\documentclass[twocolumn,tighten,times,twocolappendix]{aastex631}

\usepackage{multirow}

\usepackage[T1]{fontenc}

\begin{document}

\title{FAST Pulsar Database III. Snapshots of nulling, mode-changing and subpulse modulation of 374 pulsars
}



\author[0009-0008-1612-9948]{Yi Yan}
\affiliation{National Astronomical Observatories, Chinese Academy of Sciences, Jia-20 Datun Road, ChaoYang District, Beijing 100012, China}

\author[0000-0002-9274-3092]{J.~L. Han}\thanks{E-mail: hjl@nao.cas.cn}
\affiliation{National Astronomical Observatories, Chinese Academy of Sciences, Jia-20 Datun Road, ChaoYang District, Beijing 100012, China}
\affiliation{School of Astronomy and Space Science, University of Chinese Academy of Sciences, Beijing 100049, China}
\affiliation{State Key Laboratory of Radio Astronomy and Technology, Beijing 100101, China }

\author[0000-0003-1778-5580]{J. Xu}
\affiliation{National Astronomical Observatories, Chinese Academy of Sciences, Jia-20 Datun Road, ChaoYang District, Beijing 100012, China}
\affiliation{State Key Laboratory of Radio Astronomy and Technology, Beijing 100101, China }

\author[0000-0002-5915-5539]{N.~N. Cai}
\affiliation{National Astronomical Observatories, Chinese Academy of Sciences, Jia-20 Datun Road, ChaoYang District, Beijing 100012, China}

\author[0009-0003-2212-4792]{W.~Q. Su}
\affiliation{National Astronomical Observatories, Chinese Academy of Sciences, Jia-20 Datun Road, ChaoYang District, Beijing 100012, China}
\affiliation{School of Astronomy and Space Science, University of Chinese Academy of Sciences, Beijing 100049, China}

\author[0000-0002-6437-0487]{P.~F. Wang}
\affiliation{National Astronomical Observatories, Chinese Academy of Sciences, Jia-20 Datun Road, ChaoYang District, Beijing 100012, China}
\affiliation{School of Astronomy and Space Science, University of Chinese Academy of Sciences, Beijing 100049, China}
\affiliation{State Key Laboratory of Radio Astronomy and Technology, Beijing 100101, China }

\author[0009-0004-3433-2027]{C. Wang}
\affiliation{National Astronomical Observatories, Chinese Academy of Sciences, Jia-20 Datun Road, ChaoYang District, Beijing 100012, China}
\affiliation{School of Astronomy and Space Science, University of Chinese Academy of Sciences, Beijing 100049, China}
\affiliation{State Key Laboratory of Radio Astronomy and Technology, Beijing 100101, China }

\author[0000-0002-4704-5340]{T. Wang}
\affiliation{National Astronomical Observatories, Chinese Academy of Sciences, Jia-20 Datun Road, ChaoYang District, Beijing 100012, China} 

\author[0000-0002-6423-6106]{D.~J. Zhou}
\affiliation{National Astronomical Observatories, Chinese Academy of Sciences, Jia-20 Datun Road, ChaoYang District, Beijing 100012, China}

\author[0000-0002-1056-5895]{W.~C. Jing}
\affiliation{National Astronomical Observatories, Chinese Academy of Sciences, Jia-20 Datun Road, ChaoYang District, Beijing 100012, China}

\author[0009-0009-6590-1540]{Z.~L. Yang}
\affiliation{National Astronomical Observatories, Chinese Academy of Sciences, Jia-20 Datun Road, ChaoYang District, Beijing 100012, China}



\begin{abstract}
Based on sensitive L-band (1.0 to 1.5 GHz) observations of pulsars using the Five-hundred-meter Aperture Spherical radio Telescope (FAST), we analyzed single-pulse sequences from FAST-detected pulsars and identified nulling, mode changing, or subpulse modulation phenomena in 374 sources. Among these, nulling has been detected in 160 pulsars, with 127 cases reported for the first time. Emission mode changes have been observed in 52 pulsars, including 51 first-time detections. Subpulse modulation has been identified in 272 pulsars, 180 of which are newly revealed, with the majority displaying subpulse drifting behavior. Subpulse drifting in some pulsars exhibits distinct modes with varying drift properties, leading to mode changes and divergent mean profiles. 
Statistics on pulsar parameters show that pulsars exhibiting nulling and/or subpulse modulation tend to be older, with longer periods and lower spin-down energy-loss rates. The modulation period $P_3$ is predominantly correlated with pulsar rotation periods, magnetic field strengths, and spin-down energy loss.
\end{abstract}



\section{Introduction}          
\label{sect:intro}

Pulsars generally emit periodic single pulses. Each pulse within a period is termed an ``individual pulse'', which consists of multiple ``subpulses''. The average of many individual pulses observed in a given duration is the mean pulse or integrated pulse. For a given pulsar, the mean pulse is stable and defines the emission window in a rotation period. From single-pulse sequences of some pulsars, three phenomena have been identified: nulling, mode changing, and subpulse modulation.

Nulling refers to the absence of detectable radio pulses for one or more pulsar rotation periods. This phenomenon was first reported by \citet{BackerNull1970}. The nulling fraction ($N\!F$, in percentage) is defined as the fraction of pulse periods in the nulling state relative to the observed periods. Recent statistics from \citet{Wang2020} indicate that over 200 pulsars exhibit nulling behavior. It has been found that the spin-down rates between the emission-on and -off states of PSR B1931+24 are largely changed \citep{Kramer2006}, suggesting that nulling is physically linked to changes in the magnetosphere, either the cessation of current or the emission moving out of sightline \citep{Kramer2006, Wang2007, Gajjar2012, Basu2017}. The quasi-periodic nulling detected in some pulsars has been interpreted as the line of sight passing through the empty region of the rotating carousel beams \citep{Herfindal2007, Rankin2008, Herfindal2009, Basu2017, Wang2020}.

Some pulsars display two or more quasi-stable emission states with distinct mean profiles, a phenomenon known as emission mode changing, often characterized by abrupt transitions. Changes in radio pulsar emission mode were first reported by \citet{Backer1970_ModeChange}. \citet{Lyne2010} demonstrated that different emission states in certain pulsars correlate with varying spin-down rates. \citet{Kou2018} observed that the spin-down rate and pulse profile changes of PSR B2035+36 are triggered by glitches, i.e., the sudden changes of spin-down rates. However, it is widely accepted that different emission modes reflect distinct magnetospheric states \citep{Wang2007, Lyne2010, Ng2020}.
In addition, emission modes are sometimes accompanied by distinct polarization properties. 
Several pulsars exhibit different orthogonal polarization modes (OPMs) across emission modes. For instance, the Q and B modes of PSR J0946+0951 arise from the distinct contributions of the primary polarization mode (PPM) and secondary polarization mode (SPM), leading to differences in the linear polarization position angle curves of the integrated profiles of the two emission modes \citep{Suleymanova1998, Rankin2006}. 
The interplay between emission modes and OPM has been observed in PSR J2006$-$0807 \citep{Basu2018_J2006m0807}, J0738$-$4042 \citep{Karastergiou2011}, and PSR J0742$-$2822 and J1107$-$5907 \citep{Keith2013, Young2014}. Using FAST observation data, \citet{Yan2023_ModePol} identified bright and weak emission states in PSRs J1838+1523, J1901+0510, J1909+0007, and J1929+1844, each exhibiting distinct polarization properties.

Single pulses typically comprise one or more subpulses. For certain pulsars, subpulses in successive rotation periods appear at different longitudinal phases within the pulse phase window defined by the mean profile, such that these subpulses align along a drifting band. This subpulse drifting phenomenon was first discovered by \citet{Drake1968}. 
The most widely accepted model for subpulse drifting is that proposed by \citet{Ruderman1975}, in which drifting arises from the circulation of sparks around the magnetic axis in the polar gap due to $E \times B$ drift, leading to the circulation of carousel radio emission beams. The subpulse drifting is characterized by three parameters \citep{Backer1973}: the longitude difference in a period for two adjacent drift subpulses $P_2$, the period number for a drift-band coming back to the same phase $P_3$, and the drift rate $D= P_2 / P_3$. There are also many pulsars that display only temporal modulation, without related intensity modulation on the rotation phases \citep{Weltevrede2006,Song2023}.

\begin{table*}
\centering
\renewcommand\arraystretch{0.85}
\setlength\tabcolsep{10pt}
\scriptsize 
\caption{Pulsars exhibiting nulling, mode changing and subpulse drifting revealed by FAST observations. Summaries are presented in Sect.3 and details of each pulsar are presented in Sect.4. {\it -- to be continued}
} 
\label{Tab:all}
\begin{tabular}{lcrccrrrlll}
 \hline
PSR name  & $P$ &  DM  & Obs.date & Beam  & $T_{\rm obs}$ & $N_{\rm P}$ & $N_{\rm null}$    &  $N_{\rm Mode}$   & $D$ &  Notes.Ref.  \\
(1)       & (2)    &  (3) & (4)      & (5)    & (6)           & (7)             & (8)                & (9)               & (10)           & (11)  \\
  \hline
\multicolumn{11}{c}{Nulling merely} \\
\hline
J0209+5759            & 1.0638 & 55.3  & 20250823 & P1M01 & 16 & 883  & 613    & - & -    & \ref{subsec:J0209+5759}, [1] \\
J0244+14              & 2.1279 & 27.2  & 20210927 & P2M19 & 5  & 151  & 47     & - & -    & \ref{subsec:J0244+14}   \\
J0517+3436g (gpps0676)& 1.5958 & 191.0 & All      & -     & 30 & 1130 & 325    & - & -    & \ref{subsec:J0517+3436g} \\
                      &        &       & 20240629 & P1M01 & 15 & 565  & 144    & - & -    & -                       \\
                      &        &       & 20250205 & P1M01 & 5  & 189  & 46     & - & -    & -                       \\
\end{tabular}
\end{table*}

\begin{table*}
\centering
\renewcommand\arraystretch{0.85}
\scriptsize 
\setlength\tabcolsep{10pt}
\caption{Continued.} 
 \begin{tabular}{lcrccrrrlll}
 \hline
PSR name  & $P$ &  DM  & Obs.date & Beam  & $T_{\rm obs}$ & $N_{\rm P}$ & $N_{\rm null}$    &  $N_{\rm Mode}$   & $D$ &  Notes,Ref. \\
(1)  & (2) &  (3)  & (4) & (5)  & (6) & (7) & (8) & (9) & (10) & (11)  \\
\hline 
J1924+2037g (gpps0192)& 0.6848 & 85.3  & 20230717 & P1M01 & 96 & 8432 & 5818 & - & -     & \ref{subsec:J1924+2037g}\\
J1928-0108            & 2.3660 & 124.1 & 20211216 & P1M03 & 5  & 131  & 31   & - & -     & \ref{subsec:J1928-0108} \\
J1928+0840g (gpps0802)& 1.7835 & 121.7 & All      & -     & 20 & 675  & 115  & - & -     & \ref{subsec:J1928+0840g}\\
                      &        &       & 20250810 & P4M09 & 5  & 170  & 30   & - & -     & -                       \\
                      &        &       & 20251019 & P1M01 & 15 & 505  & 85   & - & -     & -                       \\
J1928+1923            & 0.8173 & 482.0 & 20200804 & P1M17 & 5  & 362  & 73   & - & -     & \ref{subsec:J1928+1923} \\
\end{tabular}
\end{table*}

\begin{table*}
\centering
\setlength\tabcolsep{7.5pt}
\renewcommand\arraystretch{0.85}
\scriptsize 
\caption{Continued.} 
 \begin{tabular}{lcrccrrrlll}
 \hline
PSR name  & Period &  DM  & Obs.date & Beam  & $T_{\rm obs}$ & $N_{\rm P}$ & $N_{\rm null}$    &  $N_{\rm Mode}$   & $D$ &  Notes,Ref.  \\
(1)  & (2) &  (3)  & (4) & (5)  & (6) & (7) & (8) & (9) & (10) & (11)  \\
\hline 
J0514+4009g (gpps0556)& 2.6256 & 101.0 & 20230213 & P1M01 & 15 & 339  & - & - & $+$;0     & \ref{subsec:J0514+4009g}; Table~\ref{Tab:moduPars-modu}\\
J0538+2817            & 0.1432 & 40.0  & 20190320 & P2M08 & 5  & 2107 & - & - & $-$;$+$;0 & \ref{subsec:J0538+2817}; Table~\ref{Tab:moduPars-modu}\\
J0543+2329 (B0540+23) & 0.2460 & 77.5  & 20190422 & P3M08 & 5  & 1309 & - & - & 0         & \ref{subsec:J0543+2329}; Table~\ref{Tab:moduPars-modu}, [5,16,18]\\
J0612+3721 (B0609+37) & 0.2980 & 27.2  & 20210109 & P1M01 & 9  & 1727 & - & - & $-$;0     & \ref{subsec:J0612+3721}; Table~\ref{Tab:moduPars-modu}, [16] \\
J0613+3731            & 0.6192 & 18.6  & 20210110 & P1M01 & 9  & 832  & - & - & 0         & \ref{subsec:J0613+3731}; Table~\ref{Tab:moduPars-modu} \\
J0621+0336            & 0.2700 & 72.6  & 20220307 & P3M01 & 5  & 1120 & - & - & $-$;0;$+$ & \ref{subsec:J0621+0336}; Table~\ref{Tab:moduPars-modu}, [8] \\
\end{tabular}
\end{table*}

\begin{table*}[t!]
\centering
\setlength\tabcolsep{7.5pt}
\renewcommand\arraystretch{0.85}
\scriptsize 
\caption{Continued.} 
 \begin{tabular}{lcrccrrrlll}
 \hline
PSR name  & $P$ &  DM  & Obs.date & Beam  & $T_{\rm obs}$ & $N_{\rm P}$ & $N_{\rm null}$    &  $N_{\rm Mode}$   & $D$ &  Notes,Ref.  \\
(1)  & (2) &  (3)  & (4) & (5)  & (6) & (7) & (8) & (9) & (10) & (11)  \\
\hline
J1845-0229Ag (gpps0493)&0.6578 & 848.0 & 20220907 & P4M11 & 5  & 421  & - & - & 0       &\ref{subsec:J1845-0229Ag}; Table~\ref{Tab:moduPars-modu}\\
J1845-0743            & 0.1047 & 281.1 & 20250321 & P1M06 & 6  & 3503 & - & - & 0       &\ref{subsec:J1845-0743}; Table~\ref{Tab:moduPars-modu}\\
J1846-0211g (gpps0456)& 0.7882 & 848.7 & 20220902 & P1M01 & 15 & 1127 & - & - & 0       & \ref{subsec:J1846-0211g}; Table~\ref{Tab:moduPars-modu} \\
J1846-07492           & 0.8613 & 191.1 & 20250321 & P1M01 & 6  & 426  & - & - & $-$;0   & \ref{subsec:J1846-07492}; Table~\ref{Tab:moduPars-modu}, [8] \\
J1847+0614g (gpps0165)& 1.6631 & 273.0 & 20210220 & P1M01 & 15 & 533  & - & - & $-$     & \ref{subsec:J1847+0614g}; Table~\ref{Tab:moduPars-modu} \\
J1847-0048g (gpps0337)& 0.5825 & 664.6 & All      & -     & 45 & 4593 & - & - & 0       & \ref{subsec:J1847-0048g}; Table~\ref{Tab:moduPars-modu}\\
\end{tabular}
\end{table*}

\begin{table*}
\centering
\setlength\tabcolsep{7.5pt}
\renewcommand\arraystretch{0.85}
\scriptsize 
\caption{Continued.} 
 \begin{tabular}{lcrccrrrlll}
 \hline
PSR name  & $P$ &  DM  & Obs.date & Beam  & $T_{\rm obs}$ & $N_{\rm P}$ & $N_{\rm null}$    &  $N_{\rm Mode}$   & $D$ &  Notes,Ref.  \\
(1)  & (2) &  (3)  & (4) & (5)  & (6) & (7) & (8) & (9) & (10) & (11)  \\
\hline
J1914+1122 (B1911+11) & 0.6010 & 100.2 & All      & -     & 20 & 1994 & - & - & $-$       & \ref{subsec:J1914+1122}; Table~\ref{Tab:moduPars-modu} \\
                      &        &       & 20200304 & P3M03 & 5  & 514  & - & - & $-$       & -    \\
                      &        &       & 20220513 & P1M10 & 15 & 1480 & - & - & $-$       & -    \\
J1915+0227            & 0.3173 & 192.1 & All      & -     & 20 & 3818 & - & - & $-$       & \ref{subsec:J1915+0227}; Table~\ref{Tab:moduPars-modu}, [8] \\
                      &        &       & 20201229 & P3M15 & 5  & 974  & - & - & $-$       & -  \\
                      &        &       & 20250813 & P1M02 & 15 & 2844 & - & - & $-$       & -  \\
\end{tabular}
\end{table*}

\begin{table*}[t!]
\centering
\setlength\tabcolsep{8pt}
\renewcommand\arraystretch{0.85}
\scriptsize 
\caption{Continued.} 
 \begin{tabular}{lcrccrrrlll}
 \hline
PSR name  & $P$ &  DM  & Obs.date & Beam  & $T_{\rm obs}$ & $N_{\rm P}$ & $N_{\rm null}$    &  $N_{\rm Mode}$   & $D$ &  Notes,Ref.  \\
(1)  & (2) &  (3)  & (4) & (5)  & (6) & (7) & (8) & (9) & (10) & (11)  \\
\hline
J2048+4951            & 0.5683 & 223.2 & All      & -     & 20 & 2117 & -    & - & $-$;0   & \ref{subsec:J2048+4951}; Table~\ref{Tab:moduPars-modu} \\
                      &        &       & 20240731 & P4M16 & 5  & 532  & -    & - & $-$;0   & -          \\
                      &        &       & 20250210 & P1M17 & 15 & 1585 & -    & - & $-$;0   & -          \\
J2055+3630 (B2053+36) & 0.2215 & 97.4  & 20210621 & P1M06 & 5  & 1394 & -    & - & $+$     & \ref{subsec:J2055+3630}; Table~\ref{Tab:moduPars-modu}, [16,36] \\
J2101+5028g (gpps0687)& 0.2544 & 157.2 & All      &       & 20 & 4731 & -    & - & $+$     & \ref{subsec:J2101+5028g}; Table~\ref{Tab:moduPars-modu} \\
                      &        &       & 20240731 & P4M13 & 5  & 1191 & -    & - & $+$     & -          \\
\end{tabular}
\end{table*}

\begin{table*}[t!]
\centering
\setlength\tabcolsep{8pt}
\renewcommand\arraystretch{0.85}
\scriptsize 
\caption{Continued.} 
 \begin{tabular}{lcrccrrrlll}
 \hline
PSR name  & $P$ &  DM  & Obs.date & Beam  & $T_{\rm obs}$ & $N_{\rm P}$ & $N_{\rm null}$    &  $N_{\rm Mode}$   & $D$ &  Notes,Ref.  \\
(1)  & (2) &  (3)  & (4) & (5)  & (6) & (7) & (8) & (9) & (10) & (11)  \\
\hline
J1852-0947            & 3.0195 & 236.5 & All      & -     &115 & 2286 & 1182 & - & $-$     & \ref{subsec:J1852-0947}; Table~\ref{Tab:moduPars-moduNull}, [2] \\
                      &        &       & 20230528 & P1M01 &100 & 1988 & 1020 & - & $-$     & -        \\
                      &        &       & 20251019 & P1M01 & 15 & 298  & 162  & - & $-$     & -        \\
J1853+0853            & 3.9148 & 233.9 & All      & -     & 30 & 453  & 132  & - & $+$     & \ref{subsec:J1853+0853}; Table~\ref{Tab:moduPars-moduNull} \\
                      &        &       & 20210306 & P1M09 & 15 & 227  & 69   & - & $+$     & -                       \\
                      &        &       & 20211018 & P1M09 & 15 & 226  & 63   & - & $+$     & -                       \\
\end{tabular}
\end{table*}

\begin{table*}[t!]
\centering
\setlength\tabcolsep{8pt}
\renewcommand\arraystretch{0.85}
\scriptsize 
\caption{ --- Continued and ended.} 
 \begin{tabular}{lcrccrrrlll}
 \hline
PSR name  & $P$ &  DM  & Obs.date & Beam  & $T_{\rm obs}$ & $N_{\rm P}$ & $N_{\rm null}$    &  $N_{\rm Mode}$   & $D$ &  Notes,Ref.  \\
\hline
(1)  & (2) &  (3)  & (4) & (5)  & (6) & (7) & (8) & (9) & (10) & (11)  \\
\hline
J2053+4718            & 4.9102 & 322.3 & 20200810 & P1M02 & 15 & 183  & 97   & - & $+$     & \ref{subsec:J2053+4718}; Table~\ref{Tab:moduPars-moduNull} \\
J2105+07              & 3.7468 & 50.0  & 20210118 & P1M01 & 9  & 145  & 30   & - & $+$     & \ref{subsec:J2105+07}; Table~\ref{Tab:moduPars-moduNull}   \\
\hline
\multicolumn{11}{c}{Mode changing and modulation} \\
\hline
J1811-0154            & 0.9249 & 147.5 & 20230405 & P1M01 & 64 & 4159 & -    & 2 & $+$;0   &\ref{subsec:J1811-0154}; Table~\ref{Tab:moduPars-modeModu}, [8] \\
J1818-0151            & 0.8376 & 200.7 & 20210923 & P4M02 & 6  & 397  & -    & 2 & -       & \ref{subsec:J1818-0151}; Table~\ref{Tab:moduPars-modeModu}\\
J1850+0026            & 1.0818 & 201.3 & 20200119 & P1M15 & 5  & 273  & -    & 2 & $-$     & \ref{subsec:J1850+0026}; Table~\ref{Tab:moduPars-modeModu}, [8]\\
\end{tabular}
\tablecomments{Columns (1) pulsar name (with Bname or gpps number in the bracket); (2) pulsar period $P$ in second; (3) pulsar DM in pc~cm$^{-3}$; (4) Observation date in format of ``yyyymmdd''; (5) FAST beam name; (6) observation length $T_{\rm obs}$ in minutes; (7) Period number $N_{\rm P}$ in an observation; (8) number of nulling periods $N_{\rm null}$; (9) number of emission modes $N_{\rm mode}$; (10) subpulse drifting rate $D$: ``+'' for drifting to a later phase, ``$-$ " for drifting to an earlier phase and ``0" for no drifting but temporal intensity modulation; (11) Subsection for details and references. 
References in columns (8)-(10): [1]: \citet{Good2021}; [2]: \citet{Wu2023}; [3]: \citet{Tan2020}; [4]: \citet{Deneva2016}; [5]: \citet{Weltevrede2007}; [6]: \citet{BurkeSpolaor2012}; [7]: \citet{Basu2017}; [8] \citet{Song2023}; [9]: \citet{Herfindal2009}; [10]: \citet{Redman2009}; [11]: \citet{Parent2022}; [12]: \citet{Tyulbashev2018}; [13]: \citet{Tyulbashev2018_5RRAT}; [14]: \citet{Michilli2020}; [15]: \citet{Rankin1986}; [16]: \citet{Weltevrede2006}; [17]: \citet{Basu2016}; [18]: \citet{Nowakowski1991}; [19]: \citet{Wang2021}; [20]: \citet{Zhao2023}; [21]: \citet{Brinkman2018}; [22]: \citet{Wahl2023}; [23]: \citet{Basu2019}; [24]: \citet{Smirnova2024}; [25]: \citet{Lorimer2012}; [26]: \citet{Wang2020}; [27]: \citet{Hankins1987}; [28]: \citet{Basu2020}; [29]: \citet{Wang2025}; [30]: \citet{Backer1970}; [31]: \citet{Proszynski1986}; [32]: \citet{Ritchings1976}; [33]: \citet{Vivekanand1995}; [34]: \citet{Oster1977}; [35]: \citet{Chen2024}; [36]: \citet{Weisberg1986}; [37]:\citet{Ng2020}; [38]: \citet{Backer1973}; [39]: \citet{Yan2024}; [40]: \citet{Burgay2013}; [41]: \citet{Gajjar2017}; [42]: \citet{Xu2024}; [43]: \citet{Rankin2008}; [44]: \citet{Navarro2003}; [45]: \citet{Lynch2018}; [46]: \citet{Sengar2023}; [47]: \citet{Lorimer2002}; [48]: \citet{Zhou2023}; [49]: \citet{Rankin2023}; [50]: \citet{Basu2020MNRAS}; [51]: \citet{Edwards2001}; [52]: \citet{Ord2001}; [53]: \citet{Wang2007}
}
\end{table*}

Numerous studies have investigated subpulse modulation. For example, \citet{Weltevrede2006} analyzed 187 pulsars observed with the Westerbork Synthesis Radio Telescope (WSRT) at 21 cm and found that over 55\% exhibit subpulse drifting. \citet{Basu2016} studied fluctuation spectra of 123 pulsars observed with the Giant Meterwave Radio Telescope at 333 MHz and 618 MHz, categorizing them into three groups: phase-modulated drifting, amplitude-modulated drifting, and non-drifting pulsars. These groups exhibit distinct distributions of spin-down energy loss rate $\dot{E}$. Phase-modulated drifting pulsars show an anti-correlation between $P_3$ and $\dot{E}$, supporting the ``partially screened gap" model. \citet{Basu2019} further refined this classification into four types: coherent phase-modulated drifting, switching phase-modulated drifting, diffuse phase-modulated drifting, and low-mixed phase-modulated drifting. \citet{Song2023} reported on the subpulse modulation properties of 1198 pulsars using MeerKAT. 
Subpulses of several pulsars exhibit diverse drifting behaviors, such as reversed drift or bi-drift, which challenge the traditional carousel circulation model. Drift reversals in pulsars like PSRs B2303+30, B0826$-$34, B2148+63, and B2310+42 \citep{Redman2005, Gupta2004, Esamdin2005, Weltevrede2006} can be explained by line-of-sight effects if the intrinsic drift direction remains constant. Bidirectional drifting, where different profile components exhibit opposite drift directions, is attributed to coexisting inner annular and core gaps \citep{Qiao2004}. 
This phenomenon has been observed in PSRs J0815+09 \citep{McLaughlin2004, Champion2005, Szary2017, Shang2022}, B1839$-$3224 \citep{Basu2018}. Additionally, some pulsars exhibit non-straight drift bands \citep{Weltevrede2006, Basu2018, Basu2019}, further complicating theoretical models.

Many pulsars exhibit multiple phenomena in their single-pulse sequences. Some pulsars display subpulse drifting in two or more distinct modes with different drift rates \citep[e.g.,][etc]{Deich1986, Hankins1987, van2002, Redman2005, Kloumann2010, Rankin2013, Basu2018_J1822m2256, Rahaman2021}. Some pulsars show different emission or drifting modes separated by nulls \citep{Redman2005, van2002, Wang2007, Kloumann2010, Rankin2013, Basu2018_J1822m2256}. \citet{Lyne1983} showed that the drifting rate of PSRs B0809+74 and B0818$-$13 changes at the end of emission and then relaxes back exponentially to the normal rate after null starts, with the timescale proportional to the null length. The phases of drifting subpulses of PSRs B0031$-$07 and J1840$-$0840 are related before and after nulls \citep{Vivekanand1997, Gajjar2017}. Related to drifting subpulses, the intensities at given longitude phases often show regular fluctuations, which can be analyzed by the fluctuation spectra \citep[e.g.,][]{Backer1970, EdwSta2002, Weltevrede2006, Song2023}.

Numerous statistical studies have been conducted on single-pulse phenomena. 
\citet{Ritchings1976} and \citet{Weltevrede2006} demonstrated that pulse nulling and subpulse drifting, respectively, are characteristics of older pulsars.
\citet{Wang2007} further reported that the nulling fraction correlates more strongly with pulsar age than with its spin period. \citet{Ng2020} found that nulling is linked to both pulsar period and age, with long-period old pulsars exhibiting a higher likelihood of nulling, while mode-changing is predominantly associated with longer-period pulsars. \citet{Wolszczan1980} and \citet{Rankin1986} reported the correlations between $P_3$ (the drift cycle period) and the pulsar spin-down energy loss rate $\dot{E}$, age, and pulsar magnetic fields. \citet{Basu2016, Basu2019} confirmed an anti-correlation between $\dot{E}$ and $P_3$. 

Since 2019, we have been using the Five-hundred-meter Aperture Spherical radio Telescope \citep[FAST,][]{nan06} to conduct the ``Galactic Plane Pulsar Snapshot (GPPS) survey'' \citep{Han2021,han2025} and other projects, such as probing magnetic fields in the Galactic halo through pulsar polarization observations \citep{Xu2022}. These long-term efforts have yielded a substantial dataset for numerous pulsars \citep{Wang2023, Jing2025}. Owing to the exceptionally high sensitivity of the FAST, the pulse sequences of more pulsars exhibit nulling, mode changing, or subpulse modulation phenomena. In this paper, we analyze 374 pulsars observed with FAST (see Table \ref{Tab:all}), excluding some sources already published using FAST data, such as PSRs B1859+07 \citep{WT2023}, B1929+10 \citep{Kou2021}, B2016+28 \citep{Lu2019}, and B1946+35 \citep{Chang2025ApJ}. The observational and analytical methods for studying nulling and subpulse modulation are described in Section~\ref{sect:Obse}. In Section~\ref{sect:Resu}, we discuss the nulling, mode-changing, or subpulse modulation phenomena in both previously known and newly discovered pulsars. Section~\ref{sect:Conc} presents statistical analyses of these pulsars, including parameter distributions related to these phenomena and correlations between nulling fraction or subpulse modulation period $P_3$ and pulsar parameters, and also provides conclusions. 


\section{FAST observations and analysis methods}
\label{sect:Obse}

Pulsar observations with FAST, conducted as part of the GPPS survey key project \citep{Han2021} or other co-authors' free-applied projects, together with the opened archived FAST data, all utilize the L-band 19-beam receiver \citep{jth+20}. This receiver covers a frequency band of 1.0 to 1.5~GHz and has a system temperature of approximately 22~K. Digital backend systems record data for 1024, 2048, or 4096 channels across one or all 19 beams, with each channel capturing either four polarization products ($XX$, $YY$, Re[$X^{*}Y$] and Im[$X^{*}Y$]). The data sampling rate is consistently 49.152~{\textmu}s, and recorded data are stored in FITS files.

When a pulsar is detected in any of the 19 beams, the recorded data are folded into a single-pulse sequence using previously obtained pulsar ephemeris which were often taken from the pulsar catalog \cite{Manchester+2005AJ....129.1993M}. We generally update the dispersion measure and period from the new FAST observations by using {\it DSPSR} \citep{van2011}. Subsequently, {\it PSRCHIVE} software \citep{Hotan2004} is employed to remove radio frequency interference (RFI) from the data and correct for Faraday rotation when four polarization channels are recorded.
Following the polarization calibration procedure described in \citet{Wang2023}, we obtain the total power pulse profile or polarization profiles of pulsars for which data with four polarization channels are recorded. The new rotation measure (RM) is obtained via \textsc{rmfit} \citep{Hotan2004}. The Faraday rotation of linear polarization is corrected, and the final polarization profiles are obtained independently for each mode discussed below. %
To generate a high-quality single-pulse sequence, the baseline is typically fitted with a polynomial function in off-pulse regions and subtracted. In total, we get detailed single-pulse sequences of  374 pulsars observed by FAST, see Table \ref{Tab:all} for a summary.


\section{Various behaviors of individual pulses of pulsars revealed by FAST}
\label{sect:Resu}

Analyzing single-pulse sequences can uncover phenomena such as nulling, mode changing, and subpulse modulation. FAST observations in general lasted for short duration, except for some pulsars with longer observation time. Short FAST observations can qualitatively reveal the various behaviors of individual pulses owing to the high sensitivity, but quantitative statistics of some individual pulse properties (such as the nulling fraction) are meaningful only from long FAST observations.

\begin{figure}[htbp]
\centering
\includegraphics[width=0.21\textwidth, angle=0]{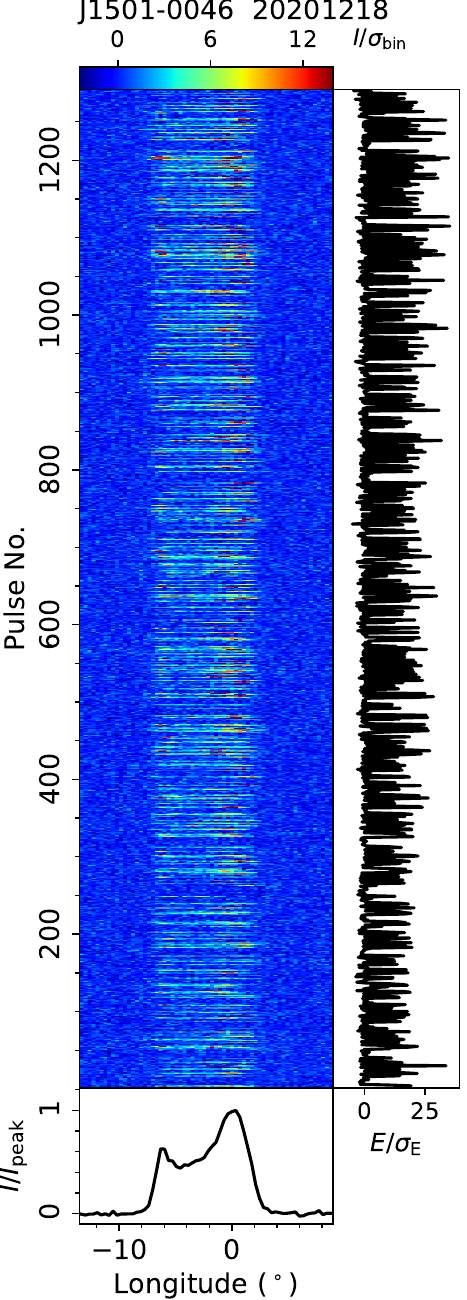}
\includegraphics[width=0.21\textwidth, angle=0]{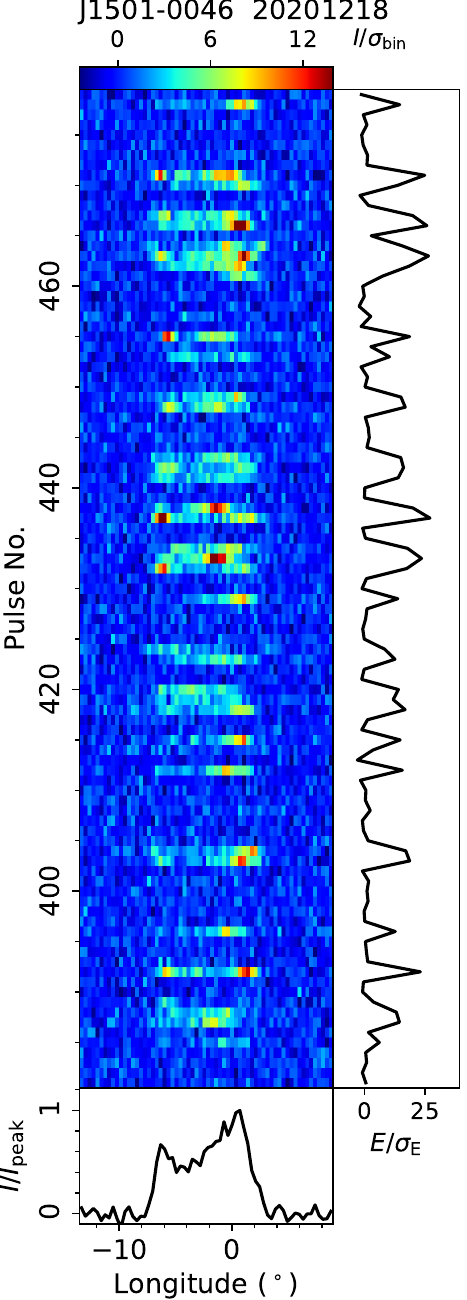}
\caption{Single-pulse sequence of PSR J1501-0046 observed by FAST on 2020 December 18 (hereafter in format of ``20201218''), accompanied by a zoom-in view in the right for the period range No. 380–480. The color is scaled by $\sigma_{\rm bin}$, which is the standard deviation of the off-pulse data sampling. The detailed variations of on-pulse energy of individual pulses are plotted in the right sub-panel, scaled to $\sigma_{\rm E}$, which is the standard deviation of off-pulse energy obtained with the same size of on-pulse  window. The mean pulse profile normalized to the peak intensity is displayed below the stack. 
\label{subfig:TP:J1501-0046}
}
\includegraphics[width=0.35\textwidth, angle=0]{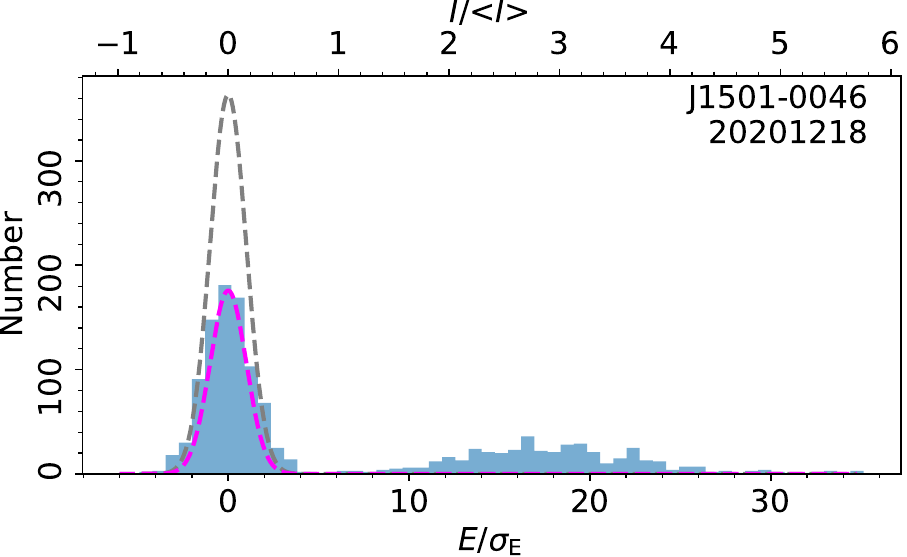} 
\caption{On-pulse energy histogram of individual pulses of PSR J1501-0046 from the FAST observation on 20201218, as shown in Fig.~\ref{subfig:TP:J1501-0046}. The grey dashed line represents a Gaussian fit to the distribution of the off-pulse energy in the same size of the on-pulse phase window. The magenta dashed line is the fitted Gaussian to the histogram section of intensities less than 0, where the Gaussian width is equivalent to that for the grey line. \label{subfig:Hist:J1501-0046}}
\end{figure}

\begin{figure}[htbp]
\centering
\includegraphics[width=0.40\textwidth, angle=0]{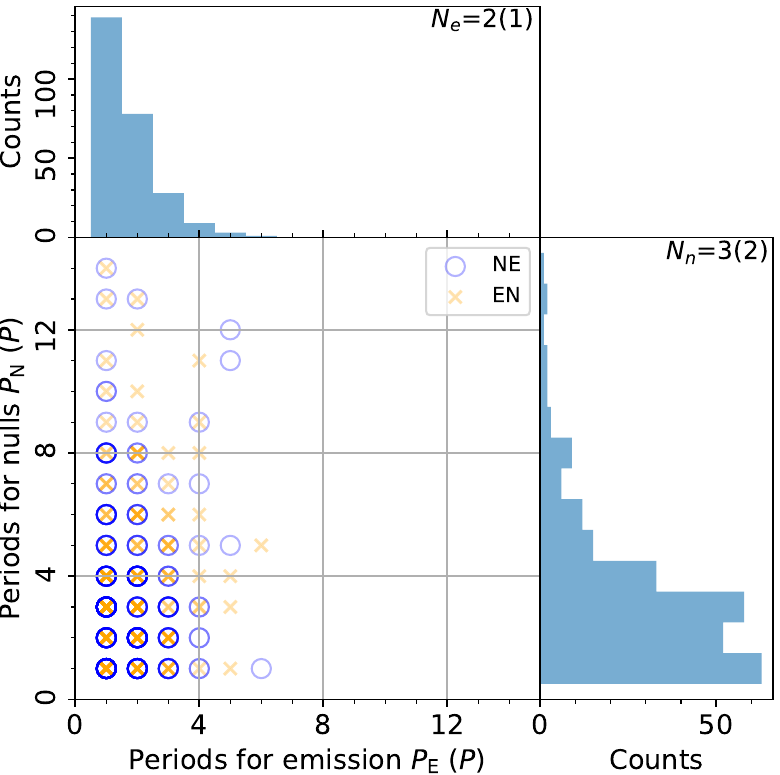}
\caption{Distribution of period numbers for continuous nulling $P_{\rm N}$ against period numbers for adjacent pulses $P_{\rm E}$ of PSR J1501$-$0046 (NE for advance nulling or EN for follow-up nulling) observed by FAST on 20201218. Symbols with deeper color in the $P_{\rm N}$ and $P_{\rm E}$ panel indicate a larger number of data with the same value. The duration histograms for the emission and nulls are shown in the top and right panels, respectively.
\label{subfig:scaleHist:J1501-0046}}
%
\includegraphics[width=0.42\textwidth, angle=0]{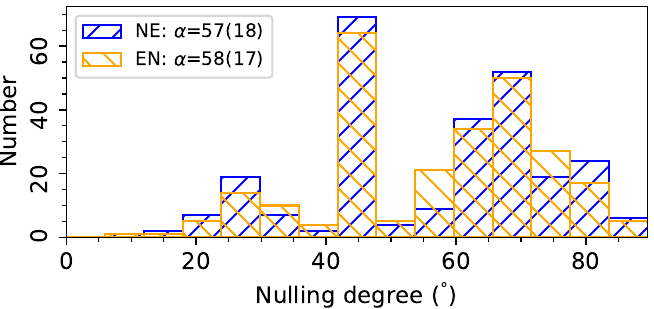}\\
\includegraphics[width=0.42\textwidth, angle=0]{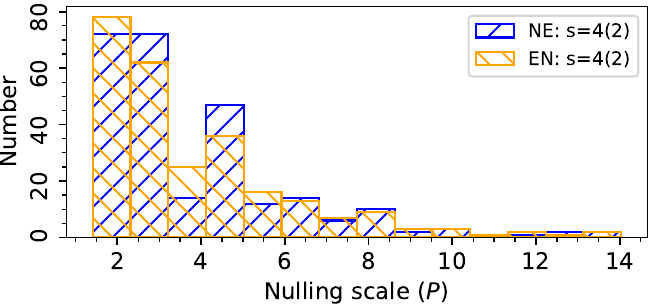}
\caption{Histogram of the nulling degree and nulling scale \citep[see][]{Yang2014, Wang2020} for PSR J1501-0046 observed by FAST on 20201218. The nulling scale and degree have the average (and standard deviation) of 57$^\circ$(18) and 4(2) pulsar period. 
\label{subfig:nullDegreeScale:J1501-0046}
}
\end{figure}


Among the 374 pulsars observed by FAST, we find from available data that 70 pulsars show merely nulling phenomenon, 31 pulsars show different modes in our observations, and 170 pulsars only show subpulse modulation. Some pulsars show two kinds of behaviors: a pulsar has nulling and mode changing; 82 pulsars show nulling and modulation; 13 pulsars show different emission modes related to subpulse modulation; 7 pulsars show nulling, mode changing and also modulation. 
We have summarized these behaviors in the Table \ref{Tab:all}. 
In the following, we will discuss the general methods that will be applied to each pulsar and work on each kind of these behaviors in a subsection.

\subsection{Nulling merely}
\label{subsec:nullPSRs}

From the sensitive FAST observations, we found 70 pulsars with merely nulling behavior as listed in Table~\ref{Tab:all}, and 59 of them have the nulls detected for the first time. The references are given in Table~\ref{Tab:all} for pulsars with previously revealed nulling. 
Furthermore, the time-limited FAST observations did not detect nulling for several previously known nulling pulsars, such as PSRs J0540+3207 (subsection~\ref{subsec:J0540+3207}), J0629+2415 (\ref{subsec:J0629+2415}) and J1921+2153 (\ref{subsec:J1921+2153}). Some previously known nulling pulsars, such as PSR J0540+3207 (see Subsection~\ref{subsec:J0540+3207}),  exhibit a weak emission state instead of a nulling state in the FAST data. Longer observations probably can reveal more nulling pulsars, known or previously unknown.
%


\begin{figure}[htbp]
\centering
\includegraphics[width=0.22\textwidth, angle=0]{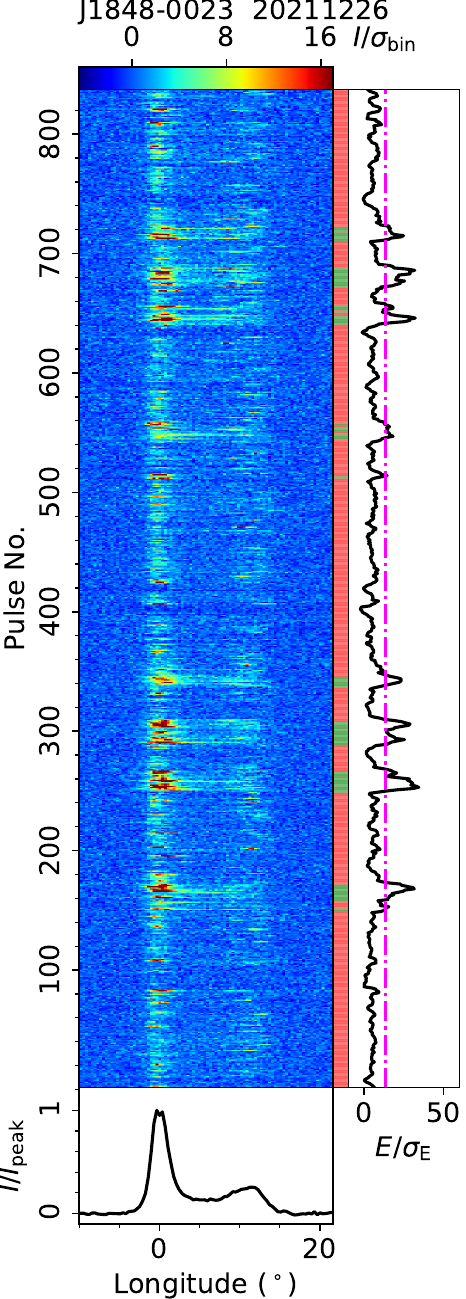} 
\includegraphics[width=0.22\textwidth, angle=0]{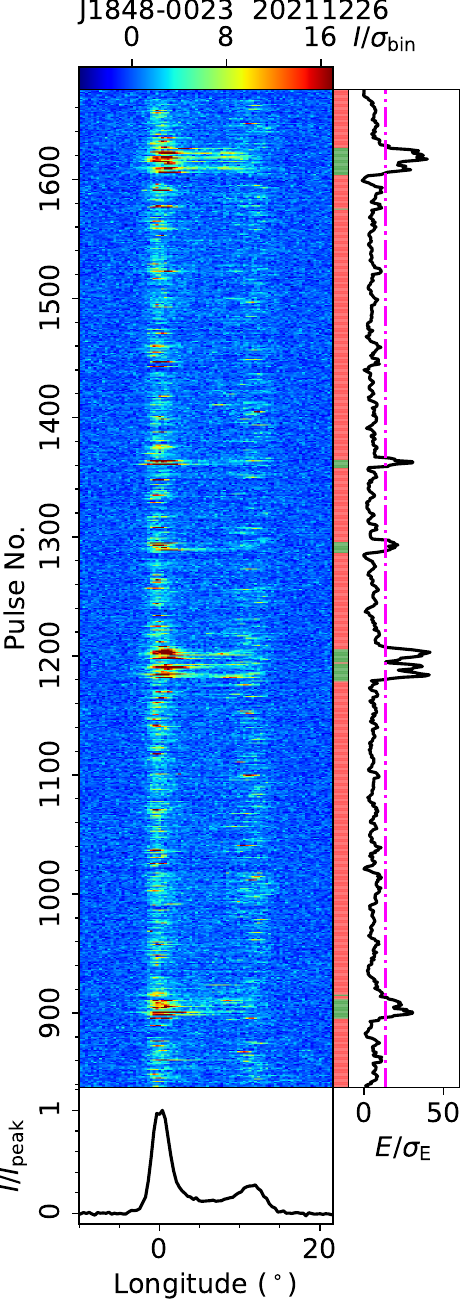}
\caption{Similar to Figure~\ref{subfig:TP:J1501-0046}, but for two segments of single-pulse sequences of PSR J1848$-$0023 observed on 20211226 with FAST. The green and red bars indicate the recognized bright or weak modes. In the right subpanels, the on-pulse energy variation is smoothed over every 5 periods, and is plotted against period, with a dashed line for the threshold to distinguish the weak and bright emission modes, as marked by the color-bar in red and green, respectively. 
}  \label{subfig:TP:J1848-0023}
%
\includegraphics[width=0.35\textwidth, angle=0]{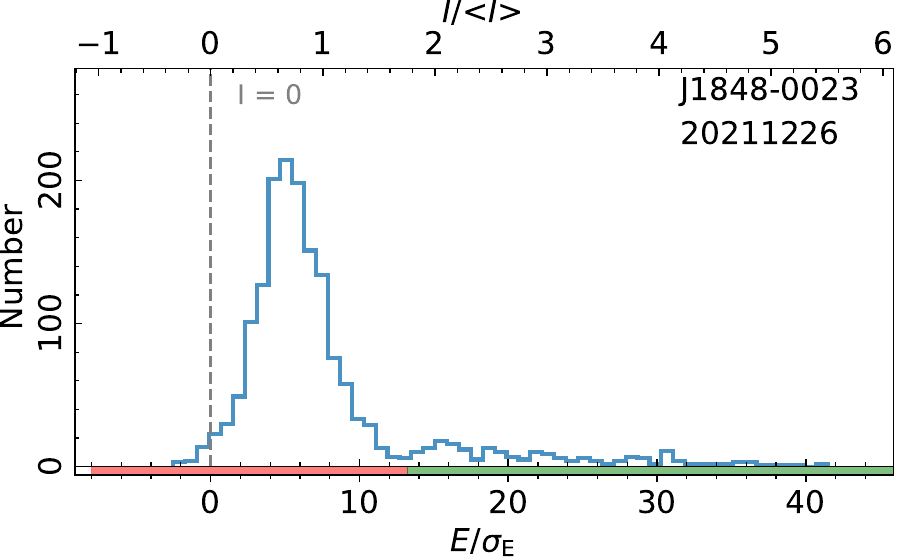}
\caption{On-pulse energy distribution for PSR J1848$-$0023 shown in Fig.~\ref{subfig:TP:J1848-0023}, with energy values smoothed over 5 periods. The green and red bars in the bottom indicate the bright and weak emission modes. \label{subfig:Hist:J1848-0023}}
\end{figure}

\begin{figure}
\centering
\includegraphics[width=0.39\textwidth, angle=0]{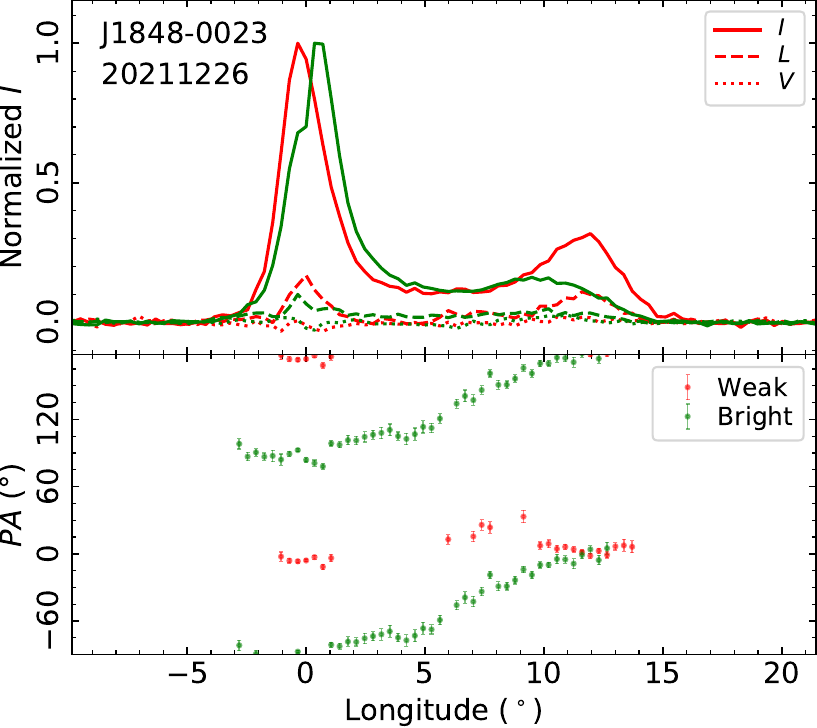}
\caption{Averaged intensity profiles ($I$, solid line), the linear polarization profiles ($L$, dashed line) and the circular polarization profiles ($V$, dotted line) for the weak (red) and bright (blue) emission modes of PSR J1848-0023 observed on 20211226 are shown in the top panel. The averaged PA curves of the two emission modes are shown in the bottom panel. For every emission mode, the profiles in the top panel are normalized to the peak of the averaged intensity profiles.
\label{subfig:PolModes:J1848-0023}}
\end{figure}

The presence of pulse nulling can be directly observed in single-pulse sequences. In general analyses, we first compute the average profile to define the on-pulse window. Subsequently, we conduct a statistical analysis of the pulse energy. The on-pulse energy distribution histogram is compared with that of an off-pulse window of the same size. Nulling is indicated by an additional distinct peak around zero in the on-pulse energy distribution, in addition to the typical on-pulse energy histogram. The nulling fraction ($N\!F$) can be calculated from these two peaks in the on-pulse energy distribution  \citep[e.g.][]{Wang2007, Basu2017, Wang2020}. Typically, the energy distribution of the off-pulse window can be approximated by a Gaussian function, as it results from random fluctuations. This Gaussian function is characterized by an amplitude $A_{\rm off}$ and a width $\sigma_{\rm off}$. If there is a peak around zero in the on-pulse energy distribution, we can fit the negative part of this distribution with a Gaussian function. This is because this part of the distribution stems from the random fluctuations during nulling periods. We expect to obtain a width similar to $\sigma_{\rm off}$, but a different amplitude $A_{\rm on}$. The widely used parameter, the nulling fraction $N\!F$, can then be calculated as the percentage given by $A_{\rm on}/A_{\rm off}$, with an uncertainty of \citep{Wang2020}
\begin{equation}
\sigma_{N\!F}=\sqrt{(\frac{\delta A_{\rm on}}{A_{\rm off}})^2+(\frac{A_{\rm on}\delta A_{\rm off}}{(A_{\rm off})^2})^2}
\end{equation}
where $\delta A_{\rm on}$ and $\delta A_{\rm off}$ represent the uncertainties obtained from the fitting. 
When the emission and null distributions are well separated, $N\!F$ reflects the true nulling fraction. In cases where there is a significant overlap of two distributions, the $N\!F$ estimate tends to represent an upper limit. Ideally, when the pulse sequence is long enough, we should determine the  duration and frequency of nulling events. 
Moreover, if the observation duration is insufficient, the nulling fraction derived from a short pulse sequence may not accurately represent the true nulling properties of a pulsar. 

The simple ``nulling fraction'' is inadequate for a comprehensive description of nulling. For instance, given a nulling fraction of 40\%, we cannot determine which 400 periods out of 1000 are in a nulling state. \citet{Yang2014} and \citet{Wang2020} introduced the ``nulling degree'' and ``nulling scale'' to characterize nulling properties when the pulse sequence is of sufficient length. The nulling degree is defined as the angle calculated from the durations of the nulling and the emission states, while the nulling scale is defined as the scale period-number for consecutive nulling and emission states. 

Here we discuss one example of nulling-merely pulsars, PSR J1501$-$0046. 
PSR J1501$-$0046 was discovered by \citet{Lynch2013} using the 100~m Robert C. Byrd Green Bank Telescope (GBT) at 350 MHz, with a period of $P=0.4641$~s. The pulsar was observed by the FAST on 20201218 for 10 minutes, over 1290 periods. The dispersion measurement (DM) determined from this observation is $DM= 22.3~{\rm cm^{-3}\,pc}$. Single pulse sequences are displayed in Figure~\ref{subfig:TP:J1501-0046}, which shows the nulling phenomenon. From the on-pulse energy histogram of single pulses, evidently the distribution around zero energy is an indication for nulling. The nulling fraction, estimated as the fraction of the on-pulse related to the off-pulse energy Gaussian distribution, is $48\pm5\%$. According to the distributions of continuous period numbers for adjacent nulling and emission duration (see Figure~\ref{subfig:scaleHist:J1501-0046}), the nulling scale is estimated to be $4\pm2$~$P_0$, and the nulling degrees are $57\pm18^\circ$ for nulling-emission (NE) pairs and $58\pm17^\circ$ for emission-nulling (EN) pairs. According to histograms in Figure~\ref{subfig:nullDegreeScale:J1501-0046}, the most probable duration for emission is 1 to 3 periods, and that for the nulling state is 1 to 4 periods, coincident with the pulse stack in Figure~\ref{subfig:TP:J1501-0046}. 

\begin{table}[htbp]
\centering
\renewcommand\arraystretch{0.85}
\scriptsize 
\setlength\tabcolsep{4pt}
\caption{Parameters of subpulse modulation (-- to be continued)} 
\label{Tab:moduPars-modu}
\begin{tabular}{llcrrrl}
 \hline
PSR name  & Obs.date  & M.N & \multicolumn{1}{c}{$P_2$} & \multicolumn{1}{c}{$P_3$} & \multicolumn{1}{c}{$D$} &  Notes     \\
\multicolumn{1}{c}{(1)}       & \multicolumn{1}{c}{(2)}        & (3)  & \multicolumn{1}{c}{(4)}   & \multicolumn{1}{c}{(5)}   & \multicolumn{1}{c}{(6)} & \multicolumn{1}{c}{(7)}   \\
\hline
J0039+35   & 20210927 & W.1  & -32(10)  & 10.4(2)  & -3(1)   & \ref{subsec:J0039+35} \\
J0058+4950 & 20201120 & W.1  & -8.7(3)  & 2.48(1)  & -3.5(1) & \ref{subsec:J0058+4950} \\
           &          & W.2  & *        & 51(1)    & -       & -                       \\
J0115+6325 & 20250811 & W.1  & *        & 10.0(2)  & -       & \ref{subsec:J0115+6325} \\
           &          & W.2  & -7.5(1)  & 3.6(1)   & -2.1(1) & -                       \\
J0453+1559 & 20210108 & L.1  & -23(3)   & 2.83(1)  & -8(1)   & \ref{subsec:J0453+1559} \\
\end{tabular}
\end{table}

\begin{table}[htbp]
\centering
\renewcommand\arraystretch{0.85}
\scriptsize 
\setlength\tabcolsep{4pt}
\caption{-- Continued.} 
 \begin{tabular}{llcrrrl}
 \hline
PSR name  & Obs.date  & M.N & \multicolumn{1}{c}{$P_2$} & \multicolumn{1}{c}{$P_3$} & \multicolumn{1}{c}{$D$} &  Notes     \\
\multicolumn{1}{c}{(1)}       & \multicolumn{1}{c}{(2)}        & (3)  & \multicolumn{1}{c}{(4)}   & \multicolumn{1}{c}{(5)}   & \multicolumn{1}{c}{(6)} & \multicolumn{1}{c}{(7)}   \\
\hline
           &          & T.1  & -21(1)   & 7.3(1)   & -2.9(2) & -                       \\
J1807-0847 & 20250319 & W.1  & *        & 14.8(3)  & -       & \ref{subsec:J1807-0847} \\
           &          & W.2  & 7.1(1)   & 4.34(5)  & 1.640(4)& -                       \\           
J1809+17   & 20240920 & W.1  & -11(1)   & 4.6(1)   & -2.4(2) & \ref{subsec:J1809+17} \\
J1809-0119 & 20210118 & L.1  & -3.6(1)  & 3.58(1)  &-1.01(4) & \ref{subsec:J1809-0119} \\
           &          & T.1  & -6.3(3)  & 3.52(1)  & -1.8(1) & -                       \\
\end{tabular}
\end{table}

\begin{table}
\centering
\renewcommand\arraystretch{0.85}
\scriptsize 
\setlength\tabcolsep{4pt}
\caption{-- Continued.} 
 \begin{tabular}{llcrrrl}
 \hline
PSR name  & Obs.date  & M.N & \multicolumn{1}{c}{$P_2$} & \multicolumn{1}{c}{$P_3$} & \multicolumn{1}{c}{$D$} &  Notes     \\
\multicolumn{1}{c}{(1)}       & \multicolumn{1}{c}{(2)}        & (3)  & \multicolumn{1}{c}{(4)}   & \multicolumn{1}{c}{(5)}   & \multicolumn{1}{c}{(6)} & \multicolumn{1}{c}{(7)}   \\
\hline 
J1845-0229Ag&20220907 & W.1  & *       & 14.1(2)  & -       &\ref{subsec:J1845-0229Ag}\\
J1845-0743 & 20250321 & W.1  & *       & 10.6(2)  & -       &\ref{subsec:J1845-0743}\\
J1846-0211g& 20220902 & W.1  & *       & 18.6(4)  & -       & \ref{subsec:J1846-0211g}\\
J1846-07492& 20250321 & W.1  & -8.8(4) & 2.7(1)   & -3.3(2) & \ref{subsec:J1846-07492}\\
           &          & W.2  & *       & 34(2)    & -       & -                       \\
J1847+0614g& 20210220 & W.1  & -7.4(2) & 3.06(1)  & -2.4(1) & \ref{subsec:J1847+0614g}\\
\end{tabular}
\end{table}

\begin{table}
\centering
\renewcommand\arraystretch{0.85}
\scriptsize 
\setlength\tabcolsep{4pt}
\caption{-- Continued.} 
 \begin{tabular}{llcrrrl}
 \hline
PSR name  & Obs.date  & M.N & \multicolumn{1}{c}{$P_2$} & \multicolumn{1}{c}{$P_3$} & \multicolumn{1}{c}{$D$} &  Notes     \\
\multicolumn{1}{c}{(1)}       & \multicolumn{1}{c}{(2)}        & (3)  & \multicolumn{1}{c}{(4)}   & \multicolumn{1}{c}{(5)}   & \multicolumn{1}{c}{(6)} & \multicolumn{1}{c}{(7)}   \\
\hline     
           & 20250522 & W.1  & *       & 2.49(1)  & -       & -                       \\
J1909+1102 & 20200824 & L.1  &  54(23) & 14.2(2)  & 4(2)    & \ref{subsec:J1909+1102}\\
           &          & T.1  & 148(74) & 17.1(1)  & 9(4)    & -                       \\
J1910-0309 & 20250317 & L.1  & *       &  8.4(3)  & -       & \ref{subsec:J1910-0309}\\
           &          & C.1  & *       &   20(1)  & -       & -                       \\
           &          & T.1  & *       & 12.1(3)  & -       & -                       \\
\end{tabular}
\end{table}

\begin{table}
\centering
\renewcommand\arraystretch{0.85}
\scriptsize 
\setlength\tabcolsep{5pt}
\caption{-- Continued.} 
 \begin{tabular}{llcrrrl}
 \hline
PSR name  & Obs.date  & M.N & \multicolumn{1}{c}{$P_2$} & \multicolumn{1}{c}{$P_3$} & \multicolumn{1}{c}{$D$} &  Notes     \\
\multicolumn{1}{c}{(1)}       & \multicolumn{1}{c}{(2)}        & (3)  & \multicolumn{1}{c}{(4)}   & \multicolumn{1}{c}{(5)}   & \multicolumn{1}{c}{(6)} & \multicolumn{1}{c}{(7)}   \\
\hline       
J1940+0239 & 20210719 & L.1  & 13(6)   & 6.5(1)   & 2.0(9)  & \ref{subsec:J1940+0239} \\ 
           &          & C.1  &   *     &  31(2)   & -       & -                       \\
           &          & T.1  & 11(1)   & 6.7(1)   & 1.6(2)  & -                       \\
J1941+0121 & 20250410 & W.1  & -248(39)& 6.28(3)  & -40(6)  & \ref{subsec:J1941+0121} \\
J1942+1743 & 20190330 & L.1  & 34(5)   & 6.4(1)   & 5.2(6)  & \ref{subsec:J1942+1743} \\
           &          & T.1  & 89(27)  & 6.6(1)   & 14(4)   & -                       \\
\end{tabular}
\end{table}

\begin{table}
\centering
\renewcommand\arraystretch{0.85}
\scriptsize 
\setlength\tabcolsep{5pt}
\caption{-- Continued and ended.} 
 \begin{tabular}{llcrrrl}
 \hline
           &          & T.1  &  *      &  8.5(1)  & -       & -                      \\
J2208+4610 & 20211124 & L.1  & 24(6)   & 6.9(1)   & 3.4(9)  & \ref{subsec:J2208+4610} \\
           &          & T.1  & 8.4(9)  & 6.8(2)   & 1.2(1)  & -                       \\
J2215+1538 & 20250317 & W.1  & *       & 5.30(5)  & -       & \ref{subsec:J2215+1538} \\
J2219+4754 & 20210608 & W.1  &-1.57(1) & *        & -       & \ref{subsec:J2219+4754} \\
           &          & W.2  & 1.56(1) & *        & -       & -                       \\
J2329+4743 & 20201028 & L.1  & -26(4)  & 4.56(2)  & -6(1)   & \ref{subsec:J2329+4743} \\
           &          & C.1  & -22(5)  & 4.48(1)  & -5(1)   & -                       \\
J2338+4818 & 20250225 & T.1  & *       & 2.50(1)  & -       & \ref{subsec:J2338+4818}\\
J2340+08   & 20210125 & L.1  & 18.2(5) & 2.80(1)  & 6.5(2)  & \ref{subsec:J2340+08}   \\
           &          & L.2  & *       & 61(4)    & -       & -                       \\
\hline 
\end{tabular}
\tablecomments{\scriptsize 
Columns: 
(1) pulsar name; 
(2) FAST observation date (in yyyymmdd); 
(3) Modulation features number (M.N) for various parts of the profiles: `W', `L', `C', and `T' denote the  phase range of the whole profile, or the leading, central, and trailing parts of a mean pulse profile, respectively. Then, the suffix number stands for the sequence number of modulation features. For instance, L.3 refers to the 3rd feature in the leading part. 
For PSR J1857+0943, M.W.1 and I.W.1 denote, respectively, the first modulation feature of the whole profile for the main pulse and interpulse (a special case of the notation); 
(4) subpulse phase interval $P_2$ in degrees, and `*' for no drifting but temporal modulation of subpulses;  
(5) subpulse modulation periodicity $P_3$, counted in pulsar period; 
(6) subpulse drifting rate $D=P_2/P_3$ in phase degrees per period. Uncertainties for the columns (4-5) are in brackets; 
(7) Subsection for details. 
}
\end{table}


Note that some pulsars show partial nulling, that is the emission absence of certain phase ranges for some periods. During the usually-recognized nulling periods, dwarf pulses could appear as weak narrow pulses detectable only by high time-resolution analyses, as shown by \citet{Chen2023} for PSR B2111+46, and later for PSR J2323+1214  \citep{Tedila2025}, PSR B1931+24 \citep{Rusul2025arXiv}, and other 10 pulsars \citep{Yan2024}. 

\begin{figure}[hbtp]
\centering
\includegraphics[width=0.22\textwidth, angle=0]{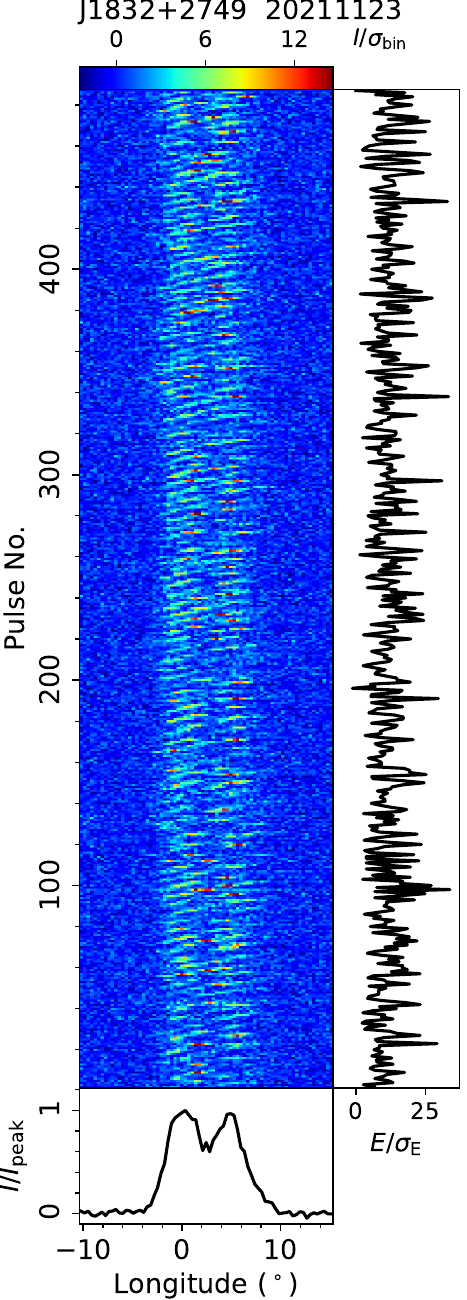}
\includegraphics[width=0.22\textwidth, angle=0]{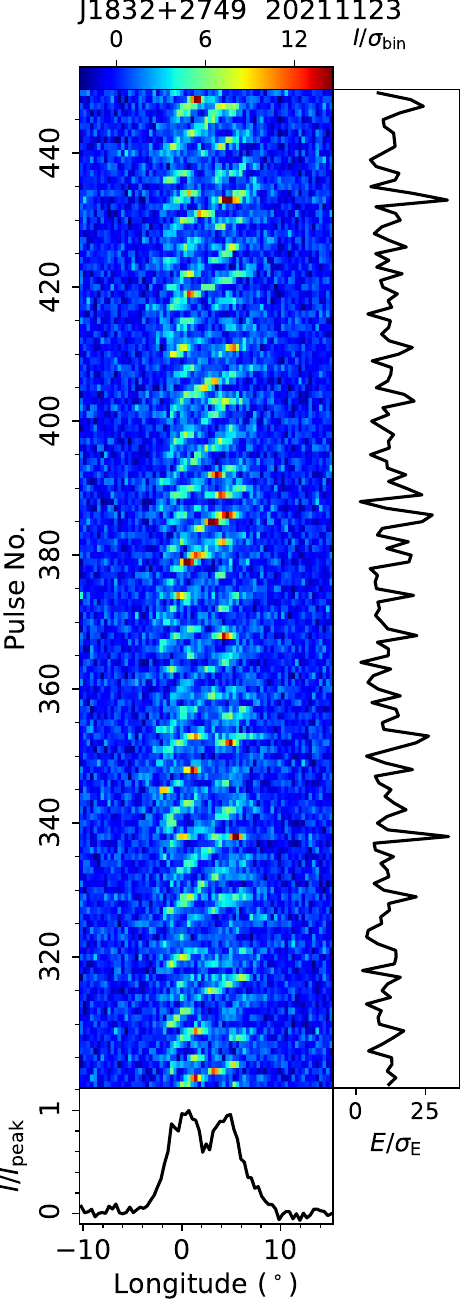}
\caption{Similar to Figure~\ref{subfig:TP:J1501-0046}, but for the observation of PSR J1832+2749 by FAST conducted on 20211123, showing subpulse drifting, with a zoomed-in view in the left.
\label{subfig:TP:J1832+2749}
}
\end{figure}

\begin{figure}
\centering
\includegraphics[width=0.44\textwidth, angle=0]{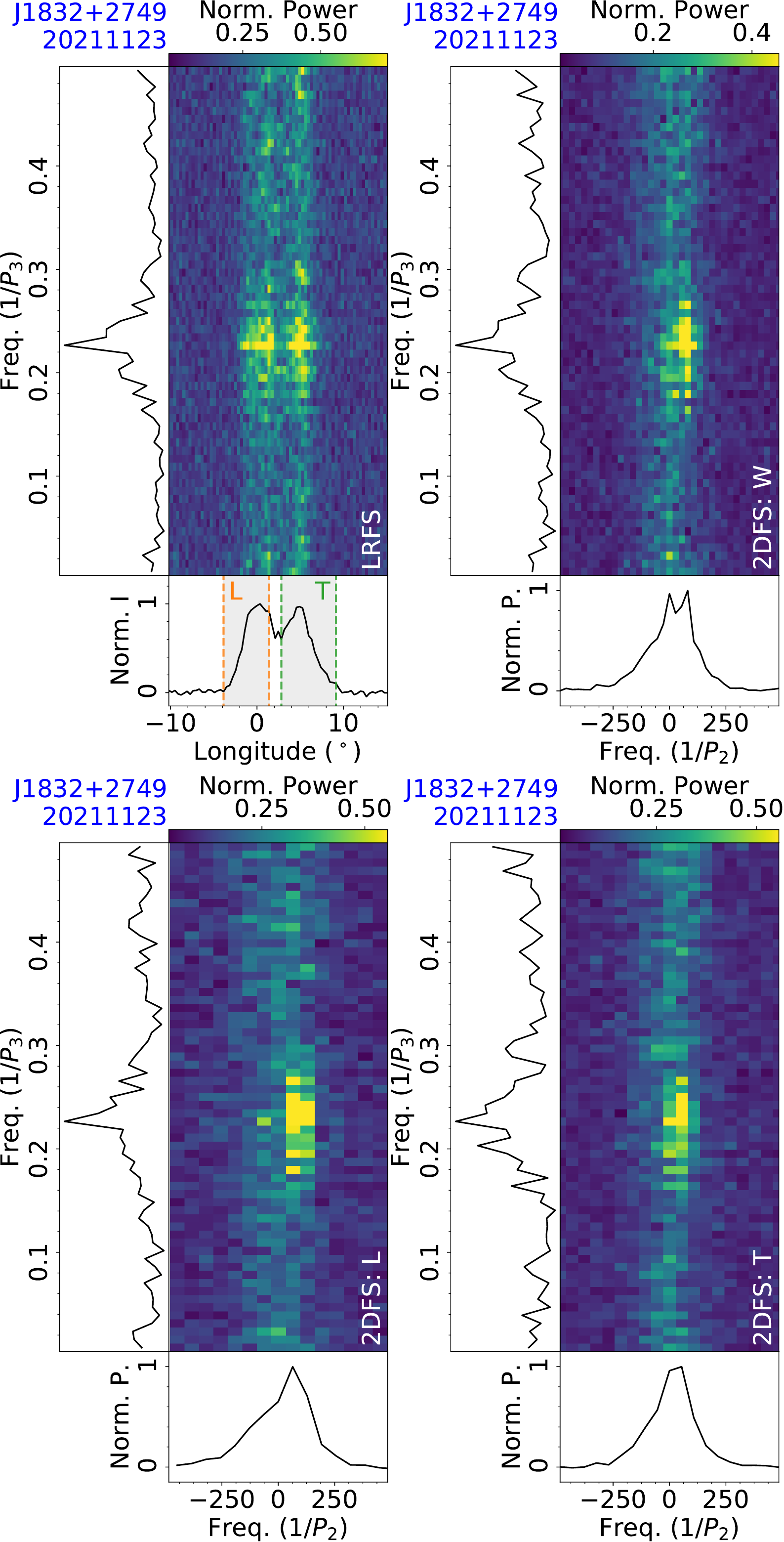}
\caption{Intensity fluctuation analysis of PSR J1832+2749 with the LRFS (top-left) and 2DFS approaches for the emission of the whole pulse phase range (top-right), the Leading half part (bottom left), and the Trailing half part (bottom right). Regular modulations for all parts exhibit a widely enhanced fluctuation spectrum with a peak frequency around $1/P_3=0.22$, corresponding to $P_3=4.4$ periods. 
The drifting frequency for the leading part is about $P_2=3.8\pm0.1^\circ$ and for the trailing part $P_2=7.0\pm0.3^\circ$. 
\label{subfig:fluctu:J1832+2749}
}
\end{figure}

\subsection{Just mode-changing}
\label{subsec:ModePSRs}

Based on FAST observation data, we found that 31 pulsars exhibit just mode-changing behavior, including one pulsar whose mode changes had been reported previously (see Table~\ref{Tab:all}). 

For pulsars exhibiting mode-changing behavior, we manually categorize emission modes from single-pulse sequences by leveraging differences in their morphological features or by analyzing histograms of specific characteristic parameters (e.g., intensity of individual profile components). Once emission modes for individual pulses are identified, one can compare single-pulse properties of different modes and properties of mean polarization profiles including the polarization position angle (PA) curves. In general, the PA curves of different emission modes are almost identical, with only a few exceptions with different PA curves 
\citep{Yan2023_ModePol}.


Here we analyse data of PSR J1848-0023 as an example, and detailed analyses of mode changing for other pulsars are presented in Sect. 4.

PSR J1848-0023 was discovered during the Parkes multibeam pulsar survey by \citet{hfs+04}. The pulsar was observed by the FAST on 20211226 for 15 minutes. The rotation period and dispersion measurement determined by this FAST observation are $P=0.5376$~s and $DM=34.8~{\rm cm^{-3}\,pc}$. Single pulse sequences presented in Figure~\ref{subfig:TP:J1848-0023} show both weak and bright emission modes. In contrast to the weak mode, the pulse intensity in the bright mode is enhanced for all components, and the duration is relatively shorter. In the left subpanels of Figure~\ref{subfig:TP:J1848-0023}, the on-pulse energy variation is smoothed over 5 periods, showing a demarcation between the two modes which is detailed in the histogram in Figure~\ref{subfig:Hist:J1848-0023}, which has been used to distinguish two emission modes. The averaged polarization profiles and polarization curves of two emission modes are shown in Figure~\ref{subfig:PolModes:J1848-0023}. The leading component is always strongest for the two emission modes. In the bright mode, the averaged profile is relatively narrow, and the trailing component is relatively weaker. In the weak mode, the bright leading component has a linear polarization in the orthogonal modes, and the bridge and trailing components have orthogonal polarization emission mixed to various extents.


Similar to PSR J1848-0023, the polarization angles of PSRs J0302+2252 and J1922+2110 are in orthogonal modes for different emission modes. We found that 4 pulsars, PSRs J1909+0912, J1919+0021, J1935+1616 and J1927+1852 have different PA curves, see Sect. 4 for details.


\begin{figure}[htbp]
\centering
\includegraphics[width=0.22\textwidth, angle=0]{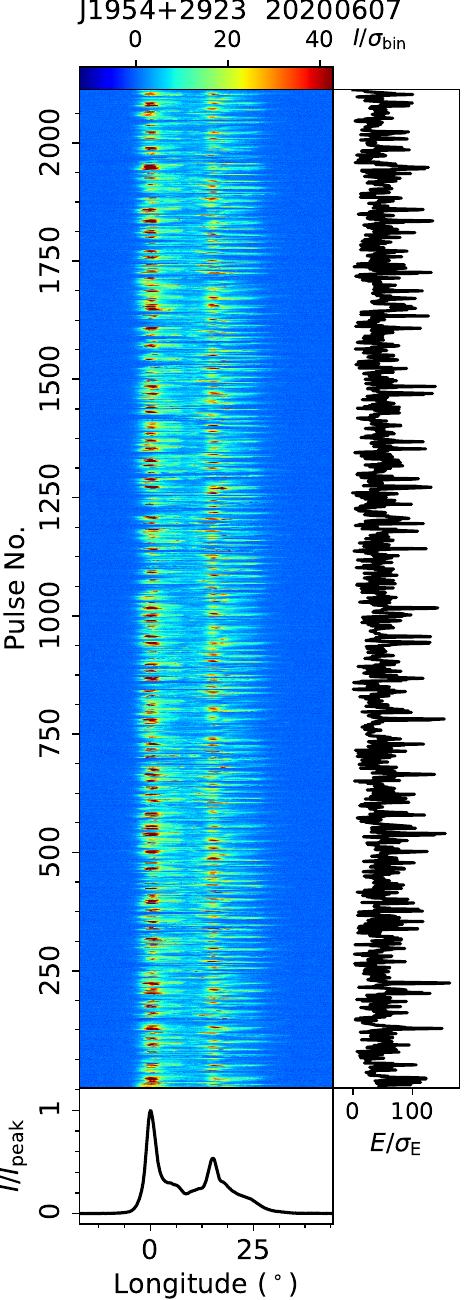}
\includegraphics[width=0.22\textwidth, angle=0]{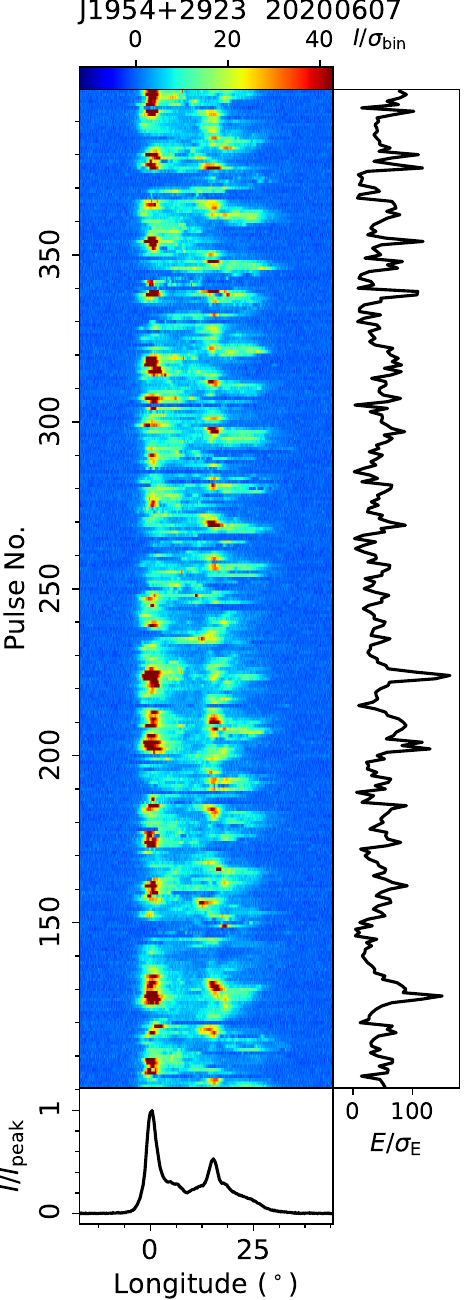}
\caption{Similar to Figure~\ref{subfig:TP:J1501-0046}, but for PSR J1954+2923 observed on 20200607 for just modulations.
\label{subfig:TP:J1954+2923}}
\end{figure}

\begin{figure}[htbp]
\centering
%
\includegraphics[width=0.44\textwidth, angle=0]{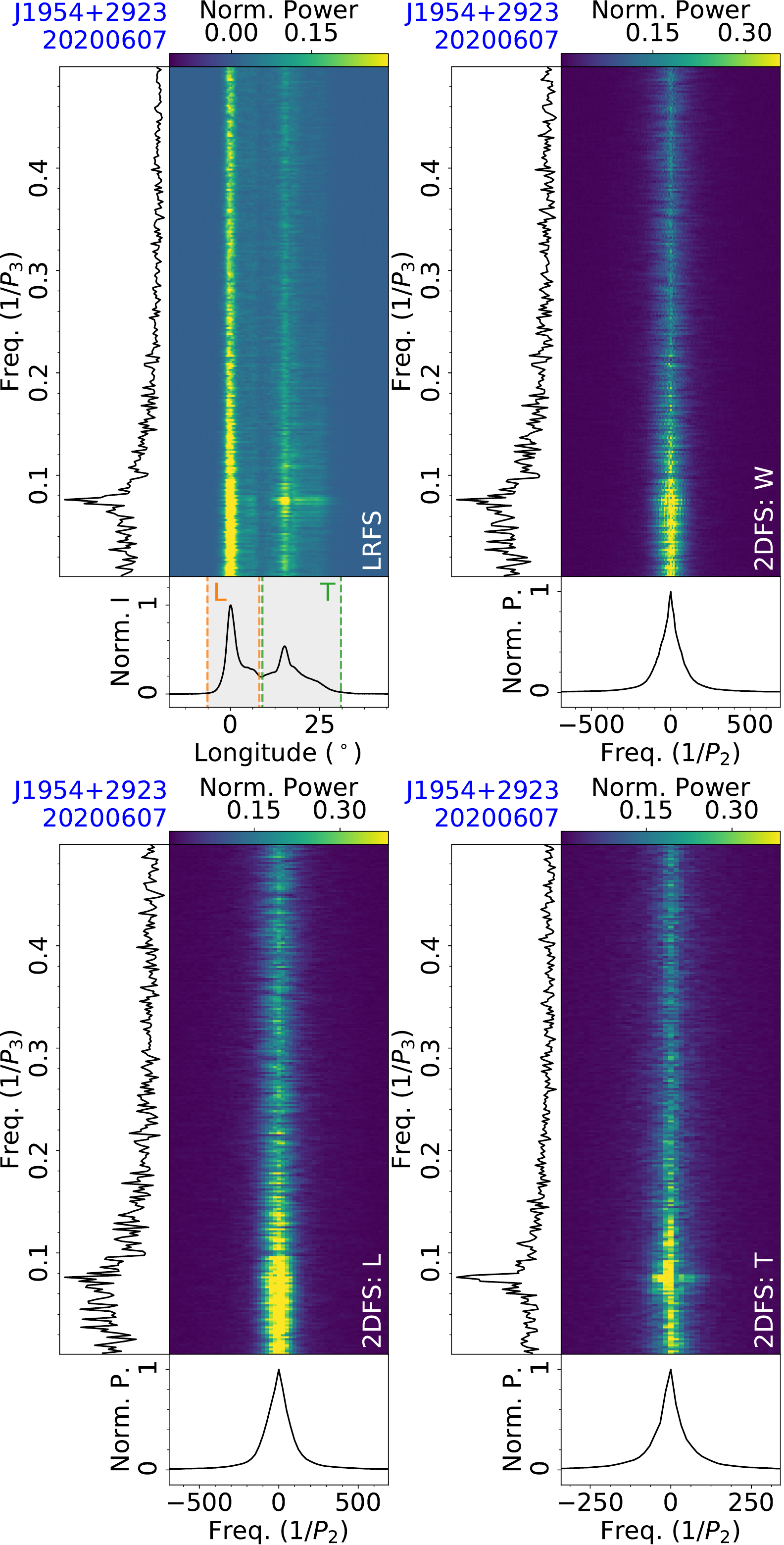}
\caption{Similar to Fig.~\ref{subfig:fluctu:J1832+2749}, here is fluctuation analysis for data in Fig~\ref{subfig:TP:J1954+2923} for PSR J1954+2923 observed on 20200607, with a distinct peak around $1/P_3 =0.076$ mainly caused by the trailing part. 
\label{subfig:fluctu:J1954+2923}}
\end{figure}

\begin{figure}[htbp]
\centering
\includegraphics[width=0.22\textwidth, angle=0]{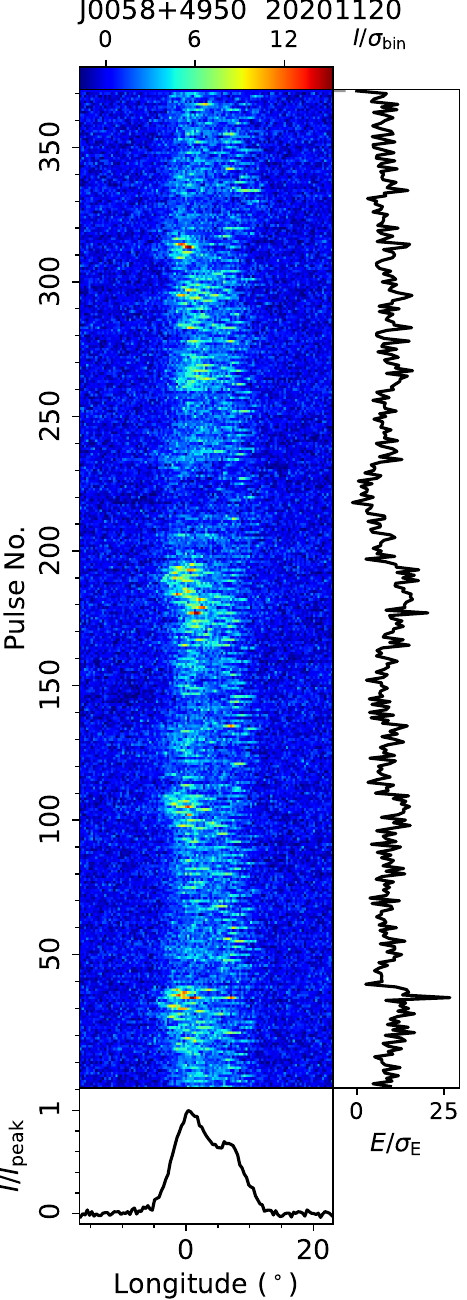} 
\includegraphics[width=0.22\textwidth, angle=0]{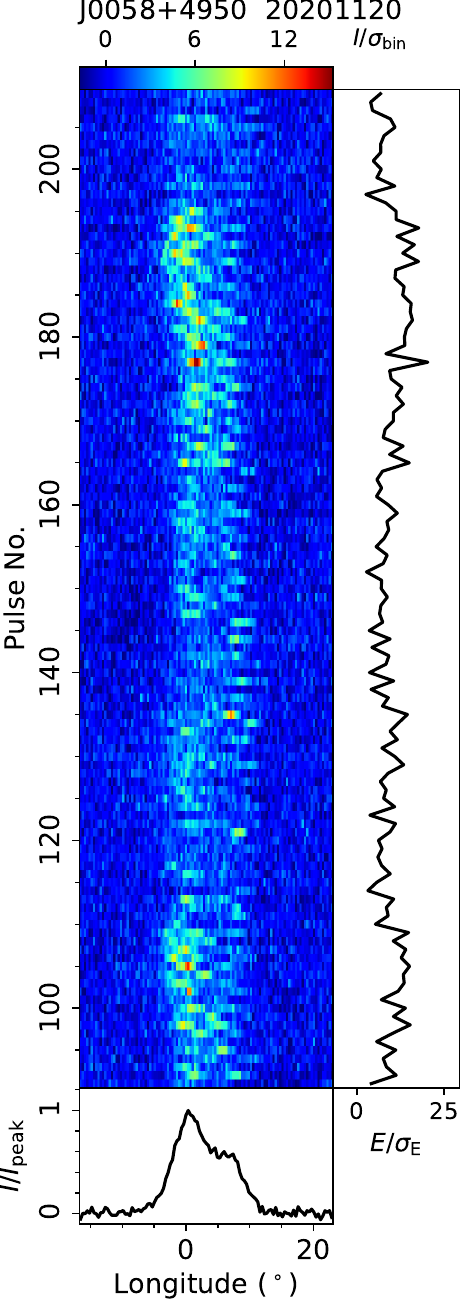}
\caption{Similar to Figure~\ref{subfig:TP:J1501-0046}, but for the single pulse stack of PSR J0058+4950 observed on 20201120, and the zoomed-in view for pulses No. 90-210. }
\label{fig:TP:J0058+4950} 
\vspace{4mm}
%
\includegraphics[width=0.22\textwidth, angle=0]{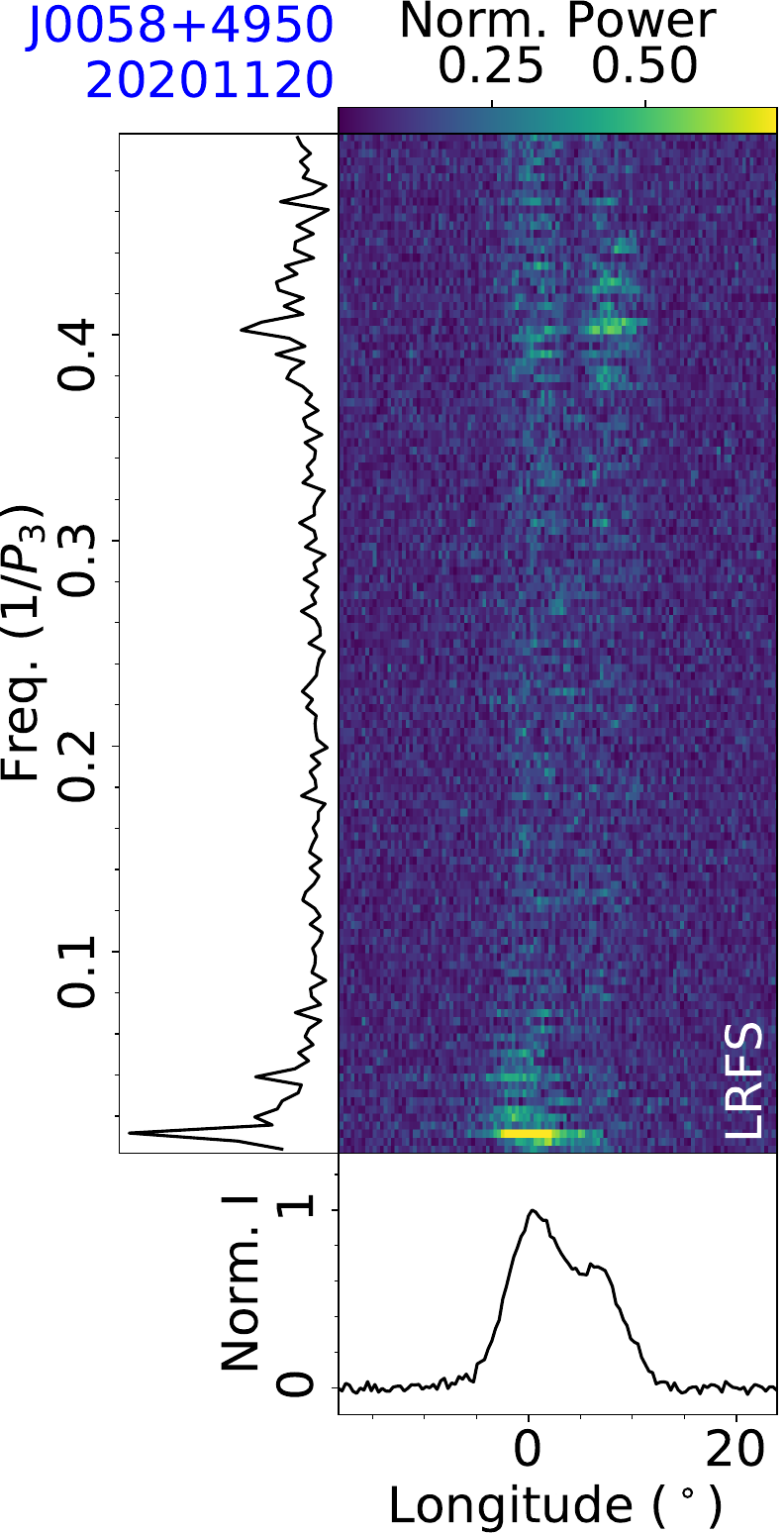} 
\includegraphics[width=0.22\textwidth, angle=0]{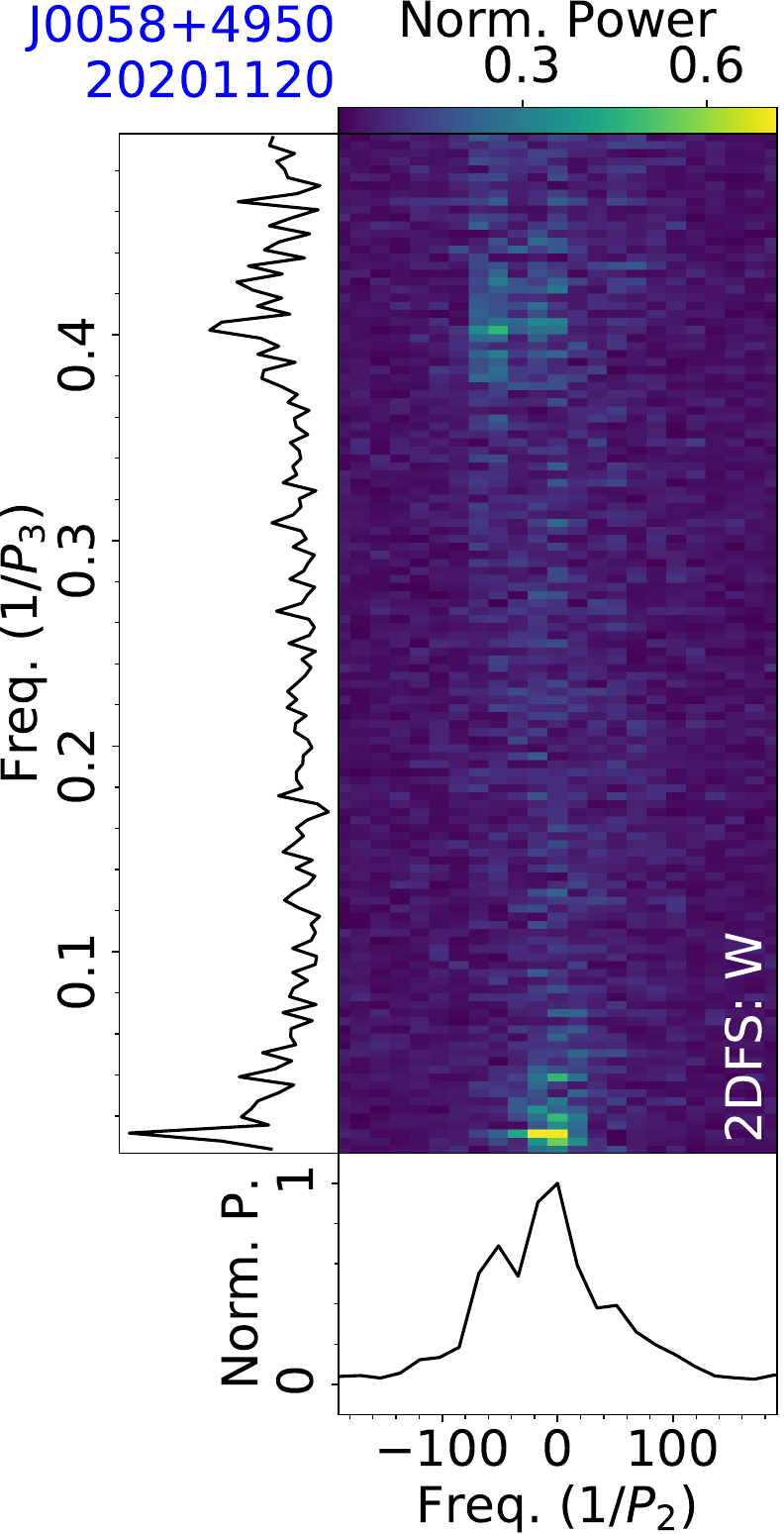}
\caption{Similar to Fig.~\ref{subfig:fluctu:J1832+2749}, but for LRFS and 2DFS of PSR J0058+4950 data in Fig.~\ref{fig:TP:J0058+4950} which show a distinct peak around $1/P_3 = 0.02$ for a very low frequency fluctuation, in addition to the fluctuation energy around $1/P_3 = 0.4$ from subpulse driftings.  
} 
\label{fig:fluctu:J0058+4950} 
\end{figure}

\subsection{Subpulse drifting and modulation}
\label{subsec:moduPSRs}

In general, continuous drifting over all profile components has the same modulations for all subpulses. 
The subpulse drifts towards the later rotation phase in the emission window, which is a positive drift with a drifting rate $D>0$, and the drifting to an earlier phase in the next period is the negative drift with a rate of $D<0$. For some pulsars, the subpulses do not drift in phase ($D \sim 0$), but pulse intensities periodically fluctuate for one or more pulse profile components, which is the so-called temporal modulation.

From the FAST folded data, some pulsars show subpulse drifting, but some pulsars do not show drifting but just temporal modulation. We detected 170 pulsars only having subpulse modulation from FAST data, and 110 of them have the drifting or modulations detected for the first time. For these pulsars with previously known drifting subpulses, sensitive FAST observations revealed more details. Some pulsars show distinct phase modulation features for different mean profile components, not necessarily co-occurring drifting features.

\begin{table}
\centering
\renewcommand\arraystretch{0.85}
\scriptsize
\setlength\tabcolsep{4pt}
\caption{Similar to Table~\ref{Tab:moduPars-modu}, but for pulsars with nulling and modulation behaviors. -- to be continued}
\label{Tab:moduPars-moduNull}
 \begin{tabular}{llcrrrl}
 \hline
PSR name  & Obs.date  & M.N & \multicolumn{1}{c}{$P_2$} & \multicolumn{1}{c}{$P_3$} & \multicolumn{1}{c}{$D$}  &  Notes     \\
\multicolumn{1}{c}{(1)}       & \multicolumn{1}{c}{(2)}        & (3)  & \multicolumn{1}{c}{(4)}   & \multicolumn{1}{c}{(5)}   & \multicolumn{1}{c}{(6)}  & \multicolumn{1}{c}{(7)} \\
\hline
J0343+5312   & 20240623 & W.1    & -5.8(4) & 2.91(2)    & -2.0(1)  & \ref{subsec:J0343+5312} \\
J0528+2200   & 20230913 & L.1    & -5.6(1) & 3.64(3)    & -1.55(4) & \ref{subsec:J0528+2200} \\
             &          & T.1    &   23(3) & 4.27(5)    &     5(1) & -      \\
J0546+2441   & 20190418 & W.1    & -1.15(3)& 4(1)       & -0.27(5) & \ref{subsec:J0546+2441} \\
J0555+3948   & 20210524 & W.1    & -       & -          & -        & \ref{subsec:J0555+3948} \\
J0623+0340   & 20240623 & C.1    & *       & 11.7(4)    & -        & \ref{subsec:J0623+0340} \\
J0711+0931   & 20210111 & W.1    & -4.9(2) & 6.19(2)    & -0.79(4) & \ref{subsec:J0711+0931} \\
J0811+37     & 20210928 & W.1    & -6.3(4) & 2.307(5)   & -2.7(2)  & \ref{subsec:J0811+37}   \\
J0944+4106   & 20201124 & L.1    & 9(2)    & 3.4(1)     & 2.7(6)   & \ref{subsec:J0944+4106} \\
             &          & T.1    & 12(3)   & 3.20(3)    & 3.9(8)   & -                       \\
J1502+4653   & 20250125 & L.1    & -8(1)   & 9.0(1)     & -0.9(1)  & \ref{subsec:J1502+4653} \\
             &          & T.1    & -2.2(1) & 9.8(1)     & -0.23(1) & -                       \\
J1549+2113   & 20250217 & W.1    & -5.4(1) & 2.466(5)   & -2.2(1)  & \ref{subsec:J1549+2113} \\
J1741-0840   & 20250521 & L.1    & 52(19)  & 5.02(5)    & 10(4)    & \ref{subsec:J1741-0840} \\
             &          & T.1    & 6.3(4)  & 4.89(3)    & 1.3(1)   & -                       \\
J1745-0129   & 20201128 & W.1    & 4.7(1)  & 2.582(4)   & 1.8(1)   & \ref{subsec:J1745-0129} \\
J1759-1029   & 20250803 & W.1    & 10.4(5) & 3.46(3)    & 3.0(1)   & \ref{subsec:J1759-1029} \\
J1802+0128   & 20210117 & W.1    & *       & 2.16(1)    & -        & \ref{subsec:J1802+0128}\\
J1802-0523   & 20250910 & W.1    & 6.3(2)  & 7.9(1)     & 0.79(2)  & \ref{subsec:J1802-0523}\\
J1810-0100g  & 20230527 & L.1    & 6.0(4)  & 2.443(5)   & 2.5(2)   & \ref{subsec:J1810-0100g} \\
J1812+0226   & 20201129 & L.1    & -12(1)  & 2.22(1)    & -6(1)    & \ref{subsec:J1812+0226}\\
             &          & L.2    & 11(1)   & 26(1)      & 0.41(3)  & -                       \\
             &          & T.2    & *       & 2.19(1)    &  -       & -                       \\
             &          & T.1    & 18(2)   & 22.3(5)    & 0.8(1)   & -                       \\
J1818+03     & 20240208 & W.1    & -5.9(1) & 4.42(4)    & -1.34(3) & \ref{subsec:J1818+03} \\
J1819+1305   & 20250217 & L.1    & 20(3)   & 2.88(1)    & 7(1)     & \ref{subsec:J1819+1305} \\
             &          & T.1    & 42(10)  & 6.56(4)    & 6(2)     & -                       \\
J1820+0006g  & 20240125 & T.1    & 34(22)  & 3.23(1)    & 10(7)    & \ref{subsec:J1820+0006g}\\
J1821+4147   & 20201031 & C.1    & *       & 25(1)      & -        & \ref{subsec:J1821+4147} \\
J1822+0044g  & 20240627 & W.1    & -3.81(2)& 2.3(1)     & -1.68(8) & \ref{subsec:J1822+0044g} \\
J1824-0127   & 20211223 & L.1    & -16(4)  & 3.24(3)    & -5(1)    & \ref{subsec:J1824-0127} \\
             &          & T.1    & -14(3)  & 3.31(3)    & -4(1)    & -    \\
             & 20220902 & L.1    & -11(3)  & 3.19(4)    & -4(1)    & -    \\
             &          & T.1    & -20(8)  & 3.28(3)    & -6(3)    & -    \\
J1825-0208g  & 20221109 & W.1    & -2.05(1)& 2.4(1)     & -0.84(4) & \ref{subsec:J1825-0208g} \\
J1826-1131   & 20250422 & L.1    & 11(1)   & 2.09(1)    & 5(1)     & \ref{subsec:J1826-1131} \\
             &          & T.1    & *       & 2.10(1)    & -        & -                        \\
J1831+0222g  & 20241012 & W.1    & -4.9(4) & 2.72(1)    & -1.8(2)  & \ref{subsec:J1831+0222g} \\
             & 20241107 & W.1    & -4.6(2) & 2.74(1)    & -1.7(1)  & -    \\
J1834+10     & 20210213 & L.1    & -30(11) & 2.79(1)    & -11(4)   & \ref{subsec:J1834+10} \\
             &          & T.1    & -12(1)  & 2.96(2)    & -3.9(3)  & -                     \\
J1837+1221   & 20210701 & L.1    & -3.0(2) & 6.3(1)     & -0.47(4) & \ref{subsec:J1837+1221}  \\
J1838+0044g  & 20200203 & L.1    & -9(1)   & 4.9(1)     & -1.9(2)  & \ref{subsec:J1838+0044g} \\
             &          & T.1    & -11(1)  & 4.01(4)    & -2.7(3)  & -               \\
J1839-1238   & 20250422 & W.1    & -3.9(1) & 5.8(1)     & -0.68(4) & \ref{subsec:J1839-1238} \\
J1840-0141g  & 20220118 & W.1    & 2.13(5) & 2.8(8)     & 0.7(2)   & \ref{subsec:J1840-0141g} \\
J1841+0912   & 20230206 & W.1    & 100(31) & 49(1)      & 2(1)     & \ref{subsec:J1841+0912} \\
J1842-0153   & 20230328 & W.1    & *       & 30.7(9)    & -        & \ref{subsec:J1842-0153} \\
J1843+0526g  & 20230218 & W.1    & 3.59(2) & 2.0(1)     & 1.79(6)  & \ref{subsec:J1843+0526g}\\
J1843-0050   & 20211009 & W.1    & -11.0(3)& 3.24(1)    & -3.4(1)  & \ref{subsec:J1843-0050} \\
             & 20241224 & W.1    & -13.5(5)& 3.25(2)    & -4.2(2)  & -    \\
J1843-0757   & 20251015 & T.1    & 15(4)   & 2.48(1)    & 6(1)     & \ref{subsec:J1843-0757} \\
J1844-0433   & 20210219 & L.1    & -5(1)   & 8.9(1)     & -0.6(1)  & \ref{subsec:J1844-0433} \\
J1849+0340g  & 20230309 & W.1    & *       & 5.40(3)    & -        & \ref{subsec:J1849+0340g}\\
J1851+0241   & 20200529 & W.1    & 1.7(2)  & 2.0(14)    & 0.9(7)   & \ref{subsec:J1851+0241} \\
J1851-0053   & 20210822 & W.1    & 4.3(1)  & 3.74(2)    & 1.16(2)  & \ref{subsec:J1851-0053} \\
             & 20210903 & W.1    & 4.3(1)  & 3.69(3)    & 1.16(2)  & -      \\
             & 20211009 & W.1    & 4.0(1)  & 3.37(2)    & 1.17(2)  & -      \\
             & 20250331 & W.1    & 3.7(1)  & 3.48(2)    & 1.07(2)  & -      \\
J1852-0947   & 20230528 & W.1    & -4.1(1) & 2.23(1)    & -1.9(1)  & \ref{subsec:J1852-0947} \\
             & 20251019 & W.1    & -4.0(2) & 2.30(2)    & -1.7(1)  & -      \\
J1853+0853   & 20210306 & W.1    & 1.7(1)  & 2.2(3)     & 0.7(1)   & \ref{subsec:J1853+0853} \\
             & 20211018 & W.1    & 2.4(1)  & 2.9(5)     & 0.8(2)   & -                    \\
J1858+0026g  & 20200426 & W.1    & -3.8(6) & 6.9(1)     & -0.6(1)  & \ref{subsec:J1858+0026g} \\
J1901+0511   & 20220102 & T.1    & 4.1(6)  & 2.19(1)    & 1.9(2)   & \ref{subsec:J1901+0511} \\
J1902+0248   & 20210119 & T.1    & 9(1)    & 2.12(1)    & 4.5(4)   & \ref{subsec:J1902+0248} \\
             & 20221018 & T.1    & 9(1)    & 2.20(1)    & 4.3(5)   & -                       \\
J1903+0415   & 20200223 & W.1    & -6.9(2) & 3.87(3)    & -1.8(1)  & \ref{subsec:J1903+0415} \\
J1904+1011   & 20230525 & L.1    & -22(2)  & 12.4(2)    & -1.7(2)  & \ref{subsec:J1904+1011}  \\
\hline 
\addtocounter{table}{-1}
\end{tabular}
\end{table}

\begin{table}[ht!]
\centering
\renewcommand\arraystretch{0.85}
\scriptsize 
\setlength\tabcolsep{4pt}
\caption{-- Continued and ended.} 
 \begin{tabular}{llcrrrl}
 \hline
PSR name  & Obs.date  & M.N & \multicolumn{1}{c}{$P_2$} & \multicolumn{1}{c}{$P_3$} & \multicolumn{1}{c}{$D$} &  Notes     \\
\multicolumn{1}{c}{(1)}       & \multicolumn{1}{c}{(2)}        & (3)  & \multicolumn{1}{c}{(4)}   & \multicolumn{1}{c}{(5)}   & \multicolumn{1}{c}{(6)}  & \multicolumn{1}{c}{(7)} \\
\hline
             &          & T.1    & -30(3)  & 13.6(2)    & -2.2(3)  & -                        \\
J1905+0935g  & 20220529 & L.1    & -4.3(2) & 2.054(2)   & -2.1(1)  & \ref{subsec:J1905+0935g} \\
J1910+0714   & 20210806 & L.1    & -4.6(3) & 3.12(2)    & -1.5(1)  & \ref{subsec:J1910+0714} \\
J1910-0219g  & 20250619 & W.1    & -4.5(3) & 2.13(1)    & -2.1(1)  & \ref{subsec:J1910-0219g} \\
J1913+3732   & 20201128 & L.1    & -30(11) & 2.266(3)   & -13(5)   & \ref{subsec:J1913+3732} \\
J1920+1040   & 20220113 & W.1    & 7.7(1)  & 2.77(1)    & 2.78(4)  & \ref{subsec:nulldrift} \\
             & 20210130 & W.1    & 7.5(2)  & 2.67(1)    & 2.82(6)  & -  \\
             & 20240106 & W.1    & 8.2(3)  & 2.73(1)    & 2.98(9)  & -  \\
J1921+0851g  & 20230510 & W.1    & -3.4(3) & -20(24)    & -0.2(2)  & \ref{subsec:J1921+0851g} \\
J1923+1706   & 20190321 & L.1    & -13(2)  & 12(1)      & -1.1(2)  & \ref{subsec:J1923+1706} \\
             &          & C.1    & *       & 20(2)      & -        & -                       \\
             &          & T.1    & *       & 16(1)      & -        & -                       \\
J1925+19     & 20221002 & L.1    & -30(10) & 9.4(2)     & -3(1)    & \ref{subsec:J1925+19} \\
             &          & T.1    & -12(2)  & 10.1(2)    & -1.2(2)  & -                       \\
J1926+0431   & 20210702 & W.1    &  6.1(3) & 2.58(3)    &  2.4(1)  & \ref{subsec:J1926+0431} \\
             &          & W.2    & -6.9(3) & 2.30(1)    & -3.0(1)  & -                       \\
J1926+1434   & 20210317 & L.1    & -57(13) & 38(1)      & -1.5(4)  & \ref{subsec:J1926+1434} \\
             &          & L.2    & -36(4)  & 7.47(4)    & -5(1)    & -                       \\
             &          & T.2    & -86(22) & 28(1)      & -3(1)    & -                       \\
J1927+2234   & 20210608 & L.1    & 26(4)   & 6.03(3)    & 4.3(7)   & \ref{subsec:J1927+2234} \\
             &          & T.1    & 11(1)   & 5.51(2)    & 2.0(2)   & -                       \\
J1930+1722   & 20250414 & W.1    & -5.2(1) & 3.81(4)    & -1.37(4) & \ref{subsec:J1930+1722} \\
J1931+1439   & 20240129 & L.1    & -9(1)   & 2.16(1)    & -4.3(5)  & \ref{subsec:J1931+1439} \\
             &          & T.1    & *       & 2.17(1)    &  -       & -                       \\
             & 20240917 & L.1    & -8(1)   & 2.14(1)    & -3.8(6)  & -                       \\
             &          & T.1    & *       & 2.16(1)    &  -       & -                       \\
J1933+1304   & 20201220 & L.1    & -10(2)  & 4.83(2)    & -2.0(4)  & \ref{subsec:J1933+1304} \\
             &          & L.2    & *       & 6.61(3)    & -        & -                       \\
             &          & T.1    & -8(1)   & 4.82(1)    & -1.7(2)  & -                       \\
             &          & T.2    & *       & 6.91(3)    & -        & -                       \\
J1935+1159   & 20210618 & L.1    & -25(3)  & 7.2(1)     & -3.5(4)  & \ref{subsec:J1935+1159} \\
J1937+1358g  & 20210814 & L.1    & -24(7)  & 13.5(5)    & -1.7(6)  & \ref{subsec:J1937+1358g} \\
             &          & T.1    & -22(5)  & 16.0(6)    & -1.4(3)  & -                       \\
J1937+1505   & 20201226 & L.1    & *       & 2.12(1)    & -        & \ref{subsec:J1937+1505} \\
J1938+0650   & 20210702 & W.1    & 2.83(4) & -          & 0.27(18) & \ref{subsec:J1938+0650} \\
J1938+14     & 20201226 & W.1    & 15(3)   & 6.8(1)     & 2.2(3)   & \ref{subsec:J1938+14} \\
J1938+14a    & 20210814 & L.1    & *       & 2.36(2)    & -        & \ref{subsec:J1938+14a} \\
J1939+10     & 20230215 & L.1    & -4.5(1) & 4.84(5)    & -0.93(4) & \ref{subsec:J1939+10} \\
             &          & T.1    & -7.6(5) & 4.8(1)     & -1.6(1)  & -         \\
J1939+2352g  & 20240317 & W.1    & 5.9(3)  & 4.1(1)     & 1.44(5)  & \ref{subsec:J1939+2352g}\\
J1944+1755   & 20191226 & L.1    & -21(1)  & 11.8(1)    & -1.8(1)  & \ref{subsec:J1944+1755} \\
             & (P3M11)  & T.1    & -17(1)  & 11.3(1)    & -1.5(1)  & -         \\
             & 20191226 & L.1    & -21(2)  & 16.5(2)    & -1.3(1)  & -         \\
             & (P4M11)  & T.1    & -34(4)  & 17.4(2)    & -1.9(3)  & -         \\
J1944+1934g  & 20240703 & W.1    & 1.93(4) & 3.88(3)    & 0.50(1)  & \ref{subsec:J1944+1934g} \\
J1945-0040   & 20250410 & L.1    & -34(8)  & 8.8(1)     & -4(1)    & \ref{subsec:J1945-0040} \\
             &          & T.1    & -10(1)  & 8.6(1)     & -1.1(1)  & -         \\
J1952+3021   & 20211009 & W.1    & -6.1(2) & 3.84(3)    & -1.6(1)  & \ref{subsec:J1952+3021} \\
J1954+1021   & 20210117 & W.1    & -5.8(2) & 3.012(4)   & -1.9(1)  & \ref{subsec:J1954+1021}  \\
J2002+4050   & 20210717 & L.1    & *       & 2.53(1)    & -        & \ref{subsec:J2002+4050}  \\
             &          & C.1    & -3.8(2) & 2.55(1)    & -1.5(1)  & -         \\
             &          & T.1    & -14(3)  & 2.52(1)    & -5(1)    & -         \\
J2011+3521g  & 20230804 & W.1    & 6.6(2)  & 2.236(2)   & 2.9(1)   & \ref{subsec:J2011+3521g} \\
J2016+3318g  & 20240909 & T.1    &  8(2)   & 2.93(2)    & 2.8(6)   & \ref{subsec:J2016+3318g} \\
             & 20241026 & L.1    & -5.2(7) & 3.67(3)    & -1.4(2)  & -                       \\
J2030+31     & 20240726 & T.1    & 5.6(4)  & 3.25(5)    & 1.7(1)   & \ref{subsec:J2030+31} \\
J2044+4614   & 20210324 & L.1    &-103(32) & 9.1(2)     & -11(4)   & \ref{subsec:J2044+4614} \\
             &          & T.1    & -24(3)  & 8.2(2)     & -2.9(5)  & -                       \\
J2053+4718   & 20200810 & W.1    & 2.04(1) & 3.1(1)     & 0.65(2)  & \ref{subsec:J2053+4718} \\
J2105+07     & 20210118 & T.1    & 11(2)   & 6.5(1)     & 1.6(3)   & \ref{subsec:J2105+07}  \\
\hline 
\end{tabular}
\end{table}

\begin{figure}[tbph]
\centering
\includegraphics[width=0.15\textwidth, angle=0]{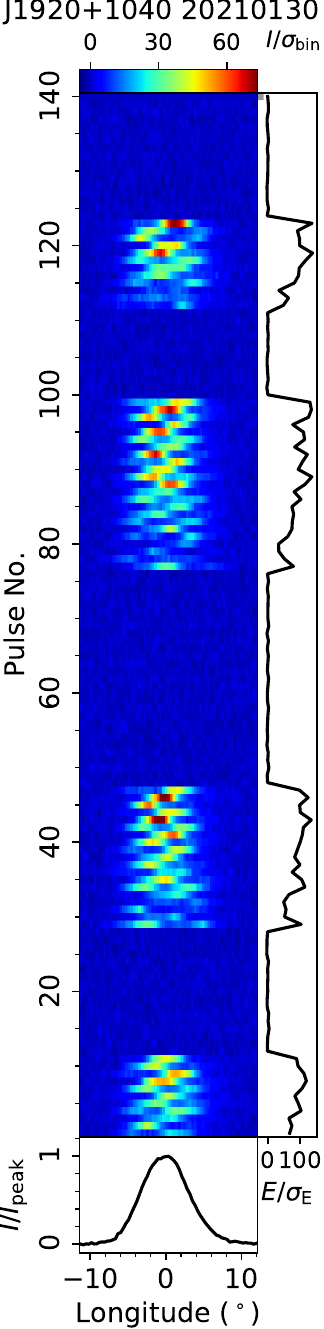}
\includegraphics[width=0.15\textwidth, angle=0]{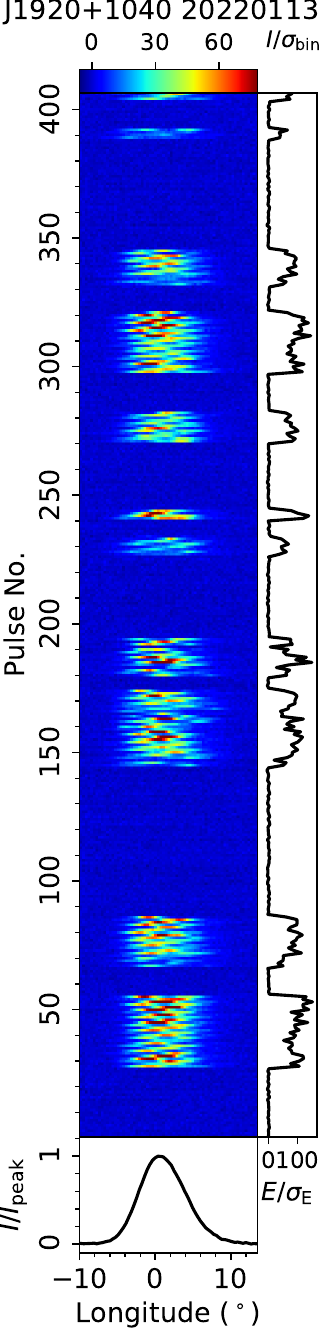}
\includegraphics[width=0.15\textwidth, angle=0]{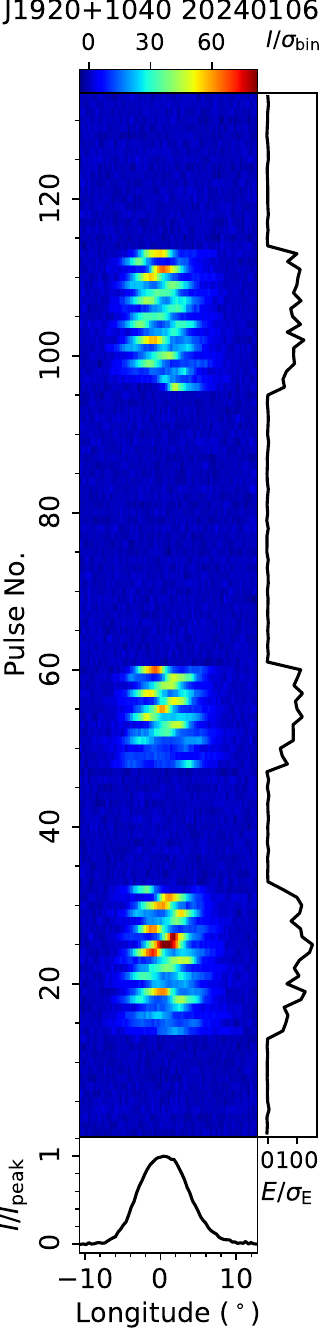}
\caption{Similar to Figure~\ref{subfig:TP:J1501-0046}, but for PSR J1920+1040 observed on 20220113, 20210130 and 20240106, showing nulling and subpulse drifting. 
\label{subfig:TP:J1920+1040}
}
%
\includegraphics[width=0.39\textwidth, angle=0]{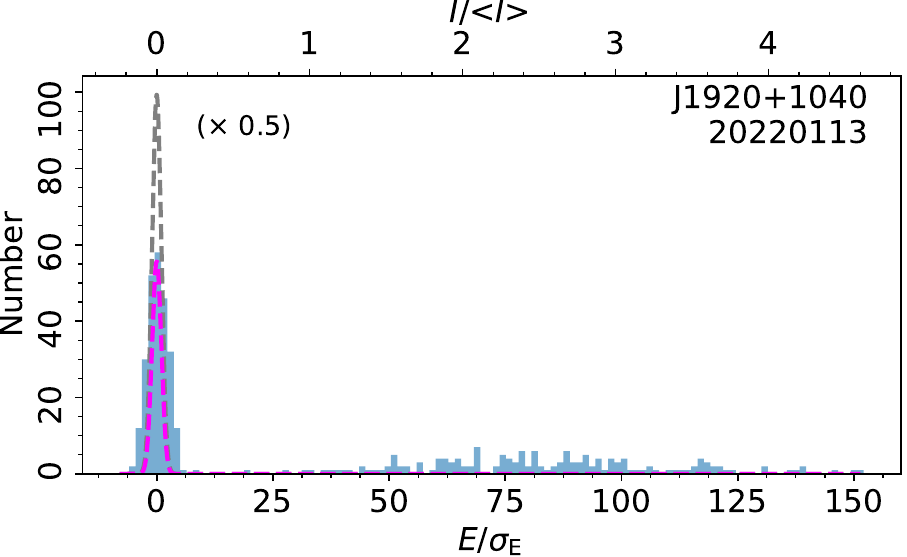}
\caption{Similar to Figure~\ref{subfig:Hist:J1501-0046}, but for the nulling of PSR J1920+1040 using data on 20220113 presented in Fig.~\ref{subfig:TP:J1920+1040}
\label{subfig:Hist:J1920+1040}
}
\end{figure}

In early days, the cross-correlation method suggested by \citet{Sieber1975} has been used to determine the drift rate $D$ that is the longitudinal shift of subpulses in degree per pulsar period, and the interval between adjacent subpulses ($P_2$, in degree of rotation phase), and also the drift cycles $P_3$ for subpulses appearing at the same longitudes that is counted in unites of the pulsar period.
The correlation coefficient $K$ in the cross-correlation method is calculated from the sum of the correlations of every two successive individual pulses as a function of phase shift $p$. The $K-p$ curve is then fitted by a multi-Gaussian function, with the Gaussian center for zero phase shift, the space between adjacent Gaussian peaks for $P_2$. 
%
%
This method is effective for a short sequence of drifting subpulses, even with a weak signal-to-noise ratio or embedded in a long non-drifting pulse sequence.

\begin{figure}[htbp]
\centering
\includegraphics[width=0.22\textwidth, angle=0]{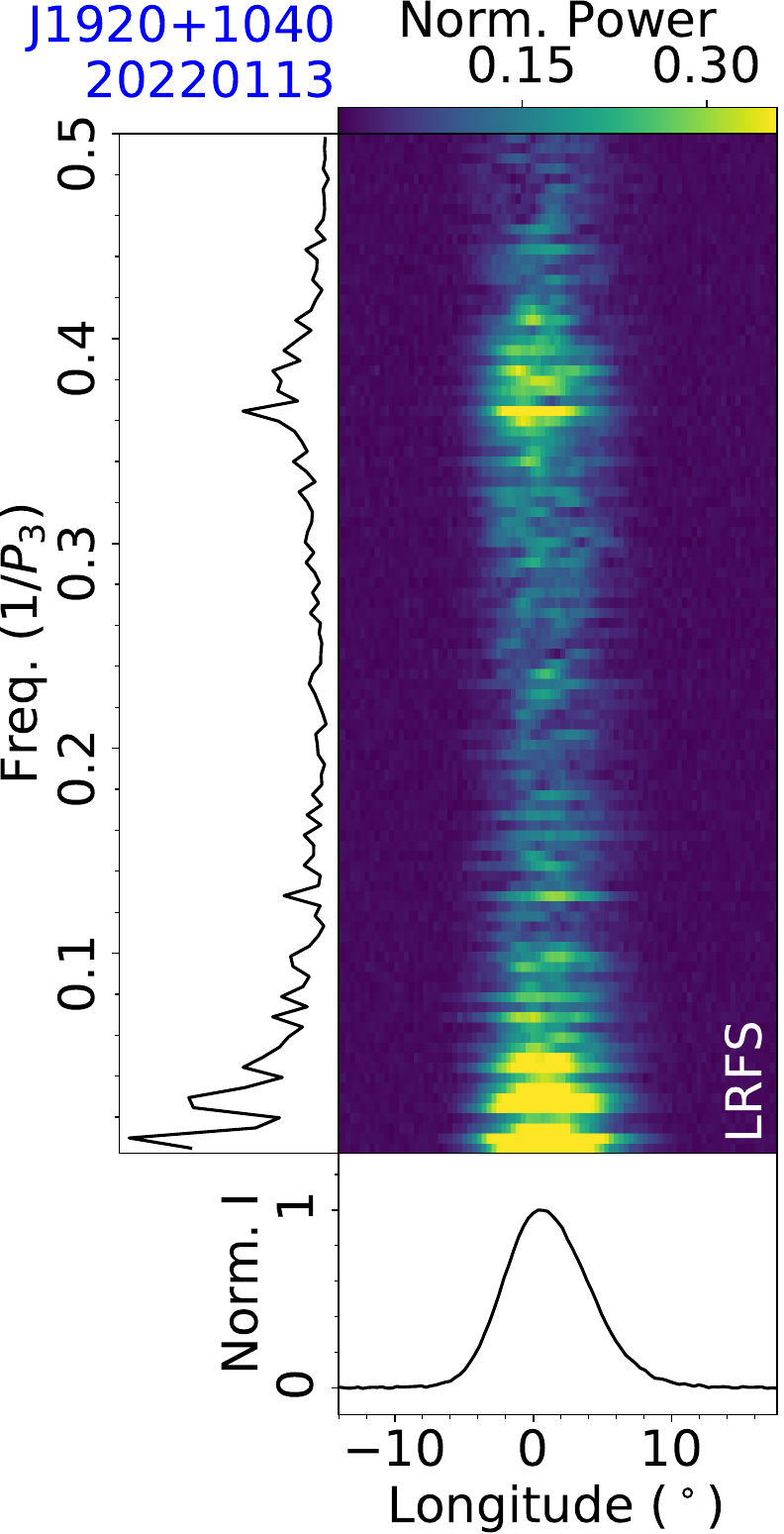}
\includegraphics[width=0.22\textwidth, angle=0]{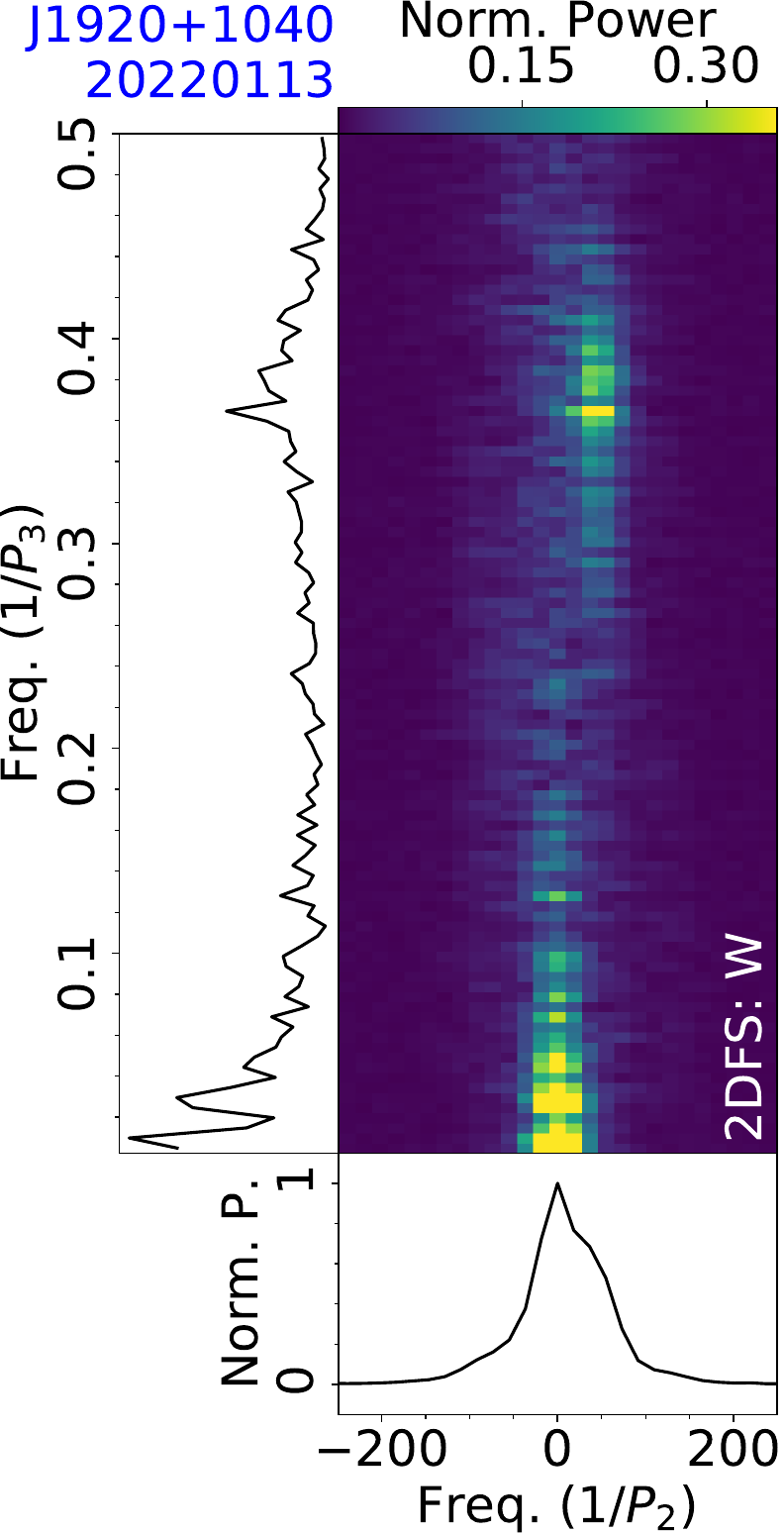}\\
\caption{Similar to Figure~\ref{subfig:fluctu:J1832+2749}, but for the LRFS and 2DFS analyses of subpulse drifting of PSR J1920+1040 observed on 20220113. 
\label{subfig:fluctu:J1920+1040}}
\end{figure}

Subpulse drifting can also be analyzed via time-phase plots or through the 1D Longitude Resolved Fluctuation Spectrum \citep[LRFS,][]{Backer1970, EdwSta2002, Weltevrede2006, Song2023}. Single-pulse sequences are either processed as a whole or divided into segments of $\rm n_s$ pulses. The results from all segments are averaged to produce a final 2D image, with the horizontal axis for pulse longitude and the vertical axis for pulse energy fluctuation frequency with the peak for periodical modulation corresponds to $1/P_3$. The LRFS can be integrated vertically to get the mean pulse profile (bottom panel) and horizontally to show the total energy fluctuations of the whole individual pulses. This technique can reveal the intensity modulation frequencies for each phase bin or each component.

To characterize the energy fluctuations along the pulse phase, a Discrete Fourier Transform (DFT) is applied to the 2D data of pulse intensities against pulsar rotation phases (in longitude), that is the Two-Dimensional Fluctuation Spectrum \citep[2DFS,][]{EdwSta2002} with the horizontal axis for the subpulse energy fluctuation frequency per pulsar ($1/P_2$), standing for the number of adjacent subpulses per period with a value given by $360^\circ/P_2$. As in LRFS, the vertical axis is for pulse energy fluctuation frequency $1/P_3$. The standard deviation of power values in the 2DFS analysis window is used to quantify the uncertainties of $P_2$ and $P_3$. Based on the mean pulse profiles, we also analyze the drifting properties for different parts or components of the profiles for some pulsars, with marked phase ranges.

%

Here we show examples of various drifting and temporal-only modulation features: PSR J1832+2749 for a standard subpulse drifting, 
PSR J1954+2923 for temporal modulation, and PSR J0058+4950 for low-frequency modulation superimposed on the negative subpulse drifting.

\subsubsection{PSR J1832+2749: Standard subpulse drifting}

PSR J1832+2749 was discovered by \citet{Tyulbashev2017} using the Large Scanning Antenna of the Pushchino Radio Astronomy Observatory at 110.25~MHz, and the final timing parameters were published by \citet{arecibo49_2025}. This pulsar has a period of $P=0.6318$~s and a dispersion measurement of $DM=47.6~{\rm cm^{-3}\,pc}$, as measured by the FAST for 5~minutes on 20211123. The single pulse sequence is shown in Figure~\ref{subfig:TP:J1832+2749}, showing a standard subpulse drifting behavior. 

We analyzed the fluctuation spectra by the 1D LRFS and 2DFS as shown in Figure~\ref{subfig:fluctu:J1832+2749}. The subpulses drifting parameters are almost the same with a slight difference for the two main profile components with a power fluctuation frequency of $1/P_3=0.22$, i.e., the $P_3$ is about 4.4 periods. However, the subpulse intervals are slightly different in the leading part and trailing part, with the centroid in spectra at $1/P_2=94\pm4$ or $1/P_2=51\pm4$, corresponding to $P_2=3.8\pm0.2^\circ$ and $P_2=7.0\pm0.5^\circ$ for the two parts, respectively.

\begin{table}
\centering
\renewcommand\arraystretch{0.85}
\scriptsize 
\setlength\tabcolsep{3pt}
\caption{Similar to Table~\ref{Tab:moduPars-modu}, but for pulsars with at least one emission mode showing  subpulse drifting.} 
\label{Tab:moduPars-modeModu}
\begin{tabular}{lllcrrrl}
 \hline
PSR name  & Obs.date & Mode  & M.N & \multicolumn{1}{c}{$P_2$} & \multicolumn{1}{c}{$P_3$} & \multicolumn{1}{c}{$D$} &  Notes     \\
\multicolumn{1}{c}{(1)}       & \multicolumn{1}{c}{(2)}  & (3)      & (4)  & \multicolumn{1}{c}{(5)}   & \multicolumn{1}{c}{(6)}   & \multicolumn{1}{c}{(7)} & \multicolumn{1}{c}{(8)}   \\
\hline
J1811-0154  & 20230405  & -  & L.1  & *         & 27.8(2)& -       & \ref{subsec:J1811-0154} \\
            &           & -  & T.1  & *         & 26.1(5)& -       & - \\
            &           & -  & T.1  & 6.8(1)    & 4.26(1)& 1.60(2) & - \\
J1818-0151  & 20210923  & -  & W.1  & -         & -      & -       & \ref{subsec:J1818-0151} \\
J1850+0026  & 20200119  & -  & L.1  & -57(7)    & 12.4(2)& -5(1)   & \ref{subsec:J1850+0026} \\
            &           & N  & T.1  & -32(6)    & 10.2(2)& -3(1)   & - \\
            &           & An & T.1  & -7.8(3)   & 3.95(4)& -2.0(1) & - \\
J1854+0319  & 20200509  & Sd & L.1  & -13(1)    & 11.8(2)& -1.1(1) & \ref{subsec:J1854+0319} \\
            &           & Sd & T.1  & -27(3)    & 11.7(1)& -2.3(3) & - \\
            &           & Fd & L.1  & -23(4)    & 10.3(2)& -2.2(4) & - \\
            &           & Fd & T.1  & -15(2)    & 7.7(1) & -2.0(2) & - \\
J1854-0230g & 20230527  & -  & L.1  & -45(2)    & 5.8(1) & -7.7(4) & \ref{subsec:J1854-0230g} \\
            &           & -  & T.1  & -61(3)    & 3.15(3)& -19(1)  & -    \\
J1856-0526  & 20250304  & Wc & W.1  &    48(6)  & 12.4(3)&  3.9(4) & \ref{subsec:J1856-0526} \\
            &           & Wc & W.2  &  -9.0(1)  & 2.47(1)& -3.6(1) & -    \\
            &           & Sc & W.1  & 447(307)  & 10.5(2)&  43(28) & -    \\
            &           & Sc & W.2  &  -8.6(1)  & 2.89(4)& -3.0(1) & -    \\
            & 20250429  & Wc & W.1  &   61(11)  & 12.7(3)&    5(1) & -    \\
            &           & Wc & W.2  &  -9.1(1)  & 2.46(2)& -3.7(1) & -    \\
            &           & Sc & W.1  & 412(247)  & 10.6(2)&  39(23) & -    \\
            &           & Sc & W.2  &  -8.6(1)  & 2.87(3)& -3.0(1) & -    \\
J1905+0616  & 20221025  & -  & W.1  &-0.838(3)  &  *     & -       & \ref{subsec:J1905+0616} \\
            &           & -  & W.2  & 0.90(1)   &  *     & -       & -                       \\
J1909+0423g & 20250209  & -  & W.1  & -84(4)    & 13.6(1)& -6.2(4) & \ref{subsec:J1909+0423g} \\
            & 20250523  & -  & W.1  & -80(4)    & 13.7(2)& -5.8(3) & -   \\
J1922+2110  & 20210706  & N  & T.1  & 42(12)    & 2.78(4)& 15(4)   & \ref{subsec:J1922+2110} \\
J1935+1616  & 20190919  & -  & L.1  & *         & 32(1)  & -       & \ref{subsec:J1935+1616} \\
            &           & -  & L.2  & *         & 7.7(1) & -       & - \\
            &           & -  & C.1  & *         & 10.6(2)& -       & - \\
            &           & -  & T.1  & *         & 3.6(1) & -       & - \\
            & 20211117  & -  & L.1  & *         & 34(1)  & -       & - \\
            &           & -  & L.2  & *         & 7.40(5)& -       & - \\ 
            &           & -  & C.1  & *         & 10.7(2)& -       & - \\
            &           & -  & T.1  & *         & 3.74(4)& -       & - \\
J1956+35    & 20240628  & WE & L.1  & -14(5)    & 18(1)  & -0.8(3) & \ref{subsec:J1956+35} \\
            &           & WE & T.1  & 13(4)     & 18(1)  & 0.7(1)  & - \\
J2051+4434g & 20230525  & Sd & L.1  & -29(1)    & 52.6(8)&-0.56(3) & \ref{subsec:modedrift} \\
            &           & Sd & T.1  & -37(1)    & 53.4(7)&-0.70(4) & -  \\
            &           & Fd & L.1  & -41(2)    & 15.9(2)&-2.6(2)  & -  \\
            &           & Fd & T.1  & -47(2)    & 16.8(2)&-2.8(2)  & -  \\
J2258+5222g & 20240723  & D1 & L.1  &   -8(1)   & 7.1(1) & -1.2(2) & \ref{subsec:J2258+5222g}\\
            &           & D1 & T.1  &  -11(2)   & 7.1(1) & -1.6(3) & -  \\
            &           & D2 & L.1  & -5.5(5)   & 3.43(2)& -1.6(1) & -  \\
            &           & D2 & T.1  &   -9(1)   & 3.48(2)& -2.6(4) & -  \\
            & 20240908  & D1 & L.1  & -10(1)    &  6.1(1)& -1.6(2) & -  \\
            &           & D1 & T.1  &  -8(1)    &  7.6(1)& -1.6(2) & -  \\
            &           & D2 & L.1  & -12(2)    & 2.92(3)& -2.7(4) & -  \\
            &           & D2 & T.1  & -10(1)    & 3.30(3)& -2.9(4) & -  \\
\hline
\end{tabular}
\tablecomments{\footnotesize Mode in Column (3): N = normal mode; An = abnormal mode; Sd = slow drifting; Fd = fast drifting mode; D1= drift mode 1; D2 = drift mode 2.; Wc = central-weak emission mode; Sc = central-strong emission mode}
\end{table}

\begin{table}
\centering
\renewcommand\arraystretch{0.8}
\scriptsize 
\setlength\tabcolsep{4pt}
\caption{Similar to Table~\ref{Tab:moduPars-modeModu}, but for pulsars with nulling, mode changing, and modulation behaviors.}
\label{Tab:moduPars-nullModuMode}
 \begin{tabular}{lllcrrrl}
 \hline
PSR name  & Obs. Date   & Mode  & M.N  & $P_2$ & $P_3$ & $D$ &  Notes  \\
          &             &       &       &       &       &     &         \\
(1)       & (2)         &  (3)  & (4)   & (5)   & (6)   & (7) & (8) \\
\hline
J1610-1322 & 20250816 & N   & W.1  & -13.1(3) & 7.35(5)    & -1.78(5) & \ref{subsec:J1610-1322} \\
           &          & Sd  & W.1  & -12.3(4) &   23(1)    & -0.52(4) & -    \\
           &          & Fd  & W.1  & -12.1(5) &  4.5(1)    &  -2.7(2) & -    \\
J1822+1120 & 20210116 & W   & L.1  & 6(2)     & 3.36(5)    & 1.9(3)   & \ref{subsec:J1822+1120} \\
           &          & W   & T.1  & 6(1)     & 3.32(3)    & 1.6(3)   & -                       \\
           &          & B   & L.1  & -7(1)    & 2.21(2)    & -3(1)    & -                       \\
J1916+1023 & 20200301 & Sd  & W.1  & -8.48(2) & 11.0(9)    & -0.77(6) & \ref{subsec:J1916+1023} \\
           & 20201230 & Sd  & W.1  & -8.65(2) & 11(1)      & -0.77(9) & -                       \\
           &          & Fd  & W.1  & -8.03(1) & 6.8(3)     & -1.19(6) & -                       \\
           & 20210420 & Sd  & W.1  & -8.71(2) & 14(1)      & -0.64(7) & -                       \\
           & 20220202 & Sd  & W.1  & -8.96(2) & 13(1)      & -0.70(8) & -                       \\
J1919+0134 & 20220310 & N   & W.1  & -7.34(1) & 6.8(3)     & -1.08(4) & \ref{subsec:J1919+0134} \\
           &          & Fd  & W.1  & -6.80(1) & 4.5(1)     & -1.51(3) & -    \\
           &          & Sd  & W.1  & -6.68(1) & 11.4(9)    & -0.59(5) & -    \\
           & 20230328 & N   & W.1  & -7.43(1) & 6.8(3)     & -1.09(6) & -    \\
           &          & Fd  & W.1  & -6.88(1) & 4.8(1)     & -1.43(4) & -    \\
           &          & Sd  & W.1  & -6.75(1) & 8.5(5)     & -0.80(5) & -    \\
J1927+1852 & 20200401 & Sd  & W.1  & -8.9(2)  & 6.39(1)    & -1.39(3) & \ref{subsec:nullmodedrift}\\ 
           &          & Fd  & W.1  & -14.6(5) & 2.49(1)    & -5.8(2)  & -    \\
J1941+4320 & 20210808 & Sd  & W.1  & 4.0(2)   & 9.9(2)     & 0.40(2)  & \ref{subsec:J1941+4320}\\
           &          & Fd  & W.1  & 3.11(8)  & 3(1)       & 1.0(4)   & -    \\
J2346-0609 & 20250223 & W   & T.1  & *        & 2.23(1)    & -        & \ref{subsec:J2346-0609} \\
%
\hline
\end{tabular}
\end{table}

\subsubsection{PSR J1954+2923: temporal-only modulation}

PSR J1954+2923 was discovered by \citet{Davies1970} using the Mark I radio telescope at the Jodrell Bank at 408 MHz. 
\citet{Weltevrede2006} has studied weak single pulses observed at the 21~cm band, and suggested the possible drifting behavior of pulses for the two main components. 

The FAST observation was conducted on 20200607 for 15 minutes with a much higher sensitivity. This pulsar has a period of $P=0.4267$ s and a dispersion measurement of $DM=7.9~{\rm cm^{-3}\,pc}$. The pulse sequence in Fig.~\ref{subfig:TP:J1954+2923} does not show subpulse drifting over the whole profile or each of the two primary components. The LRFS and 2DFS analyses of leading and trailing parts in the pulse window show different pulse energy fluctuations, as displayed in Figure~\ref{subfig:fluctu:J1954+2923}, but no drifting feature. The leading part exhibits a wide red-noise-like fluctuation feature, while the trailing part has a distinct fluctuation feature with a centroid around  $1/P_3=0.0746\pm0.0001$, corresponding to $P_3=13.40\pm0.02$ periods. 

\subsubsection{Low-frequency modulation superimposed on subpulse drifting}
\label{subsec:J0058+4950}

Multiple modulation features of some pulsars appear 
different components,  
and some exhibit low-frequency modulation in addition to the fast subpulse modulation feature, probably caused by the rotating subbeam carousel cycle. 
PSR J1857+0057 has been reported to have a low-frequency modulation of about every 50 periods 
\citep{Yan2023_drift}. \citet{Hsu2025} presented the coexistence of periodic drifting subpulses and rapid amplitude modulation 
observed in PSR J1514$-$4834.

From our FAST observations, 4 pulsars, PSRs J0058+4950, J1835-0149g, J1857+0057 and J2340+08, are found to exhibit a low-frequency modulation in addition to subpulse drifting or modulations. The details of PSR J1857+0057 have been reported by \citet{Yan2023_drift}, which has a low-frequency modulation of about every 50 periods on pulse intensity and drift parameters.
For PSRs J0453+1559 and J0621+1002, two profile components show a similar modulation periodicity, but a low-frequency modulation occurs only for one component.

Here we present the results for PSR J0058+4950 in the main text and the data of the others in the Sect.4.

\begin{figure}[htbp]
\centering
\includegraphics[width=0.215\textwidth, angle=0]{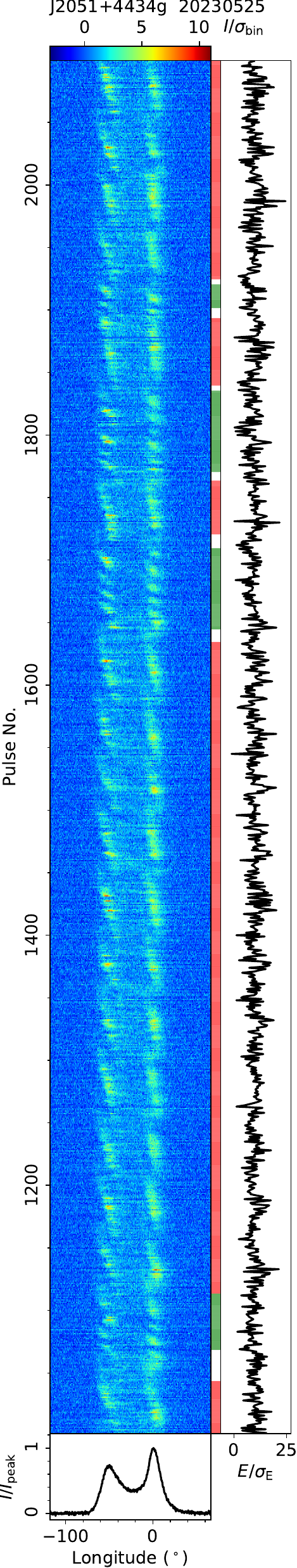}
\includegraphics[width=0.215\textwidth, angle=0]{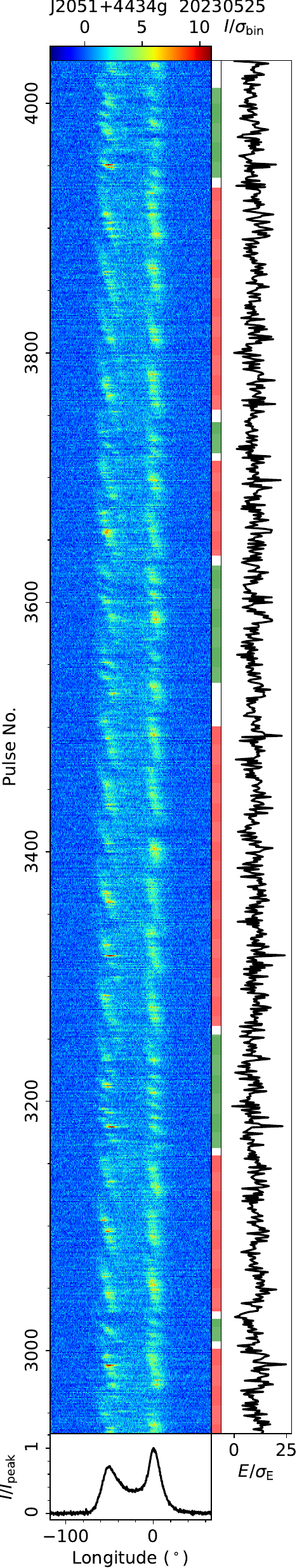}
\caption{Similar to Figure~\ref{subfig:TP:J1848-0023}, but for two segments of single pulse stacks for PSR J2051+4434g from FAST observation on 20230525. The slow-drifting mode and fast-drifting mode are labelled with red and green bars, respectively. The white ranges are segments that cannot be classified. 
\label{subfig:TP:J2051+4434g}
}
\end{figure}

\begin{figure}[htbp]
\centering
\includegraphics[width=0.44\textwidth, angle=0]{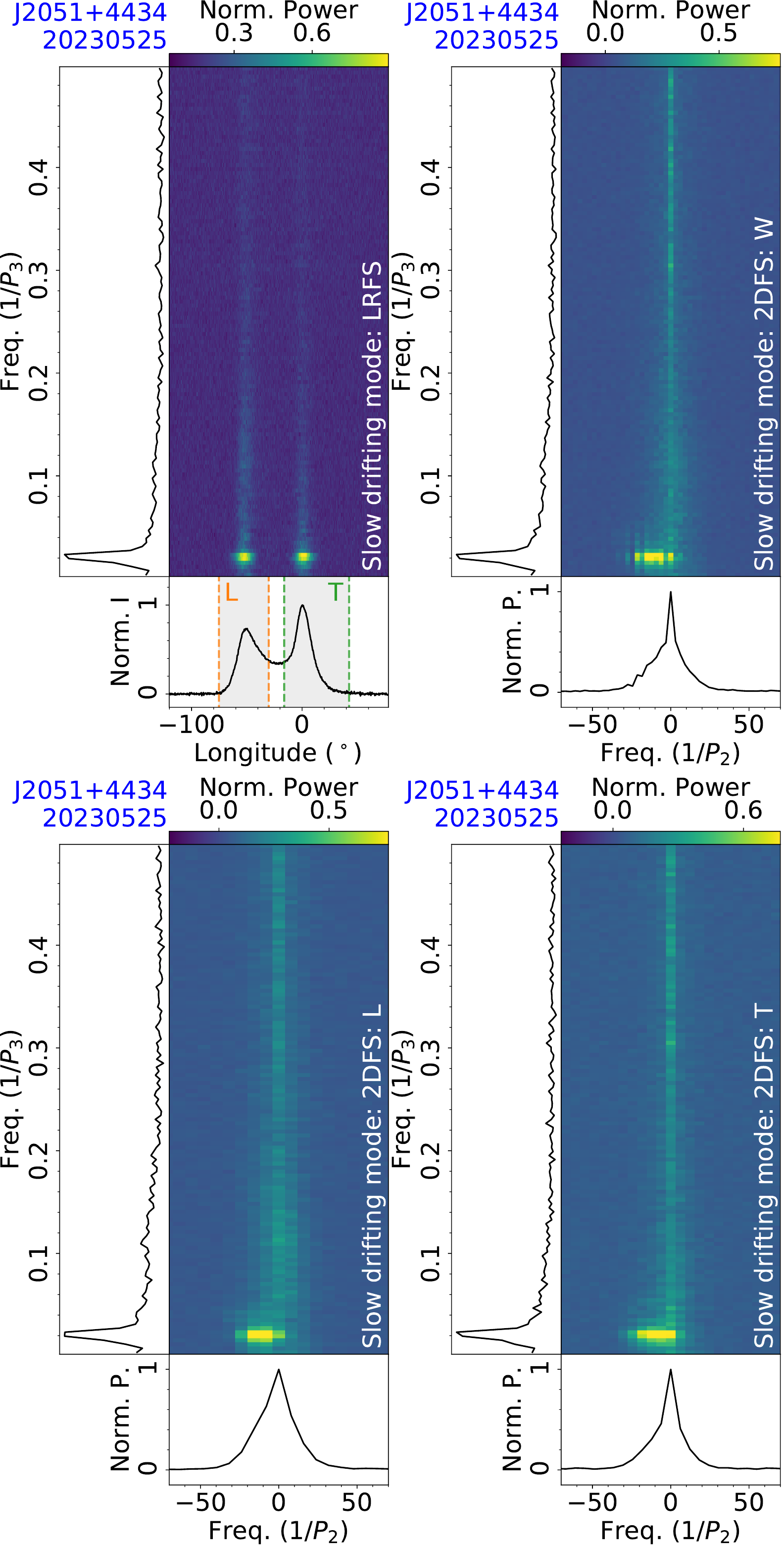}
\caption{Similar to Figure~\ref{subfig:fluctu:J1832+2749}, but for the LRFS and 2DFS for the slow drifting mode of PSR J2051+4434 from FAST observation on 20230525. }   \label{subfig:fluctu:J2051+4434g}
\addtocounter{figure}{-1}
\end{figure}

\begin{figure}[htbp]
\centering
\includegraphics[width=0.44\textwidth, angle=0]{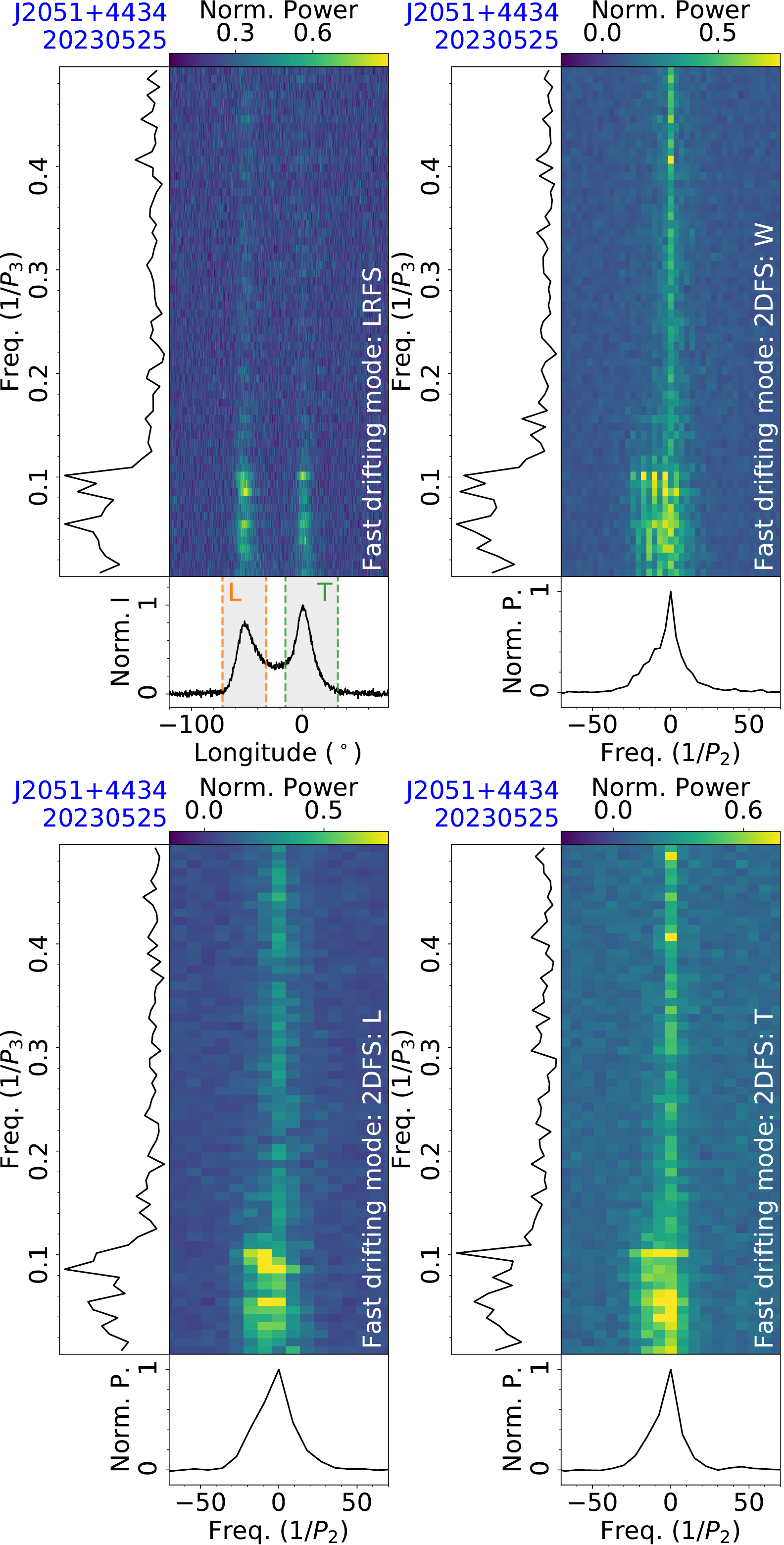}
\caption{Continued for Figure~\ref{subfig:fluctu:J2051+4434g}, but for the fast drifting mode. 
The drifting parameters of the leading ($L$) and trailing ($T$) components are almost identical in each mode.
}  
\end{figure}

\begin{figure}[htbp]
\centering
\includegraphics[width=0.35\textwidth, angle=0]{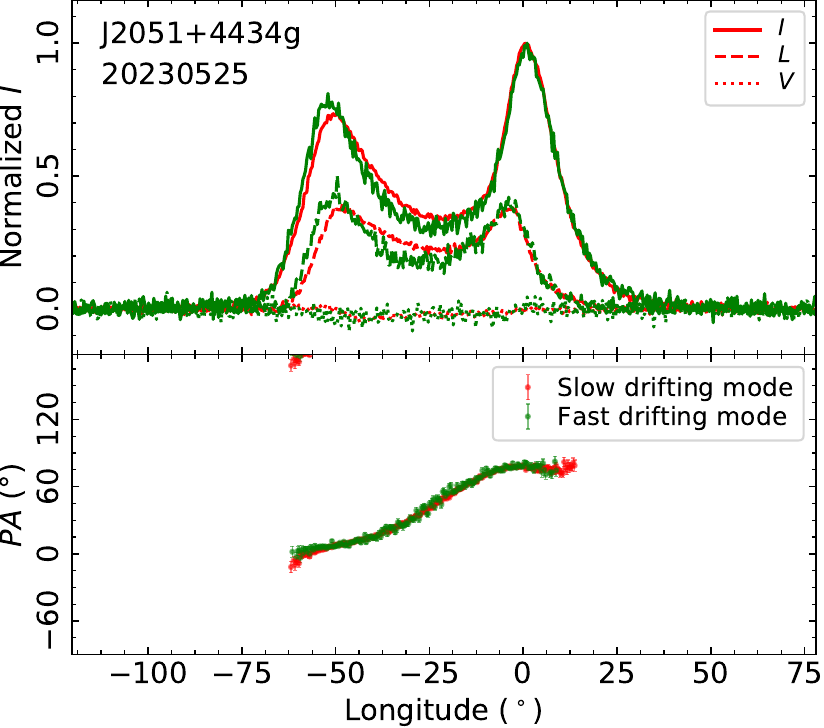}
\caption{Similar to Figure~\ref{subfig:PolModes:J1848-0023}, but for the polarization profiles for slow drifting mode and fast drifting mode of PSR J2051+4434g from FAST observation on 20230525. \label{subfig:PolModes:J2051+4434g}}
\end{figure}

PSR J0058+4950 was discovered by \citet{Stovall2014} using the Green Bank Telescope at 350 MHz. Single pulses are weak and cannot be well detected by other telescopes but FAST. We got the pulsar period $P=0.9961$~s and dispersion measurement $DM=67.1~{\rm cm^{-3}\,pc}$ from FAST data observed on 20201120. The single pulse sequences over the 6-minute FAST observation in Figure~\ref{fig:TP:J0058+4950} show the subpulse driftings and low-frequency intensity modulations. The spectral analyses in Figure~\ref{fig:fluctu:J0058+4950} show a  negatively drifting cycle frequency of $1/P_3=0.404\pm0.001$ and $P_2=-8.7^\circ\pm0.3^\circ$, which is caused by normal subpulse drifting with $P_3=2.48\pm0.01$ periods. Nevertheless, there is a distinct low-frequency modulation feature at $1/P_3= 0.020\pm0.001$, corresponding to $P_3=51\pm1$ periods without phase-shifts. Such a modulation is similar to that of PSR J1857+0057 \citep{Yan2023_drift} and is evident by the repeated patterns in Figure~\ref{fig:TP:J0058+4950}.

\begin{figure}[htbp]
\centering
\includegraphics[width=0.21\textwidth, angle=0]{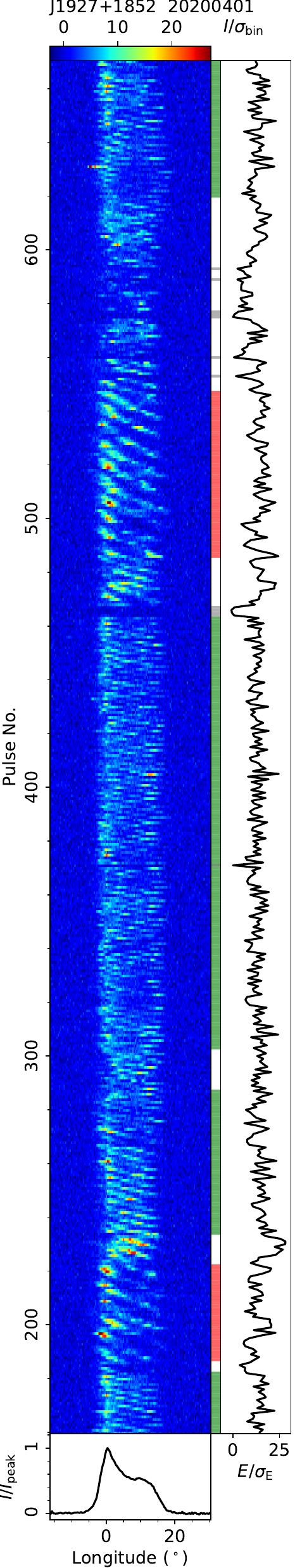}
\includegraphics[width=0.21\textwidth, angle=0]{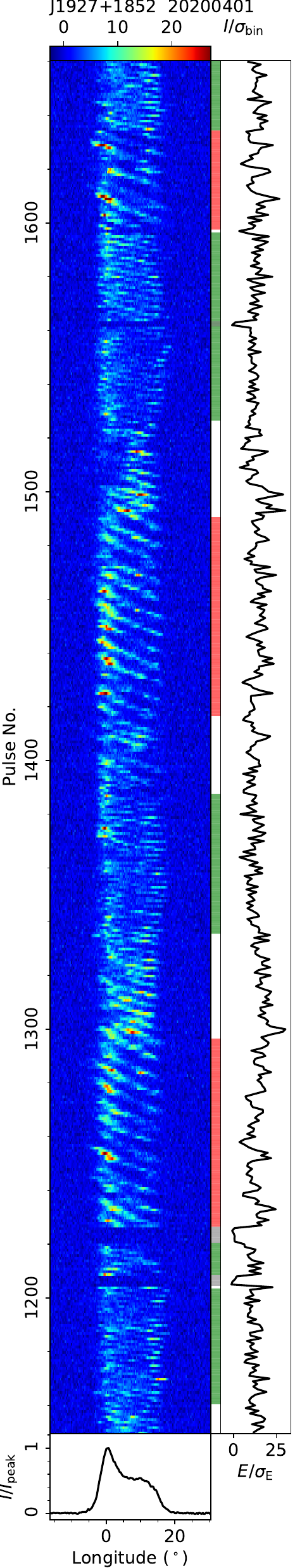}
\caption{Similar to Figure~\ref{subfig:TP:J1848-0023}, but for two segments of pulses of PSR J1927+1852 from FAST observation on 20200401. The slow drift and fast drifting modes are labelled with green and red bars, and nulling durations by grey bars. The white bars indicate the transitions between the two drifting modes with varying drifting rates.
}
\label{subfig:TP:J1927+1852}
\end{figure}

\begin{figure}
\centering
%
\includegraphics[width=0.42\textwidth, angle=0]{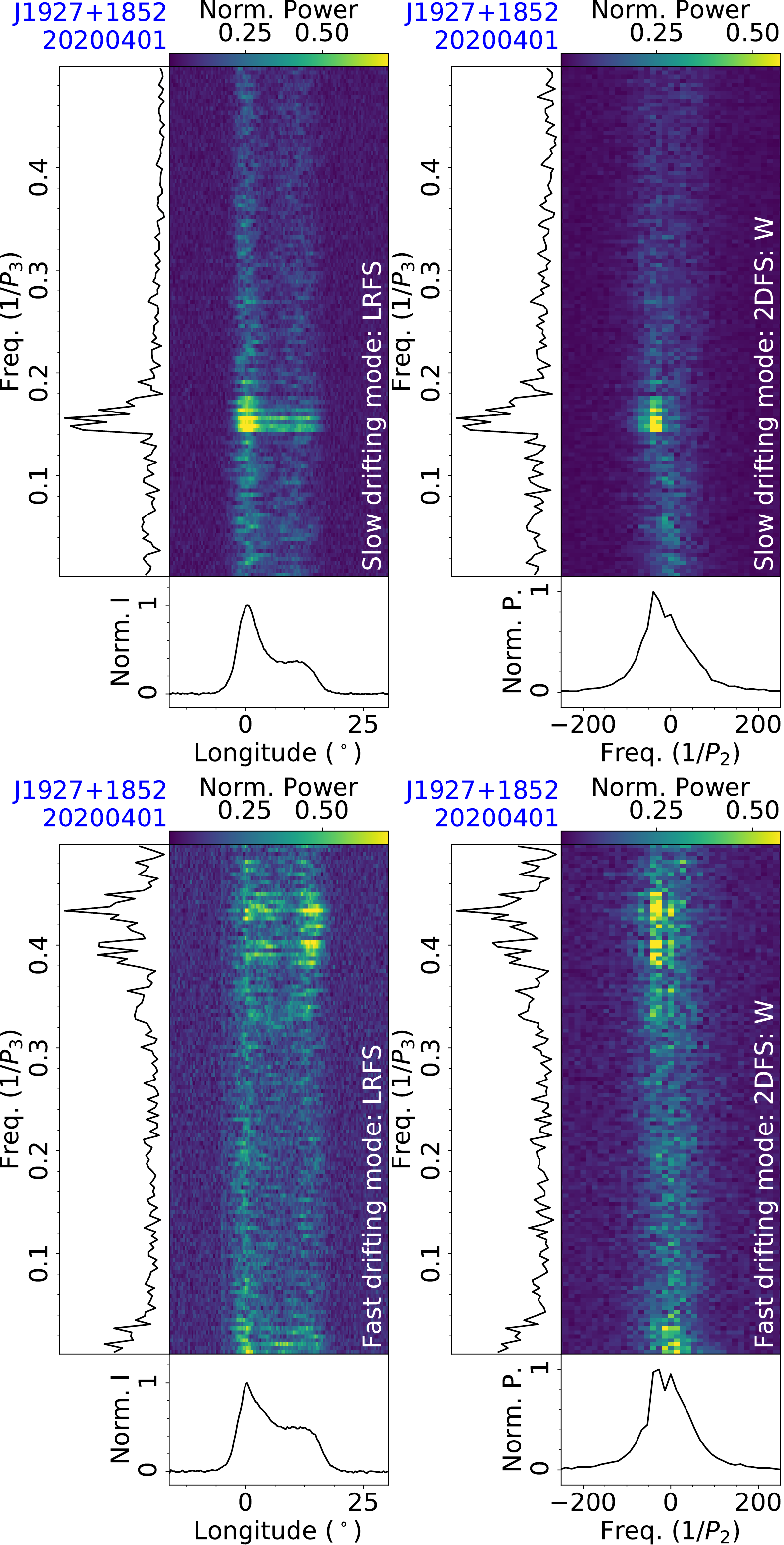}
\vspace{-0.35cm}
\caption{Similar to Figure~\ref{subfig:fluctu:J1832+2749}, but for LRFS and 2DFS of PSR J1927+1852 data observed by FAST on 20200401. The top and bottom panels are related to the slow drifting mode and fast drifting mode. The LRFS 2DFS of the on-pulse phase region are plotted in the left and right panels.
\label{subfig:fluctu:J1927+1852}}
%
\centering
\includegraphics[width=0.4\textwidth, angle=0]{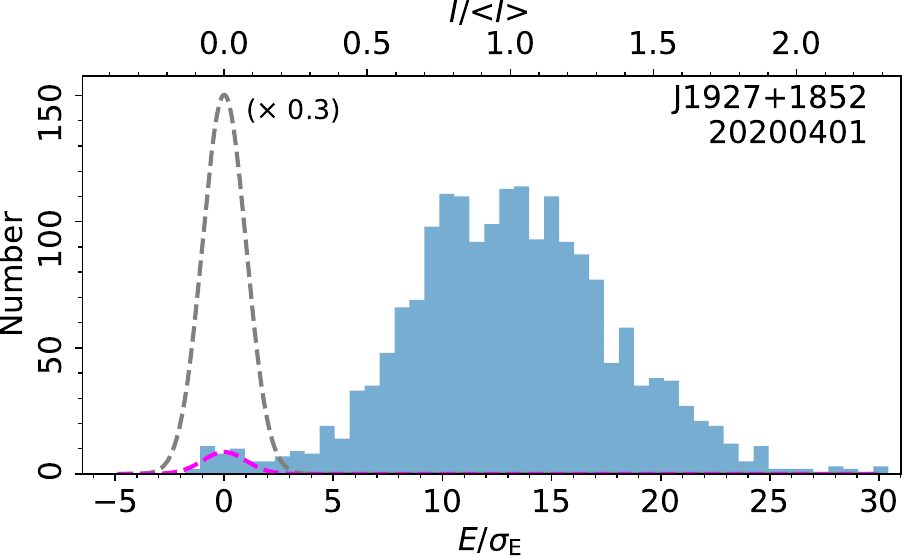}
\vspace{-0.35cm}
\caption{Similar to Figure~\ref{subfig:Hist:J1501-0046}, but for nulling analysis for PSR J1927+1852 using the FAST observation on 20200401.
\label{subfig:Hist:J1927+1852}}
\end{figure}

\begin{figure}
\centering
\includegraphics[width=0.38\textwidth, angle=0]{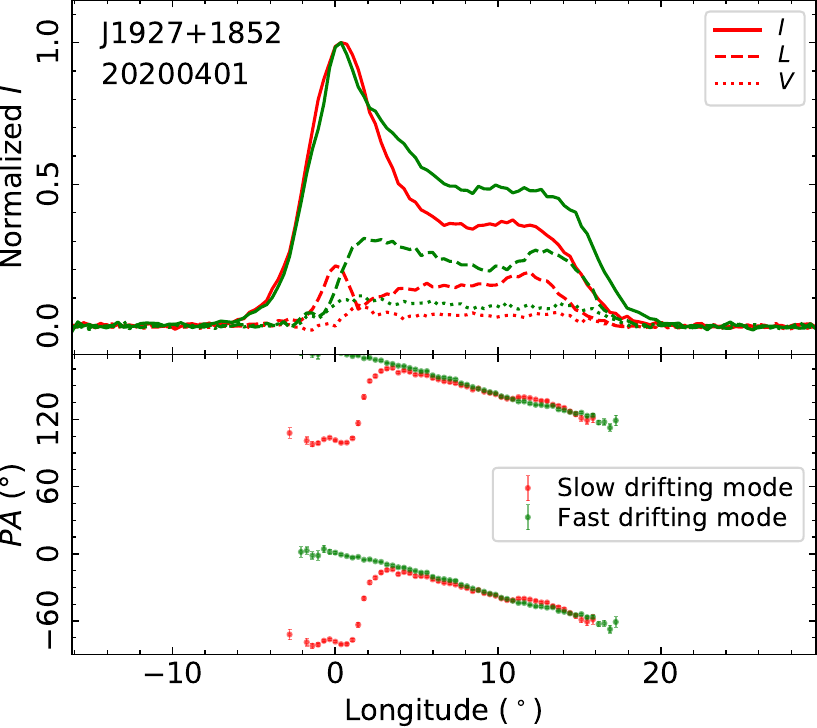}
\caption{Similar to Figure~\ref{subfig:PolModes:J1848-0023}, but for the polarization profiles of the slow drifting mode and fast drifting mode of PSR J1927+1852 observed by FAST on 20200401.
\label{subfig:PolModes:J1927+1852}}
\end{figure}

\subsection{Nulling and subpulse modulation}
\label{subsec:nulldrift}

FAST observations identified 82 pulsars exhibiting both nulling and subpulse modulation behaviors. Their nulling period numbers have been given in Table~\ref{Tab:all}, while the drifting parameters are given in Table~\ref{Tab:moduPars-moduNull}. Among them, the nulling phenomenon is revealed for the first time for 61 pulsars, and subpulse drifting is newly identified for 58 pulsars. 
Here we analyse PSR J1920+1040 with both nulling and subpulse drifting behaviors, as an example. The data of other pulsars are presented in the Sect. 4.

PSR J1920+1040 was discovered in the Parkes multibeam pulsar survey \citep{hfs+04}. With a subintegration time of 10 seconds, \citet{Wang2007} estimated the nulling fraction to be $\sim$50\% from a 2-hour observation by Parkes, but no single pulse and hence no drifting was ever detected after such a subintegration. 

The pulsar was observed three times by FAST: a 15-minute observation on 20220113, and two 5-minute observations on 20210130 and 20240106.  Figure~\ref{subfig:TP:J1920+1040} displays single pulse sequences of three observations. The rotation period and dispersion measurement from FAST observations are $P=2.2158$~s and $DM=307.2~{\rm cm^{-3}\,pc}$, respectively. Both the on-pulse energy histograms in  Figure~\ref{subfig:Hist:J1920+1040} and the period numbers in Table~\ref{Tab:all} give the nulling fraction of this pulsar to be 32$\pm$5\%.  In the emission mode, the subpulses are drifting in the one-component phase window. The LRFS and 2DFS analyses of FAST data on 20220113 (Figure~\ref{subfig:fluctu:J1920+1040}) exhibit the drifting feature for $1/P_2=46.8\pm0.8$ and $1/P_3=0.362\pm0.001$, corresponding to $P_3=2.77\pm0.01$ periods and $P_2=7.7\pm0.1^\circ$. 
%

\subsection{Mode changing and subpulse drifting}
\label{subsec:modedrift}


FAST observations revealed different emission modes for 13 pulsars (see Table~\ref{Tab:all}) with at least one mode for subpulse drifting. We present data of PSR J2051+4434g here, which has two drifting modes. Data of other pulsars are presented in the Sect. 4.

PSR J2051+4434g was discovered in the FAST GPPS survey by \citet{Han2021}, with the longest observation by FAST on 20230525 for 88 minutes. The rotation period and dispersion measurement from this FAST observation are $P=1.3031$~s and $DM=613.02~{\rm cm^{-3}\,pc}$, respectively. 
Two single pulse segments from this observation are shown in Fig.~\ref{subfig:TP:J2051+4434g}, which display two drifting modes. The slow drifting mode dominates and appears in most of the observation time, as labelled by red bars in the figure. The fast drifting mode occupies much shorter durations, as labelled by the green bars. Drifting parameters are derived from features in 2DFS of two drifting modes  (Fig.~\ref{subfig:fluctu:J2051+4434g}), as listed in Table~\ref{Tab:all}. 
For the fast drifting mode, the modulation frequencies of the leading component are $1/P_3=0.063\pm0.001$ and $1/P_2=-8.7\pm0.5$, yielding $P_3=15.9\pm0.2$ periods and $P_2=-41\pm2^\circ$. The trailing component displays values of $1/P_3=0.060\pm0.001$ and $1/P_2=-7.7\pm0.4$, corresponding to $P_3=16.8\pm0.2$ periods and $P_2=-47\pm2^\circ$.  
The slow drifting mode is characterized by longer temporally modulated periodicities: the leading component has $1/P_3=0.0190\pm0.0003$ and $1/P_2=-12.3\pm0.6$, yielding $P_3=52.6\pm0.8$ periods and $P_2=-29\pm1^\circ$; and the trailing component has a drifting feature of $1/P_3=0.0187\pm0.0002$ and $1/P_2=-9.6\pm0.4$, which correspond to $P_3=53.4\pm0.7$ periods and $P_2=-37\pm1^\circ$. As shown in Figure~\ref{subfig:PolModes:J2051+4434g}, the averaged polarization profiles are almost identical for the two drifting modes.

\subsection{Pulsars with nulling, mode changing and subpulse drifting}
\label{subsec:nullmodedrift}


From FAST observations, we found that 7 pulsars, PSRs J1610+1322, J1822+1120, J1919+0134, J1916+1023, J1927+1852, J1941+4320 and J1941+4320, have nulling, mode changing and subpulse drifting phenomena revealed in the pulse sequence. Here we present the features of PSR J1927+1852 and leave the other 6 pulsars in the Sect. 4. 

PSR J1927+1852 was discovered by the Arecibo telescope \citet{Hulse1975}. \citet{Song2023} reported two drifting features for components. J1927+1852 was observed by the FAST on 20200401 for 15 minutes, and we got the rotation period of $P=0.4827$~s and a dispersion measurement of $DM=264.5~{\rm cm^{-3}\,pc}$ from our observation. 
The single pulse sequences from FAST observation show much more detail, as shown in Figure~\ref{subfig:TP:J1927+1852}. In addition to the two regular drifting modes, the slow and fast drifting modes, the nulling or varying drifting emerges for some periods as the transitions between the two drifting modes. The drift bands of the slow drifting mode are very clear, showing a dominating leading profile component. The fluctuation analyses of the LRFS and 2DFS in Fig.~\ref{subfig:fluctu:J1927+1852} give the drifting feature of  $1/P_3=0.1565\pm0.0004$ and $1/P_2=-40.4\pm0.7$, corresponding to $P_3=6.39\pm0.01$ periods and $P_2=-8.9\pm0.2^\circ$. However, the drift bands of the fast drifting mode are less stable, sometimes with emission in the central profile phase screened for some periods. The fluctuation analysis of the LRFS and 2DFS 
give a wide frequency features around $1/P_3=0.402$ and $1/P_2=-24.6$, corresponding to $P_3=2.49\pm0.01$ periods and $P_2=-14.6\pm0.5^\circ$.

%
There are transition periods between the fast and slow drifting modes, such as the period numbers 225 to 235,  1294 to 1304, and 1492 to 1518 in Figure~\ref{subfig:TP:J1927+1852}. The drift bands appear to change the drifting rates gradually from the slow drifting mode to the fast drifting mode. 
%
In some periods, the subpulses are very disordered or chaotic, such as between the period numbers of 
280 to 300, 560 to 600, and 1520 to 1530. 
%

Generally, nulling appears after the end of fast drifting mode or for reorganizing the disordered drifting subpulses, for example, around period numbers of 465, 1205 and 1563.
To analyse the nulling phenomenon, we plot the on-pulse energy distribution from the FAST observation conducted on 20200411, as shown in Figure~\ref{subfig:Hist:J1927+1852}. The distribution around the zero energy indicates the nulling, with a nulling fraction of about $1.7\pm0.2$\%. 
%


Fig.~\ref{subfig:PolModes:J1927+1852} shows the average polarization profiles for the 
slow and fast drifting modes. The leading component of the slow drifting mode is much brighter than that in the fast drifting mode, and its linear polarization angles are in the 
orthogonal polarization modes (OPM) with a $90^{\circ}$ jump from the 
smooth average PA curve of the other part. In contrast, the PA curve of the fast drifting mode is a continuously monotonically decreasing function over all profile components. This implies  
the correlation between emission modes and orthogonal polarization modes, as previously found in PSRs J2006-0807 and J1239+2453 \citep{Basu2019,Smith2013}. 


\section{Details of individual pulses for 366 pulsars} 
\label{sec:detail}


In addition to the examples in the last Section, we present details for prominent features of individual pulses of 366 pulsars here.  
Hereafter, except for some specific cases remarked in the captions, figures in this section have the same format as in the previous section, i.e. all single pulse sequences are plotted in the same form as Fig.~\ref{subfig:TP:J1501-0046}, and the pulse energy distribution in the same form as Fig.~\ref{subfig:Hist:J1501-0046}, and the LRFS and 2DFS analyses in the same form as Fig.~\ref{subfig:fluctu:J1832+2749}, polarization profiles in the same form as Fig.~\ref{subfig:PolModes:J1848-0023}. Relevant parameters have been presented in the tables for various features in Sect. 3.

\begin{figure}[hbpt]
\centering
\includegraphics[width=0.22\textwidth, angle=0]{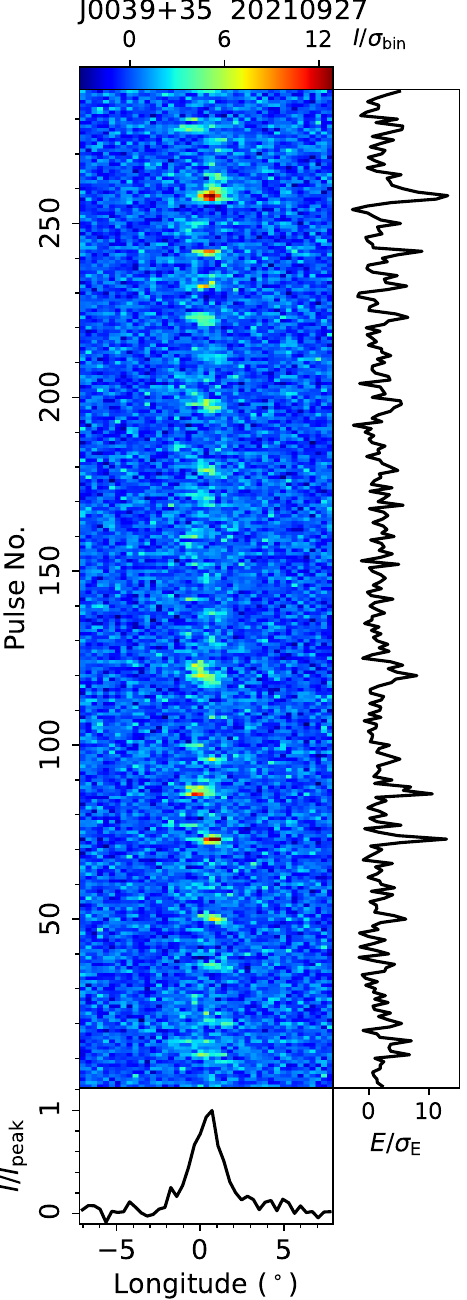}
\includegraphics[width=0.22\textwidth, angle=0]{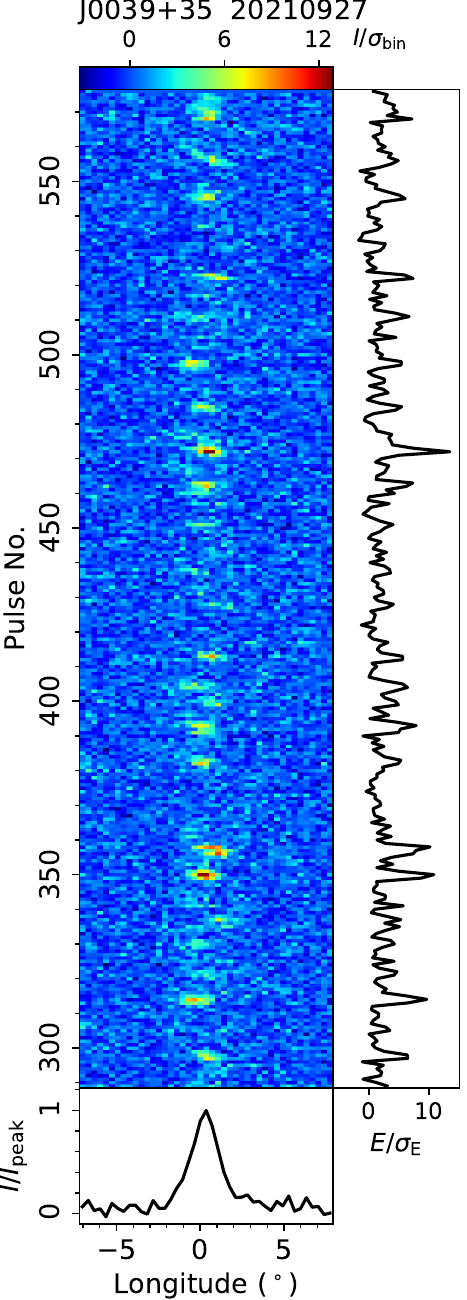}
\figcaption{Single pulse sequences of PSR J0039+35 from the FAST observation on 20210927.
\label{subfig:TP:J0039+35}}
\end{figure}

\begin{figure}[hbpt]
\centering
\includegraphics[width=0.44\textwidth, angle=0]{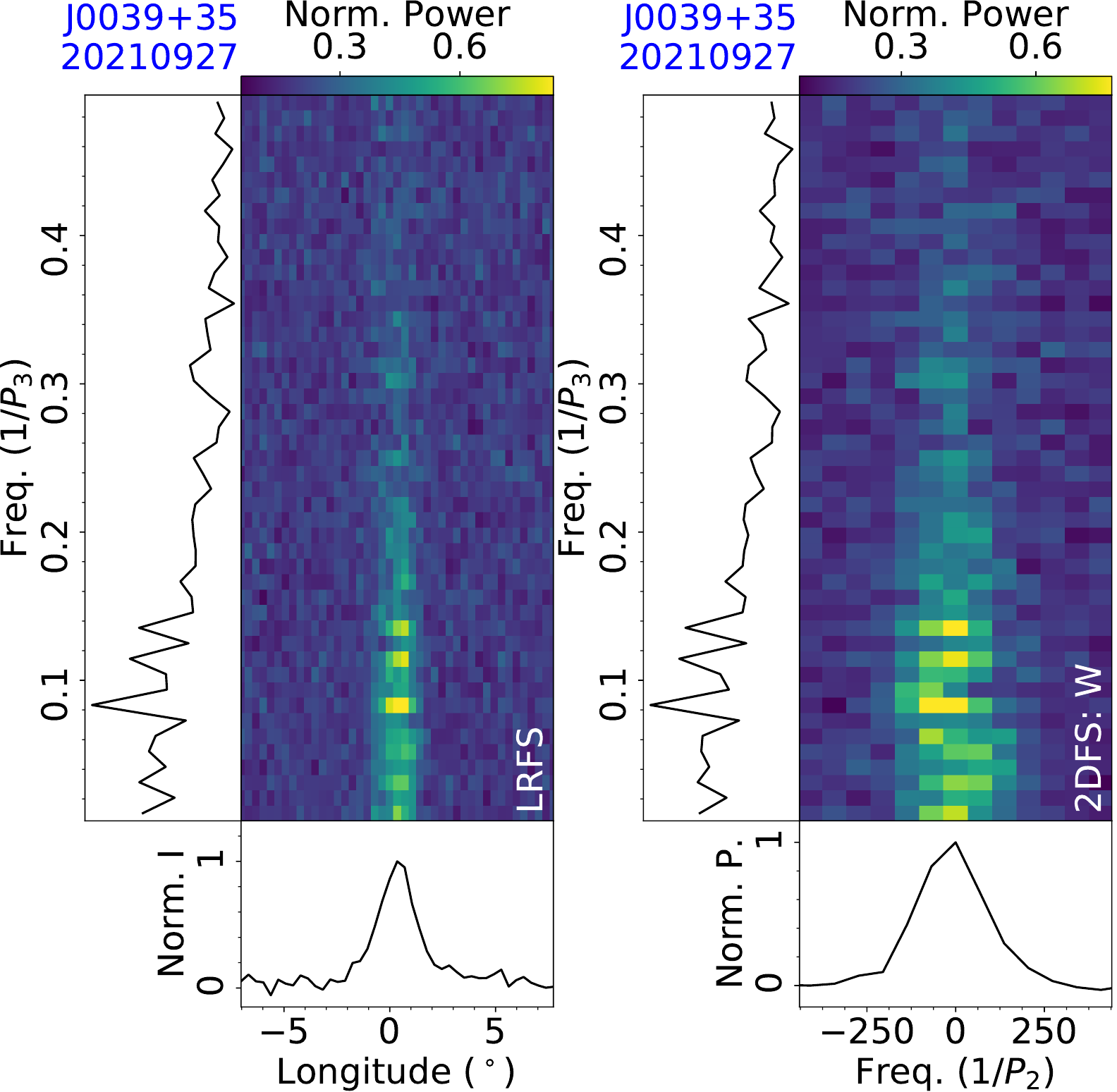}
\figcaption{Fluctuation analysis of PSR J0039+35 from the FAST observation on 20210927, with LRFS and 2DFS for the whole pulse phase range in a mean pulse profile.
\label{subfig:fluctu:J0039+35}}
%
\centering
\includegraphics[width=0.22\textwidth, angle=0]{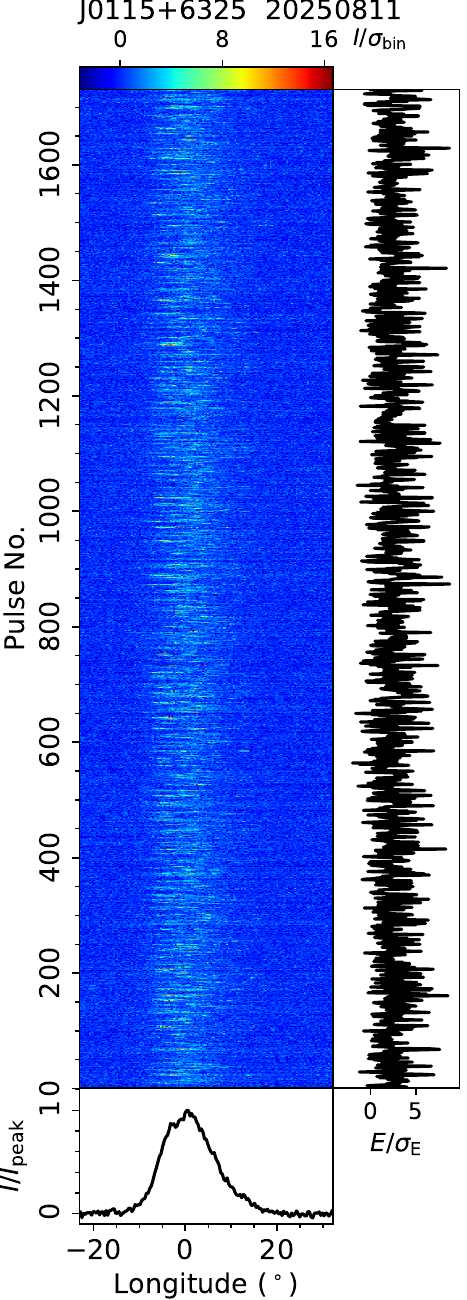}
\includegraphics[width=0.22\textwidth, angle=0]{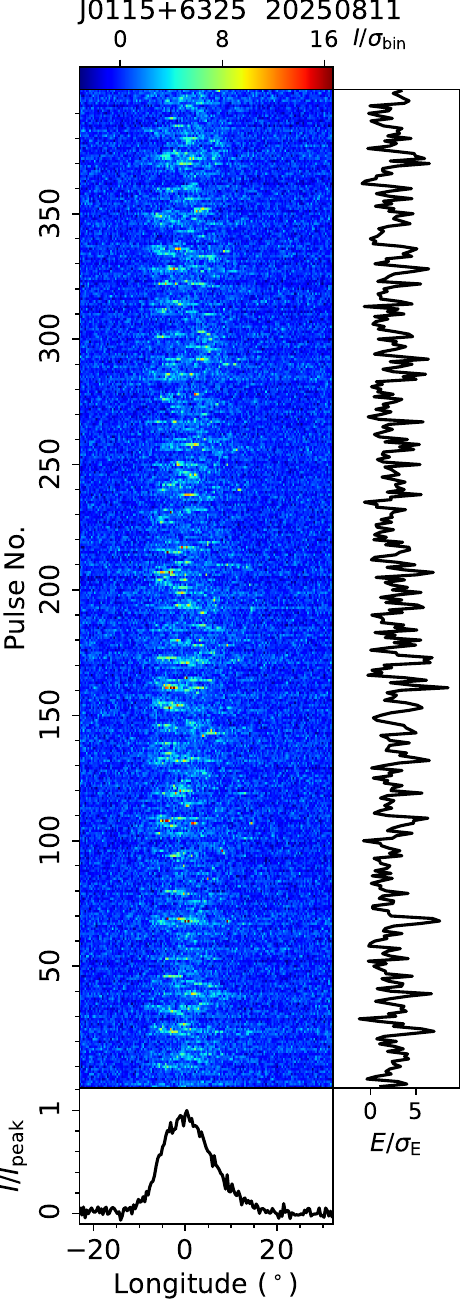}
\figcaption{Single pulse sequence of PSR J0115+6325 from the FAST observation on 20250811, and a zoomed-in view of pulses No.1-400.
\label{subfig:TP:J0115+6325}}
\end{figure}

\begin{figure}[hbpt]
\centering
\includegraphics[width=0.44\textwidth, angle=0]{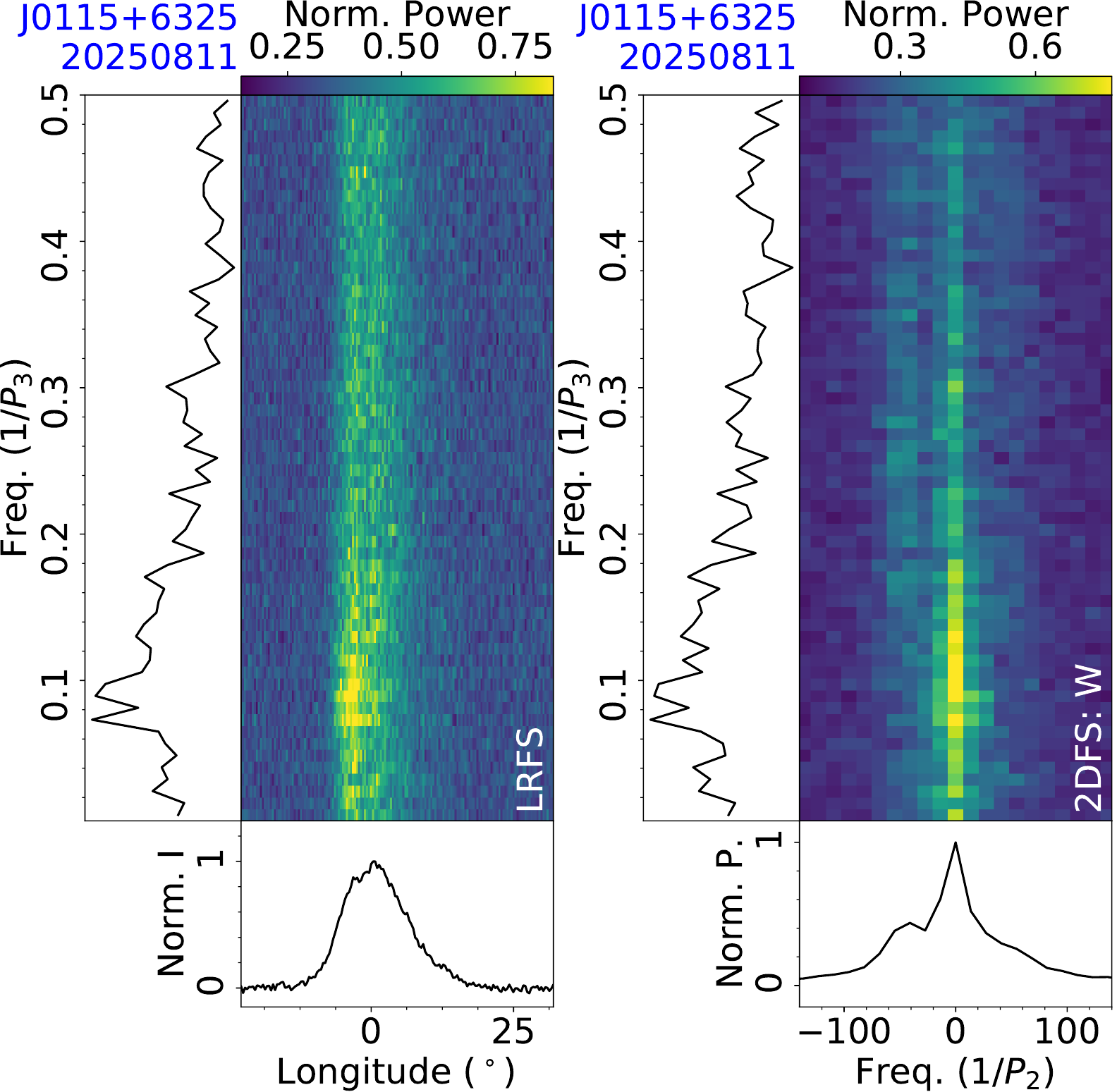}
\figcaption{Fluctuation analysis of PSR J0115+6325 from the FAST observation on 20250811, with LRFS and 2DFS for the whole pulse phase range in a mean pulse profile.
\label{subfig:fluctu:J0115+6325}}
\end{figure}

\subsection{J0039+35}
\label{subsec:J0039+35}

PSR J0039+35 was discovered in the LOFAR Tied-Array All-Sky Survey (LOTAAS) \citep{Sanidas2019}.

This pulsar was observed by FAST on 20210927 for 5 minutes, deriving a rotation period $P=0.5367$~s and a dispersion measure $D\!M=53.0~{\rm cm^{-3}\,pc}$. Single pulse sequences in Fig.~\ref{subfig:TP:J0039+35} and fluctuation spectra in Fig.~\ref{subfig:fluctu:J0039+35} illustrate the existence of subpulse drifting behavior. The negative drift feature in LRFS and 2DFS is distributed over a wide temporal range, which suggests a non-systematic modulation. The centroid of the drift feature in 2DFS is characterized by frequencies of $1/P_3=0.096\pm0.002$ and $1/P_2=-11\pm3$, corresponding to periodicities of $P_3=10.4\pm0.2$ periods and $P_2=-32\pm10$ degrees.

\begin{figure}[hbpt]
\centering
\includegraphics[width=0.22\textwidth, angle=0]{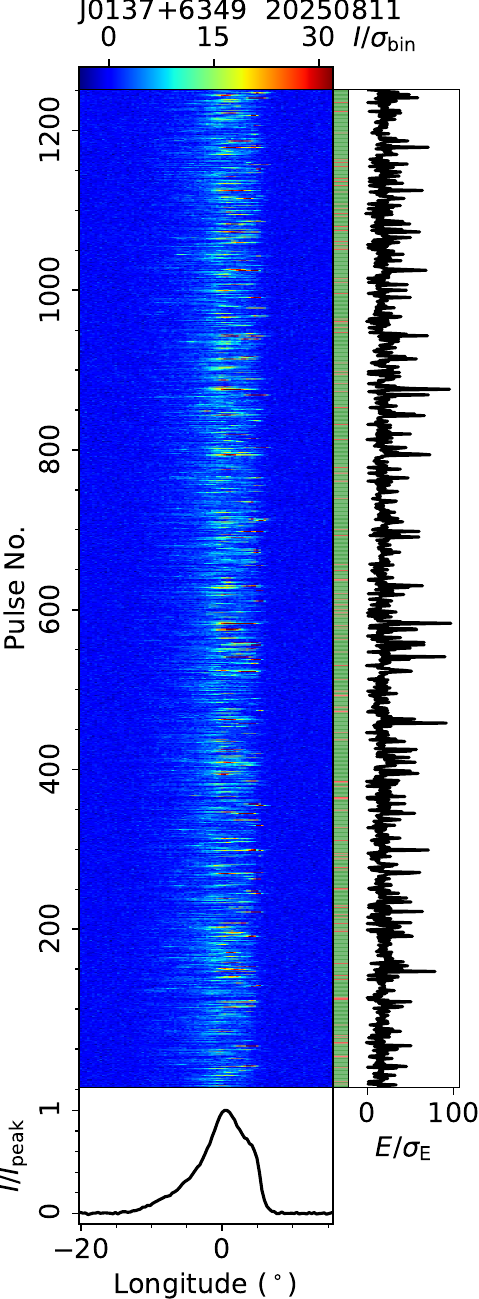}
\includegraphics[width=0.22\textwidth, angle=0]{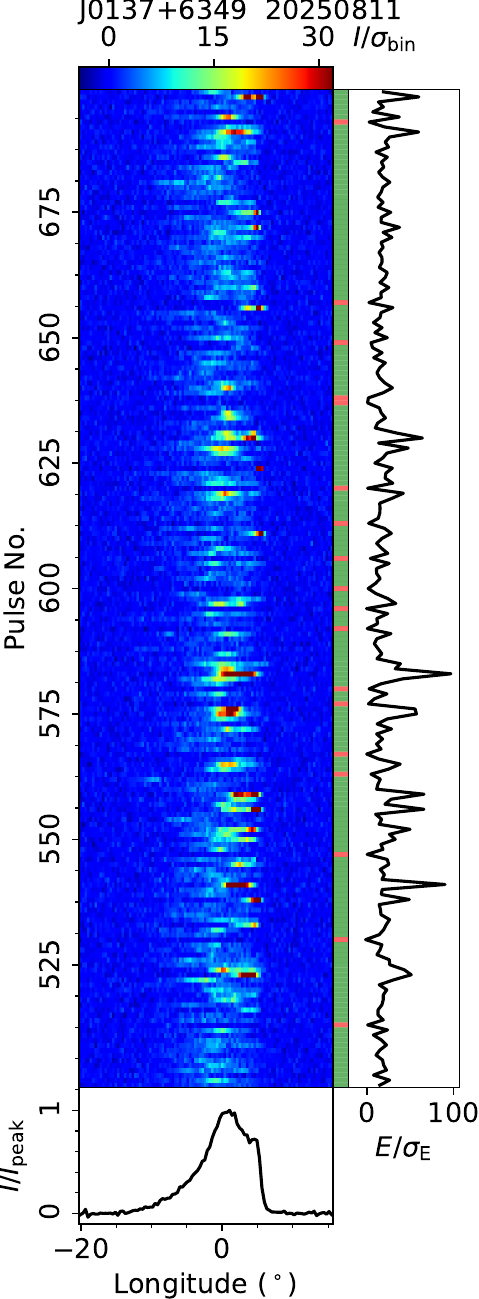}
\figcaption{Single pulse sequence of PSR J0137+6349 from the FAST observation on 20250811, and a zoomed-in view of pulses No.500-700.
\label{subfig:TP:J0137+6349}}
\end{figure}

\begin{figure}[hbpt]
\centering
\includegraphics[width=0.39\textwidth, angle=0]{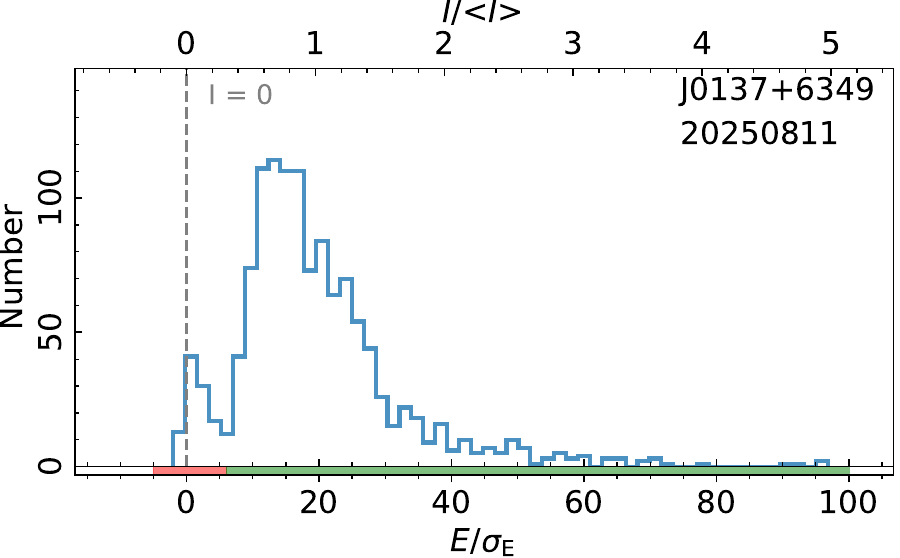}
\figcaption{On-pulse energy histogram of individual pulses of PSR J0137+6349 from the observation on 20250811.
\label{subfig:Hist:J0137+6349}}
\end{figure}

\begin{figure}[hbpt]
\centering
\includegraphics[width=0.39\textwidth, angle=0]{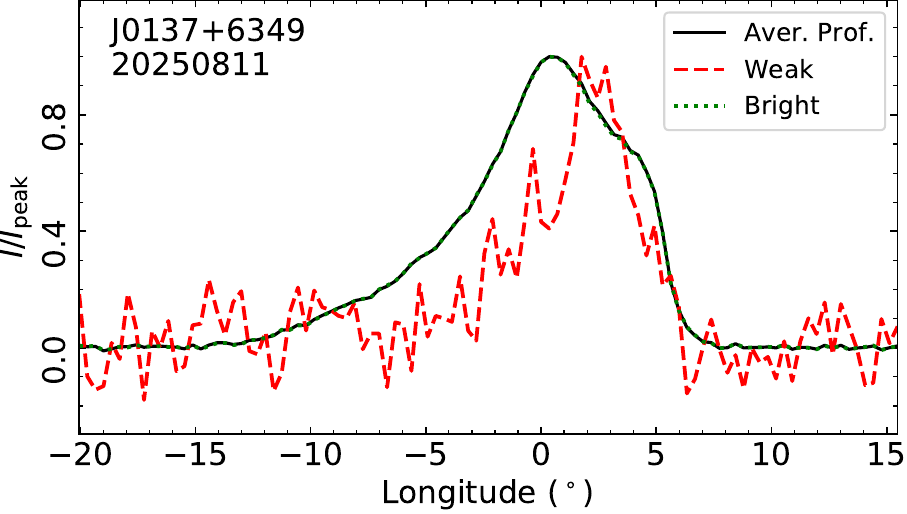}
\figcaption{Mean profiles for the weak and bright emission modes of PSR J0137+6349 from the observation on 20250811. Profiles are normalized by their respective peaks. \label{subfig:ProfModes:J0137+6349}}
\end{figure}

\subsection{J0115+6325}
\label{subsec:J0115+6325}

PSR J0115+6325 was discovered in the LOFAR Tied-Array All-Sky Survey (LOTAAS) \citep{Sanidas2019}.

This pulsar was observed by FAST on 20250811 for 15 minutes, yielding a rotation period $P=0.5214$~s and a dispersion measure $D\!M=65.1~{\rm cm^{-3}\,pc}$. The single pulse sequence and a zoomed-in view of pulses No.1-400 are shown in Fig.~\ref{subfig:TP:J0115+6325}. From fluctuation spectra in Fig.~\ref{subfig:fluctu:J0115+6325}, there are a negative drift feature and a temporal modulation feature in 2DFS. For the negative drift feature, the frequency $1/P_3$ ranges widely from 0.14 to 0.5. The centroid frequencies of the drift feature are estimated to be $1/P_3=0.28\pm0.01$ and $1/P_2=-48\pm1$, corresponding to drift periodicities of $P_3=3.6\pm0.1$ periods and $P_2=-7.5\pm0.1$ degrees. The centroid frequency of the modulation feature is $1/P_3=0.100\pm0.002$, yielding $P_3=10.0\pm0.2$.

\begin{figure}[hbpt]
\centering
\includegraphics[width=0.22\textwidth, angle=0]{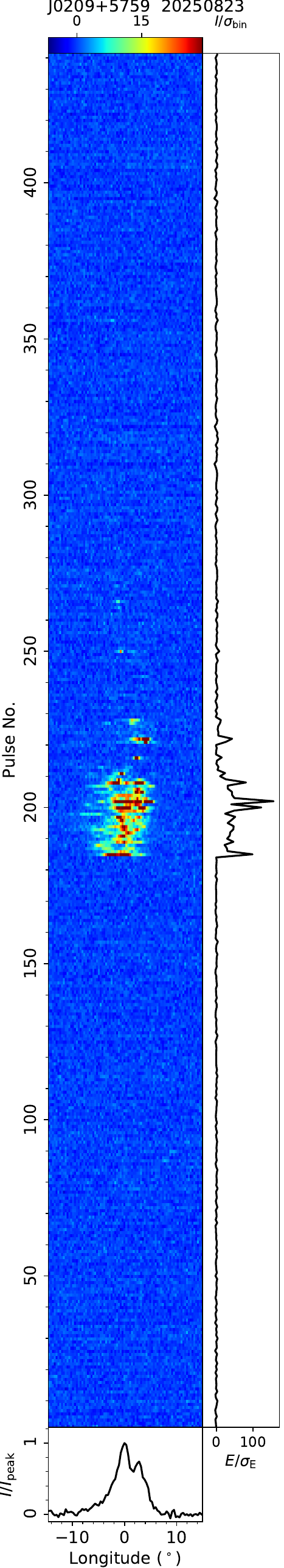}
\includegraphics[width=0.22\textwidth, angle=0]{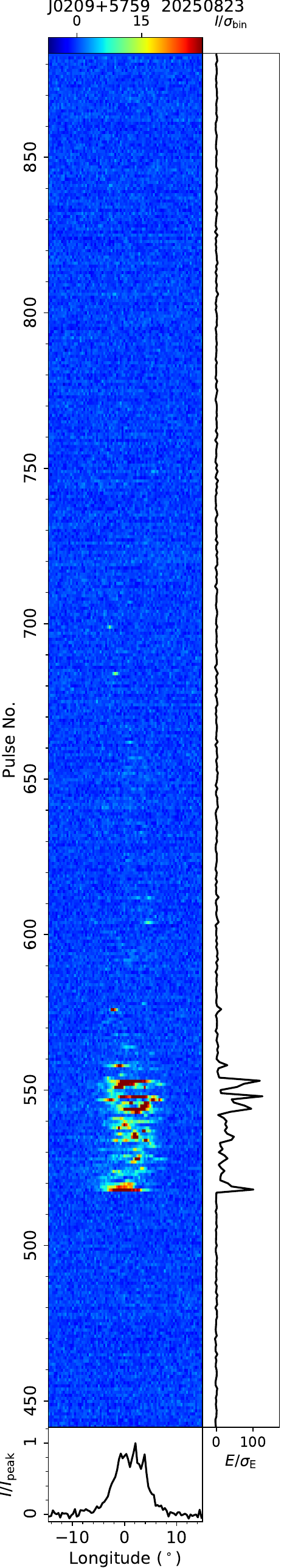}
\figcaption{Single pulse sequences of PSR J0209+5759 from the FAST observation on 20250823.
\label{subfig:TP:J0209+5759}}
\end{figure}

\begin{figure}[hbpt]
\centering
\includegraphics[width=0.39\textwidth, angle=0]{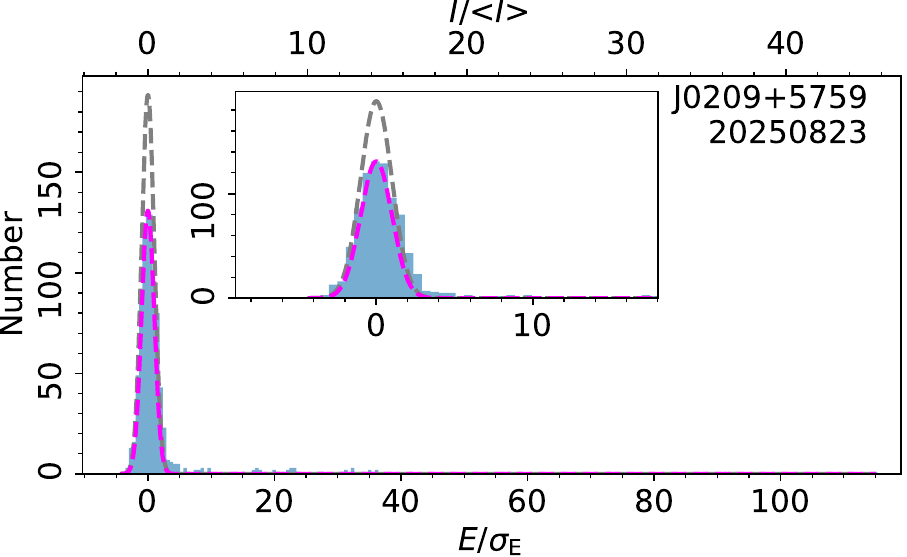}
\figcaption{On-pulse energy histogram of individual pulses of PSR J0209+5759 from the FAST observation on 20250823. The inset provides a view of the x‑axis region from -9 to 15.
\label{subfig:Hist:J0209+5759}}
\end{figure}

\begin{figure}[hbpt]
\centering
\includegraphics[width=0.22\textwidth, angle=0]{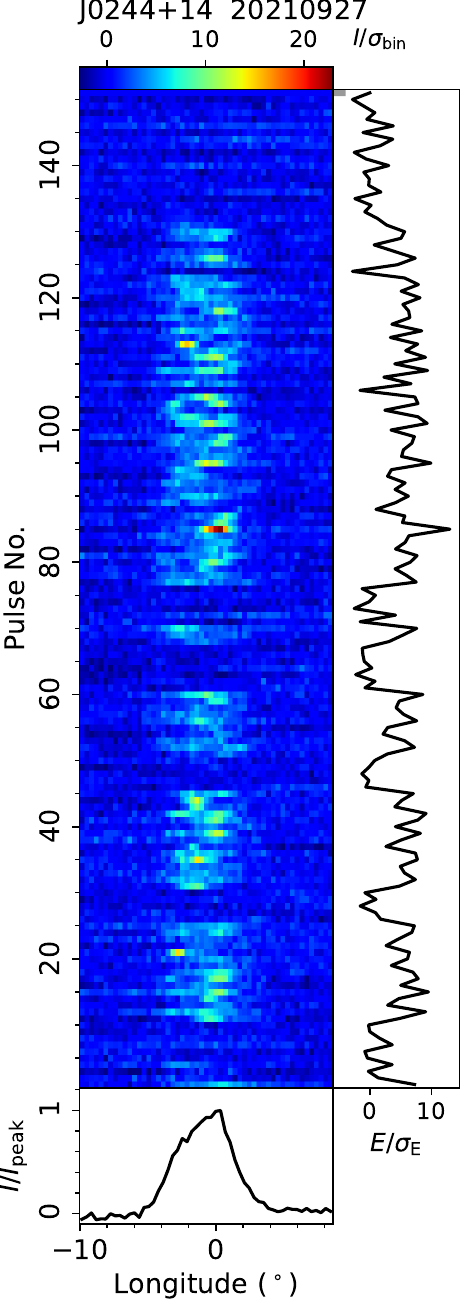}
\figcaption{Single pulse sequence of PSR J0244+14 from the FAST observation on 20210927. \label{subfig:TP:J0244+14}}
\end{figure}

\begin{figure}[hbpt]
\centering
\includegraphics[width=0.39\textwidth, angle=0]{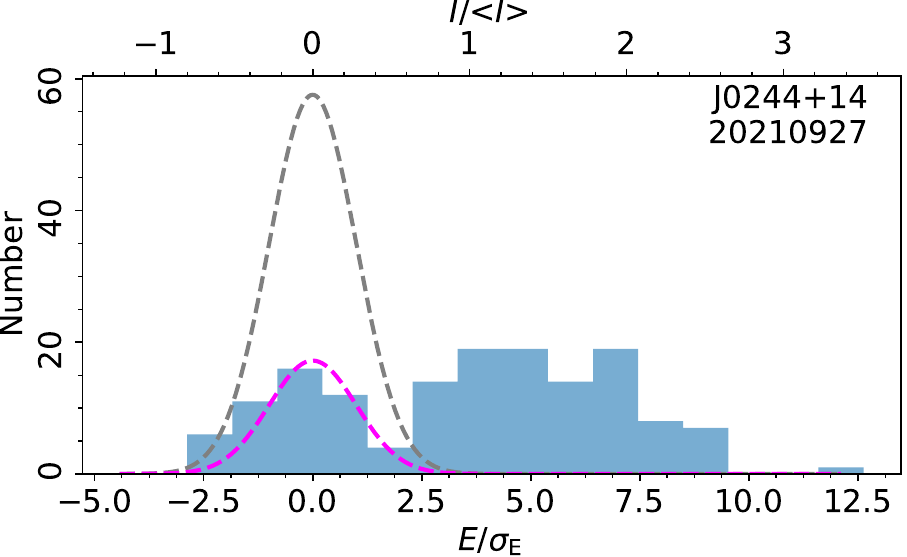}
\figcaption{On-pulse energy histogram of individual pulses of PSR J0244+14 from the FAST observation on 20210927. \label{subfig:Hist:J0244+14}}
\end{figure}

\subsection{J0137+6349}
\label{subsec:J0137+6349}

PSR J0137+6349 was discovered by the Green Bank telescope \citep{Stovall2014}.

This pulsar was observed by FAST on 20250811 for 15 minutes, deriving a rotation period $P=0.7179$~s and a dispersion measure $D\!M=285.1~{\rm cm^{-3}\,pc}$. The single pulse sequence and a zoomed-in view of pulses No.500-700 are shown in Fig.~\ref{subfig:TP:J0137+6349}. 
The on-pulse integral energy histogram in Fig.~\ref{subfig:Hist:J0137+6349} shows a distribution centered near zero energy with a slightly positive mean, indicating the weak emission mode of PSR J0137+6349. Single pulses of weak and bright emission modes are distinguished from the energy histogram, labeled using red and green. Mean profiles of two modes are displayed in Fig.~\ref{subfig:ProfModes:J0137+6349}.

\subsection{J0209+5759}
\label{subsec:J0209+5759}

PSR J0209+5759 was discovered by \citet{Good2021} with CHIME/FRB and reported to display a high degree of nulling.

This pulsar was observed by FAST on 20250823 for 16 minutes, and a rotation period $P=1.0638$~s and a dispersion measure $D\!M=55.3~{\rm cm^{-3}\,pc}$ were determined. From single pulse sequences in Fig.~\ref{subfig:TP:J0209+5759}, weak emission is observed following bright single pulses, for instance, in pulses No. 570-700. 
The nulling fraction of this observation is estimated from the on-pulse energy histogram (Fig.~\ref{subfig:Hist:J0209+5759}), which is 69.4$\pm$3.9\%. 
More observations are required to analyze the properties of the weak emission and its relationship with nulling and bright states.

\begin{figure}[hbpt]
\centering
\includegraphics[width=0.22\textwidth, angle=0]{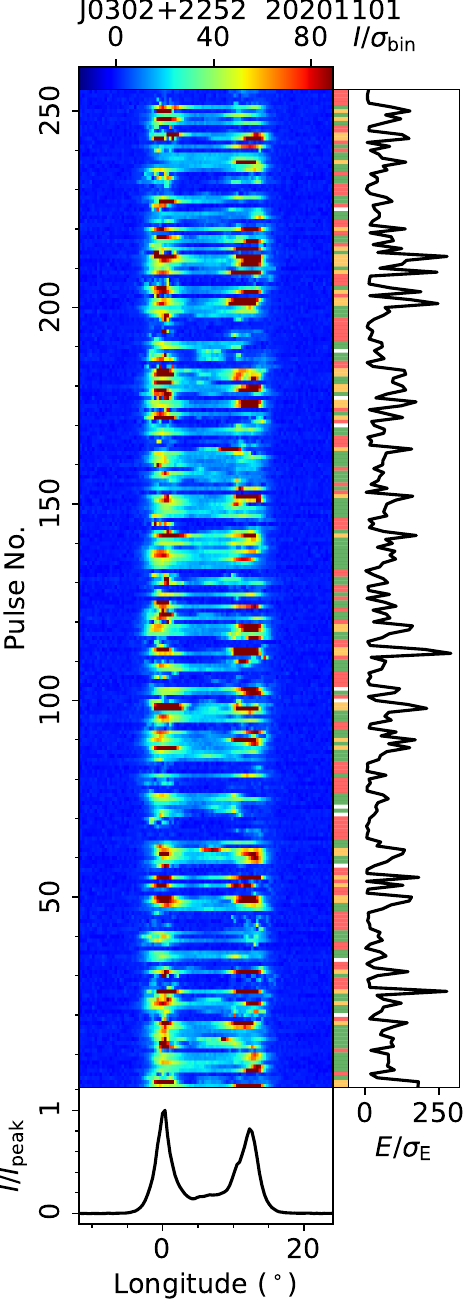} 
\includegraphics[width=0.22\textwidth, angle=0]{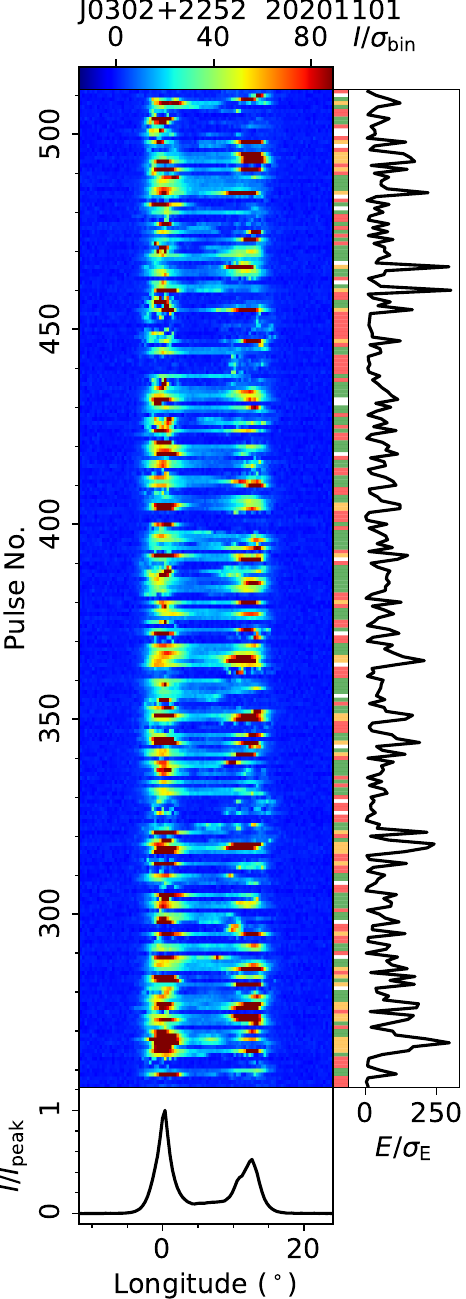}
\figcaption{Single pulse sequences of PSR J0302+2252 from the FAST observation on 20201101. \label{subfig:TP:J0302+2252}}
\end{figure}

\begin{figure}[hbpt]
\centering
\includegraphics[width=0.39\textwidth, angle=0]{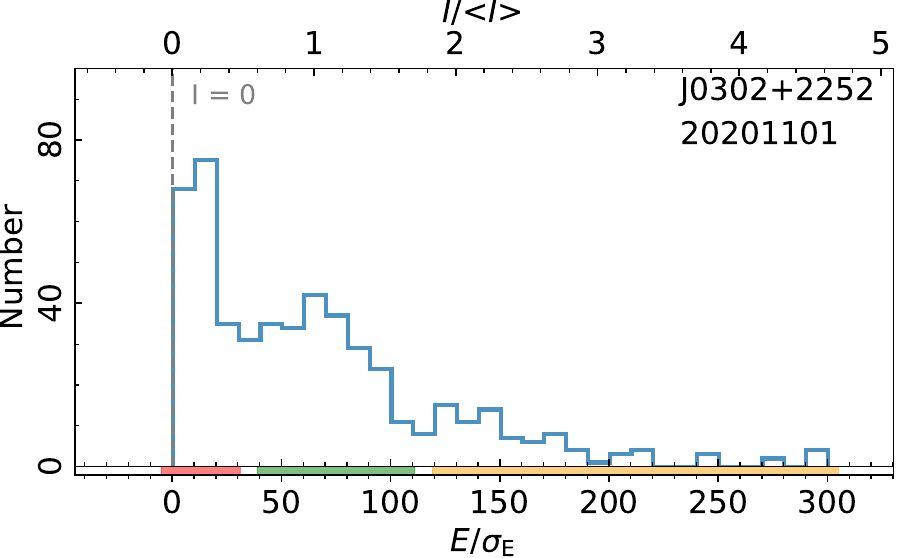}
\figcaption{On-pulse energy histogram of individual pulses of PSR J0302+2252 from the observation on 20201101. \label{subfig:Hist:J0302+2252}}
\end{figure}

\begin{figure}[hbpt]
\centering
\centering
\includegraphics[width=0.37\textwidth, angle=0]{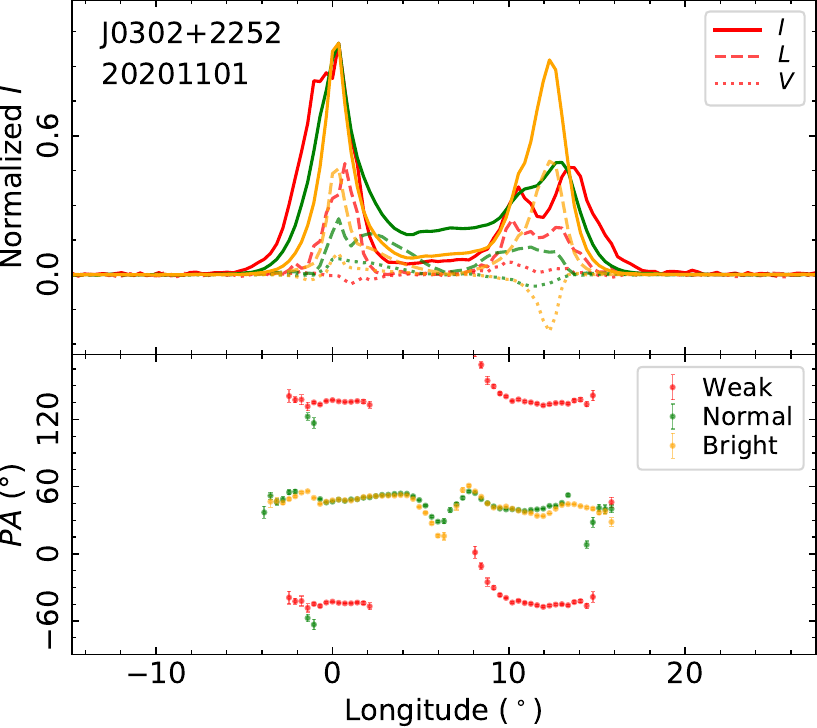}
\figcaption{Averaged polarization profiles (the top panel) for the slow drifting mode and fast drifting mode of PSR J0302+2252 from the observation on 20201101, as well as the average PA curves (the bottom panel). Profiles in the top panel are normalized to their respective peaks.
\label{subfig:PolModes:J0302+2252}}
\end{figure}

\begin{figure}[hbpt]
\centering
\includegraphics[width=0.22\textwidth, angle=0]{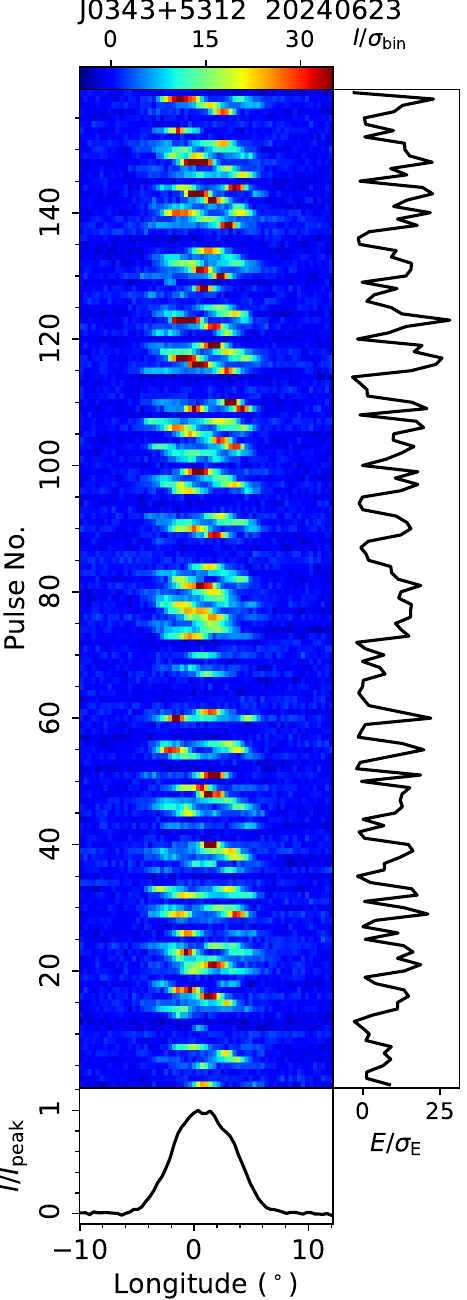}
\figcaption{Single pulse sequence of PSR J0343+5312 from the FAST observation on 20240623.
\label{subfig:TP:J0343+5312}}
\end{figure}

\begin{figure}[hbpt]
\centering
\includegraphics[width=0.39\textwidth, angle=0]{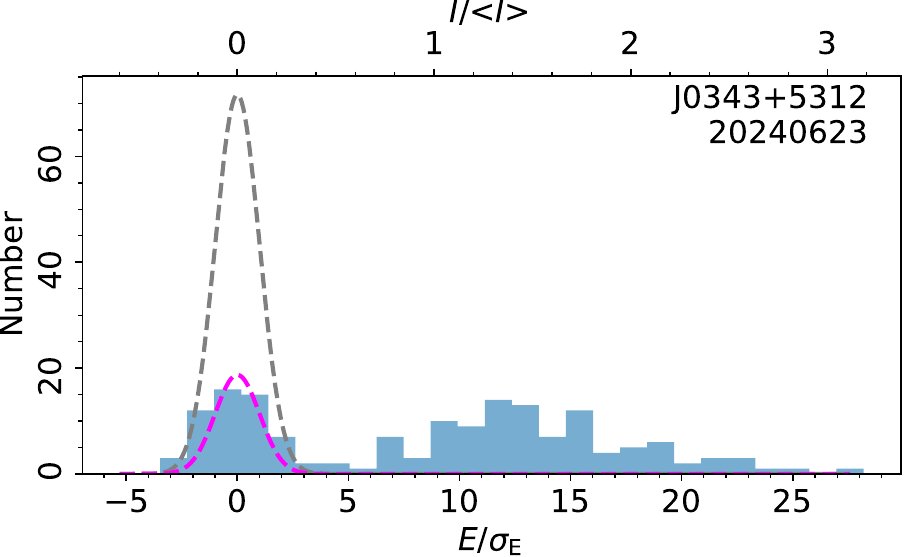}
\figcaption{On-pulse energy histogram of PSR J0343+5312 from the observation on 20240623. \label{subfig:Hist:J0343+5312}}
\end{figure}

\begin{figure}[hbpt]
\centering
\includegraphics[width=0.22\textwidth, angle=0]{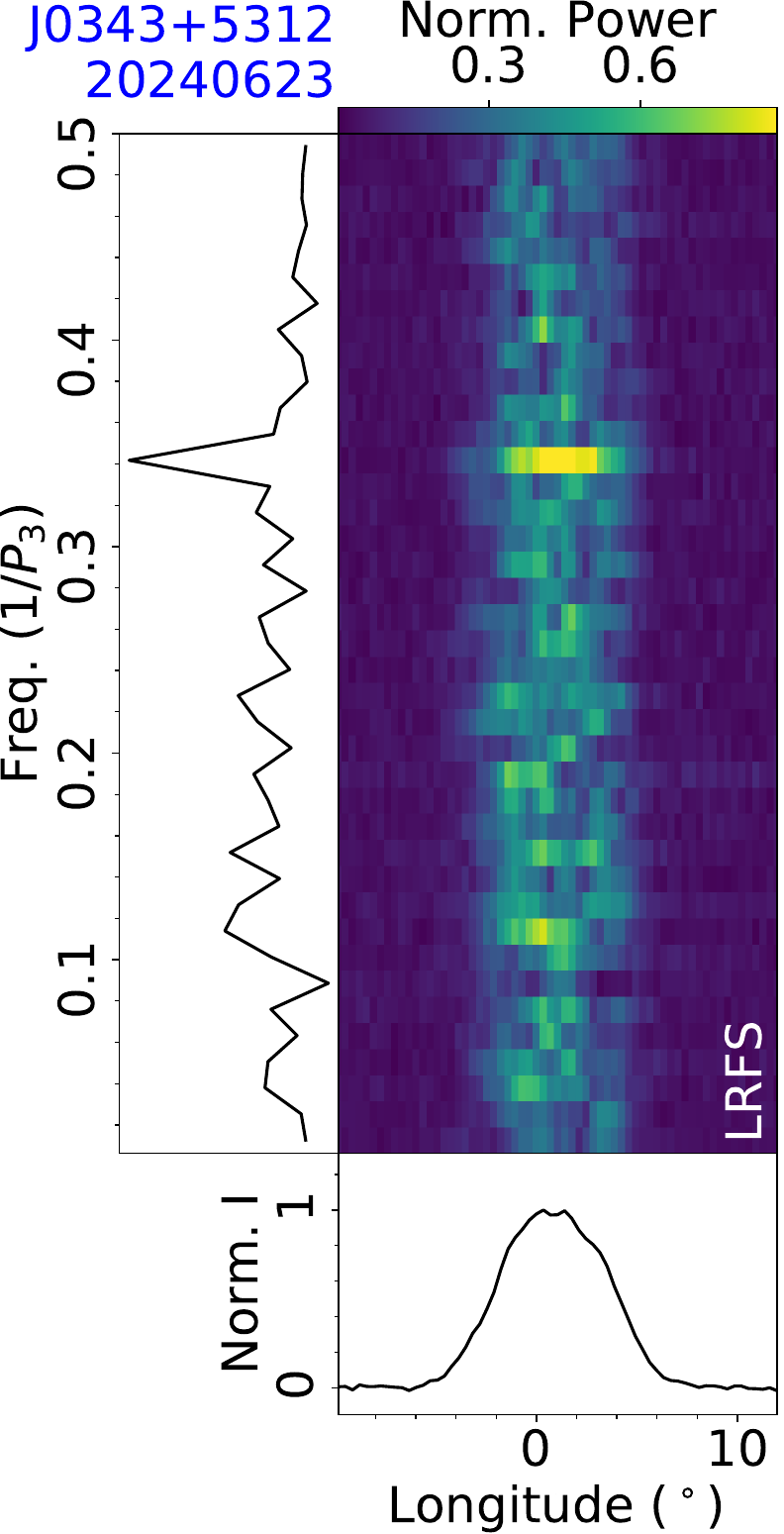}
\includegraphics[width=0.22\textwidth, angle=0]{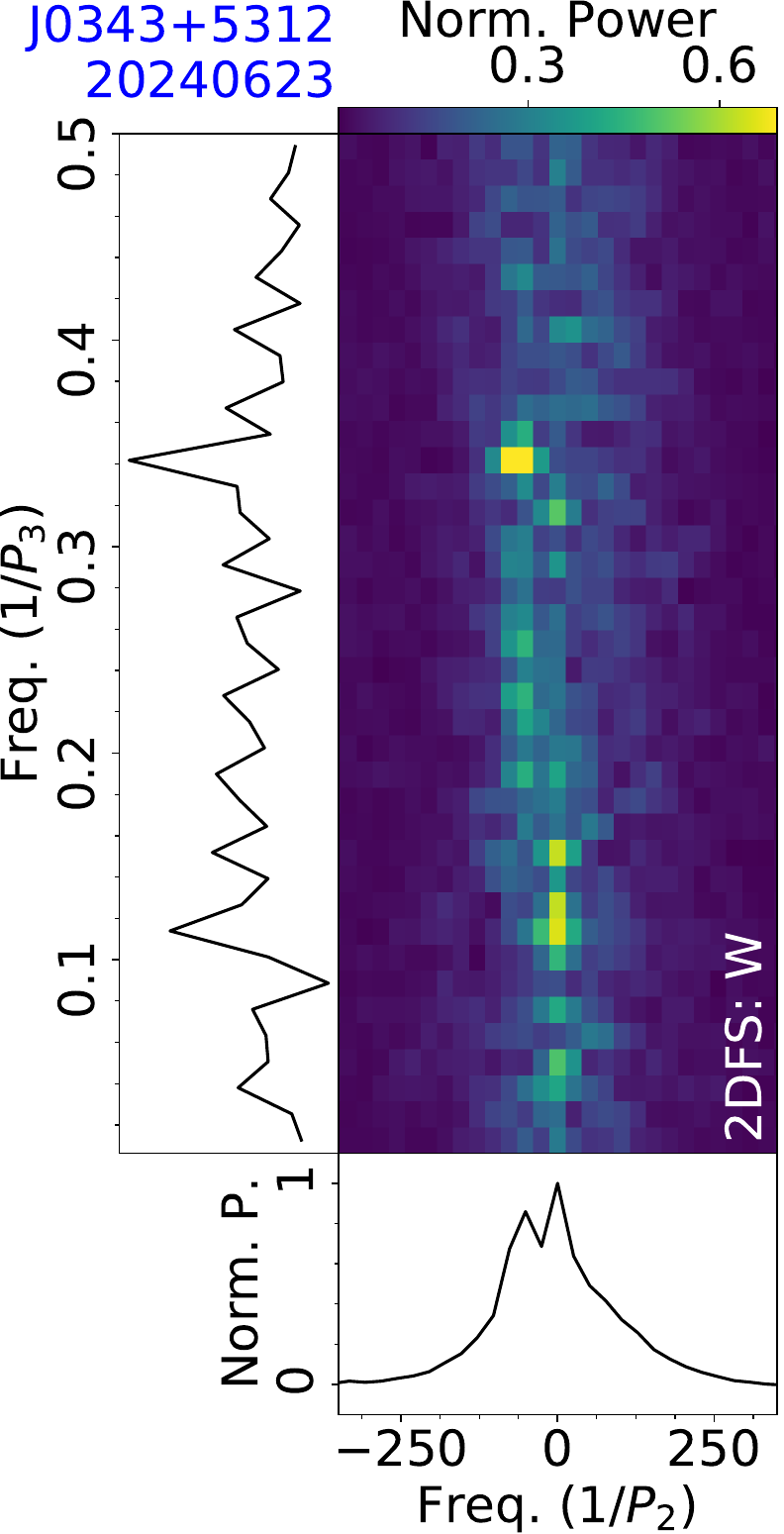}
\figcaption{Fluctuation analysis of PSR J0343+5312 from the FAST observation on 20210108, with LRFS and 2DFS for the whole pulse phase range in a mean pulse profile.  \label{subfig:fluctu:J0343+5312}}
\end{figure}

\subsection{J0244+14}
\label{subsec:J0244+14}

PSR J0244+14 was discovered by the Arecibo telescope at 327 MHz \citep{Deneva2013}. 

This pulsar was observed by FAST on 20210927 for 5 minutes, with a rotation period $P=2.1279$~s and a dispersion measure $D\!M=27.2~{\rm cm^{-3}\,pc}$ from this observation. The single pulse sequence shown in Fig.~\ref{subfig:TP:J0244+14} indicates the existence of the nulling phenomenon. 
Based on the on-pulse integral energy histogram (Fig.~\ref{subfig:Hist:J0244+14}), the nulling fraction of this data is estimated to be 31$\pm$4\%. 

\subsection{J0302+2252}
\label{subsec:J0302+2252}

PSR J0302+2252 was first reported by \citet{Tyulbashev2016} using the Large Phased Array of the Lebedev Physical Institute at 111 MHz and detected as fast transients \citep{Tyulbashev2018_5RRAT}. \citet{Michilli2020} mentioned the nulling behavior from the phase-resolved flux density plot observed with LOFAR at 149 MHz. 

This pulsar was observed by FAST on 20201101, with a rotation period $P=1.2071$~s and a dispersion measure $D\!M=19.0~{\rm cm^{-3}\,pc}$. Single pulse sequences of this observation in Fig.~\ref{subfig:TP:J0302+2252} illustrate the intensity variations as reported by the previous studies. From the tripodal shape of the on-pulse integral energy histogram of individual pulses in Fig.~\ref{subfig:TP:J0302+2252}, weak, normal, and bright emission modes are distinguished, which are labeled by red, green, and yellow, respectively. Averaged polarization profiles and PA curves of these three modes are displayed in Fig.~\ref{subfig:PolModes:J0302+2252}. From PA tracks of different emission modes, the weakest mode is dominated by the secondary polarization mode, while normal and bright modes are dominated by the primary polarization mode, resulting in a PA difference of about $90^{\circ}$ between the weak mode and the other two modes. For the leading and trailing parts in the profile, the sign of circular polarization of the weak mode is also opposite to that of the other two emission modes.





\subsection{J0343+5312}
\label{subsec:J0343+5312}

PSR J0343+5312 was discovered using the 91-m transit telescope of the National Radio Astronomy Observatory in Green Bank \citep{Damashek1978}.

This pulsar was observed by FAST on 20240623 for 5 minutes, deriving a rotation period $P=1.9344$~s and a dispersion measure $D\!M=66.3~{\rm cm^{-3}\,pc}$. The single pulse sequence in Fig.~\ref{subfig:TP:J0343+5312} displays nulling and subpulse drifting phenomena. The nulling fraction of this observation is estimated from the on-pulse integral energy histogram in Fig.~\ref{subfig:Hist:J0343+5312}, which is 26$\pm$4\%. LRFS and 2DFS of the on-pulse phase region are shown in Fig.~\ref{subfig:fluctu:J0343+5312}, where the main drift feature has $1/P_2=-62\pm4$ and $1/P_3=0.343\pm0.002$, yielding $P_2=-5.8\pm0.4^\circ$ and $P_3=2.91\pm0.02$ periods. 
In addition, there is also a drift feature with $f_3$ widely ranging from 0.18 to 0.32, indicating the variation of drifting properties, which requires a longer FAST observation to confirm.

\begin{figure}[hbpt]
\includegraphics[width=0.44\textwidth, angle=0]{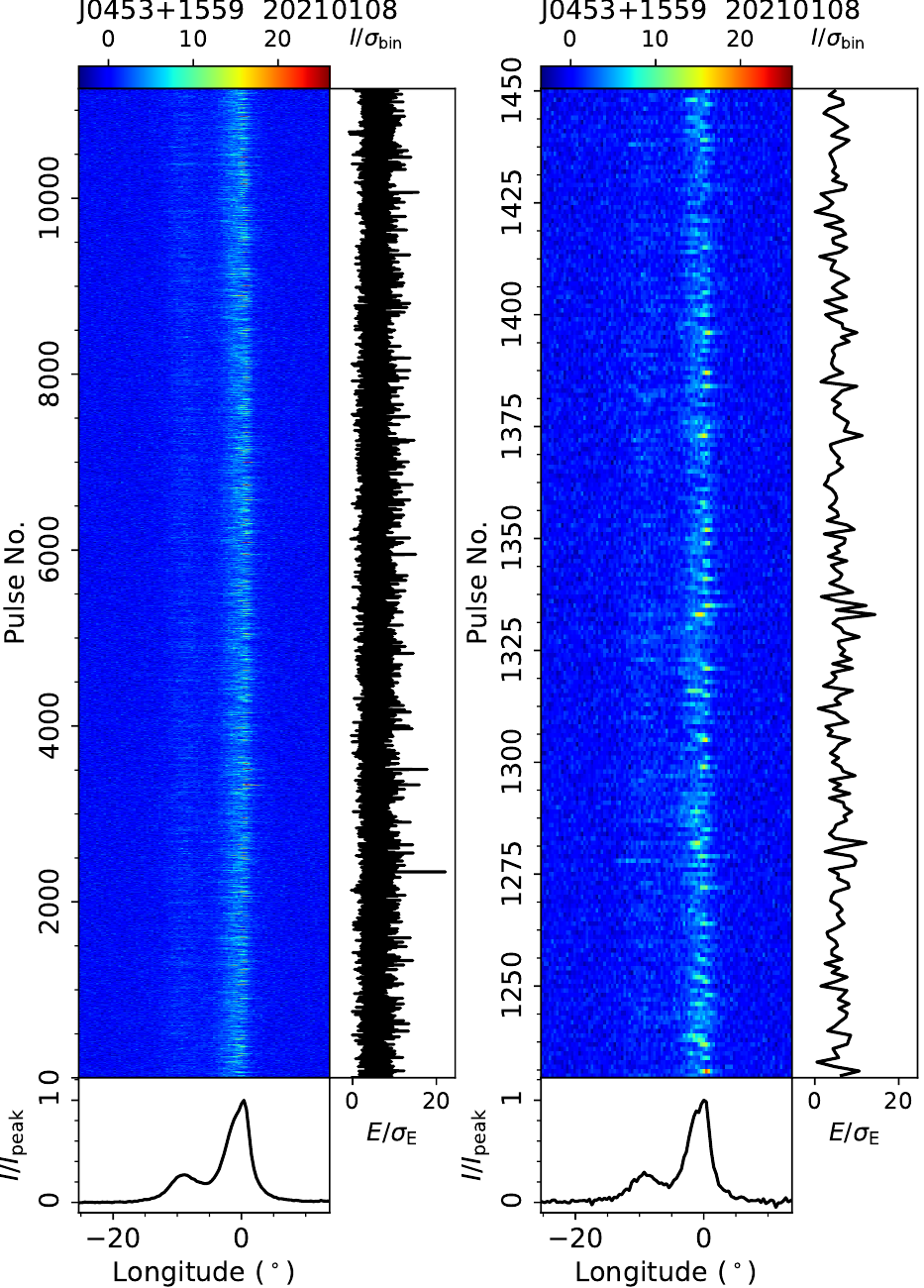} 
\figcaption{Single pulse sequence of PSR J0453+1559 from the FAST observation on 20210108, accompanied by a zoomed-in view in the right for the period range No. 1230-1450. \label{subfig:TP:J0453+1559}}
\end{figure}

\begin{figure}[hbpt]
\centering
\includegraphics[width=0.44\textwidth, angle=0]{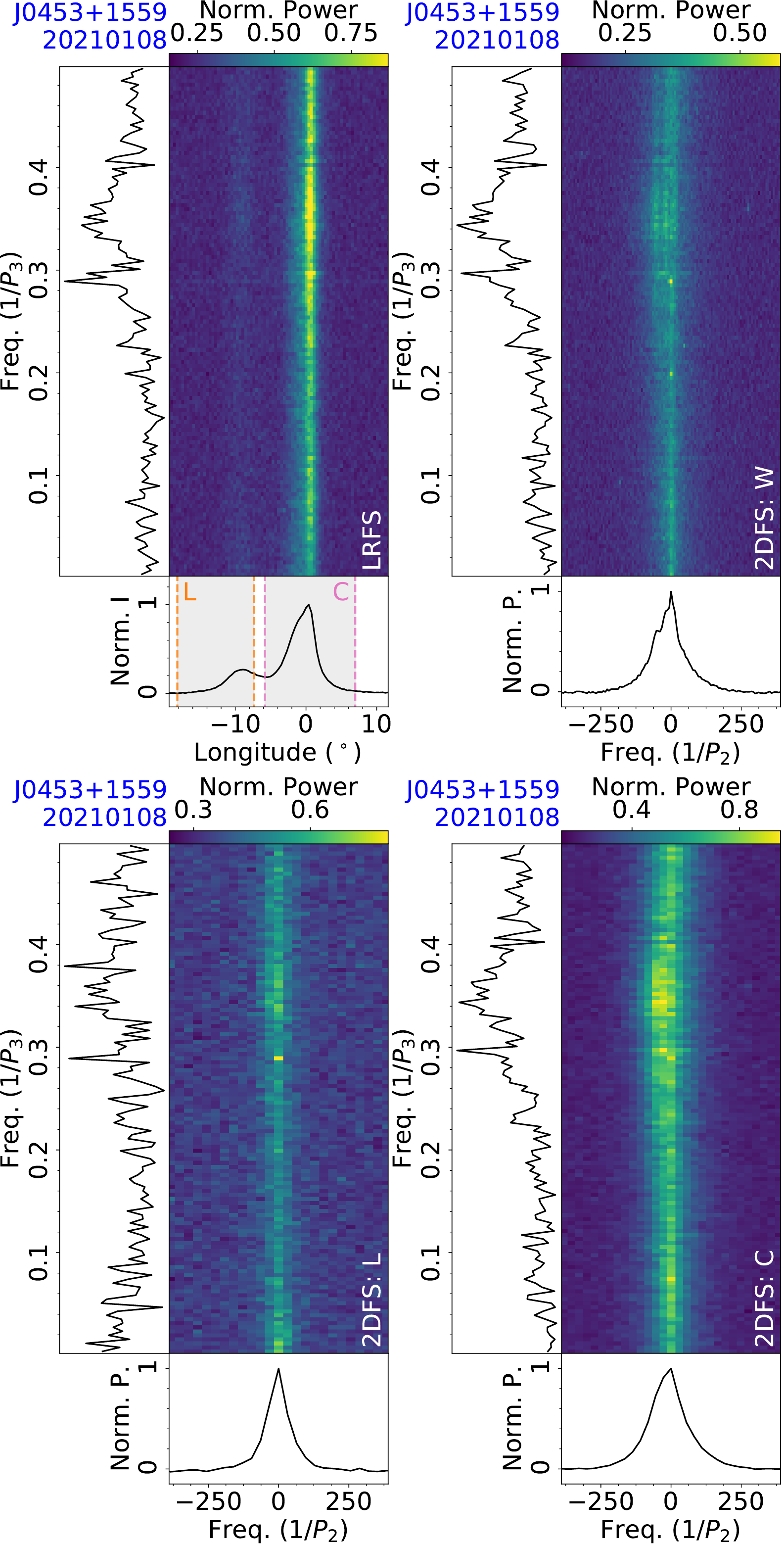}
\figcaption{Fluctuation analysis of PSR J0453+1559 from the FAST observation on 20210108, with LRFS (top-left), and 2DFS for the whole pulse phase range (top-right), the leading part (bottom left) and the central part (bottom-right) of a mean pulse profile. \label{subfig:fluctu:J0453+1559}}
\end{figure}

\begin{figure}[hbpt]
\centering
\includegraphics[width=0.22\textwidth, angle=0]{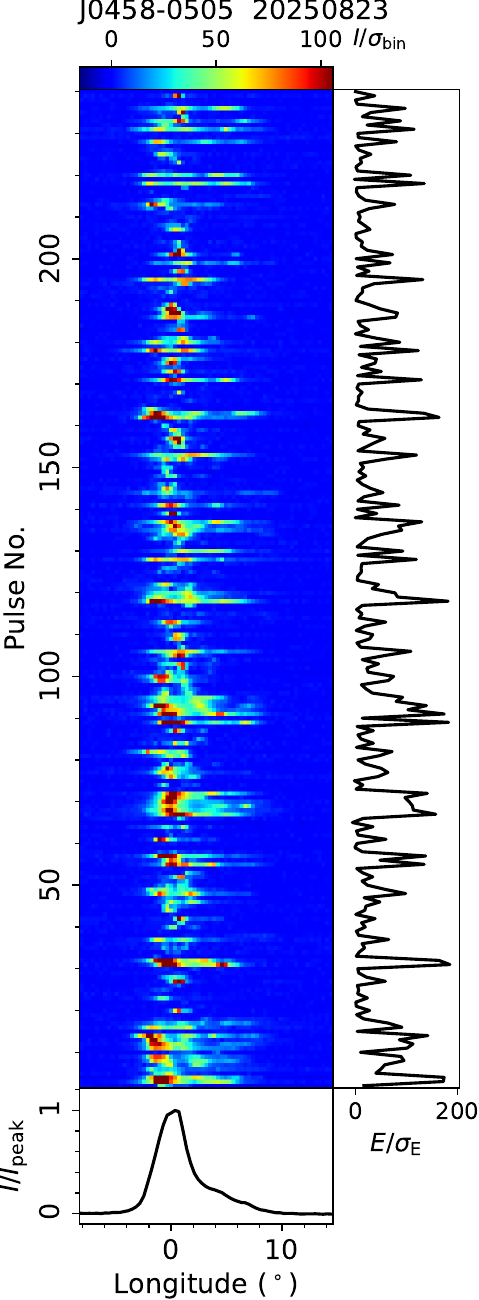}
\figcaption{Single pulse sequence of PSR J0458-0505 from the FAST observation on 20250823.
\label{subfig:TP:J0458-0505}}
\end{figure}

\begin{figure}[hbpt]
\centering
\includegraphics[width=0.44\textwidth, angle=0]{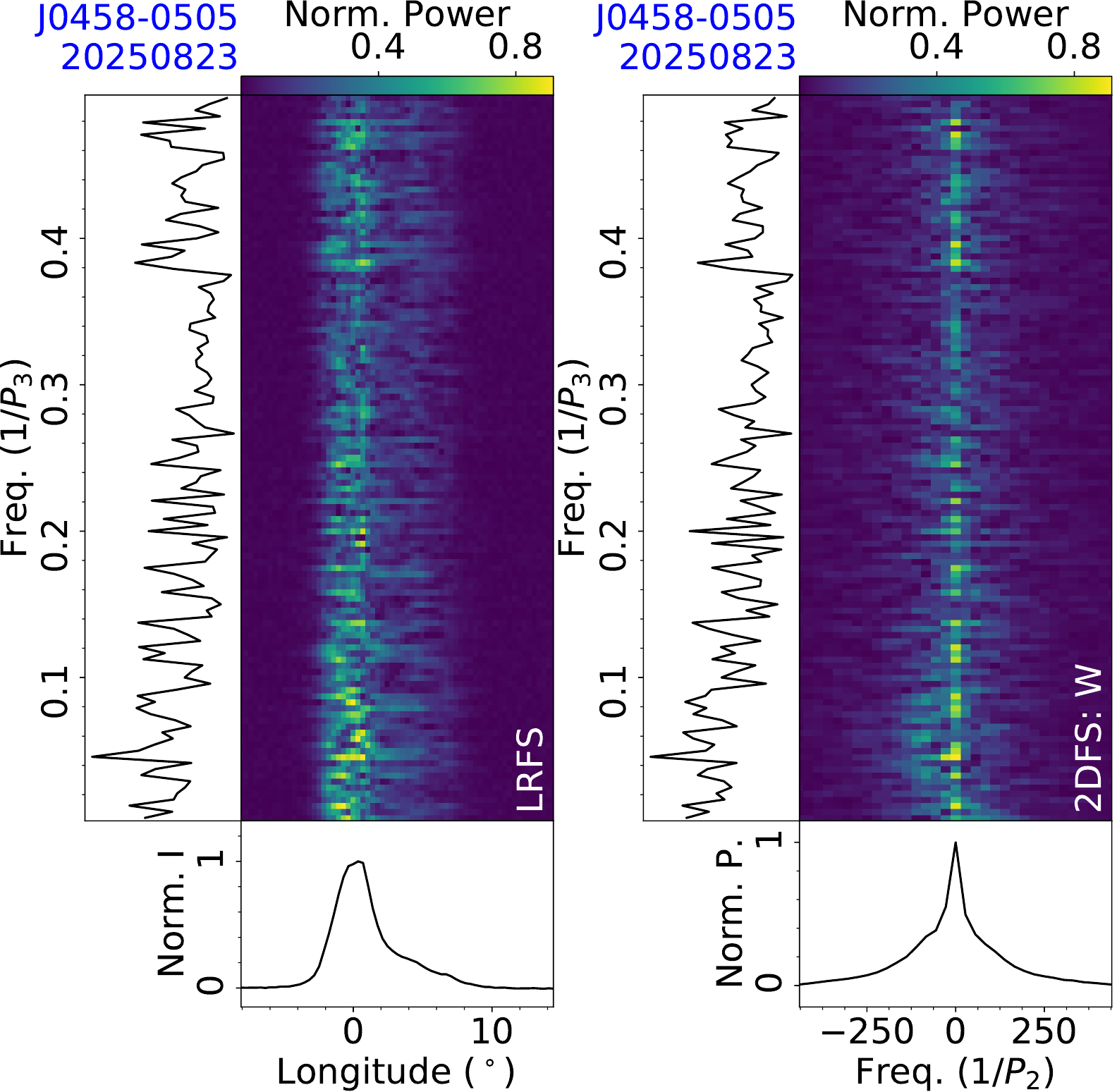}
\figcaption{Fluctuation analysis of PSR J0458-0505 from the FAST observation on 20250823, with LRFS and 2DFS for the on-pulse phase range of the mean pulse profile.
\label{subfig:fluctu:J0458-0505}}
\end{figure}

\subsection{J0453+1559}
\label{subsec:J0453+1559}


PSR J0453+1559 was discovered by the Arecibo telescope at 327 MHz \citep{Deneva2013}. 

This pulsar was observed by FAST on 20210108 for 9 minutes, with a rotation period $P=0.0458$~s and a dispersion measure $D\!M=30.4~{\rm cm^{-3}\,pc}$ from this observation. 
Single-pulse sequences of PSR J0453+1559 are displayed in Fig.~\ref{subfig:TP:J0453+1559}. 
Because of the very weak emission in the trailing part of the mean profile \citep{Wang2023}, we refer to the phase ranges related to the two brighter regions as ``leading" and ``central" parts, respectively. 
The enlarged view of pulses No.1230-1450 in the right panel of Fig.~\ref{subfig:TP:J0453+1559} illustrates the subpulse drifting for the central part of the profile.
From spectral analysis of the leading part in a mean pulse profile in Fig.~\ref{subfig:fluctu:J0453+1559}, there are two temporal modulation features: a negative drifting of $1/P_3=0.353\pm0.002$ and $1/P_2=-16\pm2$, yielding $P_3=2.83\pm0.01$ periods and $P_2=-23\pm3^\circ$; and a low-frequency modulation of $1/P_3=0.021\pm0.001$, which corresponds to $47\pm2$ periods. The 2DFS for the central part of the profile shows a clear drift feature of $1/P_3=0.338\pm0.001$ and $1/P_2=-46\pm1$, yielding $P_3=2.96\pm0.01$ periods and $P_2=-7.8\pm0.2^\circ$.


\begin{figure}[hbpt]
\centering
\includegraphics[width=0.44\textwidth, angle=0]{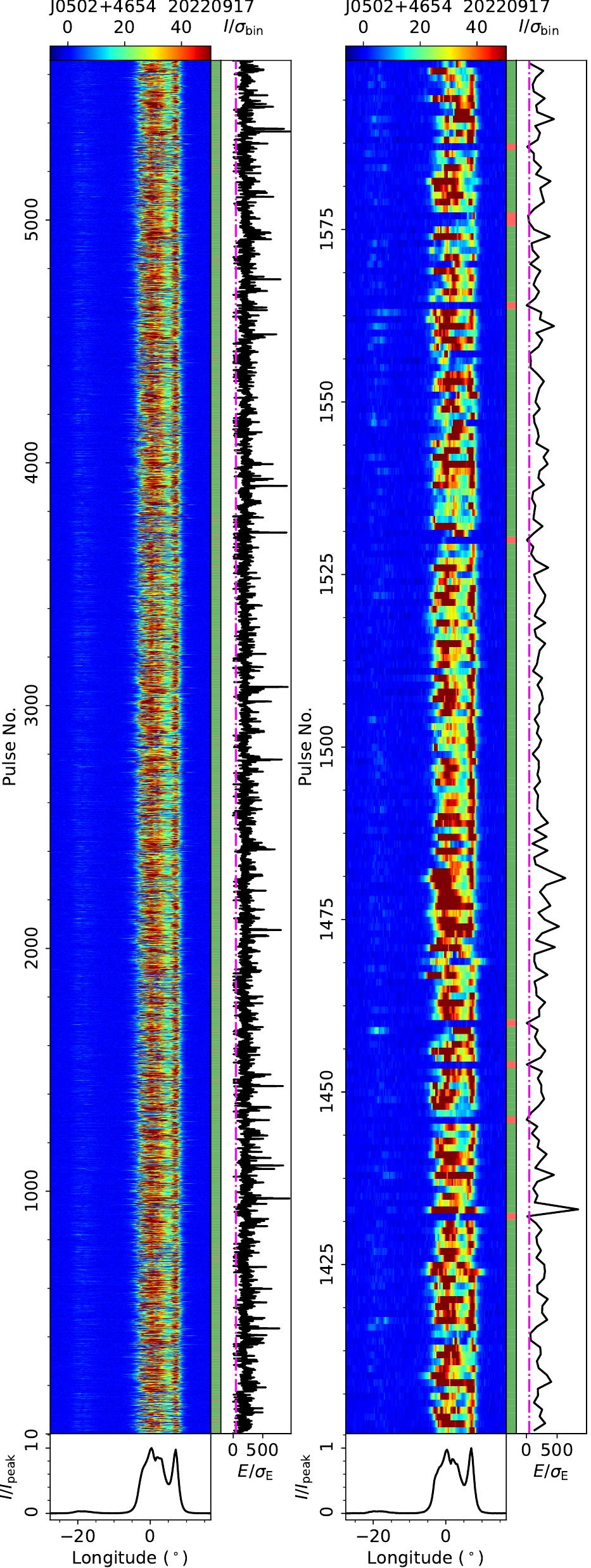}
\vspace{-0.3cm}
\figcaption{Single pulse sequence of PSR J0502+4654 from the FAST observation on 20220917, and a zoomed-in view of pulses No. 1400-1600. In the right subpanel, the energy variation is integrated over the longitude range from -6.9$^\circ$ to 11.1$^\circ$. The magenta dash-dot line represents the boundary between the two emission modes.
\label{subfig:TP:J0502+4654}}
\end{figure}

\begin{figure}[hbpt]
\centering
\includegraphics[width=0.39\textwidth, angle=0]{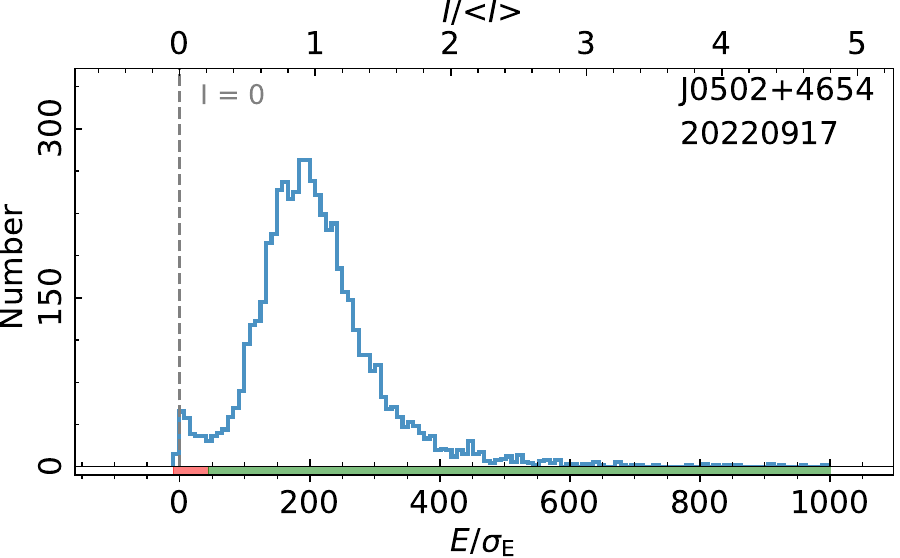}
\figcaption{Energy histograms of PSR J0502+4654 from the FAST observation on 20220917. For every single pulse, the energy is integrated over the longitude range from -6.9$^\circ$ to 11.1$^\circ$. The weak and strong emission modes are labeled in red and green, respectively. 
\label{subfig:Hist:J0502+4654}}
\end{figure}

\begin{figure}[hbpt]
\centering
\includegraphics[width=0.42\textwidth, angle=0]{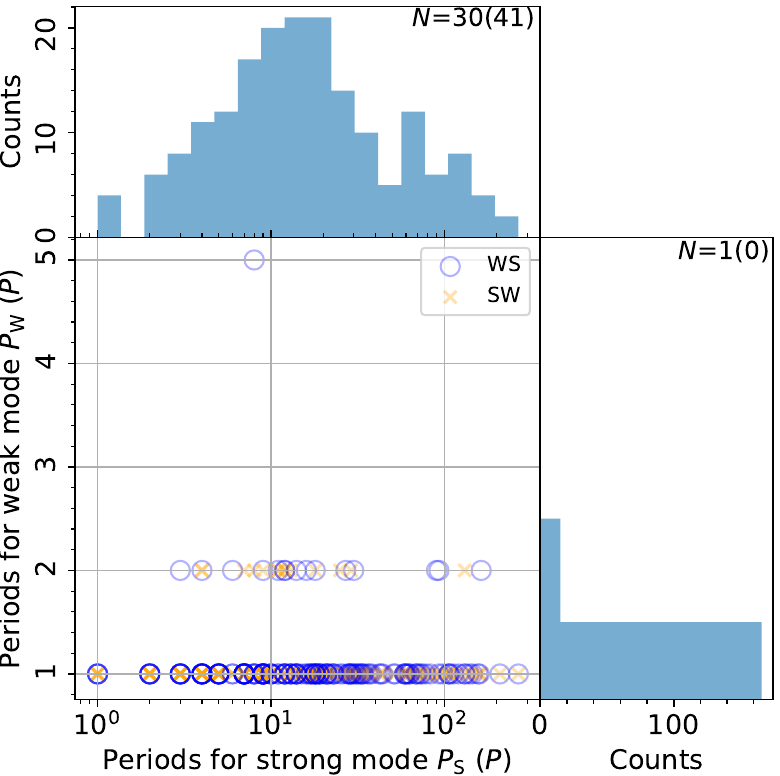}
\figcaption{Distribution of period numbers for the weak emission mode $P_{\rm W}$ against period numbers for the adjacent strong mode $P_{\rm S}$ of PSR J0502+4654 observed by FAST on 20220917, as well as the duration histograms for the strong and weak emission modes shown in the top and right panels, respectively. \label{subfig:scaleHist:J0502+4654}}
\end{figure}

\begin{figure}[htpb]
\centering
\includegraphics[width=0.37\textwidth, angle=0]{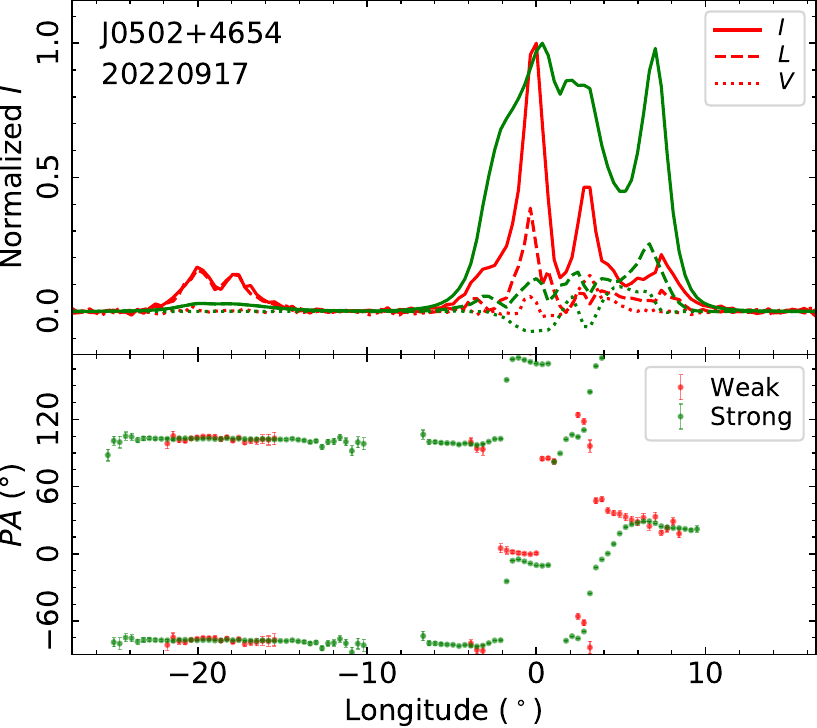}
\figcaption{Mean polarization profiles (the top panel) for the weak (red) and strong modes (green), as well as the averaged PA curves (the bottom panel) of PSR J0502+4654 from the FAST observation on 20220917.
\label{subfig:PolModes:J0502+4654}}
\end{figure}

\begin{figure}[hbpt]
\centering
\includegraphics[width=0.22\textwidth, angle=0]{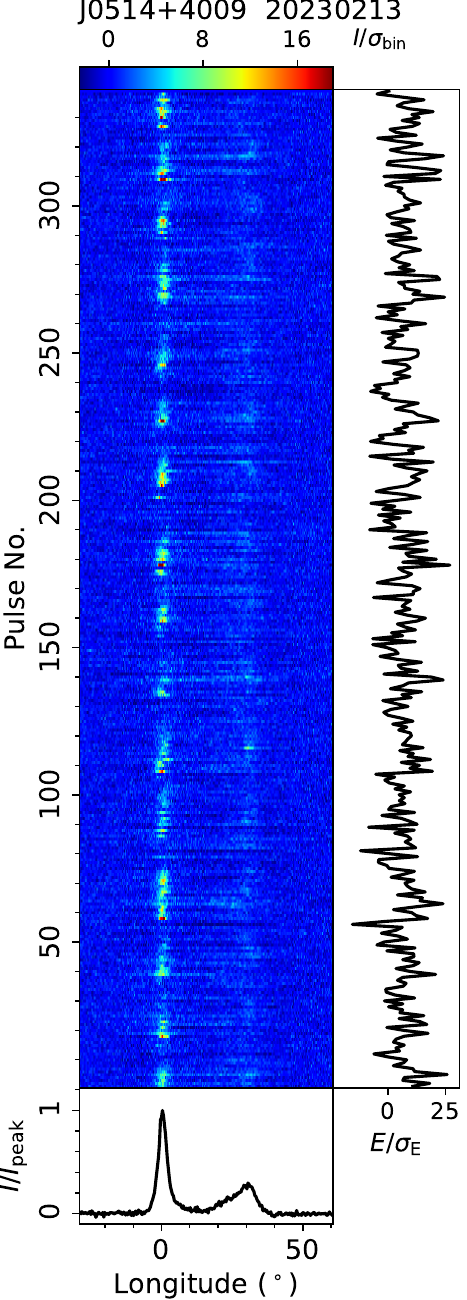}
\figcaption{Single pulse sequence of PSR J0514+4009 from the FAST observation on 20230213. \label{subfig:TP:J0514+4009}}
\end{figure}

\begin{figure}[hbpt]
\centering
\includegraphics[width=0.44\textwidth, angle=0]{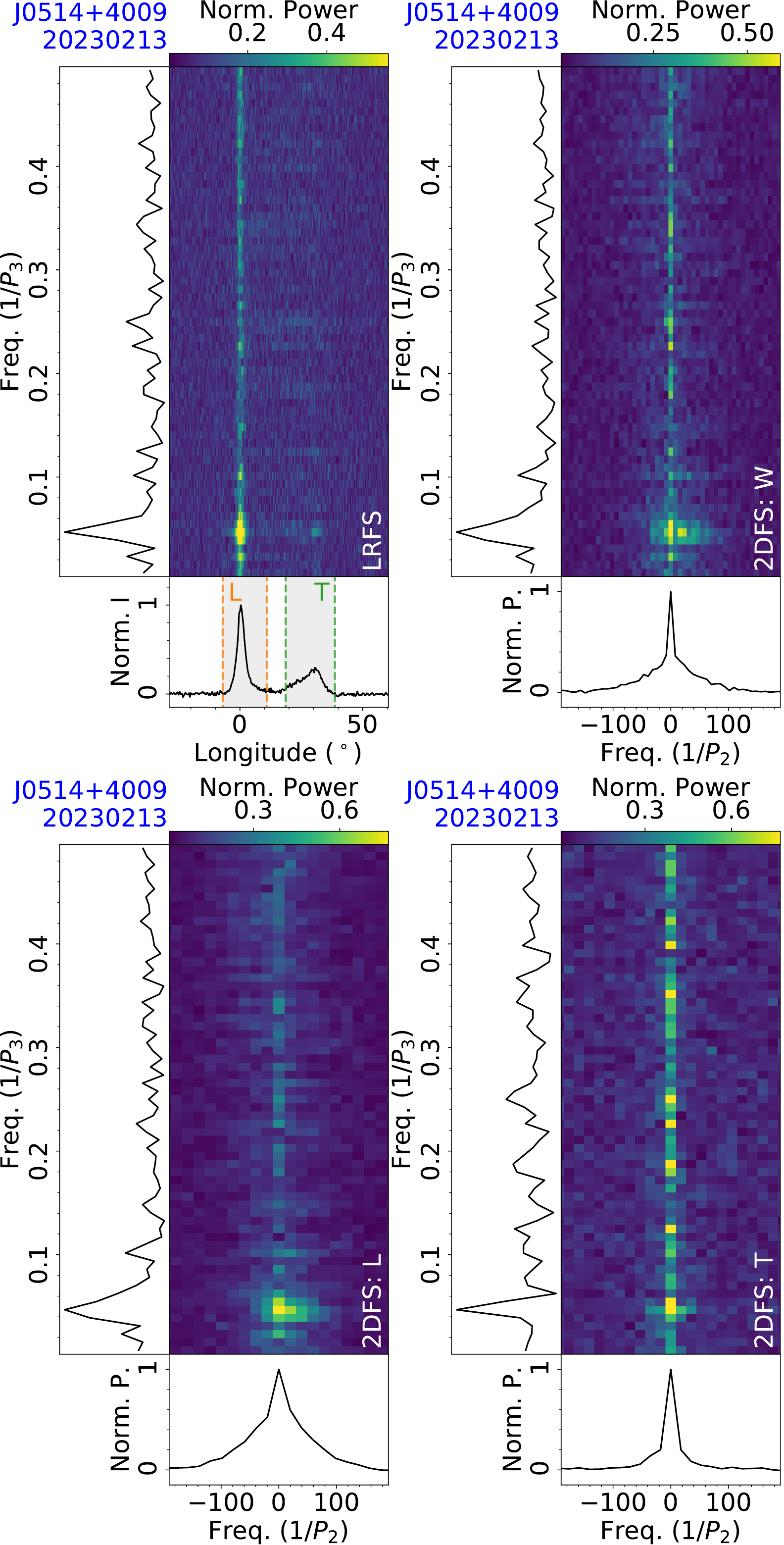}
\figcaption{Fluctuation analysis of PSR J0514+4009 from the FAST observation on 20230213, with LRFS (top-left), and 2DFS for the whole pulse phase range (top-right), the leading half (bottom left) and central half (bottom-right) of a mean pulse profile.
\label{subfig:fluctu:J0514+4009}}
\end{figure}

\subsection{J0458-0505}
\label{subsec:J0458-0505}

PSR J0458-0505 was discovered by \citet{Lynch2013} with the Green Bank Telescope. 

This pulsar was observed by FAST on 20250823 for 15 minutes, deriving a rotation period $P=3.7667$~s and a dispersion measure $D\!M=49.8~{\rm cm^{-3}\,pc}$. From the single pulse sequence and fluctuation spectra, the pulsar has a negative drift behavior. The centroid of the drift feature in 2DFS is at $1/P_3=0.052\pm0.001$ and $1/P_2=-96\pm2$, corresponding to periodicities of $P_3=19.2\pm0.4$ periods and $P_2=-3.8\pm0.1$ degrees.

\subsection{J0502+4654}
\label{subsec:J0502+4654}

PSR J0502+4654 was discovered by \citet{Damashek1978} with the 300-foot transit telescope of the National Radio Astronomy Observatory in Green Bank.

This pulsar was observed by FAST on 20220917 for 1 hour, deriving a rotation period $P=0.6385$~s and a dispersion measure $D\!M=42.1~{\rm cm^{-3}\,pc}$. The single pulse sequence and a zoomed-in view of pulses No. 1400-1600 in Fig.~\ref{subfig:TP:J0502+4654} show the intensity decrease with a short duration of $\sim$1 period for the longitude range from -6.9$^\circ$ to 11.1$^\circ$. As shown in the energy histogram integrated over this longitude range (Fig.~\ref{subfig:Hist:J0502+4654}), the weak and strong emission modes of single pulses are distinguished and labeled in red and green, respectively. 

Based on the duration statistics of two emission modes (Fig.~\ref{subfig:scaleHist:J0502+4654}), the strong mode lasts from 1 to 267 periods, whereas the weak mode lasts from 1 to 5 periods and typically only one period. 
Weak and strong modes both exhibit precursor emission over the longitude range -24$^\circ$ to -14$^\circ$. A distinct shape change in the longitude range of -10$^\circ$ to 14 $^\circ$ (main pulse) is observed between the two modes. In the weak emission mode, the leading part of the main pulse dominates in intensity, while the trailing part is weakest. In contrast, for the strong mode, the leading and trailing parts are of comparable intensity.

\begin{figure}[hbpt]
\centering
\includegraphics[width=0.22\textwidth, angle=0]{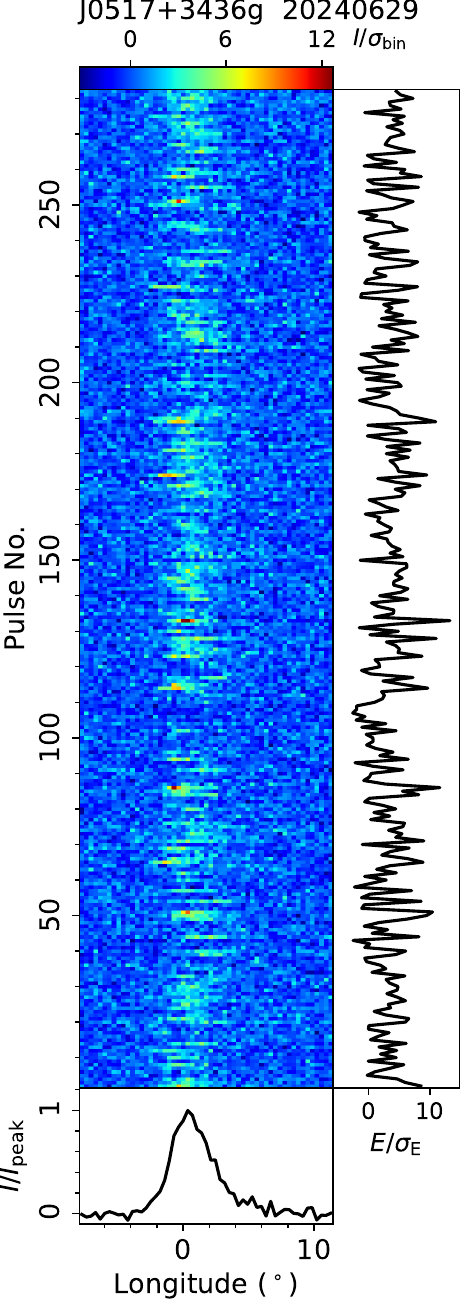}
\includegraphics[width=0.22\textwidth, angle=0]{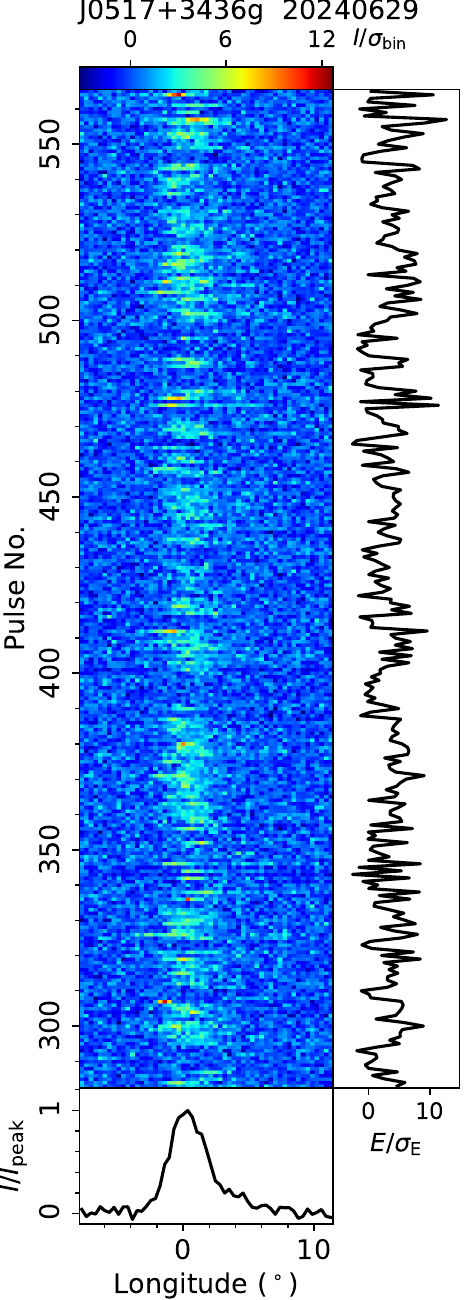}
\figcaption{Single pulse sequences of PSR J0517+3436g from the observation on 20240629.
\label{subfig:TP:J0517+3436g}}
\end{figure}

\begin{figure}[hbpt]
\centering
\includegraphics[width=0.39\textwidth, angle=0]{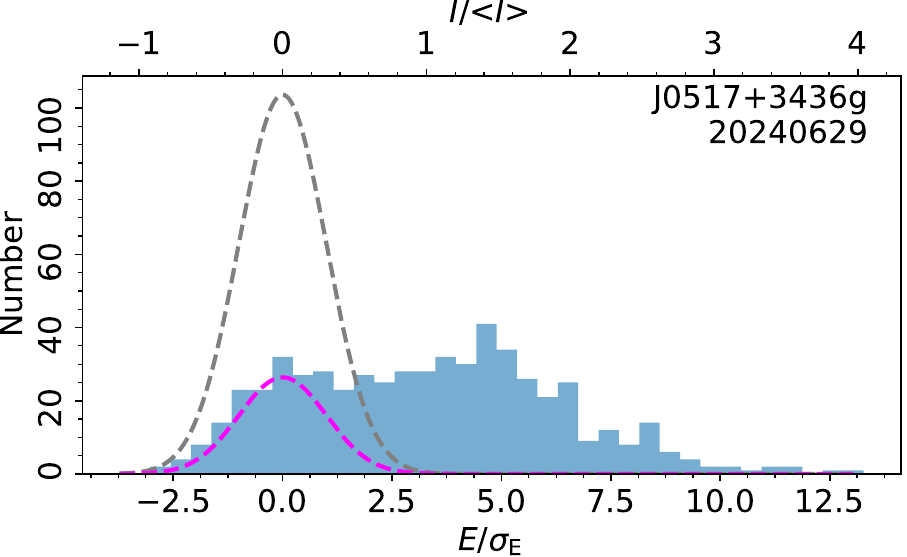}
\figcaption{On-pulse energy histogram of single pulses of PSR J0517+3436g from the FAST observation on 20240629.
\label{subfig:Hist:J0517+3436g}}
\end{figure}

\subsection{J0514+4009g}
\label{subsec:J0514+4009g}

PSR J0514+4009g was discovered in the FAST GPPS survey \citep{Han2021,han2025}. 

This pulsar was observed by FAST on 20230213 for 15 minutes, with a rotation period $P=2.6256$~s and a dispersion measure $D\!M=101.0~{\rm cm^{-3}\,pc}$. 
From the single pulse sequence in Fig.~\ref{subfig:TP:J0514+4009}, the pulsar has systematic modulation behavior. LRFS and 2DFS are displayed in Fig.~\ref{subfig:fluctu:J0514+4009}. The temporal modulation periodicity of about 20 periods is similar for both components. The leading component has a positive subpulse drifting, with modulation frequencies of $1/P_3=0.048\pm0.001$ and $1/P_2=16\pm2$, yielding drifting parameters of $P_3=20.7\pm0.3$ periods and $P_2=23\pm3^\circ$. While the trailing component is more likely to be amplitude modulated with the temporal frequency of $1/P_3=0.047\pm0.001$, which corresponds to $P_3=21.5\pm0.5$ periods. 


\begin{figure}[hbpt]
\centering
\includegraphics[width=0.22\textwidth, angle=0]{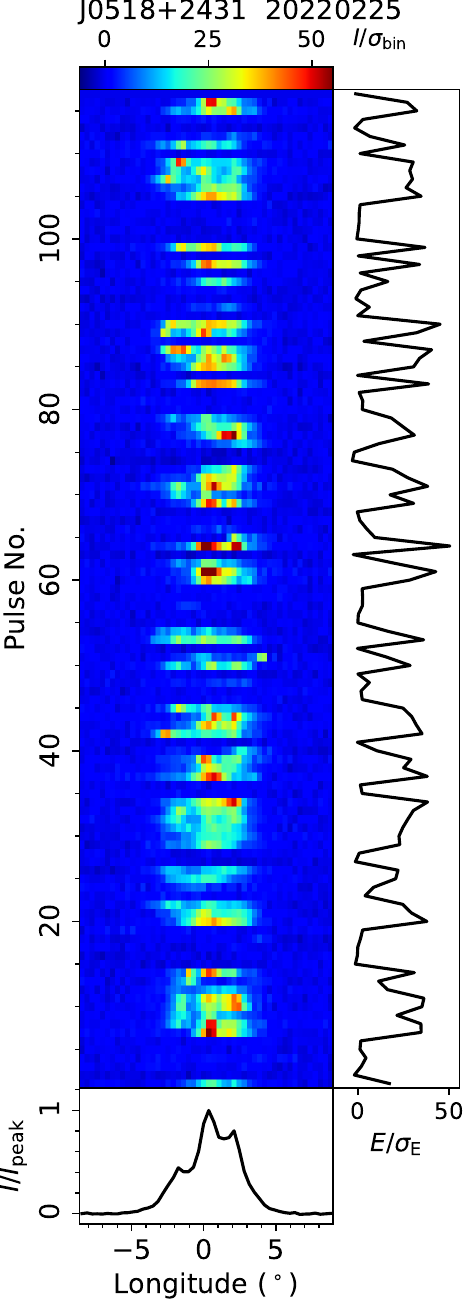}
\figcaption{Single pulse sequence of PSR J0518+2431 from the observation on 20220225. \label{subfig:TP:J0518+2431}}
\end{figure}

\begin{figure}[hbpt]
\centering
\includegraphics[width=0.39\textwidth, angle=0]{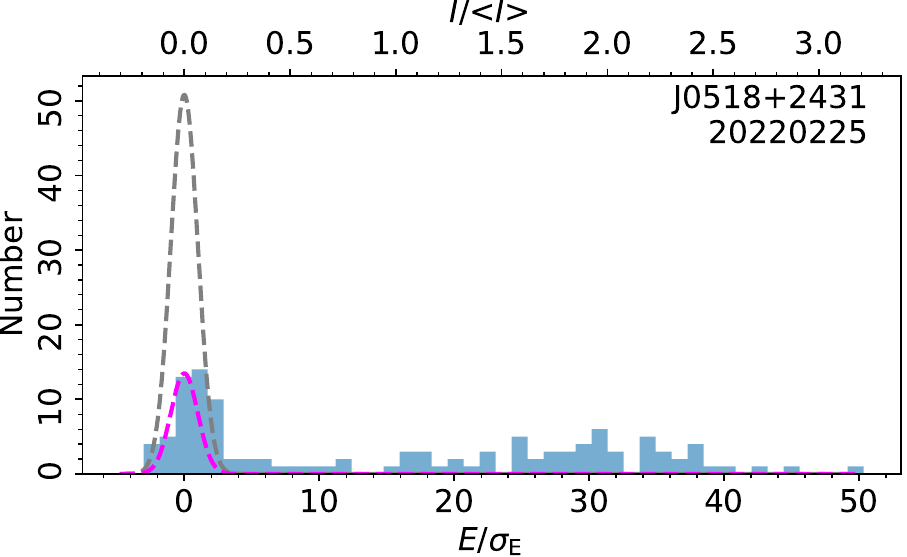}
\figcaption{On-pulse energy histogram of PSR J0518+2431 from the observation on 20220225. \label{subfig:Hist:J0518+2431}}
\end{figure}

\begin{figure}[hbpt]
\includegraphics[width=0.44\textwidth, angle=0]{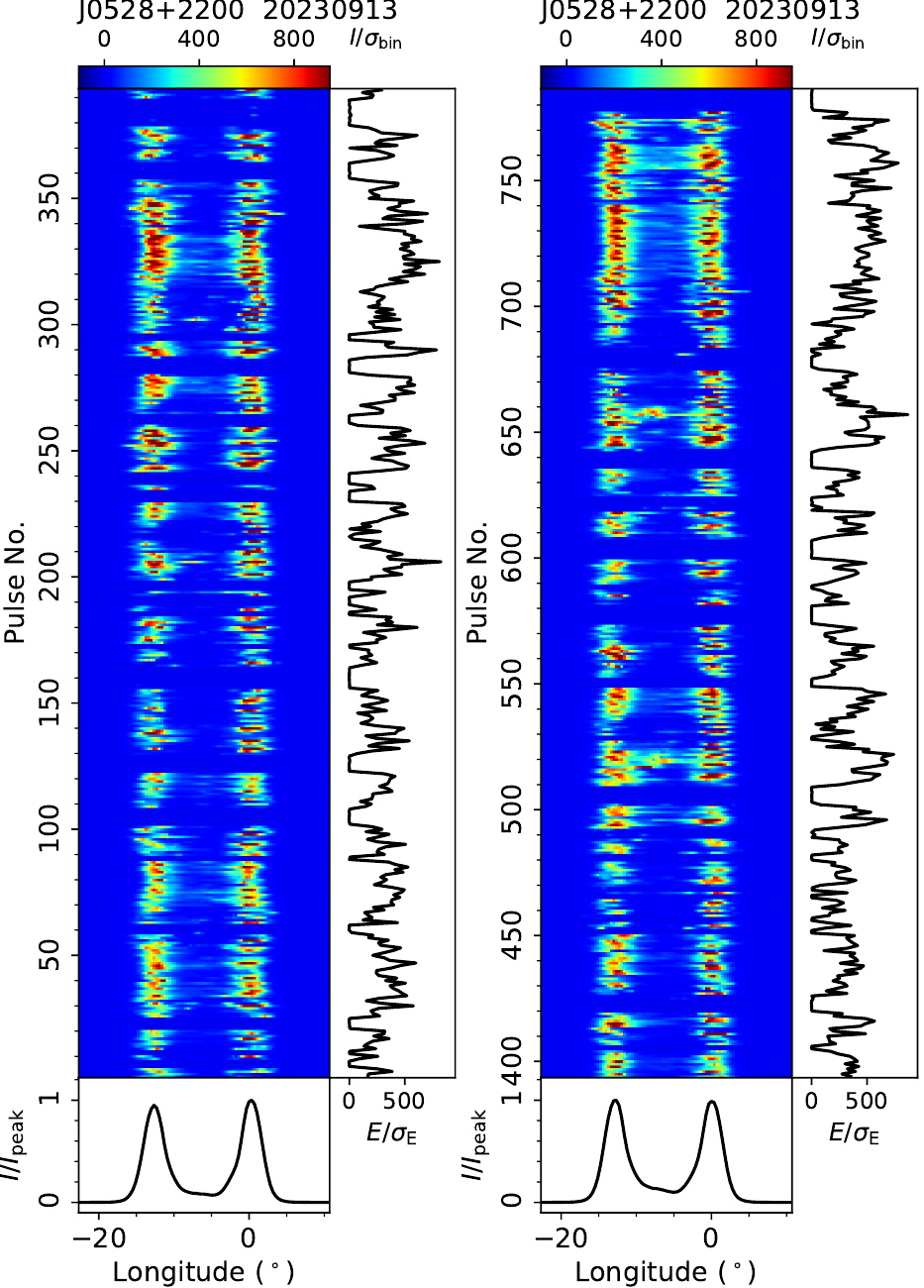}
\figcaption{Single pulse sequences of PSR J0528+2200 from the FAST observation on 20230913.
\label{subfig:TP:J0528+2200}}
\end{figure}

\begin{figure}[htpb]
\centering
\includegraphics[width=0.39\textwidth, angle=0]{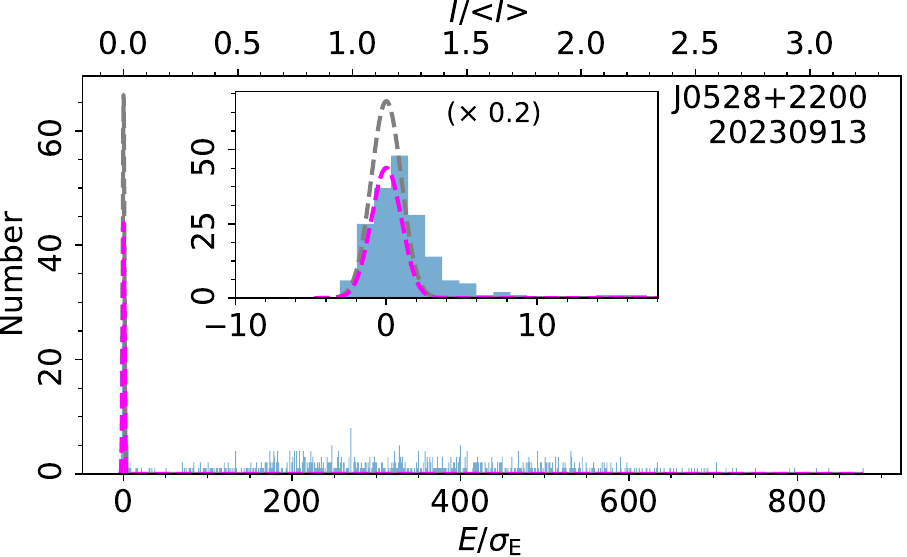}
\figcaption{On-pulse energy histogram of PSR J0528+2200 from the observation on 20230913.
\label{subfig:Hist:J0528+2200}}
\end{figure}

\begin{figure}[hbpt]
\centering
\includegraphics[width=0.44\textwidth, angle=0]{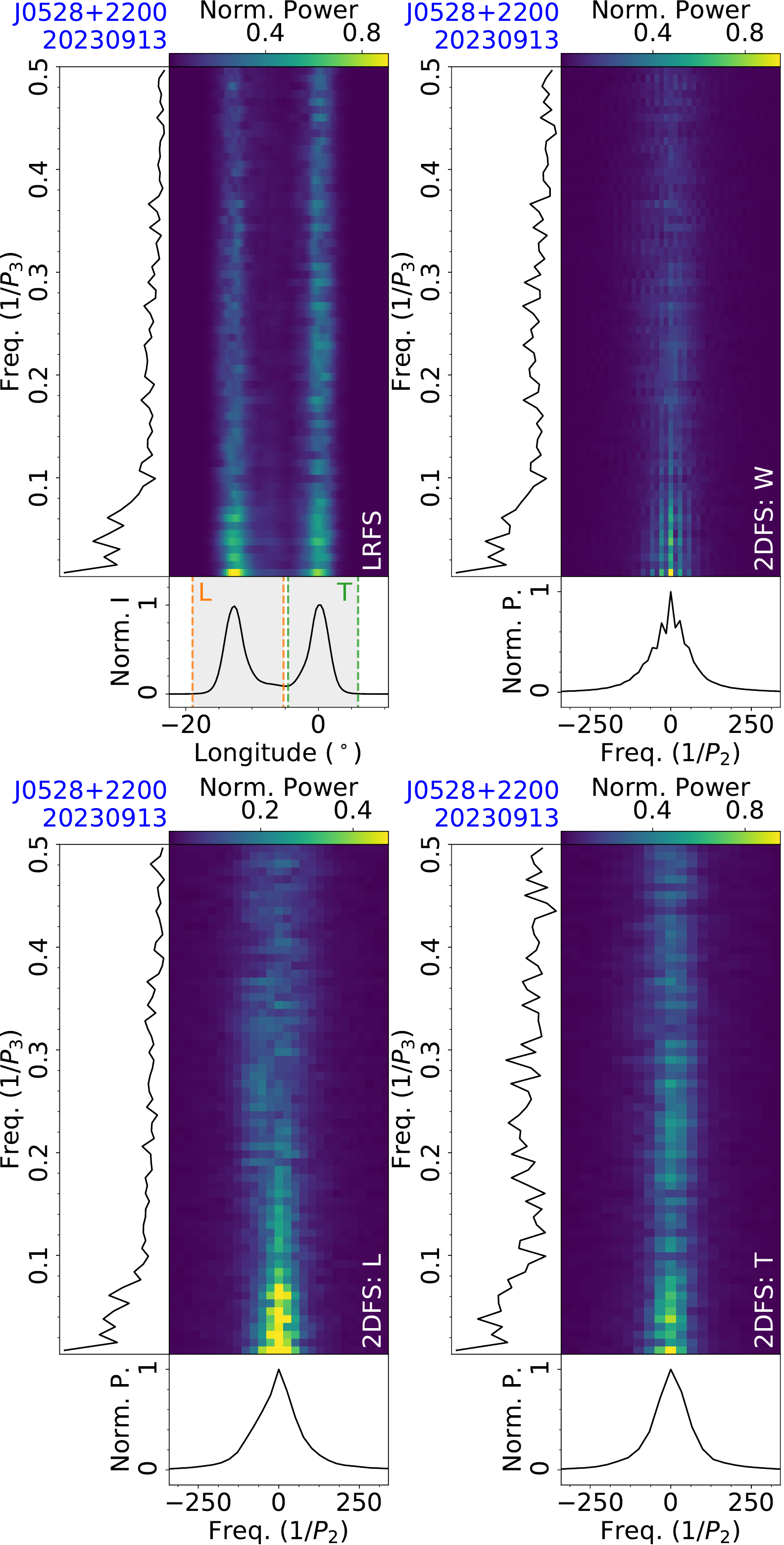}
\figcaption{Fluctuation analysis of PSR J0528+2200 from the FAST observation on 20230913, with LRFS (top-left), and 2DFS for the whole pulse phase range (top-right), the leading half (bottom left) and central half (bottom-right) of the mean pulse profile.
\label{subfig:fluctu:J0528+2200}}
\end{figure}

\subsection{J0517+3436g}
\label{subsec:J0517+3436g}

PSR J0517+3436g was discovered in the FAST GPPS survey \citep{Han2021,han2025}.

This pulsar was observed by FAST for 15 minutes on 20240629, and for 5 minutes each on 20250205, 20250308 and 20250815. From the 15-minute data, a rotation period $P=1.5958$~s and a dispersion measure $D\!M=191.0~{\rm cm^{-3}\,pc}$ were derived. 
Single pulse sequences are shown in Fig.~\ref{subfig:TP:J0517+3436g}, displaying the nulling phenomenon. From the on-pulse integral energy histogram of single pulses, the nulling fraction of this observation is estimated to be 25.5$\pm$2.3\%. 
Nulling behavior is also exhibited in the 5-minute observations.

\subsection{J0518+2431}
\label{subsec:J0518+2431}

PSR J0518+2431 was discovered by FAST, and the nulling behavior was reported in \citet{Wu2023}. 

The pulsar has also been observed in the FAST GPPS survey on 20220225 for 5 minutes, with a rotation period $P=2.6370$~s and a dispersion measure $D\!M=72.5~{\rm cm^{-3}\,pc}$. 
The single pulse sequence shown in Fig.~\ref{subfig:TP:J0518+2431} indicates the existence of nulls. The nulling fraction of this observation is estimated to be 27$\pm$2\% from the on-pulse integral energy histogram in Fig.~\ref{subfig:Hist:J0518+2431}.


\begin{figure}[hbpt]
\centering
\includegraphics[width=0.44\textwidth, angle=0]{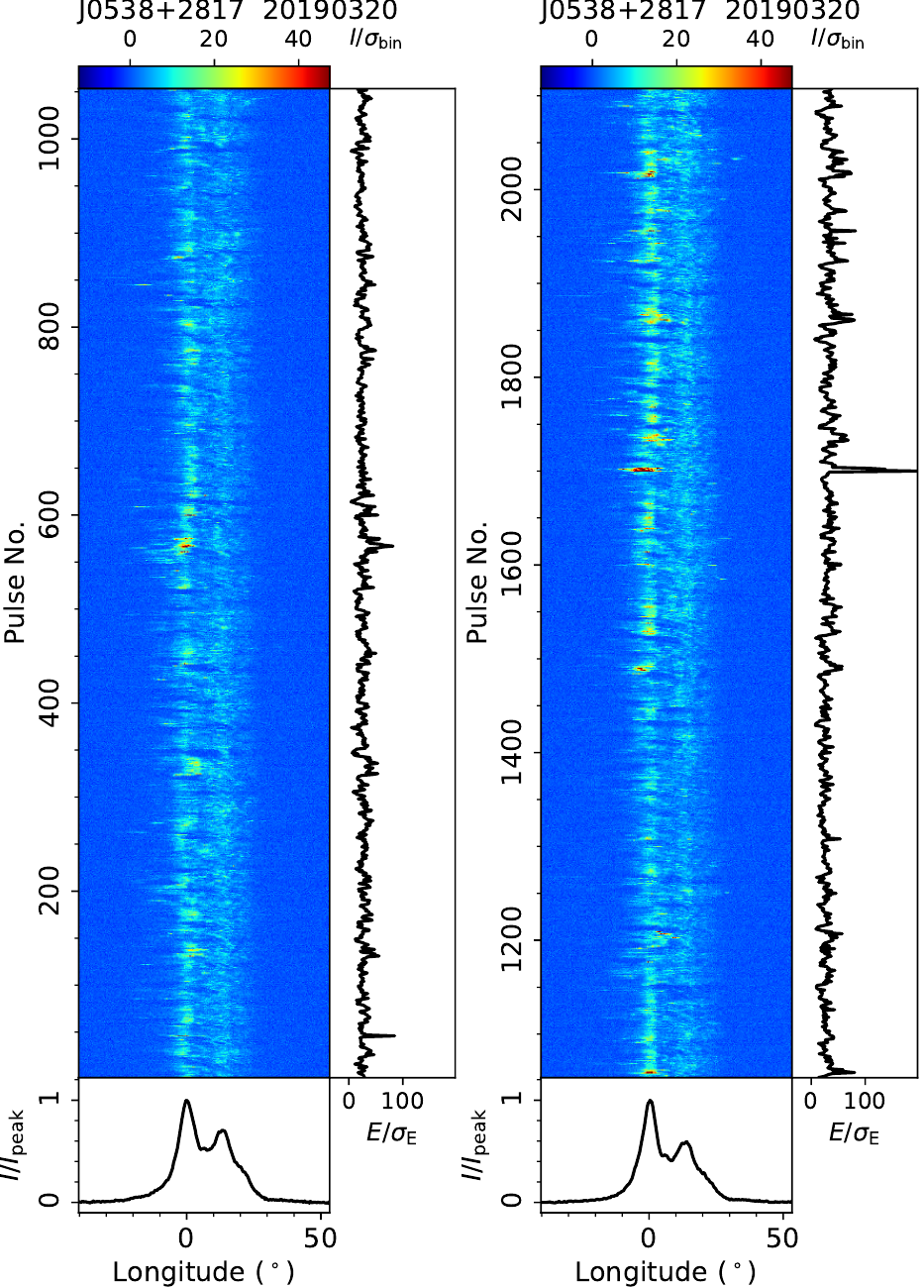}
\figcaption{Single pulse sequences of PSR J0538+2817 from the FAST observation on 20190320.
\label{subfig:TP:J0538+2817}}
\end{figure}

\begin{figure}[hbpt]
\centering
\includegraphics[width=0.44\textwidth, angle=0]{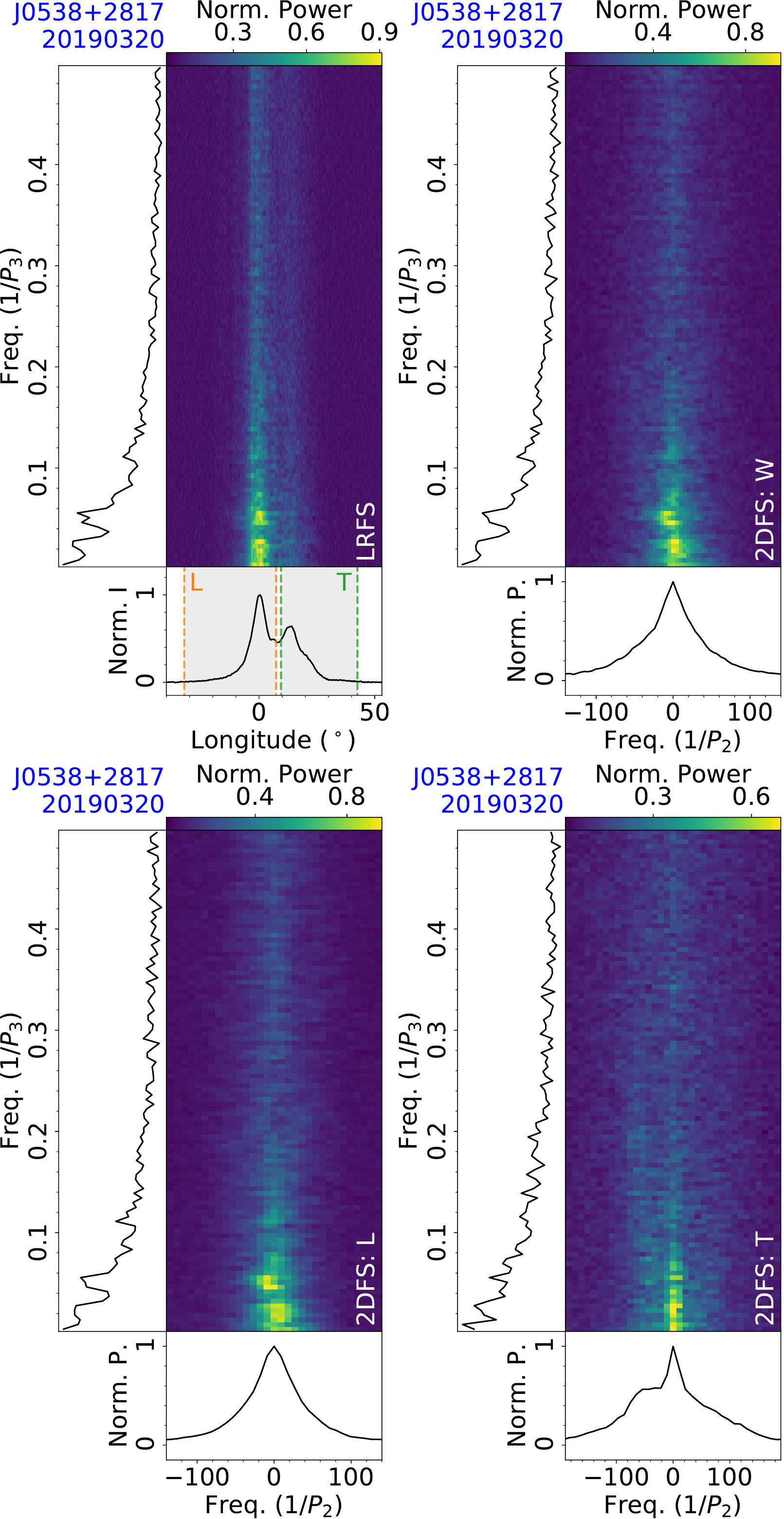}
\figcaption{Fluctuation analysis of PSR J0538+2817 from the FAST observation on 20190320, with LRFS (top-left), and 2DFS for the whole pulse phase range (top-right), the leading half (bottom left) and the central half (bottom-right) of the mean pulse profile.
\label{subfig:fluctu:J0538+2817}}
\end{figure}

\subsection{J0528+2200}
\label{subsec:J0528+2200}

PSR J0528+2200 was discovered by \citet{Staelin1968}. 
The pulsar has nulling and drifting phenomena. The nulling fraction was reported to be 25\% at 610 MHz by \citet{Ritchings1976} and at 327 MHz by \citet{Redman2009}, 28.31\% at 327 MHz by \citet{Herfindal2009}, 14.4\% at 333 MHz by \citet{Basu2017}, and 15.2\% at 2250 MHz by \citet{Wang2020}. Dwarf pulses of the pulsar have been detected by \citet{Yan2024}. 
Subpulse modulation was reported by \citet{Backer1973} and \citet{Herfindal2009} at 318 MHz and 327 MHz, and drifting behavior was reported by \citet{Weltevrede2006,Weltevrede2007} at 21 cm and 92 cm. 

This pulsar was observed by FAST on 20230913 for 49 minutes, with a rotation period $P=3.7452$~s and a dispersion measure $D\!M=51.7~{\rm cm^{-3}\,pc}$. 
Single pulse sequences in Fig.~\ref{subfig:TP:J0528+2200} display nulling and subpulse drifting behaviors. In the on-pulse energy histogram shown in Fig.~\ref{subfig:Hist:J0528+2200}, the asymmetric distribution around zero energy is caused by dwarf pulses \citep{Yan2024}, and the nulling fraction for this observation is estimated to be 13$\pm$1\%. 
From fluctuation spectra shown in Fig.~\ref{subfig:fluctu:J0528+2200}, the leading and trailing parts in the mean pulse profile have opposite drift directions, which is consistent with \citet{Weltevrede2006}. For the leading profile part, the centroid of the drift feature in 2DFS is at $1/P_3=0.275\pm0.002$ and $1/P_2=-64\pm1$, corresponding to $P_3=3.64\pm0.03$ periods and $P_2=-5.6\pm0.1$ degrees. The trailing part has a preferred positive drift feature in 2DFS, with the centroid at $1/P_3=0.234\pm0.003$ and $1/P_2=15\pm2$, yielding $P_3=4.27\pm0.05$ periods and $P_2=23\pm3$ degrees.

\begin{figure}[hbpt]
\centering
\includegraphics[width=0.21\textwidth, angle=0]{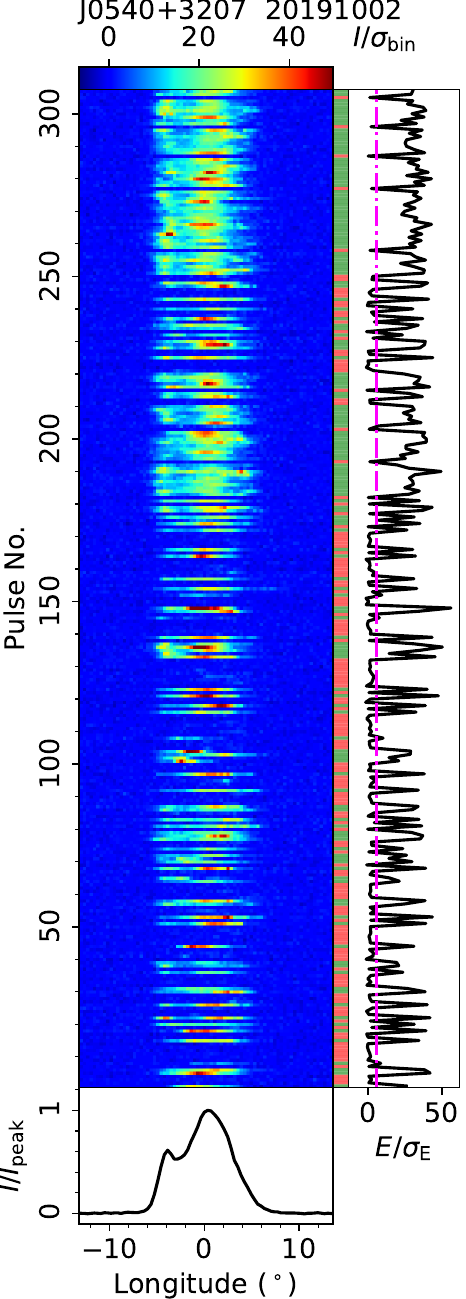}
\includegraphics[width=0.21\textwidth, angle=0]{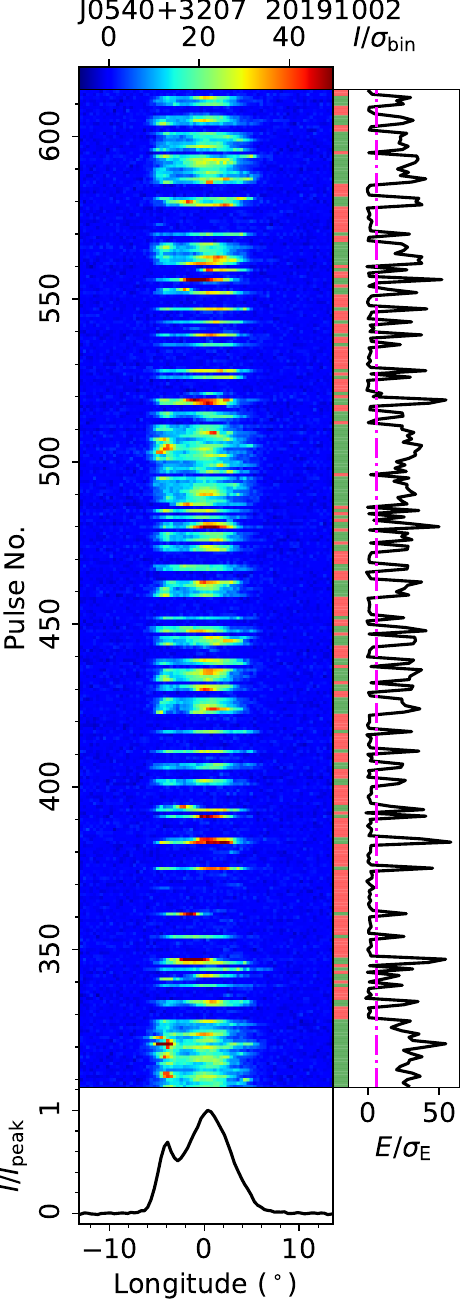}\\
\includegraphics[width=0.21\textwidth, angle=0]{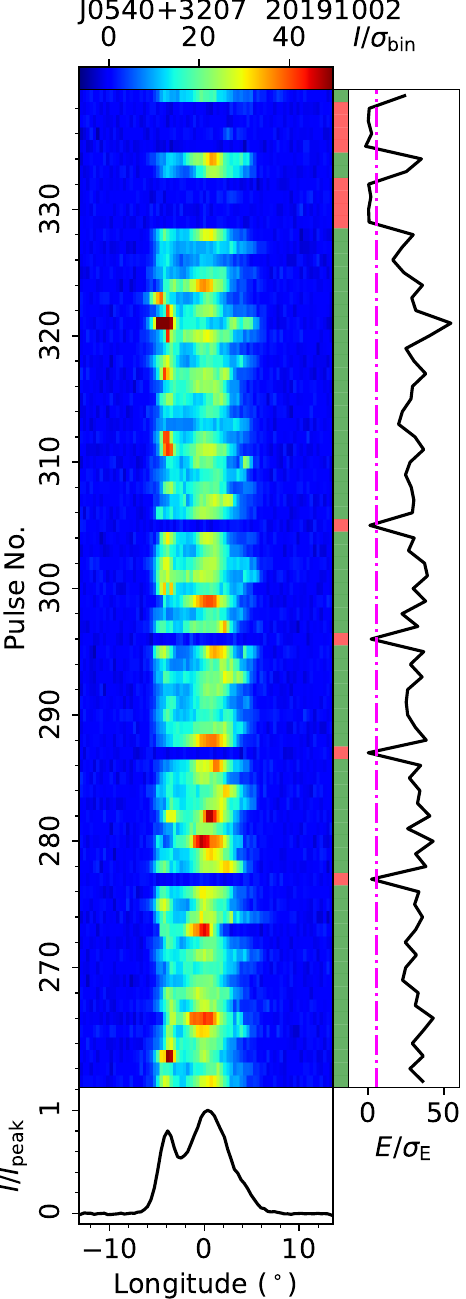}
\includegraphics[width=0.21\textwidth, angle=0]{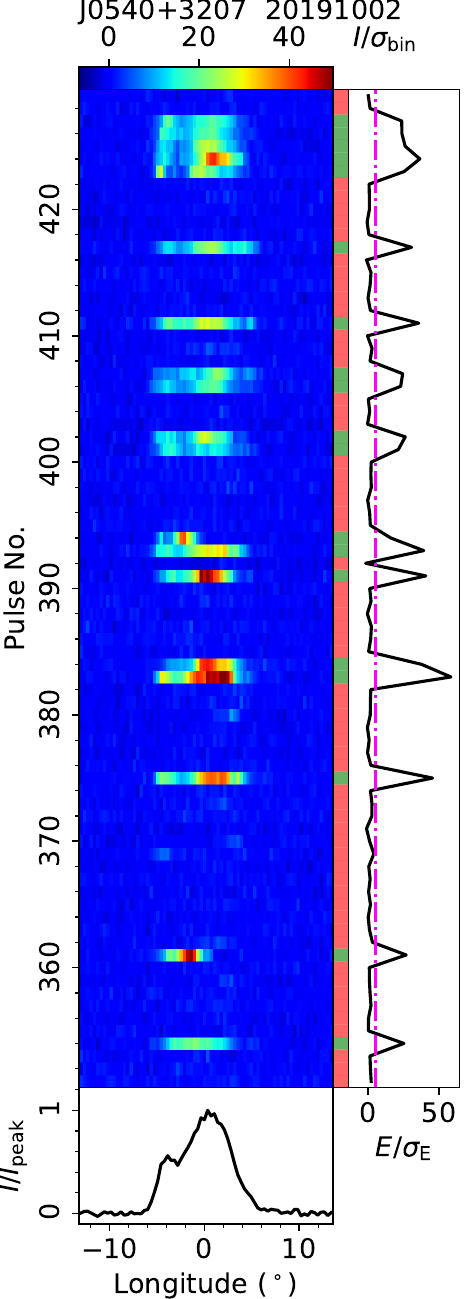}
\figcaption{Single pulse sequences of PSR J0540+3207 from the FAST observation on 20191002, as well as two zoomed-in segments in the bottom. \label{subfig:TP:J0540+3207}}
\end{figure}

\begin{figure}[hbpt]
\centering
\includegraphics[width=0.39\textwidth, angle=0]{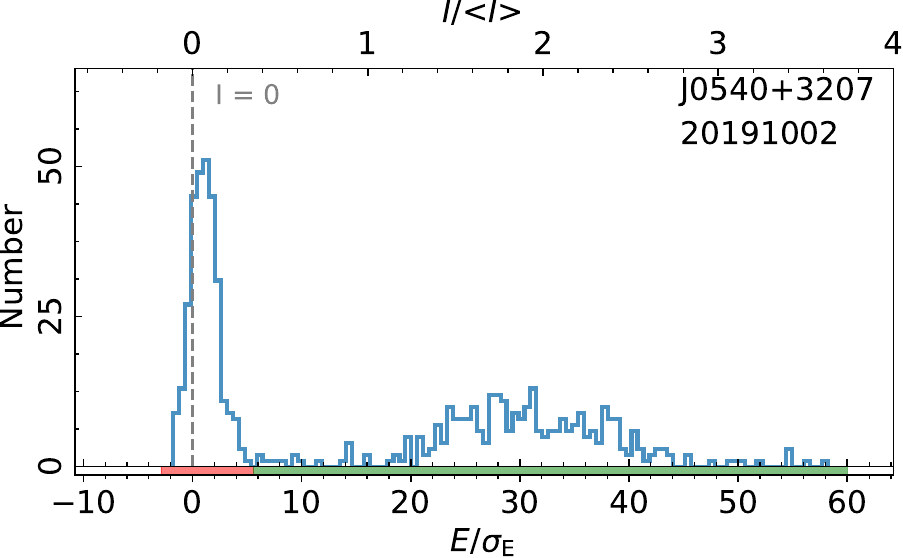}
\figcaption{On-pulse energy histograms of PSR J0540+3207 from the FAST observation on 20191002. \label{subfig:Hist:J0540+3207}}
\end{figure}

\begin{figure}[hbpt]
\centering
\includegraphics[width=0.42\textwidth, angle=0]{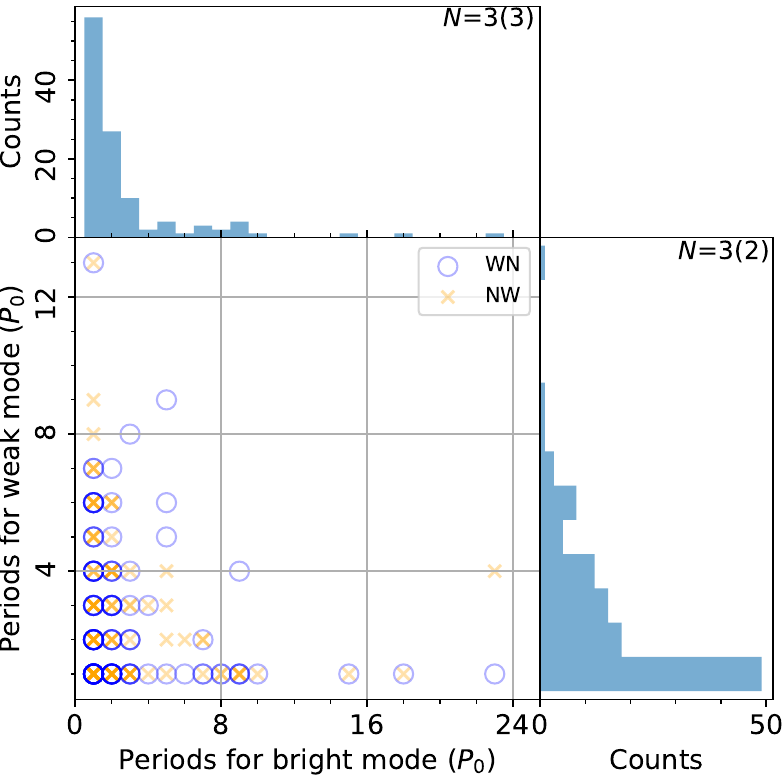}
\figcaption{Similar to Fig.~\ref{subfig:scaleHist:J0540+3207}, but for PSR J0540+3207 from the observation on 20191002. \label{subfig:scaleHist:J0540+3207}}
\end{figure}

\begin{figure}[hbpt]
\centering
\includegraphics[width=0.39\textwidth, angle=0]{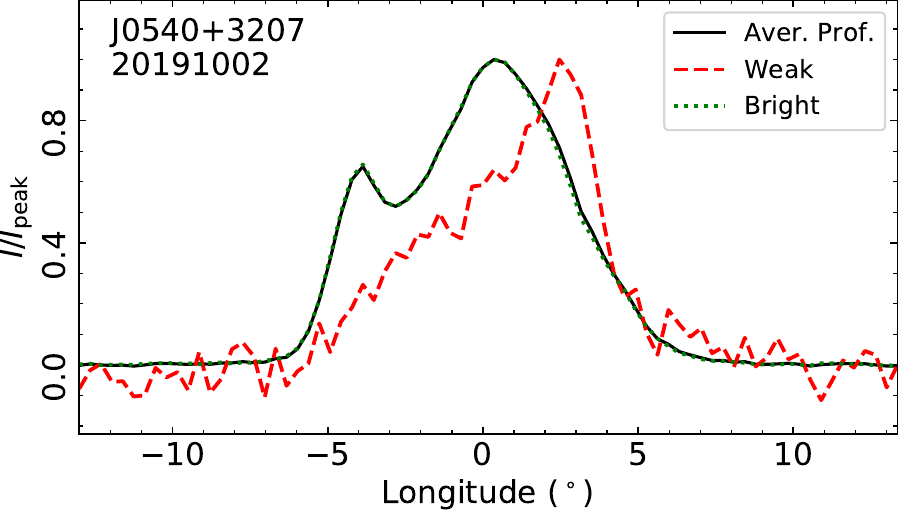}
\figcaption{Mean profiles for the weak and bright emission modes of PSR J0540+3207 from the observation on 20191002. Profiles are normalized by their respective peaks. \label{subfig:ProfModes:J0540+3207}}
\end{figure}

\subsection{J0538+2817}
\label{subsec:J0538+2817}

PSR J0538+2817 was discovered by the Arecibo telescope \citep{Foster1995}. 

This pulsar was observed by FAST on 20190320 for 5 minutes, deriving a rotation period $P=0.1432$~s and a dispersion measure $D\!M=40.0~{\rm cm^{-3}\,pc}$. Single pulse sequences in Fig.~\ref{subfig:TP:J0538+2817} show the subpulse modulation phenomenon. LRFS and 2DFS of the leading and trailing parts in the mean pulse profile are displayed in Fig.~\ref{subfig:TP:J0538+2817}. In the 2DFS of the leading profile part, there are a positive and a negative drift features and the centroids are respectively: $1/P_3=0.0167\pm0.0003$ ($P_3=60\pm1$ periods) and $1/P_2=5\pm1$ ($P_2=76\pm9$ degrees); $1/P_3=0.0507\pm0.0004$ ($P_3=19.7\pm0.2$ periods) and $1/P_2=-6\pm1$ ($P_2=-65\pm9$ degrees). The trailing part exhibits a low-frequency modulation feature in 2DFS with the centroid frequency of $1/P_3=0.0336\pm0.0003$ ($P_3=29.7\pm0.3$ periods), and a negative drift feature with the centroid at $1/P_3=0.089\pm0.001$ ($P_3=11.2\pm0.1$ periods) and $1/P_2=-51.5\pm0.4$ ($P_2=-6.99\pm0.05$ degrees).

\subsection{J0540+3207}
\label{subsec:J0540+3207}

PSR J0540+3207 was discovered by the Arecibo telescope, emitting strong, sporadic single pulses \citep{Cordes2006}. This pulsar has been reported by \citet{Herfindal2009} and \citet{Redman2009} with a nulling fraction of 53\% at 327MHz. 

The pulsar was observed by FAST on 20191002 for 5 minutes, with a rotation period $P=0.5242$~s and a dispersion measure $D\!M=62.1~{\rm cm^{-3}\,pc}$ from this observation. 
Single pulse sequences and two segments of zoomed-in views are displayed in Fig.~\ref{subfig:TP:J0540+3207}. 
From the zoomed-in view of the single pulse stack of pulses No.351-429 (the bottom panel of Fig.~\ref{subfig:TP:J0540+3207}), there is very weak emission, such as pulses No. 369, 370, and 380. Furthermore, the distribution around the 0 energy seems not symmetric about the vertical axis in the on-pulse integral energy histogram (Fig.~\ref{subfig:Hist:J0540+3207}). These evidences indicate the existence of weak emission instead of nulls. 
The on-pulse integral energy histogram in Fig.~\ref{subfig:Hist:J0540+3207} is used to distinguish weak and bright emission modes, where weak and bright emission modes are labeled by red and green colors. 

From the two bottom single pulse sequences in Fig.~\ref{subfig:TP:J0540+3207}, the duration of the weak mode or bright mode can be either one period or several periods. From the duration distributions of two emission modes in Fig.~\ref{subfig:scaleHist:J0540+3207}, the duration ranges of the weak and bright modes are 1-13 $P_0$ and 1-23 $P_0$, and their distribution peaks are both at 1 $P_0$. Profile contrast is illustrated in Fig.~\ref{subfig:ProfModes:J0540+3207}. The peak of the weak mode is lagged relative to the normal mode. 

\begin{figure}[htpb]
\includegraphics[width=0.22\textwidth, angle=0]{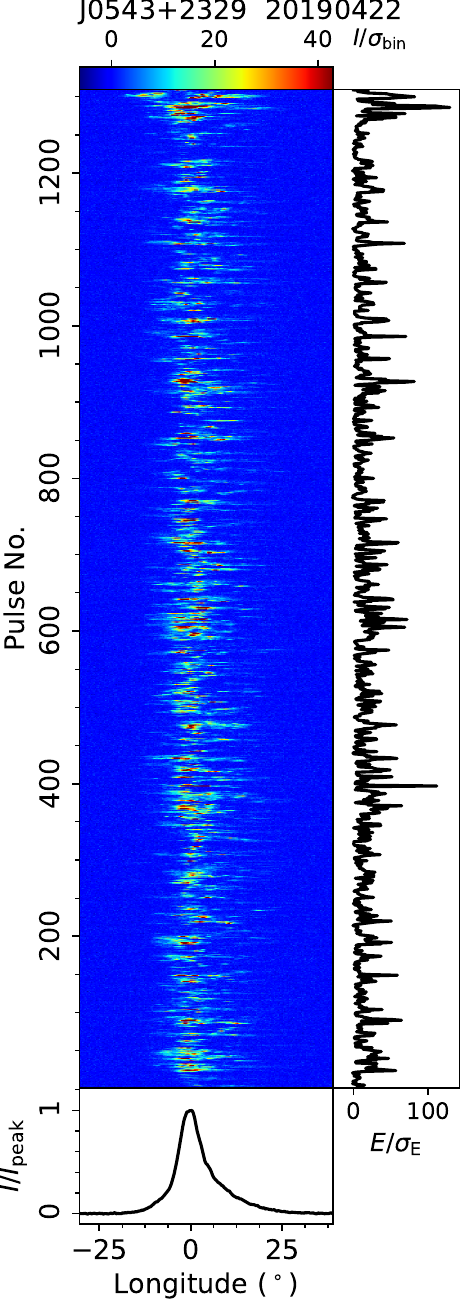}
\includegraphics[width=0.22\textwidth, angle=0]{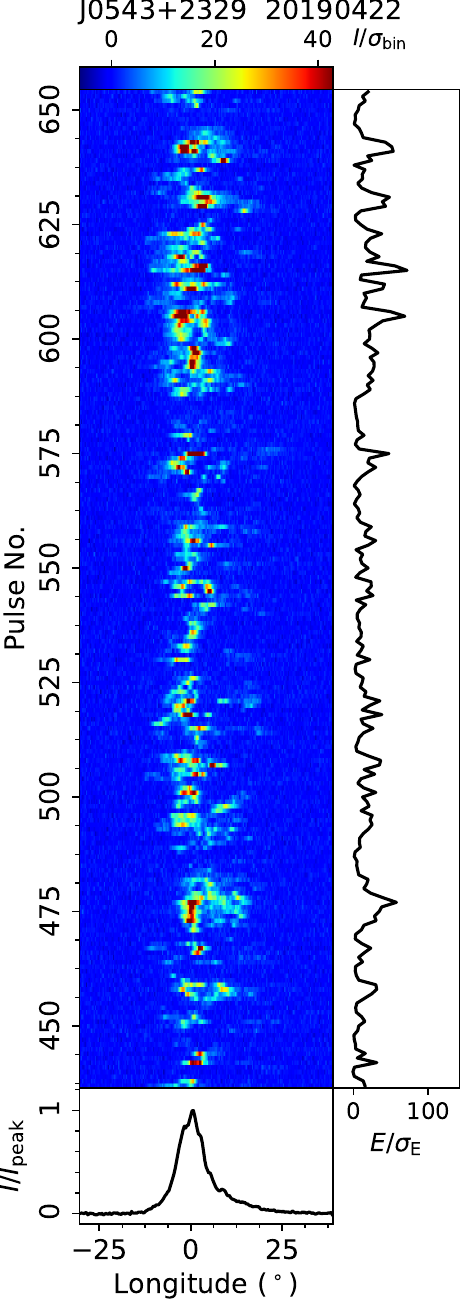}
\figcaption{Single pulse sequences of PSR J0543+2329 from the FAST observation on 20190422. \label{subfig:TP:J0543+2329}}
\end{figure}

\begin{figure}[htpb]
\centering
\includegraphics[width=0.22\textwidth, angle=0]{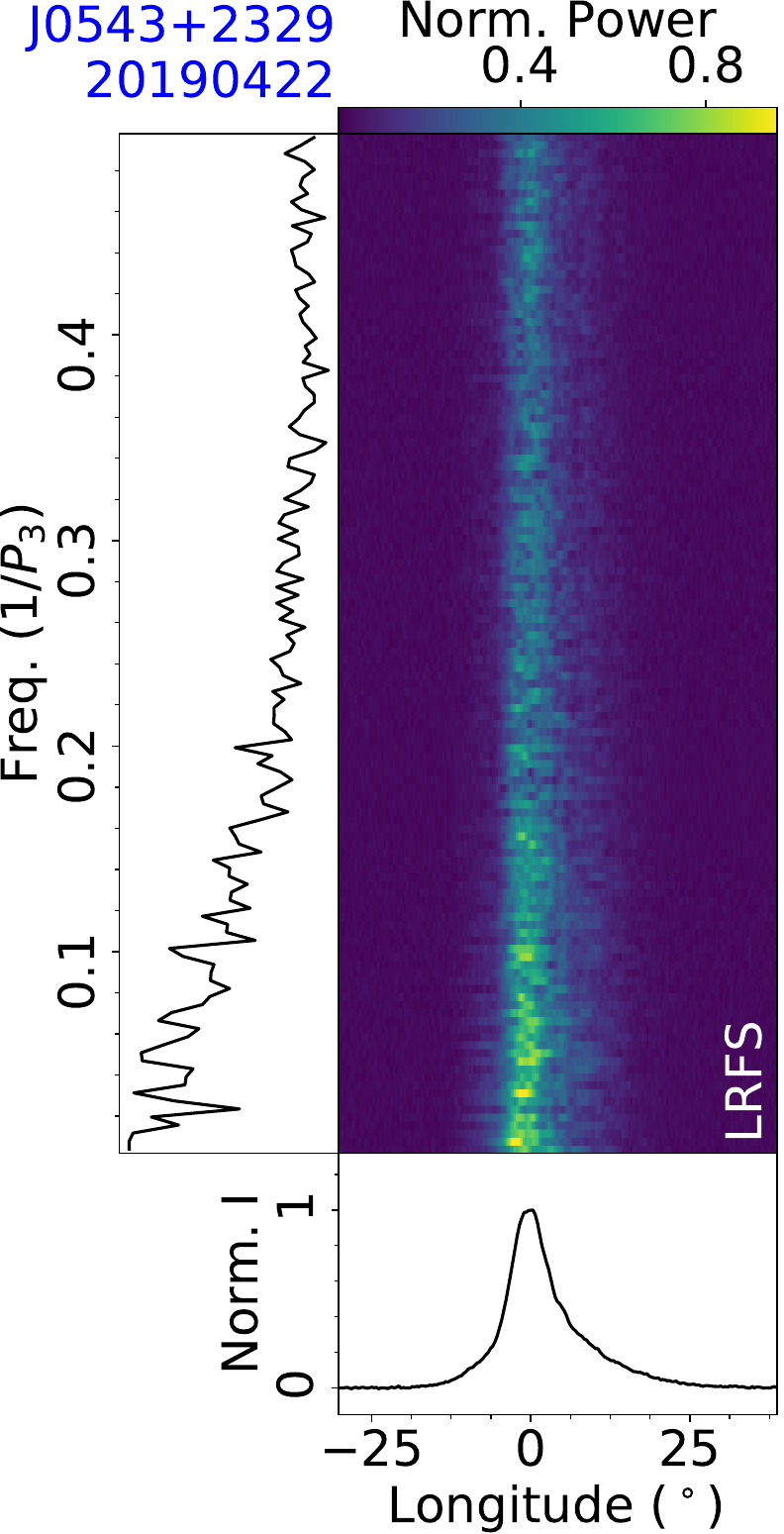}
\includegraphics[width=0.22\textwidth, angle=0]{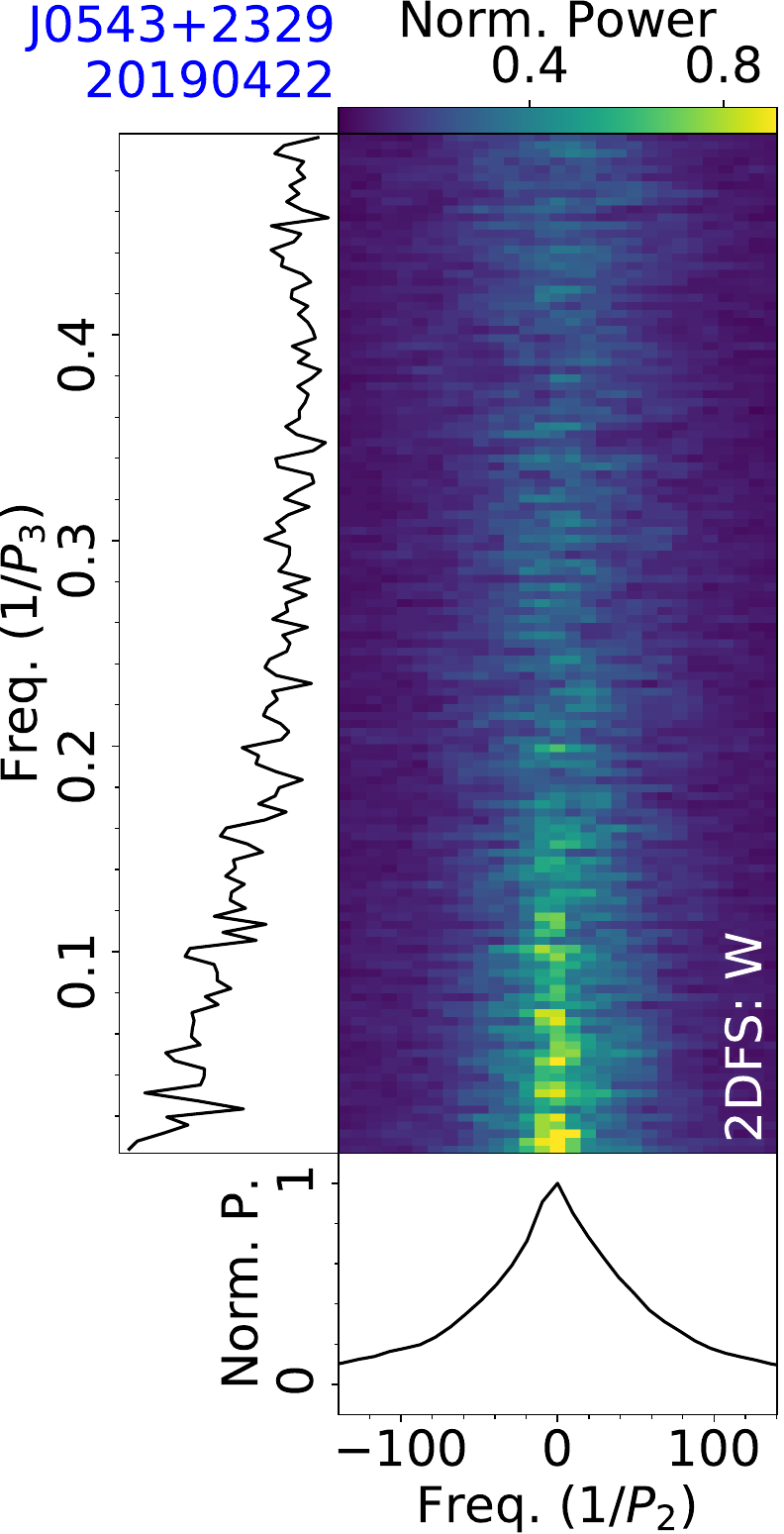}
\figcaption{Fluctuation analysis of PSR J0543+2329 from the observation on 20190422, with LRFS and 2DFS for the whole pulse phase range of a mean pulse profile.  \label{subfig:fluctu:J0543+2329}}
\end{figure}

\begin{figure}[htpb]
\centering
\includegraphics[width=0.22\textwidth, angle=0]{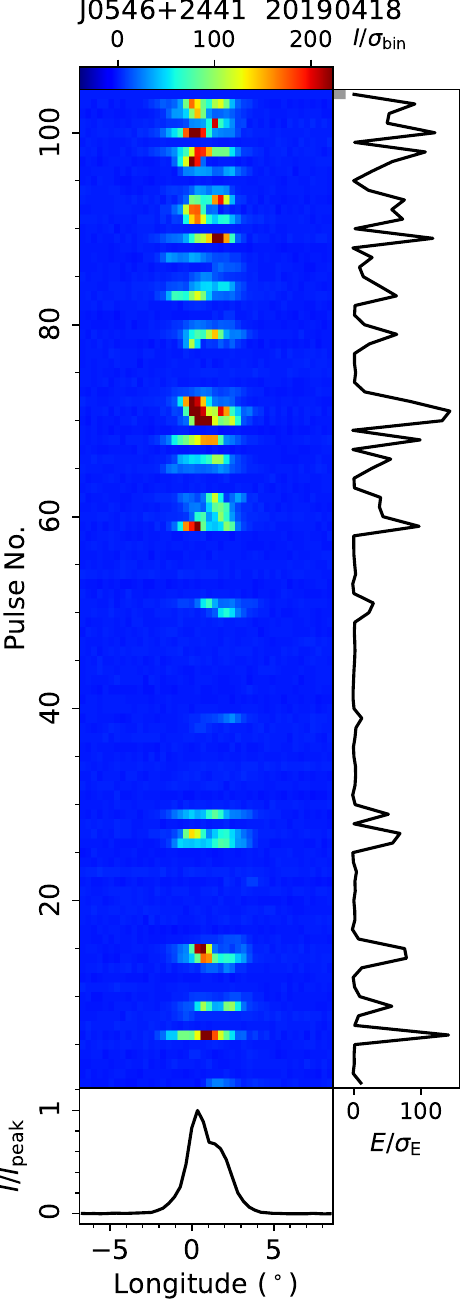}
\figcaption{Single pulse sequence of PSR J0546+2441 from the observation on 20190418.
\label{subfig:TP:J0546+2441}}
\end{figure}

\begin{figure}[htpb]
\centering
\includegraphics[width=0.39\textwidth, angle=0]{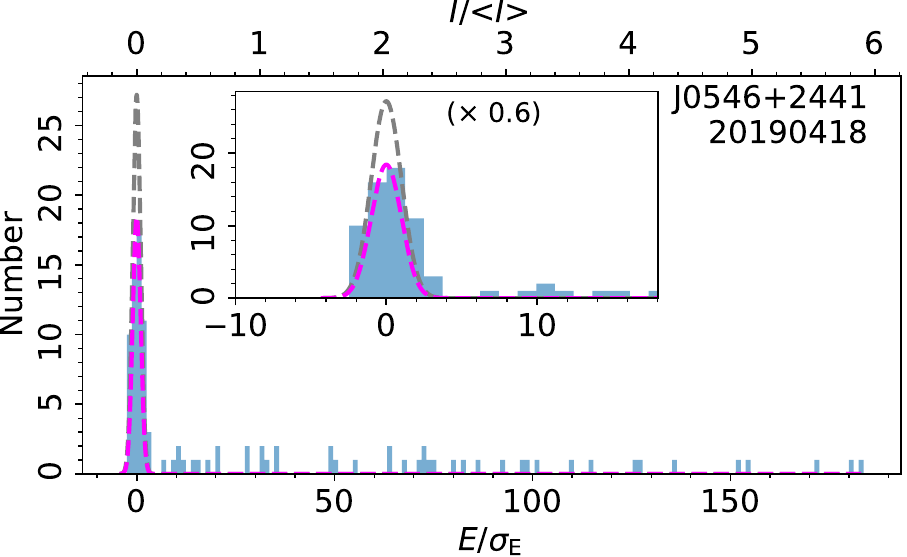}
\figcaption{On-pulse energy histogram of PSR J0546+2441 from the observation on 20190418.
\label{subfig:Hist:J0546+2441}}
\end{figure}

\begin{figure}[htpb]
\centering
\includegraphics[width=0.39\textwidth, angle=0]{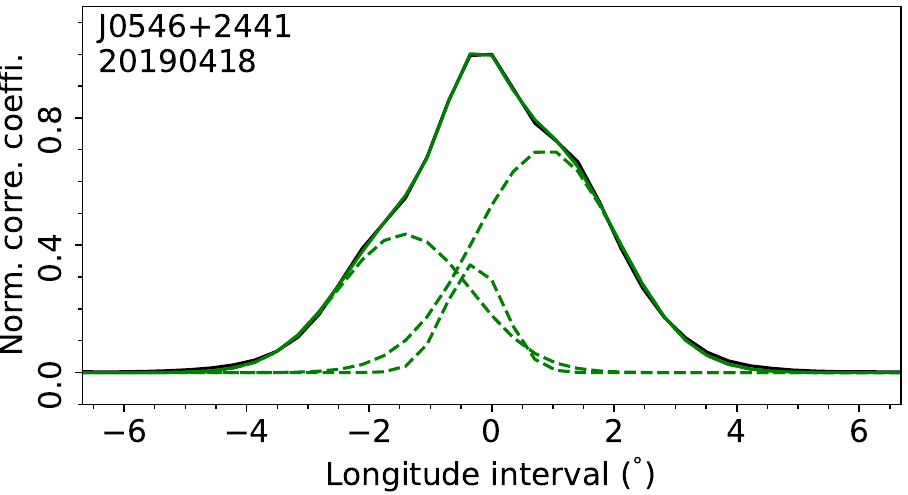}
\figcaption{Cross correlation of PSR J0546+2441 from the FAST observation on 20190418.
\label{subfig:Corre:J0546+2441}}
\end{figure}

\begin{figure}[htpb]
\centering
\includegraphics[width=0.22\textwidth, angle=0]{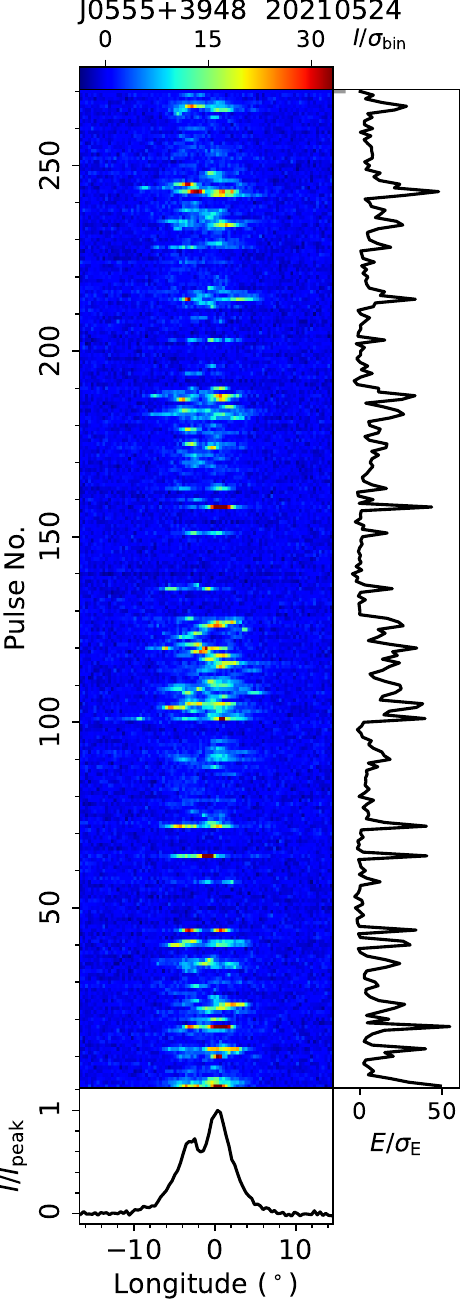}
\figcaption{Single pulse sequence of PSR J0555+3948 from the observation on 20210524. \label{subfig:TP:J0555+3948}}
\end{figure}

\begin{figure}[htpb]
\centering
\includegraphics[width=0.39\textwidth, angle=0]{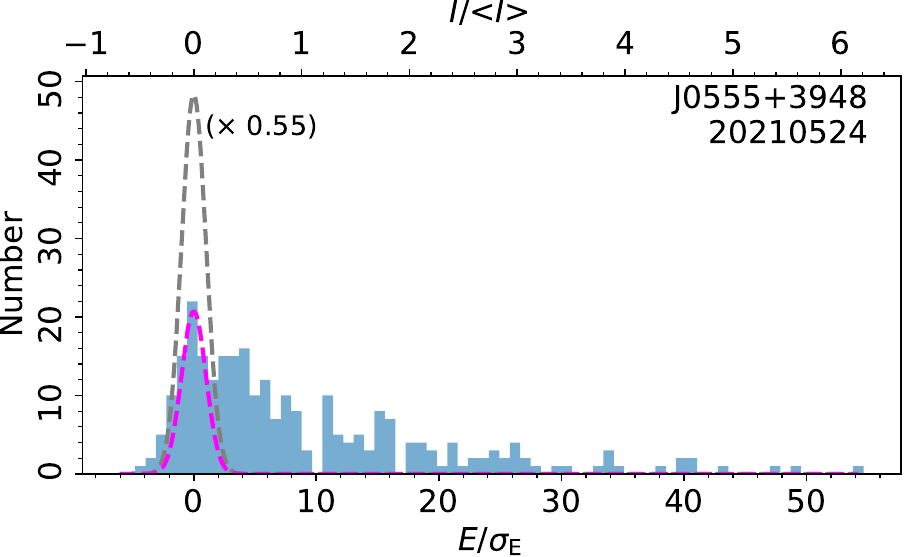}
\figcaption{On-pulse energy histogram of individual pulses of PSR J0555+39481 from the observation on 20210524. \label{subfig:Hist:J0555+3948}}
\end{figure}

\subsection{J0543+2329}
\label{subsec:J0543+2329}

PSR J0543+2329 was discovered by the Jodrell Bank Mark IA radio telescope at 408 MHz \citep{Davies1972}. This pulsar was reported to have mode changing behavior. \citet{Nowakowski1991} took repeated short segments of single pulses behaving in a similar way as different modes at 430 MHz, and the typical duration is about 5 pulses. 
Modulation behavior but no preferred drift direction were reported by \citet{Nowakowski1991} at 430 MHz and \citet{Weltevrede2006,Weltevrede2007} at 21 cm and 92 cm.

This pulsar was observed by FAST on 20190422 for 5 minutes, with a rotation period $P=0.2460$~s and a dispersion measure $D\!M=77.5~{\rm cm^{-3}\,pc}$ from this observation. The modulation behavior could be confirmed by FAST observation from single pulse sequences (Fig.~\ref{subfig:TP:J0543+2329}) and fluctuation spectra (Fig.~\ref{subfig:fluctu:J0543+2329}). For the modulation feature in 2DFS, the temporal frequency of the centroid is estimated to $1/P_3=0.080\pm0.001$, which corresponds to $P_3=P_3=12.6\pm0.1$ periods.


\begin{figure}[hbpt]
\centering
\includegraphics[width=0.22\textwidth, angle=0]{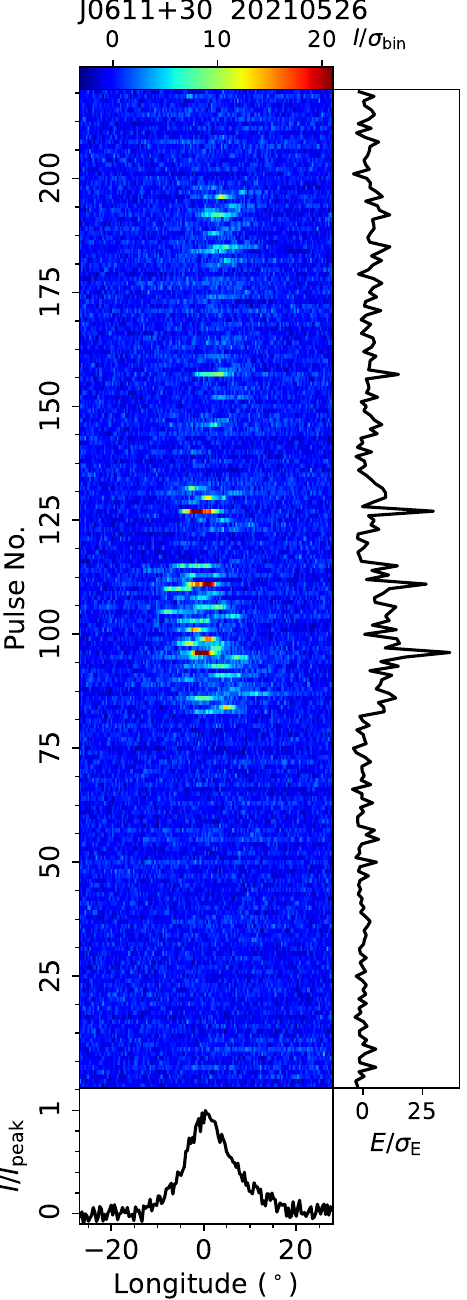}
\figcaption{Single pulse sequence of PSR J0611+30 from the FAST observation on 20210526. \label{subfig:TP:J0611+30}}
\end{figure}

\begin{figure}[hbpt]
\centering
\includegraphics[width=0.39\textwidth, angle=0]{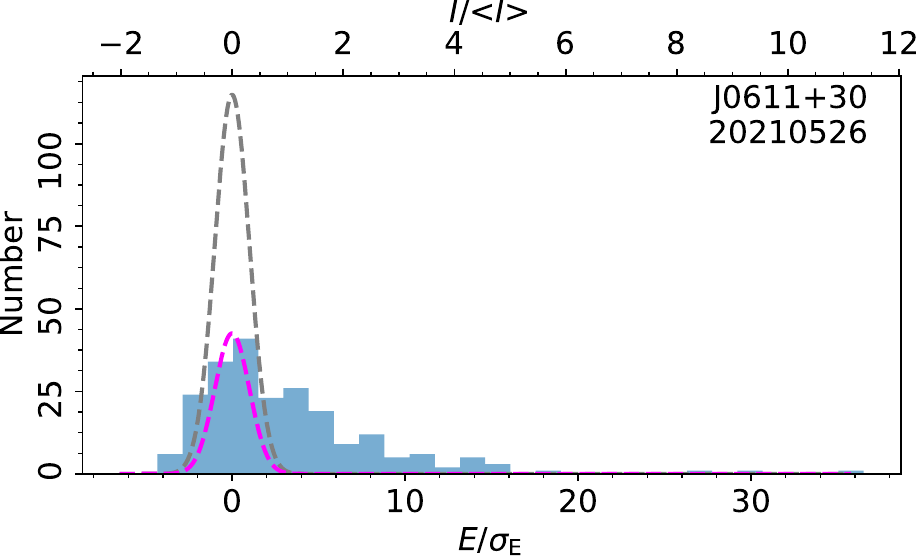}
\figcaption{On-pulse energy histogram of individual pulses of PSR J0611+30 from the FAST observation on 20210526.\label{subfig:Hist:J0611+30}}
\end{figure}

\subsection{J0546+2441}
\label{subsec:J0546+2441}

PSR J0546+2441 was discovered by \citet{Champion2005} using the Arecibo telescope at 430 MHz. 

The pulsar was observed by FAST on 20190418 for 5 minutes, with a rotation period $P=2.8438$~s and a dispersion measure $D\!M=72.1~{\rm cm^{-3}\,pc}$. 
The single pulse sequence in Fig.~\ref{subfig:TP:J0546+2441} shows the nulling and negative subpulse drifting phenomena. The nulling fraction is estimated to be 41$\pm$5\% from the on-pulse energy histogram (Fig.~\ref{subfig:Hist:J0546+2441}). The drifting parameters are derived from the cross-correlation method (Fig.~\ref{subfig:Corre:J0546+2441}), which are $P_2=1.15\pm0.03$ degrees and $D=-0.27\pm0.05$ degrees per period.

\begin{figure}[hbpt]
\centering
\includegraphics[width=0.22\textwidth, angle=0]{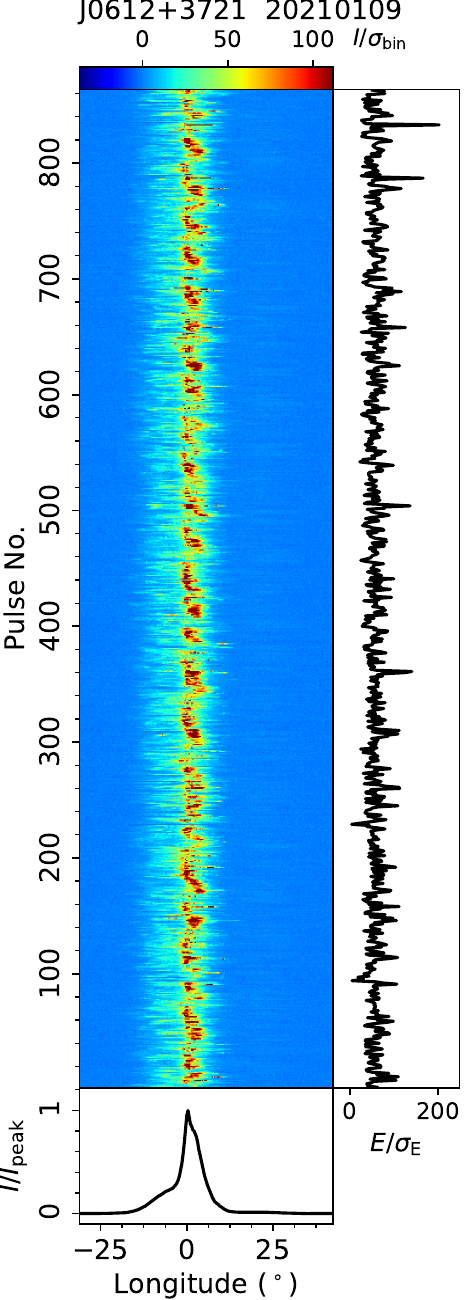} 
\includegraphics[width=0.22\textwidth, angle=0]{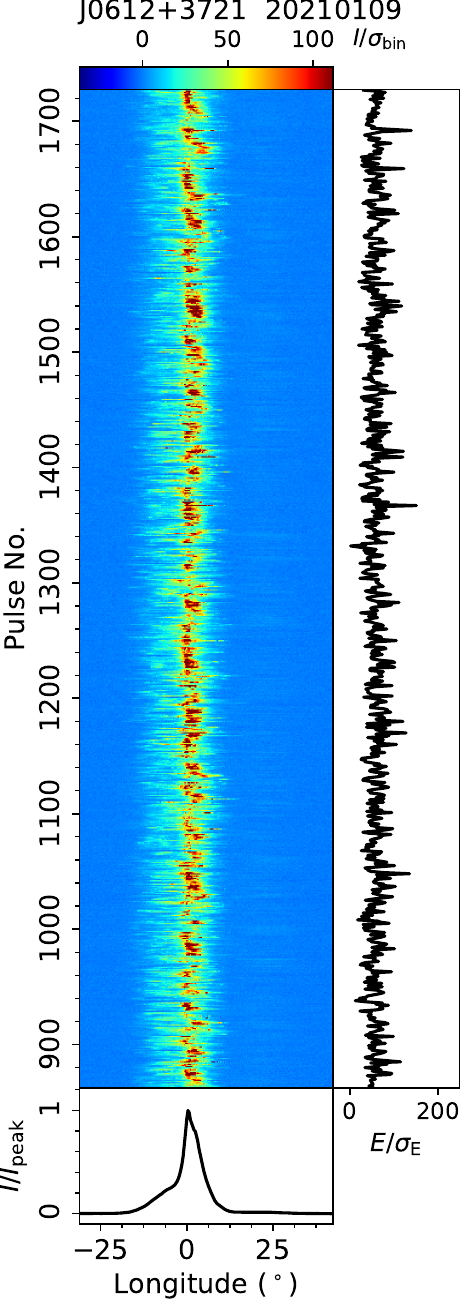}
\figcaption{Single pulse sequences of PSR J0612+3721 from the FAST observation on 20210109. \label{subfig:TP:J0612+3721}}
\end{figure}

\begin{figure}[hbpt]
\centering
\includegraphics[width=0.22\textwidth, angle=0]{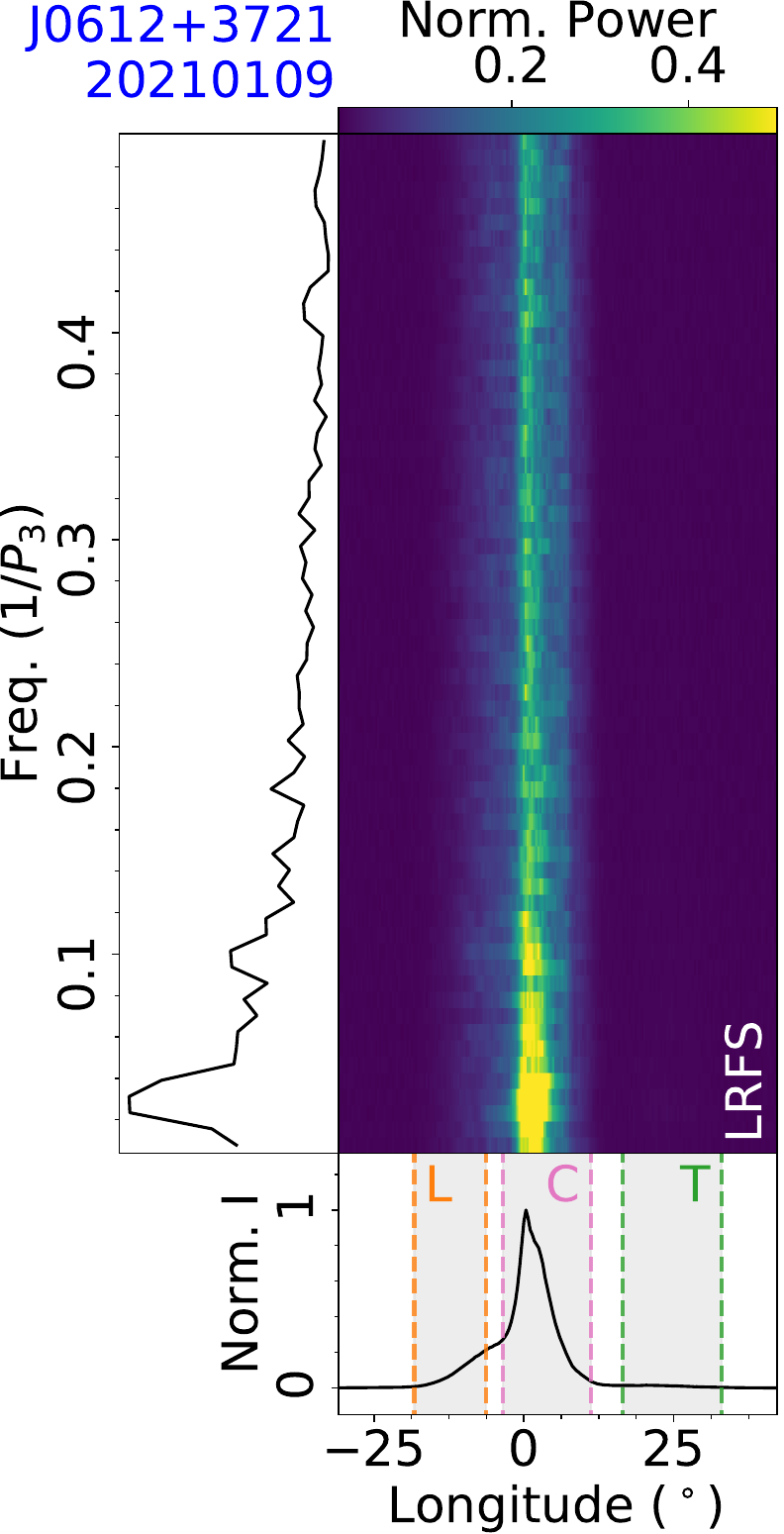}
\includegraphics[width=0.22\textwidth, angle=0]{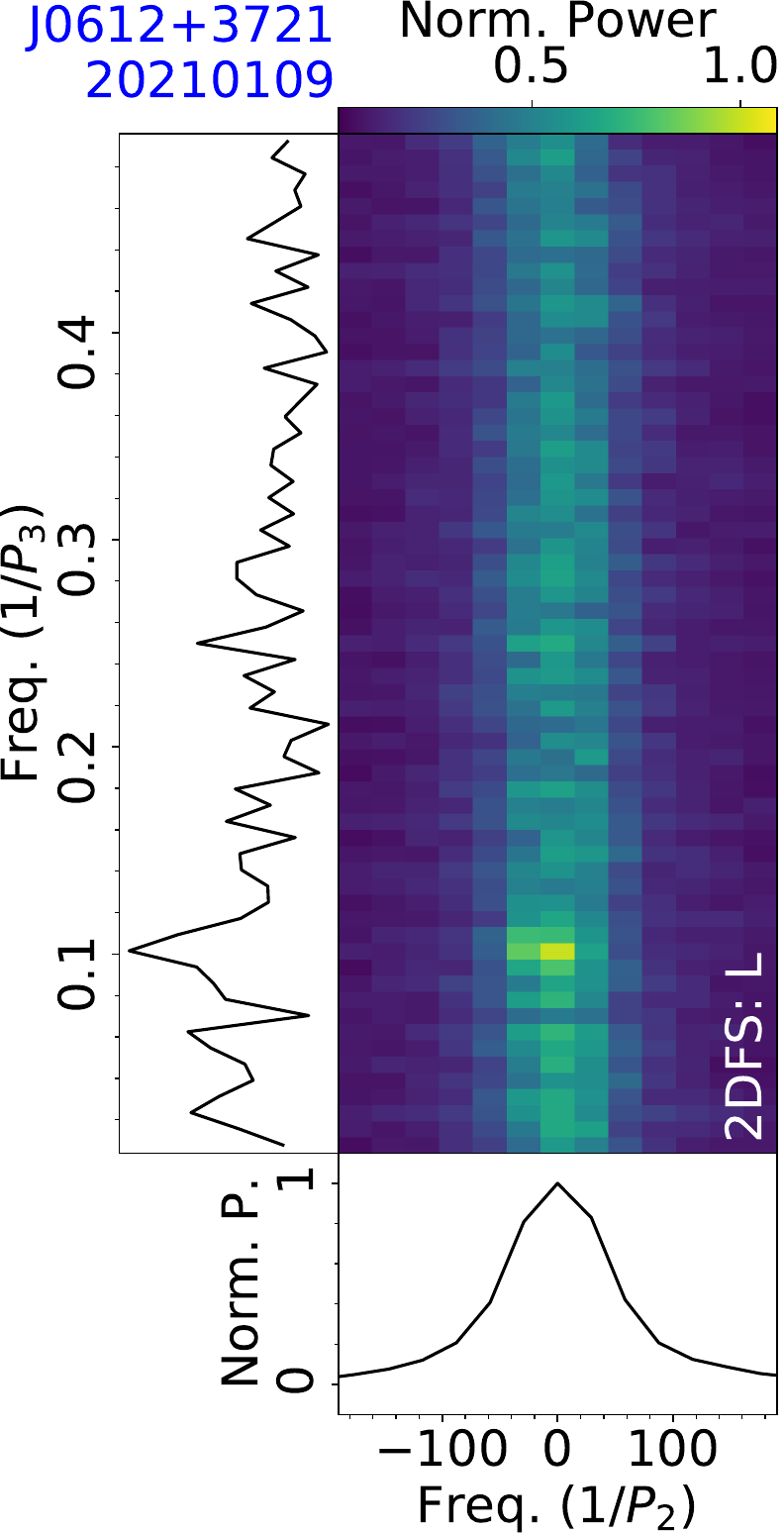}\\
\includegraphics[width=0.22\textwidth, angle=0]{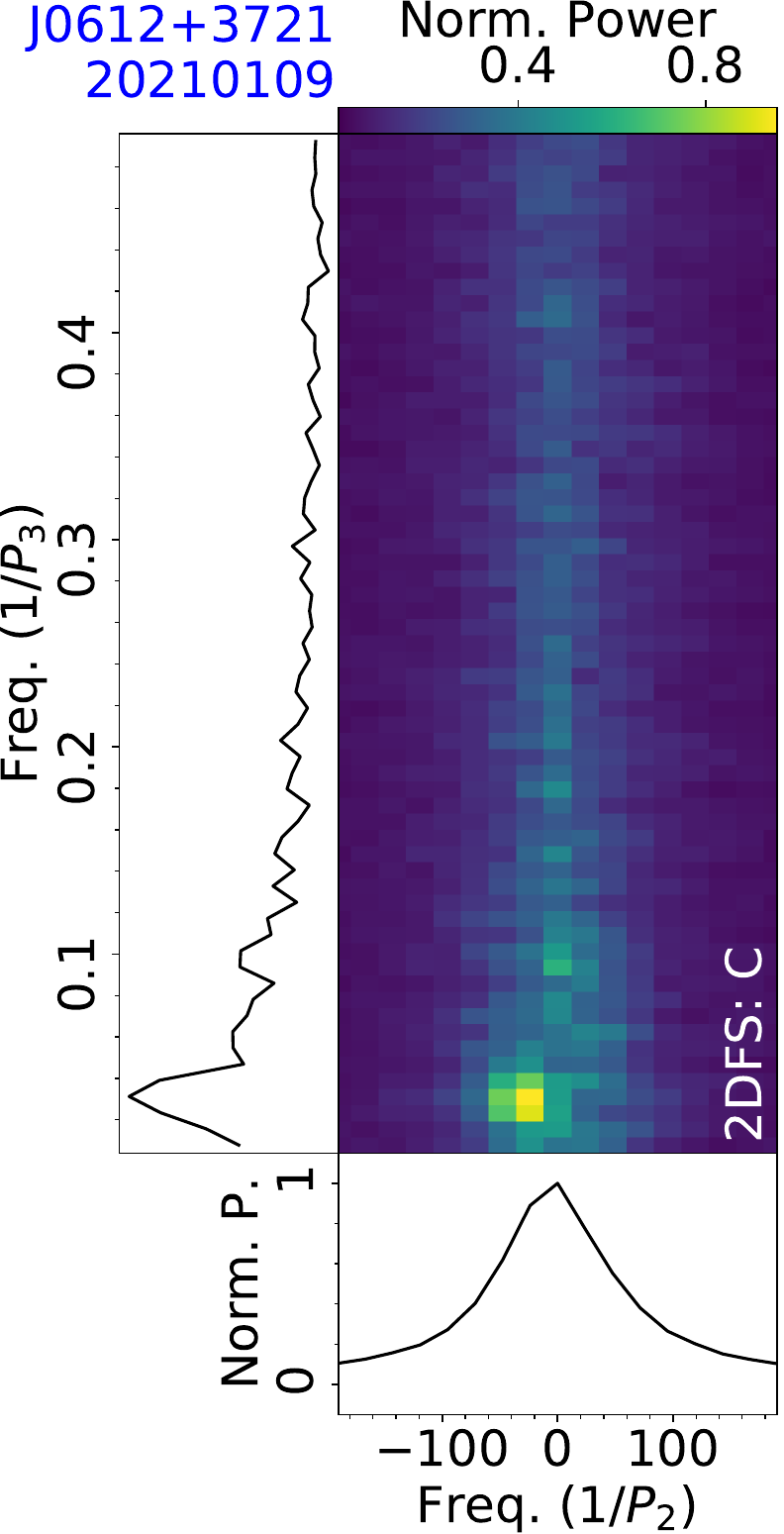} 
\includegraphics[width=0.22\textwidth, angle=0]{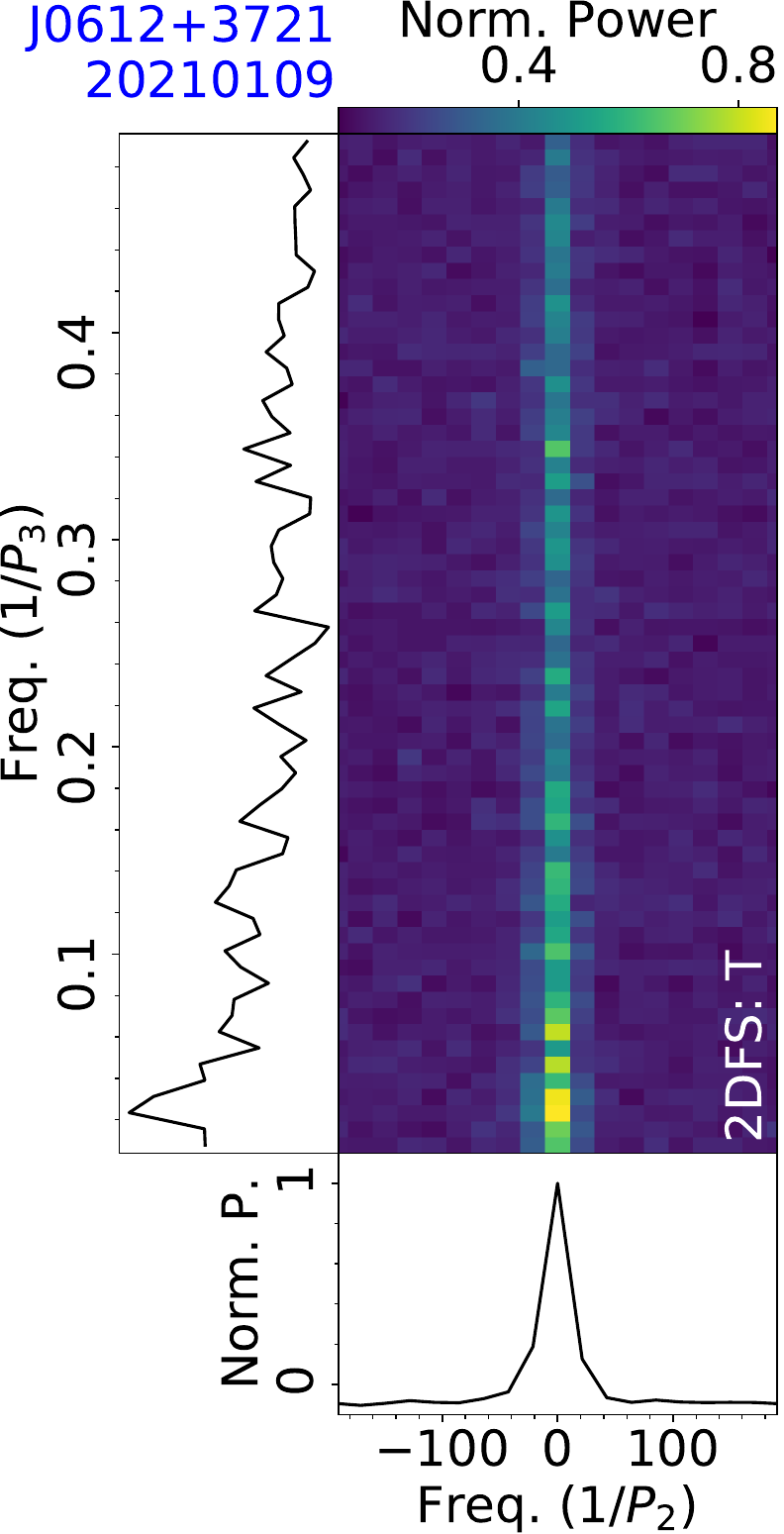}
\figcaption{Fluctuation analysis of PSR J0612+3721 from the FAST observation on 20210109, with LRFS (top-left), and 2DFS for the leading part (top-right), the central part (bottom left) and trailing part (bottom-right) of a mean pulse profile.
\label{subfig:fluctu:J0612+3721}}
\end{figure}

\subsection{J0555+3948}
\label{subsec:J0555+3948}

PSR J0555+3948 was discovered in the Northern High Time Resolution Universe survey using the Effelsberg radio telescope \citep{Barr2013}. 
The pulsar was reported by \citet{Ng2020} to be a possible new nulling pulsar. 

This pulsar was observed by FAST on 20210524 for 5 minutes, with a rotation period $P=1.1468$~s and a dispersion measure $D\!M=35.9~{\rm cm^{-3}\,pc}$. 
From this FAST observation, the nulling behavior of the pulsar is confirmed from single pulse sequences in Fig.~\ref{subfig:TP:J0555+3948} and on-pulse integral energy histogram in Fig.~\ref{subfig:Hist:J0555+3948}, with a nulling fraction estimated to be 24$\pm$2\%. In addition to nulls and normal emission, there is also weak emission, such as single pulses from No. 70 to 95, and 218 to 227, which is also consistent with that null and pulse distributions are not well separated. Single pulse stacks also show clear drift bands from single pulses from No. 112 to 122.


\begin{figure}[hbpt]
\centering
\includegraphics[width=0.22\textwidth, angle=0]{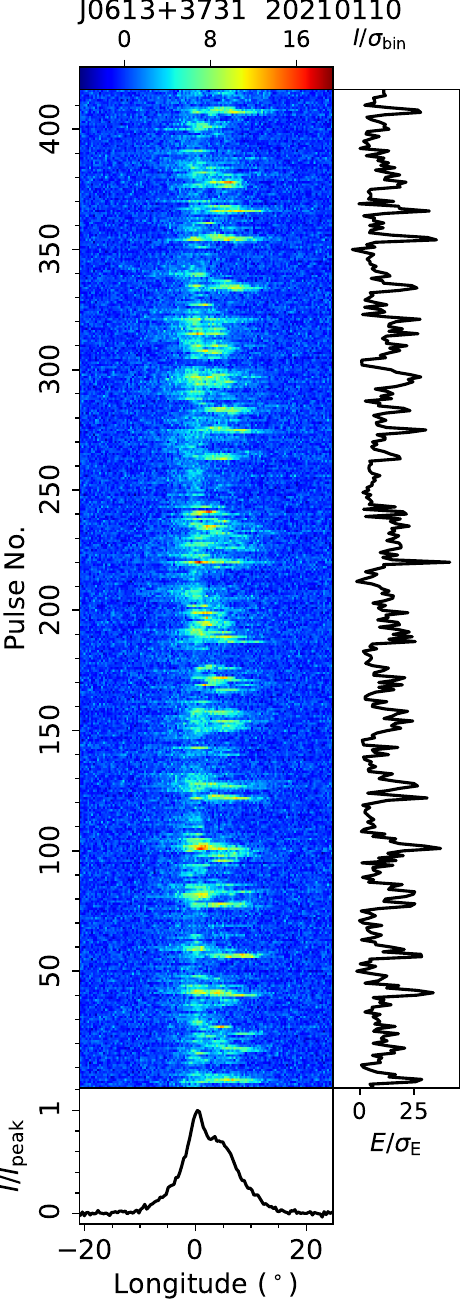}
\includegraphics[width=0.22\textwidth, angle=0]{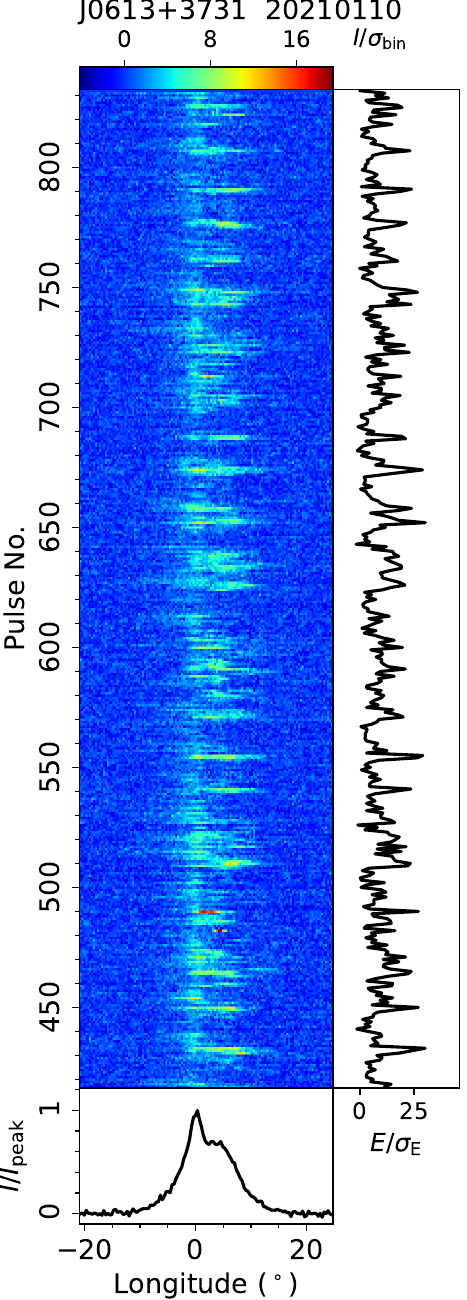}
\figcaption{Single pulse sequences of PSR J0613+3731 from the FAST observation on 20210110. \label{subfig:TP:J0613+3731}}
\end{figure}

\begin{figure}[hbpt]
\centering
\includegraphics[width=0.22\textwidth, angle=0]{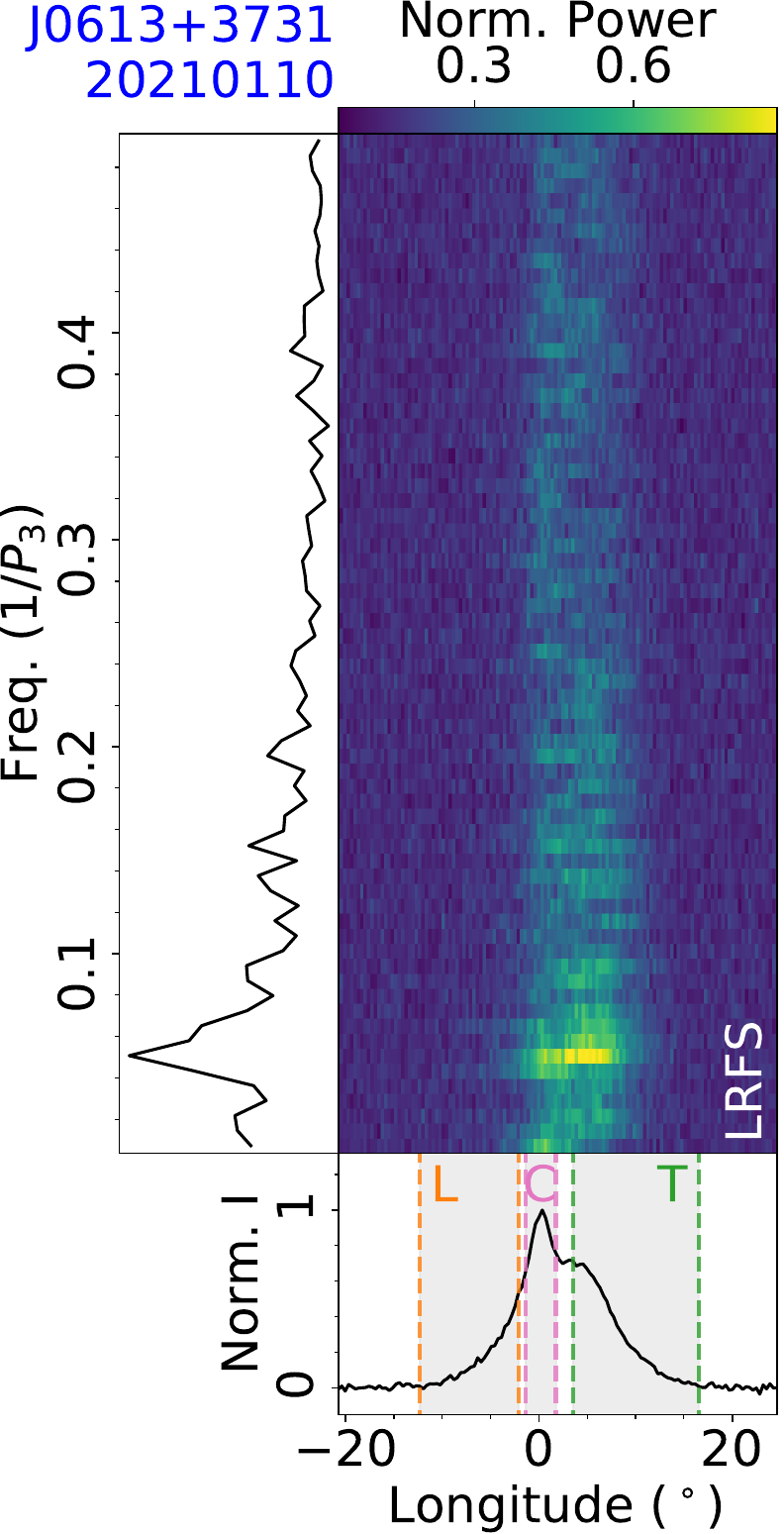}
\includegraphics[width=0.22\textwidth, angle=0]{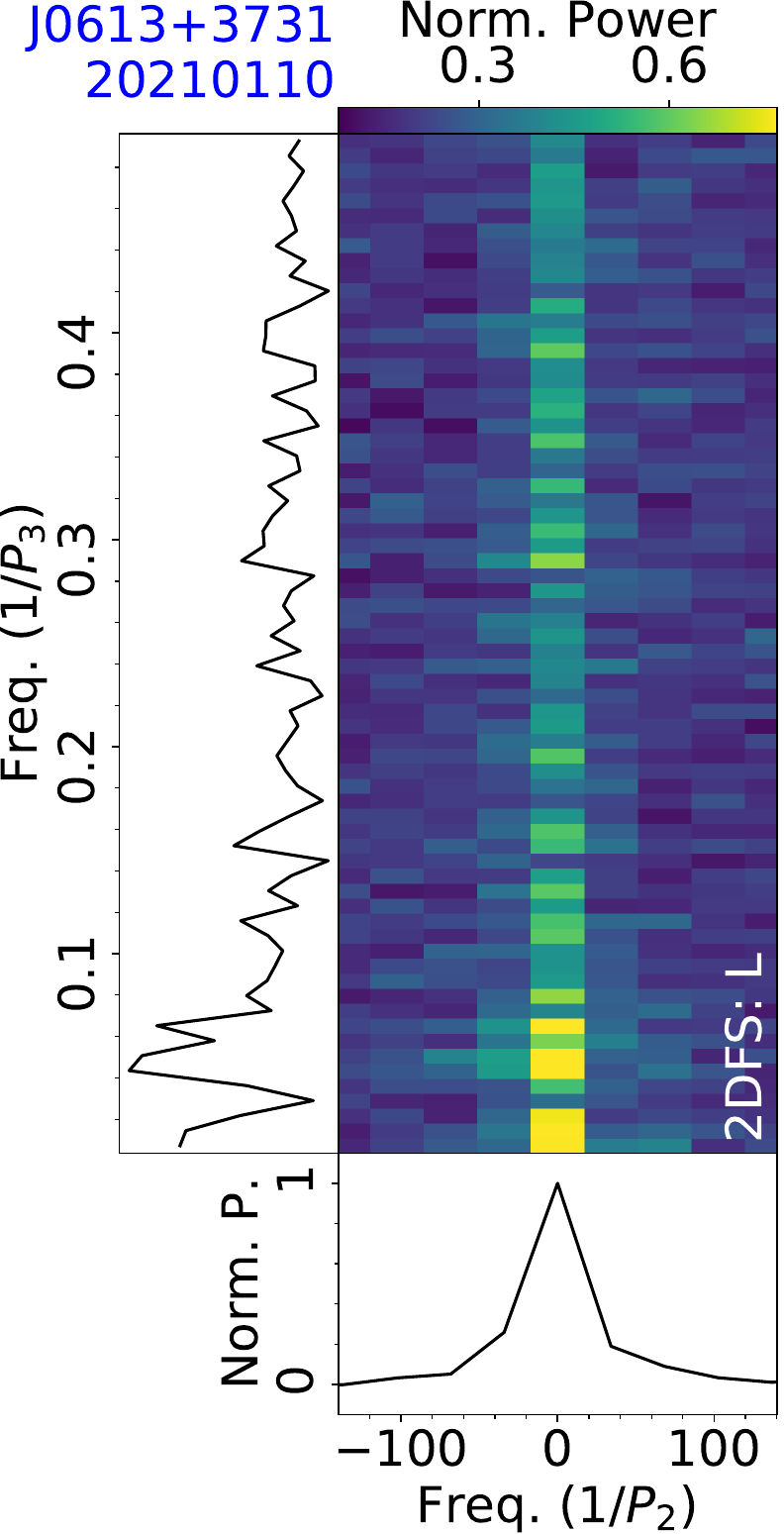}\\
\includegraphics[width=0.22\textwidth, angle=0]{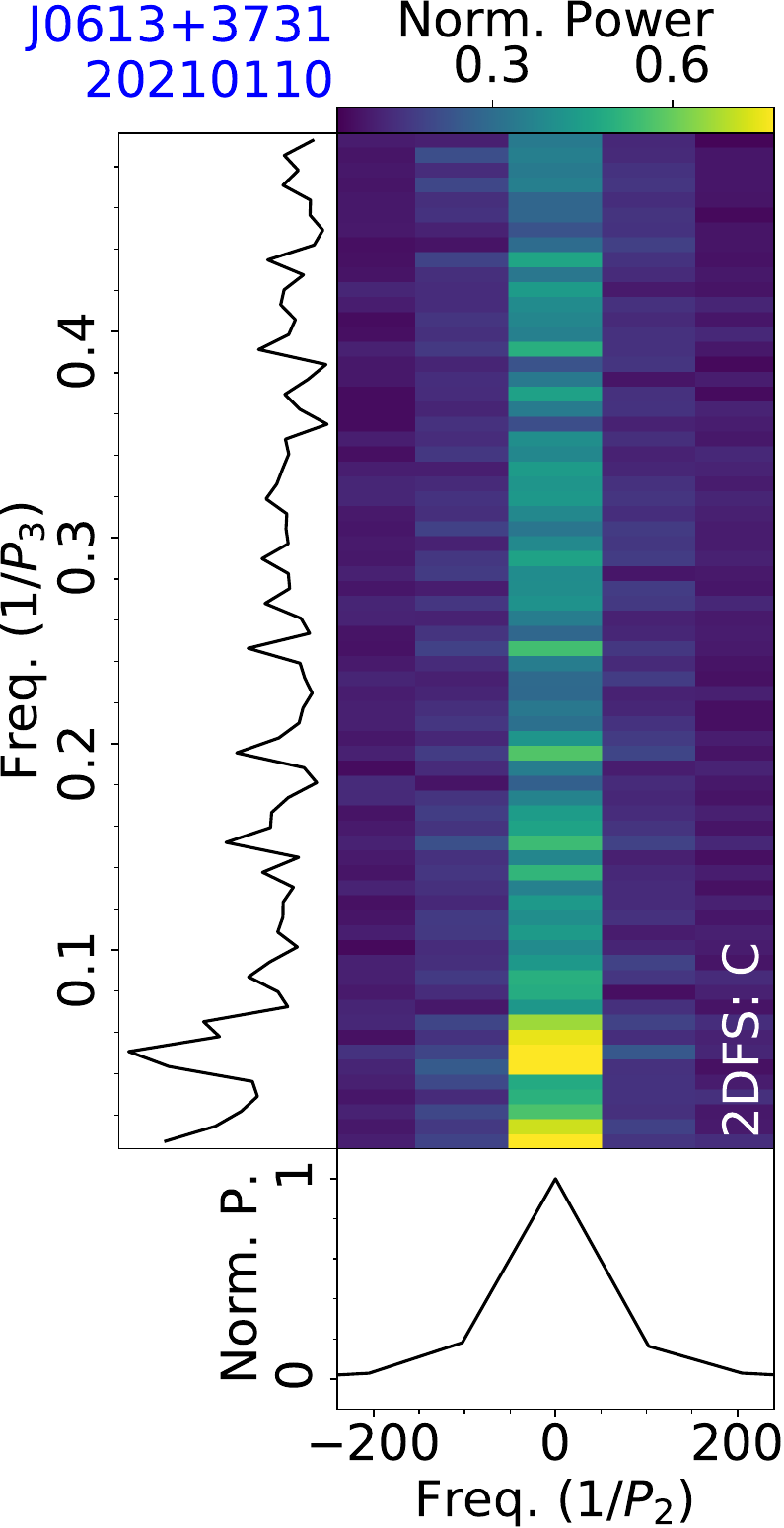}
\includegraphics[width=0.22\textwidth, angle=0]{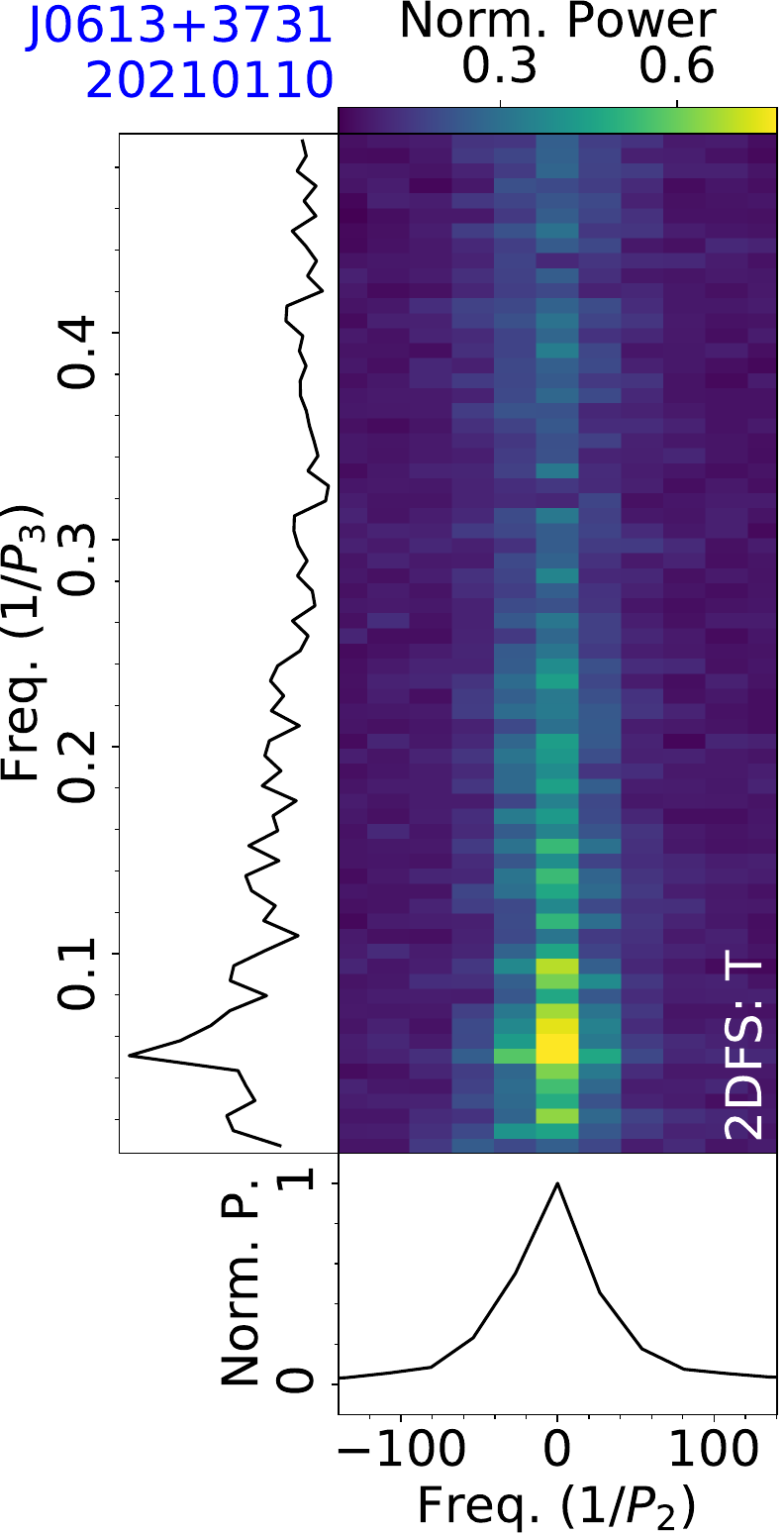}
\figcaption{Fluctuation analysis of PSR J0613+3731 from the FAST observation on 20210110, with LRFS (top-left), and 2DFS for the leading part (top-right), the central part (bottom left) and trailing part (bottom-right) of a mean pulse profile.
\label{subfig:fluctu:J0613+3731}}
\end{figure}

\subsection{J0611+30}
\label{subsec:J0611+30}

PSR J0611+30 was discovered by \citet{Camilo1996} using the Arecibo telescope at 430 MHz.

This pulsar was observed by FAST on 20210526 for 5 minutes, deriving a rotation period $P=1.4119$~s and a dispersion measure $D\!M=43.0~{\rm cm^{-3}\,pc}$. The single pulse sequence (Fig.~\ref{subfig:TP:J0611+30}) indicates the existence of nulling behavior, especially at the beginning of the observation, and the nulling fraction is estimated to be 37$\pm$7\% from the on-pulse energy histogram (Fig.~\ref{subfig:Hist:J0611+30}). Additionally, the emission longitude is variable in the single pulse sequence.

\subsection{J0612+3721}
\label{subsec:J0612+3721}

PSR J0612+3721 was discovered by \citet{Stokes1985} with the 92-m telescope at Green Bank. \citet{Weltevrede2006} reported the negative subpulse drifting of this pulsar. 

The pulsar was observed by FAST on 20210109 for 9 minutes, deriving a rotation period $P=0.2980$~s and a dispersion measure $D\!M=27.2~{\rm cm^{-3}\,pc}$. Single pulse sequences in Fig.~\ref{subfig:TP:J0612+3721} show the drifting behavior of the central component. From fluctuation spectra in Fig.~\ref{subfig:fluctu:J0612+3721}, the central component has a negative drift, and the leading and trailing components seem to have temporal modulation. The drifting parameters of the central component are estimated to be $P_3=35\pm2$ periods and $P_2=-11\pm1 ^\circ$. The trailing component has a temporal modulation of 30$\pm$4 periods. Different from other components, the leading part of the mean profile has a lower modulation periodicity of 9.8$\pm$0.2 periods.





\subsection{J0613+3731}
\label{subsec:J0613+3731}

PSR J0613+3731 was discovered by the Low-Frequency Array (LOFAR) \citep{Coenen2014}. 

This pulsar was observed by FAST on 20210110 for 9 minutes, deriving a rotation period $P=0.6192$~s and a dispersion measure $D\!M=18.6~{\rm cm^{-3}\,pc}$. 
Single pulse sequences in Fig.~\ref{subfig:TP:J0613+3731} display modulation behavior. Leading, central, and trailing phase ranges are divided in a mean profile based on the time-phase plot, and LRFS and 2DFS of these three phase ranges are shown in Fig.~\ref{subfig:fluctu:J0613+3731}. For the main modulation features in fluctuation spectra of three phase ranges, centroid temporal frequencies are estimated to be $1/P_3=0.053\pm0.001$, $0.051\pm0.001$ and $0.055\pm0.001$, yielding $P_3=19.0\pm0.4$, $19.4\pm0.5$, and $18.1\pm0.2$ periods, respectively.

\begin{figure}[hbpt]
\centering
\includegraphics[width=0.22\textwidth, angle=0]{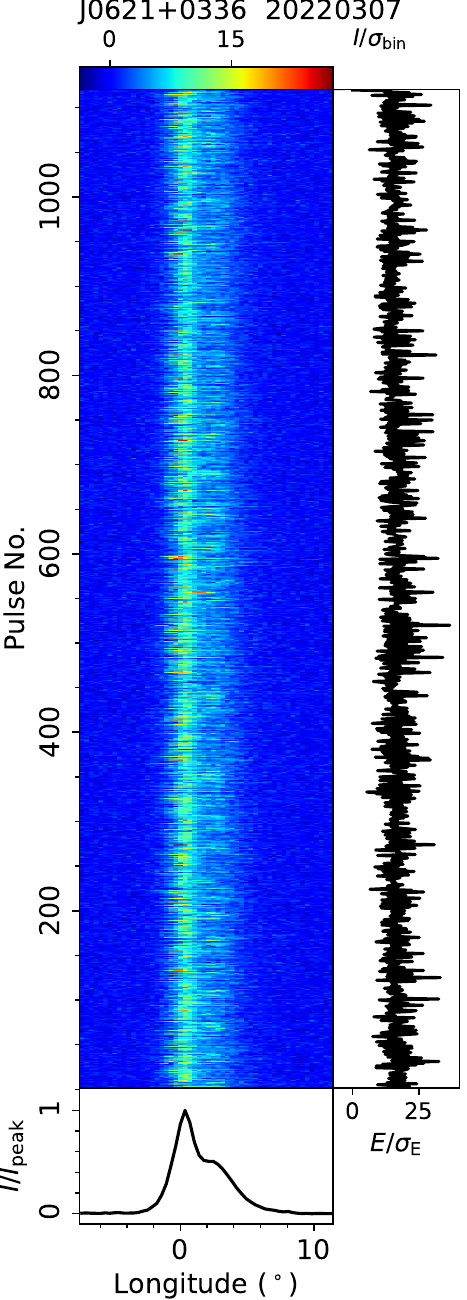} 
\includegraphics[width=0.22\textwidth, angle=0]{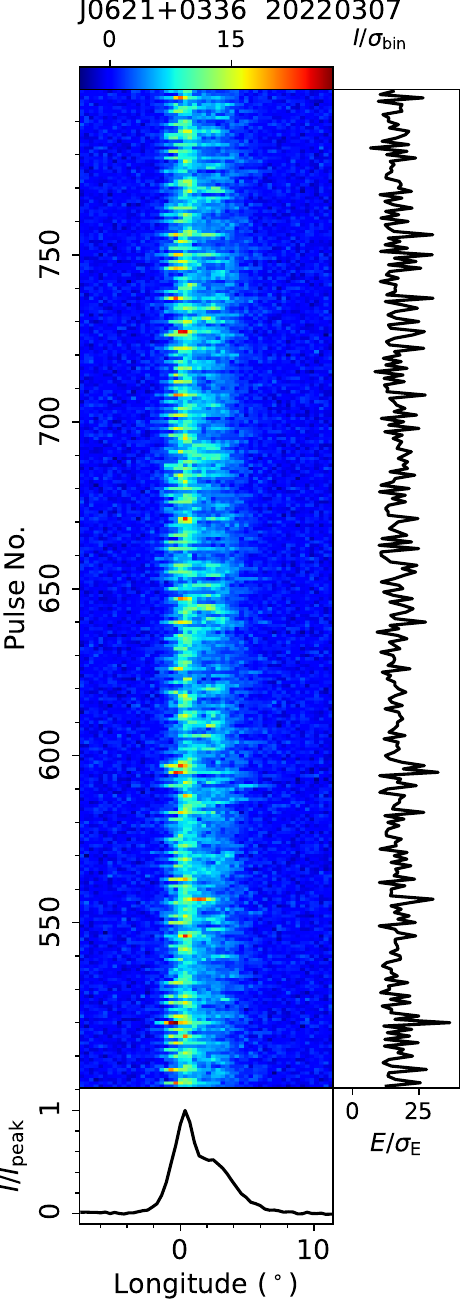}
\figcaption{Single pulse sequences of PSR J0621+0336 from the FAST observation on 20220307. \label{subfig:TP:J0621+0336}}
\end{figure}

\begin{figure}[hbpt]
\centering
\includegraphics[width=0.22\textwidth, angle=0]{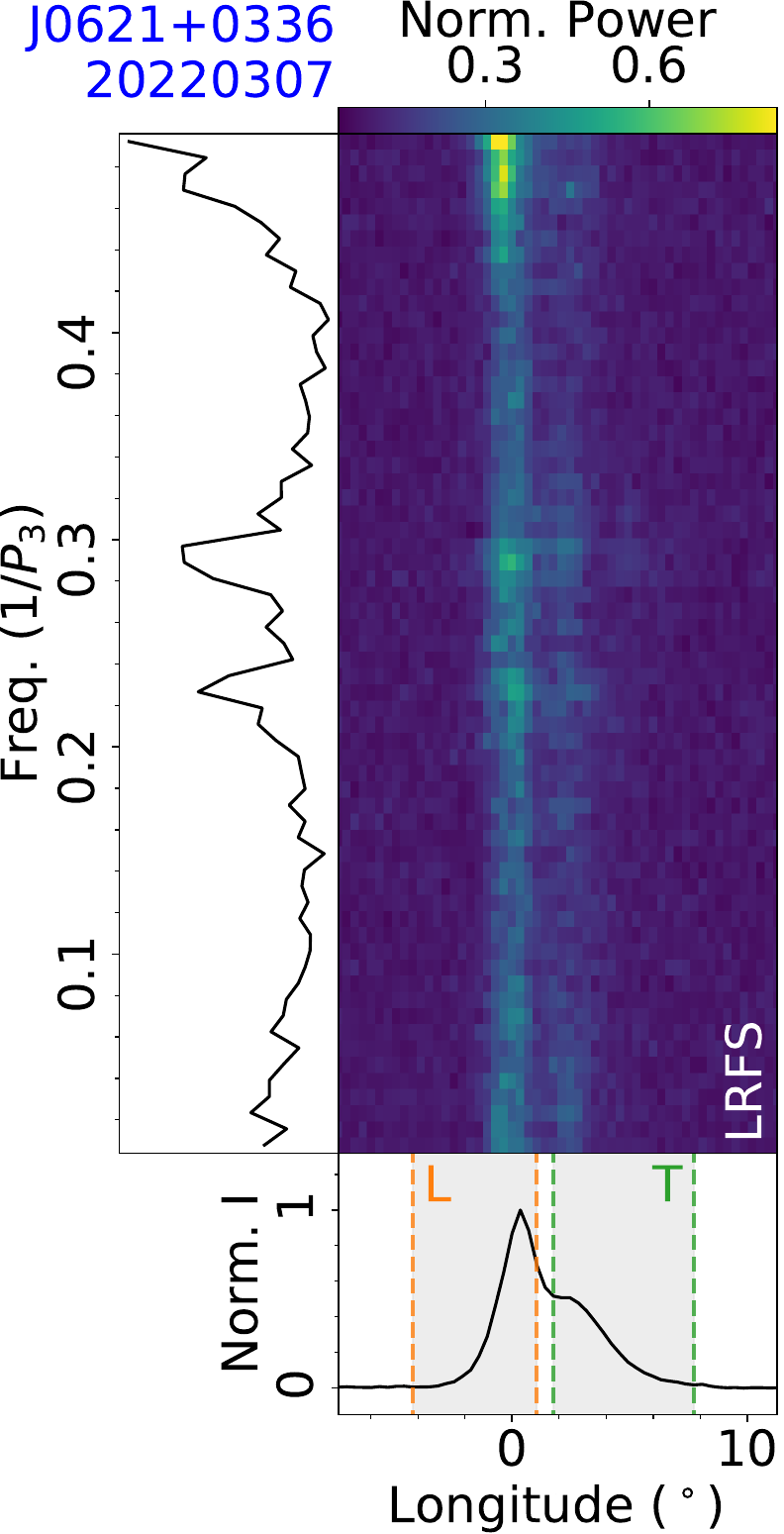} 
\includegraphics[width=0.22\textwidth, angle=0]{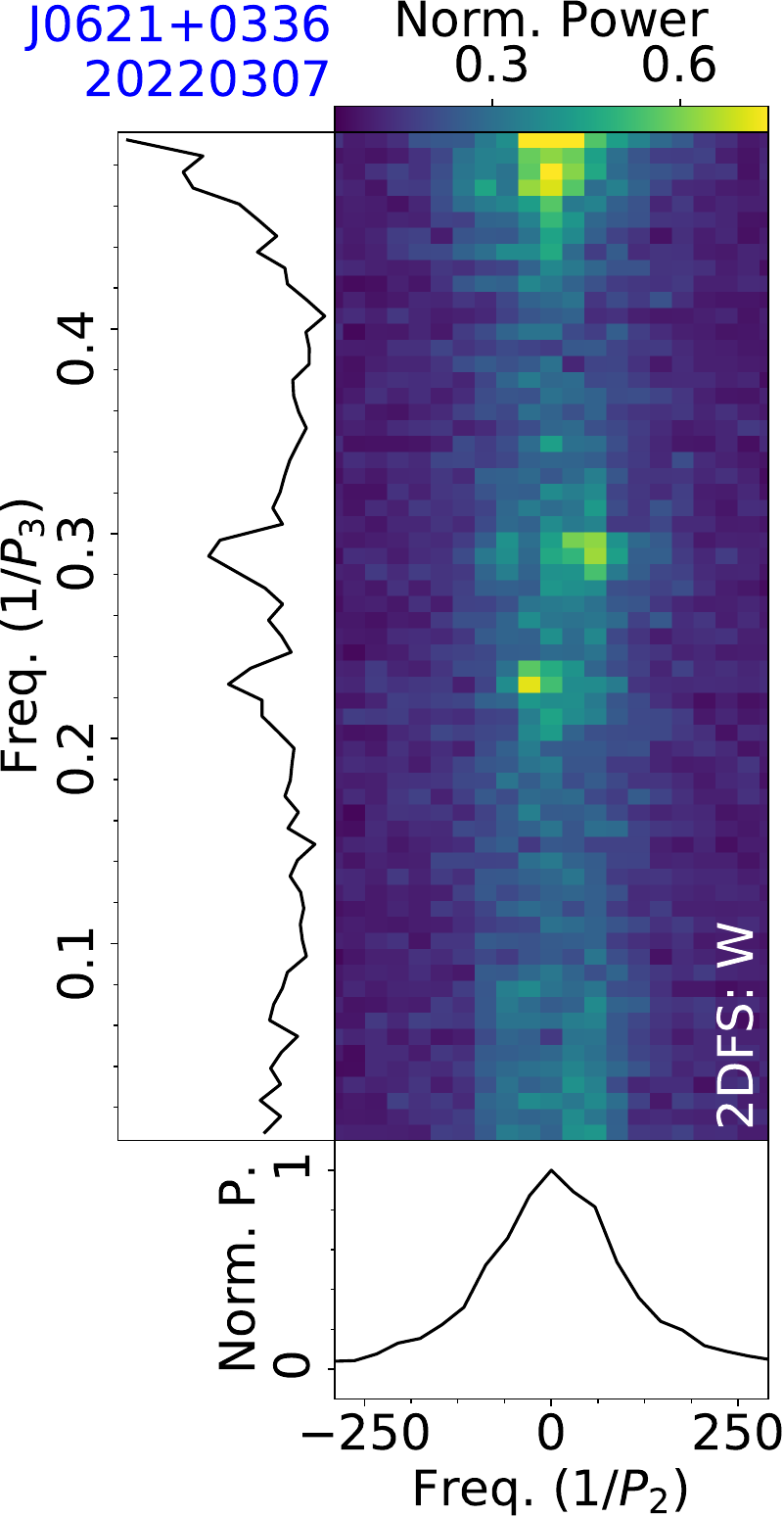}
\figcaption{Fluctuation analysis of PSR J0621+0336 from the FAST observation on 20220307, with LRFS and 2DFS for the on-pulse phase range of a mean pulse profile. \label{subfig:fluctu:J0621+0336}}
\end{figure}

\begin{figure}[htpb]
\centering
\includegraphics[width=0.22\textwidth, angle=0]{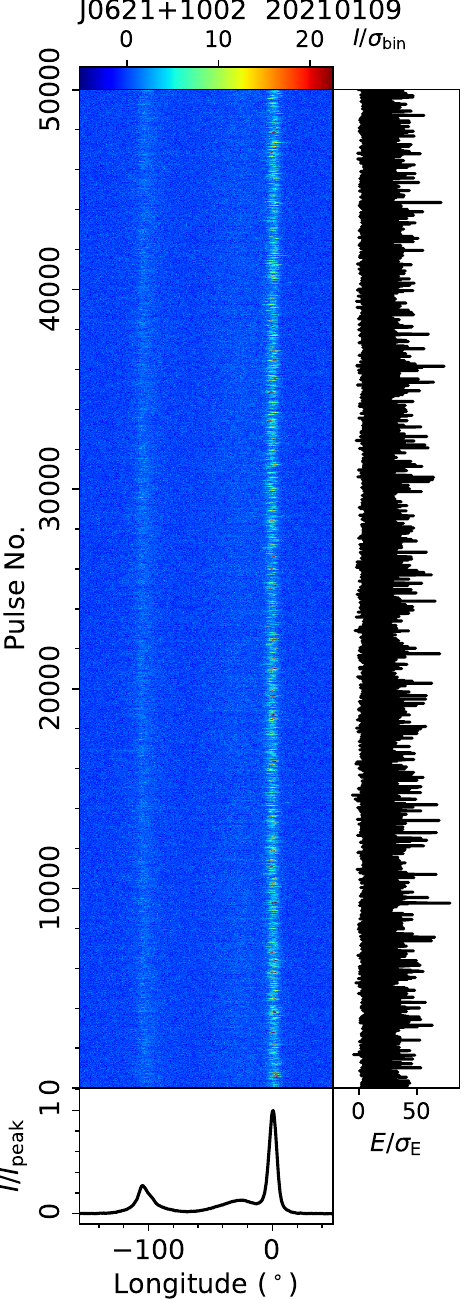} 
\includegraphics[width=0.22\textwidth, angle=0]{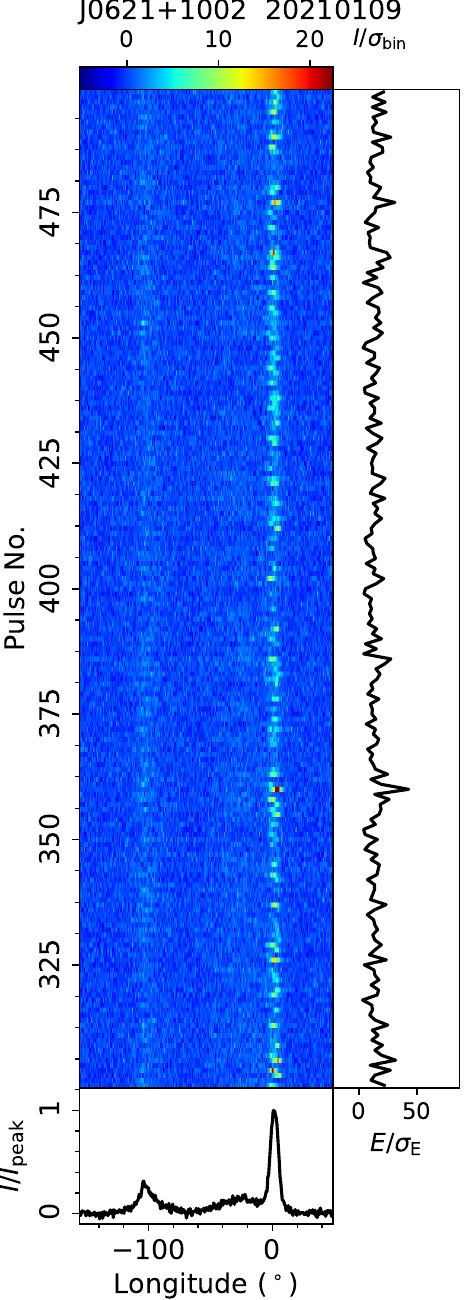}
\figcaption{Single pulse sequences of PSR J0621+1002 from the FAST observation on 20210109. \label{subfig:TP:J0621+1002}}
\end{figure}

\begin{figure}[htpb]
\centering
\includegraphics[width=0.22\textwidth, angle=0]{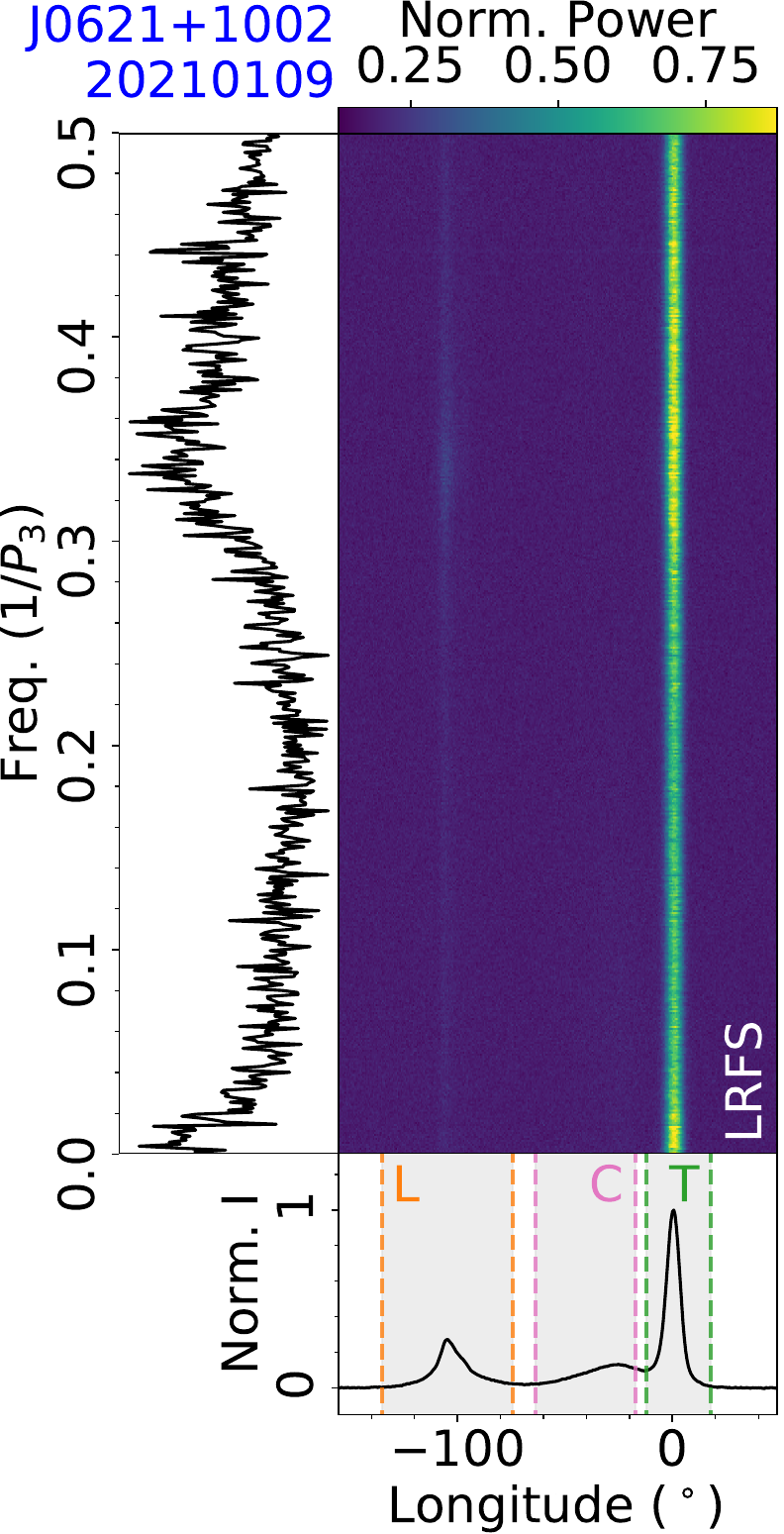} 
\includegraphics[width=0.22\textwidth, angle=0]{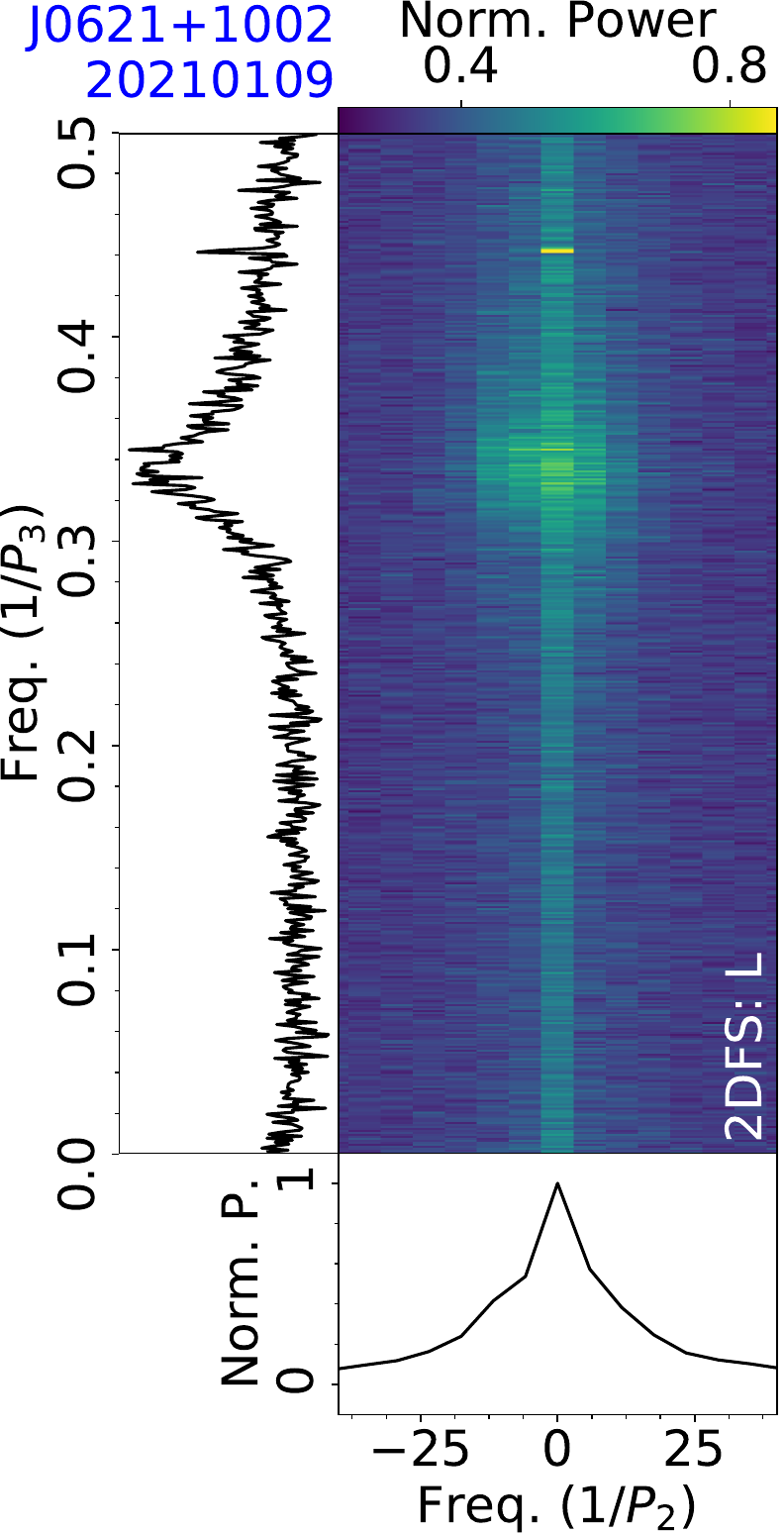}\\
\includegraphics[width=0.22\textwidth, angle=0]{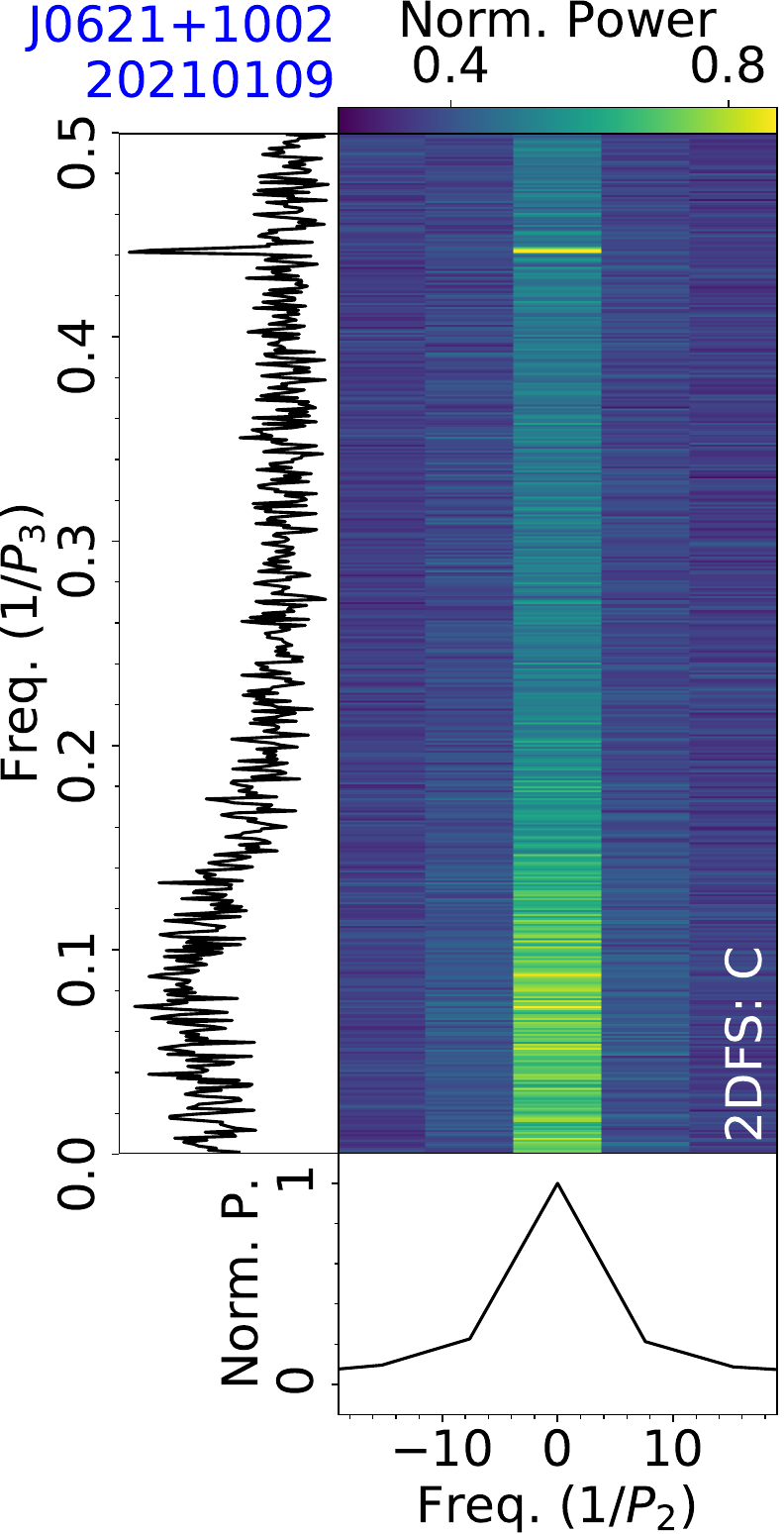} 
\includegraphics[width=0.22\textwidth, angle=0]{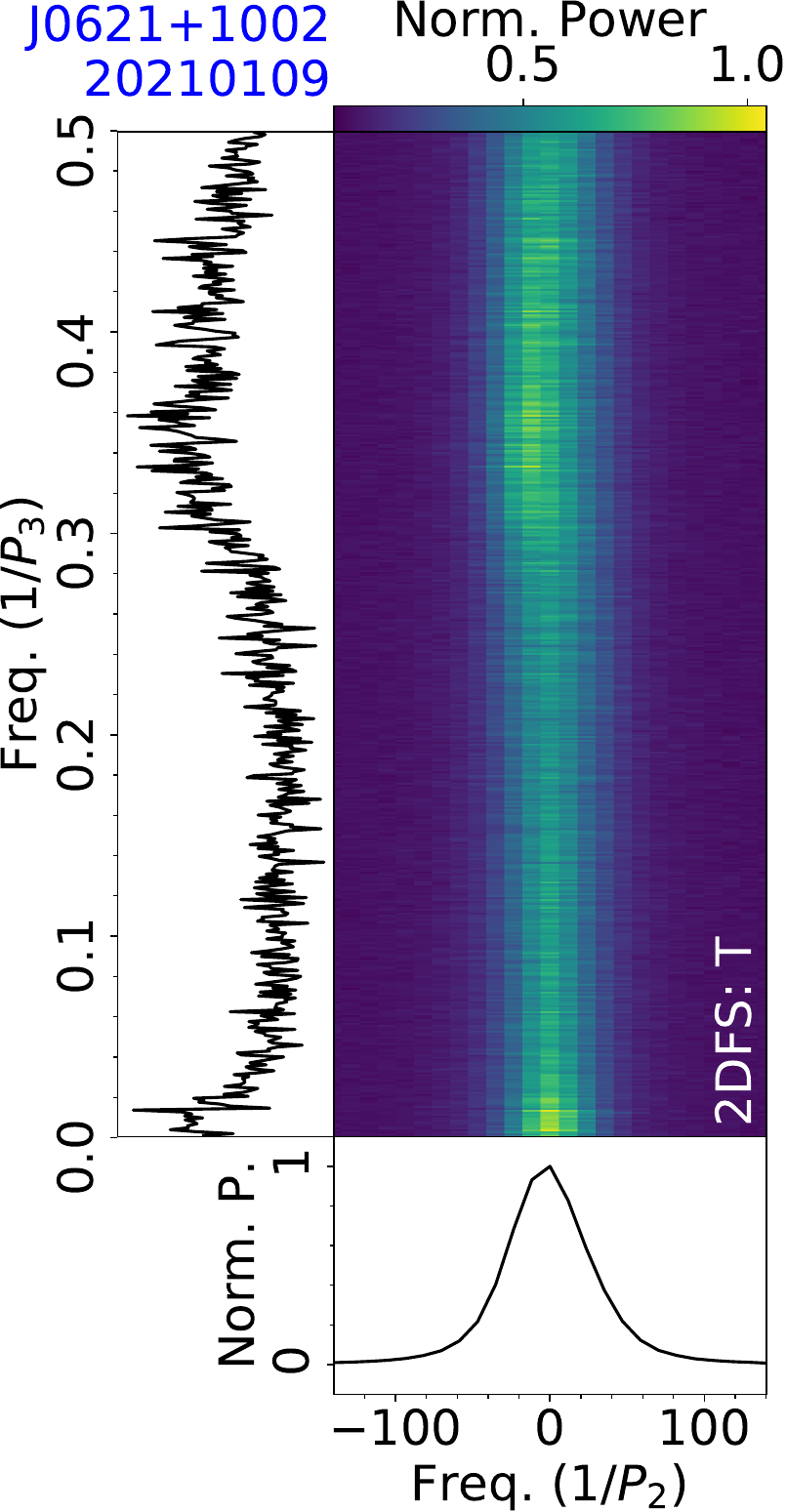}
\figcaption{Fluctuation analysis of PSR J0621+1002 from the observation on 20210109, with LRFS (top-left), and 2DFS for the leading part (top-right), the central part (bottom left) and trailing part (bottom-right) of a mean pulse profile. \label{subfig:fluctu:J0621+1002}}
\end{figure}

\subsection{J0621+0336}
\label{subsec:J0621+0336}

PSR J0621+0336 was discovered in the Perseus Arm Pulsar Survey using the Parkes 64-m radio telescope \citep{Burgay2013}. Positive drifting with $P_3$=3.4(4) $P_0$ and $P_2$=10$^{+10}_{-3}$ degrees, as well as $P_3$-only feature of 2.10(4) $P_0$, were reported by \citet{Song2023}. 

The pulsar was observed by FAST on 20220307 for 5 minutes, with a rotation period $P=0.2700$~s and a dispersion measure $D\!M=72.6~{\rm cm^{-3}\,pc}$ from this observation. Single pulse sequences and fluctuation spectra are shown in Fig.~\ref{subfig:TP:J0621+0336} and \ref{subfig:fluctu:J0621+0336}. Similar to results of \citet{Song2023}, there are positive drift feature ($P_3=3.45\pm0.02$ period and $P_2=8\pm1^\circ$) and temporal modulation features in 2DFS ($P_3=2.09\pm0.01$ period). Additionally, there is also a negative drift feature in 2DFS of $P_3=4.35\pm0.03$ periods and $P_2=-14\pm4^\circ$.


\begin{figure}[htpb]
\centering
\includegraphics[width=0.22\textwidth, angle=0]{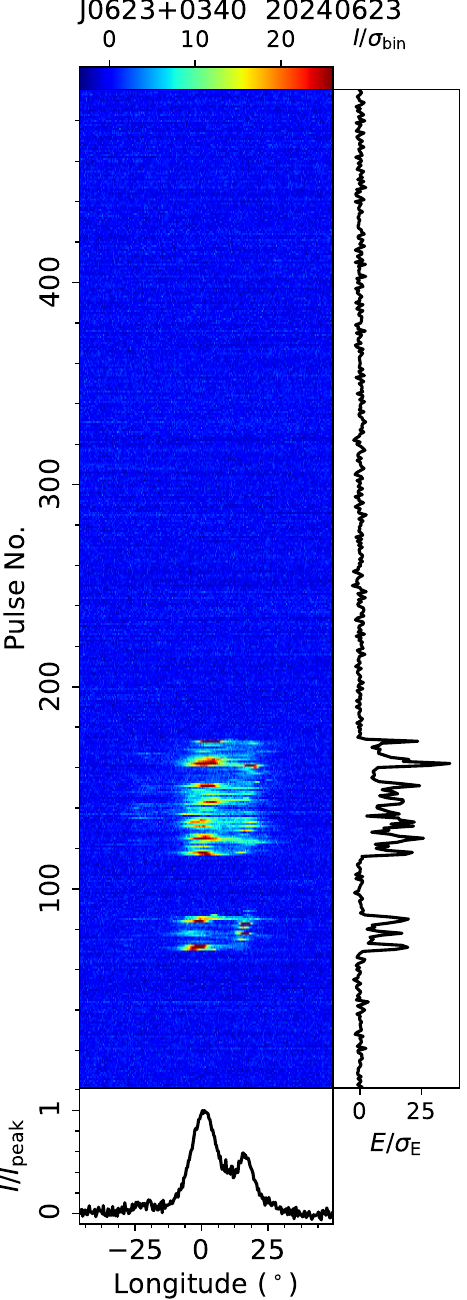} 
\includegraphics[width=0.22\textwidth, angle=0]{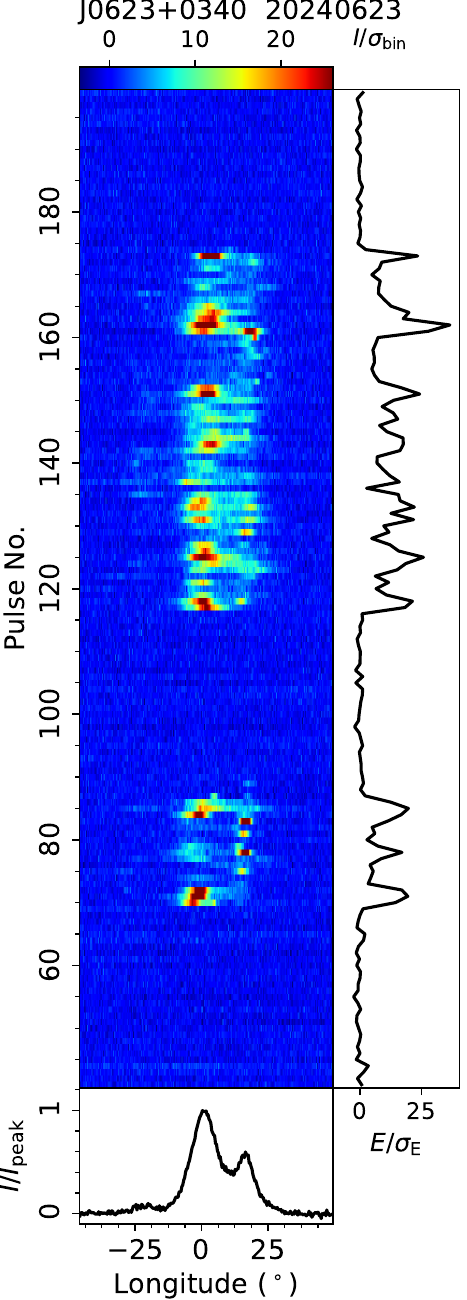}
\figcaption{Single pulse sequence of PSR J0623+0340 from the FAST observation on 20240623, and a zoomed-in view of pulses No.40-200.
\label{subfig:TP:J0623+0340}}
\end{figure}

\begin{figure}[htpb]
\centering
\includegraphics[width=0.39\textwidth, angle=0]{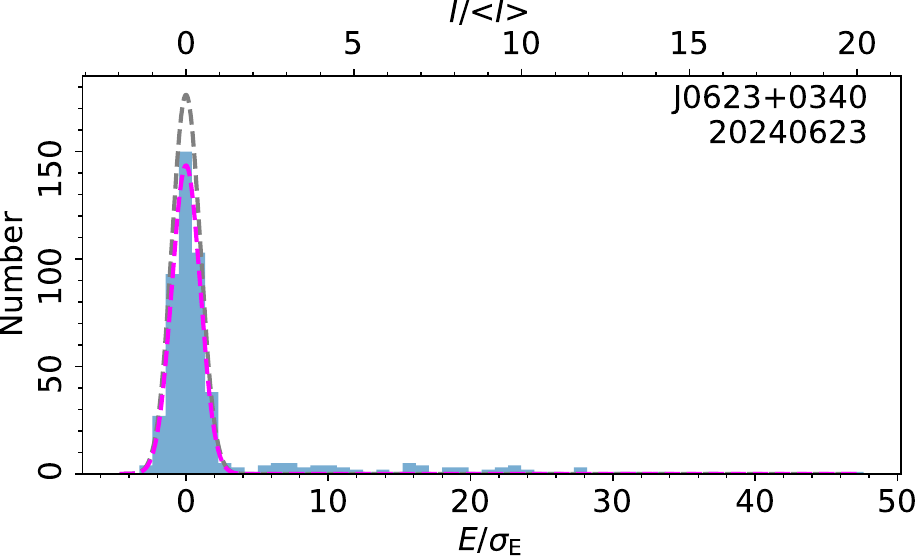}
\figcaption{On-pulse energy histogram of PSR J0623+03405 from the FAST observation on 20240623.
\label{subfig:Hist:J0623+0340}}
\end{figure}

\begin{figure}[htpb]
\centering
\includegraphics[width=0.22\textwidth, angle=0]{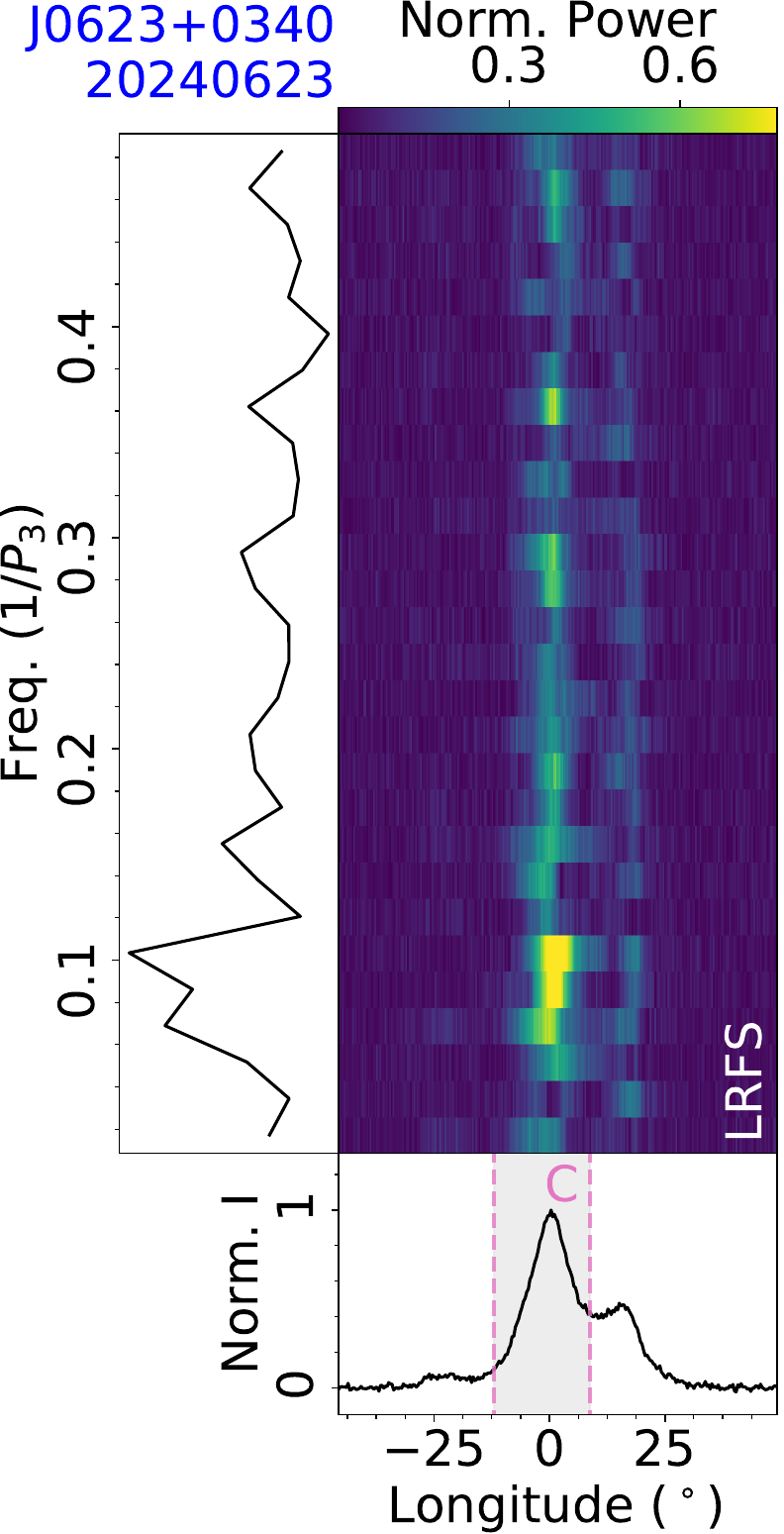}
\includegraphics[width=0.22\textwidth, angle=0]{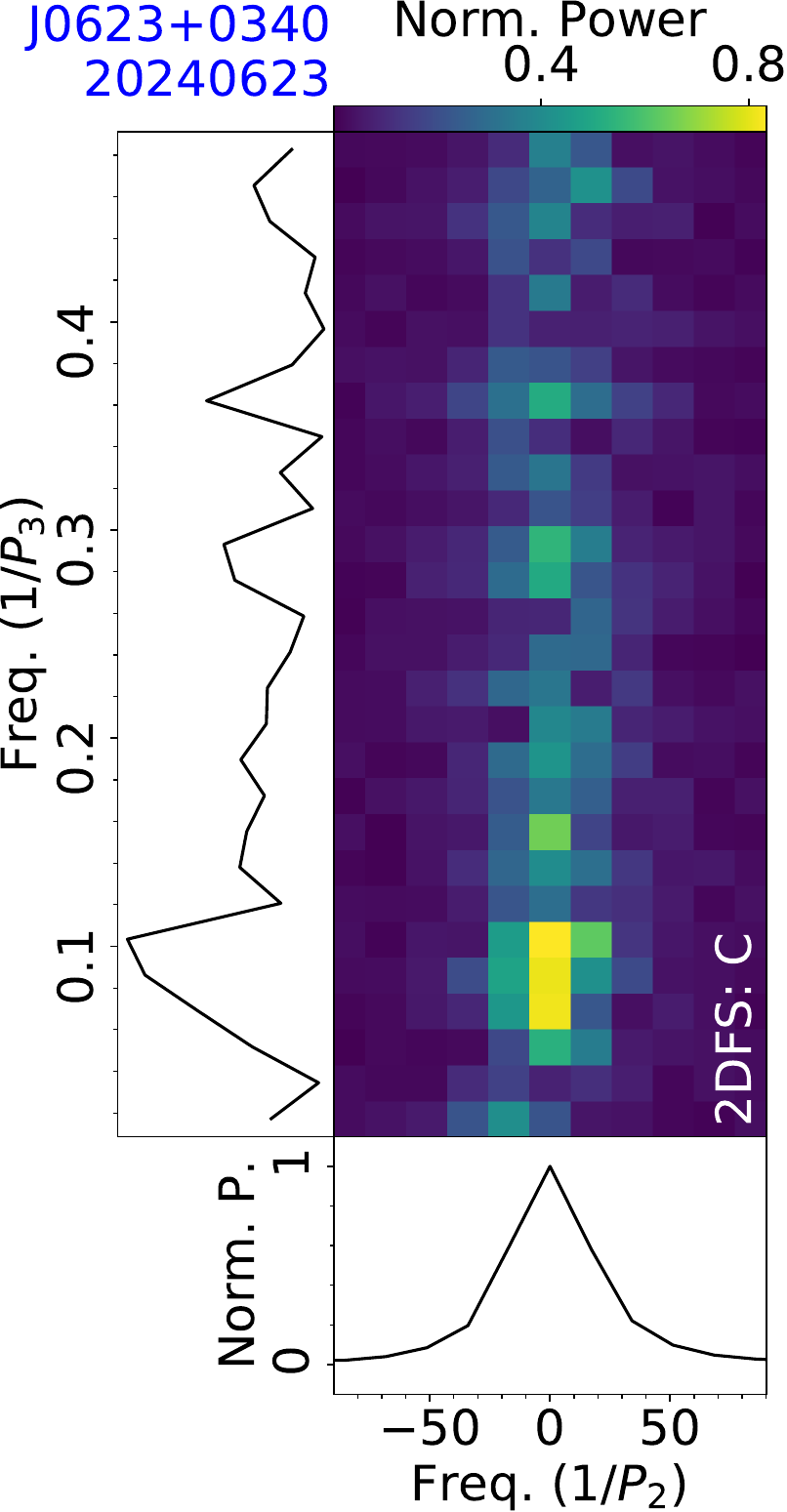}
\figcaption{Fluctuation analysis of PSR J0623+0340 for pulses No.116-174 from the observation on 20240623, with LRFS and 2DFS for the central part of a mean pulse profile. \label{subfig:fluctu:J0623+0340}}
\end{figure}

\begin{figure}[htpb]
\centering
\includegraphics[width=0.205\textwidth, angle=0]{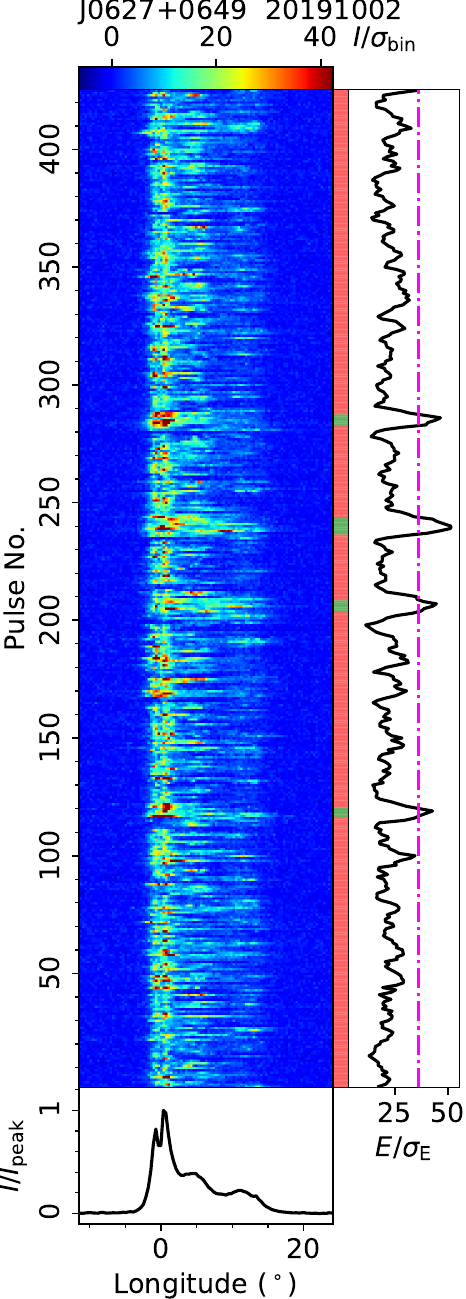} 
\includegraphics[width=0.205\textwidth, angle=0]{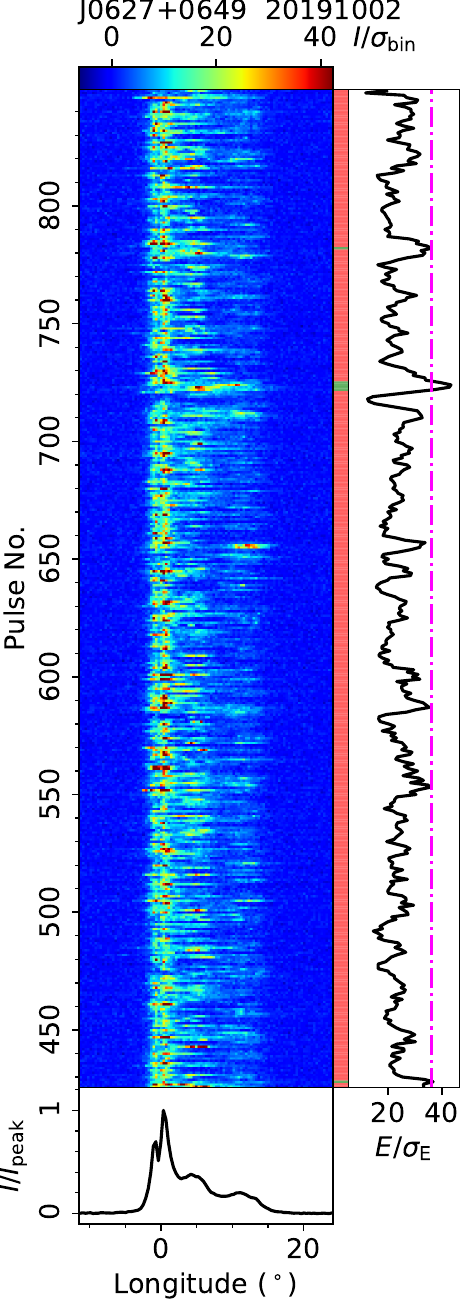}
\figcaption{Single pulse sequences of PSR J0627+0649 from the FAST observation on 20191002. \label{subfig:TP:J0627+0649}}
%
\centering
\includegraphics[width=0.36\textwidth, angle=0]{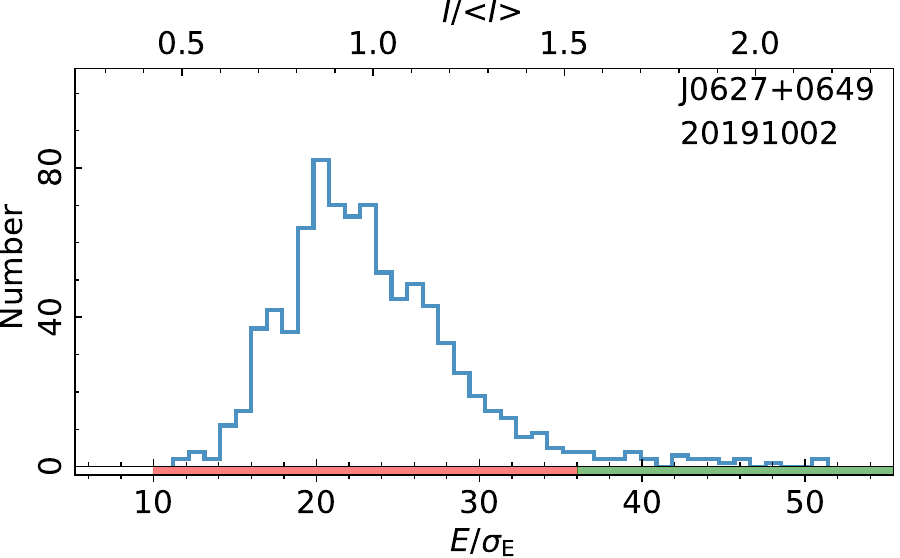}
\figcaption{On-pulse energy histogram of PSR J0627+0649 from the observation on 20210111. \label{subfig:Hist:J0627+0649}}
\end{figure}

\begin{figure}[htpb]
\centering
\includegraphics[width=0.39\textwidth, angle=0]{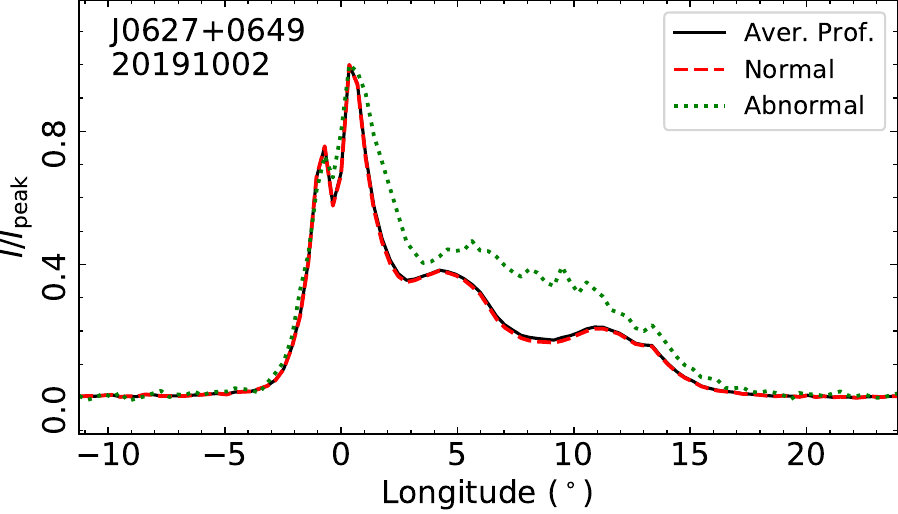}
\figcaption{Mean profiles for the normal and abnormal modes of PSR J0627+0649 from the observation on 20200401. Profiles in the plot are normalized by the respective peaks.
\label{subfig:PolModes:J0627+0649}}
\end{figure}

\begin{figure}[htpb]
\centering
\includegraphics[width=0.21\textwidth, angle=0]{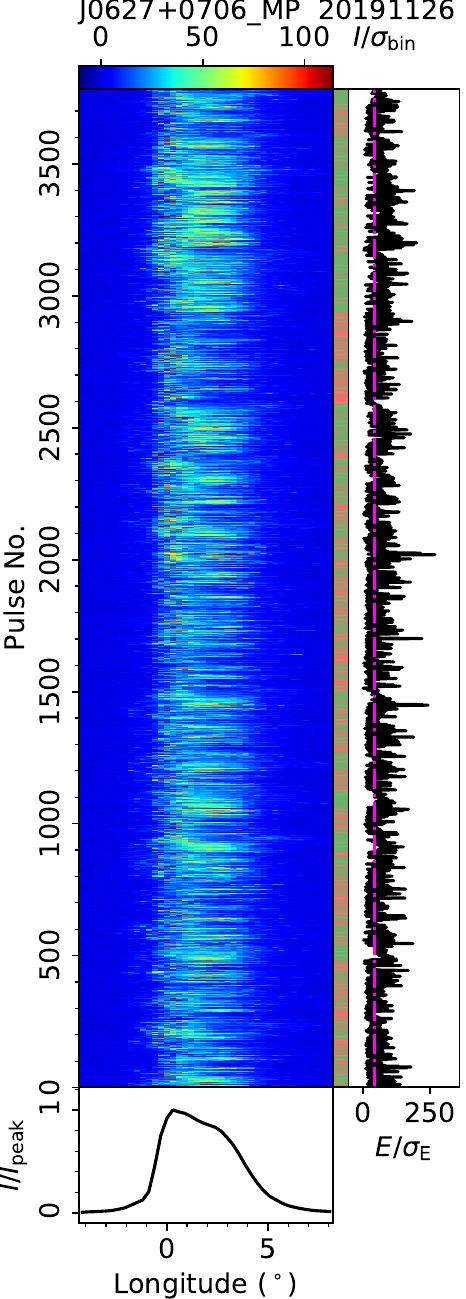}
\includegraphics[width=0.21\textwidth, angle=0]{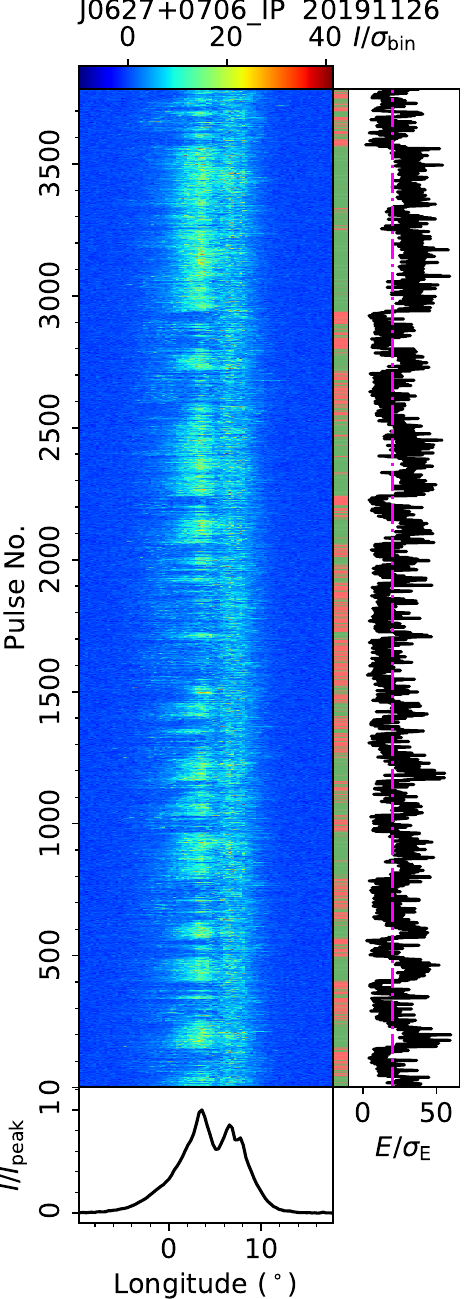}\\
\includegraphics[width=0.21\textwidth, angle=0]{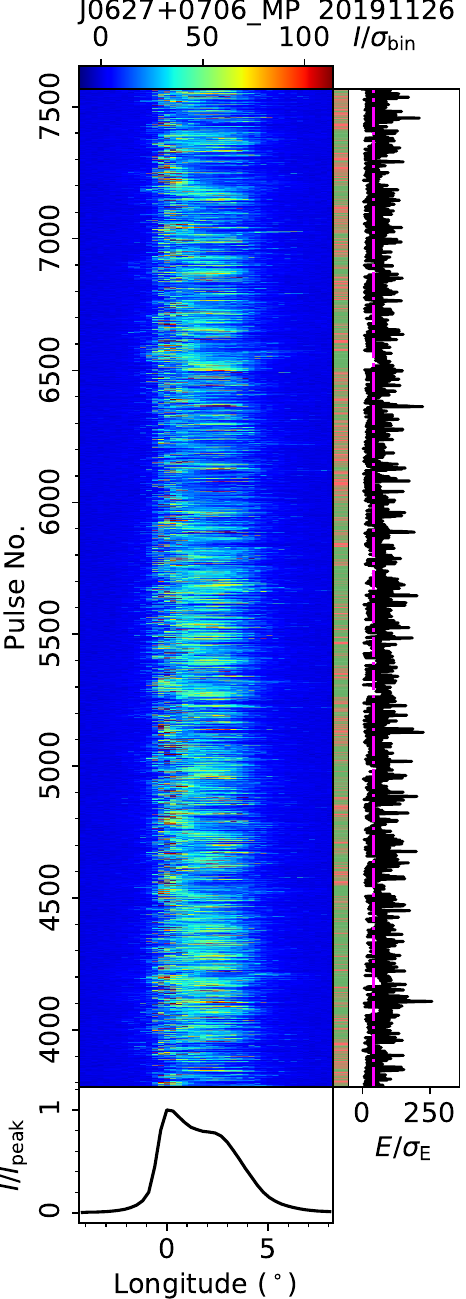}
\includegraphics[width=0.21\textwidth, angle=0]{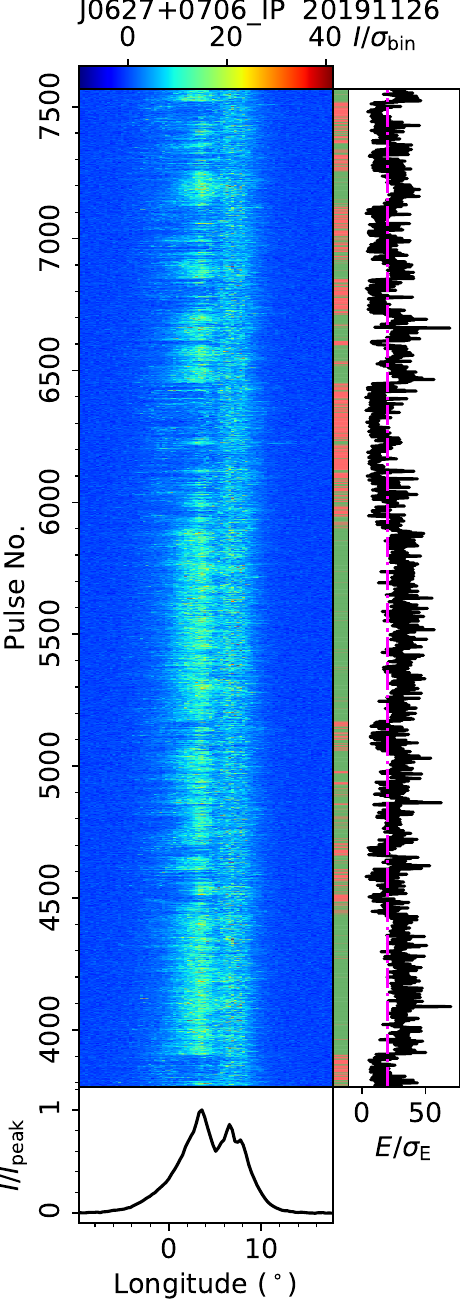}
\figcaption{Single pulse sequences of PSR J0627+0706 from the FAST observation on 20191126. \label{subfig:TP:J0627+0706}}
\end{figure}

\begin{figure}[htpb]
\centering
\includegraphics[width=0.35\textwidth, angle=0]{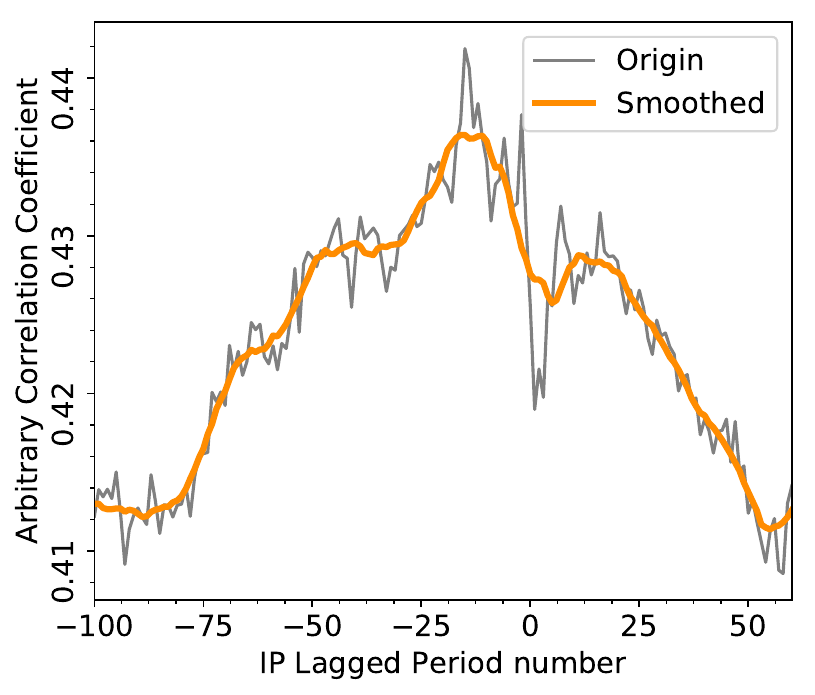}
\figcaption{Occurrence correlation between two emission modes of PSR J0627+0706 from the FAST observation on 20191126. \label{subfig:ModesCorr:J0627+0706}}
\end{figure}

\begin{figure}[htpb]
\centering
\includegraphics[width=0.35\textwidth, angle=0]{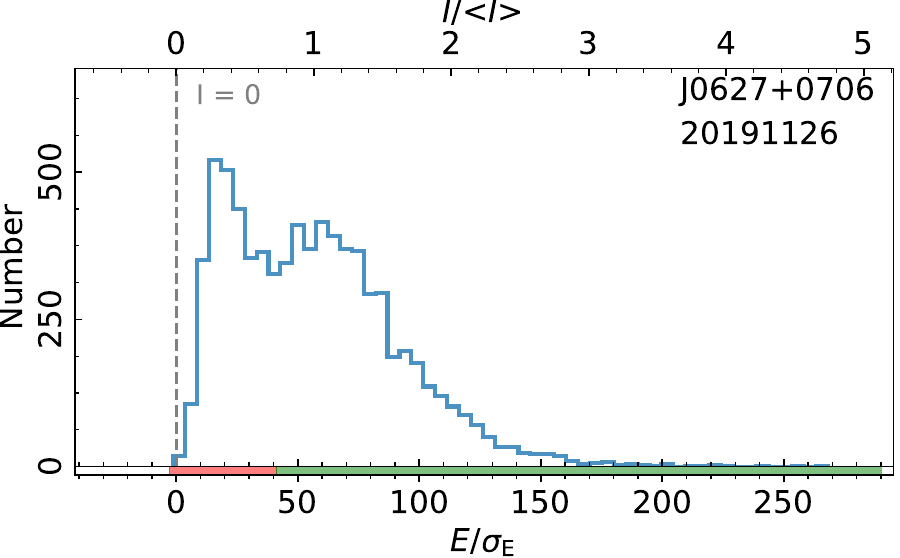}\\
\includegraphics[width=0.35\textwidth, angle=0]{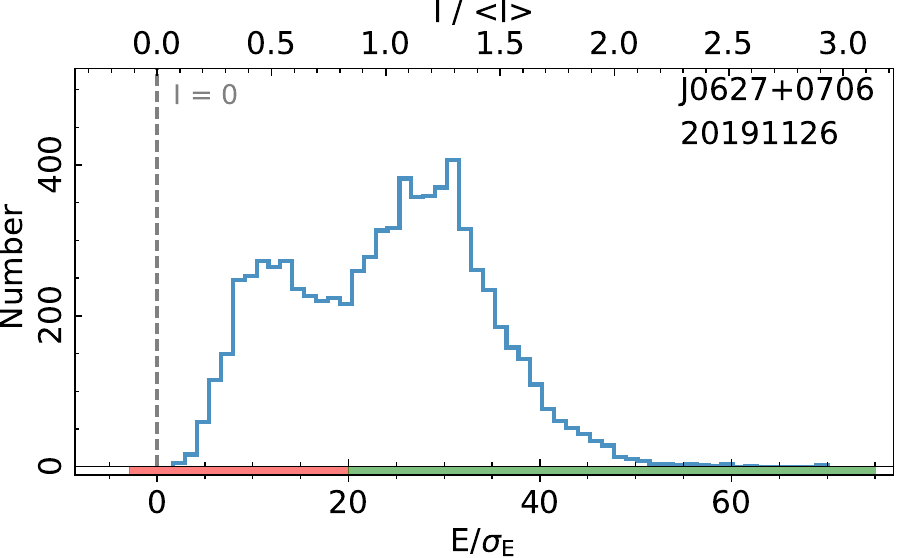}
\figcaption{On-pulse energy histogram of PSR J0627+0706 for the main pulse and inter-pulse from the observation on 20210111. \label{subfig:Hist:J0627+0706}}
\end{figure}

\begin{figure}[htpb]
\centering
\includegraphics[width=0.395\textwidth, angle=0]{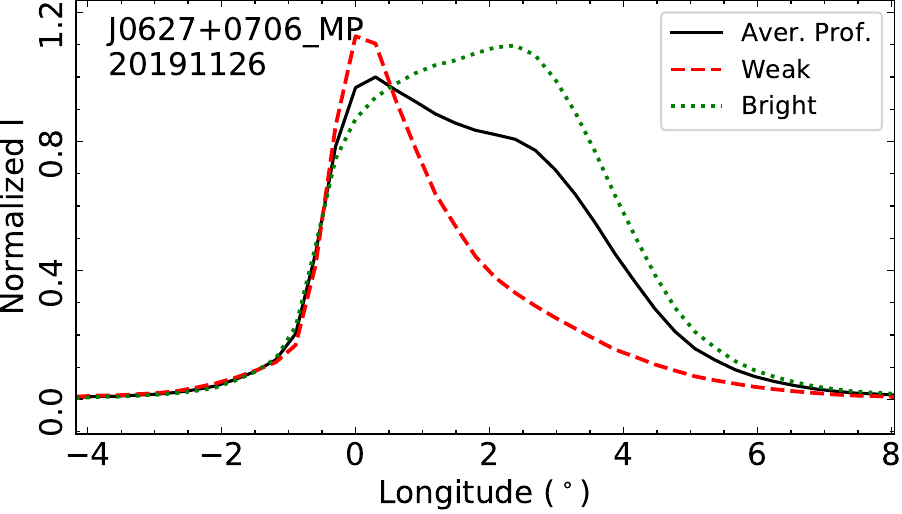}\\
\includegraphics[width=0.39\textwidth, angle=0]{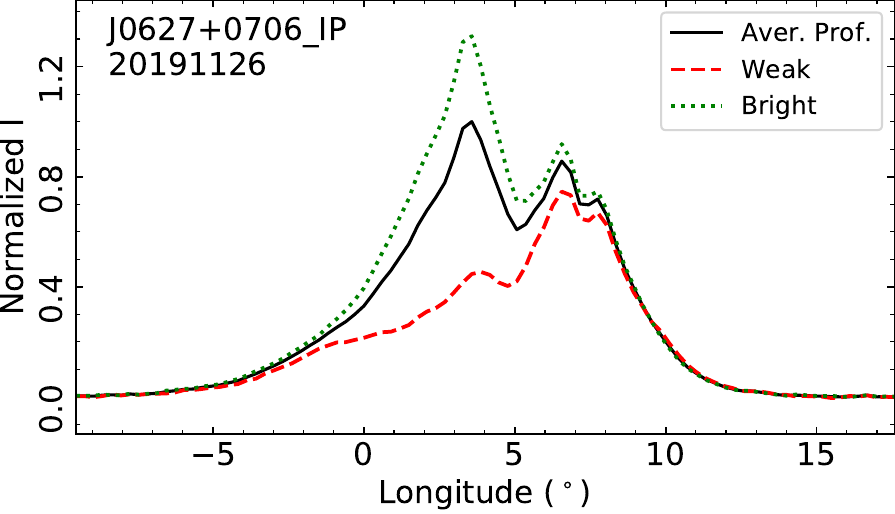}
\figcaption{Mean profiles of the weak (red dashed line) and bright emission (green dotted line) modes for the main pulse (upper plot) and interpulse (lower plot) of PSR J0627+0706, from the FAST observation on 20191126. The profiles of the two modes are normalized by the peak of the mean profile (black solid line) of all periods.
\label{subfig:ProfModes:J0627+0706}}
\end{figure}

\subsection{J0621+1002}
\label{subsec:J0621+1002}

PSR J0621+1002 was discovered using the Arecibo telescope at 430 MHz \citet{Camilo1996}. \citet{Wang2021} reported intensity modulation features in fluctuation spectra of the first component with peak of $3.0\pm0.1$ periods, and third pulse components with two diffused peaks of $3.0\pm0.1$ periods and $200\pm1$ periods.

This pulsar was observed by FAST on 20200104 for 5 minutes and 20210109 for 24 minutes, with a rotation period $P=0.0289$~s and a dispersion measure $D\!M=36.5~{\rm cm^{-3}\,pc}$ from the 24-minute observation. 
Single pulse sequences and fluctuation spectra of the observation on 20210109 are shown in Fig.~\ref{subfig:TP:J0621+1002} and \ref{subfig:fluctu:J0621+1002}. 
For the leading component, there is a hint of negative drifting with the temporal modulation centroid of $1/P_3=0.3448\pm0.0003$ ($P_3=2.88\pm0.08$ periods). There is a wide low-frequency modulation feature with the centroid of $1/P_3=0.084\pm0.001$ in 2DFS of the central component, corresponding to $P_3=11.9\pm0.1$ periods. In 2DFS of the trailing component, the drift feature is not symmetric about vertical axis, which demonstrates negative drifting instead of periodic intensity modulation. The centroid modulation frequencies are $1/P_3=0.386\pm0.001$ and $1/P_2=-14.6\pm0.2$, corresponding to periodicities of $P_3=2.591\pm0.004$ periods and $P_2=-24.7\pm0.3$. There is also a low-frequency modulation with the centroid of $1/P_3=0.028\pm0.001$, that corresponds $P_3=35\pm1$ periods for the trailing component. 
There is also a very narrow modulation frequency on time of $1/P_3=0.4422\pm0.0002$, yielding $P_3=2.262\pm0.001$ periods, especially for the leading and central components. 
The modulation properties are distinct for these three components. 

\subsection{J0623+0340}
\label{subsec:J0623+0340}

PSR J0623+0340 was discovered in the Perseus Arm Pulsar Survey using the Parkes radio telescope, exhibiting nulling behavior \citet{Burgay2013}.

This pulsar was observed by FAST on 20240623 for 5 minutes, with a rotation period $P=0.6137$~s and a dispersion measure $D\!M=57.1~{\rm cm^{-3}\,pc}$. The single pulse sequence and a zoomed-in view of pulses No.40-200 are shown in Fig.~\ref{subfig:TP:J0623+0340}, illustrating the nulling phenomenon with a long duration. The nulling fraction of this observation is estimated to be 81.4$\pm$1.8\% from the on-pulse integral energy histogram in Fig.~\ref{subfig:Hist:J0623+0340}. For the emission segment between pulses No.116 and 174, LRFS and 2DFS for the central profile part are displayed in Fig.~\ref{subfig:fluctu:J0623+0340}, where the centroid frequency of the modulation feature is $1/P_3=0.086\pm0.003$ ($P_3=11.7\pm0.4$ periods). 
More observations are required for more detailed and accurate modulation properties.

\subsection{J0627+0649}
\label{subsec:J0627+0649}

PSR J0627+0649 was discovered in the Perseus Arm Pulsar Survey using the Parkes 64-m radio telescope \citep{Burgay2013}. 

This pulsar was observed by FAST on 20191002 for 5 minutes, deriving a rotation period $P=0.3465$~s and a dispersion measure $D\!M=86.5~{\rm cm^{-3}\,pc}$. Single pulse sequences in Fig.~\ref{subfig:TP:J0627+0649} display the modes changing behavior.
The grey line in the right panel represents the variation over time of the single pulse on-pulse integral energy, and the black line shows the smoothed result obtained from every 5 single pulses of the grey line.
Based on the smoothed energy variation histogram (Fig.~\ref{subfig:Hist:J0627+0649}), normal and abnormal emission modes of single pulses are distinguished and then labeled in red and green, respectively, in the plots of the histogram and single pulse sequences. For the abnormal mode, the emission of the whole on-pulse phase region is enhanced, especially for the central and trailing components.

\begin{figure}[htpb]
\centering
\includegraphics[width=0.22\textwidth, angle=0]{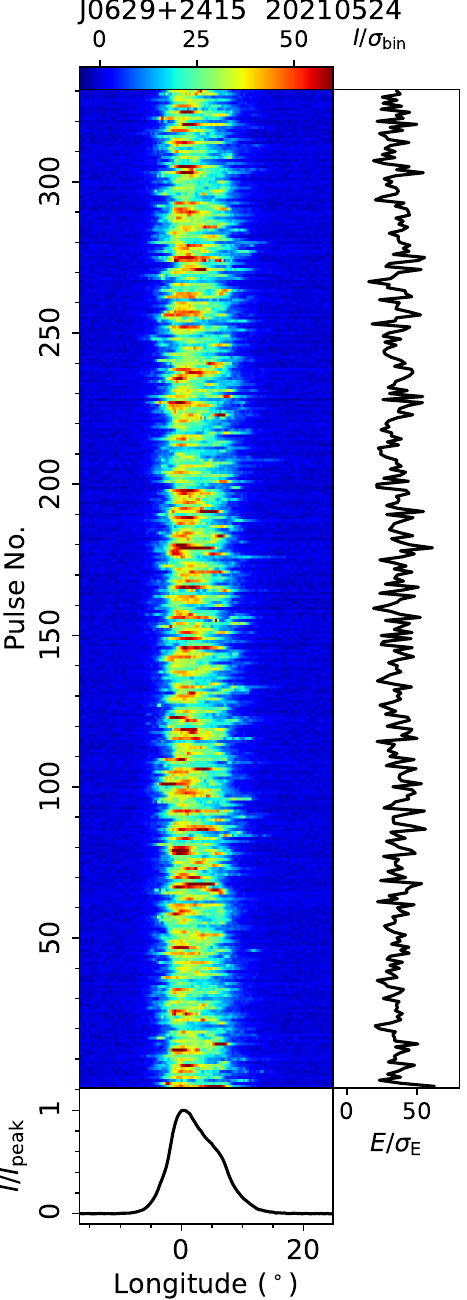}
\includegraphics[width=0.22\textwidth, angle=0]{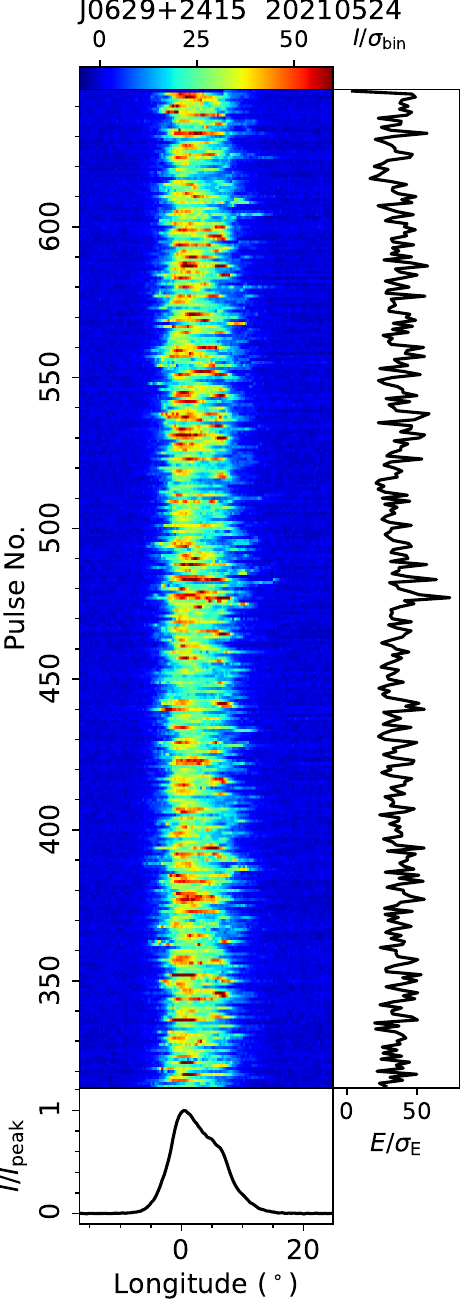}
\figcaption{Single pulse sequences of PSR J0629+2415 from the FAST observation on 20210524. \label{subfig:TP:J0629+2415}}
\end{figure}

\begin{figure}[htpb]
\centering
\includegraphics[width=0.39\textwidth, angle=0]{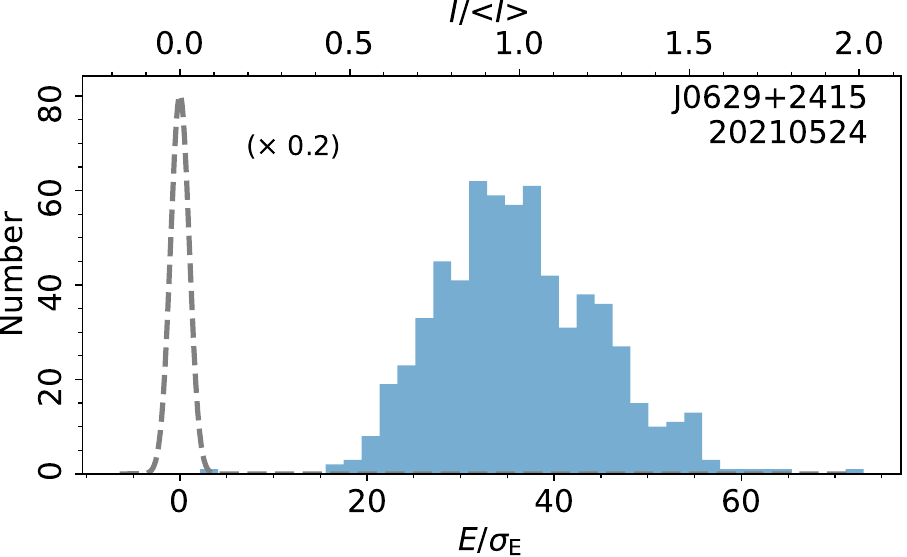}
\figcaption{On-pulse energy histogram of PSR J0629+2415 from the FAST observation on 20210111. \label{subfig:Hist:J0629+2415}}
\end{figure}

\begin{figure}[htpb]
\centering
\includegraphics[width=0.22\textwidth, angle=0]{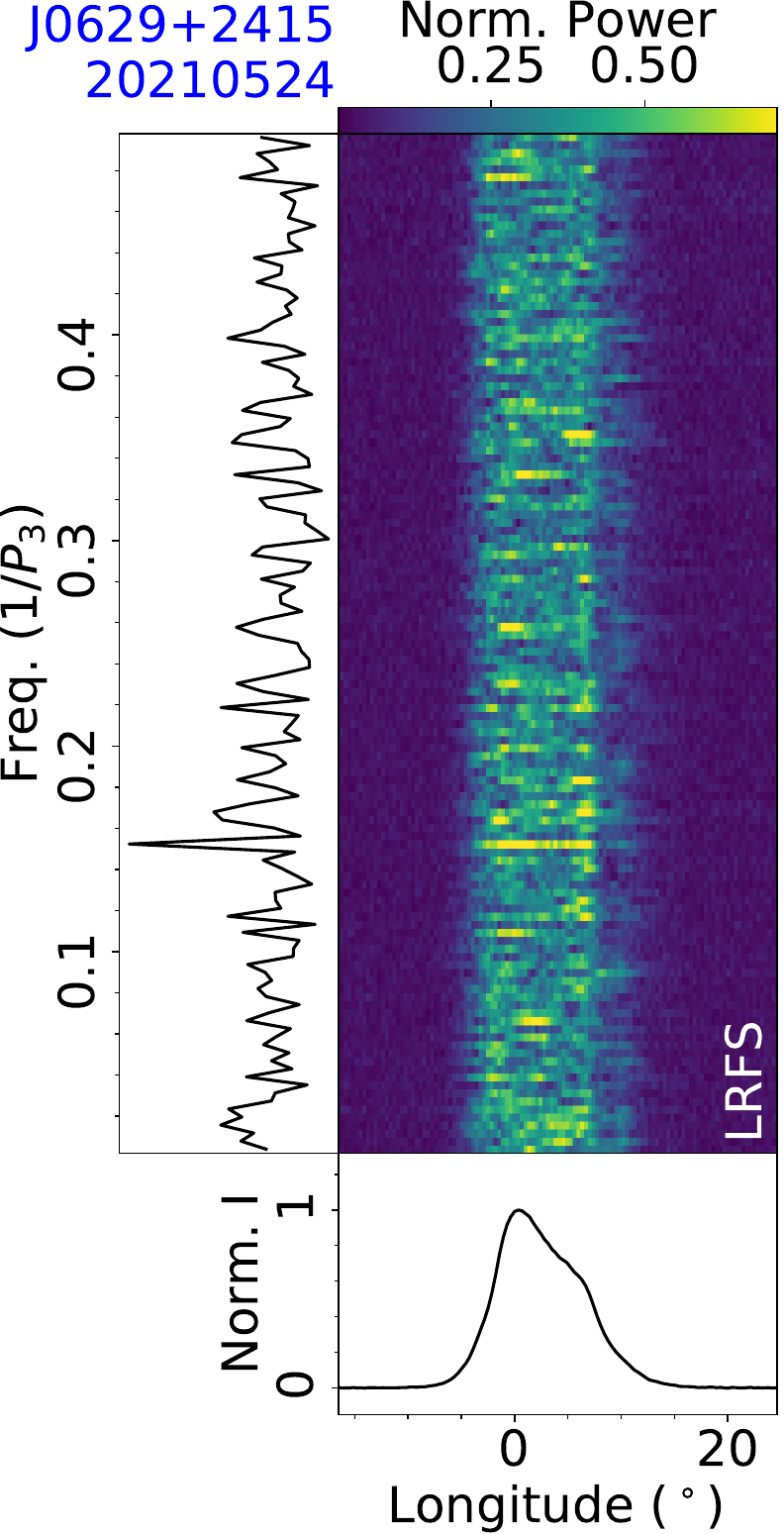}
\includegraphics[width=0.22\textwidth, angle=0]{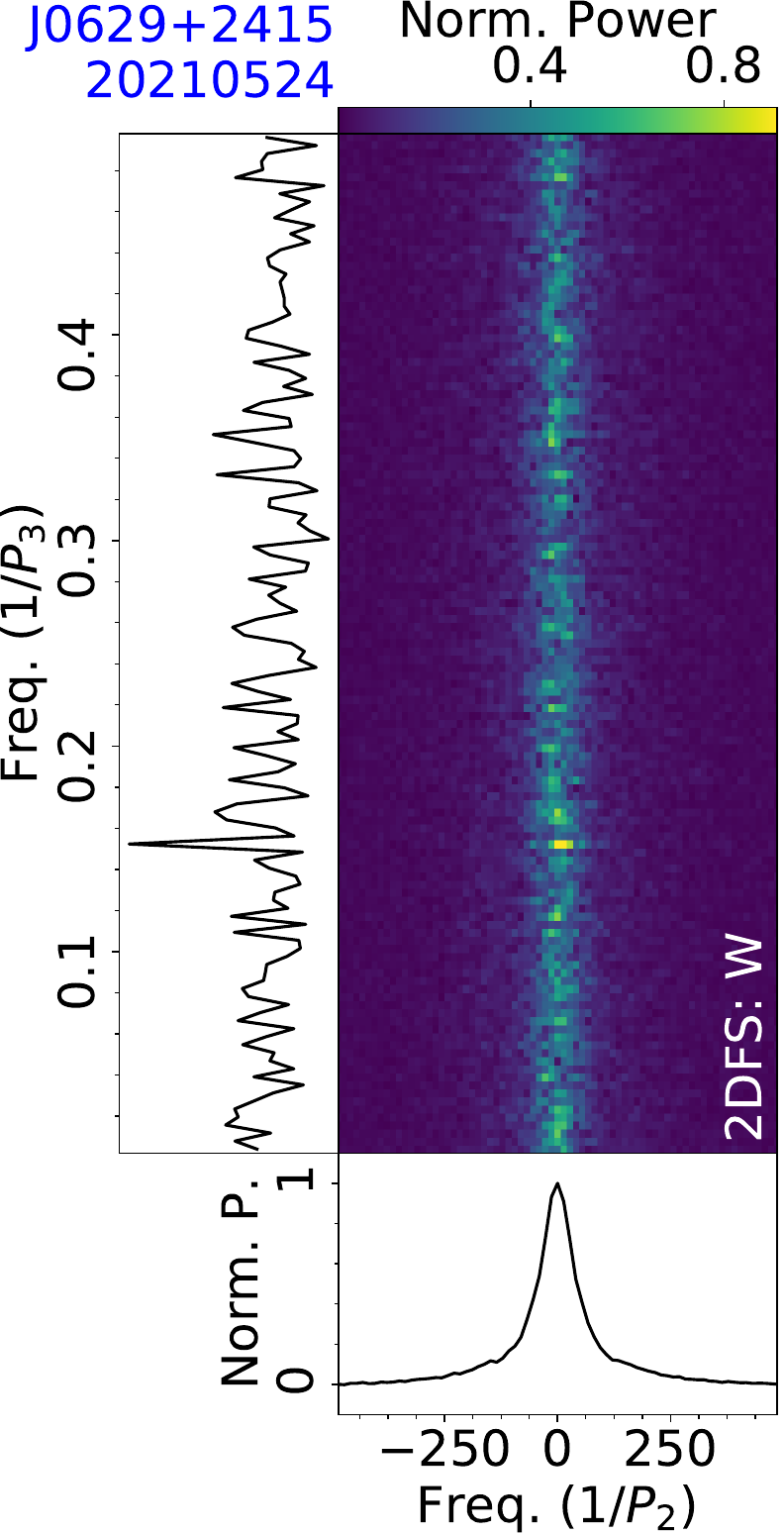}
\figcaption{Fluctuation analysis of PSR J0629+2415 from the FAST observation on 20210524, with LRFS and 2DFS for the on-pulse phase range of a mean pulse profile. 
\label{subfig:fluctu:J0629+2415}}
\end{figure}

\subsection{J0627+0706}
\label{subsec:J0627+0706}

PSR J0627+0706 was discovered in 2003 by Chandler and independently detected in the Perseus Arm Pulsar Survey \citep{Burgay2013}. 

The pulsar was observed by FAST on 20191002 for 5 minutes and 20191126 for 1 hour. A rotation period $P=0.4758$~s and a dispersion measure $D\!M=138.2~{\rm cm^{-3}\,pc}$ were yielded from the 1-hour observation. 
Single pulse sequences of the main pulse (MP) and interpulse (IP) of the FAST observation on 20191126 are shown in Fig.~\ref{subfig:TP:J0627+0706}, illustrating that both MP and IP have a mode changing phenomenon. Different emission modes are distinguished from the integral energy histogram of the trailing part for MP and the leading part for IP (Fig.~\ref{subfig:Hist:J0627+0706}), which display a bimodal shape corresponding to two emission modes. Two emission modes are labeled using red and green colors. 

To analyse the relationship between MP and IP, we set the weak mode to be "0" and the bright mode to be "1". As a result, two 0-1 sequences related to the occurrences of emission modes are obtained. Values of shifted cross-correlation between 0-1 sequences of MP and IP are calculated, which are shown in Fig.~\ref{subfig:ModesCorr:J0627+0706}. On average, the correlation coefficient reaches a maximum of 0.44 when mode changes of IP are lagged by 15 periods from MP.

Profiles contrast between emission modes of MP and IP are shown in Fig.~\ref{subfig:ProfModes:J0627+0706}. For the main pulse, the emission of the weak mode is brighter in the leading edge while weaker in the trailing edge of the profile compared to the bright mode. For the interpulse, the emission of the weak mode is weakened, especially in the leading part of the profile, and the trailing components are much brighter than the leading components, which is different from the bright mode.

\subsection{J0629+2415}
\label{subsec:J0629+2415}

The pulsar was first reported by \citet{Damashek1978} using the Green Bank telescope at 400 MHz. \citet{Weisberg1986} reported no evidence of nulling at 430 MHz, and the nulling fraction was less than 0.02\% of PSR J0629+2415. $P_3$-only feature with 2.70 periods was reported by \citet{Song2023}. 

This pulsar was observed by FAST on 20210524 for 5 minutes, deriving a rotation period $P=0.4766$~s and a dispersion measure $D\!M=84.2~{\rm cm^{-3}\,pc}$ from this observation. From single pulse sequences in Fig.~\ref{subfig:TP:J0629+2415} and the on-pulse integral energy histogram, there is no null existing in this observation. From LRFS and 2DFS (Fig.~\ref{subfig:fluctu:J0629+2415}) of the FAST observation, the temporal modulation periodicity $P_3$ is estimated to be $6.56\pm0.03$ periods with no apparent phase modulation.

\begin{figure}[htpb]
\centering
\includegraphics[width=0.22\textwidth, angle=0]{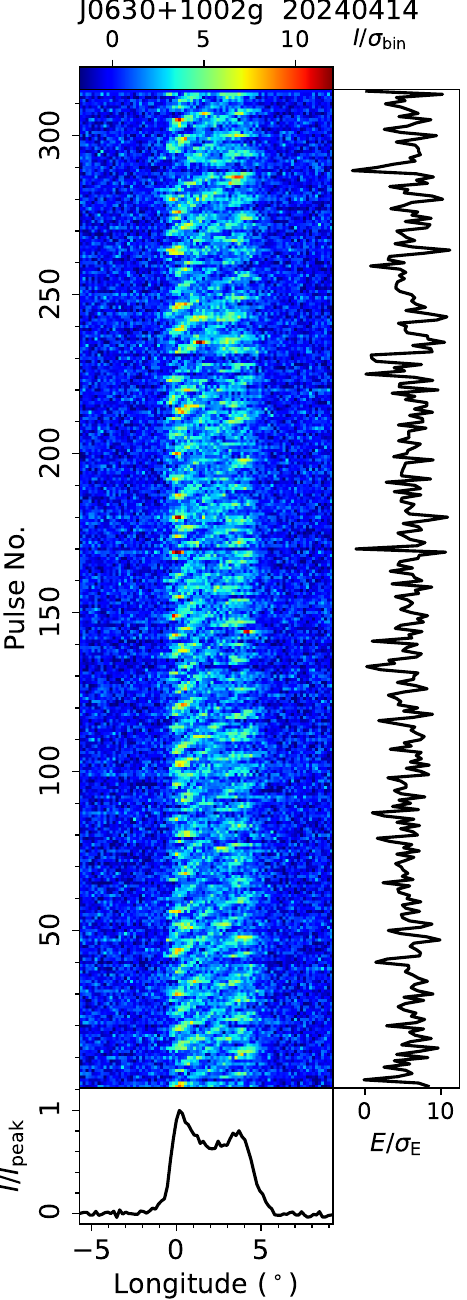} 
\figcaption{Single pulse sequence of PSR J0630+1002g from the FAST observation on 20240414. \label{subfig:TP:J0630+1002g}}
\end{figure}

\begin{figure}[htpb]
\centering
\includegraphics[width=0.22\textwidth, angle=0]{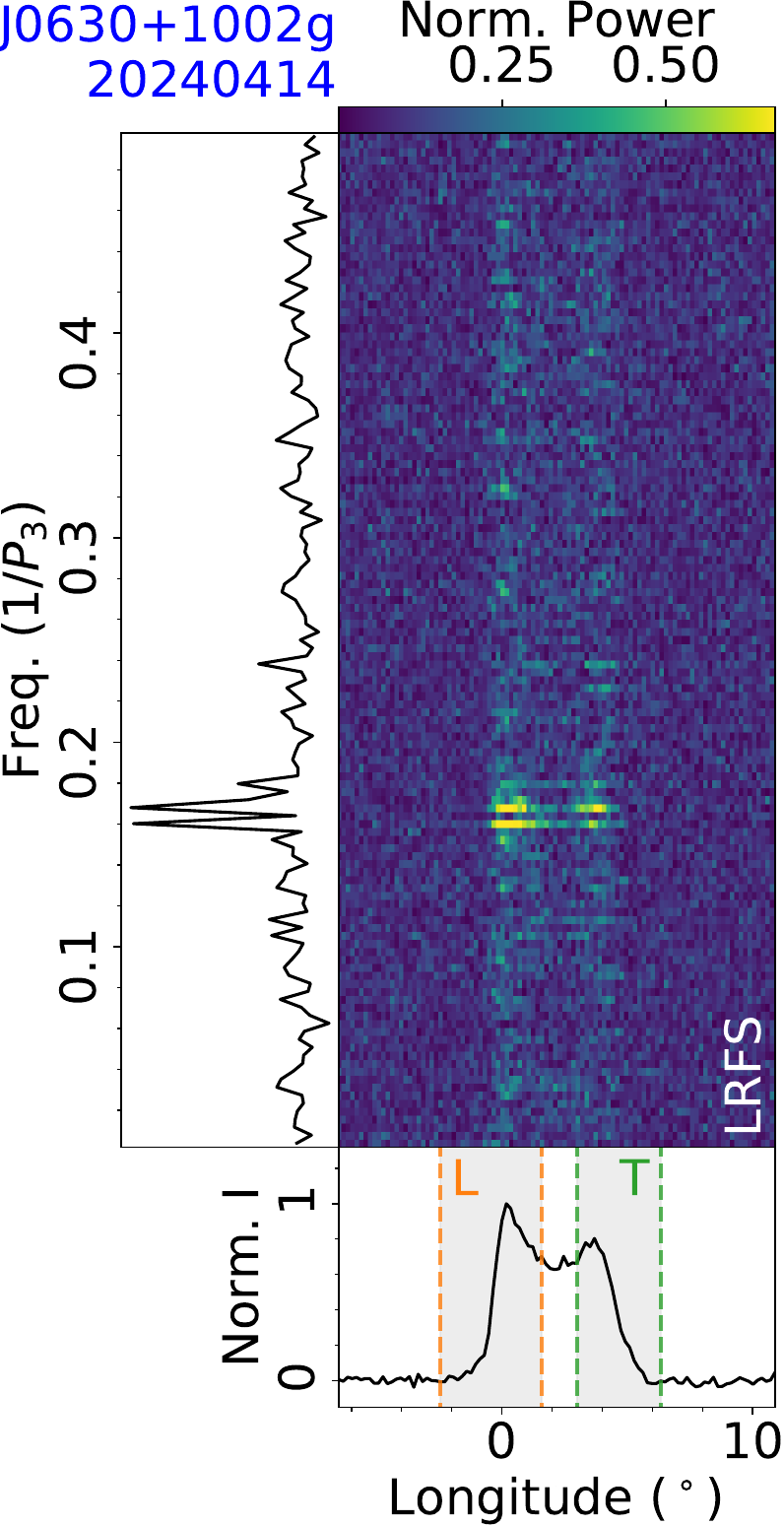}
\includegraphics[width=0.22\textwidth, angle=0]{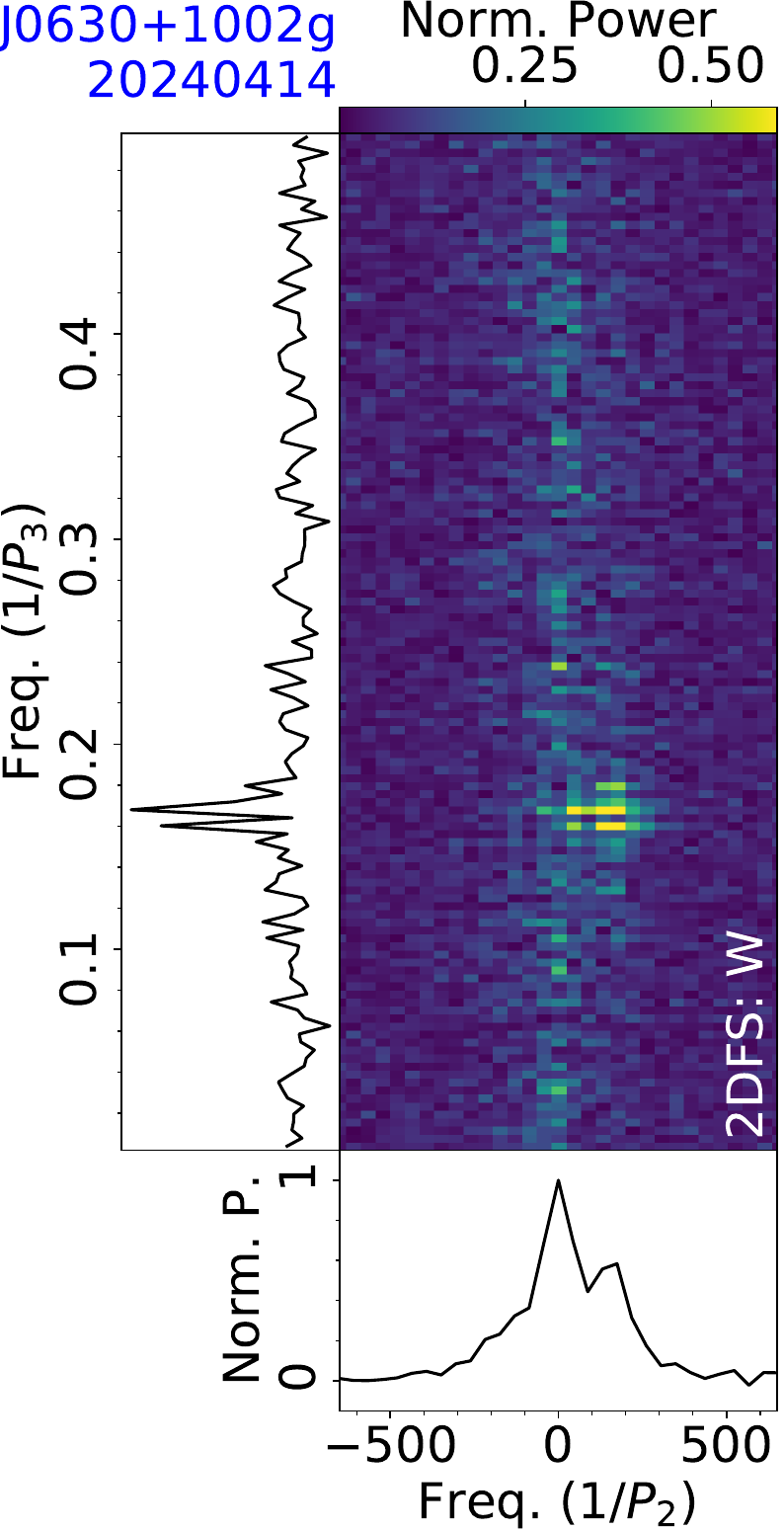}\\
\figcaption{Fluctuation analysis of PSR J0630+1002g from the FAST observation on 20240414, with LRFS and 2DFS for the on-pulse phase range of a mean pulse profile.  \label{subfig:fluctu:J0630+1002g}}
\end{figure}

\begin{figure}[hbpt]
\centering
\includegraphics[width=0.22\textwidth, angle=0]{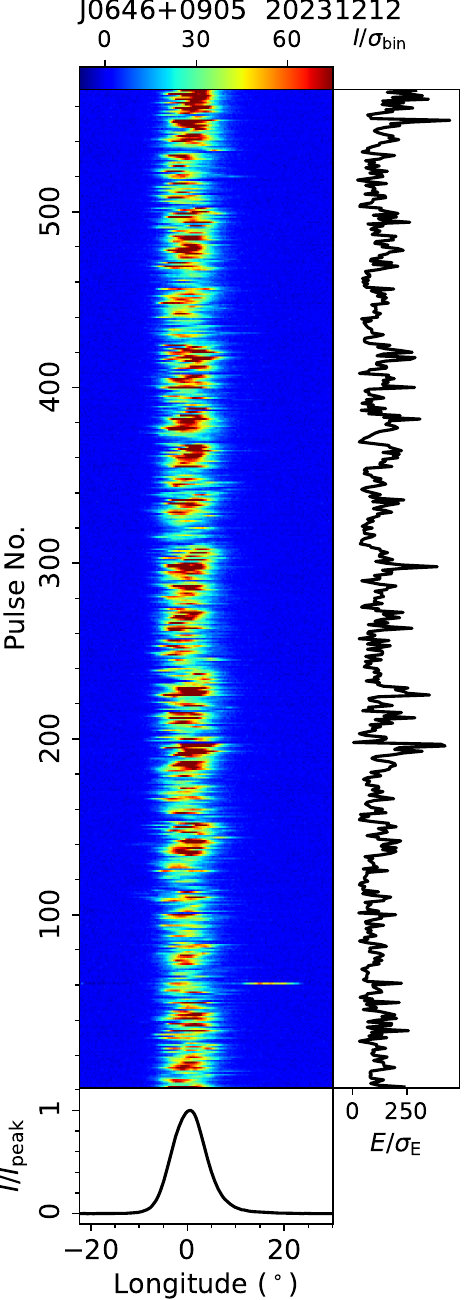}
\includegraphics[width=0.22\textwidth, angle=0]{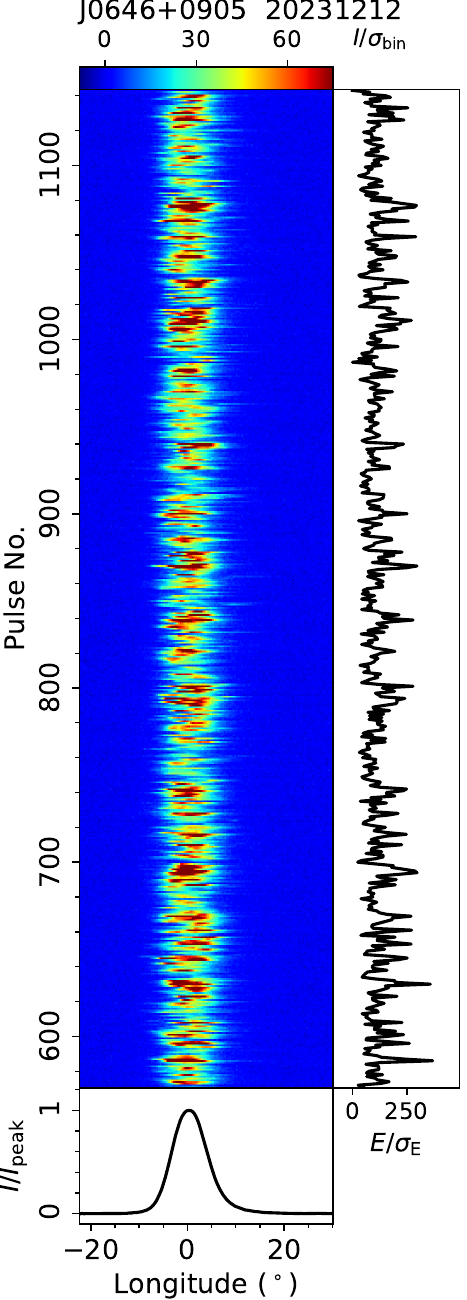}
\figcaption{Single pulse sequences of PSR J0646+0905 from the FAST observation on 20231212. \label{subfig:TP:J0646+0905}}
\end{figure}

\begin{figure}[hbpt]
\centering
\includegraphics[width=0.22\textwidth, angle=0]{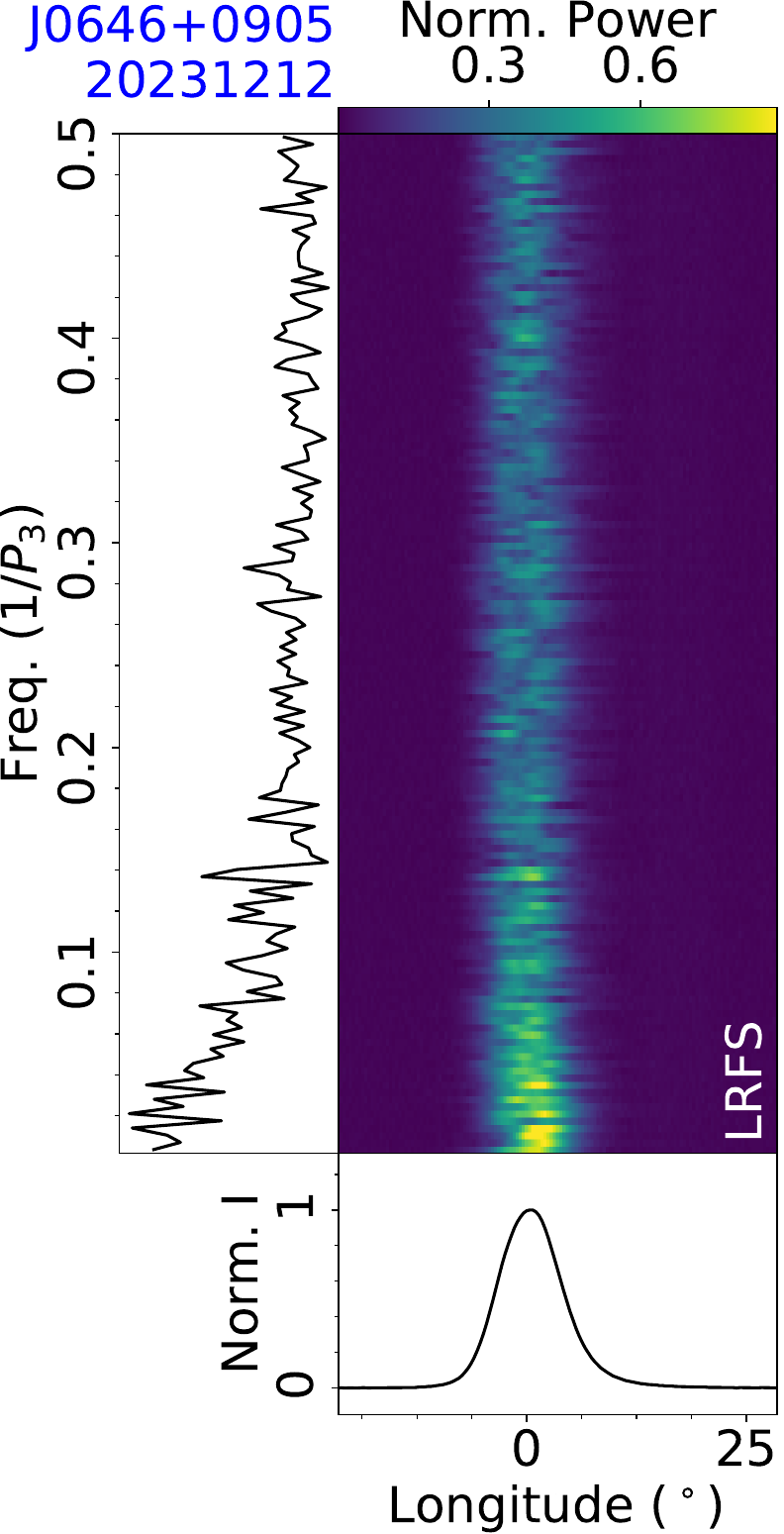}
\includegraphics[width=0.22\textwidth, angle=0]{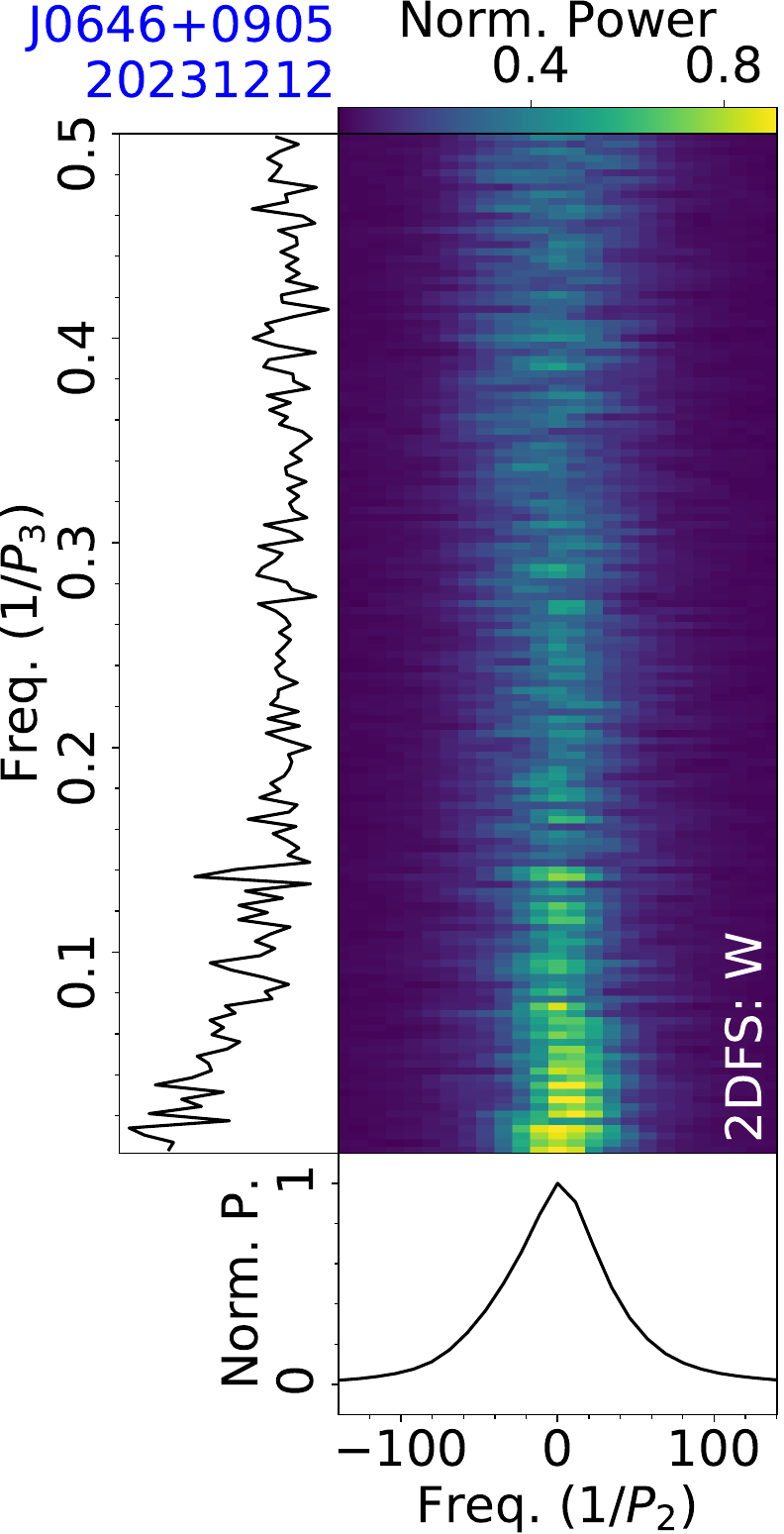}
\figcaption{Fluctuation analysis of PSR J0646+0905 from the FAST observation on 20231212, with LRFS and 2DFS for the on-pulse phase range of a mean pulse profile.
\label{subfig:fluctu:J0646+0905}}
\end{figure}

\subsection{J0630+1002g}
\label{subsec:J0630+1002g}

PSR J0630+1002g was discovered in the FAST GPPS survey \citep{Han2021,han2025}. 

FAST observed the pulsar on 20240414 for 15 minutes and on 20250125 for 46 minutes. From the longer data, a rotation period of $P=2.8815$~s and a dispersion measure of $D\!M=151.8~{\rm cm^{-3}\,pc}$ were determined. 
The single pulse sequence of the observation on 20240414 (Fig.~\ref{subfig:TP:J0630+1002g}) displays systematic subpulse drifting behavior. From LRFS and 2DFS in Fig.~\ref{subfig:fluctu:J0630+1002g}, two components have similar temporal modulation frequencies of $1/P_3=0.1657\pm0.0004$ ($P_3=6.03\pm0.02$) for the leading component and $1/P_3=0.167\pm0.001$ ($P_3=5.98\pm0.03$) for the trailing component. The leading component has the centroid phase modulation frequency of $1/P_2=166\pm7$, which corresponds to $P_2=2.2\pm0.1^\circ$. While the trailing component has a faster subpulse drifting with $1/P_2=99\pm11$, yielding $P_2=3.6\pm0.4^\circ$.

\subsection{J0646+0905}
\label{subsec:J0646+0905}

PSR J0646+0905 was discovered using the Parkes radio telescope in the Perseus Arm Pulsar Survey \citep{Burgay2013}. 

This pulsar was observed by FAST on 20231201 for 18 minutes and 20231212 for 17 minutes. From the observation on 20231212, a rotation period of $P=0.9039$~s and a dispersion measure of $D\!M=149.1~{\rm cm^{-3}\,pc}$ were derived. Single pulse sequences of this observation in Fig.~\ref{subfig:TP:J0646+0905} provide evidence for the positive drifting phenomenon. In the fluctuation spectra (Fig.~\ref{subfig:fluctu:J0646+0905}), the centroid frequencies of the positive drift feature are $1/P_3=0.0417\pm0.0004$ and $1/P_2=9.4\pm0.6$, which correspond to drifting parameters of $P_3=24.0\pm0.2$ periods and $P_3=38\pm2^\circ$.

\begin{figure}[htpb]
\centering
\includegraphics[width=0.22\textwidth, angle=0]{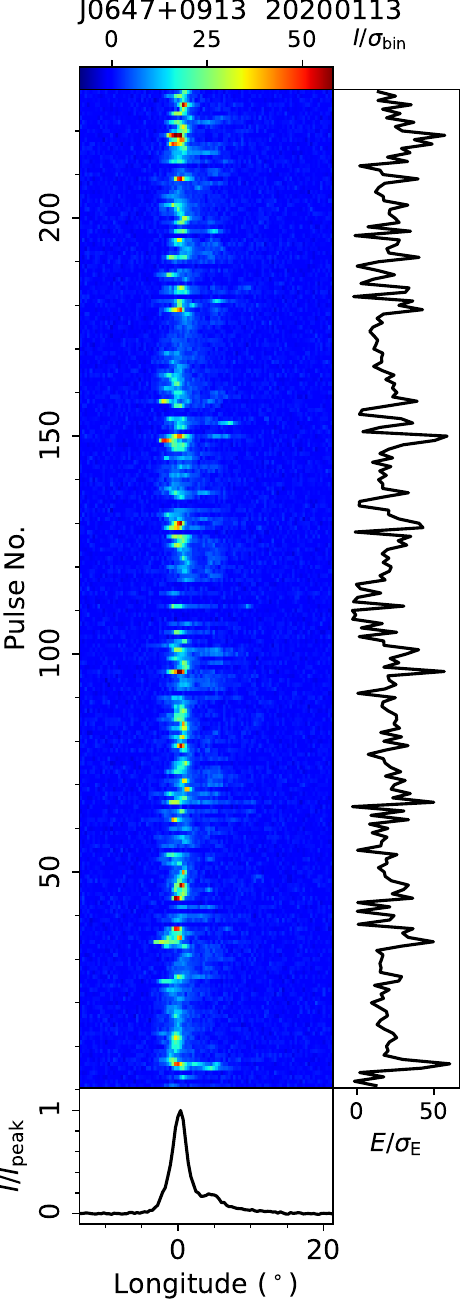}
\figcaption{Single pulse sequence of PSR J0647+0913 from the observation on 20200113. \label{subfig:TP:J0647+0913}}
\end{figure}

\begin{figure}[htpb]
\centering
\includegraphics[width=0.39\textwidth, angle=0]{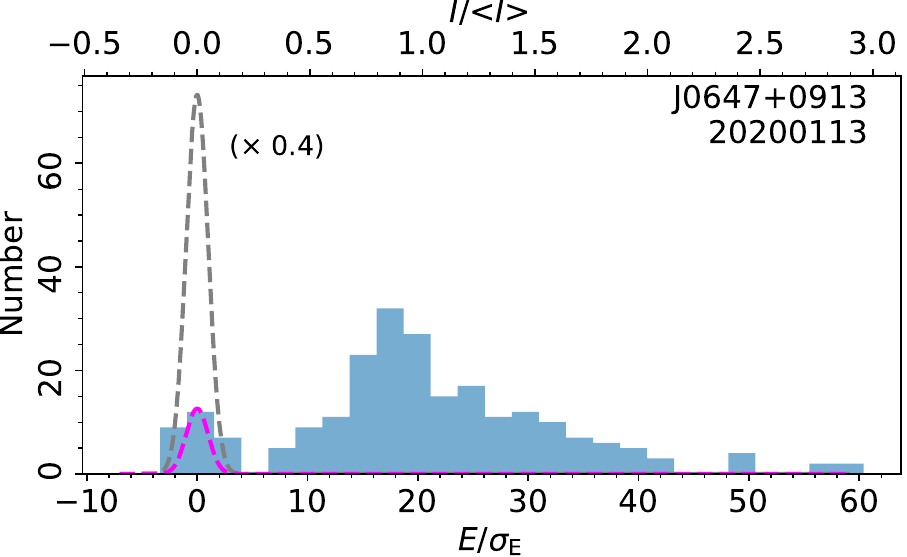}
\figcaption{On-pulse energy histogram of PSR J0647+0913 from the observation on 20200113. \label{subfig:Hist:J0647+0913}}
\end{figure}

\begin{figure}[htpb]
\centering
\includegraphics[width=0.22\textwidth, angle=0]{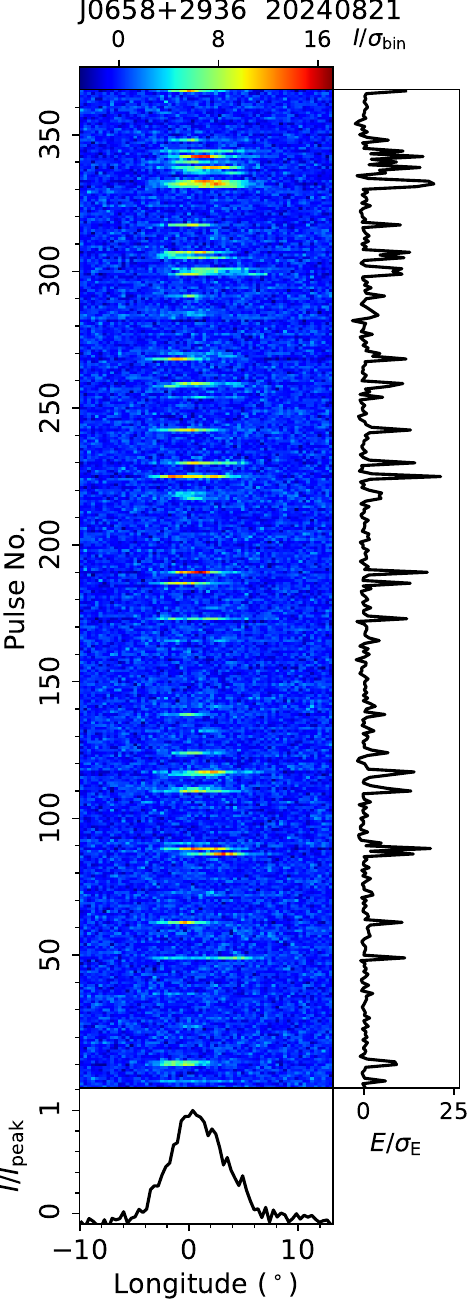}
\includegraphics[width=0.22\textwidth, angle=0]{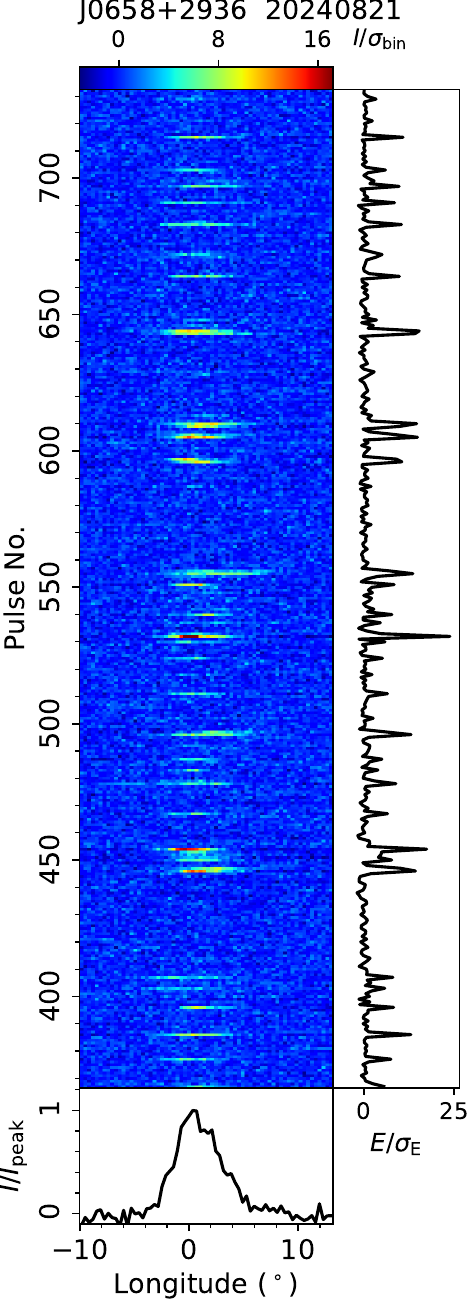}
\figcaption{Single pulse sequences of PSR J0658+2936 from the observation on 20240821.
\label{subfig:TP:J0658+2936}}
\end{figure}

\begin{figure}[htpb]
\centering
\includegraphics[width=0.39\textwidth, angle=0]{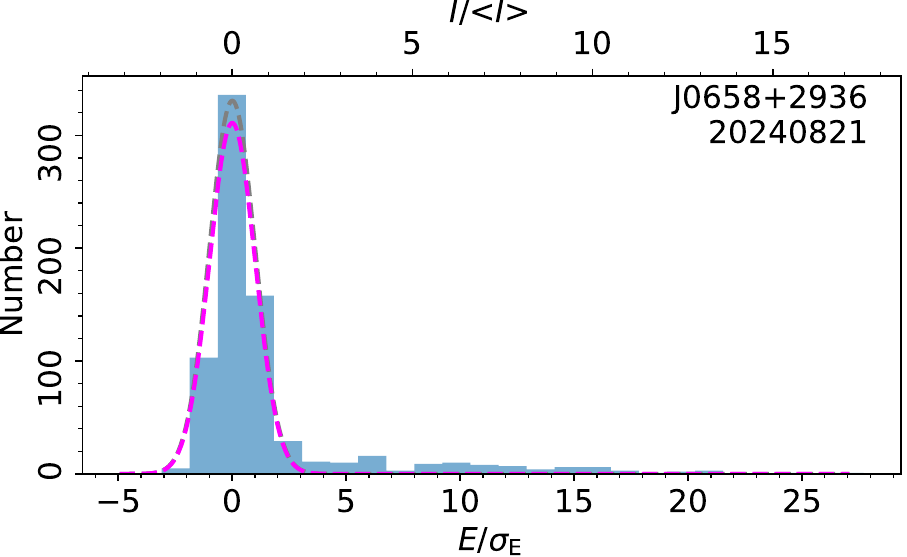}
\figcaption{On-pulse energy histogram of PSR J0658+2936 from the observation on 20240821.
\label{subfig:Hist:J0658+2936}}
\end{figure}

\subsection{J0647+0913}
\label{subsec:J0647+0913}

PSR J0647+0913 was discovered by \citet{Burgay2013} using the Parkes 64-m radio telescope at a central frequency of 1374 MHz. 

The pulsar was observed by FAST on 20200113 for 5 minutes, deriving a rotation period $P=1.2350$~s and a dispersion measure $D\!M=156.9~{\rm cm^{-3}\,pc}$. 
The single pulse sequence and on-pulse integral energy histogram of the FAST observation are shown in Fig.~\ref{subfig:TP:J0647+0913} and \ref{subfig:Hist:J0647+0913}, which illustrate the existence of the nulling behavior. There are 26 single pulses whose integral energies are less than 3$\sigma_{\rm E}$ in this observation.

\begin{figure}[htpb]
\centering
\includegraphics[width=0.22\textwidth, angle=0]{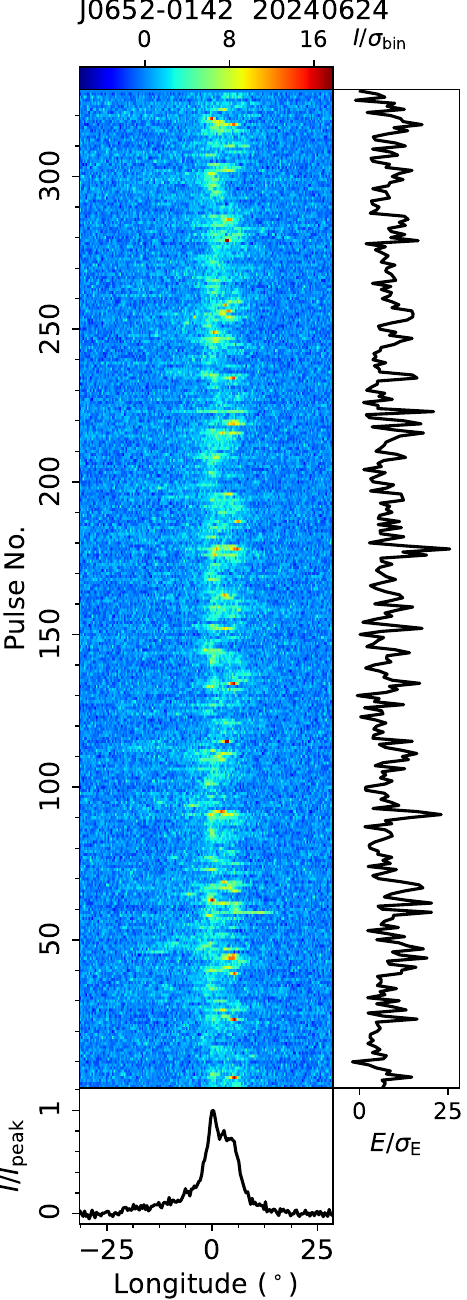}
\figcaption{Single pulse sequence of PSR J0652-0142 from the FAST observation on 20240624.
\label{subfig:TP:J0652-0142}}
\end{figure}

\begin{figure}[hbpt]
\centering
\includegraphics[width=0.44\textwidth, angle=0]{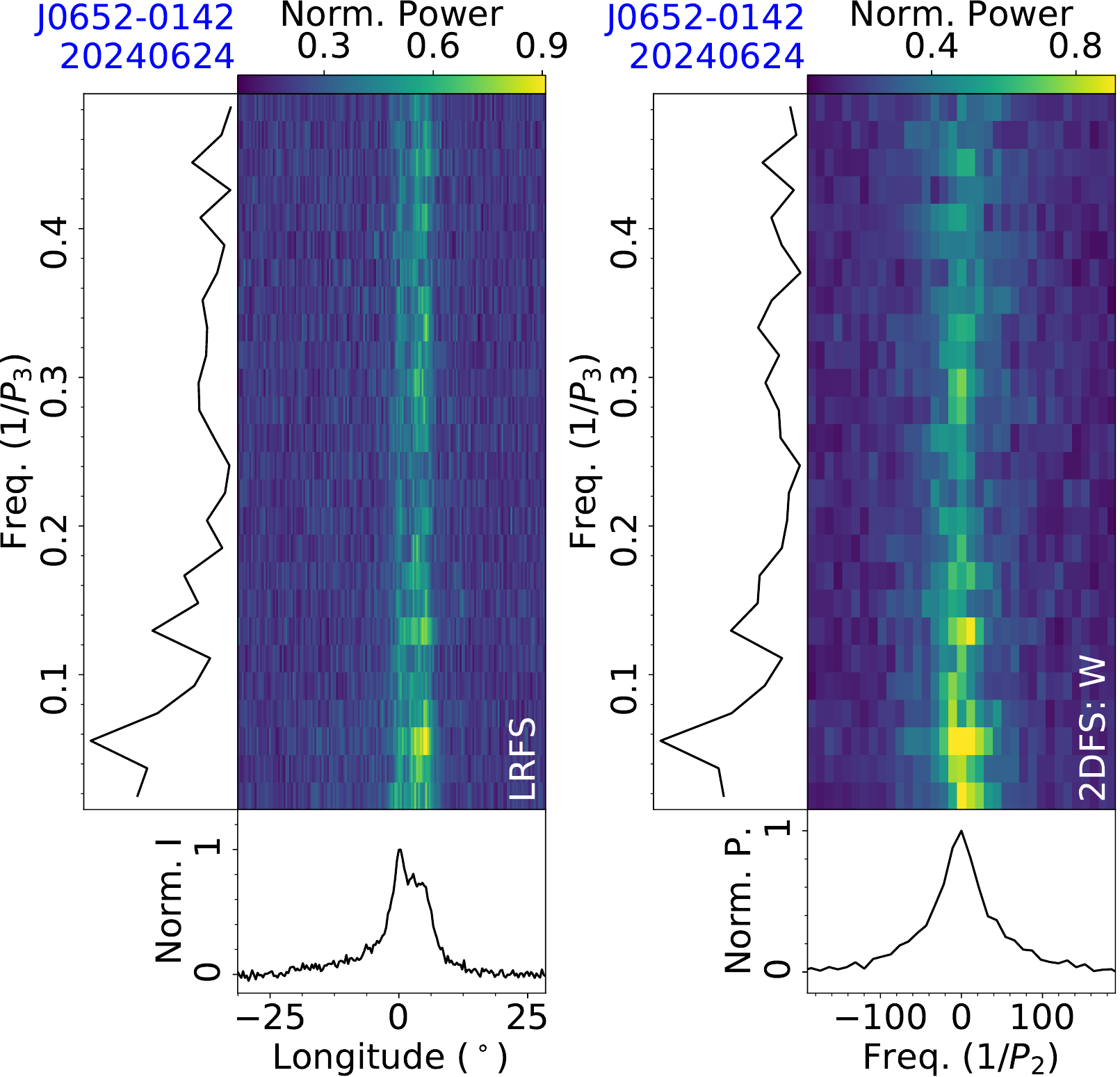}
\figcaption{Fluctuation analysis of PSR J0652-0142 from the FAST observation on 20240624, with LRFS and 2DFS for the on-pulse phase range of the mean pulse profile.
\label{subfig:fluctu:J0652-0142}}
\end{figure}

\begin{figure}[htpb]
\centering
\includegraphics[width=0.22\textwidth, angle=0]{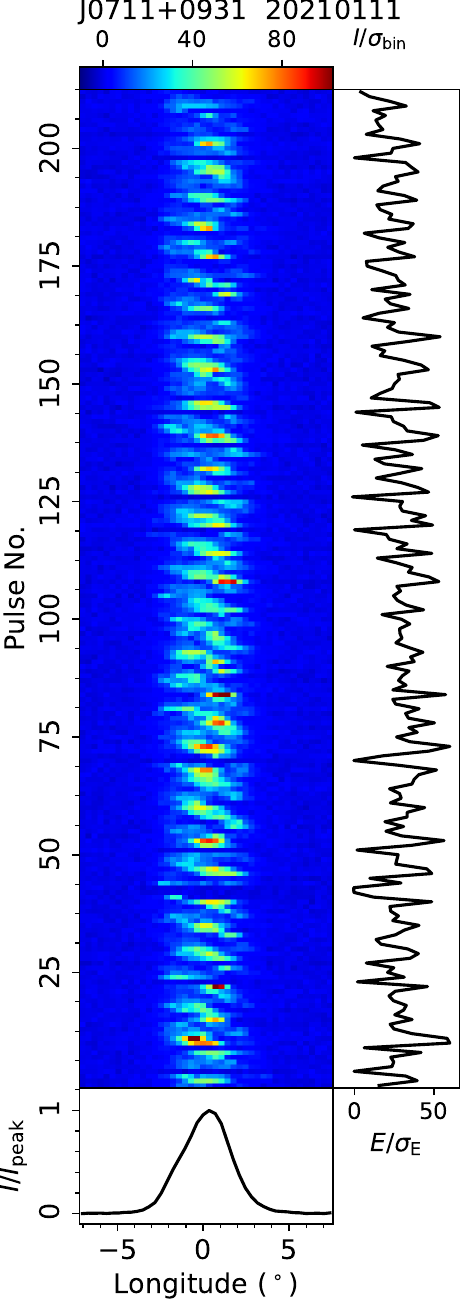}
\figcaption{Single pulse sequence of PSR J0711+0931 from the observation on 20210111.
\label{subfig:TP:J0711+0931}}
\end{figure}

\begin{figure}[htpb]
\centering
\includegraphics[width=0.39\textwidth, angle=0]{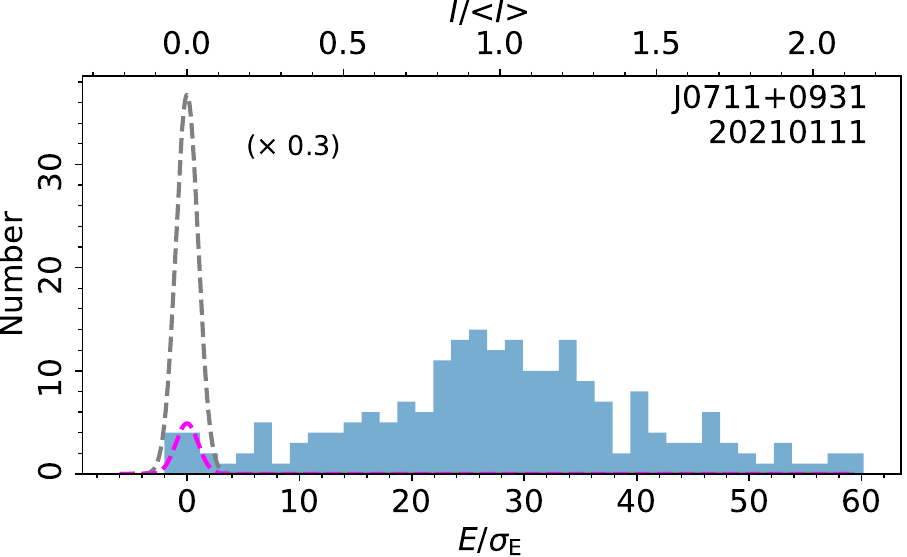}
\figcaption{On-pulse energy histogram of PSR J0711+0931 from the observation on 20210111. \label{subfig:Hist:J0711+0931}}
\end{figure}

\begin{figure}[htpb]
\centering
\includegraphics[width=0.22\textwidth, angle=0]{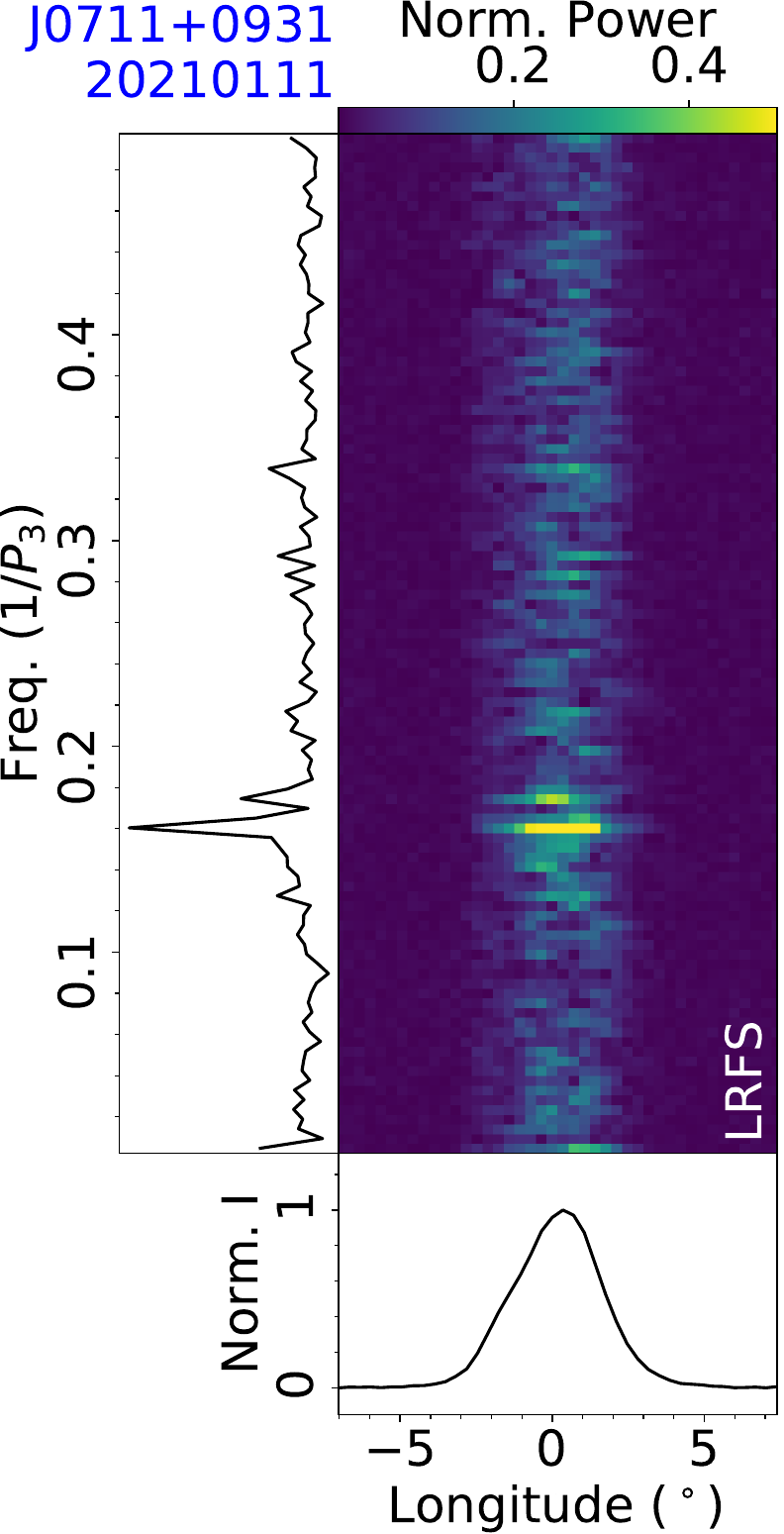}
\includegraphics[width=0.22\textwidth, angle=0]{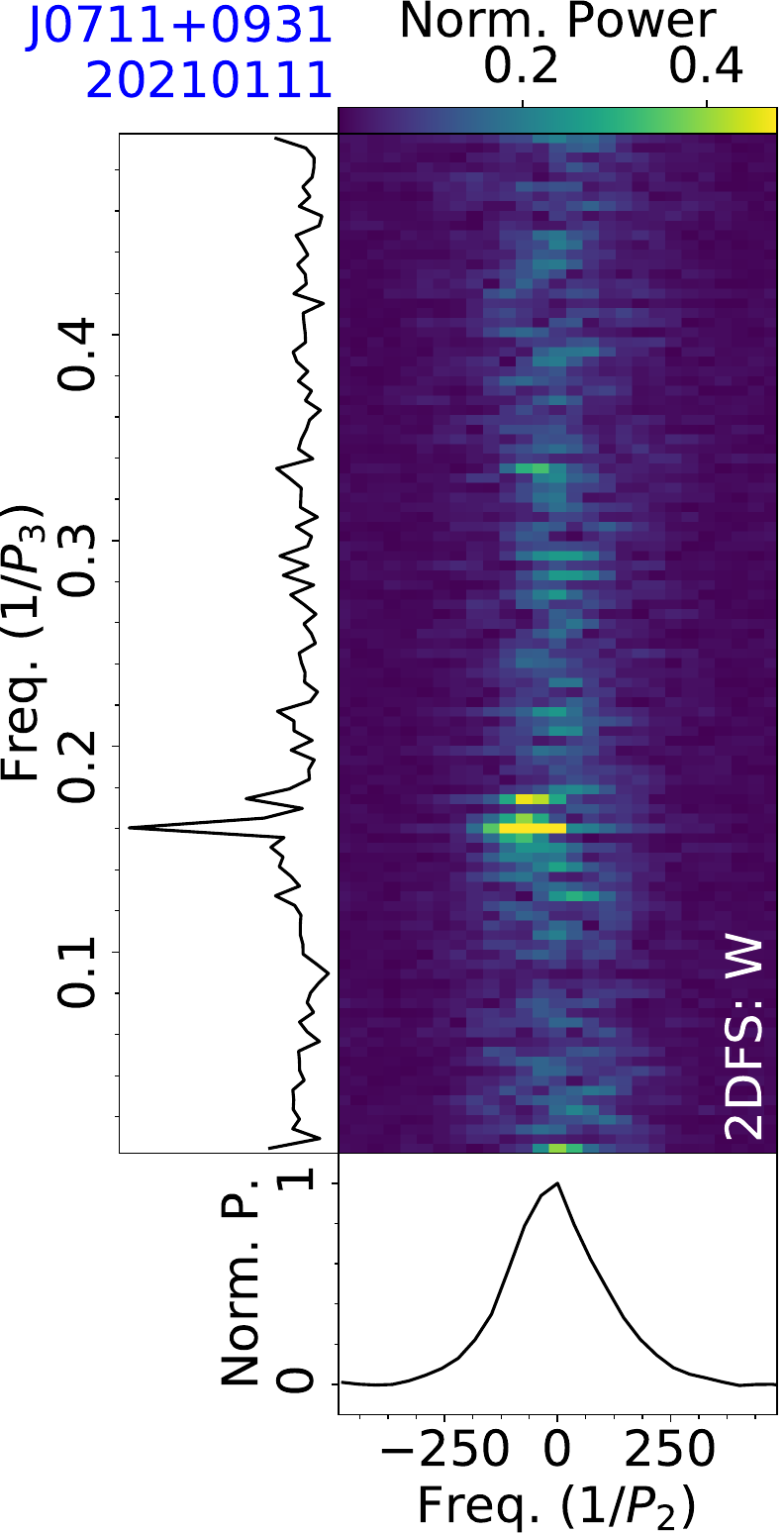}
\figcaption{Fluctuation analysis of PSR J0711+0931 from the FAST observation on 20210111, with LRFS and 2DFS for the on-pulse phase range of a mean pulse profile.
\label{subfig:fluctu:J0711+0931}}
\end{figure}

\subsection{J0652-0142}
\label{subsec:J0652-0142}

PSR J0652-0142 was discovered in the Perseus Arm Pulsar Survey with the Parkes 64-m radio telescope \citep{Burgay2013}.

This pulsar was observed by FAST on 20240624 for 5 minutes, deriving a rotation period $P=0.9241$~s and a dispersion measure $D\!M=116.4~{\rm cm^{-3}\,pc}$. The single pulse sequence is shown in Fig.~\ref{subfig:TP:J0652-0142}. From the fluctuation spectra in Fig.~\ref{subfig:fluctu:J0652-0142}, the pulsar prefers to exhibit a positive drift behavior. The centroid of the drift feature in 2DFS is at $1/P_3=0.051\pm0.002$ and $1/P_2=3\pm1$, corresponding to periodicities of $P_3=20\pm1$ periods and $P_2=115\pm45$ degrees.

\subsection{J0658+2936}
\label{subsec:J0658+2936}

PSR J0658+2936 was discovered by \citet{Dong2023} using the CHIME telescope. 

This pulsar was observed by FAST on 20240821 for 10 minutes, and a rotation period $P=0.8236$~s and a dispersion measure $D\!M=38.8~{\rm cm^{-3}\,pc}$ were determined. Single pulse sequences in Fig.~\ref{subfig:TP:J0658+2936} show the nulling phenomenon. The nulling fraction of this observation is estimated to be 94$\pm$7\%, from the on-pulse integral energy histogram in Fig.~\ref{subfig:Hist:J0658+2936}.

\subsection{J0711+0931}
\label{subsec:J0711+0931}

PSR J0711+0931 was discovered by \citet{Lommen2000} using the 305 m Arecibo telescope at 430 MHz. The drifting behavior was reported by \citet{Song2023}. 

This pulsar was observed by FAST on 20210111 for 8 minutes, deriving a rotation period $P=2.4284$~s and a dispersion measure $D\!M=44.2~{\rm cm^{-3}\,pc}$. We also detected subpulse drifting from the FAST data. The single pulse sequence in Fig.~\ref{subfig:TP:J0711+0931} shows organized negatively drifting bands and a short-scale nulling phenomenon. Temporally intensity modulation frequency in LRFS and 2DFS (Fig.~\ref{subfig:fluctu:J0711+0931}) is narrow, indicating a systematic drift feature. Centroid modulation frequencies of the drift feature in 2DFS are estimated to be $1/P_3=0.162\pm0.001$ and $1/P_2=-74\pm3$, corresponding to $P_3=6.19\pm0.02$ periods and $P_2=-4.9\pm0.2^\circ$. 
The nulling fraction is 4.0$\pm$0.7\% from the energy histogram in Fig.~\ref{subfig:Hist:J0711+0931}. 

\begin{figure}[htpb]
\centering
\includegraphics[width=0.22\textwidth, angle=0]{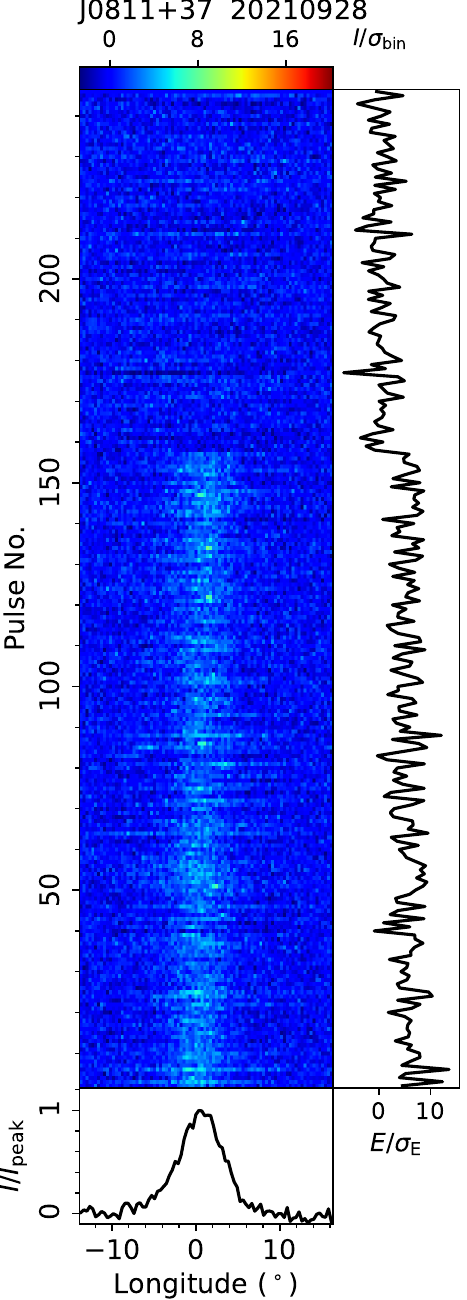}
\figcaption{Single pulse sequence of PSR J0811+37 from the observation on 20210928. \label{subfig:TP:J0811+37}}
\end{figure}

\begin{figure}[htpb]
\centering
\includegraphics[width=0.39\textwidth, angle=0]{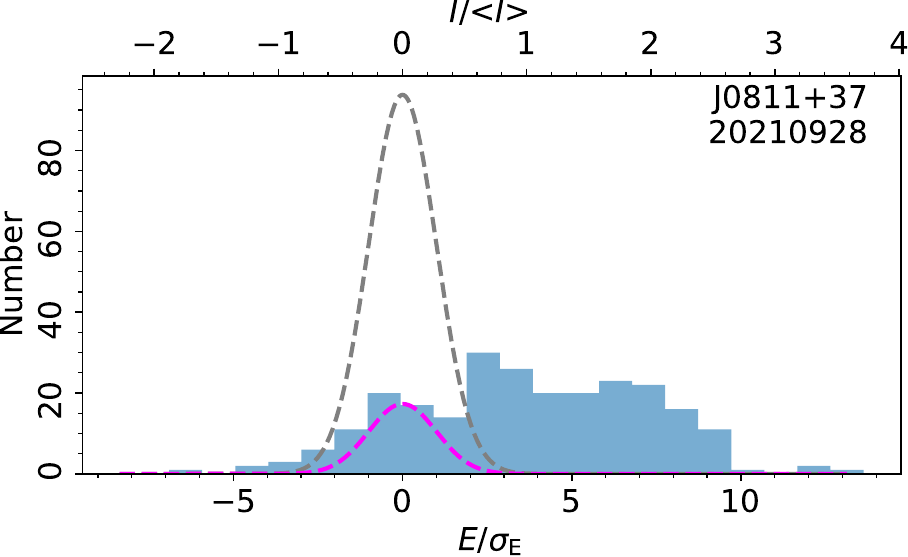}
\figcaption{On-pulse energy histogram of PSR J0811+37 for the observation on 20210928. \label{subfig:Hist:J0811+37}}
\end{figure}

\begin{figure}[htpb]
\centering
\includegraphics[width=0.22\textwidth, angle=0]{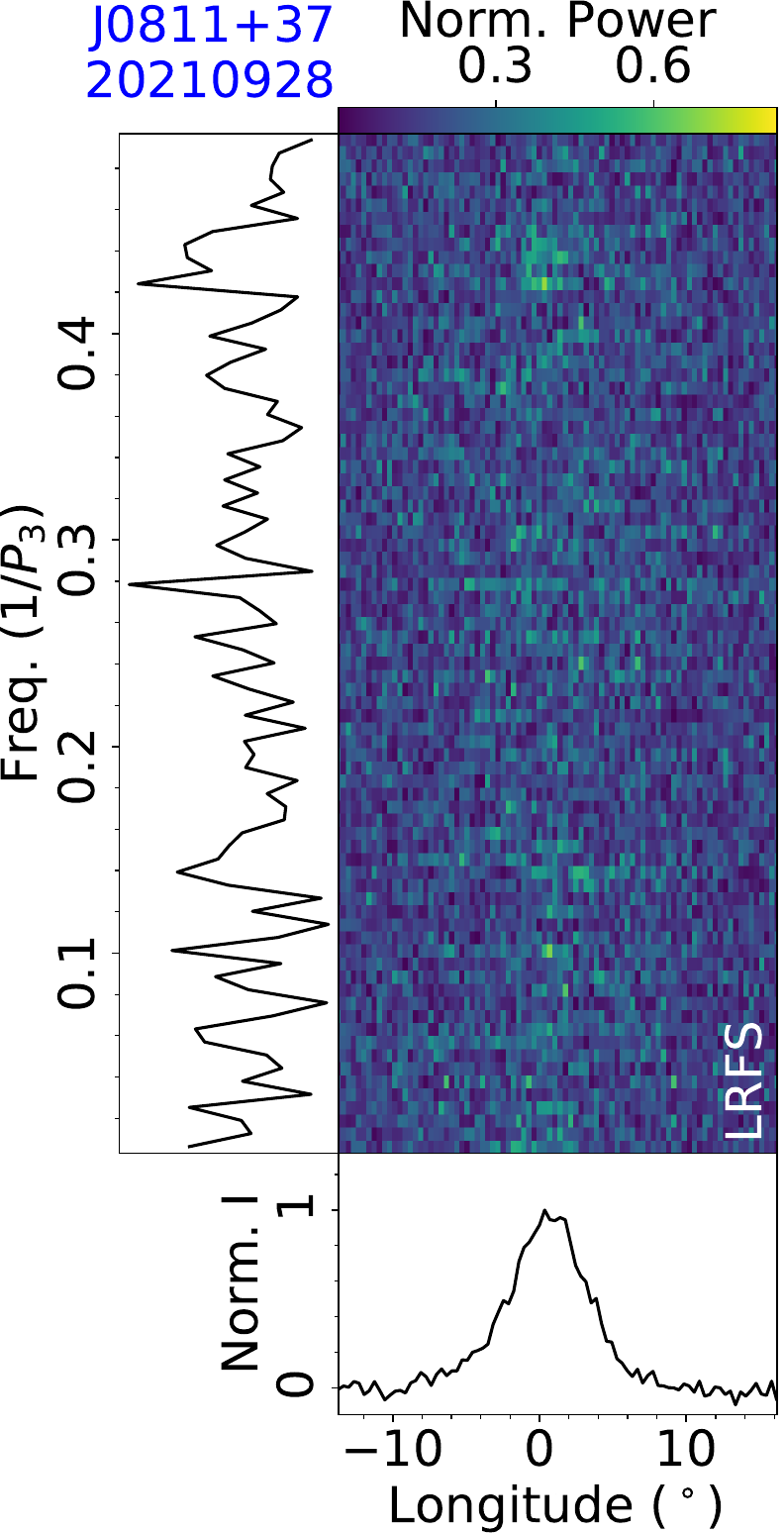}
\includegraphics[width=0.22\textwidth, angle=0]{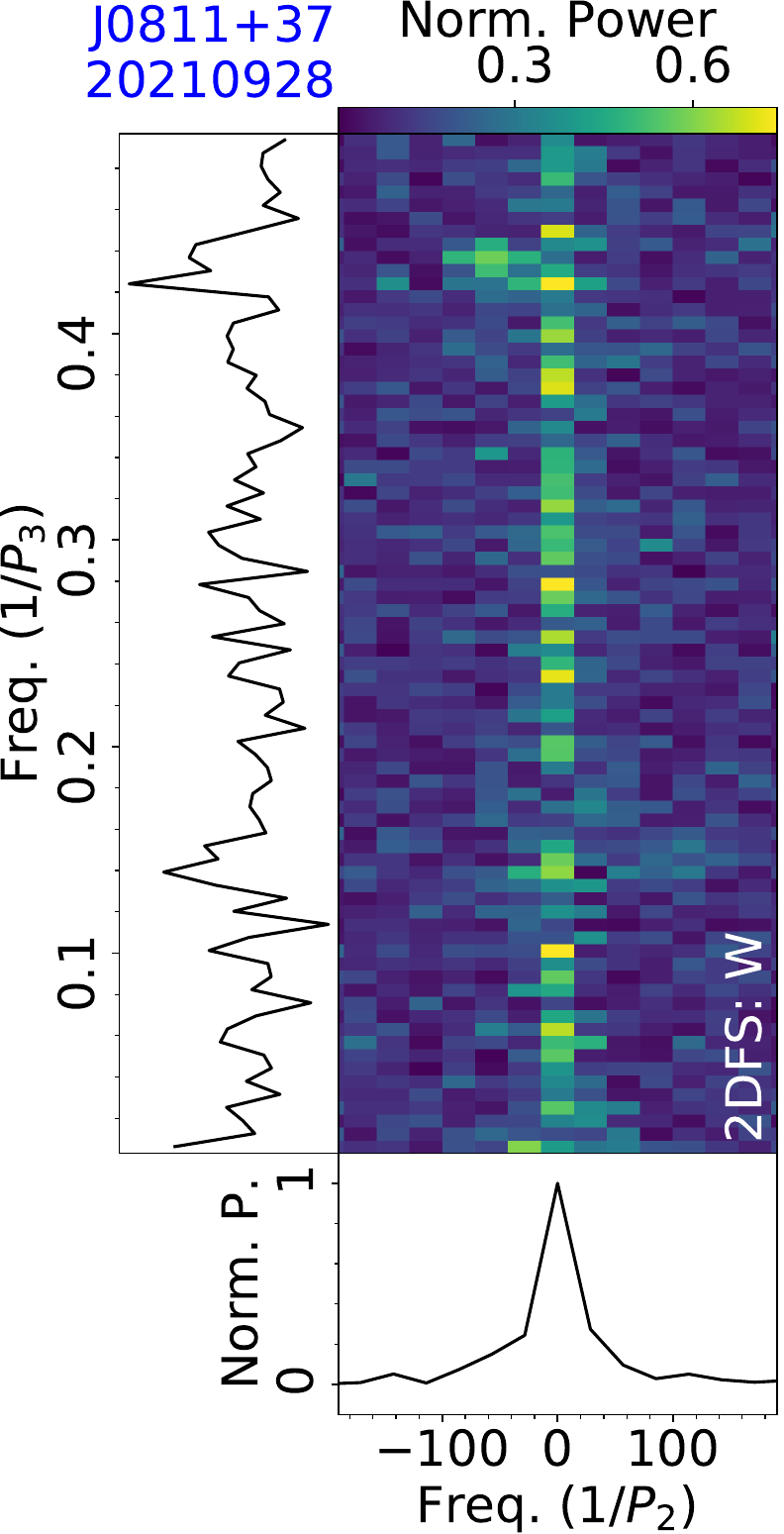}
\figcaption{Fluctuation analysis of PSR J0811+37 from the observation on 20210928, with LRFS and 2DFS for the on-pulse phase range of a mean pulse profile.
\label{subfig:fluctu:J0811+37}}
\end{figure}

\subsection{J0811+37}
\label{subsec:J0811+37}

PSR J0811+37 was discovered by \citet{Tyulbashev2017} using the Big Scanning Antenna at 111 MHz. 

This pulsar was observed by FAST on 20210928 for 5 minutes, with a rotation period $P=1.2483$~s and a dispersion measure $D\!M=17.2~{\rm cm^{-3}\,pc}$ from this observation. The single pulse sequence of the FAST observation is shown in Fig.~\ref{subfig:TP:J0811+37}, which displays nulling phenomenon. From the on-pulse integral energy histogram in Fig.~\ref{subfig:TP:J0811+37}, the nulling fraction of this observation is estimated to be 18$\pm$3\%. In 2DFS (Fig.~\ref{subfig:fluctu:J0811+37}), there is a negative drift feature with the centroid of $1/P_3=0.433\pm0.001$ and $1/P_2=-57\pm3$, corresponding to $P_3=2.307\pm0.005$ periods and $P_2=-6.3\pm0.4^\circ$. 

\begin{figure}[htpb]
\centering
\includegraphics[width=0.22\textwidth, angle=0]{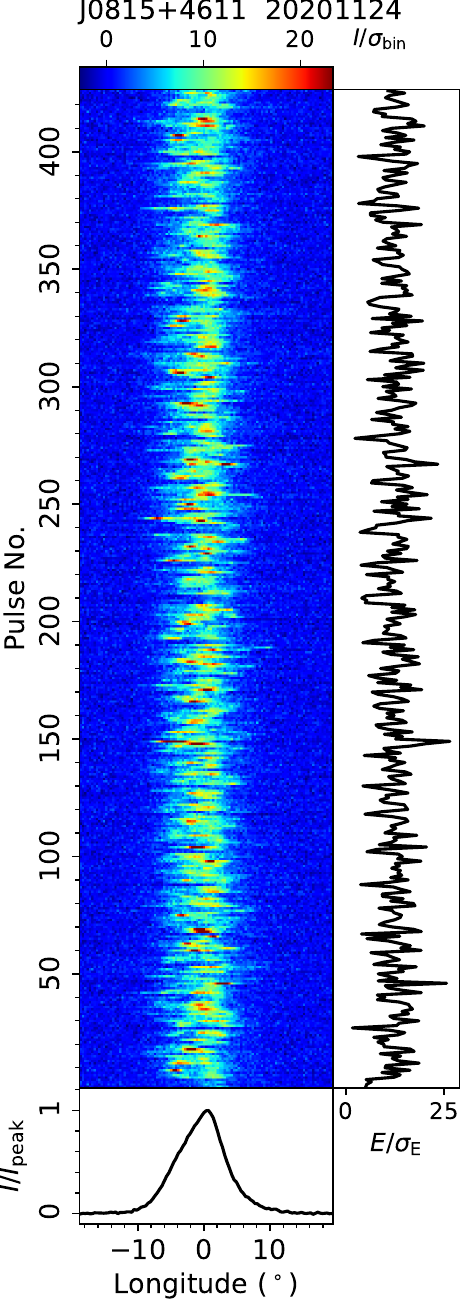}
\includegraphics[width=0.22\textwidth, angle=0]{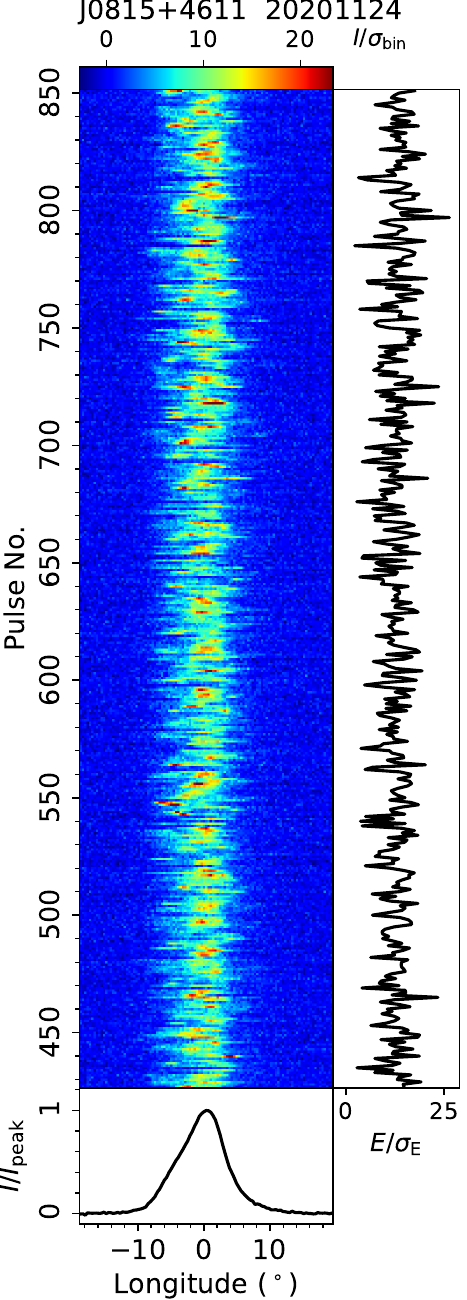}
\figcaption{Single pulse sequences of PSR J0815+4611 from the observation on 20201124.
\label{subfig:TP:J0815+4611}}
\end{figure}

\begin{figure}[htpb]
\centering
\includegraphics[width=0.22\textwidth, angle=0]{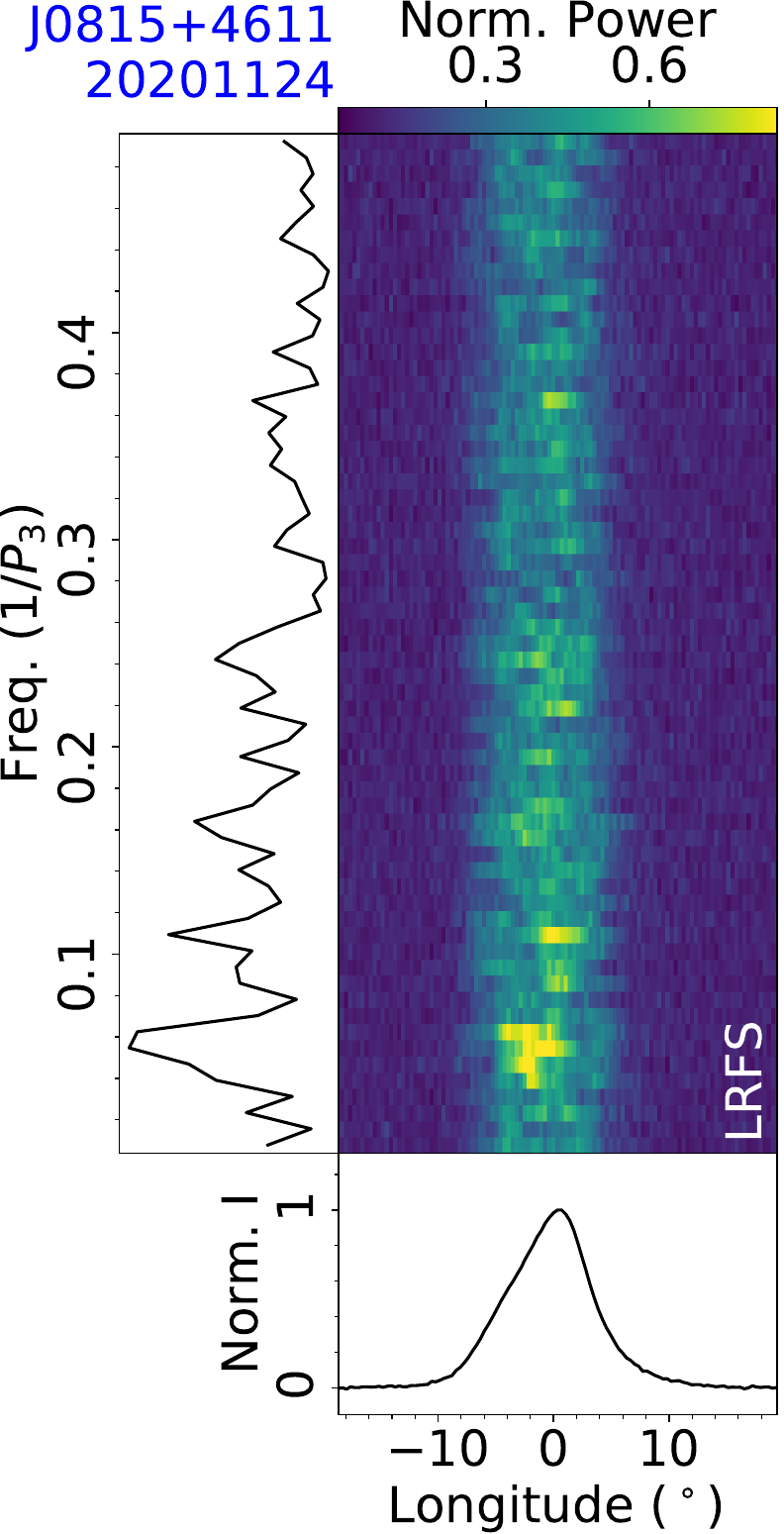} 
\includegraphics[width=0.22\textwidth, angle=0]{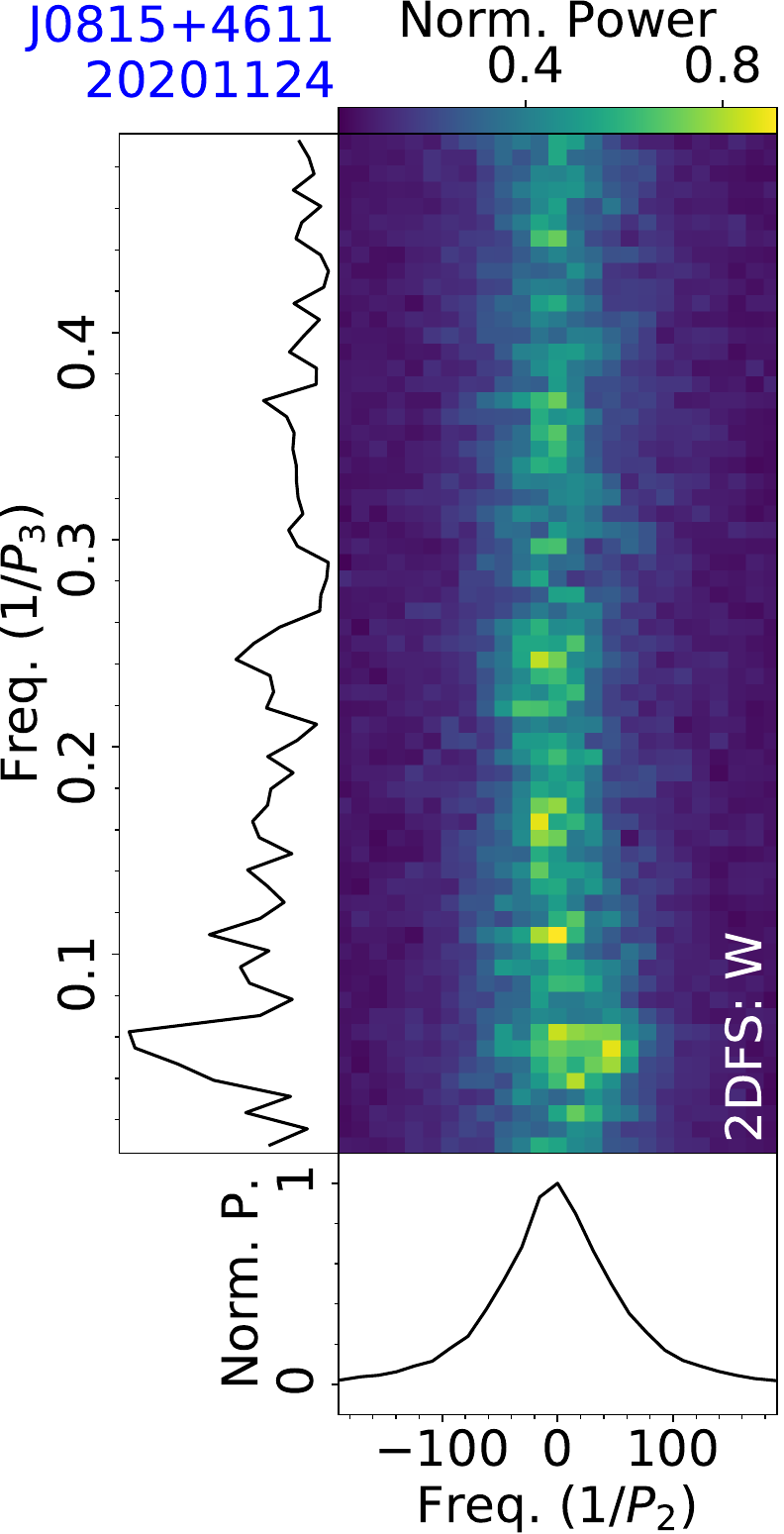}
\figcaption{Fluctuation analysis of PSR J0815+4611 from the observation on 20201124, with LRFS and 2DFS for the on-pulse phase range of a mean pulse profile. \label{subfig:fluctu:J0815+4611}}
\end{figure}

\begin{figure}[htpb]
\centering
\includegraphics[width=0.22\textwidth, angle=0]{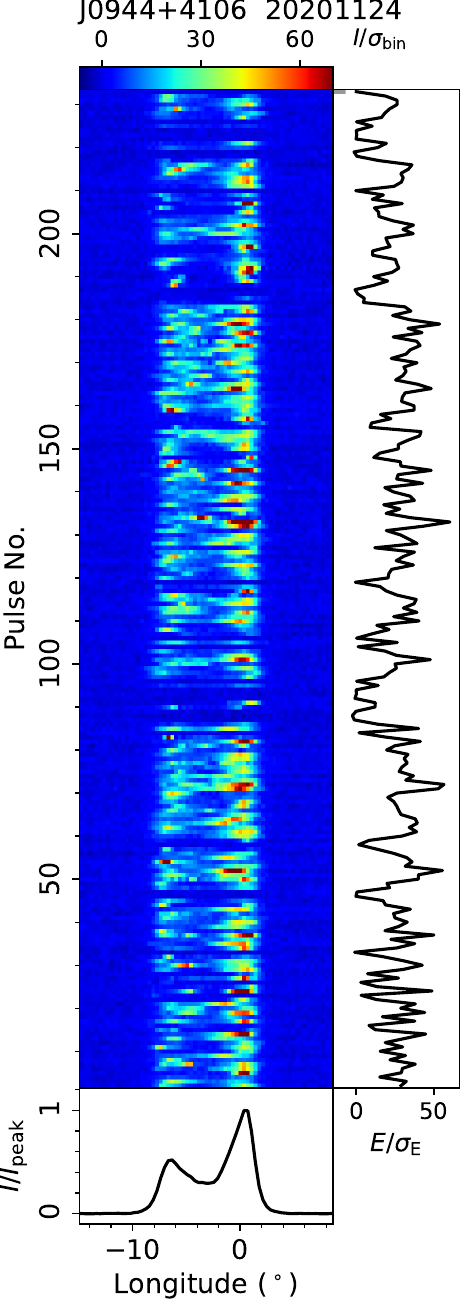}
\figcaption{Single pulse sequence of PSR J0944+4106 from the observation on 20201124. \label{subfig:TP:J0944+4106}}
\end{figure}

\begin{figure}[htpb]
\centering
\includegraphics[width=0.39\textwidth, angle=0]{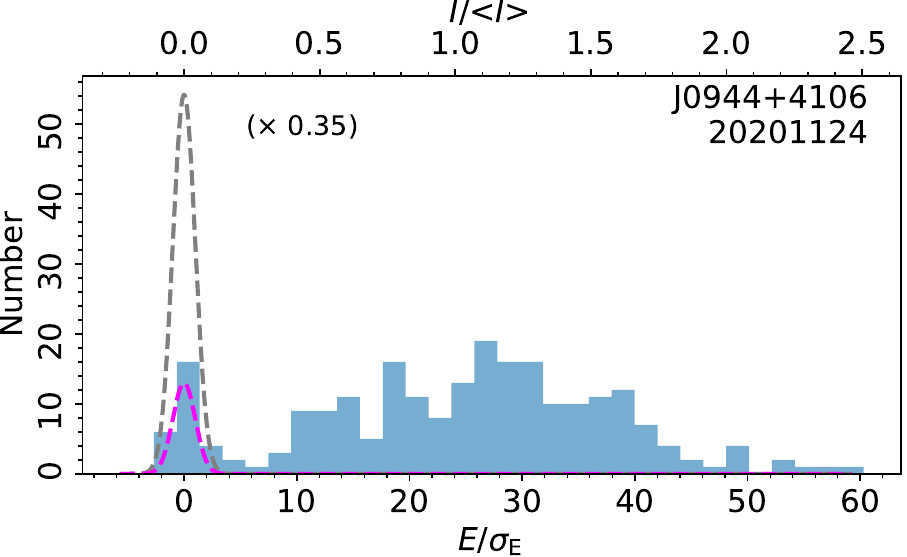}
\figcaption{On-pulse energy histogram of PSR J0944+4106 from the observation on 20201124. \label{subfig:Hist:J0944+4106}}
\end{figure}

\begin{figure}[htpb]
\centering
\includegraphics[width=0.22\textwidth, angle=0]{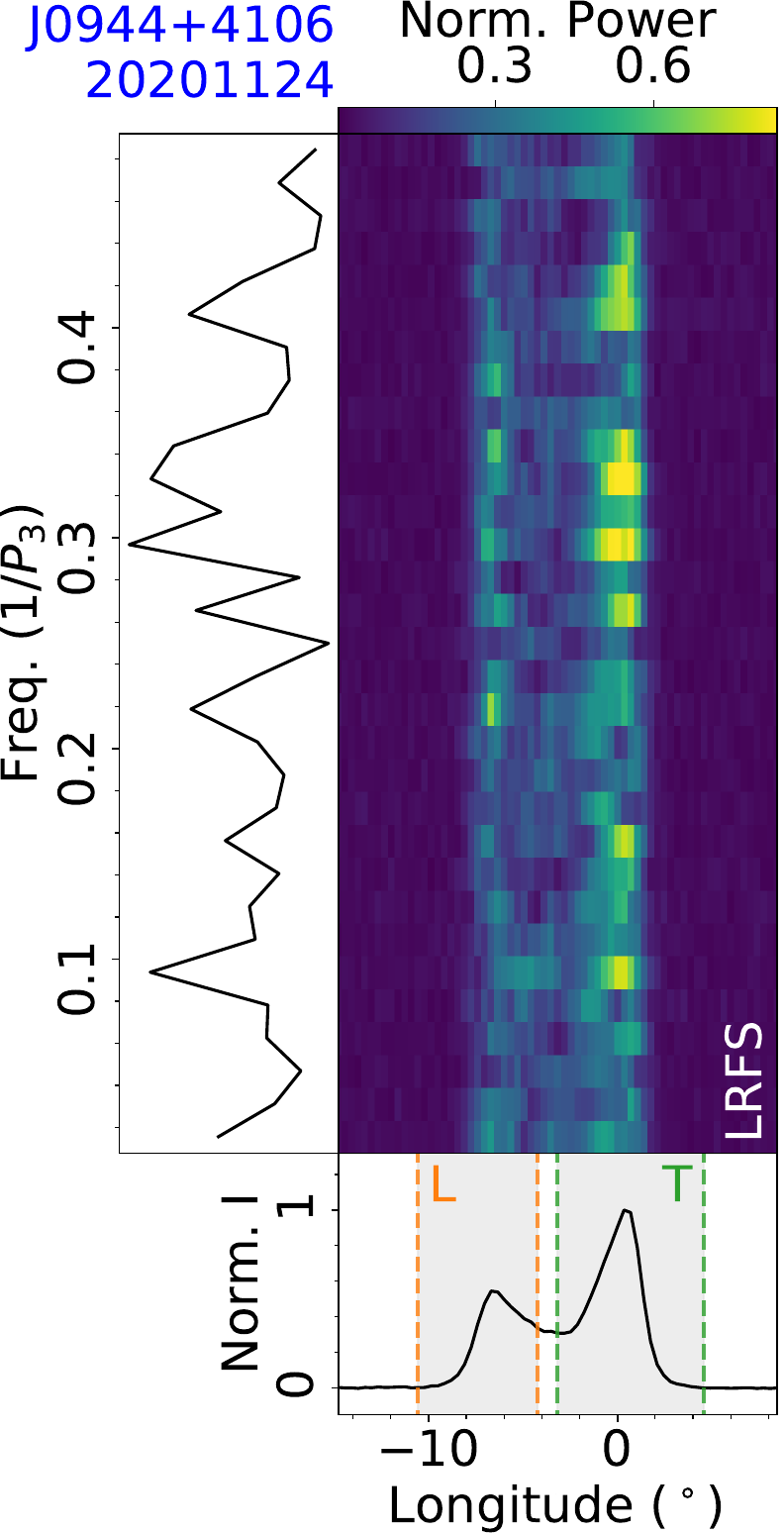}
\includegraphics[width=0.22\textwidth, angle=0]{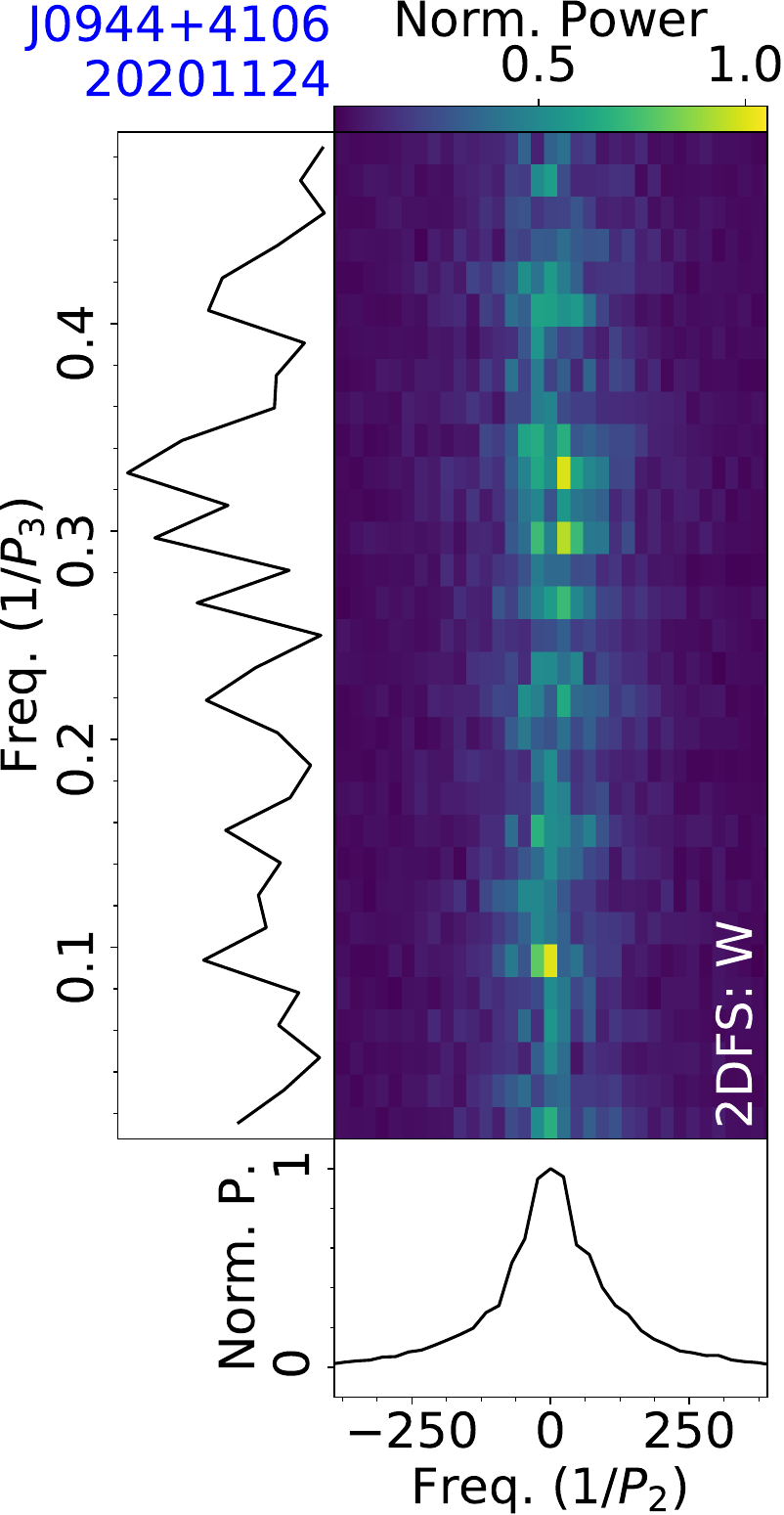}\\
\includegraphics[width=0.22\textwidth, angle=0]{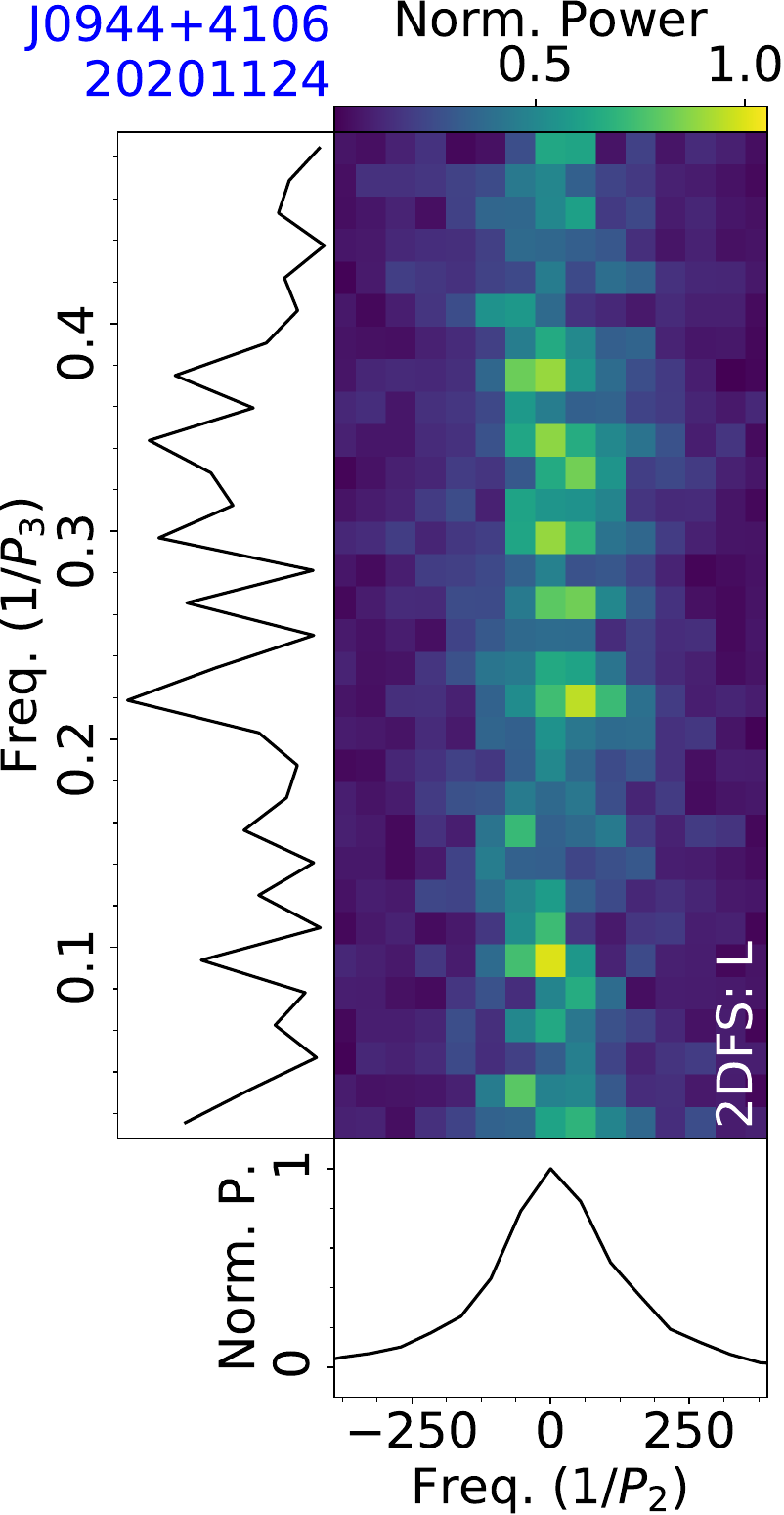}
\includegraphics[width=0.22\textwidth, angle=0]{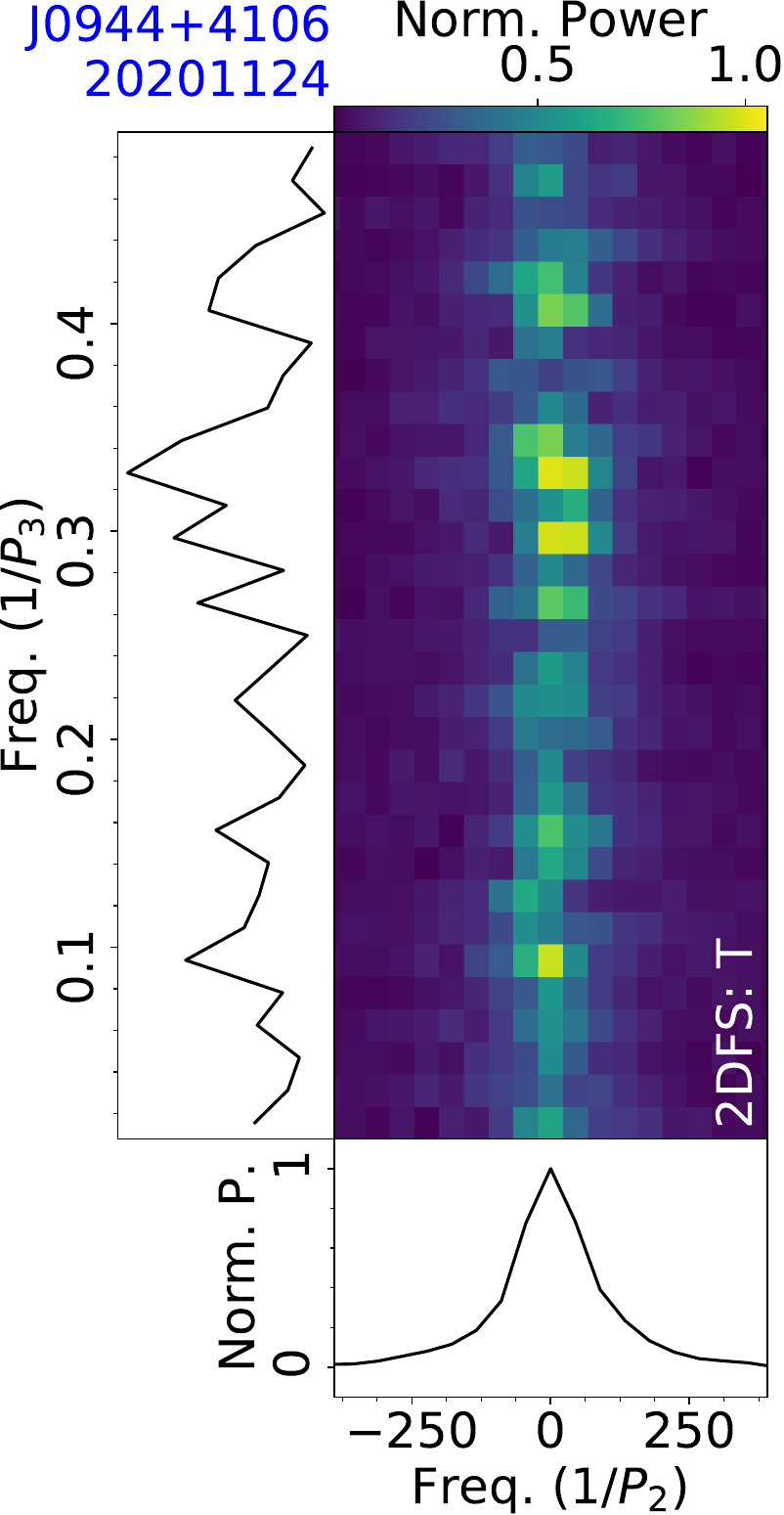}
\figcaption{Fluctuation analysis of PSR J0944+4106 for the observation on 20201124, with LRFS (top-left), and 2DFS for the leading part (top-right), central part (bottom-left) and trailing part (bottom-right) of a mean pulse profile.
\label{subfig:fluctu:J0944+4106}}
\end{figure}

\subsection{J0815+4611}
\label{subsec:J0815+4611}

PSR J0815+4611 was first reported by \citet{Sanidas2019} using the LOFAR Tied-Array All-Sky Survey (LOTAAS). \citet{Zhao2023} reported the drifting properties at 290-500 MHz using the FAST wide-band observations. 

The pulsar was also observed by FAST on 20201124 for 6 minutes, yielding a rotation period $P=0.4342$~s and a dispersion measure $D\!M=10.6~{\rm cm^{-3}\,pc}$. 
Single pulse sequences are shown in Fig.~\ref{subfig:TP:J0815+4611}, displaying a subpulse drifting behavior. In LRFS and 2DFS (Fig.~\ref{subfig:fluctu:J0815+4611}), there is a drift feature with the centroid characterized by $1/P_3=0.053\pm0.001$ and $1/P_2=29\pm2$, which correspond to $P_3=19.1\pm0.3$ periods and $P_2=12\pm2^\circ$.

\begin{figure}[htpb]
\centering
\includegraphics[width=0.22\textwidth, angle=0]{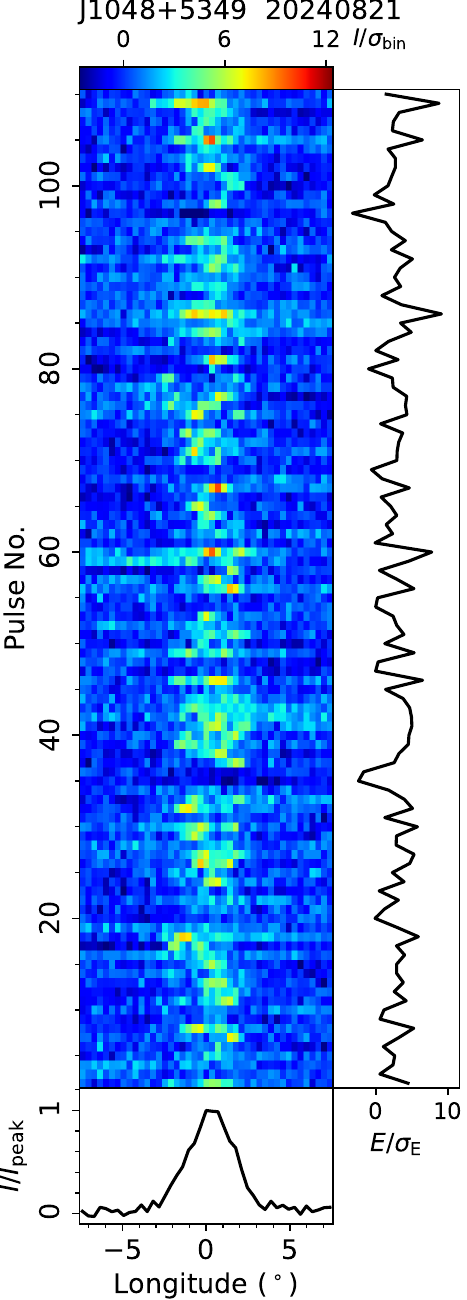}
\includegraphics[width=0.22\textwidth, angle=0]{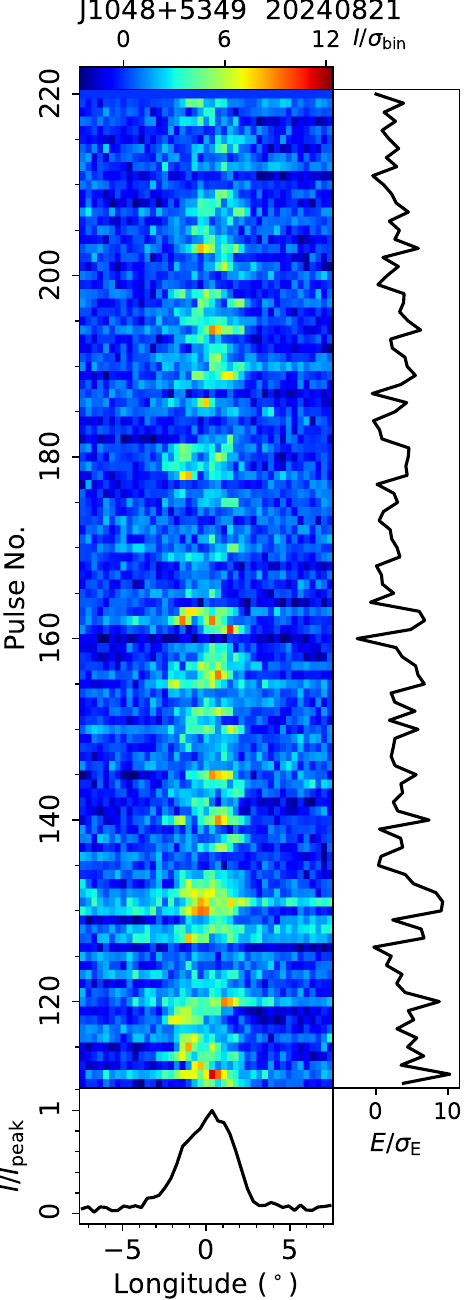}
\figcaption{Single pulse sequences of PSR J1048+5349 from the observation on 20240821.
\label{subfig:TP:J1048+5349}}
\end{figure}

\begin{figure}[htpb]
\centering
\includegraphics[width=0.22\textwidth, angle=0]{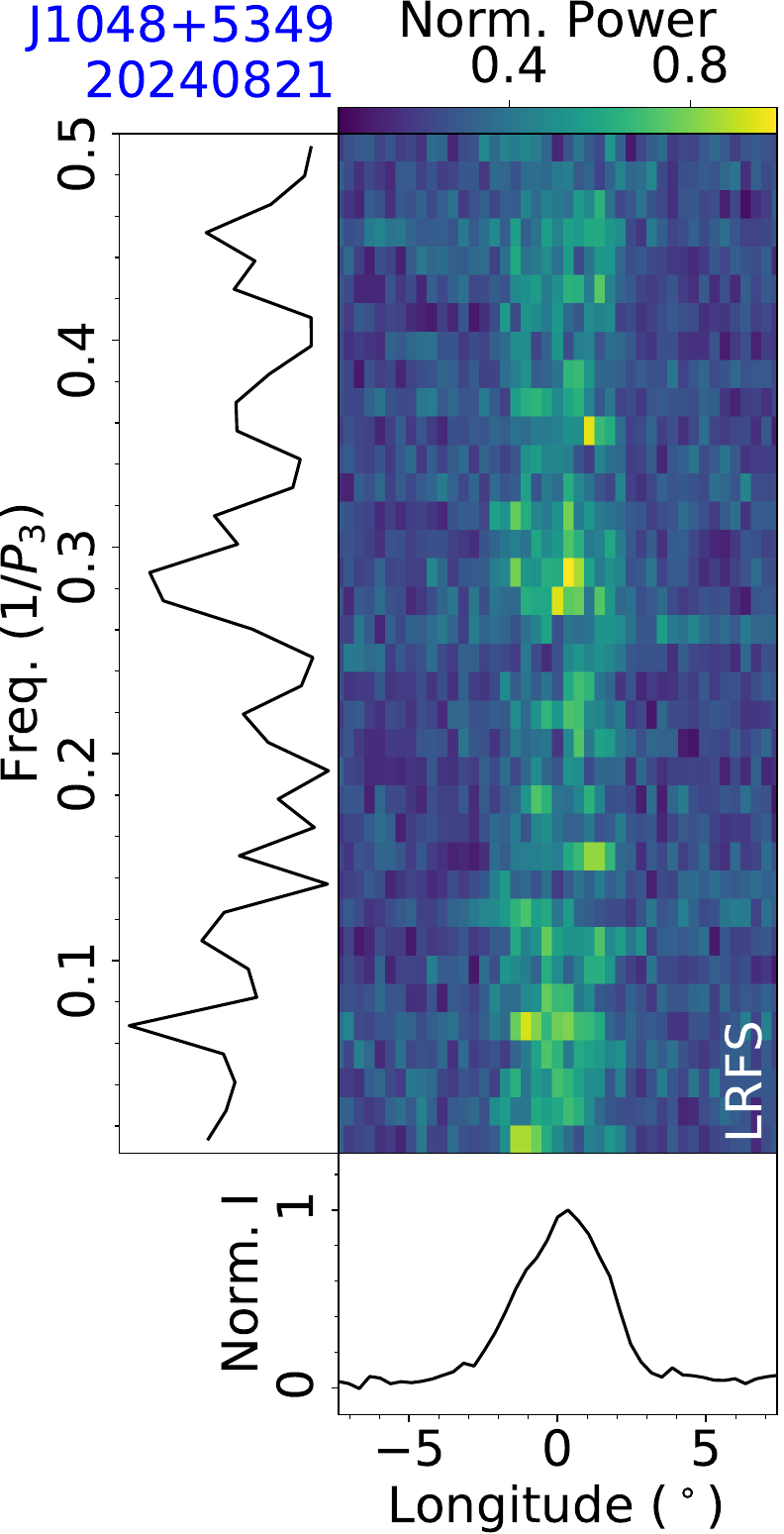}
\includegraphics[width=0.22\textwidth, angle=0]{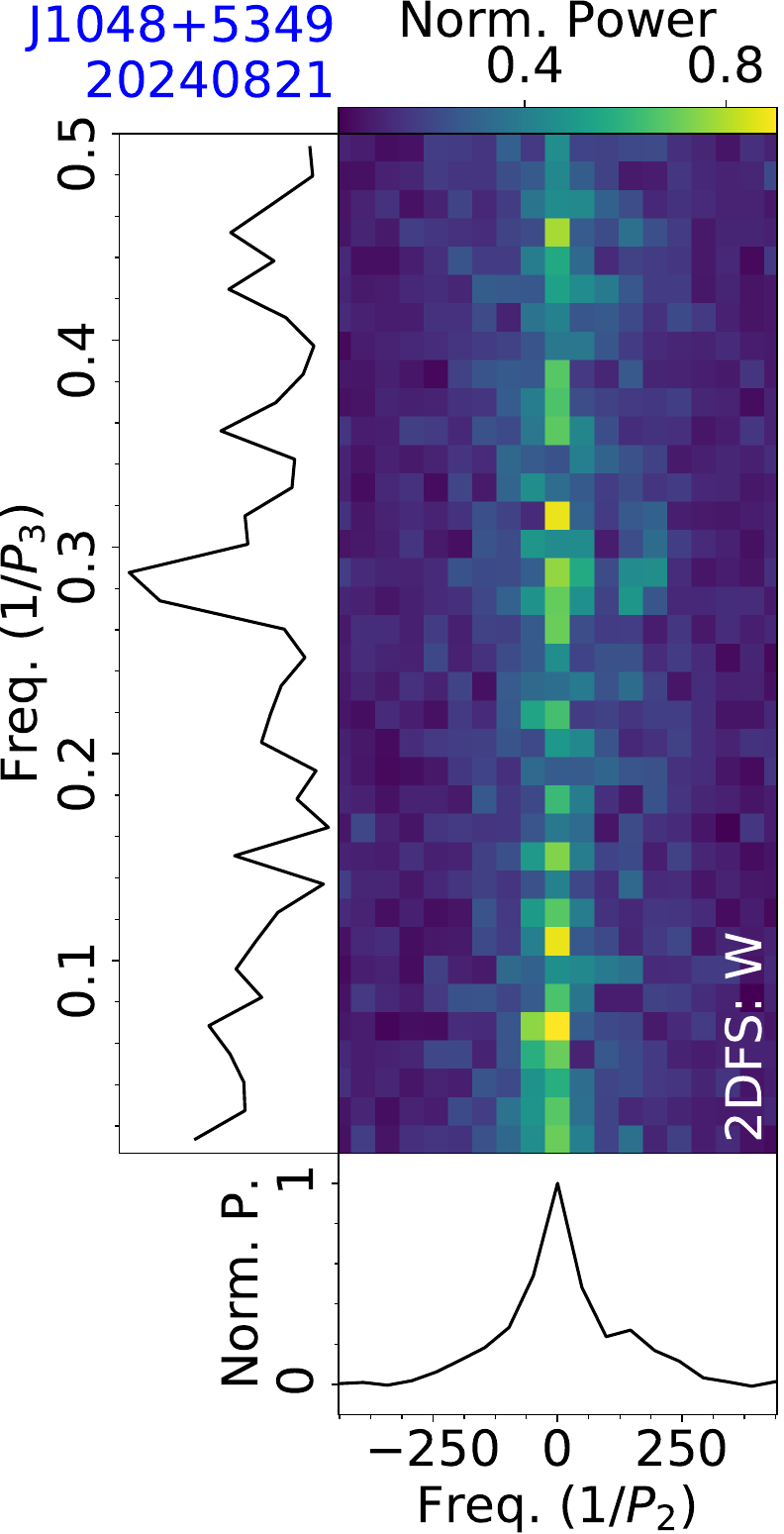}
\figcaption{Fluctuation analysis of PSR J1048+5349 for the observation on 20240821, with LRFS and 2DFS for the on-pulse phase range of a mean pulse profile.
\label{subfig:fluctu:J1048+5349}}
\end{figure}

\begin{figure}[htpb]
\centering
\includegraphics[width=0.22\textwidth, angle=0]{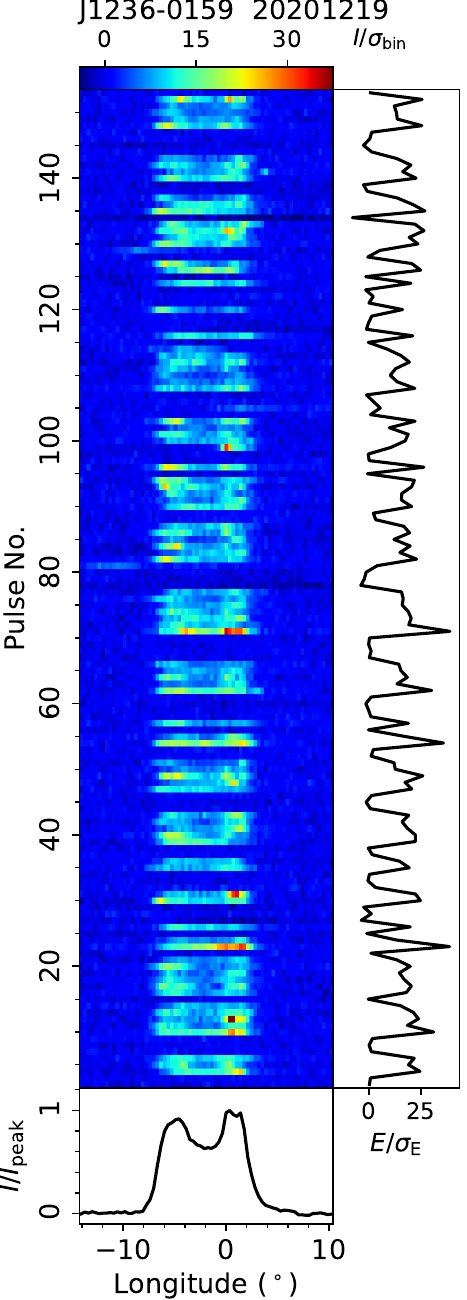}
\figcaption{Single pulse sequence of PSR J1236-0159 from the observation on 20201219. \label{subfig:TP:J1236-0159}}
\end{figure}

\begin{figure}[htpb]
\centering
\includegraphics[width=0.39\textwidth, angle=0]{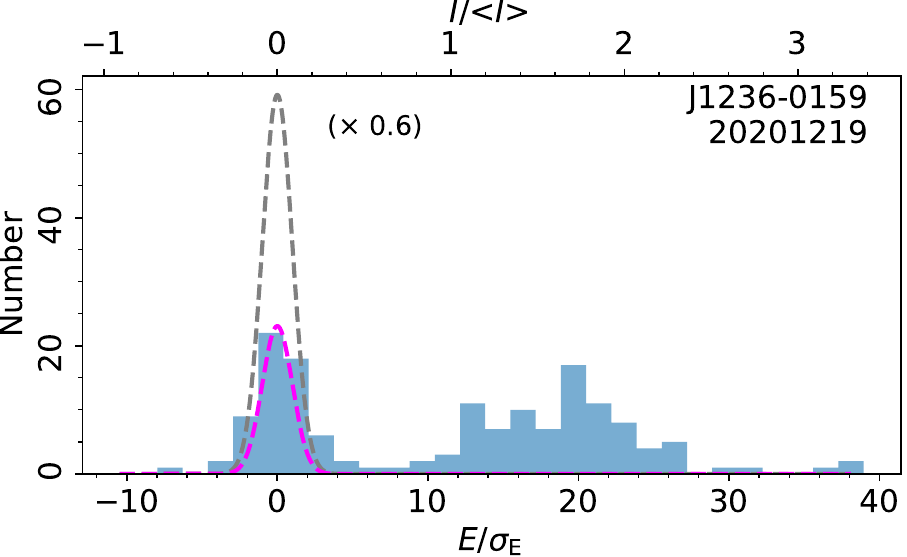}
\figcaption{On-pulse integral energy histogram of PSR J1236-0159 from the observation on 20210111. \label{subfig:Hist:J1236-0159}}
\end{figure}

\subsection{J0944+4106}
\label{subsec:J0944+4106}

PSR J0944+4106 was discovered by \citet{Stovall2014} using the Green Bank Telescope. 

The pulsar was observed by FAST on 20201124 for 9 minutes, deriving a rotation period $P=2.2296$~s and a dispersion measure $D\!M=20.6~{\rm cm^{-3}\,pc}$. 
The single pulse sequence of the observation is displayed in Fig.~\ref{subfig:TP:J0944+4106}. The histogram of single-pulse on-pulse integrated energy in Fig.~\ref{subfig:TP:J0944+4106} indicates the existence of nulls, and the nulling fraction of this observation is estimated to be 8$\pm$1\%. Fluctuation spectra are displayed in Fig.~\ref{subfig:fluctu:J0944+4106}, illustrate the weak drift features for leading and trailing parts in a mean pulse profile. 2DFS exhibits the centroid modulation frequencies of $1/P_3=0.30\pm0.01$ and $1/P_2=40\pm10$ for the leading part of the profile, corresponding to drifting parameters of $P_3=3.4\pm0.1$ periods and $P_2=9\pm2^\circ$. 
For the trailing part of the profile, the drift feature in 2DFS has the centroid frequencies of $1/P_3=0.313\pm0.003$ and $1/P_2=29\pm7$, yielding $P_3=3.20\pm0.03$ periods and $P_2=12\pm3^\circ$.

\begin{figure}[hbpt]
\centering
\includegraphics[width=0.22\textwidth, angle=0]{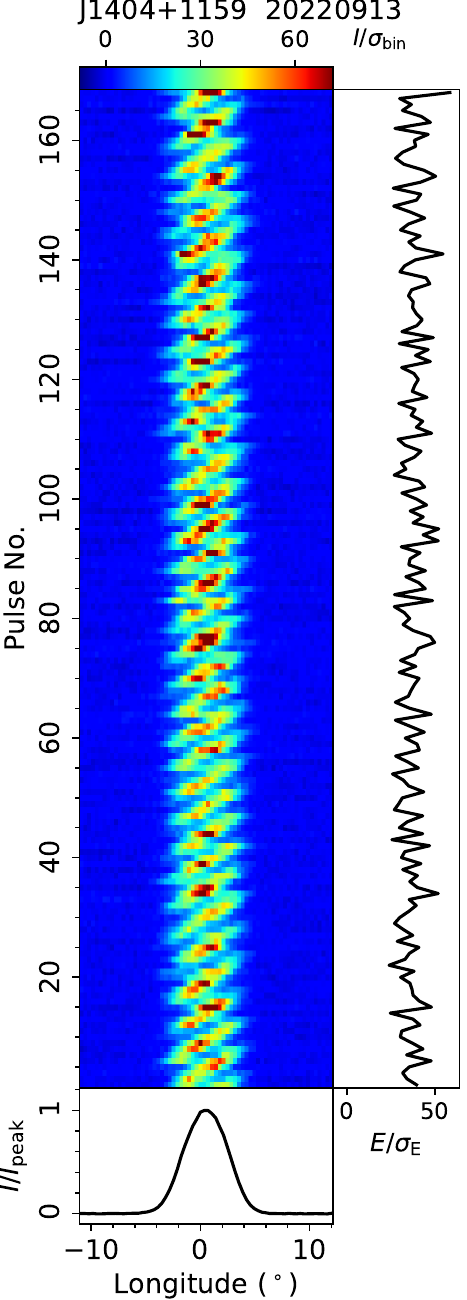}
\includegraphics[width=0.22\textwidth, angle=0]{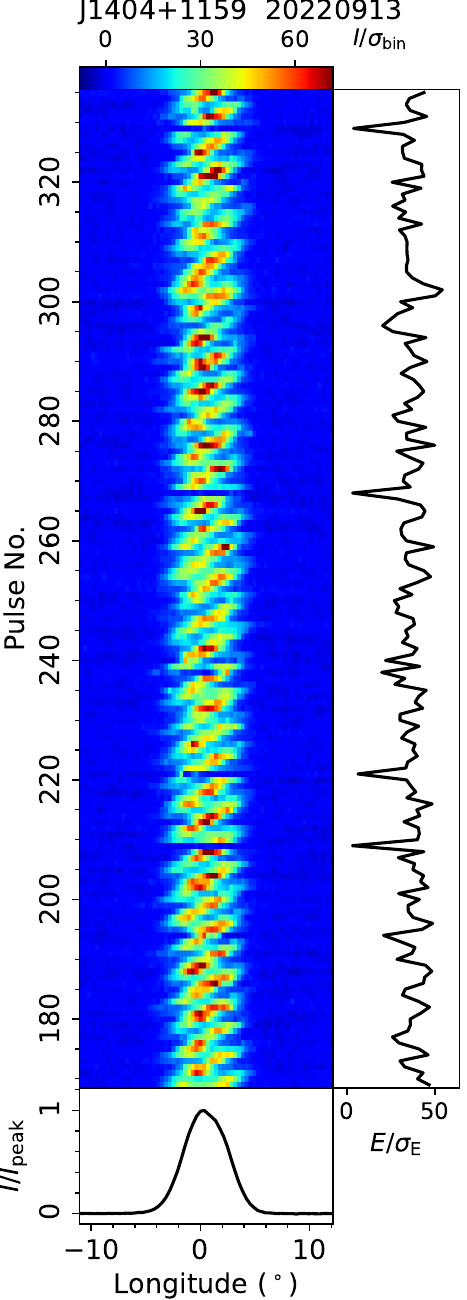}
\figcaption{Single pulse sequences of PSR J1404+1159 from the FAST observation on 20220913.
\label{subfig:TP:J1404+1159}}
\end{figure}

\begin{figure}[hbpt]
\centering
\includegraphics[width=0.22\textwidth, angle=0]{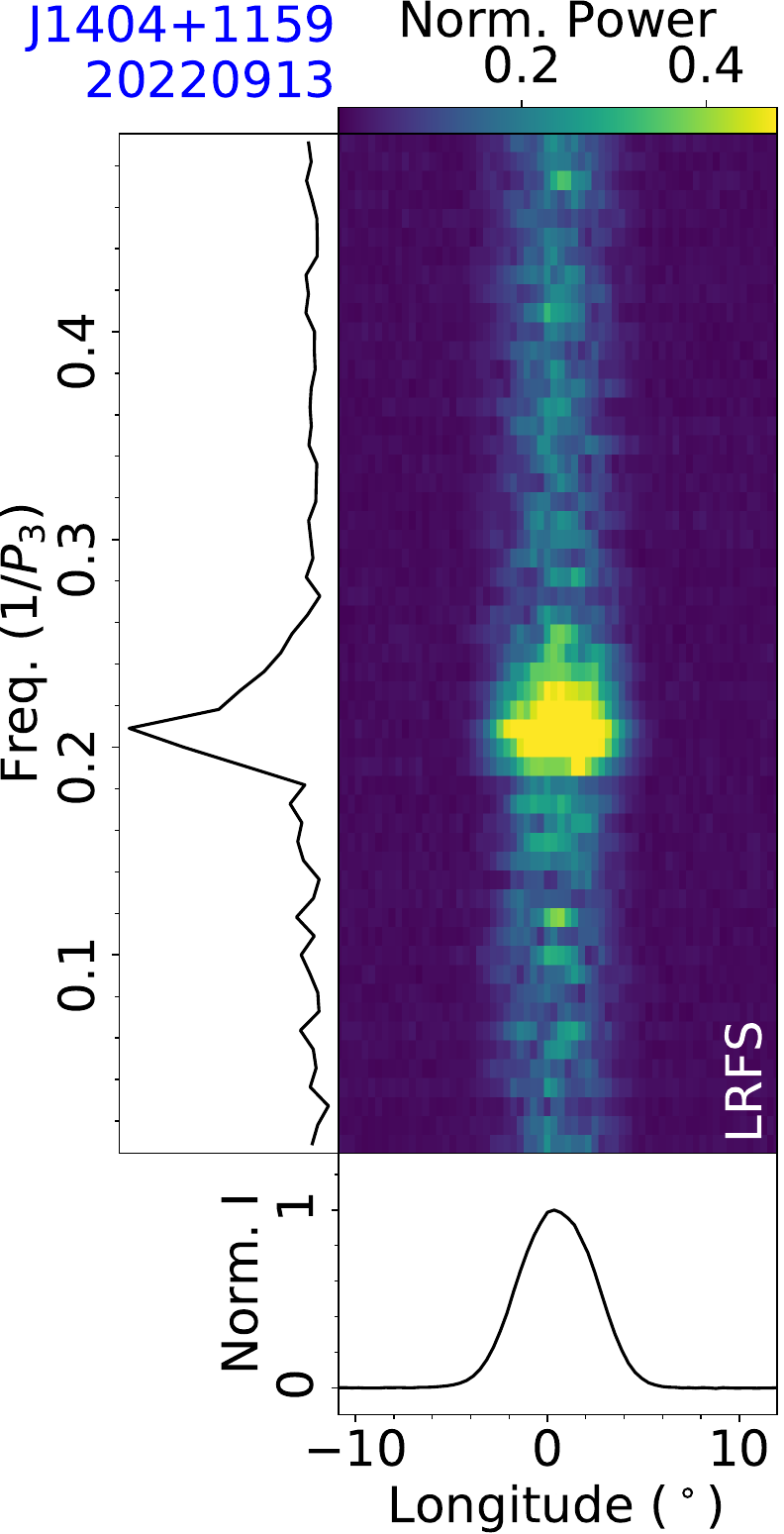}
\includegraphics[width=0.22\textwidth, angle=0]{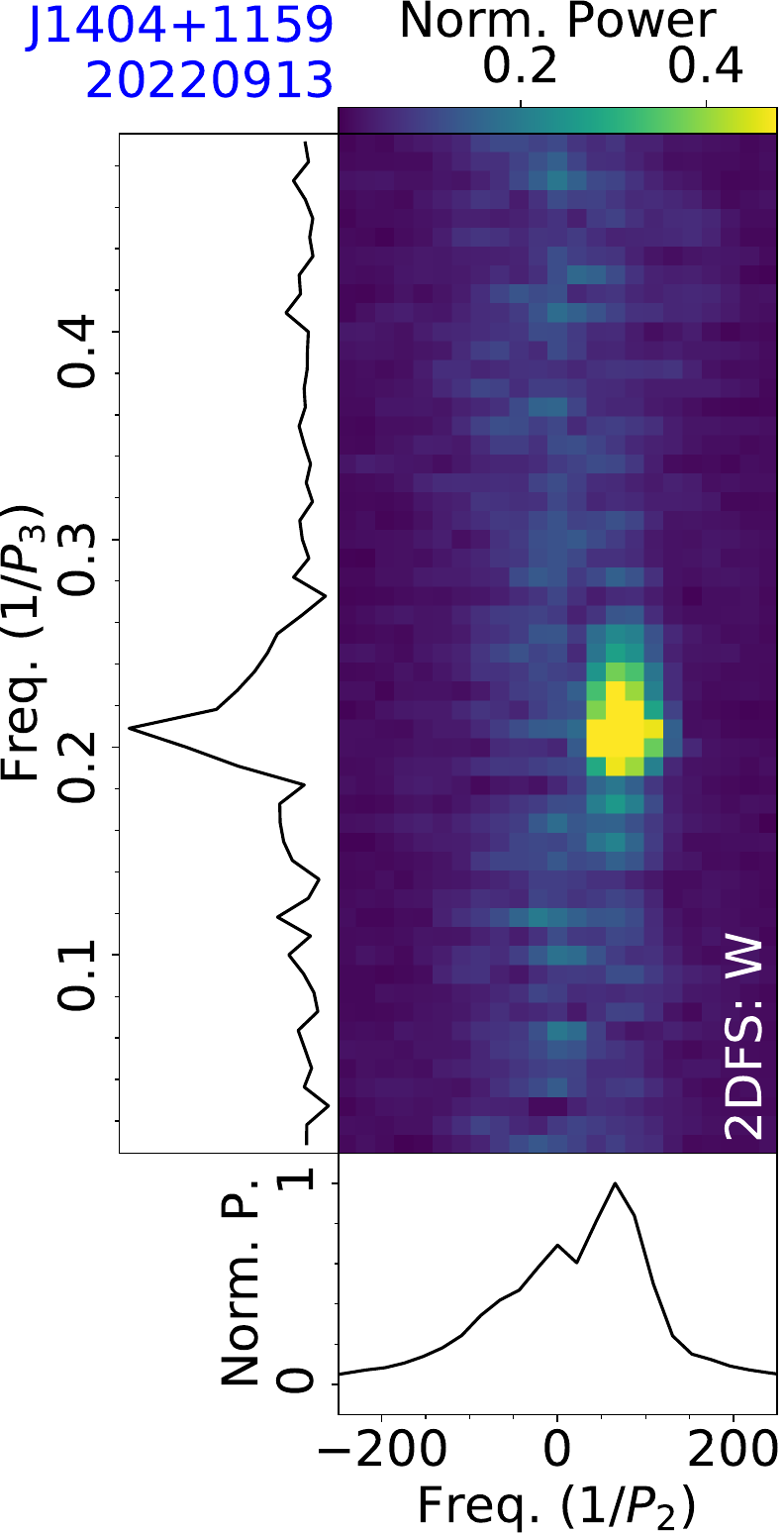}
\figcaption{Fluctuation analysis of PSR J1404+1159 from the FAST observation on 20220913, with LRFS and 2DFS for the on-pulse region of a mean pulse profile. 
\label{subfig:fluctu:J1404+1159}}
\end{figure}

\begin{figure}[htpb]
\centering
\includegraphics[width=0.39\textwidth, angle=0]{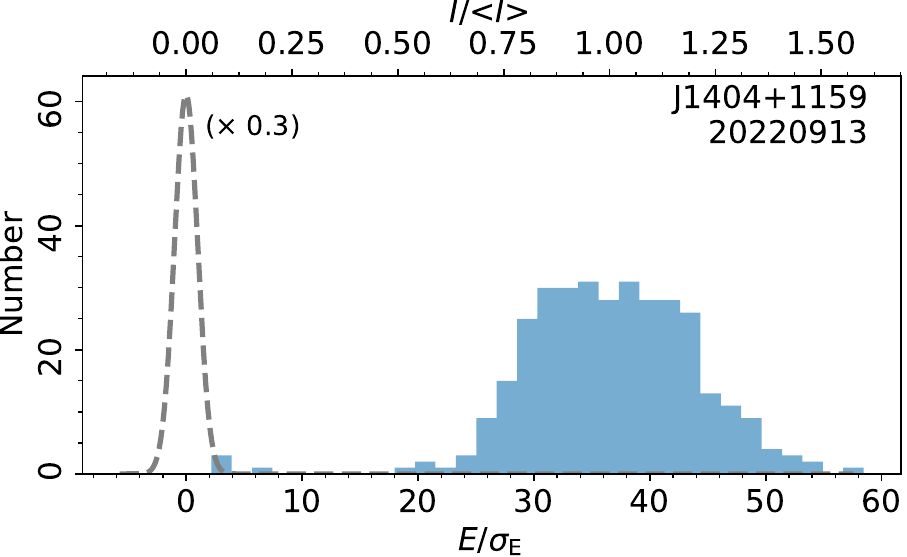}
\figcaption{On-pulse integral energy histogram of individual pulses of PSR J1404+1159 from the observation on 20220913.
\label{subfig:Hist:J1404+1159}}
\end{figure}

\subsection{J1048+5349}
\label{subsec:J1048+5349}

PSR J1048+5349 was discovered by the CHIME telescope  \citep{Dong2023}.

This pulsar was observed by FAST on 20240821 for 10 minutes, deriving a rotation period $P=2.7308$~s and a dispersion measure $D\!M=29.2~{\rm cm^{-3}\,pc}$. 
Single pulse sequences in Fig.~\ref{subfig:TP:J1048+5349} and fluctuation spectra in Fig.~\ref{subfig:fluctu:J1048+5349} illustrate the existence of subpulse drifting phenomenon. In 2DFS, the centroid frequencies of the positive drift feature are estimated to be $1/P_3=0.290\pm0.003$ and $1/P_2=173\pm7$, corresponding to periodicities of $P_3=3.45\pm0.03$ periods and $P_2=2.1\pm0.1$ degrees.

\subsection{J1236-0159}
\label{subsec:J1236-0159}

PSR J1236-0159 was discovered in the LOFAR Tied-Array All-Sky Survey (LOTAAS) \citep{Sanidas2019}. 

The pulsar is observed by FAST on 20201219 for 9 minutes, with a rotation period $P=3.5980$~s and a dispersion measure $D\!M=18.3~{\rm cm^{-3}\,pc}$ from this observation. The single pulse sequence in Fig.~\ref{subfig:TP:J1236-0159} indicates the existence of nulls. The nulling fraction of this observation is estimated to be 23$\pm$3\% from the on-pulse integral energy histogram in Fig.~\ref{subfig:Hist:J1236-0159}.

\begin{figure}[hbpt]
\centering
\includegraphics[width=0.22\textwidth, angle=0]{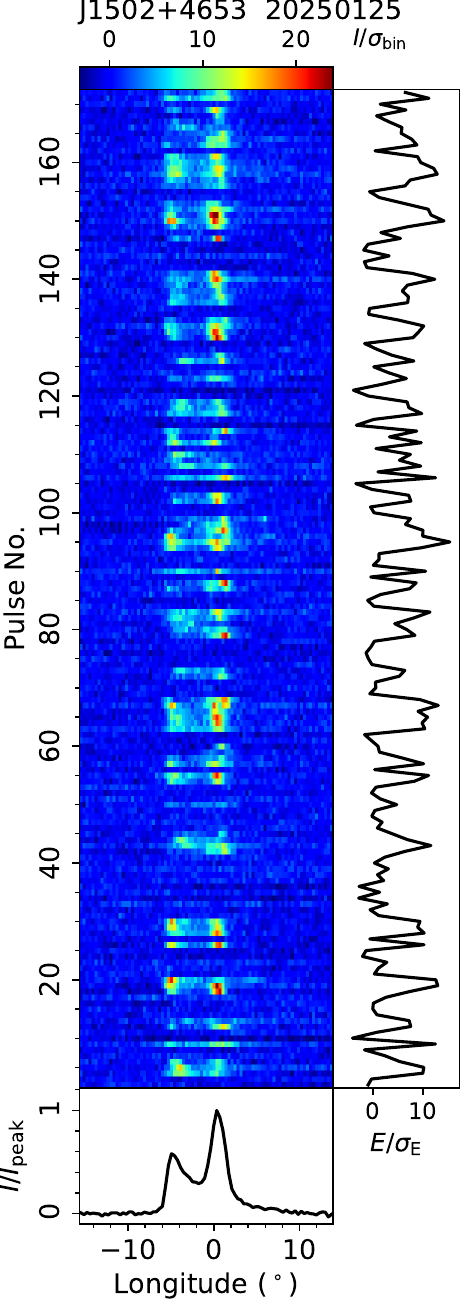}
\includegraphics[width=0.22\textwidth, angle=0]{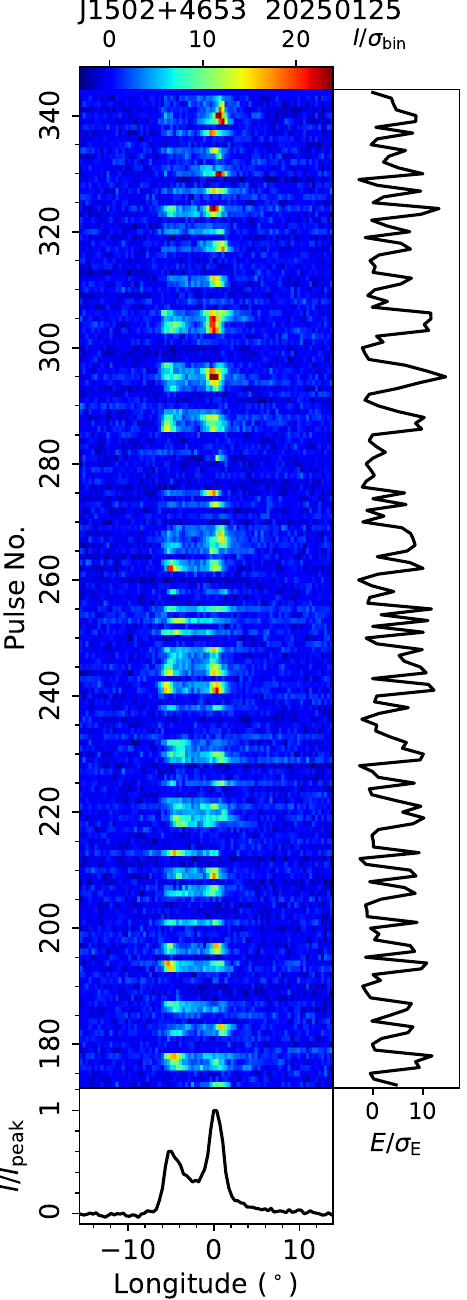}
\figcaption{Single pulse sequences of PSR J1502+4653 from the FAST observation on 20250125. \label{subfig:TP:J1502+4653}}
\end{figure}

\begin{figure}[htpb]
\centering
\includegraphics[width=0.39\textwidth, angle=0]{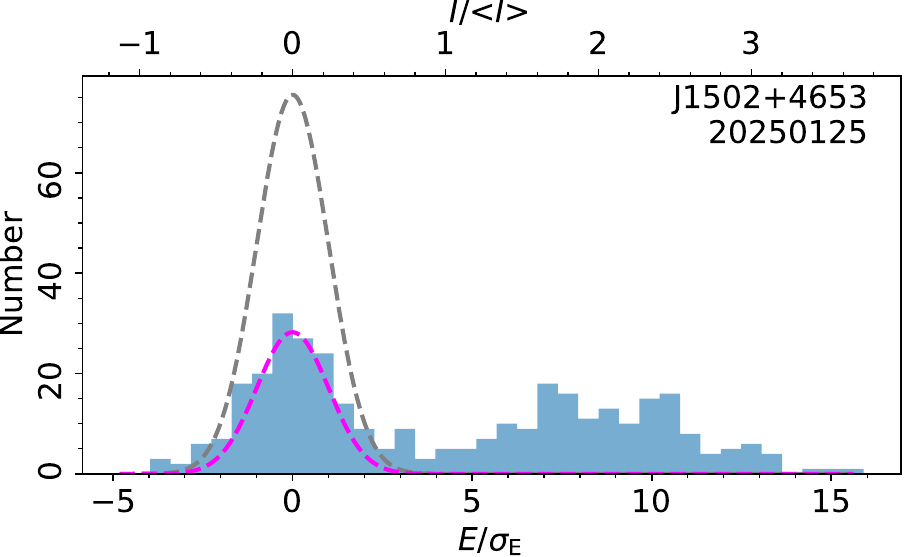}
\vspace{-0.2cm}
\figcaption{On-pulse integral energy histogram of individual pulses of PSR J1502+4653 from the observation on 20250125.
\label{subfig:Hist:J1502+4653}}
\end{figure}

\begin{figure}[hbpt]
\centering
\includegraphics[width=0.22\textwidth, angle=0]{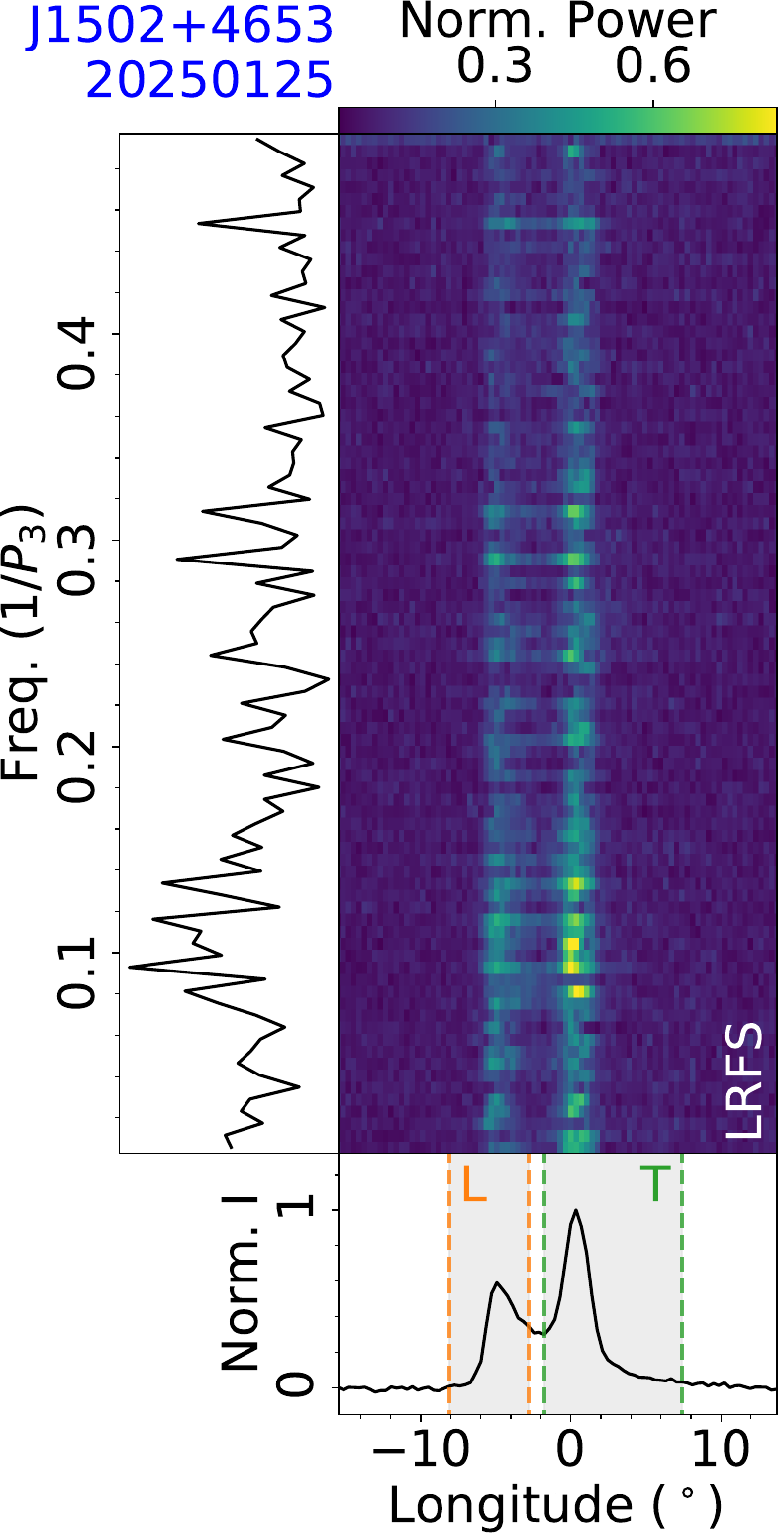}
\includegraphics[width=0.22\textwidth, angle=0]{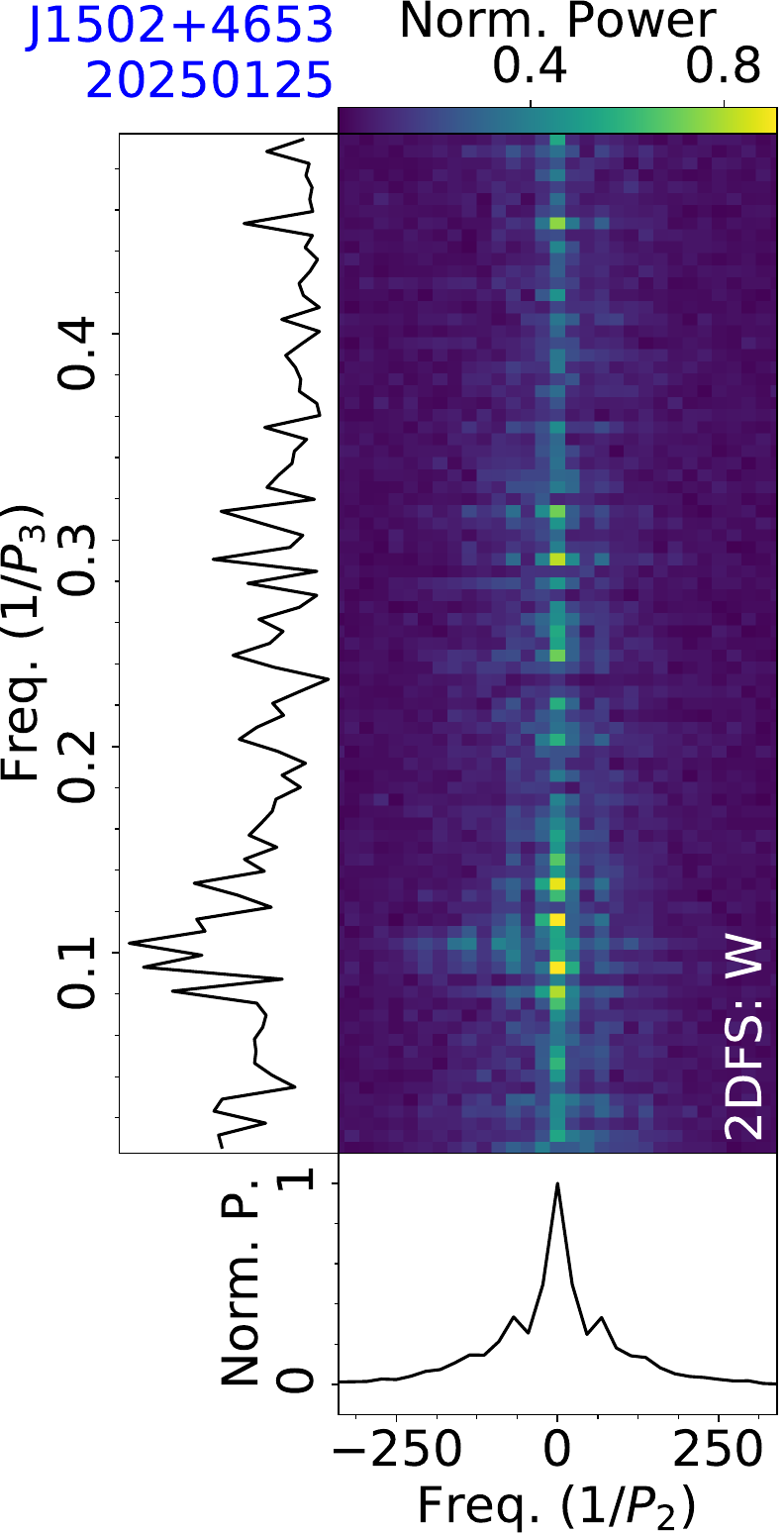}\\
\includegraphics[width=0.22\textwidth, angle=0]{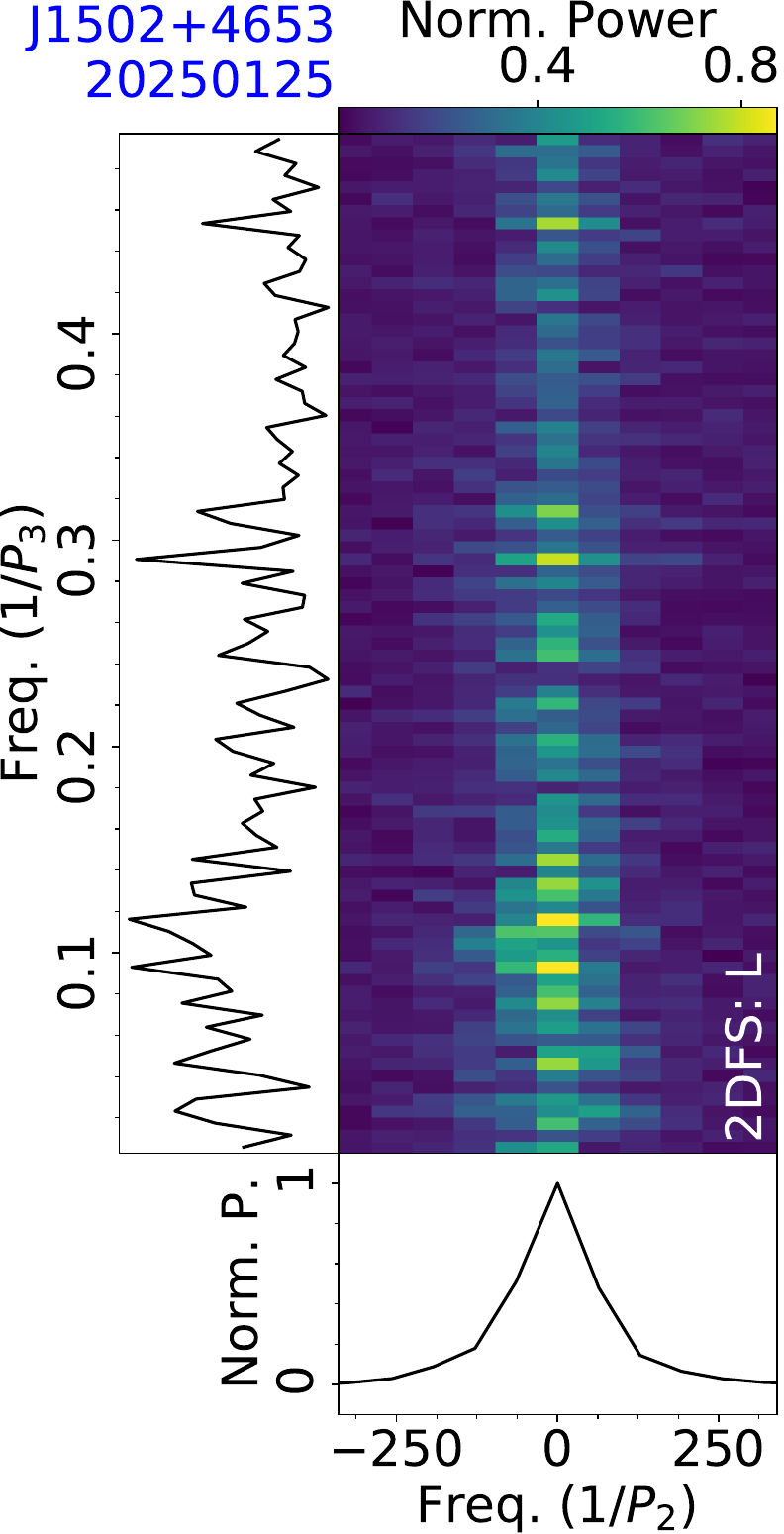}
\includegraphics[width=0.22\textwidth, angle=0]{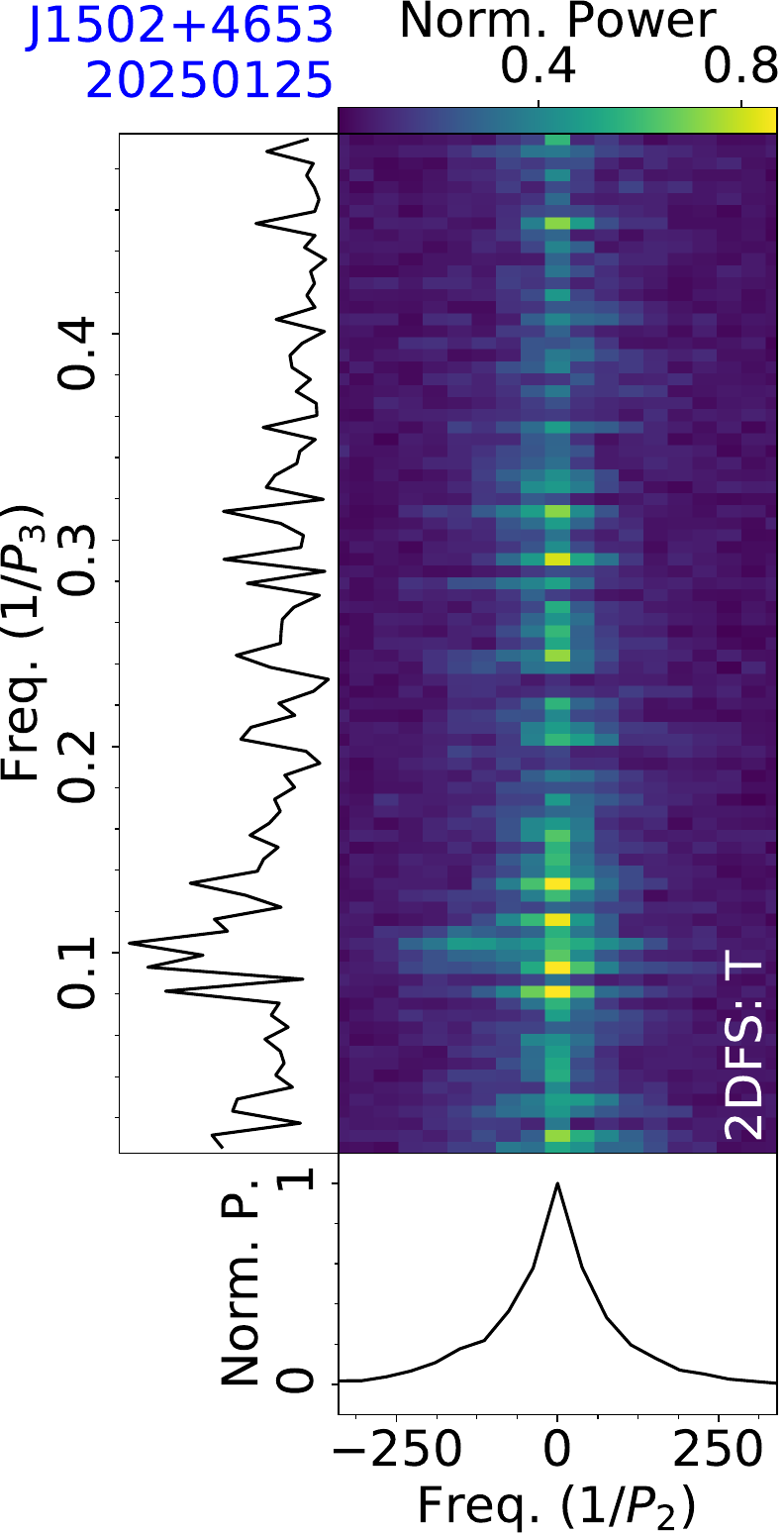}
\figcaption{Fluctuation analysis of PSR J1502+4653 from the FAST observation on 20250125, with LRFS (top-left), and 2DFS for the on-pulse region (top-right), leading part (bottom-left) and trailing part (bottom-right) of a mean pulse profile. 
\label{subfig:fluctu:J1502+4653}}
\end{figure}

\subsection{J1404+1159}
\label{subsec:J1404+1159}

PSR J1404+1159 was discovered by the Arecibo telescope \citep{Chandler2003}. Subpulse drifting behavior was reported by several studies \citep{Brinkman2018,Michilli2020,Wahl2023}, and $P_2=5.3^{+0.4}_{-0.6}$ degrees and $P_3=4.9\pm0.2$ periods were recently reported by \citet{Song2023}.

This pulsar was observed by FAST on 20220913 for 15 minutes, deriving a rotation period $P=2.6506$~s and a dispersion measure $D\!M=18.9~{\rm cm^{-3}\,pc}$. Single pulse sequences in Fig.~\ref{subfig:TP:J1404+1159} illustrate the systematic subpulse drifting behavior. The fluctuation spectra in Fig.~\ref{subfig:fluctu:J1404+1159} exhibit a positive drift feature with the centroid of $1/P_3=0.212\pm0.001$ and $1/P_2=74\pm1$, corresponding to $P_3=4.71\pm0.01$ periods and $P_2=4.9\pm0.1$ degrees. The drifting properties are consistent with previous studies. 
From single pulse sequences of this highly sensitive observation by FAST, we also detect the decrease in energy lasting for a very short duration of $\sim$1 period. Furthermore, the single pulses with an energy decrease is not nulls, as there is no single pulse whose on-pulse integral energy is less than 3$\sigma_{\rm E}$ (Fig.~\ref{subfig:Hist:J1404+1159}). 

A further high-sensitivity observation is required for the analysis of the short-duration decrease in intensity.

\subsection{J1502+4653}
\label{subsec:J1502+4653}

PSR J1502+4653 was discovered by FAST \citep{Cruces2021}. 

This pulsar was also observed by FAST on 20250125 for 10 minutes, deriving a rotation period $P=1.7524$~s and a dispersion measure $D\!M=26.5~{\rm cm^{-3}\,pc}$. Single pulse sequences of this observation in Fig.~\ref{subfig:TP:J1502+4653} illustrate the existence of nulling and subpulse drifting behaviors. A nulling fraction is estimated from the on-pulse integral energy histogram (Fig.~\ref{subfig:Hist:J1502+4653}), that is 37$\pm$4\% for this observation. Drifting parameters of the leading and trailing components are obtained from the drift features in fluctuation spectra (Fig.~\ref{subfig:fluctu:J1502+4653}). 
For the leading part in a mean pulse profile, the 2DFS has a negative drift feature with the centroid of $1/P_3=0.111\pm0.001$ and $1/P_2=-43\pm6$, corresponding to $P_3=9.0\pm0.1$ periods and $P_2=-8\pm1^\circ$. 
The negative drift feature in 2DFS of the trailing profile part is characterized by $1/P_3=0.102\pm0.001$ and $1/P_2=-163\pm4$, yielding $P_3=9.8\pm0.1$ periods and $P_2=-2.2\pm0.1^\circ$.

\begin{figure}[hbpt]
\centering
\includegraphics[width=0.22\textwidth, angle=0]{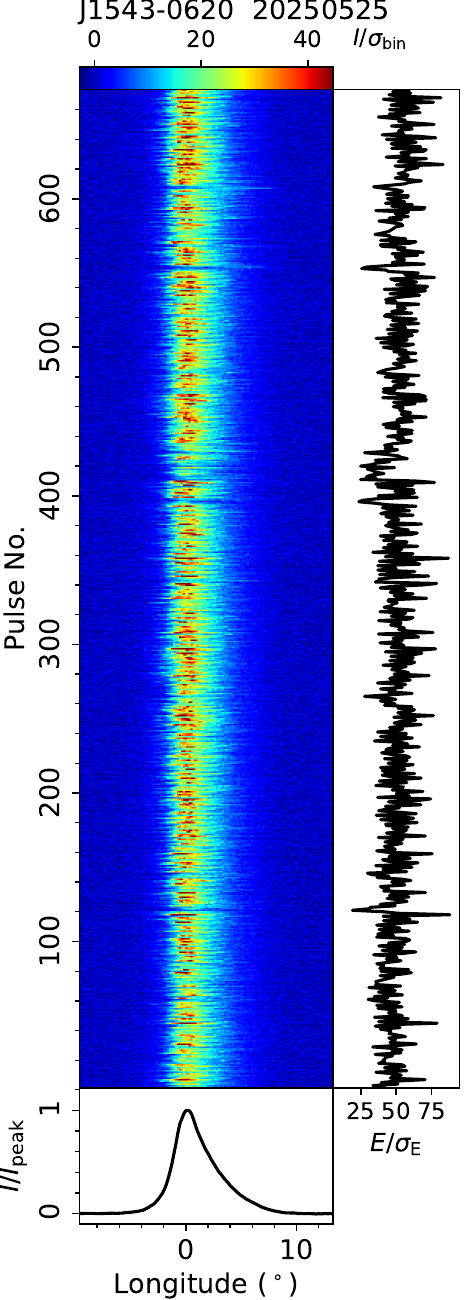}
\includegraphics[width=0.22\textwidth, angle=0]{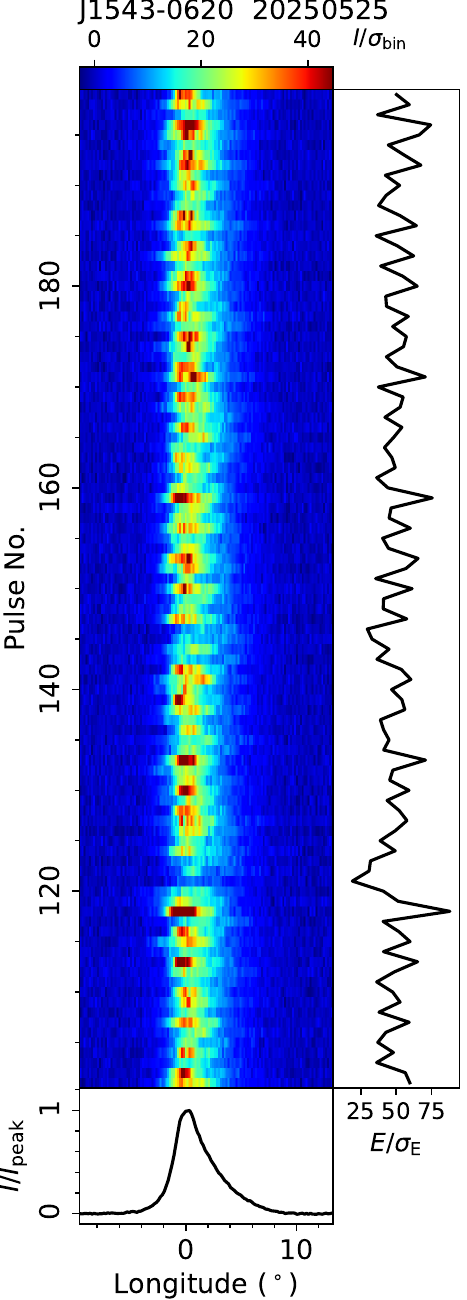}
\figcaption{Single pulse sequence of PSR J1547-0944 from the FAST observation on 20250811, and a zoomed-in view of pulses No. 100-200.
\label{subfig:TP:J1543-0620}}
\end{figure}

\begin{figure}[hbpt]
\centering
\includegraphics[width=0.22\textwidth, angle=0]{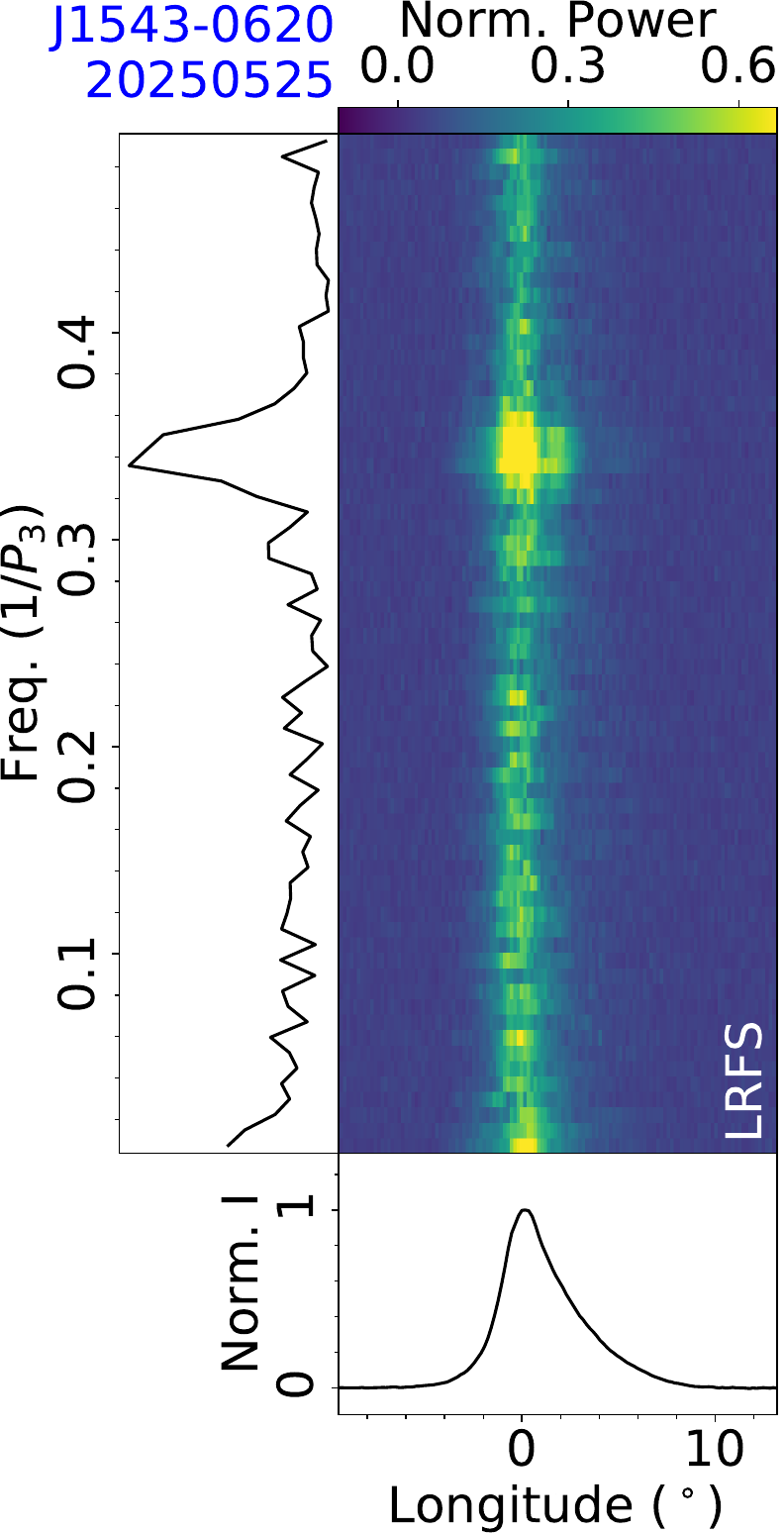}
\includegraphics[width=0.22\textwidth, angle=0]{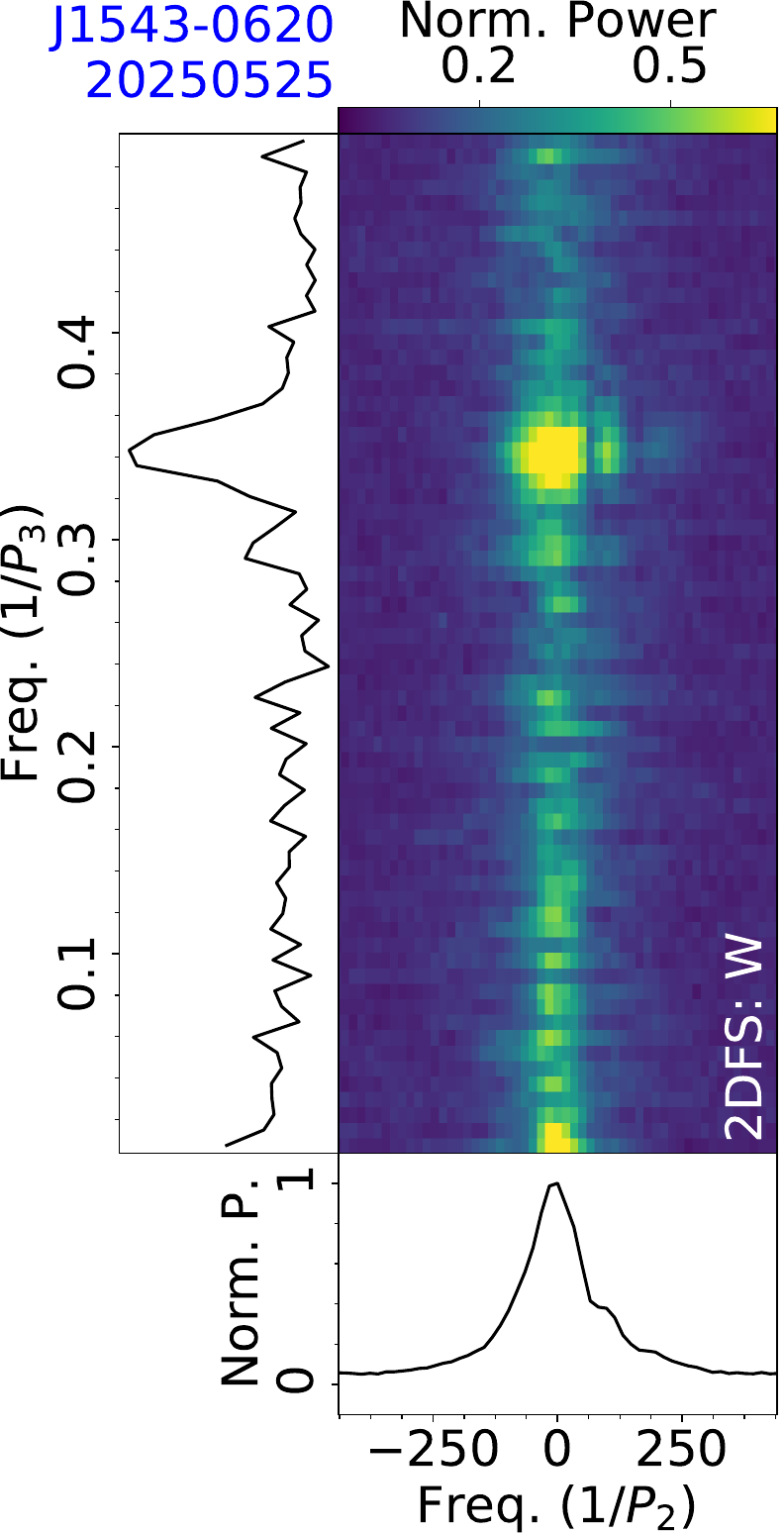}
\figcaption{Fluctuation analysis of PSR J1543-0620 from the FAST observation on 20250525, with LRFS and 2DFS for the on-pulse region of a mean pulse profile.
\label{subfig:fluctu:J1543-0620}}
\end{figure}

\begin{figure}[hbpt]
\centering
\includegraphics[width=0.22\textwidth, angle=0]{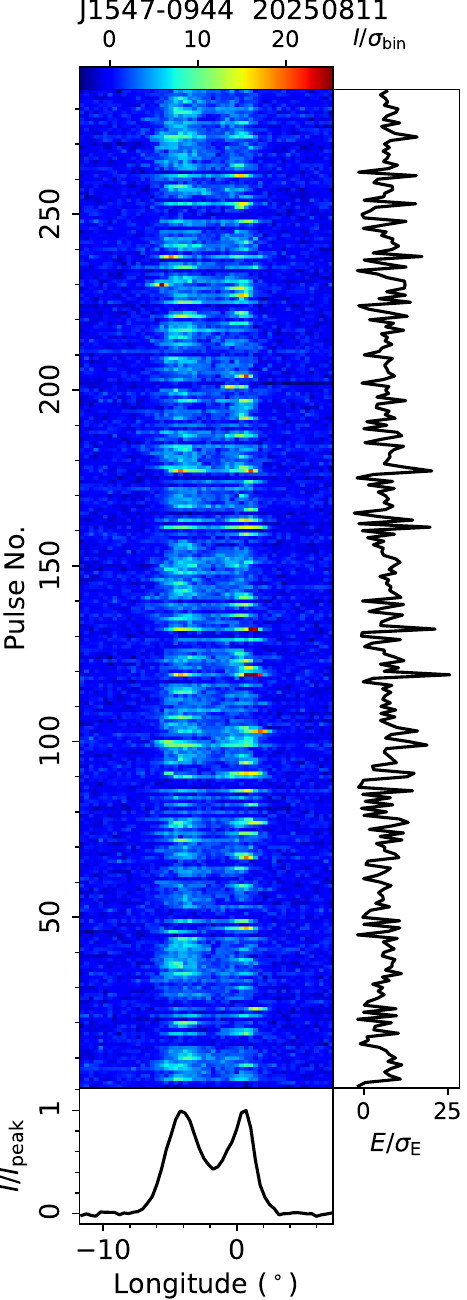}
\includegraphics[width=0.22\textwidth, angle=0]{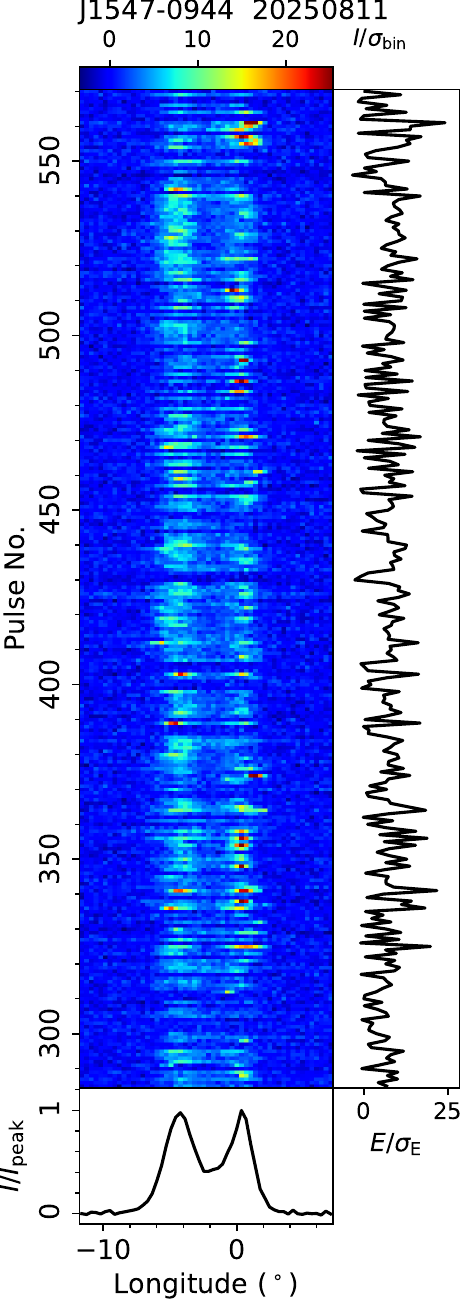}
\figcaption{Single pulse sequences of PSR J1547-0944 from the FAST observation on 20250811.
\label{subfig:TP:J1547-0944}}
\end{figure}

\begin{figure}[htpb]
\centering
\includegraphics[width=0.39\textwidth, angle=0]{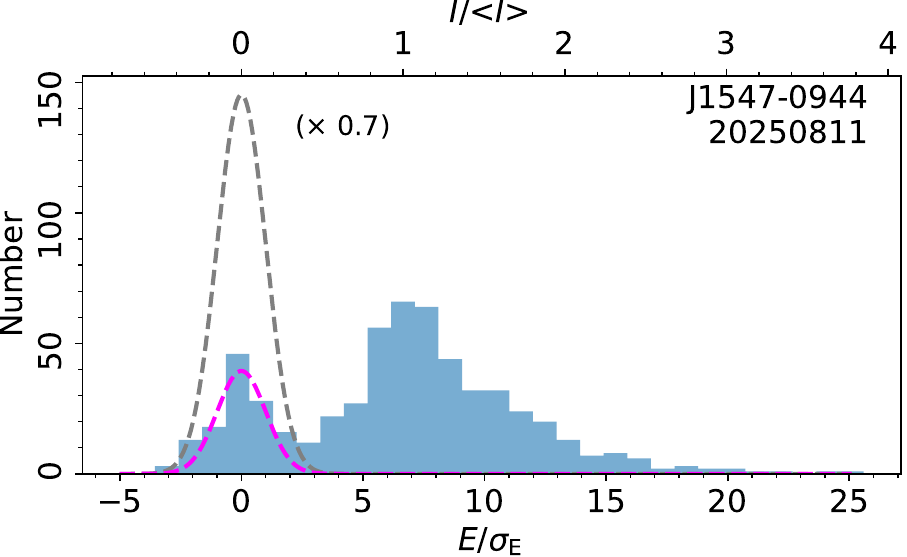}
\figcaption{On-pulse integral energy histogram of individual pulses of PSR J1547-0944 from the observation on 20250811.
\label{subfig:Hist:J1547-0944}}
\end{figure}

\begin{figure}[htpb]
\centering
\includegraphics[width=0.22\textwidth, angle=0]{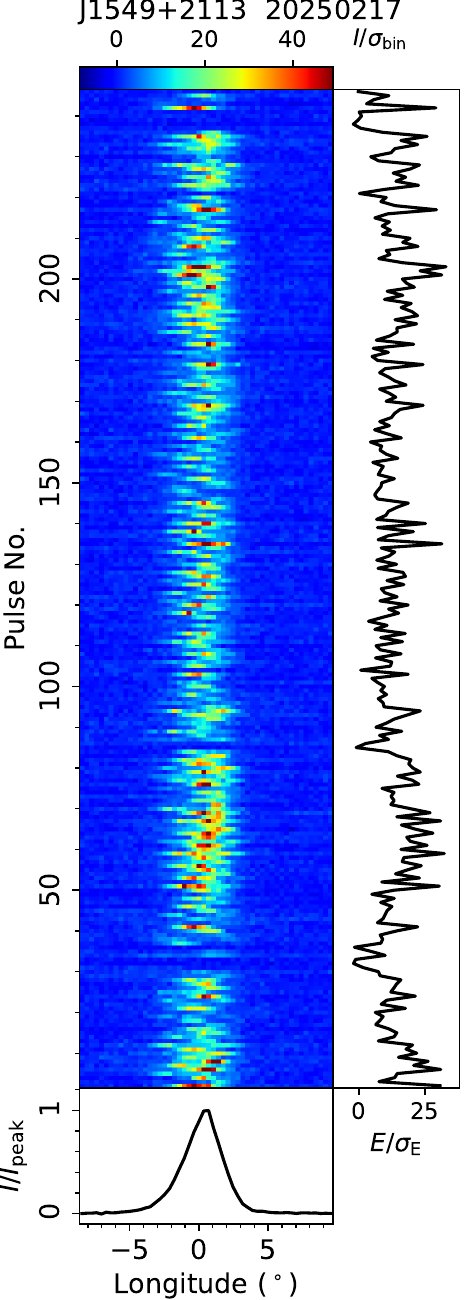}
\includegraphics[width=0.22\textwidth, angle=0]{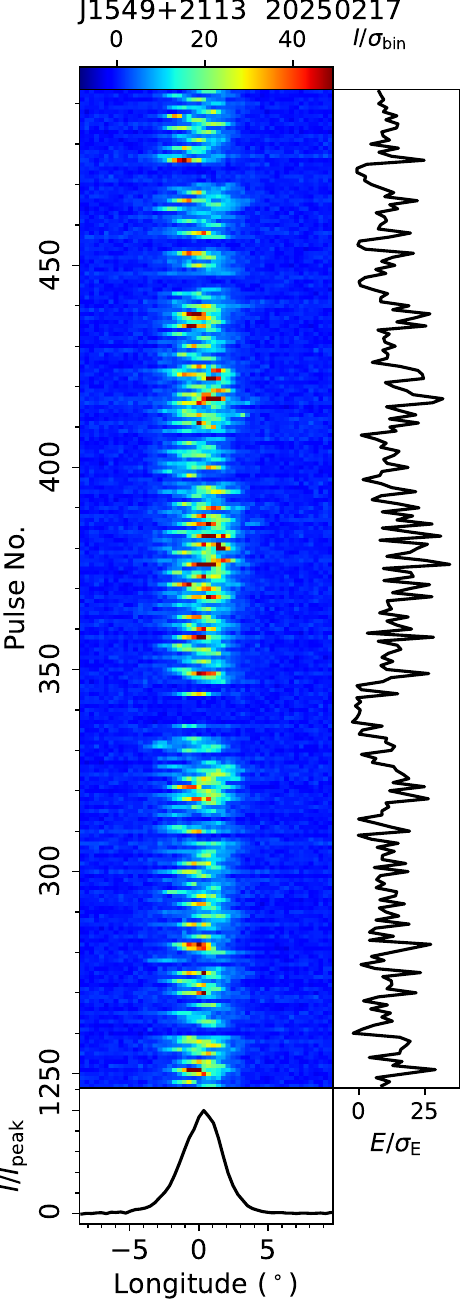}
\figcaption{Single pulse sequences of PSR J1549+2113 from the FAST observation on 20250217.
\label{subfig:TP:J1549+2113}}
\end{figure}

\begin{figure}[htpb]
\centering
\includegraphics[width=0.39\textwidth, angle=0]{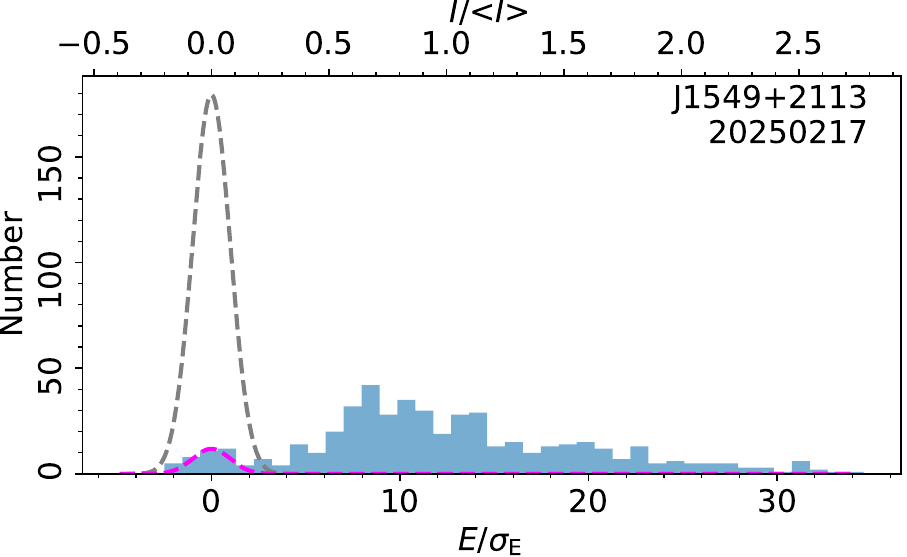}
\figcaption{On-pulse integral energy histogram of individual pulses of PSR J1549+2113 from the FAST observation on 20250217. \label{subfig:Hist:J1549+2113}}
\end{figure}

\begin{figure}[htpb]
\centering
\includegraphics[width=0.22\textwidth, angle=0]{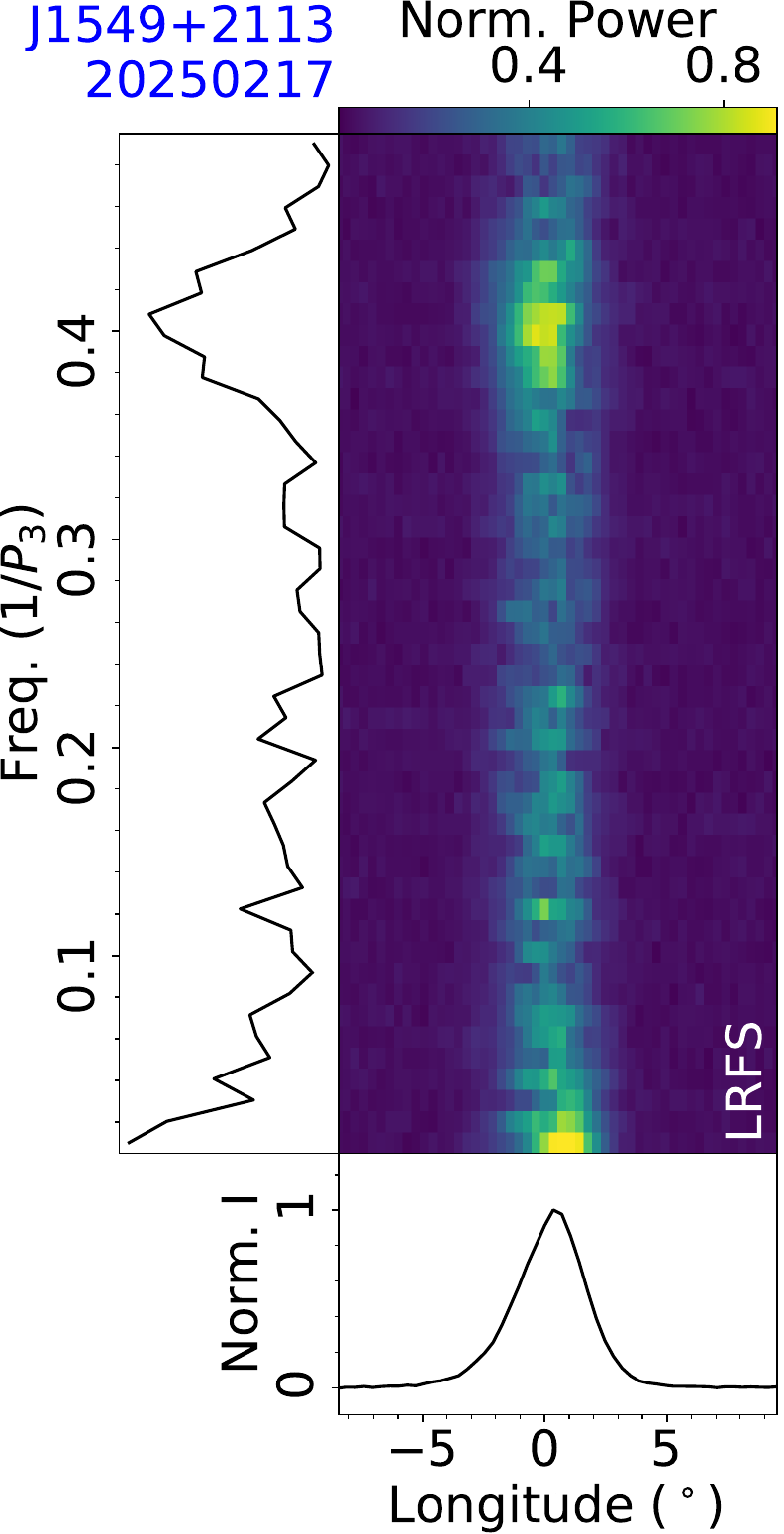}
\includegraphics[width=0.22\textwidth, angle=0]{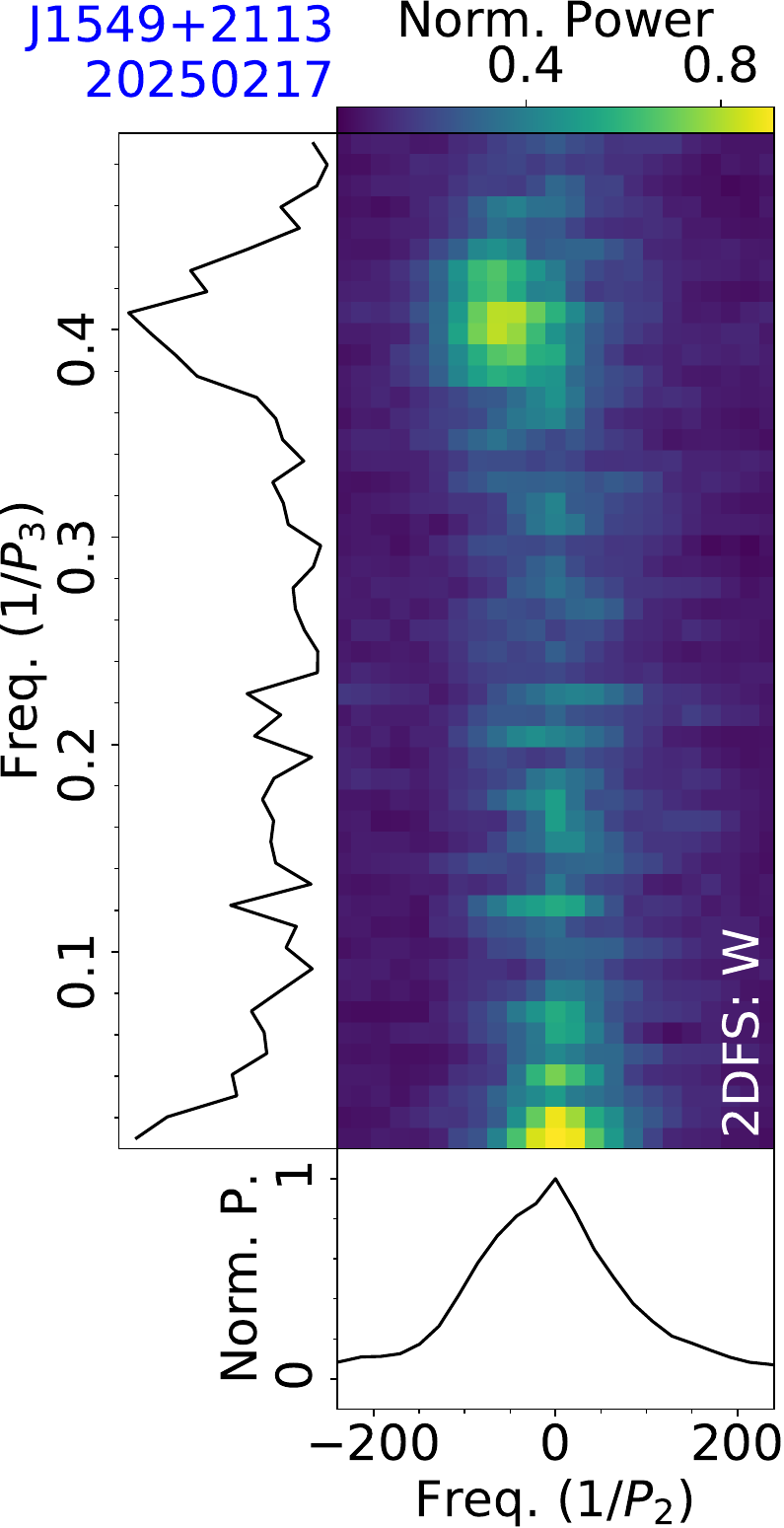}
\figcaption{Fluctuation analysis of PSR J1549+2113 from the FAST observation on 20250217, with LRFS and 2DFS for the on-pulse region of a mean pulse profile.
\label{subfig:fluctu:J1549+2113}}
\end{figure}

\subsection{J1543-0620}
\label{subsec:J1543-0620}

PSR J1543-0620 was discovered by \citet{Manchester1978} in the second Molonglo pulsar survey. Subpulse drifting behavior with $P_3\sim3$ periods of this pulsar was reported in previous studies. \citet{Weltevrede2006} observed opposite drift directions in the two pulse profile halves at 21 cm, quantifying the phase modulation periodicity of $P_2=24^{+9}_{-4}$ and $-16^{+12}_{-45}$ degrees at 21 cm. 
However, only positive drifting was reported in the work of \citet{Weltevrede2007} with $P_2=6.0^{+3}_{-0.5}$ at 92 cm, by \citet{Basu2019} at 339 MHz, and by \citet{Song2023} with $P_2=63^{+95}_{-60}$ at 1283 MHz.

This pulsar was observed by FAST on 20250525 for 8 minutes, with a rotation period $P=0.7091$~s and a dispersion measure $D\!M=18.1~{\rm cm^{-3}\,pc}$ derived. The single pulse sequence and a zoomed-in view of pulses No. 100-200 in Fig.~\ref{subfig:TP:J1543-0620} show the subpulse drifting phenomenon. LRFS and 2DFS are displayed in Fig.~\ref{subfig:fluctu:J1543-0620}, and there are three drift features in 2DFS. The negative drift feature is the most prominent, with the centroid frequencies of $1/P_3=0.341\pm0.001$ and $1/P_2=-19\pm2$, corresponding to periodicities of $P_3=2.93\pm0.01$ periods and $P_2=-19\pm2$ degrees. The feature centroid of the fast positively drifting is characterized by frequencies of $1/P_3=0.345\pm0.001$ and $1/P_2=104\pm1$, yielding $P_3=2.90\pm0.01$ periods and $P_2=3.47\pm0.04$ degrees. There is also a weak feature of positively drifting, that has a centroid of $1/P_3=0.345\pm0.001$ and $1/P_2=214\pm2$, corresponding to $P_3=2.90\pm0.01$ periods and $P_2=1.69\pm0.01$ degrees. 

\begin{figure}[htpb]
\centering
\includegraphics[width=0.44\textwidth, angle=0]{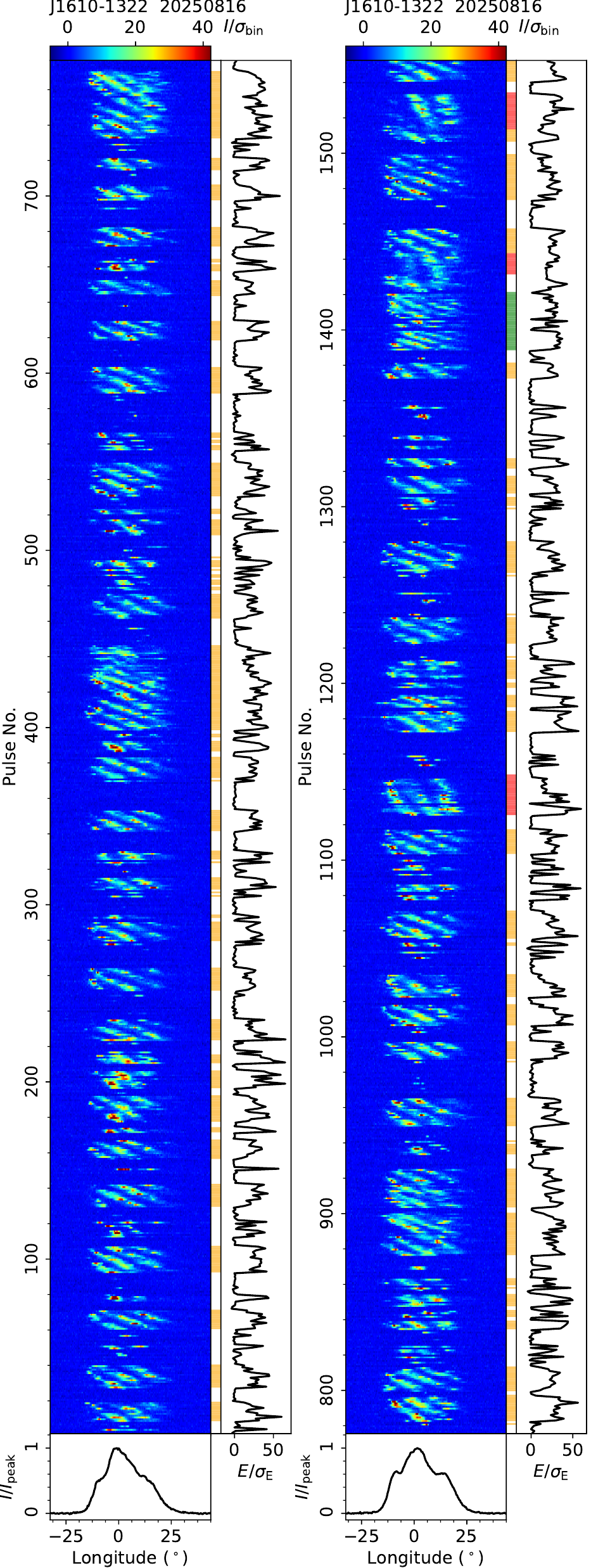}
\figcaption{Single pulse sequences of PSR J1610-1322 from the FAST observation on 20250816.
The normal, slow drifting, and fast drifting modes are labeled with orange, red, and green bars, respectively.
\label{subfig:TP:J1610-1322}}
\end{figure}

\begin{figure}[htpb]
\centering
\includegraphics[width=0.39\textwidth, angle=0]{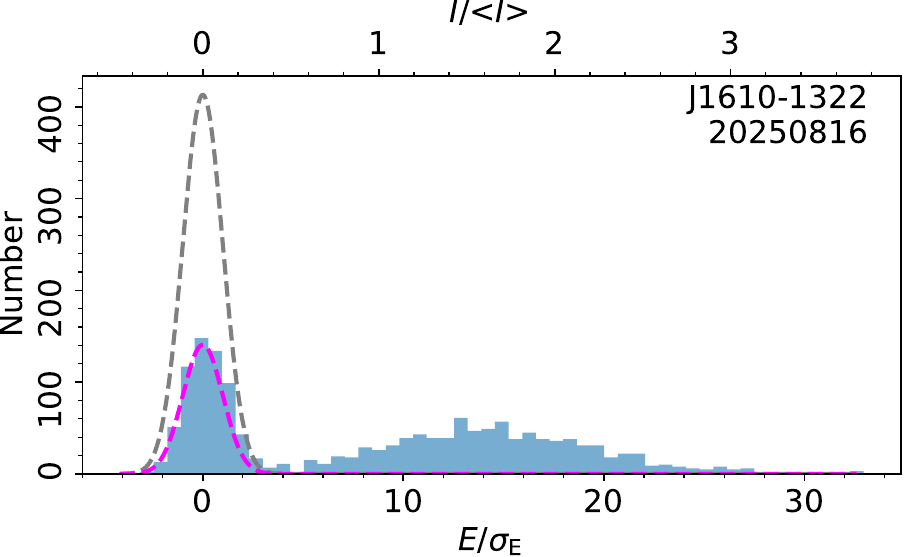}
\figcaption{On-pulse integral energy histogram of individual pulses of PSR J1610-1322 from the FAST observation on 20250816.
\label{subfig:Hist:J1610-1322}}
\end{figure}

\begin{figure}[hbpt]
\centering
\includegraphics[width=0.40\textwidth, angle=0]{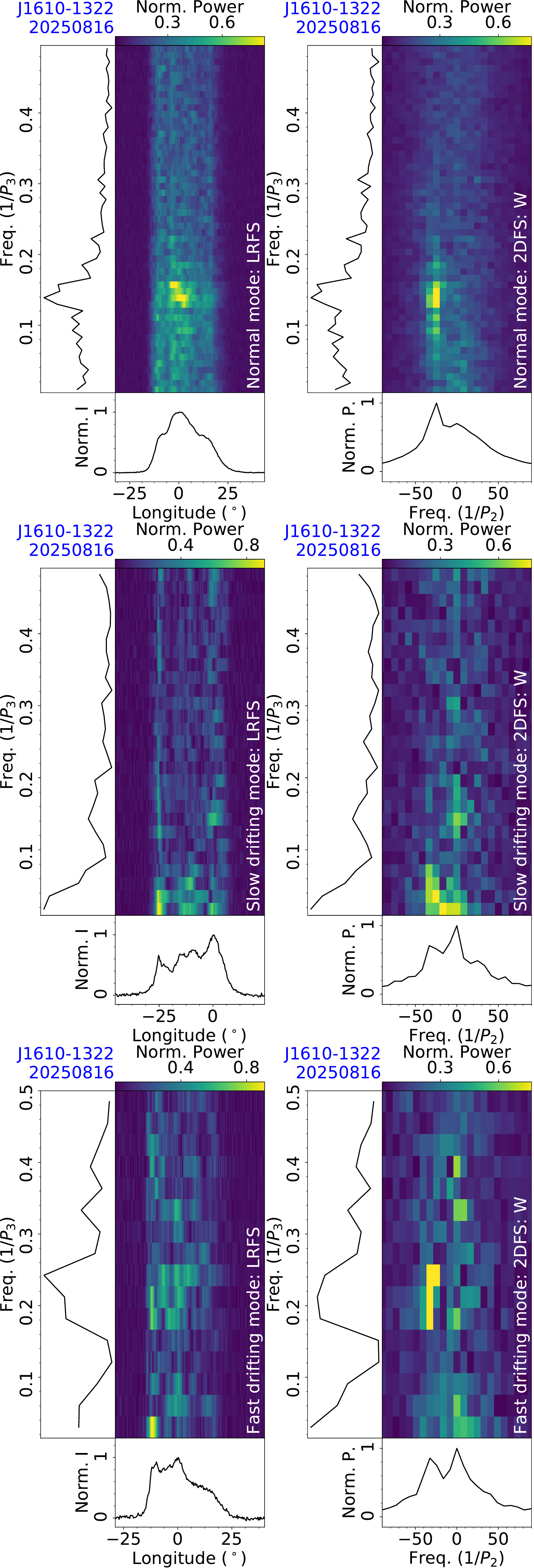}
\vspace{-0.4cm}
\figcaption{Fluctuation analysis of PSR J1610-1322 from the FAST observation on 20250816, with LRFS and 2DFS for the on-pulse region of the mean pulse profile. 
The normal, slow drifting, and fast drifting modes are shown at the top, center, and bottom, respectively.
\label{subfig:fluctu:J1610-1322}}
\end{figure}

\subsection{J1547-0944}
\label{subsec:J1547-0944}

PSR J1547-0944 was discovered by the Green Bank Telescope \citet{Lynch2013}.
This pulsar was observed by FAST on 20250811 for 15 minutes, and a rotation period $P=1.5771$~s and a dispersion measure $D\!M=36.4~{\rm cm^{-3}\,pc}$ were derived. Single pulse sequences are shown in Fig.~\ref{subfig:TP:J1547-0944}, illustrating the existence of nulls. From the on-pulse integral histogram, the nulling fraction of this observation is estimated to be 19.0$\pm$1.6\%.

\begin{figure}[htpb]
\centering
\includegraphics[width=0.22\textwidth, angle=0]{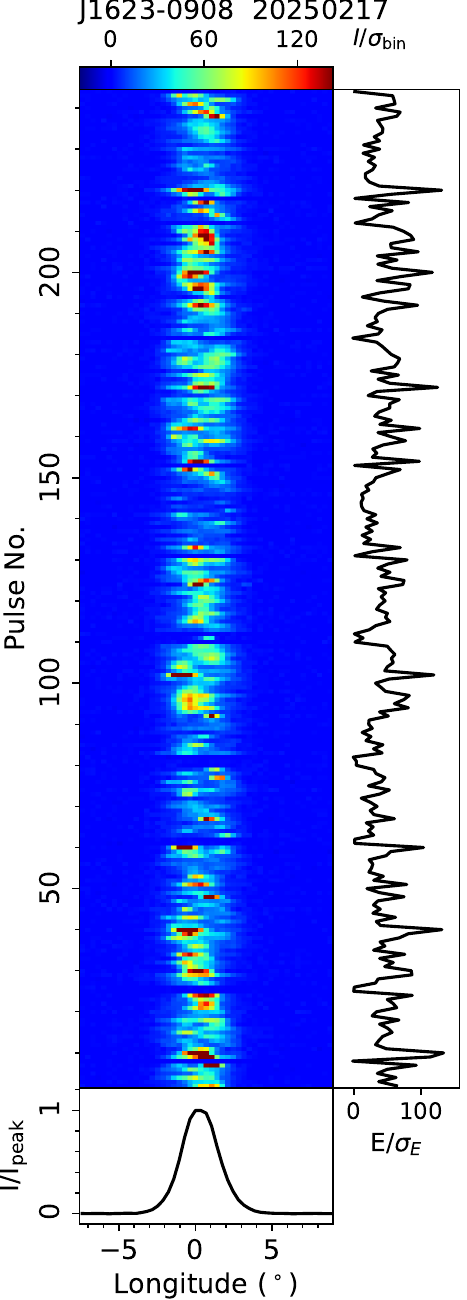}
\includegraphics[width=0.22\textwidth, angle=0]{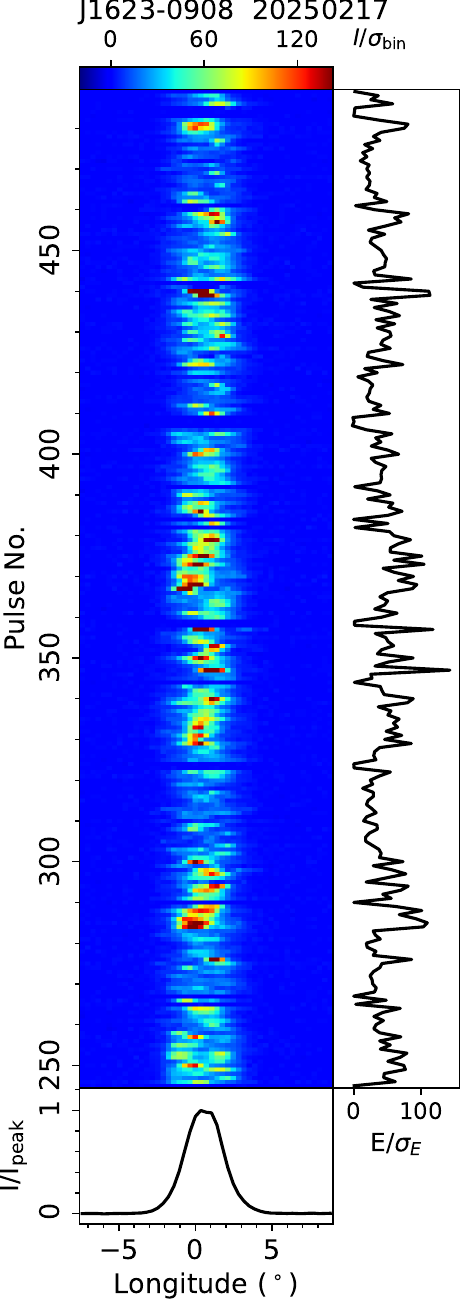}
\figcaption{Single pulse sequences of PSR J1623-0908 from the observation on 20250217.
\label{subfig:TP:J1623-0908}}
\end{figure}

\begin{figure}[htpb]
\centering
\includegraphics[width=0.39\textwidth, angle=0]{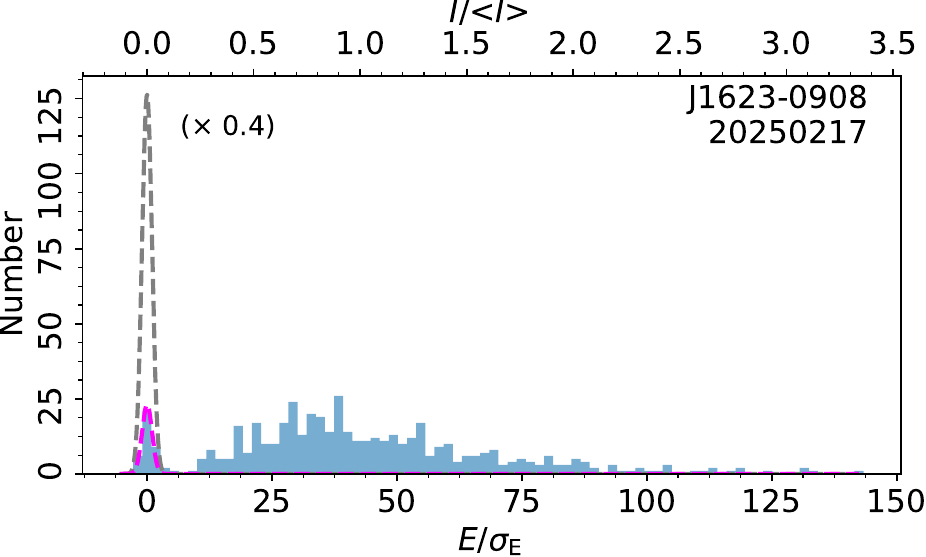}
\figcaption{On-pulse integral energy histogram of individual pulses of PSR J1623-0908 from the FAST observation on 20250217. \label{subfig:Hist:J1623-0908}}
\end{figure}

\begin{figure}[htpb]
\centering
\includegraphics[width=0.22\textwidth, angle=0]{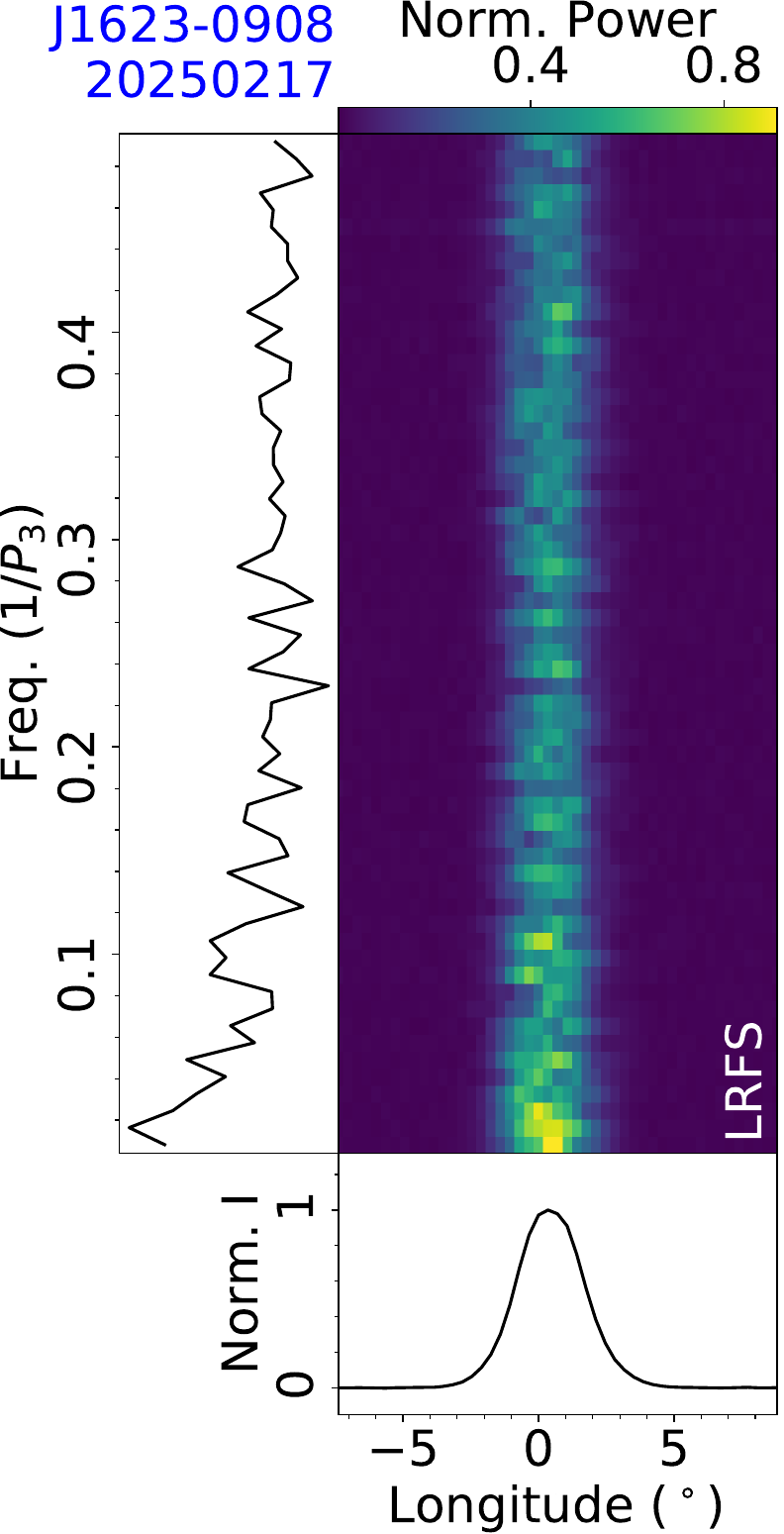}
\includegraphics[width=0.22\textwidth, angle=0]{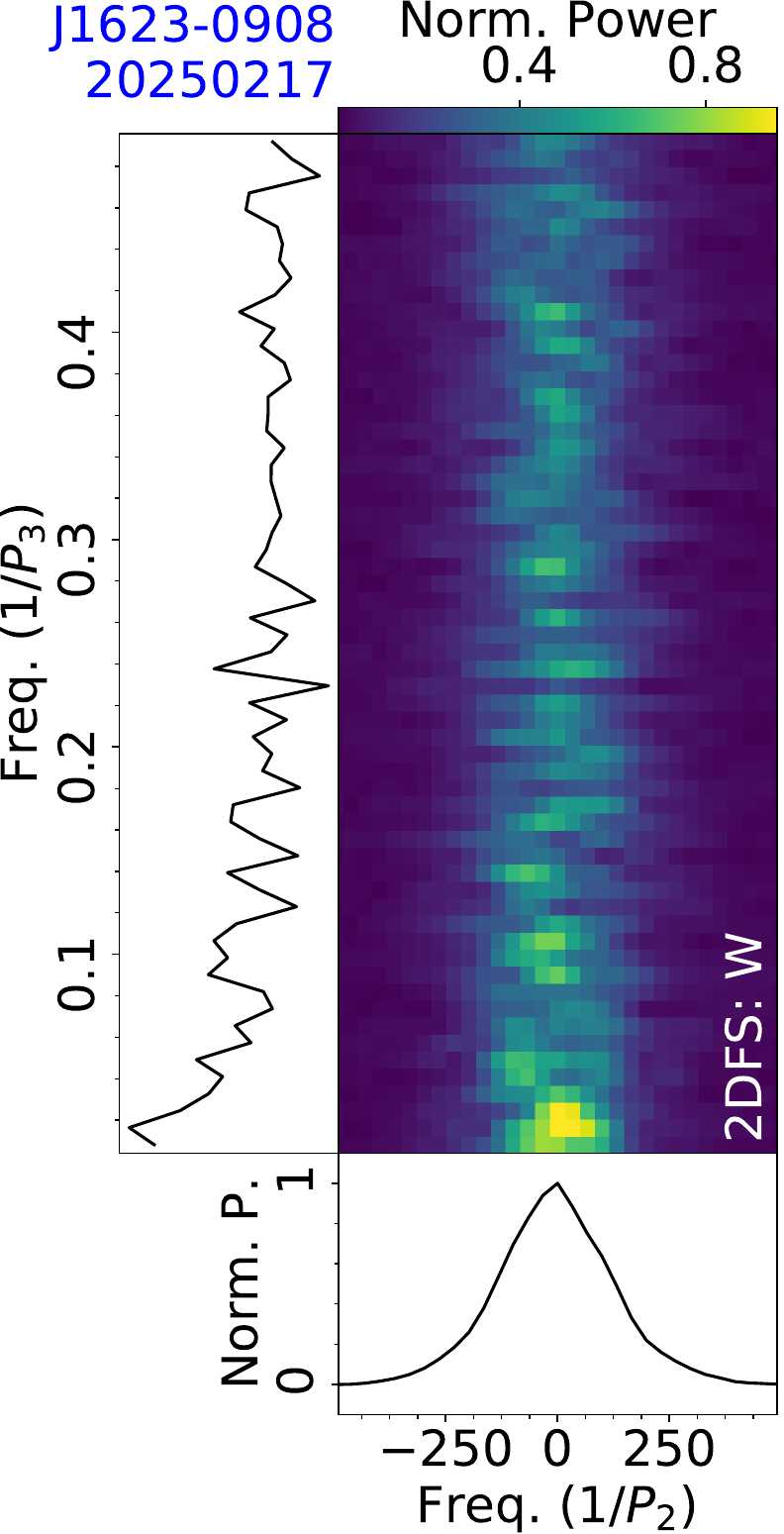}
\figcaption{Fluctuation analysis of PSR J1623-0908 from the observation on 20250217, with LRFS and 2DFS for the on-pulse region of a mean pulse profile.  \label{subfig:fluctu:J1623-0908}}
\end{figure}

\subsection{J1549+2113}
\label{subsec:J1549+2113}

PSR J1549+2113 was discovered by the Arecibo telescope \citep{Foster1995}. This pulsar was reported by citet{Song2023} to have the drifting behavior of $P_3=2.58\pm0.09$ periods and $P_2=-9.8^{+2}_{-0.9}$ degrees, as well as a $P_3$-only feature of $P_3=71\pm21$ periods.

The pulsar was observed by FAST on 20250217 for 10 minutes, deriving a rotation period $P=1.2624$~s and a dispersion measure $D\!M=24.4~{\rm cm^{-3}\,pc}$. 
Single pulse sequences shown in Fig.~\ref{subfig:TP:J1549+2113} display subpulse drifting and nulling phenomena. The nulling fraction of this observation is estimated from the on-pulse integral energy histogram in Fig.~\ref{subfig:Hist:J1549+2113}, that is 6.6$\pm$0.5\%. 
The fluctuation spectra in Fig.~\ref{subfig:fluctu:J1549+2113} display the negative drift feature, which is characterized by $1/P_3=0.406\pm0.001$ and $1/P_2=-66\pm1$, corresponding to $P_3=2.466\pm0.005$ periods and $P_2=-5.4\pm0.1^\circ$.

\begin{figure}[htpb]
\centering
\includegraphics[width=0.22\textwidth, angle=0]{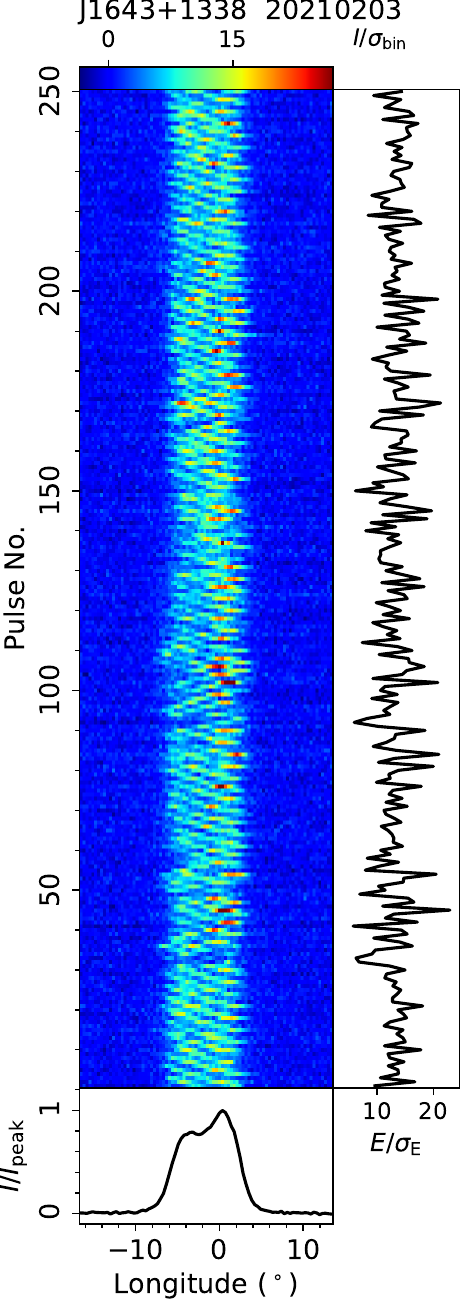}
\includegraphics[width=0.22\textwidth, angle=0]{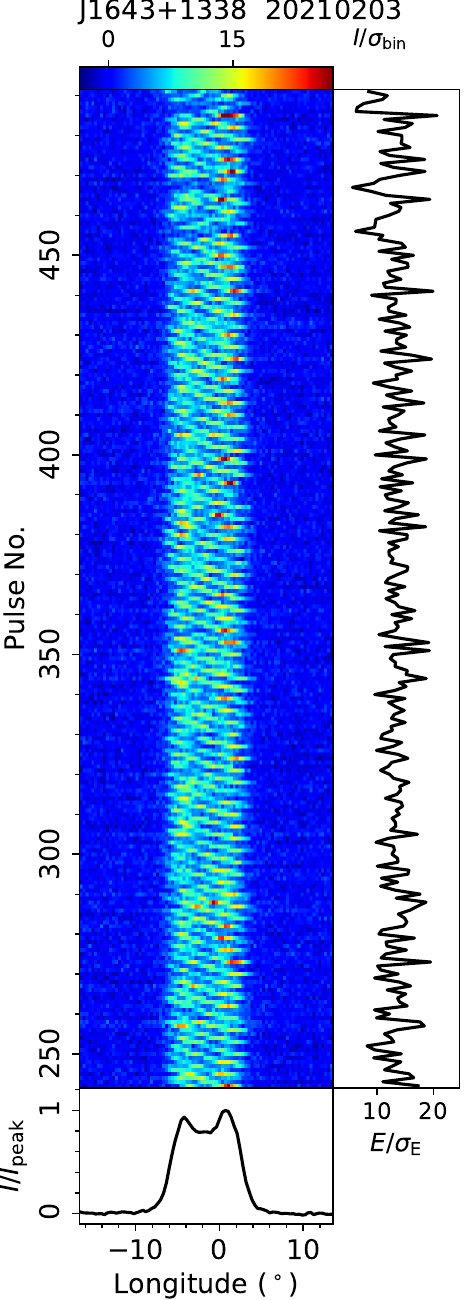}
\figcaption{Single pulse sequences of PSR J1643+1338 from the observation on 20210203. \label{subfig:TP:J1643+1338}}
\end{figure}

\begin{figure}[htpb]
\centering
\includegraphics[width=0.22\textwidth, angle=0]{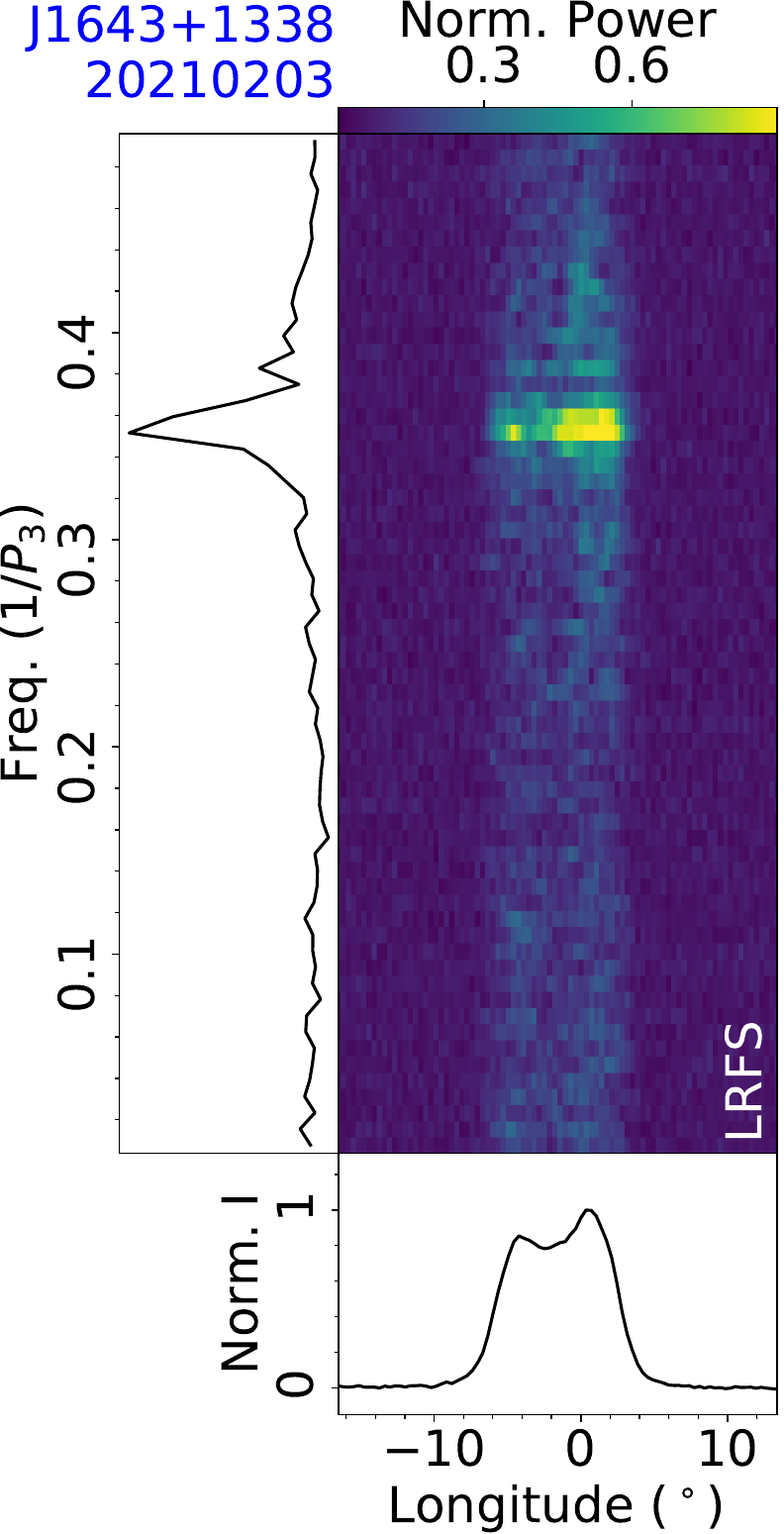}
\includegraphics[width=0.22\textwidth, angle=0]{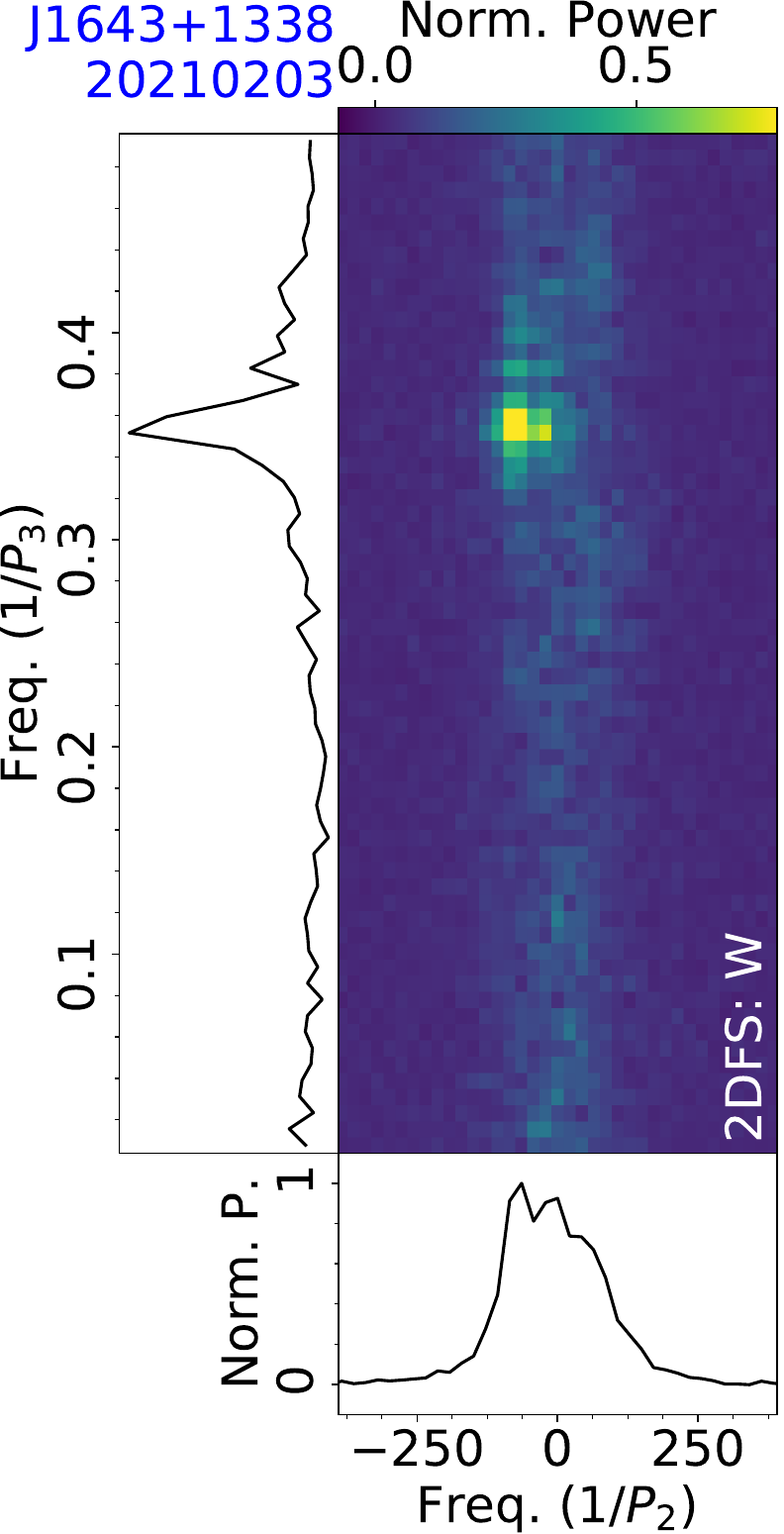}
\figcaption{Fluctuation analysis of PSR J1643+1338 from the observation on 20211123, with LRFS and 2DFS for the on-pulse region of a mean pulse profile.  \label{subfig:fluctu:J1643+1338}}
\end{figure}

\begin{figure}[htpb]
\centering
\includegraphics[width=0.22\textwidth, angle=0]{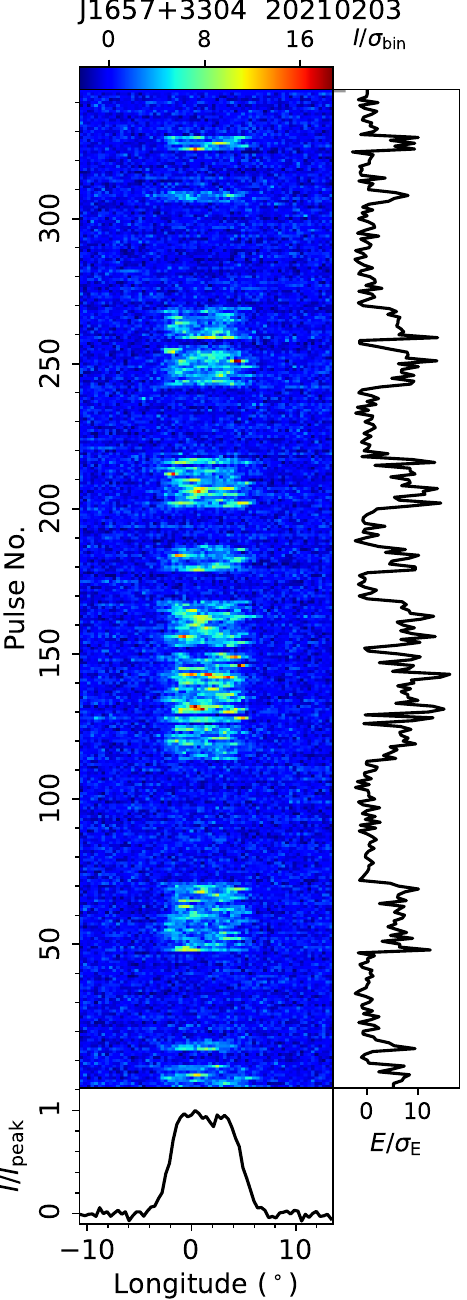}
\figcaption{Single pulse sequence of PSR J1657+3304 from the observation on 20210203. 
\label{subfig:TP:J1657+3304}
}
\end{figure}

\begin{figure}[htpb]
\centering
\includegraphics[width=0.39\textwidth, angle=0]{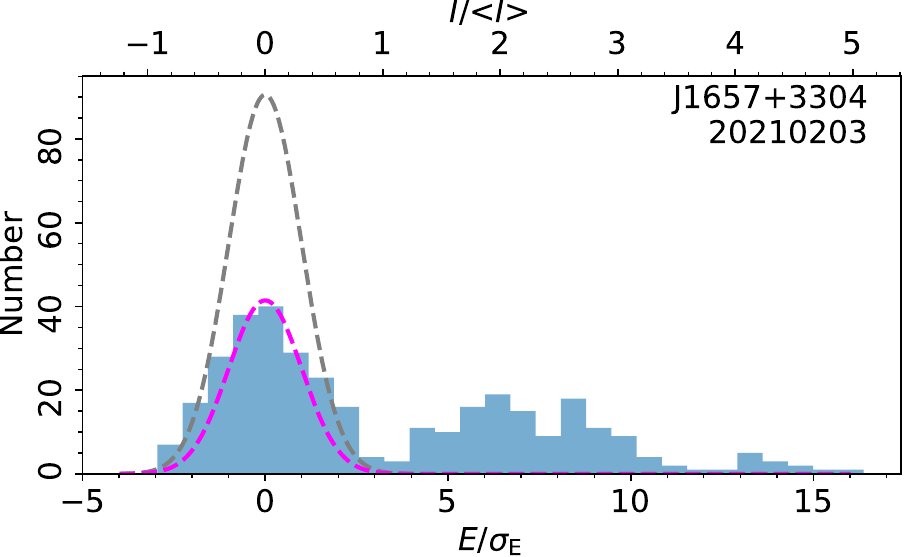}
\figcaption{On-pulse energy histogram of PSR J1657+3304 from the observation on 20210203. \label{subfig:Hist:J1657+3304}}
\end{figure}

\subsection{J1610-1322}
\label{subsec:J1610-1322}

PSR J1610-1322 was discovered in the Princeton-NRAO pulsar survey with the 92 m telescope at Green Bank \citep{Dewey1985}. Subpulse drifting with $P_3=8\pm2$ periods and $P_2=-20^{+5}_{-20}$ degrees were reported by \citet{Weltevrede2007} at 92 cm.

This pulsar was observed by FAST on 20250816 for 26 minutes, with a rotation period $P=1.0185$~s and a dispersion measure $D\!M=50.5~{\rm cm^{-3}\,pc}$ determined. Single pulse sequences in Fig.~\ref{subfig:TP:J1610-1322} illustrate the existence of the nulling, subpulse drifting, and mode changing phenomena. Three drifting modes are distinguished: normal, slow drifting, and fast drifting modes, which are labeled with orange, red, and green bars, respectively. 
The nulling fraction of this observation is estimated to be 34$\pm$1\% from the on-pulse energy histogram (Fig.~\ref{subfig:Hist:J1610-1322}). 
LRFS and 2DFS of three drifting modes, obtained from their corresponding single-pulse segments in Fig.~\ref{subfig:TP:J1610-1322}, are shown in Fig.~\ref{subfig:fluctu:J1610-1322}, and the fast drifting mode has the smallest $P_3$ value.
For the normal mode, the centroid of the negative drift feature in 2DFS is at $1/P_3=0.136\pm0.001$ and $1/P_2=-28\pm1$, corresponding to periodicities of $P_3=7.35\pm0.05$ periods and $P_2=-13.1\pm0.3$ degrees. 
The centroid of the drift feature in 2DFS of the slow drifting mode is characterized by  $1/P_3=0.043\pm0.002$ and $1/P_2=-29\pm1$, yielding $P_3=23\pm1$ periods and $P_2=-12.3\pm0.4$ degrees. 
In 2DFS of the fast drifting mode, the drift feature exhibits the centroid at $1/P_3=0.221\pm0.004$ and $1/P_2=-30\pm1$, corresponding to $P_3=4.5\pm0.1$ periods and $P_2=-12.1\pm0.5$ degrees.

\subsection{J1623-0908}
\label{subsec:J1623-0908}

PSR J1623-0908 was discovered from observations at the Molonglo Radio Observatory and the Australian National Radio Astronomy Observatory, Parkes \citep{Manchester1978}. Two positively drift features have been reported by \citet{Song2023}, which are $P_3=28\pm31$ periods and $P_2=-69^{+62}_{-35}$ degrees, and $P_3=2.5\pm0.1$ periods and $P_2=25^{+13}_{-15}$ degrees.

This pulsar was observed by FAST on 20250217 for 10 minutes, deriving a rotation period $P=1.2763$~s and a dispersion measure $D\!M=68.7~{\rm cm^{-3}\,pc}$. Single pulse sequences are displayed in Fig.~\ref{subfig:TP:J1623-0908}, illustrating the existence of the nulling phenomenon. From the on-pulse energy histogram in Fig.~\ref{subfig:Hist:J1623-0908}, the nulling fraction of this FAST observation is estimated to be 7.3$\pm$1.9\%. However, there is no obvious drift feature as previously reported in LRFS and 2DFS (Fig.~\ref{subfig:fluctu:J1623-0908}) from this data.

A longer observation is required for a detailed analysis of the drifting behavior.

\subsection{J1643+1338}
\label{subsec:J1643+1338}

The pulsar was first reported by \citet{Tyulbashev2017} using the Big Scanning Antenna at 111 MHz. 

The pulsar was observed by FAST on 20210203 for 9 minutes, deriving a rotation period $P=1.0991$~s and a dispersion measure $D\!M=35.1~{\rm cm^{-3}\,pc}$. 
The single pulse sequence is displayed in Fig.~\ref{subfig:TP:J1643+1338}. Drift bands are systematic, corresponding to the drift feature in LRFS and 2DFS (Fig.~\ref{subfig:fluctu:J1643+1338}). The centroid modulation frequencies are estimated to be $1/P_3=0.361\pm0.001$ and $1/P_2=-68\pm1$, yielding $P_3=2.77\pm0.01$ periods and $P_2=-5.3\pm0.1^\circ$. 
Moreover, apparent drift direction seems to change around pulse numbers 30, 100, and 460, before the single-pulse intensity decreases. This may be caused by drifting direction changes, or just drift rate changes and aliased effect \citep[e.g.,][]{Weltevrede2006}.

\begin{figure}[hbpt]
\centering
\includegraphics[width=0.44\textwidth, angle=0]{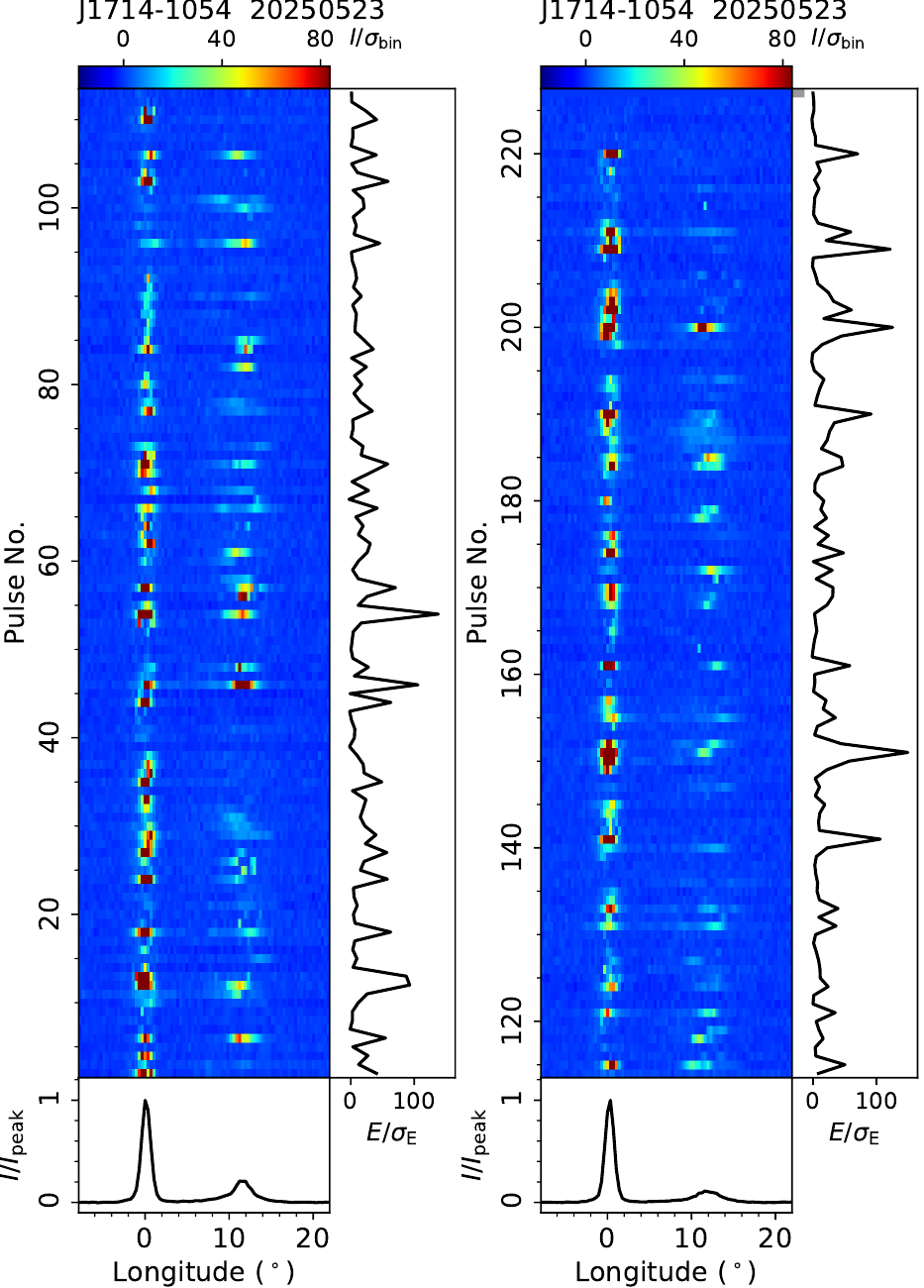}
\figcaption{Single pulse sequences of PSR J1714-1054 from the FAST observation on 20250523.
\label{subfig:TP:J1714-1054}}
\end{figure}

\begin{figure}[hbpt]
\centering
\includegraphics[width=0.44\textwidth, angle=0]{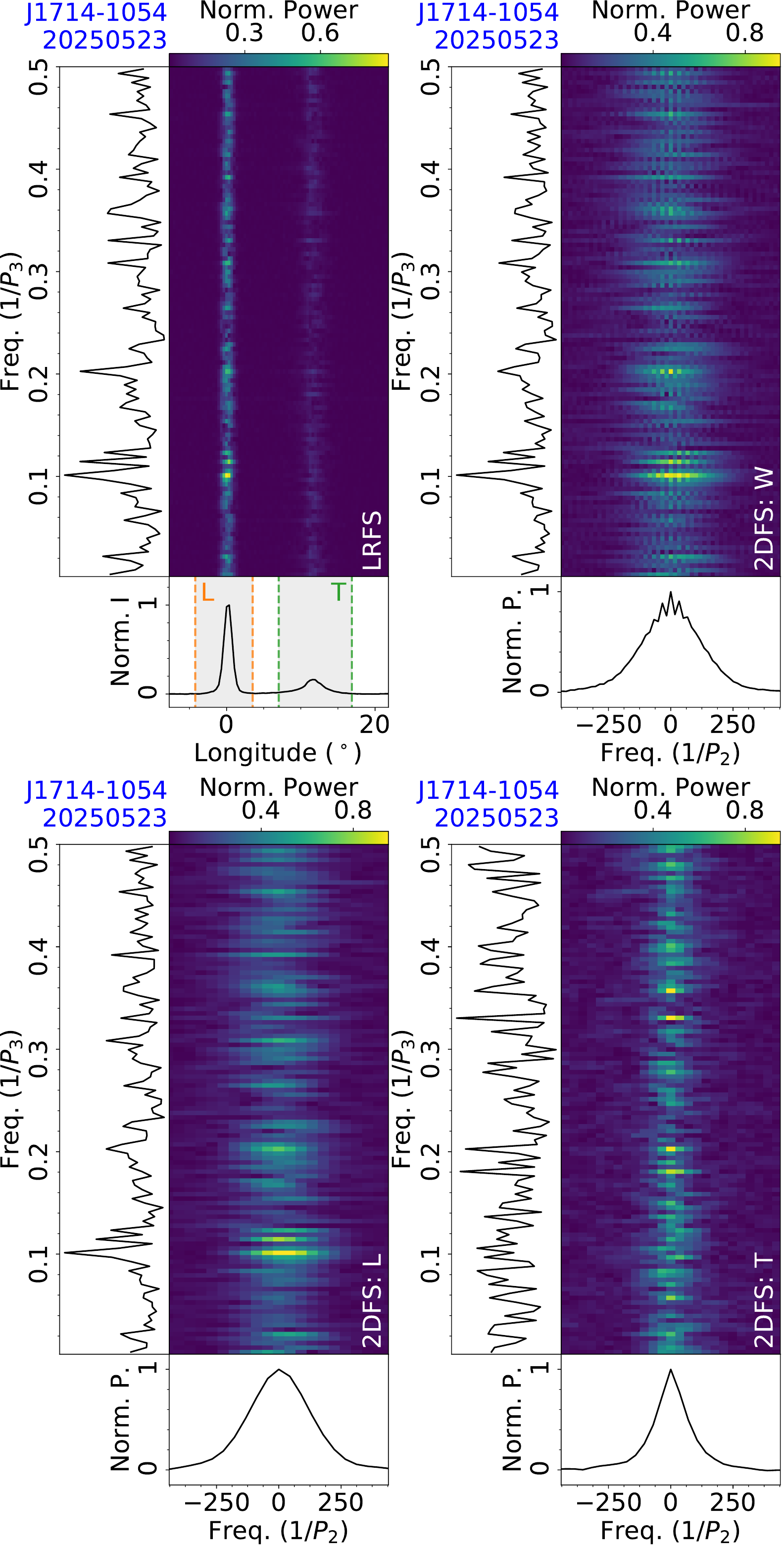}
\figcaption{Fluctuation analysis of PSR J1714-1054 from the FAST observation on 20250523, with LRFS (top-left), and 2DFS for the on-pulse region (top-right), leading part (bottom-left) and trailing part (bottom-right) of a mean pulse profile. 
\label{subfig:fluctu:J1714-1054}}
\end{figure}

\begin{figure}[htpb]
\centering
\includegraphics[width=0.22\textwidth, angle=0]{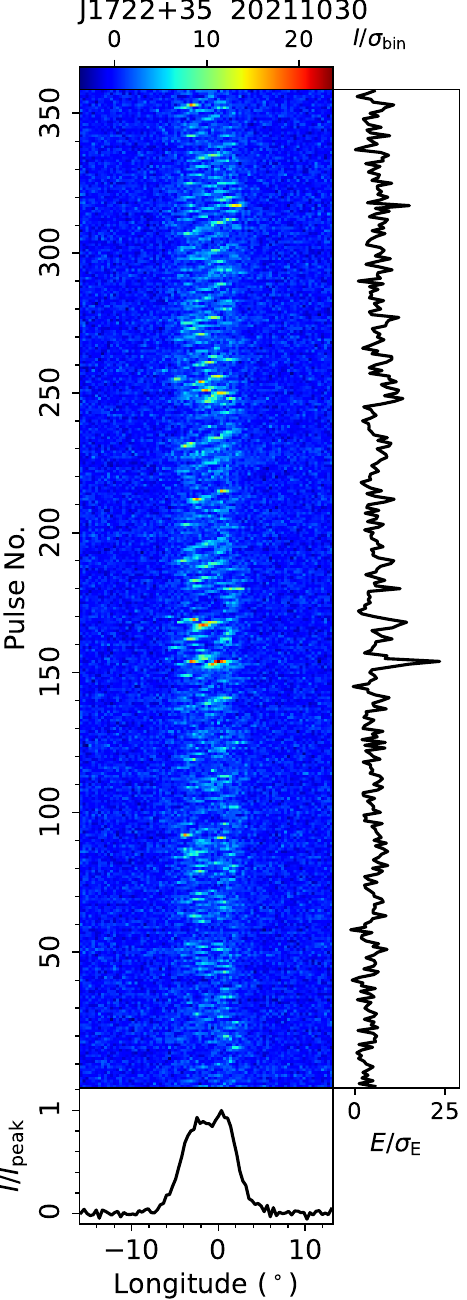}
\figcaption{Single pulse sequence of PSR J1722+35 from the observation on 20211030.
\label{subfig:TP:J1722+35}}
\end{figure}

\begin{figure}[htpb]
\centering
\includegraphics[width=0.22\textwidth, angle=0]{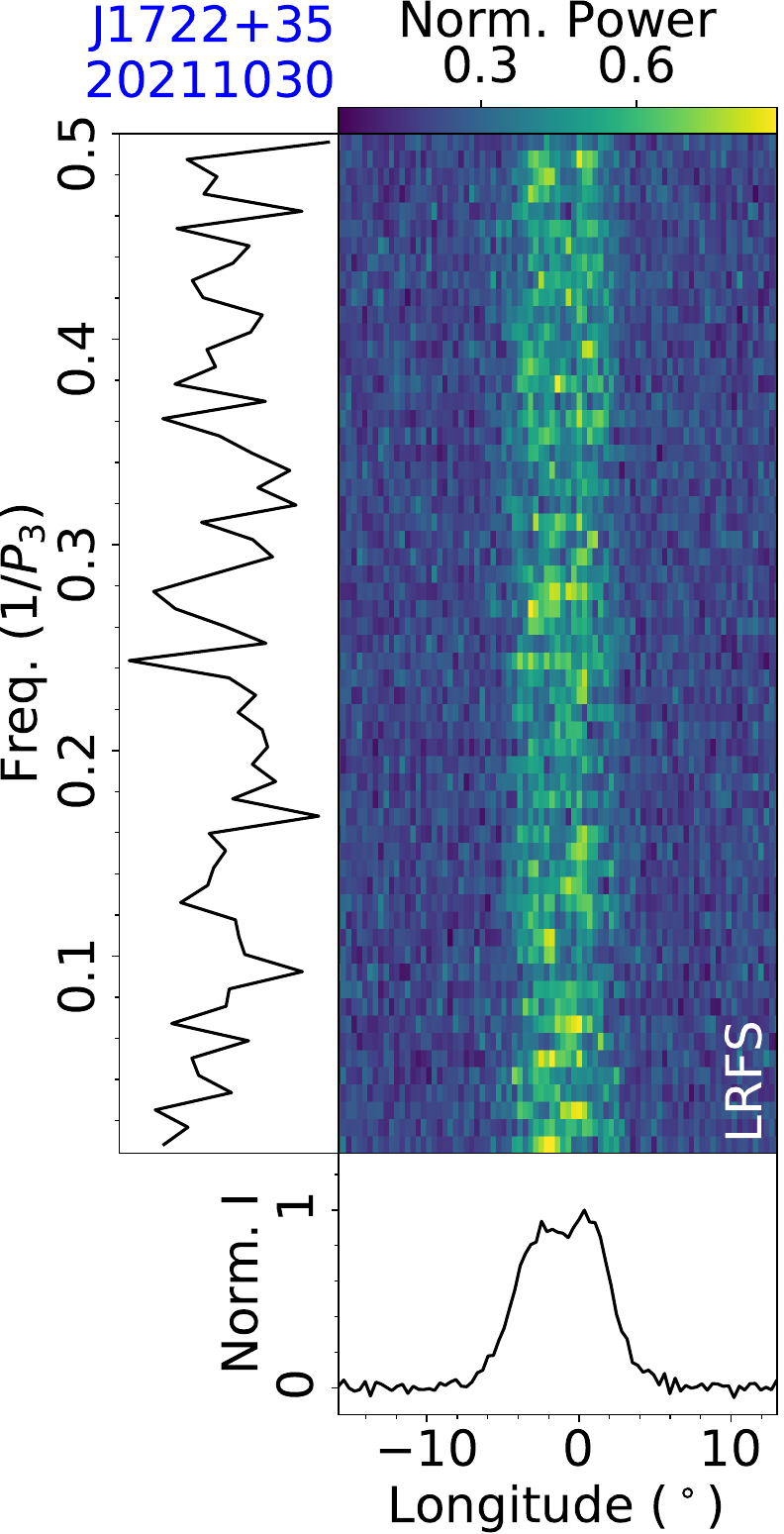}
\includegraphics[width=0.22\textwidth, angle=0]{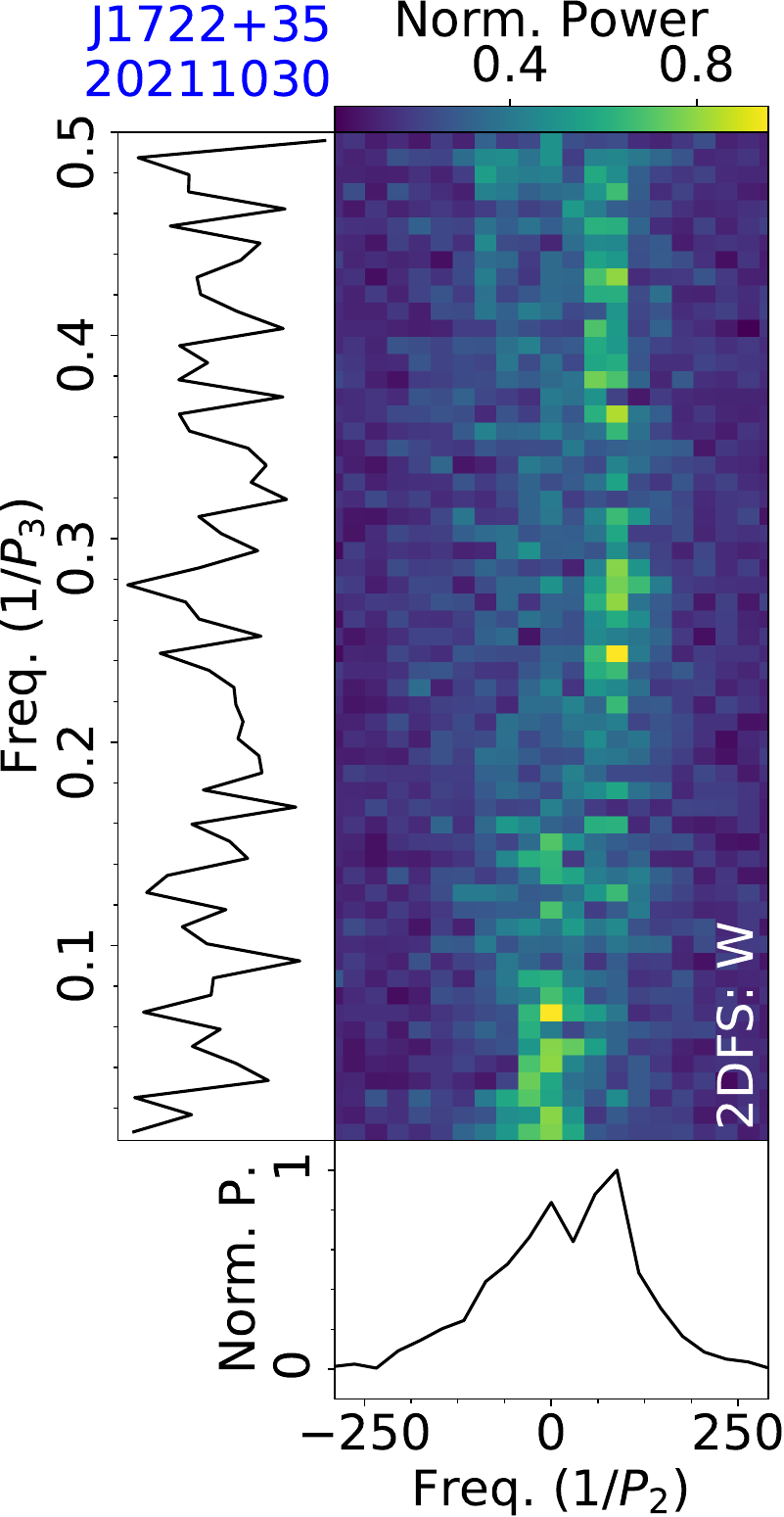}
\figcaption{Fluctuation analysis of PSR J1722+35 from the observation on 20211030, with LRFS and 2DFS for the on-pulse region of a mean pulse profile.  \label{subfig:fluctu:J1722+35}}
\end{figure}

\subsection{J1657+3304}
\label{subsec:J1657+3304}

PSR J1657+3304 was discovered by \citet{Tyulbashev2017} at 111 MHz. The nulling fraction was reported to be 27\%-43\% by \citet{Tan2020} from four observations of LOFAR at 149 MHz. 

This pulsar was observed by FAST on 20210203 for 9 minutes, deriving a rotation period $P=1.5704$~s and a dispersion measure $D\!M=23.0~{\rm cm^{-3}\,pc}$ from this observation. The single pulse sequence (Fig.~\ref{subfig:TP:J1657+3304}) also shows nulling phenomenon, and the nulling fraction is estimated to be 45$\pm$5\% from the on-pulse integral energy histogram in Fig.~\ref{subfig:Hist:J1657+3304}.

\begin{figure}[htpb]
\centering
\includegraphics[width=0.22\textwidth, angle=0]{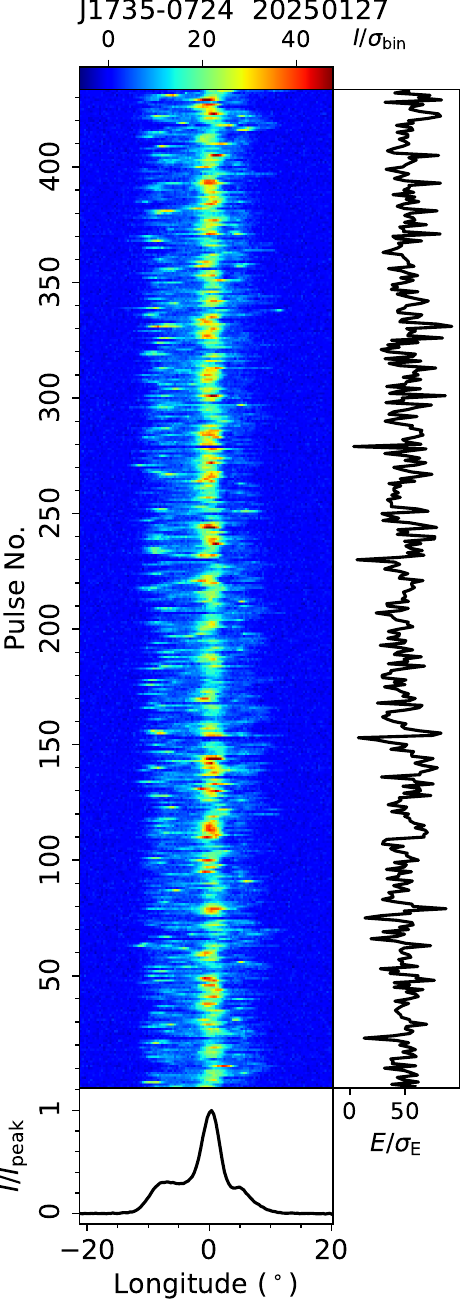}
\includegraphics[width=0.22\textwidth, angle=0]{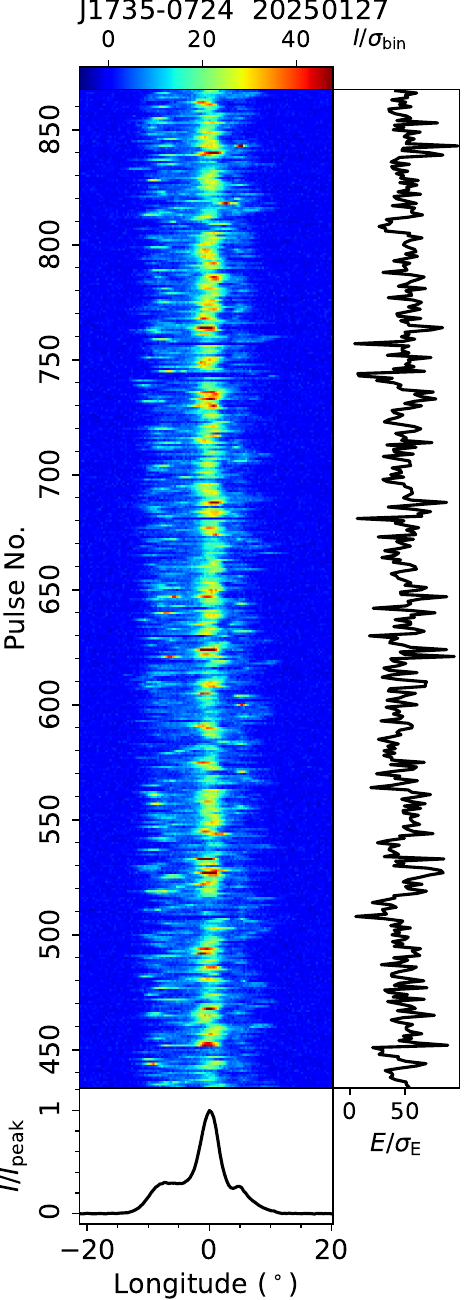}
\figcaption{Single pulse sequences of PSR J1735-0724 from the observation on 20250127.
\label{subfig:TP:J1735-0724}}
\end{figure}

\begin{figure}[htpb]
\centering
\includegraphics[width=0.22\textwidth, angle=0]{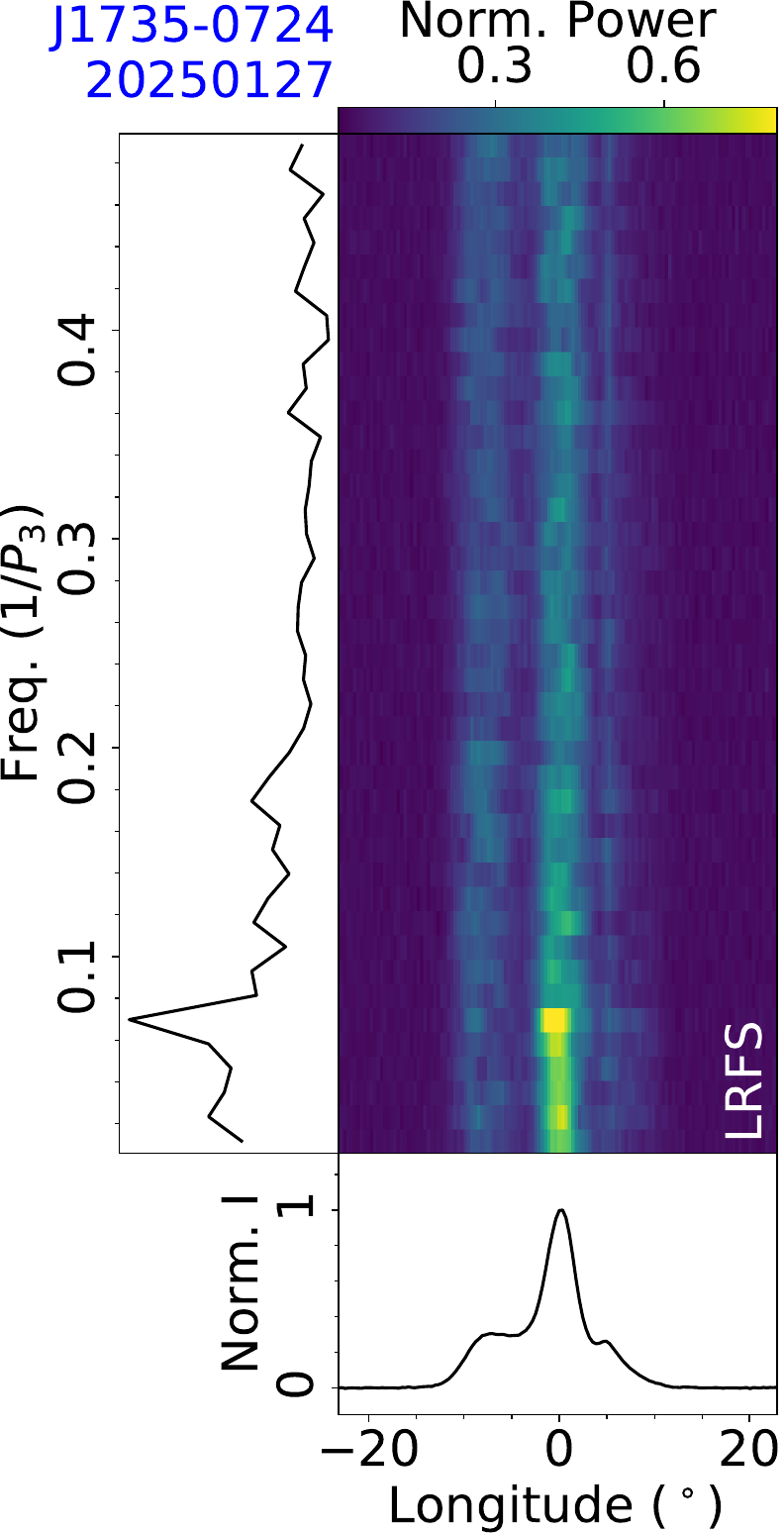}
\includegraphics[width=0.22\textwidth, angle=0]{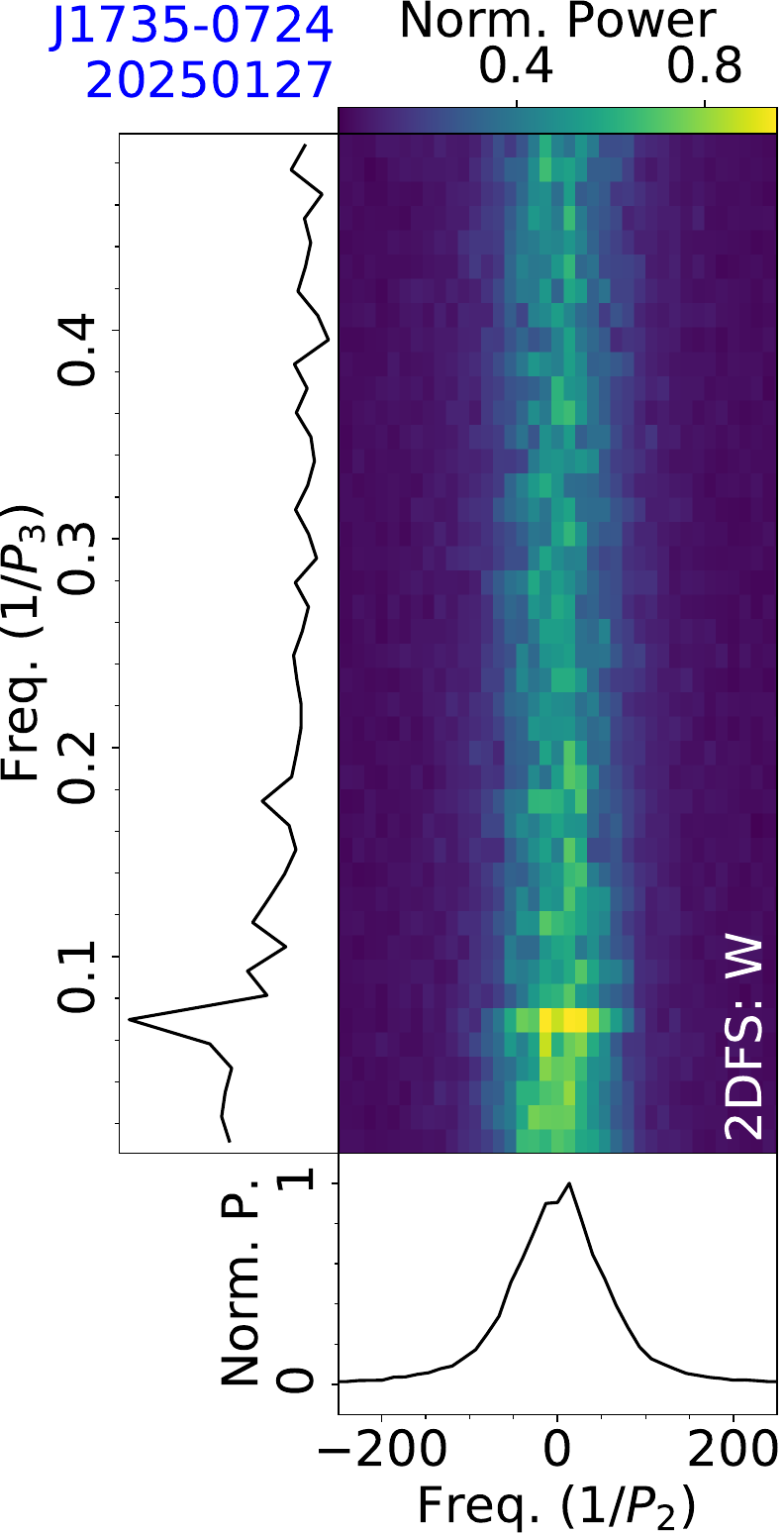}
\figcaption{Fluctuation analysis of PSR J1735-0724 from the observation on 20250127, with LRFS and 2DFS for the on-pulse region of a mean pulse profile.
\label{subfig:fluctu:J1735-0724}}
\end{figure}

\subsection{J1714-1054}
\label{subsec:J1714-1054}

PSR J1714-1054 was discovered by \citet{Jacoby2009} with the 64 m Parkes radio telescope. The drifting parameters of this pulsar were reported by \citet{Song2023}: $P_3=10.2\pm0.3$ periods and $P_2=10^{+1}_{-3}$ degrees for one component; $P_3=10.4\pm0.1$ periods and $P_2=18^{+35}_{-6}$ degrees for the other component.

This pulsar was observed by FAST on 20250523 for 8 minutes, deriving a rotation period $P=2.0888$~s and a dispersion measure $D\!M=51.9~{\rm cm^{-3}\,pc}$. Single pulse sequences are shown in Fig.~\ref{subfig:TP:J1714-1054}. From the fluctuation spectra (Fig.~\ref{subfig:fluctu:J1714-1054}), the leading part in the mean pulse profile exhibits a positive drift feature with the centroid at $1/P_3=0.1010\pm0.0004$ and $1/P_2=31\pm5$, corresponding to periodicities of $P_3=9.90\pm0.04$ periods and $P_2=11\pm2$ degrees. However, there is no obvious modulation feature in the 2DFS of the trailing profile part from this observation.

\begin{figure}[htpb]
\centering
\includegraphics[width=0.22\textwidth, angle=0]{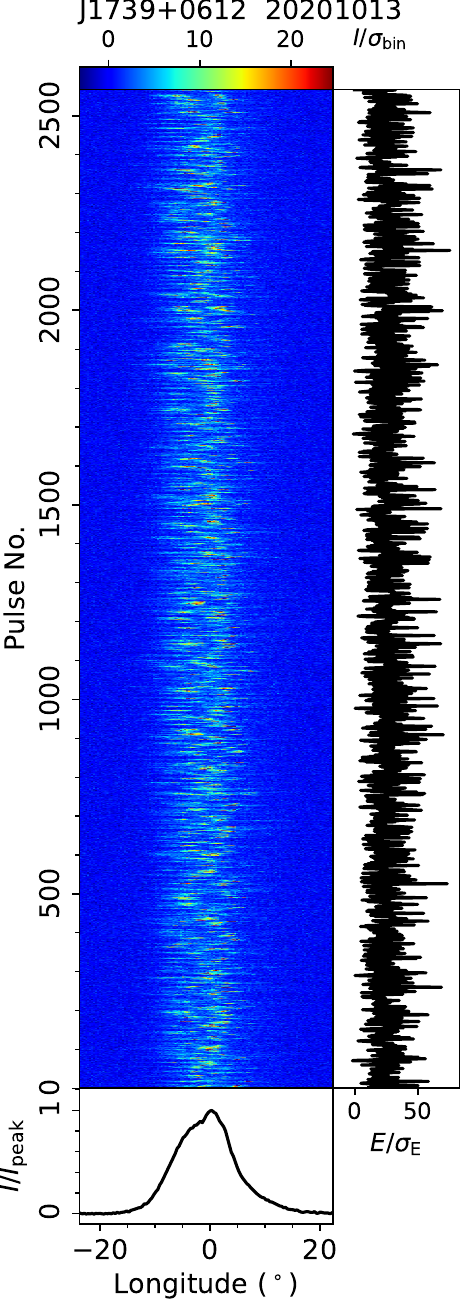}
\includegraphics[width=0.22\textwidth, angle=0]{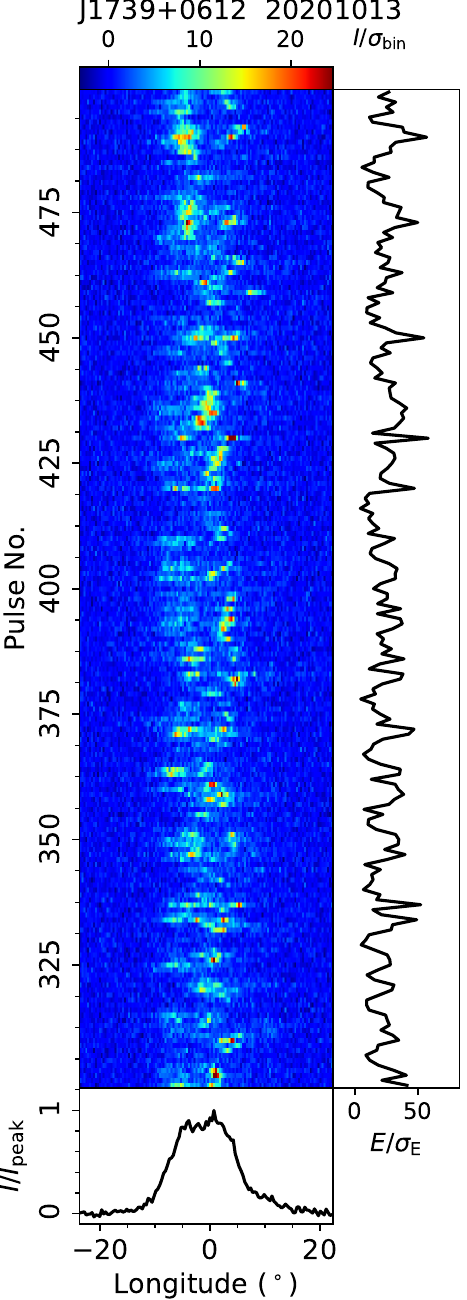}
\figcaption{Single pulse sequence of PSR J1739+0612 from the FAST observation on 20201013, and a zoomed-in view of pulses No. 300-500.
\label{subfig:TP:J1739+0612}}
\end{figure}

\begin{figure}[htpb]
\centering
\includegraphics[width=0.22\textwidth, angle=0]{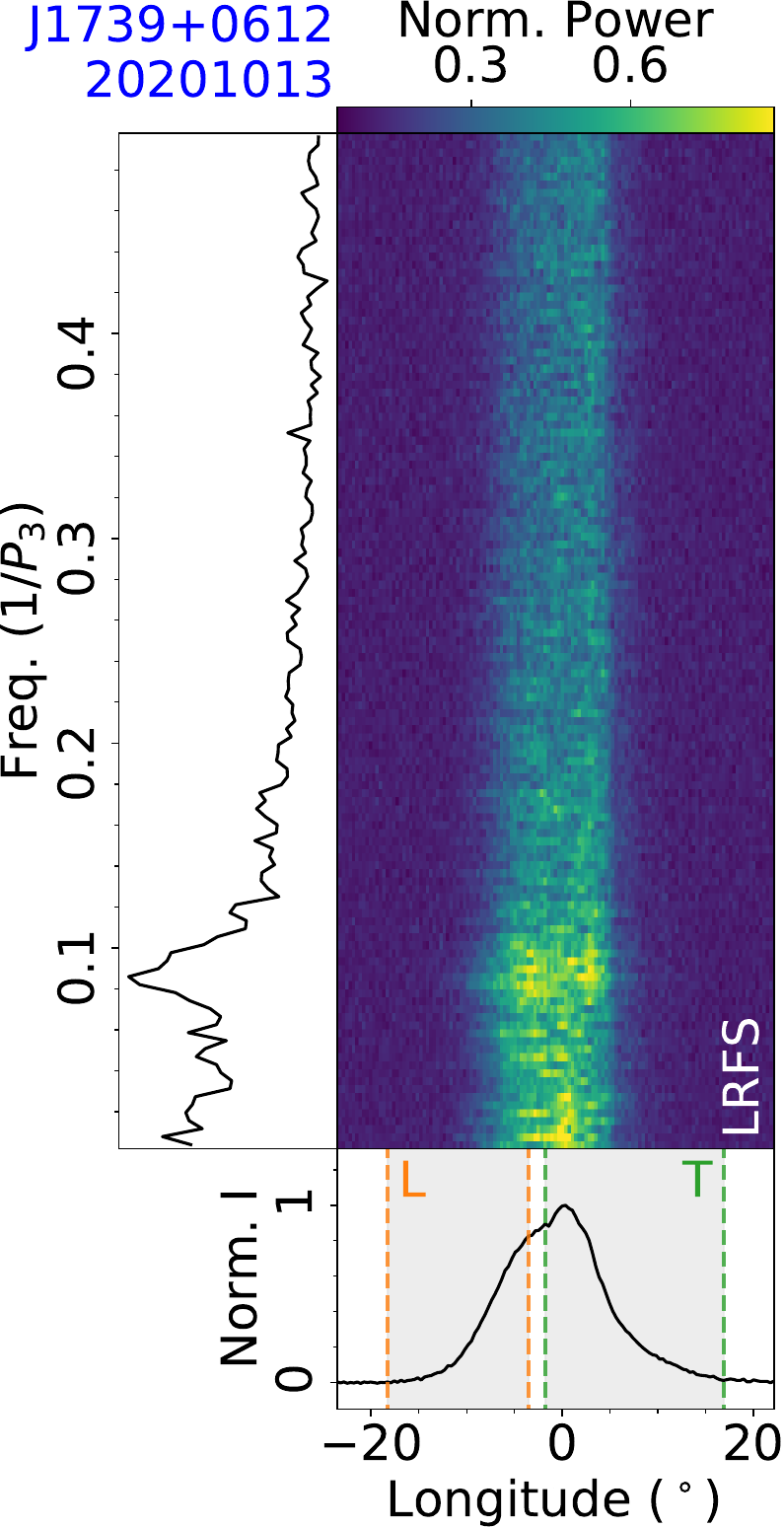}
\includegraphics[width=0.22\textwidth, angle=0]{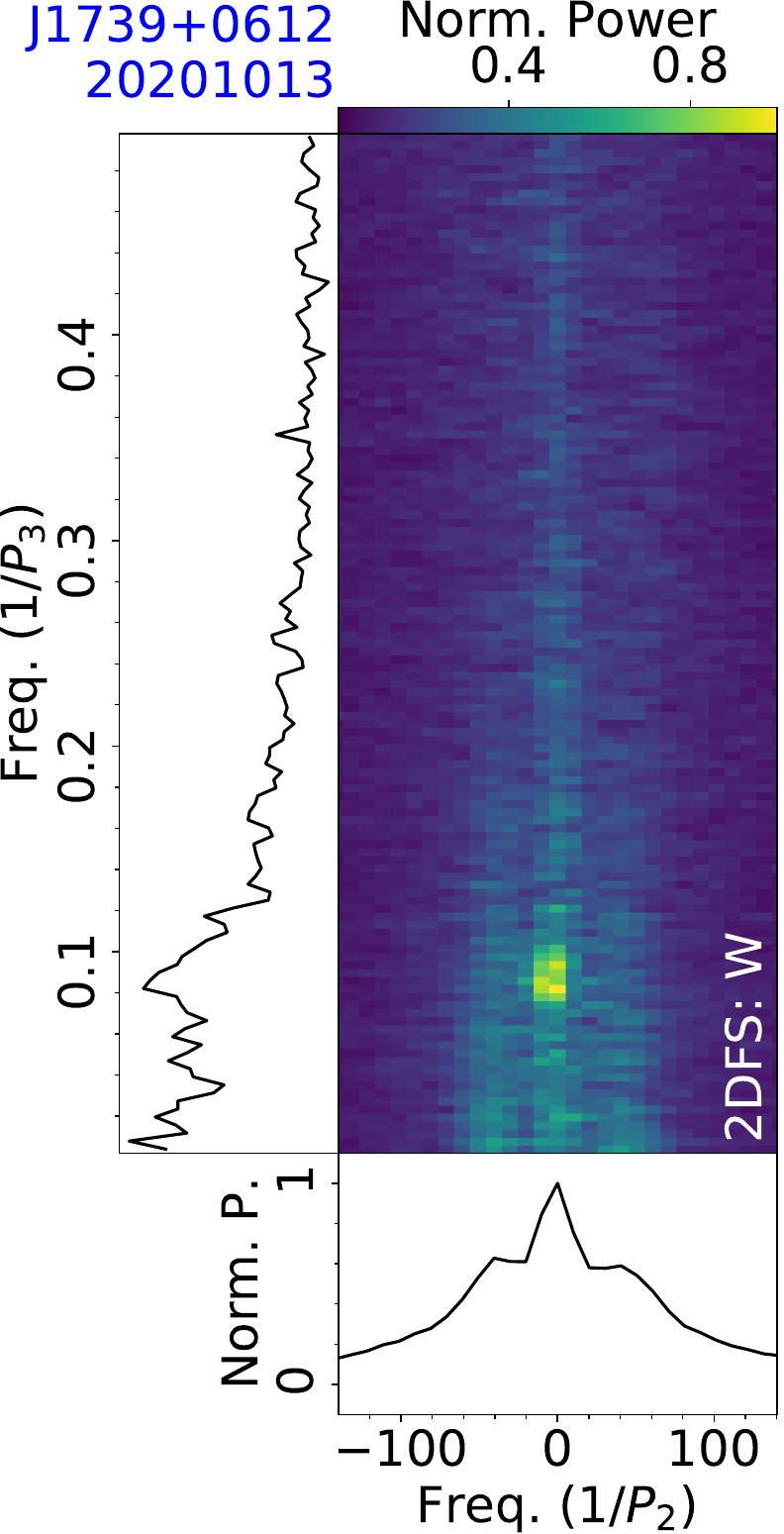}\\
\includegraphics[width=0.22\textwidth, angle=0]{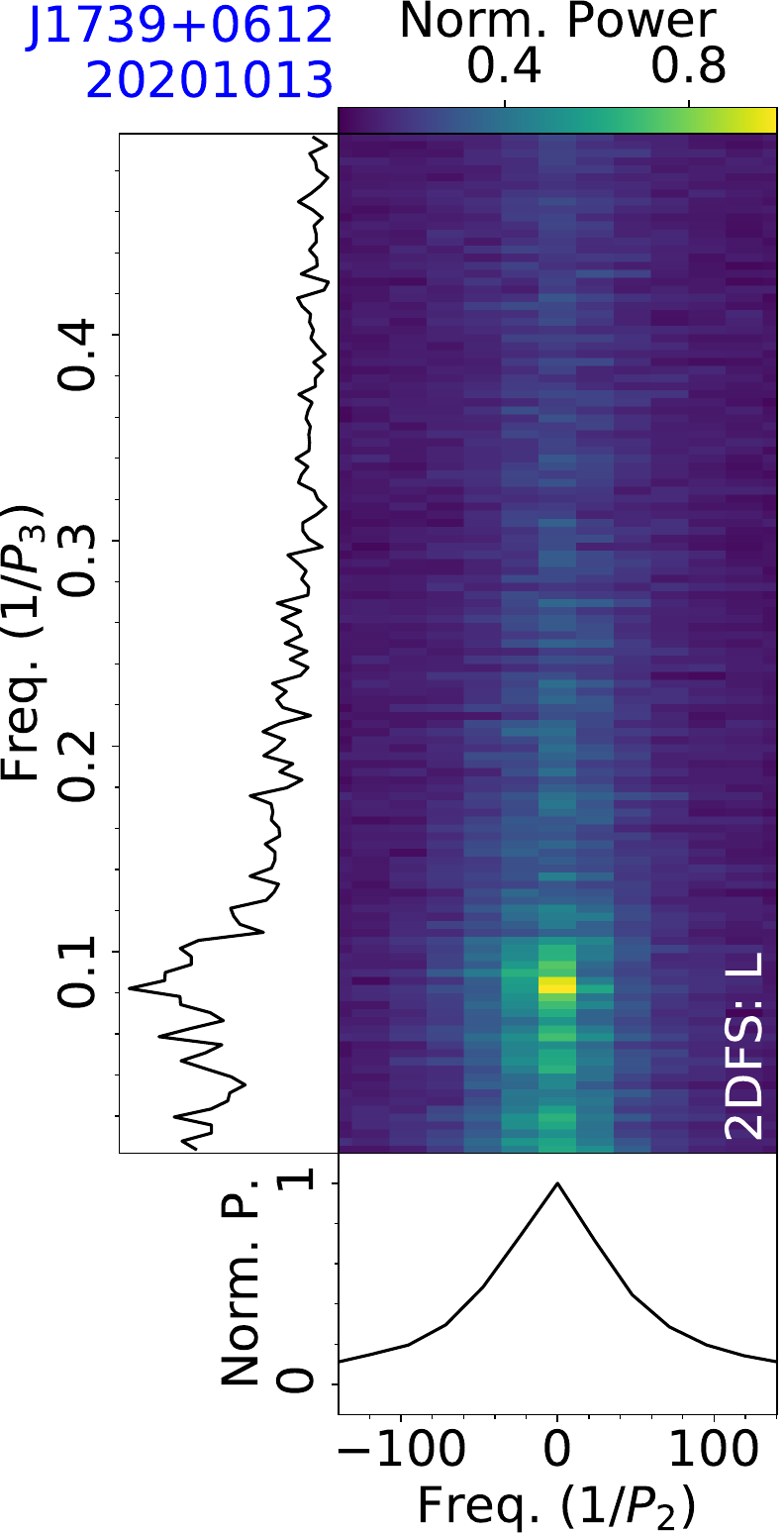}
\includegraphics[width=0.22\textwidth, angle=0]{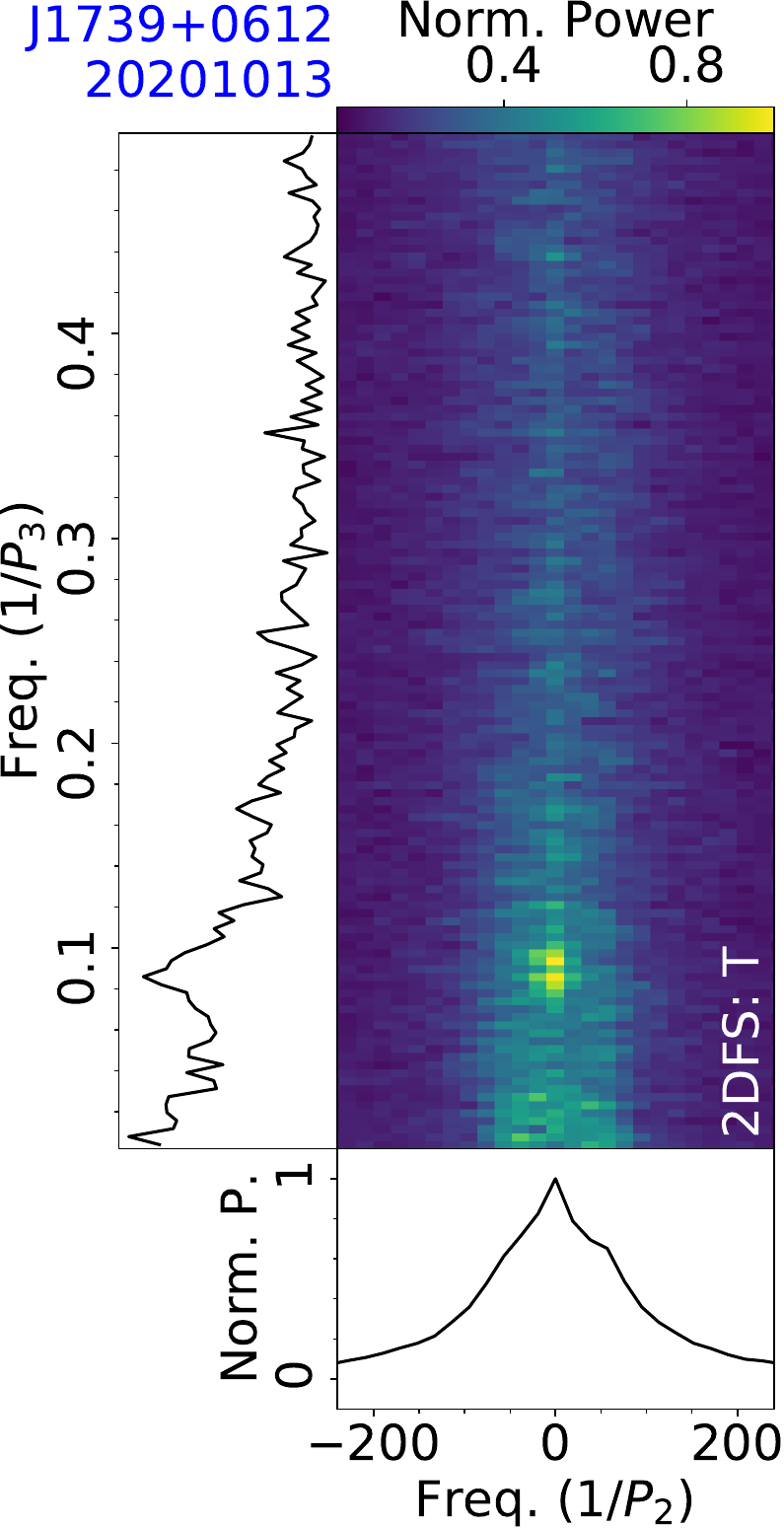}
\figcaption{Fluctuation analysis of PSR J1739+0612 from the observation on 20201013, with LRFS and 2DFS for the on-pulse region of a mean pulse profile.  \label{subfig:fluctu:J1739+0612}}
\end{figure}

\subsection{J1722+35}
\label{subsec:J1722+35}

PSR J1722+35 was discovered by \citet{Tyulbashev2016} using the Large Phased Array at 111 MHz. The drifting parameter $P_2$ at 111 MHz was reported by \citet{Smirnova2024} to be 117$^{+8}_{-11}$ ms. 

The pulsar was observed by FAST on 20211030 for 5 minutes, with a rotation period $P=0.8217$~s and a dispersion measure $D\!M=23.6~{\rm cm^{-3}\,pc}$ from this observation. The single pulse sequence is shown in Fig.~\ref{subfig:TP:J1722+35}. There are three modulation features in 2DFS (Fig.~\ref{subfig:fluctu:J1722+35}), two of them are related to positive drifting and one is temporally low-frequency modulation. For positively drift features, the centroid modulation frequencies are $1/P_3=0.265\pm0.003$ and $1/P_2=88\pm3$ ($P_3=3.78\pm0.04$ periods and $P_2=4.1\pm0.1^\circ$), and $1/P_3=0.414\pm0.004$ and $1/P_2=74\pm2$ ($P_3=2.42\pm0.0
3$ periods and $P_2=4.9\pm0.1^\circ$). 
The low-frequency modulation has $1/P_3=0.045\pm0.001$, corresponding to $P_3=22\pm1$ periods.

\subsection{J1735-0724}
\label{subsec:J1735-0724}

PSR J1735-0724 was discovered by \citet{ll76}.
%
This pulsar was reported to have the modulation behavior with $P_3$ of $19.7\pm7.5$ periods \citep{Basu2016}, and $13\pm2$ periods \citep{Weltevrede2007}. Drifting parameters of $P_3=23\pm6$ periods and $P_2=16^{+67}_{-8}$ degrees were also reported by \citet{Song2023}.

The pulsar was observed by FAST on 20250127 for 6 minutes, yielding a rotation period $P=0.4193$~s and a dispersion measure $D\!M=73.0~{\rm cm^{-3}\,pc}$. 
Single pulse sequences are shown in Fig.~\ref{subfig:TP:J1735-0724}. From the fluctuation spectra in Fig.~\ref{subfig:fluctu:J1735-0724}, there is a low-frequency modulation of $1/P_3=0.046\pm0.001$, corresponding to $P_3=21.9\pm0.4$ periods. 
There are also short-duration intensity decreases lasting for about 1-period, and longer observations are required for further analysis.

\begin{figure}[htpb]
\centering
\includegraphics[width=0.22\textwidth, angle=0]{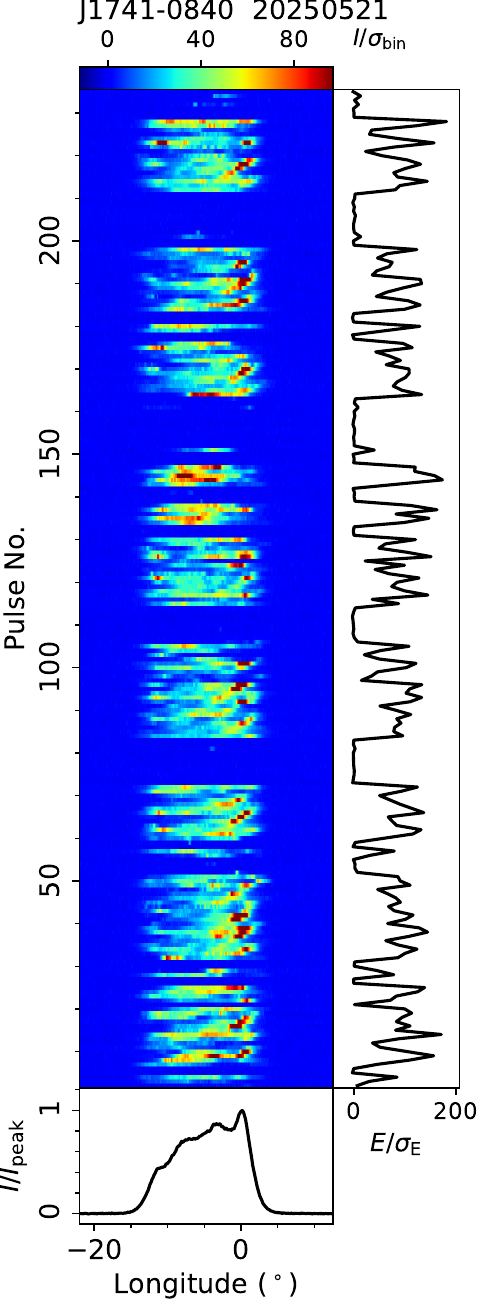}
\includegraphics[width=0.22\textwidth, angle=0]{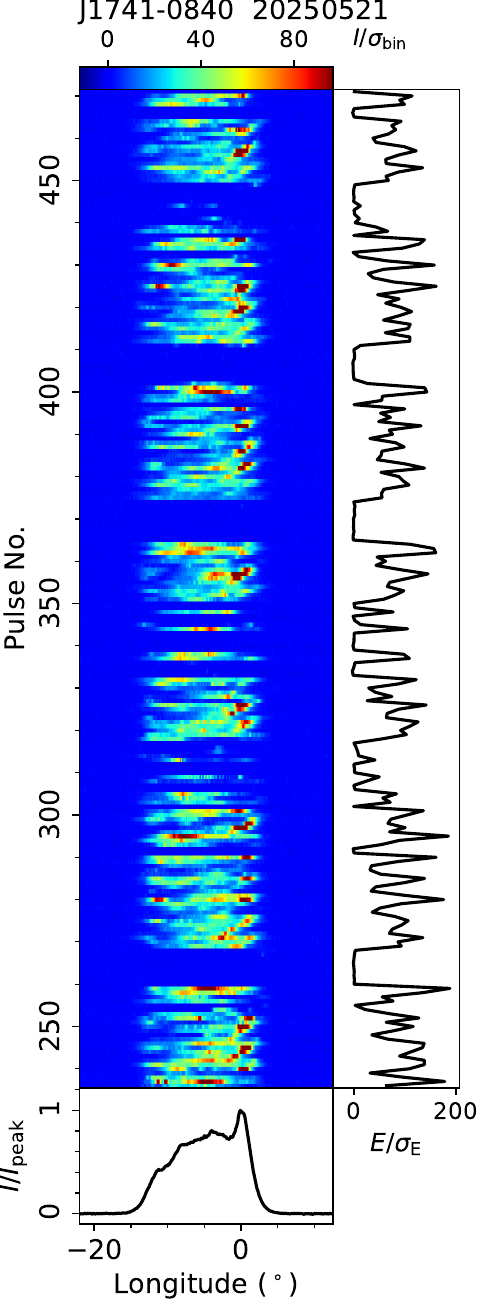}
\figcaption{Single pulse sequences of PSR J1741-0840 from the FAST observation on 20250521. 
\label{subfig:TP:J1741-0840}}
\end{figure}

\begin{figure}[htpb]
\centering
\includegraphics[width=0.39\textwidth, angle=0]{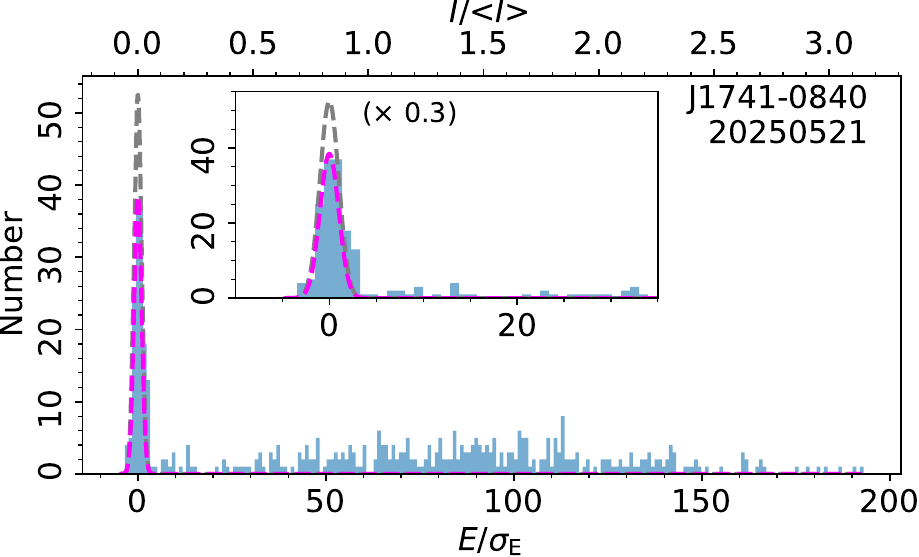}
\figcaption{On-pulse energy histogram of single pulses of PSR J1741-0840 from the FAST observation on 20250521. The inset provides a view of the x‑axis region from -10 to 35.
\label{subfig:Hist:J1741-0840}}
\end{figure}

\begin{figure}[htpb]
\centering
\includegraphics[width=0.22\textwidth, angle=0]{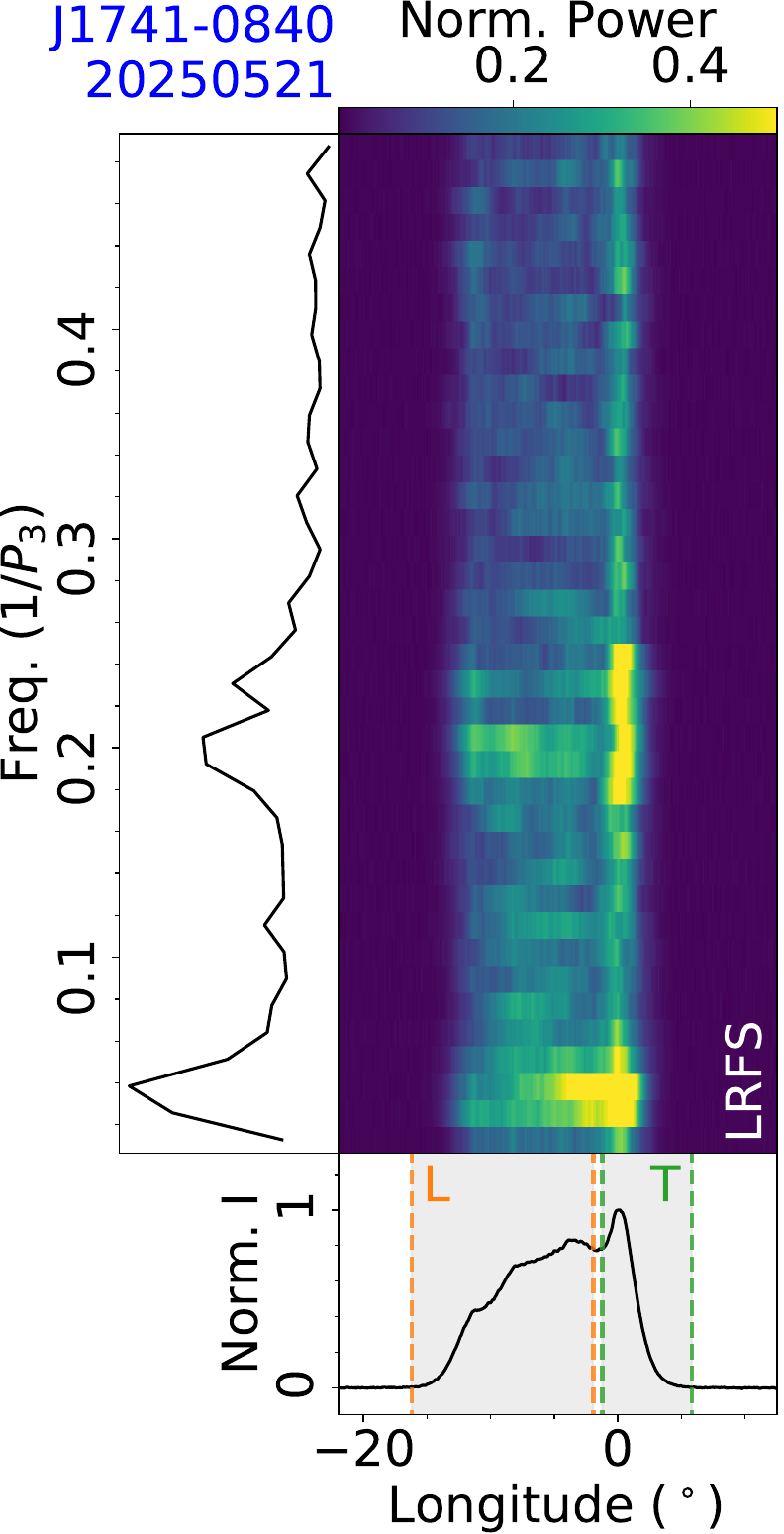}
\includegraphics[width=0.22\textwidth, angle=0]{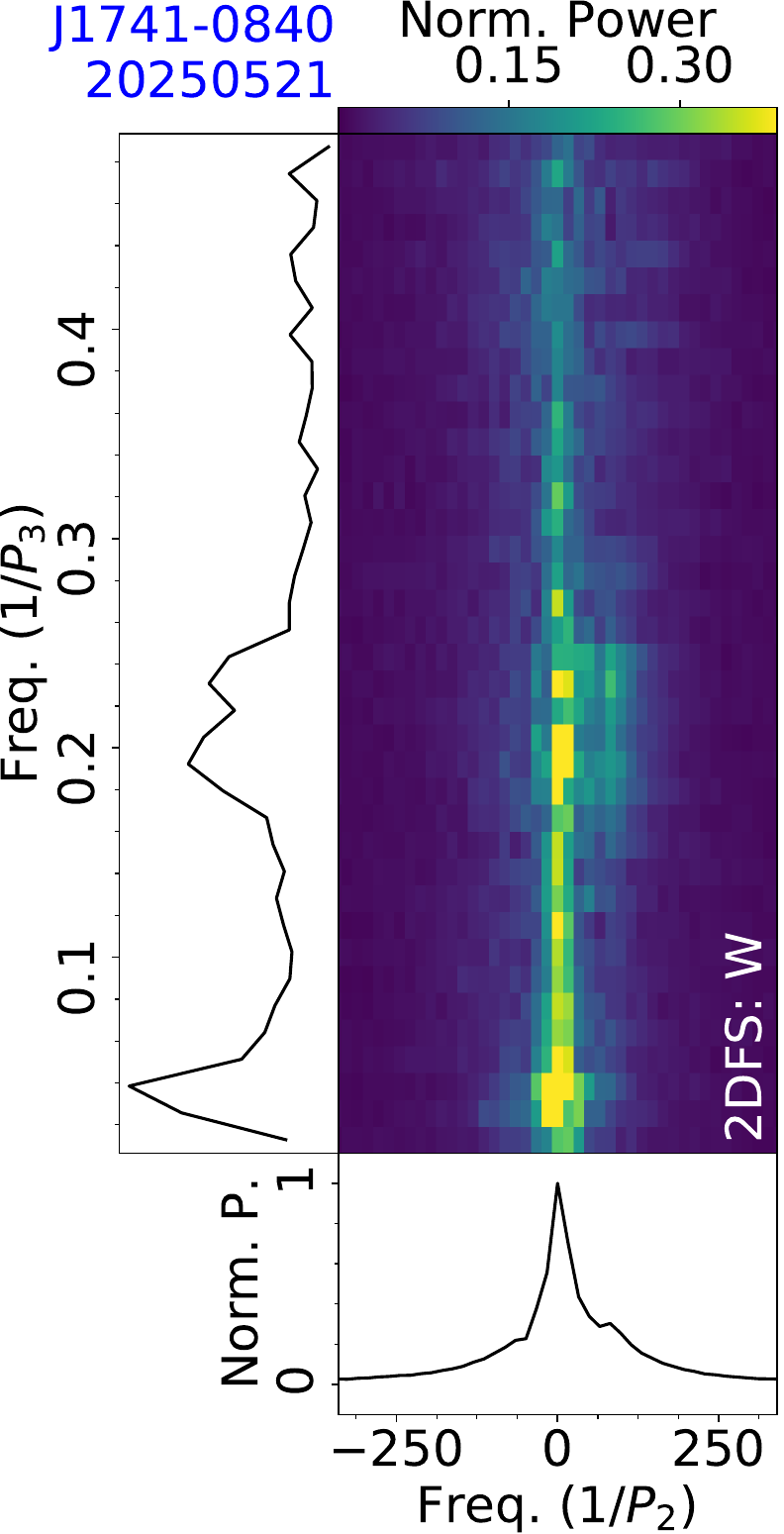}\\
\includegraphics[width=0.22\textwidth, angle=0]{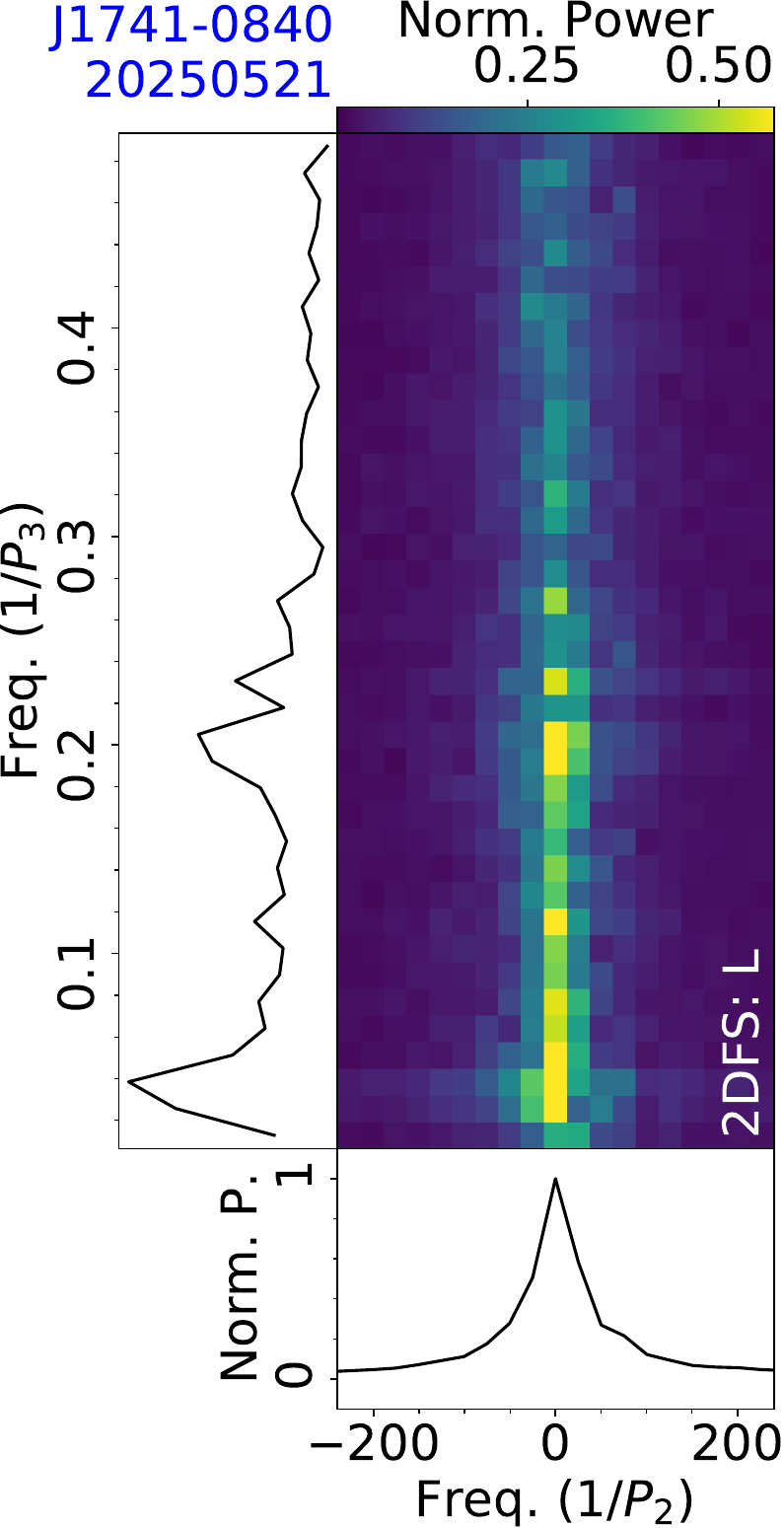}
\includegraphics[width=0.22\textwidth, angle=0]{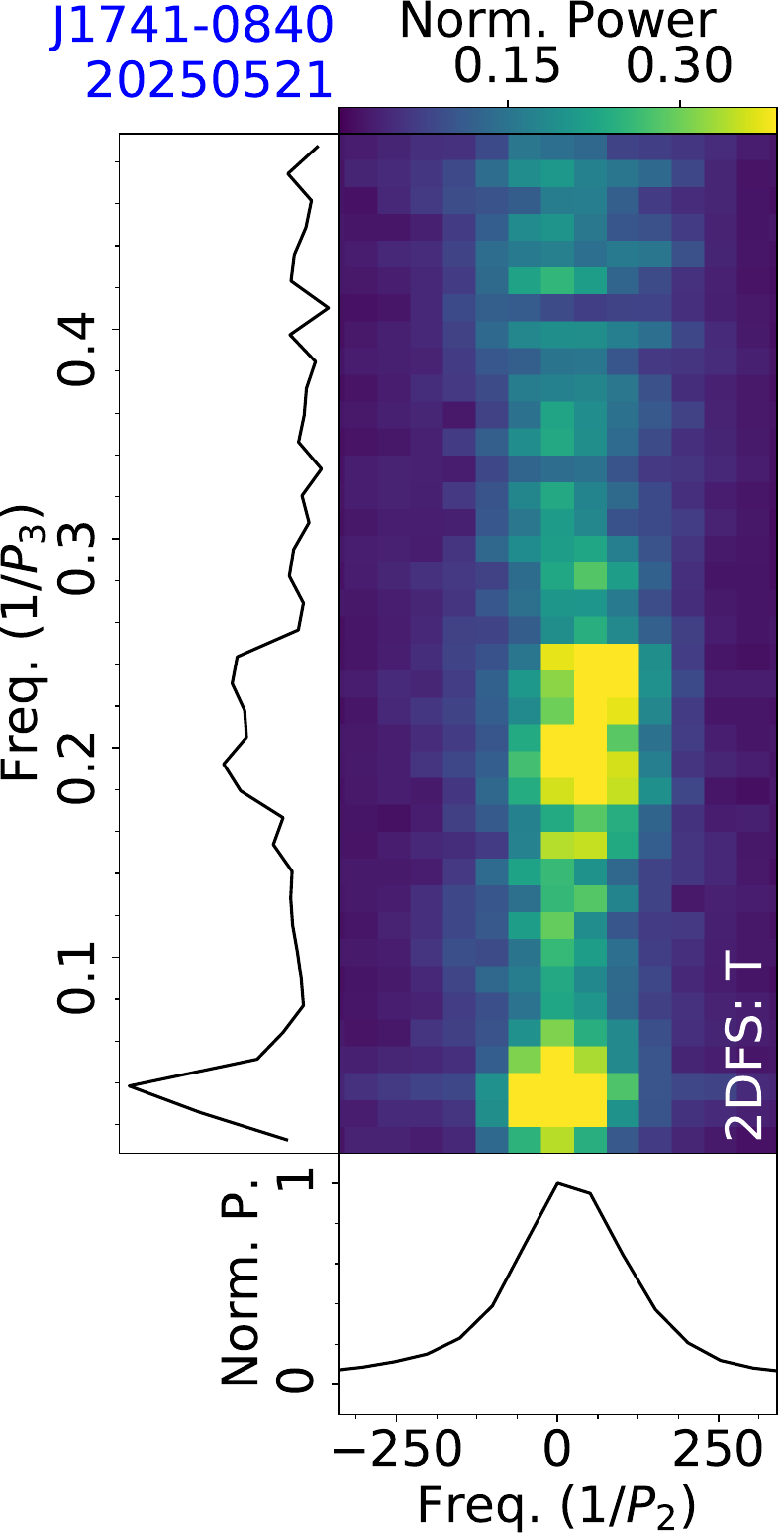}
\figcaption{Fluctuation analysis of PSR J1741-0840 from the observation on 20250521, with LRFS and 2DFS for the on-pulse region of a mean pulse profile.  \label{subfig:fluctu:J1741-0840}}
\end{figure}

\begin{figure}[htpb]
\centering
\includegraphics[width=0.21\textwidth, angle=0]{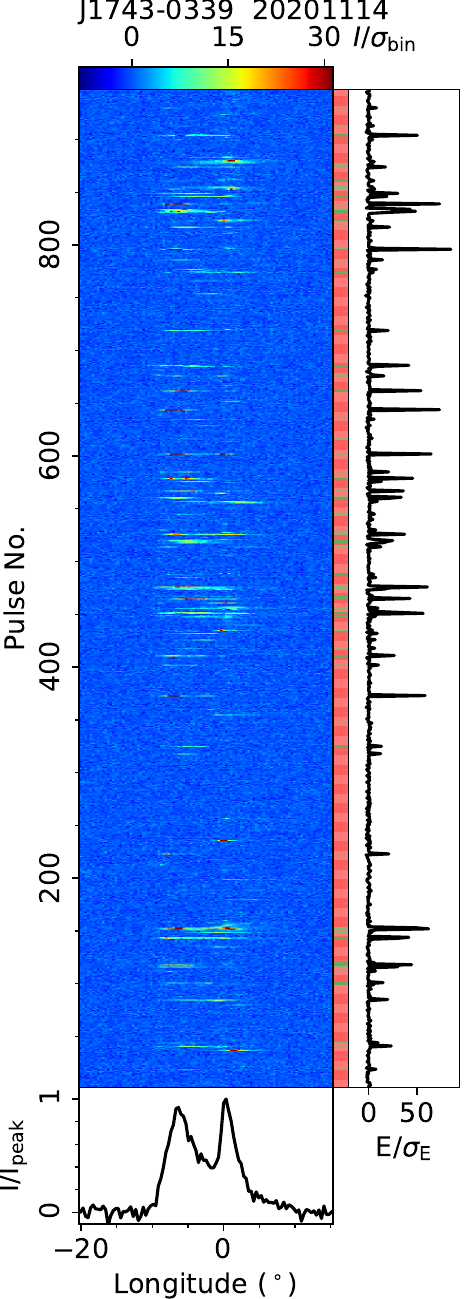}
\includegraphics[width=0.21\textwidth, angle=0]{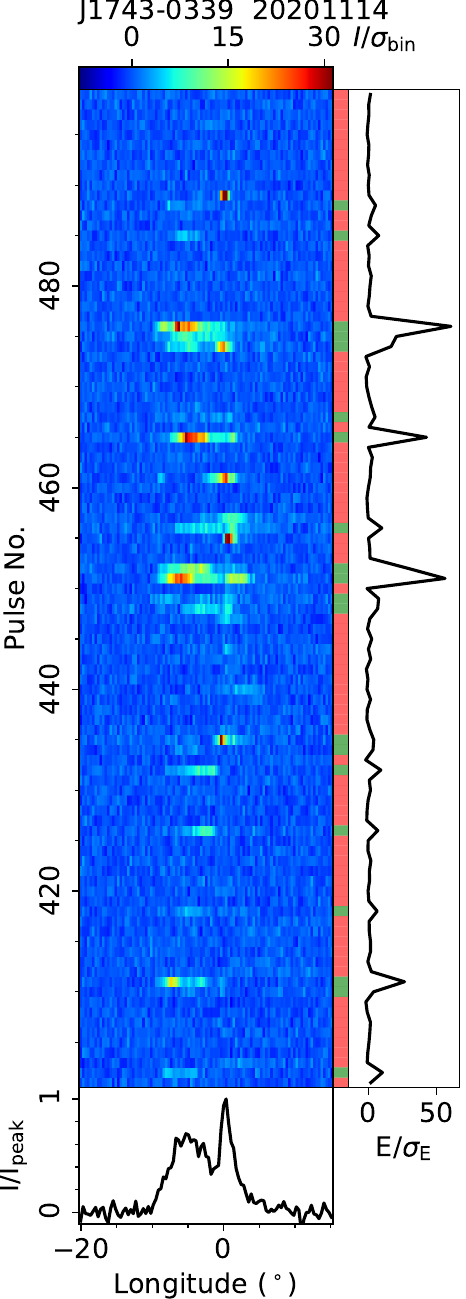}
\figcaption{Single pulse sequences of PSR J1743-0339 from the FAST observation on 20201114, as well as the energy variation of the leading profile part in the right subpanel. 
\label{subfig:TP:J1743-0339}}
\end{figure}

\begin{figure}[htpb]
\centering
\includegraphics[width=0.39\textwidth, angle=0]{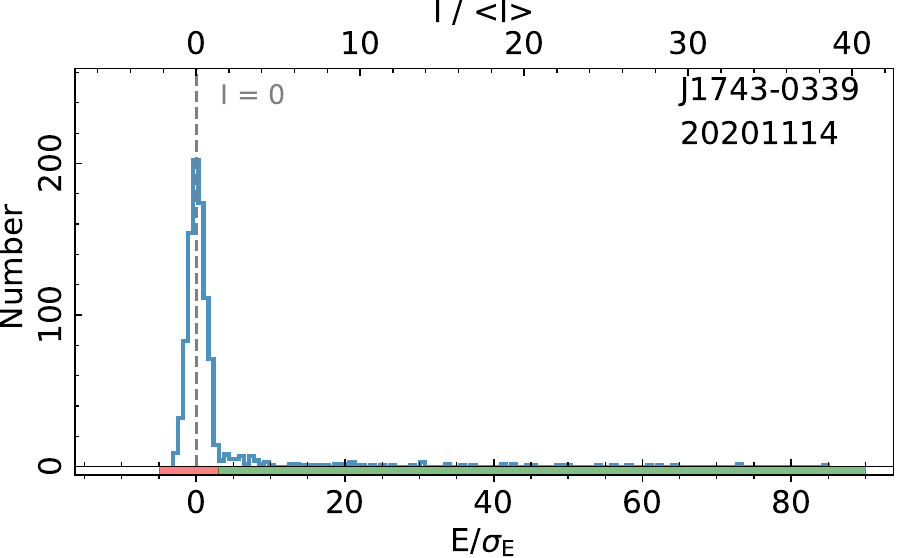}
\figcaption{Energy histogram of the leading part in the mean pulse profile of PSR J1743-0339 from the FAST observation on 20201114.
\label{subfig:Hist:J1743-0339}}
\end{figure}

\begin{figure}[htpb]
\centering
\includegraphics[width=0.39\textwidth, angle=0]{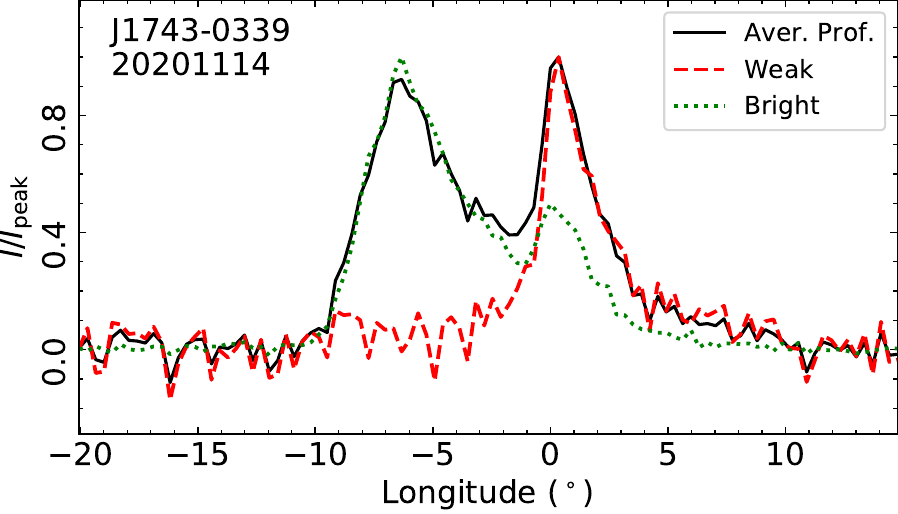}
\figcaption{Mean profiles of two emission modes of PSR J1743-0339 from the observation on 20201114.
\label{subfig:ProfModes:J1743-0339}}
\end{figure}

\begin{figure}[htpb]
\centering
\includegraphics[width=0.22\textwidth, angle=0]{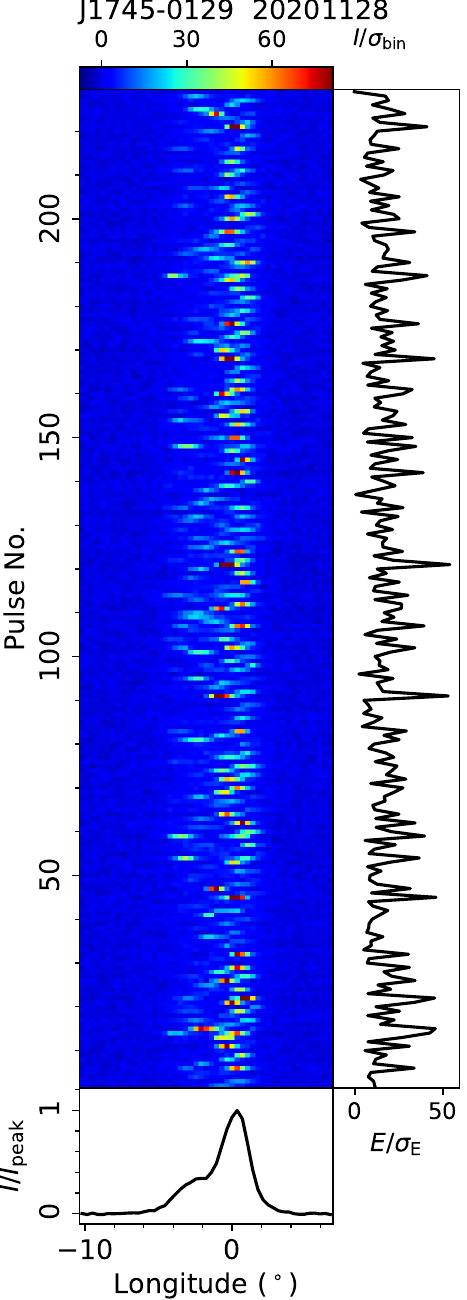}
\includegraphics[width=0.22\textwidth, angle=0]{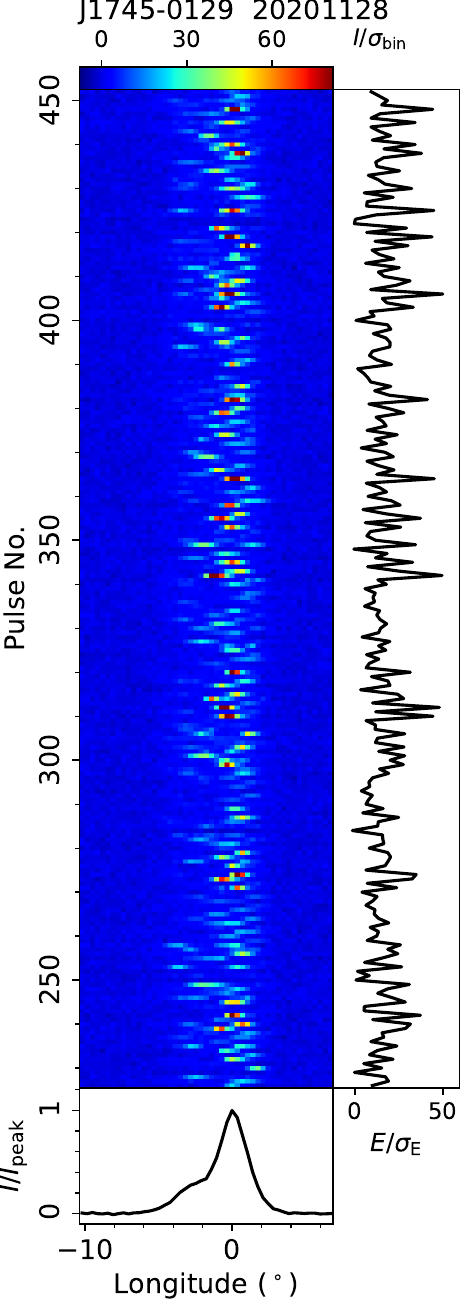}
\figcaption{Single pulse sequences of PSR J1745-0129 from the FAST observation on 20201128.
\label{subfig:TP:J1745-0129}}
\end{figure}

\begin{figure}[htpb]
\centering
\includegraphics[width=0.39\textwidth, angle=0]{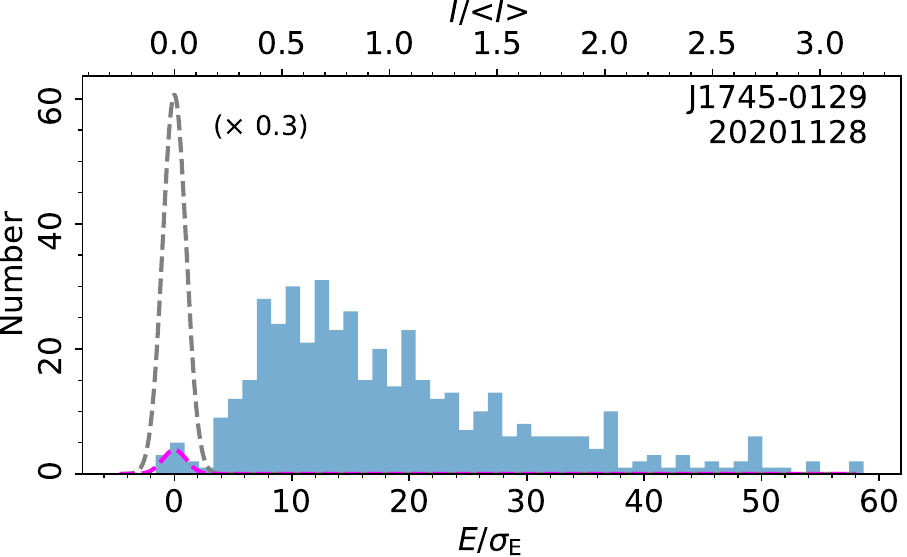}
\figcaption{On-pulse energy histogram of single pulses of PSR J1745-0129 from the FAST observation on 20201128.
\label{subfig:Hist:J1745-0129}}
\end{figure}

\begin{figure}[htpb]
\centering
\includegraphics[width=0.22\textwidth, angle=0]{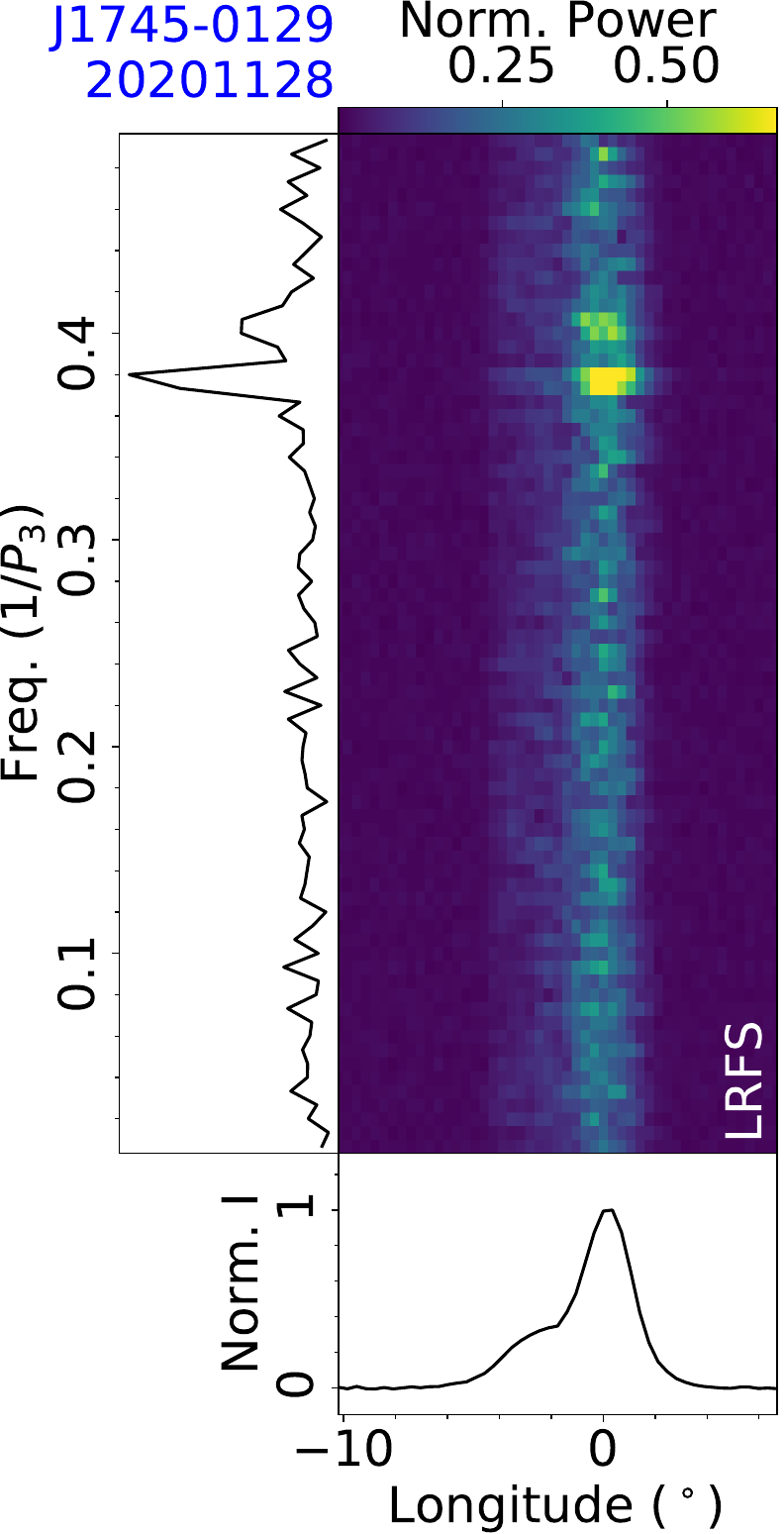}
\includegraphics[width=0.22\textwidth, angle=0]{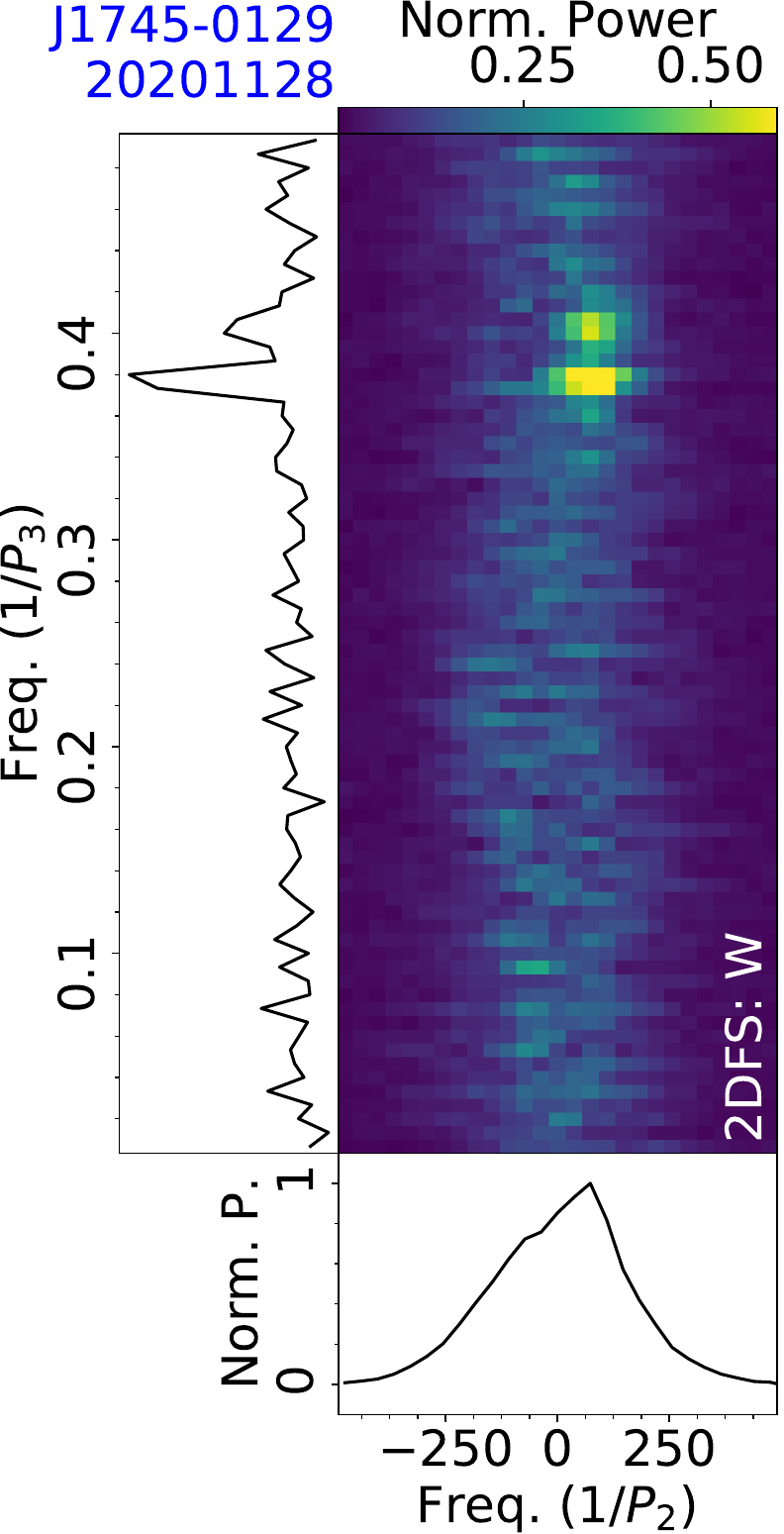}
\figcaption{Fluctuation analysis of PSR J1745-0129 from the FAST observation on 20201128, with LRFS and 2DFS for the on-pulse region of a mean pulse profile.
\label{subfig:fluctu:J1745-0129}}
\end{figure}

\subsection{J1739+0612}
\label{subsec:J1739+0612}

PSR J1739+0612 was discovered by the 64 m Parkes radio telescope \citep{Jacoby2009}. The negatively drifting parameters of this pulsar were reported by \citep{Song2023}. They are $P_3=12\pm4$ periods and $P_2=-51^{+28}_{-65}$ degrees for the leading component, and $P_3=19\pm9$ periods and $P_2=-129^{+120}_{-786}$ degrees for the trailing component. 

This pulsar was observed by FAST on 20201013 for 10 minutes, yielding a rotation period $P=0.2342$~s and a dispersion measure $D\!M=95.5~{\rm cm^{-3}\,pc}$. 
The single pulse sequence of this observation and the zoomed-in view of Nos. 300-500 are shown in Fig.~\ref{subfig:TP:J1739+0612}, where the drifting bands are not systematic. There are several modulation features in fluctuation spectra (Fig.~\ref{subfig:fluctu:J1739+0612}). For both the leading and trailing components, there is a main modulation feature, with the centroid of $1/P_3=0.0746\pm0.0003$ and $0.0897\pm0.0004$, corresponding to $P_3=13.41\pm0.06$ and $11.1\pm0.1$ periods. In the 2DFS of the on-pulse phase region, there are two drift features that are nearly symmetric about vertical axis, and widely distributed with $f_3$ ranges from 0 to 0.2. This is corresponding to the unsystematic drifting property in single pulse sequences, and reflects the varying drifting direction. The phase interval between adjacent subpulses is stable to be $P_2=8.9^\pm0.1\circ$.

\begin{figure}[htpb]
\centering
\includegraphics[width=0.44\textwidth, angle=0]{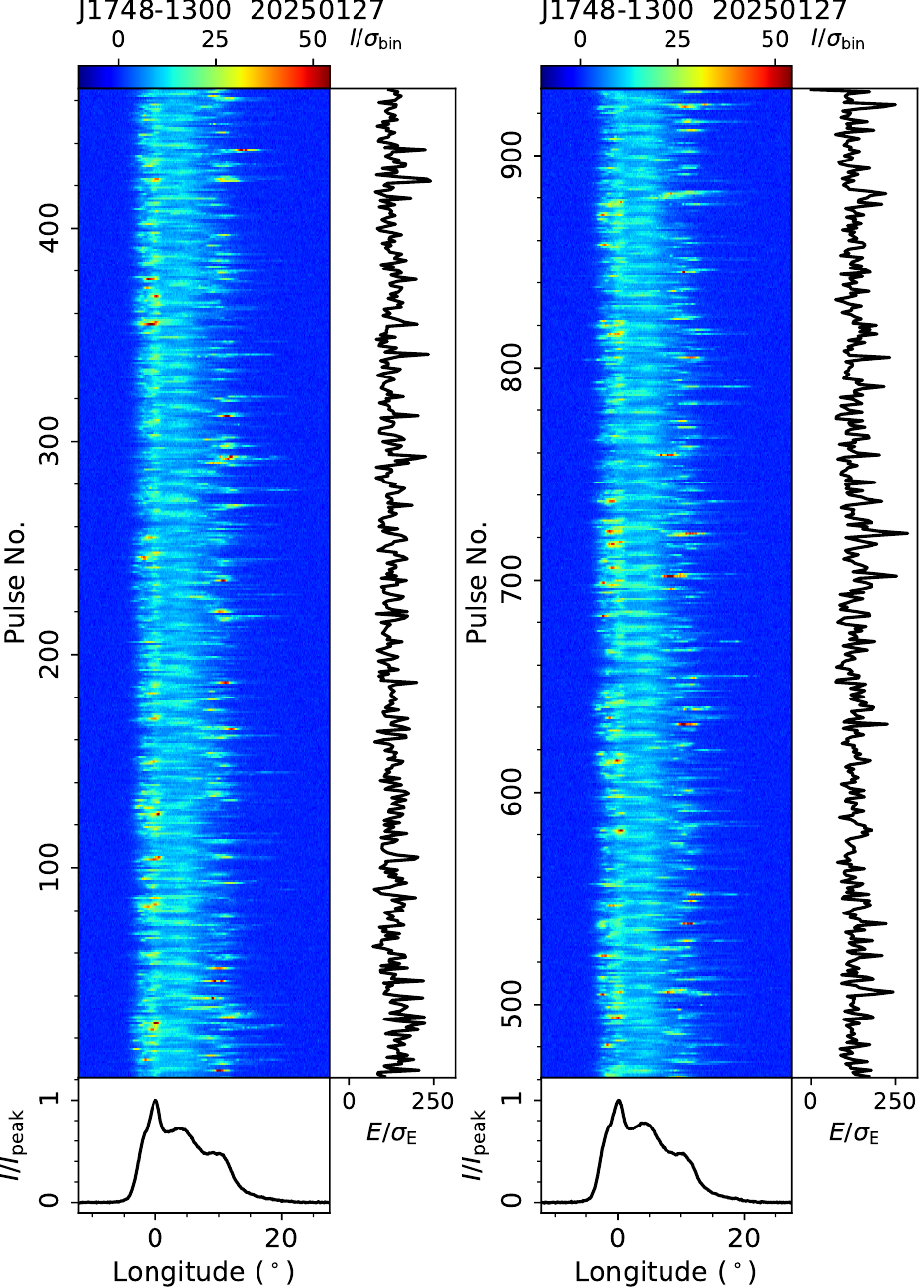}
\figcaption{Single pulse sequences of PSR J1748-1300 from the FAST observation on 20250127.
\label{subfig:TP:J1748-1300}}
\end{figure}

\begin{figure}[htpb]
\centering
\includegraphics[width=0.44\textwidth, angle=0]{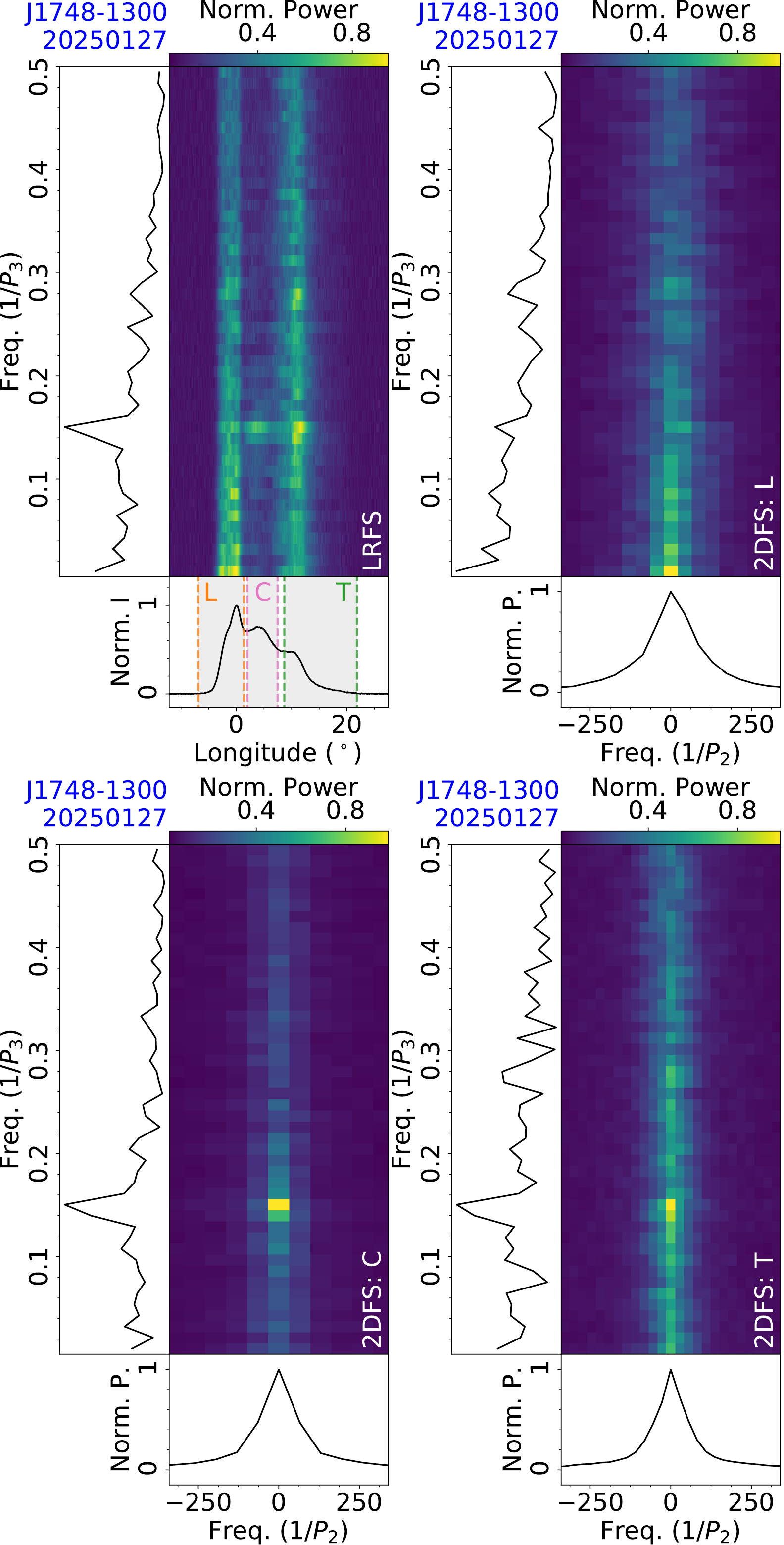}
\figcaption{Fluctuation analysis of PSR J1807+0756 from the FAST observation on 20210117, with LRFS (top-left), and 2DFS for the leading part (top-right), central part (bottom-left) and trailing part (bottom-right) of a mean pulse profile.
\label{subfig:fluctu:J1748-1300}}
\end{figure}

\begin{figure}[htpb]
\centering
\includegraphics[width=0.44\textwidth, angle=0]{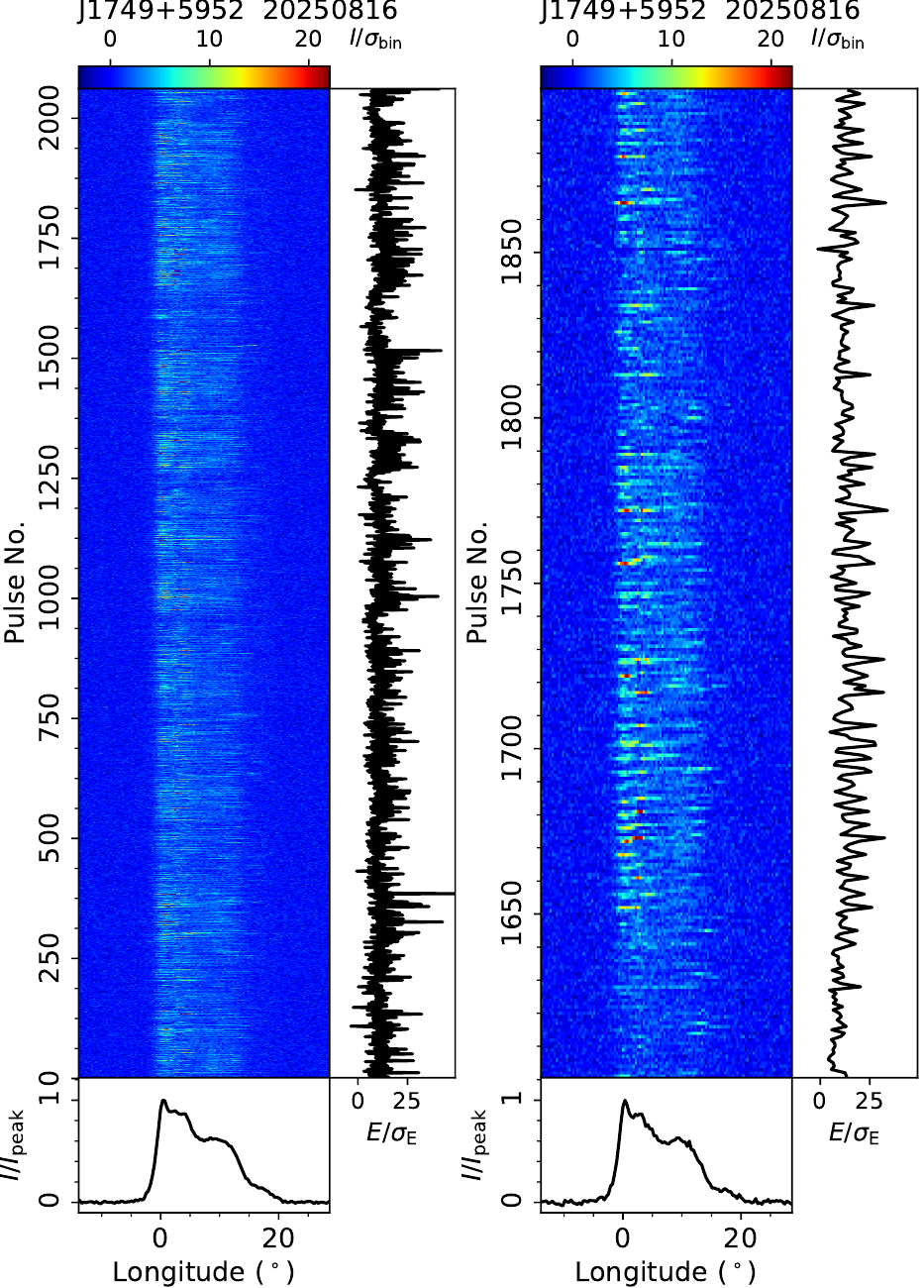}
\figcaption{Single pulse sequence of PSR J1749+5952 from the FAST observation on 20250816, and a zoomed-in view of pulses No. 1600-1900.
\label{subfig:TP:J1749+5952}}
\end{figure}

\begin{figure}[htpb]
\centering
\includegraphics[width=0.44\textwidth, angle=0]{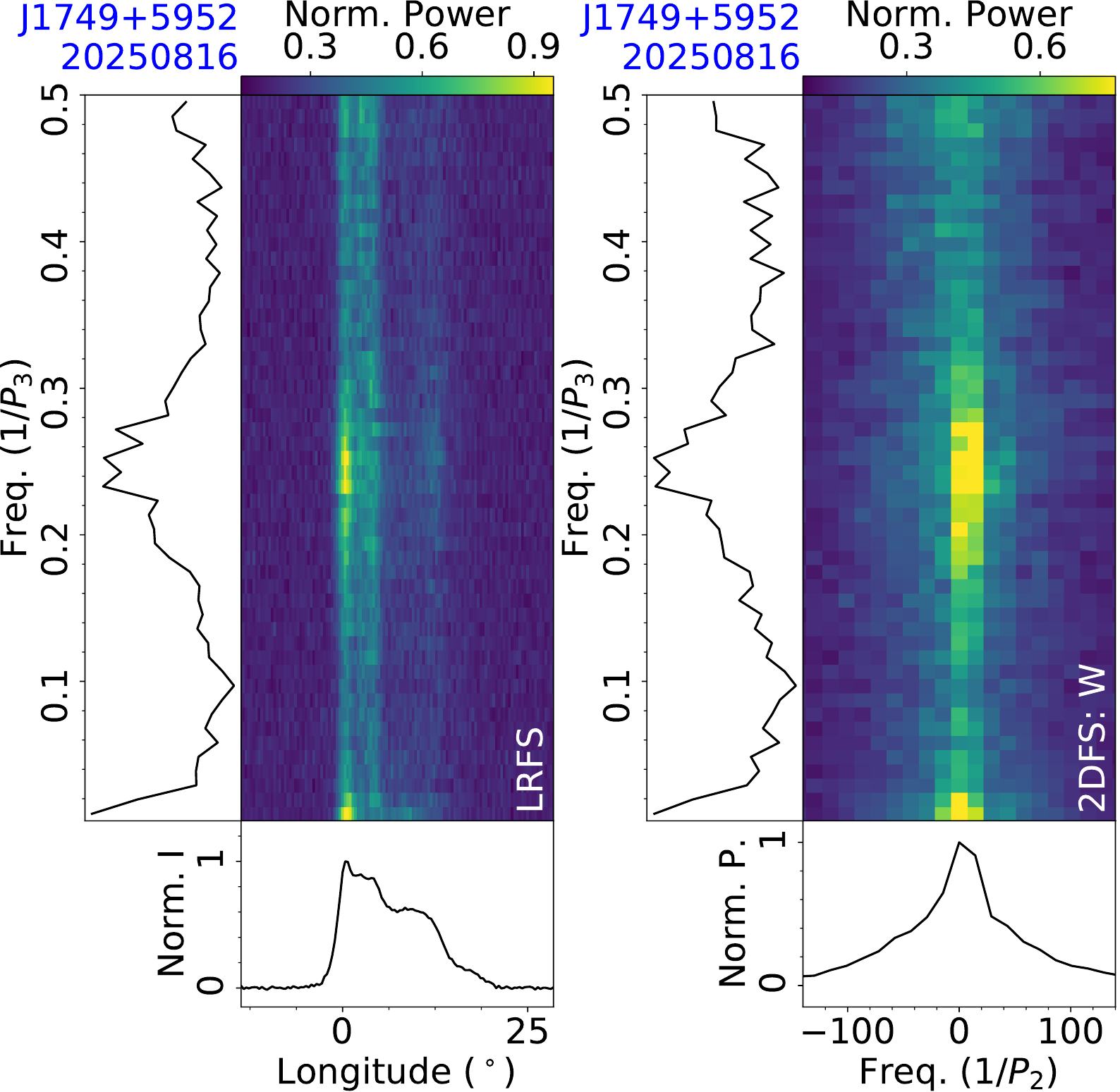}
\figcaption{Fluctuation analysis of PSR J1749+5952 from the FAST observation on 20250816, with LRFS and 2DFS for the on-pulse region of a mean pulse profile.
\label{subfig:fluctu:J1749+5952}}
\end{figure}

\subsection{J1741-0840}
\label{subsec:J1741-0840}

PSR J1741-0840 was discovered in the second Molonglo pulsar survey \citep{Manchester1978}. 
This pulsar was reported to exhibit nulling behavior by \citet{BurkeSpolaor2012} at 1352 MHz. 
The nulling fraction was estimated to be 15.7$\pm$1.7 at 333 MHz and 15.8$\pm$1.4\% at 618 MHz \citep{Basu2017}, 30$\pm$5\% at 625 MHz \citep{Gajjar2017}, and 23$\pm$2\% derived at 1347.3 MHz and 1369 MHz \citep{Xu2024}. 
Subpulse drifting and quasi‑periodicity of the nulling have also been reported \citep{Weltevrede2006,Weltevrede2007,Basu2016,Gajjar2017,Xu2024,Song2023}.

This pulsar was observed by FAST on 20250521 for 16 minutes, with a rotation period $P=2.0430$~s and a dispersion measure $D\!M=74.3~{\rm cm^{-3}\,pc}$ determined. Single pulse sequences in Fig.~\ref{subfig:TP:J1741-0840} illustrating the nulling and subpulse drifting phenomena. The nulling fraction of this observation is estimated to be 22.0$\pm$0.7\% from the on-pulse energy histogram of single pulses (Fig.~\ref{subfig:Hist:J1741-0840}). The subpulse drifting parameters are derived from the fluctuation spectra (Fig.~\ref{subfig:fluctu:J1741-0840}). For the leading part in a mean pulse profile, the centroid frequencies of the drift feature in 2DFS are $1/P_3=0.199\pm0.002$ and $1/P_2=7\pm3$, corresponding to periodicities of $P_3=5.02\pm0.05$ periods and $P_2=52\pm19$ degrees. The centroid of 2DFS for the trailing profile part is characterized by frequencies of $1/P_3=0.204\pm0.001$ and $1/P_2=57\pm4$, yielding $P_3=4.89\pm0.03$ periods and $P_2=6.3\pm0.4$ degrees. In addition, the low-frequency modulation feature in Fig.~\ref{subfig:fluctu:J1741-0840} arises from nulls.


\subsection{J1743-0339}
\label{subsec:J1743-0339}

PSR J1743-0339 was discovered in a survey using observations at the Molonglo Radio Observatory and the Australian National Radio Astronomy Observatory, Parkes \citep{Manchester1978}. 

This pulsar was observed by FAST on 20201114 for 7 minutes, deriving a rotation period $P=0.4447$~s and a dispersion measure $D\!M=30.5~{\rm cm^{-3}\,pc}$ from this observation. Single pulse sequences with the integral energy variations of the leading part in the mean pulse profile are shown in Fig.~\ref{subfig:TP:J1743-0339}. 
The integral energy histogram of the leading profile part is displayed in Fig.~\ref{subfig:Hist:J1743-0339}, and there seem to be nulls corresponding to the distribution around zero in the histogram. From the mean pulse profiles of two emission modes in Figure~\ref{subfig:ProfModes:J1743-0339}, there is only one component in the mean intensity profile of related single pulses.

\begin{figure}[htpb]
\centering
\includegraphics[width=0.22\textwidth, angle=0]{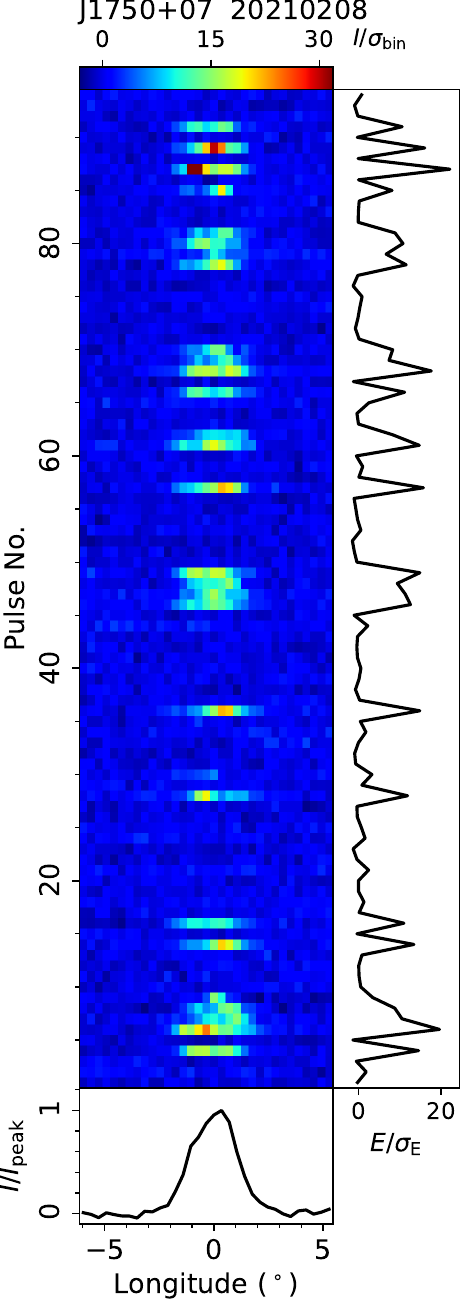}
\figcaption{Single pulse sequence of PSR J1750+07 from the FAST observation on 20210208. \label{subfig:TP:J1750+07}}
\end{figure}

\begin{figure}[htpb]
\centering
\includegraphics[width=0.39\textwidth, angle=0]{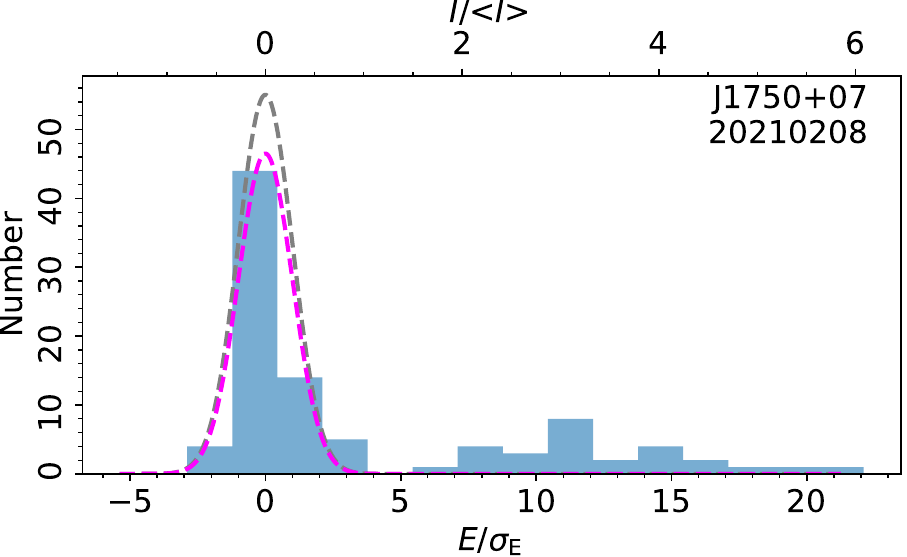}
\figcaption{On-pulse integral histogram of PSR J1750+07 from the FAST observation on 20210208. \label{subfig:Hist:J1750+07}}
\end{figure}

\begin{figure}[htpb]
\centering
\includegraphics[width=0.44\textwidth, angle=0]{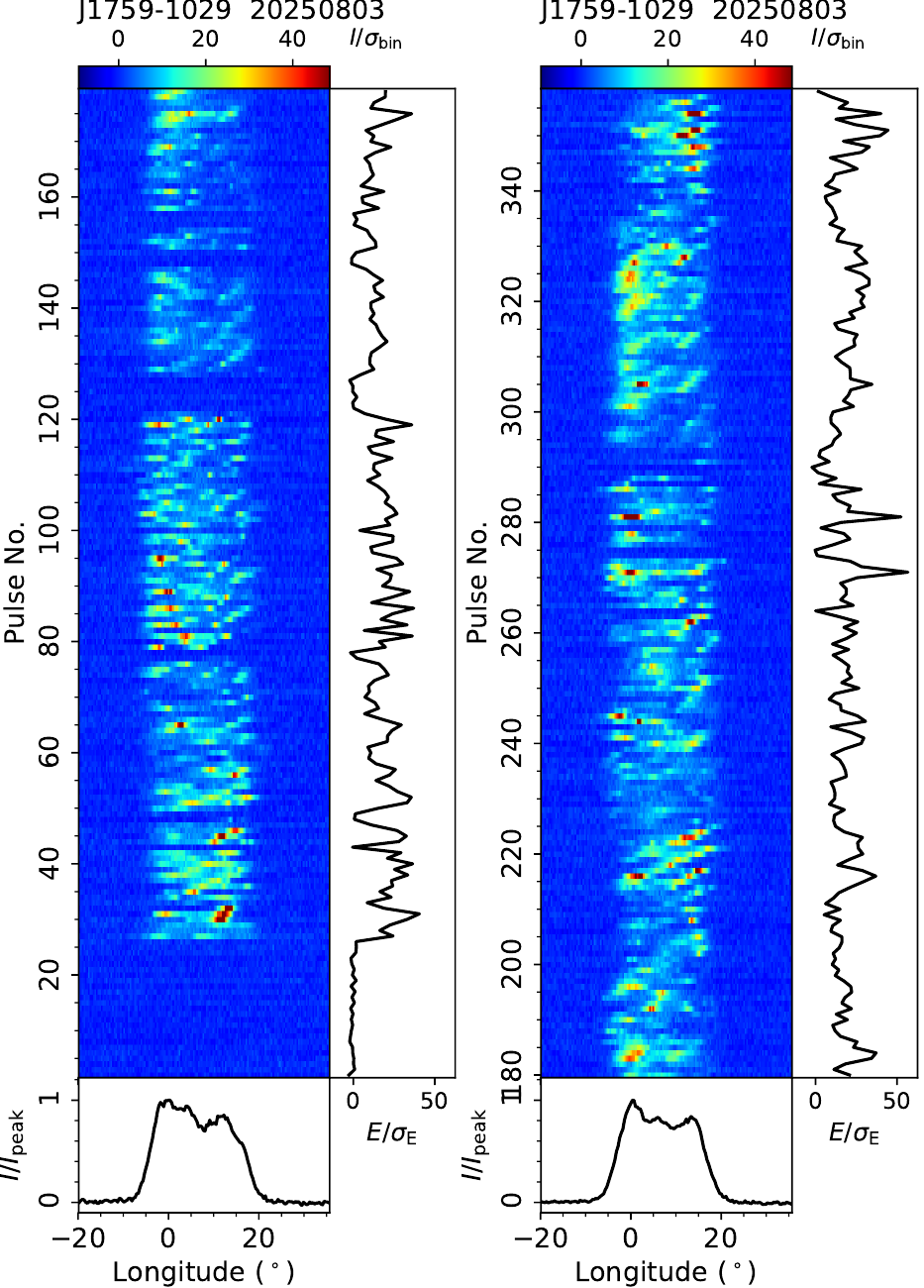}
\figcaption{Single pulse sequences of PSR J1759-1029 from the FAST observation on 20250803. 
\label{subfig:TP:J1759-1029}}
\end{figure}

\begin{figure}[htpb]
\centering
\includegraphics[width=0.39\textwidth, angle=0]{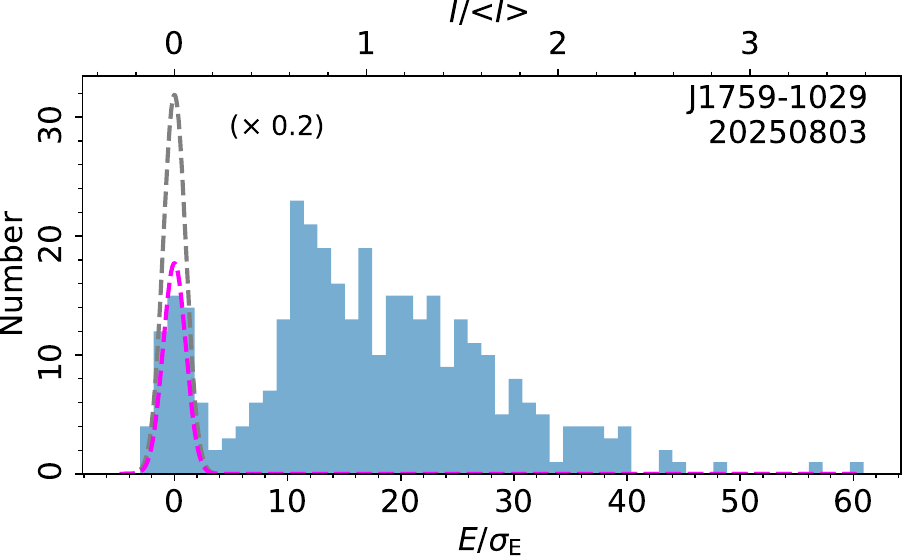}
\figcaption{On-pulse energy histogram of PSR J1759-1029 from the FAST observation on 20250803.
\label{subfig:Hist:J1759-1029}}
\end{figure}

\begin{figure}[htpb]
\centering
\includegraphics[width=0.44\textwidth, angle=0]{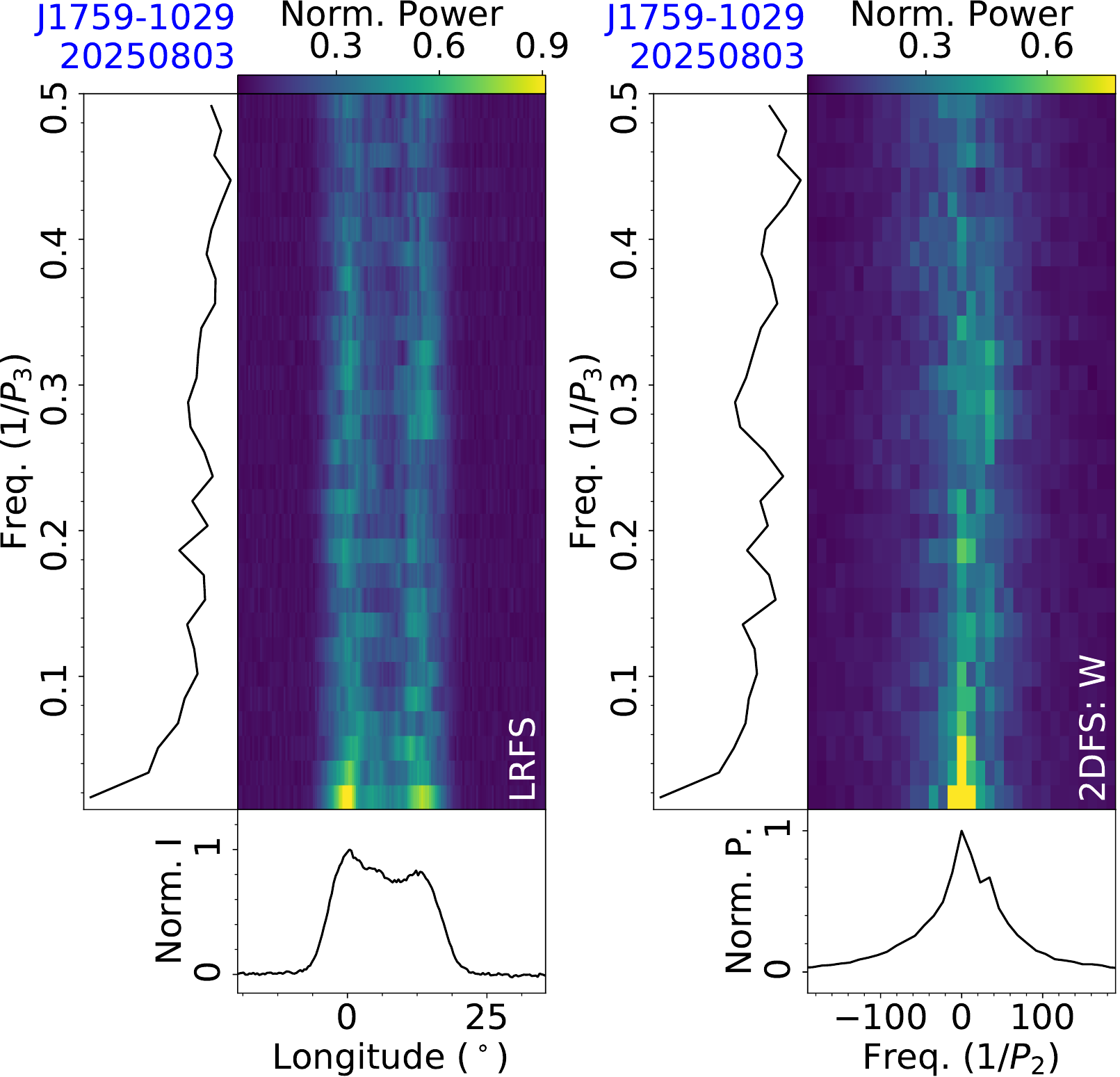} 
\figcaption{Fluctuation analysis of PSR J1759-1029 for the FAST observation on 20250803, with LRFS and 2DFS for the on-pulse region of the mean pulse profile.
\label{subfig:fluctu:J1759-1029}}
\end{figure}

\subsection{J1745-0129}
\label{subsec:J1745-0129}

The pulsar was discovered in a survey of intermediate Galactic latitudes using the Parkes 64-m radio telescope \citep{Edwards2001}. The positive drifting behavior of $P_3=2.61\pm0.04$ and $P_2=5.3^{+0.9}_{-0.5}$ degrees has been reported by \citet{Song2023}. 

This pulsar was observed by FAST on 20201128 for 8 minutes, deriving a rotation period $P=1.0455$~s and a dispersion measure $D\!M=89.4~{\rm cm^{-3}\,pc}$ from this observation. 
Single pulse sequences in Fig.~\ref{subfig:TP:J1745-0129} display the subpulse drifting phenomenon. 
In the on-pulse energy histogram (Fig.~\ref{subfig:Hist:J1745-0129}), there is a distribution around zero energy that is related to nulls. The nulling fraction of this observation is estimated to be 1.9$\pm$0.2\%. 
From fluctuation spectra in Fig.~\ref{subfig:fluctu:J1745-0129}, centroid modulation frequencies of the drift feature are $1/P_3=0.387\pm0.001$ and $1/P_2=77\pm2$, which correspond to drifting parameters of $P_3=2.582\pm0.004$ periods and $P_2=4.7\pm0.8^\circ$. Drifting properties are consistent with the result of \citet{Song2023}.

\begin{figure}[htpb]
\centering
\includegraphics[width=0.22\textwidth, angle=0]{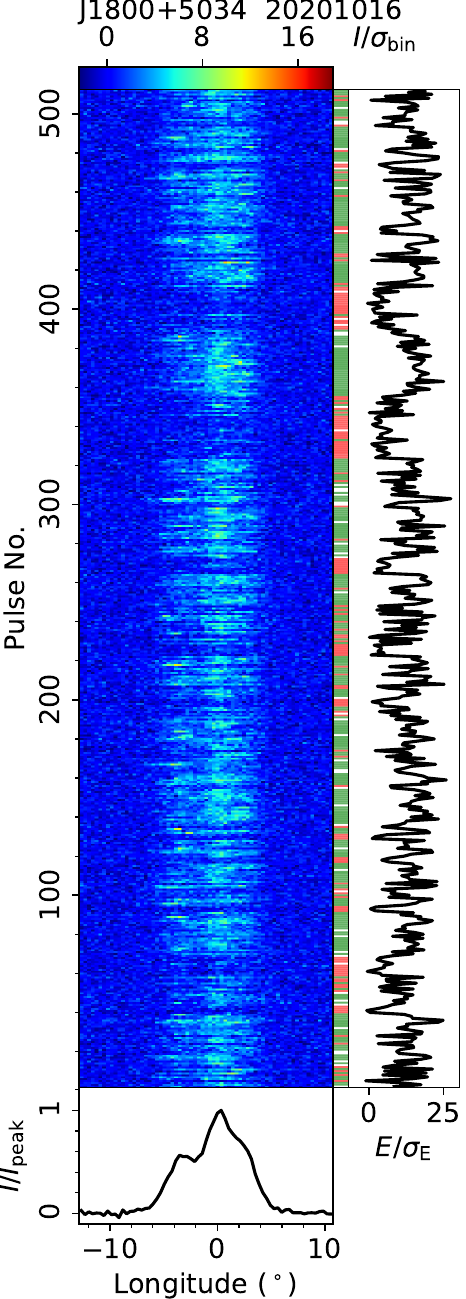}
\includegraphics[width=0.22\textwidth, angle=0]{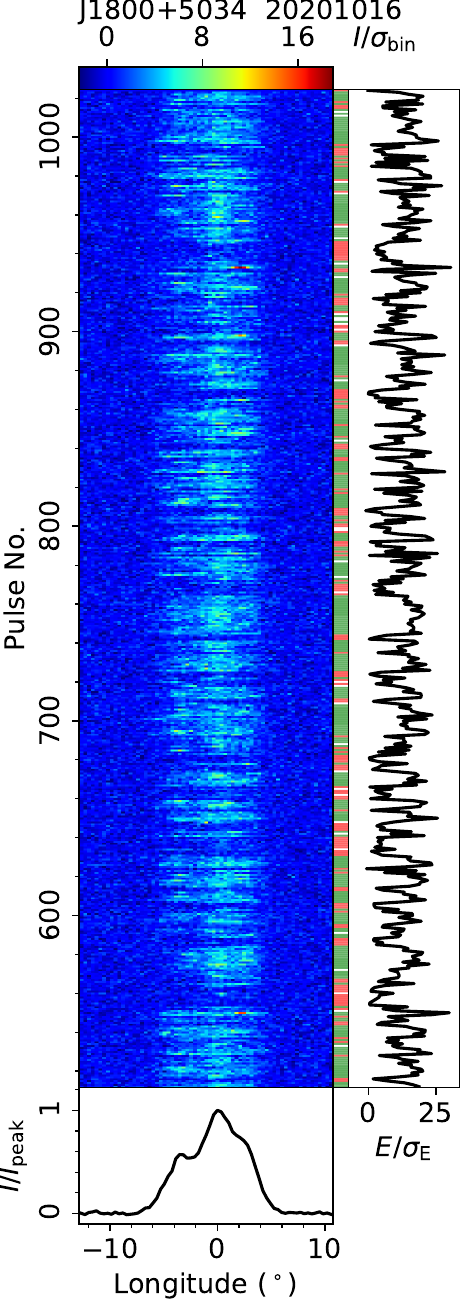}
\figcaption{Single pulse sequences of PSR J1800+5034 from the FAST observation on 20201016. \label{subfig:TP:J1800+5034}}
\end{figure}

\begin{figure}[htpb]
\centering
\includegraphics[width=0.39\textwidth, angle=0]{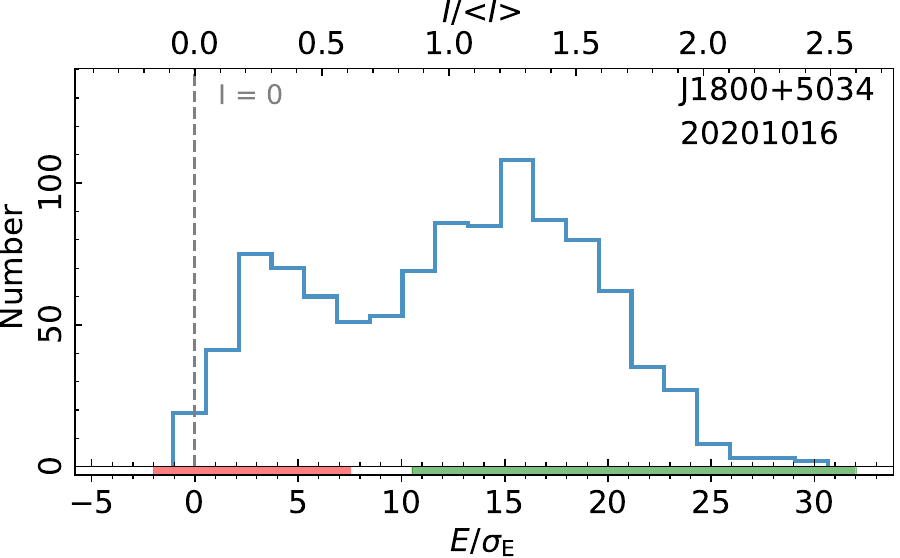}
\figcaption{On-pulse energy histogram of single pulses of PSR J1800+5034 from the FAST observation on 20201016. \label{subfig:Hist:J1800+5034}}
\end{figure}

\begin{figure}[htpb]
\centering
\includegraphics[width=0.39\textwidth, angle=0]{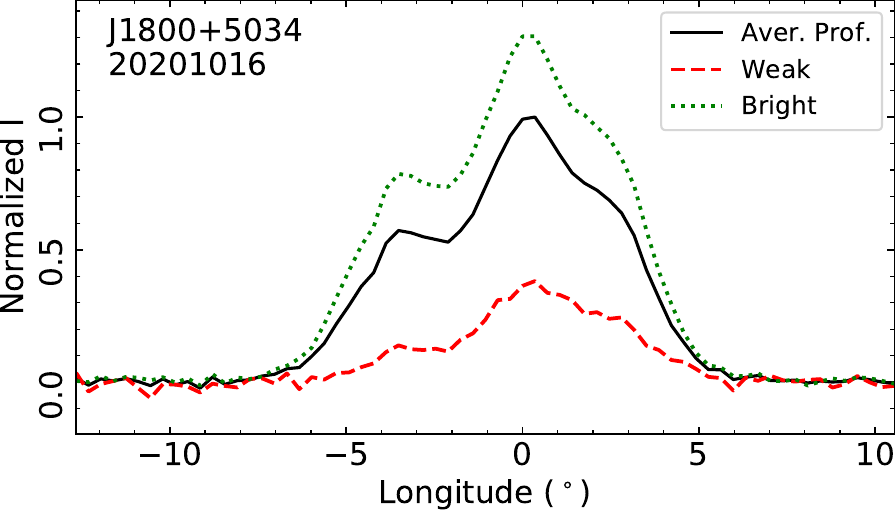}
\figcaption{Mean profiles of weak and bright emission modes of PSR J1800+5034 from the observation on 20201016.
\label{subfig:PolModes:J1800+5034}}
\end{figure}

\subsection{J1748-1300}
\label{subsec:J1748-1300}

PSR J1748-1300 was discovered in the second Molonglo pulsar survey \citep{Manchester1978}. 
\citet{Basu2016} reported the pulsar exhibiting a amplitude modulated drifting with $P_3$ of $6.7\pm0.6$ periods. \citet{Song2023} presented two features for two components, respectively: a drift feature with $P_3=7\pm2$ periods and $P_2=28^{+20}_{-7}$ degrees, and a $P_3$-only feature with $P_3=7.2\pm0.4$ periods. 

This pulsar was observed by FAST on 20250127 for 6 minutes, deriving a rotation period $P=0.3941$~s and a dispersion measure $D\!M=99.6~{\rm cm^{-3}\,pc}$. Single pulse sequences in Fig.~\ref{subfig:TP:J1748-1300} display the subpulse modulation phenomenon. LRFS and 2DFS of the leading, central, and trailing parts in the mean pulse profile are shown in Fig.~\ref{subfig:fluctu:J1748-1300}. From the 2DFS, the leading profile part tends to exhibit a positive drift feature that is widely distributed in $1/P_3$, with centroid frequencies of $1/P_3=0.100\pm0.001$ and $1/P_2=4\pm3$, corresponding to periodicities of $P_3=10.0\pm0.1$ periods and $P_2=88\pm62$ degrees. 
The 2DFS of the central and trailing profile parts show modulation features without a preferred drift direction, and the centroid frequencies are $1/P_3=0.147\pm0.001$ and $0.146\pm0.002$, yielding $P_3=6.8\pm0.1$ and $6.9\pm0.1$ periods, respectively. 
In contrast, the distribution of modulation features along the vertical axis is narrower than that of the leading part, indicating more systematic subpulse modulation.

\subsection{J1749+5952}
\label{subsec:J1749+5952}

PSR J1749+5952 was discovered by \citep{Sanidas2019} in the LOFAR Tied-Array All-Sky Survey (LOTAAS). 

This pulsar was observed by FAST on 20250816 for 15 minutes, deriving a rotation period $P=0.4360$~s and a dispersion measure $D\!M=45.1~{\rm cm^{-3}\,pc}$. The single pulse sequence and a zoomed-in view of pulses No. 1600-1900 in Fig.~\ref{subfig:TP:J1749+5952} show the positive subpulse drifting. Fluctuation spectra are displayed in Fig.~\ref{subfig:fluctu:J1749+5952}. For the positive drift feature, the centroid frequencies are $1/P_3=0.235\pm0.002$ and $1/P_2=8\pm1$, corresponding to $P_3=4.26\pm0.04$ periods and $P_2=43\pm4$ degrees. A modulation feature also appears near the $P_3$= 2 periods alias border, with a centroid at $1/P_3=0.485\pm0.001$, yielding $P_3=2.061\pm0.004$ periods.

\subsection{J1750+07}
\label{subsec:J1750+07}

PSR J1750+07 was discovered from the Arecibo 327 MHz Drift Pulsar Survey (AO327), and was reported to have the nulling phenomenon \citep{Deneva2016}. 

This pulsar was observed by FAST on 20210208 for 9 minutes, yielding a rotation period $P=5.7266$~s and a dispersion measure $D\!M=56.6~{\rm cm^{-3}\,pc}$ from this observation. Single pulse sequences in Fig.~\ref{subfig:TP:J1750+07} and on-pulse integral energy histogram in Fig.~\ref{subfig:Hist:J1750+07} show the clear nulling phenomenon, and the nulling fraction is estimated to be 85$\pm$17\% from this observation.

\begin{figure}[htpb]
\centering
\includegraphics[width=0.22\textwidth, angle=0]{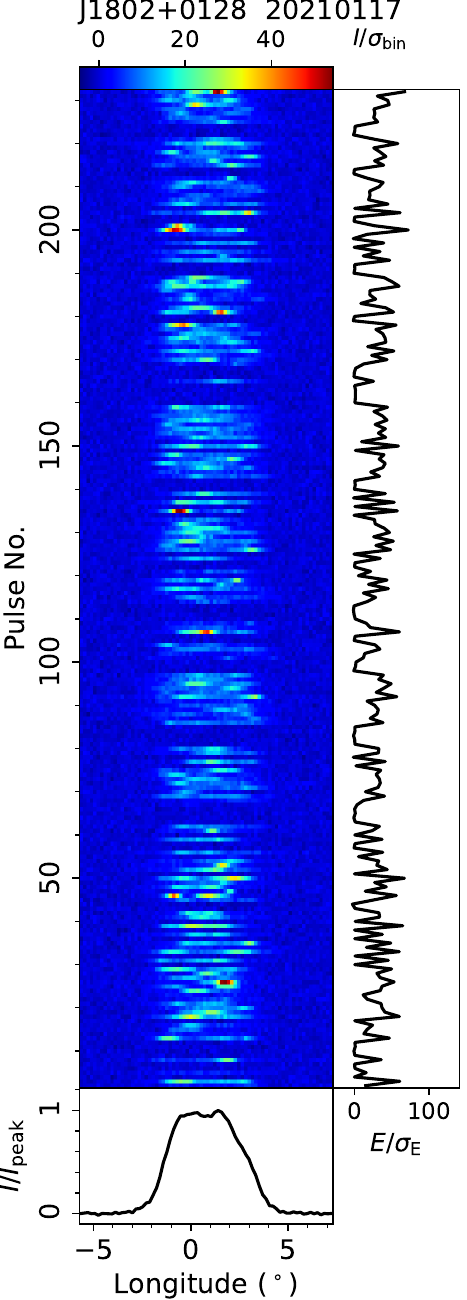}
\includegraphics[width=0.22\textwidth, angle=0]{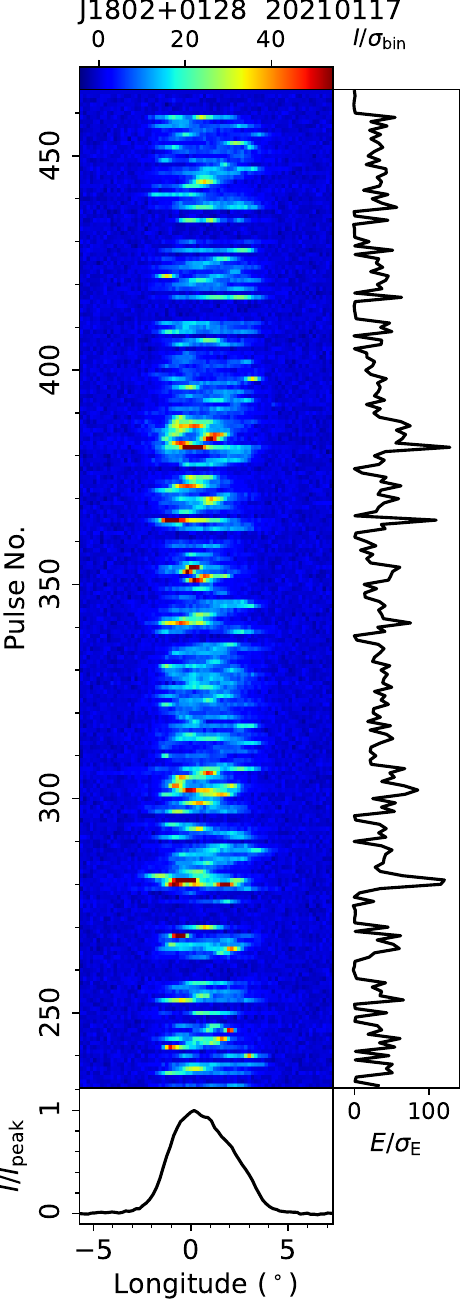}
\figcaption{Single pulse sequences of PSR J1802+0128 from the FAST observation on 20210117. \label{subfig:TP:J1802+0128}}
\end{figure}

\begin{figure}[htpb]
\centering
\includegraphics[width=0.39\textwidth, angle=0]{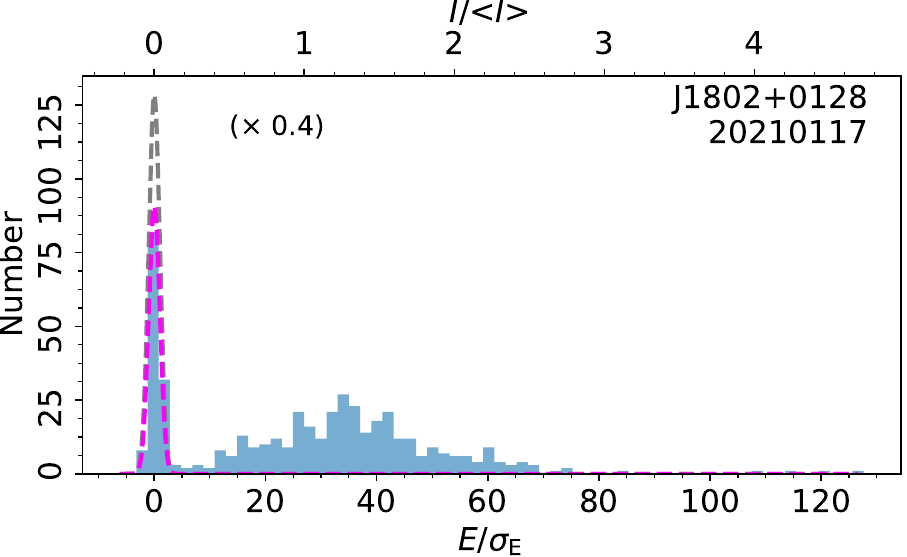}
\figcaption{On-pulse energy histogram of PSR J1802+0128 from the FAST observation on 20210117. \label{subfig:Hist:J1802+0128}}
\end{figure}

\begin{figure}[htpb]
\centering
\includegraphics[width=0.22\textwidth, angle=0]{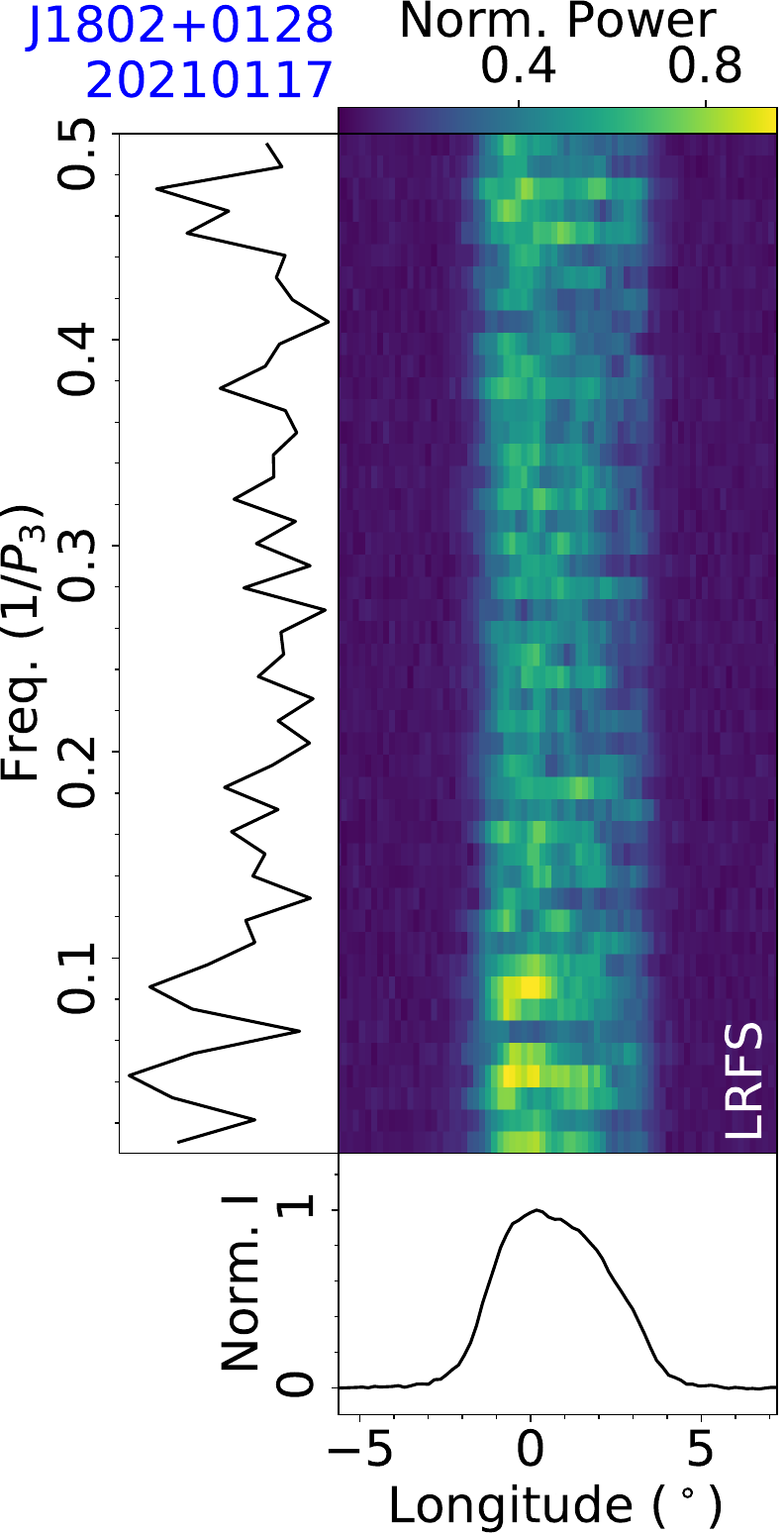} 
\includegraphics[width=0.22\textwidth, angle=0]{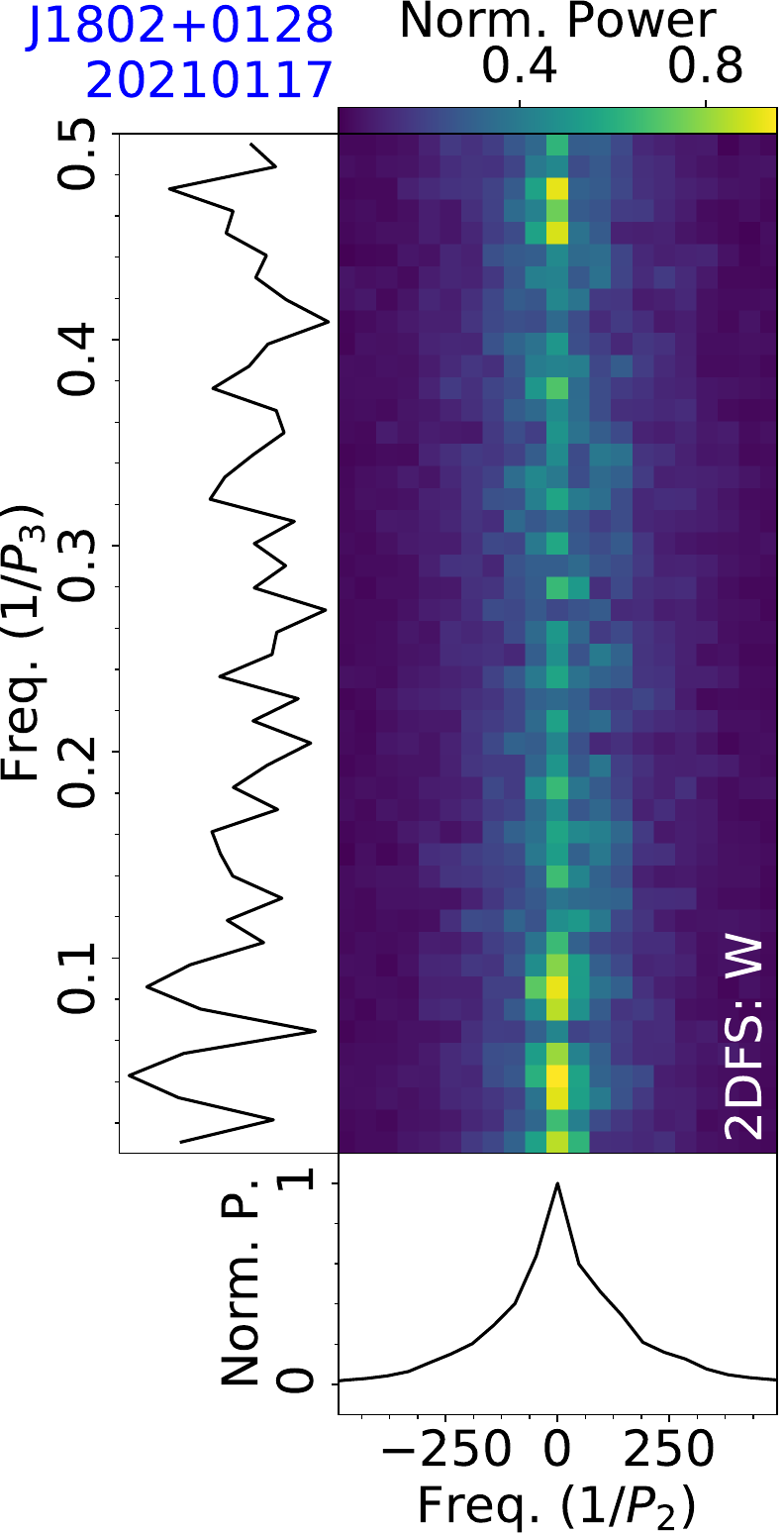}
\figcaption{Fluctuation analysis of PSR J1802+0128 for the FAST observation on 20210117, with LRFS and 2DFS for the on-pulse region of a mean pulse profile.  \label{subfig:fluctu:J1802+0128}}
\end{figure}

\subsection{J1759-1029}
\label{subsec:J1759-1029}

PSR J1759-1029 was discovered in the mid-latitude portion of the High Time Resolution
Universe survey with the Parkes 64-m radio telescope \citep{Bates2012}.

This pulsar was observed by FAST on 20250803 for 15 minutes, with a rotation period $P=2.5124$~s and a dispersion measure $D\!M=115.9~{\rm cm^{-3}\,pc}$ derived. Single pulse sequences in Fig.~\ref{subfig:TP:J1759-1029} illustrate the existence of nulling and subpulse drifting. The nulling fraction of this observation is estimated to be 11$\pm$1\% from the on-pulse energy histogram (Fig.~\ref{subfig:Hist:J1759-1029}). The fluctuation spectra are displayed in Fig.~\ref{subfig:fluctu:J1759-1029}, and the centroid of the positive drift feature in 2DFS is at $1/P_3=0.289\pm0.003$ and $1/P_2=35\pm2$, corresponding to periodicities of $P_3=3.46\pm0.03$ periods and $P_2=10.4\pm0.5$ degrees.

\subsection{J1800+5034}
\label{subsec:J1800+5034}

PSR J1800+5034 was discovered by \citet{Stovall2014} using the Green Bank Telescope at 350 MHz. 

The pulsar was observed by FAST on 20201016 for 10 minutes, deriving a rotation period $P=0.5784$~s and a dispersion measure $D\!M=22.7~{\rm cm^{-3}\,pc}$. 
Single pulse sequences of this FAST observation are shown in Fig.~\ref{subfig:TP:J1800+5034}. 
On-pulse integral energy histogram of the pulsar in Fig.~\ref{subfig:TP:J1800+5034} appears to be bimodal corresponding to two emission modes, which is used to distinguish the weak and bright emission modes. The contrast of mean profiles are displayed in Fig.~\ref{subfig:PolModes:J1800+5034}.

\subsection{J1802+0128}
\label{subsec:J1802+0128}

PSR J1802+0128 was discovered by the Parkes 64-m radio telescope \citep{Edwards2001}. 

This pulsar was observed by FAST on 20210117 for 9 minutes, deriving a rotation period $P=1.1085$~s and a dispersion measure $D\!M=98.5~{\rm cm^{-3}\,pc}$ from this observation. Single pulse sequences in Fig.~\ref{subfig:TP:J1802+0128} illustrate the existence of nulls. The nulling fraction of this observation is estimated to be 28$\pm$8\% from the on-pulse integral energy histogram in Fig.~\ref{subfig:Hist:J1802+0128}. 
In LRFS and 2DFS (Fig.~\ref{subfig:fluctu:J1802+0128}), there are two weak temporally modulated features with centroids of $1/P_3=0.054\pm0.002$ and $0.462\pm0.002$, corresponding to $P_3=18\pm1$ and $2.16\pm0.01$ periods. 
The low-frequency modulation is probably caused by the nulling behavior.

\begin{figure}[htpb]
\centering
\includegraphics[width=0.22\textwidth, angle=0]{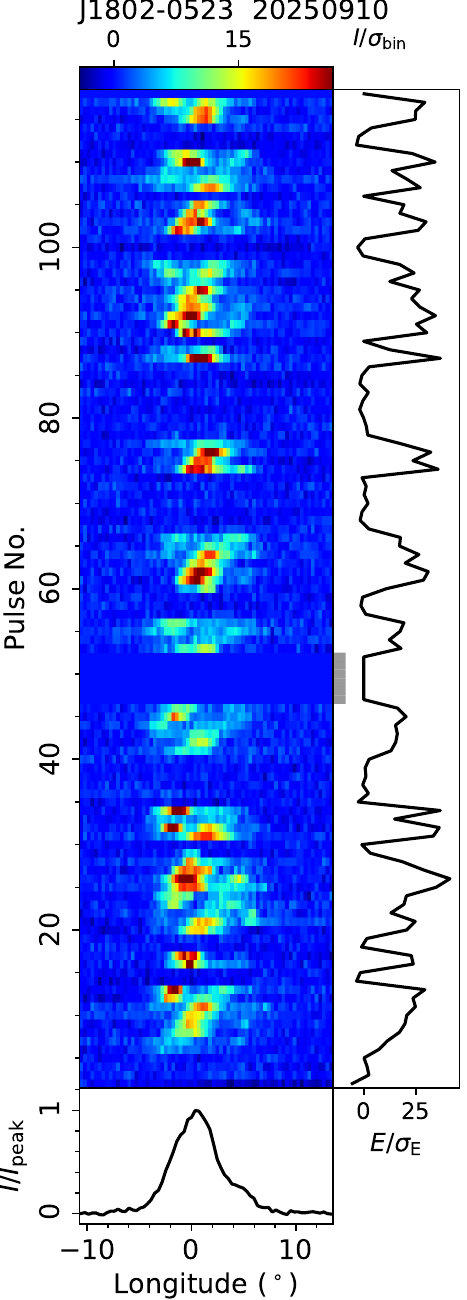}
\figcaption{Single pulse sequence of PSR J1802-0523 from the FAST observation on 20250910. \label{subfig:TP:J1802-0523}}
\end{figure}

\begin{figure}[htpb]
\centering
\includegraphics[width=0.39\textwidth, angle=0]{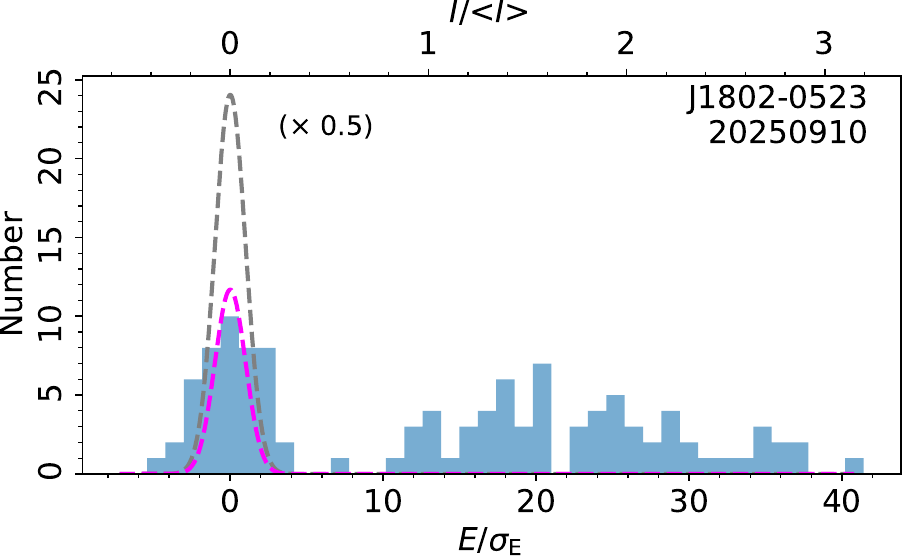}
\figcaption{On-pulse energy histogram of PSR J1802-0523 from the FAST observation on 20250910. \label{subfig:Hist:J1802-0523}}
\end{figure}

\begin{figure}[htpb]
\centering
\includegraphics[width=0.22\textwidth, angle=0]{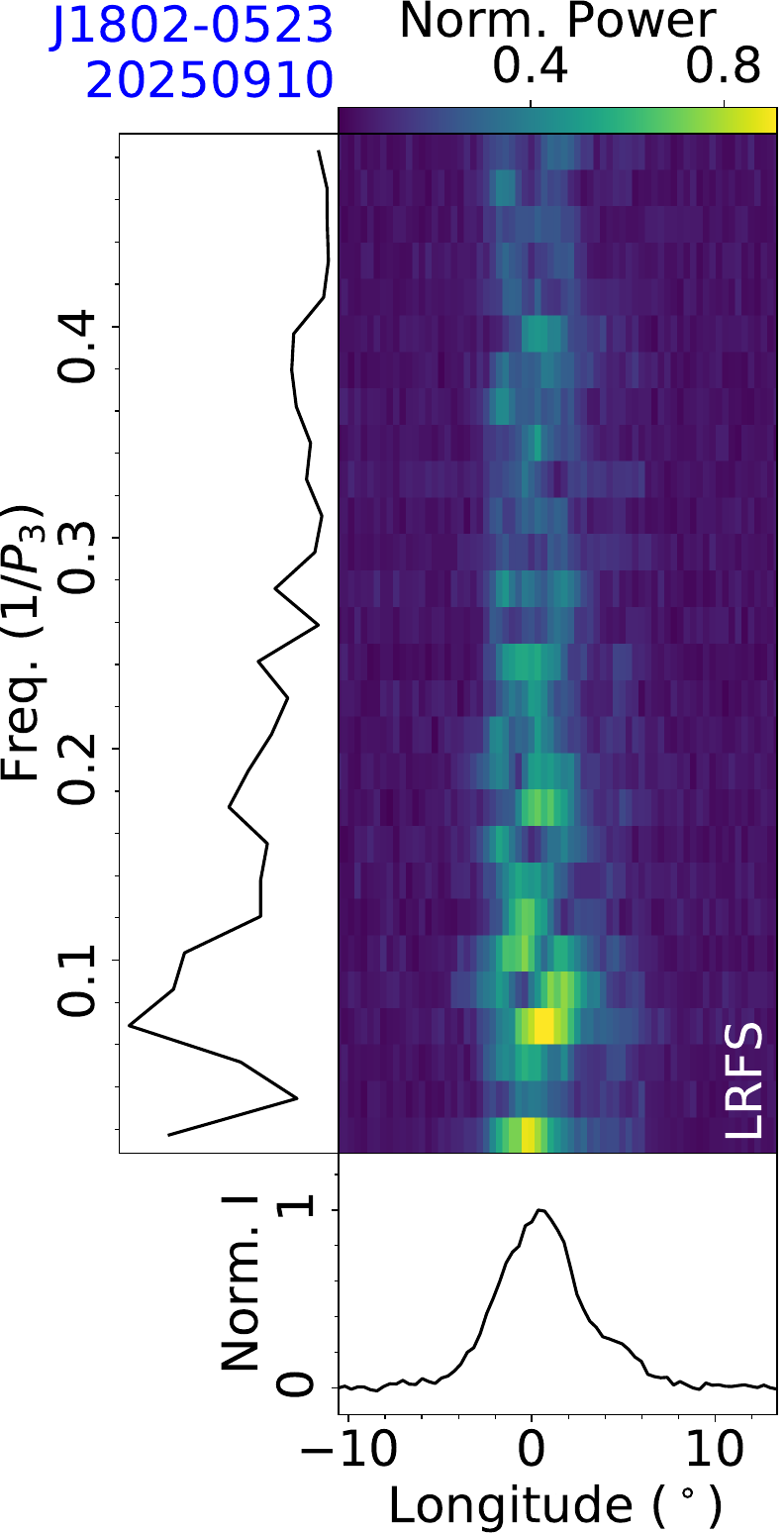} 
\includegraphics[width=0.22\textwidth, angle=0]{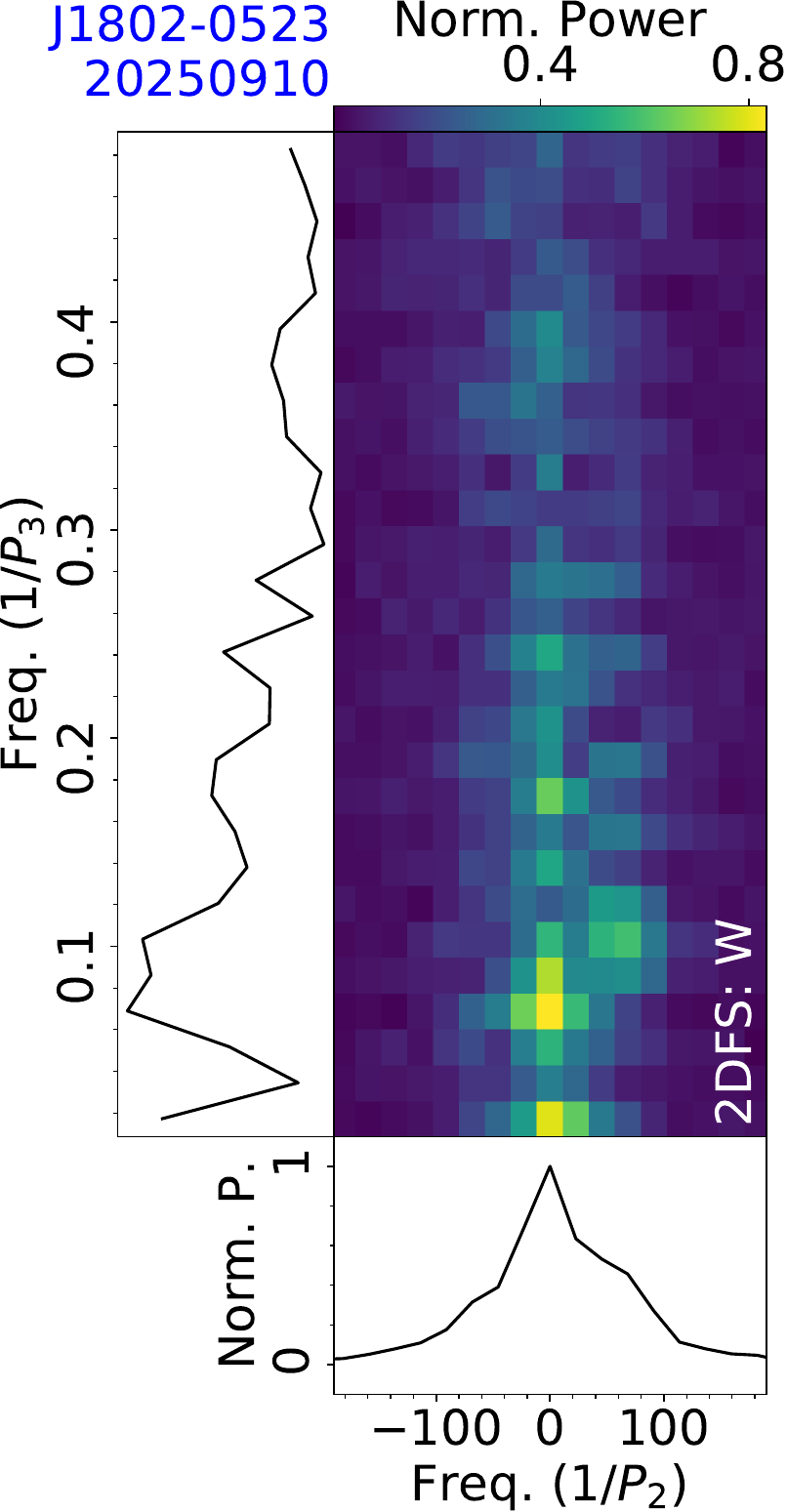}
\figcaption{Fluctuation analysis of PSR J1802-0523 for the FAST observation on 20250910, with LRFS and 2DFS for the on-pulse region of a mean pulse profile.  \label{subfig:fluctu:J1802-0523}}
\end{figure}

\begin{figure}[htpb]
\centering
\includegraphics[width=0.44\textwidth, angle=0]{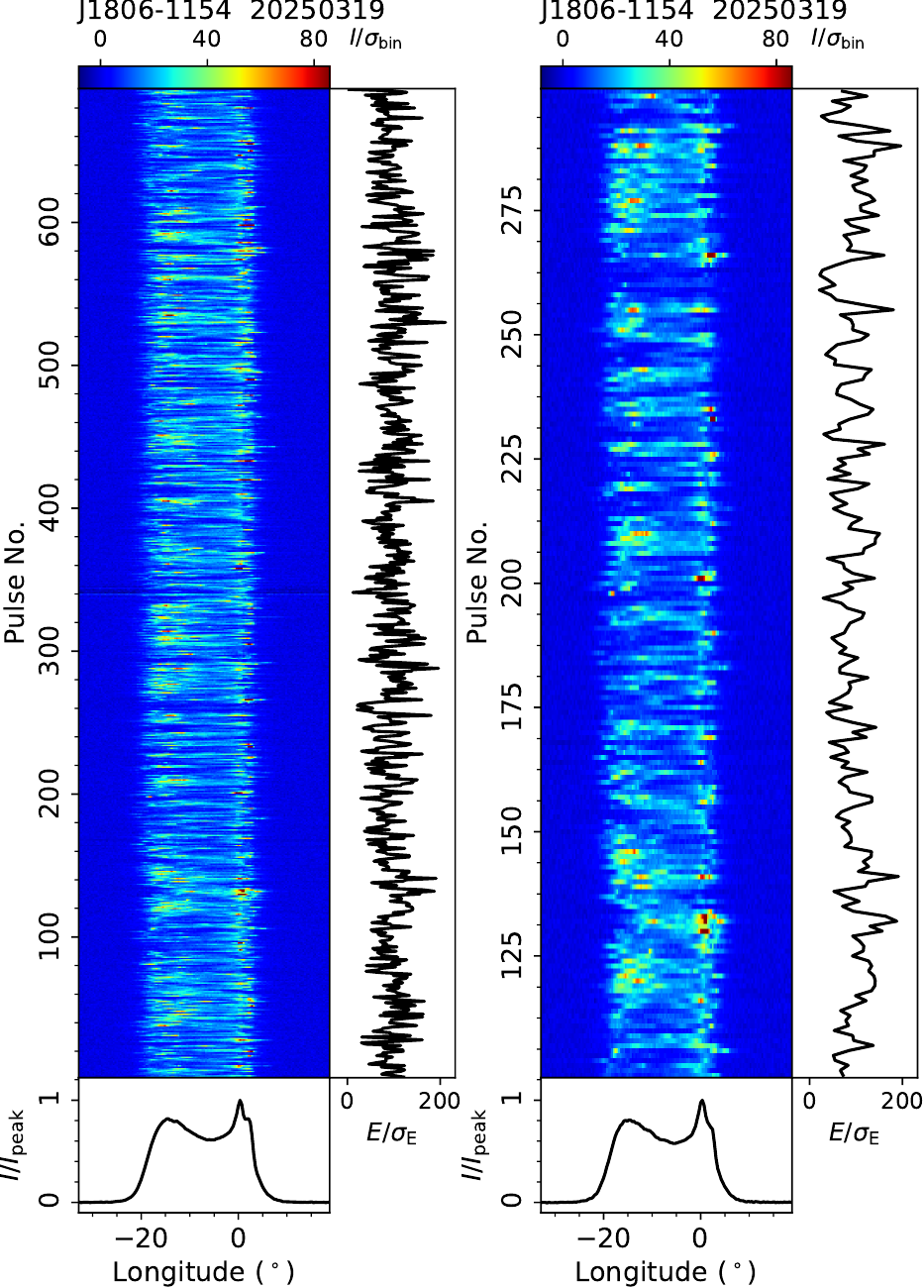}
\figcaption{Single pulse sequence of PSR J1806-1154 from the FAST observation on 20250319, and a zoomed-in view of pulses No. 100-300.
\label{subfig:TP:J1806-1154}}
\end{figure}

\begin{figure}[htpb]
\centering
\includegraphics[width=0.44\textwidth, angle=0]{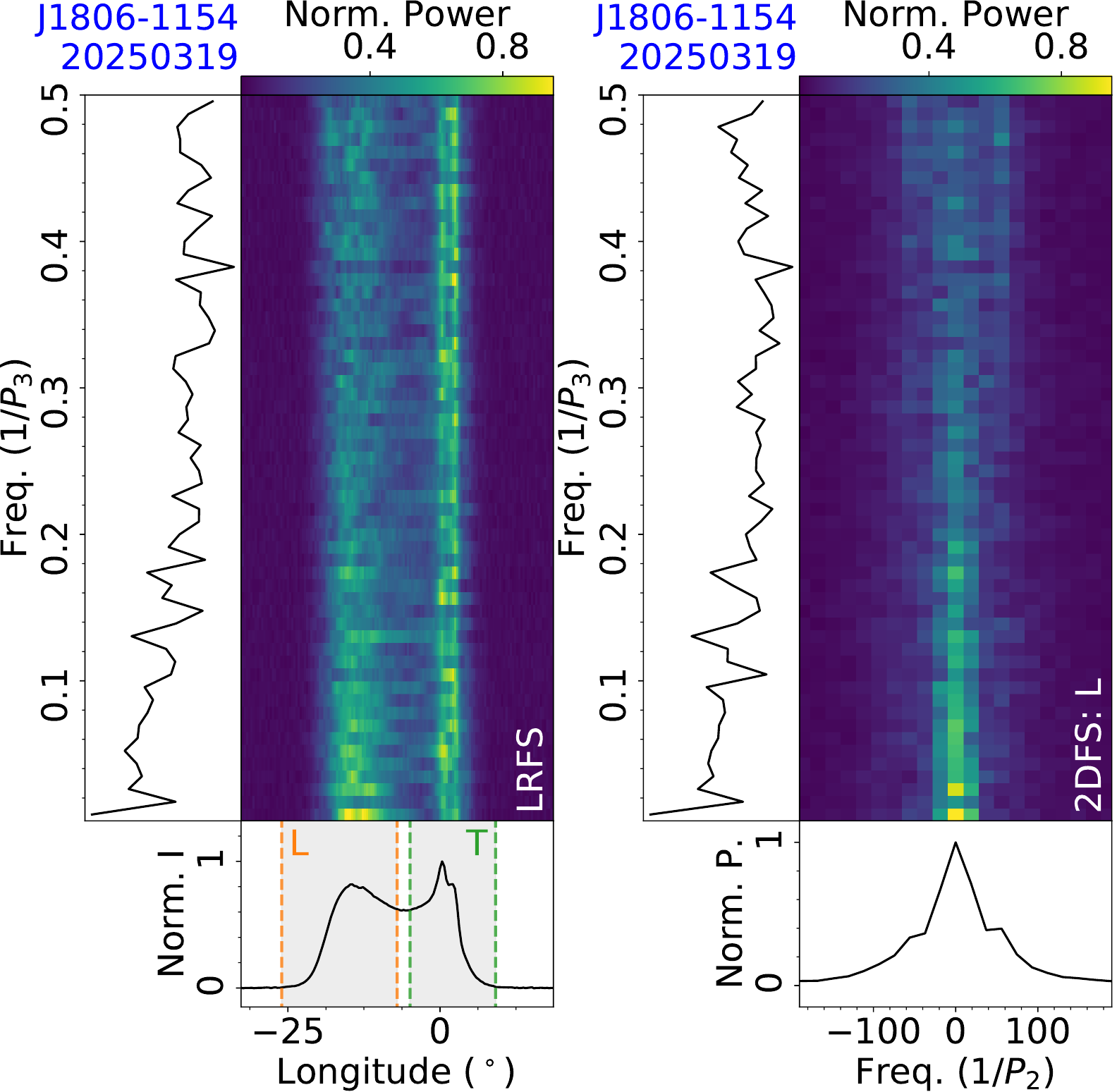}
\figcaption{Fluctuation analysis of PSR J1806-1154 from the FAST observation on 20250319, with LRFS (top-left), and 2DFS for the on-pulse region and leading part of the mean pulse profile.
\label{subfig:fluctu:J1806-1154}}
\end{figure}

\begin{figure}[htpb]
\centering
\includegraphics[width=0.22\textwidth, angle=0]{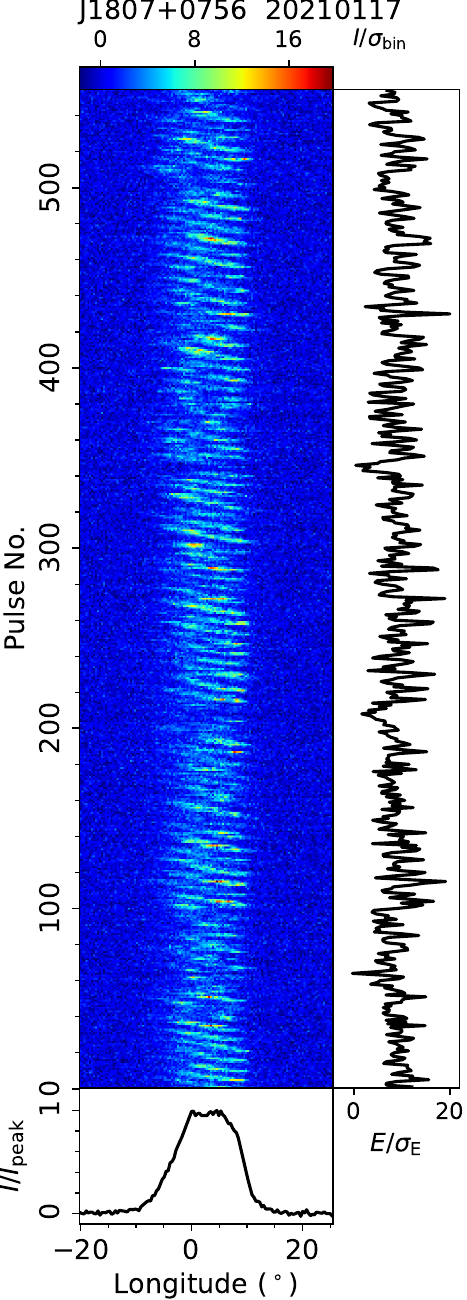} 
\includegraphics[width=0.22\textwidth, angle=0]{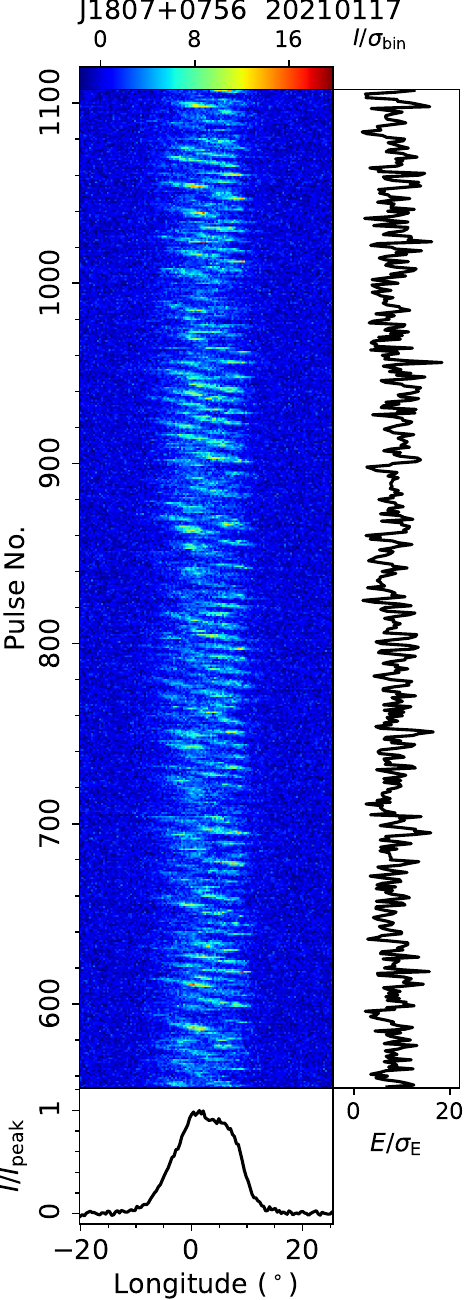}
\figcaption{Single pulse sequences of PSR J1807+0756 from the FAST observation on 20210117. \label{subfig:TP:J1807+0756}}
\end{figure}

\begin{figure}[htpb]
\centering
\includegraphics[width=0.22\textwidth, angle=0]{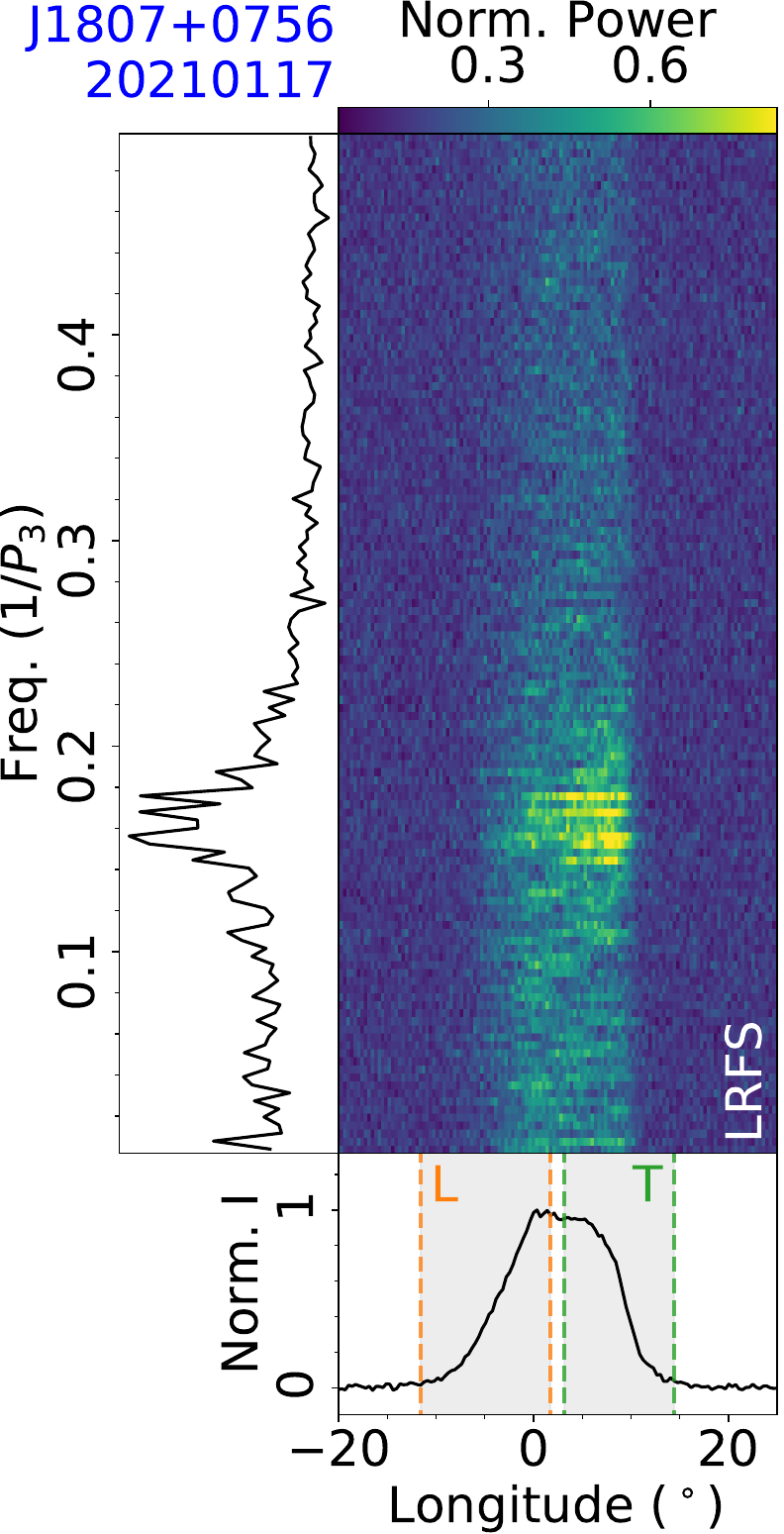}
\includegraphics[width=0.22\textwidth, angle=0]{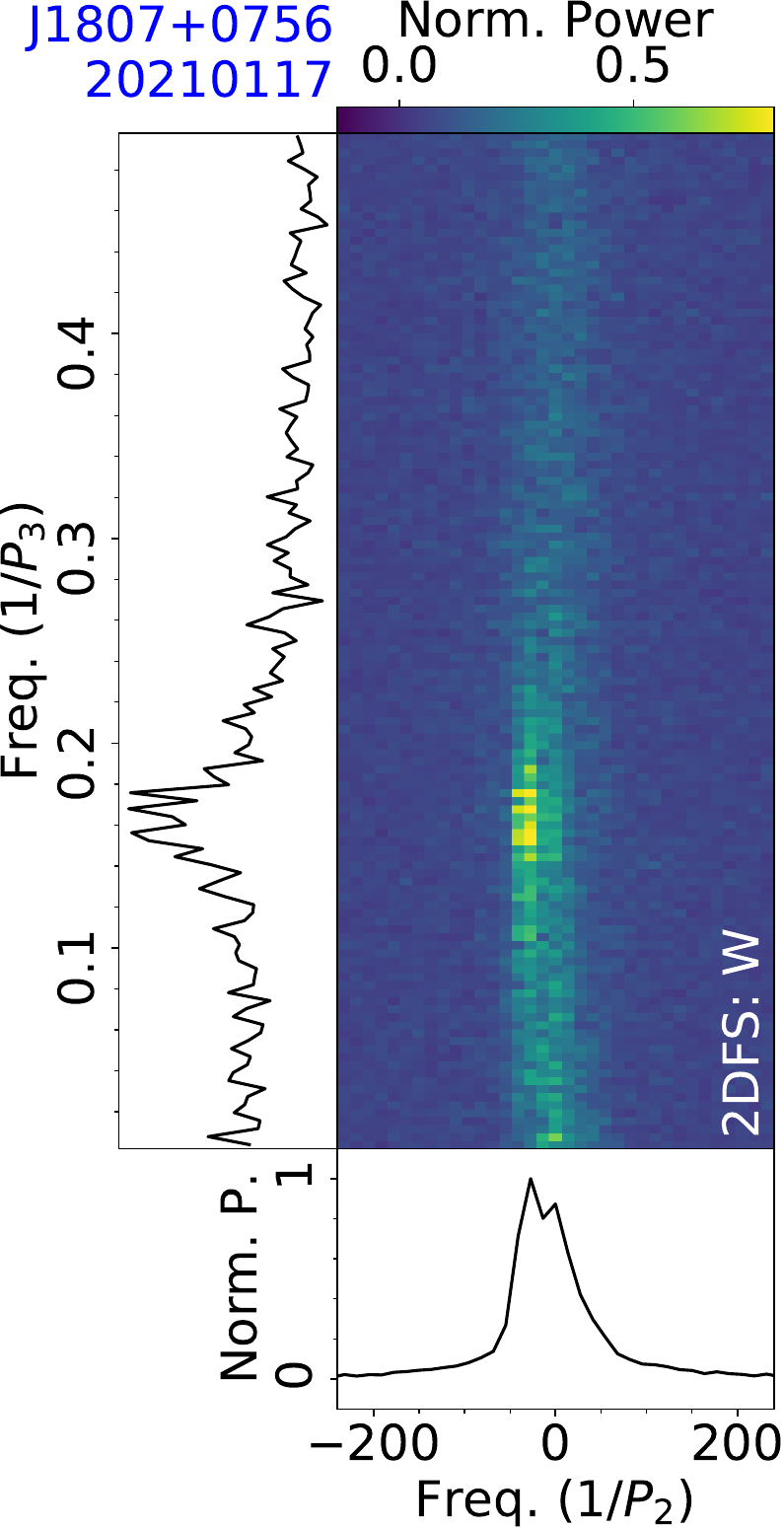}\\
\includegraphics[width=0.22\textwidth, angle=0]{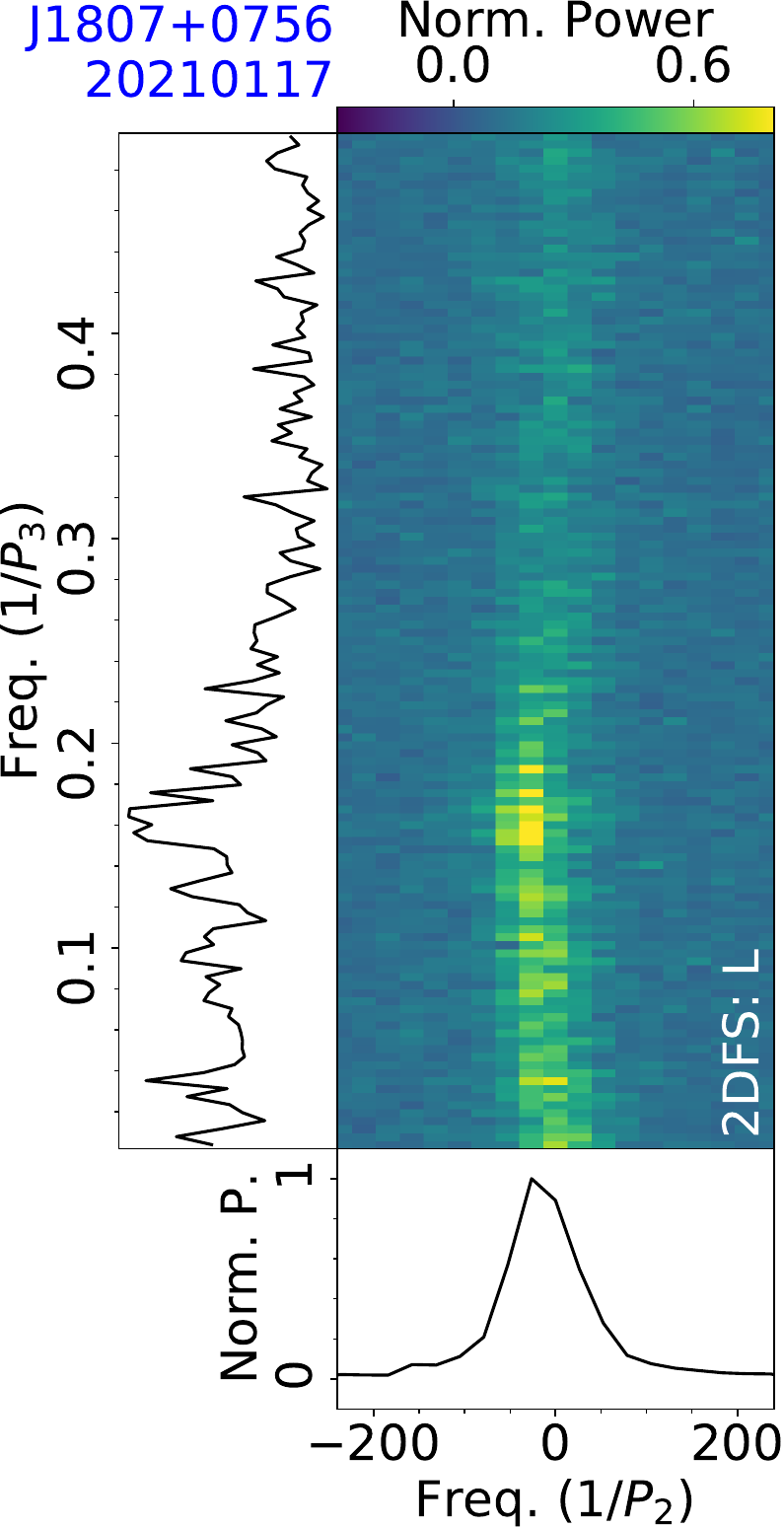}
\includegraphics[width=0.22\textwidth, angle=0]{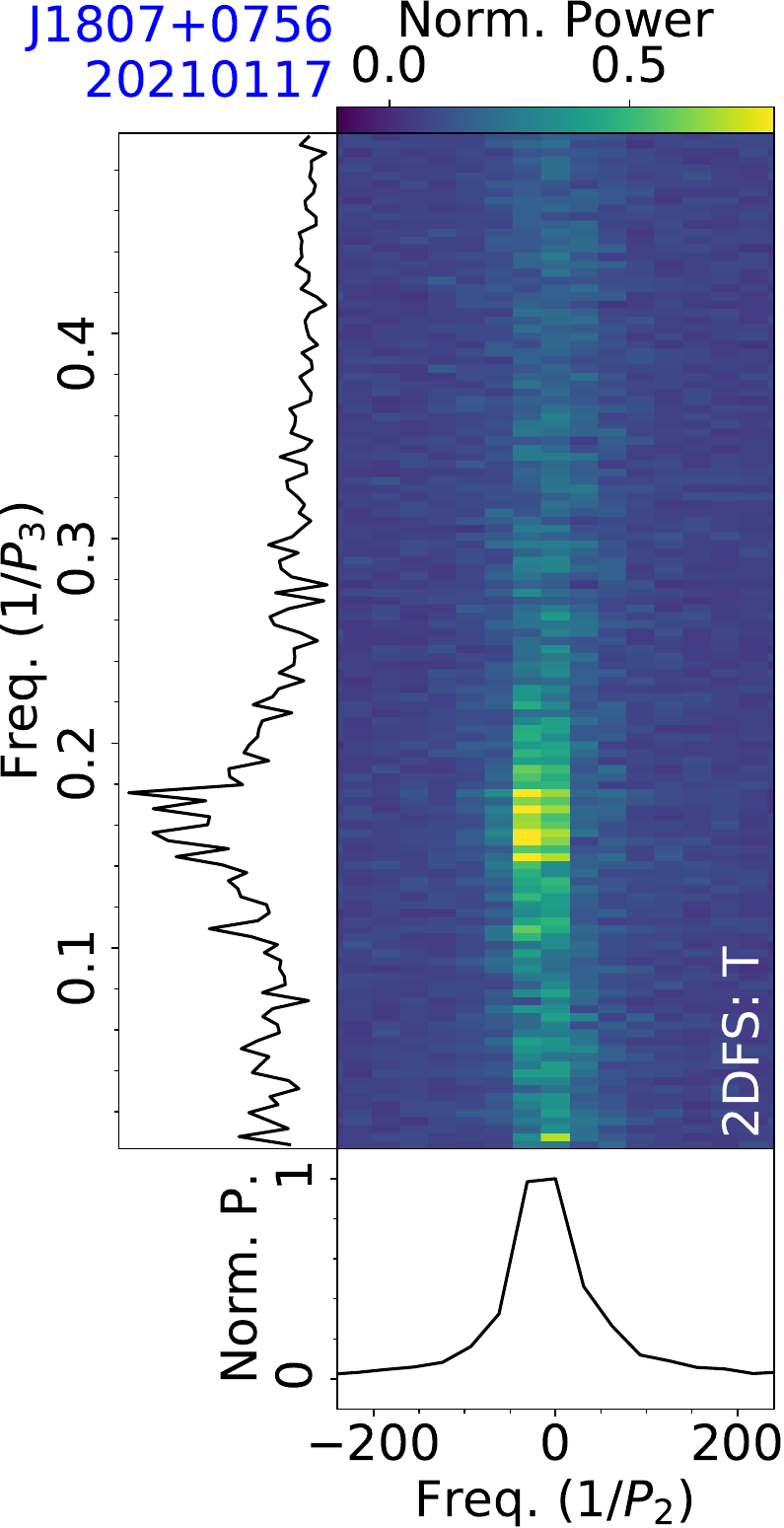}
\figcaption{Fluctuation analysis of PSR J1807+0756 from the FAST observation on 20210117, with LRFS (top-left), and 2DFS for the on-pulse region (top-right), leading part (bottom-left) and trailing part (bottom-right) of a mean pulse profile.
\label{subfig:fluctu:J1807+0756}}
\end{figure}

\subsection{J1802-0523}
\label{subsec:J1802-0523}

PSR J1802-0523 was the High Time Resolution Universe (HTRU) survey using the Parkes 64-m radio telescope \citep{Bates2012}.

This pulsar was observed by FAST on 20250910 for 10 minutes, yielding a rotation period $P=5.0422$~s and a dispersion measure $D\!M=124.7~{\rm cm^{-3}\,pc}$. 
The single pulse sequence is shown in Fig.~\ref{subfig:TP:J1802-0523}, illustrating the existence of nulling and subpulse drifting behaviors. The nulling fraction of this observation is estimated from the on-pulse energy histogram in Fig.~\ref{subfig:Hist:J1802-0523}, which is 24.3$\pm$3.6\%. In fluctuation spectra, the centroid frequencies of the positive drift feature are $1/P_3=0.126\pm0.002$ and $1/P_2=57\pm2$, corresponding to periodicities of $P_3=7.9\pm0.1$ periods and $P_2=6.3\pm0.2$ degrees.

\begin{figure}[htpb]
\centering
\includegraphics[width=0.22\textwidth, angle=0]{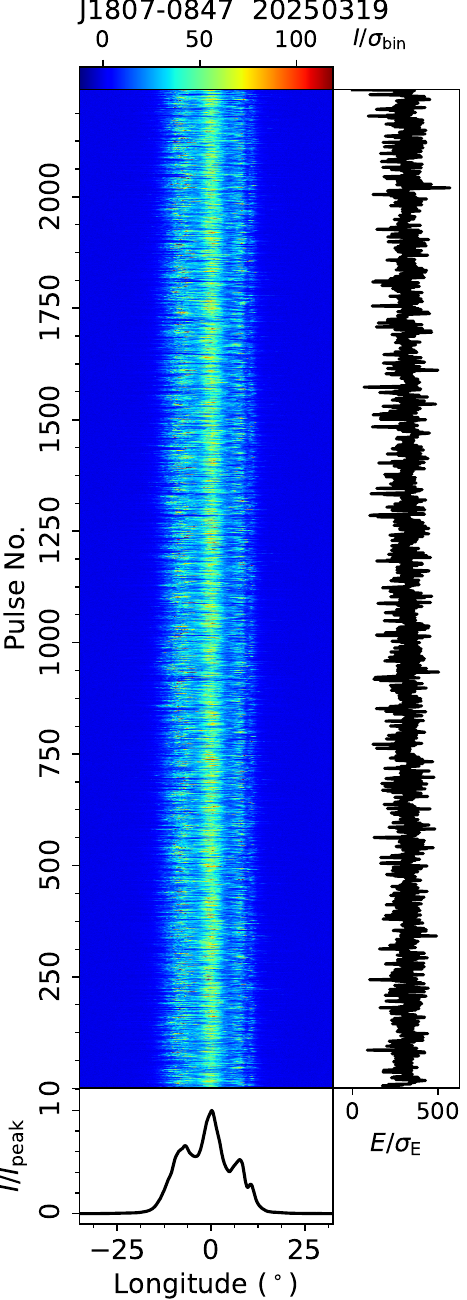}
\includegraphics[width=0.22\textwidth, angle=0]{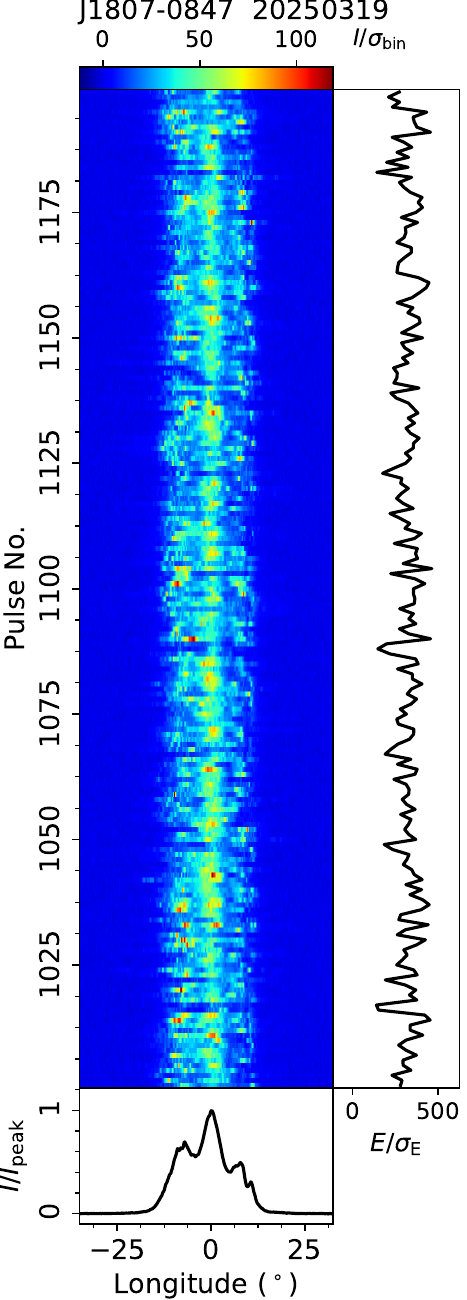}
\figcaption{Single pulse sequence of PSR J1807-0847 from the FAST observation on 20250319, and a zoomed-in view of pulses No. 1000-1200.
\label{subfig:TP:J1807-0847}}
\end{figure}

\begin{figure}[htpb]
\centering
\includegraphics[width=0.22\textwidth, angle=0]{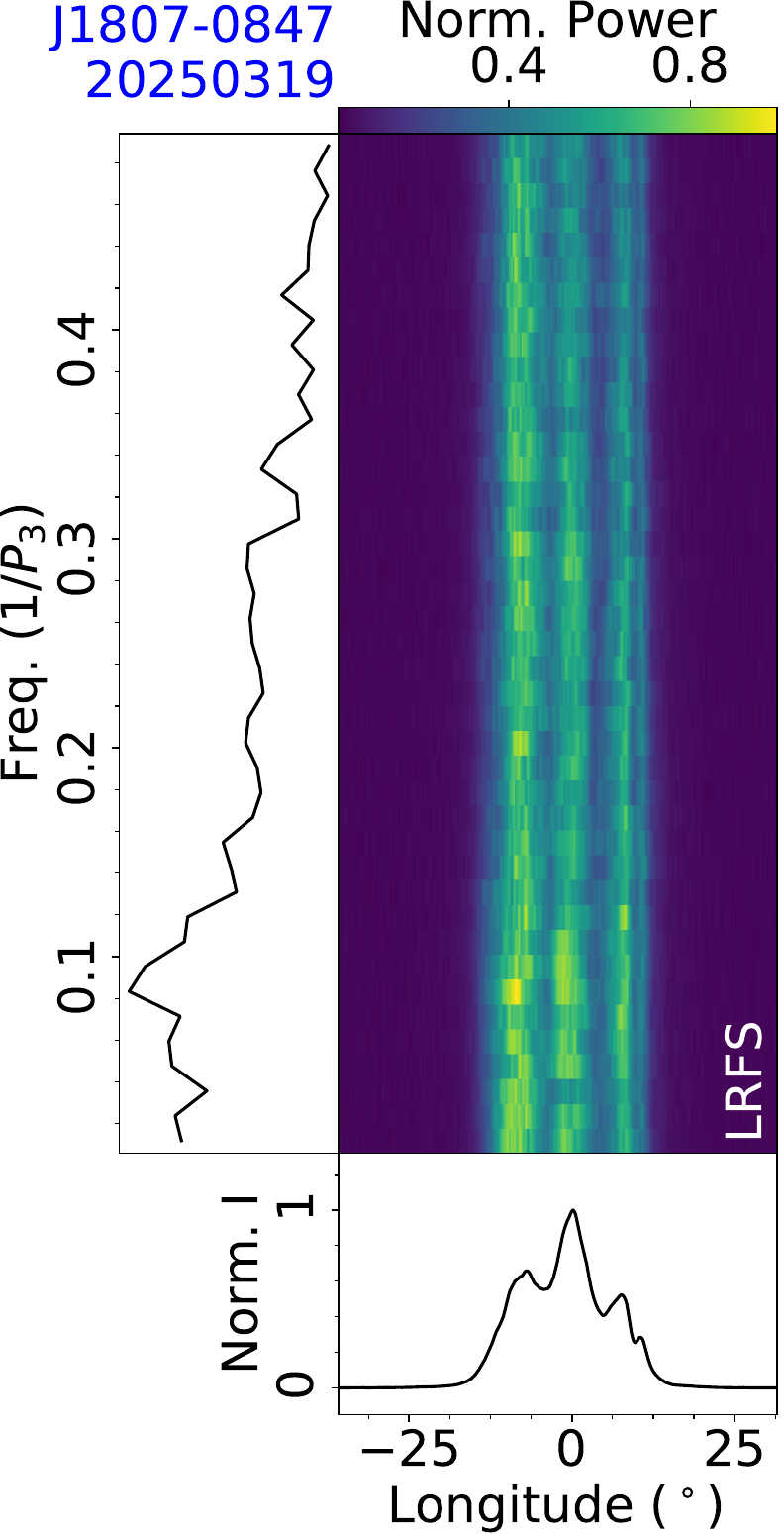}
\includegraphics[width=0.22\textwidth, angle=0]{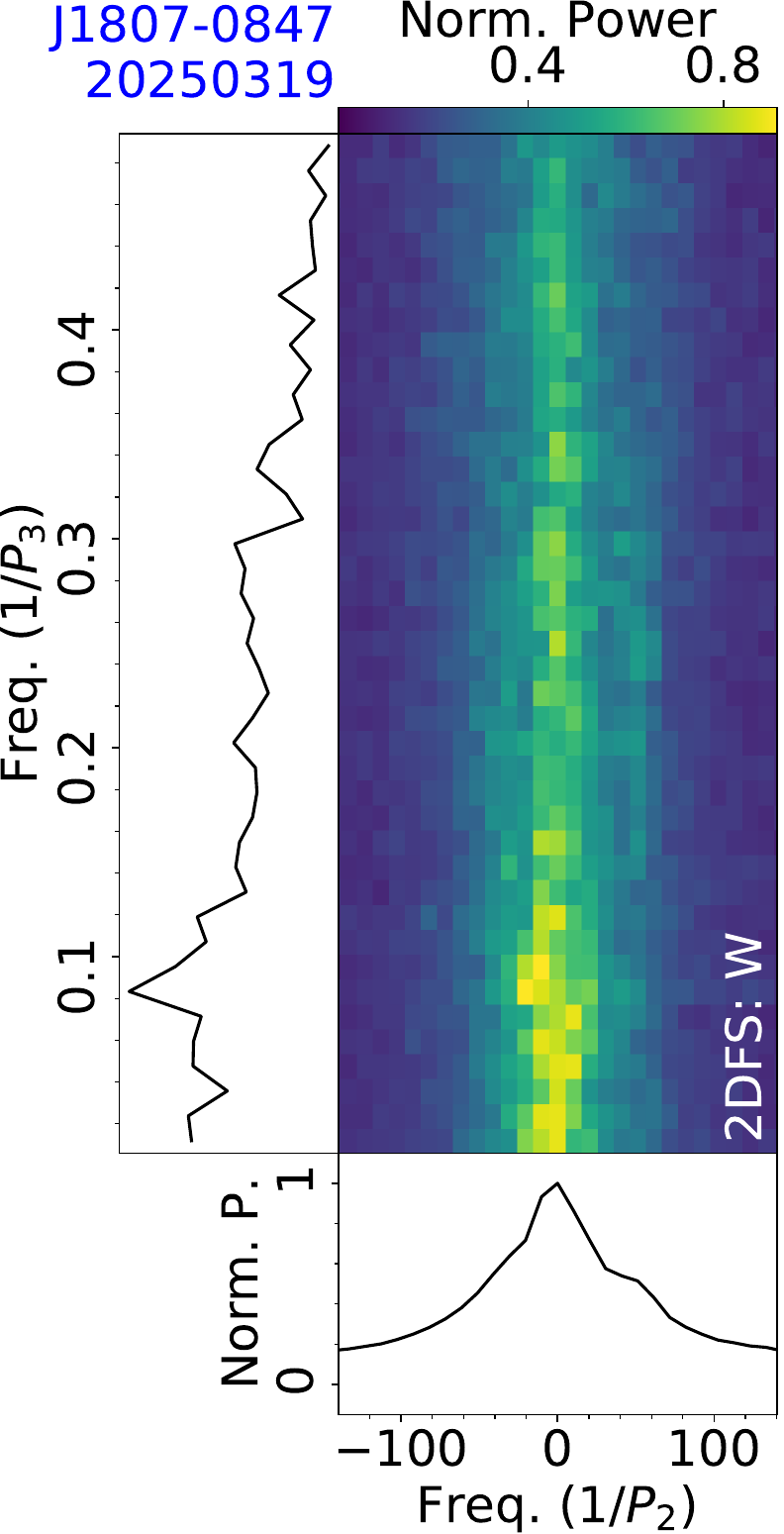}
\figcaption{Fluctuation analysis of PSR J1807-0847 for the observation on 20250319, with LRFS and 2DFS for the on-pulse region of a mean pulse profile.
\label{subfig:fluctu:J1807-0847}}
\end{figure}

\begin{figure}[htpb]
\centering
\includegraphics[width=0.22\textwidth, angle=0]{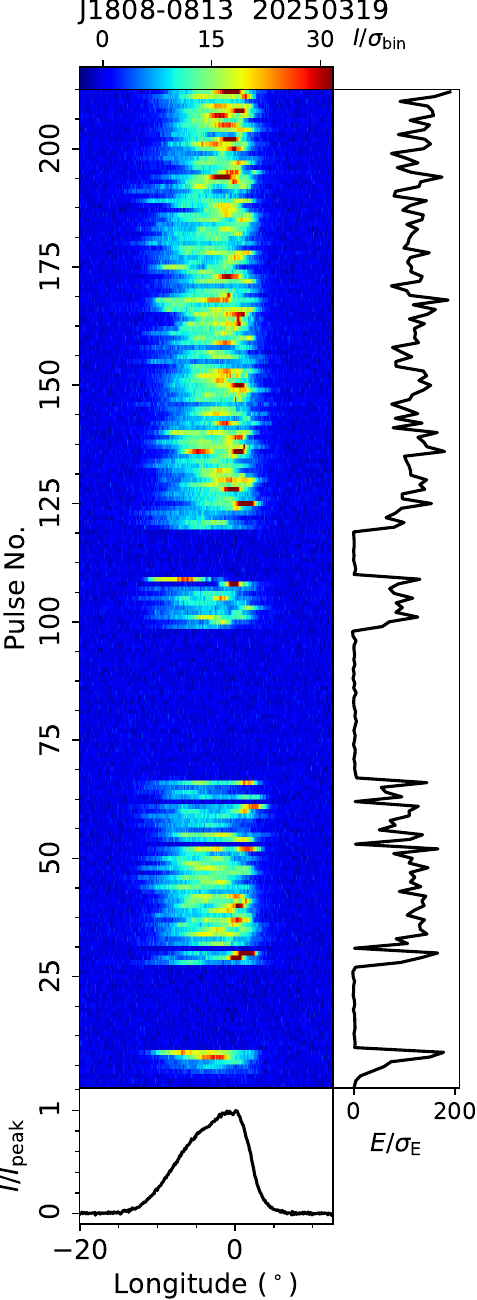}
\includegraphics[width=0.22\textwidth, angle=0]{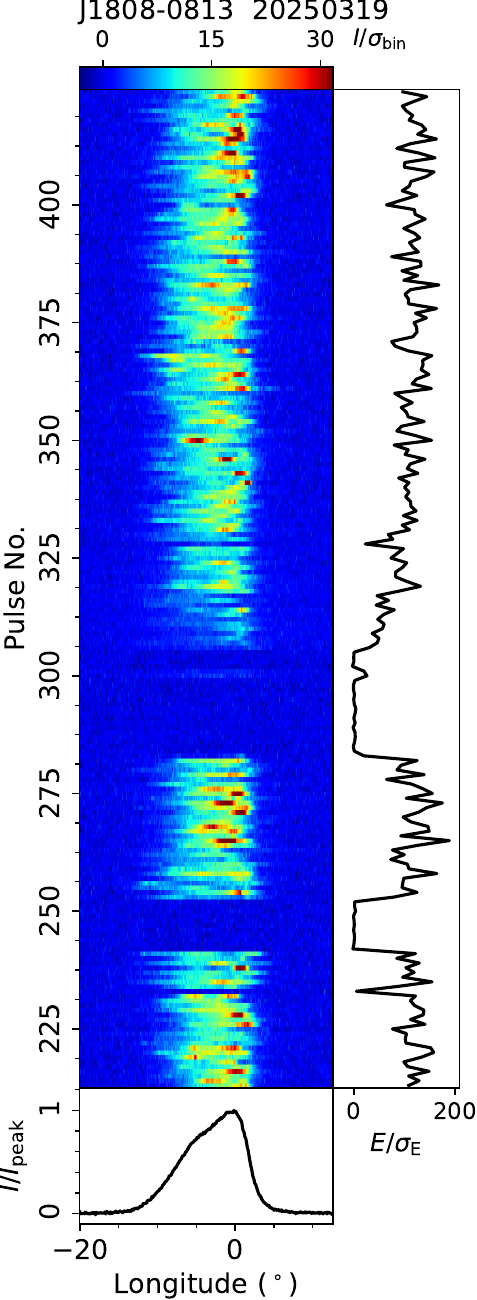}
\figcaption{Single pulse sequences of PSR J1808-0813 from the FAST observation on 20250319.
\label{subfig:TP:J1808-0813}}
\end{figure}

\begin{figure}[htpb]
\centering
\includegraphics[width=0.39\textwidth, angle=0]{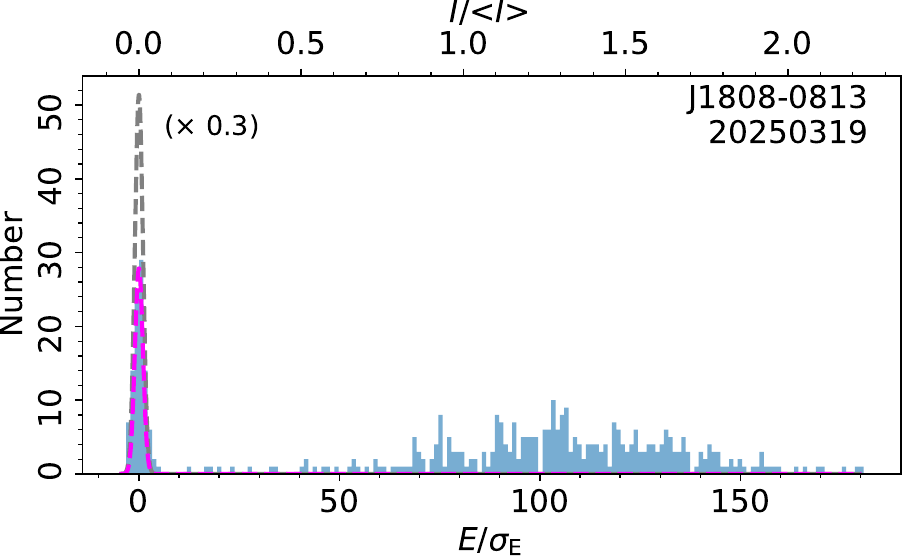}
\figcaption{On-pulse energy histogram of single pulses of PSR J1808-0813 from the FAST observation on 20250319. \label{subfig:Hist:J1808-0813}}
\end{figure}

\begin{figure}[htpb]
\centering
\includegraphics[width=0.22\textwidth, angle=0]{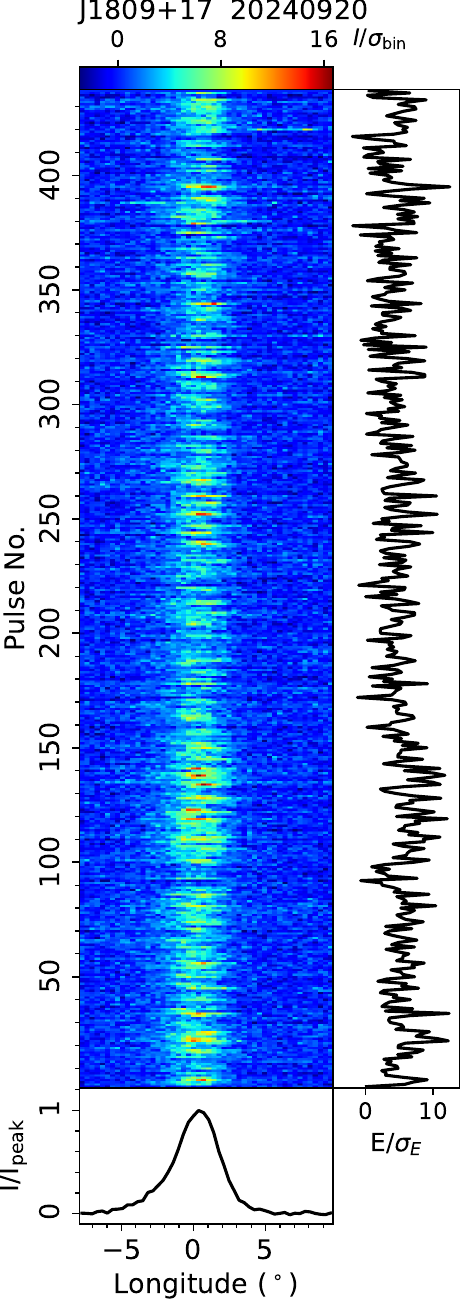}
\includegraphics[width=0.22\textwidth, angle=0]{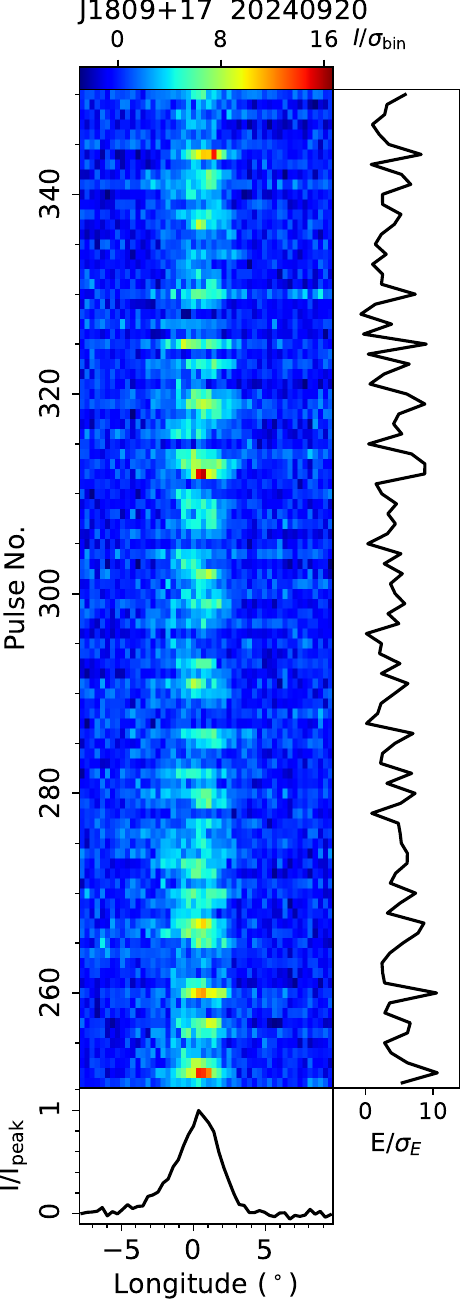}
\figcaption{Single pulse sequence of PSR J1809+17 from the FAST observation on 20240920, and a zoomed-in view of pulses No. 251-350.
\label{subfig:TP:J1809+17}}
\end{figure}

\begin{figure}[htpb]
\centering
\includegraphics[width=0.22\textwidth, angle=0]{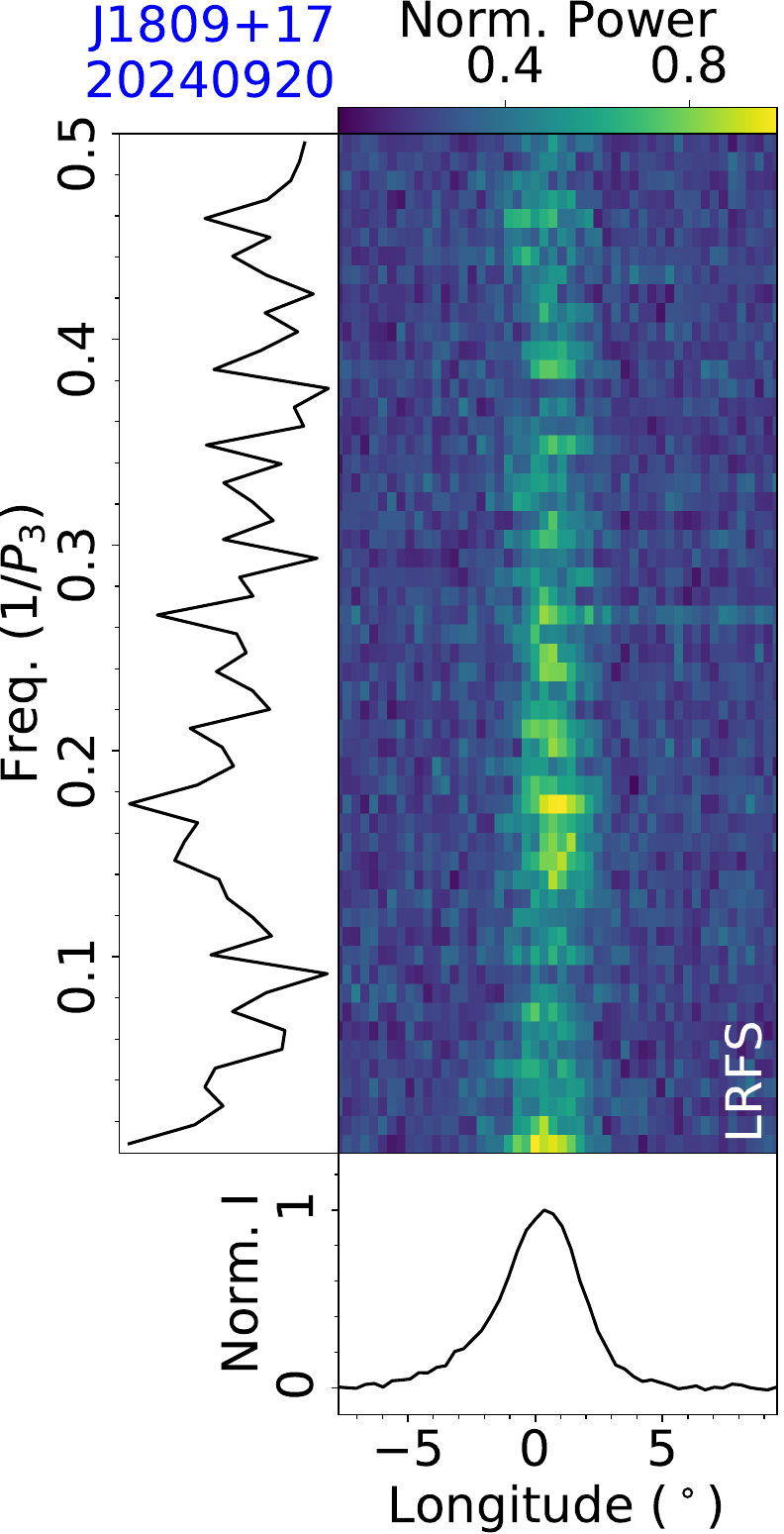}
\includegraphics[width=0.22\textwidth, angle=0]{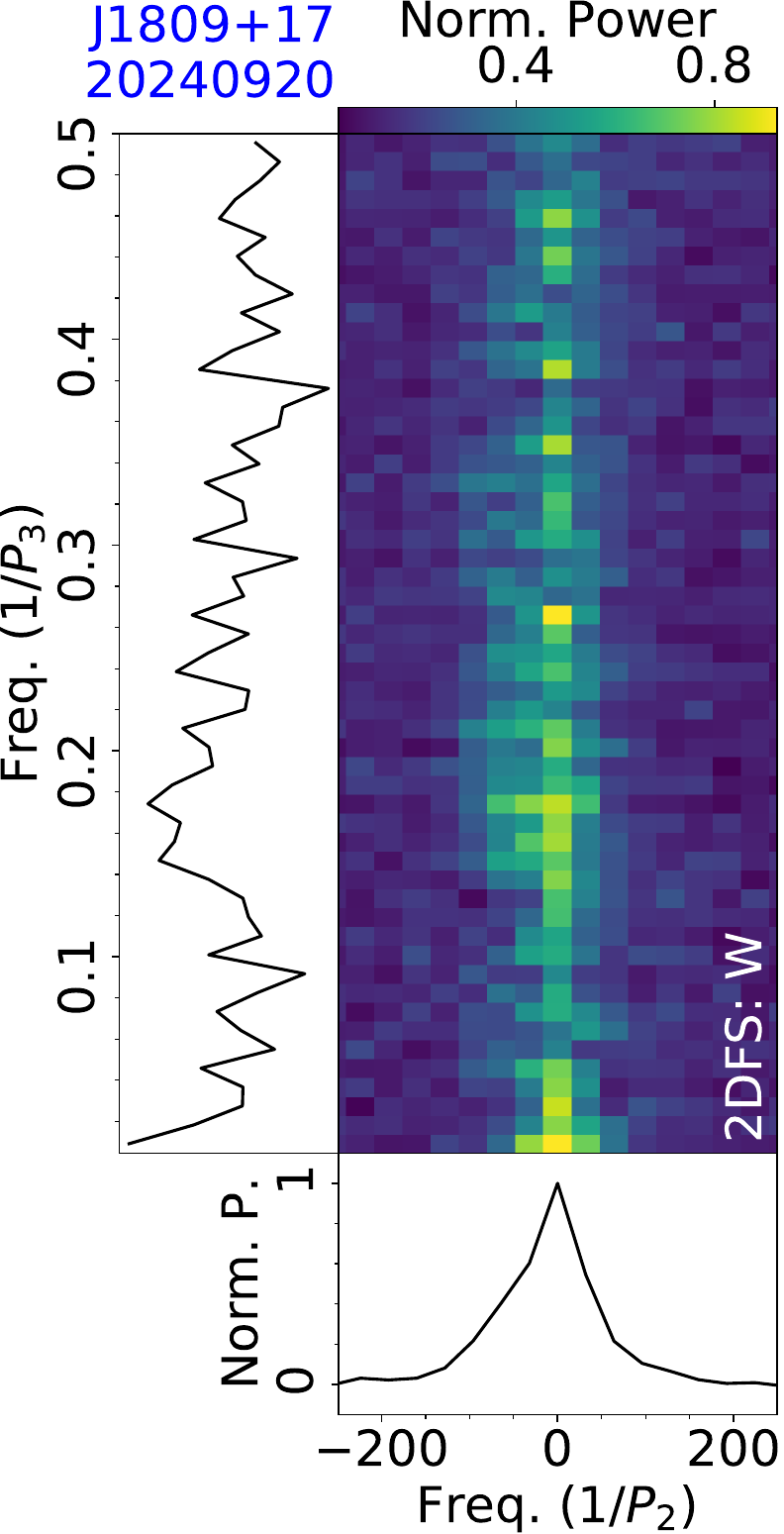}
\figcaption{Fluctuation analysis of PSR J1809+17 for the observation on 20240920, with LRFS and 2DFS for the on-pulse region of a mean pulse profile.
\label{subfig:fluctu:J1809+17}}
\end{figure}

\subsection{J1806-1154}
\label{subsec:J1806-1154}

PSR J1806-1154 was discovered by \citet{Stokes1985} with the 92-m telescope at Green Bank. \citet{Song2023} reported a drift feature with $P_3=2.3\pm0.1$ periods and $P_2=61^{+10}_{-54}$ degrees, and a $P_3$-only feature with $P_3=12\pm11$ periods.

This pulsar was observed by FAST on 20250319 for 6 minutes, with a rotation period $P=0.5226$~s and a dispersion measure $D\!M=122.4~{\rm cm^{-3}\,pc}$ derived. The single pulse sequence and a zoomed-in view of pulses No. 100-300 are shown in Fig.~\ref{subfig:TP:J1806-1154}. From the fluctuation spectra (Fig.~\ref{subfig:fluctu:J1806-1154}), the leading part in the mean pulse profile prefers to drift positively and also has a low-frequency modulation, while there is no obvious modulation feature in the 2DFS of the trailing part. There are three features in 2DFS of the leading profile part: a negative drift feature with the centroid at $1/P_3=0.42\pm0.01$ ($P_3=2.38\pm0.03$ periods) and $1/P_2=-54\pm1$ ($P_2=-6.7\pm0.2$ degrees); a positive drift feature with the centroid at $1/P_3=0.413\pm0.005$ ($P_3=2.42\pm0.03$ periods) and $1/P_2=54\pm1$ ($P_2=6.7\pm0.2$ degrees); and a low-frequency modulation feature with the centroid at $1/P_3=0.082\pm0.002$ ($P_3=12.3\pm0.3$ periods).

\subsection{J1807+0756}
\label{subsec:J1807+0756}

PSR J1807+0756 was discovered by \citet{Ray1996} using the Arecibo telescope at 430 MHz. Subpulse drifting behavior of $P_3=6.2\pm0.9$ periods and $P_2=-12^{+2}_{-9}$ degrees was reported by \citet{Song2023}. 

The pulsar has also been detected by FAST on 20210117, deriving a rotation period $P=0.4643$~s and a dispersion measure $D\!M=89.2~{\rm cm^{-3}\,pc}$ from this observation. Single pulse sequences in Fig.~\ref{subfig:TP:J1807+0756} display the subpulse drifting behavior. 
From LRFS and 2DFS in Fig.~\ref{subfig:fluctu:J1807+0756}, the temporal modulation frequency $1/P_3$ are wide, ranging from 0 to 2.8. For the leading part in a mean pulse profile, the drift feature in 2DFS has centroid frequencies of $1/P_3=0.122\pm0.001$ and $1/P_2=-24\pm1$, which correspond to $P_3=8.2\pm0.1$ periods and $P_2=-15\pm1^\circ$. The trailing part of the pulse profile has values of $1/P_3=0.137\pm0.001$ and $1/P_2=-17\pm1$, yielding $P_3=7.3\pm0.1$ periods and $P_2=-21\pm1^\circ$.



\begin{figure}[htpb]
\centering
\includegraphics[width=0.22\textwidth, angle=0]{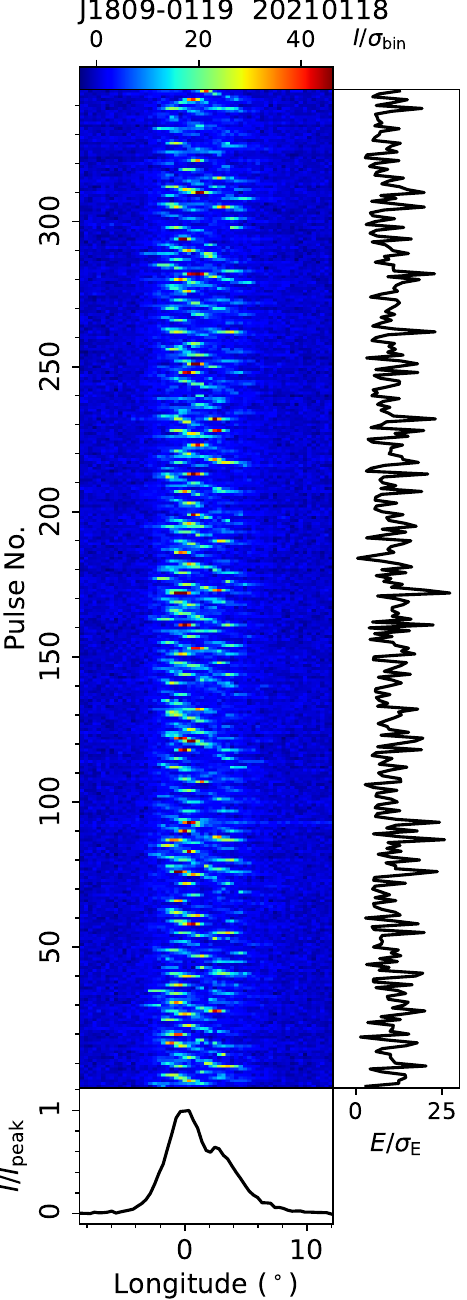}
\includegraphics[width=0.22\textwidth, angle=0]{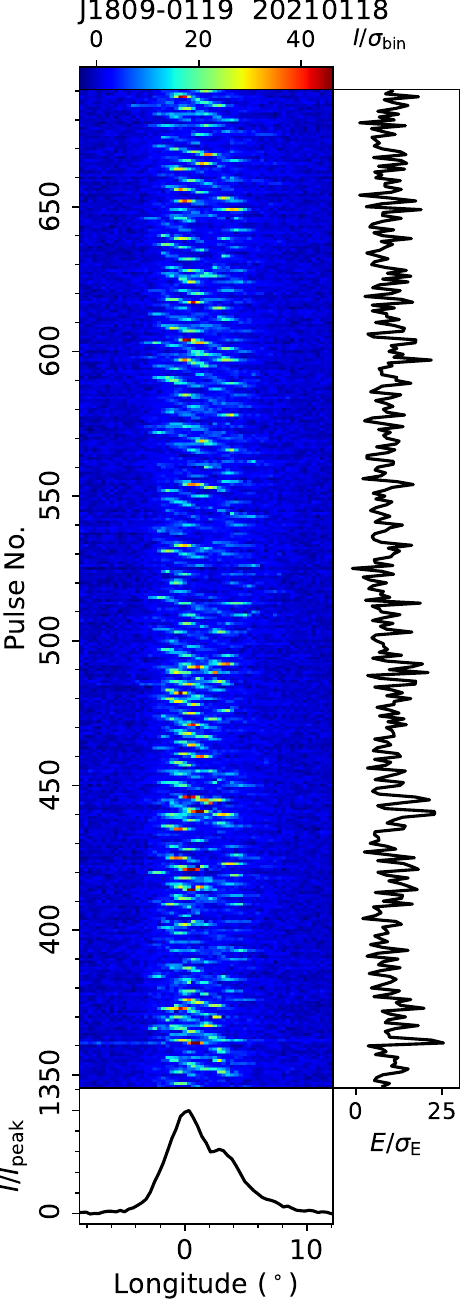}
\figcaption{Single pulse sequences of PSR J1809-0119. \label{subfig:TP:J1809-0119}}
\end{figure}

\begin{figure}[htpb]
\centering
\includegraphics[width=0.22\textwidth, angle=0]{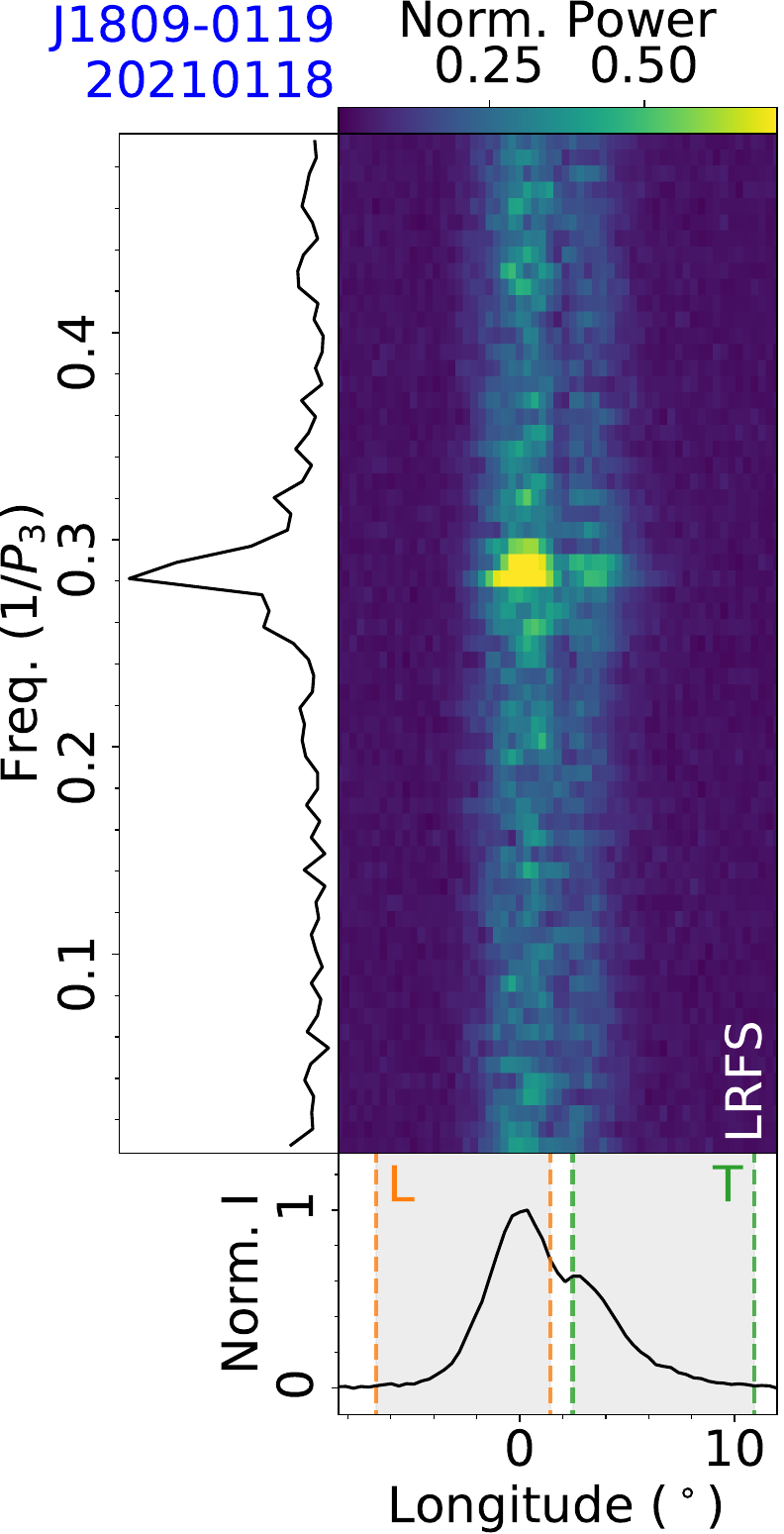}
\includegraphics[width=0.22\textwidth, angle=0]{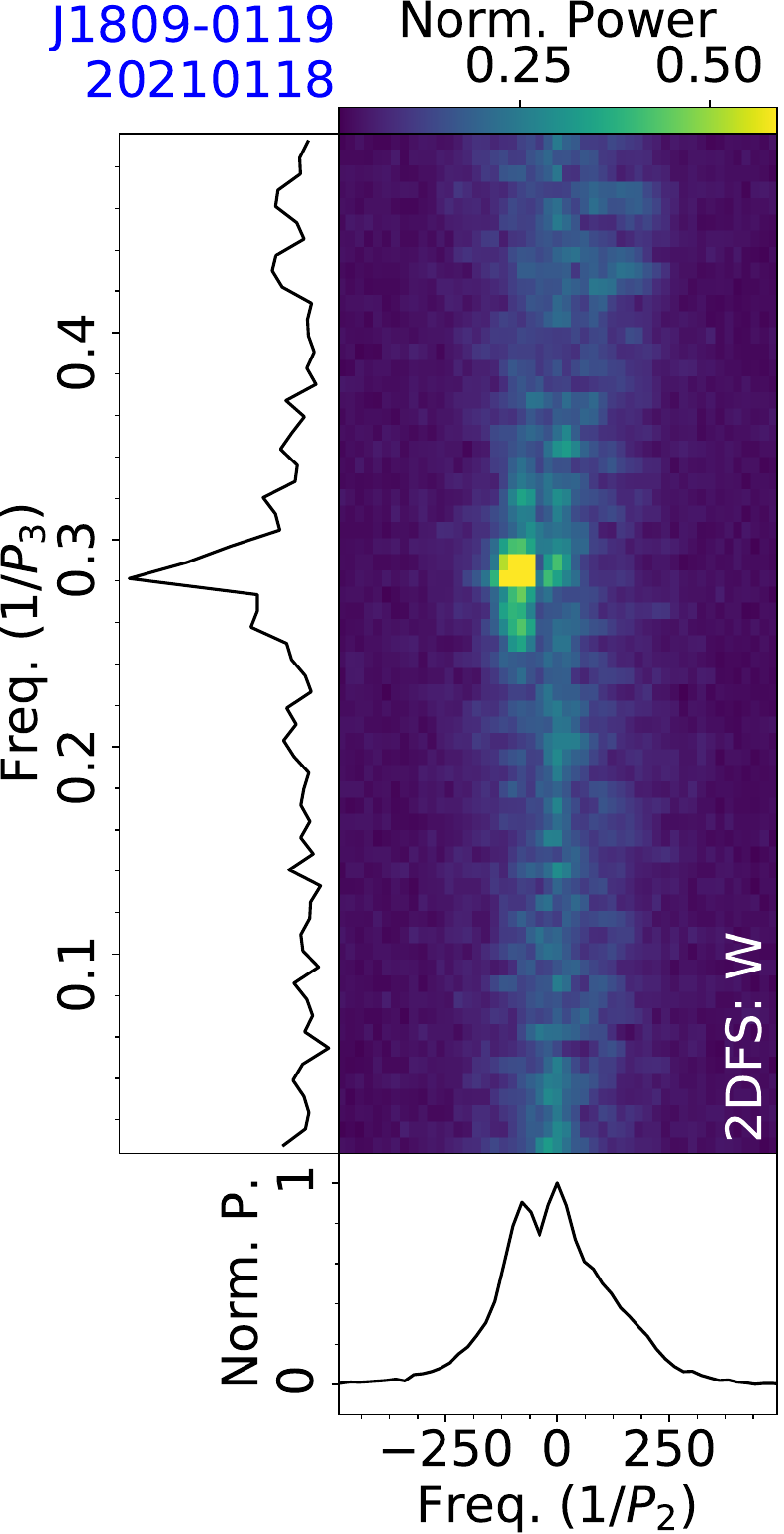}\\
\includegraphics[width=0.22\textwidth, angle=0]{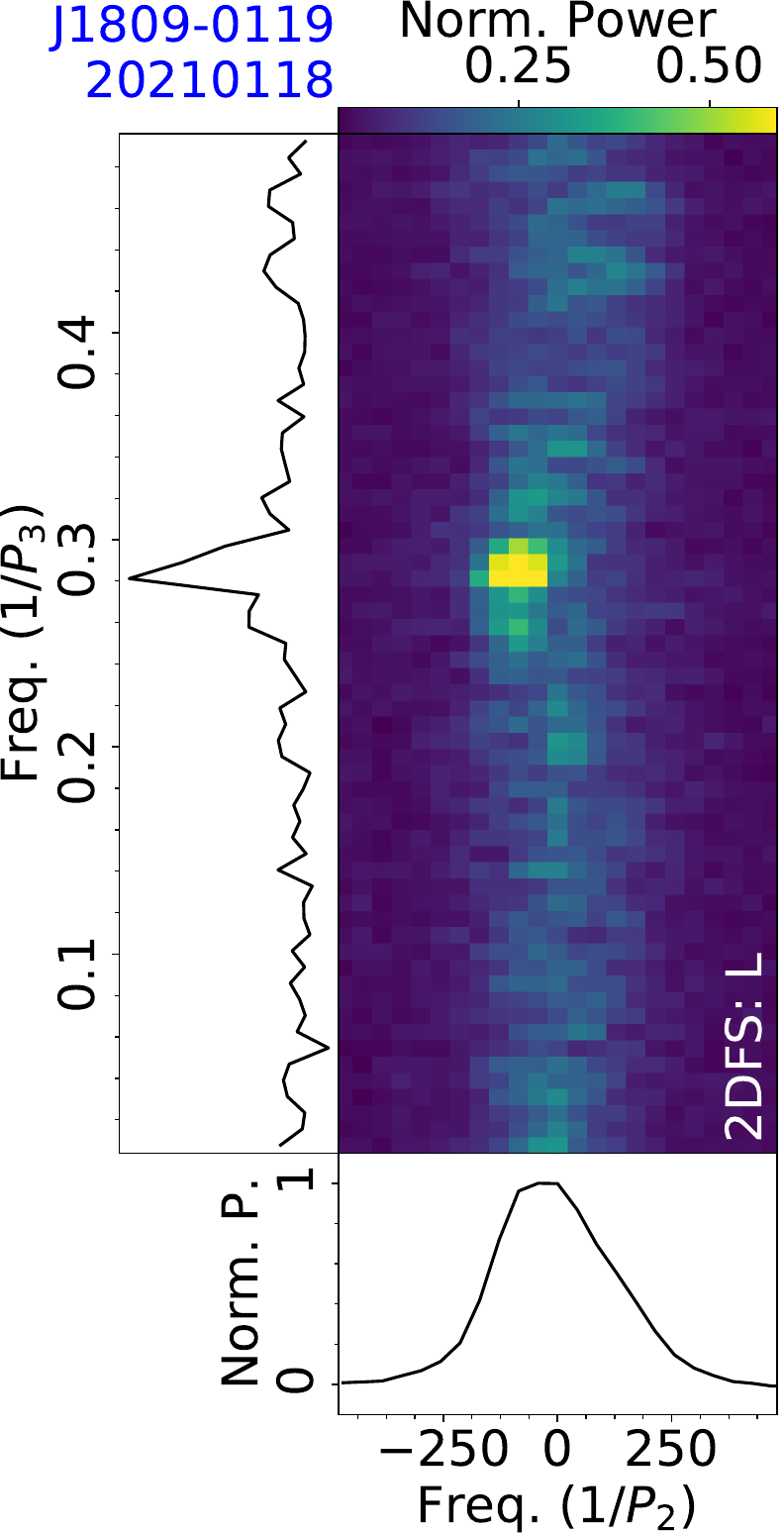}
\includegraphics[width=0.22\textwidth, angle=0]{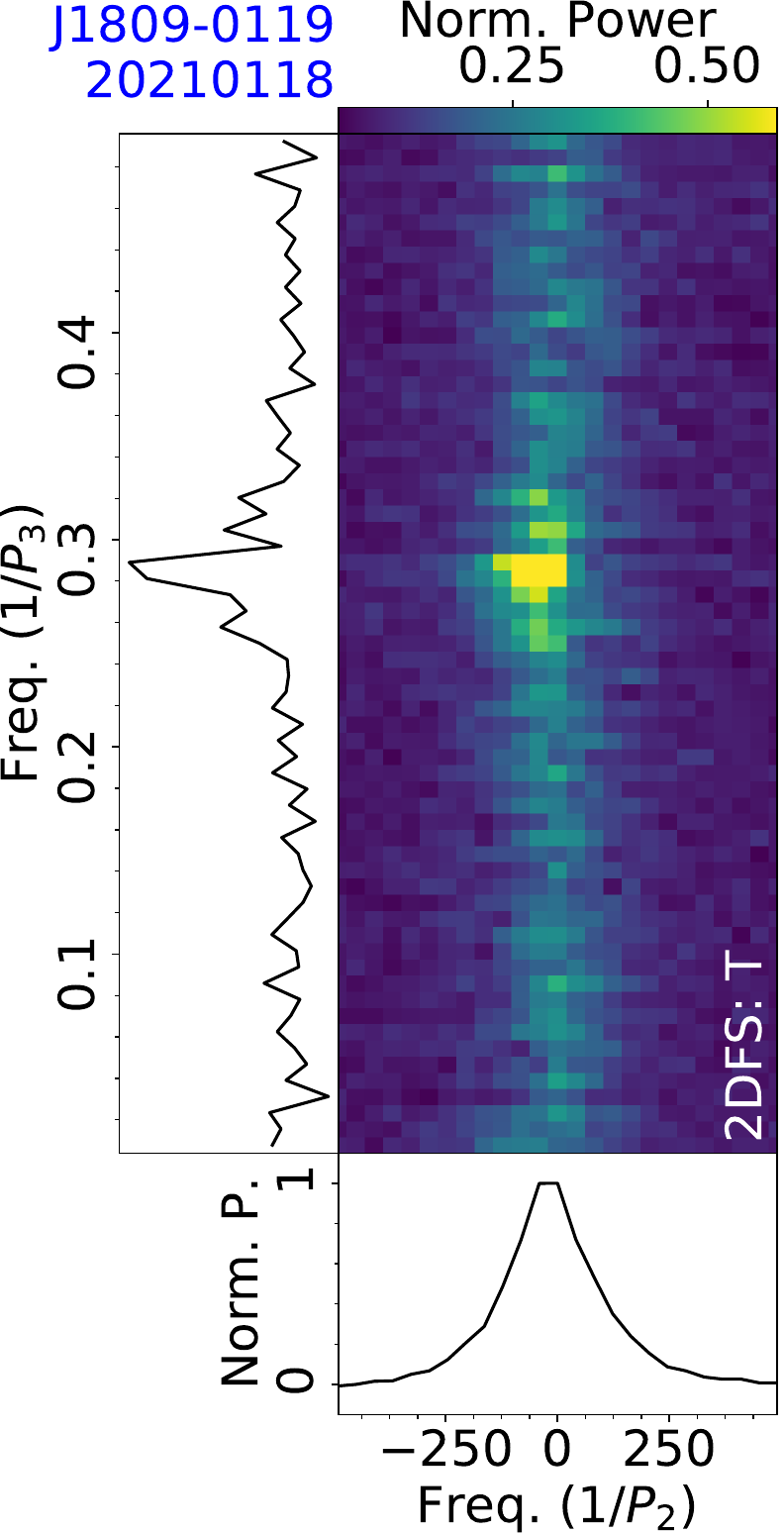}
\figcaption{Fluctuation analysis of PSR J1809-0119 from the observation on 20210118, with LRFS and 2DFS for the on-pulse region of a mean pulse profile. \label{subfig:fluctu:J1809-0119}}
\end{figure}

\subsection{J1807-0847}
\label{subsec:J1807-0847}

PSR J1807-0847 was discovered by \citet{Manchester1978} in the second Molonglo pulsar survey. \citet{Song2023} reported a positive drifting with $P_3=3.8\pm0.5$ periods and $P_2=31^{+10}_{-16}$ degrees, and a negative drifting with $P_3=4.6\pm0.6$ periods and $P_2=-35^{+19}_{-6}$ degrees, for two components respectively.

This pulsar was observed by FAST on 20250319 for 6 minutes, deriving a rotation period $P=0.1637$~s and a dispersion measure $D\!M=112.3~{\rm cm^{-3}\,pc}$. The single pulse sequence and a zoomed-in view of pulses No.1000-1200 in Fig.~\ref{subfig:TP:J1807-0847} show that the modulation behavior is not systematic. 
Fluctuation spectra are displayed in Fig.~\ref{subfig:fluctu:J1807-0847}. From this FAST observation, the 2DFS of the on-pulse region in a mean pulse profile shows a modulation feature and a positive drift feature, whereas both positive and negative drift features were reported by \citet{Song2023}. The centroid frequency of the temporal modulation feature in 2DFS is $1/P_3=0.068\pm0.001$, corresponding to $P_3=14.8\pm0.3$ periods. The centroid of the positive drift feature is characterized by frequencies of $1/P_3=0.231\pm0.003$ and $1/P_2=51\pm1$, yielding periodicities of $P_3=4.34\pm0.05$ periods and $P_2=7.1\pm0.1$ degrees.

\subsection{J1808-0813}
\label{subsec:J1808-0813}

PSR J1808-0813 was discovered by \citet{Manchester1996} using the Parkes radio telescope. Nulling behavior has been reported by \citet{Weltevrede2007} at 21 cm and by \citet{BurkeSpolaor2012} at 1352 MHz. Nulling fractions at 333MHz and 618 MHz were estimated by \citet{Basu2017} to be 12.8$\pm$1.3\% and 8.2$\pm$1.0\%, respectively. 
In addition, \citet{Song2023} presented subpulse drifting parameters of $P_3=2.4\pm0.3$ periods and $P_2=88^{+85}_{-80}$ degrees. 

This pulsar was observed by FAST on 20250319 for 6 minutes, with a rotation period $P=0.8760$~s and a dispersion measure $D\!M=151.2~{\rm cm^{-3}\,pc}$ derived. Single pulse sequences of this observation in Fig.~\ref{subfig:TP:J1808-0813} display the nulling phenomenon, and the nulling fraction is estimated to be 16.3$\pm$1.7\% from the on-pulse energy histogram in Fig.~\ref{subfig:Hist:J1808-0813}. While there is no obvious drift feature in the fluctuation spectra of this data.

\begin{figure}[htpb]
\centering
\includegraphics[width=0.22\textwidth, angle=0]{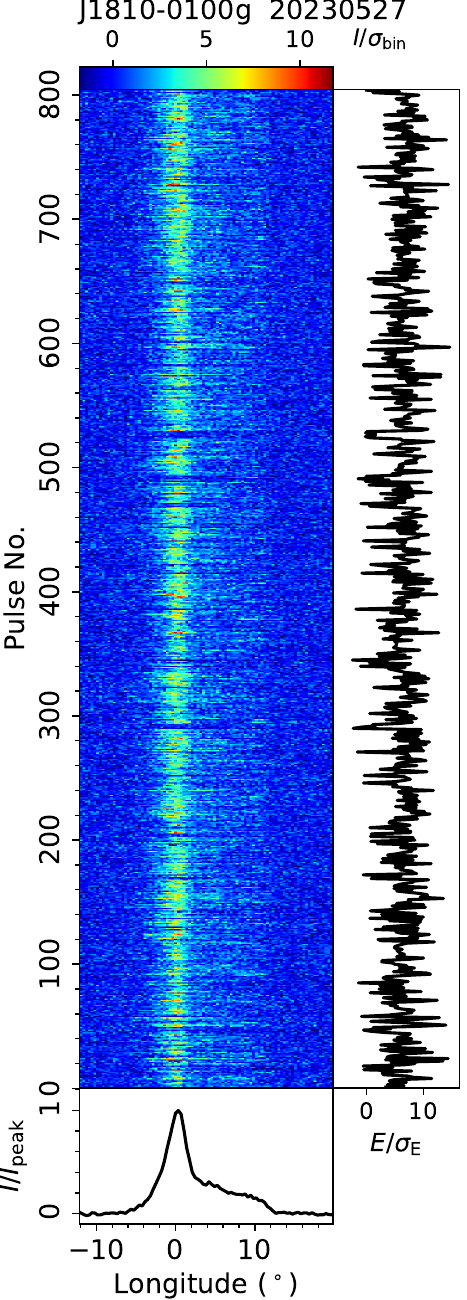}
\includegraphics[width=0.22\textwidth, angle=0]{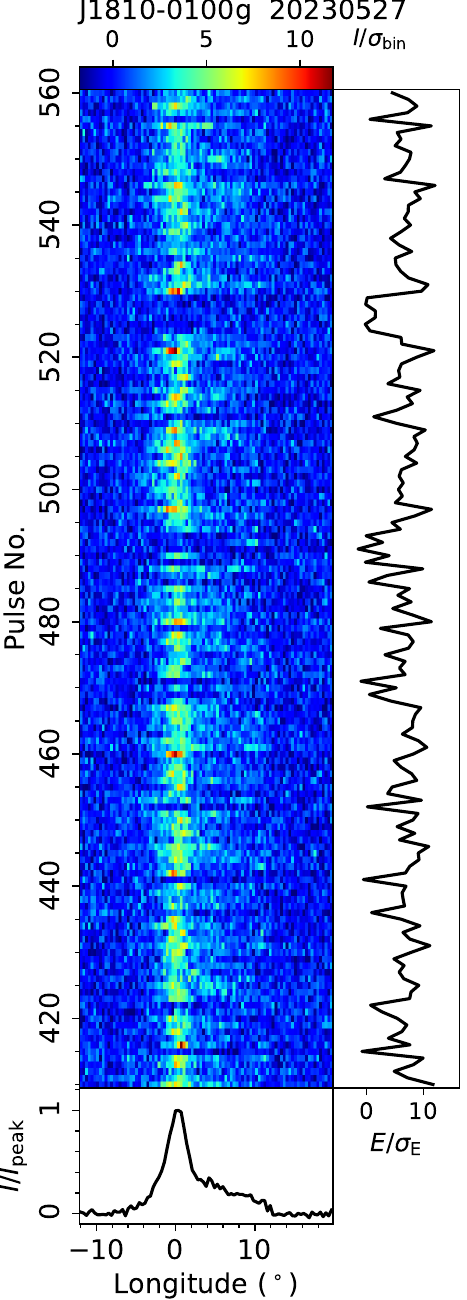}
\figcaption{Single pulse sequences of PSR J1810-0100g from the FAST observation on 20230527. \label{subfig:TP:J1810-0100g}}
\includegraphics[width=0.39\textwidth, angle=0]{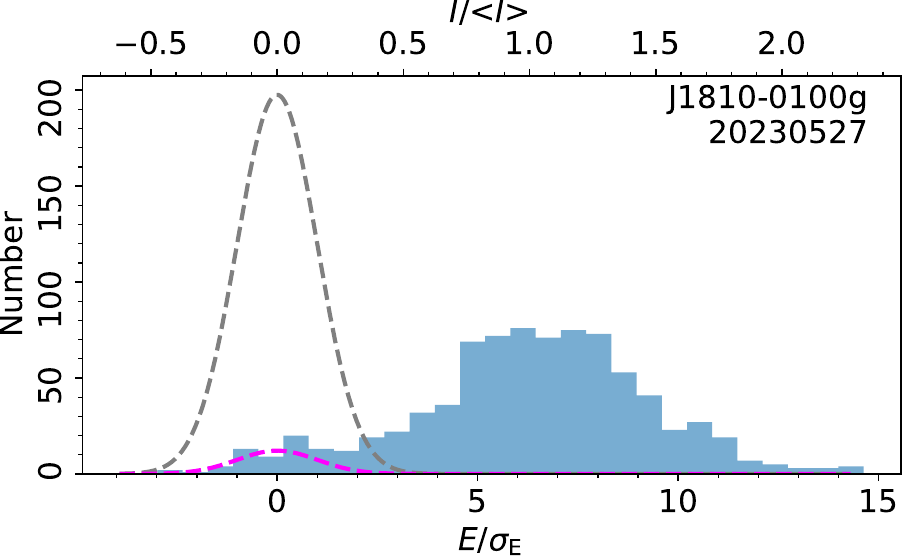}
\figcaption{On-pulse energy histogram of single pulses of PSR J1810-0100g from the FAST observation on 20230527.
\label{subfig:Hist:J1810-0100g}}
\end{figure}

\begin{figure}[htpb]
\centering
\includegraphics[width=0.22\textwidth, angle=0]{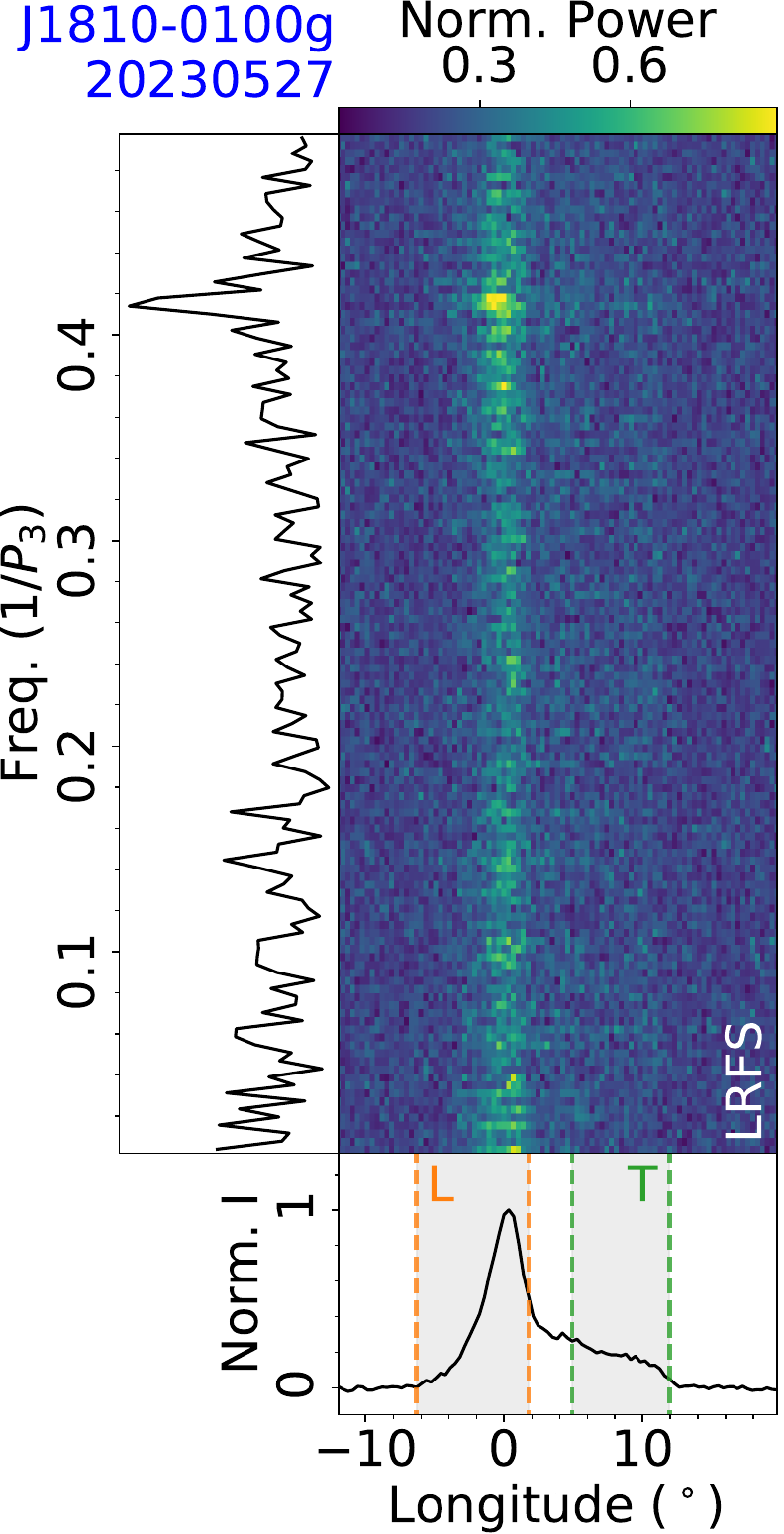}
\includegraphics[width=0.22\textwidth, angle=0]{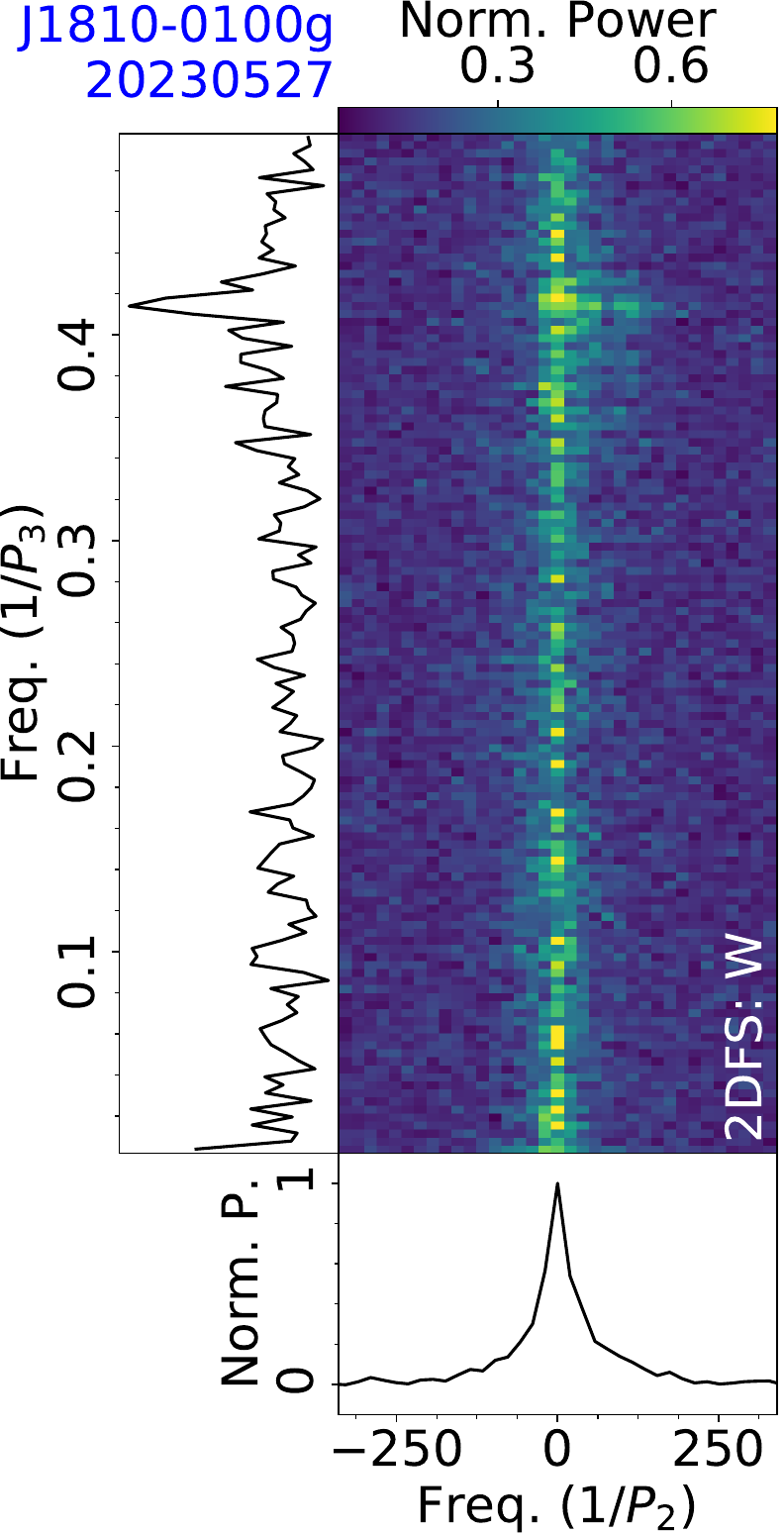}\\
\includegraphics[width=0.22\textwidth, angle=0]{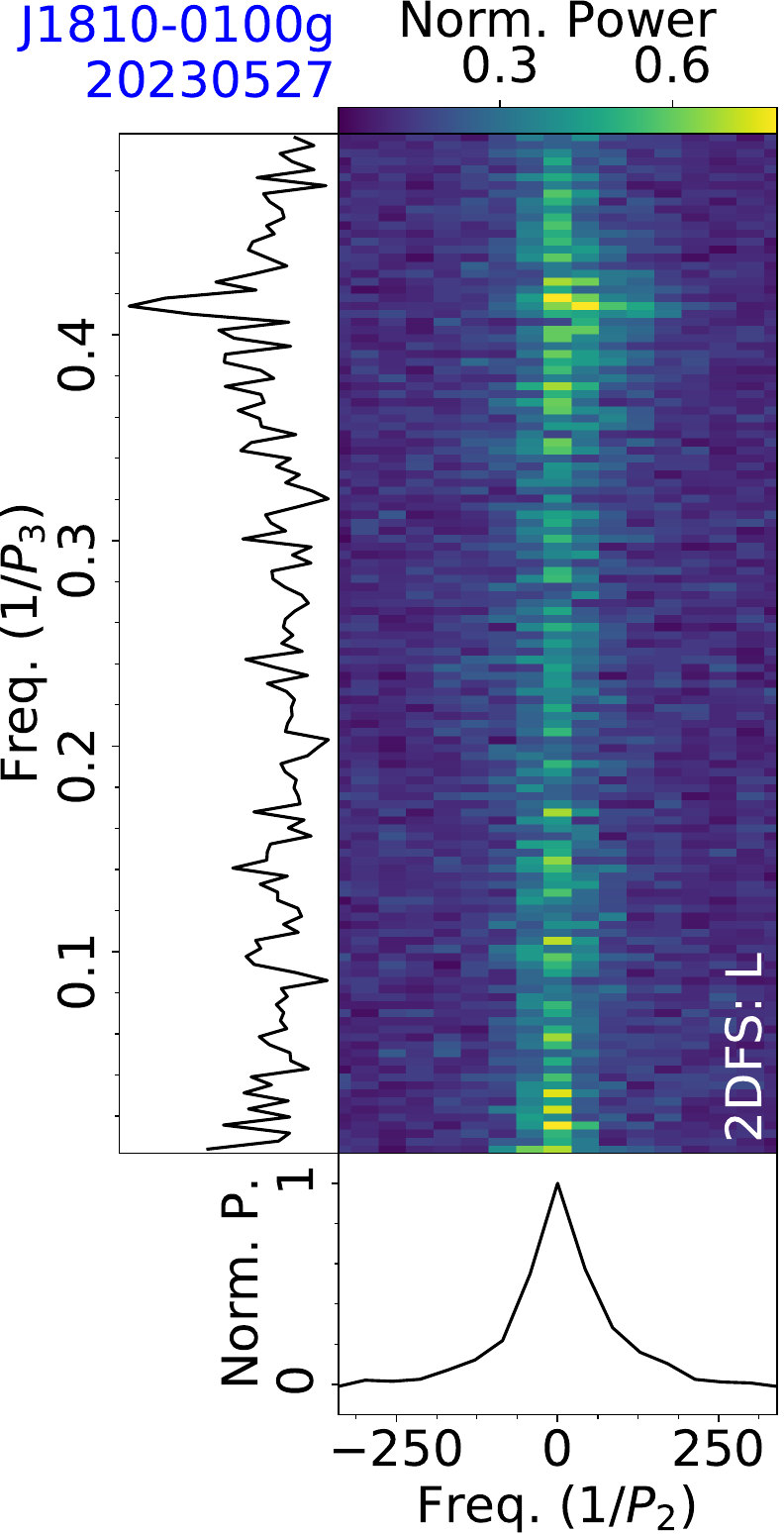}
\includegraphics[width=0.22\textwidth, angle=0]{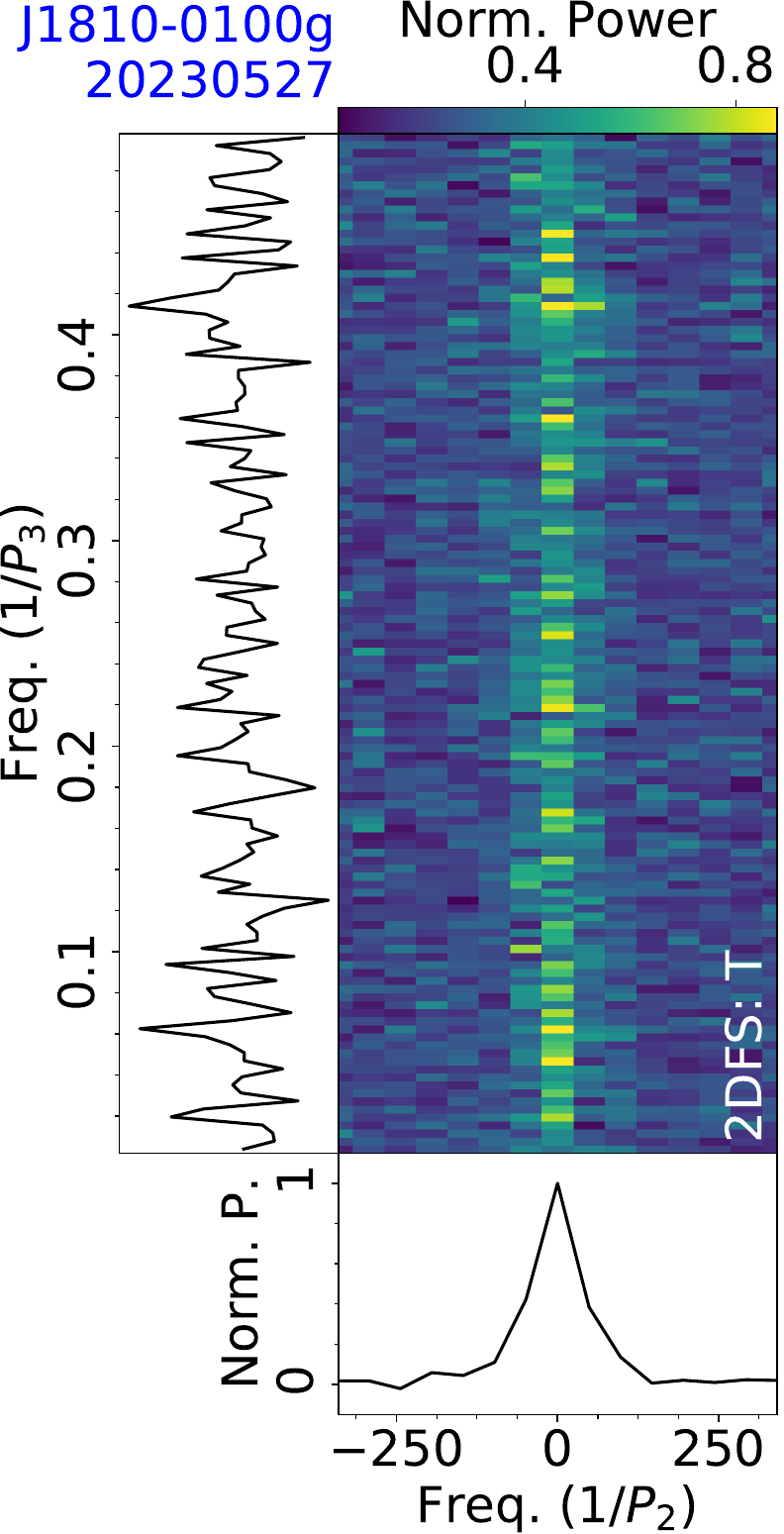}
\figcaption{Fluctuation analysis of PSR J1810-0100g from the observation on 20230527, with LRFS (top-left), and 2DFS for the on-pulse region (top-right), leading part (bottom-left) and trailing part (bottom-right) of a mean pulse profile. \label{subfig:fluctu:J1810-0100g}}
\end{figure}

\subsection{J1809+17}
\label{subsec:J1809+17}

PSR was discovered in the LOFAR Tied-Array All-Sky Survey (LOTAAS) \citep{Sanidas2019}. 

The pulsar was observed by FAST on 20240920 for 15 minutes, deriving a rotation period $P=2.0668$~s and a dispersion measure $D\!M=47.7~{\rm cm^{-3}\,pc}$ from this observation. The single pulse sequence is shown in Fig.~\ref{subfig:TP:J1809+17}.
From fluctuation spectra in Fig.~\ref{subfig:fluctu:J1809+17}, this pulsar has a negative subpulse drifting behavior. 
The centroid modulation frequencies are estimated to be $1/P_3=0.216\pm0.004$ and $1/P_2=-33\pm3$, corresponding to drifting parameters of $P_3=4.6\pm0.1$ periods and $P_2=-11\pm1^\circ$. 
The wide temporal modulation frequency range of the drift feature indicates an unsystematic modulation behavior.


\begin{figure}[htpb]
    \centering
	\includegraphics[height=0.92\textheight]{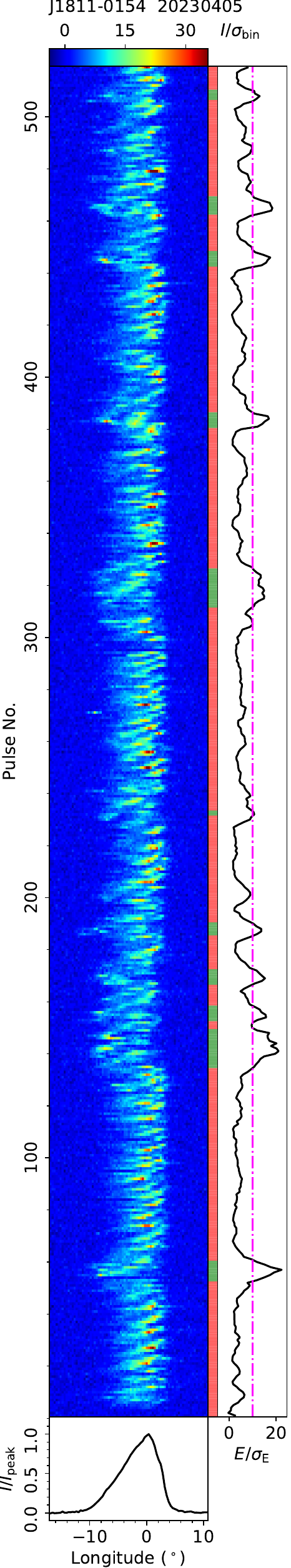}
	\includegraphics[height=0.92\textheight]{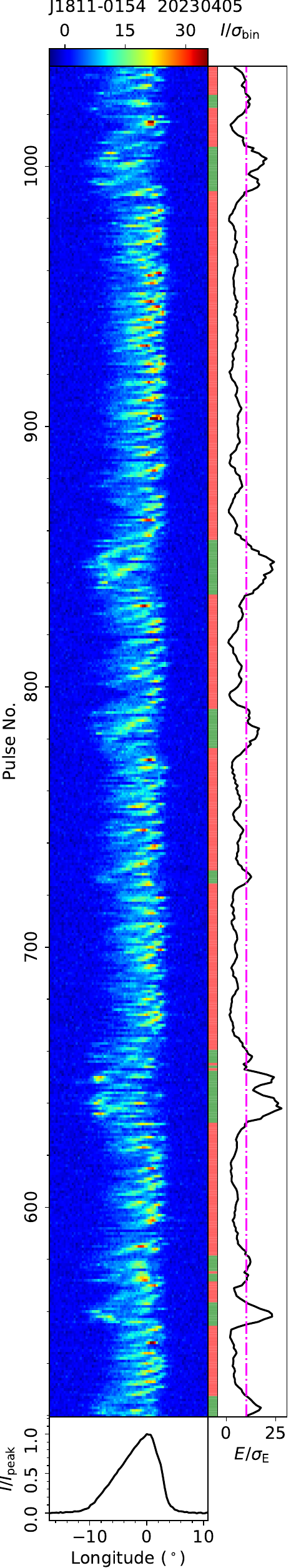} 
	\figcaption{Single pulse sequences of PSR J1811-0154 from the FAST observation on 20230405. \label{subfig:TP:J1811-0154}}
	\addtocounter{figure}{-1}
\end{figure}

\begin{figure}[htpb]
    \centering
	\includegraphics[height=0.92\textheight]{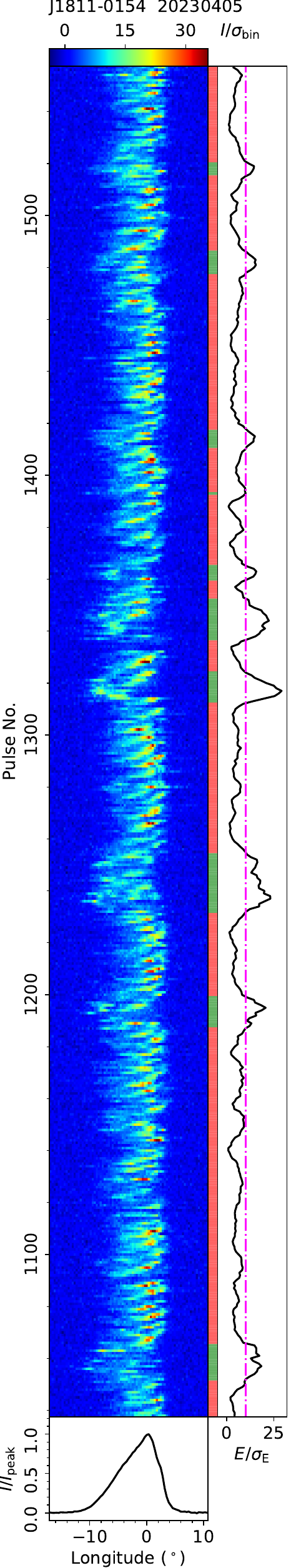}
    \includegraphics[height=0.92\textheight]{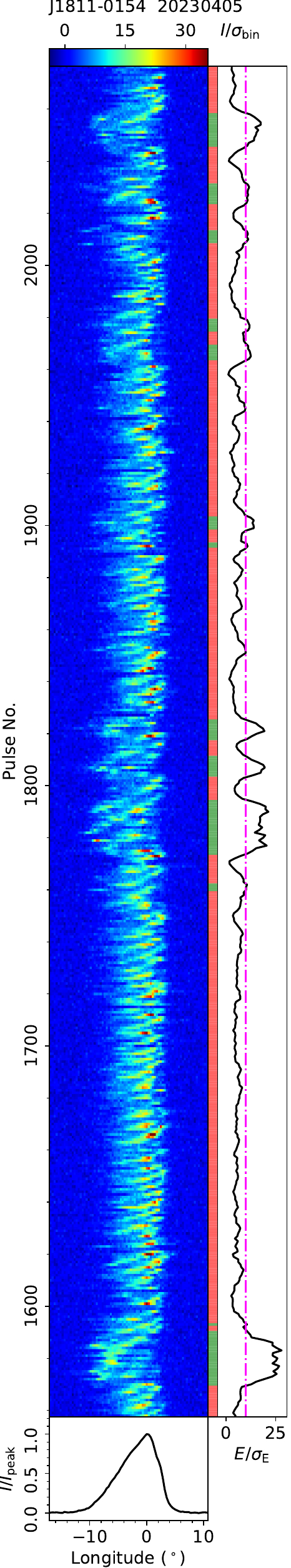} 
    \figcaption{Continued.}
    \addtocounter{figure}{-1}
\end{figure}

\begin{figure}[htpb]
    \centering
    \includegraphics[height=0.92\textheight]{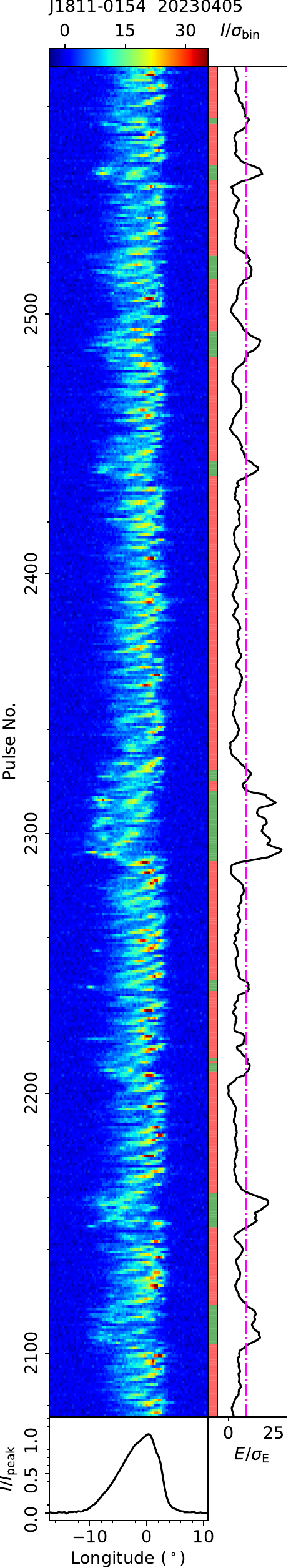}
	\includegraphics[height=0.92\textheight]{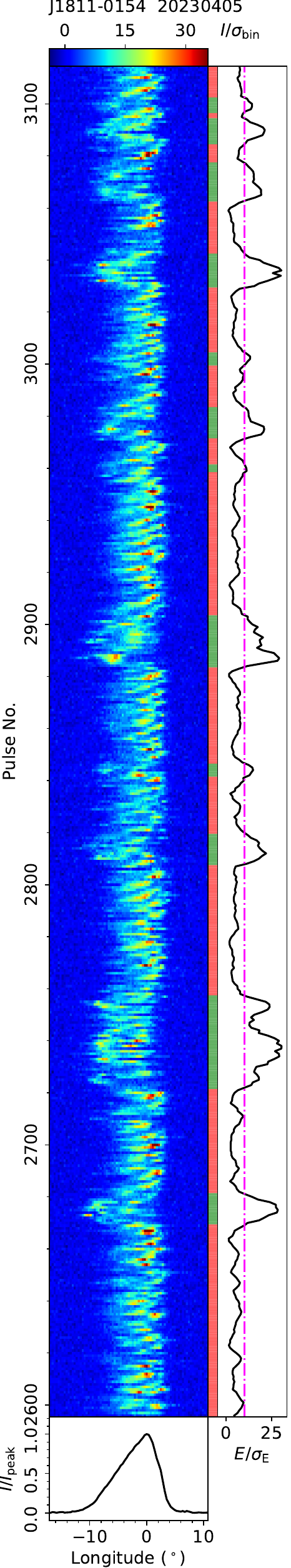}
    \figcaption{Continued.}
    \addtocounter{figure}{-1}
\end{figure}

\begin{figure}[htpb]
    \centering
	\includegraphics[height=0.92\textheight]{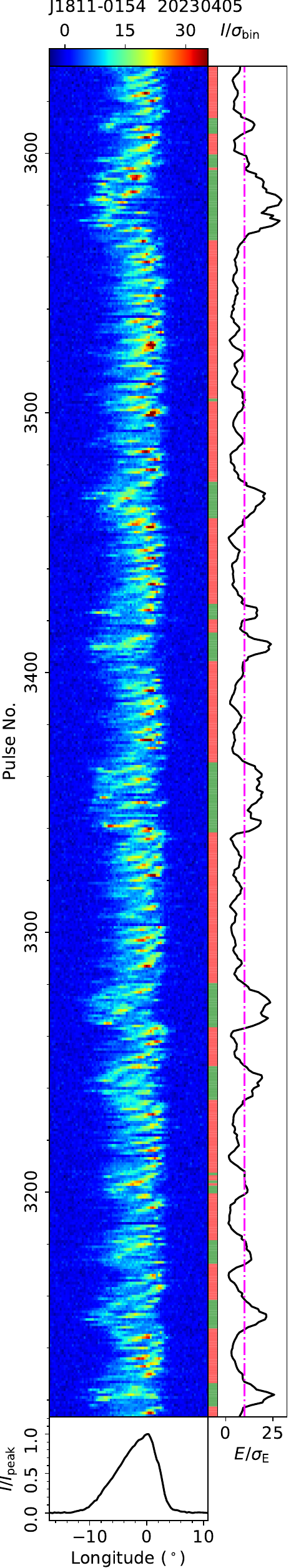}
    \includegraphics[height=0.92\textheight]{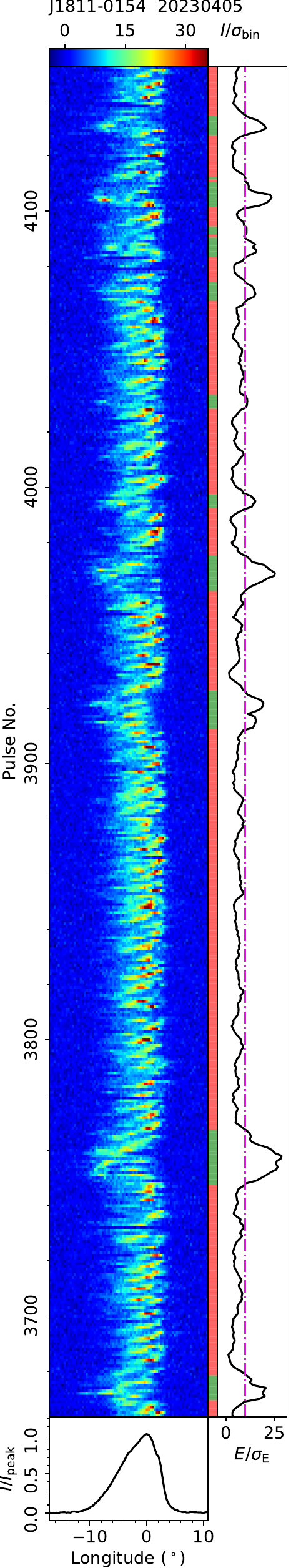}
\figcaption{Continued and ended.}
\end{figure}

\begin{figure}[htpb]
\centering
\includegraphics[width=0.22\textwidth, angle=0]{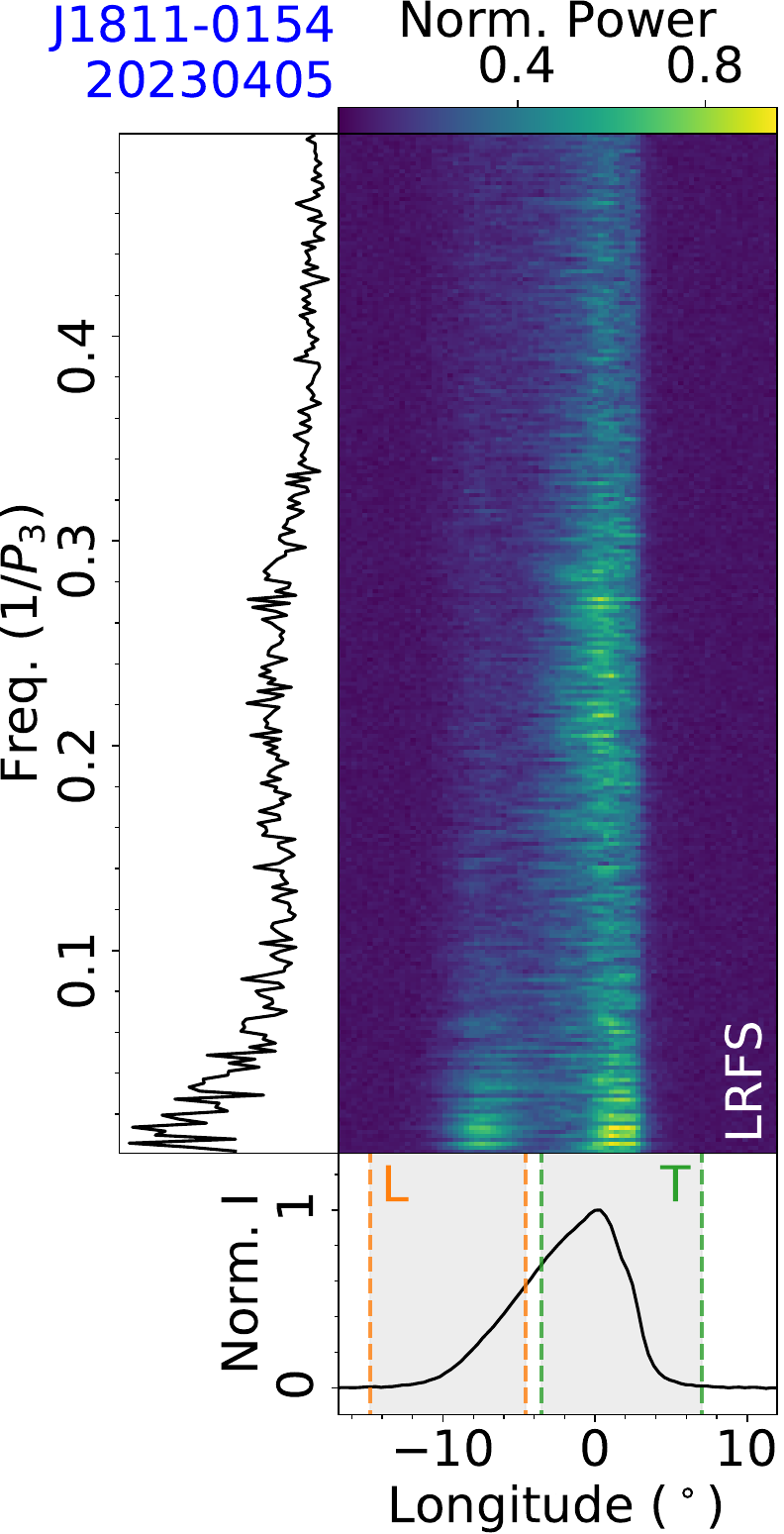}
\includegraphics[width=0.22\textwidth, angle=0]{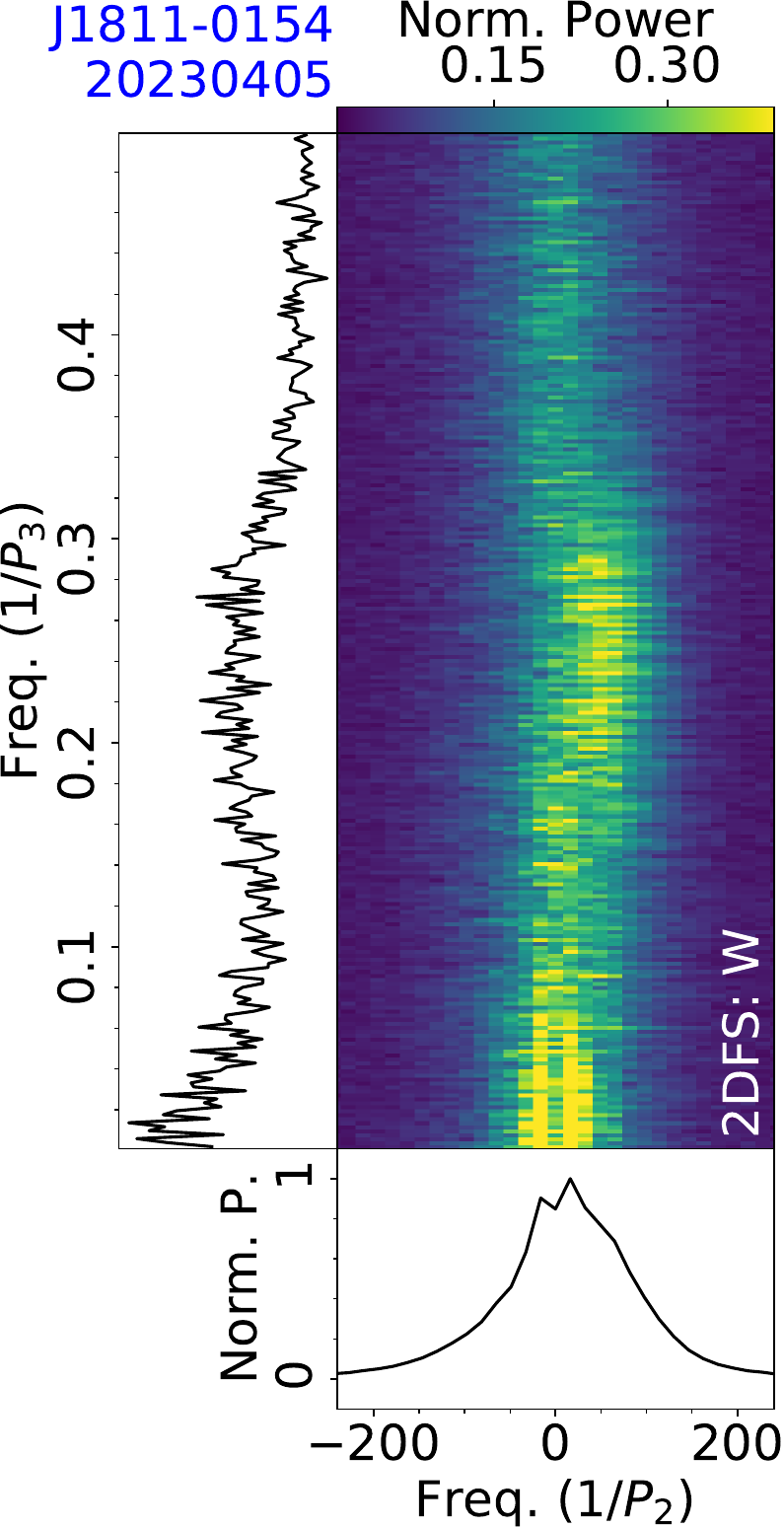}\\
\includegraphics[width=0.22\textwidth, angle=0]{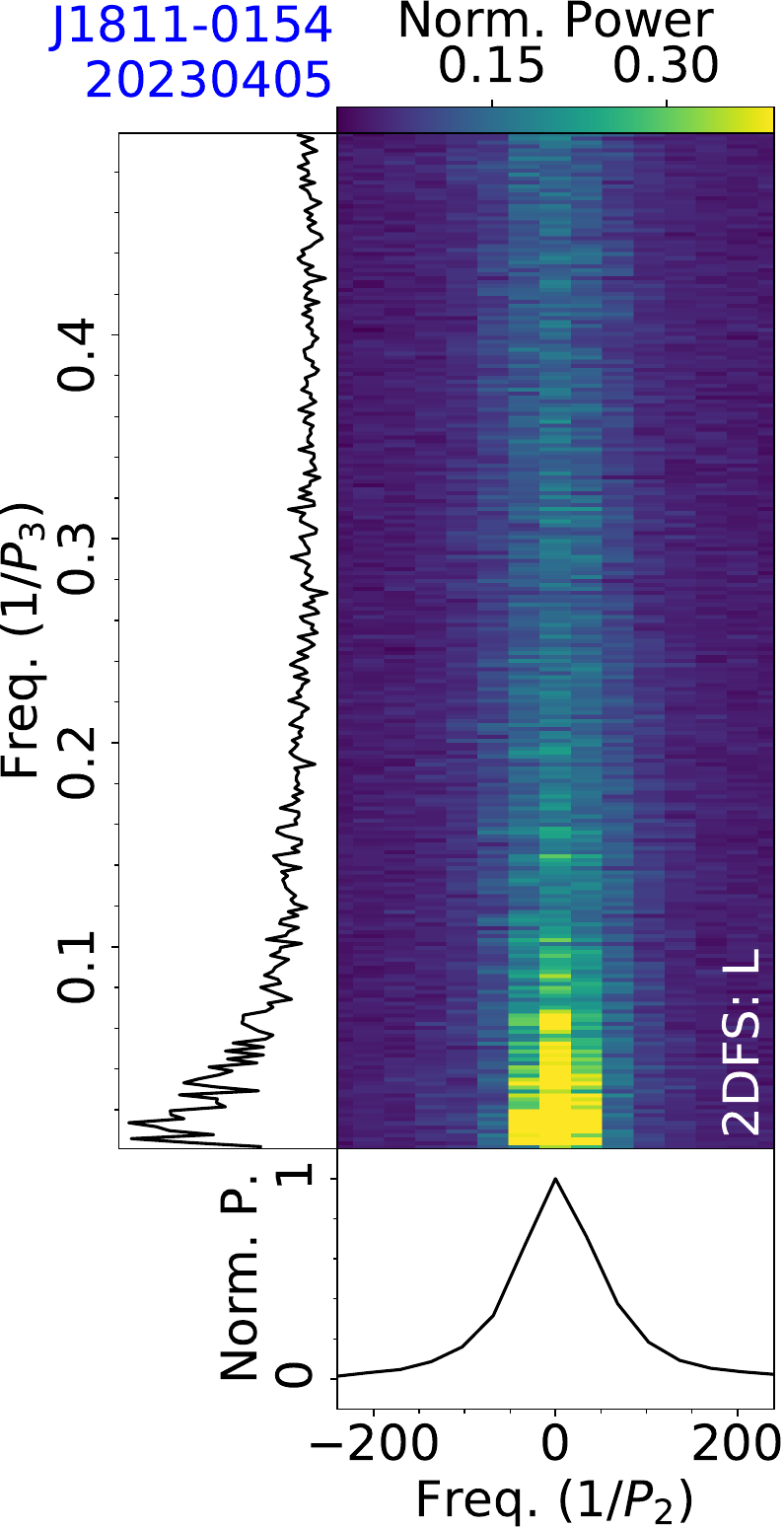}
\includegraphics[width=0.22\textwidth, angle=0]{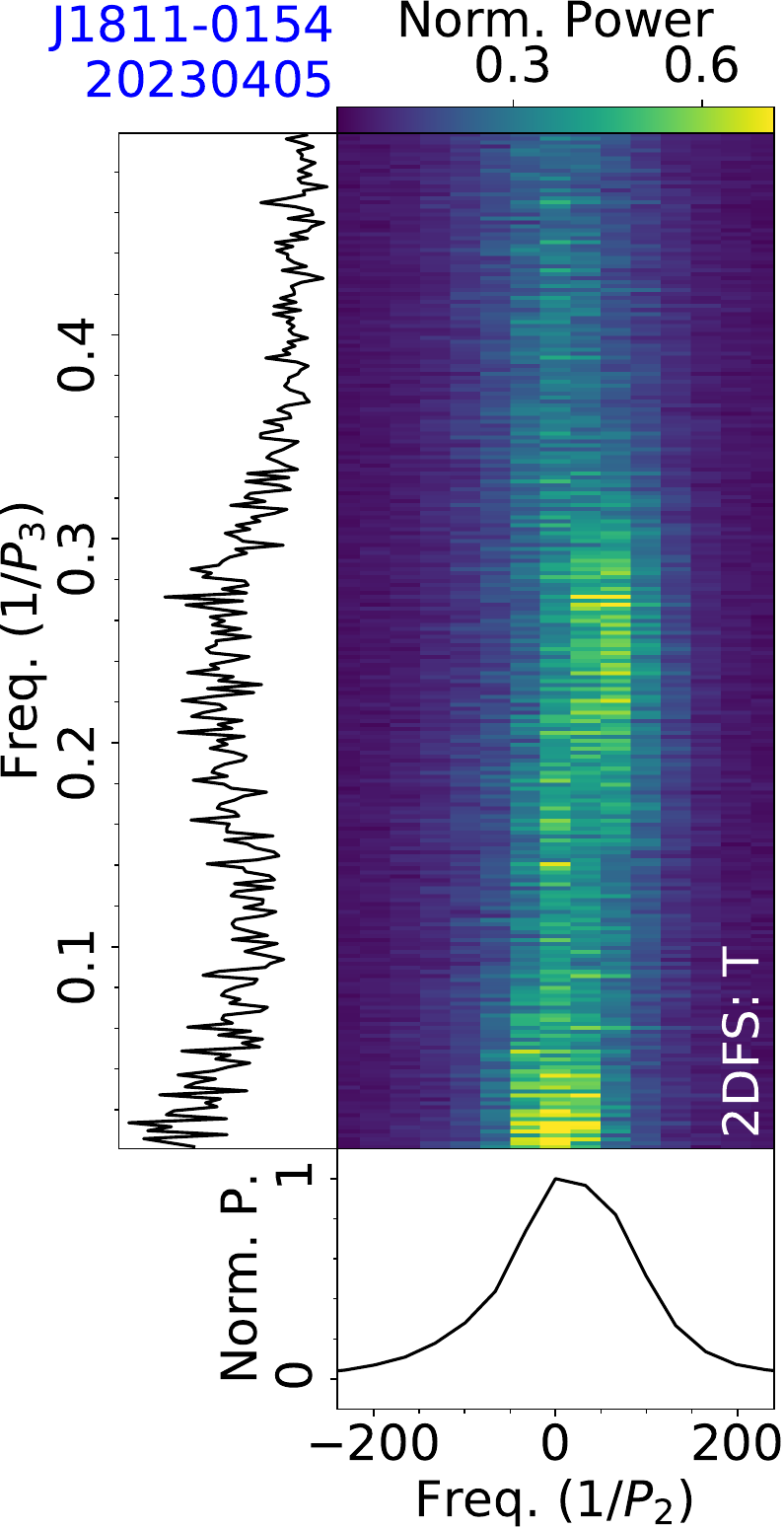}
\figcaption{Fluctuation analysis of PSR J1811-0154 for the observation on 20211123, with LRFS (top-left), and 2DFS for the on-pulse region (top-right), leading part (bottom-left) and trailing part (bottom-right) of a mean pulse profile. \label{subfig:fluctu:J1811-0154}}
\end{figure}

\begin{figure}[htpb]
\centering
\includegraphics[width=0.39\textwidth, angle=0]{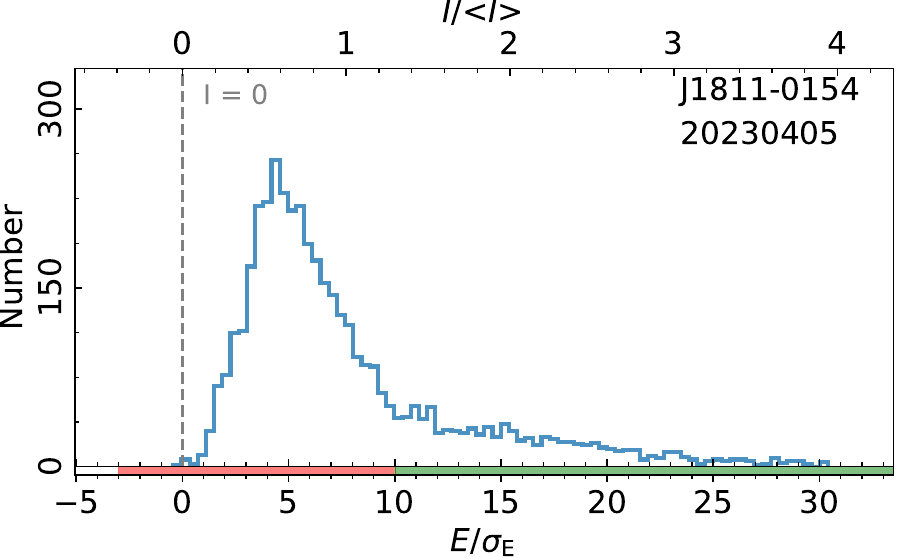}
\figcaption{On-pulse energy histogram of single pulses of PSR J1811-0154 from the FAST observation on 20210111. \label{subfig:Hist:J1811-0154}}
\end{figure}

\begin{figure}[htpb]
\centering
\includegraphics[width=0.37\textwidth, angle=0]{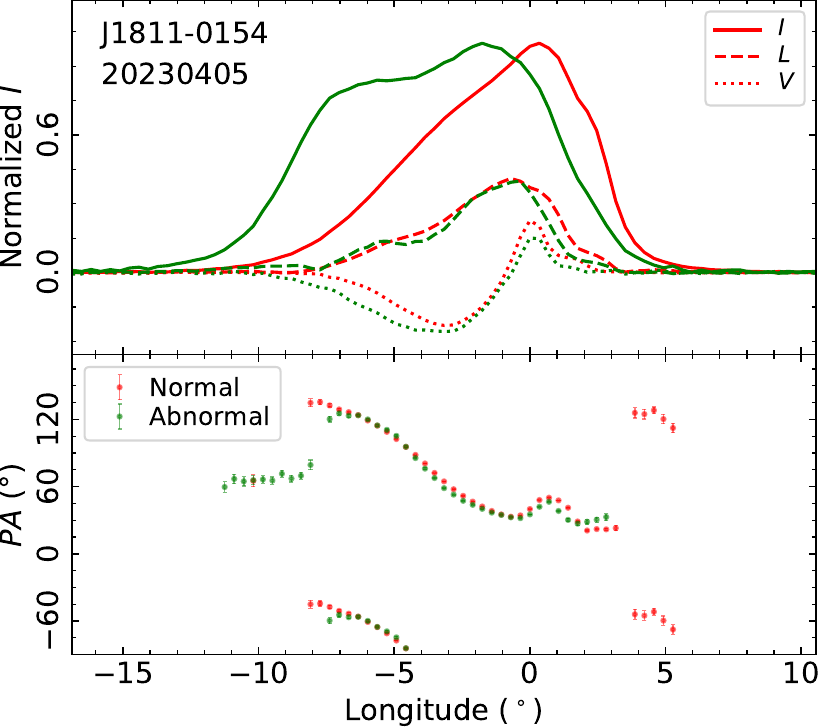}
\figcaption{Mean polarization profiles (the top panel) for the normal and abnormal emission modes of PSR J1811-0154 from the FAST observation on 20200401, as well as the averaged PA curves (the bottom panel). Profiles in the top panel are normalized by the respective peaks. \label{subfig:PolModes:J1811-0154}}
\end{figure}

\subsection{J1809-0119}
\label{subsec:J1809-0119}

PSR J1809-0119 was discovered by \citet{Keith2010} using the Parkes Radio Telescope. The subpulse drifting with parameters of $P_3=3.57\pm0.05$ periods and $P_2=-4.1^{+0.2}_{-2}$ degrees was reported by \citet{Song2023}.

The pulsar was observed by FAST on 20210118 for 9 minutes, deriving a rotation period $P=0.7450$~s and a dispersion measure $D\!M=138.8~{\rm cm^{-3}\,pc}$ from this observation. Single pulse sequences in Fig.~\ref{subfig:TP:J1809-0119} display subpulse drifting phenomenon. 
Drifting parameters are obtained from fluctuation spectra in Fig.~\ref{subfig:fluctu:J1809-0119}. 
For the leading profile part, the centroid of the drift feature in 2DFS has modulation frequencies of $1/P_3=0.279\pm0.001$ and $1/P_2=-99\pm3$, yielding $P_3=3.58\pm0.01$ periods and $P_2=-3.6\pm0.1^\circ$. 
For the trailing part in a mean pulse profile, the modulation frequencies are estimated to be $1/P_3=0.284\pm0.001$ and $1/P_2=-57\pm3$, corresponding to $P_3=3.52\pm0.01$ periods and $P_2=-6.3\pm0.3^\circ$. 
The subpulses related to the trailing component drift slightly faster than the leading component.

\subsection{J1810-0100g}
\label{subsec:J1810-0100g}

PSR J1810-0100g was discovered in the FAST GPPS survey \citep{Han2021,han2025}. 

This pulsar was observed by FAST on 20230527 for 15 minutes, deriving a rotation period $P=1.1061$~s and a dispersion measure $D\!M=79.8~{\rm cm^{-3}\,pc}$.
Single pulse sequences of this observation (Fig.~\ref{subfig:TP:J1810-0100g}) show nulling and subpulse drifting phenomena. The nulling fraction is estimated from the on-pulse integral energy histogram (Fig.~\ref{subfig:Hist:J1810-0100g}) to be 6.1$\pm$5\%. 
From LRFS and 2DFS of the leading profile part in Fig.~\ref{subfig:fluctu:J1810-0100g}, the positive drift feature has centroid modulation frequencies of $1/P_3=0.409\pm0.001$ and $1/P_2=60\pm4$, which correspond to drifting parameters $P_3=2.443\pm0.005$ periods and $P_2=6.0\pm0.4^\circ$.

\begin{figure}[htpb]
\centering
\includegraphics[width=0.22\textwidth, angle=0]{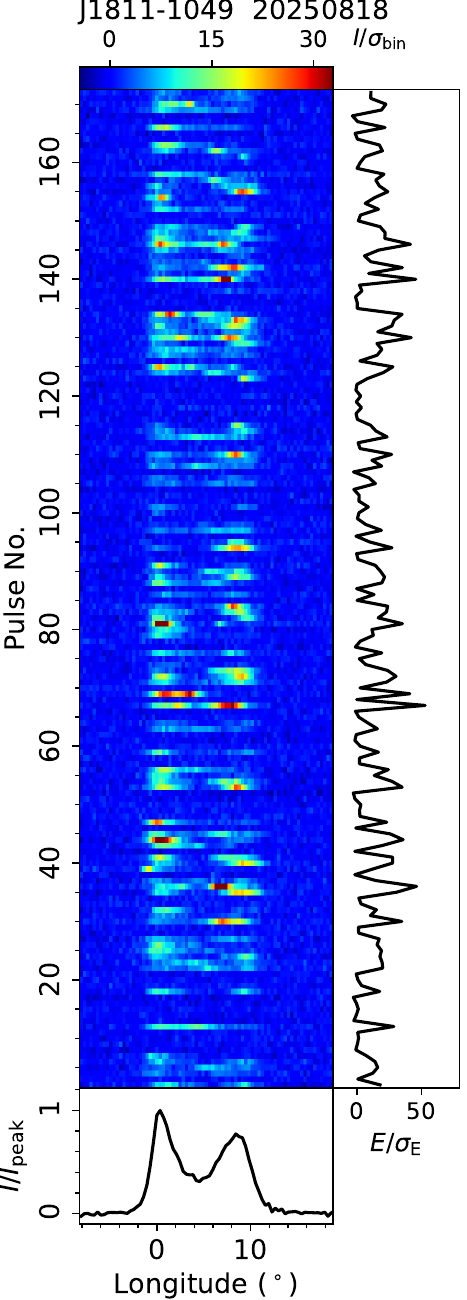}
\includegraphics[width=0.22\textwidth, angle=0]{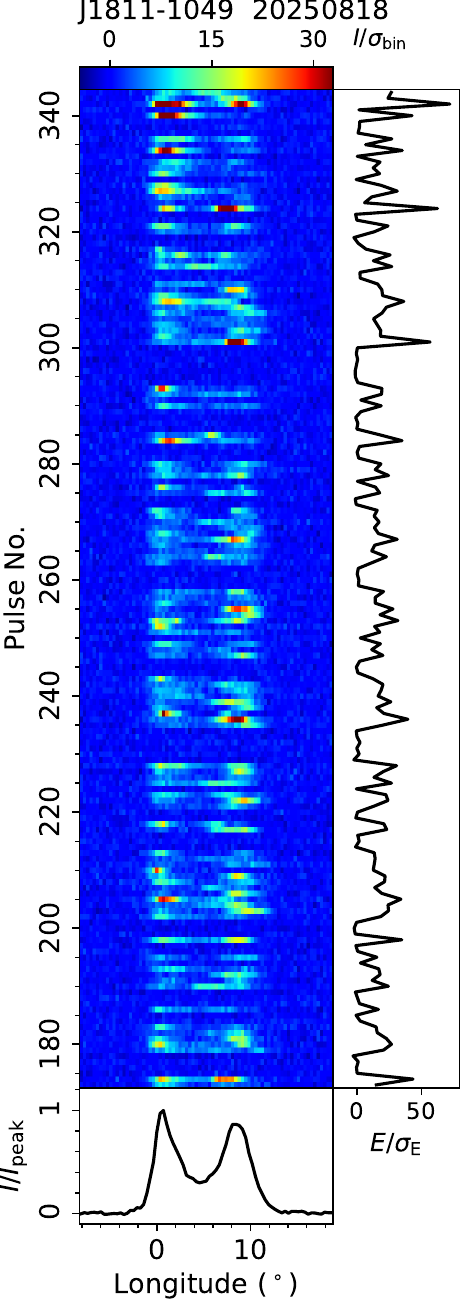}
\figcaption{Single pulse sequences of PSR J1811-1049 from the FAST observation on 20250818.
\label{subfig:TP:J1811-1049}}
\end{figure}

\begin{figure}[htpb]
\centering
\includegraphics[width=0.39\textwidth, angle=0]{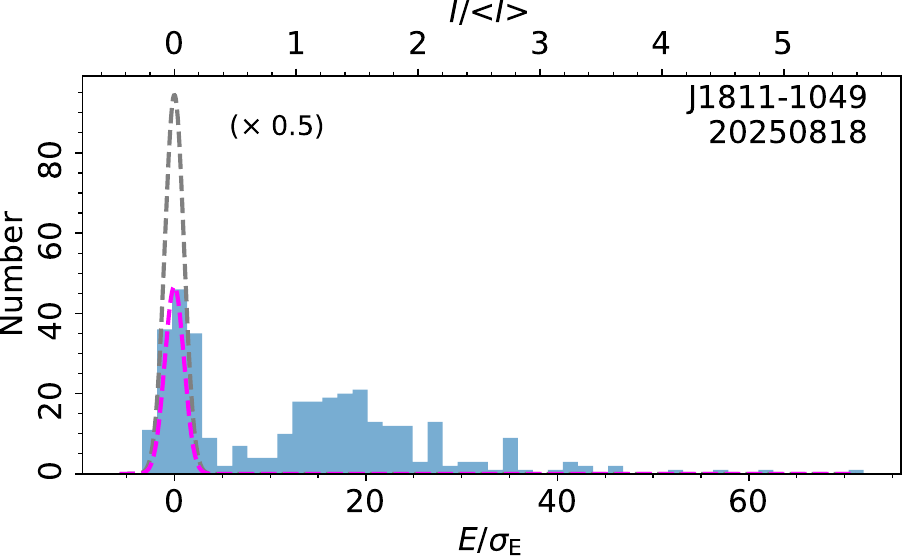}
\figcaption{On-pulse energy histogram of single pulses of PSR J1811-1049 from the FAST observation on 20250818. \label{subfig:Hist:J1811-1049}}
\end{figure}

\begin{figure}[htpb]
\centering
\includegraphics[width=0.22\textwidth, angle=0]{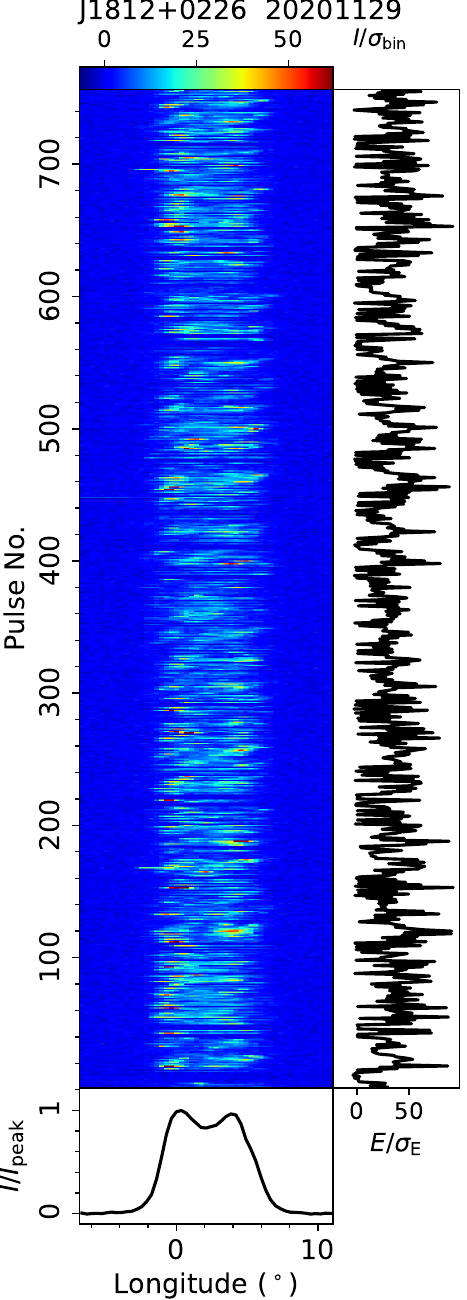}
\includegraphics[width=0.22\textwidth, angle=0]{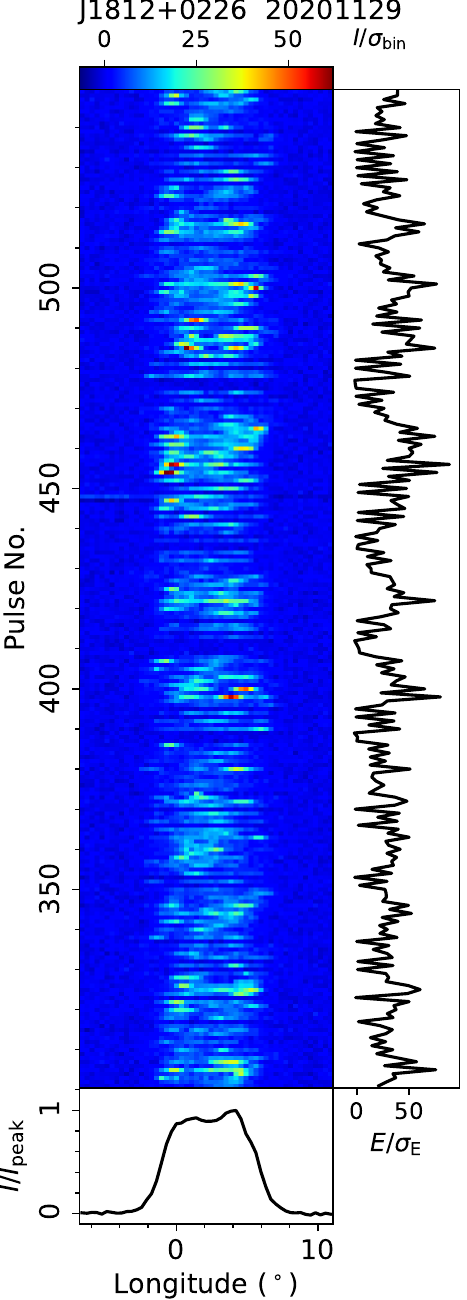}
\figcaption{Single pulse sequences of PSR J1812+0226 from the FAST observation on 20201129. \label{subfig:TP:J1812+0226}}
\end{figure}

\begin{figure}[htpb]
\centering
\includegraphics[width=0.39\textwidth, angle=0]{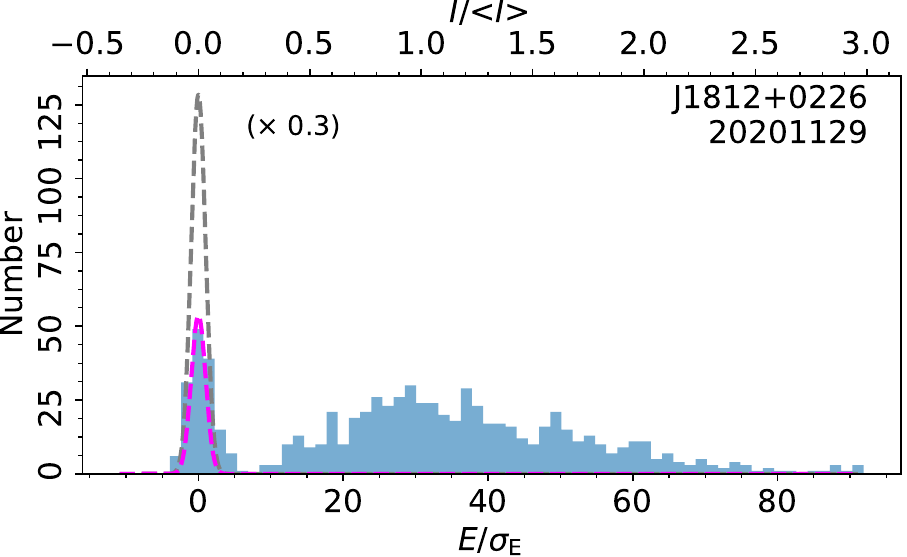}
\figcaption{On-pulse energy histogram of single pulses of PSR J1812+0226 from the FAST observation on 20201129. \label{subfig:Hist:J1812+0226}}
\end{figure}

\begin{figure}[htpb]
\centering
\includegraphics[width=0.22\textwidth, angle=0]{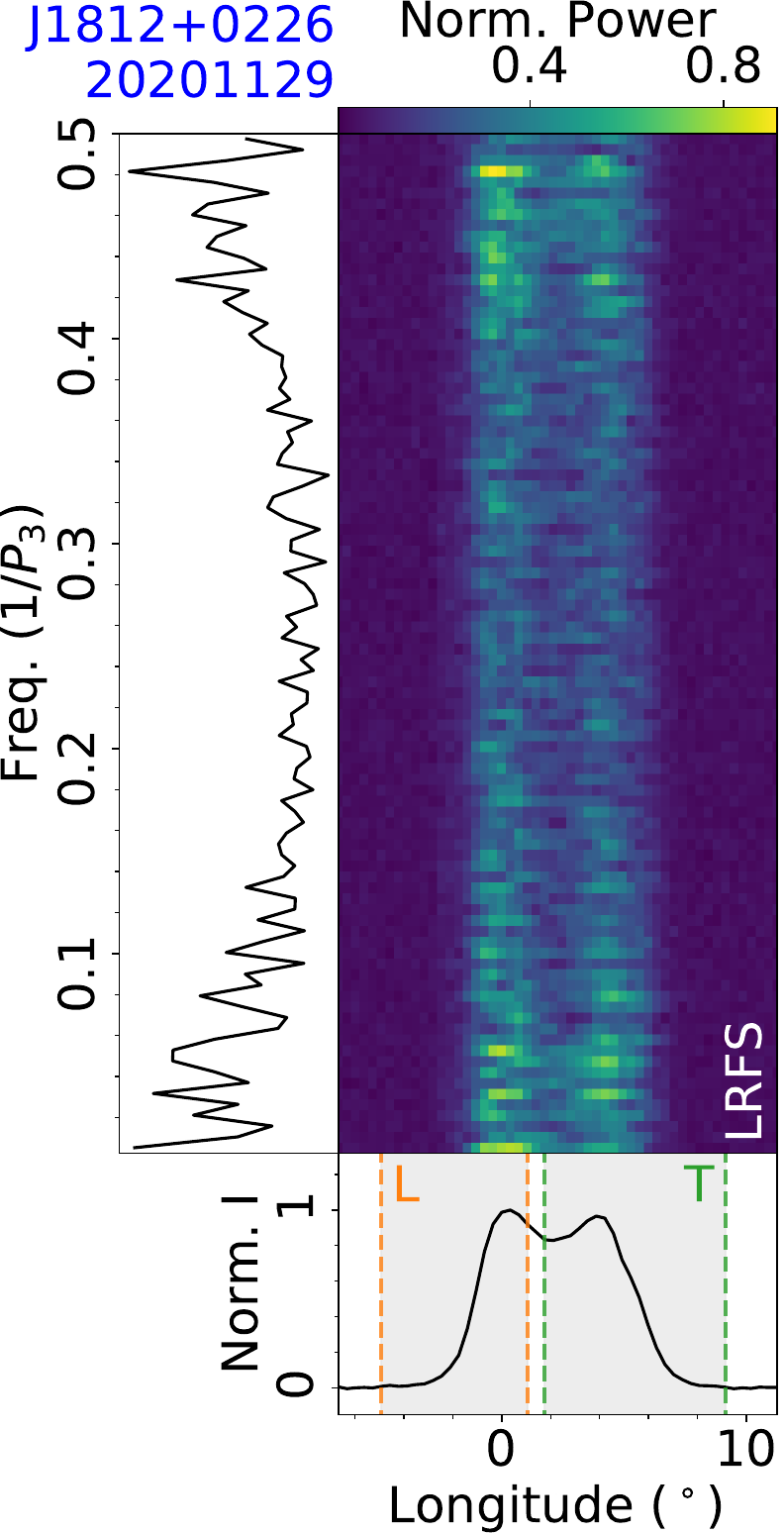}
\includegraphics[width=0.22\textwidth, angle=0]{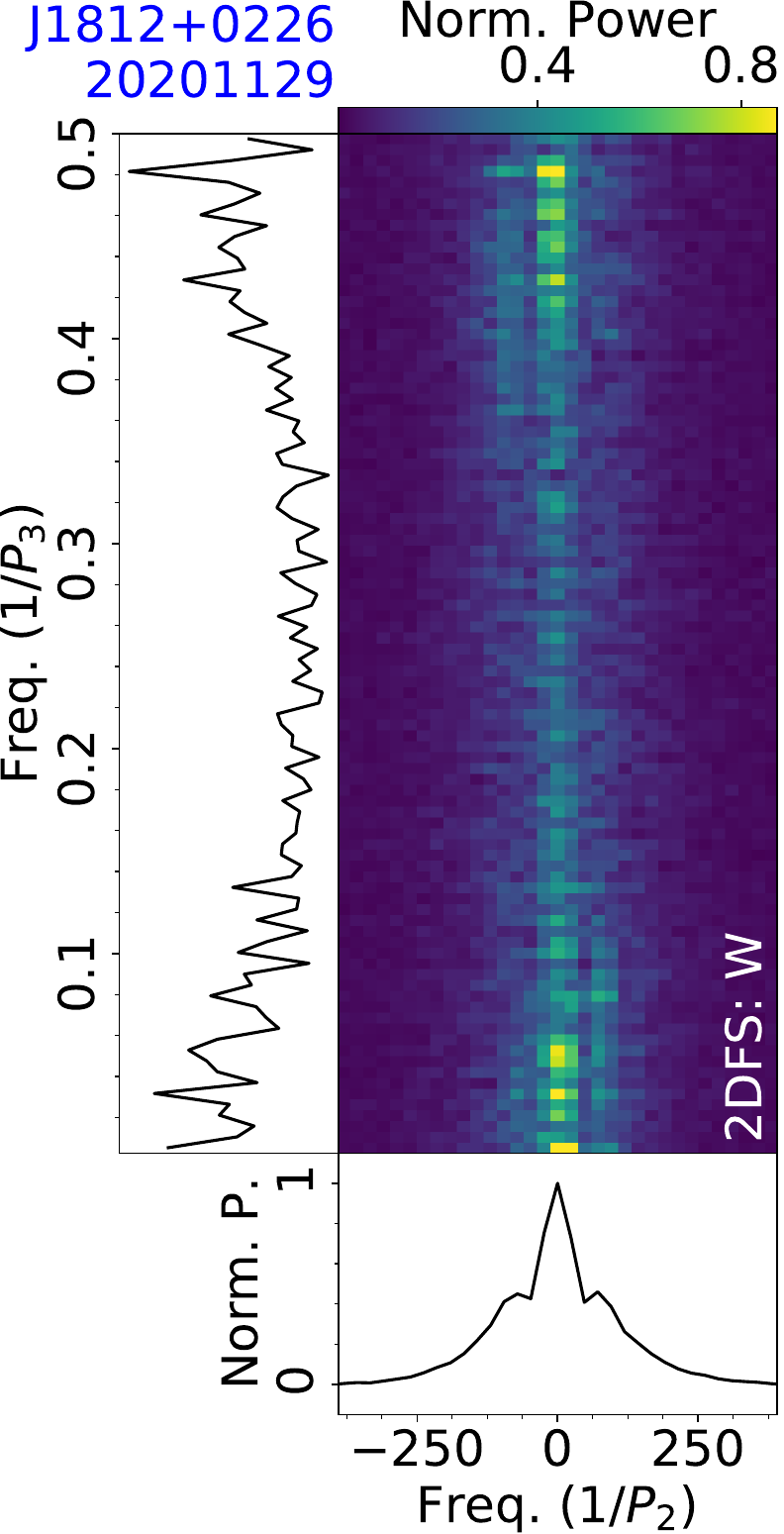}\\
\includegraphics[width=0.22\textwidth, angle=0]{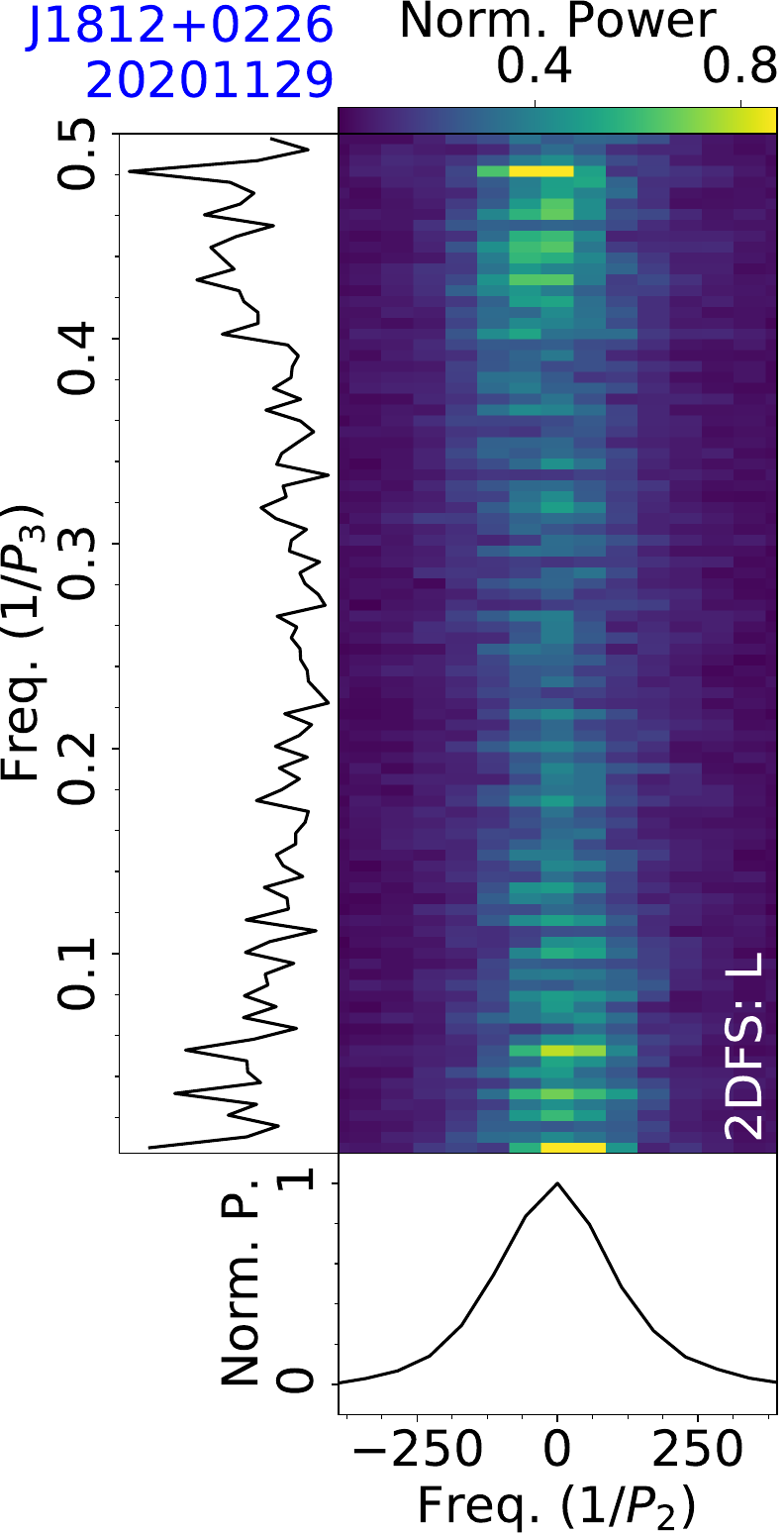}
\includegraphics[width=0.22\textwidth, angle=0]{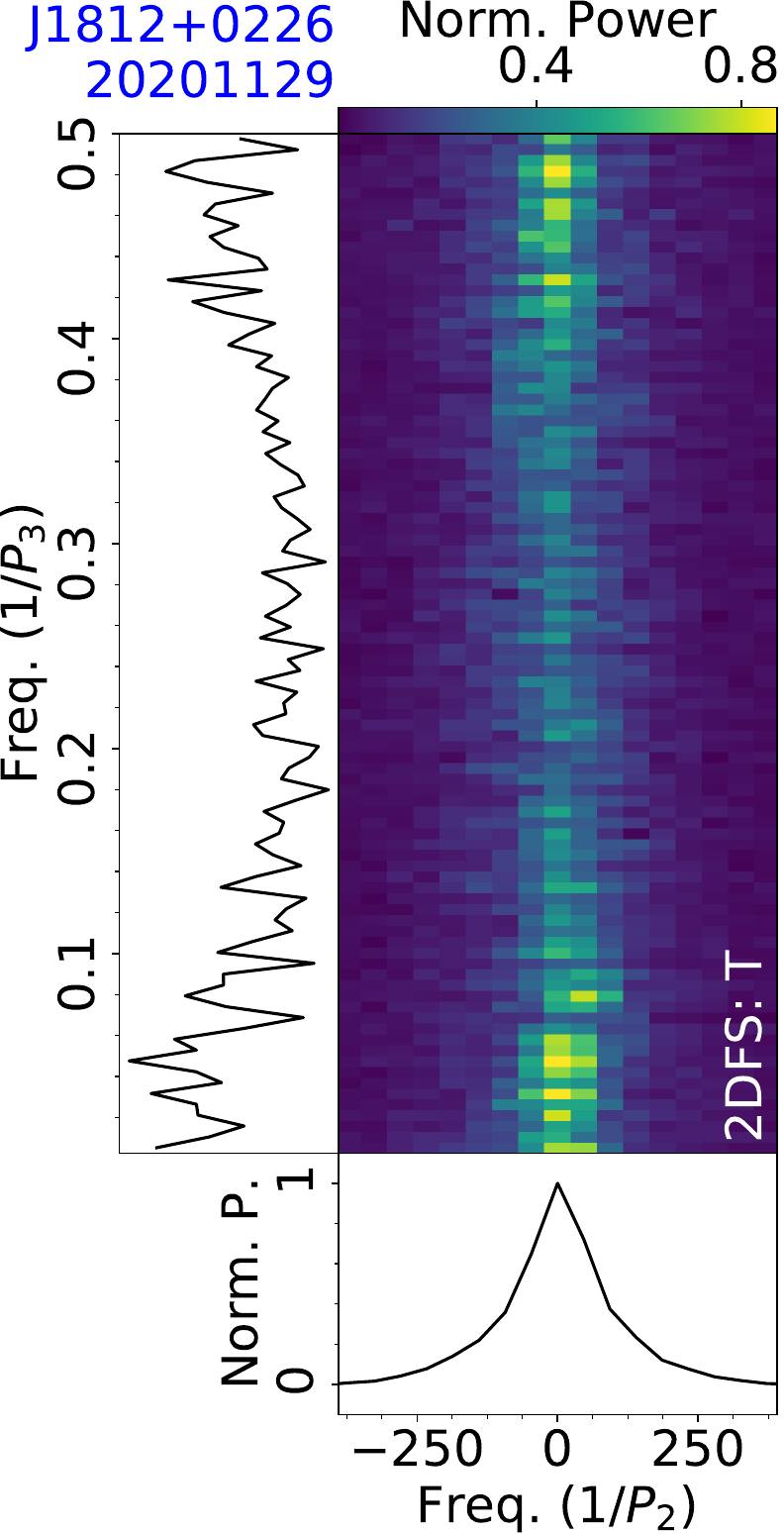}
\figcaption{Fluctuation analysis of PSR J1812+0226 from the FAST observation on 20201129, with LRFS (top-left), and 2DFS for the on-pulse region (top-right), leading part (bottom-left) and trailing part (bottom-right) of a mean pulse profile.  \label{subfig:fluctu:J1812+0226}}
\end{figure}

\subsection{J1811-0154}
\label{subsec:J1811-0154}

PSR J1811-0154 was discovered by the Parkes 64-m radio telescope \citep{Edwards2001}. Subpulse drifting was reported by \citep{Song2023} of $P_3=3.9\pm0.2$ periods and $P_2=9^{+3}_{-2}$ degrees.

This pulsar was observed by FAST on 20230405 for 64 minutes, deriving a rotation period $P=0.9249$~s and a dispersion measure $D\!M=147.5~{\rm cm^{-3}\,pc}$ from this observation. Single pulse sequences of the observation shown in Fig.~\ref{subfig:TP:J1811-0154} illustrate the existence of subpulse drifting and mode changing behavior. 
Fluctuation spectra are displayed in Fig.~\ref{subfig:fluctu:J1811-0154}. There is a low-frequency modulation of $\sim$27 periods for the on-pulse phase region. 
For the trailing part of a mean pulse profile, the positive drift feature widely temporally distributed has the centroid of $1/P_3=0.235\pm0.001$ and $1/P_2=53\pm1$, which correspond to $P_3=4.26\pm0.01$ periods and $P_2=6.8\pm0.1^\circ$. 
Single pulses of two drifting modes could be distinguished from the on-pulse energy histogram of the leading edge in the profile (Fig.~\ref{subfig:Hist:J1811-0154}). Normal and abnormal modes are labeled by red and green colors, respectively. 
From single pulse sequences (Fig.~\ref{subfig:TP:J1811-0154}) and contrast of averaged polarization profiles (Fig.~\ref{subfig:PolModes:J1811-0154}), the abnormal mode shifts to an earlier phase relative to the normal mode, with the intensity of the leading edge enhanced and the trailing edge reduced. The PA curves of two drifting modes are consistent in the longitude range of $-7^\circ$ to 0$^\circ$.

\subsection{J1811-1049}
\label{subsec:J1811-1049}

PSR J1811-1049 was discovered by \citet{Knispel2013} using the Parkes multi-beam pulsar survey (PMPS) data.

This pulsar was observed by FAST on 20250818 for 15 minutes, with a rotation period $P=2.6241$~s and a dispersion measure $D\!M=253.4~{\rm cm^{-3}\,pc}$ derived. Single pulse sequences and integral energy histogram in Fig.~\ref{subfig:TP:J1811-1049} and \ref{subfig:Hist:J1811-1049} illustrating the existence of nulling phenomenon. The nulling fraction of this observation is estimated to be 24.7$\pm$3.6\%.

\begin{figure}[htpb]
\centering
\includegraphics[width=0.21\textwidth, angle=0]{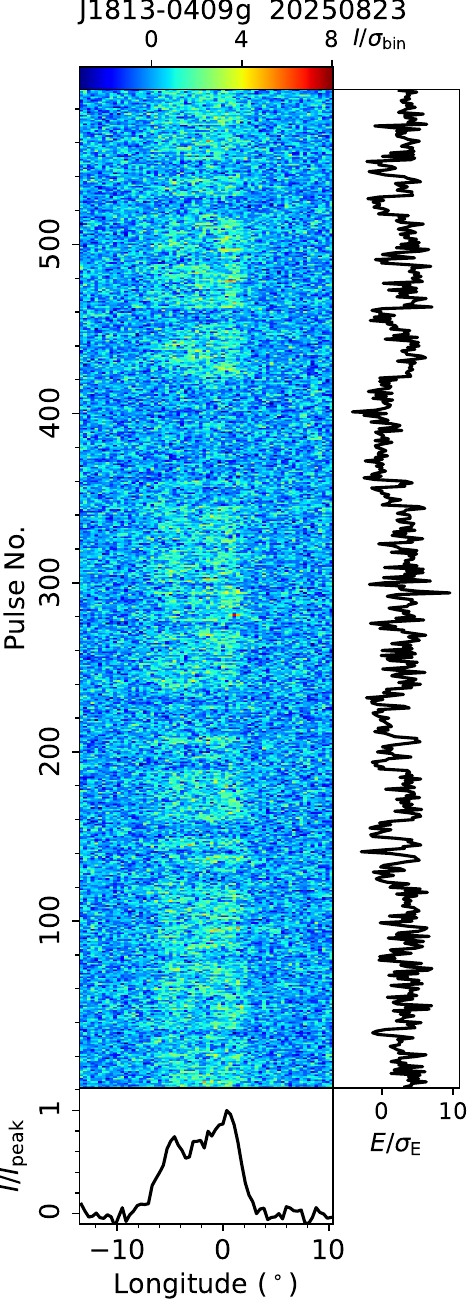}
\includegraphics[width=0.21\textwidth, angle=0]{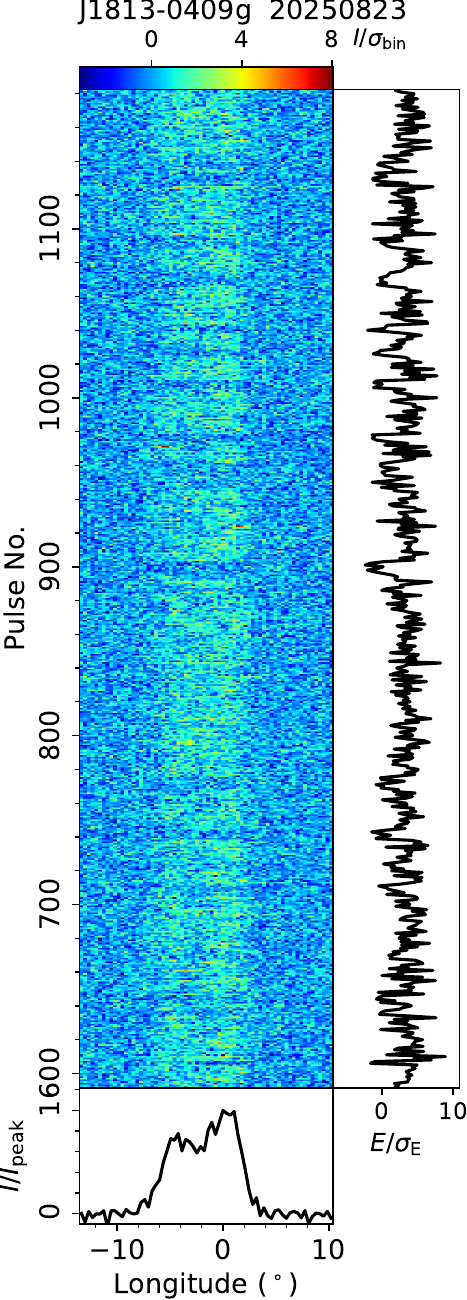}
\figcaption{Single pulse sequences of PSR J1813-0409g from the FAST observation on 20250823.
\label{subfig:TP:J1813-0409g}}
\end{figure}

\begin{figure}[htpb]
\centering
\includegraphics[width=0.39\textwidth, angle=0]{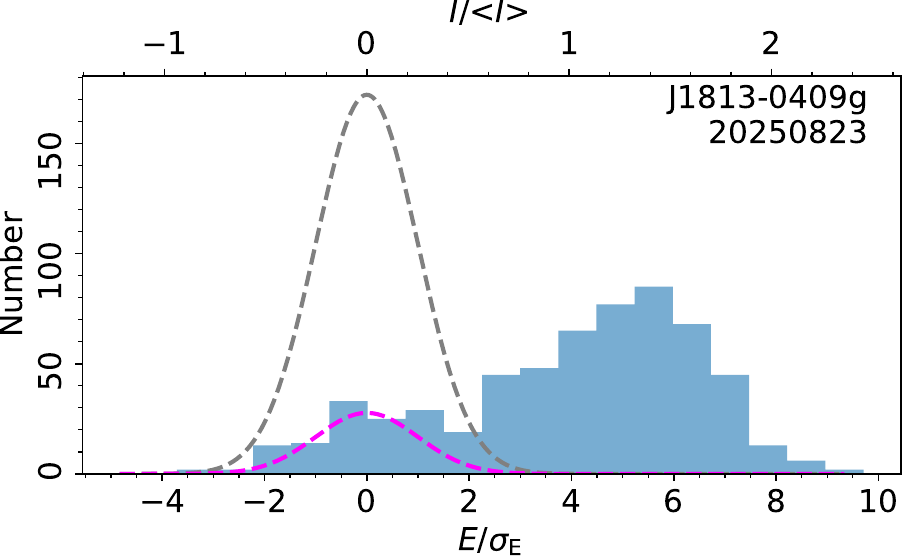}
\figcaption{On-pulse energy histogram for single pulses of PSR J1813-0409g from the FAST observation on 20250823, with energy values averaged over every 2 periods. 
\label{subfig:Hist:J1813-0409g}}
\end{figure}

\begin{figure}[htpb]
\centering
\includegraphics[width=0.44\textwidth, angle=0]{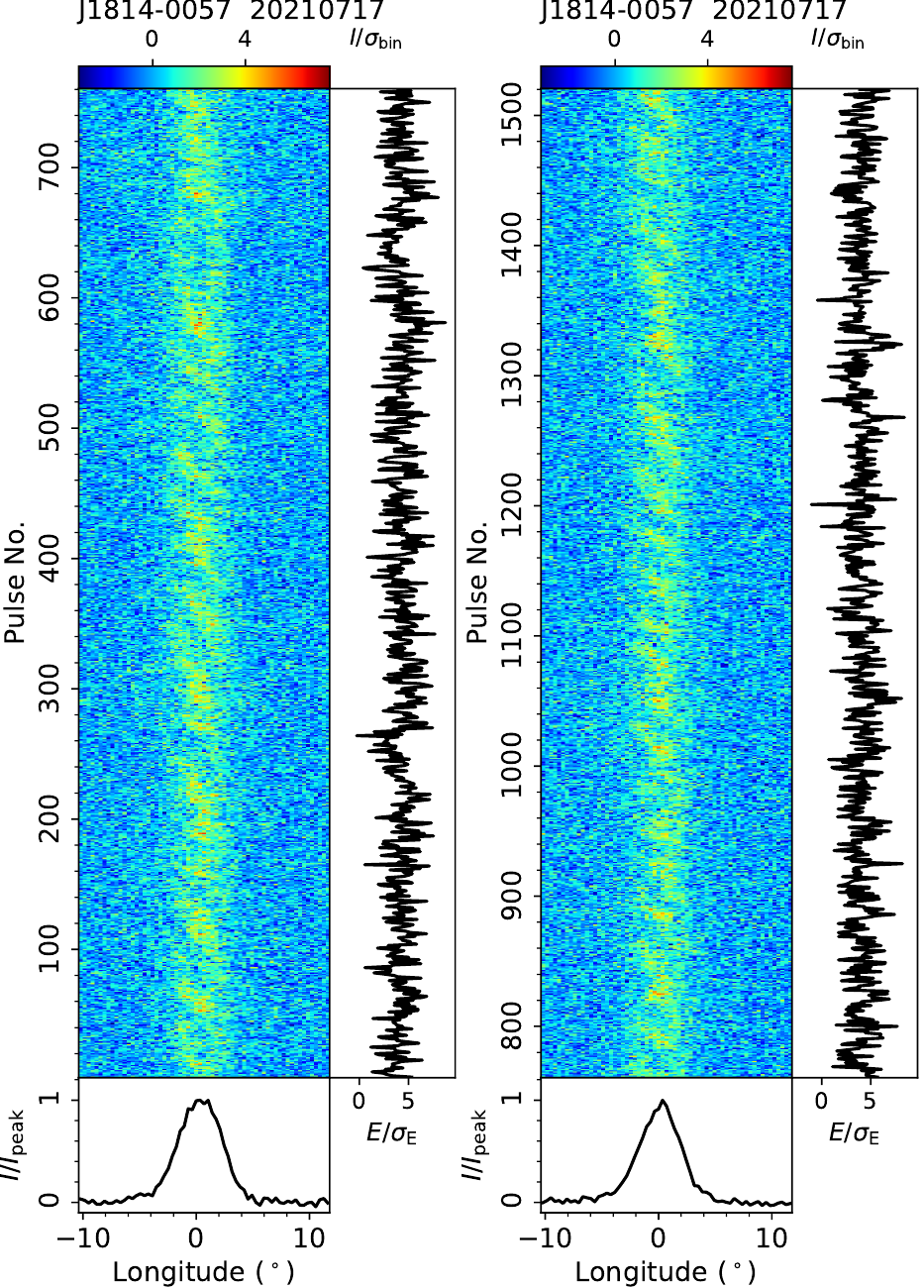}
\figcaption{Single pulse sequences of PSR J1814-0057 from the FAST observation on 20210717.
\label{subfig:TP:J1814-0057}}
\end{figure}

\begin{figure}[htpb]
\centering
\includegraphics[width=0.44\textwidth, angle=0]{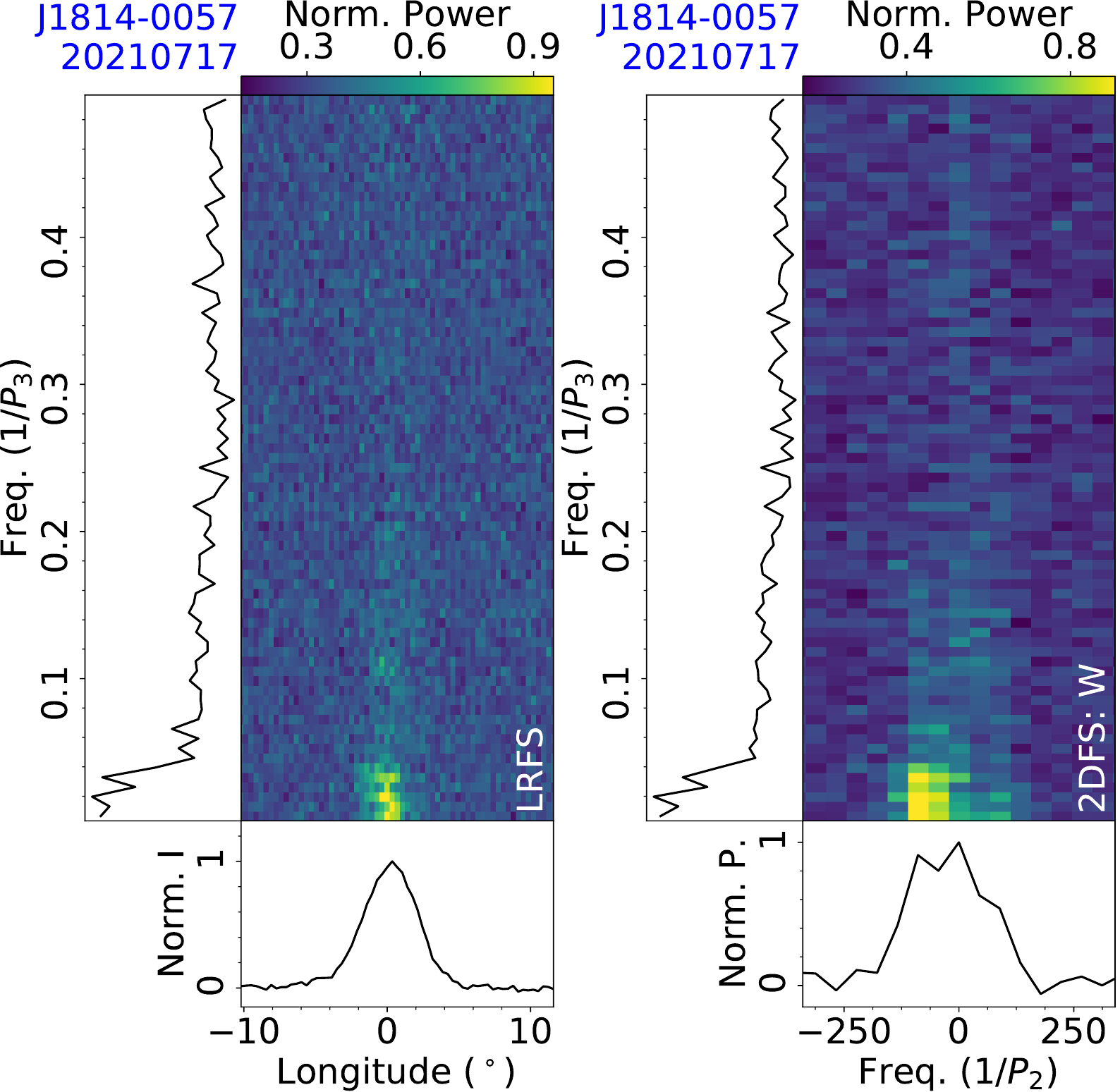}
\figcaption{Fluctuation analysis of PSR J1814-0057 for the observation on 20210717, with LRFS and 2DFS for the on-pulse region of a mean pulse profile.
\label{subfig:fluctu:J1814-0057}}
\end{figure}

\begin{figure}[htpb]
\centering
\includegraphics[width=0.21\textwidth, angle=0]{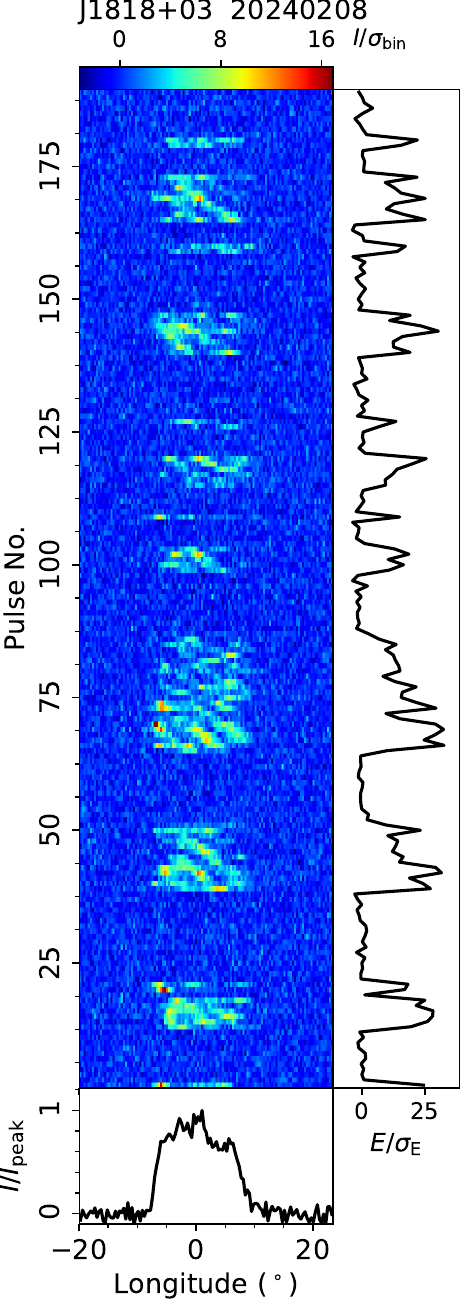}
\includegraphics[width=0.21\textwidth, angle=0]{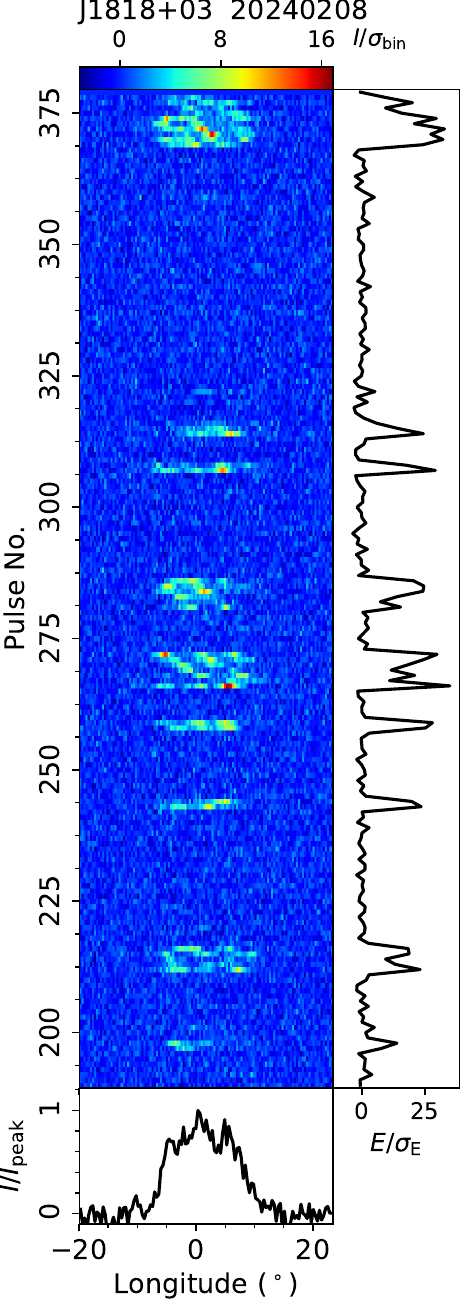}
\figcaption{Single pulse sequences of PSR J1818+03 from the FAST observation on 20240208.
\label{subfig:TP:J1818+03}}
\end{figure}

\begin{figure}[htpb]
\centering
\includegraphics[width=0.39\textwidth, angle=0]{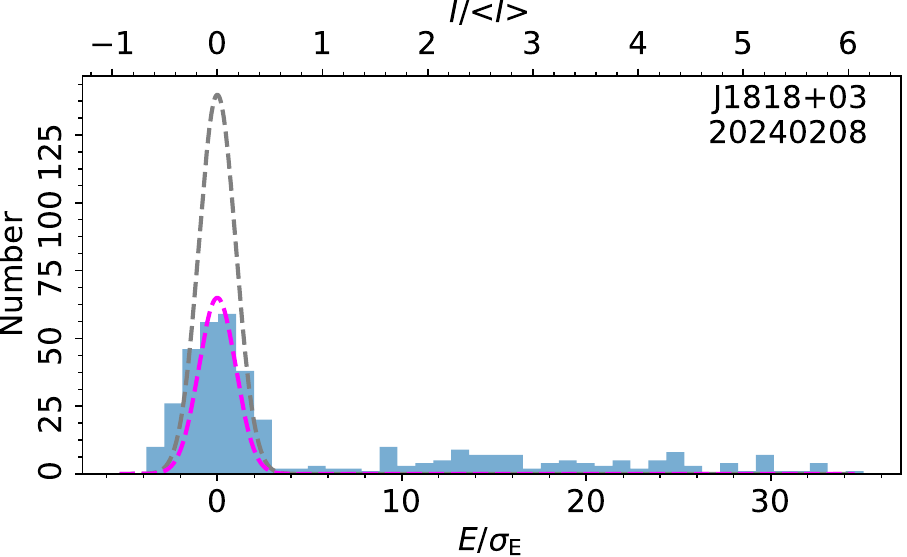}
\figcaption{On-pulse energy histogram of PSR J1818+03 from the FAST observation on 20240208.
\label{subfig:Hist:J1818+03}}
\end{figure}

\begin{figure}[htpb]
\centering
\includegraphics[width=0.22\textwidth, angle=0]{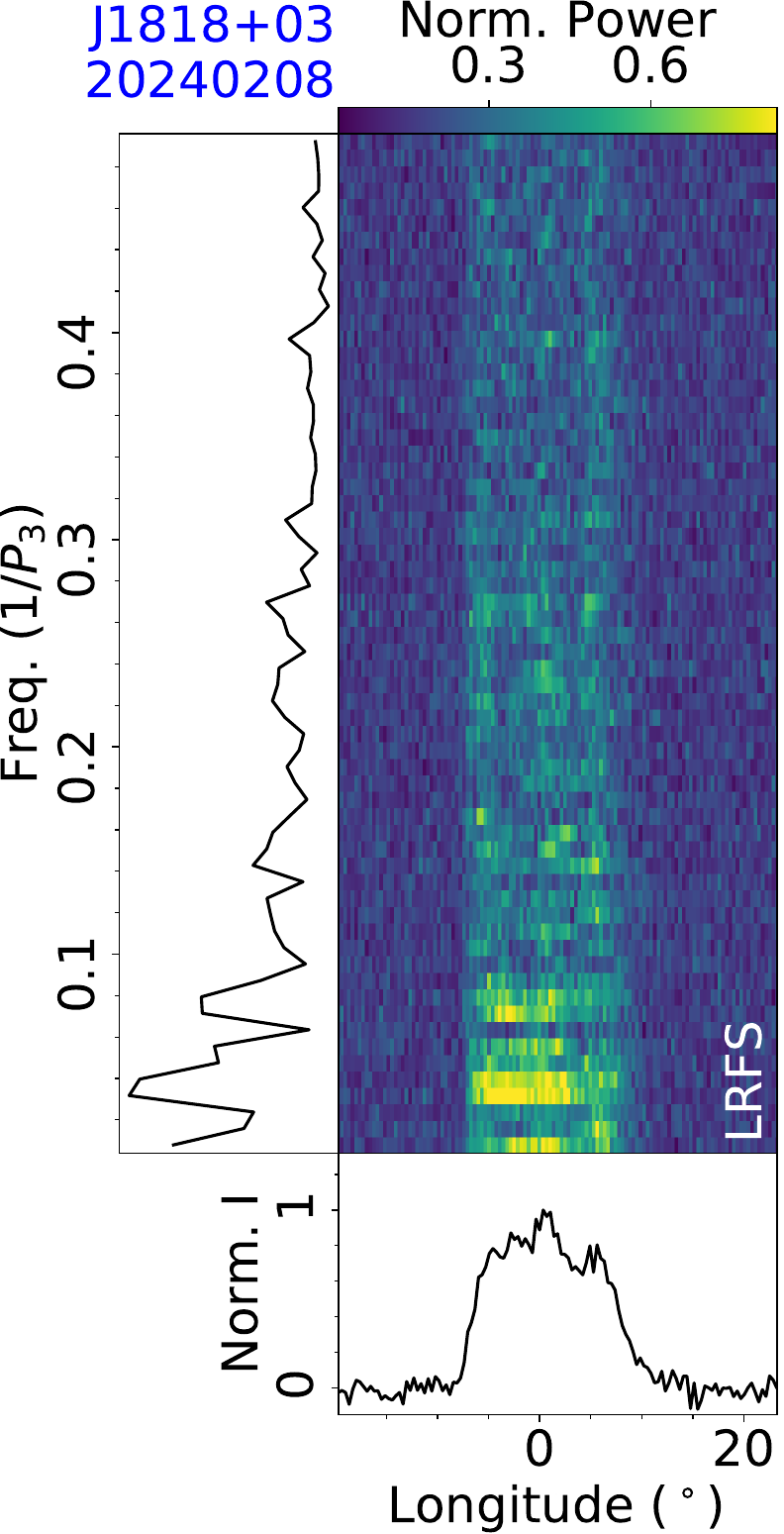}
\includegraphics[width=0.22\textwidth, angle=0]{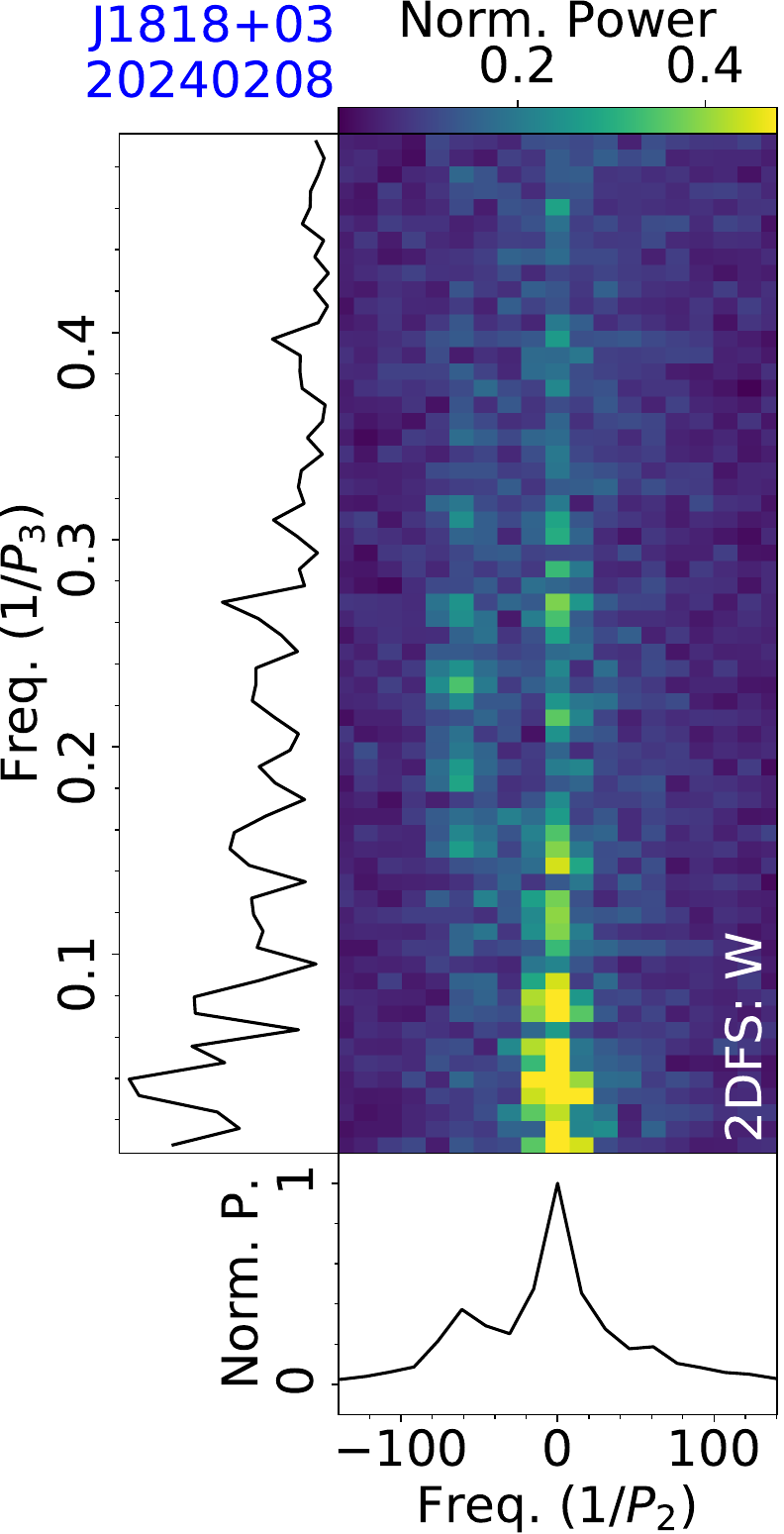}
\figcaption{Fluctuation analysis of PSR J1818+03 for the observation on 20240208, with LRFS and 2DFS for the on-pulse region of a mean pulse profile.
\label{subfig:fluctu:J1818+03}}
\end{figure}

\subsection{J1812+0226}
\label{subsec:J1812+0226}

PSR J1812+0226 was discovered by \citet{Stokes1985} using the 92-m telescope at Green Bank at 390 MHz. \citet{Song2023} has reported the drifting behaviors of two components, and both components have two modulation features. One component has a negative drift feature of $P_3=2.13\pm0.08$ periods and $P_2=-37^{+26}_{-22}$ degrees, and a positive drifting behavior of $P_3=18\pm9$ periods and $P_2=28^{+19}_{-20}$ degrees. The other component has the positive drifting behavior of $P_3=17\pm25$ periods and $P_2=27^{+5}_{-16}$, as well as a $P_3$-only feature of $P_3=2.21\pm0.07$ periods.

This pulsar was observed by FAST on 20201129 for 10 minutes, deriving a rotation period $P=0.7940$~s and a dispersion measure $D\!M=103.8~{\rm cm^{-3}\,pc}$ from this observation. 
Single pulse sequences are shown in Fig.~\ref{subfig:TP:J1812+0226}. 
The pulsar has both nulling and modulation phenomena. The nulling fraction of this observation is estimated to be 13$\pm$1\% from the on-pulse integral energy histogram (Fig.~\ref{subfig:Hist:J1812+0226}). 
Fluctuation spectra are displayed in Fig.~\ref{subfig:fluctu:J1812+0226}. Leading and trailing components both have two modulation features in 2DFS, which is similar to the results of \citet{Song2023}. 
In 2DFS of the leading component, there is a negative drift feature with the centroid of $1/P_3=0.450\pm0.001$ ($P_3=2.22\pm0.01$ periods) and $1/P_2=-29\pm3$ ($P_2=-12\pm1^\circ$), as well as a temporally low-frequency modulation of $1/P_3=0.039\pm0.001$ ($P_3=26\pm1$ periods) with the positively phase modulated frequency of $1/P_2=34\pm4$ ($P_2=11\pm1^\circ$).
The 2DFS of the trailing profile part exhibits a positive drift feature, characterized by a temporal low-frequency modulation of $1/P_3=0.045\pm0.001$ ($P_3=22.3\pm0.5$ periods) and the phase modulated frequency $1/P_2=20\pm3$ ($P_2=18\pm2^\circ$). However, there is no preferred phase modulation direction for the temporally modulation frequency of $1/P_3=0.456\pm0.001$, corresponding to $P_3=2.19\pm0.01$ periods. From the FAST observation (Fig.~\ref{subfig:TP:J1812+0226}), the temporally low-frequency modulated feature with positively phase modulated seems superimposed on the 2-period modulation, which is significant and similar to the behavior of PSR J1857+0057 \citep{Yan2023_drift}.

\subsection{J1813-0409g}
\label{subsec:J1813-0409g}

PSR J1813-0409g was discovered in the FAST GPPS survey \citep{Han2021,han2025}.

This pulsar was observed by FAST on 20250823 for 15 minutes and 20251103 for 10 minutes. From the longer data, a rotation period $P=0.7604$~s and a dispersion measure $D\!M=159.2~{\rm cm^{-3}\,pc}$ were derived. 
Single pulse sequences in Fig.~\ref{subfig:TP:J1813-0409g} exhibit the temporal variations in energy. To improve the significance of pulses, on-pulse energy values are averaged over every two periods, and the energy histogram is shown in Fig.~\ref{subfig:Hist:J1813-0409g}. From the energy histogram, the nulling fraction of this observation is estimated to be 16.1$\pm$1.7\%.

\subsection{J1814-0057}
\label{subsec:J1814-0057}

PSR J1814-0057 was discovered in the Commensal Radio Astronomy FAST Survey (CRAFTS) (http://groups.bao.ac.cn/ism/CRAFTS/).

This pulsar was observed by FAST on 20210717 for 4 minutes, with a rotation period $P=0.1601$~s and a dispersion measure $D\!M=40.2~{\rm cm^{-3}\,pc}$ derived. Single pulse sequences in Fig.~\ref{subfig:TP:J1814-0057} show the subpulse drifting phenomenon. From the fluctuation spectra in Fig.~\ref{subfig:fluctu:J1814-0057}, the main drift feature has the centroid at $1/P_3=0.022\pm0.001$ and $1/P_2=-69\pm4$, corresponding to $P_3=45\pm1$ periods and $P_2=-5.2\pm0.3$ degrees.

\begin{figure}[htpb]
\centering
\includegraphics[width=0.21\textwidth, angle=0]{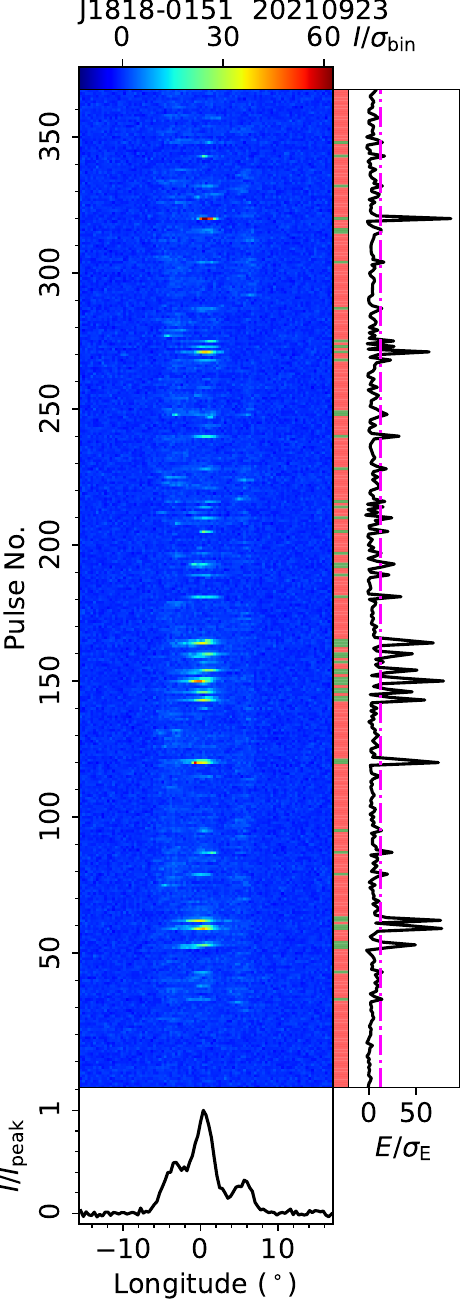}
\includegraphics[width=0.21\textwidth, angle=0]{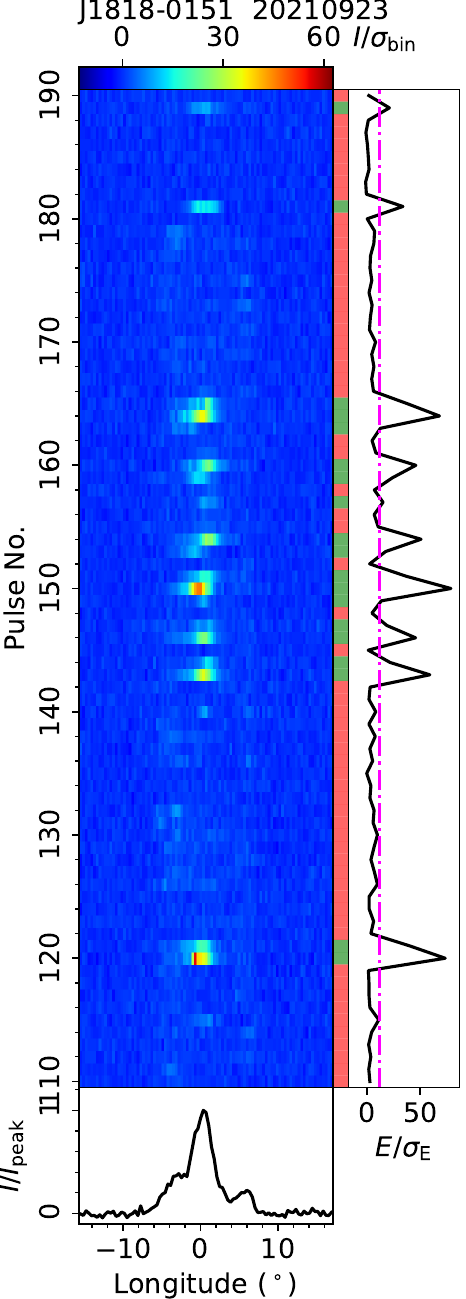}
\figcaption{Single pulse sequences of PSR J1818-0151 from the FAST observation on 20210923.
\label{subfig:TP:J1818-0151}}
\end{figure}

\begin{figure}[htpb]
\centering
\includegraphics[width=0.39\textwidth, angle=0]{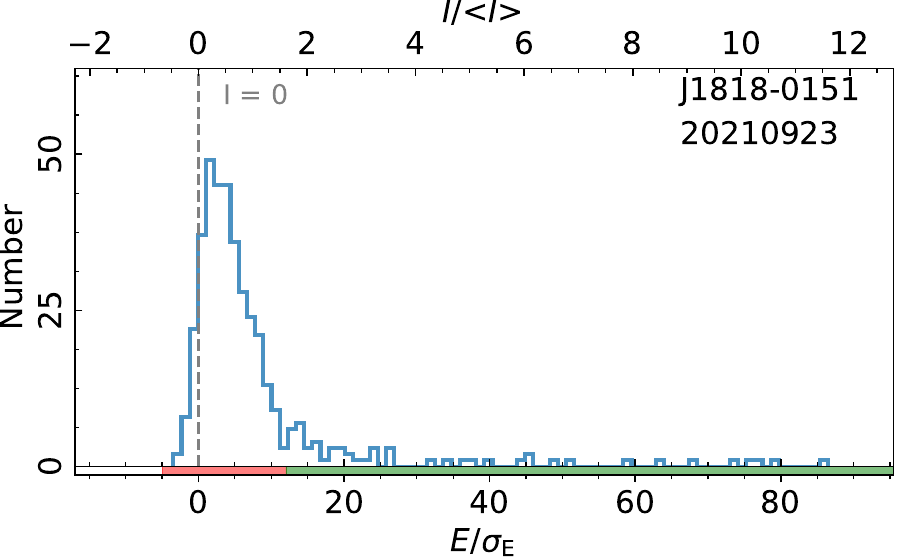}
\figcaption{Energy histogram for the central component of PSR J1818-0151 from the FAST observation on 20210923, which is used to distinguish different emission modes.
\label{subfig:Hist:J1818-0151}}
\end{figure}

\begin{figure}[htpb]
\centering
\includegraphics[width=0.37\textwidth, angle=0]{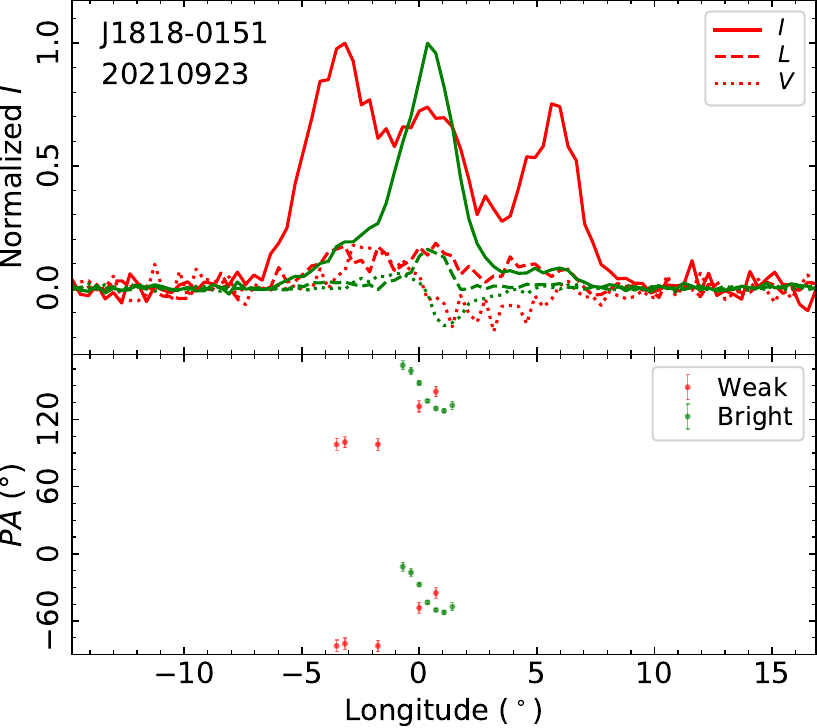}
\figcaption{Mean polarization profiles (the top panel) of the weak and bright emission modes of PSR J1818-0151 from the FAST observation on 20210923, as well as the averaged PAs (the bottom panel). \label{subfig:PolModes:J1818-0151}}
\end{figure}

\subsection{J1818+03}
\label{subsec:J1818+03}

PSR J1818+03 was discovered by the Green Bank North Celestial Cap (GBNCC) pulsar survey (http://astro.phys.wvu.edu/GBNCC/). 

This pulsar was observed by FAST on 20240208 for 5 minutes, yielding a rotation period $P=0.7991$~s and a dispersion measure $D\!M=97.9~{\rm cm^{-3}\,pc}$. Single pulse sequences shown in Fig.~\ref{subfig:TP:J1818+03} display the existence of nulling and subpulse drifting behaviors. The nulling fraction of this observation is estimated from the on-pulse energy histogram in Fig.~\ref{subfig:Hist:J1818+03}, which is 47$\pm$7\%. 
From LRFS and 2DFS in Fig.~\ref{subfig:fluctu:J1818+03}, the centroid modulation frequencies of the negative drift feature are estimated to be $1/P_3=0.226\pm0.002$ and $1/P_2=-61\pm1$, which correspond to drifting parameters of $P_3=4.42\pm0.04$ periods and $P_2=-5.9\pm0.1^\circ$.

\subsection{J1818-0151}
\label{subsec:J1818-0151}

PSR J1818-0151 was discovered by \citet{Bates2012} in the High Time Resolution Universe survey using the Parkes radio telescope. 

The FAST observation of this pulsar was conducted on 20210923 for 6 minutes, deriving a rotation period $P=0.8376$~s and a dispersion measure $D\!M=200.7~{\rm cm^{-3}\,pc}$ from this observation. 
From single pulse sequences (Fig.~\ref{subfig:TP:J1818-0151}), mode changes occur occasionally, and the central component is greatly enhanced in the bright mode compared to that of the normal mode (Fig.~\ref{subfig:PolModes:J1818-0151}). The integral energy histogram of the central phase range in the profile is used to distinguish the weak and bright emission modes, as Fig.~\ref{subfig:Hist:J1818-0151} shows. In addition, the central component of the bright mode also shows obvious modulation behavior.

\begin{figure}[htpb]
\centering
\includegraphics[width=0.22\textwidth, angle=0]{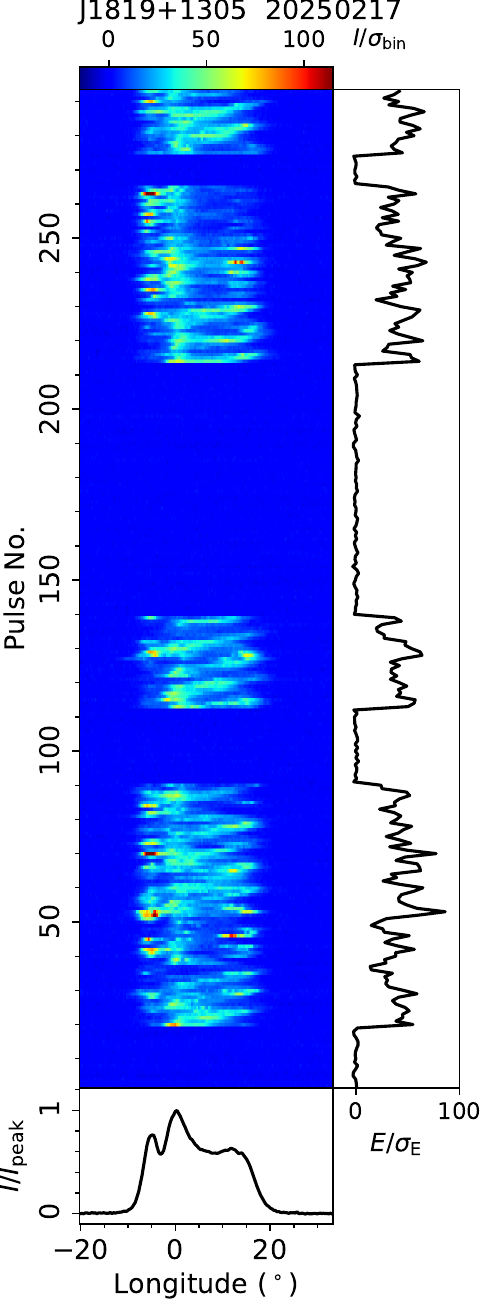}
\includegraphics[width=0.22\textwidth, angle=0]{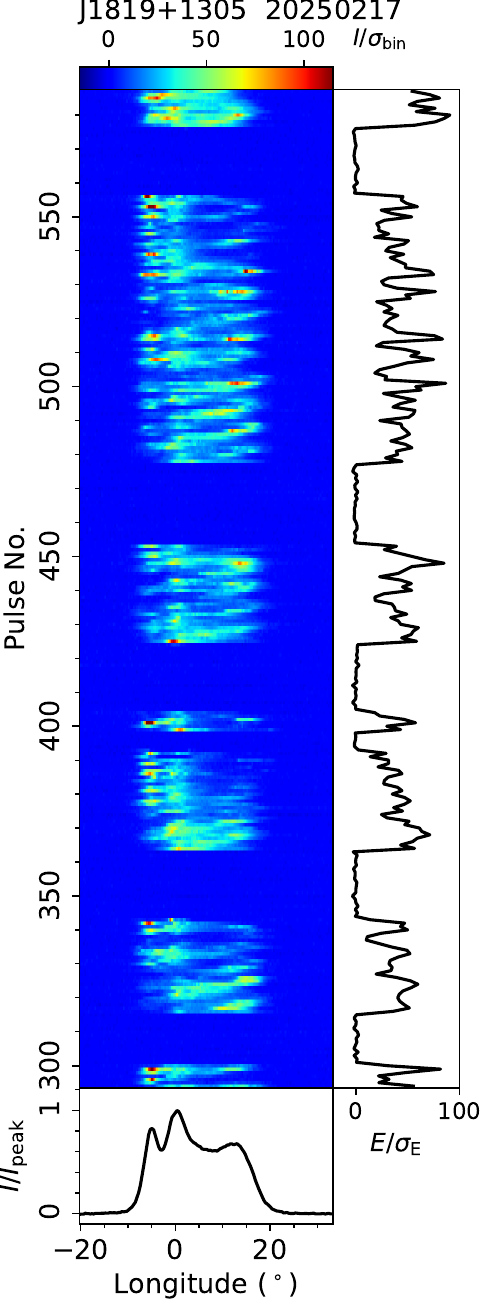}
\figcaption{Single pulse sequences of PSR J1819+1305 from the FAST observation on 20250217.
\label{subfig:TP:J1819+1305}}
\end{figure}

\begin{figure}[htpb]
\centering
\includegraphics[width=0.39\textwidth, angle=0]{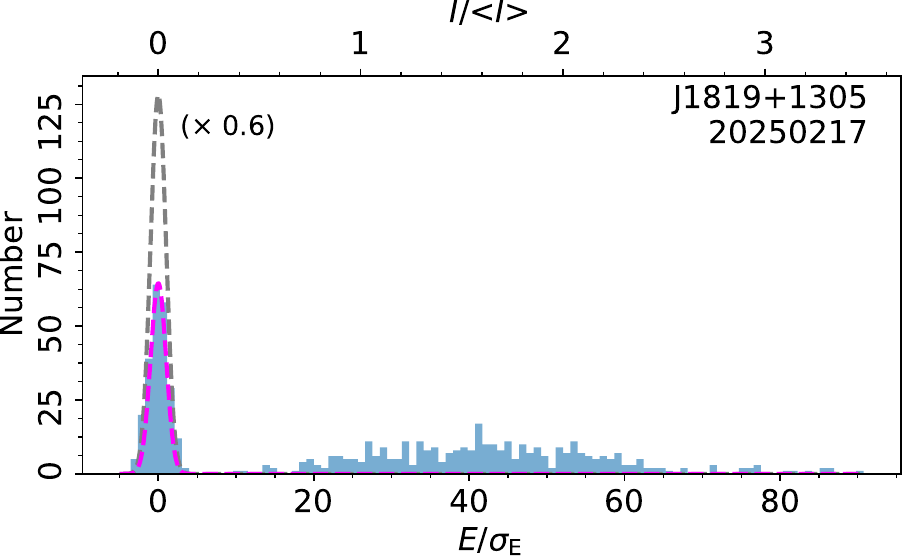}
\figcaption{On-pulse energy histogram of PSR J1819+1305 from the FAST observation on 20250217.
\label{subfig:Hist:J1819+1305}}
\end{figure}

\begin{figure}[htpb]
\centering
\includegraphics[width=0.22\textwidth, angle=0]{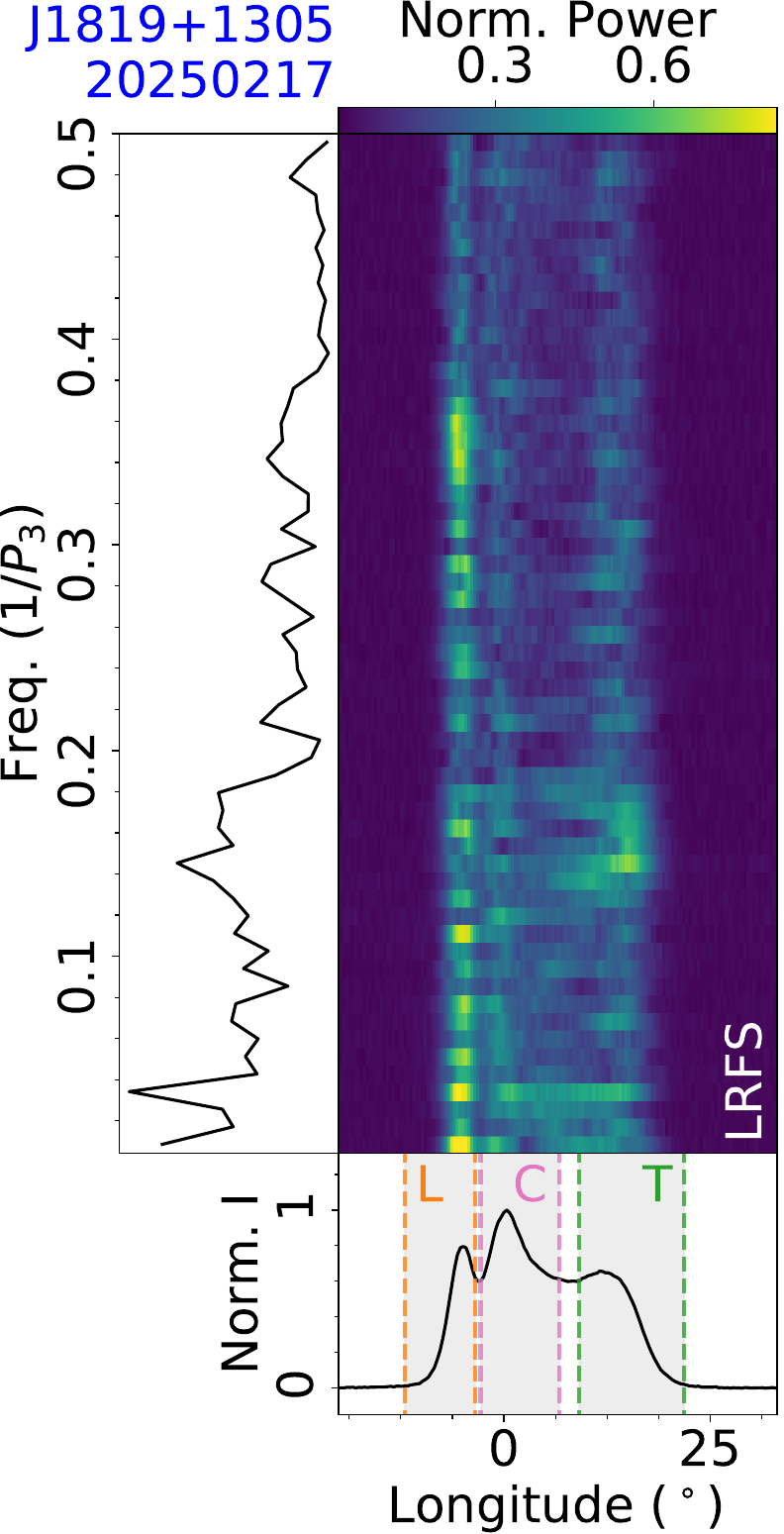}
\includegraphics[width=0.22\textwidth, angle=0]{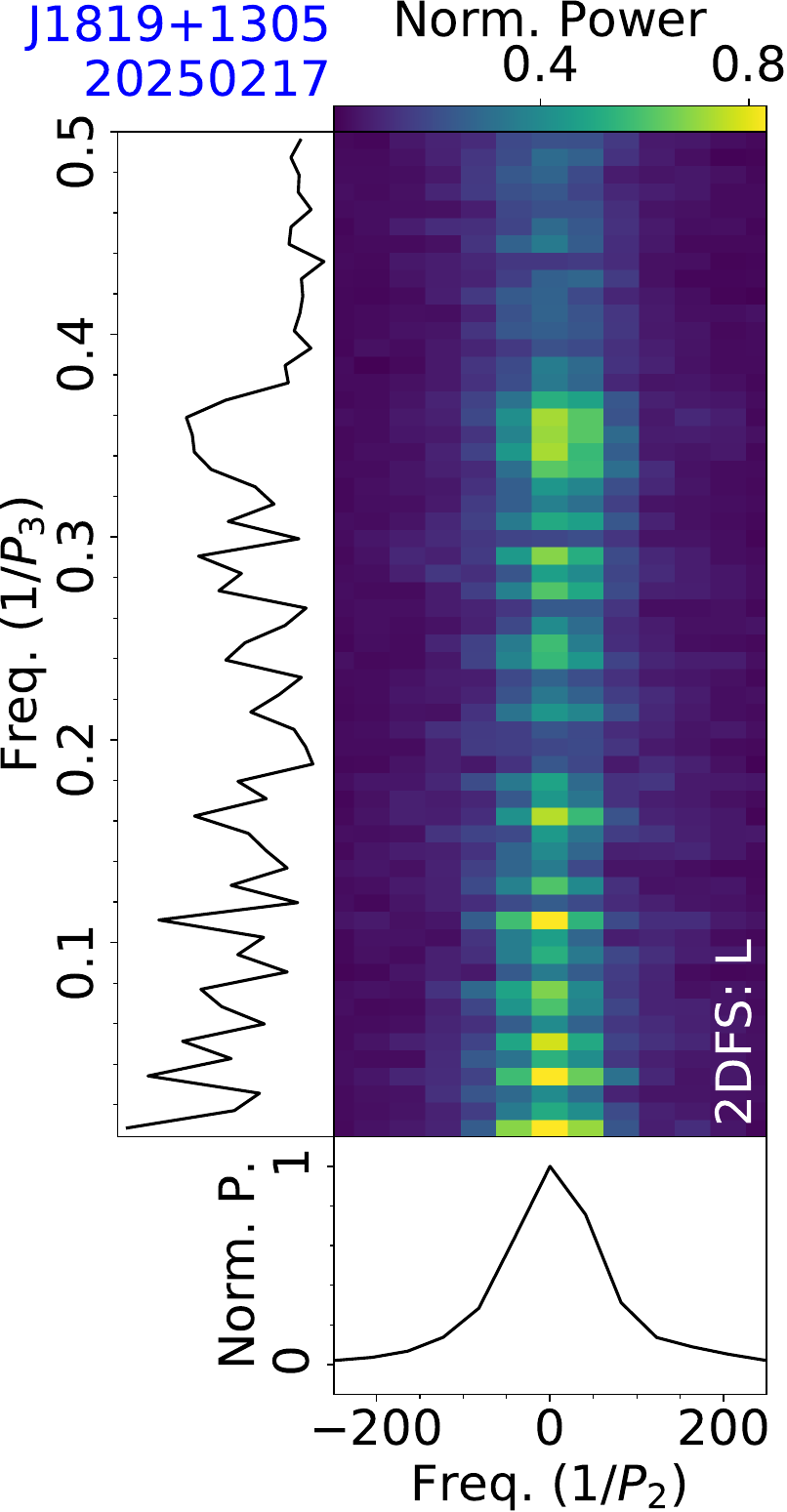}\\
\includegraphics[width=0.22\textwidth, angle=0]{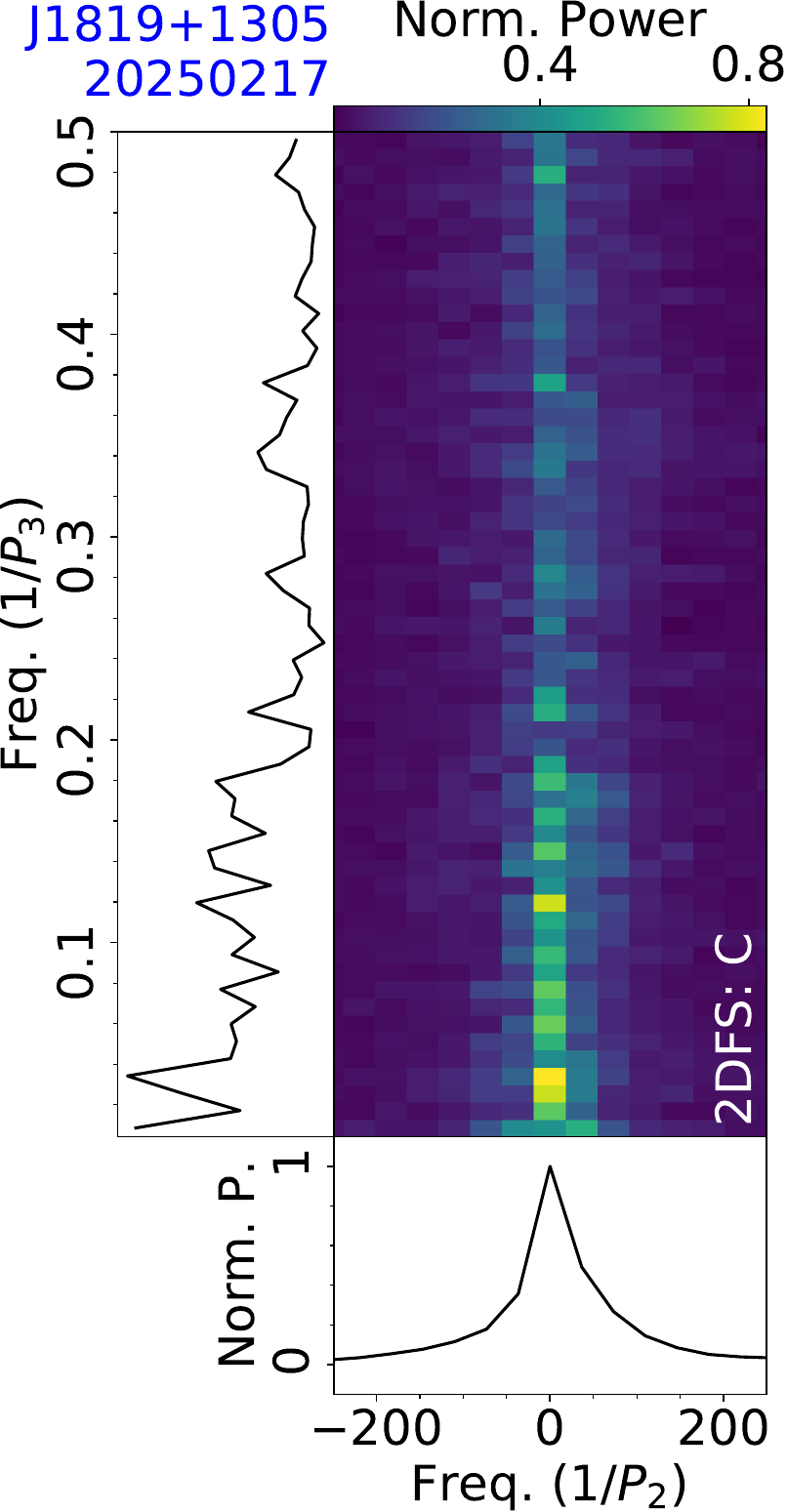}
\includegraphics[width=0.22\textwidth, angle=0]{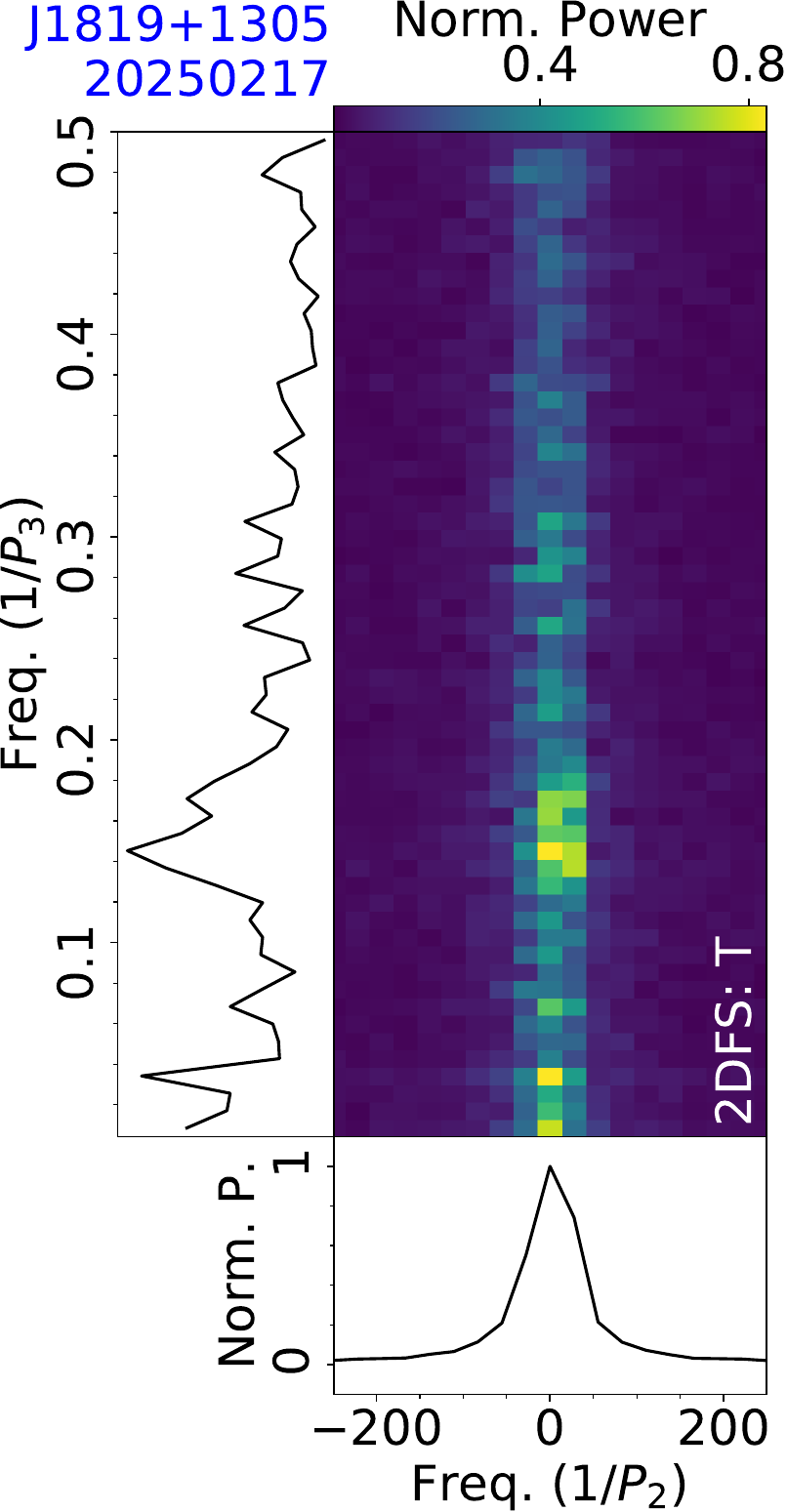}
\figcaption{Fluctuation analysis of PSR J1819+1305 for the observation on 20250217, with LRFS (top-left), and 2DFS for the on-pulse region (top-right), leading part (bottom-left) and trailing part (bottom-right) of a mean pulse profile.
\label{subfig:fluctu:J1819+1305}}
\end{figure}

\subsection{J1819+1305}
\label{subsec:J1819+1305}

PSR J1819+1305 was discovered in a survey of intermediate Galactic latitudes using the Parkes 64-m radio telescope \citet{Edwards2001}. Periodic nulling with a fraction of 41$\pm$6\% has been reported previously \citep{Navarro2003,Rankin2008}. This pulsar also has drifting behaviors of about 6 and 3 periods \citep{Rankin2008,Song2023}. 

This pulsar was observed by FAST on 20250217 for 10 minutes, deriving a rotation period $P=1.0603$~s and a dispersion measure $D\!M=65.0~{\rm cm^{-3}\,pc}$. Single pulse sequences in Fig.~\ref{subfig:TP:J1819+1305} display nulling, subpulse drifting, as well as the intensity weakening of the trailing component (pulses No.250-265 and 380-390) as \citet{Rankin2008} described. 
The nulling fraction of this FAST observation is estimated from the on-pulse energy histogram in Fig.~\ref{subfig:Hist:J1819+1305}, that is 30$\pm$2\%. 
To reduce the effect of nulls, LRFS and 2DFS for the single pulse sequence after the removal of nulls are obtained (Fig.~\ref{subfig:fluctu:J1819+1305}). 
For the leading part in a mean pulse profile, the temporal modulation frequency is widely distributed, and the main drift feature in 2DFS is positive and with the centroid modulation frequencies of $1/P_3=0.348\pm0.001$ ($P_3=2.88\pm0.01$ periods) and $1/P_2=18\pm3$ ($P_2=20\pm3^\circ$). 
For the central component, the drift feature is very weak from this observation. 
In 2DFS of the trailing component, there is a positive drift feature with the centroid of $1/P_3=0.153\pm0.001$ and $1/P_2=9\pm2$, which correspond to $P_3=6.56\pm0.04$ periods and $P_2=42\pm10^\circ$. 
$P_3=6.5\pm0.1$ periods and $P_2=61\pm54^\circ$.

\begin{figure}[htpb]
\centering
\includegraphics[width=0.22\textwidth, angle=0]{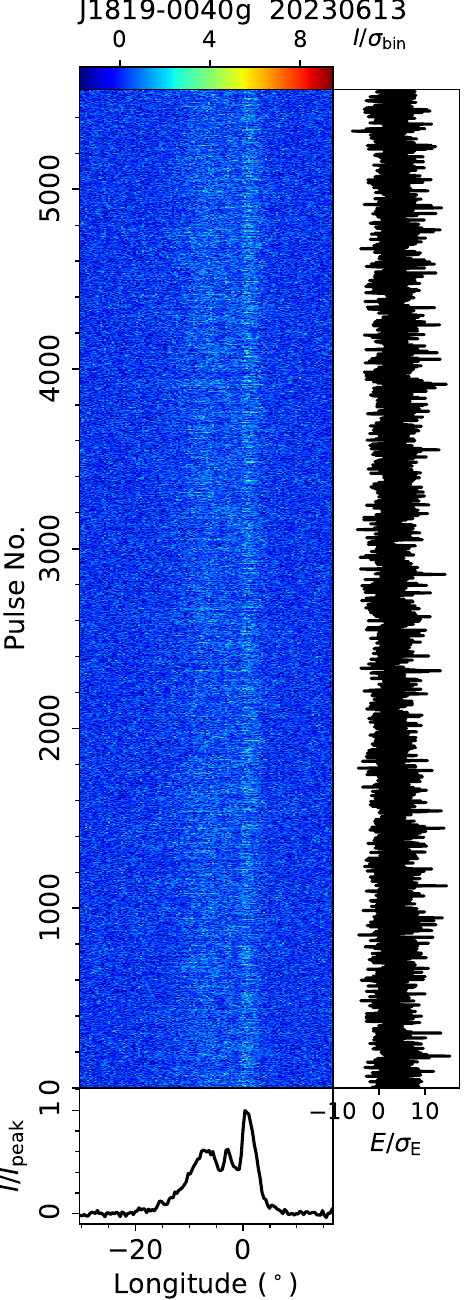}
\includegraphics[width=0.22\textwidth, angle=0]{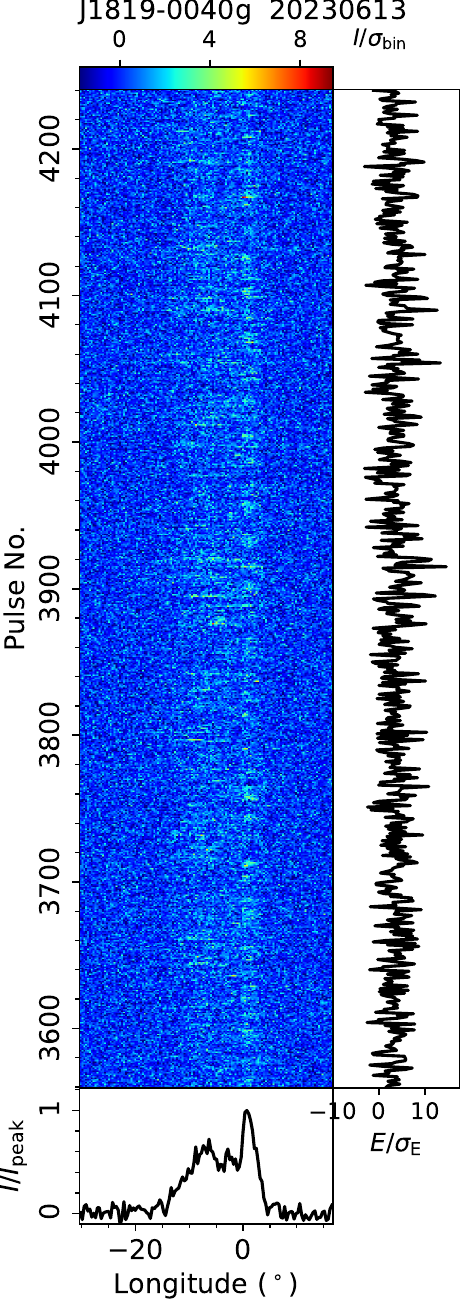}
\figcaption{Single pulse sequences of PSR J1819-0040g from the FAST observation on 20230613. \label{subfig:TP:J1819-0040g}}
\end{figure}

\begin{figure}[htpb]
\centering
\includegraphics[width=0.22\textwidth, angle=0]{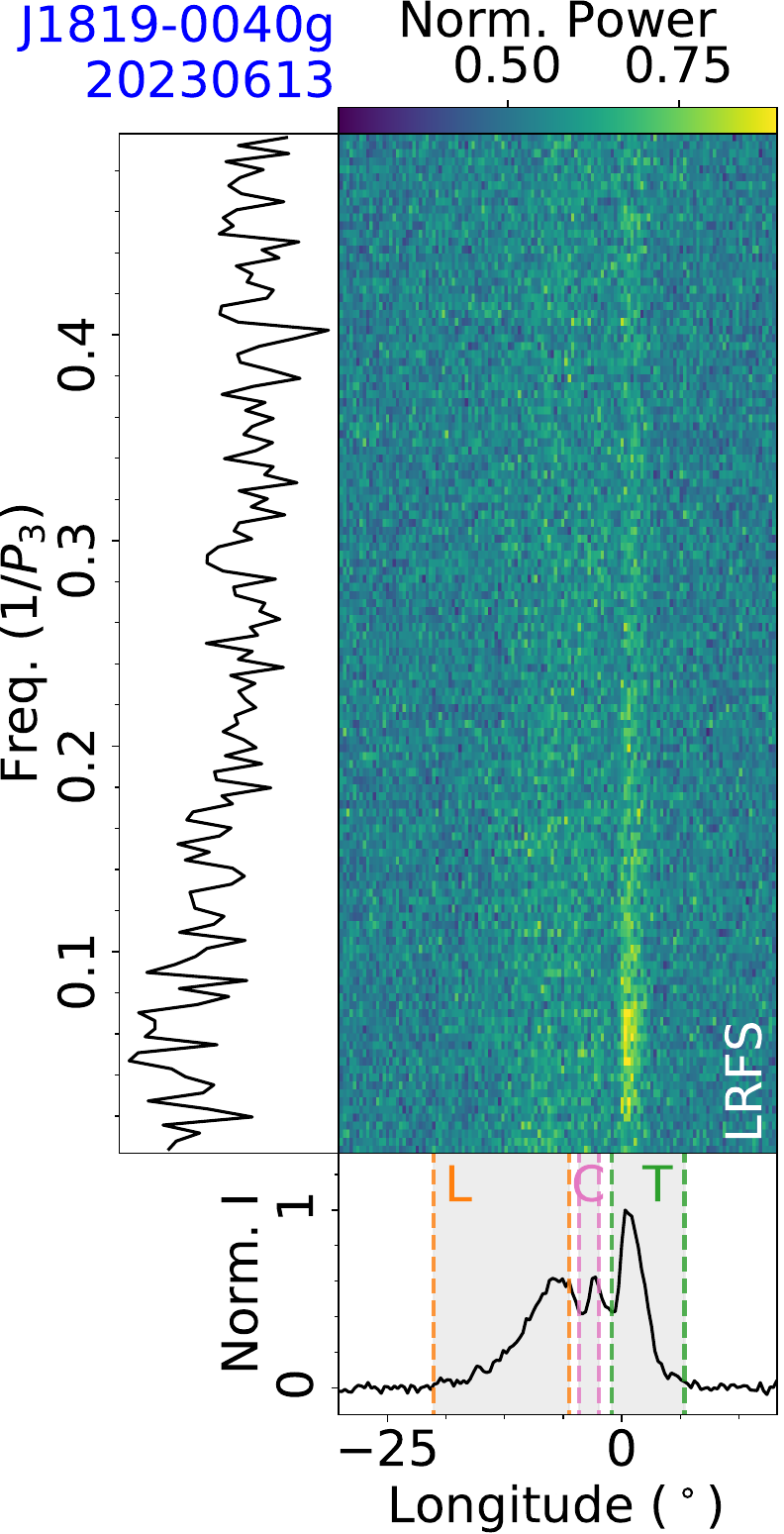}
\includegraphics[width=0.22\textwidth, angle=0]{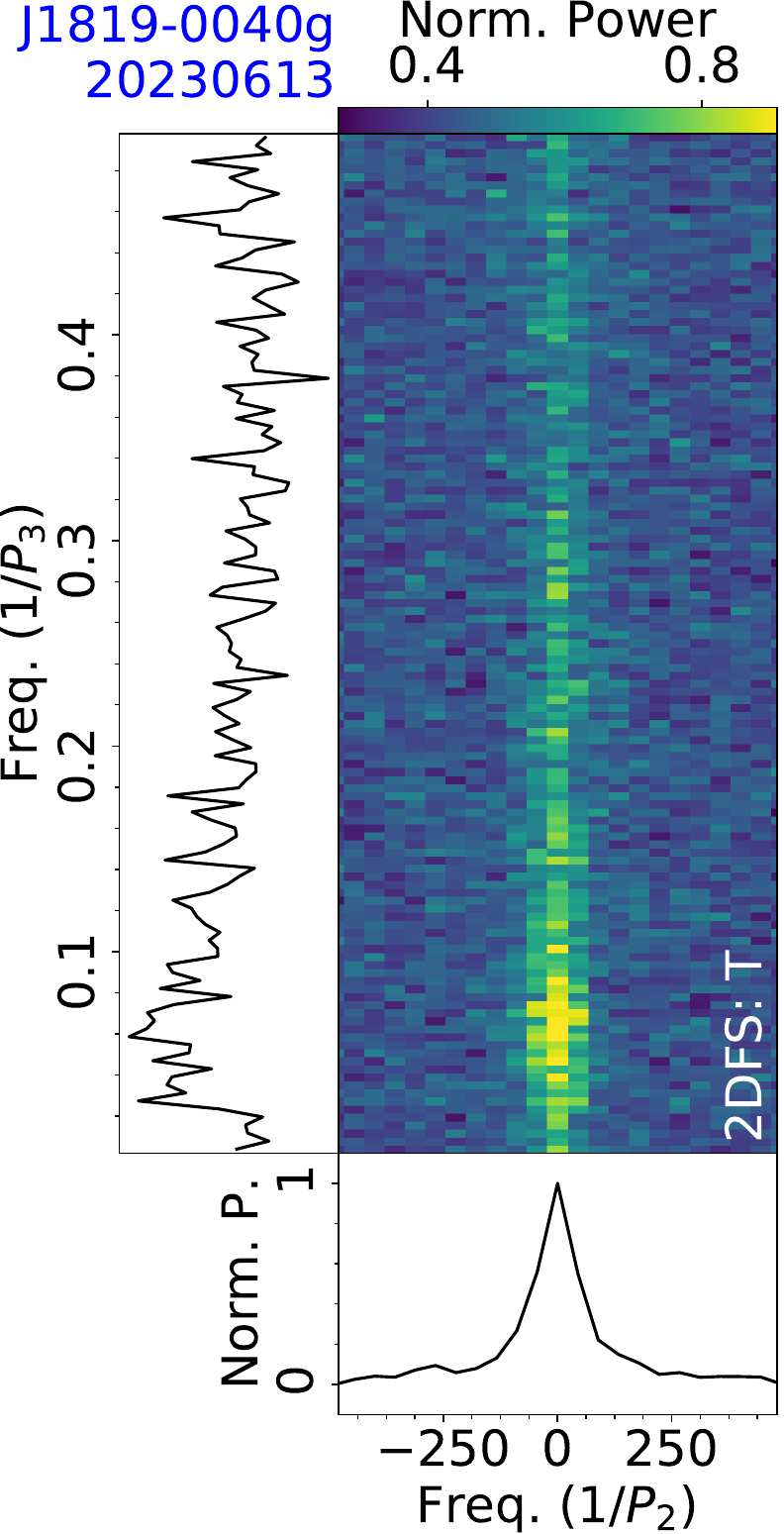}
\figcaption{Fluctuation analysis of PSR J1819-0040g for the observation on 20230613, with LRFS and 2DFS for the on-pulse region of a mean pulse profile.  \label{subfig:fluctu:J1819-0040g}}
\end{figure}

\begin{figure}[htpb]
\centering
\includegraphics[width=0.22\textwidth, angle=0]{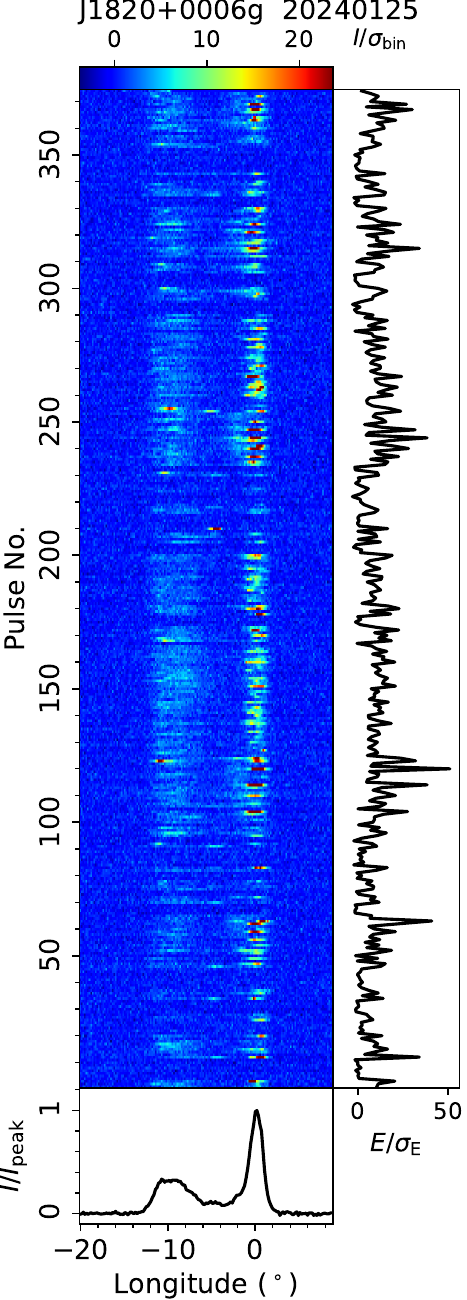}
\figcaption{Single pulse sequence of PSR J1820+0006g from the FAST observation on 20240125. \label{subfig:TP:J1820+0006g}}
\end{figure}

\begin{figure}[htpb]
\centering
\includegraphics[width=0.39\textwidth, angle=0]{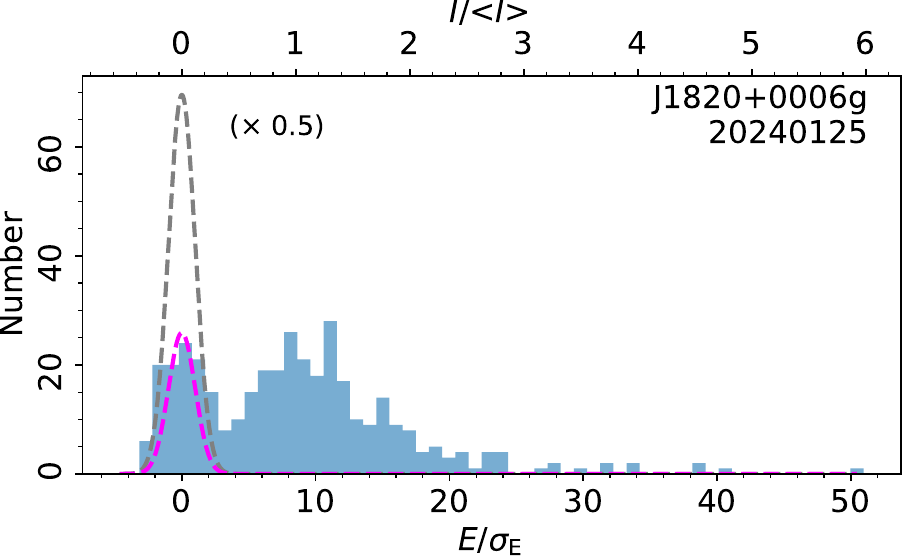}
\figcaption{On-pulse energy histogram of single pulses of PSR J1820+0006g from the FAST observation on 20240125. \label{subfig:Hist:J1820+0006g}}
\end{figure}

\begin{figure}[htpb]
\centering
\includegraphics[width=0.22\textwidth, angle=0]{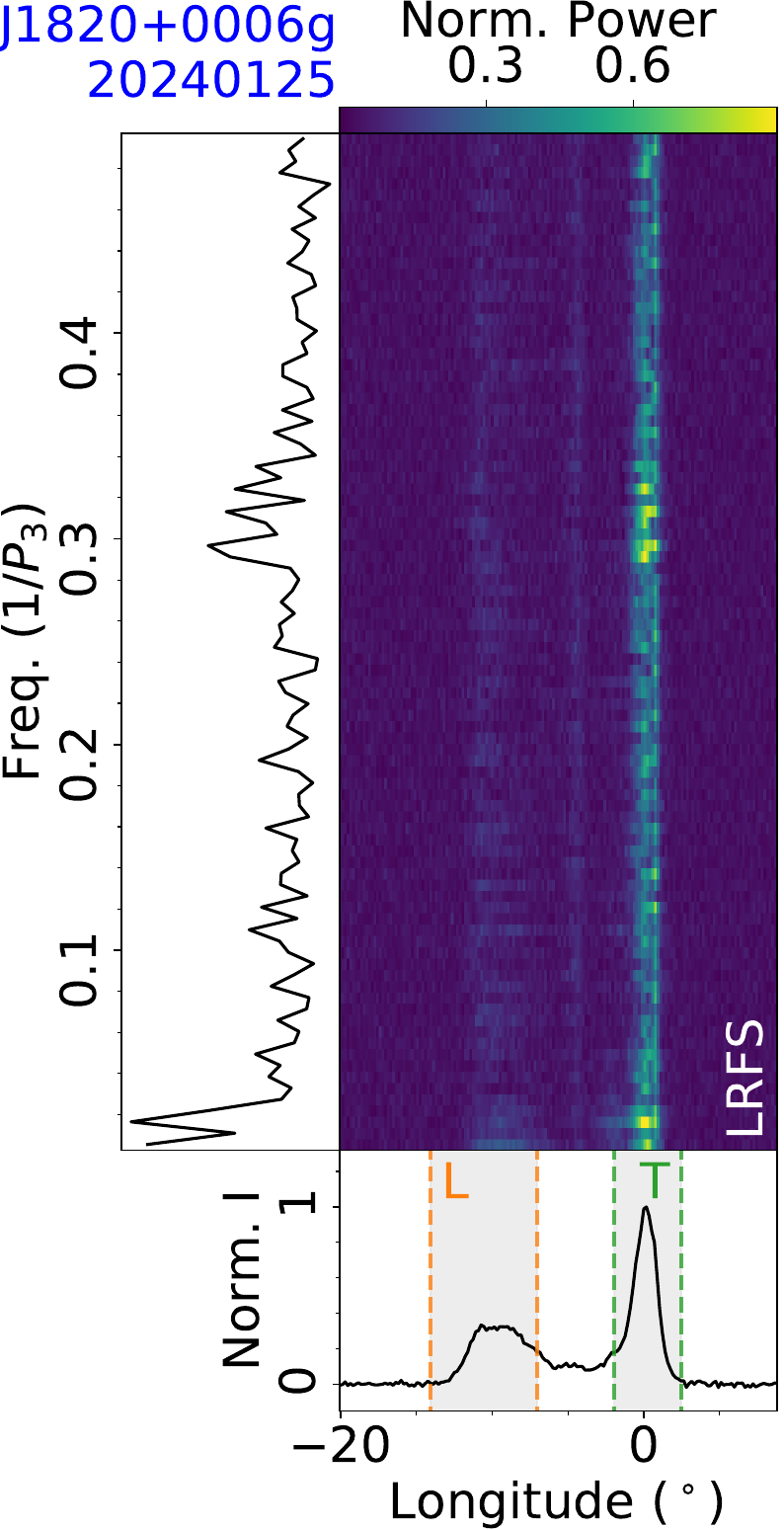}
\includegraphics[width=0.22\textwidth, angle=0]{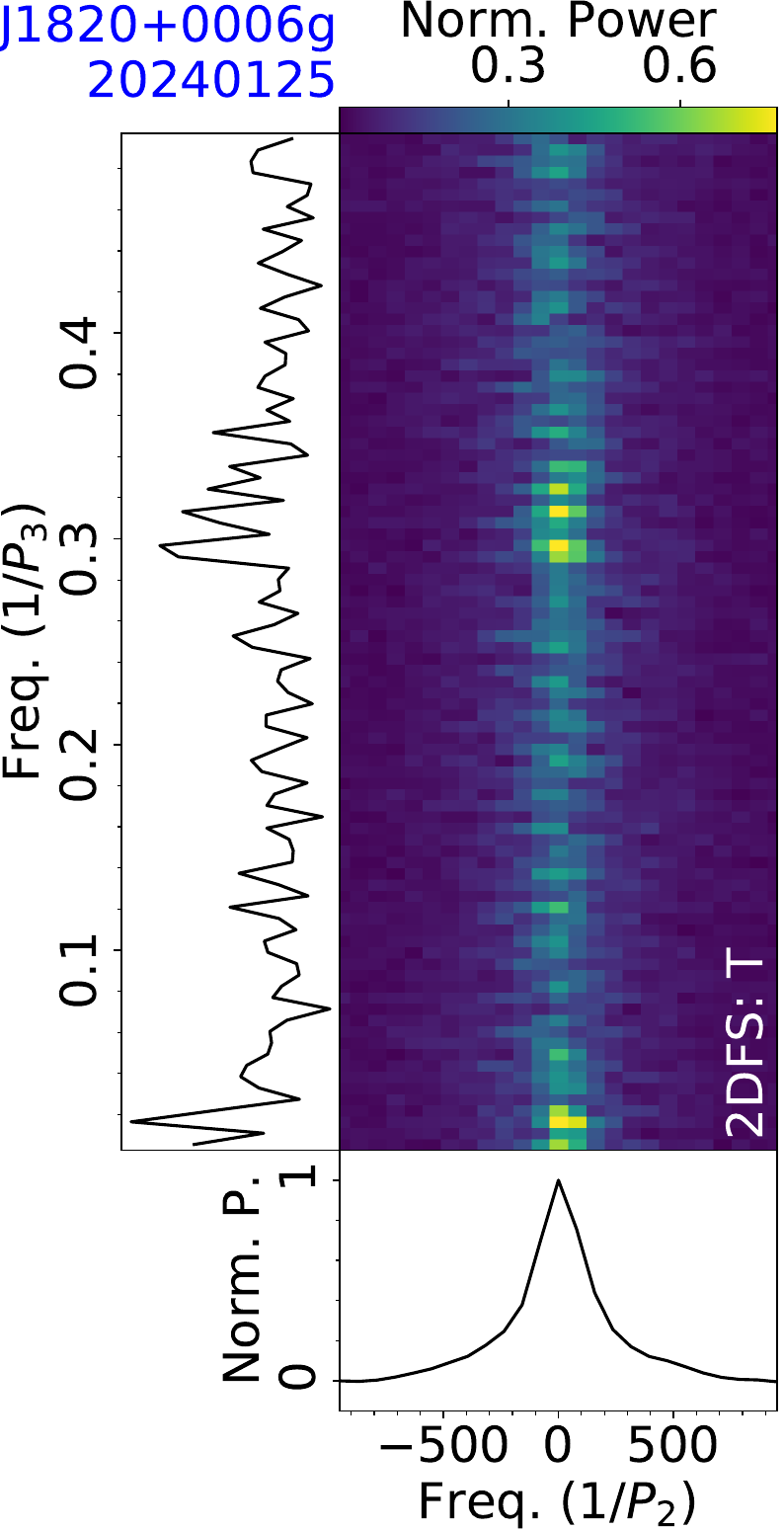}
\figcaption{Fluctuation analysis of PSR J1820+0006g for the observation on 20240125, with LRFS and 2DFS for the trailing part of a mean pulse profile.  \label{subfig:fluctu:J1820+0006g}}
\end{figure}

\subsection{J1819-0040g}
\label{subsec:J1819-0040g}

PSR J1819-0040g was discovered in the FAST GPPS survey \citep{Han2021,han2025}. 

This pulsar was observed by FAST on 20230613 for 15 minutes, yielding a rotation period $P=0.1601$~s and a dispersion measure $D\!M=58.2~{\rm cm^{-3}\,pc}$ from this observation. 
Single pulse sequences displayed in Fig.~\ref{subfig:TP:J1819-0040g} show modulation behavior. From LRFS and 2DFS in Fig.~\ref{subfig:fluctu:J1819-0040g}, the trailing component has a centroid temporally modulated frequency of $1/P_3=0.064\pm0.001$, corresponding to $P_3=15.5\pm0.2$ periods. 

\subsection{J1820+0006g}
\label{subsec:J1820+0006g}

PSR J1820+0006g was discovered in the FAST GPPS survey \citep{Han2021,han2025}. 

This pulsar was observed by FAST on 20240125 for 15 minutes, deriving a rotation period $P=2.4049$~s and a dispersion measure $D\!M=186.3~{\rm cm^{-3}\,pc}$. The single pulse sequence of the observation is displayed in Fig.~\ref{subfig:TP:J1820+0006g}. The pulsar is found to have nulling and subpulse drifting behaviors. The nulling fraction is estimated to be 19$\pm$4\% from the on-pulse energy histogram (Fig.~\ref{subfig:Hist:J1820+0006g}). 
For the trailing component, 2DFS exhibits a preferred positive drift feature with the centroid modulation frequencies of $1/P_3=0.310\pm0.001$ and $1/P_2=11\pm7$, which correspond to $P_3=3.23\pm0.01$ periods and $P_2=34\pm22^\circ$. 
In addition, bright subpulses occur in the central phase part of the profile from the single pulse sequence.

\begin{figure}[htpb]
\centering
\includegraphics[width=0.22\textwidth, angle=0]{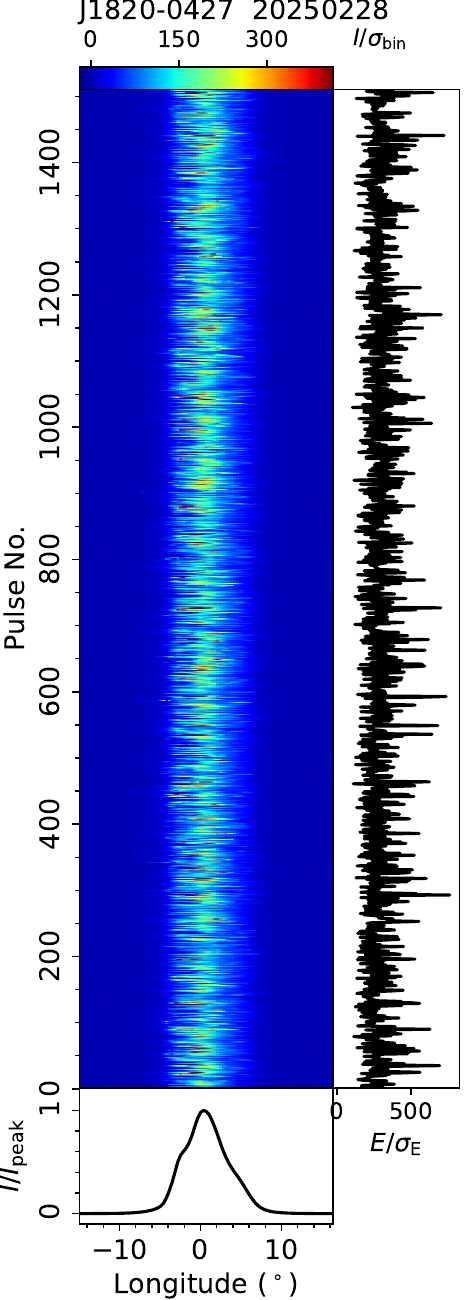}
\includegraphics[width=0.22\textwidth, angle=0]{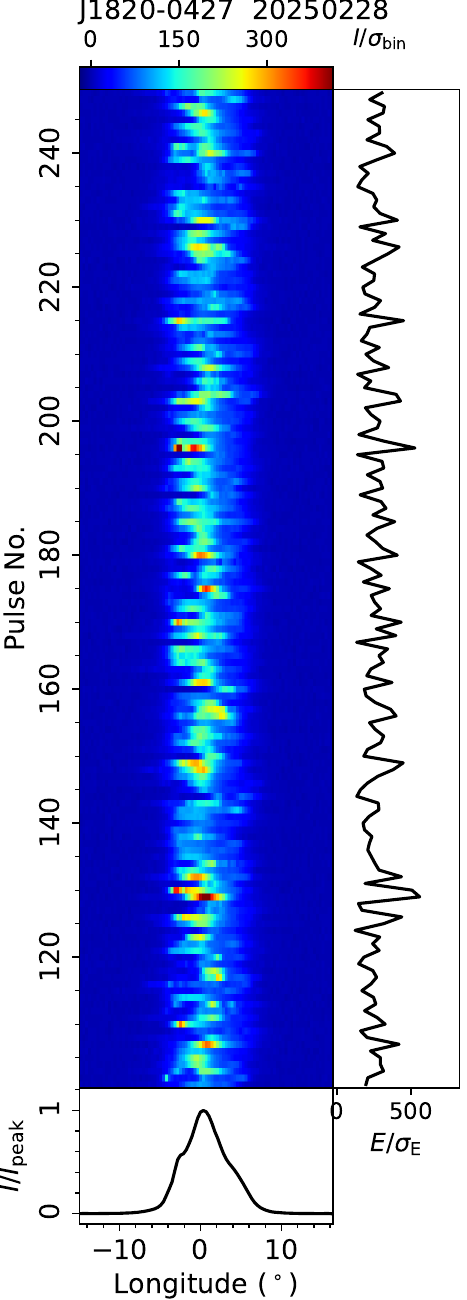}
\figcaption{The single pulse sequence of PSR J1820-0427 from the FAST observation on 20250228, and a zoomed-in view of pulses No. 101-250.
\label{subfig:TP:J1820-0427}}
\end{figure}

\begin{figure}[htpb]
\centering
\includegraphics[width=0.22\textwidth, angle=0]{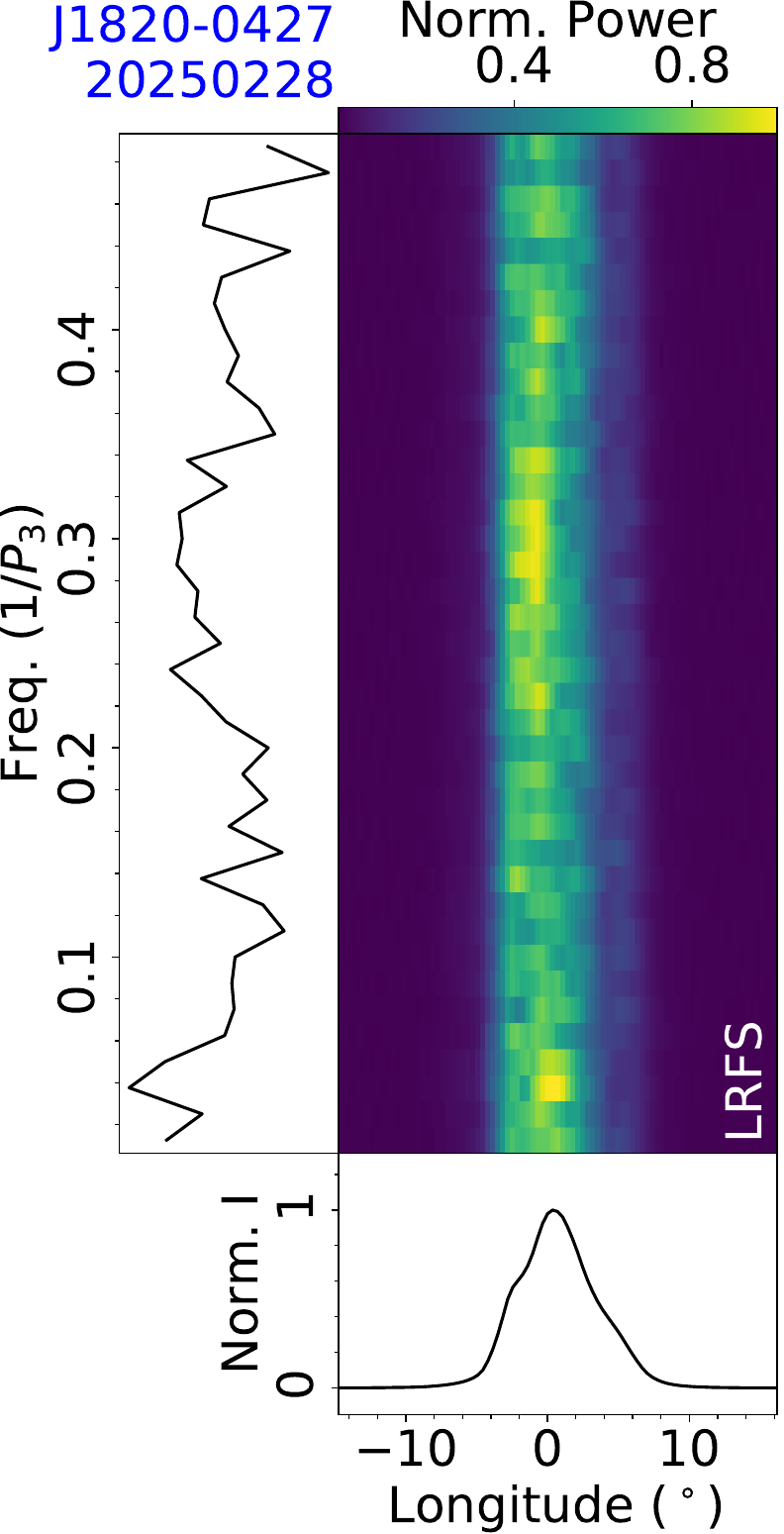}
\includegraphics[width=0.22\textwidth, angle=0]{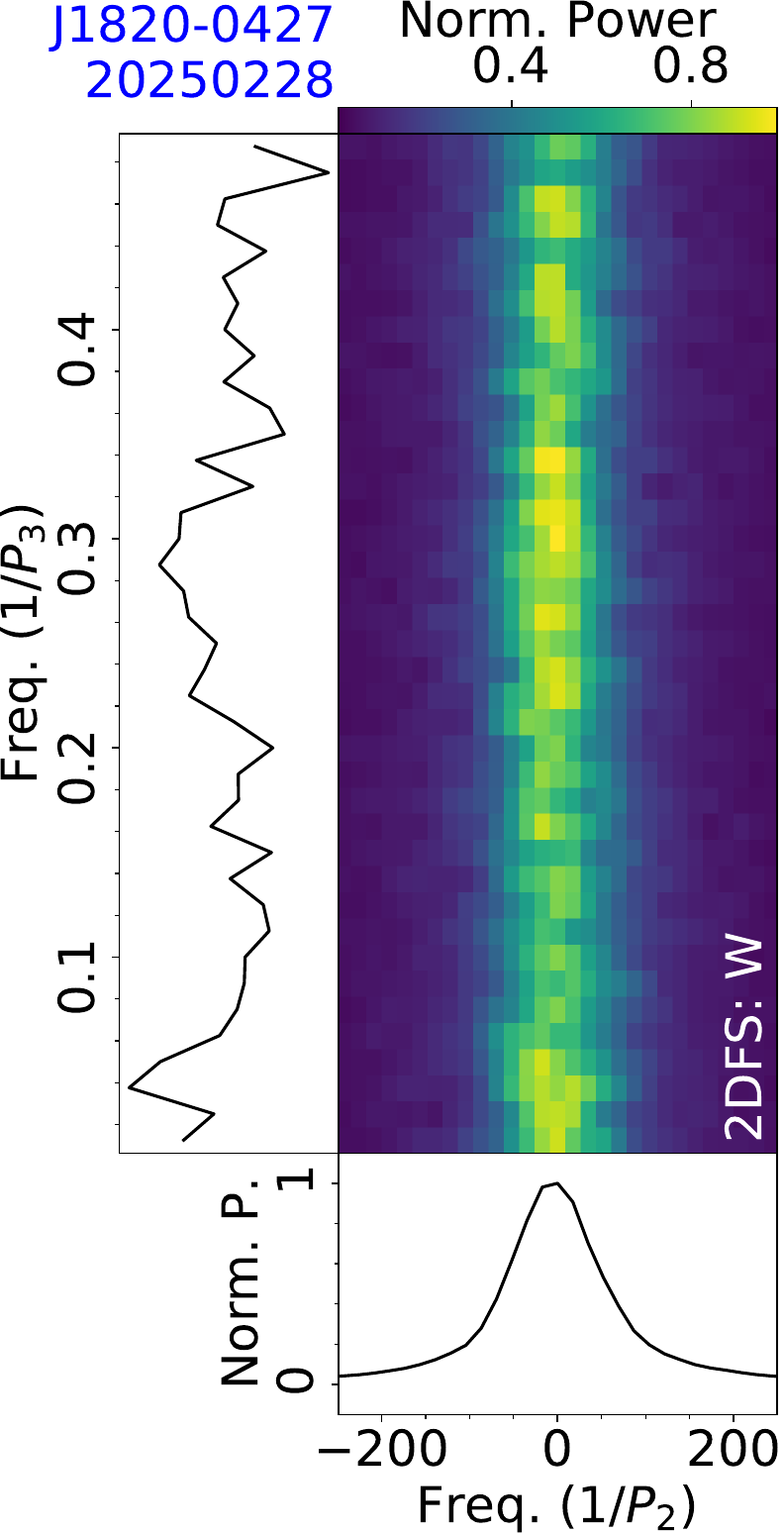}
\figcaption{Fluctuation analysis of PSR J1820-0427 for the FAST observation on 20250228, with LRFS and 2DFS for the on-pulse phase region of a mean pulse profile.
\label{subfig:fluctu:J1820-0427}}
\end{figure}




\begin{figure}[htpb]
\centering
\includegraphics[width=0.22\textwidth, angle=0]{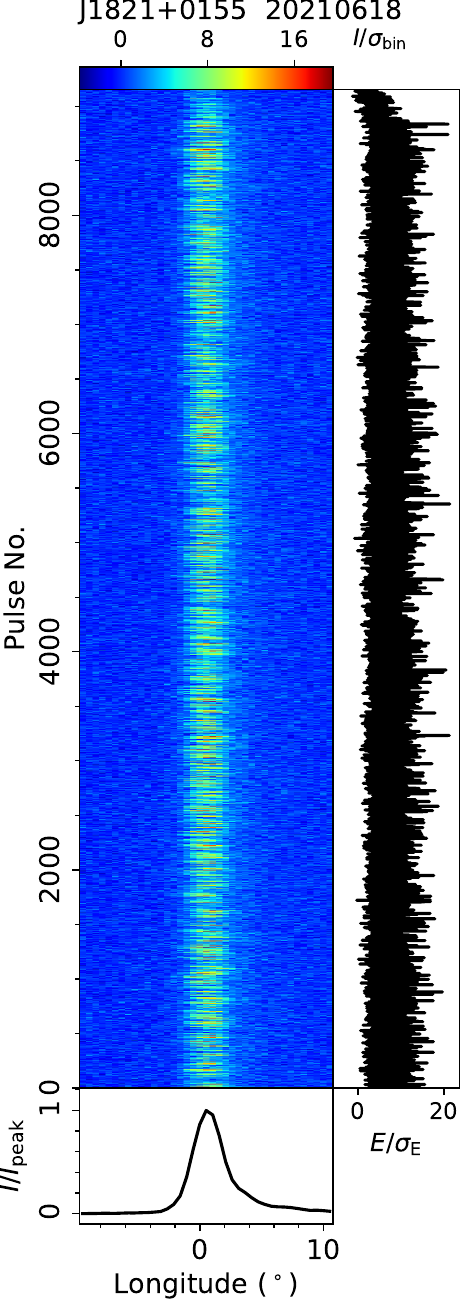}
\includegraphics[width=0.22\textwidth, angle=0]{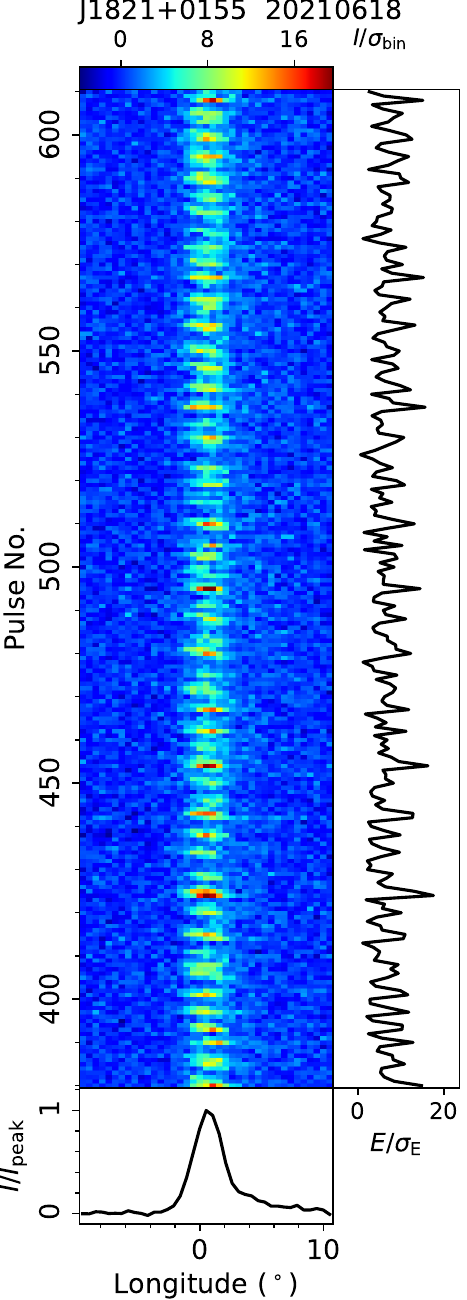}
\figcaption{Single pulse sequences of PSR J1821+0155 from the FAST observation on 20210618. \label{subfig:TP:J1821+0155}}
\end{figure}

\begin{figure}[htpb]
\centering
\includegraphics[width=0.22\textwidth, angle=0]{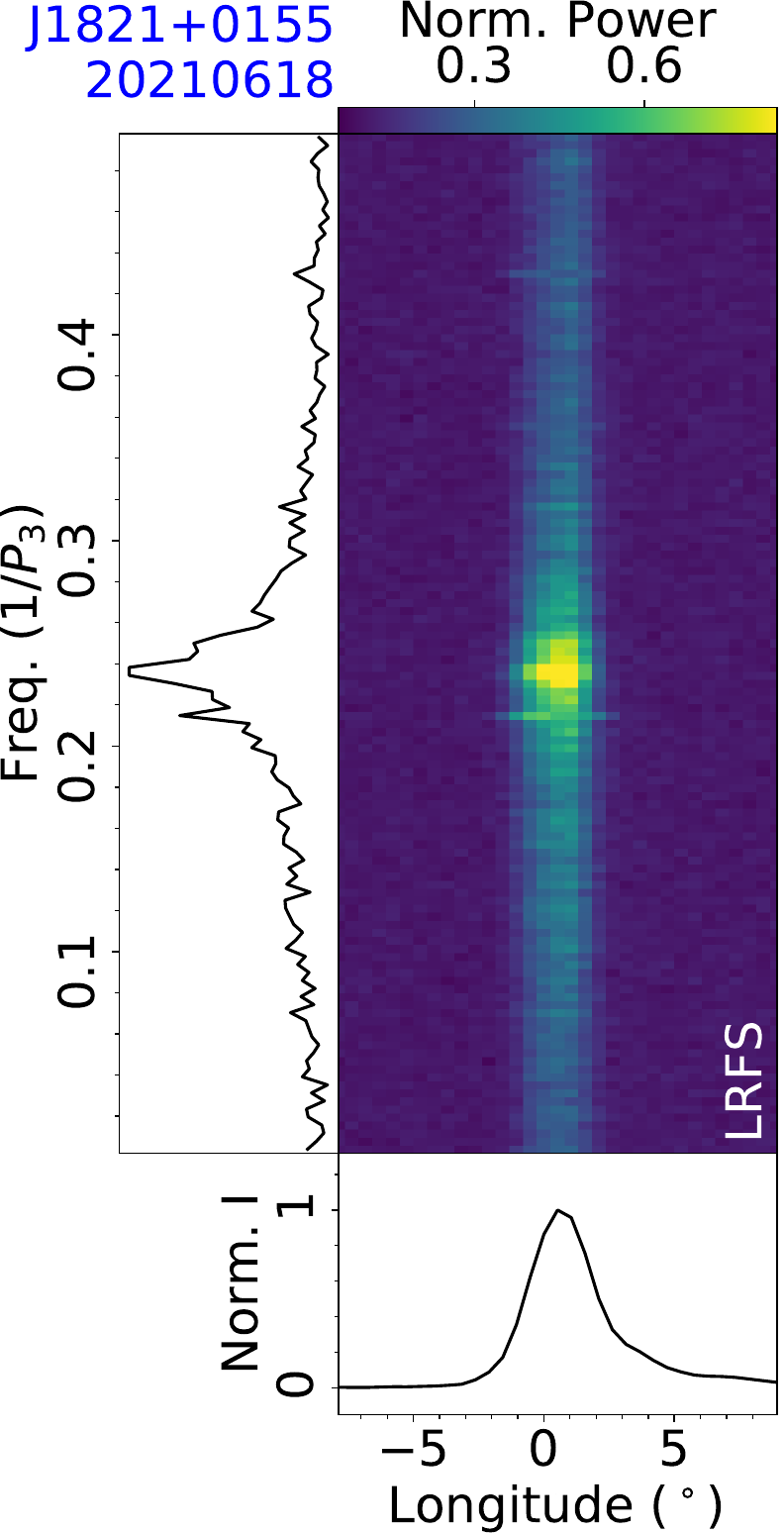}
\includegraphics[width=0.22\textwidth, angle=0]{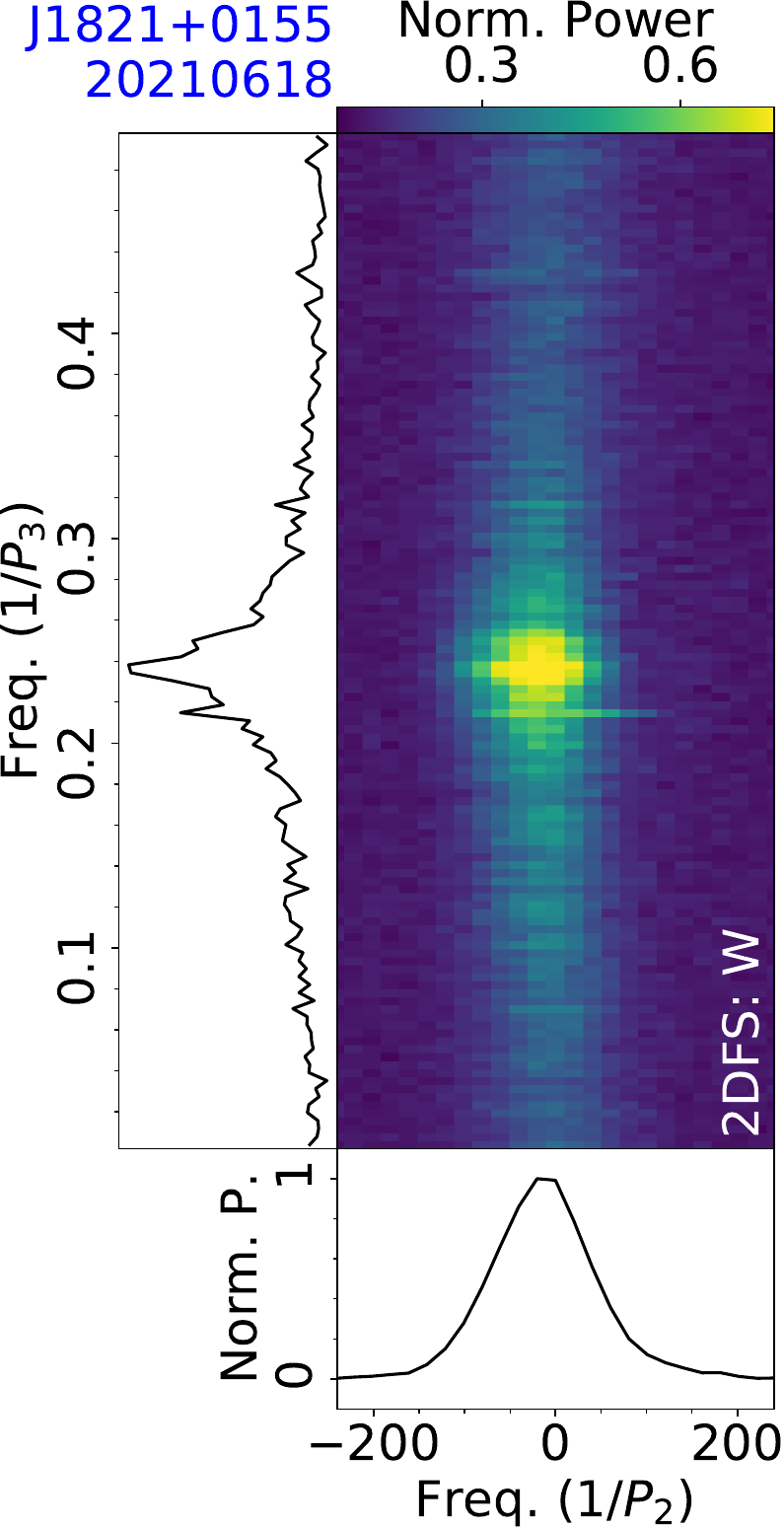}
\figcaption{Fluctuation analysis of PSR J1821+0155 for the FAST observation on 20210618, with LRFS and 2DFS for the on-pulse phase region of a mean pulse profile.
\label{subfig:fluctu:J1821+0155}}
\end{figure}

\begin{figure}[htpb]
\includegraphics[width=0.22\textwidth, angle=0]{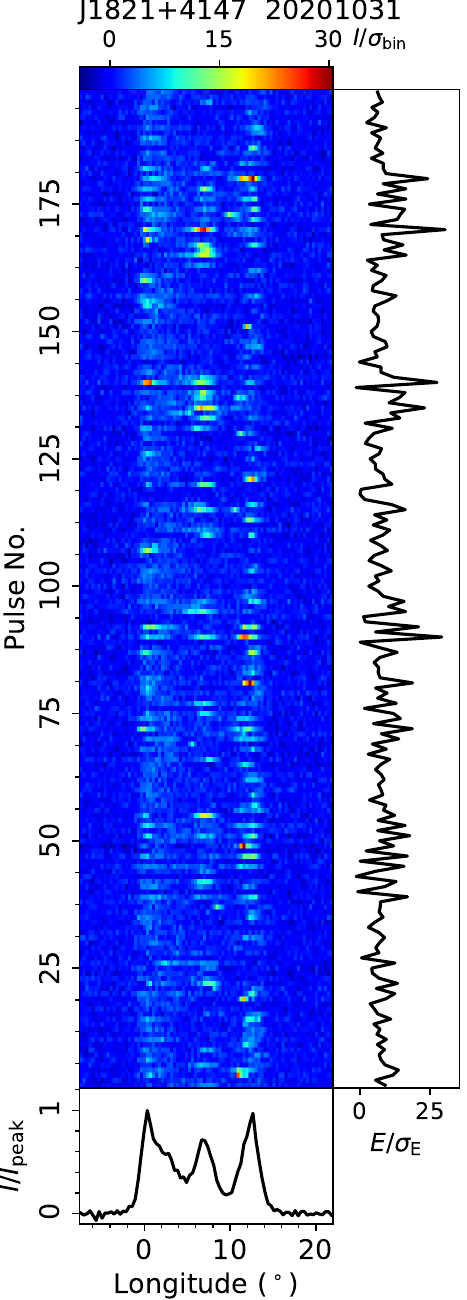}
\includegraphics[width=0.22\textwidth, angle=0]{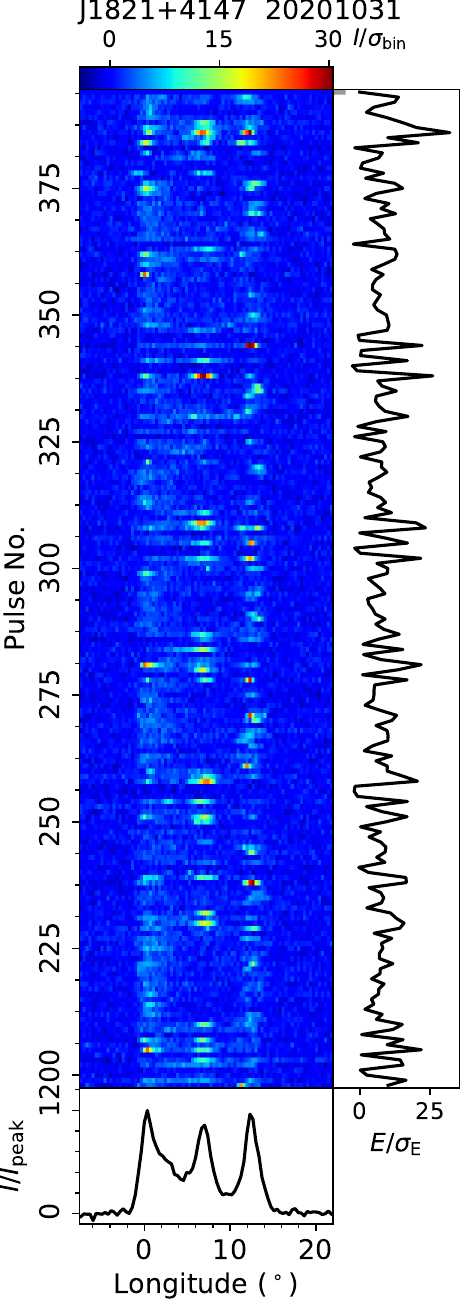}
\figcaption{Single pulse sequences of PSR J1821+4147 from the FAST observation on 20201031. \label{subfig:TP:J1821+4147}}
\end{figure}

\begin{figure}[htpb]
\centering
\includegraphics[width=0.39\textwidth, angle=0]{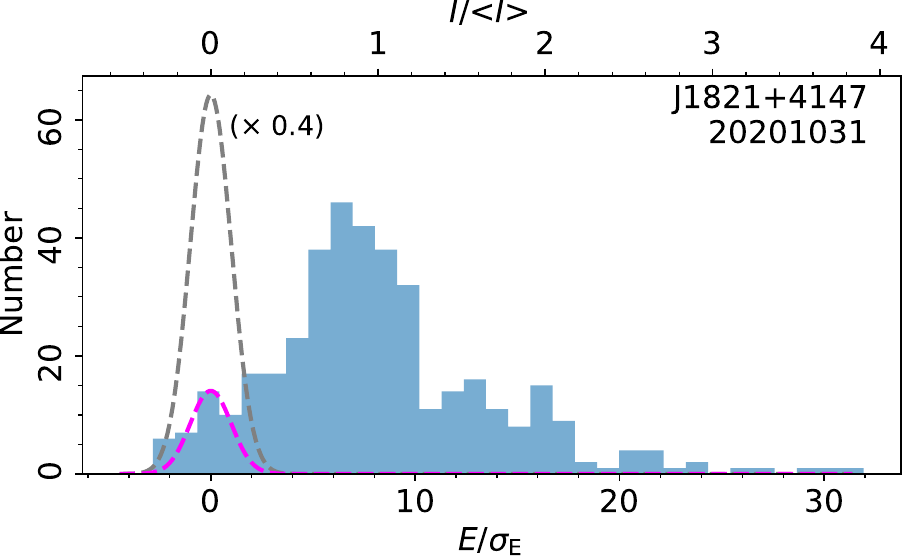}
\figcaption{On-pulse energy histogram of single pulses of PSR J1821+4147 from the FAST observation on 20201031.
\label{subfig:Hist:J1821+4147}}
\end{figure}

\begin{figure}[htpb]
\centering
\includegraphics[width=0.22\textwidth, angle=0]{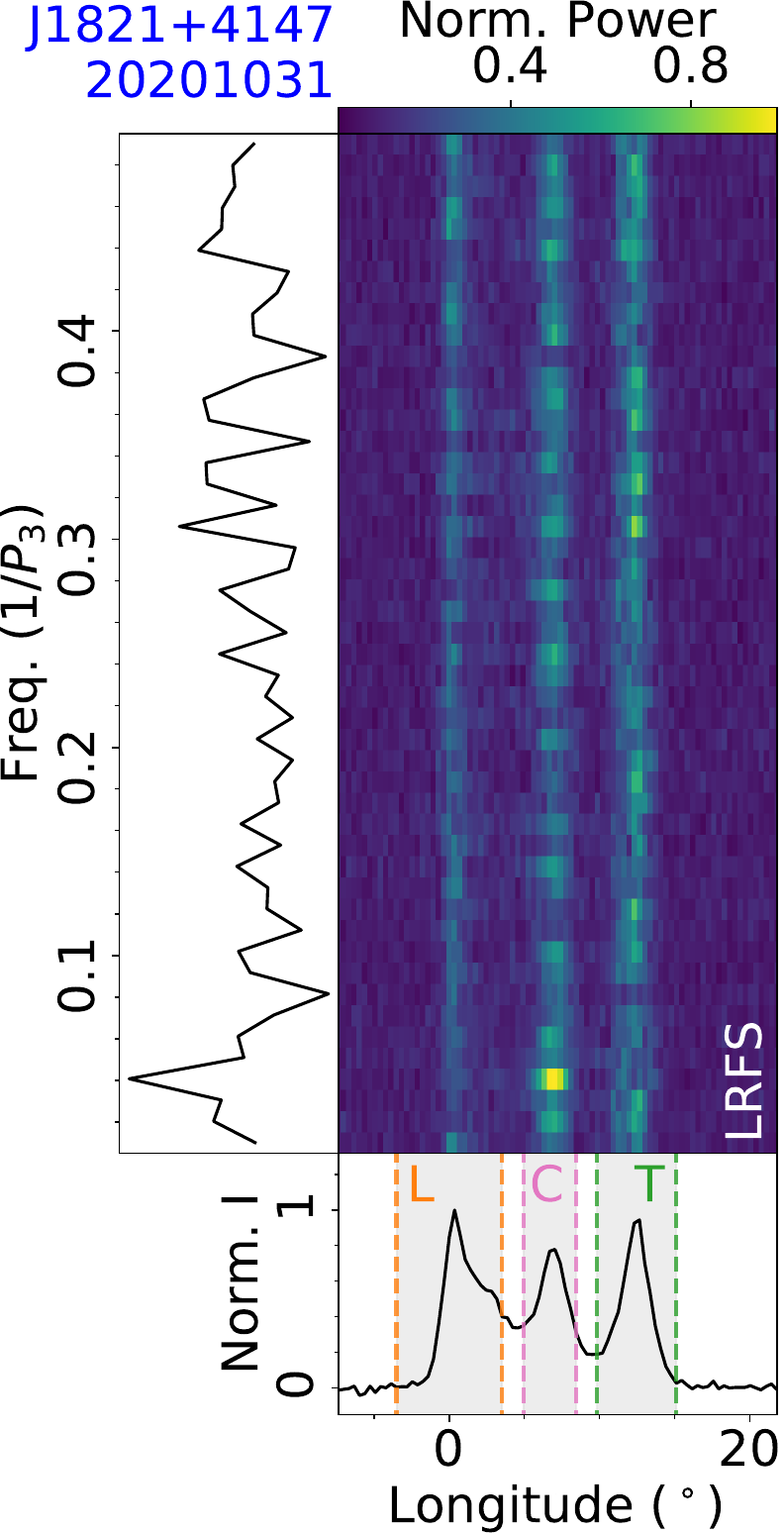}
\includegraphics[width=0.22\textwidth, angle=0]{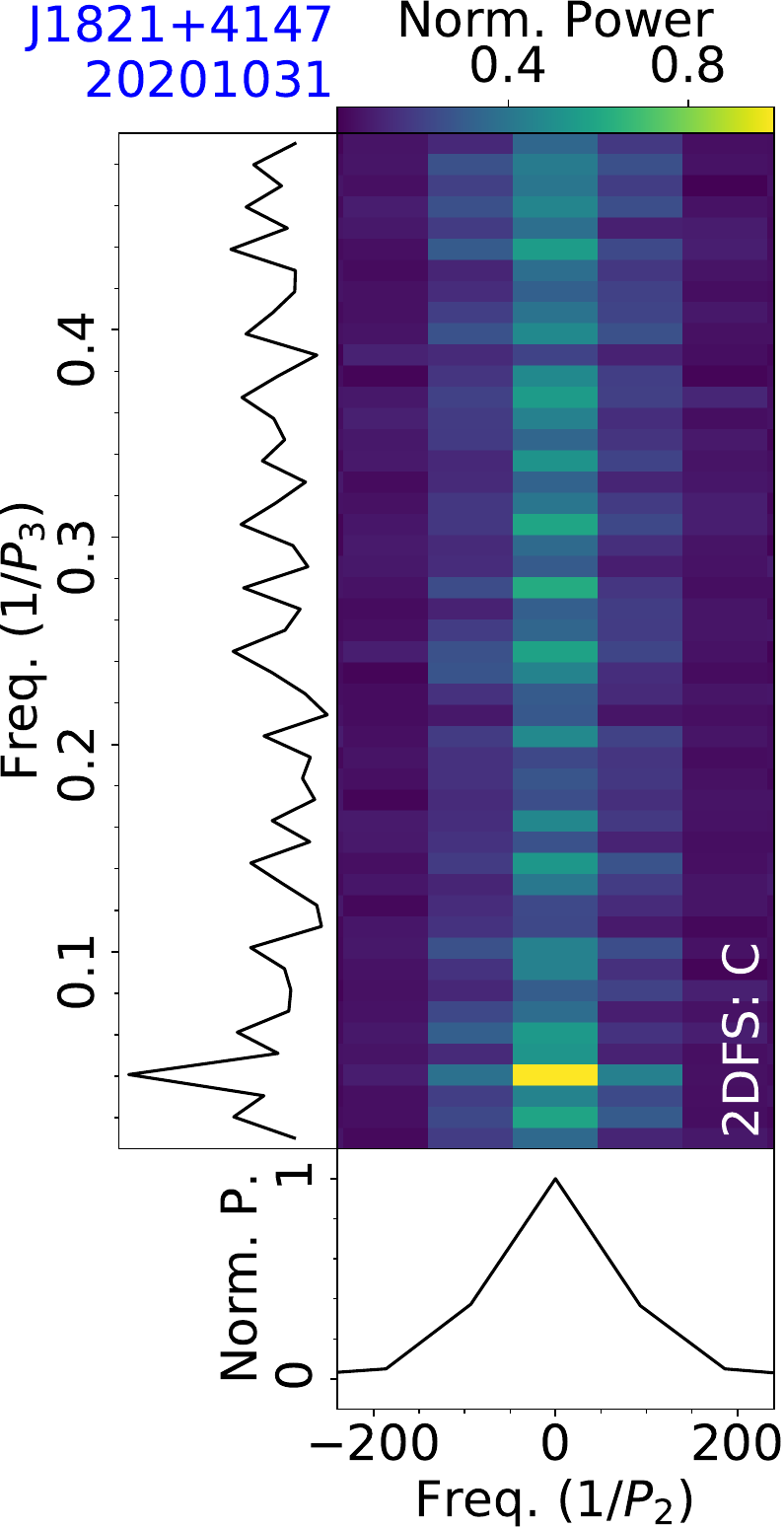}
\figcaption{Fluctuation analysis of PSR J1821+4147 for the observation on 20211123, with LRFS and 2DFS for the central phase region of a mean pulse profile.  \label{subfig:fluctu:J1821+4147}}
\end{figure}

\subsection{J1820-0427}
\label{subsec:J1820-0427}

PSR J1820-0427 was discovered by \citet{Vaughan1969} using the Molonglo radio telescope. In previous studies of \citet{Weltevrede2007}, the pulsar has a very broad feature with a preferred drift direction of $P_2=-300^{+50}_{-100}$ degrees and $P_3=3\pm1$ periods at 92 cm, as well as a longitude stationary feature of $P_3=22\pm3$ periods. \citet{Song2023} reported the drifting parameters of $P_3=3.4\pm0.3$ periods and $P_2=-23^{+7}_{-78}$ degrees.

This pulsar was observed by FAST on 20250201 for 6 minutes and on 20250228 for 15 minutes. From the longer observation, a rotation period $P=0.5980$~s and a dispersion measure $D\!M=84.6~{\rm cm^{-3}\,pc}$ were derived. The single pulse sequence and a zoomed-in view are shown in Fig.~\ref{subfig:TP:J1820-0427}, in which subpulses show no organized drift pattern. From LRFS and 2DFS in Fig.~\ref{subfig:fluctu:J1820-0427}, the pulsar has a preferred negative drift feature with the centroid modulation frequencies of $1/P_3=0.284\pm0.002$ ($P_3=3.52\pm0.03$ periods) and $1/P_2=-9\pm2$ ($P_2=-42\pm8^\circ$), as well as a low-frequency modulation feature of $1/P_3=0.033\pm0.001$ ($P_3=30\pm1$ periods).




\subsection{J1821+0155}
\label{subsec:J1821+0155}

PSR J1821+0155 was discovered from data for the GBT 350 MHz Drift Scan Survey \citep{Rosen2013}.

This pulsar was observed by FAST on 20210618 for 5 minutes, with a rotation period $P=0.0338$~s and a dispersion measure $D\!M=51.8~{\rm cm^{-3}\,pc}$ from this observation. Single pulse sequences in Fig.~\ref{subfig:TP:J1821+0155} show subpulse drifting behavior. From fluctuation spectra in Fig.~\ref{subfig:fluctu:J1821+0155}, the centroid modulation frequencies of the negative drift feature are estimated to be $1/P_3=0.2345\pm0.0003$ and $1/P_2=-25\pm1$, yielding the drifting parameters of $P_3=4.26\pm0.01$ periods and $P_2=-14.2\pm0.3^\circ$.

\begin{figure}[htpb]
\centering
\includegraphics[width=0.22\textwidth, angle=0]{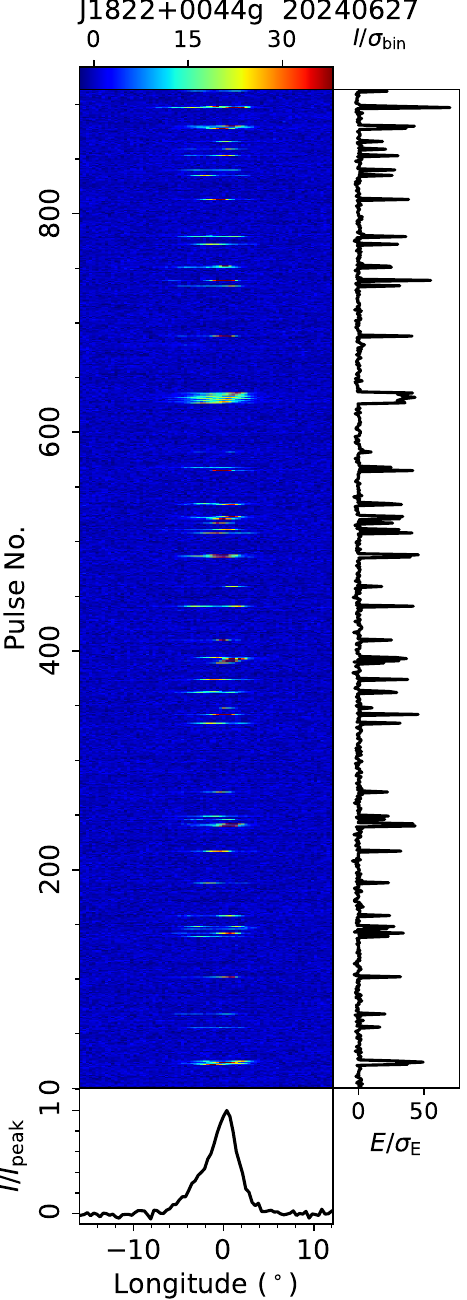}
\includegraphics[width=0.22\textwidth, angle=0]{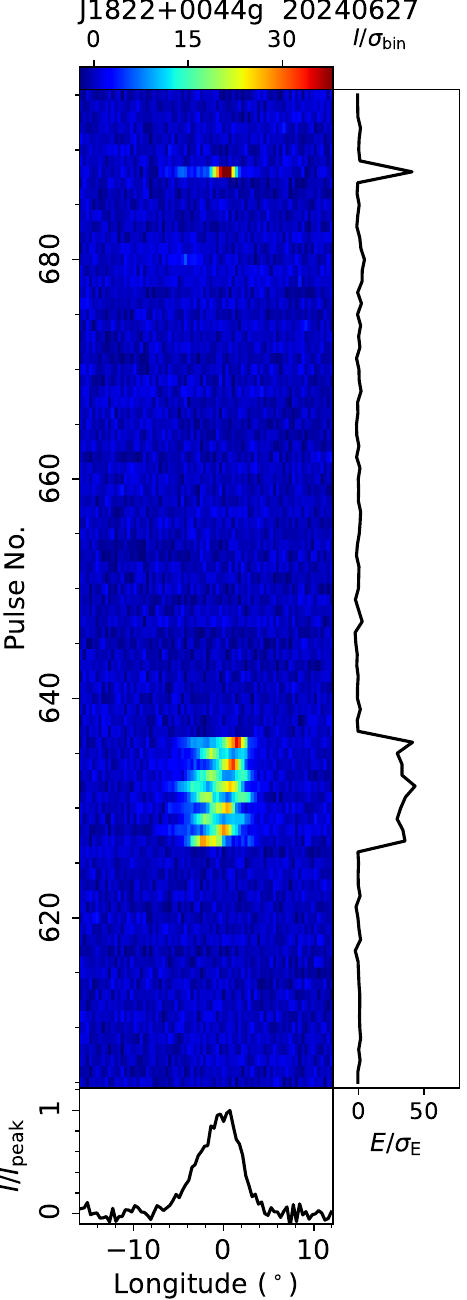}
\figcaption{Single pulse sequence of PSR J1822+0044g from the FAST observation on 20240627, and a zoomed-in view of pulses No. 605-695.
\label{subfig:TP:J1822+0044g}}
%
\centering
\includegraphics[width=0.39\textwidth, angle=0]{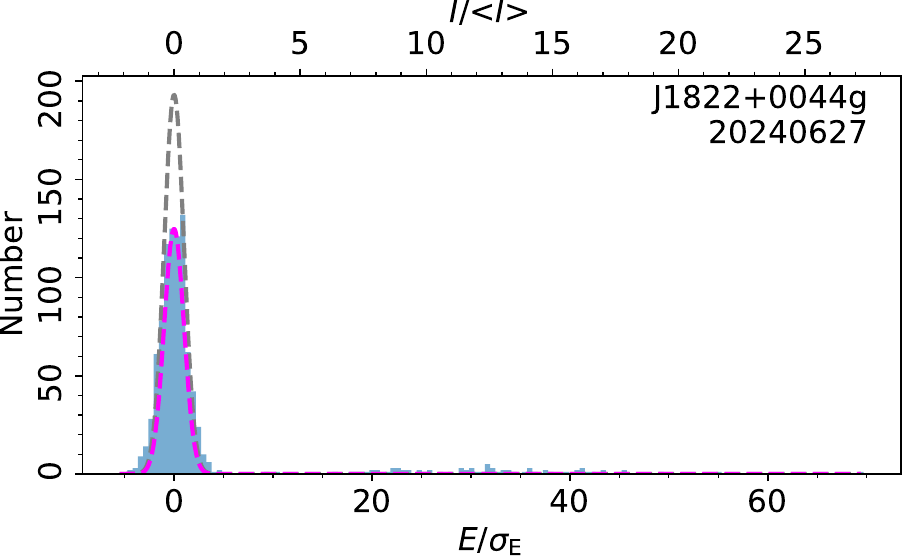}
\figcaption{On-pulse integral energy histograms of single pulses of PSR J1822+0044g from the FAST observation on 20240627. \label{subfig:Hist:J1822+0044g}}
%
\centering
\includegraphics[width=0.39\textwidth, angle=0]{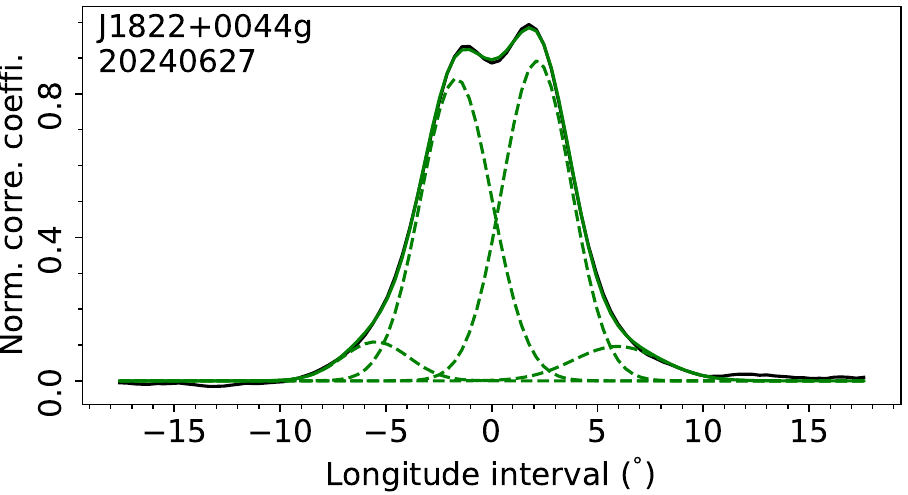}
\figcaption{Cross correlation of PSR J1822+0044g from the FAST observation on 20240627.
\label{subfig:Corre:J1822+0044g}}
\end{figure}

\begin{figure}[htpb]
\centering
\includegraphics[width=0.22\textwidth, angle=0]{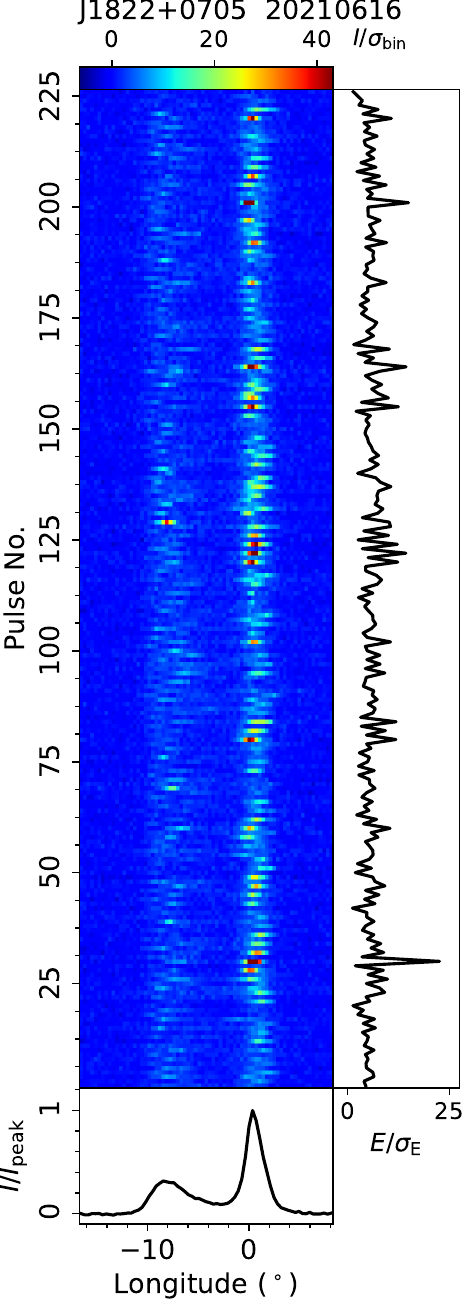}
\includegraphics[width=0.22\textwidth, angle=0]{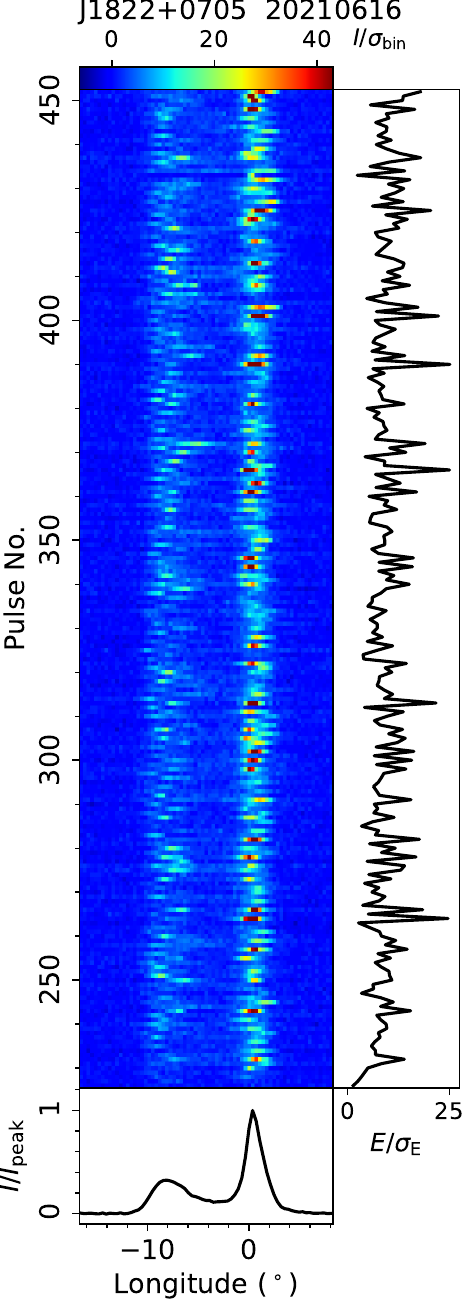}
\figcaption{Single pulse sequences of PSR J1822+0705 from the FAST observation on 20210616. \label{subfig:TP:J1822+0705}}
\end{figure}

\begin{figure}[htpb]
\centering
\includegraphics[width=0.22\textwidth, angle=0]{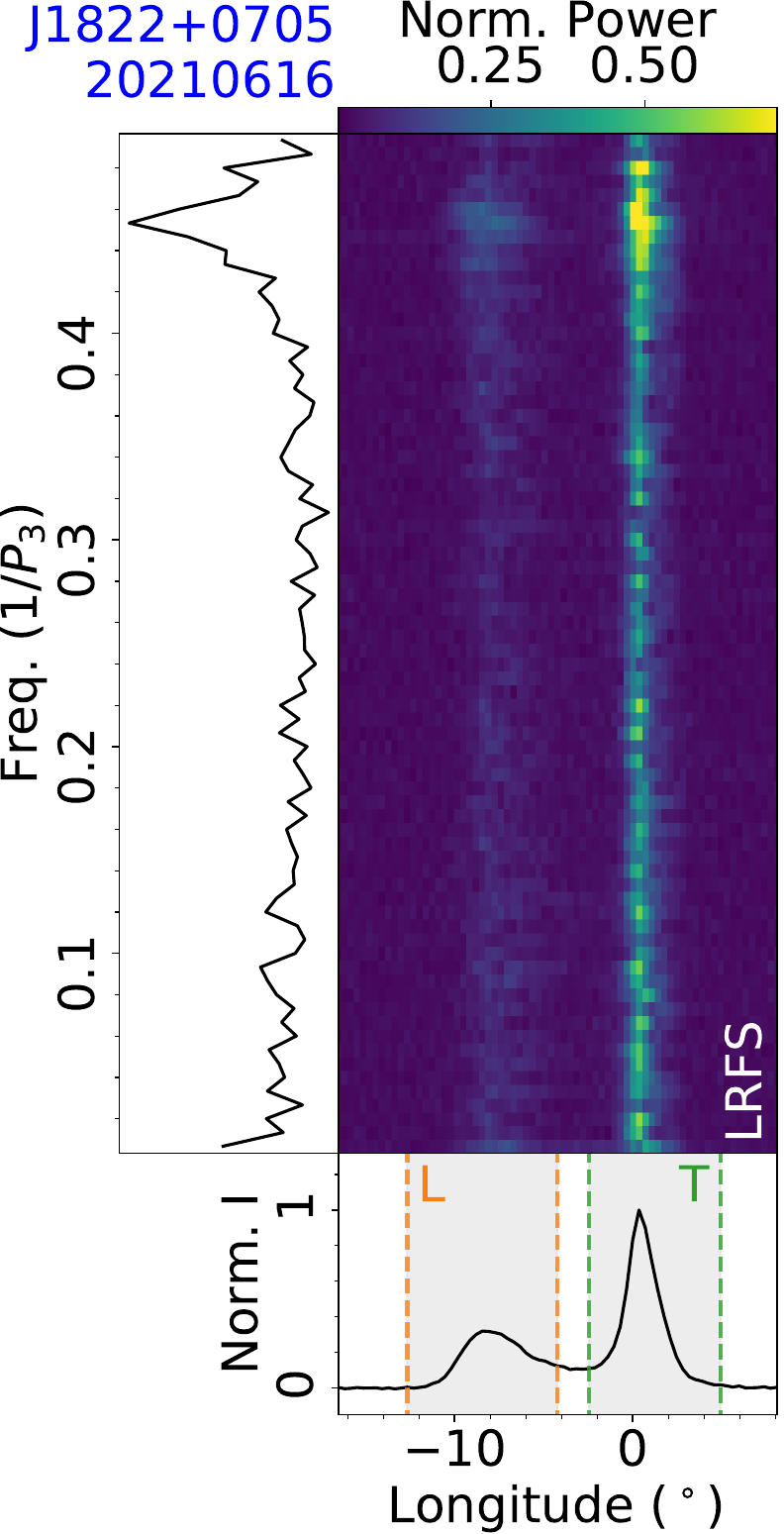}
\includegraphics[width=0.22\textwidth, angle=0]{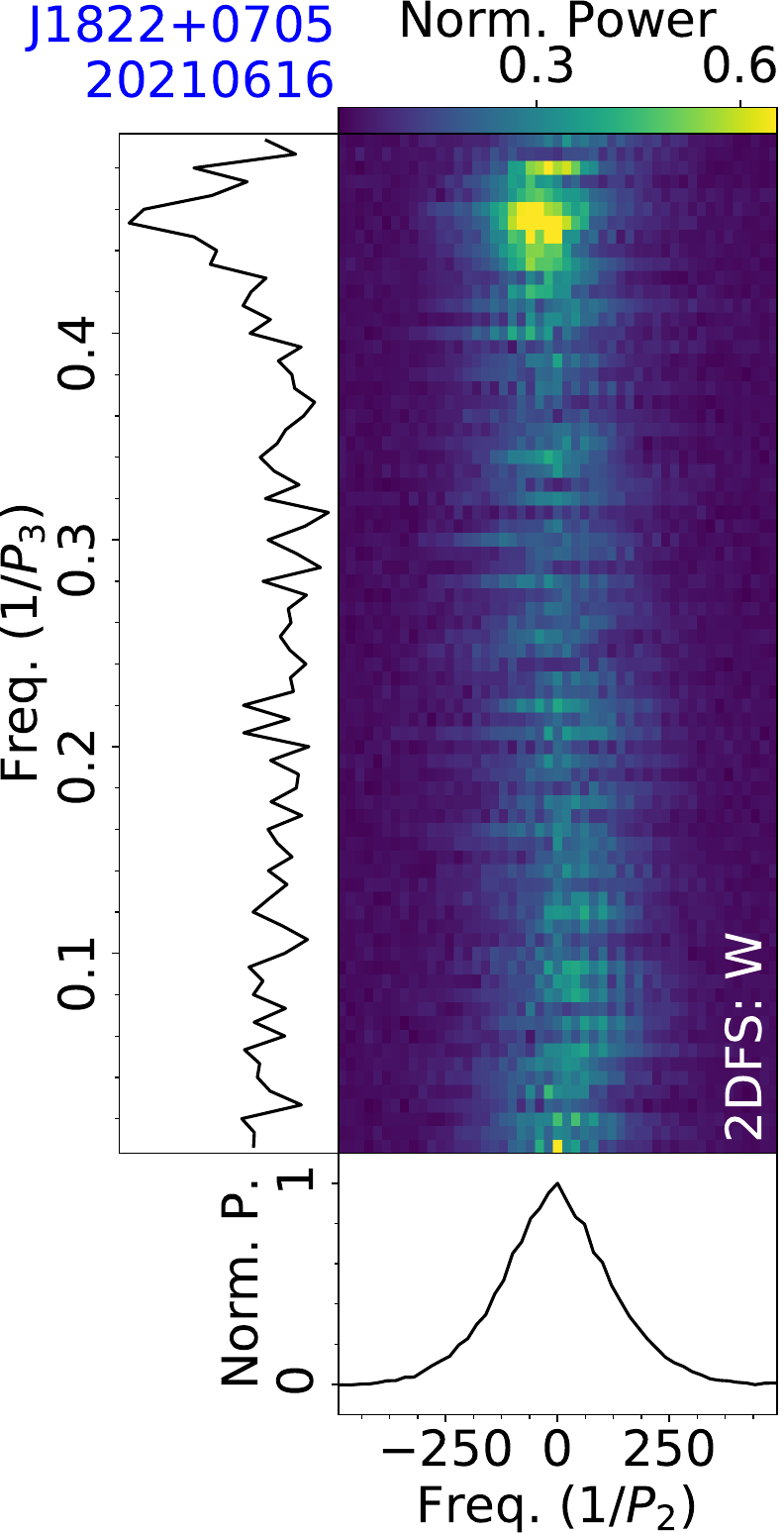}\\
\includegraphics[width=0.22\textwidth, angle=0]{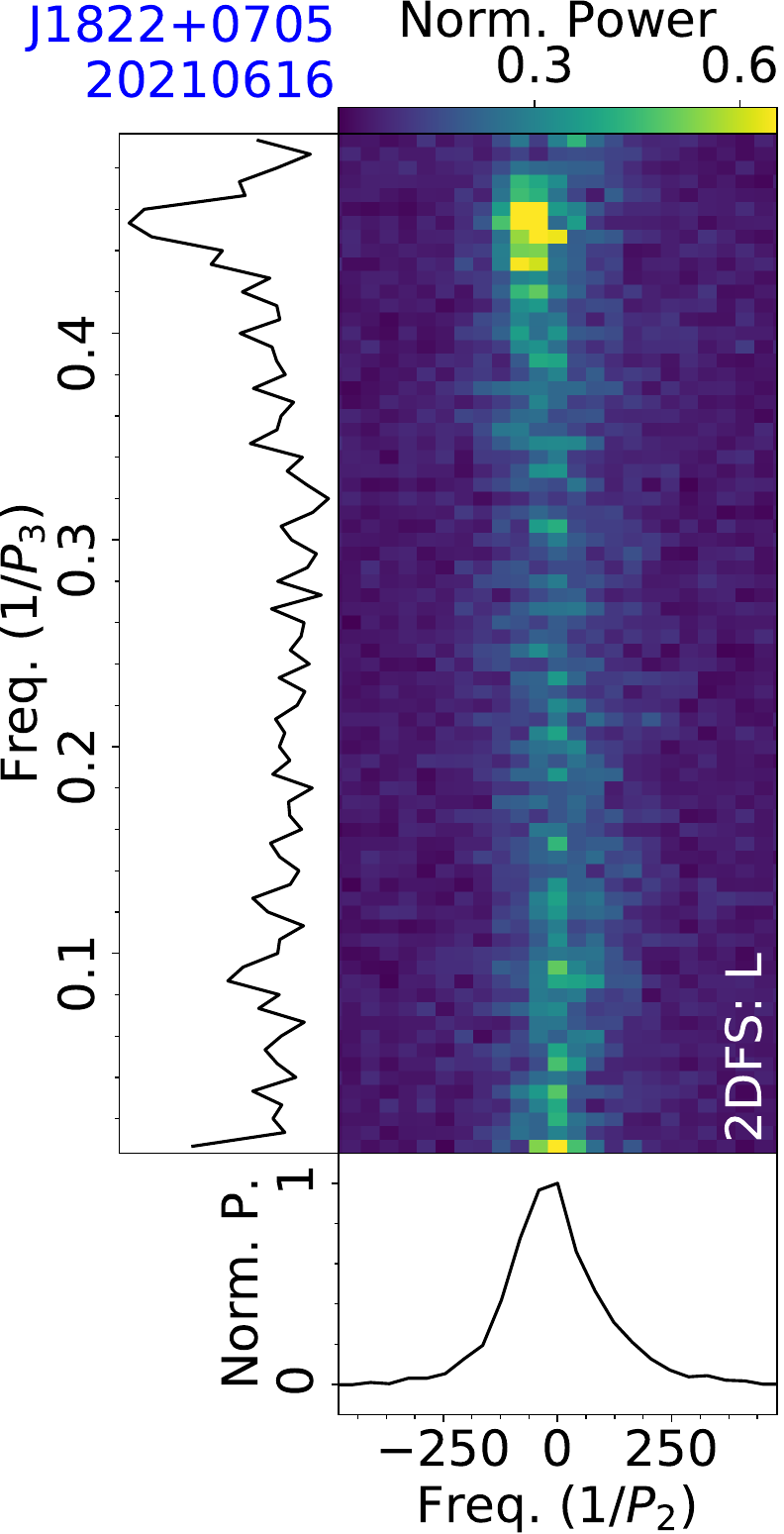}
\includegraphics[width=0.22\textwidth, angle=0]{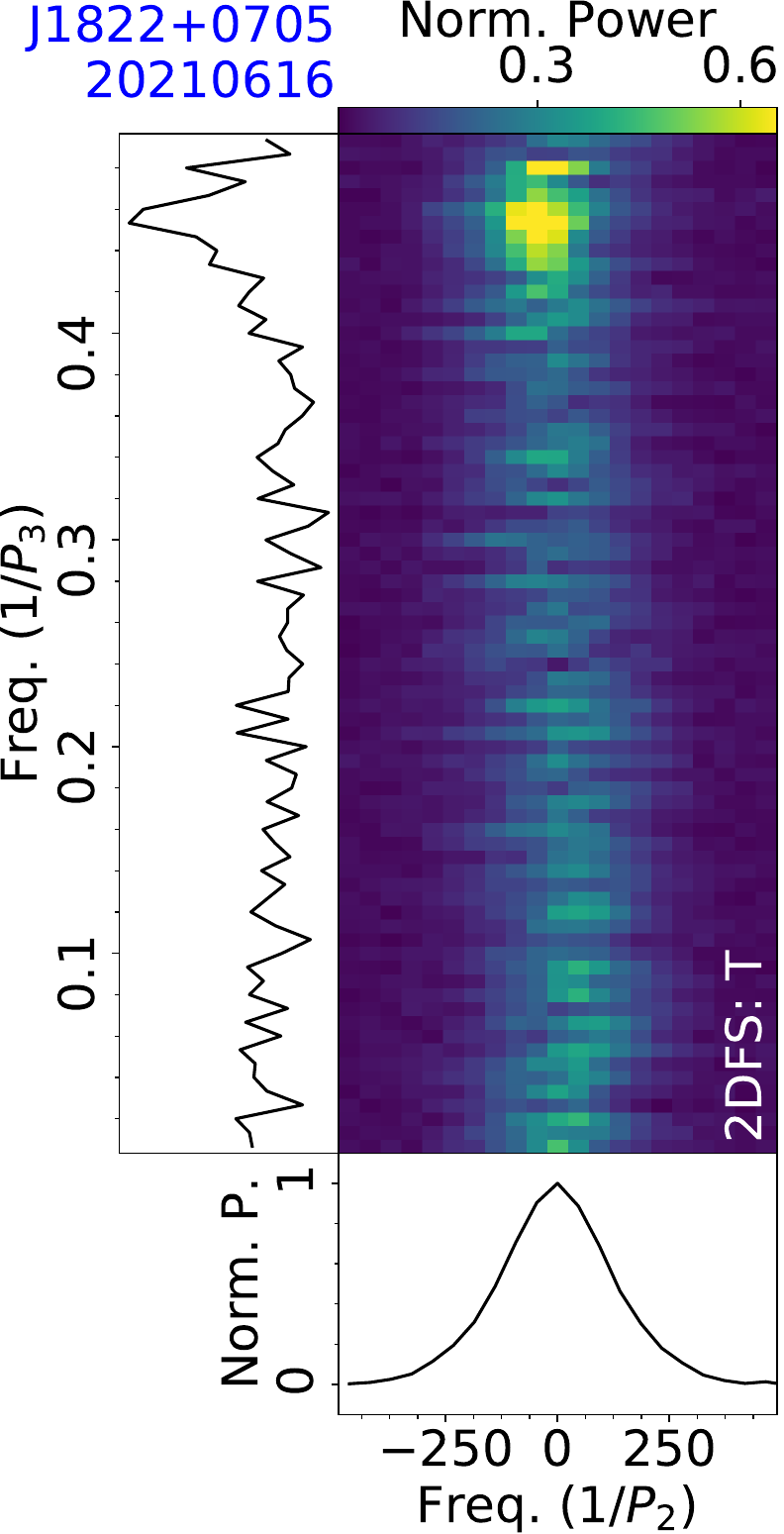}
\figcaption{Fluctuation analysis of PSR J1822+0705 for the observation on 20210616, with LRFS (top-left), and 2DFS for the on-pulse region (top-right), leading part (bottom-left) and trailing part (bottom-right) of a mean pulse profile.   \label{subfig:fluctu:J1822+0705}}
\end{figure}

\subsection{J1821+4147}
\label{subsec:J1821+4147}

PSR J1821+4147 was found in the Green Bank Northern Celestial Cap Pulsar Survey \citep{Stovall2014}. Nulling behavior of the pulsar was identified by \citet{Lynch2018}. 

This pulsar was observed by FAST on 20201031 for 8 minutes, deriving a rotation period $P=1.2619$~s and a dispersion measure $D\!M=41.1~{\rm cm^{-3}\,pc}$ from this observation.
Single pulse sequences are shown in Fig.~\ref{subfig:TP:J1821+4147}. The on-pulse integral energy histogram shown in Fig.~\ref{subfig:Hist:J1821+4147} indicates the existence of nulls, with the nulling fraction estimated to be 8.8$\pm$0.9\%. From single pulse sequences, there is a modulation of dual rotation periods related to changes between weak and bright intensities. The central component also exhibits a low-frequency modulation in intensity, with periodic enhancements and weakenings at a centroid frequency of $1/P_3=0.040\pm0.002$ (Fig.~\ref{subfig:fluctu:J1821+4147}), corresponding to $P_3=25\pm1$ periods.

\subsection{J1822+0044g}
\label{subsec:J1822+0044g}

PSR J1822+0044g was discovered in the FAST GPPS survey \citep{Han2021,han2025}. 

This pulsar was observed by FAST on 20240627 for 15 minutes, yielding a rotation period $P=0.9989$~s and a dispersion measure $D\!M=55.3~{\rm cm^{-3}\,pc}$ from this observation. 
Single pulse sequences of PSR J1822+0044g in Fig.~\ref{subfig:TP:J1822+0044g} illustrate that there are both nulling and subpulse drifting behaviors. From the on-pulse energy histogram in Fig.~\ref{subfig:Hist:J1822+0044g}, the nulling fraction is 65\%. 
The emission duration is typically short, lasting for one to two periods. While the single pulse segment of pulse No.627-636 shows clear subpulse drifting, with $P_2=-3.81\pm0.02^\circ$ and drifting rate $D=-1.68\pm0.08$ degrees per period, which are determined by the cross-correlation method (Fig.~\ref{subfig:Corre:J1822+0044g}).

\subsection{J1822+0705}
\label{subsec:J1822+0705}

PSR J1822+0705 was discovered by \citet{Foster1995} using the Arecibo telescope. Subpulse drifting behavior was reported by \citet{Song2023} previously. 

The pulsar was observed by FAST on 20210616 for 10 minutes, with a rotation period $P=1.3627$~s and a dispersion measure $D\!M=61.4~{\rm cm^{-3}\,pc}$ from this observation. 
Fig.~\ref{subfig:TP:J1822+0705} shows the clear drifting phenomenon of the pulsar. 2DFS (Fig.~\ref{subfig:fluctu:J1822+0705}) of the leading and trailing parts in a mean pulse profile have drift features with similar temporal modulation frequencies. For the leading part, the centroid frequencies are estimated to be $1/P_3=0.449\pm0.001$ and $1/P_2=-66\pm3$, corresponding to $P_3=2.227\pm0.004$ periods and $P_2=-5.4\pm0.2^\circ$. In 2DFS of the trailing profile part, the centroid drift feature is characterized by $1/P_3=0.456\pm0.001$ and $1/P_2=-47\pm3$, yielding $P_3=2.191\pm0.003$ periods and $P_2=-8\pm1$.

\begin{figure}[htpb]
\centering
\includegraphics[width=0.21\textwidth, angle=0]{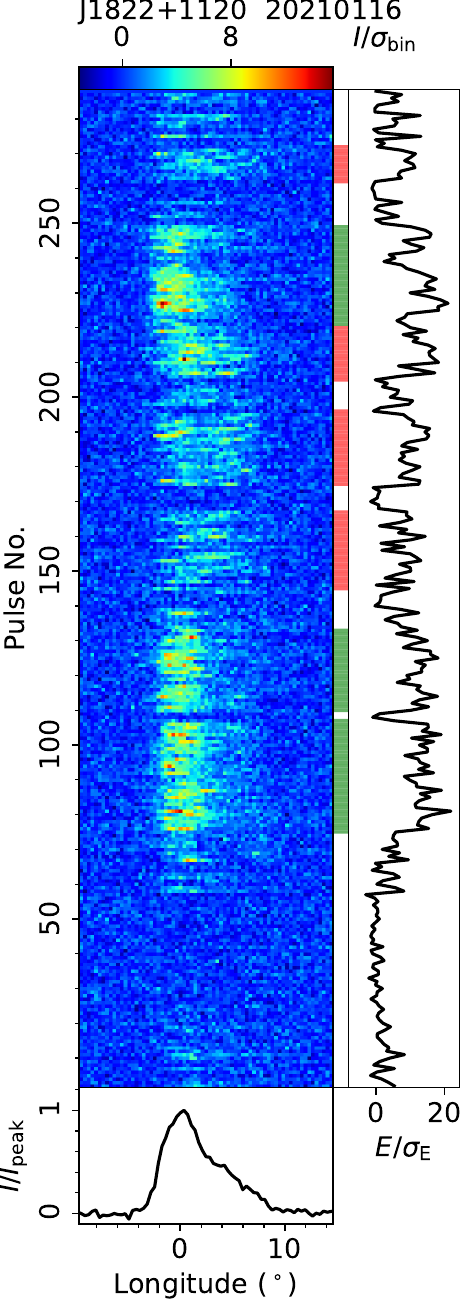}
\figcaption{Single pulse sequence of PSR J1822+1120 from the FAST observation on 20210116. \label{subfig:TP:J1822+1120}}
\end{figure}

\begin{figure}[htpb]
\centering
\includegraphics[width=0.39\textwidth, angle=0]{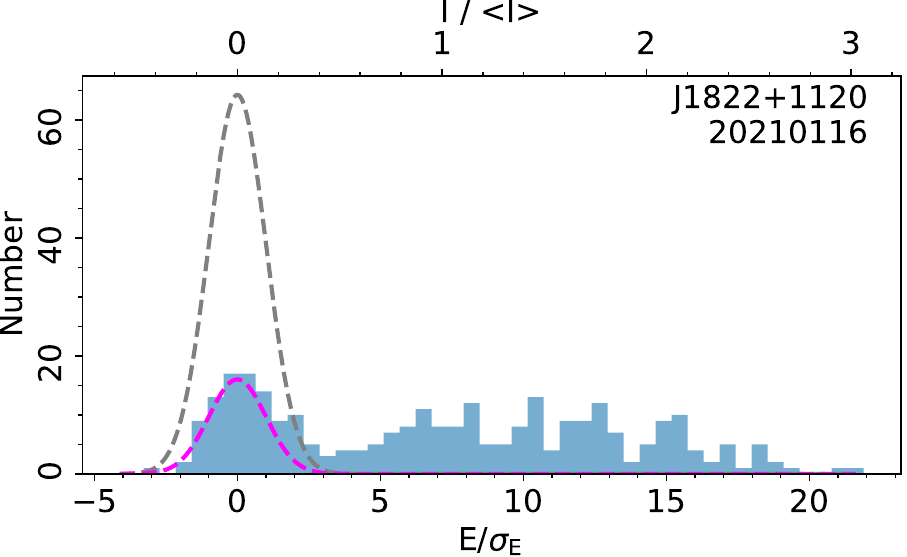}
\figcaption{On-pulse energy histogram of single pulses of PSR J1822+1120 from the FAST observation on 20210116.
\label{subfig:Hist:J1822+1120}}
\end{figure}

\begin{figure}[htpb]
\centering
\includegraphics[width=0.39\textwidth, angle=0]{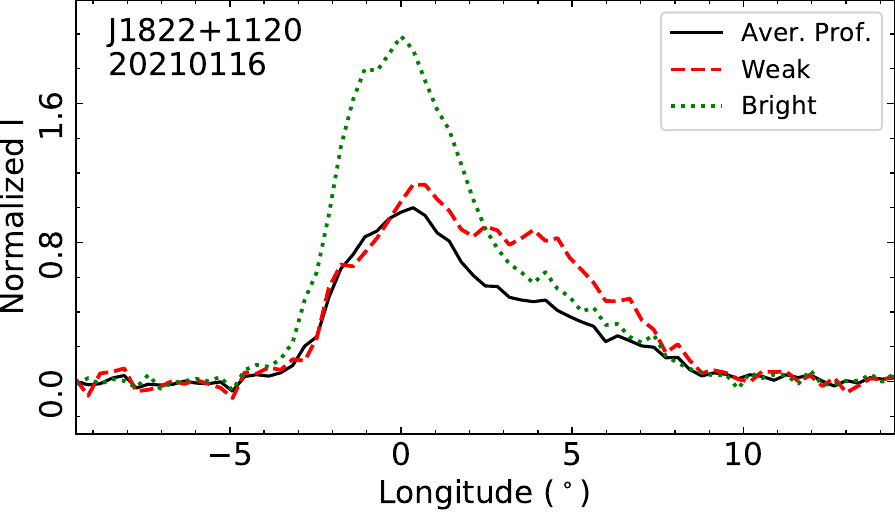} 
\figcaption{Mean profiles of weak and bright emission modes of PSR J1822+1120 from the FAST observation on 20210116.
\label{subfig:PolModes:J1822+1120}}
\end{figure}

\begin{figure}[htpb]
\centering
\includegraphics[width=0.22\textwidth, angle=0]{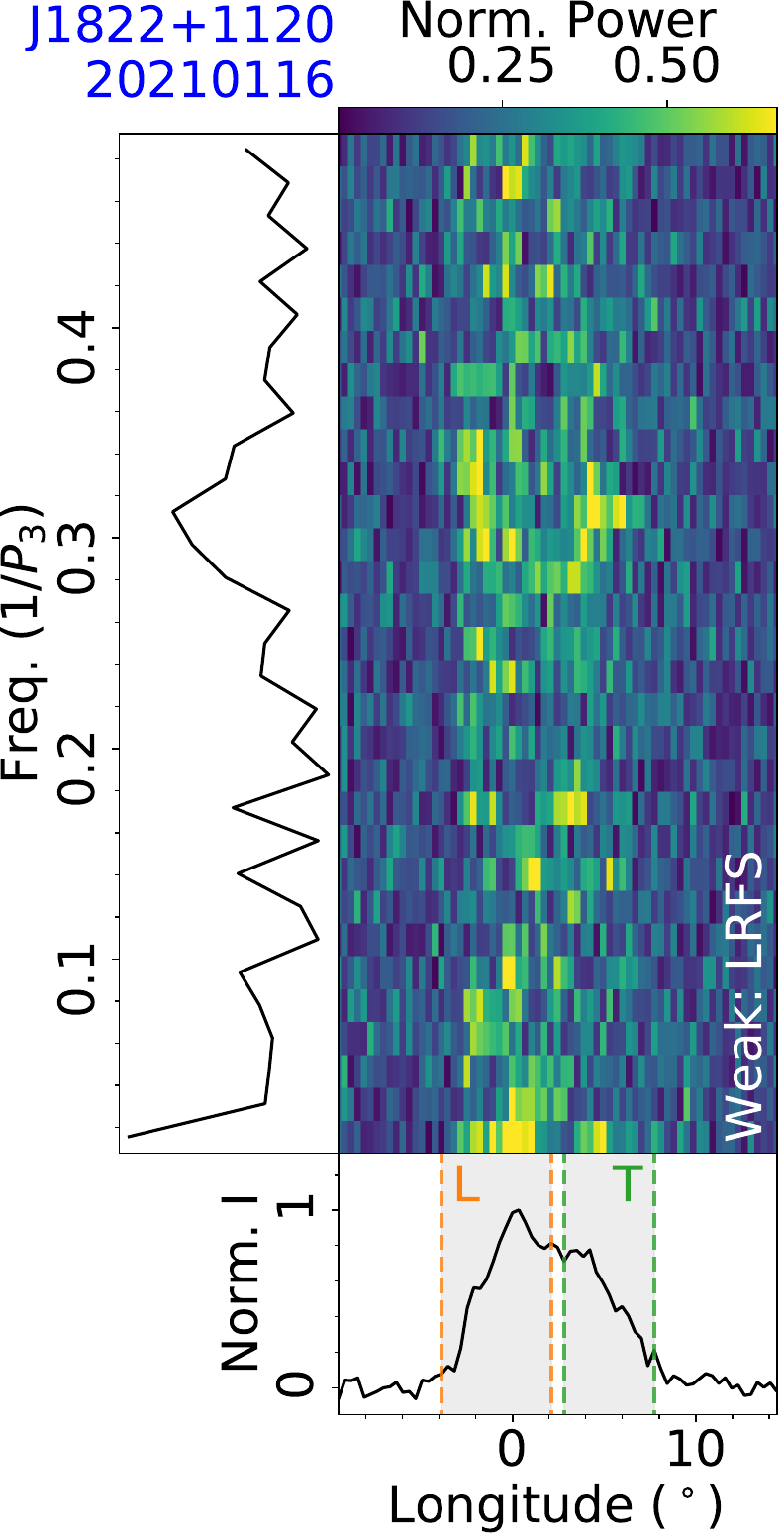}
\includegraphics[width=0.22\textwidth, angle=0]{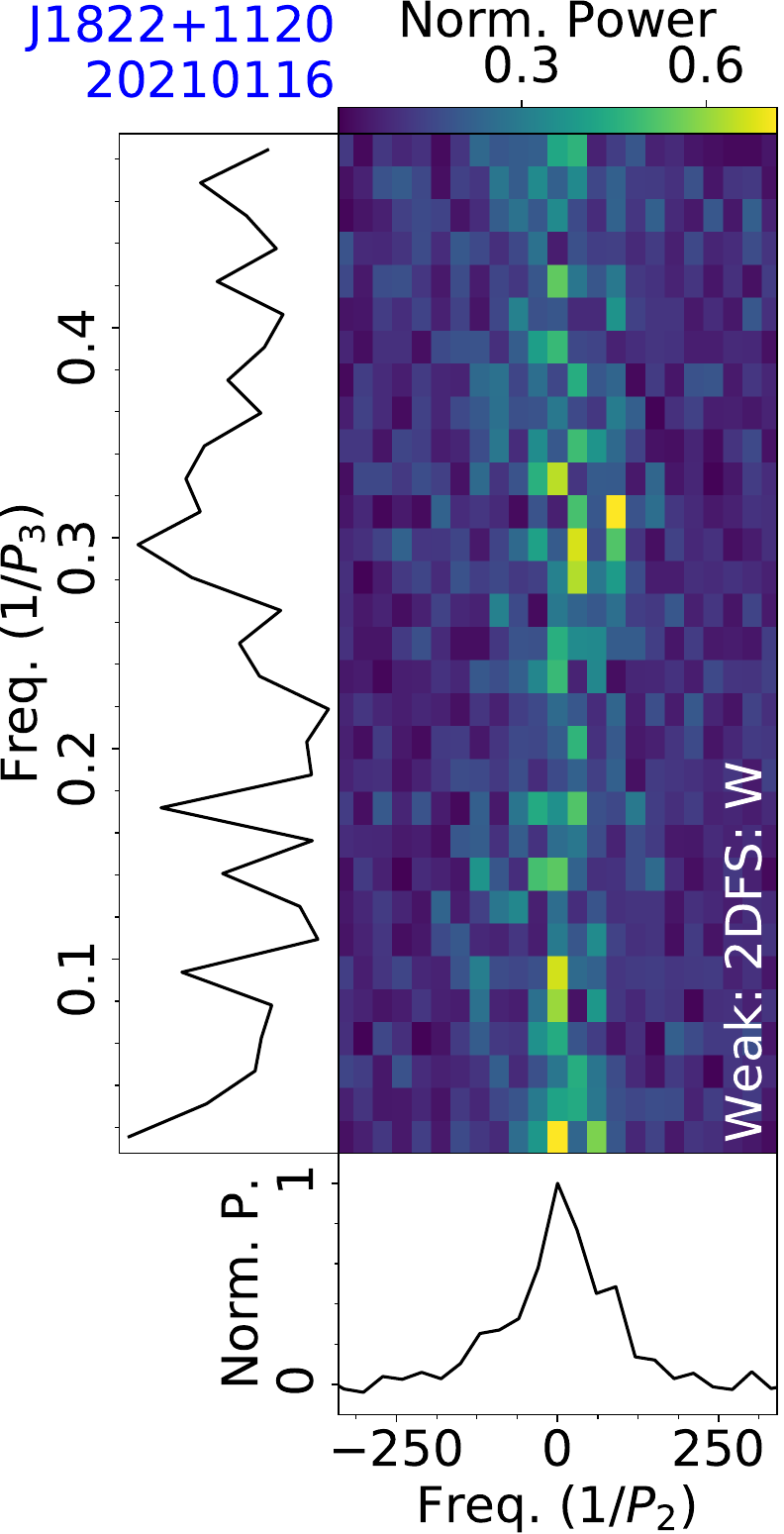}
\figcaption{Fluctuation analysis of PSR J1822+1120 for the observation on 20210116, with LRFS 
and 2DFS for the on-pulse region 
for the weak emission mode. 
\label{subfig:fluctu:J1822+1120}}
\end{figure}

\begin{figure}[htpb]
\centering
\includegraphics[width=0.22\textwidth, angle=0]{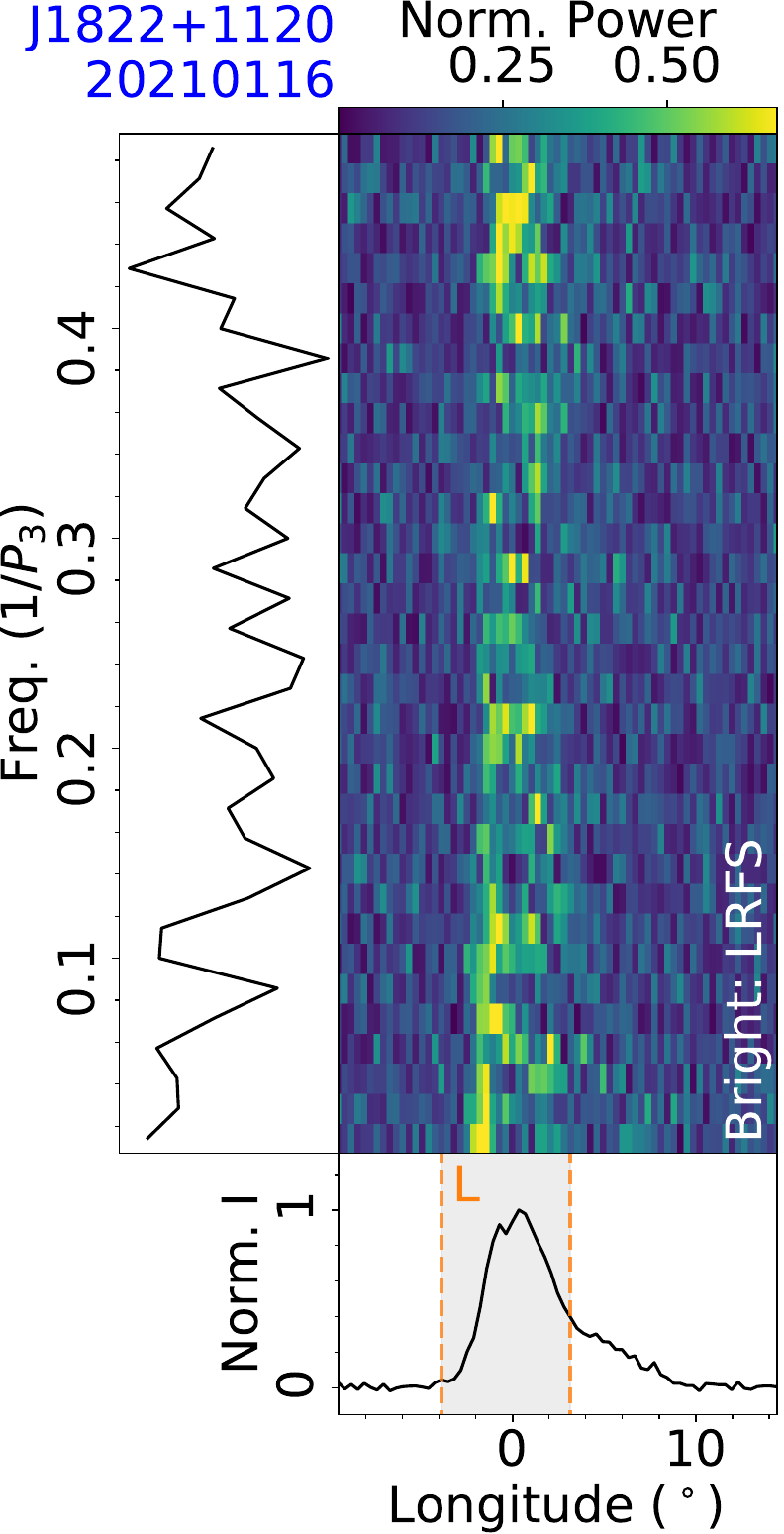}
\includegraphics[width=0.22\textwidth, angle=0]{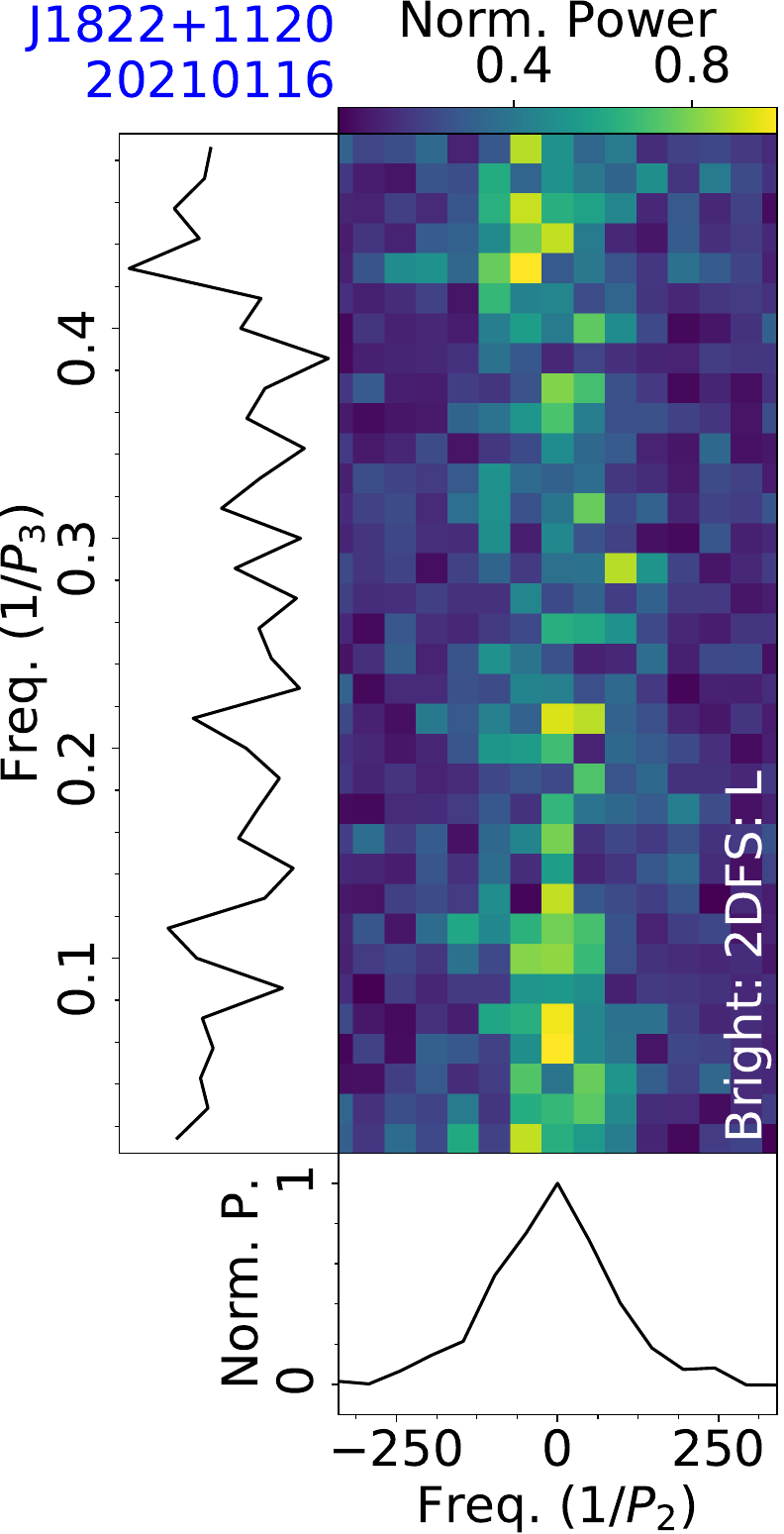}
\figcaption{Similar to Fig.~\ref{subfig:fluctu:J1822+1120}, this figure shows LRFS and 2DFS of the leading part of the profile for the bright emission mode.}
\end{figure}

\begin{figure}[htpb]
\centering
\includegraphics[width=0.22\textwidth, angle=0]{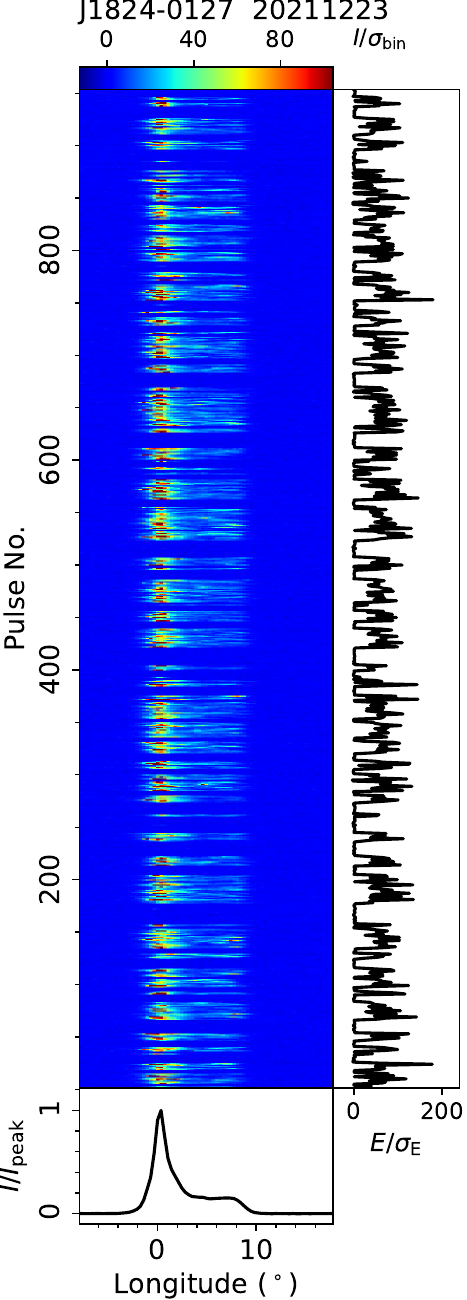}
\includegraphics[width=0.22\textwidth, angle=0]{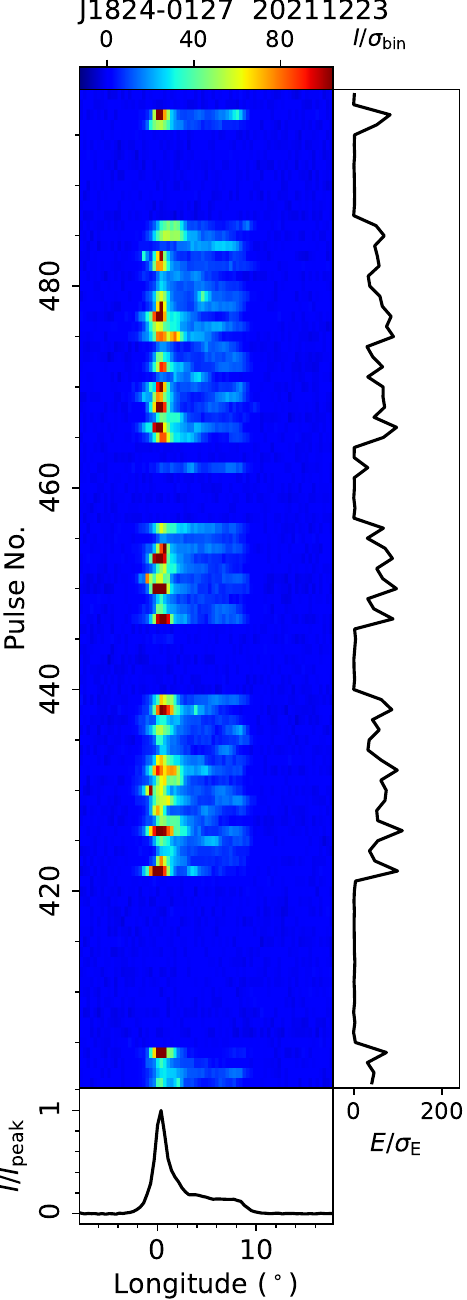}
\figcaption{Single pulse sequence of PSR J1824-0127 from the FAST observation on 20211223, and a zoomed-in view of pulses No. 400-500.
\label{subfig:TP:J1824-0127}}
\end{figure}

\begin{figure}[htpb]
\centering
\includegraphics[width=0.39\textwidth, angle=0]{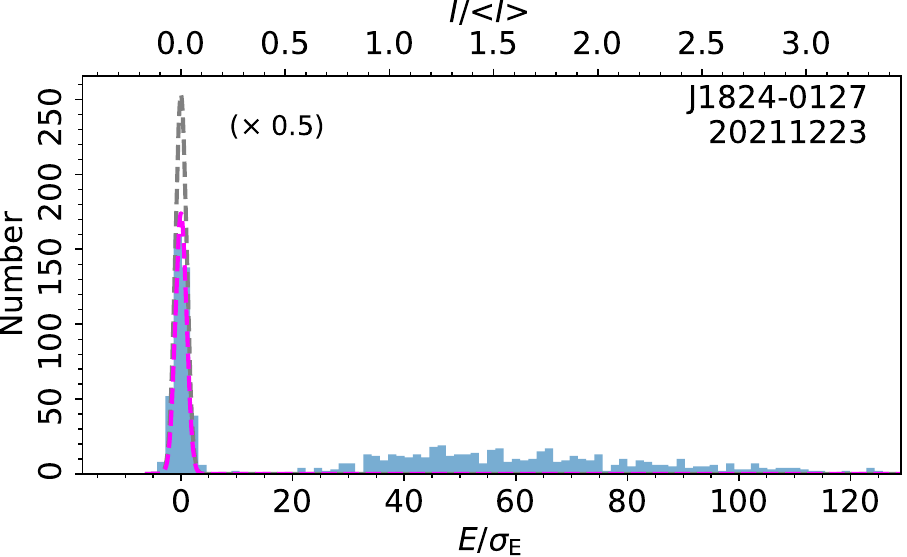}
\vspace{-0.4cm}
\figcaption{On-pulse energy histogram of PSR J1824-0127 from the FAST observation on 20211223. \label{subfig:Hist:J1824-0127}}
%
\centering
\includegraphics[width=0.22\textwidth, angle=0]{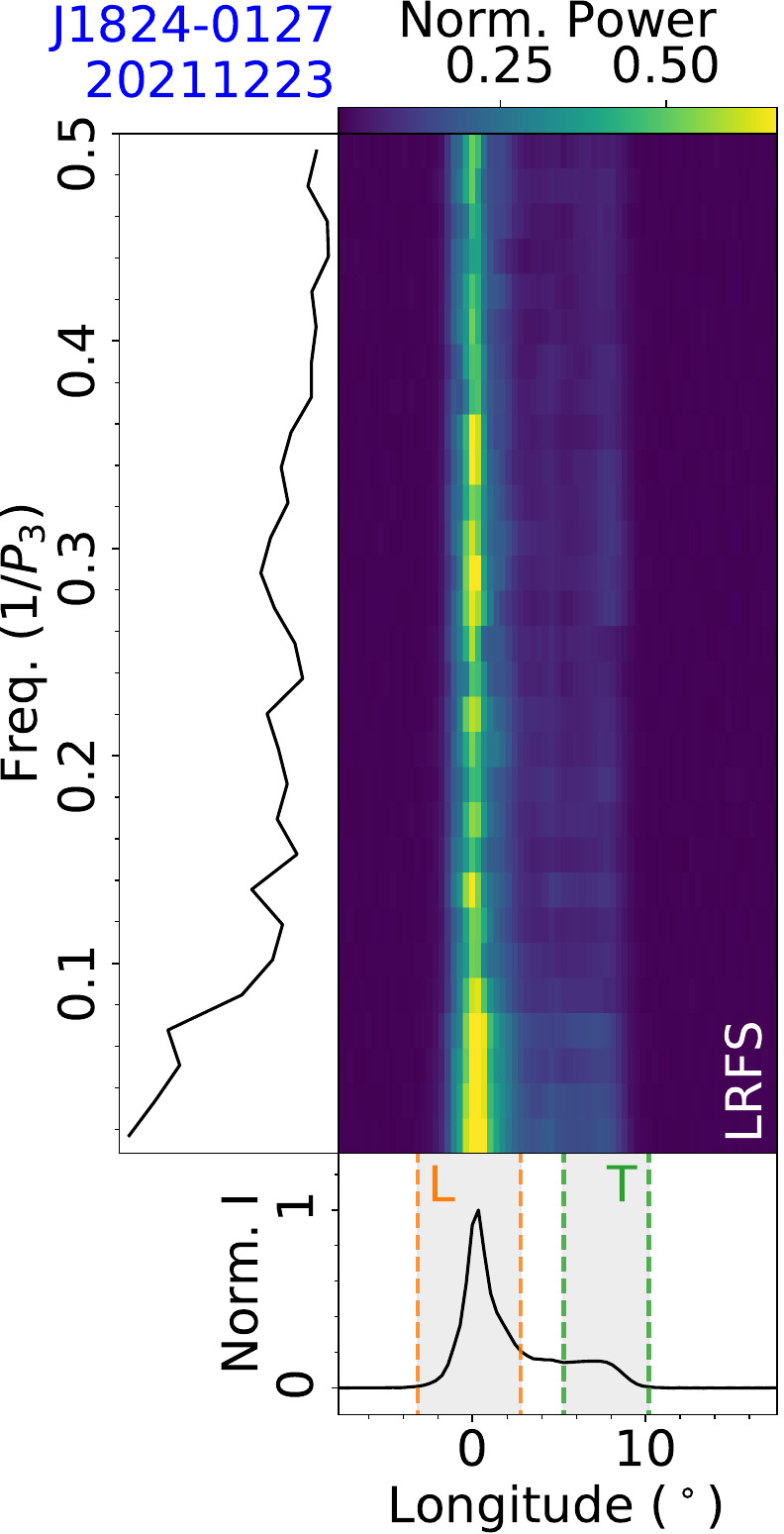}
\includegraphics[width=0.22\textwidth, angle=0]{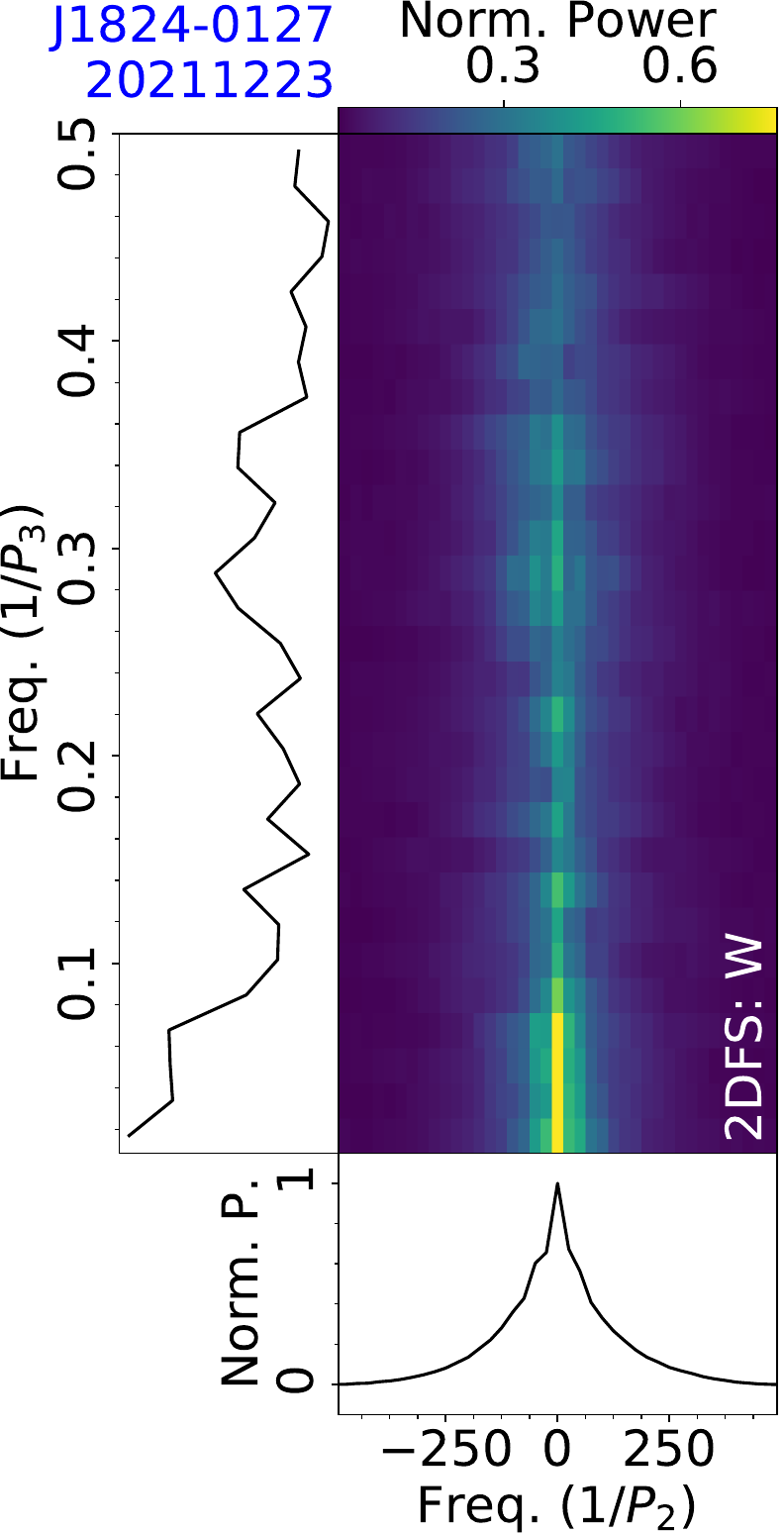}\\
\includegraphics[width=0.22\textwidth, angle=0]{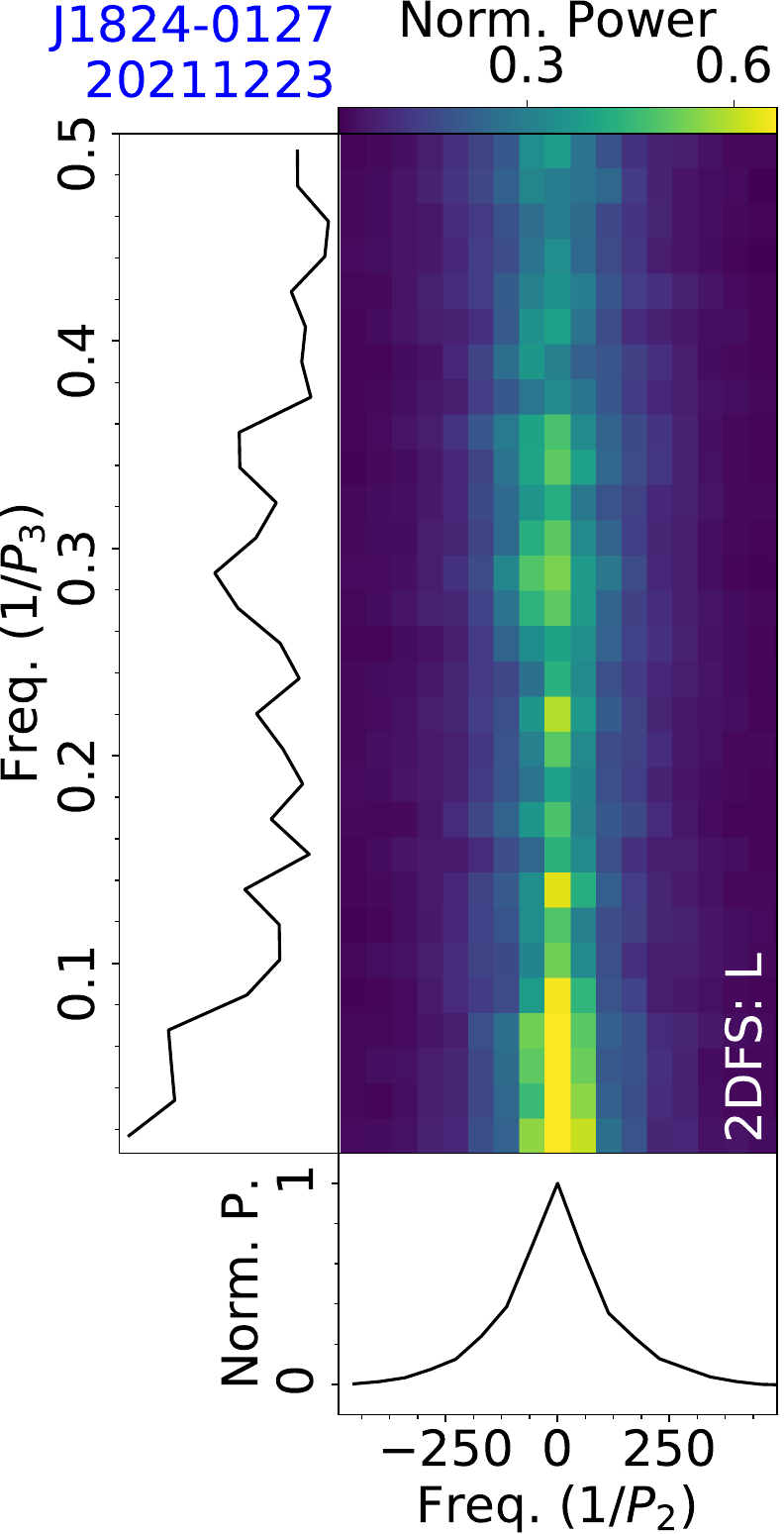}
\includegraphics[width=0.22\textwidth, angle=0]{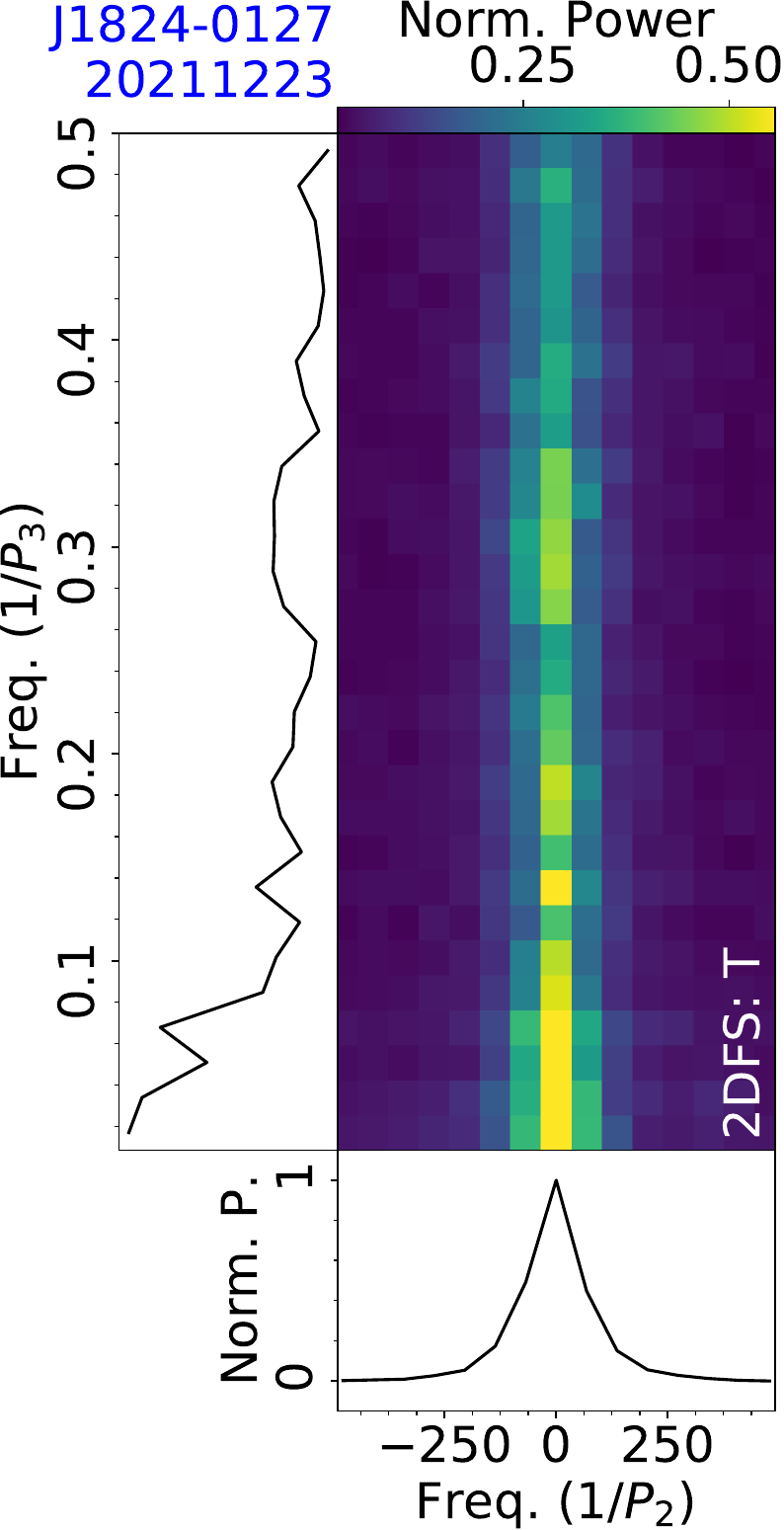}
\figcaption{Fluctuation analysis of PSR J1824-0127 for the observation on 20211223, with LRFS (top-left), and 2DFS for the on-pulse region (top-right), leading part (bottom-left) and trailing part (bottom-right) of a mean pulse profile. \label{subfig:fluctu:J1824-0127}}
\end{figure}

\subsection{J1822+1120}
\label{subsec:J1822+1120}

PSR J1822+1120 was discovered by the 305-m Arecibo radio telescope in a 430-MHz survey for pulsars \citep{Lorimer2005}.

This pulsar was observed by FAST on 20210116 for 9 minutes, deriving a rotation period $P=1.7872$~s and a dispersion measure $D\!M=93.4~{\rm cm^{-3}\,pc}$ from this observation. 
The single pulse sequence of this observation is shown in Fig.~\ref{subfig:TP:J1822+1120}, illustrating the existence of nulling, mode changing, as well as subpulse drifting phenomena for the first time. 

The nulling fraction is estimated from the on-pulse integral energy histogram in Fig.~\ref{subfig:Hist:J1822+1120}. The distribution around the zero energy indicates the nulling behavior of this pulsar. From fitting with the Gaussian function obtained using the off-pulse integral energy, the nulling fraction of this observation is estimated to be 25$\pm$2\%. 

Additionally, there seem to be 2 emission modes from the single pulse sequence, which are labeled using red and green colors, and defined to be the bright and weak modes on the basis of the intensity of the leading component. From averaged profiles of two emission modes shown in Fig.~\ref{subfig:PolModes:J1822+1120}, the intensity ratio of the leading component to the trailing component of the bright mode are lager than the weak mode. The trailing part in the profile of the longitude 2.5$^\circ$ to 9$^\circ$ of the weak mode is stronger than that of the bright mode. 

Weak and bright emission modes both have subpulse drifting behavior. To obtain the drifting parameters, fluctuation spectra of two components for the weak emission mode and the leading component for the bright emission mode are displayed in Fig.~\ref{subfig:fluctu:J1822+1120}. 
For the weak emission mode, drift bands are clear in the single pulse sequence (Fig.~\ref{subfig:TP:J1822+1120}). There are positive drift features in 2DFS of the leading and trailing profile parts. 
For the weak emission mode, the centroid frequencies of the drift feature in 2DFS are $1/P_3=0.298\pm0.004$ ($P_3=3.36\pm0.05$ periods) and $1/P_2=57\pm10$ ($P_2=6\pm2^\circ$) for the leading part, and $1/P_3=0.301\pm0.003$ ($P_3=3.32\pm0.03$ periods) and $1/P_2=68\pm12$ ($P_2=6\pm1^\circ$) for the trailing part of a mean pulse profile. 
For the bright emission mode, 2DFS of the leading part of the profile exhibits a negative drift feature, with the centroid frequencies of $1/P_3=0.452\pm0.003$ and $1/P_2=-52\pm8$, corresponding to $P_3=2.21\pm0.02$ periods and $P_2=-7\pm1^\circ$.
%

More observations are needed for a more accurate analysis of this pulsar.

\begin{figure}[htpb]
\centering
\includegraphics[width=0.22\textwidth, angle=0]{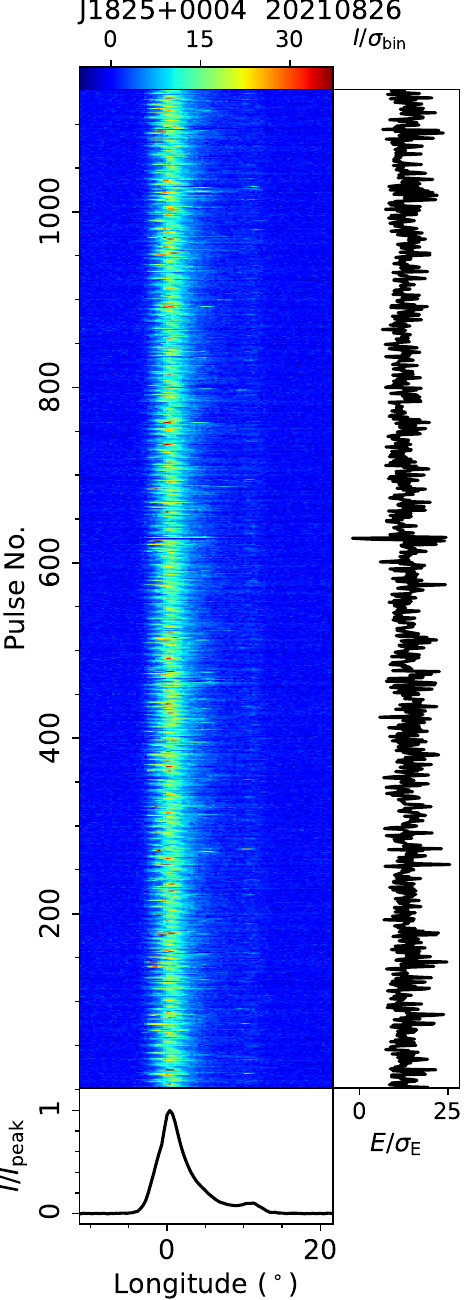}
\includegraphics[width=0.22\textwidth, angle=0]{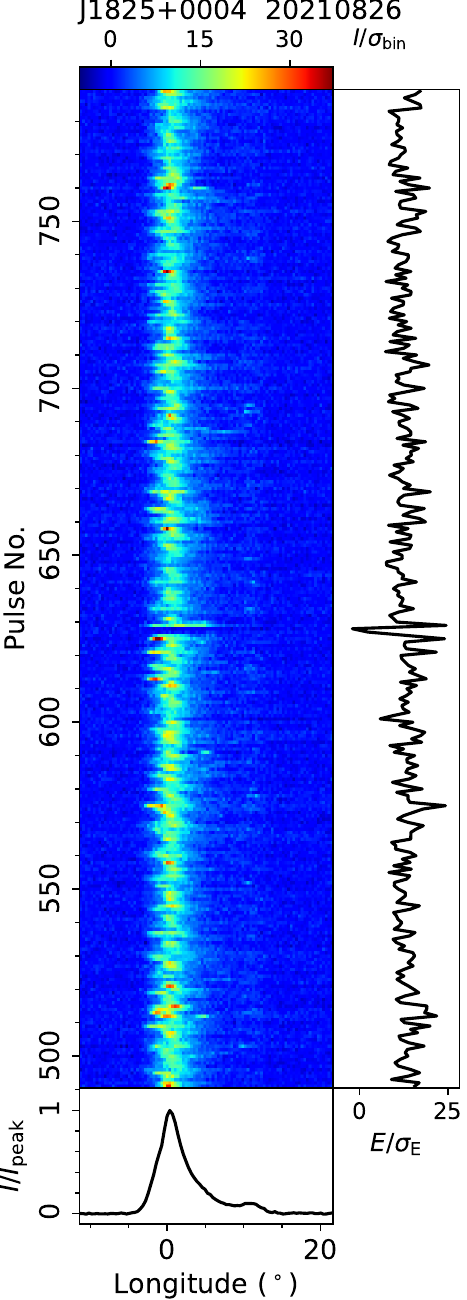}
\figcaption{Single pulse sequence of PSR J1825+0004 from the FAST observation on 20210826, and a zoomed-in view of pulses No. 490-790.
\label{subfig:TP:J1825+0004}}
\end{figure}

\begin{figure}[htpb]
\centering
\includegraphics[width=0.39\textwidth, angle=0]{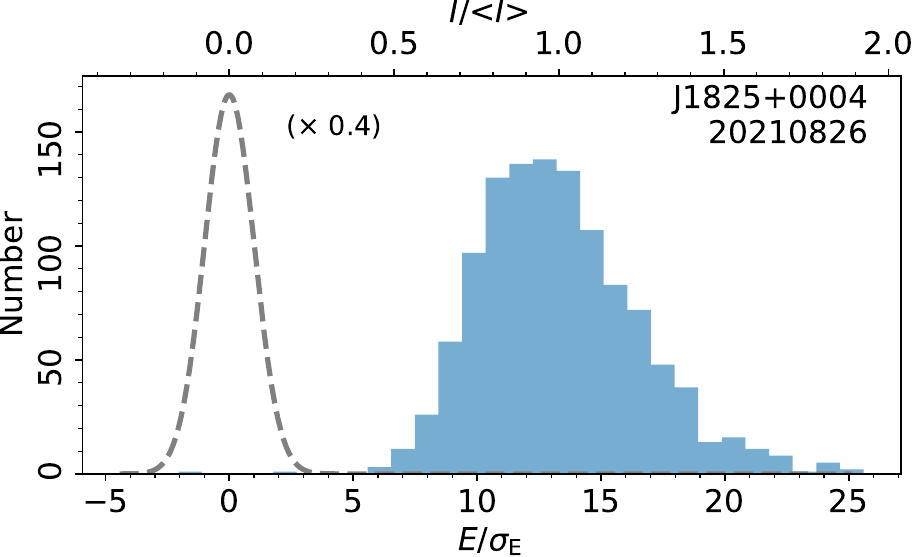}
\figcaption{On-pulse energy histogram of single pulses of PSR J1825+0004 from the FAST observation on 20210826. \label{subfig:Hist:J1825+0004}}
\end{figure}

\begin{figure}[htpb]
\centering
\includegraphics[width=0.22\textwidth, angle=0]{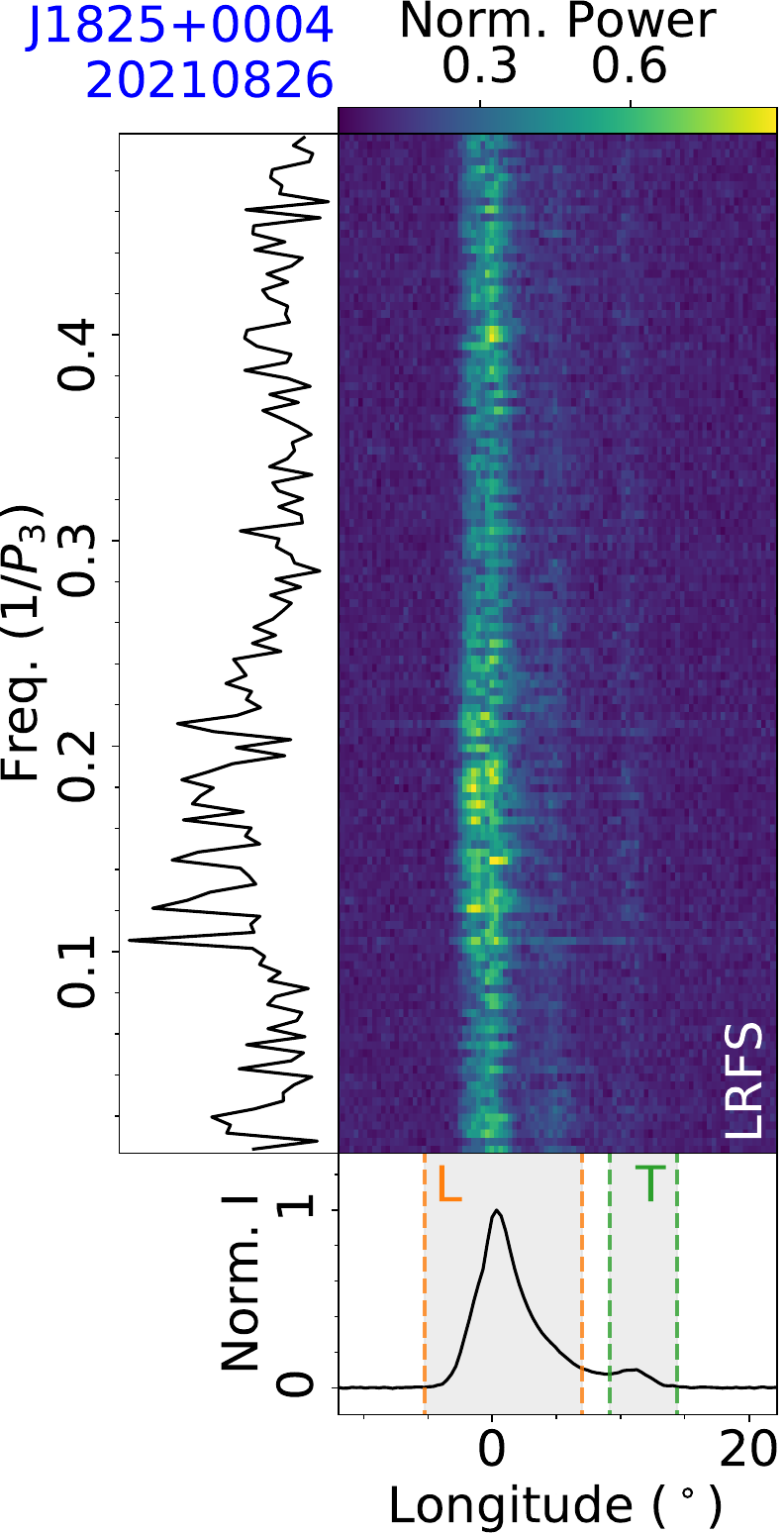}
\includegraphics[width=0.22\textwidth, angle=0]{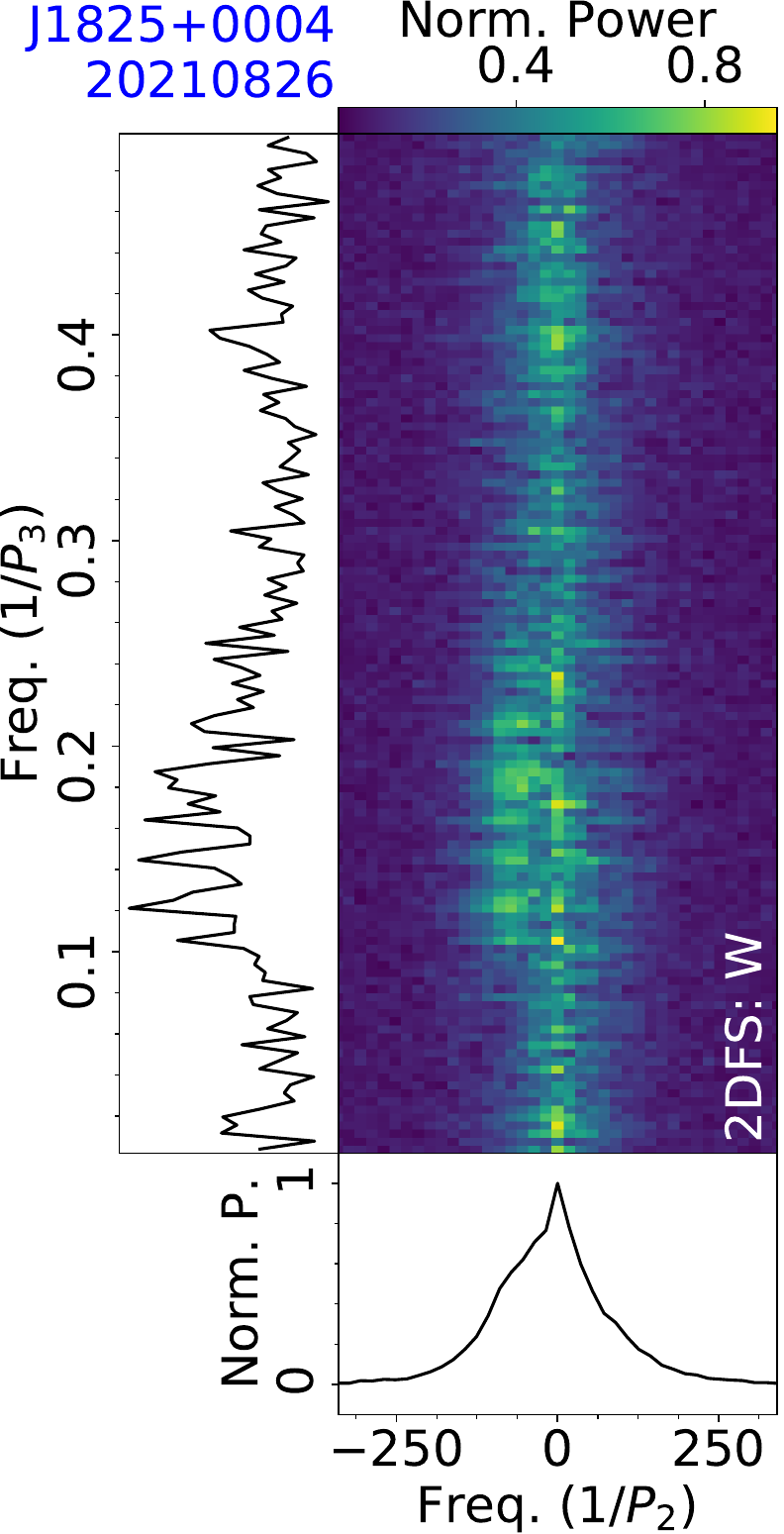}\\
\includegraphics[width=0.22\textwidth, angle=0]{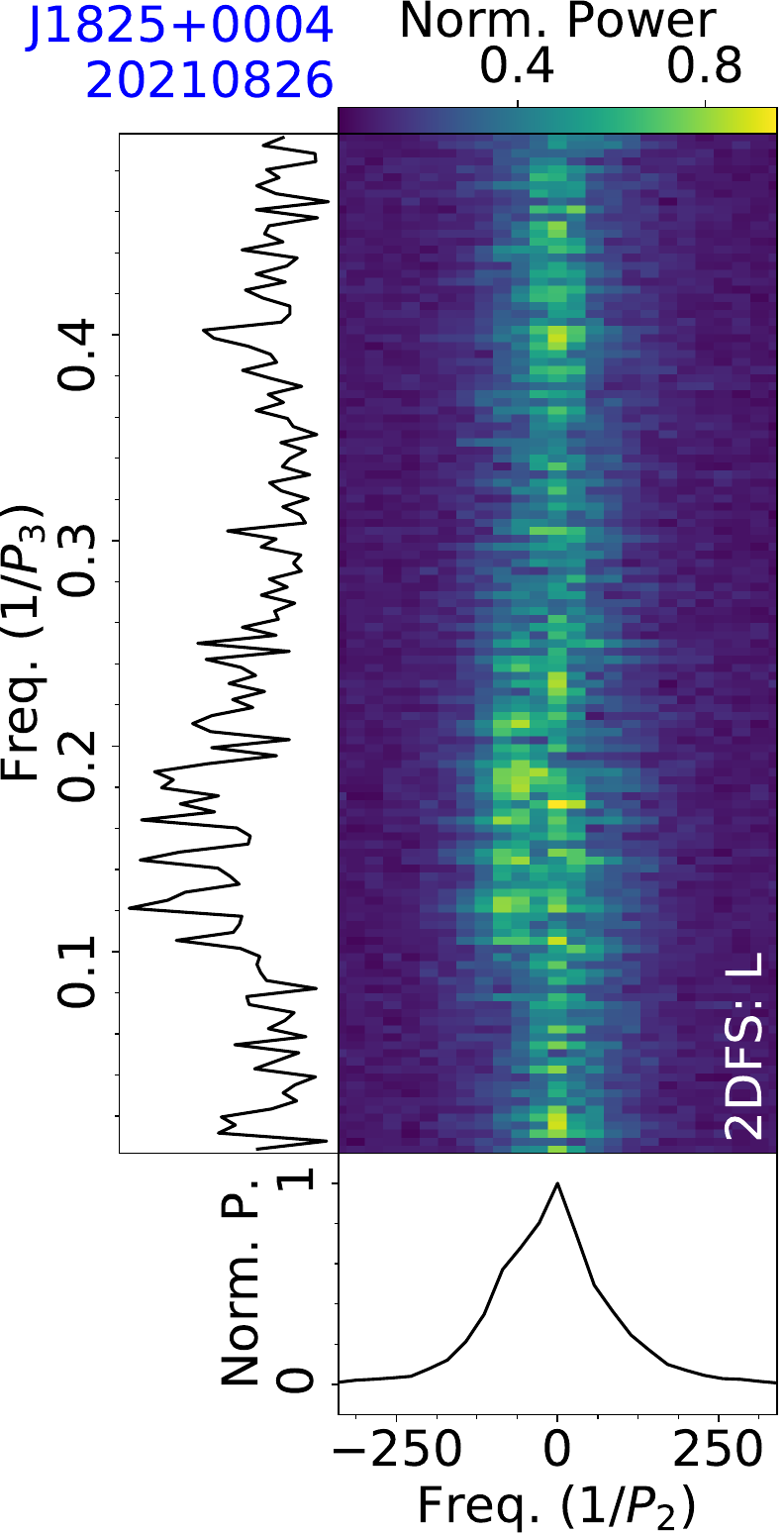}
\includegraphics[width=0.22\textwidth, angle=0]{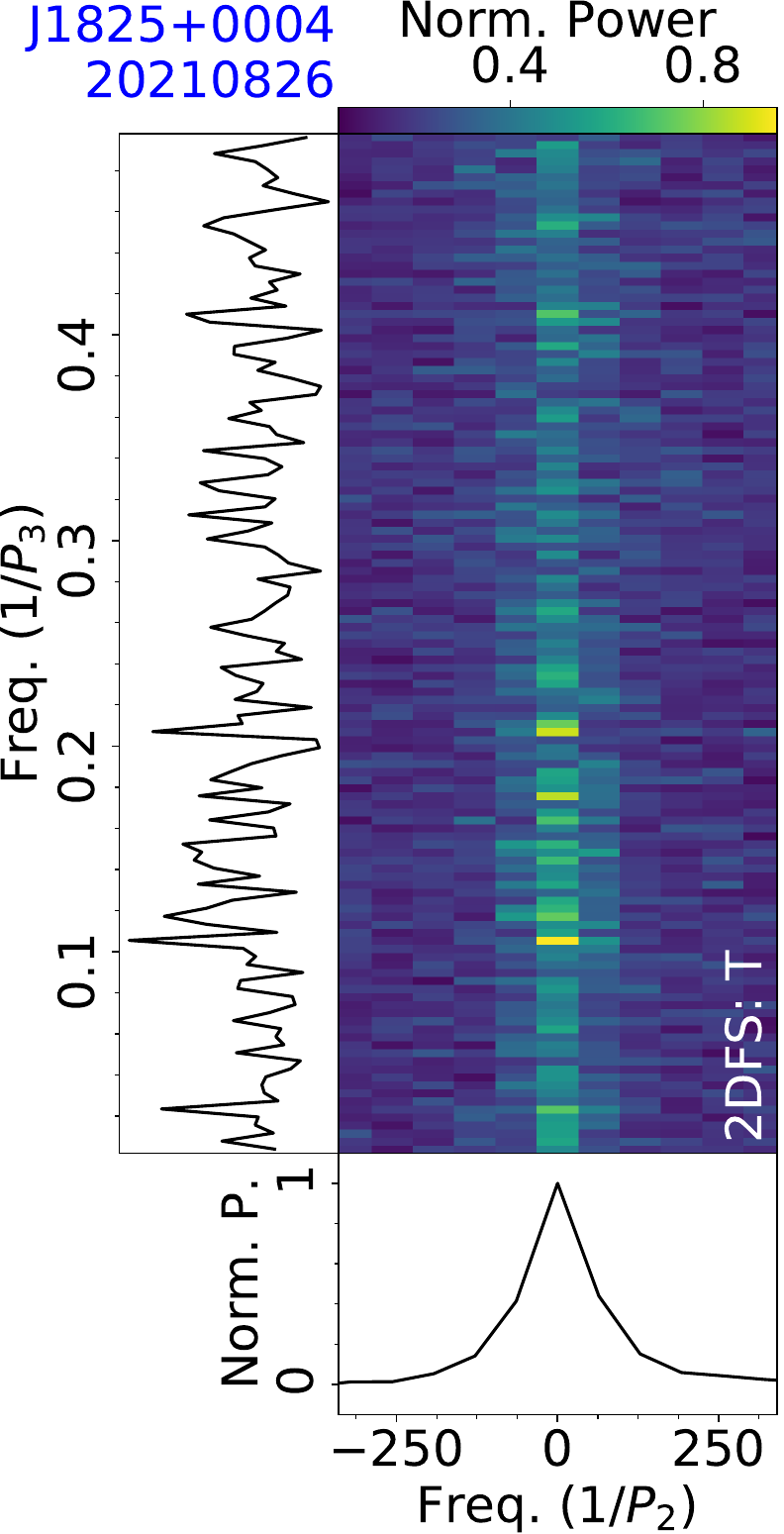}
\figcaption{Fluctuation analysis of PSR J1825+0004 for the observation on 20210826, with LRFS (top-left), and 2DFS for the on-pulse region (top-right), leading part (bottom-left) and trailing part (bottom-right) of a mean pulse profile. \label{subfig:fluctu:J1825+0004}}
\end{figure}

\begin{figure}[htpb]
\centering
\includegraphics[width=0.22\textwidth, angle=0]{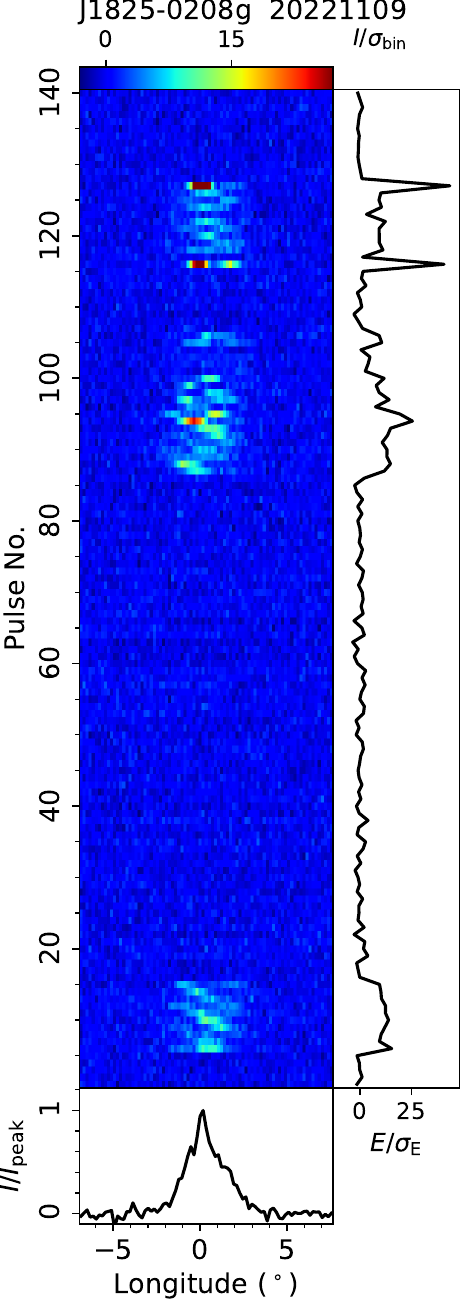}
\includegraphics[width=0.22\textwidth, angle=0]{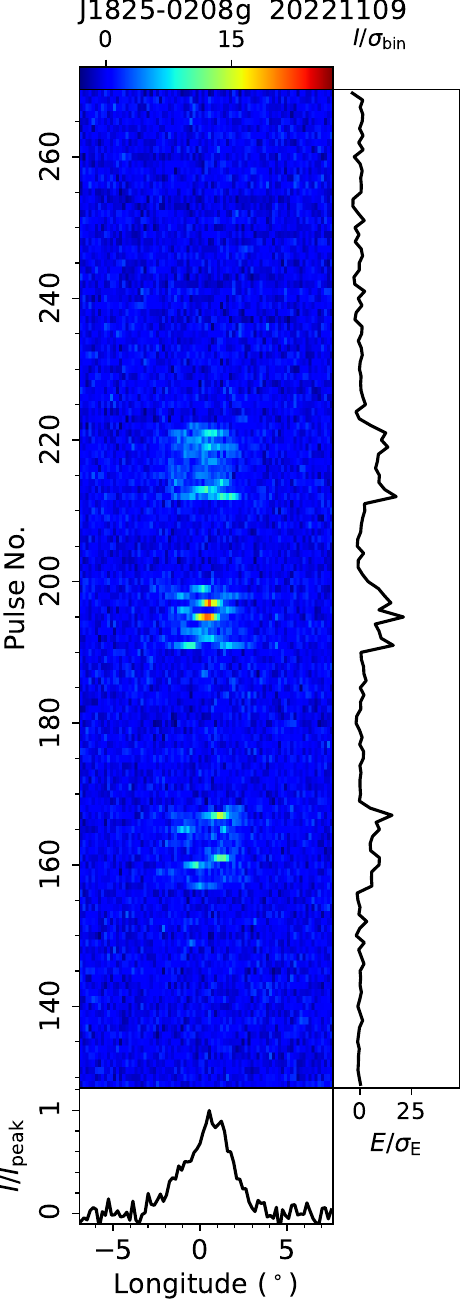}
\figcaption{Single pulse sequences of PSR J1825-0208g from the FAST observation on 20221109.
\label{subfig:TP:J1825-0208g}}
\end{figure}

\begin{figure}[htpb]
\centering
\includegraphics[width=0.39\textwidth, angle=0]{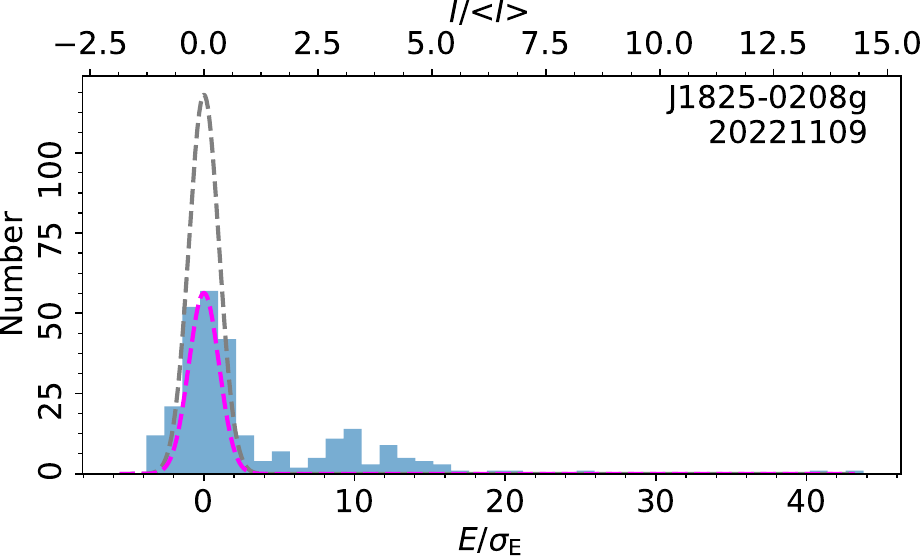}
\figcaption{On-pulse energy histogram of single pulses of PSR J1825-0208g from the FAST observation on 20221109.
\label{subfig:Hist:J1825-0208g}}
\end{figure}

\begin{figure}[htpb]
\centering
\includegraphics[width=0.39\textwidth, angle=0]{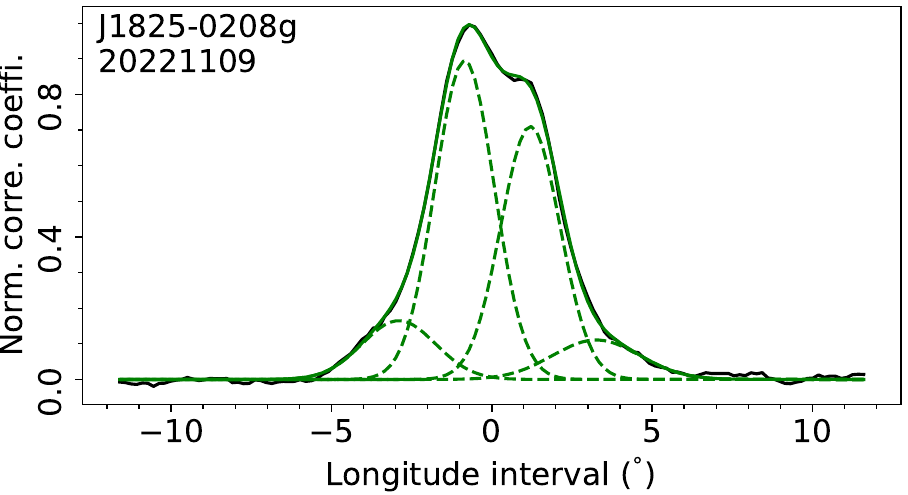}
\figcaption{Cross correlation of PSR J1825-0208g from the FAST observation on 20221109.
\label{subfig:Corre:J1825-0208g}}
\end{figure}

\subsection{J1824-0127}
\label{subsec:J1824-0127}

PSR J1824-0127 was discovered by \citep{Lorimer2006} in the Parkes Multibeam Pulsar Survey. \citet{Yan2024} showed this pulsar exhibiting dwarf pulses during the nulling state. \citet{Song2023} reported negative drifting parameters of two components, they are $P_3=3.5\pm0.1$ and $P_2=-49^{+40}_{-2}$ degrees for the leading component, and $P_3=3.3\pm0.1$ and $P_2=-44^{+25}_{-39}$ degrees for the trailing component. There is also a $P_3$-only feature of 26 periods. 

This pulsar was observed by FAST on 20211223 for 40 minutes and on 20220902 for 15 minutes. From the 40-minute data, a rotation period $P=2.4995$~s and a dispersion measure $D\!M=63.9~{\rm cm^{-3}\,pc}$ were determined. 
Here we show the result of the observation on 20211223. 
The nulling fraction is estimated to be 34$\pm$3\% from the on-pulse integral energy histogram in Fig.~\ref{subfig:Hist:J1824-0127}. 
Fluctuation spectra of this observation are shown in Fig.~\ref{subfig:fluctu:J1824-0127}, from which the negative drifting parameters are estimated. For leading part of a mean pulse profile, the drift feature in 2DFS has centroid frequencies of $1/P_3=0.309\pm0.003$ and $1/P_2=-23\pm6$, corresponding to $P_3=3.24\pm0.03$ periods and $P_2=-16\pm4^\circ$. For the trailing part, the frequencies are $1/P_3=0.302\pm0.002$ and $1/P_2=-27\pm7$, yielding $P_3=3.31\pm0.03$ periods and $P_2=-14\pm3^\circ$.
The $P_3$-only feature reported by \citet{Song2023} is probably caused by the nulling behavior.

\begin{figure}[htpb]
\centering
\includegraphics[width=0.22\textwidth, angle=0]{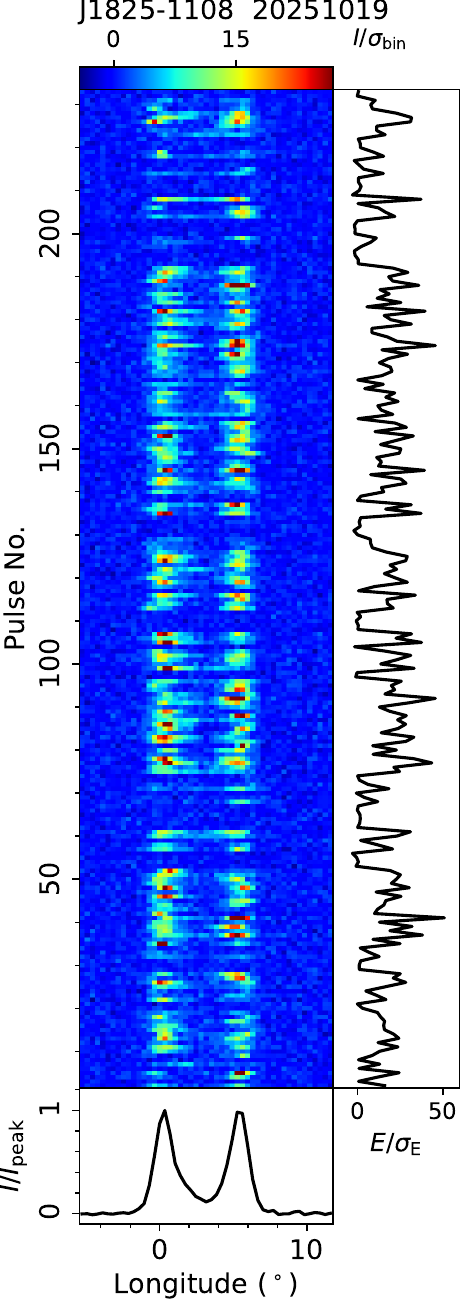}
\includegraphics[width=0.22\textwidth, angle=0]{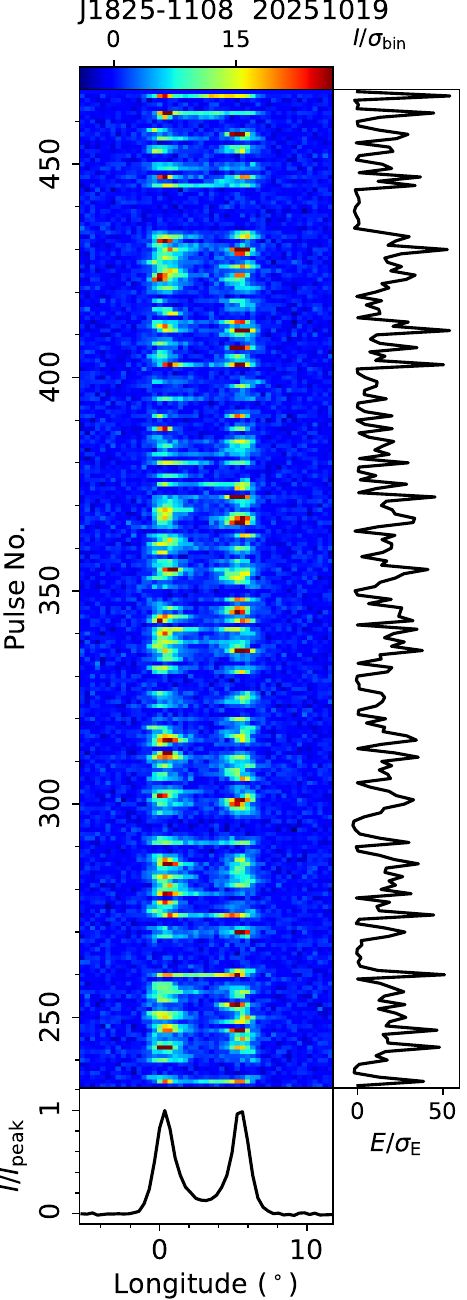}
\figcaption{Single pulse sequences of PSR J1825-1108 from the FAST observation on 20251019.
\label{subfig:TP:J1825-1108}}
\end{figure}

\begin{figure}[htpb]
\centering
\includegraphics[width=0.39\textwidth, angle=0]{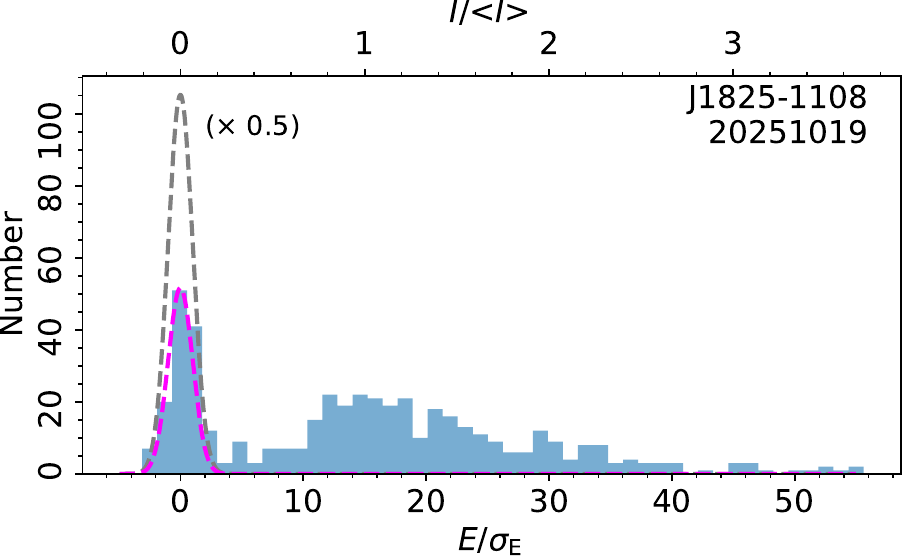}
\figcaption{On-pulse energy histogram of single pulses of PSR J1825-1108 from the FAST observation on 20251019.
\label{subfig:Hist:J1825-1108}}
\end{figure}

\subsection{J1825+0004}
\label{subsec:J1825+0004}

PSR J1825+0004 was discovered by \citet{Dewey1985} using the 92 m telescope at Green Bank at 390 MHz. $P_3$-only feature of 14.1$\pm$0.9 periods was reported by \citet{Song2023}. 

The pulsar was observed by FAST on 20210826 for 15 minutes, deriving a rotation period $P=0.7790$~s and a dispersion measure $D\!M=56.3~{\rm cm^{-3}\,pc}$ from this observation. 
Single pulse sequences are shown in Fig.~\ref{subfig:TP:J1825+0004}, displaying the subpulse drifting phenomenon. 
For the leading part in a mean pulse profile, 2DFS in Fig.~\ref{subfig:fluctu:J1825+0004} exhibit a negative drift feature with the centroid frequencies of $1/P_3=0.169\pm0.001$ and $1/P_2=-77\pm1$, which correspond to $P_3=5.90\pm0.04$ periods and $P_2=-4.6\pm0.1^\circ$. 
For the weak trailing part of the profile, there is also a temporal modulation feature in 2DFS, with the centroid frequency of $1/P_3=0.151\pm0.003$ ($P_3=6.6\pm0.1$ periods).
Additionally, on-pulse integral energies of two single pulses (Nos. 627 and 628) are less than 3$\sigma_{E}$.

\begin{figure}[htpb]
\centering
\includegraphics[width=0.22\textwidth, angle=0]{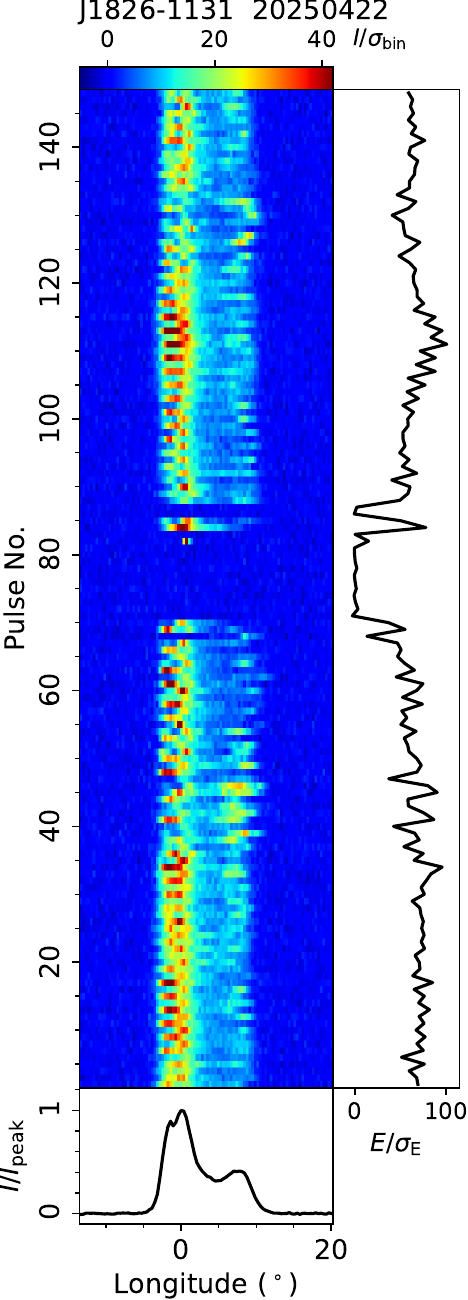}
\includegraphics[width=0.22\textwidth, angle=0]{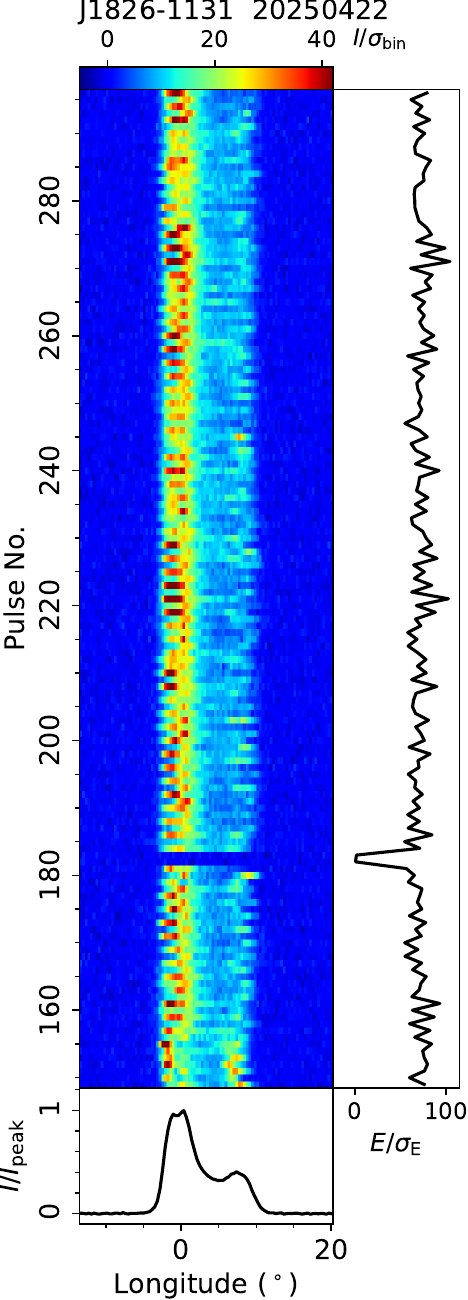}
\figcaption{Single pulse sequences of PSR J1826-1131 from the FAST observation on 20250422.
\label{subfig:TP:J1826-1131}}
\end{figure}

\begin{figure}[htpb]
\centering
\includegraphics[width=0.39\textwidth, angle=0]{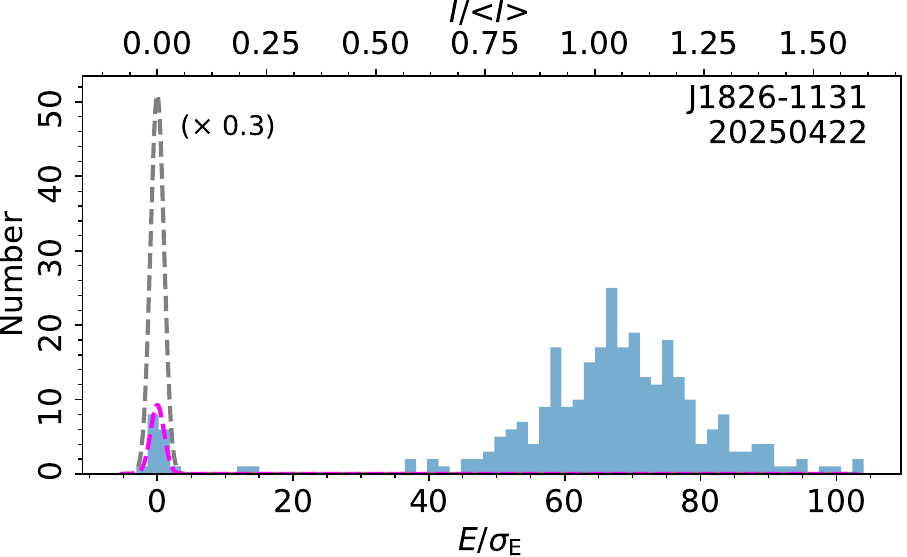}
\vspace{-0.3cm}
\figcaption{On-pulse energy histogram of single pulses of PSR J1826-1131 from the FAST observation on 20250422.
\label{subfig:Hist:J1826-1131}}
\end{figure}

\begin{figure}[htpb]
\centering
\includegraphics[width=0.22\textwidth, angle=0]{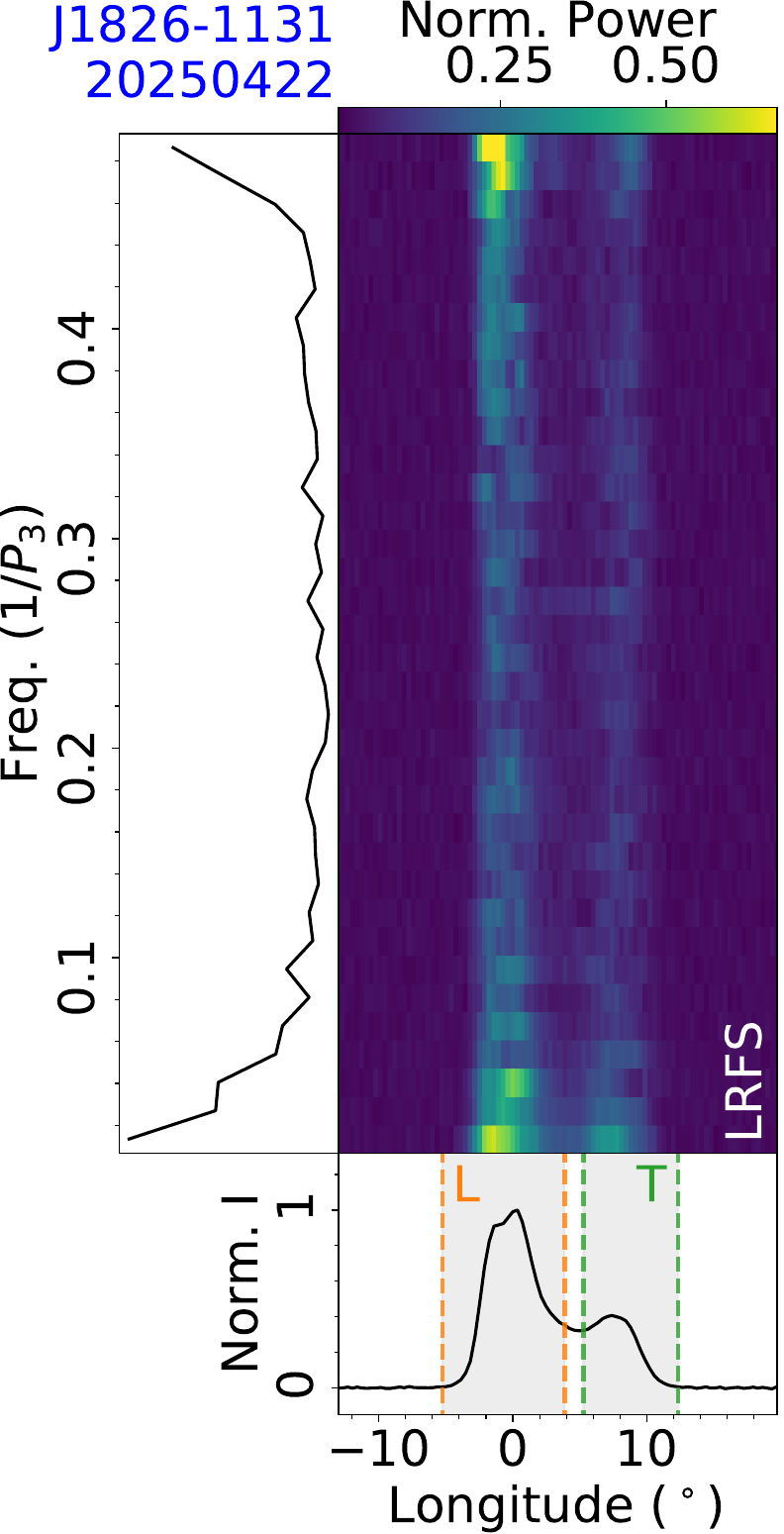}
\includegraphics[width=0.22\textwidth, angle=0]{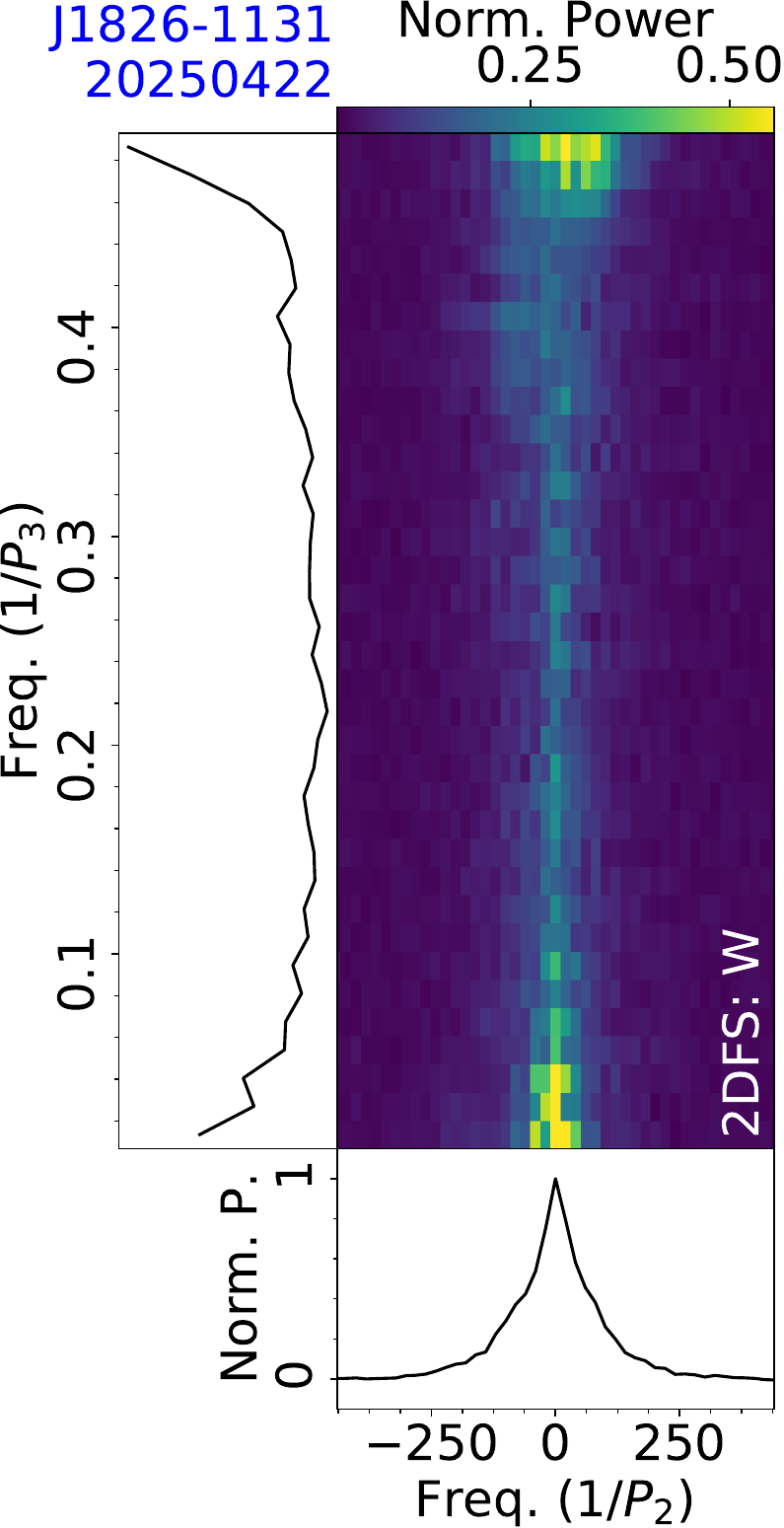}\\
\includegraphics[width=0.22\textwidth, angle=0]{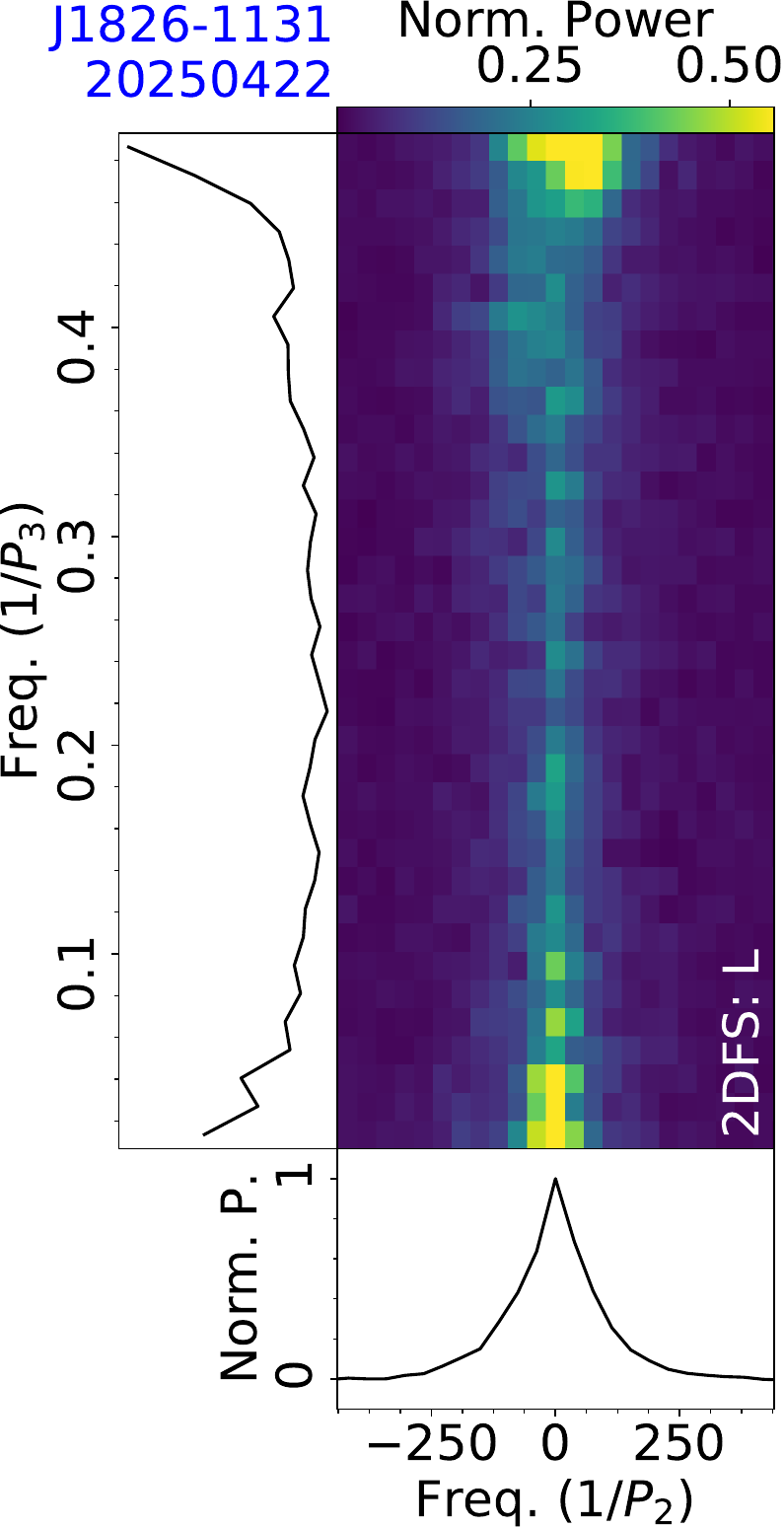}
\includegraphics[width=0.22\textwidth, angle=0]{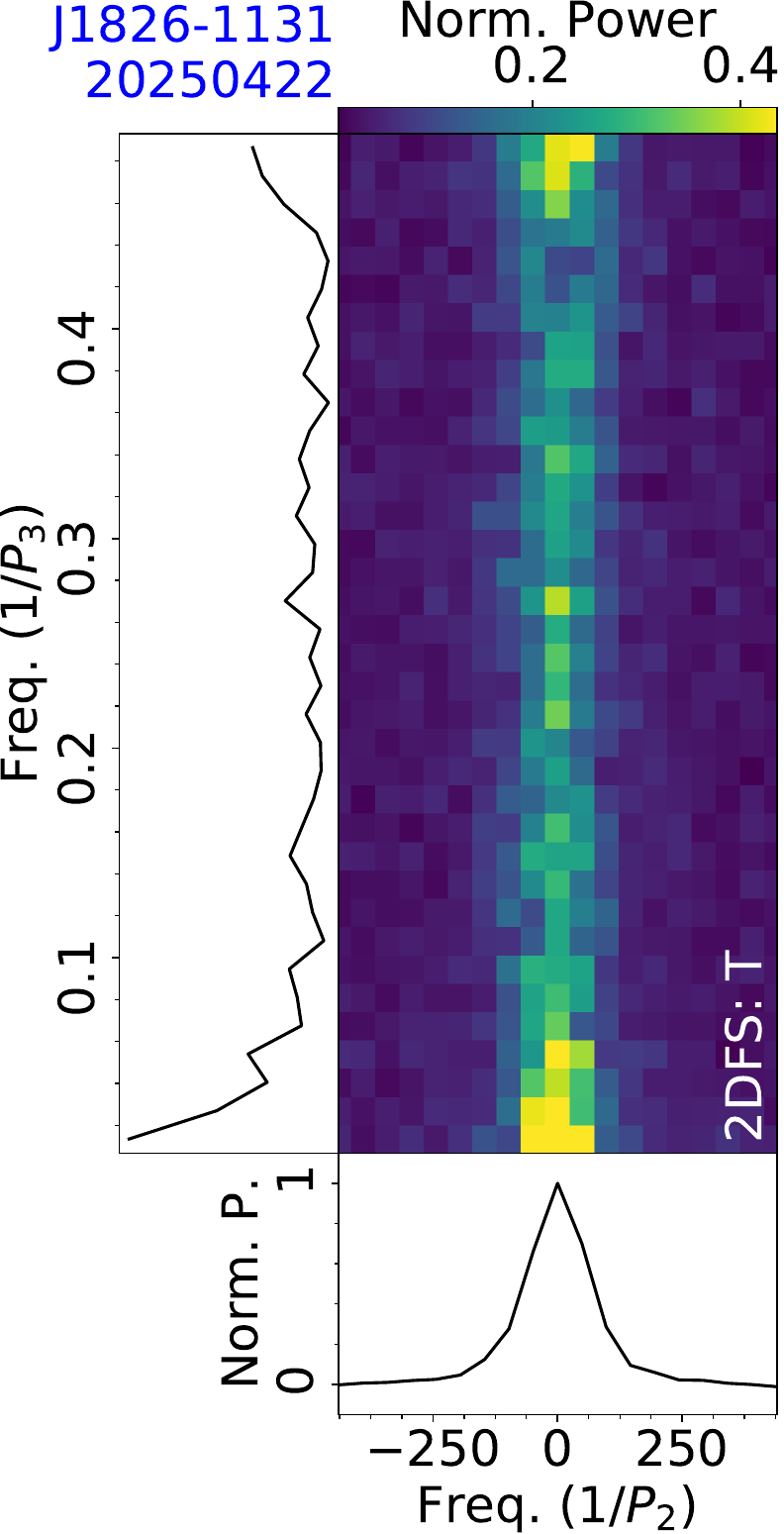}
\figcaption{Fluctuation analysis of PSR J1826-1131 for the observation on 20250422, with LRFS (top-left), and 2DFS for the on-pulse region (top-right), leading part (bottom-left) and trailing part (bottom-right) of a mean pulse profile. \label{subfig:fluctu:J1826-1131}}
\end{figure}

\begin{figure}[htpb]
\centering
\includegraphics[width=0.22\textwidth, angle=0]{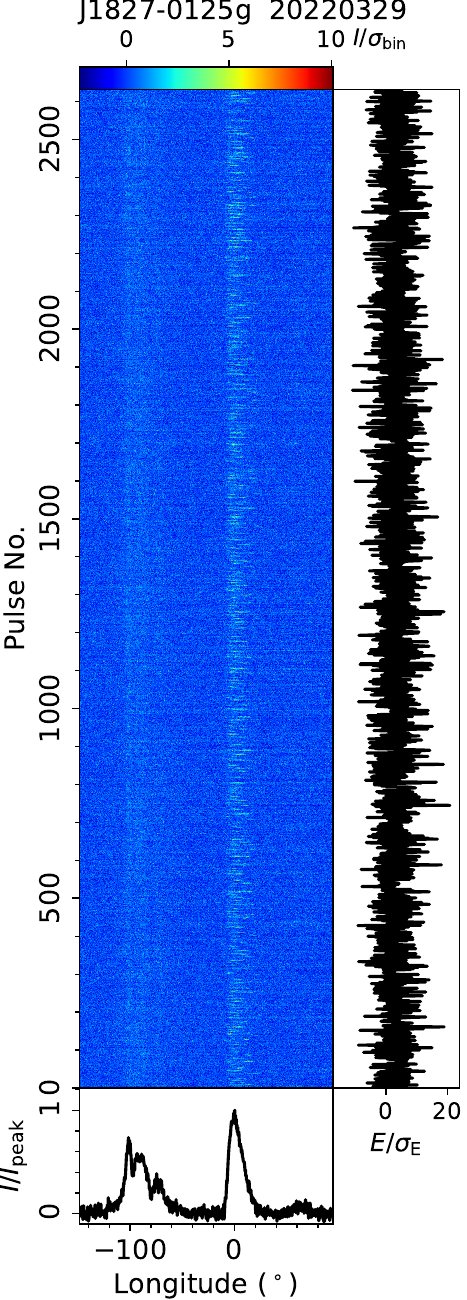}
\includegraphics[width=0.22\textwidth, angle=0]{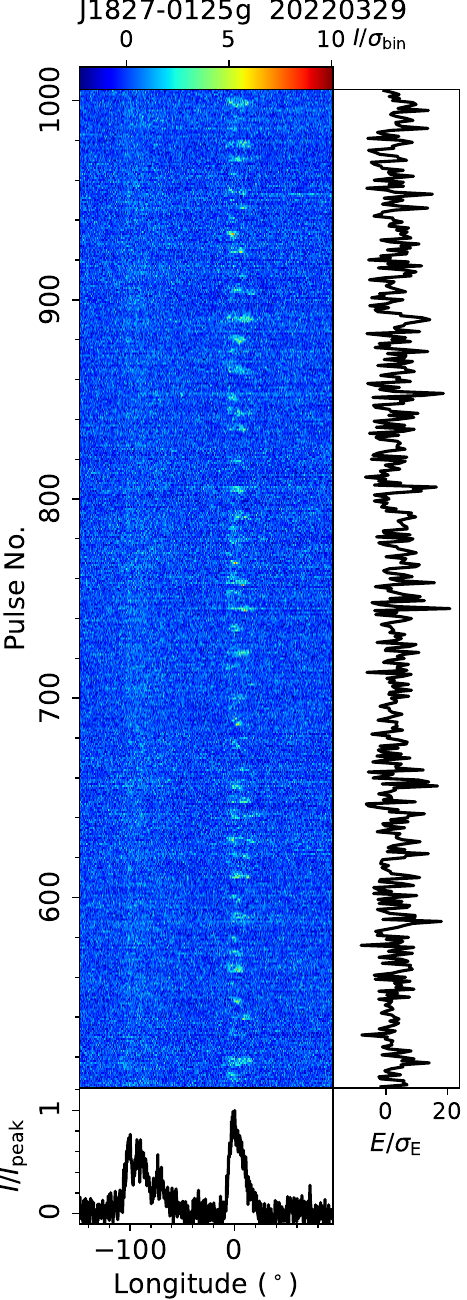}
\figcaption{Single pulse sequences of PSR J1827-0125g from the FAST observation on 20220329, and a zoomed-in view of pulses No. 505-1005.
\label{subfig:TP:J1827-0125g}}
\end{figure}

\begin{figure}[htpb]
\centering
\includegraphics[width=0.22\textwidth, angle=0]{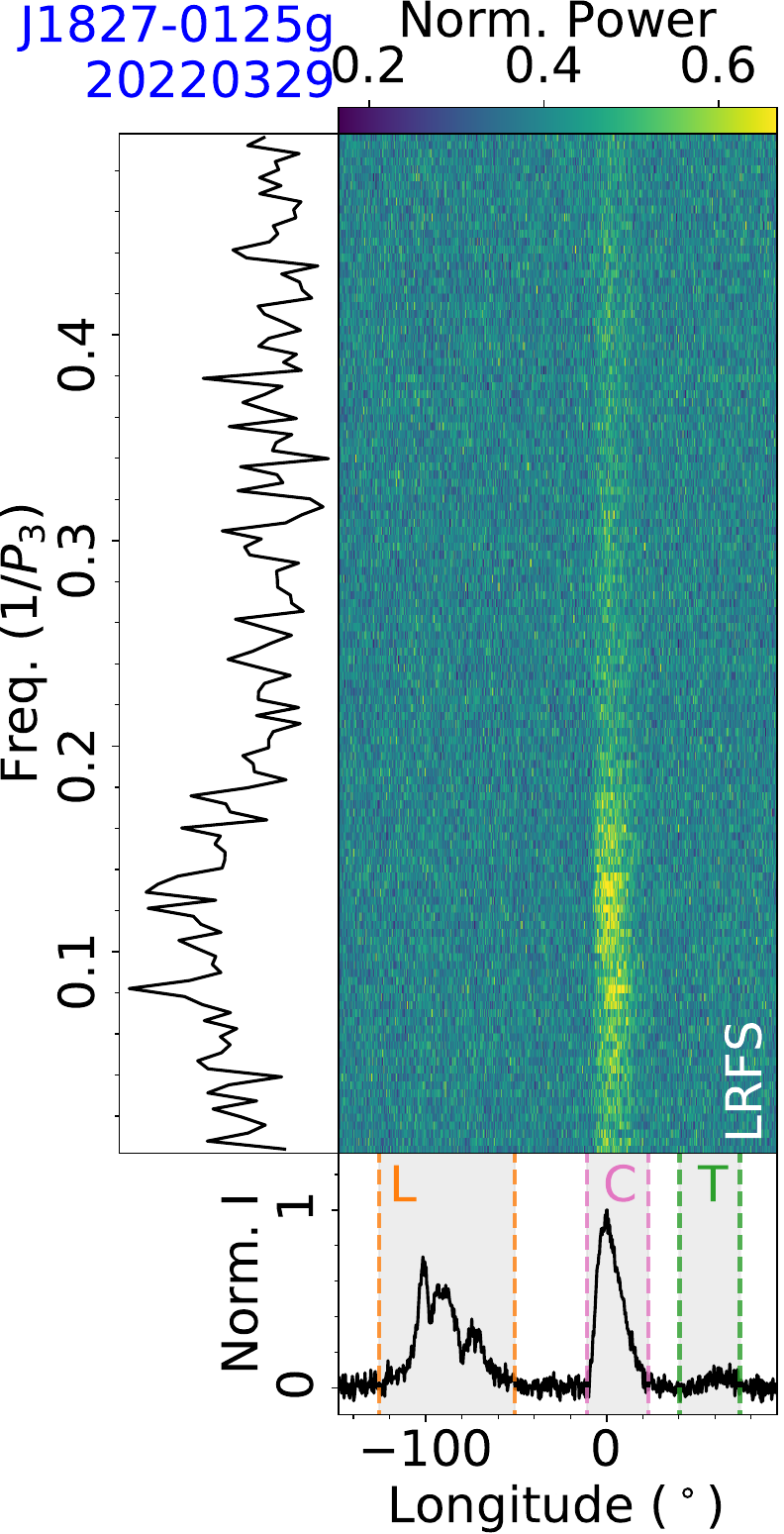}
\includegraphics[width=0.22\textwidth, angle=0]{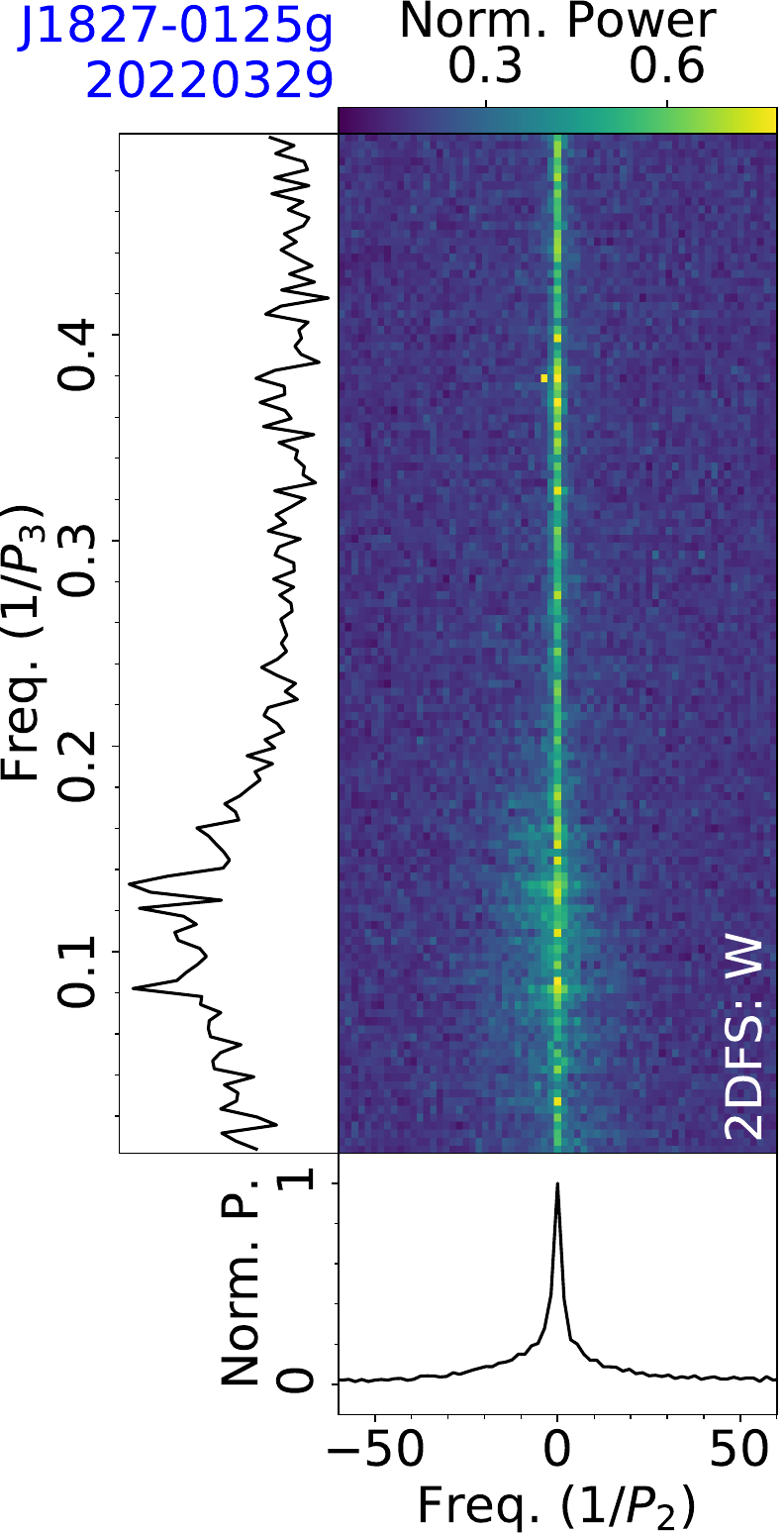}\\
\includegraphics[width=0.22\textwidth, angle=0]{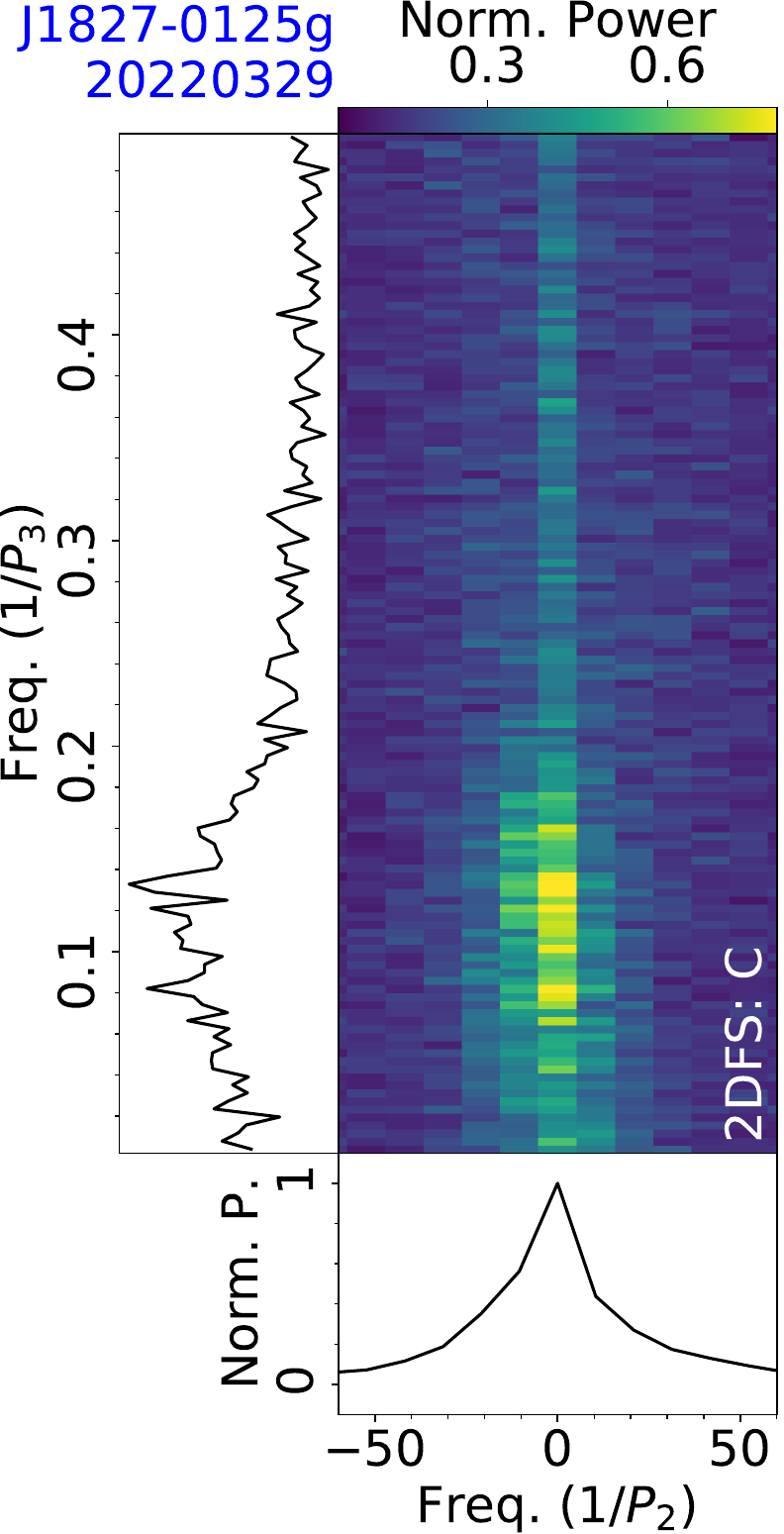}
\includegraphics[width=0.22\textwidth, angle=0]{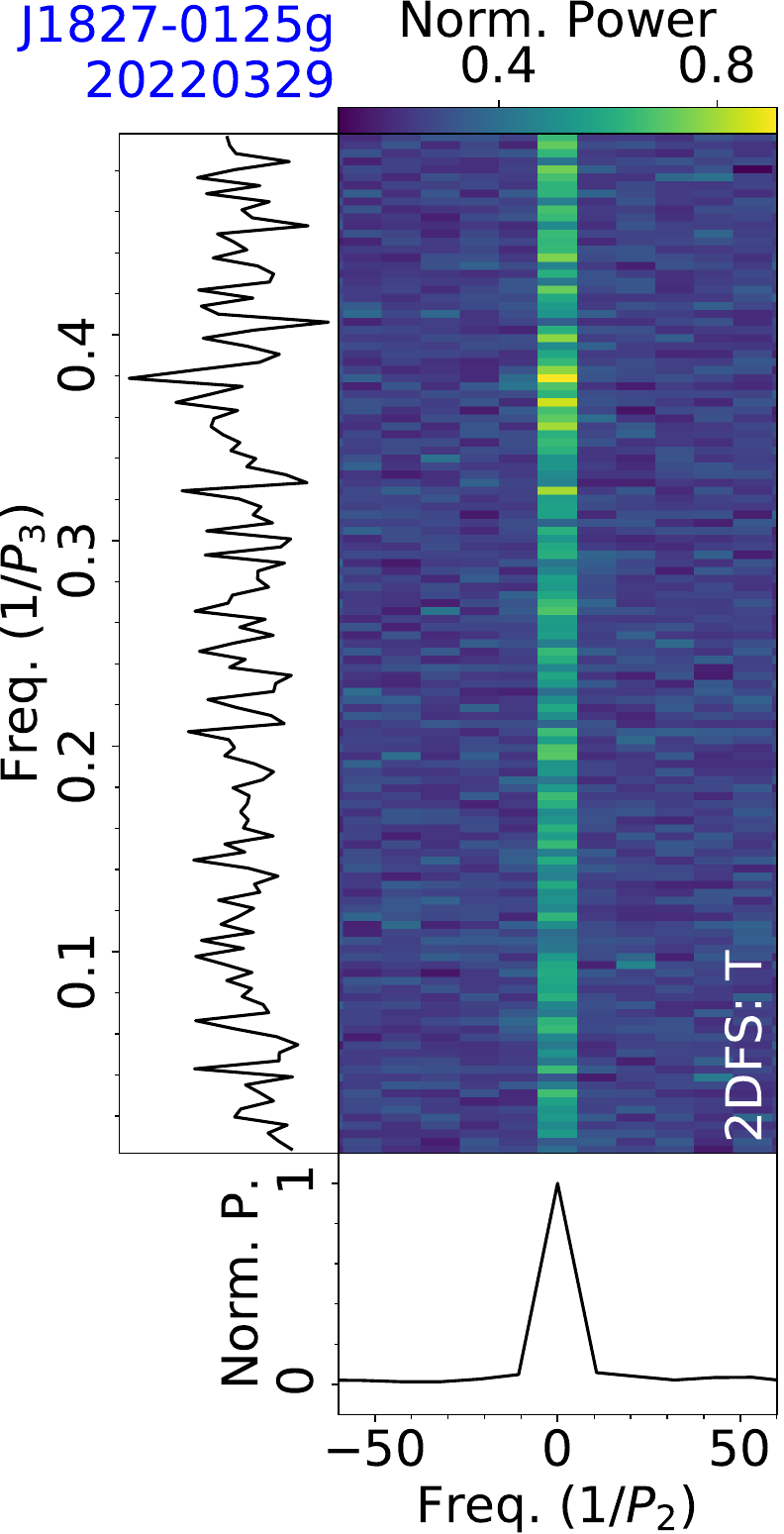}
\figcaption{Fluctuation analysis of PSR J1827-0125g for the observation on 20220329, with LRFS (top-left), and 2DFS for the on-pulse region (top-right), central part (bottom-left) and trailing part (bottom-right) of a mean pulse profile. \label{subfig:fluctu:J1827-0125g}}
\end{figure}

\subsection{J1825-0208g}
\label{subsec:J1825-0208g}

PSR J1825-0208g was discovered in the FAST GPPS survey \citep{Han2021,han2025}. 

This pulsar was observed by FAST on 20221109 for 16 minutes, yielding a rotation period $P=3.3076$~s and a dispersion measure $D\!M=192.6~{\rm cm^{-3}\,pc}$. 
Single pulse sequences in Fig.~\ref{subfig:TP:J1825-0208g} display nulling and subpulse drifting phenomena. The on-pulse integral energy histogram (Fig.~\ref{subfig:Hist:J1825-0208g}) indicates the nulling fraction to be 48$\pm$8\%. From the correlation between adjacent single pulses in Fig.~\ref{subfig:Corre:J1825-0208g}, the pulsar has negative subpulse drifting with a drift rate of $D=-0.84\pm0.04$ degrees per period.

\subsection{J1825-1108}
\label{subsec:J1825-1108}

PSR J1825-1108 was discovered by the Parkes 64-m radio telescope \citep{Ng2015}.

This pulsar was observed by FAST on 20251019 for 15 minutes, with a rotation period $P=1.9260$~s and a dispersion measure $D\!M=131.4~{\rm cm^{-3}\,pc}$ determined. Single pulse sequences in Fig.~\ref{subfig:TP:J1825-1108} and on-pulse integral energy histogram in Fig.~\ref{subfig:Hist:J1825-1108} illustrating the existence of the nulling phenomenon. The nulling fraction of this observation is estimated to be 24.5$\pm$1.5\%.

\begin{figure}[htpb]
\centering
\includegraphics[width=0.435\textwidth, angle=0]{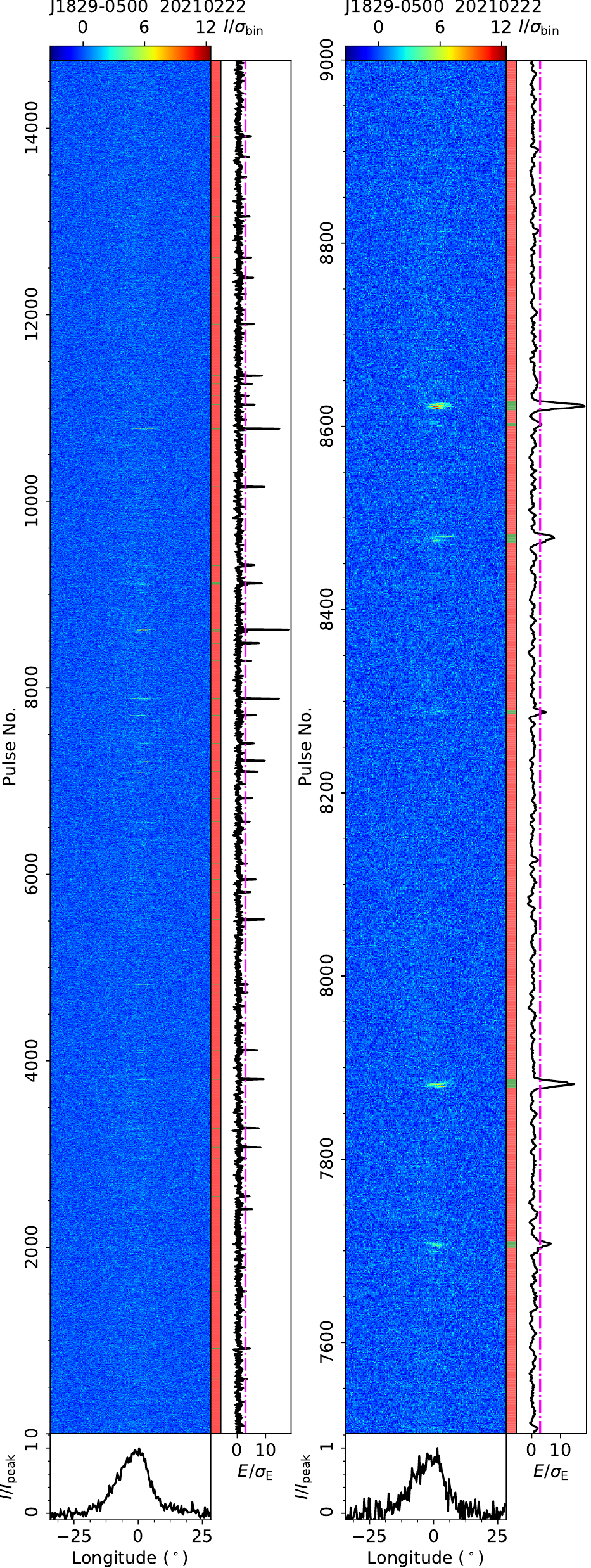}
\vspace{-0.3cm}
\figcaption{Single pulse sequence of PSR J1829-0500 from the FAST observation on 20210222, and a zoomed-in view of pulses No. 7500-9000. 
The right subplot shows the time series of single-pulse energy integrated over the longitude range from -1.58$^\circ$ to 8.26$^\circ$, smoothed with a 5-period moving average.
\label{subfig:TP:J1829-0500}}
\end{figure}

\begin{figure}[htpb]
\centering
\includegraphics[width=0.39\textwidth, angle=0]{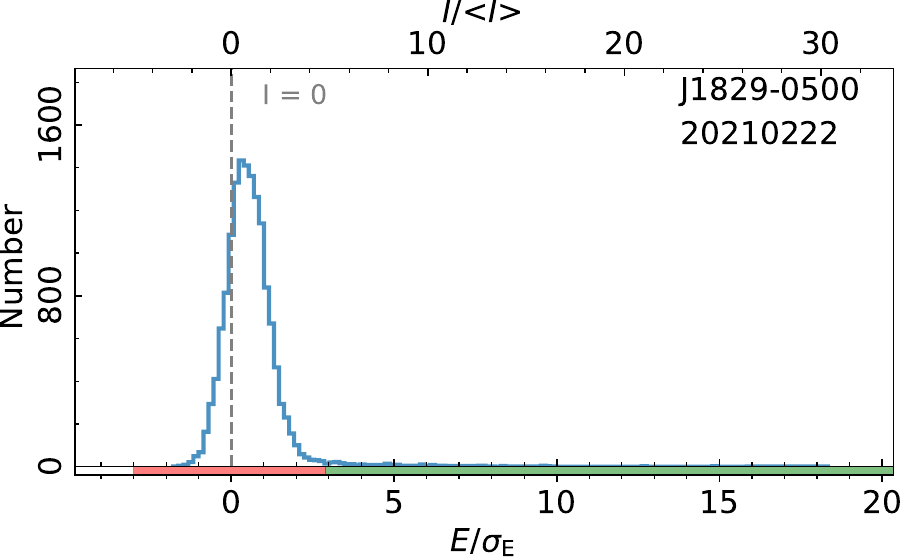}
\figcaption{Histogram of integrated single-pulse energies (longitude -1.58$^\circ$ to 8.26$^\circ$) for PSR J1829-0500 from the FAST observation on 20210222, with energy values smoothed using a 5-period moving average. 
\label{subfig:Hist:J1829-0500}}
\end{figure}

\begin{figure}[htpb]
\centering
\includegraphics[width=0.39\textwidth, angle=0]{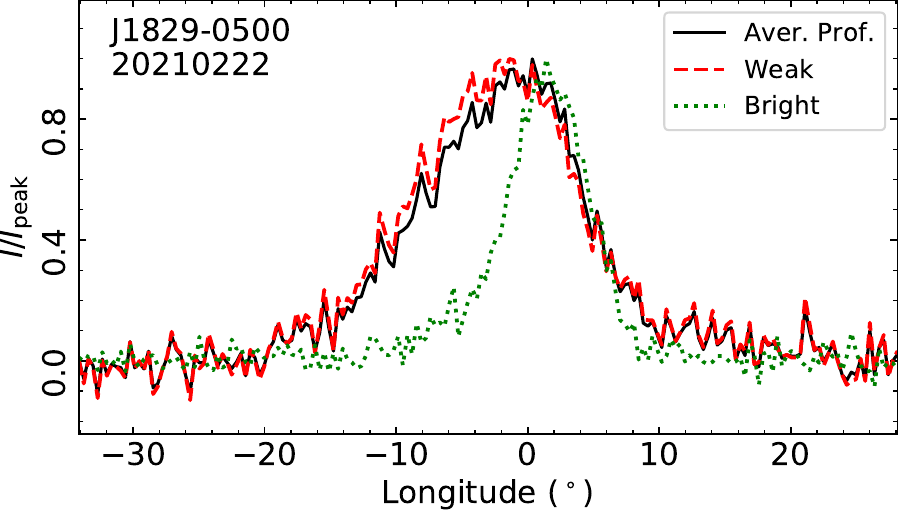}
\figcaption{Mean profiles of weak and bright emission modes of PSR J1829-0500 from the FAST observation on 20210222, normalized by their respective peaks.
\label{subfig:PolModes:J1829-0500}}
\end{figure}

\subsection{J1826-1131}
\label{subsec:J1826-1131}

PSR J1826-1131 was discovered in a survey carried out at Jodrell Bank \citep{Clifton1986}. \citet{Song2023} reported modulation parameters for both components. For the leading component, two drift features were identified: $P_3=2.09\pm0.01$ periods and $P_2=16^{+2}_{-8}$ degrees; and $P_3=2.92\pm0.05$ periods and $P_2=-4.1^{+0.5}_{-4}$ degrees. For the trailing component, a $P_3$-only feature was found with  $P_3=2.07\pm0.02$ periods.

This pulsar was observed by FAST on 20250422 for 10 minutes, with a rotation period $P=2.0929$~s and a dispersion measure $D\!M=319.4~{\rm cm^{-3}\,pc}$ determined.  
The single-pulse sequences in Fig.~\ref{subfig:TP:J1826-1131} reveal that this pulsar exhibits nulling and subpulse drifting, along with a single pulse (No.~82) with a narrow emission longitude range during the nulling state. The nulling fraction of this observation is estimated to be 5.5$\pm$0.7\% from the on-pulse energy histogram of single pulses (Fig.~\ref{subfig:Hist:J1826-1131}). The leading and trailing parts in a mean pulse profile have different modulation properties, and the fluctuation spectra are displayed in Fig.~\ref{subfig:fluctu:J1826-1131}. For the leading profile part, the centroid frequencies of the main drift feature in 2DFS are $1/P_3=0.478\pm0.001$ ($P_3=2.09\pm0.01$ periods) and $1/P_2=31\pm3$ ($P_2=11\pm1$ degree). 2DFS of the trailing profile part exhibits a modulation feature with the centroid frequency of $1/P_3=0.477\pm0.002$, corresponding to $P_3=2.10\pm0.01$ periods. For the negative drifting behavior of the leading component reported by \citet{Song2023}, as well as for the cause of the low-frequency modulation feature in the fluctuation spectra (Fig.~\ref{subfig:fluctu:J1826-1131}), more observations are necessary.

\begin{figure}[htpb]
\centering
\includegraphics[width=0.44\textwidth, angle=0]{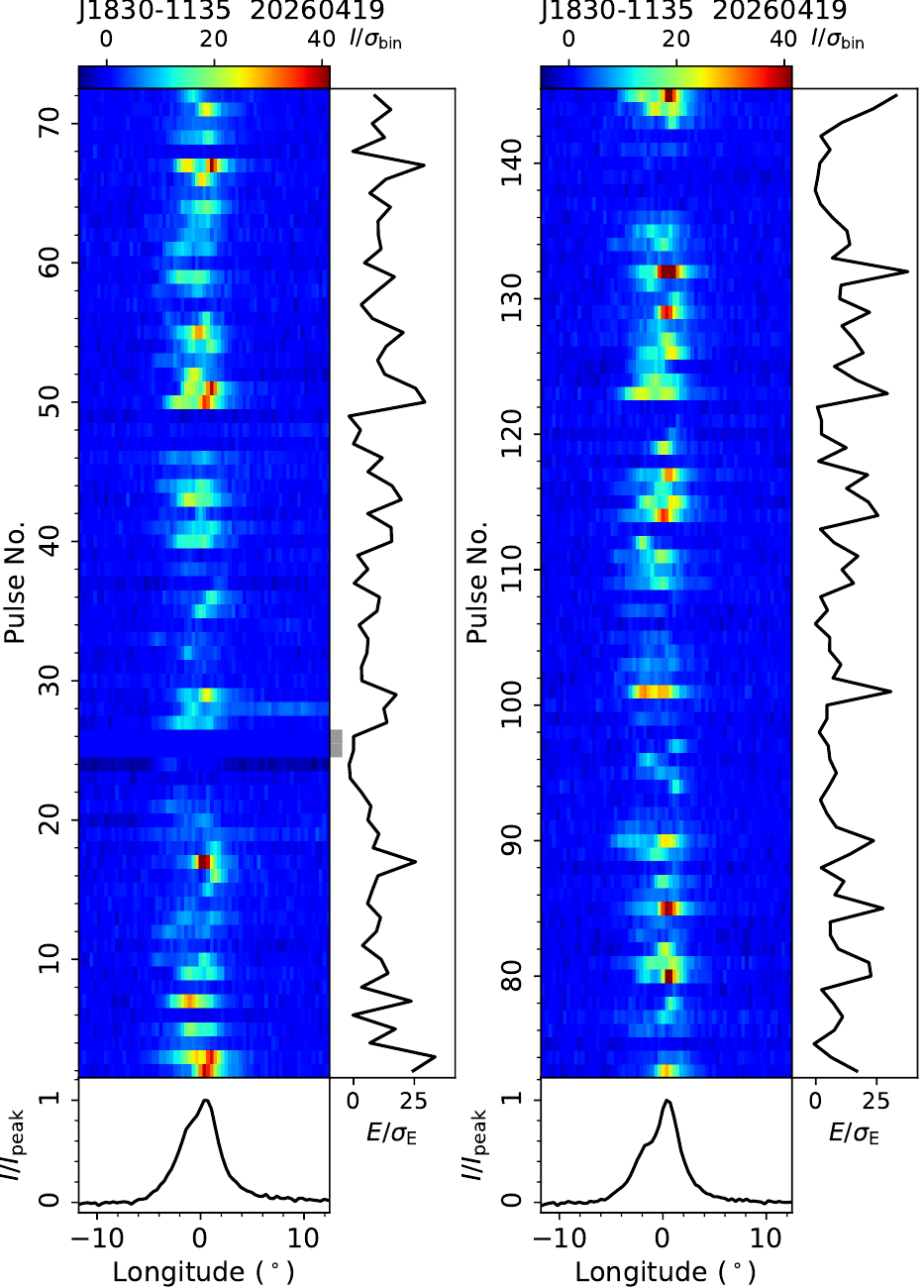}
\figcaption{Single pulse sequences of PSR J1830-1135 from the FAST observation on 20260419. \label{subfig:TP:J1830-1135}}
\end{figure}

\begin{figure}[htpb]
\centering
\includegraphics[width=0.44\textwidth, angle=0]{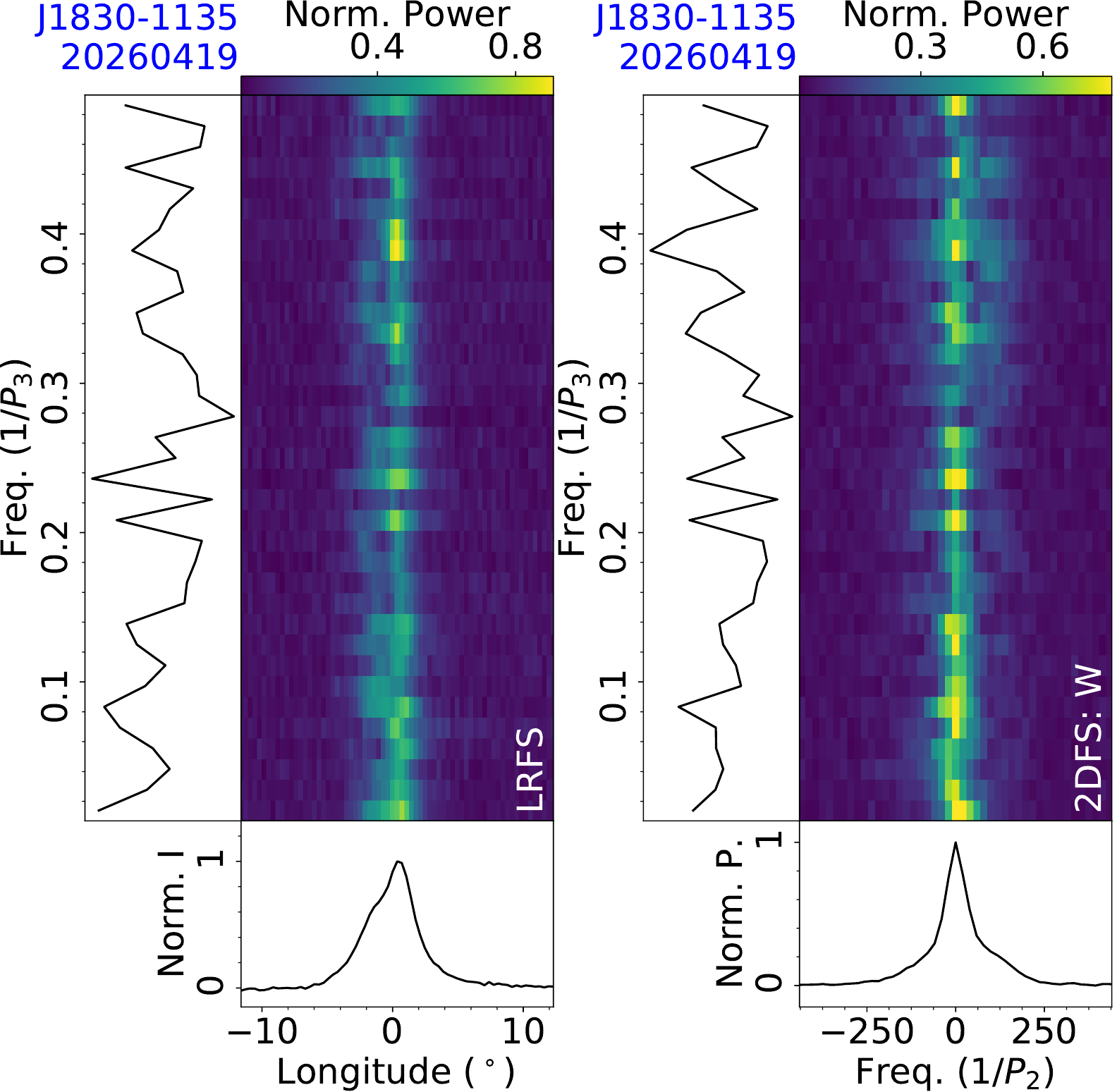}
\figcaption{Fluctuation analysis of PSR J1830-1135 for the observation on 20260419, with LRFS and 2DFS for the on-pulse region of the mean pulse profile. 
\label{subfig:fluctu:J1830-1135}}
\end{figure}

\subsection{J1827-0125g}
\label{subsec:J1827-0125g}

PSR J1827-0125g was discovered in the FAST GPPS survey \citep{Han2021,han2025}. 

This pulsar was observed by FAST on 20220329 for 15 minutes, deriving a rotation period $P=0.3379$~s and a dispersion measure $D\!M=97.7~{\rm cm^{-3}\,pc}$. 
Single pulse sequences are displayed in Fig.~\ref{subfig:TP:J1827-0125g}, where the central part in the profile has subpulse drifting behavior. Fluctuation spectra in Fig.~\ref{subfig:fluctu:J1827-0125g} illustrate that the central and trailing profile parts have different modulation properties. 
For the central part in a mean pulse profile, 2DFS exhibits a preferred negative drift feature with the centroid frequencies of $1/P_3=0.105\pm0.001$ and $1/P_2=-2.1\pm0.3$, corresponding to $P_3=9.5\pm0.1$ periods and $P_2=-175\pm27^\circ$. The centroid temporal modulation frequency in 2DFS of the trailing part is $1/P_3=0.375\pm0.003$, yielding $P_3=2.66\pm0.02$ periods.

\begin{figure}[htpb]
\centering
\includegraphics[width=0.22\textwidth, angle=0]{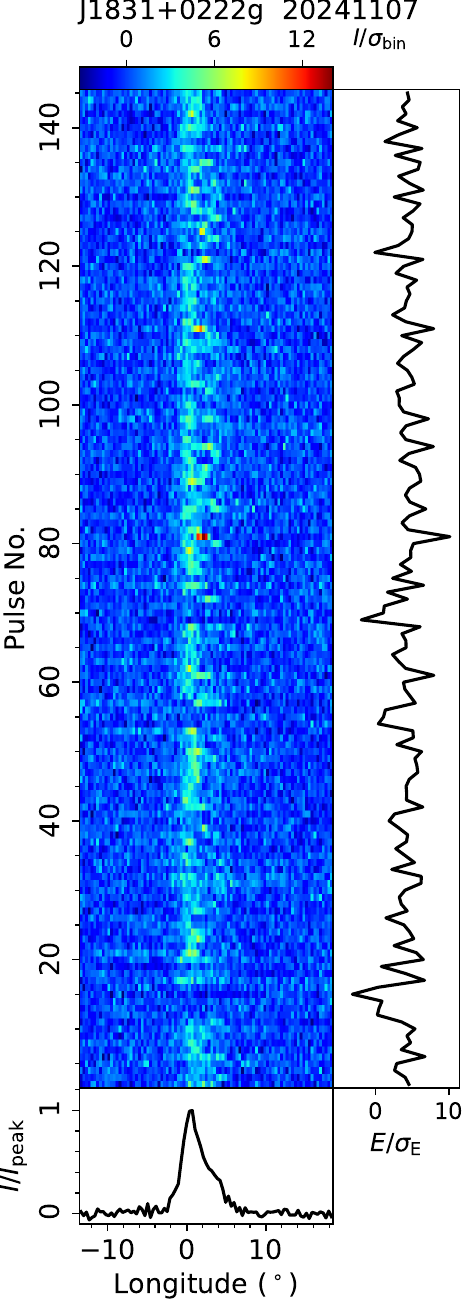}
\includegraphics[width=0.22\textwidth, angle=0]{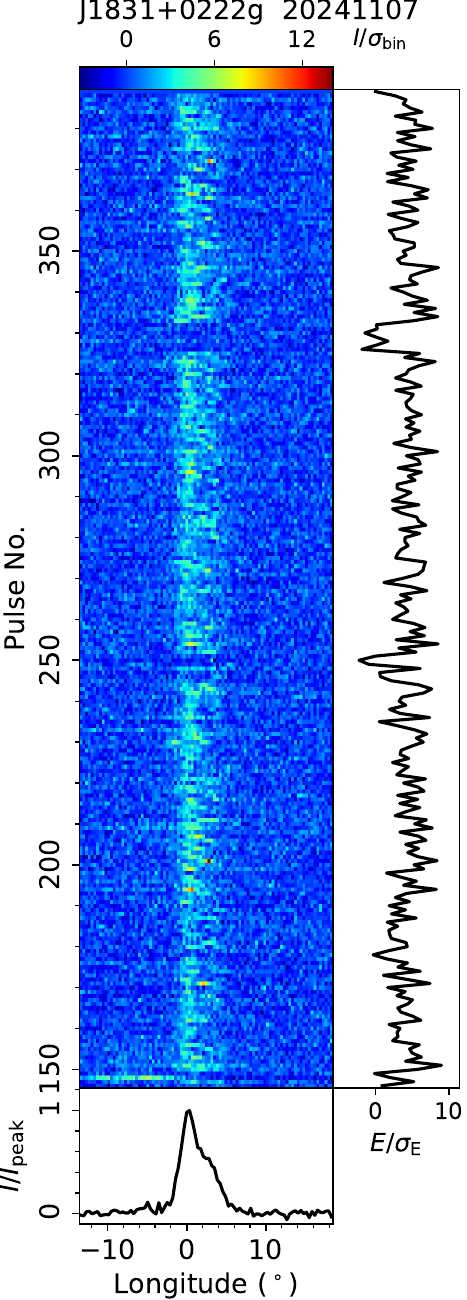}
\figcaption{Single pulse sequences of PSR J1831+0222g from the FAST observation on 20241107. \label{subfig:TP:J1831+0222g}}
\end{figure}

\begin{figure}[htpb]
\centering
\includegraphics[width=0.39\textwidth, angle=0]{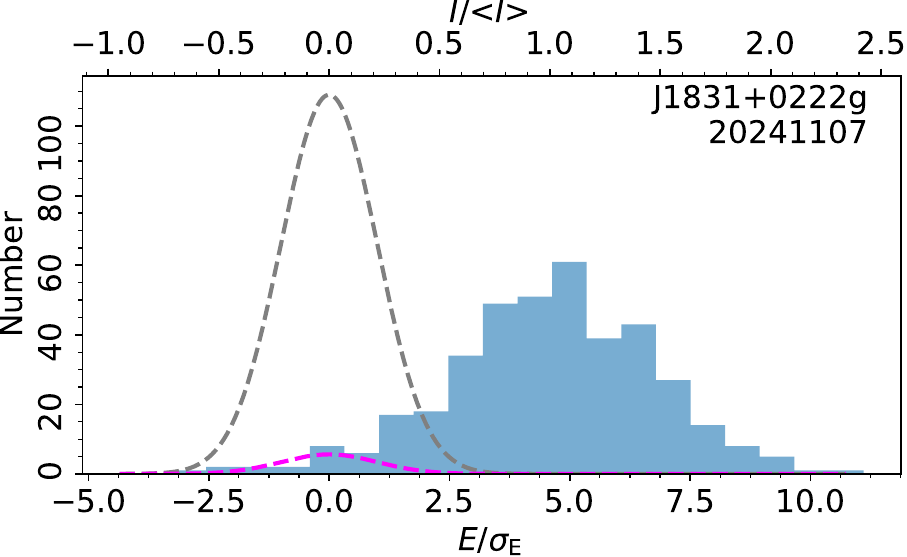}
\figcaption{On-pulse energy histogram of single pulses of PSR J1831+0222g from the FAST observation on 20241107.
\label{subfig:Hist:J1831+0222g}}
\end{figure}

\begin{figure}[htpb]
\centering
\includegraphics[width=0.22\textwidth, angle=0]{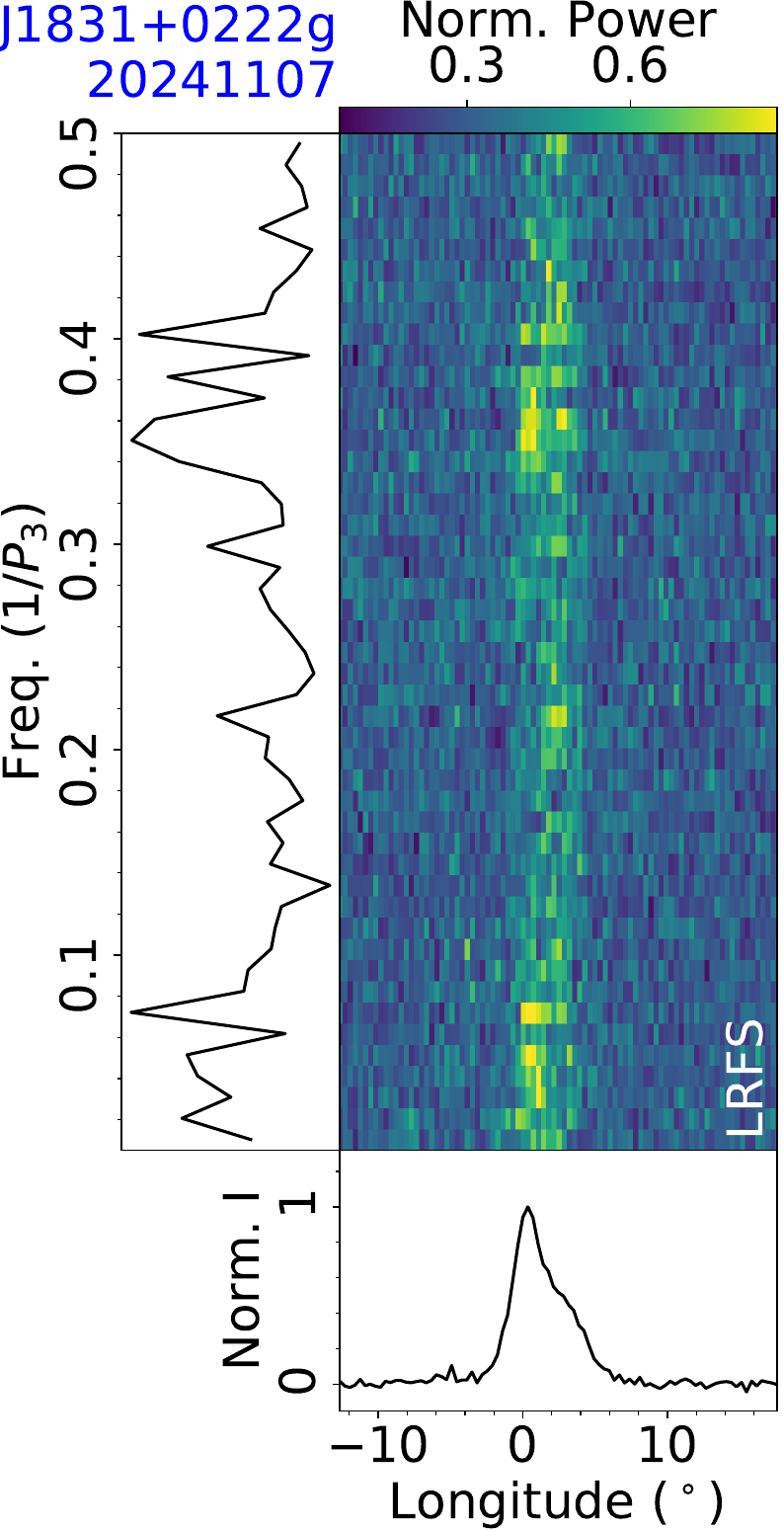}
\includegraphics[width=0.22\textwidth, angle=0]{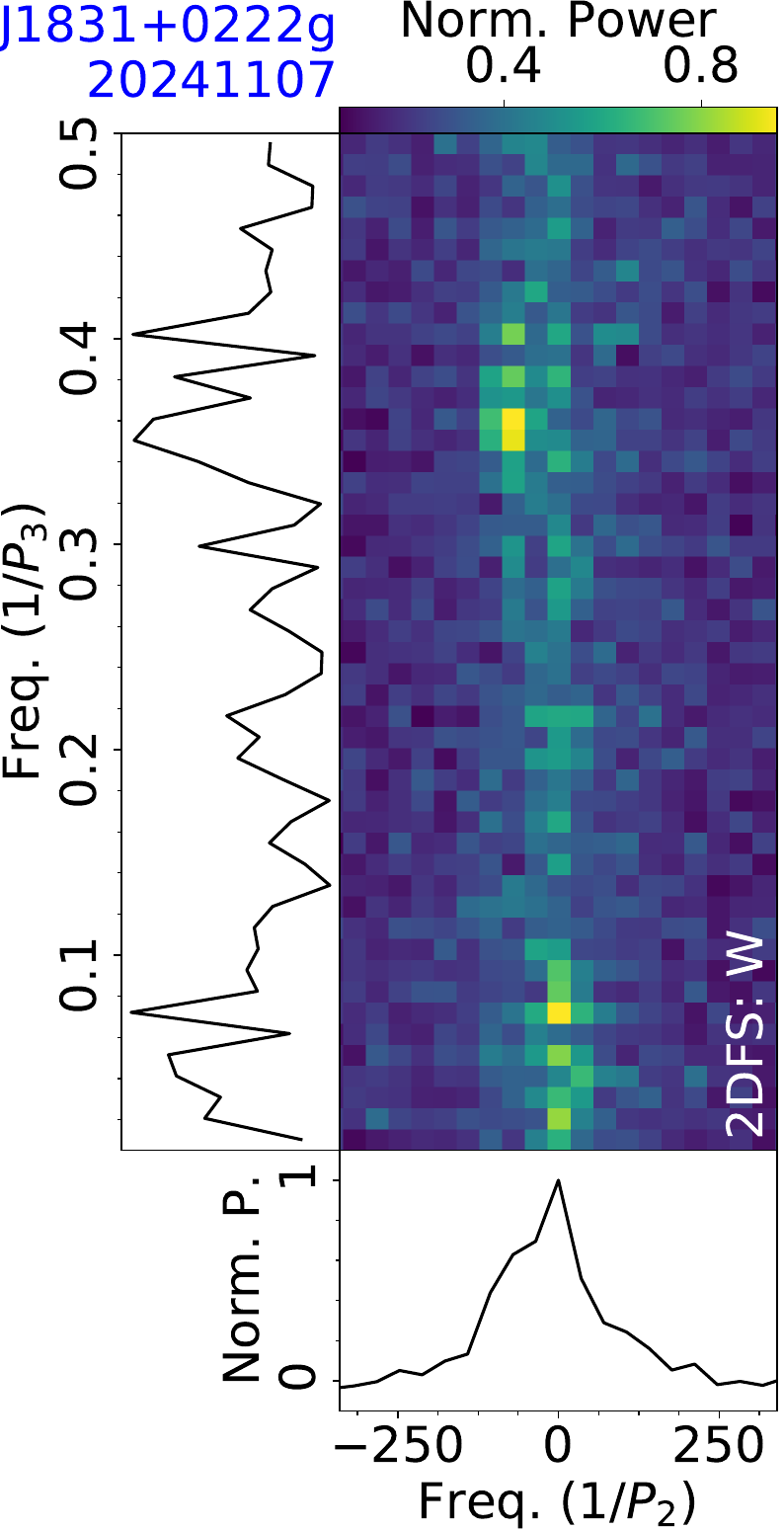}
\figcaption{Fluctuation analysis of PSR J1831+0222g from the FAST observation on 20241107, with LRFS and 2DFS for the central phase region of a mean pulse profile.  \label{subfig:fluctu:J1831+0222g}}
\end{figure}

\begin{figure}[htpb]
\centering
\includegraphics[width=0.22\textwidth, angle=0]{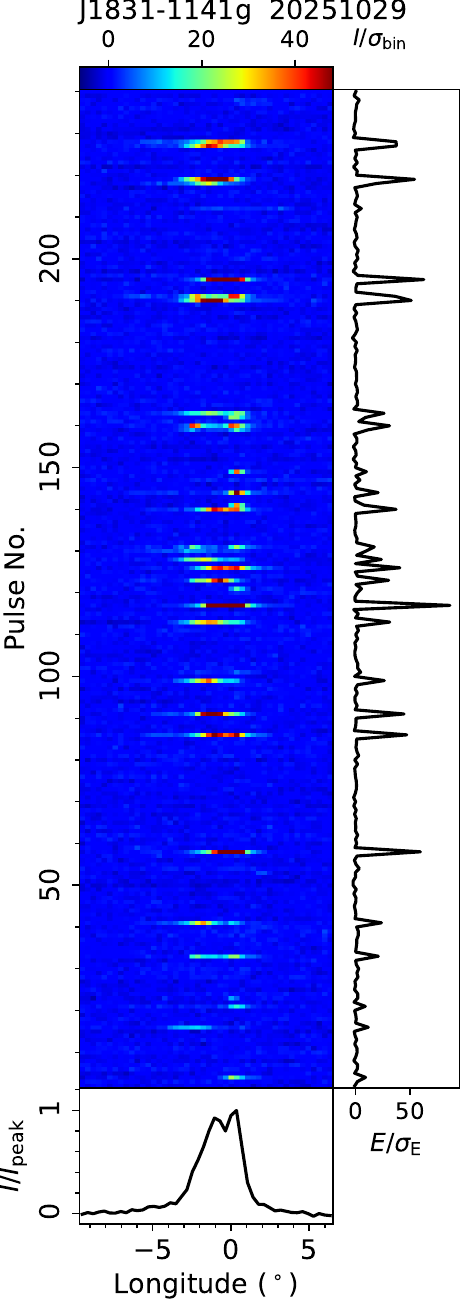}
\figcaption{Single pulse sequence of PSR J1831-1141g from the FAST observation on 20251029.
\label{subfig:TP:J1831-1141g}}
\end{figure}

\begin{figure}[htpb]
\centering
\includegraphics[width=0.39\textwidth, angle=0]{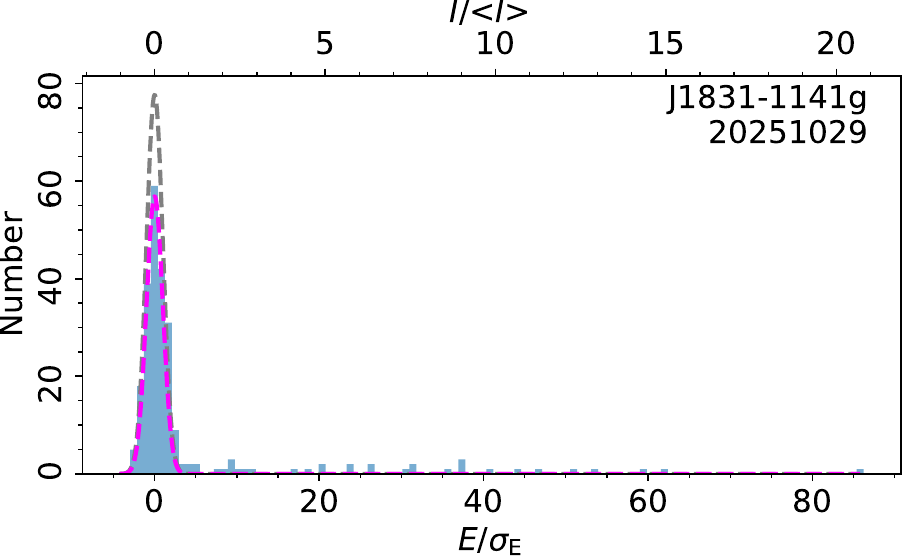}
\figcaption{On-pulse energy histogram of single pulses of PSR J1831-1141g from the FAST observation on 20251029.
\label{subfig:Hist:J1831-1141g}}
\end{figure}

\begin{figure}[htpb]
\includegraphics[width=0.22\textwidth, angle=0]{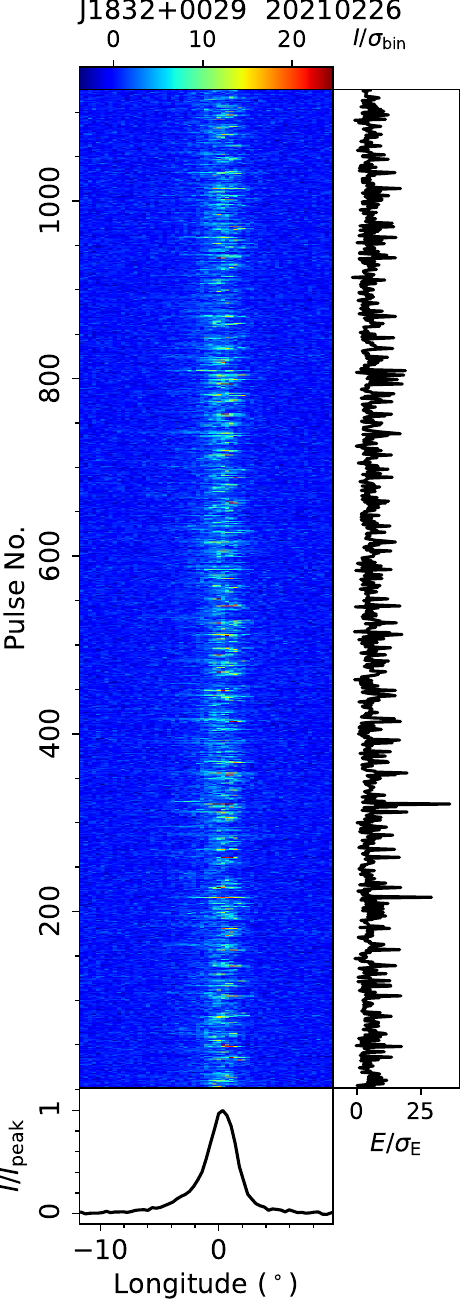}
\includegraphics[width=0.22\textwidth, angle=0]{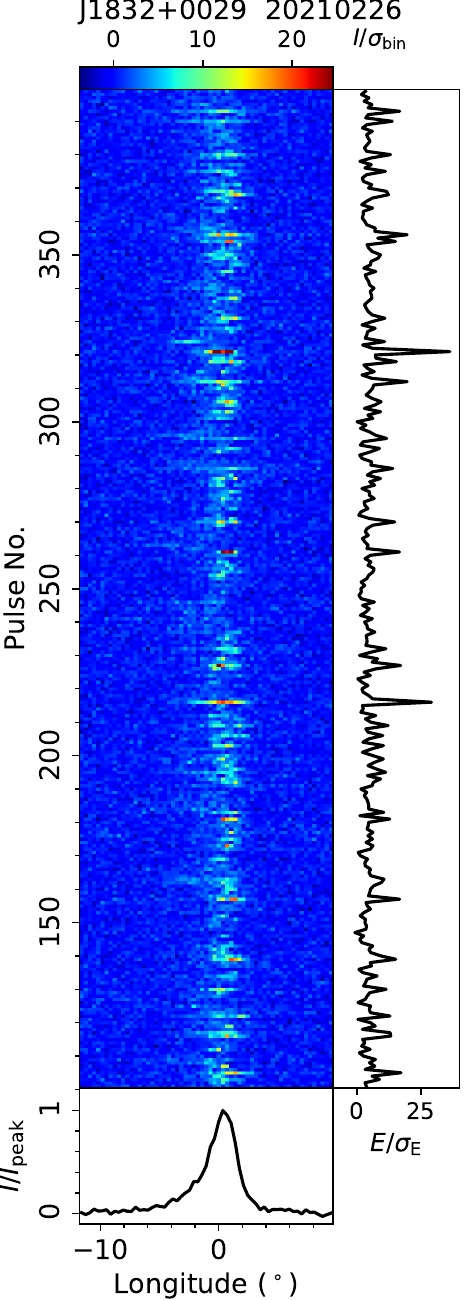}
\figcaption{Single pulse sequences of PSR J1832+0029 from the FAST observation on 20210226, and a zoomed-in view of pulses No. 100-400.
\label{subfig:TP:J1832+0029}}
\end{figure}

\begin{figure}[htpb]
\centering
\includegraphics[width=0.22\textwidth, angle=0]{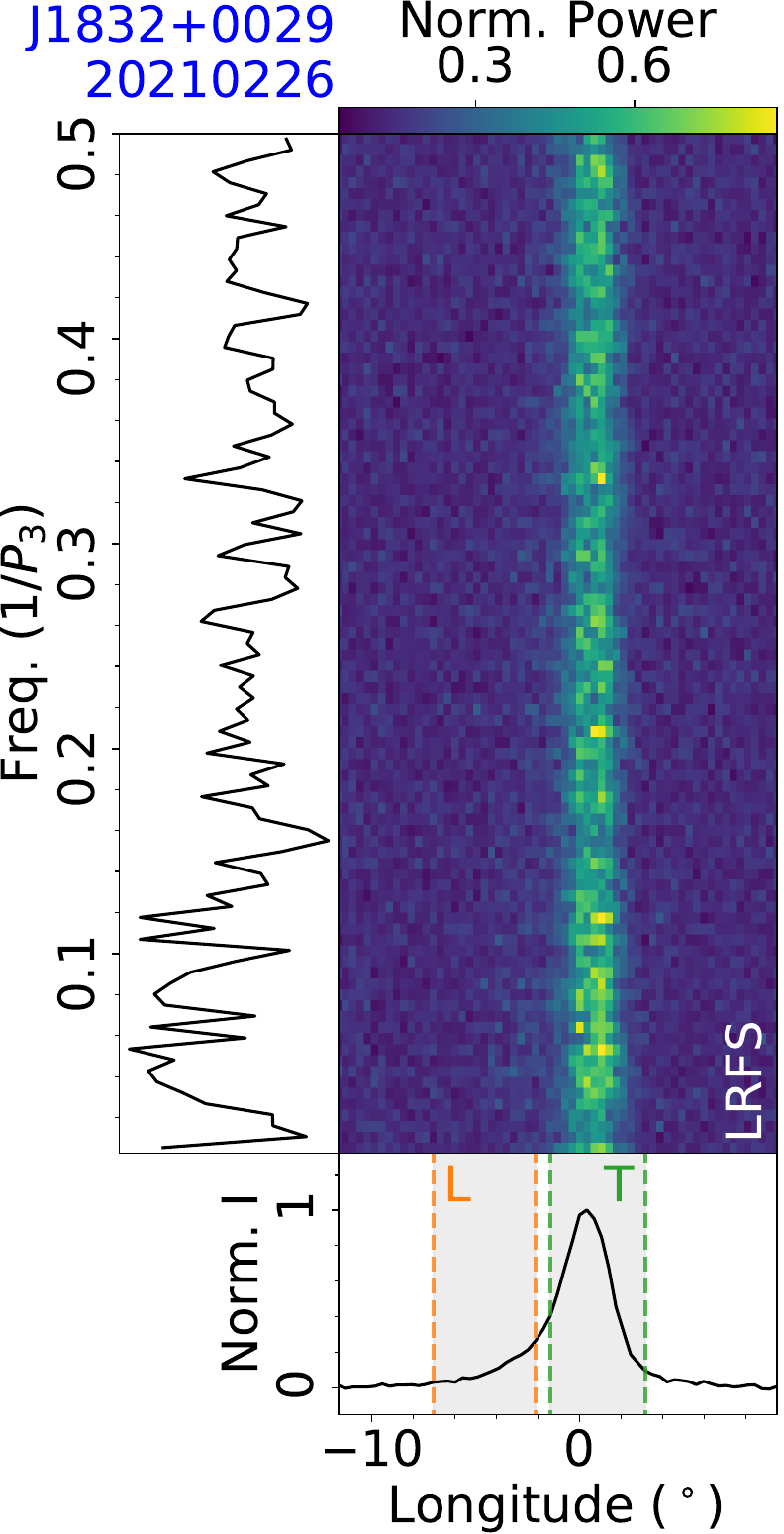}
\includegraphics[width=0.22\textwidth, angle=0]{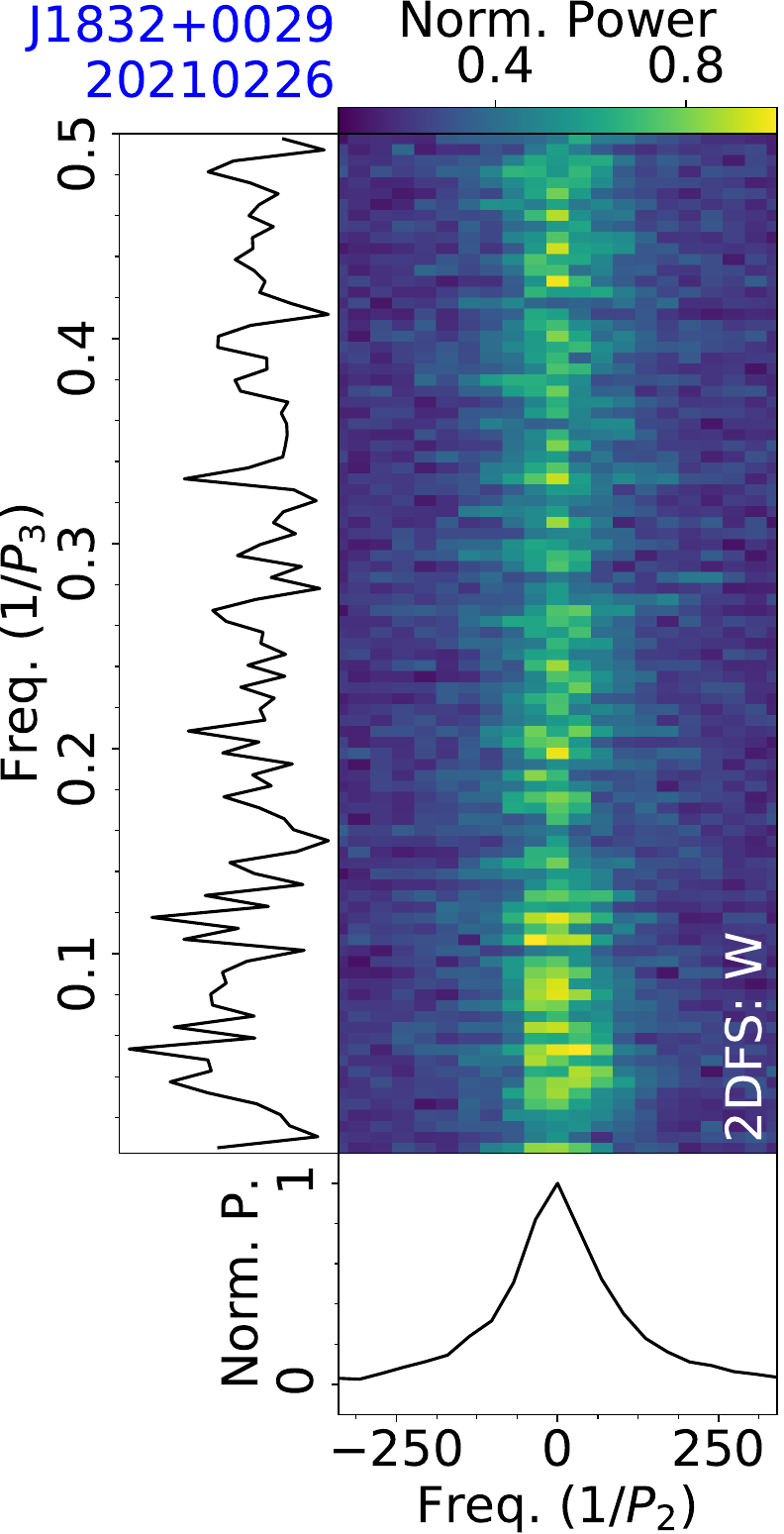}\\
\includegraphics[width=0.22\textwidth, angle=0]{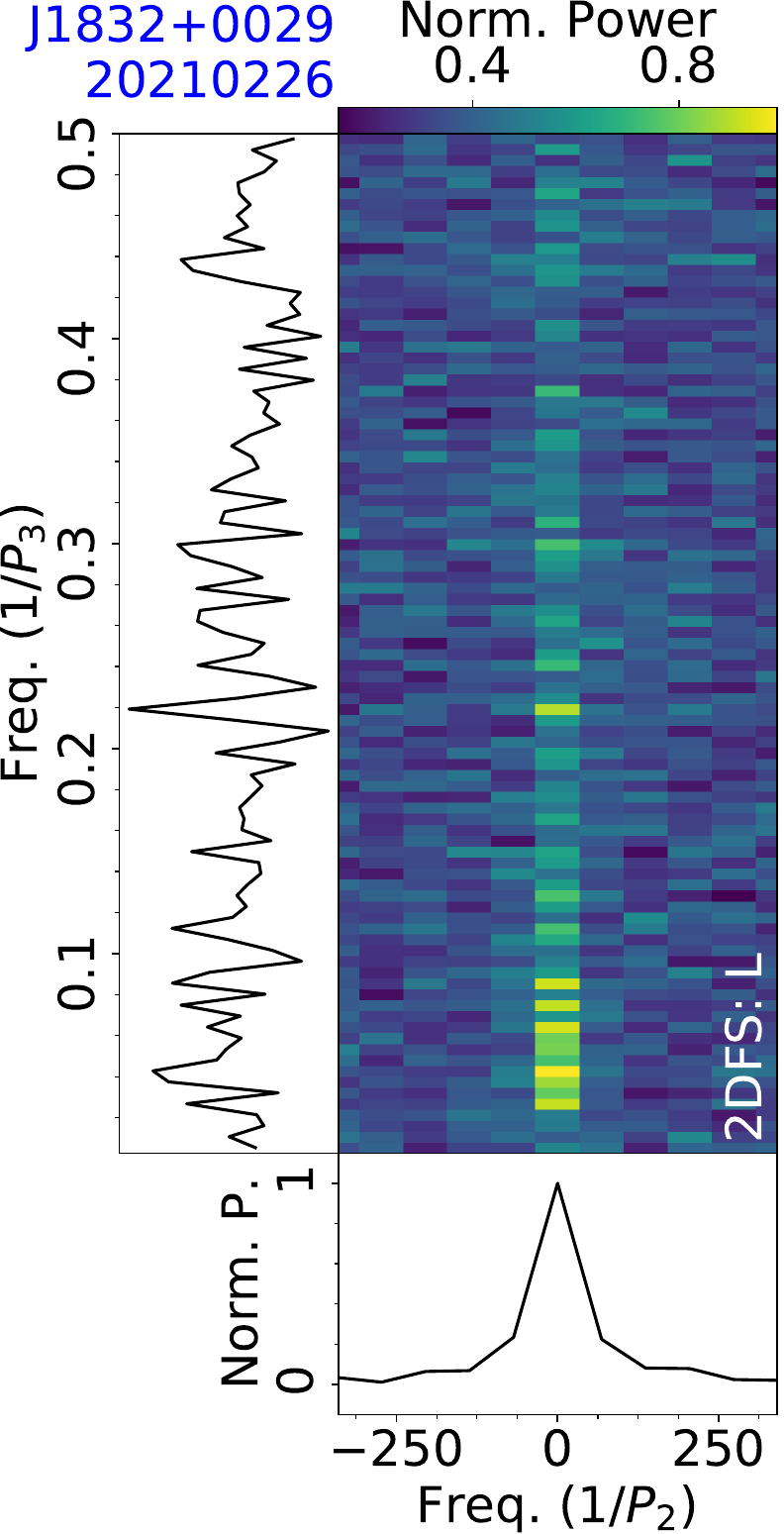}
\includegraphics[width=0.22\textwidth, angle=0]{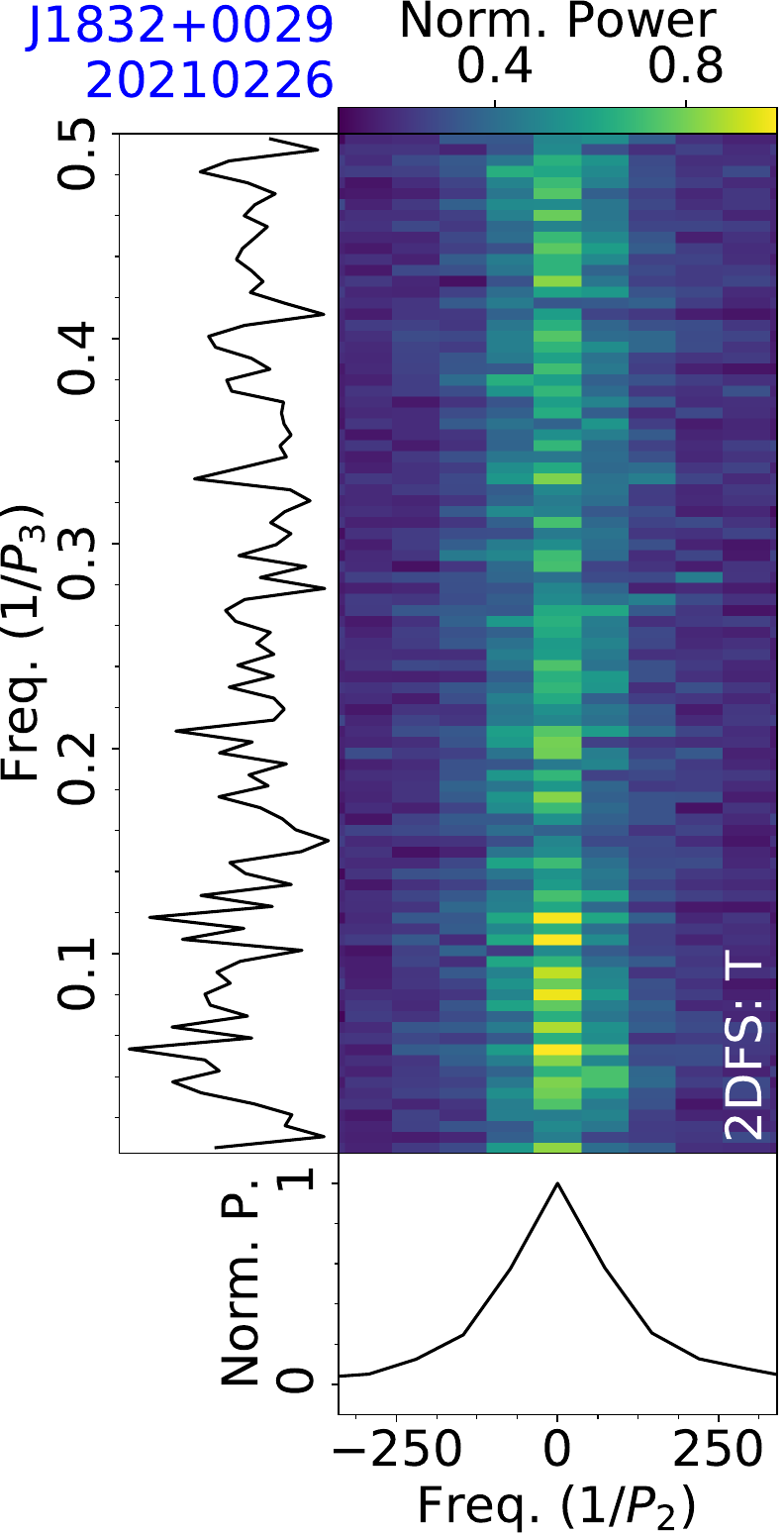}
\figcaption{Fluctuation analysis of PSR J1832+0029 for the observation on 20210226, with LRFS (top-left), and 2DFS for the on-pulse region (top-right), leading part (bottom-left) and trailing part (bottom-right) of a mean pulse profile. \label{subfig:fluctu:J1832+0029}}
\end{figure}

\subsection{J1829-0500}
\label{subsec:J1829-0500}

PSR J1829-0500 was discovered in the Commensal Radio Astronomy FAST Survey (CRAFTS) (http://groups.bao.ac.cn/ism/CRAFTS/).

This pulsar was observed by FAST on 20210222 for 59 minutes, deriving a rotation period $P=0.2406$~s and a dispersion measure $D\!M=274.0~{\rm cm^{-3}\,pc}$. The single pulse sequence and a zoomed-in view of pulses No. 7500-9000 in Fig.~\ref{subfig:TP:J1829-0500} show the emission occasionally brightened. 
For every single pulse, the energy is integrated over the longitude range from -1.58$^\circ$ to 8.26$^\circ$, and the resulting energy time series is smoothed with a 5-period moving average. Weak and bright emission modes of single pulses are then distinguished from the energy histogram shown in Fig.~\ref{subfig:Hist:J1829-0500}. 
The bright emission mode has a narrower profile and appears on the trailing side of the weak mode, as the profile contrast in Fig.~\ref{subfig:PolModes:J1829-0500} shows.

\begin{figure}[htpb]
\centering
\includegraphics[width=0.22\textwidth, angle=0]{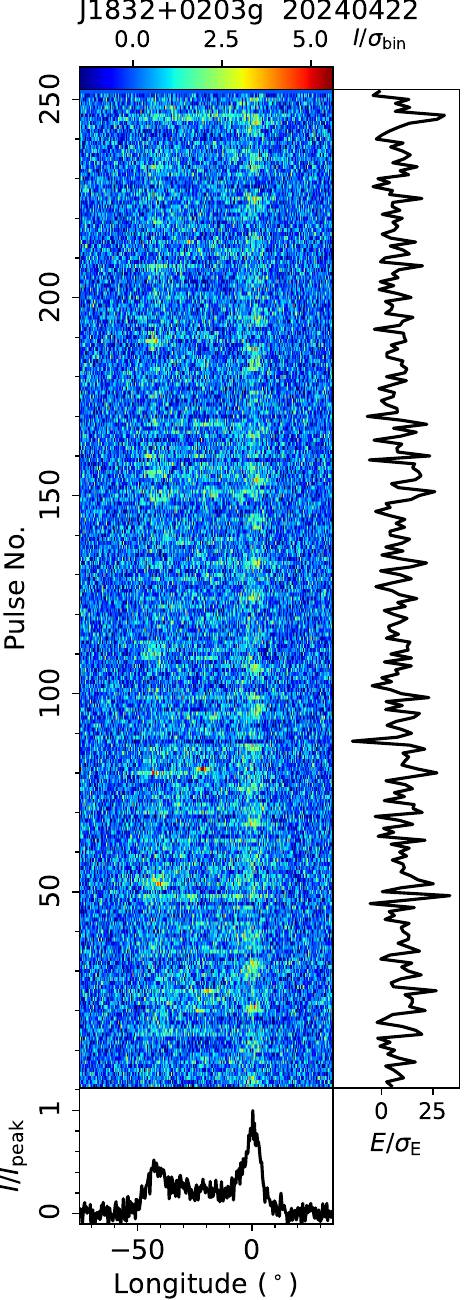}
\figcaption{Single pulse sequence of PSR J1832+0203g from the FAST observation on 20240422. \label{subfig:TP:J1832+0203g}}
\end{figure}

\begin{figure}[htpb]
\includegraphics[width=0.44\textwidth, angle=0]{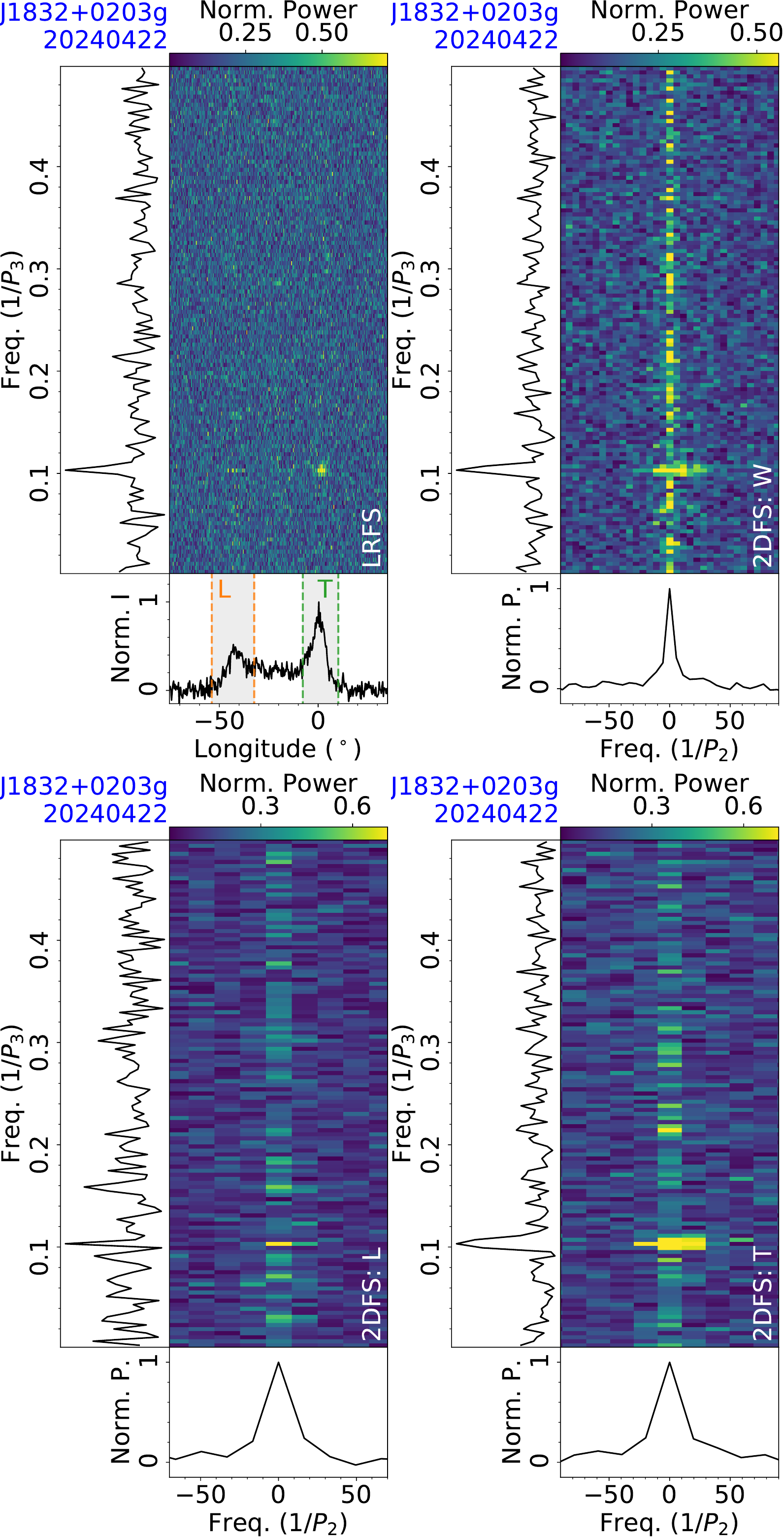}
\figcaption{Fluctuation analysis of PSR J1832+0203g for the observation on 20240422, with LRFS (top-left), and 2DFS for the on-pulse region (top-right), leading part (bottom-left) and trailing part (bottom-right) of a mean pulse profile. \label{subfig:fluctu:J1832+0203g}}
\end{figure}

\subsection{J1830-1135}
\label{subsec:J1830-1135}

PSR J1830-1135 was discovered in the Parkes Multibeam Pulsar Survey \citep{Morris2002}. A drift feature in 2DFS with $P_2=25^{+25}_{-7}$ degrees and $P_3=2.1\pm0.2$ periods was reported by \citet{Weltevrede2007} at 21 cm.

This pulsar was observed by FAST on 20260419 for 15 minutes, deriving a rotation period $P=6.2210$~s and a dispersion measure $D\!M=250.6~{\rm cm^{-3}\,pc}$. Single pulse sequences are shown in Fig.~\ref{subfig:TP:J1830-1135}. From the fluctuation spectra in Fig.~\ref{subfig:fluctu:J1830-1135}, the drift feature in 2DFS has the centroid at $1/P_3=0.391\pm0.002$ and $1/P_2=105\pm2$, corresponding to periodicities of $P_3=2.56\pm0.02$ periods and $P_2=3.4\pm0.1$ degrees.

\subsection{J1831+0222g}
\label{subsec:J1831+0222g}

PSR J1831+0222g was discovered in the FAST GPPS survey \citep{Han2021,han2025}. 
 
This pulsar was observed by FAST on 20241012 for 5 minutes and 20241107 for 15 minutes, with a rotation period $P=2.3153$~s and a dispersion measure $D\!M=158.8~{\rm cm^{-3}\,pc}$ from the 15-minute observation. 
Single pulse sequences of the observation on 20241107 in Fig.~\ref{subfig:TP:J1831+0222g} display the existence of nulling and subpulse drifting phenomena. 
The nulling fraction of this observation is estimated from the on-pulse integral energy histogram (Fig.~\ref{subfig:Hist:J1831+0222g}) to be 5.2$\pm$0.7\%. 
The centroid frequencies, corresponding to the negative drift feature in 2DFS (Fig.~\ref{subfig:fluctu:J1831+0222g}), are estimated to be $1/P_3=0.364\pm0.002$ and $1/P_2=-78\pm3$, corresponding to $P_3=2.74\pm0.01$ periods and $P_2=-4.6\pm0.2^\circ$.

\subsection{J1831-1141g}
\label{subsec:J1831-1141g}

PSR J1831-1141g was discovered in the FAST GPPS survey \citep{Han2021,han2025}. 

This pulsar was observed by FAST on 20251029 for 10 minutes, with a rotation period and a dispersion measure determined to be  $P=2.4722$~s and a dispersion measure $D\!M=43.8~{\rm cm^{-3}\,pc}$. The single pulse sequence shown in Fig.~\ref{subfig:TP:J1831-1141g} illustrating the existence of the nulling phenomenon. The nulling fraction is estimated from the on-pulse integral energy histogram (Fig.~\ref{subfig:Hist:J1831-1141g}), which is 73.2$\pm$2.3\% for this observation.

\begin{figure}[htpb]
\centering
\includegraphics[width=0.22\textwidth, angle=0]{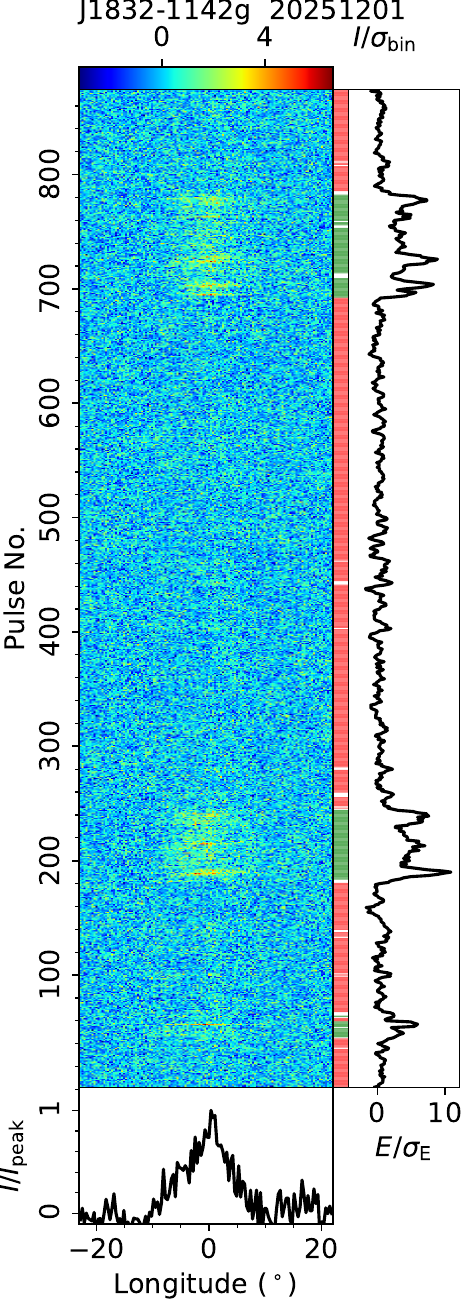}
\figcaption{Single pulse sequence of PSR J1832-1142g from the FAST observation on 20251201. 
The right subplot shows the energy variation integrated over the on-pulse phase region of the mean pulse profile, smoothed with a 5-period moving average.
\label{subfig:TP:J1832-1142g}}
\end{figure}

\begin{figure}[htpb]
\centering
\includegraphics[width=0.39\textwidth, angle=0]{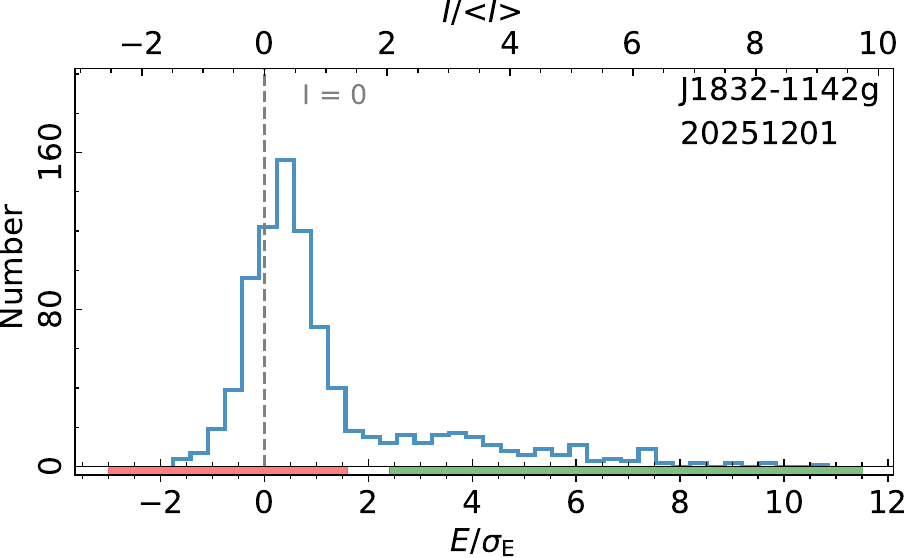}
\figcaption{On-pulse energy histogram of single pulses of PSR J1832-1142g from the FAST observation on 20251201, with energy values smoothed using a 5-period moving average. 
\label{subfig:Hist:J1832-1142g}}
\end{figure}

\begin{figure}[htpb]
\centering
\includegraphics[width=0.39\textwidth, angle=0]{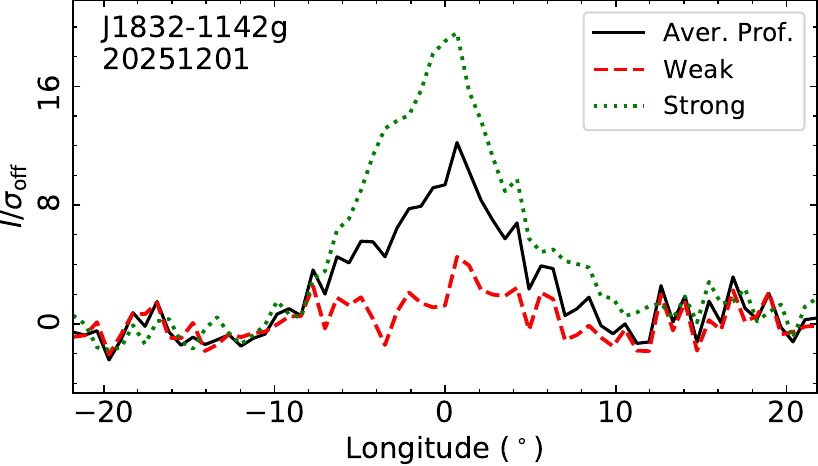}
\figcaption{Mean profiles of weak and bright emission modes of PSR J1832-1142g from the FAST observation on 20251201, with each profile normalized separately by its corresponding off-pulse standard deviation $\sigma_{\rm off}$. 
\label{subfig:profModes:J1832-1142g}}
\end{figure}

\subsection{J1832+0029}
\label{subsec:J1832+0029}

PSR J1832+0029 was discovered in the Parkes multibeam pulsar survey \citep{Lorimer2006}. 
The pulsar is known as an intermittent pulsar, with off-state time scale fraction of about 50\% reported by \citet{Lorimer2012}. 

This pulsar was observed by FAST on 20191009 for 5 minutes and 20210226 for 10 minutes, and on on-state during these two observations. Single pulse sequences of the observation on 20210226 are shown in Fig.~\ref{subfig:TP:J1832+0029}. 2DFS of the leading and trailing parts of the profile in Fig.~\ref{subfig:fluctu:J1832+0029} illustrate the existence of the low-frequency modulation, with the centroid frequencies of $1/P_3=0.055\pm0.002$ ($P_3=18\pm1$ periods) and $1/P_3=0.073\pm0.003$ ($P_3=14\pm1$ periods), respectively.

\begin{figure}[htpb]
\centering
\includegraphics[width=0.22\textwidth, angle=0]{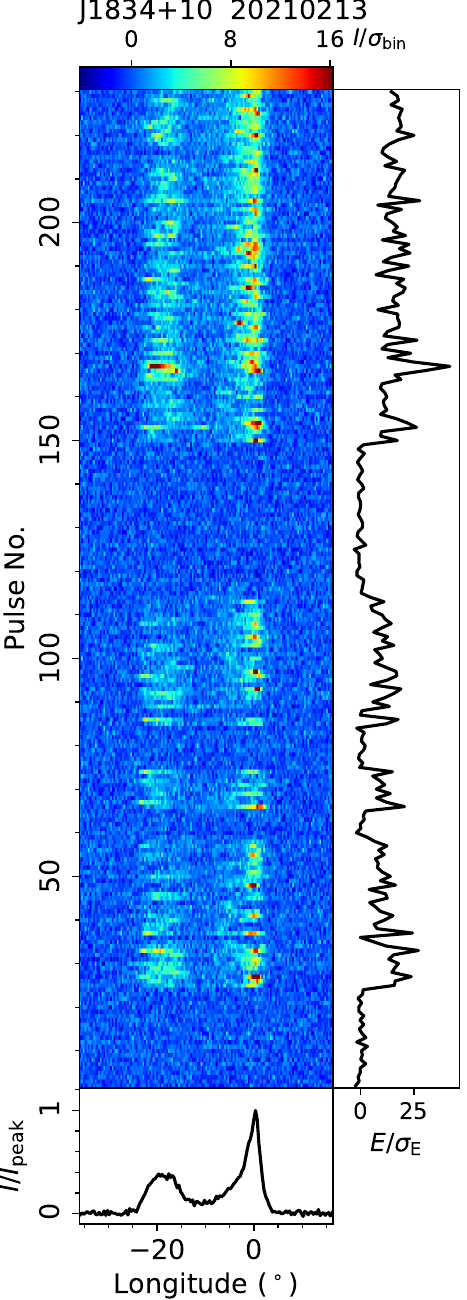}
\includegraphics[width=0.22\textwidth, angle=0]{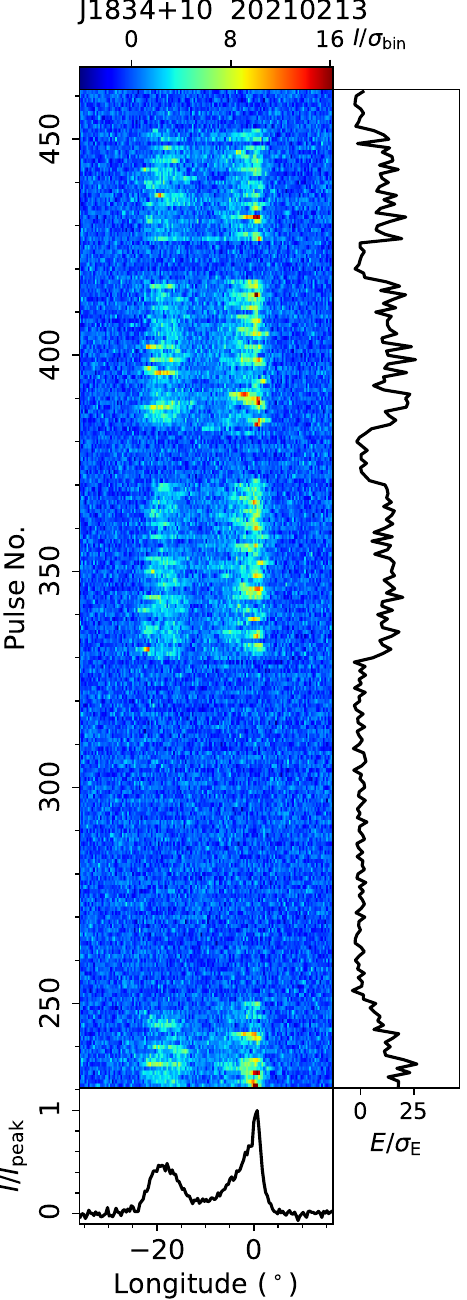}
\figcaption{Single pulse sequences of PSR J1834+10 from the FAST observation on 20210213. \label{subfig:TP:J1834+10}}
\end{figure}

\begin{figure}[htpb]
\centering
\includegraphics[width=0.39\textwidth, angle=0]{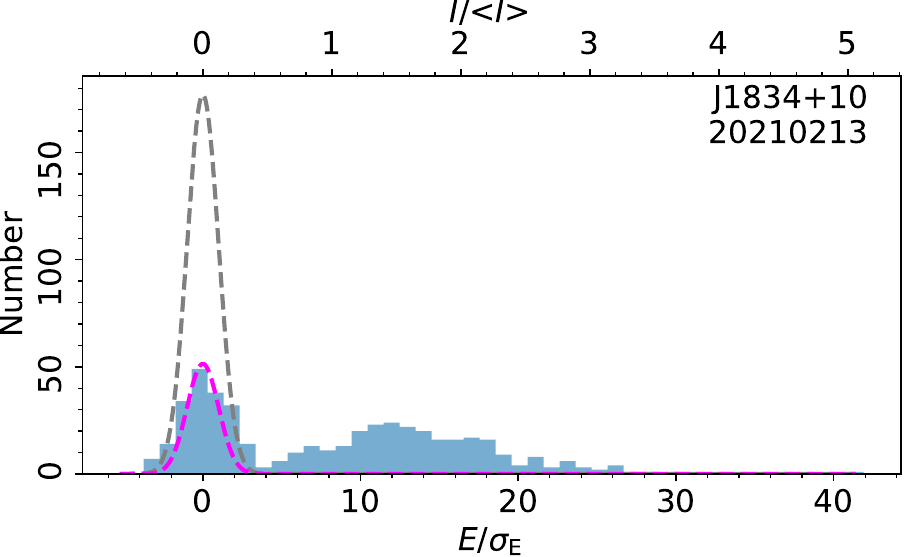}
\vspace{-0.3cm}
\figcaption{On-pulse integral energy histogram of single pulses of PSR J1834+10 from the FAST observation on 20210213.
\label{subfig:Hist:J1834+10}}
%
\centering
\includegraphics[width=0.44\textwidth, angle=0]{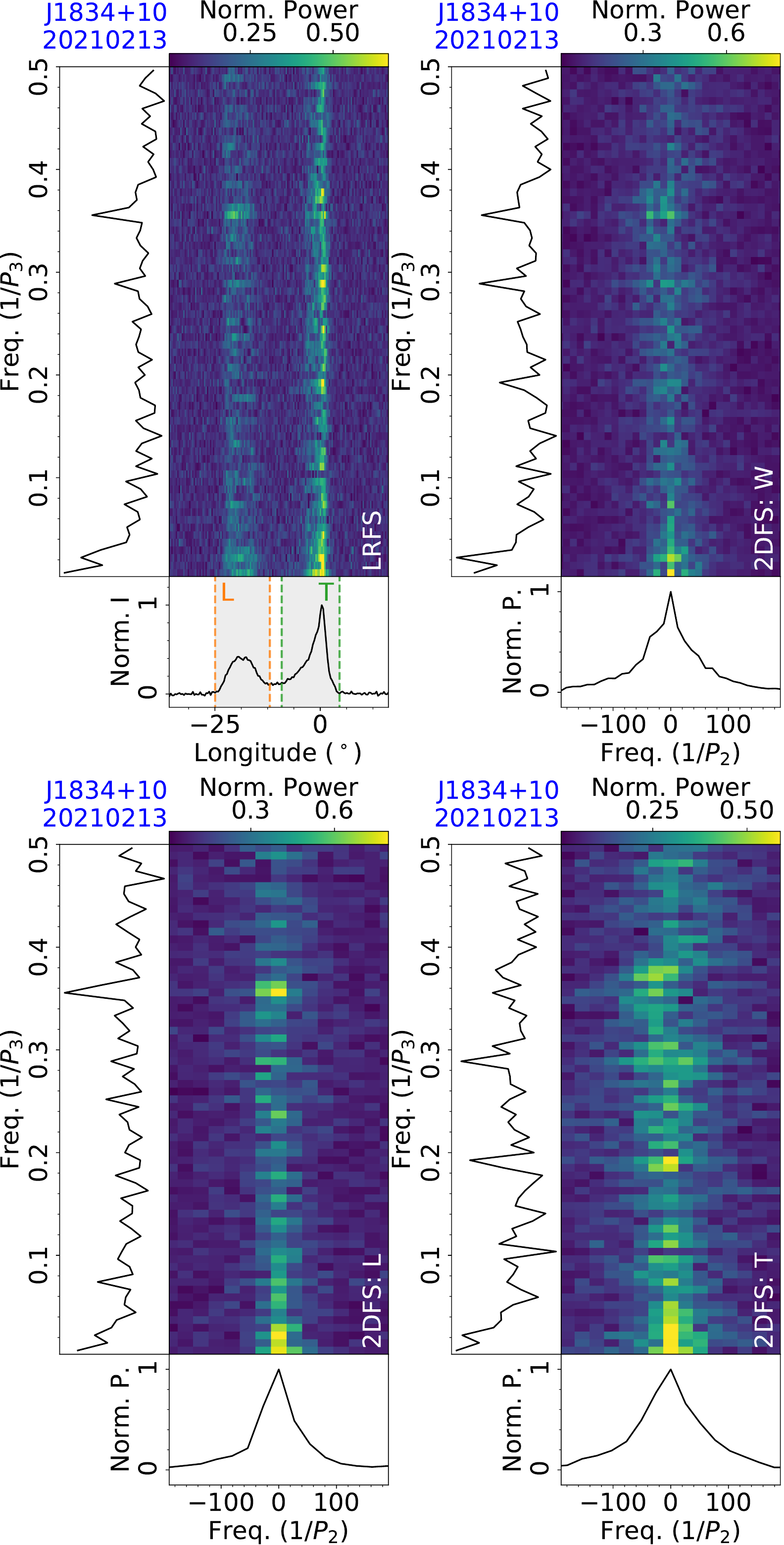}
\vspace{-0.2cm}
\figcaption{Fluctuation analysis of PSR J1834+10 for the observation on 20210213, with LRFS (top-left), and 2DFS for the on-pulse region (top-right), leading part (bottom-left) and trailing part (bottom-right) of a mean pulse profile. \label{subfig:fluctu:J1834+10}}
\end{figure}

\begin{figure}[htpb]
\centering
\includegraphics[width=0.22\textwidth, angle=0]{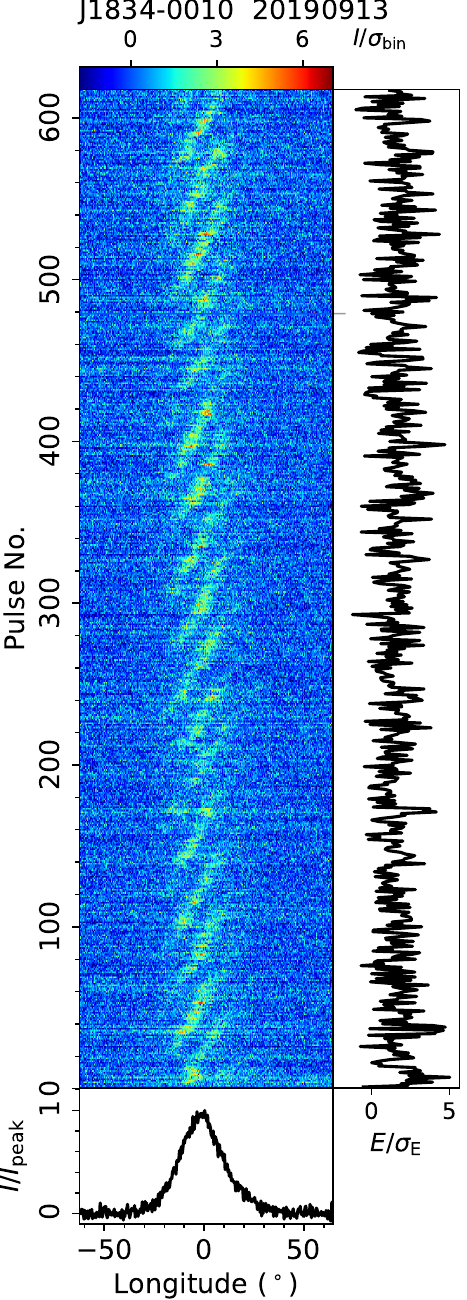}
\figcaption{Single pulse sequence of PSR J1834-0010 from the FAST observation on 20190913.
\label{subfig:TP:J1834-0010}}
\end{figure}

\begin{figure}[htpb]
\centering
\includegraphics[width=0.22\textwidth, angle=0]{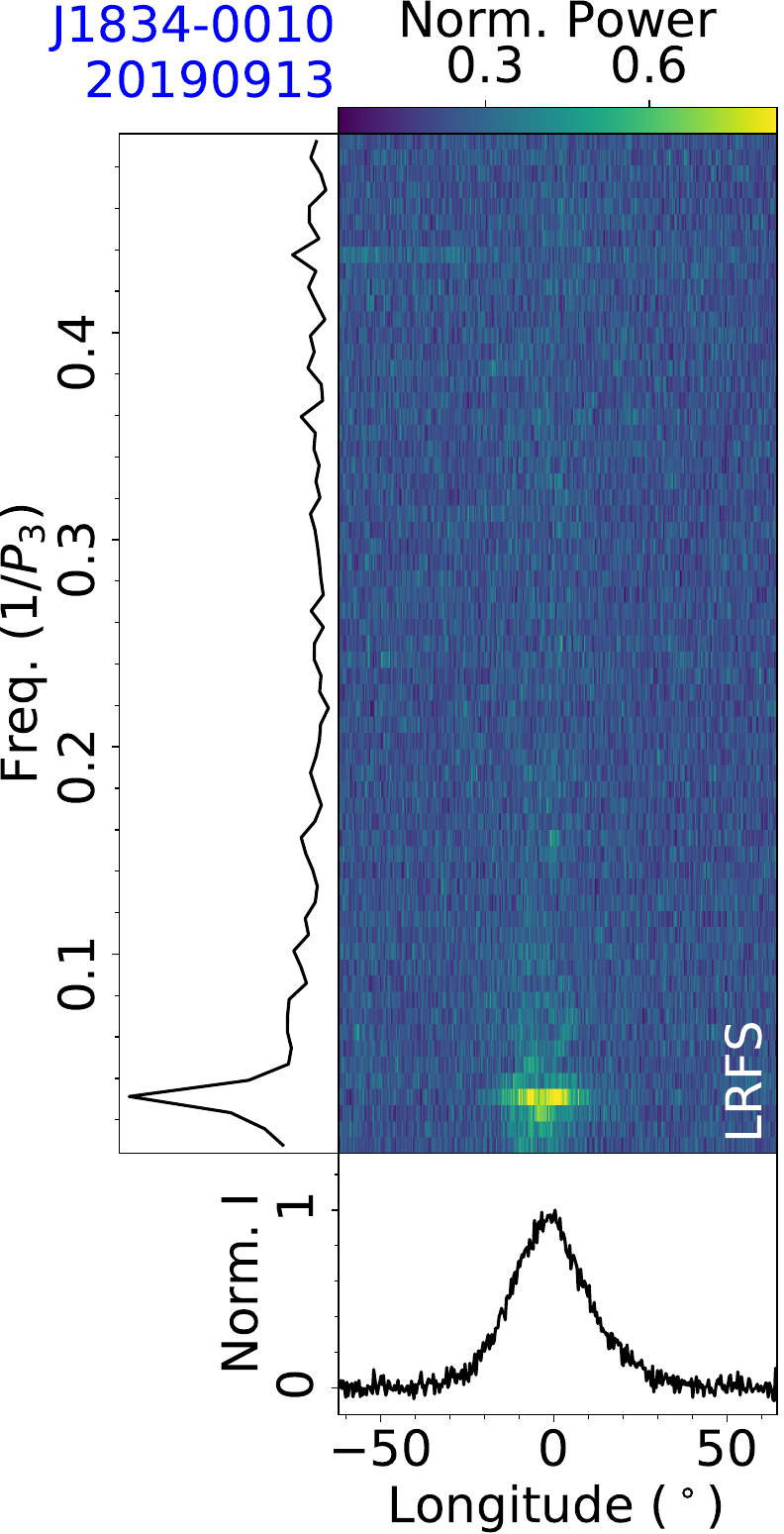}
\includegraphics[width=0.22\textwidth, angle=0]{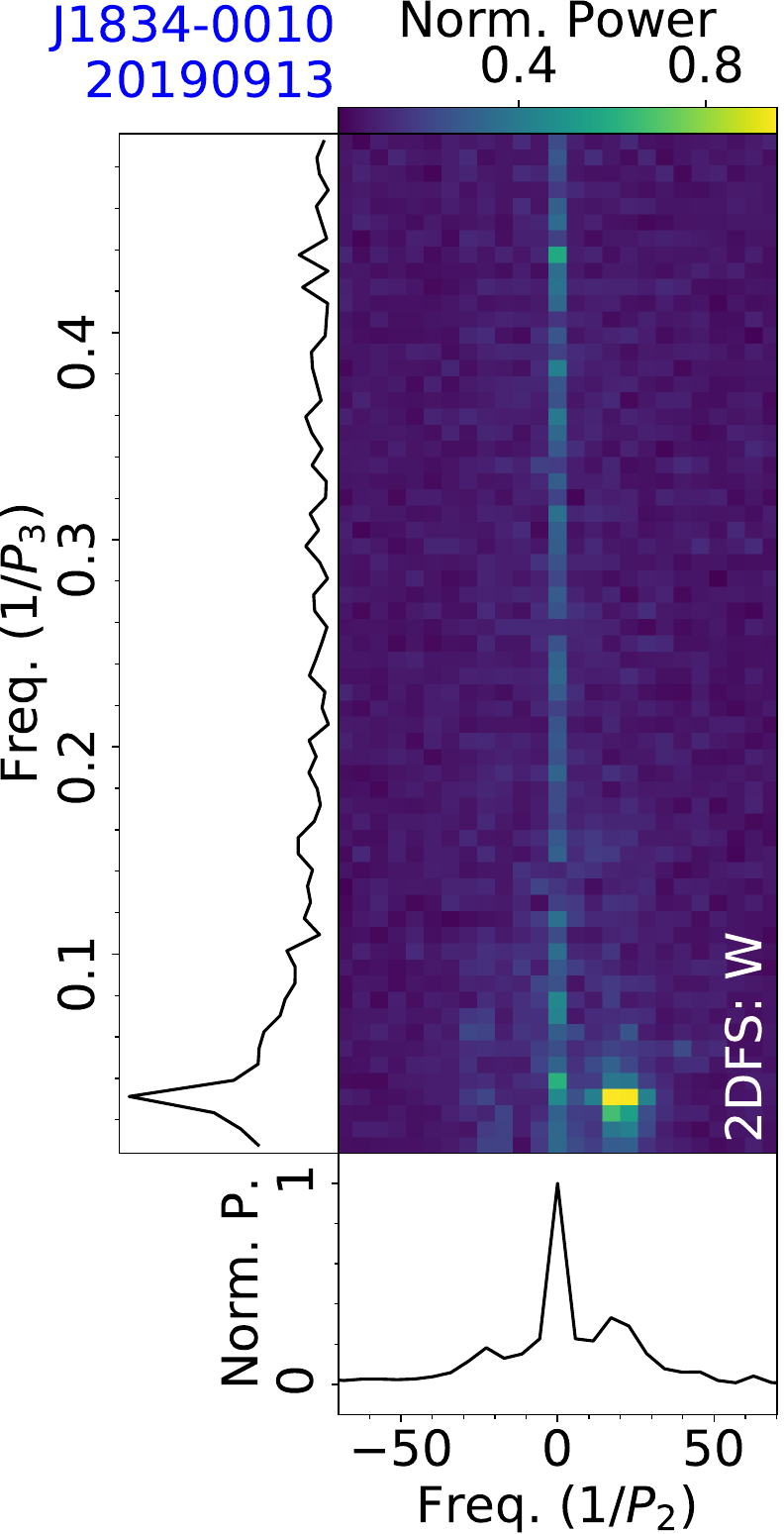}
\figcaption{Fluctuation analysis of PSR J1834-0010 for the observation on 20190913, with LRFS and 2DFS for the on-pulse region of a mean pulse profile.
\label{subfig:fluctu:J1834-0010}}
\end{figure}

\begin{figure}[htpb]
\centering
\includegraphics[width=0.44\textwidth, angle=0]{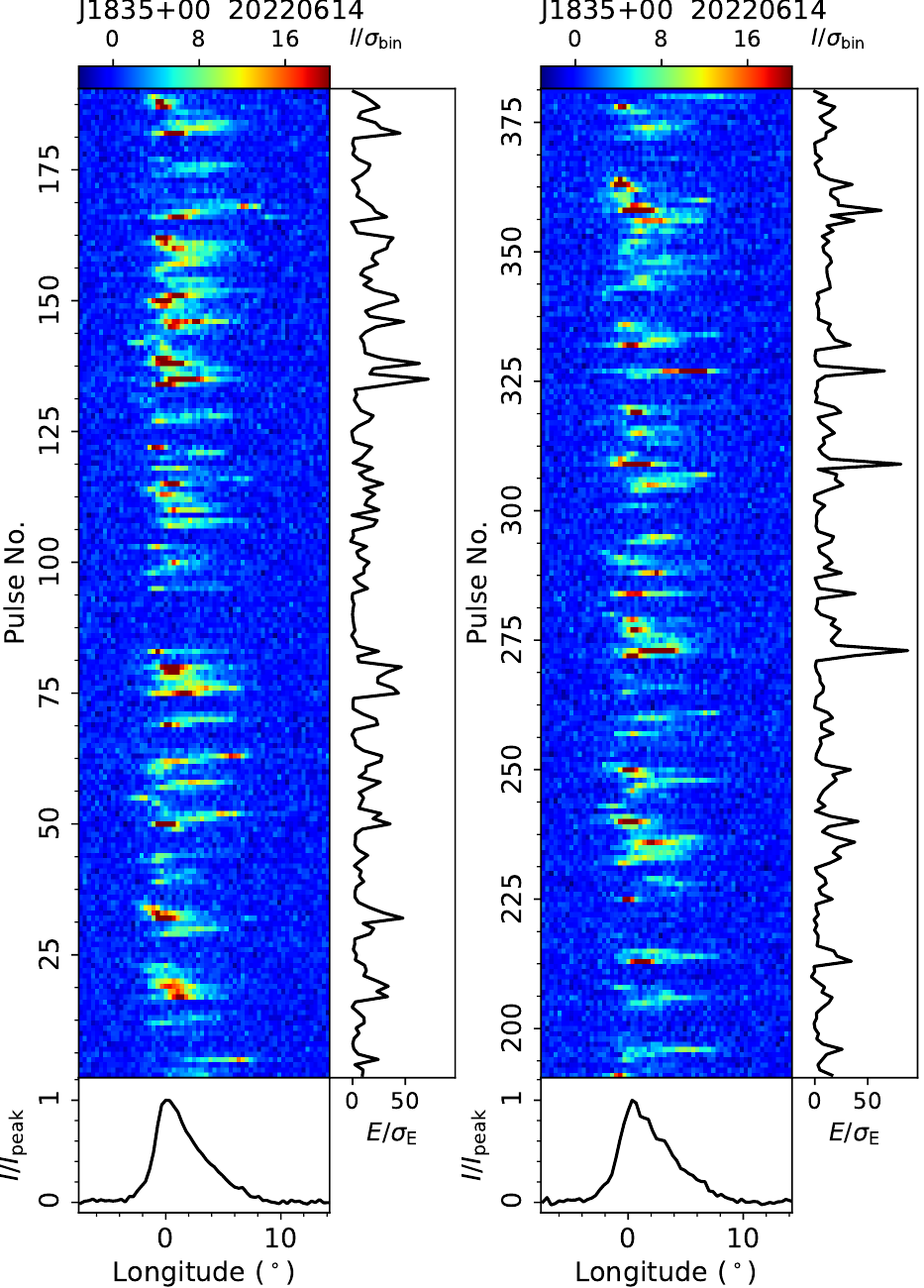}
\figcaption{Single pulse sequences of PSR J1835+00 from the FAST observation on 20220614.
\label{subfig:TP:J1835+00}}
\end{figure}

\begin{figure}[htpb]
\centering
\includegraphics[width=0.44\textwidth, angle=0]{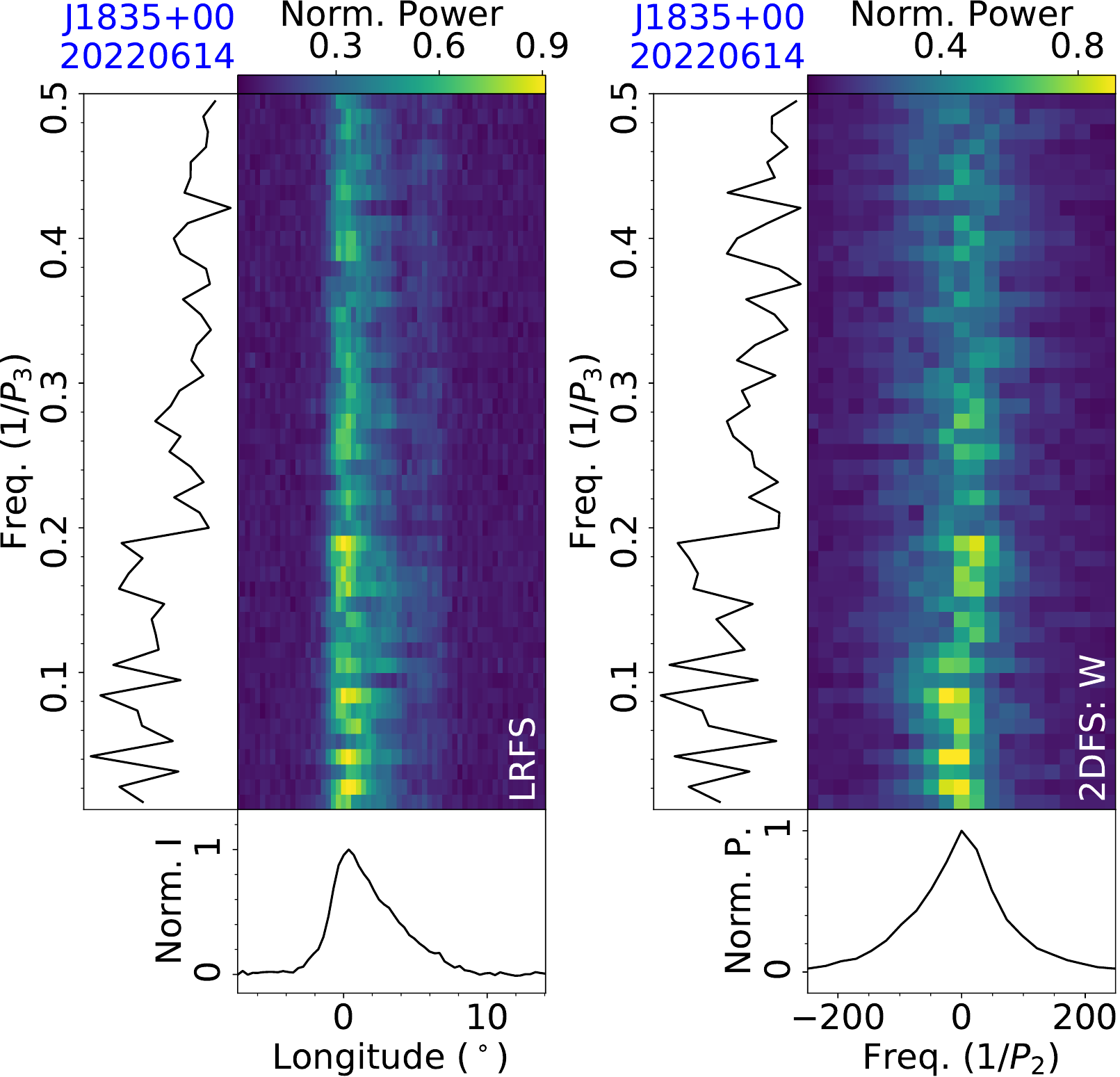}
\figcaption{Fluctuation analysis of PSR J1835+00 for the observation on 20220614, with LRFS and 2DFS for the on-pulse region of the mean pulse profile.
\label{subfig:fluctu:J1835+00}}
\end{figure}
 
\subsection{J1832+0203g}
\label{subsec:J1832+0203g}

PSR J1832+0203g was discovered in the FAST GPPS survey \citep{Han2021,han2025}. 

This pulsar was observed by FAST on 20240422 for 5 minutes, deriving a rotation period $P=1.2127$~s and a dispersion measure $D\!M=226.2~{\rm cm^{-3}\,pc}$. 
The single pulse sequence of this observation is displayed in Fig.~\ref{subfig:TP:J1832+0203g}. The LRFS and 2DFS in Fig.~\ref{subfig:fluctu:J1832+0203g} indicate that the leading and trailing parts in a mean pulse profile have similar temporal modulation frequencies of $1/P_3=0.103\pm0.001$, corresponding to $P_3=9.7\pm0.1$ periods. 
2DFS of the trailing part exhibits a preferred positive subpulse drift feature, with the centroid of $1/P_2=5\pm2$, yielding $P_2=73\pm34^\circ$.

\subsection{J1832-1142g}
\label{subsec:J1832-1142g}

PSR J1832-1142g was discovered in the FAST GPPS survey \citep{Han2021,han2025}. 

This pulsar was observed by FAST on 20251201 for 15 minutes, deriving a rotation period $P=1.0308$~s and a dispersion measure $D\!M=704.0~{\rm cm^{-3}\,pc}$. The observation indicates that this pulsar has weak and strong emission modes. The single pulse sequence in Fig.~\ref{subfig:TP:J1832-1142g} displays three segments of enhanced emission. Single pulses of two emission modes are distinguished from the on-pulse energy histogram with energy values smoothed using a 5-period moving average (Fig.~\ref{subfig:Hist:J1832-1142g}). Weak and strong emission modes are labeled in red and green respectively. The mean pulse profiles of two modes are shown in Fig.~\ref{subfig:profModes:J1832-1142g}.

\subsection{J1834+10}
\label{subsec:J1834+10}

PSR J1834+10 was discovered by the Arecibo telescope at 430 MHz \citep{Camilo1996}.

The pulsar was observed by FAST on 20210213 for 9 minutes and 20210621 for 5 minutes. From the 9-minute data, the rotation period is $P=1.1728$~s and the dispersion measure is $D\!M=78.1~{\rm cm^{-3}\,pc}$. 
Single pulse sequences of the observation on 20210213 in Fig.~\ref{subfig:TP:J1834+10} display nulling and modulation phenomena. From on-pulse integral energy histogram in Fig.~\ref{subfig:Hist:J1834+10}, the nulling fraction of this observation is estimated to be 29$\pm$3\%. Drifting parameters is calculated from 2DFS (Fig.~\ref{subfig:fluctu:J1834+10}).
In 2DFS of the leading part in a mean pulse profile, the centroid frequencies of the negative drift feature are $1/P_3=0.359\pm0.001$ and $1/P_2=-12\pm4$, corresponding to $P_3=2.79\pm0.01$ periods and $P_2=-30\pm11^\circ$. For the trailing profile part, the negative drift feature exhibits the centroid frequencies of $1/P_3=0.337\pm0.003$ and $1/P_2=-31\pm2$, yielding $P_3=2.96\pm0.02$ periods and $P_2=-12\pm1^\circ$.
While there is no obvious modulation feature in 2DFS of the data on 20210621.

%

\begin{figure}[htpb]
\centering
\includegraphics[width=0.22\textwidth, angle=0]{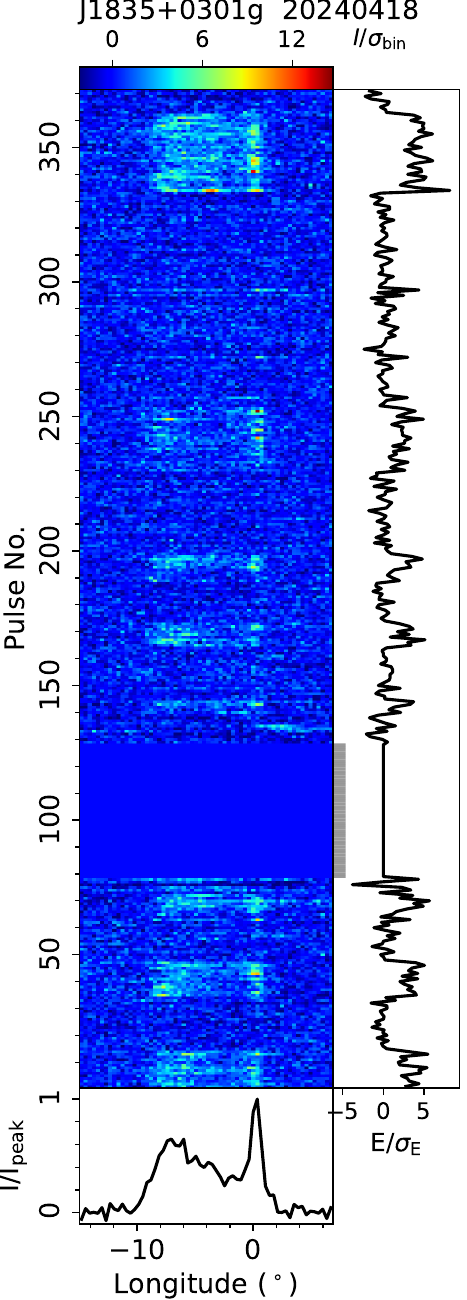}
\figcaption{Single pulse sequence of PSR J1835+0301g from the FAST observation on 20240418.
\label{subfig:TP:J1835+0301g}}
\end{figure}

\begin{figure}[htpb]
\centering
\includegraphics[width=0.39\textwidth, angle=0]{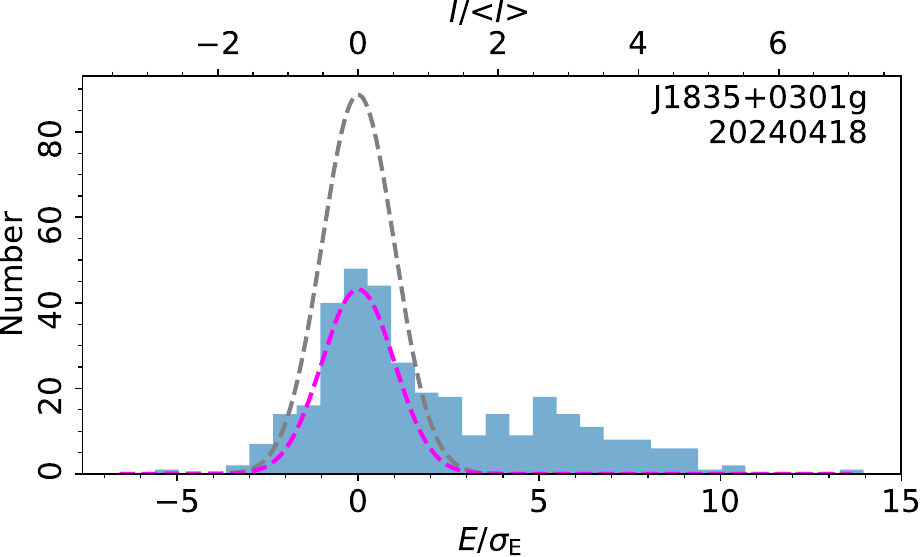}
\figcaption{On-pulse energy histograms of single pulses of PSR J1835+0301g from the FAST observation on 20240418.
\label{subfig:Hist:J1835+0301g}}
\end{figure}

\begin{figure}[htpb]
\centering
\includegraphics[width=0.22\textwidth, angle=0]{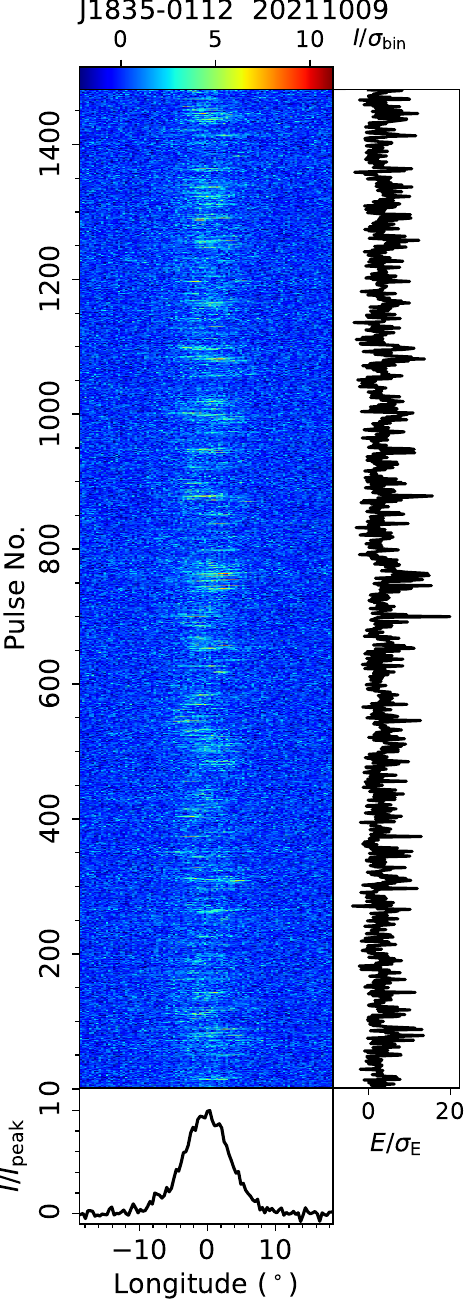}
\includegraphics[width=0.22\textwidth, angle=0]{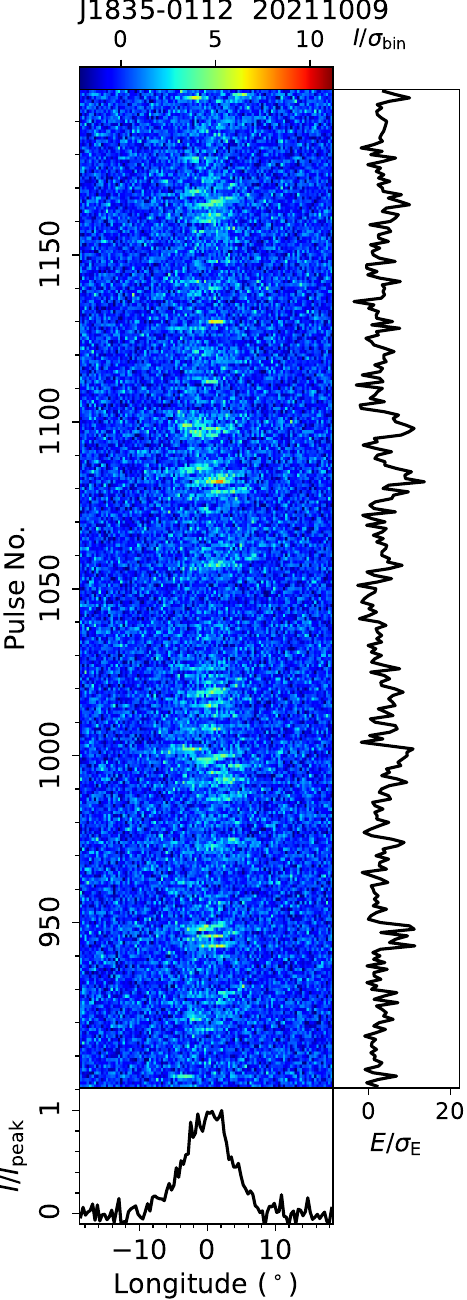}
\figcaption{Single pulse sequence of PSR J1835-0112 from the FAST observation on 20211009, and a zoomed-in view of pulses No. 900-1200.
\label{subfig:TP:J1835-0112}}
\end{figure}

\begin{figure}[htpb]
\centering
\includegraphics[width=0.22\textwidth, angle=0]{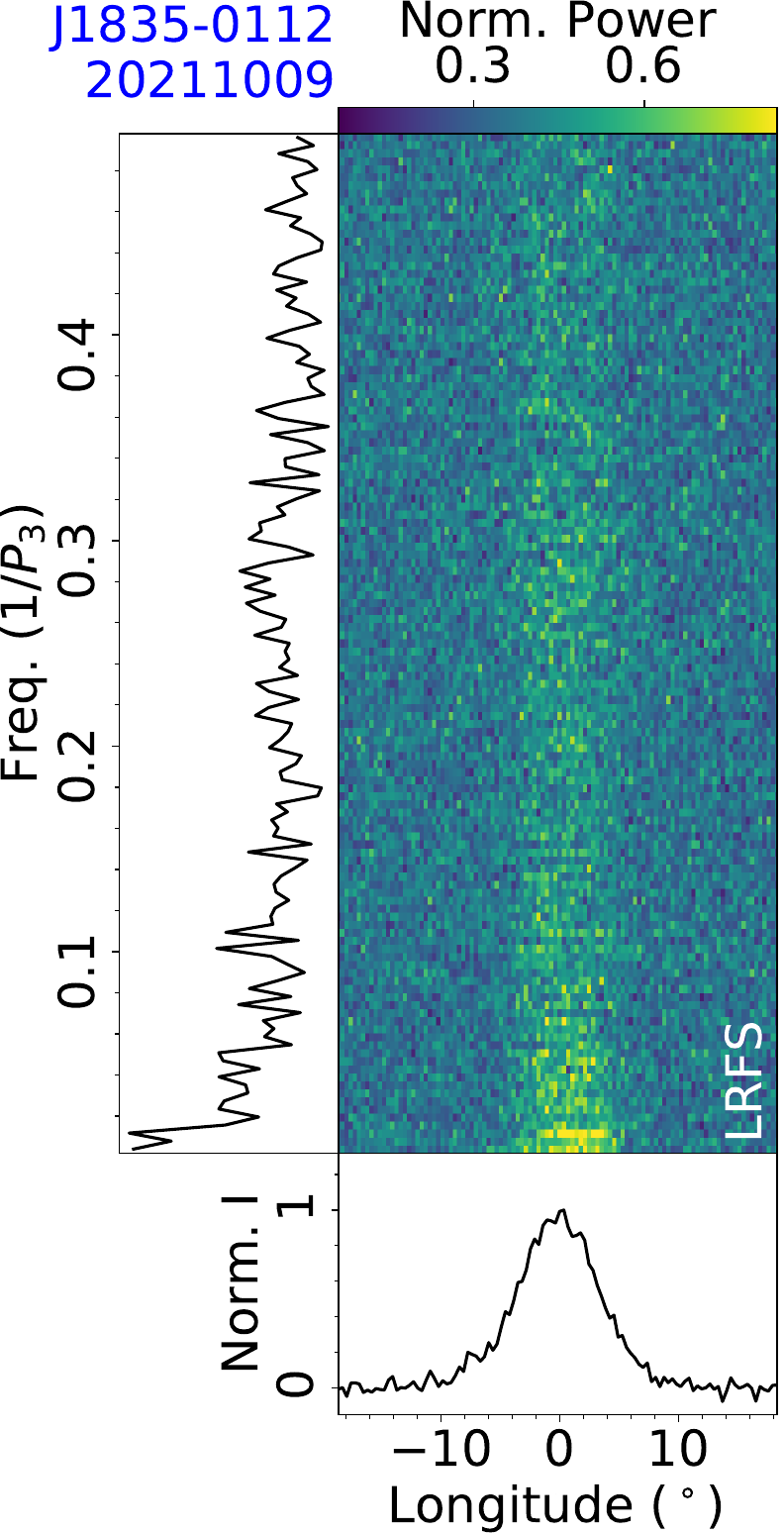}
\includegraphics[width=0.22\textwidth, angle=0]{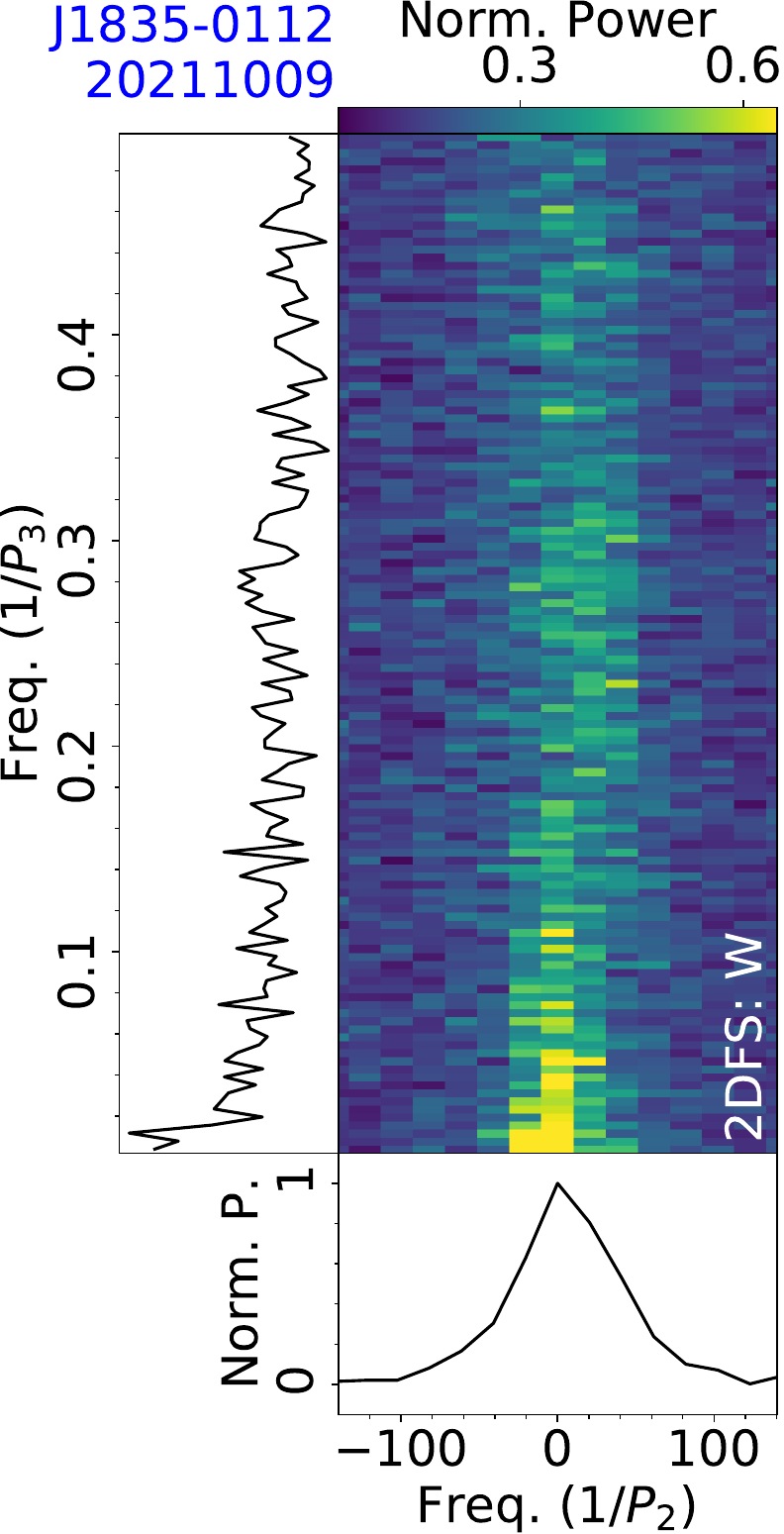}
\figcaption{Fluctuation analysis of PSR J1835-0112 from the FAST observation on 20211009, with LRFS and 2DFS for the on-pulse region of a mean pulse profile.
\label{subfig:fluctu:J1835-0112}}
\end{figure}

\begin{figure}[htpb]
\centering
\includegraphics[width=0.22\textwidth, angle=0]{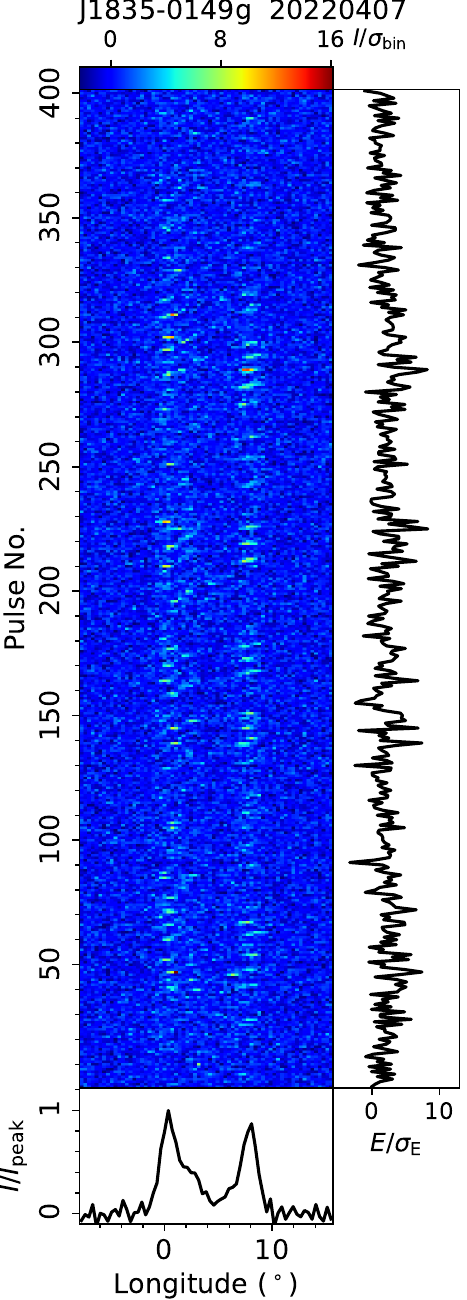}
\includegraphics[width=0.22\textwidth, angle=0]{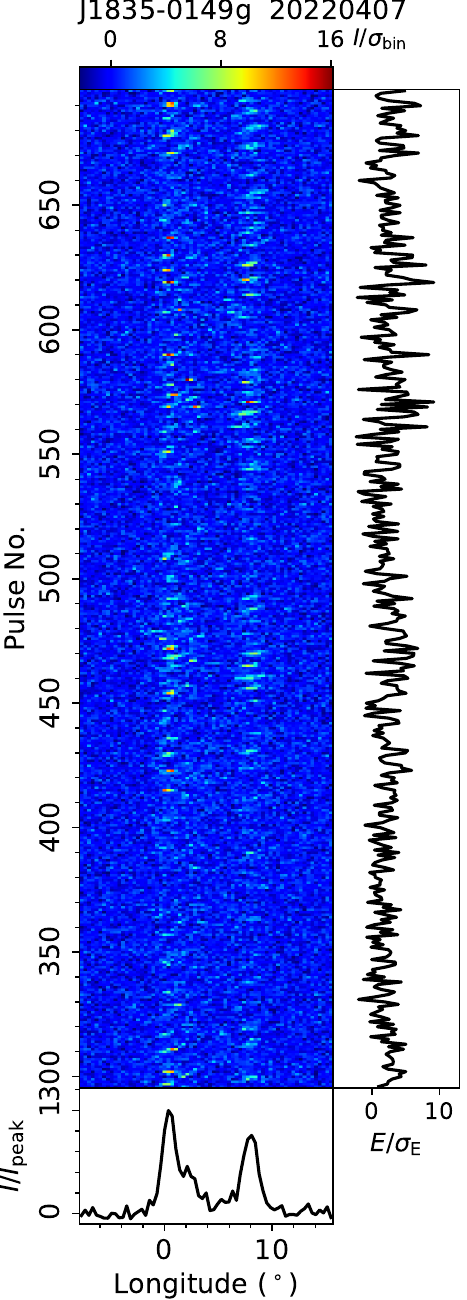}
\figcaption{Single pulse sequences of PSR J1835-0149g from the FAST observation on 20220407. \label{subfig:TP:J1835-0149g}}
\end{figure}

\begin{figure}[htpb]
\centering
\includegraphics[width=0.44\textwidth, angle=0]{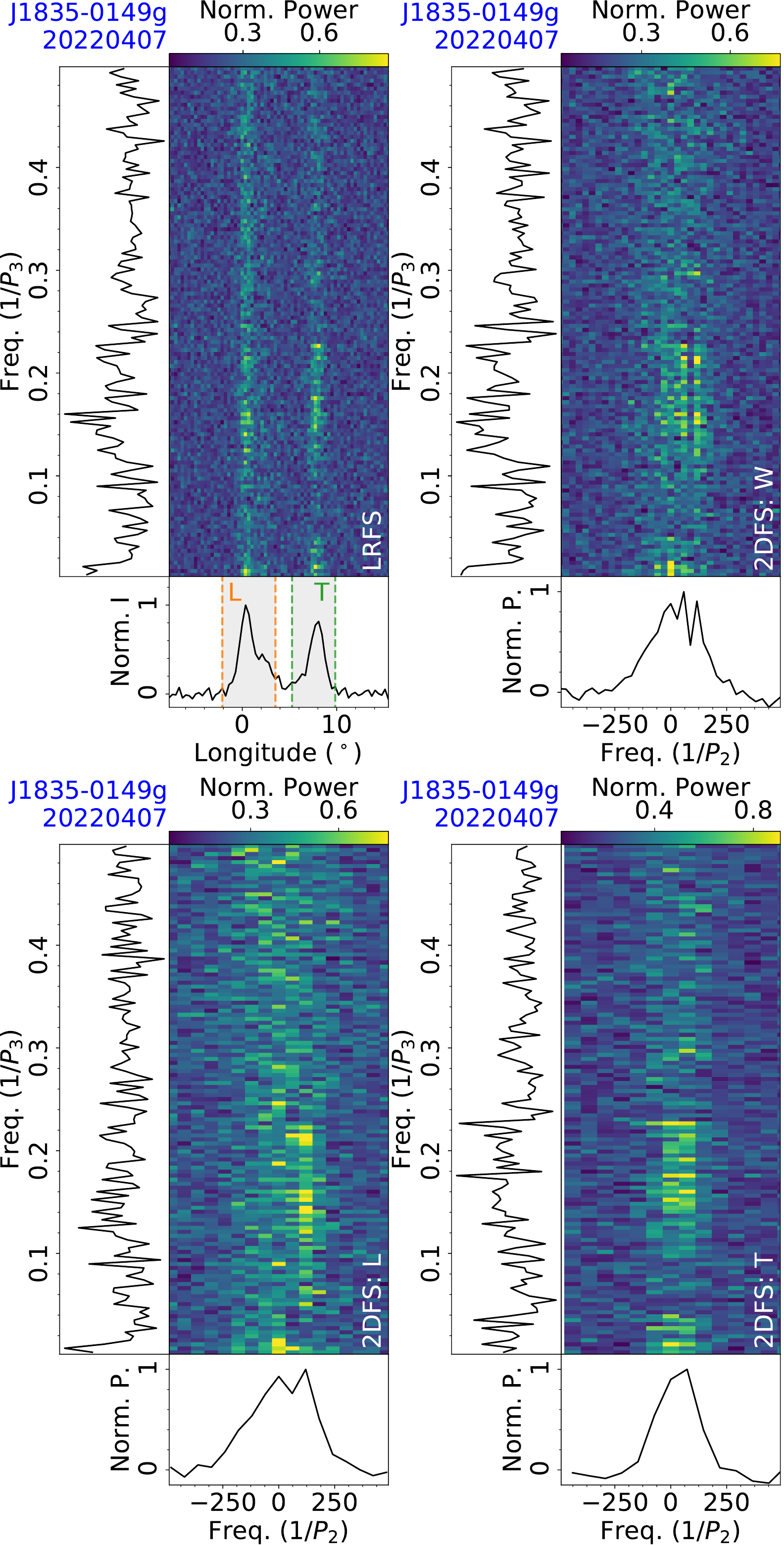}
\figcaption{Fluctuation analysis of PSR J1835-0149g for the observation on 20220407, with LRFS (top-left), and 2DFS for the on-pulse region (top-right), leading part (bottom-left) and trailing part (bottom-right) of a mean pulse profile. \label{subfig:fluctu:J1835-0149g}}
\end{figure}

\subsection{J1834-0010}
\label{subsec:J1834-0010}

PSR J1834-0010 was discovered by \citet{Dewey1985} at 390 MHz using the 92 m telescope at Green Bank. In previous studies, \citet{Herfindal2009} reported this pulsar with a nulling fraction of 2.618\% at 327MHz. \citet{Song2023} discovered that the pulsar has drifting behavior of $P_3=37.6\pm0.7$ periods and $P_2=18.5^{+0.5}_{-0.3}$ degrees.

This pulsar was observed by FAST on 20190913 for 5 minutes, deriving a rotation period $P=0.5210$~s and a dispersion measure $D\!M=87.8~{\rm cm^{-3}\,pc}$. 
The single pulse sequence in Fig.~\ref{subfig:TP:J1835+0301g} displays subpulse drifting bands. 
%
Fluctuation spectra are displayed in Fig.~\ref{subfig:fluctu:J1834-0010}, where the centroid frequencies of the positive drift feature are estimated to be $1/P_3=0.026\pm0.001$ and $1/P_2=20.1\pm0.5$, corresponding to $P_3=38\pm1$ periods and $P_2=17.9\pm0.4^\circ$. 
Drifting properties are consistent with the results of \citet{Song2023}.

\begin{figure}[htpb]
\centering
\includegraphics[width=0.44\textwidth, angle=0]{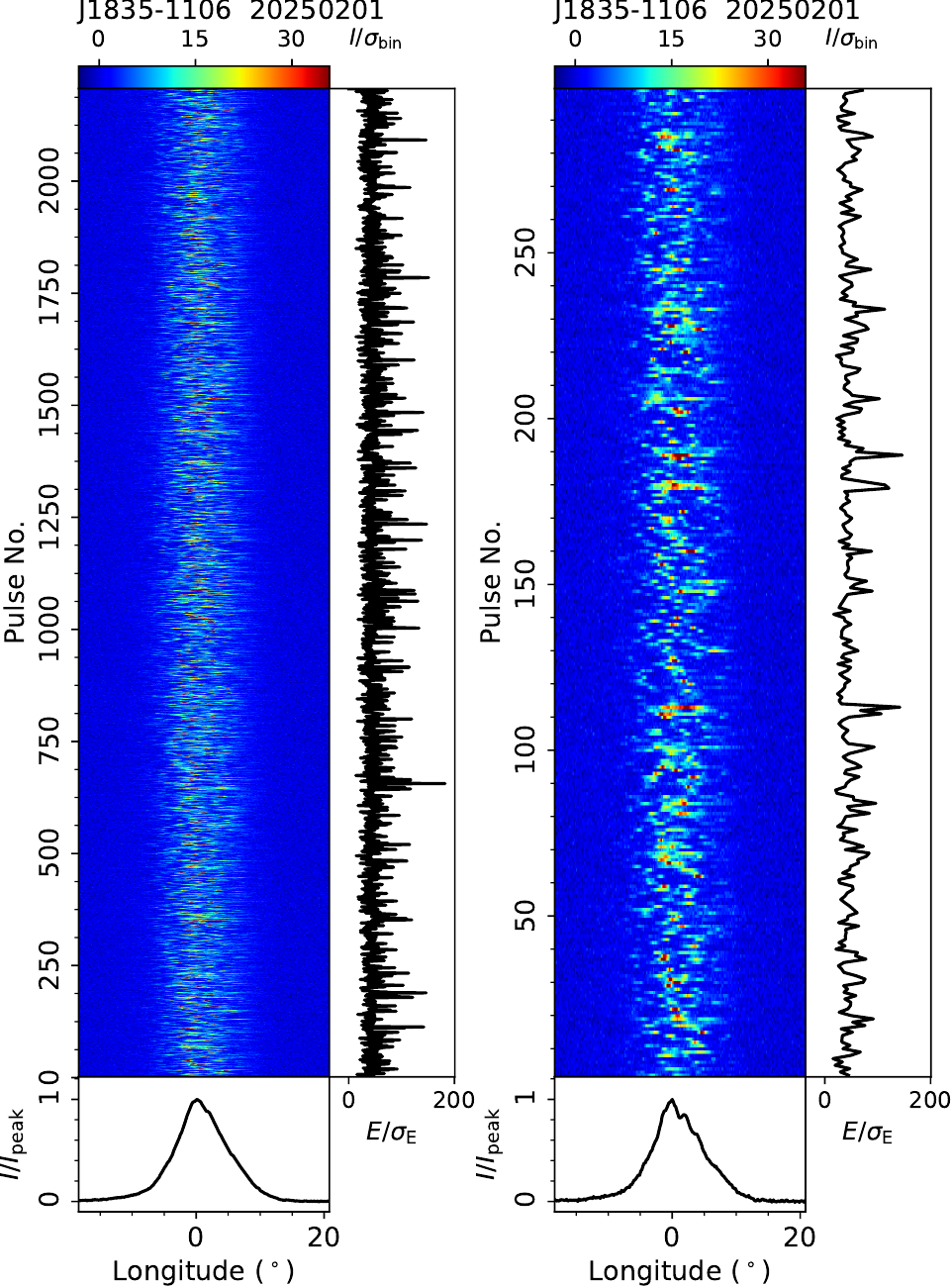}
\figcaption{Single pulse sequence of PSR J1835-1106 from the FAST observation on 20250201, and a zoomed-in view of pulses No. 1-300.
\label{subfig:TP:J1835-1106}}
\end{figure}

\begin{figure}[htpb]
\centering
\includegraphics[width=0.44\textwidth, angle=0]{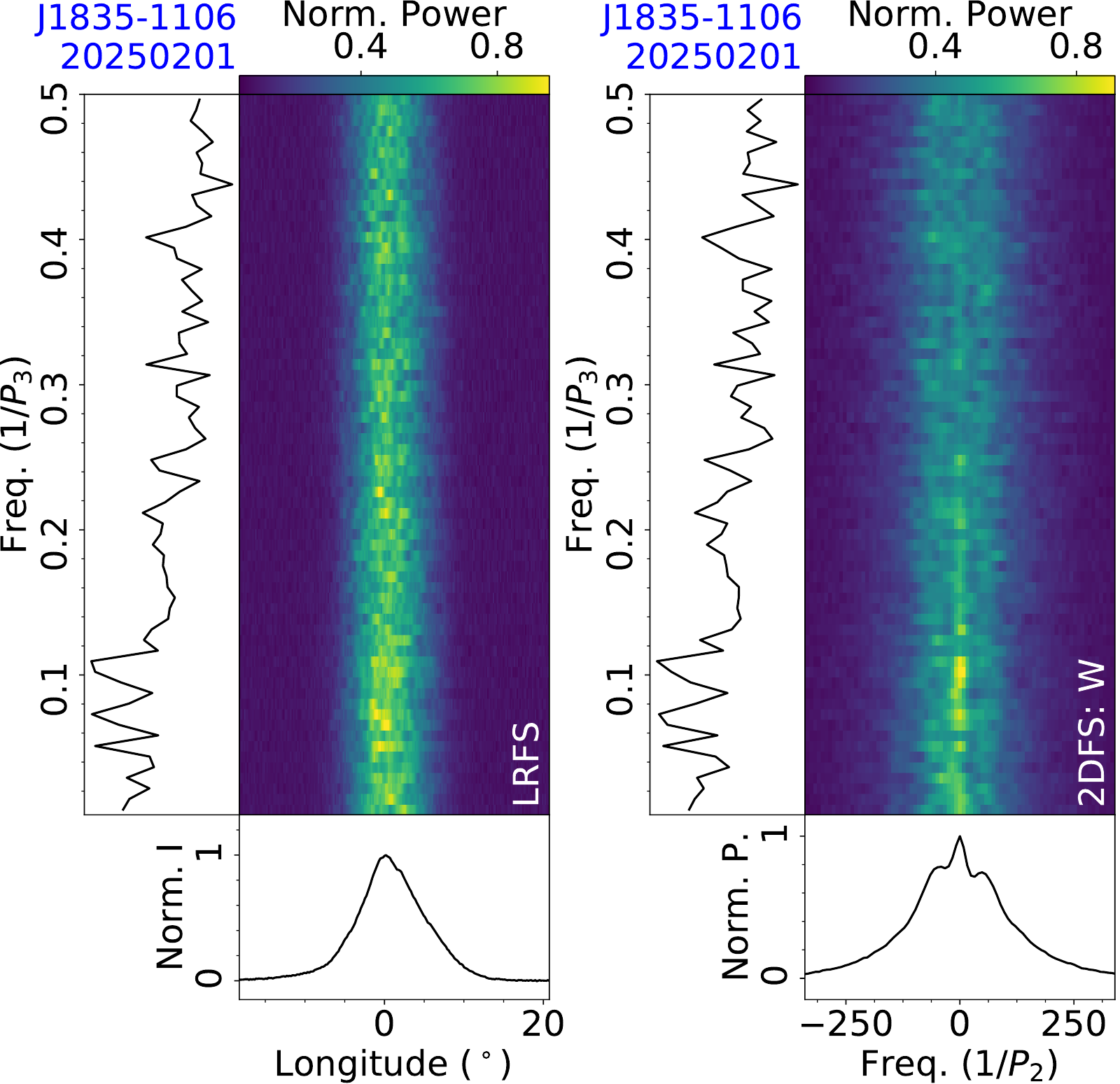}
\figcaption{Fluctuation analysis of PSR J1835-1106 from the FAST observation on 20250201, with LRFS and 2DFS for the on-pulse region of a mean pulse profile.
\label{subfig:fluctu:J1835-1106}}
\end{figure}

\subsection{J1835+00}
\label{subsec:J1835+00}

PSR J1835+00 was discovered in  the Pulsar Arecibo L-band Feed Array (PALFA) survey \citep{Parent2022}.

This pulsar was observed by FAST on 20220614 for 5 minutes, with a rotation period $P=0.7900$~s and a dispersion measure $D\!M=134.6~{\rm cm^{-3}\,pc}$ determined. Single pulse sequences in Fig.~\ref{subfig:TP:J1835+00} display the unsystematic subpulse drifting phenomenon. Fluctuation spectra are shown in Fig.~\ref{subfig:fluctu:J1835+00}, and both negative and positive drift features are present in the 2DFS. The centroid of the negative drift feature is at $1/P_3=0.060\pm0.002$ and $1/P_2=-13\pm2$, corresponding to periodicities of $P_3=16.7\pm0.5$ periods and $P_2=-28\pm4$ degrees. For the positive drift feature, the centroid frequencies are $1/P_3=0.164\pm0.002$ and $1/P_2=18\pm2$, yielding $P_3=6.1\pm0.1$ periods and $P_2=20\pm3$ degrees.

\subsection{J1835+0301g}
\label{subsec:J1835+0301g}

PSR J1835+0301g was discovered in the FAST GPPS survey \citep{Han2021,han2025}. 

This pulsar was observed by FAST on 20240418 for 15 minutes, deriving a rotation period $P=2.2987$~s and a dispersion measure $D\!M=206.6~{\rm cm^{-3}\,pc}$ from this observation. 
The single pulse sequence in Fig.~\ref{subfig:TP:J1835+0301g} shows the nulling phenomenon. After the removal of RFI, the nulling fraction of this observation is determined to be 49\% from the on-pulse integrated energy histogram (Fig.~\ref{subfig:Hist:J1835+0301g}).

\begin{figure}[htpb]
\centering
\includegraphics[width=0.22\textwidth, angle=0]{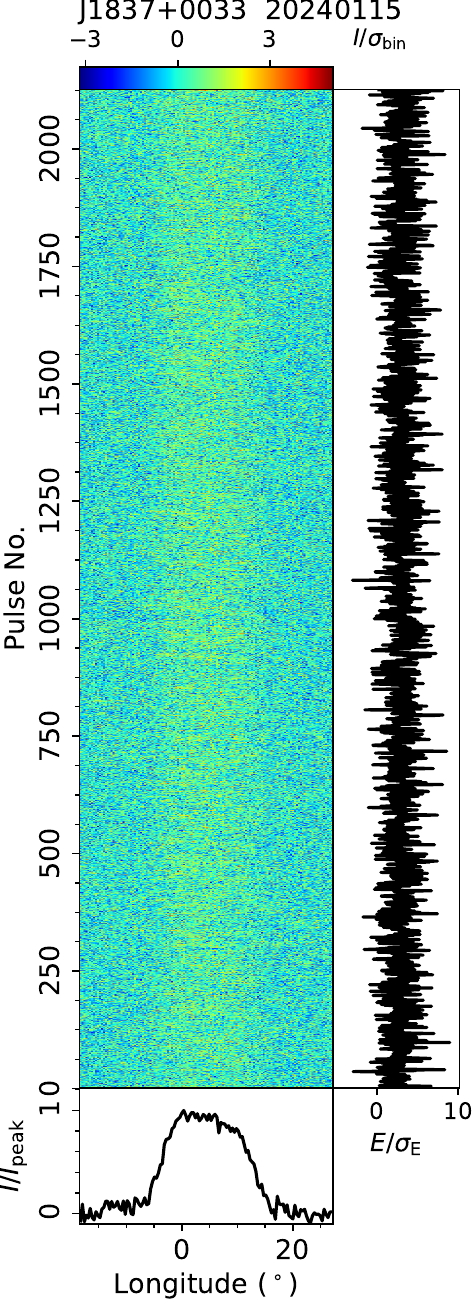}
\includegraphics[width=0.22\textwidth, angle=0]{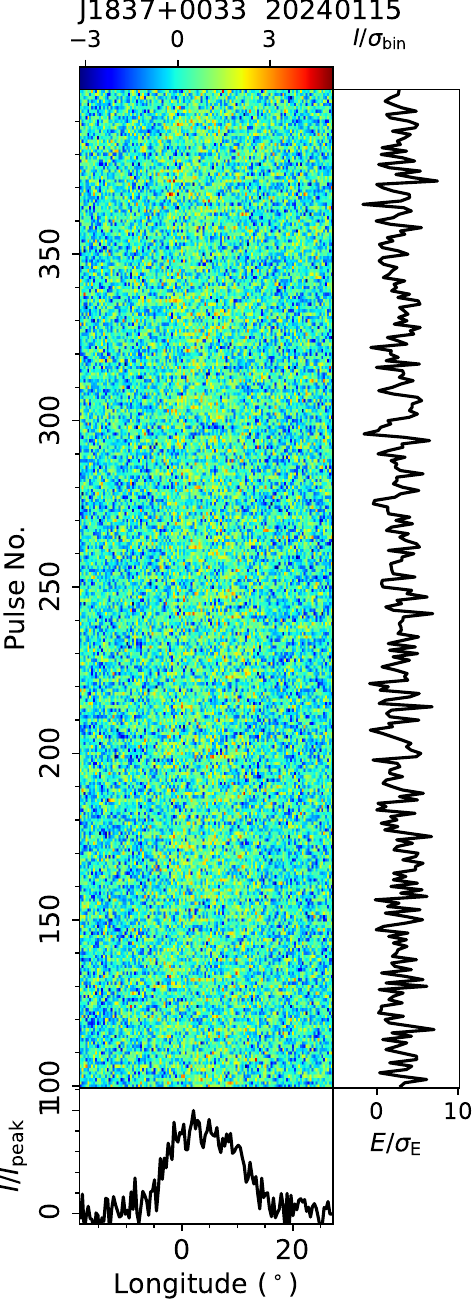}
\figcaption{Single pulse sequence of PSR J1837+0033 from the FAST observation on 20240115, and a zoomed-in view of pulses No. 100-400.
\label{subfig:TP:J1837+0033}}
\end{figure}

\begin{figure}[htpb]
\centering
\includegraphics[width=0.22\textwidth, angle=0]{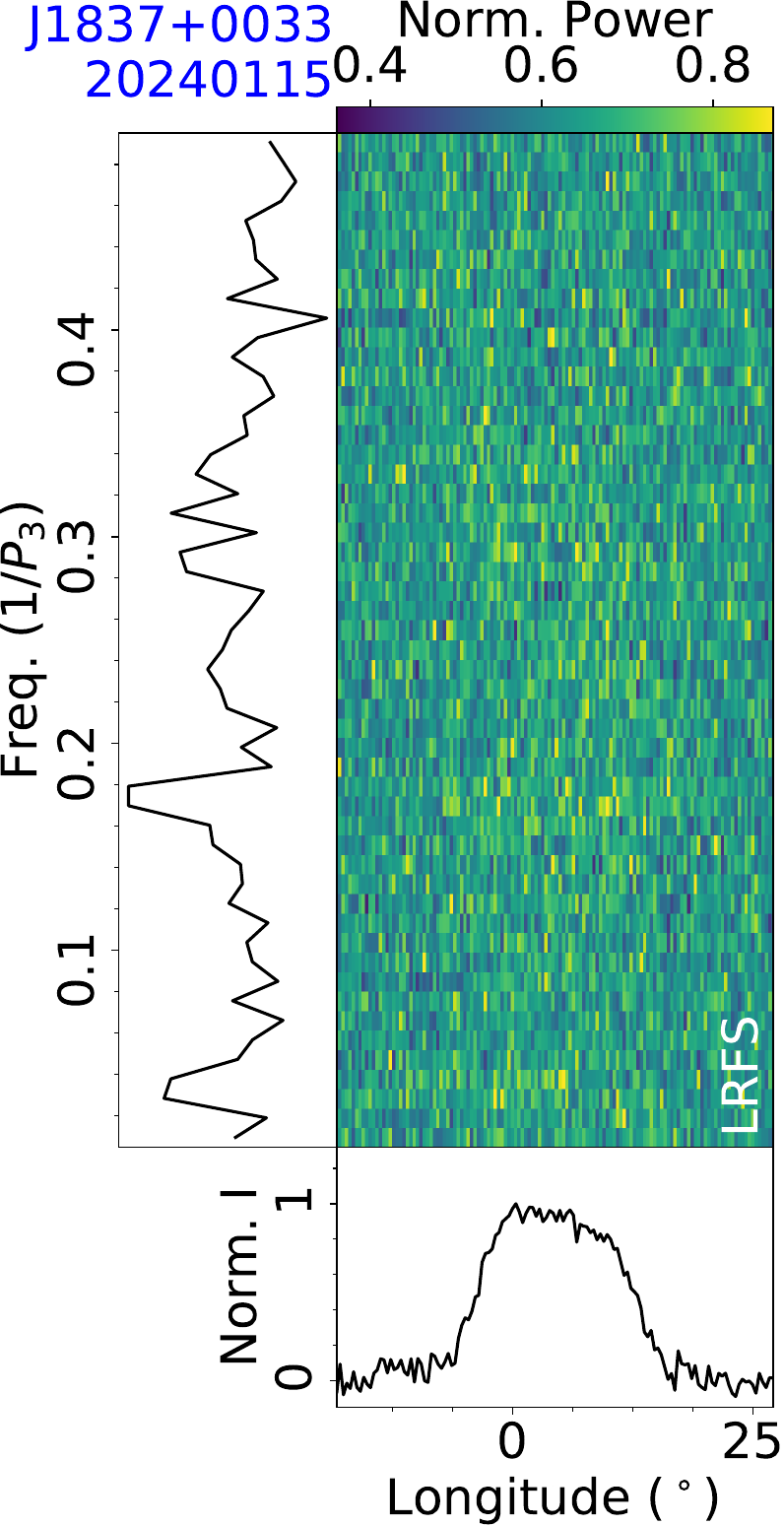}
\includegraphics[width=0.22\textwidth, angle=0]{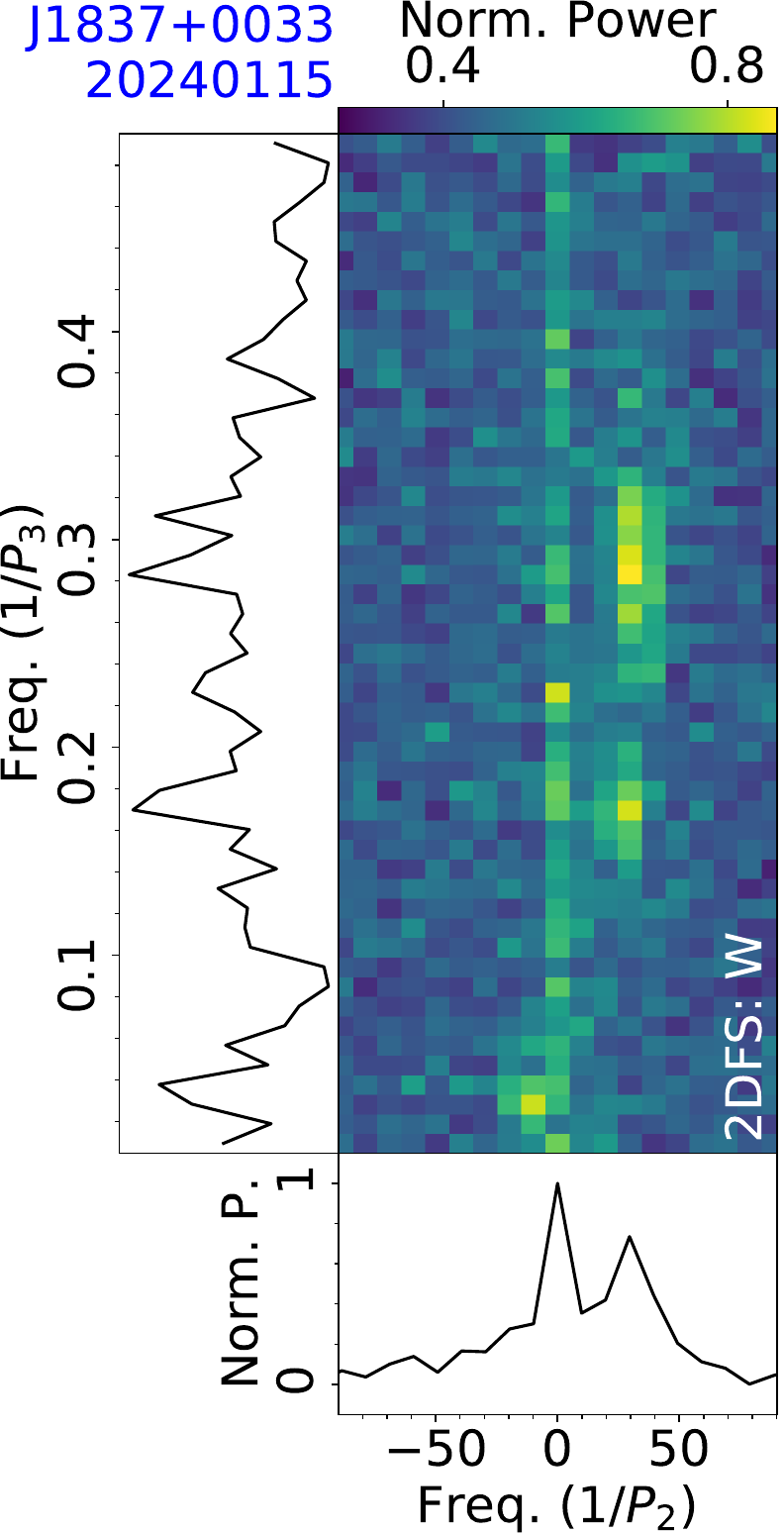}
\figcaption{Fluctuation analysis of PSR J1837+0033 from the FAST observation on 20240115, with LRFS and 2DFS for the on-pulse region of a mean pulse profile.
\label{subfig:fluctu:J1837+0033}}
\end{figure}

\subsection{J1835-0112g}
\label{subsec:J1835-0112g}

PSR J1835-0112g was discovered in the FAST GPPS survey \citep{Han2021,han2025}. 

This pulsar was observed by FAST on 20211009 for 15 minutes, deriving a rotation period $P=0.5996$~s and a dispersion measure $D\!M=316.7~{\rm cm^{-3}\,pc}$ from this observation. 
Single pulse sequence and a zoomed-in view of pulses No. 900-1200 in Fig.~\ref{subfig:TP:J1835-0112} display positive drift bands. The fluctuation spectra are shown in Fig.~\ref{subfig:fluctu:J1835-0112}, and the 2DFS reveals two modulation features. For the positive drift feature in the 2DFS, the centroid is at $1/P_3=0.258\pm0.004$ and $1/P_2=24\pm2$, corresponding to periodicities of $P_3=3.9\pm0.1$ periods and $P_2=15\pm1$ degrees. 
In addition, there is also a low-frequency modulation on the drifting behavior, and the centroid frequency of the modulation feature in 2DFS is $1/P_3=0.046\pm0.002$ ($P_3=22\pm1$ periods).

\begin{figure}[htpb]
\centering
\includegraphics[width=0.22\textwidth, angle=0]{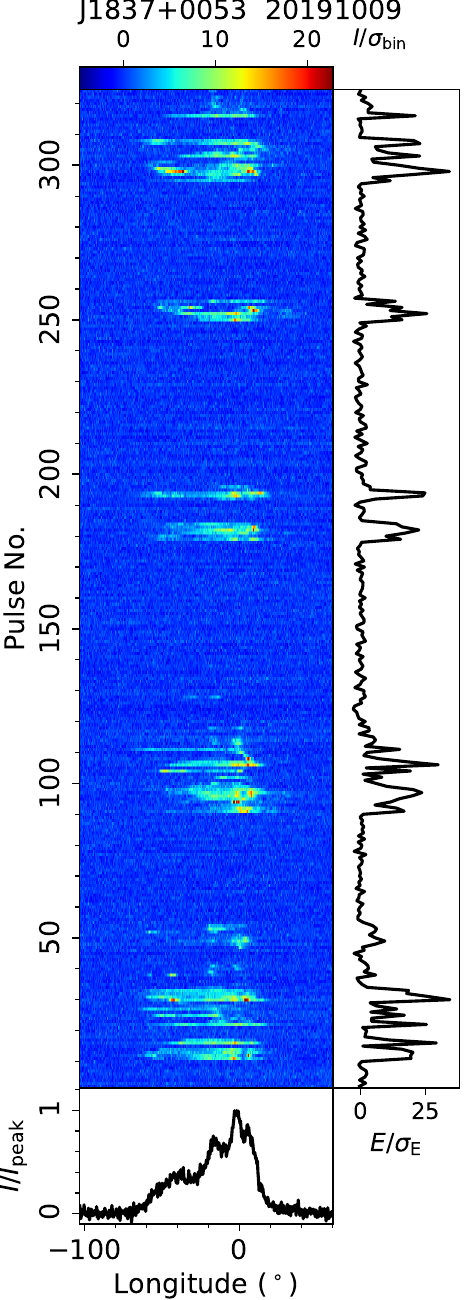}
\includegraphics[width=0.22\textwidth, angle=0]{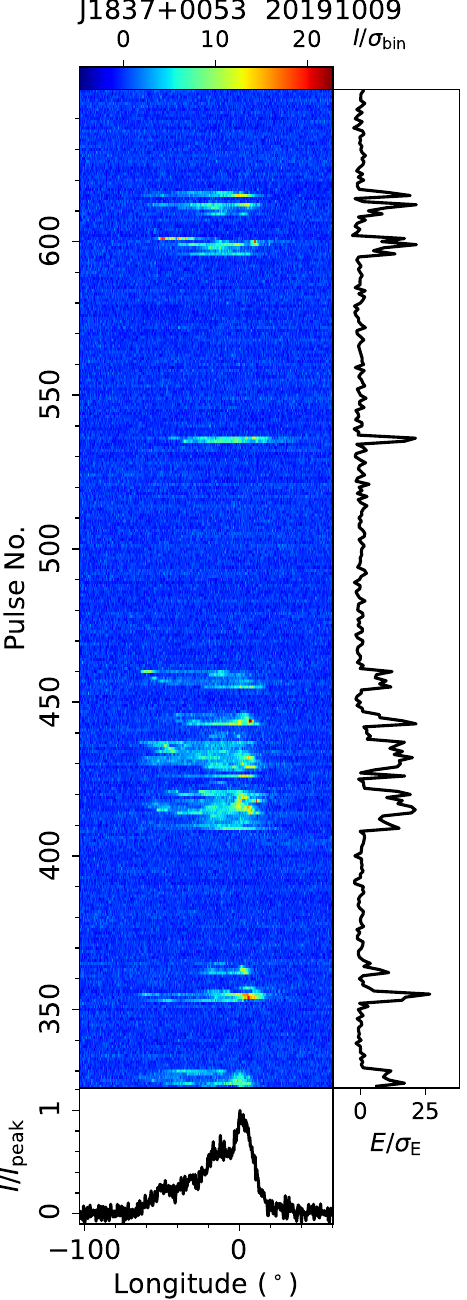}
\figcaption{Single pulse sequences of PSR J1837+0053 from the FAST observation on 20191009.
\label{subfig:TP:J1837+0053}}
\end{figure}

\begin{figure}[htpb]
\centering
\includegraphics[width=0.39\textwidth, angle=0]{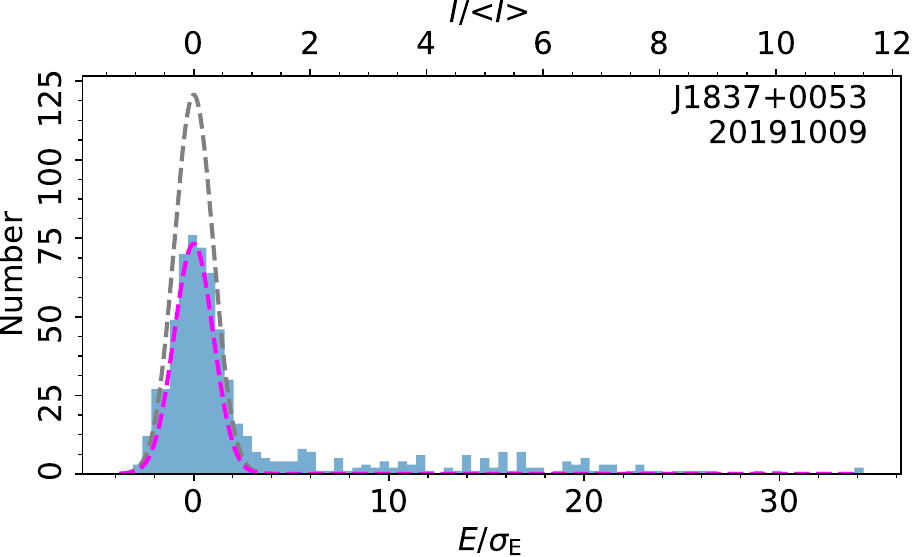}
\figcaption{On-pulse energy histogram of single pulses of PSR J1837+0053 from the FAST observation on 20191009. \label{subfig:Hist:J1837+0053}}
\end{figure}

\subsection{J1835-0149g}
\label{subsec:J1835-0149g}

PSR J1835-0149g was discovered in the FAST GPPS survey \citep{Han2021,han2025}. 

This pulsar was observed by FAST on 20220407 for 15 minutes, deriving a rotation period $P=1.2775$~s and a dispersion measure $D\!M=104.2~{\rm cm^{-3}\,pc}$ from this observation. 
Single pulse sequences and fluctuation spectra of the observation are displayed in Fig.~\ref{subfig:TP:J1835-0149g} and \ref{subfig:fluctu:J1835-0149g}. The pulsar has subpulse drifting and low-frequency modulation behaviors. Leading and trailing longitude parts both have a $P_3$ of 6$\pm$1 periods, with $P_2$ of 3.1$\pm$0.9 and 9$\pm$6 degrees. Additionally, there are also low-frequency modulations for two parts, with periods of 89$\pm$36 and 50$\pm$17 periods.

\begin{figure}[htpb]
\centering
\includegraphics[width=0.22\textwidth, angle=0]{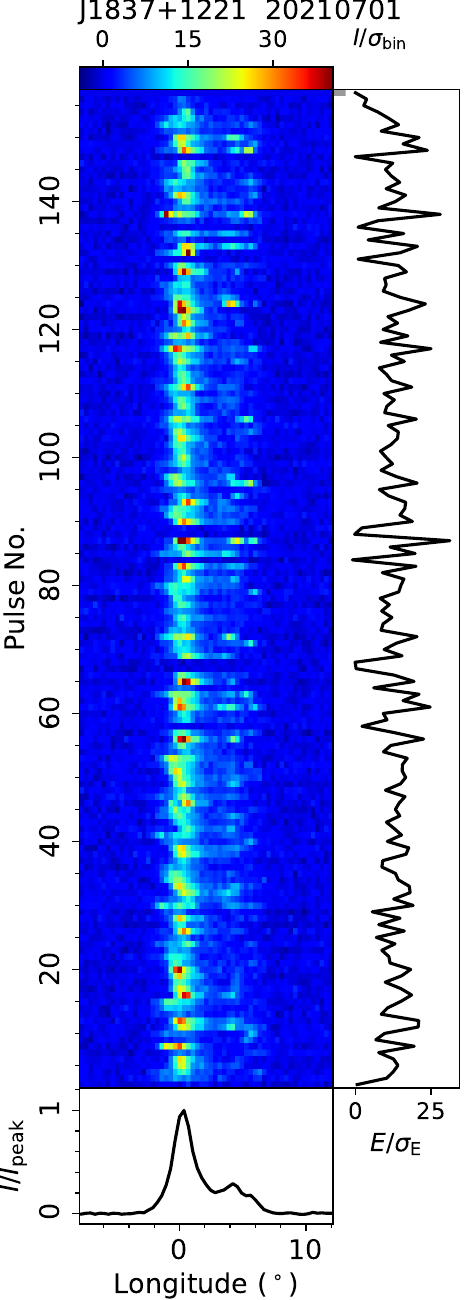}
\figcaption{Single pulse sequence of PSR J1837+1221 from the FAST observation on 20210701.
\label{subfig:TP:J1837+1221}}
\end{figure}

\begin{figure}[htpb]
\centering
\includegraphics[width=0.39\textwidth, angle=0]{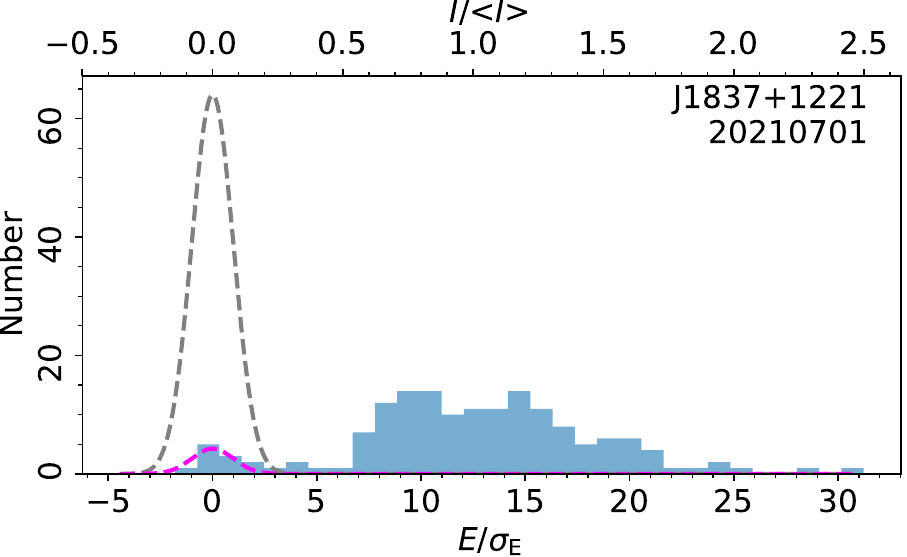}
\figcaption{On-pulse energy histogram of single pulses of PSR J1837+1221 from the FAST observation on 20210701.
\label{subfig:Hist:J1837+1221}}
\end{figure}

\begin{figure}[htpb]
\centering
\includegraphics[width=0.22\textwidth, angle=0]{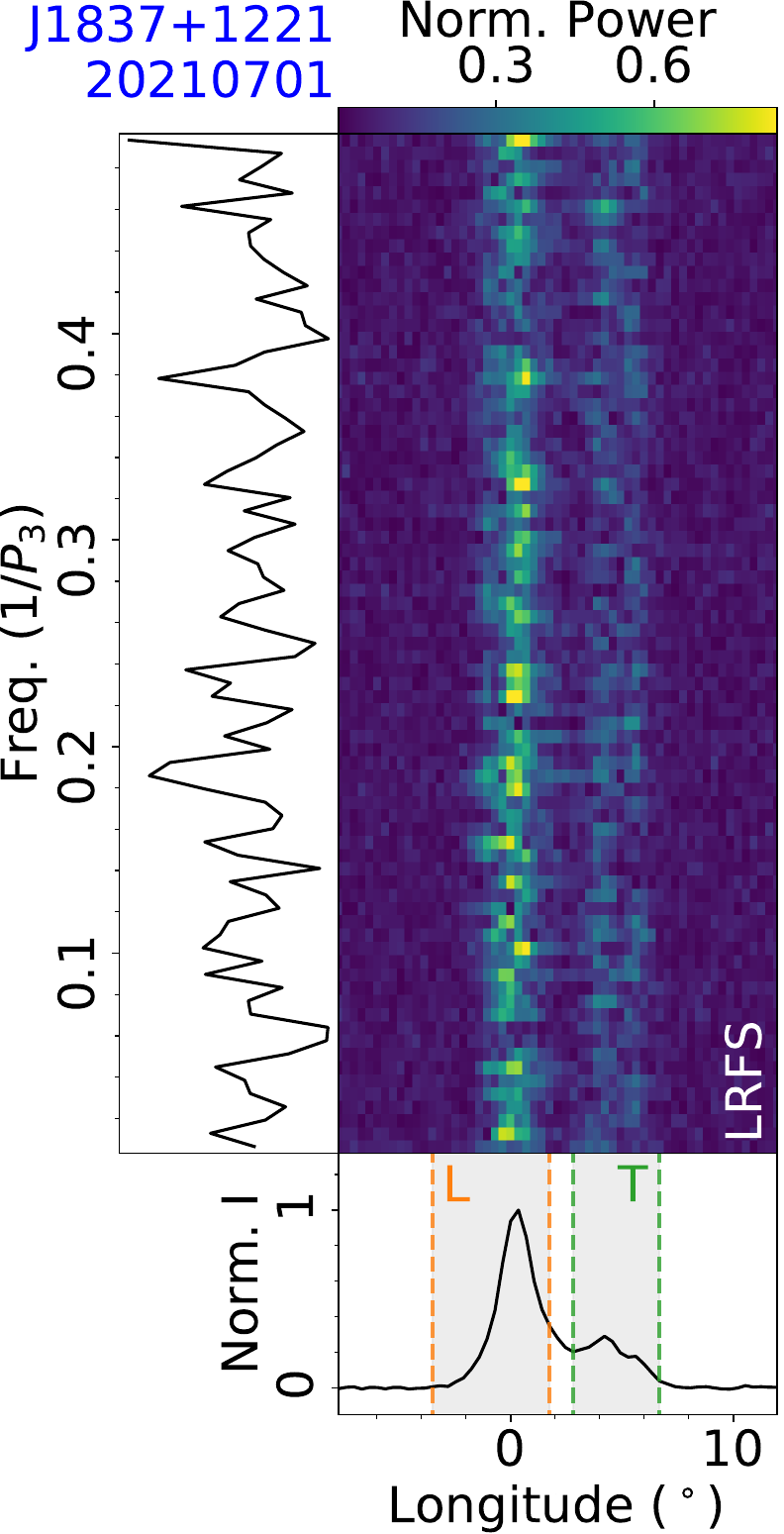}
\includegraphics[width=0.22\textwidth, angle=0]{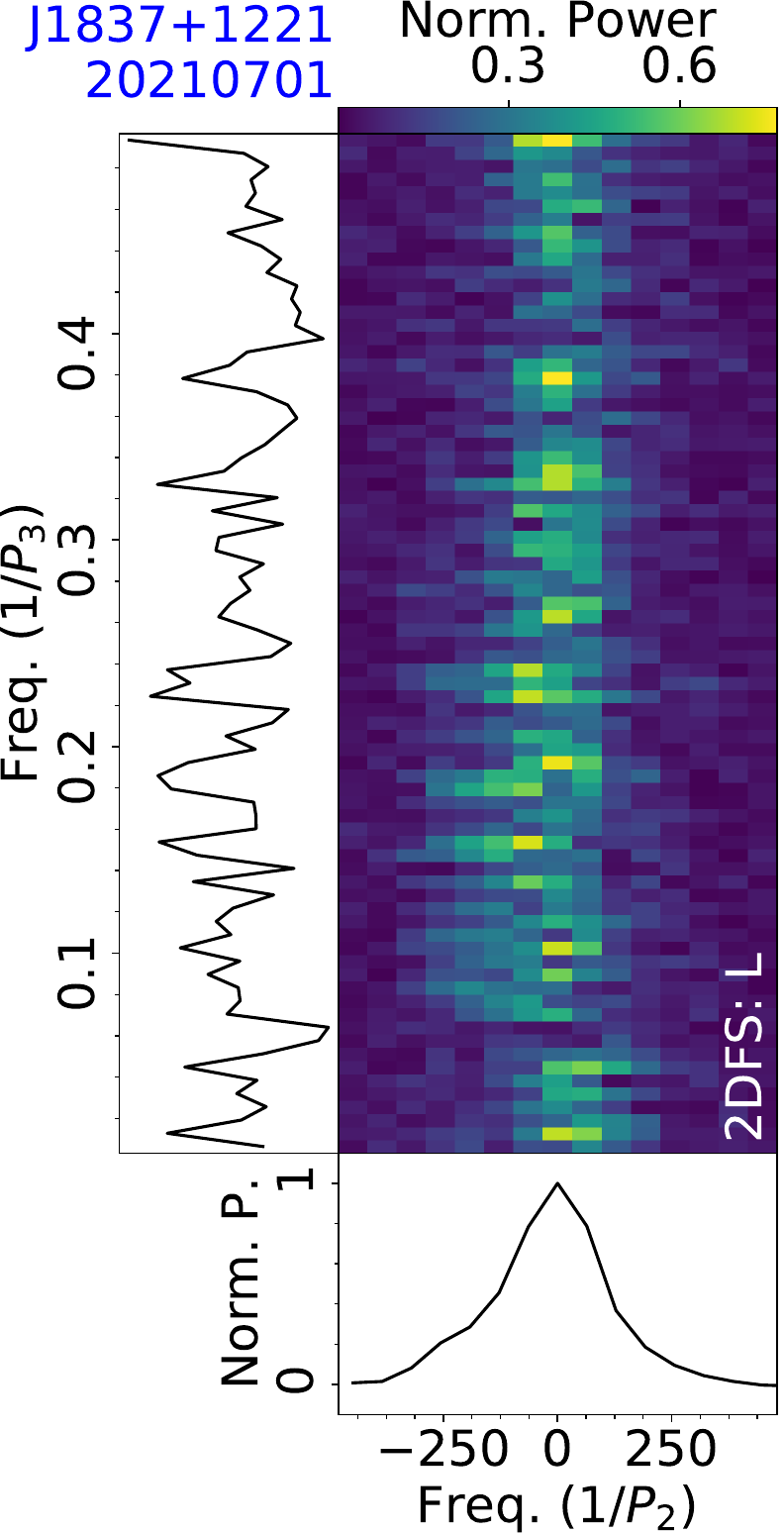}
\figcaption{Fluctuation analysis of PSR J1837+1221 from the FAST observation on 20210701, with LRFS and 2DFS for the on-pulse region of a mean pulse profile. \label{subfig:fluctu:J1837+1221}}
\end{figure}

\subsection{J1835-1106}
\label{subsec:J1835-1106}

PSR J1835-1106 was discovered in the Parkes Southern Pulsar Survey \citet{Manchester1996}. 
A $P_2$-only feature was reported by \citet{Song2023} with $P_2=5.4^{+0.3}_{-0.7}$ degrees. 

This pulsar was observed by FAST on 20250201 for 6 minutes, deriving a rotation period $P=0.1659$~s and a dispersion measure $D\!M=132.8~{\rm cm^{-3}\,pc}$. The single pulse sequence and a zoomed-in view of pulses No. 1-300 are shown in Fig.~\ref{subfig:TP:J1835-1106}. 
The fluctuation spectra (Fig.~\ref{subfig:fluctu:J1835-1106}) reveal both negative and positive drift features with constant $P_2$ that are widely distributed over $1/P_3$ from 0 to 0.5. 
The centroids of these two features in $1/P_2$ are at $-51\pm1$ ($P_2=-7.1\pm0.1$ degrees) and $1/P_2=53\pm1$ ($P_2=6.9\pm0.1$ degrees), respectively. 
These indicate variable subpulse drifting rates but a relatively stable subpulse phase interval. 
In addition, there is a low-frequency modulation feature in 2DFS with the centroid frequency of $1/P_3=0.077\pm0.001$, corresponding to $P_3=13.0\pm0.2$ periods.

\begin{figure}[htpb]
\centering
\includegraphics[width=0.44\textwidth, angle=0]{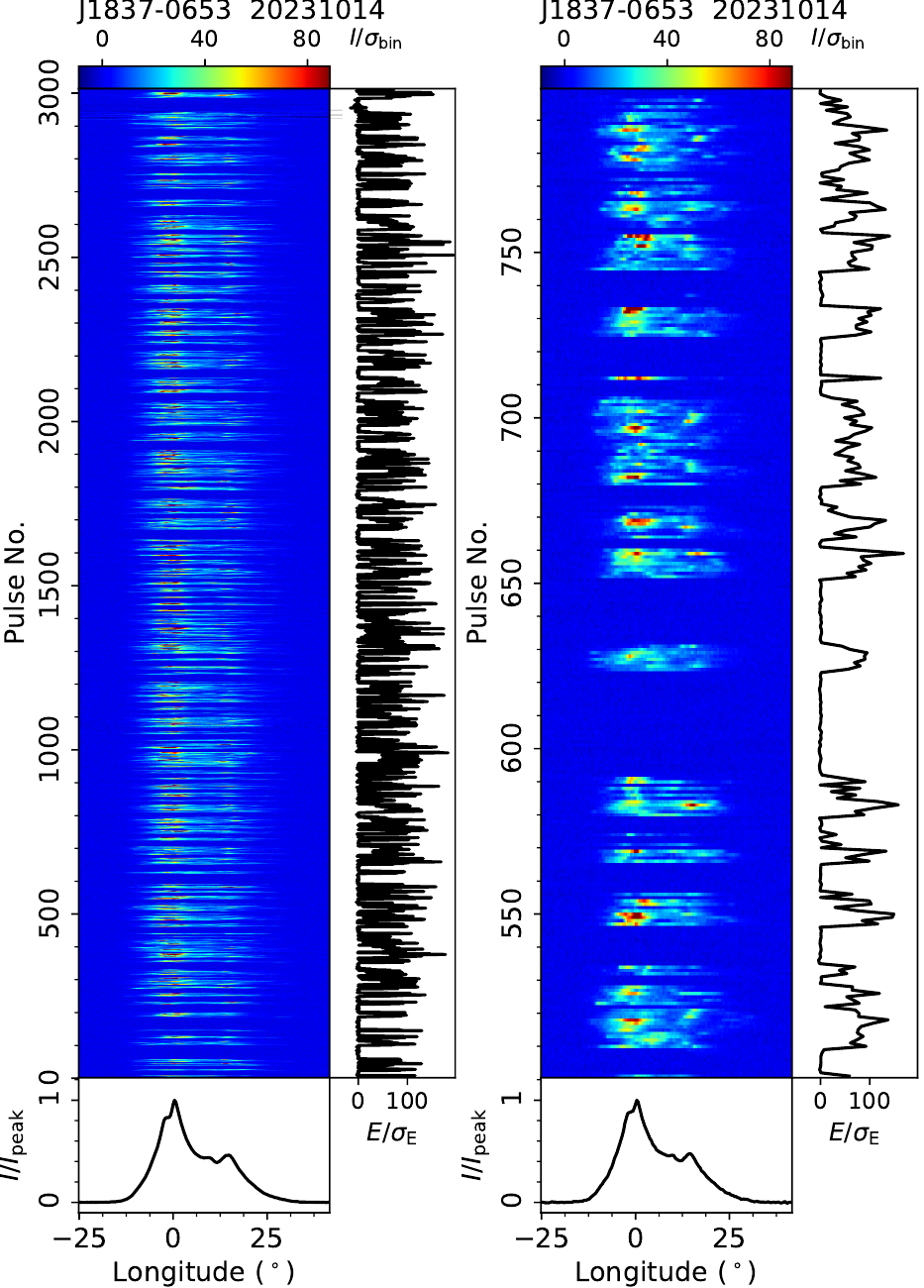}
\figcaption{Single pulse sequence of PSR J1837-0653 from the FAST observation on 20231014, and a zoomed-in view of pulses No. 500-800.
\label{subfig:TP:J1837-0653}}
\end{figure}

\begin{figure}[htpb]
\centering
\includegraphics[width=0.39\textwidth, angle=0]{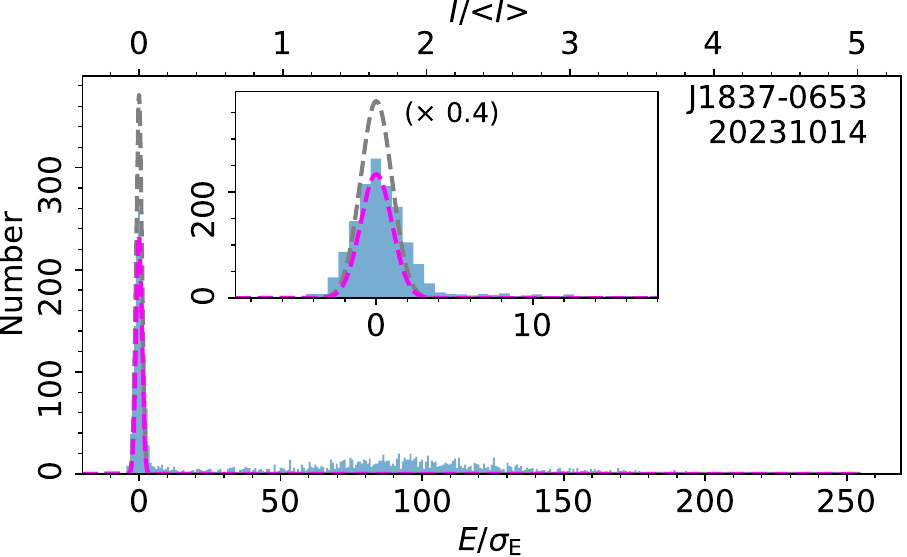}
\figcaption{On-pulse energy histogram of single pulses of PSR J1837-0653 from the FAST observation on 20231014. The inset provides a view of the x‑axis region from -10 to 18.
\label{subfig:Hist:J1837-0653}}
\end{figure}

\subsection{J1837+0033}
\label{subsec:J1837+0033}

PSR J1837+0033 was discovered in the FAST GPPS survey \citep{Han2021,han2025}.

This pulsar was observed by FAST on 20240115 and 20240329 each for 15 minutes. From the observation on 20240115, a rotation period $P=0.4181$~s and a dispersion measure $D\!M=187.7~{\rm cm^{-3}\,pc}$ were derived. The single pulse sequence and a zoomed-in view of pulses No. 100-400 are displayed in Fig.~\ref{subfig:TP:J1837+0033}, and the fluctuation spectra are shown in Fig.~\ref{subfig:fluctu:J1837+0033}. 
Three drift features are identified in the 2DFS. 
The low-frequency negative drift feature has a centroid of $1/P_3=0.030\pm0.001$ ($P_3=33\pm1$ periods) and $1/P_2=-12\pm1$ ($P_2=-31\pm3$ degrees). Two positive features are also present, with centroids at $1/P_3=0.171\pm0.002$ ($P_3=5.8\pm0.1$ periods), $1/P_2=26\pm1$ ($P_2=14\pm1$ degrees), and at $1/P_3=0.284\pm0.002$ ($P_3=3.52\pm0.02$ periods), $1/P_2=34\pm1$ ($P_2=10.7\pm0.2$ degrees). 
The drifting properties derived from the observation on 20240329 are consistent with those obtained on 20240115.

\subsection{J1837+0053}
\label{subsec:J1837+0053}

PSR J1837+0053 was found in the Parkes 20-cm multibeam pulsar survey of the Galactic plane \citep{Lorimer2006}. 

The pulsar was observed by FAST on 20191009 for 5 minutes, deriving a rotation period $P=0.4735$~s and a dispersion measure $D\!M=123.6~{\rm cm^{-3}\,pc}$. 
Single pulse sequences are shown in Fig.~\ref{subfig:TP:J1837+0053}, illustrating the existence of nulls. From the on-pulse integrated energy histogram in Fig.~\ref{subfig:Hist:J1837+0053}, the nulling fraction of this observation is estimated to be 61$\pm$4\%. There are also some weak emissions such as pulses around Nos. 40 and 320, which result in distributions of nulling and emission not well separated in the histogram.

\begin{figure}[htpb]
\centering
\includegraphics[width=0.22\textwidth, angle=0]{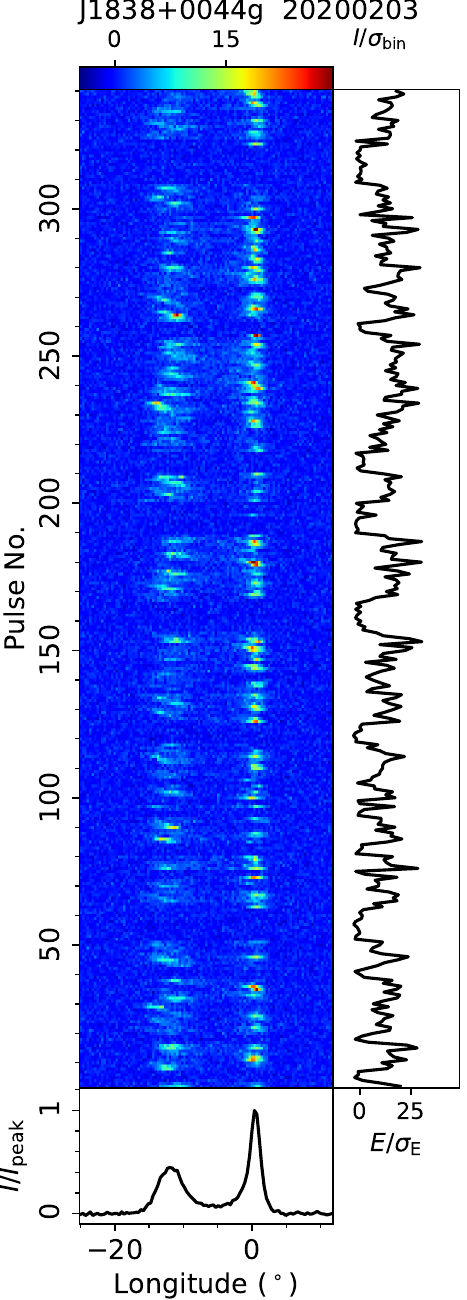}
\includegraphics[width=0.22\textwidth, angle=0]{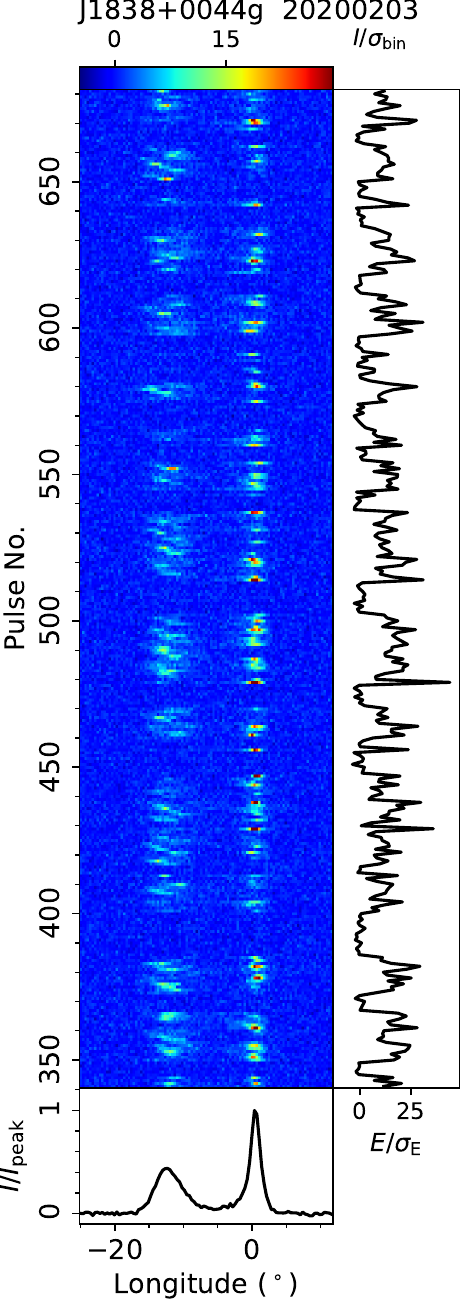}
\figcaption{Single pulse sequences of PSR J1838+0044g from the FAST observation on 20200203.
\label{subfig:TP:J1838+0044g}}
\end{figure}

\begin{figure}[htpb]
\centering
\includegraphics[width=0.39\textwidth, angle=0]{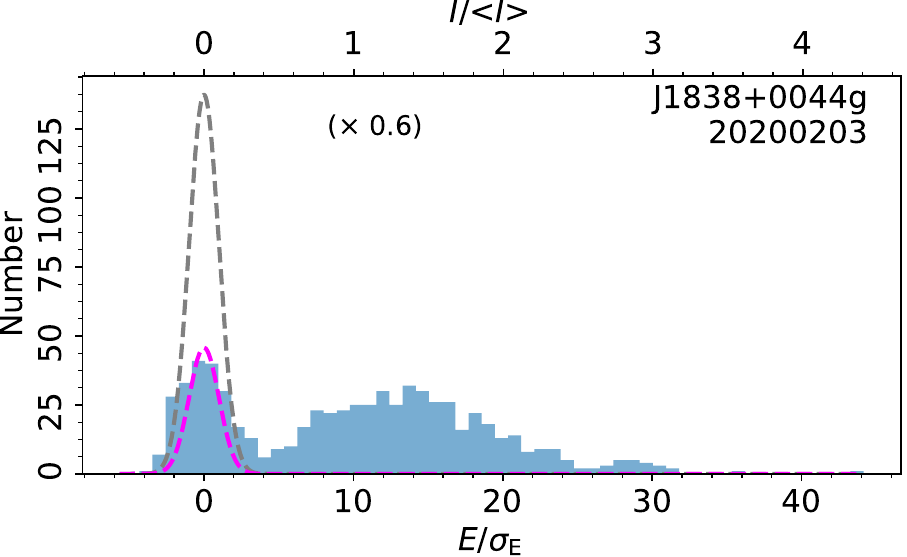}
\figcaption{On-pulse energy histogram of single pulses of PSR J1838+0044g from the FAST observation on 20200203.
\label{subfig:Hist:J1838+0044g}}
\end{figure}

\begin{figure}[htpb]
\centering
\includegraphics[width=0.21\textwidth, angle=0]{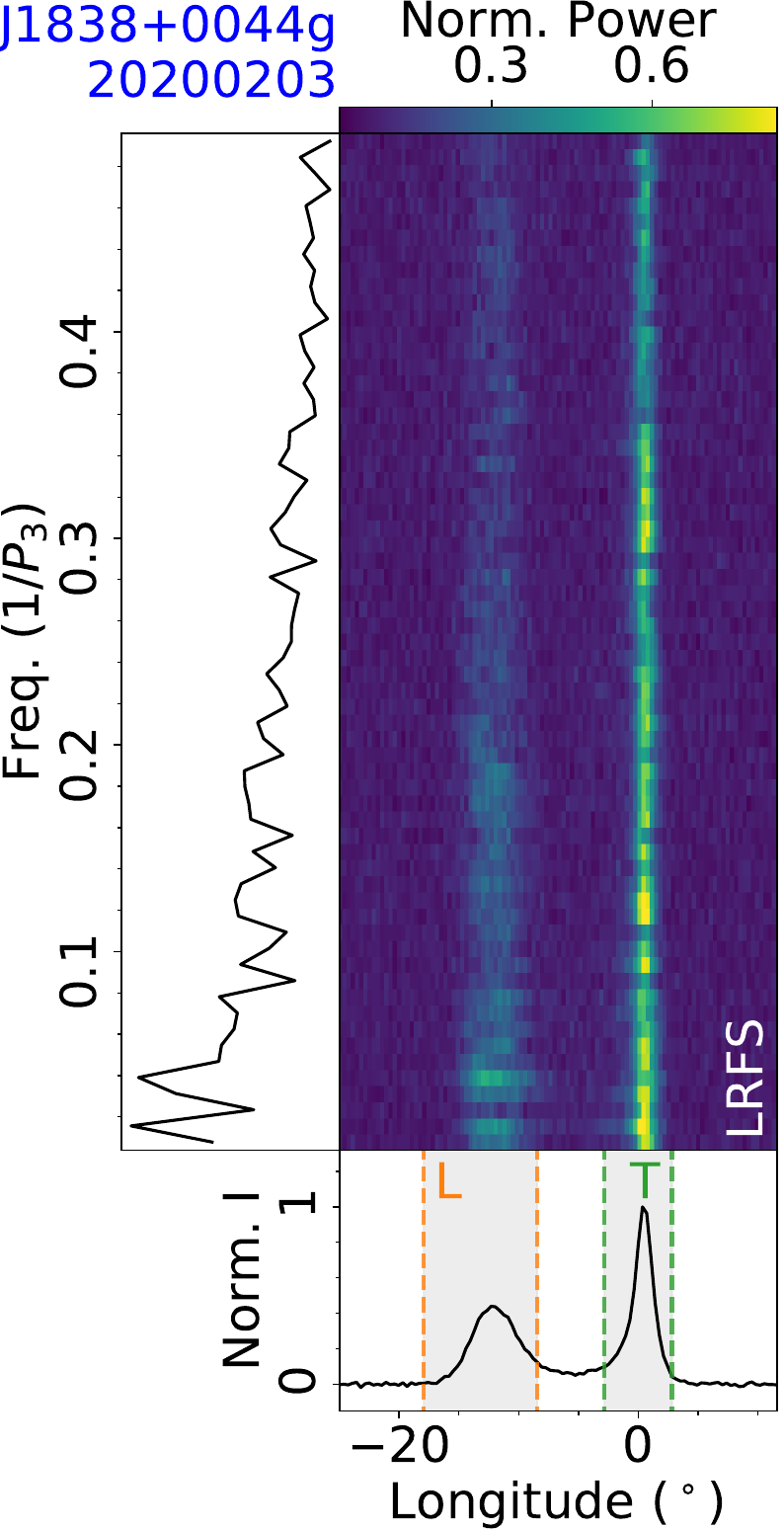}
\includegraphics[width=0.21\textwidth, angle=0]{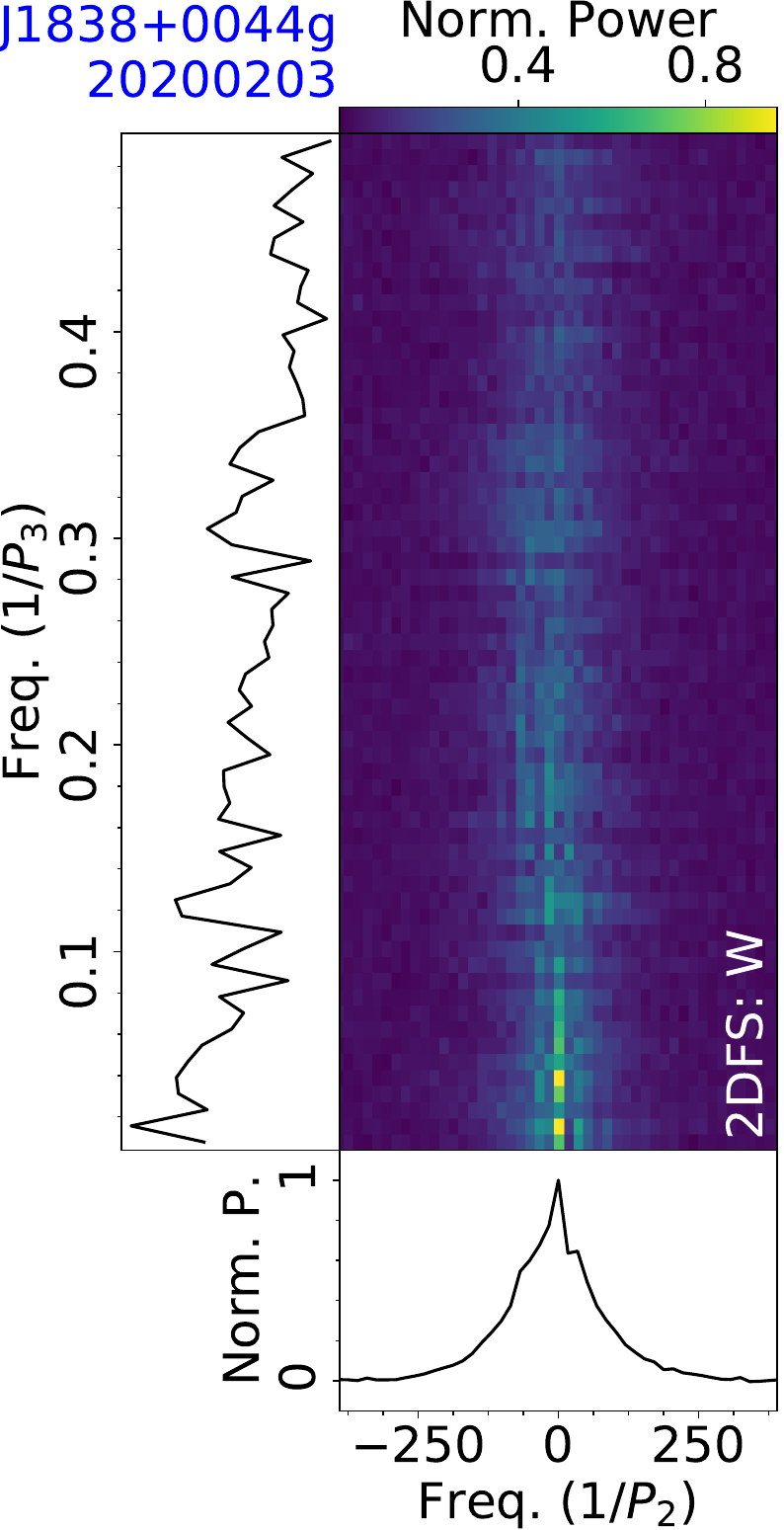}\\
\includegraphics[width=0.21\textwidth, angle=0]{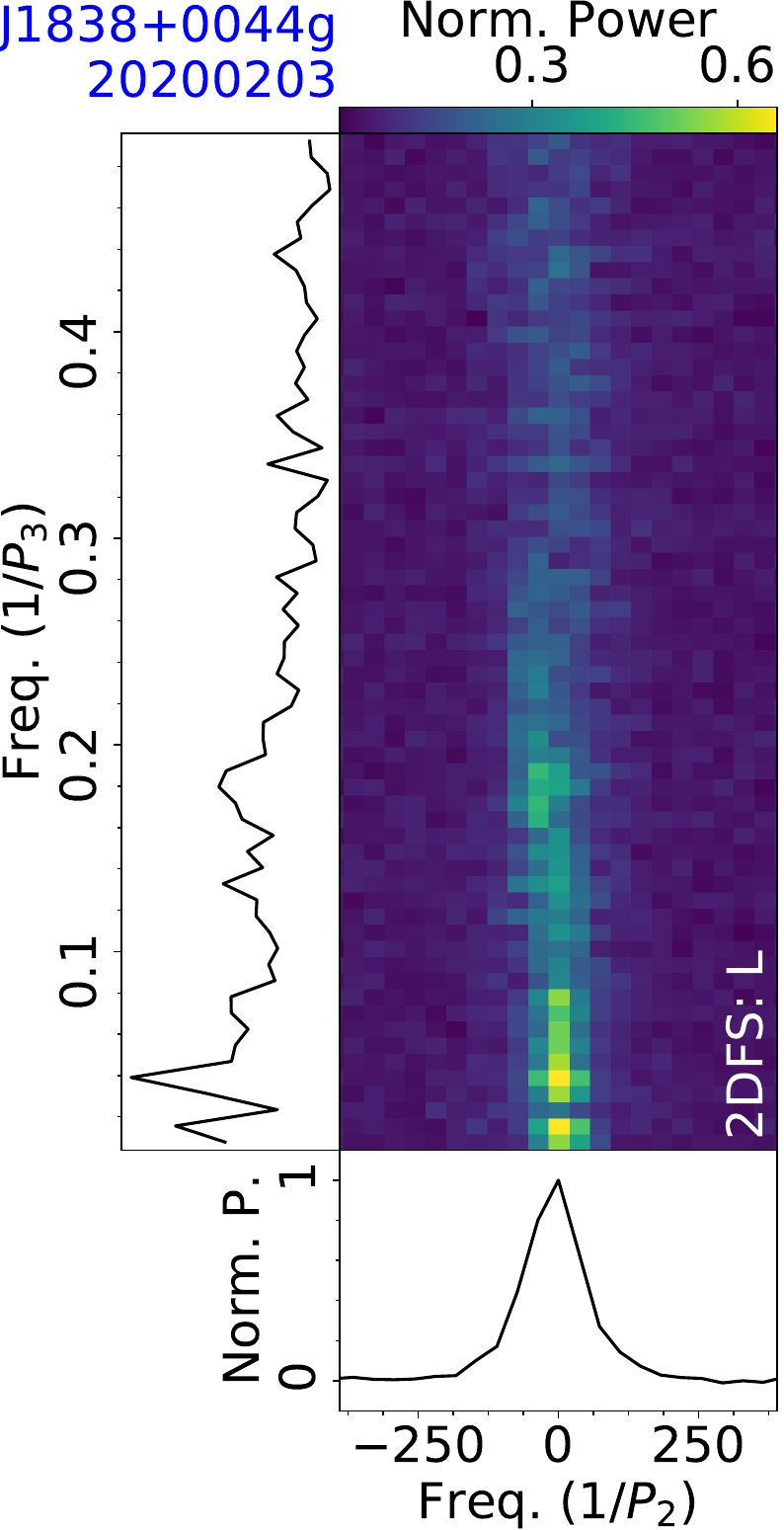}
\includegraphics[width=0.21\textwidth, angle=0]{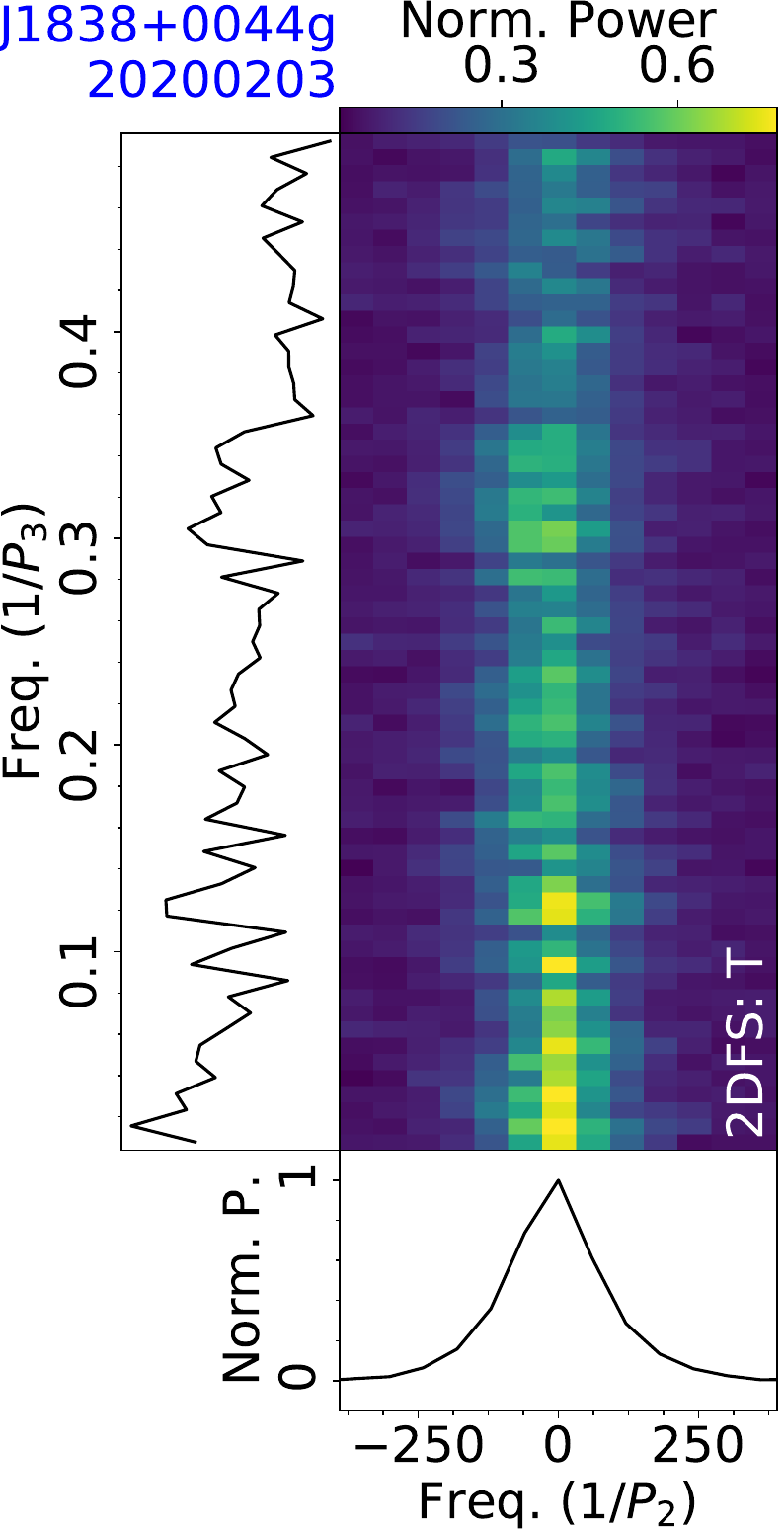}
\figcaption{Fluctuation analysis of PSR J1838+0044g for the observation on 20200203, with LRFS (top-left), and 2DFS for the on-pulse region (top-right), leading part (bottom-left) and trailing part (bottom-right) of a mean pulse profile. \label{subfig:fluctu:J1838+0044g}}
\end{figure}

\subsection{J1837+1221}
\label{subsec:J1837+1221}

PSR J1837+1221 was discovered in a survey of intermediate Galactic latitudes using the Parkes 64-m radio telescope \citep{Edwards2001}. 
The subpulse drifting of $P_3$=7$\pm$1 periods and $P_2$=-9$^{+6}_{-5}$ degrees was previously reported by \citet{Song2023}.

This pulsar was observed by FAST on 20210701 for 5 minutes, deriving a rotation period $P=1.9633$~s and a dispersion measure $D\!M=101.0~{\rm cm^{-3}\,pc}$ from this observation. 
The time-phase graph in Fig.~\ref{subfig:TP:J1837+1221} indicates the existence of nulls. The nulling fraction of this observation is estimated to be 6.7$\pm$0.6\% from the histogram of on-pulse integrated energy of single pulses in Fig.~\ref{subfig:Hist:J1837+1221}. 
Subpulse drifting of the leading phase part in a mean pulse profile is confirmed by this observation. From fluctuation spectra in Fig.~\ref{subfig:fluctu:J1837+1221}, there is a negative drift feature with the temporal modulation frequency widely distributed. 
The centroid frequencies of the negative drift feature are estimated to be $1/P_3=0.158\pm0.004$ and $1/P_2=-121\pm7$, corresponding to drifting parameters of $P_3=6.3\pm0.1$ periods and $P_2=-3.0\pm0.2^\circ$.

\begin{figure}[htpb]
\centering
\includegraphics[width=0.22\textwidth, angle=0]{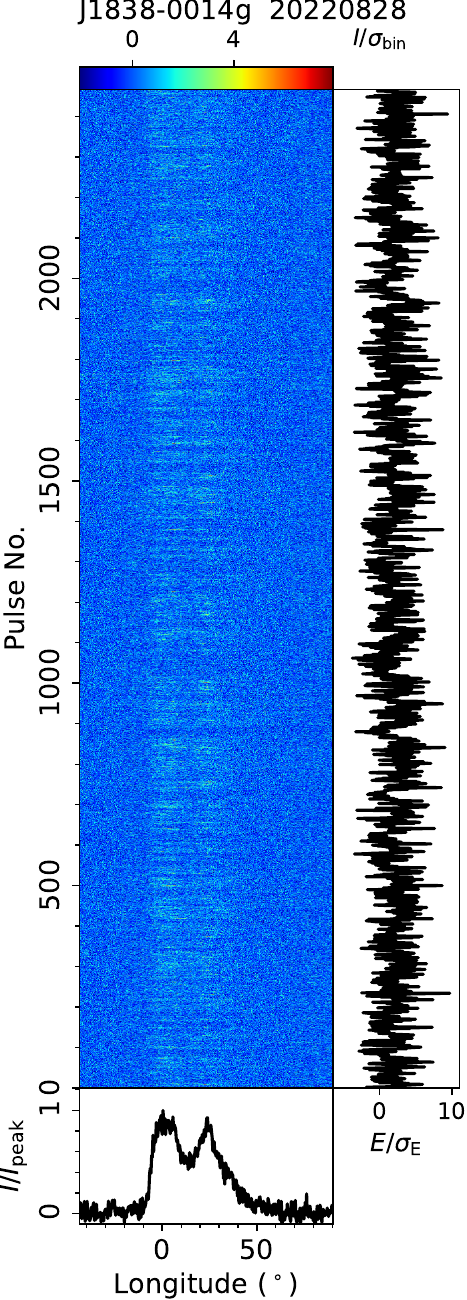}
\includegraphics[width=0.22\textwidth, angle=0]{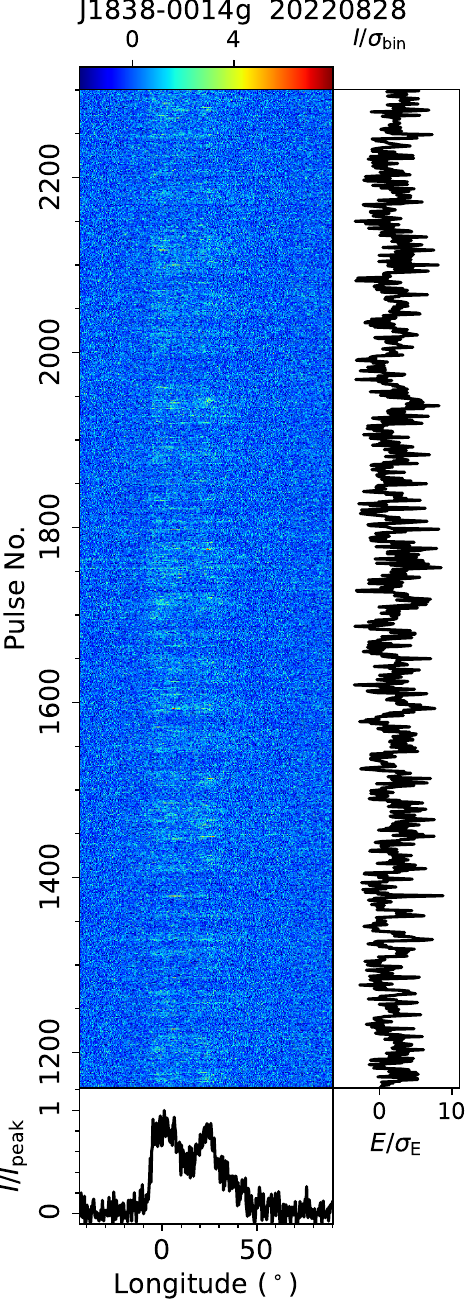}
\figcaption{Single pulse sequence of PSR J1838-0014g from the FAST observation on 20220828, and a zoomed-in view of pulses No. 1160-2300.
\label{subfig:TP:J1838-0014g}}
\end{figure}

\begin{figure}[htpb]
\centering
\includegraphics[width=0.22\textwidth, angle=0]{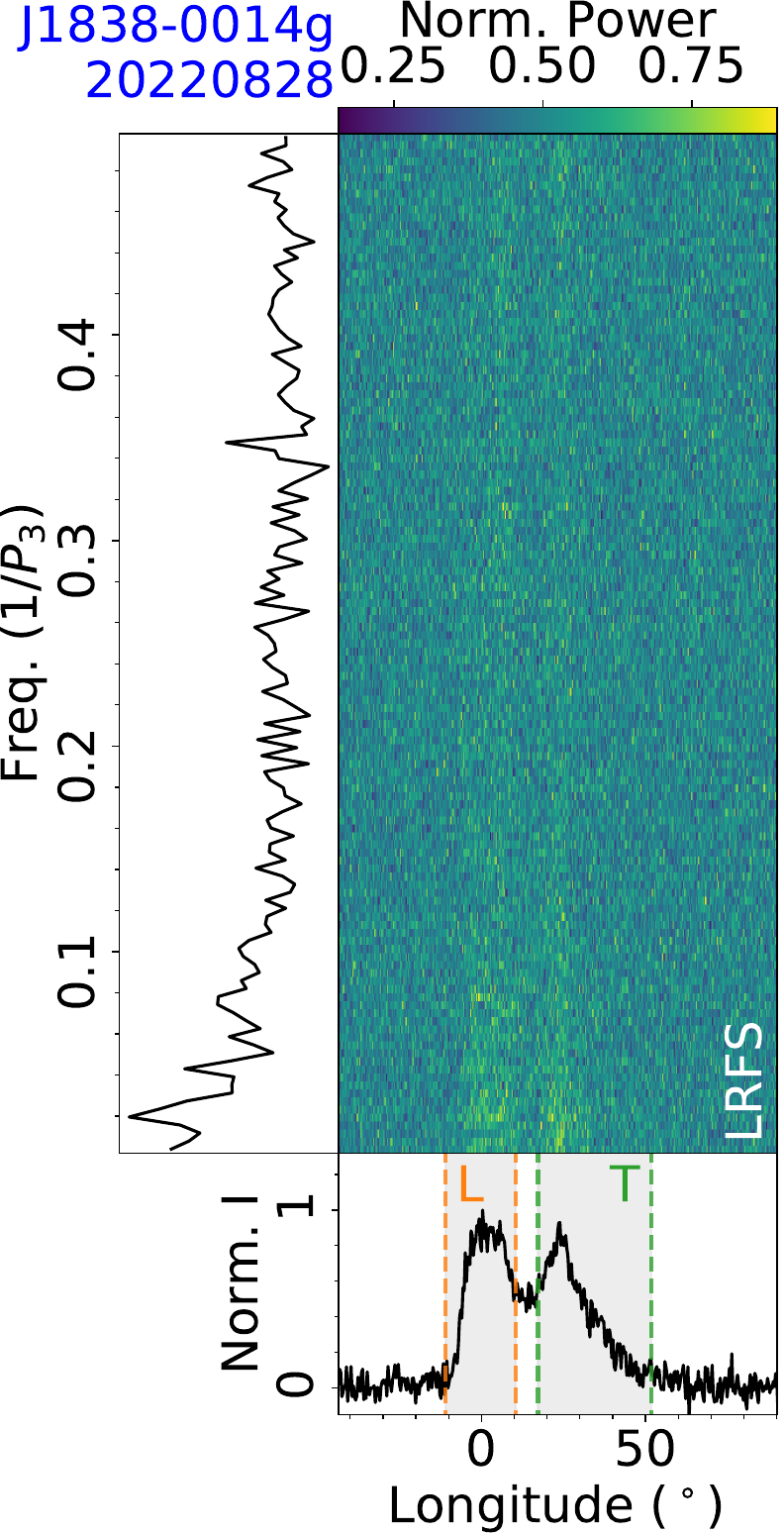}
\includegraphics[width=0.22\textwidth, angle=0]{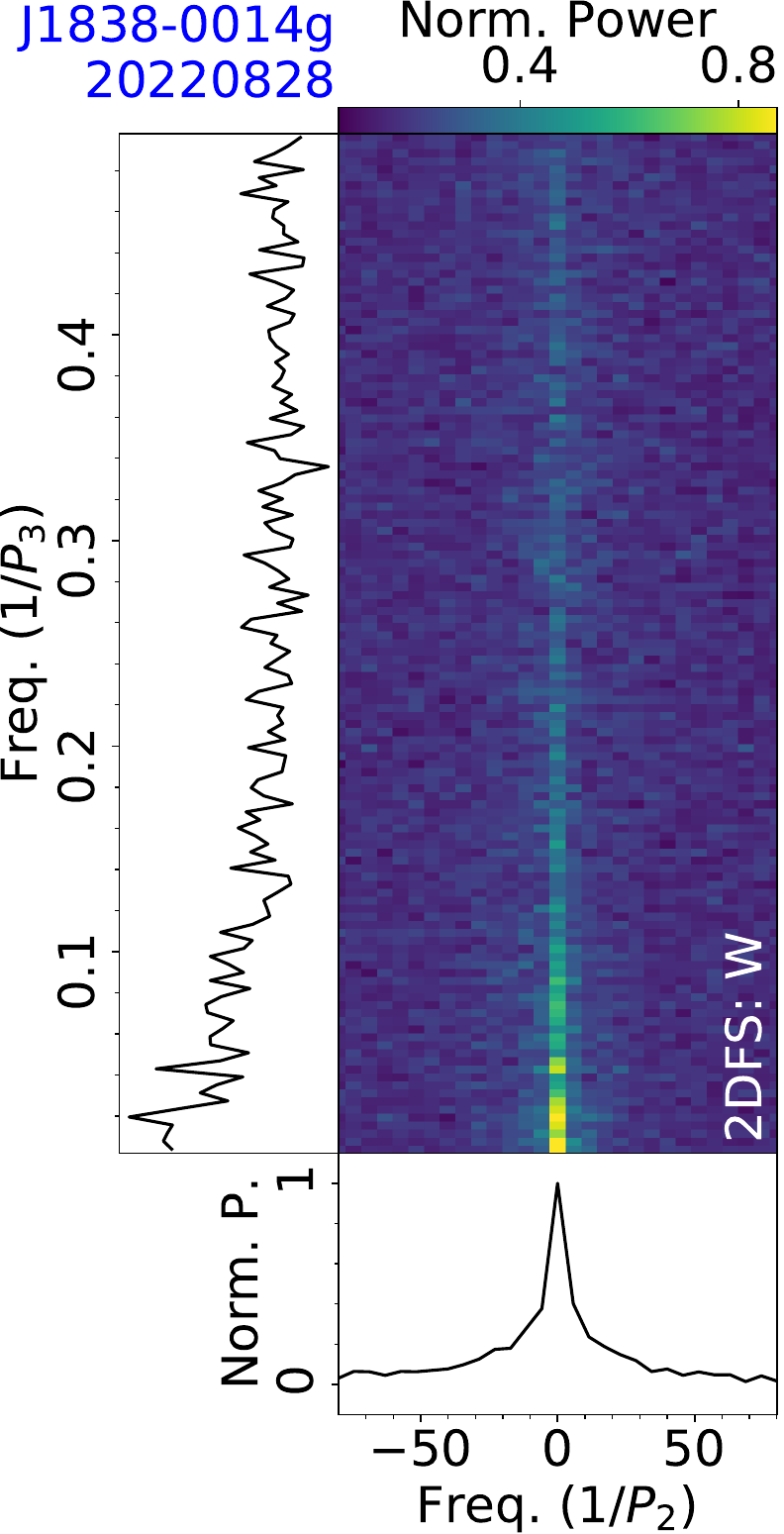}\\
\includegraphics[width=0.22\textwidth, angle=0]{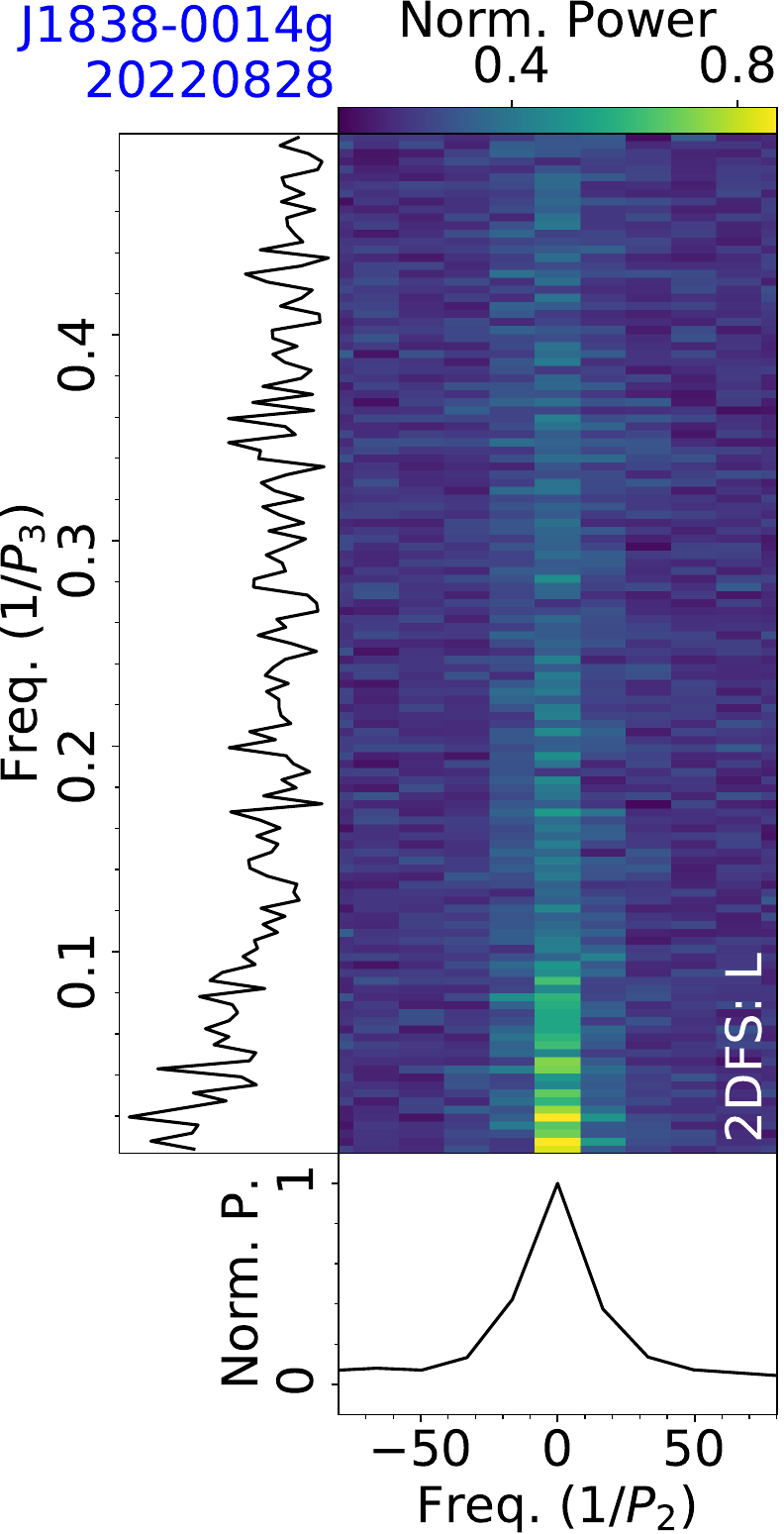}
\includegraphics[width=0.22\textwidth, angle=0]{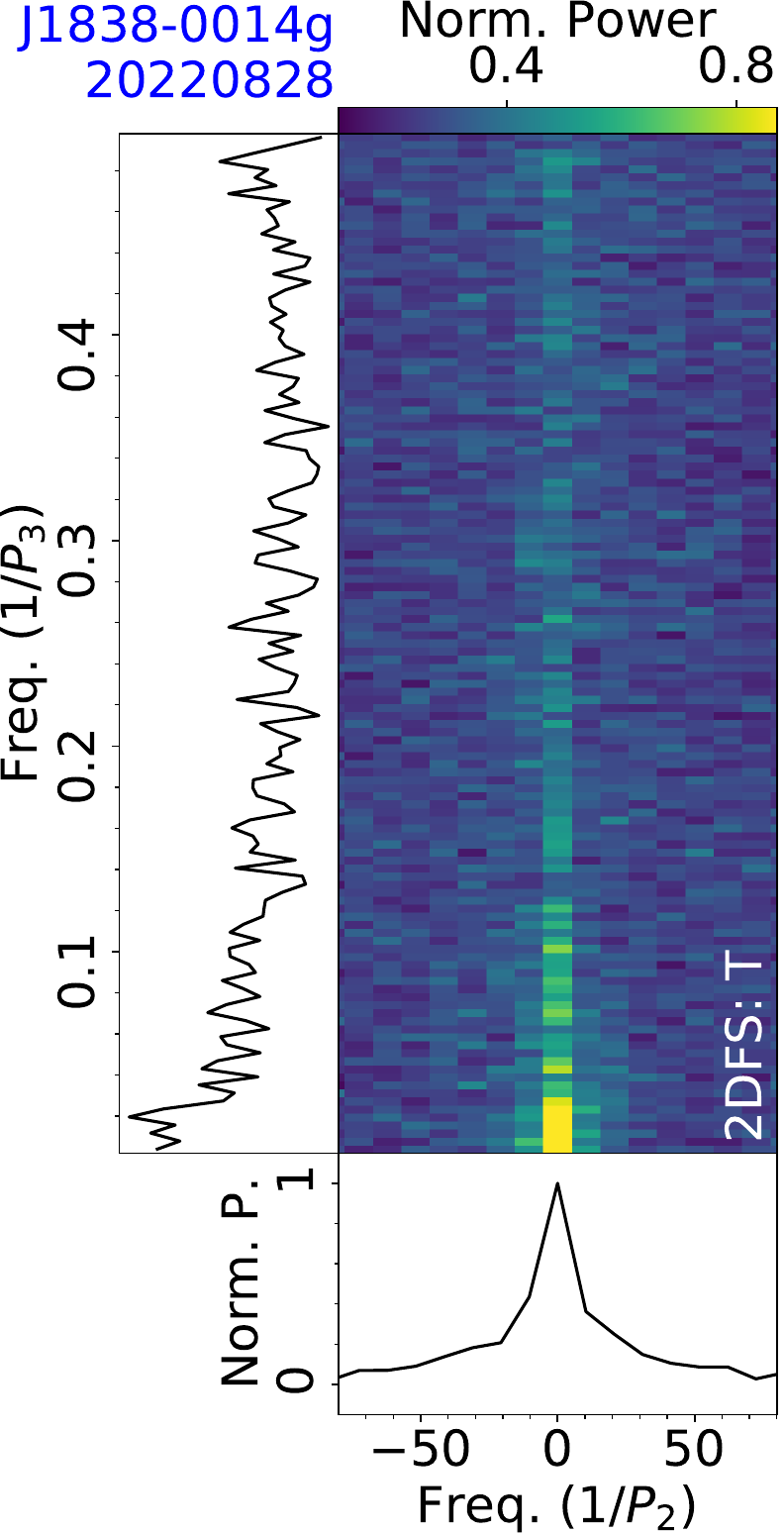}
\figcaption{Fluctuation analysis of PSR J1838-0014g from the FAST observation on 20220828, with LRFS (top-left), and 2DFS for the on-pulse region (top-right), leading part (bottom-left) and trailing part (bottom-right) of a mean pulse profile. \label{subfig:fluctu:J1838-0014g}}
\end{figure}

\begin{figure}[htpb]
\centering
\includegraphics[width=0.22\textwidth, angle=0]{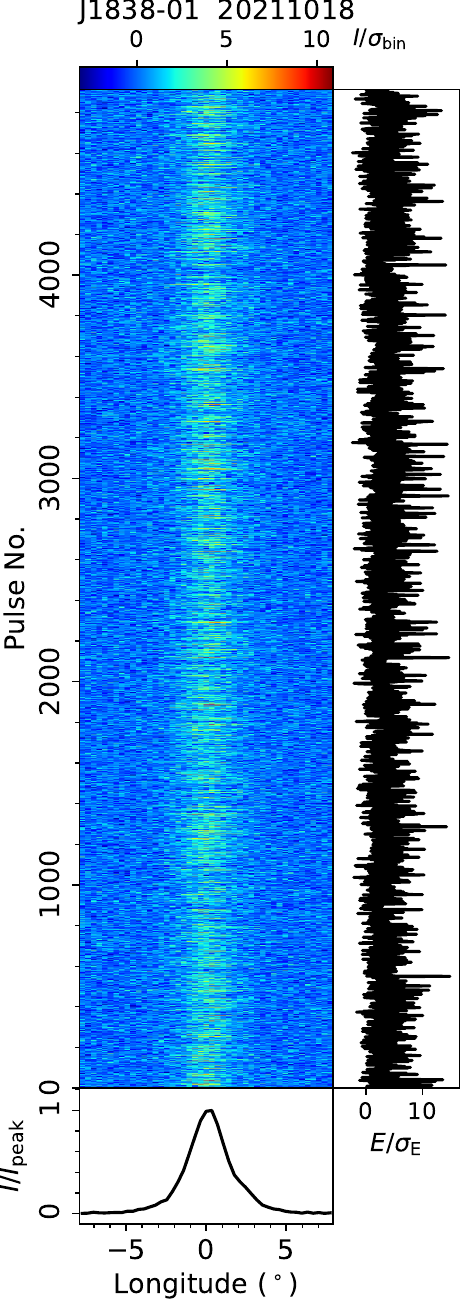}
\includegraphics[width=0.22\textwidth, angle=0]{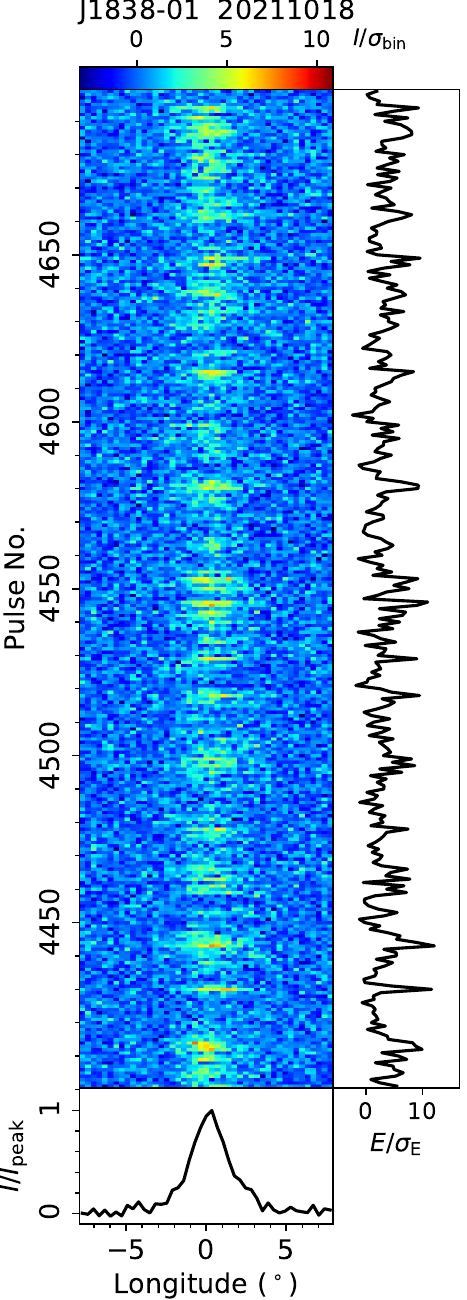}
\figcaption{Single pulse sequence of PSR J1838-01 from the FAST observation on 20211018, and a zoomed-in view of pulses No. 4400-4700.
\label{subfig:TP:J1838-01}}
\end{figure}

\begin{figure}[htpb]
\centering
\includegraphics[width=0.22\textwidth, angle=0]{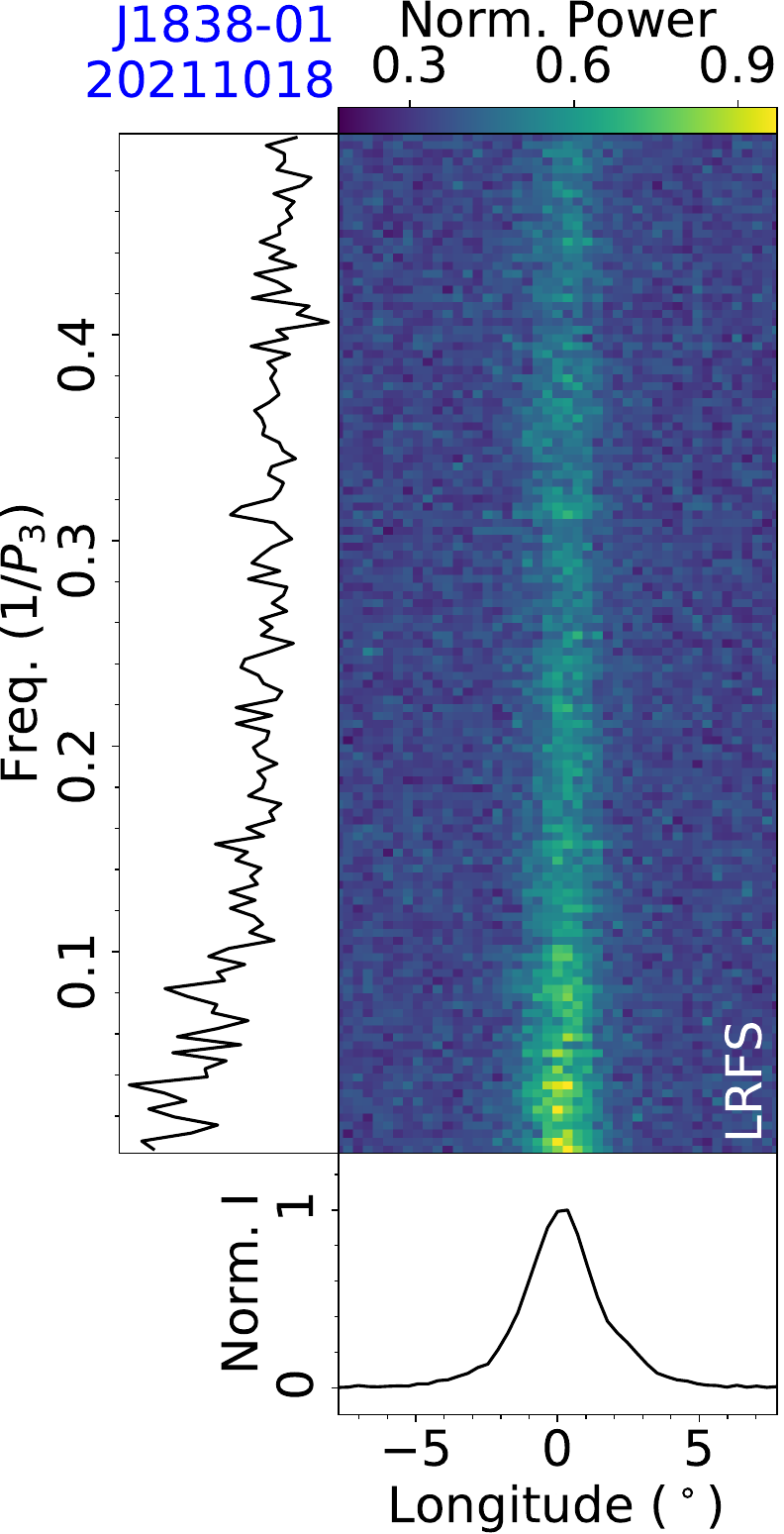}
\includegraphics[width=0.22\textwidth, angle=0]{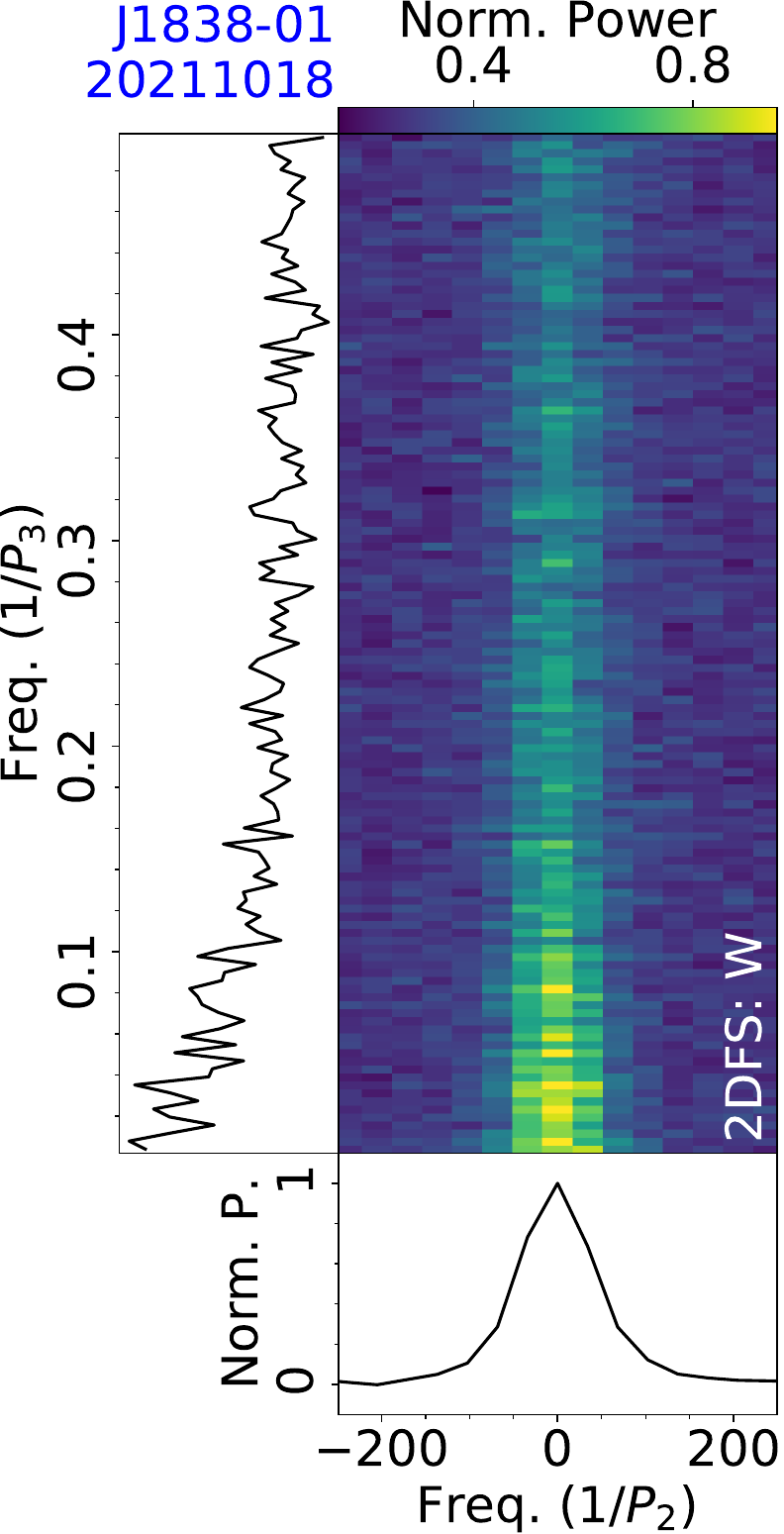}
\figcaption{Fluctuation analysis of PSR J1838-01 for the observation on 20211018, with LRFS and 2DFS for the on-pulse region of a mean pulse profile. \label{subfig:fluctu:J1838-01}}
\end{figure}

\begin{figure}[htpb]
\centering
\includegraphics[width=0.44\textwidth, angle=0]{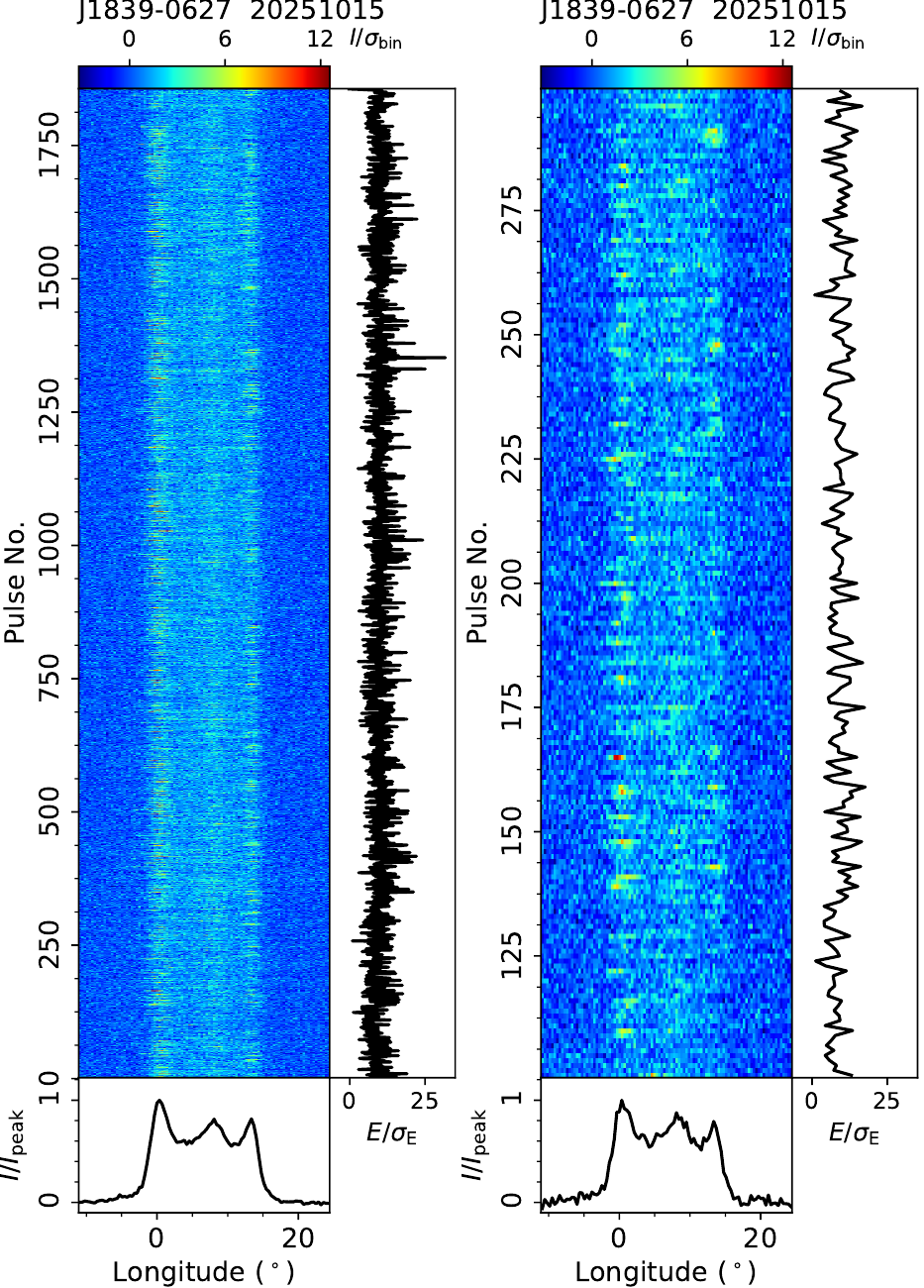}
\figcaption{Single pulse sequence of PSR J1839-0627 from the FAST observation on 20251015, and a zoomed-in view of pulses No. 100-300.
\label{subfig:TP:J1839-0627}}
\end{figure}

\begin{figure}[htpb]
\centering
\includegraphics[width=0.44\textwidth, angle=0]{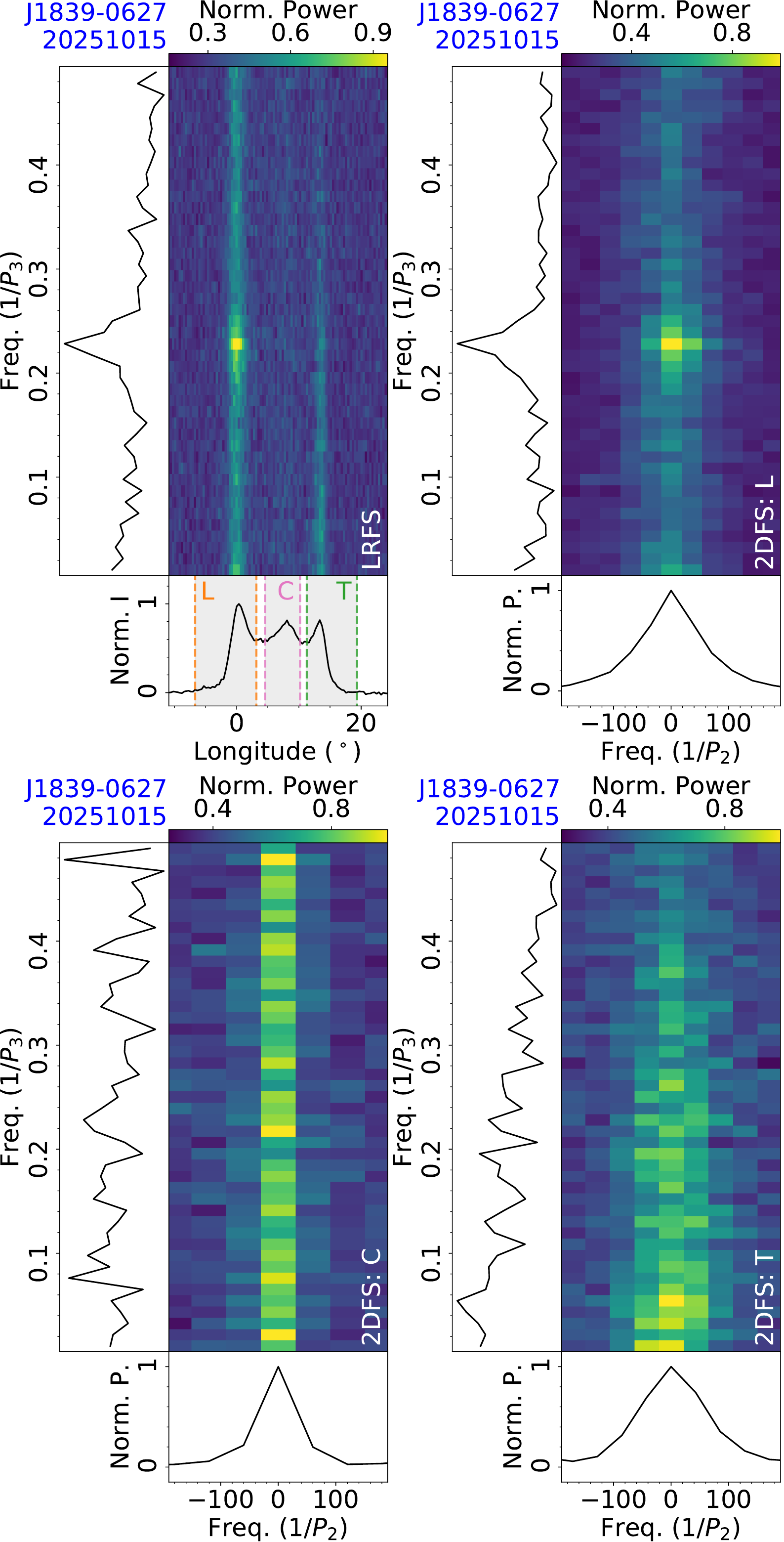}
\figcaption{Fluctuation analysis of PSR J1839-0627 from the FAST observation on 20251015, with LRFS (top-left), and 2DFS for the leading part (top-right), central part (bottom-left) and trailing part (bottom-right) of the mean pulse profile.
\label{subfig:fluctu:J1839-0627}}
\end{figure}

\begin{figure}[htpb]
\centering
\includegraphics[width=0.22\textwidth, angle=0]{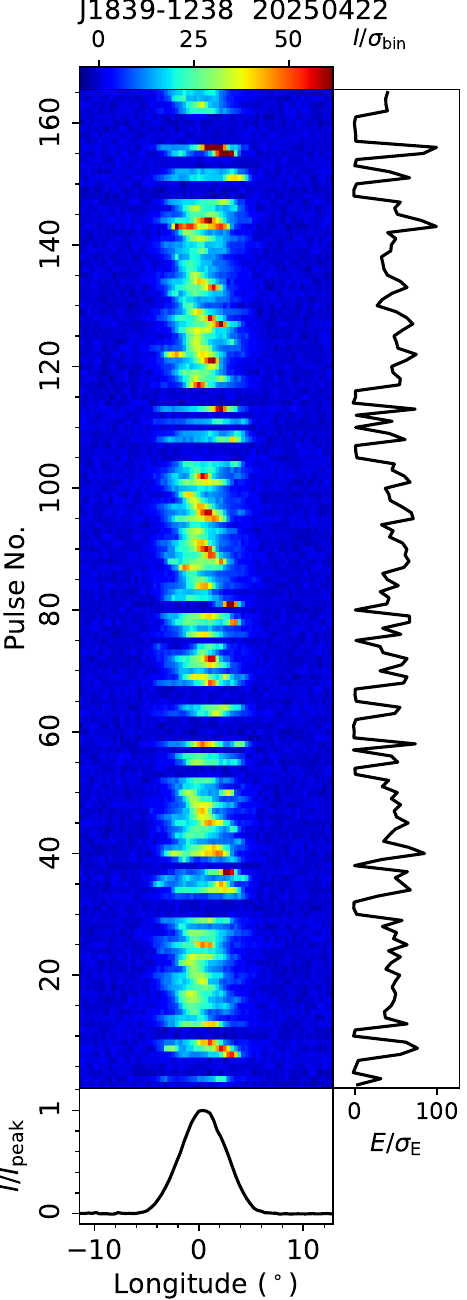}
\includegraphics[width=0.22\textwidth, angle=0]{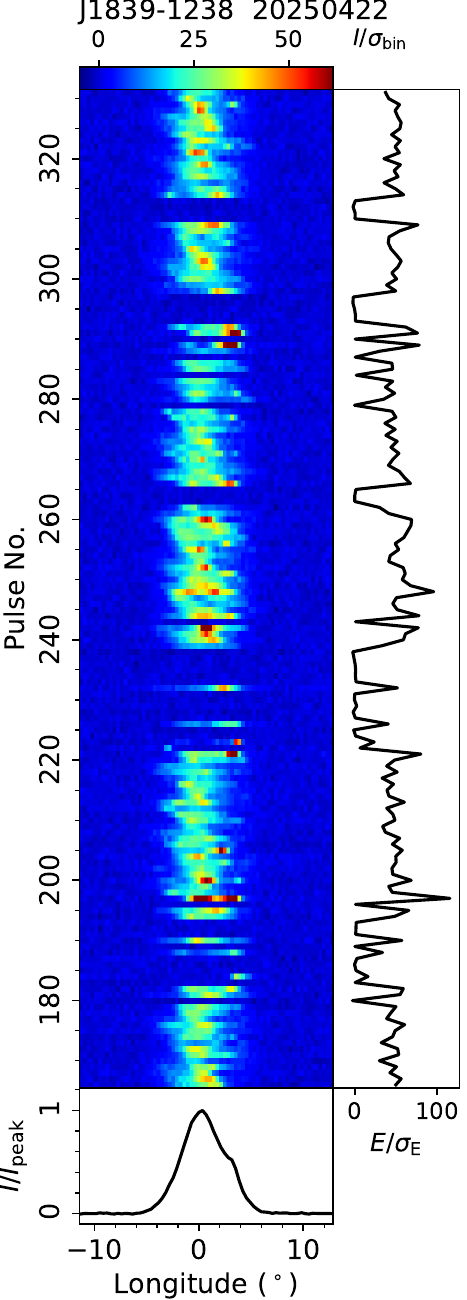}
\figcaption{Single pulse sequences of PSR J1839-1238 from the FAST observation on 20250422.
\label{subfig:TP:J1839-1238}}
\end{figure}

\begin{figure}[htpb]
\centering
\includegraphics[width =0.39\textwidth, angle=0]{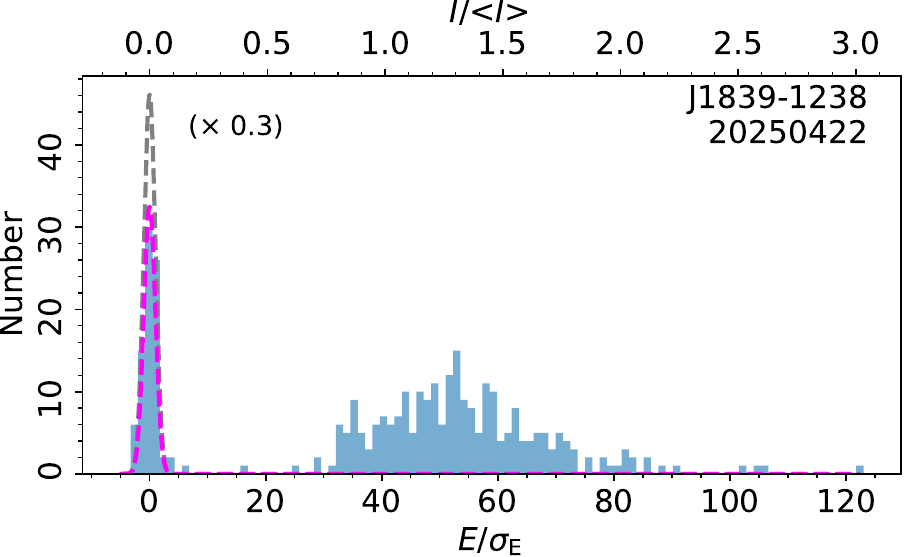}
\figcaption{On-pulse energy histogram of single pulses of PSR J1839-1238 from the FAST observation on 20250422. \label{subfig:Hist:J1839-1238}}
\end{figure}

\begin{figure}[htpb]
\centering
\includegraphics[width=0.22\textwidth, angle=0]{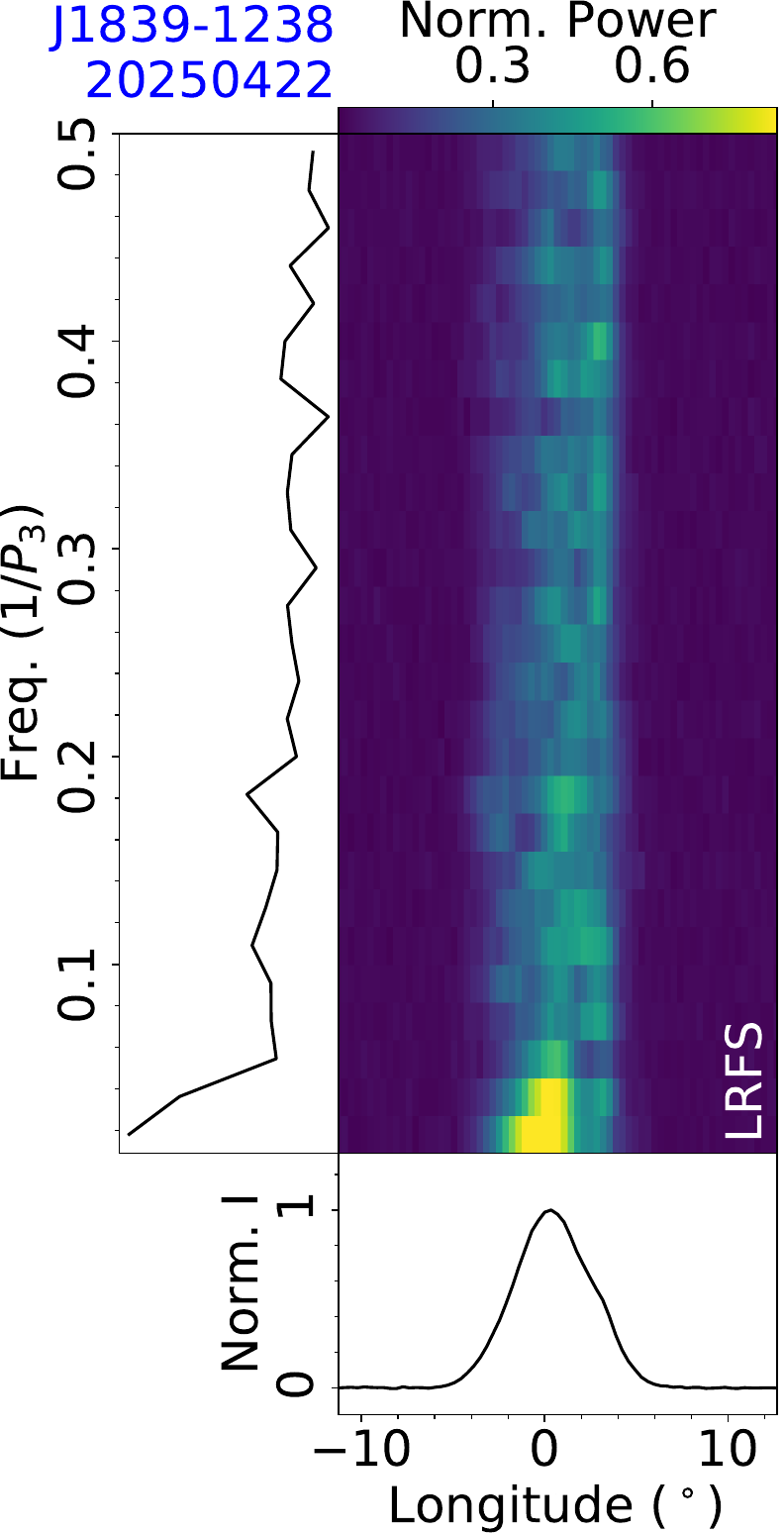}
\includegraphics[width=0.22\textwidth, angle=0]{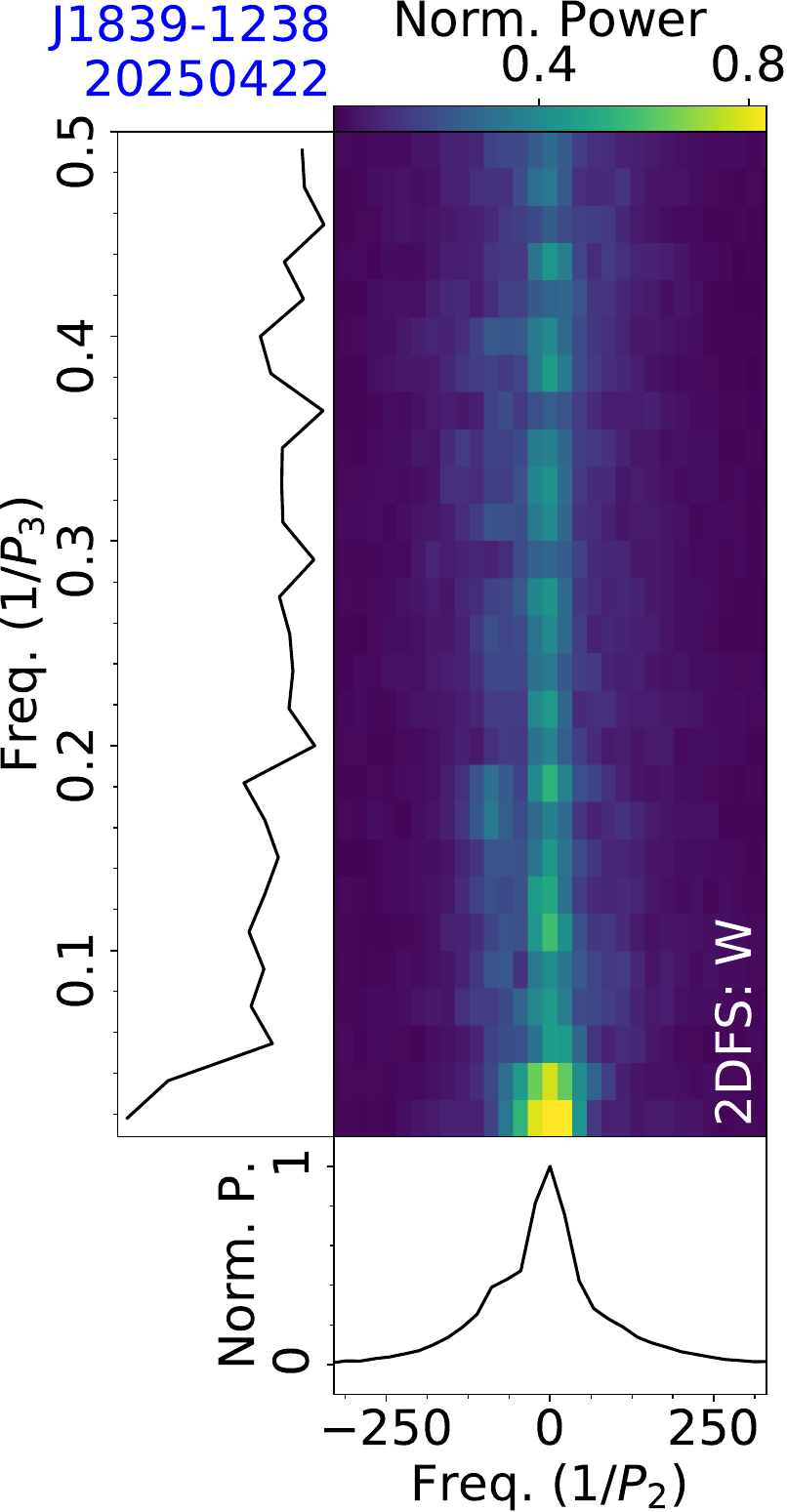}
\figcaption{Fluctuation analysis of PSR J1839-1238 for the observation on 20250422, with LRFS and 2DFS for the on-pulse region of a mean pulse profile.
\label{subfig:fluctu:J1839-1238}}
\end{figure}

\subsection{J1837-0653}
\label{subsec:J1837-0653}

PSR J1837-0653 was discovered during a survey conducted at Jodrell Bank with the MKIA telescope \citep{Clifton1986}.

This pulsar was observed by FAST on 20231014 for 96 minutes, with a rotation period $P=1.9060$~s and a dispersion measure $D\!M=313.3~{\rm cm^{-3}\,pc}$ derived. The single pulse sequence and a zoomed-in view of pulses No. 500-800 in Fig.~\ref{subfig:TP:J1837-0653} display the nulling phenomenon. Based on the histogram of on-pulse integrated energies of single pulses (Fig.~\ref{subfig:Hist:J1837-0653}), the nulling fraction of this observation is estimated to be 25$\pm$2\%.

\subsection{J1838+0044g}
\label{subsec:J1838+0044g}

PSR J1838+0044g was discovered in the FAST GPPS survey with both nulling and drifting phenomena \citep{Han2021,han2025}.

This pulsar was observed by FAST on 20200203 for 25 minutes, deriving a rotation period $P=2.2032$~s and a dispersion measure $D\!M=229.6~{\rm cm^{-3}\,pc}$ from this observation. 
Single pulse sequences of this observation are displayed in Fig.~\ref{subfig:TP:J1838+0044g}. 
On-pulse energy histogram in Fig.~\ref{subfig:Hist:J1838+0044g} indicates the nulling behavior with a nulling fraction of 20$\pm$3\%. 
In addition, fluctuation spectra (Fig.~\ref{subfig:fluctu:J1838+0044g}) of leading and trailing parts in a mean pulse profile both show negative drift features, with the temporally modulated frequency widely distributed. 
For the leading profile part, centroid frequencies of the drift feature are estimated to be $1/P_3=0.204\pm0.002$ and $1/P_2=-39\pm2$, corresponding to $P_3=4.9\pm0.1$ periods and $P_2=-9\pm1^\circ$. 
The drift feature in 2DFS of the trailing profile part exhibits centroid frequencies of $1/P_3=0.249\pm0.003$ and $1/P_2=-33\pm3$, yielding $P_3=4.01\pm0.04$ periods and $P_2=-11\pm1^\circ$.
The low-frequency modulation in spectra is related to nulls.

\begin{figure}[htpb]
\centering
\includegraphics[width=0.22\textwidth, angle=0]{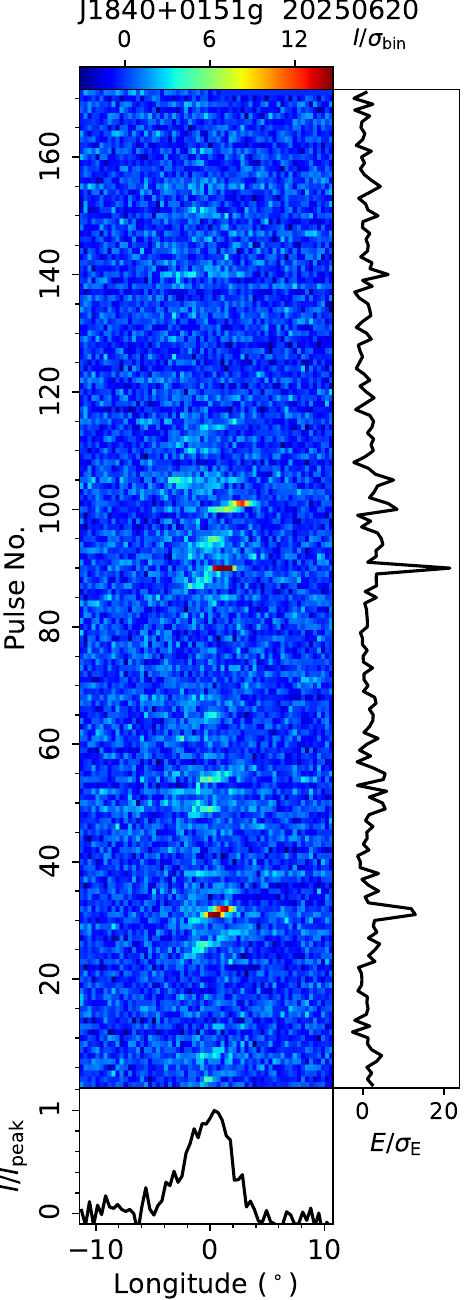}
\figcaption{Single pulse sequence of PSR J1840+0151g from the FAST observation on 20250620.
\label{subfig:TP:J1840+0151g}}
\end{figure}

\begin{figure}[htpb]
\centering
\includegraphics[width=0.22\textwidth, angle=0]{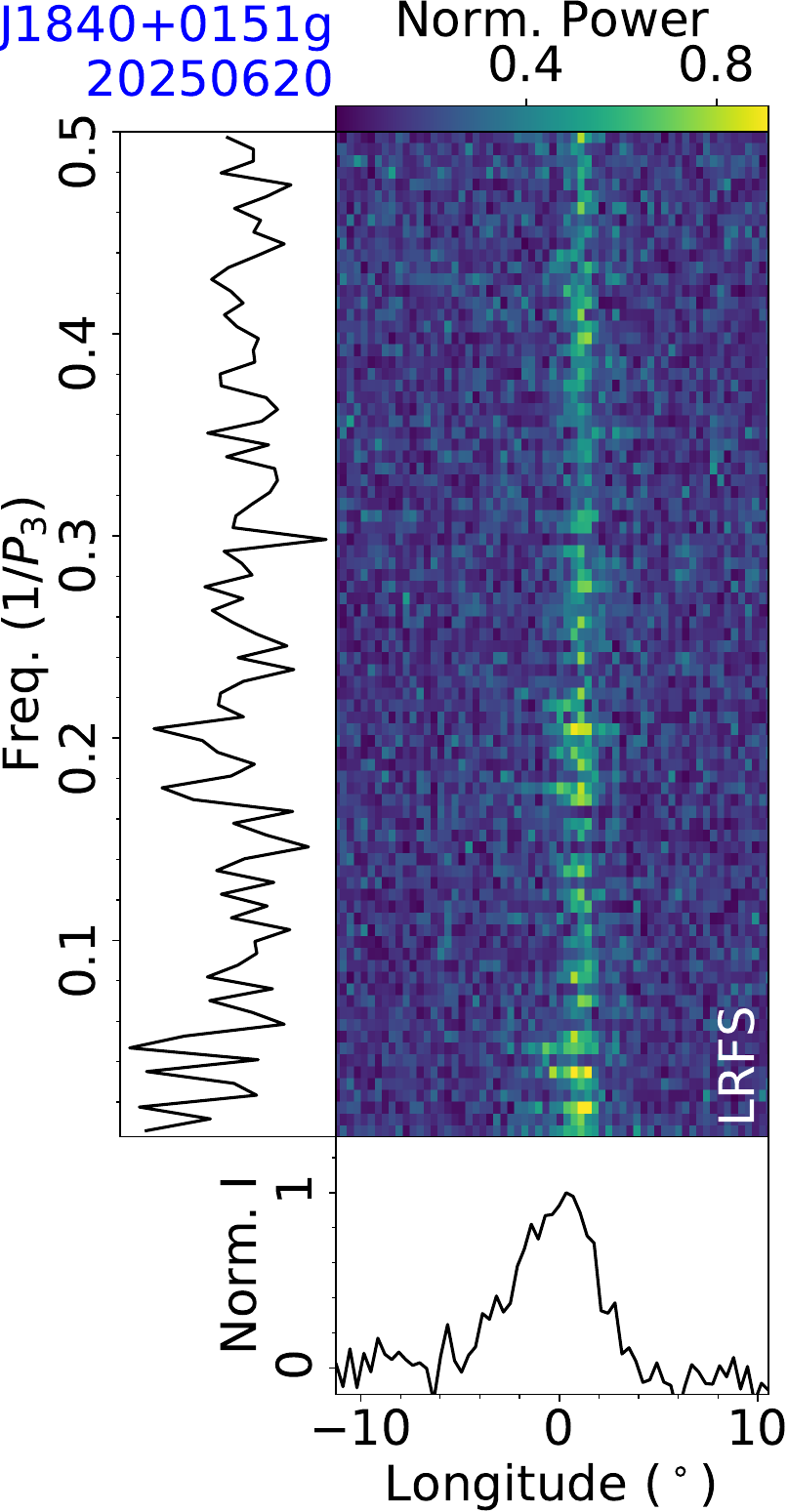}
\includegraphics[width=0.22\textwidth, angle=0]{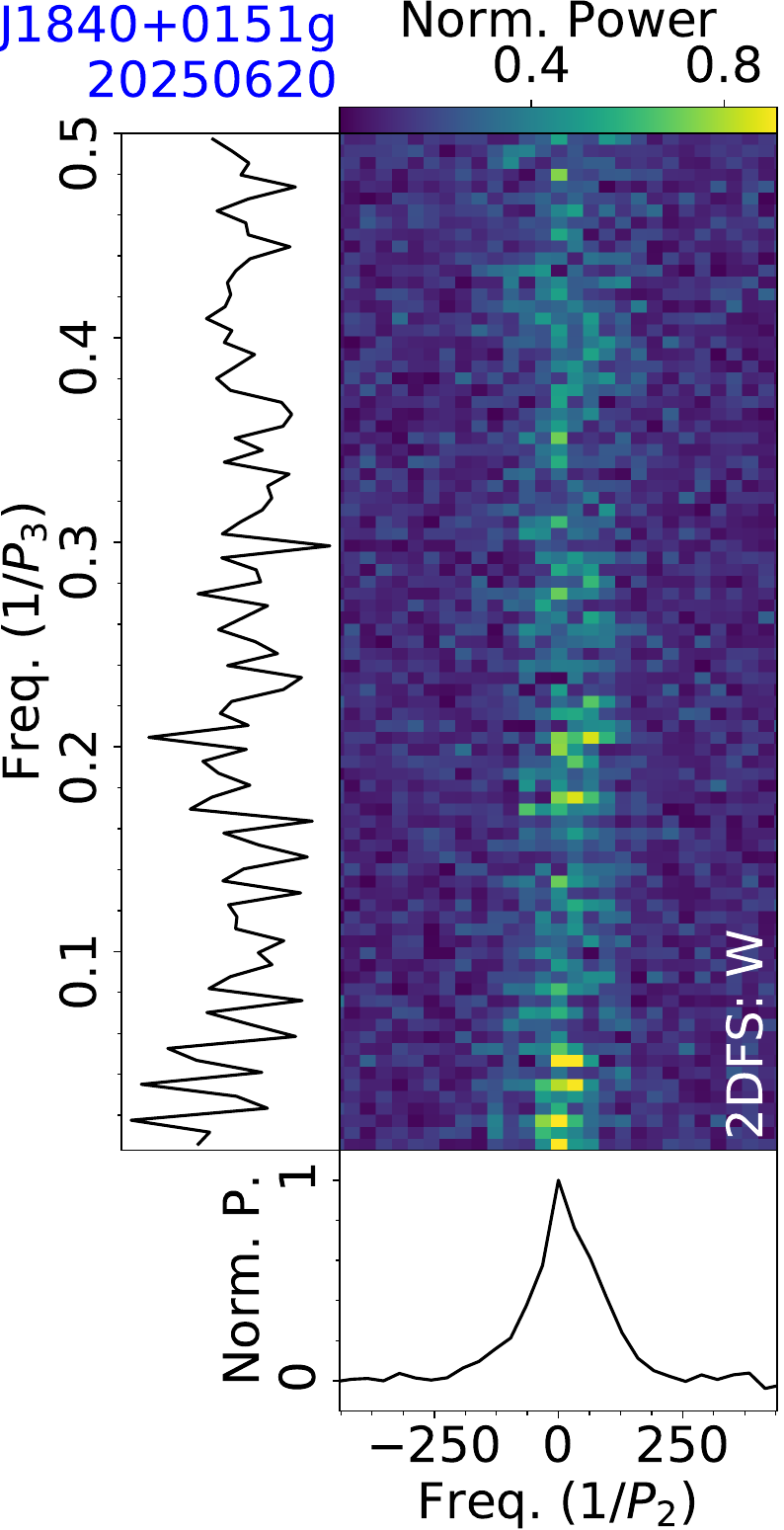}
\figcaption{Fluctuation analysis of PSR J1840+0151g for the observation on 20250620, with LRFS and 2DFS for the on-pulse region of a mean pulse profile.
\label{subfig:fluctu:J1840+0151g}}
\end{figure}

\begin{figure}[htpb]
\centering
\includegraphics[width=0.22\textwidth, angle=0]{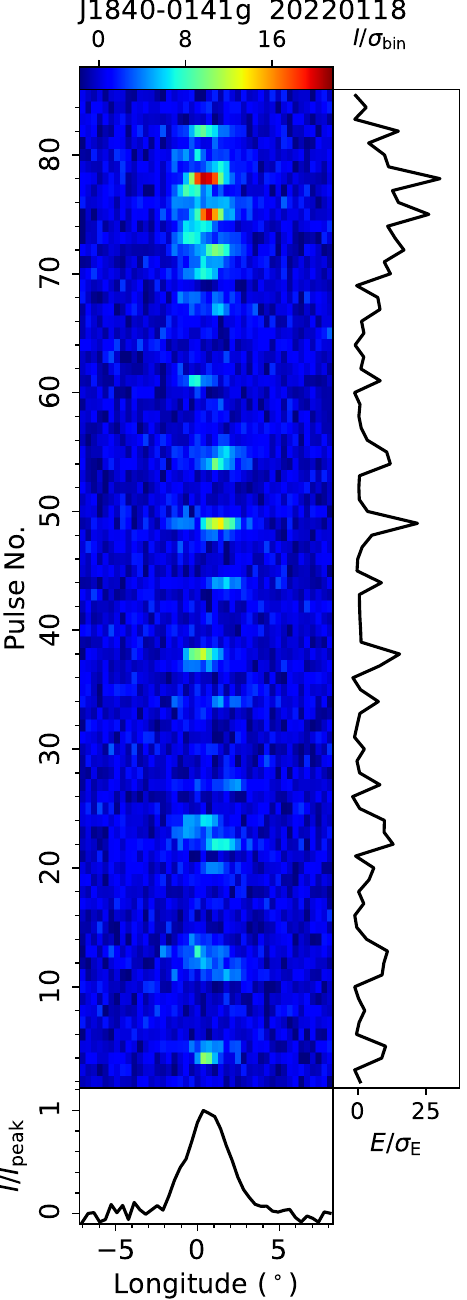}
\includegraphics[width=0.22\textwidth, angle=0]{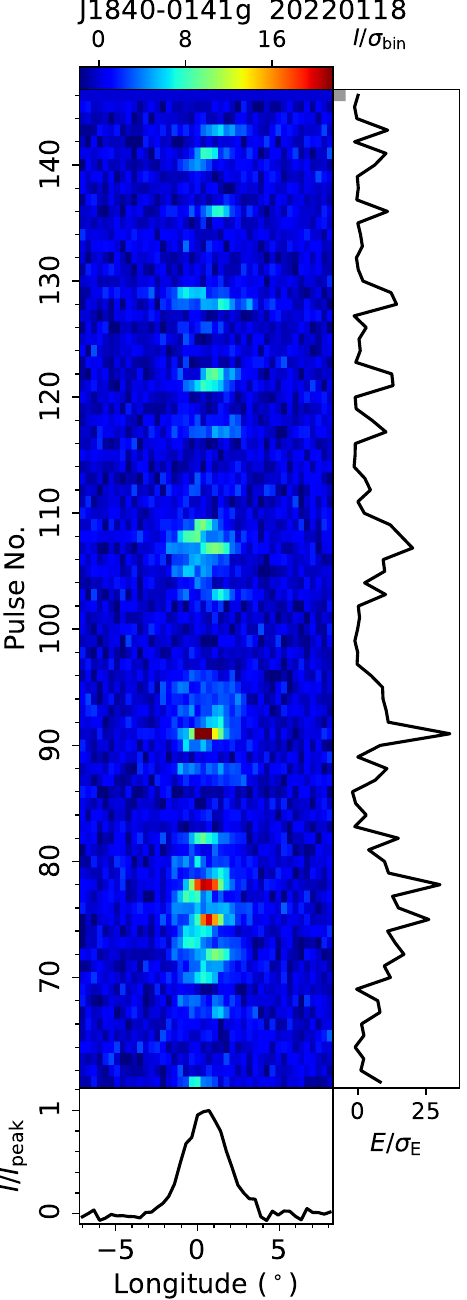}
\figcaption{Single pulse sequences of PSR J1840-0141g from the FAST observation on 20220118. \label{subfig:TP:J1840-0141g}}
\end{figure}

\begin{figure}[htpb]
\centering
\includegraphics[width =0.39\textwidth, angle=0]{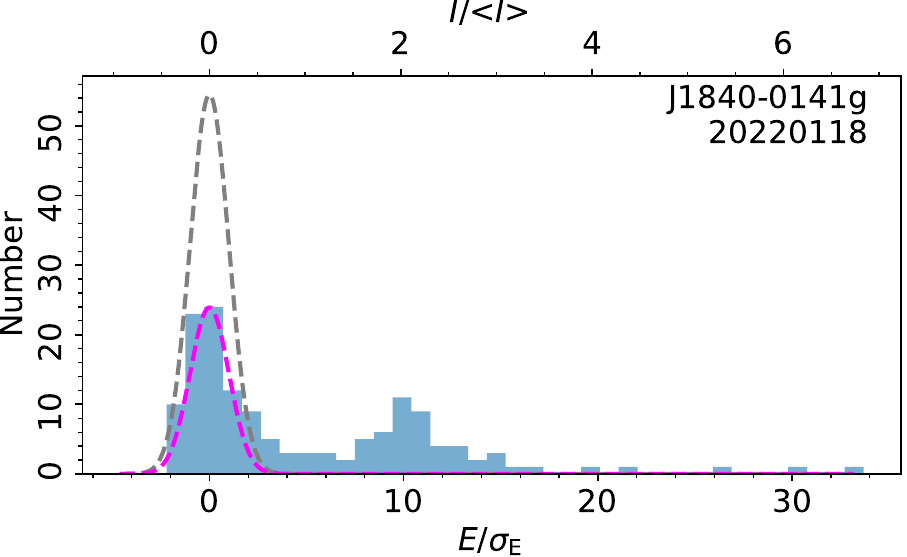}
\figcaption{On-pulse energy histogram of single pulses of PSR J1840-0141g from the FAST observation on 20220118. \label{subfig:Hist:J1840-0141g}}
\end{figure}

\begin{figure}[htpb]
\centering
\includegraphics[width=0.39\textwidth, angle=0]{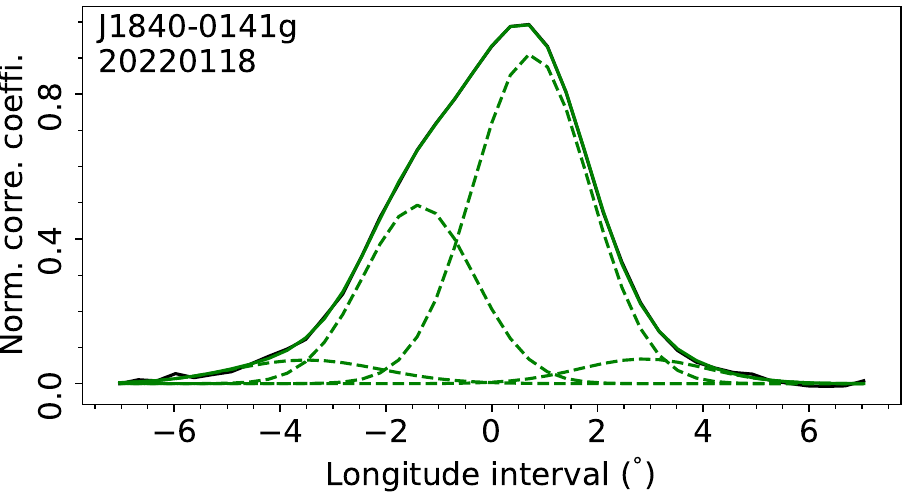}
\figcaption{Cross correlation of PSR J1843+0526g from the FAST observation on 20220118. \label{subfig:Corre:J1840-0141g}}
\end{figure}

\subsection{J1838-0014g}
\label{subsec:J1838-0014g}

PSR J1838-0014g was discovered in the FAST GPPS survey \citep{Han2021,han2025}. 

This pulsar was observed by FAST on 20220828 for 15 minutes, deriving a rotation period $P=0.3605$~s and a dispersion measure $D\!M=283.0~{\rm cm^{-3}\,pc}$ from this observation. 
Single pulse sequences and fluctuation spectra are shown in Fig.~\ref{subfig:TP:J1838-0014g} and \ref{subfig:fluctu:J1838-0014g}. 
Leading and trailing parts in the mean profile all have a temporal modulation, with the frequency centroid of $1/P_3=0.046\pm0.001$ ($P_3=22\pm1$ periods), without any preferred phase modulation.





\begin{figure}[htpb]
\centering
\includegraphics[width=0.21\textwidth, angle=0]{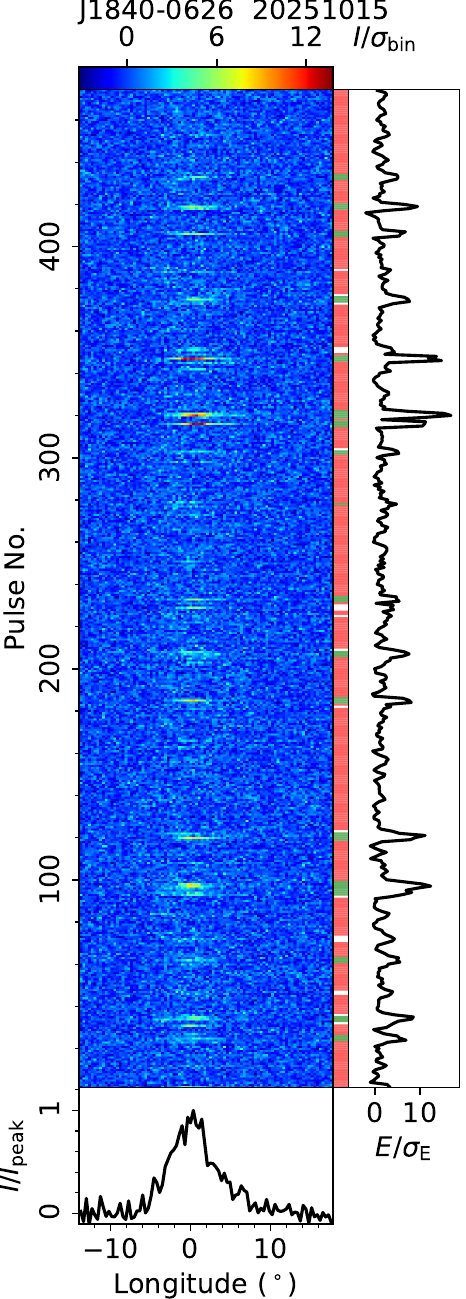}
\figcaption{Single pulse sequence of PSR J1840-0626 from the FAST observation on 20251015. In the right subpanel, the on-pulse energy variation is smoothed over every 3 periods.
\label{subfig:TP:J1840-0626}}
\end{figure}

\begin{figure}[htpb]
\centering
\includegraphics[width=0.39\textwidth, angle=0]{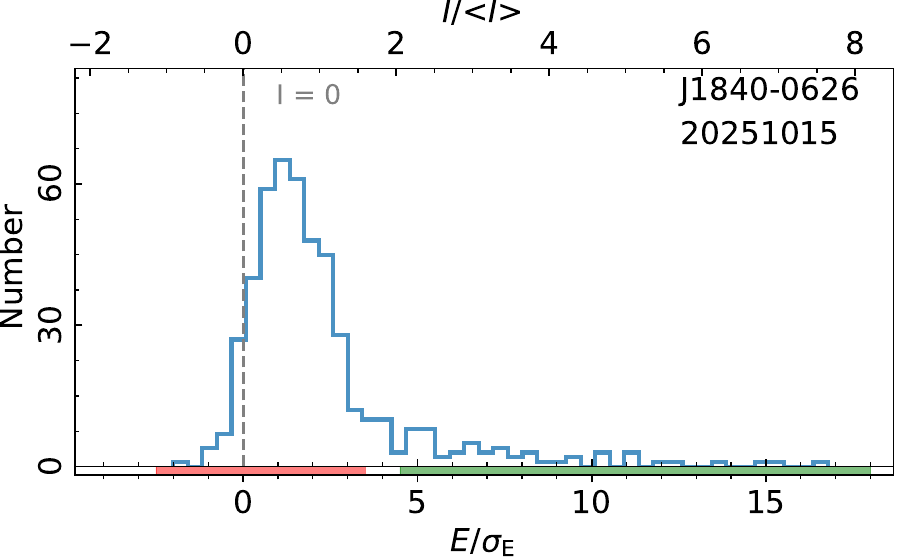}
\figcaption{On-pulse energy histogram of single pulses of PSR J1840-0626 from the FAST observation on 20251015, with energy values smoothed over 3 periods. 
\label{subfig:Hist:J1840-0626}}
\end{figure}

\begin{figure}[htpb]
\centering
\includegraphics[width=0.22\textwidth, angle=0]{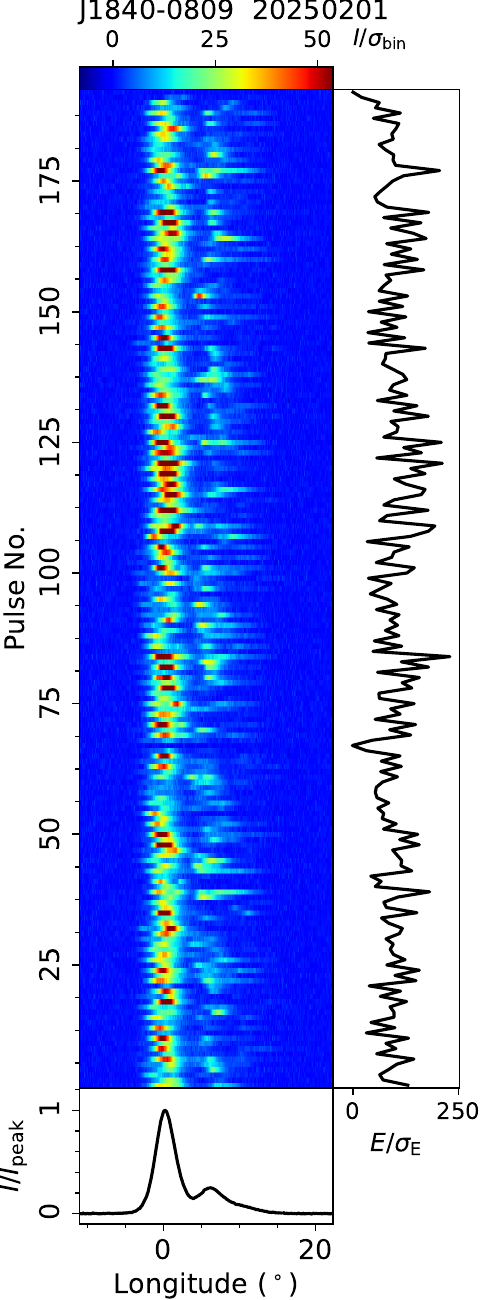}
\includegraphics[width=0.22\textwidth, angle=0]{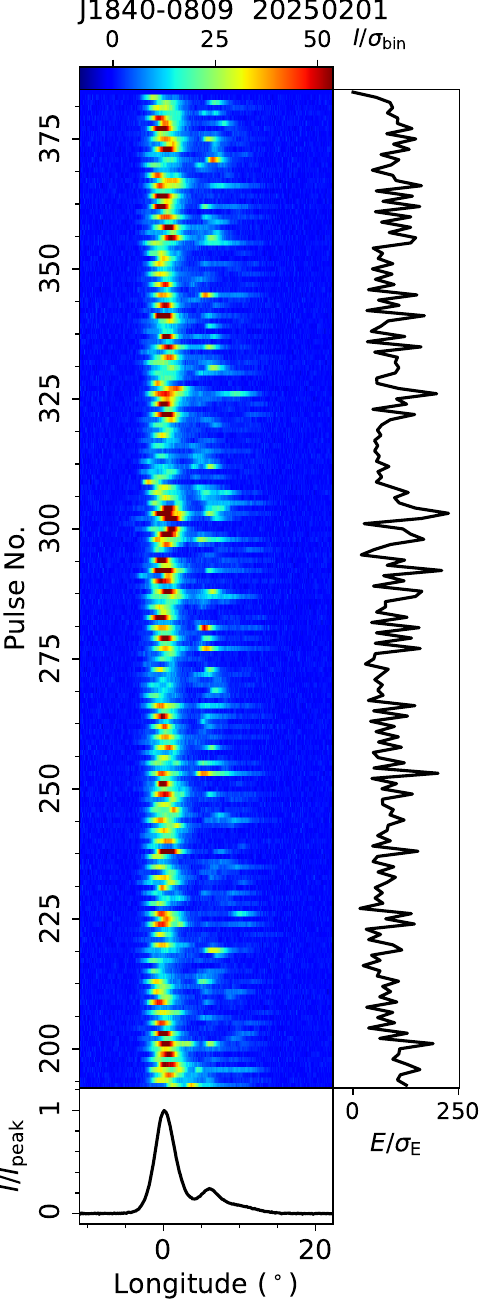}
\figcaption{Single pulse sequences of PSR J1840-0809 from the FAST observation on 20250201.
\label{subfig:TP:J1840-0809}}
\end{figure}

\begin{figure}[htpb]
\centering
\includegraphics[width=0.21\textwidth, angle=0]{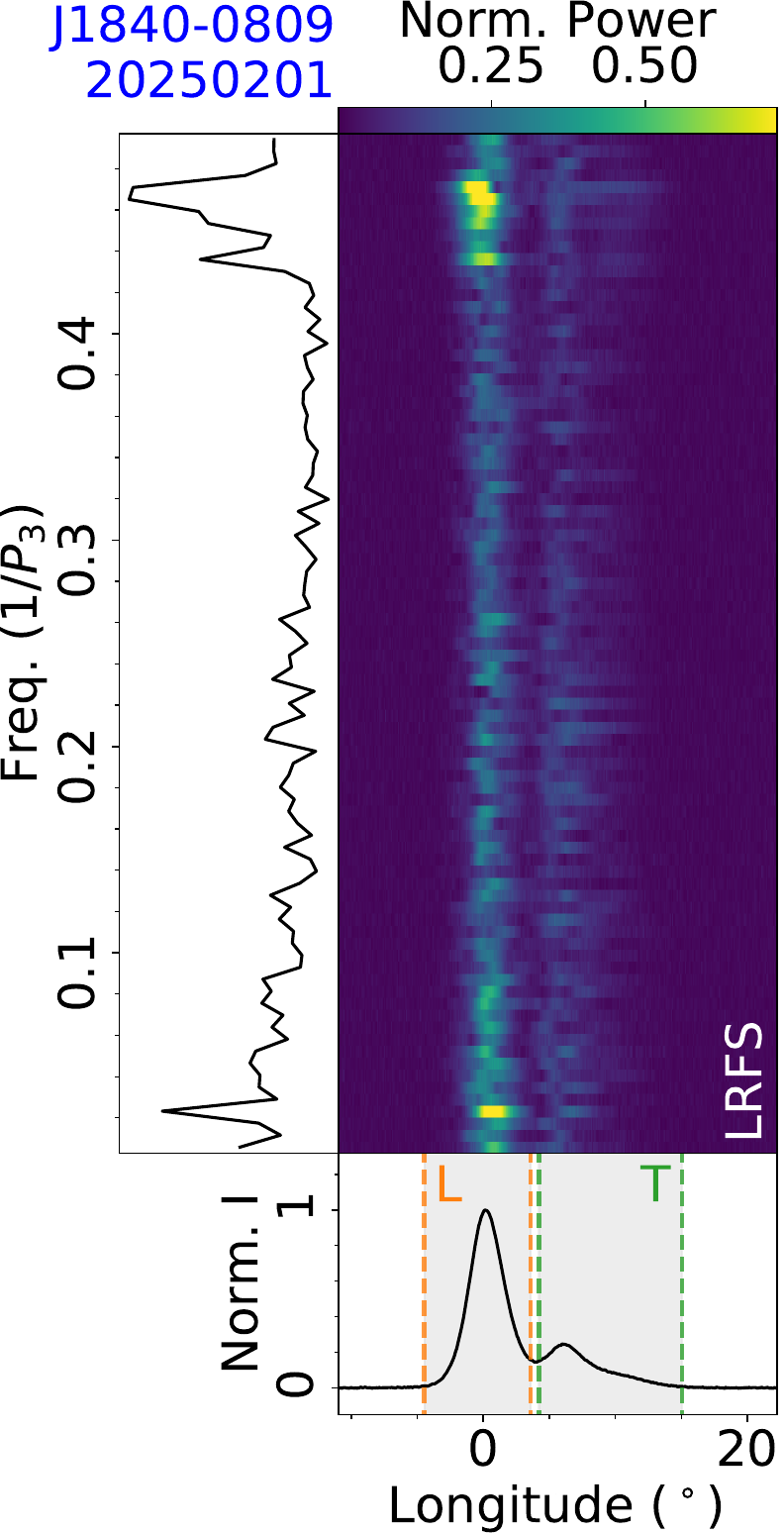}
\includegraphics[width=0.21\textwidth, angle=0]{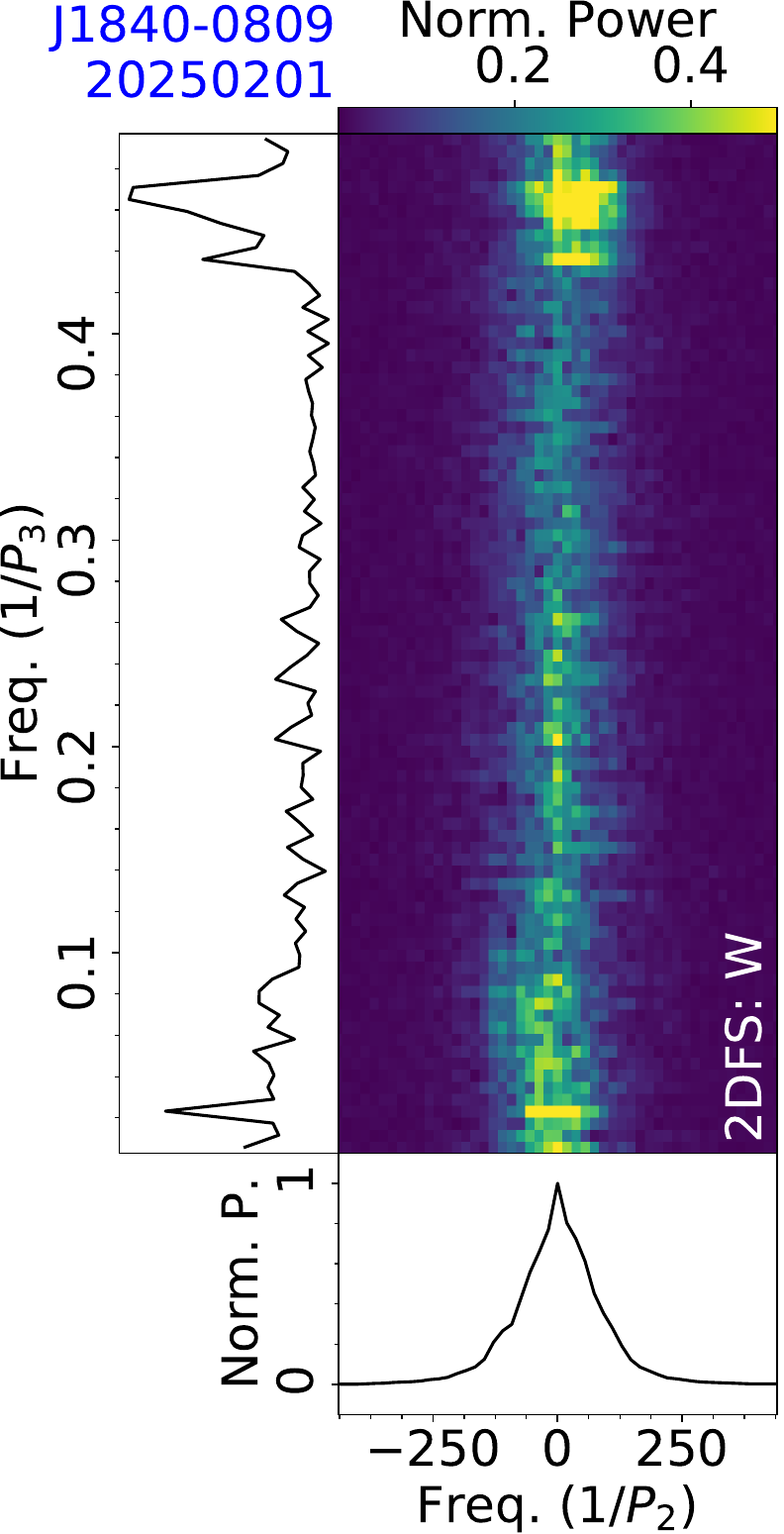}\\
\includegraphics[width=0.21\textwidth, angle=0]{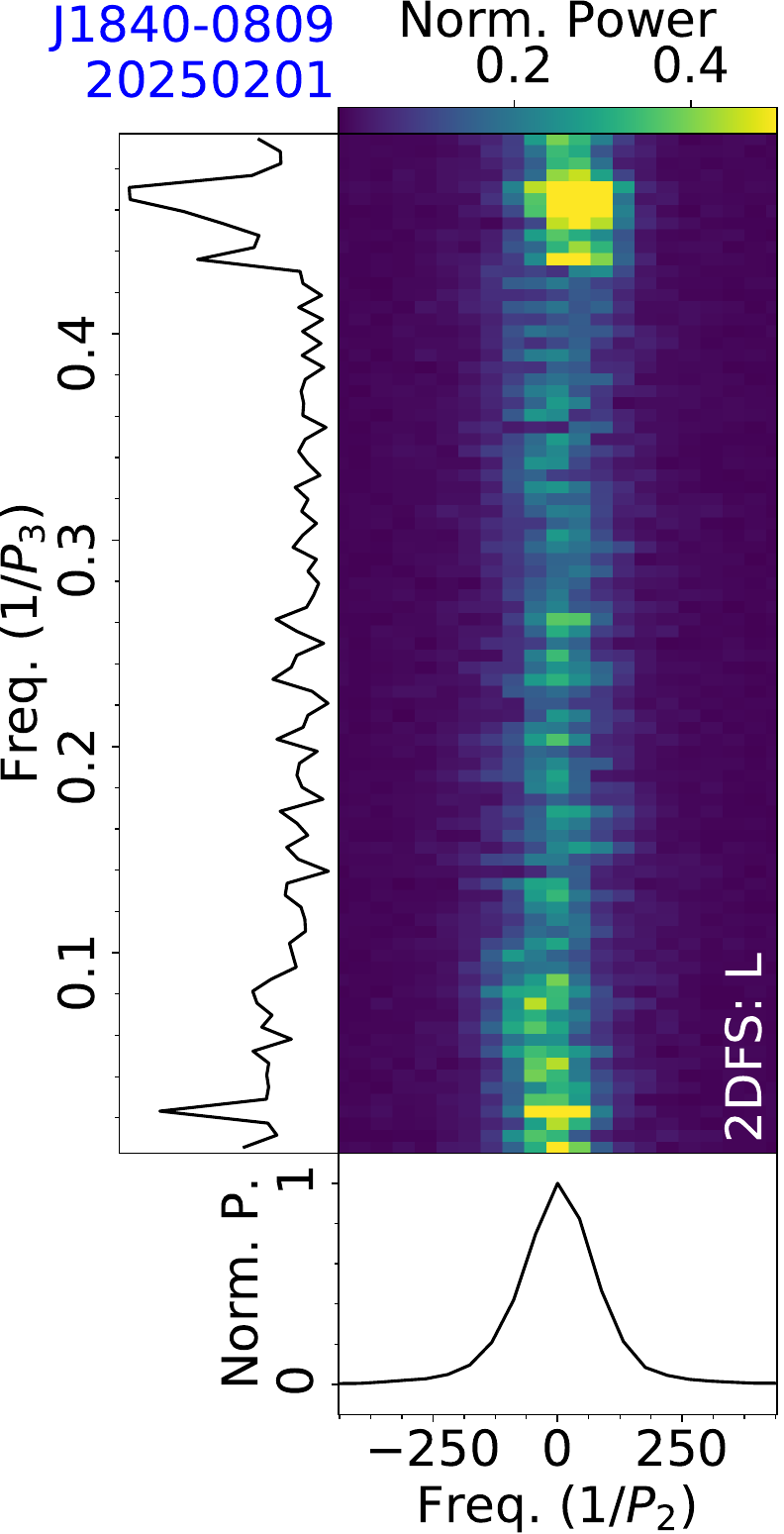}
\includegraphics[width=0.21\textwidth, angle=0]{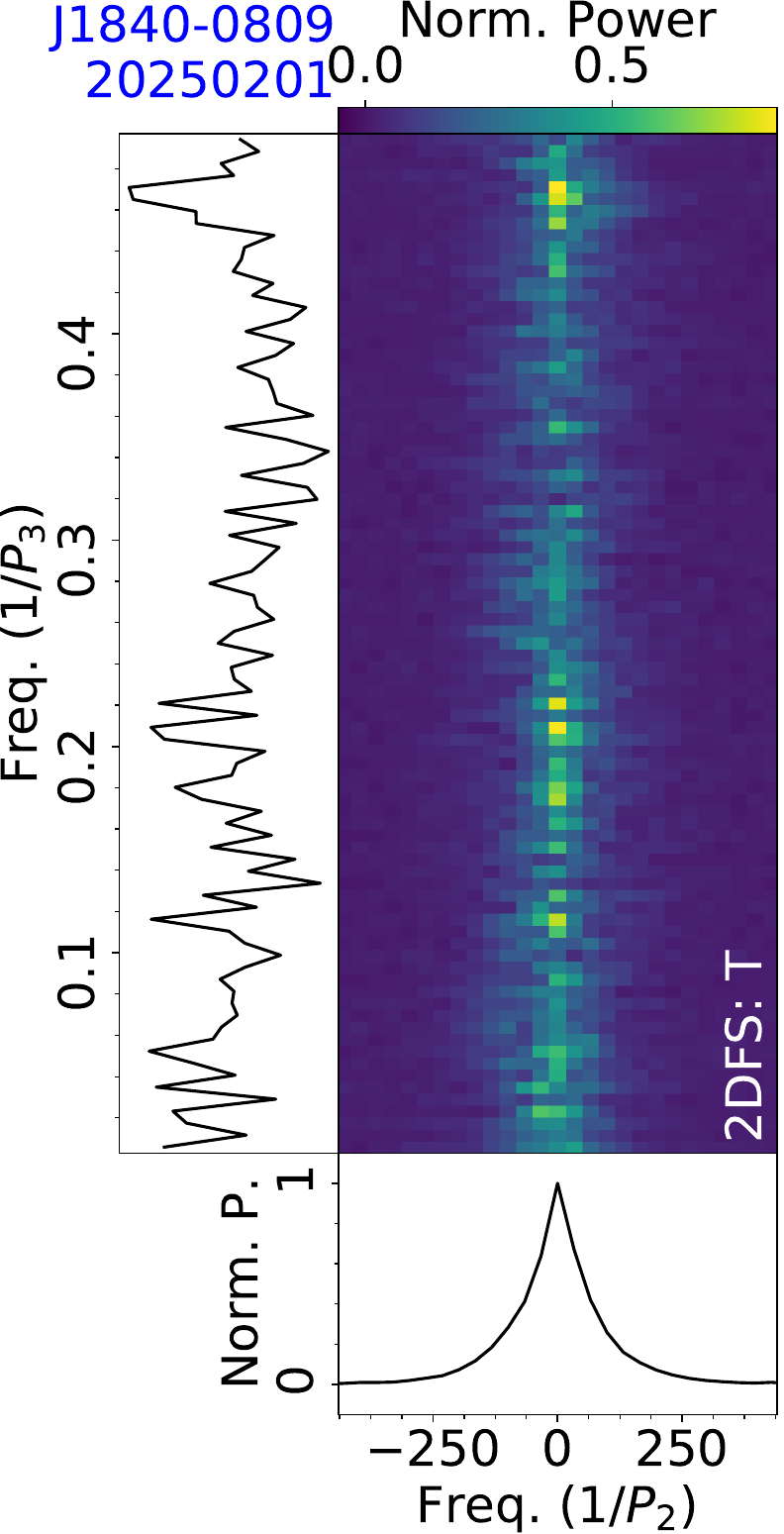}
\figcaption{Fluctuation analysis of PSR J1840-0809 for the observation on 20250201, with LRFS (top-left), and 2DFS for the on-pulse region (top-right), leading part (bottom-left) and trailing part (bottom-right) of a mean pulse profile.
\label{subfig:fluctu:J1840-0809}}
\end{figure}

\begin{figure}[htpb]
\centering
\includegraphics[width=0.22\textwidth, angle=0]{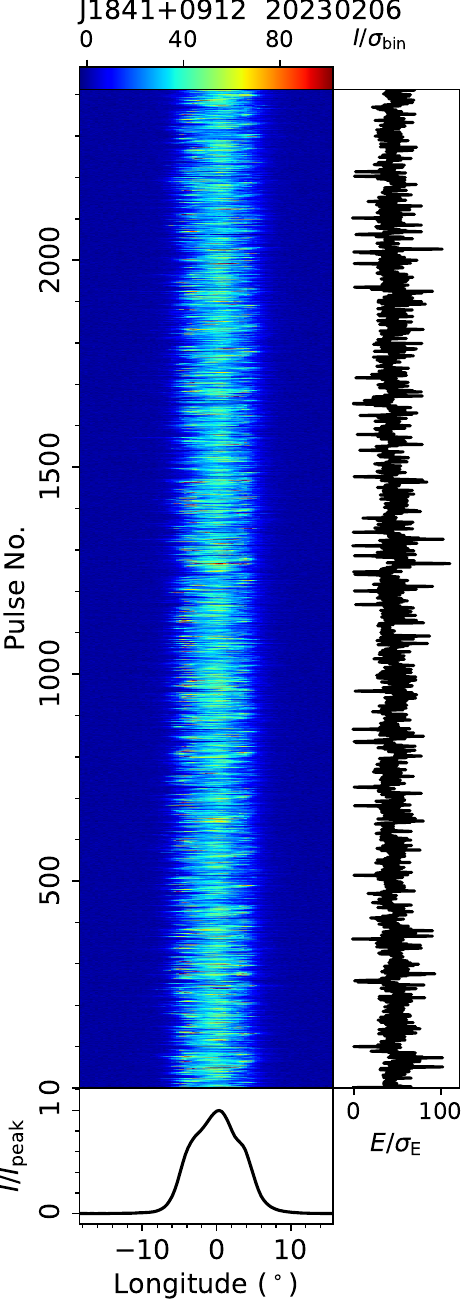}
\includegraphics[width=0.22\textwidth, angle=0]{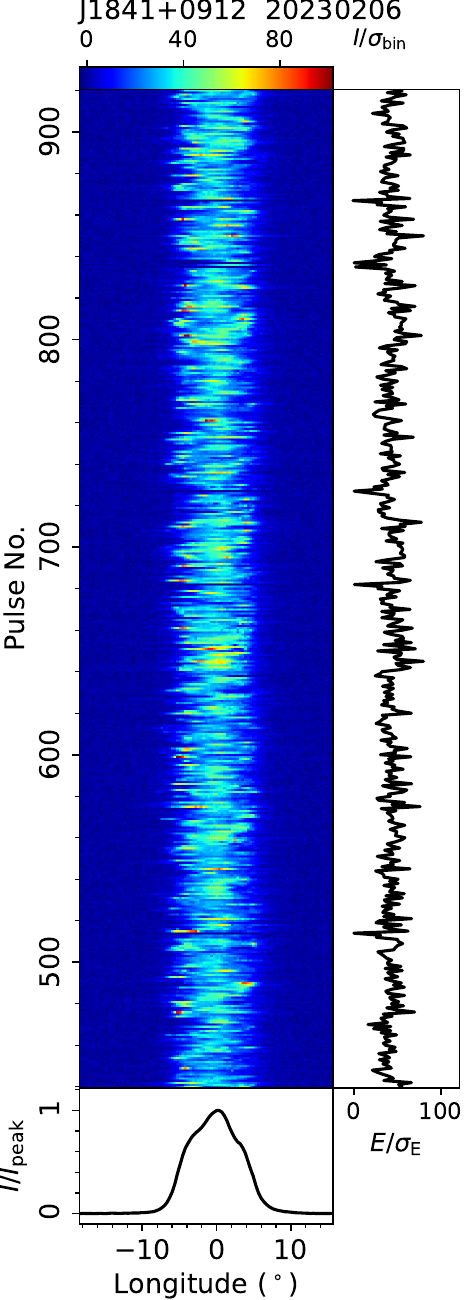}
\figcaption{Single pulse sequence of PSR J1841+0912 from the FAST observation on 20230206, and a zoomed-in view of pulses No. 440-920.
\label{subfig:TP:J1841+0912}}
\end{figure}

\begin{figure}[htpb]
\centering
\includegraphics[width=0.22\textwidth, angle=0]{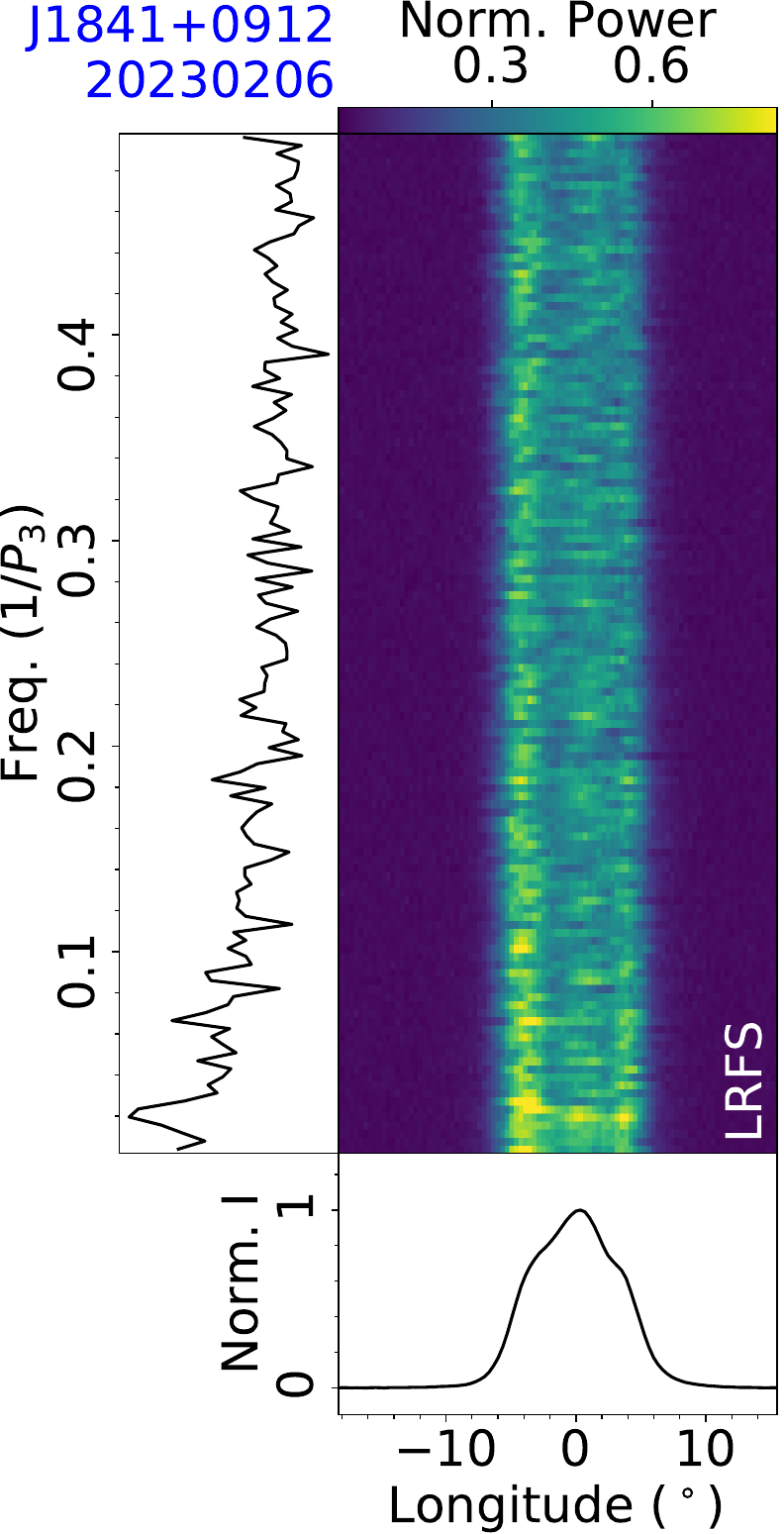}
\includegraphics[width=0.22\textwidth, angle=0]{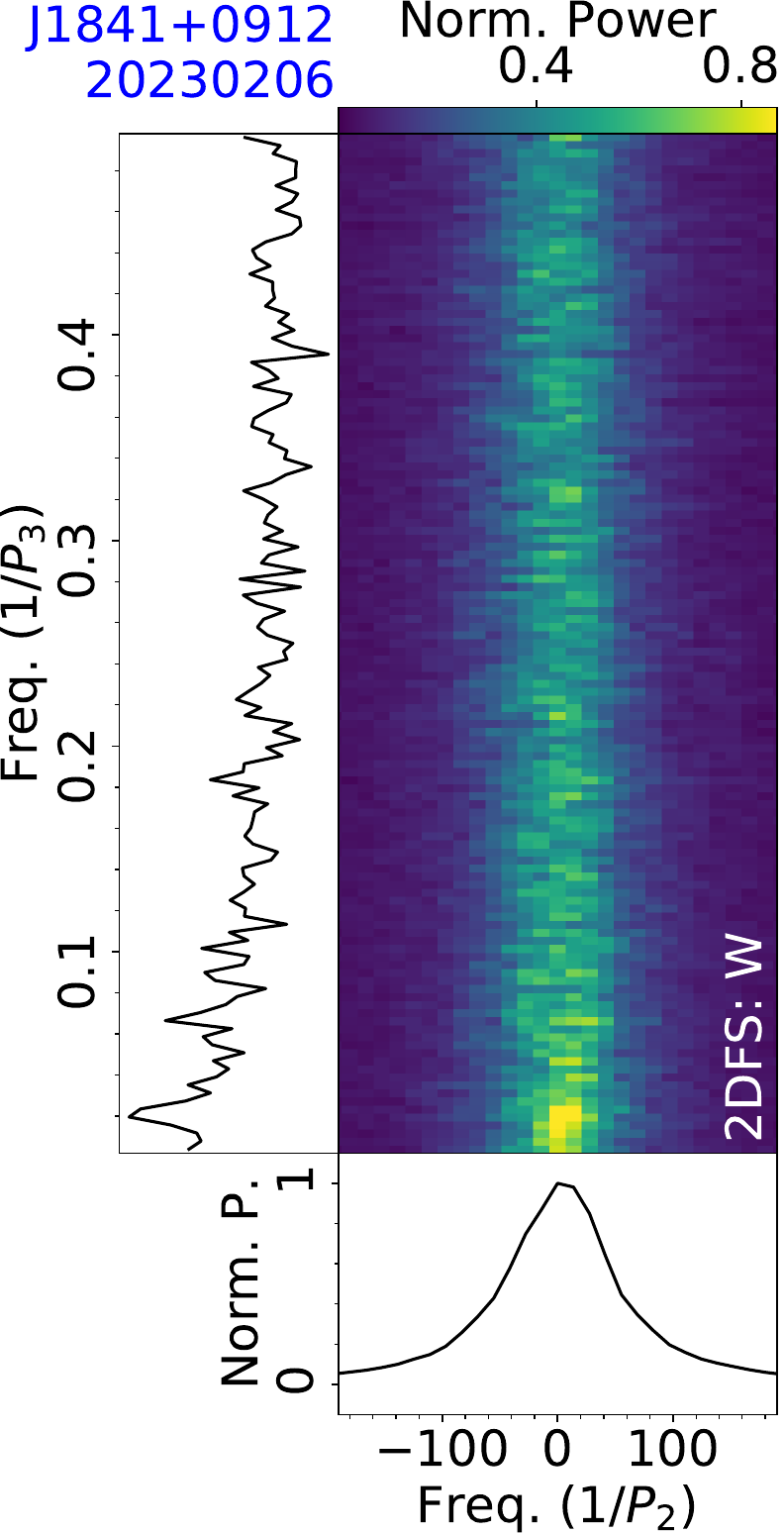}
\figcaption{Fluctuation analysis of PSR J1841+0912, with LRFS and 2DFS for the on-pulse region of a mean pulse profile.
\label{subfig:fluctu:J1841+0912}}
\end{figure}

\begin{figure}[htpb]
\centering
\includegraphics[width=0.39\textwidth, angle=0]{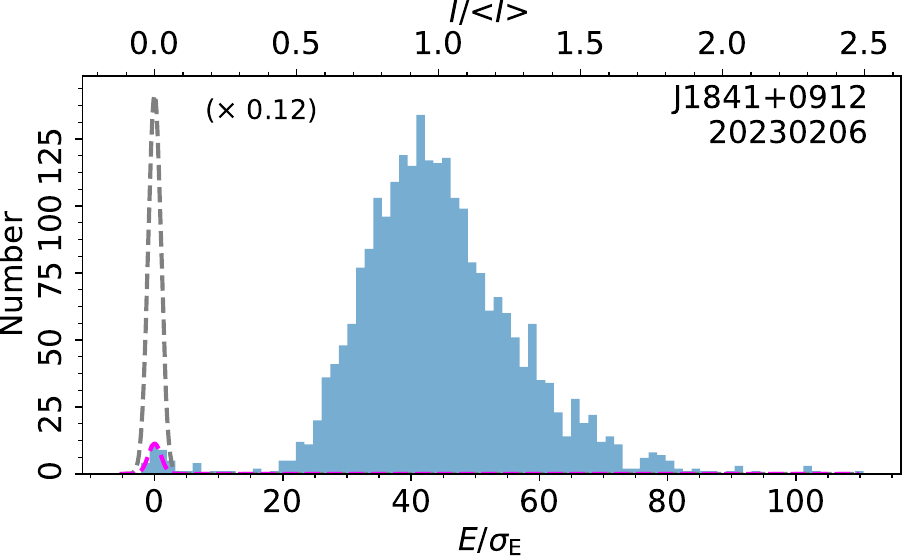}
\figcaption{On-pulse energy histograms of single pulses of PSR J1841+0912 from the FAST observation on 20230206.
\label{subfig:Hist:J1841+0912}}
\end{figure}

\subsection{J1838-01}
\label{subsec:J1838-01}

PSR J1838$-$01 was discovered from the Parkes Multi-beam Pulsar Survey Data \citep{Knispel2013}.

This pulsar was observed by FAST on 20211018 for 15 minutes, yielding a rotation period $P=0.1833$~s and a dispersion measure $D\!M=311.9~{\rm cm^{-3}\,pc}$. 
Single pulse sequences in Fig.~\ref{subfig:TP:J1838-01} display modulation behavior. In LRFS and 2DFS (Fig.~\ref{subfig:fluctu:J1838-01}), the centroid frequency of the temporally modulated feature is estimated to be $1/P_3=0.061\pm0.001$, which corresponds to $P_3=16.5\pm0.3$ periods.

\subsection{J1839-0627}
\label{subsec:J1839-0627}

PSR J1839-0627 was discovered in the Parkes Multibeam Pulsar Survey \citep{Morris2002}. 

This pulsar was observed by FAST on 20251015 for 15 minutes, driving a rotation period $P=0.4850$~s and a dispersion measure $D\!M=92.2~{\rm cm^{-3}\,pc}$. The single pulse sequence and a zoomed-in view of pulses No. 100-300 are shown in Fig.~\ref{subfig:TP:J1839-0627}. The fluctuation spectra in Fig.~\ref{subfig:fluctu:J1839-0627} display a preferred positive drift feature of the leading profile part, a flat featureless spectrum of the central part, and a temporally wide modulation feature of the trailing part. In 2DFS of the leading profile part, the centroid of the positive drift feature is at $1/P_3=0.228\pm0.001$ and $1/P_2=9\pm4$, corresponding to $P_3=4.38\pm0.02$ periods and $P_2=39\pm17$ degrees. The modulation feature in 2DFS of the trailing part exhibits a centroid at $1/P_3=0.132\pm0.002$, yielding $P_3=7.6\pm0.1$ periods.

\subsection{J1839-1238}
\label{subsec:J1839-1238}

PSR J1839-1238 was discovered in the Parkes multibeam pulsar survey \citep{Lorimer2006}. \citet{Song2023} reported a drifting behavior with $P_3=34\pm6$ periods and $P_2=-92^{+69}_{-78}$ degrees.

This pulsar was observed by FAST on 20250422 for 11 minutes, driving a rotation period $P=1.9113$~s and a dispersion measure $D\!M=174.2~{\rm cm^{-3}\,pc}$. 
Single pulse sequences in Fig.~\ref{subfig:TP:J1839-1238} illustrate the existence of nulling and subpulse drifting phenomena. The nulling fraction is estimated to be 21.1$\pm$1.6\% from the on-pulse energy histogram of single pulses (Fig.~\ref{subfig:Hist:J1839-1238}), and nulling gives rise to the low‑frequency modulation feature in the fluctuation spectra in Fig.~\ref{subfig:fluctu:J1839-1238}. In addition, the negative drift feature in 2DFS exhibits the centroid frequencies of $1/P_3=0.172\pm0.003$ and $1/P_2=-91\pm3$, corresponding to periodicities of $P_3=5.8\pm0.1$ periods and $P_2=-3.9\pm0.1$ degrees.

\subsection{J1840+0151g}
\label{subsec:J1840+0151g}

PSR J1840+0151g was discovered in the FAST GPPS survey \citep{Han2021,han2025}. 

This pulsar was observed by FAST on 20210513 and 20250620 for 15 and 5 minutes, respectively. From the data of 20250620, a rotation period $P=1.7504$~s and a dispersion measure $D\!M=69.4~{\rm cm^{-3}\,pc}$ were determined. The single pulse sequence in Fig.~\ref{subfig:TP:J1840+0151g} shows the subpulse drifting phenomenon. From fluctuation spectra in Fig.~\ref{subfig:fluctu:J1840+0151g}, centroid frequencies of the positive drift feature are estimated to be $1/P_3=0.196\pm0.002$ and $1/P_2=35\pm4$, corresponding to periodicities of $P_3=5.1\pm0.1$ periods and $P_2=10\pm1$ degrees.

\subsection{J1840-0141g}
\label{subsec:J1840-0141g}

PSR J1840-0141g was discovered in the FAST GPPS survey \citep{Han2021,han2025}. 

This pulsar was observed by FAST on 20220118 for 5 minutes, yielding a rotation period $P=1.9269$~s and a dispersion measure $D\!M=278.2~{\rm cm^{-3}\,pc}$ from this observation. 
Single pulse sequences in Fig.~\ref{subfig:TP:J1840-0141g} indicate the existence of nulling of subpulse drifting behaviors. The nulling fraction of this observation is estimated to be 44$\pm$5\% from the on-pulse energy histogram in Fig.~\ref{subfig:Hist:J1840-0141g}. The drifting parameters are obtained from the intensity correlation of adjacent single pulses of Nos. 69-82 and 102-108 (Fig.~\ref{subfig:Corre:J1840-0141g}), that are $D=0.7\pm0.2$ degrees per period and $P_2=2.13\pm0.05^\circ$.

\begin{figure}[htpb]
\centering
\includegraphics[width=0.21\textwidth, angle=0]{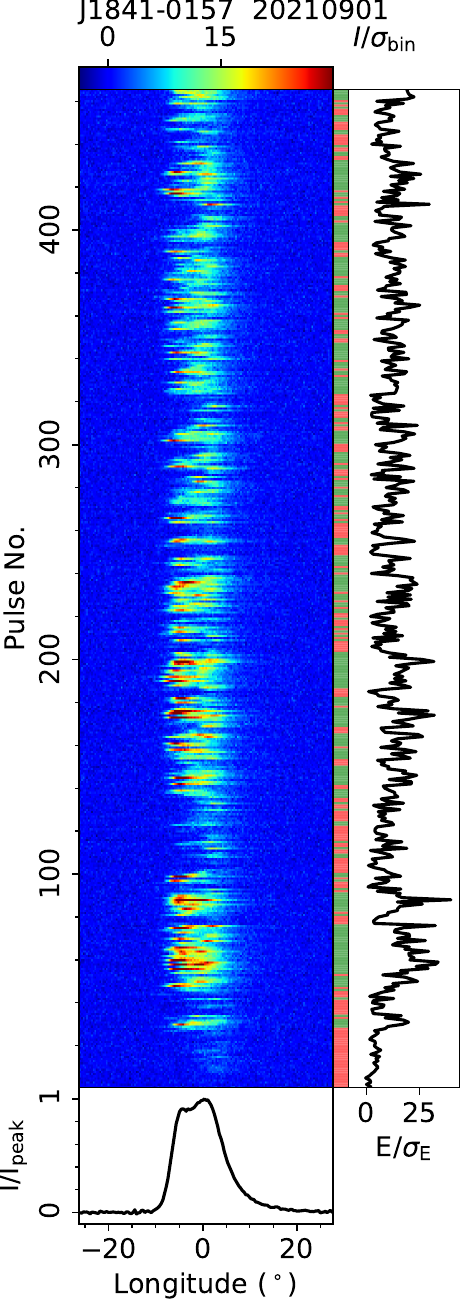}
\figcaption{Single pulse sequence of PSR J1841-0157 from the FAST observation on 20210901. 
\label{subfig:TP:J1841-0157}}
\end{figure}

\begin{figure}[htpb]
\centering
\includegraphics[width=0.39\textwidth, angle=0]{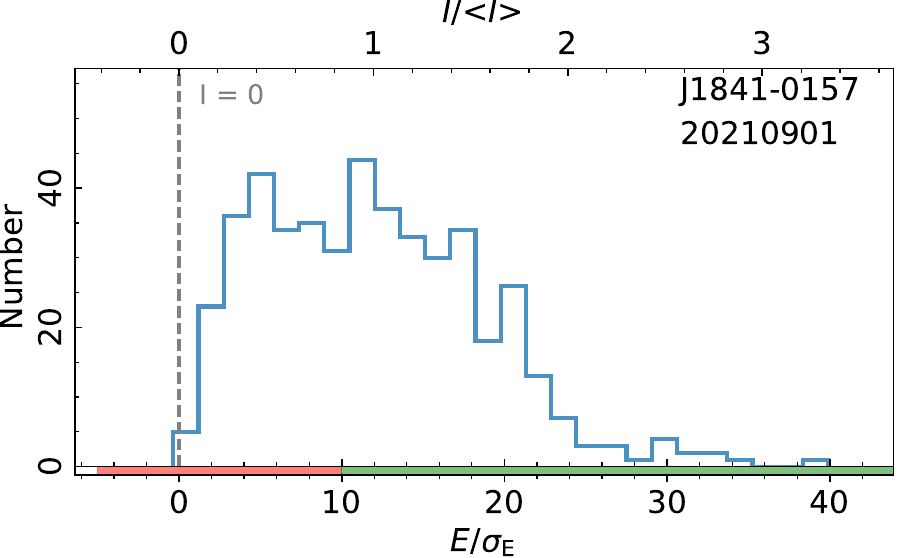}
\figcaption{Integral energy histogram of the leading part in a mean pulse profile of PSR J1841-0157 from the FAST observation on 20210901. 
\label{subfig:Hist:J1841-0157}}
\end{figure}

\begin{figure}[htpb]
\centering
\includegraphics[width=0.37\textwidth, angle=0]{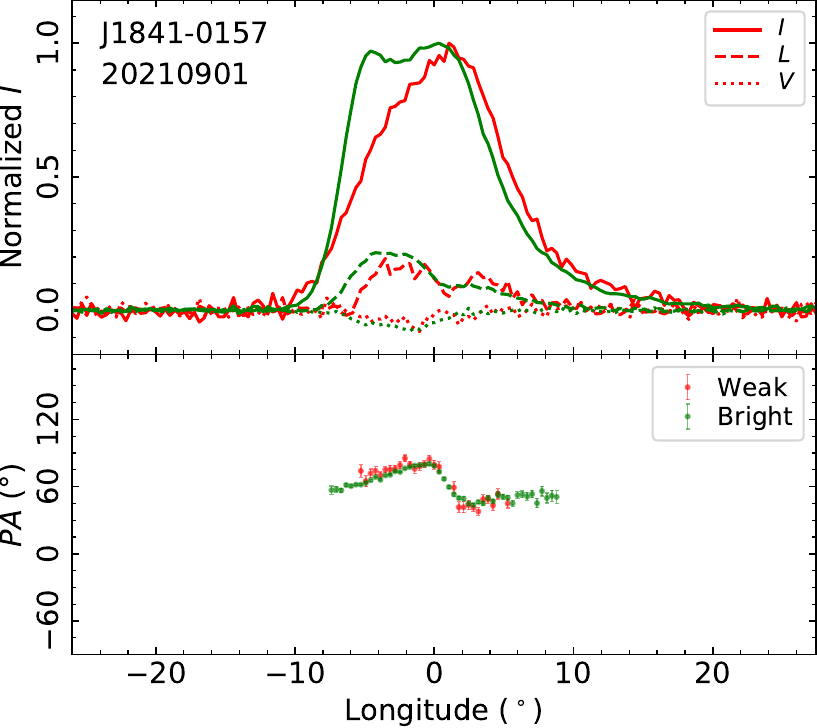}
\figcaption{Mean polarization profiles (the top panel) for the two emission modes of PSR J1841-0157 from the FAST observation on 20210901, as well as the averaged PA curves (the bottom panel). 
\label{subfig:PolModes:J1841-0157}}
\end{figure}

\subsection{J1840-0626}
\label{subsec:J1840-0626}

PSR J1840-0626 was discovered by \citet{Keith2009} in the data from the Parkes Multi-beam Pulsar Survey. 

This pulsar was observed by FAST on 20251015 for 15 minutes, and a rotation period $P=1.8935$~s and a dispersion measure $D\!M=760.5~{\rm cm^{-3}\,pc}$  were determined. The single pulse sequence in Fig.~\ref{subfig:TP:J1840-0626} shows mode changes between the weak and bright emission modes. Single pulses of two emission modes are distinguished from the 3-period smoothed energy variation histogram (Fig.~\ref{subfig:Hist:J1840-0626}). In the figures, weak and bright modes are labeled with red and green colors, respectively. 

\subsection{J1840-0809}
\label{subsec:J1840-0809}

PSR J1840-0809 was discovered in the Parkes multibeam pulsar survey \citep{hfs+04}. This pulsar exhibits subpulse drifting behavior \citep{Song2023}. For the leading component, they measured a positive drift with $P_3=2.18\pm0.04$ periods and $P_2=10.1^{+0.5}_{-0.6}$ degrees, as well as a negative drifting with $P_3=12\pm3$ periods and $P_2=-3.2^{+0.5}_{-3}$ degrees. The trailing component was reported to have a positive drifting characterized by $P_3=2.17\pm0.04$ periods and $P_2=100^{+85}_{-76}$ degrees.

This pulsar was observed by FAST on 20250201 for 6 minutes, deriving a rotation period $P=0.9556$~s and a dispersion measure $D\!M=350.5~{\rm cm^{-3}\,pc}$. Single pulse sequences in Fig.~\ref{subfig:TP:J1840-0809} reveal a positive drift with $P_3\sim$2 periods, as well as a low-frequency negative drift superimposed on it. 
The drifting properties are estimated from the fluctuation spectra displayed in Fig.~\ref{subfig:fluctu:J1840-0809}. There are two features in 2DFS of the leading profile part: a positive drift feature with the centroid of $1/P_3=0.461\pm0.001$ and $1/P_2=45\pm3$, corresponding to $P_3=2.169\pm0.003$ periods and $P_2=8.0\pm0.5$ degree; and a low-frequency modulated negative drift feature with the centroid of $1/P_3=0.045\pm0.001$ and $1/P_2=-33\pm3$, corresponding to $P_3=22\pm1$ periods and $P_2=-11\pm1$ degrees. For the trailing profile part, the centroid of the main drift feature is characterized by $1/P_3=0.463\pm0.002$ and $1/P_2=42\pm6$, yielding $P_3=2.16\pm0.01$ periods and $P_2=9\pm1$ degrees.

\subsection{J1841+0912}
\label{subsec:J1841+0912}

PSR J1841+0912 was discovered by \citet{Manchester1978} using observations at the Molonglo Radio Observatory and the Australian National Radio Astronomy Observatory, Parkes. 
From previous studies, the nulling fraction was estimated to be no more than 5\% at 430 MHz \citep{Weisberg1986} and less than 2\% at 327 MHz \citep{Herfindal2009}. 
This pulsar also has modulation phenomena \citep{Weltevrede2007,Herfindal2009}. \citet{Song2023} reported the drifting parameters of $P_3=51\pm39$ periods and $P_2=68^{+37}_{-44}$ degrees.

The pulsar was observed by FAST on 20230206 for 15 minutes, yielding a rotation period $P=0.3813$~s and a dispersion measure $D\!M=49.2~{\rm cm^{-3}\,pc}$. 
The nulling fraction of this observation is about 1.0$\pm$0.1\% from the on-pulse energy histogram in Fig.~\ref{subfig:Hist:J1841+0912}, with the duration of nulls being no more than 2 periods (Fig.~\ref{subfig:TP:J1841+0912}). In fluctuation spectra, there is a low-frequency drift feature with the centroid frequencies of $1/P_3=0.020\pm0.001$ and $1/P_2=4\pm1$, which correspond to $P_3=49\pm1$ periods and $P_2=100\pm31^\circ$.

\begin{figure}[htpb]
\centering
\includegraphics[width=0.22\textwidth, angle=0]{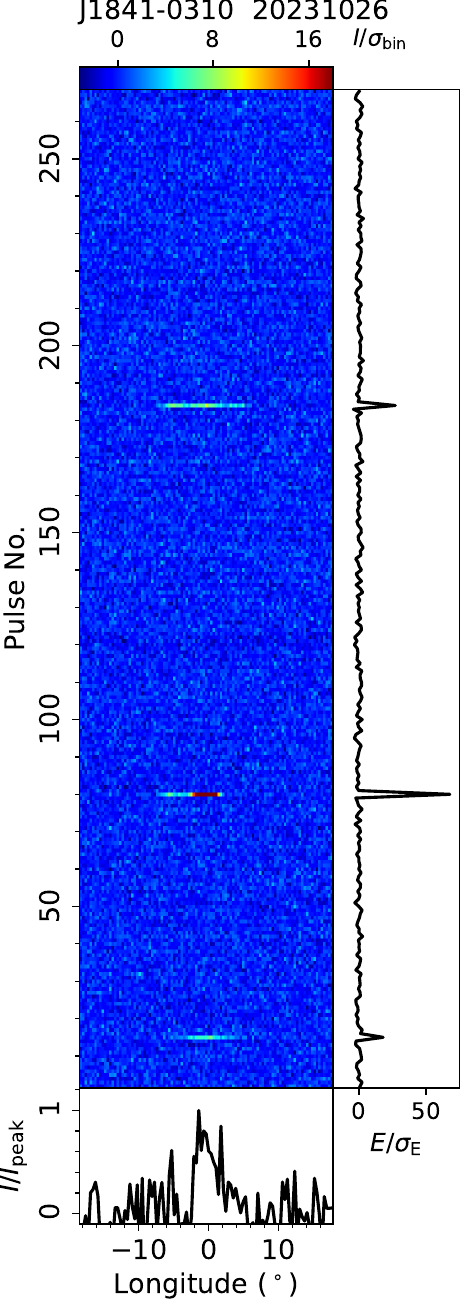}
\includegraphics[width=0.22\textwidth, angle=0]{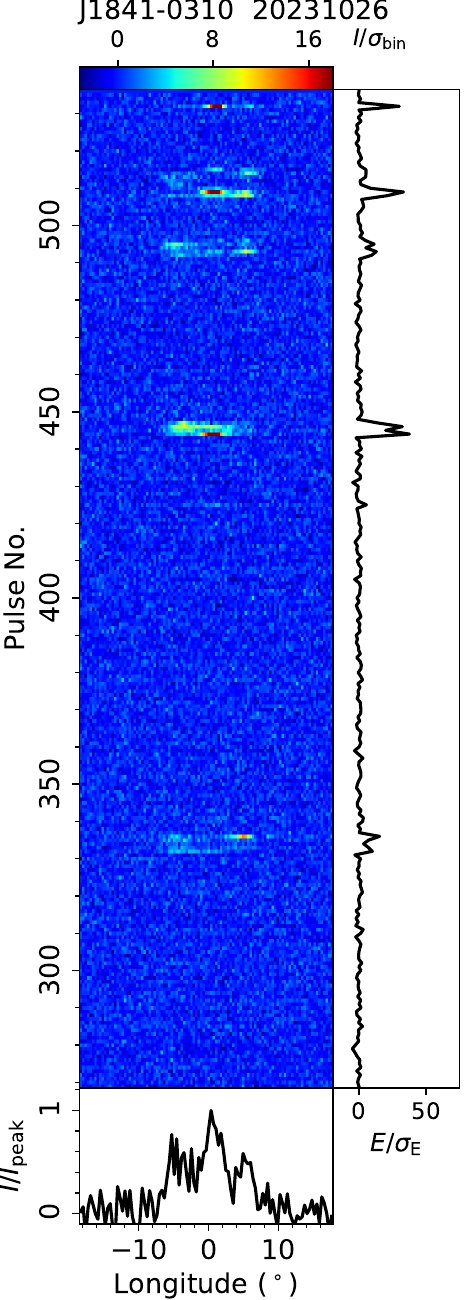}
\figcaption{Single pulse sequences of PSR J1841-0310 from the FAST observation on 20231026. 
\label{subfig:TP:J1841-0310}}
\end{figure}

\begin{figure}[htpb]
\centering
\includegraphics[width=0.39\textwidth, angle=0]{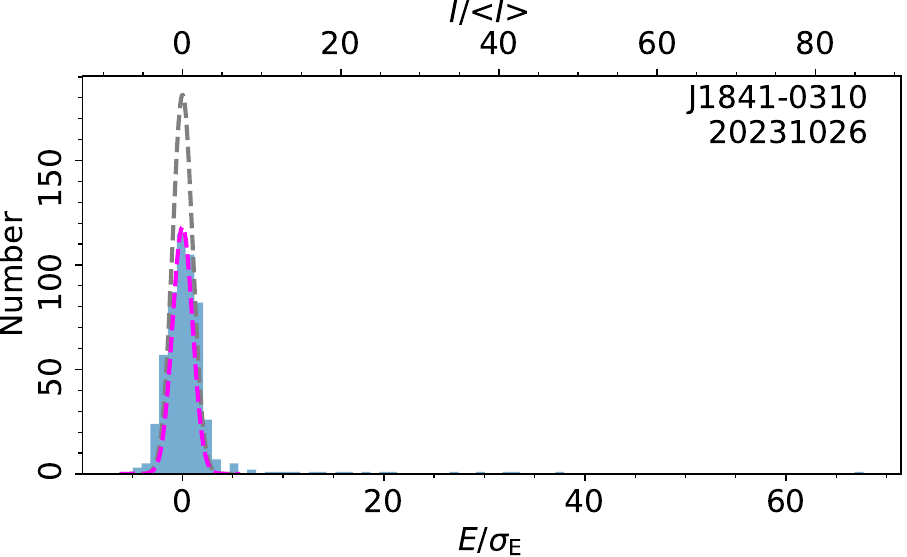}
\figcaption{On-pulse energy histogram of single pulses of PSR J1841-0310 from the FAST observation on 20231026.
\label{subfig:Hist:J1841-0310}}
\end{figure}

\begin{figure}[htpb]
\centering
\includegraphics[width=0.44\textwidth, angle=0]{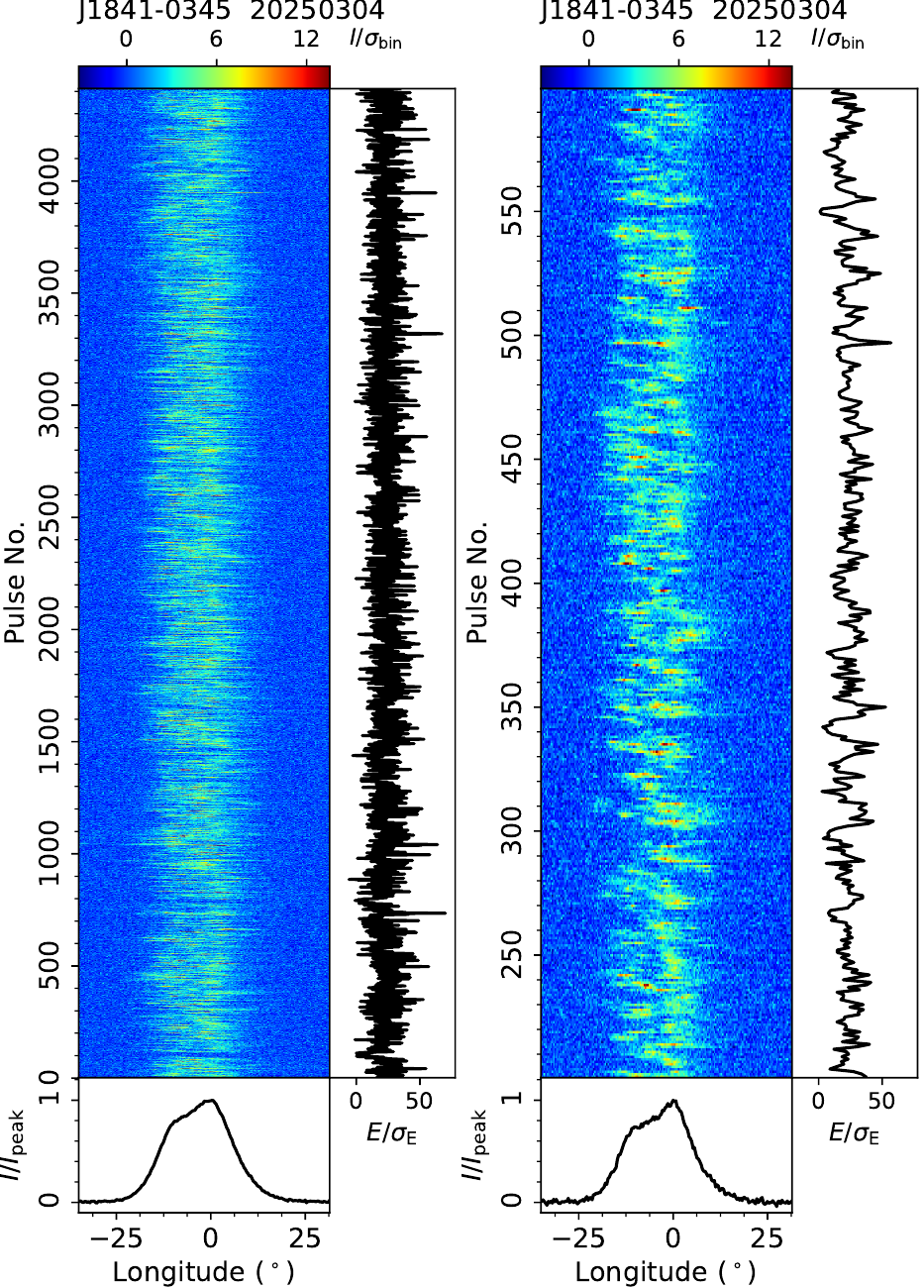}
\figcaption{Single pulse sequence of PSR J1841-0345 from the FAST observation on 20250304, and a zoomed-in view of pulses No. 200-600.
\label{subfig:TP:J1841-0345}}
\end{figure}

\begin{figure}[htpb]
\centering
\includegraphics[width=0.44\textwidth, angle=0]{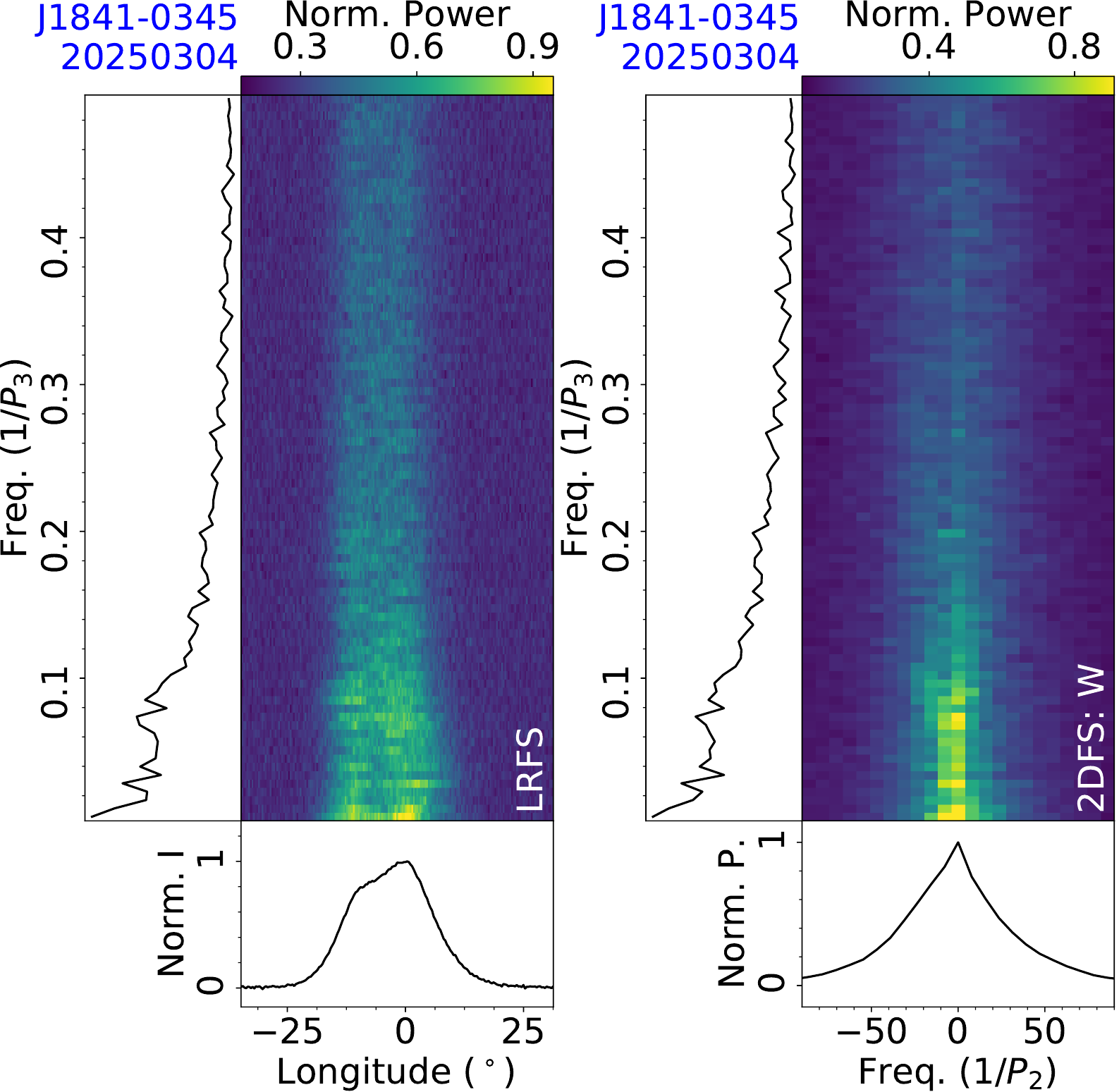}
\figcaption{Fluctuation analysis of PSR J1841-0345 for the observation on 20250304, with LRFS and 2DFS for the on-pulse region of a mean pulse profile.
\label{subfig:fluctu:J1841-0345}}
\end{figure}

\begin{figure}[htpb]
\centering
\includegraphics[width=0.22\textwidth, angle=0]{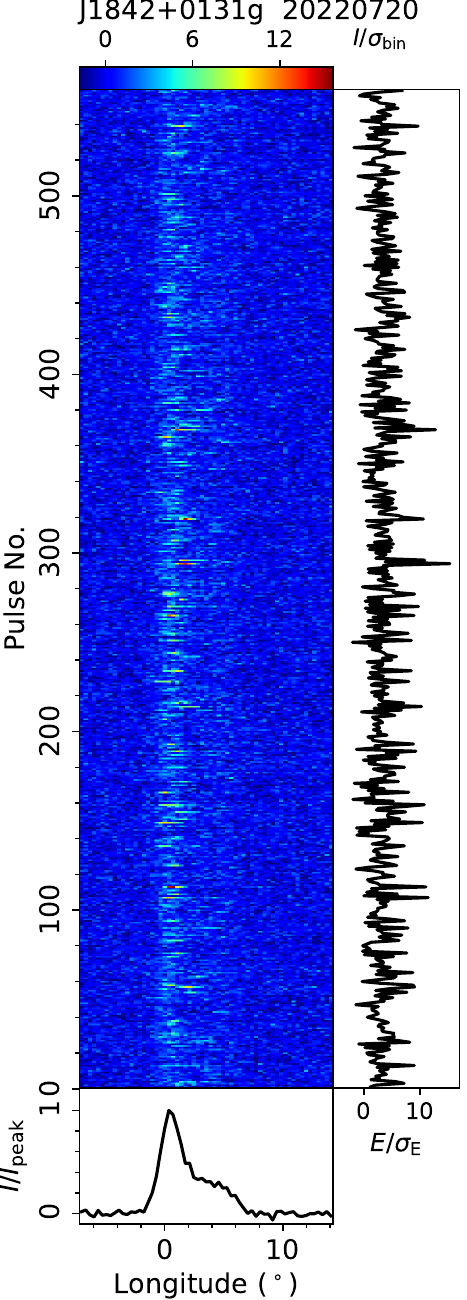}
\includegraphics[width=0.22\textwidth, angle=0]{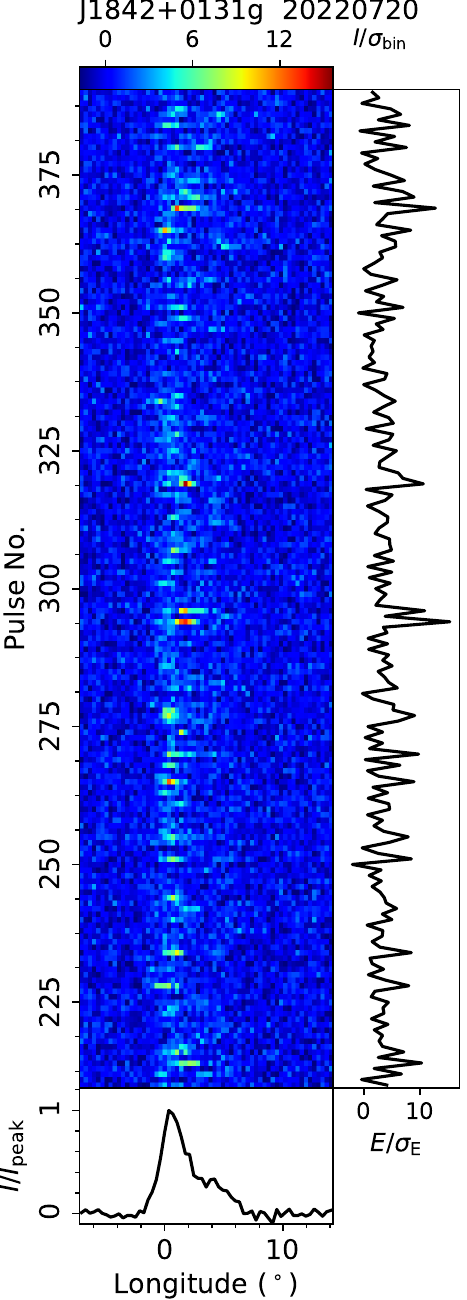}
\figcaption{Single pulse sequence of PSR J1842+0131g from the FAST observation on 20220720, and a zoomed-in view of pulses No. 210-390.
\label{subfig:TP:J1842+0131g}}
\end{figure}

\begin{figure}[htpb]
\centering
\includegraphics[width=0.22\textwidth, angle=0]{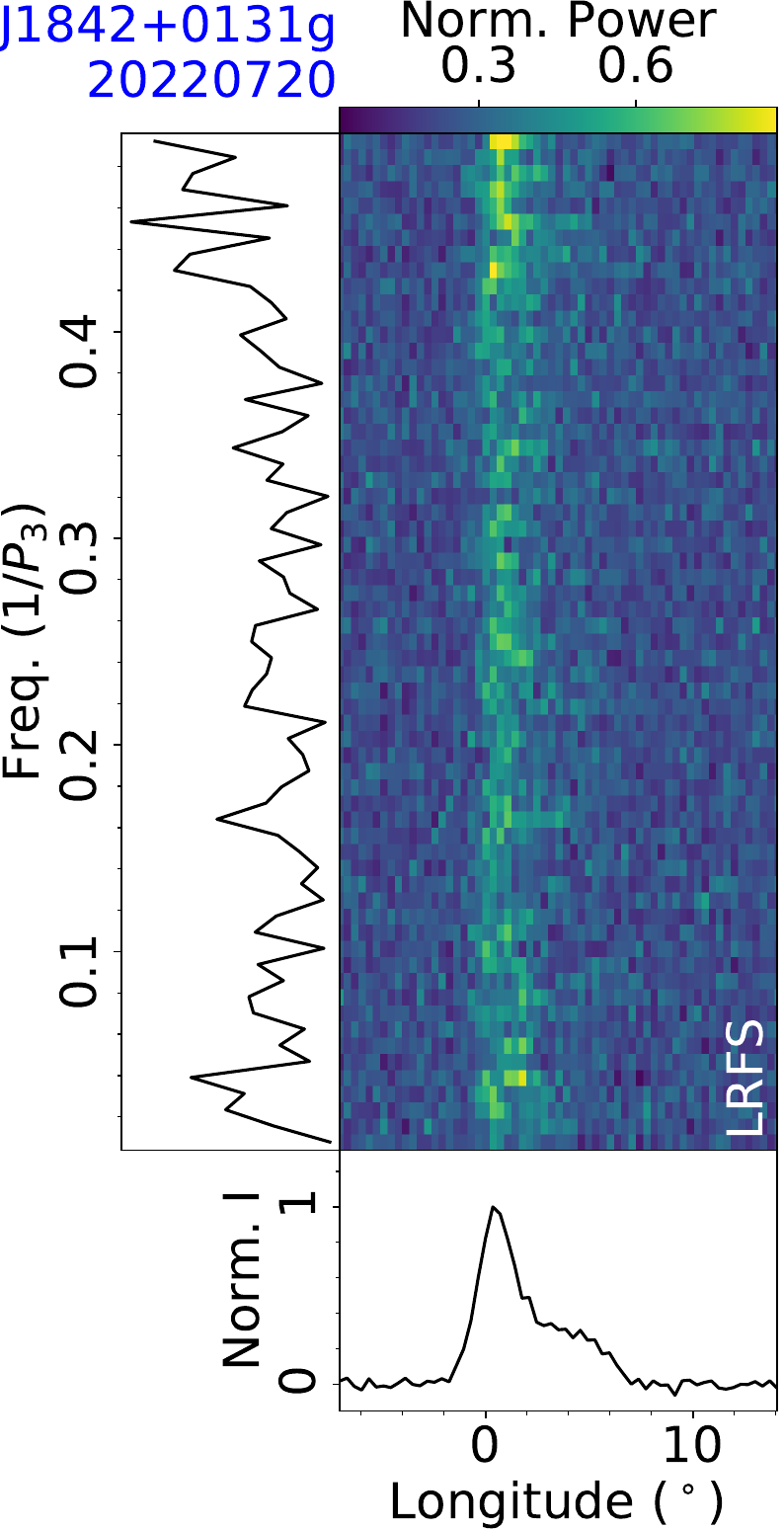}
\includegraphics[width=0.22\textwidth, angle=0]{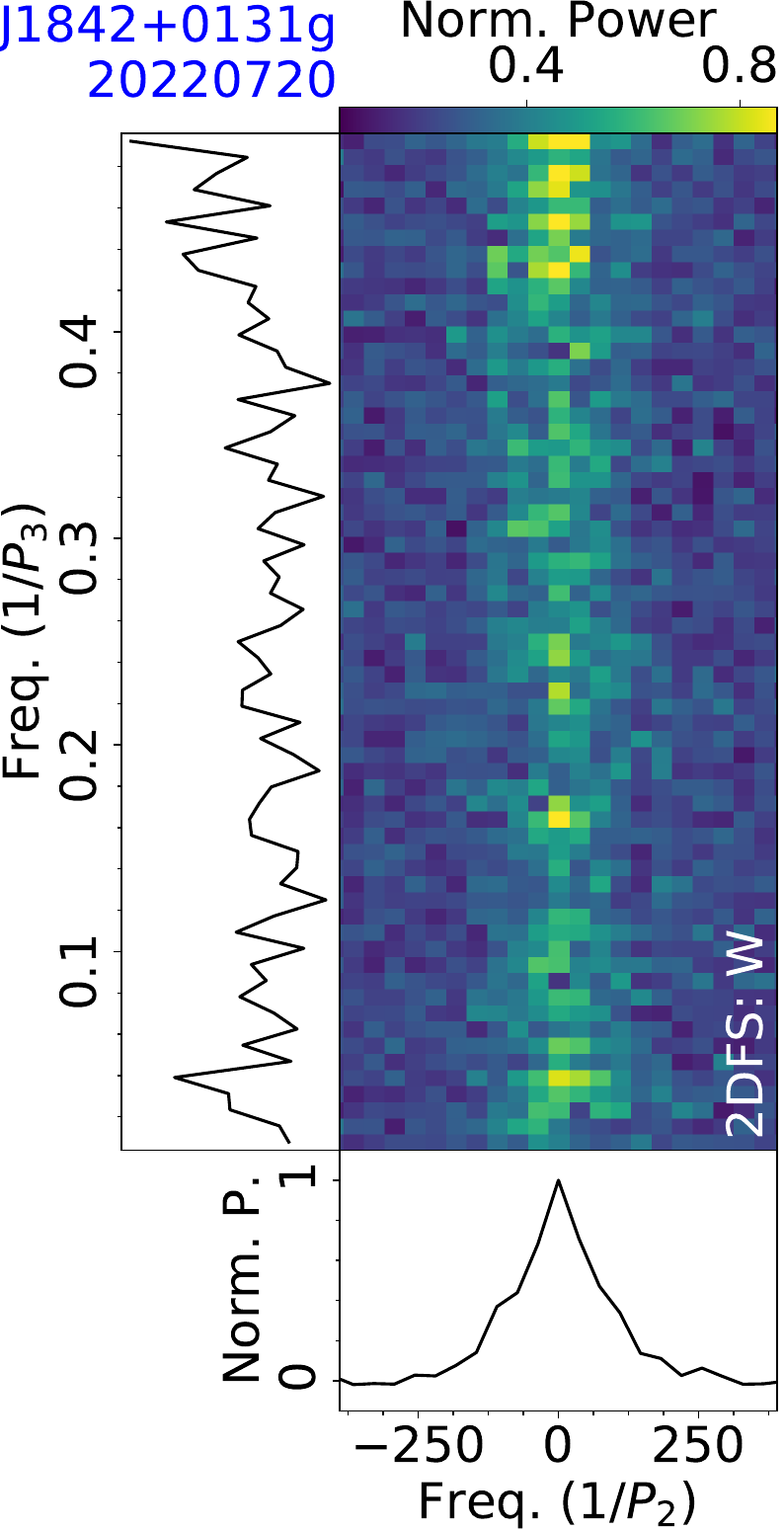}
\figcaption{Fluctuation analysis of PSR J1842+0131g for the observation on 20220720, with LRFS and 2DFS for the on-pulse region of a mean pulse profile.
\label{subfig:fluctu:J1842+0131g}}
\end{figure}

\subsection{J1841-0157}
\label{subsec:J1841-0157}

PSR J1841-0157 was discovered in the Parkes multibeam pulsar survey \citep{hfs+04}. 

The pulsar was observed by FAST on 20210901 for 5 minutes, deriving a rotation period $P=0.6634$~s and a dispersion measure $D\!M=474.7~{\rm cm^{-3}\,pc}$ from this observation. 
The single pulse sequence is displayed in Fig.~\ref{subfig:TP:J1841-0157}. The histogram in Fig.~\ref{subfig:Hist:J1841-0157} of integral energy from the leading part of the profile is used to distinguish the weak or bright mode of single pulses. 
For the bright emission mode, the leading part is stronger than the trailing component. However, the leading component seems not to be obvious in the average pulse profile in the weak mode (Fig.~\ref{subfig:PolModes:J1841-0157}).

\begin{figure}[htpb]
    \centering
	\includegraphics[height=0.82\textheight]{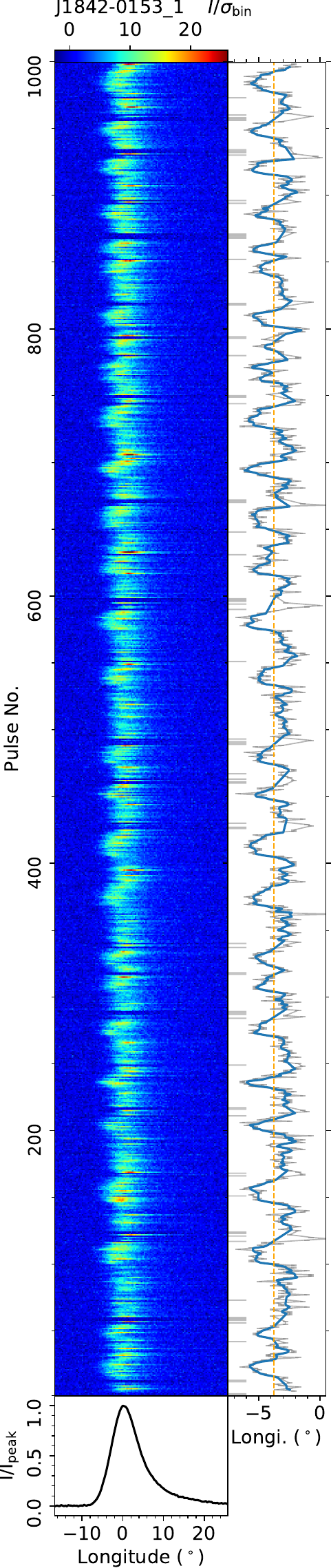}
	\includegraphics[height=0.82\textheight]{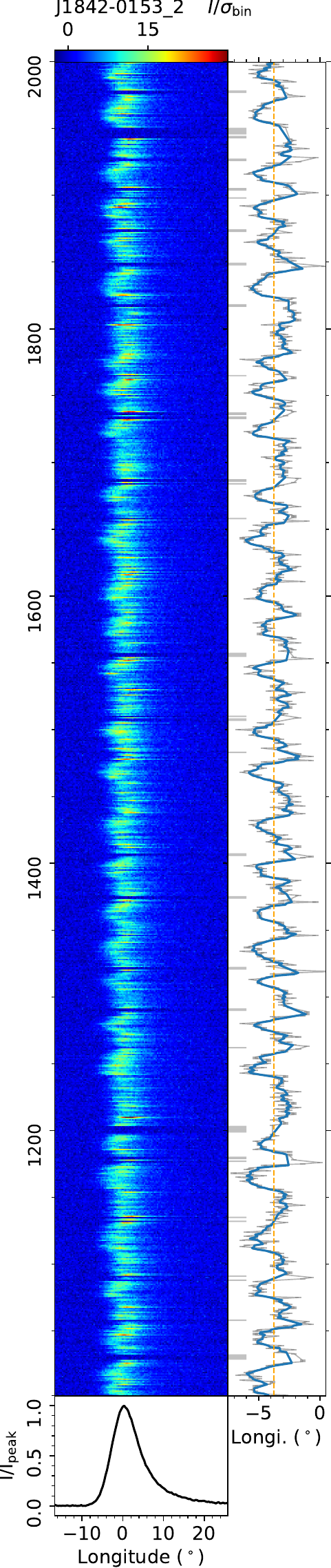}
    \figcaption{{\it The left sub-panel} displays single-pulse sequences of PSR J1842-0153 from the FAST observation on 20230328. {\it The central sub-panel} shows the variation of starting longitude of single pulses (grey line) and the smoothed line (blue line), with orange splitting dashed line from Fig.~\ref{fig:J1842-0153:longiHist}. Horizontal grey lines correspond to nulls. {\it The right sub-panel} shows the variance between adjacent single pulses on smoothed starting longitudes in the central sub-panel. -- to be continued.}
    \label{fig:J1842-0153:sinPulSequ}
    \addtocounter{figure}{-1}
\end{figure}

\begin{figure}[htpb]
    \centering
	\includegraphics[height=0.82\textheight]{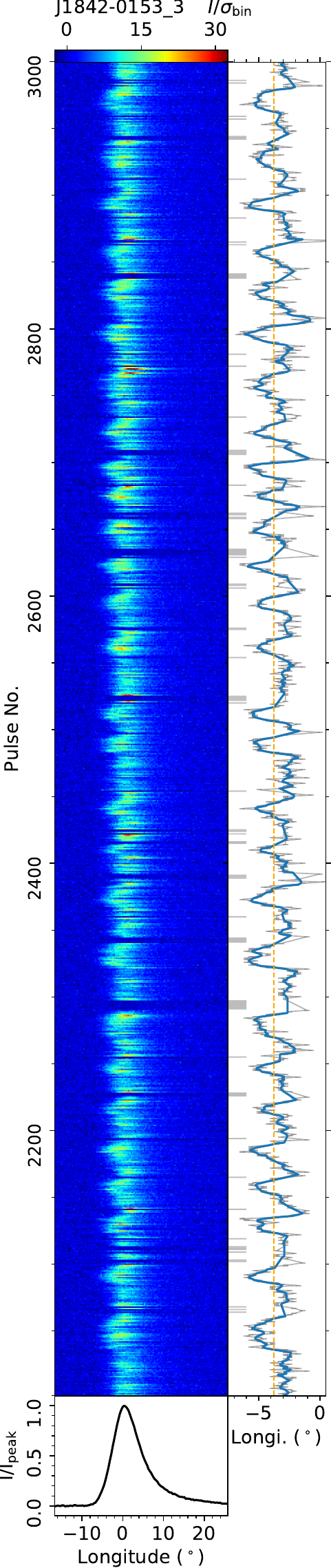}
	\includegraphics[height=0.82\textheight]{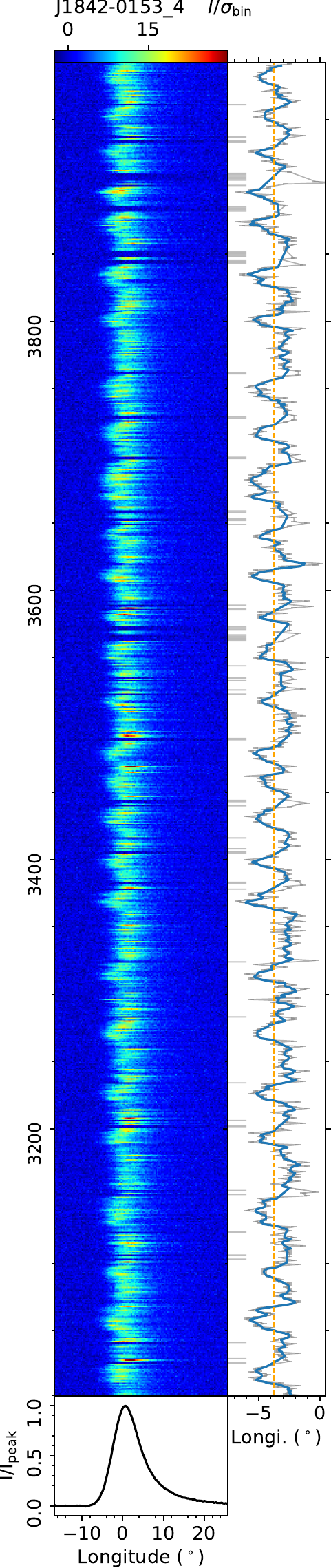}
    \figcaption{Continued and ended.}
\end{figure}

\begin{figure}[htpb]
\centering
\includegraphics[width=0.22\textwidth, angle=0]{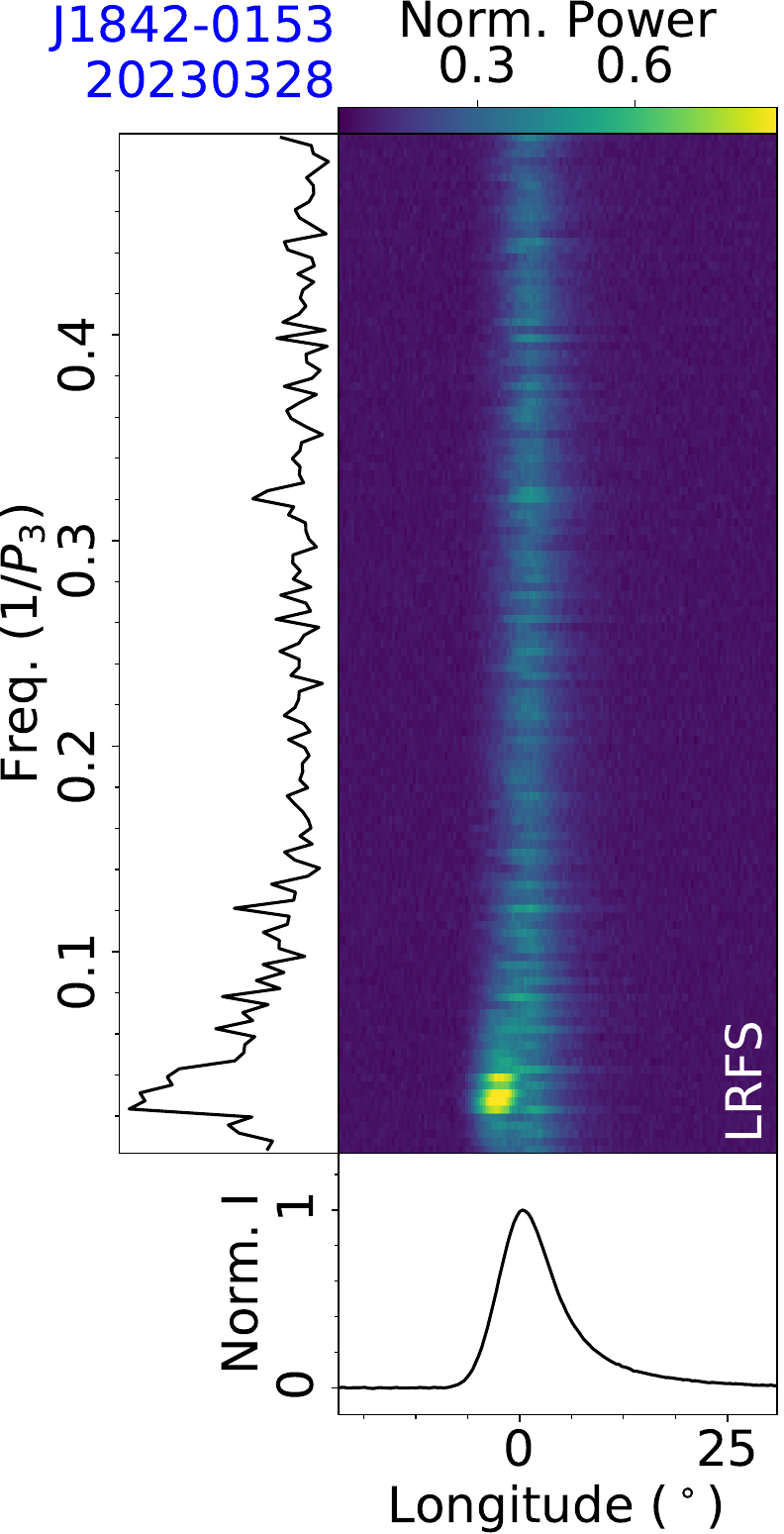}
\includegraphics[width=0.22\textwidth, angle=0]{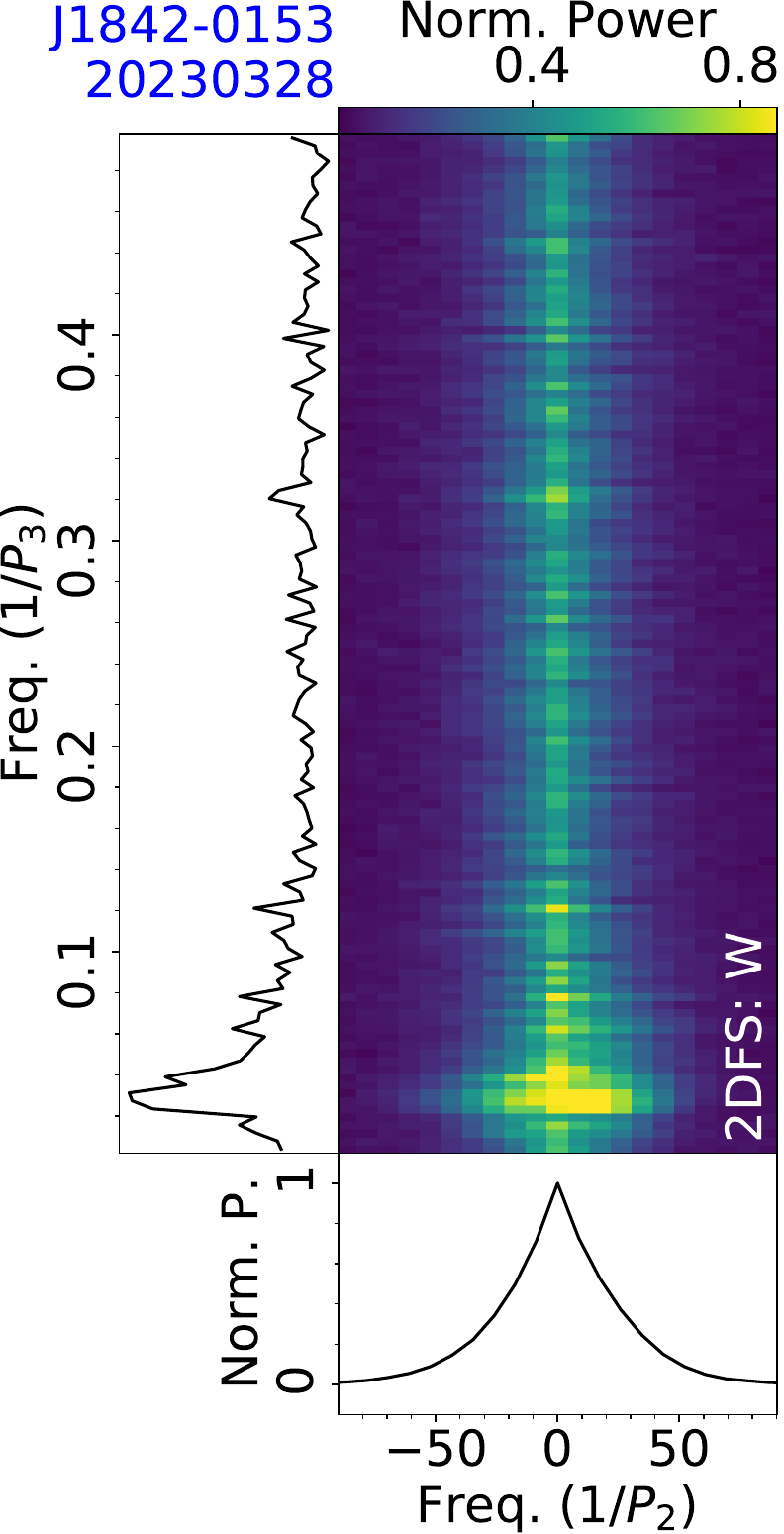}
\figcaption{Fluctuation analysis of PSR J1842-0153 from the FAST observation on 20230328, with LRFS and 2DFS for the on-pulse region of a mean pulse profile.
\label{subfig:fluctu:J1842-0153}}
\end{figure}

\begin{figure}[htpb]
    \centering
    \includegraphics[width=0.39\textwidth]{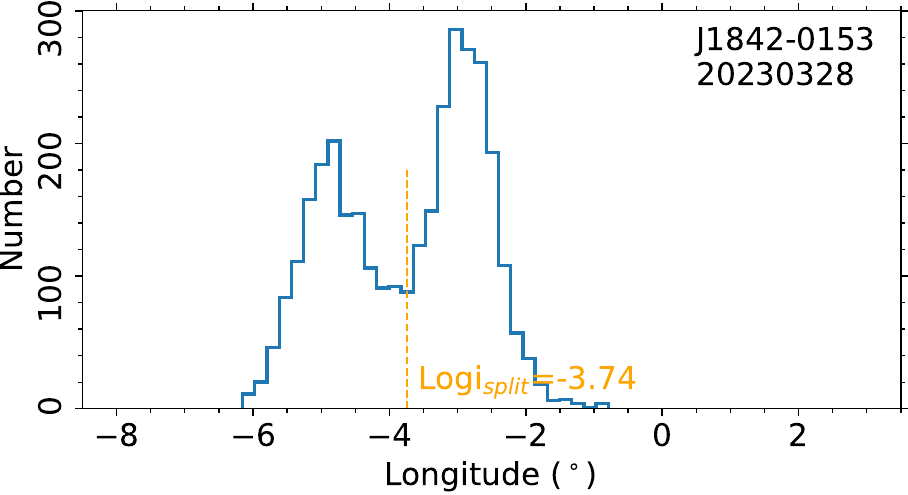}
    \figcaption{Histogram of smoothed single-pulse starting longitudes, which is blue line in the central sub-panel of Fig.~\ref{fig:J1842-0153:sinPulSequ}. 
   }
   \label{fig:J1842-0153:longiHist}
\end{figure}
   
\begin{figure}[htpb]
    \centering
    \includegraphics[width=0.39\textwidth]{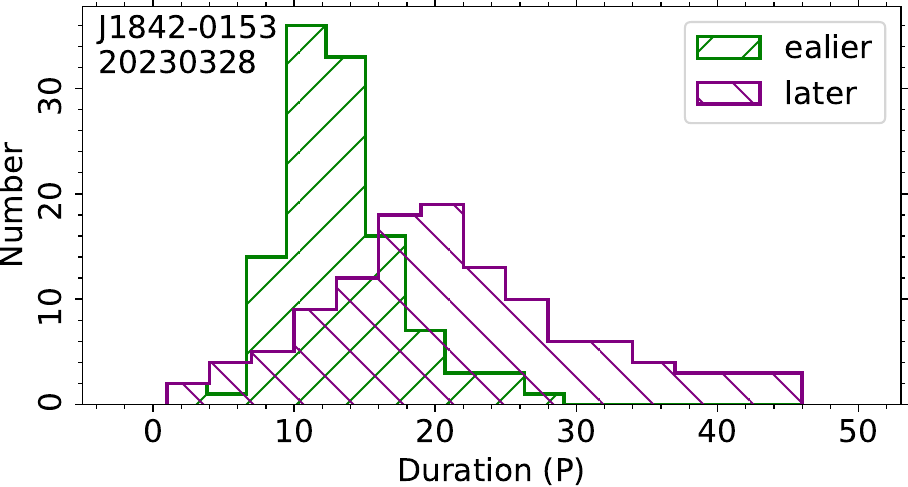}
   \figcaption{Duration periods of emission in the earlier longitude region (green) and later longitude region (purple), that are distinguished from Fig.~\ref{fig:J1842-0153:longiHist}.
   }
   \label{fig:J1842-0153:duraDiffLongi}
\end{figure}



\begin{figure}[htpb]
\centering
\includegraphics[width=0.39\textwidth, angle=0]{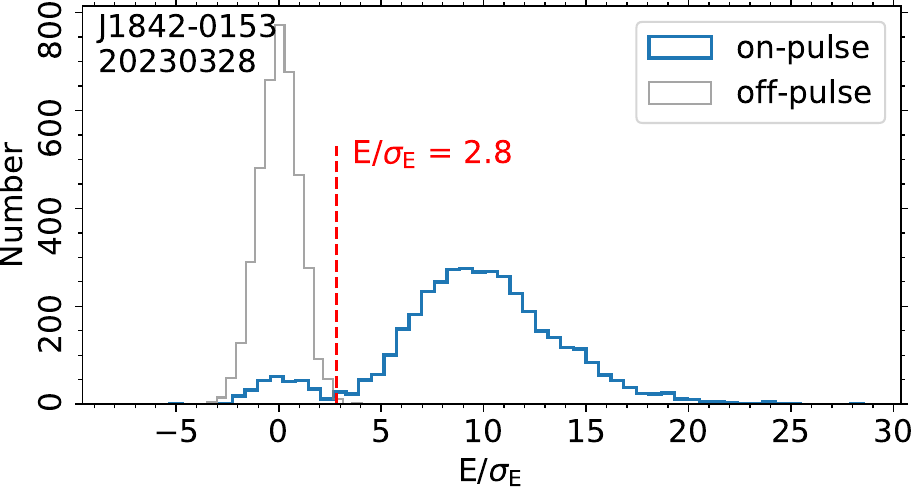}
\figcaption{On-pulse energy histogram of PSR J1842-0153 from the FAST observation on 20230328. \label{subfig:Hist:J1842-0153}}
\end{figure}

\begin{figure}[htpb]
\centering
\includegraphics[width=0.42\textwidth, angle=0]{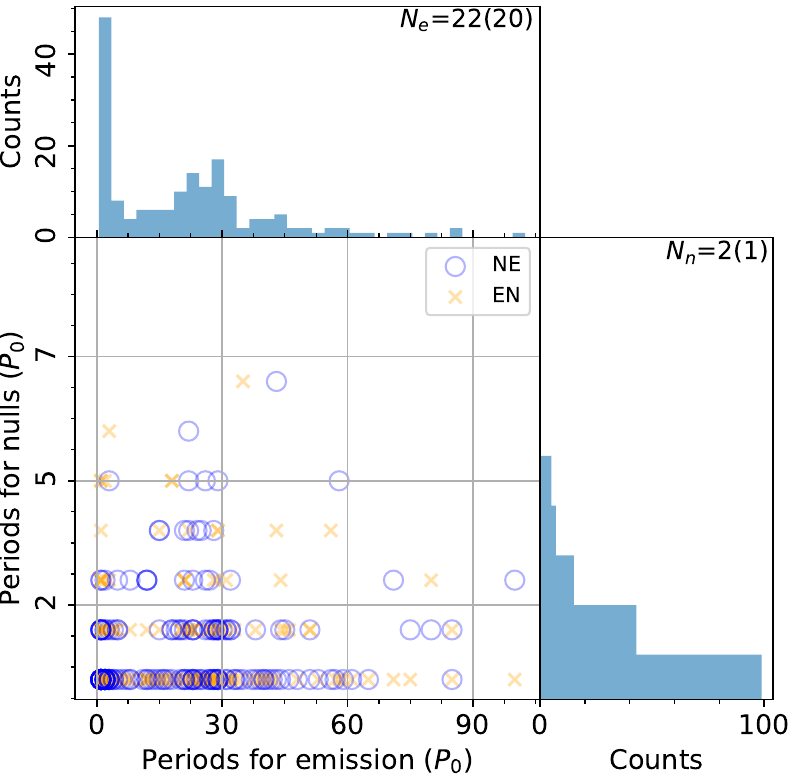}
\figcaption{Distribution of period numbers for continuous nulling $P_{\rm N}$ against period numbers for adjacent pulses $P_{\rm E}$ of PSR J1842-0153 observed by FAST on 20201218, as well as the duration histograms for the emission and null shown in the top and right panels, respectively. 
\label{subfig:scaleHist:J1842-0153}}
\end{figure}

\begin{figure}[htpb]
\centering
\includegraphics[width=0.39\textwidth, angle=0]{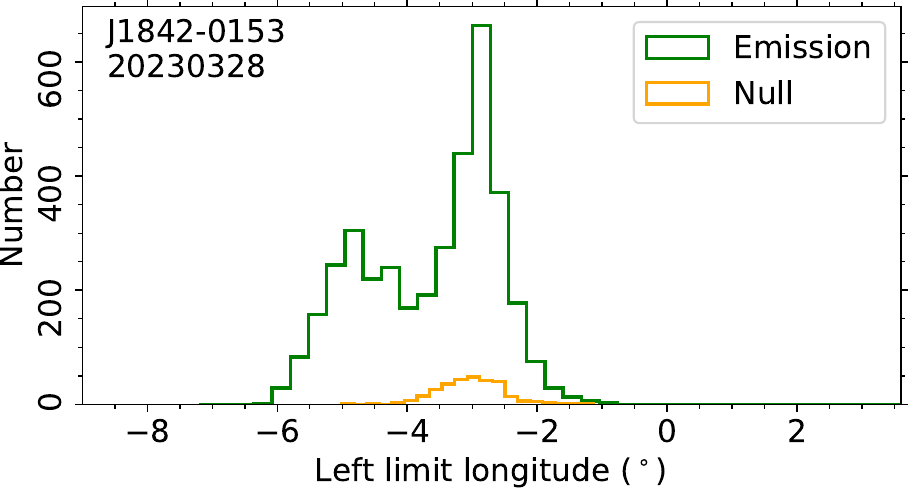}
\figcaption{Histogram of single-pulse starting longitudes which are smoothed and interpolated at nulls. Orange and blue lines correspond respectively to nulls and emissions. \label{subfig:longiHist:J1842-0153}}
\end{figure}

\subsection{J1841-0310}
\label{subsec:J1841-0310}

PSR J1841-0310 was first reported by \citet{hfs+04} from the Parkes multibeam pulsar survey. \citet{BurkeSpolaor2012} revealed that the pulsar appears to be nulling on 1352MHz.

This pulsar was observed by FAST on 20231026 for 15 minutes, 20240327 for 2 minutes and 20250721 for 5 minutes. From the 15-minute data, a rotation period $P=1.6578$~s and a dispersion measure $D\!M=223.8~{\rm cm^{-3}\,pc}$ were derived. 
Single pulse sequences of the observation on 20231026 are displayed in Fig.~\ref{subfig:TP:J1841-0310}, illustrating that the pulsar has the nulling phenomenon and the emission could last for just one period. The nulling fraction of this observation is estimated to be 65.2$\pm$6.0\% from the on-pulse integral energy histogram (Fig.~\ref{subfig:Hist:J1841-0310}).

\begin{figure}[htpb]
\centering
\includegraphics[width=0.22\textwidth, angle=0]{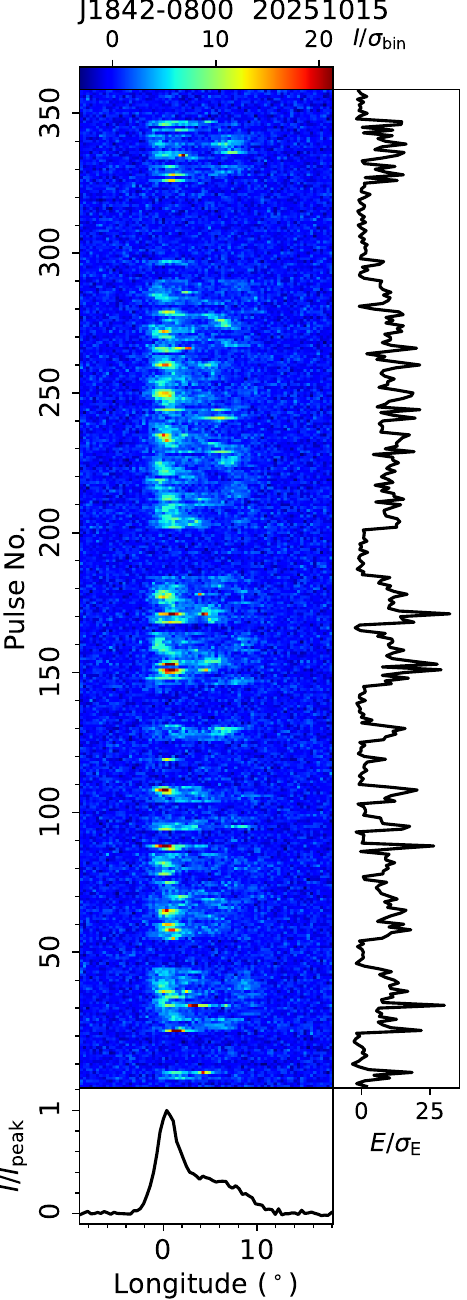}
\includegraphics[width=0.22\textwidth, angle=0]{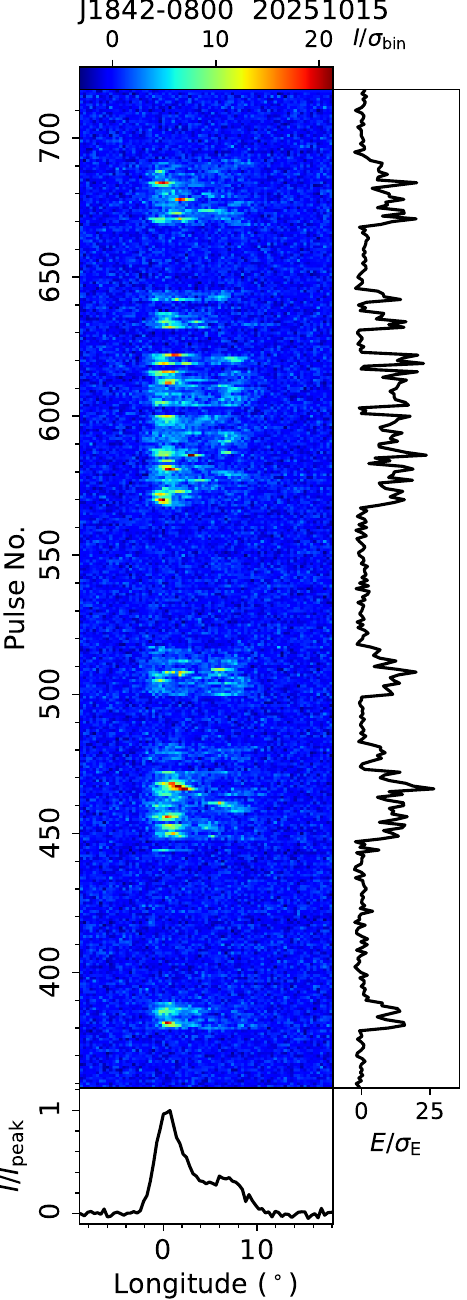}
\figcaption{Single pulse sequences of PSR J1842-0800 from the FAST observation on 20251015.
\label{subfig:TP:J1842-0800}}
\end{figure}

\begin{figure}[htpb]
\includegraphics[width=0.39\textwidth, angle=0]{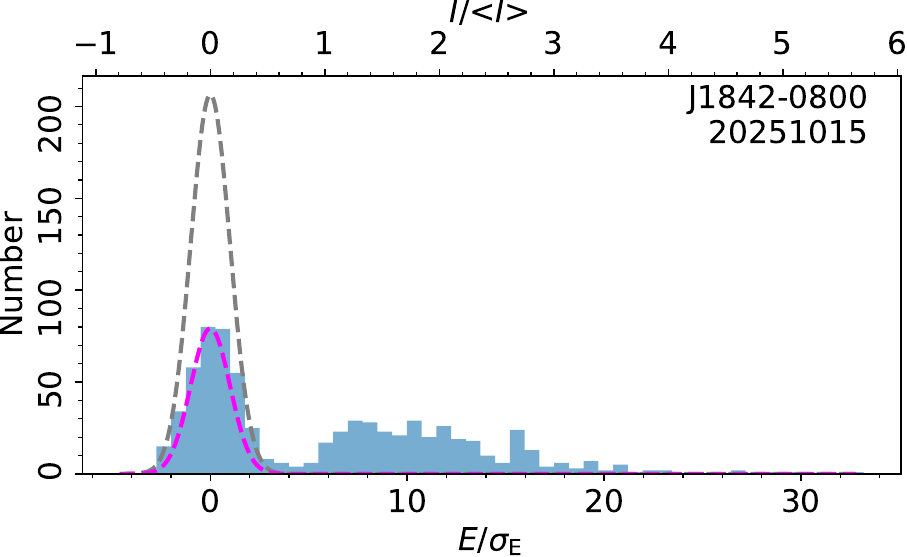}
\figcaption{On-pulse integral energy histogram of single pulses of PSR J1842-0800 from the FAST observation on 20251015.
\label{subfig:Hist:J1842-0800}}
\end{figure}

\subsection{J1841-0345}
\label{subsec:J1841-0345}

PSR J1841-0345 was discovered by \citet{Lorimer2000} with the Effelsberg 100-m radio telescope. Negative drifting periodicities of $P_3=27\pm18$ periods and $P_2=-258^{+160}_{-134}$ degrees were reported by \citet{Song2023}.

This pulsar was observed by FAST on 20250304 for 15 minutes, deriving a rotation period $P=0.2041$~s and a dispersion measure $D\!M=196.3~{\rm cm^{-3}\,pc}$. The single pulse sequence and a zoomed-in view of pulses No. 200-600 in Fig.~\ref{subfig:TP:J1841-0345} show an unsystematic negative drifting behavior. Fluctuation spectra in Fig.~\ref{subfig:fluctu:J1841-0345} display that the negative drift feature is widely distributed in $1/P_3$, consistent with the unsystematic drifting. The centroid of the feature in 2DFS is at $1/P_3=0.0528\pm0.0003$ and $1/P_2=-3.7\pm0.3$, corresponding to $P_3=18.9\pm0.1$ periods and $P_2=-99\pm8$ degrees.

\subsection{J1842+0131g}
\label{subsec:J1842+0131g}

PSR J1842+0131g was discovered in the FAST GPPS survey \citep{Han2021,han2025}. 

This pulsar was observed by FAST on 20220720 for 15 minutes, deriving a rotation period $P=1.5903$~s and a dispersion measure $D\!M=115.8~{\rm cm^{-3}\,pc}$ from this observation. 
Single pulse sequences in Fig.~\ref{subfig:TP:J1842+0131g} and fluctuation spectra in Fig.~\ref{subfig:fluctu:J1842+0131g} illustrate that this pulsar has a temporal modulation feature with a frequency of $1/P_3=0.461\pm0.003$, corresponding to $P_3=2.17\pm0.01$ periods.

\subsection{J1842-0153}
\label{subsec:J1842-0153}

PSR J1842-0153 was discovered in the Parkes Multibeam Pulsar Survey \citep{Morris2002}. $P_3$-only feature of $P_3=28\pm10$ periods has been reported by \citep{Song2023}.

This pulsar was observed by FAST on 20230328 for 70 minutes, deriving a rotation period $P=1.0542$~s and a dispersion measure $D\!M=426.5~{\rm cm^{-3}\,pc}$. 
Segments of single pulses are shown in Fig.~\ref{fig:J1842-0153:sinPulSequ}, which reveal two phenomena: nulling and subpulse drifting.

Significantly, the drifting direction changes and there is no preferred direction, resulting in 2DFS (Fig.~\ref{subfig:fluctu:J1842-0153}) almost symmetric about the vertical axis. The temporally modulated periodicity is estimated to be $P_3=30.7\pm0.9$ periods, which is the strongest for the leading edge in the profile (LRFS in Fig.~\ref{subfig:fluctu:J1842-0153}). The variation sequence of the single-pulse starting longitudes is displayed in grey in the central sub-panel of Fig.~\ref{fig:J1842-0153:sinPulSequ}, together with the smoothed line in blue. There are two distributions in the histogram of smoothed values of single-pulse starting longitudes in Fig.~\ref{fig:J1842-0153:longiHist}. Single pulses corresponding to these two distributions are distinguished using $-3.77^\circ$, labeled in the orange dashed line. Histograms of earlier and later starting longitude of single pulses are displayed in Fig.~\ref{fig:J1842-0153:duraDiffLongi}, illustrating that single pulses with earlier starting longitude tend to have a shorter duration, and the duration distribution is narrower. 

This pulsar also has nulling behavior. The nulling fraction is estimated to be 6.1\% from the intensity histogram in Fig.~\ref{subfig:Hist:J1842-0153}. 
From the distribution of continuous period numbers for adjacent nulling and emission states in Fig.~\ref{subfig:scaleHist:J1842-0153}, the duration of the nulling state is short, ranging from 1 to 7 periods. The duration of the emission state peaks at a short value, while a significant portion of the values cluster at a longer duration.
Single-pulse starting longitudes are smoothed and interpolated at nulls, and the histogram of longitudes related to nulling and emission states in Fig.~\ref{subfig:longiHist:J1842-0153} indicates that nulls tend to appear at later longitudes, reflecting the relation between nulling and drifting direction changes.

\begin{figure}[htpb]
\centering
\includegraphics[width=0.22\textwidth, angle=0]{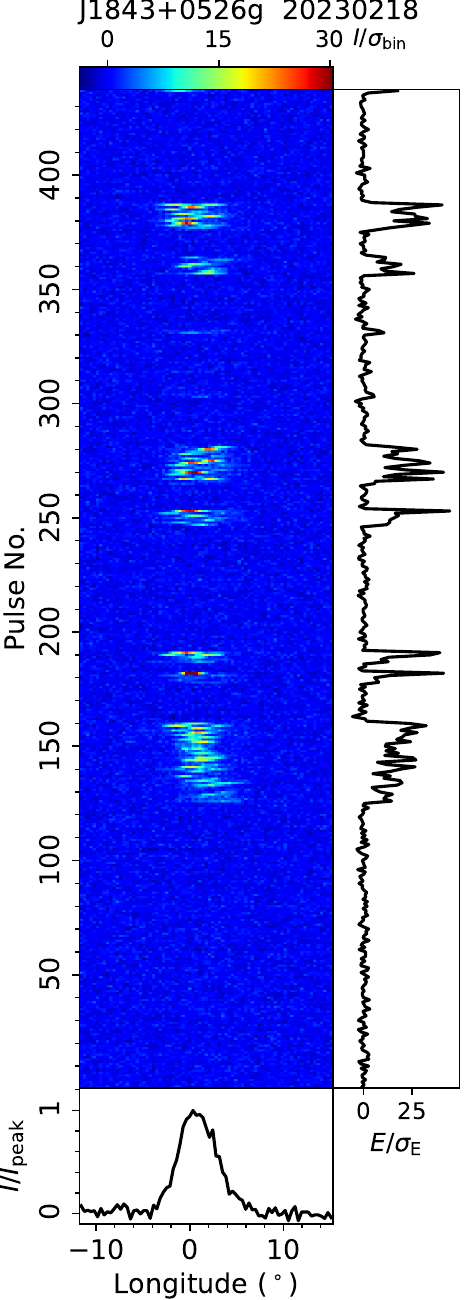}
\includegraphics[width=0.22\textwidth, angle=0]{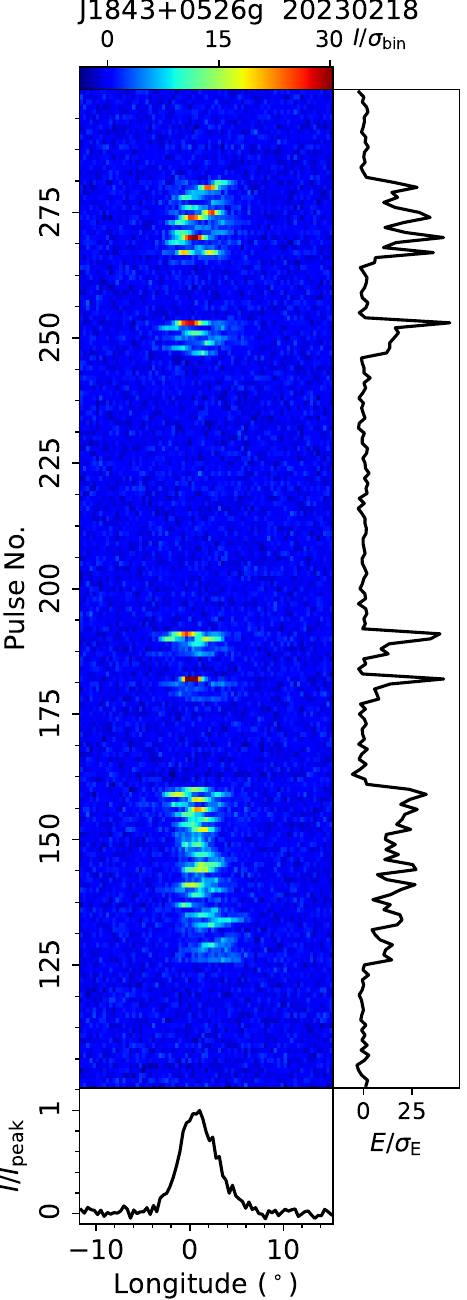}
\figcaption{Single pulse sequences of PSR J1843+0526g from the FAST observation on 20230218. \label{subfig:TP:J1843+0526g}}
\end{figure}

\begin{figure}[htpb]
\centering
\includegraphics[width=0.39\textwidth, angle=0]{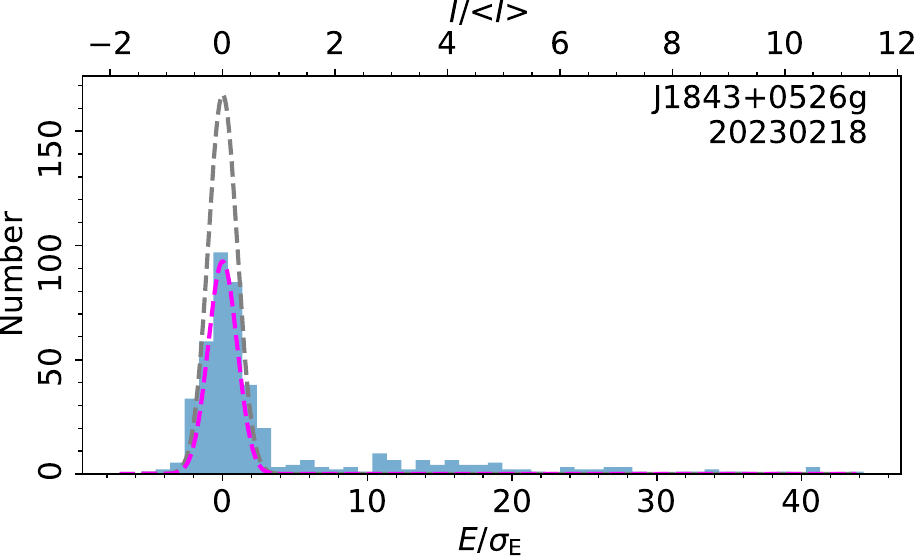}
\figcaption{On-pulse energy histogram of single pulses of PSR J1843+0526g from the FAST observation on 20230218. \label{subfig:Hist:J1843+0526g}}
\end{figure}

\begin{figure}[htpb]
\centering
\includegraphics[width=0.39\textwidth, angle=0]{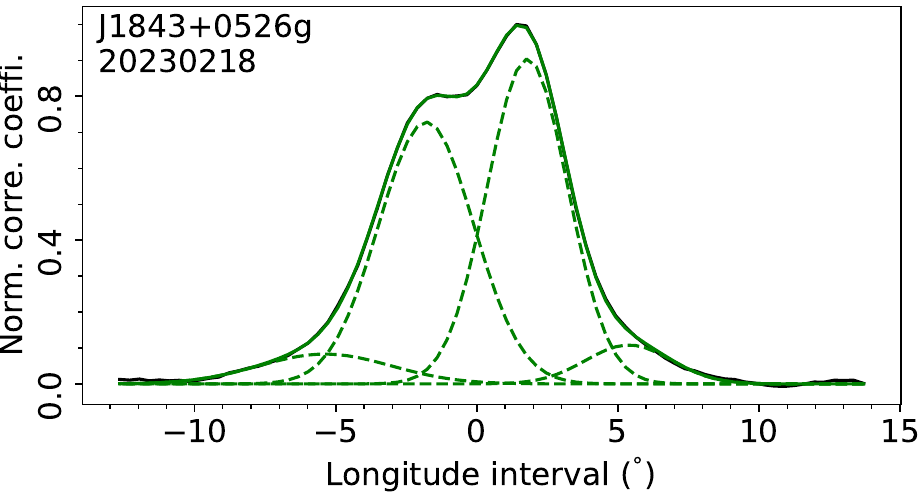}
\figcaption{Cross correlation of PSR J1843+0526g from the FAST observation on 20230218. \label{subfig:Corre:J1843+0526g}}
\end{figure}

\begin{figure}[htpb]
\centering
\includegraphics[width=0.22\textwidth, angle=0]{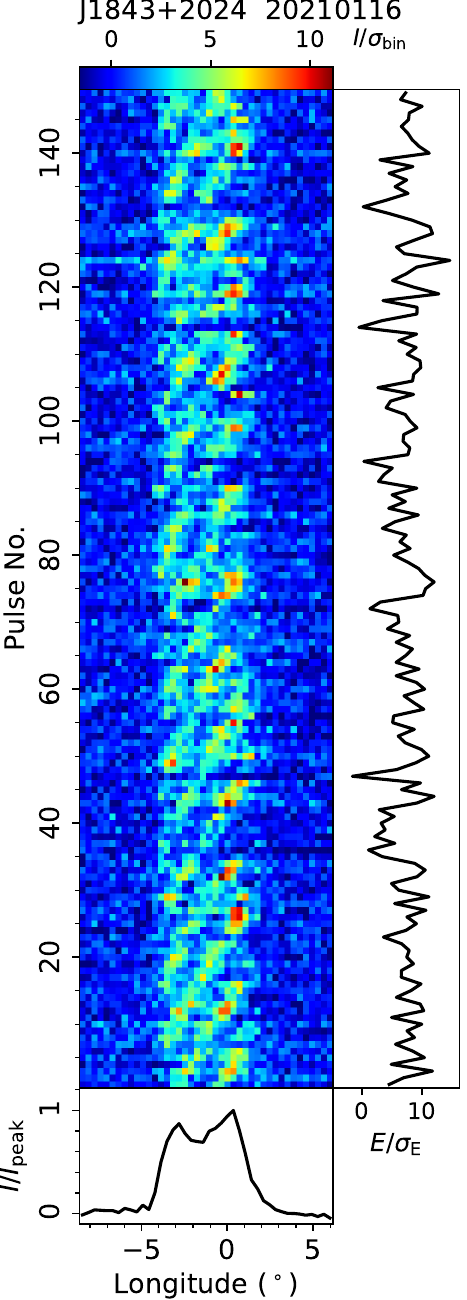}
\figcaption{Single pulse sequence of PSR J1843+2024 from the FAST observation on 20210116. \label{subfig:TP:J1843+2024}}
\end{figure}

\begin{figure}[htpb]
\centering
\includegraphics[width=0.22\textwidth, angle=0]{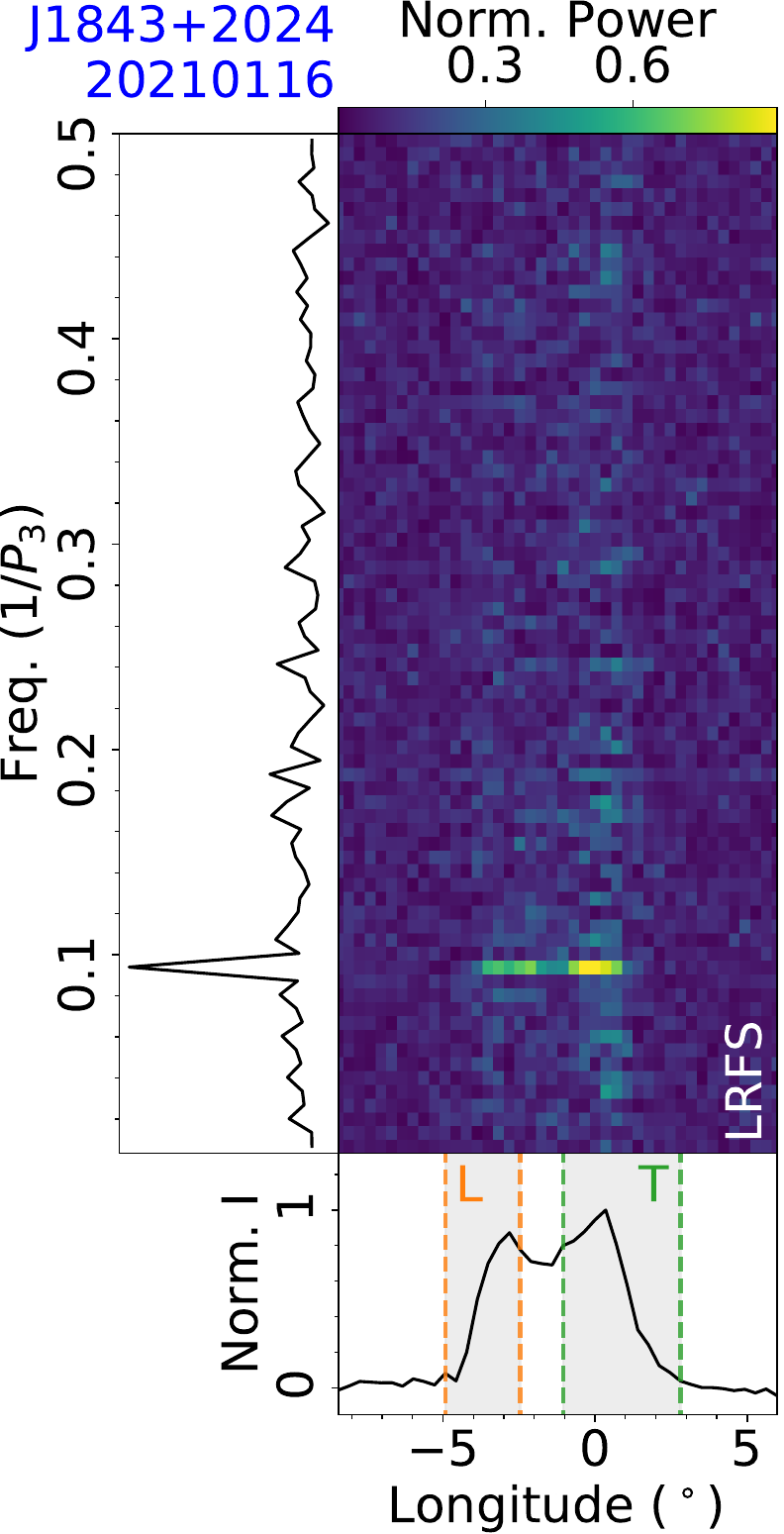}
\includegraphics[width=0.22\textwidth, angle=0]{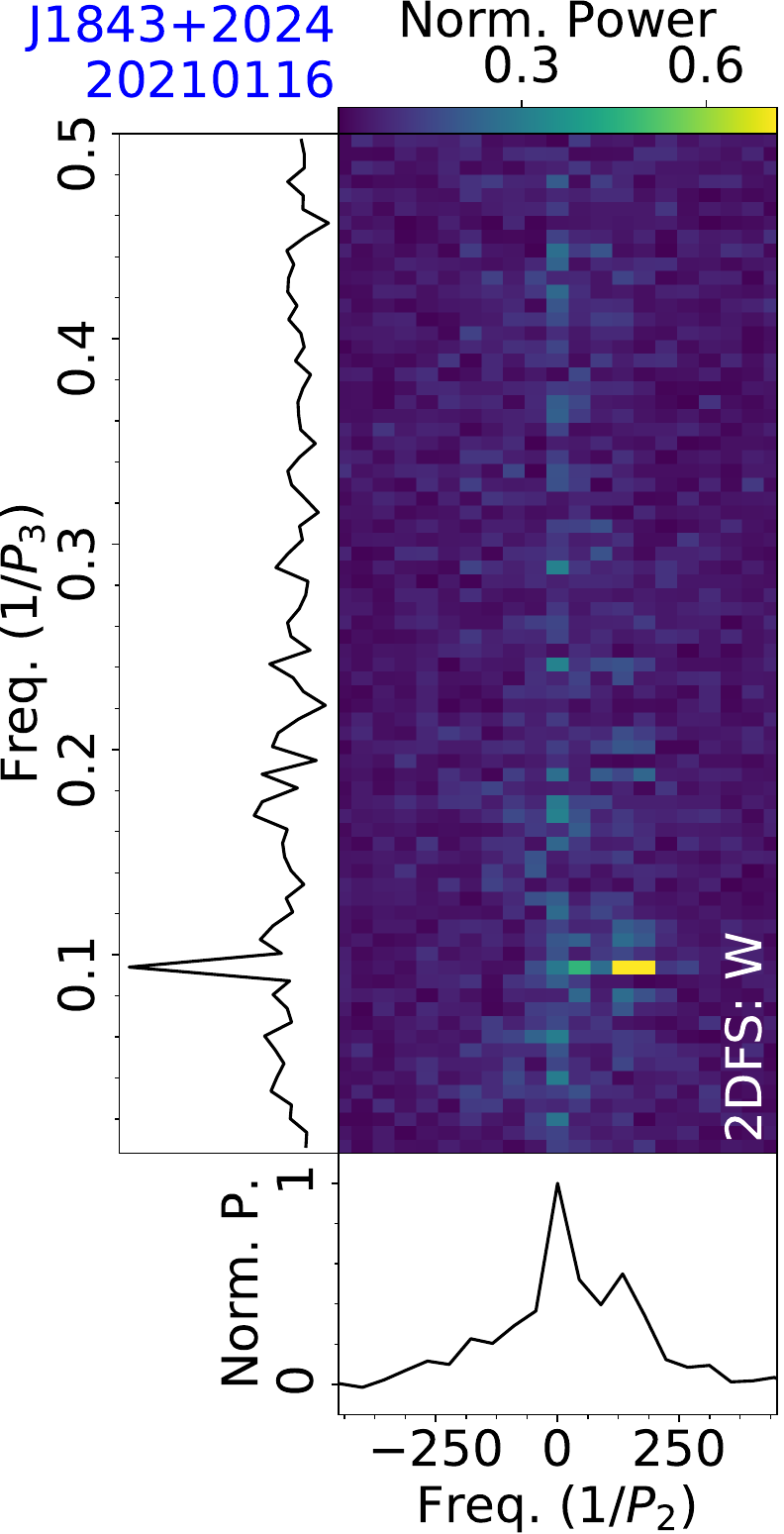}\\
\includegraphics[width=0.22\textwidth, angle=0]{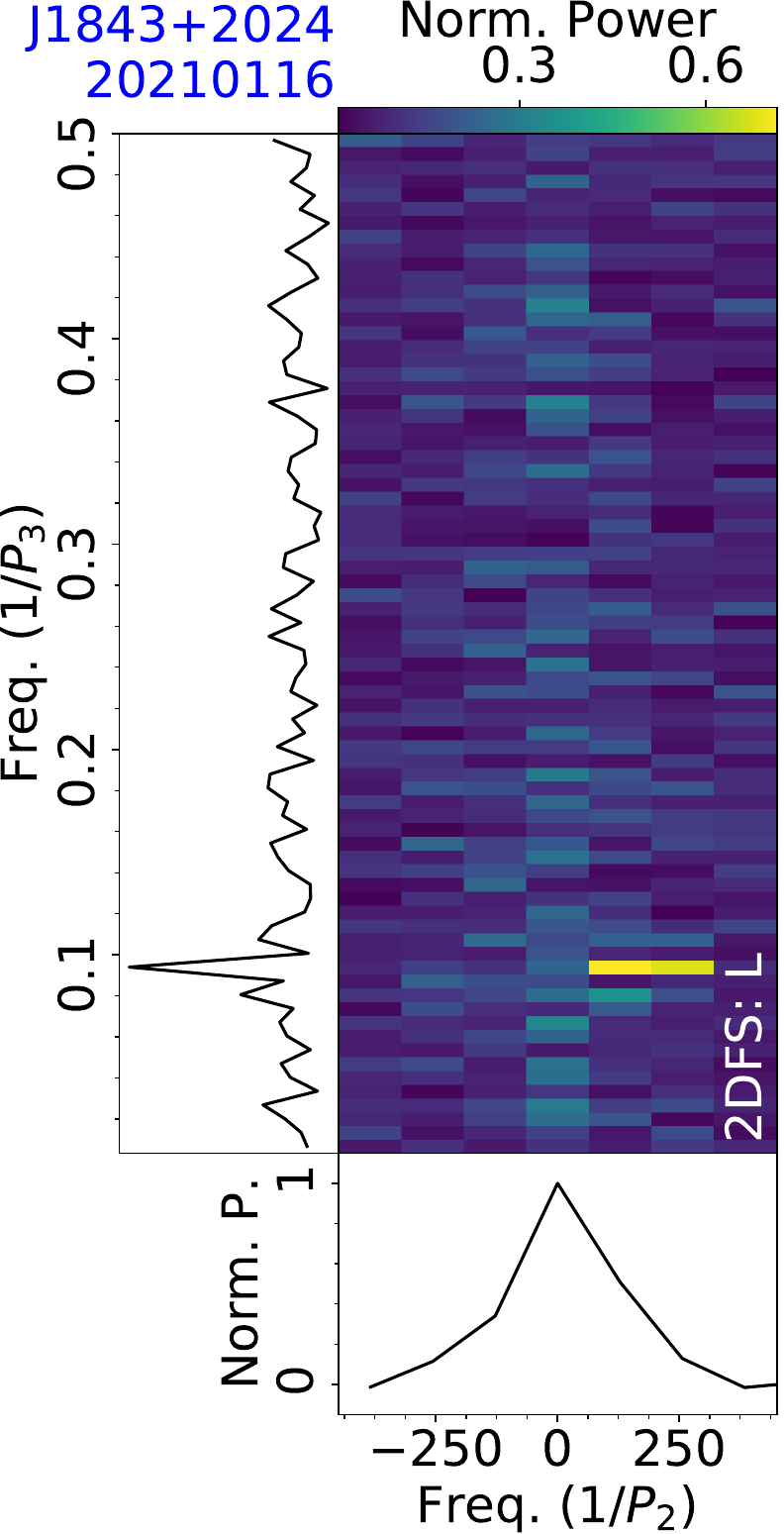}
\includegraphics[width=0.22\textwidth, angle=0]{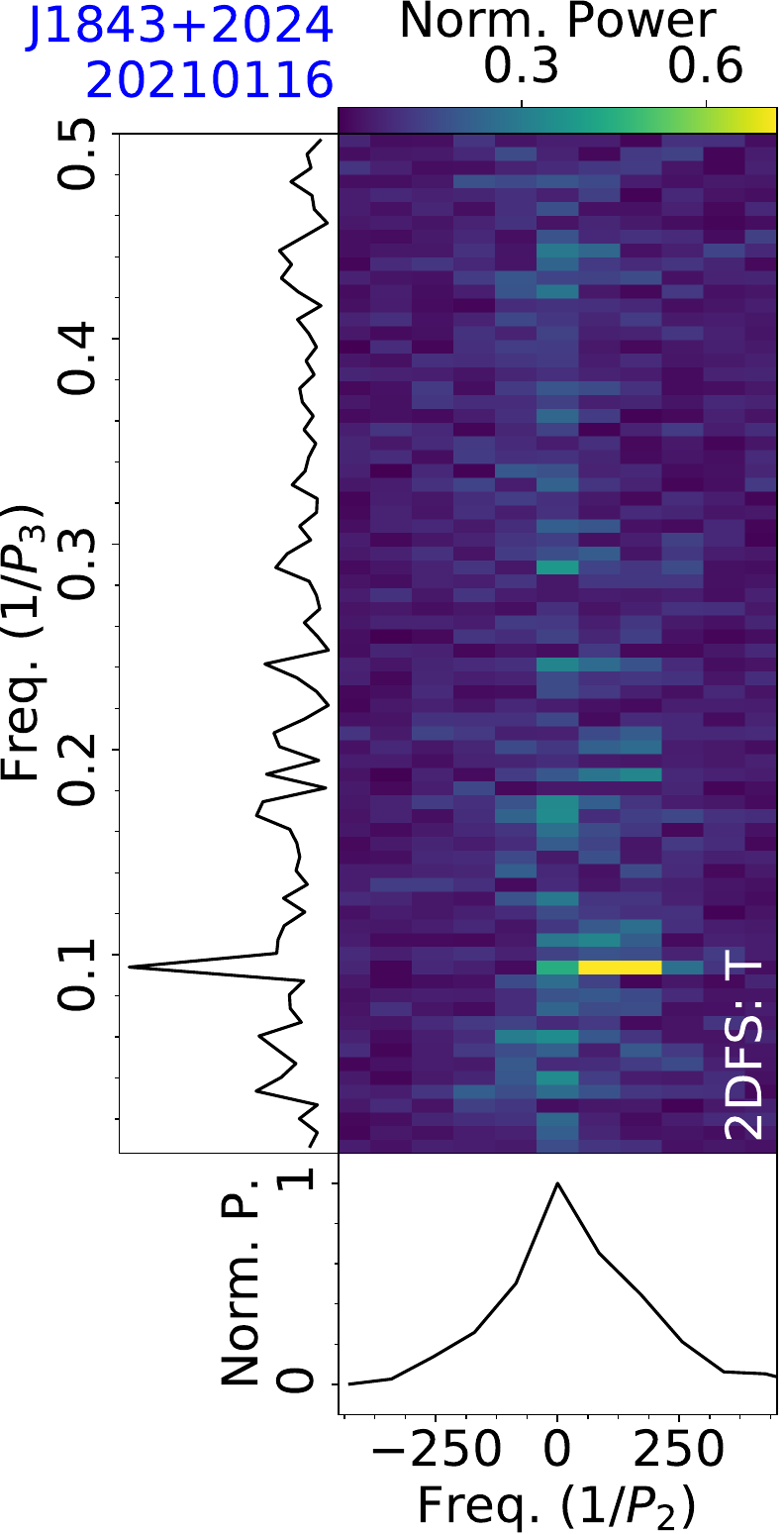}
\figcaption{Fluctuation analysis of PSR J1843+2024 from the FAST observation on 20210116, with LRFS (top-left), and 2DFS for the on-pulse region (top-right), leading part (bottom-left) and trailing part (bottom-right) of a mean pulse profile. \label{subfig:fluctu:J1843+2024}}
\end{figure}

\subsection{J1842-0800}
\label{subsec:J1842-0800}

PSR J1842-0800 was discovered in the High Time Resolution Universe pulsar survey with the Parkes 64-m radio telescope \citep{Ng2015}. 

This pulsar was observed by FAST on 20251015 for 15 minutes, and a rotation period $P=1.2556$~s and a dispersion measure $D\!M=187.9~{\rm cm^{-3}\,pc}$ were determined. Single pulse sequences in Fig.~\ref{subfig:TP:J1842-0800} illustrate the exhibition of the nulling phenomenon. The nulling fraction of this observation is estimated to be 38.4$\pm$2.3\% from the on-pulse integral energy histogram (Fig.~\ref{subfig:Hist:J1842-0800}).

\subsection{J1843+0526g}
\label{subsec:J1843+0526g}

PSR J1843+0526g was discovered in the GPPS survey \citep{Han2021,han2025} using the FAST. 

The pulsar was observed by FAST on 20230218 for 15 minutes, deriving a rotation period $P=2.0348$~s and a dispersion measure $D\!M=259.0~{\rm cm^{-3}\,pc}$ from this observation. 
From single pulse sequences displayed in Fig.~\ref{subfig:TP:J1843+0526g}, this pulsar is found to exhibit nulling and subpulse drifting behaviors. In particular, the emission phase regions of single pulses vary instead of remaining stable. The nulling fraction of the observation on PSR J1843+0526g is estimated to be 56$\pm$5\% from the energy histogram (Fig.~\ref{subfig:Hist:J1843+0526g}). From the cross-correlation analysis of single pulses with emission (Fig.~\ref{subfig:Corre:J1843+0526g}), the drifting parameters are estimated to be $P_2=3.59\pm0.02\pm^\circ$ and $D=1.79\pm0.06$ degrees per period.

\begin{figure}[htpb]
\centering
\includegraphics[width=0.22\textwidth, angle=0]{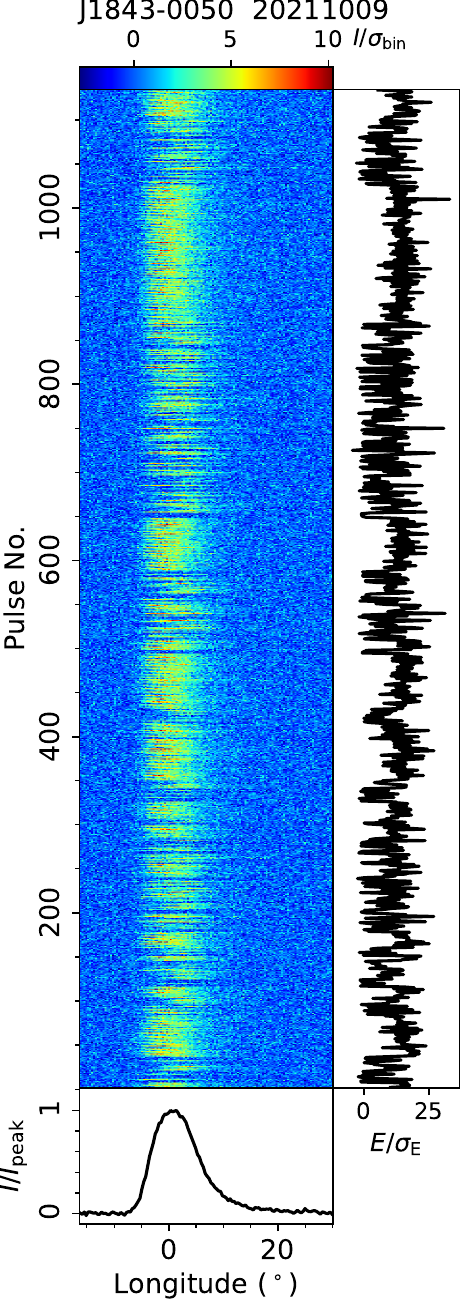}
\includegraphics[width=0.22\textwidth, angle=0]{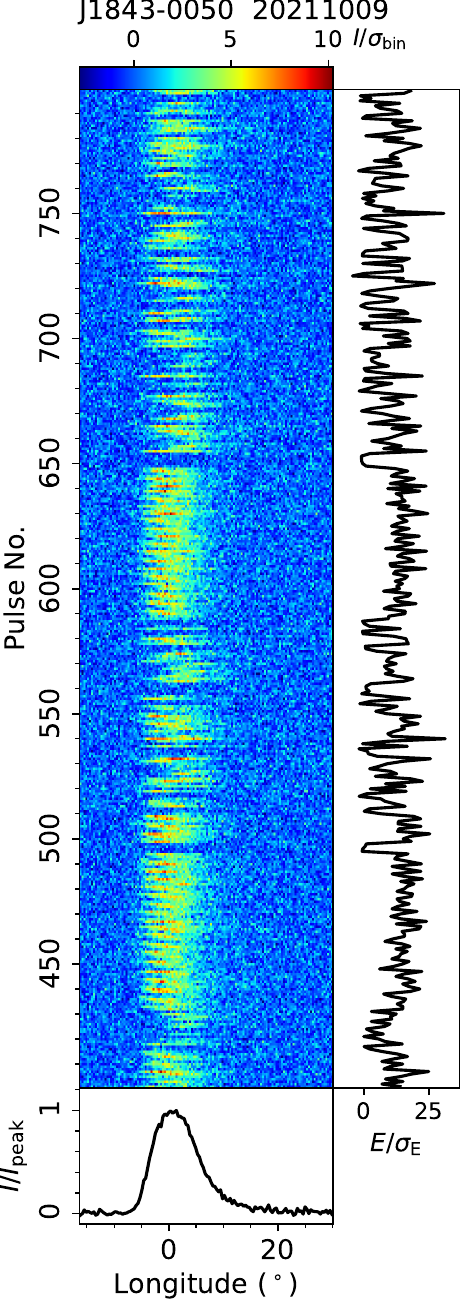}
\figcaption{Single pulse sequences of PSR J1843-0050 from the FAST observation on 20211009. \label{subfig:TP:J1843-0050}}
\end{figure}

\begin{figure}[htpb]
\includegraphics[width=0.39\textwidth, angle=0]{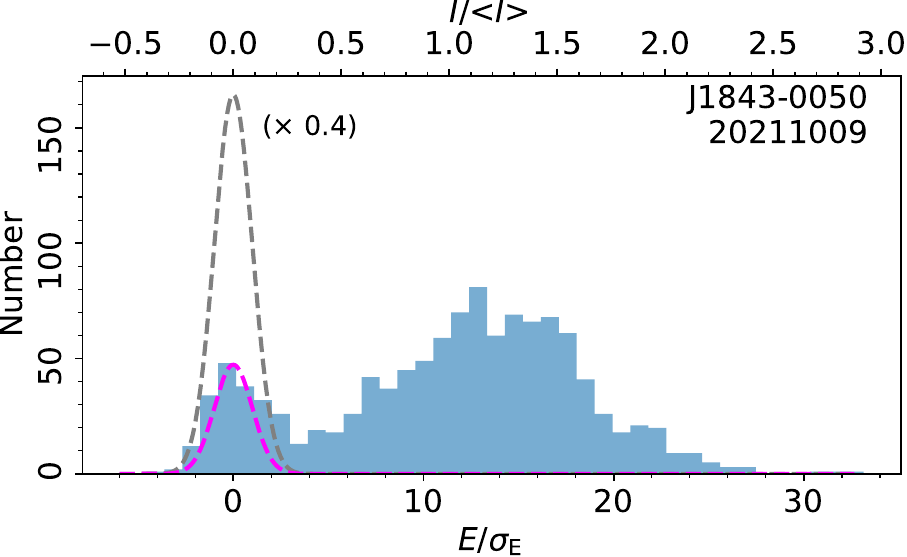}
\figcaption{On-pulse integral energy histogram of single pulses of PSR J1843-0050 from the FAST observation on 20211009. \label{subfig:Hist:J1843-0050}}
\end{figure}

\begin{figure}[htpb]
\centering
\includegraphics[width=0.22\textwidth, angle=0]{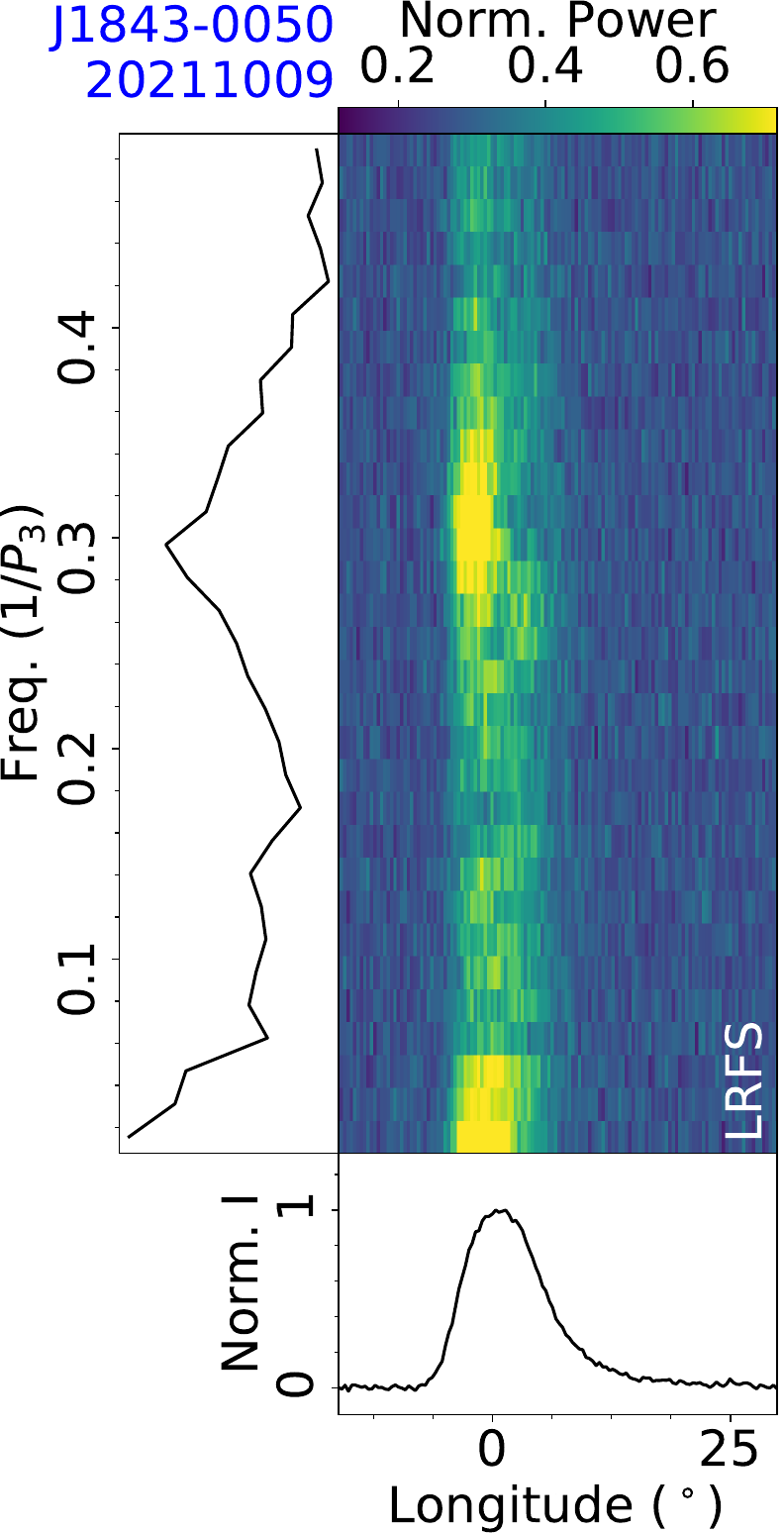}
\includegraphics[width=0.22\textwidth, angle=0]{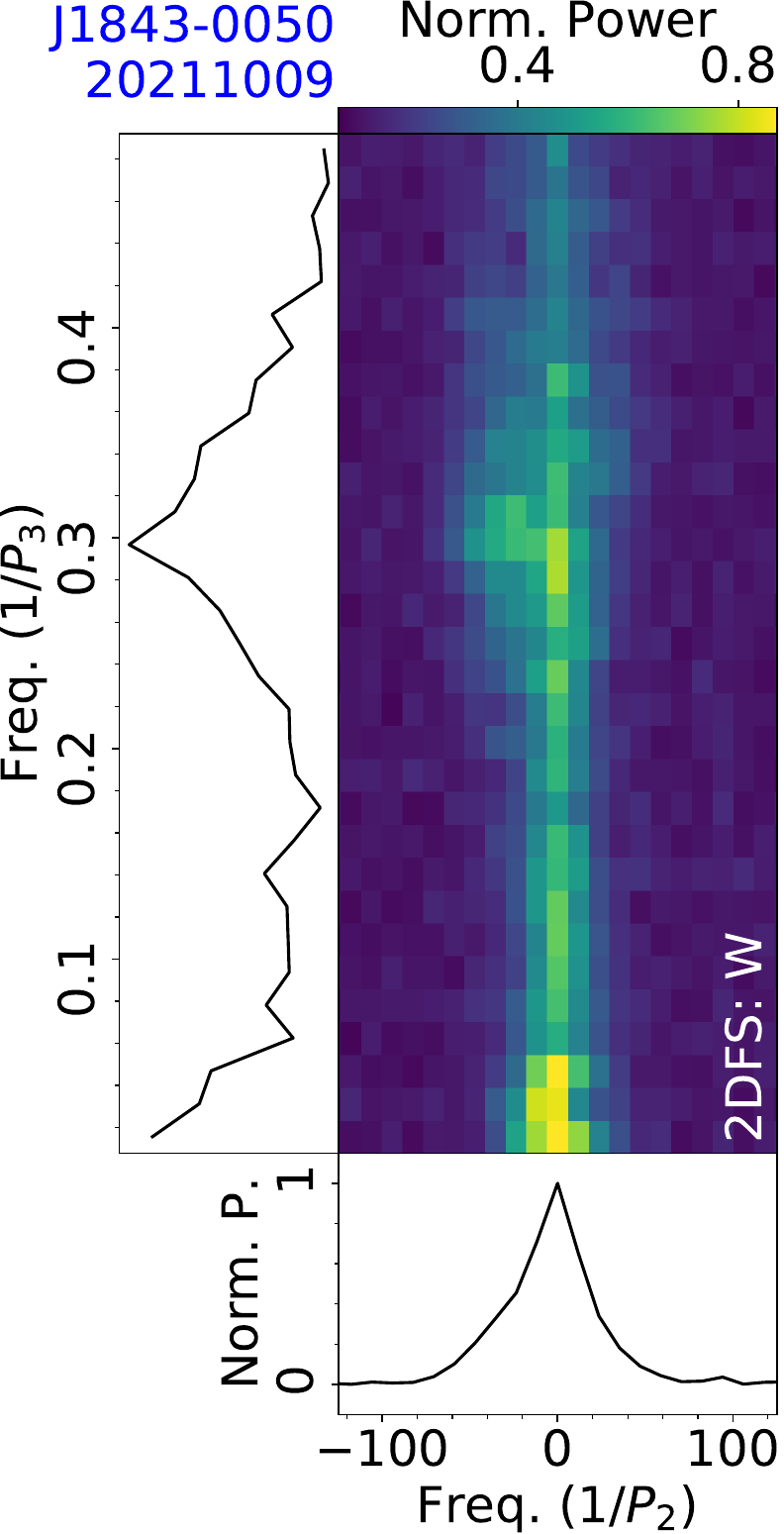}
\figcaption{Fluctuation spectra of PSR J1843-0050 from the FAST observation on 20211009, with LRFS and 2DFS for the on-pulse region of a mean pulse profile. \label{subfig:fluctu:J1843-0050}}
\end{figure}

\subsection{J1843+2024}
\label{subsec:J1843+2024}

PSR J1843+2024 was discovered by \citet{Zepka1996}. For two components of the pulsar, \citet{Song2023} reported drifting parameters with values of $P_3=10.4\pm0.5$ periods, $P_2=2.3^{+1}_{-0.2}$ degrees and $P_3=10.5\pm0.3$ periods, $P_2=3^{+2}_{-1}$ degrees, respectively.

This pulsar was observed by FAST on 20210116 for 9 minutes, deriving a rotation period $P=3.4069$~s and a dispersion measure $D\!M=86.5~{\rm cm^{-3}\,pc}$. 
The single pulse sequence of this observation is shown in Fig.~\ref{subfig:TP:J1843+2024}. From the LRFS and 2DFS in Fig.~\ref{subfig:fluctu:J1843+2024}, we obtain the drifting properties $P_3=10.6\pm0.1$ periods and $P_2=2.0\pm0.3^\circ$ for the leading component, and $P_3=10.6\pm0.1$ periods and $P_2=3.0\pm0.4^\circ$ for the trailing component.

\begin{figure}[htpb]
\centering
\includegraphics[width=0.22\textwidth, angle=0]{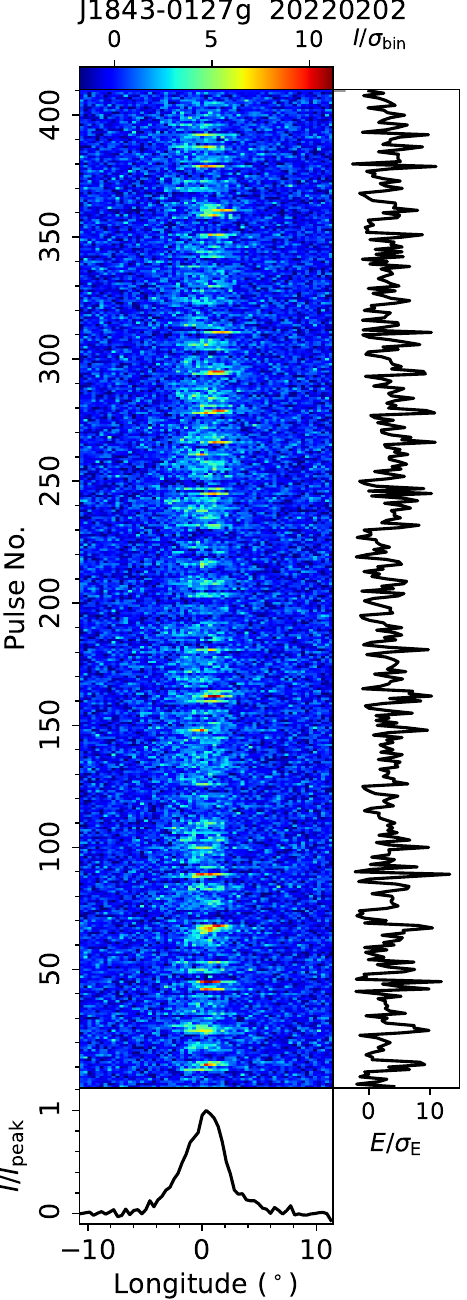}
\figcaption{Single pulse sequence of PSR J1843-0127g from the FAST observation on 20220202.
\label{subfig:TP:J1843-0127g}}
\end{figure}

\begin{figure}[htpb]
\centering
\includegraphics[width=0.39\textwidth, angle=0]{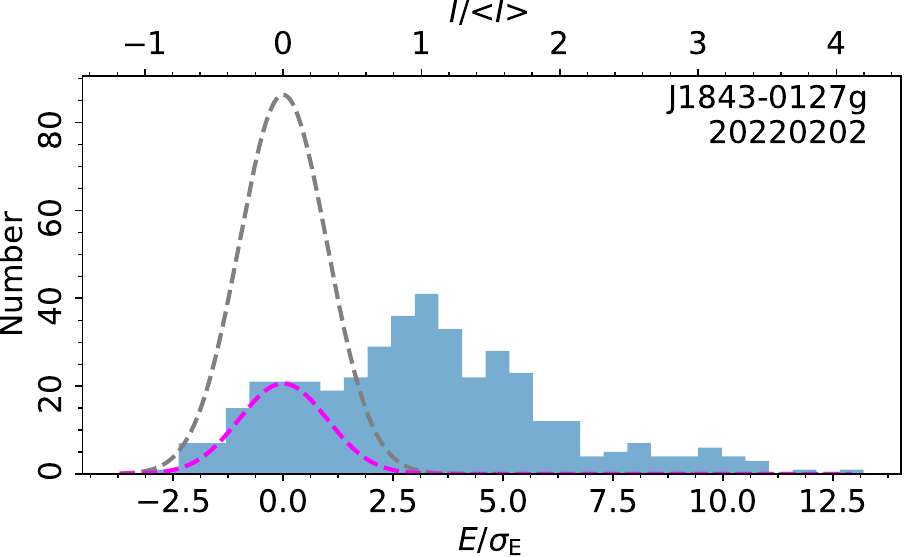}
\figcaption{On-pulse energy histogram of single pulses of PSR J1843-0127g from the FAST observation on 20220202.
\label{subfig:Hist:J1843-0127g}}
\end{figure}


\begin{figure}[htpb]
\centering
\includegraphics[width=0.44\textwidth, angle=0]{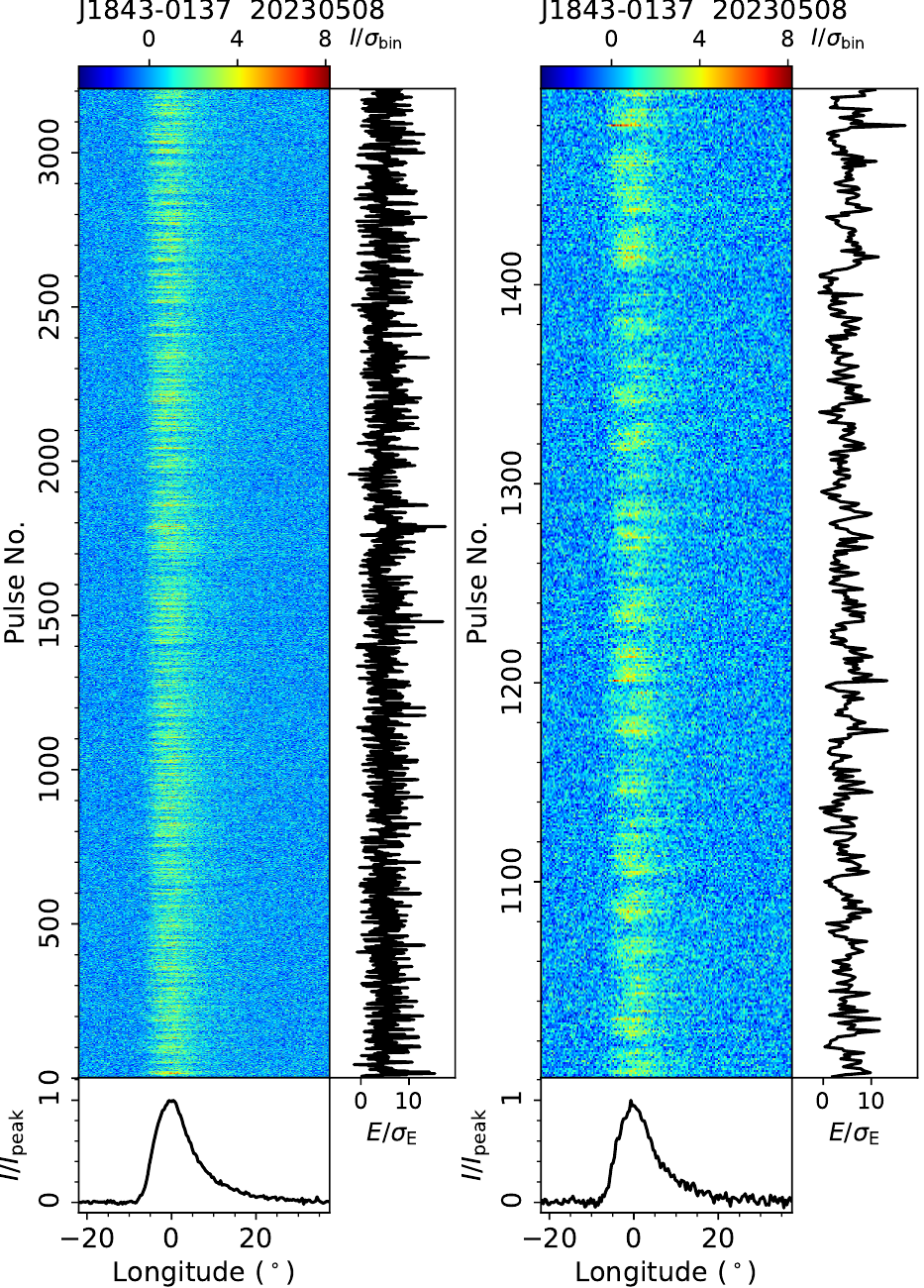}
\figcaption{Single pulse sequence of PSR J1843-0137 from the FAST observation on 20230508, and a zoomed-in view of pulses No. 1000-1500.
\label{subfig:TP:J1843-0137}}
\end{figure}

\begin{figure}[htpb]
\centering
\includegraphics[width=0.44\textwidth, angle=0]{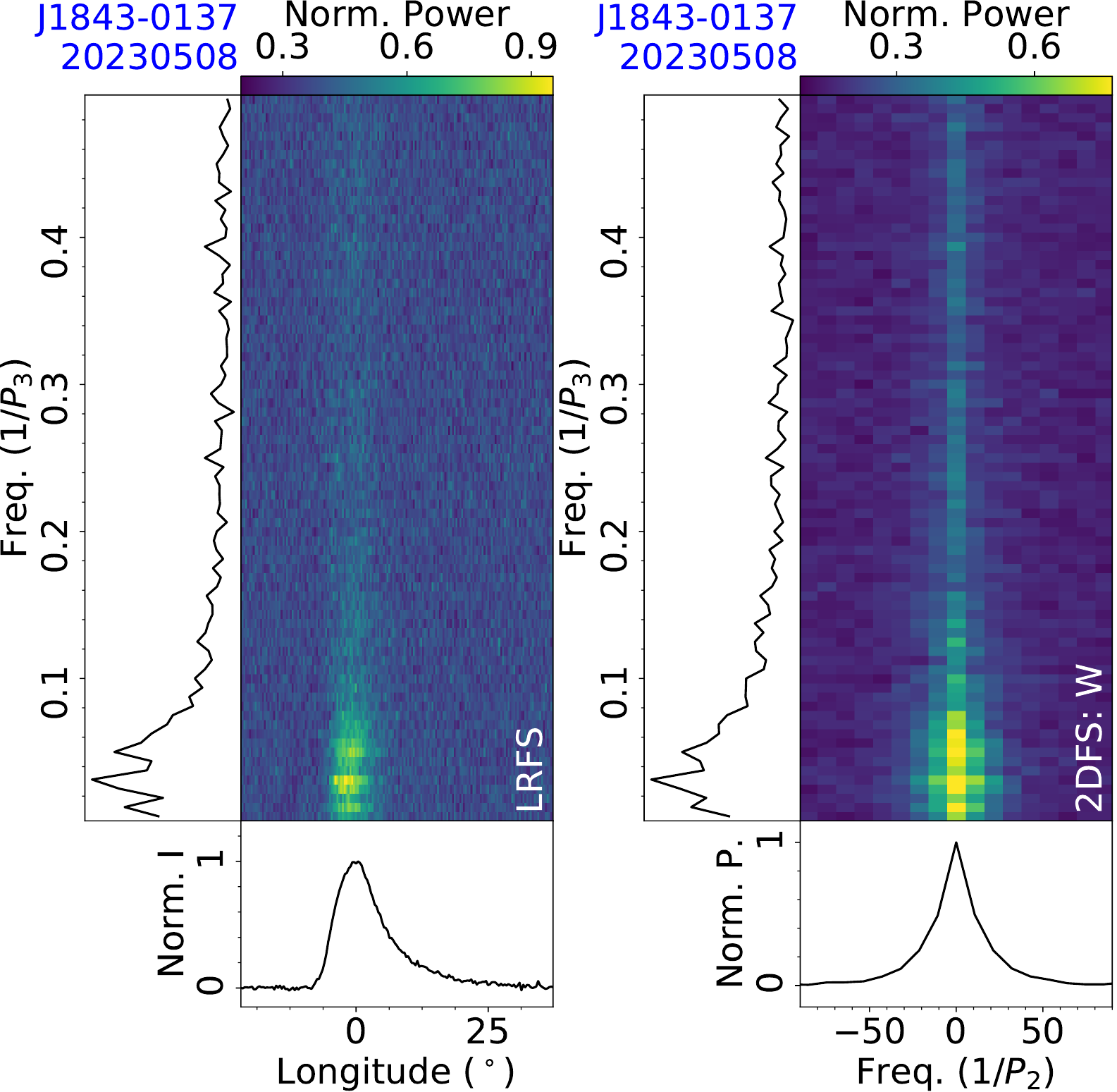}
\figcaption{Fluctuation spectra of PSR J1843-0137 from the FAST observation on 20230508, with LRFS and 2DFS for the on-pulse region of the mean pulse profile.
\label{subfig:fluctu:J1843-0137}}
\end{figure}

\begin{figure}[htpb]
\centering
\includegraphics[width=0.22\textwidth, angle=0]{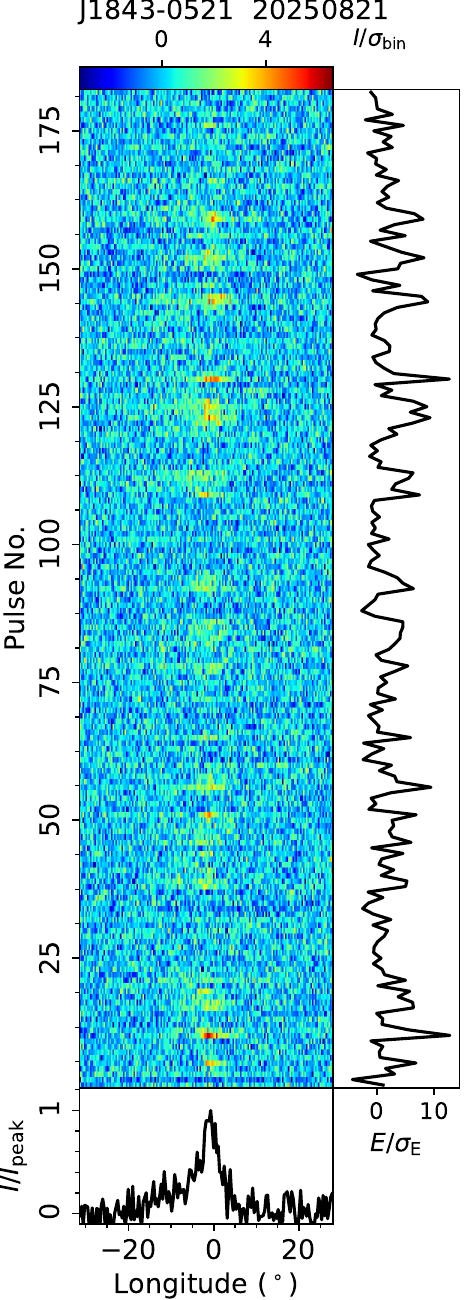}
\includegraphics[width=0.22\textwidth, angle=0]{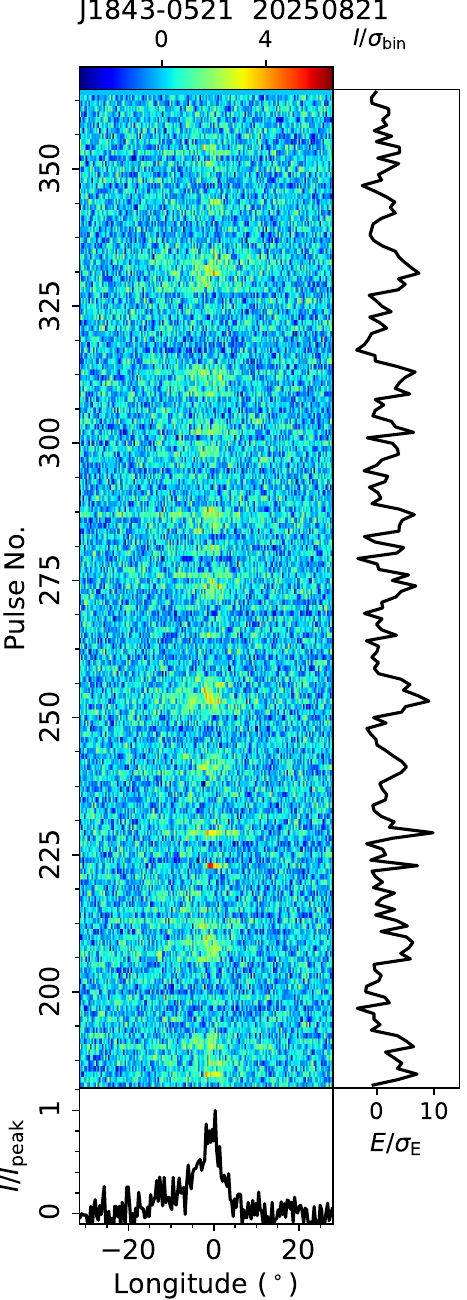}
\figcaption{Single pulse sequences of PSR J1843-0521 from the FAST observation on 20250821.
\label{subfig:TP:J1843-0521}}
\end{figure}

\begin{figure}[htpb]
\centering
\includegraphics[width=0.39\textwidth, angle=0]{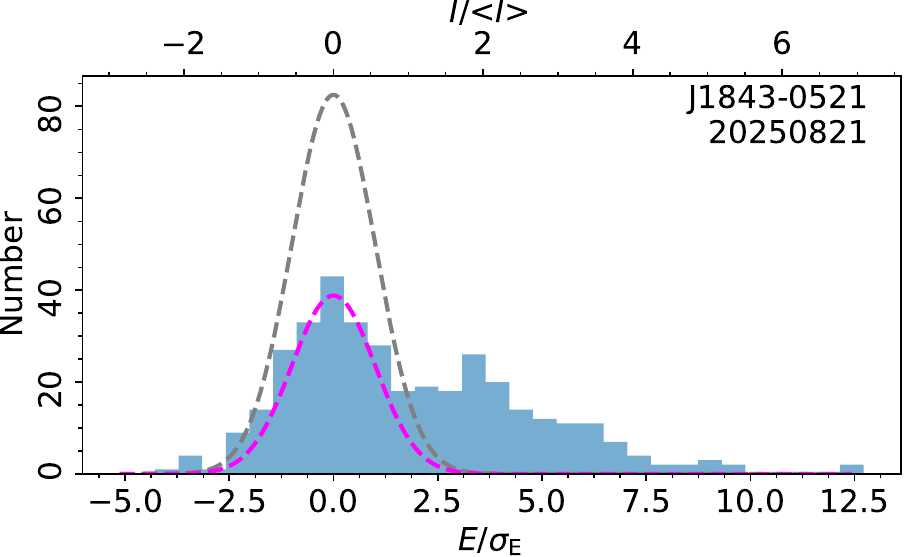}
\figcaption{On-pulse energy histogram of single pulses of PSR J1843-0521 from the FAST observation on 20250821.
\label{subfig:Hist:J1843-0521}}
\end{figure}

\subsection{J1843-0050}
\label{subsec:J1843-0050}

PSR J1843-0050 was discovered in the Parkes Multibeam Pulsar Survey \citep{Morris2002}. 
This pulsar was observed by FAST on 20211009 and 20241224, each for 15 minutes. From the observation on 20211009, a rotation period and a dispersion measure are $P=0.7827$~s and a dispersion measure $D\!M=512.3~{\rm cm^{-3}\,pc}$. Single pulse sequences of this data shown in Fig.~\ref{subfig:TP:J1843-0050} illustrate nulling and subpulse drifting phenomena. The nulling fraction is estimated to be 12$\pm$1\% from the on-pulse integral energy histogram in Fig.~\ref{subfig:Hist:J1843-0050}. LRFS and 2DFS in Fig.~\ref{subfig:fluctu:J1843-0050} display a negative drift feature, with centroid frequencies of $1/P_3=0.308\pm0.001$ and $1/P_2=-33\pm1$, which correspond to $P_3=3.24\pm0.01$ periods and $P_2=-11.0\pm0.3^\circ$.

\begin{figure}[htpb]
\centering
\includegraphics[width=0.22\textwidth, angle=0]{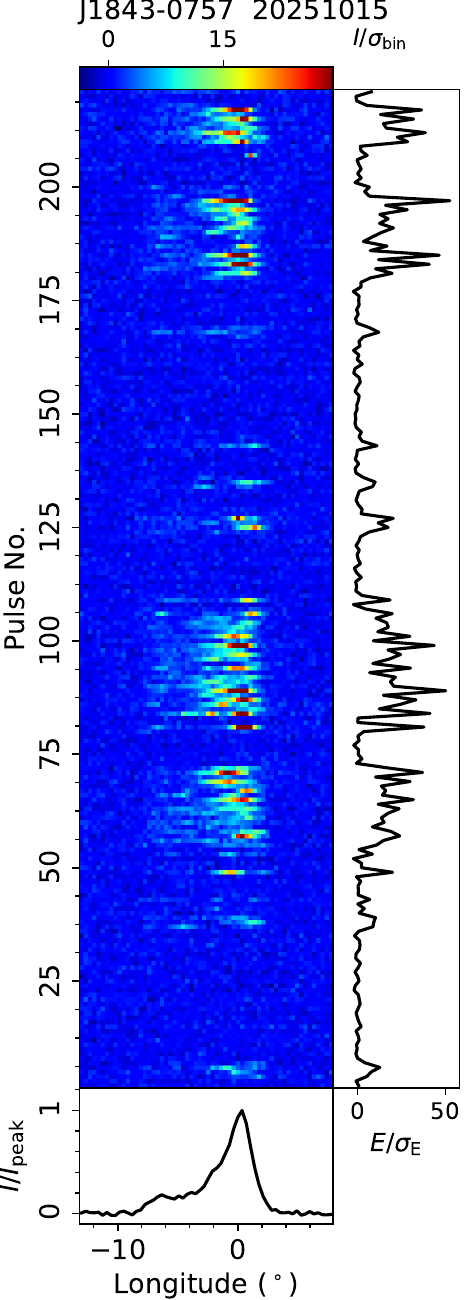}
\includegraphics[width=0.22\textwidth, angle=0]{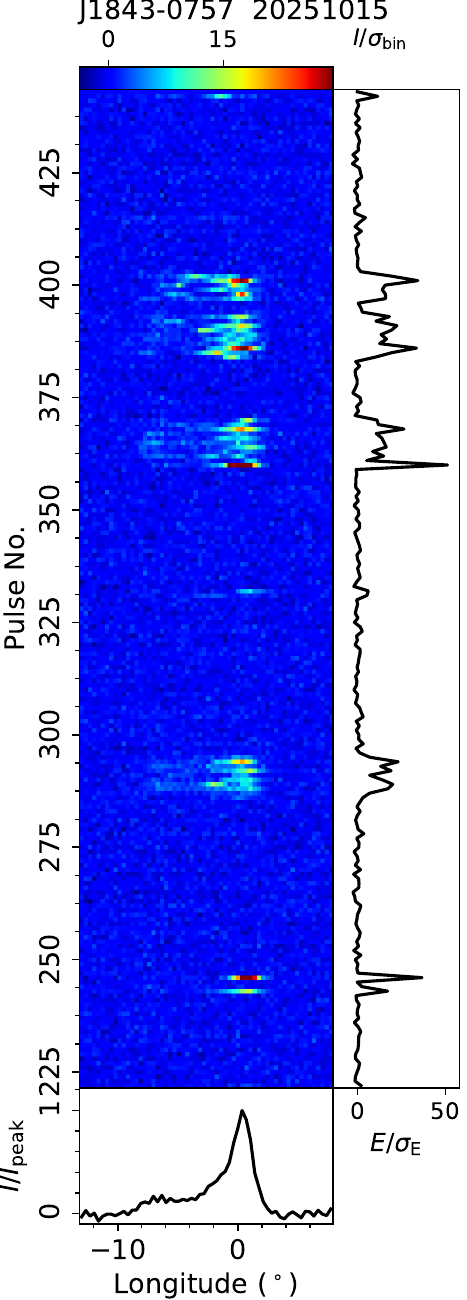}
\figcaption{Single pulse sequences of PSR J1843-0757 from the FAST observation on 20251015.
\label{subfig:TP:J1843-0757}}
\end{figure}

\begin{figure}[htpb]
\centering
\includegraphics[width=0.39\textwidth, angle=0]{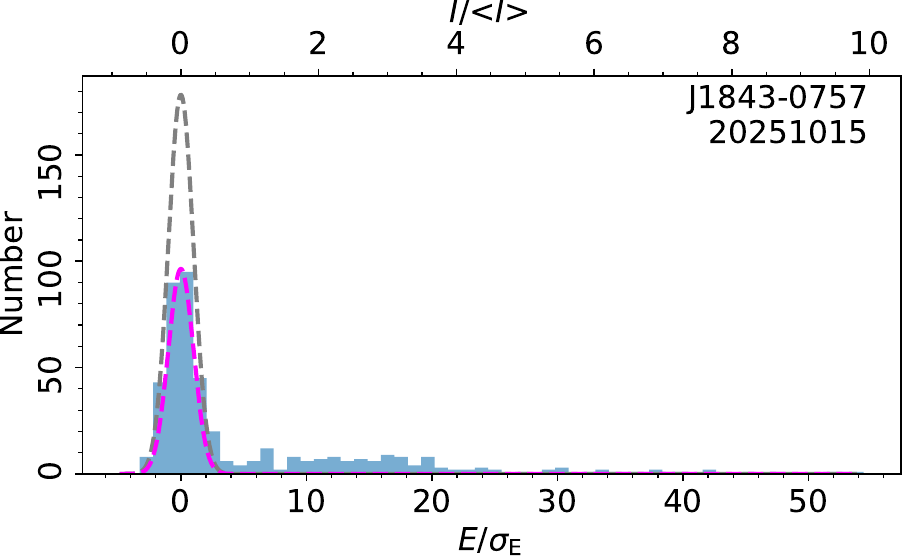}
\figcaption{On-pulse energy histogram of single pulses of PSR J1843-0757 from the FAST observation on 20251015.
\label{subfig:Hist:J1843-0757}}
\end{figure}

\begin{figure}[htpb]
\centering
\includegraphics[width=0.22\textwidth, angle=0]{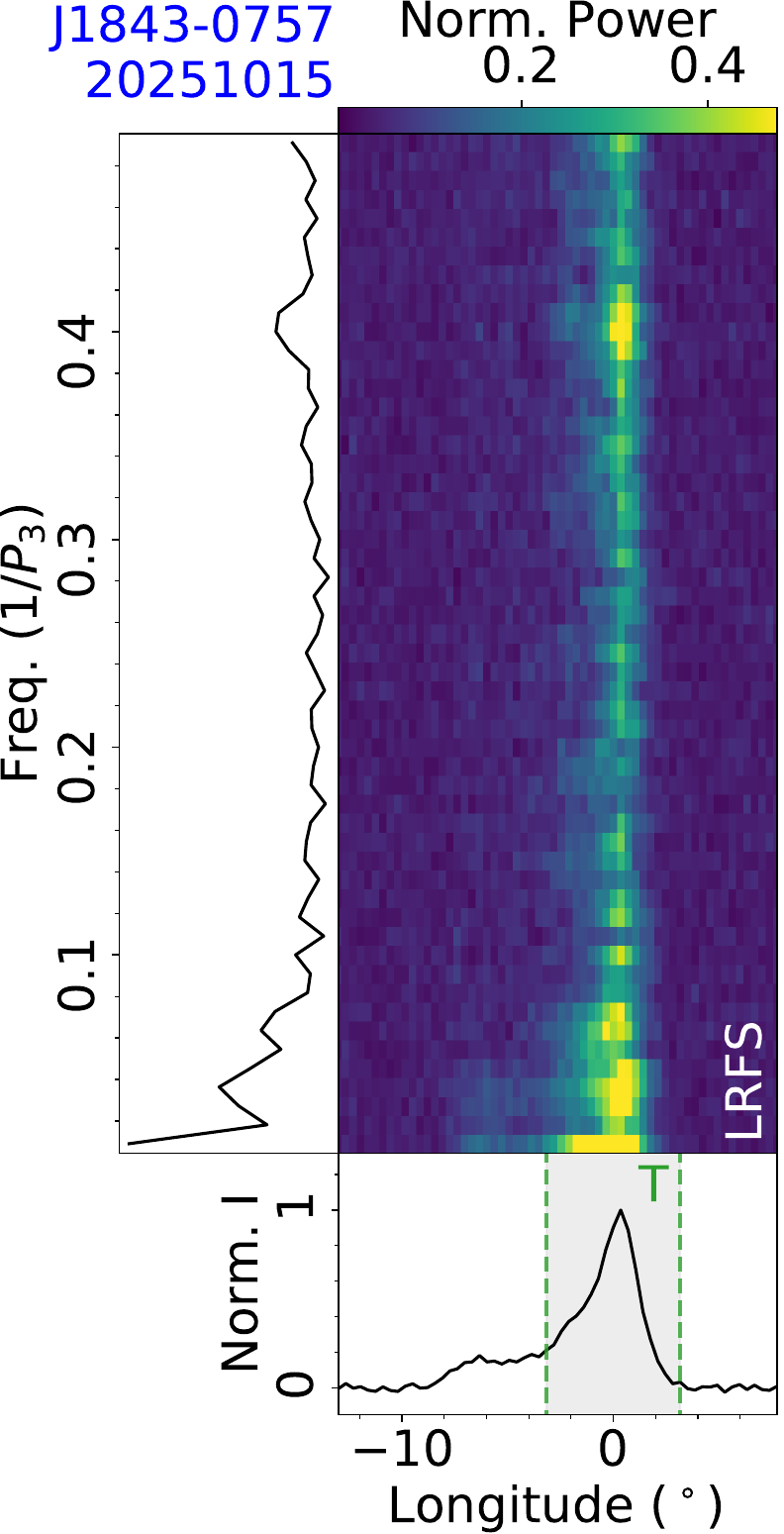}
\includegraphics[width=0.22\textwidth, angle=0]{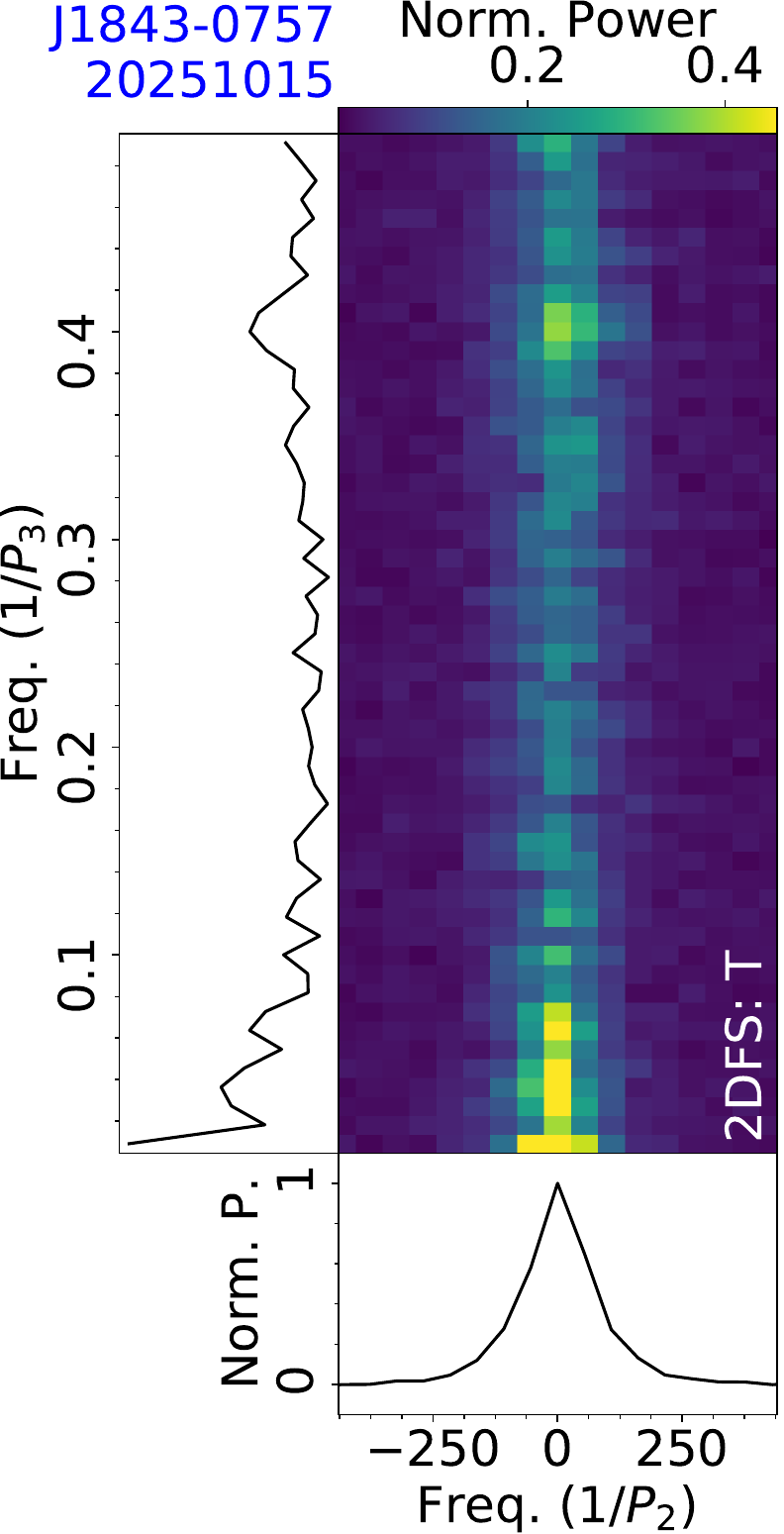}
\figcaption{Fluctuation analysis of PSR J1843-0757 for the observation on 20251015, with LRFS and 2DFS for the trailing phase region of a mean pulse profile.
\label{subfig:fluctu:J1843-0757}}
\end{figure}

\subsection{J1843-0127g}
\label{subsec:J1843-0127g}

PSR J1843-0127g was discovered in the FAST GPPS survey \citep{Han2021,han2025}. 

This pulsar was observed by FAST on 20220202 for 15 minutes, deriving a rotation period $P=2.1648$~s and a dispersion measure $D\!M=449.0~{\rm cm^{-3}\,pc}$. 
The single pulse sequence in Figures~\ref{subfig:TP:J1843-0127g} shows intensity variation. 
From the on-pulse energy histogram in Fig.~\ref{subfig:Hist:J1843-0127g}, there is evidence that the pulsar has nulling behavior, with a nulling fraction estimated to be 24(2)\% for this observation.

\subsection{J1843-0137}
\label{subsec:J1843-0137}

PSR J1843-0137 was discovered in the Parkes multibeam pulsar survey \citep{hfs+04}. 

This pulsar was observed by FAST on 20230508 for 36 minutes, with a rotation period $P=0.6698$~s and a dispersion measure $D\!M=484.7~{\rm cm^{-3}\,pc}$ determined. The single pulse sequence and a zoomed-in view of pulses No. 1000-1500 in Fig.~\ref{subfig:TP:J1843-0137} show the modulation phenomenon. In the fluctuation spectra in Fig.~\ref{subfig:fluctu:J1843-0137}, the centroid frequency of the modulation feature is $1/P_3=0.0453\pm0.0004$, corresponding to the periodicity of $P_3=22.1\pm0.2$ periods.

\subsection{J1843-0521}
\label{subsec:J1843-0521}

PSR J1930+2836 was discovered by FAST in the Commensal Radio Astronomy FAST Survey (CRAFTS) (http://groups.bao.ac.cn/ism/CRAFTS/).

This pulsar was observed by FAST on 20250821 for 15 minutes, deriving a rotation period $P=2.4705$~s and a dispersion measure $D\!M=771.0~{\rm cm^{-3}\,pc}$. Single pulse sequences in Fig.~\ref{subfig:TP:J1843-0521} show the nulling phenomenon. From the on-pulse integral energy histogram (Fig.~\ref{subfig:Hist:J1843-0521}), the nulling fraction of this observation is estimated to be 47.1$\pm$3.4\%.

\begin{figure}[htpb]
\centering
\includegraphics[width=0.22\textwidth, angle=0]{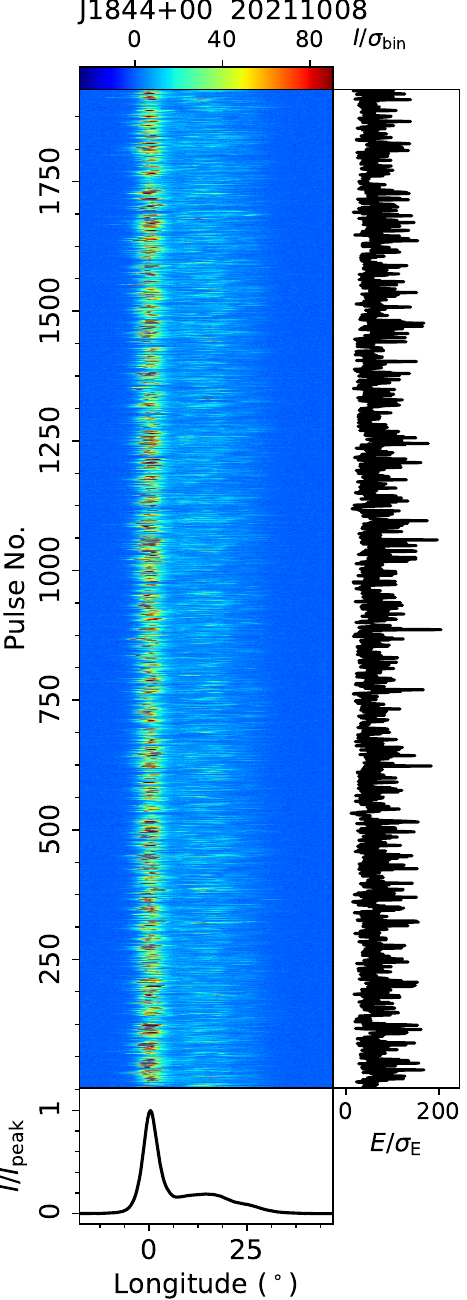}
\includegraphics[width=0.22\textwidth, angle=0]{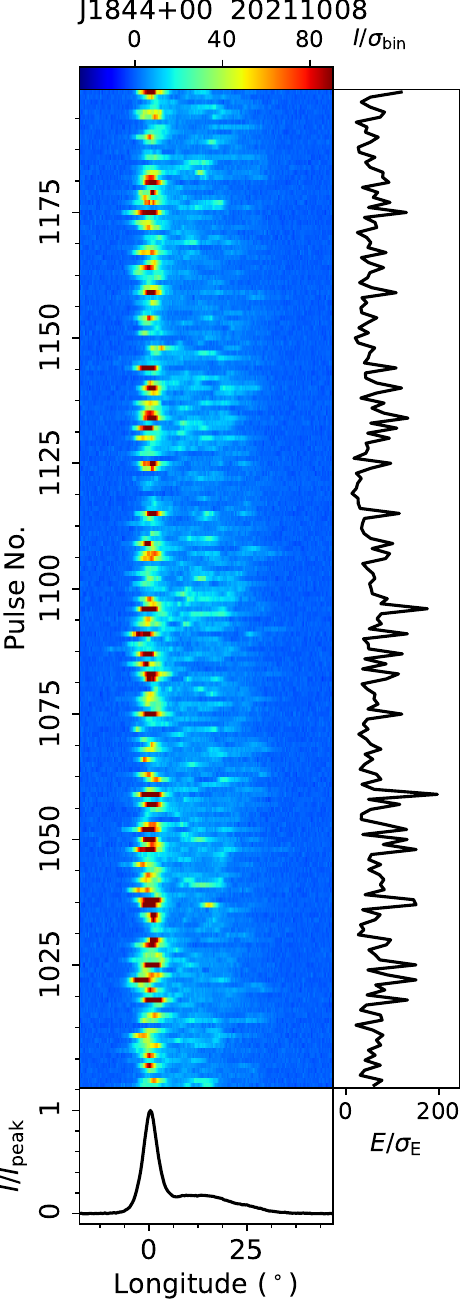}\\
\includegraphics[width=0.22\textwidth, angle=0]{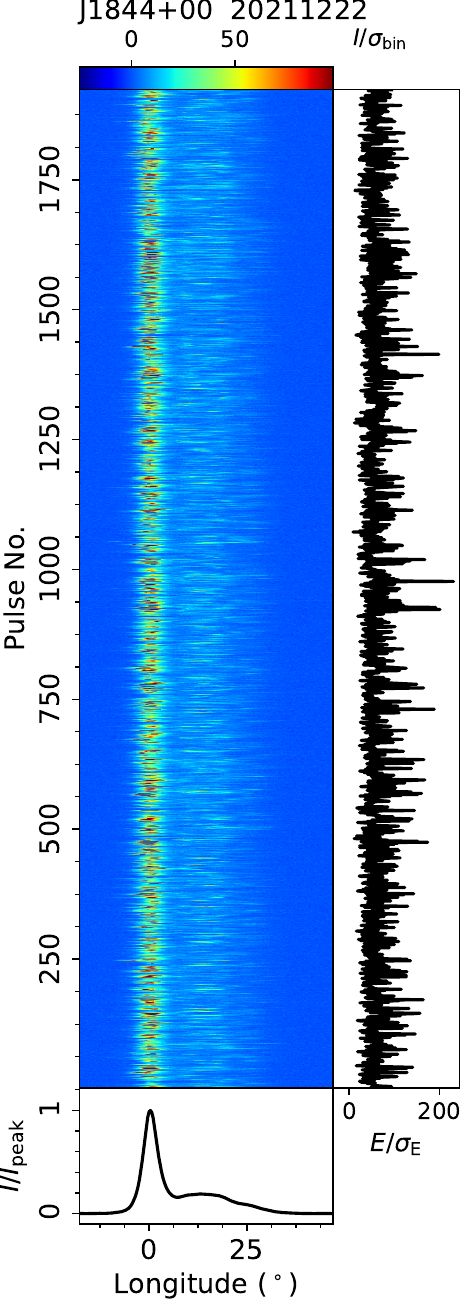}
\includegraphics[width=0.22\textwidth, angle=0]{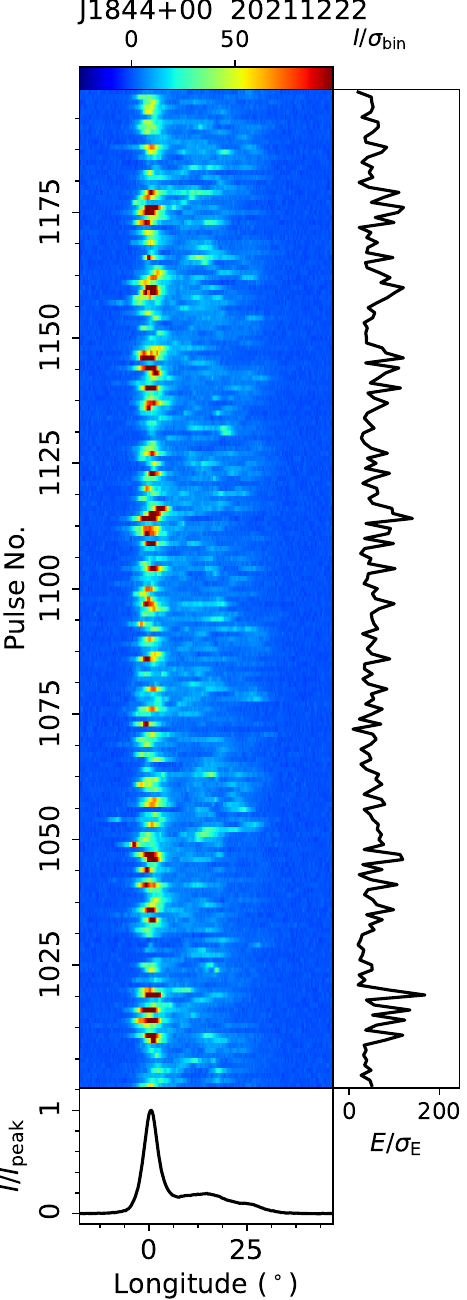}
\vspace{-0.3cm}
\figcaption{Single pulse sequences of PSR J1844+00 from the FAST observation on 20211008 and 20211222. 
\label{subfig:TP:J1844+00}}
\end{figure}

\begin{figure}[htpb]
\centering
\includegraphics[width=0.22\textwidth, angle=0]{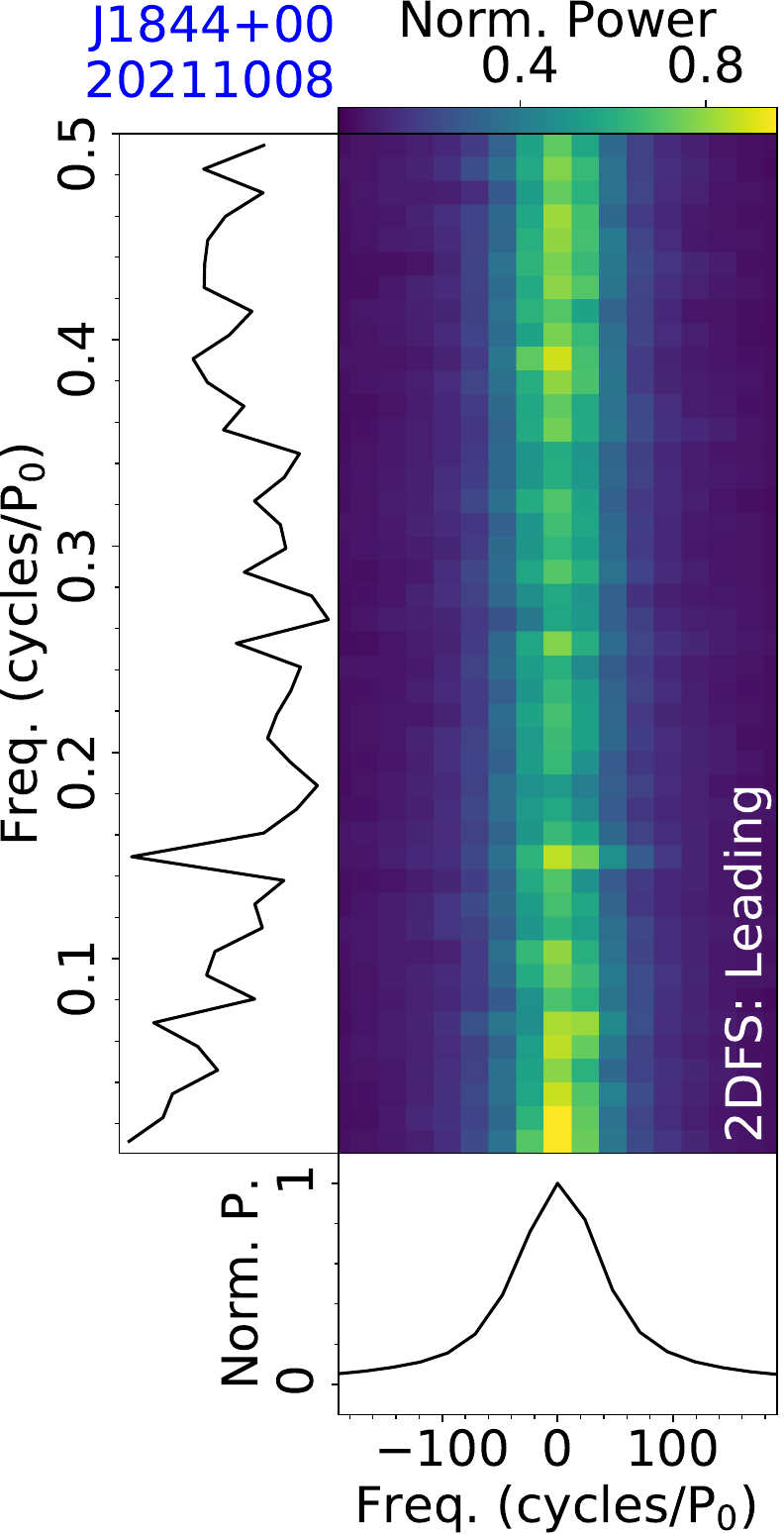}
\includegraphics[width=0.22\textwidth, angle=0]{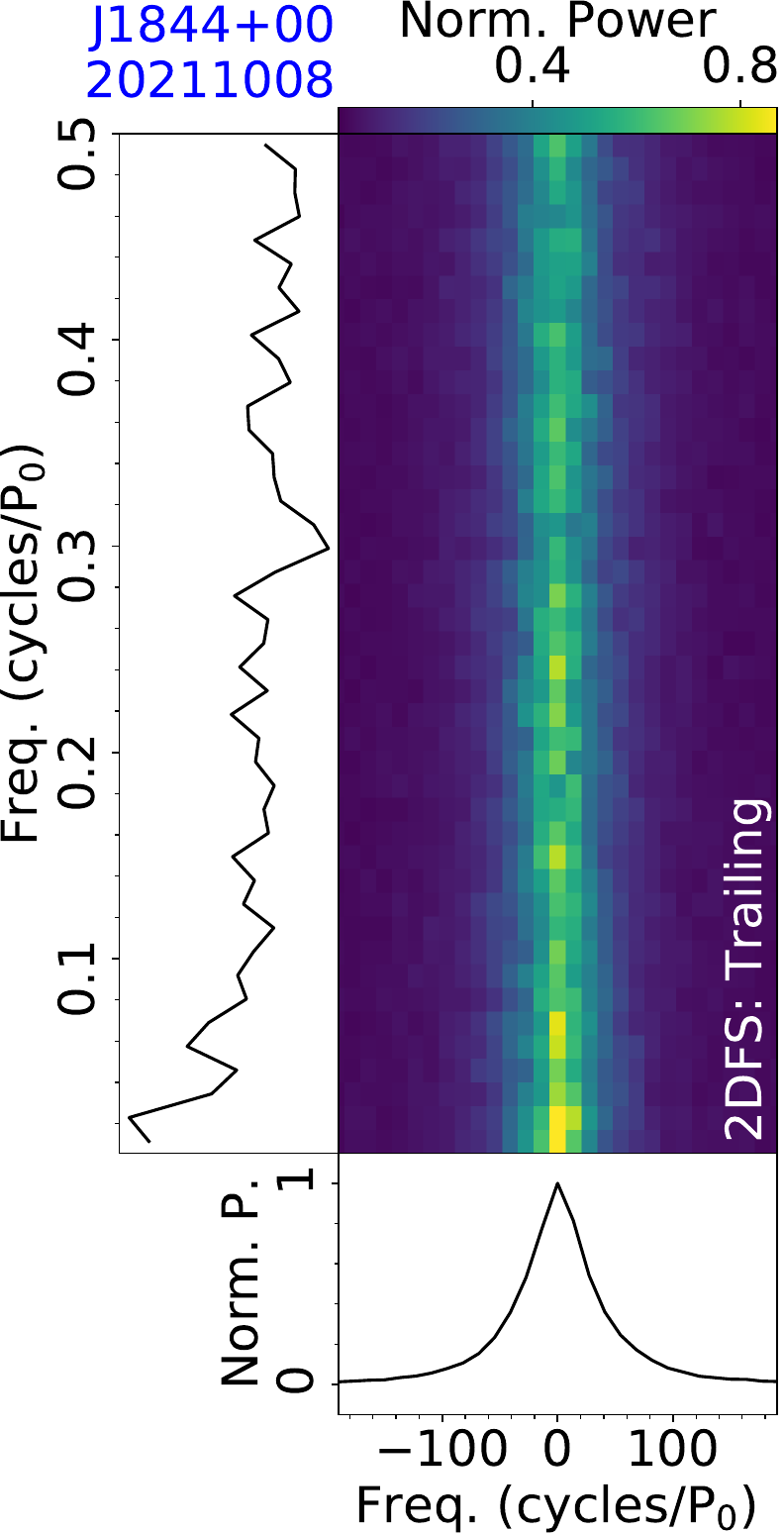}\\
\includegraphics[width=0.22\textwidth, angle=0]{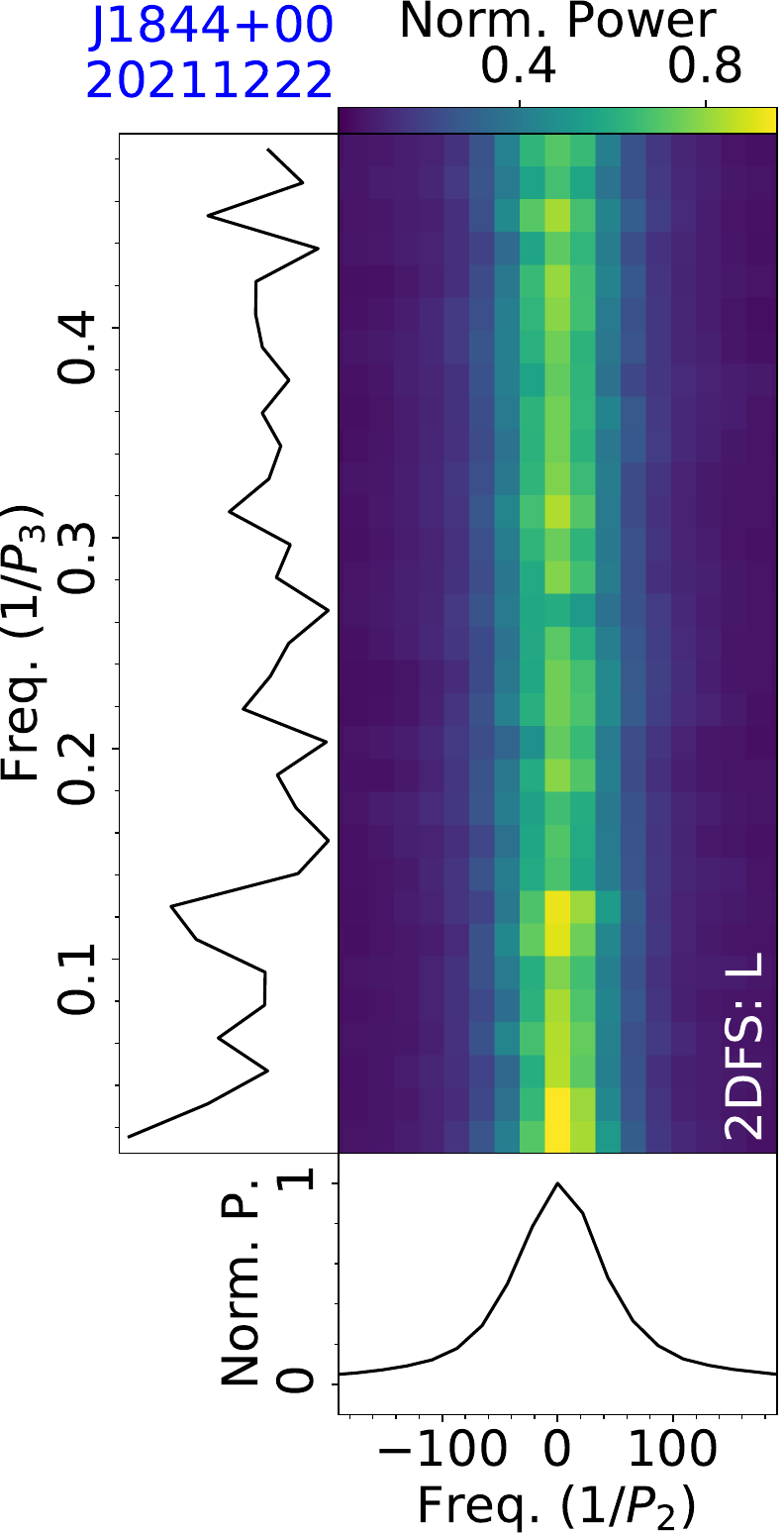}
\includegraphics[width=0.22\textwidth, angle=0]{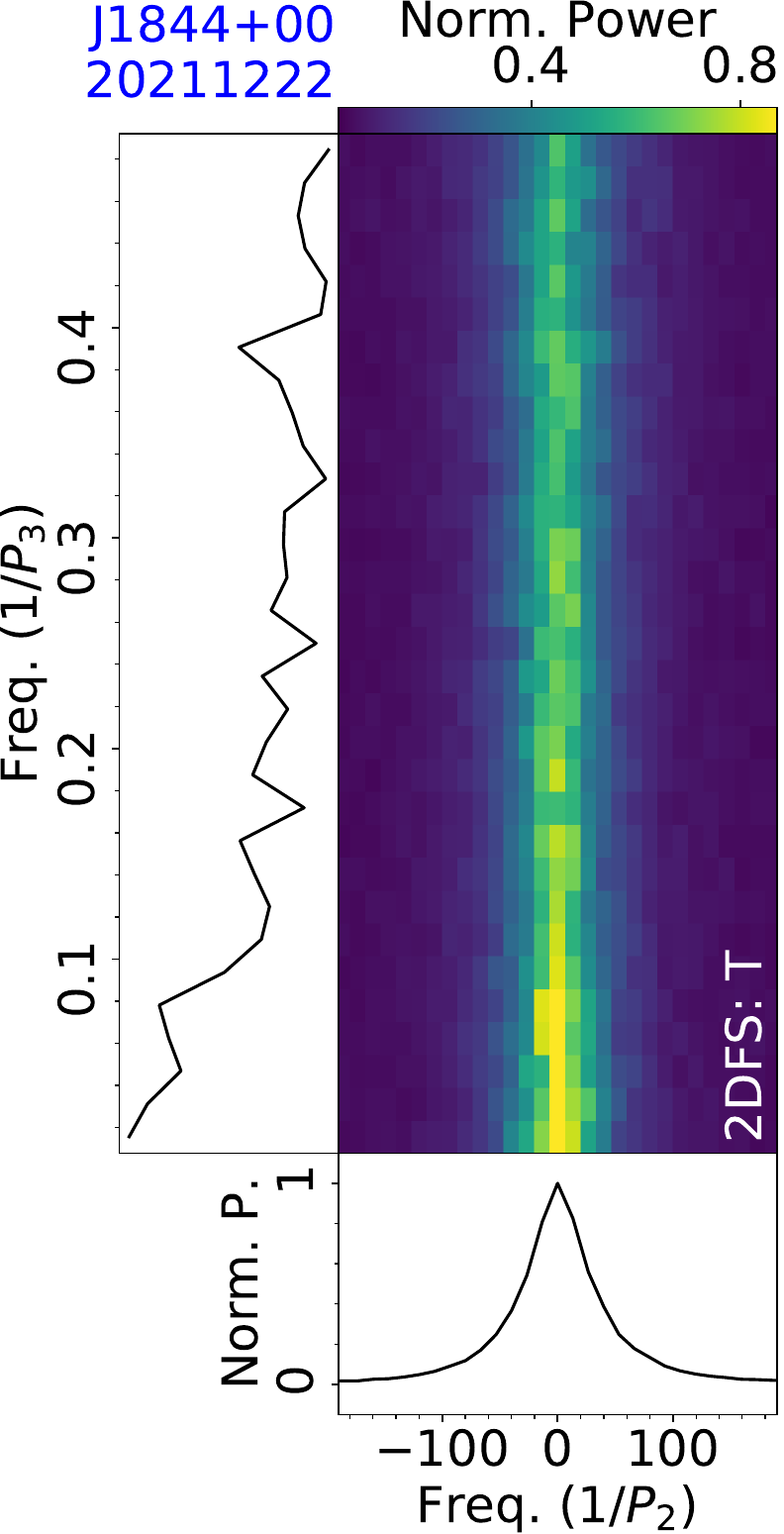}
\figcaption{Fluctuation analysis of PSR J1844+00 from the observation on 20211008 and 20211222, with LRFS and 2DFS for the on-pulse region of a mean pulse profile.
\label{subfig:fluctu:J1844+00}}
\end{figure}

\subsection{J1843-0757}
\label{subsec:J1843-0757}

PSR J1843-0757 was discovered by MeerTRAP with the MeerKAT telescope \citep{Bezuidenhout2022}. A nulling fraction of around 35 percent was reported by \citet{Sengar2023}.

This pulsar was observed by FAST on 20251015 for 15 minutes, deriving a rotation period $P=2.0321$~s and a dispersion measure $D\!M=254.2~{\rm cm^{-3}\,pc}$. Single pulse sequences in Fig.~\ref{subfig:TP:J1843-0757} show nulling and subpulse drifting phenomena. The nulling fraction of this observation is estimated to be 54.1$\pm$5.3\% from the on-pulse integral energy histogram (Fig.~\ref{subfig:Hist:J1843-0757}). Fluctuation spectra are displayed in Fig.~\ref{subfig:fluctu:J1843-0757}. For 2DFS of the trailing profile part, the centroid frequencies of the positive drift feature are estimated to be $1/P_3=0.403\pm0.001$ and $1/P_2=25\pm6$, corresponding to periodicities of $P_3=2.48\pm0.01$ periods and $P_2=15\pm4$ degrees.

\begin{figure}[htpb]
\centering
\includegraphics[height=0.94\textheight]{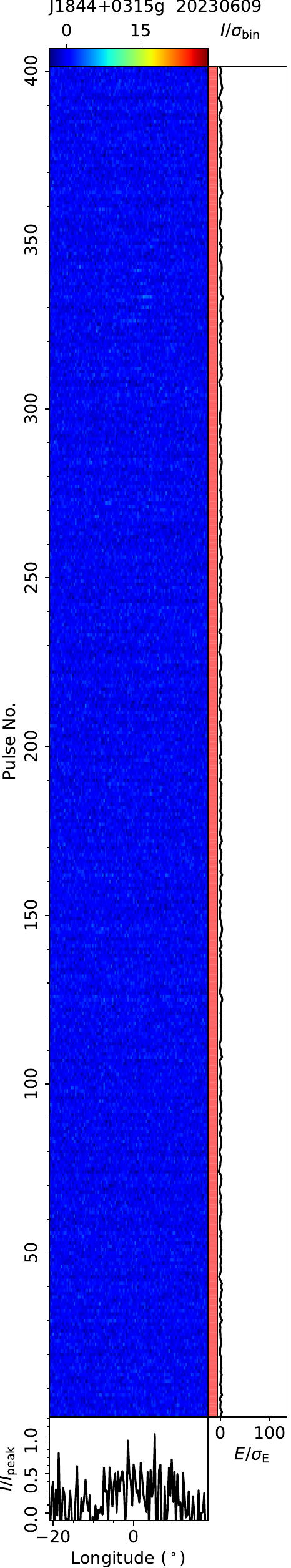}
\includegraphics[height=0.94\textheight]{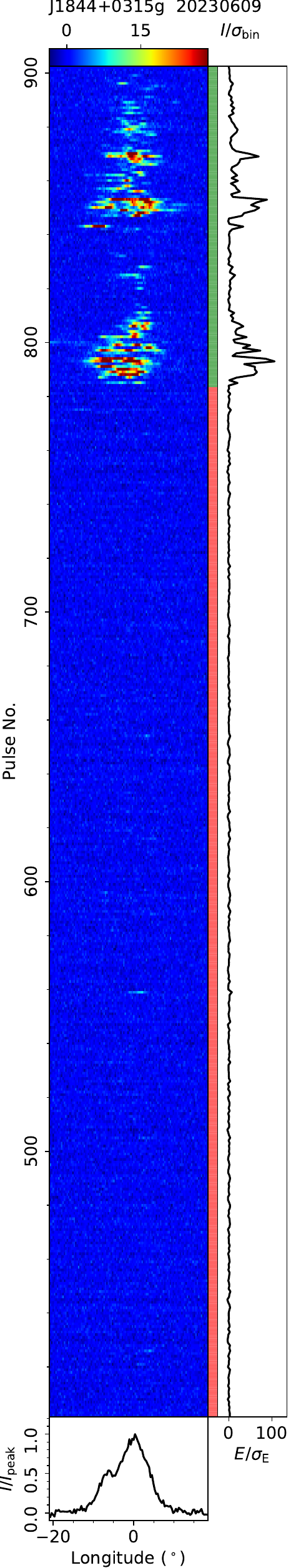} 
\figcaption{Single pulse sequences of PSR J1844+0315g from the FAST observation on 20230609. \label{subfig:TP:J1844+0315g}}
\end{figure}

\begin{figure}[htpb]
\centering
\includegraphics[width=0.39\textwidth, angle=0]{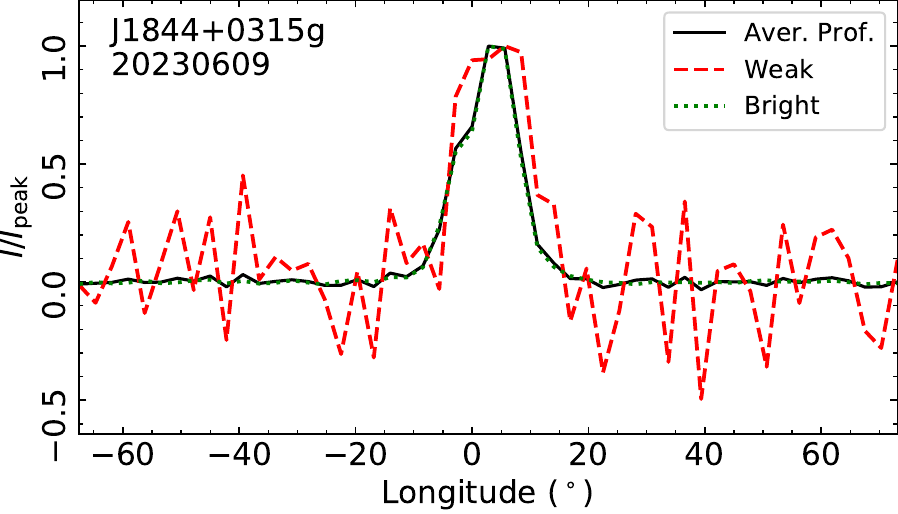}
\figcaption{Mean profiles of weak and bright emission modes of PSR J1844+0315g from the FAST observation on 20230609, normalized by their respective peaks.
\label{subfig:PolModes:J1844+0315g}}
\end{figure}

\subsection{J1844+00}
\label{subsec:J1844+00}

PSR J1844+00 was discovered using the Arecibo radio telescope at 430 MHz \citep{Camilo1996}. The nulling phenomenon of this pulsar was first reported by \citet{Wang2020} using the Jiamusi 66 m telescope at a central frequency of 2250 MHz. The pulsar was observed for 180 minutes, and 256 nulling pulses were detected by integrating 32 pulses to form a subintegration. The nulling fraction was estimated to be between 0.3\% and 34.0\%. 

This pulsar was observed by FAST on 20210713, 20211008, and 20211222, with each observation conducted for 15 minutes. From the data of 20211222, a rotation period $P=0.4605$~s and a dispersion measure $D\!M=345.3~{\rm cm^{-3}\,pc}$ were determined. 
Single pulse sequences of the observation on 20211008 and 20211222 are shown in Fig.~\ref{subfig:TP:J1844+00}, and 2DFS of the leading and trailing parts in the profiles are displayed in Fig.~\ref{subfig:fluctu:J1844+00}. For the leading part in the profile, there are two modulation features in 2DFS with the centroid frequencies of $1/P_3=0.068\pm0.003$ ($P_3=15\pm1$ periods) and $1/P_3=0.422\pm0.003$ ($P_3=2.37\pm0.02$ periods) from the observation on 20211008. The temporal modulation feature in 2DFS of the trailing part is more widely distributed, with the centroid frequency of $1/P_3=0.120\pm0.004$ ($P_3=8.3\pm0.3$ periods). However, the $\sim$2-period feature is not obvious in the 2DFS of the leading component from the observation on 20211222. The modulation features of the observation on 20210713 are consistent with those on 20211008. 
There is no single pulse whose intensity is less than 3$\sigma_{\rm E}$, indicating no nulling behavior during these FAST observations. 



\begin{figure}[htpb]
\centering
\includegraphics[width=0.22\textwidth, angle=0]{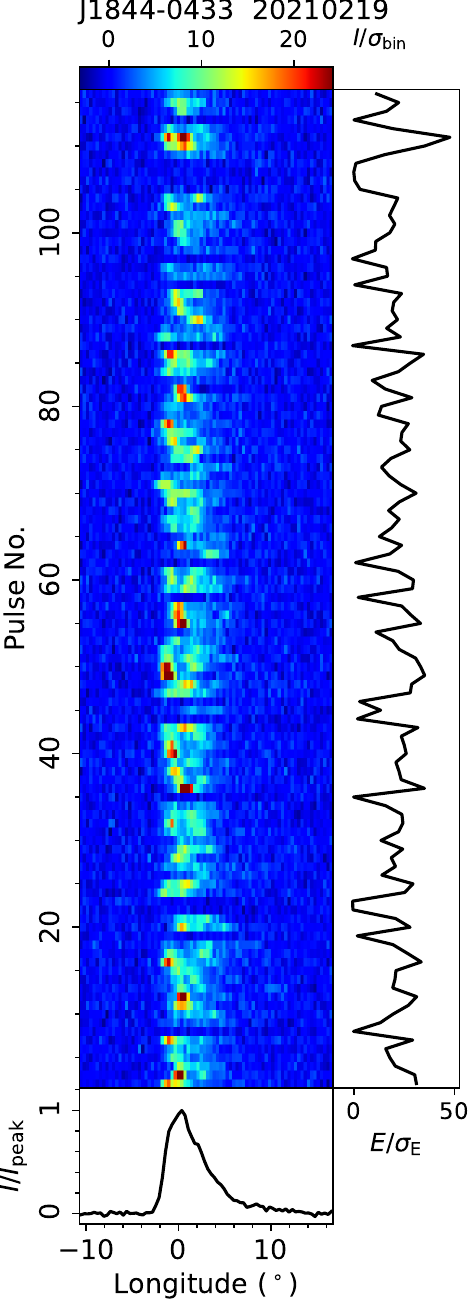}
\figcaption{Single pulse sequence of PSR J1844-0433 from the FAST observation on 20210219.
\label{subfig:TP:J1844-0433}}
\end{figure}

\begin{figure}[htpb]
\centering
\includegraphics[width=0.39\textwidth, angle=0]{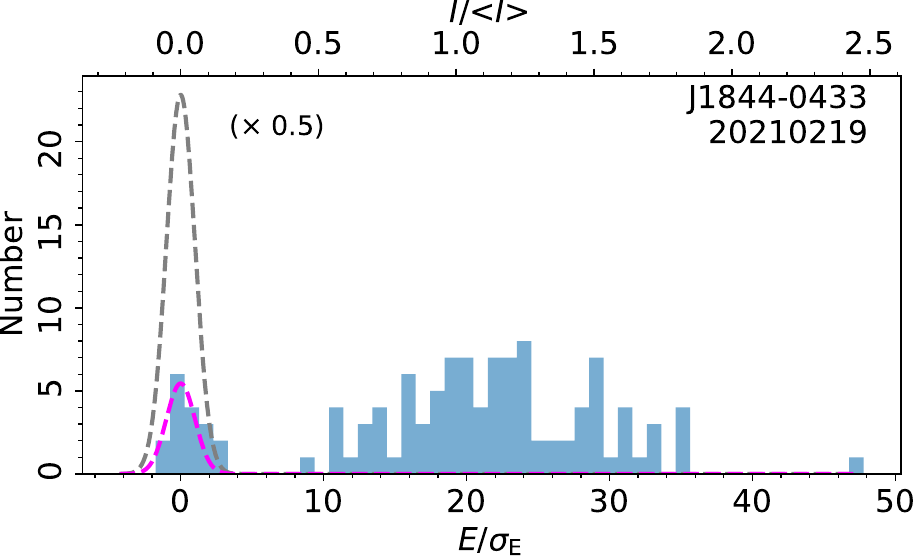}
\figcaption{On-pulse energy histogram of single pulses of PSR J1844-0433 from the FAST observation on 20210219.
\label{subfig:Hist:J1844-0433}}
\end{figure}

\begin{figure}[htpb]
\centering
\includegraphics[width=0.44\textwidth, angle=0]{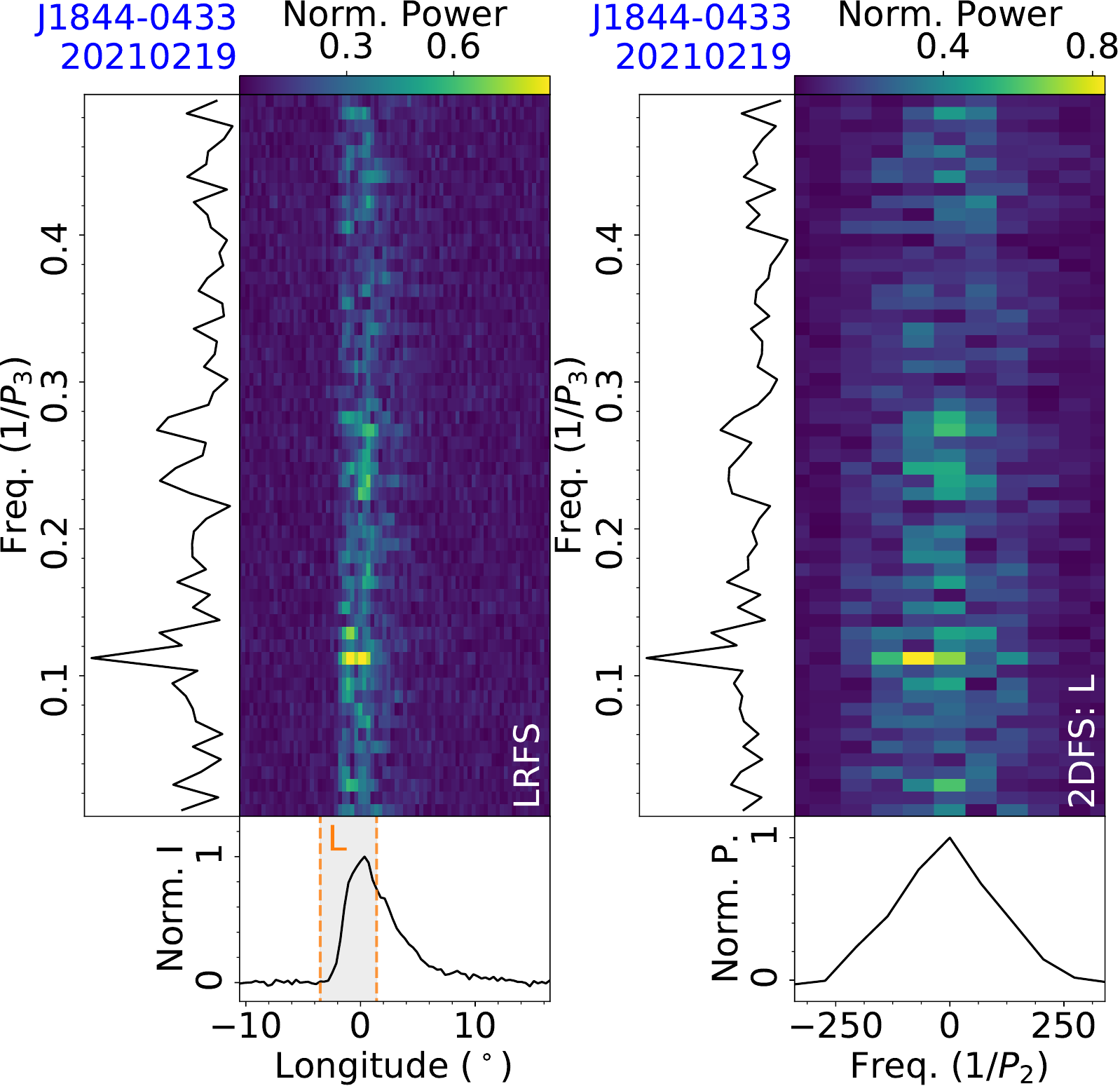}
\figcaption{Fluctuation analysis of PSR J1844-0433 for the observation on 20210219, with LRFS and 2DFS for the leading phase region of a mean pulse profile.
\label{subfig:fluctu:J1844-0433}}
\end{figure}

\subsection{J1844+0315g}
\label{subsec:J1844+0315g}

PSR J1844+0315g was discovered in the FAST GPPS survey \citep{Han2021,han2025}.

This pulsar was observed by FAST on 20230609 for 5 minutes, with a rotation period $P=0.3189$~s and a dispersion measure $D\!M=23.3~{\rm cm^{-3}\,pc}$. 
Single pulse sequences are shown in Fig.~\ref{subfig:TP:J1844+0315g}, where the weak and bright emission modes are divided and labeled using red and green colors. For the bright emission mode, the single-pulse width and integral energy are temporally variable. The weak mode instead of nulling is proved by the mean pulse profile shown in Fig.~\ref{subfig:PolModes:J1844+0315g}. For the weak emission mode in single pulse sequences, the intensity is occasionally enhanced lasting for only a single period, such as pulse No. 559.

More detailed single-pulse properties require longer observations.

\begin{figure}[htpb]
\centering
\includegraphics[width=0.22\textwidth, angle=0]{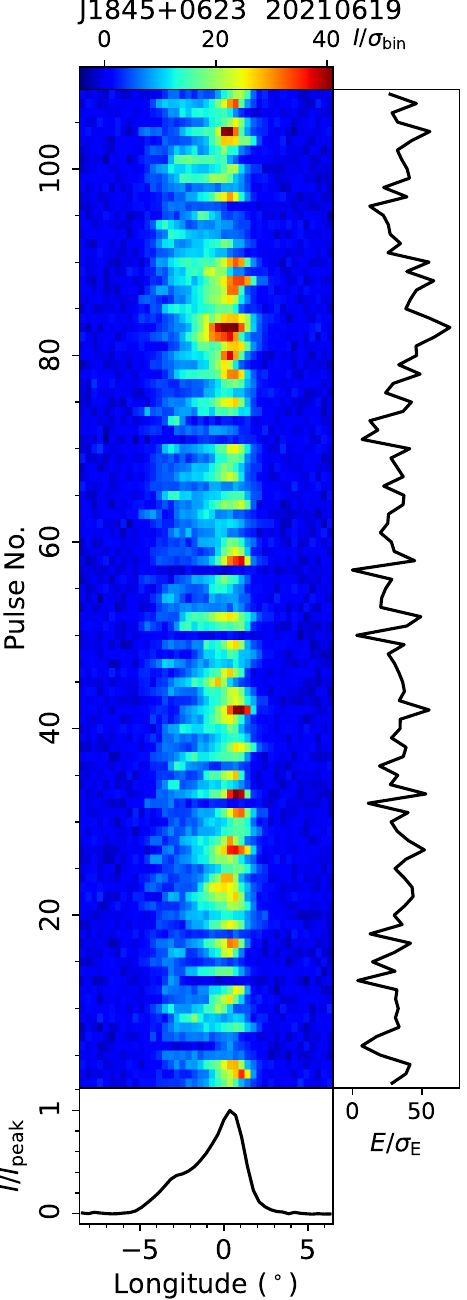}
\includegraphics[width=0.22\textwidth, angle=0]{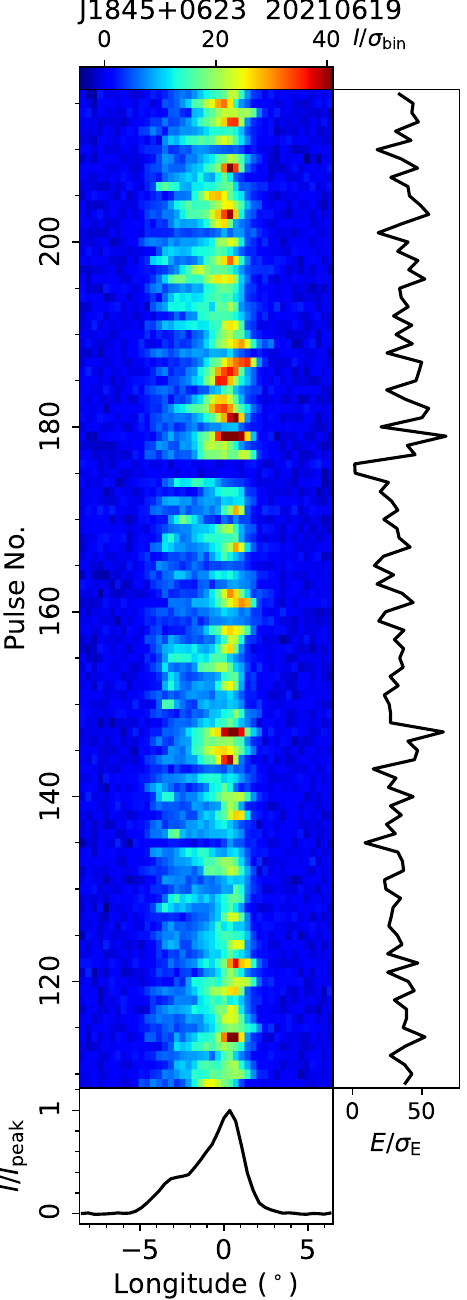}
\figcaption{Single pulse sequences of PSR J1845+0623 from the FAST observation on 20210619.
\label{subfig:TP:J1845+0623}}
\end{figure}

\begin{figure}[htpb]
\centering
\includegraphics[width=0.39\textwidth, angle=0]{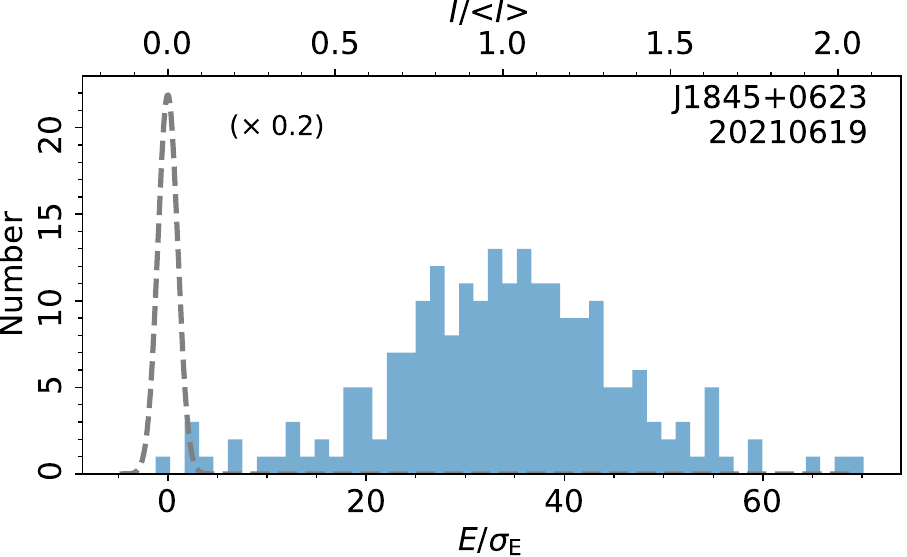}
\figcaption{On-pulse energy histogram of single pulses of PSR J1845+0623 from the FAST observation on 20210619.
\label{subfig:Hist:J1845+0623}}
\end{figure}

\begin{figure}[htpb]
\centering
\includegraphics[width=0.22\textwidth, angle=0]{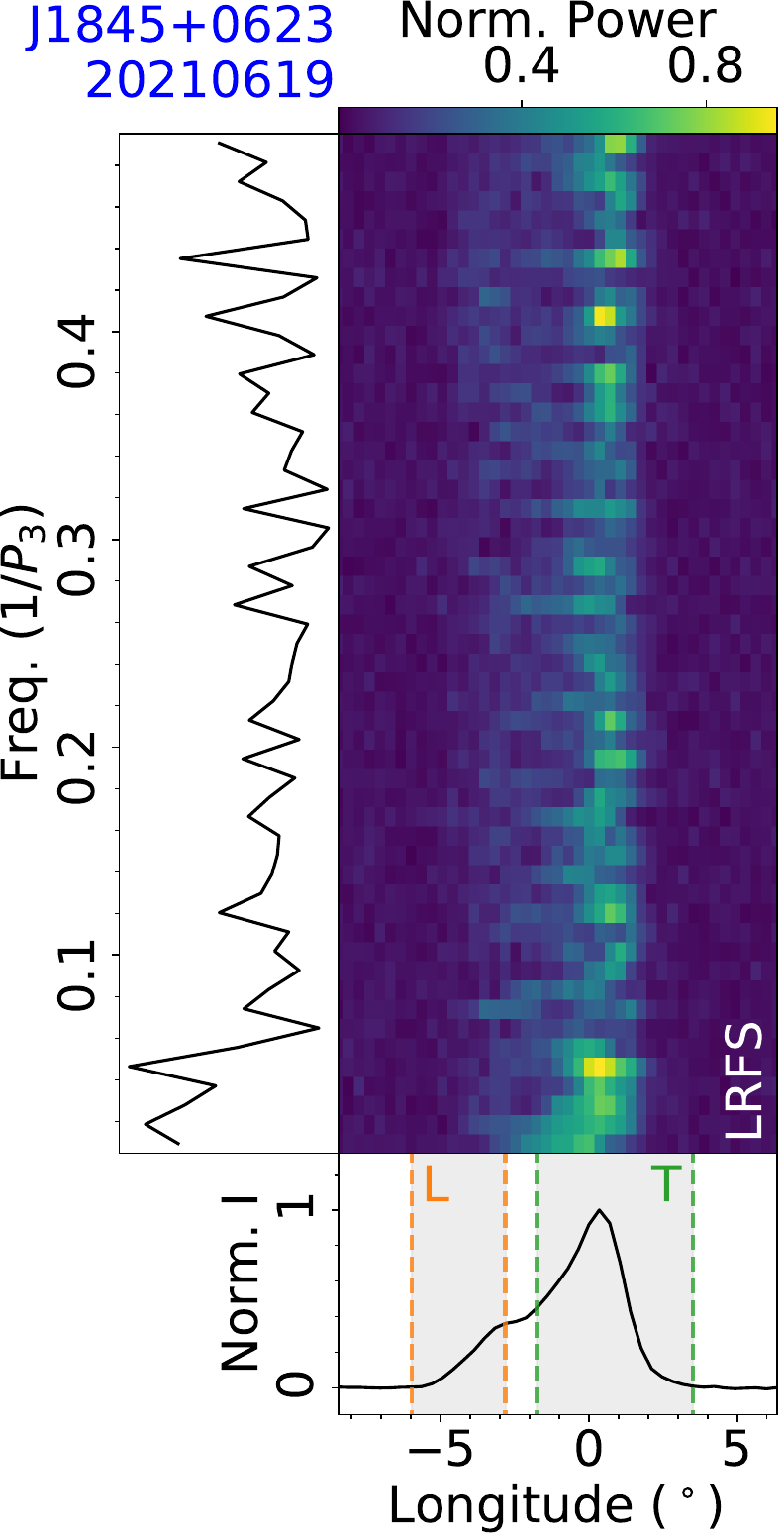}
\includegraphics[width=0.22\textwidth, angle=0]{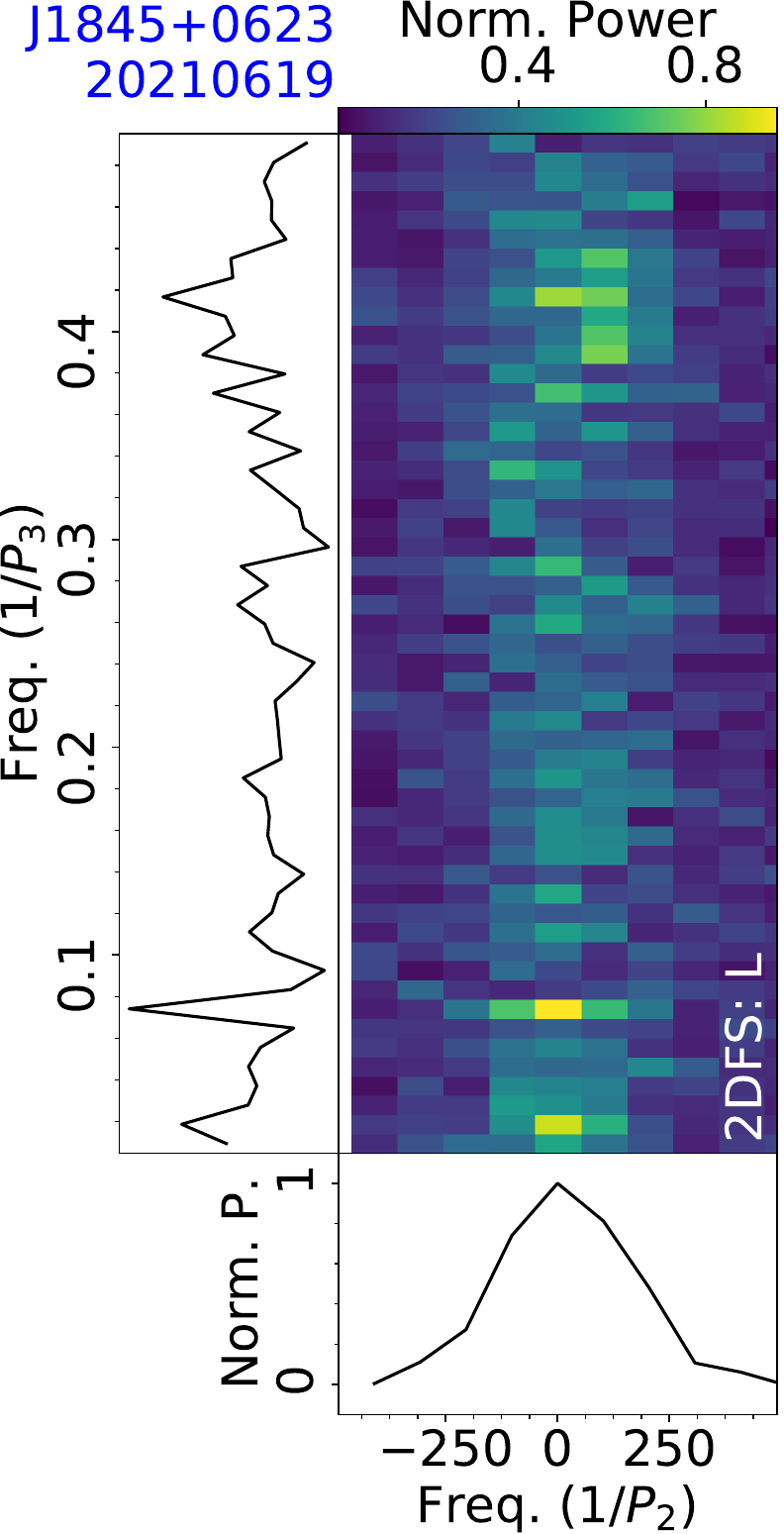}
\figcaption{Fluctuation analysis of PSR J1845+0623 for the observation on 20210619, with LRFS and 2DFS for the leading phase region of a mean pulse profile.
\label{subfig:fluctu:J1845+0623}}
\end{figure}




\begin{figure}[hbt]
\centering
\includegraphics[width=0.22\textwidth, angle=0]{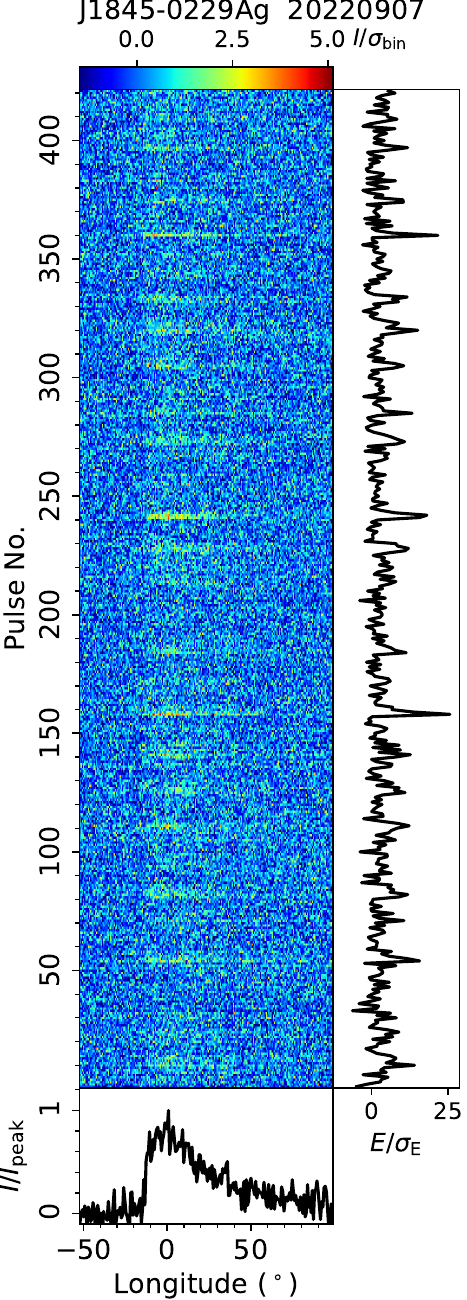}
\figcaption{Single pulse sequence of PSR J1845-0229Ag from the FAST observation on 20220907. 
\label{subfig:TP:J1845-0229Ag}}
\end{figure}

\begin{figure}[htpb]
\centering
\includegraphics[width=0.22\textwidth, angle=0]{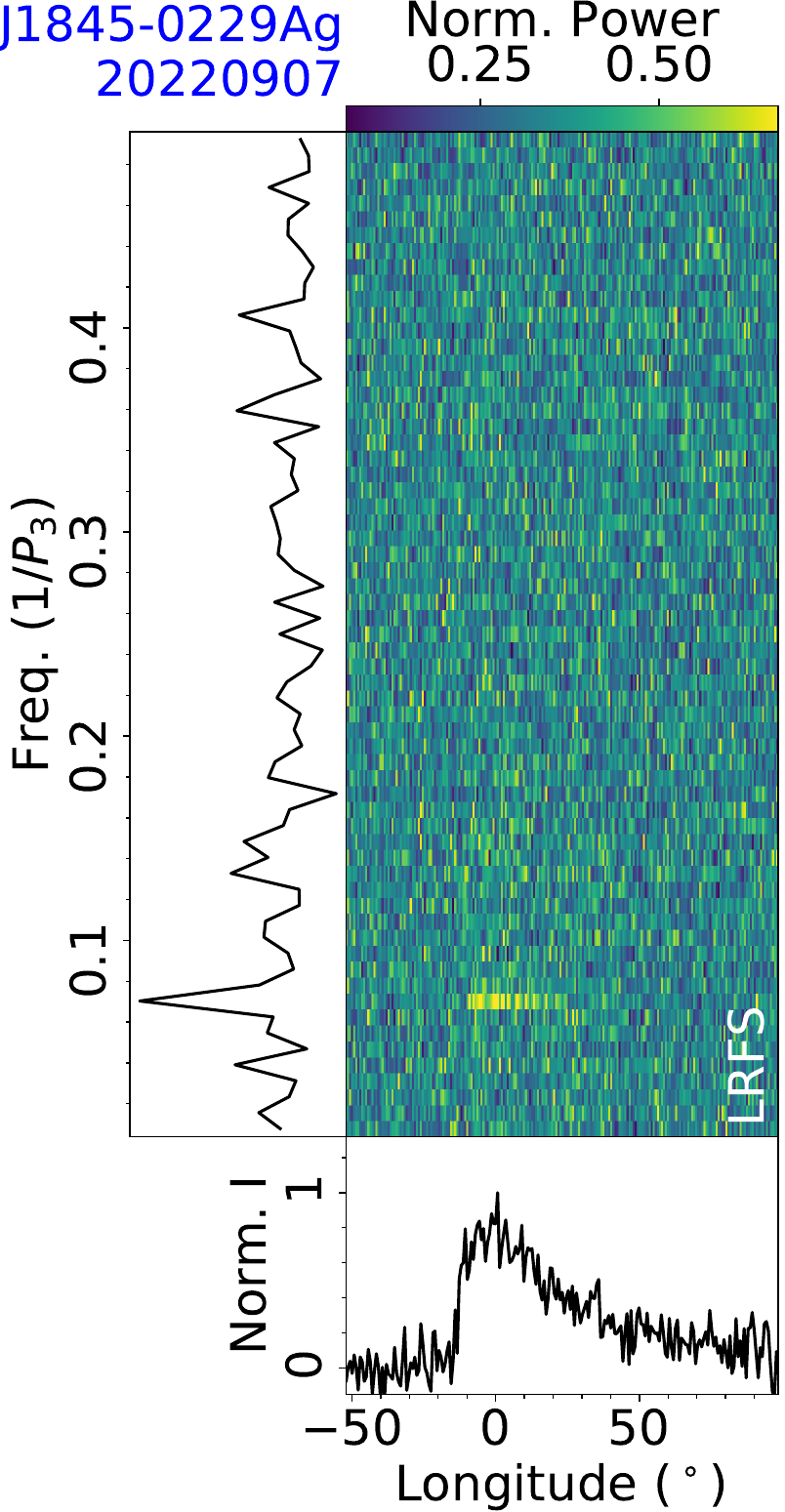}
\includegraphics[width=0.22\textwidth, angle=0]{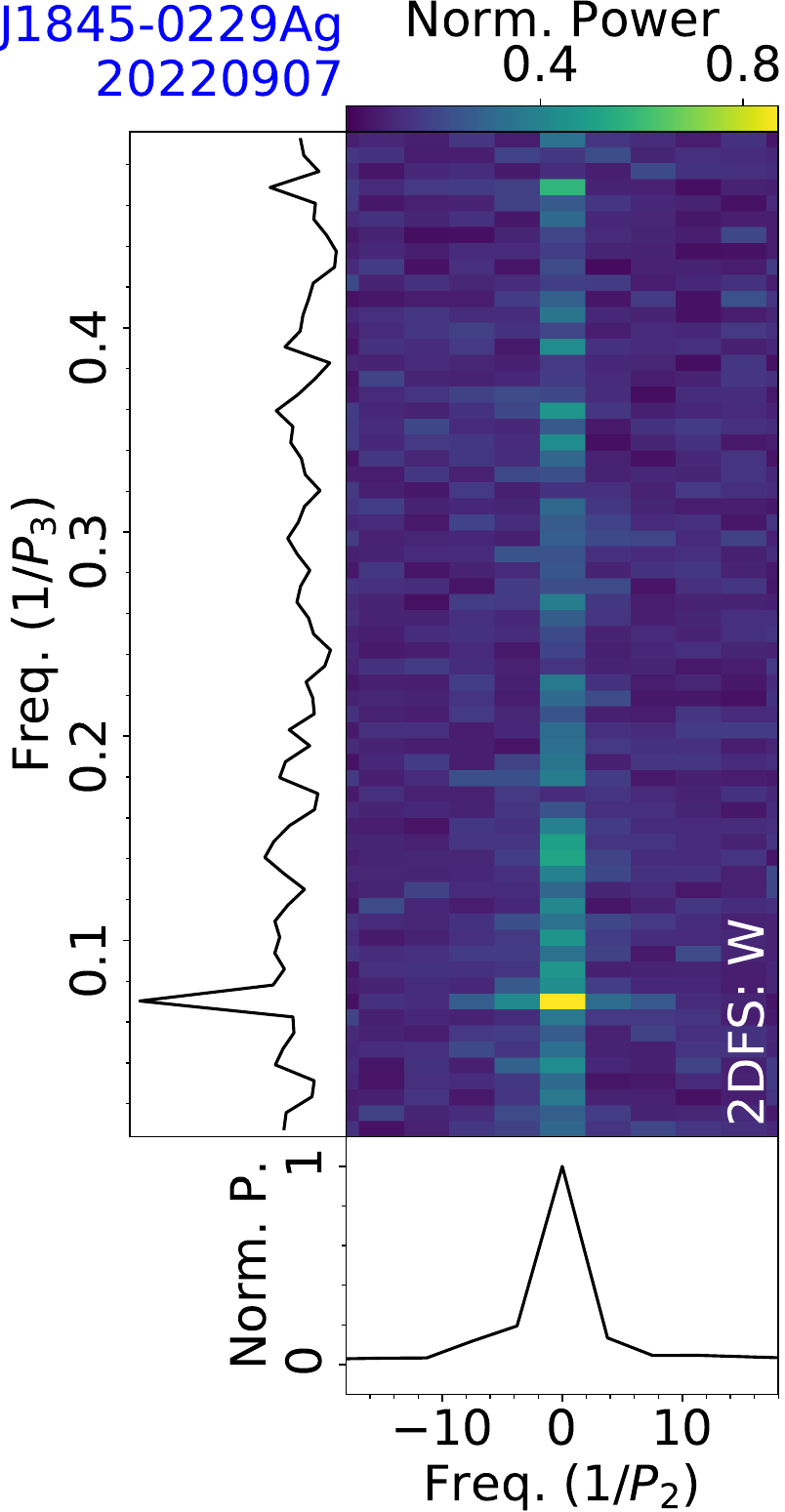}
\figcaption{Fluctuation analysis of PSR J1845-0229Ag from the observation on 20220907, with LRFS and 2DFS for the on-pulse region of a mean pulse profile.
\label{subfig:fluctu:J1845-0229Ag}}
\end{figure}

\begin{figure}[htpb]
\centering
\includegraphics[width=0.21\textwidth, angle=0]{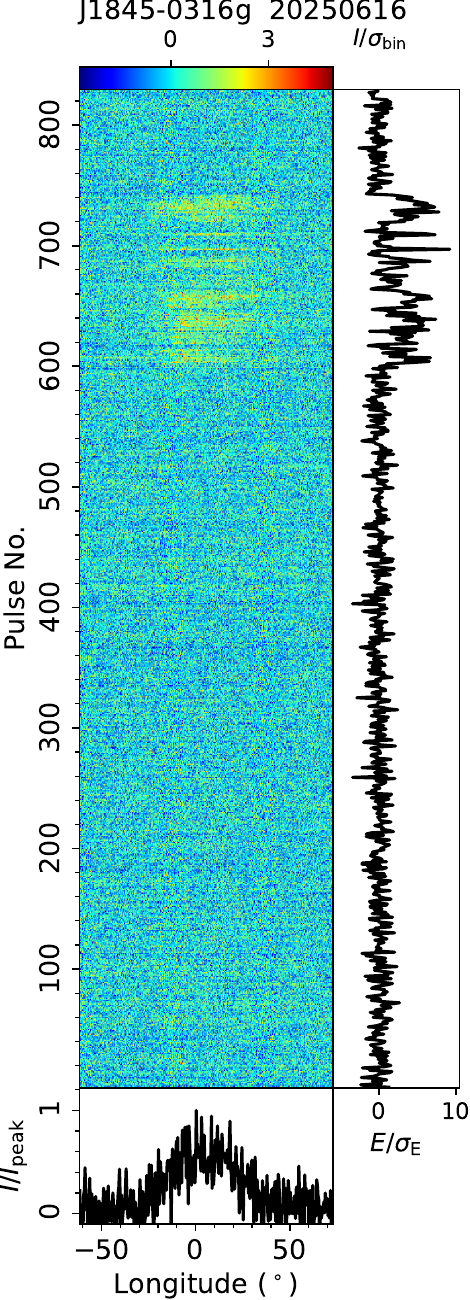}
\includegraphics[width=0.21\textwidth, angle=0]{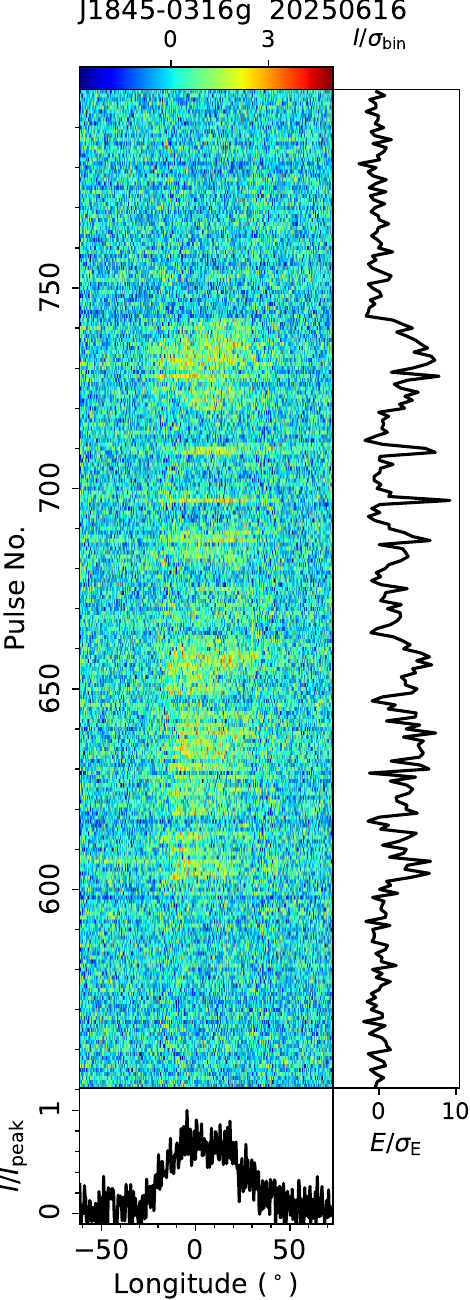}
\figcaption{Single pulse sequence of PSR J1845-0316g from the FAST observation on 20250616, and a zoomed-in view of pulses No. 550-800.
\label{subfig:TP:J1845-0316g}}
\end{figure}

\begin{figure}[htpb]
\centering
\includegraphics[width=0.39\textwidth, angle=0]{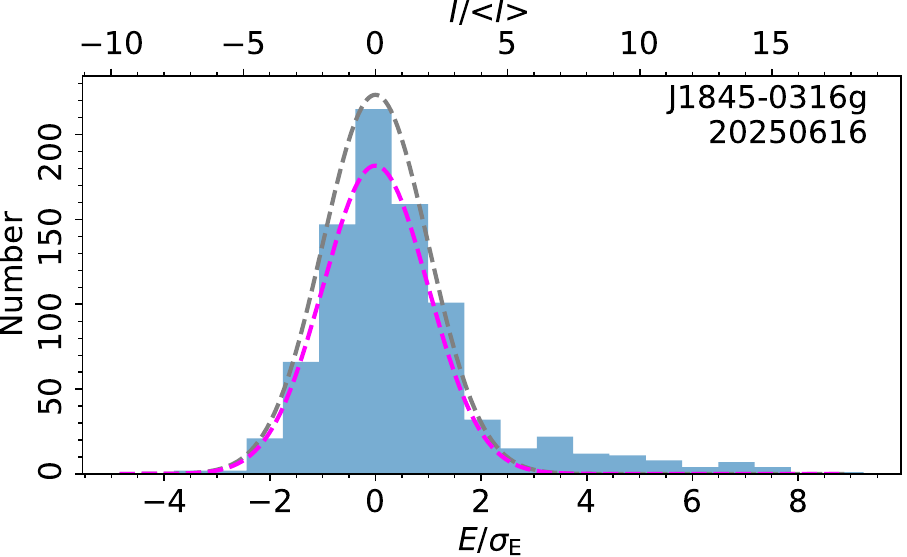}
\figcaption{On-pulse energy histogram of single pulses of PSR J1845-0316g from the FAST observation on 20250616.
\label{subfig:Hist:J1845-0316g}}
\end{figure}

\subsection{J1844-0433}
\label{subsec:J1844-0433}

PSR J1844-0433 was discovered by the MKIA telescope \citep{Clifton1987}. \citet{Weltevrede2006} reported the negative drift behavior at 21 cm with $P_3=8.5\pm0.1$ periods and $P_2=-13^{+11}_{−5}$ degrees. \citet{Song2023} presented the drift periodicities of $P_3=8.1\pm0.3$ periods and $P_2=-5^{+2}_{-4}$ degrees measured at 1280 MHz with the MeerKAT telescope. 

This pulsar was observed by FAST on 20210219 for 2 minutes, deriving a rotation period $P=0.9910$~s and a dispersion measure $D\!M=124.1~{\rm cm^{-3}\,pc}$. The single pulse sequence in Fig.~\ref{subfig:TP:J1844-0433} displays negative drifting for the leading part in the mean pulse profile and nulling. The nulling fraction is estimated to be 12$\pm$1\% from the on-pulse energy histogram of single pulses (Fig.~\ref{subfig:Hist:J1844-0433}). From LRFS and 2DFS of the leading profile part in Fig.~\ref{subfig:fluctu:J1844-0433}, the centroid of the negative drift feature is at $1/P_3=0.112\pm0.002$ and $1/P_2=-66\pm13$, corresponding to $P_3=8.9\pm0.1$ periods and $P_2=-5\pm1$ degrees.

\begin{figure}[hbt]
\centering
\includegraphics[width=0.44\textwidth, angle=0]{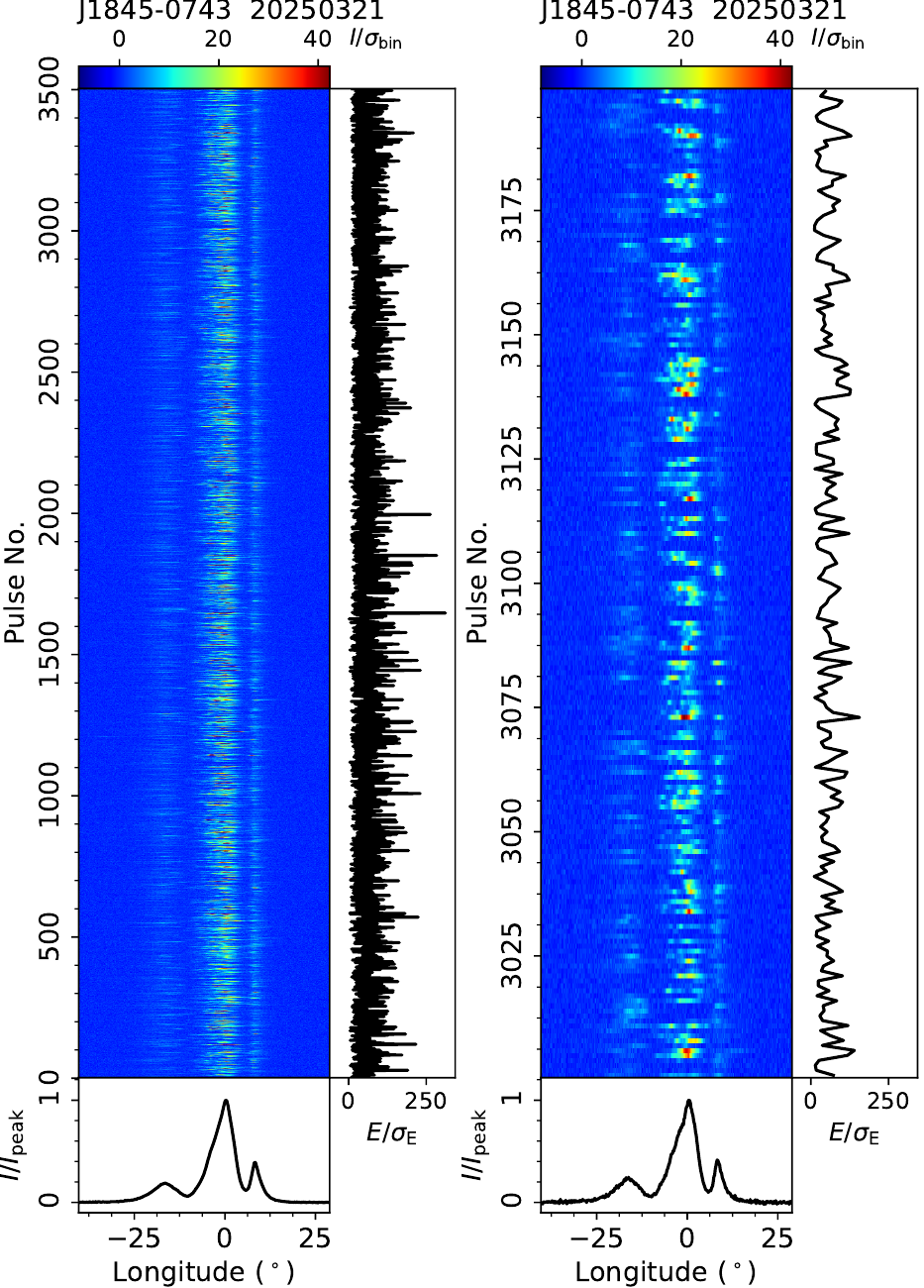}
\figcaption{Single pulse sequence of PSR J1845-0743 from the FAST observation on 20240321, and a zoomed-in view of pulses No. 3000-3200. 
\label{subfig:TP:J1845-0743}}
\end{figure}

\begin{figure}[htpb]
\centering
\includegraphics[width=0.44\textwidth, angle=0]{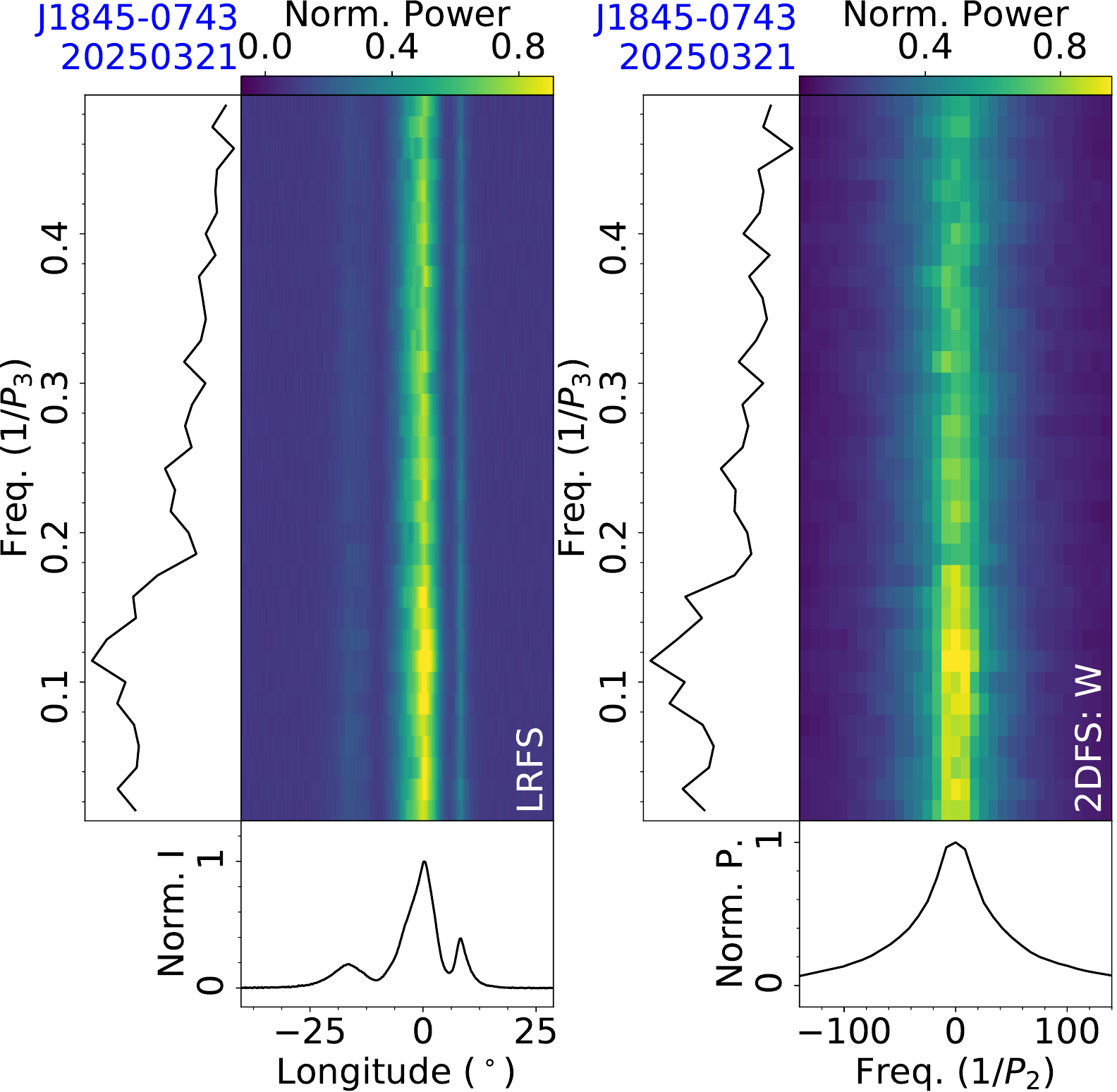}
\figcaption{Fluctuation analysis of PSR J1845-0743 from the observation on 20240321, with LRFS and 2DFS for the on-pulse region of a mean pulse profile.
\label{subfig:fluctu:J1845-0743}}
\end{figure}

\subsection{J1845+0623}
\label{subsec:J1845+0623}

PSR J1845+0623 was discovered in the Parkes Multibeam Pulsar Survey \citep{Lorimer2006}. \citet{Song2023} reported this pulsar with drifting behavior of $P_3=2.35(8)$ periods and $P_2=23_{-17}^{+24}$ degrees. 

This pulsar was observed by FAST for 5 minutes, with a rotation period $P=1.4216$~s and a dispersion measure $D\!M=114.7~{\rm cm^{-3}\,pc}$ from this observation. 
Single pulse sequences in Fig.~\ref{subfig:TP:J1845+0623} display occasional and short-duration decreases in energy. There are three single pulses whose on-pulse integral energy is less than 3$\sigma_{\rm E}$ (Fig.~\ref{subfig:Hist:J1845+0623}). In the 2DFS of the leading component (Fig.~\ref{subfig:fluctu:J1845+0623}), there is a positive drift feature with the centroid frequencies of $1/P_3=0.413\pm0.002$ and $1/P_2=113\pm12$, which correspond to $P_3=2.42\pm0.01$ periods and $P_2=3.2\pm0.4$ degrees. 
Further observations are needed for the causes of the intensity decrease and the low-frequency modulation in LRFS.

\begin{figure}[htpb]
\centering
\includegraphics[width=0.21\textwidth, angle=0]{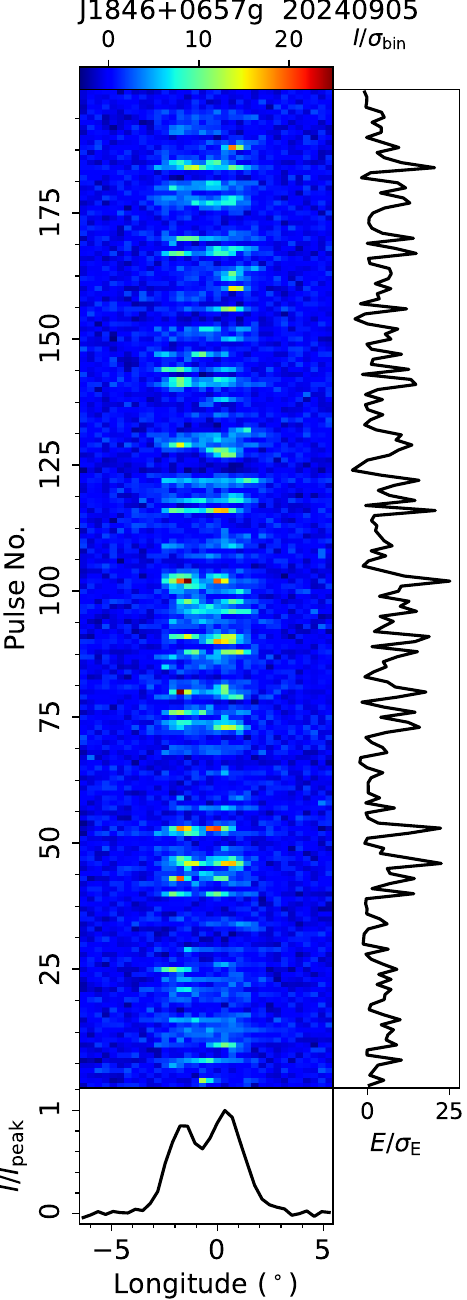}
\includegraphics[width=0.21\textwidth, angle=0]{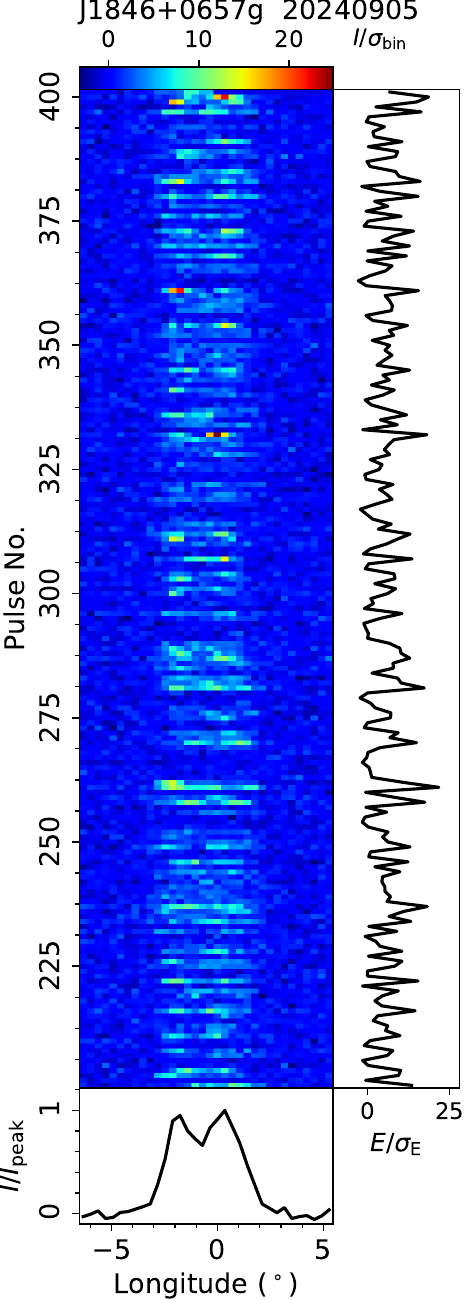}
\figcaption{Single pulse sequences of PSR J1846+0657g from the FAST observation on 20240905.
\label{subfig:TP:J1846+0657g}}
\end{figure}

\begin{figure}[htpb]
\centering
\includegraphics[width=0.39\textwidth, angle=0]{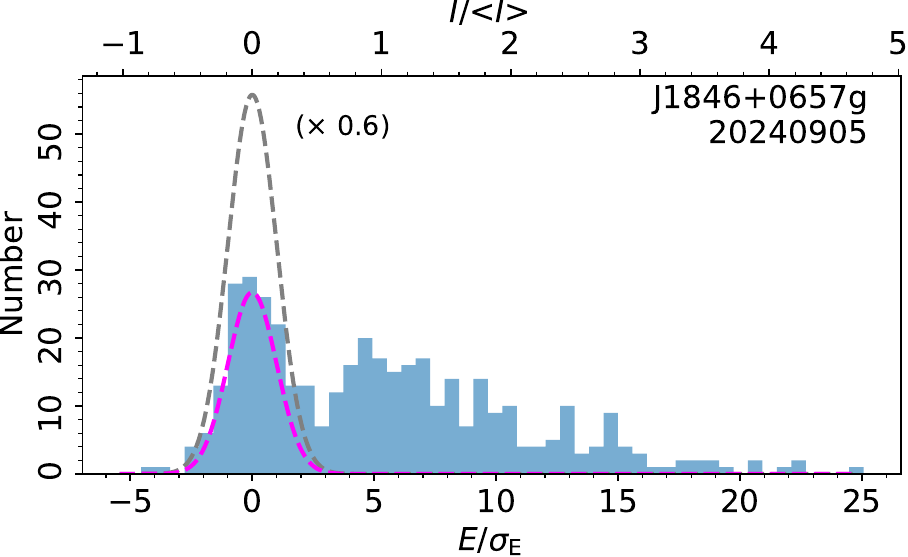}
\figcaption{On-pulse energy histogram of single pulses of PSR J1846+0657g from the FAST observation on 20240905.
\label{subfig:Hist:J1846+0657g}}
\end{figure}

\subsection{J1845-0229Ag}
\label{subsec:J1845-0229Ag}

PSR J1845-0229Ag was discovered in the FAST GPPS survey \citep{Han2021,han2025}.

This pulsar was observed by FAST on 20220907 for 5 minutes, yielding a rotation period $P=0.6578$~s and a dispersion measure $D\!M=848.0~{\rm cm^{-3}\,pc}$. 
The single pulse sequence in Fig.~\ref{subfig:TP:J1845-0229Ag} displays the modulation behavior. In the fluctuation spectrum shown in Fig.~\ref{subfig:fluctu:J1845-0229Ag}, there is a modulation feature with the centroid frequency of $1/P_3=0.071\pm0.001$, which corresponds to $P_3=14.1\pm0.2$ periods.

\begin{figure}[htpb]
\centering
\includegraphics[width=0.22\textwidth, angle=0]{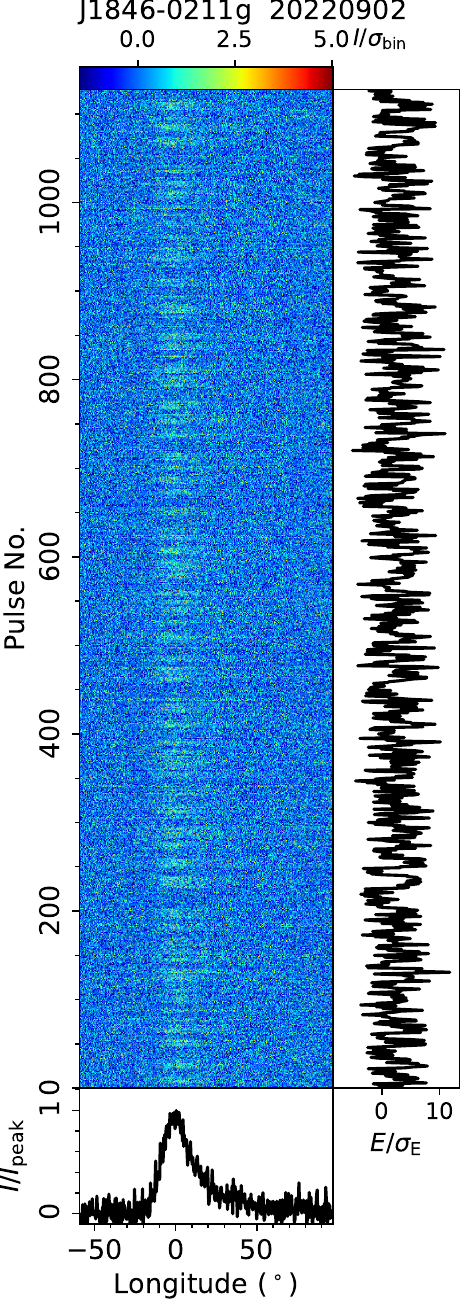}
\figcaption{Single pulse sequence of PSR J1846-0211g from the FAST observation on 20220902.
\label{subfig:TP:J1846-0211g}}
\end{figure}

\begin{figure}[htpb]
\centering
\includegraphics[width=0.22\textwidth, angle=0]{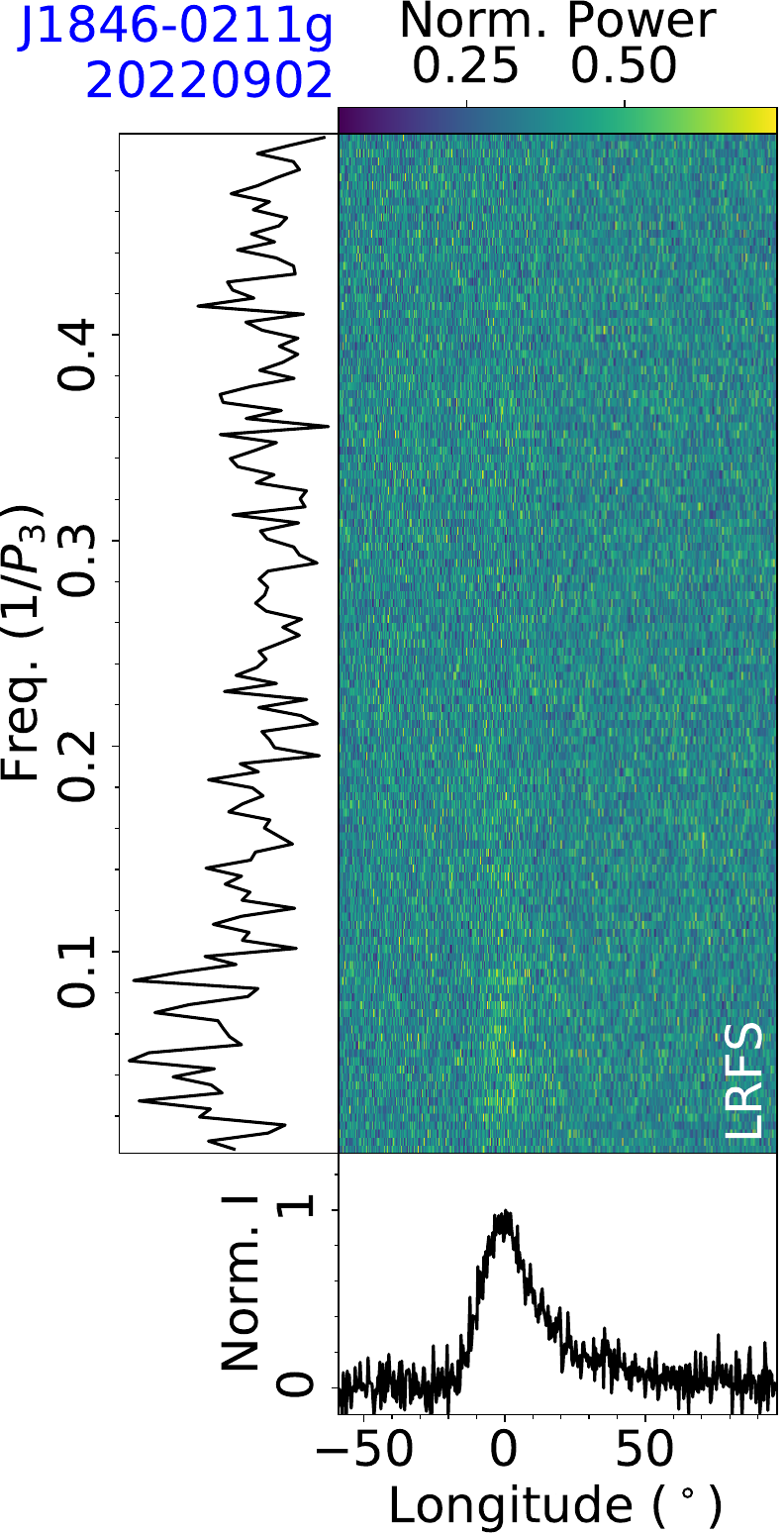}
\includegraphics[width=0.22\textwidth, angle=0]{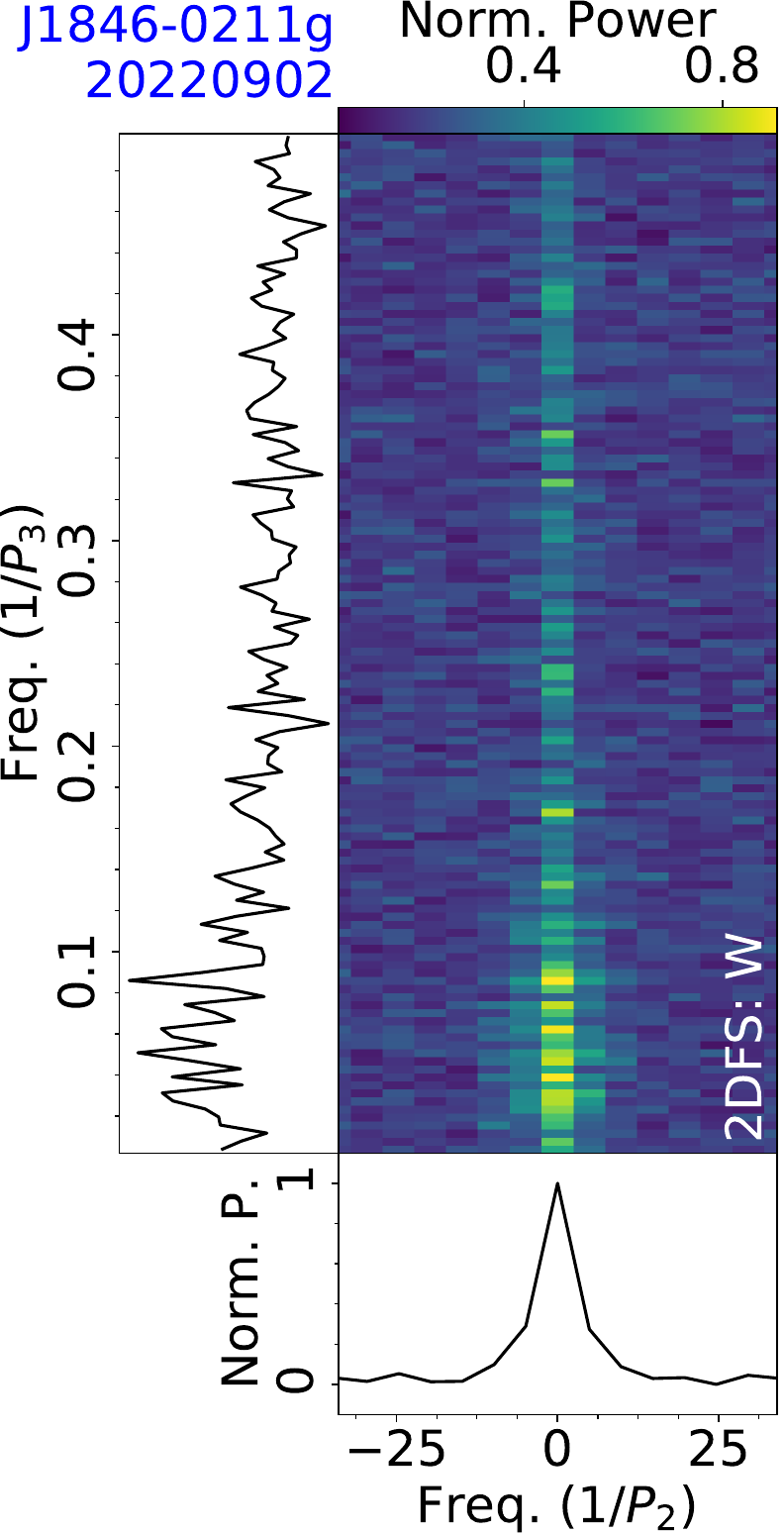}
\figcaption{Fluctuation analysis of PSR J1846-0211g from the observation on 20220902, with LRFS and 2DFS for the on-pulse region of a mean pulse profile.
\label{subfig:fluctu:J1846-0211g}}
\end{figure}

\subsection{J1845-0316g}
\label{subsec:J1845-0316g}

PSR J1845-0316g was discovered in the FAST GPPS survey \citep{Han2021,han2025}.

This pulsar was observed by FAST on 20250616 and 20250829, both for 15 minutes. From the data of 20250616, a rotation period of $P=1.0822$~s and a dispersion measure of $D\!M=913.0~{\rm cm^{-3}\,pc}$ were derived. 
The single pulse sequence and a zoomed-in view of pulses No. 550-800 in Fig.~\ref{subfig:TP:J1845-0316g} display a substantial fraction of  nulls. The nulling fraction is estimated from the on-pulse integral energy histogram (Fig.~\ref{subfig:Hist:J1845-0316g}), which is 81.3$\pm$5.5\% for this observation.

\begin{figure}[htpb]
\centering
\includegraphics[width=0.44\textwidth, angle=0]{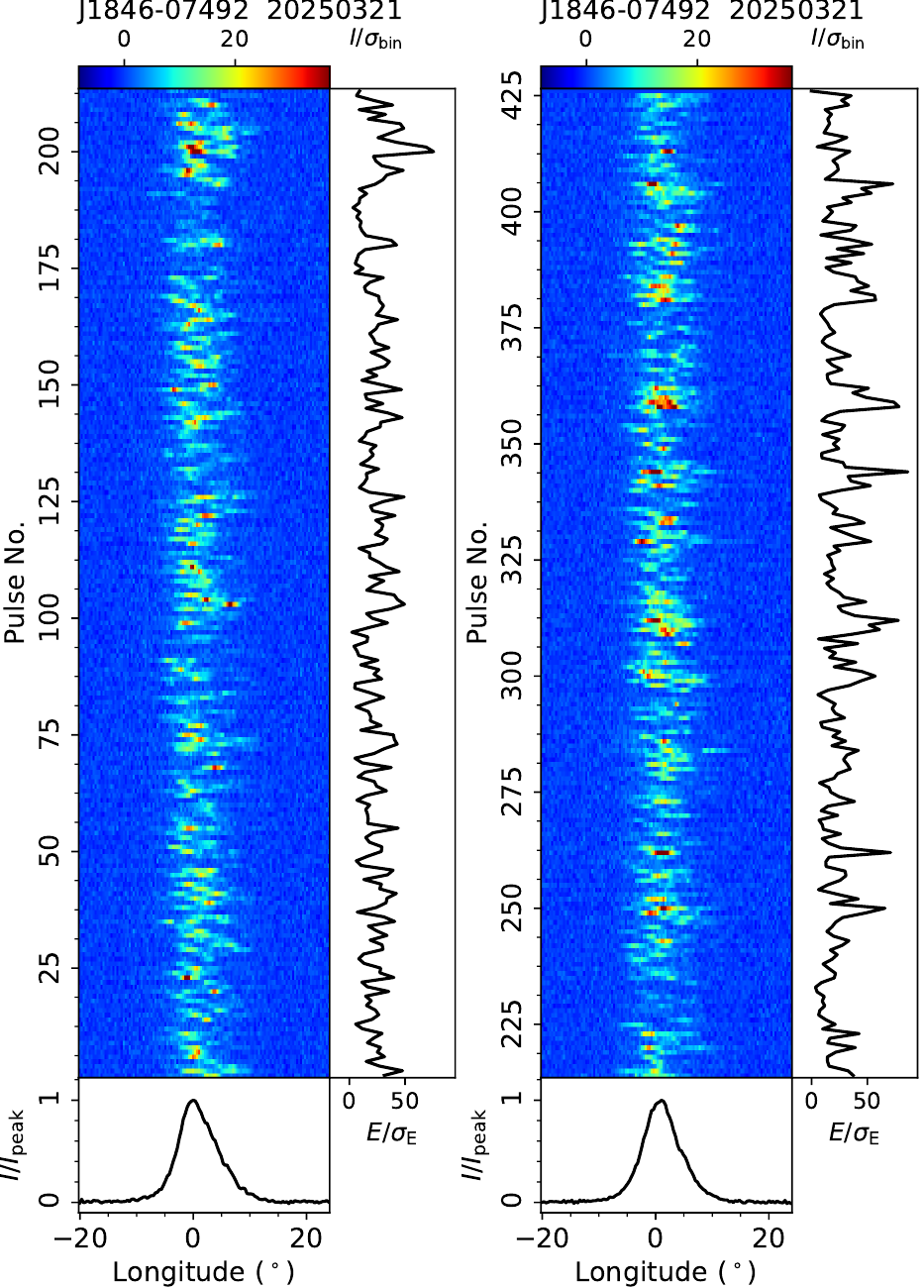}
\figcaption{Single pulse sequences of PSR J1846-07492 from the FAST observation on 20250321.
\label{subfig:TP:J1846-07492}}
\end{figure}

\begin{figure}[htpb]
\centering
\includegraphics[width=0.44\textwidth, angle=0]{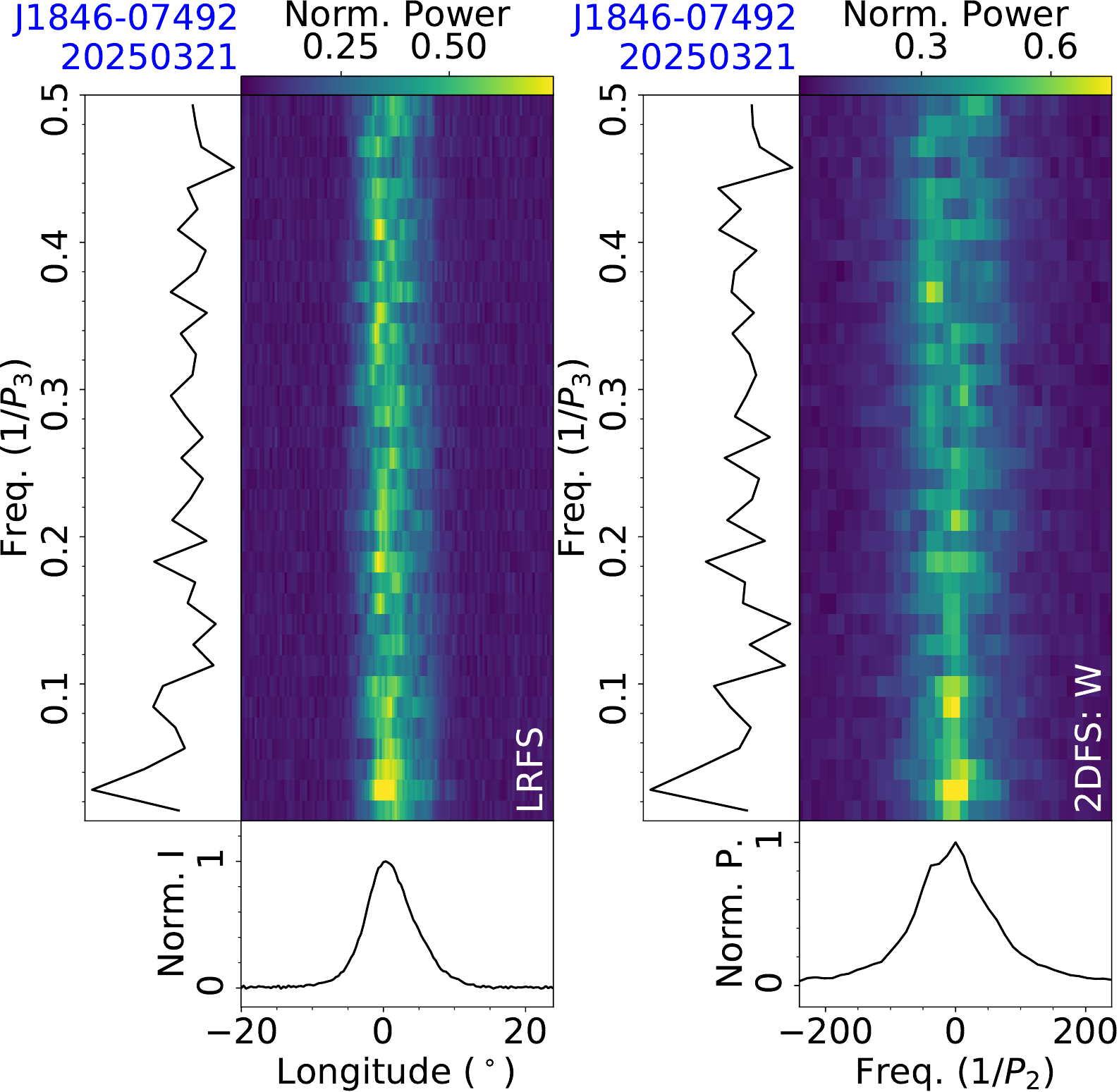}
\figcaption{Fluctuation analysis of PSR J1846-07492 from the FAST observation on 20250321, with LRFS and 2DFS for the on-pulse region of the mean pulse profile.  \label{subfig:fluctu:J1846-07492}}
\end{figure}

\subsection{J1845-0743}
\label{subsec:J1845-0743}

PSR J1845-0743 was discovered in the Parkes Multibeam Pulsar Survey \citep{Kramer2003}. 

This pulsar was observed by FAST on 20250321 for 6 minutes, driving a rotation period of $P=0.1047$~s and a dispersion measure of $D\!M=281.1~{\rm cm^{-3}\,pc}$. The single pulse sequence and a zoomed-in view of pulses No. 3000-3200 are shown in Fig.~\ref{subfig:TP:J1845-0743}. From LRFS and 2DFS in Fig.~\ref{subfig:fluctu:J1845-0743}, this pulsar exhibits the modulation behavior. The modulation feature has a centroid at $1/P_3=0.095\pm0.002$, corresponding to $P_3=10.6\pm0.2$ periods.

\begin{figure}[htpb]
\centering
\includegraphics[width=0.22\textwidth, angle=0]{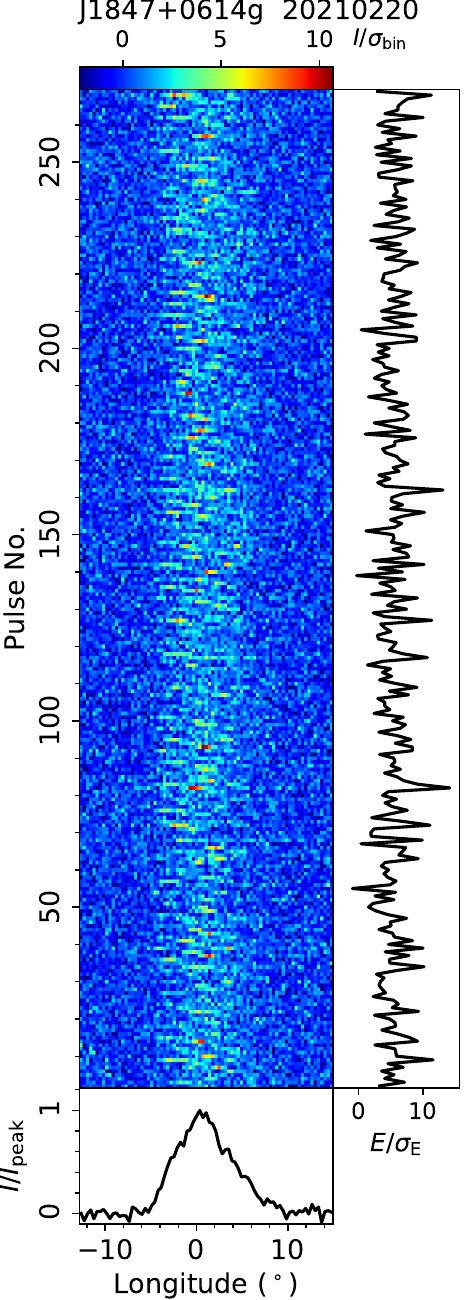}
\includegraphics[width=0.22\textwidth, angle=0]{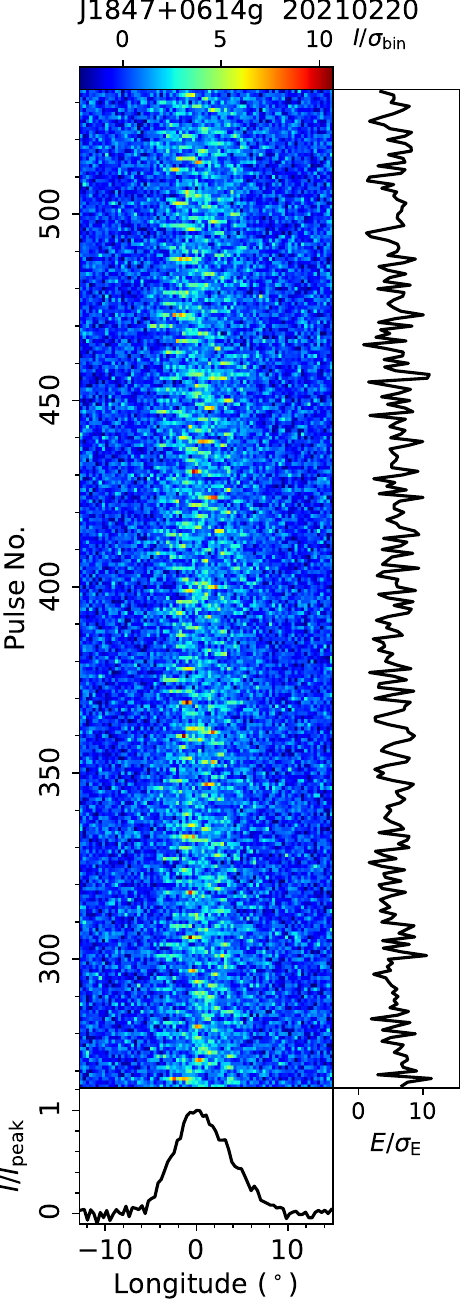}
\figcaption{Single pulse sequences of PSR J1847+0614g from the FAST observation on 20210220.
\label{subfig:TP:J1847+0614g}}
\end{figure}

\begin{figure}[htpb]
\centering
\includegraphics[width=0.21\textwidth, angle=0]{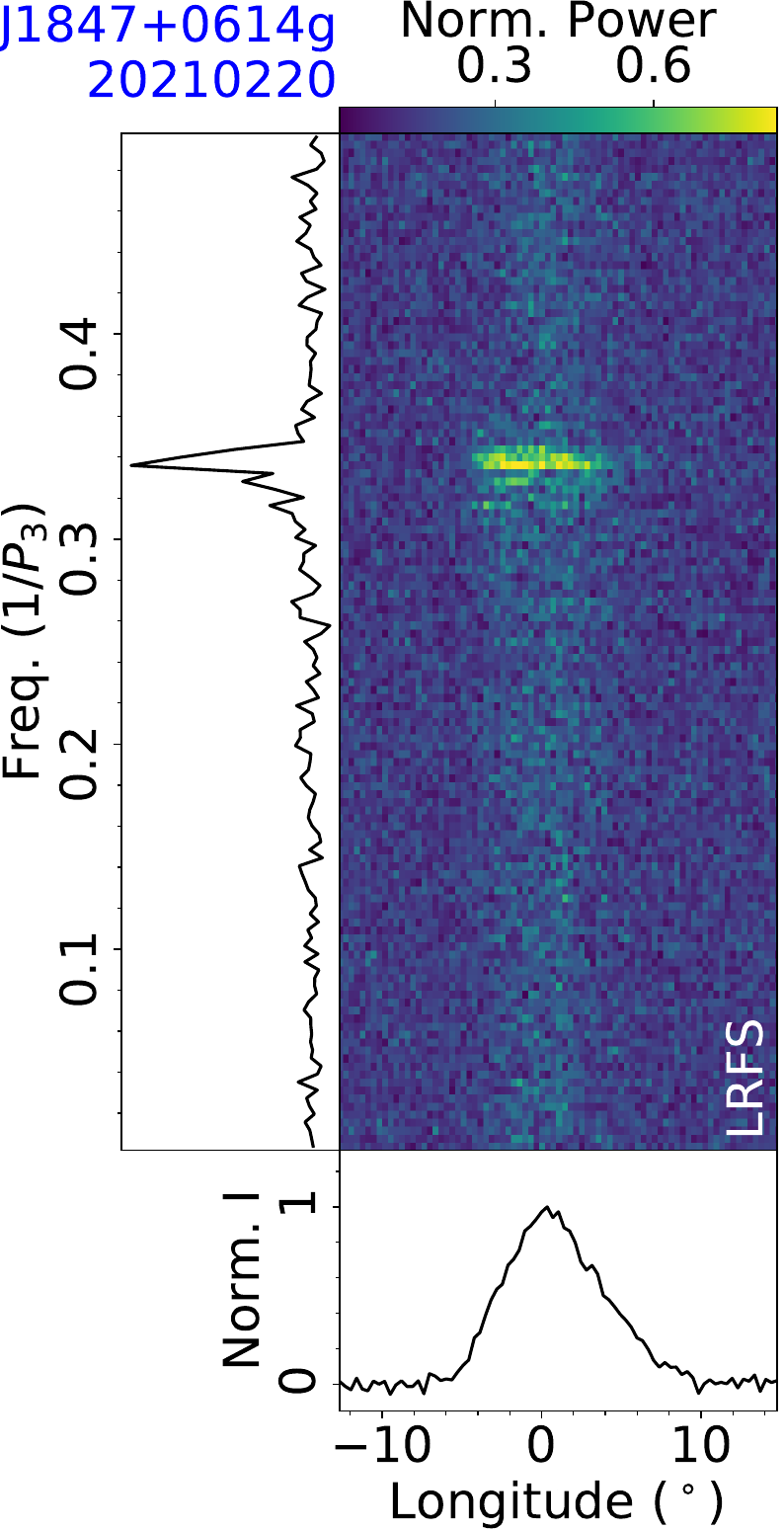}
\includegraphics[width=0.21\textwidth, angle=0]{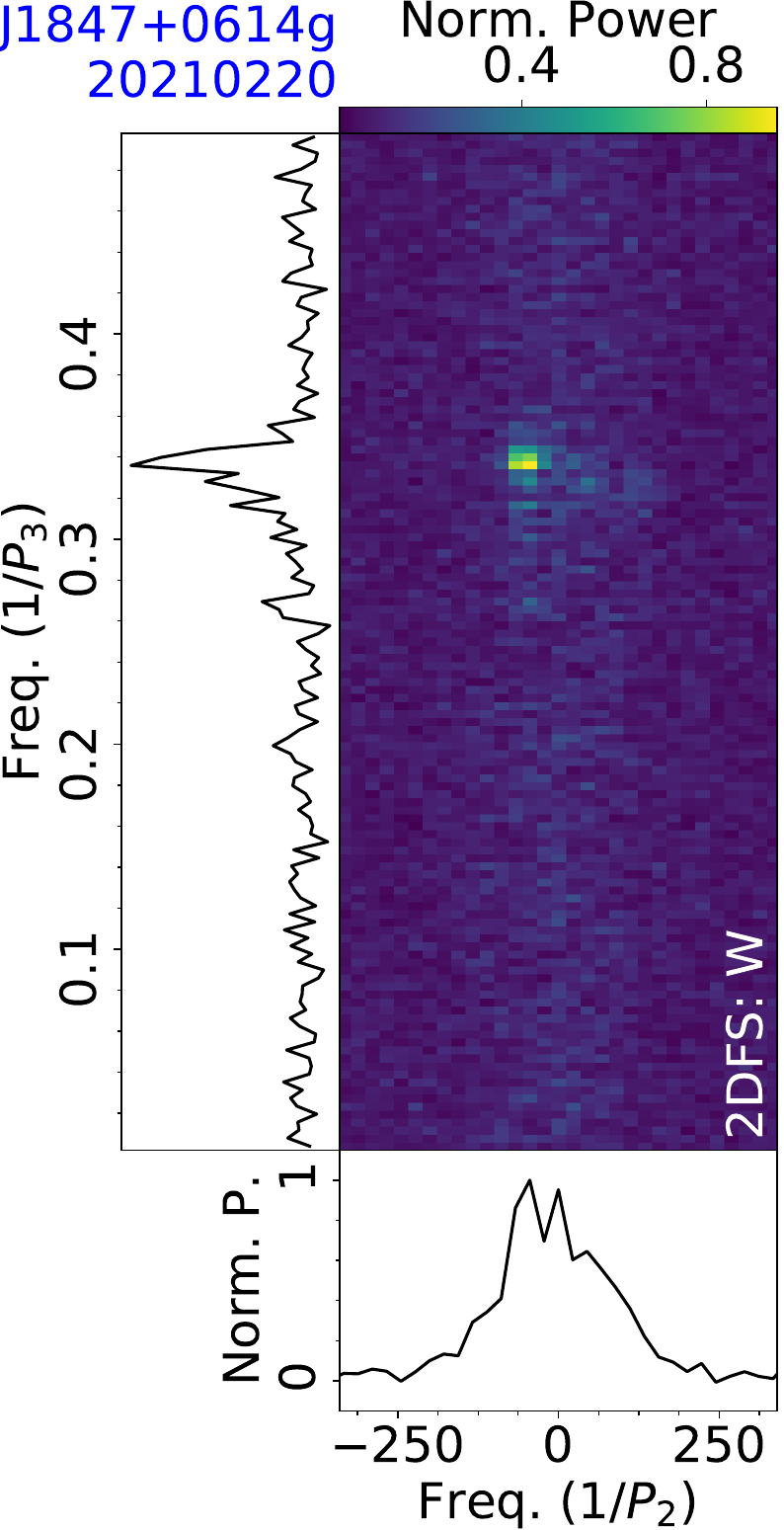}
\figcaption{Fluctuation analysis of PSR J1847+0614g from the FAST observation on 20210220, with LRFS and 2DFS for the on-pulse region of a mean pulse profile.  \label{subfig:fluctu:J1847+0614g}}
\end{figure}

\subsection{J1846+0657g}
\label{subsec:J1846+0657g}

PSR J1846+0657g was discovered in the FAST GPPS survey \citep{Han2021,han2025}.

This pulsar was observed by FAST on 20240905 for 15 minutes, driving a rotation period of $P=2.2442$~s and a dispersion measure of $D\!M=223.8~{\rm cm^{-3}\,pc}$. Single pulse sequences  and the on-pulse integral energy histogram of this observation are shown in Fig.~\ref{subfig:TP:J1846+0657g} and \ref{subfig:Hist:J1846+0657g}, illustrating the existence of the nulling phenomenon. The nulling fraction of this data is estimated to be 28.7$\pm$2.8\% from Fig.~\ref{subfig:Hist:J1846+0657g}.

\subsection{J1846-0211g}
\label{subsec:J1846-0211g}

PSR J1846-0211g was discovered in the FAST GPPS survey \citep{Han2021,han2025}. 

This pulsar was observed by FAST on 20220902 for 15 minutes, yielding a rotation period $P=0.7882$~s and a dispersion measure $D\!M=848.7~{\rm cm^{-3}\,pc}$. 
The single pulse sequence and fluctuation spectra of this observation are displayed in Fig.~\ref{subfig:TP:J1846-0211g} and \ref{subfig:fluctu:J1846-0211g}. The pulsar has a temporal modulation with a frequency of $1/P_3=0.054\pm0.001$ ($P_3=18.6\pm0.4$ periods), but without an obvious phase modulation.

\begin{figure}[htpb]
\centering
\includegraphics[width=0.22\textwidth, angle=0]{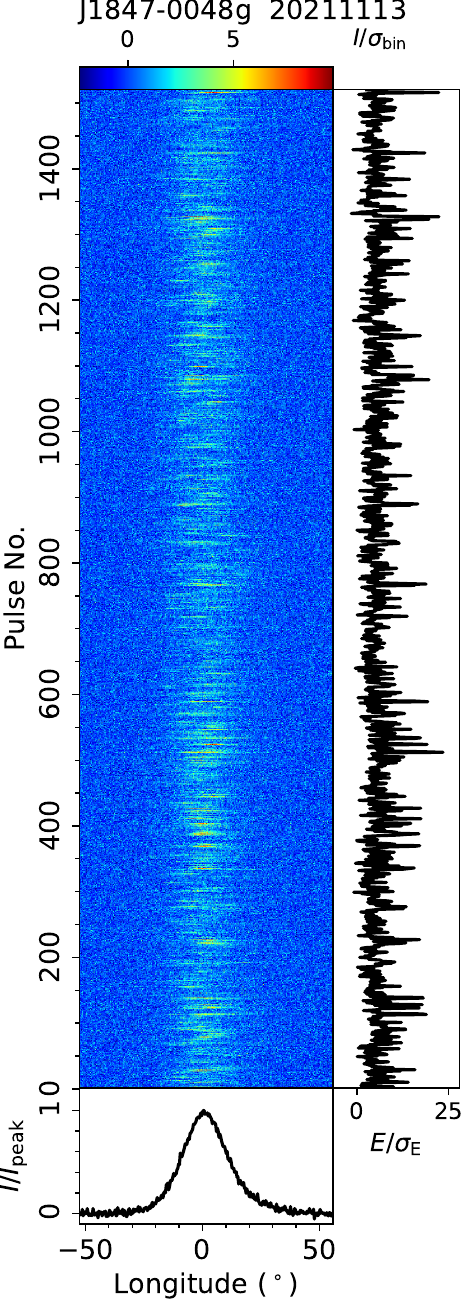}
\includegraphics[width=0.22\textwidth, angle=0]{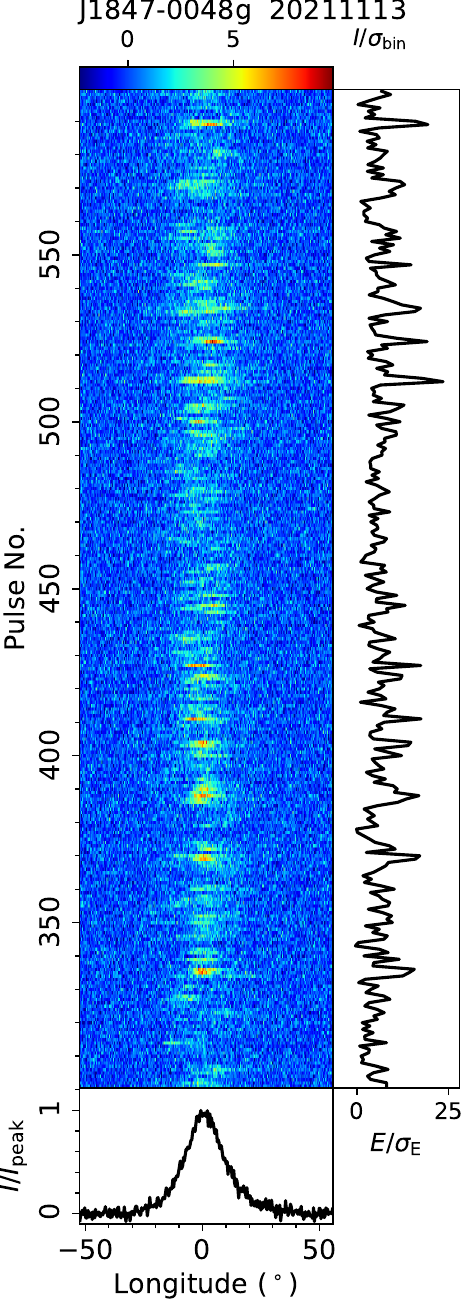}
\figcaption{Single pulse sequence of PSR J1847-0048g from the FAST observation on 20211113, and a zoomed-in view of pulses No. 300-600.
\label{subfig:TP:J1847-0048g}}
\end{figure}

\begin{figure}[htpb]
\centering
\includegraphics[width=0.22\textwidth, angle=0]{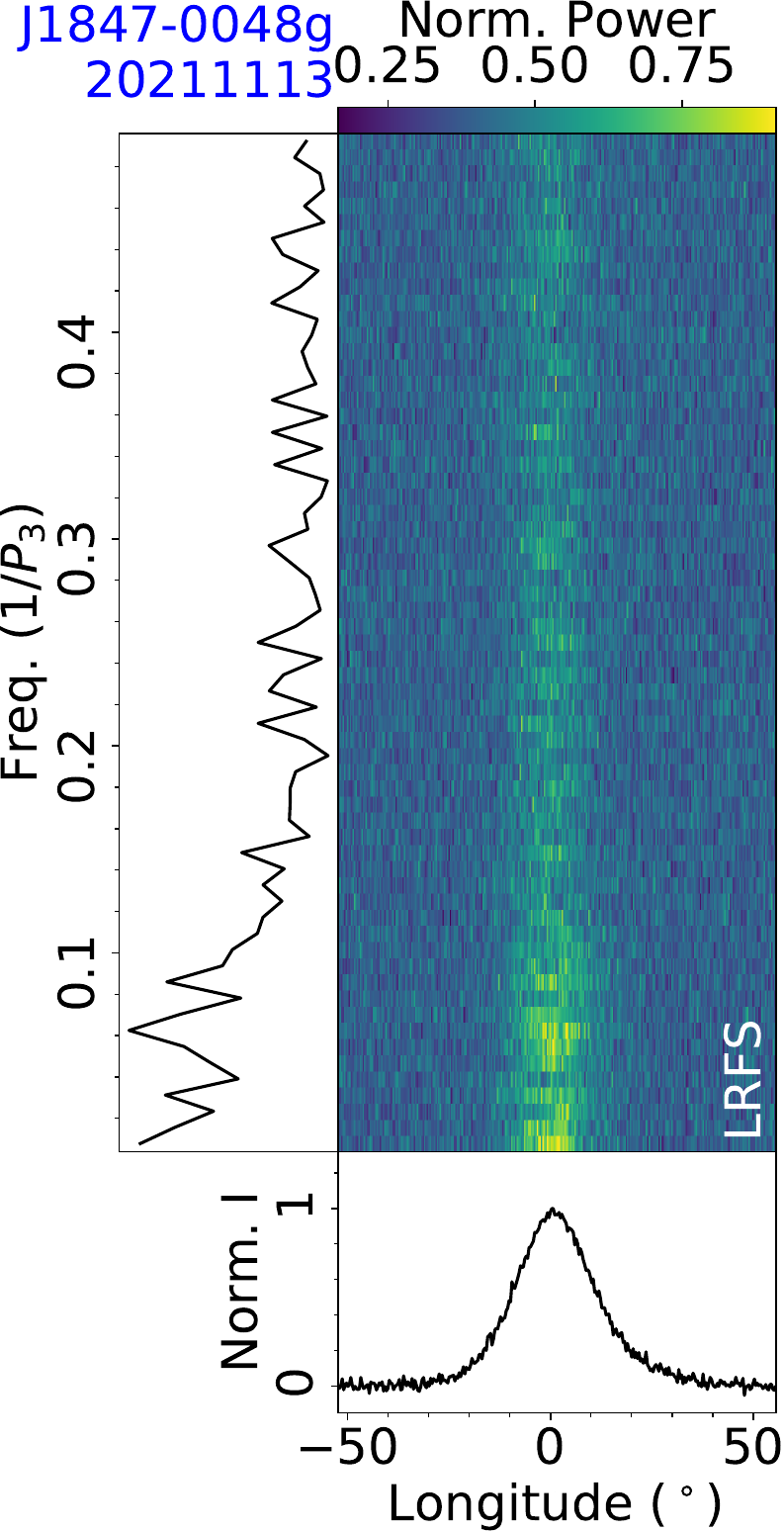}
\includegraphics[width=0.22\textwidth, angle=0]{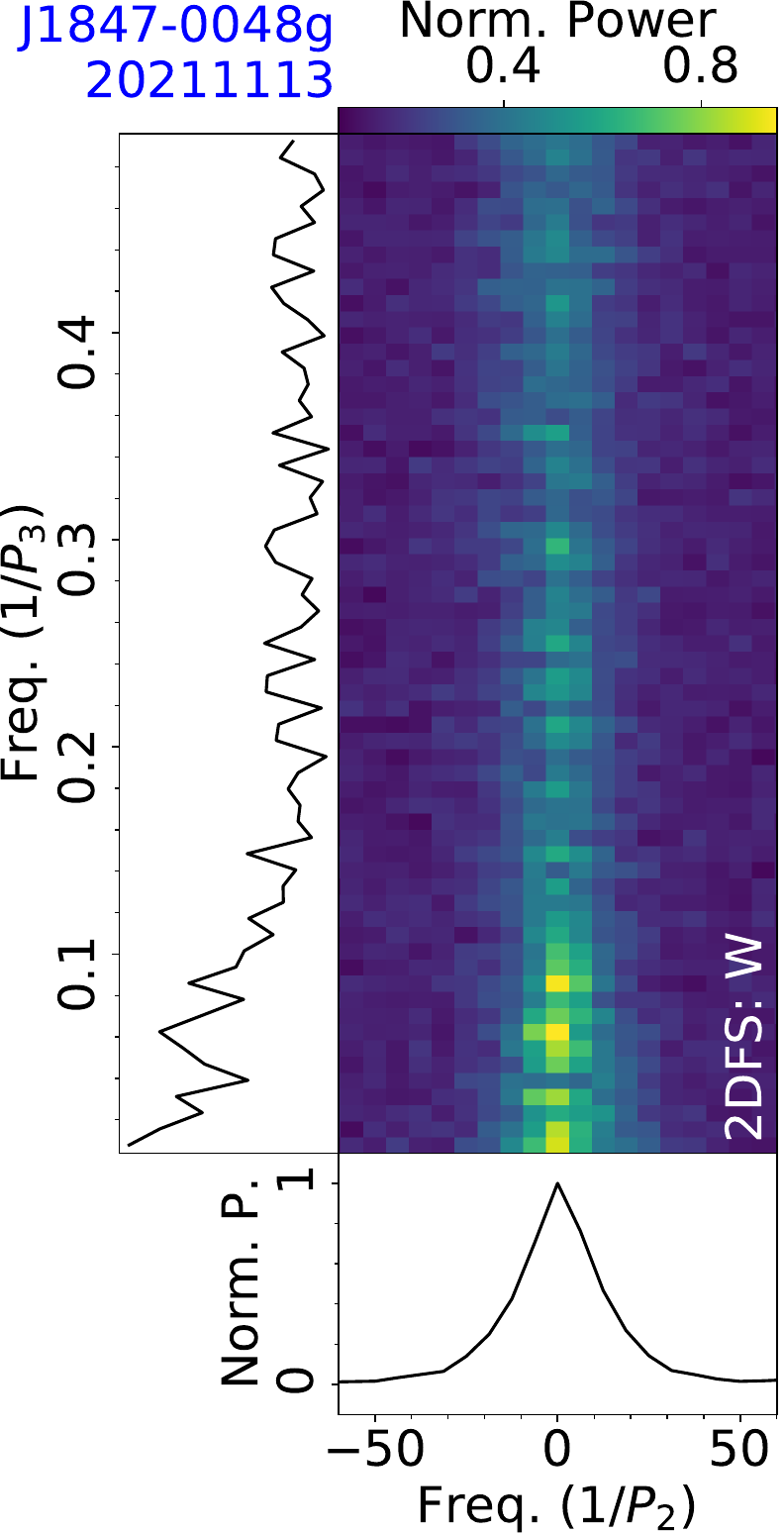}
\figcaption{Fluctuation analysis of PSR J1847-0048g for the observation on 20211113, with LRFS and 2DFS for the on-pulse region of a mean pulse profile.
\label{subfig:fluctu:J1847-0048g}}
\end{figure}

\subsection{J1846-07492}
\label{subsec:J1846-07492}

PSR J1846-07492 was discovered in data from the Parkes Multi-beam Pulsar Survey \citep{Keith2009}. \citet{Song2023} reported two drift features: one with $P_3=2.7\pm0.4$ periods and $P_2=-98^{+89}_{-35}$ degrees, and the other with $P_3=21\pm9$ periods and $P_2=-83^{+48}_{-53}$ degrees.

This pulsar was observed by FAST on 20250321 for 6 minutes, deriving a rotation period $P=0.8613$~s and a dispersion measure $D\!M=191.1~{\rm cm^{-3}\,pc}$. Single pulse sequences in Fig.~\ref{subfig:TP:J1846-07492} display the negative drifting behavior and a low-frequency modulation superimposed on it. Fluctuation spectra are shown in Fig.~\ref{subfig:fluctu:J1846-07492}. The negative drift feature in 2DFS is widely spread in $1/P_3$, with the centroid frequencies of $1/P_3=0.38\pm0.01$ and $1/P_2=-41\pm2$, corresponding to $P_3=2.7\pm0.1$ periods and $P_2=-8.8\pm0.4$ degrees. The low-frequency modulation feature has the centroid at $1/P_3=0.029\pm0.002$, yielding $P_3=34\pm2$ periods.

\subsection{J1847+0614g}
\label{subsec:J1847+0614g}

PSR J1847+0614g was discovered in the FAST GPPS survey \citep{Han2021,han2025}.

This pulsar was observed by FAST on 20210220 for 15 minutes, yielding a rotation period $P=1.6631$~s and a dispersion measure $D\!M=273.0~{\rm cm^{-3}\,pc}$. 
Single pulse sequences of this observation displayed in Fig.~\ref{subfig:TP:J1847+0614g} show the existence of a negative subpulse drifting behavior. 
In fluctuation spectra (Fig.~\ref{subfig:fluctu:J1848+1245g}), the centroid frequencies of the main drift feature are $1/P_3=0.327\pm0.001$ and $1/P_2=-49\pm1$, which correspond to $P_3=3.06\pm0.01$ periods and $P_2=-7.4\pm0.2^\circ$.

\begin{figure}[htpb]
\centering
\includegraphics[width=0.44\textwidth, angle=0]{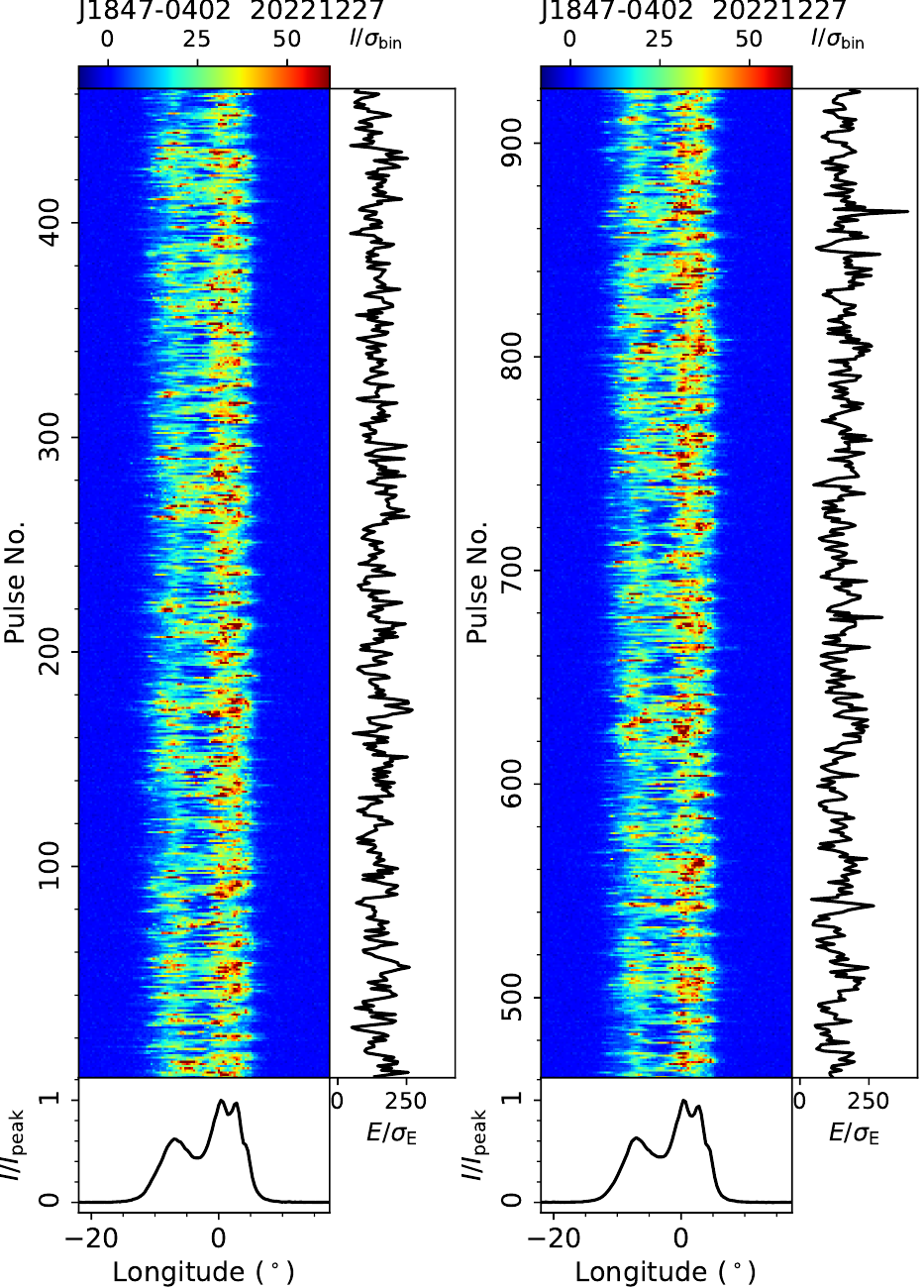}
\figcaption{Single pulse sequences of PSR J1847-0402 from the FAST observation on 20221227.
\label{subfig:TP:J1847-0402}}
\end{figure}

\begin{figure}[htpb]
\centering
\includegraphics[width=0.44\textwidth, angle=0]{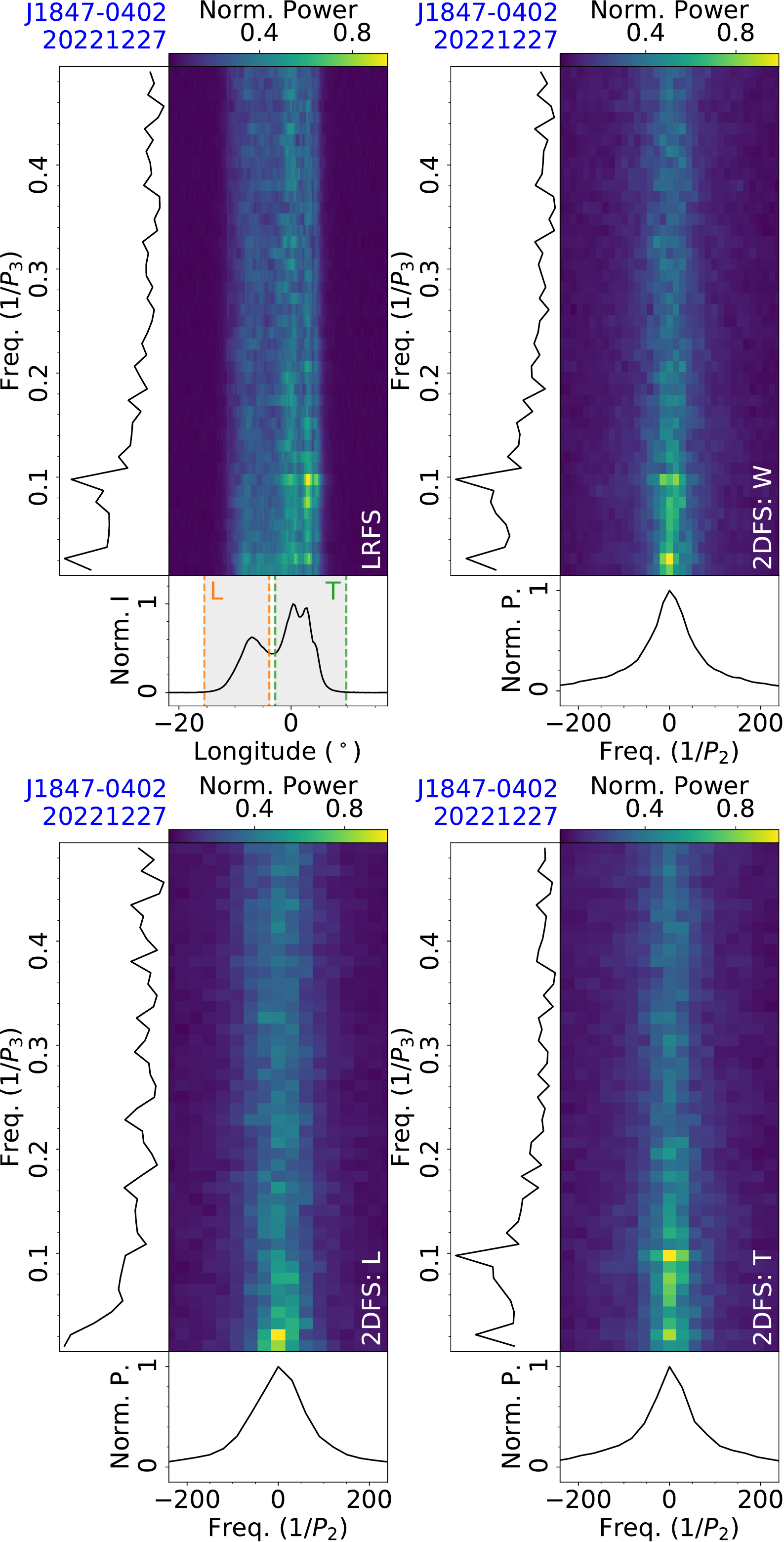}
\figcaption{Fluctuation analysis of PSR J1847-0402 from the FAST observation on 20221227, with LRFS (top-left), and 2DFS for the on-pulse region (top-right), leading part (bottom-left) and trailing part (bottom-right) of a mean pulse profile.
\label{subfig:fluctu:J1847-0402}}
\end{figure}

\begin{figure}[htpb]
\centering
\includegraphics[width=0.21\textwidth, angle=0]{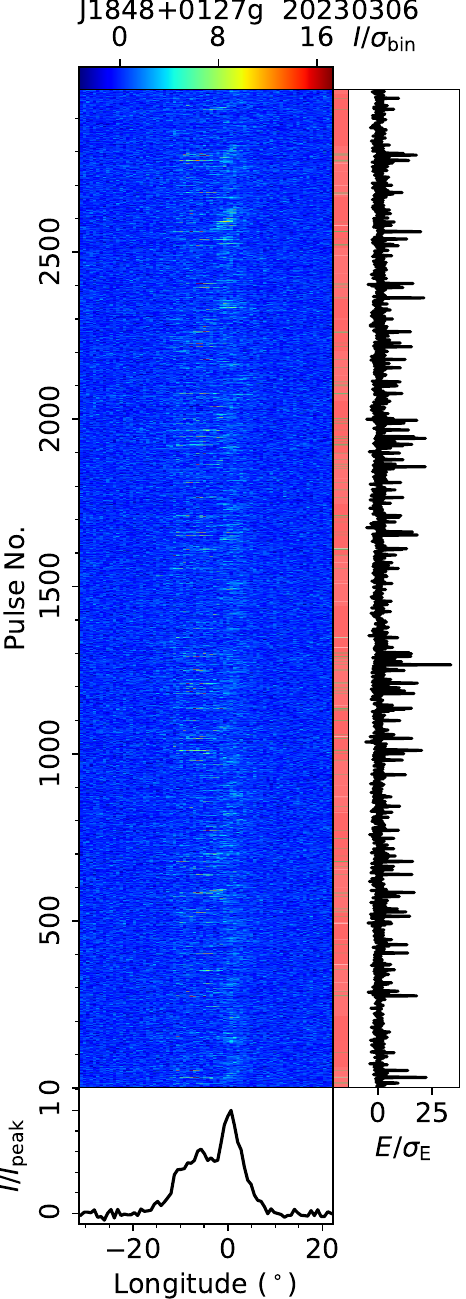}
\includegraphics[width=0.21\textwidth, angle=0]{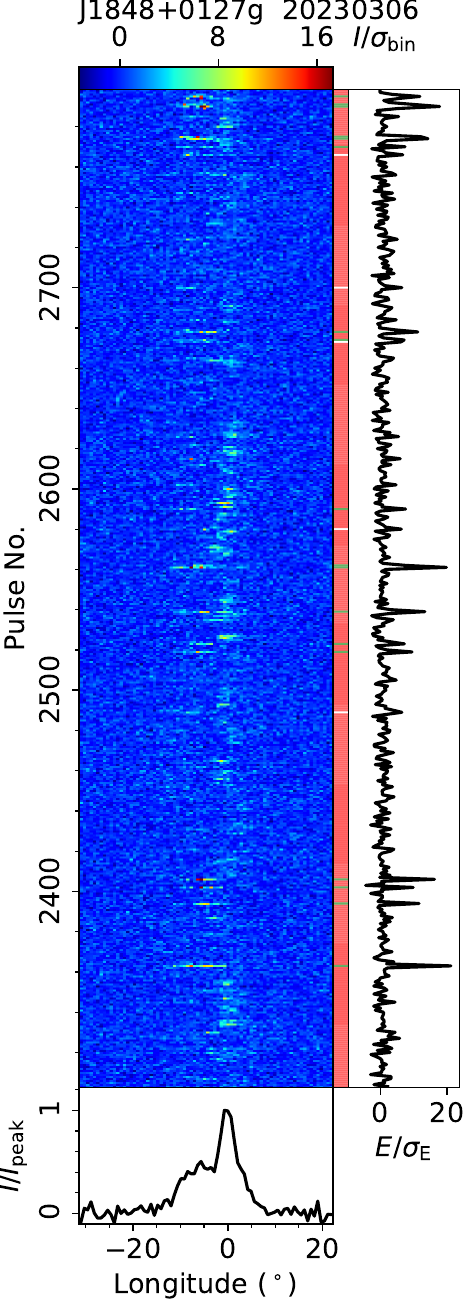}
\figcaption{Single pulse sequences of PSR J1848+0127g from the FAST observation on 20230306.
\label{subfig:TP:J1848+0127g}}
\end{figure}

\begin{figure}[htpb]
\centering
\includegraphics[width=0.39\textwidth, angle=0]{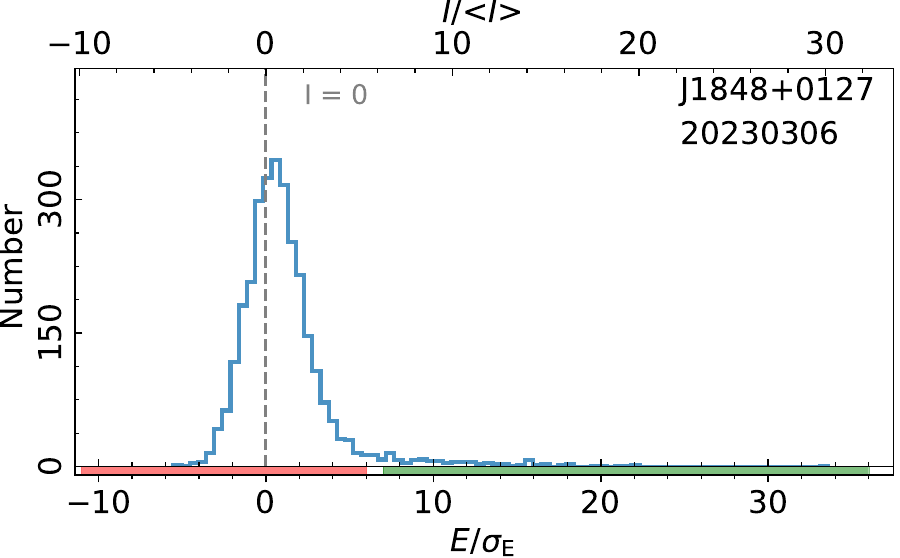}
\figcaption{Energy histogram of single pulses of PSR J1848+0127g for the leading part in a mean pulse profile, from the FAST observation on 20230306.
\label{subfig:Hist:J1848+0127g}}
\end{figure}

\begin{figure}[htpb]
\centering
\includegraphics[width=0.37\textwidth, angle=0]{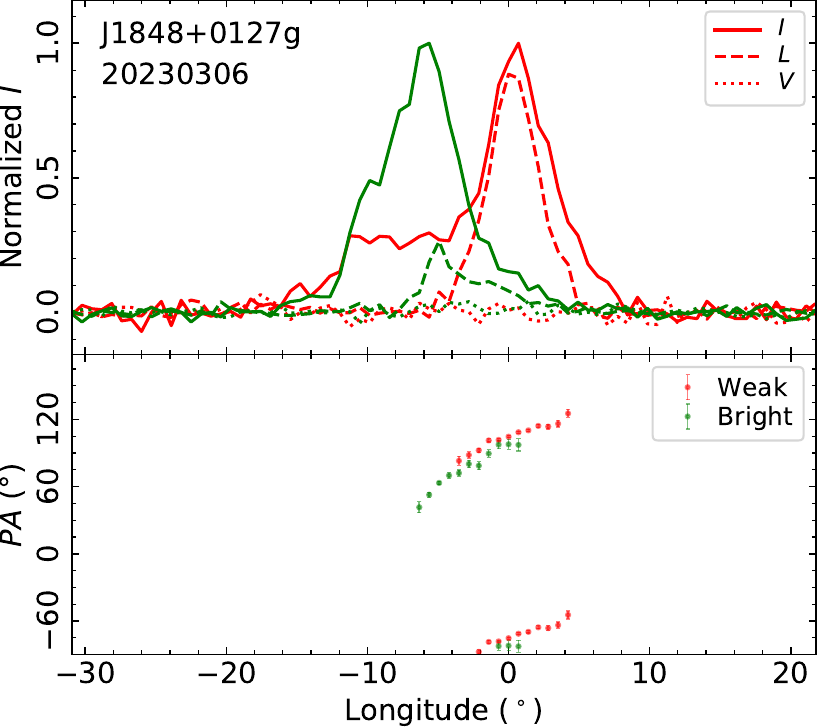}
\figcaption{Mean polarization profiles (the top panel) for weak and bright emission modes of PSR J1848+0127g observed on 20230306, as well as the averaged PA curves (the bottom panel). \label{subfig:PolModes:J1848+0127g}}
\end{figure}

\subsection{J1847-0048g}
\label{subsec:J1847-0048g}

PSR J1847-0048g was discovered in the FAST GPPS survey \citep{Han2021,han2025}. 

The pulsar was observed by FAST on 20211113 for 15 minutes and 20220606 for 30 minutes. From the data of 20211113, a rotation period $P=0.5825$~s and a dispersion measure $D\!M=664.6~{\rm cm^{-3}\,pc}$ were determined. 
Single pulse sequences of the observation on 20211113 are shown in Fig.~\ref{subfig:TP:J1847-0048g}, which display the modulation behavior. In the fluctuation spectra (Fig.~\ref{subfig:fluctu:J1847-0048g}), the centroid frequency of the low-frequency feature is estimated to be $1/P_3=0.056\pm0.001$, corresponding to $P_3=17.9\pm0.4$ periods.

The temporal modulation periodicity is  16$\pm2$ periods.
Single pulse behavior of the observation on 20220606 is consistent with 20211113.

\subsection{J1847-0402}
\label{subsec:J1847-0402}

PSR J1847-0402 was discovered by \citet{Davies1970} using the Mark I radio telescope at Jodrell Bank. \citet{Weltevrede2006} reported the drifting behavior at 21 cm with $P_3=12\pm1$ periods and $P_2=80^{+70}_{-45}$ degrees. At 1280 MHz, \citet{Song2023} presented two modulation features for two components: a $P_3$-only feature with $P_3=71\pm39$ periods for one component, and a drift feature with $P_3=12\pm2$ periods and $P_2=57^{+15}_{-25}$ degrees for the other.

This pulsar was observed by FAST on 20221227 for 9 minutes and on 20250321 for 6 minutes. From the longer data, a rotation period $P=0.5978$~s and a dispersion measure $D\!M=141.5~{\rm cm^{-3}\,pc}$ were determined. 
Single pulse sequences of the observation on 20221227 in Fig.~\ref{subfig:TP:J1847-0402} show the subpulse modulation phenomenon. Fluctuation spectra are shown in Fig.~\ref{subfig:fluctu:J1847-0402}. For the leading phase part in the mean pulse profile, the main modulation feature is at low frequency, with a centroid frequency of $1/P_3=0.024\pm0.001$, corresponding to $P_3=42\pm2$ periods. In the 2DFS, the trailing profile part also exhibits this low-frequency feature, with a centroid frequency of $1/P_3=0.026\pm0.001$ ($P_3=39\pm2$ periods), as well as an additional feature with the centroid at $1/P_3=0.080\pm0.001$ ($P_3=12.4\pm0.2$ periods). 
The modulation properties observed on 20250321 are consistent with those observed on 20221227.

\begin{figure}[htpb]
\centering
\includegraphics[width=0.22\textwidth, angle=0]{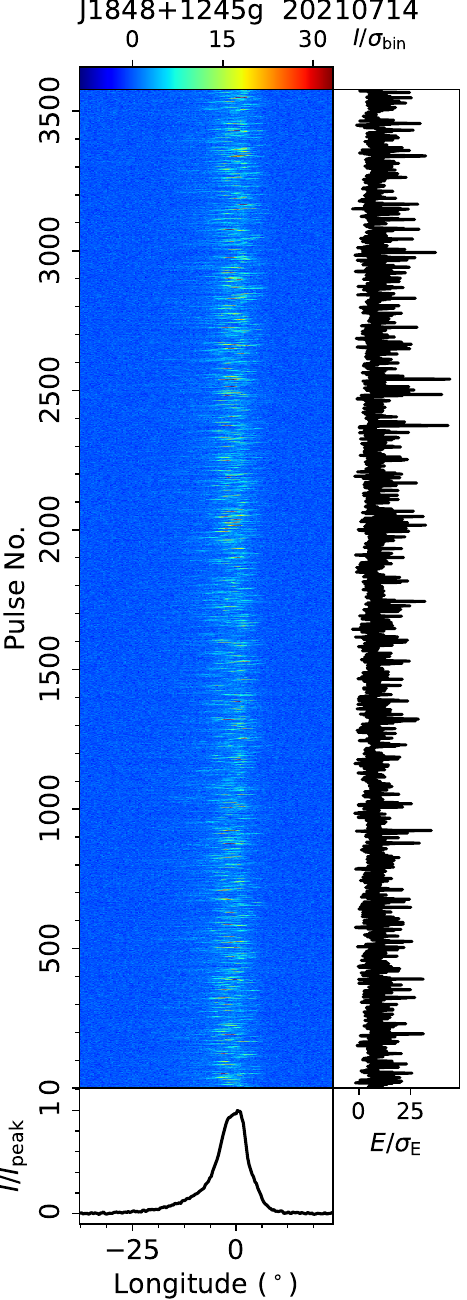}
\includegraphics[width=0.22\textwidth, angle=0]{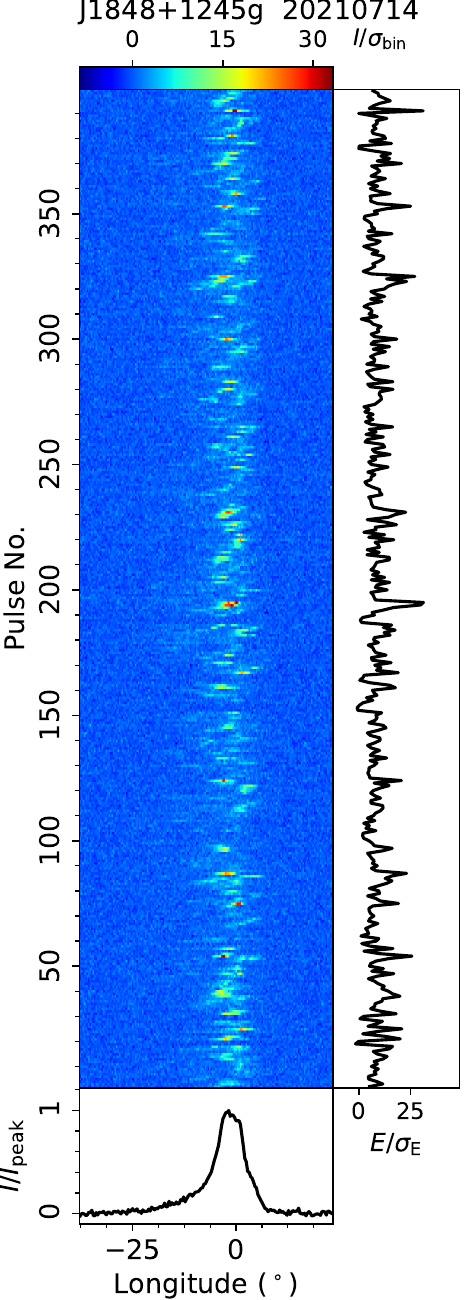}
\figcaption{Single pulse sequence of PSR J1848+1245g from the FAST observation on 20210714, and a zoomed-in view of pulses No. 1-400.
\label{subfig:TP:J1848+1245g}}
\end{figure}

\begin{figure}[htpb]
\centering
\includegraphics[width=0.22\textwidth, angle=0]{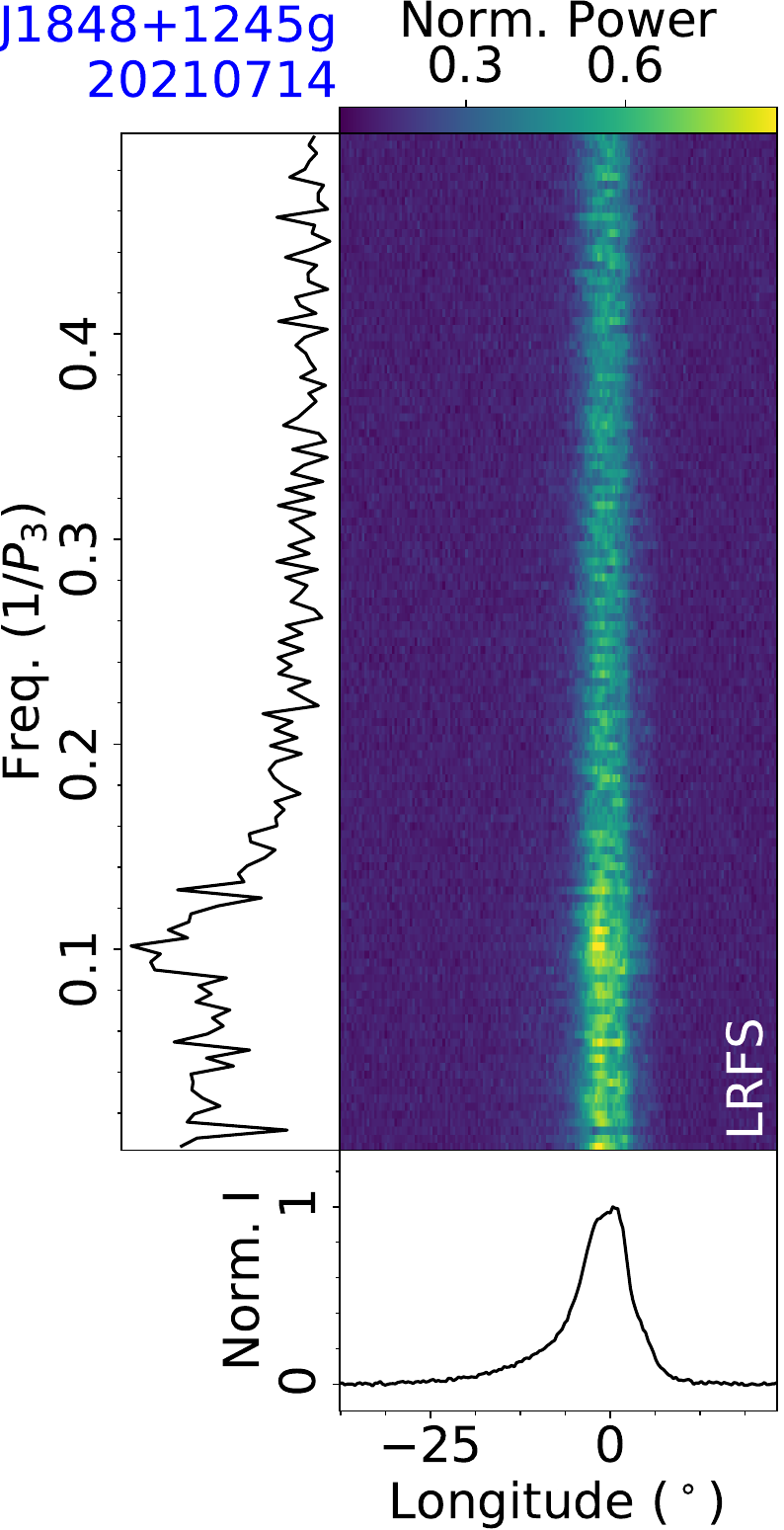}
\includegraphics[width=0.22\textwidth, angle=0]{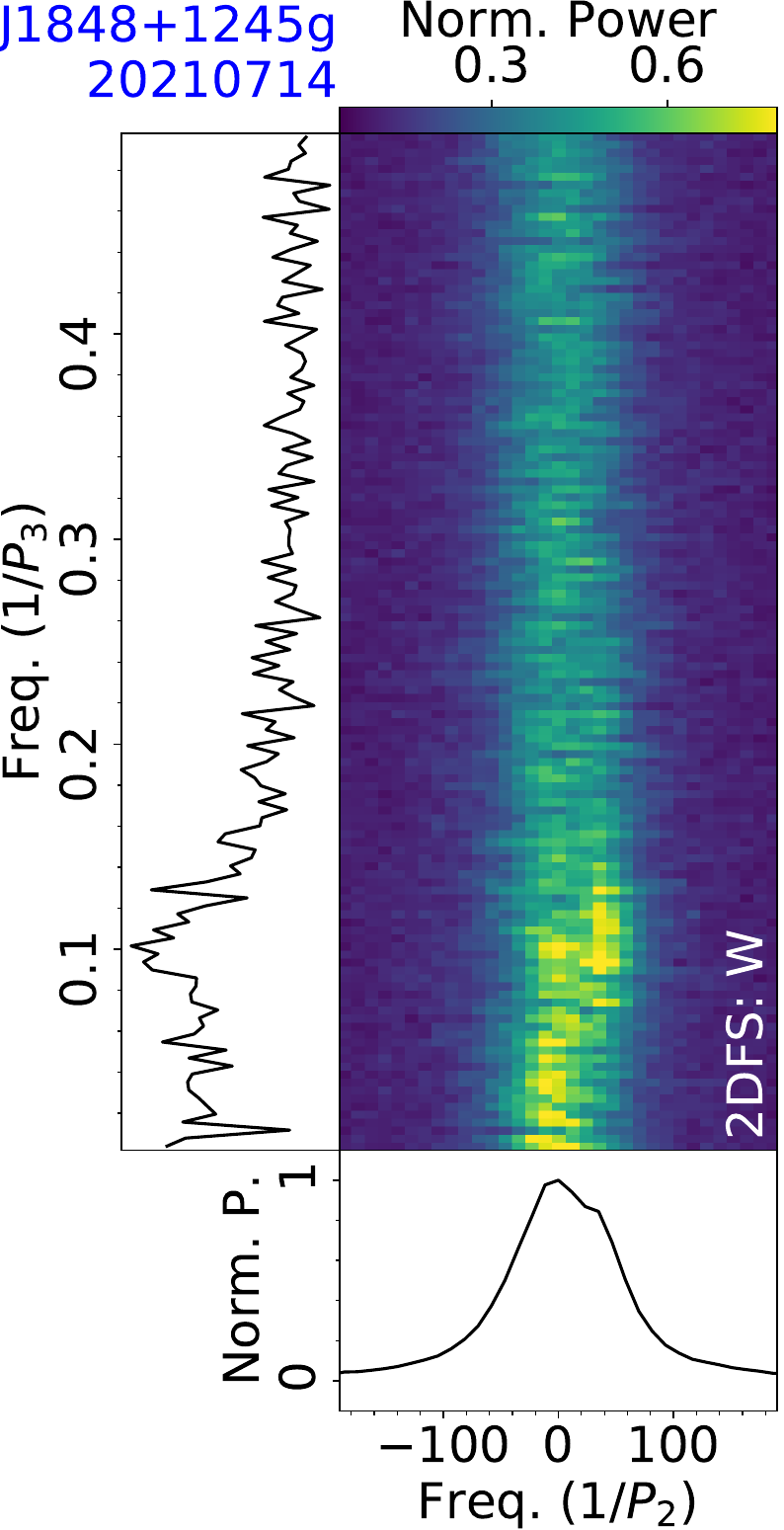}
\figcaption{Fluctuation analysis of PSR J1848+1245g from the FAST observation on 20210714, with LRFS and 2DFS for the on-pulse region of a mean pulse profile.
\label{subfig:fluctu:J1848+1245g}}
\end{figure}

\begin{figure}[htpb]
\centering
\includegraphics[width=0.22\textwidth, angle=0]{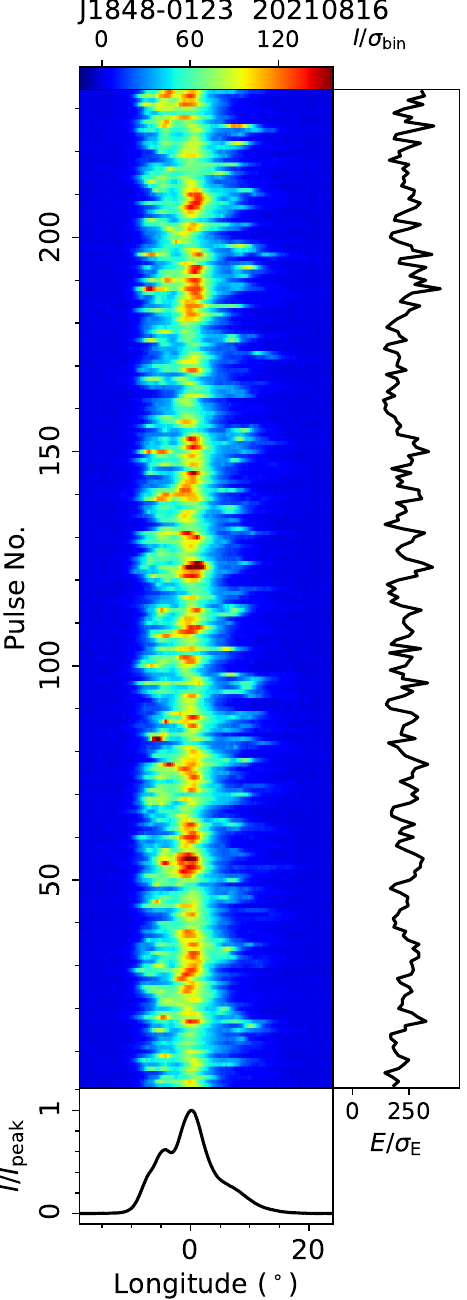}
\includegraphics[width=0.22\textwidth, angle=0]{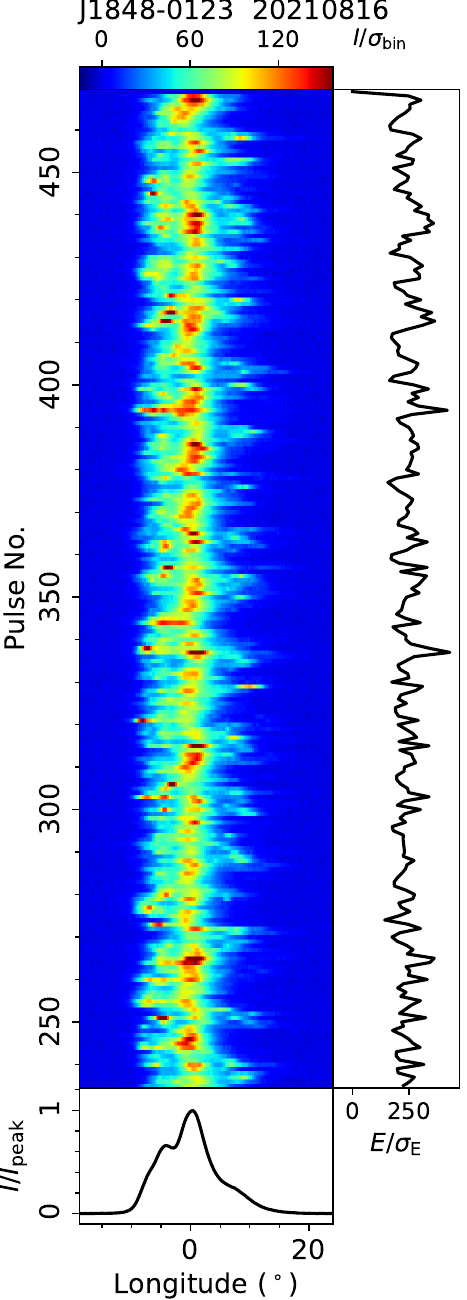}
\figcaption{Single pulse sequences of PSR J1848-0123 from the FAST observation on 20210816. \label{subfig:TP:J1848-0123}}
\end{figure}

\begin{figure}[htpb]
\centering
\includegraphics[width=0.22\textwidth, angle=0]{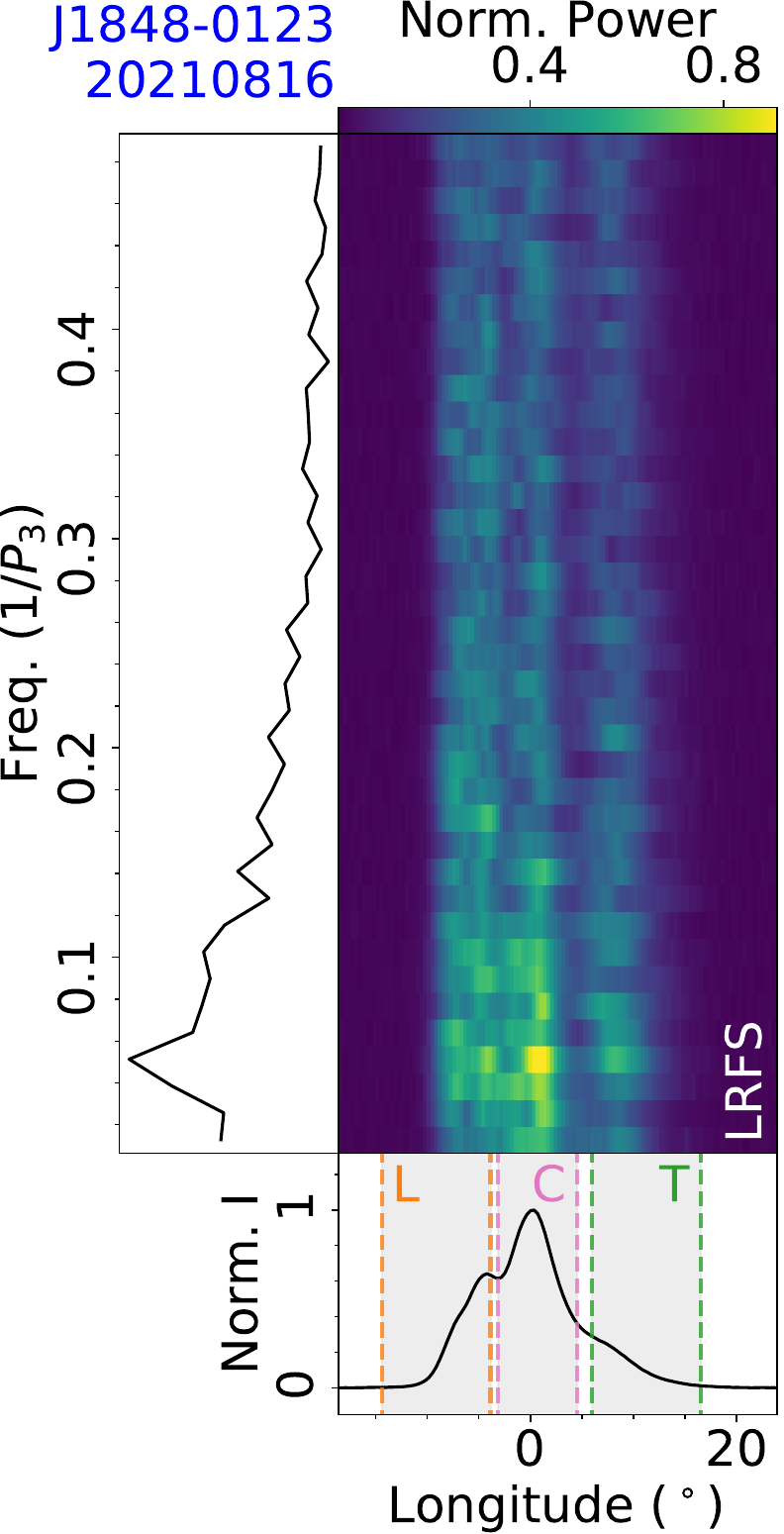}
\includegraphics[width=0.22\textwidth, angle=0]{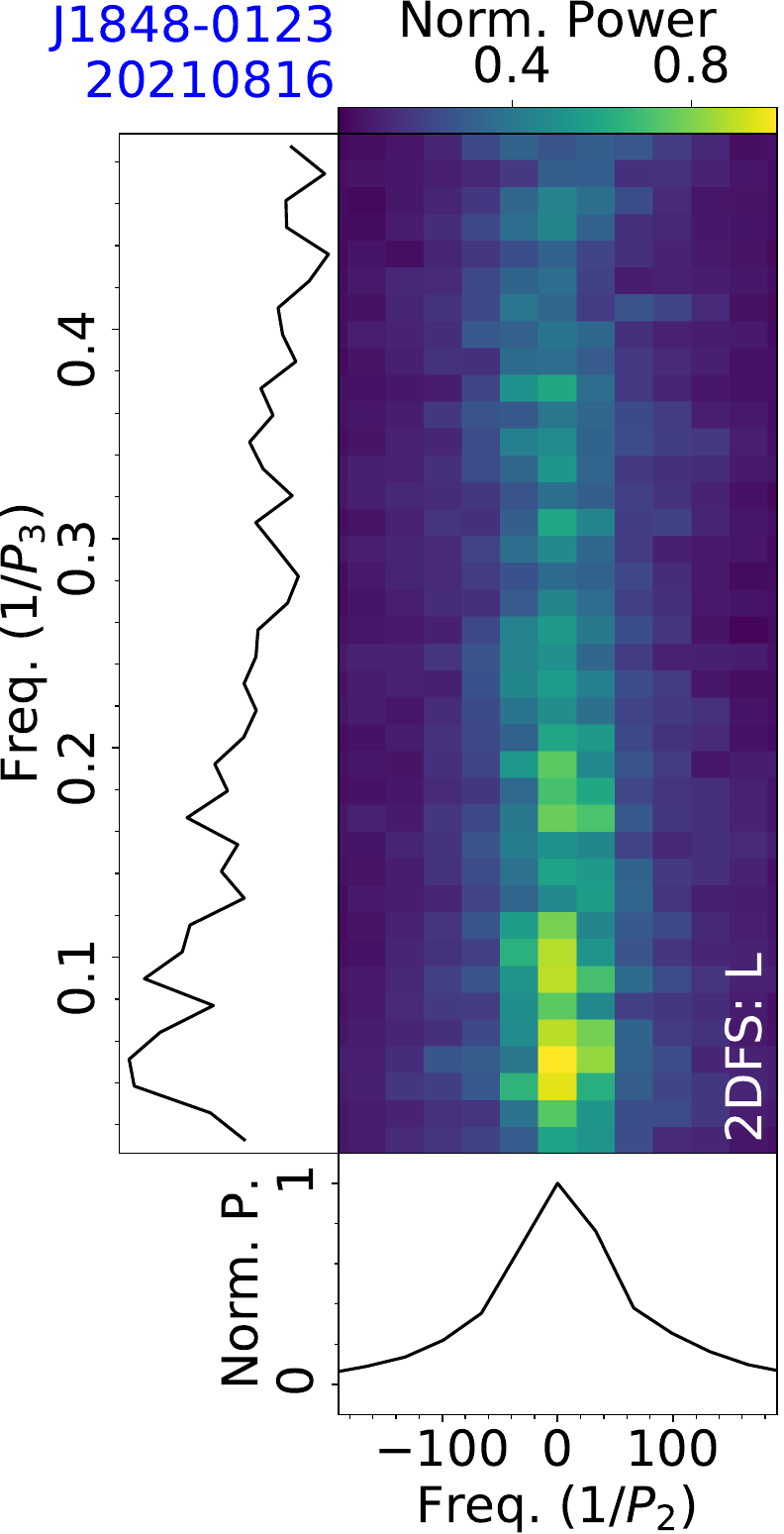}\\
\includegraphics[width=0.22\textwidth, angle=0]{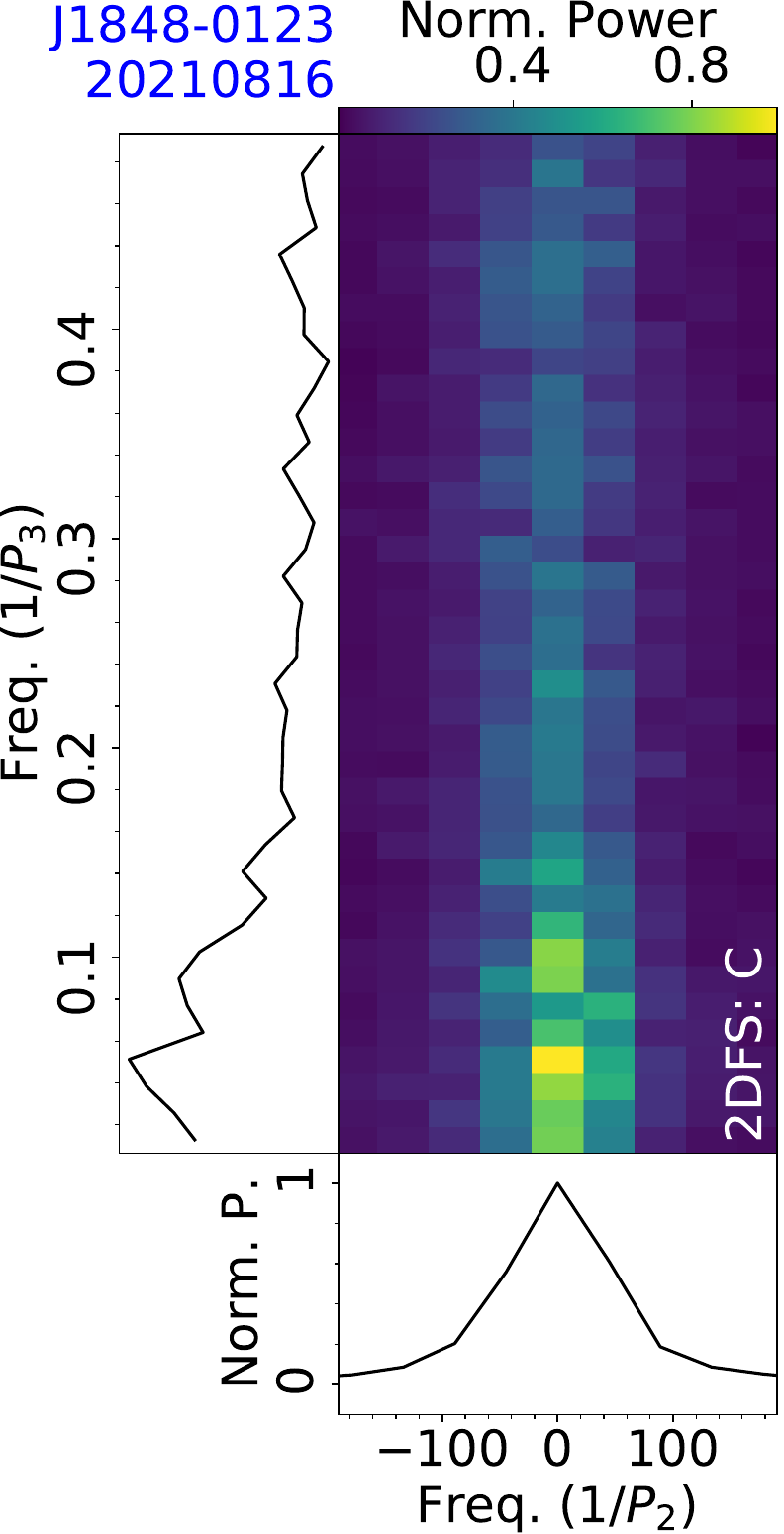}
\includegraphics[width=0.22\textwidth, angle=0]{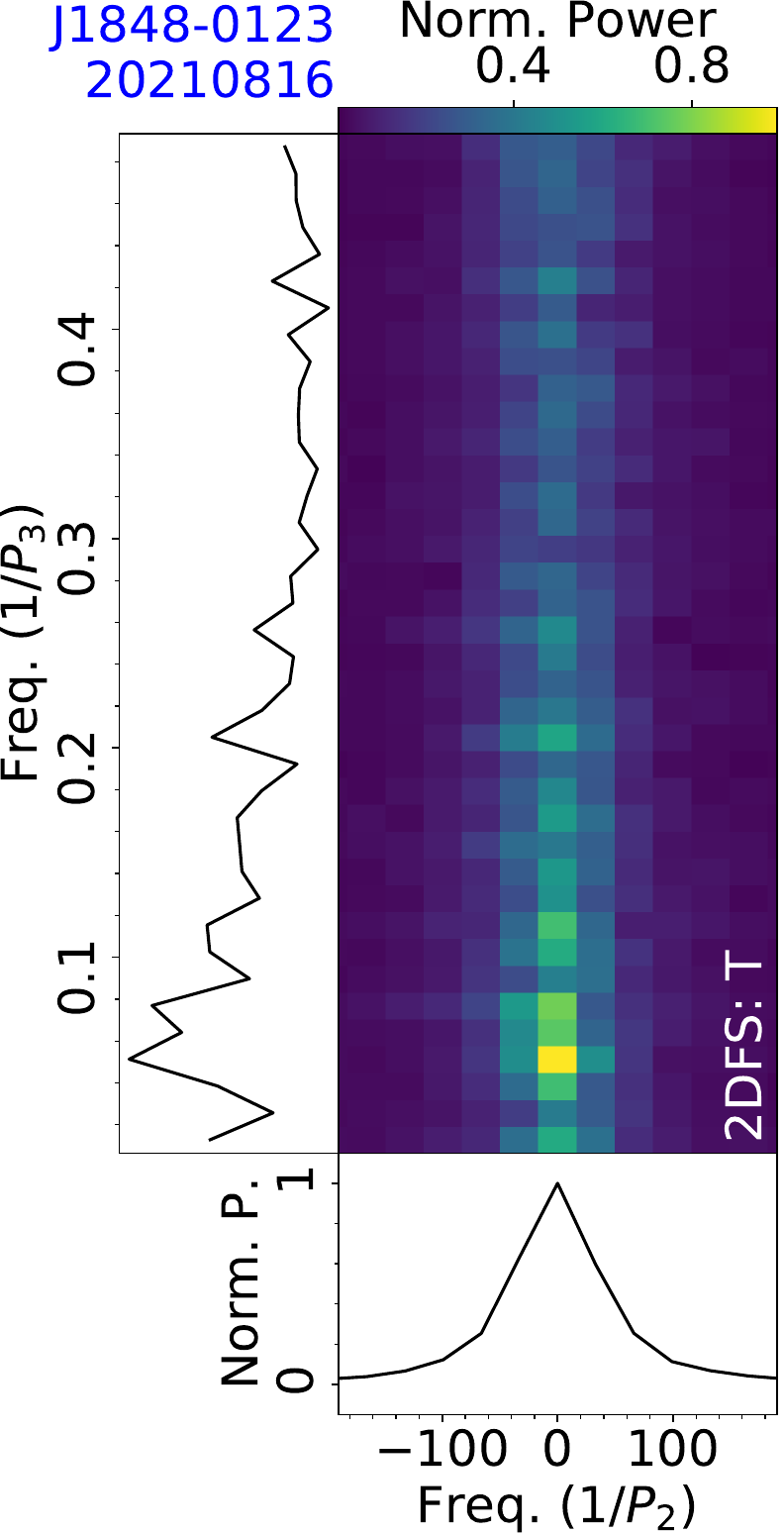}
\figcaption{Fluctuation analysis of PSR J1848-0123 for the observation on 20210816, with LRFS (top-left), and 2DFS for the leading part (top-right), central part (bottom-left) and trailing part (bottom-right) of a mean pulse profile. \label{subfig:fluctu:J1848-0123}}
\end{figure}

\subsection{J1848+0127g}
\label{subsec:J1848+0127g}

PSR J1848+0127g was discovered in the FAST GPPS survey \citep{Han2021,han2025}. 

This pulsar was observed by FAST on 20230306 for 17 minutes, deriving a rotation period $P=0.5339$~s and a dispersion measure $D\!M=77.8~{\rm cm^{-3}\,pc}$. 
Single pulse sequences of this observation are shown in Fig.~\ref{subfig:TP:J1848+0127g}, where the leading component displays weak and bright emission states. From the energy histogram for the leading part in a mean pulse profile in Fig.~\ref{subfig:Hist:J1848+0127g}, we distinguish the weak and bright modes for single pulses. The integral polarization profiles and PA curves of two modes are shown in Fig.~\ref{subfig:PolModes:J1848+0127g}.

\begin{figure}[htpb]
\centering
\includegraphics[width=0.44\textwidth, angle=0]{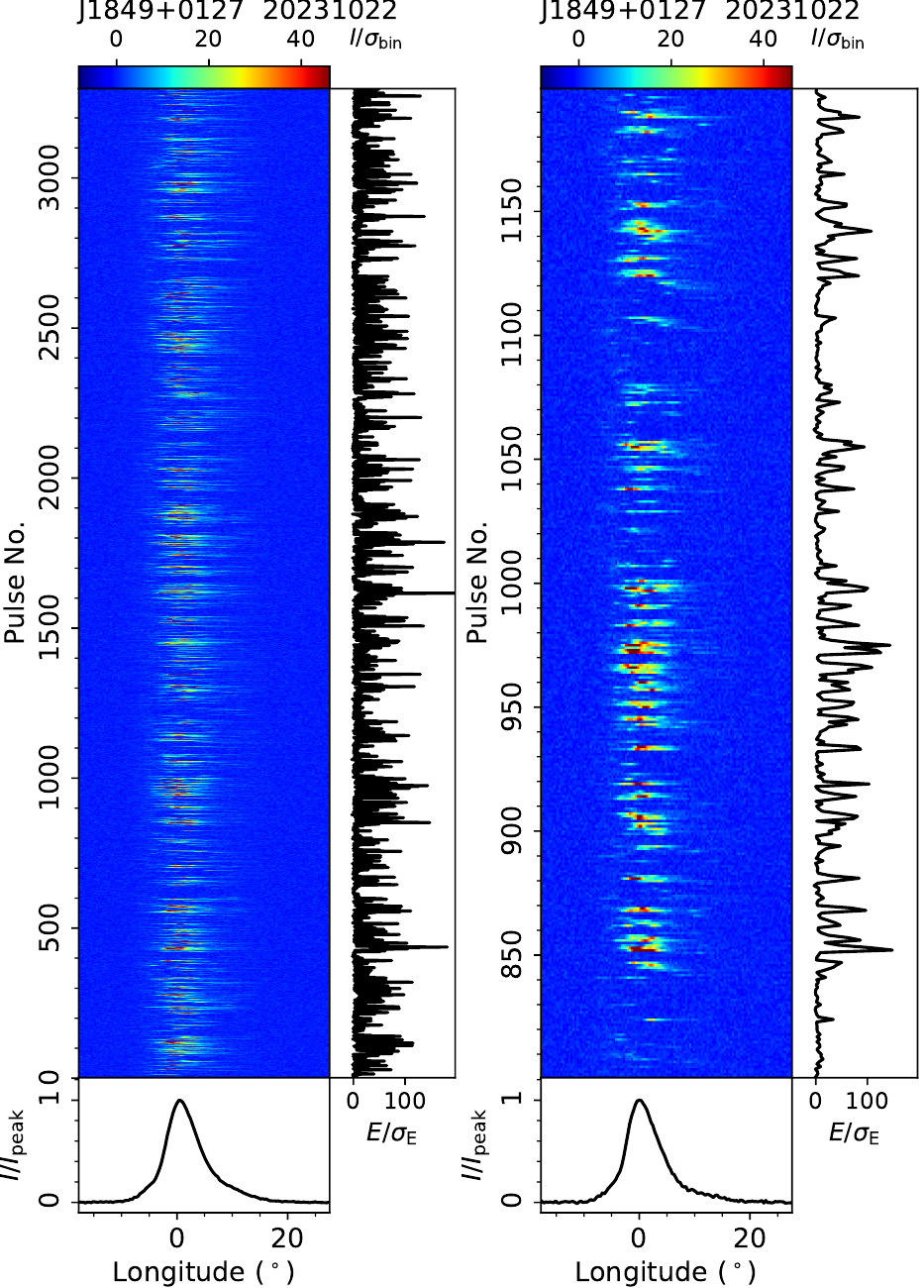}
\figcaption{Single pulse sequence of PSR J1849+0127 from the FAST observation on 20231022, and a zoomed-in view of pulses No. 800-1200.
\label{subfig:TP:J1849+0127}}
\end{figure}

\begin{figure}[htpb]
\centering
\includegraphics[width=0.44\textwidth, angle=0]{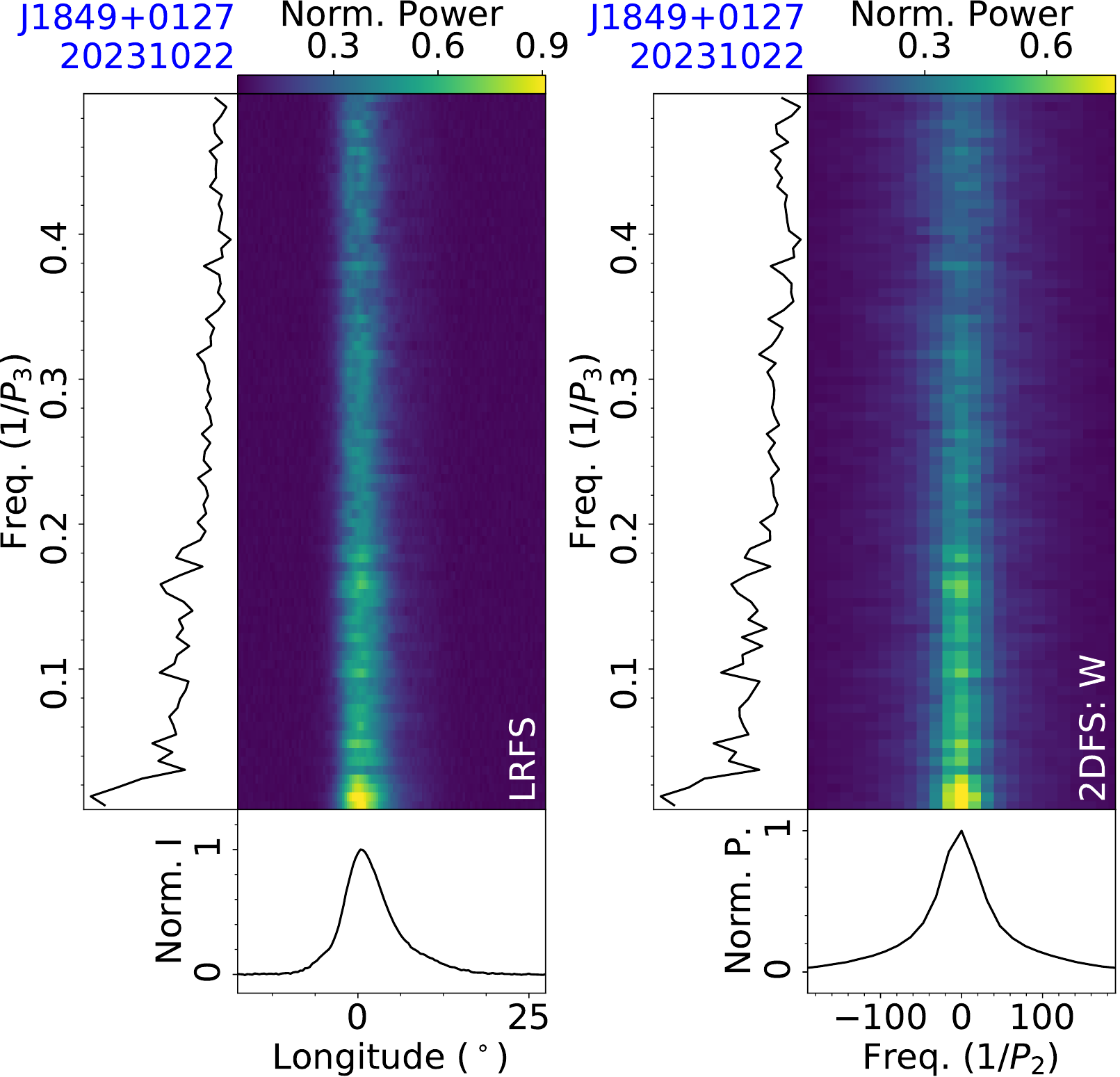}
\figcaption{Fluctuation analysis of PSR J1849+0127 from the FAST observation on 20231022, with LRFS and 2DFS for the on-pulse region of a mean pulse profile.
\label{subfig:fluctu:J1849+0127}}
\end{figure}

\subsection{J1848+1245g}
\label{subsec:J1848+1245g}

PSR J1848+1245g was discovered in the FAST GPPS survey \citep{Han2021,han2025}. 

This pulsar was observed by FAST on 20210714 for 15 minutes, deriving a rotation period $P=0.2482$~s and a dispersion measure $D\!M=99.6~{\rm cm^{-3}\,pc}$. 
Single pulse sequences shown in Fig.~\ref{subfig:TP:J1848+1245g} display unsystematic drifting bands. In fluctuation spectra (Fig.~\ref{subfig:fluctu:J1848+1245g}), there are two features. They are the drift feature of $1/P_3=0.1035\pm0.0004$ ($P_3=9.66\pm0.04$ periods) and $1/P_2=41\pm1$ ($P_2=8.8\pm0.1^\circ$), and the temporally low-frequency modulation feature of $1/P_3=0.059\pm0.001$ ($P_3=17.0\pm0.2$ periods).

\begin{figure}[htpb]
\centering
\includegraphics[width=0.21\textwidth, angle=0]{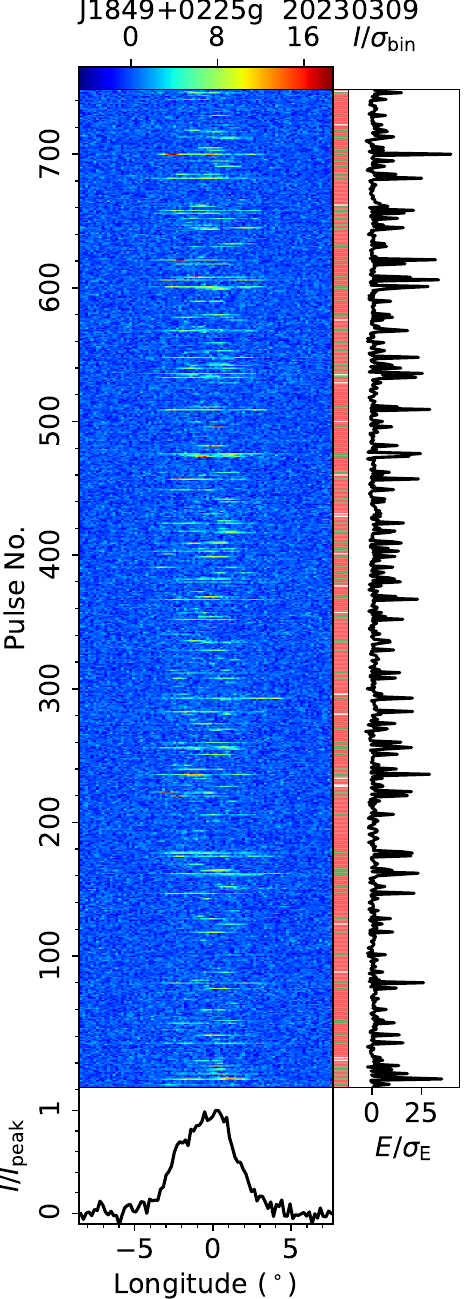}
\includegraphics[width=0.21\textwidth, angle=0]{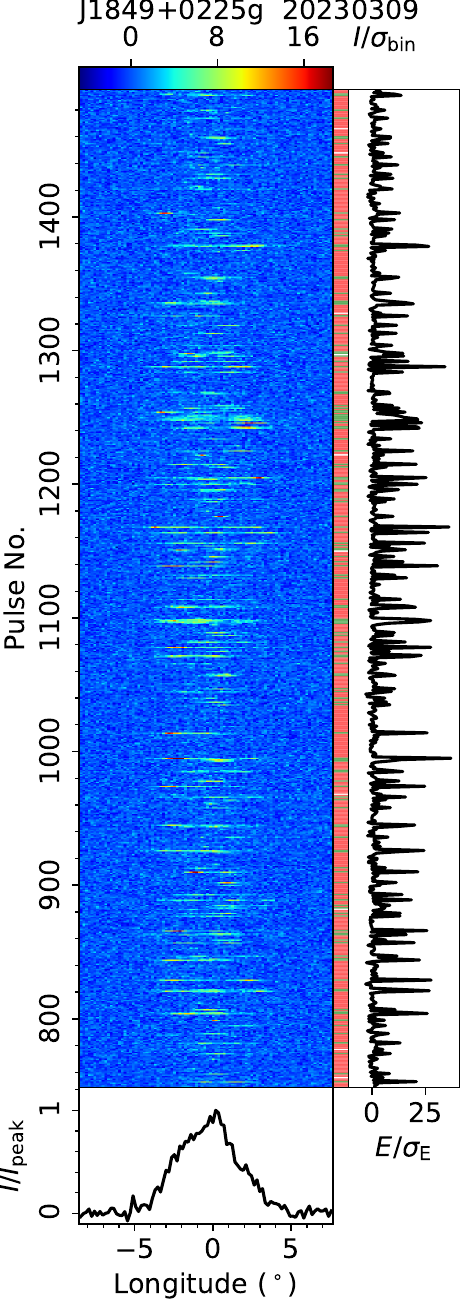}
\figcaption{Single pulse sequences of PSR J1849+0225g from the FAST observation on 20230309.
\label{subfig:TP:J1849+0225g}}
\end{figure}

\begin{figure}[htpb]
\centering
\includegraphics[width=0.39\textwidth, angle=0]{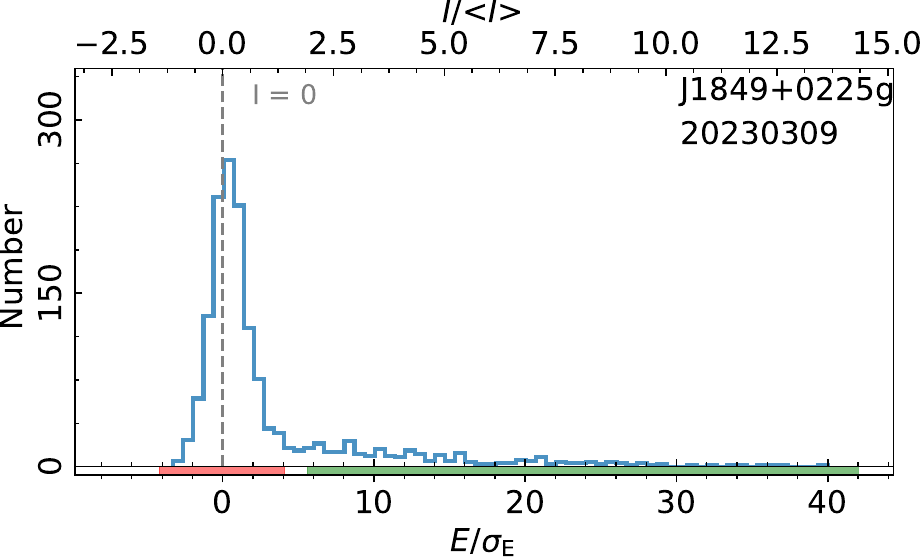}
\figcaption{On-pulse energy histogram of single pulses of PSR J1849+0225g from the FAST observation on 20230309. \label{subfig:Hist:J1849+0225g}}
\end{figure}

\begin{figure}[htpb]
\centering
\includegraphics[width=0.37\textwidth, angle=0]{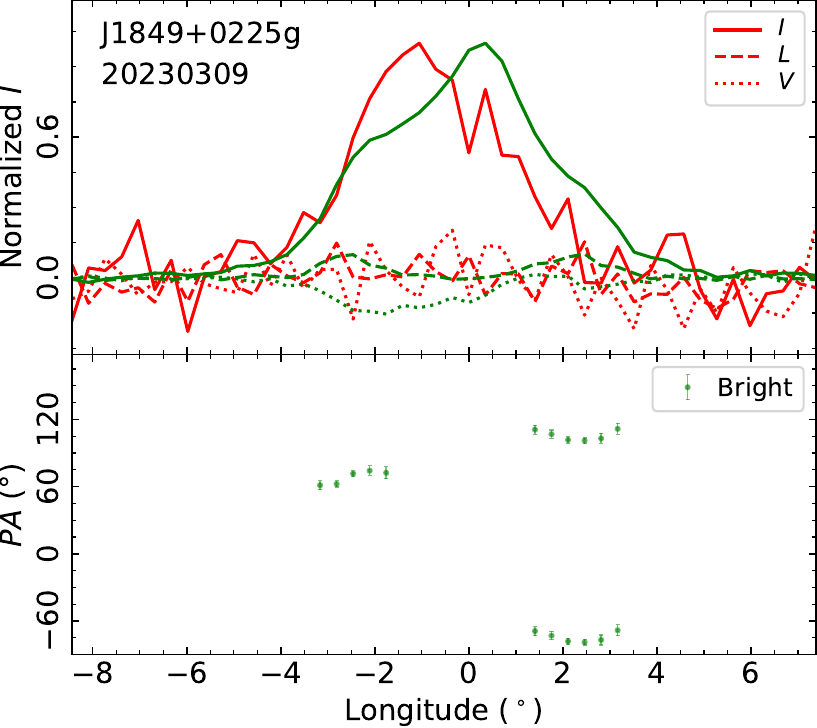}
\figcaption{Mean polarization profiles (the top panel) for weak and bright emission modes of PSR J1849+0225g observed on 20230309, as well as the averaged PA curves (the bottom panel).
\label{subfig:PolModes:J1849+0225g}}
\end{figure}

\subsection{J1848-0123}
\label{subsec:J1848-0123}

PSR J1848-0123 was reported with the drifting phenomenon by \citet{Hankins1987} at 1414 MHz and \citet{Weltevrede2006} at 21 cm. In addition, \citet{Basu2016} reported amplitude modulation at 618 MHz. 

This pulsar was observed by FAST on 20210816 for 5 minutes, deriving a rotation period $P=0.6594$~s and a dispersion measure $D\!M=159.9~{\rm cm^{-3}\,pc}$. 
Single pulse sequences of this observation in Fig.~\ref{subfig:TP:J1848-0123} display the modulation behavior. From fluctuation spectra of three longitude ranges in the mean pulse profile (Fig.~\ref{subfig:fluctu:J1848-0123}), the modulation frequencies over time are $1/P_3=0.070\pm0.001$, $0.060\pm0.001$, and $0.062\pm0.002$, which correspond to the periodicities of $P_3=14.3\pm0.3$, $16.8\pm0.4$, and $16.0\pm0.4$ periods. 
Longer observations are required for the property of phase modulation.





\begin{figure}[htpb]
\centering
\includegraphics[width=0.22\textwidth, angle=0]{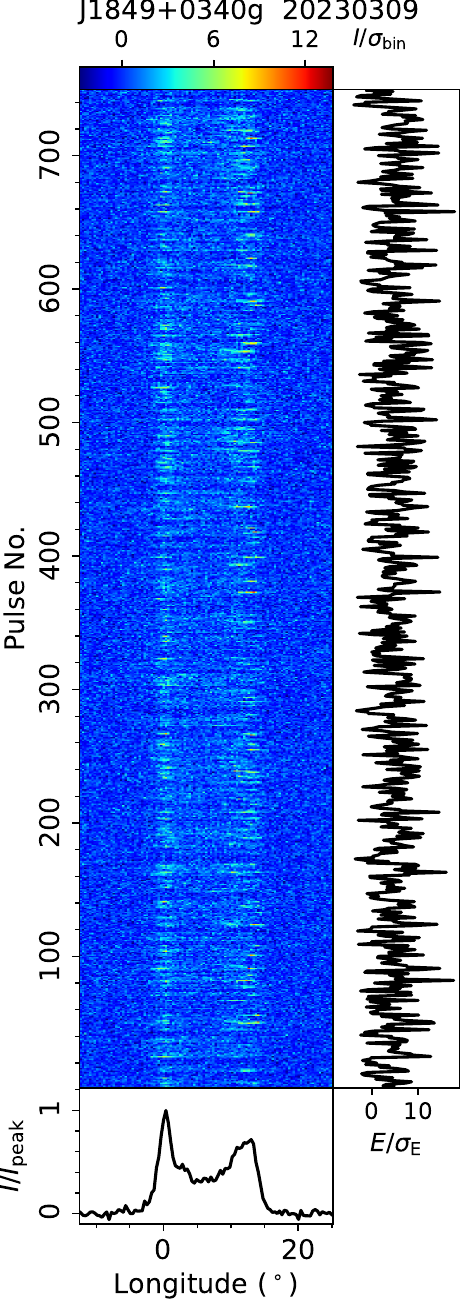}
\includegraphics[width=0.22\textwidth, angle=0]{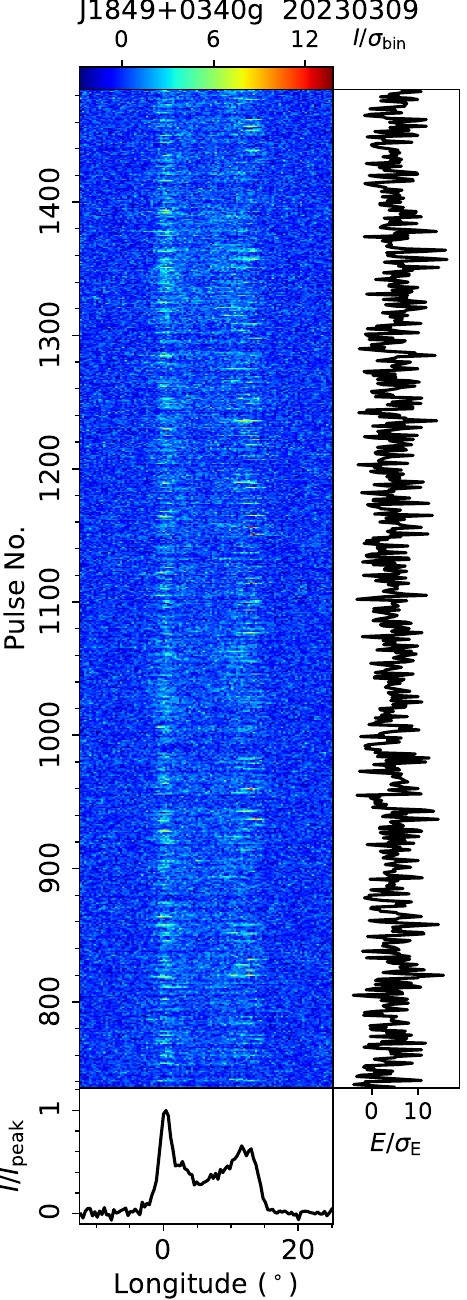}
\figcaption{Single pulse sequences of PSR J1849+0340g from the FAST observation on 20230309.
\label{subfig:TP:J1849+0340g}}
\end{figure}

\begin{figure}[htpb]
\centering
\includegraphics[width=0.39\textwidth, angle=0]{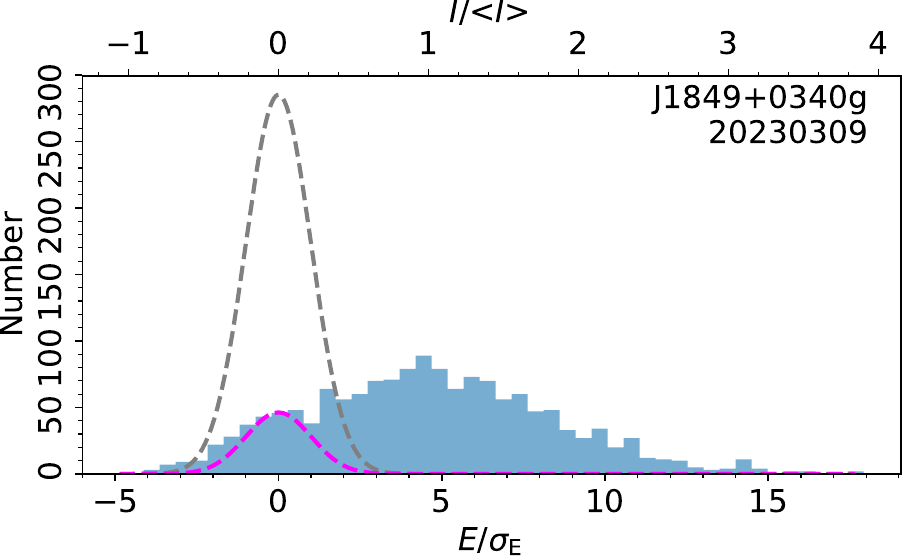}
\figcaption{On-pulse energy histogram of single pulses of PSR J1849+0340g from the FAST observation on 20230309.
\label{subfig:Hist:J1849+0340g}}
\end{figure}

\begin{figure}[htpb]
\centering
\includegraphics[width=0.22\textwidth, angle=0]{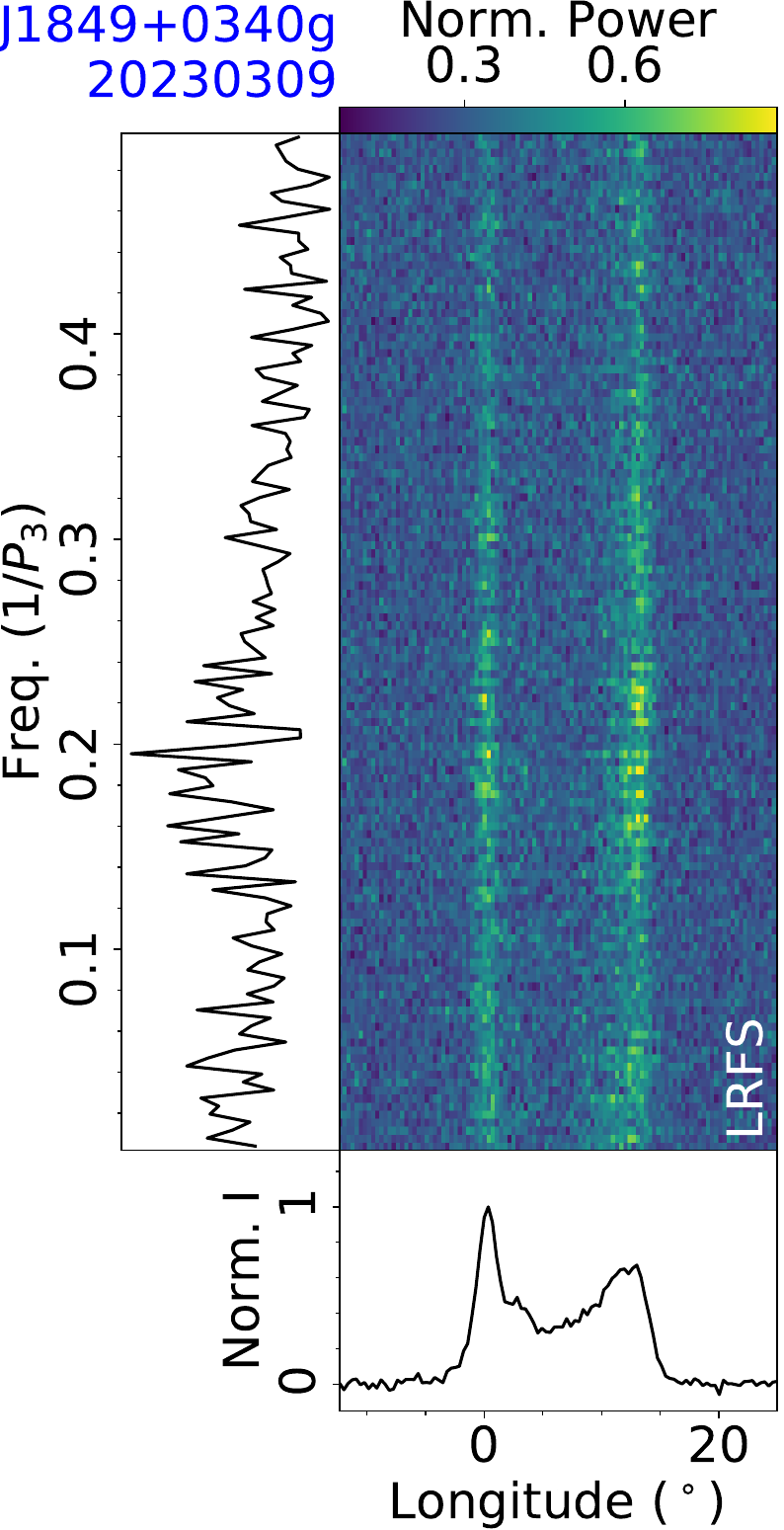}
\includegraphics[width=0.22\textwidth, angle=0]{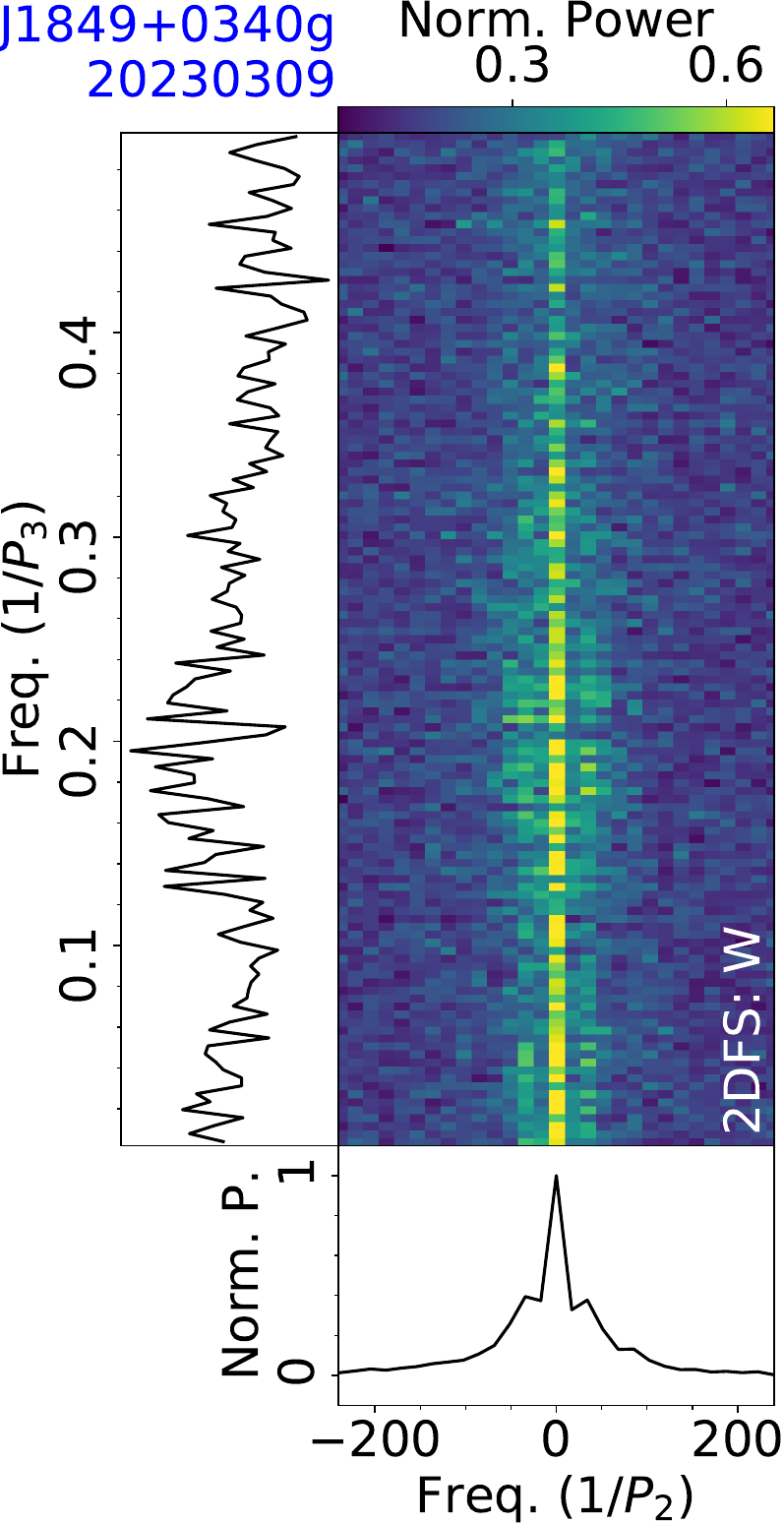}
\figcaption{Fluctuation analysis of PSR J1849+0340g for the observation on 20230309, with LRFS and 2DFS for the on-pulse region of a mean pulse profile.
\label{subfig:fluctu:J1849+0340g}}
\end{figure}

\subsection{J1849+0127}
\label{subsec:J1849+0127}

PSR J1849+0127 was discovered in the Parkes Multibeam Pulsar Survey \citep{Morris2002}. 

This pulsar was observed by FAST on 20231022 for 30 minutes, with a rotation period $P=0.5422$~s and a dispersion measure $D\!M=203.0~{\rm cm^{-3}\,pc}$ derived. The single pulse sequence and a zoomed-in view of pulses No. 800-1200 in Fig.~\ref{subfig:TP:J1849+0127} illustrate the subpulse modulation behavior. From the fluctuation spectra in Fig.~\ref{subfig:fluctu:J1849+0127}, the pulsar prefers to drift negatively. The centroid of the drift feature in 2DFS is at $1/P_3=0.105\pm0.001$ and $1/P_2=-2.1\pm0.6$, corresponding to periodicities of $P_3=9.5\pm0.1$ periods and $P_2=-172\pm48$ degrees.

\subsection{J1849+0225g}
\label{subsec:J1849+0225g}

PSR J1849+0225g was discovered in the FAST GPPS survey \citep{Han2021,han2025}. 

This pulsar was observed by FAST on 20230309 for 37 minutes, yielding a rotation period $P=1.4744$~s and  a dispersion measure $D\!M=266.4~{\rm cm^{-3}\,pc}$. 
Single pulse sequences of this observation are shown in Fig.~\ref{subfig:TP:J1849+0225g}. In the energy histogram in Fig.~\ref{subfig:Hist:J1849+0225g}, the distribution around zero intensity is not symmetric, which illustrates that the pulsar has a weak emission state instead of the nulling state. Emission modes of single pulses are distinguished from the histogram. Averaged profiles of two emission modes in Fig.~\ref{subfig:PolModes:J1849+0225g} show that the peak of the weak mode is earlier compared to the bright mode.

\begin{figure}[htpb]
\centering
\includegraphics[width=0.22\textwidth, angle=0]{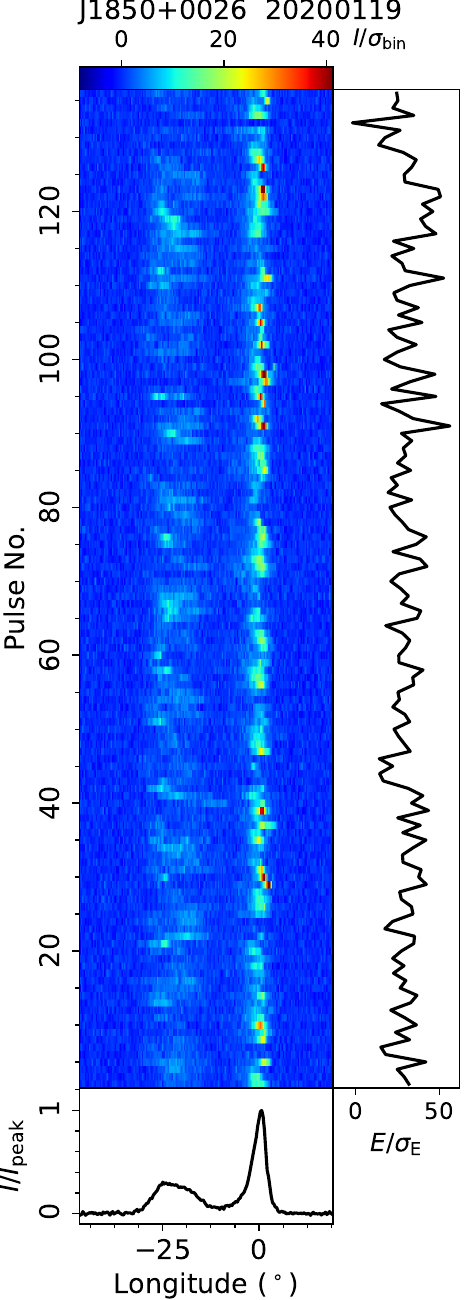}
\includegraphics[width=0.22\textwidth, angle=0]{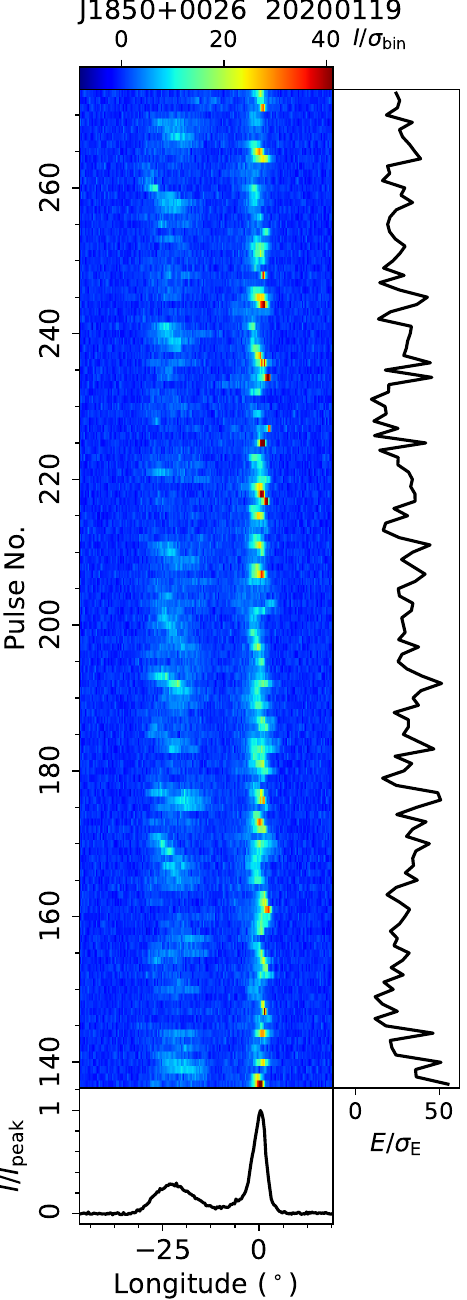}
\figcaption{Single pulse sequences of PSR J1850+0026 from the FAST observation on 20200119. \label{subfig:TP:J1850+0026}}
\end{figure}

\begin{figure}[htpb]
\centering
\includegraphics[width=0.22\textwidth, angle=0]{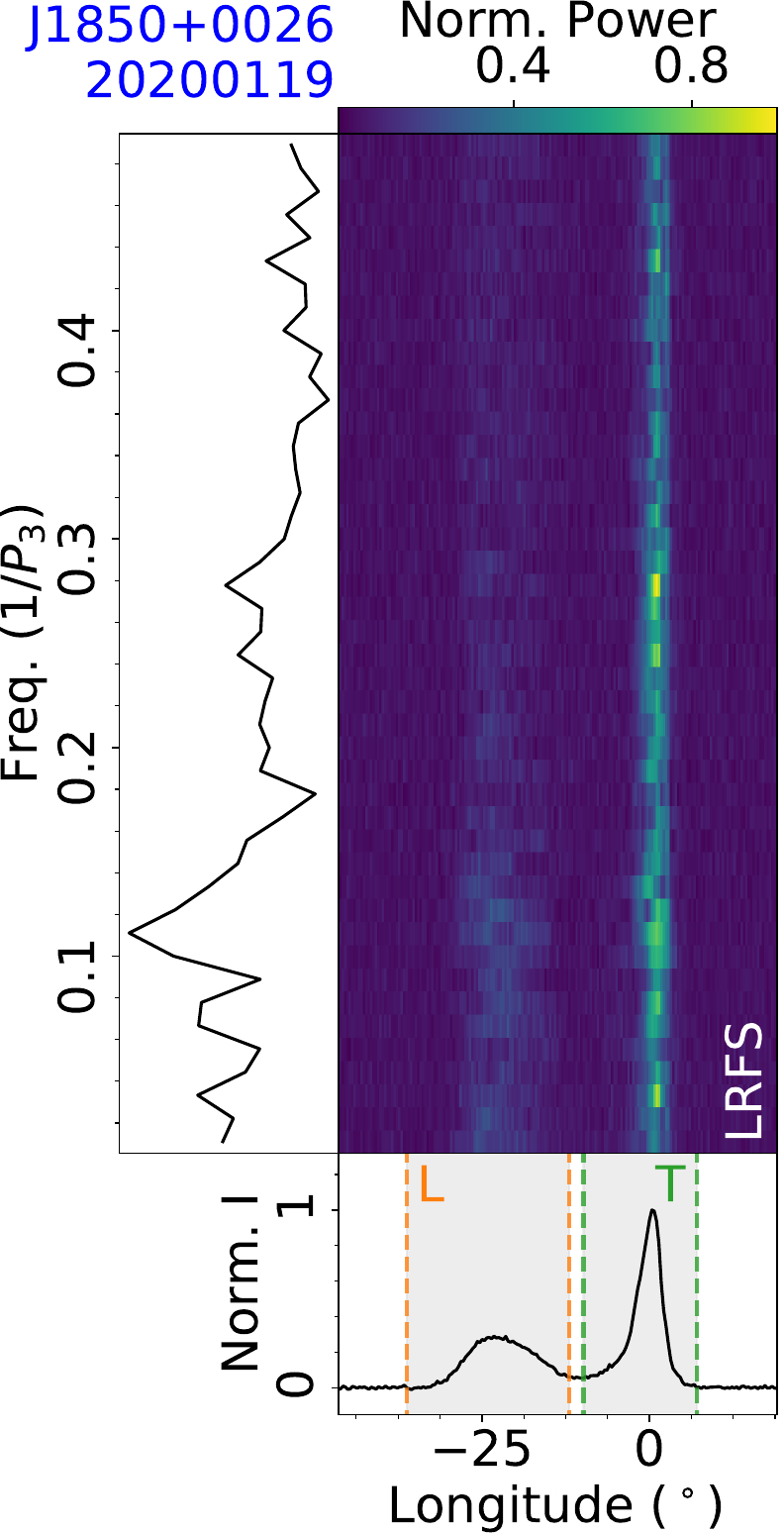}
\includegraphics[width=0.22\textwidth, angle=0]{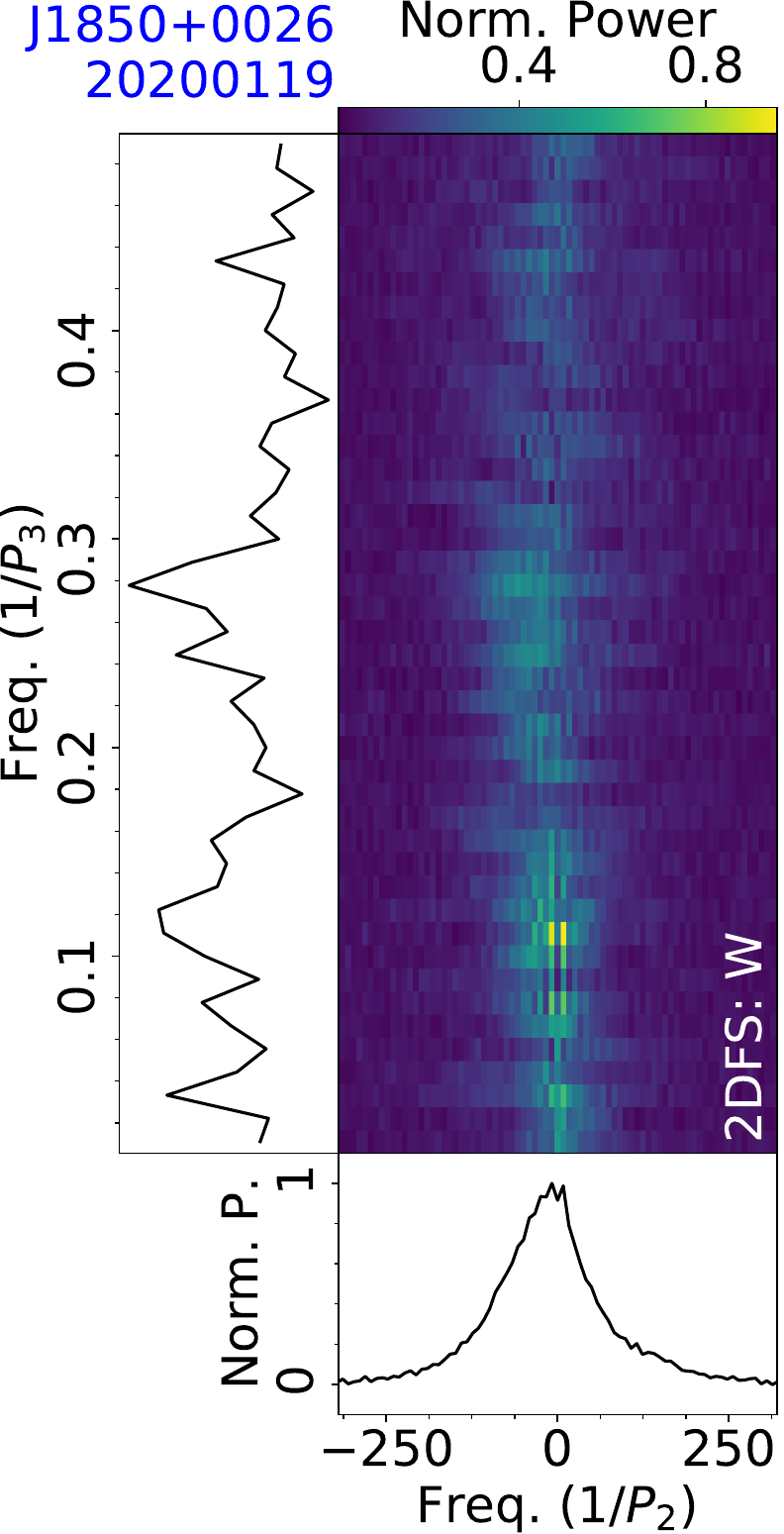}\\
\includegraphics[width=0.22\textwidth, angle=0]{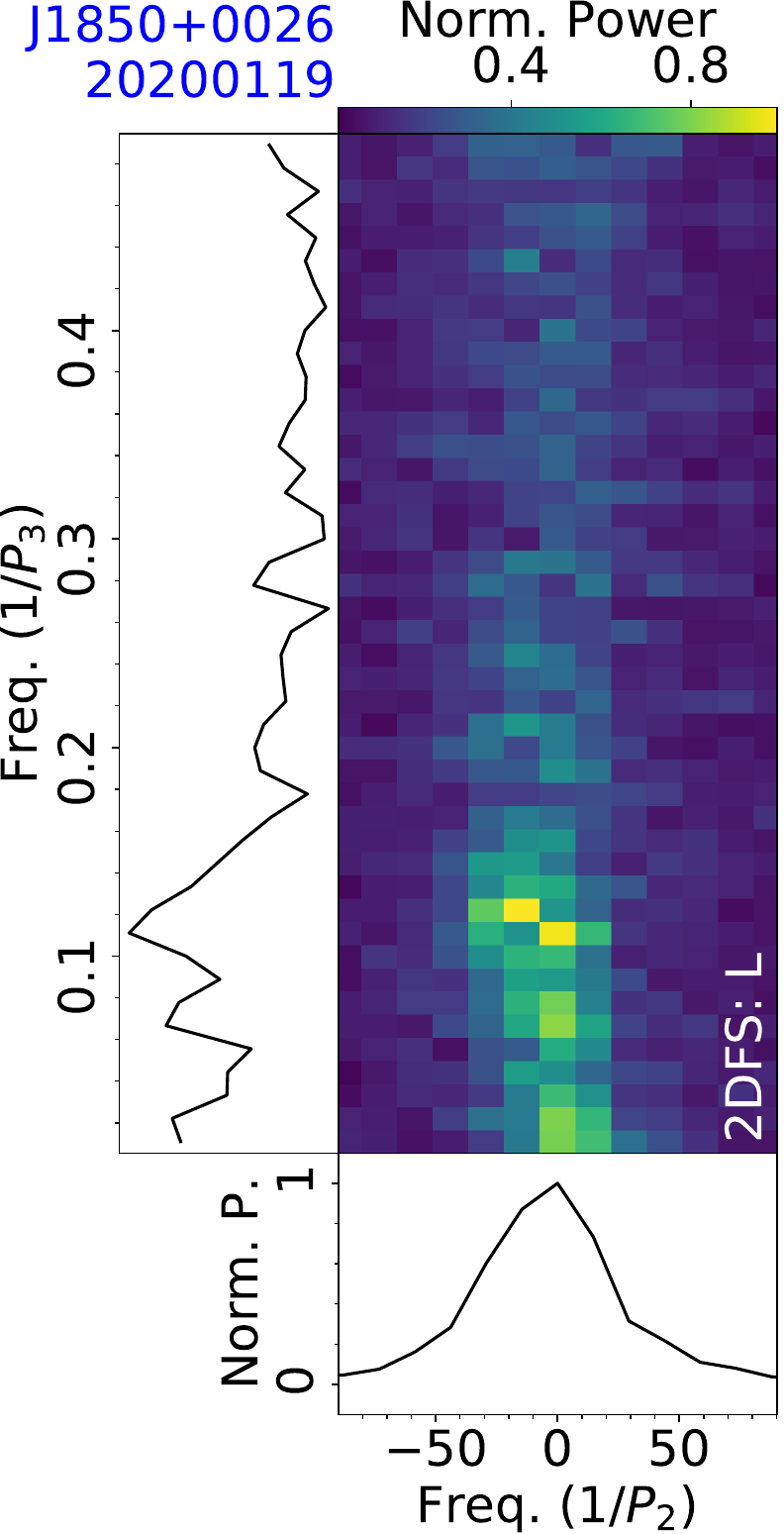}
\includegraphics[width=0.22\textwidth, angle=0]{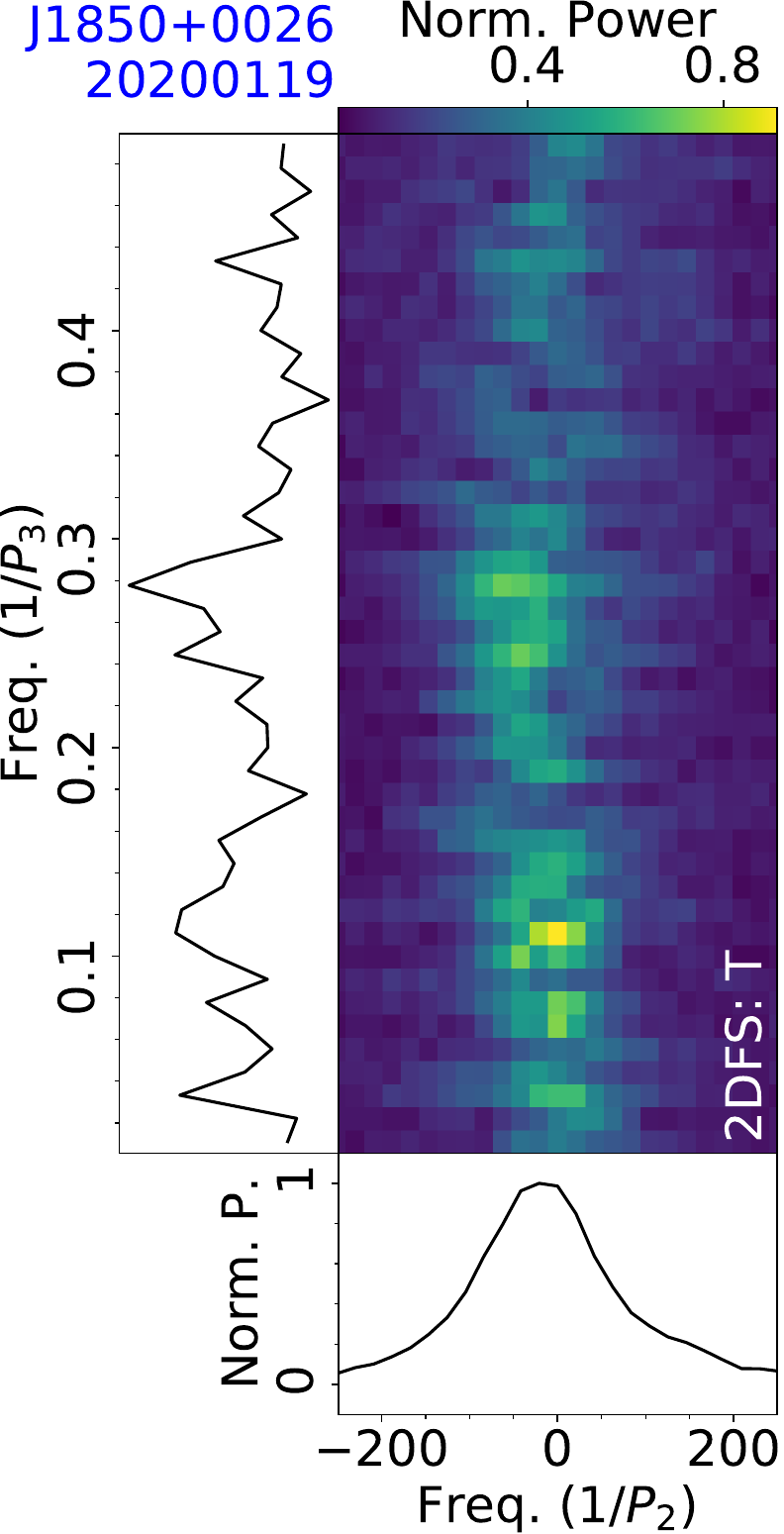}
\figcaption{Fluctuation analysis of PSR J1850+0026 from the FAST observation on 20200119, with LRFS (top-left), and 2DFS for the on-pulse region (top-right), leading part (bottom-left) and trailing part (bottom-right) of a mean pulse profile.
\label{subfig:fluctu:J1850+0026}}
\end{figure}

\subsection{J1849+0340g}
\label{subsec:J1849+0340g}

PSR J1849+0340g was discovered in the FAST GPPS survey \citep{Han2021,han2025}.

This pulsar was observed by FAST on 20230309 for 41 minutes, deriving a rotation period $P=1.6666$~s and a dispersion measure $D\!M=350.6~{\rm cm^{-3}\,pc}$. 
Single pulse sequences of this observation are shown in Fig.~\ref{subfig:TP:J1849+0340g}. The on-pulse energy histogram in Fig.~\ref{subfig:Hist:J1849+0340g} indicates the existence of the nulling phenomenon with a nulling fraction of 16$\pm$2\%. From the fluctuation spectra of the on-pulse phase region (Fig.~\ref{subfig:Hist:J1849+0340g}), there is a temporal modulation with the centroid frequency of $1/P_3=0.185\pm0.001$, corresponding to $P_3=5.40\pm0.03$ periods.


\begin{figure}[htpb]
\centering
\includegraphics[width=0.22\textwidth, angle=0]{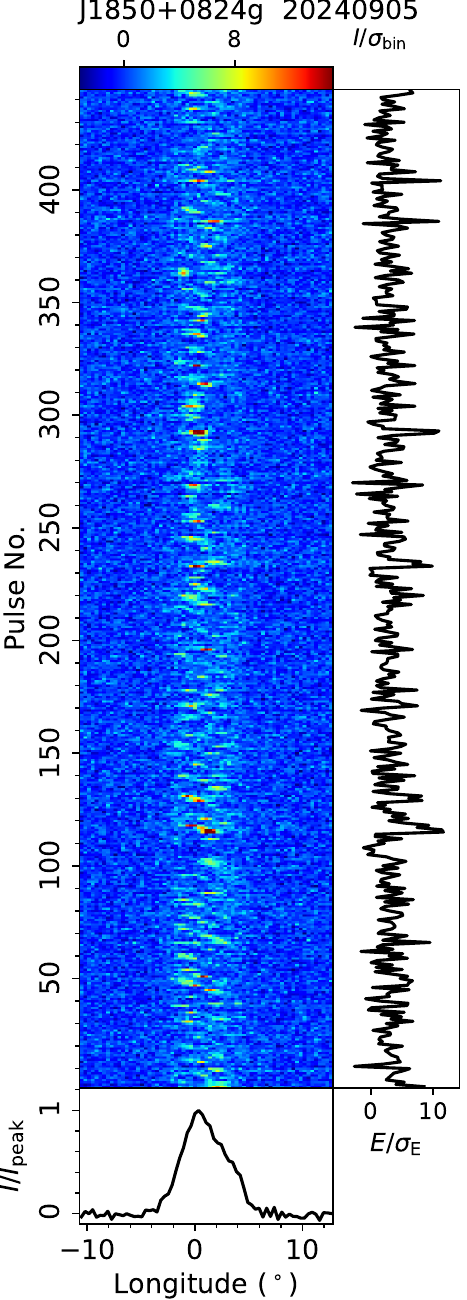}
\includegraphics[width=0.22\textwidth, angle=0]{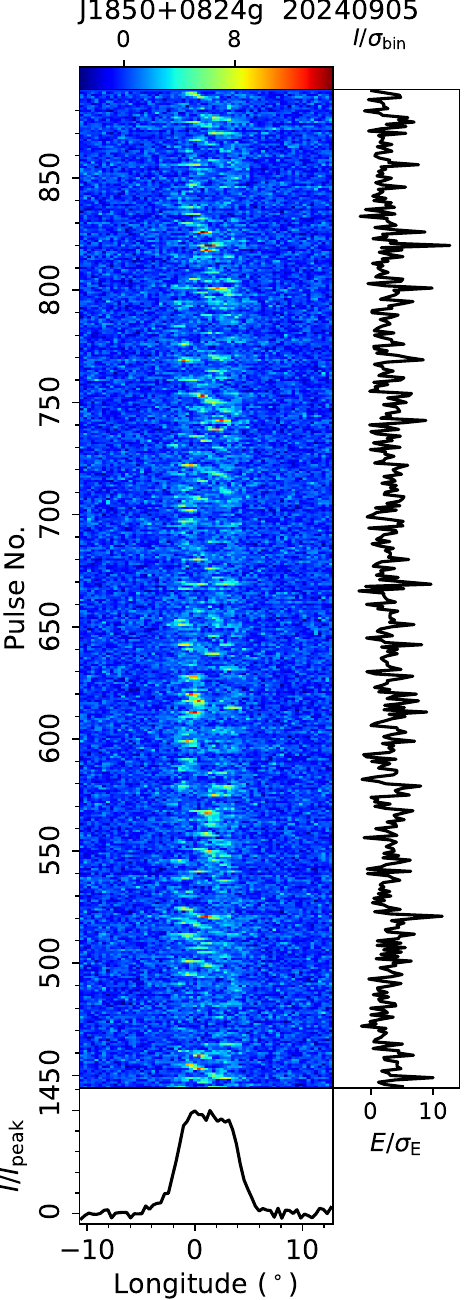}
\figcaption{Single pulse sequences of PSR J1850+0824g from the FAST observation on 20240905. \label{subfig:TP:J1850+0824g}}
\end{figure}

\begin{figure}[htpb]
\centering
\includegraphics[width=0.22\textwidth, angle=0]{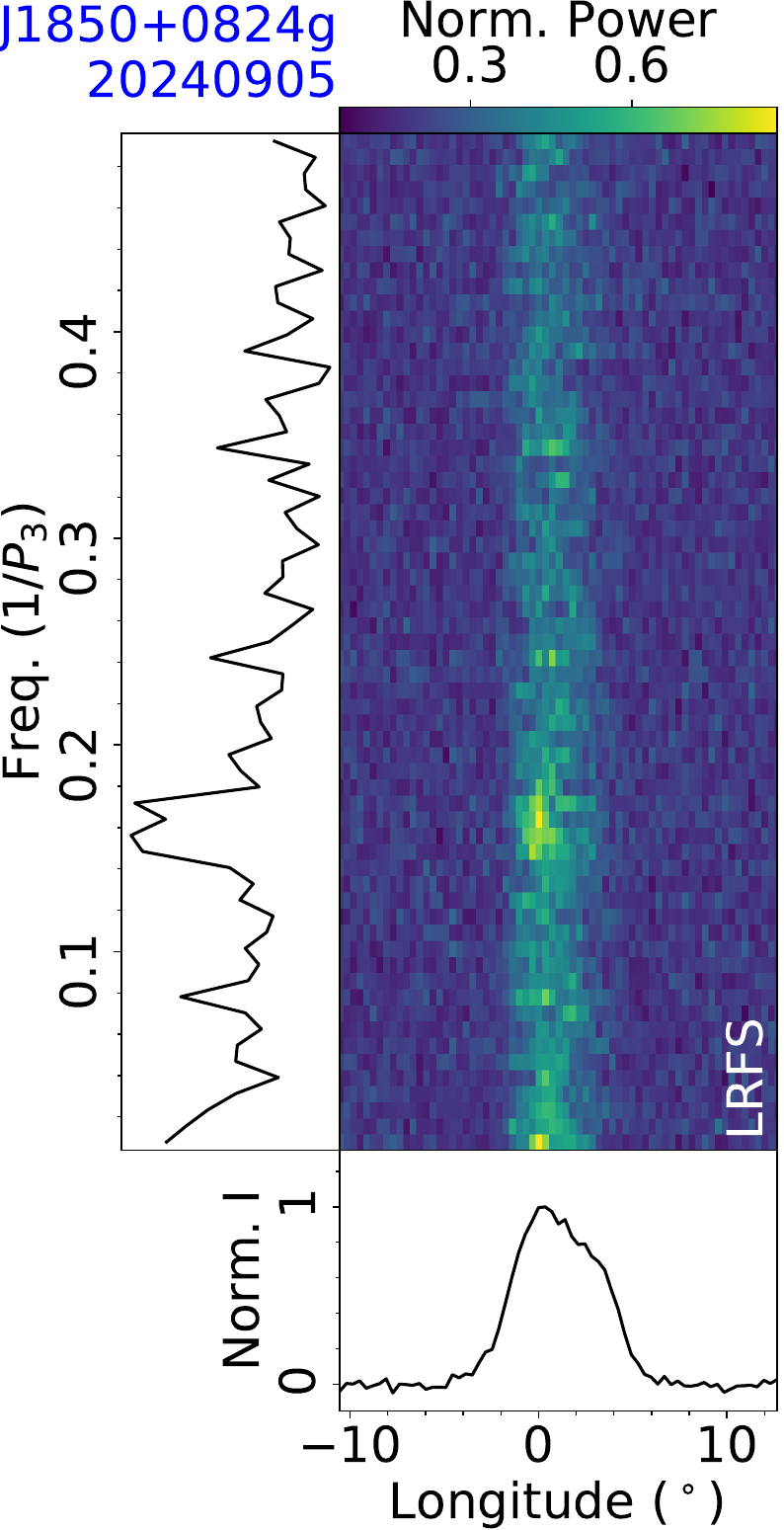}
\includegraphics[width=0.22\textwidth, angle=0]{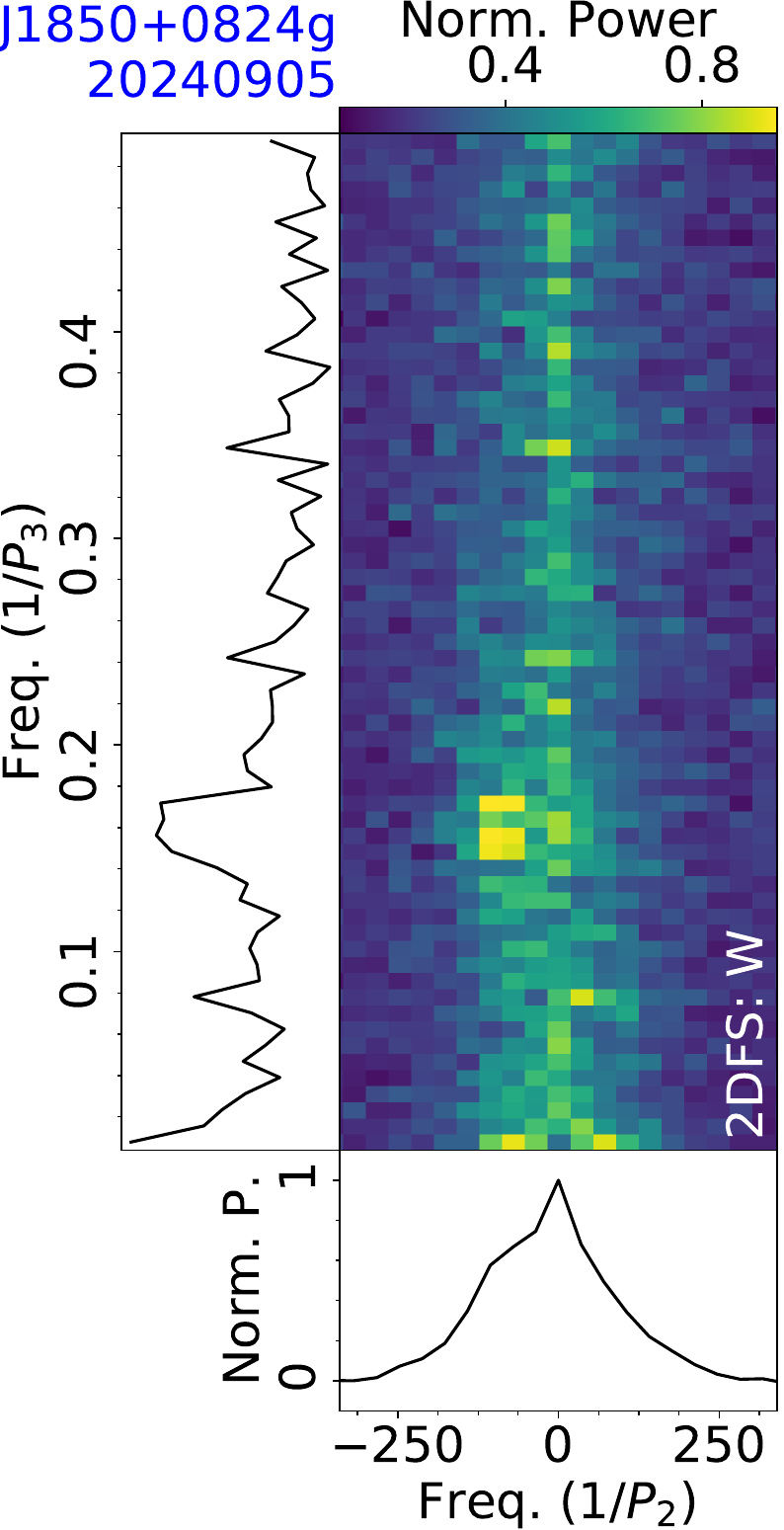}
\figcaption{Fluctuation analysis of PSR J1850+0824g from the FAST observation on 20240905, with LRFS and 2DFS for the on-pulse region of a mean pulse profile.
\label{subfig:fluctu:J1850+0824g}}
\end{figure}

\begin{figure}[htpb]
\centering
\includegraphics[width=0.22\textwidth, angle=0]{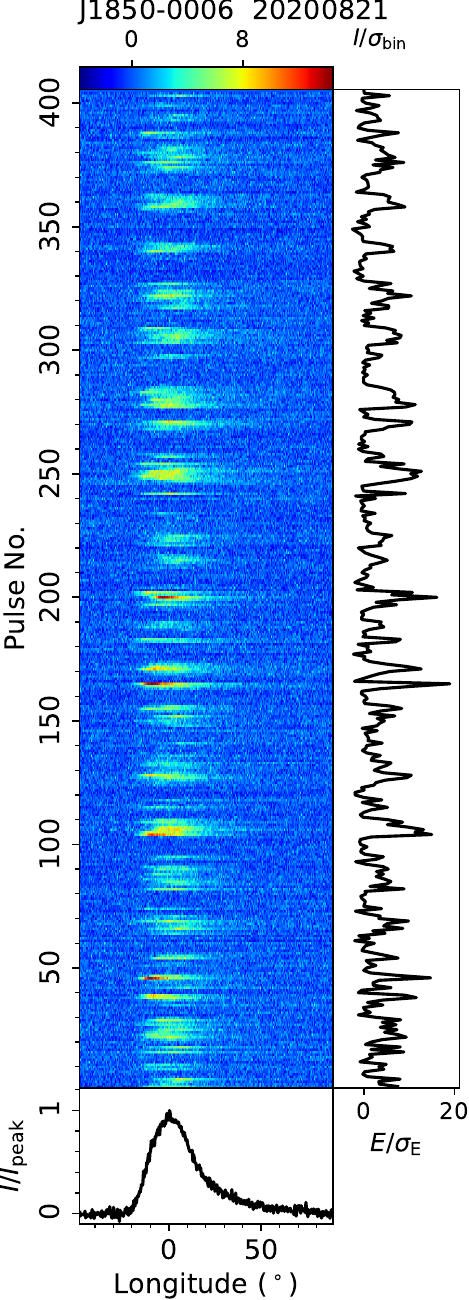}
\figcaption{Single pulse sequence of PSR J1850-0006 from the FAST observation on 20200821.
\label{subfig:TP:J1850-0006}}
\end{figure}

\begin{figure}[htpb]
\centering
\includegraphics[width=0.39\textwidth, angle=0]{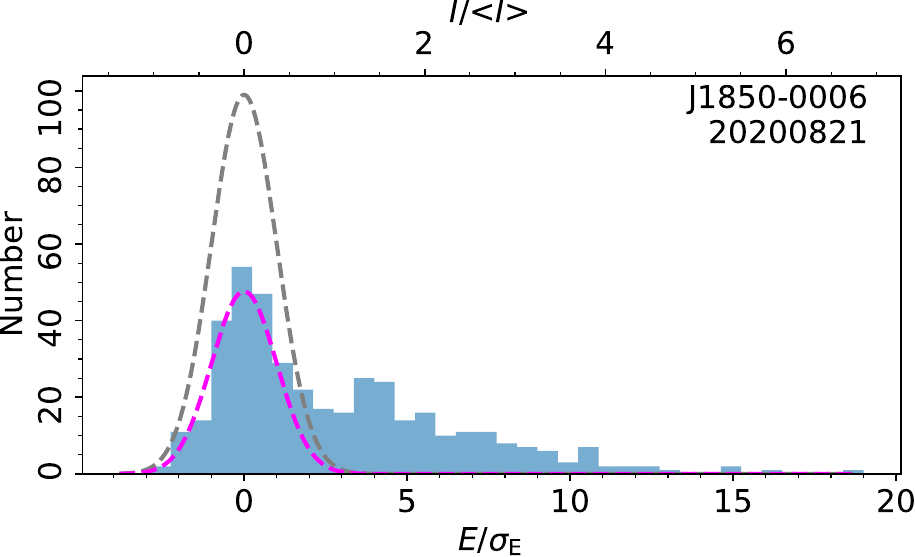}
\figcaption{On-pulse energy histogram of single pulses of PSR J1850-0006 from the FAST observation on 20200821. \label{subfig:Hist:J1850-0006}}
\end{figure}

\begin{figure}[htpb]
\centering
\includegraphics[width=0.22\textwidth, angle=0]{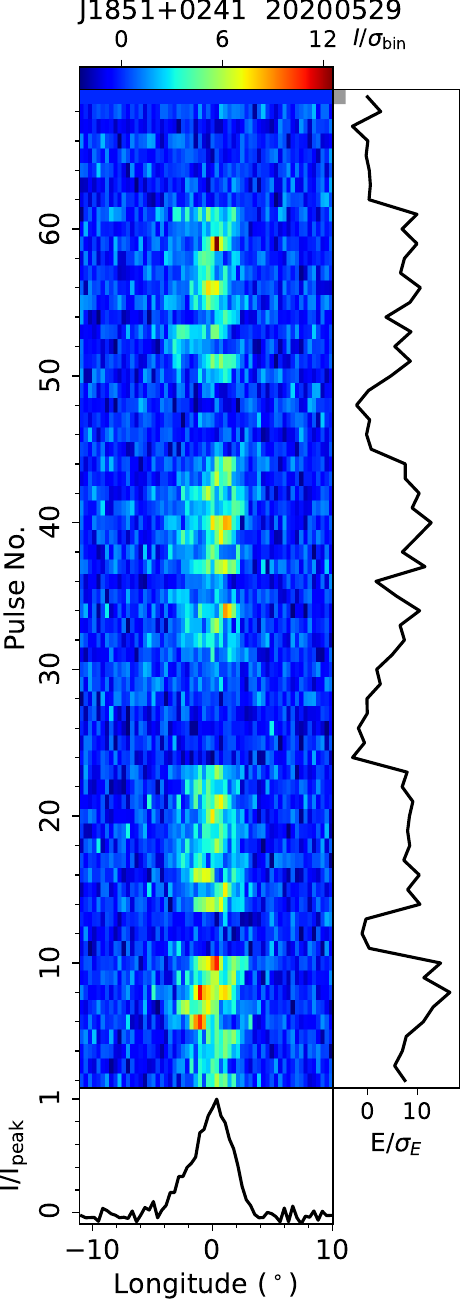}
\figcaption{Single pulse sequence of PSR J1851+0241 from the FAST observation on 20200529.
\label{subfig:TP:J1851+0241}}
\end{figure}

\begin{figure}[htpb]
\centering
\includegraphics[width=0.39\textwidth, angle=0]{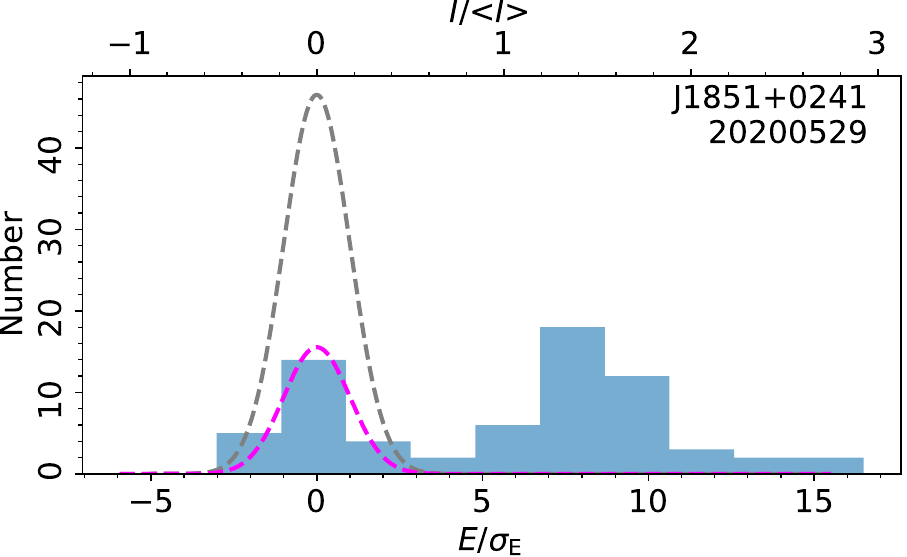}
\figcaption{On-pulse energy histogram of single pulses of PSR J1851+0241 from the FAST observation on 20200529.
\label{subfig:Hist:J1851+0241}}
\end{figure}

\begin{figure}[htpb]
\centering
\includegraphics[width=0.39\textwidth, angle=0]{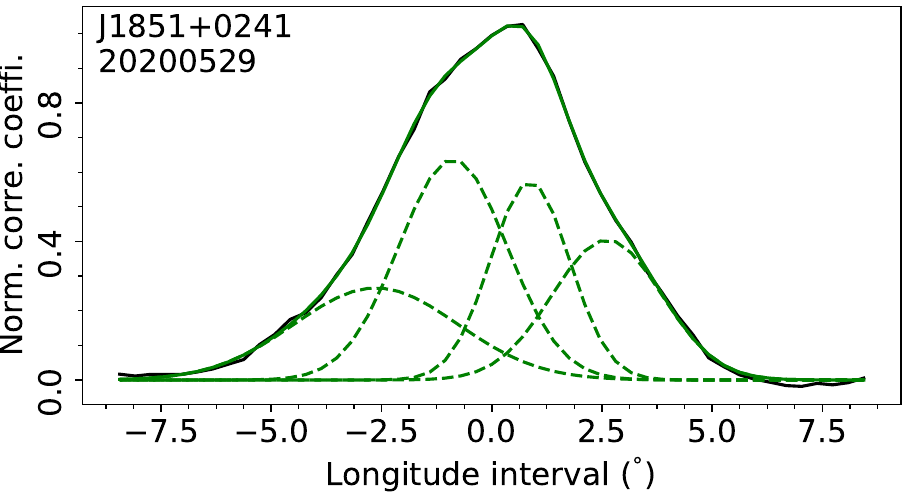}
\figcaption{Cross correlation of PSR J1851+0241 from the FAST observation on 20200529.
\label{subfig:Corre:J1851+0241}}
\end{figure}

\subsection{J1850+0026}
\label{subsec:J1850+0026}

PSR J1850+0026 was discovered in the Parkes Multibeam Pulsar Survey \citep{Morris2002}. \citet{Song2023} reported parameters of three negative drift features: one from the leading component of $P_3=9\pm2$ periods and $P_2=-65^{+31}_{-18}$ degree; two from the trailing component, that are $P_3=8.2\pm0.8$ periods and $P_2=-11^{+1}_{-28}$ degrees, and $P_3=4.1\pm0.2$ periods and $P_2=-10^{+2}_{-4}$ degrees.

The pulsar was observed by FAST on 20200119 for 5 minutes, deriving a rotation period $P=1.0818$~s and a dispersion measure $D\!M=201.3~{\rm cm^{-3}\,pc}$. 
Single pulse sequences of this observation are displayed in Fig.~\ref{subfig:TP:J1850+0026}. Fluctuation spectra are shown in Fig.~\ref{subfig:fluctu:J1850+0026}. 
There is a negative drift feature in 2DFS of the leading component, with the centroid frequencies of $1/P_3=0.081\pm0.001$ and $1/P_2=-6\pm1$, corresponding to drifting parameters of $P_3=12.4\pm0.2$ periods and $P_2=-57\pm7^\circ$.
In 2DFS of the trailing component, two negatively drifting features are related to different drifting modes in the single pulse sequences (Fig.~\ref{subfig:TP:J1850+0026}). For the normal mode, the centroid frequencies of the drift feature are $1/P_3=0.099\pm0.002$ ($P_3=10.2\pm0.2$ periods) and $1/P_2=-11\pm2$ ($P_2=-32\pm6^\circ$). 
For the abnormal drifting mode, the drift feature in 2DFS has centroid frequencies of $1/P_3=0.253\pm0.002$ and $1/P_2=-46\pm2$.
drifting parameters of 
($P_3=3.95\pm0.04$ periods) and ($P_2=-7.8\pm0.3^\circ$).

\begin{figure}[htpb]
\centering
\includegraphics[width=0.22\textwidth, angle=0]{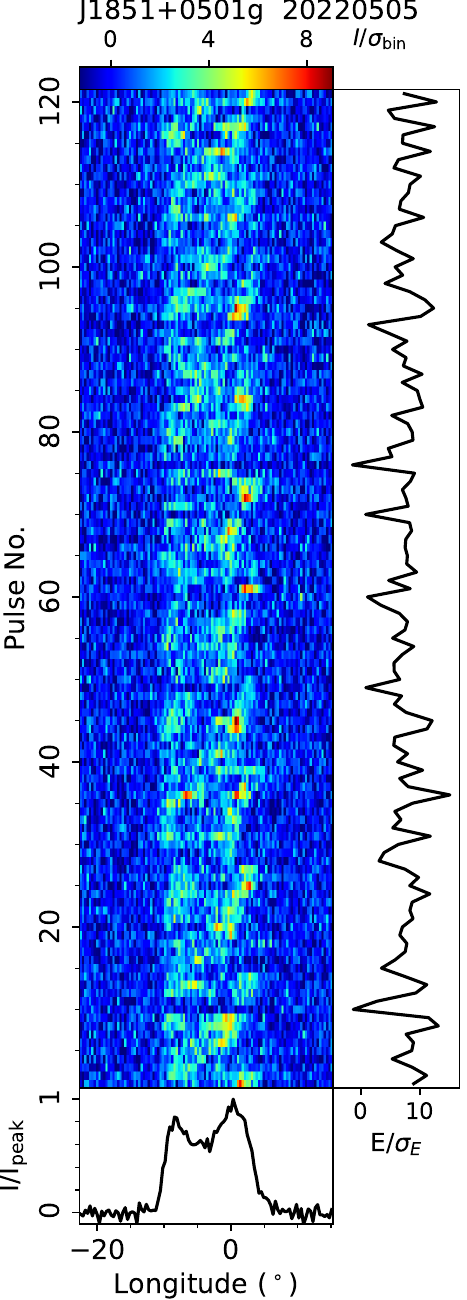}
\figcaption{Single pulse sequence of PSR J1851+0501g from the FAST observation on 20220505.
\label{subfig:TP:J1851+0501g}}
\end{figure}

\begin{figure}[htpb]
\centering
\includegraphics[width=0.22\textwidth, angle=0]{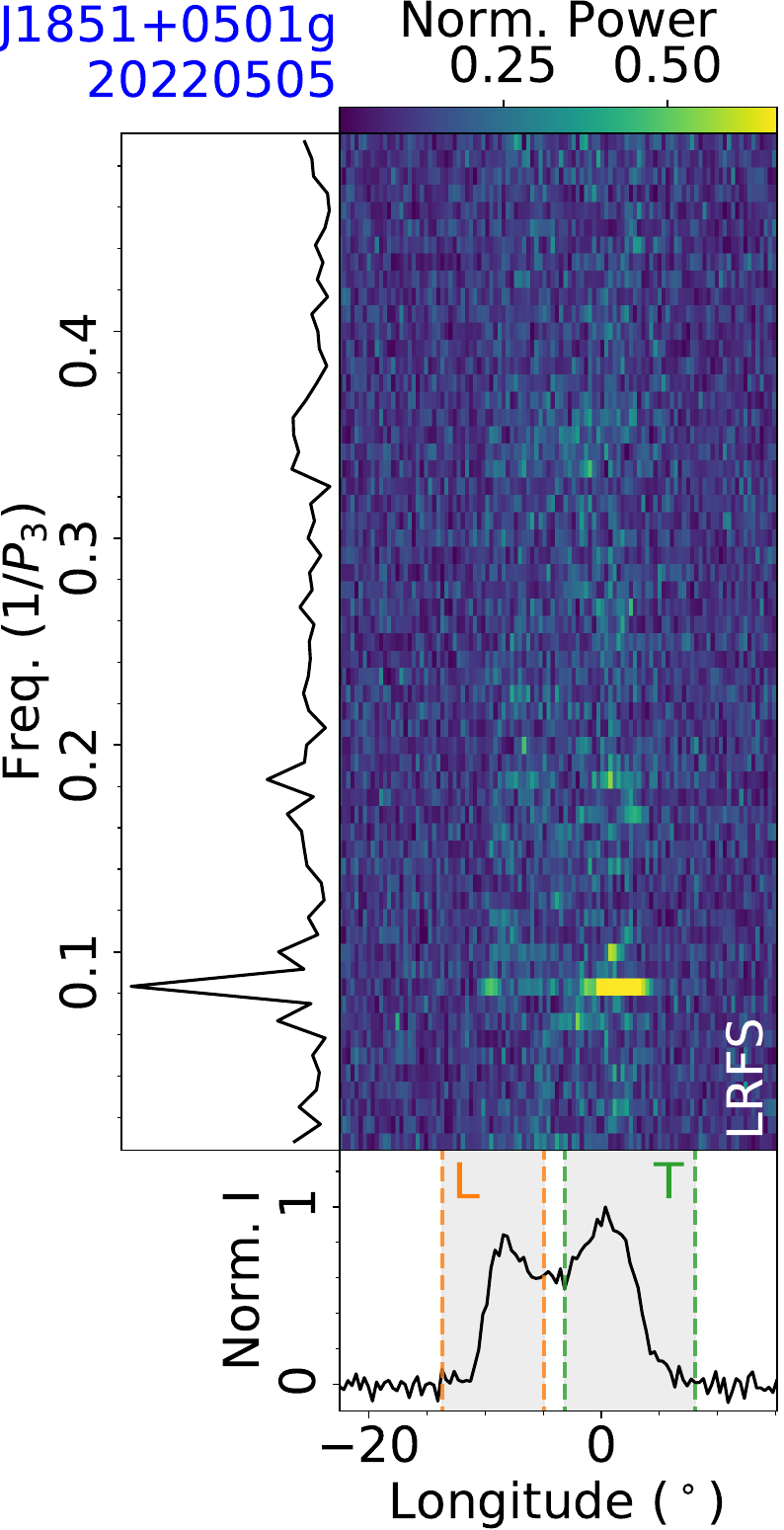}
\includegraphics[width=0.22\textwidth, angle=0]{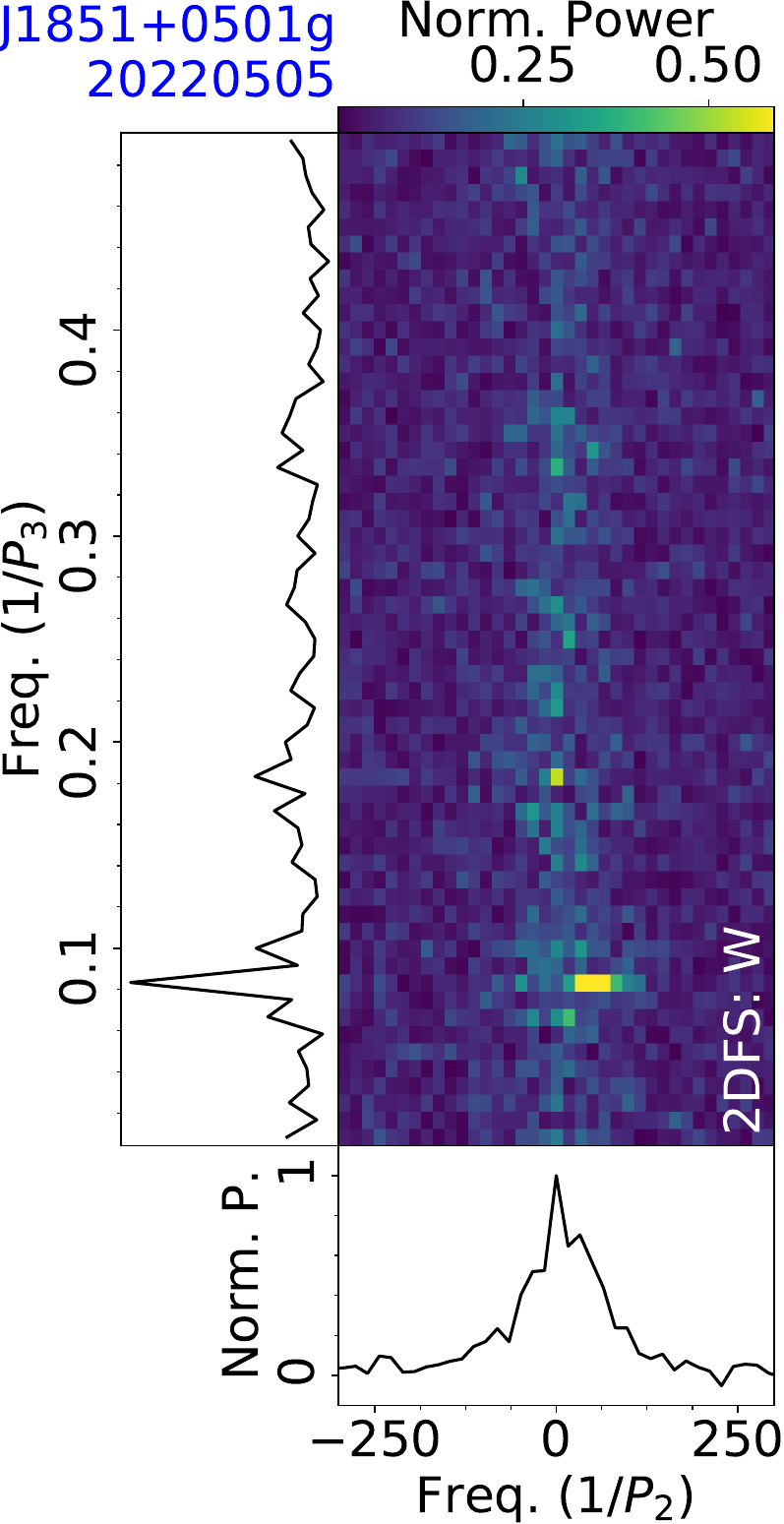}\\
\includegraphics[width=0.22\textwidth, angle=0]{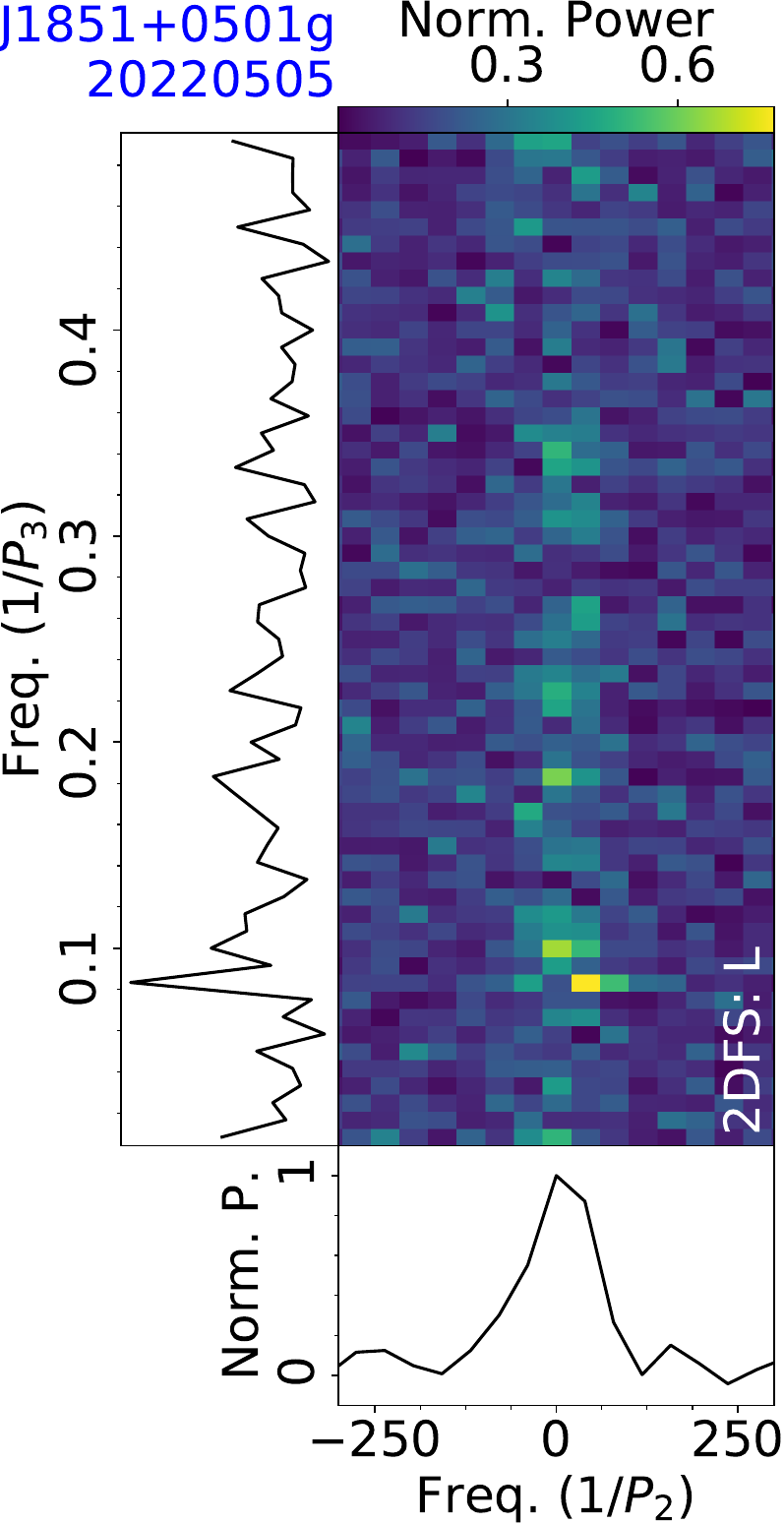}
\includegraphics[width=0.22\textwidth, angle=0]{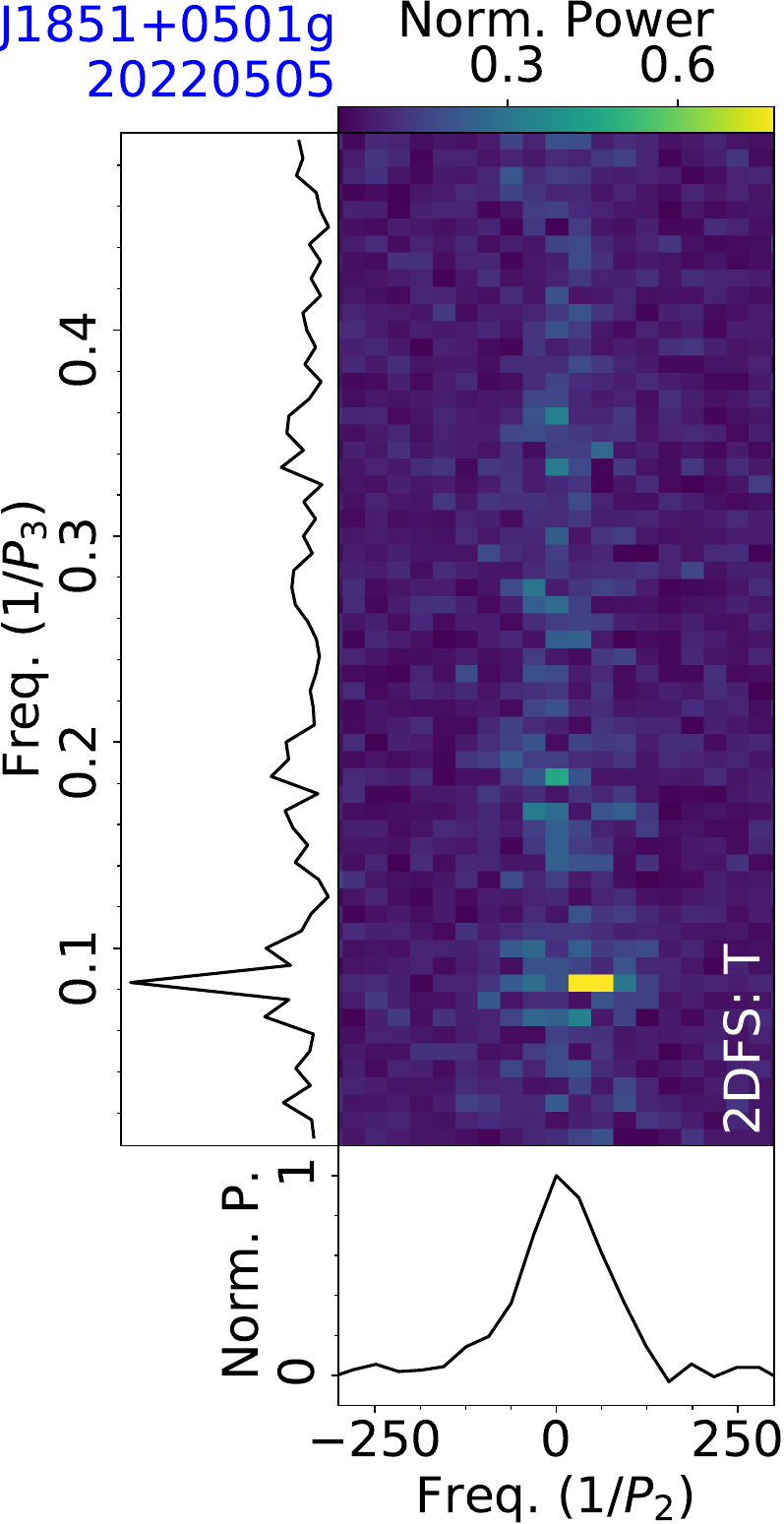}
\figcaption{Fluctuation analysis of PSR J1851+0501g for the observation on 20220505, with LRFS (top-left), and 2DFS for the on-pulse region (top-right), leading part (bottom-left) and trailing part (bottom-right) of a mean pulse profile.
\label{subfig:fluctu:J1851+0501g}}
\end{figure}

\begin{figure}[htpb]
\centering
\includegraphics[width=0.22\textwidth, angle=0]{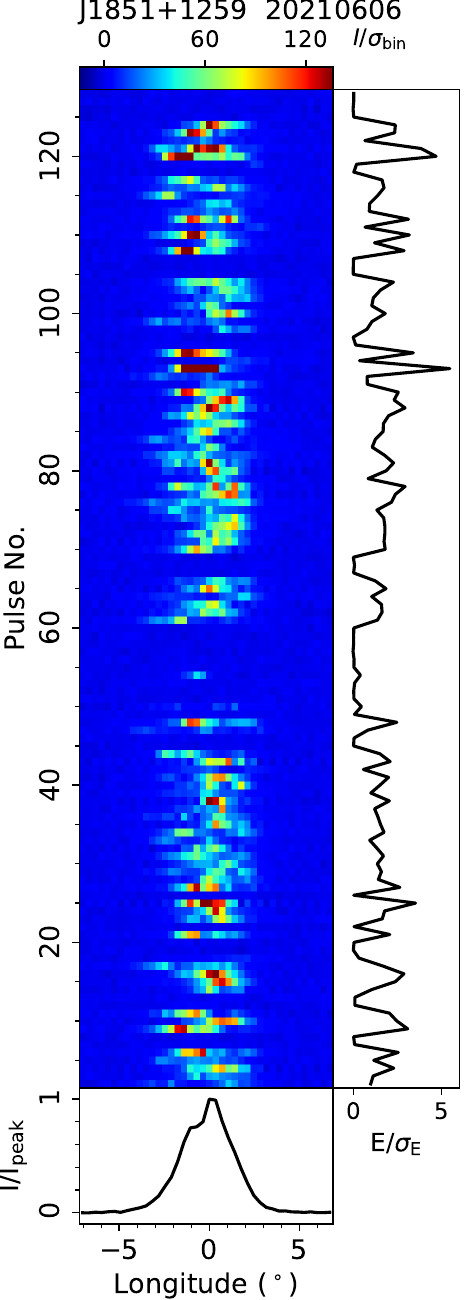}
\includegraphics[width=0.22\textwidth, angle=0]{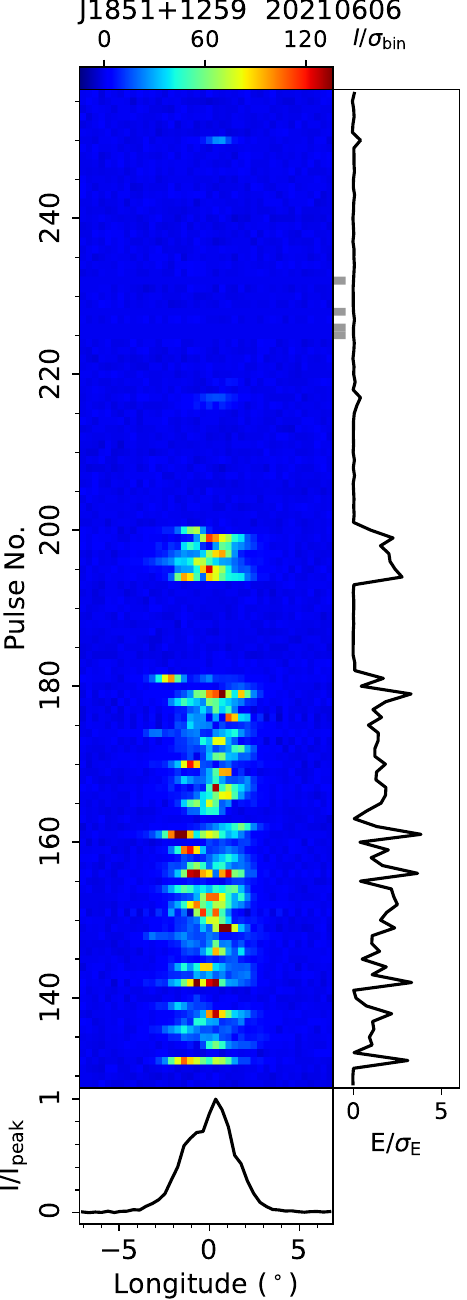}
\figcaption{Single pulse sequences of PSR J1851+1259 from the FAST observation on 20210606.
\label{subfig:TP:J1851+1259}}
\end{figure}

\begin{figure}[htpb]
\centering
\includegraphics[width=0.39\textwidth, angle=0]{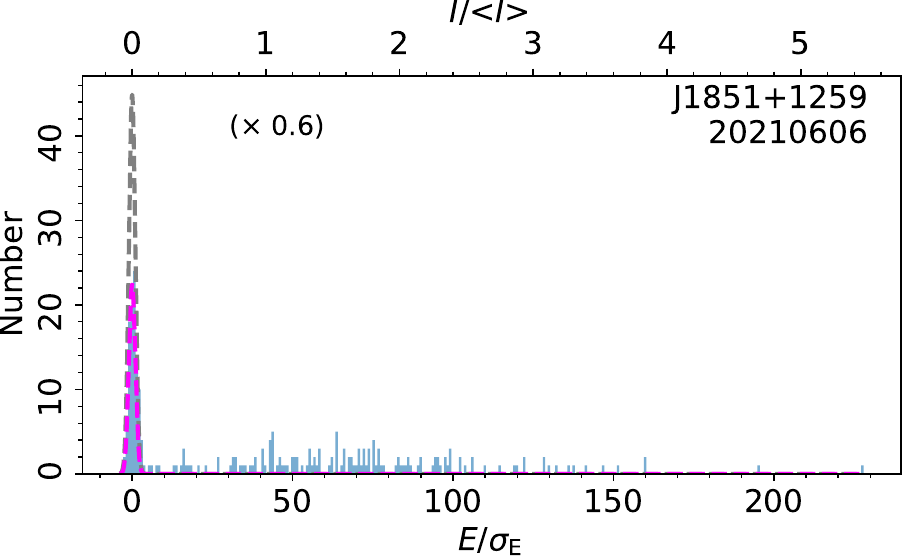}
\figcaption{On-pulse energy histogram of single pulses of PSR J1851+1259 from the FAST observation on 20210111.
\label{subfig:Hist:J1851+1259}}
\end{figure}

\begin{figure}[htpb]
\centering
\includegraphics[width=0.22\textwidth, angle=0]{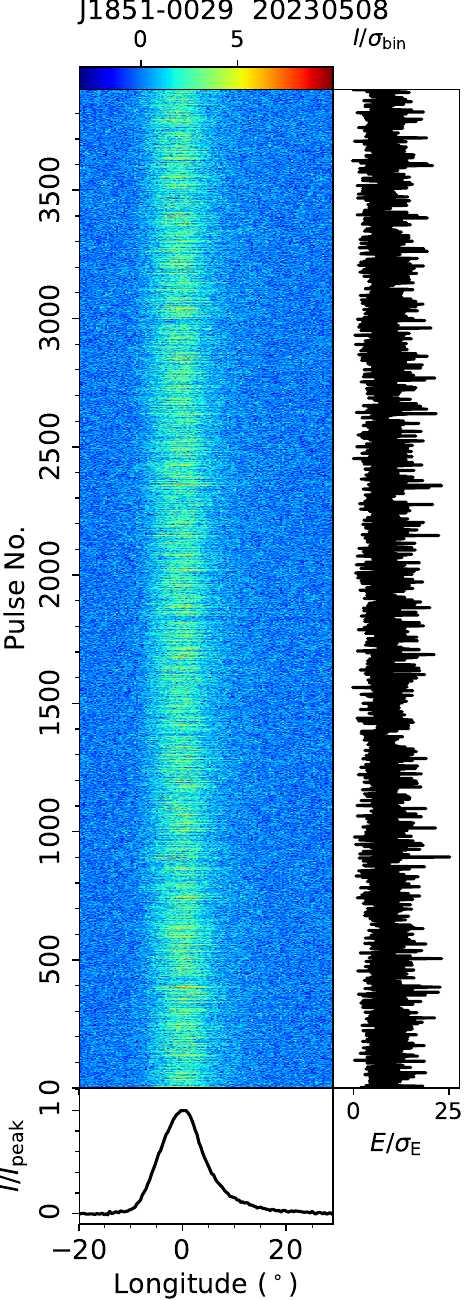}
\includegraphics[width=0.22\textwidth, angle=0]{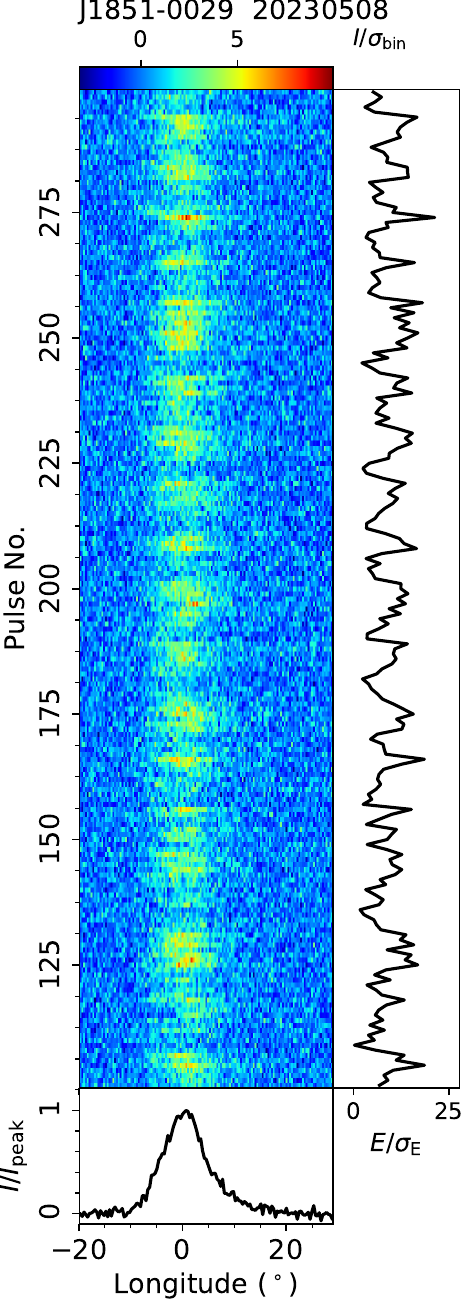}
\figcaption{Single pulse sequence of PSR J1851-0029 from the FAST observation on 20230508, and a zoomed-in view of pulses No.100-300.
\label{subfig:TP:J1851-0029}}
\end{figure}

\begin{figure}[htpb]
\centering
\includegraphics[width=0.22\textwidth, angle=0]{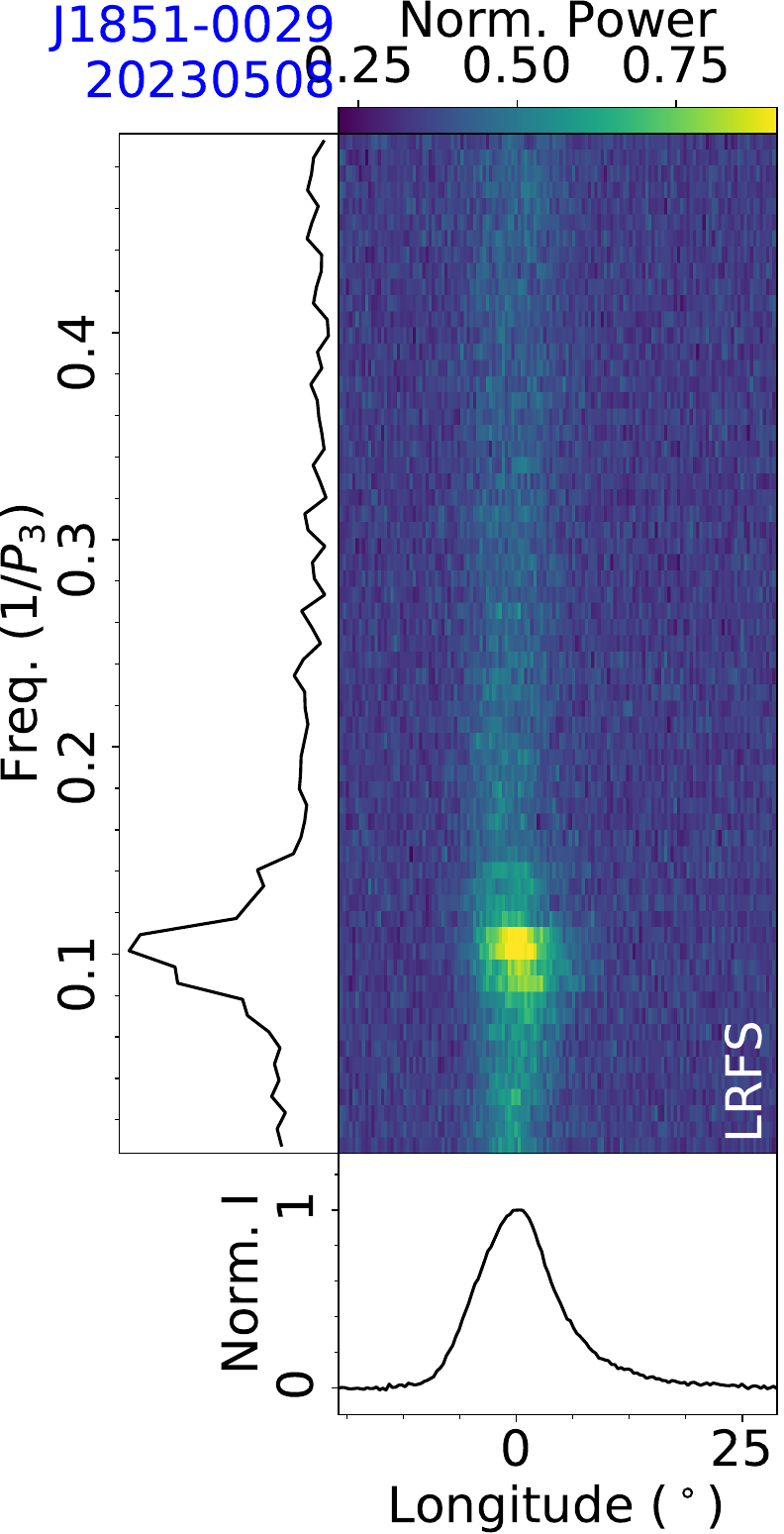}
\includegraphics[width=0.22\textwidth, angle=0]{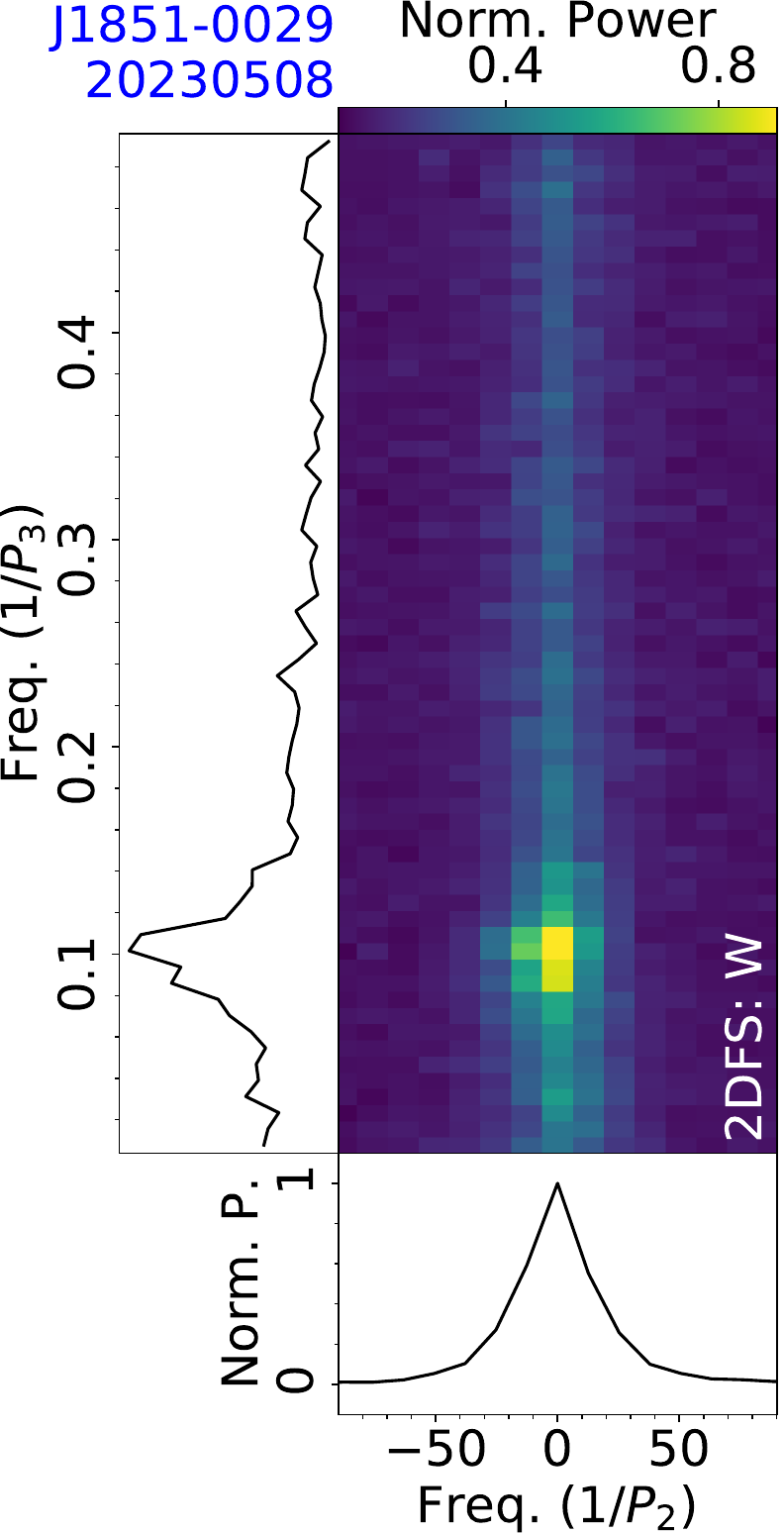}
\figcaption{Fluctuation analysis of PSR J1851-0029 for the observation on 20230508, with LRFS and 2DFS for the on-pulse region of a mean pulse profile.
\label{subfig:fluctu:J1851-0029}}
\end{figure}

\subsection{J1850+0824g}
\label{subsec:J1850+0824g}

PSR J1850+0824g was discovered in the FAST GPPS survey \citep{Han2021,han2025}.

The pulsar was observed by FAST on 20240905, deriving a rotation period of 1.0132~s and a dispersion measure of 72.8 $\rm cm^{-3}$pc from this observation. Single pulse sequences are shown in Fig.~\ref{subfig:TP:J1850-0006}, which display subpulse drifting behavior. 
In fluctuation spectra (Fig.~\ref{subfig:fluctu:J1850+0824g}), there is a negative drift feature with the centroid frequencies of $1/P_3=0.160\pm0.001$ and $1/P_2=-90\pm4$, corresponding to drifting parameters of $P_3=6.25\pm0.04$ periods and $P_2=-4.0\pm0.2^\circ$.

\begin{figure}[htpb]
\centering
\includegraphics[width=0.22\textwidth, angle=0]{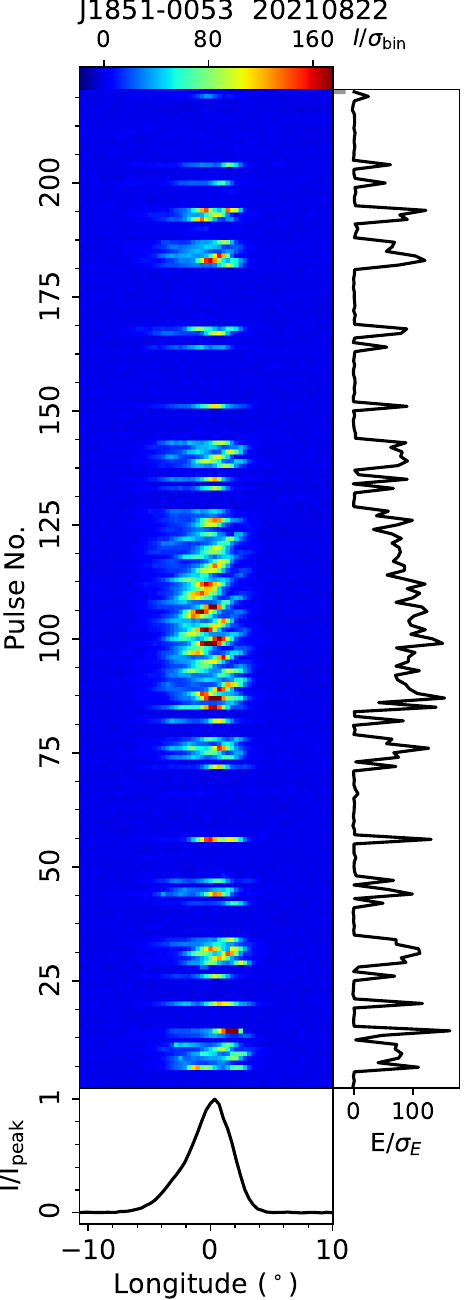}
\includegraphics[width=0.22\textwidth, angle=0]{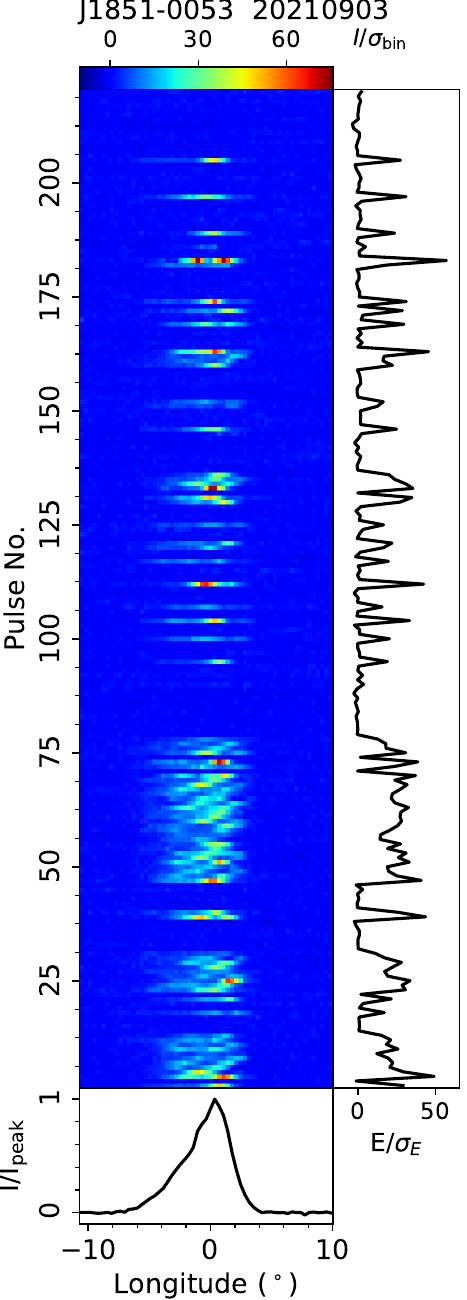}\\
\includegraphics[width=0.22\textwidth, angle=0]{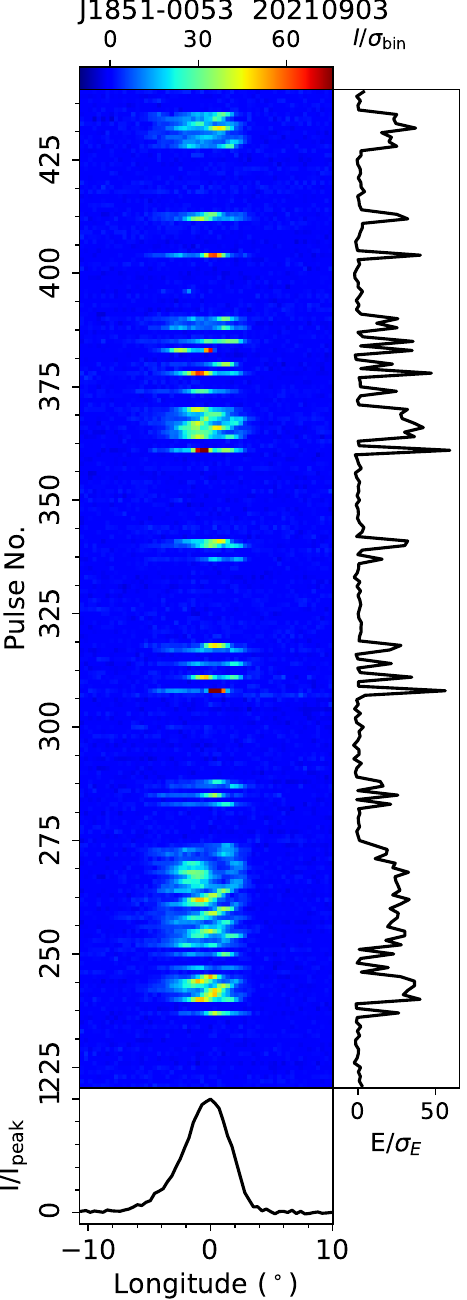}
\includegraphics[width=0.22\textwidth, angle=0]{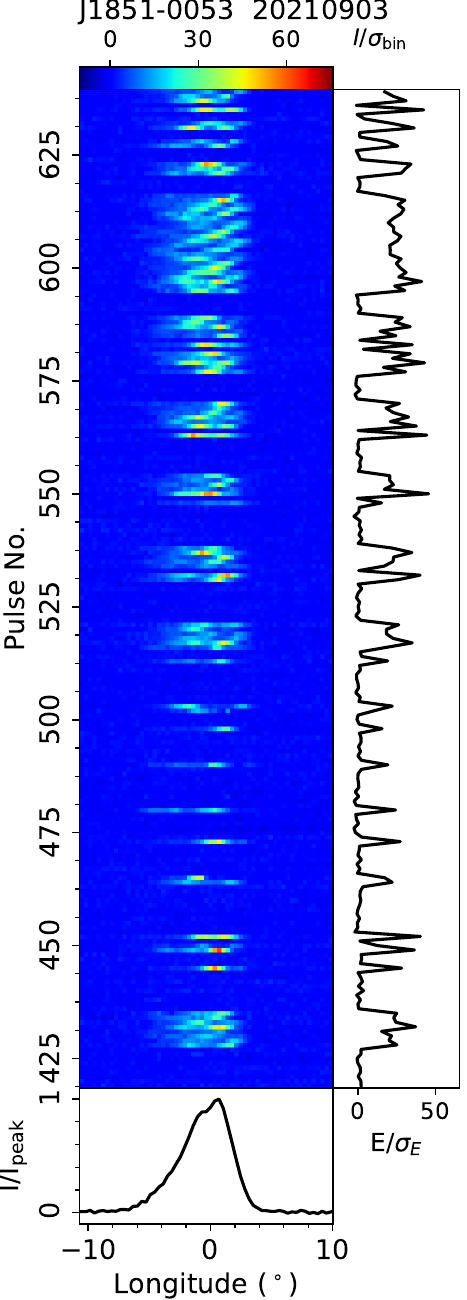}
\vspace{-0.3cm}
\figcaption{Single pulse sequences of PSR J1851-0053 from the FAST observation on 20210903.
\label{subfig:TP:J1851-0053}}
\end{figure}

\begin{figure}[htpb]
\centering
\includegraphics[width=0.39\textwidth, angle=0]{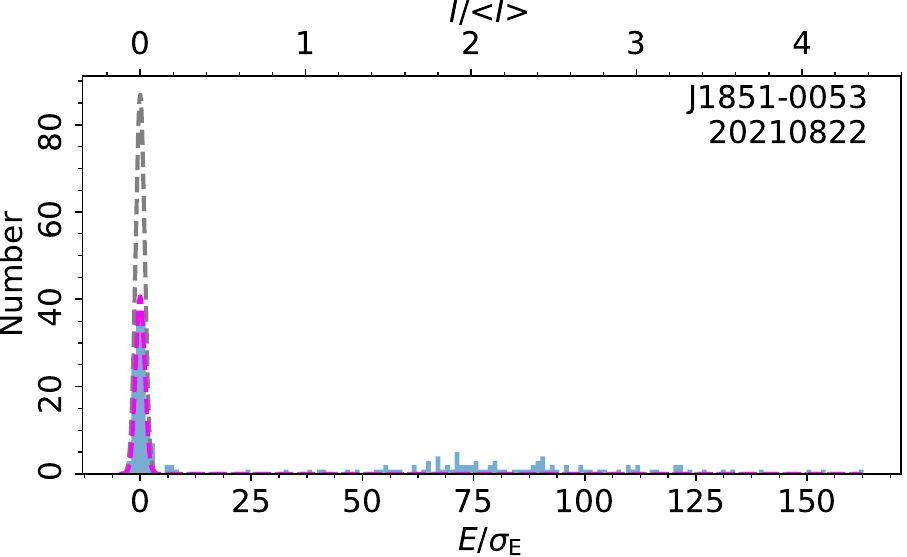}
\includegraphics[width=0.39\textwidth, angle=0]{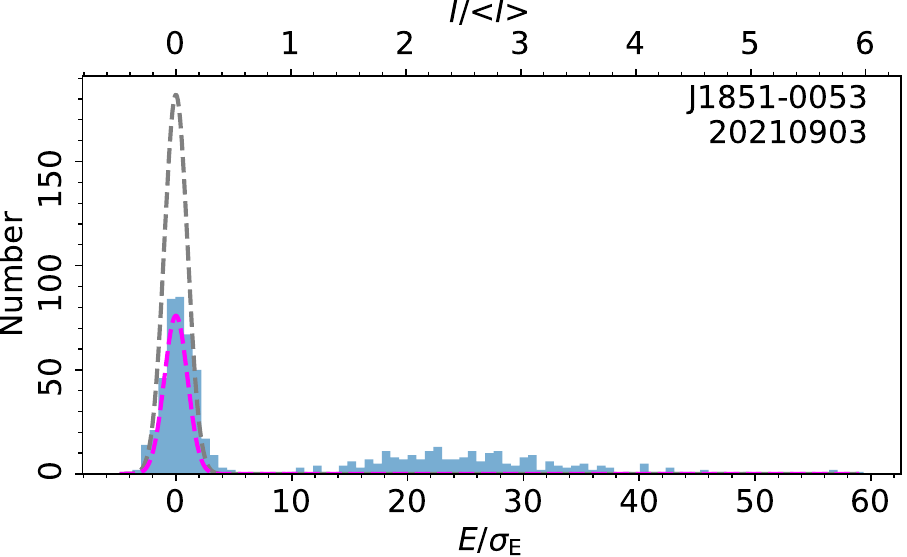}
\figcaption{On-pulse energy histograms of single pulses of PSR J1851-0053 from the FAST observation on 20210822 and 20210903. \label{subfig:Hist:J1851-0053}}
\end{figure}

\begin{figure}[htpb]
\centering
\includegraphics[width=0.22\textwidth, angle=0]{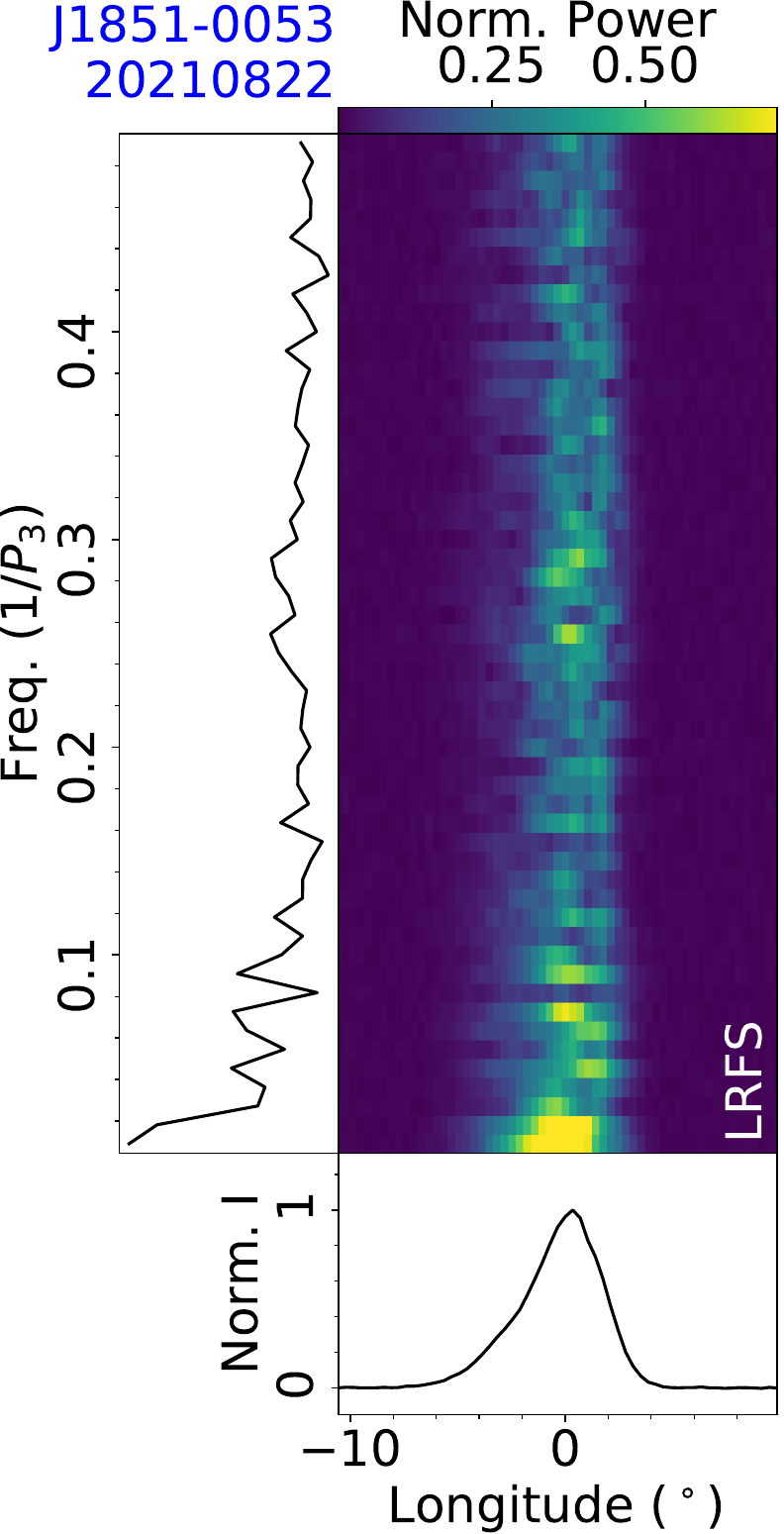}
\includegraphics[width=0.22\textwidth, angle=0]{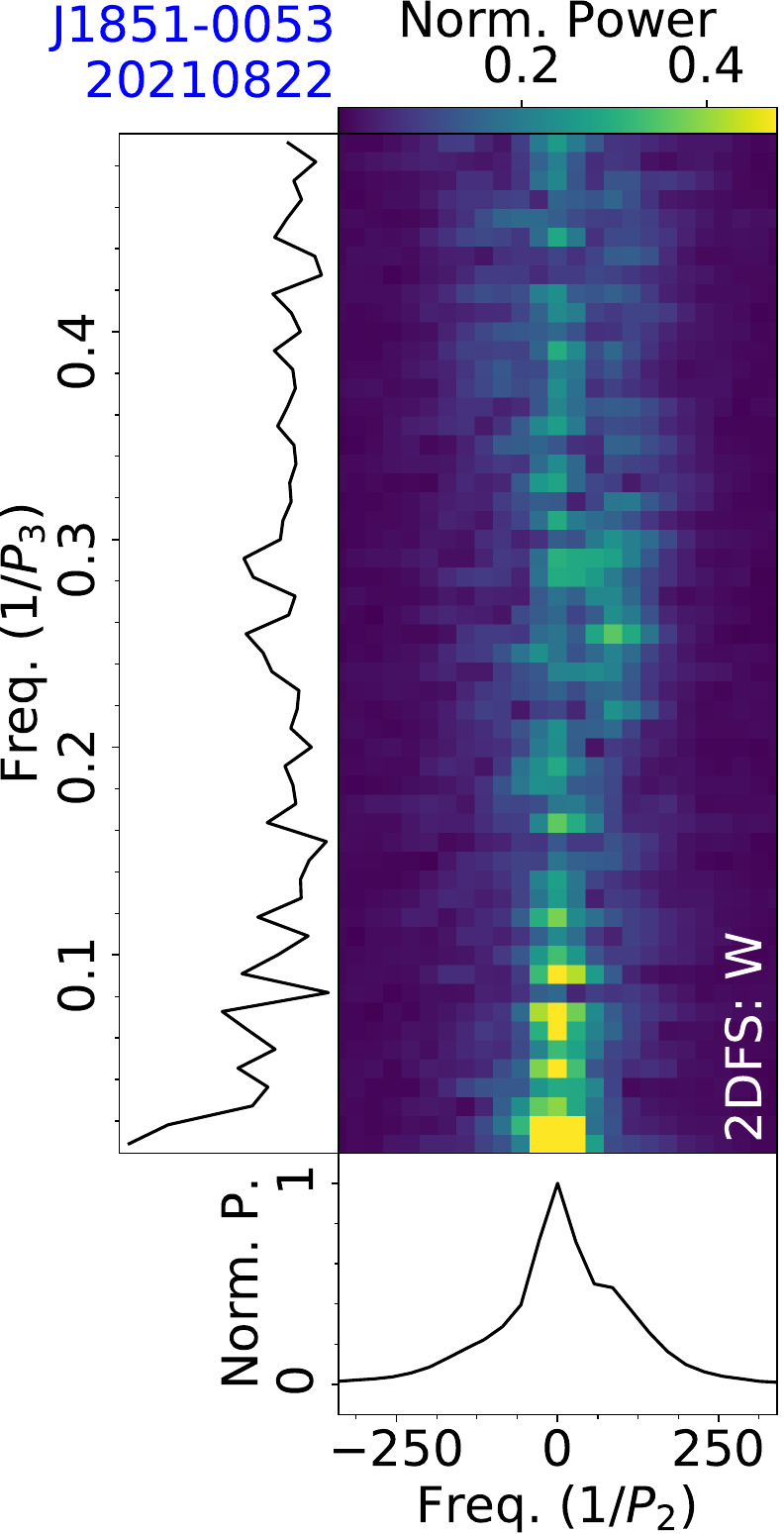}\\
\includegraphics[width=0.22\textwidth, angle=0]{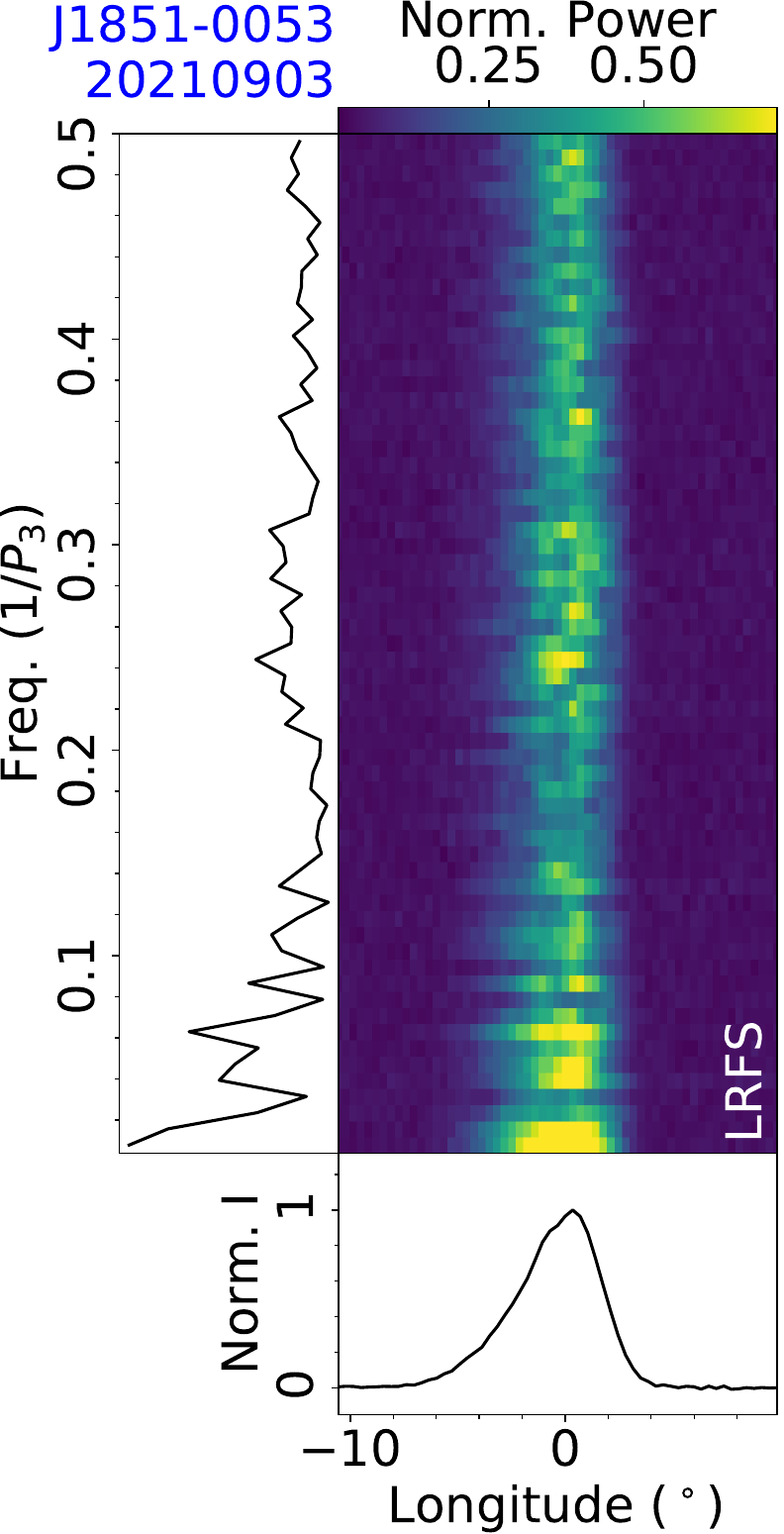}
\includegraphics[width=0.22\textwidth, angle=0]{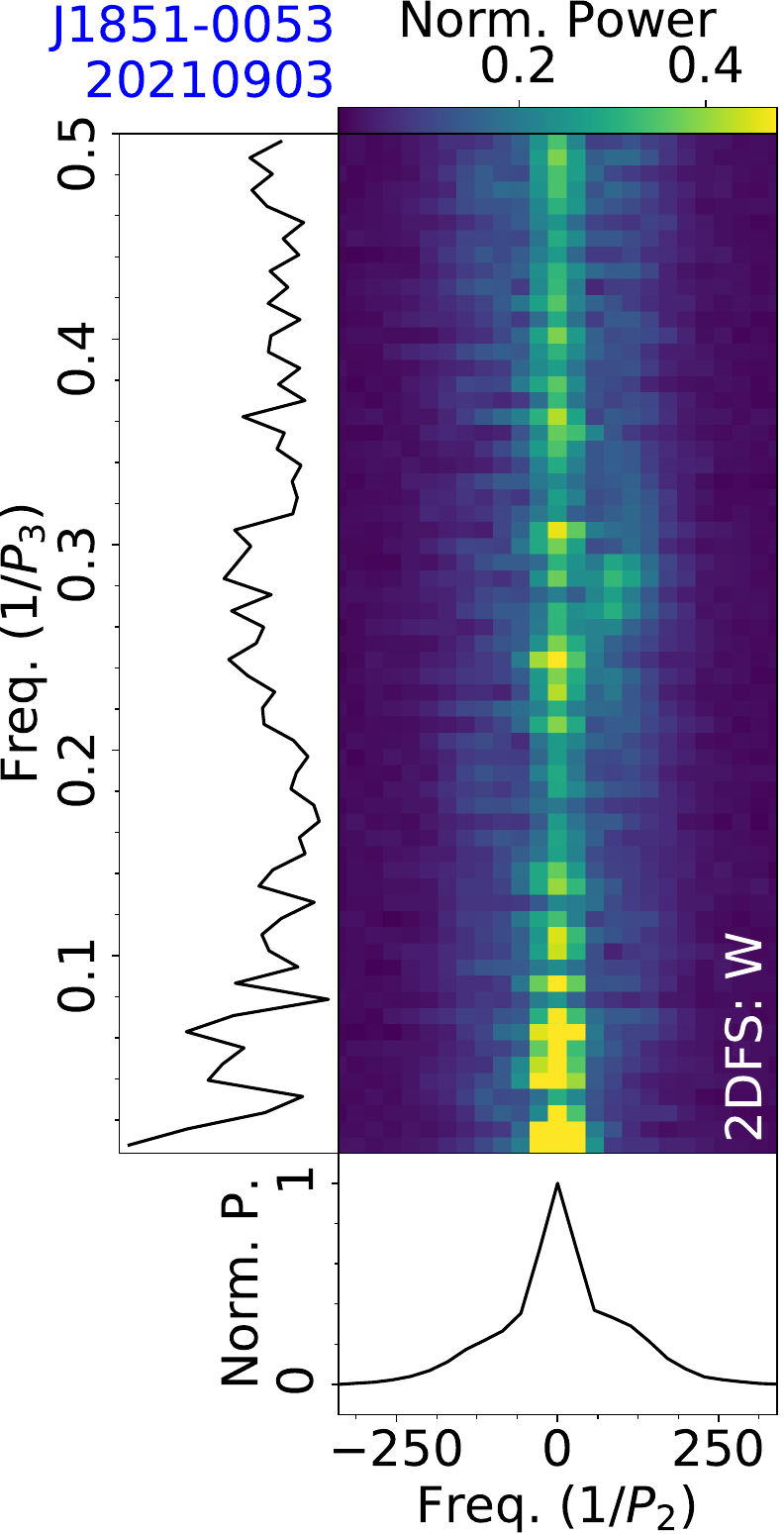}
\figcaption{Fluctuation analysis of PSR J1851-0053 from the FAST observation on 20210822 and 20210903, with LRFS and 2DFS for the on-pulse region of a mean pulse profile.
\label{subfig:fluctu:J1851-0053}}
\end{figure}

\subsection{J1850-0006}
\label{subsec:J1850-0006}

PSR J1850-0006 was discovered by \citet{Keith2009} using data of the Parkes Multi-beam Pulsar Survey. 

The pulsar was observed by FAST on 20200821 for 15 minutes, with a rotation period $P=2.1917$~s and a dispersion measure $D\!M=653.0~{\rm cm^{-3}\,pc}$ from this observation. 
The single pulse sequence shown in Fig.~\ref{subfig:TP:J1850-0006} indicates the existence of the nulling phenomenon. From the on-pulse integral energy histogram (Fig.~\ref{subfig:Hist:J1850-0006}), the nulling fraction of the observation is estimated to be 48$\pm$3\%.


\subsection{J1851+0241}
\label{subsec:J1851+0241}

PSR J1851+0241 was discovered in the Pulsar Arecibo L-band Feed Array (PALFA) survey \citep{Parent2022}. 

The pulsar was observed by FAST on 20200529 for 5 minutes, deriving a rotation period $P=4.4908$~s and a dispersion measure $D\!M=525.3~{\rm cm^{-3}\,pc}$ from this observation. 
The single pulse sequence is shown in Fig.~\ref{subfig:TP:J1851+0241}. The on-pulse integral energy histogram in Fig.~\ref{subfig:Hist:J1851+0241} shows a bimodal shape, indicating the existence of nulls. There are 24 single pulses whose on-pulse integral energies are less than 3$\sigma_{\rm E}$ in this observation. From the cross-correlation method (Fig.~\ref{subfig:Corre:J1851+0241}), the drifting parameters are estimated to be $D=0.9
\pm0.7$ degrees per period and $P_2=1.7\pm0.2^\circ$.

\subsection{J1851+0501g}
\label{subsec:J1851+0501g}

PSR J1851+0501g was discovered in the FAST GPPS survey \citep{Han2021,han2025}. 

This pulsar was observed by FAST on 20220505 for 5 minutes, deriving a rotation period $P=2.3271$~s and a dispersion measure $D\!M=340.8~{\rm cm^{-3}\,pc}$. 
The single pulse sequence of this observation is displayed in Fig.~\ref{subfig:TP:J1851+0501g}, illustrating that the pulsar has subpulse drifting behavior. For drift features in fluctuation spectra (Fig.~\ref{subfig:fluctu:J1851+0501g}), the centroid frequencies are $1/P_3=0.083\pm0.002$ ($P_3=12.0\pm0.3$ periods), and $1/P_2=52\pm9$ and $46\pm6$ ($P_2=7\pm1^\circ$ and $8\pm1^\circ$) for the leading and trailing components, respectively.

\subsection{J1851+1259}
\label{subsec:J1851+1259}

PSR was discovered by the 92-m telescope at Green Bank \citep{Stokes1985}. 
This pulsar has been reported to have the nulling phenomenon in previous studies. The nulling fraction was estimated to be 51\% and 54\% at 327 MHz by \citet{Herfindal2009} and \citet{Redman2009}, respectively.

The pulsar was observed by FAST on 20210111 for 5 minutes, yielding a rotation period $P=2.3271$~s and a dispersion measure $D\!M=340.8~{\rm cm^{-3}\,pc}$. 
Single pulse sequences of this observation, shown in Fig.~\ref{subfig:TP:J1851+1259}, display the nulling phenomenon. 
The nulling fraction is estimated from the on-pulse integral energy histogram (Fig.~\ref{subfig:Hist:J1851+1259}) to be 30$\pm$1\%. 
Besides nulls and normal emission, weak emission also seems to be present in the single-pulse stacks.

\begin{figure}[htpb]
\centering
\includegraphics[width=0.22\textwidth, angle=0]{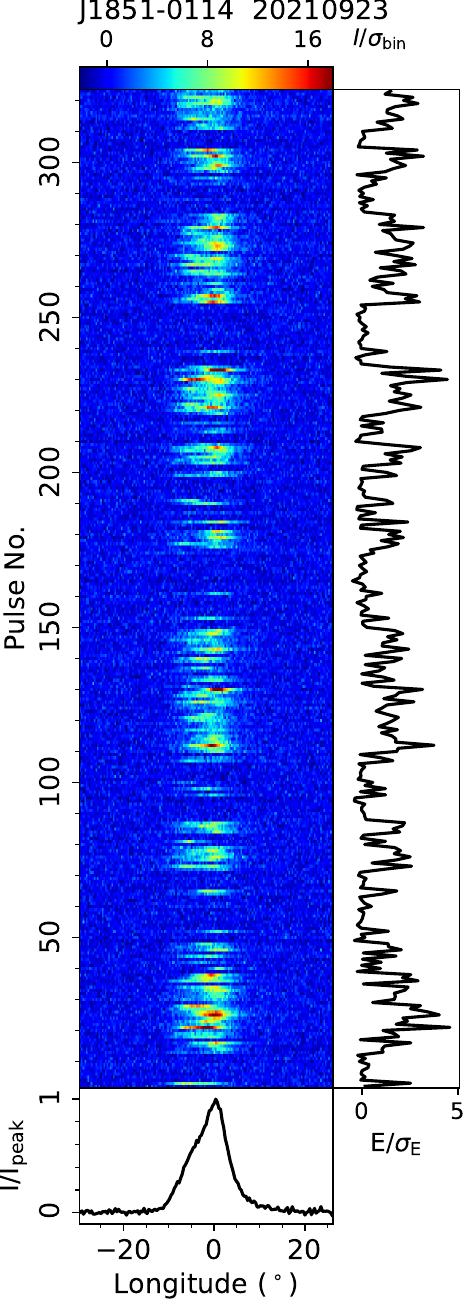}
\includegraphics[width=0.22\textwidth, angle=0]{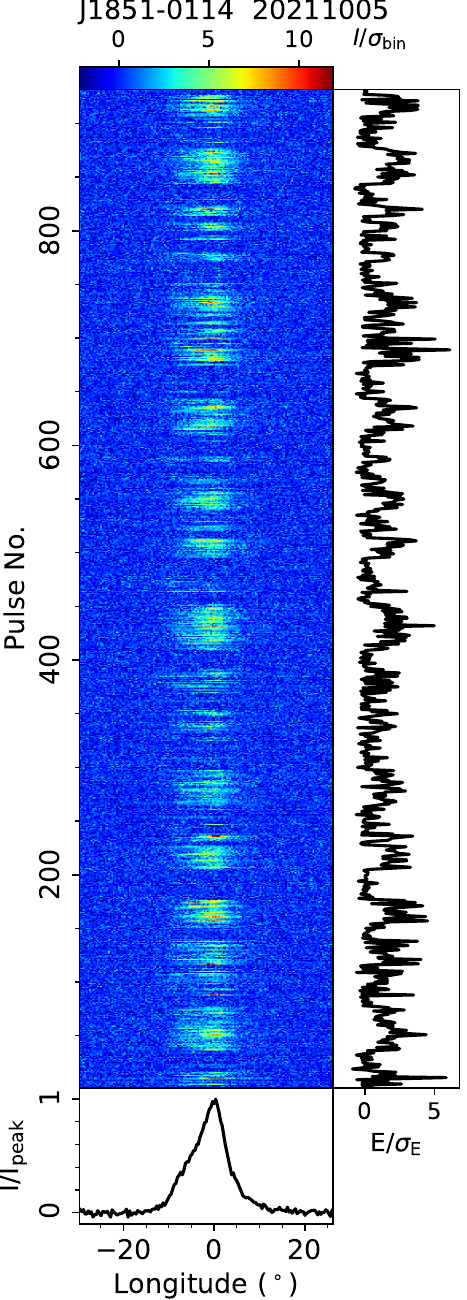}
\figcaption{Single pulse sequences of PSR J1851-0114 from the FAST observations on 20210923 and 20211005. \label{subfig:TP:J1851-0114}}
\end{figure}

\begin{figure}[htpb]
\centering
\includegraphics[width=0.39\textwidth, angle=0]{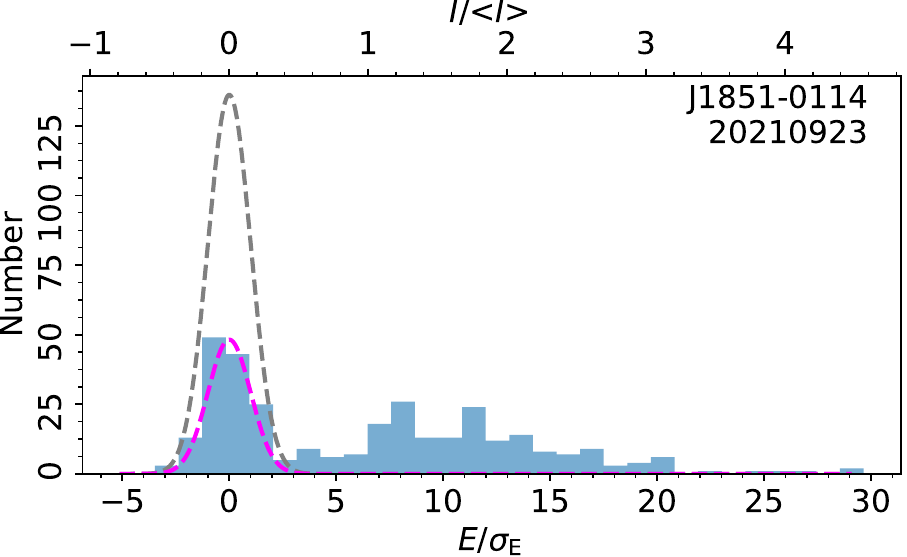}
\includegraphics[width=0.39\textwidth, angle=0]{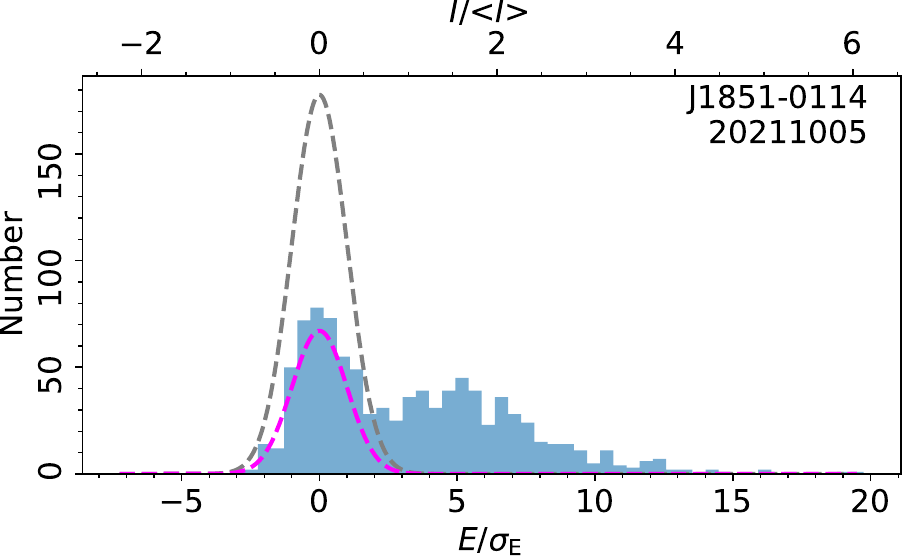}
\figcaption{On-pulse energy histograms of single pulses of PSR J1851-0114 from the FAST observations on 20210923 and 20211005.
\label{subfig:Hist:J1851-0114}}
\end{figure}

\begin{figure}[htpb]
\centering
\includegraphics[width=0.22\textwidth, angle=0]{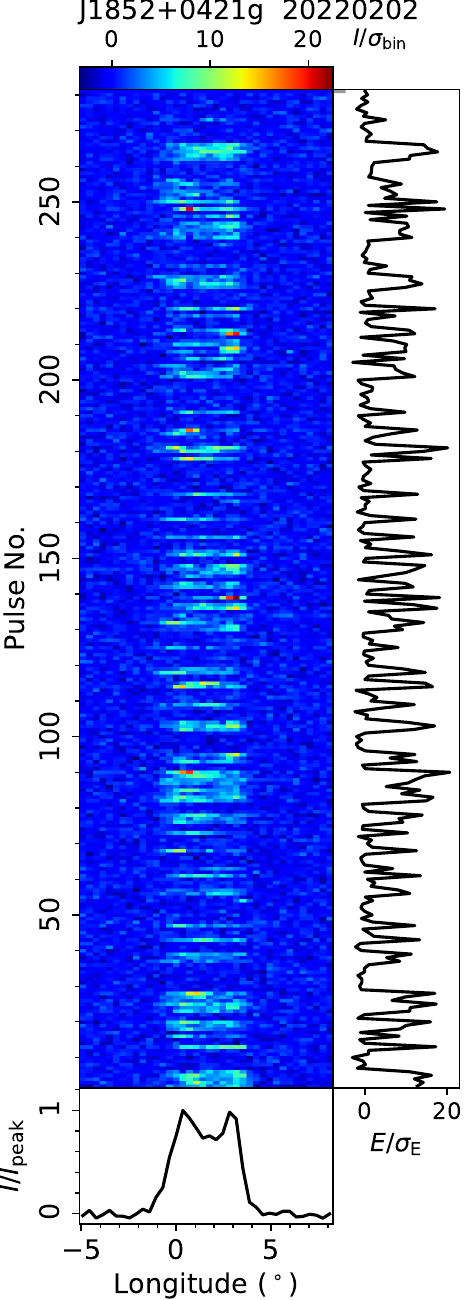}
\figcaption{Single pulse sequence of PSR J1852+0421g from the FAST observation on 20220202.
\label{subfig:TP:J1852+0421g}}
\end{figure}

\begin{figure}[htpb]
\centering
\includegraphics[width=0.39\textwidth, angle=0]{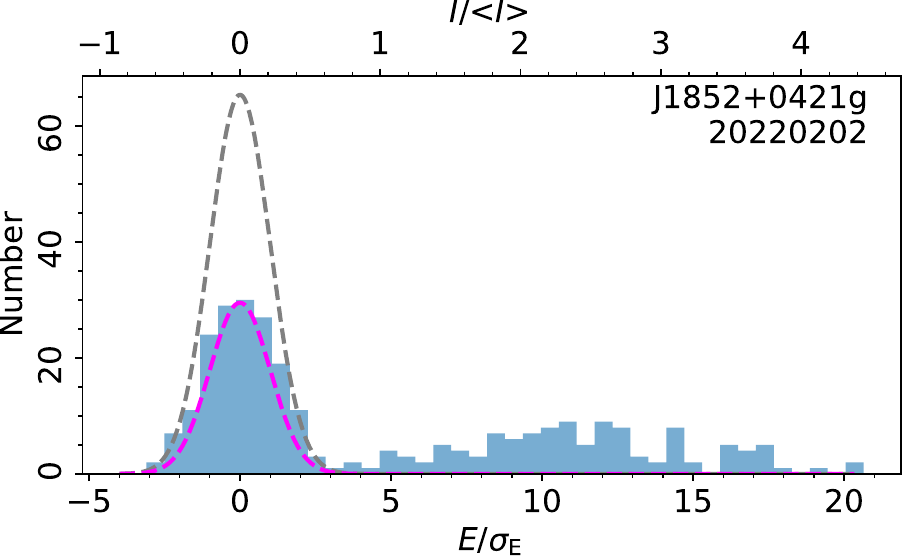}
\figcaption{On-pulse energy histogram of single pulses of PSR J1852+0421g from the FAST observation on 20220202.
\label{subfig:Hist:J1852+0421g}}
\end{figure}

\begin{figure}[htpb]
\centering
\includegraphics[width=0.22\textwidth, angle=0]{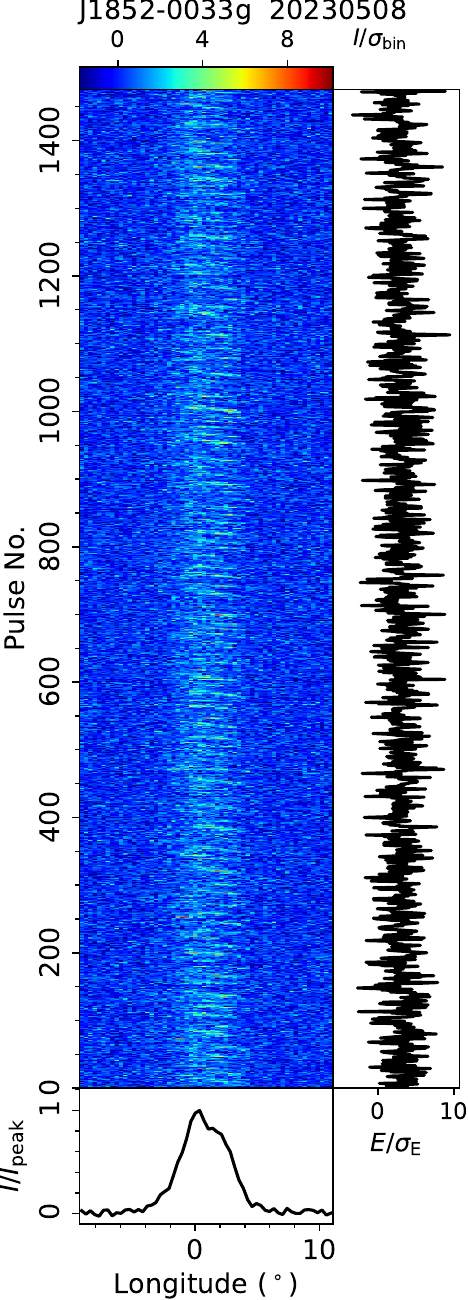}
\includegraphics[width=0.22\textwidth, angle=0]{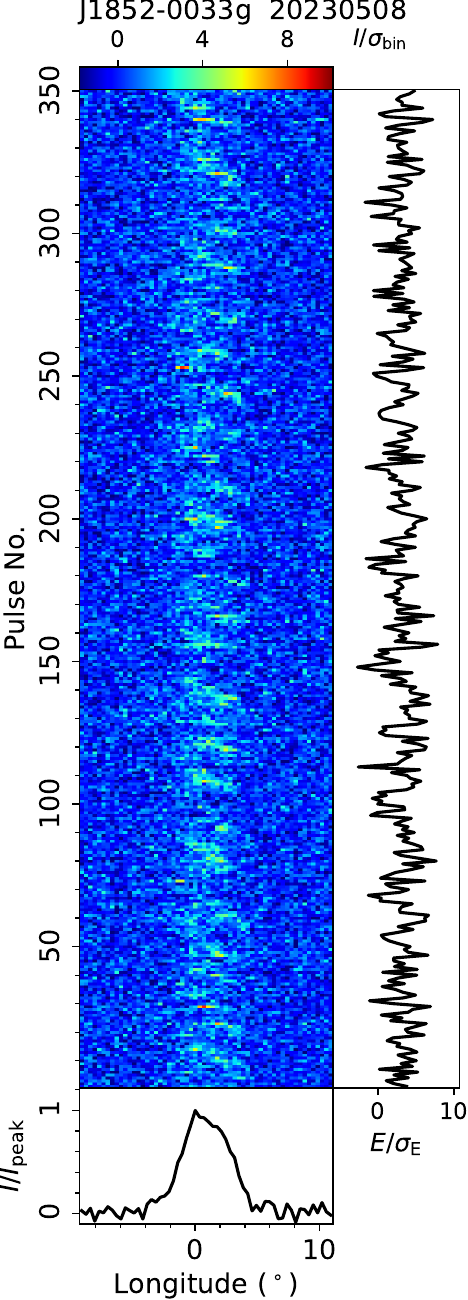}
\figcaption{Single pulse sequence of PSR J1852-0033g from the FAST observation on 20230508, and a zoomed-in view of pulses No. 1-350.
\label{subfig:TP:J1852-0033g}}
\end{figure}

\begin{figure}[htpb]
\centering
\includegraphics[width=0.22\textwidth, angle=0]{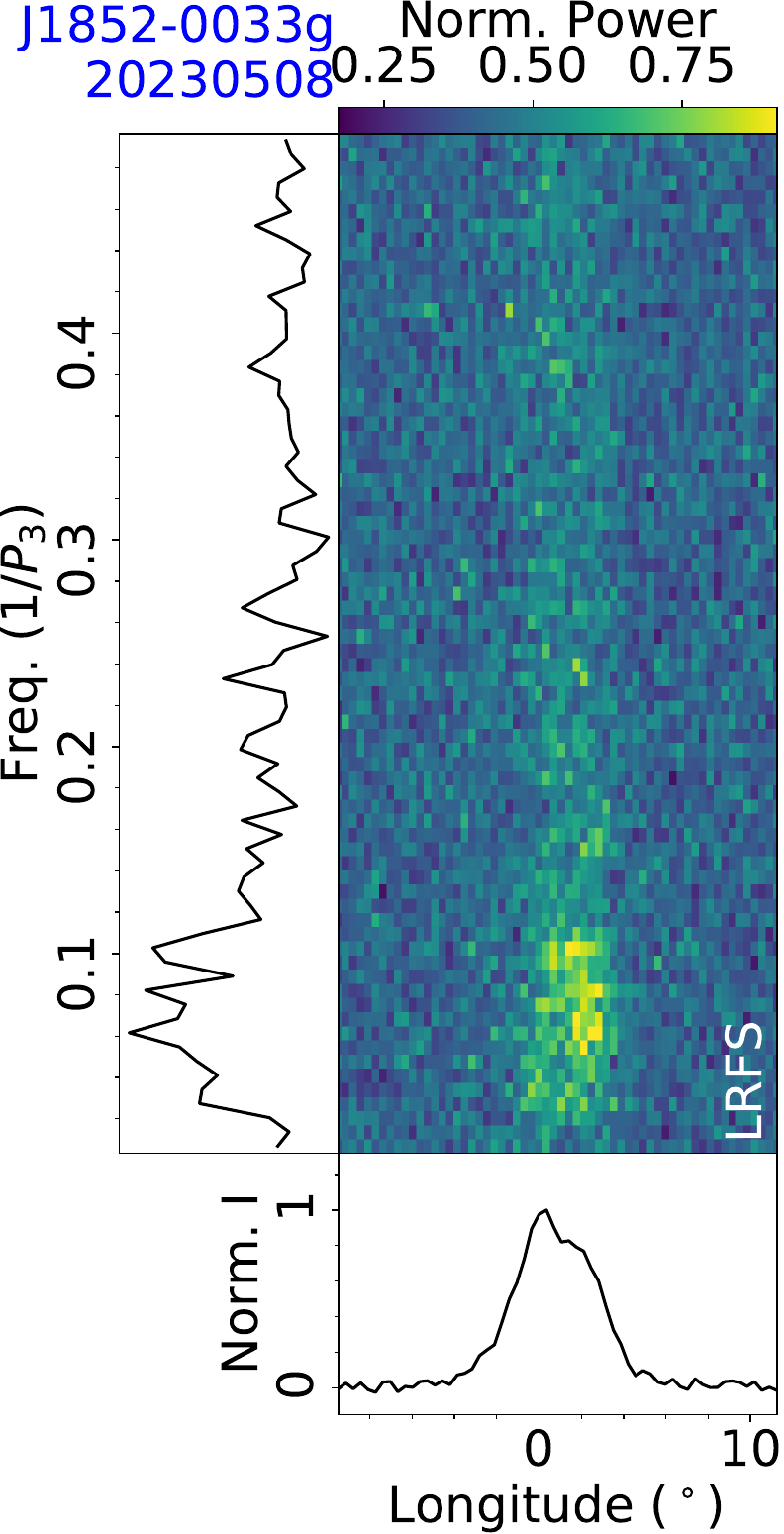}
\includegraphics[width=0.22\textwidth, angle=0]{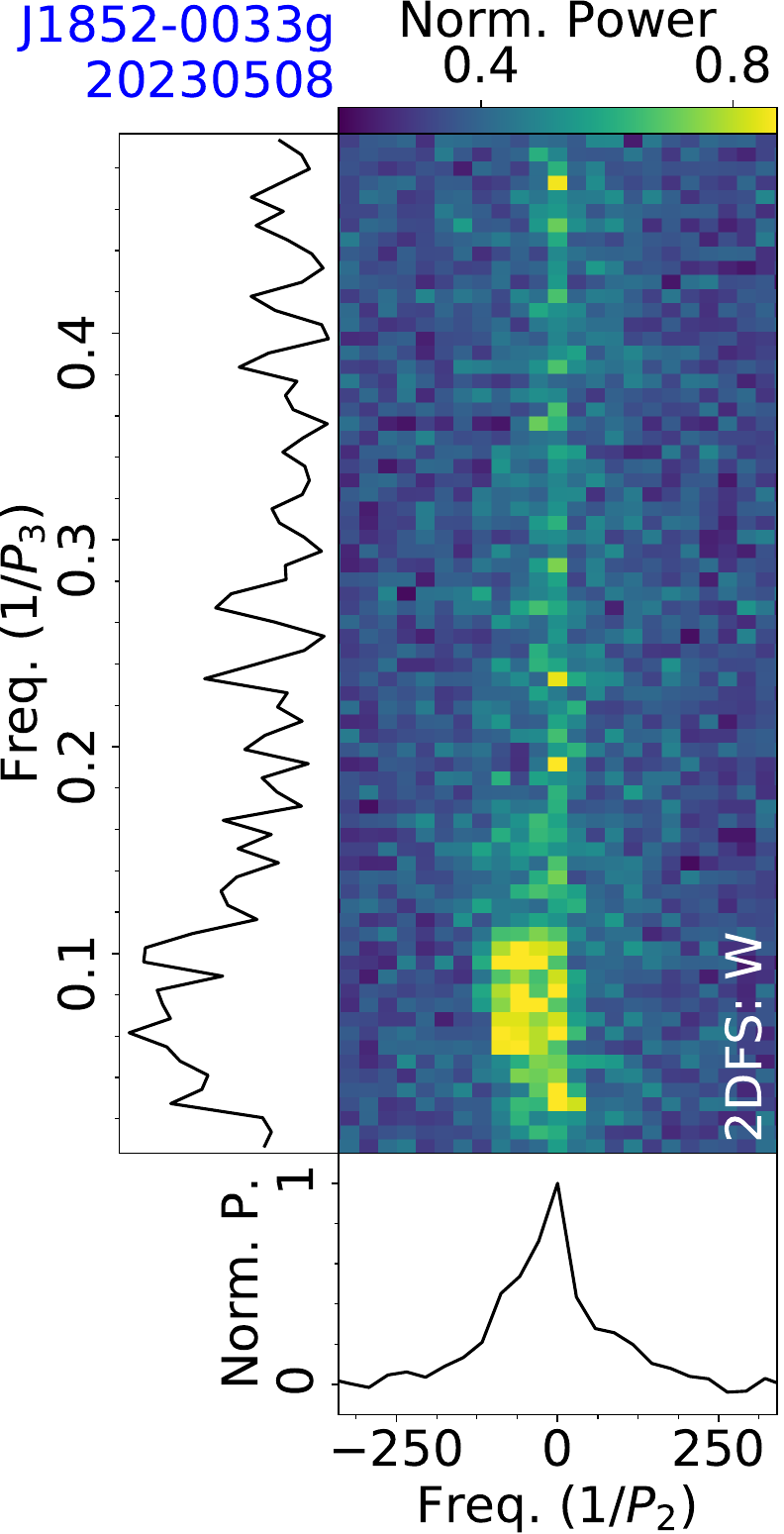}
\figcaption{Fluctuation analysis of PSR J1852-0033g from the FAST observation on 20230508, with LRFS and 2DFS for the on-pulse region of a mean pulse profile.
\label{subfig:fluctu:J1852-0033g}}
\end{figure}

\subsection{J1851-0029}
\label{subsec:J1851-0029}

PSR J1851-0029 was discovered by \citet{Keith2009} using data of the Parkes Multi-beam Pulsar Survey. $P_3$-only feature with the modulation periodicity of 10.0$\pm$0.5 has been reported by \citet{Song2023}.

This pulsar was observed by FAST on 20220121 for 15 minutes and 20230508 for 34 minutes. From the longer data, a rotation period and a dispersion measure are derived to be $P=0.5187$~s and a dispersion measure $D\!M=521.5~{\rm cm^{-3}\,pc}$. The single pulse sequence and a zoomed-in view of the observation on 20230508 are shown in Fig.~\ref{subfig:TP:J1851-0029}, displaying the modulation behavior. From the fluctuation spectra in Fig.~\ref{subfig:fluctu:J1851-0029}, the centroid modulation frequency is estimated to be $1/P_3=0.087\pm0.001$, corresponding to $P_3=11.5\pm0.1$ periods.

\begin{figure}[htpb]
\centering
\includegraphics[width=0.21\textwidth, angle=0]{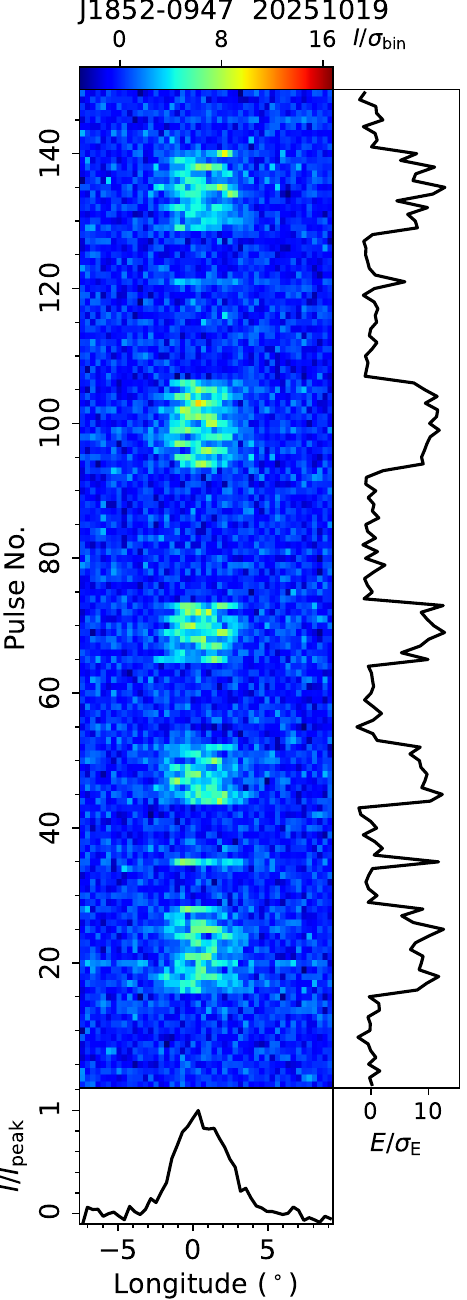}
\includegraphics[width=0.21\textwidth, angle=0]{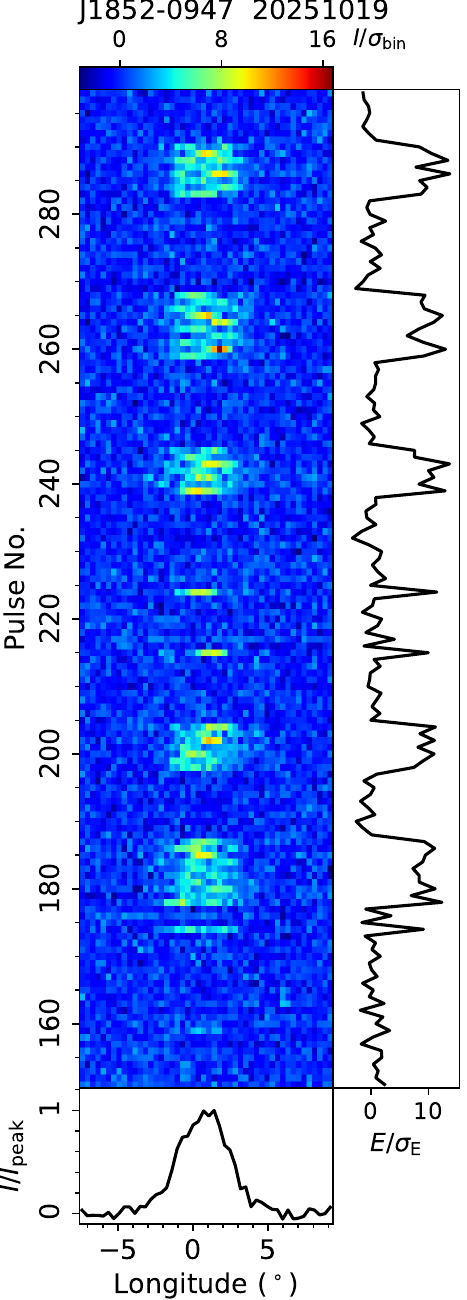}
\figcaption{Single pulse sequences of PSR J1852-0947 from the FAST observation on 20251019.
\label{subfig:TP:J1852-0947}}
\end{figure}

\begin{figure}[htpb]
\centering
\includegraphics[width=0.39\textwidth, angle=0]{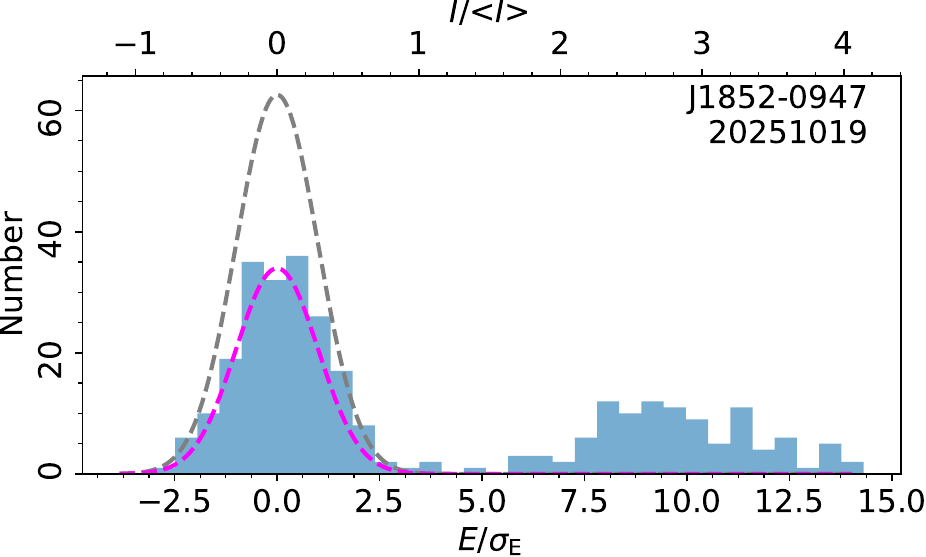}
\figcaption{On-pulse energy histogram of single pulses of PSR J1852-0947 from the FAST observation on 20251019.
\label{subfig:Hist:J1852-0947}}
\end{figure}

\begin{figure}[htpb]
\centering
\includegraphics[width=0.22\textwidth, angle=0]{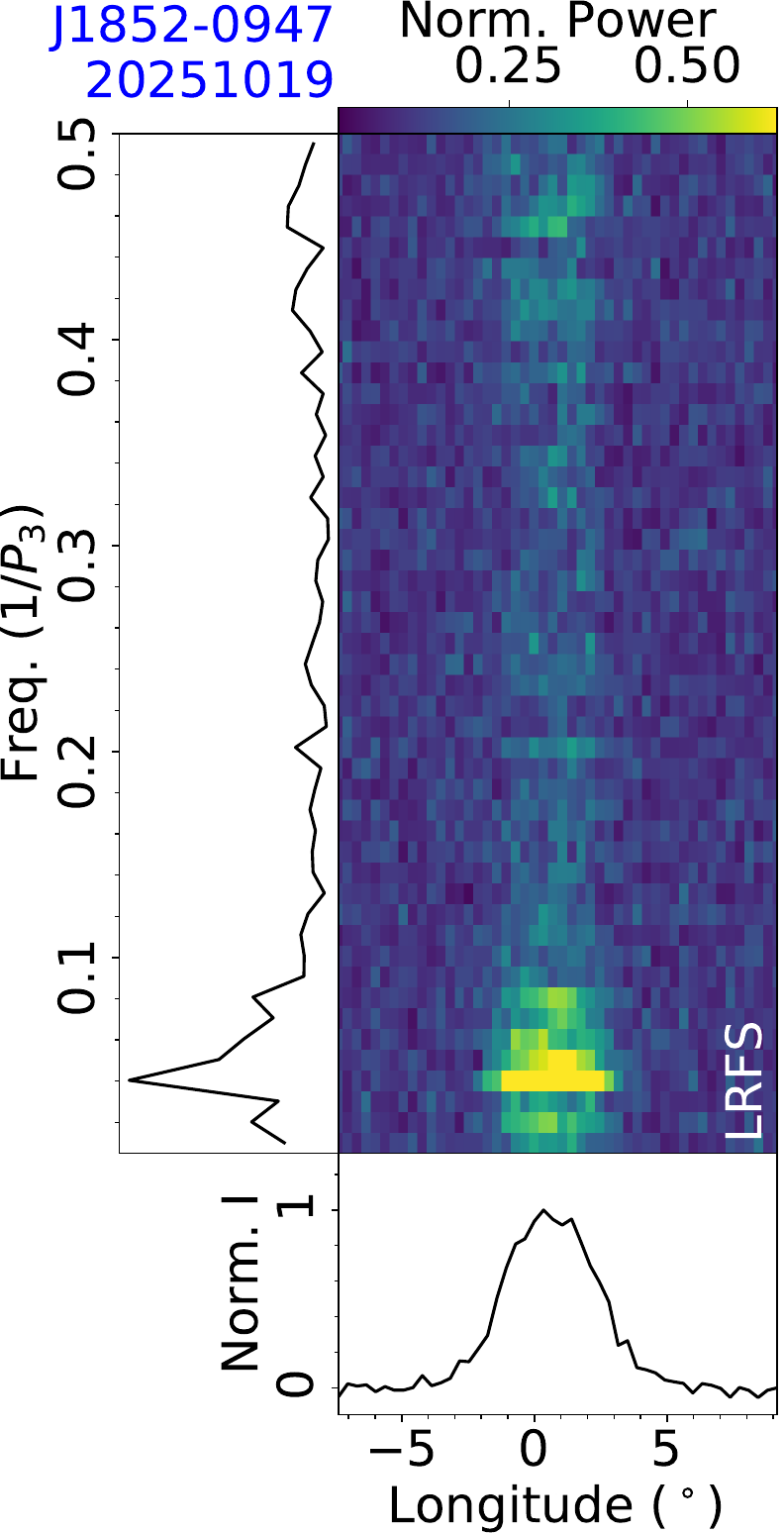}
\includegraphics[width=0.22\textwidth, angle=0]{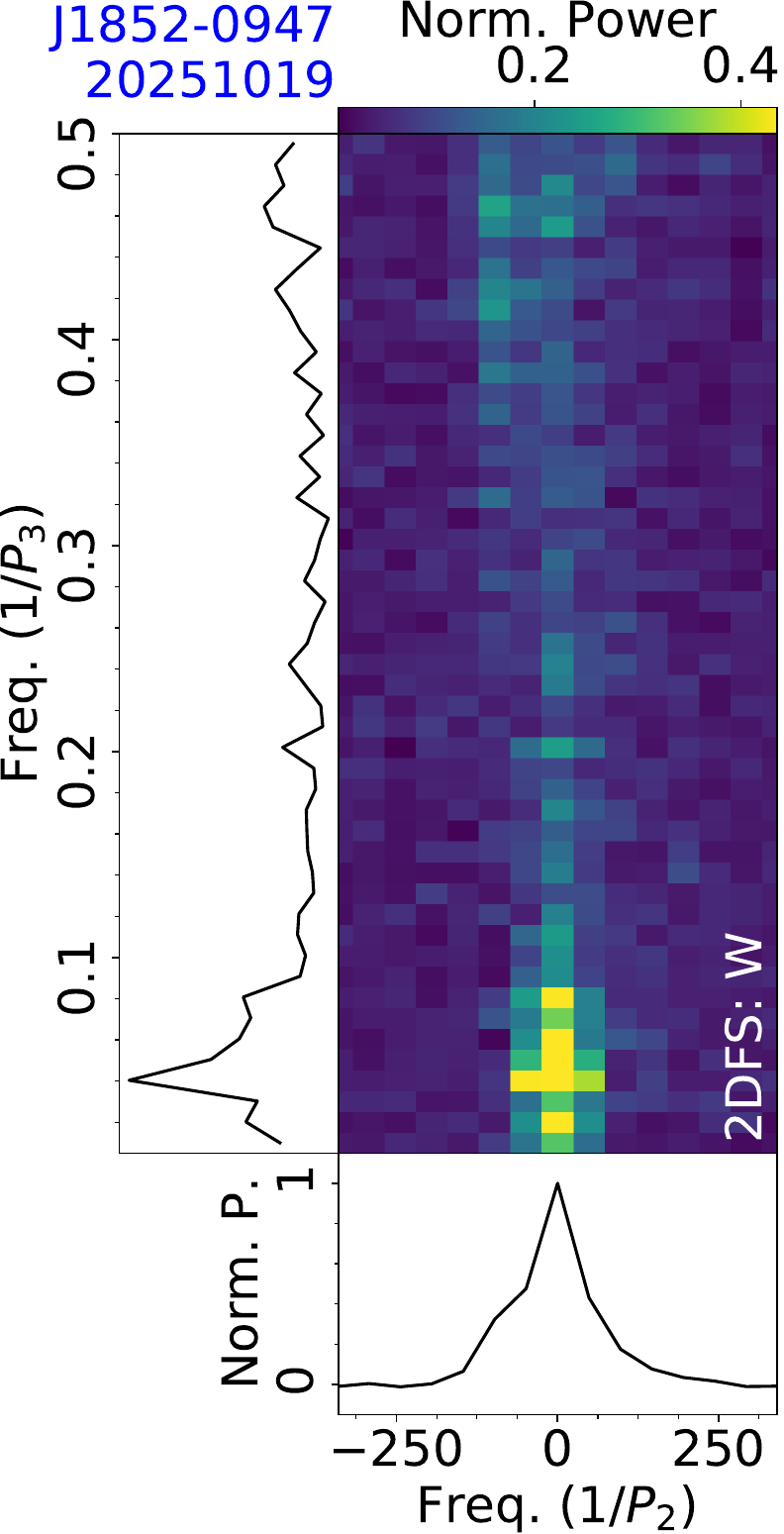}
\figcaption{Fluctuation analysis of PSR J1852-0947 from the FAST observation on 20251019, with LRFS and 2DFS for the on-pulse region of a mean pulse profile.
\label{subfig:fluctu:J1852-0947}}
\end{figure}

\subsection{J1851-0053}
\label{subsec:J1851-0053}

PSR J1851-0053 was discovered in the Parkes multibeam pulsar survey \citep{hfs+04}. \citep{Song2023} reported the drifting behavior of $P_3=3.5(3)$ periods and $P_2=3.70^{+13}_{-0.08}$ degrees.

The pulsar was observed by FAST for 5 minutes on 20210822 and for 15 minutes each on 20210903, 20211009, and 20250331. From the observation on 20210822, a rotation period and a dispersion measure are determined to be $P=1.4092$~s and a dispersion measure $D\!M=24.5~{\rm cm^{-3}\,pc}$. 
Single pulse sequences of two observations on 20210822 and 20210903 are shown in Fig.~\ref{subfig:TP:J1851-0053}, illustrating the existence of nulling and subpulse drifting phenomena. The pulsar also has emission with a duration of only a few periods. From the on-pulse integral energy histograms (Fig.~\ref{subfig:Hist:J1851-0053}), the nulling fractions of these two observations are 47$\pm$4\% and 42$\pm$4\%. In single pulse sequences, the subpulse drifting bands are not systematic, such as pulses Nos. 75-130 and 240-275. The 2DFS (Fig.~\ref{subfig:fluctu:J1851-0053}) exhibits drift features with centroids at $1/P_3=0.267\pm0.002$ ($P_3=3.74\pm0.02$ periods) and $1/P_2=83\pm2$ ($P_2=4.3\pm0.1^\circ$) on 20210822 and $1/P_3=0.271\pm0.002$ ($P_3=3.69\pm0.03$ periods) and $1/P_2=84\pm2$ ($P_2=4.3\pm0.1^\circ$) on 20210903. Single-pulse features are consistent across observations. Longer observations are necessary for detailed drifting analysis.

\begin{figure}[htpb]
\centering
\includegraphics[width=0.42\textwidth, angle=0]{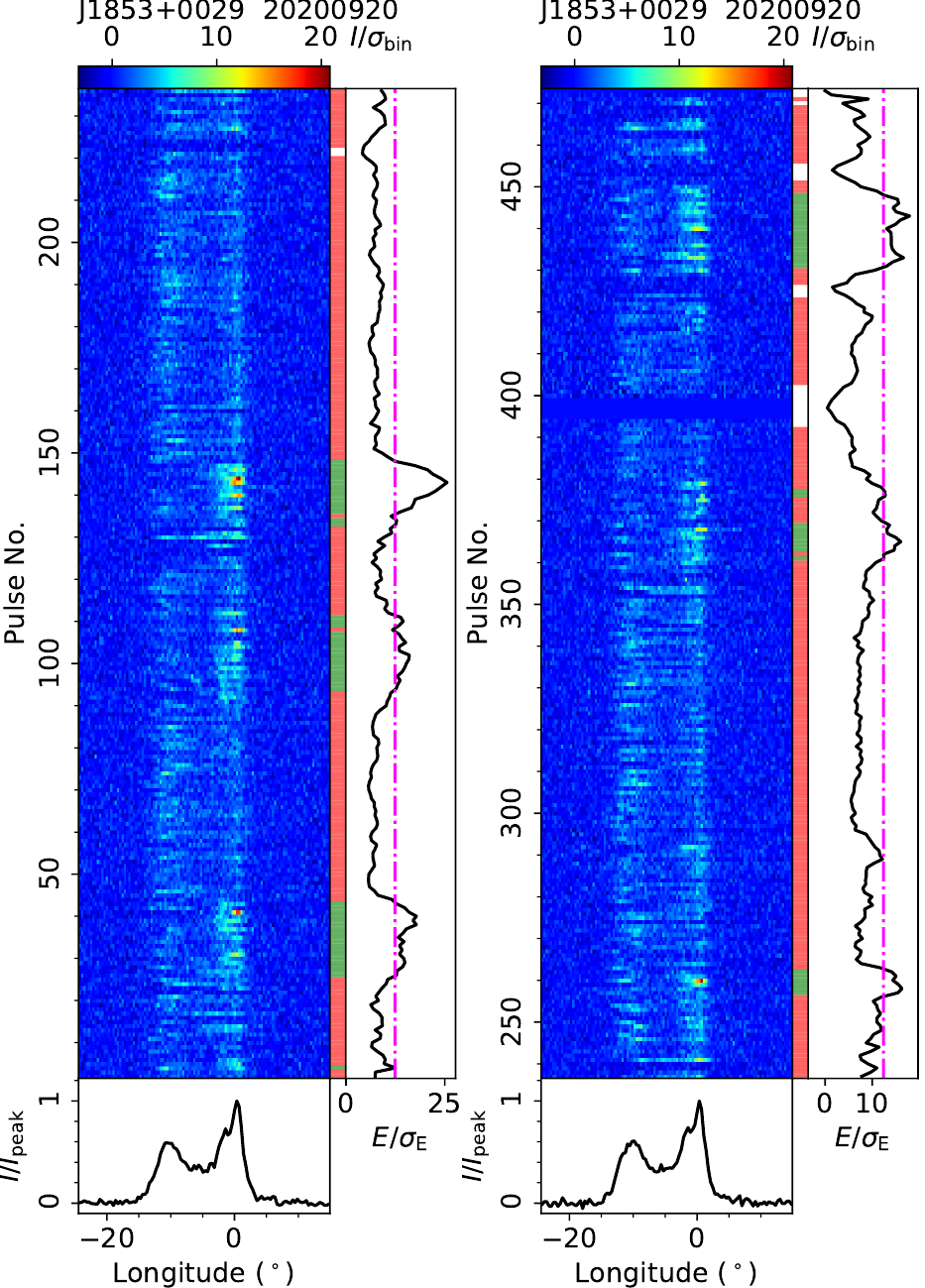}
\figcaption{Single pulse sequences of PSR J1853+0029 from the FAST observation on 20200920. In the right subpanel, the energy variation of the trailing profile part is smoothed over every 7 periods.
\label{subfig:TP:J1853+0029}}
\end{figure}

\begin{figure}[htpb]
\centering
\includegraphics[width=0.39\textwidth, angle=0]{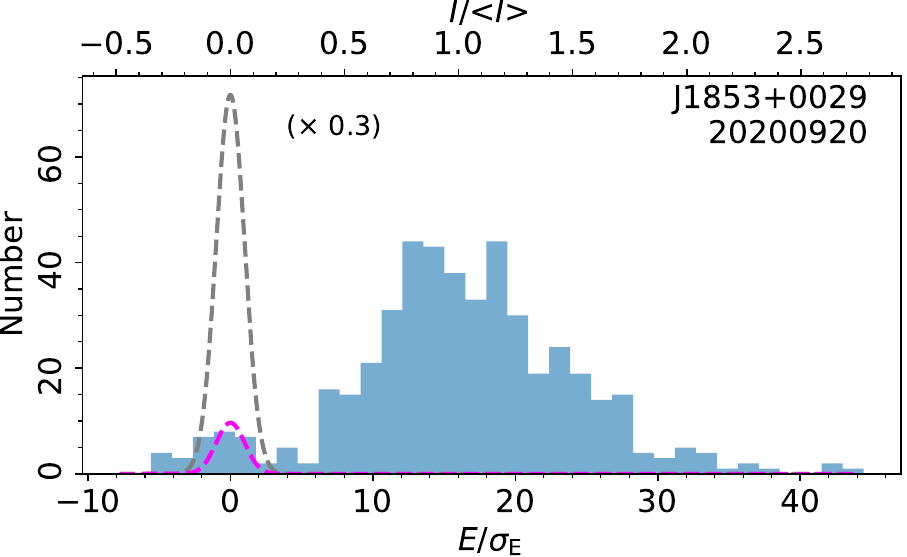}
\figcaption{On-pulse energy histogram of single pulses of PSR J1853+0029 from the FAST observation on 20200920.
\label{subfig:Hist:J1853+0029}}
\end{figure}

\begin{figure}[htpb]
\centering
\includegraphics[width=0.39\textwidth, angle=0]{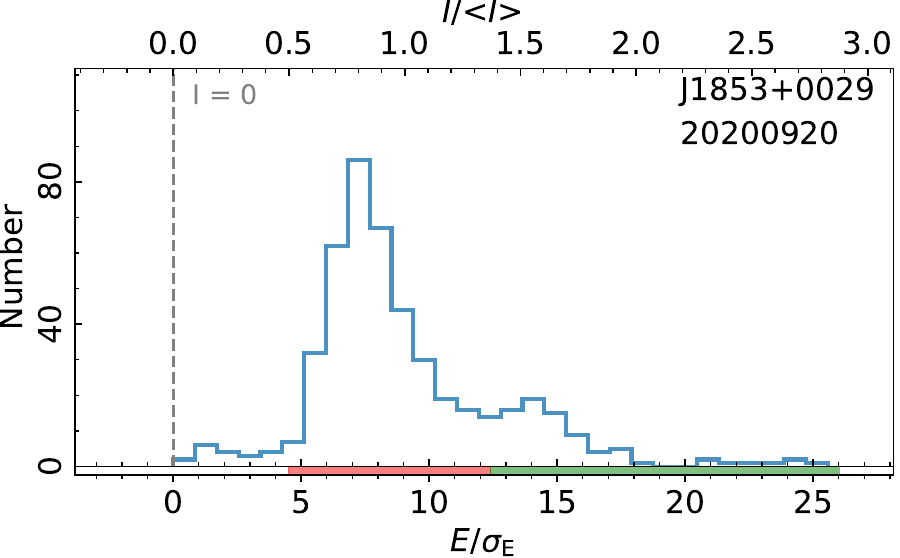}
\figcaption{Energy histogram of single pulses of PSR J1853+0029 from the FAST observation on 20200920. The energy values are integrated over the trailing part of the pulse profile and smoothed over 7 periods. 
\label{subfig:HistModes:J1853+0029}}
\end{figure}

\begin{figure}[htpb]
\centering
\includegraphics[width=0.39\textwidth, angle=0]{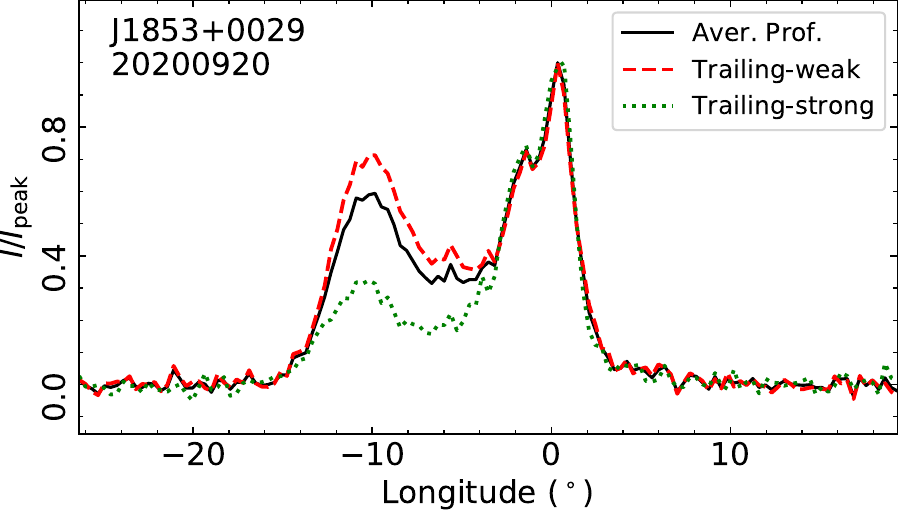}
\figcaption{Mean profiles for the trailing-weak and trailing-strong emission modes of PSR J1853+0029 from the FAST observation on 20200920, which are normalized by their respective peaks.
\label{subfig:ProfModes:J1853+0029}}
\end{figure}

\subsection{J1851-0114}
\label{subsec:J1851-0114}

PSR J1851-0114 was discovered by \citet{Lorimer2006} in the Parkes Multibeam Pulsar Survey. 

This pulsar was observed by FAST on 20210923 for 5 minutes and 20211005 for 15 minutes. From the observation on 20210923, a rotation period and a dispersion measure are determined to be $P=0.9532$~s and a dispersion measure $D\!M=432.8~{\rm cm^{-3}\,pc}$. 
Single pulses sequences of these two observations are displayed in Fig.~\ref{subfig:TP:J1851-0114}. The nulling fractions of two observations are estimated to be 35$\pm$4\% and 38$\pm$3\% from energy histograms in Fig.~\ref{subfig:Hist:J1851-0114}.




\subsection{J1852+0421g}
\label{subsec:J1852+0421g}

PSR J1852+0421g was discovered in the FAST GPPS survey \citep{Han2021,han2025}. 

The pulsar was observed by FAST observation on 20220202 for 15 minutes, yielding a rotation period $P=3.1612$~s and a dispersion measure $D\!M=300.3~{\rm cm^{-3}\,pc}$. 
The single pulse sequence of this observation is displayed in Fig.~\ref{subfig:Hist:J1852+0421g}, and the nulling fraction is estimated to be 45$\pm$4\% from the energy histogram (Fig.~\ref{subfig:Hist:J1852+0421g}).

\begin{figure}[htpb]
\centering
\includegraphics[width=0.22\textwidth, angle=0]{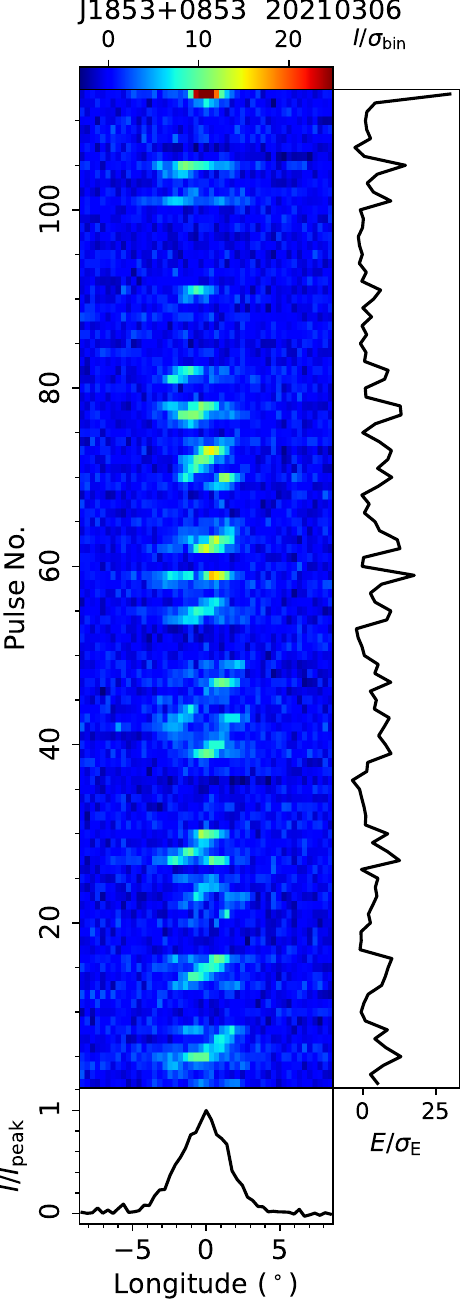}
\includegraphics[width=0.22\textwidth, angle=0]{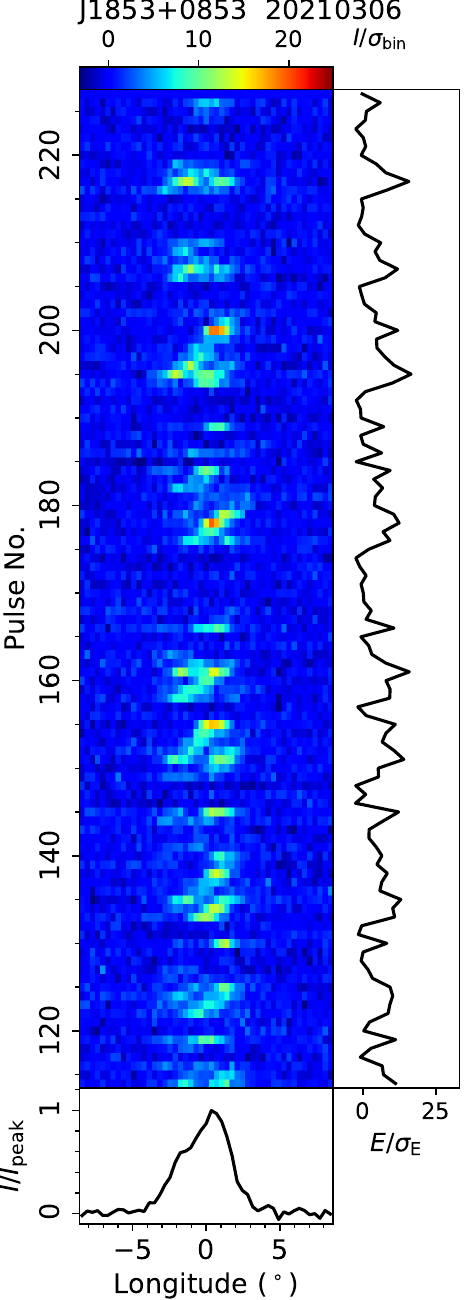}
\figcaption{Single pulse sequences of PSR J1853+0853 from the FAST observation on 20210306.
\label{subfig:TP:J1853+0853}}
\end{figure}

\begin{figure}[htpb]
\centering
\includegraphics[width=0.39\textwidth, angle=0]{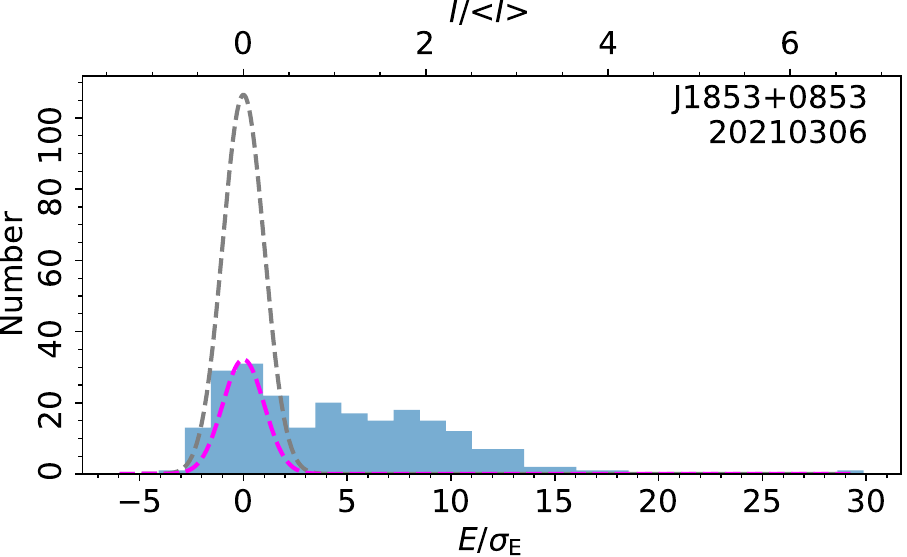}
\figcaption{On-pulse energy histogram of single pulses of PSR J1853+0853 from the FAST observation on 20210306.
\label{subfig:Hist:J1853+0853}}
\end{figure}

\begin{figure}[htpb]
\centering
\includegraphics[width=0.39\textwidth, angle=0]{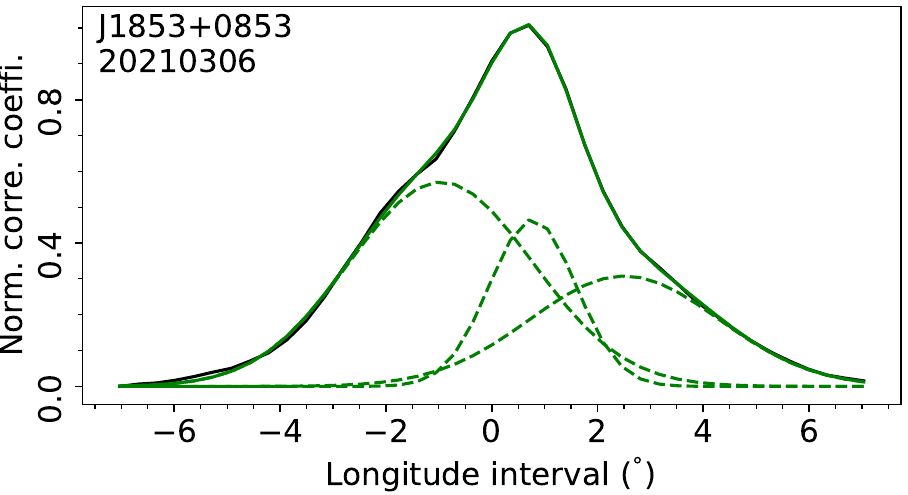}
\figcaption{Cross correlation of PSR J1853+0853 from the FAST observation on 20210306.
\label{subfig:Corre:J1853+0853}}
\end{figure}

\subsection{J1852-0033g}
\label{subsec:J1852-0033g}

PSR J1852-0033g was discovered in the FAST GPPS survey \citep{Han2021,han2025}. 

The pulsar was observed by FAST on 20230508 for 34 minutes and 20240316 for 30 minutes. From the 34-minute data, a rotation period $P=1.3689$~s and a dispersion measure $D\!M=321.3~{\rm cm^{-3}\,pc}$ were determined. 
The single-pulse behavior of these two observations is consistent. Single pulse sequences shown in Fig.~\ref{subfig:TP:J1852-0033g} display the negative subpulse drifting phenomenon. From the fluctuation in Fig.~\ref{subfig:fluctu:J1852-0033g}, the centroid frequencies of the drift feature are $1/P_3=0.073\pm0.001$ and $1/P_2=-48\pm2$, corresponding to periodicities of $P_3=13.8\pm0.2$ periods and $P_2=-7.5\pm0.3^\circ$.





\subsection{J1852-0947}
\label{subsec:J1852-0947}

PSR J1852-0947 was discovered by FAST in the Commensal Radio Astronomy FAST Survey, exhibiting the nulling phenomenon \citep{Wu2023}.

This pulsar was observed by FAST on 20230528 for 1.7 hours and 20251019 for 15 minutes. From the observation with a longer duration, the rotation period and a dispersion measure are derived to be $P=3.0195$~s and a dispersion measure $D\!M=236.5~{\rm cm^{-3}\,pc}$. 
Single pulse sequences of the observation on 20251019 are shown in Fig.~\ref{subfig:TP:J1852-0947}, illustrating the existence of nulling and subpulse drifting behaviors. The emission could last for only one period, such as pulse Nos. 215 and 224. From the on-pulse integral energy histogram, the nulling fraction of this data is estimated to be 54.2$\pm$3.0\%. LRFS and 2DFS in Fig.~\ref{subfig:fluctu:J1852-0947} display two modulation features. The low-frequency modulation feature with the centroid frequency of $1/P_3=0.046\pm0.001$ is caused by the nulling behavior, yielding $P_3=21.7\pm$ periods. The negative drift feature is characterized by the centroid frequencies of $1/P_3=0.435\pm0.003$ and $1/P_2=-89\pm3$, which correspond to the modulation frequencies of $P_3=2.30\pm0.02$ periods and $P_2=-4.0\pm0.2$ degrees.

\begin{figure}[htpb]
\centering
\includegraphics[width=0.21\textwidth, angle=0]{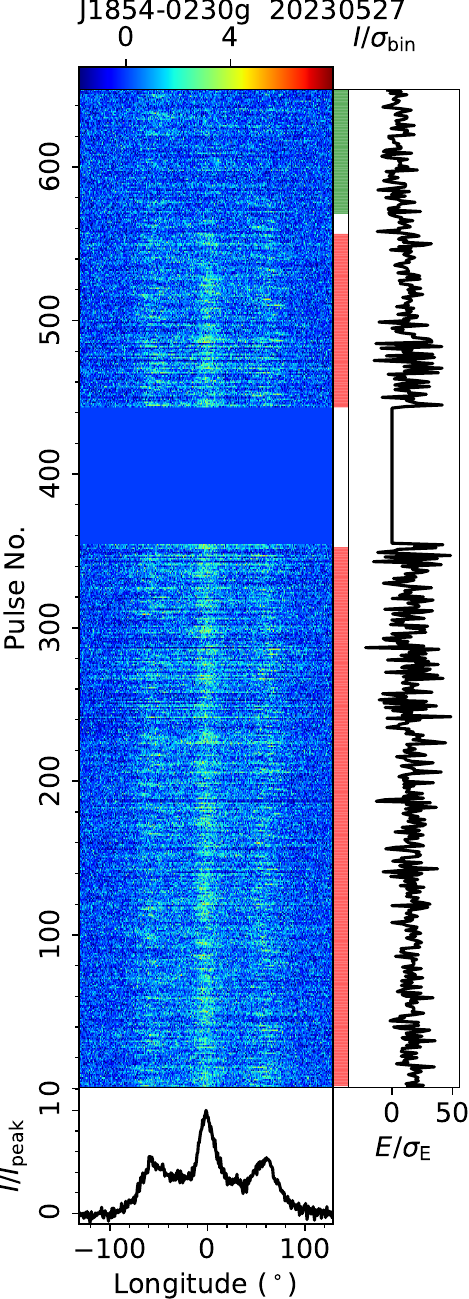}
\includegraphics[width=0.21\textwidth, angle=0]{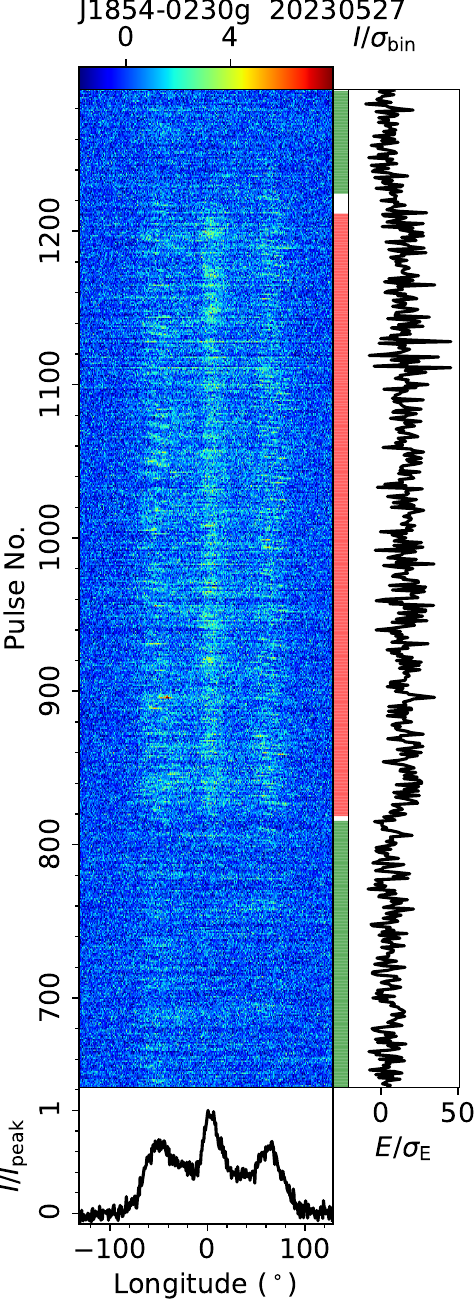}
\figcaption{Single pulse sequences of PSR J1854-0230g from the FAST observation on 20230527.
\label{subfig:TP:J1854-0230g}}
\end{figure}

\begin{figure}[htpb]
\centering
\includegraphics[width=0.39\textwidth, angle=0]{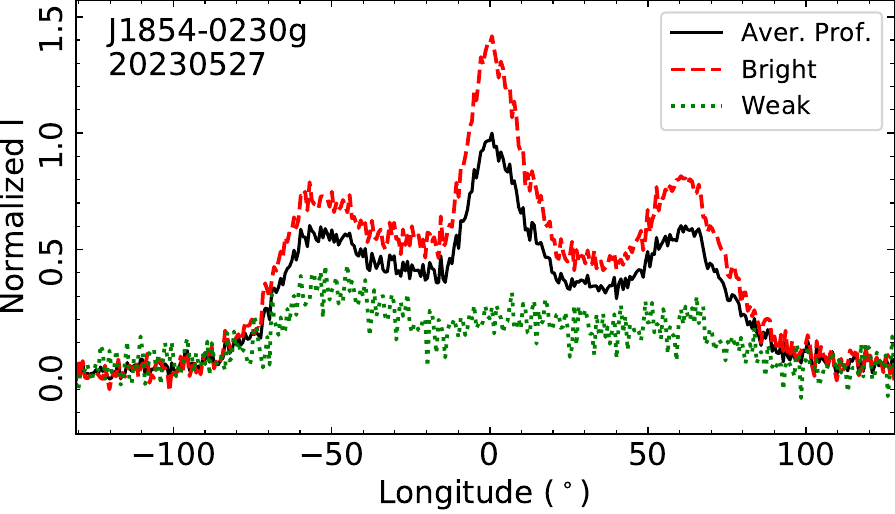}
\figcaption{Mean profiles of the weak (blue dotted) and bright (red dashed) emission modes for PSR J1854-0230g from the FAST observation on 20230527. Profiles of modes are normalized by the peak of the mean profile (black solid) of all periods.
\label{subfig:PolModes:J1854-0230g}}
\end{figure}

\begin{figure}[htpb]
\centering
\includegraphics[width=0.21\textwidth, angle=0]{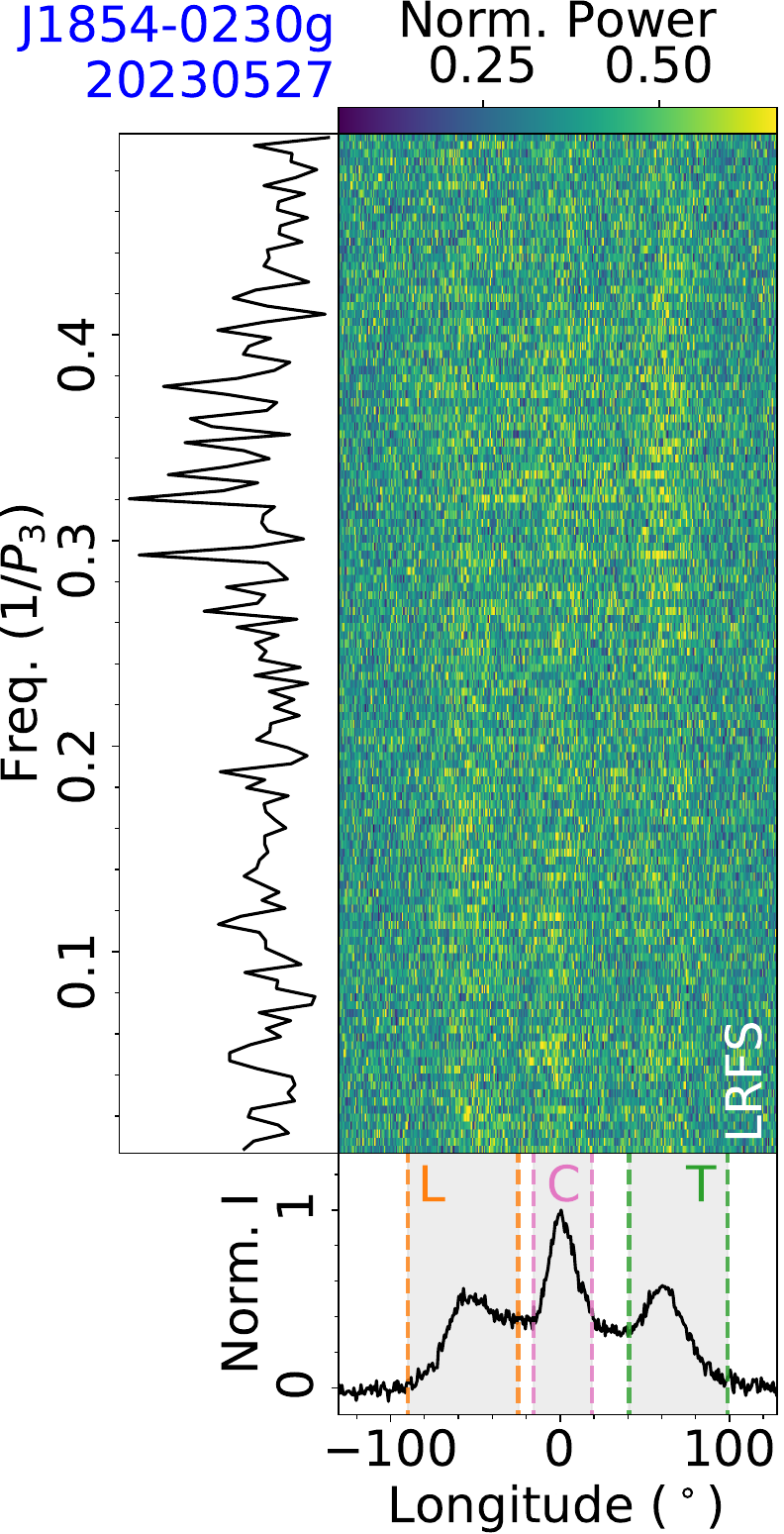}
\includegraphics[width=0.21\textwidth, angle=0]{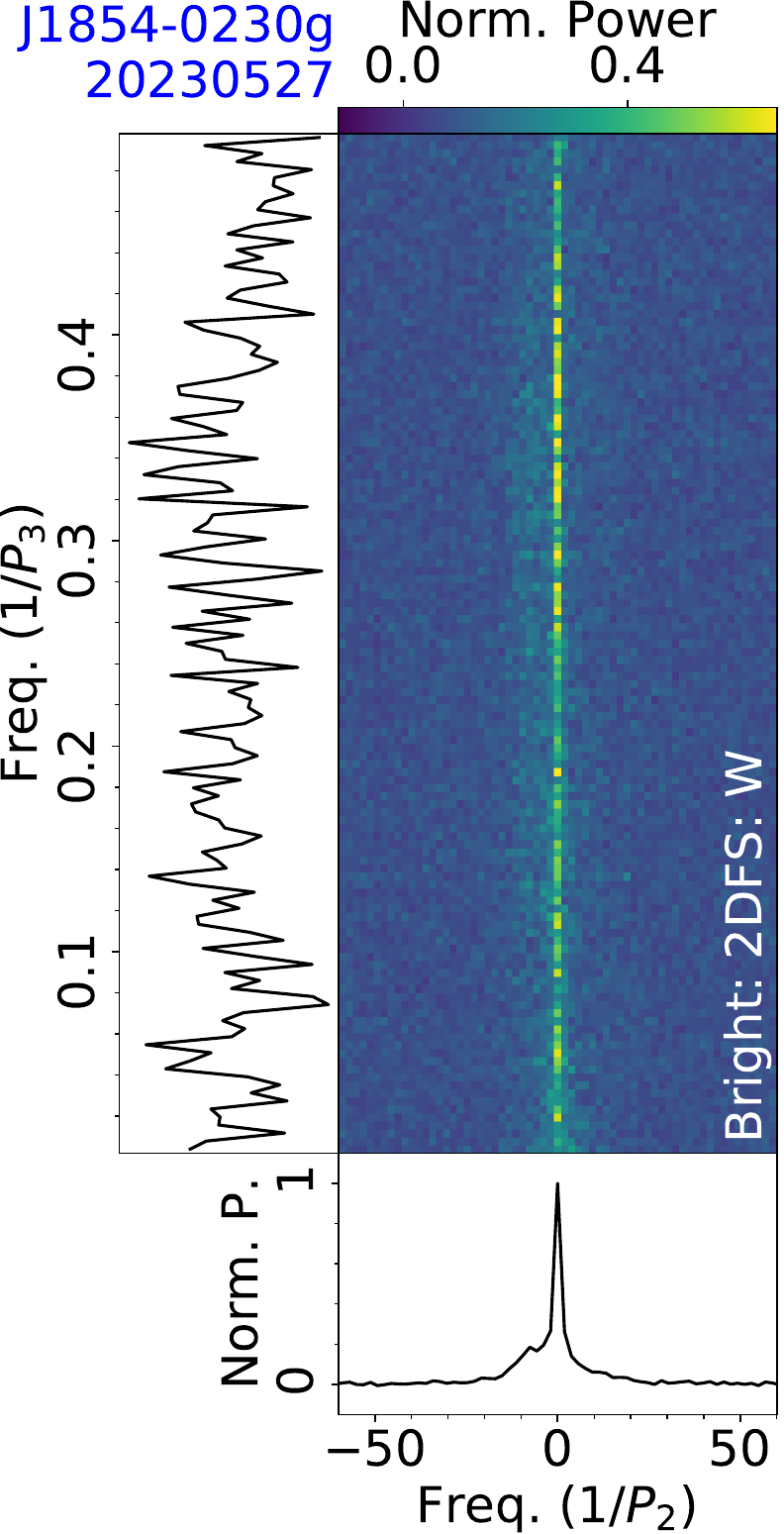}\\
\includegraphics[width=0.21\textwidth, angle=0]{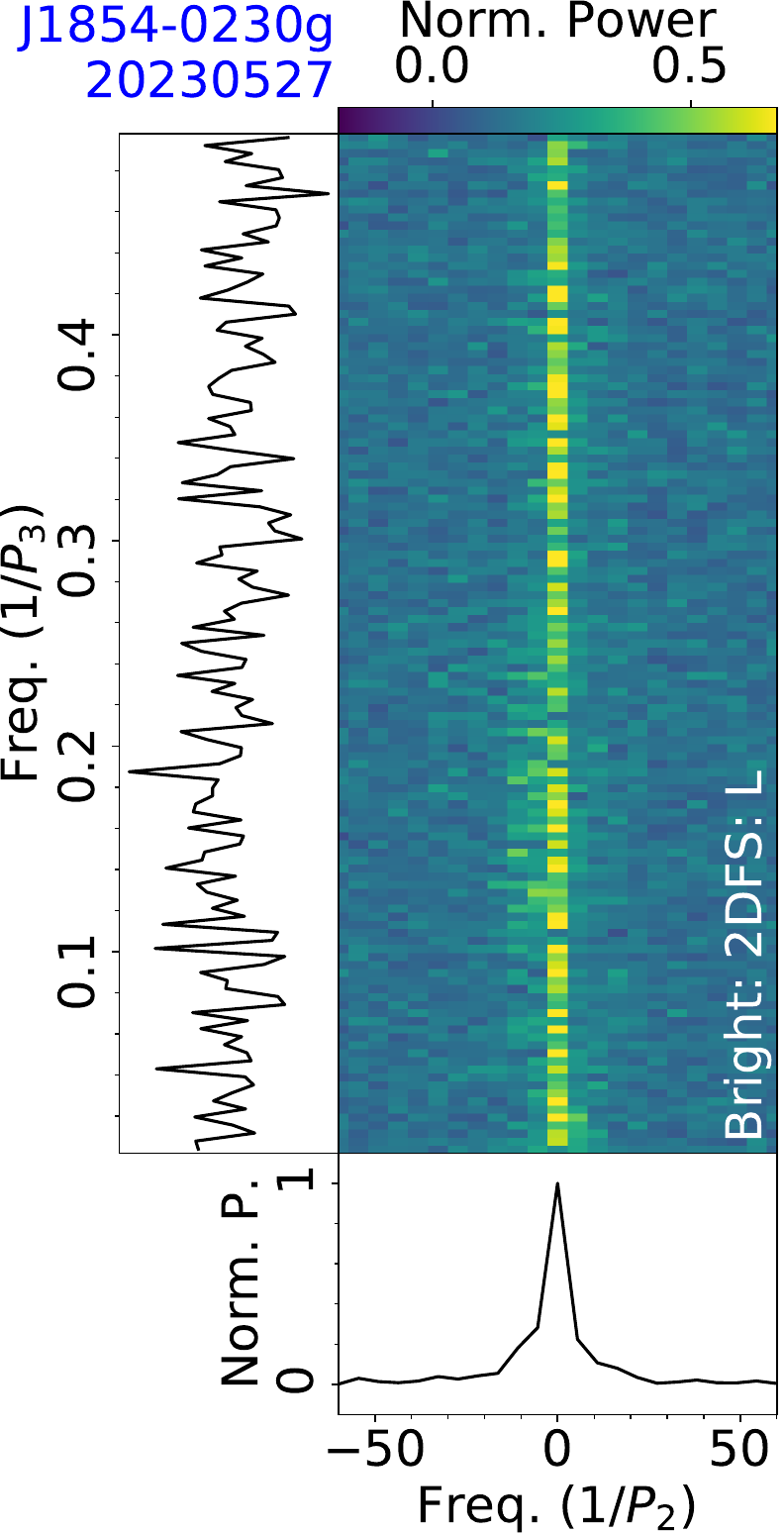}
\includegraphics[width=0.21\textwidth, angle=0]{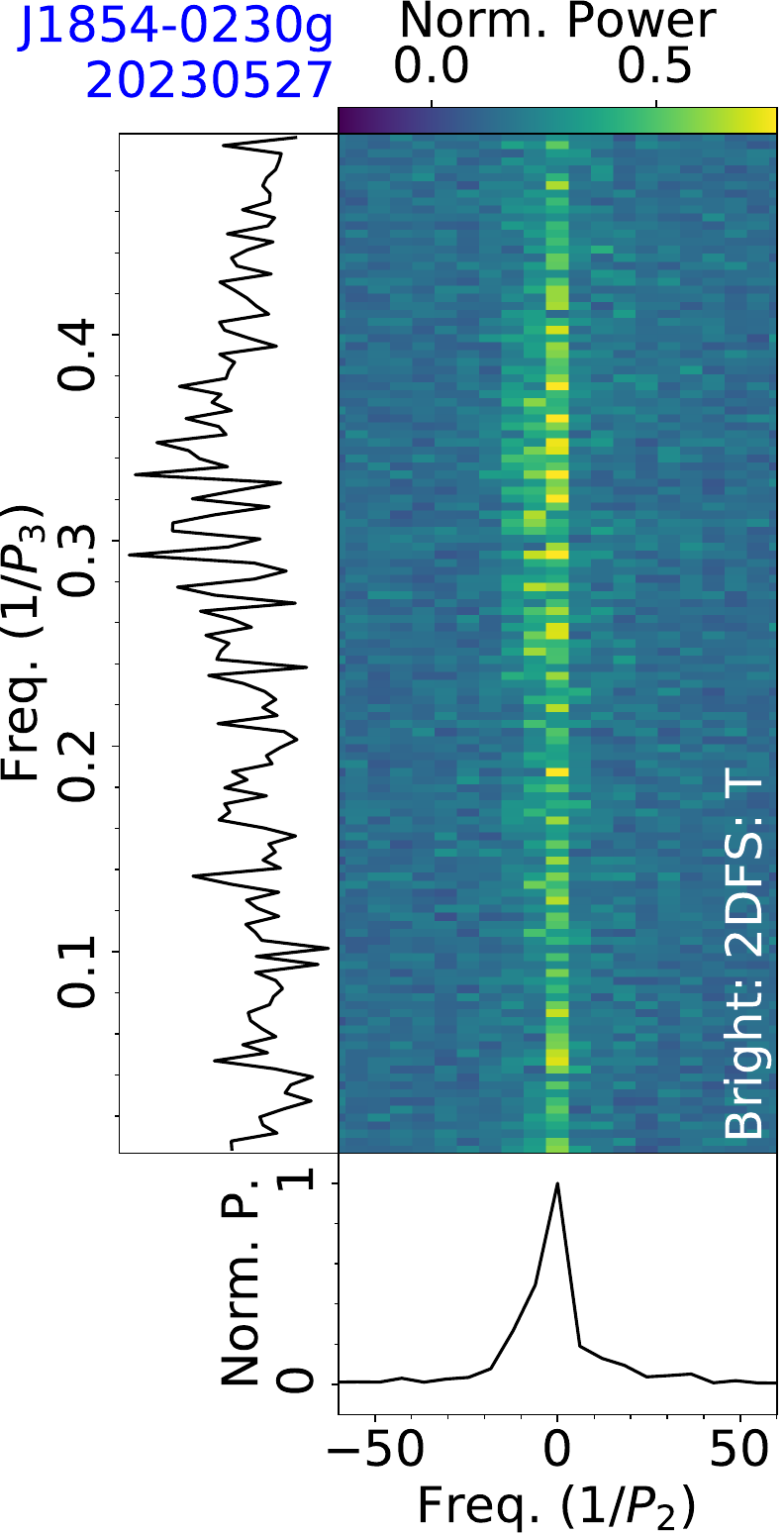}
\figcaption{Fluctuation analysis of the bright emission mode of PSR J1854-0230g for the observation on 20230527, with LRFS and 2DFS for the on-pulse region of a mean pulse profile.
\label{subfig:fluctu:J1854-0230g}}
\end{figure}

\subsection{J1853+0029}
\label{subsec:J1853+0029}

PSR J1853+0029 was discovered in the Pulsar Arecibo L-band Feed Array (PALFA) survey \citep{Parent2022}. 

This pulsar was observed by FAST on 20200920 for 15 minutes, deriving a rotation period $P=1.8769$~s and a dispersion measure $D\!M=227.7~{\rm cm^{-3}\,pc}$. Single pulse sequences in Fig.~\ref{subfig:TP:J1853+0029} display nulling and mode changing with the emission of the trailing profile part occasionally enhanced. 
The nulling fraction of this observation is estimated to be 4$\pm$1\% from the on-pulse energy histogram in Fig.~\ref{subfig:Hist:J1853+0029}. 
Single pulses of two emission modes are distinguished from the energy histogram (Fig.~\ref{subfig:HistModes:J1853+0029}), where each energy value is obtained by integrating over the trailing profile part and then smoothing the resulting time series over 7 periods. The trailing-weak and trailing-strong modes are labeled in red and green, respectively. The mean profiles of the two emission modes (Fig.~\ref{subfig:ProfModes:J1853+0029}) both exhibit a stronger trailing part, while the contrast between trailing and leading is much more pronounced in the trailing-strong mode.




\begin{figure}[htpb]
\centering
\includegraphics[width=0.21\textwidth, angle=0]{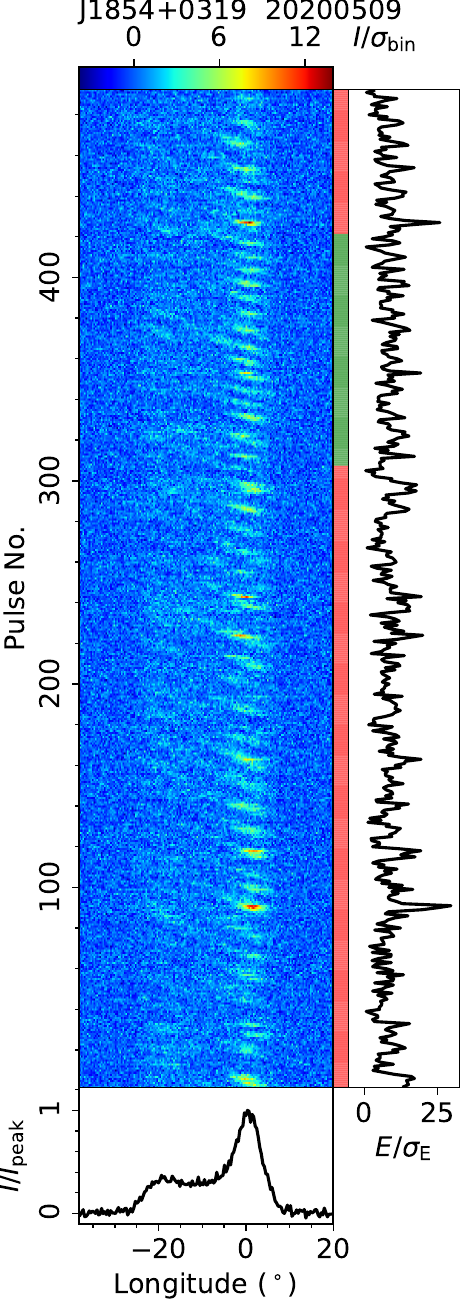}
\figcaption{Single pulse sequence of PSR J1854+0319 from the FAST observation on 20200509.
\label{subfig:TP:J1854+0319}}
\end{figure}

\begin{figure}[htpb]
\centering
\includegraphics[width=0.22\textwidth, angle=0]{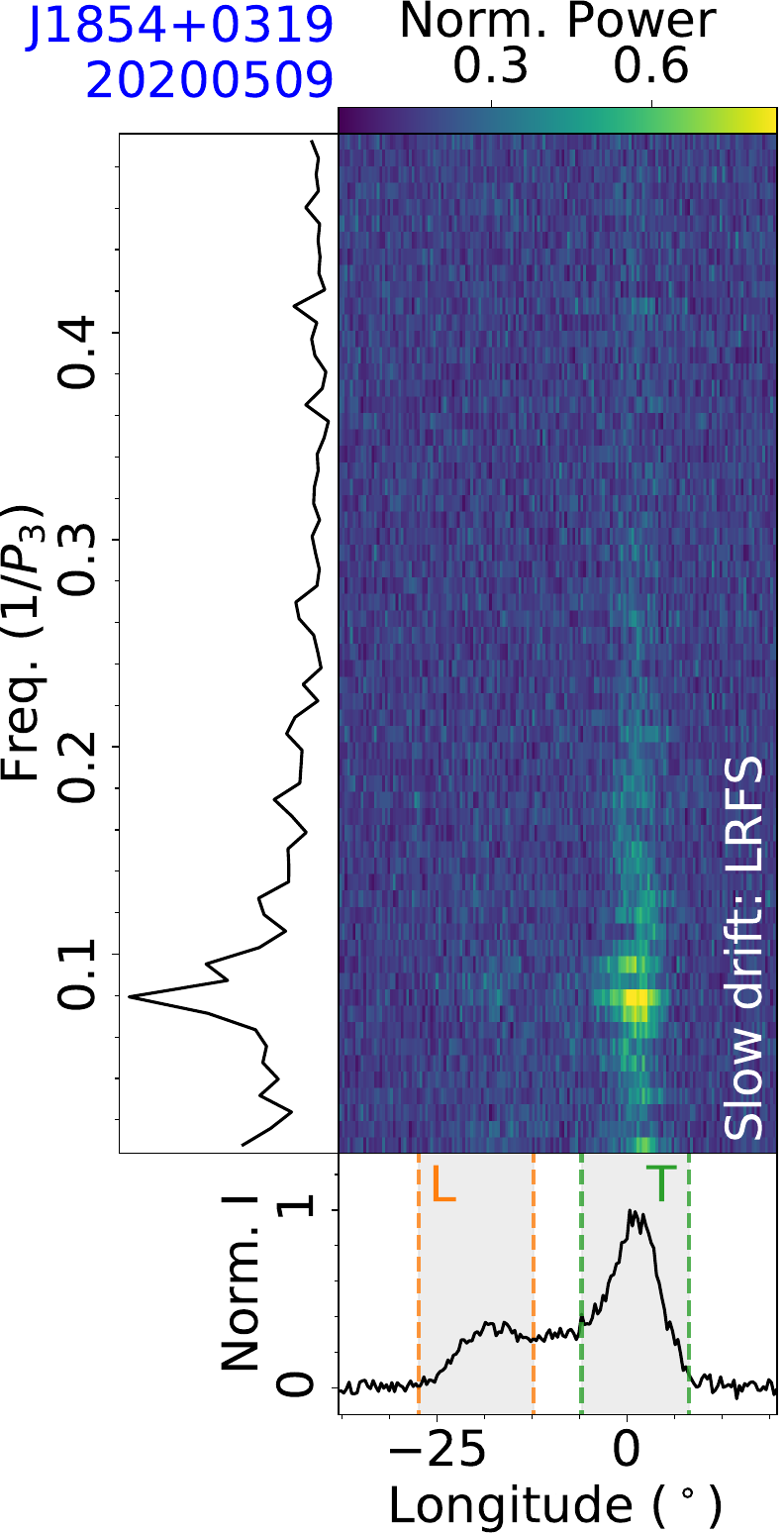}
\includegraphics[width=0.22\textwidth, angle=0]{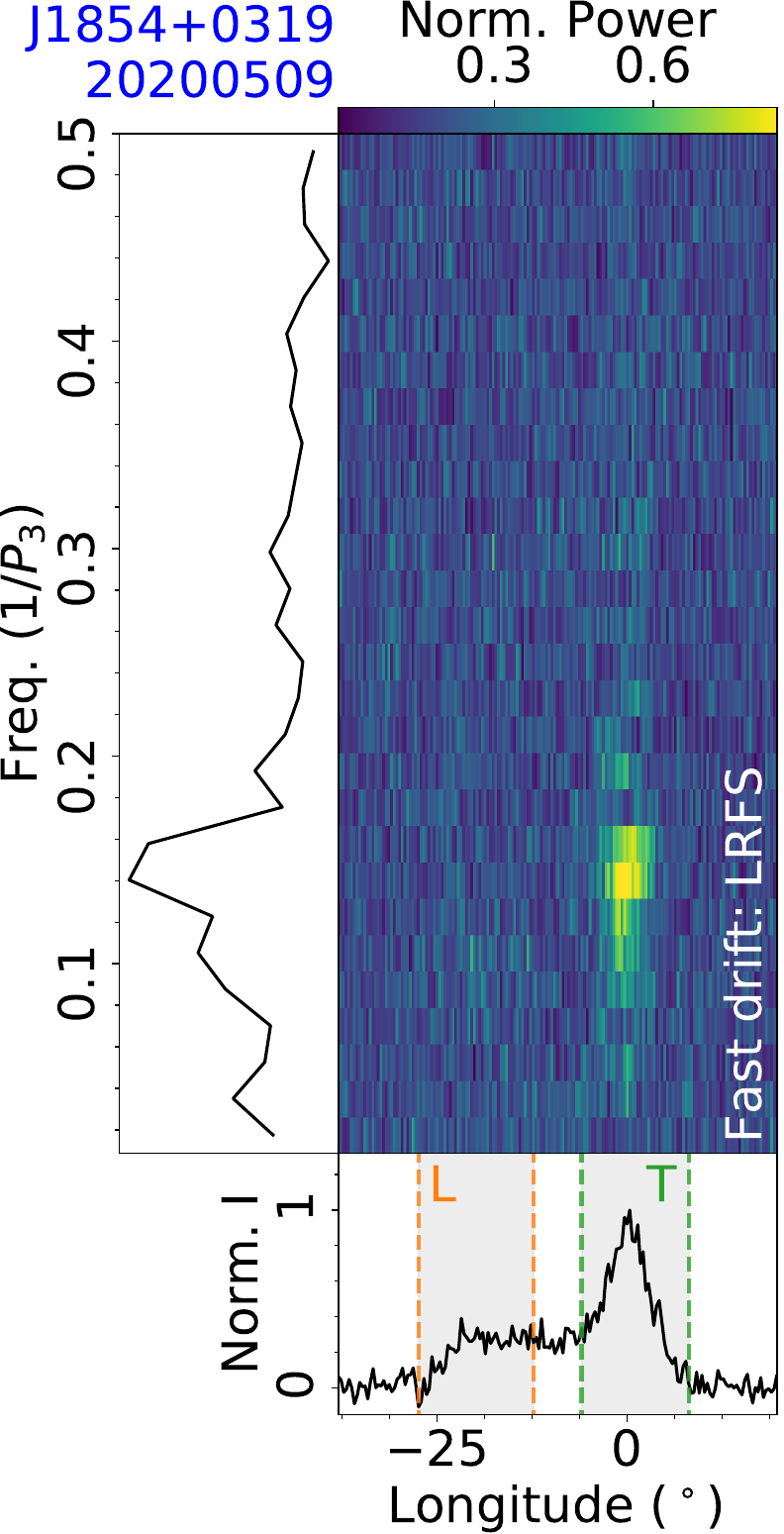}\\
\includegraphics[width=0.22\textwidth, angle=0]{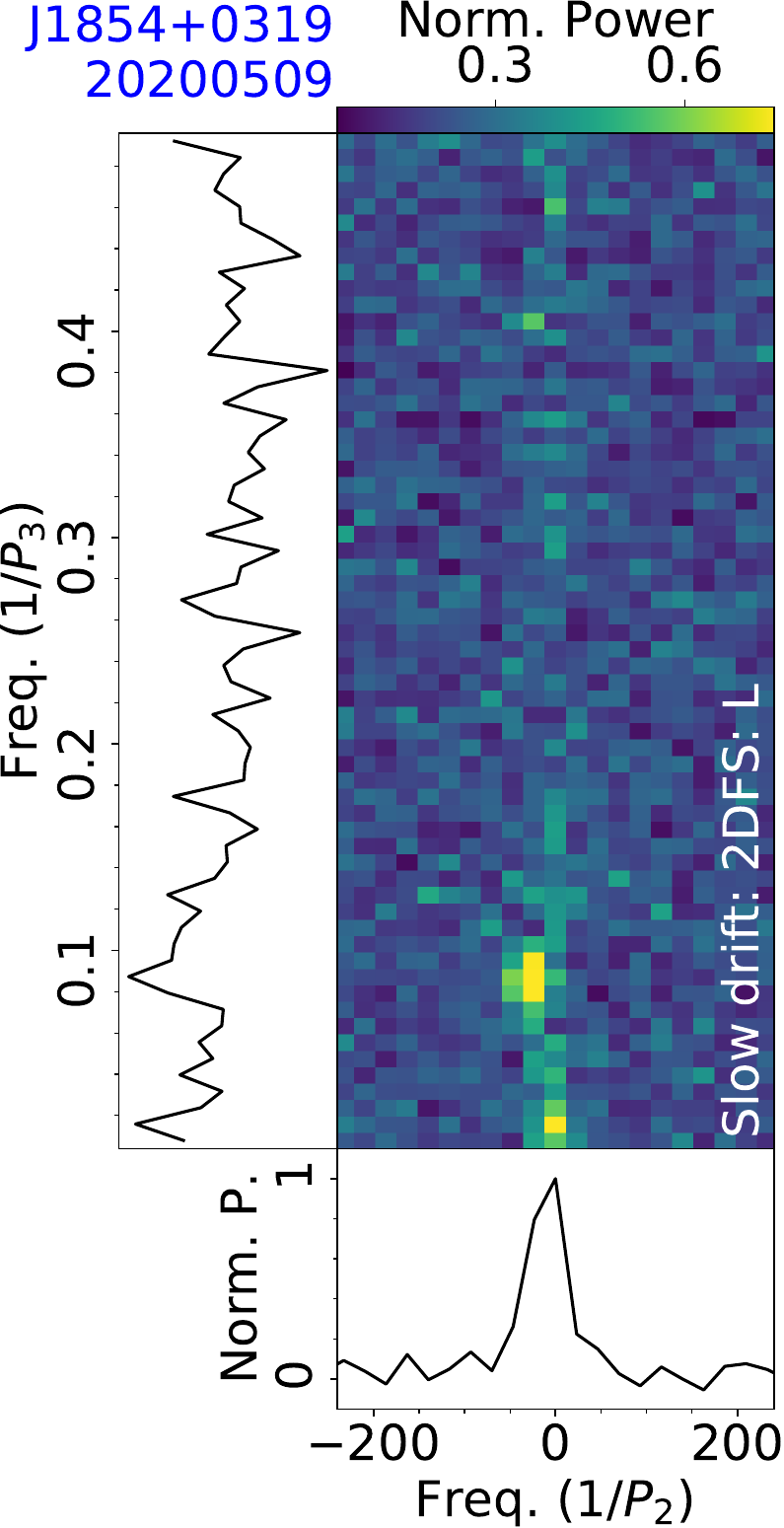}
\includegraphics[width=0.22\textwidth, angle=0]{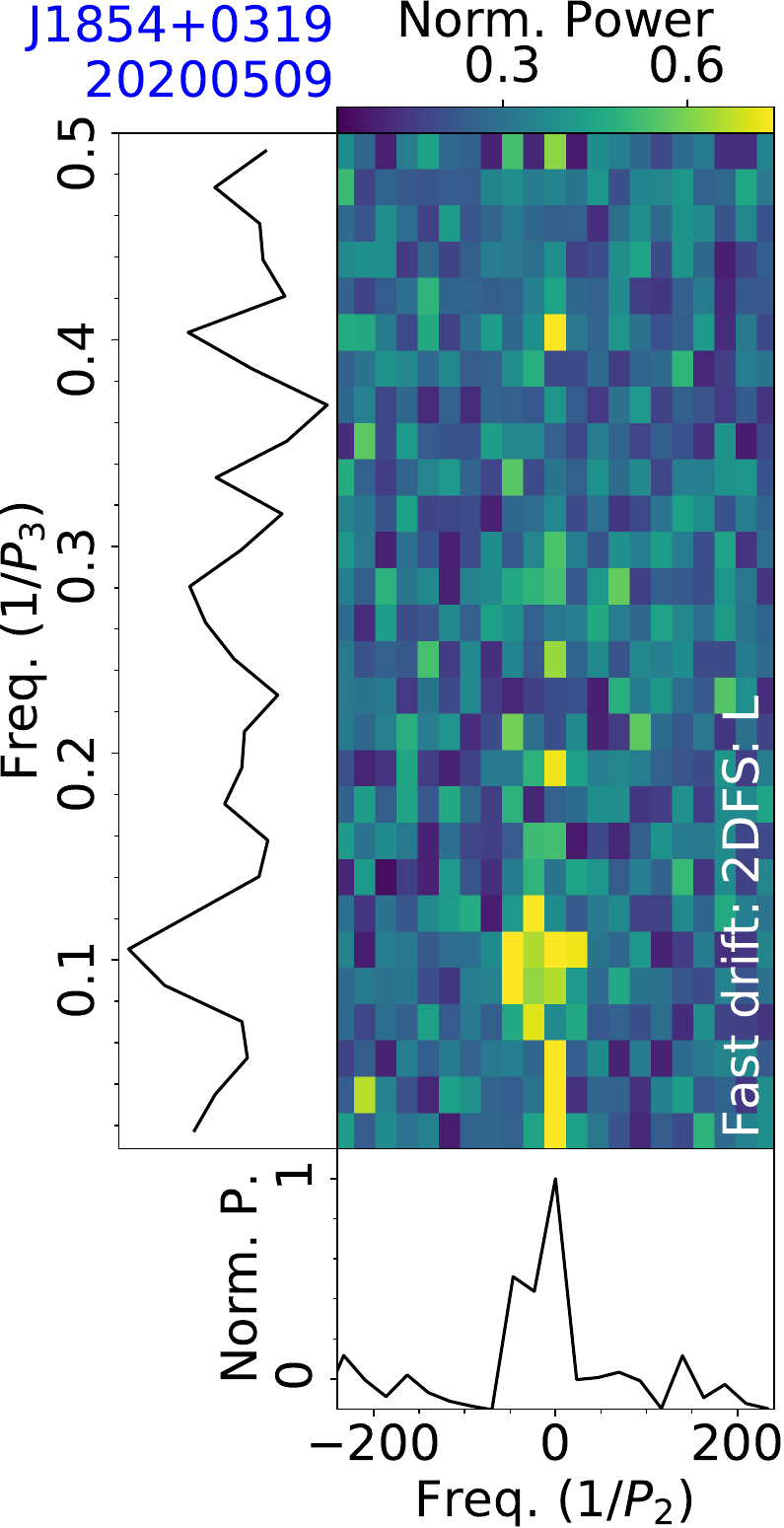}
\figcaption{Fluctuation analysis of slow drifting mode (left panels) and fast drifting mode (right panels) of PSR J1854+0319 from the FAST observation on 20200509, with LRFS (top), and 2DFS for the leading part of a mean pulse profile (bottom). 
-- to be continued.
\label{subfig:fluctu:J1854+0319}}
\addtocounter{figure}{-1}
\end{figure}

\begin{figure}[htpb]
\centering
\includegraphics[width=0.22\textwidth, angle=0]{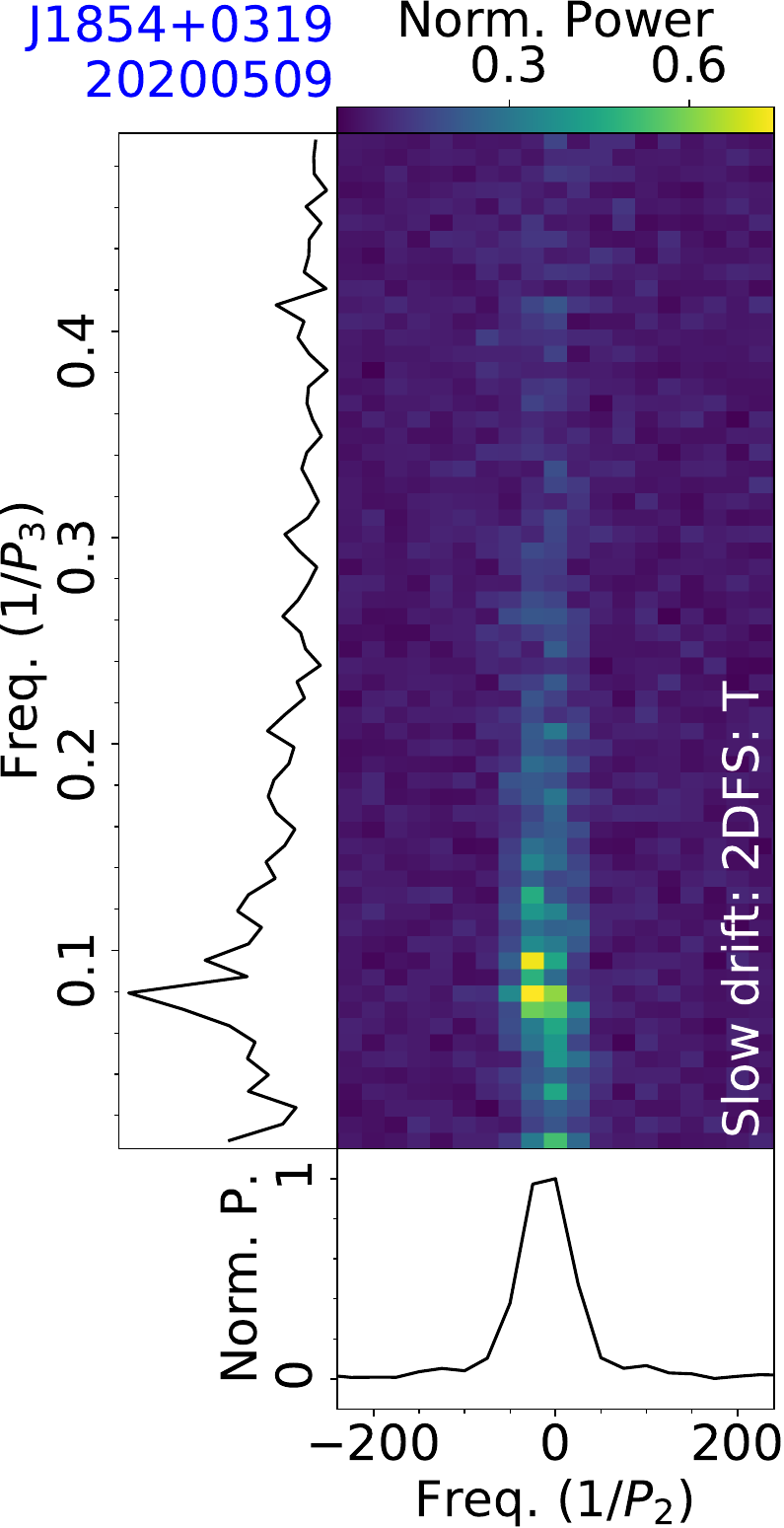}
\includegraphics[width=0.22\textwidth, angle=0]{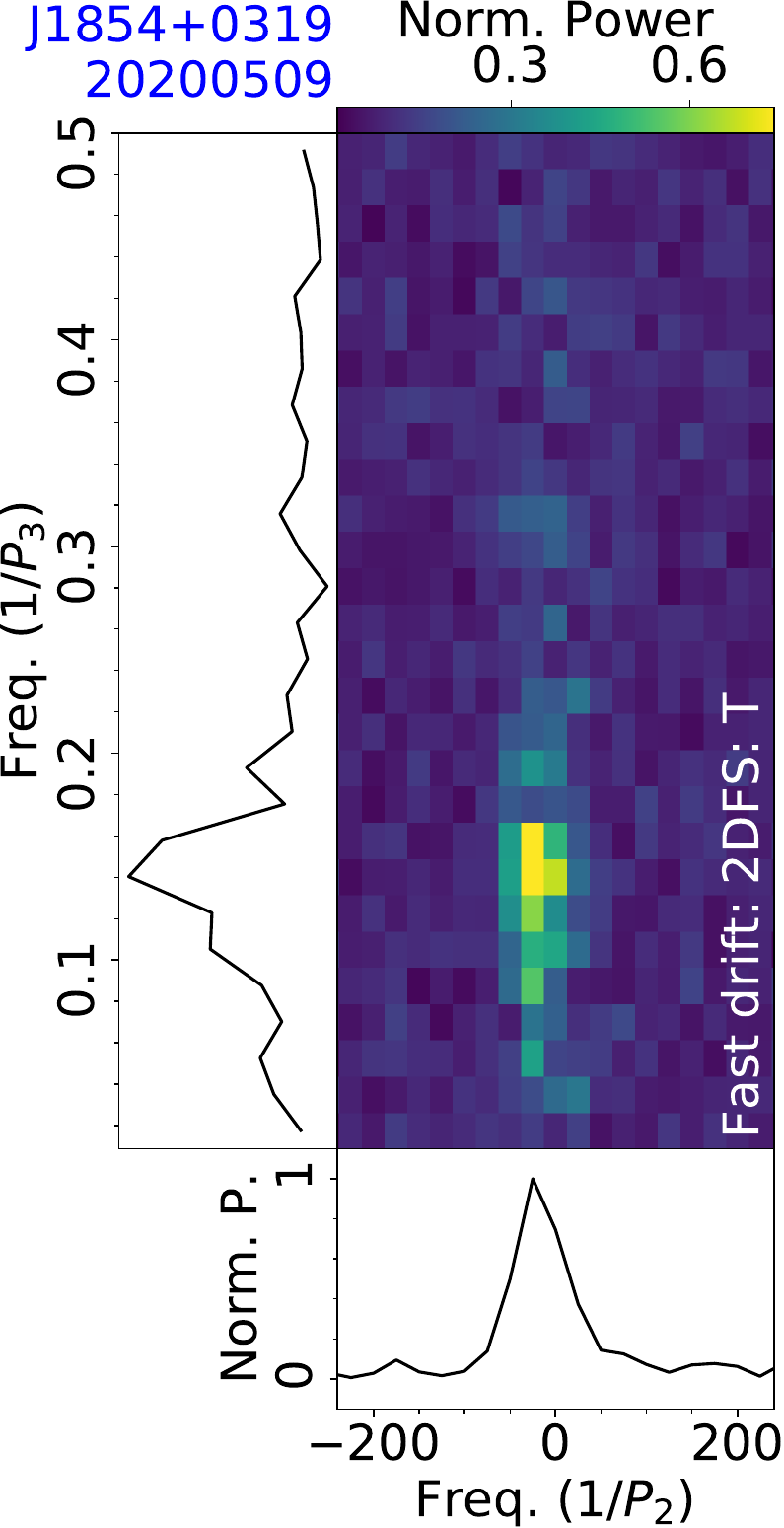}
\figcaption{Continued for Fig.~\ref{subfig:fluctu:J1854+0319}, but for 2DFS of the trailing part of a mean pulse profile.
}
\end{figure}

\subsection{J1853+0853}
\label{subsec:J1853+0853}

PSR J1853+0853 was found in the Parkes multibeam pulsar survey \citep{Lorimer2006}. 

This pulsar was observed by FAST on 20210306 and 20211018,  both for 15 minutes. From the observation on 20211018, a rotation period $P=3.9148$~s and a dispersion measure $D\!M=233.9~{\rm cm^{-3}\,pc}$ were derived.
Single pulse sequences of the observation on 20210306 are displayed in Fig.~\ref{subfig:TP:J1853+0853}, illustrating the existence of both nulling and subpulse drifting behaviors. The nulling fraction of the observation is estimated from the on-pulse integral energy histogram (Fig.~\ref{subfig:Hist:J1853+0853}) to be 30$\pm$5\%. Because of the frequent nulling and few single pulses of the observation, the drifting parameters are obtained from the cross-correlation method (Fig.~\ref{subfig:Corre:J1853+0853}), which are $P_2=1.7\pm0.1^\circ$ and $D=0.8\pm0.1$ degrees per period.

\begin{figure}[htpb]
\centering
\includegraphics[width=0.22\textwidth, angle=0]{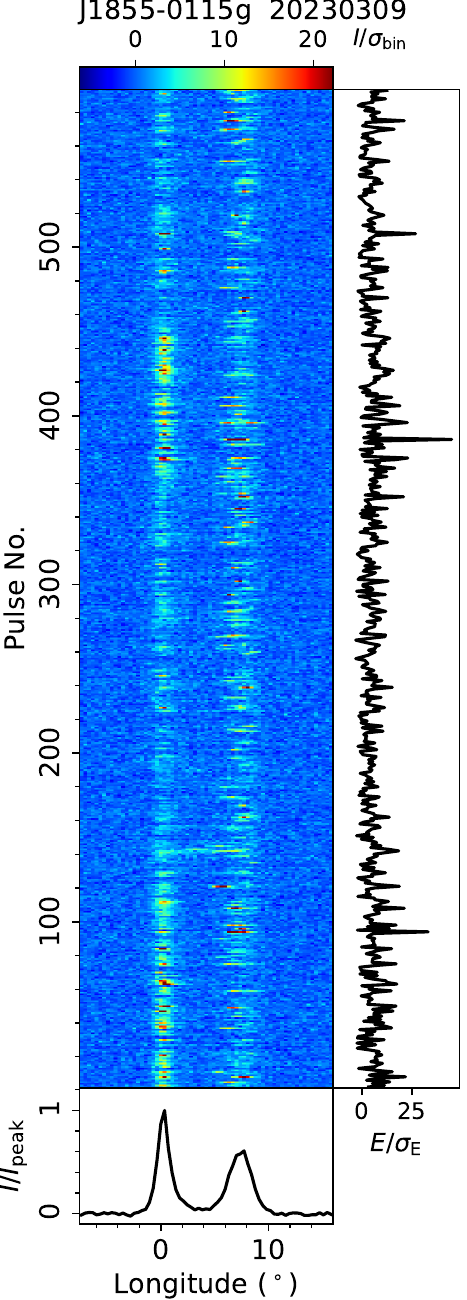}
\includegraphics[width=0.22\textwidth, angle=0]{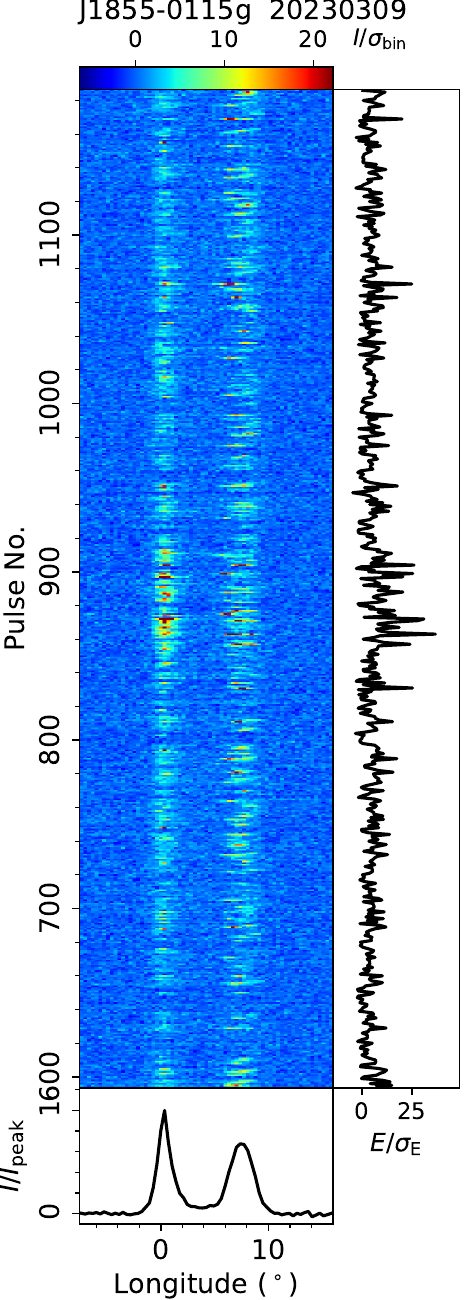}
\figcaption{Single pulse sequences of PSR J1855-0115g from the FAST observation on 20230309.
\label{subfig:TP:J1855-0115g}}
\end{figure}

\begin{figure}[htpb]
\centering
\includegraphics[width=0.39\textwidth, angle=0]{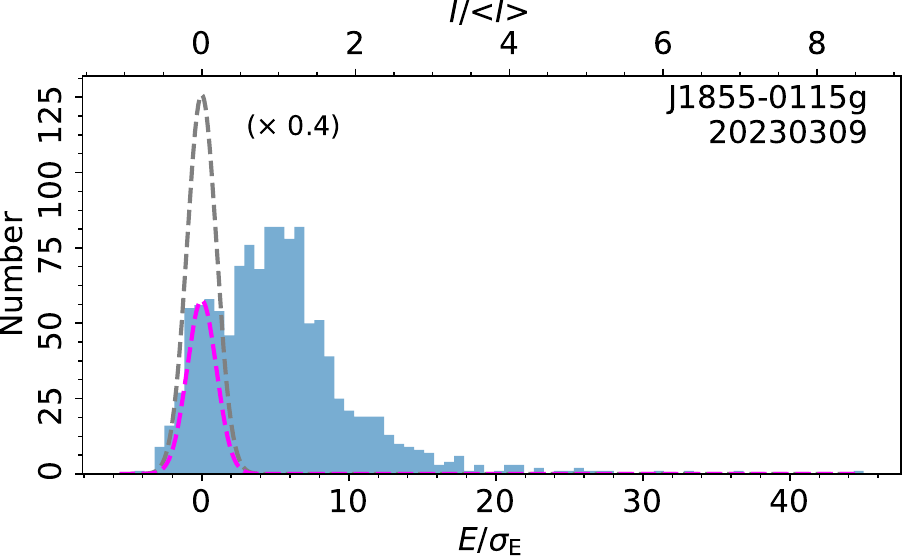}
\figcaption{On-pulse energy histogram of single pulses of PSR J1855-0115g from the FAST observation on 20230309. \label{subfig:Hist:J1855-0115g}}
\end{figure}

\begin{figure}[htpb]
\centering
\includegraphics[width=0.22\textwidth, angle=0]{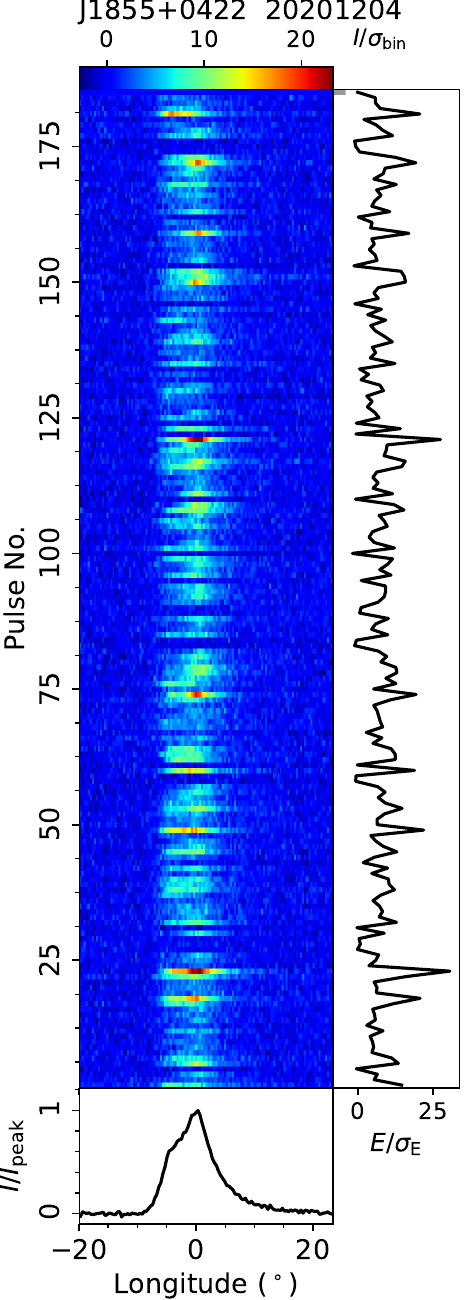}
\figcaption{Single pulse sequence of PSR J1855+0422 from the FAST observation on 20201204.
\label{subfig:TP:J1855+0422}}
\end{figure}

\begin{figure}[htpb]
\centering
\includegraphics[width=0.39\textwidth, angle=0]{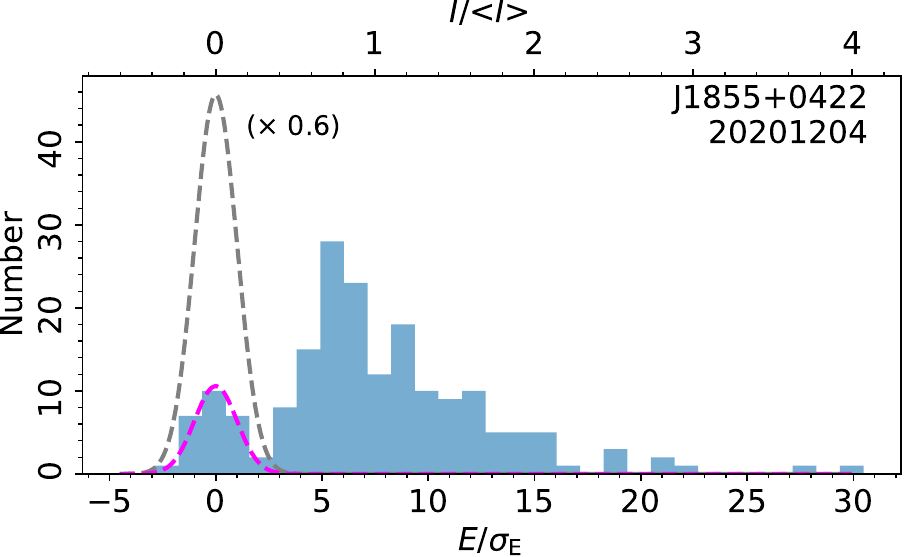}
\figcaption{On-pulse energy histogram of single pulses of PSR J1855+0422 from the FAST observation on 20201204.
\label{subfig:Hist:J1855+0422}}
\end{figure}

\subsection{J1854-0230g}
\label{subsec:J1854-0230g}

PSR J1854-0230g was discovered in the FAST GPPS survey \citep{Han2021,han2025}. 

The pulsar was observed by FAST on 20230527 for 14 minutes, yielding a rotation period $P=0.6874$~s and a dispersion measure $D\!M=548.6~{\rm cm^{-3}\,pc}$. 
The single pulse sequence of this observation (Fig.~\ref{subfig:TP:J1854-0230g}) shows that there are weak and bright emission modes. The polarization profiles of two emission modes are displayed in Fig.~\ref{subfig:PolModes:J1854-0230g}, where the profile shape changes a lot. In the bright mode, the central component is the strongest. While in the weak emission mode, the intensity of the leading component is the highest. Additionally, the bright emission mode displays subpulse drifting behavior. 
From fluctuation spectra of the leading component (Fig.~\ref{subfig:fluctu:J1854-0230g}), the centroid frequencies of the drift feature are $1/P_3=0.173\pm0.003$ and $1/P_2=-8.1\pm0.3$, corresponding to the periodicities of $P_3=5.8\pm0.1$ periods and $P_2=-45\pm2^\circ$. 2DFS of the trailing component exhibit a drift feature with the centroid of $1/P_3=0.318\pm0.003$ and $1/P_2=-5.9\pm0.3$, yielding $P_3=3.15\pm0.03$ periods and $P_2=-61\pm3^\circ$.

\subsection{J1854+0319}
\label{subsec:J1854+0319}

PSR J1854+0319 was discovered in the PALFA survey \citet{Lyne2017}. 
The drifting behavior of the trailing component was published by \citet{Song2023}. 

This pulsar was observed by FAST on 20200509 for 5 minutes, deriving a rotation period $P=0.6285$~s and a dispersion measure $D\!M=481.0~{\rm cm^{-3}\,pc}$. 
From this observation, we newly find that the pulsar exhibits changes between different drifting states, and the weaker leading component also has drifting behavior. 
Single pulse sequences are displayed in Fig.~\ref{subfig:TP:J1854+0319}, with two drifting modes labeled using bars of different colors. Drifting parameters of two drifting modes are obtained from fluctuation spectra in Fig.~\ref{subfig:fluctu:J1854+0319}. 
For the slow drifting mode, the centroid frequencies of the drift feature in 2DFS are $1/P_3=0.085\pm0.001$ ($P_3=11.8\pm0.2$ periods) and $1/P_2=-28\pm3$ ($P_2=-13\pm1^\circ$) for the leading part in the mean pulse profile, and $1/P_3=0.086\pm0.001$ ($P_3=11.7\pm0.1$ periods) and $1/P_2=-13\pm1$ ($P_2=-27\pm3^\circ$) for the trailing part.
While for the fast drifting mode, they are $1/P_3=0.097\pm0.002$ ($P_3=10.3\pm0.2$ periods) and $1/P_2=-16\pm3$ ($P_2=-23\pm4^\circ$) for the leading part in the mean pulse profile, and $1/P_3=0.131\pm0.002$ ($P_3=7.7\pm0.1$ periods) and $1/P_2=-23\pm2$ ($P_2=-15\pm2^\circ$) for the trailing part.

\subsection{J1855-0115g}
\label{subsec:J1855-0115g}

PSR J1855-0115g was discovered in the FAST GPPS survey \citep{Han2021,han2025}. 

This pulsar was observed by FAST on 20230309 for 50 minutes, deriving a rotation period $P=2.5618$~s and a dispersion measure $D\!M=261.7~{\rm cm^{-3}\,pc}$. 
From the single pulse sequence in Fig.~\ref{subfig:TP:J1855-0115g}, there is a modulation on the leading component that becomes brighter every several hundred periods. The pulsar has nulling behavior with a nulling fraction estimated to be 18$\pm$2\% from the on-pulse energy histogram in Fig.~\ref{subfig:Hist:J1855-0115g}. There are also very bright single pulses whose intensity could be up to 9 times the averaged intensity <I>.

\begin{figure}[htpb]
\centering
\includegraphics[width=0.21\textwidth, angle=0]{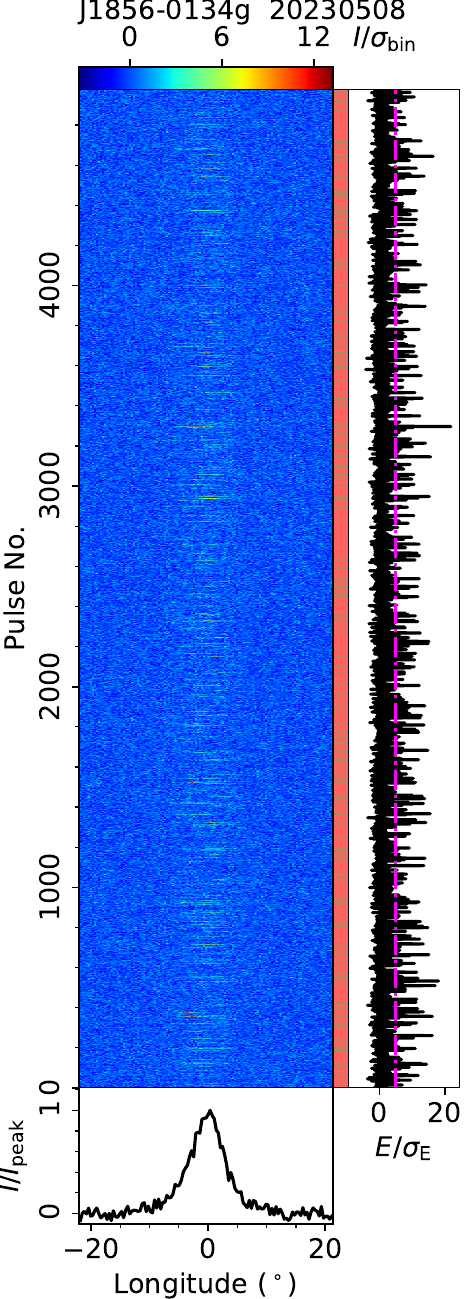}
\includegraphics[width=0.21\textwidth, angle=0]{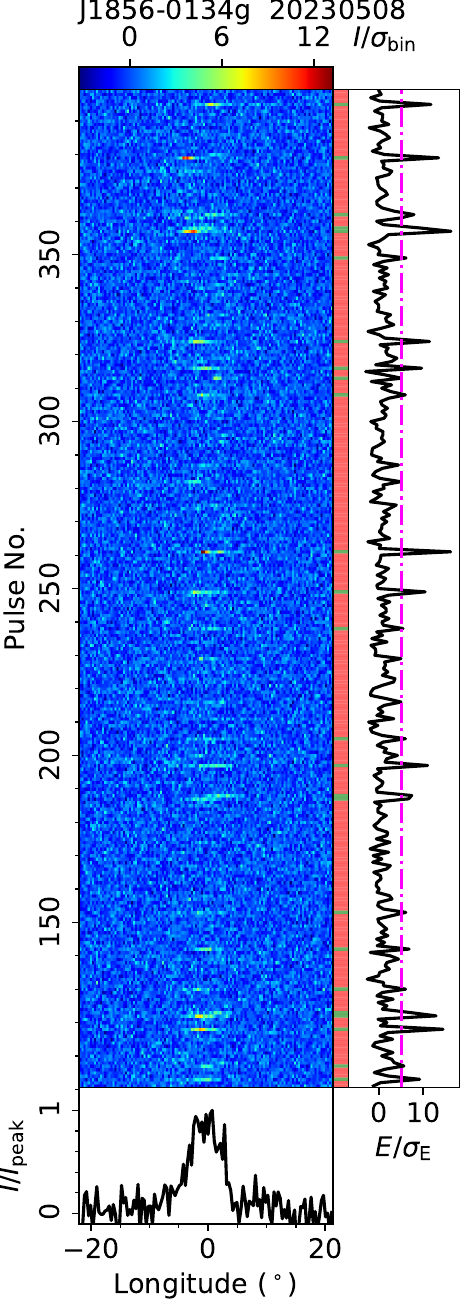}
\figcaption{Single pulse sequence of PSR J1856-0134g from the FAST observation on 20230508, and a zoomed-in view of pulses No. 100-400.
\label{subfig:TP:J1856-0134g}}
\end{figure}

\begin{figure}[htpb]
\centering
\includegraphics[width=0.39\textwidth, angle=0]{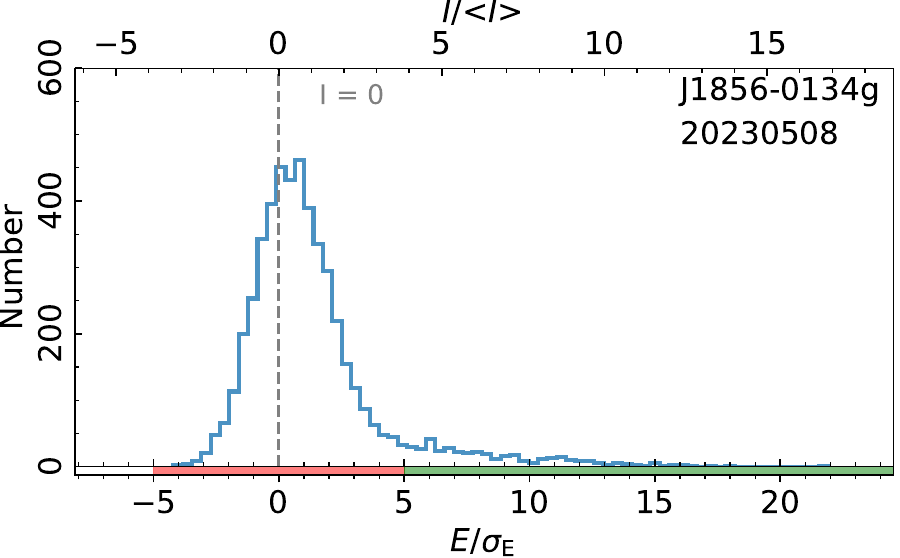}
\figcaption{On-pulse energy histogram of single pulses of PSR J1856-0134g from the FAST observation on 20230508.
\label{subfig:Hist:J1856-0134g}}
\end{figure}

\begin{figure}[htpb]
\centering
\includegraphics[width=0.39\textwidth, angle=0]{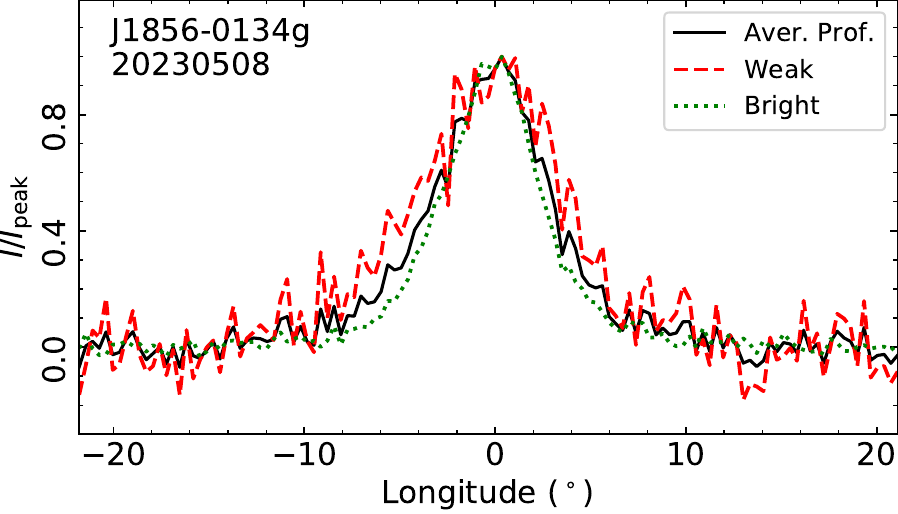}
\figcaption{Mean profiles for the weak and bright emission modes of PSR J1856-0134g from the FAST observation on 20230508, which are normalized by their respective peaks. 
\label{subfig:ProfModes:J1856-0134g}}
\end{figure}

\begin{figure}[htpb]
\centering
\includegraphics[width=0.42\textwidth, angle=0]{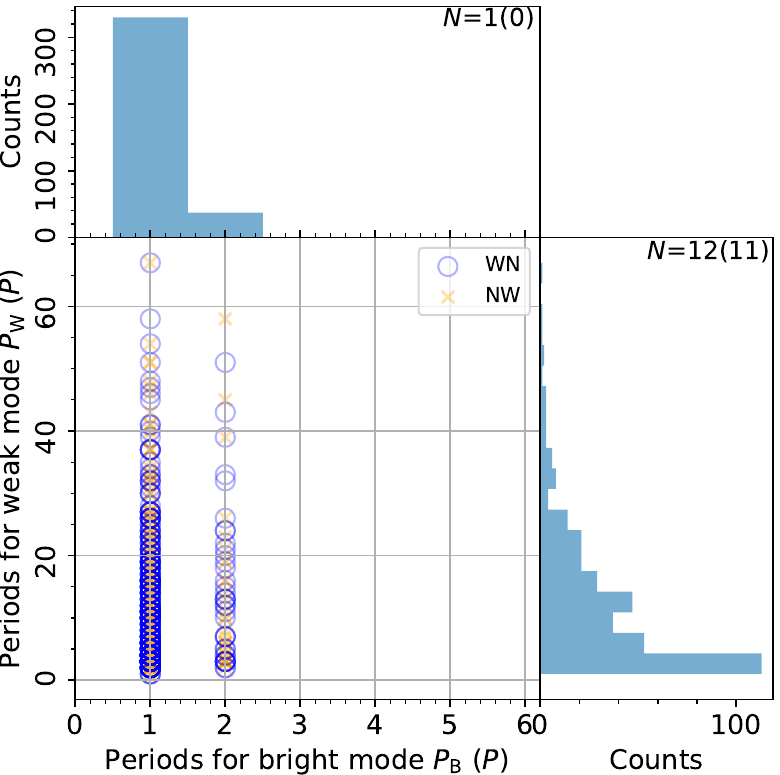}
\figcaption{Distribution of period numbers for the weak emission mode $P_{\rm W}$ against period numbers for adjacent bright mode $P_{\rm B}$ of PSR J1856-0134g observed by FAST on 20230508, as well as the duration histograms for the bright and weak emission modes shown in the top and right panels, respectively. 
\label{subfig:scaleHist:J1856-0134g}}
\end{figure}

\begin{figure}[htpb]
\centering
\includegraphics[height=0.88\textheight]{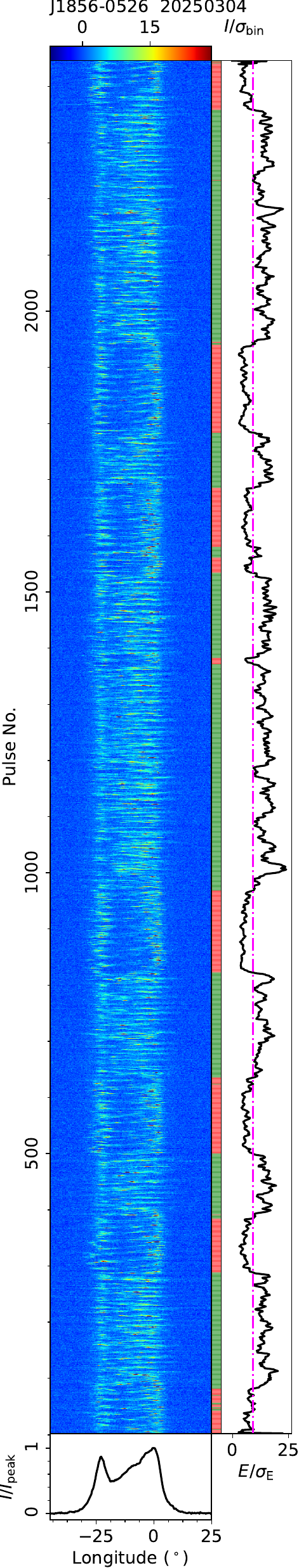}
\includegraphics[height=0.88\textheight]{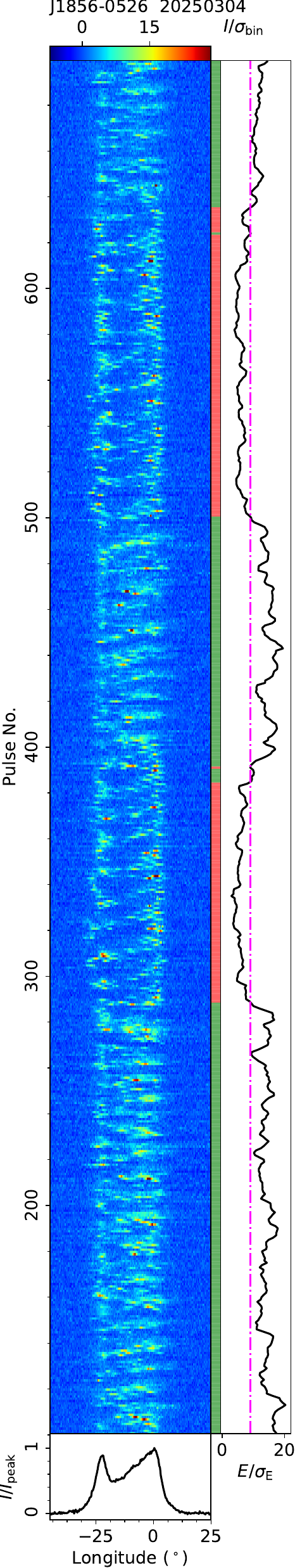} 
\vspace{-0.3cm}
\figcaption{Single pulse sequence of PSR J1856-0526 from the FAST observation on 20250304, and a zoomed-in view of pulses No. 100-700. 
The right subplot shows the energy variation integrated over the central phase interval of the mean pulse profile, smoothed with a 17-period moving average. The magenta dash-dot line represents the boundary between the two emission modes.
\label{subfig:TP:J1856-0526}}
\end{figure}

\begin{figure}[htpb]
\centering
\includegraphics[width=0.39\textwidth, angle=0]{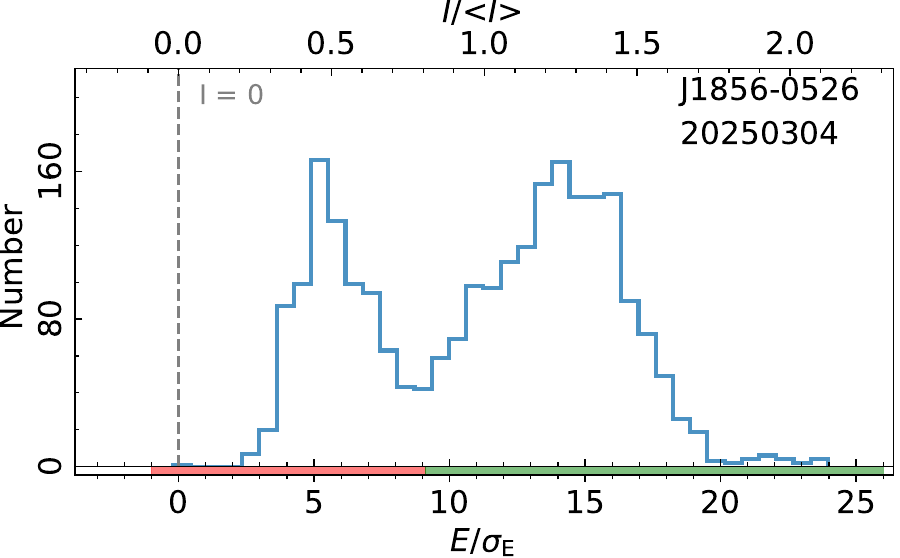}
\figcaption{Energy histogram of single pulses from PSR J1856-0526, obtained from the FAST observation on 20250304. 
For every single pulse, the energy is integrated over the central phase interval of the mean pulse profile, and the energy sequence is smoothed with a 17-period moving average. 
The central-weak and central-strong modes, distinguished by their relative integrated energy in this interval, are shown in red and green, respectively.
\label{subfig:Hist:J1856-0526}}
\end{figure}

\begin{figure}[htpb]
\centering
\includegraphics[width=0.37\textwidth, angle=0]{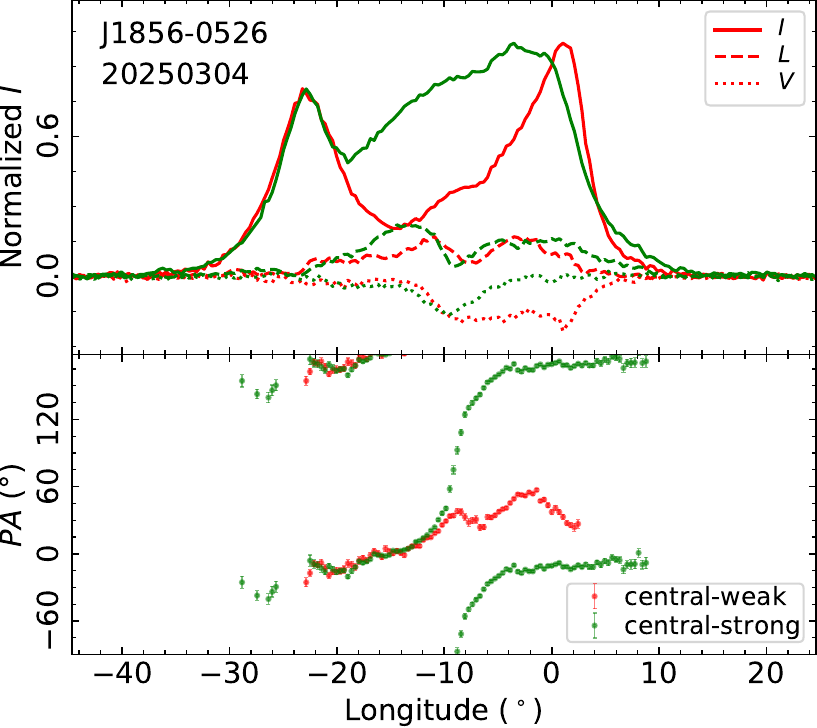}
\figcaption{Mean polarization profiles (the top panel) for the central-weak (red) and central-strong modes (green), as well as the averaged PA curves (the bottom panel) of PSR J1856-0526 from the FAST observation on 20250304.
\label{subfig:PolModes:J1856-0526}}
\end{figure}

\begin{figure}[htpb]
\centering
\includegraphics[width=0.22\textwidth, angle=0]{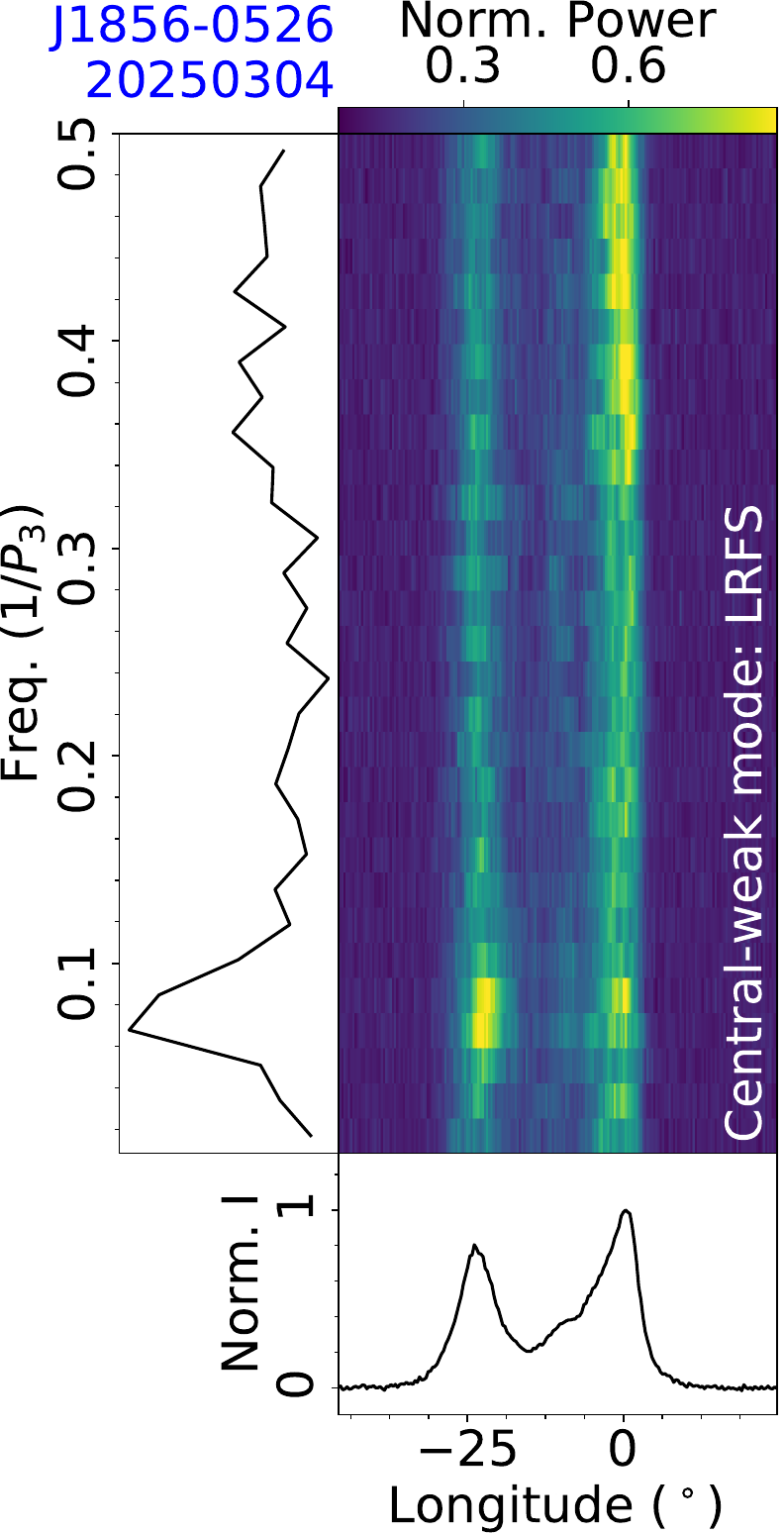}
\includegraphics[width=0.22\textwidth, angle=0]{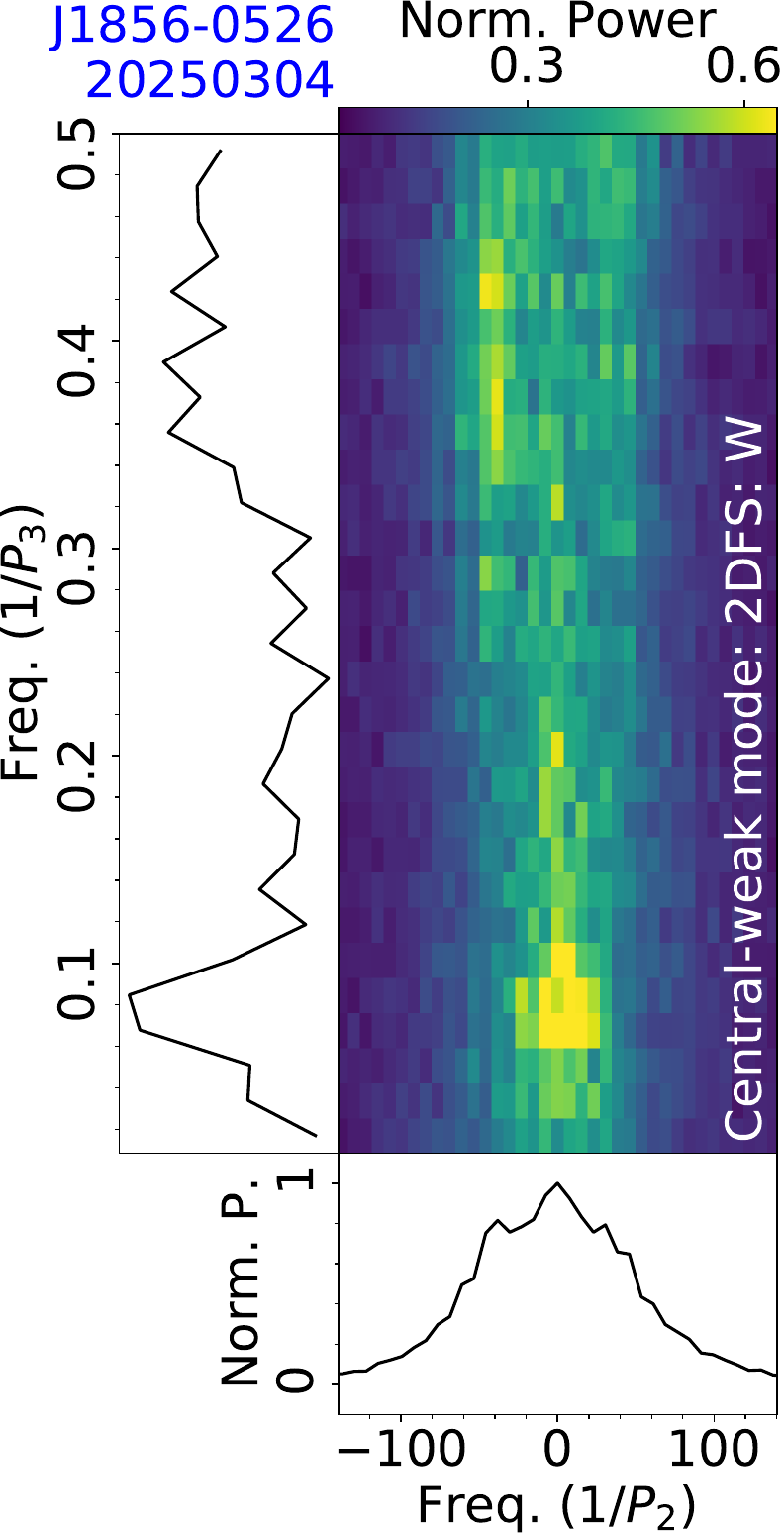}
\includegraphics[width=0.22\textwidth, angle=0]{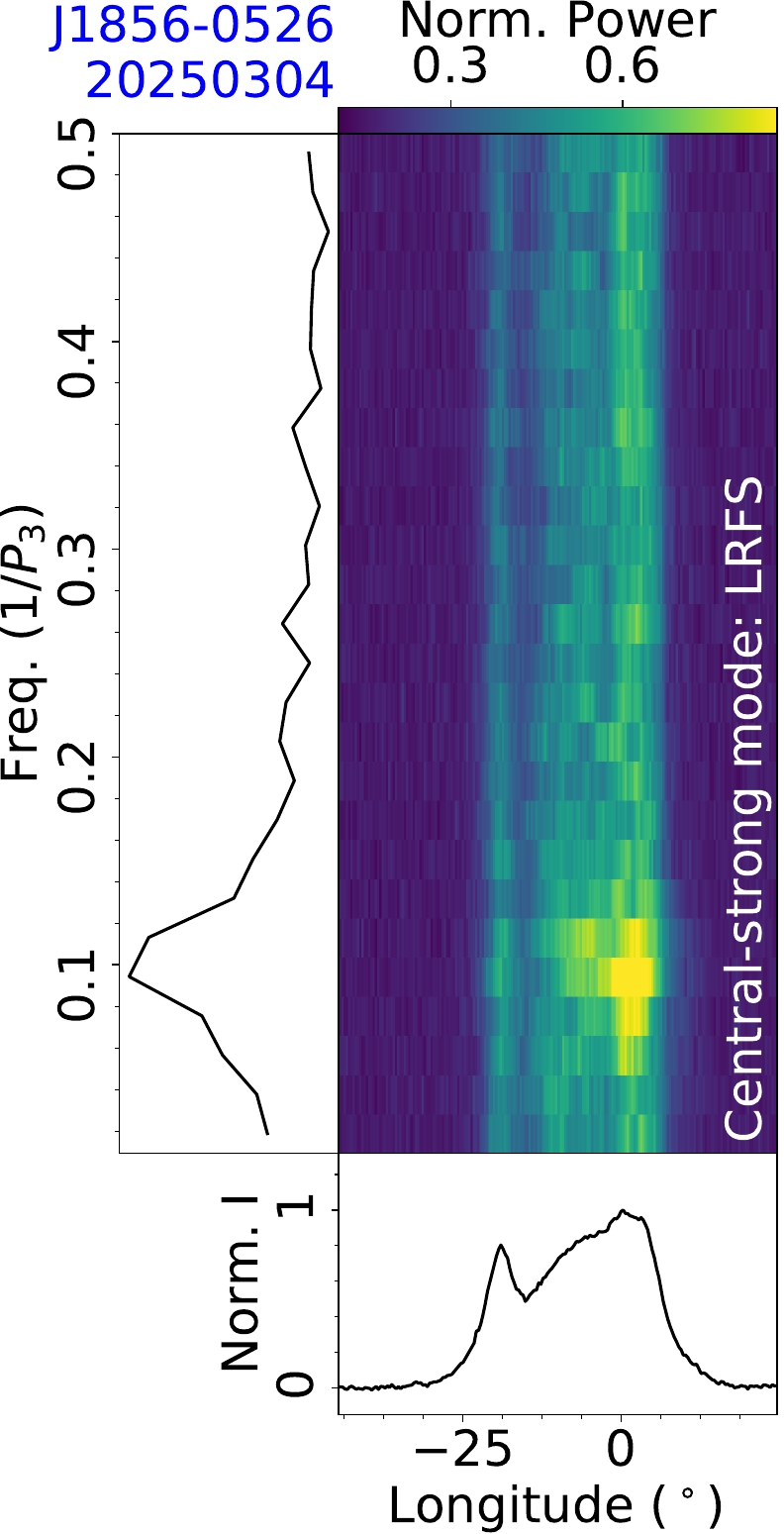}
\includegraphics[width=0.22\textwidth, angle=0]{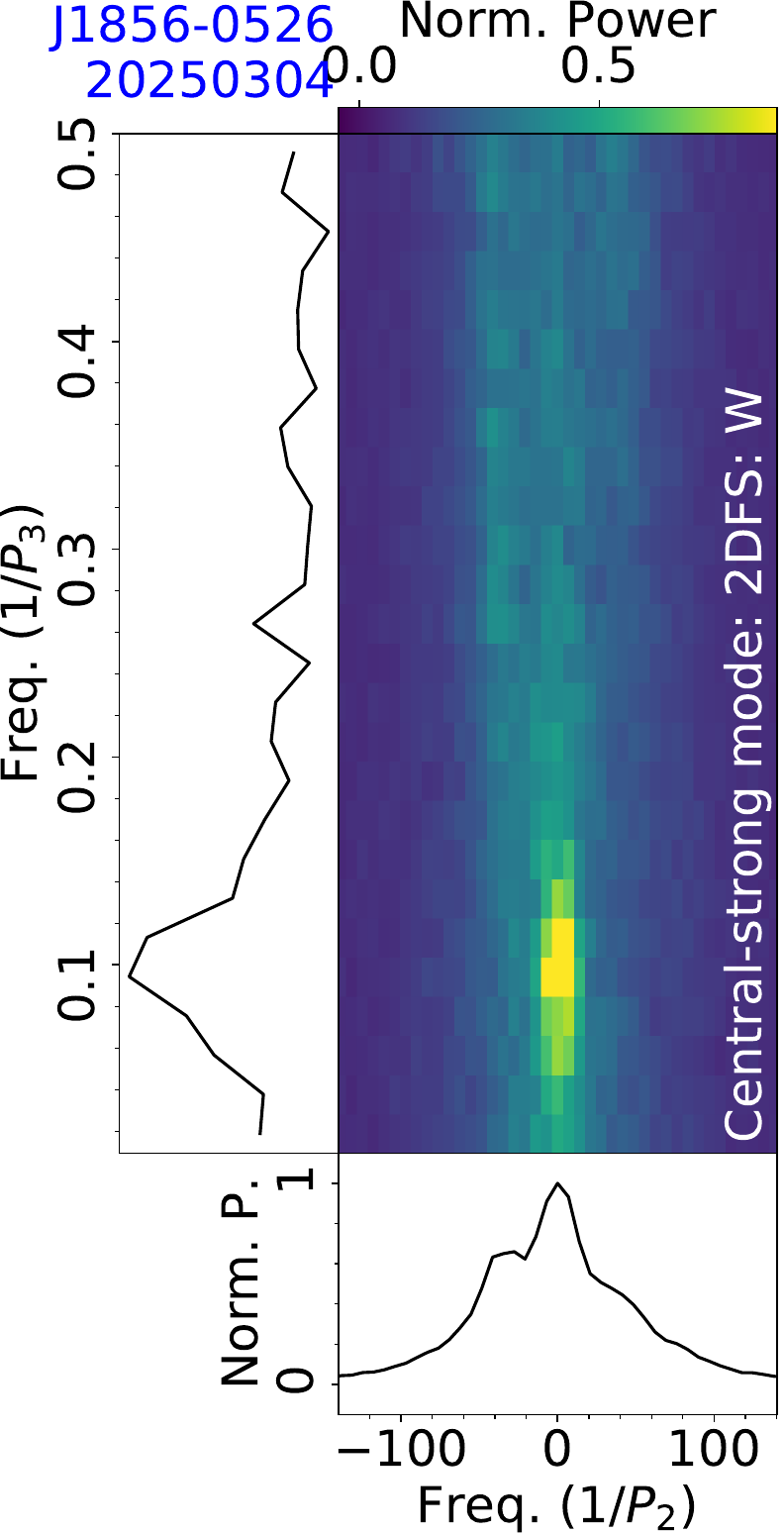}
\figcaption{Fluctuation analysis for the central-weak (upper) and central-strong (bottom) modes of PSR J1856-0526 for the observation on 20250304, with LRFS and 2DFS for the on-pulse region of the mean pulse profile.
\label{subfig:fluctu:J1856-0526}}
\end{figure}

\subsection{J1855+0422}
\label{subsec:J1855+0422}

PSR J1855+0422 was discovered by \citet{Morris2002} in the Parkes Multibeam Pulsar Survey. 

The pulsar was observed by FAST on 20201204 for 5 minutes, deriving a rotation period $P=1.6783$~s and a dispersion measure $D\!M=456.8~{\rm cm^{-3}\,pc}$. 
The single pulse sequence in Fig.~\ref{subfig:TP:J1855+0422} and on-pulse integral energy histogram in Fig.~\ref{subfig:Hist:J1855+0422} illustrate that the pulsar has short-duration nulls. The nulling fraction of this observation is estimated from Fig.~\ref{subfig:Hist:J1855+0422} to be 14$\pm$1 percent.

\begin{figure}[htpb]
\centering
\includegraphics[width=0.22\textwidth, angle=0]{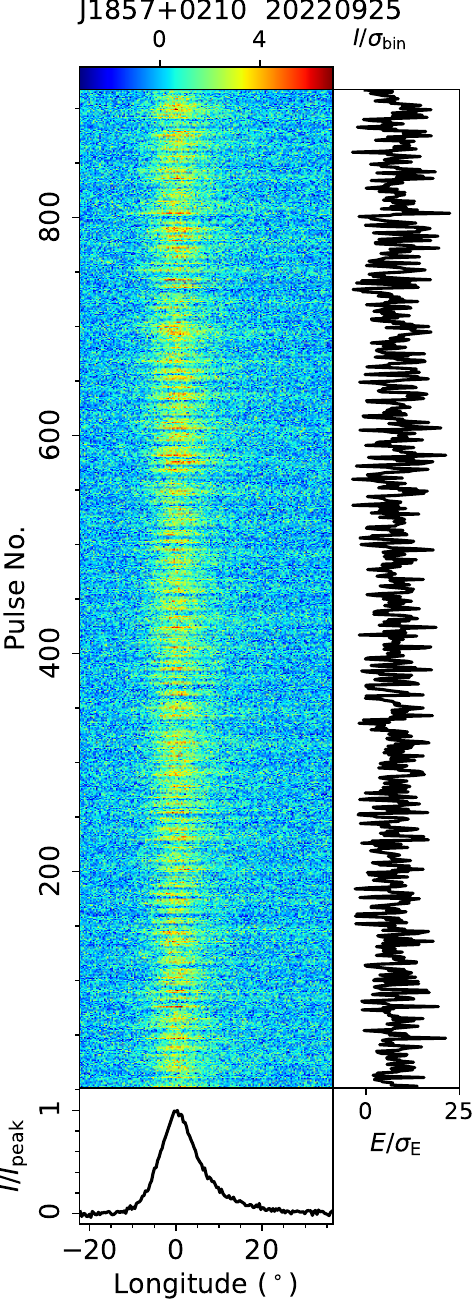}
\includegraphics[width=0.22\textwidth, angle=0]{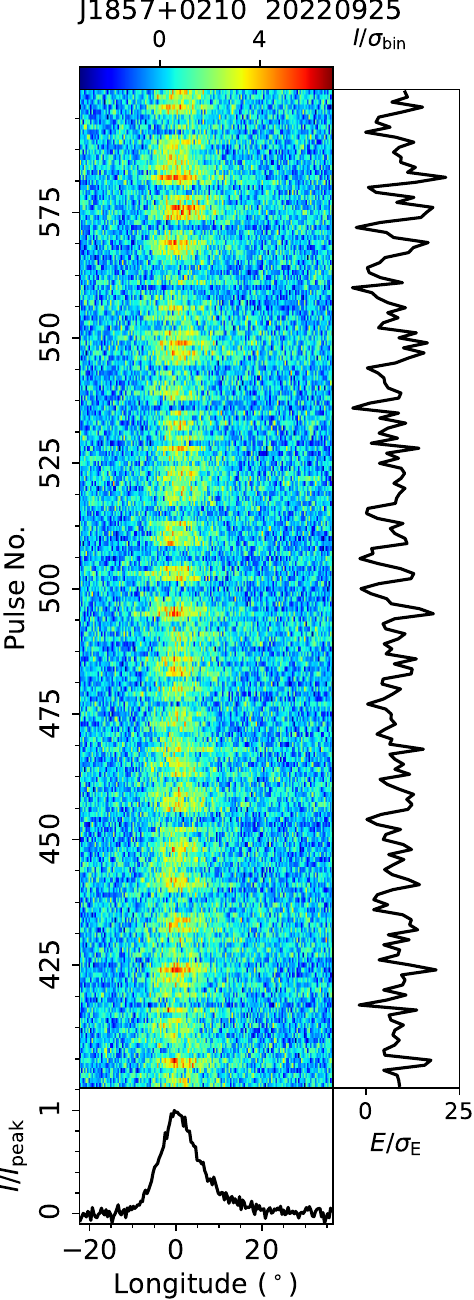}
\figcaption{Single pulse sequence of PSR J1857+0210 from the FAST observation on 20220925, and a zoomed-in view of pulses No. 400-600.
\label{subfig:TP:J1857+0210}}
\end{figure}

\begin{figure}[htpb]
\centering
\includegraphics[width=0.22\textwidth, angle=0]{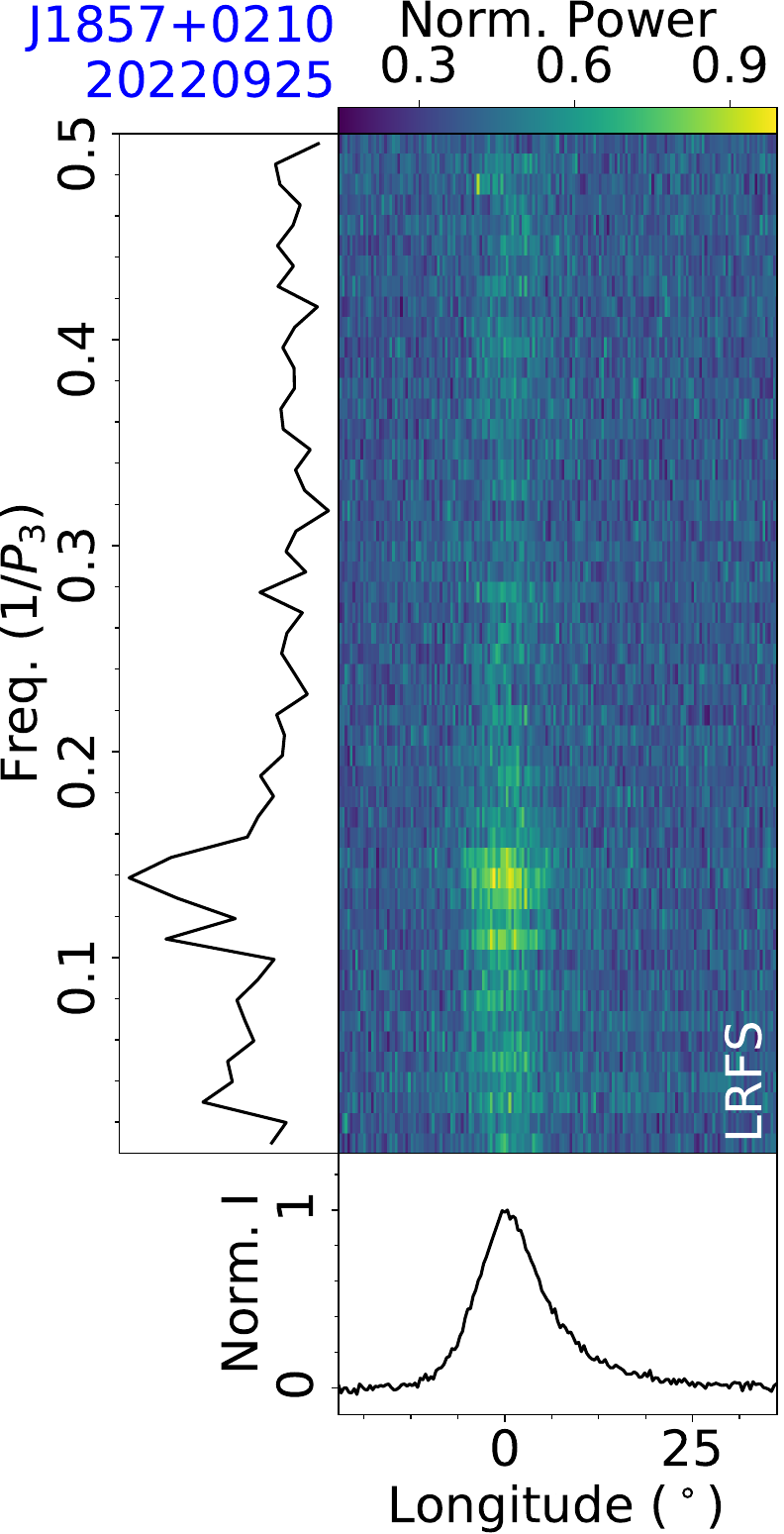}
\includegraphics[width=0.22\textwidth, angle=0]{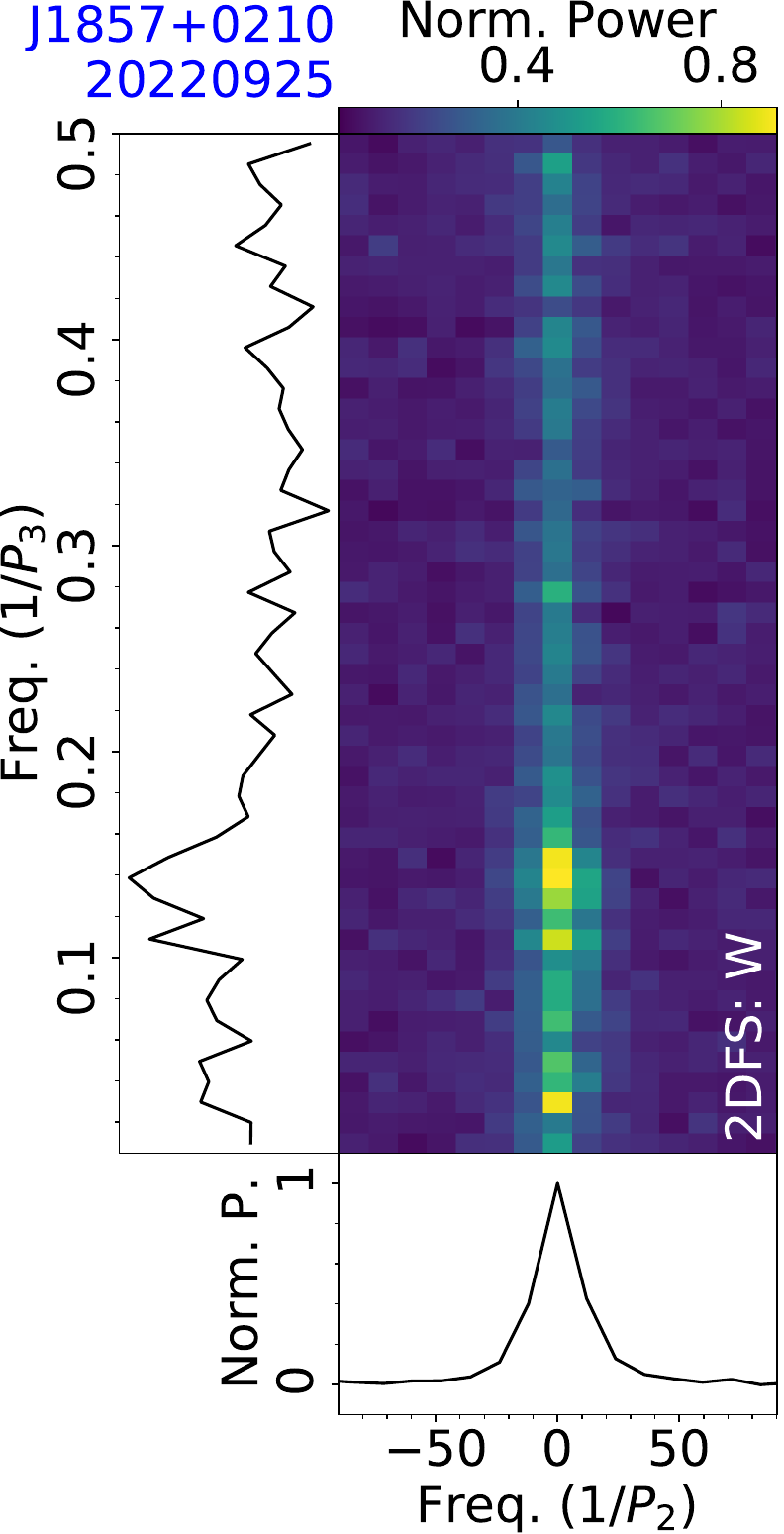}
\figcaption{Fluctuation analysis of PSR J1857+0210 for the observation on 20220925, with LRFS and 2DFS for the on-pulse region of a mean pulse profile. \label{subfig:fluctu:J1857+0210}}
\end{figure}

\begin{figure}[htpb]
\centering
\includegraphics[width=0.21\textwidth, angle=0]{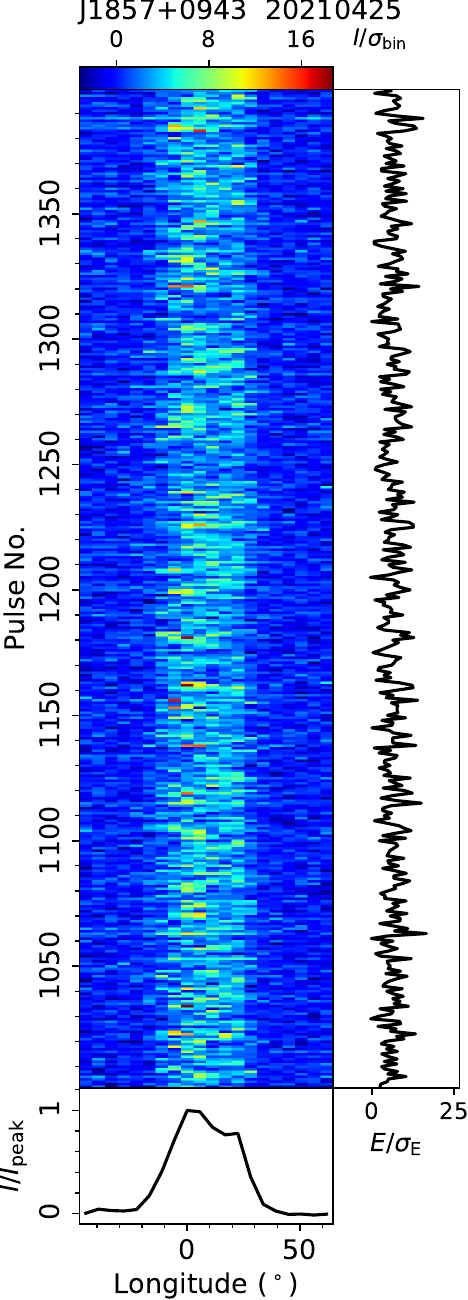}
\includegraphics[width=0.21\textwidth, angle=0]{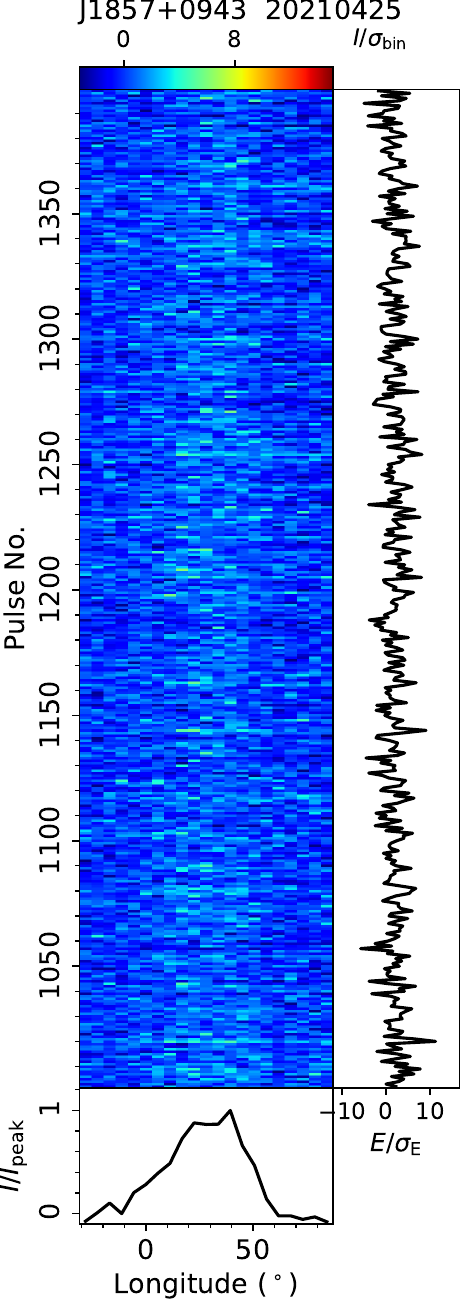}\\
\includegraphics[width=0.21\textwidth, angle=0]{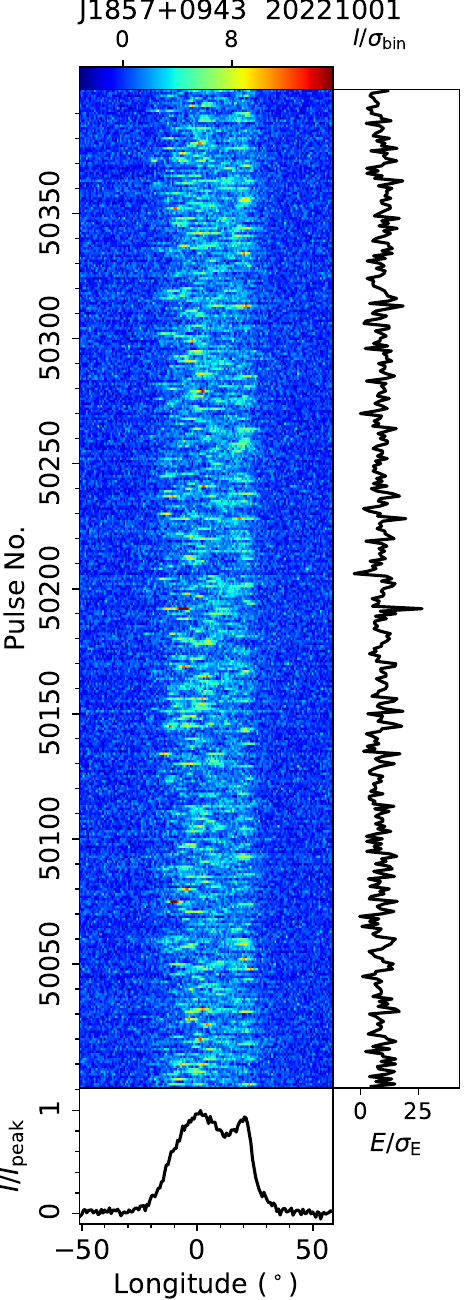}
\includegraphics[width=0.21\textwidth, angle=0]{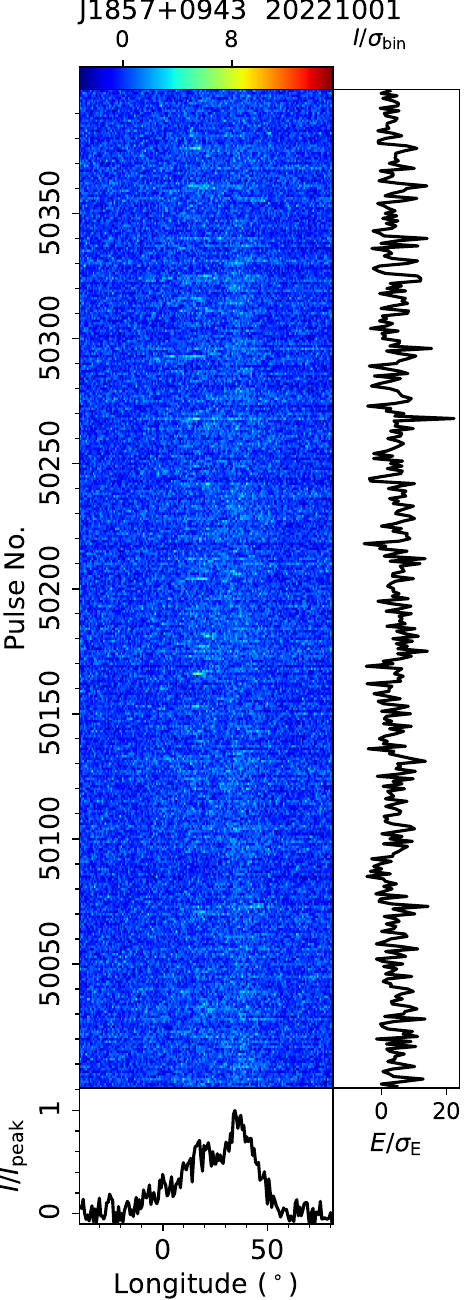}
\vspace{-0.3cm}
\figcaption{Zoomed-in segments of 400 pulses each for main pulse (left) and interpulse (right) of PSR J1857+0943, from the FAST observation on 20210425 (top) and 20221001 (bottom). The data from 20210425 and 20221001 were folded with 64 and 512 bins per pulse period, respectively.
\label{subfig:TP:J1857+0943}}
\end{figure}

\begin{figure}[htpb]
\centering
\includegraphics[width=0.22\textwidth, angle=0]{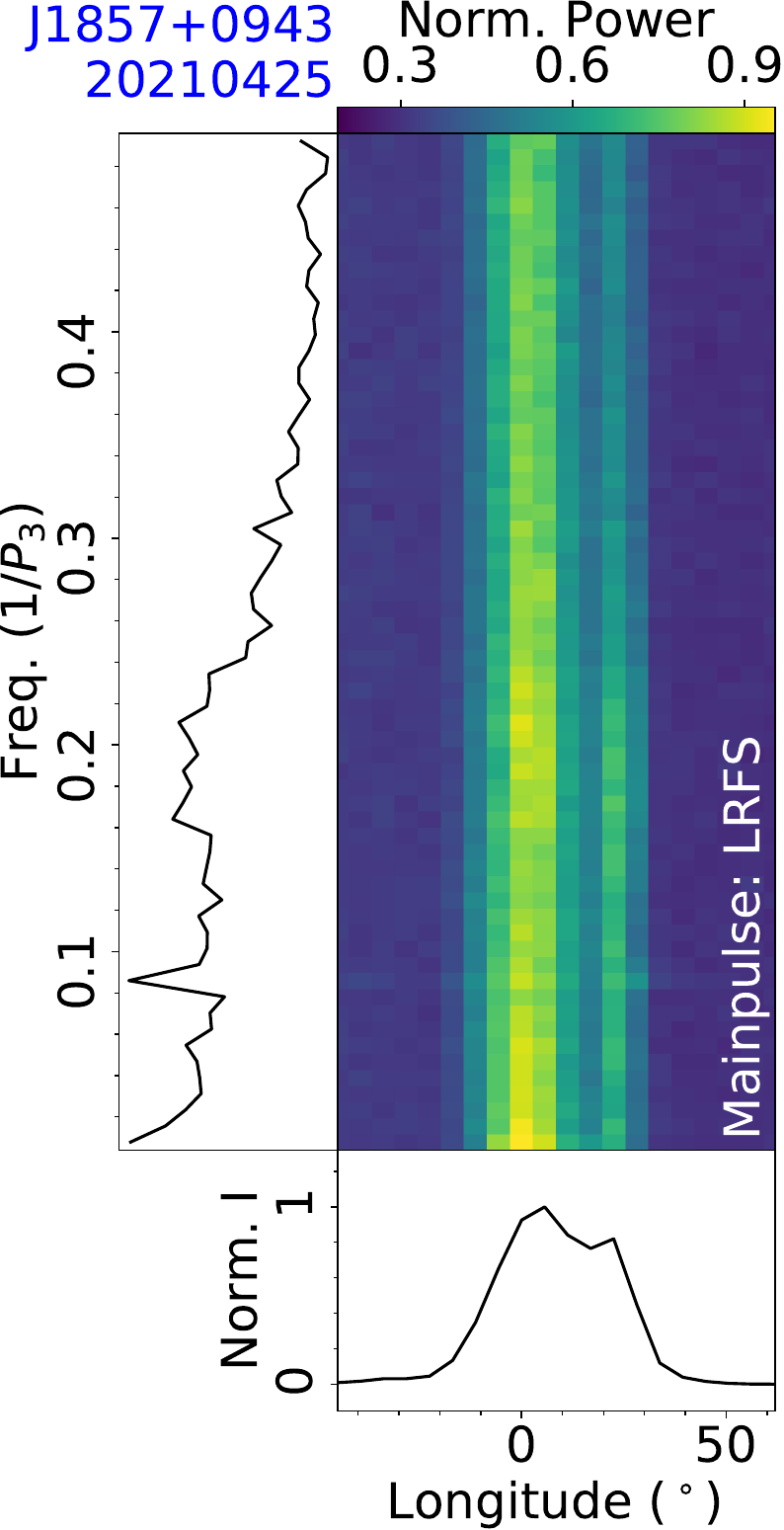}
\includegraphics[width=0.22\textwidth, angle=0]{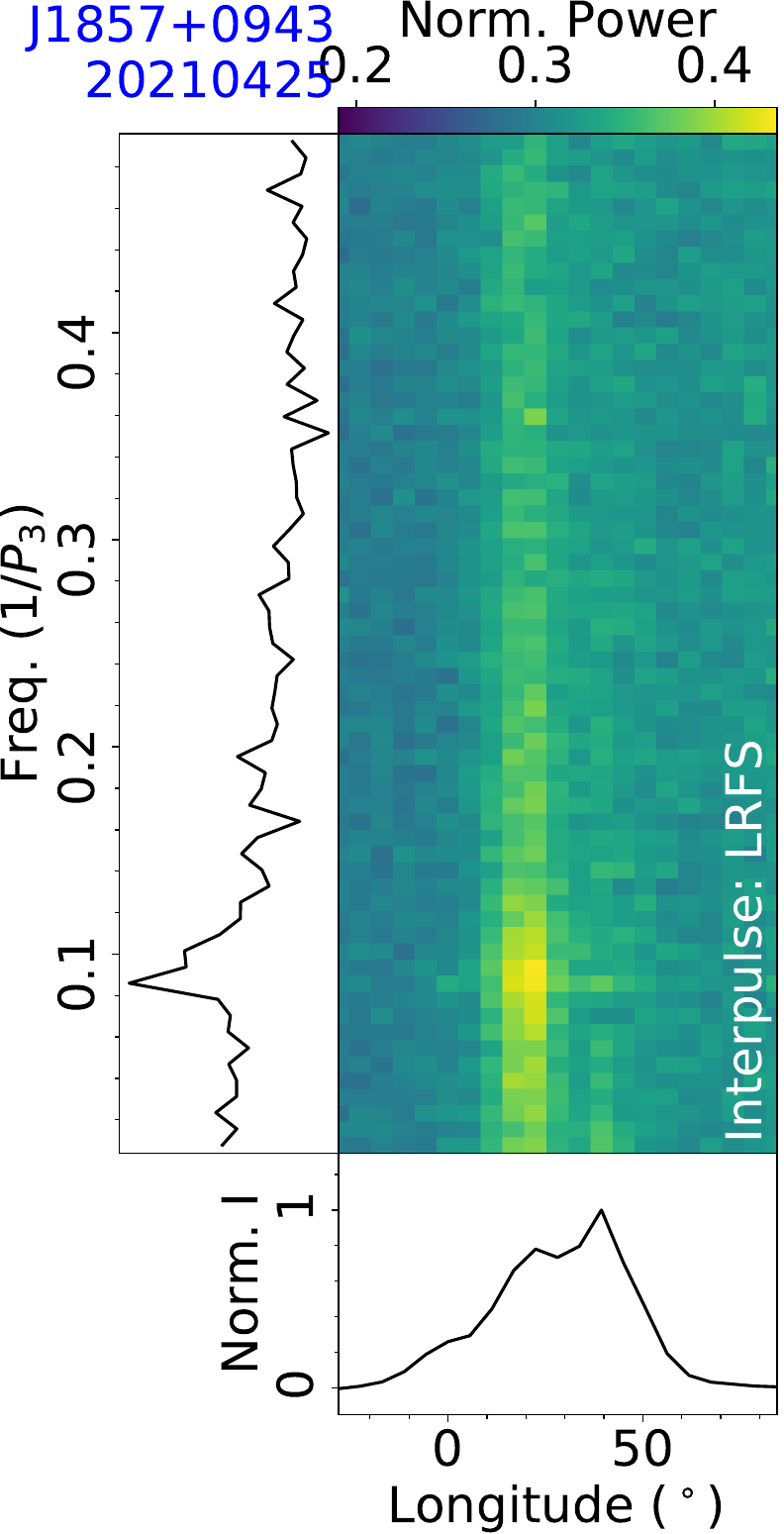}\\
\includegraphics[width=0.22\textwidth, angle=0]{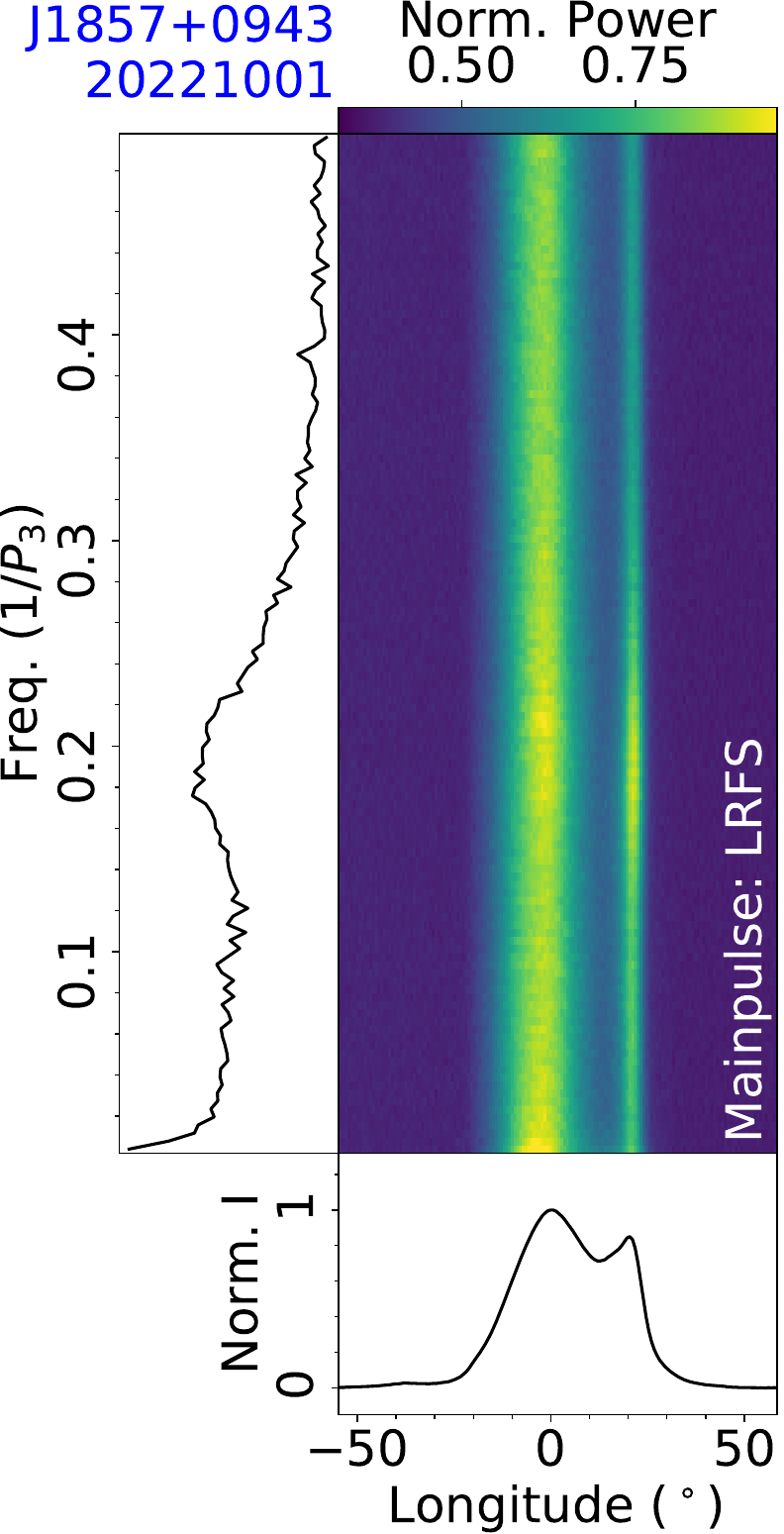}
\includegraphics[width=0.22\textwidth, angle=0]{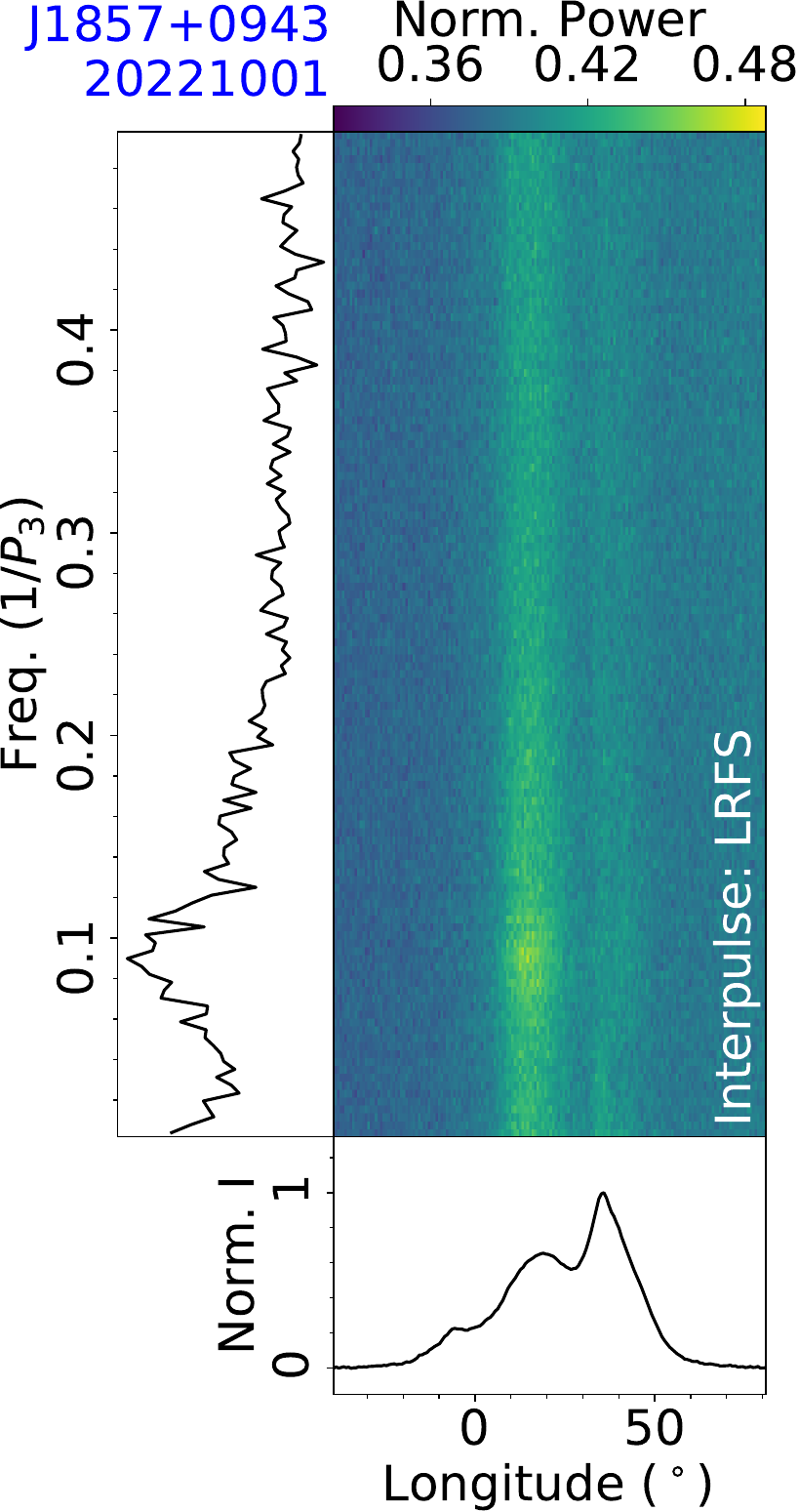}
\figcaption{LRFS of PSR J1857+0943 for the main pulse (left panels) and interpulse (right panels), from FAST observations on 20210425 (top) and 20221001 (bottom). 
\label{subfig:fluctu:J1857+0943}}
\end{figure}

\subsection{J1856-0134g}
\label{subsec:J1856-0134g}

PSR J1856-0134g was discovered in the FAST GPPS survey \citep{Han2021,han2025}. 

The pulsar was observed by FAST on 20230508 for 32 minutes, yielding a rotation period $P=0.3818$~s and a dispersion measure $D\!M=237.1~{\rm cm^{-3}\,pc}$. 
Single pulse sequences of this observation are shown in Fig.~\ref{subfig:TP:J1856-0134g}, illustrating the changes between the weak and bright emission states. Emission modes of single pulses are distinguished from the energy histogram of the on-pulse phase region (Fig.~\ref{subfig:Hist:J1856-0134g}), and averaged profiles of two states are contrasted in Fig.~\ref{subfig:ProfModes:J1856-0134g}. From the distribution of continuous period numbers for adjacent weak and bright modes in Fig.~\ref{subfig:scaleHist:J1856-0134g}, the duration of the bright emission mode is very short and mostly 1 period, while that of the weak mode is 12$\pm$11 periods.

\begin{figure}[htpb]
\centering
\includegraphics[width=0.22\textwidth, angle=0]{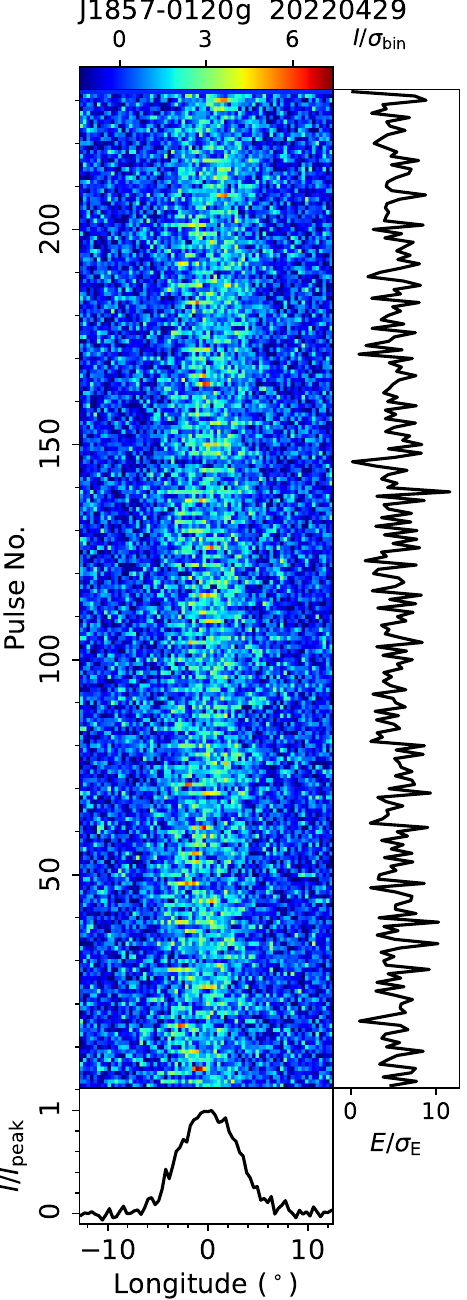}
\figcaption{Single pulse sequence of PSR J1857-0120 from the FAST observation on 20220429.
\label{subfig:TP:J1857-0120}}
\end{figure}

\begin{figure}[htpb]
\centering
\includegraphics[width=0.22\textwidth, angle=0]{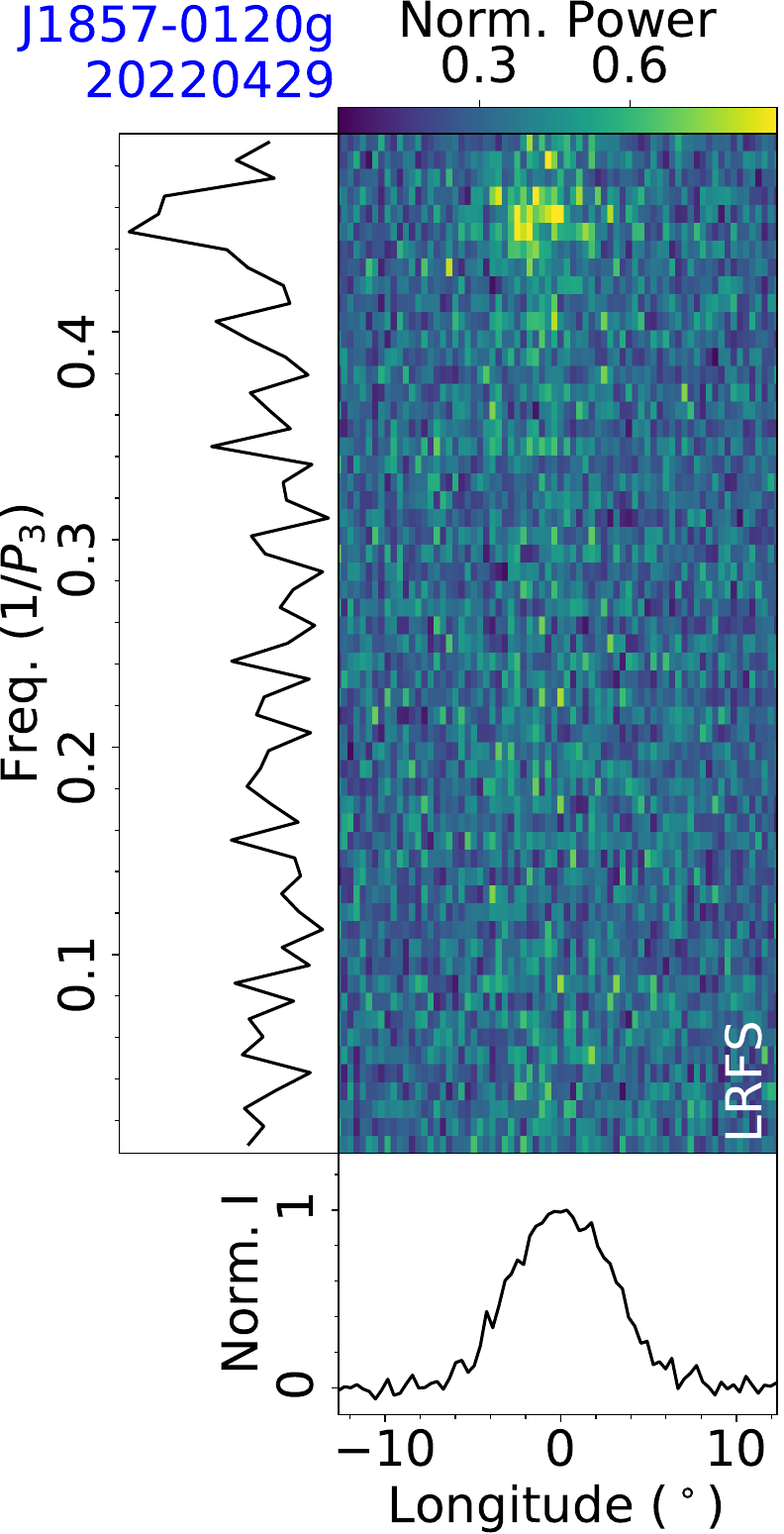}
\includegraphics[width=0.22\textwidth, angle=0]{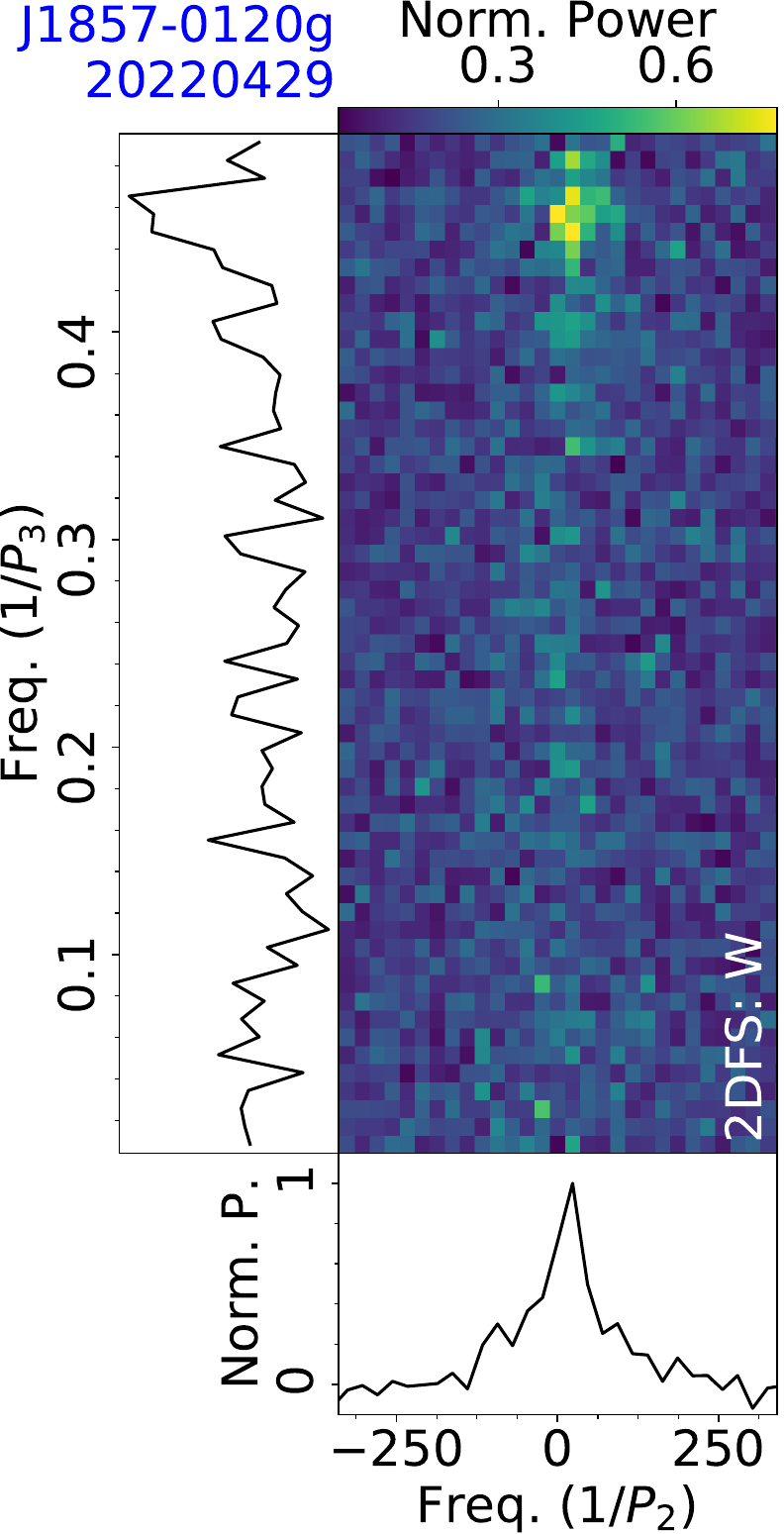}
\figcaption{Fluctuation analysis of PSR J1857-0120 for the observation on 20220429, with LRFS and 2DFS for the on-pulse region of a mean pulse profile. \label{subfig:fluctu:J1857-0120}}
\end{figure}

\subsection{J1856-0526}
\label{subsec:J1856-0526}

PSR J1856-0526 was discovered in the Parkes multibeam pulsar survey \citep{Lorimer2006}. \citet{Song2023} subsequently analysed the subpulse modulation properties. 
For one component, they found a $P_3$-only modulation with $P_3=13\pm2$ periods. The other component exhibits a negative drift feature characterized by $P_3=2.5\pm0.1$ periods and $P_2=-68^{+58}_{-21}$ degrees, as well as a $P_3$-only modulation with $P_3=11\pm1$ periods. 

This pulsar was observed by FAST on 20250304 and 20250429, each for 15 minutes. From the data on 20250304, a rotation period $P=0.3705$~s and a dispersion measure $D\!M=132.2~{\rm cm^{-3}\,pc}$ were determined. 
In the single pulse sequence and a zoomed-in view of pulses No. 100-700 from the FAST observation on 20250304 (Fig.~\ref{subfig:TP:J1856-0526}), two drifting modes are evident. These are designated as the central-weak and central-strong modes, based on the relative energy integrated over the central phase interval of the mean pulse profile. 
The corresponding energy histogram in Fig.~\ref{subfig:Hist:J1856-0526} then distinguishes single pulses of the two drifting modes, labeled in red and green, respectively. 
A comparison of the mean polarization profiles and averaged PA curves for the two emission modes is shown in Fig.~\ref{subfig:PolModes:J1856-0526}. The PA curves of the two modes diverge in the trailing profile, starting at the longitude of around $-10^\circ$. 
In addition, a $\sim$2-period modulation was also observed from Fig.~\ref{subfig:TP:J1856-0526}, nested within the lower-frequency drifting pattern, especially for the trailing profile part of the central-weak mode. 
Figure~\ref{subfig:fluctu:J1856-0526} displays the fluctuation spectra for the two drifting modes, characterizing their subpulse modulation properties. 
For the central-weak drifting mode, the lower-frequency positive drift feature has a centroid at $1/P_3=0.081\pm0.002$ and $1/P_2=7\pm1$ ($P_3=12.4\pm0.3$ periods, $P_2=48\pm6$ degrees). 
Near the temporally high frequency boundary, both negative and positive drift features appear, widely spread in $1/P_3$, which is most prominent in the trailing part of the mean pulse profile. 
The preferred negative drift feature is with a centroid of $1/P_3=0.404\pm0.002$ and $1/P_2=-40\pm1$ ($P_3=2.47\pm0.01$ periods, $P_2=-9.0\pm0.1$ degrees). 
For the central-strong drifting mode, the centroid of the positive drift feature is at $1/P_3=0.095\pm0.002$ and $1/P_2=0.8\pm0.6$ ($P_3=10.5\pm0.2$ periods, $P_2=447\pm307$ degrees). The centroid of the preferred negative drift feature is at $1/P_3=0.346\pm0.004$ and $1/P_2=-42.0\pm0.4$ ($P_3=2.89\pm0.04$ periods, $P_2=-8.6\pm0.1$ degrees). 
The single pulse behaviors from the observation on 20250429 are consistent with those on 20250304.

\subsection{J1857+0210}
\label{subsec:J1857+0210}

PSR J1857+0210 was discovered in the Parkes Multibeam Pulsar Survey \citep{Morris2002}. 

This pulsar was observed by FAST on 20220925 for 10 minutes, and a rotation period $P=0.6310$~s and a dispersion measure $D\!M=782.8~{\rm cm^{-3}\,pc}$ were derived. The single pulse sequence and a zoomed-in view of pulses No. 400-600  in Fig.~\ref{subfig:TP:J1857+0210} show the intensity modulation behavior. LRFS and 2DFS are displayed in Fig.~\ref{subfig:fluctu:J1857+0210}, and the centroid frequency of the modulation feature is estimated to be $1/P_3=0.125\pm0.002$, corresponding to the periodicity of $P_3=8.0\pm0.1$ periods.

\subsection{J1857+0943}
\label{subsec:J1857+0943}

PSR J1857+0943 was discovered by the Arecibo telescope \citep{Segelstein1986}. 

This pulsar was observed by FAST on 20210425 for 5 minutes and on 20221001 for 26 minutes, with a rotation period $P=0.0054$~s and a dispersion measure $D\!M=13.3~{\rm cm^{-3}\,pc}$ derived. 
Zoomed-in views of 400 pulses for the main pulse and interpulse from two observations are shown in Fig.~\ref{subfig:TP:J1857+0943}. From the LRFS of the main pulse and interpulse in Fig.~\ref{subfig:fluctu:J1857+0943}, the temporal modulation frequency is widely distributed, and the main modulation properties are different for the main pulse and interpulse. 
For the data on 20210425, the centroid frequency of the main modulation feature is $1/P_3=0.1951\pm0.0005$ ($P_3=5.13\pm0.01$ periods) for the main pulse, and $0.093\pm0.001$ ($P_3=10.8\pm0.1$ periods) for the interpulse. For the data on 20221001, the derived parameters are $1/P_3=0.1934\pm0.0003$ ($P_3=5.17\pm0.01$ periods) for the main pulse, and $1/P_3=0.0926\pm0.0003$ ($P_3=10.80\pm0.03$ periods) for the interpulse.

\begin{figure}[htpb]
\centering
\includegraphics[width=0.22\textwidth, angle=0]{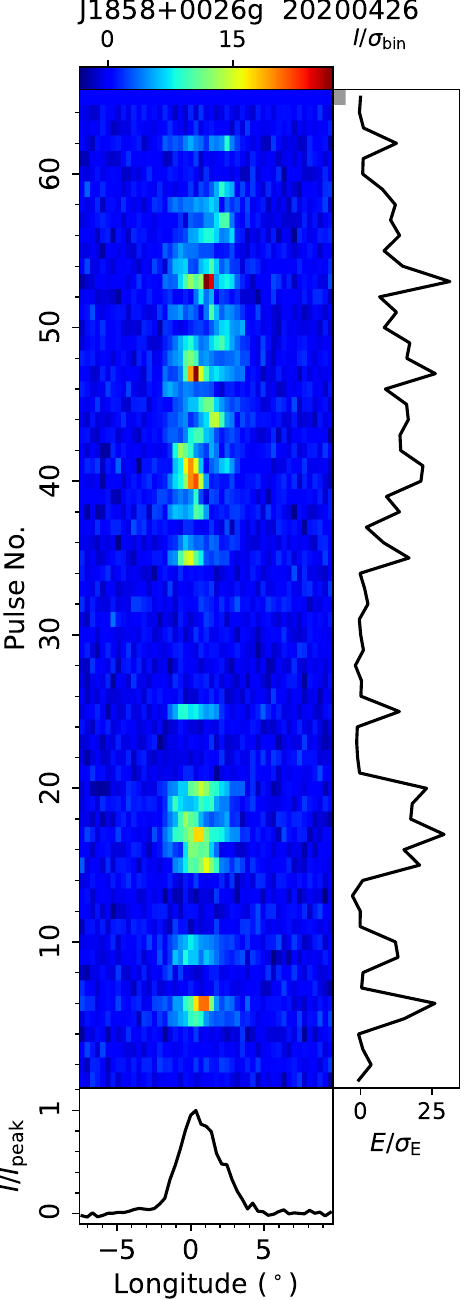}
\includegraphics[width=0.22\textwidth, angle=0]{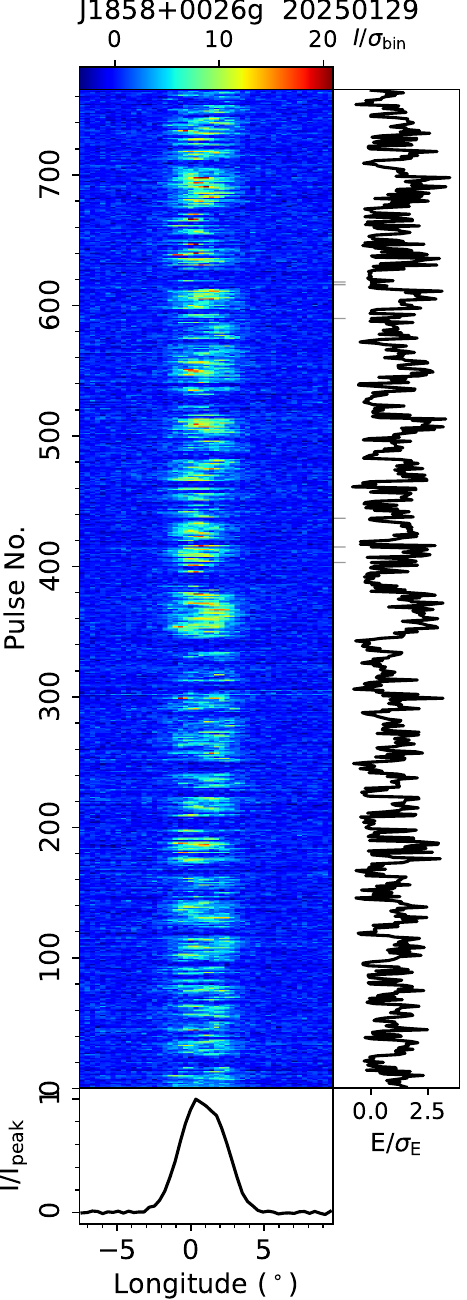}\\
\includegraphics[width=0.22\textwidth, angle=0]{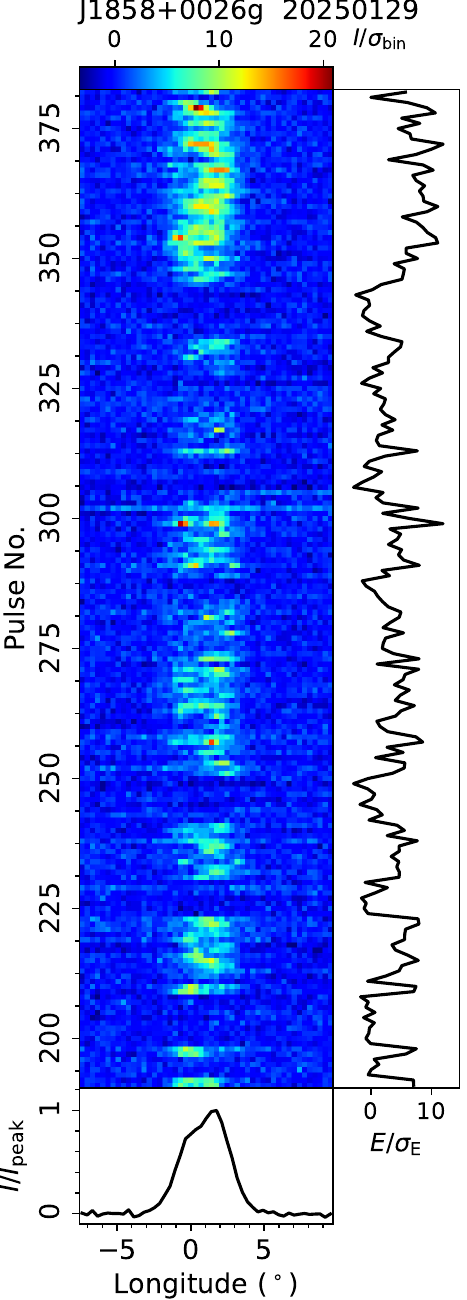}
\includegraphics[width=0.22\textwidth, angle=0]{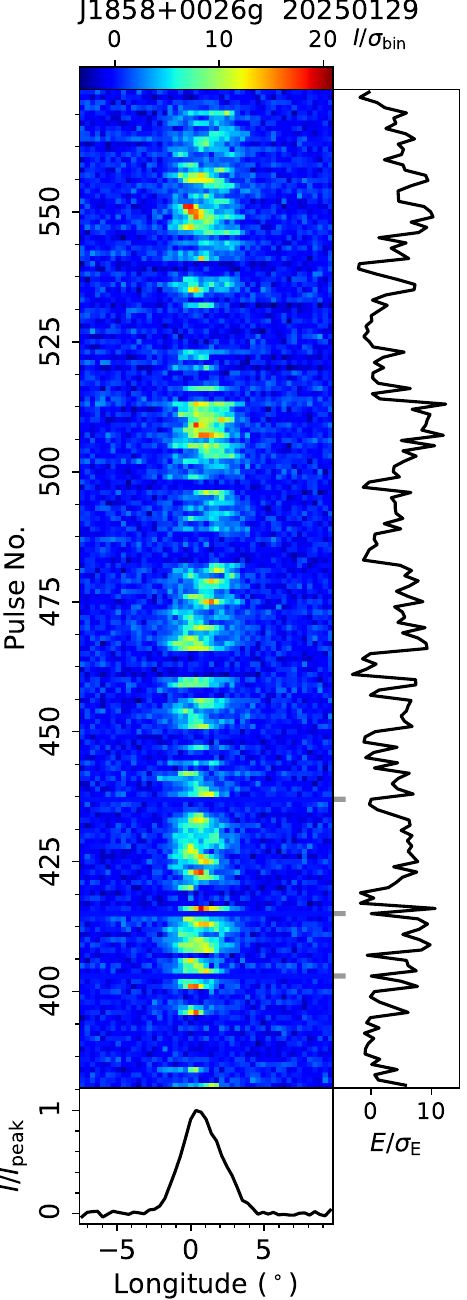}
\vspace{-0.3cm}
\figcaption{Single pulse sequences of PSR J1858+0026g from the FAST observation on 20250129.
\label{subfig:TP:J1858+0026g}}
\end{figure}

\begin{figure}[htpb]
\centering
\includegraphics[width=0.39\textwidth, angle=0]{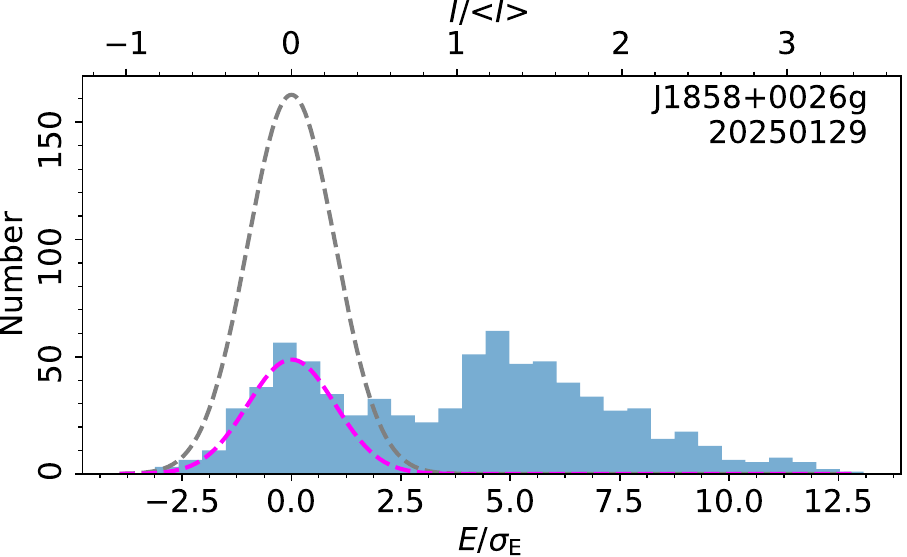}
\figcaption{On-pulse energy histogram of single pulses of PSR J1858+0026g from the FAST observation on 20250129.
\label{subfig:Hist:J1858+0026g}}
\end{figure}

\begin{figure}[htpb]
\centering
\includegraphics[width=0.22\textwidth, angle=0]{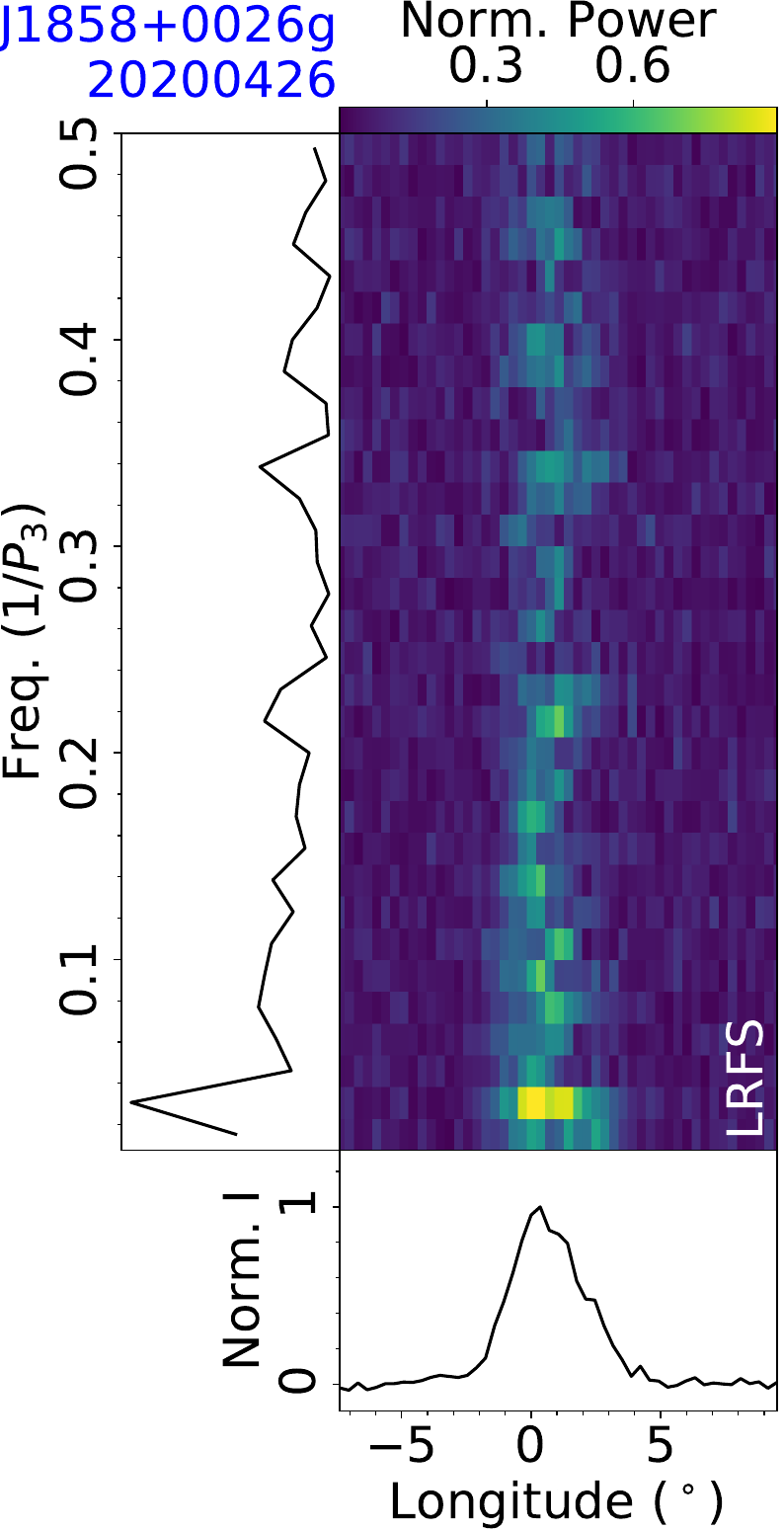}
\includegraphics[width=0.22\textwidth, angle=0]{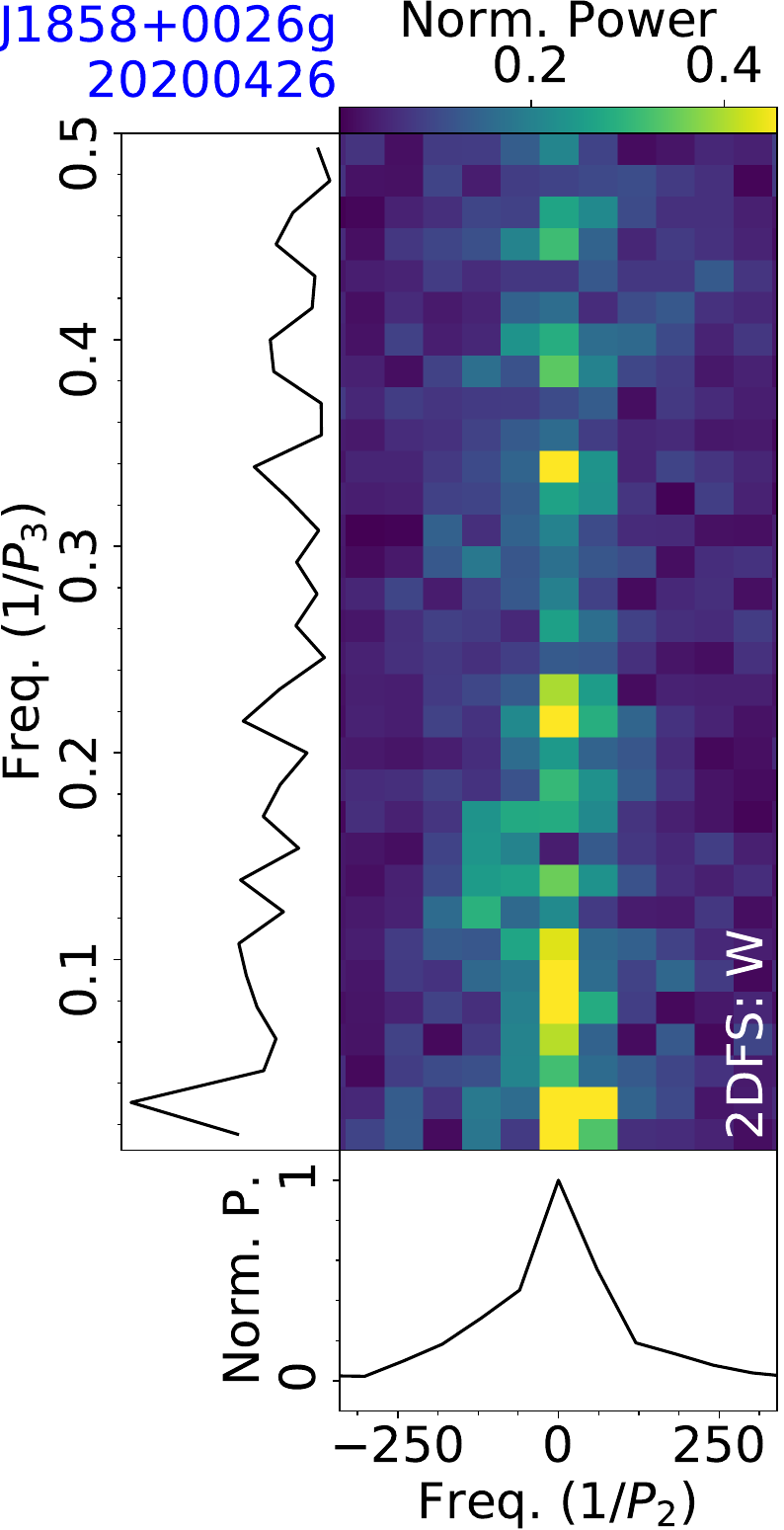}
\figcaption{Fluctuation analysis of PSR J1858+0026g for the observation on 20200426, with LRFS and 2DFS for the on-pulse region of a mean pulse profile.
\label{subfig:fluctu:J1858+0026g}}
\end{figure}

\subsection{J1857-0120g}
\label{subsec:J1857-0120g}

PSR J1857-0120g was discovered in the FAST GPPS survey \citep{Han2021,han2025}. 

This pulsar was observed by FAST on 20220429 for 5 minutes and on 20250129 for 20 minutes. From the 20-minute data, a rotation period $P=1.2225$~s and a dispersion measure $D\!M=352.6~{\rm cm^{-3}\,pc}$ were determined. 
The single pulse sequence of the observation on 20220429 is displayed in Fig.~\ref{subfig:TP:J1857-0120}. From LRFS and 2DFS in Fig.~\ref{subfig:fluctu:J1857-0120}, the centroid of the drift feature is characterized by $1/P_3=0.459\pm0.001$ and $1/P_2=31\pm2$, corresponding to drifting parameters of $P_3=2.18\pm0.01$ periods and $P_2=11\pm1^\circ$.

\begin{figure}[htpb]
\centering
\includegraphics[width=0.22\textwidth, angle=0]{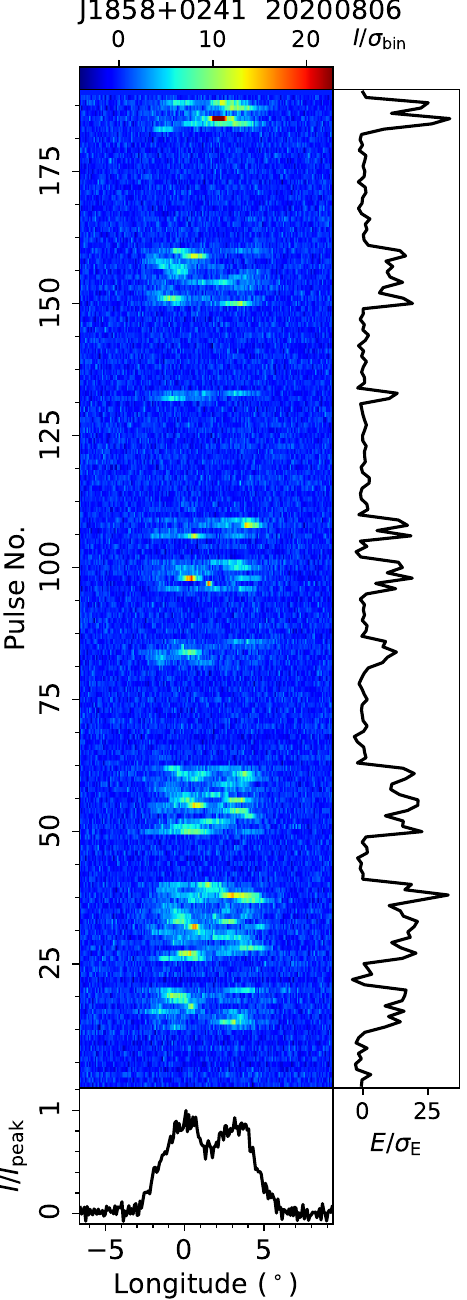}
\includegraphics[width=0.22\textwidth, angle=0]{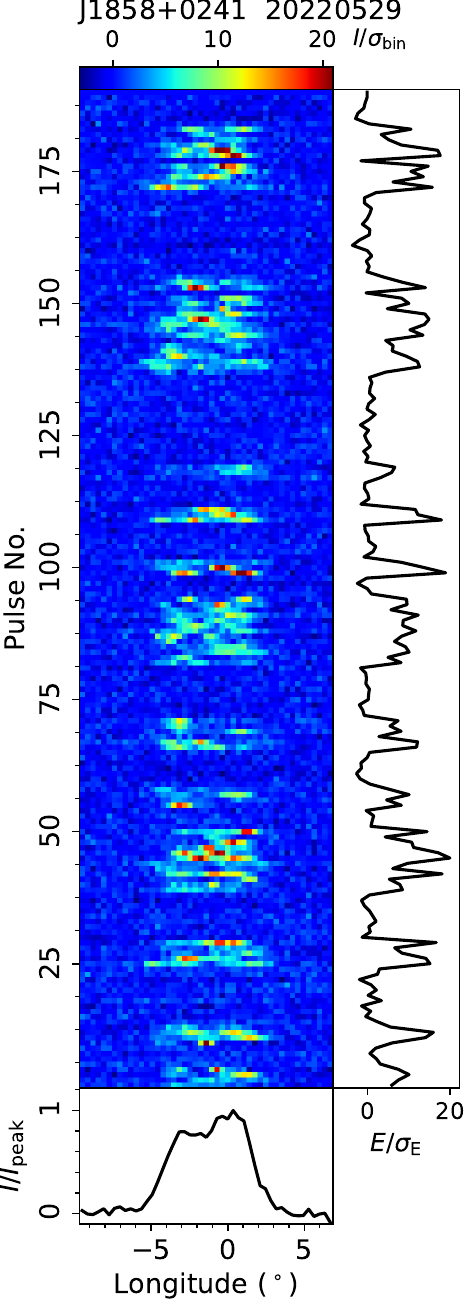}
\figcaption{Single pulse sequences of PSR J1858+0241 from the FAST observation on 20220529.
\label{subfig:TP:J1858+0241}}
\end{figure}

\begin{figure}[htpb]
\centering
\includegraphics[width=0.39\textwidth, angle=0]{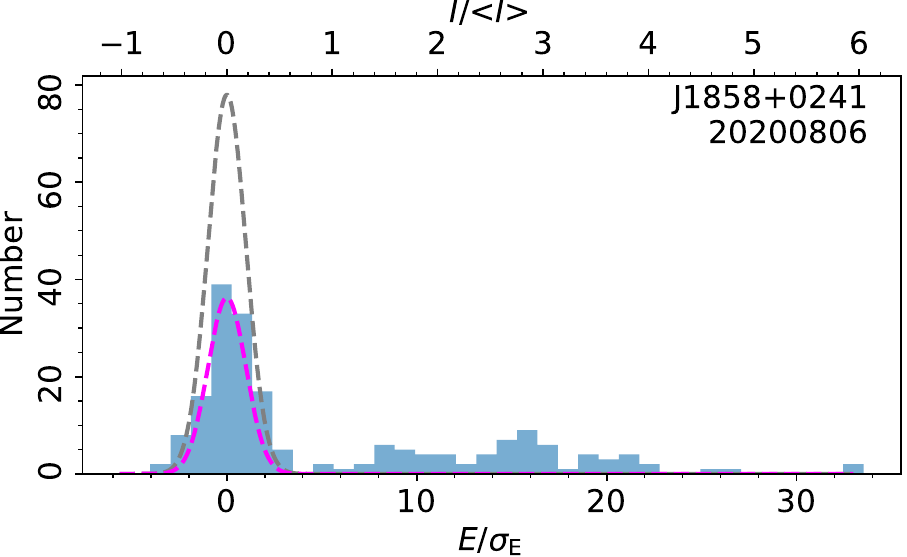}
\includegraphics[width=0.39\textwidth, angle=0]{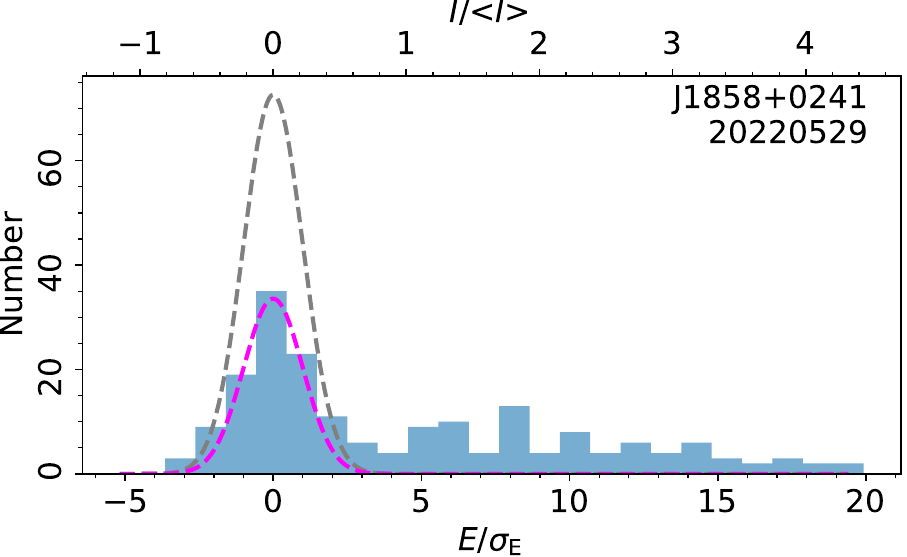}
\figcaption{On-pulse energy histograms of single pulses of PSR J1858+0241 from the FAST observations on 20200806 and 20220529.
\label{subfig:Hist:J1858+0241}}
\end{figure}

\begin{figure}[htpb]
\centering
\includegraphics[width=0.22\textwidth, angle=0]{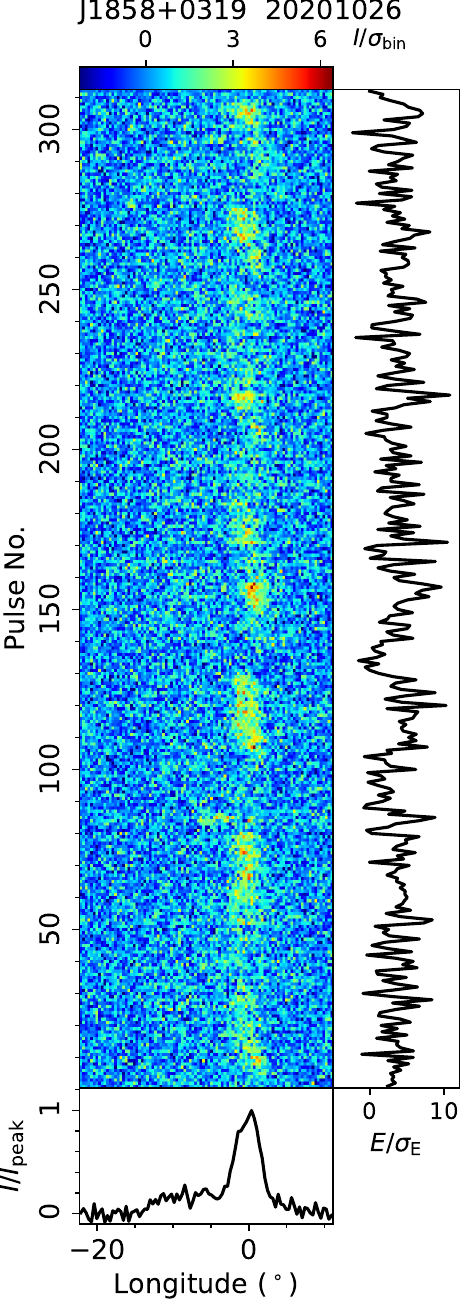}
\figcaption{Single pulse sequence of PSR J1858+0319 from the FAST observation on 20201026.
\label{subfig:TP:J1858+0319}}
\end{figure}

\begin{figure}[htpb]
\centering
\includegraphics[width=0.22\textwidth, angle=0]{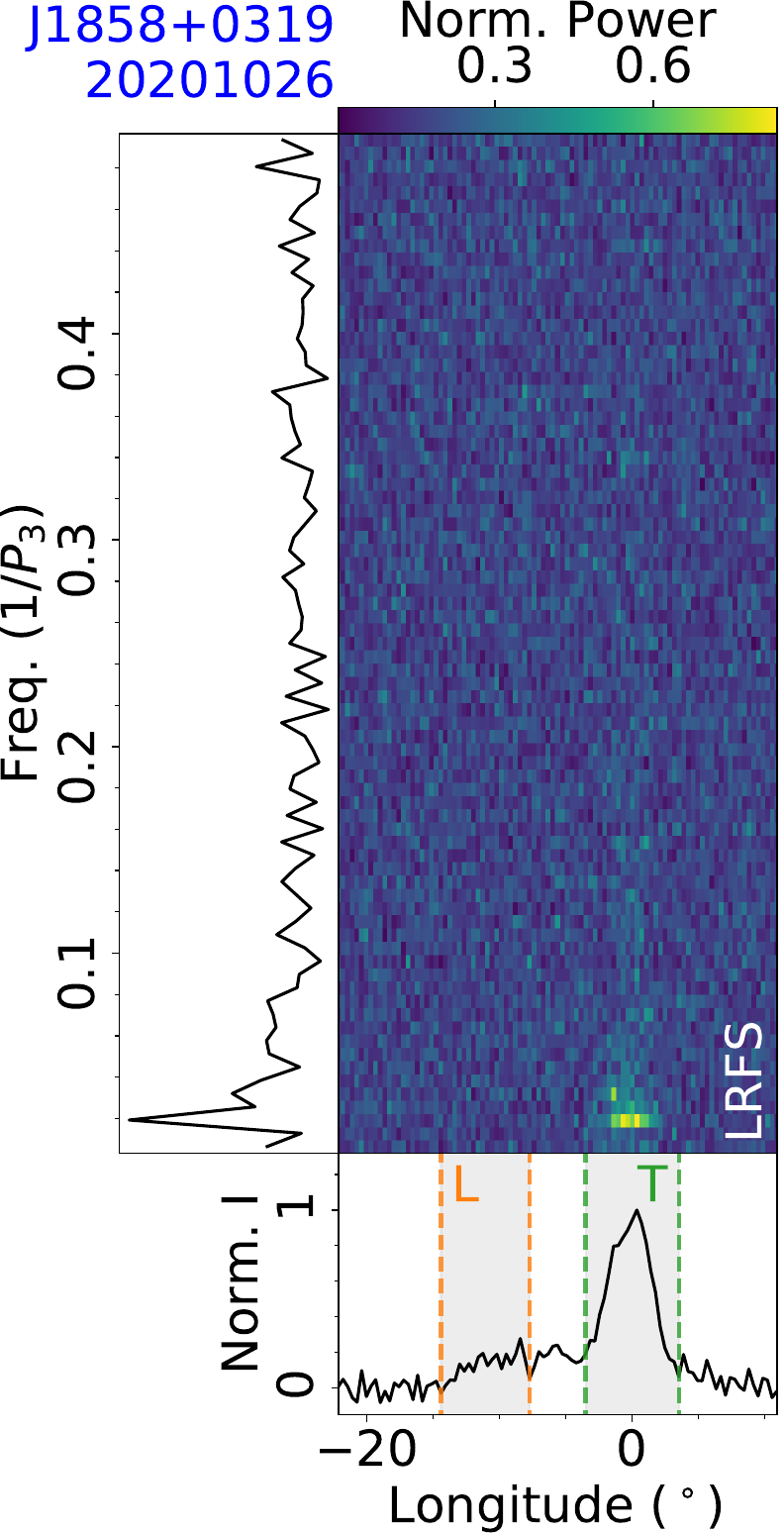}
\includegraphics[width=0.22\textwidth, angle=0]{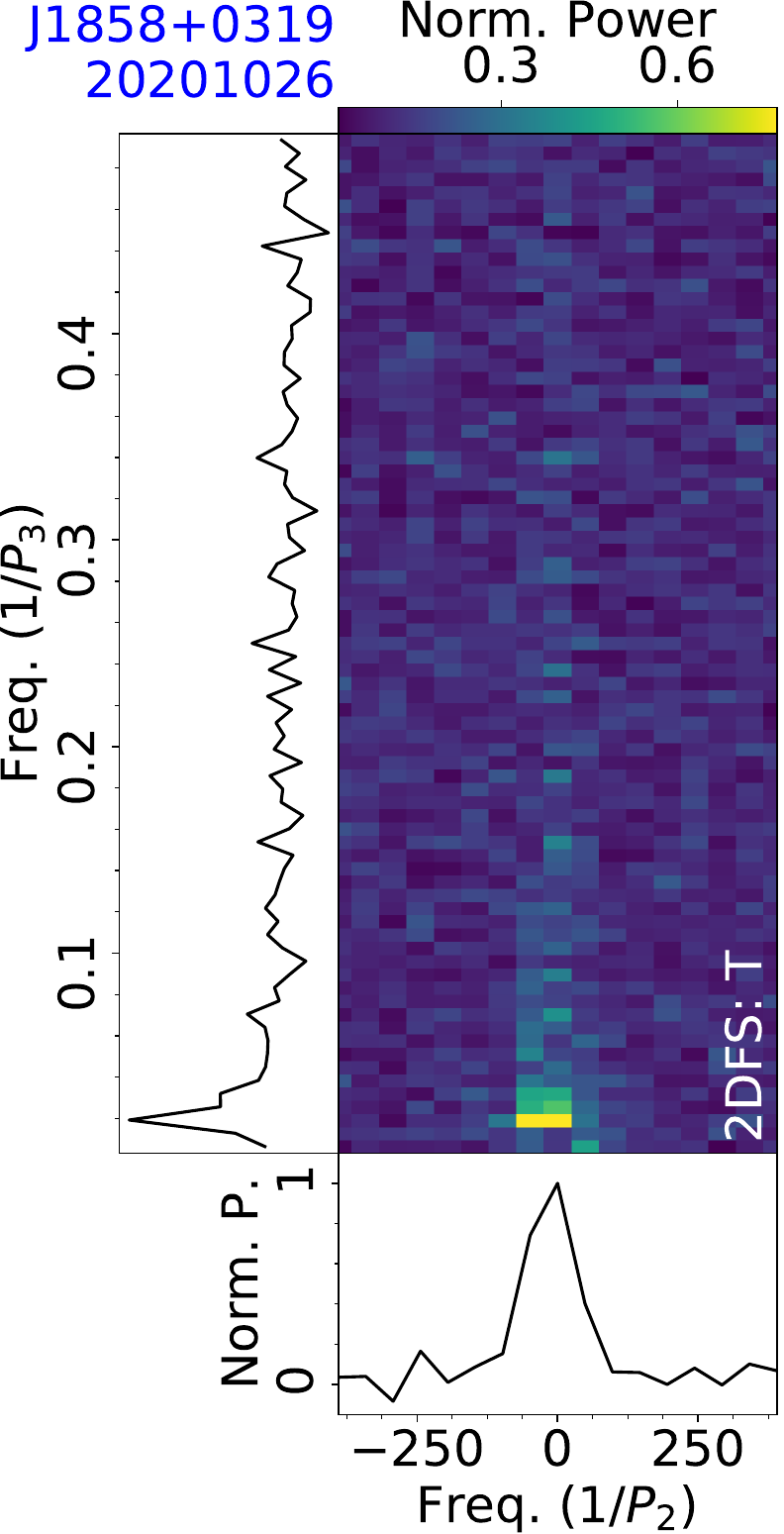}
\figcaption{Fluctuation analysis of PSR J1858+0319 from the FAST observation on 20201026, with LRFS and 2DFS for the trailing part of a mean pulse profile.
\label{subfig:fluctu:J1858+0319}}
\end{figure}

\begin{figure}[htpb]
\centering
\includegraphics[width=0.44\textwidth, angle=0]{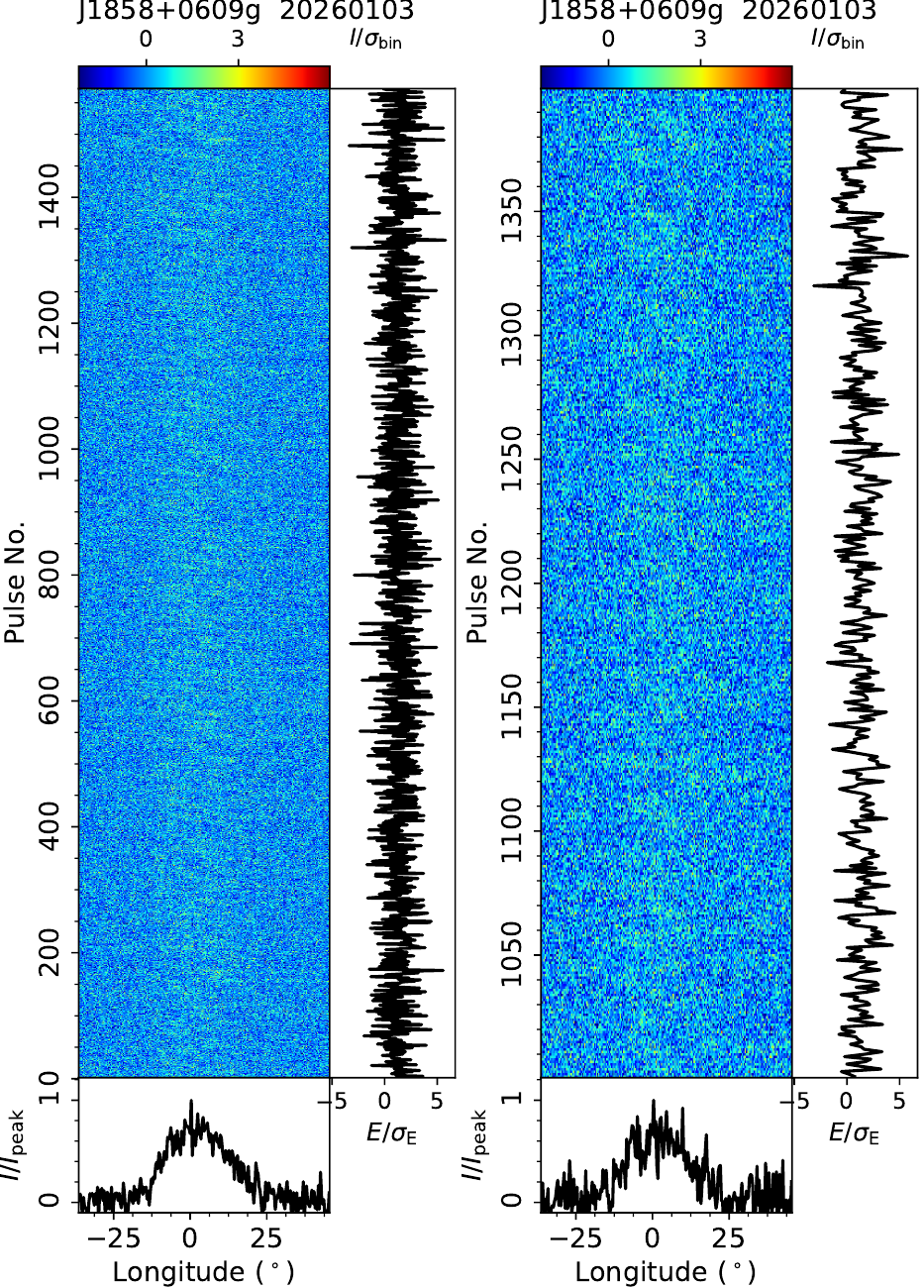}
\figcaption{Single pulse sequence of PSR J1858+0609g from the FAST observation on 20260103, and a zoomed in view of pulses No. 1000-1400.
\label{subfig:TP:J1858+0609g}}
\end{figure}

\begin{figure}[htpb]
\centering
\includegraphics[width=0.44\textwidth, angle=0]{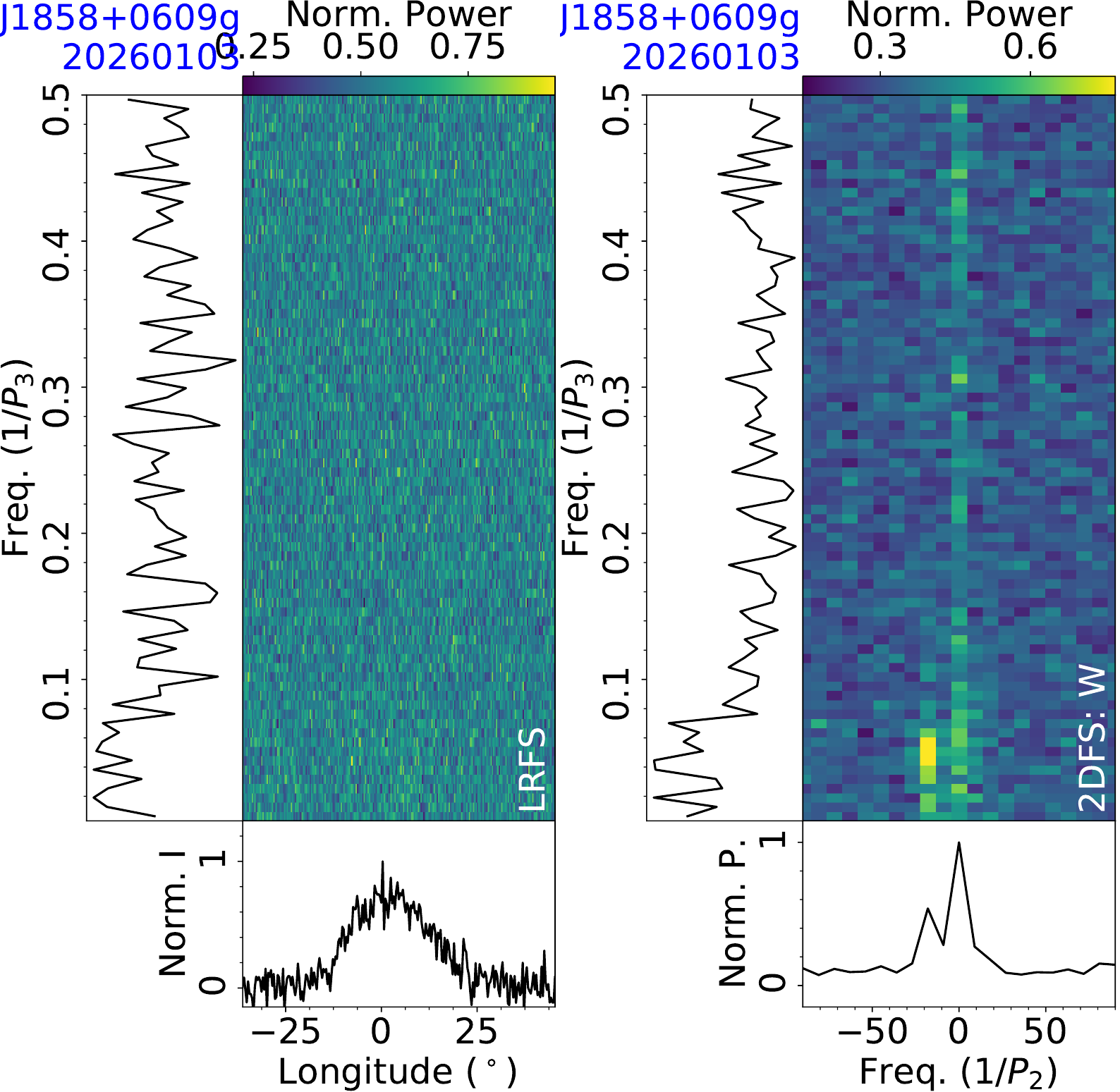}
\figcaption{Fluctuation analysis of PSR J1858+0609g for the observation on 20260103, with LRFS and 2DFS for the on-pulse region of the mean pulse profile.
\label{subfig:fluctu:J1858+0609g}}
\end{figure}

\begin{figure}[htpb]
\centering
\includegraphics[width=0.22\textwidth, angle=0]{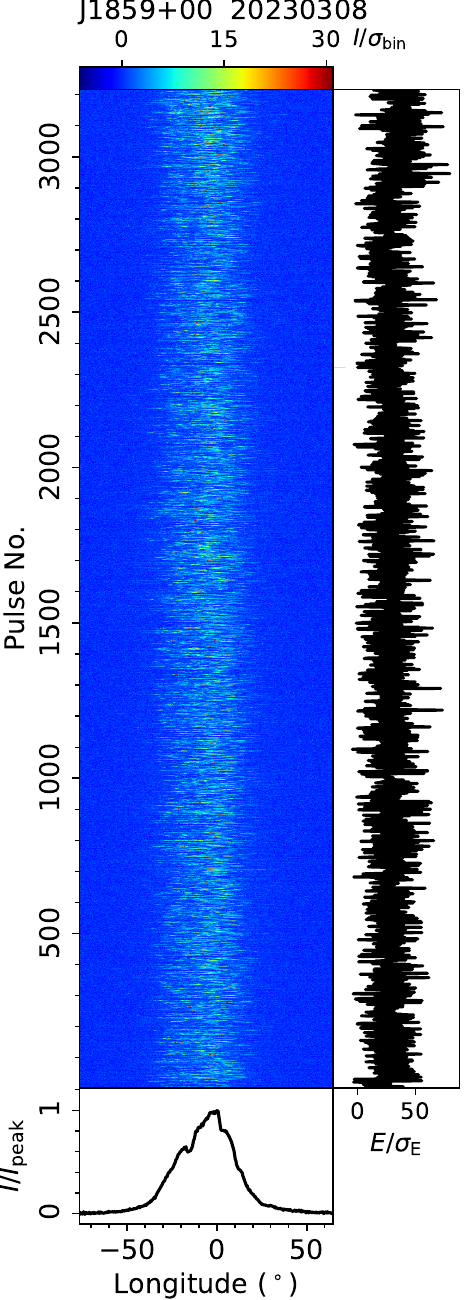}
\includegraphics[width=0.22\textwidth, angle=0]{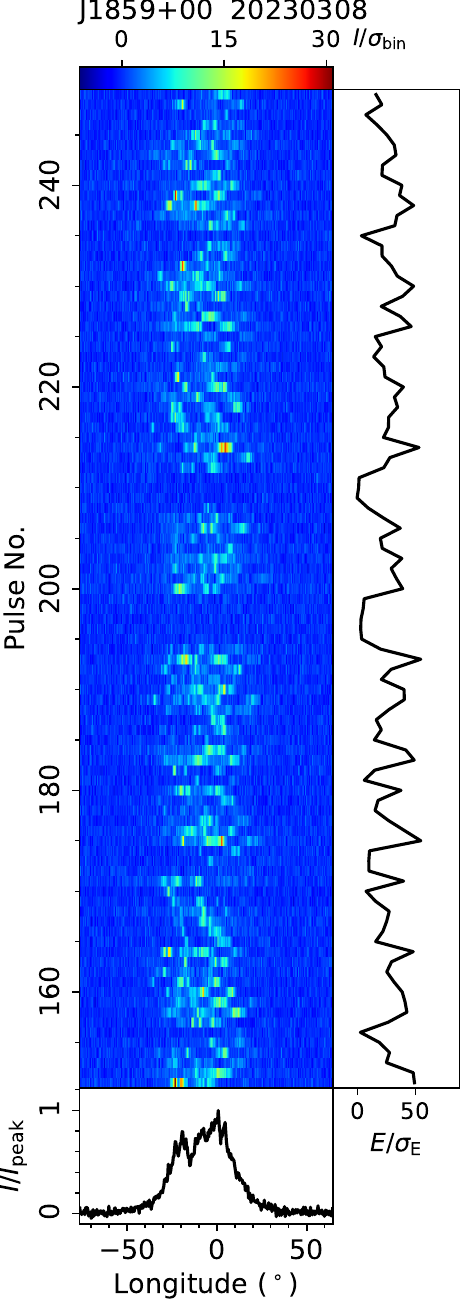}
\figcaption{Single pulse sequence of PSR J1859+00 from the FAST observation on 20230308, and a zoomed-in view of pulses No. 150-250.
\label{subfig:TP:J1859+00}}
\end{figure}

\begin{figure}[htpb]
\centering
\includegraphics[width=0.22\textwidth, angle=0]{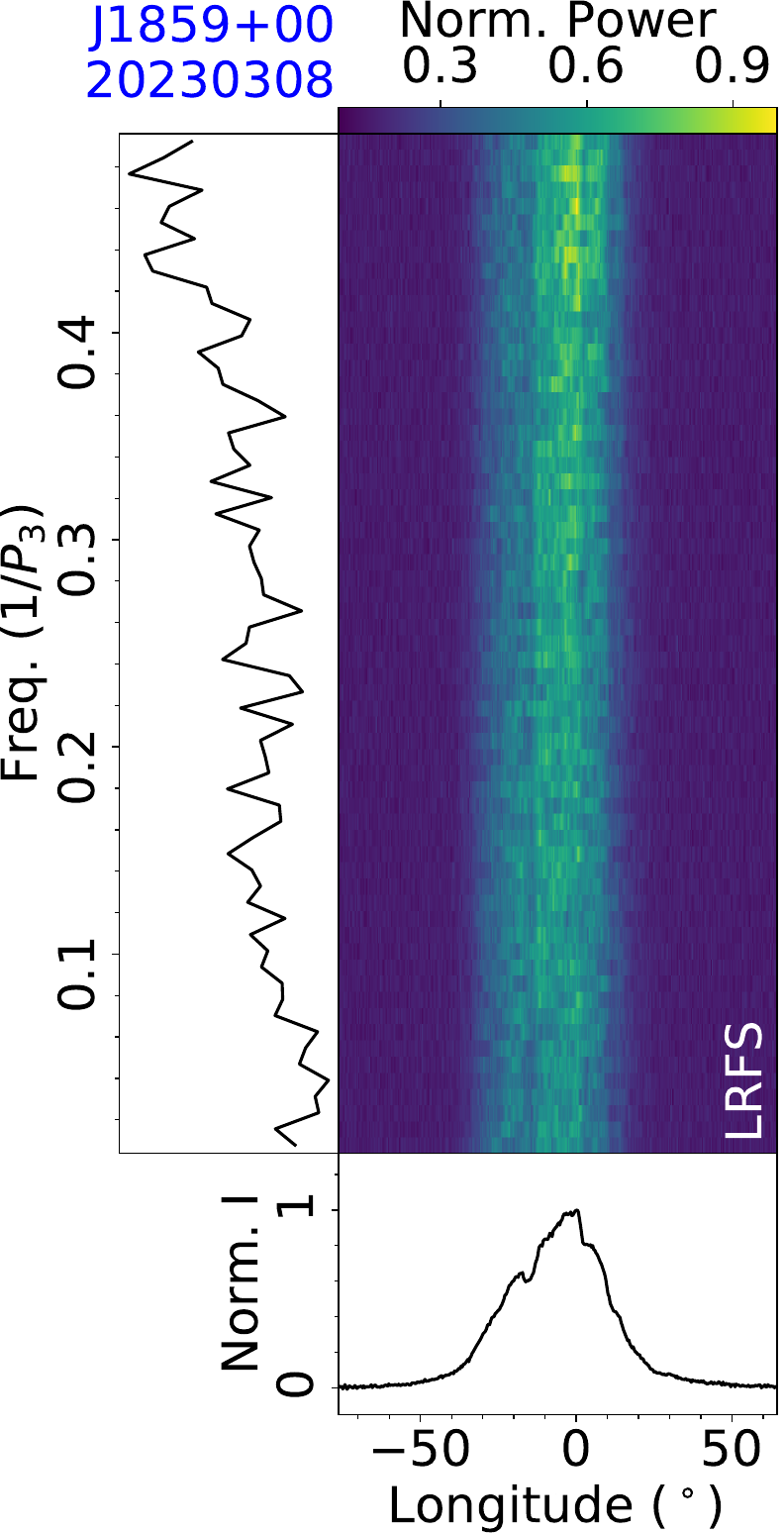}
\includegraphics[width=0.22\textwidth, angle=0]{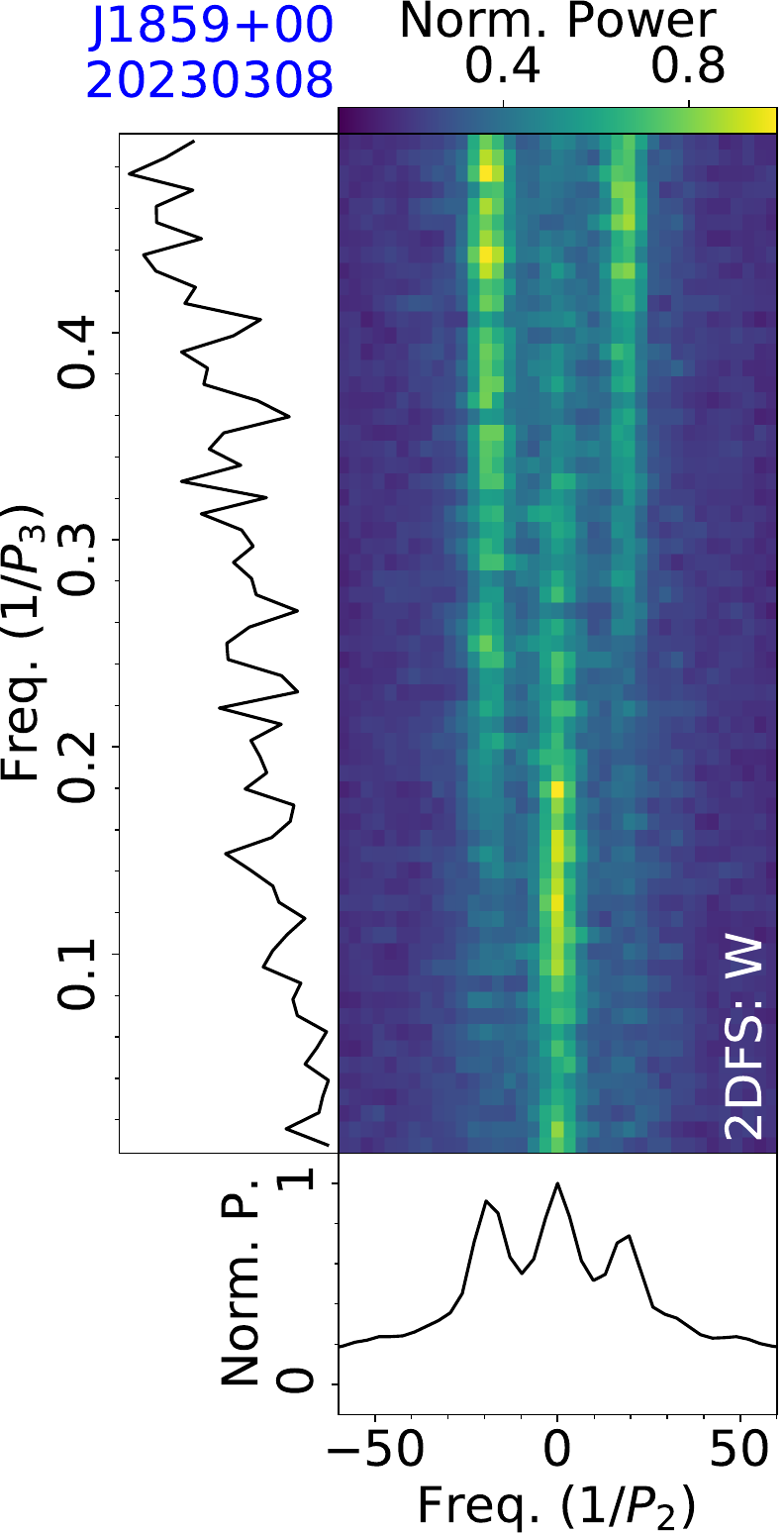}
\figcaption{Fluctuation analysis of PSR J1859+00 for the observation on 20230308, with LRFS and 2DFS for the on-pulse region of a mean pulse profile.
\label{subfig:fluctu:J1859+00}}
\end{figure}

\subsection{J1858+0026g}
\label{subsec:J1858+0026g}

PSR J1858+0026g was discovered in the FAST GPPS survey \citep{Han2021,han2025}. 

This pulsar was observed by FAST on 20200426 for 5 minutes and 20250129 for 60 minutes. From the data on 20200426, a rotation period $P=4.7144$~s and a dispersion measure $D\!M=421.7~{\rm cm^{-3}\,pc}$ were determined. 
Single pulse sequences of two observations and two zoomed-in views of the observation on 20250129 are shown in Fig.~\ref{subfig:TP:J1858+0026g}, illustrating the existence of unsystematic subpulse drifting and nulling phenomena. 
The on-pulse integral energy histogram for the data of 20250129 is displayed in Fig.~\ref{subfig:Hist:J1858+0026g}, from which the nulling fraction of this observation is estimated to be 30$\pm$2\%. Although fluctuation spectra (Fig.~\ref{subfig:fluctu:J1858+0026g}) of the observation on 20200426 indicate the negative drift feature of $1/P_3=0.146\pm0.003$ ($P_3=6.9\pm0.1$ periods) and $1/P_2=-94\pm14$ ($P_2=-3.8\pm0.6^\circ$), there is no obvious drift feature in 2DFS of the observation on 20250129, which may be caused by the unsystematic drifting behavior.

\subsection{J1858+0241}
\label{subsec:J1858+0241}

PSR J1858+0241 was discovered in the Parkes multibeam pulsar survey \citep{hfs+04}. 

This pulsar was observed by FAST on 20200806 and 20220529, both for 15 minutes. 
The rotation period and a dispersion measure were determined to be $P=4.6929$~s and $D\!M=327.2~{\rm cm^{-3}\,pc}$ from the observation on 20200806. 
Single pulse sequences shown in Fig.~\ref{subfig:TP:J1858+0241} illustrate the existence of the nulling phenomenon. From the on-pulse integral energy histograms (Fig.~\ref{subfig:Hist:J1858+0241}), the nulling fractions of two observations are estimated to be 47$\pm$4\% and 46$\pm$3\%, respectively. 
The subpulse drifting behavior appears to be unstable in the single-pulse sequences. However, more observations of longer duration are required for a detailed analysis.

\subsection{J1858+0319}
\label{subsec:J1858+0319}

PSR J1858+0319 was discovered with the Arecibo Telescope \citep{Lyne2017}. 

The pulsar was observed by FAST on 20201026 for 5 minutes, yielding a rotation period $P=0.8675$~s and a dispersion measure $D\!M=282.9~{\rm cm^{-3}\,pc}$. 
The single pulse sequence of the observation is shown in Fig.~\ref{subfig:TP:J1858+0319}, which illustrates the drifting bands temporally modulated at a low frequency. From the LRFS and 2DFS in Fig.~\ref{subfig:fluctu:J1858+0319}, the centroid frequencies of the drift feature are $1/P_3=0.024\pm0.001$ and $1/P_2=-26\pm6$, corresponding to periodicities of the subpulse drifting $P_3=42\pm1$ periods and $P_2=-14\pm3^\circ$.

\begin{figure}[htpb]
\centering
\includegraphics[width=0.22\textwidth, angle=0]{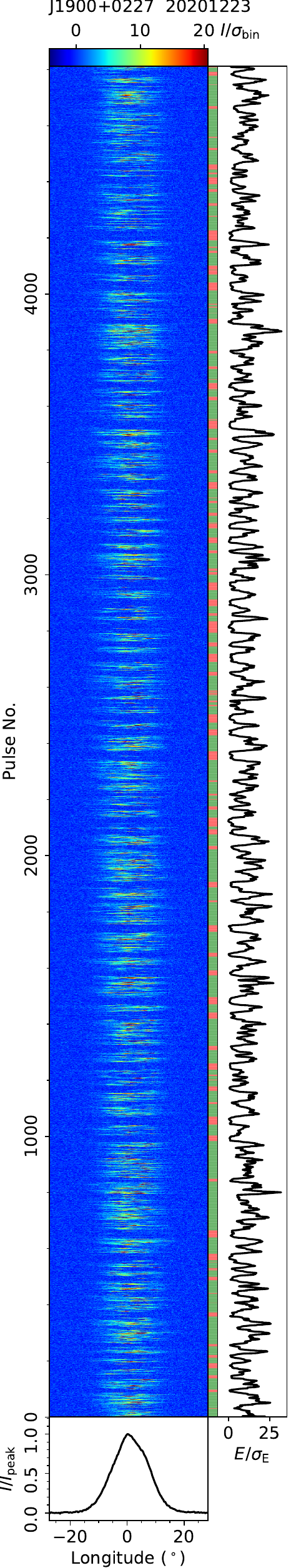}
\includegraphics[width=0.22\textwidth, angle=0]{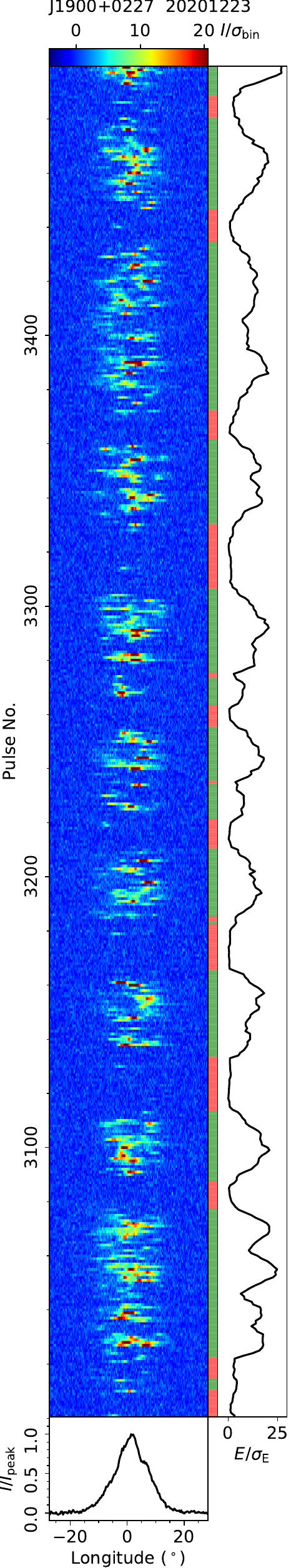}
\vspace{-0.38cm}
\figcaption{Single pulse sequence of PSR J1900+0227 from the FAST observation on 20201223, and a zoomed-in view of pulses No.3000-3500. In the right subpanel, the on-pulse energy variation is smoothed over every 9 periods.
\label{subfig:TP:J1900+0227}}
\end{figure}

\begin{figure}[htpb]
\centering
\includegraphics[width=0.39\textwidth, angle=0]{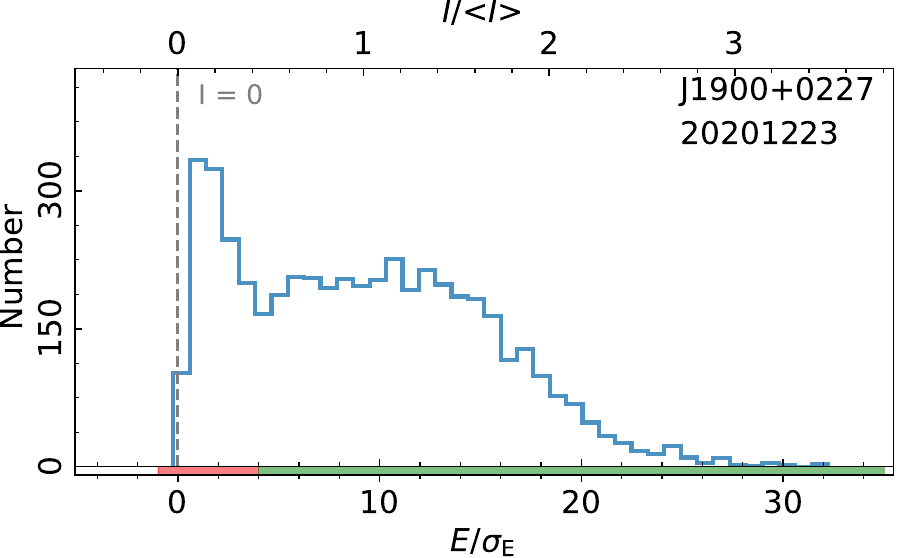}
\figcaption{On-pulse energy histogram of single pulses of PSR J1900+0227 from the FAST observation on 20201223, with energy values smoothed over 9 periods. 
\label{subfig:Hist:J1900+0227}}
\end{figure}

\begin{figure}[htpb]
\centering
\includegraphics[width=0.39\textwidth, angle=0]{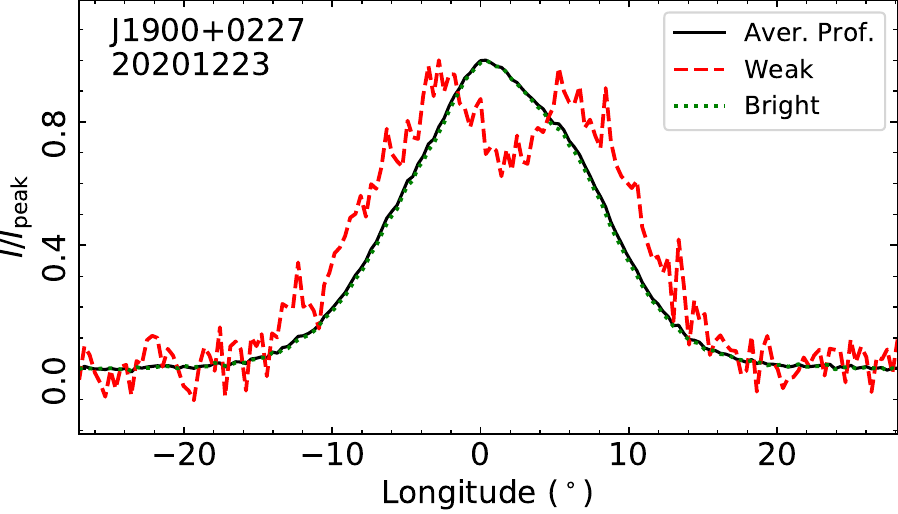}
\figcaption{Mean profiles for the weak and bright emission modes of PSR J1900+0227 from the FAST observation on 20201223, which are normalized by their respective peaks.
\label{subfig:ProfModes:J1900+0227}}
\end{figure}

\subsection{J1858+0609g}
\label{subsec:J1858+0609g}

PSR J1858+0609g was discovered in the FAST GPPS survey \citep{Han2021,han2025}.

This pulsar was observed by FAST on 20210822 and 20260103 for 5 and 13 minutes, respectively. From the longer data, a rotation period $P=0.5596$~s and a dispersion measure $D\!M=422.3~{\rm cm^{-3}\,pc}$ were derived.
The single pulse sequence observed on 20260103, along with the zoomed-in view of pulses No. 1000-1400 presented in Fig.~\ref{subfig:TP:J1858+0609g}, indicates the existence of subpulse drifting. 
Fluctuation spectra are shown in Fig.~\ref{subfig:fluctu:J1858+0609g}. The centroid of the negative drift feature is at $1/P_3=0.041\pm0.001$ and $1/P_2=-18\pm1$, corresponding to periodicities of $P_3=24\pm1$ periods and $P_2=-20\pm1$ degrees. 
drift features observed on 20210822 are consistent with those on 20260103.

\begin{figure}[htpb]
\centering
\includegraphics[width=0.22\textwidth, angle=0]{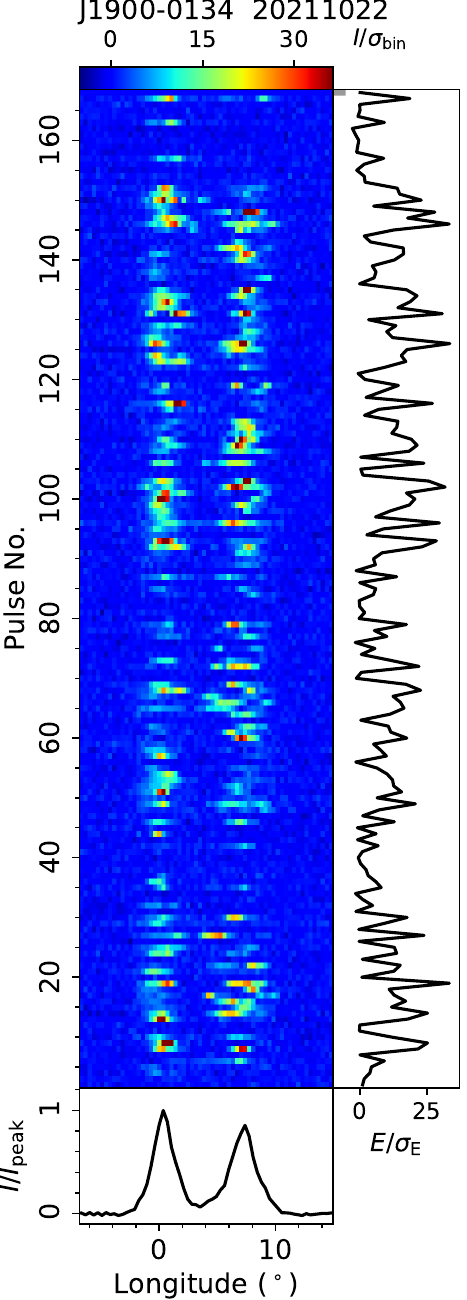}
\figcaption{Single pulse sequence of PSR J1900-0134 from the FAST observation on 20211022. 
\label{subfig:TP:J1900-0134}}
\end{figure}

\begin{figure}[htpb]
\centering
\includegraphics[width=0.39\textwidth, angle=0]{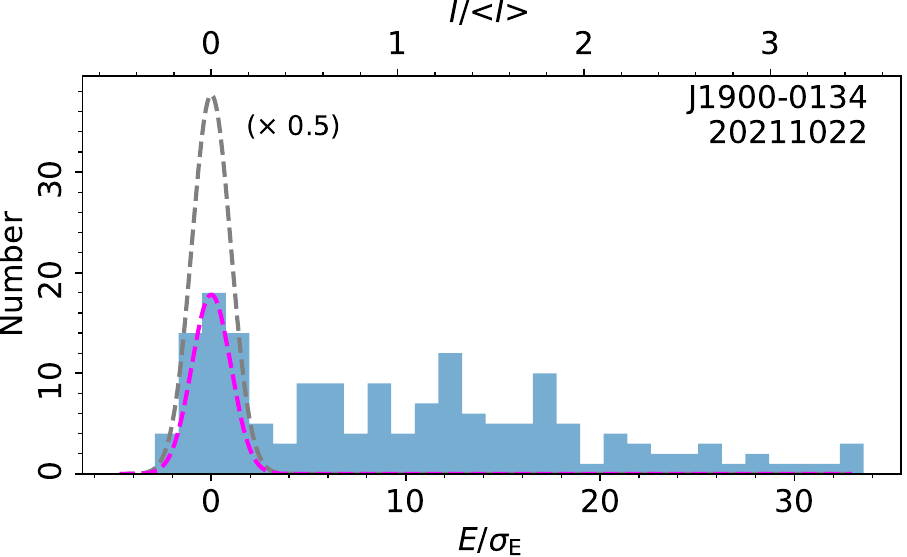}
\figcaption{On-pulse energy histogram of single pulses of PSR J1900-0134 from the FAST observation on 20211022.
\label{subfig:Hist:J1900-0134}}
\end{figure}

\subsection{J1859+00}
\label{subsec:J1859+00}

PSR J1859+00 was discovered by \citet{Camilo1996} using the Arecibo radio telescope at 430 MHz. 

This pulsar was observed by FAST on 20230308 for 30 minutes, yielding a rotation period $P=0.5596$~s and a dispersion measure $D\!M=422.3~{\rm cm^{-3}\,pc}$. 
Single pulse sequences are shown in Fig.~\ref{subfig:TP:J1859+00}. 
There are three modulation features in 2DFS. One feature has a low temporal frequency without phase modulation, while the other two features are phase modulated in opposite directions. The centroid frequency of the low-frequency modulation feature in 2DFS is $1/P_3=0.160\pm0.002$, corresponding to $P_3=6.2\pm0.1$ periods. 
The temporal modulation frequencies of the two oppositely drift features are widely distributed in 2DFS. Meanwhile, the phase modulation frequencies of these features are opposite in value, with the negative drifting being more dominant. 
The negative drift feature is characterized by the centroid frequencies of $1/P_3=0.356\pm0.002$ and $1/P_2=-18.4\pm0.1$, yielding $P_3=2.81\pm0.01$ periods and $P_2=-19.6\pm0.1^\circ$. For the positive drift feature, the centroid frequencies are $1/P_3=0.404\pm0.001$ and $1/P_2=18.1\pm0.1$, which correspond to $P_3=2.48\pm0.01$ periods and $P_2=19.8\pm0.1^\circ$. 
These indicate that widths between subpulses are constant, but the drifting rate varies with time. Oppositely drift feature in 2DFS may be caused by variable drifting rate and the aliased effect.

\begin{figure}[htpb]
\centering
\includegraphics[width=0.215\textwidth, angle=0]{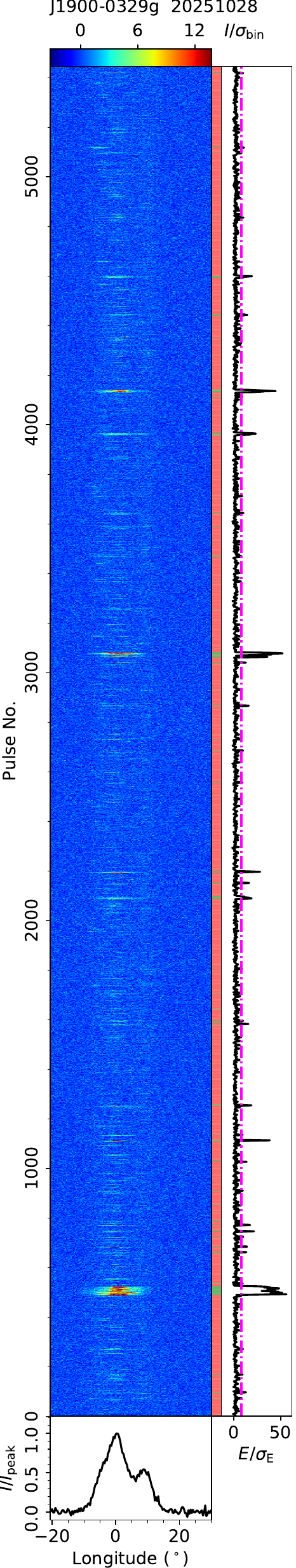}
\includegraphics[width=0.215\textwidth, angle=0]{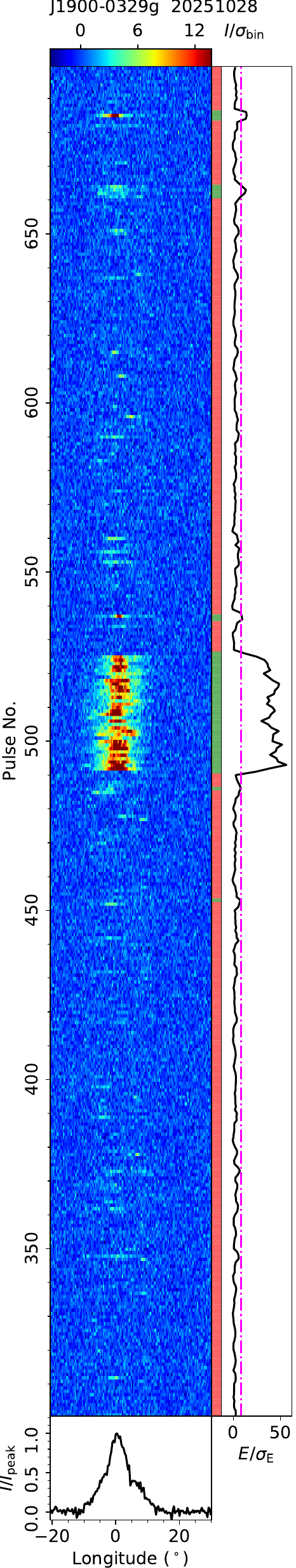}
\figcaption{Single pulse sequence of PSR J1900-0329g from the FAST observation on 20251028, and a zoomed-in view of pulses No.300-700. In the right subpanel, the on-pulse energy variation is smoothed over every 3 periods, with a dashed line for the threshold to distinguish the weak and bright emission modes.
\label{subfig:TP:J1900-0329g}}
\end{figure}

\begin{figure}[htpb]
\centering
\includegraphics[width=0.39\textwidth, angle=0]{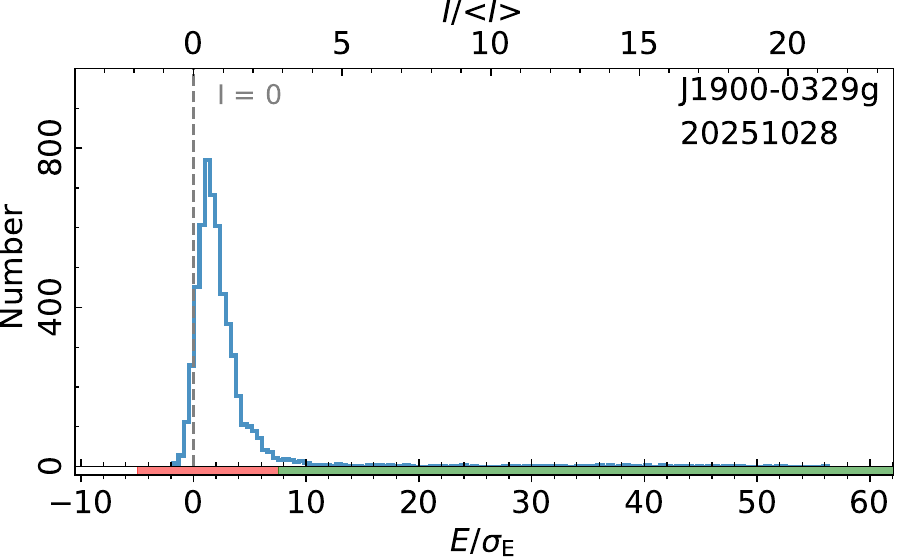}
\figcaption{On-pulse energy histogram of single pulses of PSR J1900-0329g from the FAST observation on 20251028, with energy values smoothed over 3 periods. 
\label{subfig:Hist:J1900-0329g}}
\end{figure}

\begin{figure}[htpb]
\centering
\includegraphics[width=0.39\textwidth, angle=0]{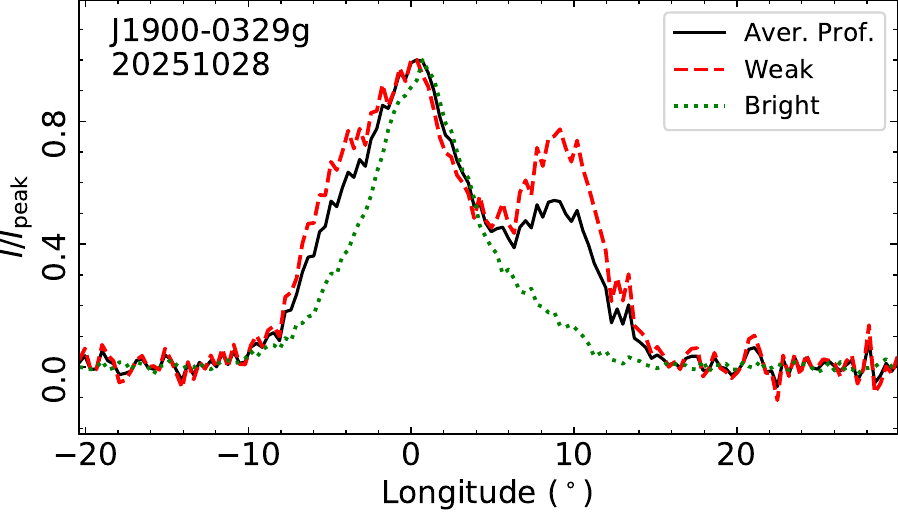}
\figcaption{Mean profiles for the weak and bright emission modes of PSR J1900-0329g from the FAST observation on 20251028, which are normalized by their respective peaks.
\label{subfig:ProfModes:J1900-0329g}}
\end{figure}

\subsection{J1900+0227}
\label{subsec:J1900+0227}

PSR J1900+0227 was discovered in the Parkes Multibeam Pulsar Survey \citep{Morris2002}. 

This pulsar was observed by FAST on 20200901 for 5 minutes and 20201223 for 30 minutes. From the 30-minute data, a rotation period $P=0.3743$~s and a dispersion measure $D\!M=202.7~{\rm cm^{-3}\,pc}$ were derived. The single pulse sequence and a zoomed-in view of pulses No. 3000-3500 in Fig.~\ref{subfig:TP:J1900+0227} display changes between weak and bright emission modes. From the smoothed energy variation histogram (Fig.~\ref{subfig:Hist:J1900+0227}), single pulses of weak and bright emission modes are distinguished, which are labeled using red and green colors, respectively. Mean pulse profiles of two emission modes are displayed in Fig.~\ref{subfig:ProfModes:J1900+0227}, and there seem to be two components for the weak mode.

\subsection{J1900-0134}
\label{subsec:J1900-0134}

PSR J1900-0134 is the first pulsar discovered by FAST \citep{Qian2019}. 

This pulsar was also observed by FAST on 20211022 for 5 minutes, with a rotation period $P=1.8324$~s and a dispersion measure $D\!M=178.9~{\rm cm^{-3}\,pc}$. 
The single pulse sequence of the observation on 20211022 in Fig.~\ref{subfig:TP:J1900-0134} illustrates the existence of nulls. From the on-pulse integral energy histogram (Fig.~\ref{subfig:Hist:J1900-0134}), the nulling fraction of this observation is estimated to be 24$\pm$4\%. 

\begin{figure}[htpb]
\includegraphics[width=0.21\textwidth, angle=0]{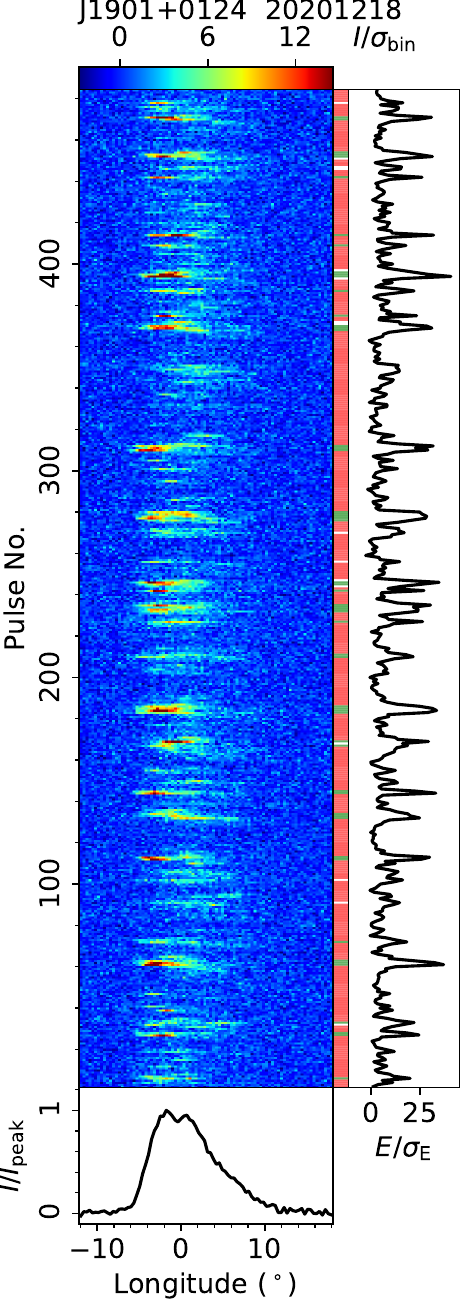}
\includegraphics[width=0.21\textwidth, angle=0]{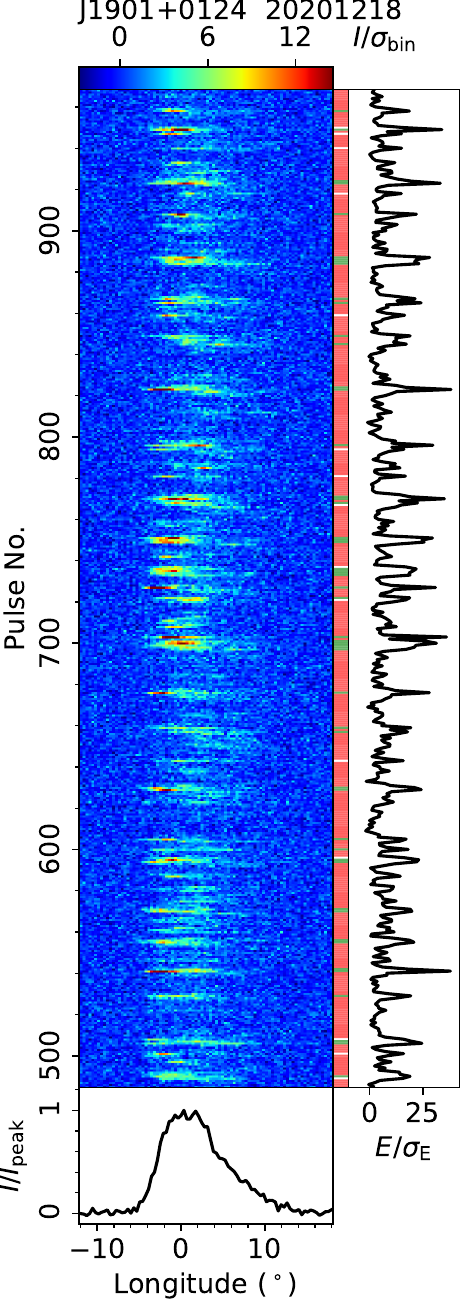}
\figcaption{Single pulse sequences of PSR J1901+0124 from the FAST observation on 20201218.
\label{subfig:TP:J1901+0124}}
\end{figure}


\begin{figure}[htpb]
\centering
\includegraphics[width=0.22\textwidth, angle=0]{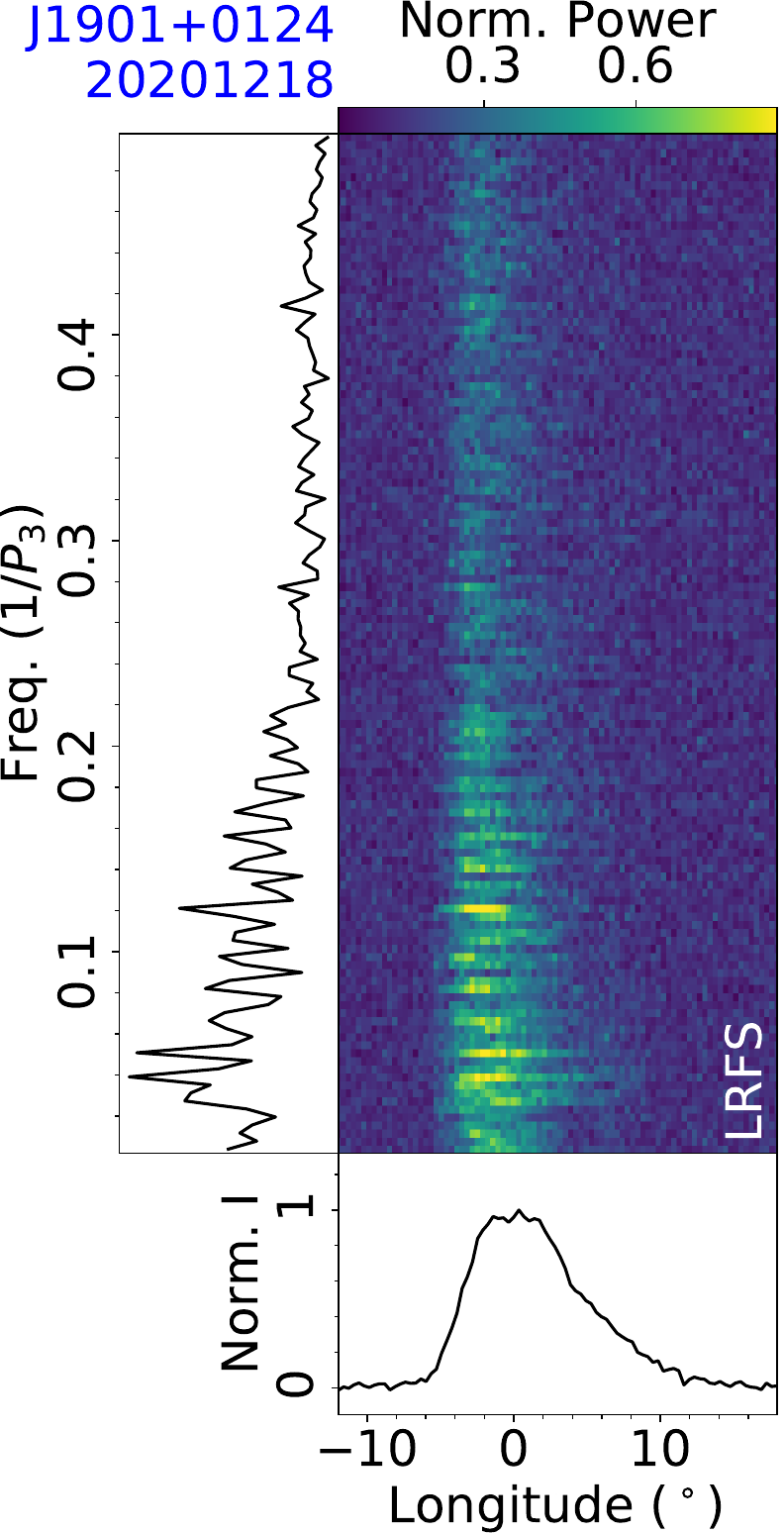}
\includegraphics[width=0.22\textwidth, angle=0]{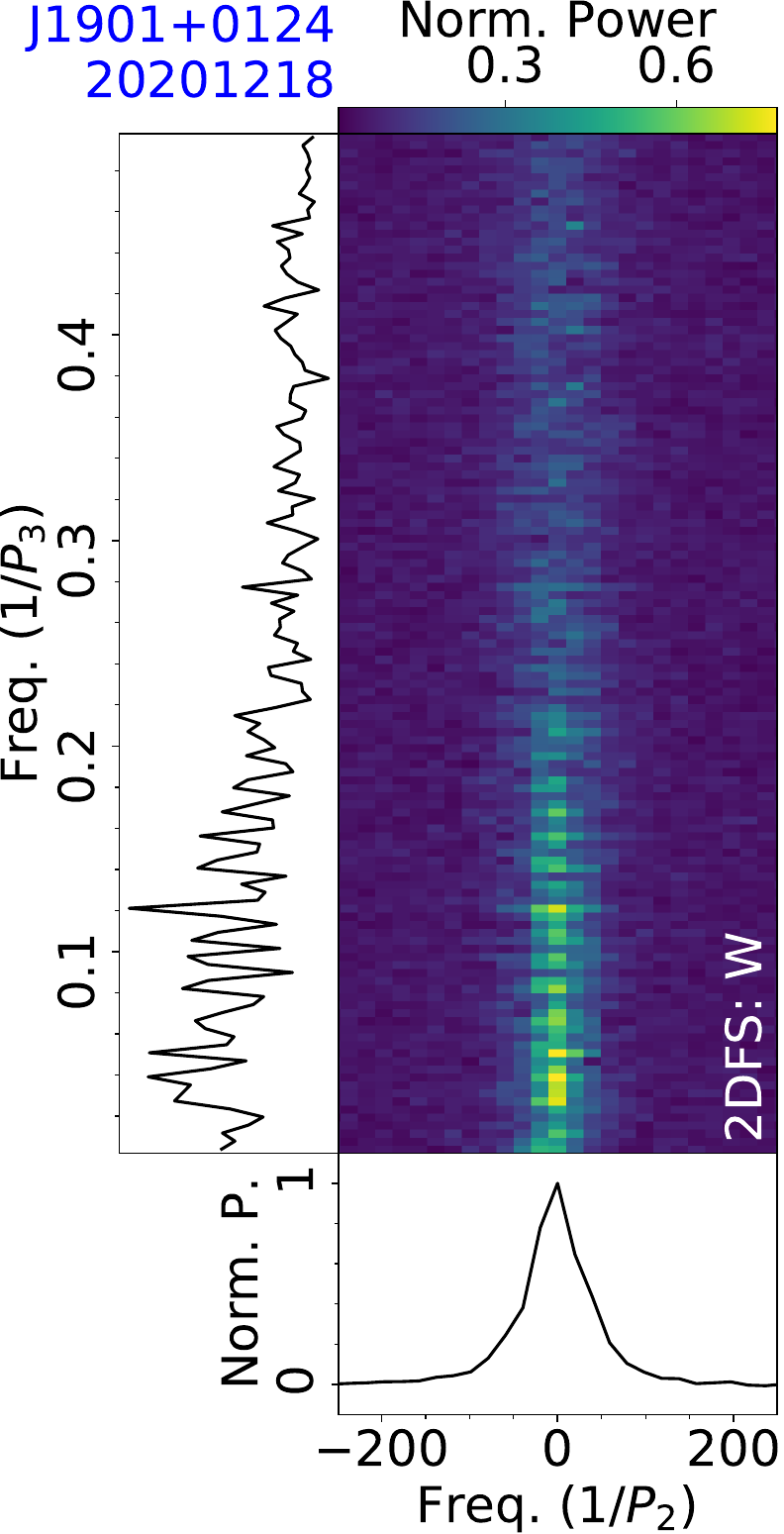}
\figcaption{Fluctuation analysis of PSR J1901+0124 for the observation on 20201218, with LRFS and 2DFS for the on-pulse region of a mean pulse profile.
\label{subfig:fluctu:J1901+0124}}
\end{figure}


\begin{figure}[htpb]
\centering
\includegraphics[width=0.22\textwidth, angle=0]{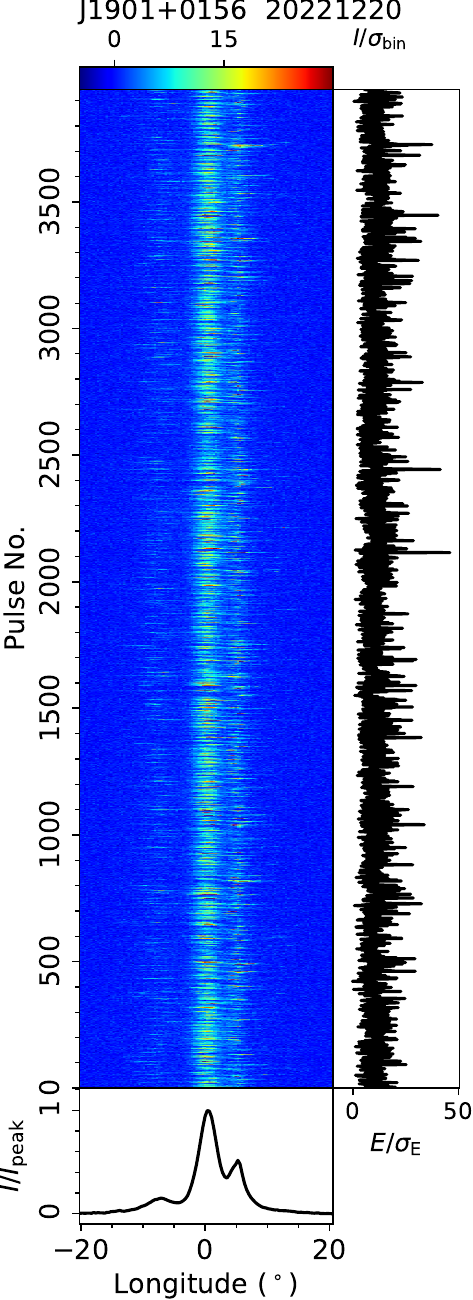}
\includegraphics[width=0.22\textwidth, angle=0]{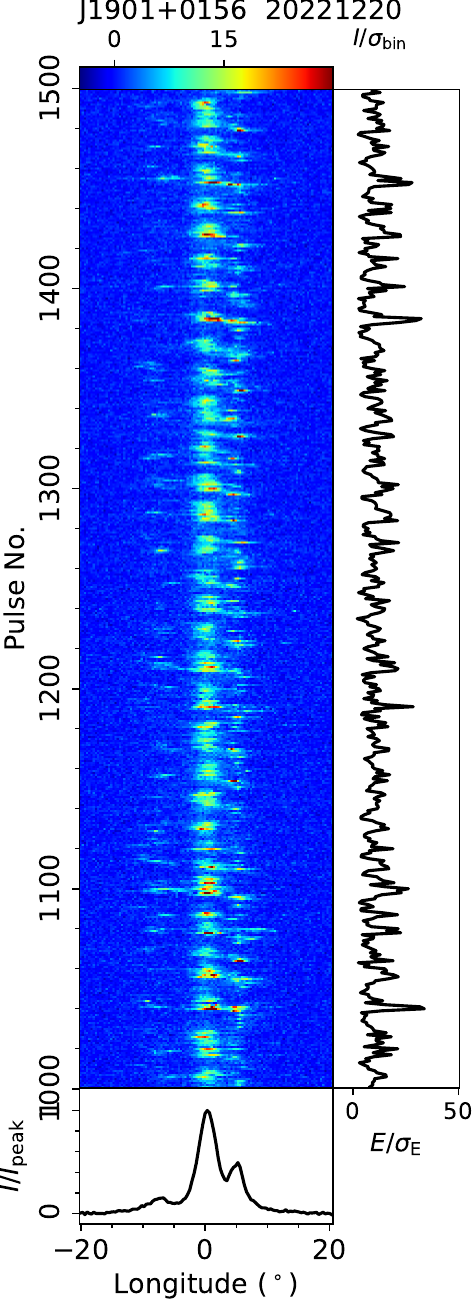}
\figcaption{The single pulse sequence of PSR J1901+0156 from the FAST observation on 20221220, and a zoomed-in view of pulses No.1000-1500.
\label{subfig:TP:J1901+0156}}
\end{figure}

\begin{figure}[htpb]
\centering
\includegraphics[width=0.22\textwidth, angle=0]{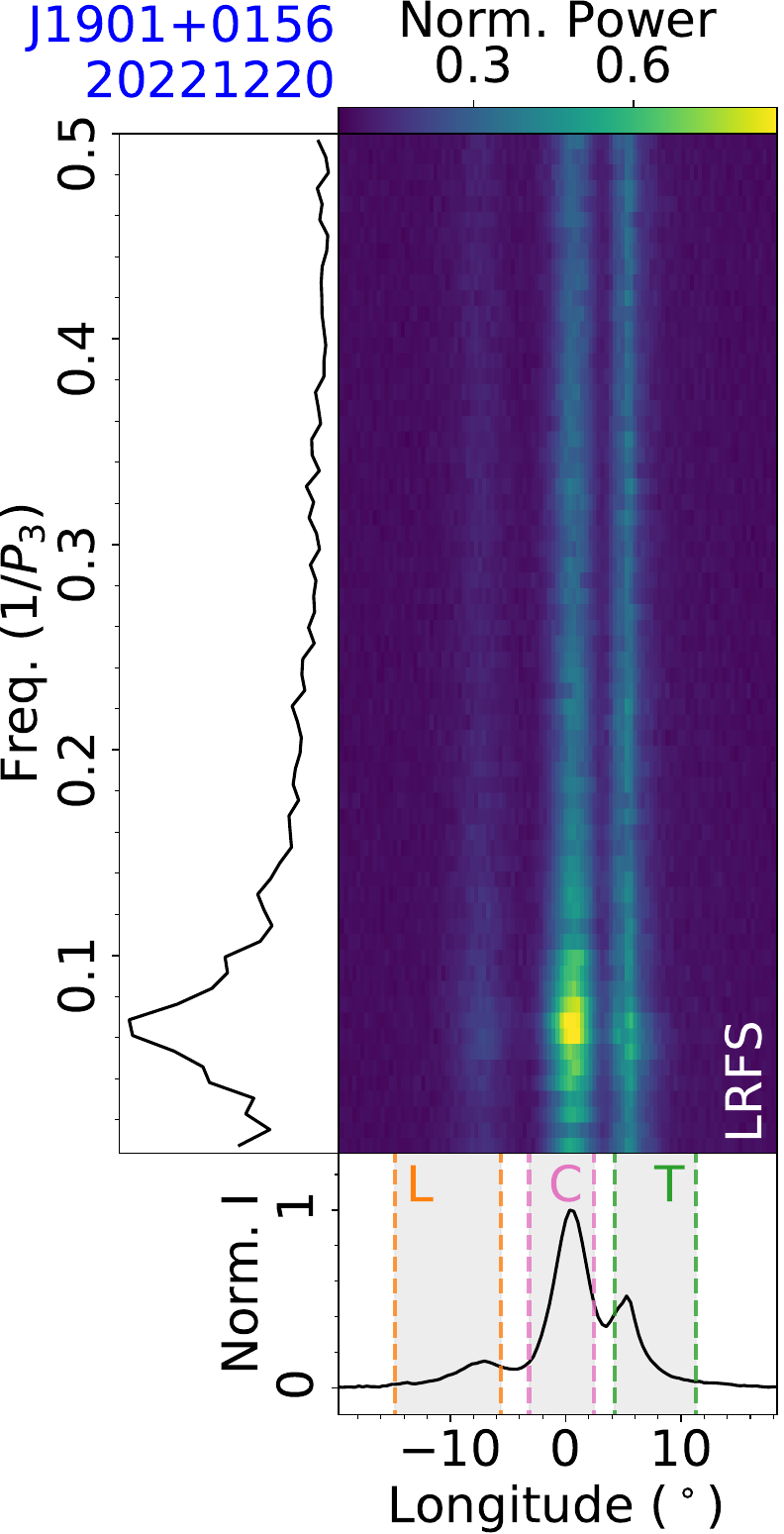}
\includegraphics[width=0.22\textwidth, angle=0]{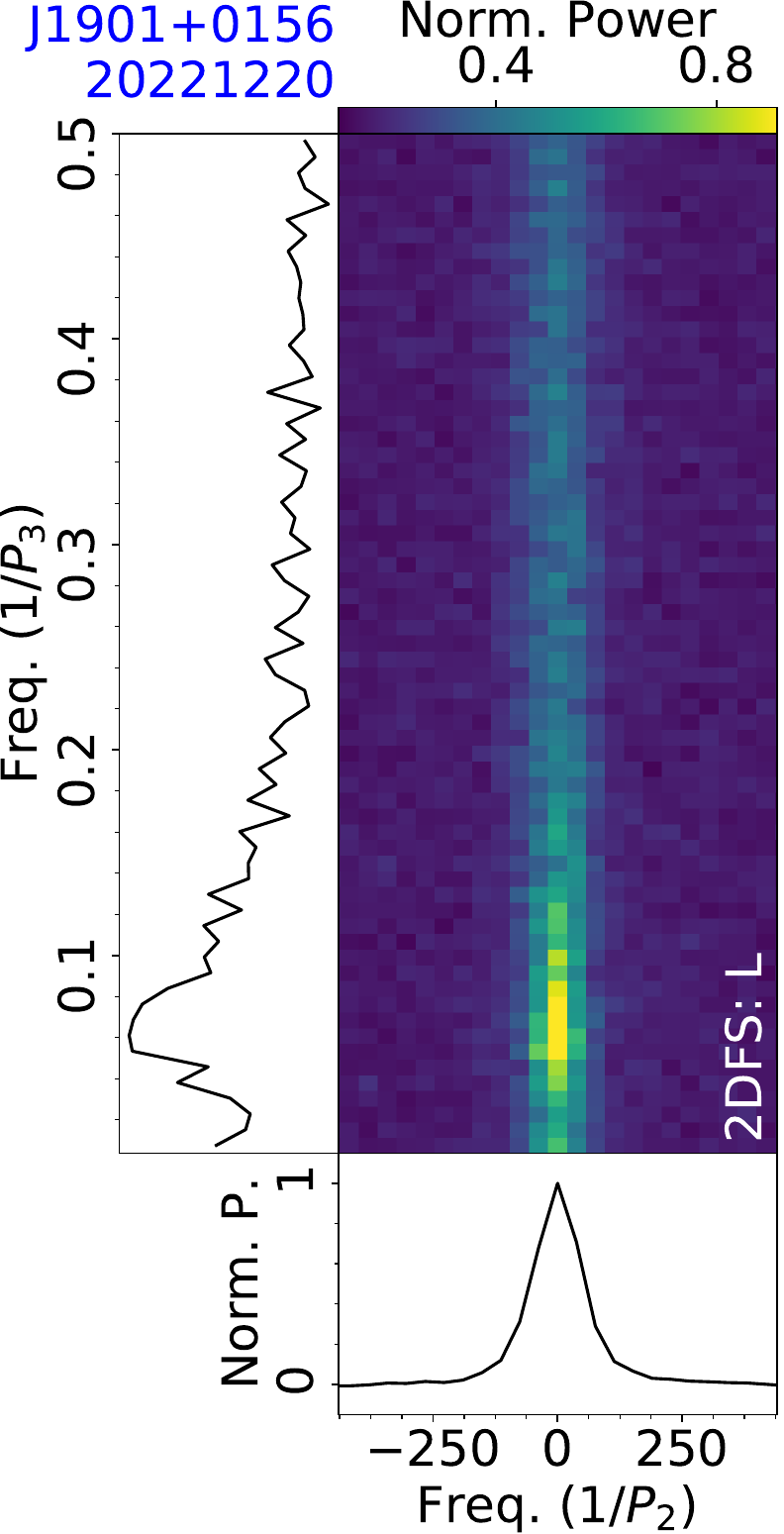}
\includegraphics[width=0.22\textwidth, angle=0]{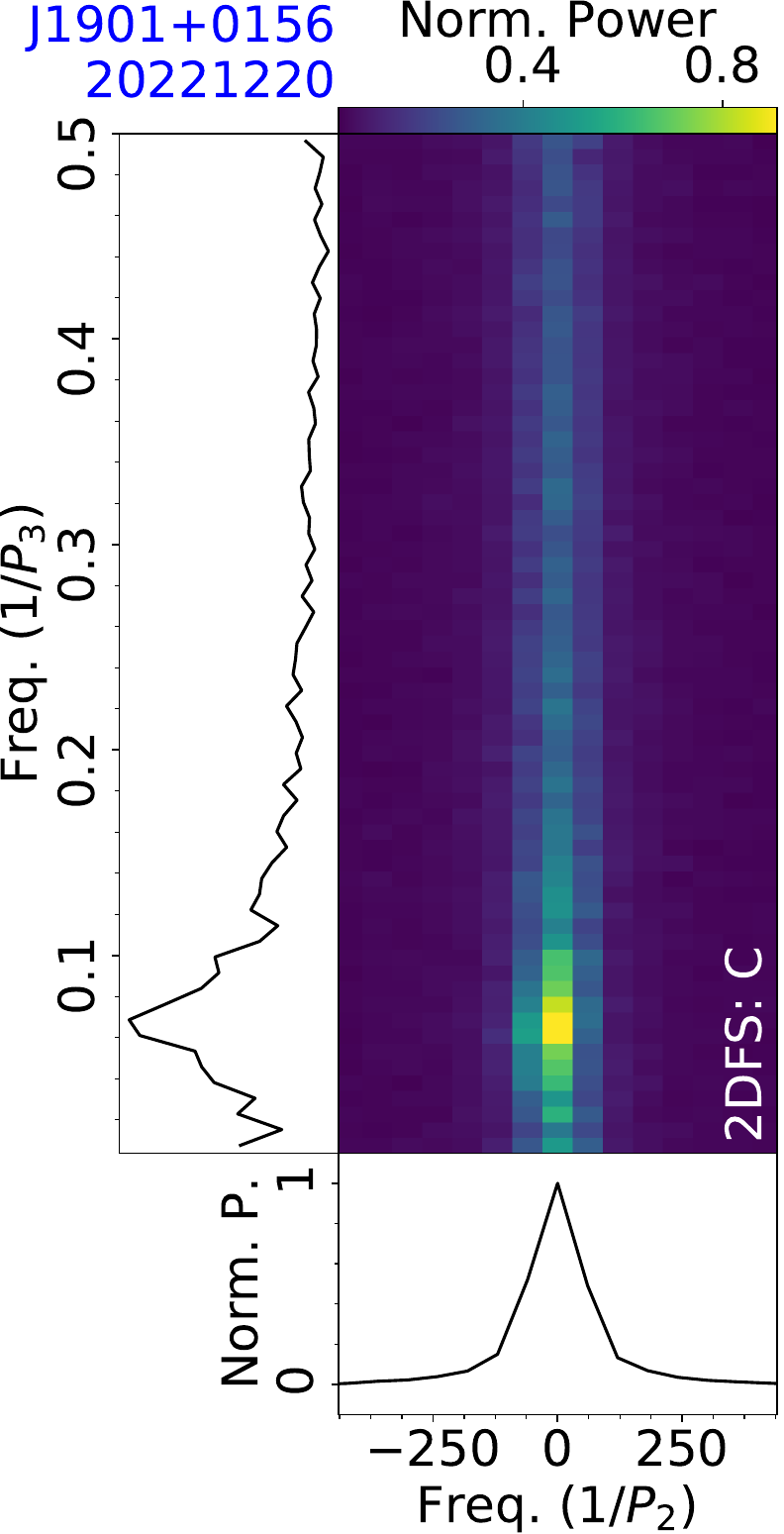}
\includegraphics[width=0.22\textwidth, angle=0]{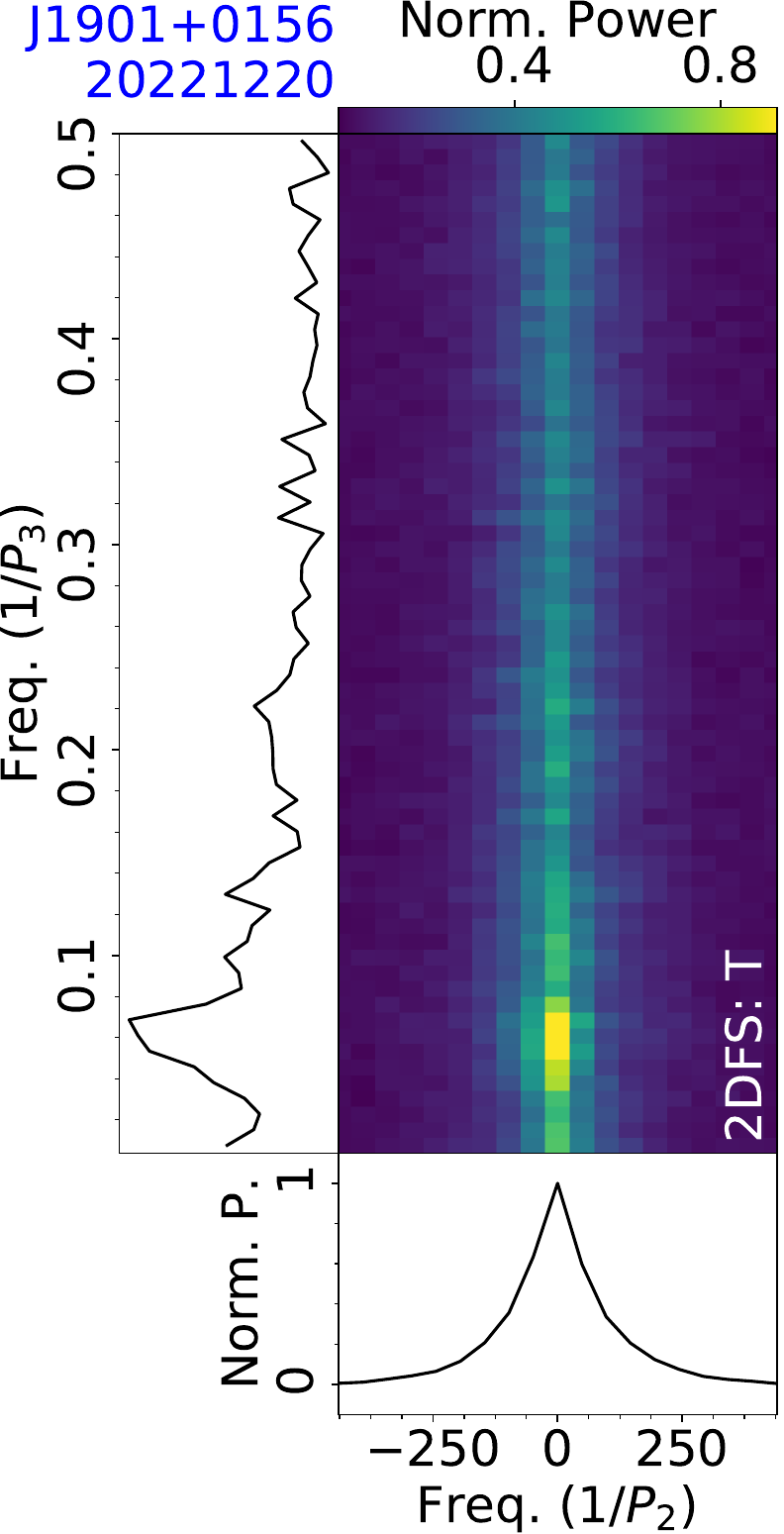}
\figcaption{Fluctuation analysis of PSR J1901+0156 for the observation on 20221220, with LRFS (top-left), and 2DFS for the leading part (top-right), central part (bottom-left) and trailing part (bottom-right) of a mean pulse profile.
\label{subfig:fluctu:J1901+0156}}
\end{figure}

\subsection{J1900-0329g}
\label{subsec:J1900-0329g}

PSR J1900-0329g was discovered in the FAST GPPS survey \citep{Han2021,han2025}. 

This pulsar was observed by FAST on 20251028 for 15 minutes, yielding a rotation period $P=0.1654$~s and a dispersion measure $D\!M=104.6~{\rm cm^{-3}\,pc}$. The single pulse sequence and a zoomed-in view of pulses No. 300-700 in Fig.~\ref{subfig:TP:J1900-0329g} show mode changes between bright and weak emission modes, and the on-pulse integral energy sequence of single pulses smoothed using a 3-period moving average is displayed in the right subpanel. 
Two emission modes of single pulses are distinguished from the smoothed on-pulse energy histogram in Fig.~\ref{subfig:Hist:J1900-0329g}, with the weak and bright emission modes labeled in red and green, respectively. From the mean pulse profiles of two emission modes in Fig.~\ref{subfig:ProfModes:J1900-0329g}, the weak mode exhibits a double-peaked profile, whereas the bright mode shows only a single peak in the leading profile part. 
For the bright emission mode, the energy is enhanced for both the leading and trailing profile parts compared to the weak mode.

\subsection{J1901+0124}
\label{subsec:J1901+0124}

PSR J1901+0124 was discovered in the Parkes multibeam pulsar survey \citep{hfs+04}. \citet{Song2023} reported that the pulsar has the negative drifting behavior with $P_3=10\pm10$ periods and $P_2=-77^{+38}_{-163}$ degrees.

This pulsar is observed by FAST on 20201218 for 5 minutes, with a rotation period $P=0.3188$~s and a dispersion measure $D\!M=314.4~{\rm cm^{-3}\,pc}$. 
Single pulse sequences in Fig.~\ref{subfig:TP:J1901+0124} show the modulation phenomenon. In the fluctuation spectra in Fig.~\ref{subfig:fluctu:J1901+0124}, the temporal modulation frequency is wide. The centroid of the modulation feature is at $1/P_3=0.095\pm0.001$ and $1/P_2=-1.8\pm0.7$, corresponding to periodicities of $P_3=10.5\pm0.1$ periods and $P_2=-201\pm78$ degrees.

\begin{figure}[htpb]
\centering
\includegraphics[width=0.22\textwidth, angle=0]{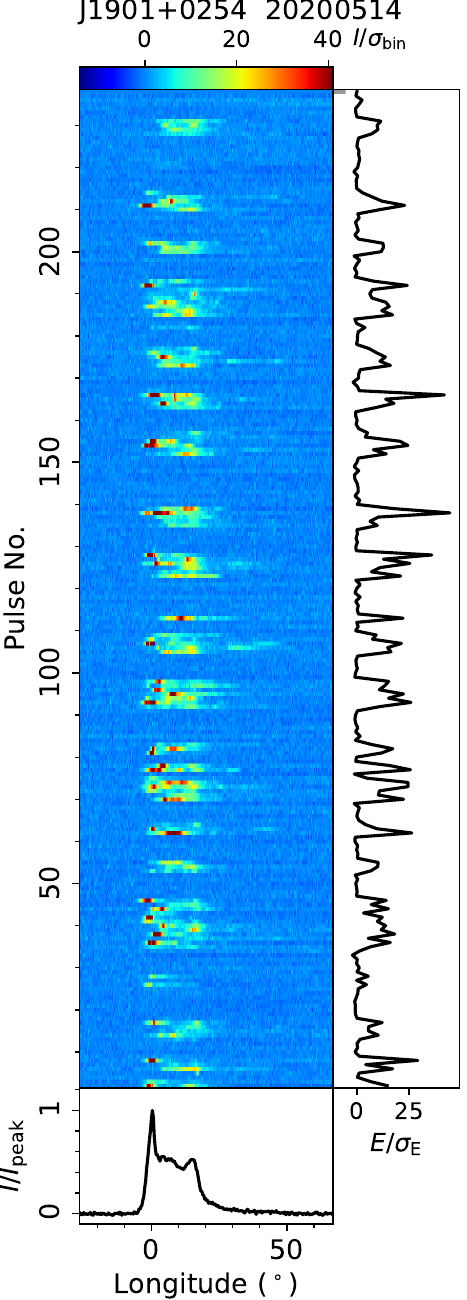}
\figcaption{Single pulse sequence of PSR J1901+0254 from the FAST observation on 20200514.
\label{subfig:TP:J1901+0254}}
\end{figure}

\begin{figure}[htpb]
\centering
\includegraphics[width=0.39\textwidth, angle=0]{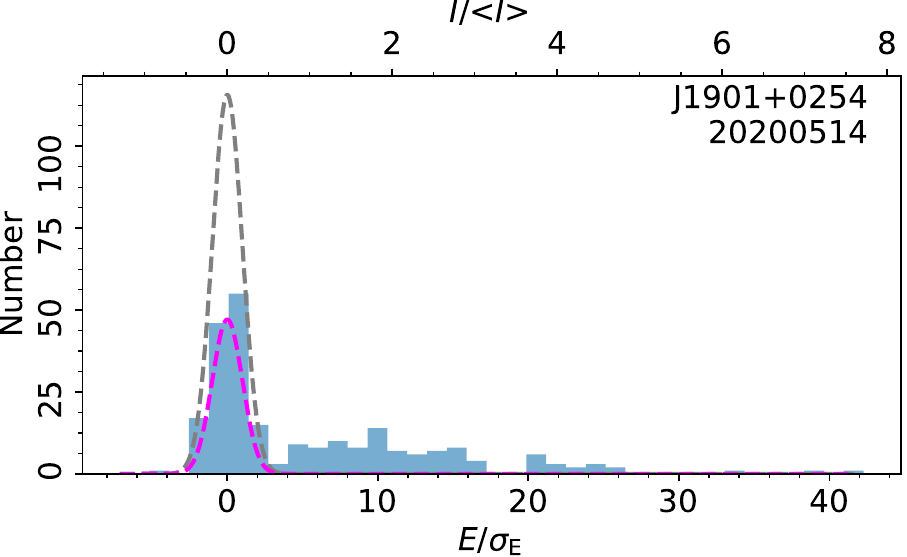}
\figcaption{On-pulse energy histogram of single pulses of PSR J1901+0254 from the FAST observation on 20200514.
\label{subfig:Hist:J1901+0254}}
\end{figure}

\begin{figure}[htpb]
\centering
\includegraphics[width=0.22\textwidth, angle=0]{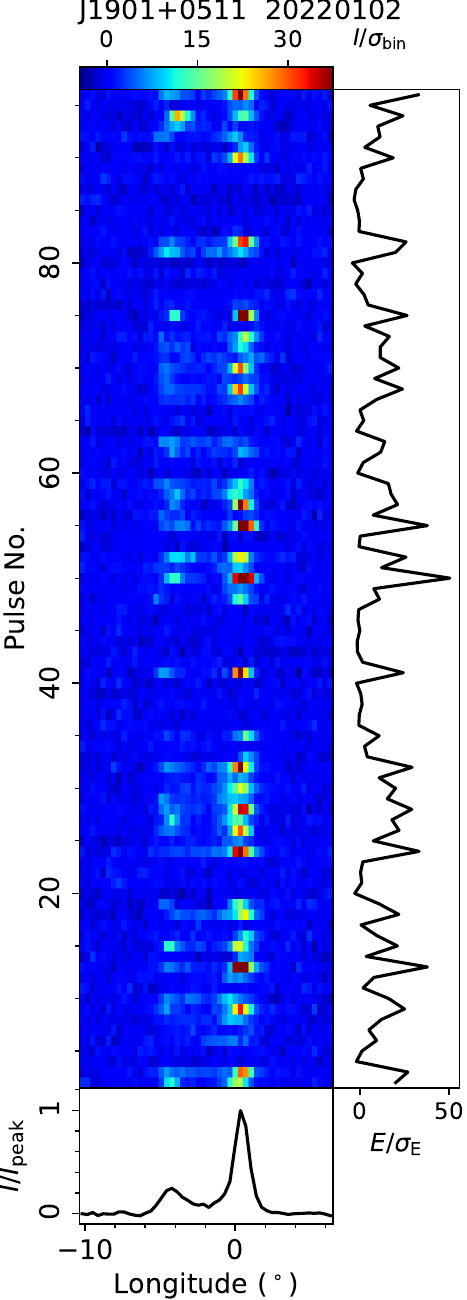}
\includegraphics[width=0.22\textwidth, angle=0]{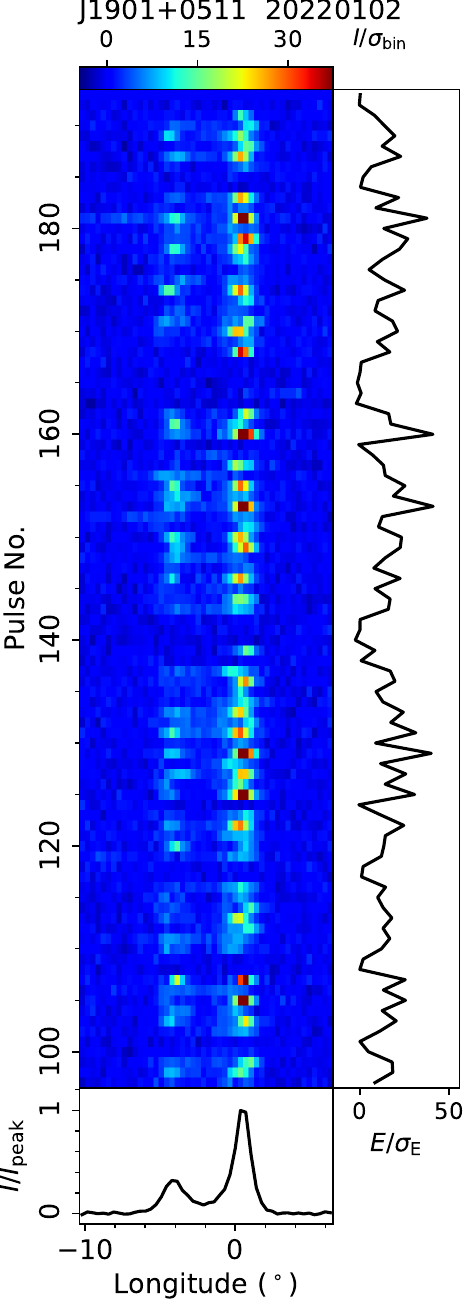}
\figcaption{Single pulse sequences of PSR J1901+0511 from the FAST observation on 20220102.
\label{subfig:TP:J1901+0511}}
\end{figure}

\begin{figure}[htpb]
\centering
\includegraphics[width=0.39\textwidth, angle=0]{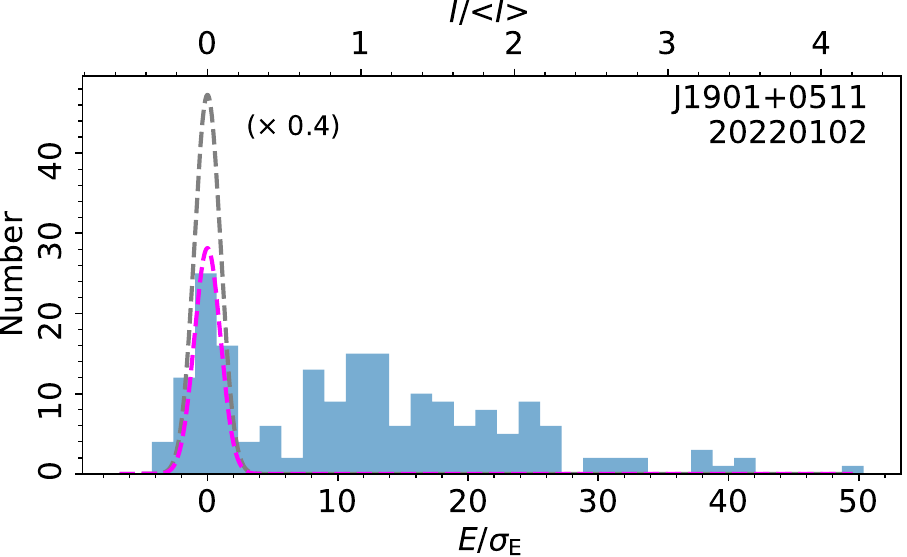}
\figcaption{On-pulse energy histogram of single pulses of PSR J1901+0511 from the FAST observation on 20220102.
\label{subfig:Hist:J1901+0511}}
\end{figure}

\begin{figure}[htpb]
\centering
\includegraphics[width=0.22\textwidth, angle=0]{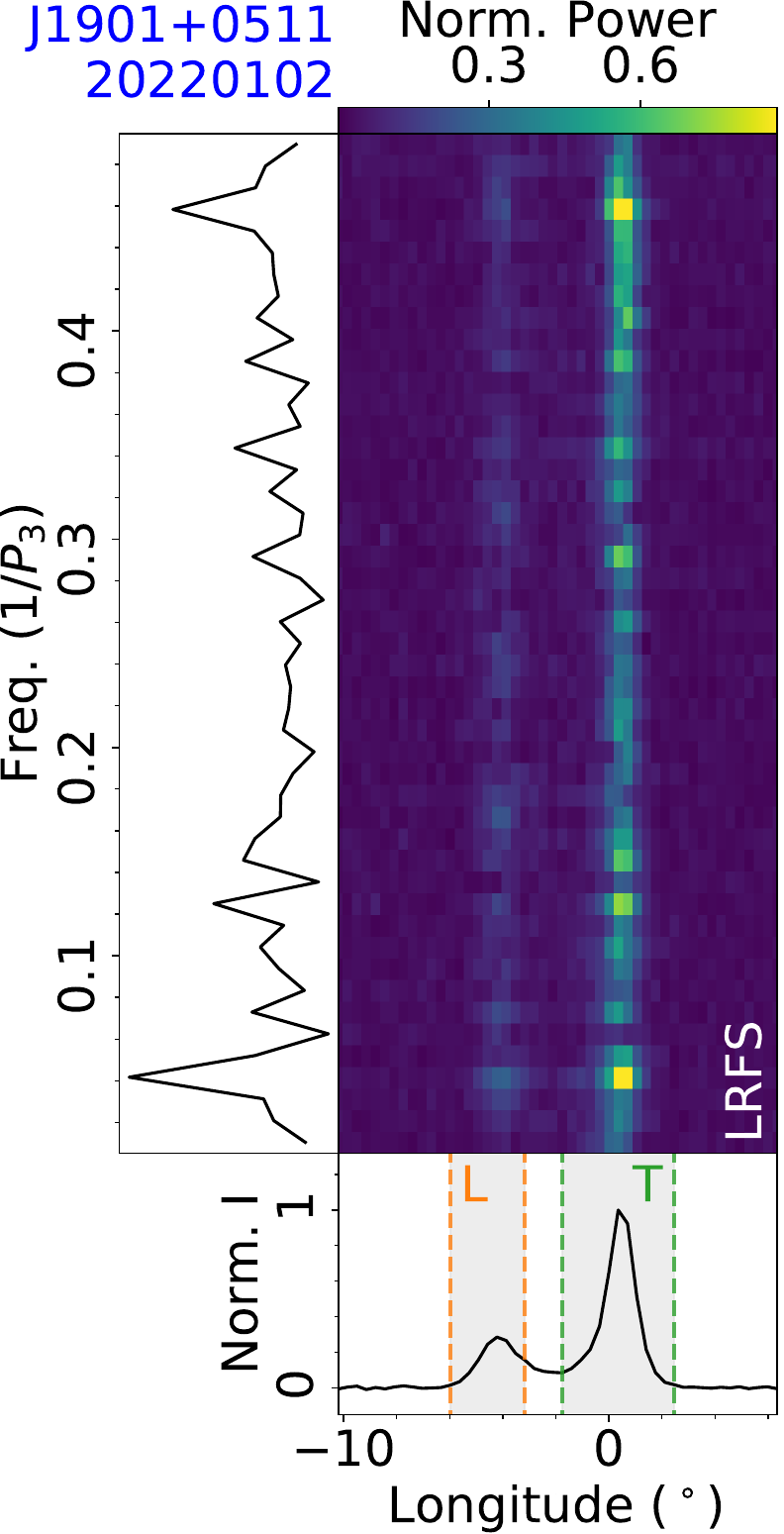}
\includegraphics[width=0.22\textwidth, angle=0]{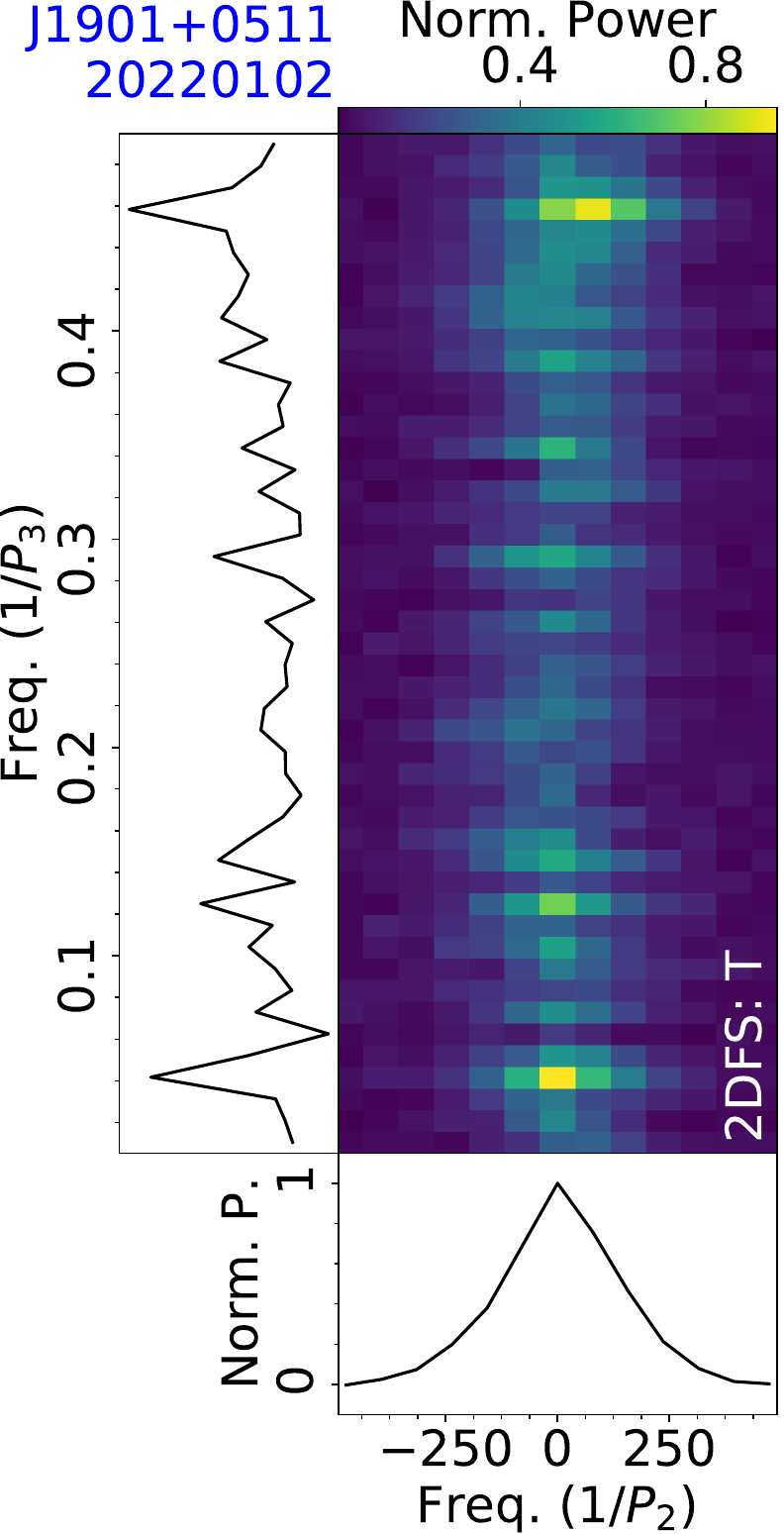}
\figcaption{Fluctuation analysis of PSR J1901+0511 for the observation on 20220102, with LRFS and 2DFS for the trailing part of a mean pulse profile.
\label{subfig:fluctu:J1901+0511}}
\end{figure}

\subsection{J1901+0156}
\label{subsec:J1901+0156}

PSR J1901+0156 was discovered by the 92-m telescope at Green Bank \citep{Stokes1985}. Periodic modulation behavior of $P_3=14\pm1$ periods was detected by \citet{Weltevrede2007,Basu2020}. \citet{Song2023} reported the negative drift feature for two components: $P_3=14\pm3$ periods and $P_2=-52^{+13}_{-29}$ degrees for the first, and $P_3=15\pm2$ periods and $P_2=-56^{+36}_{-22}$ degrees for the second.

This pulsar was observed by FAST on 20191226 for 5 minutes and 20221220 for 19 minutes. From the 19-minute data, a rotation period and a dispersion measure were determined to be $P=0.2882$~s and a dispersion measure $D\!M=105.3~{\rm cm^{-3}\,pc}$. The single pulse sequence of this observation and a zoomed-in view in Fig.~\ref{subfig:TP:J1901+0156} show subpulse modulation behavior for three components. Fluctuation spectra are shown in Fig.~\ref{subfig:fluctu:J1901+0156}. The leading part in a mean pulse profile has a temporal modulation feature with the centroid frequency of $1/P_3=0.076\pm0.001$, corresponding to $P_3=13.2\pm0.1$ periods. The central and trailing profile parts are more likely to exhibit the negative drift feature. The centroids are characterized by $1/P_3=0.073\pm0.001$ ($P_3=13.6\pm0.1$ periods) and $1/P_2=-10\pm3$ ($P_2=-36\pm11$ degree) for the central part, and $1/P_3=0.073\pm0.001$ ($P_3=13.8\pm0.2$ periods) and $1/P_2=-11\pm3$ ($P_2=-34\pm8$ degree) for the trailing part.



\subsection{J1901+0254}
\label{subsec:J1901+0254}

PSR J1901+0254 was discovered in the Parkes multibeam pulsar survey \citep{hfs+04}. 

The pulsar was observed on 20200514 for 5 minutes by FAST, with a rotation period $P=1.2996$~s and a dispersion measure $D\!M=182.3~{\rm cm^{-3}\,pc}$. 
The single pulse sequence shown in Fig.~\ref{subfig:TP:J1901+0254}  illustrates the existence of nulls, and the nulling fraction is estimated to be 41$\pm$5\% from the on-pulse integral energy histogram in Fig.~\ref{subfig:Hist:J1901+0254}.




\begin{figure}[htpb]
\centering
\includegraphics[width=0.22\textwidth, angle=0]{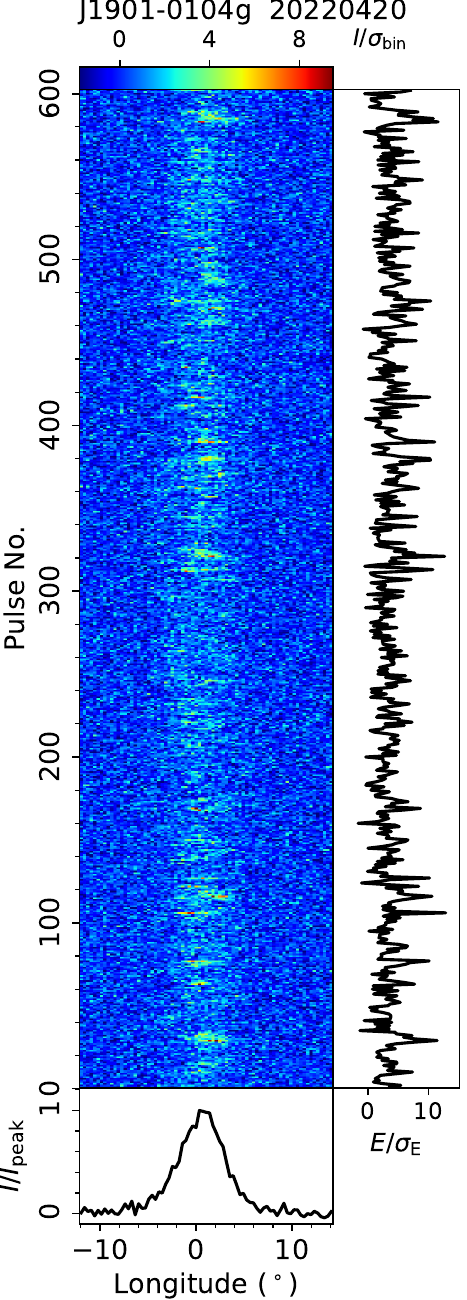}
\includegraphics[width=0.22\textwidth, angle=0]{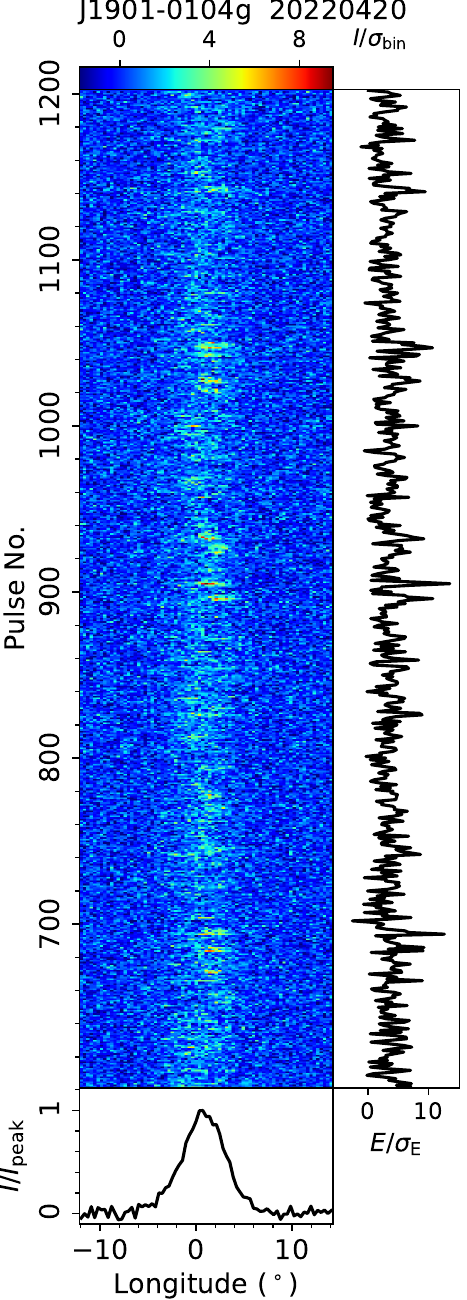}
\figcaption{Single pulse sequences of PSR J1901-0104g from the FAST observation on 20220420.
\label{subfig:TP:J1901-0104g}}
\end{figure}

\begin{figure}[htpb]
\centering
\includegraphics[width=0.22\textwidth, angle=0]{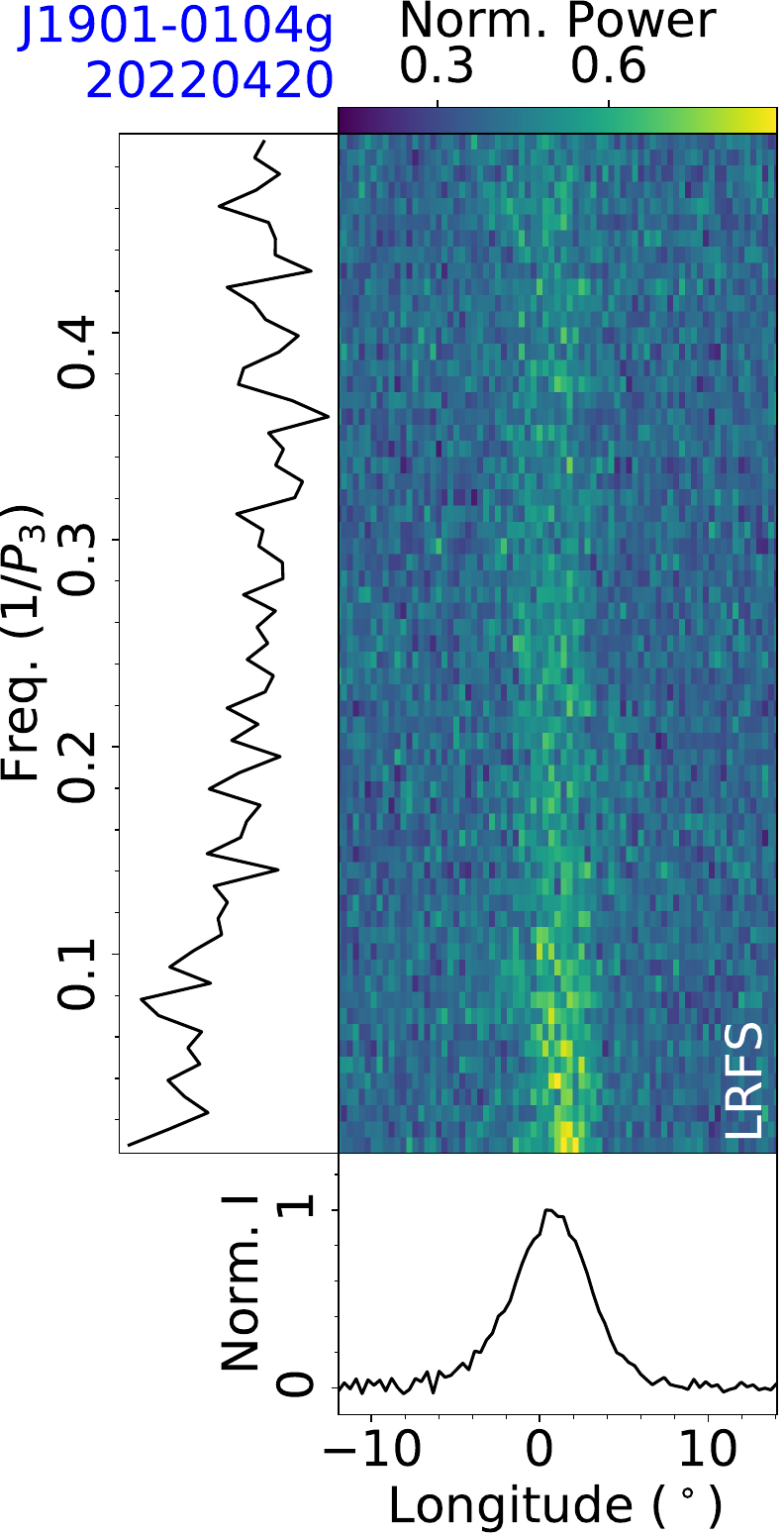}
\includegraphics[width=0.22\textwidth, angle=0]{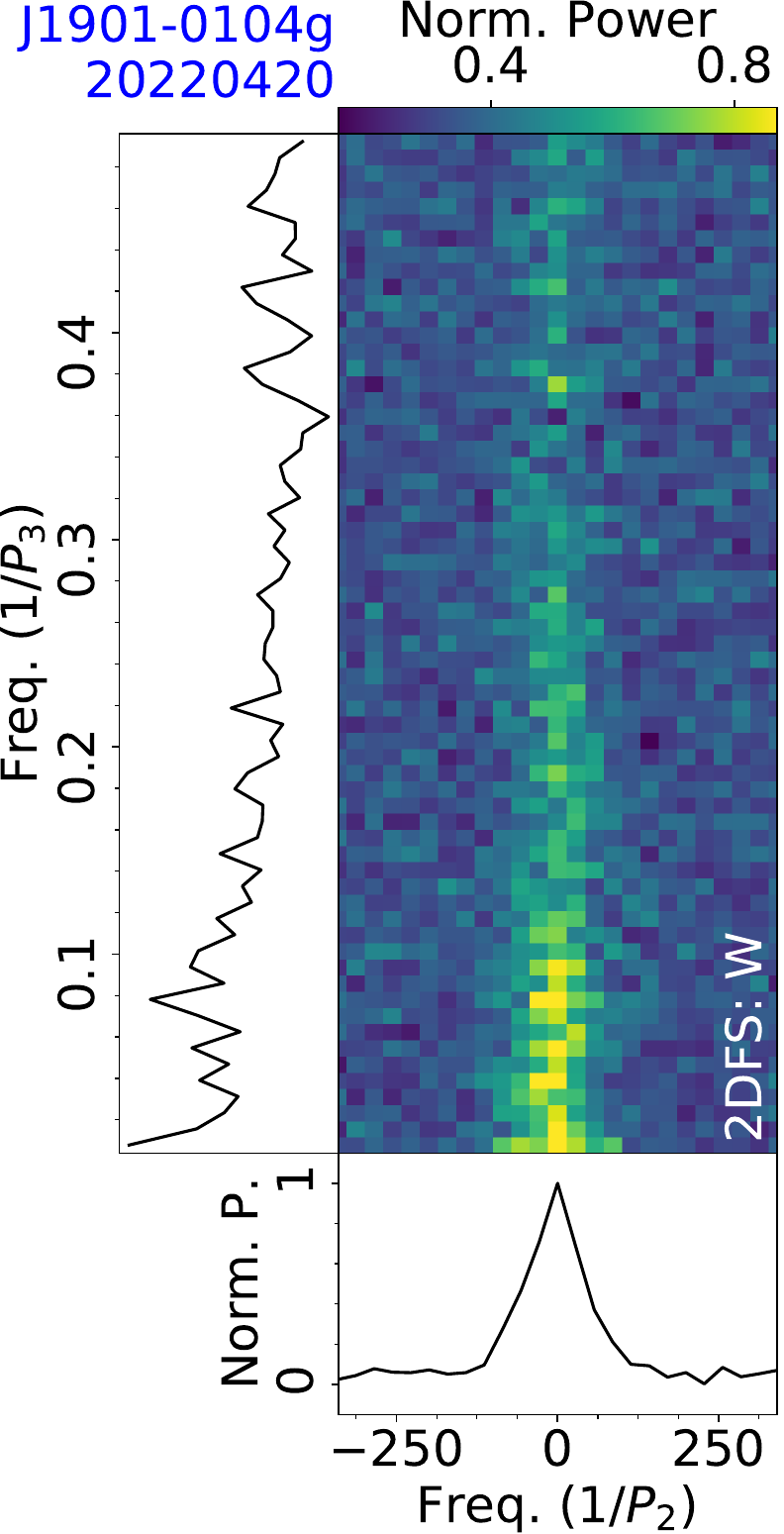}
\figcaption{Fluctuation analysis of PSR J1901-0104g for the observation on 20220420, with LRFS and 2DFS for the on-pulse region of a mean pulse profile.
\label{subfig:fluctu:J1901-0104g}}
\end{figure}

\subsection{J1901+0511}
\label{subsec:J1901+0511}

PSR J1901+0511 was found in the Arecibo L-band Feed Array pulsar survey \citep{Lyne2017}. Positive subpulse drifting with a 2.2-period temporal modulation periodicity for two components was reported by \citet{Song2023}. 

The pulsar was observed by FAST on 20220102 for 15 minutes, deriving a rotation period $P=4.6008$~s and a dispersion measure $D\!M=406.6~{\rm cm^{-3}\,pc}$. 
Single pulse sequences are shown in Fig.~\ref{subfig:TP:J1901+0511}. The pulsar has nulling and subpulse drifting behaviors. The nulling fraction of this FAST observation is estimated to be 24$\pm$3\% from the on-pulse integral energy histogram in Fig.~\ref{subfig:Hist:J1901+0511}. 2DFS of the trailing component (Fig.~\ref{subfig:fluctu:J1901+0511}) shows a positive drift feature, with centroid frequencies of $1/P_3=0.457\pm0.002$ and $1/P_2=88\pm12$, corresponding to drifting periodicities of $P_3=2.19\pm0.01$ periods and $P_2=4.1\pm0.6^\circ$.


\subsection{J1901-0104g}
\label{subsec:J1901-0104g}

PSR J1901-0104g was found in the FAST GPPS survey \citep{Han2021,han2025}. 

The pulsar was observed by FAST on 20220420 for 15 minutes, yielding a rotation period $P=0.7394$~s and a dispersion measure $D\!M=261.4~{\rm cm^{-3}\,pc}$. 
Single pulse sequences are shown in Fig.~\ref{subfig:TP:J1901-0104g}.  The fluctuation spectra in Fig.~\ref{subfig:fluctu:J1901-0104g} exhibit a temporal modulation feature with the centroid frequency of $1/P_3=0.061\pm0.002$, which corresponds to $P_3=16.3\pm0.4$ periods.

\begin{figure}[htpb]
\centering
\includegraphics[width=0.22\textwidth, angle=0]{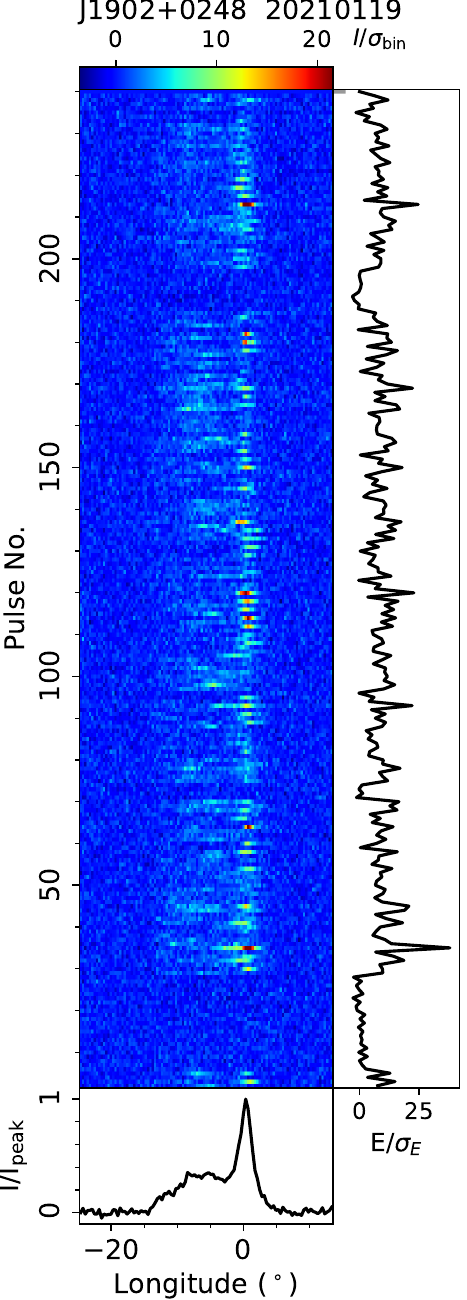}
\includegraphics[width=0.22\textwidth, angle=0]{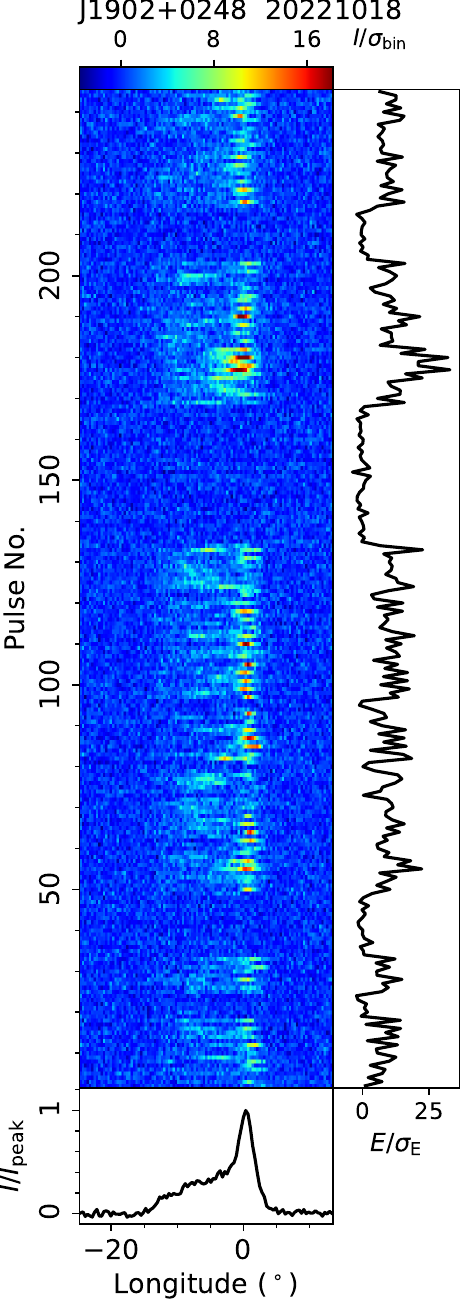}
\figcaption{Single pulse sequences of PSR J1902+0248 from the FAST observation on 20221018.
\label{subfig:TP:J1902+0248}}
\end{figure}

\begin{figure}[htpb]
\centering
\includegraphics[width=0.39\textwidth, angle=0]{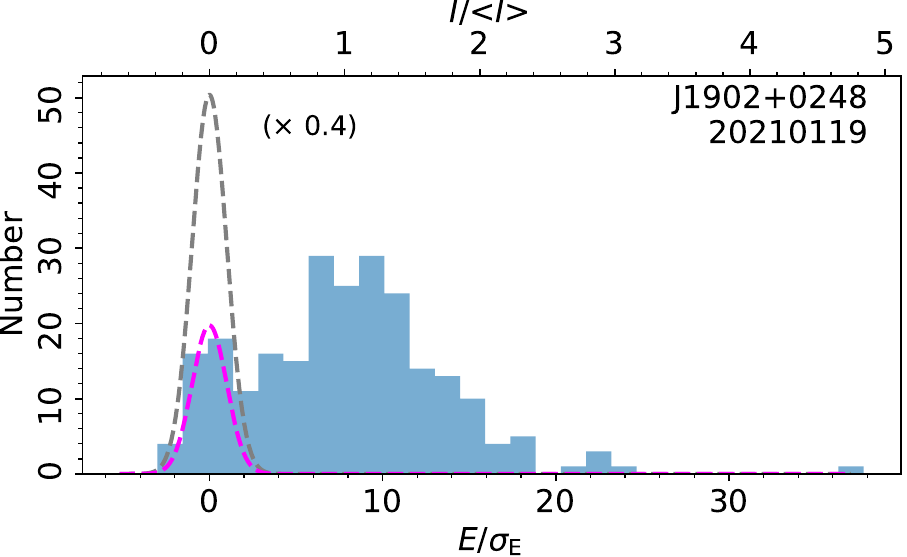}
\includegraphics[width=0.39\textwidth, angle=0]{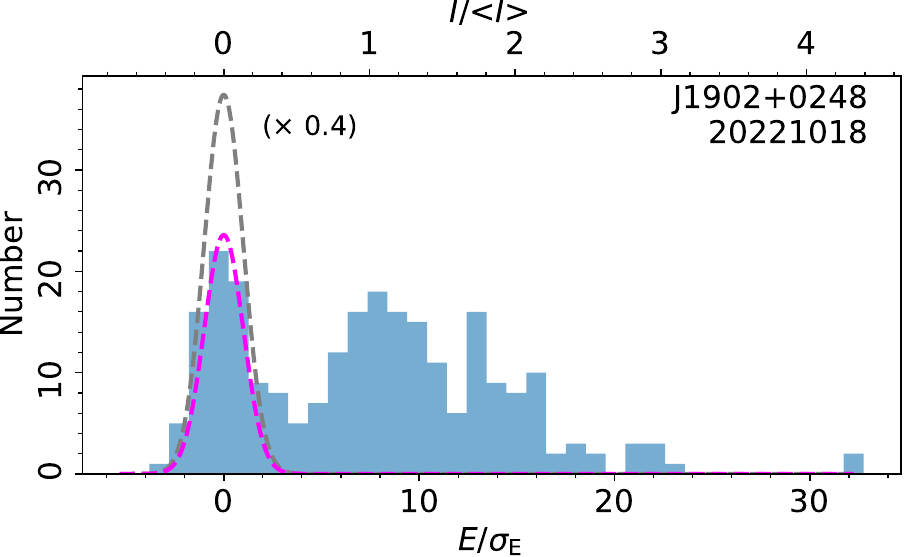}
\figcaption{On-pulse energy histograms of single pulses of PSR J1902+0248 from FAST observations on 20210119 and 20221018.
\label{subfig:Hist:J1902+0248}}
\end{figure}

\begin{figure}[htpb]
\centering
\includegraphics[width=0.22\textwidth, angle=0]{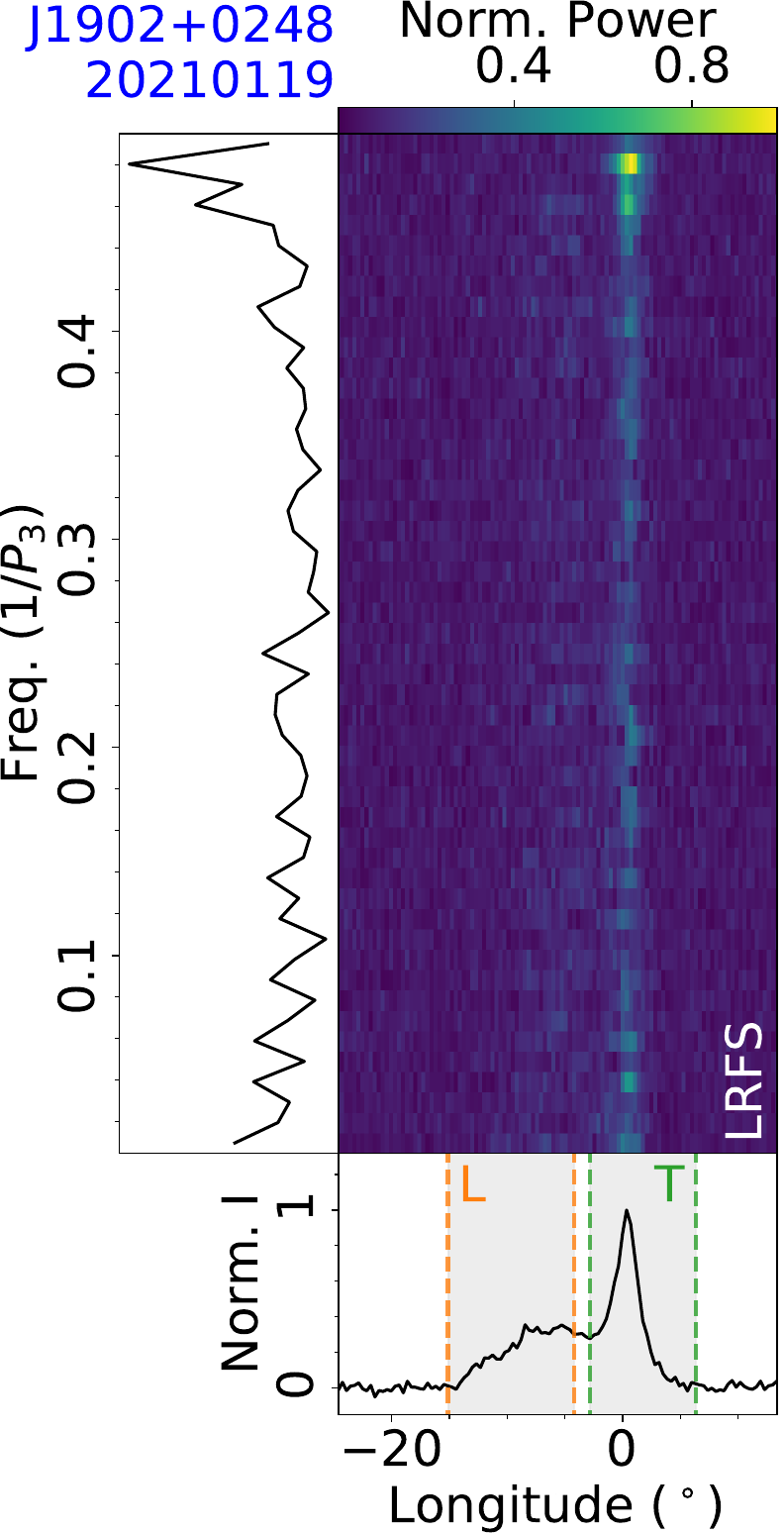}
\includegraphics[width=0.22\textwidth, angle=0]{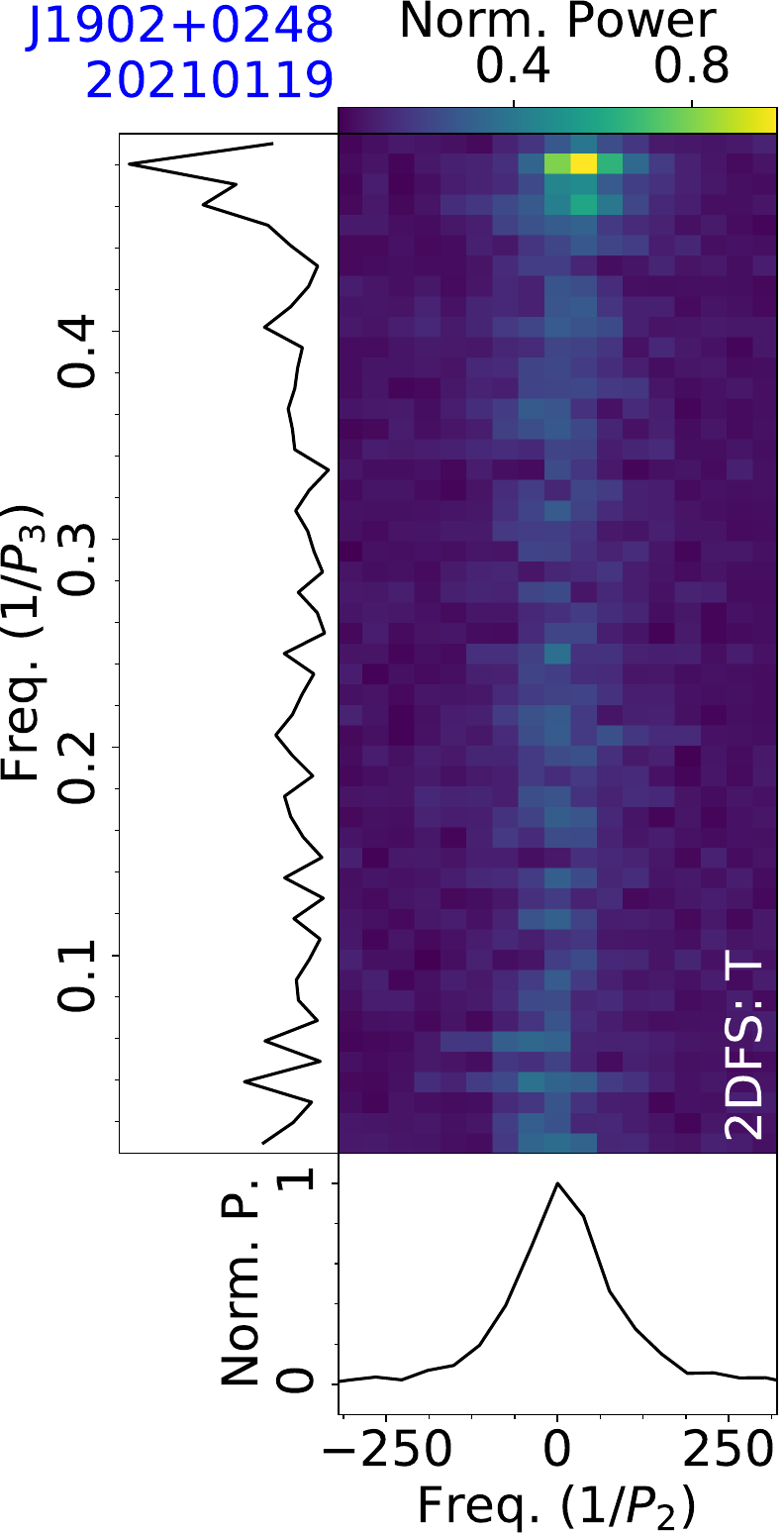}\\
\includegraphics[width=0.22\textwidth, angle=0]{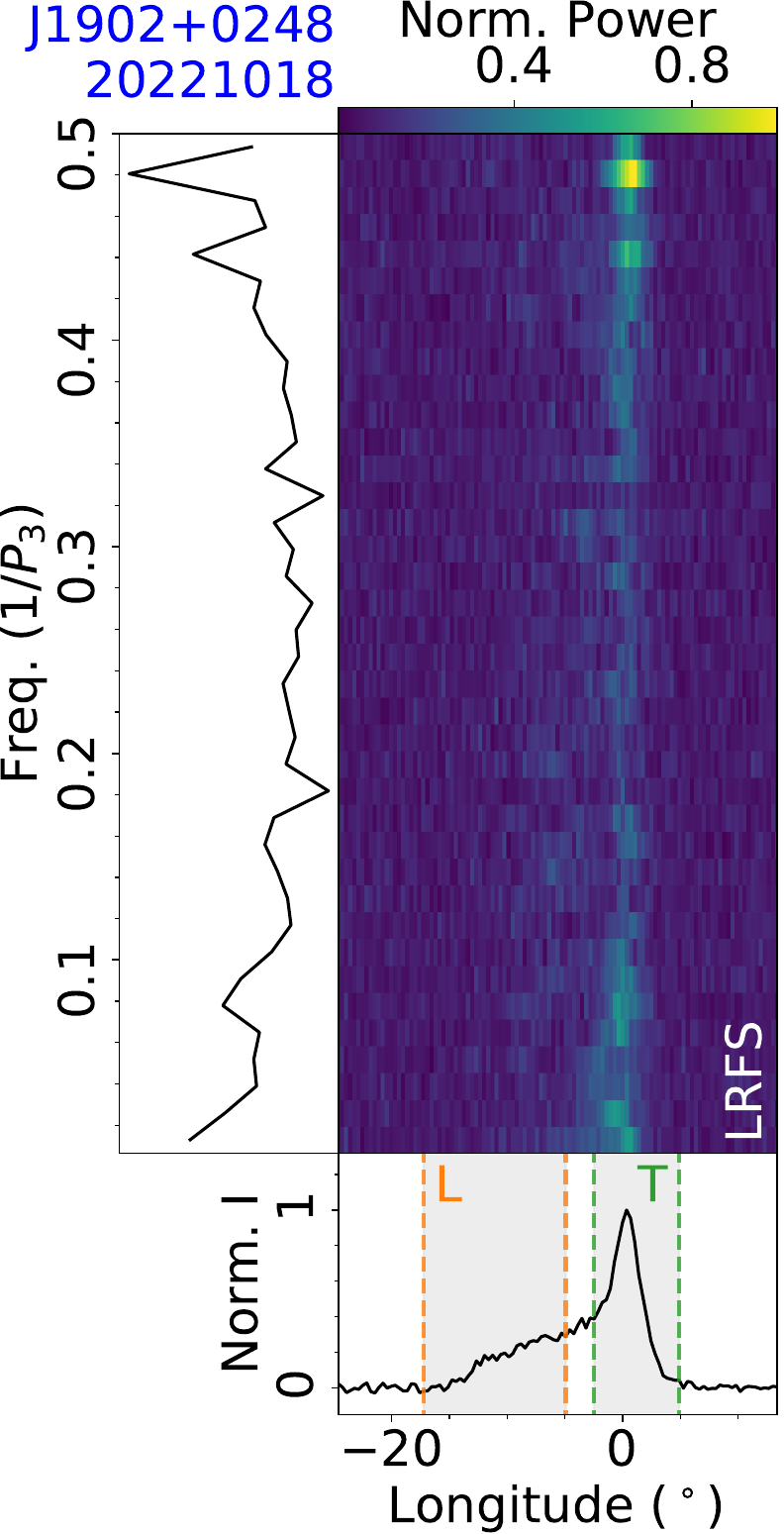}
\includegraphics[width=0.22\textwidth, angle=0]{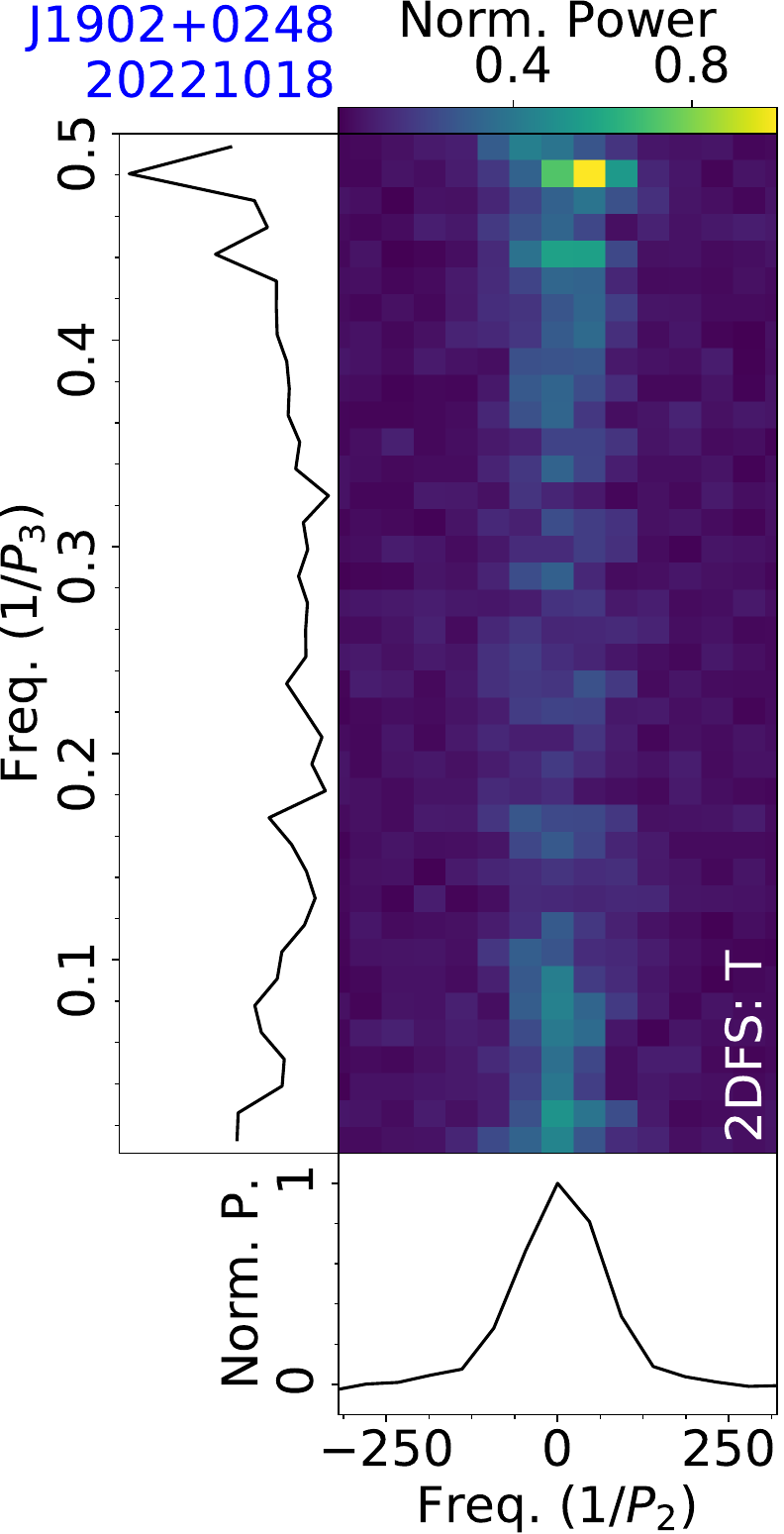}
\figcaption{Fluctuation analysis of PSR J1902+0248 for FAST observations on 20210119 and 20221018, with LRFS and 2DFS for the on-pulse region of mean pulse profiles.
\label{subfig:fluctu:J1902+0248}}
\end{figure}

\begin{figure}[htpb]
\centering
\includegraphics[width=0.22\textwidth, angle=0]{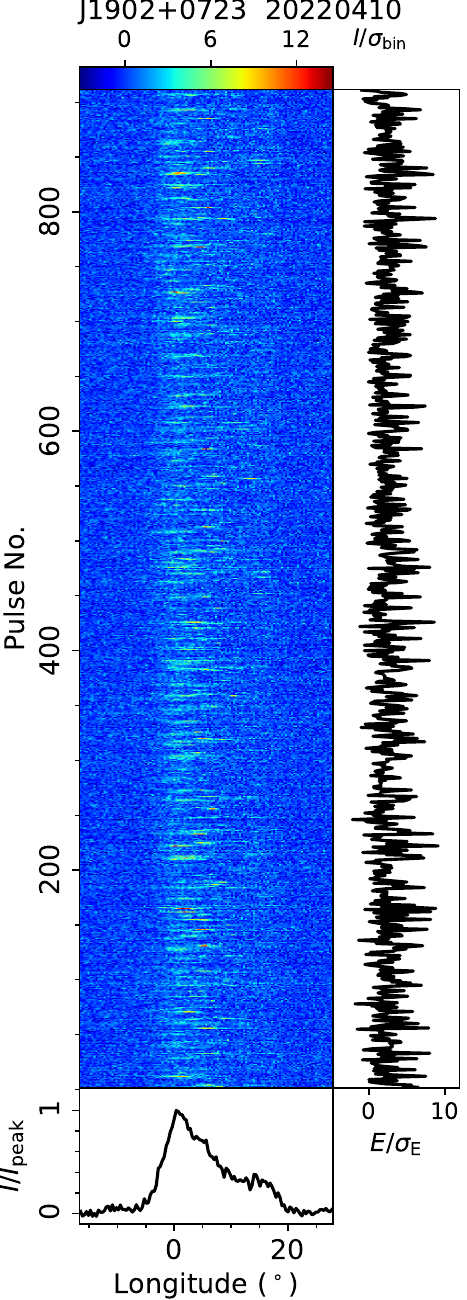}
\includegraphics[width=0.22\textwidth, angle=0]{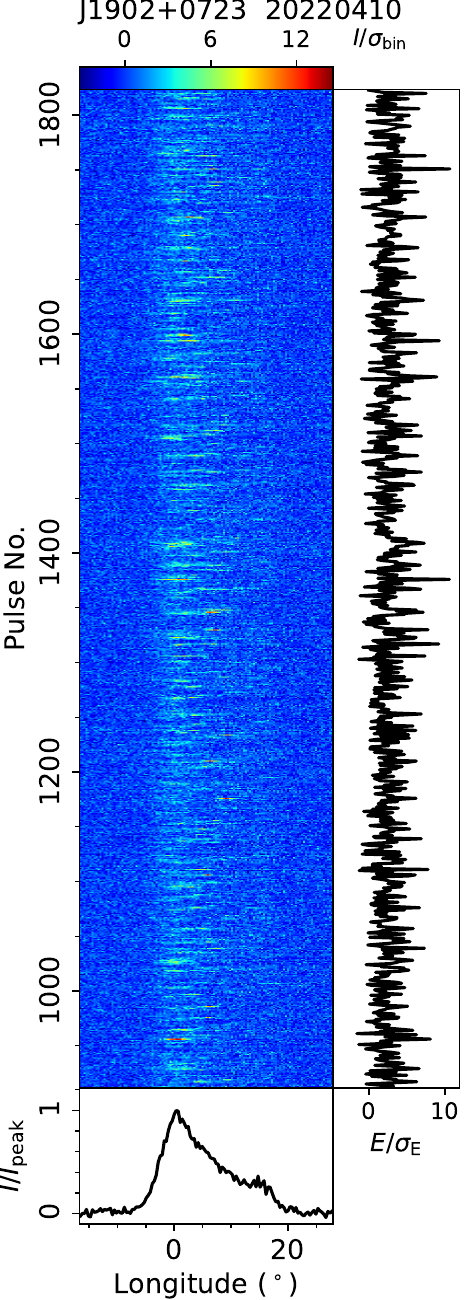}
\figcaption{Single pulse sequences of PSR J1902+0723 from the FAST observation on 20220410.
\label{subfig:TP:J1902+0723}}
\end{figure}

\begin{figure}[htpb]
\centering
\includegraphics[width=0.22\textwidth, angle=0]{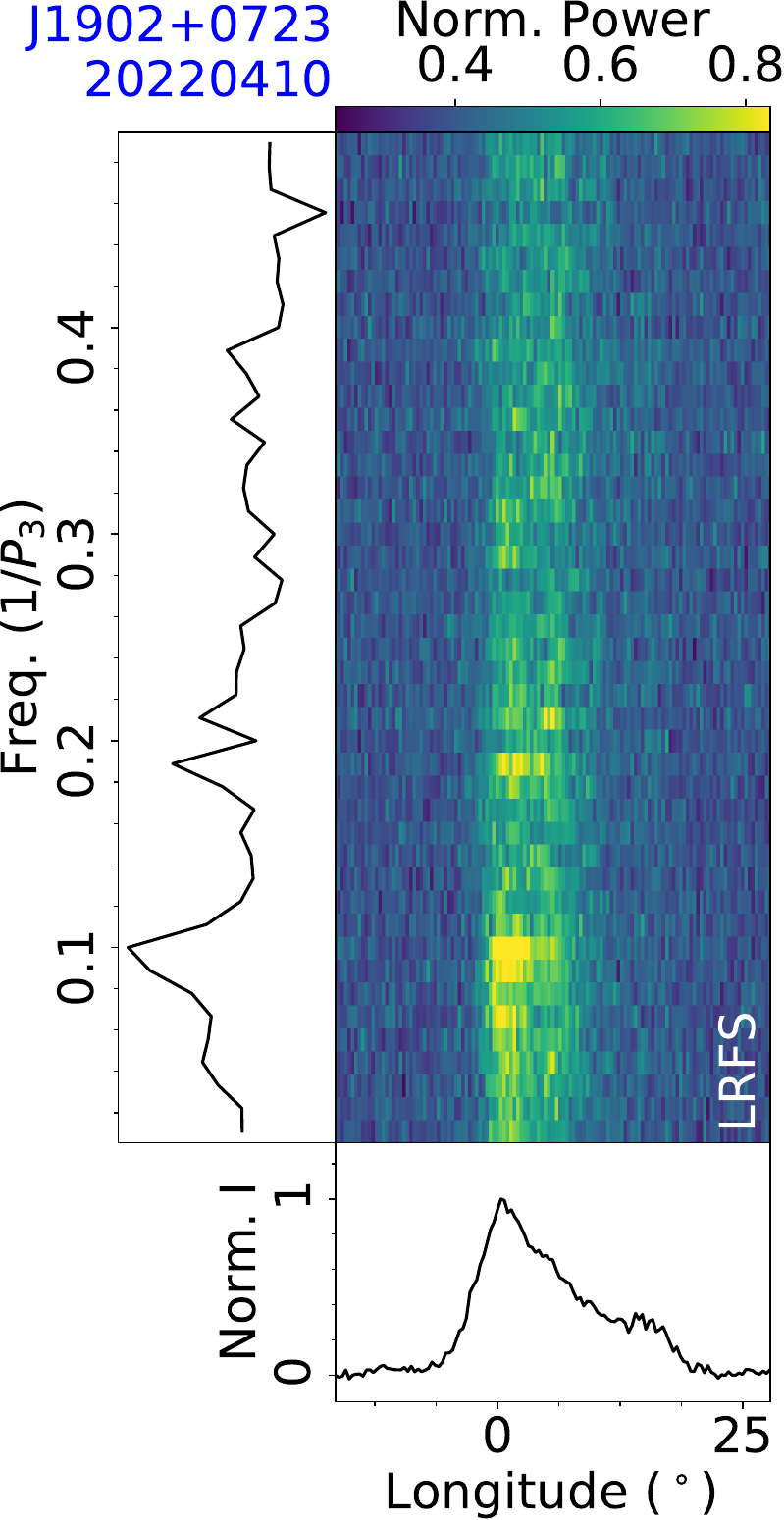}
\includegraphics[width=0.22\textwidth, angle=0]{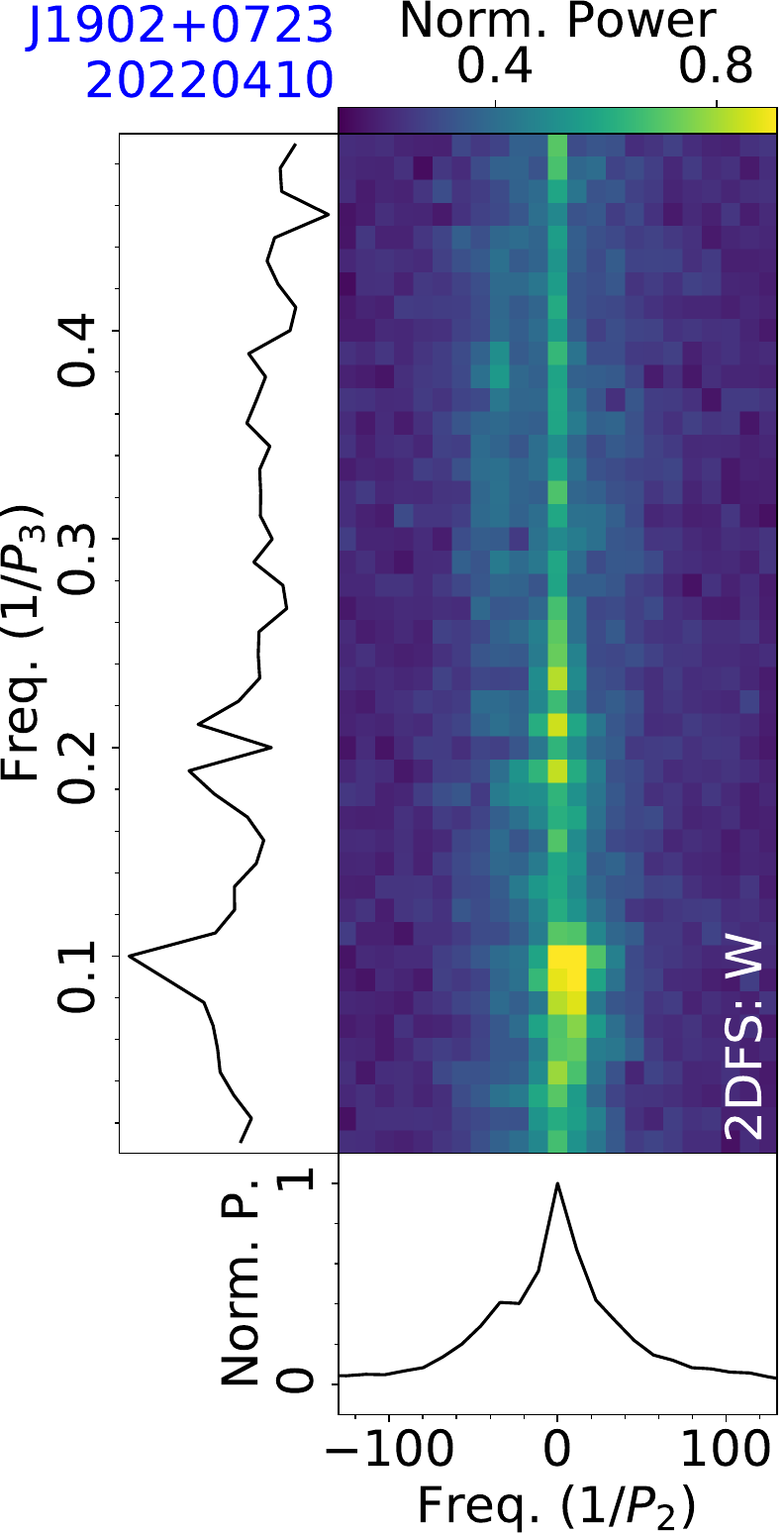}
\figcaption{Fluctuation analysis of PSR J1902+0723 for the observation on 20220410, with LRFS and 2DFS for the on-pulse region of a mean pulse profile.
\label{subfig:fluctu:J1902+0723}}
\end{figure}

\begin{figure}[htpb]
\centering
\includegraphics[width=0.21\textwidth, angle=0]{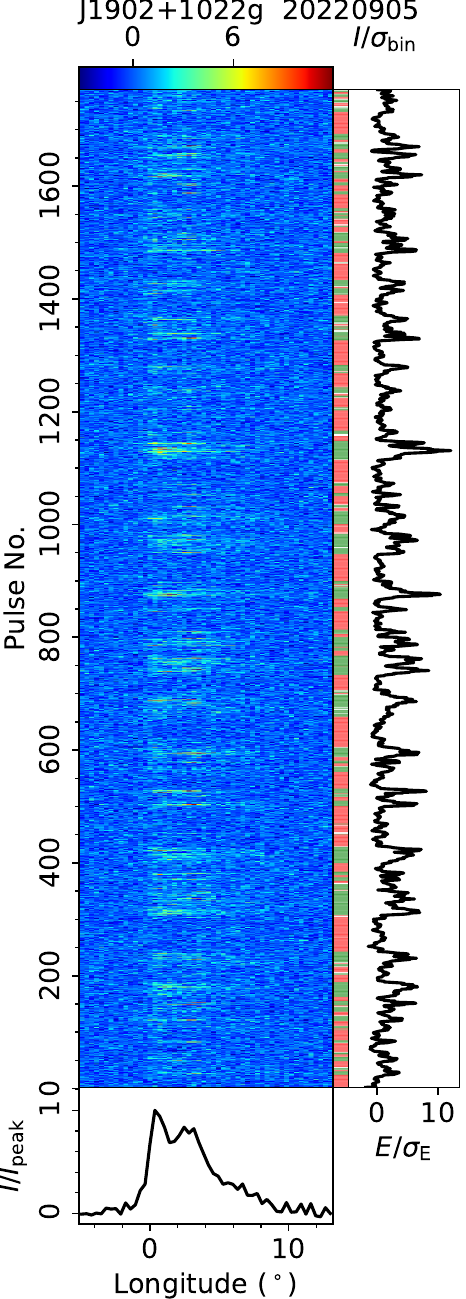}
\includegraphics[width=0.21\textwidth, angle=0]{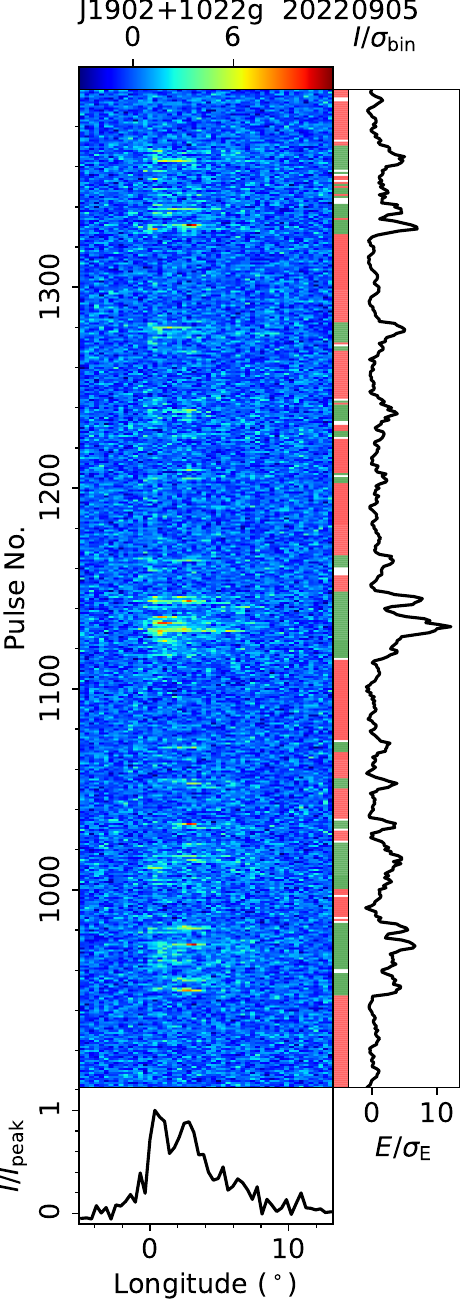}
\figcaption{Single pulse sequence of PSR J1902+1022g from the FAST observation on 20220905, as well as a zoomed-in view of pulses No.900-1400.
\label{subfig:TP:J1902+1022g}}
\end{figure}

\begin{figure}[htpb]
\centering
\includegraphics[width=0.39\textwidth, angle=0]{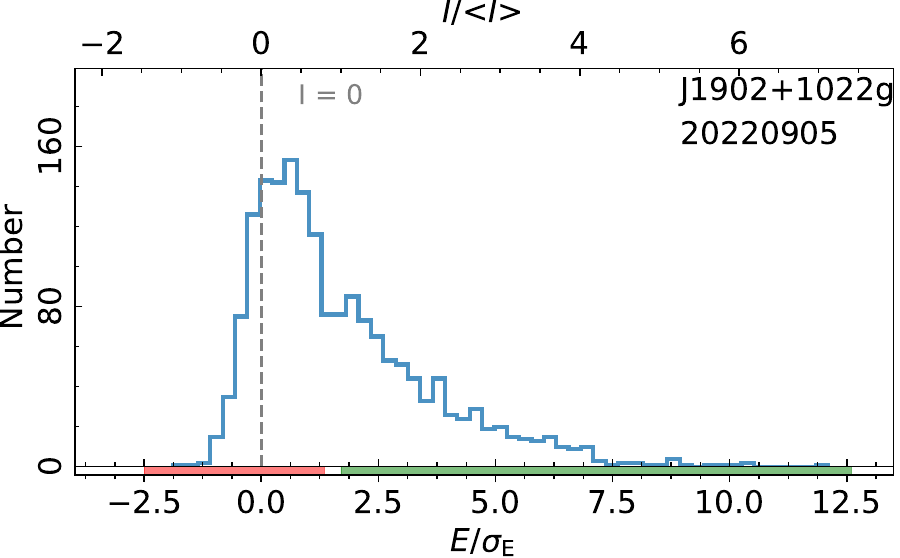}
\figcaption{Histogram of smoothed on-pulse energies for single pulses of PSR J1902+1022g from the FAST observation on 20220905.
\label{subfig:Hist:J1902+1022g}}
\end{figure}

\subsection{J1902+0248}
\label{subsec:J1902+0248}

PSR J1902+0248 was discovered in the Parkes multibeam pulsar survey \citep{hfs+04}.

The pulsar is observed by FAST on 20210119 and 20221018 for both 5 minutes. 
Single pulse sequences in Fig.~\ref{subfig:TP:J1902+0248} display that the pulsar has nulling phenomenon, and the trailing component shows clear modulation. Nulling fractions of two observations are estimated to be 16$\pm$2\% and 25$\pm$2\% from the on-pulse integral energy histograms (Fig.~\ref{subfig:Hist:J1902+0248}). The fluctuation spectra of two data are shown in Fig.~\ref{subfig:fluctu:J1902+0248}, 2DFS of the trailing parts in mean pulse profiles display positive drift features, with the centroid phase modulated frequency of $1/P_2=38\pm4$, corresponding to $P_2=9\pm1^\circ$. 
For the observation on 20210119, the centroid temporal frequency is $1/P_3=0.472\pm0.001$, corresponding to the drifting parameter of $P_3=2.12\pm0.01$ periods. The centroid temporal frequency of the data on 20221018 is $1/P_3=0.455\pm0.002$, yielding $P_3=2.20\pm0.01$ periods.

\subsection{J1902+0723}
\label{subsec:J1902+0723}

PSR J1902+0723 was discovered by the Arecibo telescope \citep{Nice1995}. 

This pulsar was observed by FAST on 20220410 for 15 minutes, and a rotation period of $P=0.4878$~s and a dispersion measure of $D\!M=105.5~{\rm cm^{-3}\,pc}$ were determined. Single pulse sequences in Fig.~\ref{subfig:TP:J1902+0723} show the subpulse drifting phenomenon. LRFS and 2DFS are displayed in Fig.~\ref{subfig:fluctu:J1902+0723}. The main feature in 2DFS is related to the positively drifting behavior, which is characterized by the centroid frequencies of $1/P_3=0.081\pm0.001$ and $1/P_2=8\pm1$, corresponding to periodicities of $1/P_3=12.4\pm0.2$ periods and $1/P_2=46\pm6$ degrees.

\subsection{J1902+1022g}
\label{subsec:J1902+1022g}

PSR J1902+1022g was discovered by the FAST GPPS survey \citep{Han2021,han2025}.

The pulsar was observed by the FAST on 20220905 for 15 minutes, deriving a rotation period of $P=0.5019$~s and a dispersion measure of $D\!M=288.8~{\rm cm^{-3}\,pc}$. The single pulse sequence and a zoomed-in view of this observation in Fig.~\ref{subfig:TP:J1902+1022g} illustrate the changes between the weak and bright emission modes. 
The on-pulse integral energy sequence of  single pulses is smoothed using a 5-period moving average. 
Single pulses of two modes are distinguished from the smoothed on-pulse energy histogram in Figure~\ref{subfig:Hist:J1902+1022g}. The weak and bright emission modes are labeled using red and green, respectively.

\subsection{J1902+1141}
\label{subsec:J1902+1141}

PSR J1902+1141 was discovered in the PALFA survey by \citep{Patel2018}, and the long-term timing results were lately presented by \citet{Parent2022}.

\begin{figure}[htpb]
\centering
\includegraphics[width=0.21\textwidth, angle=0]{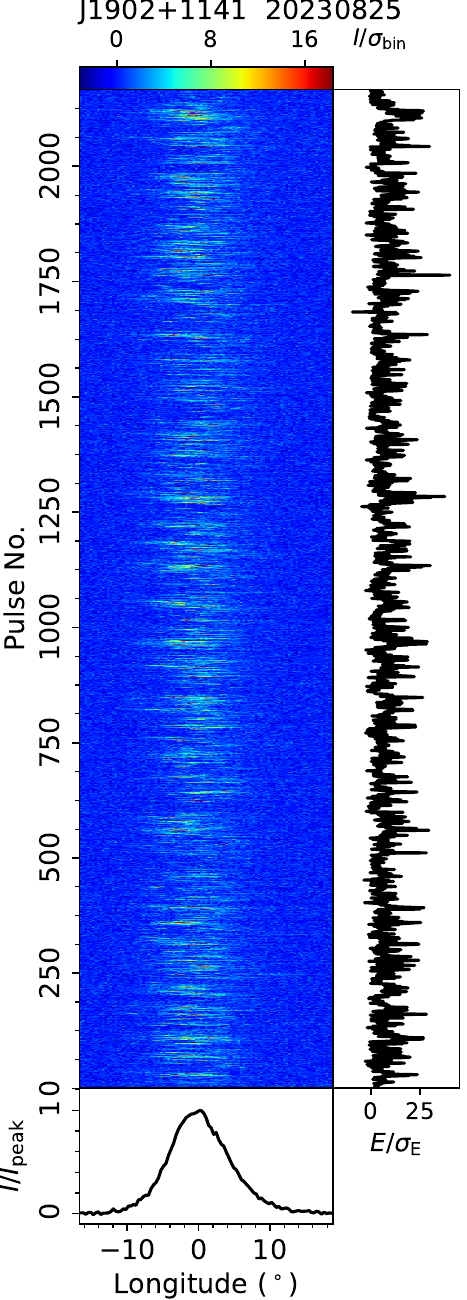}
\includegraphics[width=0.21\textwidth, angle=0]{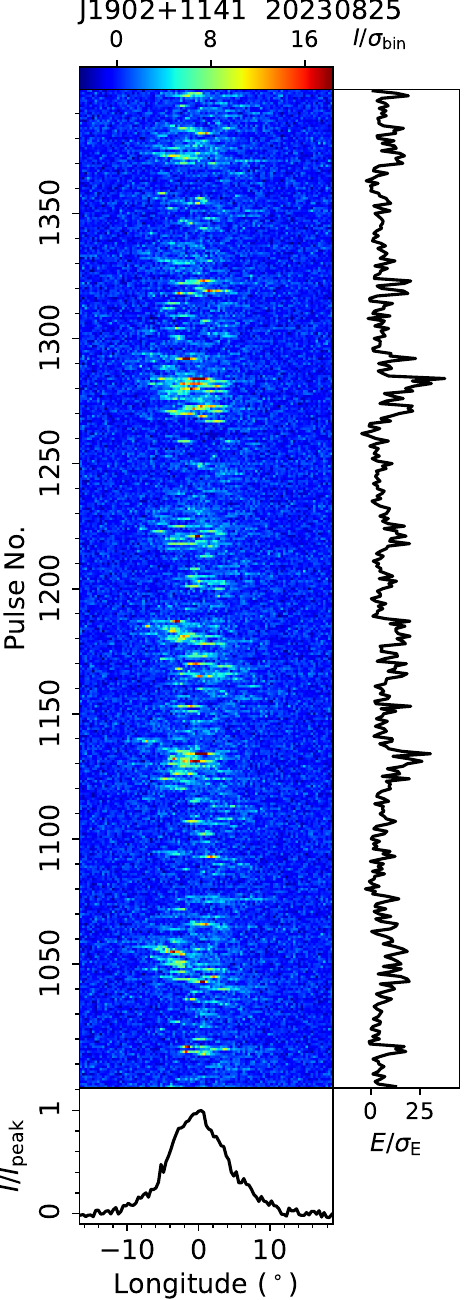}
\figcaption{Single pulse sequence of PSR J1902+1141 from the FAST observation on 20230825, as well as a zoomed-in view of pulses No.1000-1400.
\label{subfig:TP:J1902+1141}}
\end{figure}

\begin{figure}[htpb]
\centering
\includegraphics[width=0.22\textwidth, angle=0]{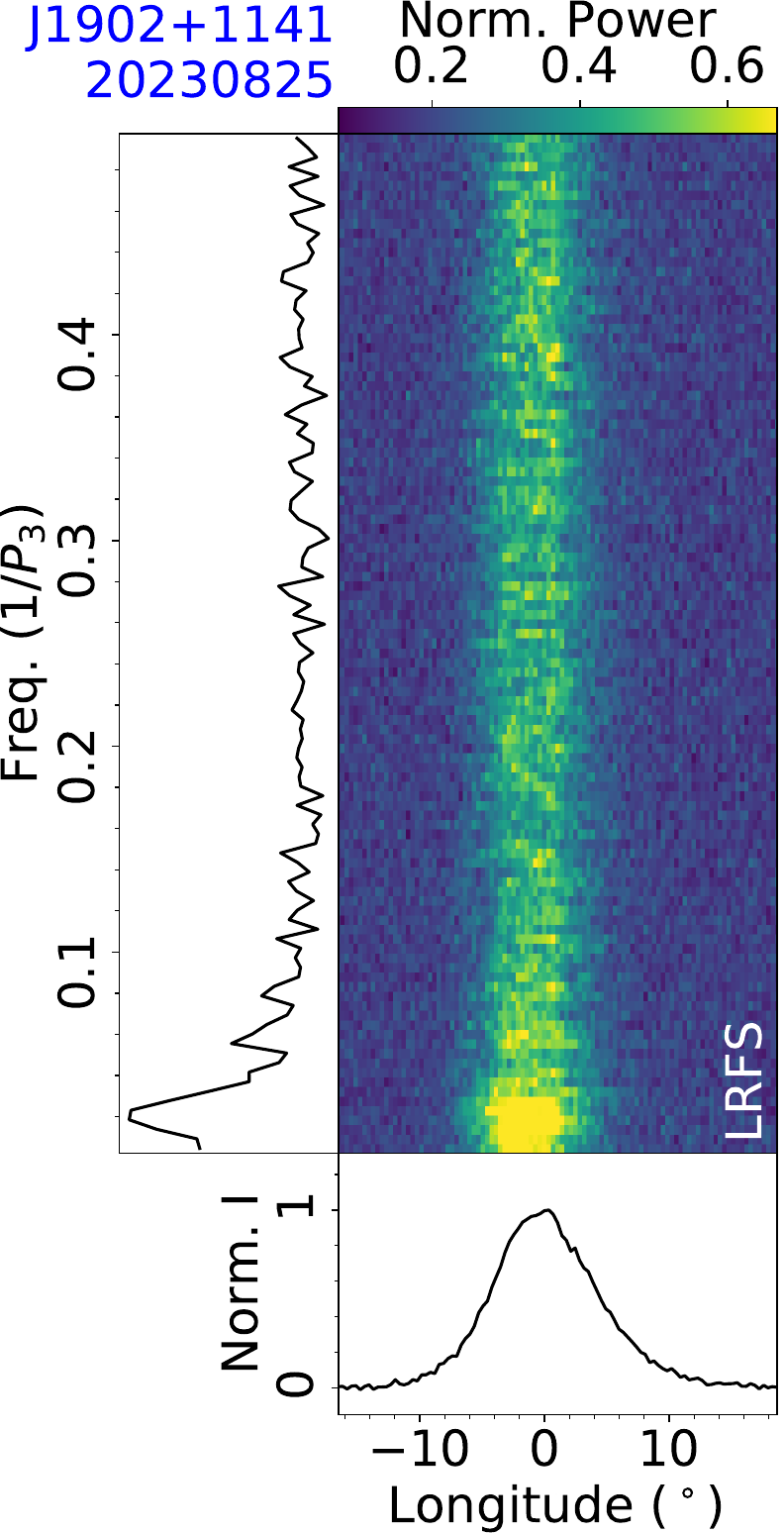}
\includegraphics[width=0.22\textwidth, angle=0]{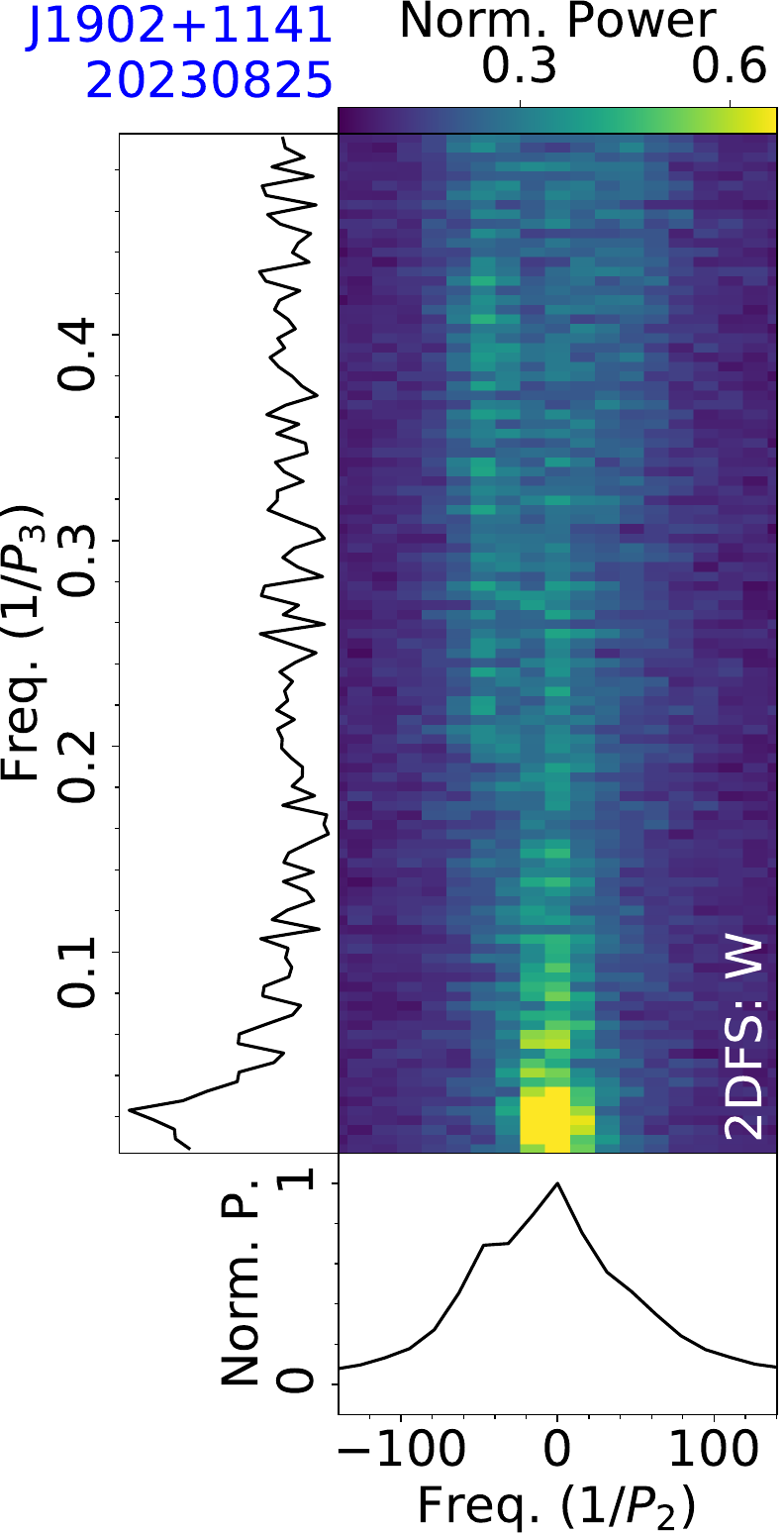}
\figcaption{Fluctuation analysis of PSR J1902+1141 for the observation on 20230825, with LRFS and 2DFS for the on-pulse region of a mean pulse profile.
\label{subfig:fluctu:J1902+1141}}
\end{figure}

This pulsar was observed by FAST on 20230825 for 15 minutes, with a rotation period $P=0.4092$~s and a dispersion measure $D\!M=269.1~{\rm cm^{-3}\,pc}$ derived. The single pulse sequence of this observation and a zoomed-in view of pulses No. 1000-1400 are shown in Figure~\ref{subfig:TP:J1902+1141}, displaying a negative drifting phenomenon, as well as a low-frequency modulation behavior. From LRFS and 2DFS in Figure~\ref{subfig:fluctu:J1902+1141}, the negative drift feature, with temporally modulated frequency widely distributed from 0.2 to 0.5, has centroid frequencies of $1/P_3=0.345\pm0.003$ and $1/P_2=-48\pm1$, corresponding to periodicities of $P_3=2.90\pm0.02$ periods and $P_2=-7.4\pm0.1$ degrees. The low-frequency modulation feature, which seems superimposed on the negative drifting, is characterized by a centroid frequency of $1/P_3=0.028\pm0.001$ ($P_3=36\pm1$ periods) with a tendency of the negative phase modulation property.

\begin{figure}[htpb]
\centering
\includegraphics[width=0.22\textwidth, angle=0]{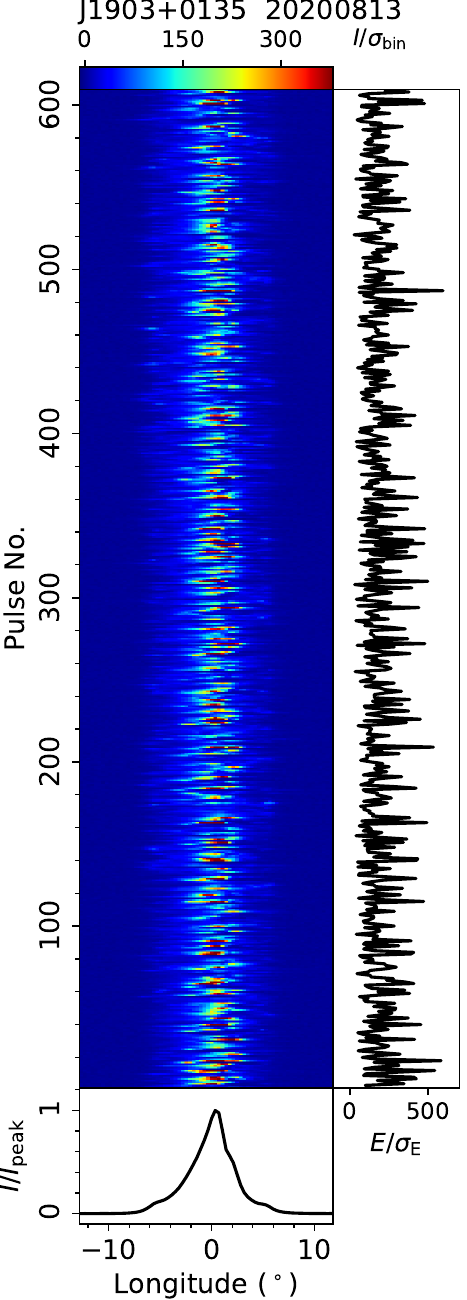}
\includegraphics[width=0.22\textwidth, angle=0]{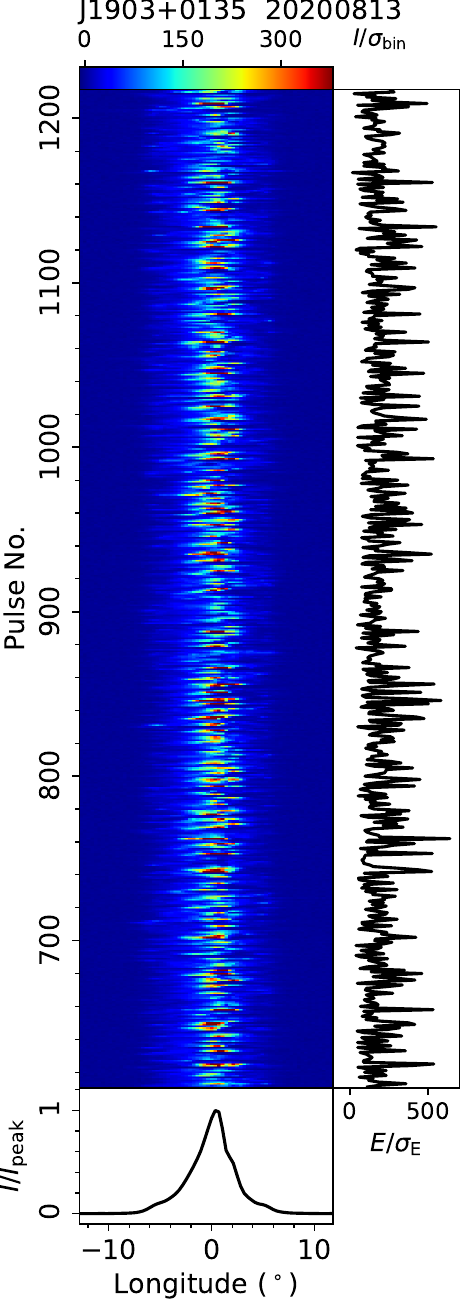}
\figcaption{Single pulse sequences of PSR J1903+0135 from the FAST observation on 20200813.
\label{subfig:TP:J1903+0135}}
\end{figure}

\begin{figure}[htpb]
\centering
\includegraphics[width=0.22\textwidth, angle=0]{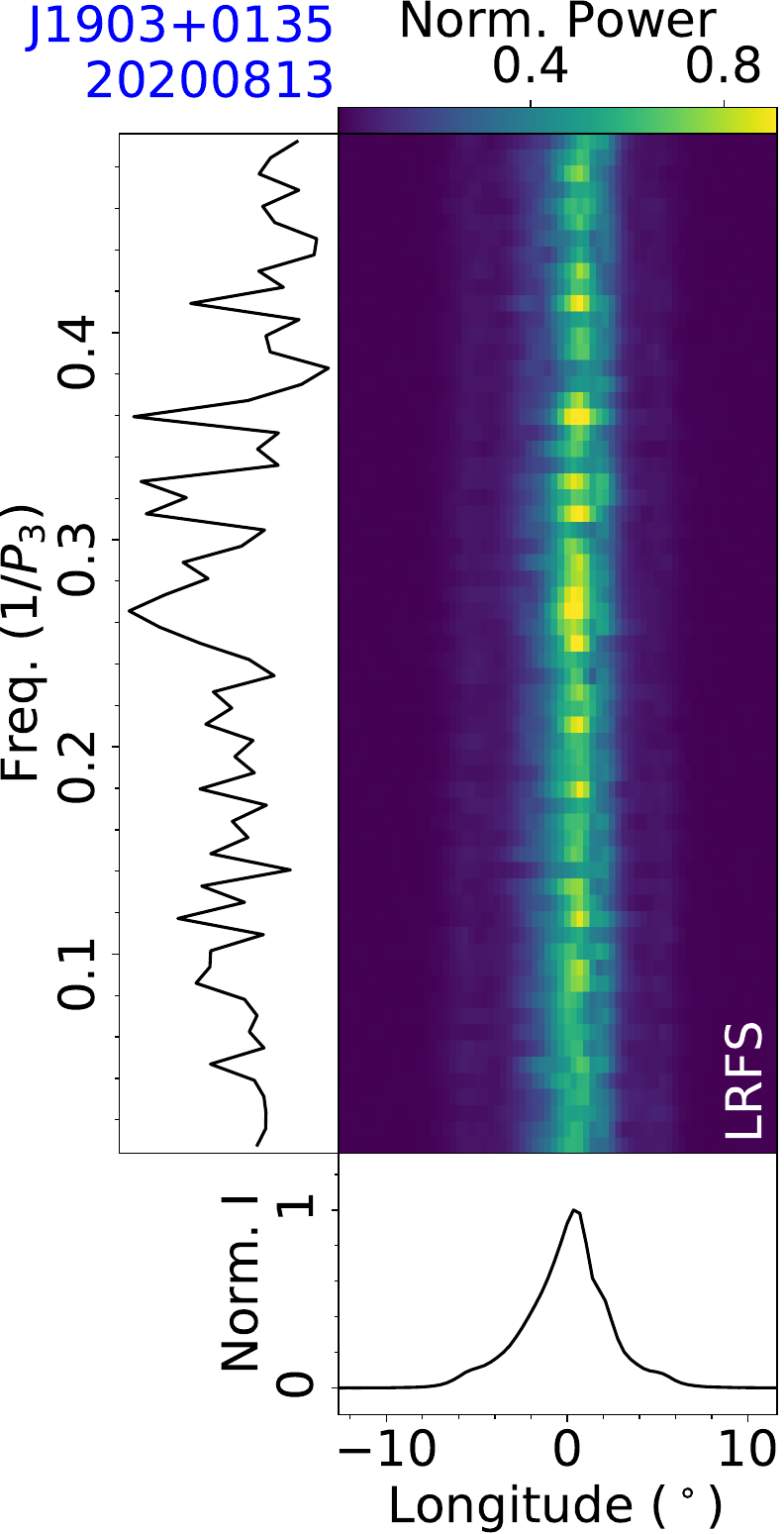}
\includegraphics[width=0.22\textwidth, angle=0]{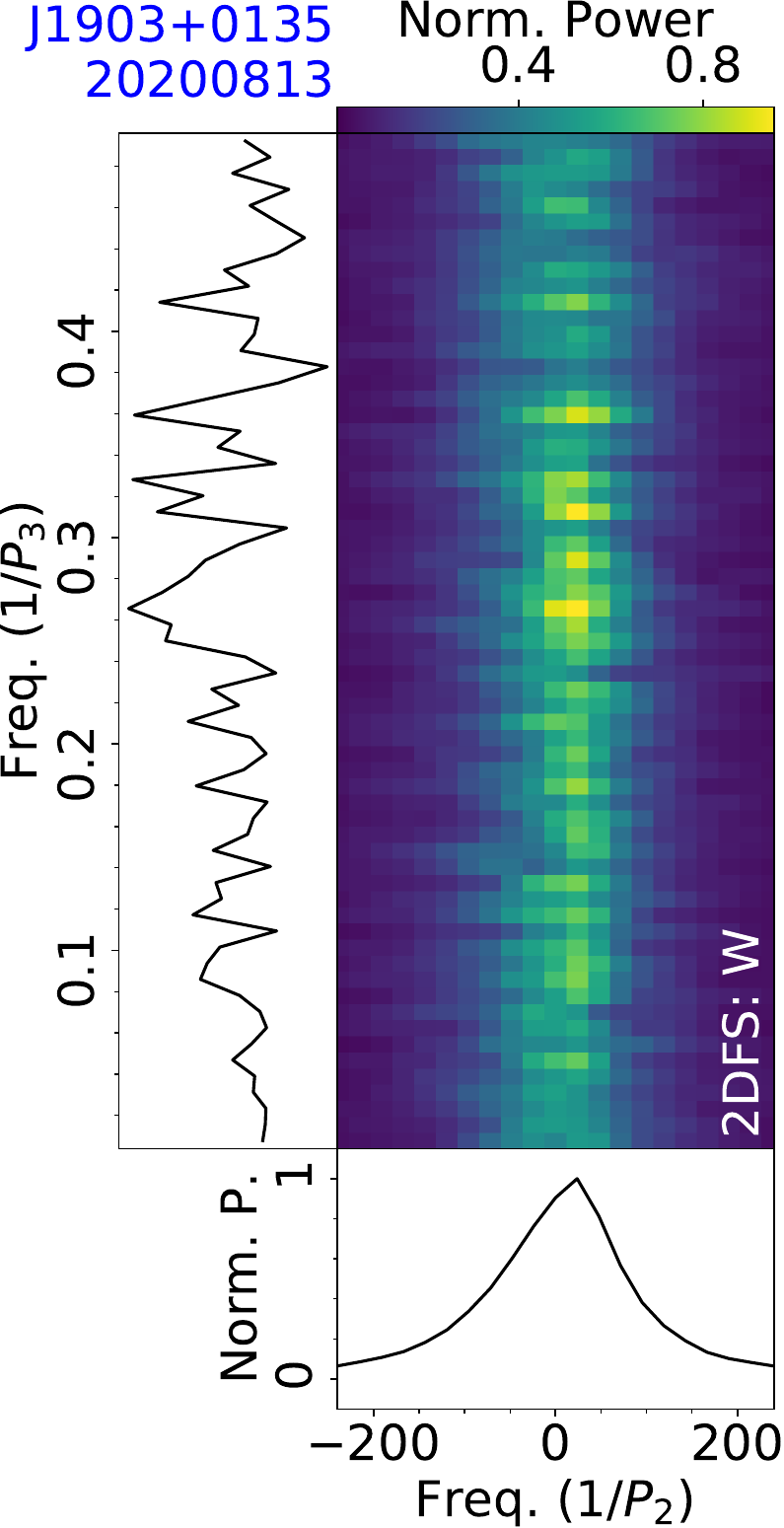}
\figcaption{Fluctuation analysis of PSR J1903+0135 from the FAST observation on 20200813, with LRFS and 2DFS for the on-pulse region of a mean pulse profile.
\label{subfig:fluctu:J1903+0135}}
\end{figure}

\begin{figure}[htpb]
\centering
\includegraphics[width=0.22\textwidth, angle=0]{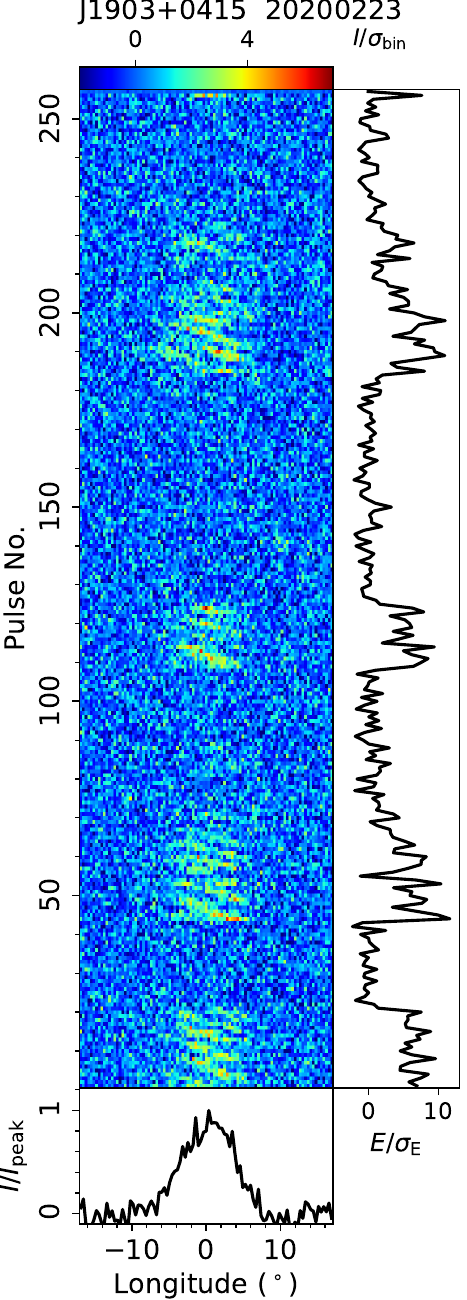}
\figcaption{Single pulse sequence of PSR J1903+0415 from the FAST observation on 20200223.
\label{subfig:TP:J1903+0415}}
\end{figure}

\begin{figure}[htpb]
\centering
\includegraphics[width=0.39\textwidth, angle=0]{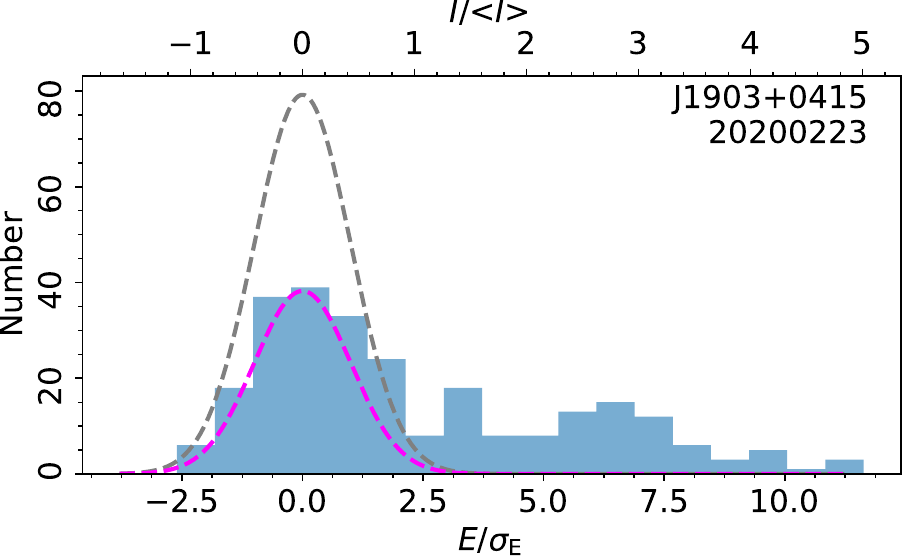}
\figcaption{On-pulse energy histogram of single pulses of PSR J1903+0415 from the FAST observation on 20200223.
\label{subfig:Hist:J1903+0415}}
\end{figure}

\begin{figure}[htpb]
\centering
\includegraphics[width=0.22\textwidth, angle=0]{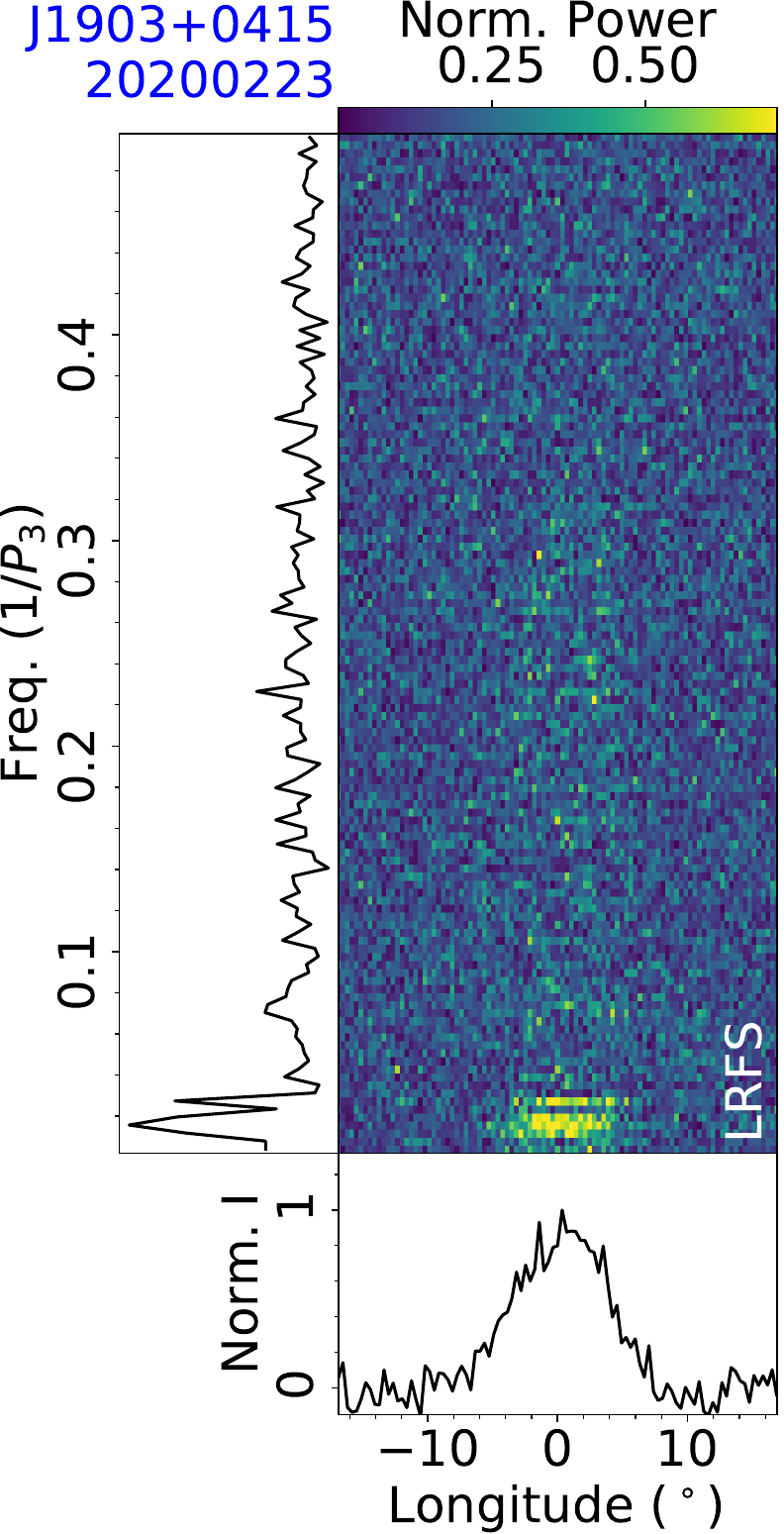}
\includegraphics[width=0.22\textwidth, angle=0]{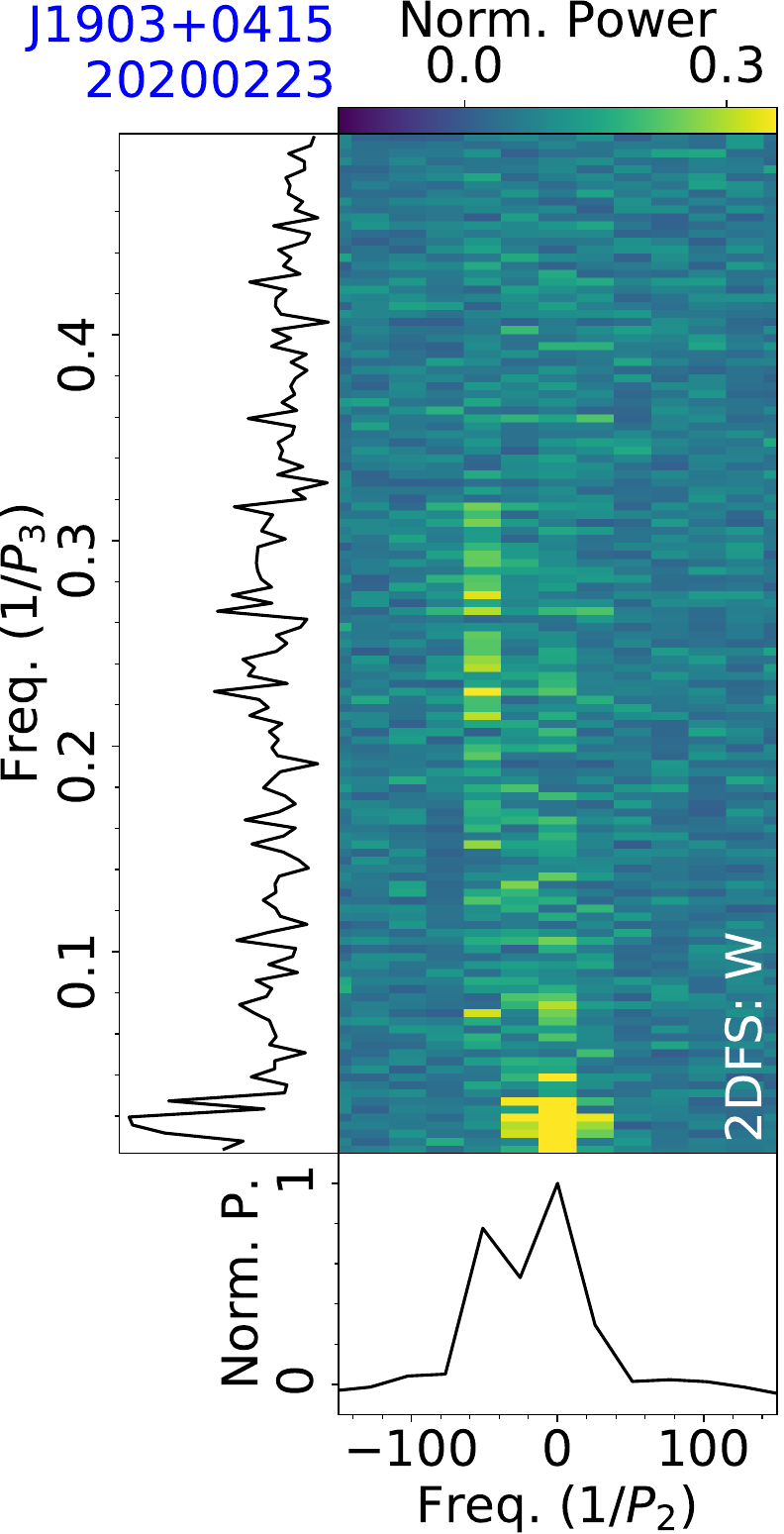}
\figcaption{Fluctuation analysis of PSR J1903+0415 from the FAST observation on 20200223, with LRFS and 2DFS for the on-pulse region of a mean pulse profile.
\label{subfig:fluctu:J1903+0415}}
\end{figure}

\subsection{J1903+0135}
\label{subsec:J1903+0135}

PSR J1903+0135 was discovered by \citet{Davies1973} using the Mark 1A radio telescope at Jodrell Bank. Subpulse drifting behavior has been reported by \citet{Weltevrede2006} and \citet{Song2023}. 

This pulsar was observed by FAST on 20200813 for 15 minutes, yielding a rotation period $P=0.7293$~s and a dispersion measure $D\!M=245.0~{\rm cm^{-3}\,pc}$. 
Single pulse sequences of this observation are displayed in Fig.~\ref{subfig:TP:J1903+0135}. 
2DFS shows a clear drift feature (Fig.~\ref{subfig:fluctu:J1903+0135}), and the modulation frequency with time is wide, which may be caused by the variable drifting rate, consistent with results in \citet{Weltevrede2006}. Centroid frequencies of the drift feature are $1/P_3=0.23\pm0.01$ and $1/P_2=24\pm1$, corresponding to drifting periodicities of $P_3=4.3\pm0.1$ periods and $P_2=15\pm1^\circ$. However, the top part of the 2DFS is not double-peaked as described earlier.

\begin{figure}[htpb]
\centering
\includegraphics[width=0.44\textwidth, angle=0]{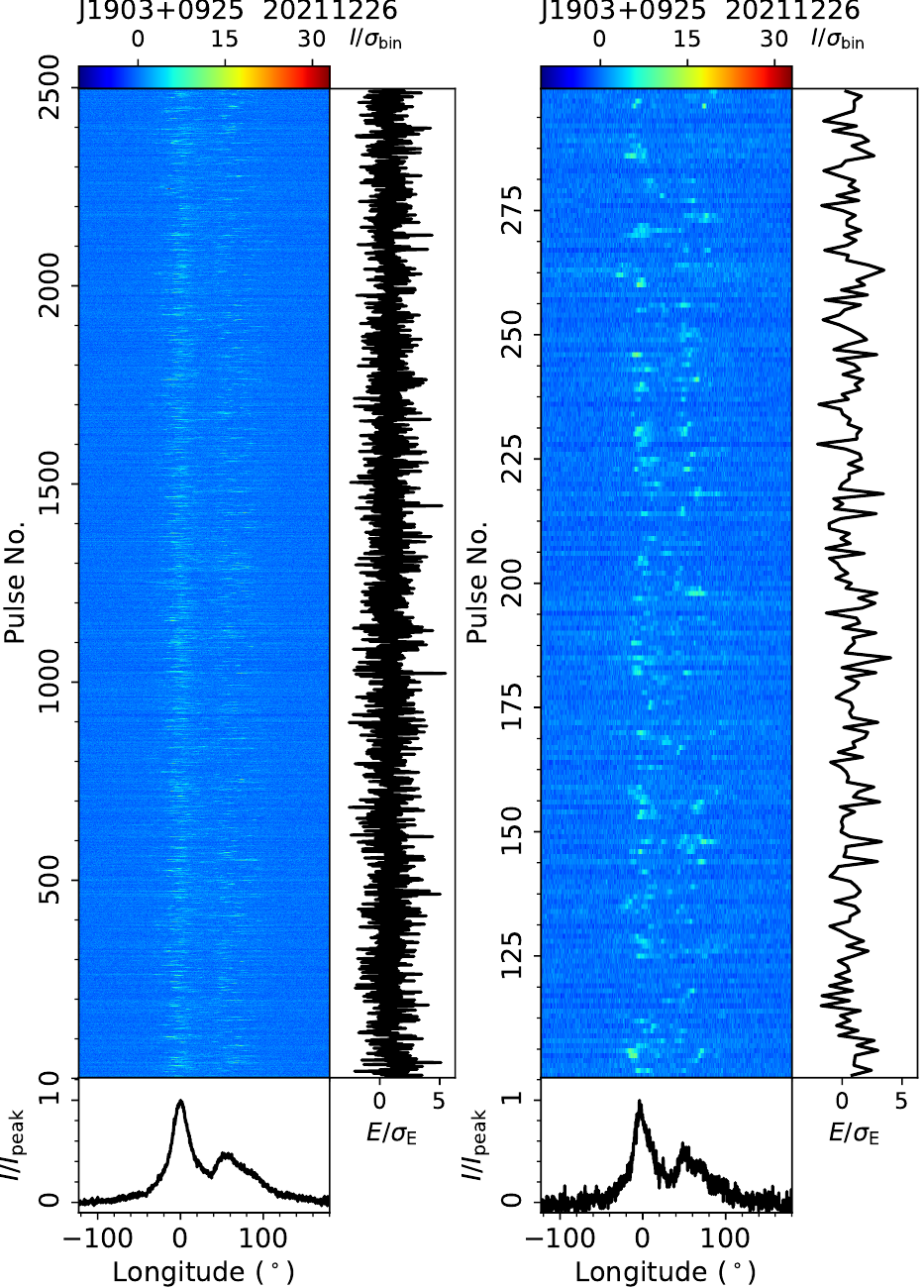}
\figcaption{Single pulse sequence of PSR J1903+0925 from the FAST observation on 20211226, and a zoomed-in view of pulses No.100-300.
\label{subfig:TP:J1903+0925}}
\end{figure}

\begin{figure}[htpb]
\centering
\includegraphics[width=0.44\textwidth, angle=0]{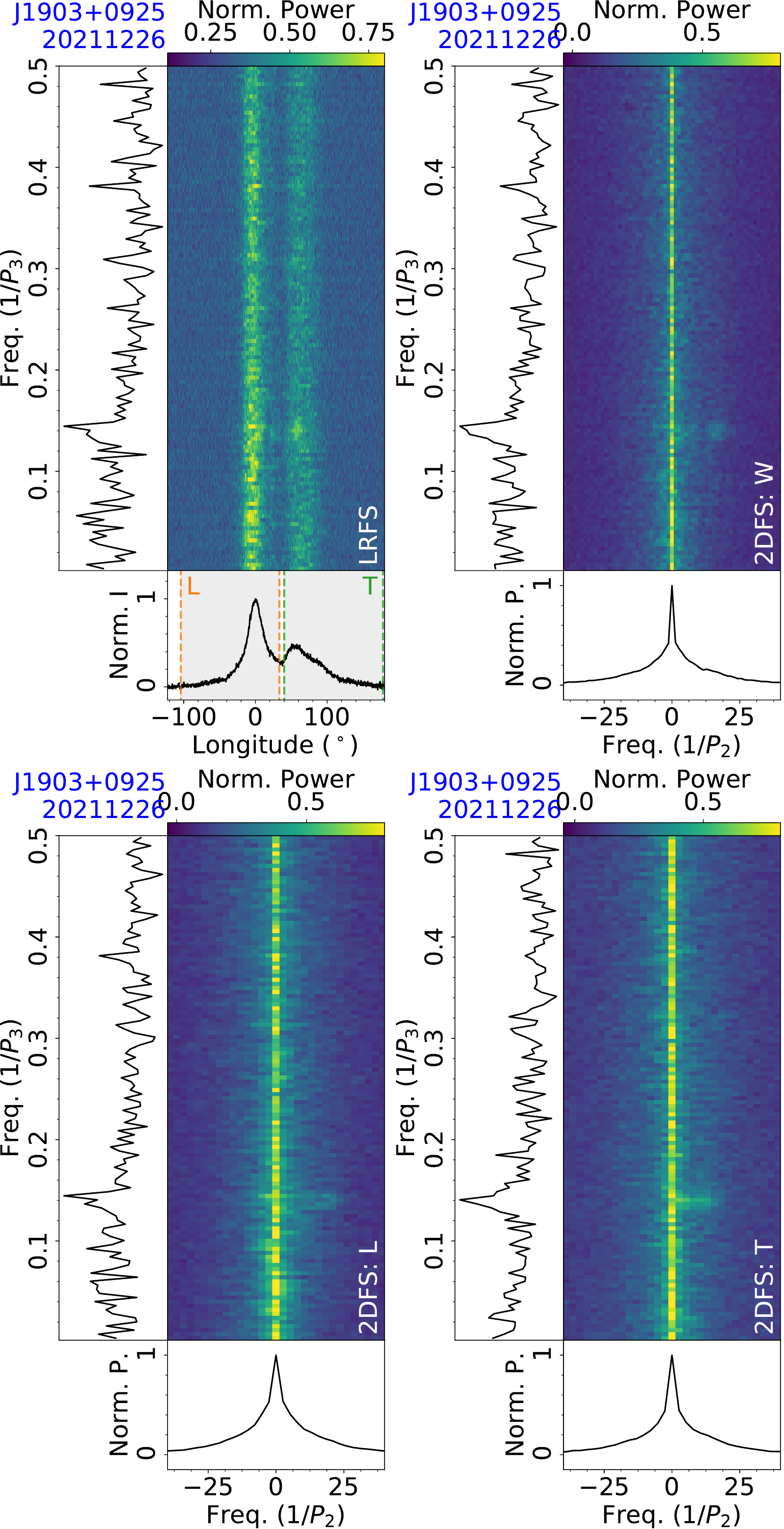}
\figcaption{Fluctuation analysis of PSR J1903+0925 for the observation on 20211226, with LRFS (top-left), and 2DFS for the on-pulse region (top-right), leading part (bottom-left) and trailing part (bottom-right) of a mean pulse profile.
\label{subfig:fluctu:J1903+0925}}
\end{figure}

\begin{figure}[htpb]
\centering
\includegraphics[width=0.22\textwidth, angle=0]{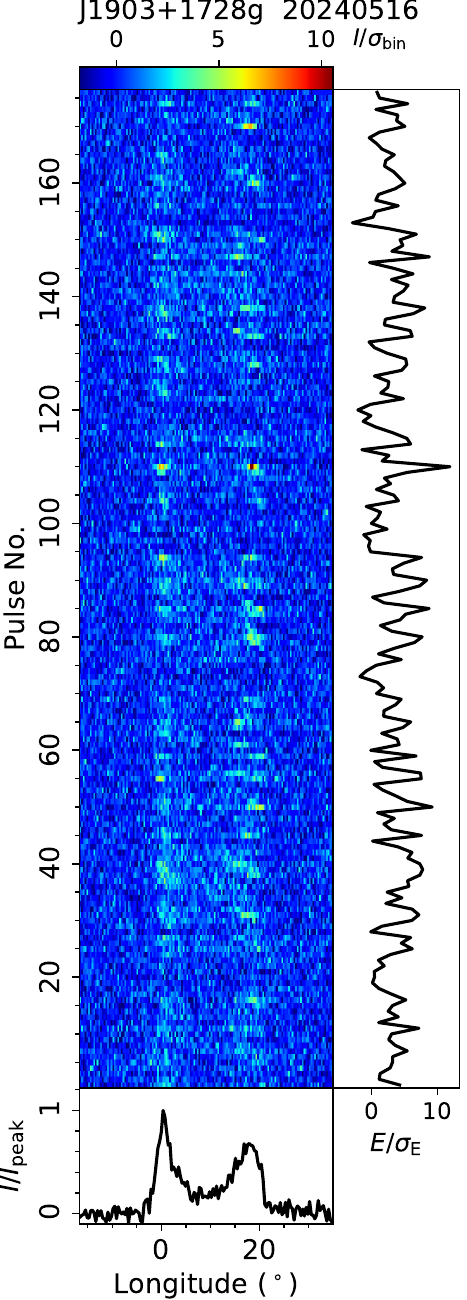}
\figcaption{Single pulse sequence of PSR J1903+1728g from the FAST observation on 20240516.
\label{subfig:TP:J1903+1728g}}
\end{figure}

\begin{figure}[htpb]
\centering
\includegraphics[width=0.22\textwidth, angle=0]{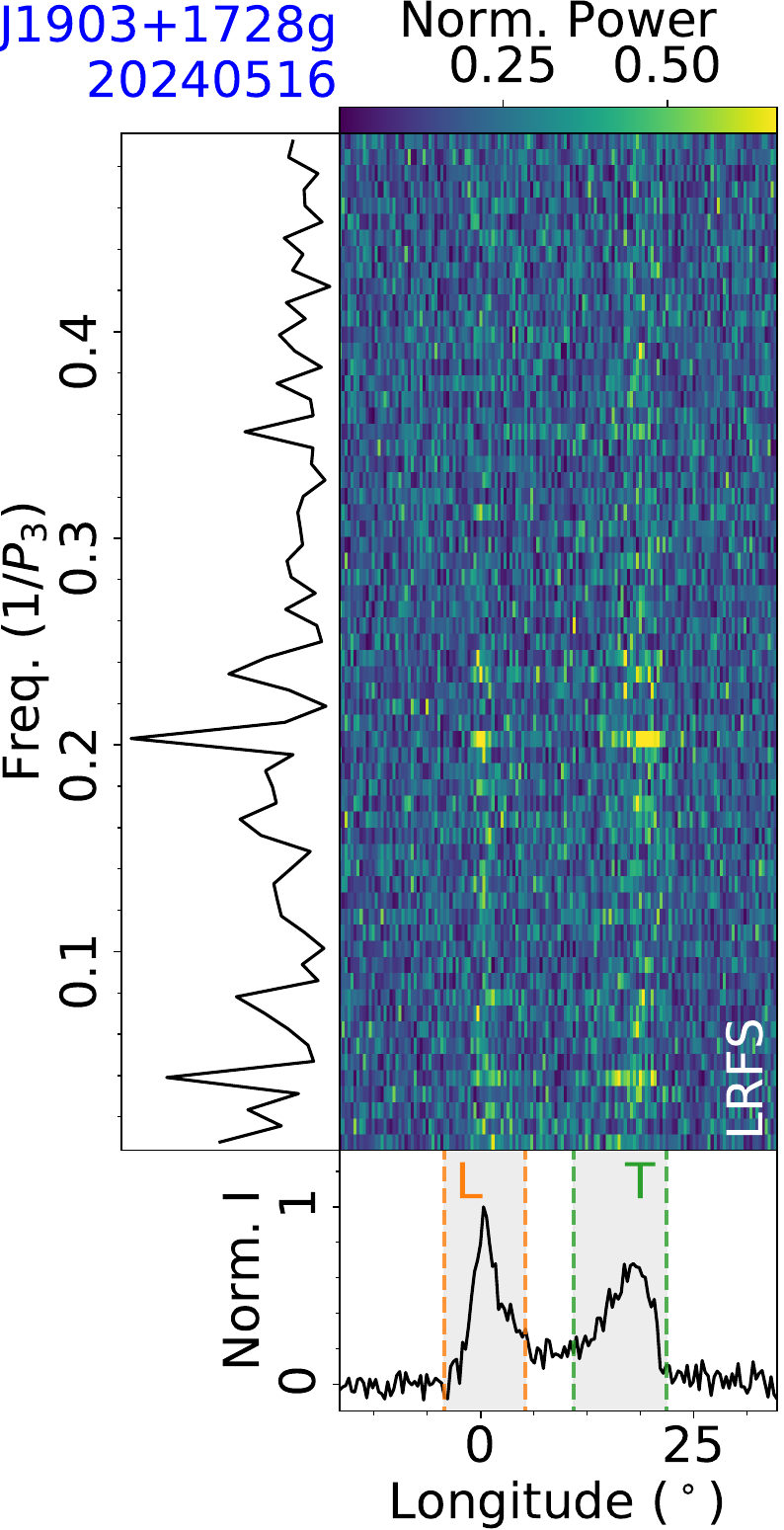}
\includegraphics[width=0.22\textwidth, angle=0]{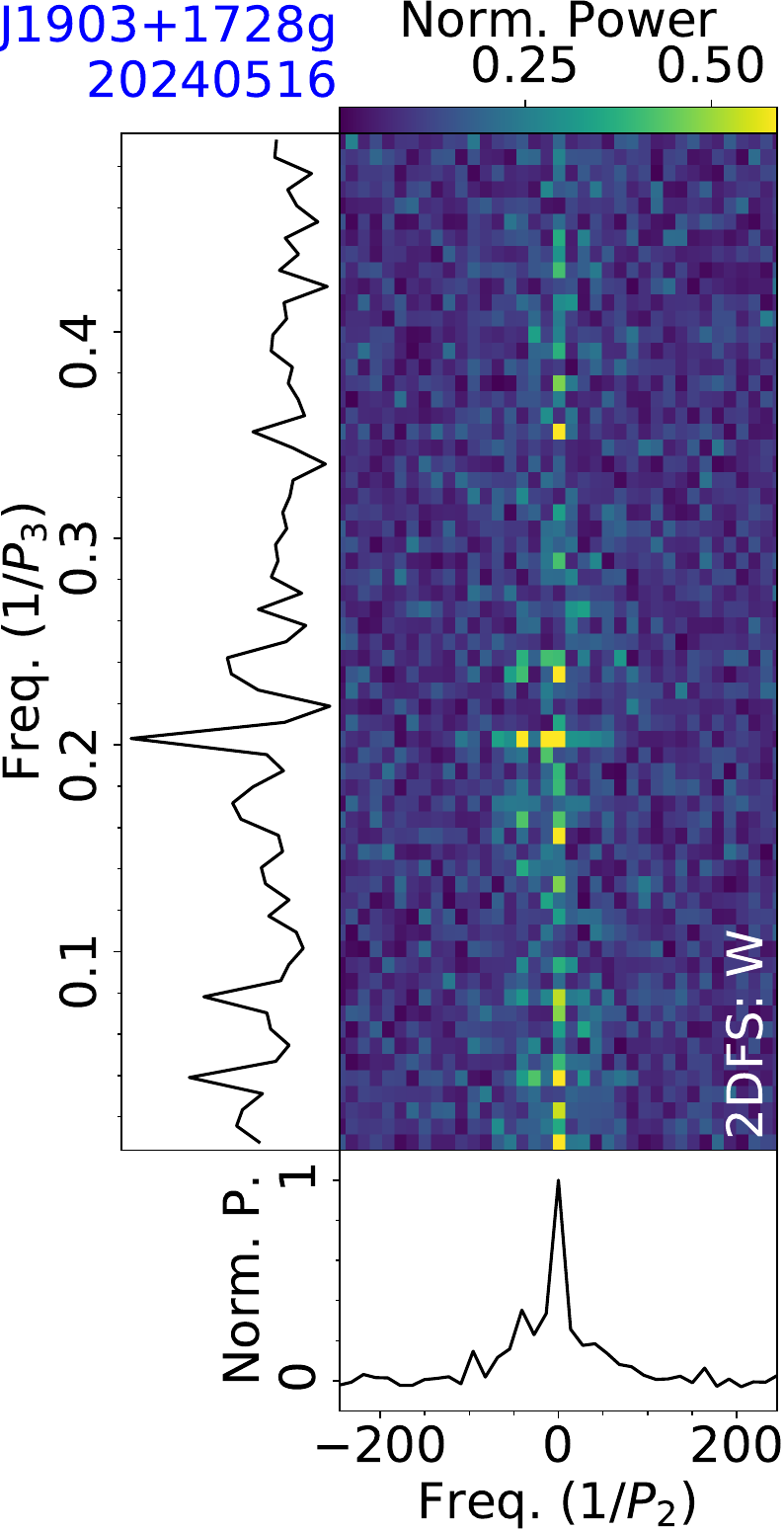}\\
\includegraphics[width=0.22\textwidth, angle=0]{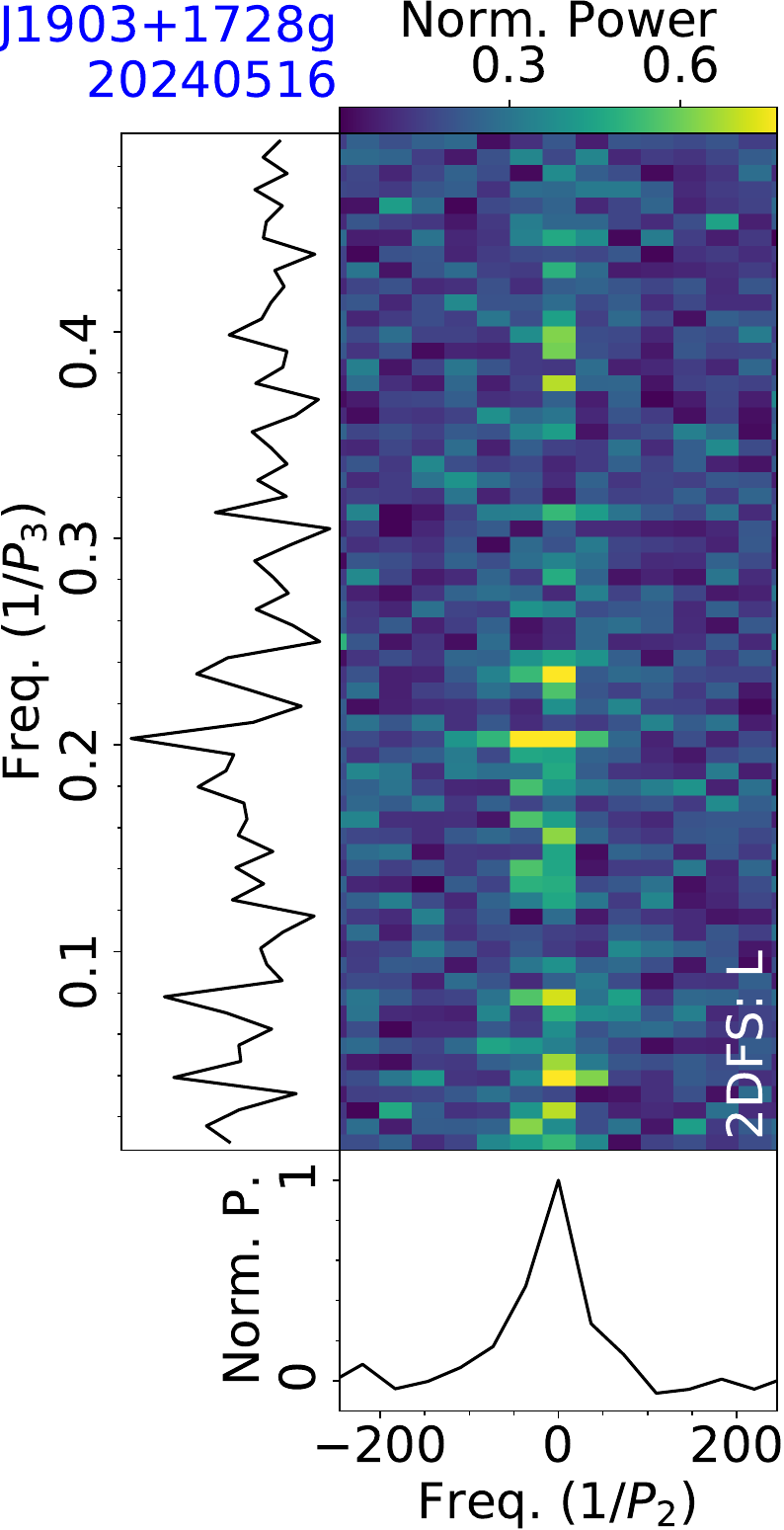}
\includegraphics[width=0.22\textwidth, angle=0]{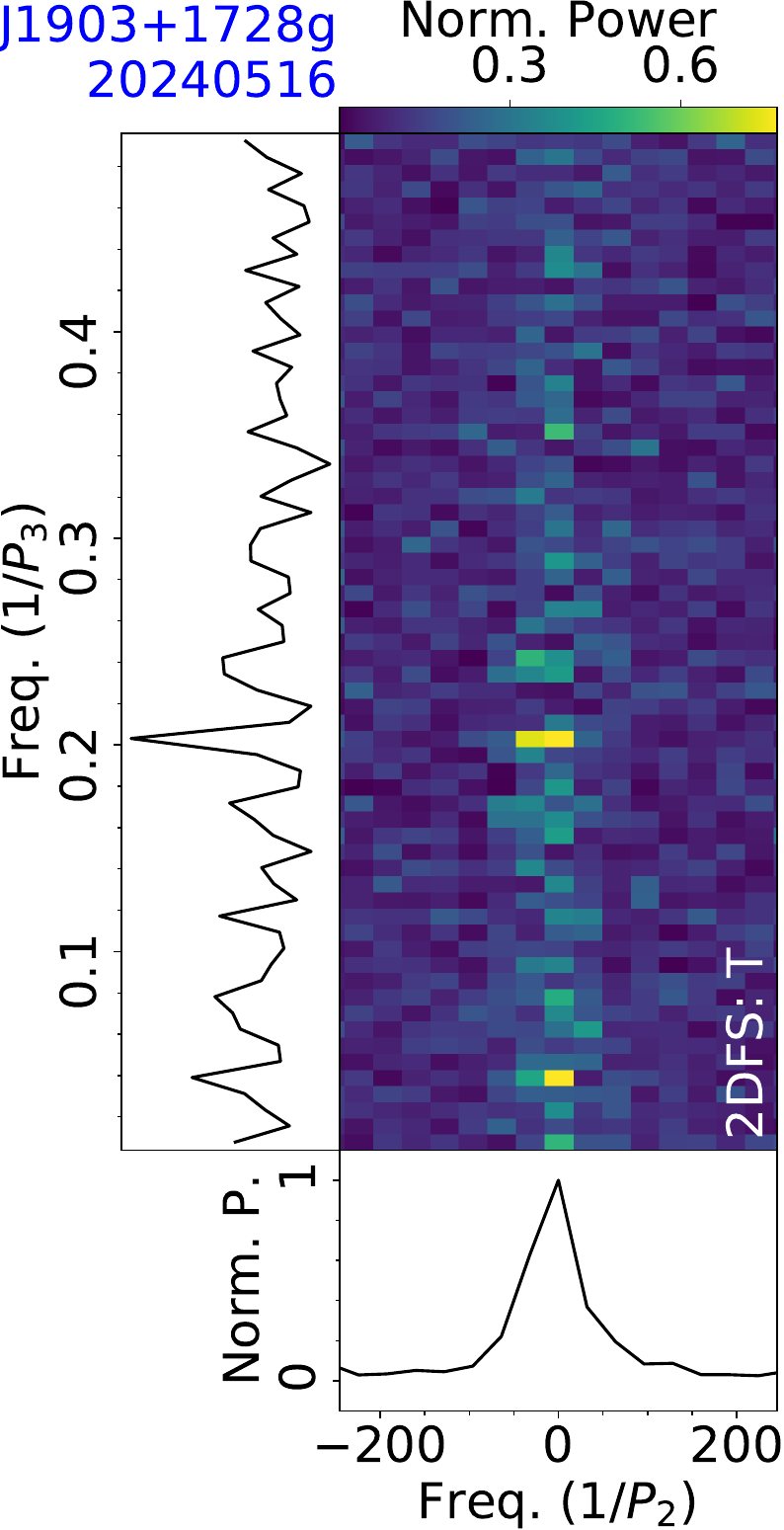}
\figcaption{Fluctuation analysis of PSR J1903+1728g for the observation on 20240516, with LRFS (top-left), and 2DFS for the on-pulse region (top-right), leading part (bottom-left) and trailing part (bottom-right) of a mean pulse profile.
\label{subfig:fluctu:J1903+1728g}}
\end{figure}

\begin{figure}[htpb]
\centering
\includegraphics[width=0.22\textwidth, angle=0]{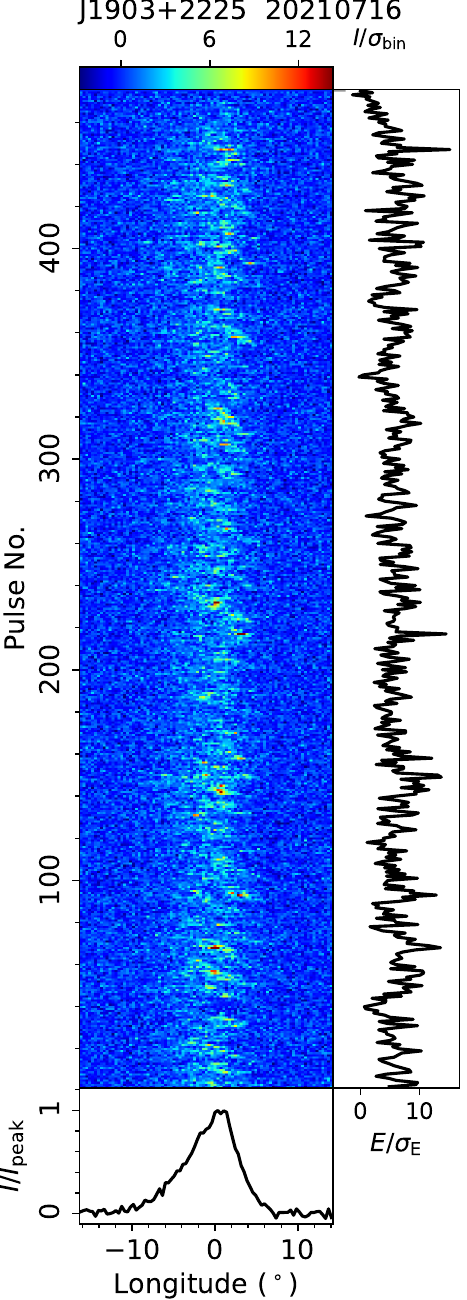}
\figcaption{Single pulse sequence of PSR J1903+2225 from the FAST observation on 20210716.
\label{subfig:TP:J1903+2225}}
\end{figure}

\begin{figure}[htpb]
\centering
\includegraphics[width=0.22\textwidth, angle=0]{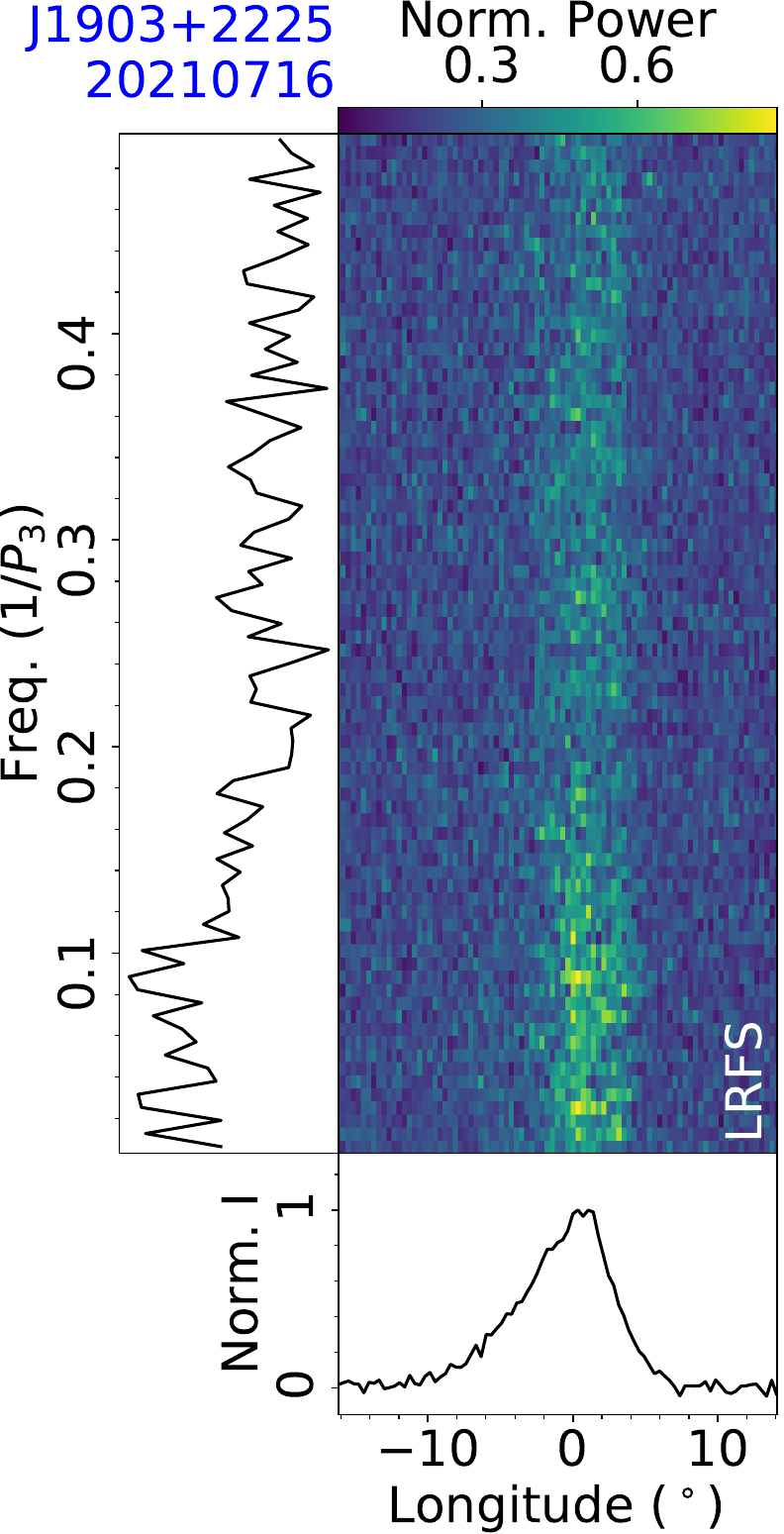}
\includegraphics[width=0.22\textwidth, angle=0]{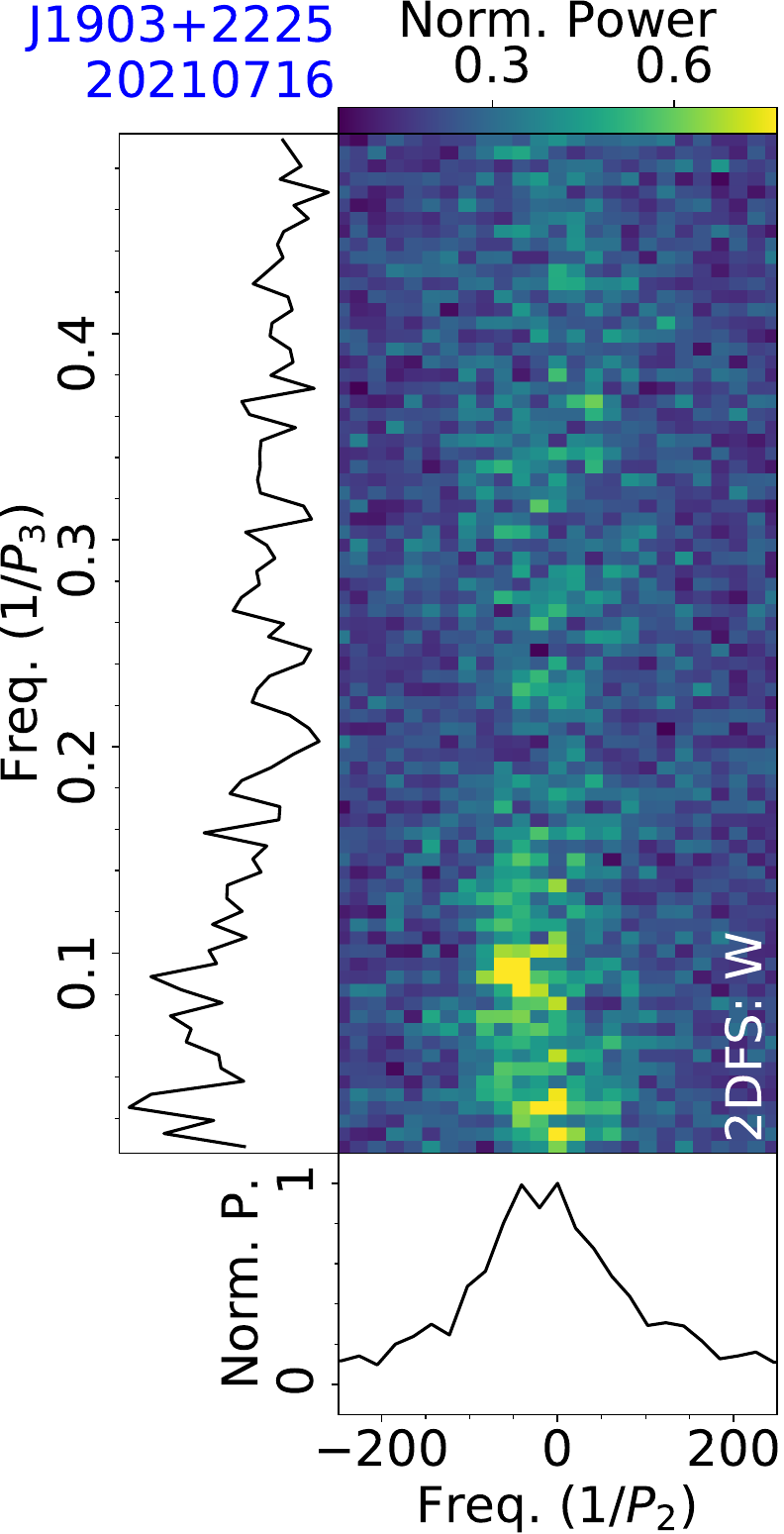}
\figcaption{Fluctuation analysis of PSR J1903+2225 for the observation on 20210716, with LRFS and 2DFS for the on-pulse region of a mean pulse profile.
\label{subfig:fluctu:J1903+2225}}
\end{figure}

\subsection{J1903+0415}
\label{subsec:J1903+0415}

PSR J1903+0415 was discovered by \citet{Lazarus2015} in the Arecibo Pulsar-ALFA (PALFA) survey at 1.4 GHz. The nulling behavior was reported by \citet{Parent2022}.

This pulsar was observed by FAST on 20200223 for 5 minutes, deriving a rotation period $P=1.1515$~s and a dispersion measure $D\!M=482.5~{\rm cm^{-3}\,pc}$. 
The single pulse sequence in Fig.~\ref{subfig:TP:J1903+0415} displays both nulling and subpulse drifting phenomena. 
Nulling is confirmed by the on-pulse integral energy histogram in Fig.~\ref{subfig:Hist:J1903+0415}, with the nulling fraction estimated to be 48$\pm$4\% for this observation. In LRFS and 2DFS in Fig.~\ref{subfig:fluctu:J1903+0415}, there is a negative drift feature with the centroid frequencies of $1/P_3=0.258\pm0.002$ ($P_3=3.87\pm0.03$ periods) and $1/P_2=-52\pm1$ ($P_2=-6.9\pm0.2^\circ$), as well as a low-frequency modulation from nulls with $P_3=59\pm1$ periods.

\begin{figure}[htpb]
\centering
\includegraphics[width=0.22\textwidth, angle=0]{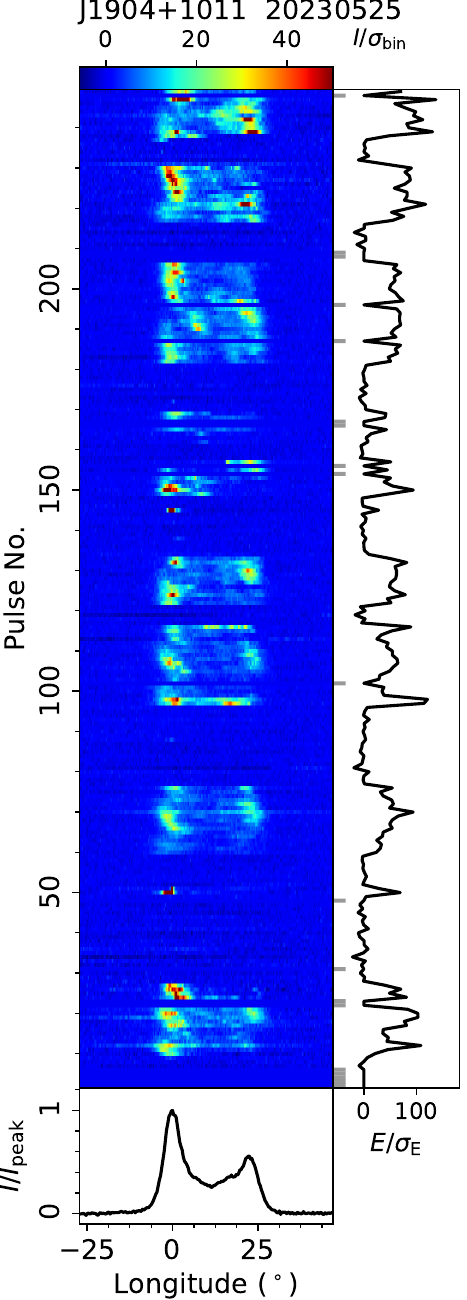}
\includegraphics[width=0.22\textwidth, angle=0]{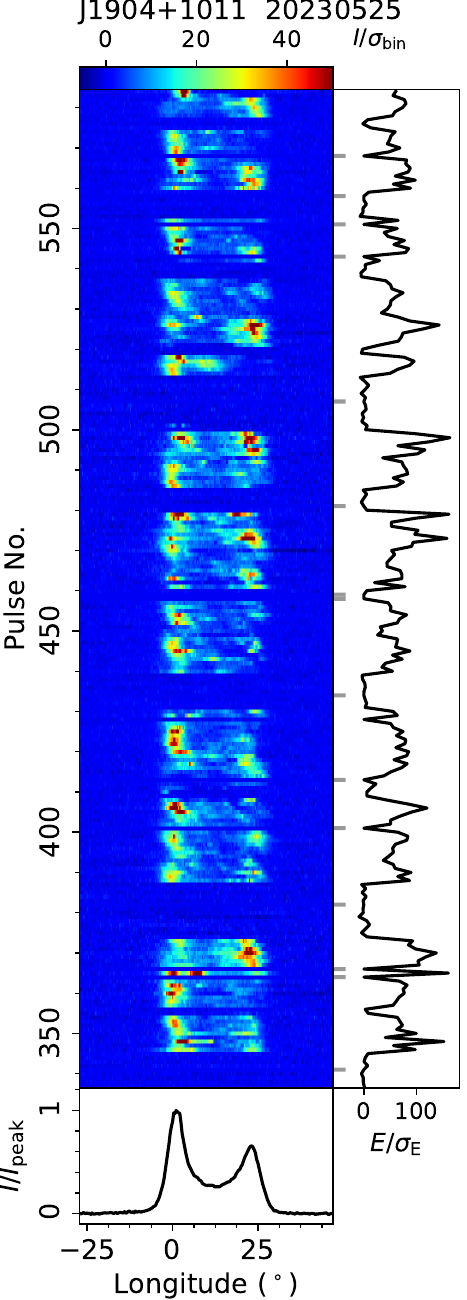}
\figcaption{Single pulse sequences of PSR J1904+1011 from the FAST observation on 20230525.
\label{subfig:TP:J1904+1011}}
\end{figure}

\begin{figure}[htpb]
\centering
\includegraphics[width=0.39\textwidth, angle=0]{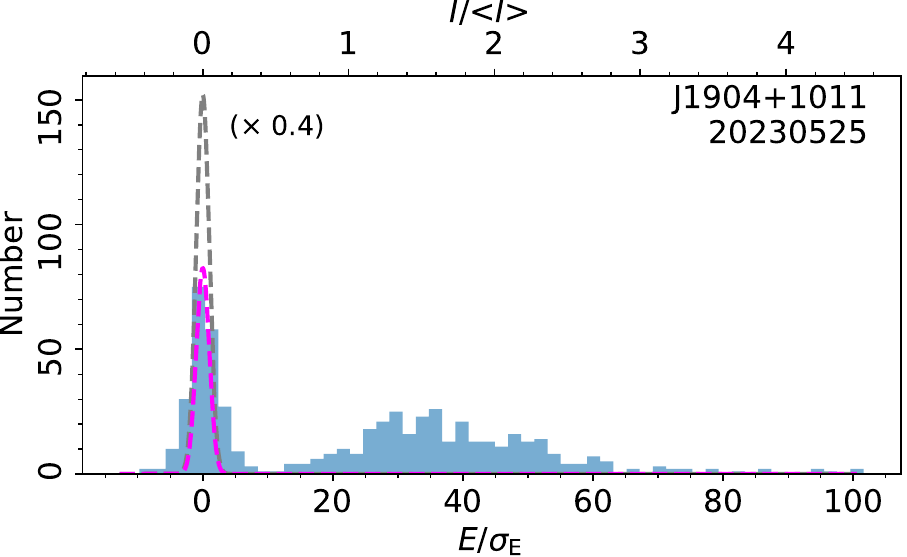}
\figcaption{On-pulse energy histogram of single pulses of PSR J1904+1011 from the FAST observation on 20230525.
\label{subfig:Hist:J1904+1011}}
\end{figure}

\begin{figure}[htpb]
\centering
\includegraphics[width=0.22\textwidth, angle=0]{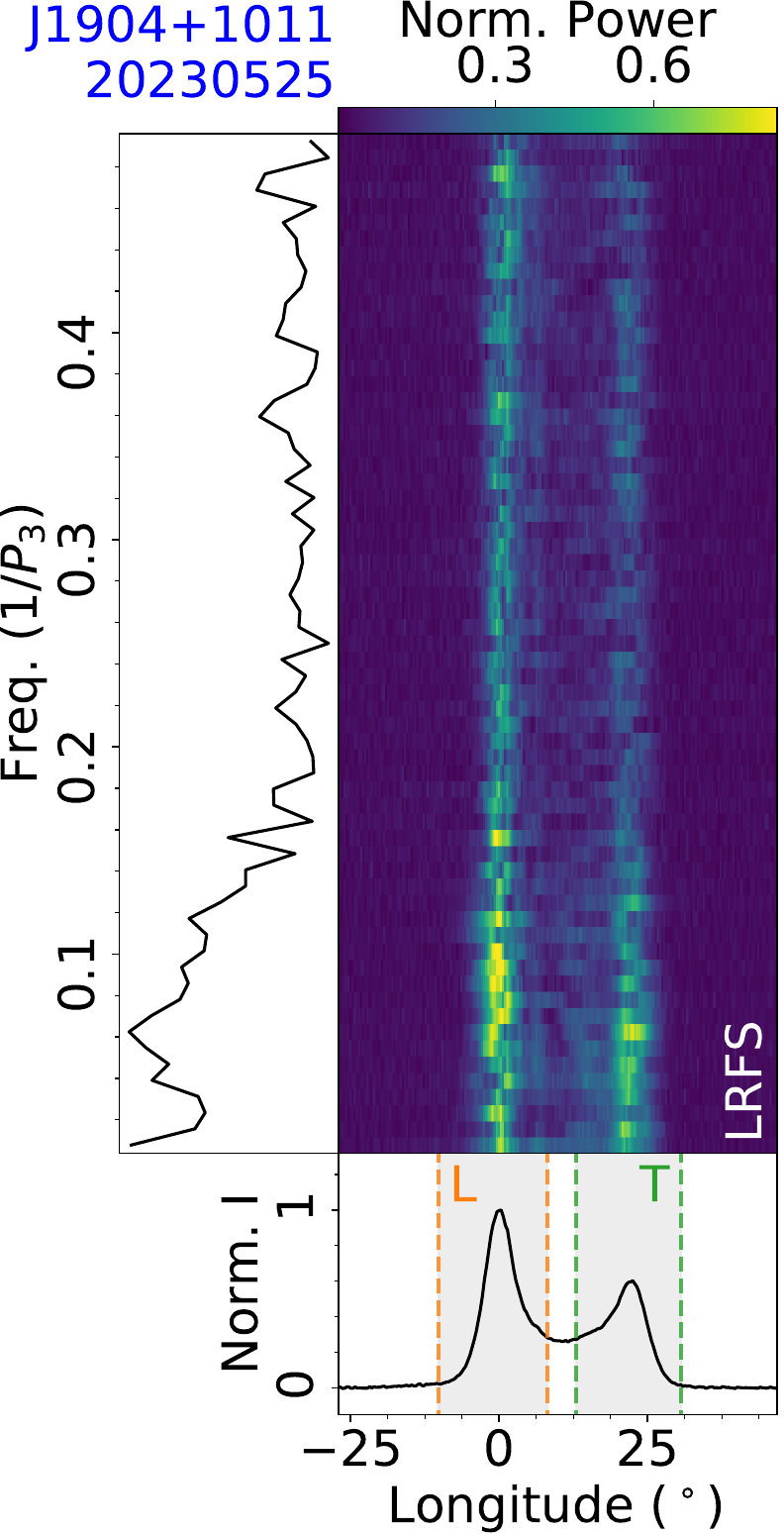}
\includegraphics[width=0.22\textwidth, angle=0]{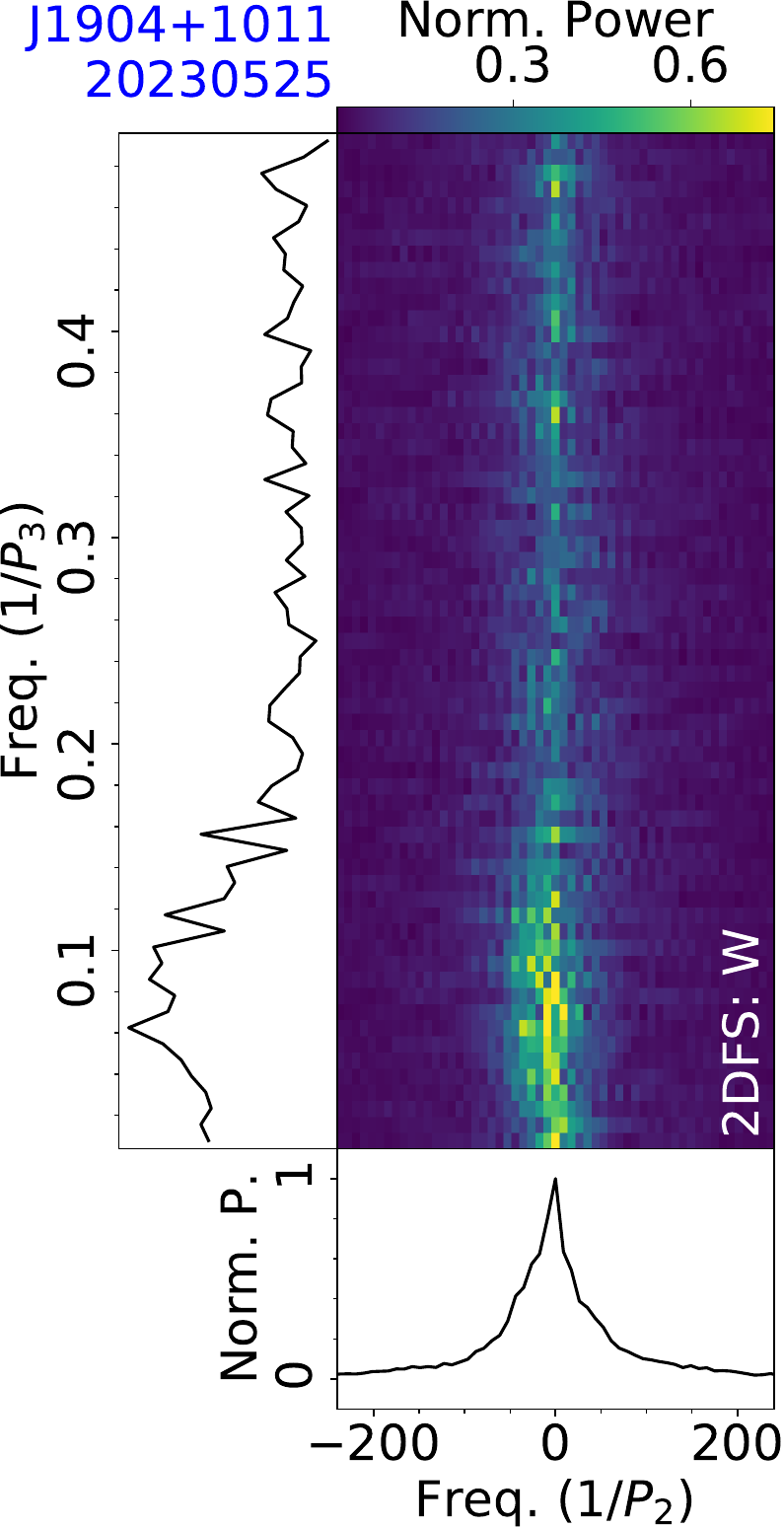}\\
\includegraphics[width=0.22\textwidth, angle=0]{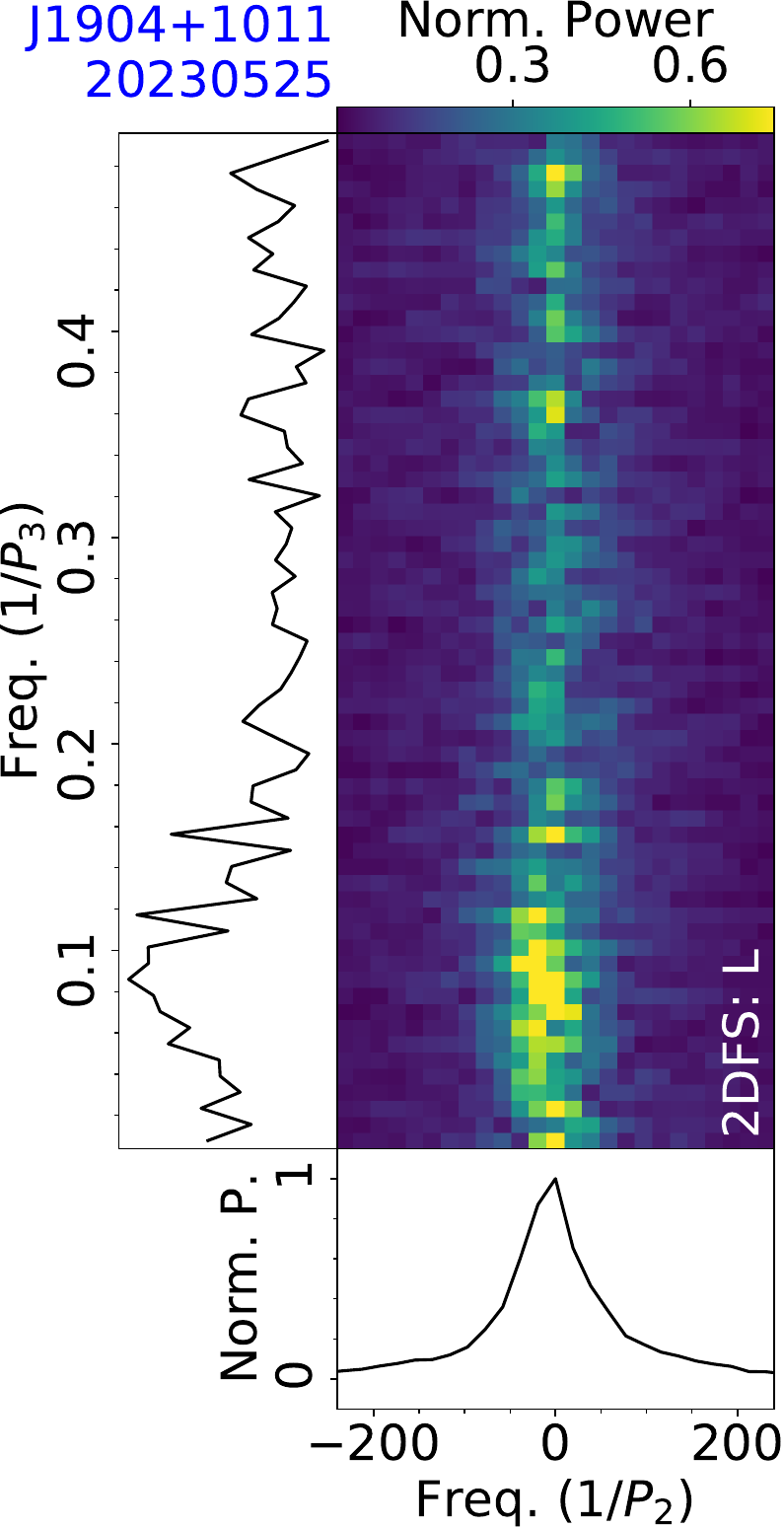}
\includegraphics[width=0.22\textwidth, angle=0]{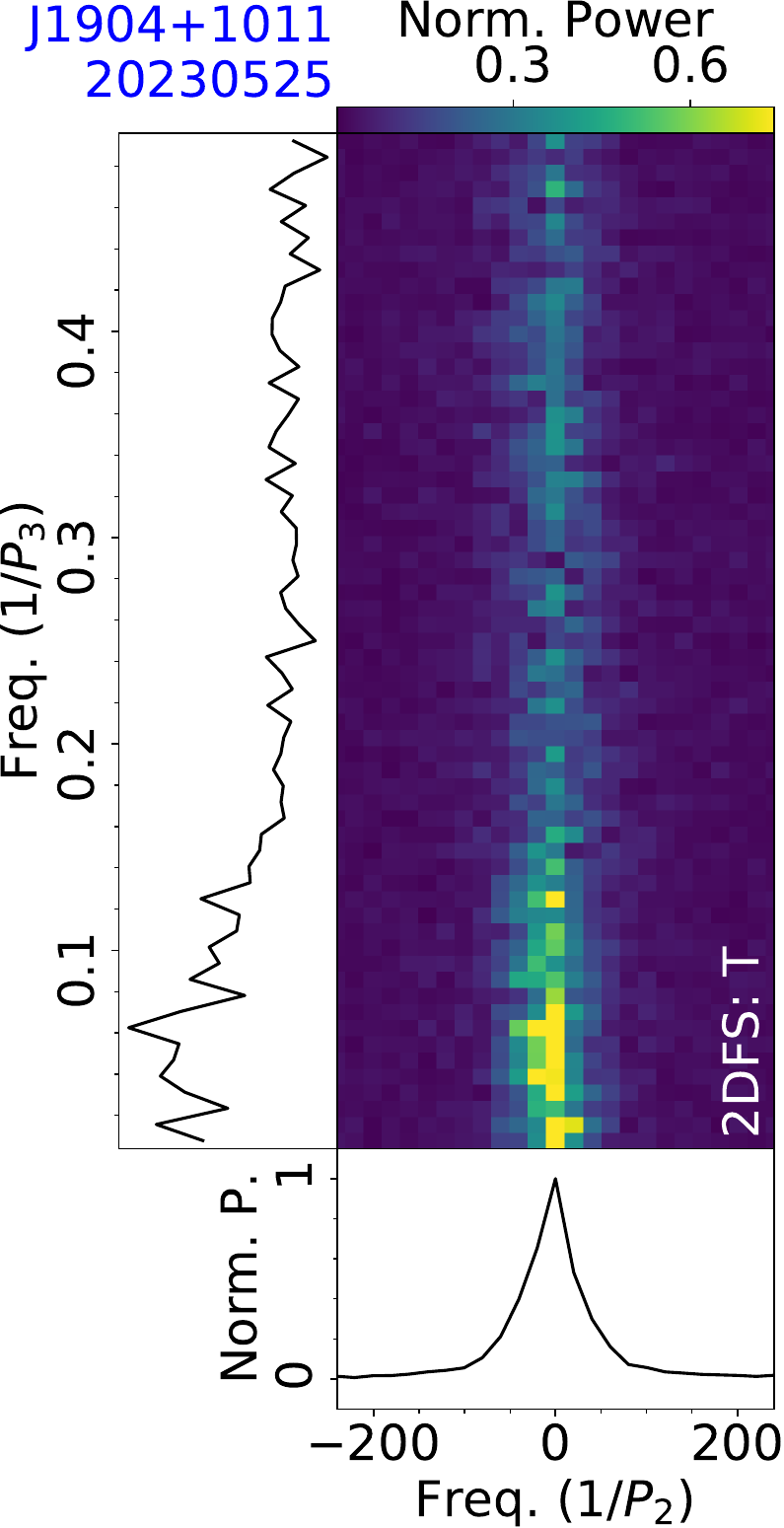}
\figcaption{Fluctuation analysis of PSR J1904+1011 from the FAST observation on 20230525, with LRFS (top-left), and 2DFS for the on-pulse region (top-right), leading part (bottom-left), and trailing part (bottom-right) of a mean pulse profile.
\label{subfig:fluctu:J1904+1011}}
\end{figure}

\begin{figure}[htpb]
\centering
\includegraphics[width=0.44\textwidth, angle=0]{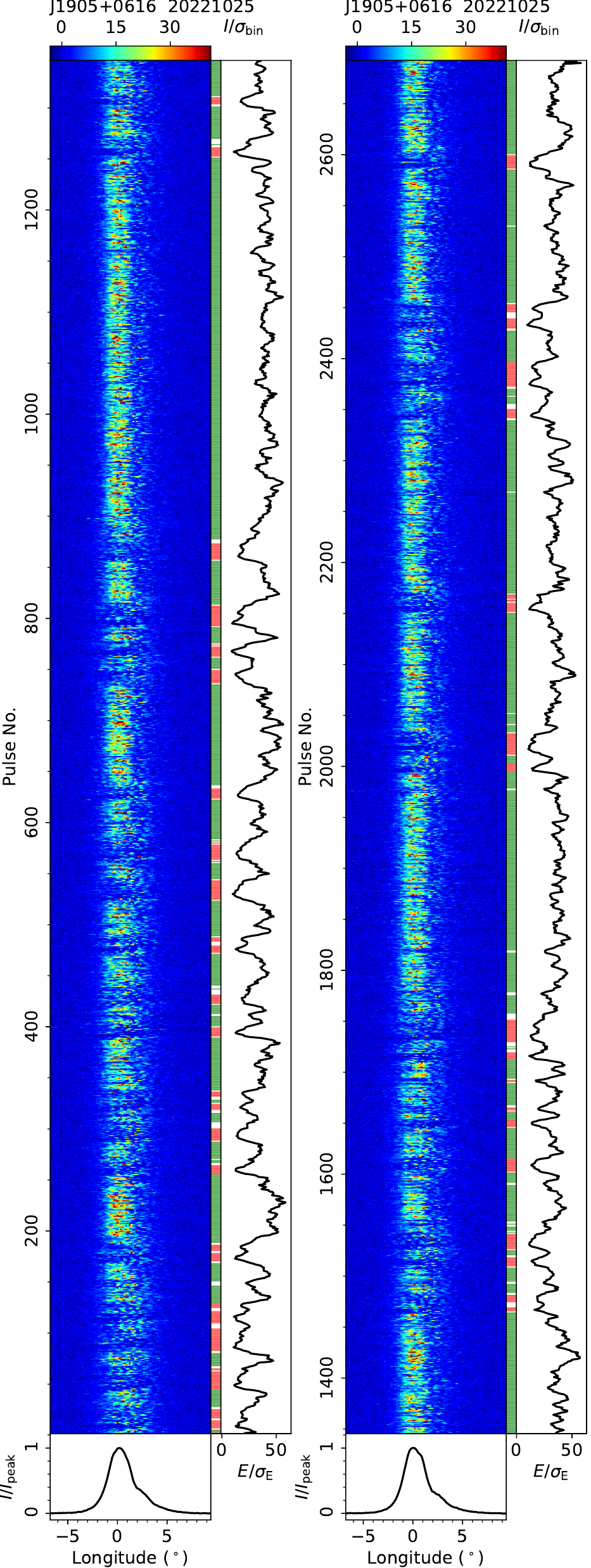}
\figcaption{Single pulse sequences of PSR J1905+0616 from the FAST observation on 20221025. In the right subpanel, the energy variation is integrated over the leading phase interval from -3.9$^\circ$ to 1.4$^\circ$ in longitude of the mean pulse profile, smoothed with a 7-period moving average.
\label{subfig:TP:J1905+0616}}
\end{figure}

\begin{figure}[htpb]
\centering
\includegraphics[width=0.39\textwidth, angle=0]{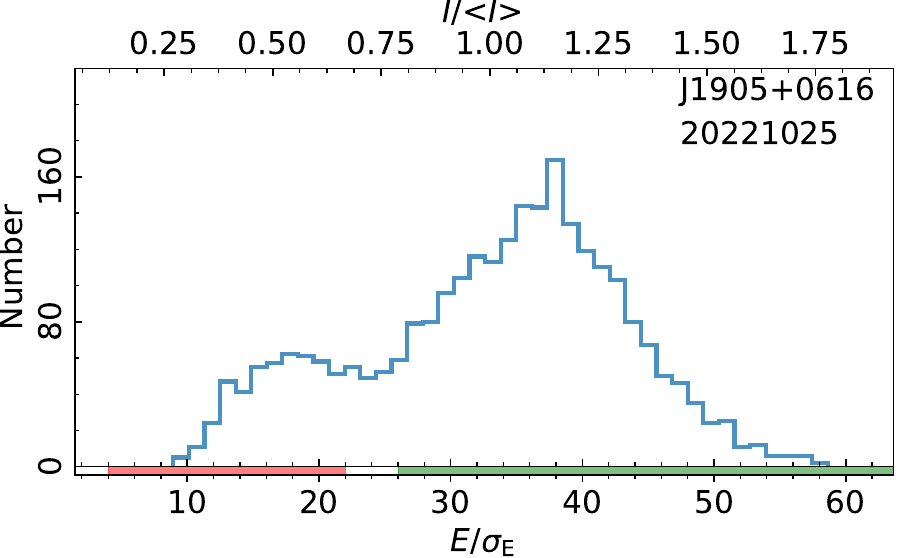}
\figcaption{Energy histogram of single pulses of PSR J1905+0616 from the FAST observation on 20221025. 
For every single pulse, the energy is integrated over the leading phase interval of the mean pulse profile, and the energy sequence is smoothed with a 7-period moving average.
The leading-weak and leading-strong modes, distinguished by their relative energy integrated in this phase interval, are shown in red and green, respectively.
\label{subfig:Hist:J1905+0616}}
\end{figure}

\begin{figure}[htpb]
\centering
\includegraphics[width=0.37\textwidth, angle=0]{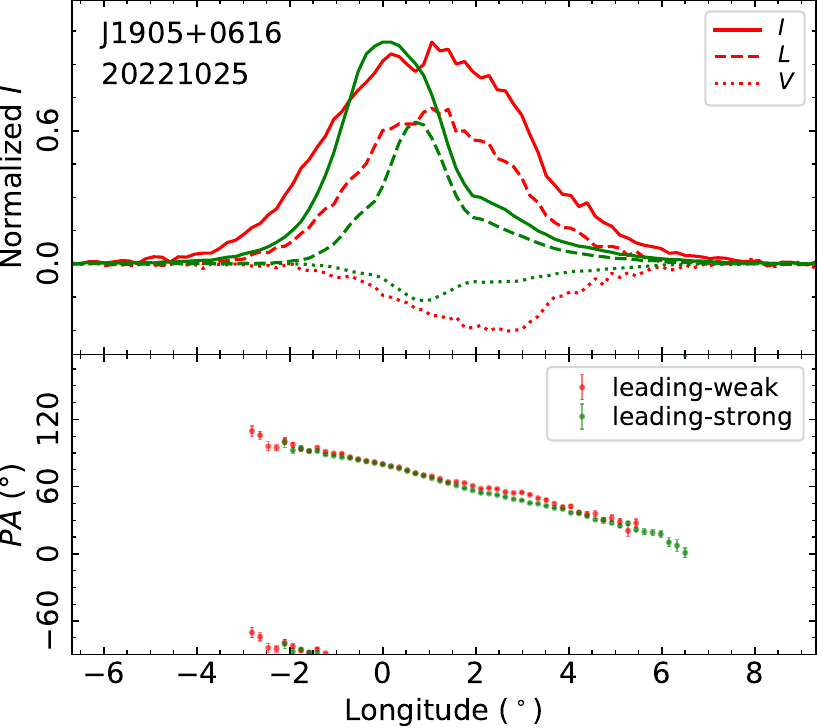}
\figcaption{Mean polarization profiles (the top panel) for the leading-weak and leading-strong emission modes of PSR J1905+0616 from the FAST observation on 20221025, as well as the averaged PA curves (the bottom panel). Profiles in the top panel are normalized by the respective peaks.
\label{subfig:PolModes:J1905+0616}}
\end{figure}

\begin{figure}[htpb]
\centering
\includegraphics[width=0.44\textwidth, angle=0]{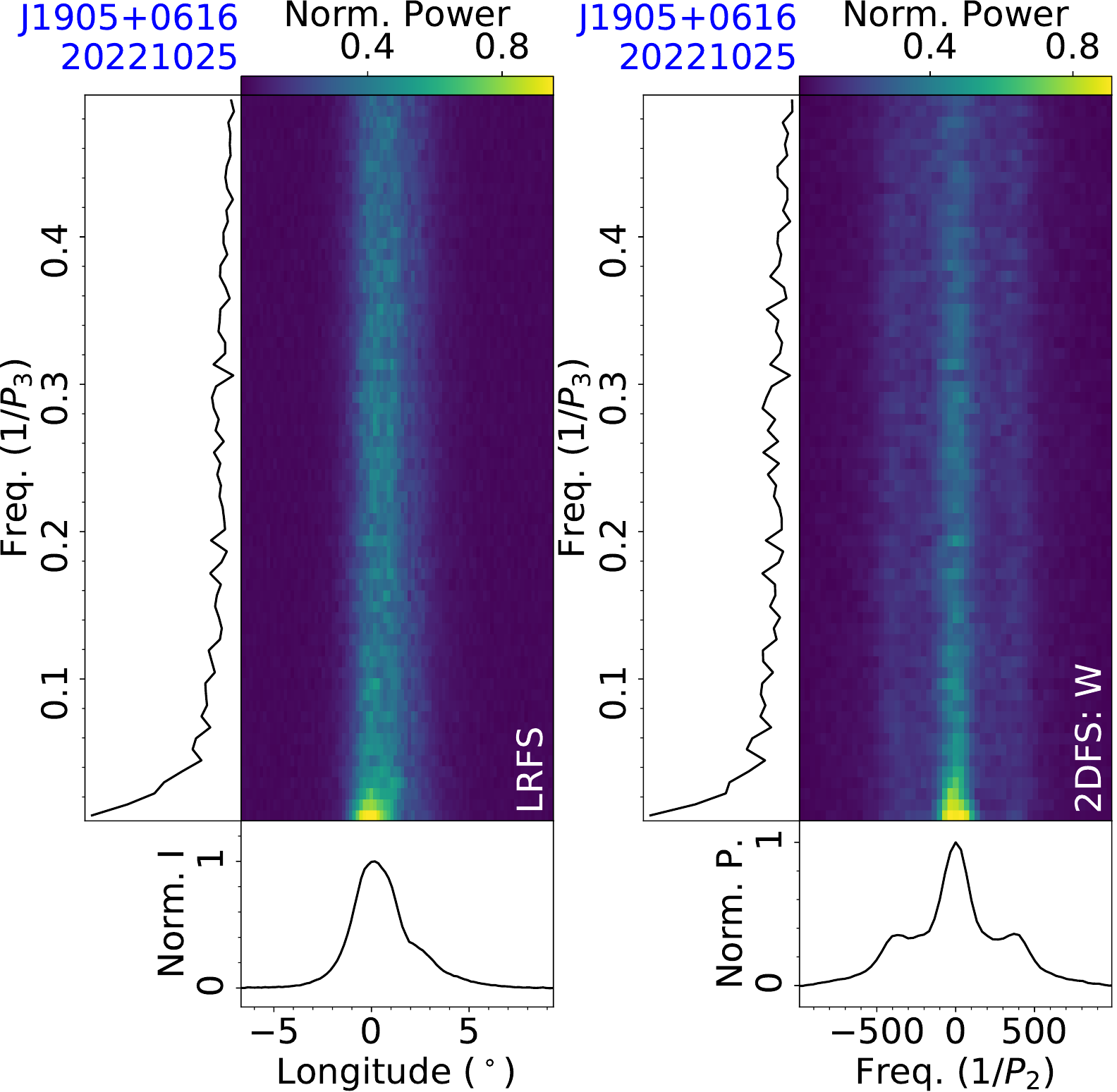}
\figcaption{Fluctuation analysis of PSR J1905+0616 for the observation on 20221025, with LRFS and 2DFS for the on-pulse region of a mean pulse profile.
\label{subfig:fluctu:J1905+0616}}
\end{figure}

\subsection{J1903+0925}
\label{subsec:J1903+0925}

PSR J1903+0925 was discovered in the Parkes multibeam pulsar survey \citep{Lorimer2006}.

This pulsar was observed by FAST on 20211226 and 20231230, each for 15 minutes. From the data on 20211226, a rotation period $P=0.3572$~s and a dispersion measure $D\!M=158.7~{\rm cm^{-3}\,pc}$ were derived. The single pulse sequence and a zoomed-in view of pulses No.100-300 are shown in Fig.~\ref{subfig:TP:J1903+0925}. Fluctuation spectra in Fig.~\ref{subfig:fluctu:J1903+0925} illustrate the existence of positive drifting for both the leading and trailing parts in the mean pulse profile. For the leading profile part, the centroid of the drift feature in 2DFS is at $1/P_3=0.1400\pm0.0004$ and $1/P_2=17.8\pm0.2$, corresponding to $1/P_3=7.14\pm0.02$ periods and $1/P_2=20.3\pm0.3$ degrees. In 2DFS of the trailing part, the drift feature has the centroid at $1/P_3=0.1403\pm0.0004$ and $1/P_2=13.3\pm0.2$, yielding $P_3=7.13\pm0.02$ periods and $P_2=27\pm1$ degrees. 
The two observations are in agreement in drift properties.

\subsection{J1903+1728g}
\label{subsec:J1903+1728g}

PSR J1903+1728g was discovered in the FAST GPPS survey \citep{Han2021,han2025}. 

This pulsar was observed by FAST on 20240516 for 5 minutes, deriving a rotation period $P=1.7164$~s and a dispersion measure $D\!M=153.2~{\rm cm^{-3}\,pc}$. 
The single pulse sequence of this observation in Fig.~\ref{subfig:TP:J1903+1728g} displays subpulse drifting behavior. In fluctuation spectra (Fig.~\ref{subfig:fluctu:J1903+1728g}), centroid frequencies of the drift feature are estimated to be $1/P_3=0.203\pm0.002$ ($P_3=4.92\pm0.04$ periods) and $1/P_2=-17\pm8$ ($P_2=-22\pm10^\circ$) for the leading part in a mean pulse profile, and $1/P_3=0.203\pm0.002$ ($P_3=4.92\pm0.04$ periods) and $1/P_2=-13\pm7$ ($P_2=-27\pm13^\circ$) for the trailing part.

\subsection{J1903+2225}
\label{subsec:J1903+2225}

PSR J1903+2225 was discovered by \citet{Nice1995} using the Arecibo telescope. Subpulse drifting of the pulsar was reported by \citet{Song2023}. 

The pulsar was observed by FAST on 20210716 for 5 minutes, deriving a rotation period $P=0.6511$~s and a dispersion measure $D\!M=109.3~{\rm cm^{-3}\,pc}$. 
The single pulse sequence is shown in Fig.~\ref{subfig:TP:J1903+2225}, which displays drifting bands. Fluctuation spectra are shown in Fig.~\ref{subfig:fluctu:J1903+2225}, and there is a negative drift feature in 2DFS. The centroid frequencies of the feature are estimated to be $1/P_3=0.090\pm0.002$ and $1/P_2=-50\pm1$, which correspond to drifting periodicities of $P_3=11.2\pm0.2$ periods and $P_2=-7.2\pm0.2^\circ$.

\begin{figure}[htpb]
\centering
\includegraphics[width=0.22\textwidth, angle=0]{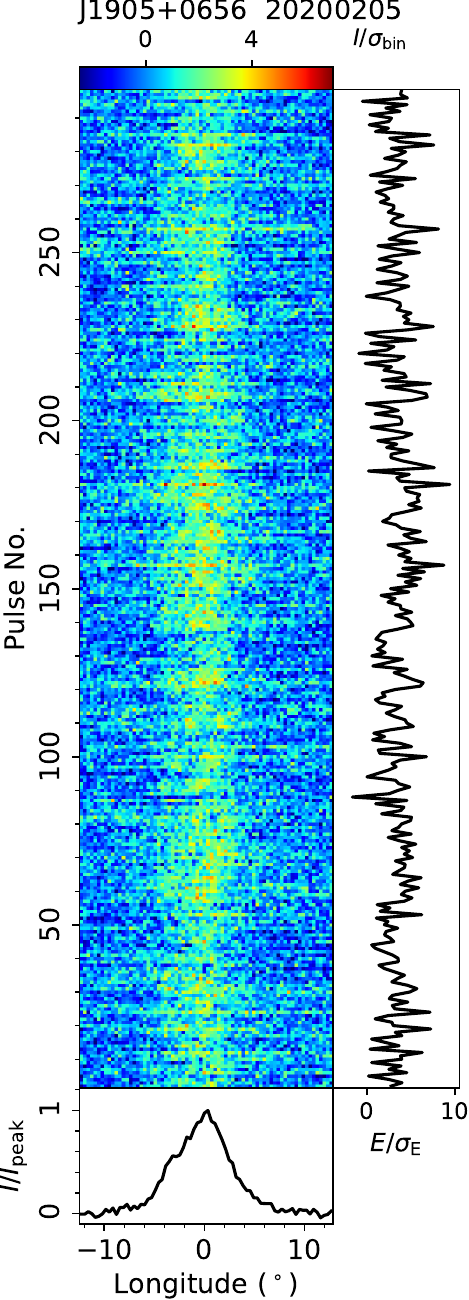}
\includegraphics[width=0.22\textwidth, angle=0]{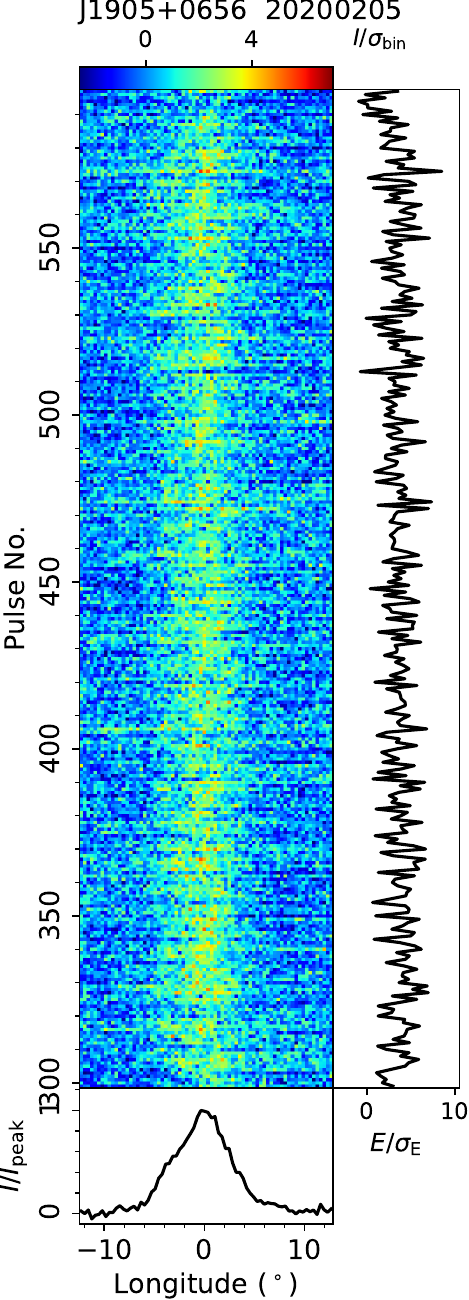}
\figcaption{Single pulse sequences of PSR J1905+0656 from the FAST observation on 20200205.
\label{subfig:TP:J1905+0656}}
\end{figure}

\begin{figure}[htpb]
\centering
\includegraphics[width=0.22\textwidth, angle=0]{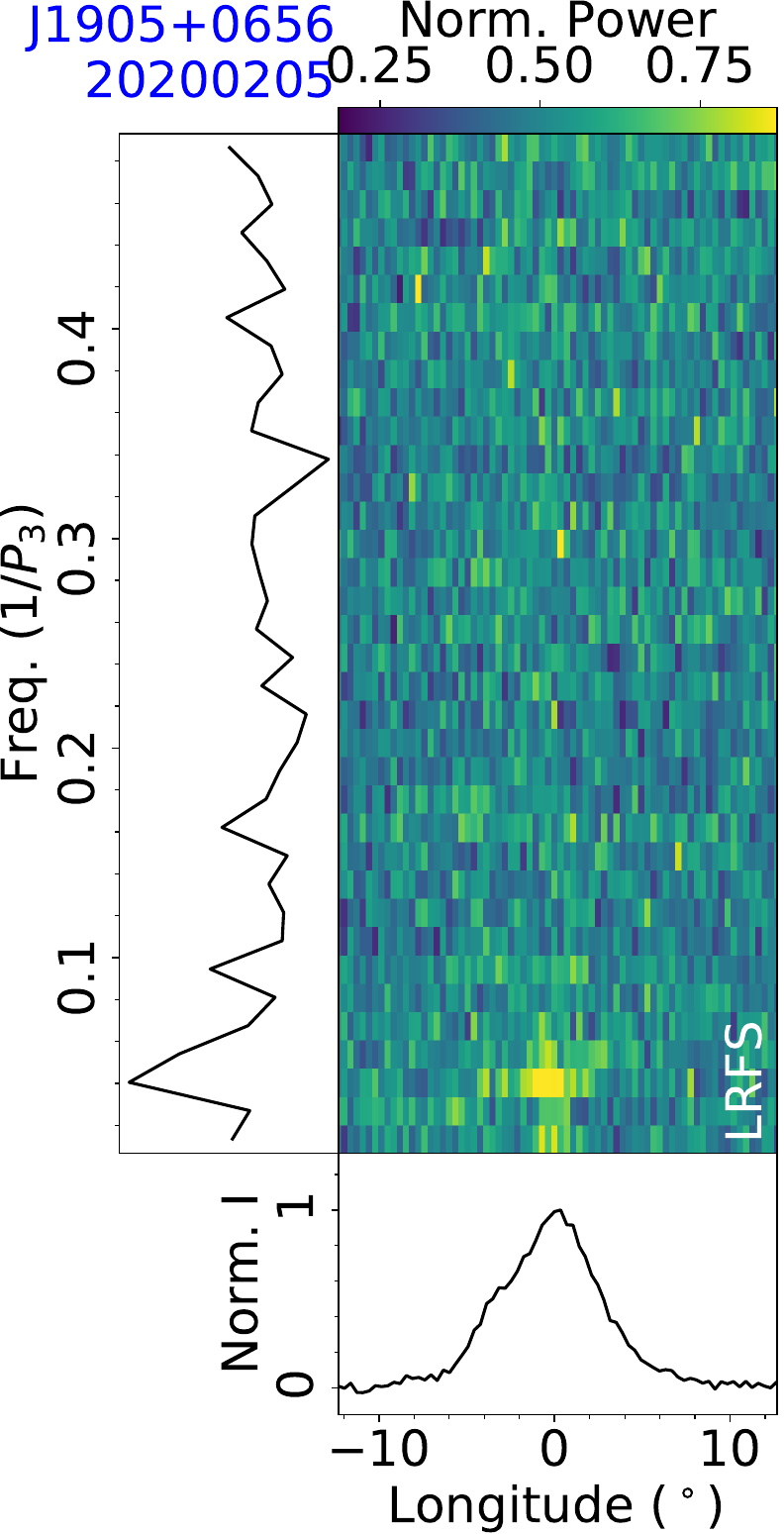}
\includegraphics[width=0.22\textwidth, angle=0]{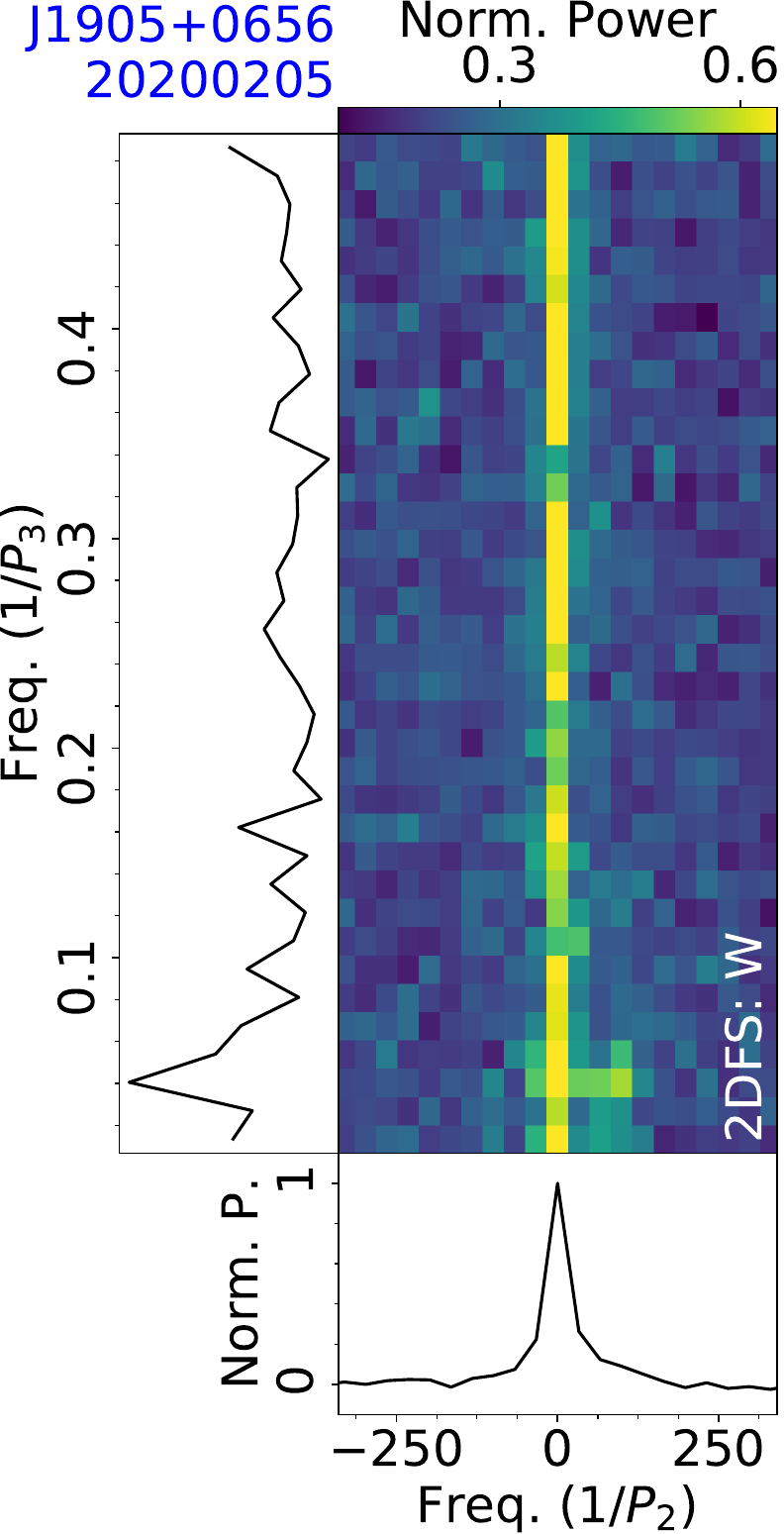}
\figcaption{Fluctuation analysis of PSR J1905+0656 for the observation on 20200205, with LRFS and 2DFS for the on-pulse region of a mean pulse profile.
\label{subfig:fluctu:J1905+0656}}
\end{figure}

\subsection{J1904+1011}
\label{subsec:J1904+1011}

PSR J1904+1011 was discovered by \citet{Hulse1975} using the Arecibo telescope. The pulsar was reported to have nulling \citep{Lorimer2002}, subpulse drifting \citep{Song2023}, as well as dwarf pulses \citep{Yan2024}. 

Here we report the result of the observation conducted by FAST on 20230525 for 18 minutes. From this data, a rotation period $P=1.8564$~s and a dispersion measure $D\!M=137.0~{\rm cm^{-3}\,pc}$ were determined. The single pulse sequences in Fig.~\ref{subfig:TP:J1904+1011} and the histogram in Fig.~\ref{subfig:Hist:J1904+1011} all indicate the existence of the nulling phenomenon with a nulling fraction estimated to be 22$\pm$3\%. As reported previously, there is also subpulse drifting behavior, and drifting parameters of the leading and trailing parts in a mean pulse profile are estimated from fluctuation spectra in Fig.~\ref{subfig:fluctu:J1904+1011}. 
For the leading profile part, 2DFS exhibits a negative drift feature with the centroid of $1/P_3=0.081\pm0.001$ and $1/P_2=-17\pm2$, which correspond to $P_3=12.4\pm0.2$ periods and $P_2=-22\pm2^\circ$. In 2DFS of the trailing part, the centroid frequencies of the drift feature are $1/P_3=0.073\pm0.001$ and $1/P_2=-12\pm1$, yielding $P_3=13.6\pm0.2$ periods and $P_2=-30\pm3^\circ$. 
Sometimes the drift band corresponding to the trailing component is missed. In addition to the weak and narrow dwarf pulses appearing in the nulling state, there are also very narrow and strong emission in pulses No. 50, 51 and 145.

\begin{figure}[htpb]
\centering
\includegraphics[width=0.44\textwidth, angle=0]{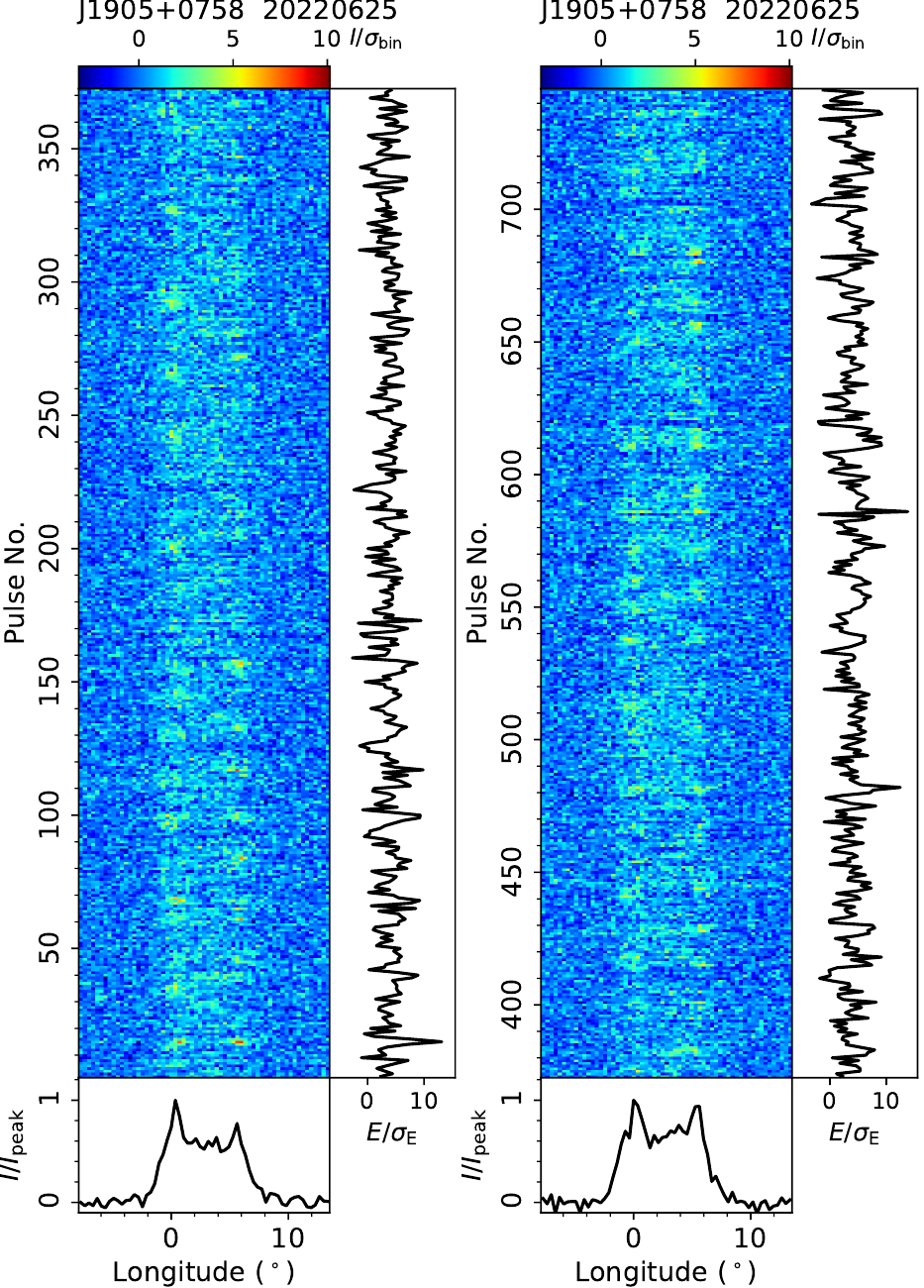}
\figcaption{Single pulse sequences of PSR J1905+0758 from the FAST observation on 20220625.
\label{subfig:TP:J1905+0758}}
\end{figure}

\begin{figure}[htpb]
\centering
\includegraphics[width=0.44\textwidth, angle=0]{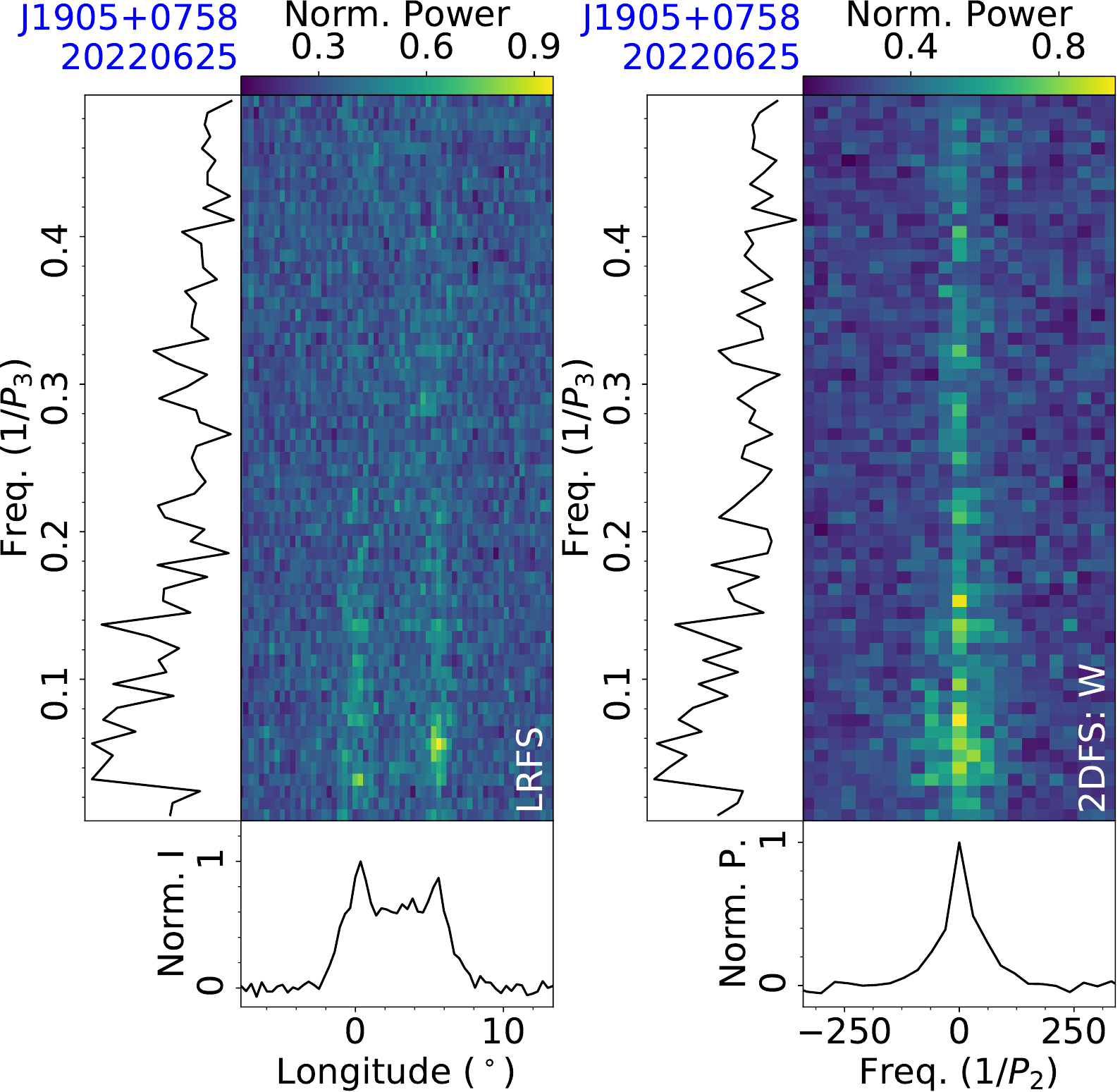}
\figcaption{Fluctuation analysis of PSR J1905+0758 for the observation on 20220625, with LRFS and 2DFS for the on-pulse region of a mean pulse profile.
\label{subfig:fluctu:J1905+0758}}
\end{figure}

\subsection{J1905+0616}
\label{subsec:J1905+0616}

PSR J1905+0616 was independently discovered in the Parkes multibeam pulsar survey \citep{Morris2002} and in an Arecibo drift-scan survey \citep{Lorimer2005}. \citet{Song2023} reported a positive drift feature of $P_3=37\pm9$ periods and $P_2=34^{+42}_{-20}$ degrees. 

This pulsar was observed by FAST on 20221025 for 44 minutes, with a rotation period $P=0.9899$~s and a dispersion measure $D\!M=255.8~{\rm cm^{-3}\,pc}$ derived. Single pulse sequences in Fig.~\ref{subfig:TP:J1905+0616} show mode changes, especially in the energy of the leading phase part of the mean pulse profile. 
Single pulses of two emission modes are distinguished from the energy histogram in Fig.~\ref{subfig:Hist:J1905+0616}. where for every single pulse the energy is integrated over the leading phase interval from $-3.9^\circ$ to 1.4$^\circ$ in longitude and the resulting energy sequence is smoothed with a 7-period moving average. The two modes are named the leading-weak and leading-strong modes, labeled in red and green colors, respectively. 
The polarization profiles and average PA curves for the two emission modes are compared in Fig.~\ref{subfig:PolModes:J1905+0616}. 
In the leading-strong mode, the leading component is stronger than the trailing component. 
The two modes have the same sense of circular polarization, and their PA curves are similar. 
The fluctuation spectra (Fig.~\ref{subfig:fluctu:J1905+0616}) show both negative and positive drift features, widely distributed over a $1/P_3$ range of 0 to 0.5. The centroid frequencies in $1/P_2$ are $-429\pm2$ and $399\pm4$, corresponding to $P_2$ values of $-0.838\pm0.003$ degrees and $0.90\pm0.01$ degrees, respectively. These results suggest variable drifting rates but a relatively stable subpulse phase interval. The low-frequency modulation feature is likely due to mode changes.

\begin{figure}[htpb]
\centering
\includegraphics[width=0.22\textwidth, angle=0]{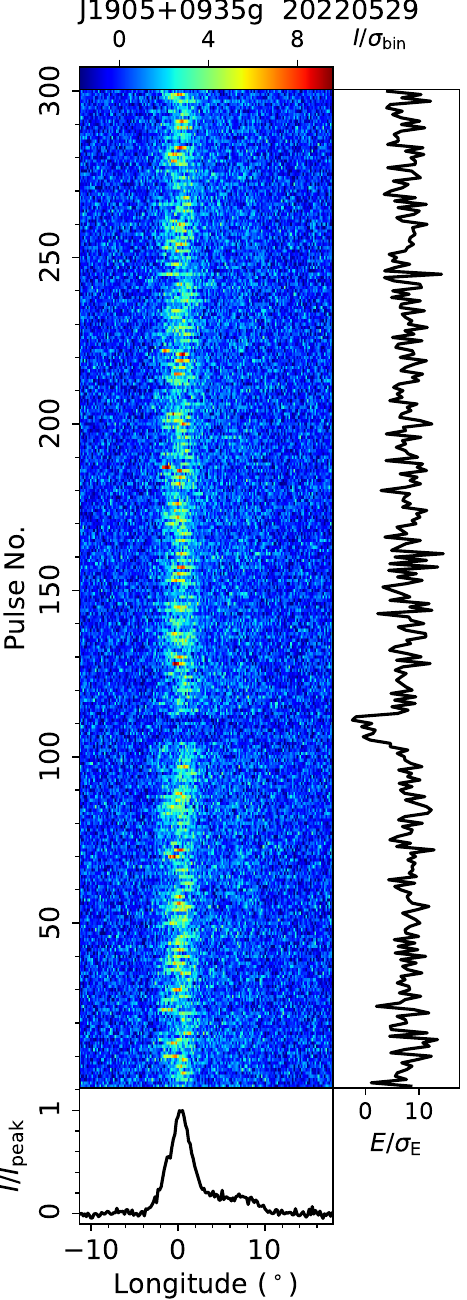}
\includegraphics[width=0.22\textwidth, angle=0]{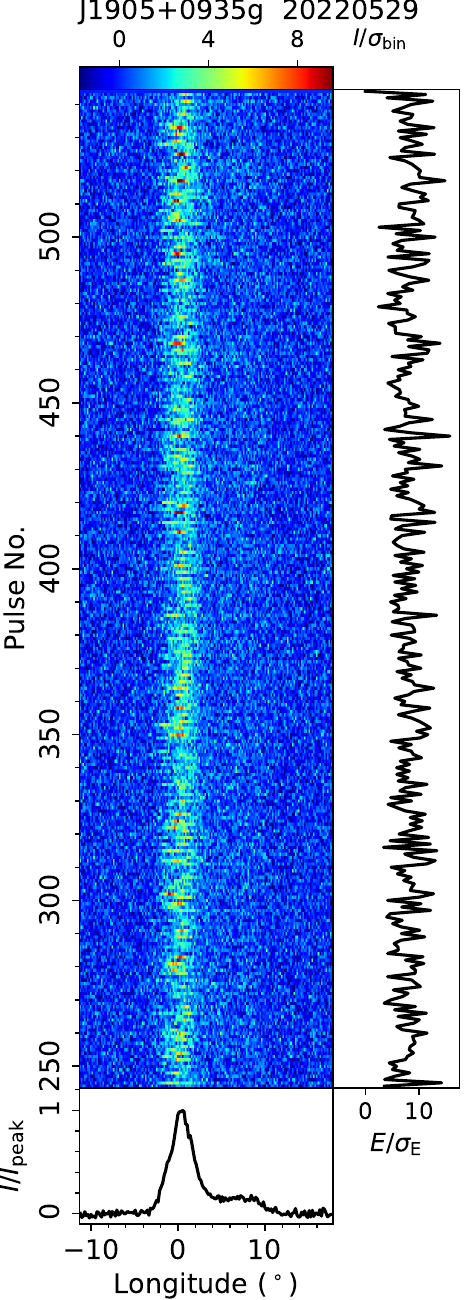}
\figcaption{Single pulse sequences of PSR J1905+0935g from the FAST observation on 20220529.
\label{subfig:TP:J1905+0935g}}
\end{figure}

\begin{figure}[htpb]
\centering
\includegraphics[width=0.39\textwidth, angle=0]{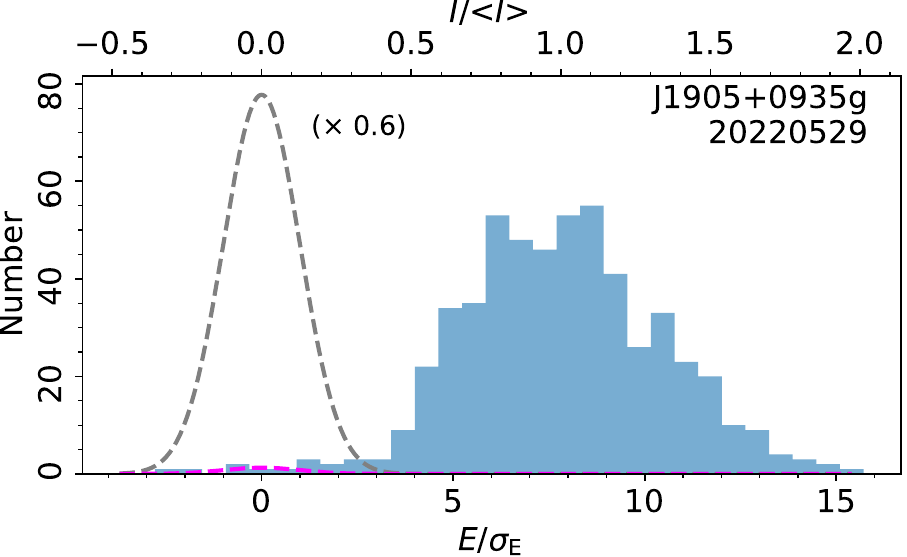}
\figcaption{On-pulse energy histogram of single pulses of PSR J1905+0935g from the FAST observation on 20220529.
\label{subfig:Hist:J1905+0935g}}
\end{figure}

\begin{figure}[htpb]
\centering
\includegraphics[width=0.22\textwidth, angle=0]{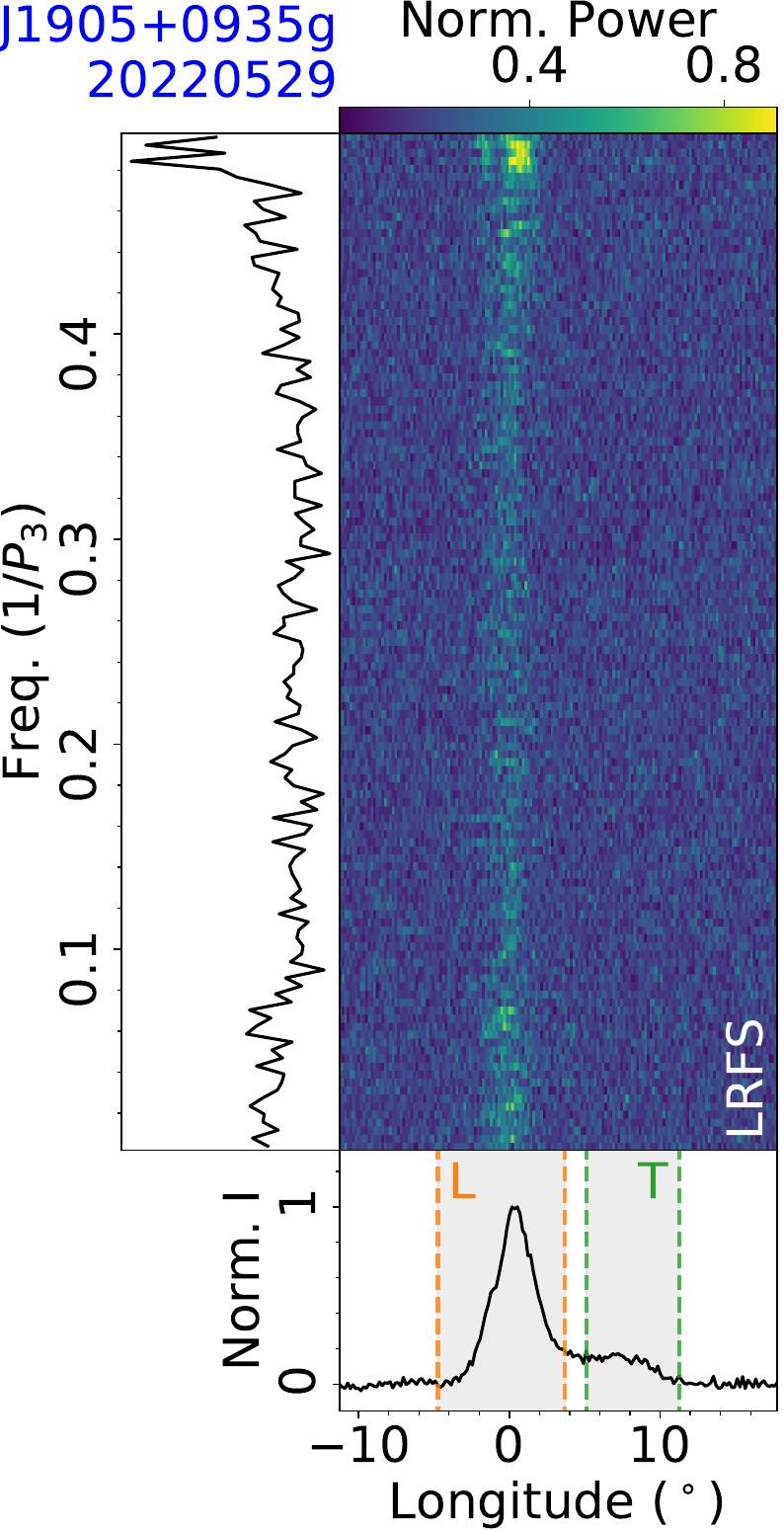}
\includegraphics[width=0.22\textwidth, angle=0]{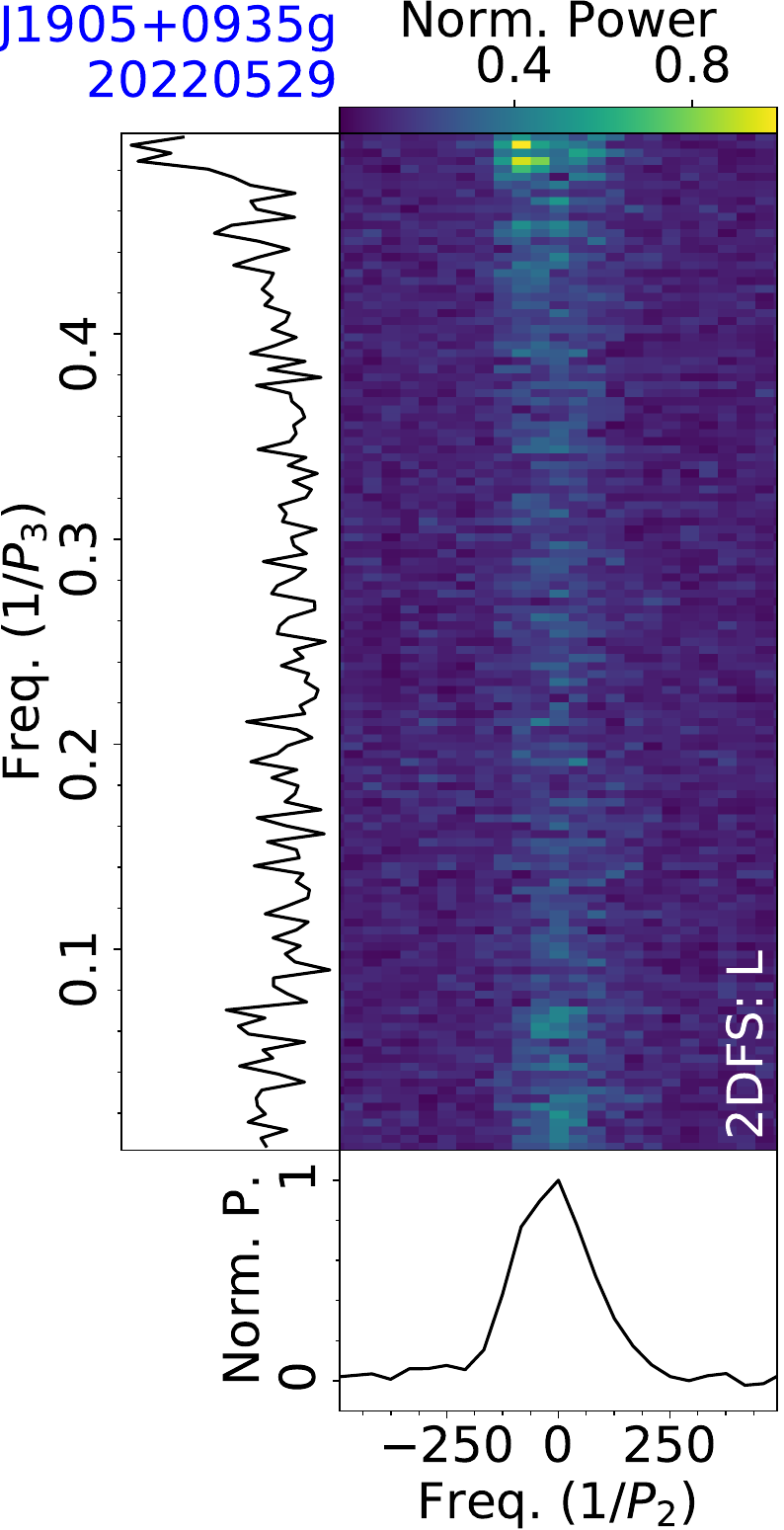}
\figcaption{Fluctuation analysis of PSR J1905+0935g for the observation on 20220529, with LRFS and 2DFS for the leading part of a mean pulse profile.
\label{subfig:fluctu:J1905+0935g}}
\end{figure}

\begin{figure}[htpb]
\centering
\includegraphics[width=0.22\textwidth, angle=0]{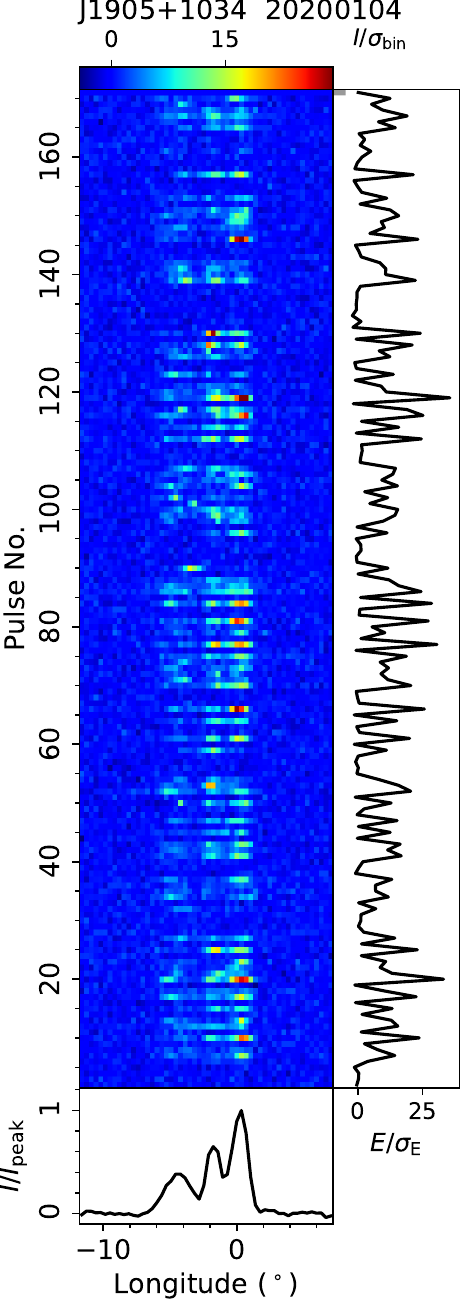}
\figcaption{Single pulse sequence of PSR J1905+1034 from the FAST observation on 20200104.
\label{subfig:TP:J1905+1034}}
\end{figure}

\begin{figure}[htpb]
\centering
\includegraphics[width=0.39\textwidth, angle=0]{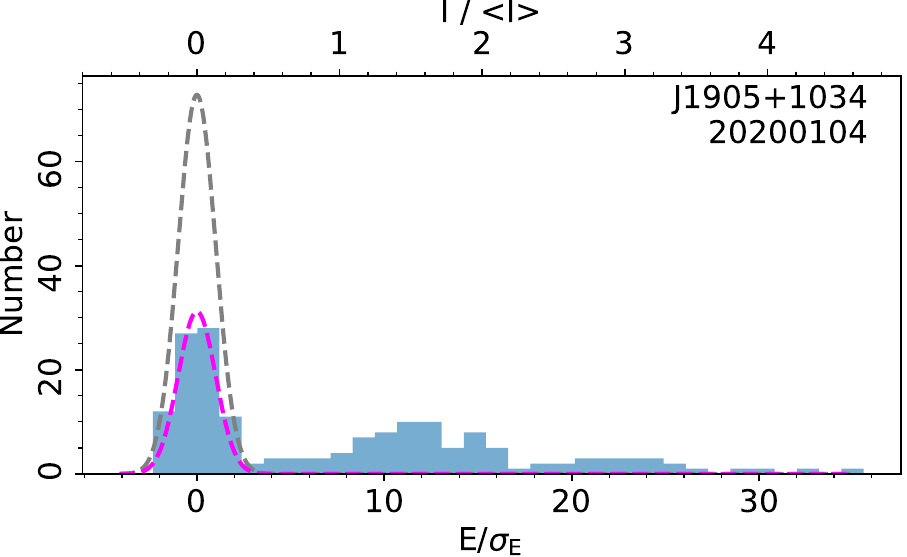}
\figcaption{On-pulse energy histogram of single pulses of PSR J1905+1034 from the FAST observation on 20200104.
\label{subfig:Hist:J1905+1034}}
\end{figure}

\begin{figure}[htpb]
\centering
\includegraphics[width=0.22\textwidth, angle=0]{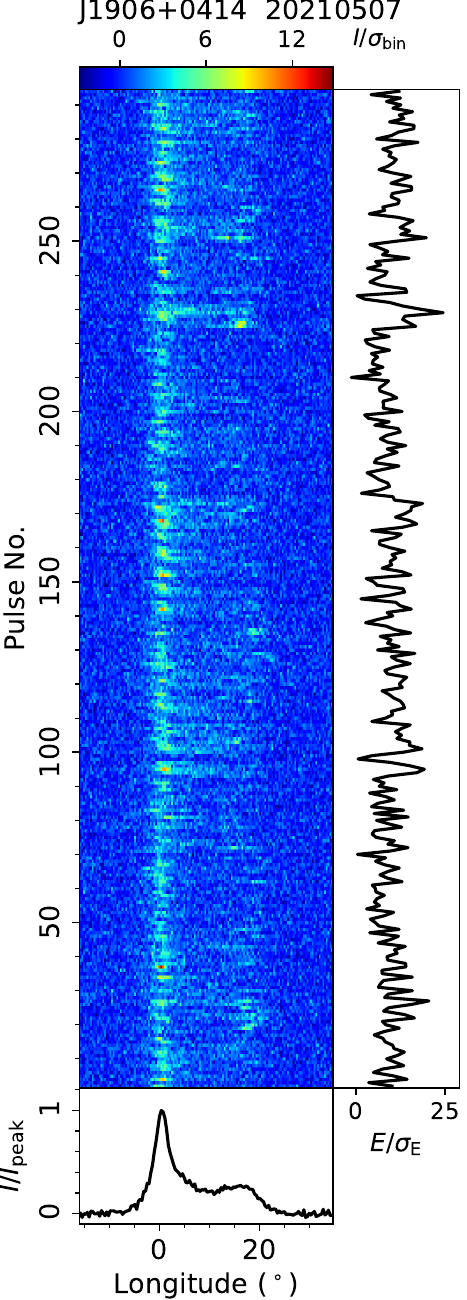}
\figcaption{Single pulse sequence of PSR J1906+0414 from the FAST observation on 20210507.
\label{subfig:TP:J1906+0414}}
\end{figure}

\begin{figure}[htpb]
\centering
\includegraphics[width=0.22\textwidth, angle=0]{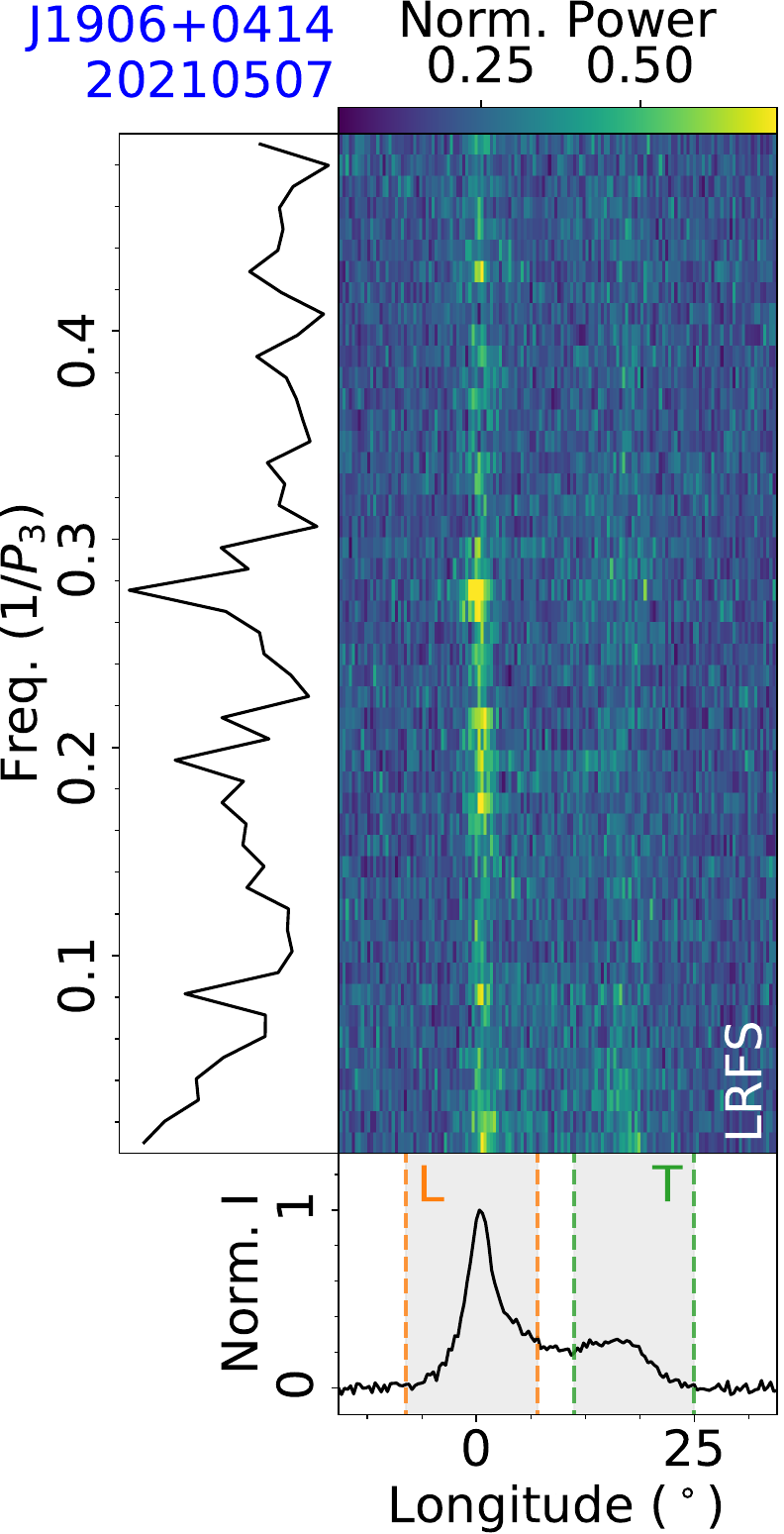}
\includegraphics[width=0.22\textwidth, angle=0]{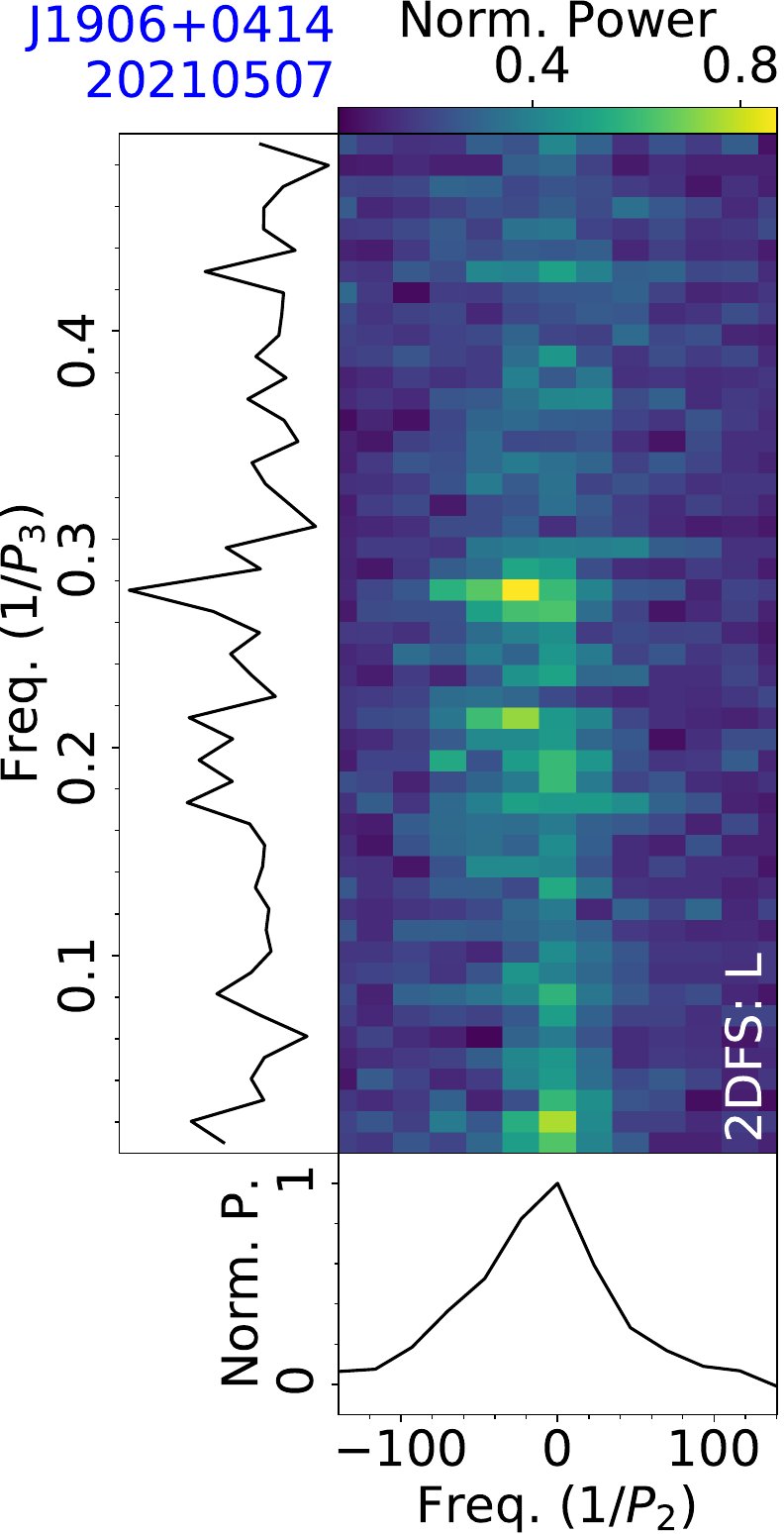}
\figcaption{Fluctuation analysis of PSR J1906+0414 for the observation on 20210507, with LRFS and 2DFS for the leading part of a mean pulse profile.
\label{subfig:fluctu:J1906+0414}}
\end{figure}

\subsection{J1905+0656}
\label{subsec:J1905+0656}

PSR J1905+0656 was discovered in the FAST GPPS survey \citep{Han2021,han2025}. 

This pulsar was observed by FAST on 20200205 for 25 minutes and on 20231001 for 15 minutes. From the 25-minute data, a rotation period $P=2.5117$~s and a dispersion measure $D\!M=24.8~{\rm cm^{-3}\,pc}$ were determined. 
Single pulse sequences of the observation on 20200205 are shown in Fig.~\ref{subfig:TP:J1905+0656}. Fluctuation spectra in Fig.~\ref{subfig:fluctu:J1905+0656} reveal a positive drift feature centered at $1/P_3=0.041\pm0.002$ and $1/P_2=86\pm5$, corresponding to $P_3=25\pm1$ periods and $P_2=4.2\pm0.2$ degrees. The drifting properties obtained from the observation on 20231001 are consistent with those from 20200205.

\begin{figure}[htpb]
\centering
\includegraphics[width=0.22\textwidth, angle=0]{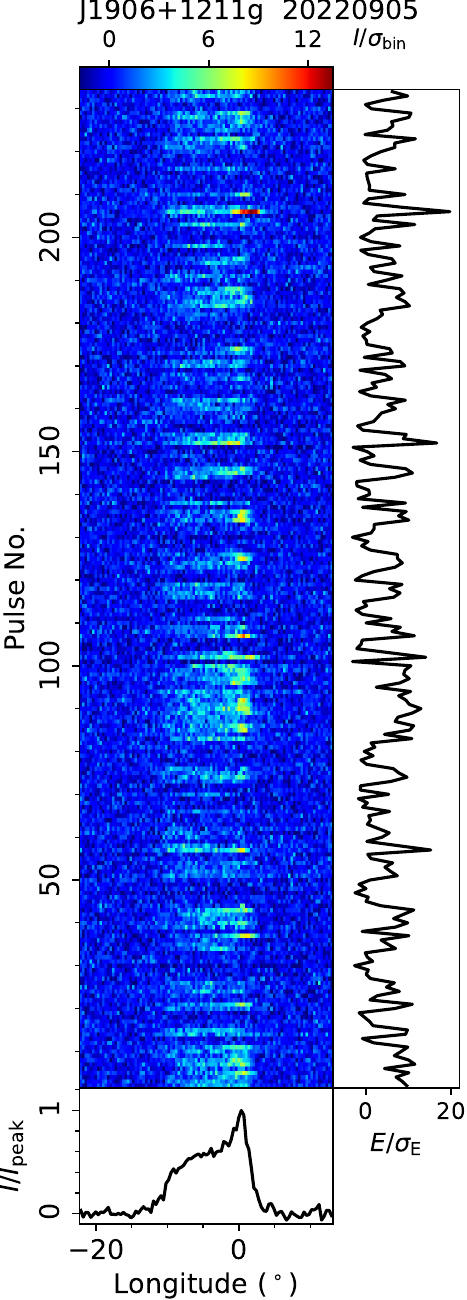}
\figcaption{Single pulse sequence of PSR J1906+1211g from the FAST observation on 20220905.
\label{subfig:TP:J1906+1211g}}
\end{figure}

\begin{figure}[htpb]
\centering
\includegraphics[width=0.39\textwidth, angle=0]{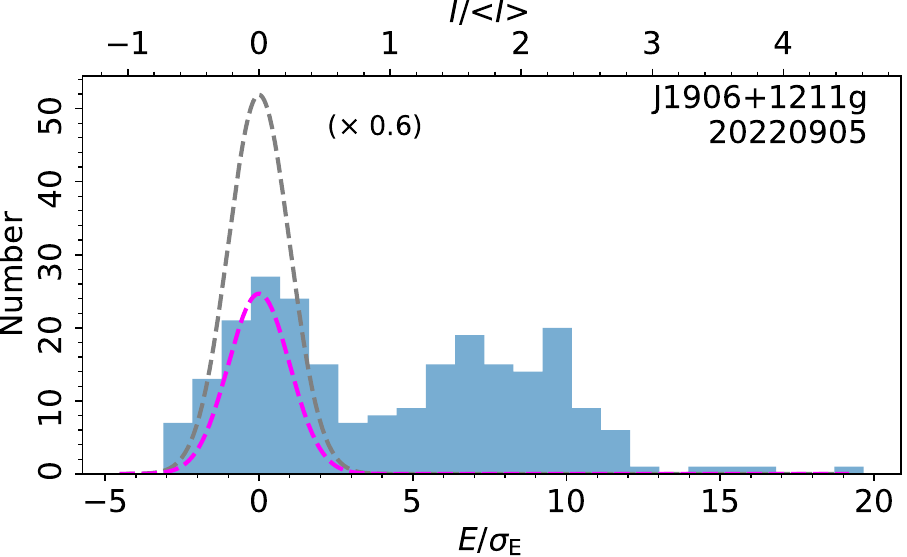}
\figcaption{On-pulse energy histogram of single pulses of PSR J1906+1211g from the FAST observation on 20220905.
\label{subfig:Hist:J1906+1211g}}
\end{figure}

\begin{figure}[htpb]
\centering
\includegraphics[width=0.22\textwidth, angle=0]{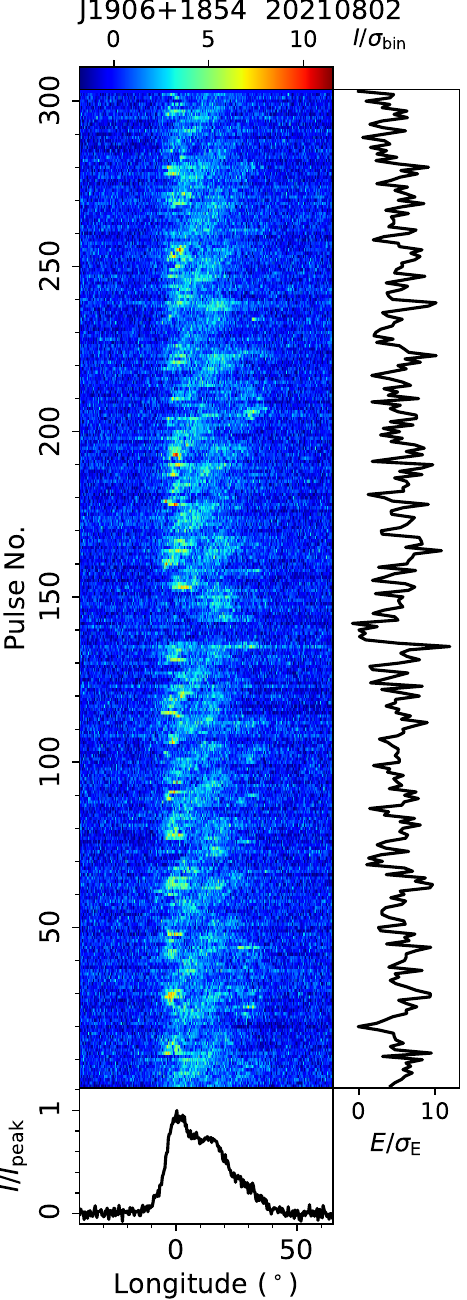}
\figcaption{Single pulse sequence of PSR J1906+1854 from the FAST observation on 20210802.
\label{subfig:TP:J1906+1854}}
\end{figure}

\begin{figure}[htpb]
\centering
\includegraphics[width=0.22\textwidth, angle=0]{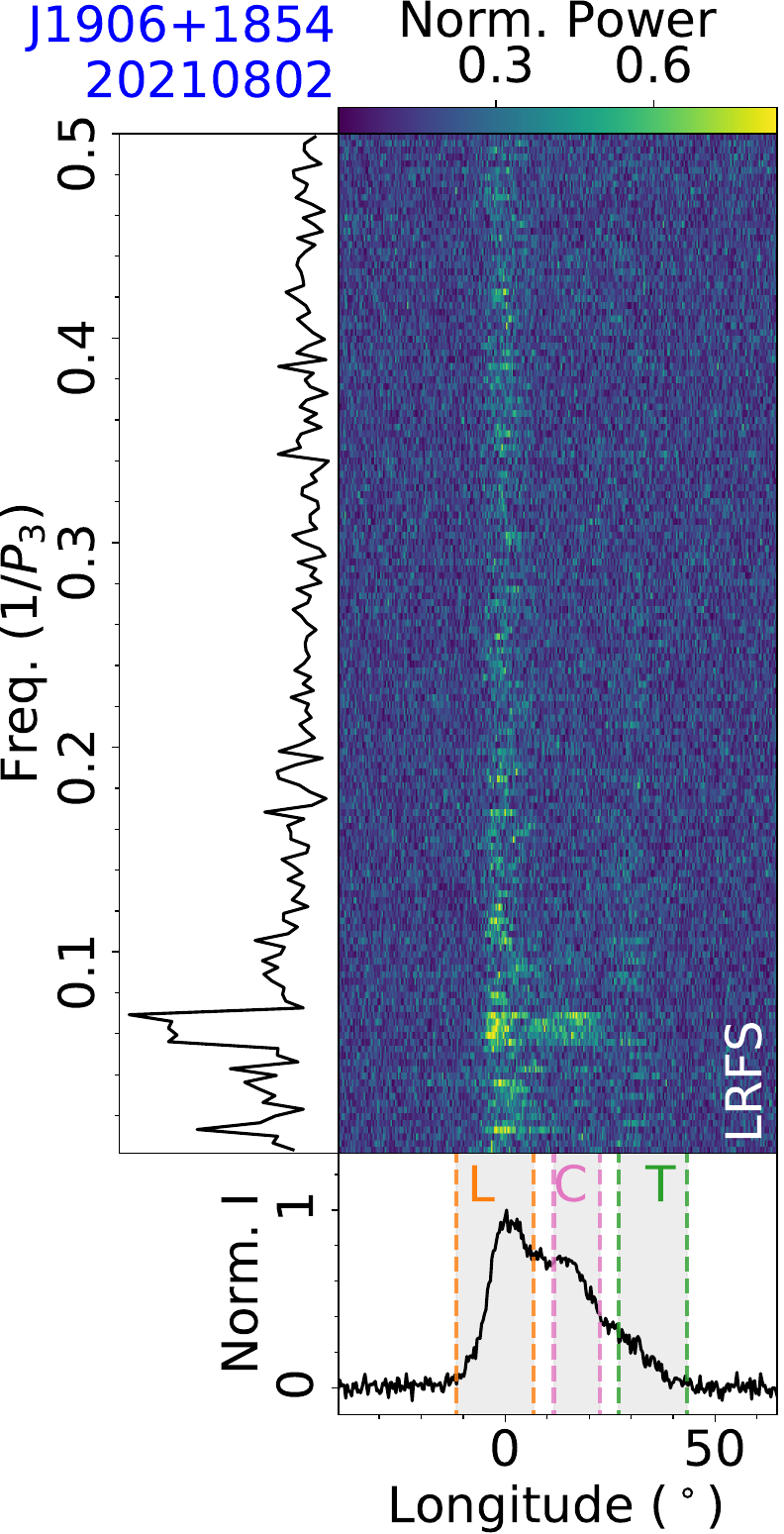}
\includegraphics[width=0.22\textwidth, angle=0]{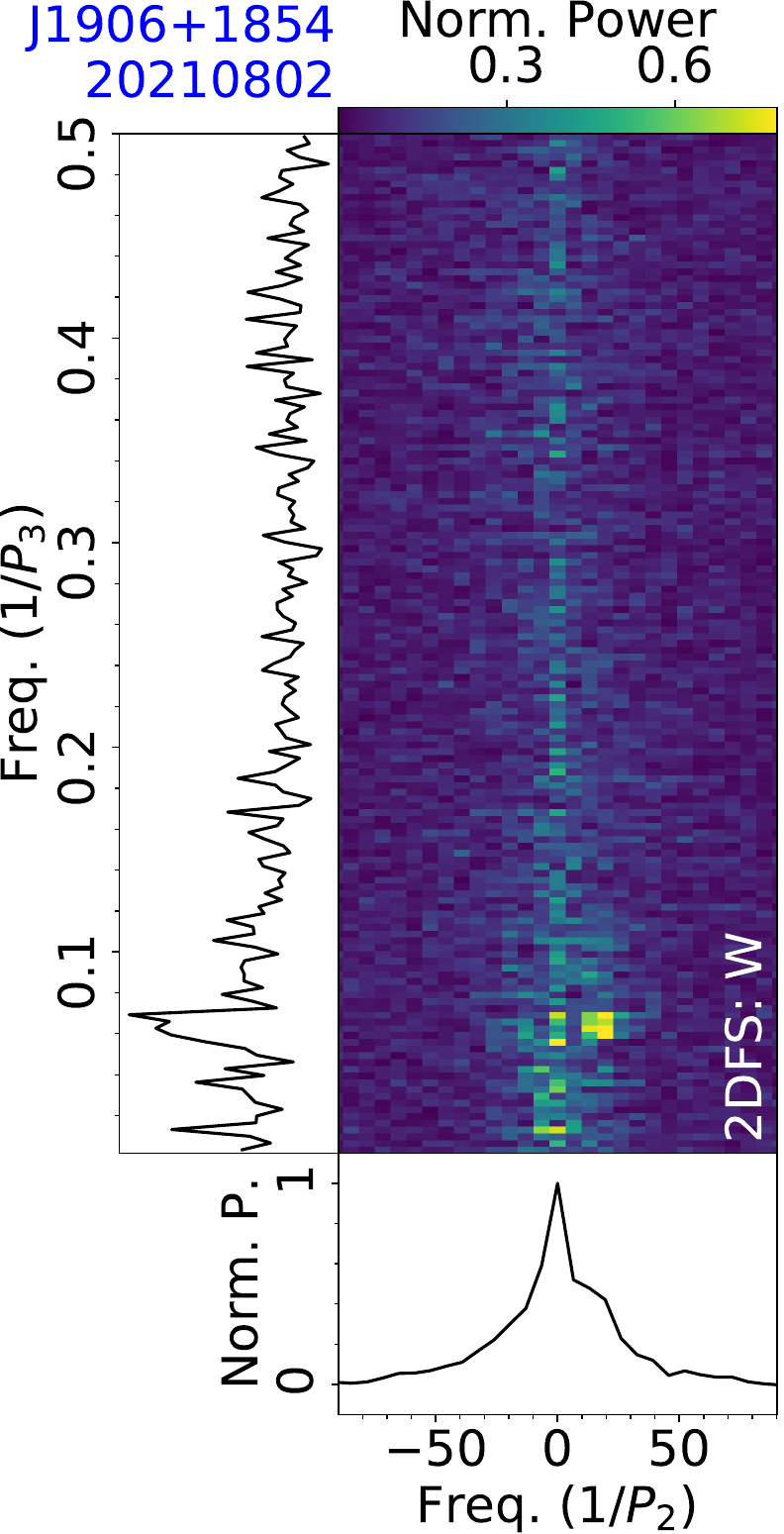}\\
\includegraphics[width=0.22\textwidth, angle=0]{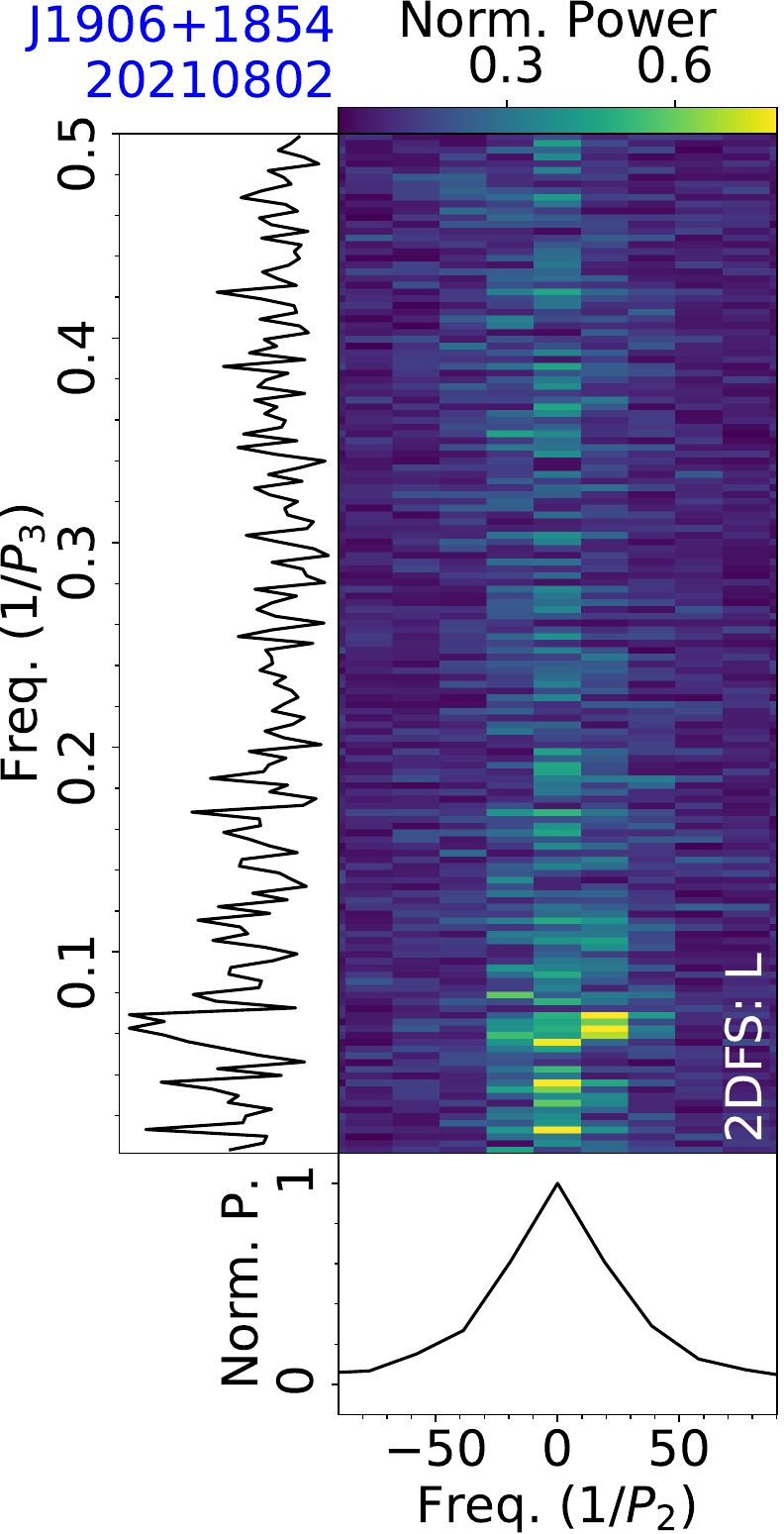}
\includegraphics[width=0.22\textwidth, angle=0]{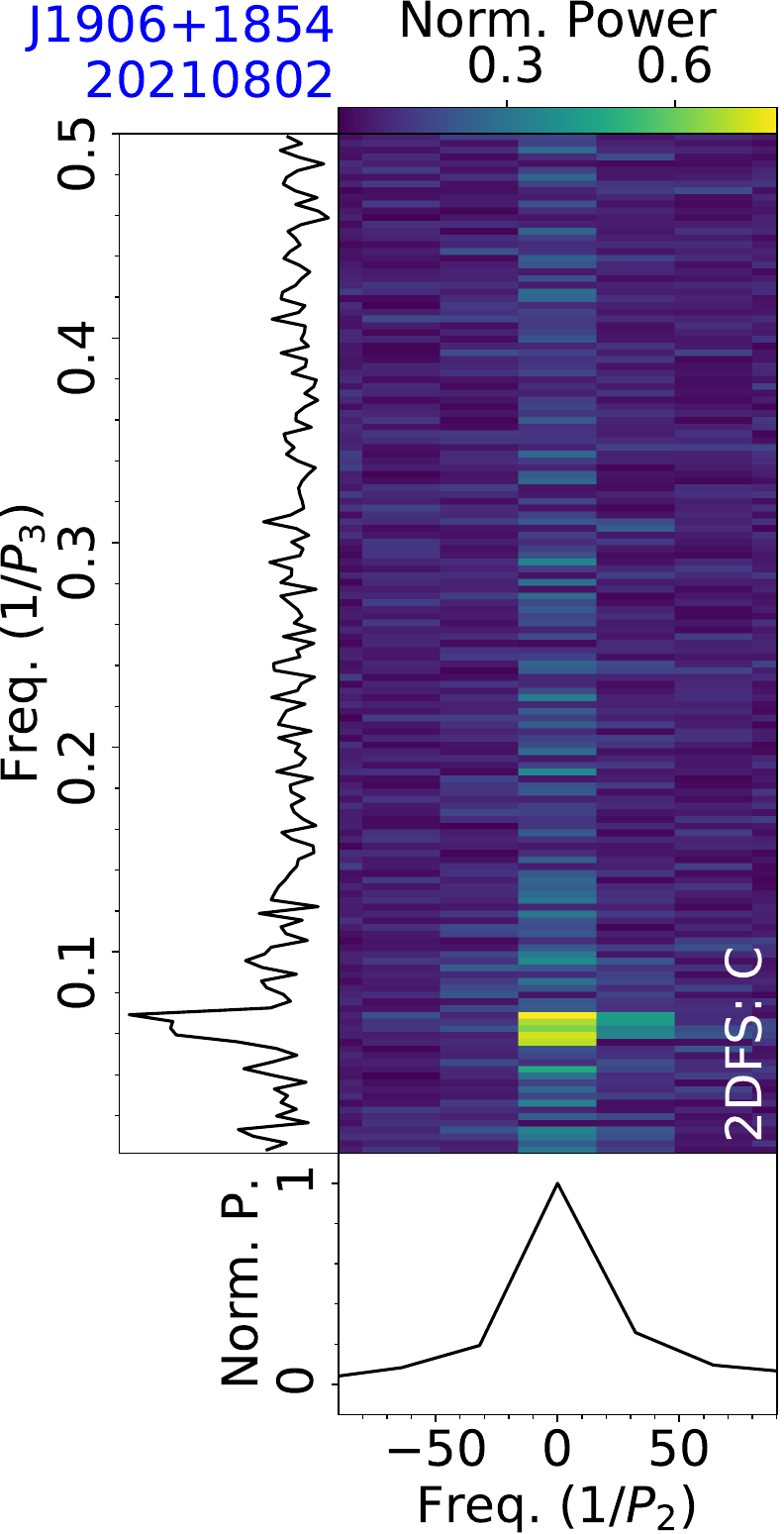}
\figcaption{Fluctuation analysis of PSR J1906+1854 for the observation on 20211123, with LRFS (top-left), and 2DFS for the on-pulse region (top-right), leading part (bottom-left) and central part (bottom-right) of a mean pulse profile.
\label{subfig:fluctu:J1906+1854}}
\end{figure}

\subsection{J1905+0758}
\label{subsec:J1905+0758}

PSR J1905+0758 was discovered in the FAST GPPS survey \citep{Han2021,han2025}. 

This pulsar was observed by FAST on 20220625 for 15 minutes, deriving a rotation period $P=1.1926$~s and a dispersion measure $D\!M=197.3~{\rm cm^{-3}\,pc}$. Single pulse sequences in Fig.~\ref{subfig:TP:J1905+0758} show subpulse modulation in both the leading and trailing profile parts. From the fluctuation spectra in Fig.~\ref{subfig:fluctu:J1905+0758}, the subpulse modulation feature for the on-pulse region of the mean pulse
profile has the centroid at $1/P_3=0.077\pm0.002$, corresponding to $P_3=13.0\pm0.3$ periods. 

Further FAST observations are required to study the subpulse modulation parameters and drifting directions for different profile parts.

\subsection{J1905+0935g}
\label{subsec:J1905+0935g}

PSR J1905+0935g was discovered in the FAST GPPS survey \citep{Han2021,han2025}. 

This pulsar was observed by FAST on 20220529 for 15 minutes, deriving a rotation period $P=1.6344$~s and a dispersion measure $D\!M=417.9~{\rm cm^{-3}\,pc}$. 
Single pulse sequences in Fig.~\ref{subfig:TP:J1905+0935g} of the observation indicate the nulling behavior, and there are 15 single pulses with the on-pulse integral energies less than 3$\sigma_{\rm E}$. In 2DFS of the leading part in a mean pulse profile, there is a 2-period temporal modulation feature with centroid frequencies of $1/P_3=0.4869\pm0.0004$ and $1/P_2=-83\pm3$, corresponding to periodicities of $P_3=2.054\pm0.002$ periods and $P_2=-4.3\pm0.2^\circ$.

\subsection{J1905+1034}
\label{subsec:J1905+1034}

PSR was discovered in the PALFA Survey, and previously reported to have nulling behavior \citet{Parent2021}. 

This pulsar was observed by FAST on 20200104 for 5 minutes. 
The single pulse sequence in Fig.~\ref{subfig:TP:J1905+1034} illustrates the existence of nulls. Nulling fraction of this observation is estimated from the on-pulse integral energy histogram in Fig.~\ref{subfig:Hist:J1905+1034} to be 43$\pm$3\%.

\subsection{J1906+0414}
\label{subsec:J1906+0414}

PSR J1906+0414 was discovered in the Parkes multibeam pulsar survey \citep{Lorimer2006}. 

This pulsar was observed by FAST on 20210507 for 5 minutes, deriving a rotation period $P=1.0433$~s and a dispersion measure $D\!M=348.0~{\rm cm^{-3}\,pc}$. 
The single pulse sequence of this observation is shown in Fig.~\ref{subfig:TP:J1906+0414}. Fluctuation spectra in Fig.~\ref{subfig:fluctu:J1906+0414} indicate that the leading component has two drift features, with centroid frequencies of $1/P_3=0.202\pm0.002$ ($P_3=4.95\pm0.04$ periods) and $1/P_2=-31\pm3$ ($P_2=-12\pm1^\circ$), and $1/P_3=0.272\pm0.001$ ($P_3=3.68\pm0.02$ periods) and $1/P_2=-31\pm3$ ($P_2=-12\pm1^\circ$), respectively.
A longer observation is required for the robust analysis of drift features.

\begin{figure}[htpb]
\centering
\includegraphics[width=0.44\textwidth, angle=0]{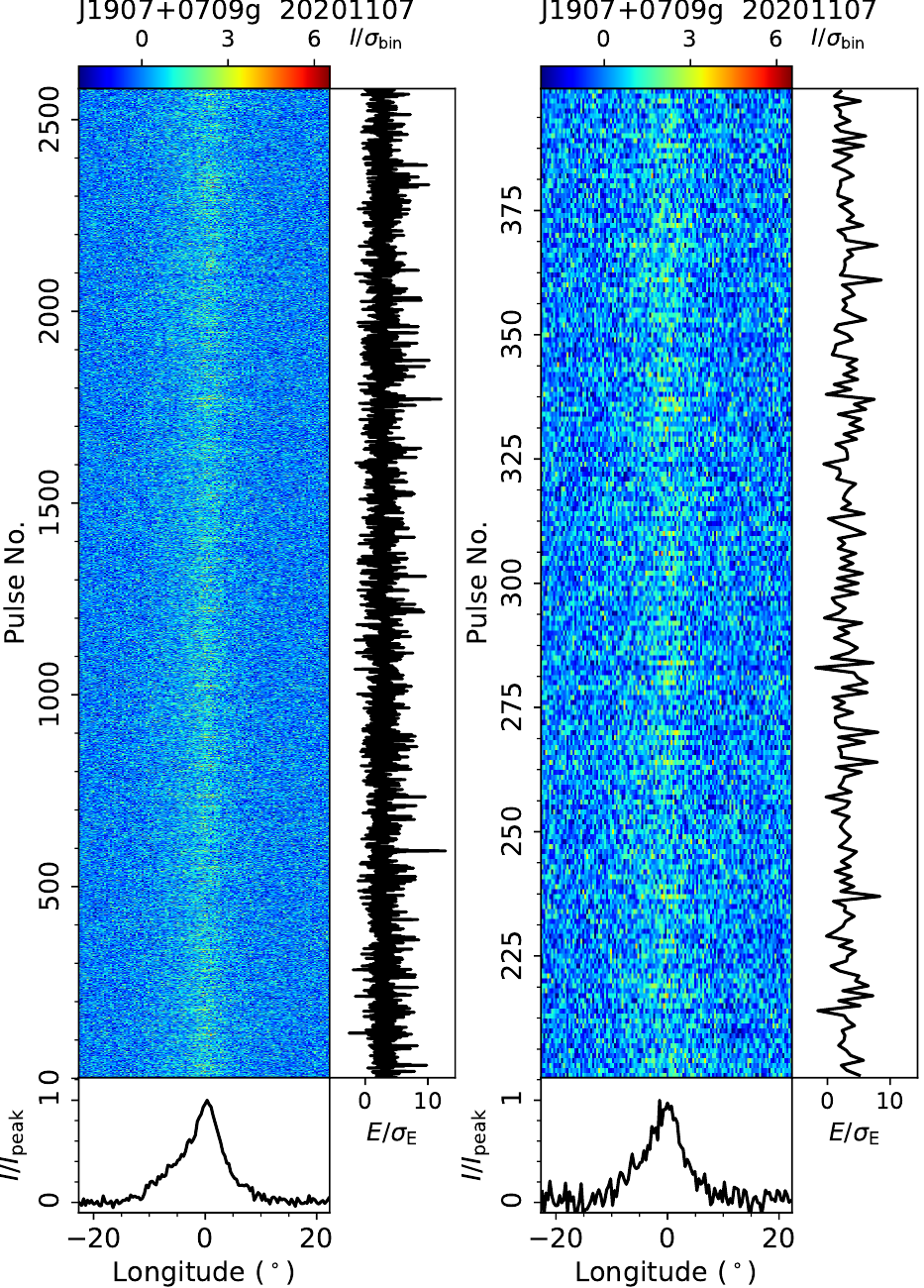}
\figcaption{Single pulse sequence of PSR J1907+0709g from the FAST observation on 20201107, as well as a zoomed-in view of pulses No.100-400.
\label{subfig:TP:J1907+0709g}}
\end{figure}

\begin{figure}[htpb]
\centering
\includegraphics[width=0.44\textwidth, angle=0]{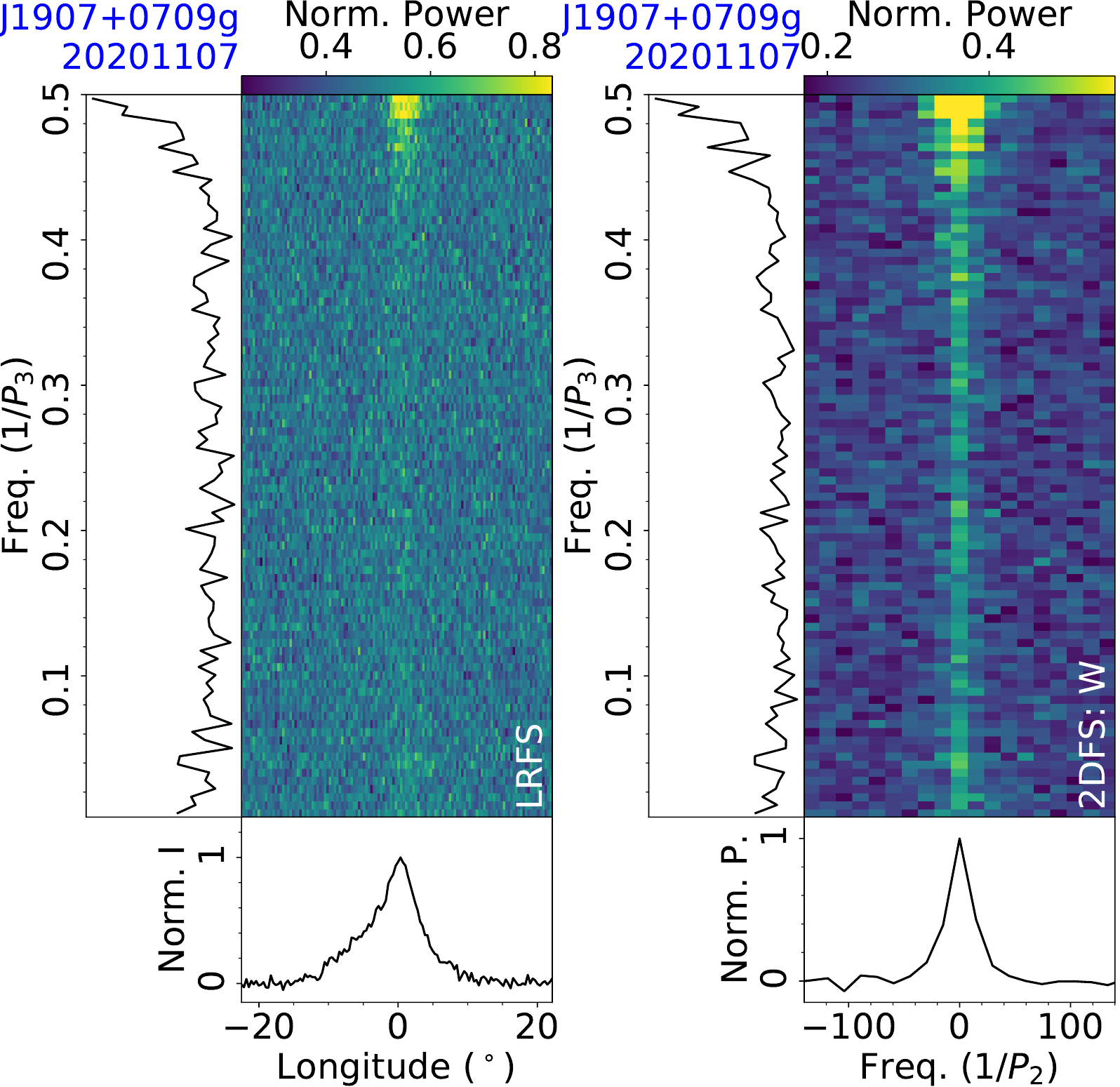}
\figcaption{Fluctuation analysis of PSR J1907+0709g for the observation on 20201107, with LRFS and 2DFS for the on-pulse region of a mean pulse profile.
\label{subfig:fluctu:J1907+0709g}}
\end{figure}

\subsection{J1906+1211g}
\label{subsec:J1906+1211g}

PSR J1906+1211g was discovered in the FAST GPPS survey \citep{Han2021,han2025}. 

This pulsar was observed by FAST on 20220905 for 15 minutes. 
The single pulse sequence in Fig.~\ref{subfig:TP:J1906+1211g} displays the nulling phenomenon. The nulling fraction is estimated to be 29$\pm$4\% from the on-pulse integral energy histogram (Fig.~\ref{subfig:Hist:J1906+1211g}).

\begin{figure}[htpb]
\centering
\includegraphics[width=0.22\textwidth, angle=0]{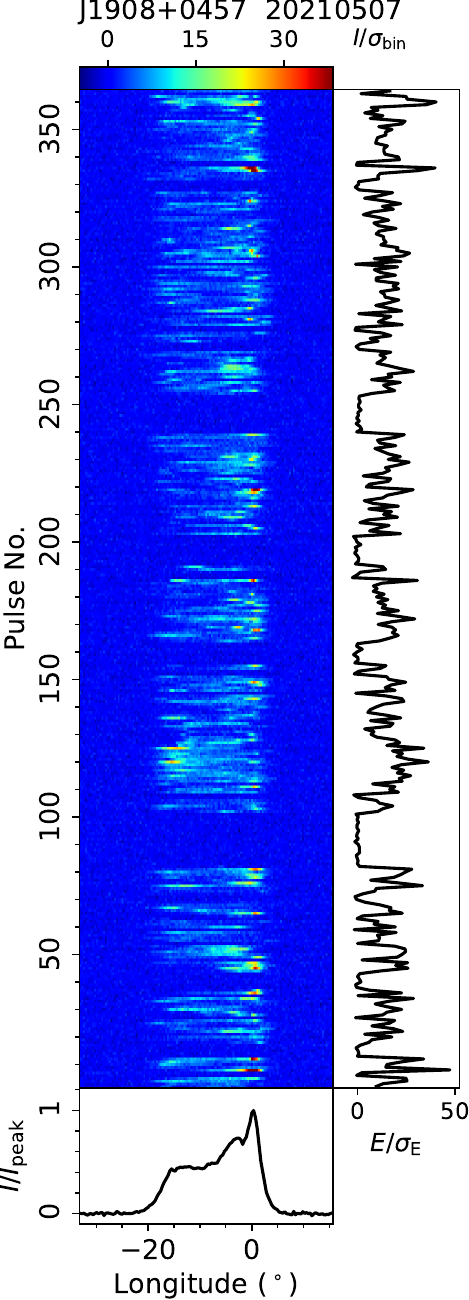}
\figcaption{Single pulse sequence of PSR J1908+0457 from the FAST observation on 20210507.
\label{subfig:TP:J1908+0457}}
\end{figure}

\begin{figure}[htpb]
\centering
\includegraphics[width=0.39\textwidth, angle=0]{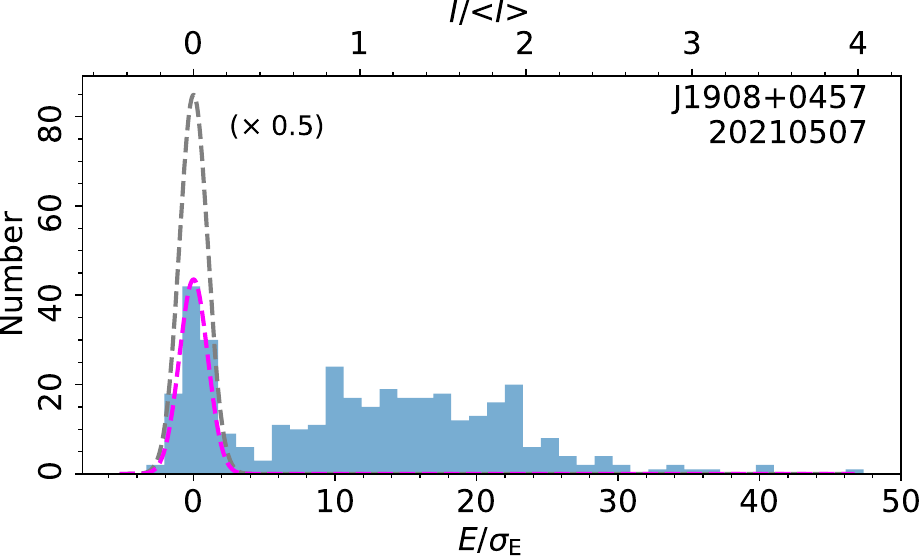}
\figcaption{On-pulse energy histogram of single pulses of PSR J1908+0457 from the FAST observation on 20210507.
\label{subfig:Hist:J1908+0457}}
\end{figure}

\subsection{J1906+1854}
\label{subsec:J1906+1854}

PSR J1906+1854 was discovered by \citet{Nice1995} using the Arecibo telescope. The subpulse drifting behavior was reported by \citet{Song2023}. 

This pulsar was observed by FAST on 20210802 for 5 minutes. 
The single pulse sequence displayed in Fig.~\ref{subfig:TP:J1906+1854} shows drifting bands. From the LRFS and 2DFS in Fig.~\ref{subfig:fluctu:J1906+1854}, the leading and central parts in the mean pulse profile both have positive subpulse drifting behaviors. 
For the leading profile part, 2DFS has a drift feature with centroid frequencies of $1/P_3=0.063\pm0.001$ and $1/P_2=9\pm2$, corresponding to $P_3=16.0\pm0.1$ periods and $P_2=39\pm8^\circ$. For 2DFS of the central part, the centroid of the drift feature is characterized by $1/P_3=0.0637\pm0.0004$ and $1/P_2=9\pm3$, yielding drifting periodicities of $P_3=15.7\pm0.1$ periods and $P_2=39\pm14^\circ$. 
However, the drifting property of the trailing part is not clear. 
It is likely to be mode changing or nulling of pulses No.135-140. Further observations are needed to confirm this behavior.

\begin{figure}[htpb]
\centering
\includegraphics[width=0.22\textwidth, angle=0]{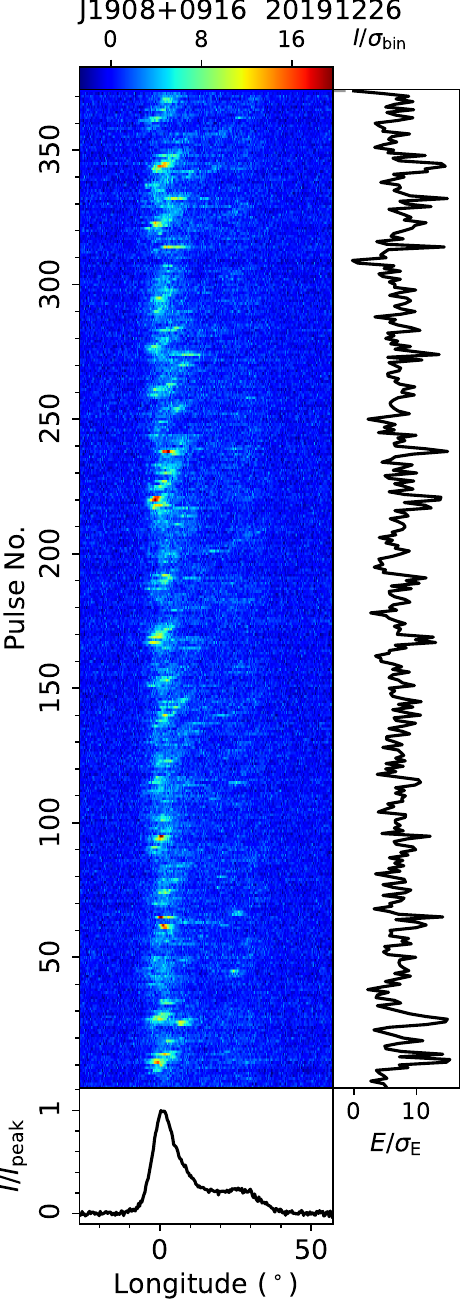}
\figcaption{Single pulse sequence of PSR J1908+0916 from the FAST observation on 20191226.
\label{subfig:TP:J1908+0916}}
\end{figure}

\begin{figure}[htpb]
\centering
\includegraphics[width=0.22\textwidth, angle=0]{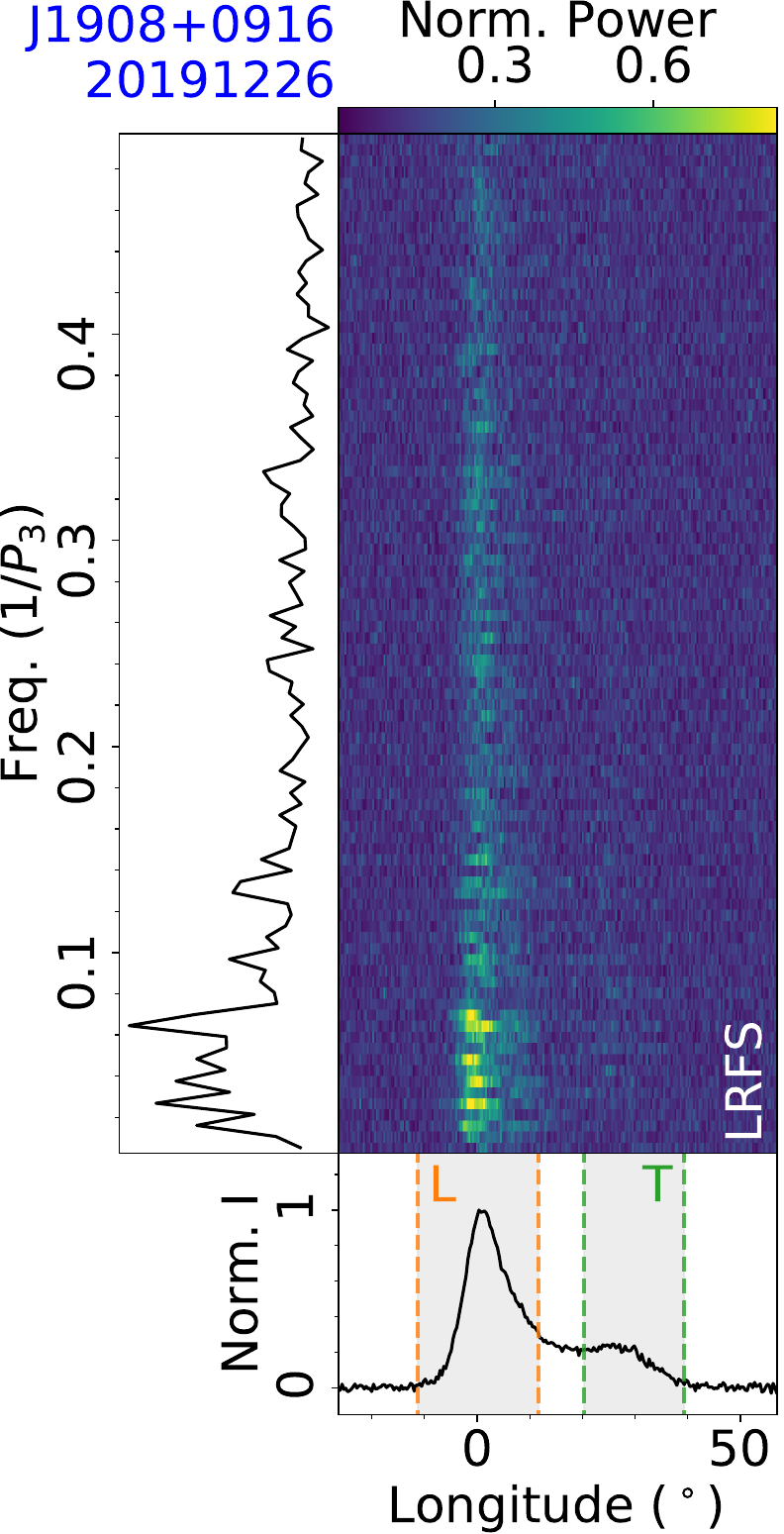}
\includegraphics[width=0.22\textwidth, angle=0]{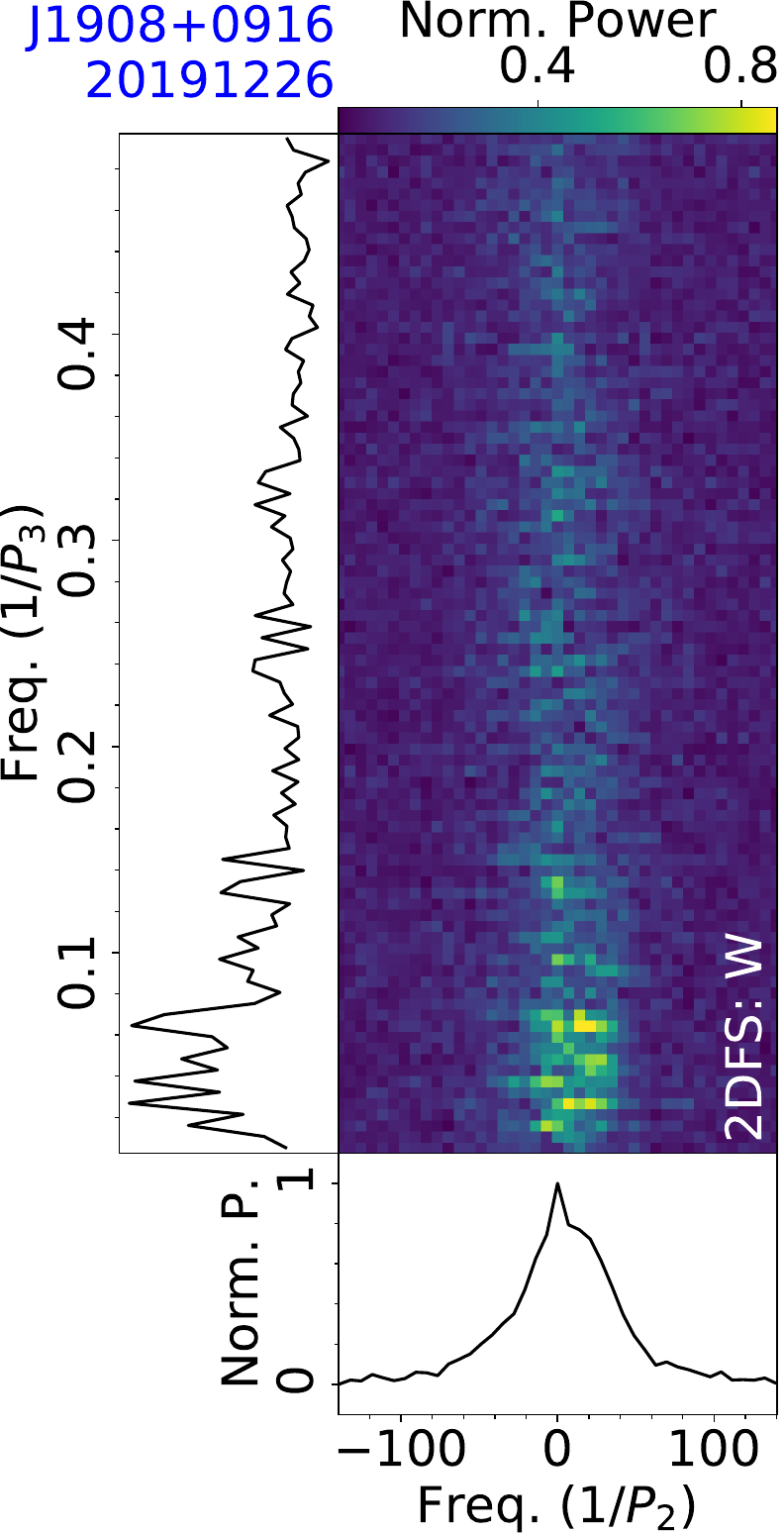}\\
\includegraphics[width=0.22\textwidth, angle=0]{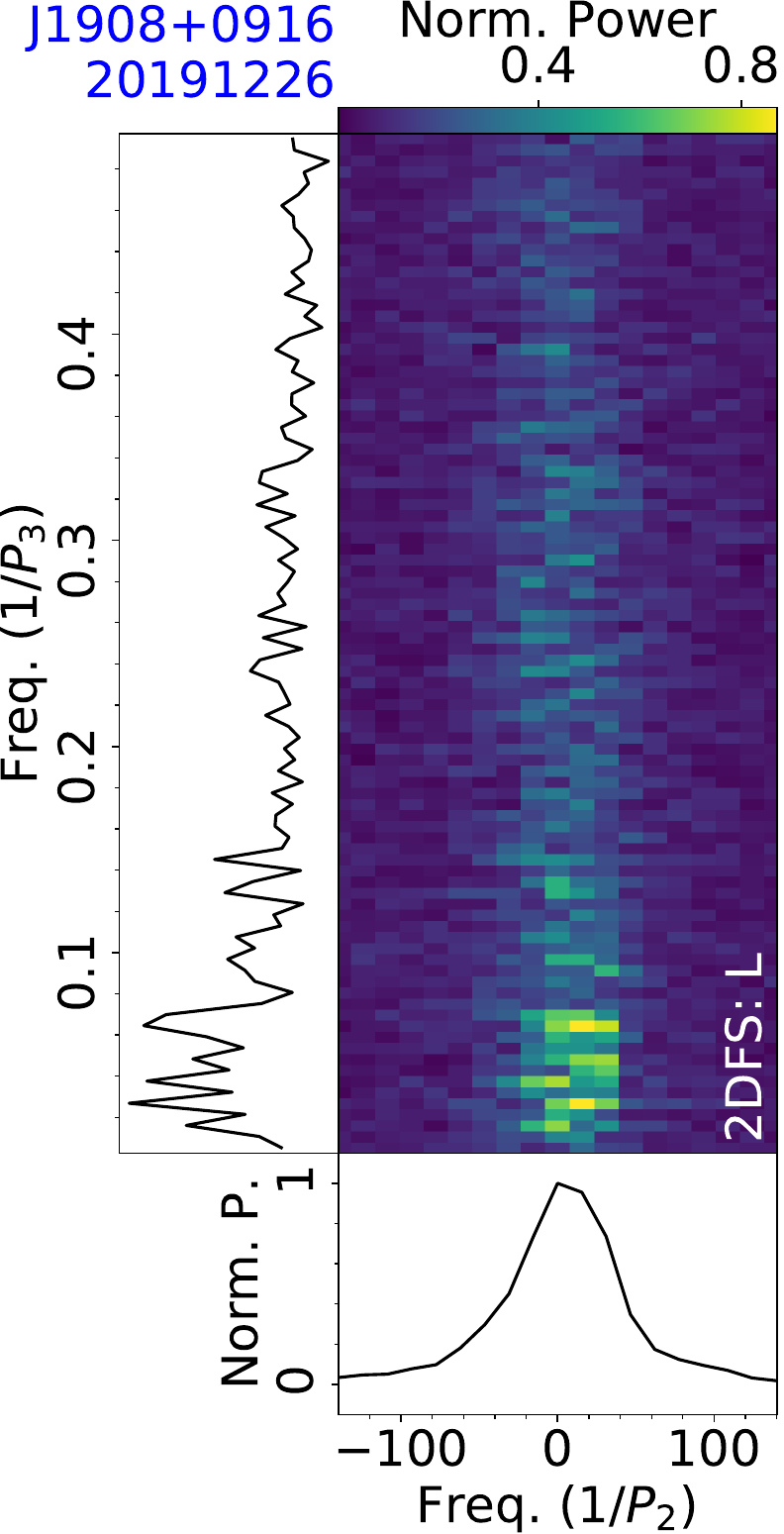}
\includegraphics[width=0.22\textwidth, angle=0]{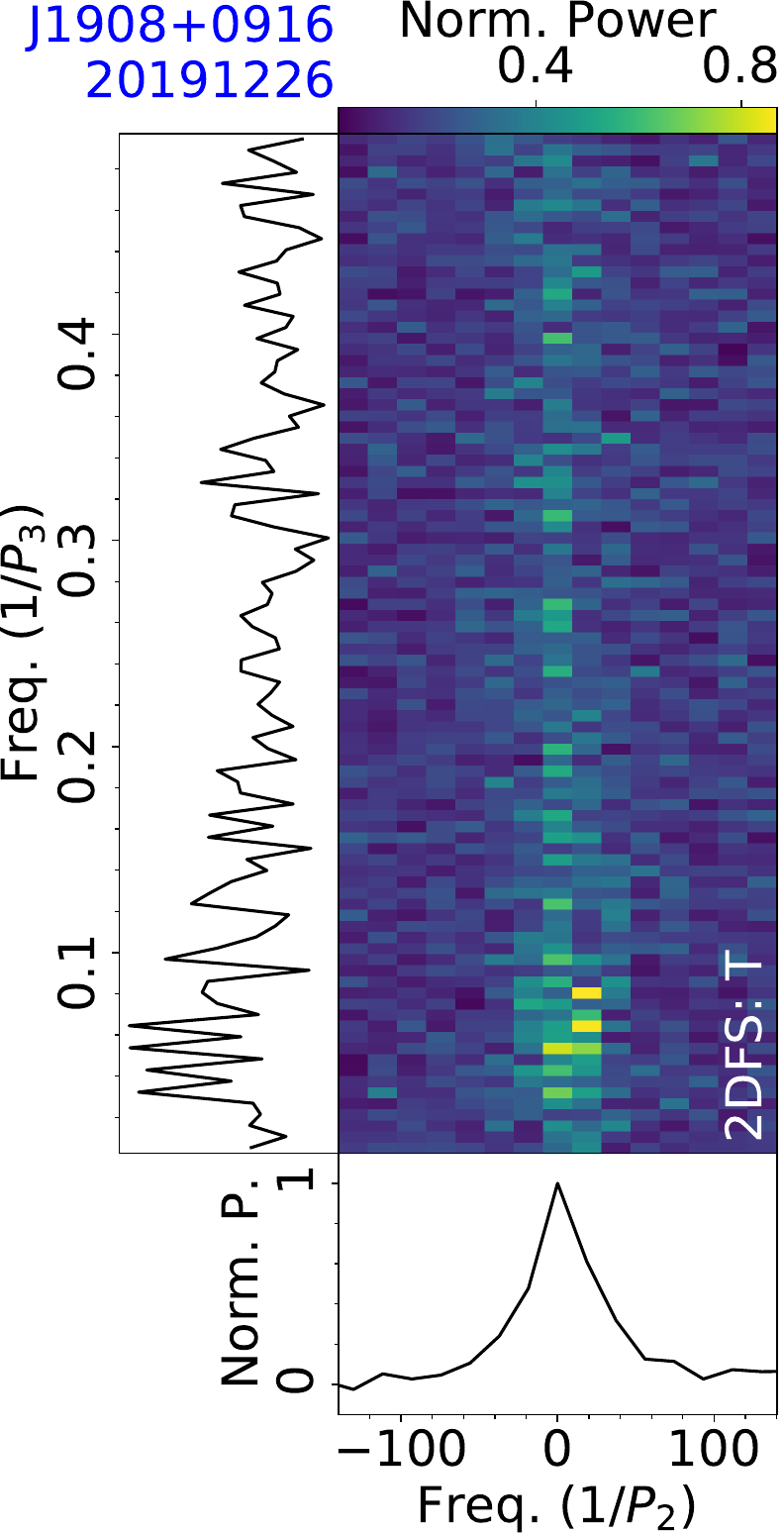}
\figcaption{Fluctuation analysis of PSR J1908+0916 for the observation on 20191226, with LRFS (top-left), and 2DFS for the on-pulse region (top-right), leading part (bottom-left) and trailing part (bottom-right) of a mean pulse profile.
\label{subfig:fluctu:J1908+0916}}
\end{figure}

\subsection{J1907+0709g}
\label{subsec:J1907+0709g}

PSR J1907+0709g was discovered in the FAST GPPS survey \citep{Han2021,han2025}. 

This pulsar was observed by FAST on 20201107 and 20250807, each for 15 minutes. From the observation on 20201107, a rotation period $P=0.3441$~s and a dispersion measure $D\!M=280.1~{\rm cm^{-3}\,pc}$ were derived. The single pulse sequence and a zoomed-in view of pulses No. 200-400 from the data on 20201107 are shown in Fig.~\ref{subfig:TP:J1907+0709g}. In the fluctuation spectra in Fig.~\ref{subfig:fluctu:J1907+0709g}, there is a modulation feature with the centroid at $1/P_3=0.479\pm0.001$, corresponding to $P_3=2.089\pm0.003$ periods. The modulation feature observed on 20250807 is consistent with that on 20201107.

\begin{figure}[htpb]
\centering
\includegraphics[width=0.22\textwidth, angle=0]{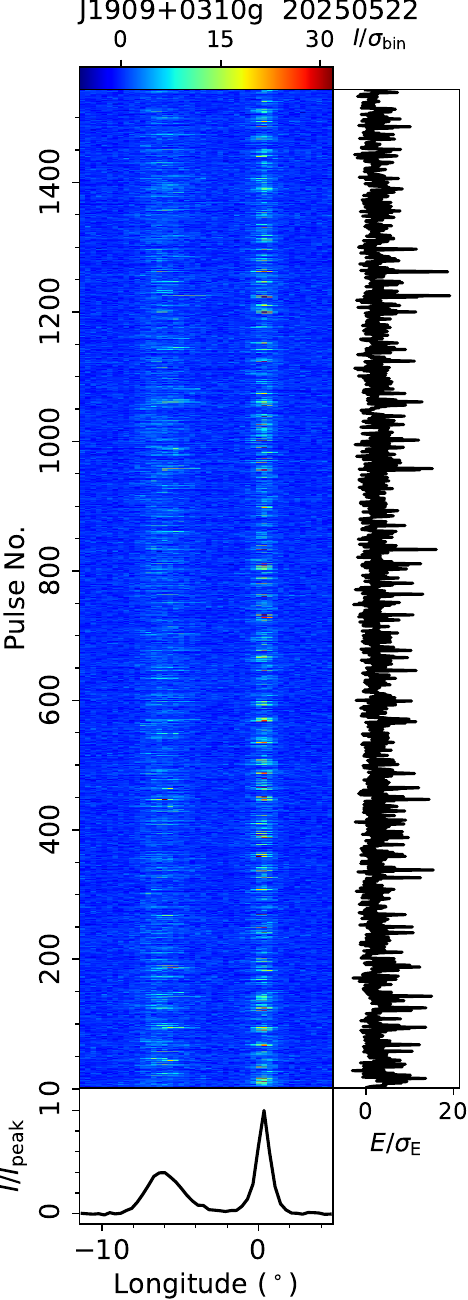}
\includegraphics[width=0.22\textwidth, angle=0]{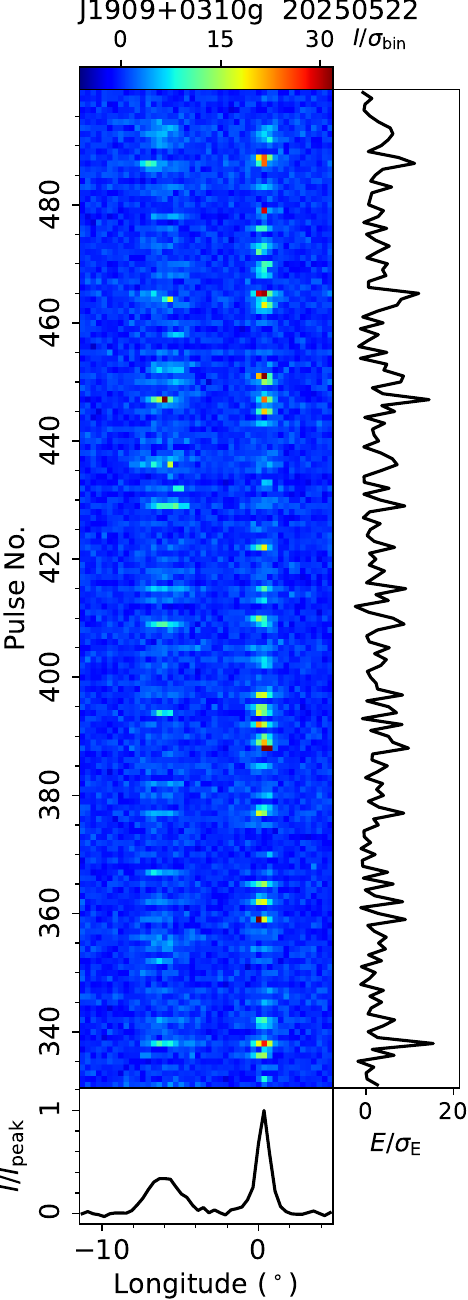}
\figcaption{Single pulse sequence of PSR J1909+0310g from the FAST observation on 20250522, and a zoomed-in view of pulses No. 330-500.
\label{subfig:TP:J1909+0310g}}
\end{figure}

\begin{figure}[htpb]
\centering
\includegraphics[width=0.22\textwidth, angle=0]{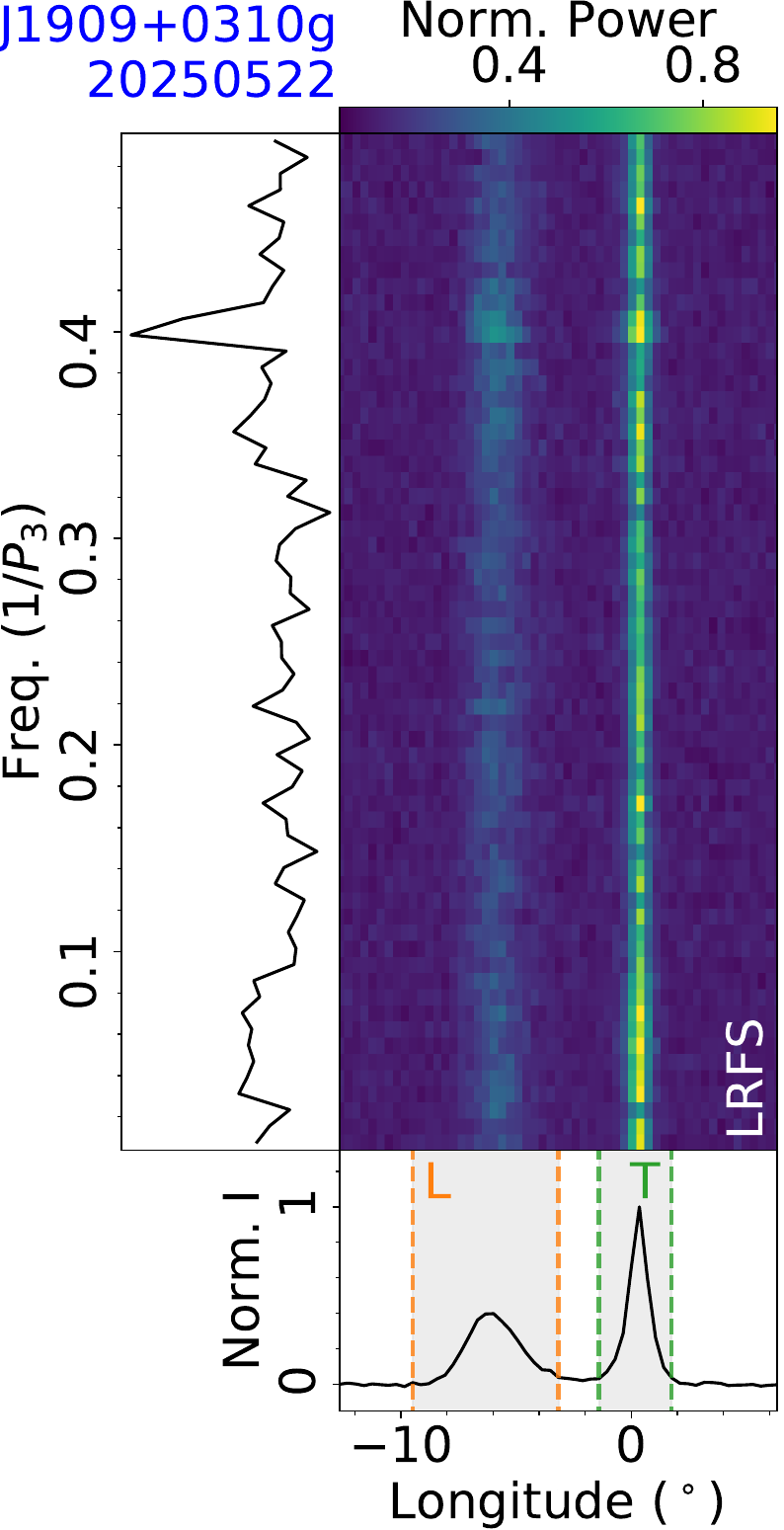}
\includegraphics[width=0.22\textwidth, angle=0]{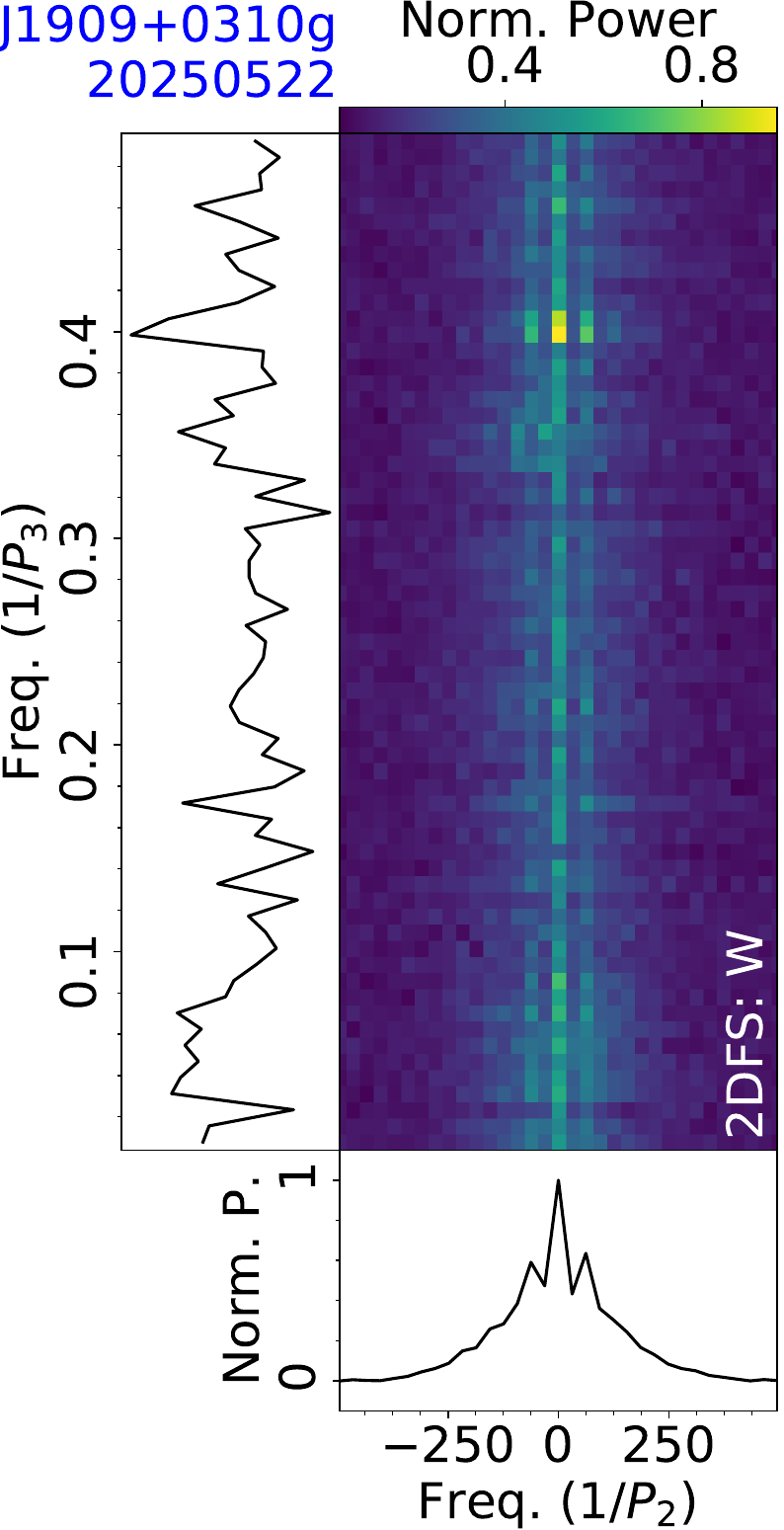}\\
\includegraphics[width=0.22\textwidth, angle=0]{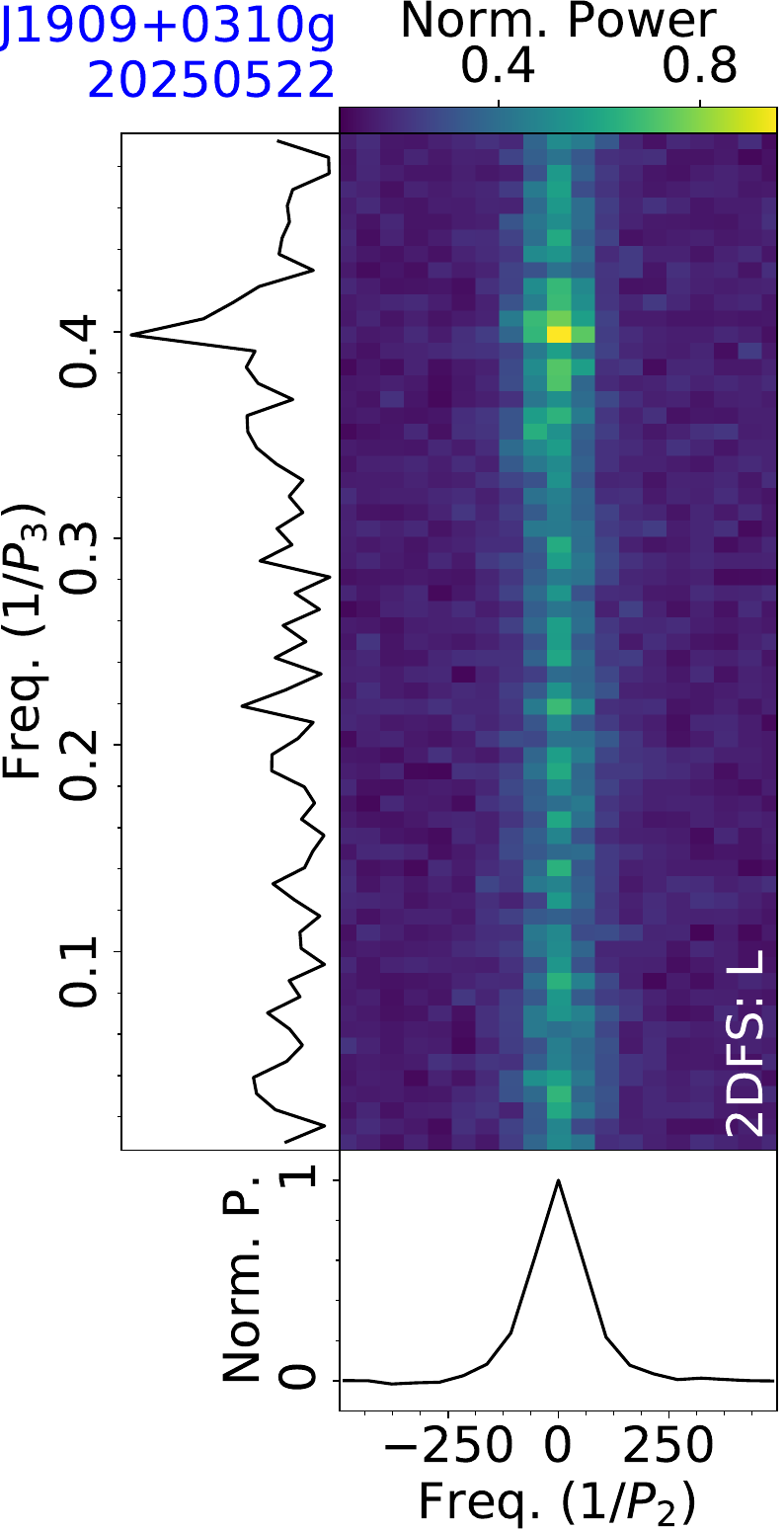}
\includegraphics[width=0.22\textwidth, angle=0]{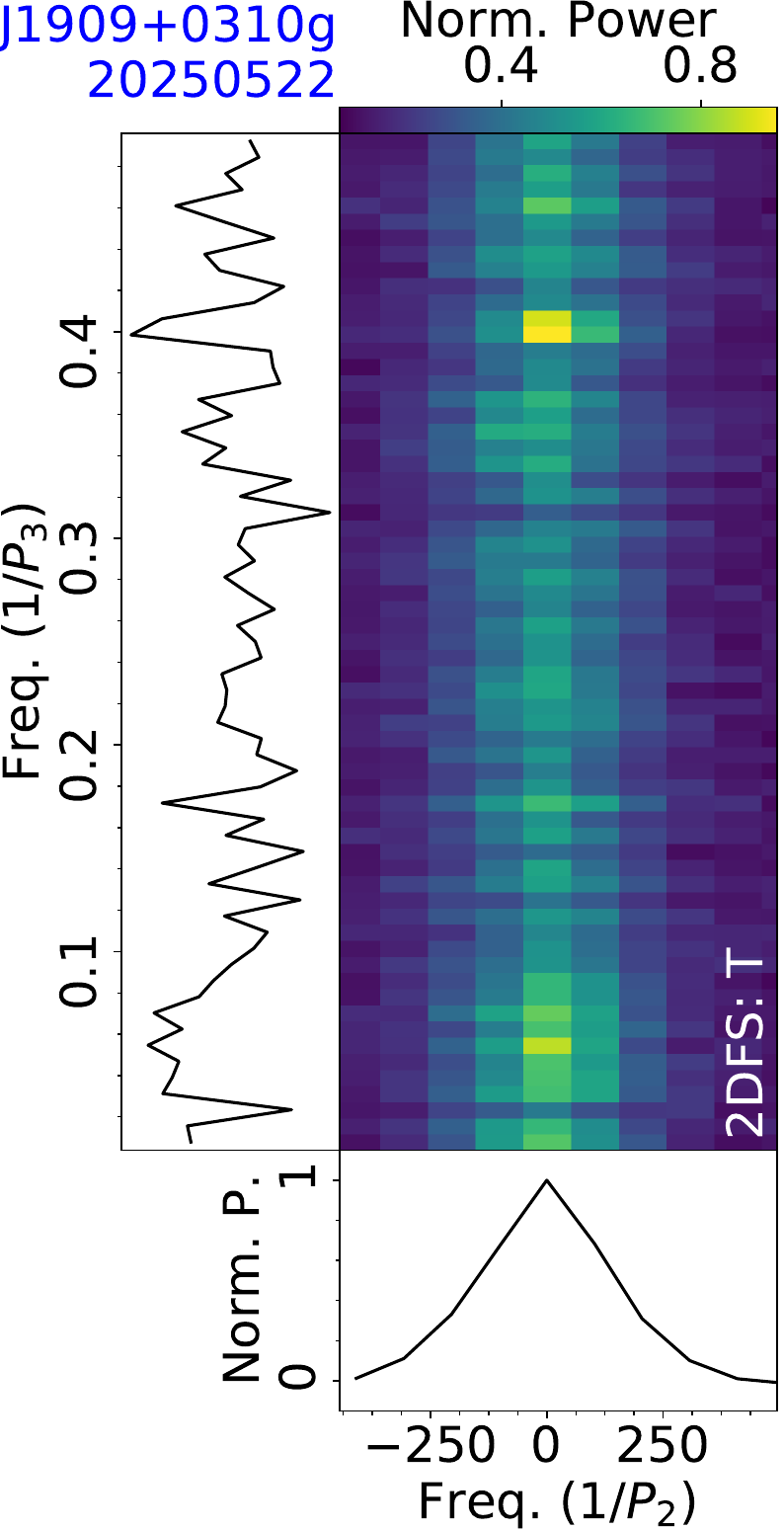}
\figcaption{Fluctuation analysis of PSR J1909+0310g for the observation on 20250522, with LRFS (top-left), and 2DFS for the on-pulse region (top-right), leading part (bottom-left) and trailing part (bottom-right) of a mean pulse profile.
\label{subfig:fluctu:J1909+0310g}}
\end{figure}

\subsection{J1908+0457}
\label{subsec:J1908+0457}

PSR J1908+0457 was discovered in a search targeted at Einstein IPC X-ray sources using the Arecibo telescope \citep{Zepka1996}. 

This pulsar was observed by FAST on 20210507 for 5 minutes, deriving a rotation period $P=0.8467$~s and a dispersion measure $D\!M=349.3~{\rm cm^{-3}\,pc}$. 
The single pulse sequence is displayed in Fig.~\ref{subfig:TP:J1908+0457}, which illustrates the existence of nulls. The nulling fraction is estimated from the on-pulse integral energy histogram (Fig.~\ref{subfig:Hist:J1908+0457}) to be 26$\pm$1\%.

\subsection{J1908+0916}
\label{subsec:J1908+0916}

PSR J1908+0916 was discovered by \citet{Hulse1975} using the Arecibo telescope. The subpulse drifting of the pulsar was reported by \citet{Song2023} to be $P_3$=24(5) periods and $P_2$=17$^{+66}_{-4}$ degrees. 

The pulsar was observed by FAST on 20191226 for 5 minutes, deriving a rotation period $P=0.8303$~s and a dispersion measure $D\!M=250.8~{\rm cm^{-3}\,pc}$. 
The single pulse sequence is displayed in Fig.~\ref{subfig:TP:J1908+0916}. 
Drifting parameters of the leading and trailing parts in a mean pulse profile are estimated from fluctuation spectra (Fig.~\ref{subfig:fluctu:J1908+0916}). 
The centroid frequencies of the drift feature in 2DFS are $1/P_3=0.044\pm0.001$ ($P_3=22.8\pm0.4$ periods) and $1/P_2=11\pm1$ ($P_2=33\pm3^\circ$) for the leading profile part, and $1/P_3=0.058\pm0.002$ ($P_3=17.4\pm0.5$ periods) and $1/P_2=12\pm2$ ($P_2=31\pm4^\circ$) for the trailing part.

\subsection{J1909+0310g}
\label{subsec:J1909+0310g}

PSR J1909+0310g was discovered in the FAST GPPS survey \citep{Han2021,han2025}.

This pulsar was observed by FAST on 20220828 for 15 minutes and 20250522 for 51 minutes. From the data of 20220828, a rotation period $P=1.9720$~s and a dispersion measure $D\!M=108.7~{\rm cm^{-3}\,pc}$ were determined. 
Single pulse sequences of the observation on 20250522 are shown in Fig.~\ref{subfig:TP:J1909+0310g}. From fluctuation spectra in Fig.~\ref{subfig:fluctu:J1909+0310g}, leading and trailing parts in a mean pulse profile both have a temporal modulation feature with the centroid frequency of $1/P_3=0.402\pm0.001$, corresponding to the periodicity of $P_3=2.49\pm0.01$ periods. Modulation property from the observation on 20220828 is consistent with that on 20250522.

\begin{figure}[htpb]
    \centering
	\includegraphics[width=0.22\textwidth]{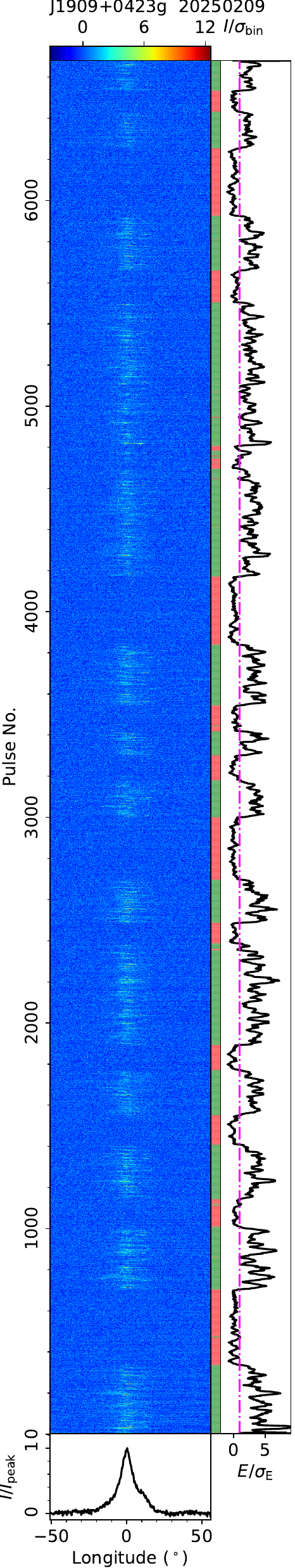}
	\includegraphics[width=0.22\textwidth]{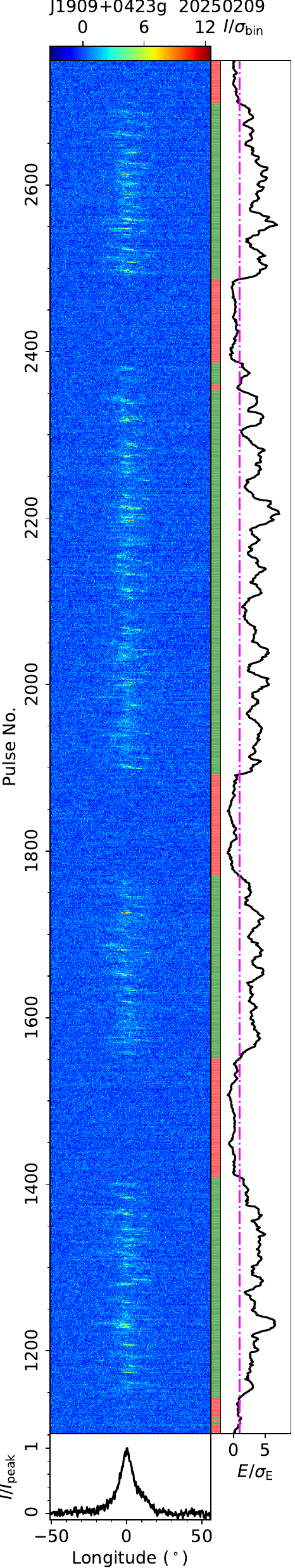}
	\figcaption{Single pulse sequence of PSR J1909+0423g from the FAST observation on 20250209, and the zoomed-in view of pulses No.1101-2750.
	\label{subfig:TP:J1909+0423g}}
\end{figure}

\begin{figure}[htpb]
\centering
\includegraphics[width=0.39\textwidth, angle=0]{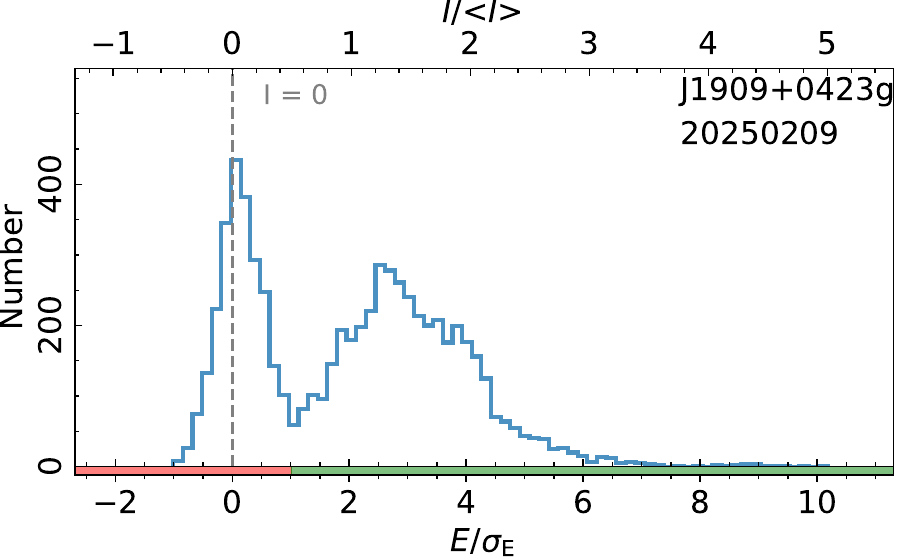}
\figcaption{On-pulse energy histogram of single pulses of PSR J1909+0423g from the FAST observation on 20250209.
\label{subfig:Hist:J1909+0423g}}
\end{figure}

\begin{figure}[htpb]
\centering
\includegraphics[width=0.39\textwidth, angle=0]{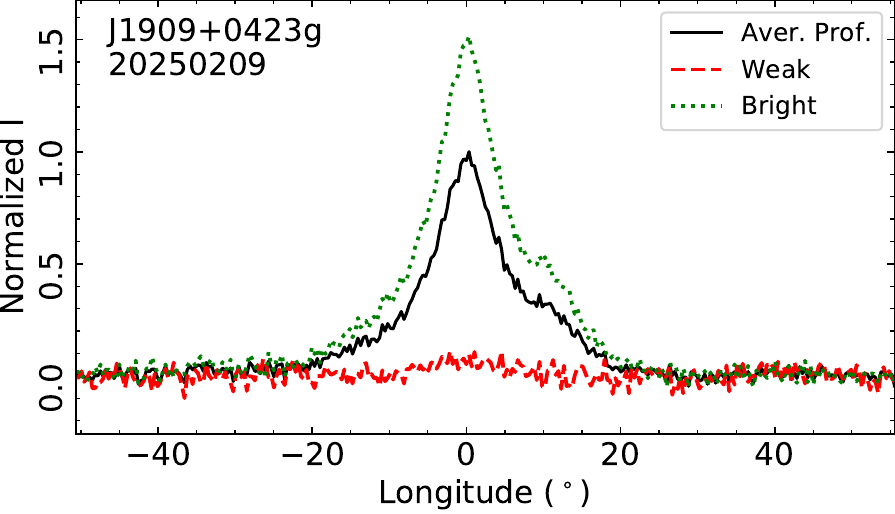}
\figcaption{Mean profiles for the weak (red dashed) and bright (green dotted) emission modes of PSR J1909+0423g from the FAST observation on 20250209, which are normalized by the peak of the mean profile (black solid) of all periods.
\label{subfig:ProfModes:J1909+0423g}}
\end{figure}

\begin{figure}[htpb]
\centering
\includegraphics[width=0.22\textwidth, angle=0]{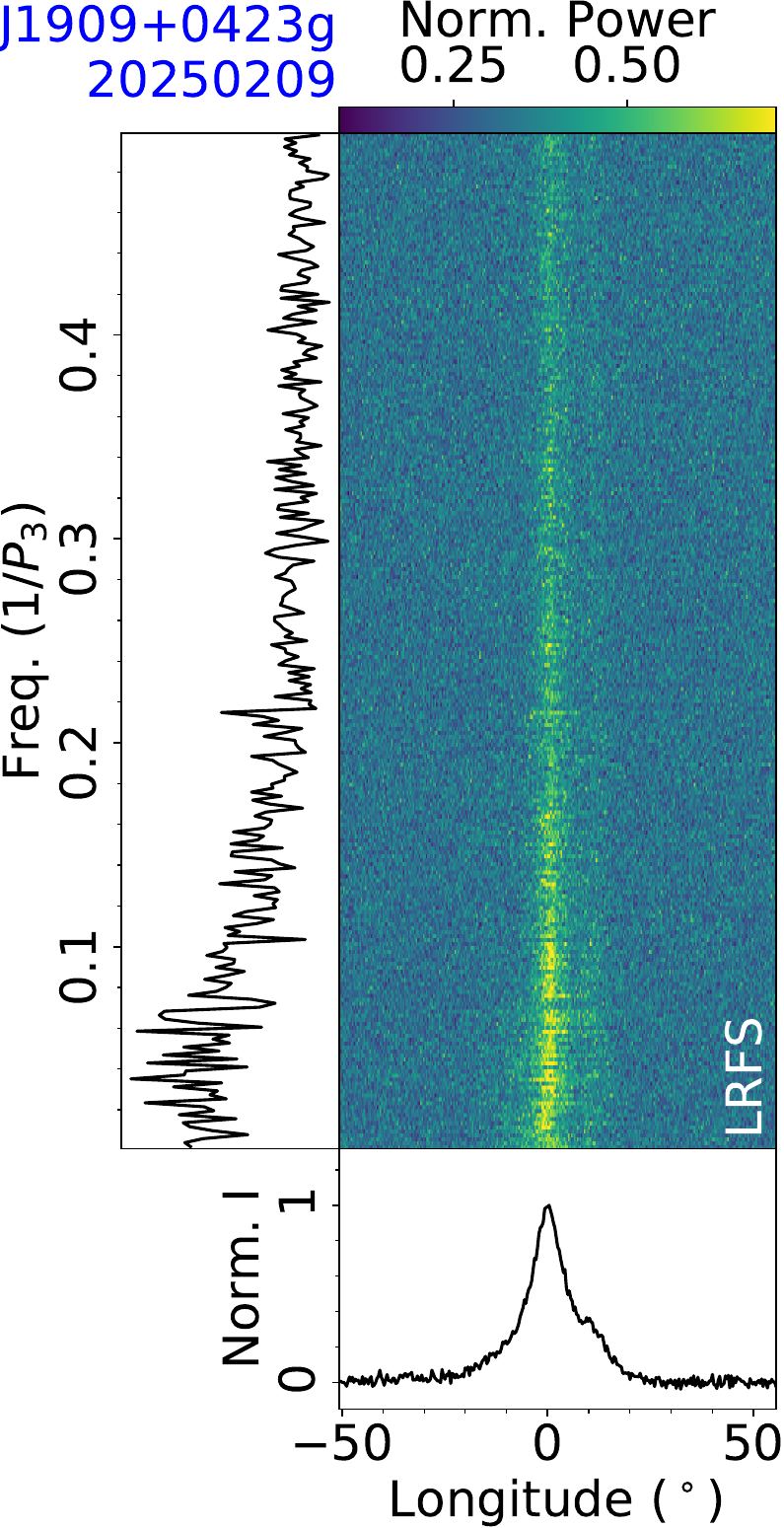}
\includegraphics[width=0.22\textwidth, angle=0]{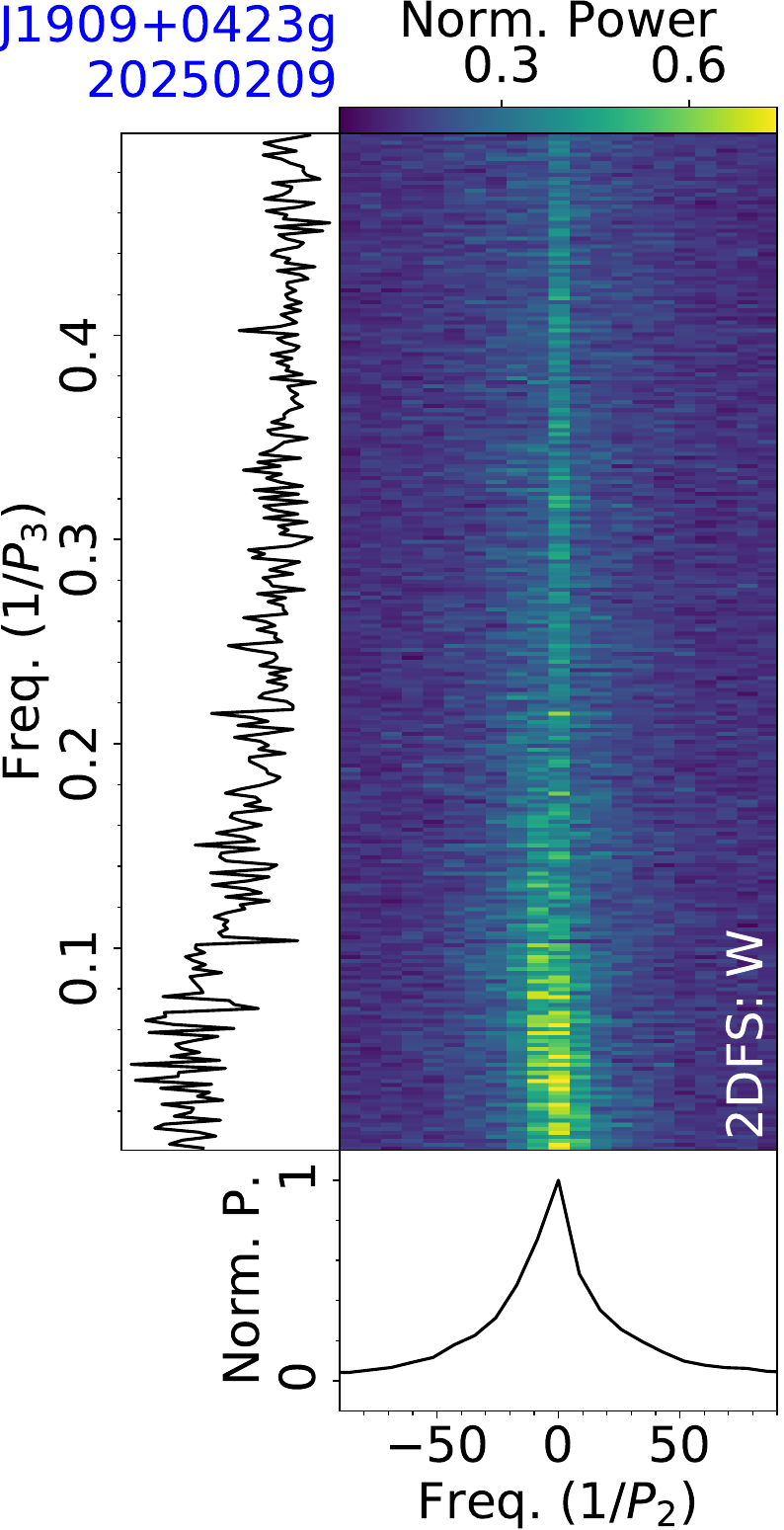}
\figcaption{Fluctuation analysis of PSR J1909+0423g for the observation on 20250209, with LRFS and 2DFS for the on-pulse region of a mean pulse profile.
\label{subfig:fluctu:J1909+0423g}}
\end{figure}

\begin{figure}[htpb]
\centering
\includegraphics[width=0.21\textwidth, angle=0]{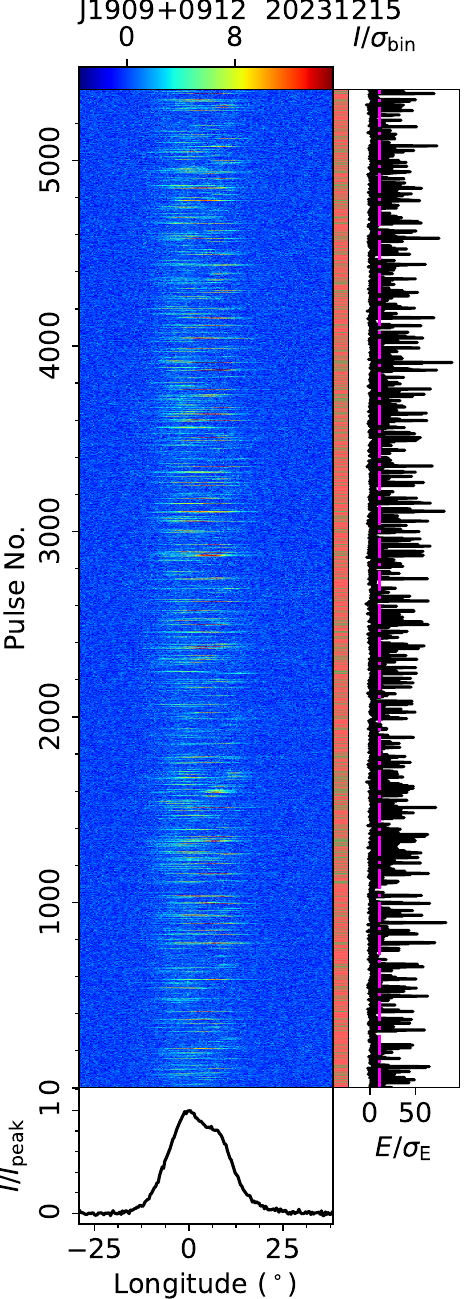}
\includegraphics[width=0.21\textwidth, angle=0]{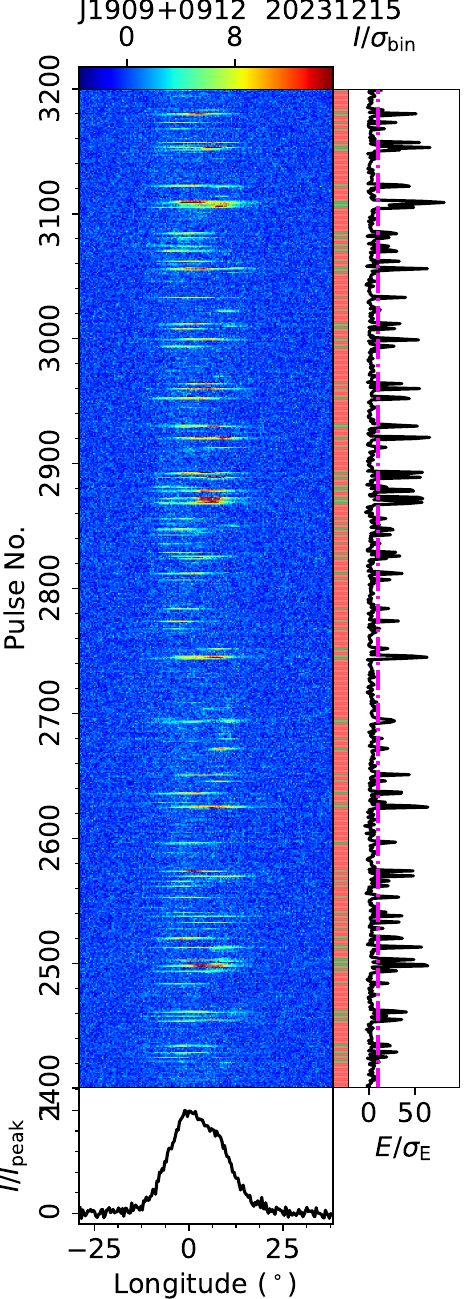}
\figcaption{Single pulse sequence of PSR J1909+0912 from the FAST observation on 20231215, and a zoomed-in view of pulses No. 2400-3200.
\label{subfig:TP:J1909+0912}}
\end{figure}

\begin{figure}[htpb]
\centering
\includegraphics[width=0.39\textwidth, angle=0]{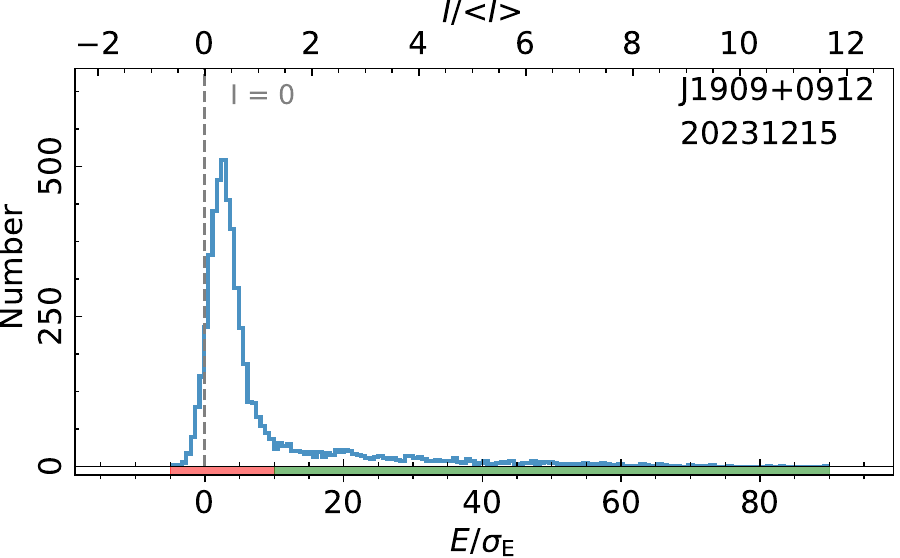}
\figcaption{On-pulse energy histogram of single pulses of PSR J1909+0912 from the FAST observation on 20231215.
\label{subfig:Hist:J1909+0912}}
\end{figure}

\begin{figure}[htpb]
\centering
\includegraphics[width=0.37\textwidth, angle=0]{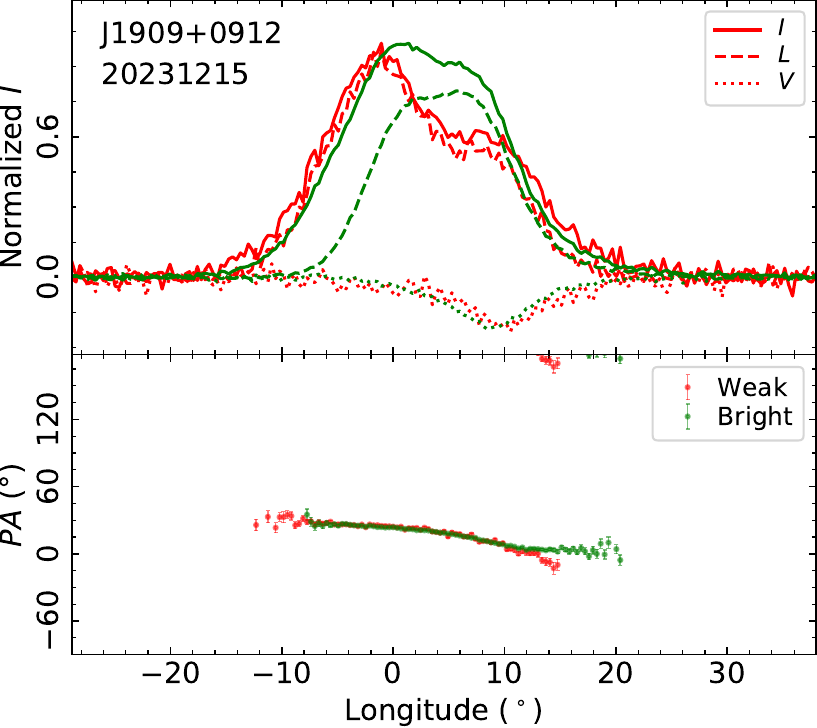}
\figcaption{Mean polarization profiles (the top panel) for the weak (red) and bright (green) emission modes of PSR J1909+0912 from the FAST observation on 20231215, as well as the averaged PA curves (the bottom panel).
\label{subfig:PolModes:J1909+0912}}
\end{figure}


\begin{figure}[htpb]
\centering
\includegraphics[width=0.44\textwidth, angle=0]{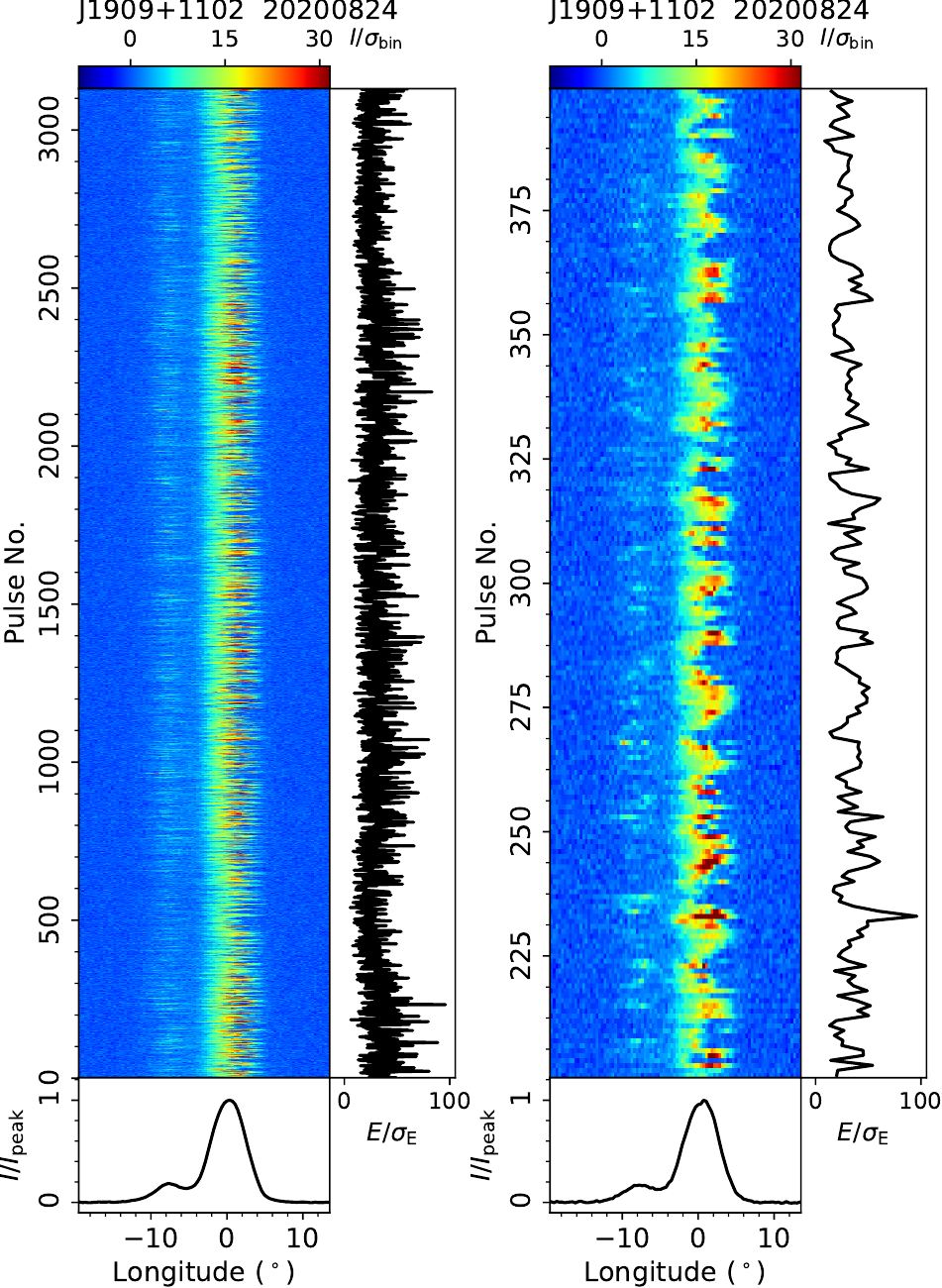}
\figcaption{Single pulse sequence of PSR J1909+1102 from the FAST observation on 20200824, and a zoomed-in view of pulses No. 200-400.
\label{subfig:TP:J1909+1102}}
\end{figure}

\begin{figure}[htpb]
\centering
\includegraphics[width=0.44\textwidth, angle=0]{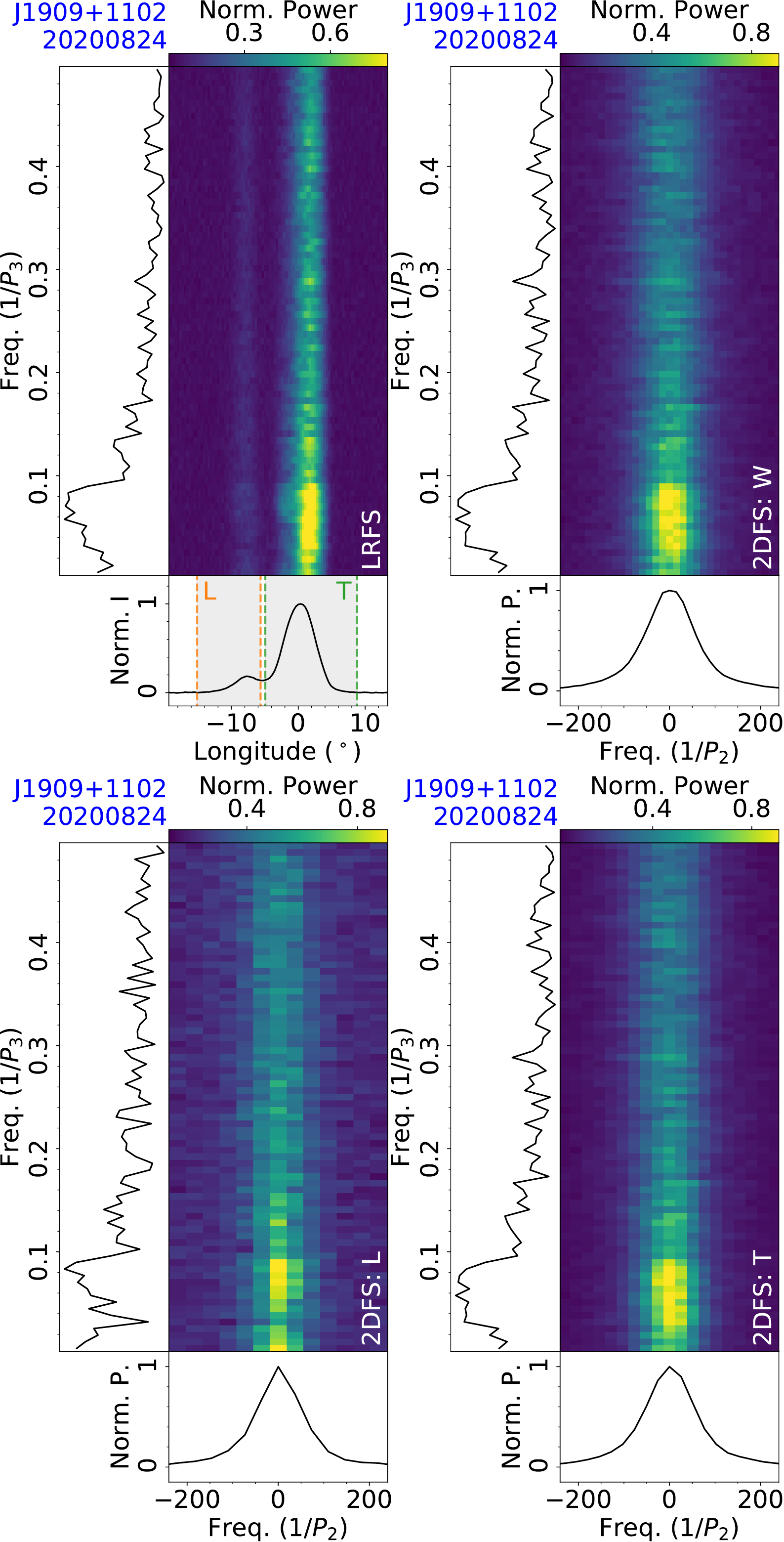}
\figcaption{Fluctuation analysis of PSR J1909+1102 for the observation on 20200824, with LRFS (top-left), and 2DFS for the on-pulse region (top-right), leading part (bottom-left) and trailing part (bottom-right) of a mean pulse profile.
\label{subfig:fluctu:J1909+1102}}
\end{figure}

\subsection{J1909+0423g}
\label{subsec:J1909+0423g}

PSR J1909+0423g was discovered in the FAST GPPS survey \citep{Han2021,han2025}.

This pulsar was observed by FAST on 20250209 for 57 minutes and 20250523 for 78 minutes. From the data of 20250209, a rotation period $P=0.5116$~s and a dispersion measure $D\!M=255.1~{\rm cm^{-3}\,pc}$ were determined. 
Single pulse sequences of the observation on 20250209 in Fig.~\ref{subfig:TP:J1909+0423g} display changes between different states, as well as negative subpulse drifting in the bright emission state. To distinguish the two states and reduce the influence of intensity variations in the bright emission state, the sequence of on-pulse integral energies for single pulses is smoothed using a 17-period moving average. The histogram of smoothed energy sequence is shown in Fig.~\ref{subfig:Hist:J1909+0423g}, where the weak and bright emission modes of single pulses are distinguished, with the averaged profile of two emission states displayed in Fig.~\ref{subfig:ProfModes:J1909+0423g}. The negative drift feature in 2DFS (Fig.~\ref{subfig:fluctu:J1909+0423g}) is temporally widely distributed, with the centroid frequencies of $1/P_3=0.074\pm0.001$ and $1/P_2=-4.3\pm0.2$, corresponding to periodicities of $P_3=13.6\pm0.1$ periods and $P_2=-84\pm4^\circ$. The single pulse behavior of the observation on 20250523 is consistent with those on 20250209.

\begin{figure}[htpb]
\centering
\includegraphics[width=0.22\textwidth, angle=0]{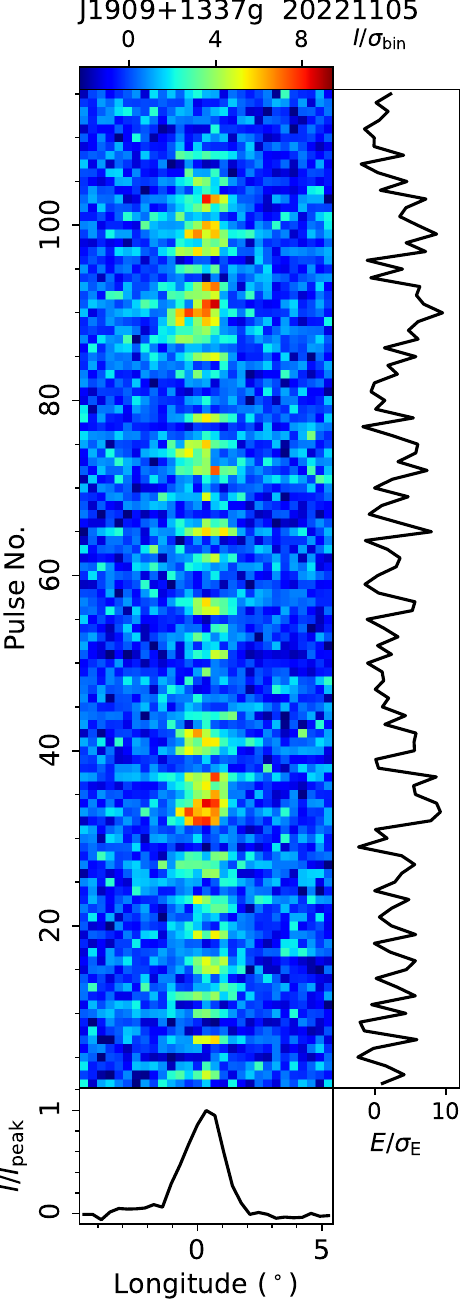}
\includegraphics[width=0.22\textwidth, angle=0]{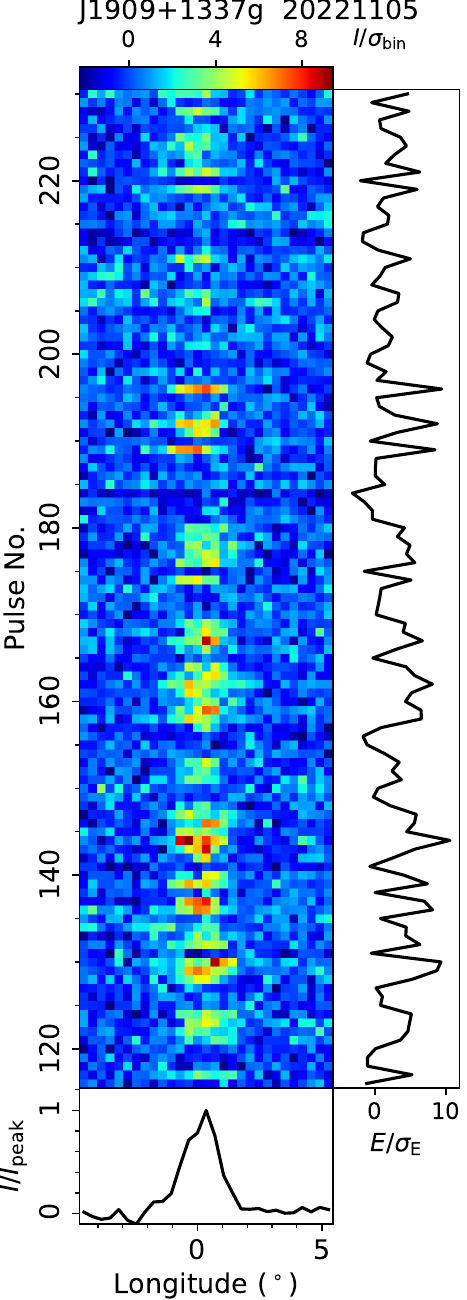}
\figcaption{Single pulse sequences of PSR J1909+1337g from the FAST observation on 20221105.
\label{subfig:TP:J1909+1337g}}
\end{figure}

\begin{figure}[htpb]
\centering
\includegraphics[width=0.39\textwidth, angle=0]{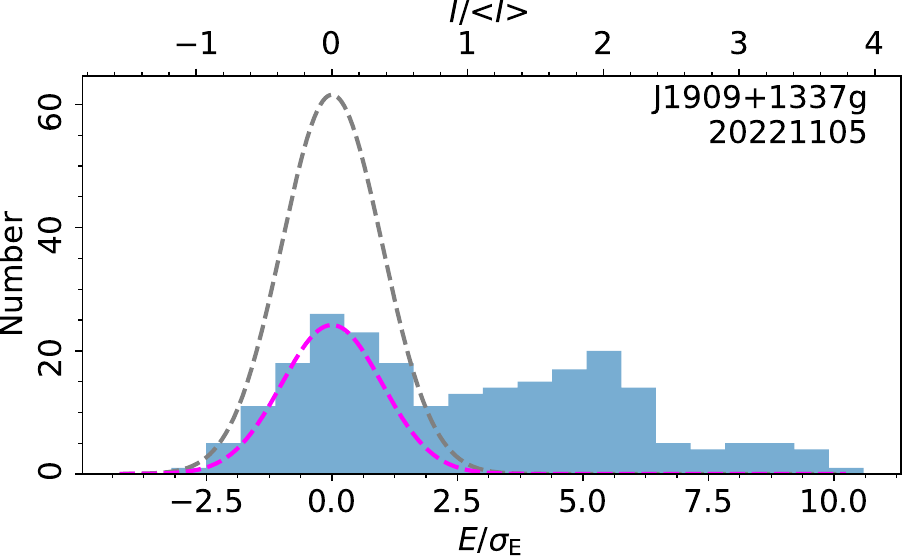}
\figcaption{On-pulse energy histogram of single pulses of PSR J1909+1337g from the FAST observation on 20221105.
\label{subfig:Hist:J1909+1337g}}
\end{figure}

\begin{figure}[htpb]
\centering
\includegraphics[width=0.21\textwidth, angle=0]{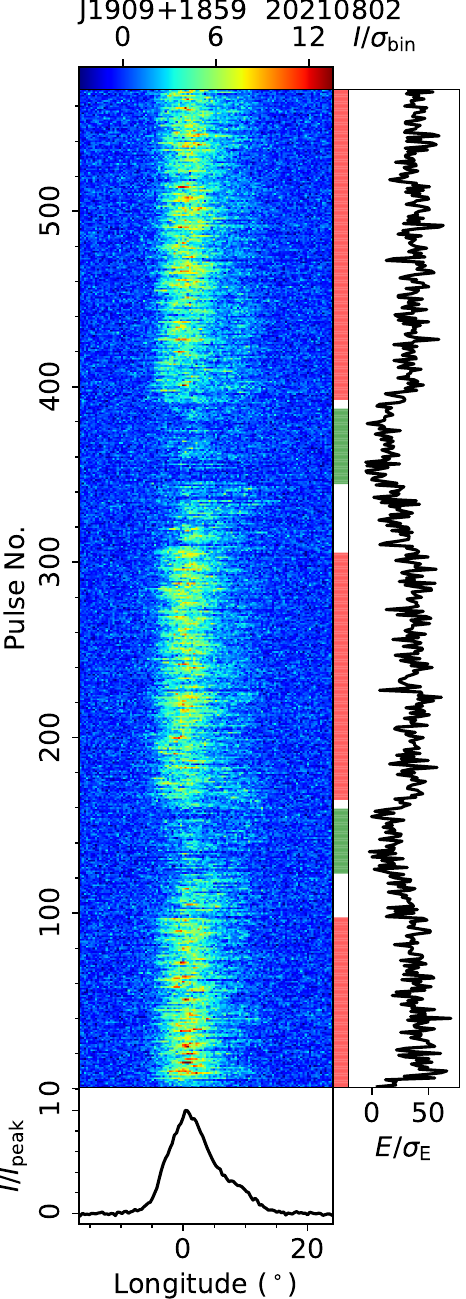}
\figcaption{Single pulse sequence of PSR J1909+1859 from the FAST observation on 20210802.
\label{subfig:TP:J1909+1859}}
\end{figure}

\begin{figure}[htpb]
\centering
\includegraphics[width=0.37\textwidth, angle=0]{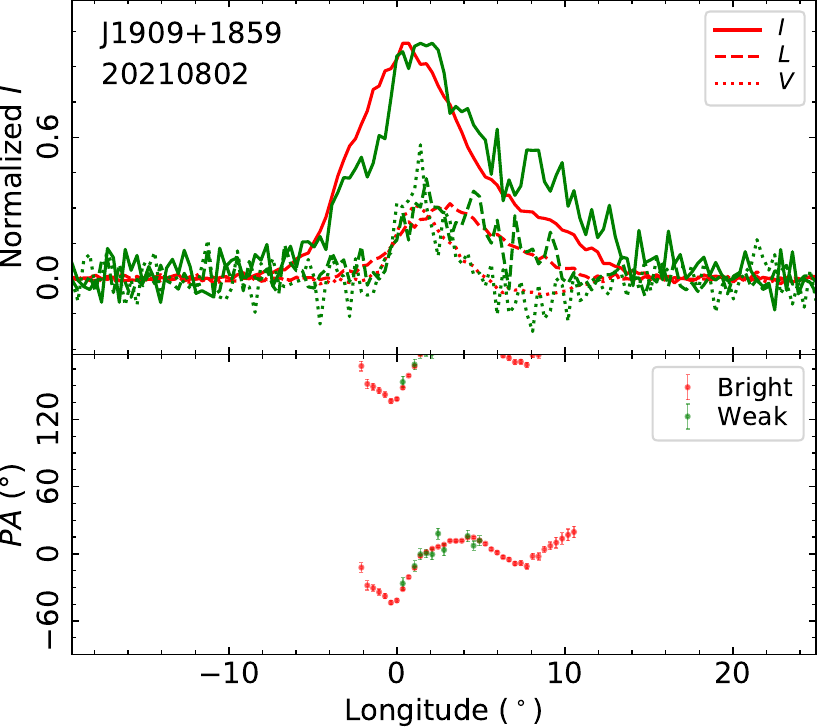}
\figcaption{Mean polarization profiles (the top panel) for the weak (green) and bright (red) emission modes of PSR J1909+1859 from the FAST observation on 20210802, as well as the averaged PA curves (the bottom panel).
\label{subfig:PolModes:J1909+1859}}
\end{figure}

\begin{figure}[htpb]
\centering
\includegraphics[width=0.22\textwidth, angle=0]{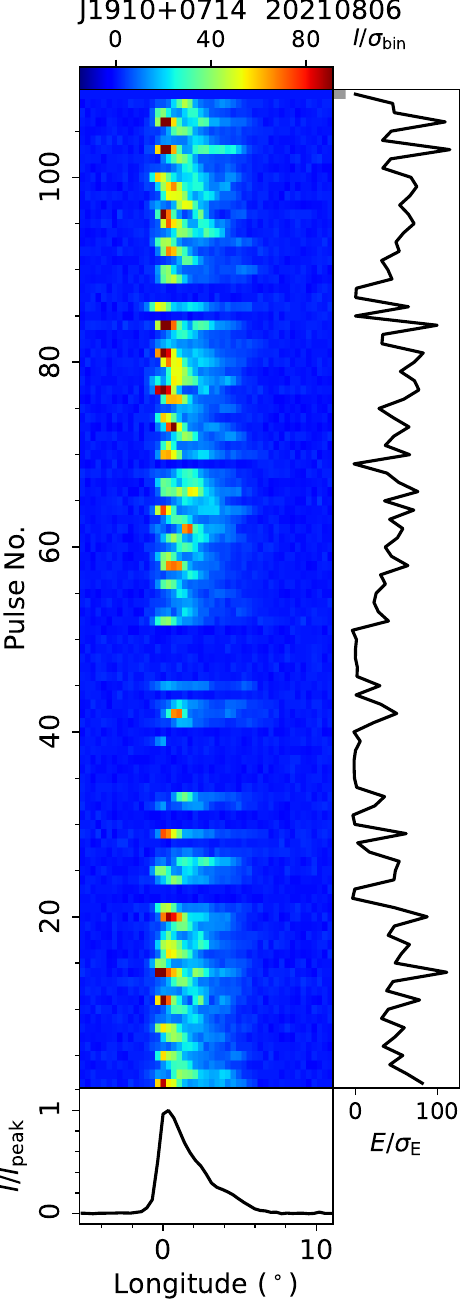}
\figcaption{Single pulse sequence of PSR J1910+0714 from the FAST observation on 20210806.
\label{subfig:TP:J1910+0714}}
\end{figure}

\begin{figure}[htpb]
\centering
\includegraphics[width=0.39\textwidth, angle=0]{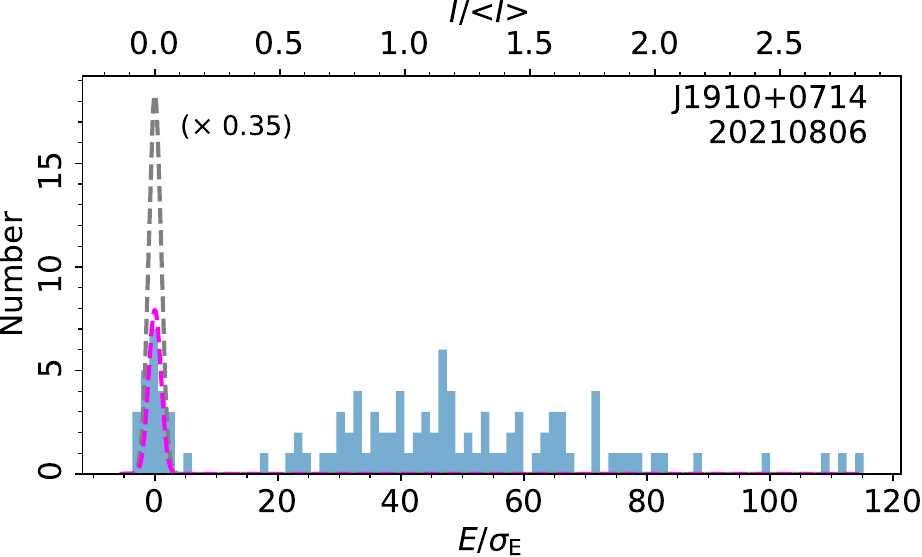}
\figcaption{On-pulse energy histogram of single pulses of PSR J1910+0714 from the FAST observation on 20210806.
\label{subfig:Hist:J1910+0714}}
\end{figure}

\begin{figure}[htpb]
\centering
\includegraphics[width=0.22\textwidth, angle=0]{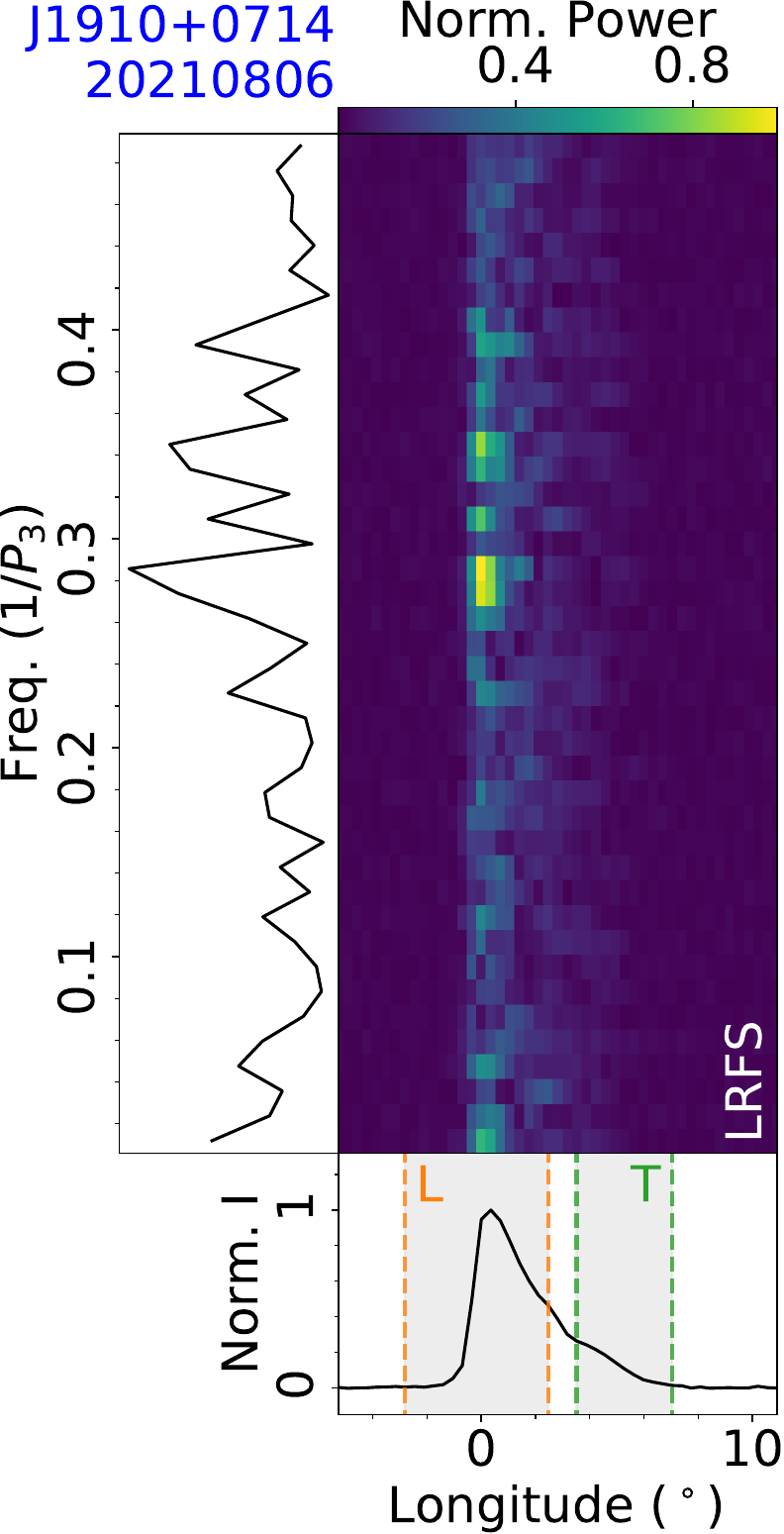}
\includegraphics[width=0.22\textwidth, angle=0]{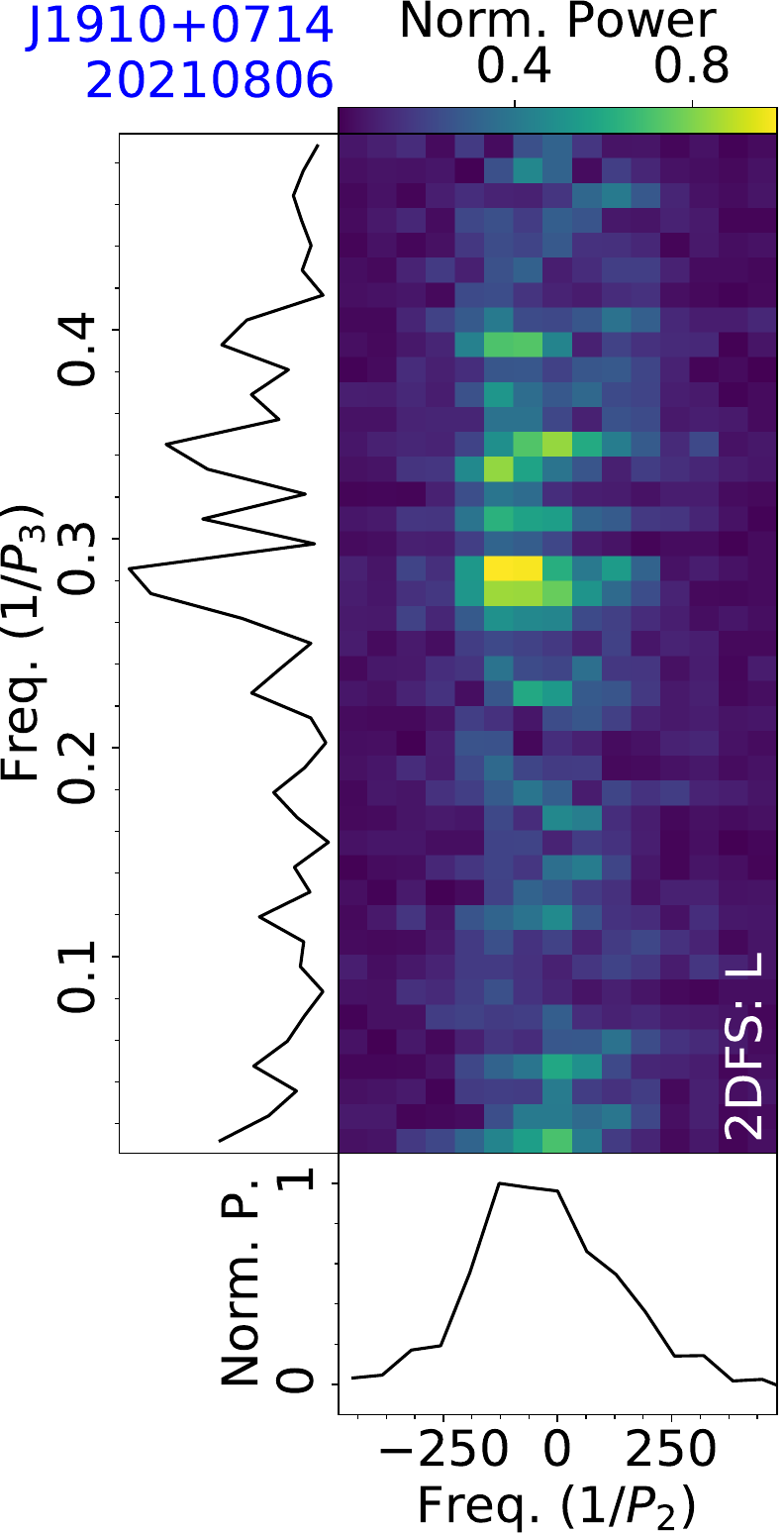}
\figcaption{Fluctuation analysis of PSR J1910+0714 for the observation on 20210806, with LRFS and 2DFS for the leading part of a mean pulse profile.
\label{subfig:fluctu:J1910+0714}}
\end{figure}

\subsection{J1909+0912}
\label{subsec:J1909+0912}

PSR J1909+0912 was discovered in the Parkes Multibeam Pulsar Survey \citep{Morris2002}. 

This pulsar was observed by FAST on 20231215 for 20 minutes, deriving a rotation period $P=0.2230$~s and a dispersion measure $D\!M=423.1~{\rm cm^{-3}\,pc}$. 
Single pulse sequences of this observation in Fig.~\ref{subfig:TP:J1909+0912} display changes between weak and bright emission states. Emission modes of single pulses, labeled in red and blue, are distinguished from the on-pulse energy histogram (Fig.~\ref{subfig:Hist:J1909+0912}). Averaged polarization profiles and PA curves are displayed in Fig.~\ref{subfig:PolModes:J1909+0912}. Relative to the bright mode, the leading component of the weak mode is much stronger than the trailing component. PA curves between two emission modes are slightly different for the longitude later than 10$^\circ$.

\subsection{J1909+1102}
\label{subsec:J1909+1102}

PSR J1909+1102 was discovered by the Mark 1A radio telescope at Jodrell Bank \citep{Davies1973}. The modulation behavior of this pulsar was reported in previous studies \citep{Weltevrede2007, Basu2016}, and for two components \citet{Song2023} presented a positive drift feature ($P_3=12.2\pm0.5$ periods and $P_2=65^{+15}_{-50}$ degrees) and a $P_3$-only feature ($P_3=16\pm2$ periods). 

This pulsar was reported by FAST on 20200824 for 15 minutes, deriving a rotation period $P=0.2837$~s and a dispersion measure $D\!M=150.0~{\rm cm^{-3}\,pc}$. The single-pulse sequence and a zoomed-in view of pulses No. 200-400 in Fig.~\ref{subfig:TP:J1909+1102} show the subpulse modulation feature in both the leading and trailing parts of the mean pulse profile. Fluctuation spectra of this observation in Fig.~\ref{subfig:fluctu:J1909+1102} illustrate that the leading and trailing profile parts both prefer to have the positive drift direction. In 2DFS of the leading profile part, the centroid of the drift feature is at $1/P_3=0.070\pm0.001$ and $1/P_2=7\pm3$, corresponding to periodicities of $P_3=14.2\pm0.2$ periods and $P_2=54\pm23$ degrees. For the trailing part, the centroid of the drift feature in 2DFS is characterized by $1/P_3=0.0586\pm0.0004$ and $1/P_2=2\pm1$, yielding $P_3=17.1\pm0.1$ periods and $P_2=148\pm74$ degrees.

\subsection{J1909+1337g}
\label{subsec:J1909+1337g}

PSR J1909+1337g was discovered in the FAST GPPS survey \citep{Han2021,han2025}.

This pulsar was observed by FAST on 20221105 for 15 minutes and 20250321 for 10 minutes. From the 15-minute data, a rotation period and a dispersion measure are estimated to be $P=3.8651$~s and $D\!M=232.4~{\rm cm^{-3}\,pc}$. 
Single pulse sequences are shown in Figure~\ref{subfig:TP:J1909+1337g}, illustrating the existence of nulls. From the on-pulse integral energy histogram, the nulling fraction of this observation is estimated to be 39.3$\pm$2.2\%.

\begin{figure}[htpb]
\centering
\includegraphics[width=0.44\textwidth, angle=0]{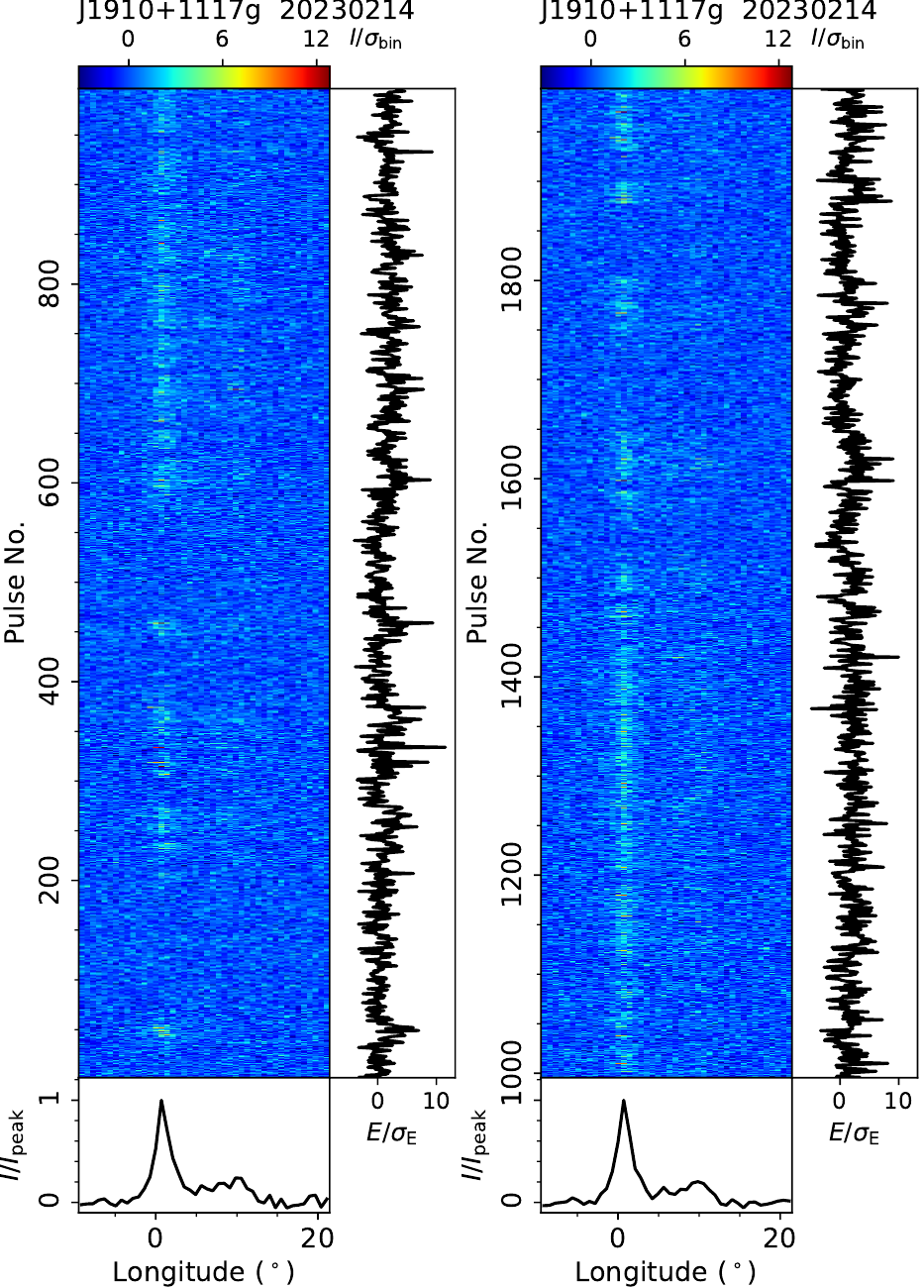}
\figcaption{Single pulse sequences of PSR J1910+1117g from the FAST observation on 20230214.
\label{subfig:TP:J1910+1117g}}
\end{figure}

\begin{figure}[htpb]
\centering
\includegraphics[width=0.39\textwidth, angle=0]{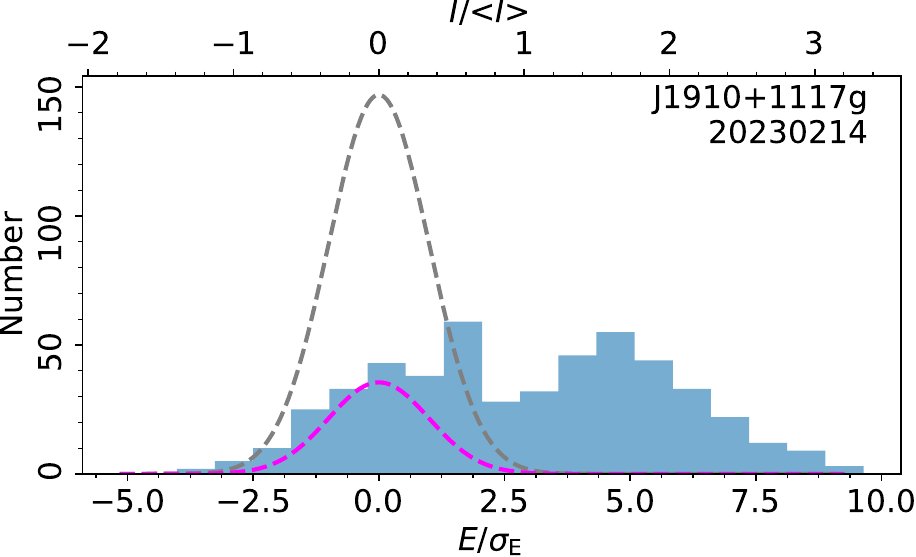}
\figcaption{On-pulse energy histogram of single pulses of PSR J1910+1117g from the FAST observation on 20230214, with energy values averaged for every 4 periods.
\label{subfig:Hist:J1910+1117g}}
\end{figure}

\subsection{J1909+1859}
\label{subsec:J1909+1859}

PSR J1909+1859 was discovered by \citet{Nice1995} using the Arecibo telescope. 

This pulsar was observed by FAST on 20210802 for 5 minutes, deriving a rotation period $P=0.5425$~s and a dispersion measure $D\!M=64.5~{\rm cm^{-3}\,pc}$. 
The single pulse sequence in Fig.~\ref{subfig:TP:J1909+1859} illustrates the existence of the bright and weak emission modes, which are labeled using the red and blue bars. The pulsar switches from the bright mode to the weak mode, with the intensity decreased gradually and the emission phase delayed, and then changes to the bright mode. 
The averaged polarization profiles and the PA curves of the two emission modes are displayed in Fig.~\ref{subfig:PolModes:J1909+1859}.




\begin{figure}[htpb]
\centering
\includegraphics[width=0.22\textwidth, angle=0]{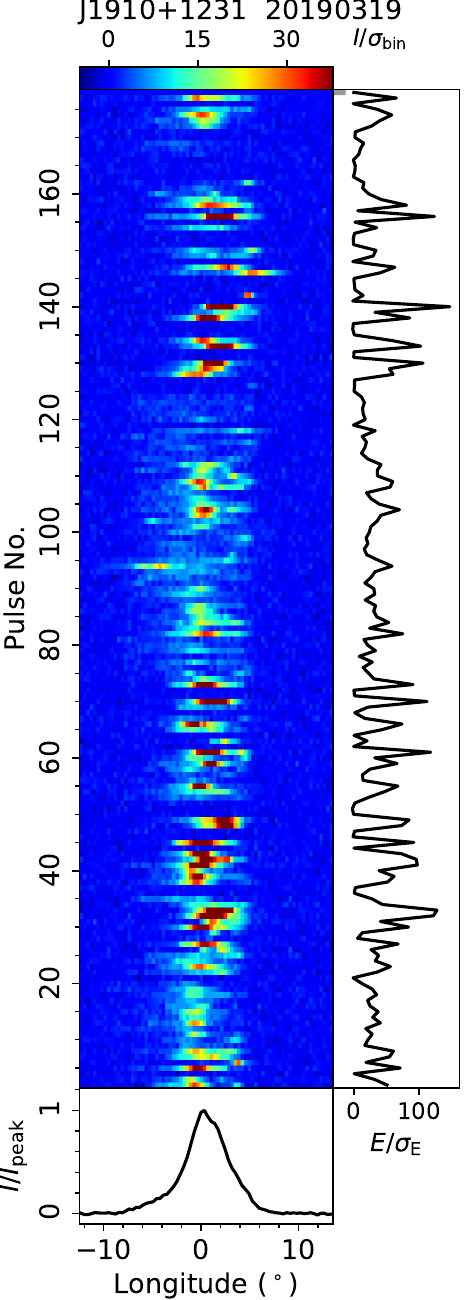}
\figcaption{Single pulse sequence of PSR J1910+1231 from the FAST observation on 20190319.
\label{subfig:TP:J1910+1231}}
\end{figure}

\begin{figure}[htpb]
\centering
\includegraphics[width=0.39\textwidth, angle=0]{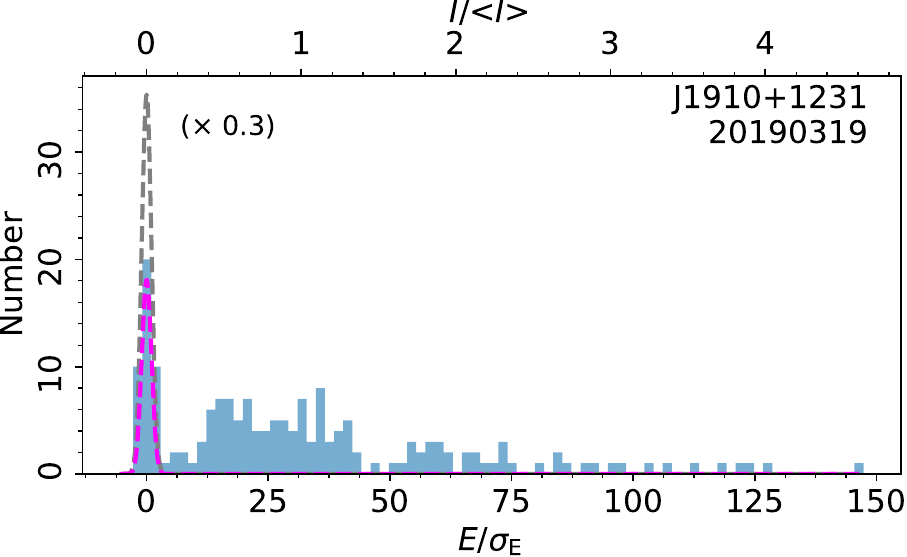}
\figcaption{On-pulse energy histogram of single pulses of PSR J1910+1231 from the FAST observation on 20190319.
\label{subfig:Hist:J1910+1231}}
\end{figure}

\begin{figure}[htpb]
\centering
\includegraphics[width=0.22\textwidth, angle=0]{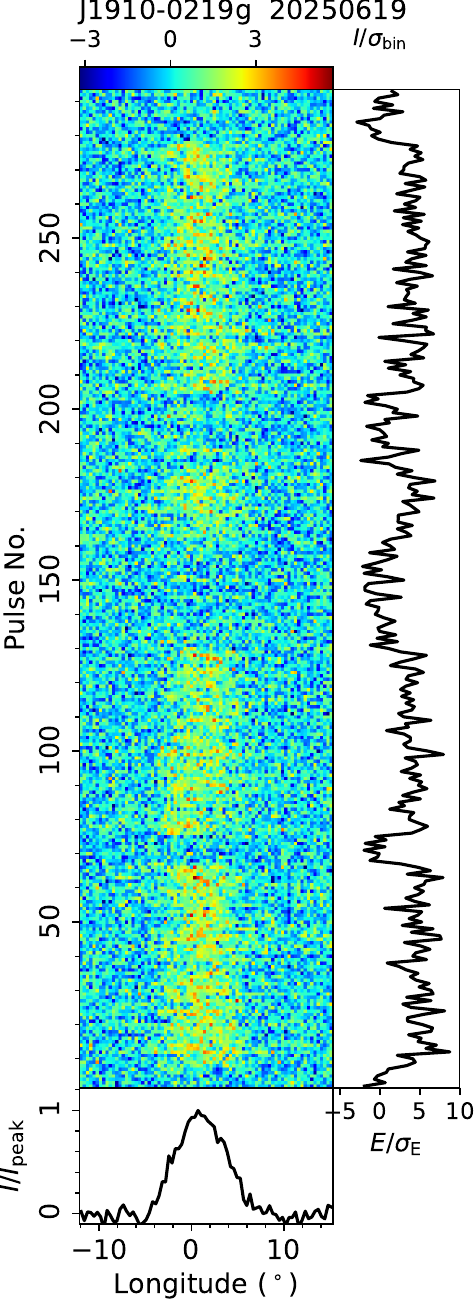}
\includegraphics[width=0.22\textwidth, angle=0]{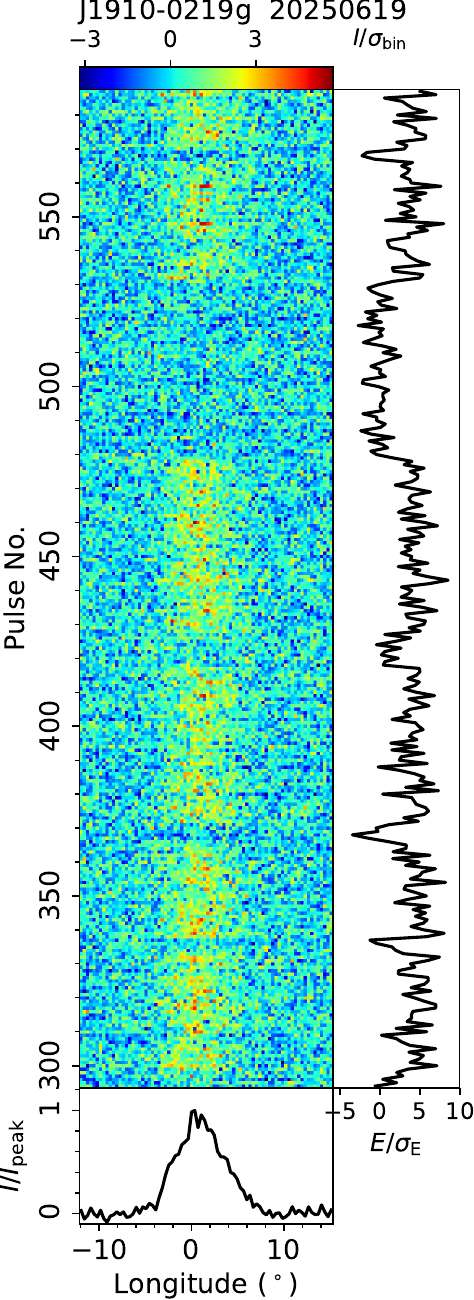}
\figcaption{Single pulse sequences of PSR J1910-0219g from the FAST observation on 20250619.
\label{subfig:TP:J1910-0219g}}
\end{figure}

\begin{figure}[htpb]
\centering
\includegraphics[width=0.39\textwidth, angle=0]{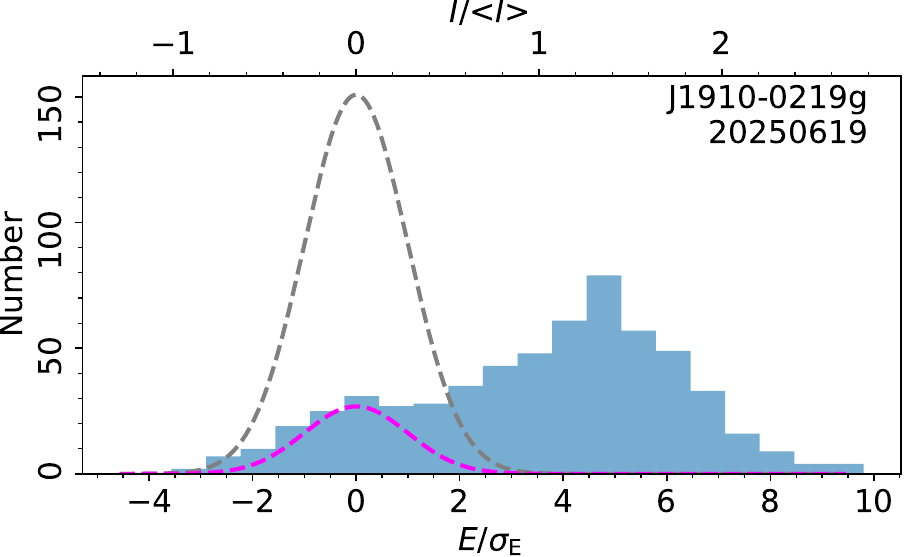}
\figcaption{On-pulse energy histogram of single pulses of PSR J1910-0219g from the FAST observation on 20250619.
\label{subfig:Hist:J1910-0219g}}
\end{figure}

\begin{figure}[htpb]
\centering
\includegraphics[width=0.22\textwidth, angle=0]{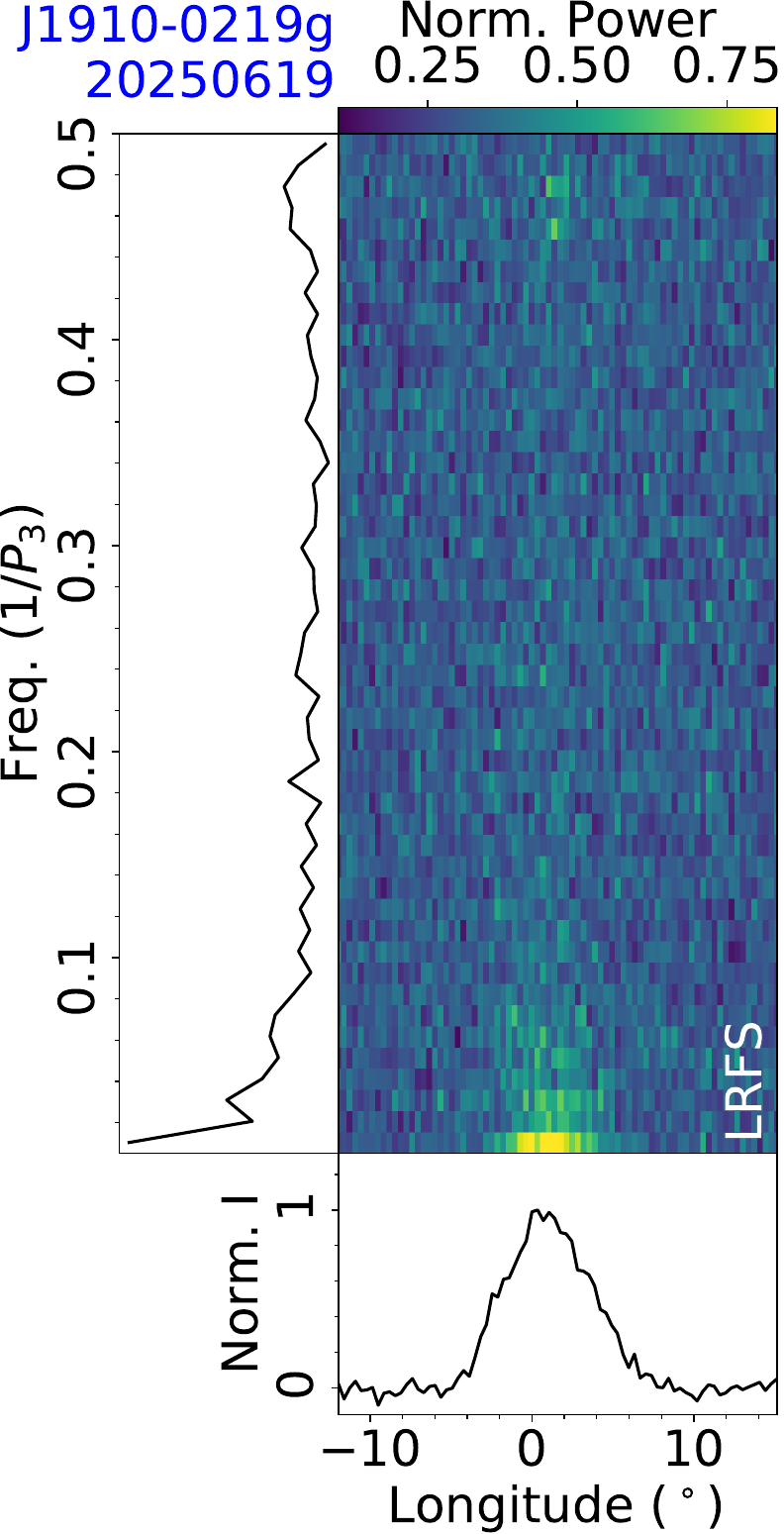}
\includegraphics[width=0.22\textwidth, angle=0]{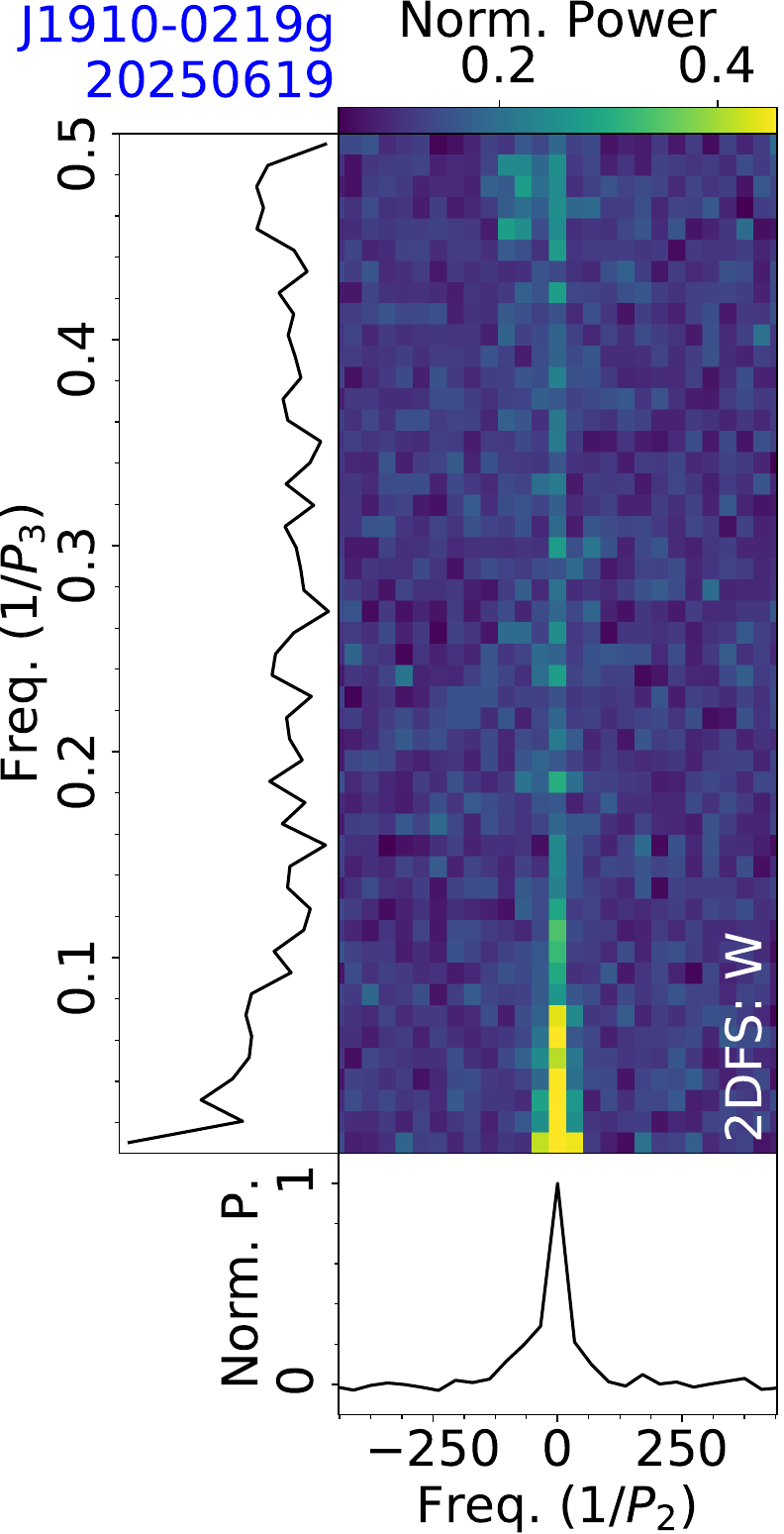}
\figcaption{Fluctuation analysis of PSR J1910-0219g for the observation on 20250619, with LRFS and 2DFS for the on-pulse region of a mean pulse profile.
\label{subfig:fluctu:J1910-0219g}}
\end{figure}

\subsection{J1910+0714}
\label{subsec:J1910+0714}

PSR J1910+0714 was discovered using the 305 m radio telescope at Arecibo \citep{Nice1995}. Drifting behavior was reported by \citet{Song2023}. 

The FAST observed this pulsar on 20210806 for 5 minutes, deriving a rotation period $P=2.7122$~s and a dispersion measure $D\!M=124.5~{\rm cm^{-3}\,pc}$. 
The single pulse sequence of this observation is shown in Fig.~\ref{subfig:TP:J1910+0714}, where subpulses related to the leading part in the profile have clear drifting behavior, while the trailing part does not. 2DFS in Fig.~\ref{subfig:fluctu:J1910+0714} of the leading part of a mean pulse profile exhibits a negative drift feature with the centroid of $1/P_3=0.321\pm0.002$ and $1/P_2=-78\pm6$, corresponding to drifting parameters of $P_3=3.12\pm0.02$ periods and $P_2=-4.6\pm0.3^\circ$. There is also nulling behavior, with the nulling fraction of this observation estimated to be 15$\pm$3\% from the on-pulse energy histogram (Fig.~\ref{subfig:Hist:J1910+0714}). From the single pulse sequence, both the duration of nulling and emission could be one period.

\begin{figure}[htpb]
\centering
\includegraphics[width=0.22\textwidth, angle=0]{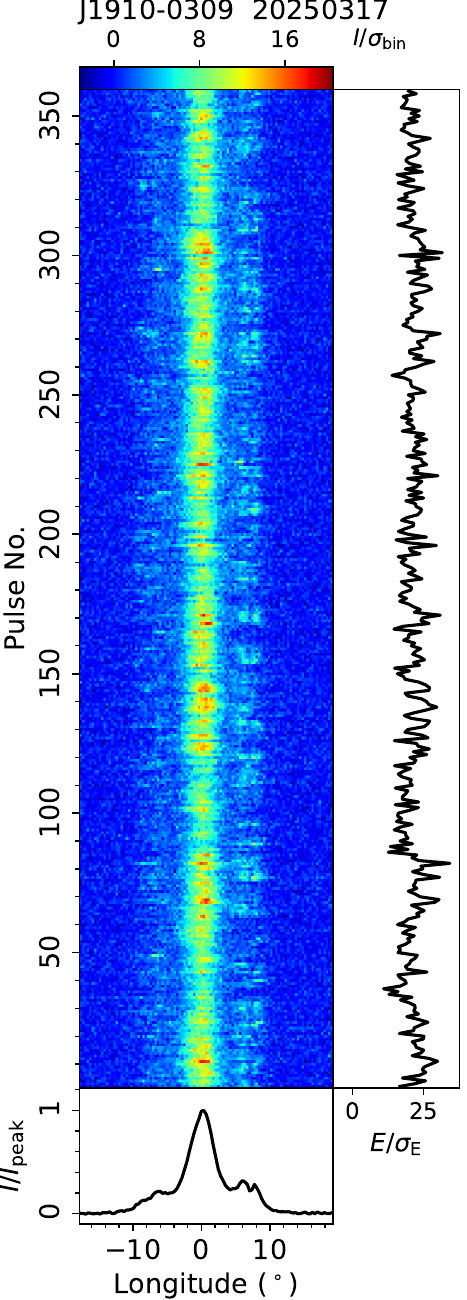}
\includegraphics[width=0.22\textwidth, angle=0]{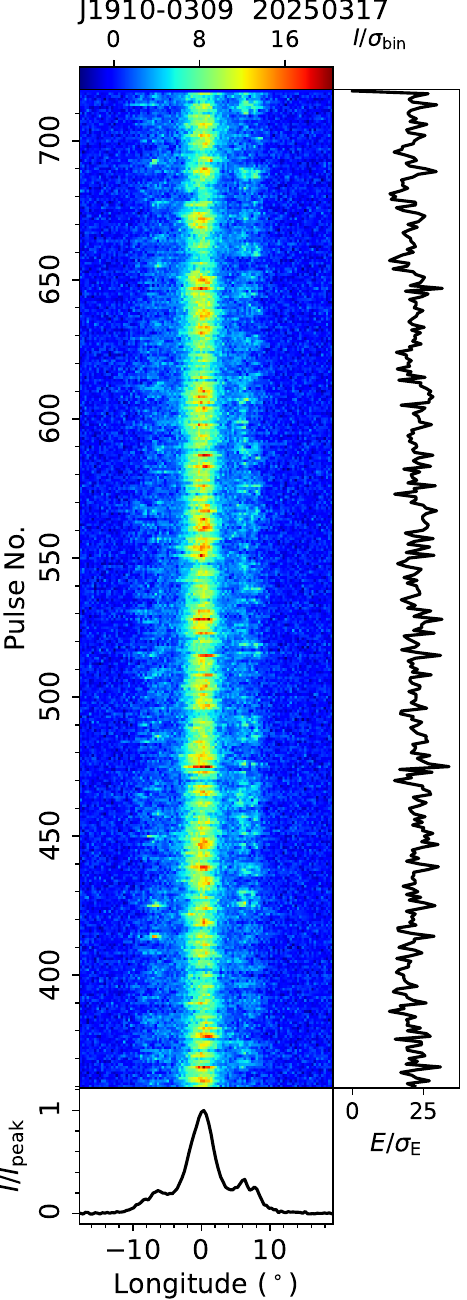}
\figcaption{Single pulse sequences of PSR J1910-0309 from the FAST observation on 20250317.
\label{subfig:TP:J1910-0309}}
\end{figure}

\begin{figure}[htpb]
\centering
\includegraphics[width=0.22\textwidth, angle=0]{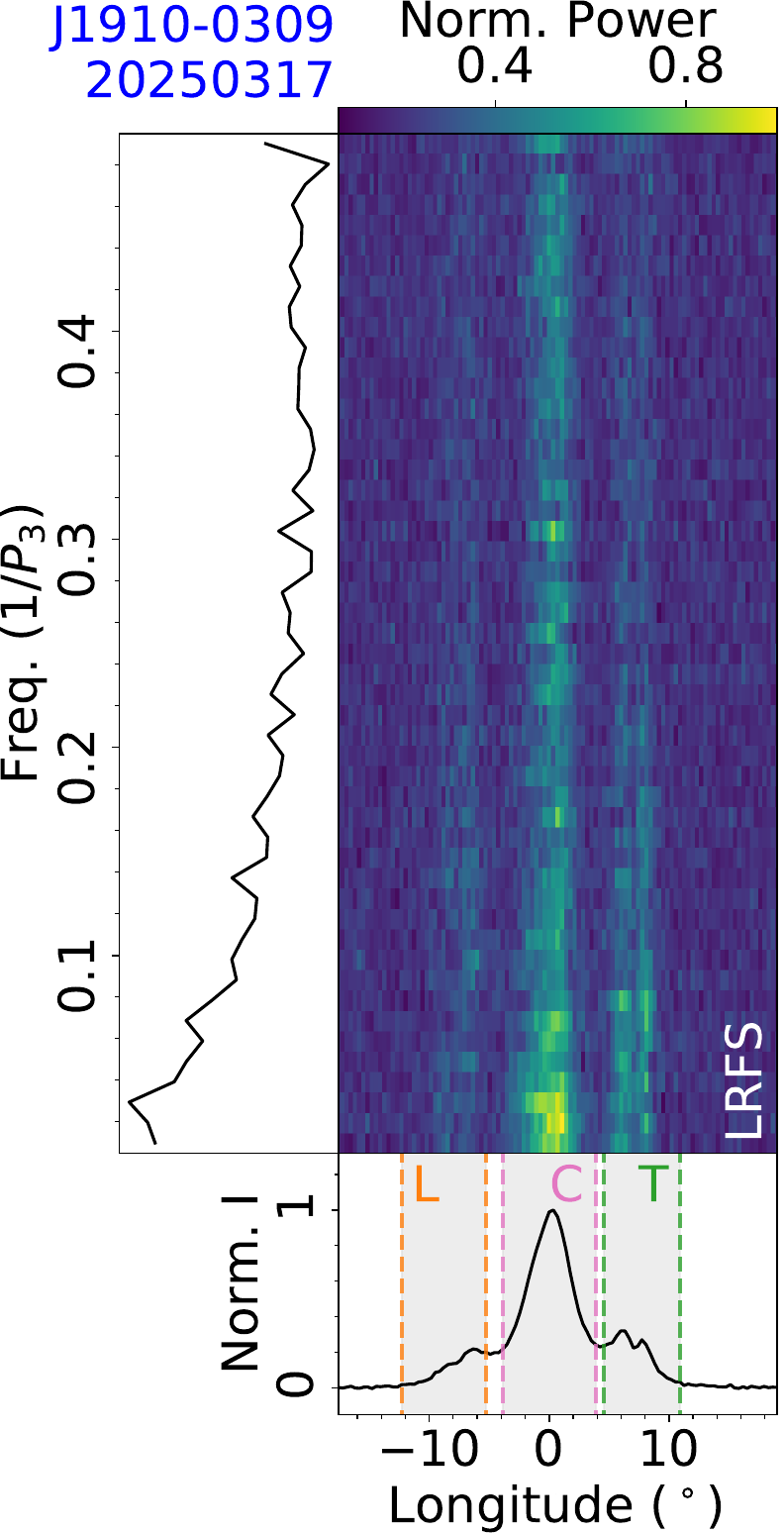}
\includegraphics[width=0.22\textwidth, angle=0]{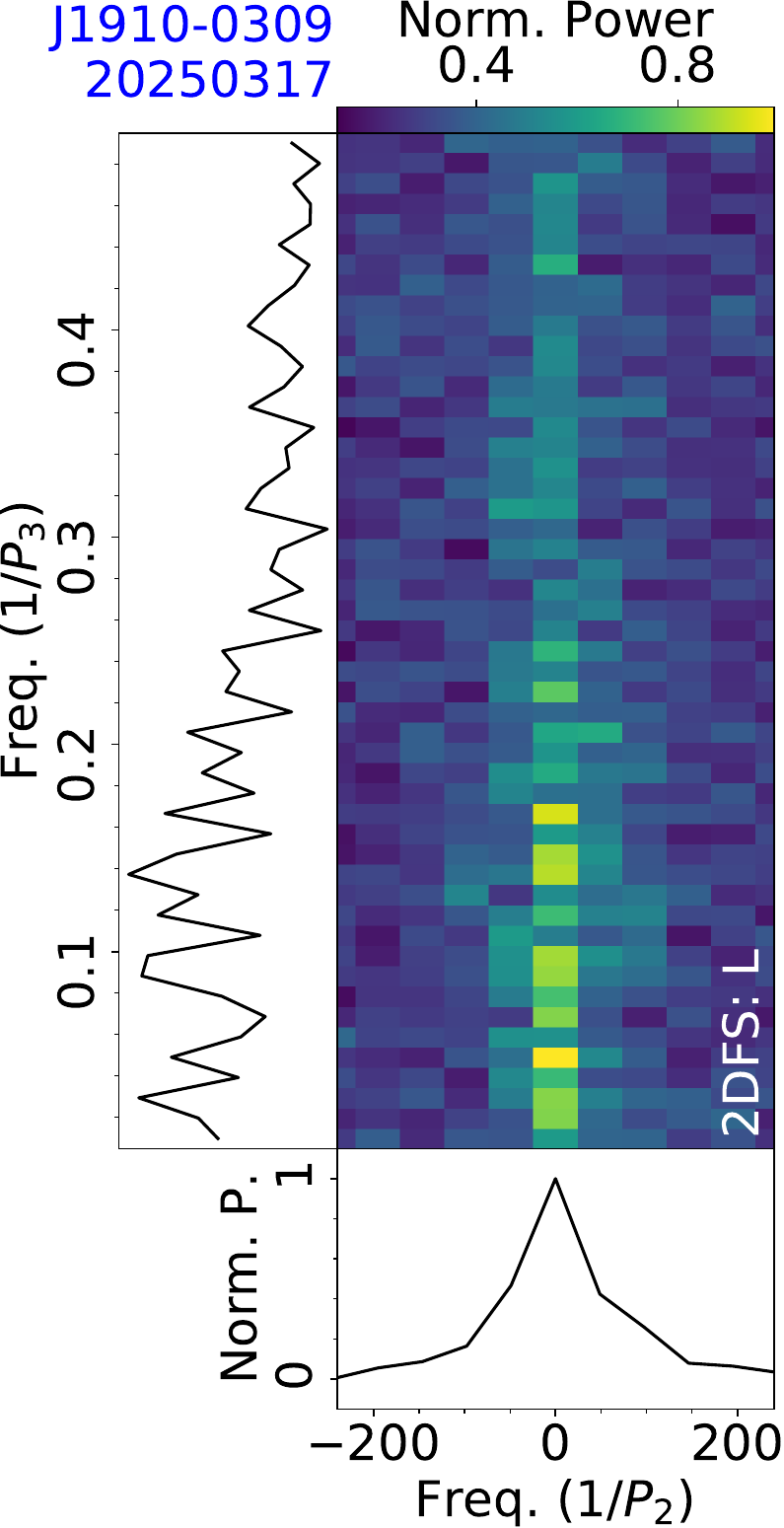}\\
\includegraphics[width=0.22\textwidth, angle=0]{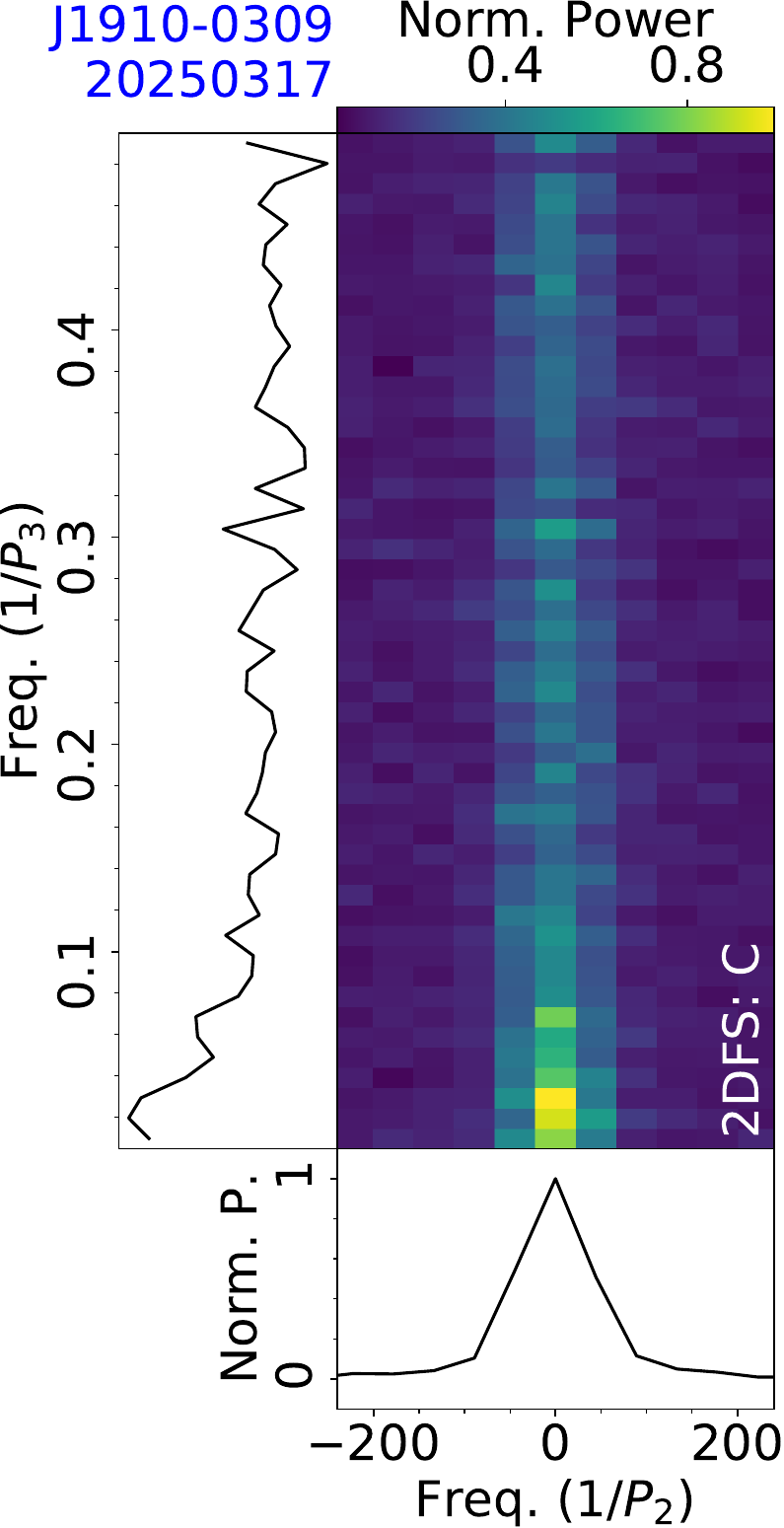}
\includegraphics[width=0.22\textwidth, angle=0]{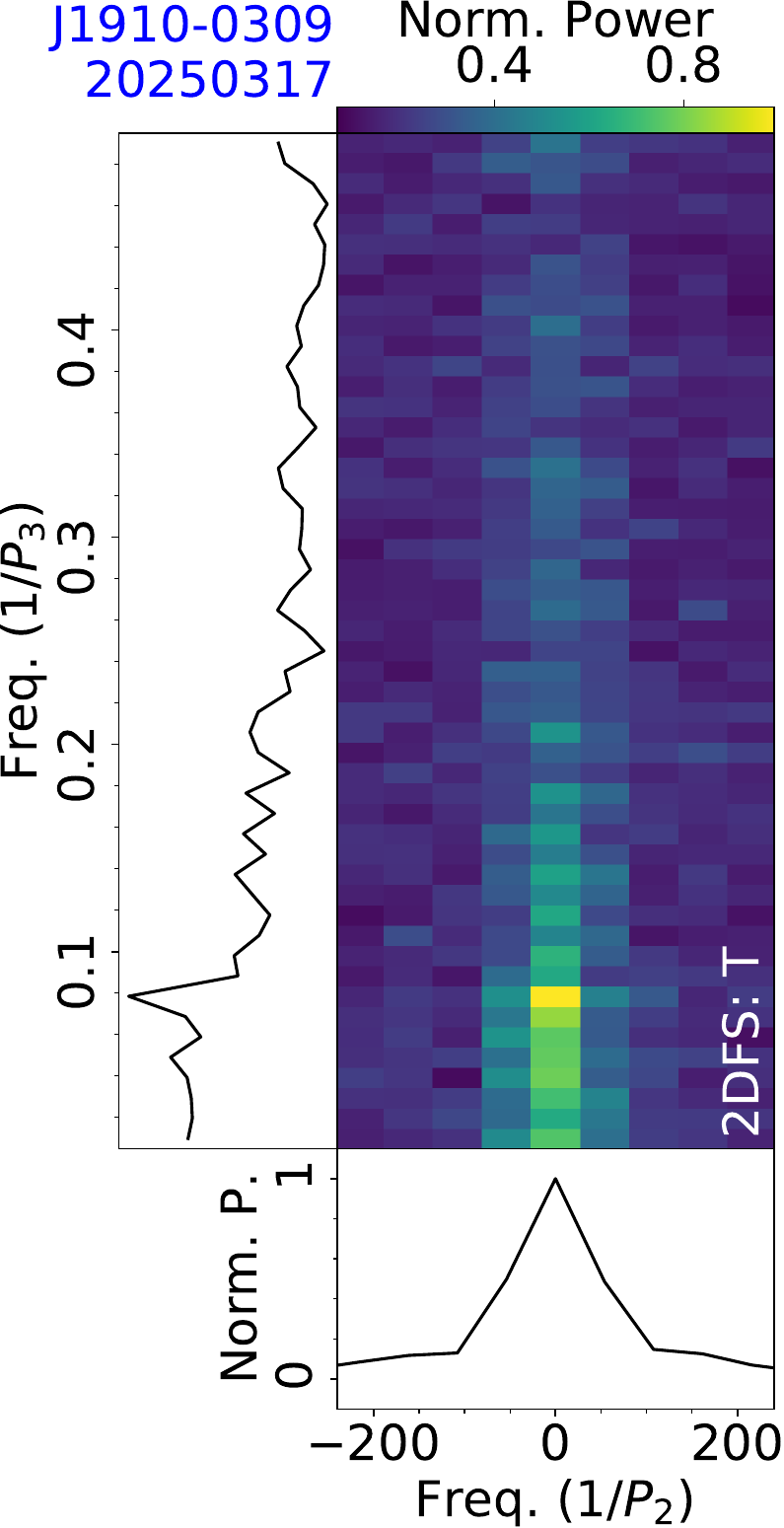}
\figcaption{Fluctuation analysis of PSR J1910-0309 for the observation on 20250317, with LRFS (top-left), and 2DFS for the leading (top-right), central (bottom-left) and trailing (bottom-right) parts of a mean pulse profile.
\label{subfig:fluctu:J1910-0309}}
\end{figure}

\begin{figure}[htpb]
\centering
\includegraphics[width=0.22\textwidth, angle=0]{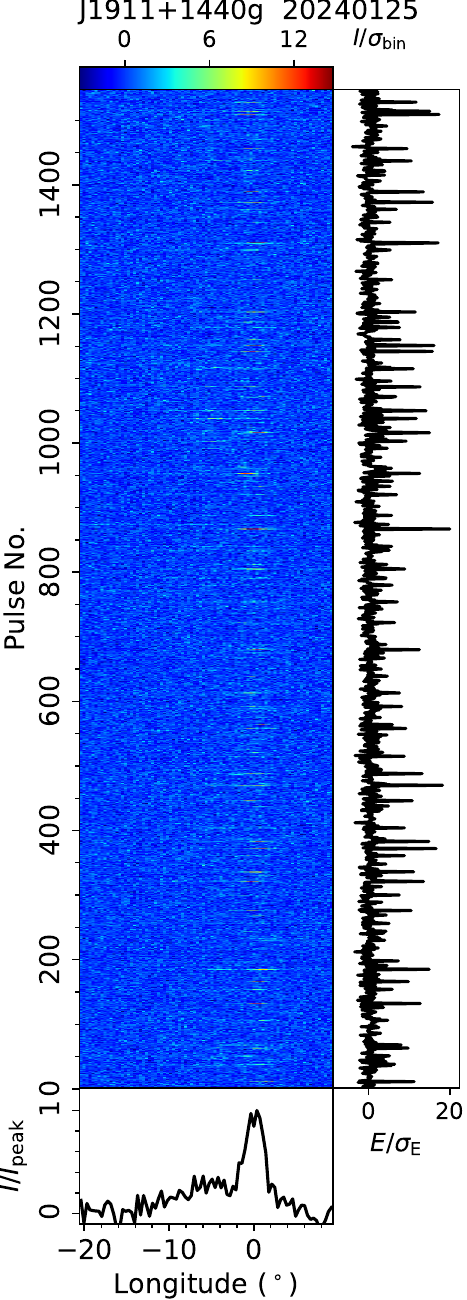}
\includegraphics[width=0.22\textwidth, angle=0]{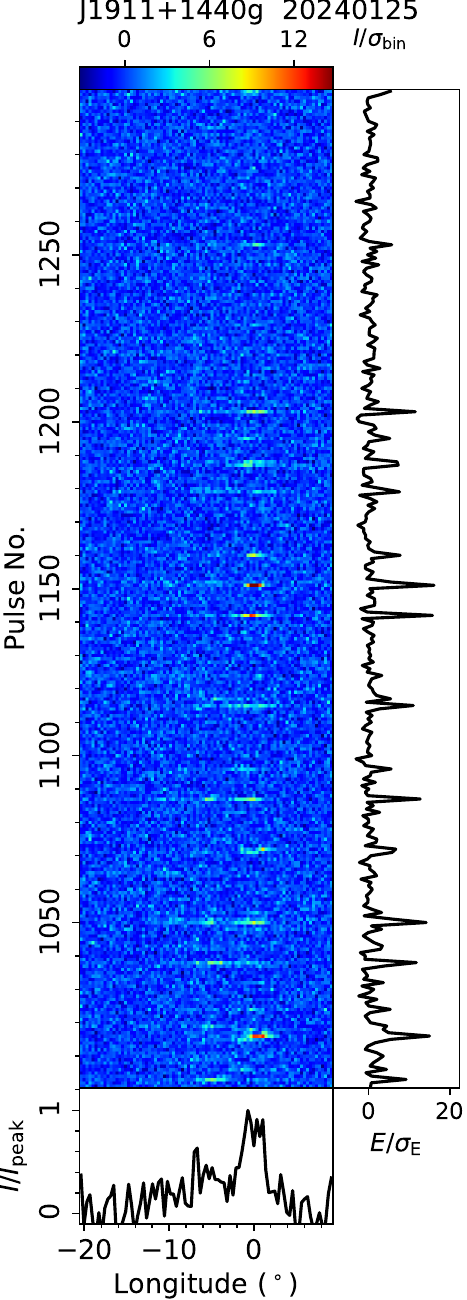}
\figcaption{Single pulse sequence of PSR J1911+1440g from the FAST observation on 20240125, and a zoomed-in view of pulses No.1001-1299.
\label{subfig:TP:J1911+1440g}}
\end{figure}

\begin{figure}[htpb]
\centering
\includegraphics[width=0.39\textwidth, angle=0]{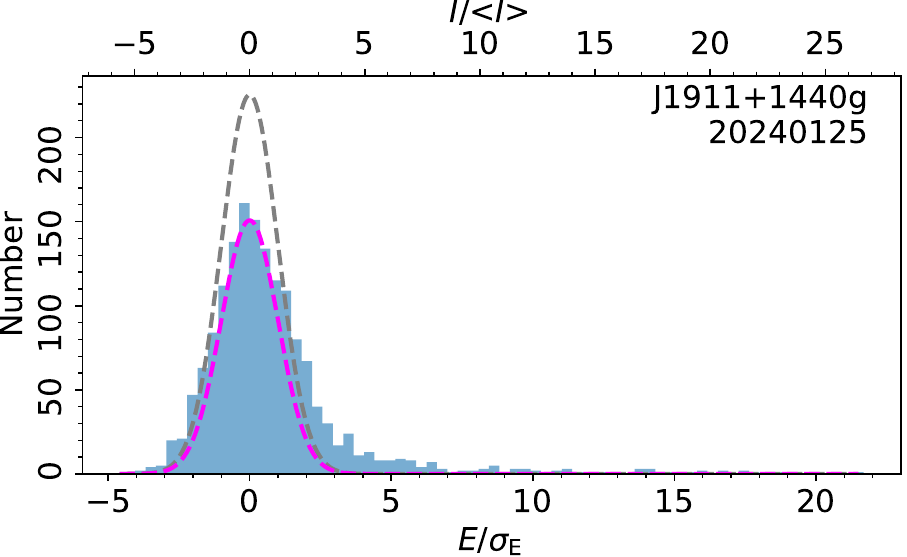}
\figcaption{On-pulse energy histogram of single pulses of PSR J1911+1440g from the FAST observation on 20240125.
\label{subfig:Hist:J1911+1440g}}
\end{figure}

\begin{figure}[htpb]
\centering
\includegraphics[width=0.42\textwidth, angle=0]{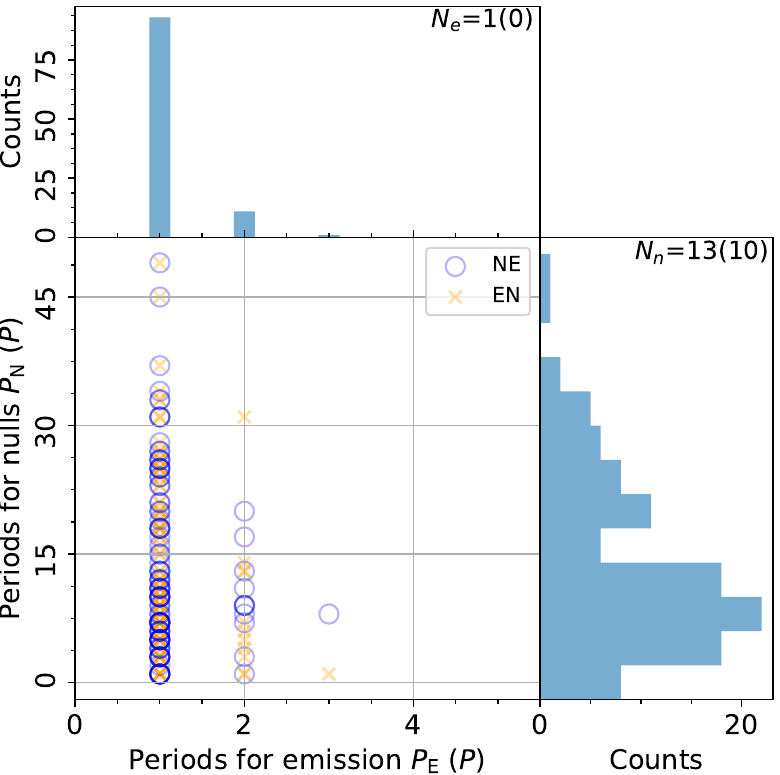}
\figcaption{Distribution of period numbers for continuous nulling $P_N$ against period numbers for adjacent pulses $P_E$ of PSR J1911+1440g observed by FAST on 20240125, as well as the duration histograms for the emission and null shown in the top and right panels, respectively. 
\label{subfig:scaleHist:J1911+1440g}}
\end{figure}

\begin{figure}[htpb]
\centering
\includegraphics[width=0.39\textwidth, angle=0]{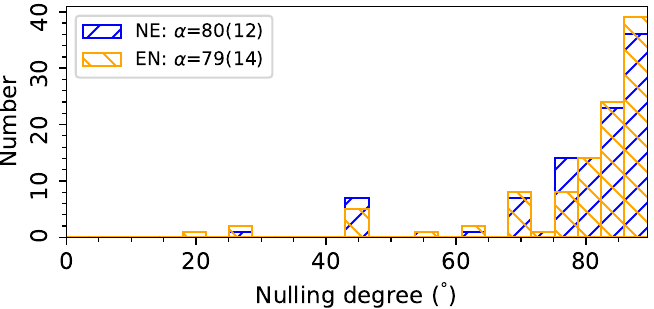}\\
\includegraphics[width=0.39\textwidth, angle=0]{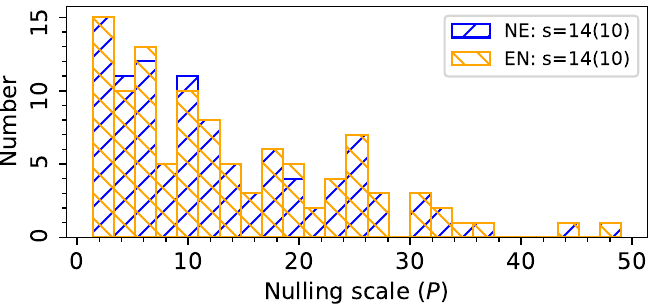}
\figcaption{Histograms of the nulling degree and nulling scale for PSR J1911+1440g observed by FAST on 20240125.
\label{subfig:nullDegreeScale:J1911+1440g}}
\end{figure}

\begin{figure}[htpb]
\centering
\includegraphics[width=0.22\textwidth, angle=0]{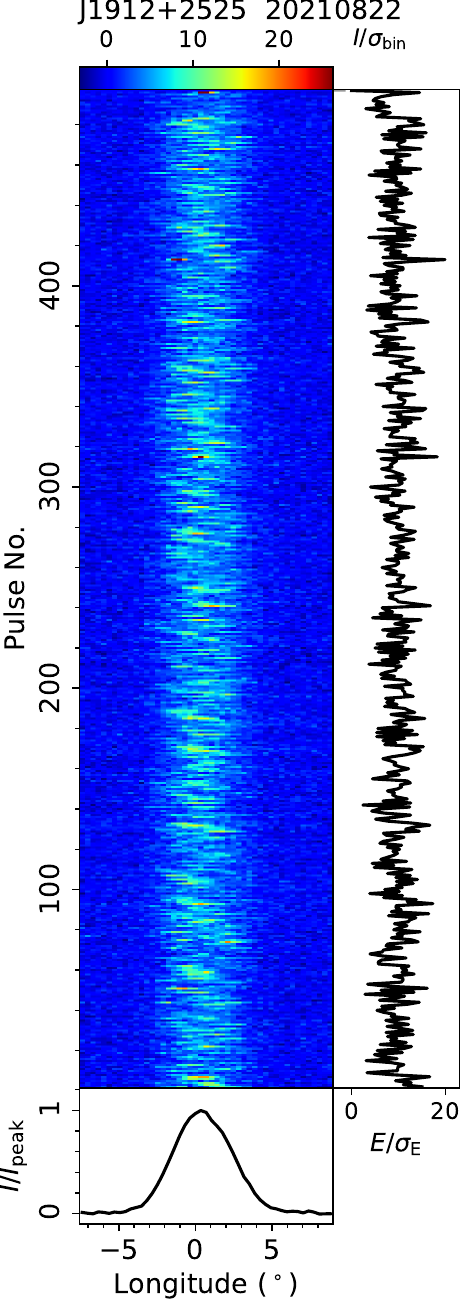}
\includegraphics[width=0.22\textwidth, angle=0]{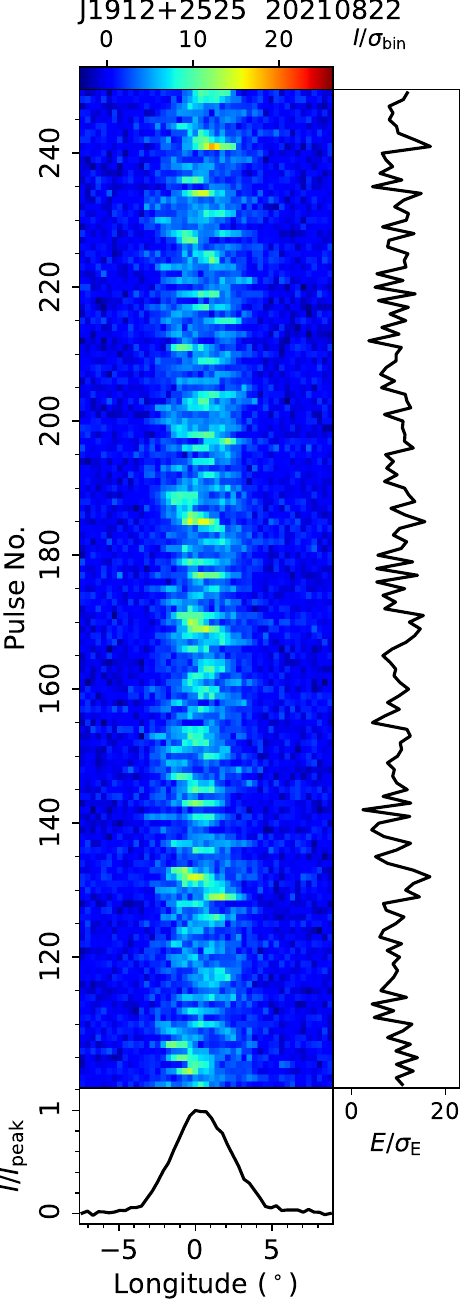}
\figcaption{Single pulse sequences of PSR J1912+2525 from the FAST observation on 20210822.
\label{subfig:TP:J1912+2525}}
\end{figure}

\begin{figure}[htpb]
\centering
\includegraphics[width=0.22\textwidth, angle=0]{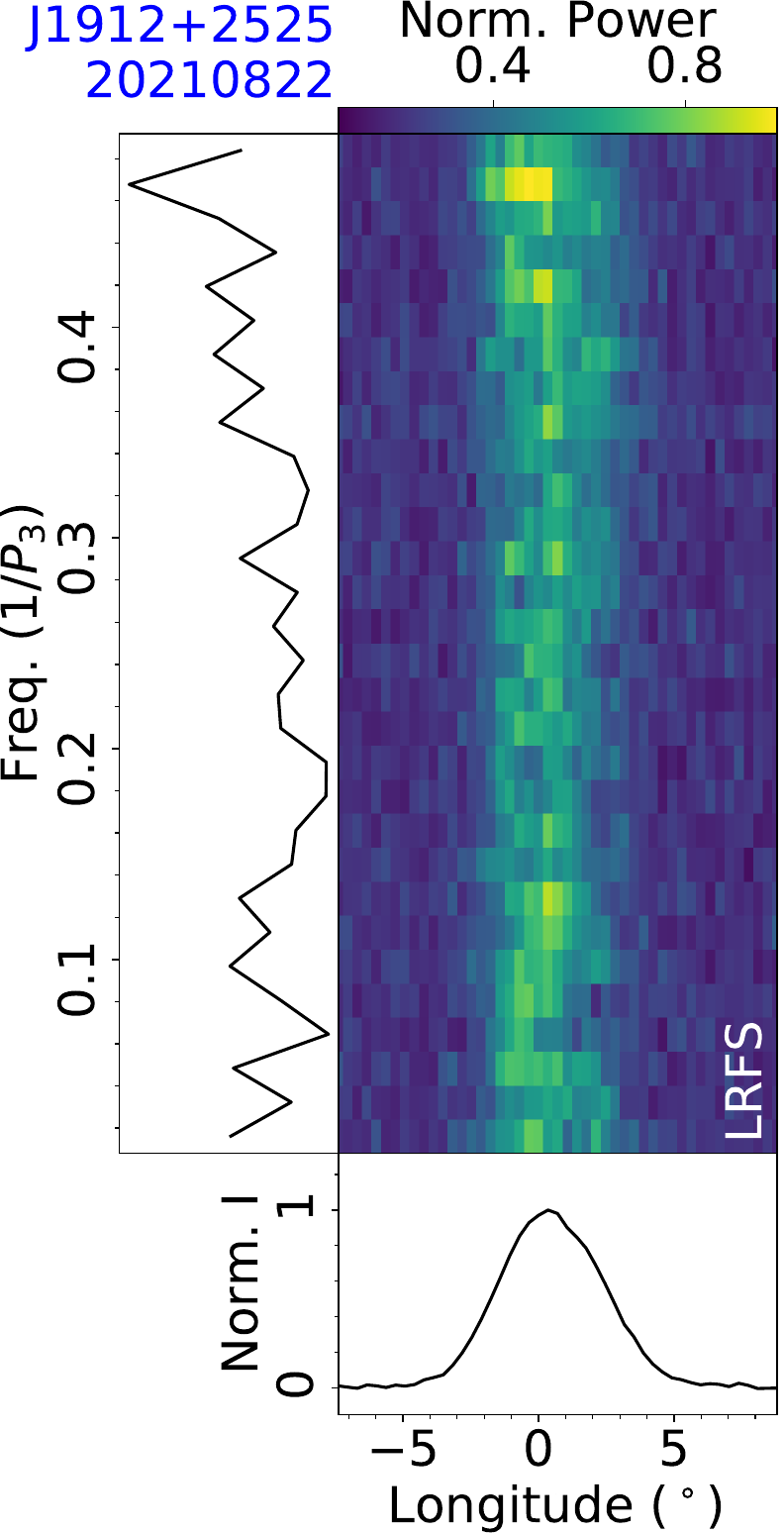}
\includegraphics[width=0.22\textwidth, angle=0]{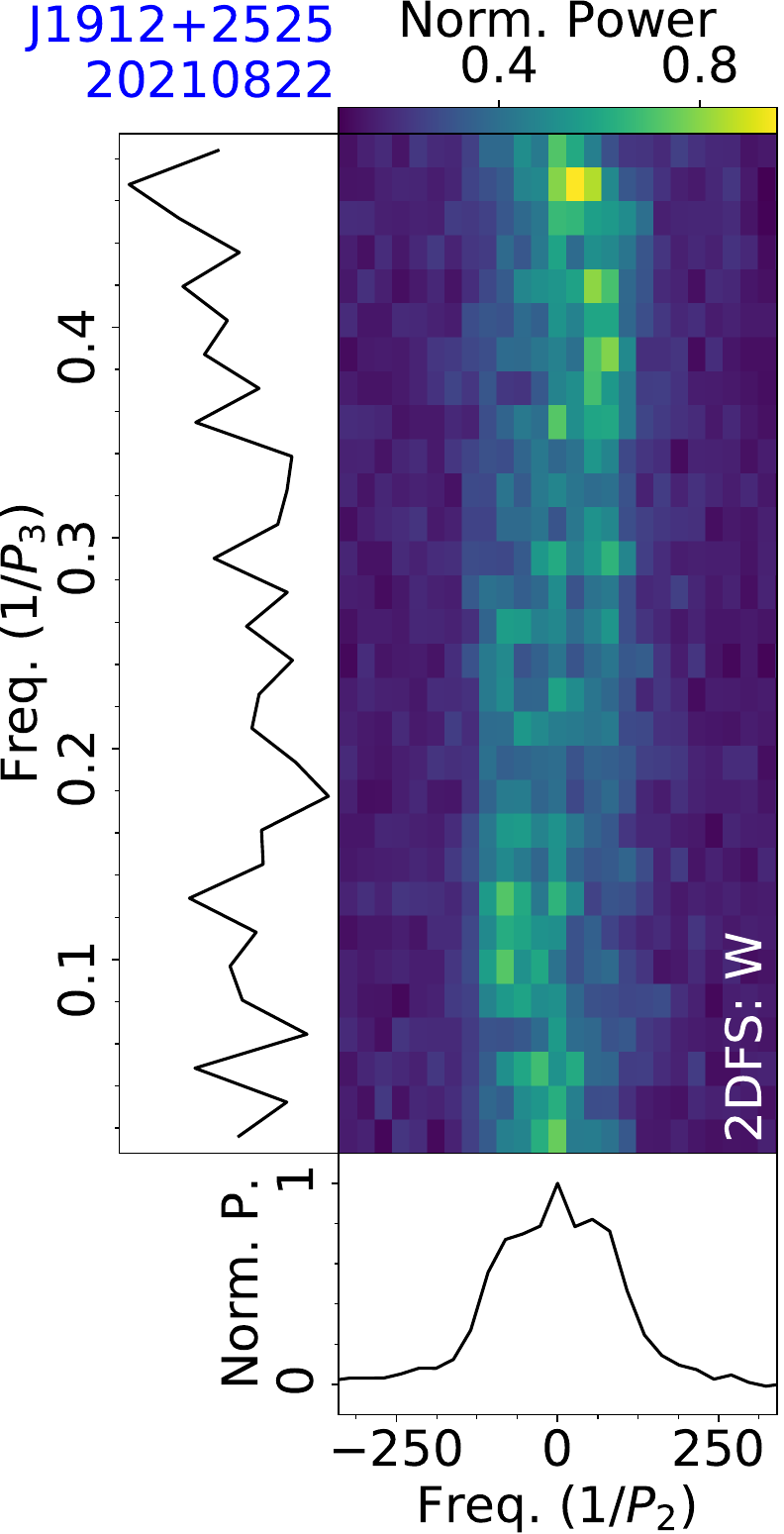}
\figcaption{Fluctuation analysis of PSR J1912+2525 for the observation on 20210822, with LRFS and 2DFS for the on-pulse region of a mean pulse profile.
\label{subfig:fluctu:J1912+2525}}
\end{figure}

\begin{figure}[htpb]
\centering
\includegraphics[width=0.22\textwidth, angle=0]{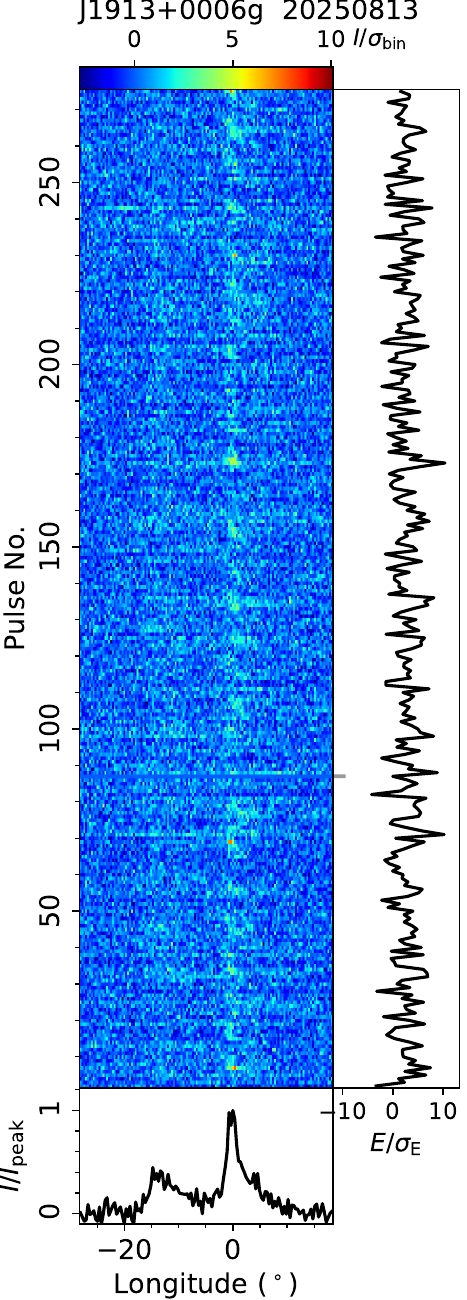}
\includegraphics[width=0.22\textwidth, angle=0]{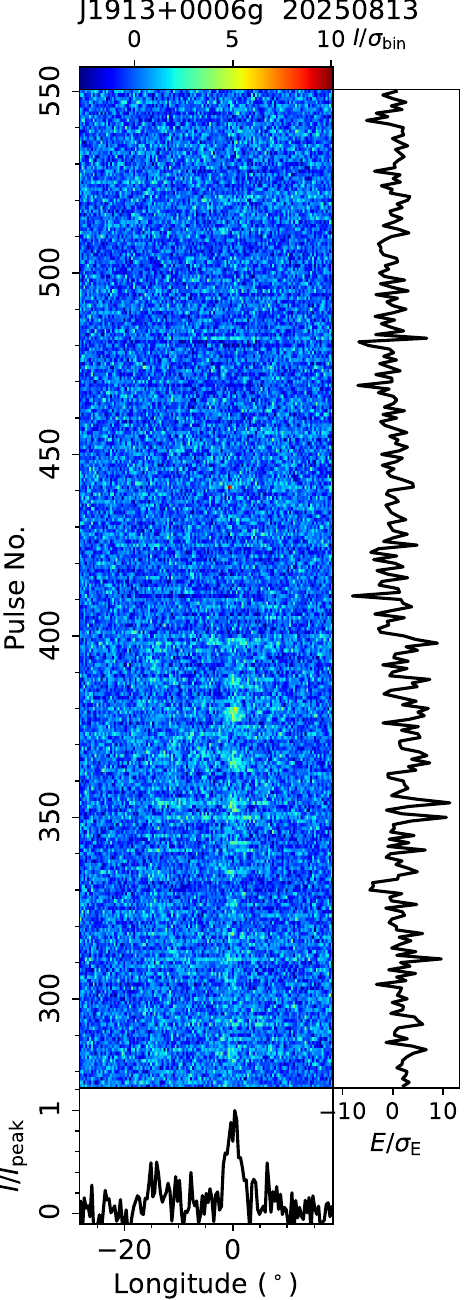}
\figcaption{Single pulse sequences of PSR J1913+0006g from the FAST observation on 20250813.
\label{subfig:TP:J1913+0006g}}
\end{figure}

\begin{figure}[htpb]
\centering
\includegraphics[width=0.22\textwidth, angle=0]{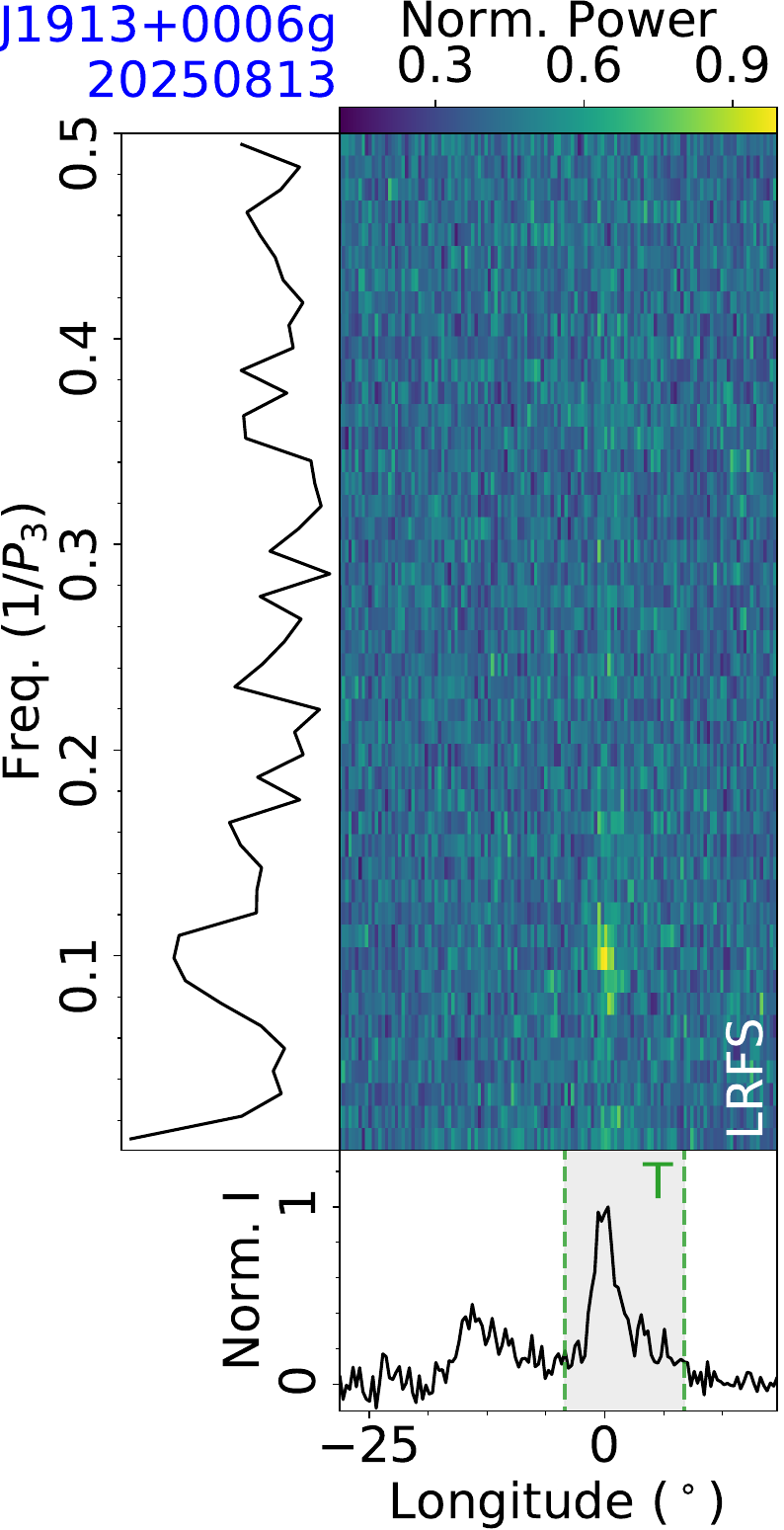}
\includegraphics[width=0.22\textwidth, angle=0]{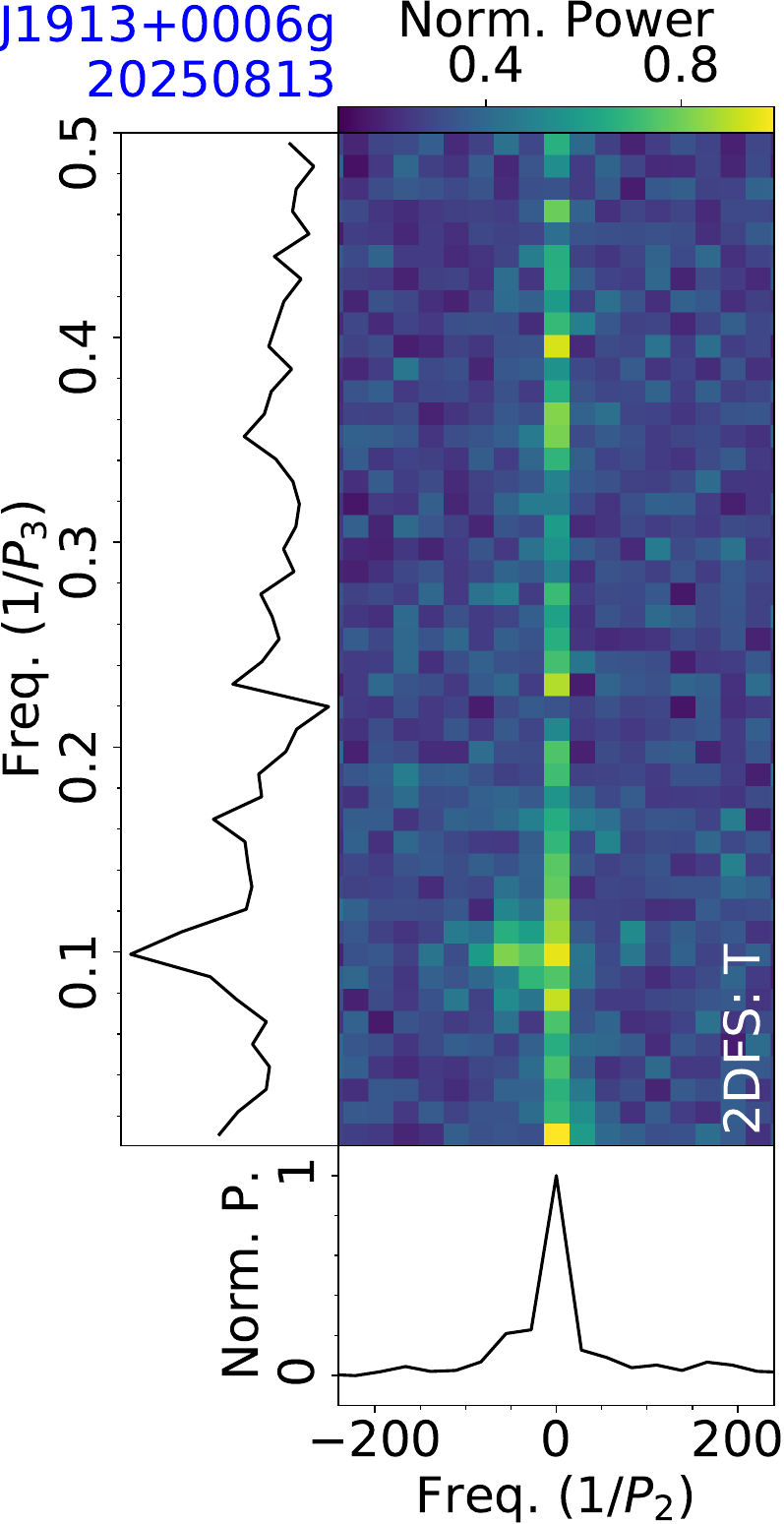}
\figcaption{Fluctuation analysis of PSR J1913+0006g for the FAST observation on 20250813, with LRFS and 2DFS for the trailing profile part of a mean pulse profile.
\label{subfig:fluctu:J1913+0006g}}
\end{figure}

\subsection{J1910+1117g}
\label{subsec:J1910+1117g}

PSR J1910+1117g was discovered in the FAST GPPS survey \citep{Han2021,han2025}.

This pulsar was observed by FAST on 20230214 for 44 minutes, deriving a rotation period $P=1.3213$~s and a dispersion measure $D\!M=296.7~{\rm cm^{-3}\,pc}$. Single pulse sequences in Fig.~\ref{subfig:TP:J1910+1117g} show the existence of the nulling phenomenon. The on-pulse energy histogram is displayed in Fig.~\ref{subfig:Hist:J1910+1117g}, where the energy values have been averaged over every four periods to improve the emission significance \citep{Wang2020}. From the energy histogram, the nulling fraction is estimated to be 24$\pm$3\%.

\subsection{J1910+1231}
\label{subsec:J1910+1231}

PSR J1910+1231 was discovered by the Arecibo telescope \citep{Hulse1975}. 

This pulsar was observed by FAST for 4 minutes, deriving a rotation period $P=1.4416$~s and a dispersion measure $D\!M=257.8~{\rm cm^{-3}\,pc}$. 
The single pulse sequence in Fig.~\ref{subfig:TP:J1910+1231} illustrates that there is the nulling phenomenon. The nulling fraction of this observation is estimated to be 15$\pm$4\% from the on-pulse integral energy histogram (Fig.~\ref{subfig:Hist:J1910+1231}).

\begin{figure}[htpb]
\centering
\includegraphics[width=0.22\textwidth, angle=0]{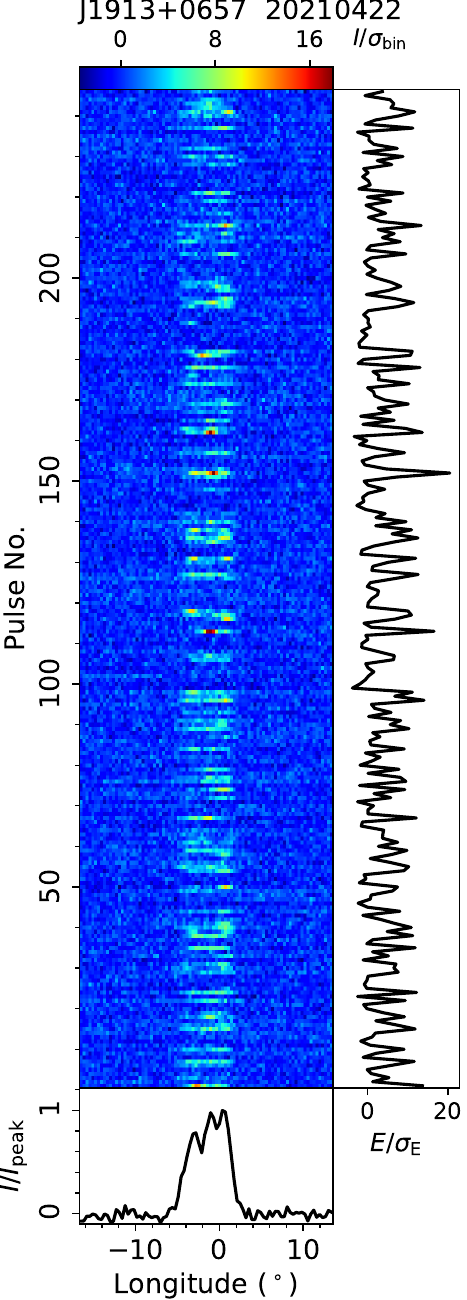}
\figcaption{Single pulse sequences of PSR J1913+0657 from the FAST observation on 20210422.
\label{subfig:TP:J1913+0657}}
\end{figure}

\begin{figure}[htpb]
\centering
\includegraphics[width=0.39\textwidth, angle=0]{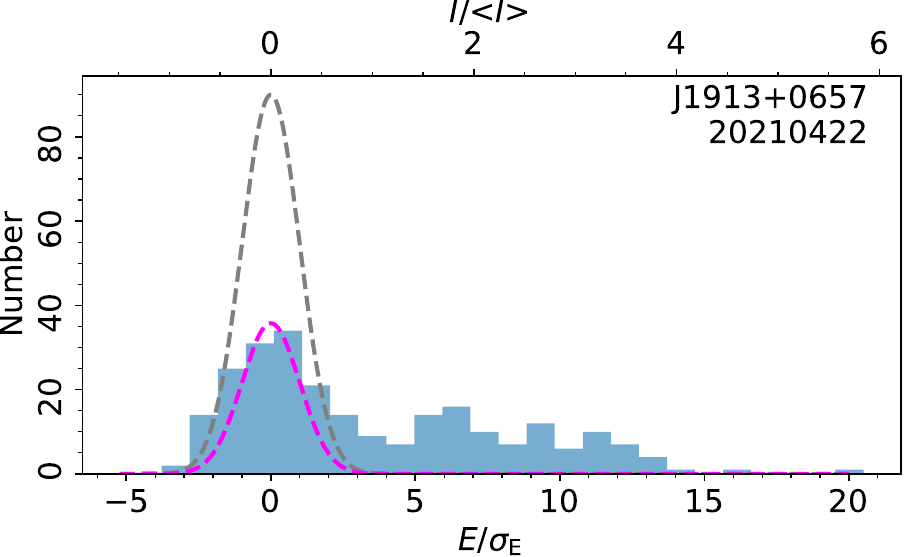}
\figcaption{On-pulse energy histogram of single pulses of PSR J1913+0657 from the FAST observation on 20210422.
\label{subfig:Hist:J1913+0657}}
\end{figure}

\begin{figure}[htpb]
\centering
\includegraphics[width=0.44\textwidth, angle=0]{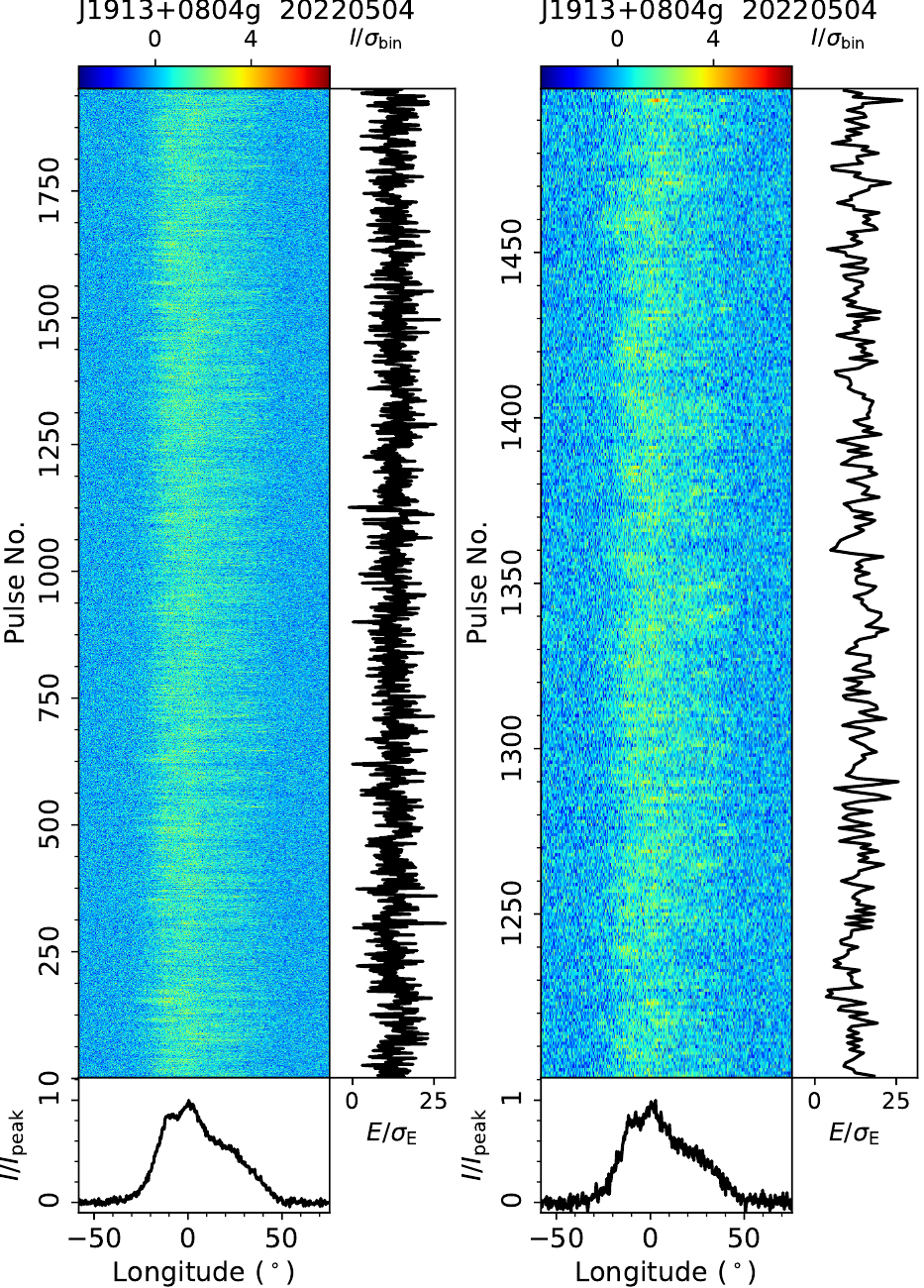}
\figcaption{Single pulse sequence of PSR J1913+0804g from the FAST observation on 20220504, and a zoomed-in view of pulses No. 1200-1500.
\label{subfig:TP:J1913+0804g}}
\end{figure}

\begin{figure}[htpb]
\centering
\includegraphics[width=0.44\textwidth, angle=0]{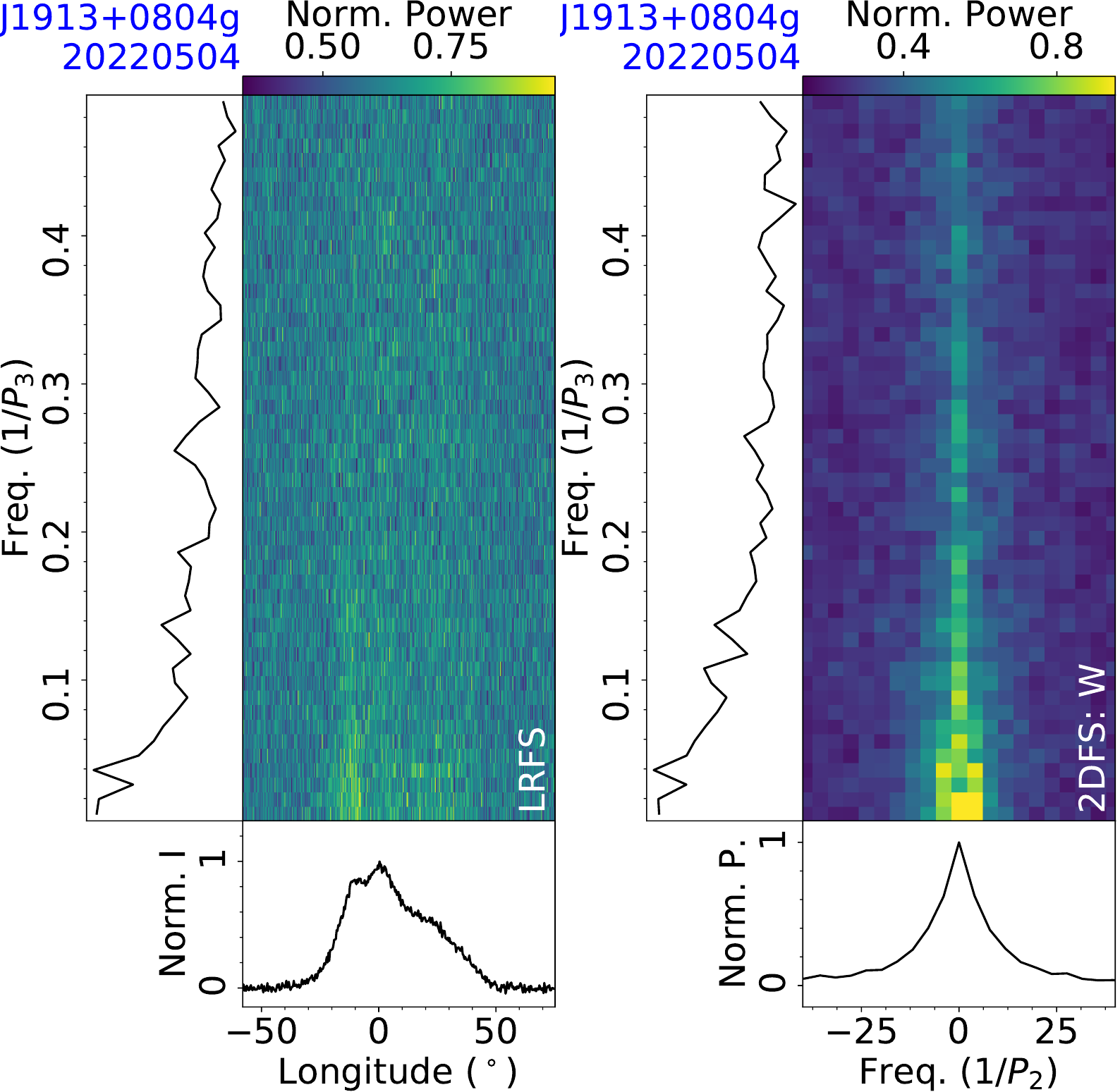}
\figcaption{Fluctuation analysis of PSR J1913+0804g for the FAST observation on 20220504, with LRFS and 2DFS for the on-pulse region of a mean pulse profile.
\label{subfig:fluctu:J1913+0804g}}
\end{figure}

\subsection{J1910-0219g}
\label{subsec:J1910-0219g}

PSR J1910-0219g was discovered in the FAST GPPS survey \citep{Han2021,han2025}.

This pulsar was observed by FAST on 20250619 for 15 minutes, yielding a rotation period $P=1.5316$~s and a dispersion measure $D\!M=185.9~{\rm cm^{-3}\,pc}$. Single pulse sequences in Fig.~\ref{subfig:TP:J1910-0219g} show the nulling phenomenon. The nulling fraction of this observation is 17.8$\pm$2.2\%, which is estimated from the on-pulse energy histogram in Fig.~\ref{subfig:Hist:J1910-0219g}. Fluctuation spectra in Fig.~\ref{subfig:fluctu:J1910-0219g} illustrate the existence of negative drifting. Centroid frequencies of the negative drift feature in 2DFS are estimated to be $1/P_3=0.470\pm0.002$ and $1/P_2=-80\pm5$, corresponding to periodicities of $P_3=2.13\pm0.01$ periods and $P_2=-4.5\pm0.3$ degrees.

\begin{figure}[htpb]
\centering
\includegraphics[width=0.22\textwidth, angle=0]{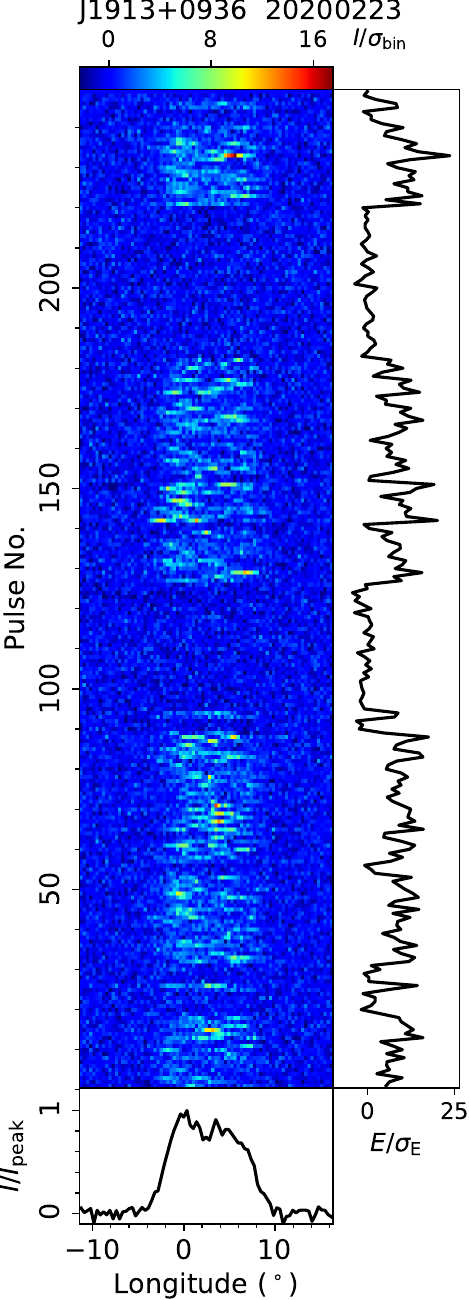}
\figcaption{Single pulse sequence of PSR J1913+0936 from the FAST observation on 20200223.
\label{subfig:TP:J1913+0936}}
\end{figure}

\begin{figure}[htpb]
\centering
\includegraphics[width=0.39\textwidth, angle=0]{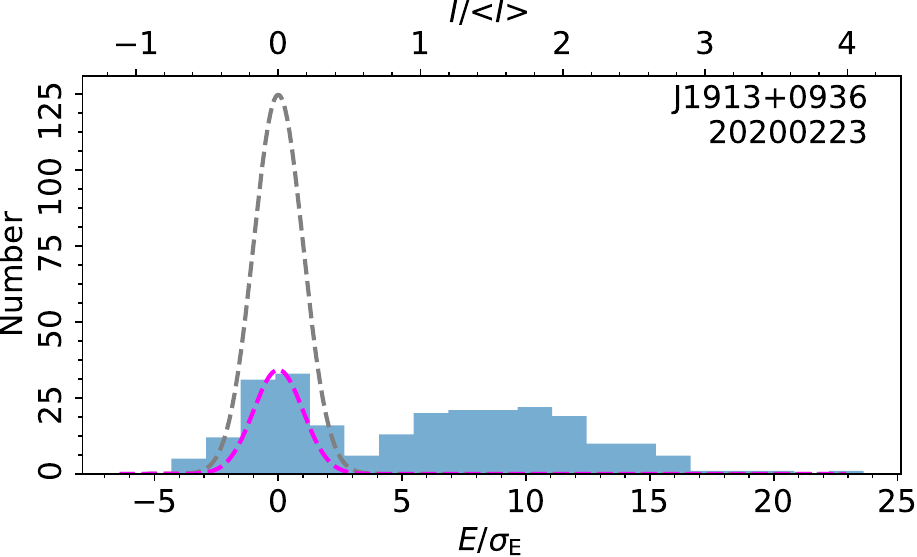}
\figcaption{On-pulse energy histogram of single pulses of PSR J1913+0936 from the FAST observation on 20200223.
\label{subfig:Hist:J1913+0936}}
\end{figure}

\begin{figure}[htpb]
\includegraphics[width=0.22\textwidth, angle=0]{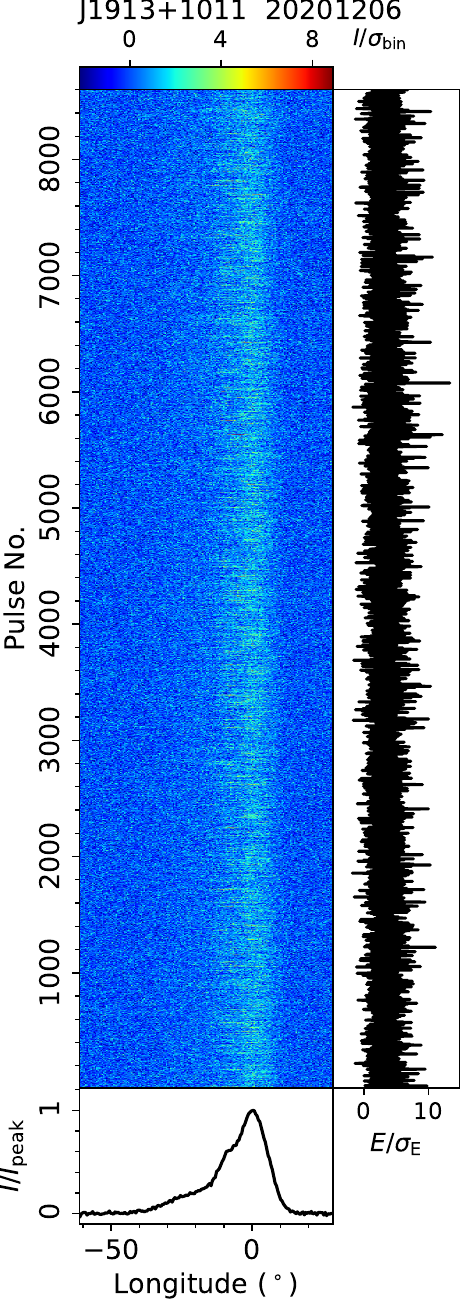}
\includegraphics[width=0.22\textwidth, angle=0]{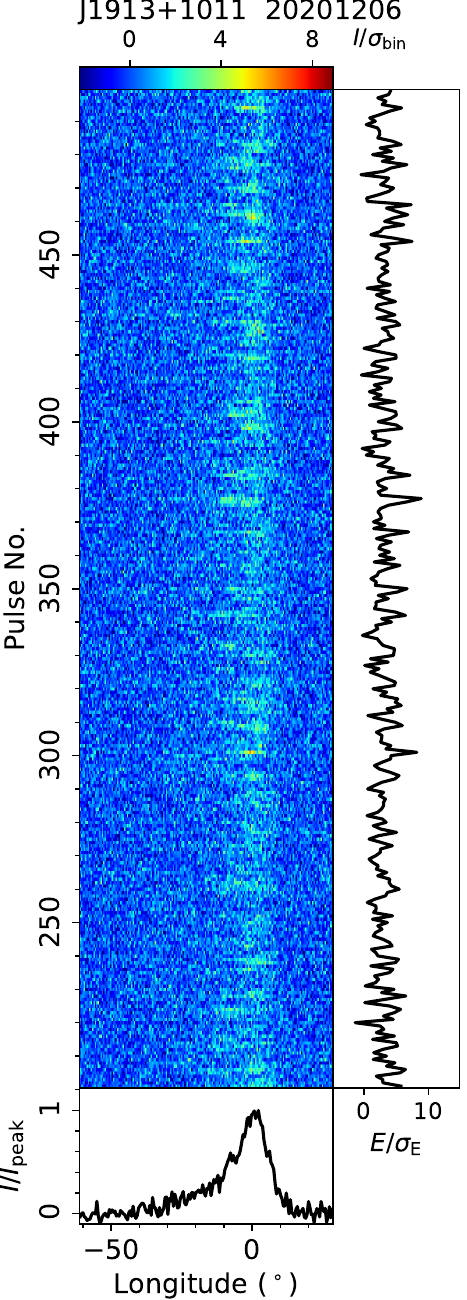}
\figcaption{Single pulse sequence of PSR J1913+1011 from the FAST observation on 20201206, and a zoomed-in view of pulses No. 200-500.
\label{subfig:TP:J1913+1011}}
\end{figure}

\begin{figure}[htpb]
\centering
\includegraphics[width=0.22\textwidth, angle=0]{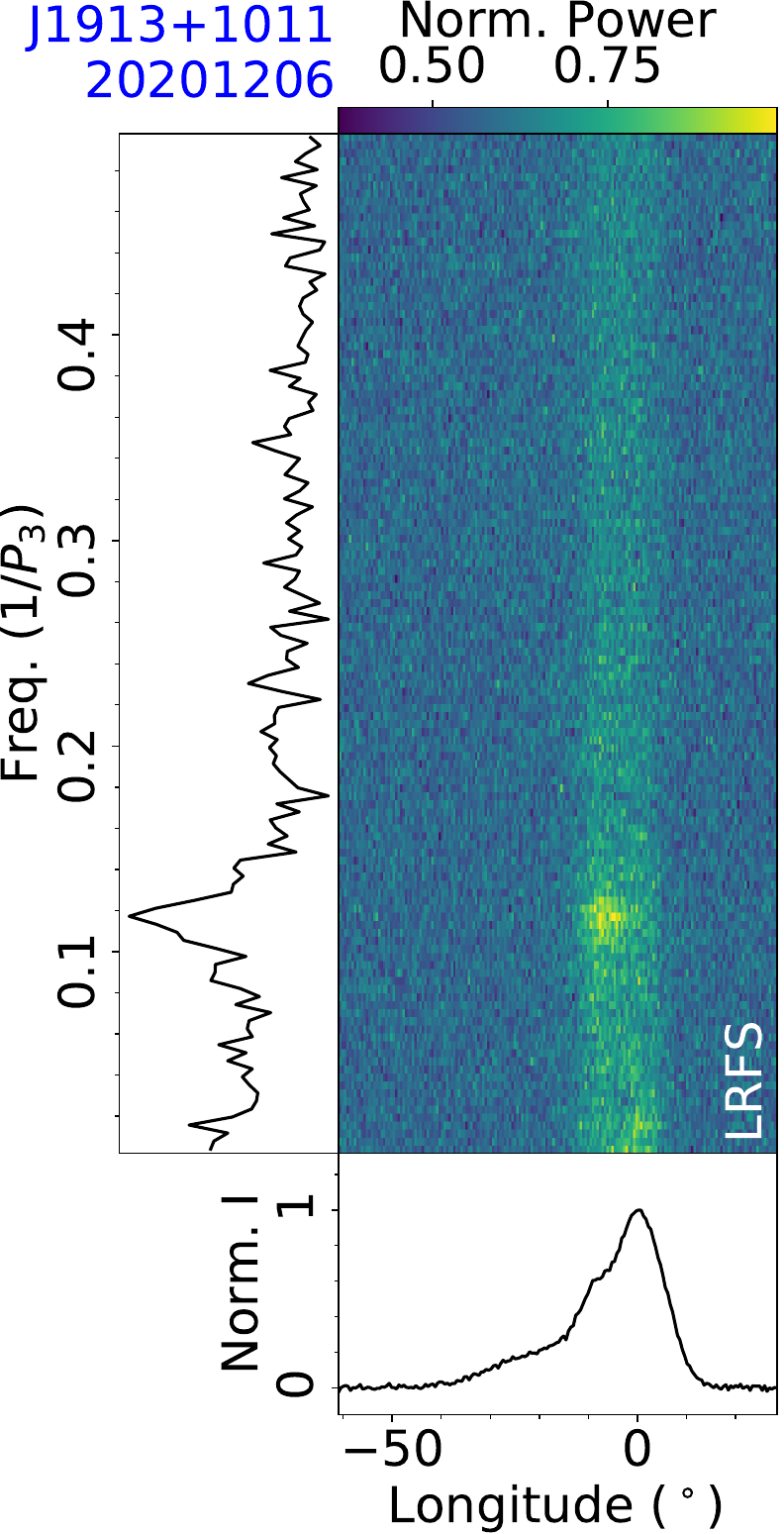}
\includegraphics[width=0.22\textwidth, angle=0]{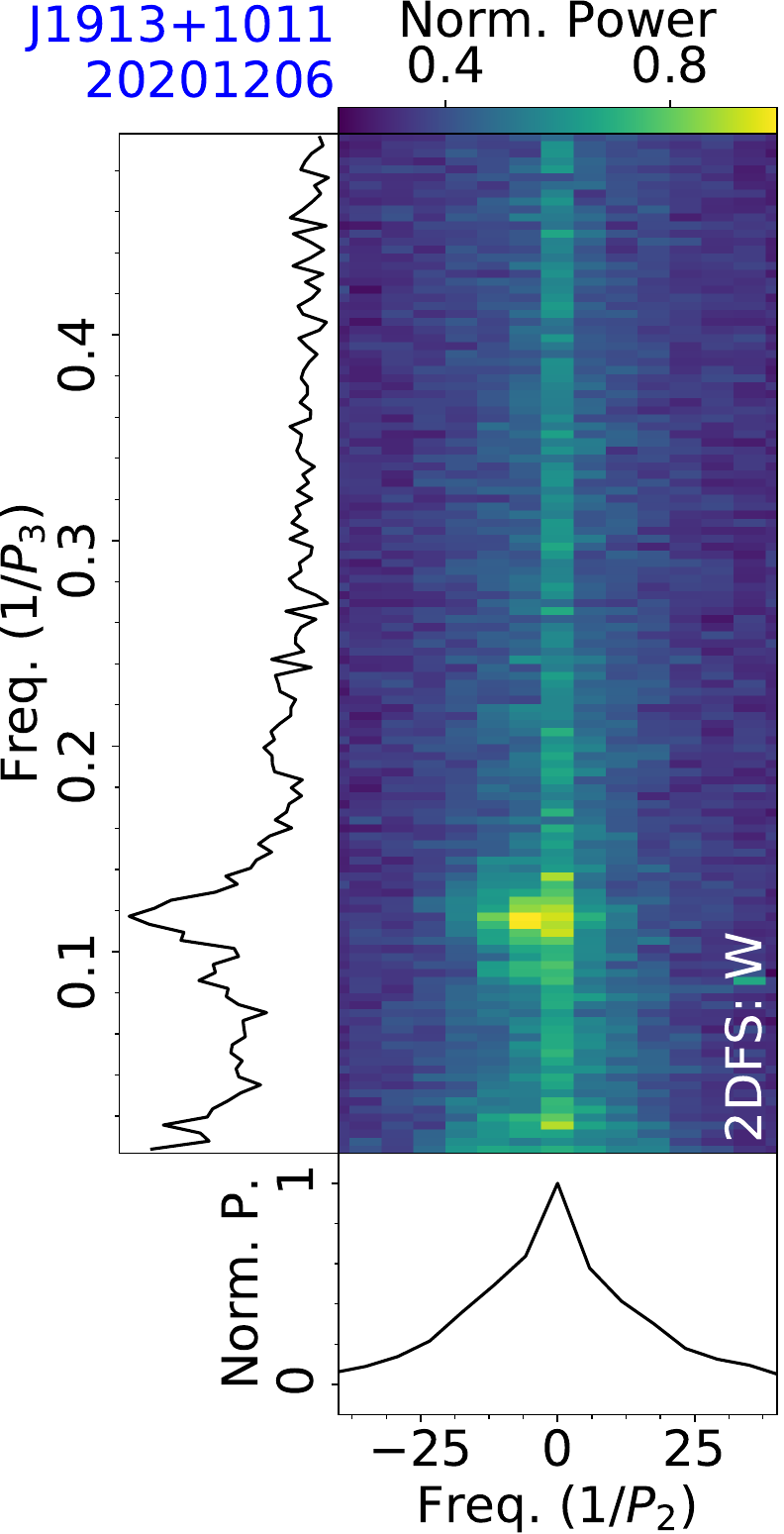}
\figcaption{Fluctuation analysis of PSR J1913+1011 for the observation on 20201206, with LRFS and 2DFS for the on-pulse region of a mean pulse profile.
\label{subfig:fluctu:J1913+1011}}
\end{figure}

\subsection{J1910-0309}
\label{subsec:J1910-0309}

PSR J1910-0309 was discovered in the second Molonglo pulsar survey \citep{Manchester1978}, exhibiting the subpulse drifting phenomenon of $P_3=22\pm14$ periods and $P_2=-127^{+107}_{-86}$ degrees. 

This pulsar was observed by FAST on 20250317 for 6 minutes, with a rotation period $P=0.5046$~s and a dispersion measure $D\!M=205.5~{\rm cm^{-3}\,pc}$ derived. Single pulse sequences are shown in Fig.~\ref{subfig:TP:J1910-0309}. LRFS and 2DFS of the leading, central, and trailing parts in a mean pulse profile (Fig.~\ref{subfig:fluctu:J1910-0309}) illustrate that the temporally modulated frequencies are widely distributed. The centroid frequencies for the modulation features of three profile parts are estimated to be $1/P_3=0.119\pm0.004$, $0.051\pm0.003$, and $0.083\pm0.002$, corresponding to periodicities of $P_3=8.4\pm0.3$, $20\pm1$, and $12.1\pm0.3$ periods.

\begin{figure}[htpb]
\centering
\includegraphics[width=0.22\textwidth, angle=0]{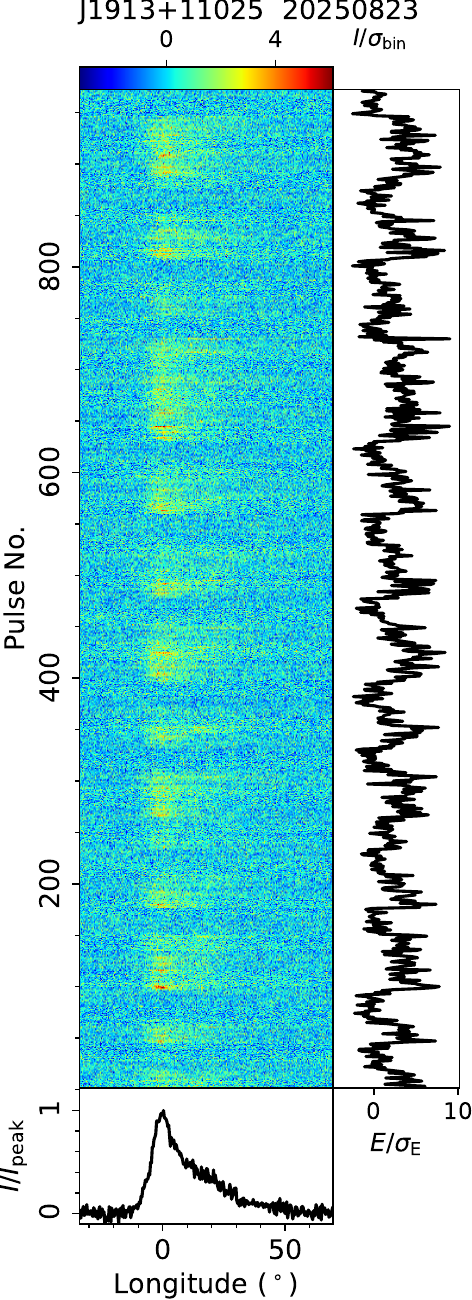}
\figcaption{Single pulse sequence of PSR J1913+11025 from the FAST observation on 20250823.
\label{subfig:TP:J1913+11025}}
\end{figure}

\begin{figure}[htpb]
\centering
\includegraphics[width=0.39\textwidth, angle=0]{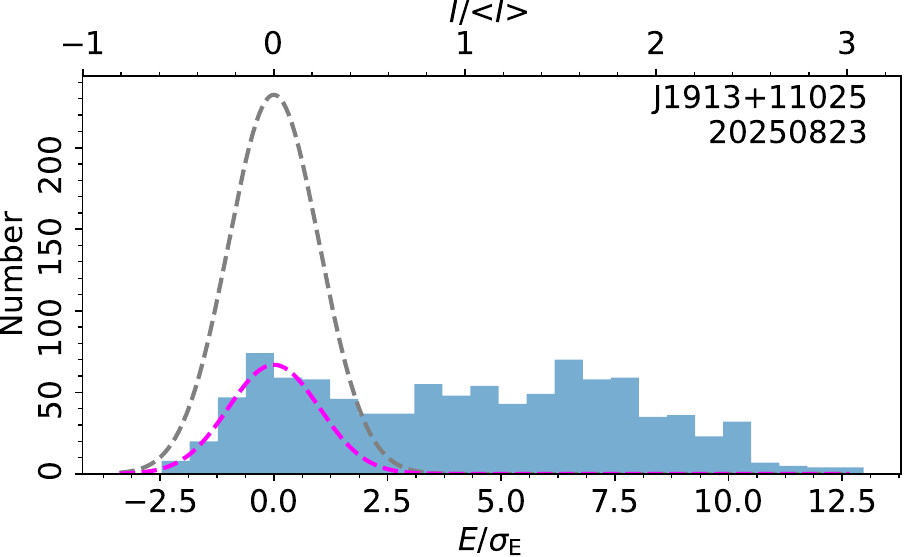}
\figcaption{On-pulse energy histogram of single pulses of PSR J1913+11025 from the FAST observation on 20250823, with energy values smoothed over every 4 periods.
\label{subfig:Hist:J1913+11025}}
\end{figure}

\subsection{J1911+1440g}
\label{subsec:J1911+1440g}

PSR J1911+1440g was discovered in the FAST GPPS survey \citep{Han2021,han2025}.

This pulsar was observed by FAST on 20240125 and 20250326, each for a duration of 15 minutes. From the data of 20240125, a rotation period $P=0.5825$~s and a dispersion measure $D\!M=86.9~{\rm cm^{-3}\,pc}$ were derived. The single pulse sequence and a zoomed-in view of this observation in Fig.~\ref{subfig:TP:J1911+1440g} display the behavior of short-duration emission enhanced. 
From the on-pulse integral energy histogram in Fig.~\ref{subfig:Hist:J1911+1440g}, the nulling fraction is estimated to be 66.9$\pm$4.4\%. 
The emission duration is short, typically 1 period but ranging up to 3, while nulls last from 1 to 49 periods (Fig.~\ref{subfig:scaleHist:J1911+1440g}). 
Histograms in Fig.~\ref{subfig:nullDegreeScale:J1911+1440g} show that the nulling degree is large, to be 80$\pm$12 degrees. The nulling scale is estimated to be 14$\pm$10 periods for NE pairs, and they are similar for EN pairs.

\begin{figure}[htpb]
\centering
\includegraphics[width=0.22\textwidth, angle=0]{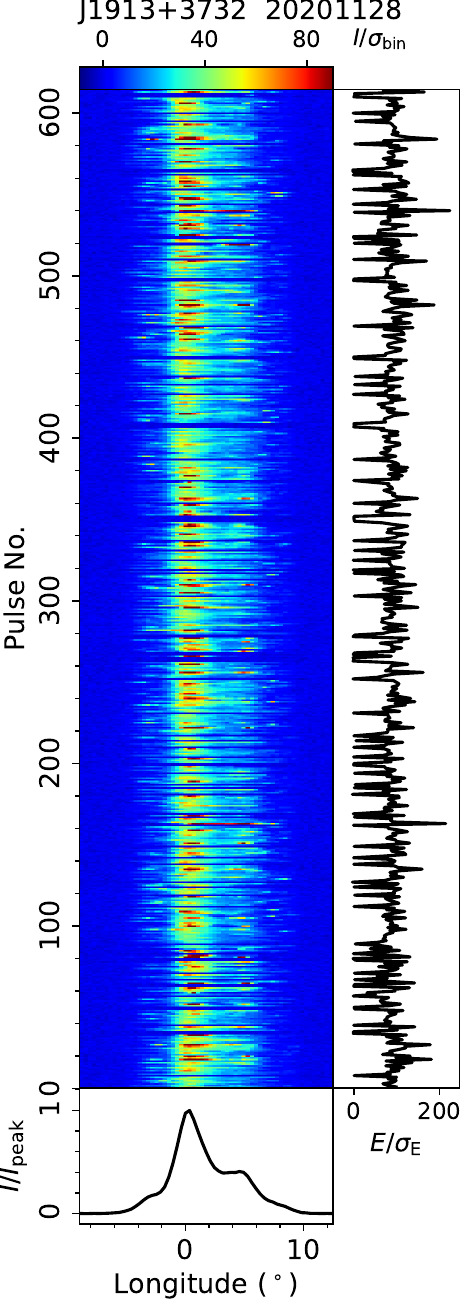}
\includegraphics[width=0.22\textwidth, angle=0]{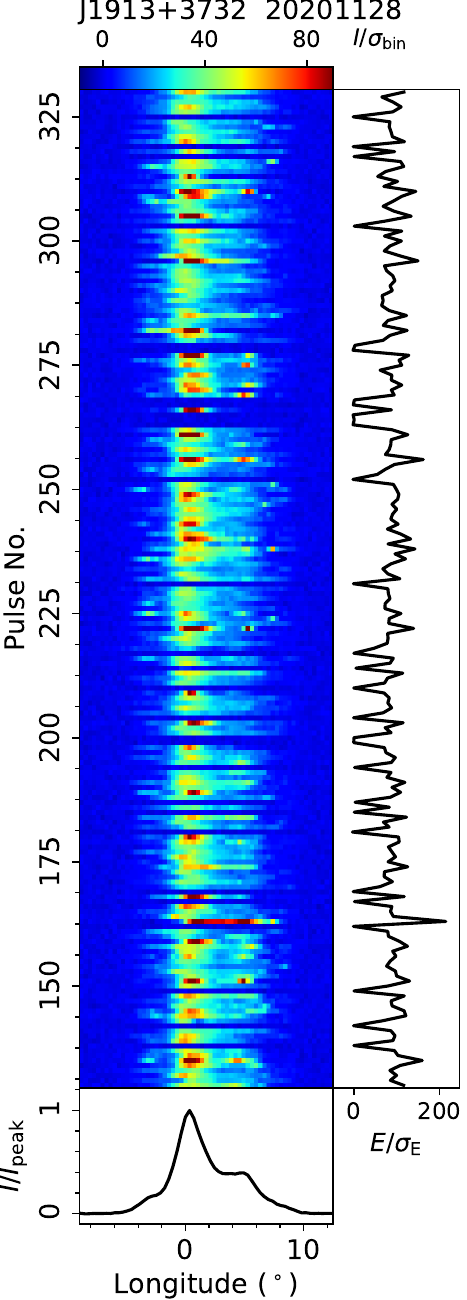}
\figcaption{Single pulse sequences of PSR J1913+3732 from the FAST observation on 20201128.
\label{subfig:TP:J1913+3732}}
\end{figure}

\begin{figure}[htpb]
\centering
\includegraphics[width=0.39\textwidth, angle=0]{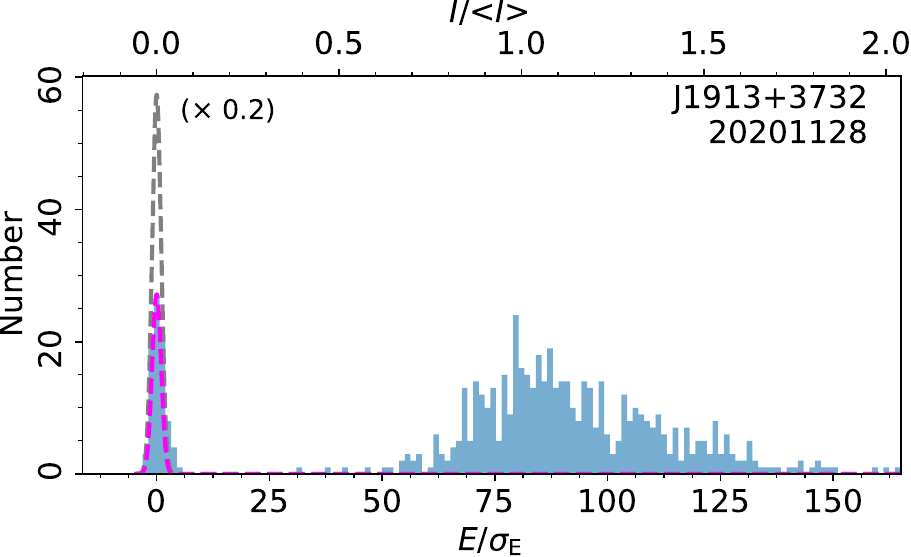}
\figcaption{On-pulse energy histogram of single pulses of PSR J1913+3732 from the FAST observation on 20201128.
\label{subfig:Hist:J1913+3732}}
\end{figure}

\begin{figure}[htpb]
\centering
\includegraphics[width=0.22\textwidth, angle=0]{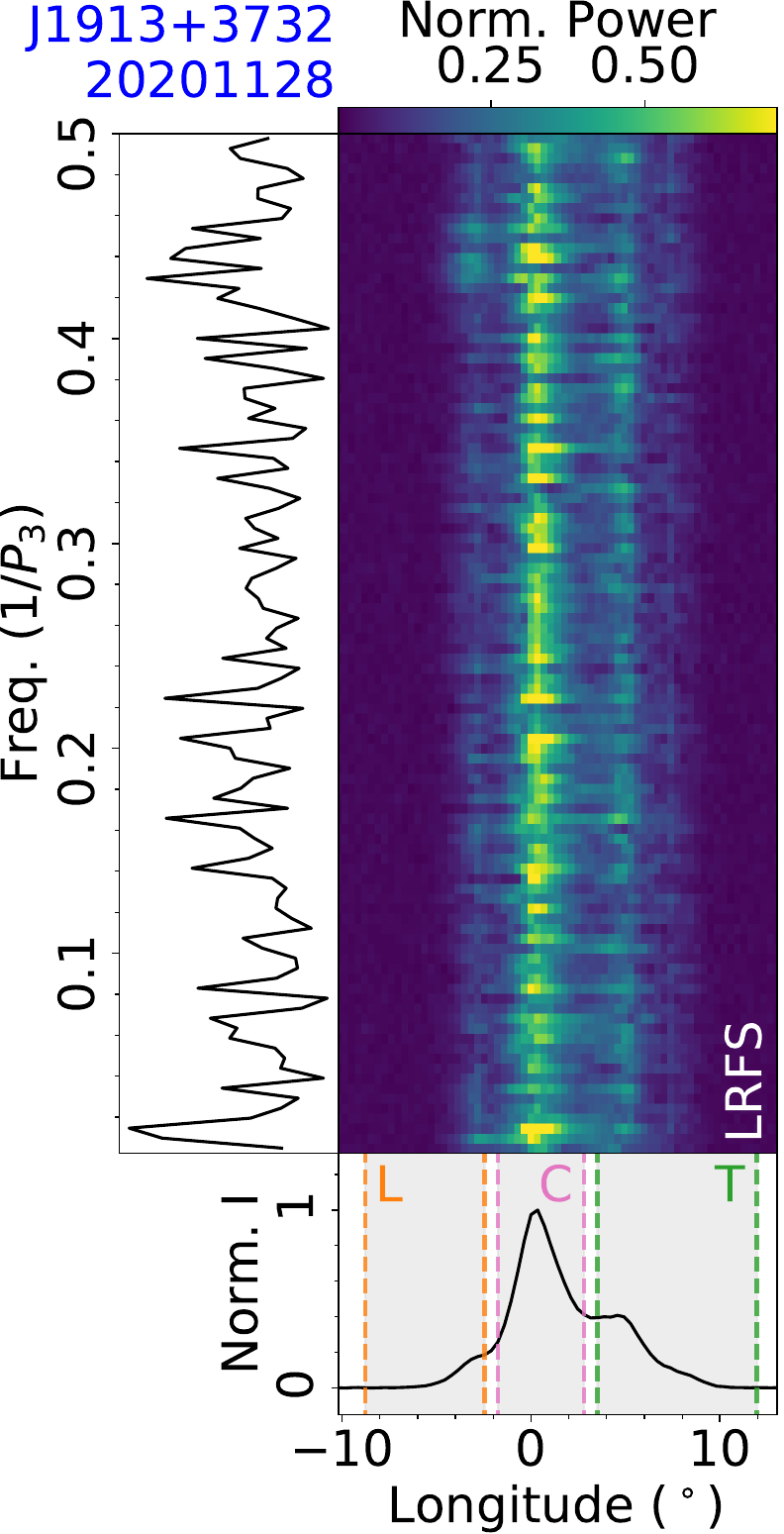}
\includegraphics[width=0.22\textwidth, angle=0]{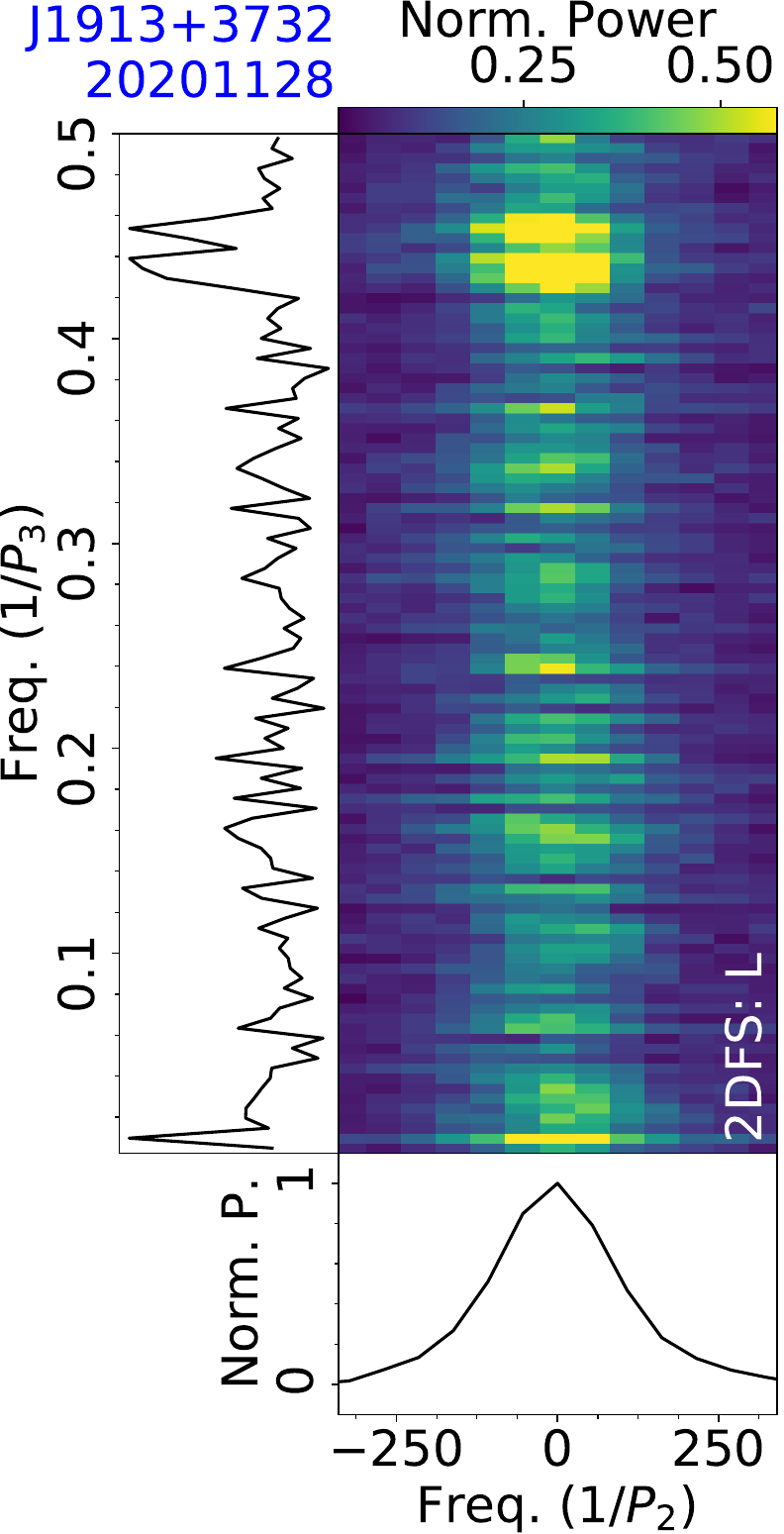}
\figcaption{Fluctuation analysis of PSR J1913+3732 for the observation on 20201128, with LRFS and 2DFS for the leading part of a mean pulse profile.
\label{subfig:fluctu:J1913+3732}}
\end{figure}

\subsection{J1912+2525}
\label{subsec:J1912+2525}

PSR J1912+2525 was discovered in a survey for fast pulsars at the Arecibo Observatory \citep{Nice1995}. 

This pulsar was observed by FAST on 20210822 for 5 minutes, yielding a rotation period $P=0.6219$~s and a dispersion measure $D\!M=37.9~{\rm cm^{-3}\,pc}$. 
From single pulse sequences in Fig.~\ref{subfig:TP:J1912+2525} and fluctuation spectra in Fig.~\ref{subfig:fluctu:J1912+2525}, there is a positive drift feature with the centroid of $1/P_3=0.418\pm0.004$ and $1/P_2=52\pm3$, corresponding to  $P_3=2.39\pm0.02$ periods and $P_2=7.0\pm0.4^\circ$. 
There is also a lower-frequency negative drift feature, with the centroid frequencies of $1/P_3=0.106\pm0.003$ and $1/P_2=-72\pm4$, yielding $P_3=9.4\pm0.2$ periods and $P_2=-5.0\pm0.3$, which seem to be superimposed on the positive drifting from single pulse sequences.

\subsection{J1913+0006g}
\label{subsec:J1913+0006g}

PSR J1913+0006g was discovered in the FAST GPPS survey \citep{Han2021,han2025}. 

This pulsar was observed by FAST on 20250709 and 20250813 for 5 and 15 minutes, respectively. From the longer observation, a rotation period $P=1.6325$~s and a dispersion measure $D\!M=128.7~{\rm cm^{-3}\,pc}$ were derived. Single pulse sequences in Fig.~\ref{subfig:TP:J1913+0006g} display the subpulse drifting of the trailing profile part and the intensity decreases in pulses No. 400-550. Fluctuation spectra are shown in Fig.~\ref{subfig:fluctu:J1913+0006g}. In 2DFS of the trailing part in a mean pulse profile, centroid frequencies of the negative drift feature are estimated to be $1/P_3=0.099\pm0.001$ and $1/P_2=-50\pm3$, corresponding to periodicities of $P_3=10.1\pm0.1$ periods and $P_2=-7.2\pm0.5$ degrees. 
Further observations are required in order to analyze the intensity decrease.

\subsection{J1913+0657}
\label{subsec:J1913+0657}

PSR J1913+0657 was discovered in the PALFA survey using the Arecibo Telescope \citep{Lyne2017}. 

The pulsar was observed by FAST on 20210422 for 5 minutes. The single pulse sequence in Fig.~\ref{subfig:TP:J1913+0657} shows state changes between nulling and emission states. The nulling fraction is estimated to be 40$\pm$5\% from the on-pulse integral energy histogram (Figures~\ref{subfig:Hist:J1913+0657}).

\subsection{J1913+0804g}
\label{subsec:J1913+0804g}

PSR J1913+0804g was discovered in the FAST GPPS survey \citep{Han2021,han2025}. 

This pulsar was observed by FAST on 20220313 and 20220504, each for 15 minutes. From the observation on 20220504, a rotation period $P=0.4555$~s and a dispersion measure $D\!M=344.2~{\rm cm^{-3}\,pc}$ were derived. The single pulse sequence, along with a zoomed-in view of pulses No. 1200-1500, is shown in Fig.~\ref{subfig:TP:J1913+0804g}. The LRFS and 2DFS of the on-pulse phase region (Fig.~\ref{subfig:fluctu:J1913+0804g}) present a modulation feature with the centroid at $1/P_3=0.056\pm0.001$, yielding a periodicity of $P_3=17.8\pm0.4$ periods.

\subsection{J1913+0936}
\label{subsec:J1913+0936}

PSR J1913+0936 was discovered by \citet{Hulse1975} at 430 MHz using the Arecibo telescope. 

The pulsar was observed by FAST on 20200223 for 5 minutes, yielding a rotation period $P=1.2419$~s and a dispersion measure $D\!M=156.2~{\rm cm^{-3}\,pc}$. 
The single pulse sequence shown in Fig.~\ref{subfig:TP:J1913+0936} illustrates the existence of the emission state and nulling state. From the on-pulse integral energy histogram in Fig.~\ref{subfig:TP:J1913+0936}, the nulling fraction is estimated to be 28$\pm$4\%.

\begin{figure}[htpb]
\includegraphics[width=0.22\textwidth, angle=0]{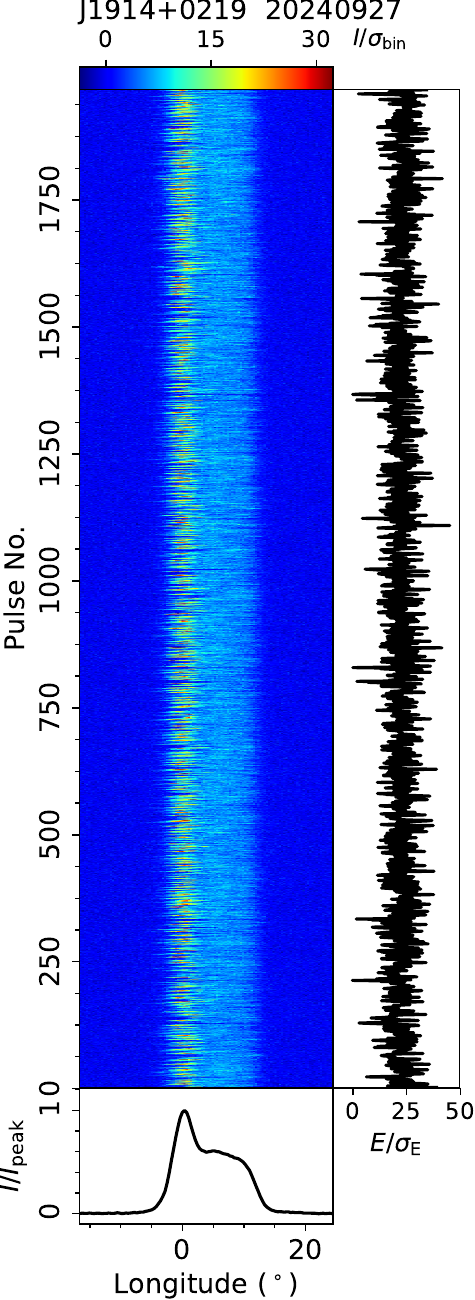}
\includegraphics[width=0.22\textwidth, angle=0]{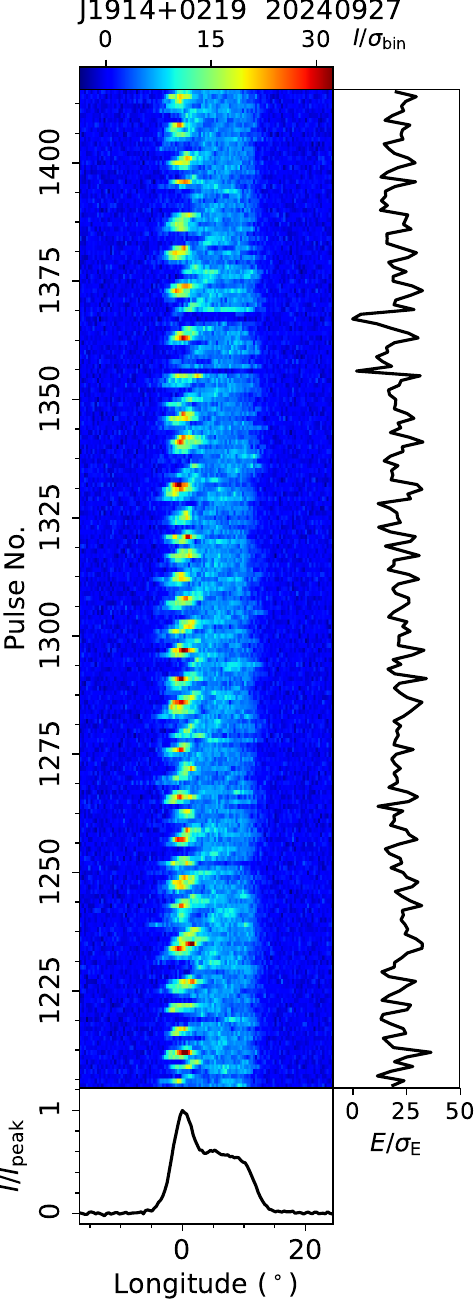}
\figcaption{Single pulse sequence of PSR J1914+0219 from the FAST observation on 20240927, and zoomed-in view of pulses No. 1205-1415.
\label{subfig:TP:J1914+0219}}
\end{figure}

\begin{figure}[htpb]
\centering
\includegraphics[width=0.39\textwidth, angle=0]{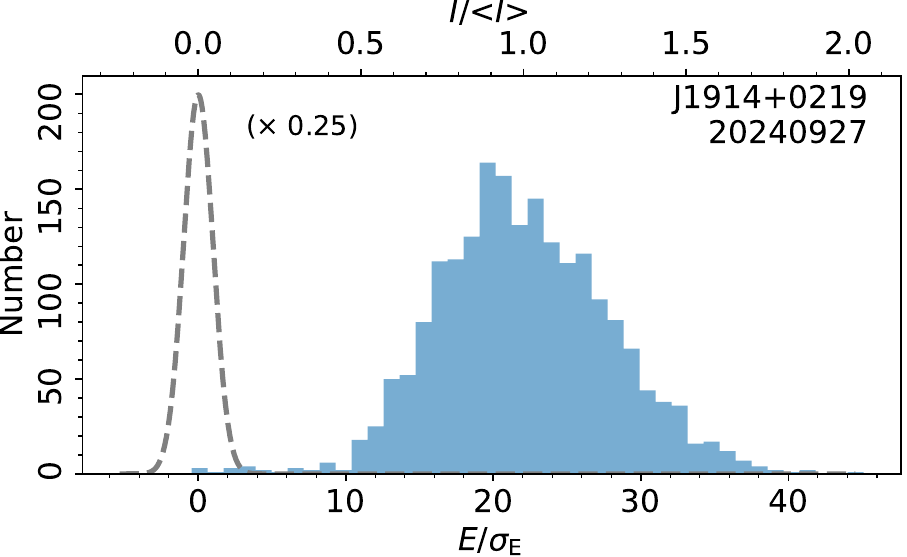}
\figcaption{On-pulse energy histogram of single pulses of PSR J1914+0219 from the FAST observation on 20240927.
\label{subfig:Hist:J1914+0219}}
\end{figure}

\begin{figure}[htpb]
\centering
\includegraphics[width=0.22\textwidth, angle=0]{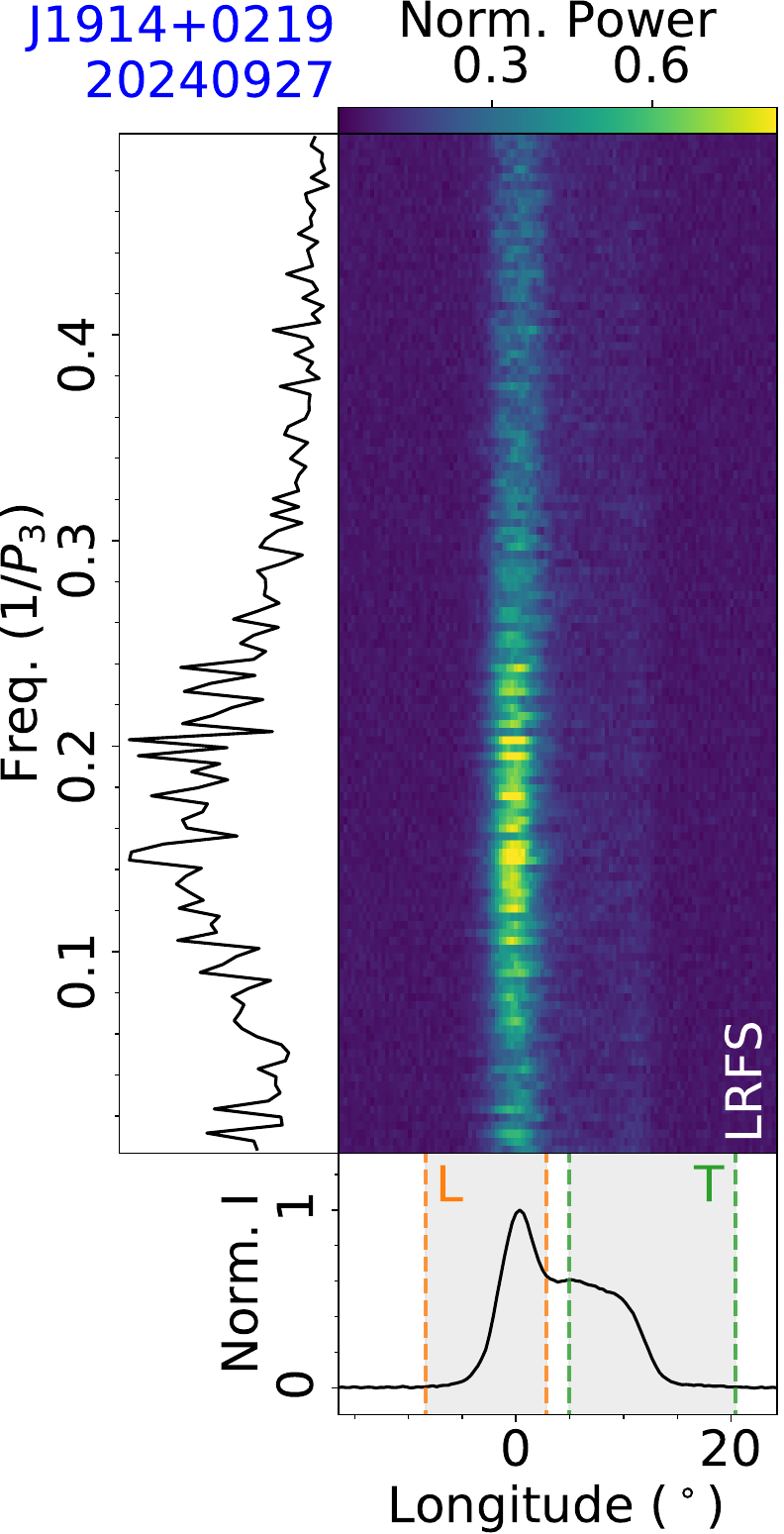}
\includegraphics[width=0.22\textwidth, angle=0]{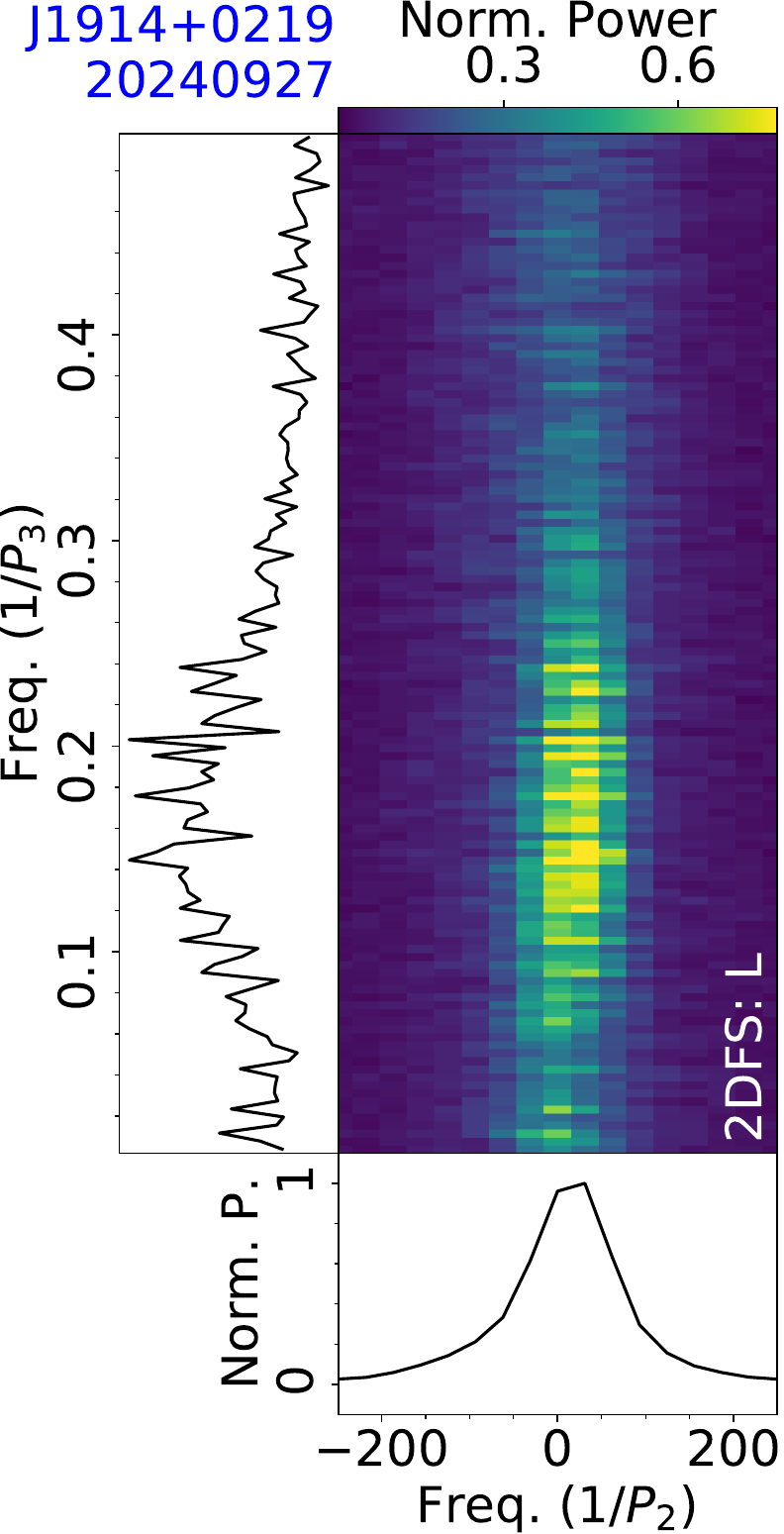}
\figcaption{Fluctuation analysis of PSR J1914+0219 for the observation on 20240927, with LRFS and 2DFS for the leading part of a mean pulse profile.
\label{subfig:fluctu:J1914+0219}}
\end{figure}

\subsection{J1913+1011}
\label{subsec:J1913+1011}

PSR J1913+1011 was discovered in the Parkes Multibeam Pulsar Survey \citet{Morris2002}. 

This pulsar was observed by FAST on 20201206 for 5 minutes, deriving a rotation period $P=0.0359$~s and a dispersion measure $D\!M=178.6~{\rm cm^{-3}\,pc}$. 
Single pulse sequences are shown in Fig.~\ref{subfig:TP:J1913+1011}. 
From LRFS and 2DFS of the on-pulse region (Fig.~\ref{subfig:fluctu:J1913+1011}), there is a temporally low-frequency modulation feature with the centroid frequency of $1/P_3=0.0183\pm0.0004$ ($P_3=55\pm1$ periods), as well as a negative drift feature with the centroid frequencies of $1/P_3=0.1119\pm0.0004$ ($P_3=8.94\pm0.03$ periods) and $1/P_2=-5.3\pm0.3$ ($P_2=-67\pm4^\circ$).

\subsection{J1913+11025}
\label{subsec:J1913+11025}

PSR J1913+11025 was discovered by the Arecibo telescope \citep{Lazarus2015}. \citet{Parent2022} reported that this pulsar displays nulling behavior at 1.4 GHz.

This pulsar was observed by FAST on 20250823 for 15 minutes, and a rotation period $P=0.9239$~s and a dispersion measure $D\!M=624.2~{\rm cm^{-3}\,pc}$ were determined. The single pulse sequence in Fig.~\ref{subfig:TP:J1913+11025} displays the nulling phenomenon. 
To improve the significance of pulses in the histogram (Fig.~\ref{subfig:Hist:J1913+11025}), the on-pulse energies are averaged for every four pulses using the method described in \citet{Wang2020}. The nulling fraction of this observation is estimated to be 28.8$\pm$1.5\%.

\begin{figure}[htpb]
\includegraphics[width=0.22\textwidth, angle=0]{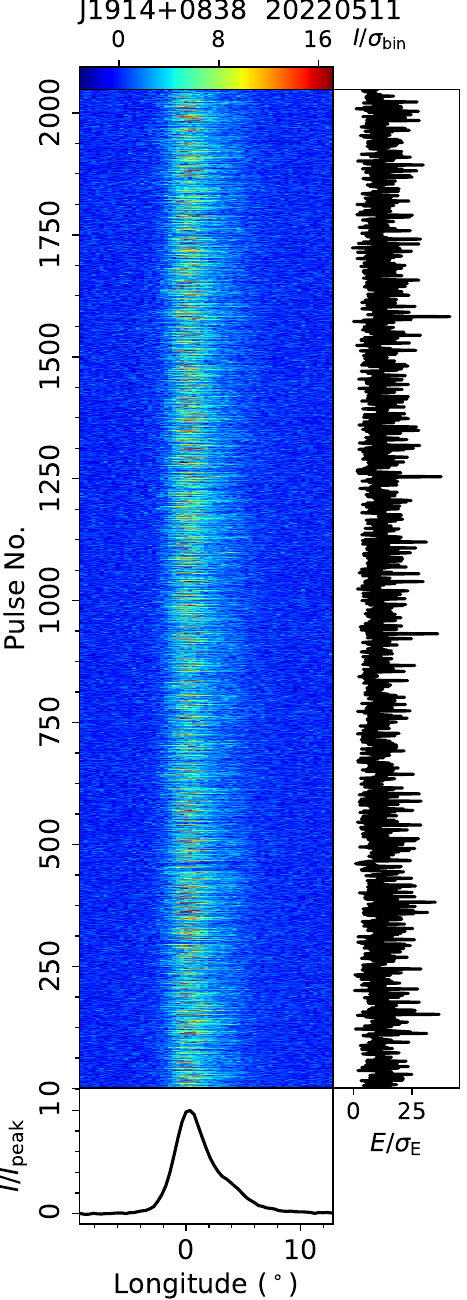}
\includegraphics[width=0.22\textwidth, angle=0]{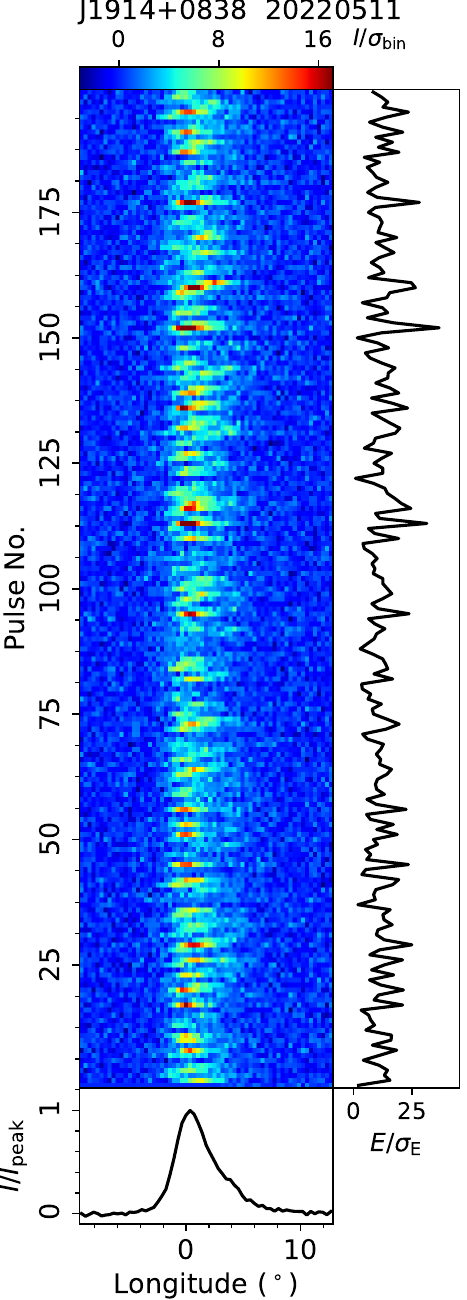}
\figcaption{Single pulse sequence of PSR J1914+0838 from the FAST observation on 20220511, and a zoomed-in view of pulses No. 1-200.
\label{subfig:TP:J1914+0838}}
\end{figure}

\begin{figure}[htpb]
\centering
\includegraphics[width=0.22\textwidth, angle=0]{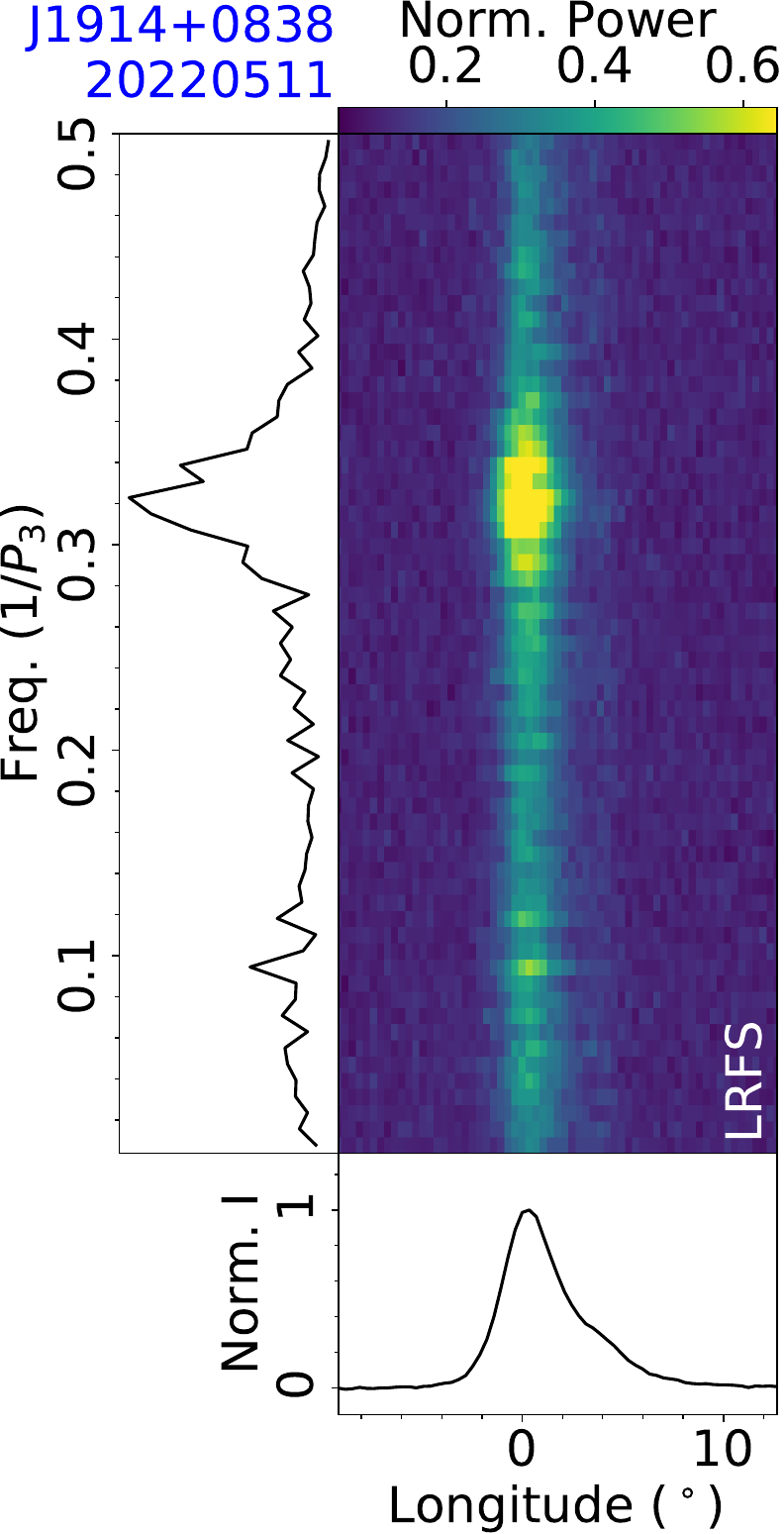}
\includegraphics[width=0.22\textwidth, angle=0]{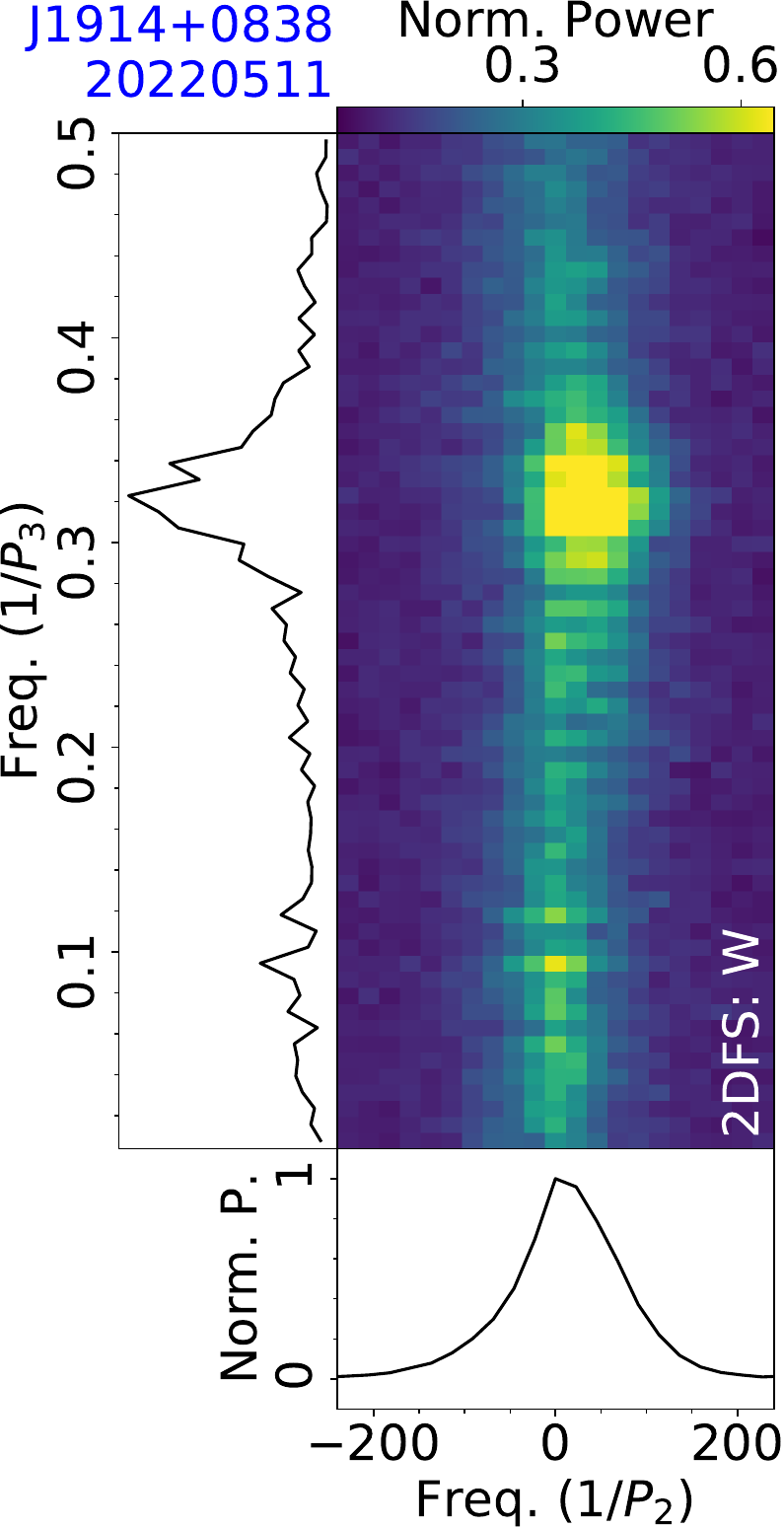}
\figcaption{Fluctuation analysis of PSR J1914+0838 for the observation on 20220511, with LRFS and 2DFS for the on-pulse region of a mean pulse profile.
\label{subfig:fluctu:J1914+0838}}
\end{figure}

\subsection{J1913+3732}
\label{subsec:J1913+3732}

This pulsar was discovered by the Northern High Time Resolution Universe survey conducted with the 100-m Effelsberg radio telescope \citep{Barr2013}. 

This pulsar was observed by FAST on 20201128 for 9 minutes, deriving a rotation period $P=0.8511$~s and a dispersion measure $D\!M=72.0~{\rm cm^{-3}\,pc}$. 
Single pulse sequences in Fig.~\ref{subfig:fluctu:J1913+3732} show occasional nulling behavior. The nulling fraction is estimated to be 10$\pm$2\% from the on-pulse integral energy histogram in Fig.~\ref{subfig:Hist:J1913+3732}. 
In addition, there is a preferred negative drifting for the leading part in the mean pulse profile, from LRFS and 2DFS in Fig.~\ref{subfig:fluctu:J1913+3732}. The drift feature in 2DFS is characterized by the centroid frequencies of $1/P_3=0.441\pm0.001$ and $1/P_2=-12\pm4$, corresponding to drifting parameters of $P_3=2.266\pm0.003$ periods and $P_2=-30\pm11$ degrees.

\begin{figure}[htpb]
\includegraphics[width=0.22\textwidth, angle=0]{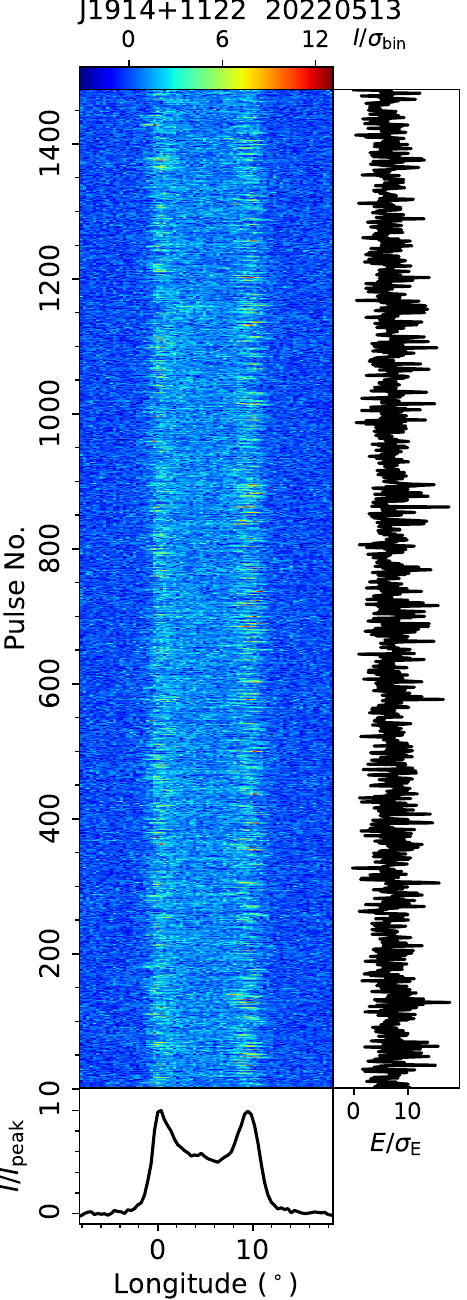}
\includegraphics[width=0.22\textwidth, angle=0]{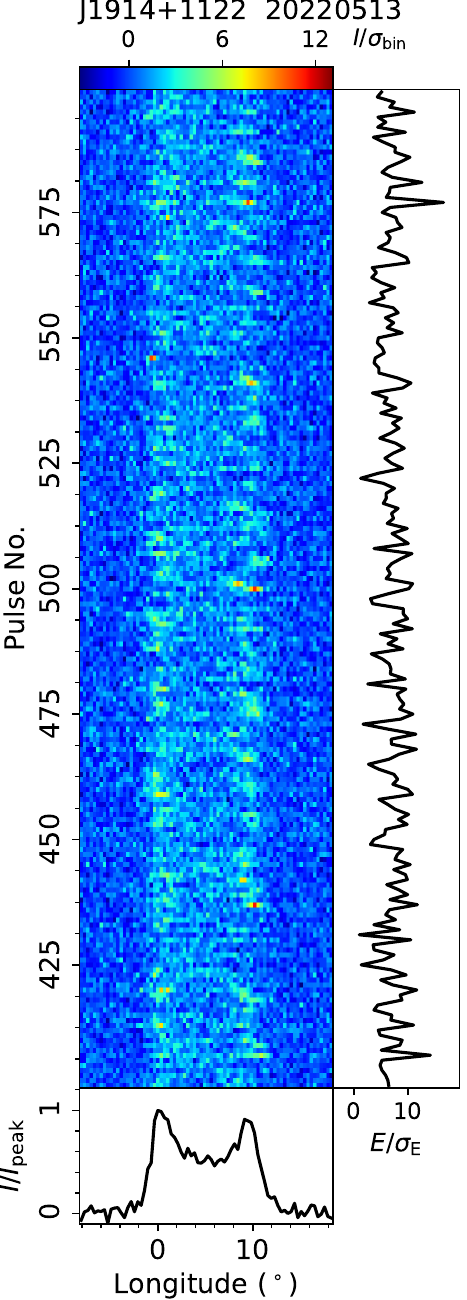}
\figcaption{Single pulse sequence of PSR J1914+1122 from the FAST observation on 20220513, and a zoomed-in view of pulses No.400-600.
\label{subfig:TP:J1914+1122}}
\end{figure}

\begin{figure}[htpb]
\centering
\includegraphics[width=0.22\textwidth, angle=0]{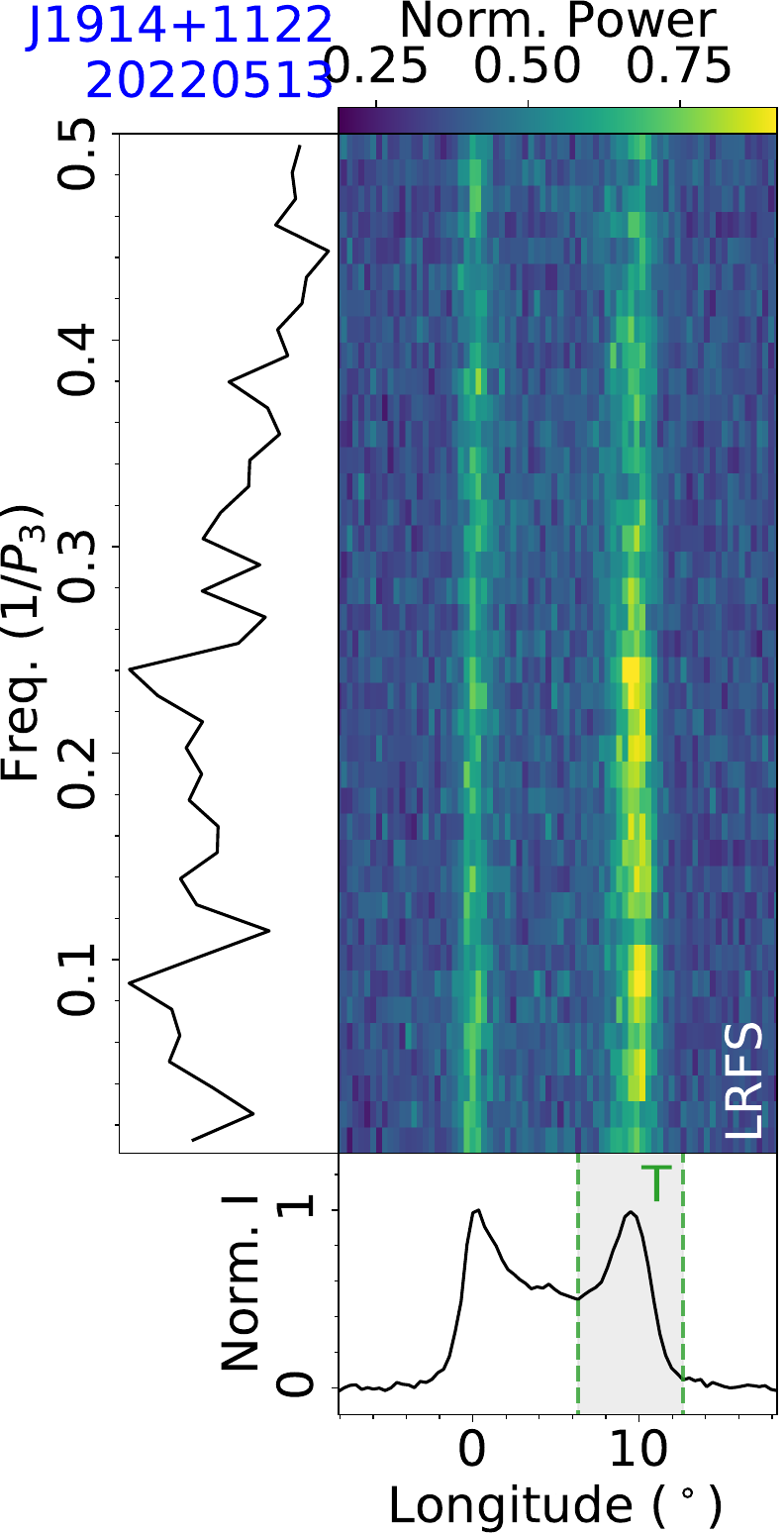}
\includegraphics[width=0.22\textwidth, angle=0]{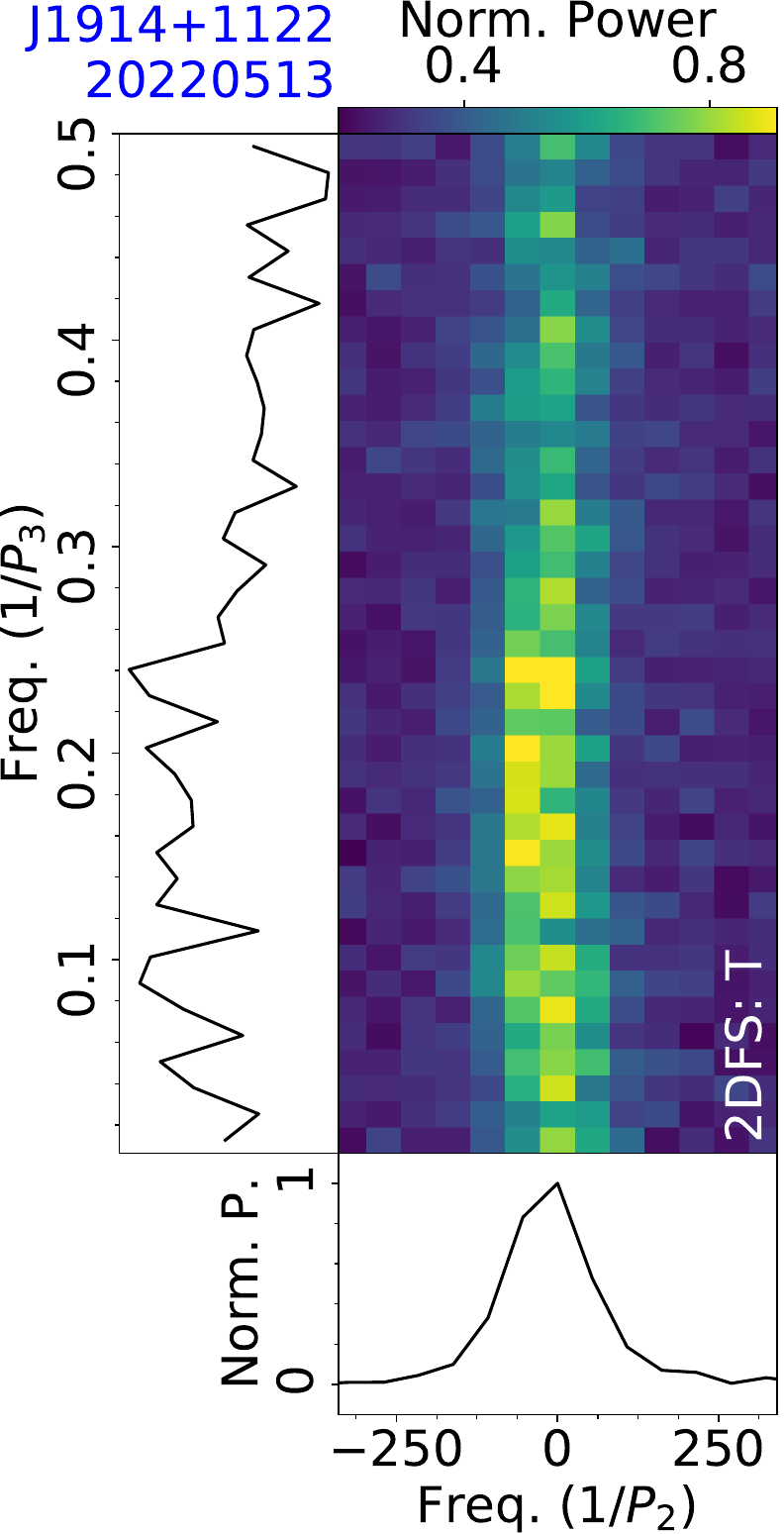}
\figcaption{Fluctuation analysis of PSR J1914+1122 for the observation on 20220513, with LRFS and 2DFS for the trailing part of a mean pulse profile.
\label{subfig:fluctu:J1914+1122}}
\end{figure}

\begin{figure}[htpb]
\centering
\includegraphics[width=0.44\textwidth, angle=0]{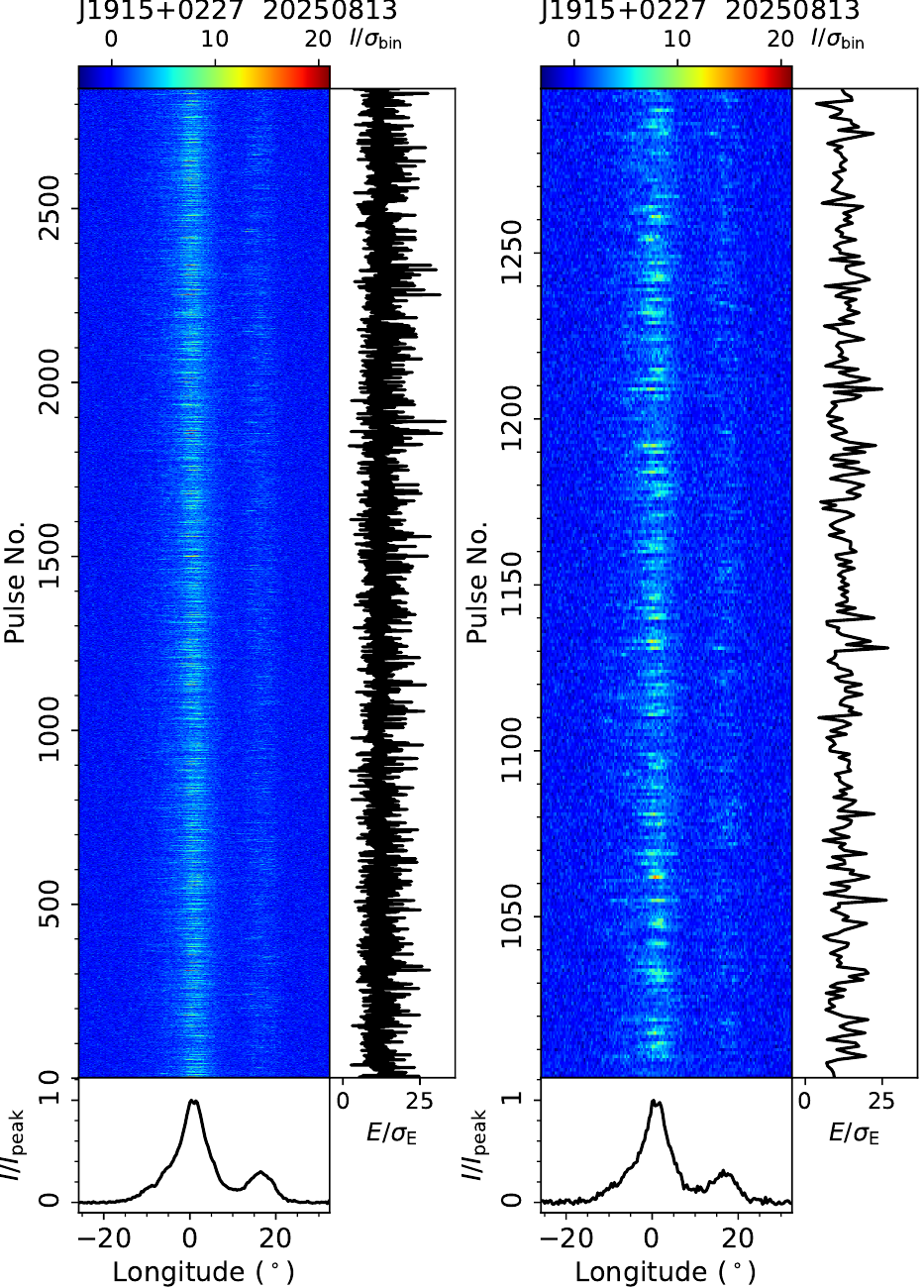}
\figcaption{Single pulse sequence of PSR J1915+0227 from the FAST observation on 20250813, and a zoomed-in view of pulses No. 1000-1300.
\label{subfig:TP:J1915+0227}}
\end{figure}

\begin{figure}[htpb]
\centering
\includegraphics[width=0.44\textwidth, angle=0]{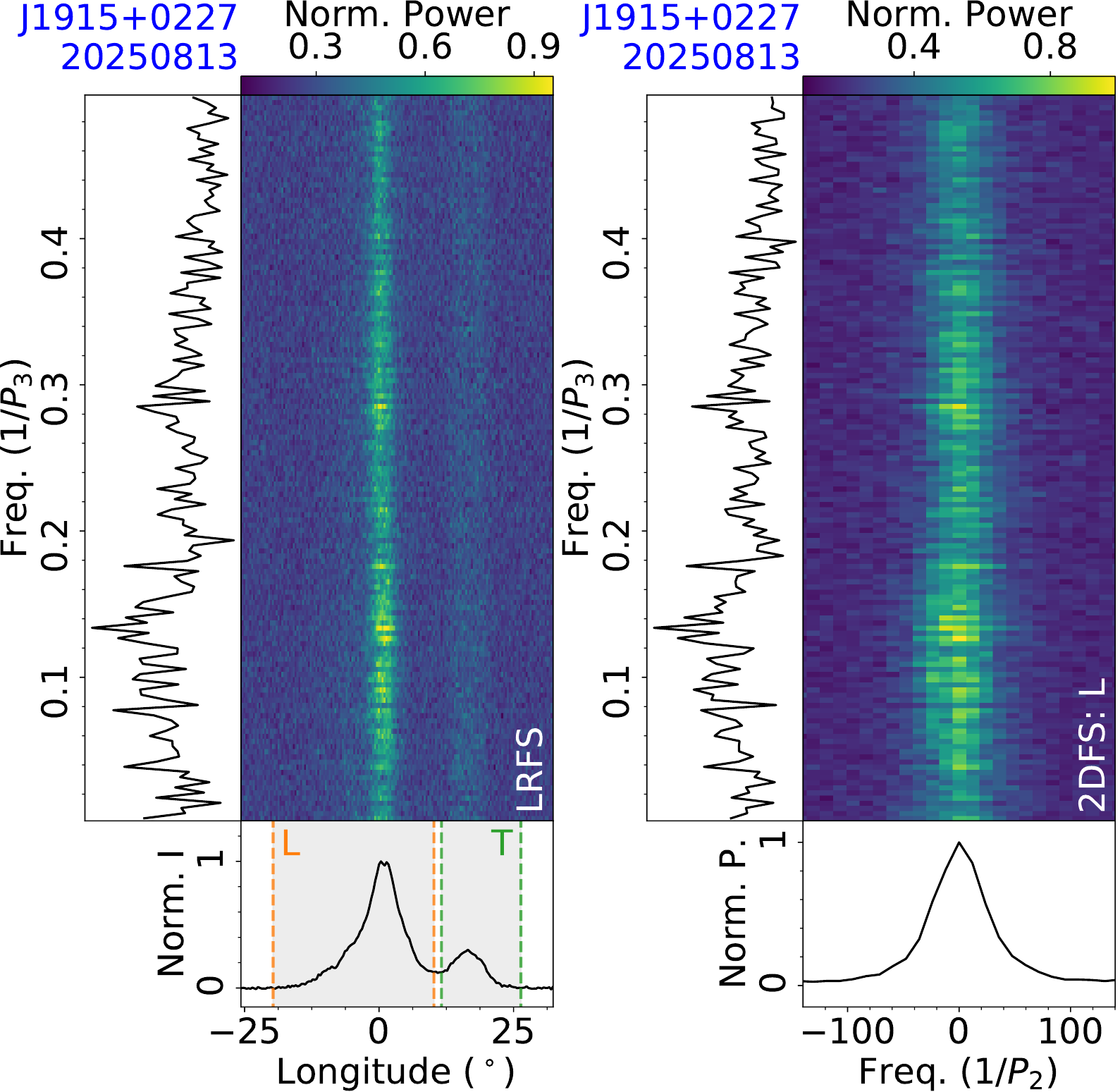}
\figcaption{Fluctuation analysis of PSR J1915+0227 for the observation on 20250813, with LRFS and 2DFS for leading part of the mean pulse profile.
\label{subfig:fluctu:J1915+0227}}
\end{figure}

\subsection{J1914+0219}
\label{subsec:J1914+0219}

PSR J1914+0219 was discovered by \citet{Lorimer2013} in the Parkes Multibeam Pulsar Survey. Subpulse drifting was reported by \citet{Song2023} with $P_3=6.2\pm0.8$ periods and $P_2=18^{+2}_{-3}$ degrees.

The pulsar was observed by FAST on 20240927 for 15 minutes, with a rotation period $P=0.4576$~s and a dispersion measure $D\!M=235.3~{\rm cm^{-3}\,pc}$ determined. 
Single pulse sequences in Fig.~\ref{subfig:TP:J1914+0219} display subpulse drifting and the short-duration decreases in pulse energy. In 2DFS of the leading part in the mean pulse profile, there is a positive drift feature with centroid frequencies of $1/P_3=0.166\pm0.001$ and $1/P_2=19\pm1$, corresponding to drift periodicities of $P_3=6.03\pm0.03$ periods and $P_2=19\pm1$ degrees. 
While there is no obvious evidence of drifting for the right profile part from this data.

\subsection{J1914+0838}
\label{subsec:J1914+0838}

PSR J1914+0838 was discovered in the Survey for Pulsars and Extragalactic Radio Bursts (SUPERB) with the Parkes radio telescope \citep{Keane2018}.

This pulsar was observed by FAST on 20220511 for 15 minutes, deriving a rotation period $P=0.4400$~s and a dispersion measure $D\!M=290.5~{\rm cm^{-3}\,pc}$. The single pulse sequence and a zoomed-in view of pulses No. 1-200 in Fig.~\ref{subfig:TP:J1914+0838} display the positive drifting phenomenon. Fluctuation spectra are shown in Fig.~\ref{subfig:fluctu:J1914+0838}, and the positive drift feature in 2DFS has a centroid at $1/P_3=0.322\pm0.001$ and $1/P_2=39\pm1$, corresponding to $P_3=3.11\pm0.01$ periods and $P_2=9.2\pm0.2$ degrees.

\subsection{J1914+1122}
\label{subsec:J1914+1122}

PSR J1914+1122 was discovered by the Arecibo telescope \citep{Hulse1975}. The drift feature of the second component has been reported by \citet{Song2023}, with $P_3=6\pm1$ periods and $P_2=-15^{+9}_{-9}$ degrees.

This pulsar was observed by FAST on 20200304 for 5 minutes and 20220513 for 15 minutes. From the 15-minute data, a rotation period $P=0.6010$~s and a dispersion measure $D\!M=100.2~{\rm cm^{-3}\,pc}$ are derived. The single pulse sequence and a zoomed-in view are shown in Fig.~\ref{subfig:TP:J1914+1122}. From the fluctuation in Fig.~\ref{subfig:fluctu:J1914+1122}, the trailing part in a mean pulse profile has a negative drift feature, with the centroid frequencies estimated to be $1/P_3=0.169\pm0.004$ and $1/P_2=-29\pm3$. Periodicities of $P_3=5.9\pm0.1$ periods and $P_2=-13\pm1$ degrees are consistent with previous studies \citep{Song2023}.

\begin{figure}[htpb]
\centering
\includegraphics[width=0.22\textwidth, angle=0]{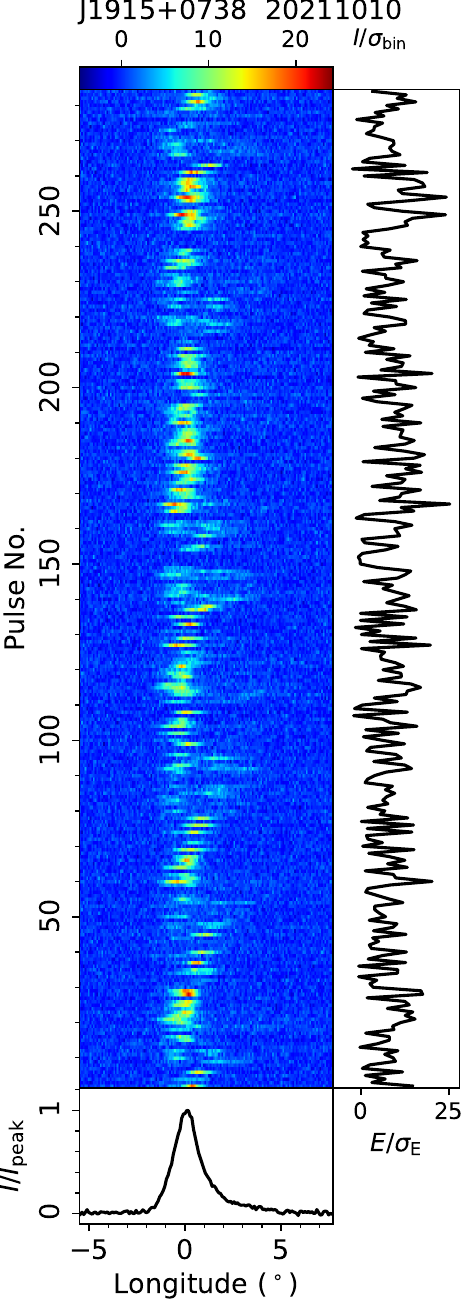}
\includegraphics[width=0.22\textwidth, angle=0]{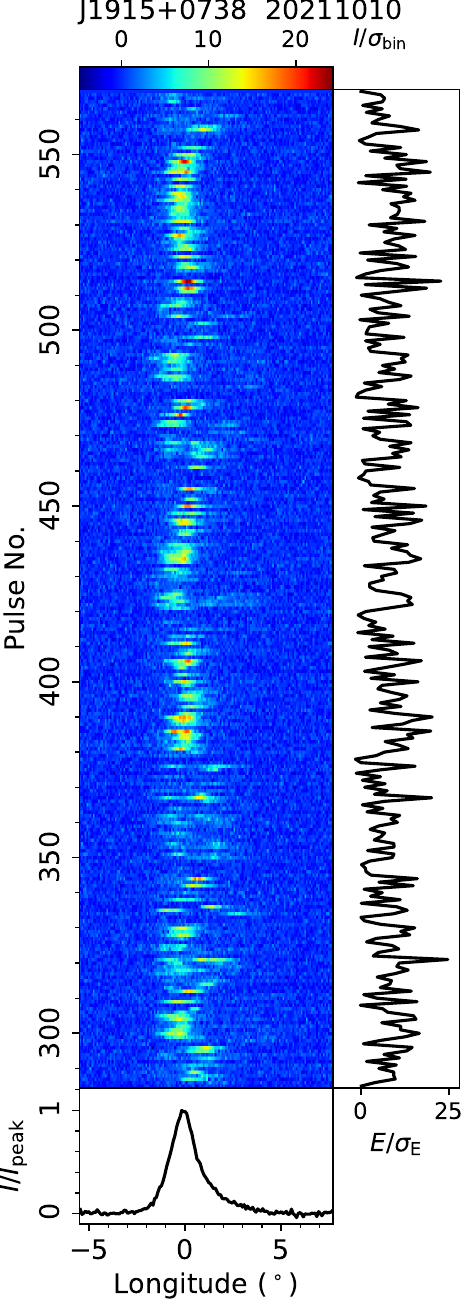}
\figcaption{Single pulse sequences of PSR J1915+0738 from the FAST observation on 20211010.
\label{subfig:TP:J1915+0738}}
\end{figure}

\begin{figure}[htpb]
\centering
\includegraphics[width=0.22\textwidth, angle=0]{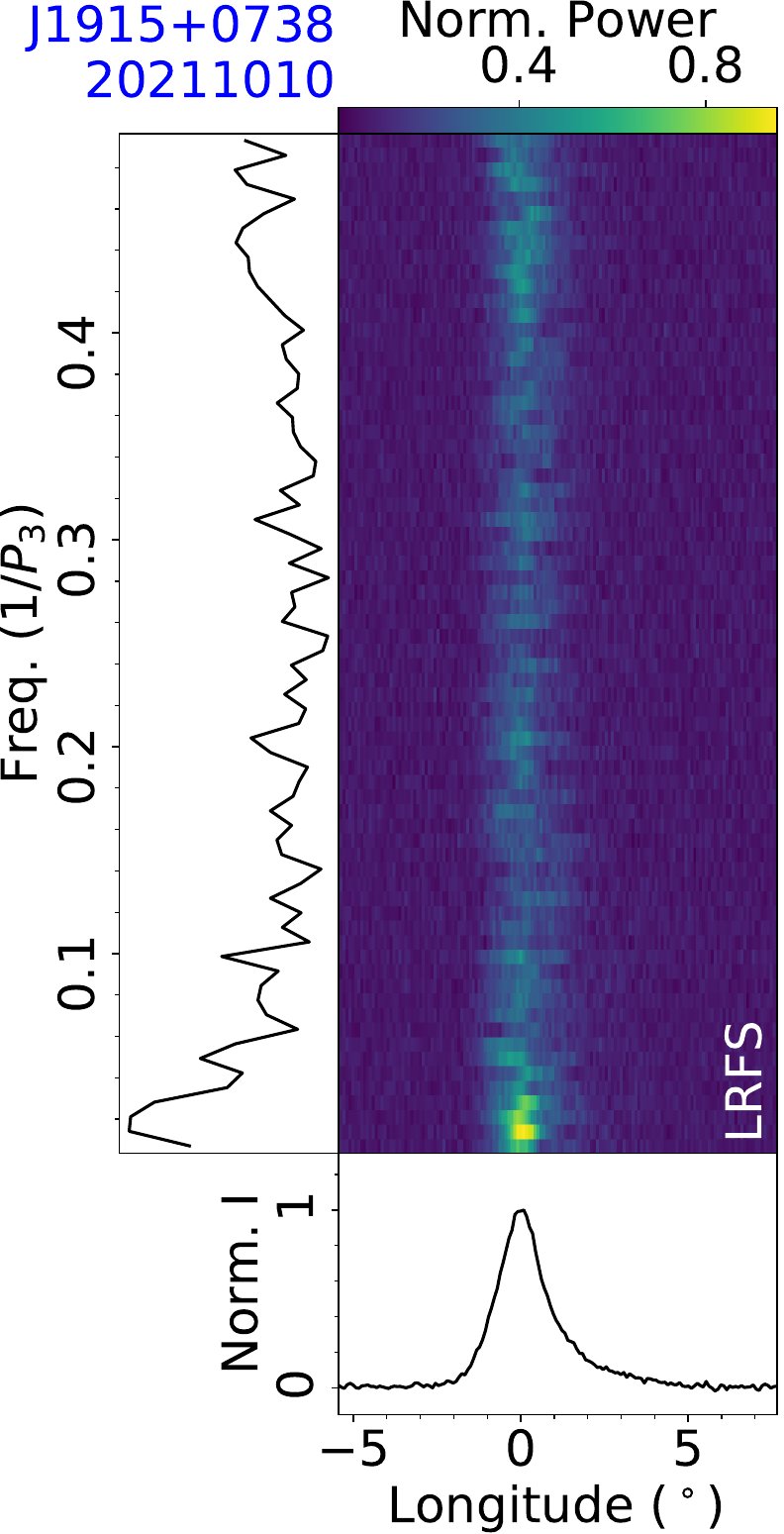}
\includegraphics[width=0.22\textwidth, angle=0]{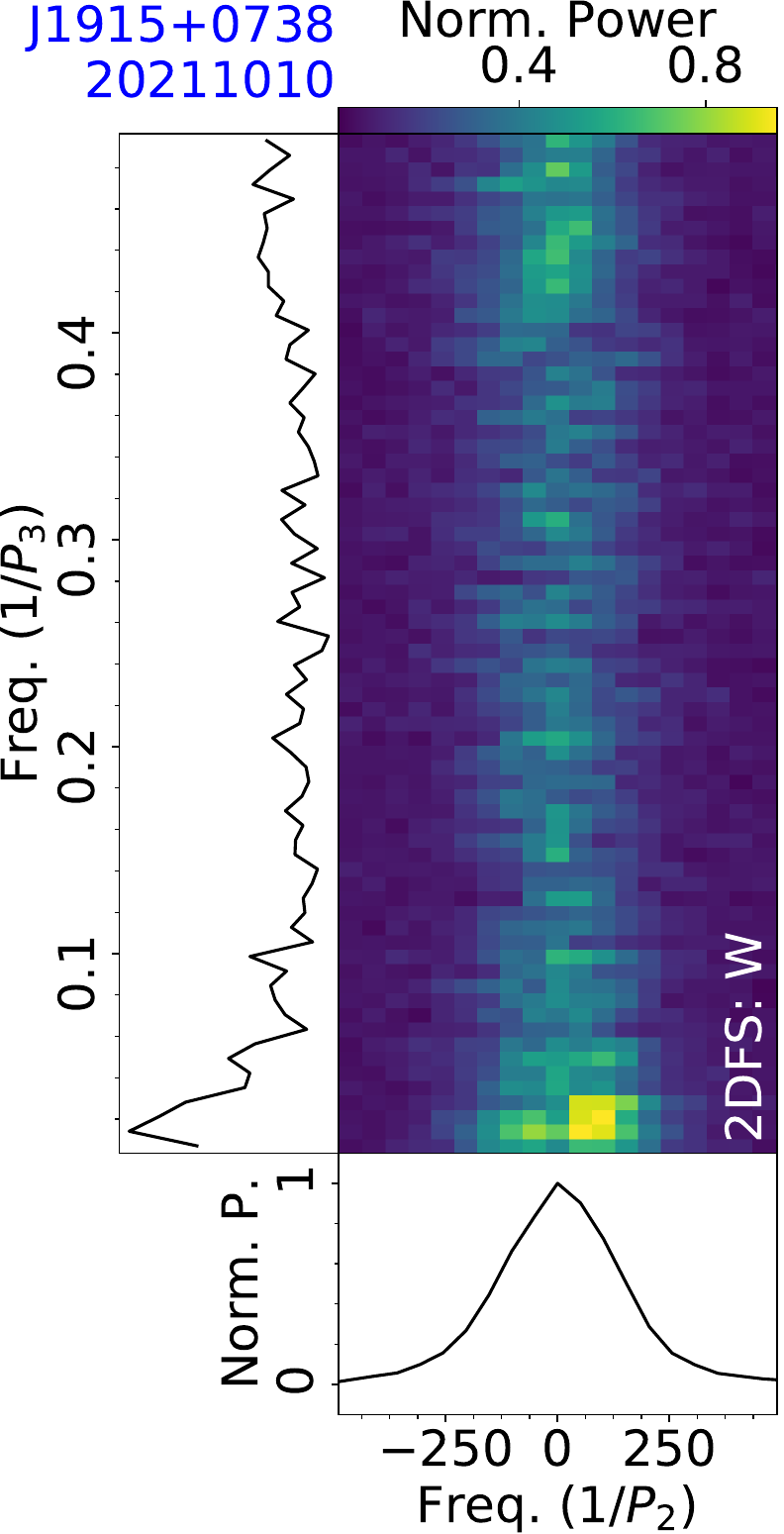}
\figcaption{Fluctuation analysis of PSR J1915+0738 for the observation on 20211010, with LRFS and 2DFS for the on-pulse region of a mean pulse profile.
\label{subfig:fluctu:J1915+0738}}
\end{figure}

\subsection{J1915+0227}
\label{subsec:J1915+0227}

PSR J1915+0227 was discovered in the Parkes Multibeam Pulsar Survey \citep{Lorimer2006}. \citet{Song2023} reported the parameters of the negative drifting behavior as $P_3=9\pm2$ periods and $P_2=163^{+114}_{-70}$ degrees.

This pulsar was observed by FAST on 20201229 for 5 minutes and on 20250813 for 15 minutes. From the longer data, a rotation period $P=0.3173$~s and a dispersion measure $D\!M=192.1~{\rm cm^{-3}\,pc}$ were derived. The single pulse sequence and a zoomed-in view of pulses No. 1000-1300 from the observation on 20250813 (Fig.~\ref{subfig:TP:J1915+0227}) illustrate the presence of subpulse modulation for the leading profile part. LRFS and 2DFS of the leading part are shown in Fig.~\ref{subfig:fluctu:J1915+0227}. The centroid of the negative drift feature in 2DFS is at $1/P_3=0.110\pm0.001$ and $1/P_2=-2.5\pm0.6$, corresponding to $P_3=9.1\pm0.1$ periods and $P_2=-146\pm35$ degrees.

\subsection{J1915+0738}
\label{subsec:J1915+0738}

PSR J1915+0738 was discovered using the 305 m radio telescope at Arecibo \citep{Nice1995}. Drifting parameters of $P_3=37(40)$ periods and $P_2=14^{+2}_{-10}$ degrees were reported by \citet{Song2023}.

This pulsar was observed by FAST on 20211010, 20211208, and 20220117, with each observation conducted for a duration of 15 minutes. From the data of 20211010, a rotation period $P=1.5428$~s and a dispersion measure $D\!M=38.0~{\rm cm^{-3}\,pc}$ were determined. 
Single pulse sequences of the observation on 20211010 are shown in Fig.~\ref{subfig:TP:J1915+0738}, where subpulse drifting is not systematic and sometimes changes directions. There is also a modulation of about 2 periods. Fluctuation spectra are displayed in Fig.~\ref{subfig:fluctu:J1915+0738}, and there are two drift features as well as a modulation feature. The main feature in 2DFS is the positive drift feature, which has the centroid frequencies $1/P_3=0.027\pm0.001$ and $1/P_2=87\pm4$, yielding $P_3=37\pm1$ periods and $P_2=4.1\pm0.2^\circ$. The negative drift feature exhibits the centroid of $1/P_3=0.025\pm0.002$ and $1/P_2=-71\pm6$, corresponding to $P_3=40\pm2$ periods and $P_2=-5.0\pm0.4^\circ$. 
There is also a modulation feature with the centroid of $1/P_3=0.450\pm0.002$, that corresponds to the periodicity of $P_3=2.22\pm0.01$ periods. For the other two observations, the drifting properties are consistent with 20211010, while there is no obvious modulation feature of ~2 periods in 2DFS.

\begin{figure}[htpb]
\centering
\includegraphics[width=0.22\textwidth, angle=0]{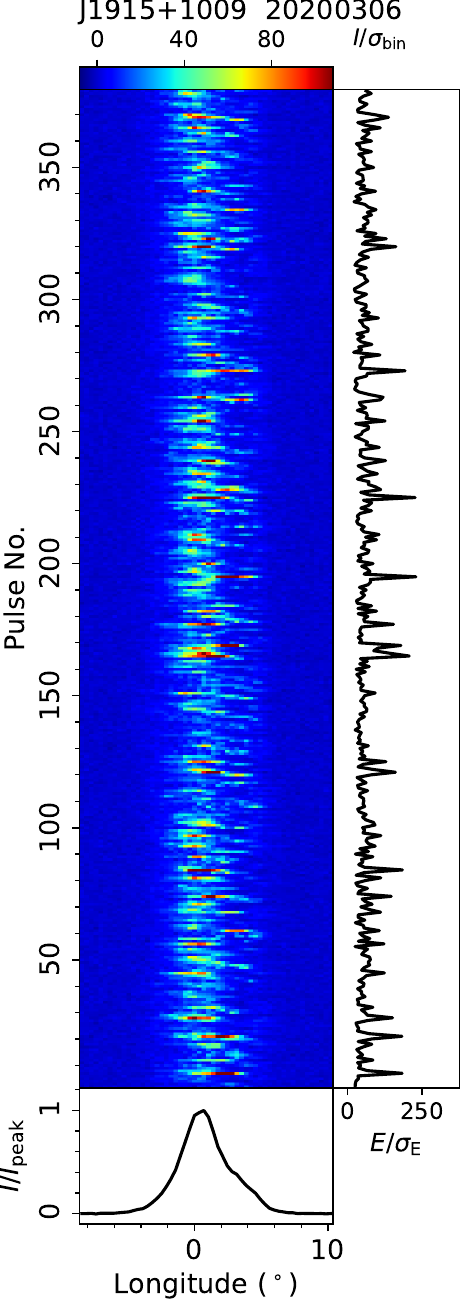}
\includegraphics[width=0.22\textwidth, angle=0]{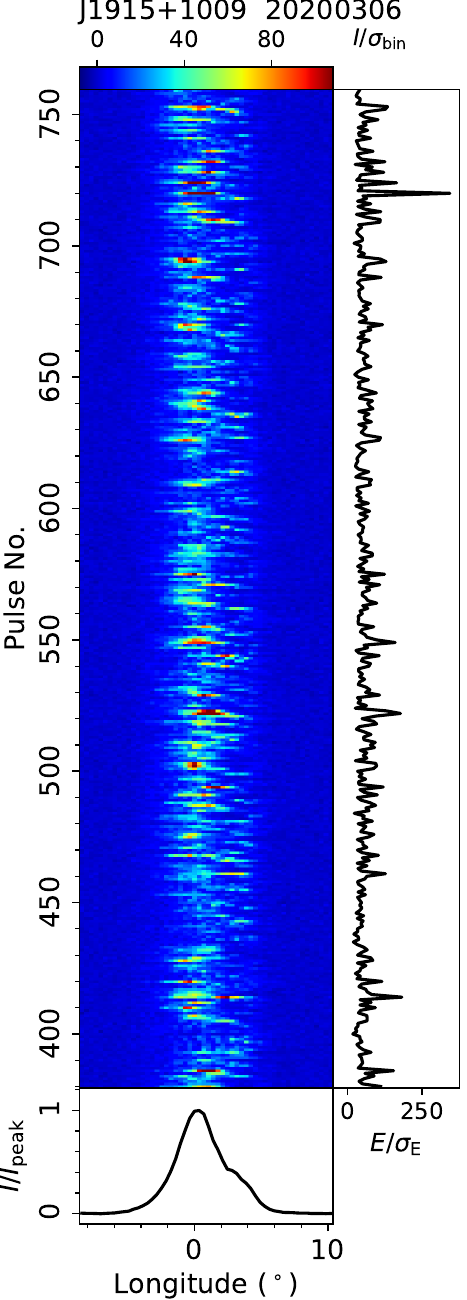}
\figcaption{Single pulse sequences of PSR J1915+1009 from the FAST observation on 20200306.
\label{subfig:TP:J1915+1009}}
\end{figure}

\begin{figure}[htpb]
\centering
\includegraphics[width=0.22\textwidth, angle=0]{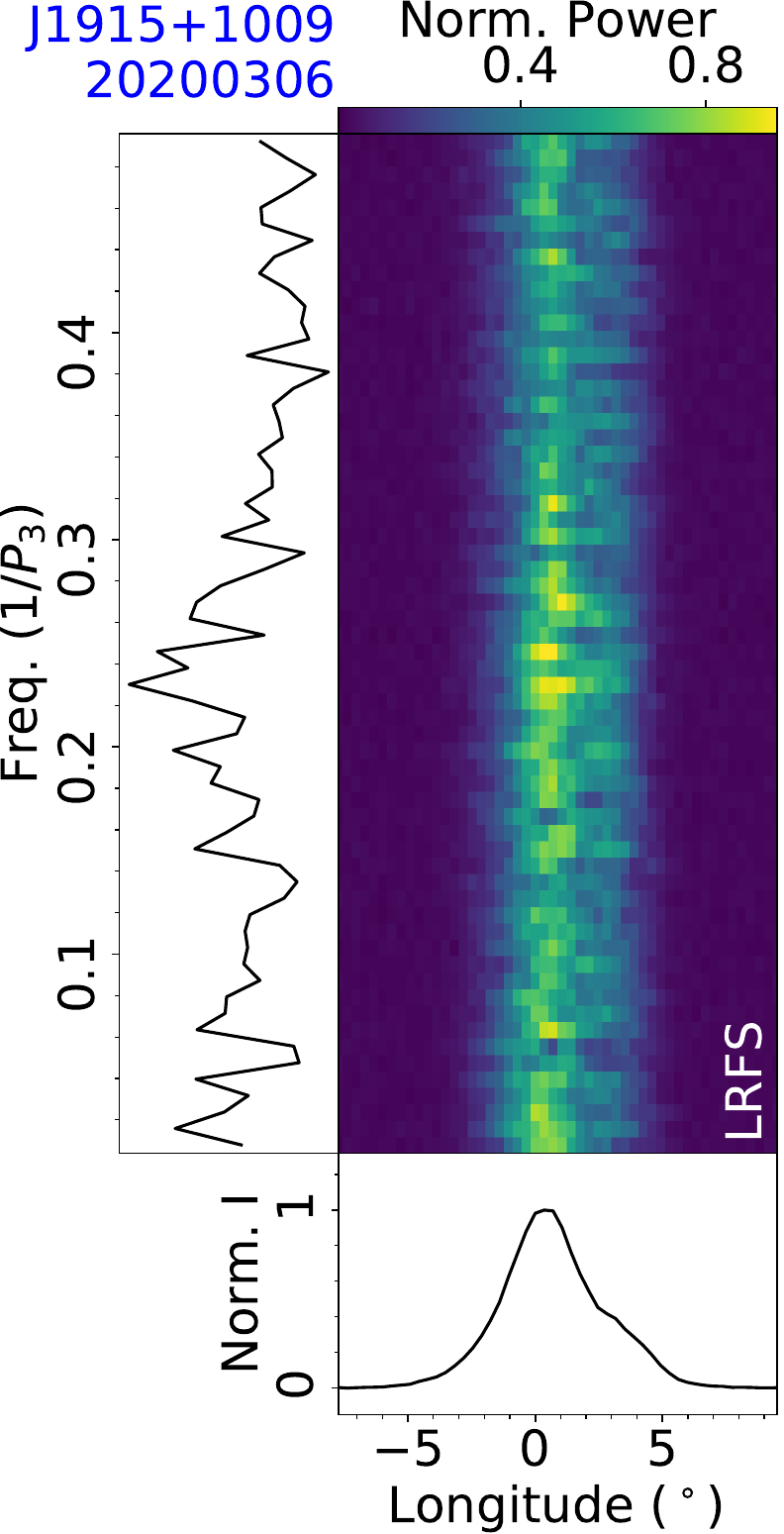}
\includegraphics[width=0.22\textwidth, angle=0]{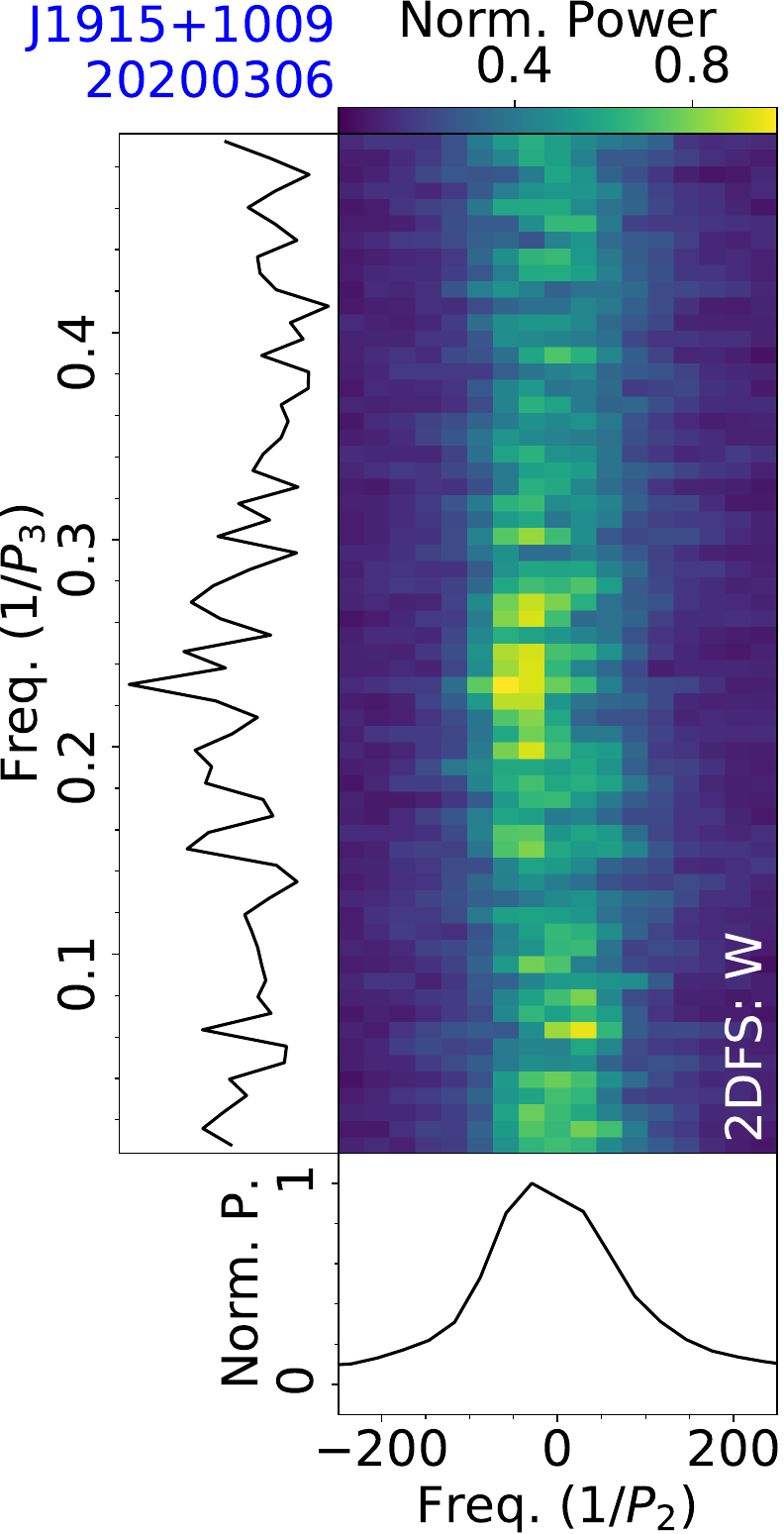}
\figcaption{Fluctuation analysis of PSR J1915+1009 for the observation on 20200306, with LRFS and 2DFS for the on-pulse region of a mean pulse profile.
\label{subfig:fluctu:J1915+1009}}
\end{figure}

\begin{figure}[htpb]
\centering
\includegraphics[width=0.21\textwidth, angle=0]{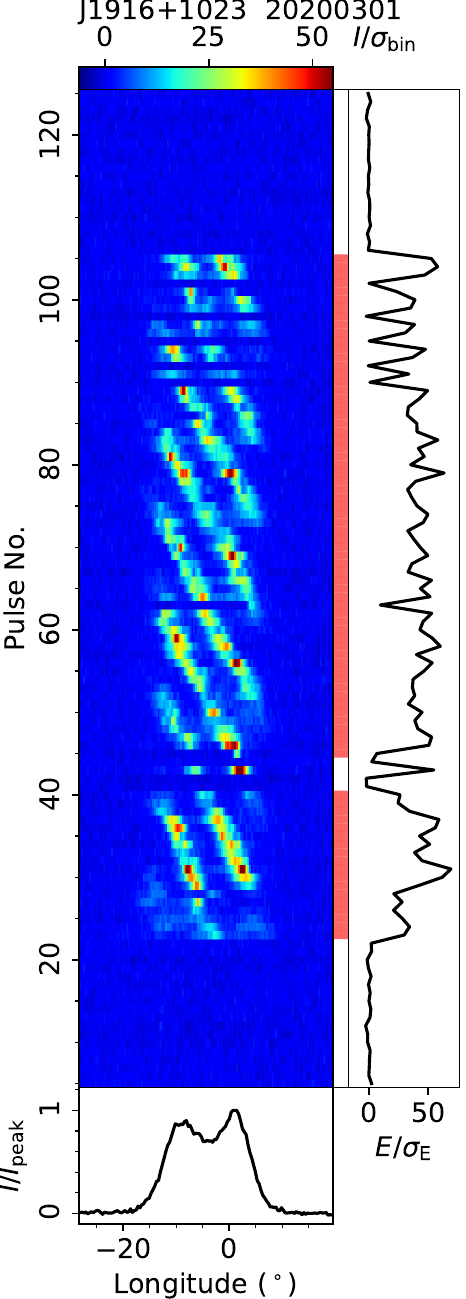}
\includegraphics[width=0.21\textwidth, angle=0]{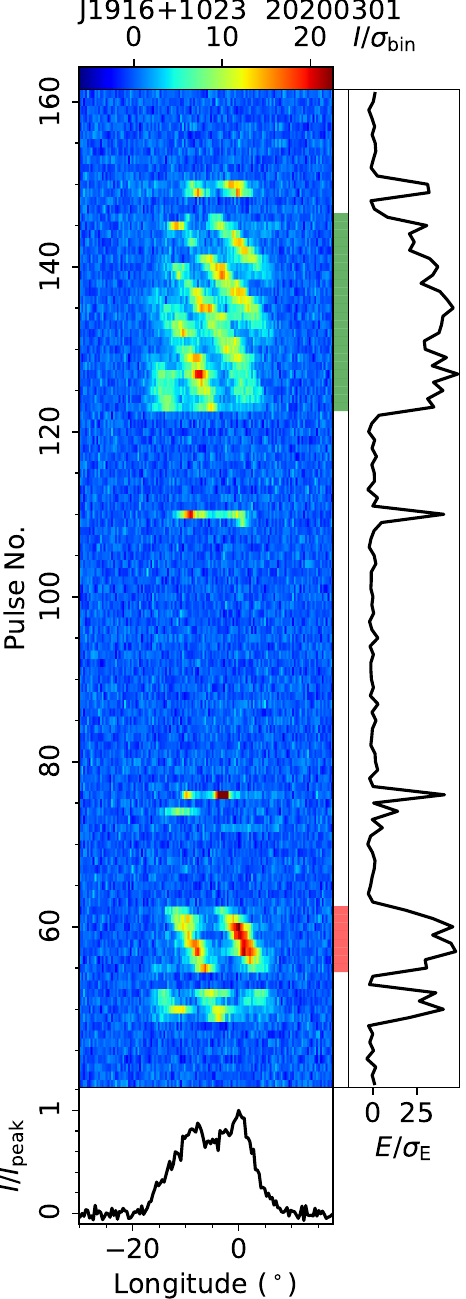}\\
\includegraphics[width=0.21\textwidth, angle=0]{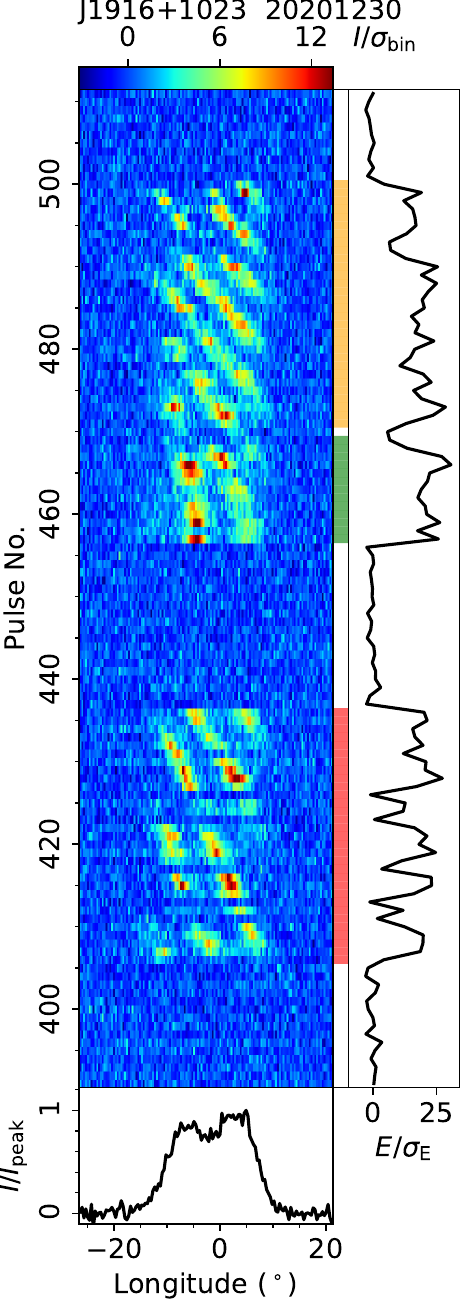}
\includegraphics[width=0.21\textwidth, angle=0]{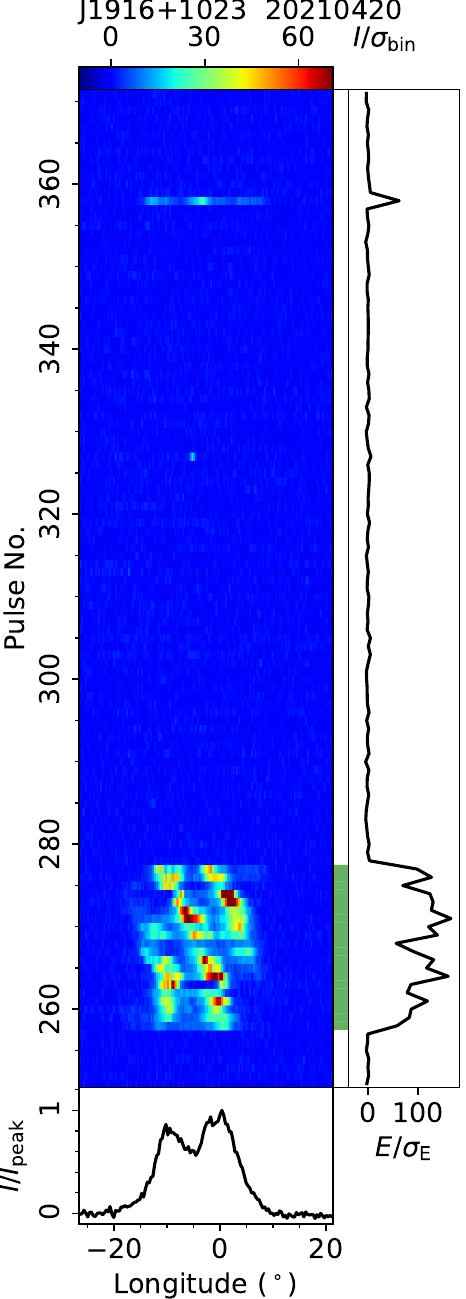}
\figcaption{Single pulse sequences of PSR J1916+1023 from FAST observations on 20200301, 20201230 and 20220202.
\label{subfig:TP:J1916+1023}}
\end{figure}

\begin{figure}[htpb]
\centering
\includegraphics[width=0.39\textwidth, angle=0]{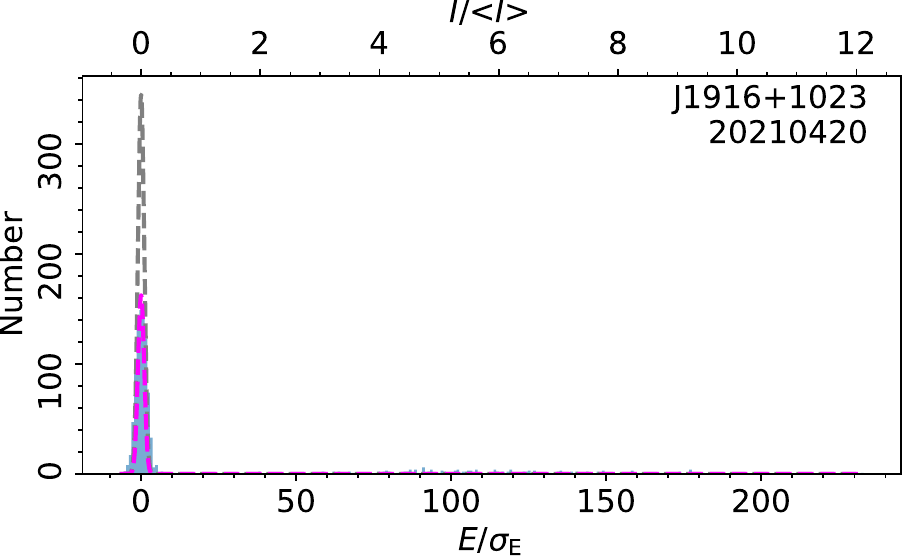}
\figcaption{On-pulse energy histogram of single pulses of PSR J1916+1023 from the FAST observation on 20210420.
\label{subfig:Hist:J1916+1023}}
\end{figure}

\begin{figure}[htpb]
\centering
\includegraphics[width=0.42\textwidth, angle=0]{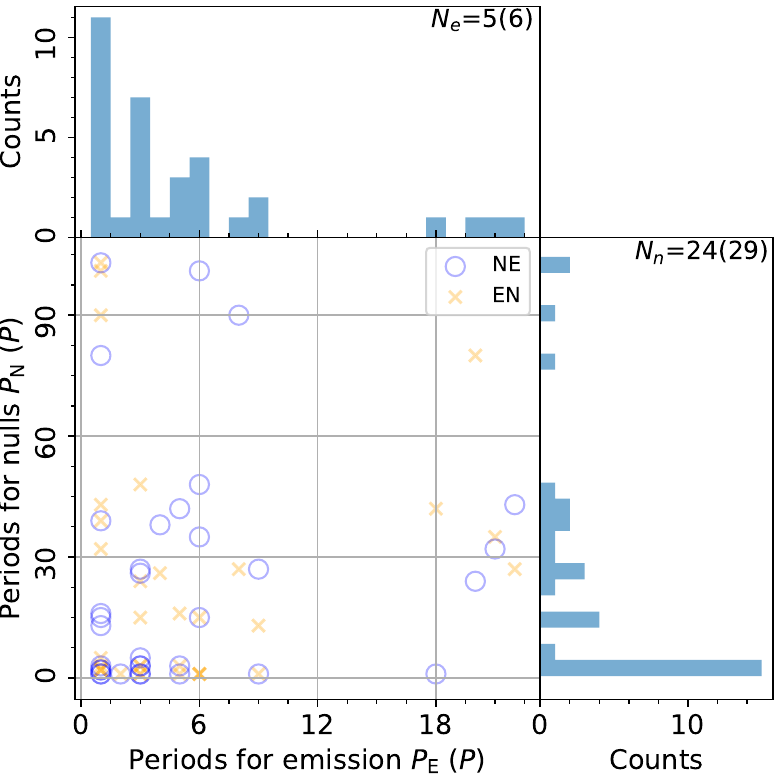}
\figcaption{Distribution of period numbers for continuous nulling $P_N$ against period numbers for adjacent pulses $P_E$ of PSR J1916+1023 observed by FAST on 20210420, as well as the duration histograms for the emission and null shown in the top and right panels, respectively. 
\label{subfig:scaleHist:J1916+1023}}
\end{figure}

\begin{figure}[htpb]
\centering
\includegraphics[width=0.39\textwidth, angle=0]{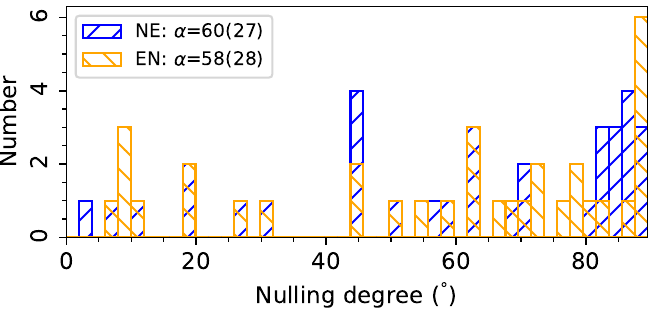}\\
\includegraphics[width=0.39\textwidth, angle=0]{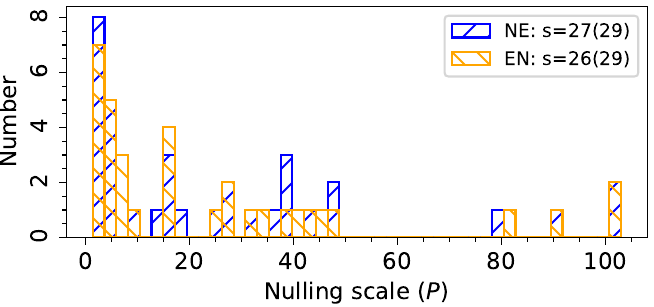}
\figcaption{Histograms of the nulling degree and nulling scale for PSR J1916+1023 observed by FAST on 20210420.
\label{subfig:nullDegreeScale:J1916+1023}}
\end{figure}

\begin{figure}[htpb]
\centering
\includegraphics[width=0.39\textwidth, angle=0]{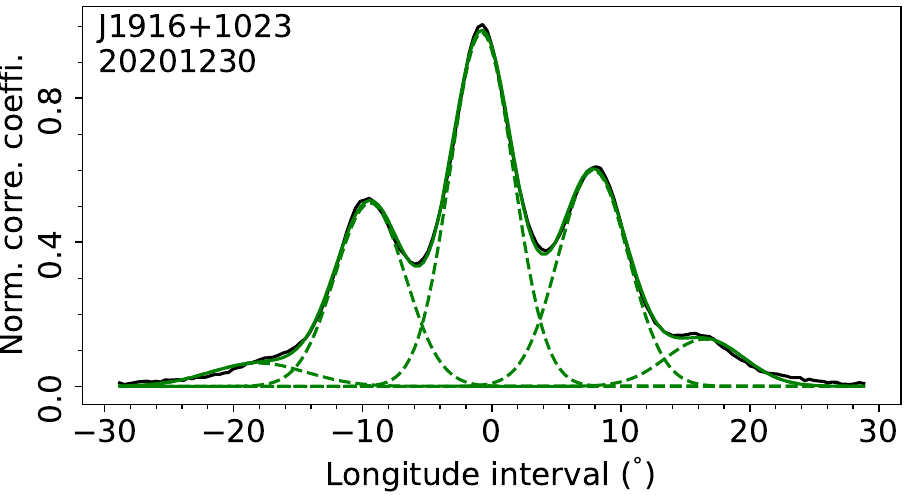}
\includegraphics[width=0.39\textwidth, angle=0]{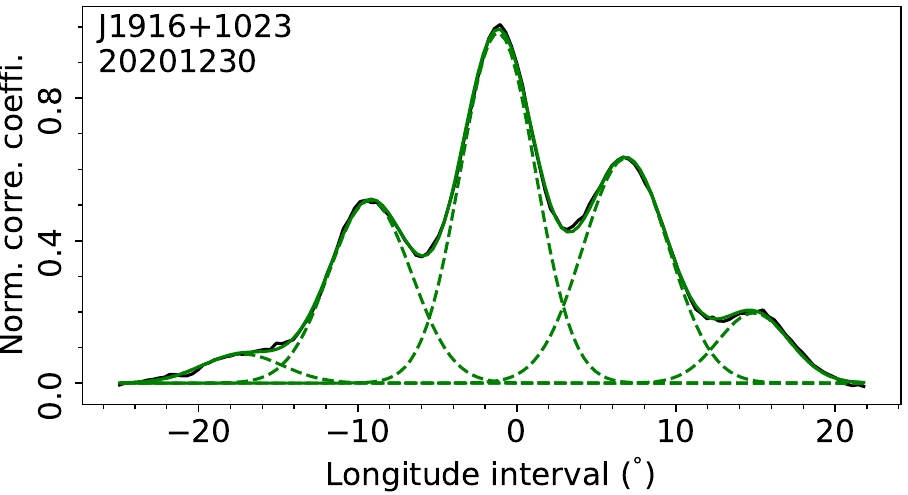}
\figcaption{Cross correlations for the slow drifting mode and fast drifting mode of PSR J1916+1023 from the FAST observation on 20201230. \label{subfig:Corre:J1916+1023}}
\end{figure}

\begin{figure}[htpb]
\centering
\includegraphics[width=0.42\textwidth, angle=0]{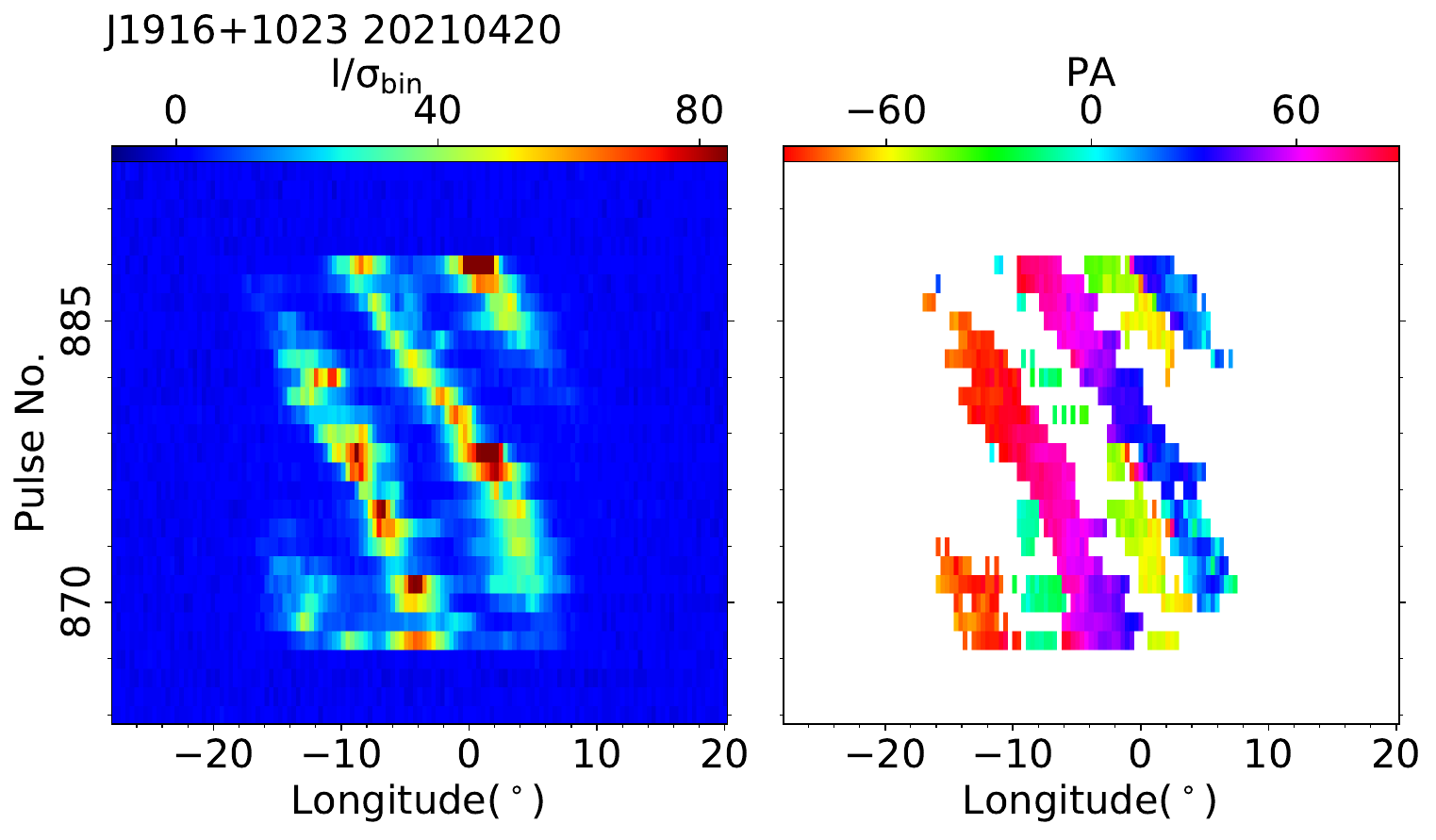}
\figcaption{A segment of single pulses of intensity I and PA from the FAST observation on 20210420 of PSR J1916+1023.
\label{subfig:PolSeg:J1916+1023}}
\end{figure}

\subsection{J1915+1009}
\label{subsec:J1915+1009}

PSR J1915+1009 was discovered by the Arecibo telescope \citep{Hulse1975}. Negative drift feature with $P_3=4.2\pm0.4$ periods and $P_2=-18_{-9}^{+7}$ degrees was reported by \citet{Song2023}.

This pulsar was observed by FAST on 20200306 and 20211118 for 5 minutes. From the data of 20200306, a rotation period $P=0.4045$~s and a dispersion measure $D\!M=241.7~{\rm cm^{-3}\,pc}$ were determined. 
Single pulse sequences of the observation on 20200306 are displayed in Fig.~\ref{subfig:TP:J1915+1009}. 
Fluctuation spectra in Fig.~\ref{subfig:fluctu:J1915+1009} exhibit a negative drift feature, with the centroid frequencies of $1/P_3=0.233\pm0.002$ and $1/P_2=-44\pm2$, corresponding to drifting periodicities of $P_3=4.30\pm0.03$ periods and $P_2=-8.2\pm0.4^\circ$.

\subsection{J1916+1023}
\label{subsec:J1916+1023}

PSR J1916+1023 was discovered in the Parkes multibeam pulsar survey \citep{hfs+04}. The nulling of this pulsar was first reported by \citet{Wang2007} with a nulling fraction of 47\% at 1518MHz. 

This pulsar was observed by FAST on 20200301 for 5 minutes, 20201230 for 20 minutes, 20210420 for 28 minutes, and 20220202 for 15 minutes. From data of 20210420, a rotation period $P=1.6182$~s and a dispersion measure $D\!M=341.0~{\rm cm^{-3}\,pc}$ were determined. 
Four segments are shown in Fig.~\ref{subfig:TP:J1916+1023}. From these sensitive FAST observations, the pulsar is found to have different drifting modes, in addition to the nulling behavior. 

\subsubsection{Nulling}

The nulling fraction of the observation on 20210420 is estimated to be 47$\pm$5\% from the on-pulse energy histogram (Fig.~\ref{subfig:Hist:J1916+1023}). Distributions of duration for adjacent nulling and emission of this observation are shown in Fig.~\ref{subfig:scaleHist:J1916+1023}. From single pulse sequences in Fig.~\ref{subfig:TP:J1916+1023}, there are short-duration nulls occurring in drifting bands, as well as short-duration emission in the nulling state. 
The 63th single pulse of the observation on 20200301 (Fig.~\ref{subfig:TP:J1916+1023}) displays partial null. The moment when nulling ends and emission begins corresponds to a position within the third subpulse, indicating that the short-duration nulls of this pulsar are not caused by the carousel rotation. 
The duration of emission ranging from 1 to 22 periods with an average of 5 periods, that is shorter than the duration of nulling state ranging from 1 to 103 periods with an average of 24 periods. 
From the histogram in Fig.~\ref{subfig:nullDegreeScale:J1916+1023}, the nulling degree and scale are estimated to be 55$\pm$32 degrees and 27$\pm$29 periods for NE pairs. For EN pairs, they are 58$\pm$28 degrees and 26$\pm$29 periods.

There is a also dwarf pulse appearing in the nulling state, that is the pulse No.327 in the observation on 20210420 (Fig.~\ref{subfig:TP:J1916+1023}).

\subsubsection{Drifting and mode changing}

There are three drifting modes from the FAST observations, labeled by bars with different colors in Fig.~\ref{subfig:TP:J1916+1023}. Drifting of two modes is relatively stable, and drifting parameters are calculated from the correlation method (Fig.~\ref{subfig:Corre:J1916+1023}). The drifting rate of one mode changes continuously from slow to fast.
For most drifting bands, the drifting rate is stable and slow with drifting rate of $-$0.7 degrees per period and $P_2$ of 8.7$^\circ$ on average, which is defined as Sd mode (red bars in Fig.~\ref{subfig:TP:J1916+1023}). 
The segment of pulse Nos. 457-550 in the left bottom of Fig.~\ref{subfig:TP:J1916+1023} from the observation on 20201230 shows that the emission begins with a very slow drifting rate, slower than that of the Sd mode. The drifting rate is faster and faster continually, eventually stable at the fastest drifting rate. The state with stable fast drifting rate is defined as Fd mode in yellow color, with the drifting parameters $P_2=-8.03\pm0.01^\circ$ and $D=-1.19\pm0.06$ degrees per period estimated from the correlation method. 
The state of drifting rate increasing continually is named Tr mode in green color. The Tr mode also appears without following Fd mode in other data, such as Nos. 123-145 of the observation on 20200301, and Nos. 210-232 of the observation on 20220202.

Fig.~\ref{subfig:PolSeg:J1916+1023} shows single pulse sequences (pulses No.864-893) of intensity and PA from the observation on 20210420, illustrating that 90$^\circ$ PA jumps occur within subpulses and orthogonal polarization modes of subpulses are along with drifting bands.

\begin{figure}[htpb]
\centering
\includegraphics[width=0.22\textwidth, angle=0]{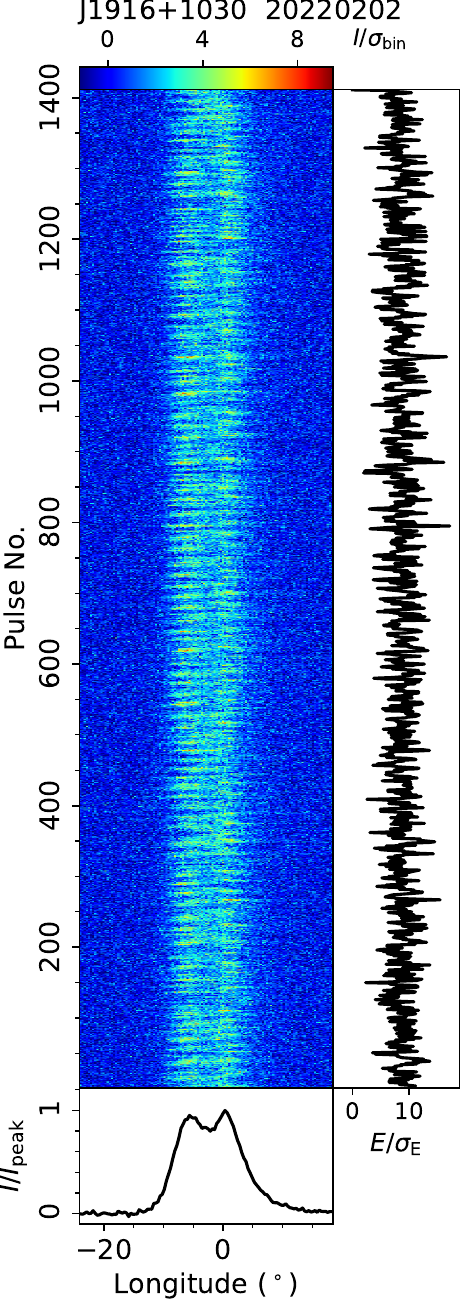}
\includegraphics[width=0.22\textwidth, angle=0]{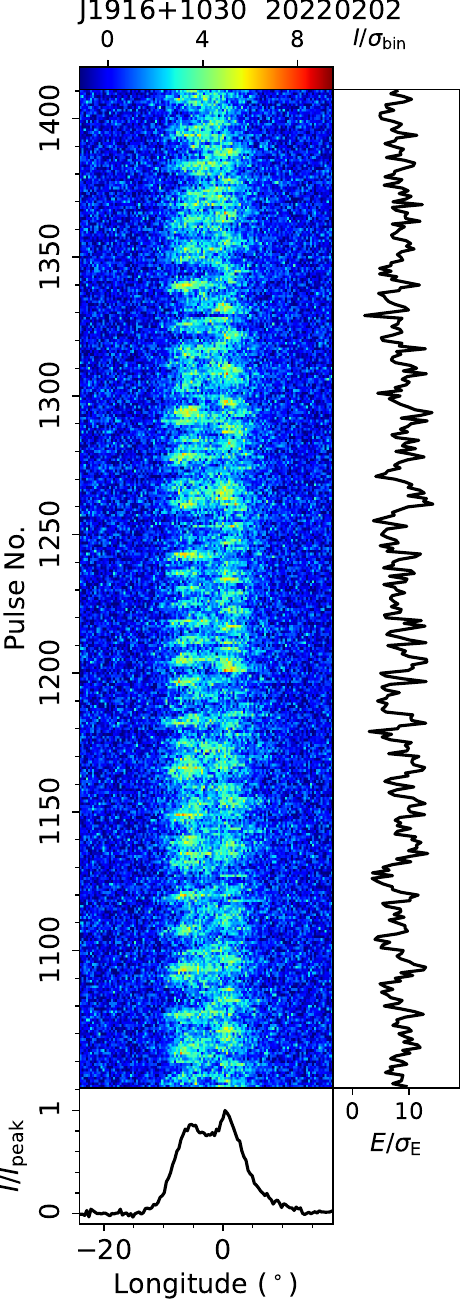}
\figcaption{Single pulse sequence of PSR J1916+1030 from the FAST observation on 20220202, and a zoomed-in view of pulses No. 1050-1410.
\label{subfig:TP:J1916+1030}}
\end{figure}

\begin{figure}[htpb]
\centering
\includegraphics[width=0.22\textwidth, angle=0]{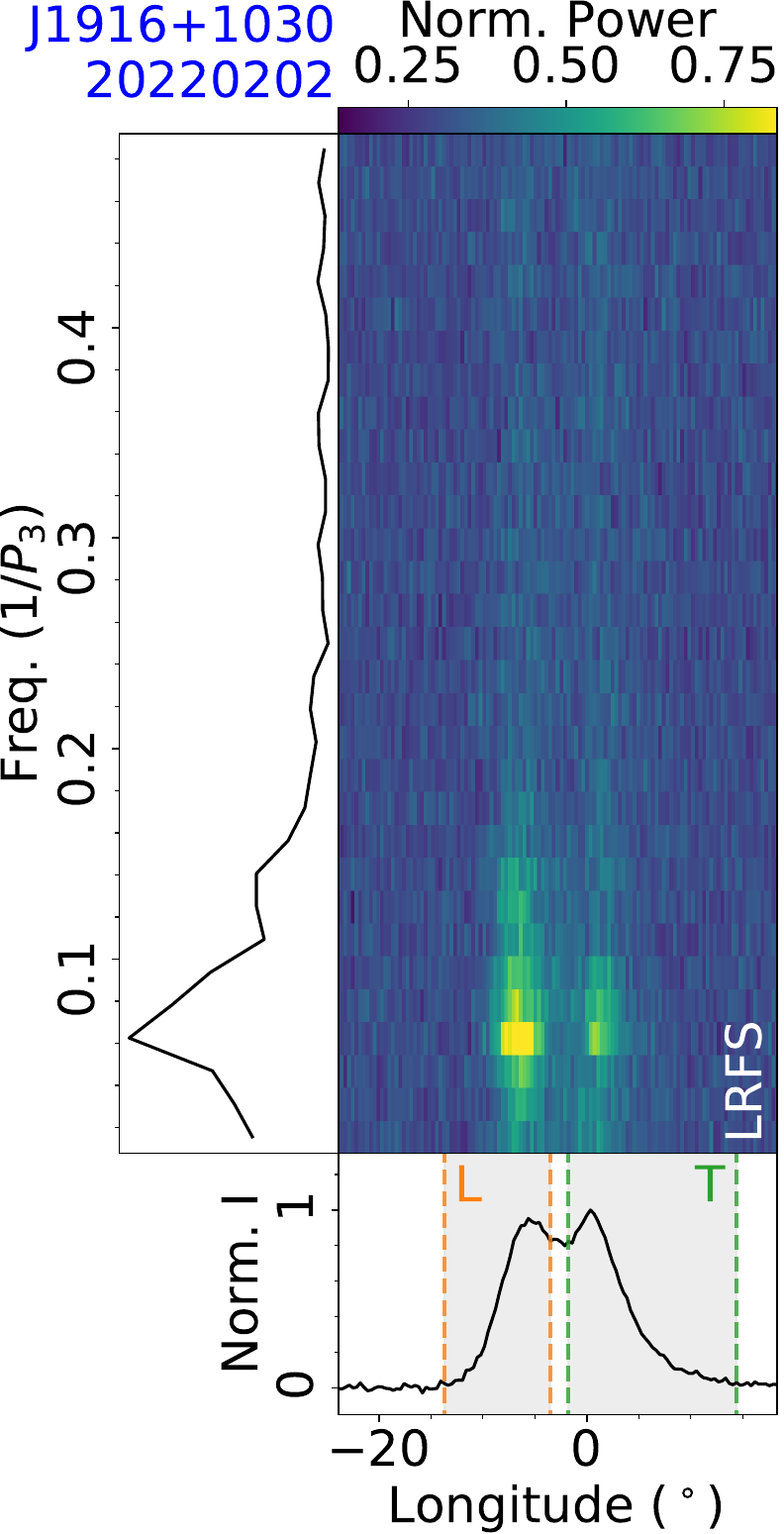}
\includegraphics[width=0.22\textwidth, angle=0]{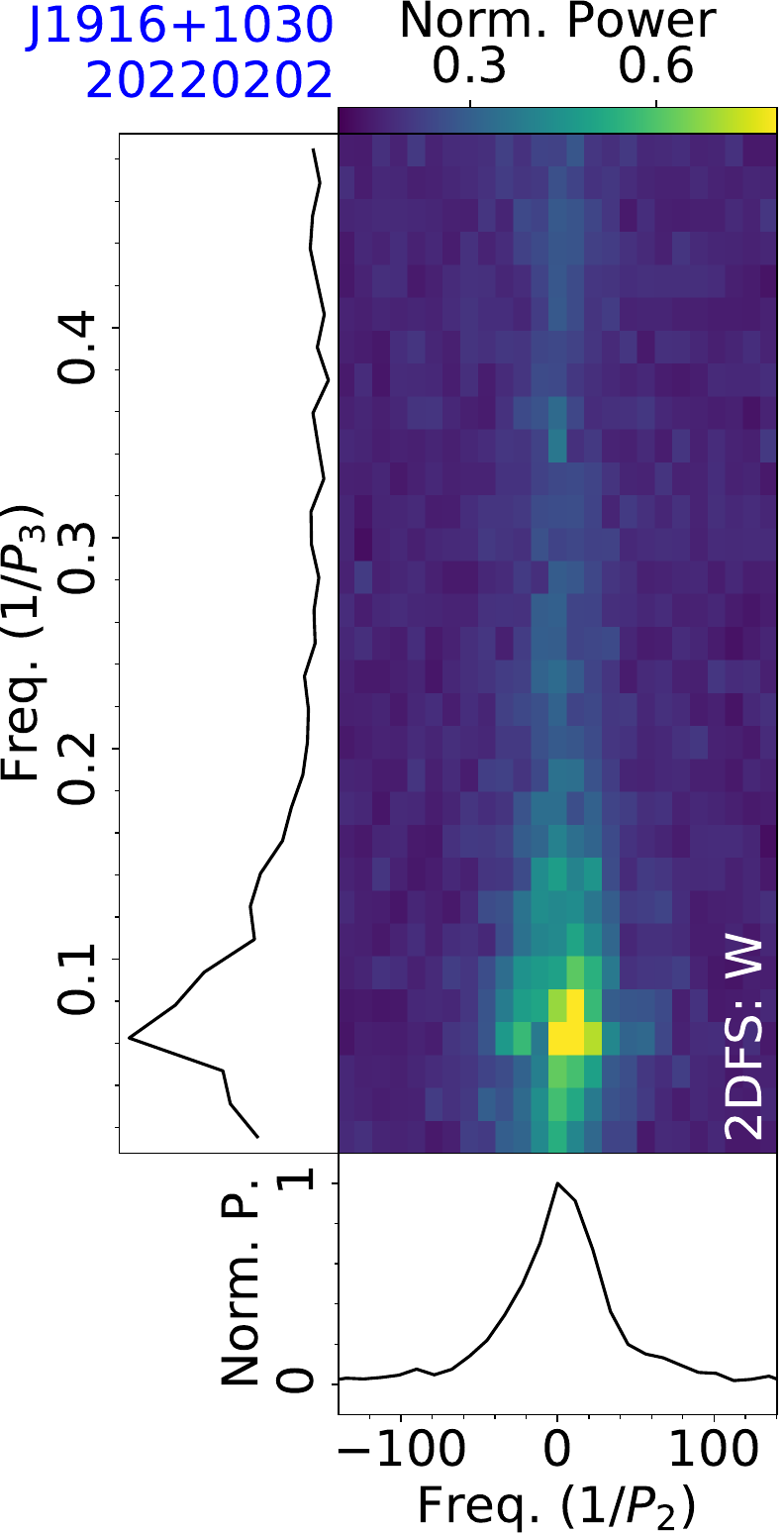}\\
\includegraphics[width=0.22\textwidth, angle=0]{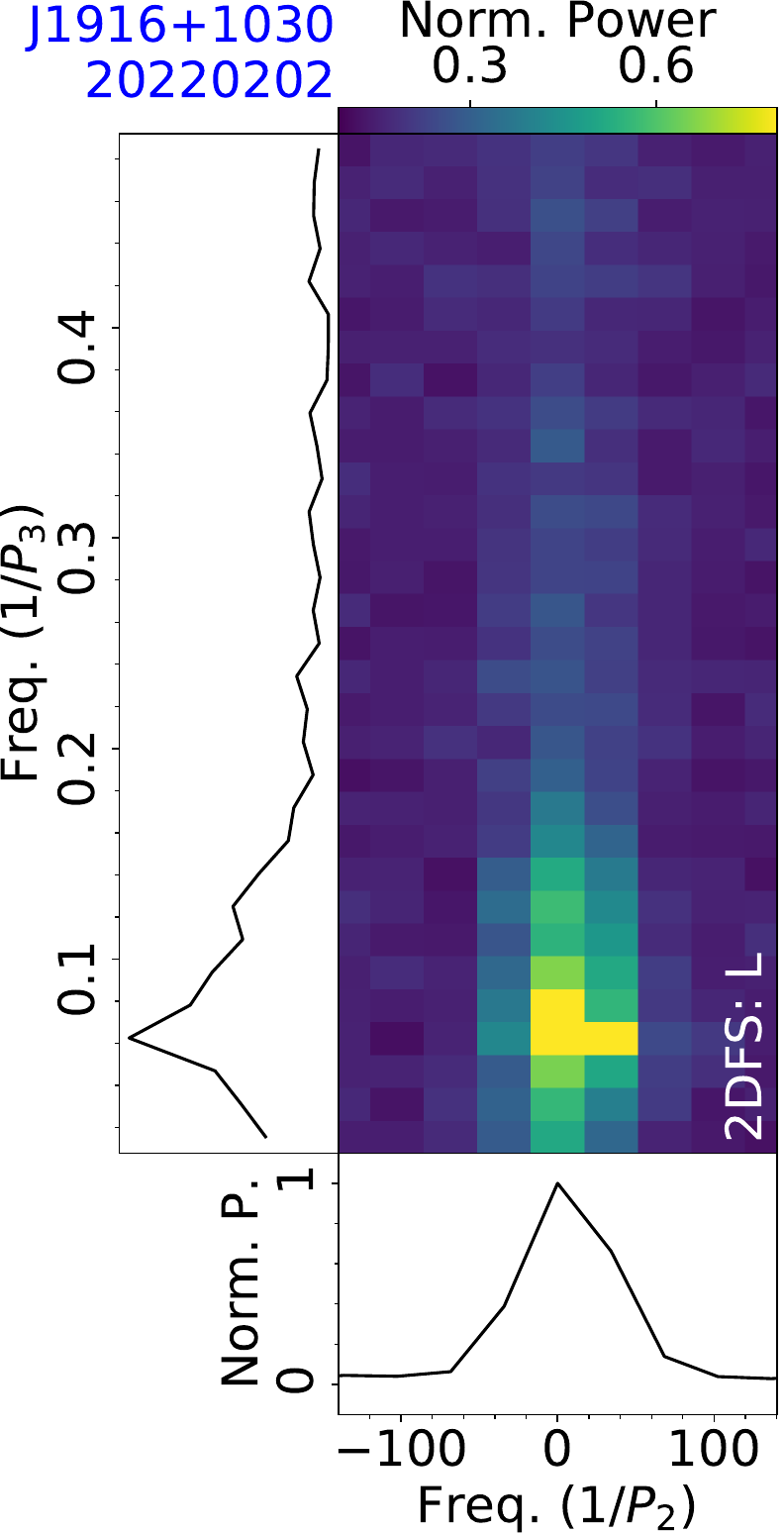}
\includegraphics[width=0.22\textwidth, angle=0]{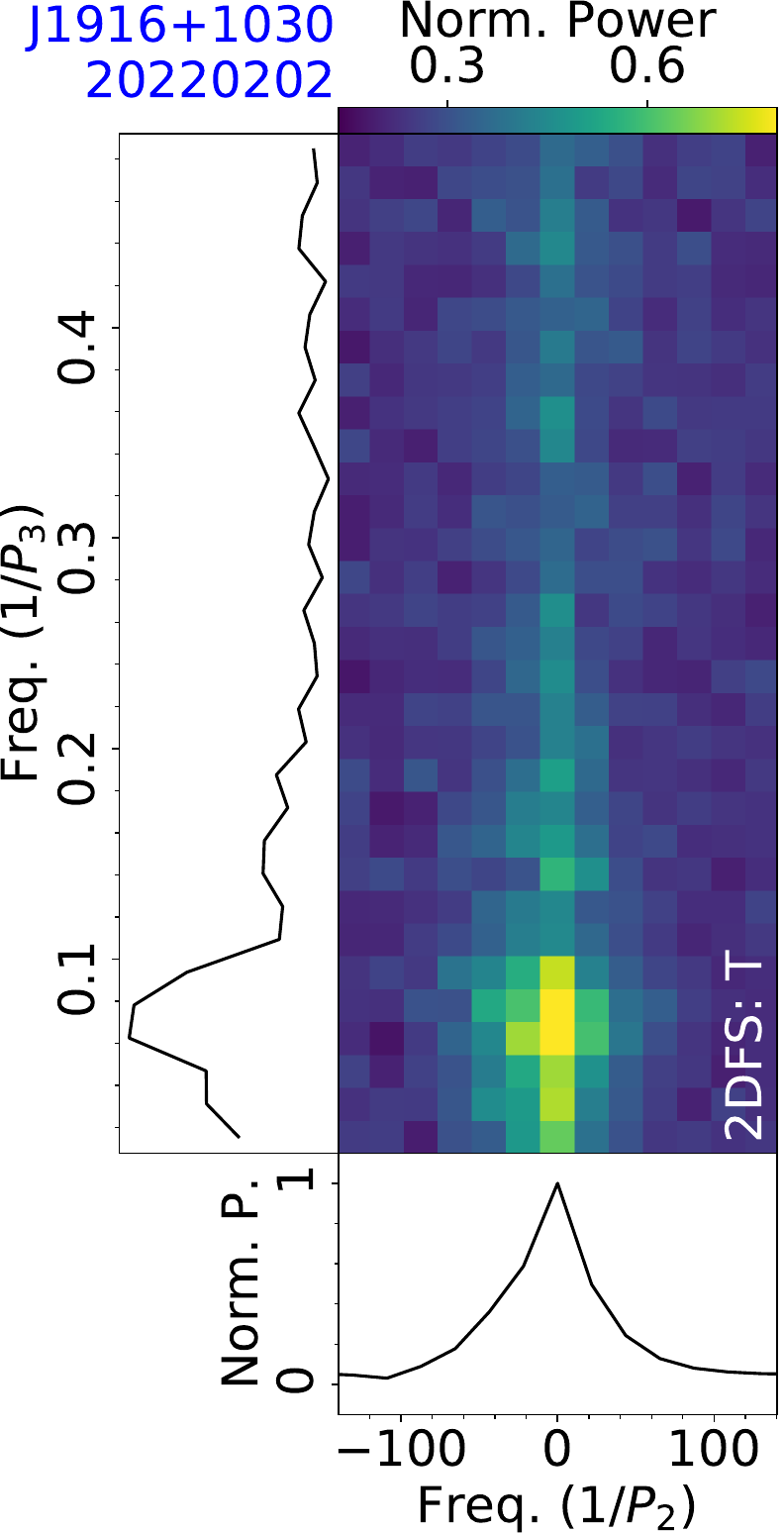}
\figcaption{Fluctuation analysis of PSR J1916+1030 for the observation on 20220202, with LRFS (top-left), and 2DFS for the on-pulse region (top-right), leading part (bottom-left) and trailing part (bottom-right) of a mean pulse profile.
\label{subfig:fluctu:J1916+1030}}
\end{figure}

\begin{figure}[htpb]
\centering
\includegraphics[width=0.44\textwidth, angle=0]{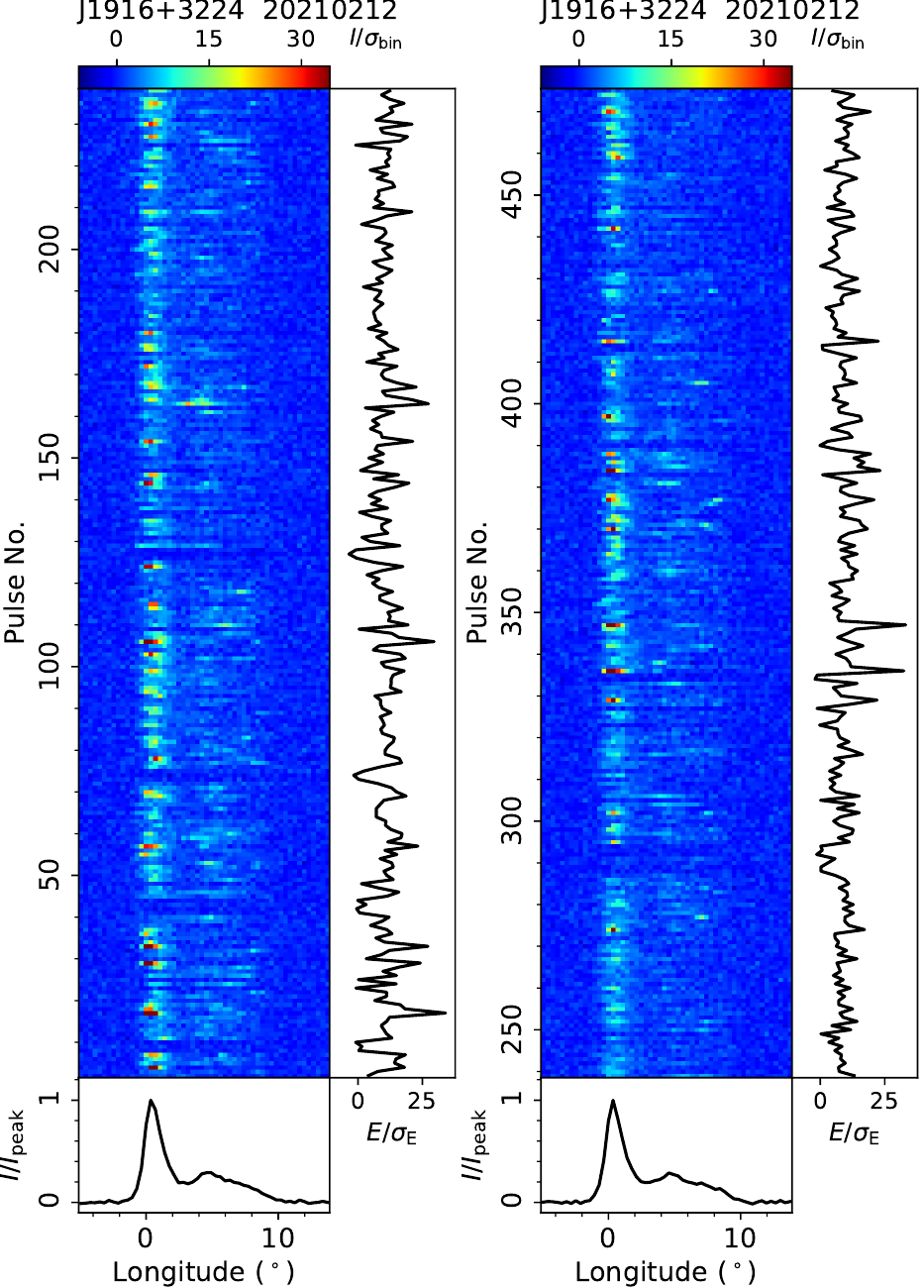}
\figcaption{Single pulse sequences of PSR J1916+3224 from the FAST observation on 20210212.
\label{subfig:TP:J1916+3224}}
\end{figure}

\begin{figure}[htpb]
\centering
\includegraphics[width=0.39\textwidth, angle=0]{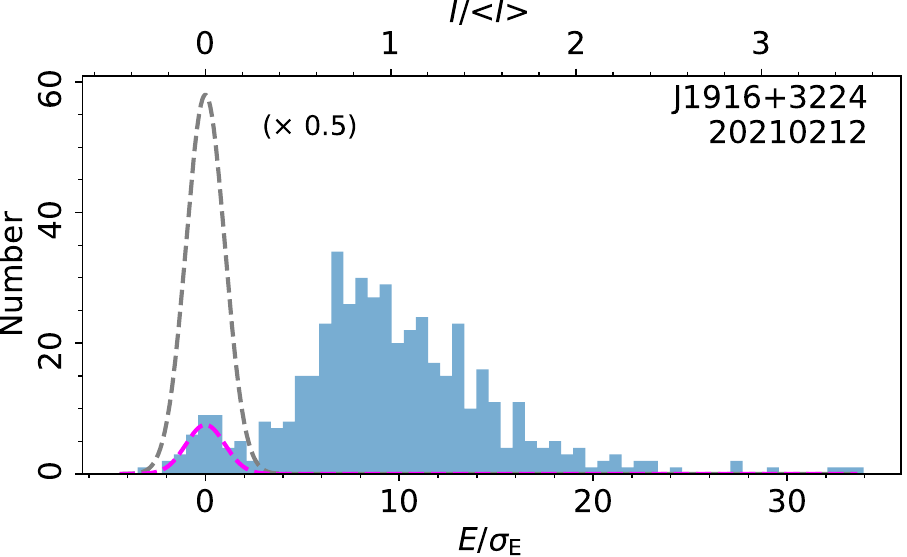}
\figcaption{On-pulse energy histogram of single pulses of PSR J1916+3224 from the FAST observation on 20210212.
\label{subfig:Hist:J1916+3224}}
\end{figure}

\subsection{J1916+1030}
\label{subsec:J1916+1030}

PSR J1916+1030 was first published by \citet{Taylor1993}. The subpulse drifting was reported by \citet{Song2023}, with $P_3=16\pm1$ periods and $P_2$=33$^{+75}_{-10}$ degrees. 

This pulsar was observed by FAST on 20220202 for 15 minutes, deriving a rotation period $P=0.6290$~s and a dispersion measure $D\!M=388.4~{\rm cm^{-3}\,pc}$. 
Single pulse sequences and fluctuation spectra of leading and trailing parts in the mean pulse profile are displayed in Figures~\ref{subfig:TP:J1916+1030} and \ref{subfig:fluctu:J1916+1030}. The leading profile part has a tendency of positive drifting from 2DFS, with the centroid frequencies of $1/P_3=0.076\pm0.001$ periods and $1/P_2=6\pm2$, corresponding to periodicities of $P_3=13.1\pm0.2$ periods and $P_2=55\pm21^\circ$. 
For the trailing part, 2DFS exhibits the modulation feature with the centroid of $1/P_3=0.059\pm0.001$, yielding the temporal periodicity of $P_3=16.9\pm0.4$ periods.

\begin{figure}[htpb]
\centering
\includegraphics[width=0.22\textwidth, angle=0]{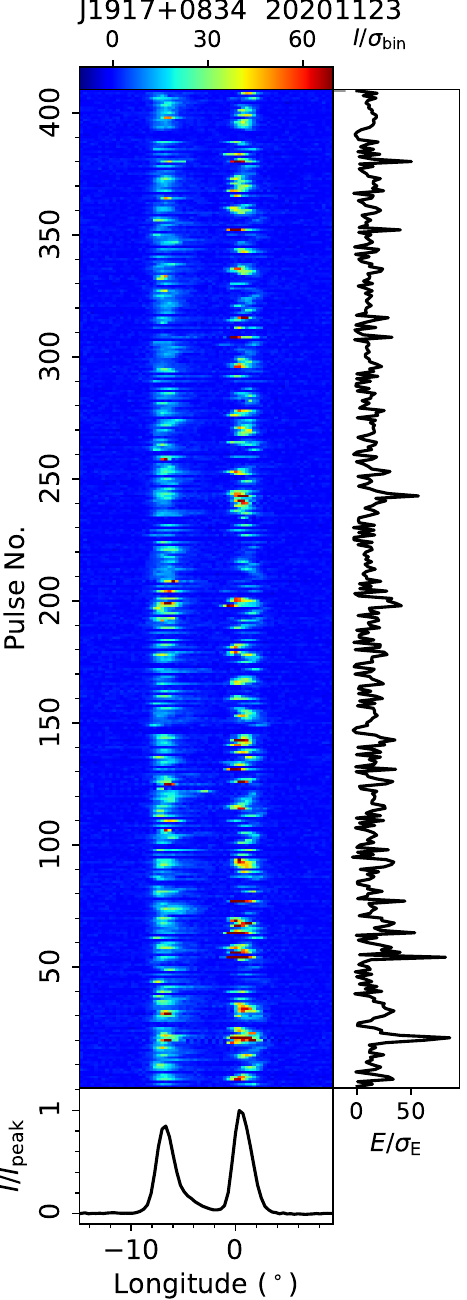}
\figcaption{Single pulse sequence of PSR J1917+0834 from the FAST observation on 20201123.
\label{subfig:TP:J1917+0834}}
\end{figure}

\begin{figure}[htpb]
\centering
\includegraphics[width=0.39\textwidth, angle=0]{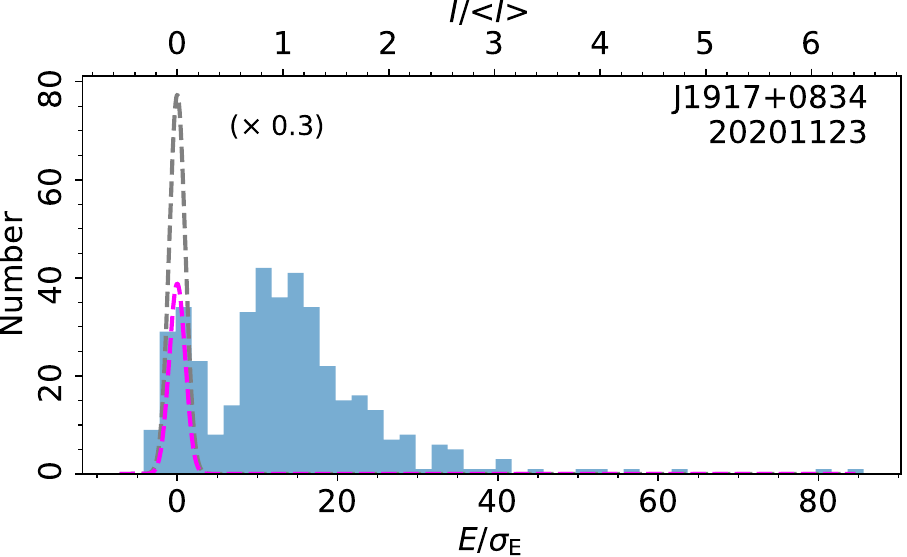}
\figcaption{On-pulse energy histogram of single pulses of PSR J1917+0834 from the FAST observation on 20201123.
\label{subfig:Hist:J1917+0834}}
\end{figure}

\begin{figure}[htpb]
\centering
\includegraphics[width=0.22\textwidth, angle=0]{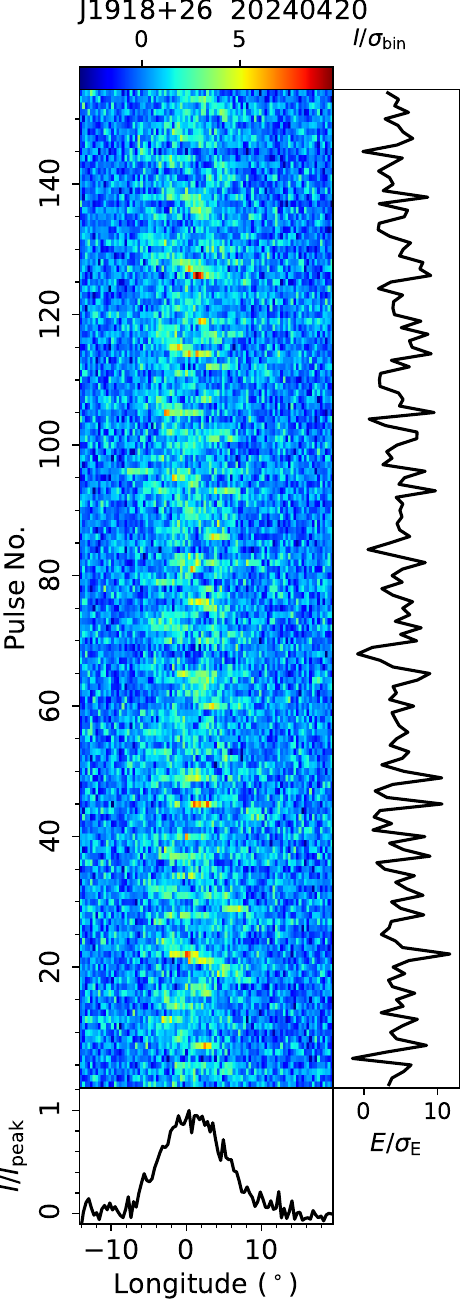}
\figcaption{Single pulse sequence of PSR J1918+26 from the FAST observation on 20240420.
\label{subfig:TP:J1918+26}}
\end{figure}

\begin{figure}[htpb]
\centering
\includegraphics[width=0.22\textwidth, angle=0]{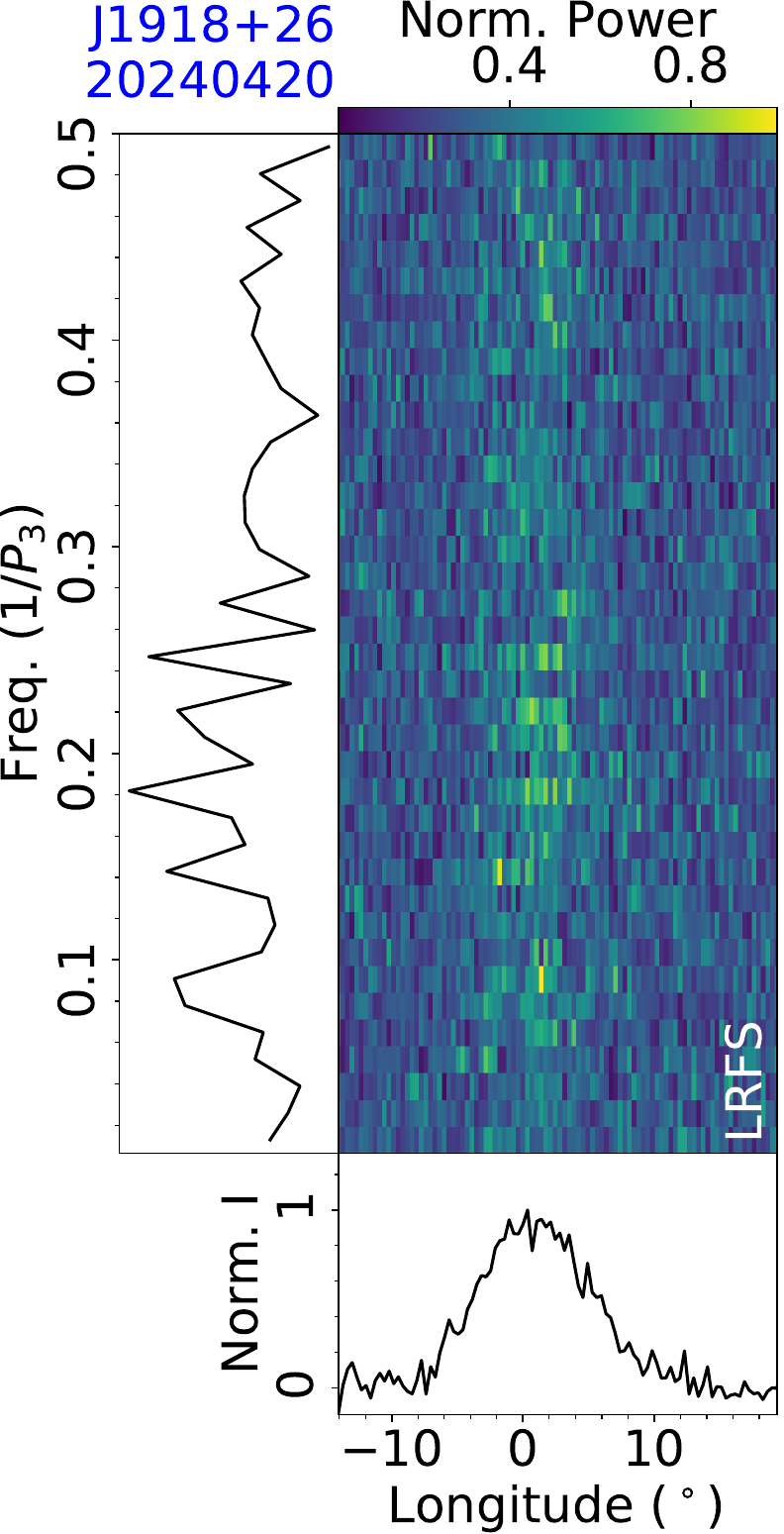}
\includegraphics[width=0.22\textwidth, angle=0]{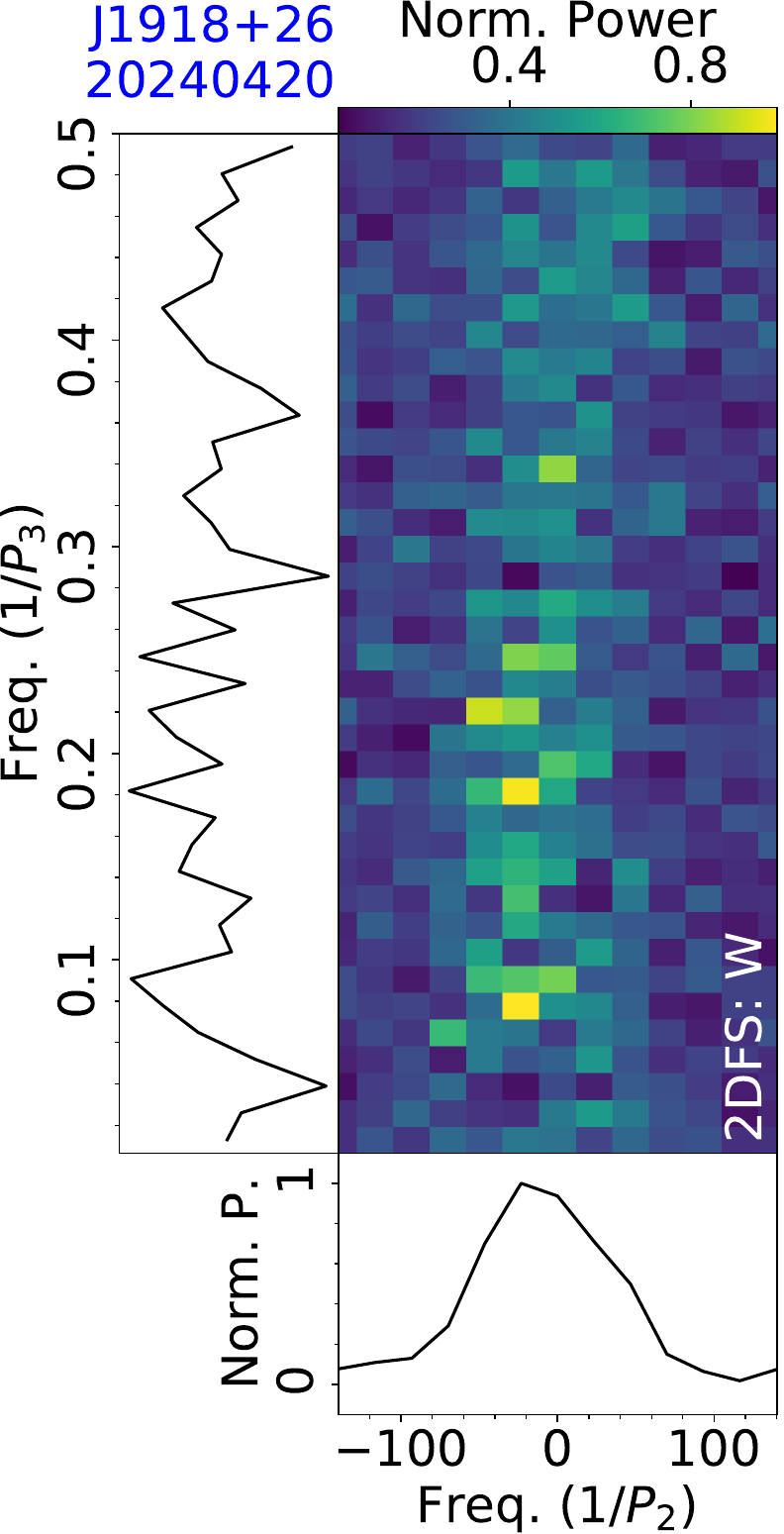}
\figcaption{Fluctuation analysis of PSR J1918+26 for the observation on 20240420, with LRFS and 2DFS for the on-pulse region of a mean pulse profile.
\label{subfig:fluctu:J1918+26}}
\end{figure}

\subsection{J1916+3224}
\label{subsec:J1916+3224}

PSR J1916+3224 was discovered in the LOFAR Tied-Array All-Sky Survey \citep{Sanidas2019}.

This pulsar was observed by FAST on 20210212 for 9 minutes, with a rotation period $P=1.1374$~s and a dispersion measure $D\!M=84.3~{\rm cm^{-3}\,pc}$ determined. Single pulse sequences in Fig.~\ref{subfig:TP:J1916+3224} show the nulling behavior. From the on-pulse integral energy histogram (Fig.~\ref{subfig:Hist:J1916+3224}), the nulling fraction of this observation is estimated to be 6.8$\pm$0.5\%.

\subsection{J1917+0834}
\label{subsec:J1917+0834}

PSR J1917+0834 was discovered by \citet{Nice1999} using the Arecibo telescope at 430 MHz. 

The pulsar was observed by FAST on 20201123 for 14 minutes, with a rotation period $P=2.1299$~s and a dispersion measure $D\!M=27.8~{\rm cm^{-3}\,pc}$ derived. 
The single pulse sequence is shown in Fig.~\ref{subfig:TP:J1917+0834}, and the on-pulse integral energy histogram in Fig.~\ref{subfig:Hist:J1917+0834} illustrates the existence of the nulling phenomenon. The nulling fraction is estimated to be 15$\pm$3\%. As previously reported \citep{Nice1999}, the pulsar have strong emission, and the single-pulse intensity could be up to 7 times than the average of this FAST observation.

\begin{figure}[htpb]
\centering
\includegraphics[width=0.21\textwidth, angle=0]{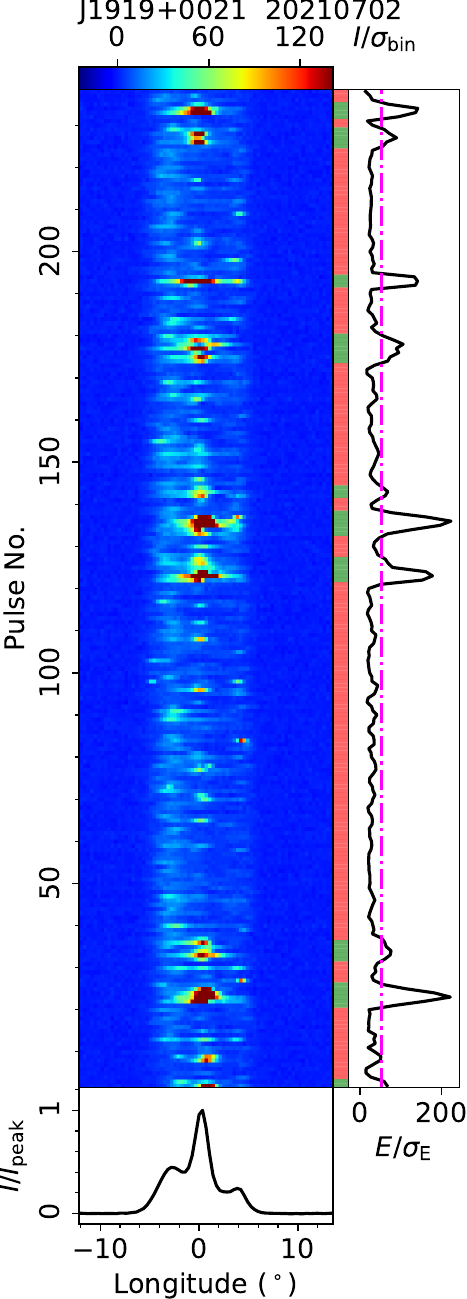}
\figcaption{Single pulse sequence of PSR J1919+0021 from the FAST observation on 20210702. The red and green bars represent normal or abnormal modes. In the right subpanel, the on-pulse energy variation smoothed over every 3 periods is plotted against period, with a dashed line for the threshold to distinguish the normal and abnormal emission modes.
\label{subfig:TP:J1919+0021}}
\end{figure}

\begin{figure}[htpb]
\centering
\includegraphics[width=0.39\textwidth, angle=0]{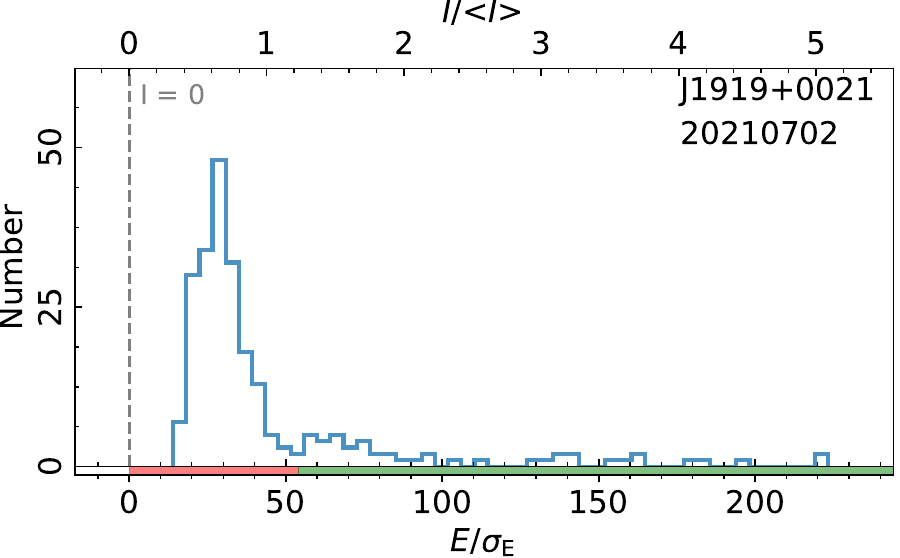}
\figcaption{On-pulse energy histogram of single pulses of PSR J1919+0021 from the FAST observation on 20210111, with energy values smoothed over 3 periods. 
The red and green bars indicate the normal and abnormal modes.
\label{subfig:Hist:J1919+0021}}
\end{figure}

\begin{figure}[htpb]
\centering
\includegraphics[width=0.37\textwidth, angle=0]{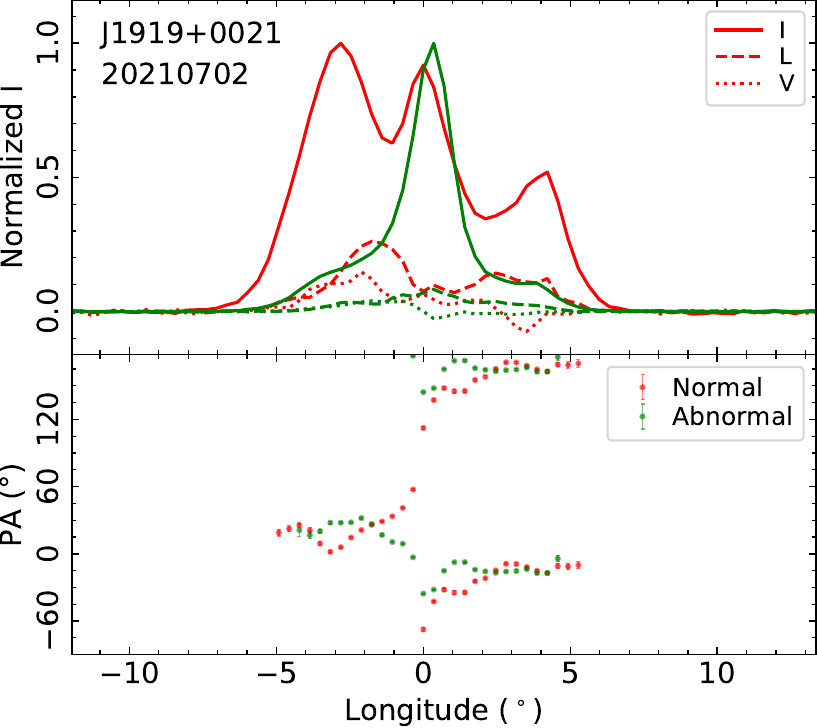}
\figcaption{Mean polarization profiles (the top panel) for the normal (red) and abnormal (green) emission modes of PSR J1919+0021 observed on 20210702, as well as the averaged PA curves (the bottom panel). Profiles in the top panel are normalized by their respective peaks.
\label{subfig:PolModes:J1919+0021}}
\end{figure}

\subsection{J1918+26}
\label{subsec:J1918+26}

PSR J1918+26 was discovered by the radio telescope Large Phased Array (LPA) \citep{Tyulbashev2024}.

This pulsar was observed by FAST on 20240420 for 2 minutes, deriving a rotation period $P=0.7219$~s and a dispersion measure $D\!M=113.4~{\rm cm^{-3}\,pc}$. The single pulse sequence in Fig.~\ref{subfig:TP:J1918+26} and fluctuation spectra in Fig.~\ref{subfig:fluctu:J1918+26} illustrate the existence of the negative drifting phenomenon. The centroid of the main drift feature in 2DFS is characterized by frequencies of $1/P_3=0.17\pm0.01$ and $1/P_2=-24\pm2$, corresponding to periodicities of $P_3=5.9\pm0.2$ periods and $P_2=-15\pm1$ degrees.

\subsection{J1919+0021}
\label{subsec:J1919+0021}

PSR J1919+0021 was discovered by the Mark 1A radio telescope at Jodrell Bank \citep{Davies1973}. 
\citet{Rankin1986} reported the pulsar with the mode-changing phenomenon that the core component was enhanced predominantly relative to the outer component. \citet{Weltevrede2006,Weltevrede2007} detected broad drifting behavior at 21 cm and 92 cm, and amplitude modulation was reported by \citet{Basu2016} at 333 MHz.

This pulsar was observed by FAST on 20210702 for 6 minutes, yielding a rotation period $P=1.2723$~s and a dispersion measure $D\!M=89.6~{\rm cm^{-3}\,pc}$. 
The single-pulse sequence in Fig.~\ref{subfig:TP:J1919+0021} shows that the pulsar exhibits a mode changing phenomenon in which the emission is sometimes enhanced.
In the histogram of smoothed on-pulse integral energy variation every 3 single pulses (Fig.~\ref{subfig:Hist:J1919+0021}), two emission modes can be easily distinguished, which are labeled by red and green colors. In the normal mode, the leading component is the brightest. The emission on-pulse phase region, especially for the central component, is enhanced in the abnormal mode (Fig.~\ref{subfig:PolModes:J1919+0021}), which is consistent with the result of \citet{Rankin1986}.

Further observations are needed for the modulation analysis.

\begin{figure}[htpb]
    \centering
	\includegraphics[width=0.22\textwidth]{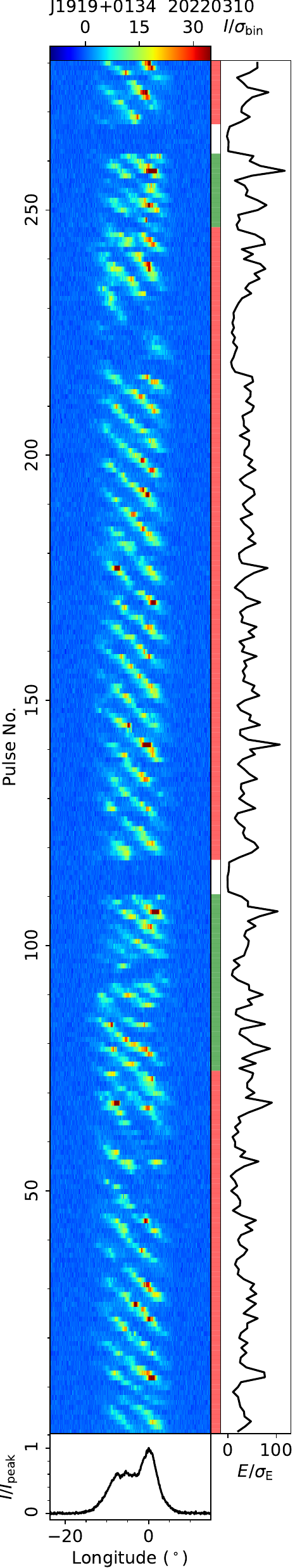}
	\includegraphics[width=0.22\textwidth]{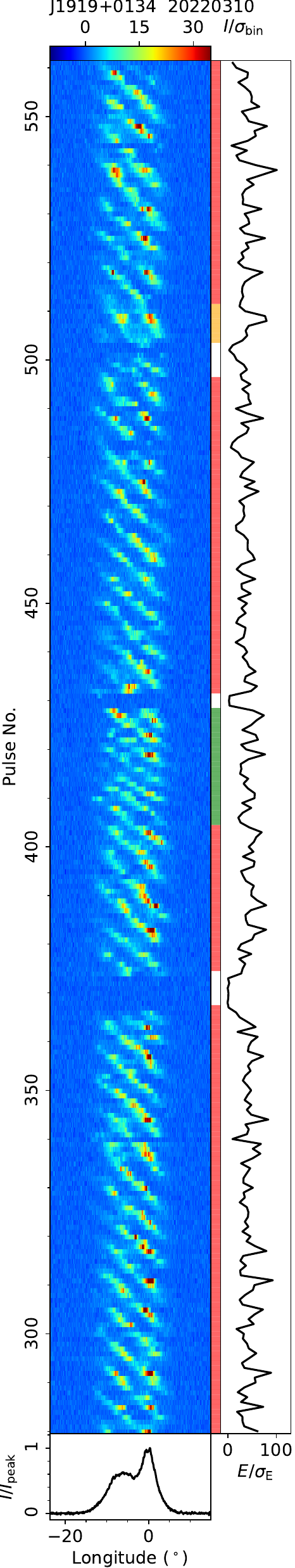}
\figcaption{Single pulse sequences of PSR J1919+0134 from FAST observations on 20220310 and 20230328. -- to be continued
	\label{subfig:TP:J1919+0134}}
    \addtocounter{figure}{-1}
\end{figure}

\begin{figure}[htpb]
    \centering
	\includegraphics[width=0.22\textwidth]{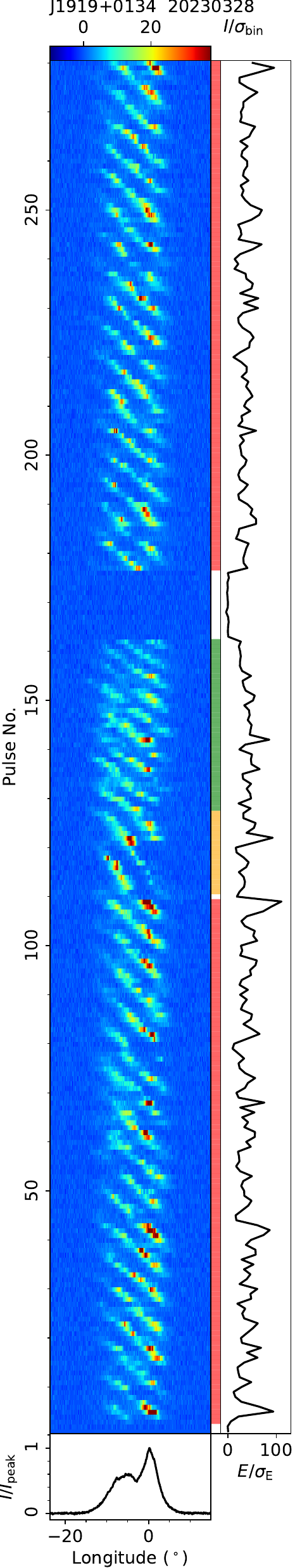}
	\includegraphics[width=0.22\textwidth]{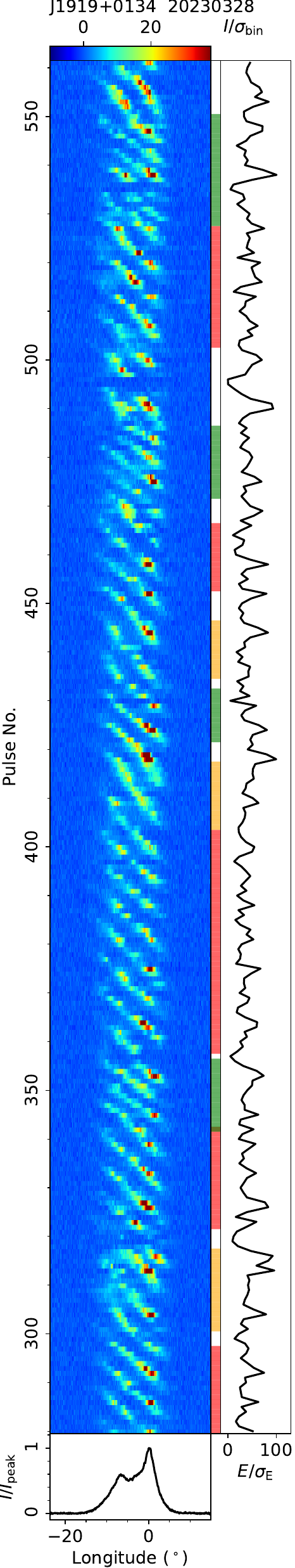}
	\figcaption{Continued.
	}
\end{figure}

\begin{figure}[htpb]
\centering
\includegraphics[width=0.39\textwidth, angle=0]{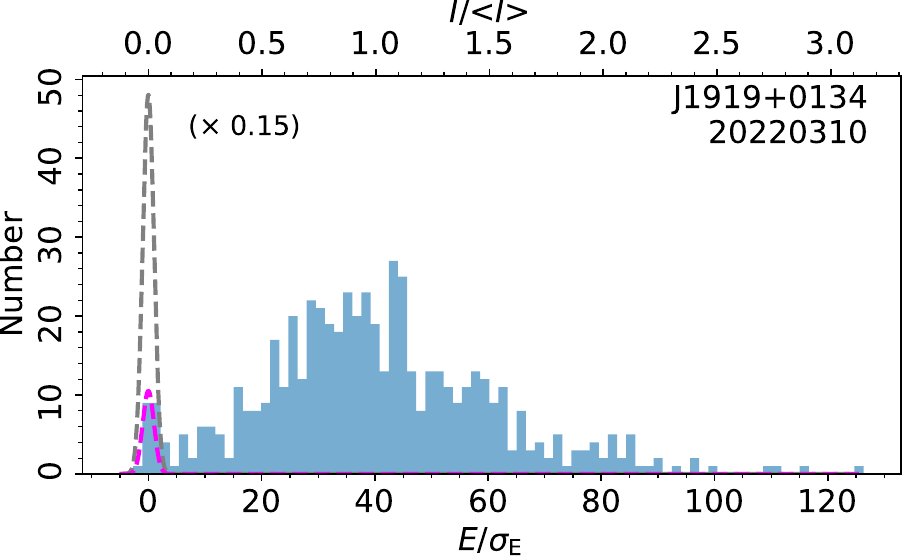}
\includegraphics[width=0.39\textwidth, angle=0]{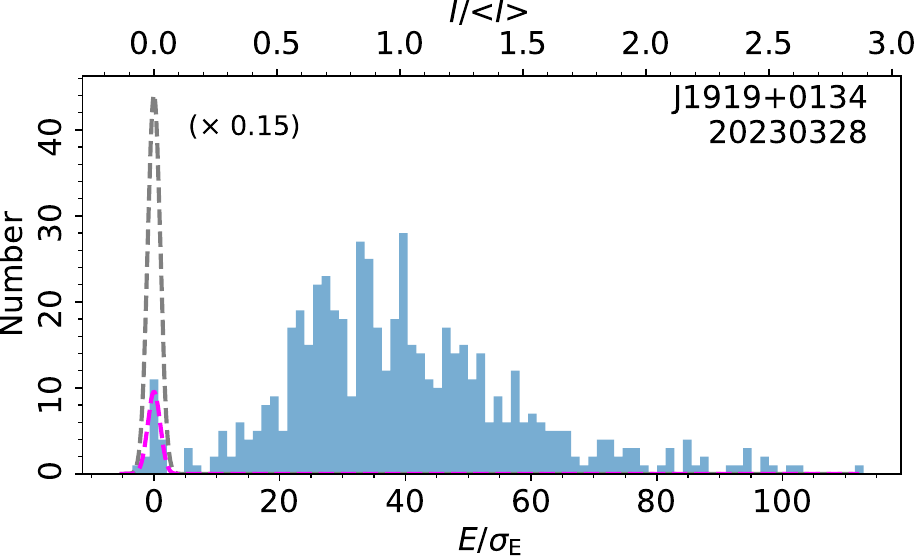}
\figcaption{On-pulse energy histograms of single pulses of PSR J1916+1023 from FAST observations on 20220310 and 2023032.
\label{subfig:Hist:J1919+0134}}
\end{figure}

\begin{figure}[htpb]
\centering
\includegraphics[width=0.365\textwidth, angle=0]{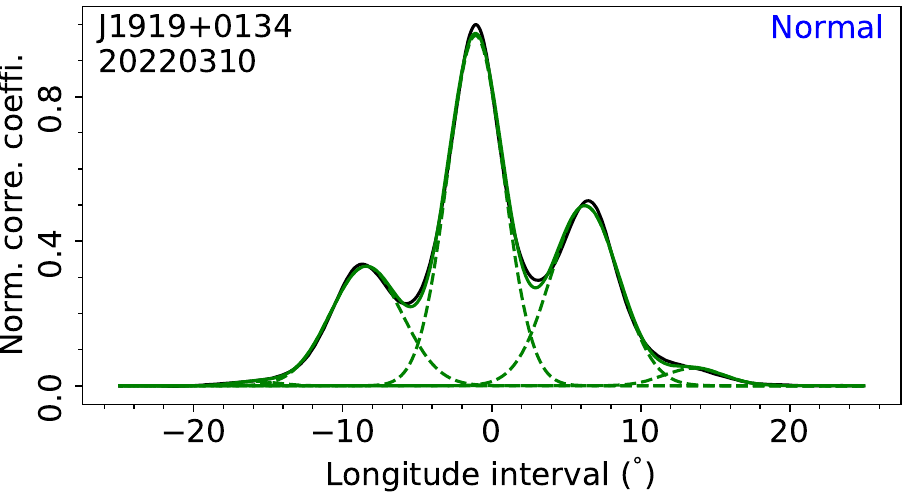}
\includegraphics[width=0.365\textwidth, angle=0]{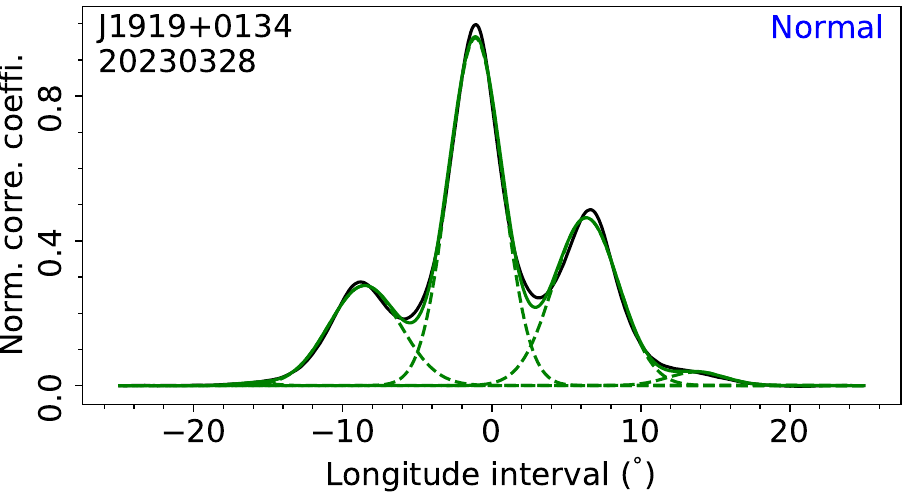}\\
\includegraphics[width=0.365\textwidth, angle=0]{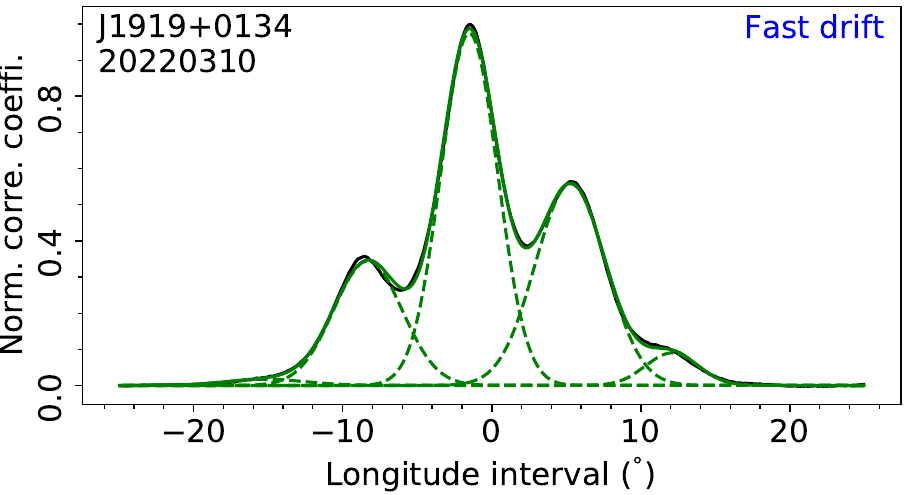}
\includegraphics[width=0.365\textwidth, angle=0]{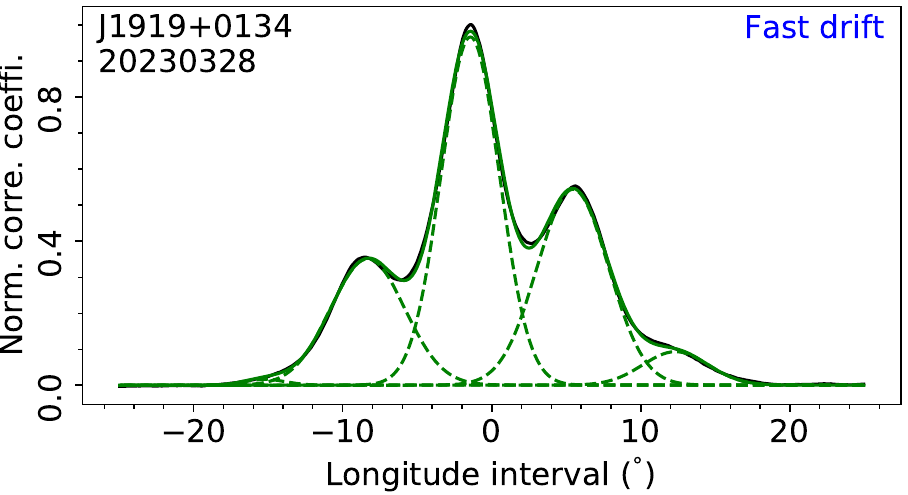}
\includegraphics[width=0.365\textwidth, angle=0]{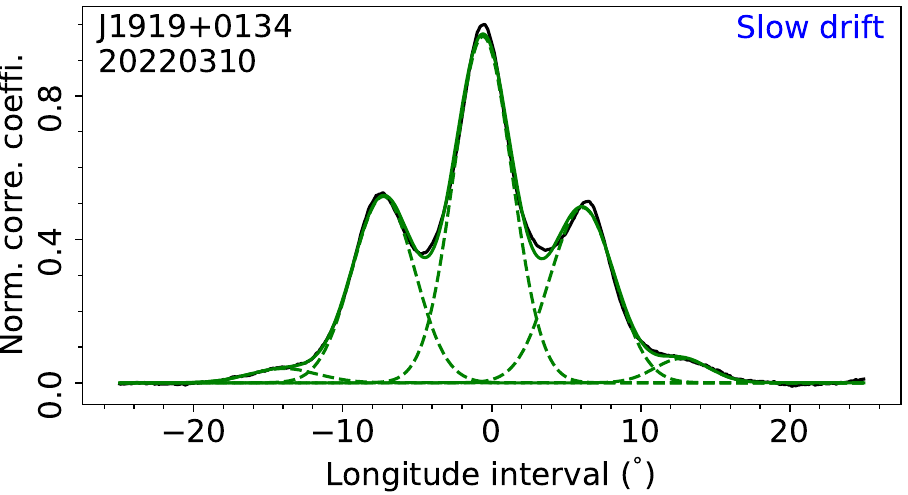}
\includegraphics[width=0.365\textwidth, angle=0]{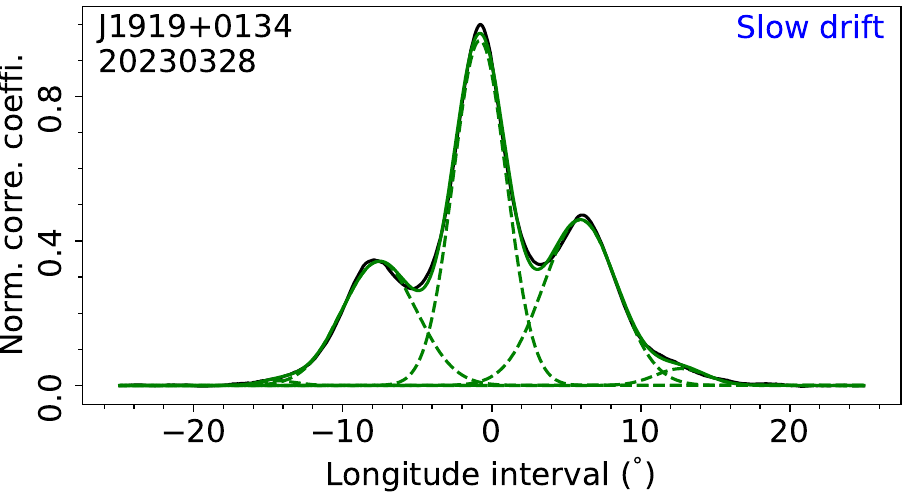}
\vspace{-0.2cm}
\figcaption{Cross correlations of normal, fast drifting and slow drifting modes of PSR J1919+0134 from FAST observations on 20220310 and 20230328.
\label{subfig:Corre:J1919+0134}}
\end{figure}

\subsection{J1919+0134}
\label{subsec:J1919+0134}

PSR J1919+0134 was discovered in a survey of intermediate Galactic latitudes using the Parkes 64-m radio telescope and shown to have regular drift \citep{Edwards2001}. Drifting parameters were previously reported to be $P_3 \sim 6.5$ periods and $P_2 \sim 9^\circ$ \citep{Ord2001,Basu2016,Song2023}. 

This pulsar was observed by FAST on 20220310 and 20230328 each for 15 minutes. A rotation period $P=1.6038$~s and a dispersion measure $D\!M=190.6~{\rm cm^{-3}\,pc}$ were determined from the observation on 20230328. 
Single pulse sequences of two observations in Fig.~\ref{subfig:TP:J1919+0134} display nulling and mode changing between different subpulse drifting modes. It's the first time revealing nulling and the drift rate changingfrom the highly sensitive FAST observations. 

Based on the on-pulse integral energy histograms in Fig.~\ref{subfig:Hist:J1919+0134}, the nulling fractions of the observations on 20220310 and 20230328 are 3.3$\pm$0.5\% and 3.3$\pm$0.3\%, respectively.

We distinguish three drifting modes on the basis of different subpulse drift features and calculate the drifting parameters using the correlation method. They are normal drifting mode, fast drifting (Fd) mode, and slow drifting (Sd) mode, where Fd and Sd modes are newly distinguished. 
The most frequent drifting mode is the normal mode, labeled in red. The drifting parameters of this mode are estimated to be $P_2=-7.34\pm0.01^\circ$ and $D=-1.08\pm0.04$ degrees per period for the observation on 20220310, and $P_2=-7.43\pm0.01^\circ$ and $D=-1.09\pm0.06$ degrees per period for the observation on 20230328, which is consistent with previous studies \citep{Ord2001,Basu2016,Song2023}. The fast drifting mode is labeled by blue color, with $P_2=-6.80\pm0.01^\circ$ and $D=-1.51\pm0.03$ degrees per period (20220310), and $P_2=-6.88\pm0.01^\circ$ and $D=-1.43\pm0.04$ degrees per period (20230328). The slow drifting mode labeled with yellow color has $P_2=-6.78\pm0.01^\circ$ and $D=-0.59\pm0.05$ degrees per period (20220310), and $P_2=-6.75\pm0.01^\circ$ and $D=-0.80\pm0.05$ degrees per period (20230328). The slow drifting mode is minimally occurring in these two observations.

\begin{figure}[htpb]
\centering
\includegraphics[width=0.22\textwidth, angle=0]{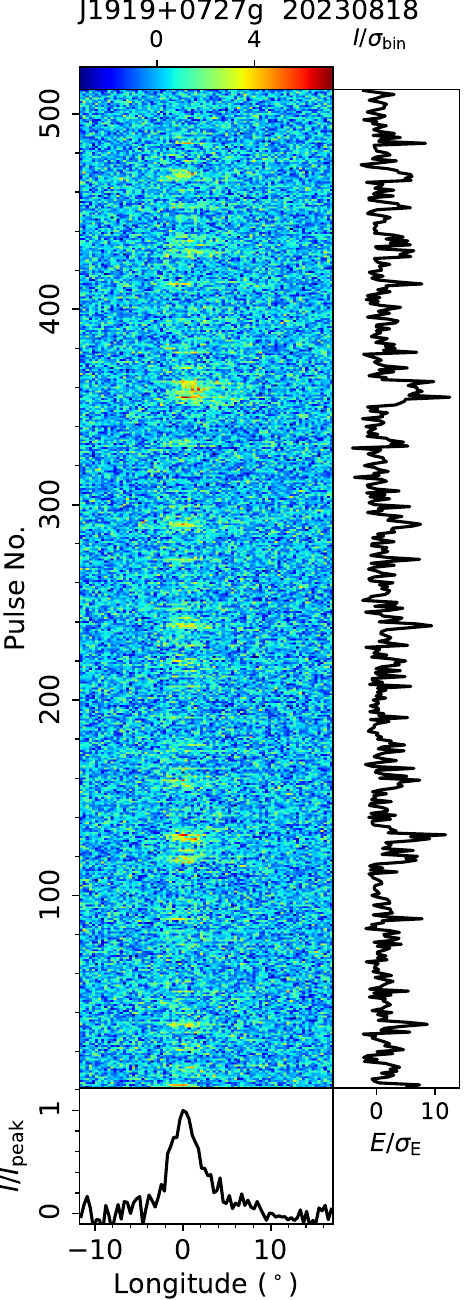}
\includegraphics[width=0.22\textwidth, angle=0]{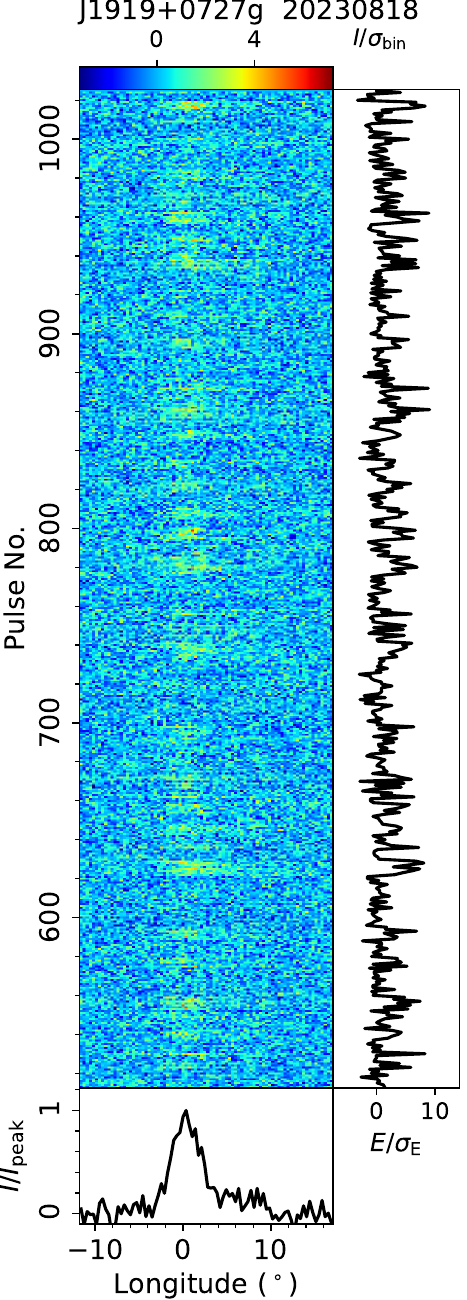}
\figcaption{Single pulse sequences of PSR J1919+0727g from the FAST observation on 20230818. 
\label{subfig:TP:J1919+0727g}}
\end{figure}

\begin{figure}[htpb]
\centering
\includegraphics[width=0.22\textwidth, angle=0]{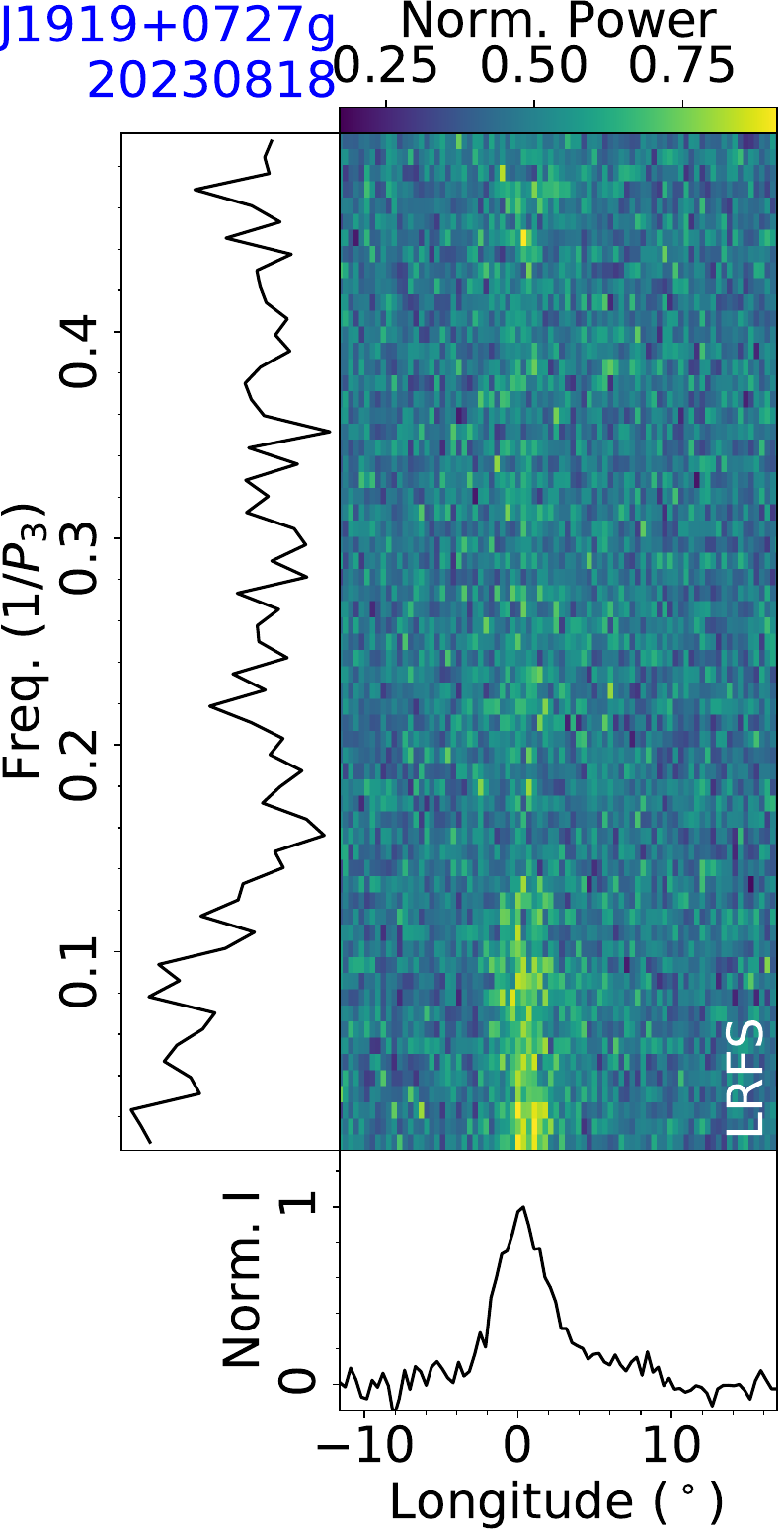}
\includegraphics[width=0.22\textwidth, angle=0]{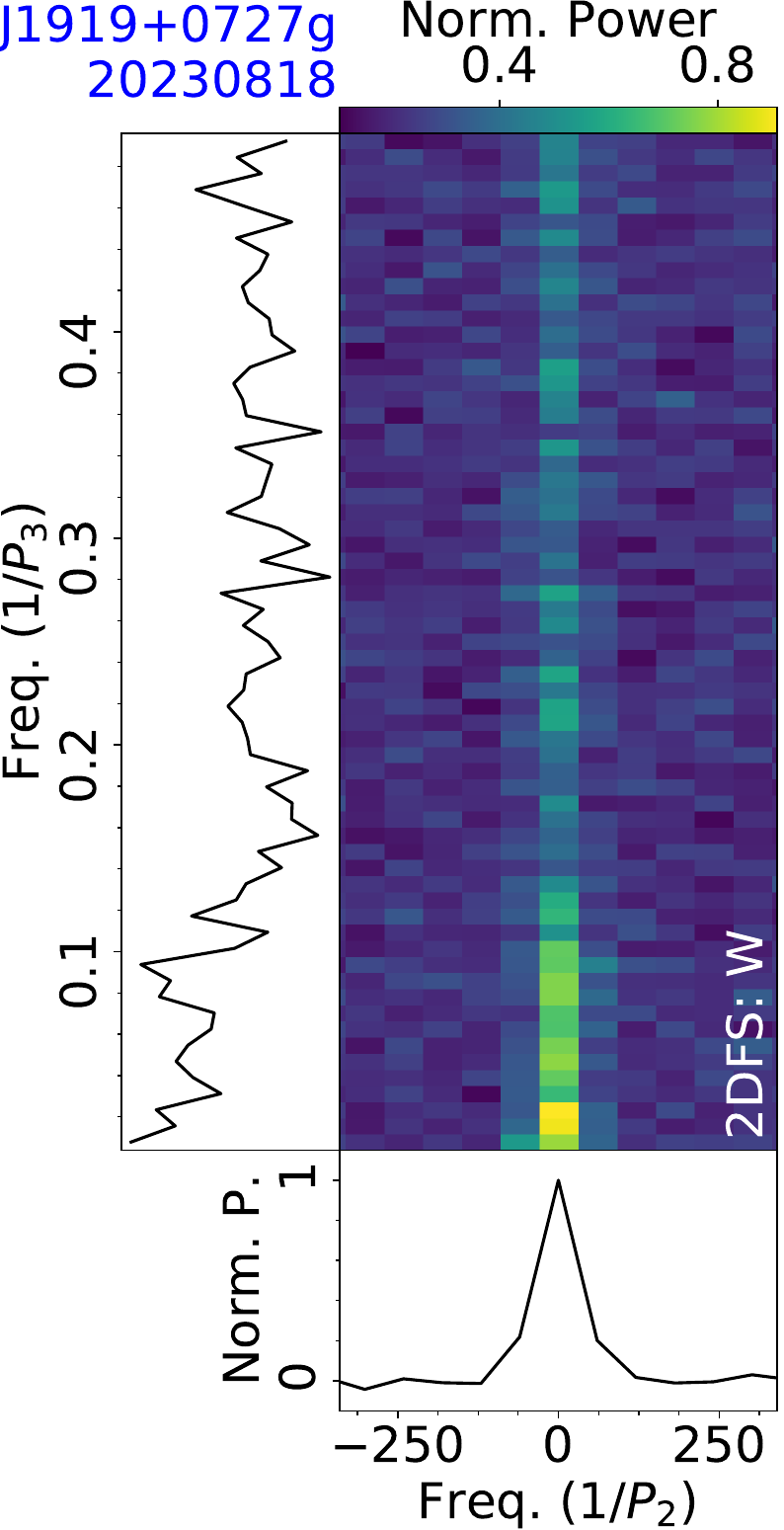}
\figcaption{Fluctuation analysis of PSR J1919+0727g for the observation on 20230818, with LRFS and 2DFS for the on-pulse region of a mean pulse profile.
\label{subfig:fluctu:J1919+0727g}}
\end{figure}

\begin{figure}[htpb]
\centering
\includegraphics[width=0.22\textwidth, angle=0]{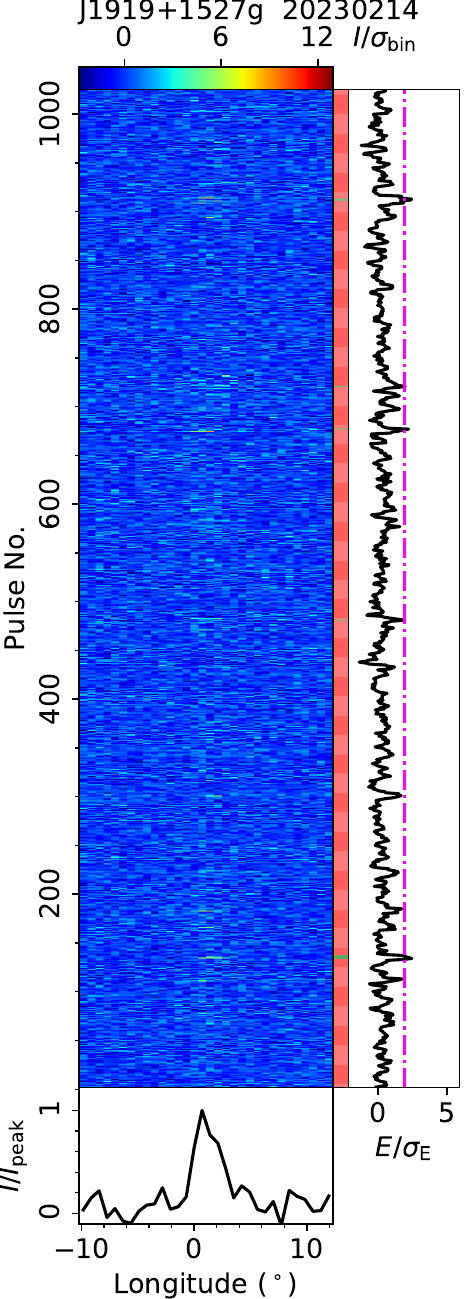}
\includegraphics[width=0.22\textwidth, angle=0]{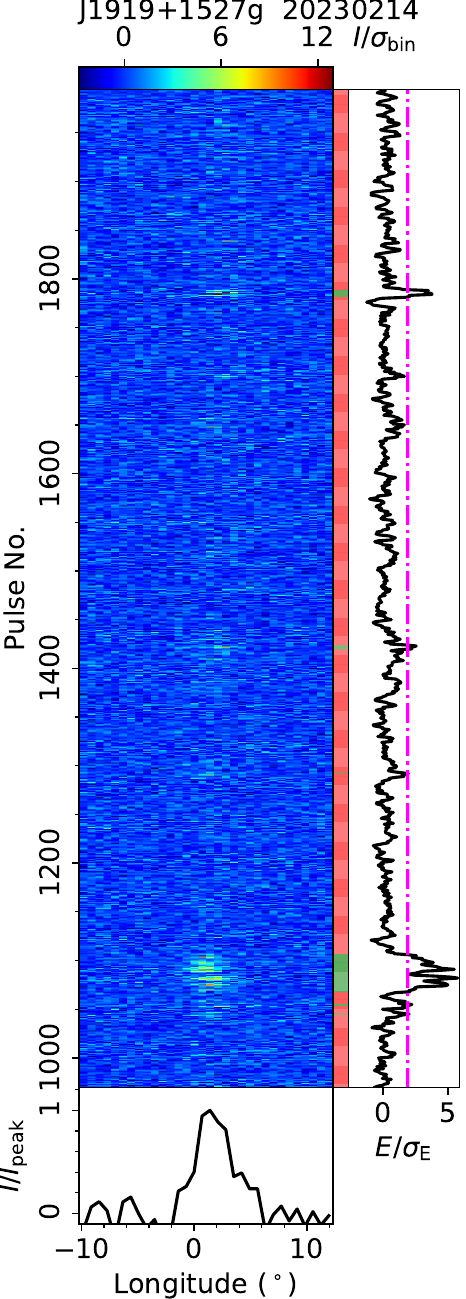}
\figcaption{Single pulse sequences of PSR J1919+1527g from the FAST observation on 20230214. 
The red and green bars represent weak or bright emission modes. In the right subpanel, the on-pulse energy variation smoothed over every 5 periods is plotted against period, with a dashed line for the threshold to distinguish the weak and bright emission modes.
\label{subfig:TP:J1919+1527g}}
\end{figure}

\begin{figure}[htpb]
\centering
\includegraphics[width=0.39\textwidth, angle=0]{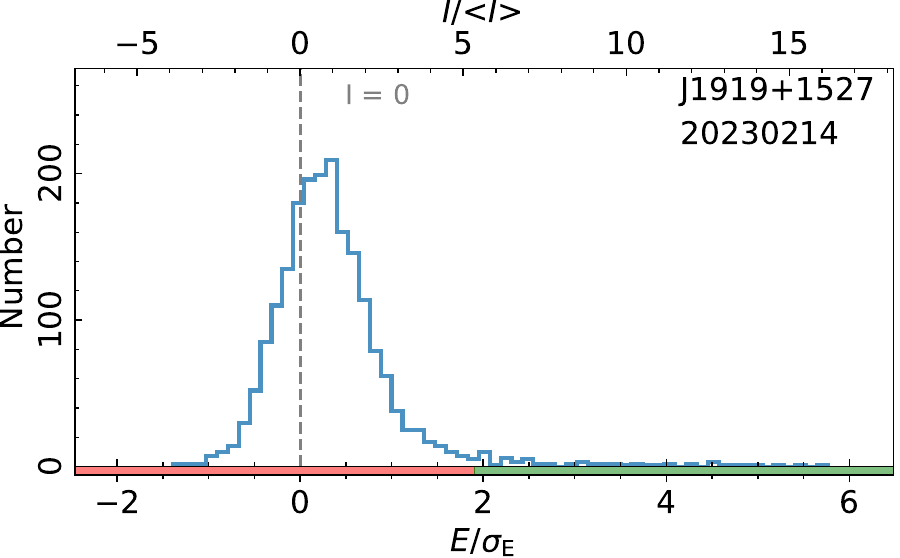}
\figcaption{On-pulse energy histogram of single pulses of PSR J1919+1527g from the FAST observation on 20230214, with energy values smoothed over 5 periods. 
The red and green bars indicate the weak and bright emission modes.
\label{subfig:Hist:J1919+1527g}}
\end{figure}

\begin{figure}[htpb]
\centering
\includegraphics[width=0.37\textwidth, angle=0]{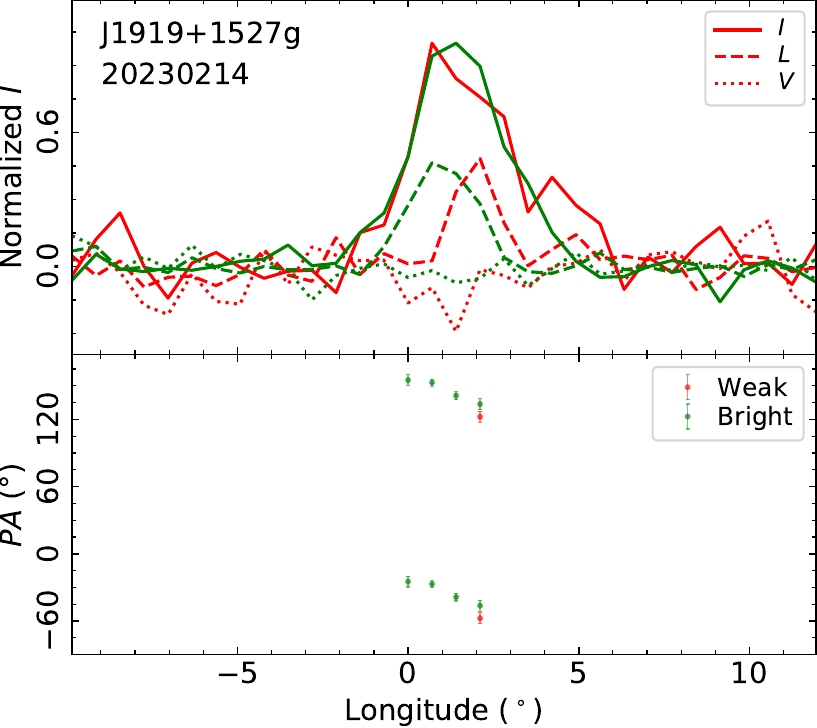}
\figcaption{Mean polarization profiles (the top panel) for the weak and bright emission modes of PSR J1919+1527g observed on 20230214. Profiles in the top panel are normalized by their respective peaks. \label{subfig:PolModes:J1919+1527g}}
\end{figure}

\subsection{J1919+0727g}
\label{subsec:J1919+0727g}

PSR J1919+0727g was discovered in the FAST GPPS survey \citep{Han2021,han2025}. 

This pulsar was observed by FAST on 20230818 for 15 minutes, and a rotation period $P=0.8669$~s and a dispersion measure $D\!M=183.1~{\rm cm^{-3}\,pc}$ were derived. Single pulse sequences in Fig.~\ref{subfig:TP:J1919+0727g} display the temporal intensity variation. From the fluctuation spectra (Fig.~\ref{subfig:fluctu:J1919+0727g}), the intensity variation is temporally low-frequency modulated, with a centroid frequency of $f_3=0.059\pm0.003$ ($P_3=17.0\pm0.8$ periods).

\subsection{J1919+1527g}
\label{subsec:J1919+1527g}

PSR J1919+1527g was discovered in the FAST GPPS survey \citep{Han2021,han2025}. 

This pulsar was observed by FAST on 20230214 for 46 minutes, yielding a rotation period $P=1.3713$~s and a dispersion measure $D\!M=695.0~{\rm cm^{-3}\,pc}$. 
From single pulse sequences in Fig.~\ref{subfig:TP:J1919+1527g}, the pulsar has the bright and weak emission modes, as demonstrated from the asymmetric distribution around the zero in the on-pulse integral energy histogram (Fig.~\ref{subfig:Hist:J1919+1527g}). Bright and weak emission modes are labeled using green and red colors. Averaged polarization profiles and PA curves of two emission modes are shown in Fig.~\ref{subfig:PolModes:J1919+1527g}. 

\begin{figure}[htpb]
\centering
\includegraphics[width=0.21\textwidth, angle=0]{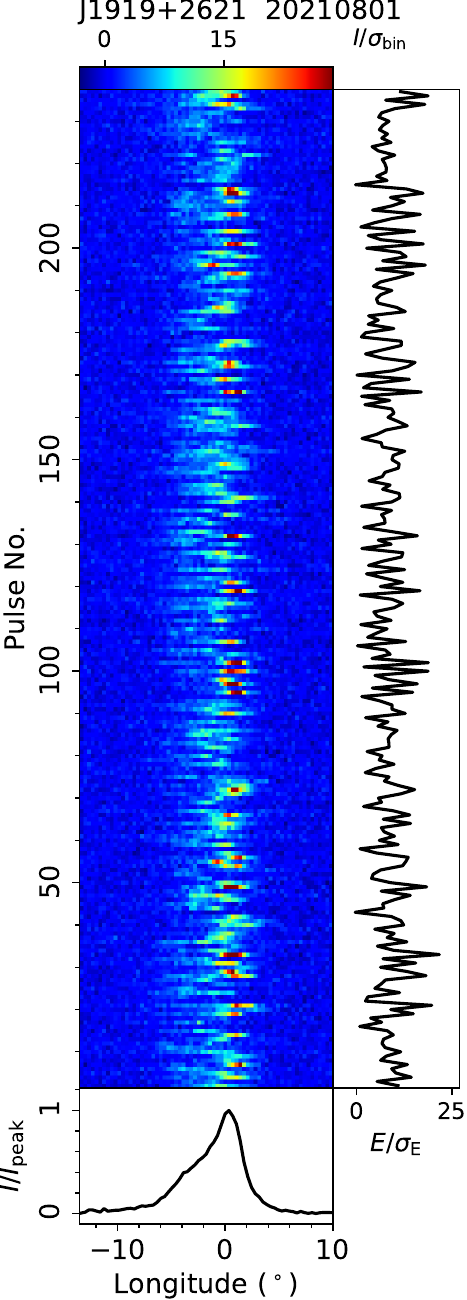}
\includegraphics[width=0.21\textwidth, angle=0]{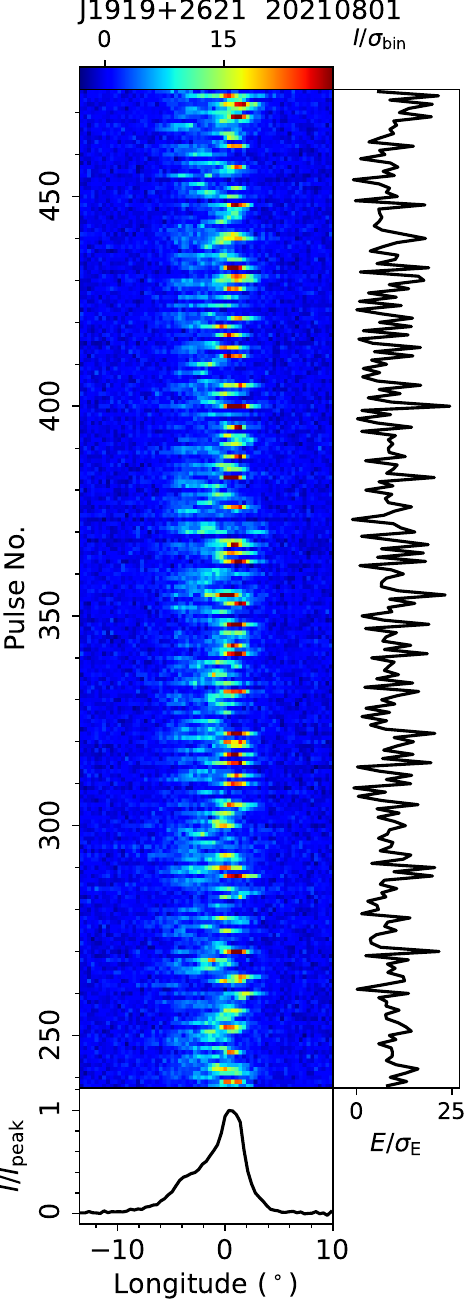}
\figcaption{Single pulse sequences of PSR J1919+2621 from the FAST observation on 20210801.
\label{subfig:TP:J1919+2621}}
\end{figure}

\begin{figure}[htpb]
\centering
\includegraphics[width=0.22\textwidth, angle=0]{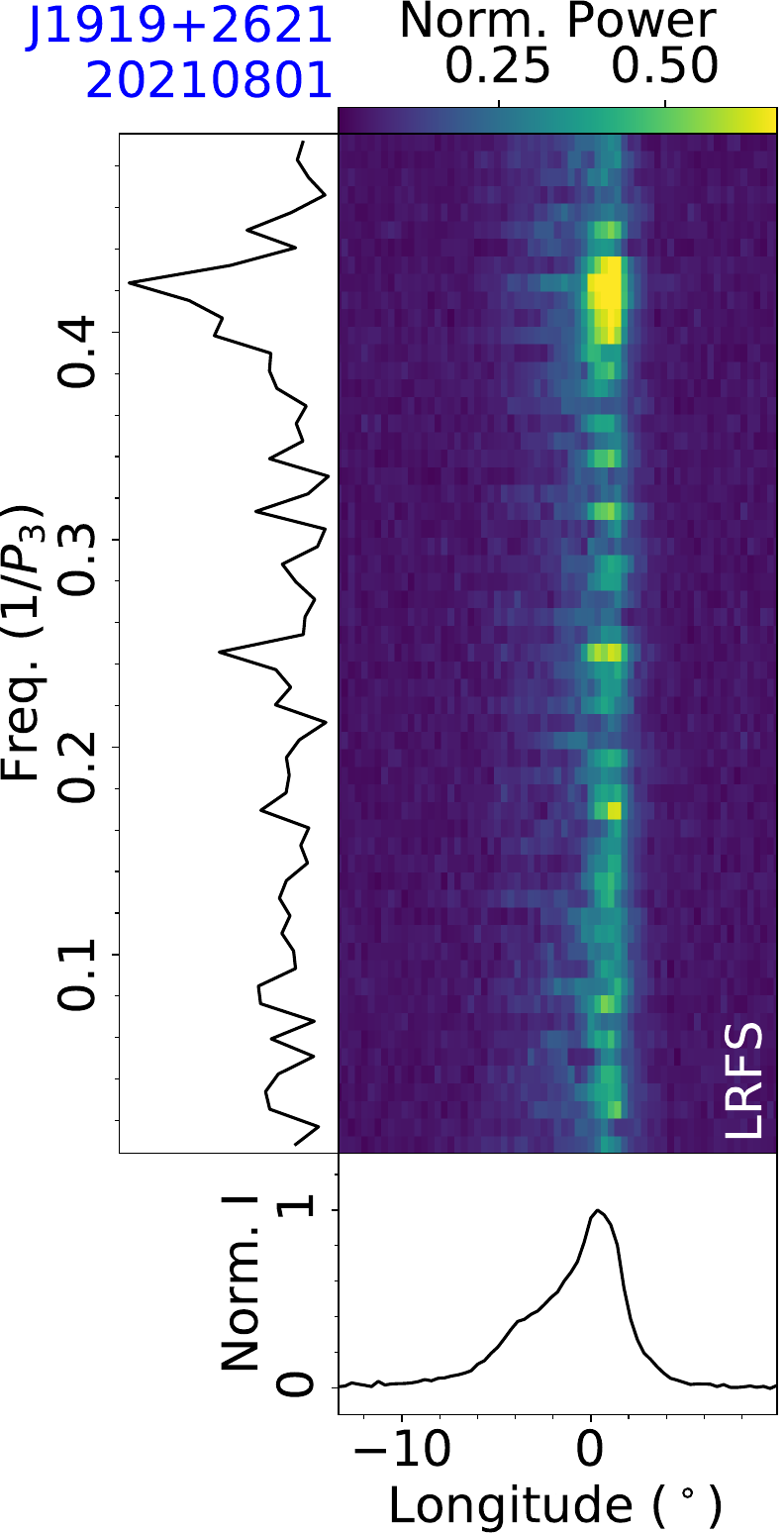}
\includegraphics[width=0.22\textwidth, angle=0]{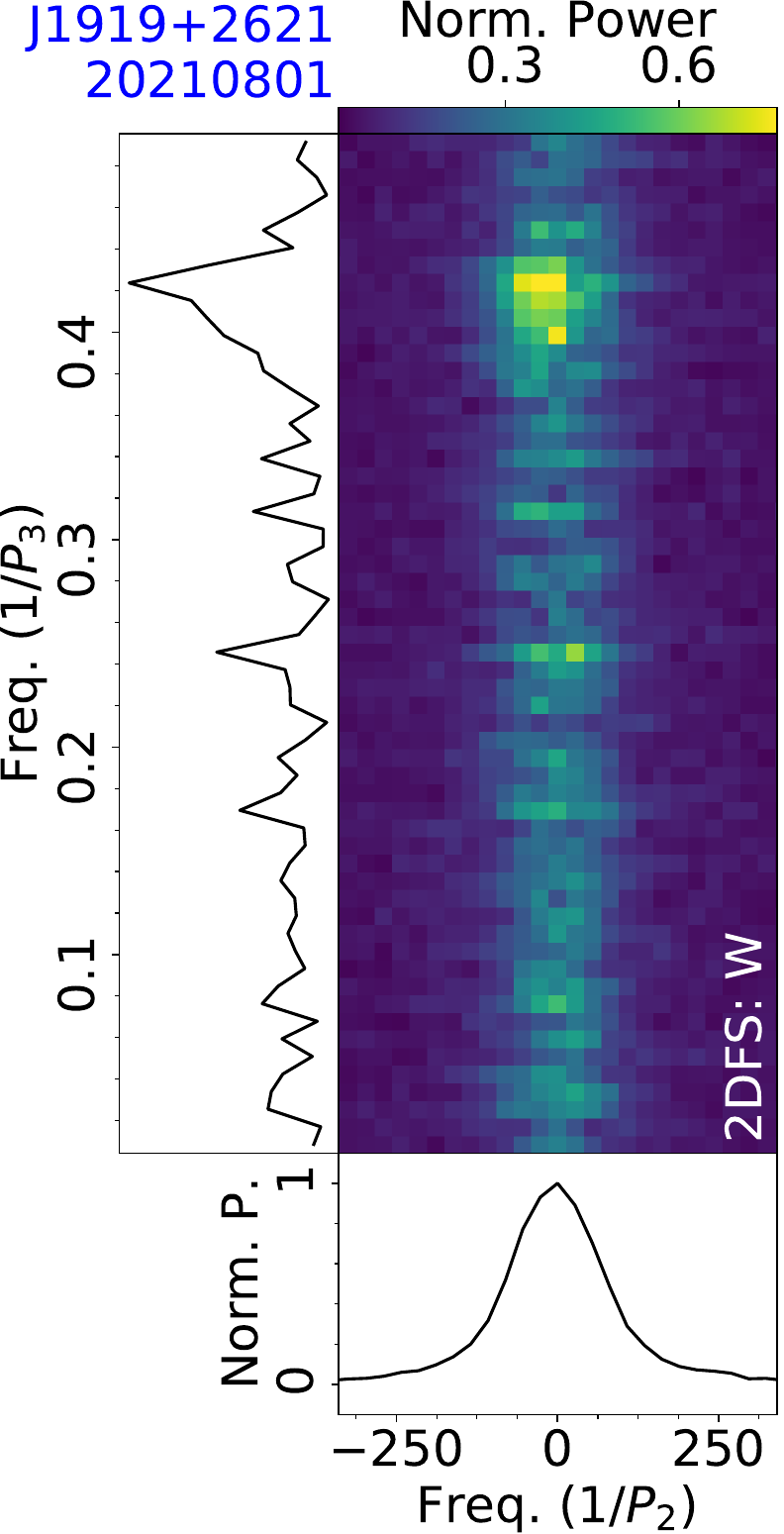}
\figcaption{Fluctuation analysis of PSR J1919+2621 for the observation on 20210801, with LRFS and 2DFS for the on-pulse region of a mean pulse profile. \label{subfig:fluctu:J1919+2621}}
\end{figure}

\begin{figure}[htpb]
\centering
\includegraphics[width=0.22\textwidth, angle=0]{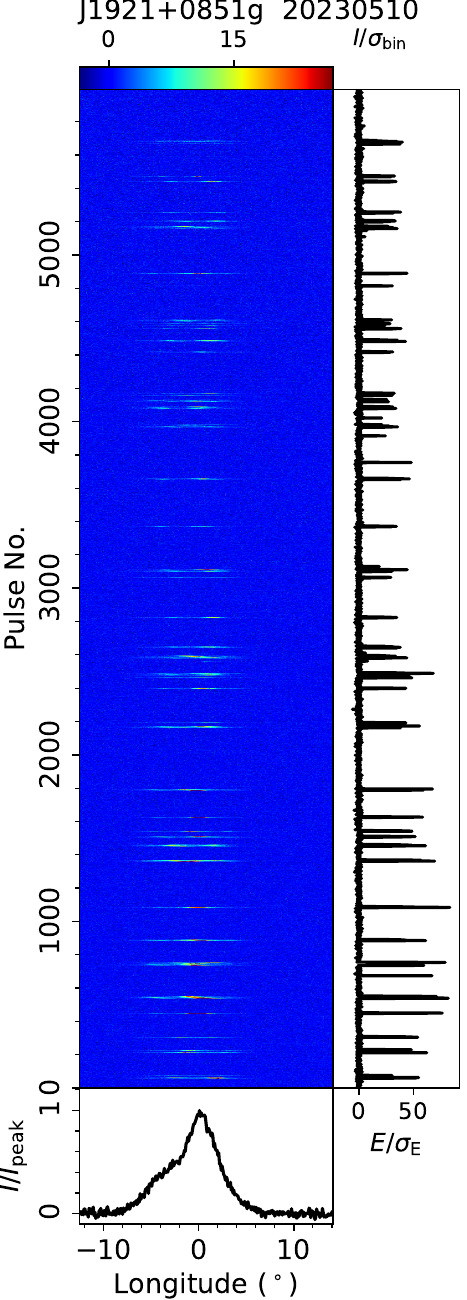}
\includegraphics[width=0.22\textwidth, angle=0]{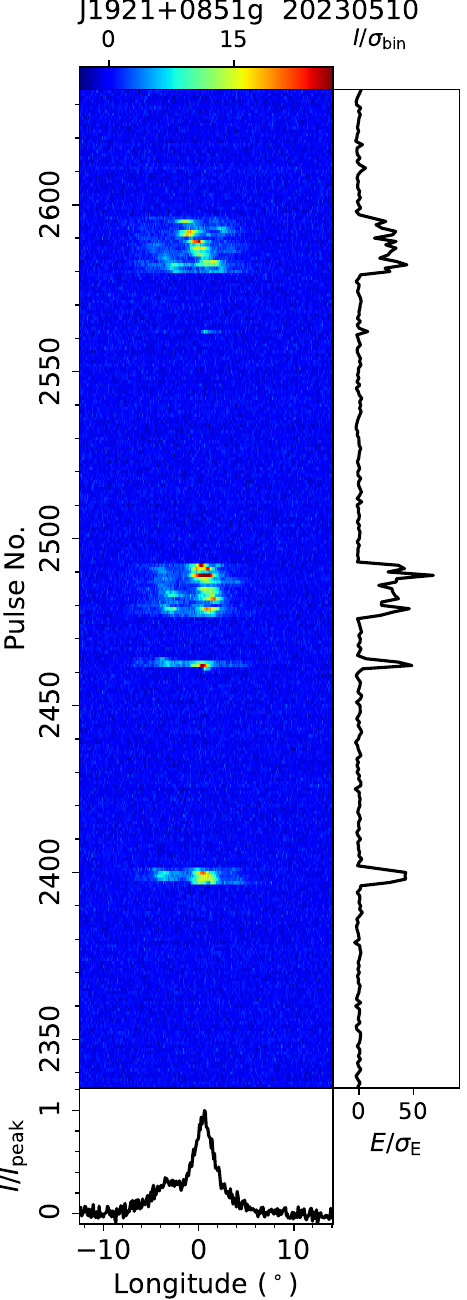}
\figcaption{Single pulse sequence of PSR J1921+0851g from the FAST observation on 20230510, and three zoomed-in views of pulses No. 2335-2635, 2550-2850 and 3950-4250. -- to be continued
\label{subfig:TP:J1921+0851g}}
\end{figure}

\begin{figure}[htpb]
\centering
\includegraphics[width=0.22\textwidth, angle=0]{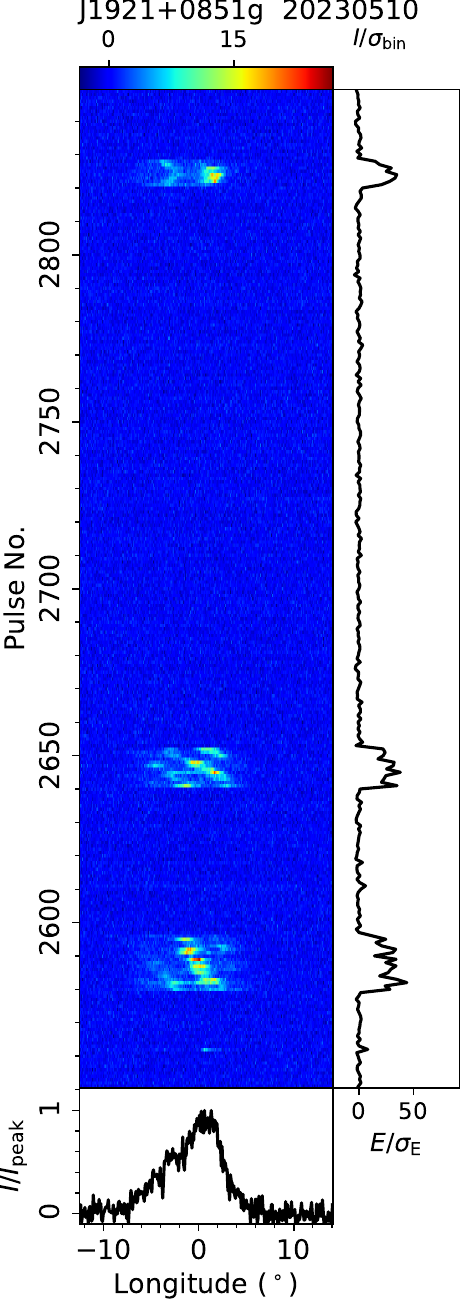}
\includegraphics[width=0.22\textwidth, angle=0]{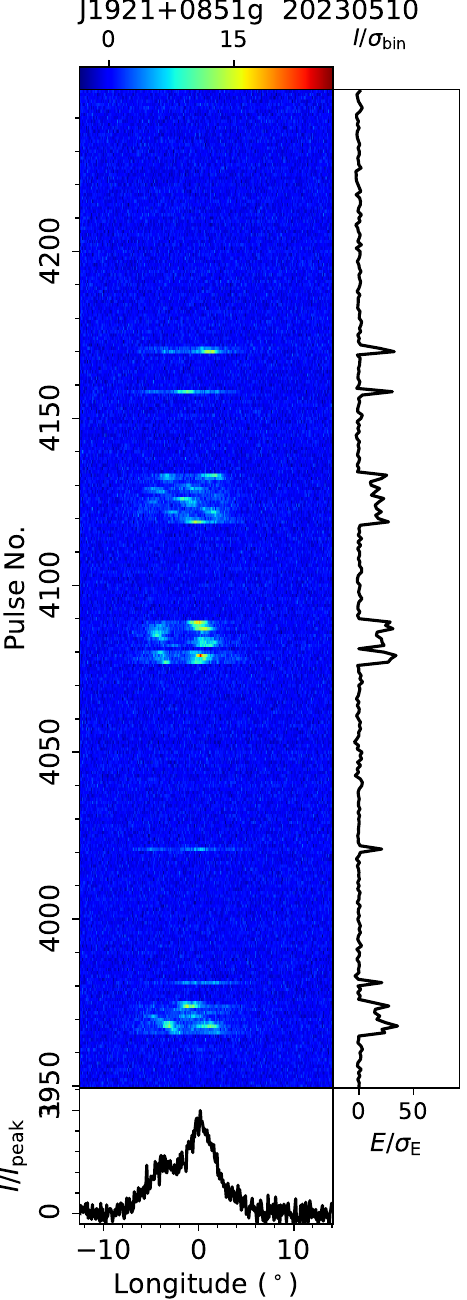}
\figcaption{Continued.
}
\end{figure}

\begin{figure}[htpb]
\centering
\includegraphics[width=0.39\textwidth, angle=0]{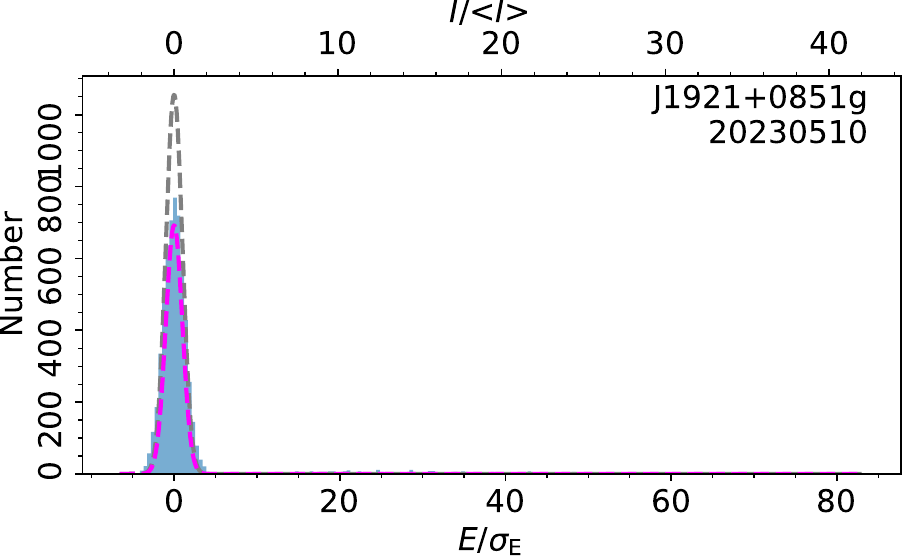}
\figcaption{On-pulse energy histogram of PSR J1921+0851g from the FAST observations on 20230510.
\label{subfig:Hist:J1921+0851g}}
\end{figure}

\begin{figure}[htpb]
\centering
\includegraphics[width=0.42\textwidth, angle=0]{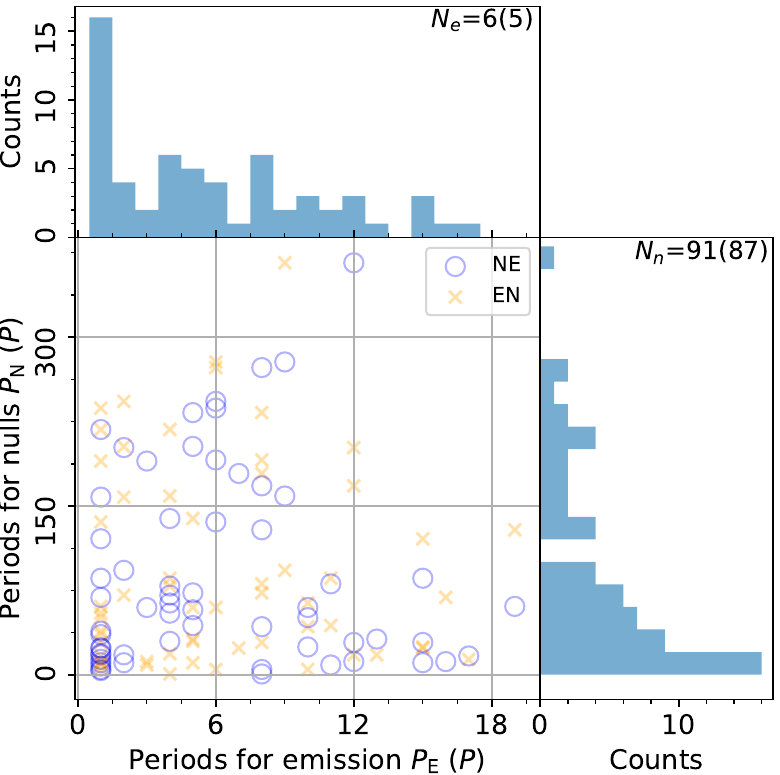}
\figcaption{Distribution of period numbers for continuous nulling $P_N$ against period numbers for adjacent pulses $P_E$ of PSR J1921+0851g observed by FAST on 20230510, as well as the duration histograms for the emission and null shown in the top and right panels, respectively. 
\label{subfig:scaleHist:J1921+0851g}}
\end{figure}

\begin{figure}[htpb]
\centering
\includegraphics[width=0.39\textwidth, angle=0]{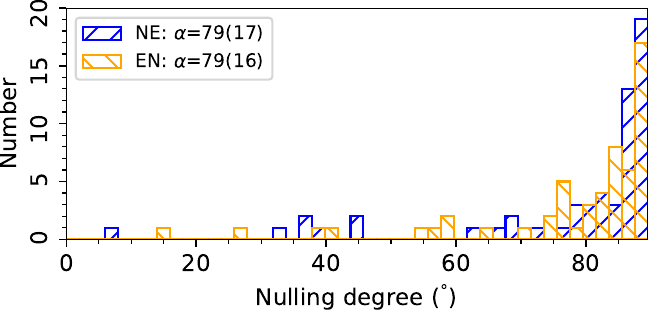}\\
\includegraphics[width=0.39\textwidth, angle=0]{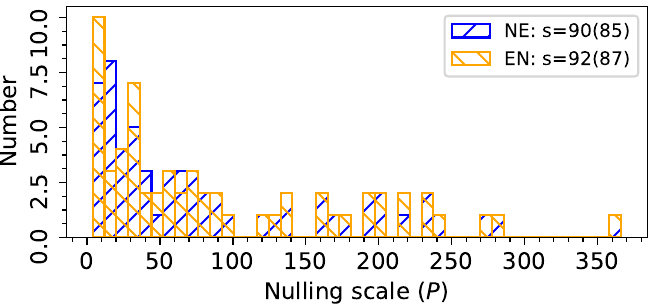}
\figcaption{Histograms of the nulling degree and nulling scale for PSR J1921+0851g observed by FAST on 20230510.
\label{subfig:nullDegreeScale:J1921+0851g}}
\end{figure}

\begin{figure}[htpb]
\centering
\includegraphics[width=0.36\textwidth, angle=0]{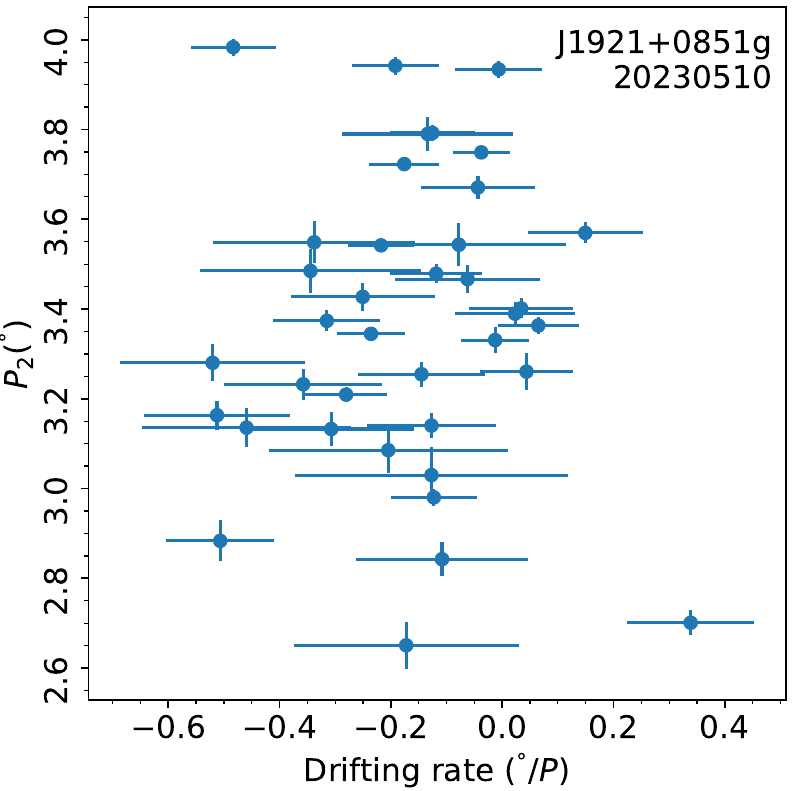}
\figcaption{Scatter plot of $P_2$ versus drifting rate of PSR J1921+0851g from the FAST observation on 20230510.
\label{subfig:DriftDis:J1921+0851g}}
\end{figure}

\subsection{J1919+2621}
\label{subsec:J1919+2621}

PSR J1919+2621 was discovered by FAST \citep{Cameron2020}. 

The pulsar was observed by FAST on 20210801 for 5 minutes, yielding a rotation period $P=0.6515$~s and a dispersion measure $D\!M=96.3~{\rm cm^{-3}\,pc}$. 
Single pulse sequences of the observation are shown in Fig.~\ref{subfig:TP:J1919+2621}, which display unsystematic subpulse drifting phenomenon. From fluctuation spectra in Fig.~\ref{subfig:fluctu:J1919+2621}, this pulsar has a preferred negative drifting direction. The centroid frequencies of the drift feature in 2DFS are estimated to be $1/P_3=0.417\pm0.001$ and $1/P_2=-25\pm2$, which correspond to periodicities of $P_3=2.40\pm0.01$ periods and $P_2=-14\pm1^\circ$.



\begin{figure}[htpt]
\centering
\includegraphics[width=0.22\textwidth, angle=0]{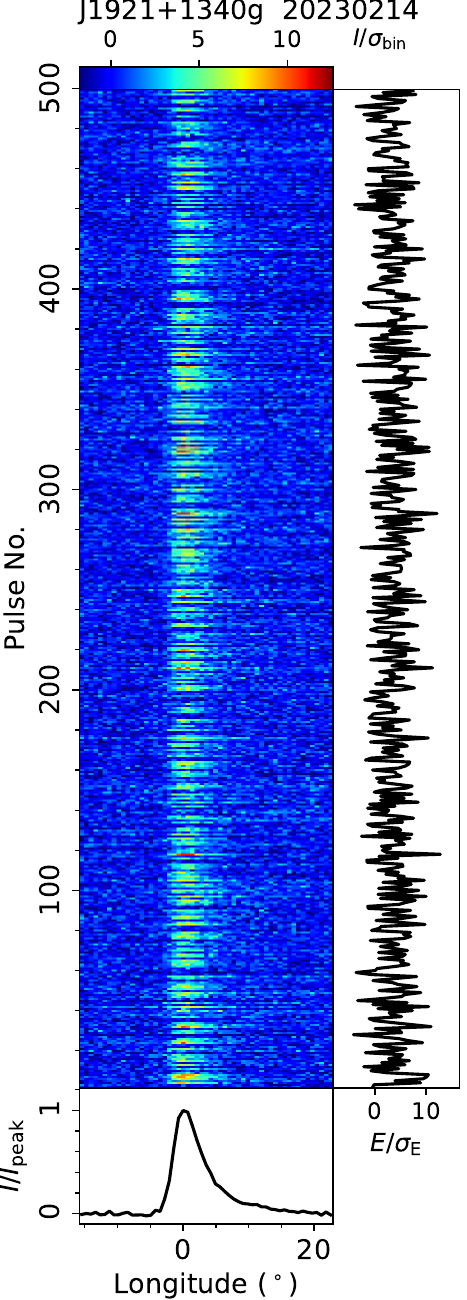}
\includegraphics[width=0.22\textwidth, angle=0]{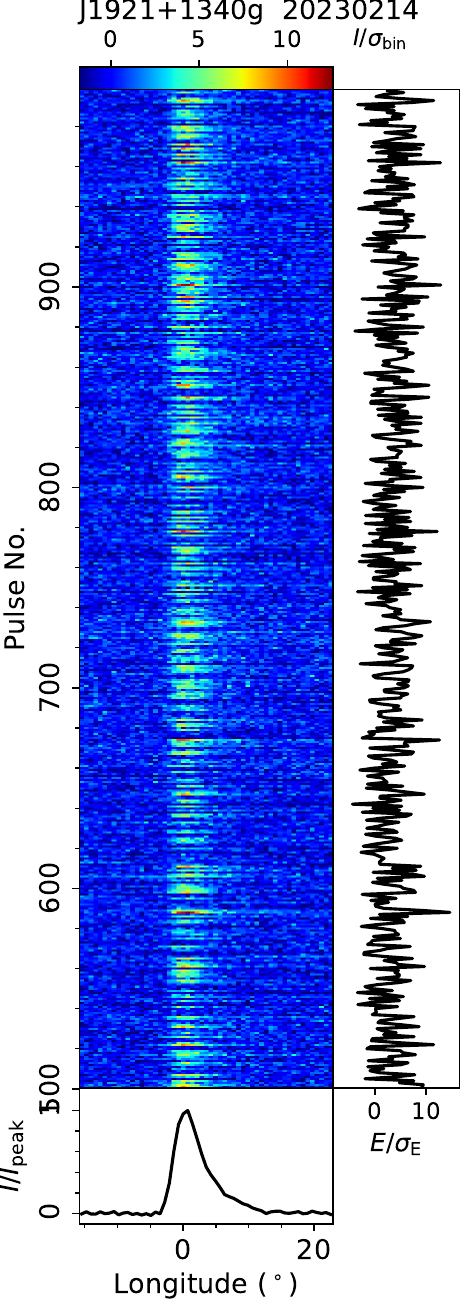}
\figcaption{Single pulse sequences of PSR J1921+1340g from the FAST observation on 20230214.
\label{subfig:TP:J1921+1340g}}
\end{figure}

\begin{figure}[htpt]
\centering
\includegraphics[width=0.39\textwidth, angle=0]{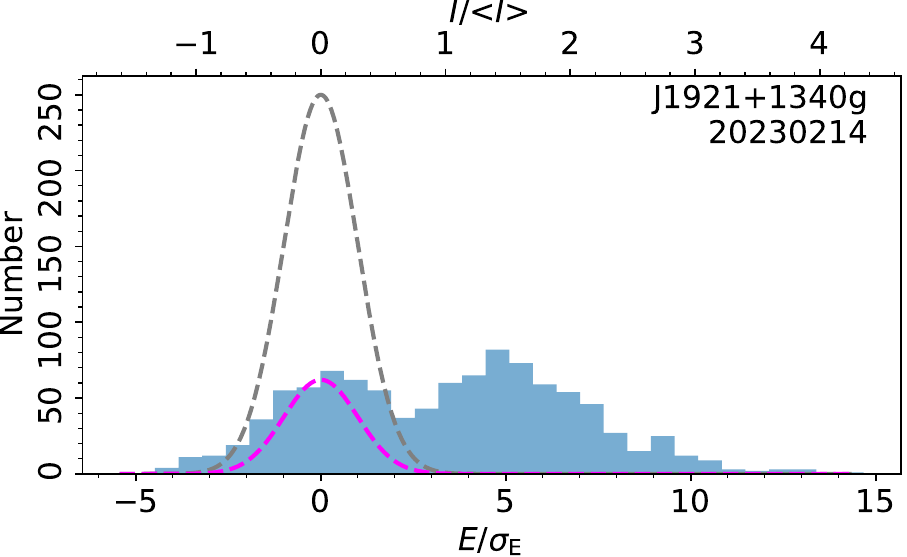}
\vspace{-0.2cm}
\figcaption{On-pulse energy histogram of single pulses of PSR J1921+1340g from the FAST observation on 20230214.
\label{subfig:Hist:J1921+1340g}}
\end{figure}

\begin{figure}[htpb]
\centering
\includegraphics[width=0.44\textwidth, angle=0]{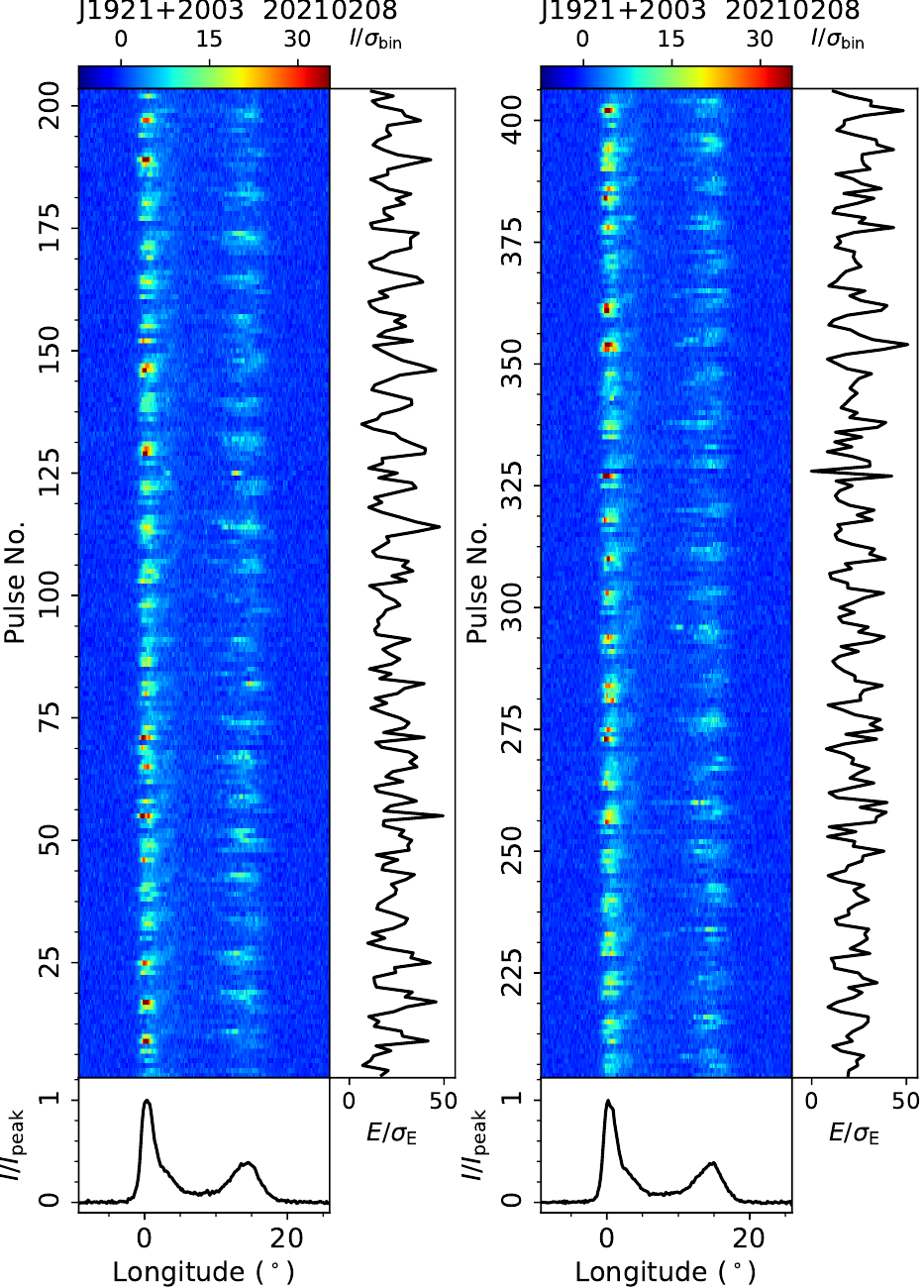}
\figcaption{Single pulse sequences of PSR J1921+2003 from the FAST observation on 20200514.
\label{subfig:TP:J1921+2003}}
\end{figure}

\begin{figure}[htpb]
\centering
\includegraphics[width=0.44\textwidth, angle=0]{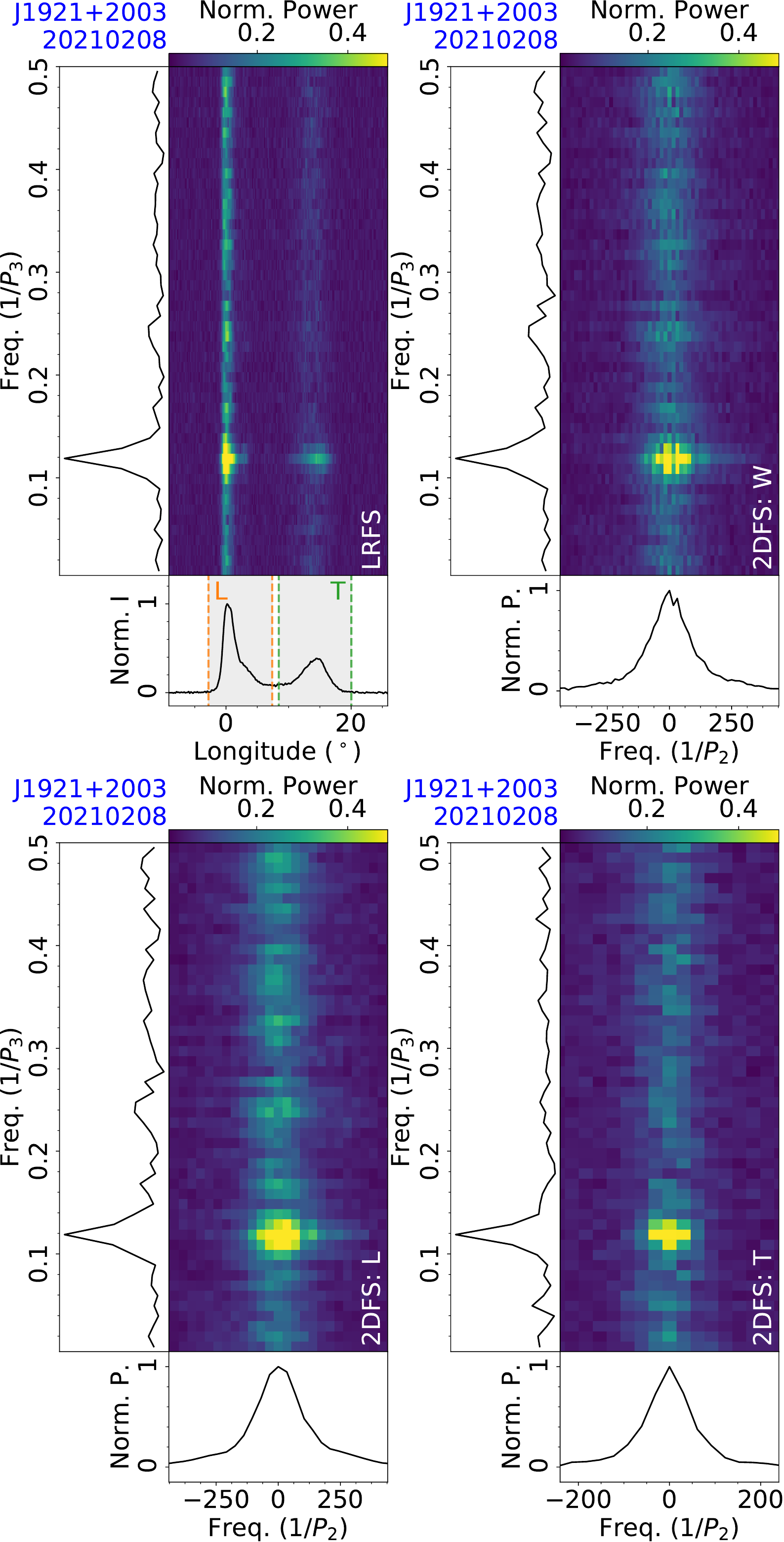}
\figcaption{Fluctuation analysis of PSR J1921+2003 for the observation on 20200514, with LRFS (top-left), and 2DFS for the on-pulse region (top-right), leading part (bottom-left) and trailing part (bottom-right) of a mean pulse profile.
\label{subfig:fluctu:J1921+2003}}
\end{figure}

\begin{figure}[htpb]
\centering
\includegraphics[width=0.22\textwidth, angle=0]{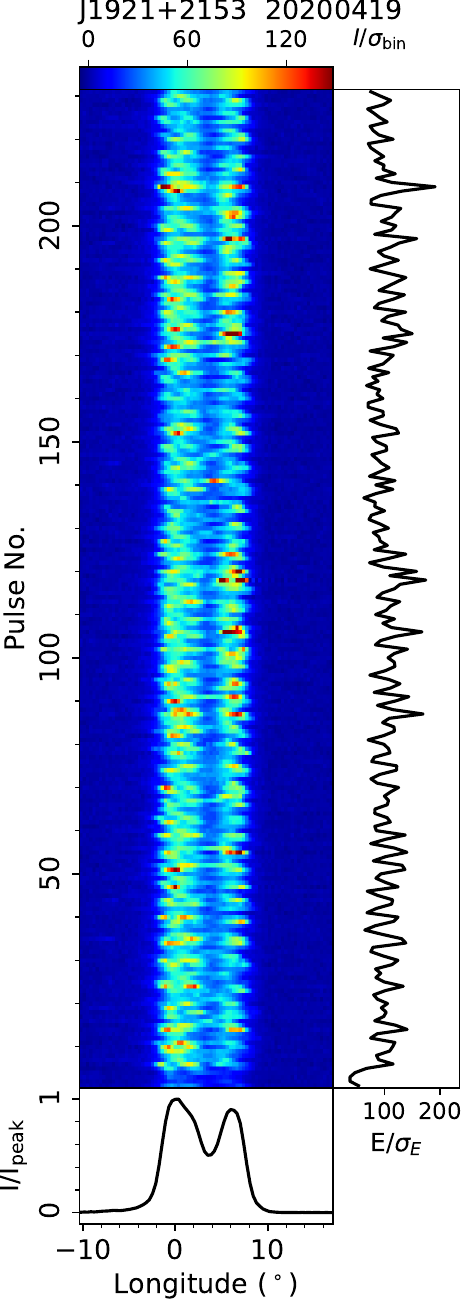}
\figcaption{Single pulse sequence of PSR J1921+2153 from the FAST observation on 20200419.
\label{subfig:TP:J1921+2153}}

\centering
\includegraphics[width=0.39\textwidth, angle=0]{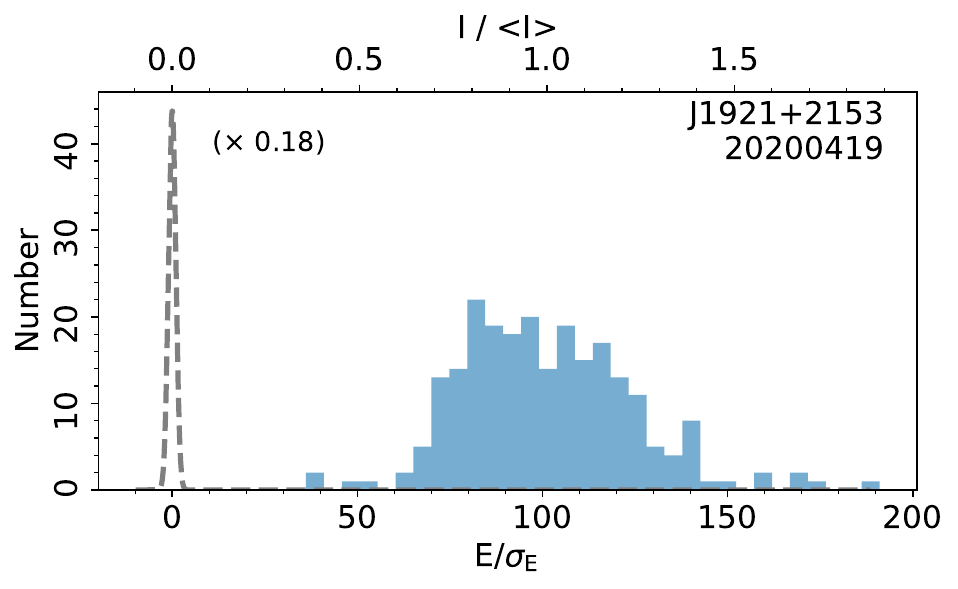}
\vspace{-0.2cm}
\figcaption{On-pulse energy histogram of single pulses of PSR J1921+2153 from the FAST observation on 20210111.
\label{subfig:Hist:J1921+2153}}
\end{figure}

\begin{figure}[htpb]
\centering
\includegraphics[width=0.22\textwidth, angle=0]{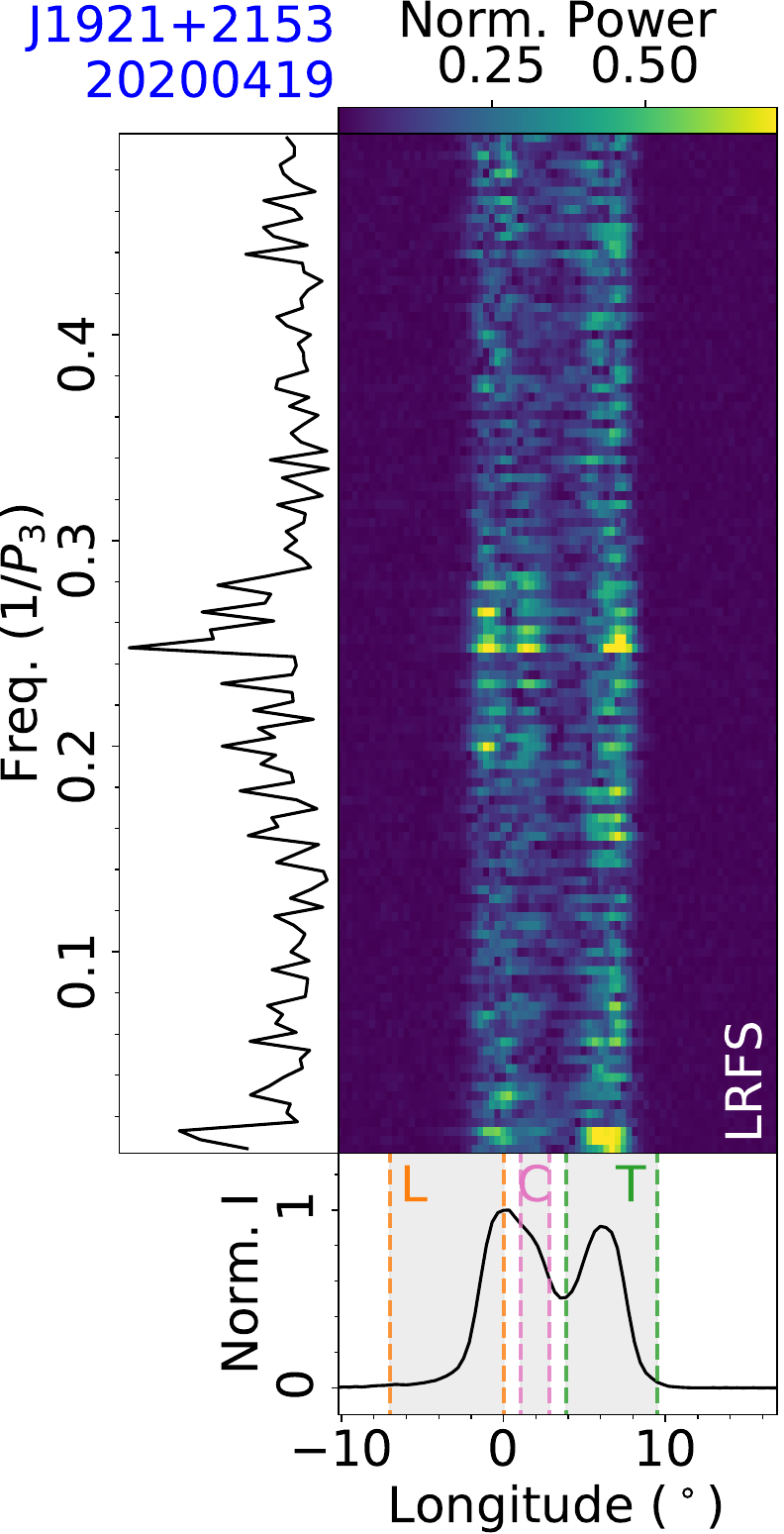}
\includegraphics[width=0.22\textwidth, angle=0]{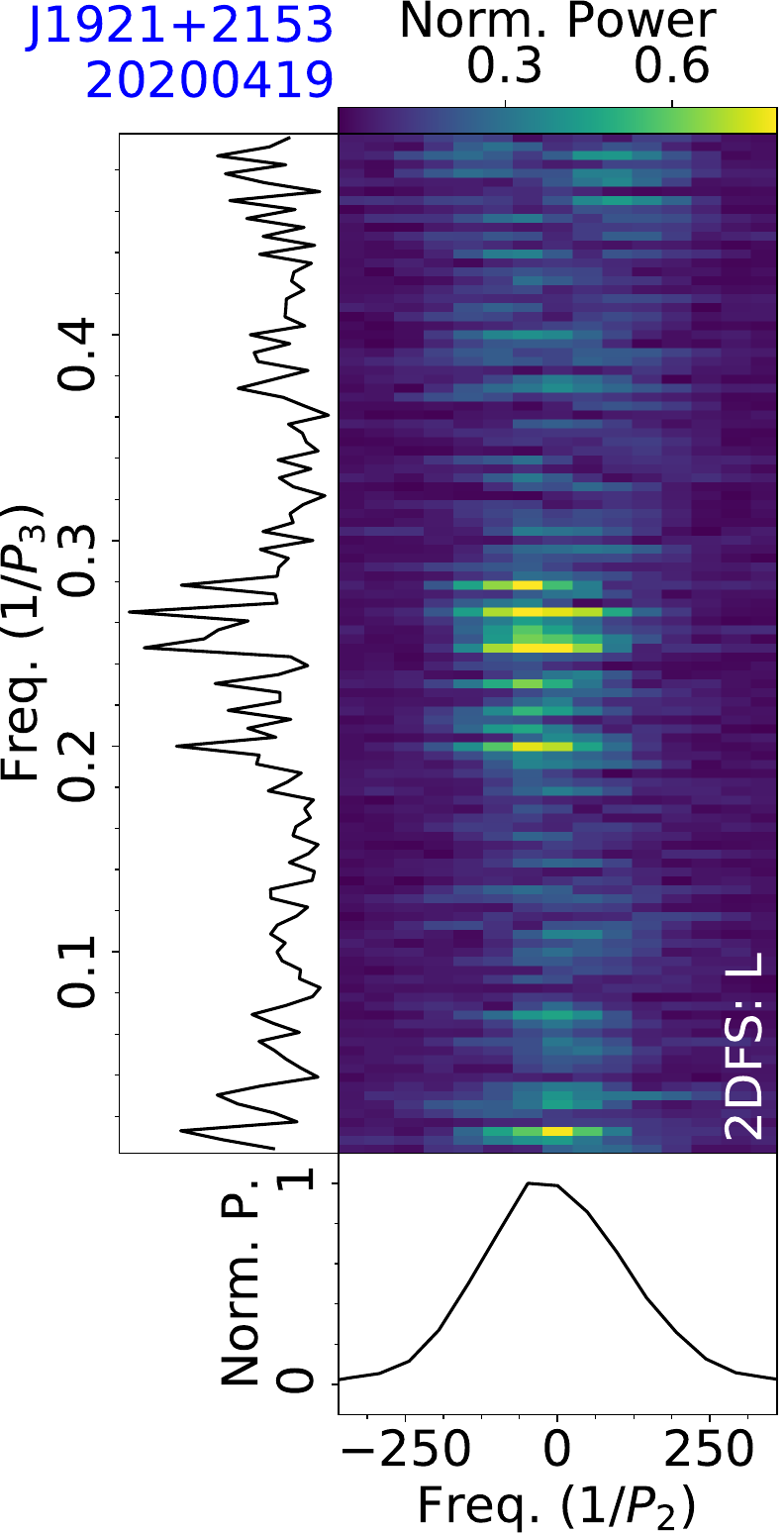}\\
\includegraphics[width=0.22\textwidth, angle=0]{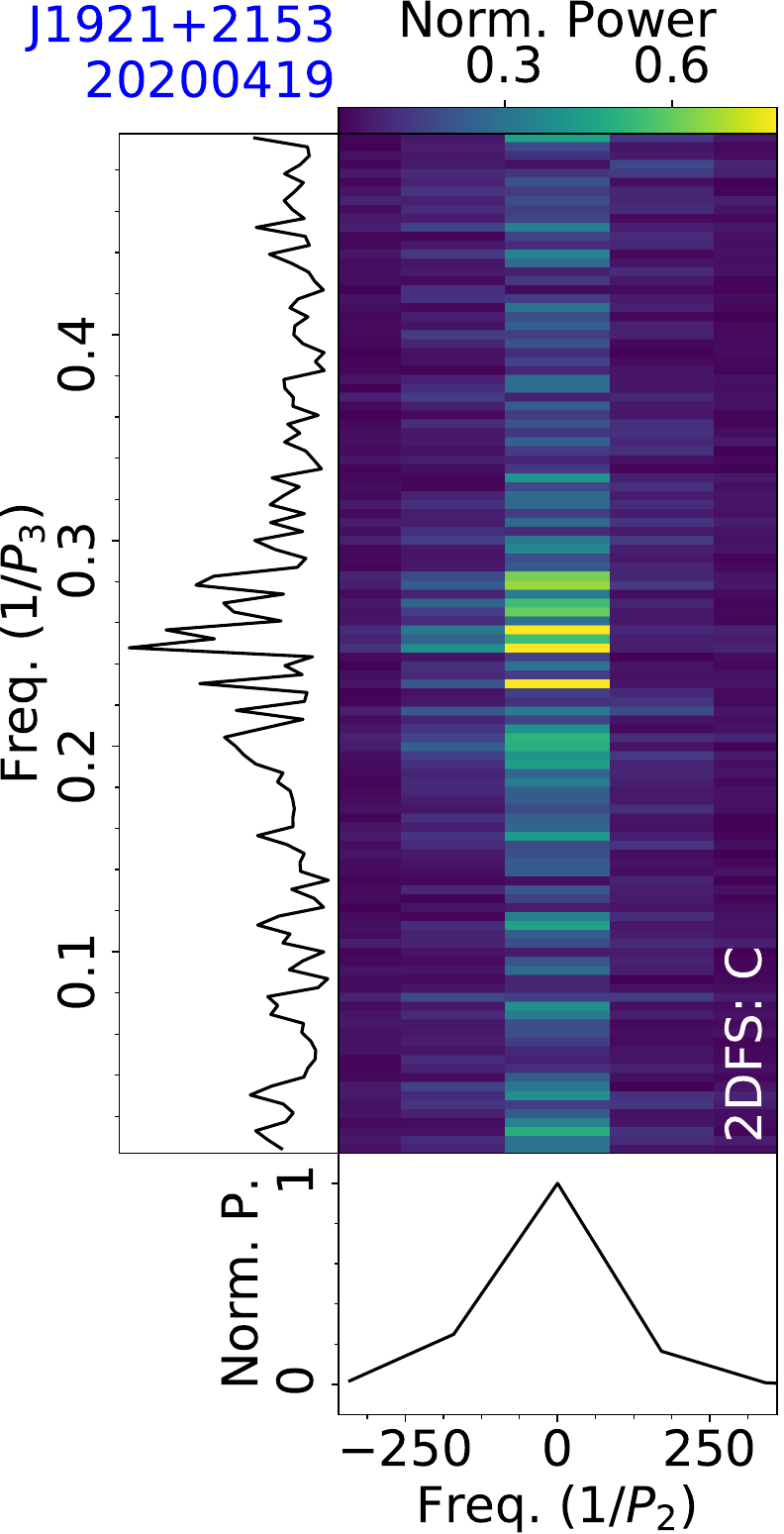}
\includegraphics[width=0.22\textwidth, angle=0]{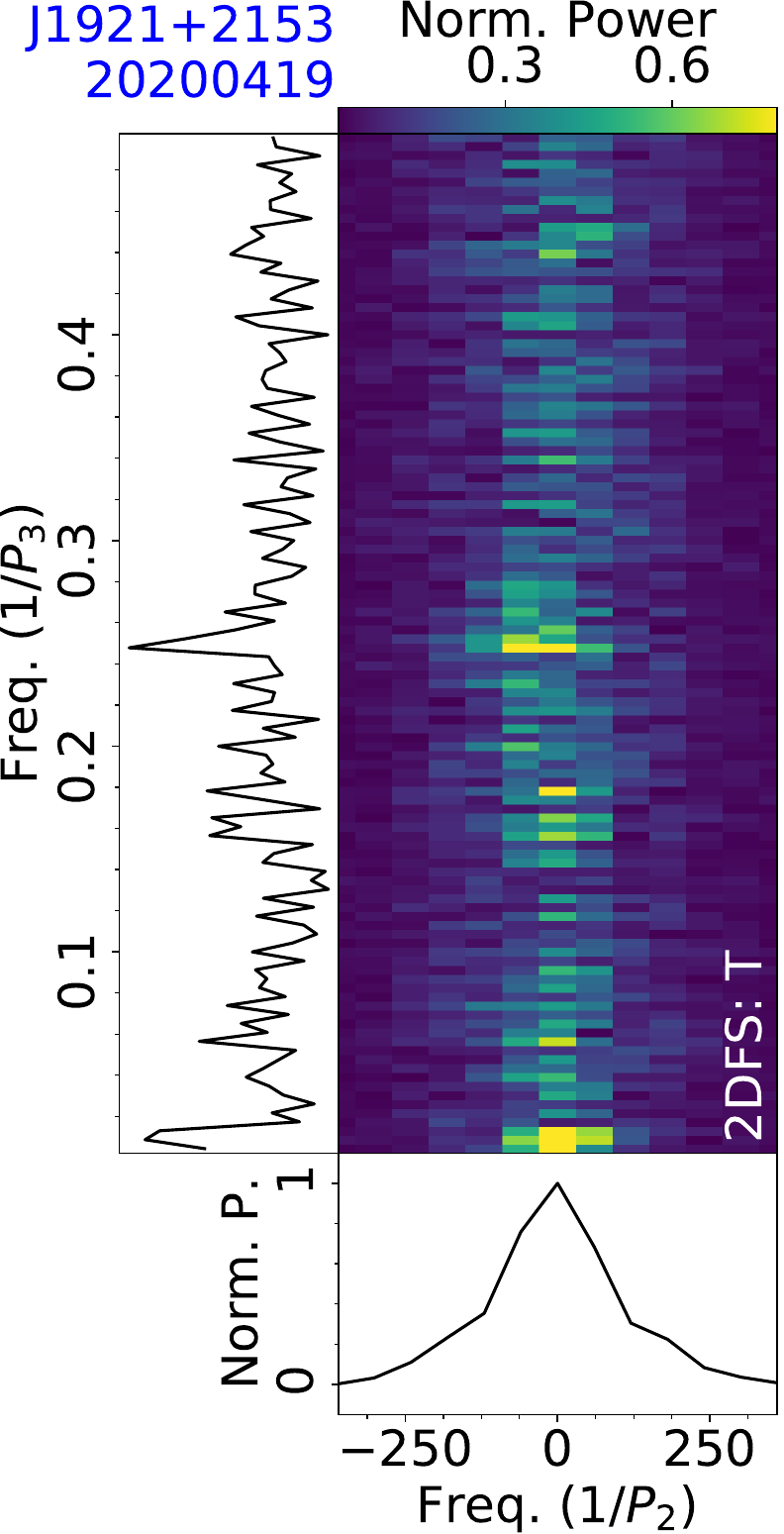}
\figcaption{Fluctuation analysis of PSR J1921+2153 for the observation on 20200419, with LRFS (top-left), and 2DFS for the leading part (top-right), central part (bottom-left) and trailing part (bottom-right) of a mean pulse profile.
\label{subfig:fluctu:J1921+2153}}
\end{figure}

\begin{figure}[htpb]
\centering
\includegraphics[width=0.22\textwidth, angle=0]{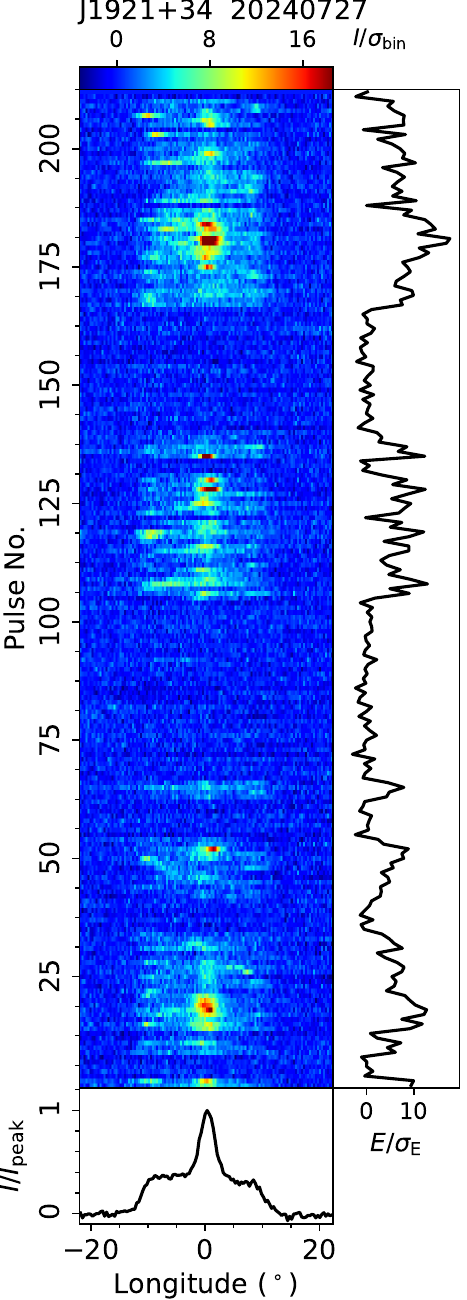}
\figcaption{Single pulse sequence of PSR J1921+34 from the FAST observation on 20240727.
\label{subfig:TP:J1921+34}}
\end{figure}

\begin{figure}[htpb]
\centering
\includegraphics[width=0.39\textwidth, angle=0]{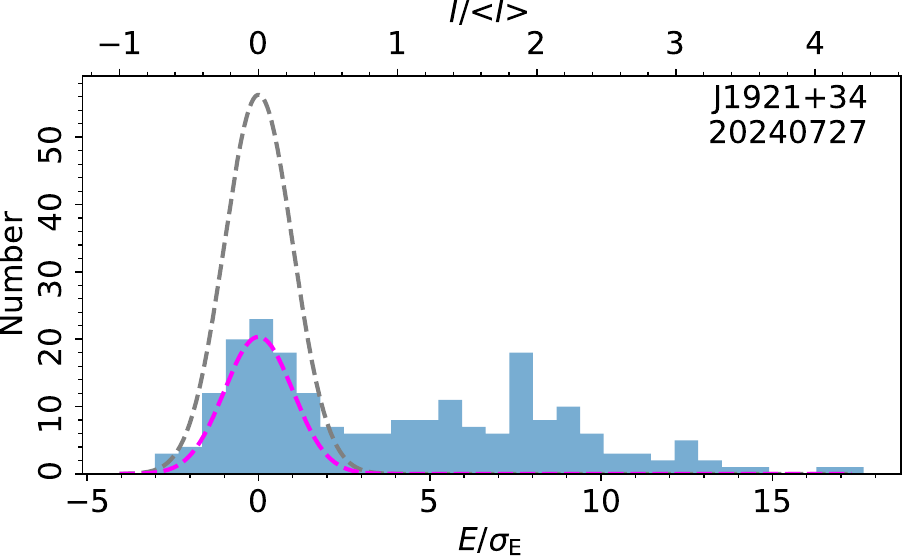}
\figcaption{On-pulse energy histogram of single pulses of PSR J1921+34 from the FAST observation on 20240727.
\label{subfig:Hist:J1921+34}}
\end{figure}

\subsection{J1921+0851g}
\label{subsec:J1921+0851g}

PSR J1921+0851g was discovered in the FAST GPPS survey \citep{Han2021,han2025}, and reported as an extremely nulling pulsar \citep{Zhou2023}.

This pulsar was observed by FAST on 20230510 for 96 minutes, deriving a rotation period $P=0.9566$~s and a dispersion measure $D\!M=102.4~{\rm cm^{-3}\,pc}$. 
The single pulse sequence of this observation and three zoomed-in views are shown in Fig.~\ref{subfig:TP:J1921+0851g}, illustrating the existence of nulling, as well as subpulse drifting behaviors. 

From the on-pulse integral energy histogram in Fig.~\ref{subfig:Hist:J1921+0851g}, the nulling fraction of this observation is estimated to be 66$\pm$6\%. The distribution of continuous period numbers for adjacent nulling and emission states is shown in Fig.~\ref{subfig:scaleHist:J1921+0851g}, where the duration of the nulling state is 91$\pm$87 periods, and that of the emission state is 6$\pm$5 periods. From histograms in Fig.~\ref{subfig:nullDegreeScale:J1921+0851g}, nulling degree and scale for NE pairs are 79$\pm$17 degrees and 90$\pm$85 periods, and those for EN pairs are 79$\pm$16 degrees and 92$\pm$87 periods, indicating a significantly longer duration of the nulling state compared to the emission state. 

This pulsar also displays subpulse drifting behavior, as zoomed-in views of the single pulse sequence in Fig.~\ref{subfig:TP:J1921+0851g}. Notably, the drifting properties change in different continuous emission segments, and the drifting direction also changes occasionally. For instance, the drifting rate and direction change in the segment between pulses No. 2819 and 2827. The drifting rates of four segments of pulses No. 2477-2492, 2580-2596, 2641-2652, and 4119-4133 are estimated from the correlation method to be $-0.17\pm0.06$, $-0.12\pm0.09$, $-0.48\pm0.08$, and $-0.55\pm0.19$ degrees per period. For emission segments that continued for more than 4 periods, drifting parameters $D$ and $P_2$ are estimated from the correlation method and plotted in Fig.~\ref{subfig:DriftDis:J1921+0851g}. In this distribution, the mean and standard deviation of $D$ are $-$0.17 and 0.19 degrees per period, and those of $P_2$ are 3.36 and 0.33 degrees.

\begin{figure}[htpb]
\centering
\includegraphics[width=0.22\textwidth, angle=0]{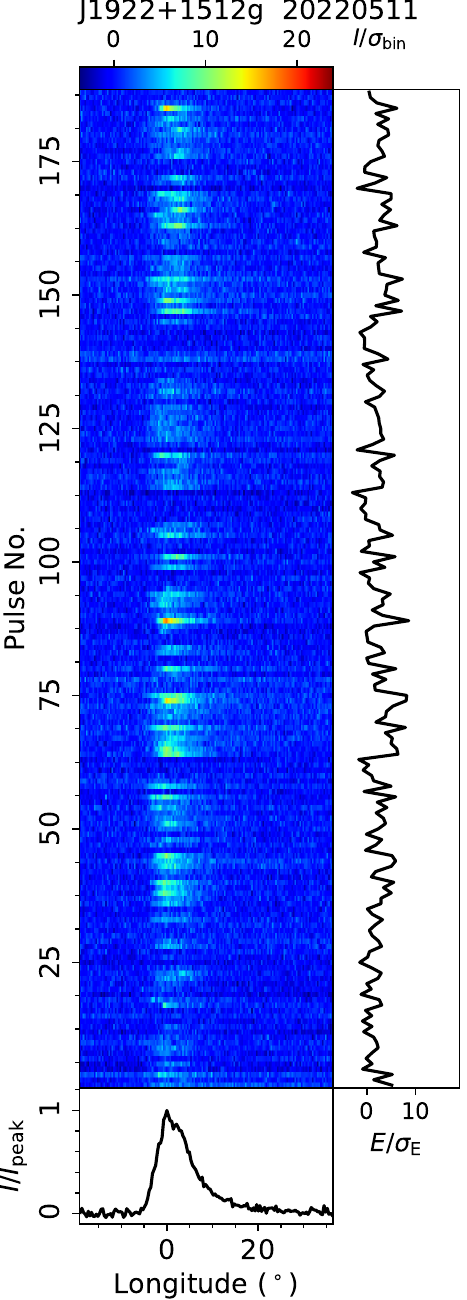}
\includegraphics[width=0.22\textwidth, angle=0]{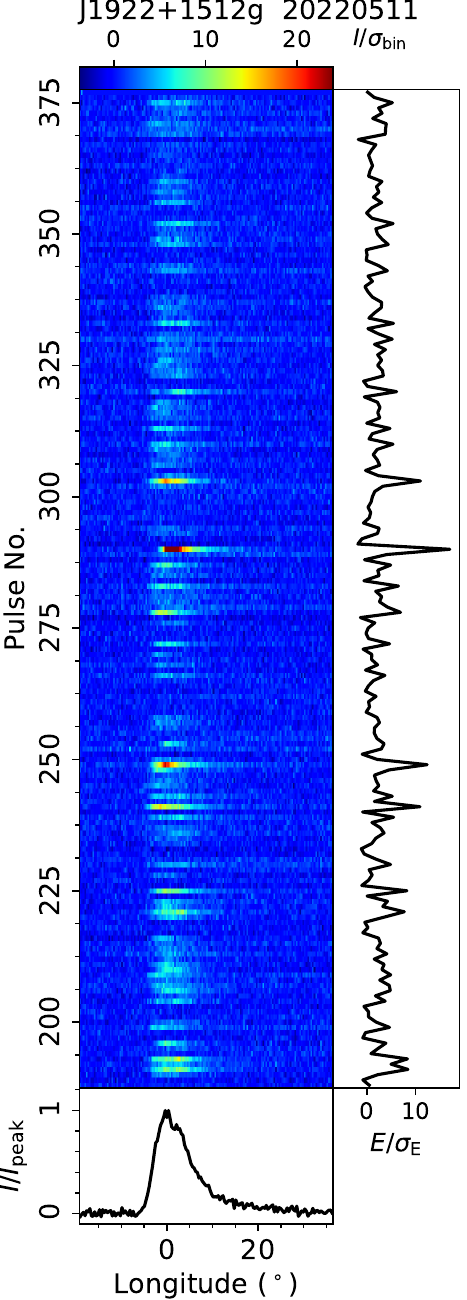}
\figcaption{Single pulse sequences of PSR J1922+1512g from the FAST observation on 20220511.
\label{subfig:TP:J1922+1512g}}
\end{figure}

\begin{figure}[htpb]
\centering
\includegraphics[width=0.39\textwidth, angle=0]{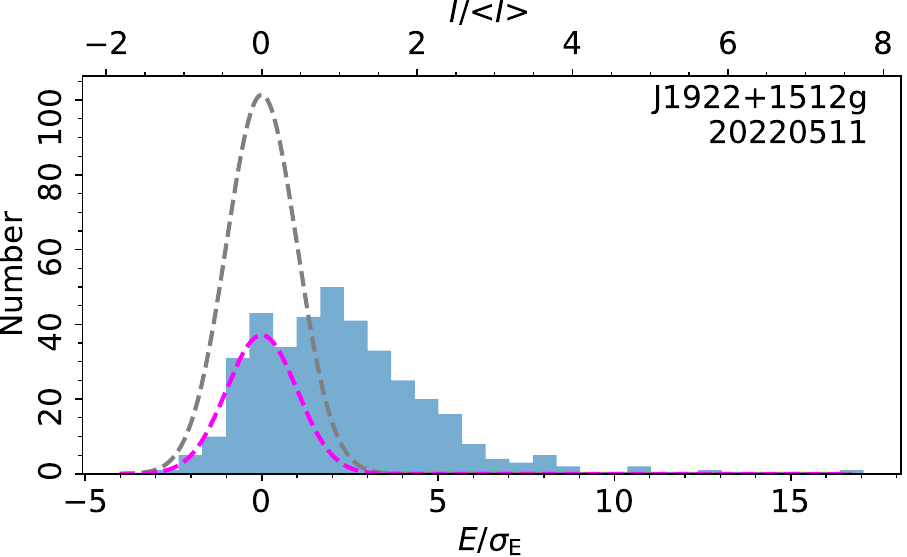}
\figcaption{On-pulse energy histogram of single pulses of PSR J1922+1512g from the FAST observation on 20220511.
\label{subfig:Hist:J1922+1512g}}
\end{figure}

\subsection{J1921+1340g}
\label{subsec:J1921+1340g}

PSR J1921+1340g was discovered in the FAST GPPS survey \citep{Han2021,han2025}. 

This pulsar was observed by FAST on 20230214 for 76 minutes, deriving a rotation period $P=4.6029$~s and a dispersion measure $D\!M=735.3~{\rm cm^{-3}\,pc}$. 
Single pulse sequences in Fig.~\ref{subfig:TP:J1921+1340g} show nulling behavior. From the on-pulse integral energy histogram in Fig.~\ref{subfig:Hist:J1921+1340g}, the nulling fraction is estimated to be 25$\pm$3\%.

\begin{figure}[htpb]
\centering
\includegraphics[width=0.22\textwidth, angle=0]{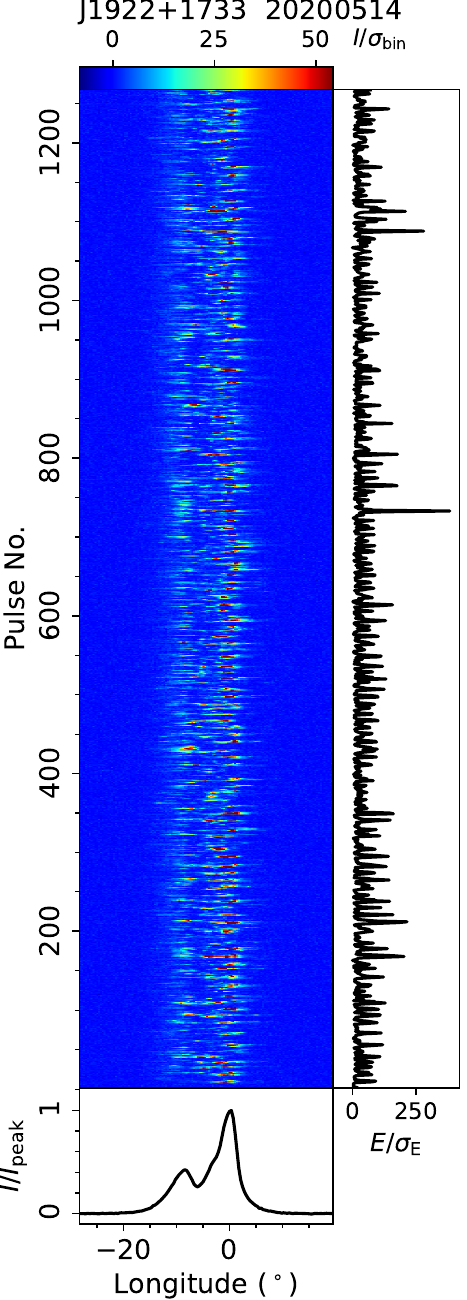}
\includegraphics[width=0.22\textwidth, angle=0]{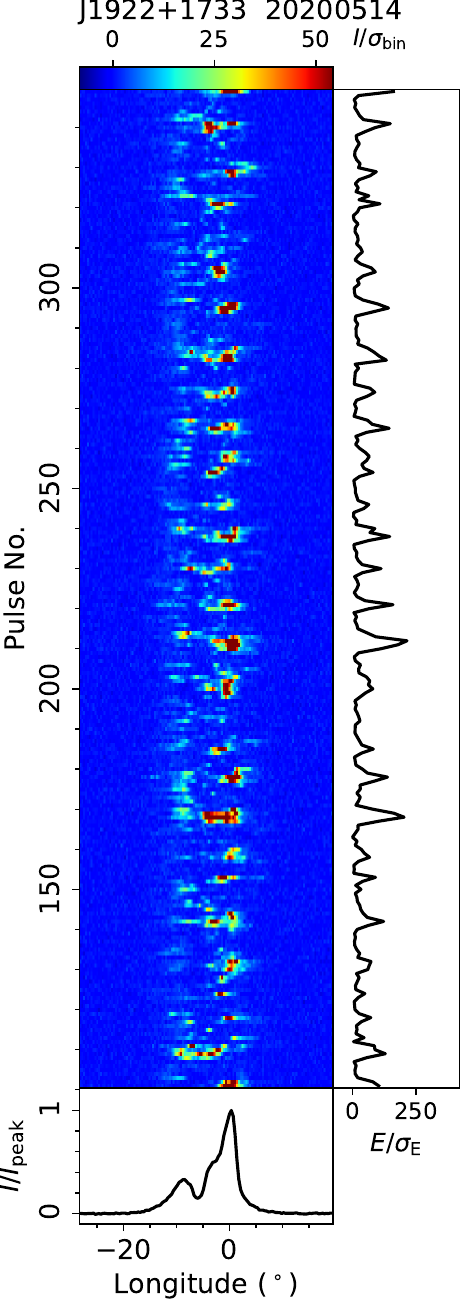}
\figcaption{Single pulse sequence of PSR J1922+1733 from the FAST observation on 20200514, and a zoomed-in view of pulses No. 100-350. 
\label{subfig:TP:J1922+1733}}
\end{figure}

\begin{figure}[htpb]
\centering
\includegraphics[width=0.22\textwidth, angle=0]{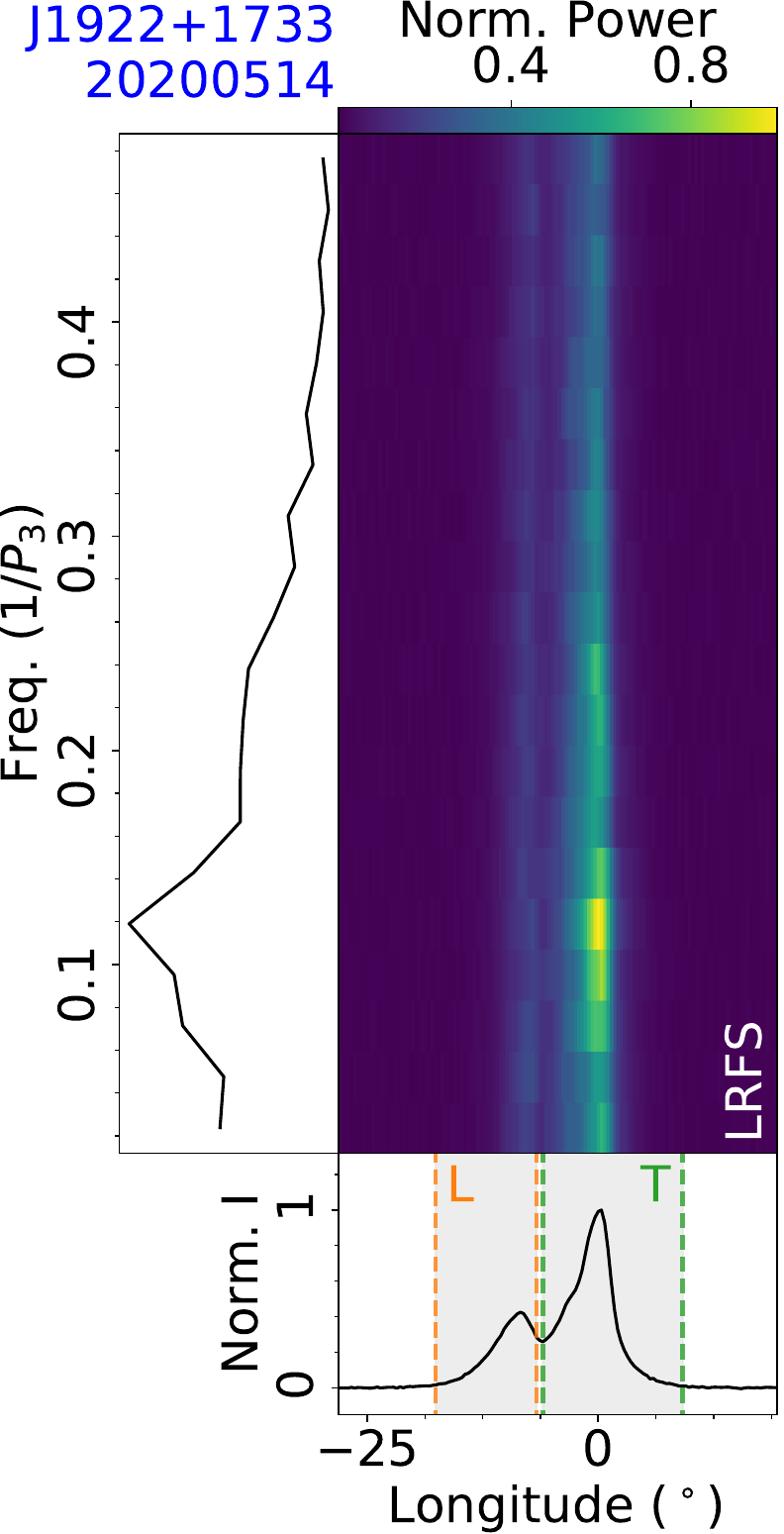}
\includegraphics[width=0.22\textwidth, angle=0]{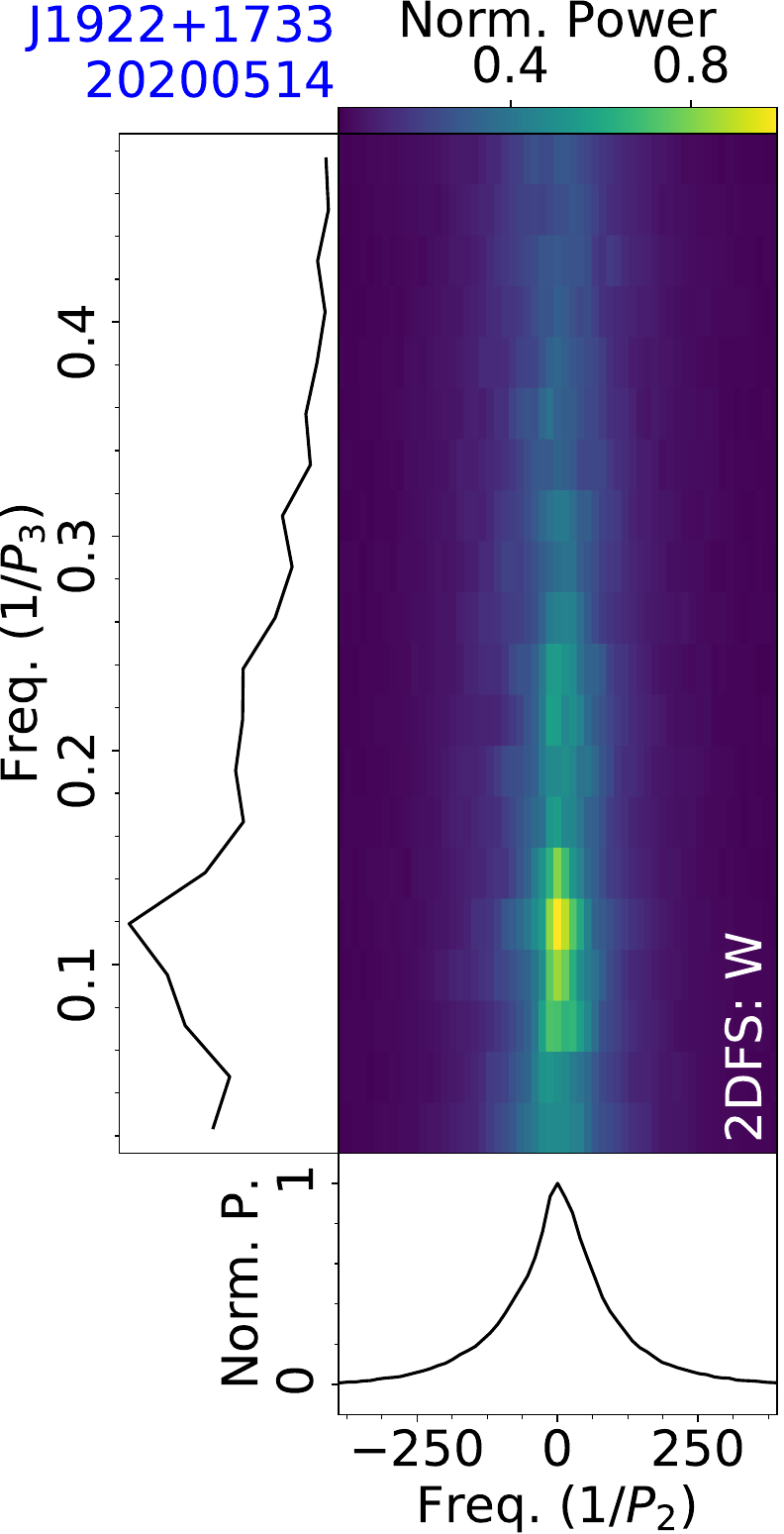}\\
\includegraphics[width=0.22\textwidth, angle=0]{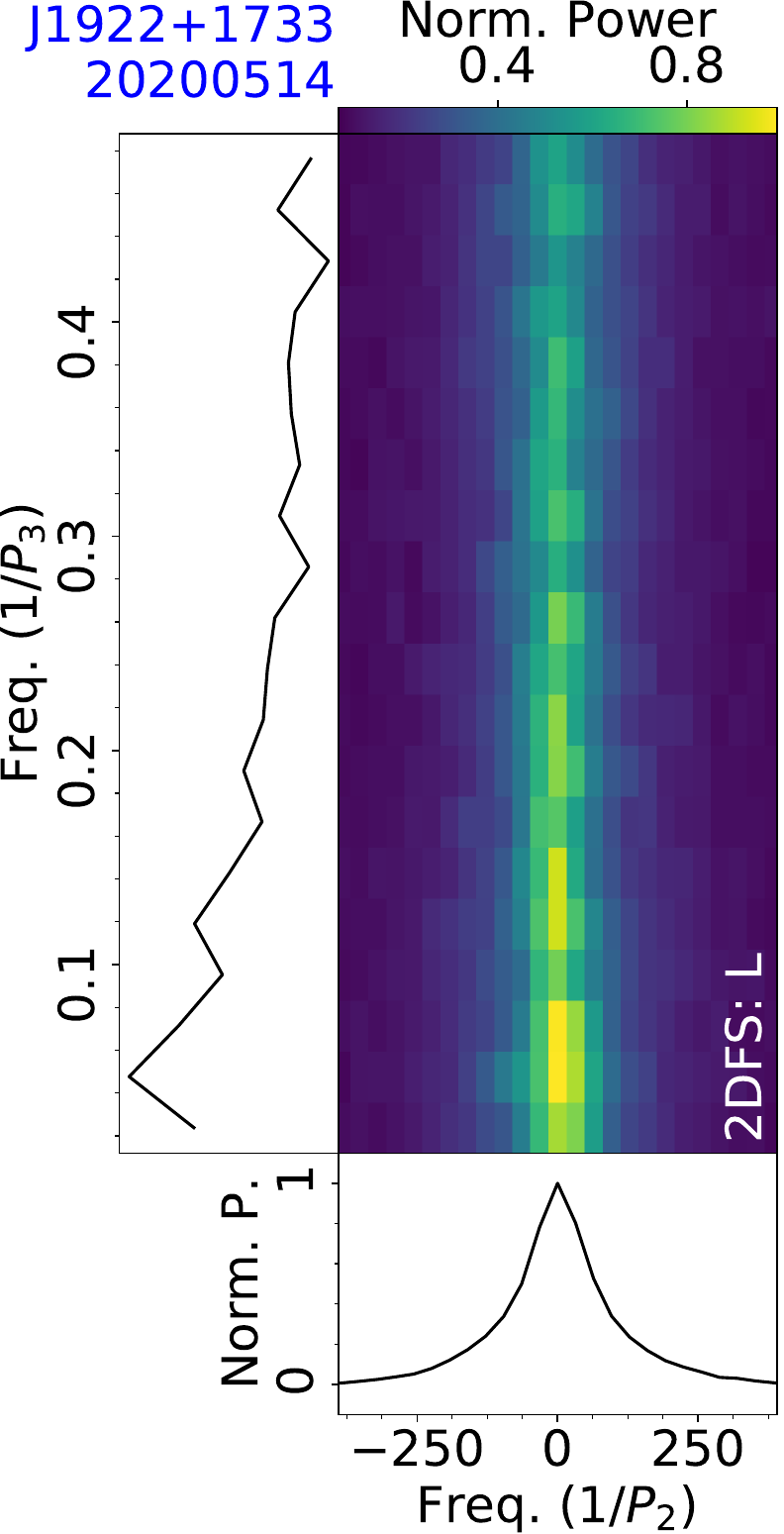}
\includegraphics[width=0.22\textwidth, angle=0]{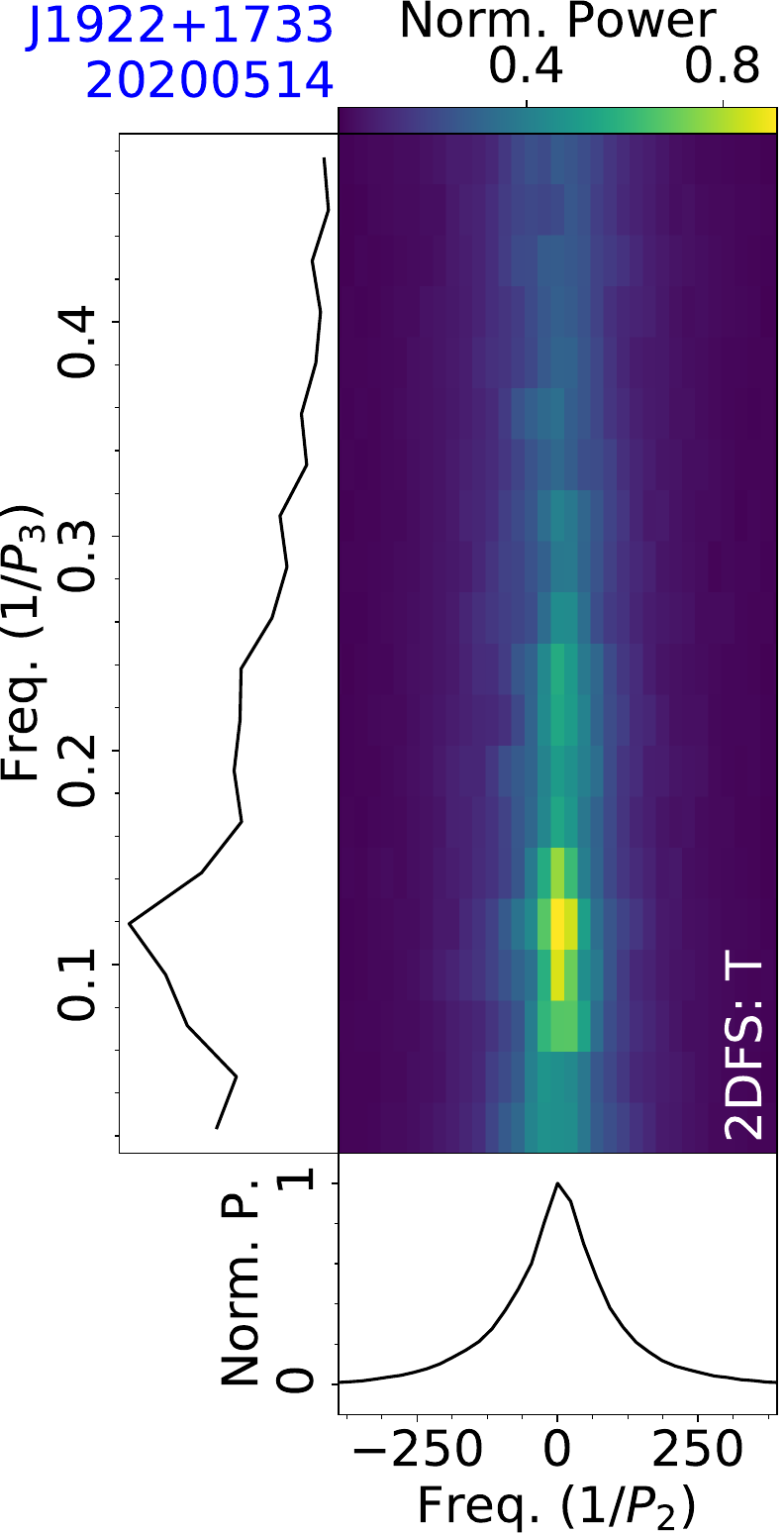}
\figcaption{Fluctuation analysis of PSR J1922+1733 for the observation on 20200514, with LRFS (top-left), and 2DFS for the on-pulse region (top-right), leading part (bottom-left) and trailing part (bottom-right) of a mean pulse profile.
\label{subfig:fluctu:J1922+1733}}
\end{figure}

\begin{figure}[htpb]
\centering
\includegraphics[width=0.22\textwidth, angle=0]{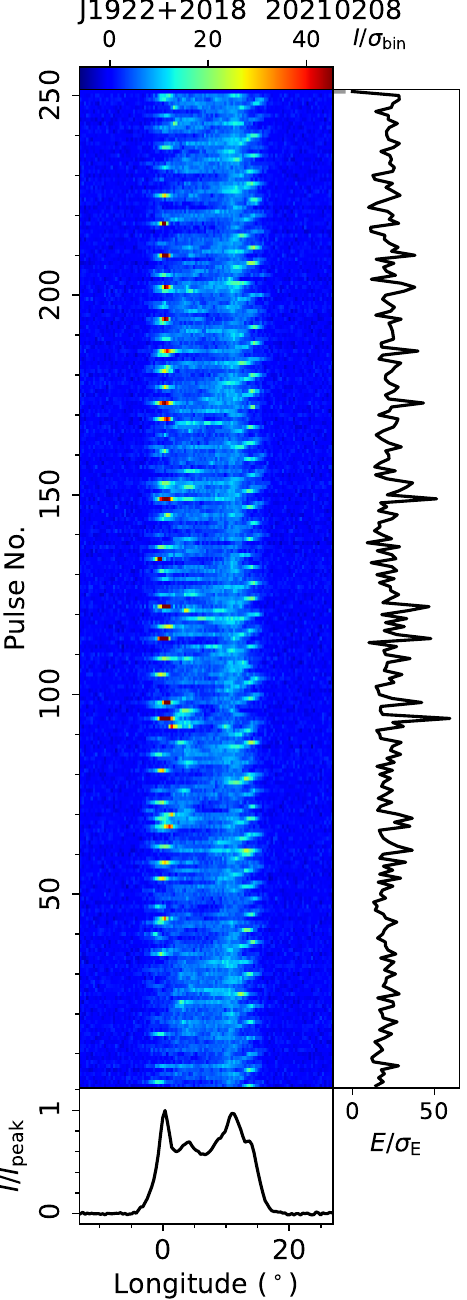}
\figcaption{Single pulse sequence of PSR J1922+2018 from the FAST observation on 20210208.
\label{subfig:TP:J1922+2018}}
\end{figure}

\begin{figure}[htpb]
\centering
\includegraphics[width=0.22\textwidth, angle=0]{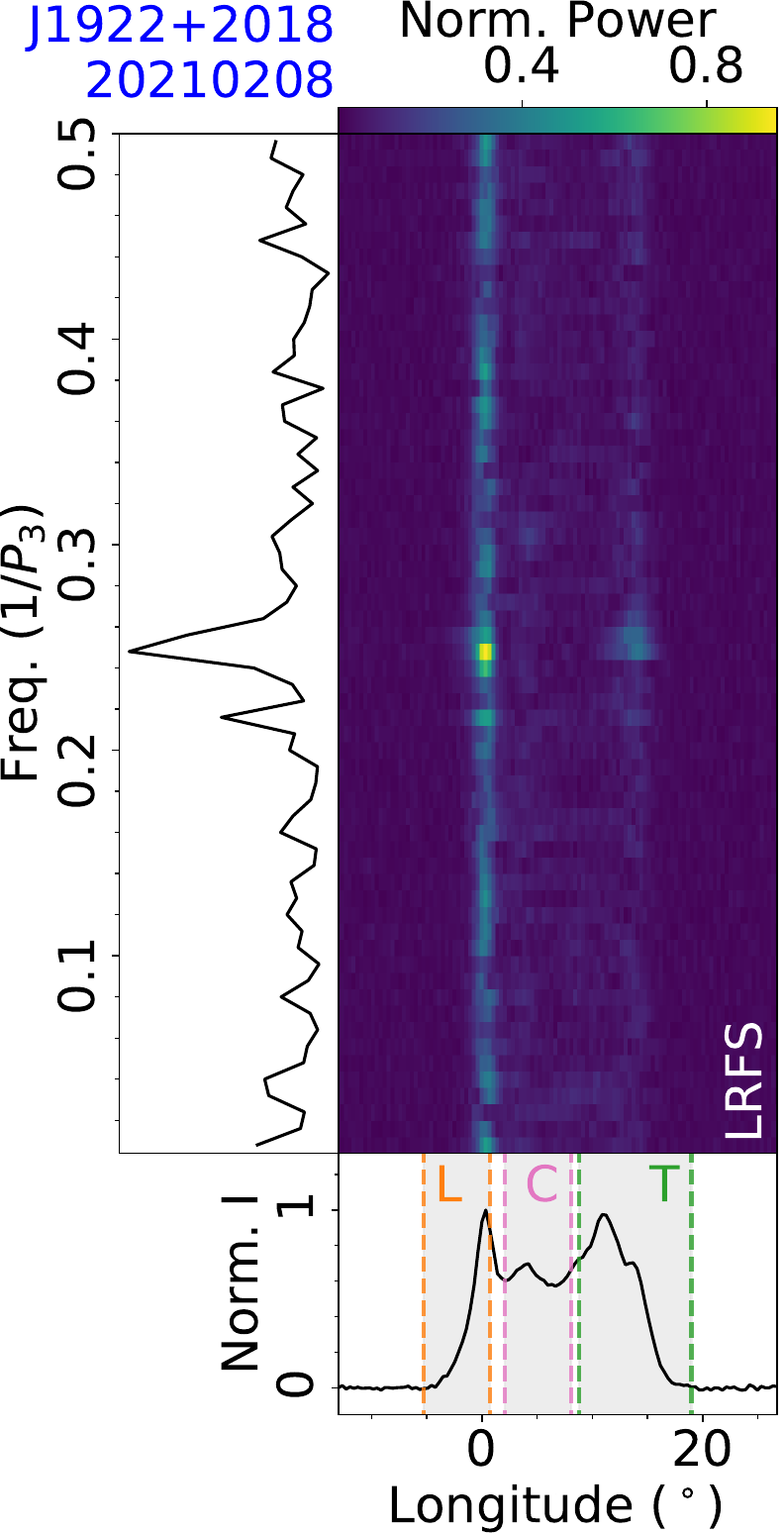}
\includegraphics[width=0.22\textwidth, angle=0]{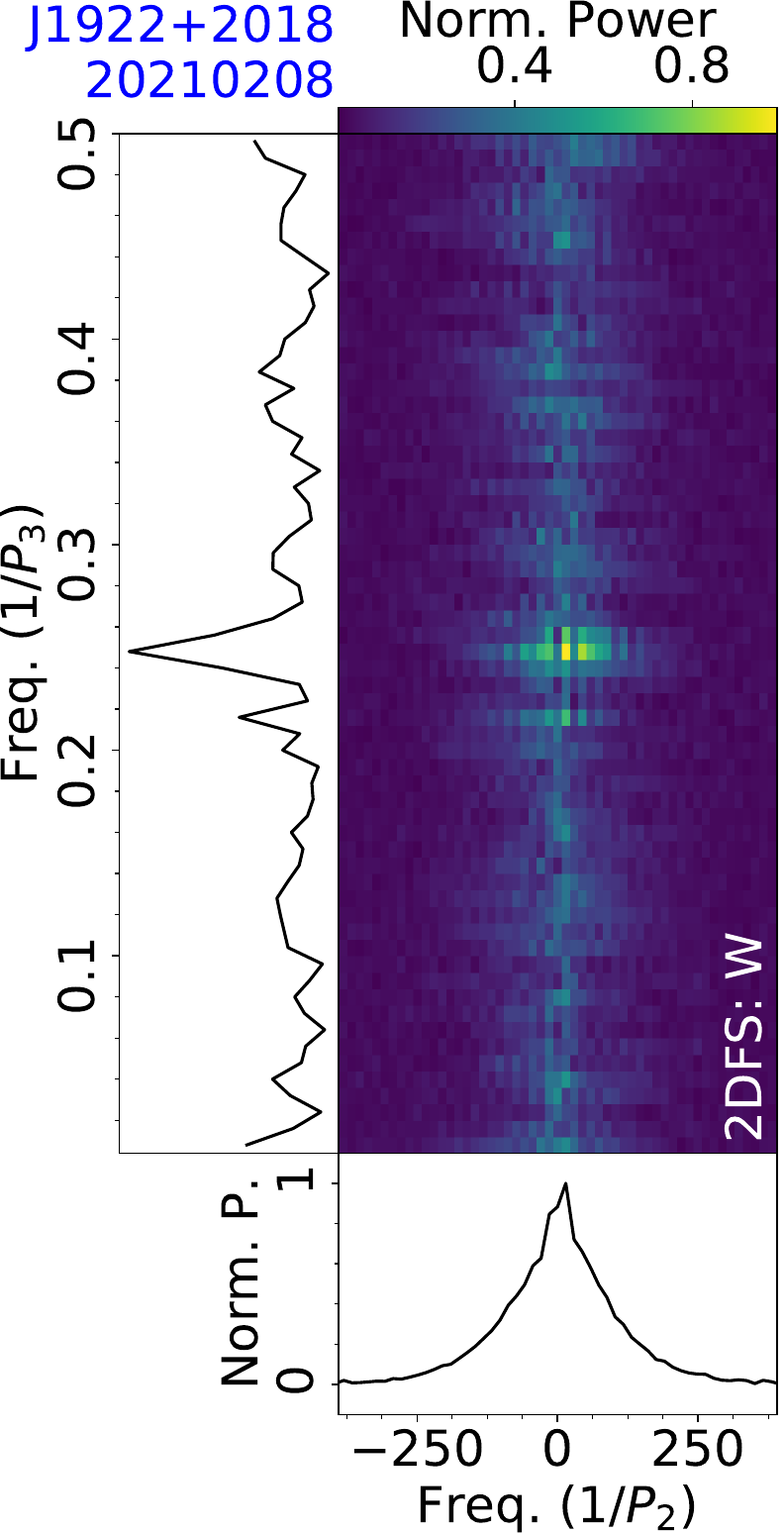}\\
\includegraphics[width=0.22\textwidth, angle=0]{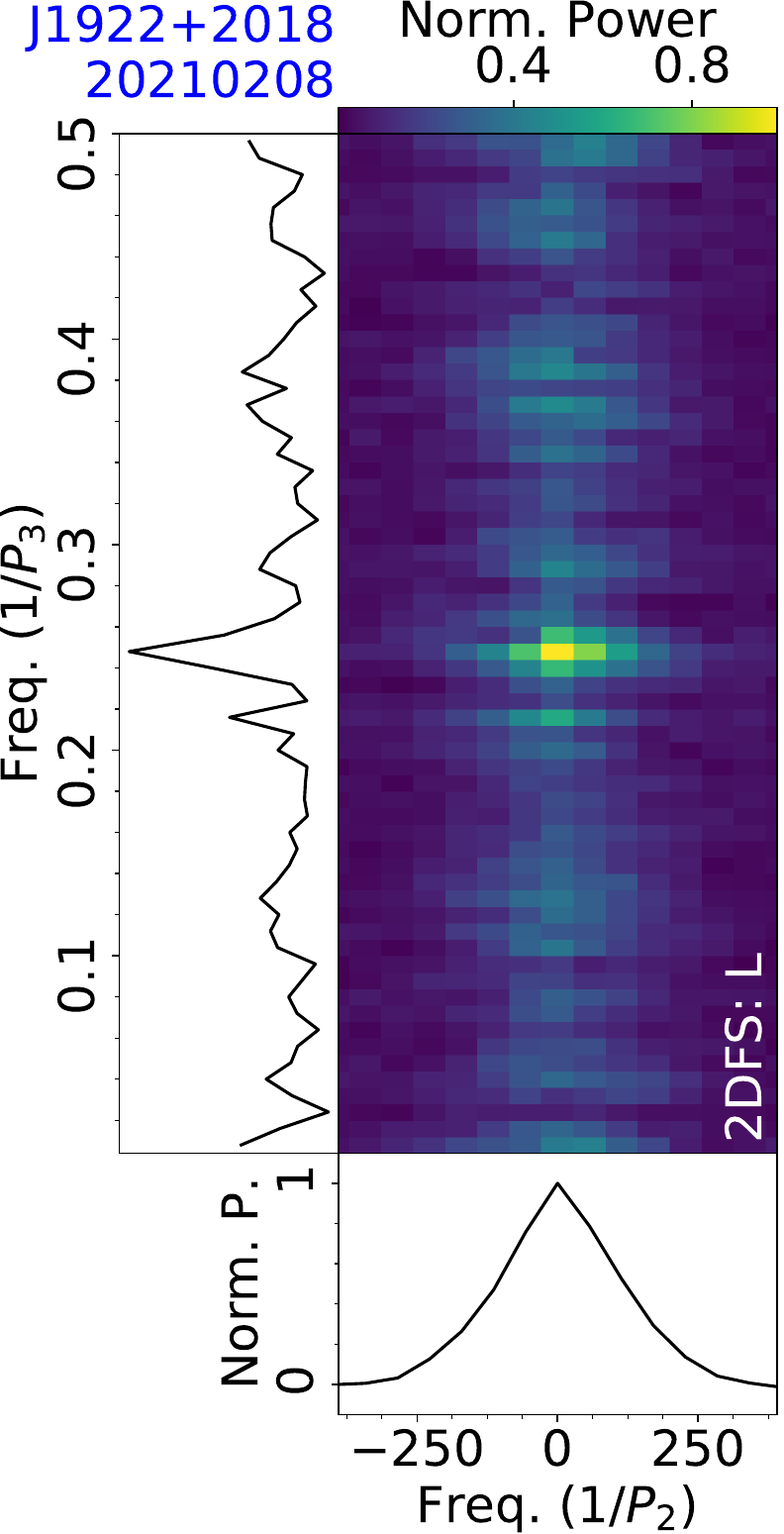}
\includegraphics[width=0.22\textwidth, angle=0]{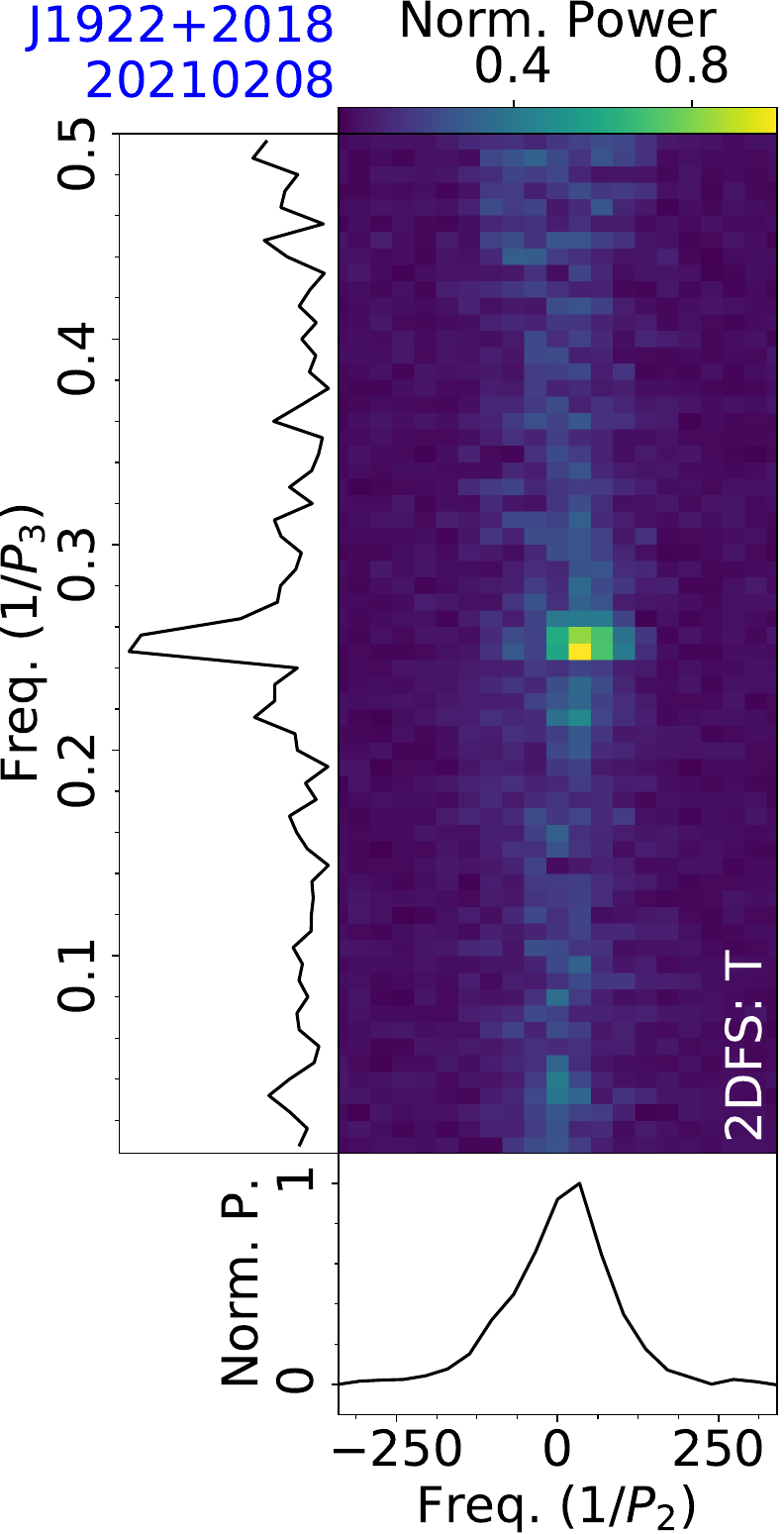}
\figcaption{Fluctuation analysis of PSR J1922+2018 for the observation on 20210208, with LRFS (top-left), and 2DFS for the on-pulse region (top-right), leading part (bottom-left) and trailing part (bottom-right) of a mean pulse profile.
\label{subfig:fluctu:J1922+2018}}
\end{figure}

\begin{figure}[hbpt]
\centering
\includegraphics[width=0.21\textwidth, angle=0]{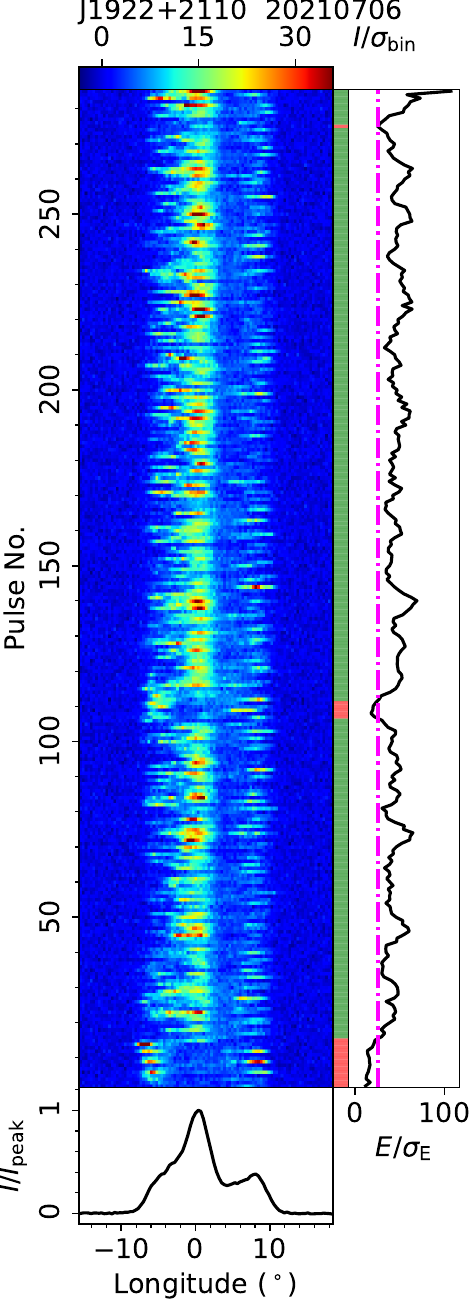}
\figcaption{Single pulse sequence of PSR J1922+2110 from the FAST observation on 20210706. 
The green and red bars represent normal or abnormal modes. In the right subpanel, the on-pulse energy variation smoothed over every 5 periods is plotted against period, with a dashed line for the threshold to distinguish the two emission modes.
\label{subfig:TP:J1922+2110}}
\end{figure}

\begin{figure}[htpb]
\centering
\includegraphics[width=0.39\textwidth, angle=0]{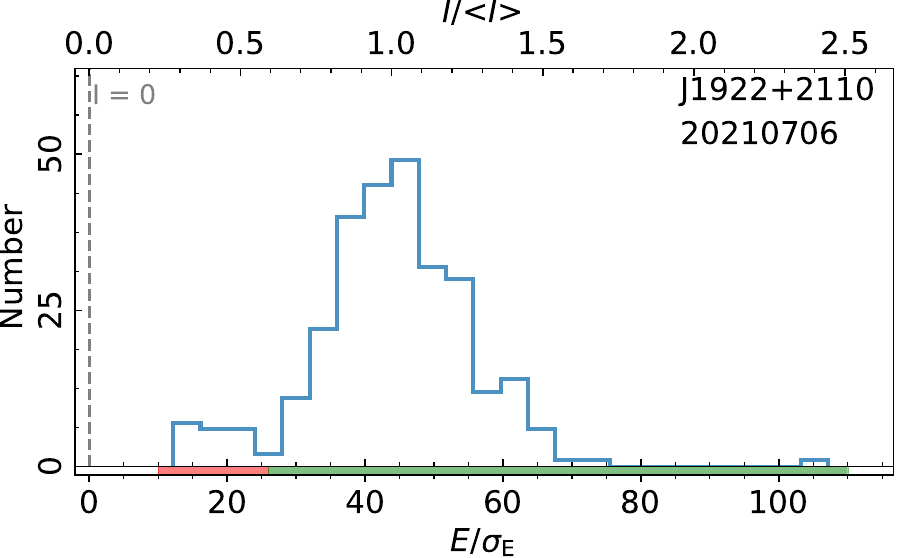}
\figcaption{On-pulse energy histogram of single pulses of PSR J1922+2110 from the FAST observation on 20210706, with energy values smoothed over 5 periods. 
The red and green bars indicate the abnormal and normal modes.
\label{subfig:Hist:J1922+2110}}
\end{figure}

\begin{figure}[hbpt]
\centering
\includegraphics[width=0.37\textwidth, angle=0]{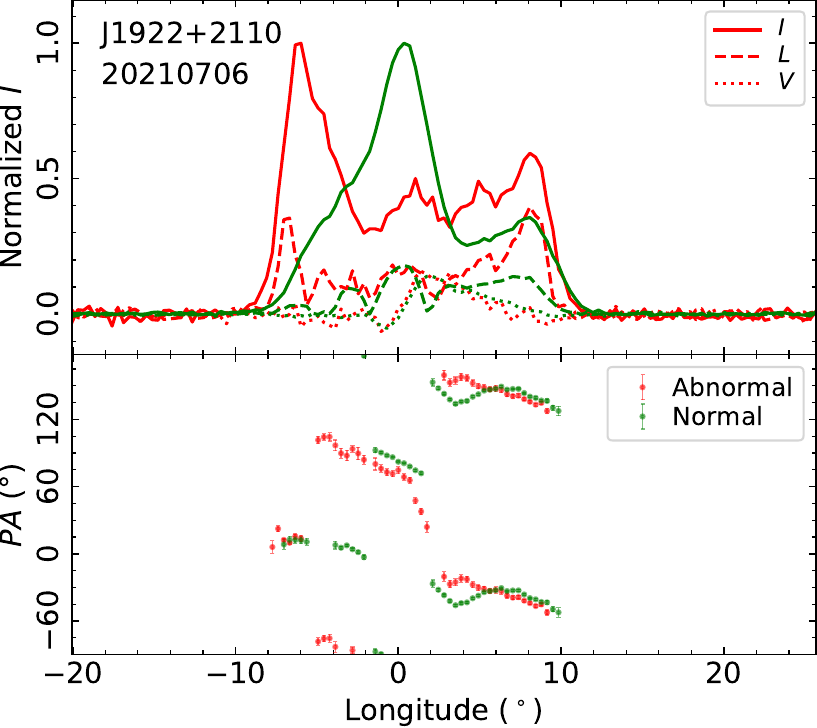}
\figcaption{Mean polarization profiles (the top panel) for the normal (green) and abnormal (red) emission modes of PSR J1922+2110 observed on 20210706, as well as the averaged PA curves (the bottom panel). Profiles in the top panel are normalized by their respective peaks.
\label{subfig:PolModes:J1922+2110}}
\end{figure}

\begin{figure}[hbpt]
\centering
\includegraphics[width=0.22\textwidth, angle=0]{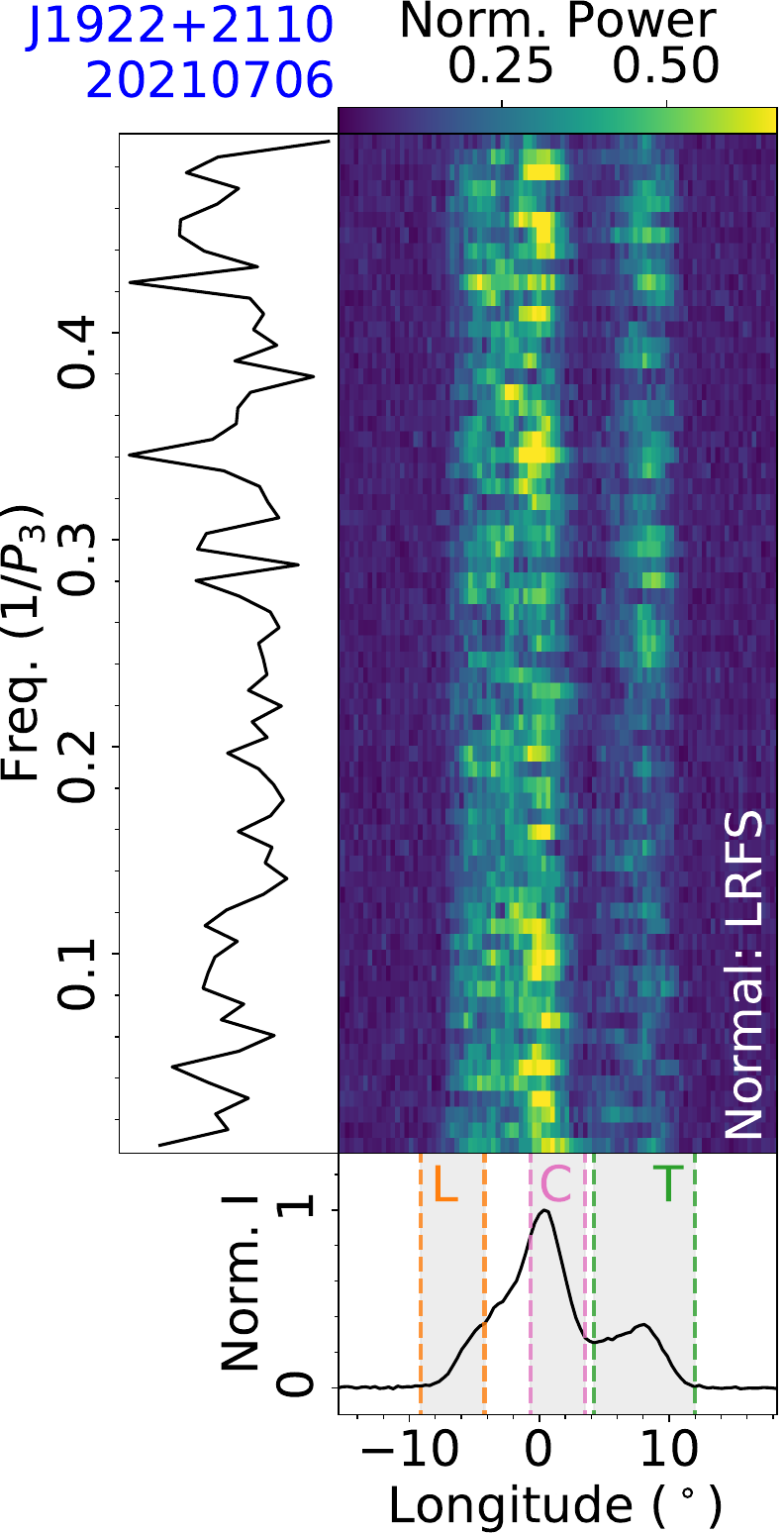}
\includegraphics[width=0.22\textwidth, angle=0]{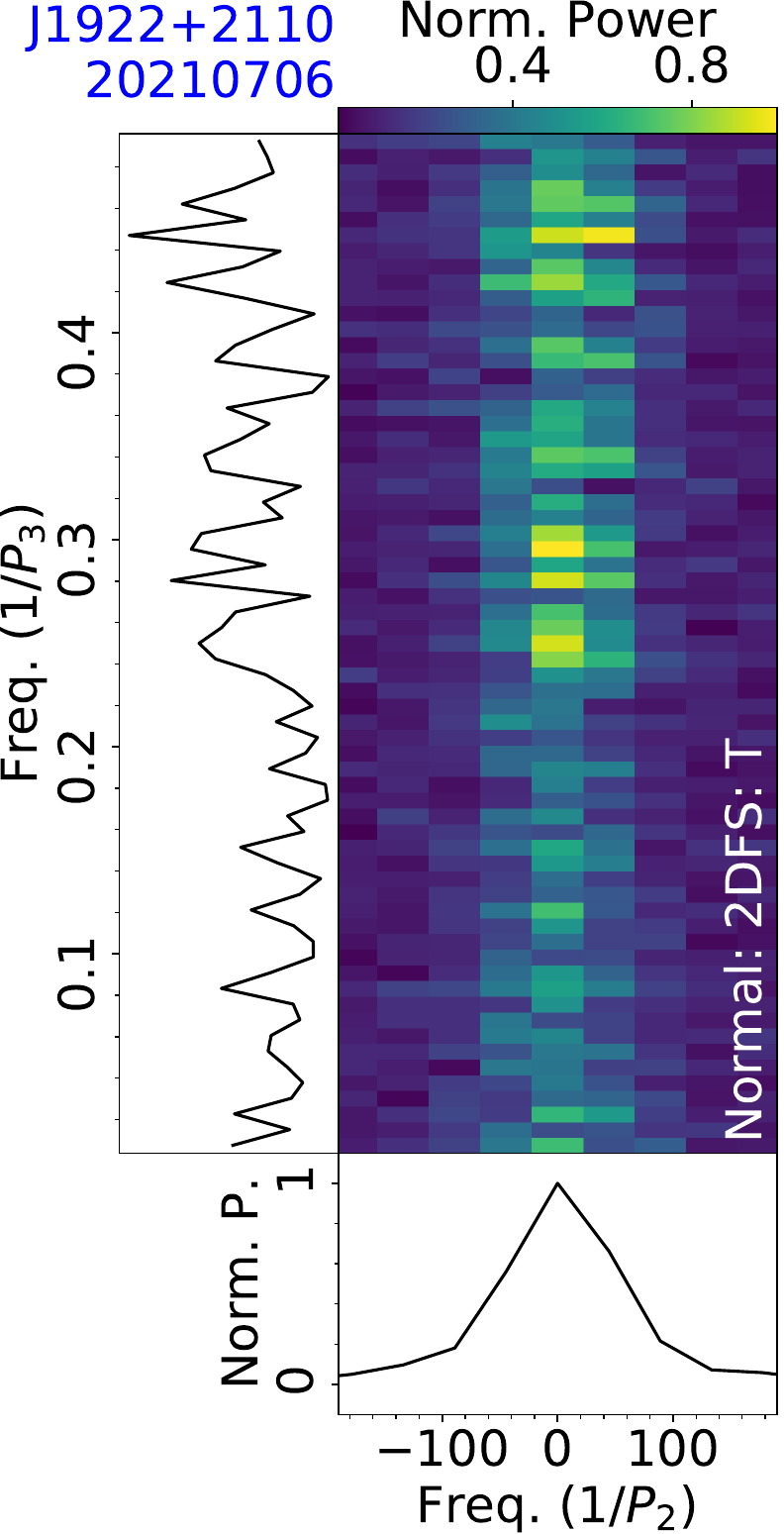}
\figcaption{Fluctuation analysis of the normal emission mode of PSR J1922+2110 for the observation on 20210706, with LRFS and 2DFS for the trailing part of a mean pulse profile.
\label{subfig:fluctu:J1922+2110}}
\end{figure}

\subsection{J1921+2003}
\label{subsec:J1921+2003}

PSR J1921+2003 was discovered by the Arecibo telescope \citep{Hulse1975}. \citet{Song2023} presented the subpulse drifting periodicities at 1280 MHz using the MeerKAT telescope: $P_3=8.4\pm0.1$ periods and $P_2=22^{+1}_{-11}$ degrees for the leading component; $P_3=8.4\pm0.2$ periods and $P_2=-105^{+51}_{-27}$ degrees for the trailing component. The bi-drifting phenomenon was reported by \citet{Wang2025} based on the MeerKAT observation at 1284 MHz and the GMRT observation at 650 MHz.

This pulsar was observed by FAST on 20210208 for 5 minutes, deriving a rotation period $P=0.7607$~s and a dispersion measure $D\!M=97.9~{\rm cm^{-3}\,pc}$. Single pulse sequences in Fig.~\ref{subfig:TP:J1921+2003} show the subpulse modulation behavior for both the leading and trailing parts of the mean pulse profile. Fluctuation spectra are presented in Fig.~\ref{subfig:fluctu:J1921+2003}. The leading profile part exhibits a positive drift feature in 2DFS with a centroid at $1/P_3=0.118\pm0.001$ and $1/P_2=20\pm3$, yielding $P_3=8.4\pm0.1$ periods and $P_2=18\pm3$ degrees. In the 2DFS of the trailing profile part, the modulation feature is centered at $1/P_3=0.119\pm0.001$ ($P_3=8.4\pm0.1$ periods),  without an obvious phase modulation direction from this 5-minute FAST data.

\subsection{J1921+2153}
\label{subsec:J1921+2153}

PSR J1921+2153 was discovered by \citet{Hewish1968} at 81.5 MHz at the Mullard Radio Astronomy Observatory. 
The pulsar has been reported with a nulling fraction of $\leq0.25\%$ by \citet{Ritchings1976} at 408 MHz and $\leq1.2\%$ by \citet{Vivekanand1995} at 326.5 MHz. Drifting phenomenon and phase step of this pulsar have been reported \citep[e.g.,][etc]{Backer1970,Proszynski1986,Weltevrede2006,Weltevrede2007,Basu2016,Basu2019}.

This pulsar was observed by FAST on 20200419 for 5 minutes, deriving a rotation period $P=1.3372$~s and a dispersion measure $D\!M=11.6~{\rm cm^{-3}\,pc}$. 
However, there is no nulling phenomenon from the single pulse sequences in Fig.~\ref{subfig:TP:J1921+2153}, and the histogram in Fig.~\ref{subfig:Hist:J1921+2153} consists of one distribution far from zero. 2DFS of three phase ranges (leading, central, and trailing parts) in the mean pulse profile in Fig.~\ref{subfig:fluctu:J1921+2153} display negative drift features. 
2DFS of the leading profile part exhibit a negative drift feature, with the centroid frequencies of $1/P_3=0.242\pm0.001$ and $1/P_2=-33\pm3$, corresponding to periodicities of $P_3=4.12\pm0.02$ periods and $P_2=-11\pm1^\circ$. For the central profile part, the centroid frequencies of the feature are estimated to be $1/P_3=0.242\pm0.002$ and $1/P_2=-34\pm10$, yielding $P_3=4.13\pm0.03$ periods and $P_2=-11\pm3^\circ$. 
The centroid of the drift feature in 2DFS is characterized by $1/P_3=0.253\pm0.002$ and $1/P_2=-49\pm6$, which correspond to $P_3=3.95\pm0.03$ periods and $P_2=-7\pm1^\circ$.

\subsection{J1921+34}
\label{subsec:J1921+34}

PSR J1921+34 was discovered by \citet{Tyulbashev2022} with the Large Phased Array of the Lebedev Physical Institute (LPA LPI). 

This pulsar was observed by FAST on 20240727 for 5 minutes, deriving a rotation period $P=1.4390$~s and a dispersion measure $D\!M=88.3~{\rm cm^{-3}\,pc}$. The single pulse sequence in Fig.~\ref{subfig:TP:J1921+34} shows the nulling phenomenon. The nulling fraction of this observation is estimated to be 36.2$\pm$3.8\%.

\subsection{J1922+1512g}
\label{subsec:J1922+1512g}

PSR J1922+1512g was discovered in the FAST GPPS survey \citep{Han2021,han2025}. 

This pulsar was observed by FAST on 20220511 for 15 minutes, and a rotation period $P=2.3559$~s and a dispersion measure $D\!M=400.3~{\rm cm^{-3}\,pc}$ were derived. 
Single pulse sequences in Fig.~\ref{subfig:TP:J1922+1512g} indicate the existence of nulling behavior as well as bright single pulses. The nulling fraction is estimated to be 37$\pm$2\% from the on-pulse integral energy histogram in Fig.~\ref{subfig:Hist:J1922+1512g}.

\begin{figure}[htpb]
\centering
\includegraphics[width=0.22\textwidth, angle=0]{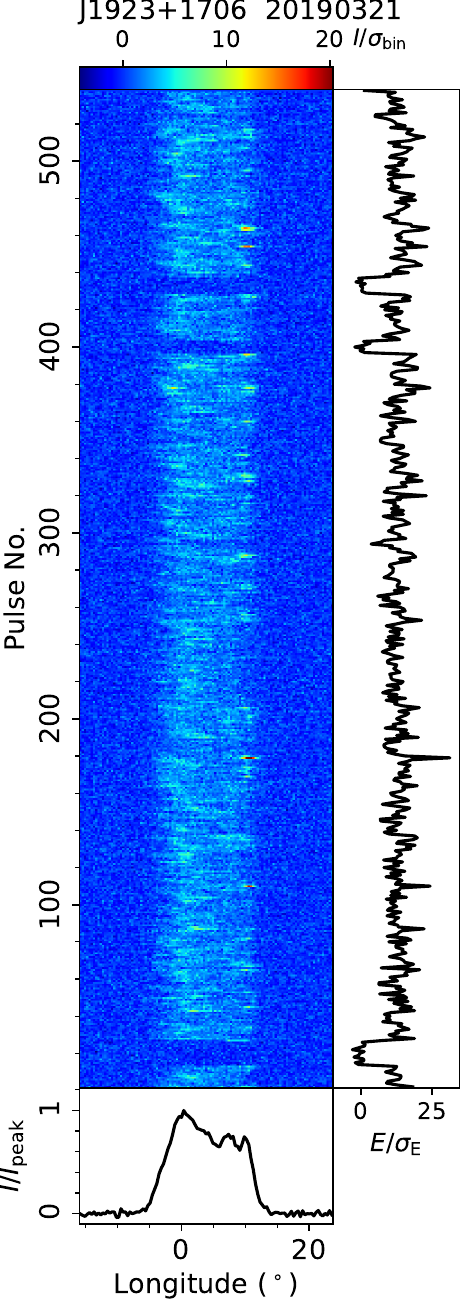}
\figcaption{Single pulse sequence of PSR J1923+1706 from the FAST observation on 20190321.
\label{subfig:TP:J1923+1706}}
\end{figure}

\begin{figure}[htpb]
\centering
\includegraphics[width=0.39\textwidth, angle=0]{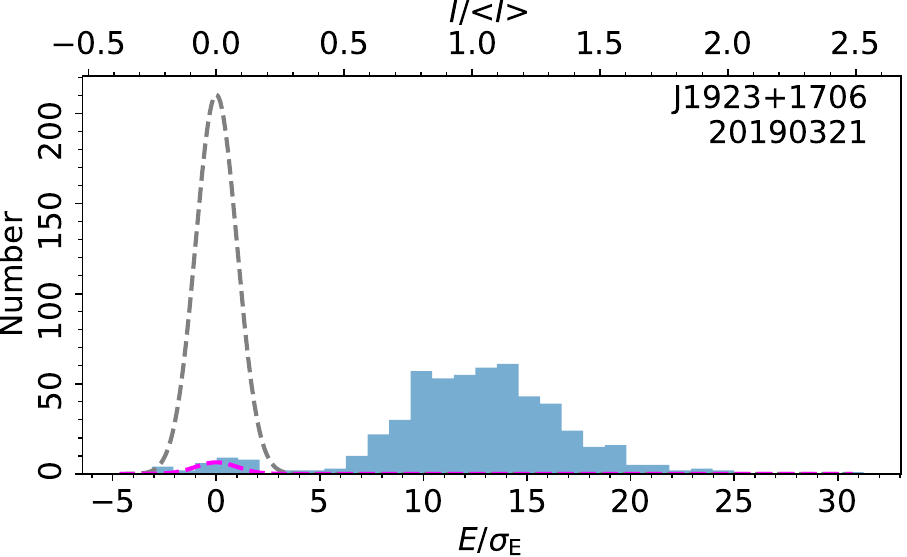}
\figcaption{On-pulse energy histogram of single pulses of PSR J1923+1706 from the FAST observation on 20190321.
\label{subfig:Hist:J1923+1706}}
\end{figure}

\begin{figure}[htpb]
\centering
\includegraphics[width=0.22\textwidth, angle=0]{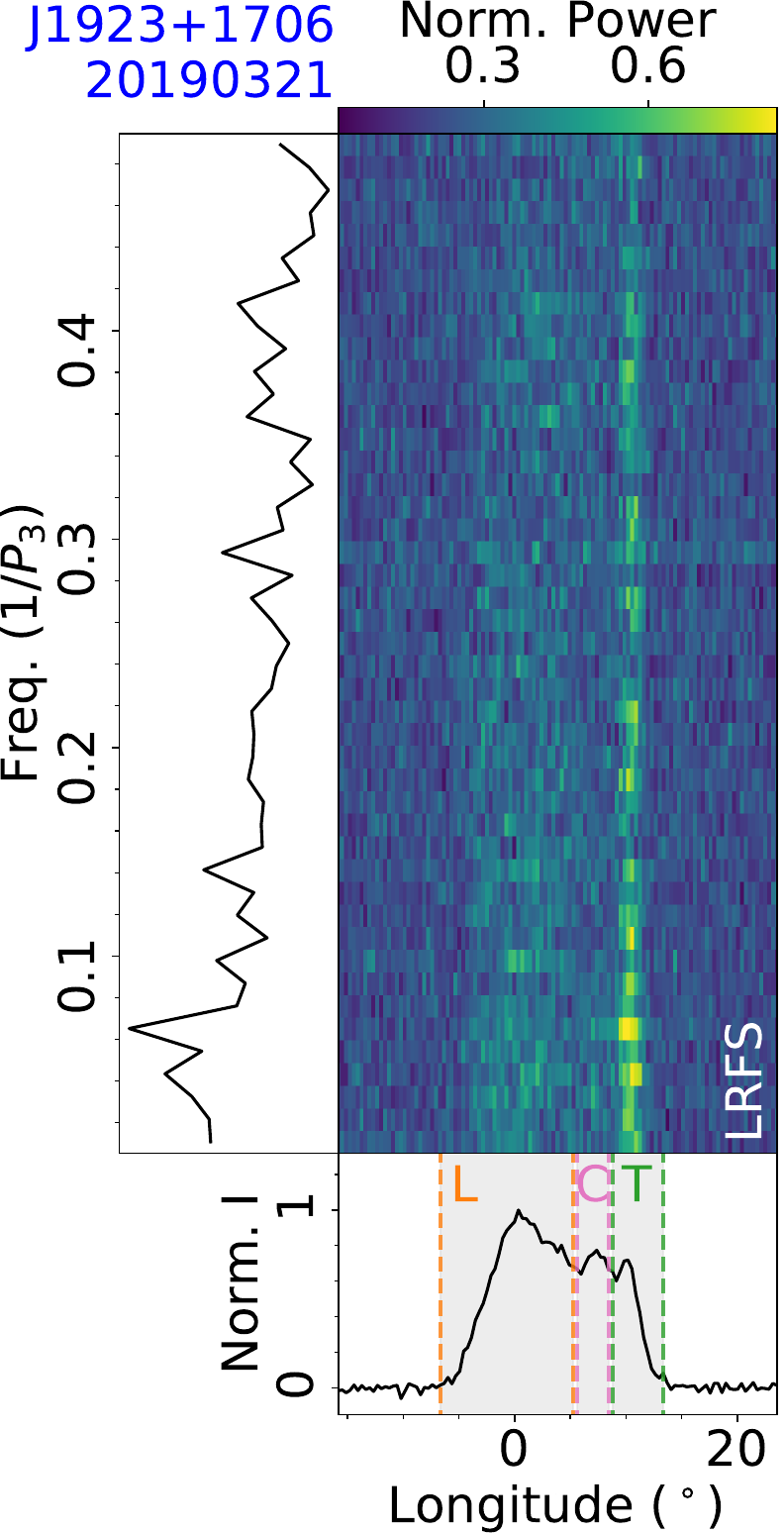}
\includegraphics[width=0.22\textwidth, angle=0]{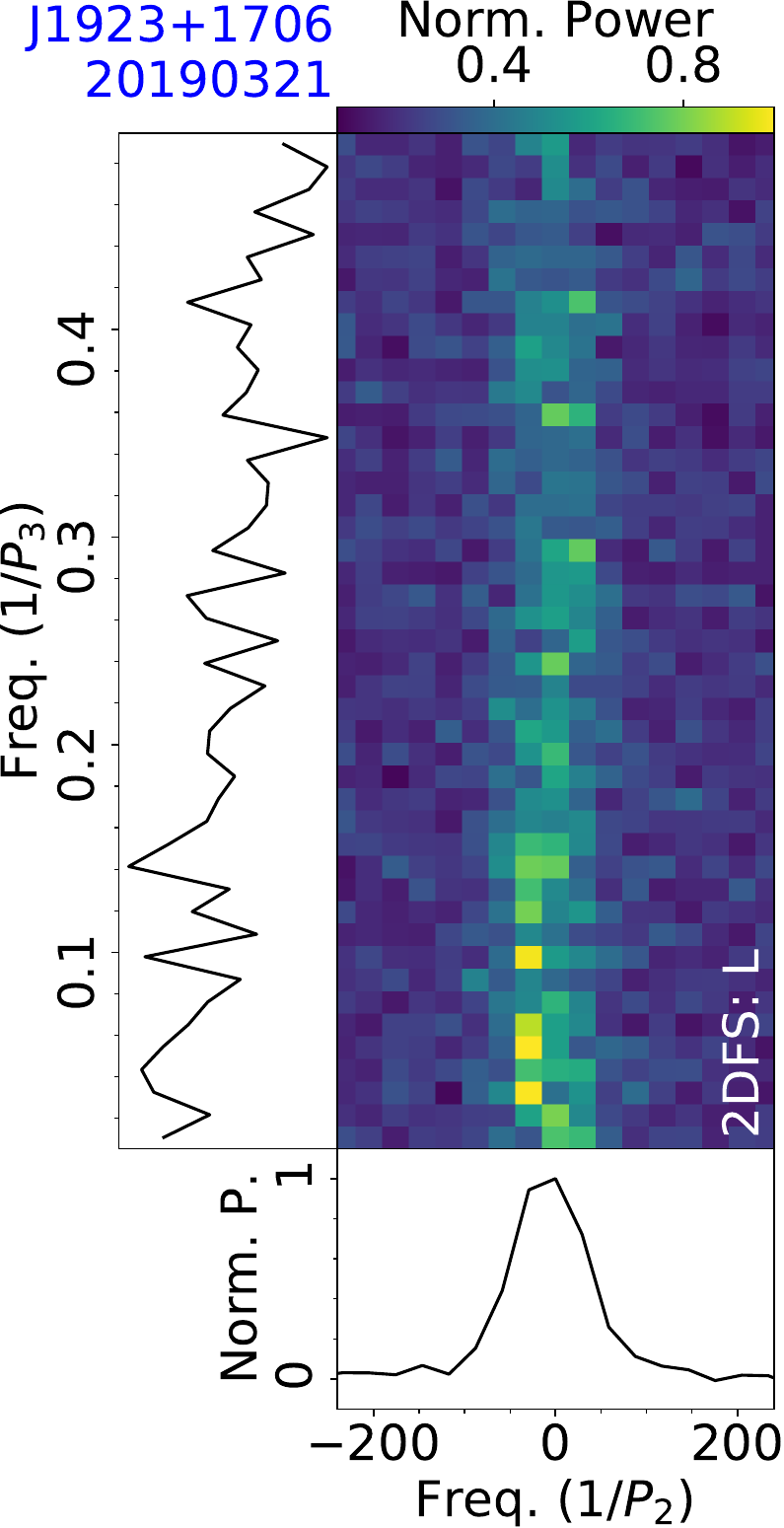}\\
\includegraphics[width=0.22\textwidth, angle=0]{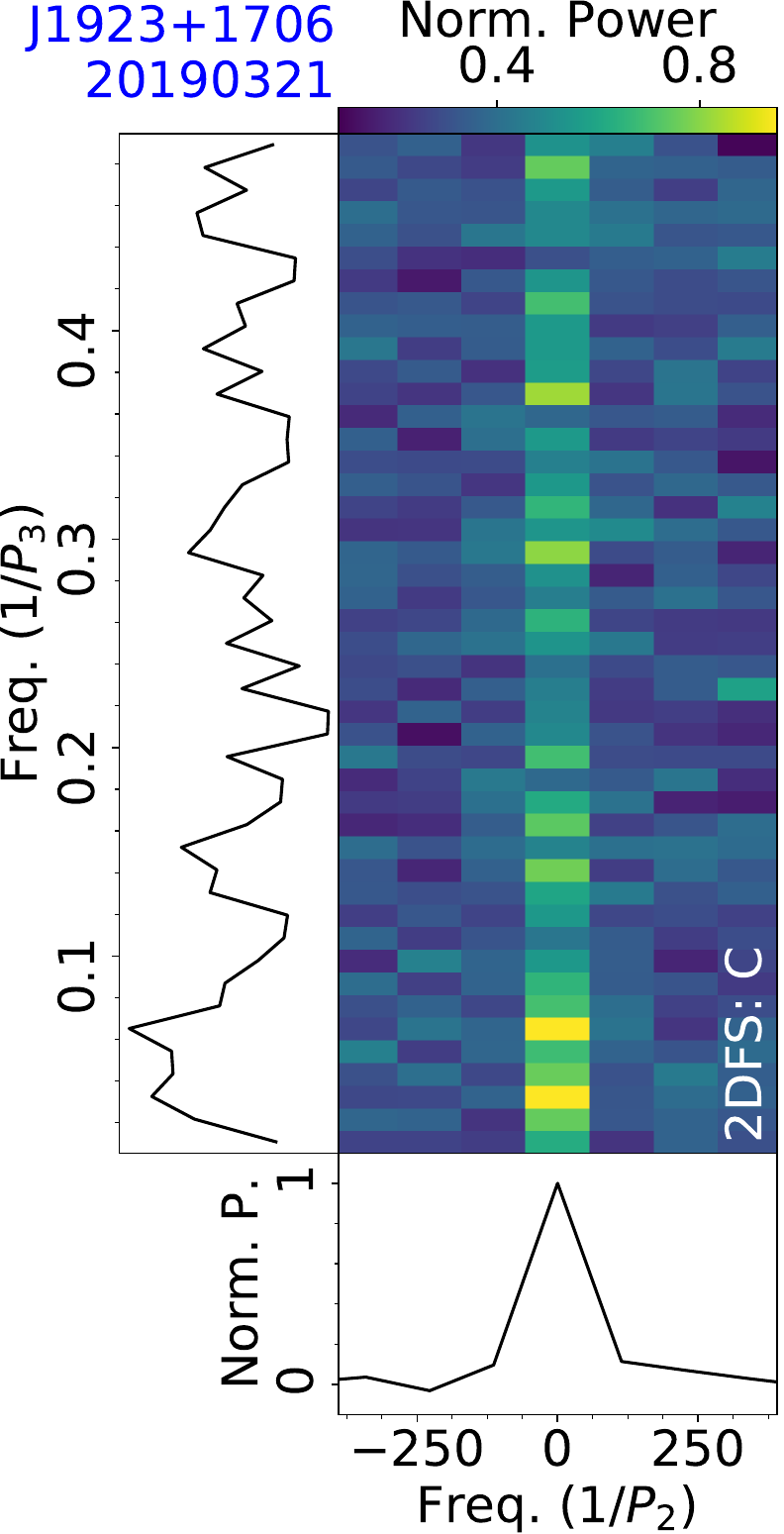}
\includegraphics[width=0.22\textwidth, angle=0]{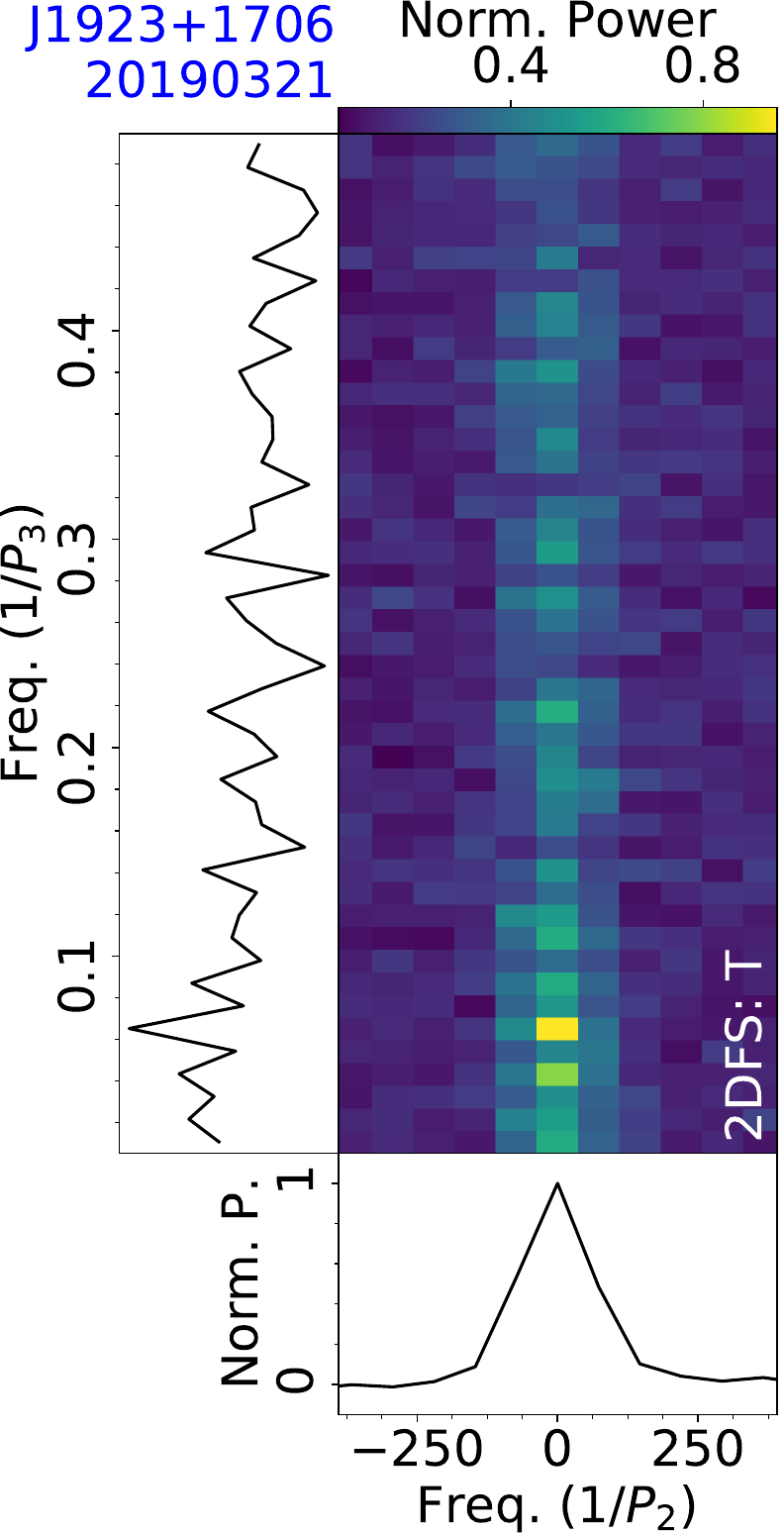}
\figcaption{Fluctuation analysis of PSR J1923+1706 for the observation on 20190321, with LRFS (top-left), and 2DFS for the leading part (top-right), central part (bottom-left) and trailing part (bottom-right) of a mean pulse profile.
\label{subfig:fluctu:J1923+1706}}
\end{figure}

\subsection{J1922+1733}
\label{subsec:J1922+1733}

PSR J1922+1733 was discovered in a deep Parkes multibeam survey \citep{Lorimer2013}. For two components, \citet{Song2023} reported a $P_3$-only feature with $P_3=17\pm12$ periods and a positive drift feature with $P_3=10\pm1$ periods and $P_2=144^{+45}_{-121}$ degrees.

This pulsar was observed by FAST on 20200514 for 5 minutes, deriving a rotation period $P=0.2362$~s and a dispersion measure $D\!M=234.1~{\rm cm^{-3}\,pc}$. The single pulse sequence and a zoomed-in view of pulses No. 100-350 in Fig.~\ref{subfig:TP:J1922+1733} show that the leading and trailing parts of a mean pulse profile exhibit the intensity modulation and the positive drifting behavior, respectively. Fluctuation spectra are displayed in Fig.~\ref{subfig:fluctu:J1922+1733}. For the leading profile part, the centroid frequency of the modulation feature in 2DFS is estimated to be $1/P_3=0.106\pm0.003$, corresponding to $P_3=9.4\pm0.3$ periods. 2DFS of the trailing profile part exhibits a positive drift feature, and the centroid is characterized by frequencies of $1/P_3=0.124\pm0.002$ and $1/P_2=3\pm1$, yielding the periodicities of $P_3=8.0\pm0.1$ periods and $P_2=130\pm55$ degrees.

\subsection{J1922+2018}
\label{subsec:J1922+2018}

PSR J1922+2018 was discovered by the Arecibo telescope \citep{Hulse1975}. 

This pulsar was observed by FAST on 20210208 and 20210220, both for 5 minutes. From the observation on 20210208, a rotation period $P=1.1728$~s and a dispersion measure $D\!M=203.8~{\rm cm^{-3}\,pc}$ were determined. 
Here we present the analysis of the observation on 20210208, which is similar to that on 20210220. 
The single pulse sequence observed by FAST on 20210208 is displayed in Fig.~\ref{subfig:TP:J1922+2018}. 
The leading and trailing parts in the profile seem to have systematic modulation. The former is more likely to have a temporal modulation, while the latter has a positive drifting behavior.  
LRFS and 2DFS of the leading part (-5$^\circ$ to 1$^\circ$) and trailing part (9$^\circ$ to 19$^\circ$) in the profile shown in Fig.~\ref{subfig:fluctu:J1922+2018}. 
The leading profile part has a centroid temporally modulated frequency of $1/P_3=0.248\pm0.001$, corresponding to $P_3=4.04\pm0.02$ periods. 2DFS of the trailing profile part exhibits a positive drift feature with centroid frequencies of $1/P_3=0.252\pm0.001$ and $1/P_2=39\pm4$, yielding the periodicities of $P_3=3.97\pm0.02$ periods and $P_2=9\pm1^\circ$.

\begin{figure}[htpb]
\centering
\includegraphics[width=0.21\textwidth, angle=0]{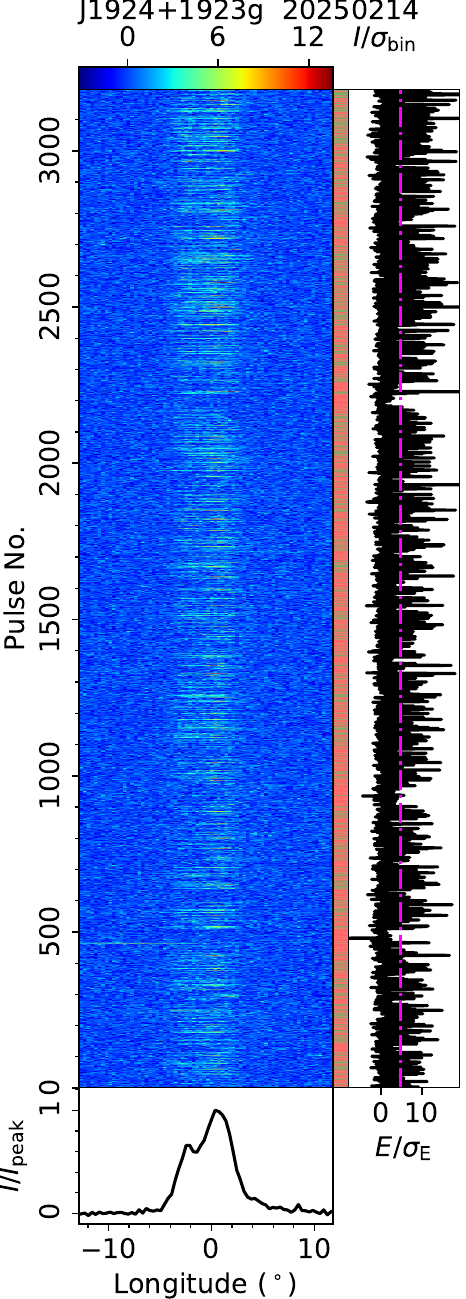}
\includegraphics[width=0.21\textwidth, angle=0]{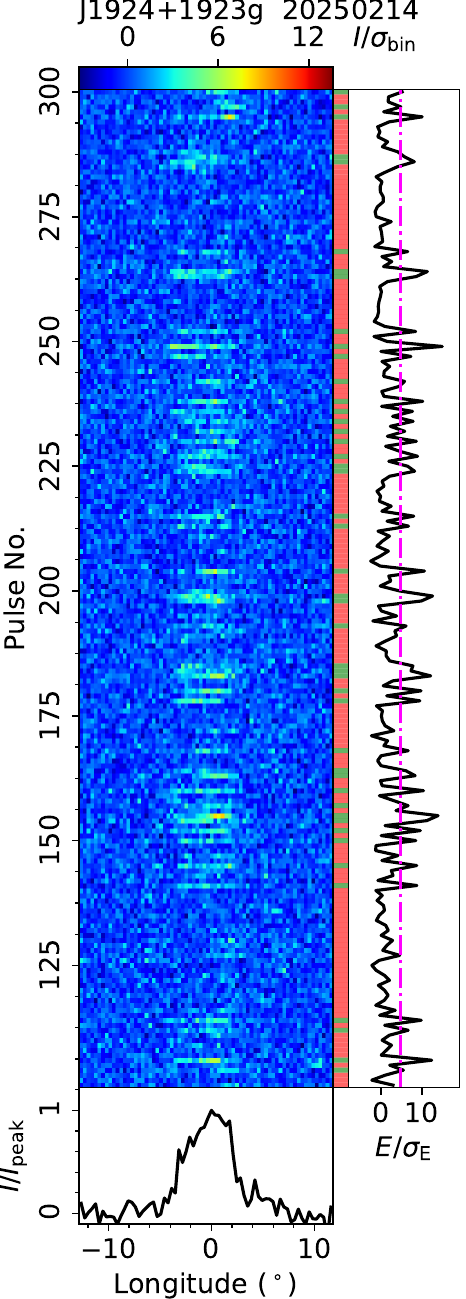}
\figcaption{Single pulse sequence of PSR J1924+1923g from the FAST observation on 20250214, and a zoomed-in view of pulses No. 101-300. The red and green bars represent weak or bright emission modes. In the right subpanel, the on-pulse energy variation is plotted against period, with a dashed line for the threshold to distinguish the two emission modes.
\label{subfig:TP:J1924+1923g}}
\end{figure}

\begin{figure}[htpb]
\centering
\includegraphics[width=0.39\textwidth, angle=0]{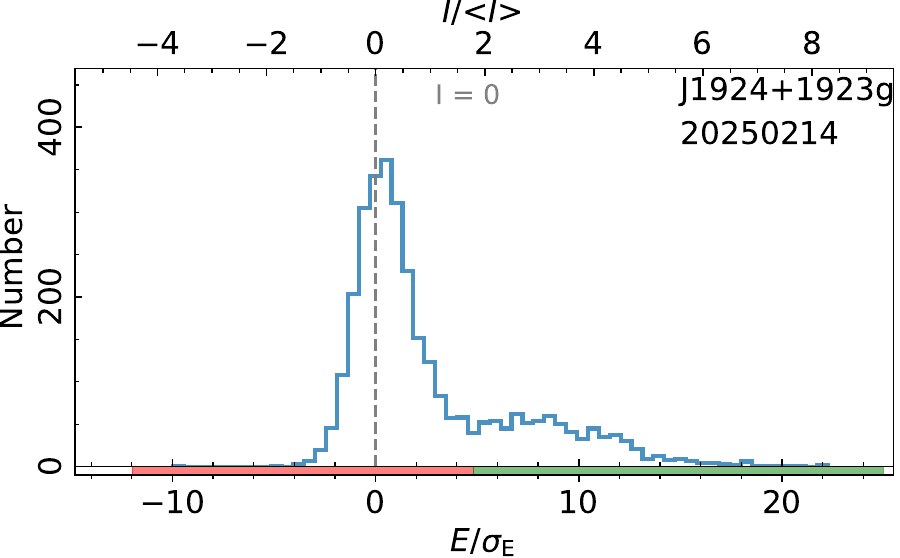}
\figcaption{On-pulse energy histogram of single pulses of PSR J1924+1923g from the FAST observation on 20250214. The red and green bars indicate the weak and bright emission modes.
\label{subfig:Hist:J1924+1923g}}
\end{figure}

\begin{figure}[htpb]
\centering
\includegraphics[width=0.42\textwidth, angle=0]{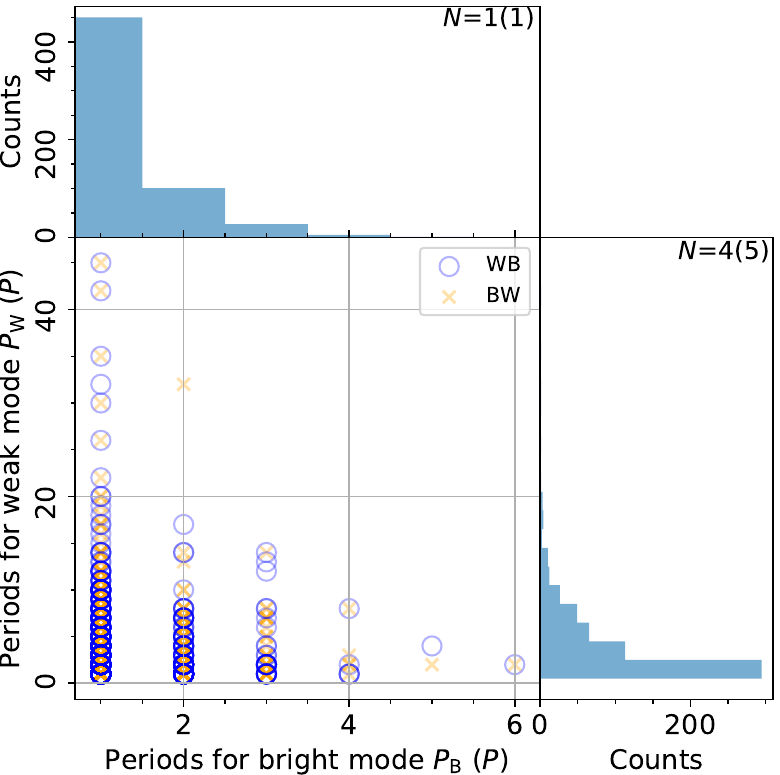}
\figcaption{Distribution of period numbers for the weak emission mode $P_{\rm W}$ against period numbers for adjacent bright mode $P_{\rm B}$ of PSR J1924+1923g observed by FAST on 20250214, as well as the duration histograms for the bright and weak emission modes shown in the top and right panels, respectively. 
\label{subfig:scaleHist:J1924+1923g}}
\end{figure}

\begin{figure}[htpb]
\centering
\includegraphics[width=0.39\textwidth, angle=0]{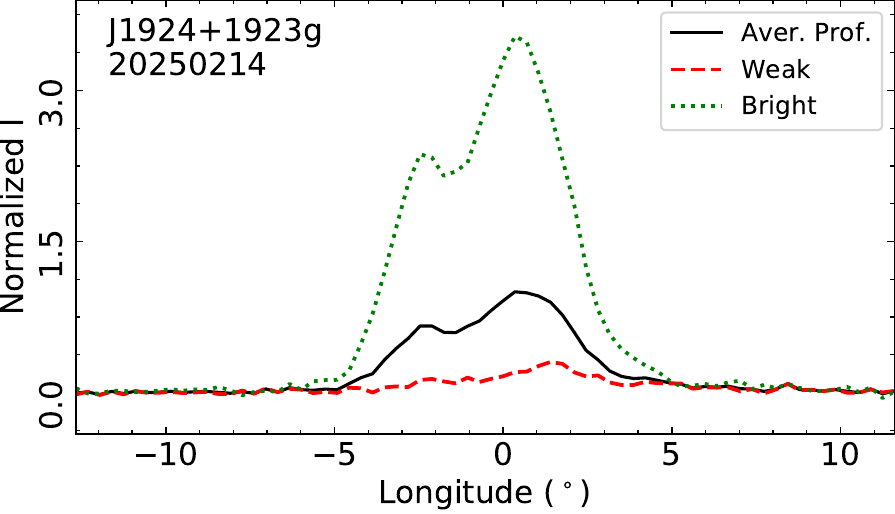}
\figcaption{Mean profiles for the weak (red dashed) and bright (green dotted) emission modes of PSR J1924+1923g observed on 20250214, which are normalized by the peak of the mean pulse profile of all periods.
\label{subfig:ProfModes:J1924+1923g}}
\end{figure}

\begin{figure}[htpb]
\centering
\includegraphics[width=0.22\textwidth, angle=0]{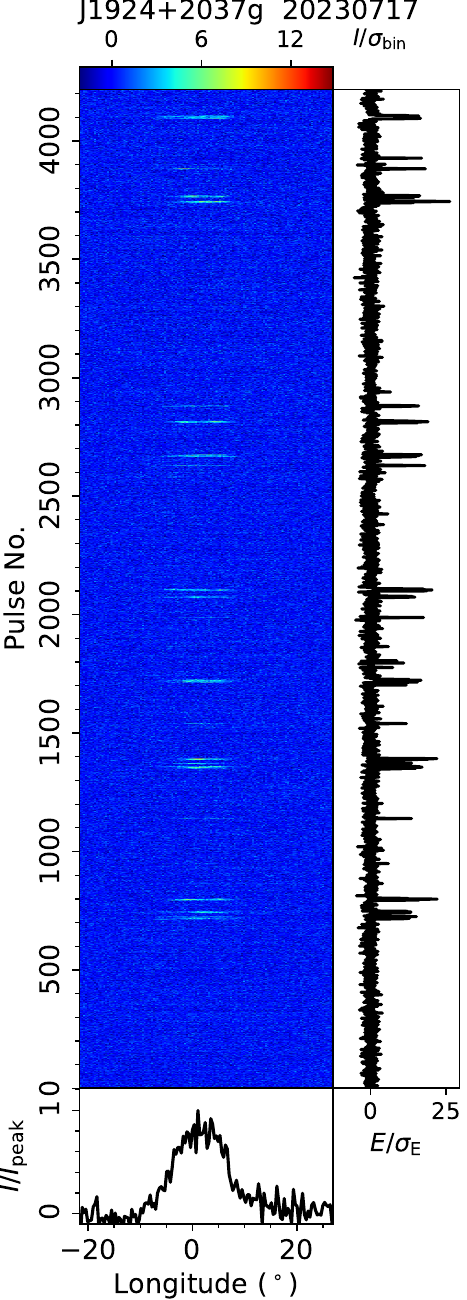}
\includegraphics[width=0.22\textwidth, angle=0]{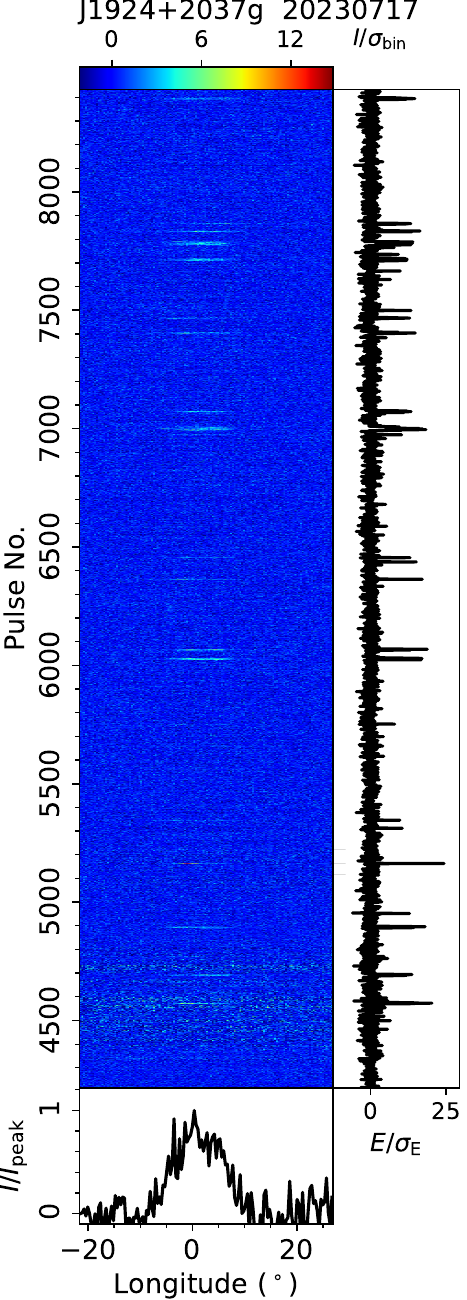}
\figcaption{Single pulse sequences of PSR J1924+2037g from the FAST observation on 20230717. 
\label{subfig:TP:20230717}}
\end{figure}

\begin{figure}[htpb]
\centering
\includegraphics[width=0.39\textwidth, angle=0]{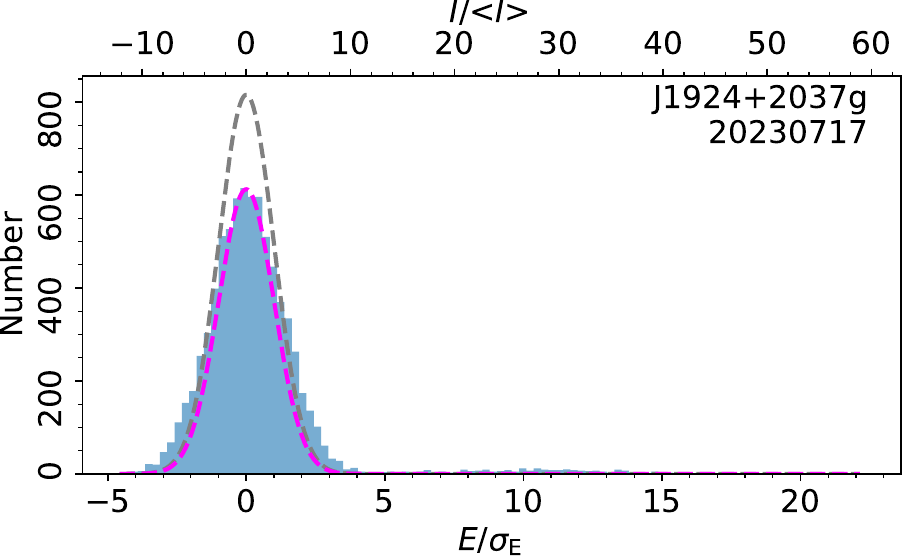}
\figcaption{On-pulse energy histogram of single pulses of PSR J1924+2037g from the FAST observation on 20230717.
\label{subfig:Hist:J1924+2037g}}
\end{figure}

\begin{figure}[htpb]
\centering
\includegraphics[width=0.42\textwidth, angle=0]{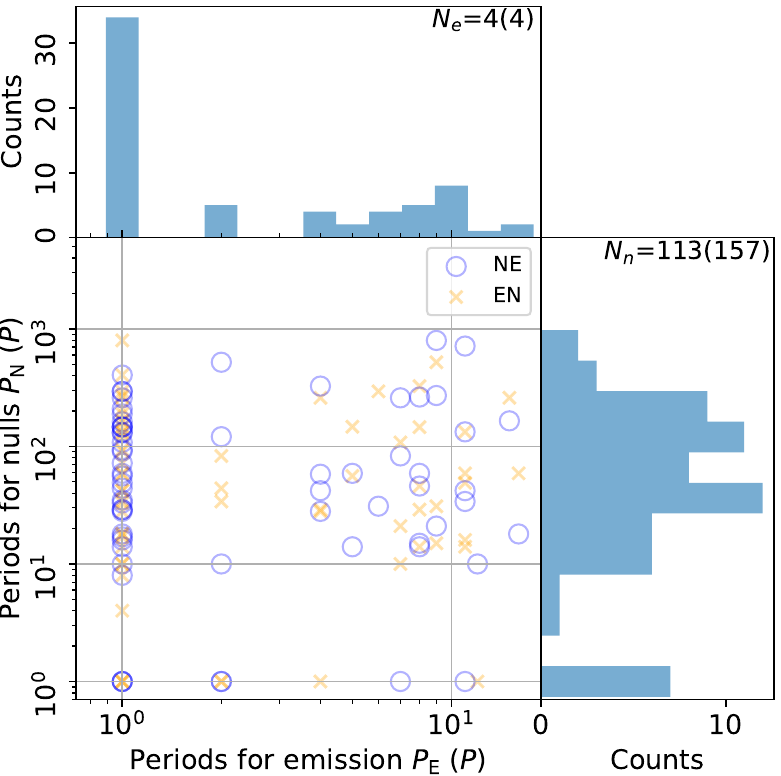}
\figcaption{Distribution of period numbers for continuous nulling $P_{\rm N}$ against period numbers for adjacent pulses $P_{\rm E}$ of PSR J1924+2037g observed by FAST on 20230717, as well as the duration histograms for the emission and null shown in the top and right panels, respectively. 
\label{subfig:scaleHist:J1924+2037g}}
\end{figure}

\begin{figure}[htpb]
\centering
\includegraphics[width=0.39\textwidth, angle=0]{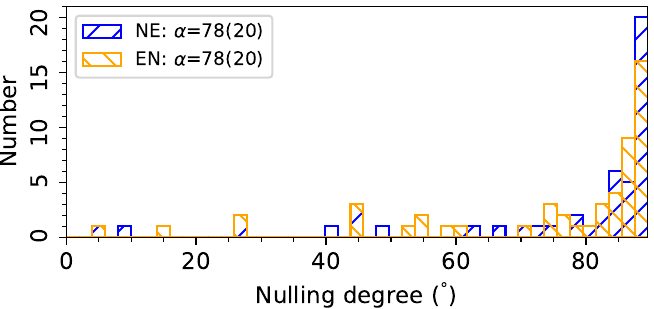}
\includegraphics[width=0.39\textwidth, angle=0]{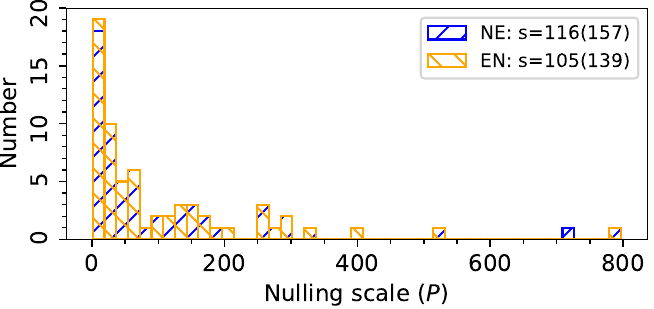}
\figcaption{Histograms of the nulling degree and nulling scale for PSR J1924+2037g observed by FAST on 20230717.
\label{subfig:nullDegreeScale:J1924+2037g}}
\end{figure}

\begin{figure}[htpb]
\centering
\includegraphics[width=0.44\textwidth, angle=0]{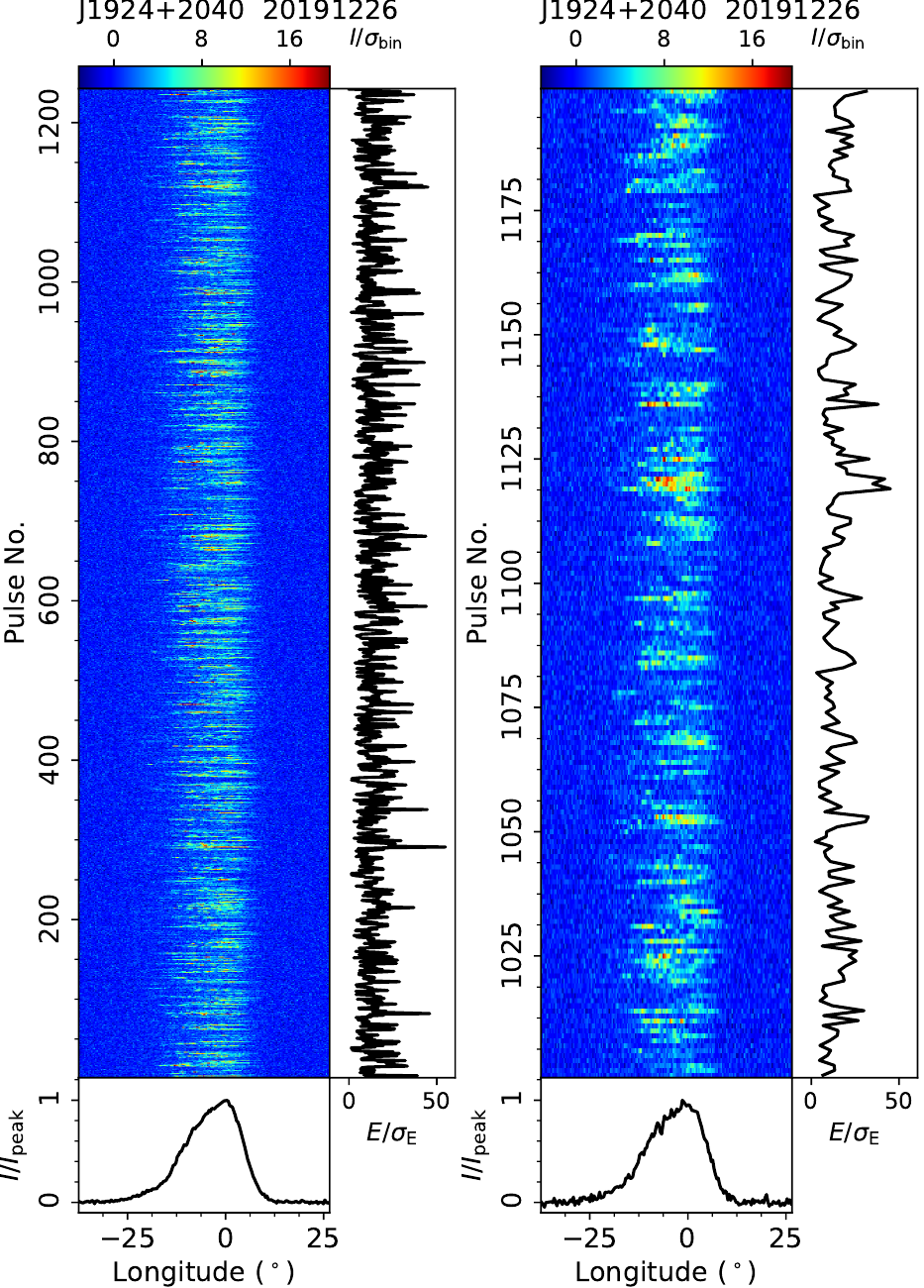}
\figcaption{Single pulse sequence of PSR J1924+2040 from the FAST observation on 20191226, and a zoomed-in view of pulses No. 1000-1200.
\label{subfig:TP:J1924+2040}}
\end{figure}

\begin{figure}[htpb]
\centering
\includegraphics[width=0.44\textwidth, angle=0]{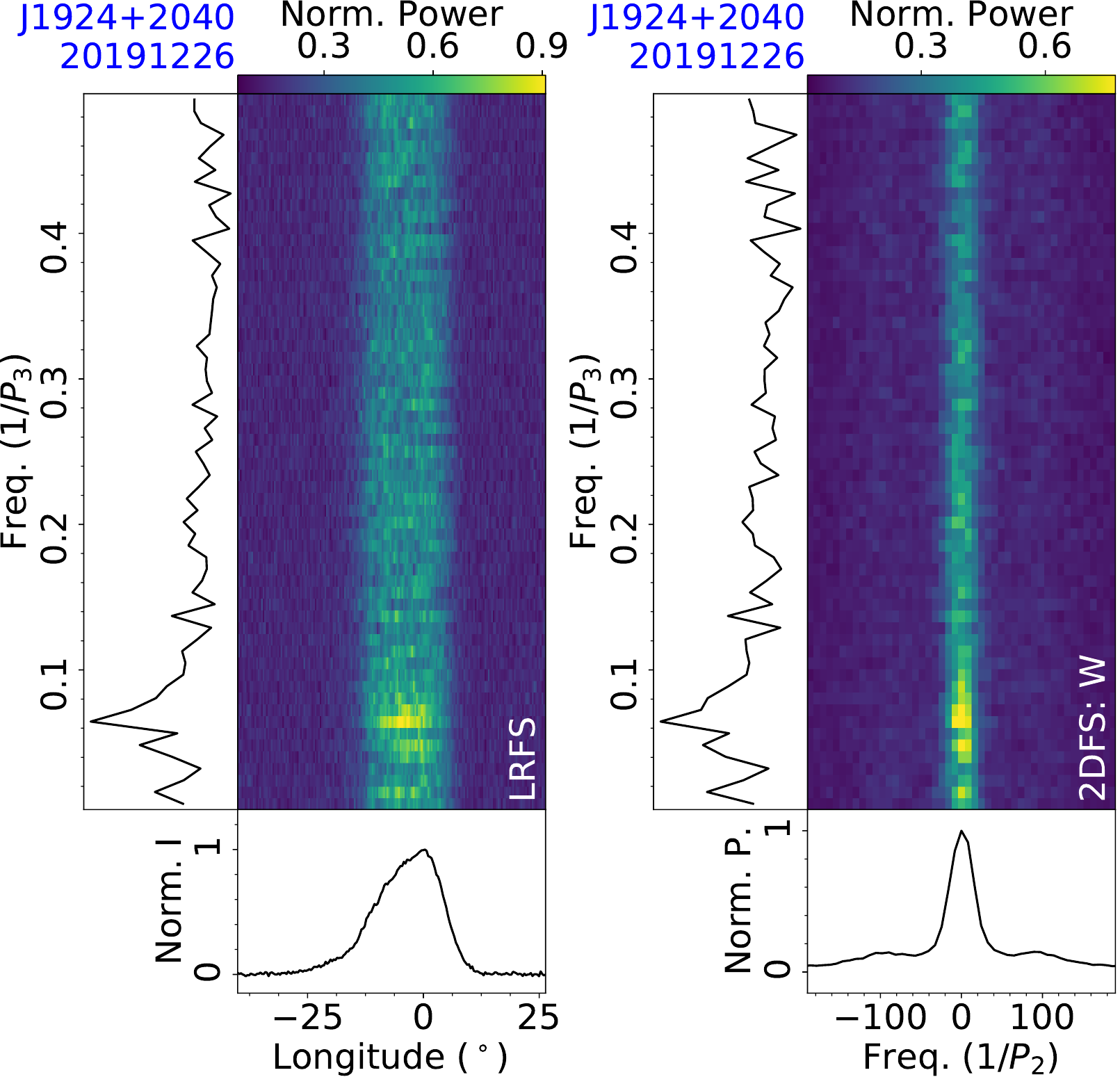}
\figcaption{Fluctuation analysis of PSR J1924+2040 for the observation on 20191226, with LRFS and 2DFS for the on-pulse phase region of the mean pulse profile.
\label{subfig:fluctu:J1924+2040}}
\end{figure}

\subsection{J1922+2110}
\label{subsec:J1922+2110}

PSR J1922+2110 was discovered by the Mark 1A radio telescope at Jodrell Bank \citep{Davies1973}. \citet{Weltevrede2007} reported three sharp features with 0.19, 0.24, and 0.28 cycle per period (cpp) in LRFS and 2DFS at 92cm, and they found the 0.24 cpp feature is in the first half of the data and the other two in the last half. 

This pulsar was observed by FAST on 20210706 for 5 minutes, deriving a rotation period $P=1.0779$~s and a dispersion measure $D\!M=217.1~{\rm cm^{-3}\,pc}$. 
The data of two beams, P1M07 and P4M07, are used to do the analysis. Single pulse sequences in Fig.~\ref{subfig:TP:J1922+2110} show that there are two emission modes, as well as the modulation phenomenon. 

The energy series of the central component is smoothed every 5 periods. 
The normal and abnormal modes are distinguished from this smoothed energy histogram of the central component in Fig.~\ref{subfig:Hist:J1922+2110}, which are labeled in red and green, respectively. There are two segments of the abnormal mode in Fig.~\ref{subfig:TP:J1922+2110}. 
Averaged polarization profiles of two emission modes, as well as PA curves, are displayed in Fig.~\ref{subfig:PolModes:J1922+2110}. The intensity of the central part is strongest in the normal mode, while the leading component is the brightest in the abnormal mode. The most prominent difference between PA curves of two emission modes is the orthogonal modes in the longitude phase range between -4 and -2 degrees. 
In addition, there is a preferred positive drifting for the trailing profile part of the normal mode. 2DFS shown in Fig.~\ref{subfig:fluctu:J1922+2110} displays a positive drift feature with centroid frequencies of $1/P_3=0.360\pm0.005$ and $1/P_2=9\pm3$, corresponding to periodicities of $P_3=2.78\pm0.04$ periods and $P_2=42\pm12^\circ$.


\begin{figure}[htpb]
\centering
\includegraphics[width=0.22\textwidth, angle=0]{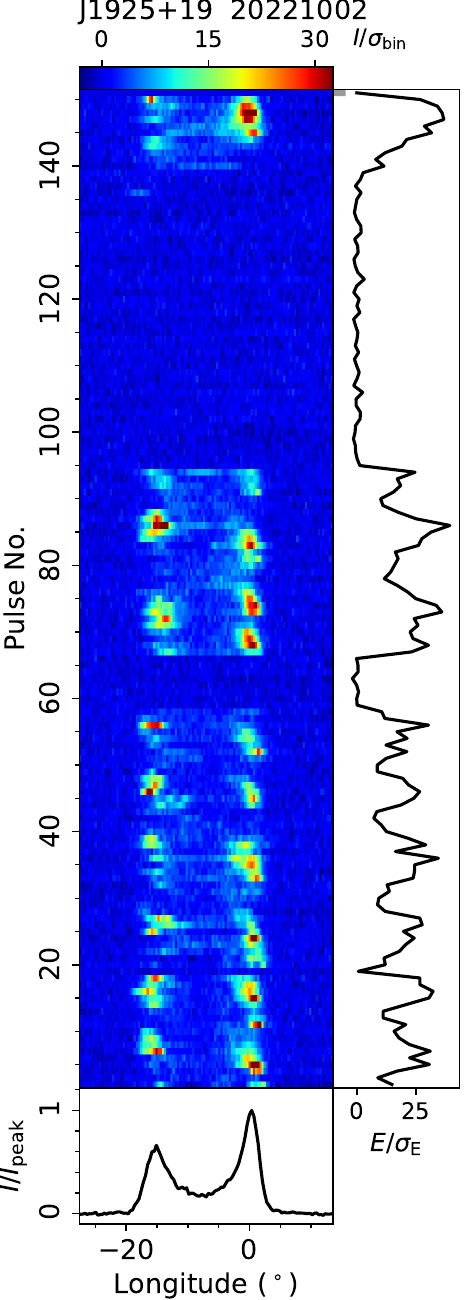}
\figcaption{Single pulse sequence of PSR J1925+19 from the FAST observation on 20221002.
\label{subfig:TP:J1925+19}}
\end{figure}

\begin{figure}[htpb]
\centering
\includegraphics[width=0.39\textwidth, angle=0]{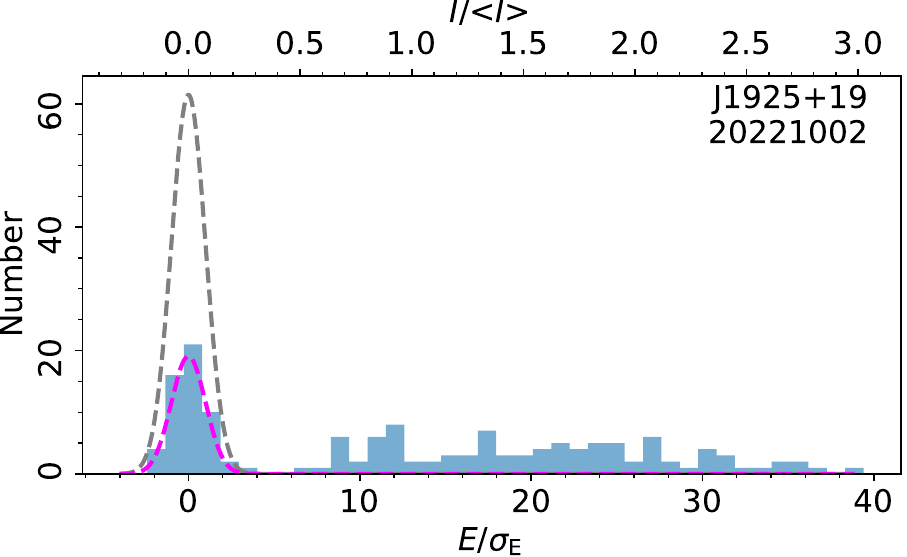}
\figcaption{On-pulse energy histogram of single pulses of PSR J1925+19 from the FAST observation on 20221002.
\label{subfig:Hist:J1925+19}}
\end{figure}

\begin{figure}[htpb]
\centering
\includegraphics[width=0.22\textwidth, angle=0]{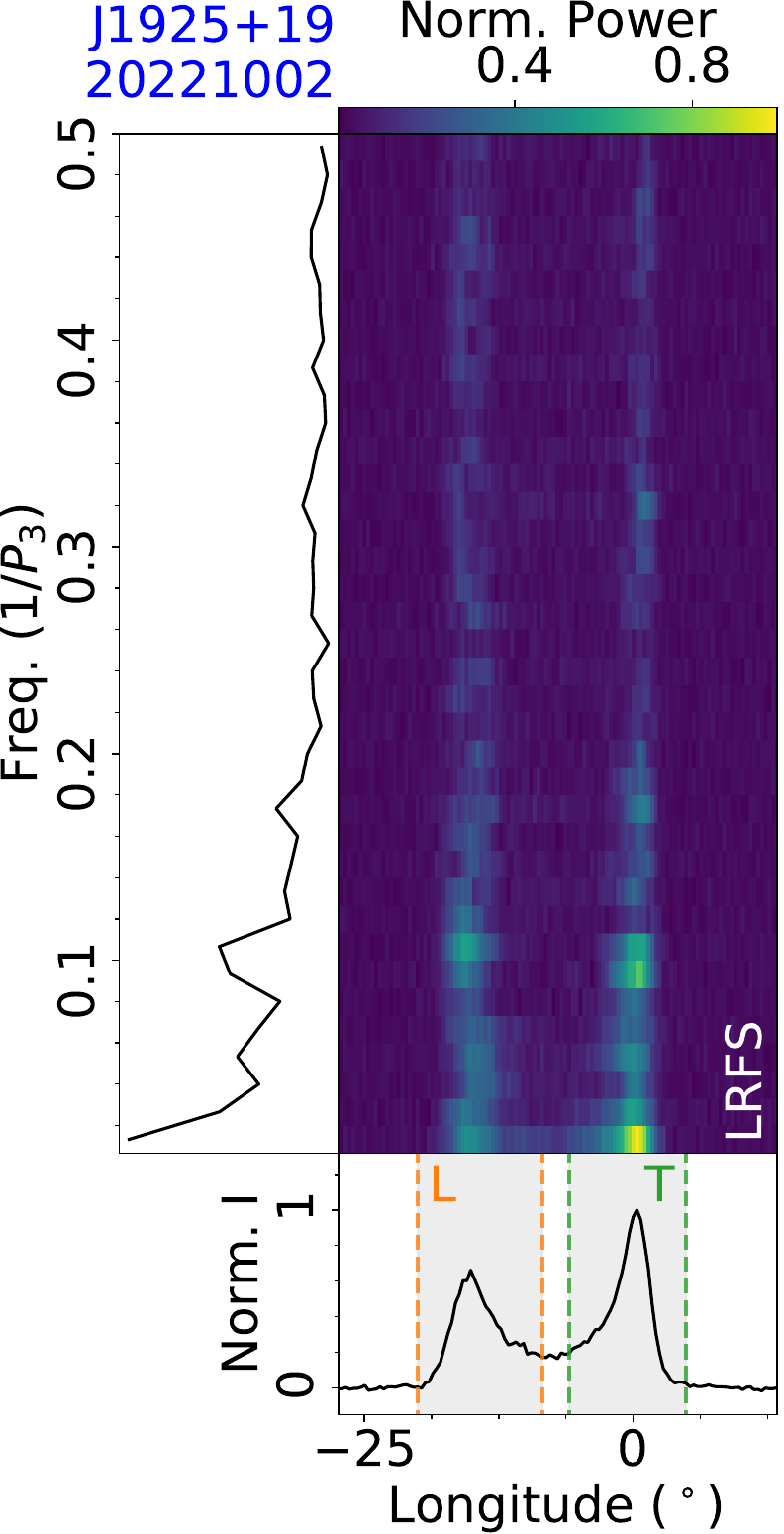}
\includegraphics[width=0.22\textwidth, angle=0]{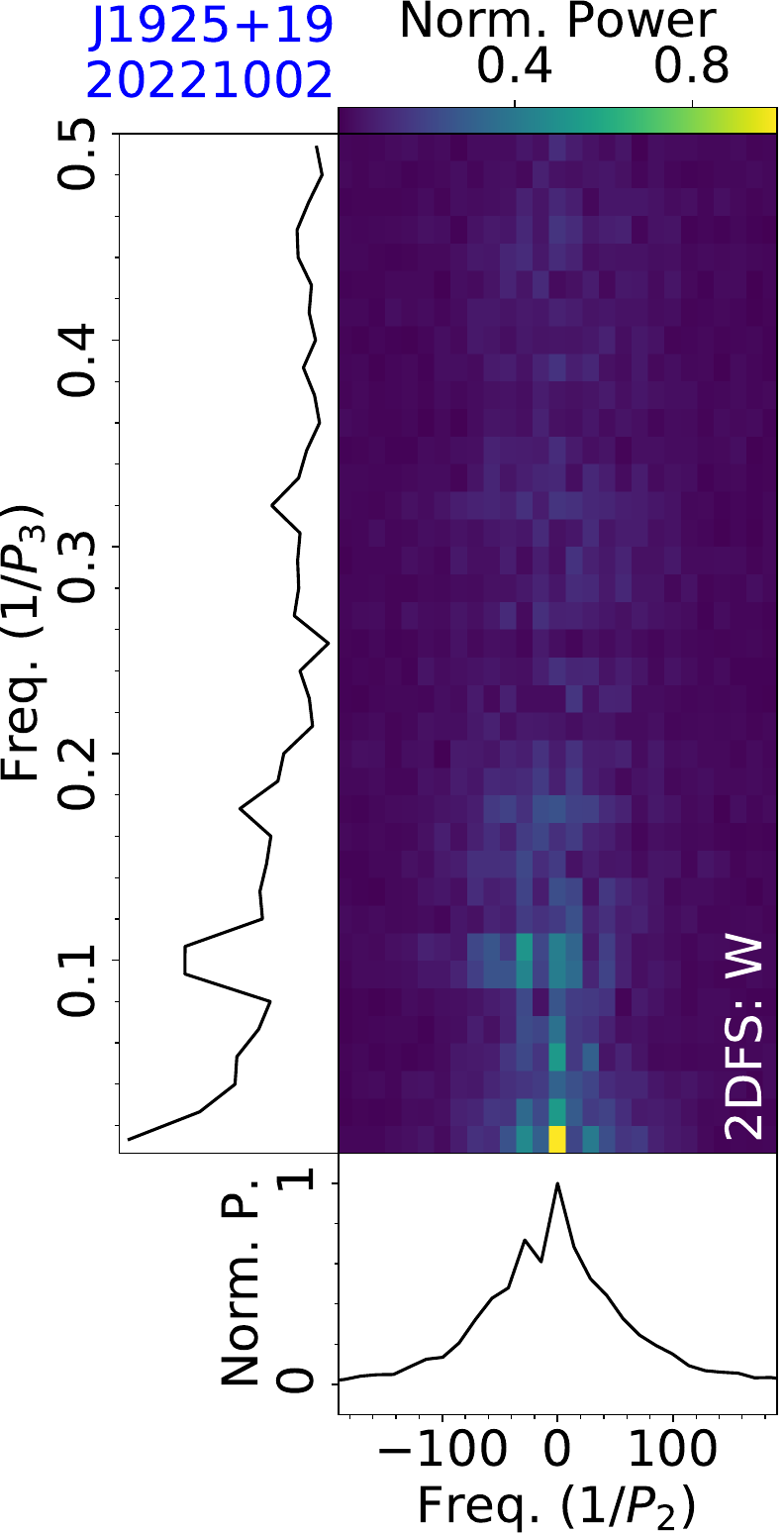}\\
\includegraphics[width=0.22\textwidth, angle=0]{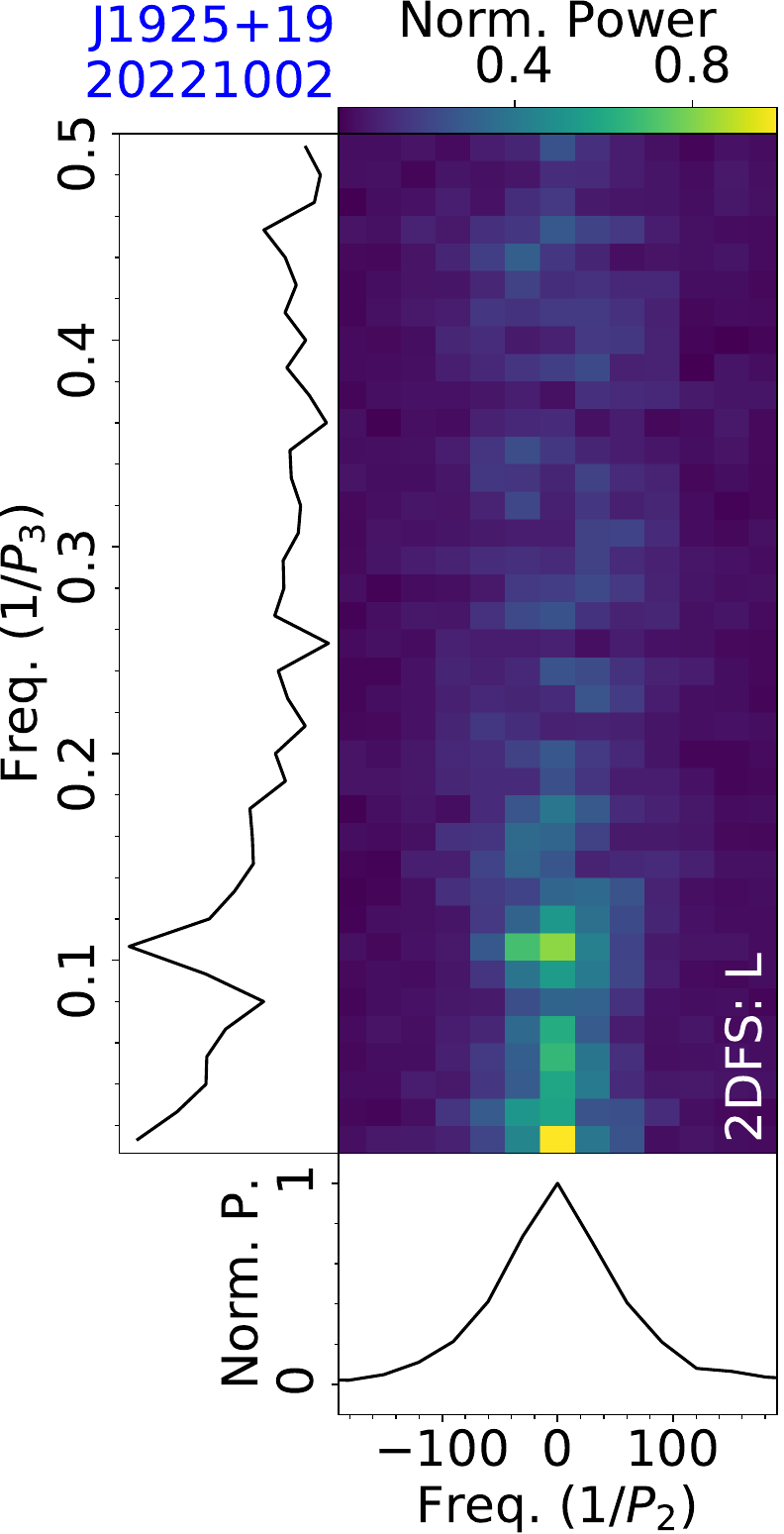}
\includegraphics[width=0.22\textwidth, angle=0]{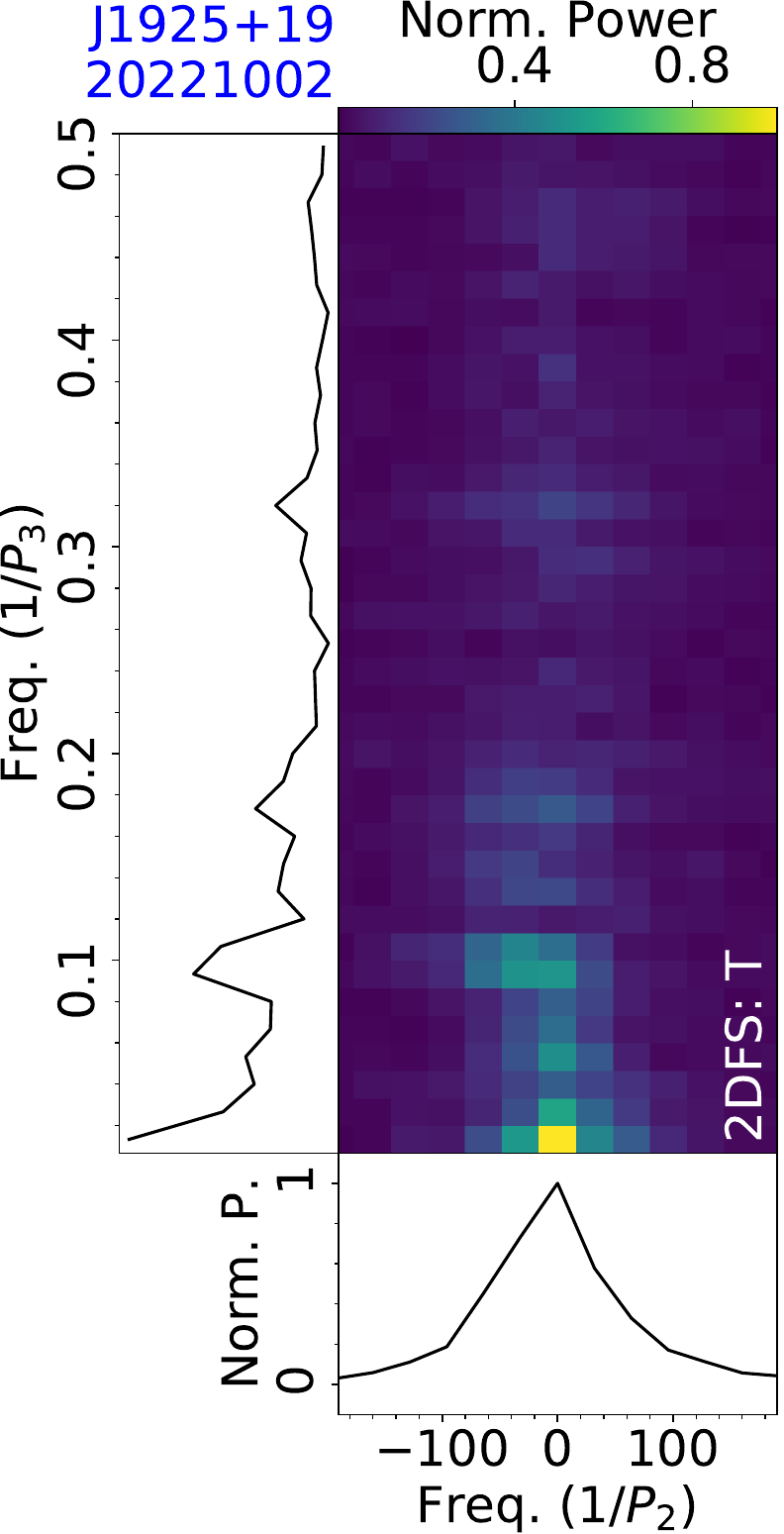}
\figcaption{Fluctuation analysis of PSR J1925+19 for the observation on 20221002, with LRFS (top-left), and 2DFS for the on-pulse phase region (top-right), leading part (bottom-left), and trailing part (bottom-right) of a mean pulse profile.
\label{subfig:fluctu:J1925+19}}
\end{figure}

\subsection{J1923+1706}
\label{subsec:J1923+1706}

PSR J1923+1706 was discovered by \citet{Hulse1975} using the Arecibo telescope. In the previous work, \citet{Song2023} reported the pulsar with $P_3$=14$\pm$5 periods and $P_2$=-66$^{+49}_{-16}$ degrees. 
This pulsar was observed by FAST on 20190321 for 5 minutes, deriving a rotation period $P=0.5472$~s and a dispersion measure $D\!M=141.5~{\rm cm^{-3}\,pc}$. 
Single pulse sequences in Fig.~\ref{subfig:TP:J1923+1706} display the nulling phenomenon. The nulling fraction of this observation is estimated to be 3.1$\pm$0.5\% from the on-pulse integral energy histogram in Fig.~\ref{subfig:Hist:J1923+1706}.
LRFS and 2DFS of three phase ranges in the mean pulse profile are displayed in Fig.~\ref{subfig:fluctu:J1923+1706}. There is a negative phase modulation feature in 2DFS of the leading profile part, with the centroid frequencies of $1/P_3=0.086\pm0.004$ and $1/P_2=-27\pm3$, corresponding to periodicities of $P_3=12\pm1$ periods and $P_2=-13\pm1^\circ$. 
2DFS of the central and trailing profile parts exhibit a temporal modulation feature, with the centroid frequencies of $1/P_3=0.05\pm0.01$ ($P_3=20\pm2$ periods) and $0.063\pm0.003$ ($16\pm1$ periods), respectively.

\subsection{J1924+1923g}
\label{subsec:J1924+1923g}

PSR J1924+1923g was observed in the FAST GPPS survey \citep{Han2021,han2025}. 

This pulsar was observed by FAST on 20250214 for 37 minutes, deriving a rotation period $P=0.6892$~s and a dispersion measure $D\!M=384.8~{\rm cm^{-3}\,pc}$. 
Single pulse sequences are shown in Fig.~\ref{subfig:TP:J1924+1923g}. In the on-pulse integral energy histogram (Fig.~\ref{subfig:Hist:J1924+1923g}), the distribution around 0 intensity is related to the weak emission. Single pulses of weak and bright emission modes are distinguished from the histogram and labeled with red and green colors. The successive length of the weak mode ranges from 1 to 6 periods, and that of the bright modes ranges from 1 to 45 periods (Fig.~\ref{subfig:scaleHist:J1924+1923g}). Emission of the weak or bright mode mostly lasts for 1 period. 
Averaged profiles of two emission modes are compared in Fig.~\ref{subfig:ProfModes:J1924+1923g}, where the peak of the weak mode is lagged relative to the bright mode.

\begin{figure}[htpb]
\centering
\includegraphics[width=0.22\textwidth, angle=0]{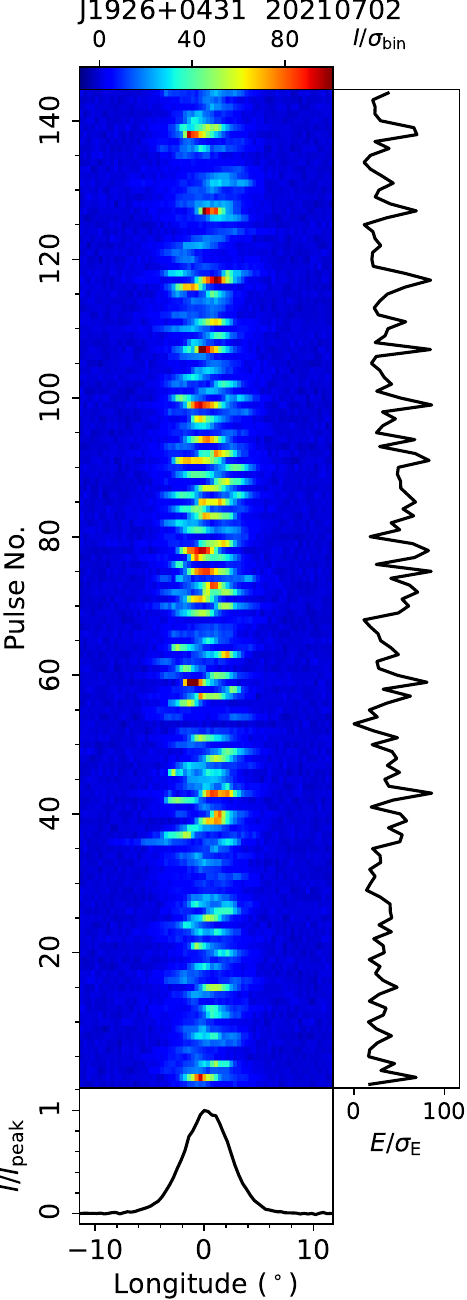}
\includegraphics[width=0.22\textwidth, angle=0]{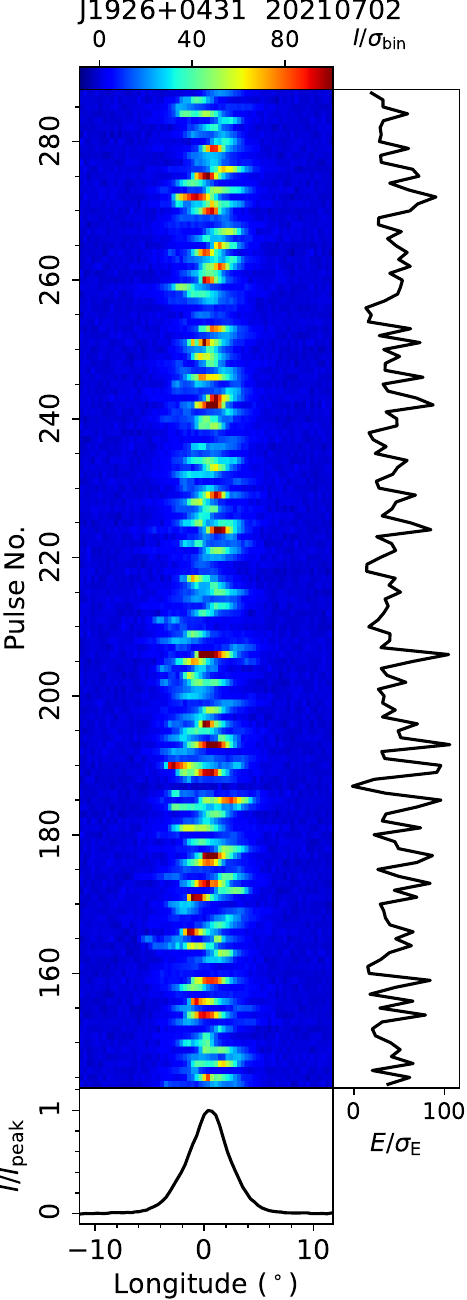}
\figcaption{Single pulse sequences of PSR J1926+0431 from the FAST observation on 20210702.
\label{subfig:TP:J1926+0431}}
\end{figure}

\begin{figure}[htpb]
\centering
\includegraphics[width=0.39\textwidth, angle=0]{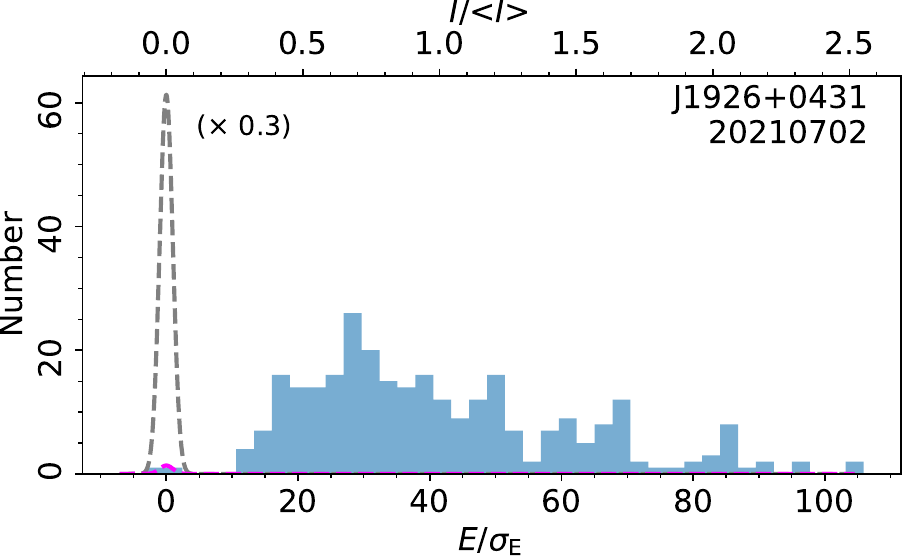}
\figcaption{On-pulse energy histogram of single pulses of PSR J1926+0431 from the FAST observation on 20210702.
\label{subfig:Hist:J1926+0431}}
\end{figure}

\begin{figure}[htpb]
\centering
\includegraphics[width=0.22\textwidth, angle=0]{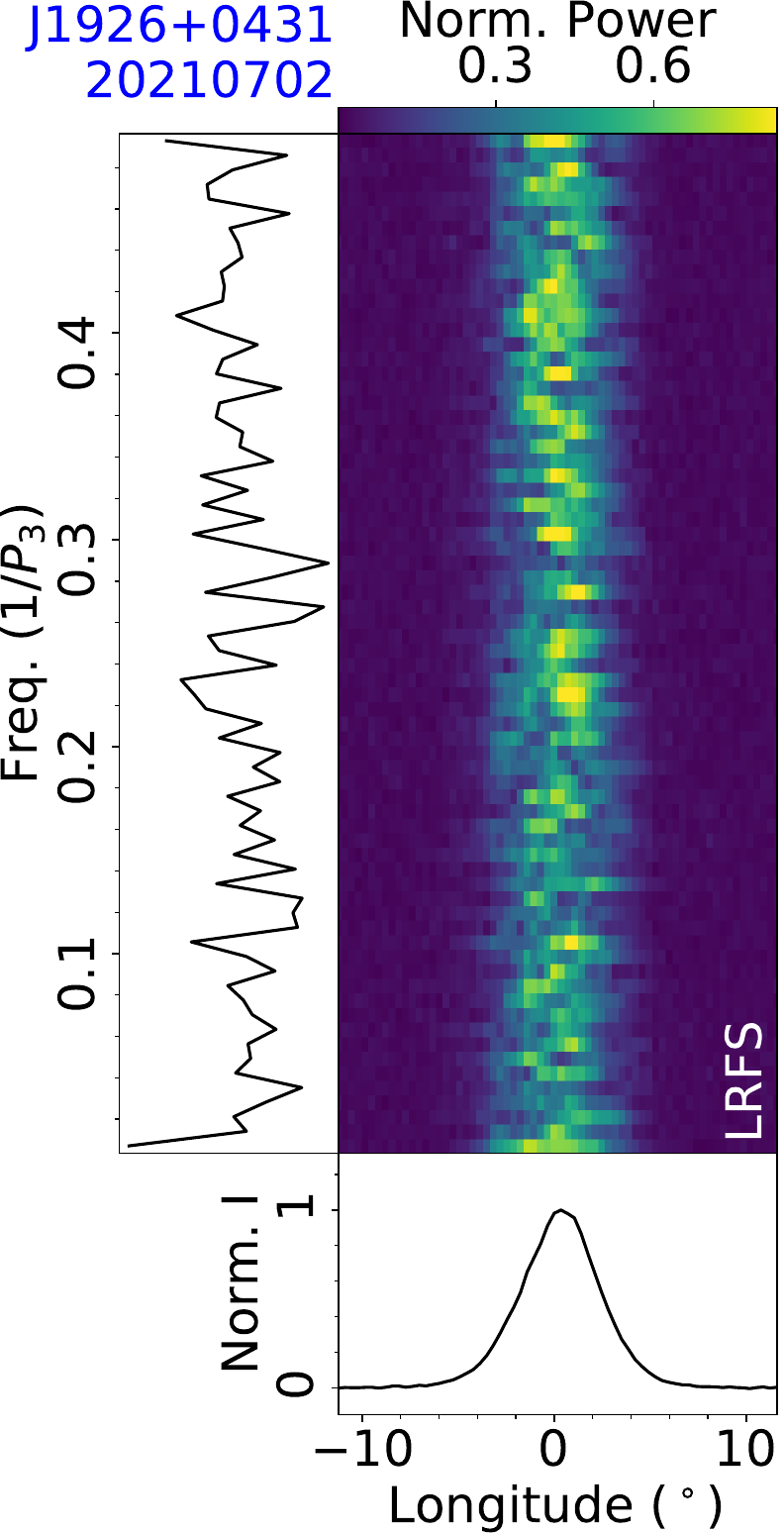}
\includegraphics[width=0.22\textwidth, angle=0]{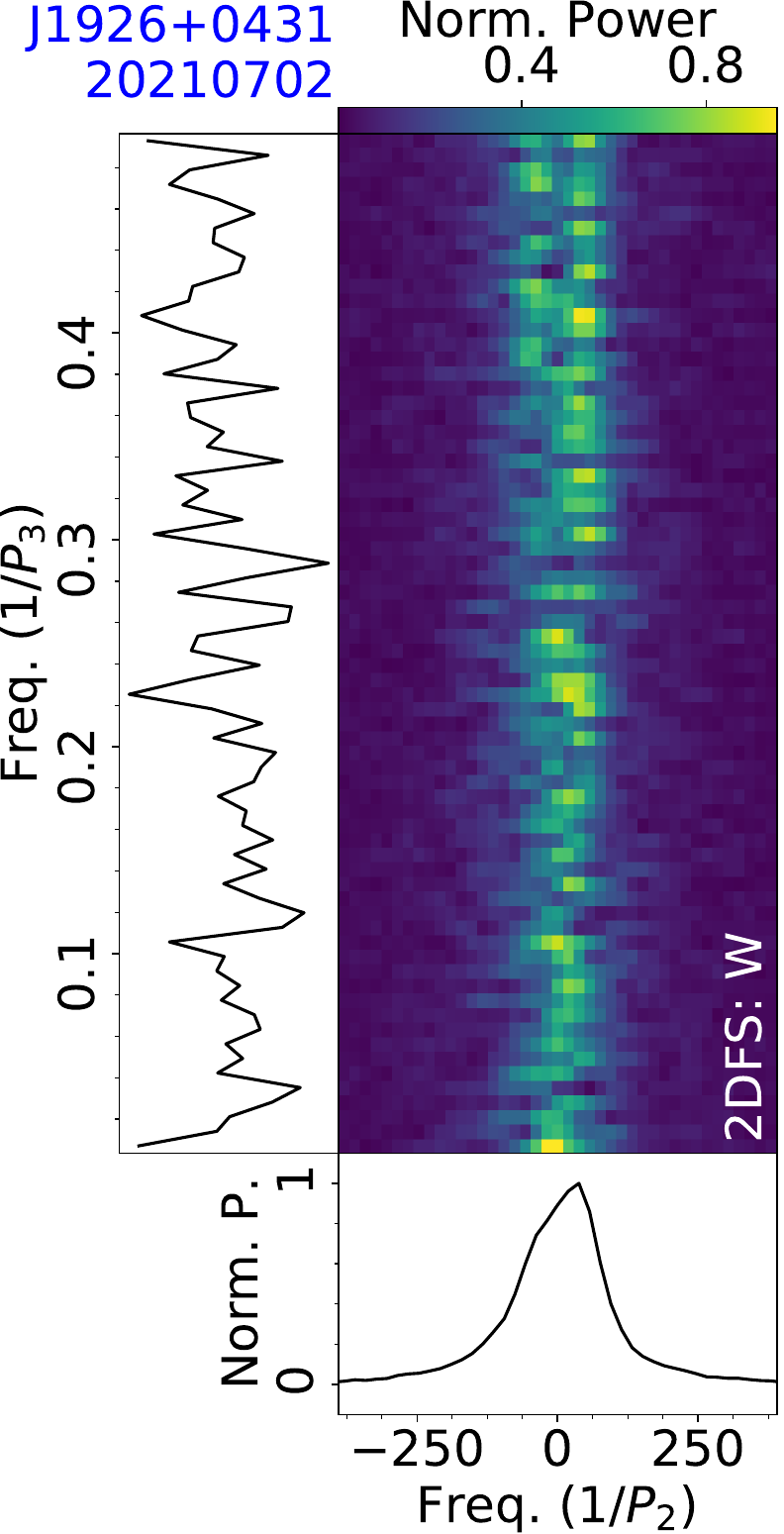}
\figcaption{Fluctuation analysis of PSR J1926+0431 from the FAST observation on 20210702, with LRFS and 2DFS for the on-pulse region of a mean pulse profile.
\label{subfig:fluctu:J1926+0431}}
%
\centering
\includegraphics[width=0.21\textwidth, angle=0]{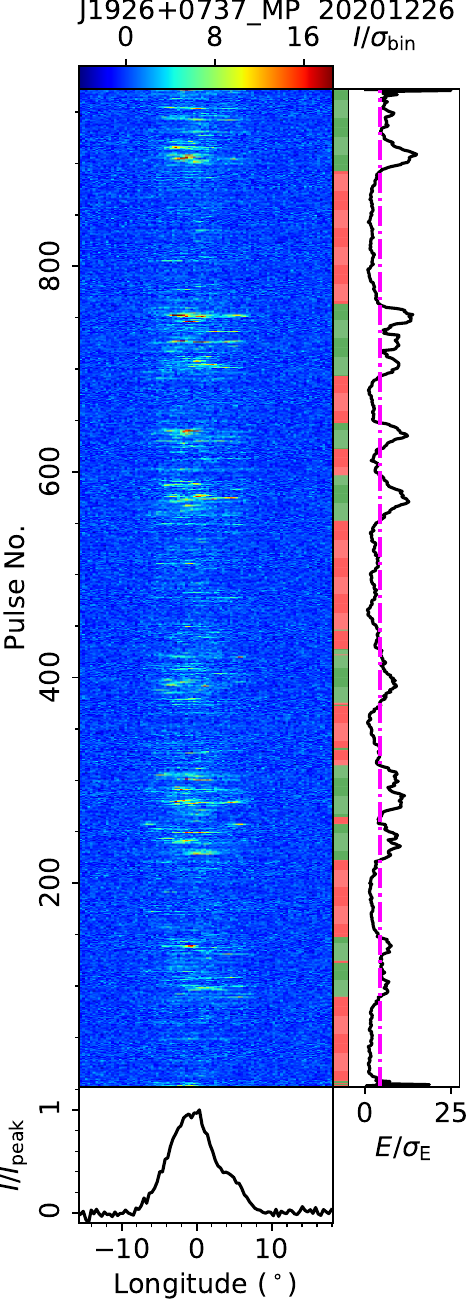}
\includegraphics[width=0.21\textwidth, angle=0]{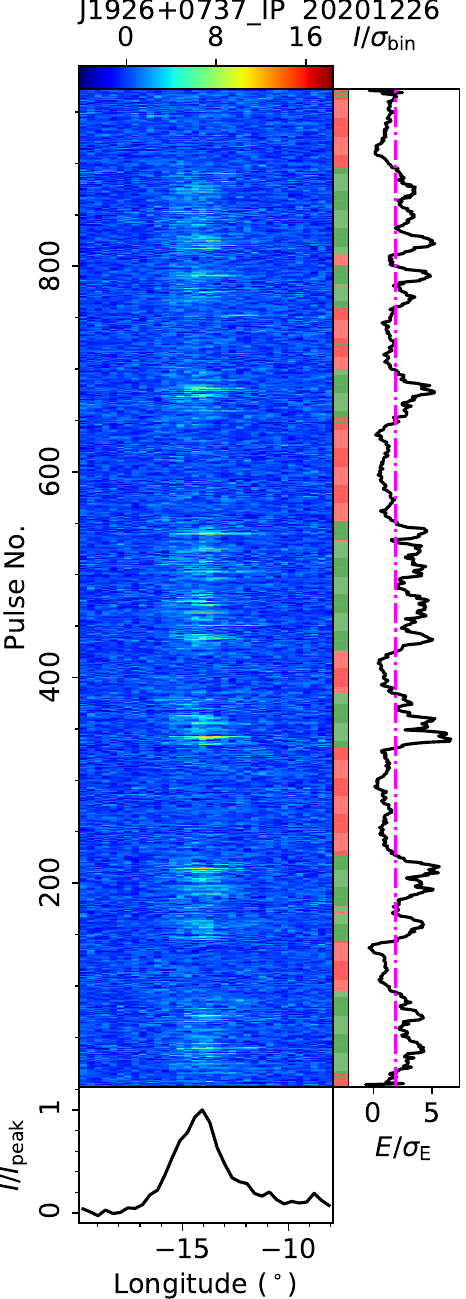}
\figcaption{Single pulse sequence of the main pulse and interpulse of PSR J1926+0737 from the FAST observation on 20201226. 
The green and red bars represent bright or weak emission modes. In the right subpanels of both plots, the on-pulse energy variations smoothed over every 15 periods for the main pulse and 11 periods for the interpulse are plotted against period, with dashed lines for thresholds to distinguish the weak and bright emission states.
\label{subfig:TP:J1926+0737}}
\end{figure}

\begin{figure}[htpb]
\centering
\includegraphics[width=0.39\textwidth, angle=0]{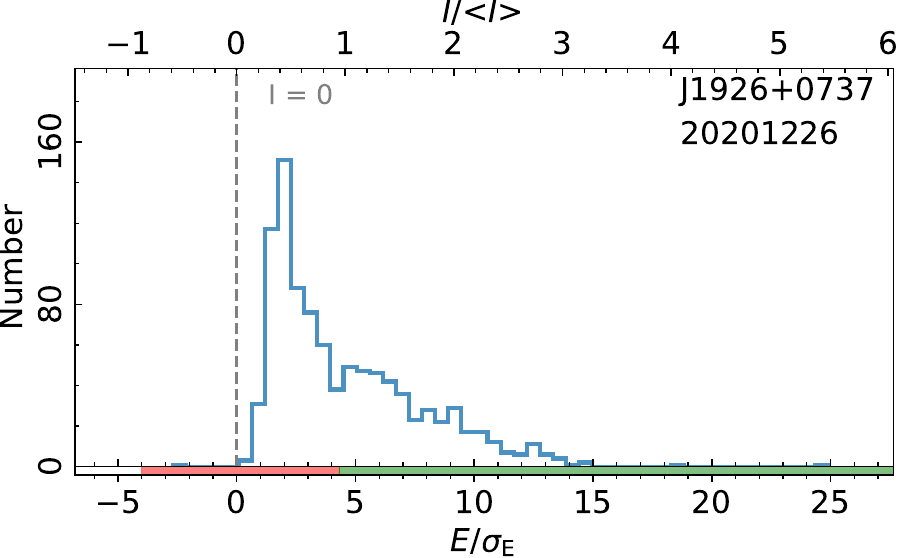}\\
\includegraphics[width=0.39\textwidth, angle=0]{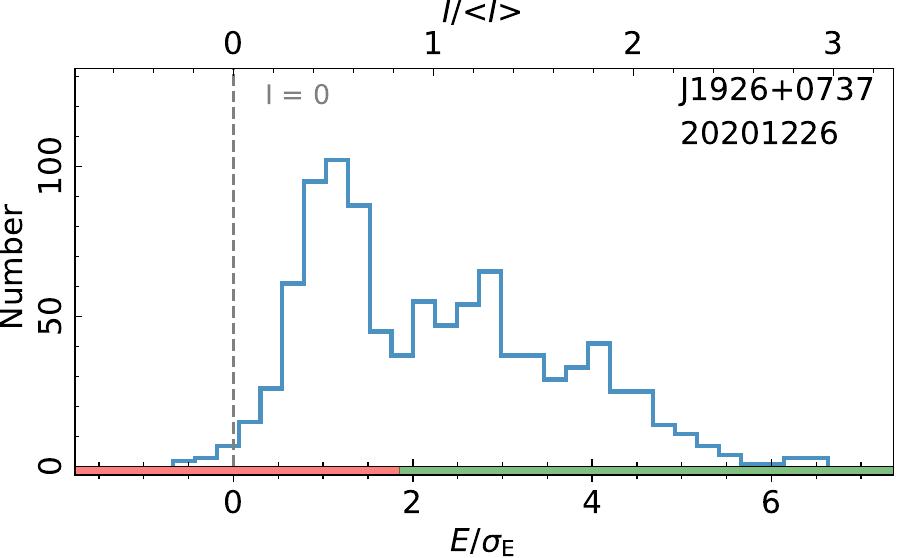}
\figcaption{On-pulse energy histograms of single pulses of the main pulse and interpulse for PSR J1926+0737 from the FAST observation on 20201226, with energy values smoothed over 15 periods for the main pulse and 11 periods for the interpulse. The red and green bars indicate the weak and bright emission states.
\label{subfig:Hist:J1926+0737}}
\end{figure}

\begin{figure}[htpb]
\centering
\includegraphics[width=0.35\textwidth, angle=0]{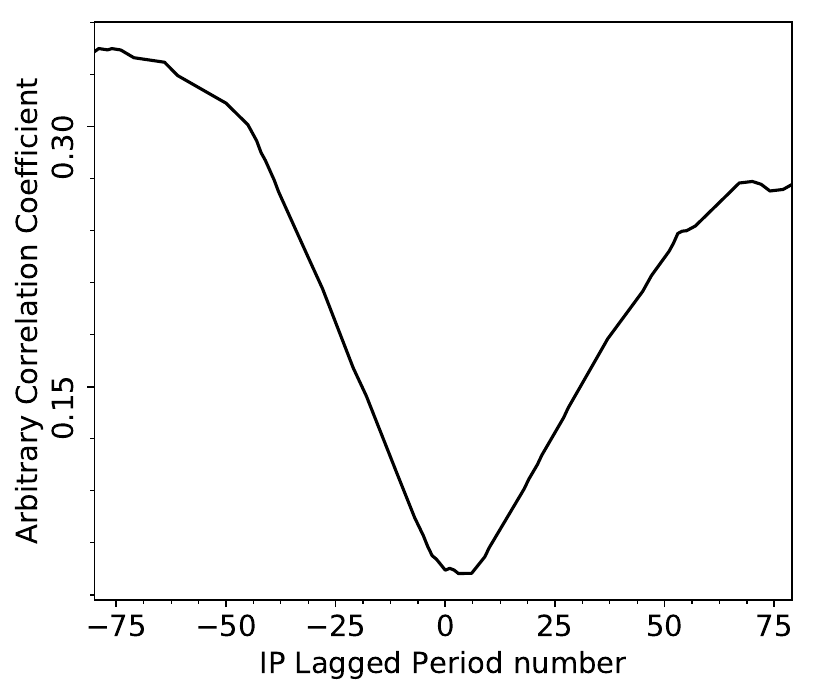}
\figcaption{Correlation of occurrence of weak and bright emission states between main pulse and interpulse of PSR J1926+0737 from the FAST observation on 20201226.
\label{subfig:ModesCorr:J1926+0737}}
\end{figure}

\begin{figure}[htpb]
\centering
\includegraphics[width=0.39\textwidth, angle=0]{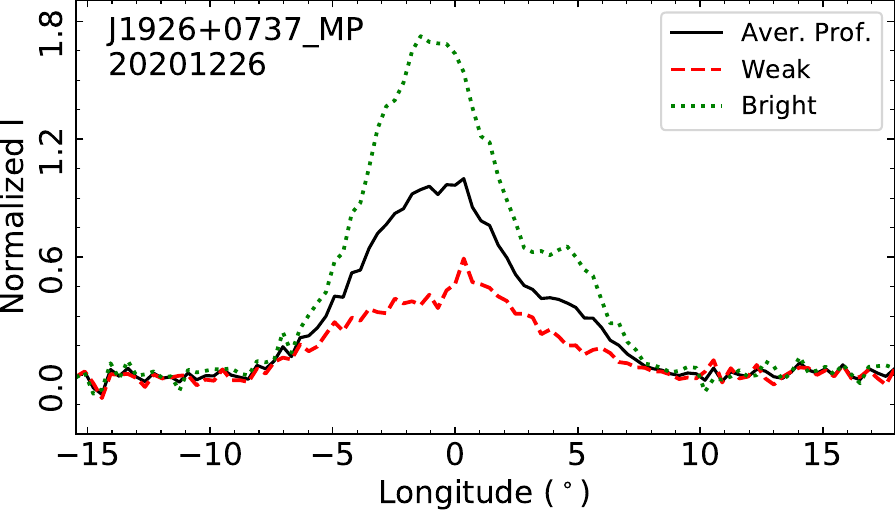}\\
\includegraphics[width=0.39\textwidth, angle=0]{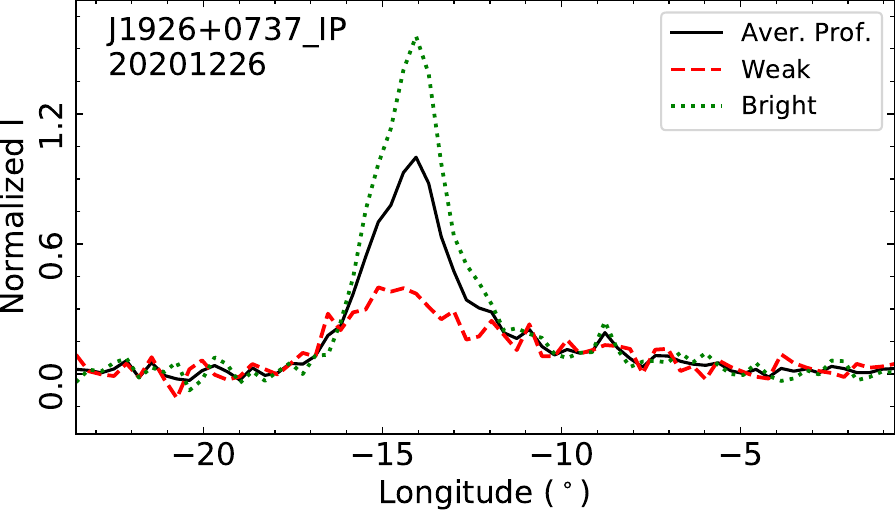}
\figcaption{Mean profiles of the weak (red dashed) and bright (green dotted) emission modes of PSR J1926+0737 for main pulse (upper) and interpulse (lower) observed on 20201226. Profiles are normalized by the peak of the mean profile of all periods. \label{subfig:ProfModes:J1926+0737}}
\end{figure}

\subsection{J1924+2037g}
\label{subsec:J1924+2037g}

PSR J1924+2037g was discovered in the FAST GPPS survey with a very nulling behavior \citep{Han2021,han2025}.

This pulsar was observed by FAST on 20230717 for 96 minutes, deriving a rotation period $P=0.6848$~s and a dispersion measure $D\!M=85.3~{\rm cm^{-3}\,pc}$. 
Single pulse sequences of this observation are shown in Fig.~\ref{subfig:TP:20230717}. From the intensity distribution in Fig.~\ref{subfig:Hist:J1924+2037g}, the nulling fraction is estimated to be 75$\pm$4\%.  
From the distributions of continuous period numbers for adjacent nulling and emission, the duration of emission is short, ranging from 1 to 16 periods. In contrast, the duration of nulling is longer, ranging from 1 to 797 periods, with an average of 113 periods. From the histogram in Fig.~\ref{subfig:nullDegreeScale:J1924+2037g}, the nulling degree and scale are estimated to be 75$\pm$25 degrees and 116$\pm$157 periods for NE pairs. For EN pairs, they are 78$\pm$20 degrees and 105$\pm$139 periods.

\begin{figure}[htpb]
\centering
\includegraphics[width=0.22\textwidth, angle=0]{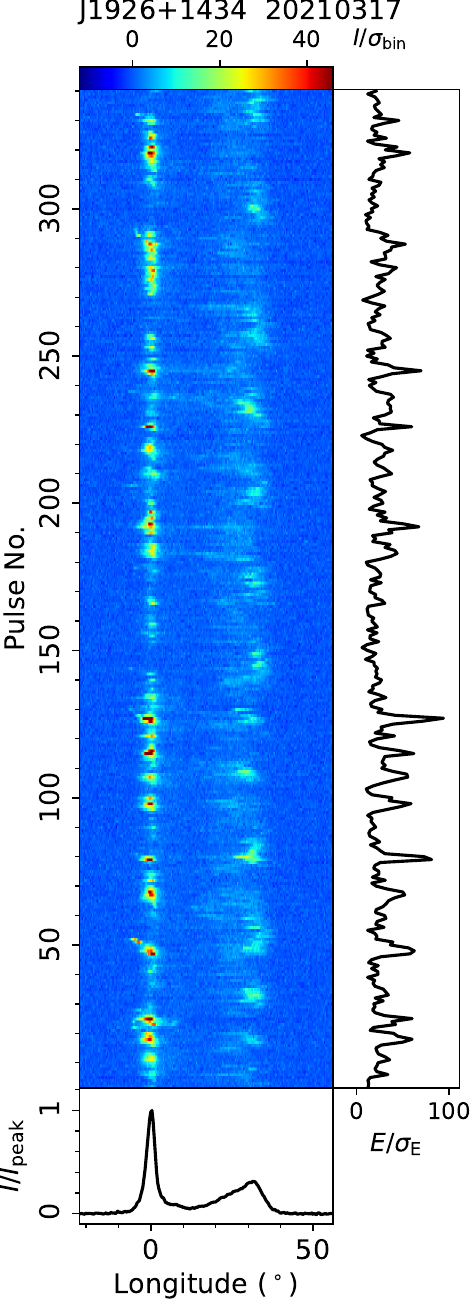}
\includegraphics[width=0.22\textwidth, angle=0]{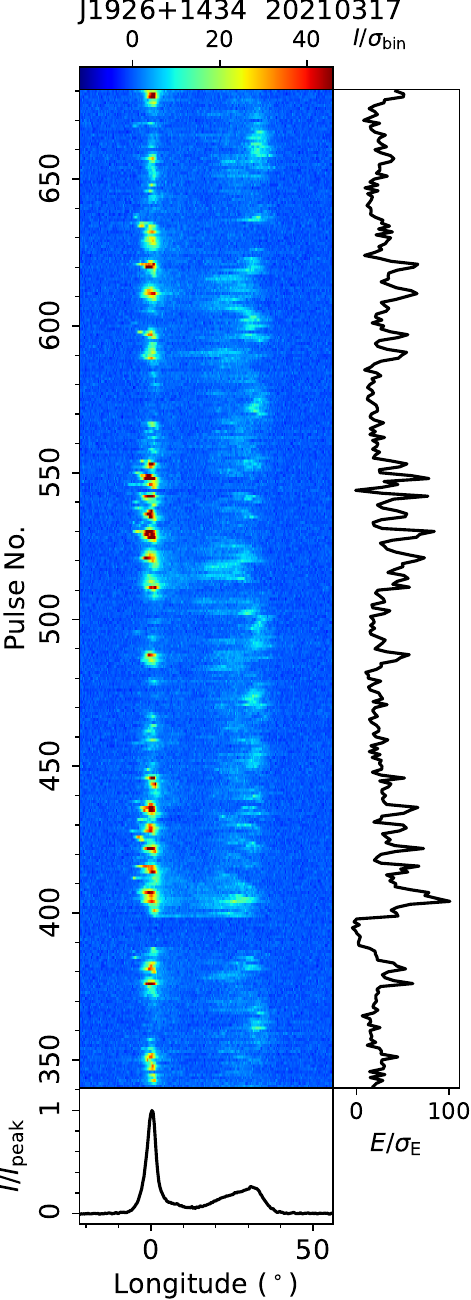}
\figcaption{Single pulse sequences of PSR J1926+1434 from the FAST observation on 20210317.
\label{subfig:TP:J1926+1434}}
\end{figure}

\begin{figure}[htpt]
\centering
\includegraphics[width=0.39\textwidth, angle=0]{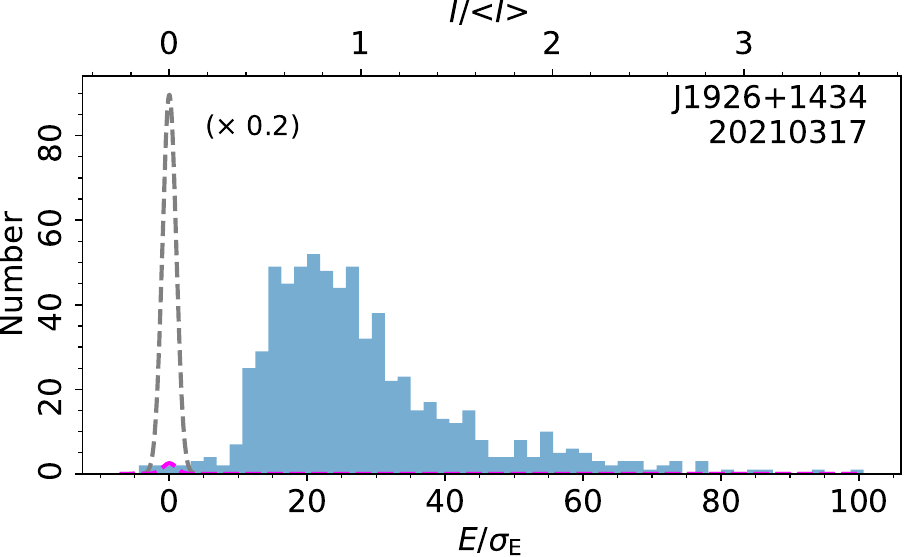}
\figcaption{On-pulse energy histogram of single pulses of PSR J1926+1434 from the FAST observation on 20210317.
\label{subfig:Hist:J1926+1434}}
\end{figure}

\begin{figure}[htpb]
\centering
\includegraphics[width=0.22\textwidth, angle=0]{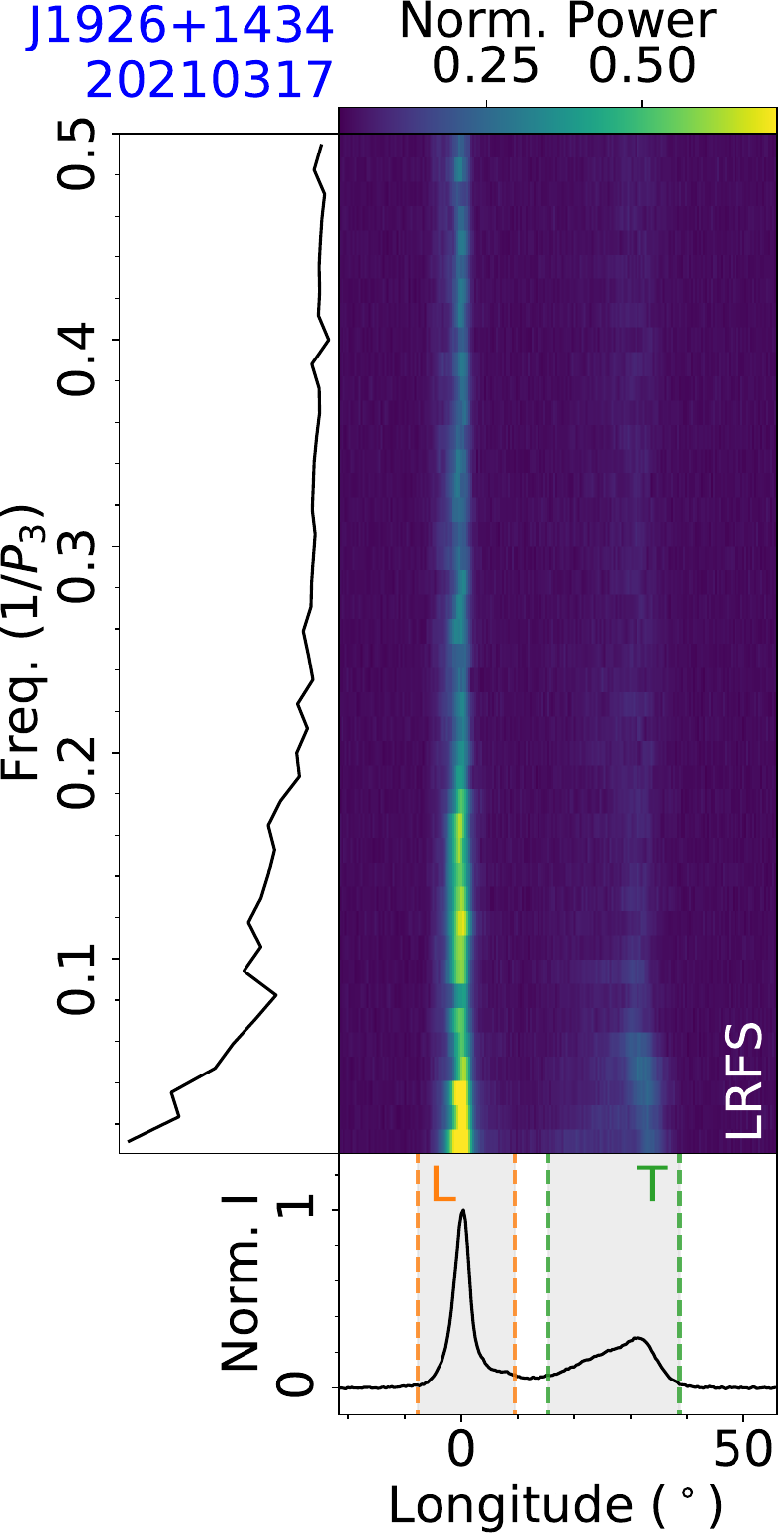}
\includegraphics[width=0.22\textwidth, angle=0]{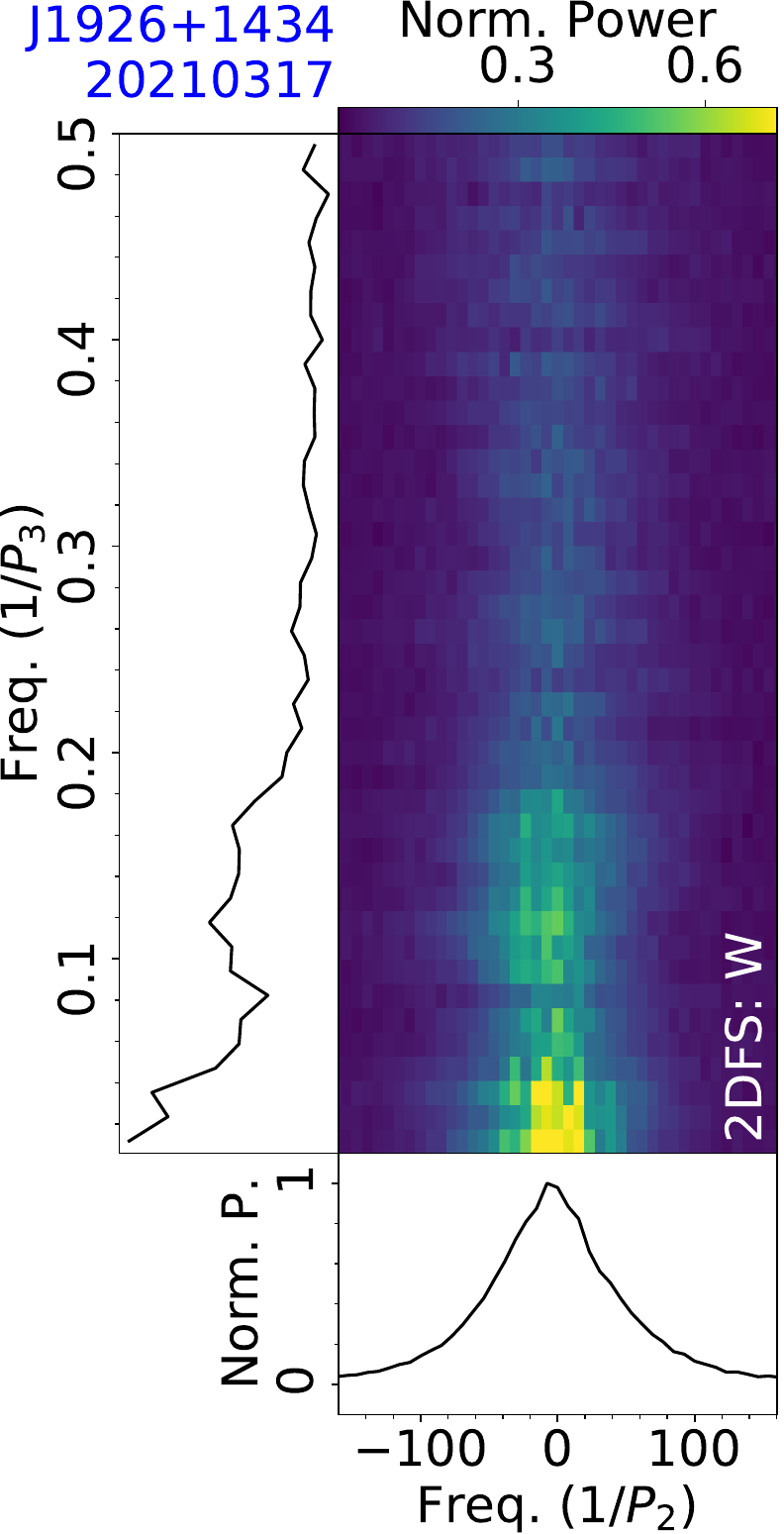}\\
\includegraphics[width=0.22\textwidth, angle=0]{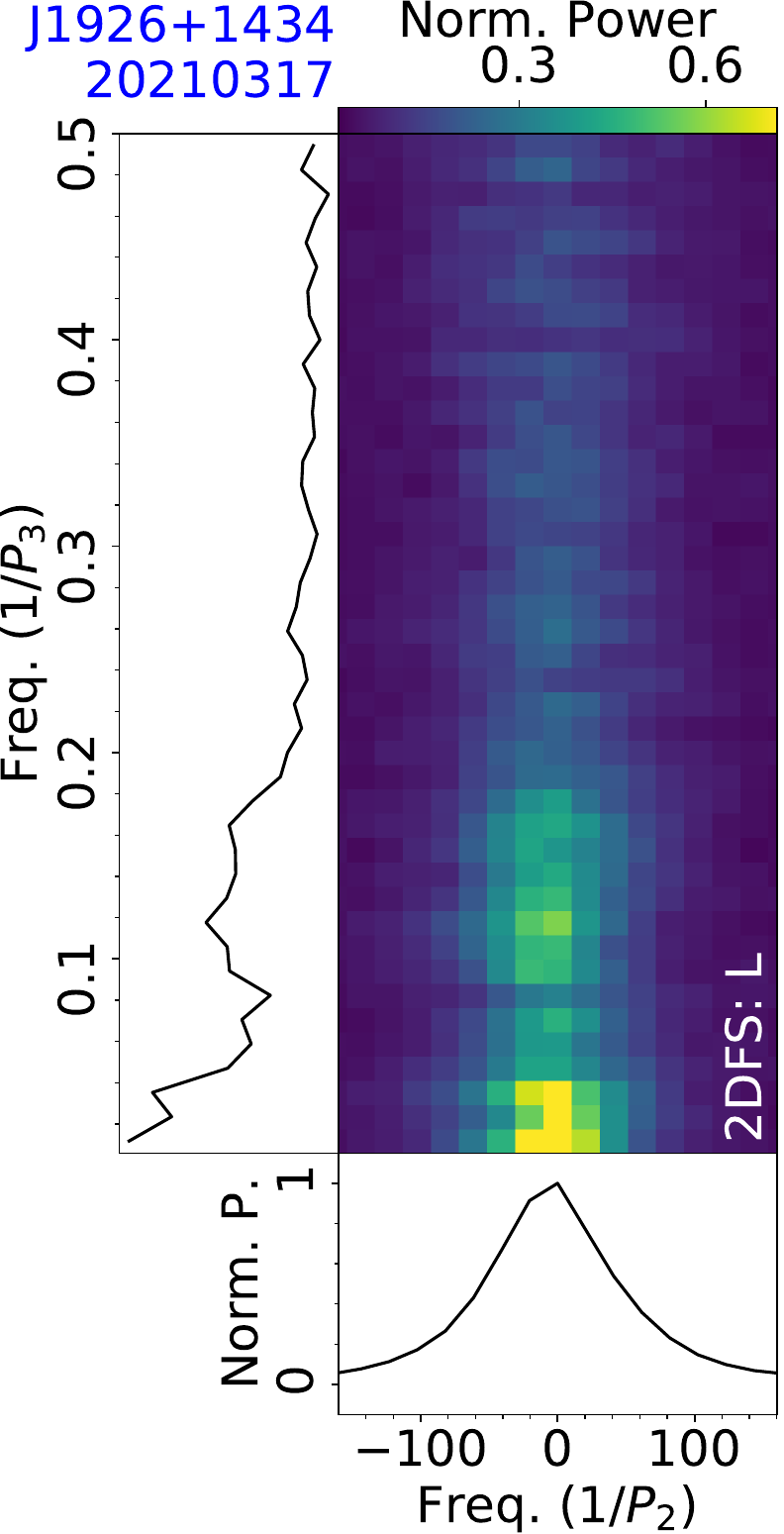}
\includegraphics[width=0.22\textwidth, angle=0]{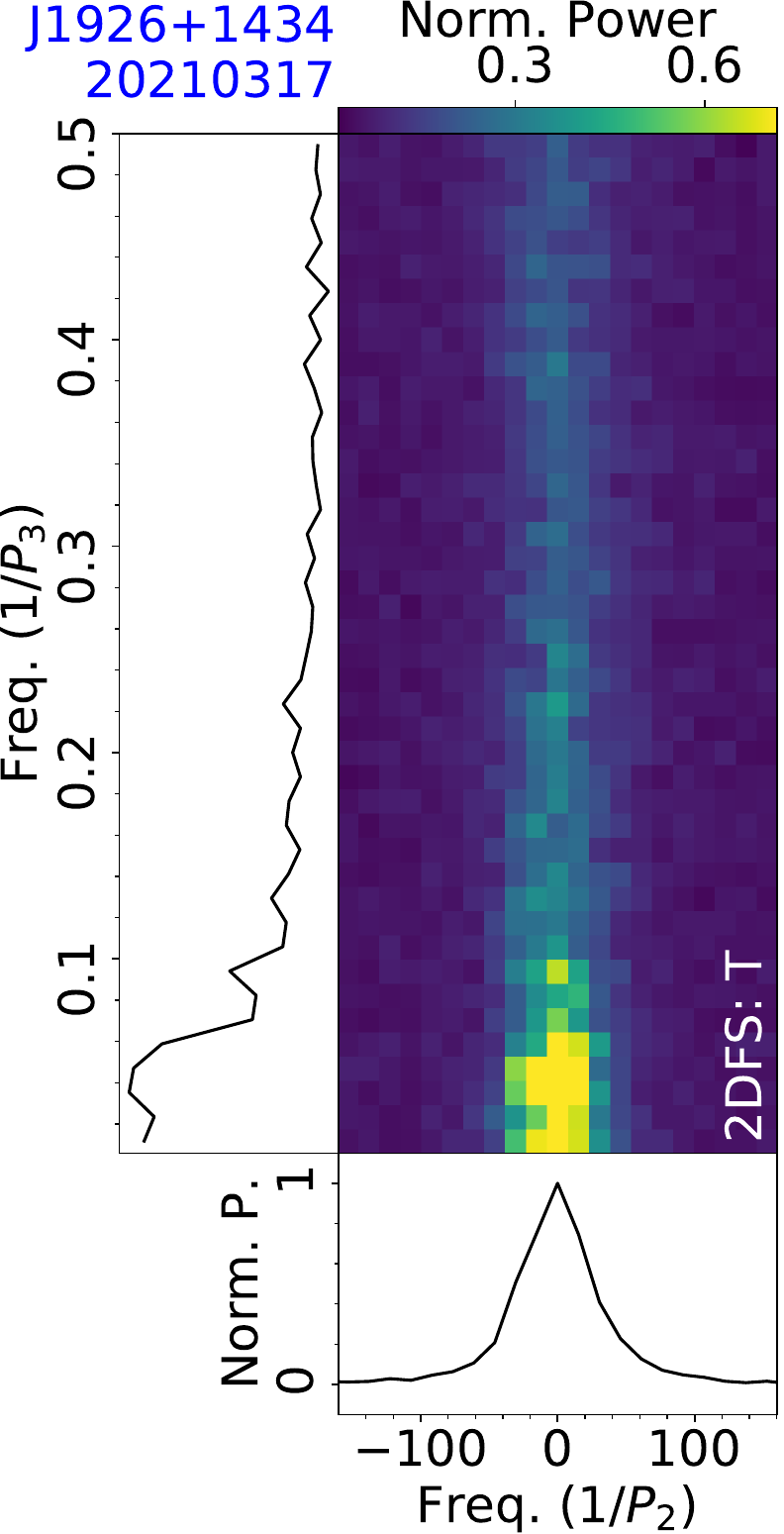}
\figcaption{Fluctuation analysis of PSR J1926+1434 for the observation on 20210317, with LRFS (top-left), and 2DFS for the on-pulse phase region (top-right), leading part (bottom-left) and trailing part (bottom-right) of a mean pulse profile.
\label{subfig:fluctu:J1926+1434}}
\end{figure}

\subsection{J1924+2040}
\label{subsec:J1924+2040}

PSR J1924+2040 was discovered by \citet{Hulse1975} with the Arecibo telescope. 

This pulsar was observed by FAST on 20191226 for 5 minutes, deriving a rotation period $P=0.2378$~s and a dispersion measure $D\!M=226.6~{\rm cm^{-3}\,pc}$. The single pulse sequence and a zoomed-in view of pulses No. 1000-1200 in Fig.~\ref{subfig:TP:J1924+2040} illustrate the modulation behavior and the presence of quasi-periodic subpulses. Fluctuation spectra are displayed in Fig.~\ref{subfig:fluctu:J1924+2040}, and the centroid frequency of the main modulation feature is $1/P_3=0.071\pm0.001$, corresponding to $P_3=14.2\pm0.3$ periods. In addition, symmetrically located secondary peaks are present in the bottom subpanel of the 2DFS. The $P_2$ values of $-3.9\pm0.1$ and $3.8\pm0.1$ degrees are derived from the centroid frequencies $1/P_2$ of $-93\pm1$ and $95\pm2$, respectively. These values correspond to the phase interval of the quasi-periodic subpulses.

\subsection{J1925+19}
\label{subsec:J1925+19}

PSR J1925+19 was discovered in a deep Parkes multibeam survey \citep{Lorimer2013}.

This pulsar was observed by FAST on 20221002 for 5 minutes, with a rotation period $P=1.9165$~s and a dispersion measure $D\!M=332.2~{\rm cm^{-3}\,pc}$ derived. The single pulse sequence in Fig.~\ref{subfig:TP:J1925+19} shows nulling and subpulse drifting phenomena. The nulling fraction of this observation is 31$\pm$2\%, estimated from the on-pulse integral energy histogram (Fig.~\ref{subfig:Hist:J1925+19}). Fluctuation spectra are shown in Fig.~\ref{subfig:fluctu:J1925+19}. For the leading part in a mean pulse profile, the centroid frequencies of the negative drift feature are $1/P_3=0.106\pm0.002$ and $1/P_2=-12\pm4$, corresponding to periodicities of $P_3=9.4\pm0.2$ periods and $P_2=-30\pm10$ degrees. The drift feature in 2DFS of the trailing profile part is characterized by frequencies of $1/P_3=0.099\pm0.002$ and $1/P_2=-31\pm4$, yielding $P_3=10.1\pm0.2$ periods and $P_2=-12\pm2$ degrees.

\begin{figure}[htpb]
\centering
\includegraphics[width=0.22\textwidth, angle=0]{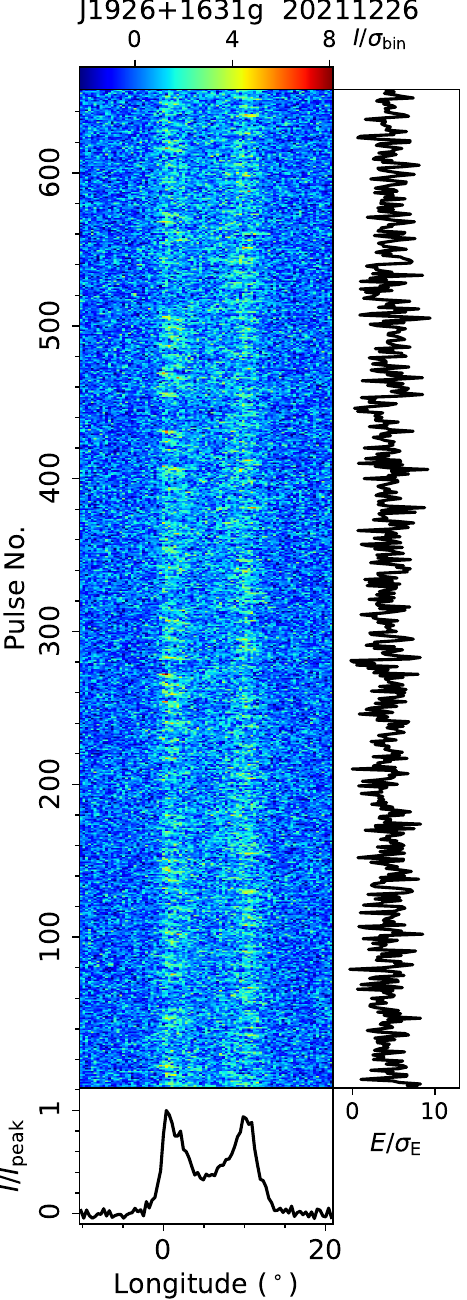}
\includegraphics[width=0.22\textwidth, angle=0]{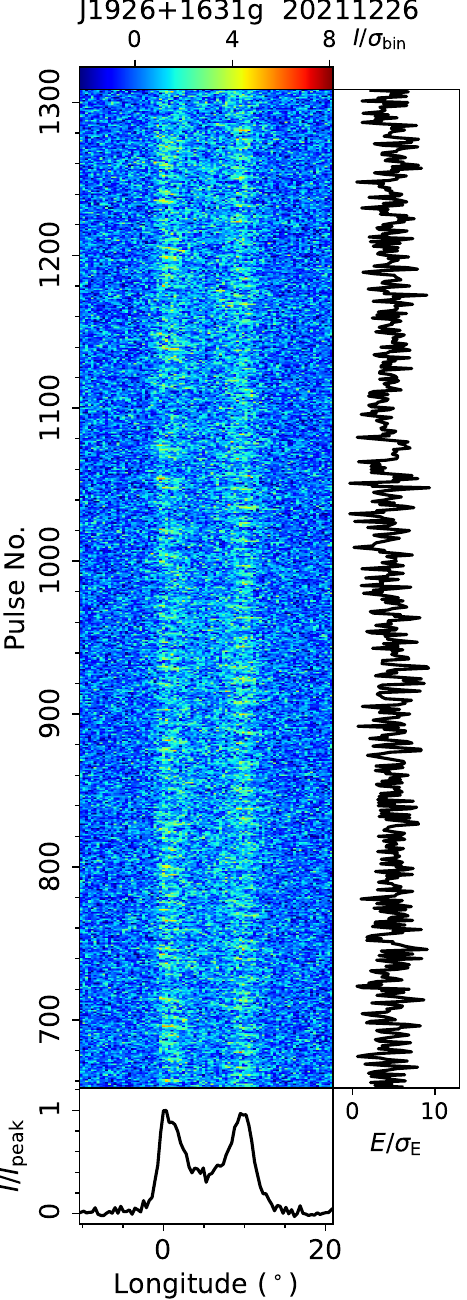}
\figcaption{Single pulse sequences of PSR J1926+1631g from the FAST observation on 20211226.
\label{subfig:TP:J1926+1631g}}
\end{figure}

\begin{figure}[htpb]
\centering
\includegraphics[width=0.22\textwidth, angle=0]{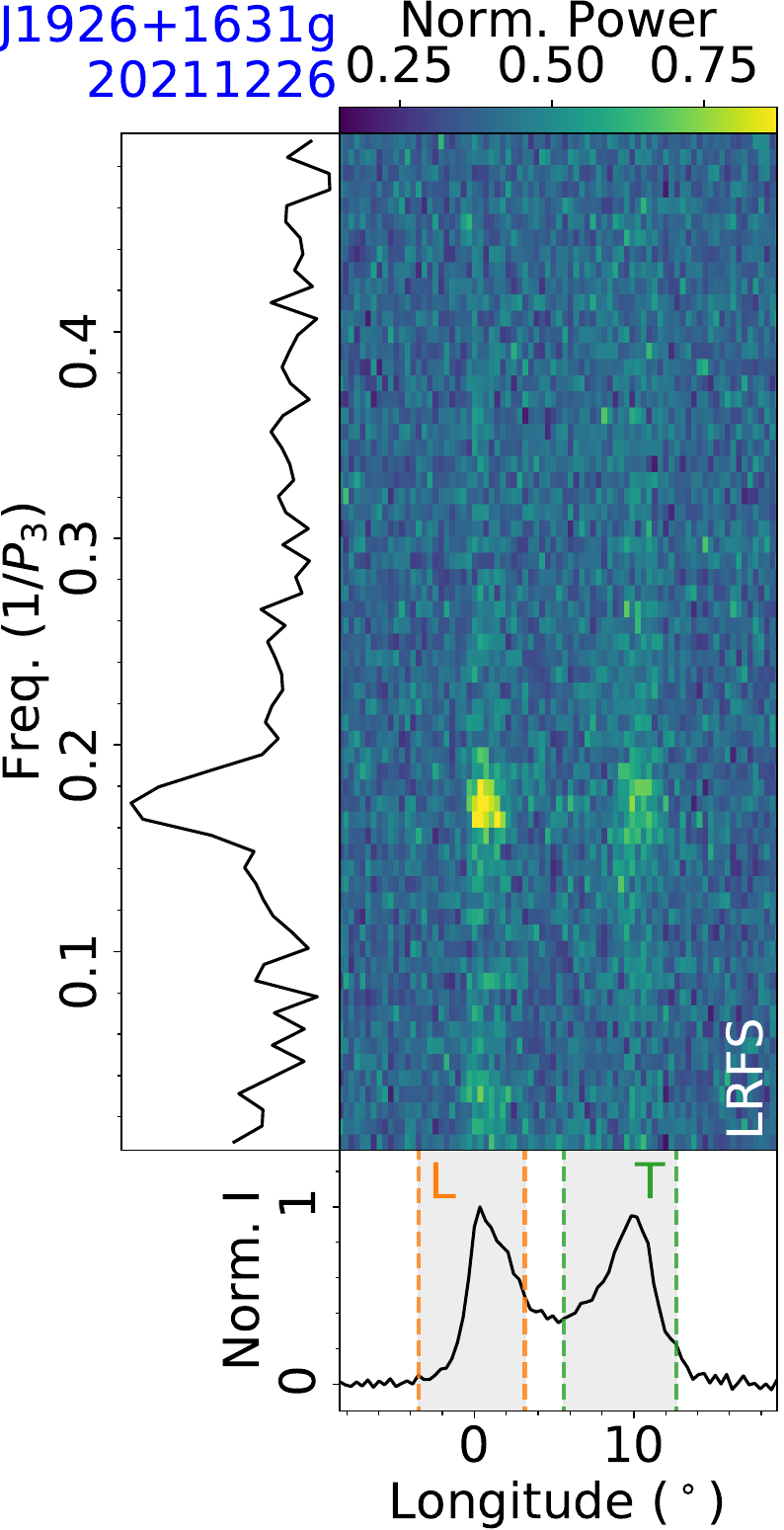}
\includegraphics[width=0.22\textwidth, angle=0]{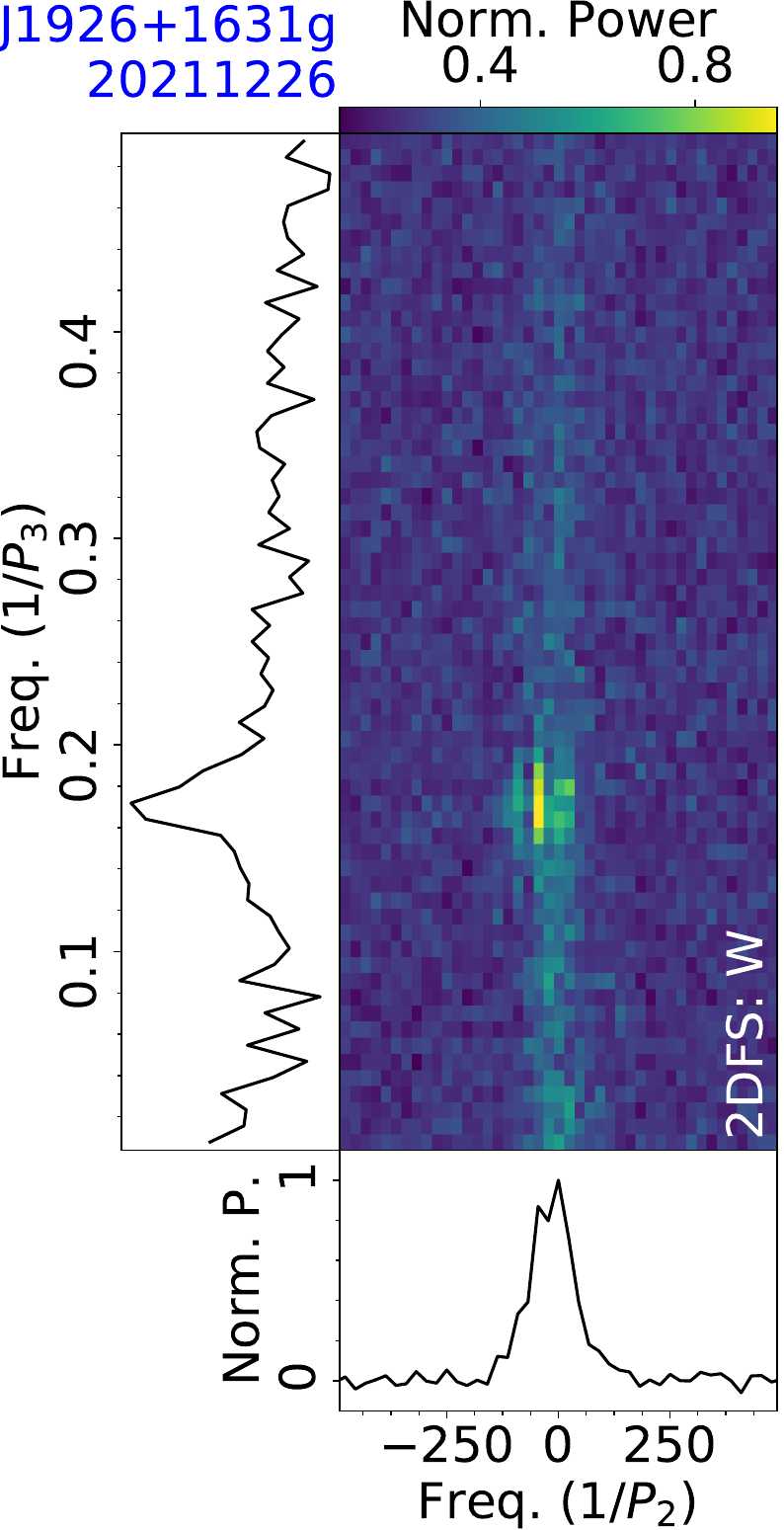}\\
\includegraphics[width=0.22\textwidth, angle=0]{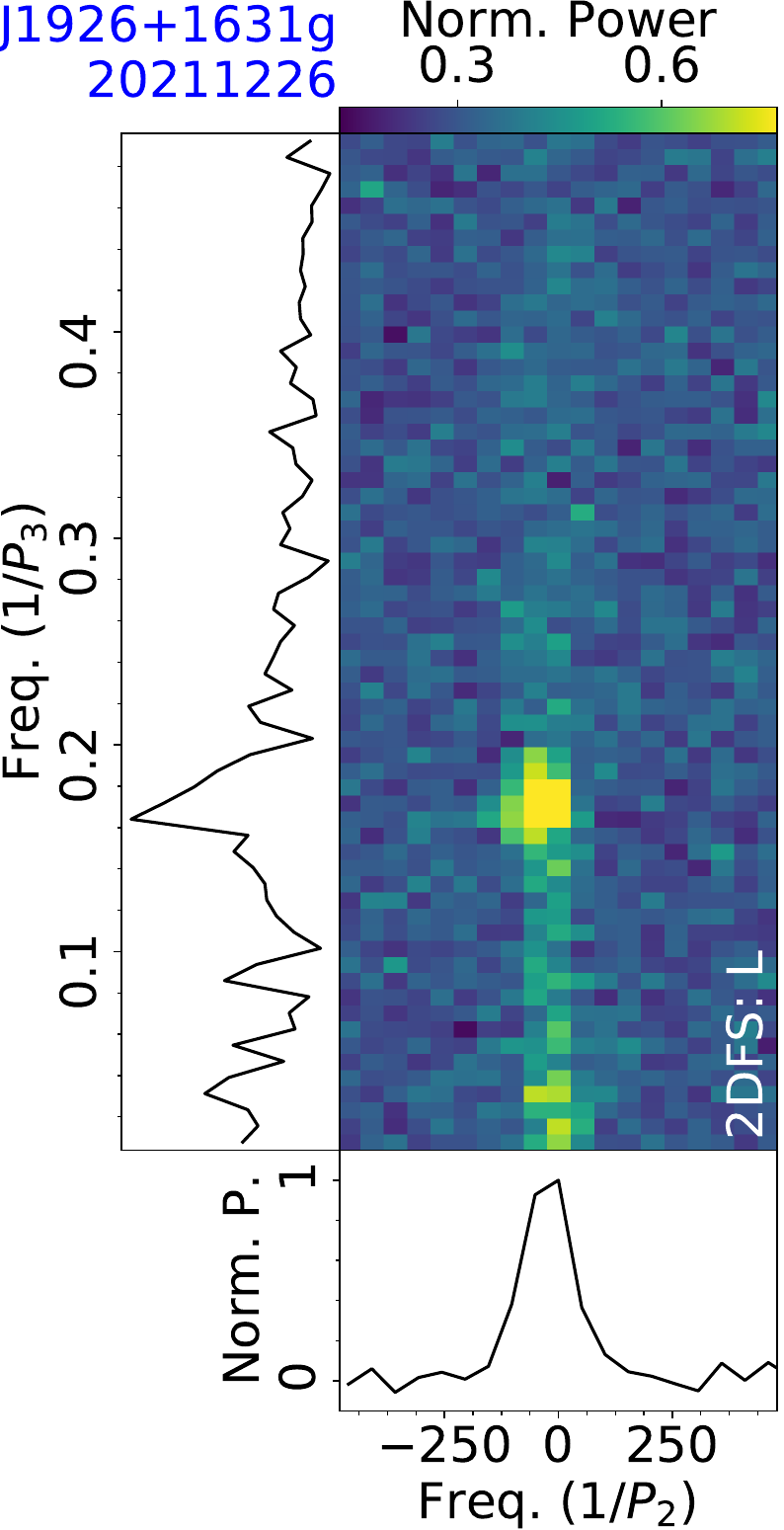}
\includegraphics[width=0.22\textwidth, angle=0]{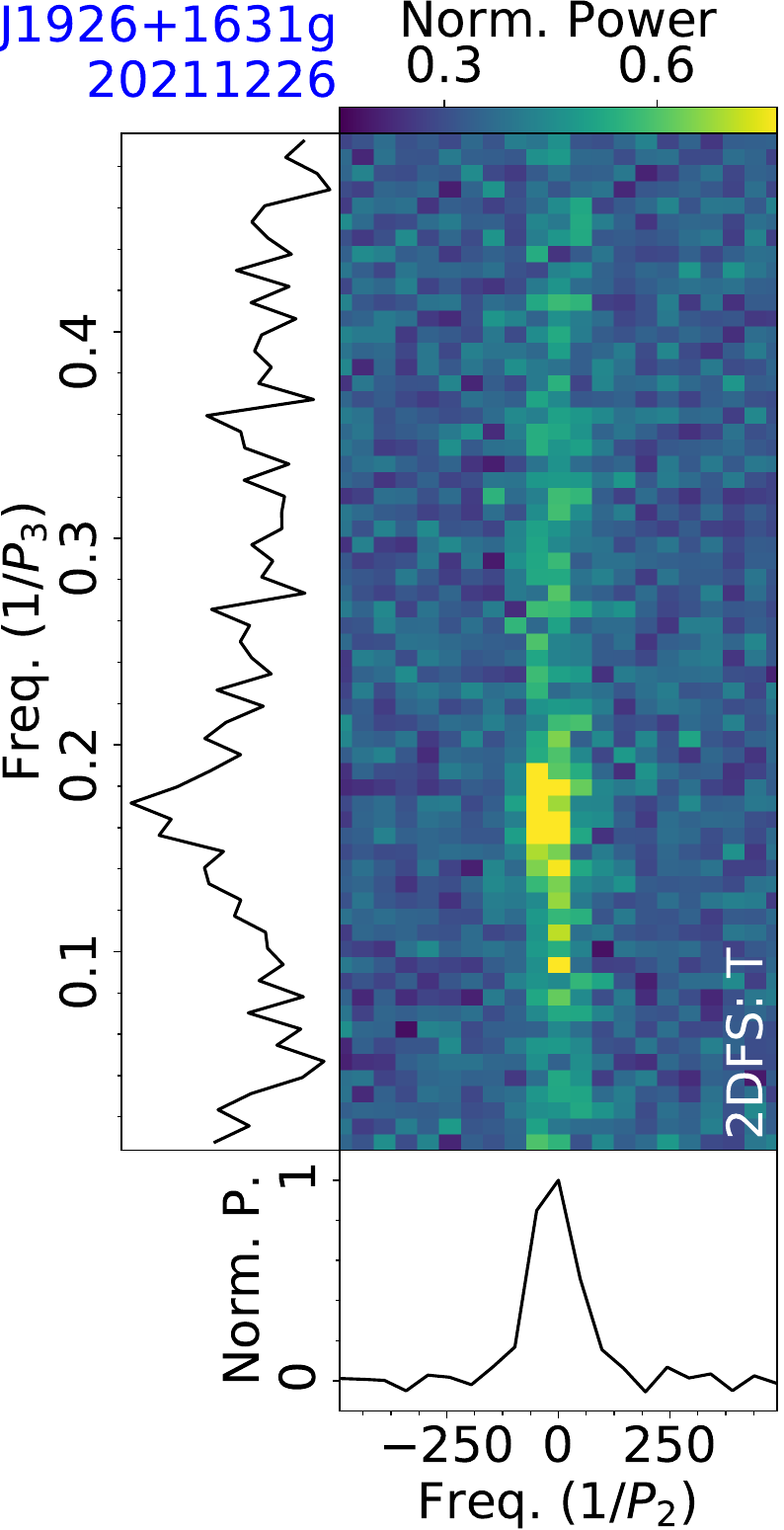}
\figcaption{Fluctuation analysis of PSR J1926+1631g for the observation on 20211226, with LRFS (top-left), and 2DFS for the on-pulse phase region (top-right), leading part (bottom-left) and trailing part (bottom-right) of a mean pulse profile.
\label{subfig:fluctu:J1926+1631g}}
\end{figure}

\begin{figure}[htpb]
\centering
\includegraphics[width=0.44\textwidth, angle=0]{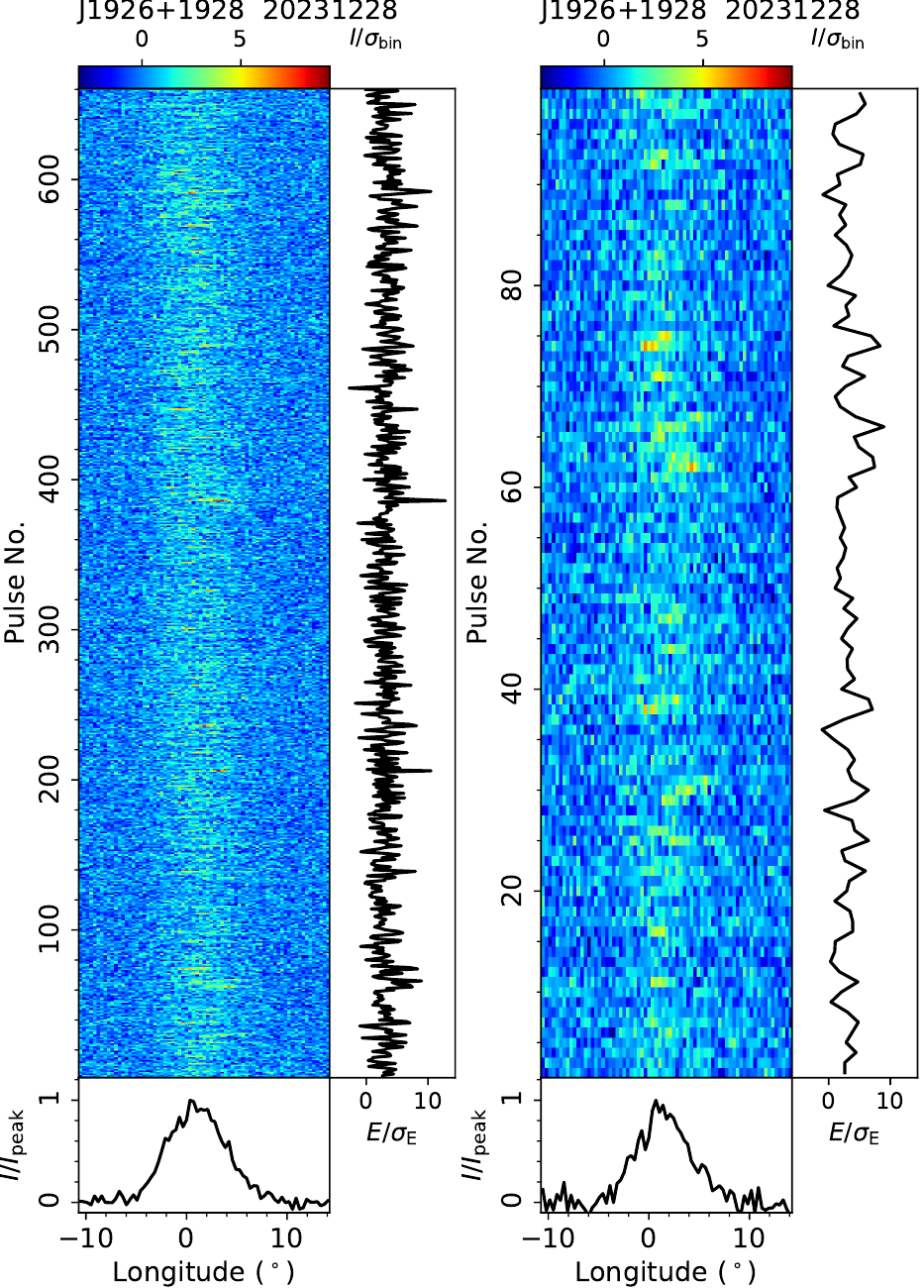}
\figcaption{Single pulse sequence of PSR J1926+1928 from the FAST observation on 20231228, and a zoomed-in view of pulses No. 1-100. 
\label{subfig:TP:J1926+1928}}
\end{figure}

\begin{figure}[htpb]
\centering
\includegraphics[width=0.44\textwidth, angle=0]{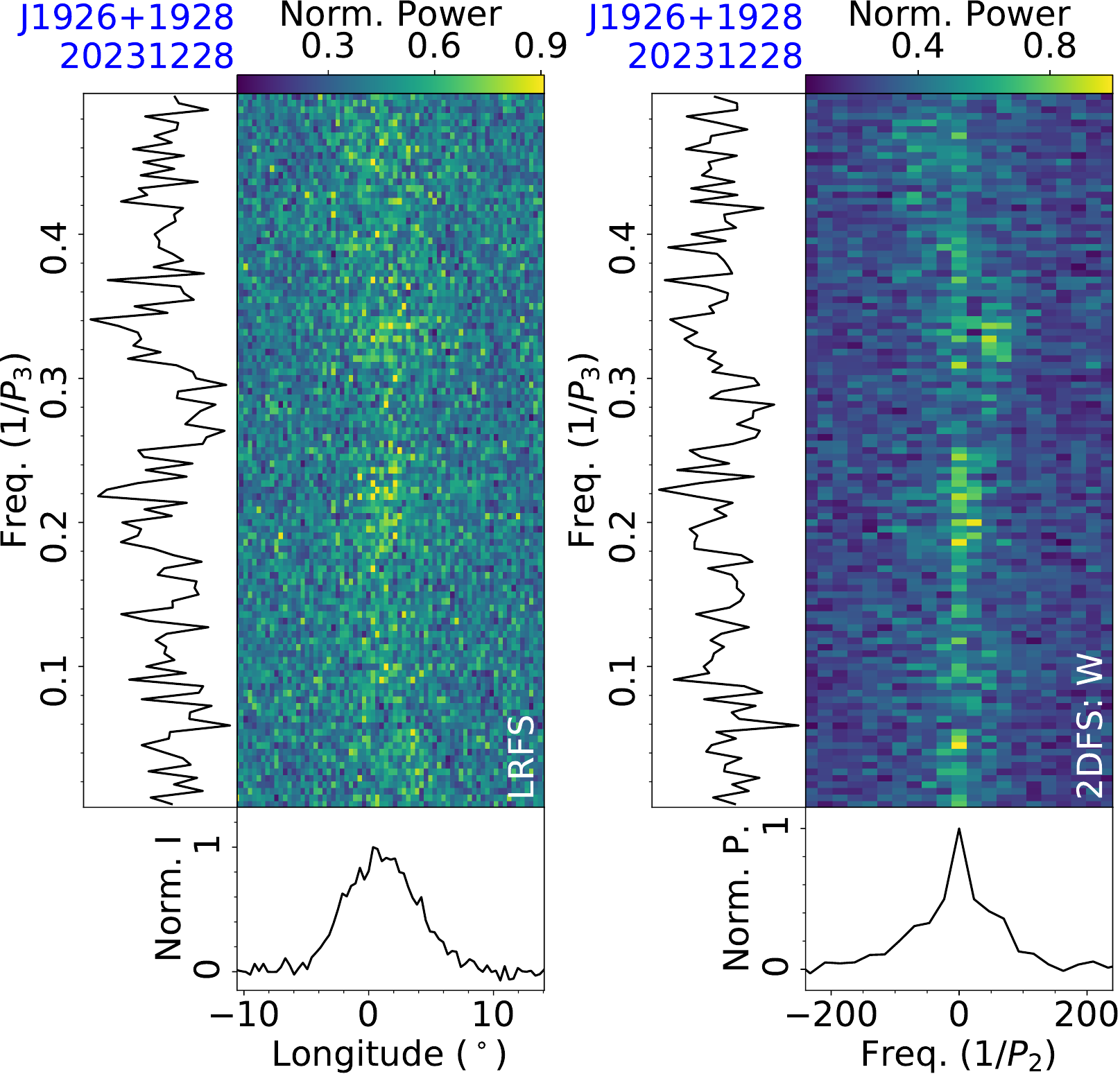}
\figcaption{Fluctuation analysis of PSR J1926+1928 for the observation on 20231228, with LRFS and 2DFS for the on-pulse phase region of the mean pulse profile.
\label{subfig:fluctu:J1926+1928}}
\end{figure}

\subsection{J1926+0431}
\label{subsec:J1926+0431}

The pulsar was discovered by \citet{Manchester1978} in the second Molonglo pulsar survey, which was reported to have nulling and drifting behaviors. Null fraction was reported to be less than 5\% at 430 MHz by \citet{Weisberg1986}. Drifting behavior has been detected by \citet{Weisberg1986} and \citet{Weltevrede2007}. 

The pulsar has also been observed by FAST on 20210702 for 5 minutes, deriving a rotation period $P=1.0740$~s and a dispersion measure $D\!M=102.7~{\rm cm^{-3}\,pc}$. 
The single pulse sequences are shown in Fig.~\ref{subfig:TP:J1926+0431}. We estimate the null fraction to be 0.7$\pm$0.2\% from the on-pulse energy histogram (Fig.~\ref{subfig:Hist:J1926+0431}). There are two modulation features in 2DFS (Fig.~\ref{subfig:fluctu:J1926+0431}), with broad high modulation frequency on time and opposite modulation frequency on phase, consistent with the result of \citet{Weltevrede2007}. In the 2DFS, the main drift feature is positive with the centroid of $1/P_3=0.388\pm0.005$ ($P_3=2.58\pm0.03$ periods) and $1/P_2=59\pm3$ ($P_2=6.1\pm0.3^\circ$), which has a wider temporal modulation frequency than the negative feature. The centroid frequencies of the negative phase modulated feature are $1/P_3=0.435\pm0.003$ and $1/P_2=-52\pm2$, corresponding to periodicities of $P_3=2.30\pm0.01$ periods and $P_2=-6.9\pm0.3^\circ$.

\begin{figure}[hbpt]
\centering
\includegraphics[width=0.21\textwidth, angle=0]{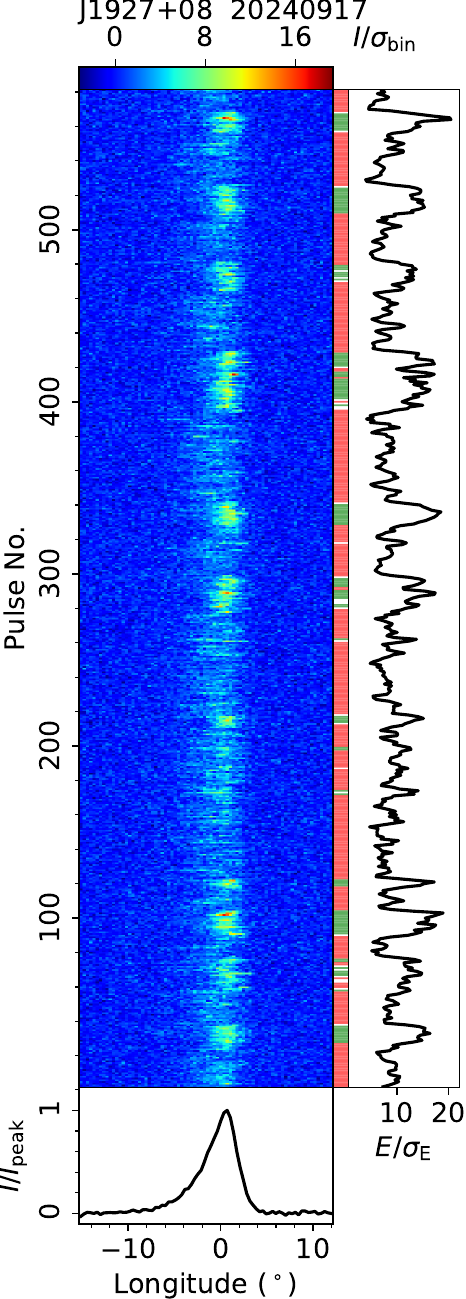}
\includegraphics[width=0.21\textwidth, angle=0]{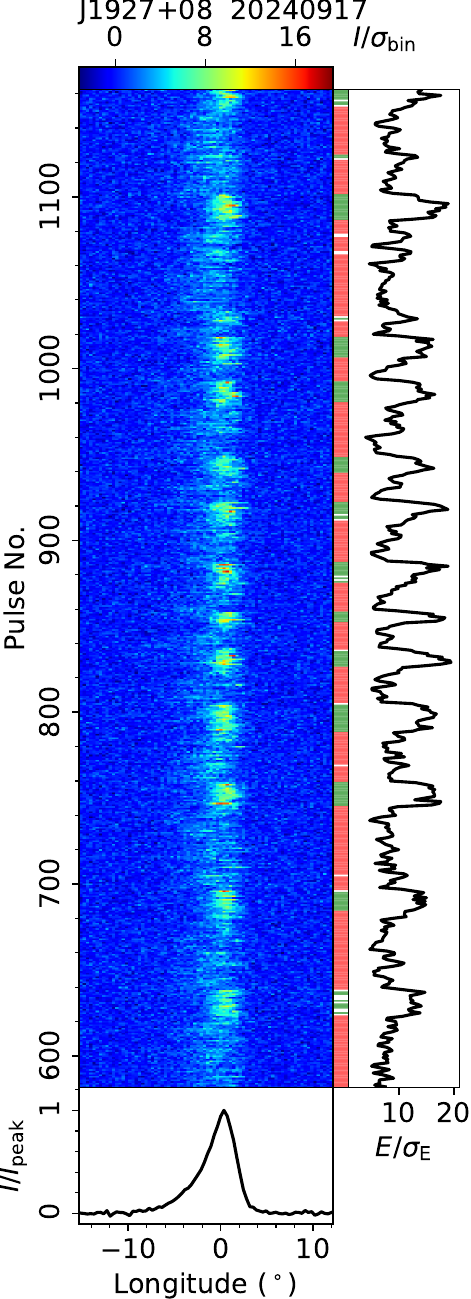}
\figcaption{Single pulse sequence of PSR J1927+08 from the FAST observation on 20210706. 
The green and red bars represent bright or weak emission modes. In the right subpanel, the on-pulse energy variation smoothed over every 3 periods is plotted against period.
\label{subfig:TP:J1927+08}}
\end{figure}

\begin{figure}[htpb]
\centering
\includegraphics[width=0.39\textwidth, angle=0]{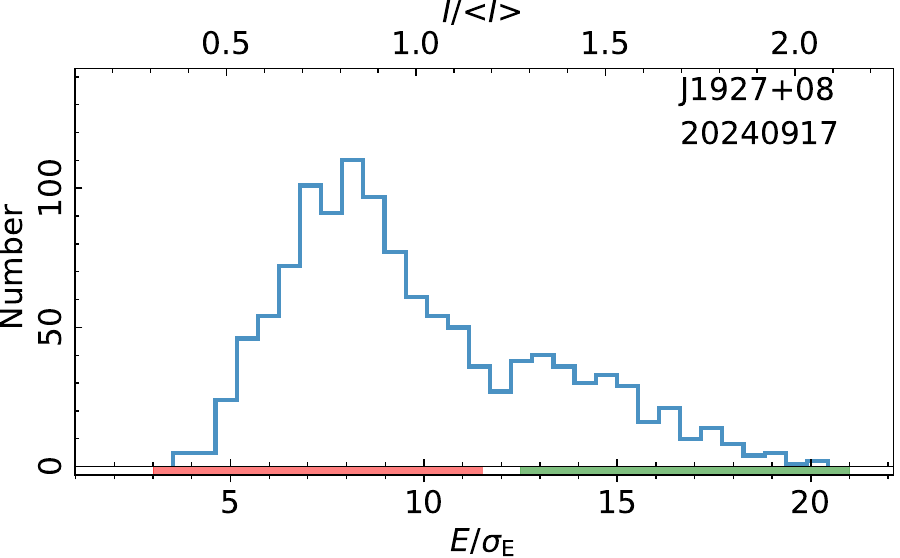}
\figcaption{On-pulse energy histogram of single pulses of PSR J1927+08 from the FAST observation on 20240917, with energy values smoothed over 3 periods. 
The red and green bars indicate the weak and bright emission modes.
\label{subfig:Hist:J1927+08}}
\end{figure}

\begin{figure}[hbpt]
\centering
\includegraphics[width=0.37\textwidth, angle=0]{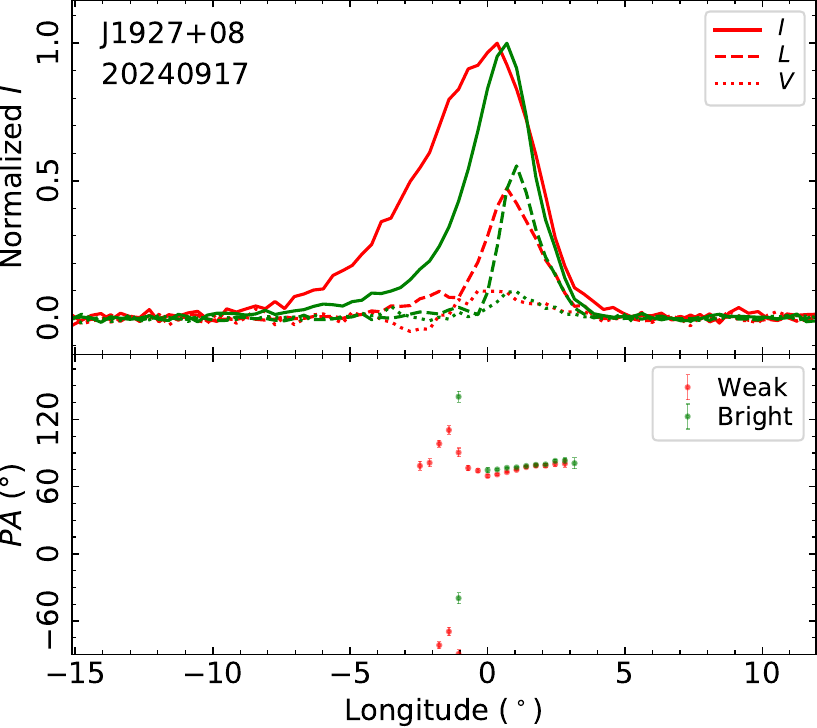}
\figcaption{Mean polarization profiles (the top panel) for the bright (green) and weak (red) emission modes of PSR J1927+08 observed on 20240917, as well as the averaged PA curves (the bottom panel). Profiles in the top panel are normalized by their respective peaks.
\label{subfig:PolModes:J1927+08}}
\end{figure}

\begin{figure}[htpb]
\centering
\includegraphics[width=0.22\textwidth, angle=0]{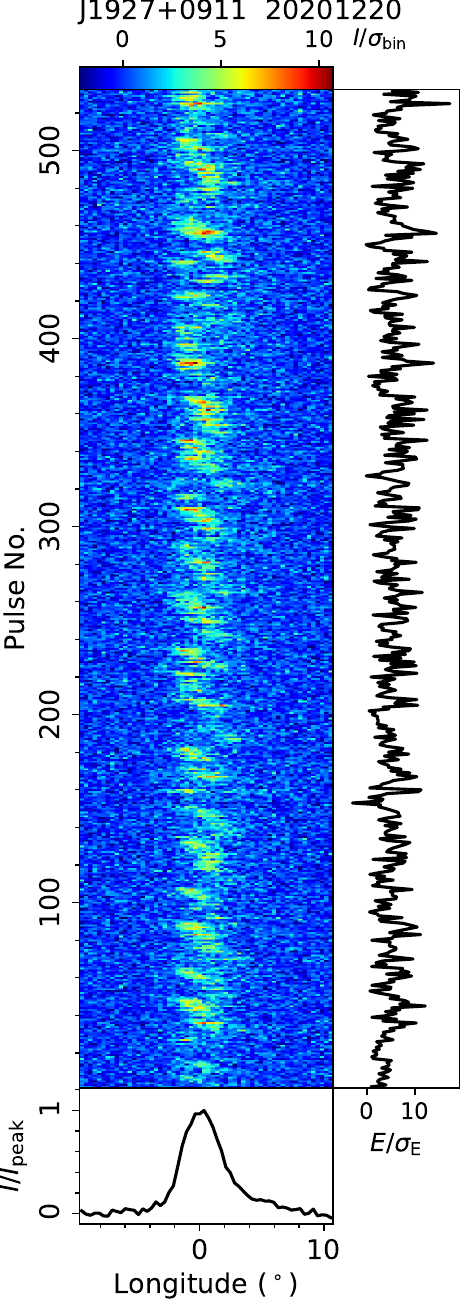}
\includegraphics[width=0.22\textwidth, angle=0]{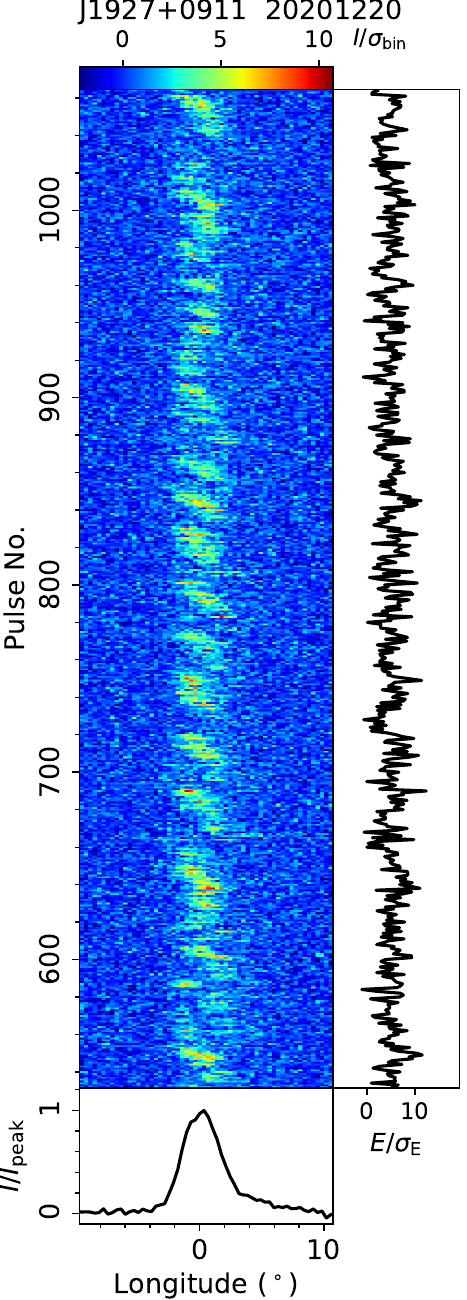}
\figcaption{Single pulse sequences of PSR J1927+0911 from the FAST observation on 20201220.
\label{subfig:TP:J1927+0911}}
\end{figure}

\begin{figure}[htpb]
\centering
\includegraphics[width=0.22\textwidth, angle=0]{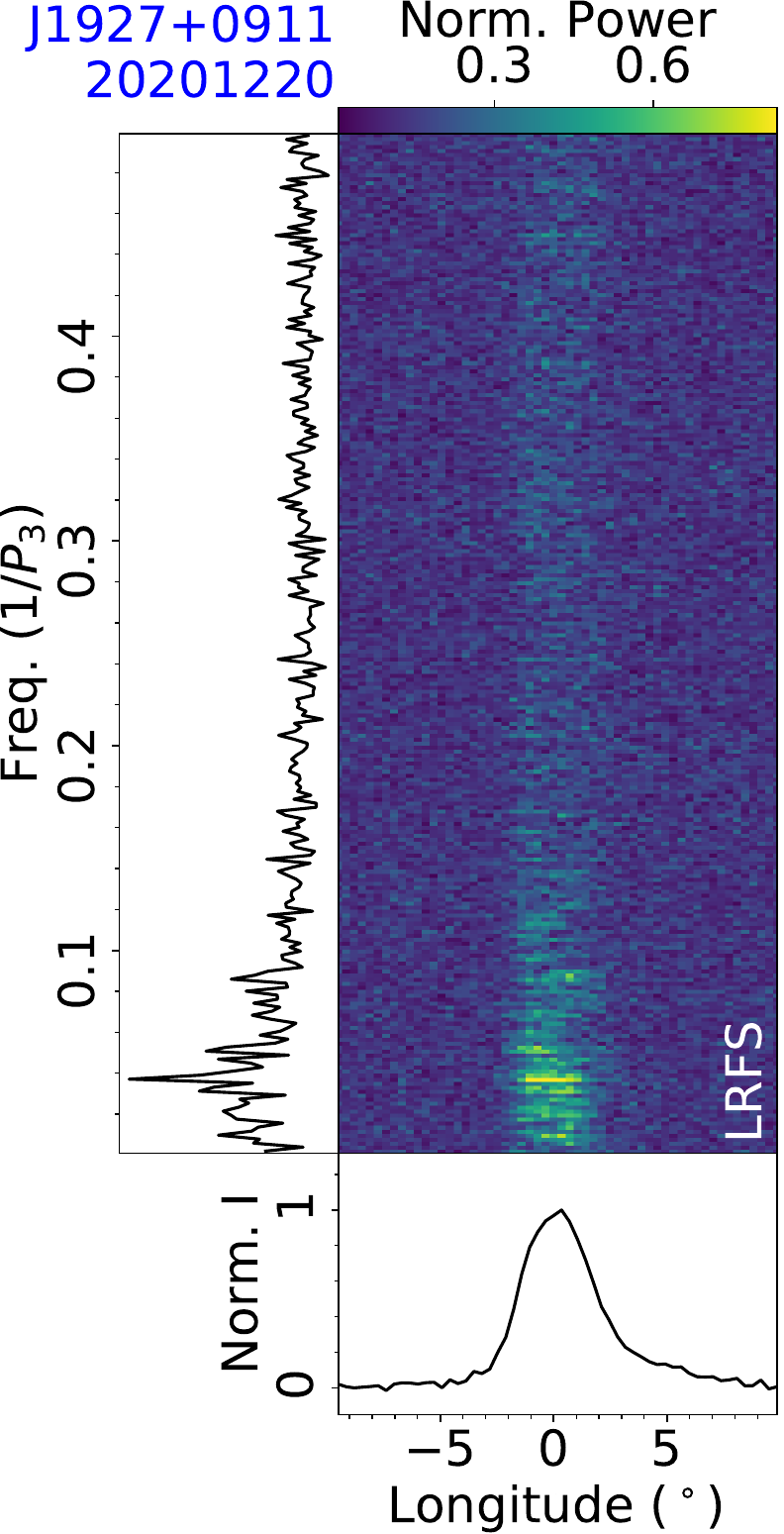}
\includegraphics[width=0.22\textwidth, angle=0]{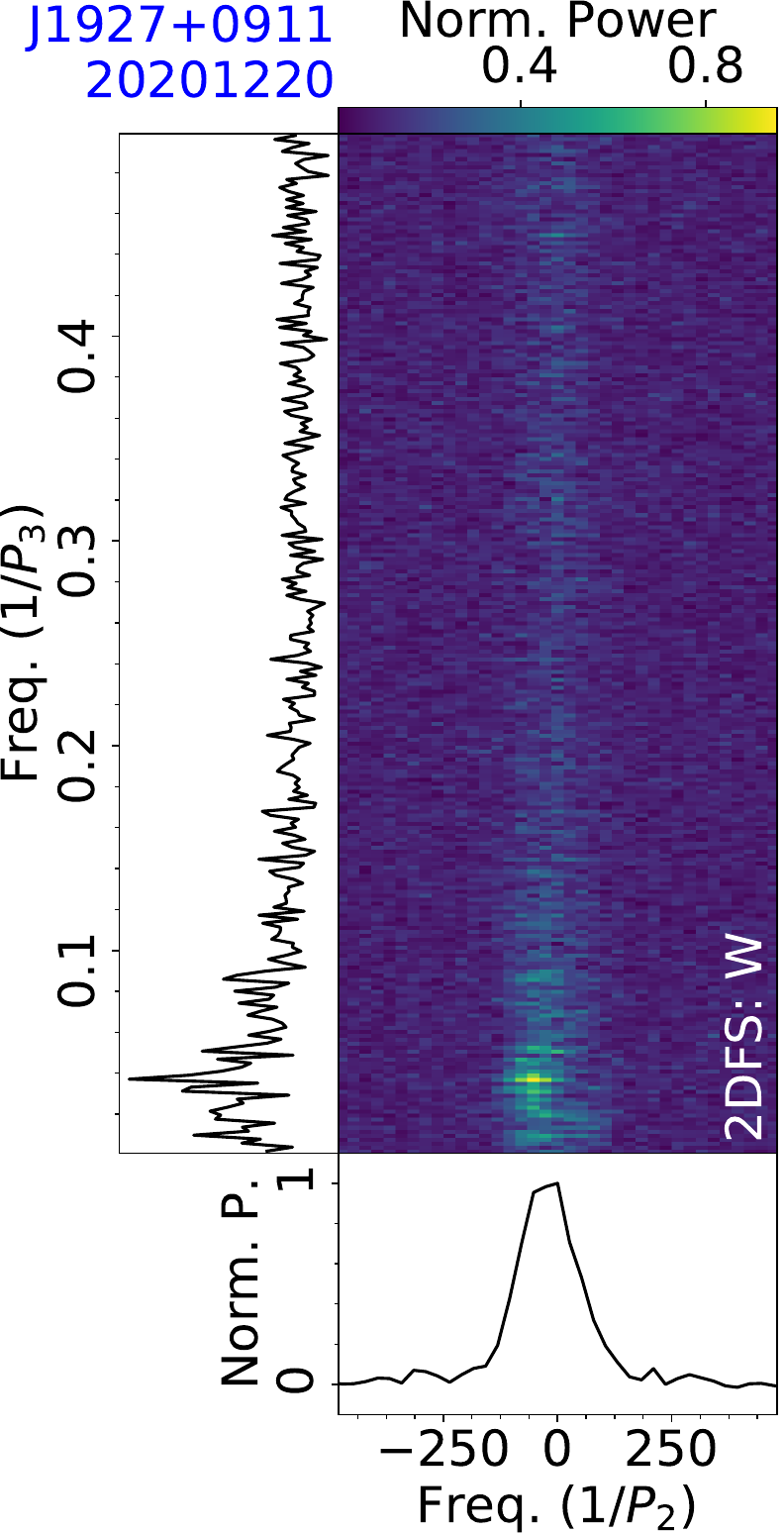}
\figcaption{Fluctuation analysis of PSR J1927+0911 for the observation on 20201220, with LRFS and 2DFS for the on-pulse region of a mean pulse profile.
\label{subfig:fluctu:J1927+0911}}
\end{figure}

\begin{figure}[htpb]
\centering
\includegraphics[width=0.22\textwidth, angle=0]{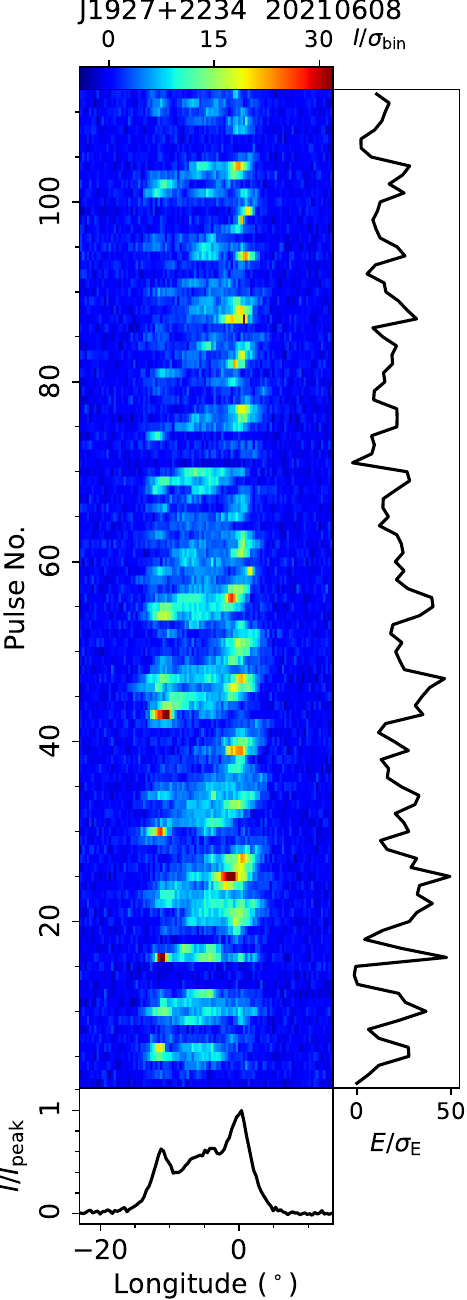}
\includegraphics[width=0.22\textwidth, angle=0]{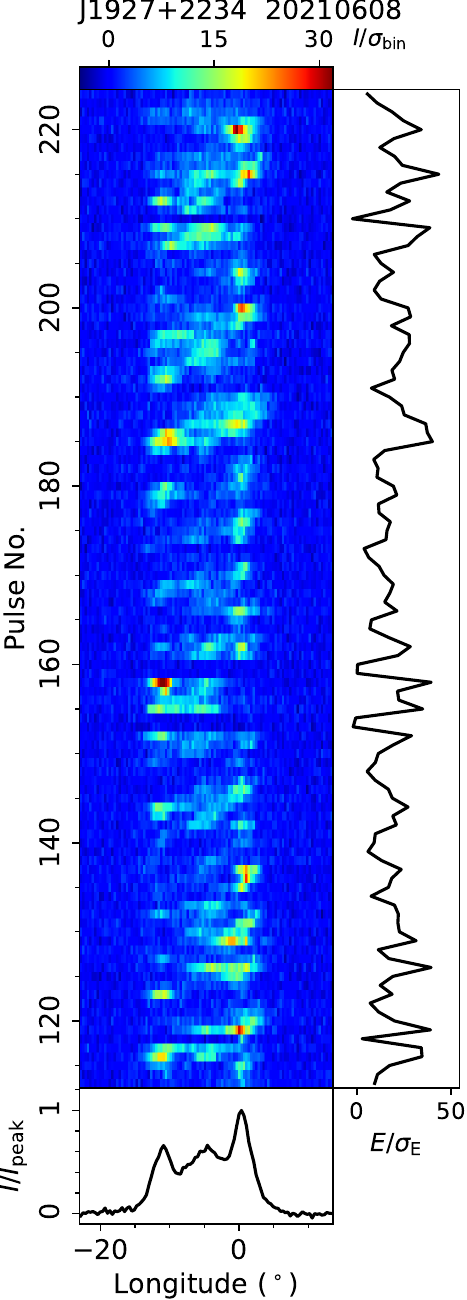}
\figcaption{Single pulse sequences of PSR J1927+2234 from the FAST observation on 20210608.
\label{subfig:TP:J1927+2234}}
\end{figure}

\begin{figure}[htpb]
\centering
\includegraphics[width=0.39\textwidth, angle=0]{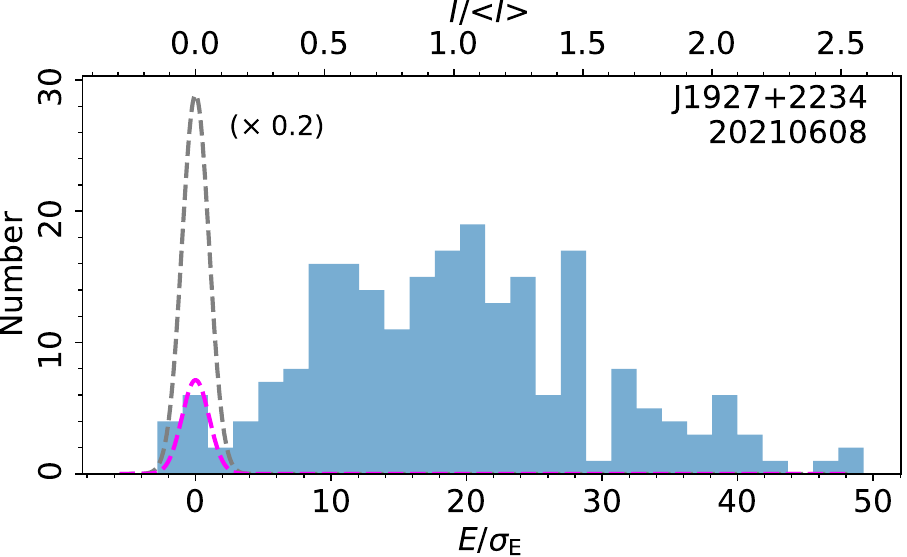}
\figcaption{On-pulse energy histogram of single pulses of PSR J1927+2234 from the FAST observation on 20210608.
\label{subfig:Hist:J1927+2234}}
%
\centering
\includegraphics[width=0.22\textwidth, angle=0]{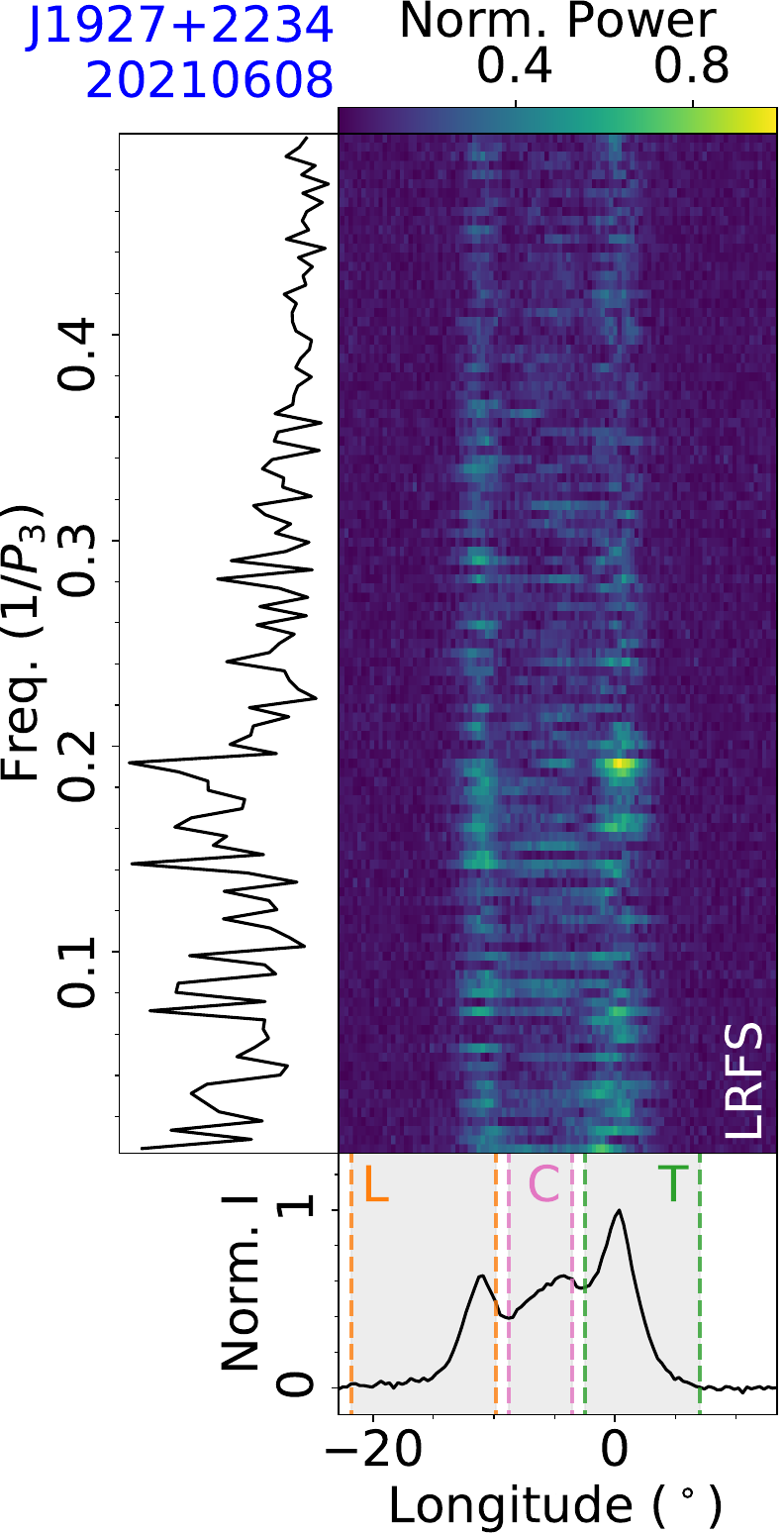}
\includegraphics[width=0.22\textwidth, angle=0]{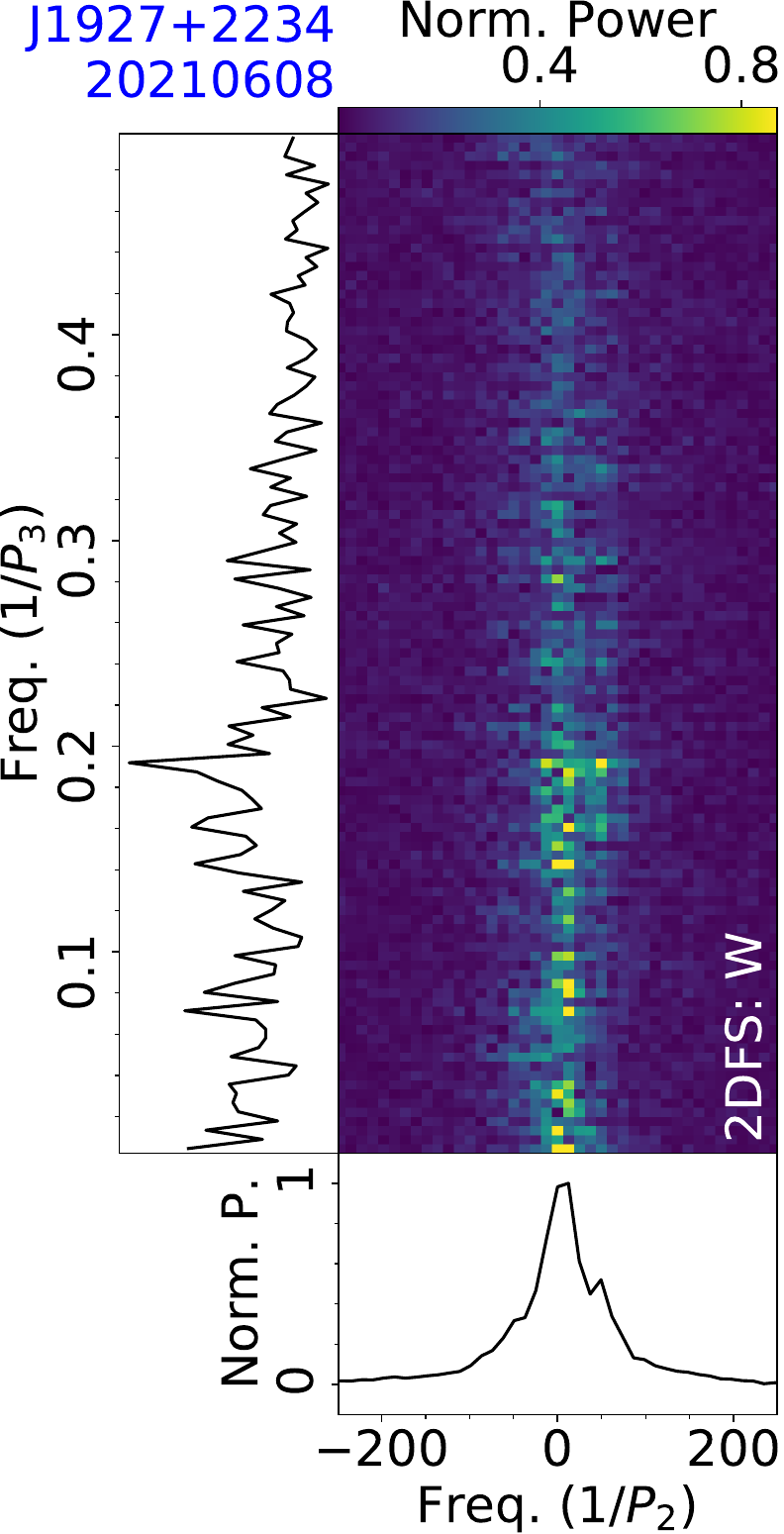}\\
\includegraphics[width=0.22\textwidth, angle=0]{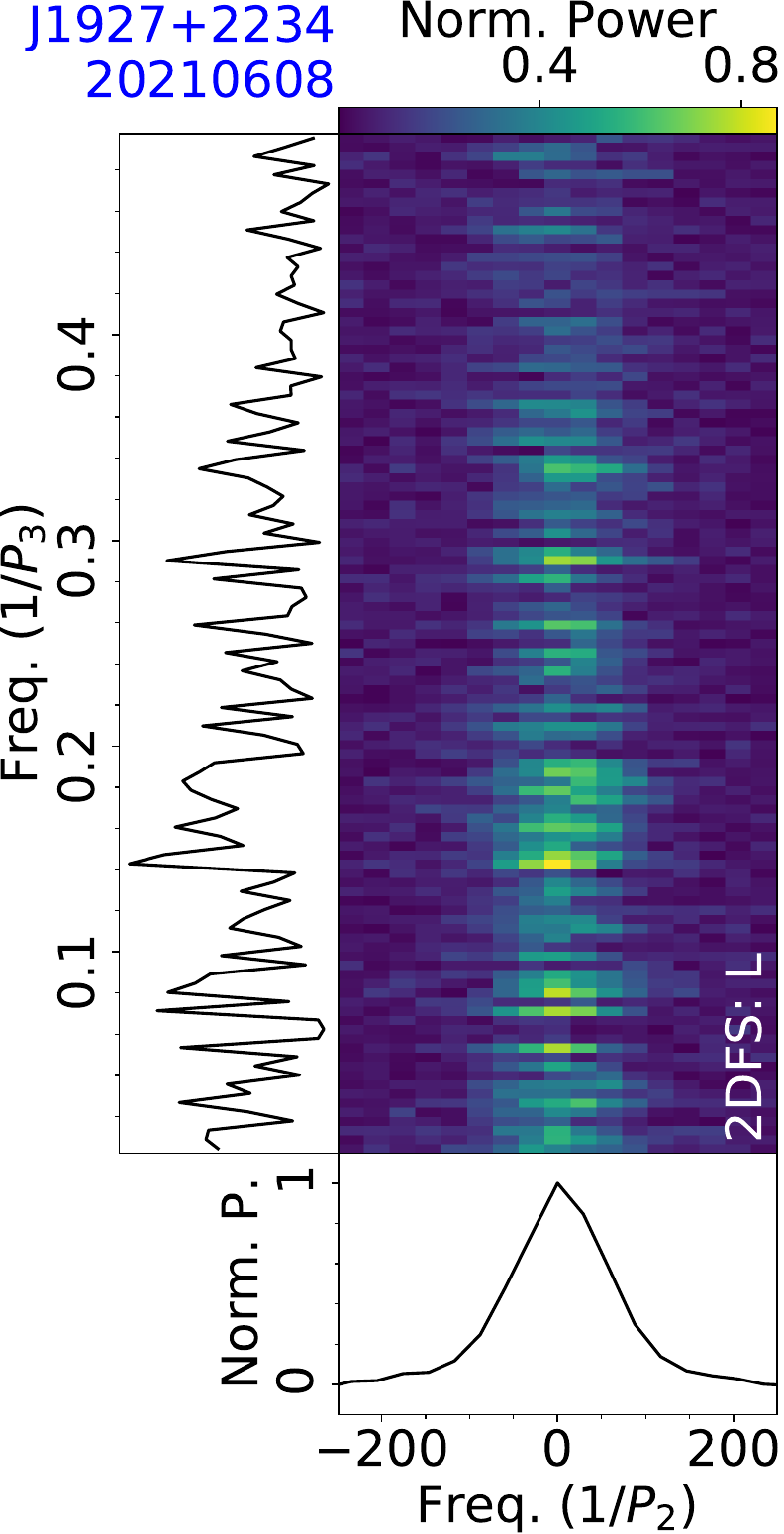}
\includegraphics[width=0.22\textwidth, angle=0]{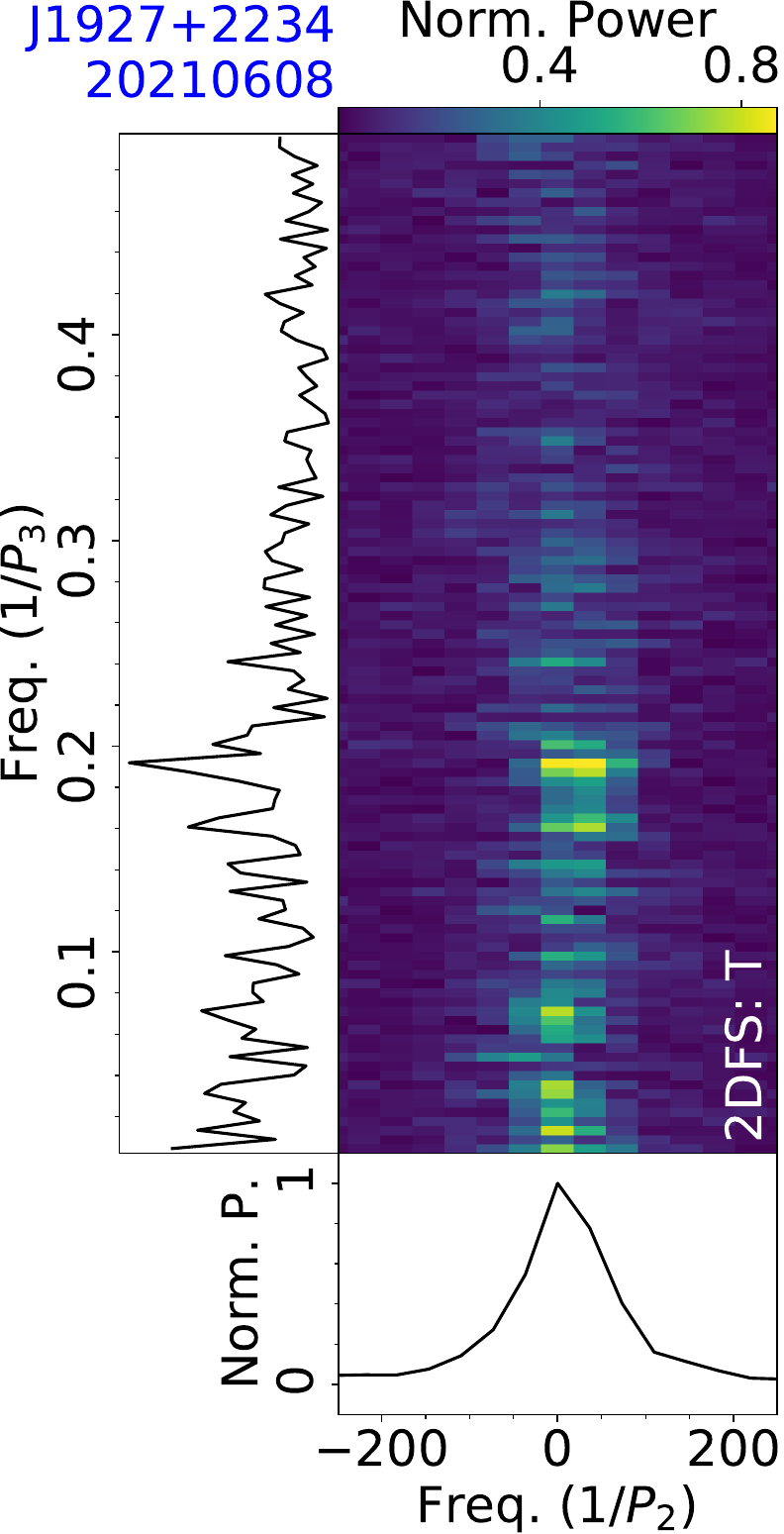}
\figcaption{Fluctuation analysis of PSR J1927+2234 for the observation on 20210608, with LRFS (top-left), and 2DFS for the on-pulse phase region (top-right), leading part (bottom-left) and trailing part (bottom-right) of a mean pulse profile.
\label{subfig:fluctu:J1927+2234}}
\end{figure}

\subsection{J1926+0737}
\label{subsec:J1926+0737}

PSR J1926+0737 was discovered from a re-analysis of the Parkes multi-beam pulsar survey using artificial neural networks \citep{Eatough2010}. 

This pulsar was observed by FAST on 20201226 for 5 minutes, deriving a rotation period $P=0.3181$~s and a dispersion measure $D\!M=159.0~{\rm cm^{-3}\,pc}$. 
Single pulse sequences of the main pulse and interpulse of this observation are shown in Fig.~\ref{subfig:TP:J1926+0737}, illustrating that both the main pulse (MP) and the interpulse (IP) have the mode changing phenomenon. 
Weak or bright emission states of single pulses of the main pulse and the interpulse are distinguished from the on-pulse energy histogram shown in Fig.~\ref{subfig:Hist:J1926+0737}. In plots of single pulse sequences and energy histograms, the weak mode and bright mode are labeled using red and green bars, respectively. 
To analyse the relationship of emission states between the main pulse and interpulse, 0-1 sequences representing the occurrences of weak or bright states are obtained. From the correlation of two occurrence sequences shown in Fig.~\ref{subfig:ModesCorr:J1926+0737}, mode changes between main pulse and interpulse are relevant, and their emission states are anti-correlated. 
For both main pulse and interpulse, the prominent difference between the two modes is in intensity, with averaged profiles of the two emission modes displayed in Fig.~\ref{subfig:ProfModes:J1926+0737}.

\subsection{J1926+1434}
\label{subsec:J1926+1434}

PSR J1926+1434 was discovered by the Arecibo telescope \citep{Hulse1974}. Drifting parameters were presented by \citet{Song2023}: $P_3=17\pm11$ periods and $P_2=-100^{+52}_{-28}$ degrees for the leading component, and $P_3=23\pm4$ periods and $P_2=-149^{+69}_{-22}$ degrees for the trailing component. 

This pulsar was observed by FAST on 20210317 for 15 minutes, driving a rotation period $P=1.3248$~s and a dispersion measure $D\!M=211.9~{\rm cm^{-3}\,pc}$. 
Single pulse sequences in Fig.~\ref{subfig:TP:J1926+1434} show nulling and unsystematic subpulse drifting phenomena. 
The nulling fraction of this observation is estimated to be 0.6$\pm$0.3\% from the on-pulse energy histogram in Fig.~\ref{subfig:Hist:J1926+1434}. 
Fluctuation spectra are shown in Fig.~\ref{subfig:fluctu:J1926+1434}. 
In the 2DFS of the leading component of the mean pulse profile, two negative drift features are present: one with centroid frequencies $1/P_3=0.026\pm0.001$ ($P_3=38\pm1$ periods) and $1/P_2=-6\pm1$ ($P_2=-57\pm13$ degrees), and the other with $1/P_3=0.134\pm0.001$ ($P_3=7.47\pm0.04$ periods) and $1/P_2=-10\pm1$ ($P_2=-36\pm4$ degrees). 
2DFS of the trailing profile part exhibits a preferred negative drift feature, with the centroid frequencies of $1/P_3=0.035\pm0.001$ and $1/P_2=-4\pm1$, corresponding to periodicities of $P_3=28\pm1$ periods and $P_2=-86\pm22$ degrees.

\subsection{J1926+1631g}
\label{subsec:J1926+1631g}

PSR J1926+1631g was discovered in the FAST GPPS survey \citep{Han2021,han2025}. 

This pulsar was observed by FAST on 20211226 for 15 minutes and on 20250214 for 20 minutes. From the observation on 20211226, a rotation period $P=0.6784$~s and a dispersion measure $D\!M=196.7~{\rm cm^{-3}\,pc}$ were determined. 
Single-pulse features of two observations are consistent. Single pulse sequences of the observation on 20211226 in Fig.~\ref{subfig:TP:J1926+1631g} illustrate that the pulsar has subpulse drifting behavior. From 2DFS in Fig.~\ref{subfig:fluctu:J1926+1631g}, the centroid frequencies are $1/P_3=0.174\pm0.001$ ($P_3=5.76\pm0.02$ periods) and $1/P_2=-45\pm4$ ($P_2=-8\pm1^\circ$) for the leading part of a mean pulse profile, and $1/P_3=0.153\pm0.002$ ($P_3=6.6\pm0.1$ periods) and $1/P_2=-24\pm4$ ($P_2=-15\pm2^\circ$) for the trailing profile part.
In addition to the subpulse drifting, 2DFS of the leading component also has a low-frequency modulation feature with the centroid of $1/P_3=0.023\pm0.001$, yielding $P_3=43\pm2$ periods.

\subsection{J1926+1928}
\label{subsec:J1926+1928}

PSR J1926+1928 was discovered by \citet{Hulse1975} with the Arecibo telescope.

This pulsar was observed by FAST on 20190327 for 5 minutes and on 20231228 for 15 minutes. From the 15-minute data, a rotation period $P=1.3461$~s and a dispersion measure $D\!M=451.4~{\rm cm^{-3}\,pc}$ were derived. Single pulse sequence and a zoomed-in view of pulses No. 1-100 of the observation on 20231228 are shown in Fig.~\ref{subfig:TP:J1926+1928}. From the fluctuation spectra in Fig.~\ref{subfig:fluctu:J1926+1928}, the pulsar has a positive drifting behavior. The centroid frequencies of the positive drift feature in 2DFS are $1/P_3=0.325\pm0.001$ and $1/P_2=54\pm2$, corresponding to periodicities of $P_3=3.08\pm0.01$ periods and $P_2=6.6\pm0.2$ degrees.

\subsection{J1927+08}
\label{subsec:J1927+08}

PSR J1927+08 was discovered in the Pulsar Arecibo L-band Feed Array (PALFA) survey \citep{Parent2022}.

This pulsar was observed by FAST on 20240917 for 5 minutes, and a rotation period $P=0.2534$~s and a dispersion measure $D\!M=223.8~{\rm cm^{-3}\,pc}$ were derived. Single pulse sequences in Fig.~\ref{subfig:TP:J1927+08} display switches between the weak and bright emission modes, and the on-pulse integral energies versus period are smoothed over every 3 periods. The weak and bright emission modes of single pulses are distinguished from the histogram of smoothed energies (Fig.~\ref{subfig:Hist:J1927+08}), which are labeled in red and green. Mean polarization profiles and the averaged PA curves of two modes are shown in Fig.~\ref{subfig:PolModes:J1927+08}, and the weak emission mode has a wider profile.

\begin{figure}[htpb]
\centering
\includegraphics[width=0.22\textwidth, angle=0]{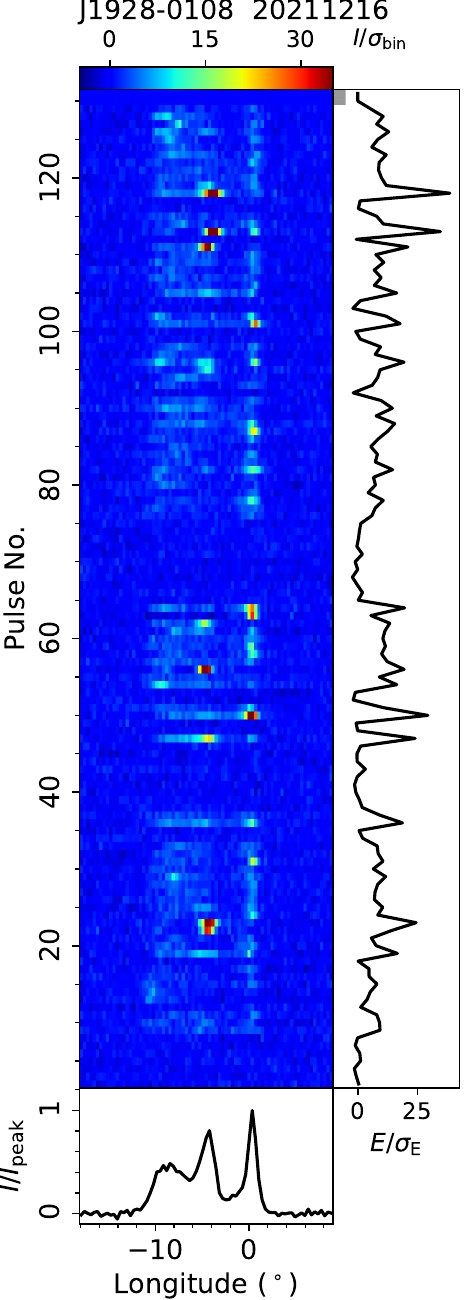}
\figcaption{Single pulse sequence of PSR J1928-0108 from the FAST observation on 20211216. 
\label{subfig:TP:J1928-0108}}
\end{figure}

\begin{figure}[htpb]
\centering
\includegraphics[width=0.39\textwidth, angle=0]{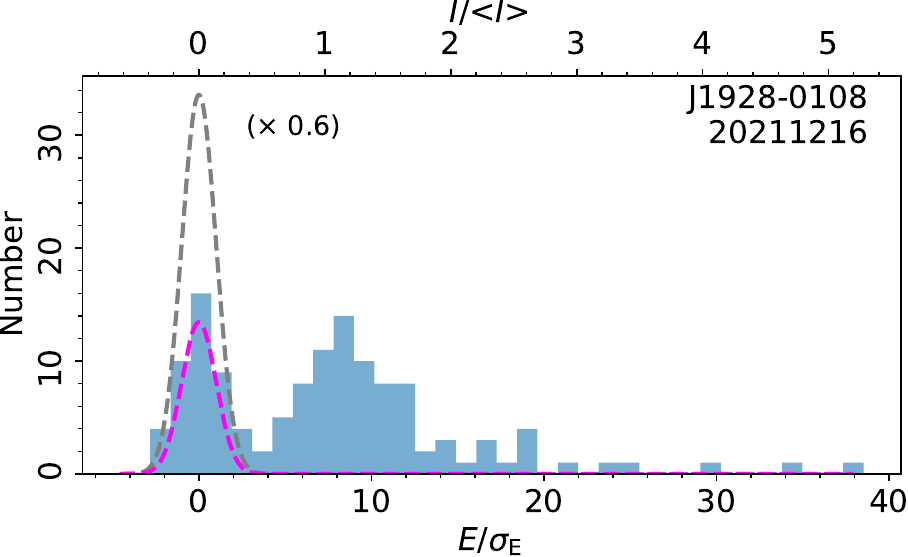}
\figcaption{On-pulse energy histogram of single pulses of PSR J1928-0108 from the FAST observation on 20211216.
\label{subfig:Hist:J1928-0108}}
\end{figure}

\begin{figure}[htpb]
\centering
\includegraphics[width=0.22\textwidth, angle=0]{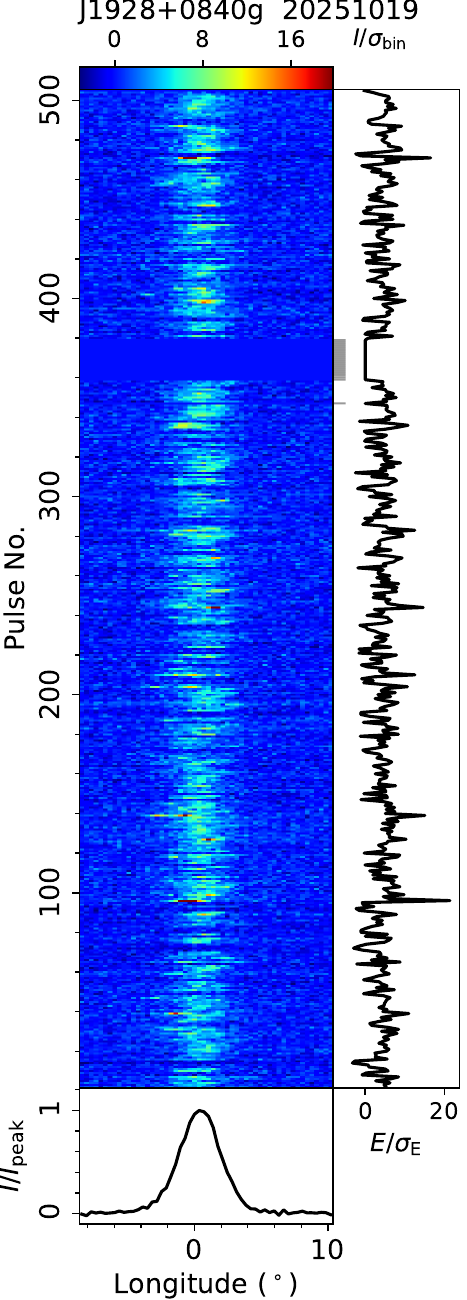}
\includegraphics[width=0.22\textwidth, angle=0]{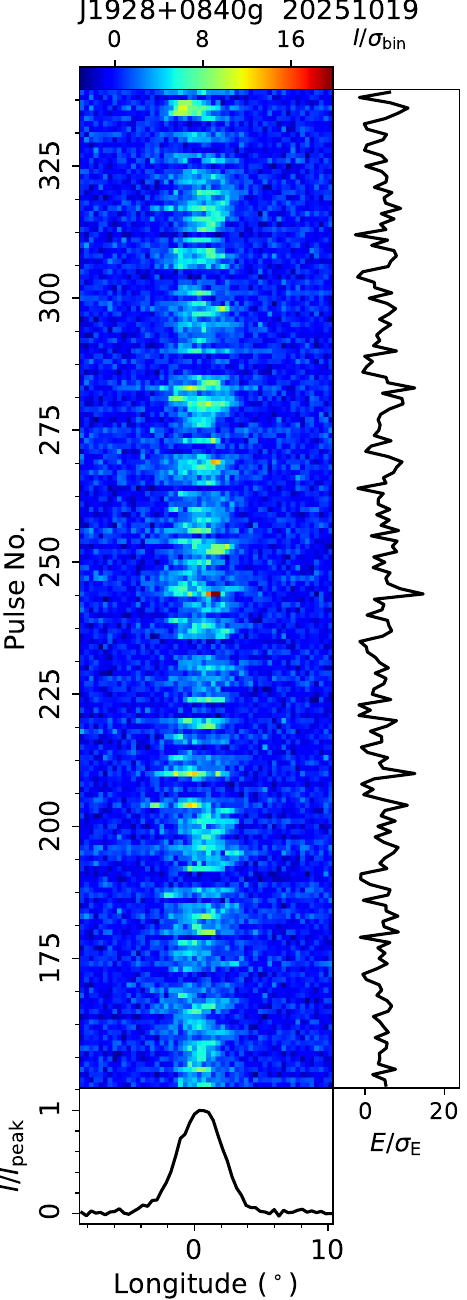}
\figcaption{Single pulse sequence of PSR J1928+0840g from the FAST observation on 20251019, and a zoomed-in view of pulses No. 150-340.
\label{subfig:TP:J1928+0840g}}
\end{figure}

\begin{figure}[htpb]
\centering
\includegraphics[width=0.39\textwidth, angle=0]{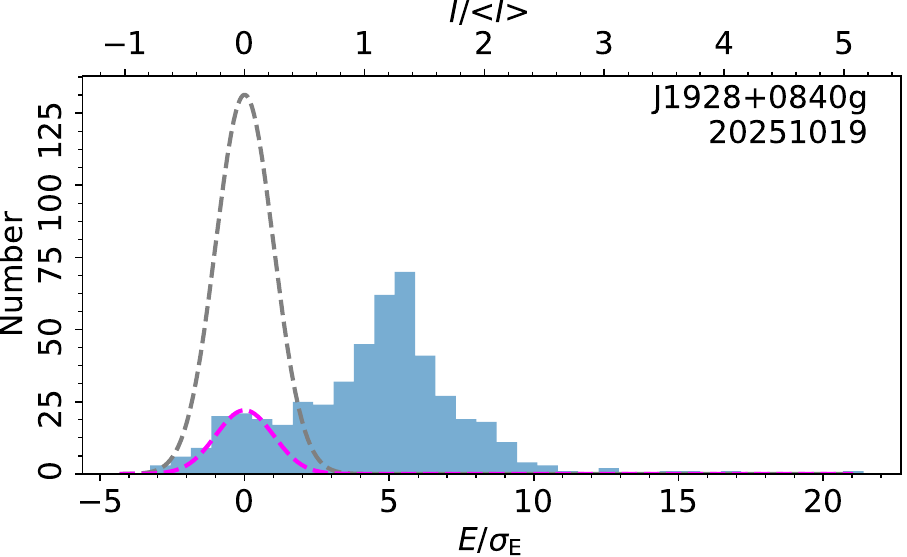}
\figcaption{On-pulse energy histogram of single pulses of PSR J1928+0840g from the FAST observation on 20251019.
\label{subfig:Hist:J1928+0840g}}
\end{figure}

\subsection{J1927+0911}
\label{subsec:J1927+0911}

PSR J1927+0911 was discovered in the Parkes 20-cm multibeam pulsar survey of the Galactic plane \citep{Lorimer2006}. The negative subpulse drifting was reported by \citet{Song2023} with $P_3$=23$\pm$5 periods and $P_2$=-7.0$^{+0.6}_{-5}$ degrees. 

This pulsar was observed by FAST on 20201220 for 5 minutes, deriving a rotation period $P=0.2903$~s and a dispersion measure $D\!M=202.7~{\rm cm^{-3}\,pc}$. 
Single pulse sequences in Fig.~\ref{subfig:TP:J1927+0911} display drifting bands. 
Fluctuation spectra shown in Fig.~\ref{subfig:fluctu:J1927+0911} exhibit a negative drift feature, with the centroid frequencies of $1/P_3=0.042\pm0.001$ and $1/P_2=-51\pm1$, that correspond to $P_3=23.6\pm0.3$ periods and $P_2=-7.1\pm0.1^\circ$. 
From the single pulse sequences in Fig.~\ref{subfig:TP:J1927+0911}, single pulses around pulses No. 440 and 920 are in a different drift state from the normal state.

\begin{figure}[htpb]
\centering
\includegraphics[width=0.22\textwidth, angle=0]{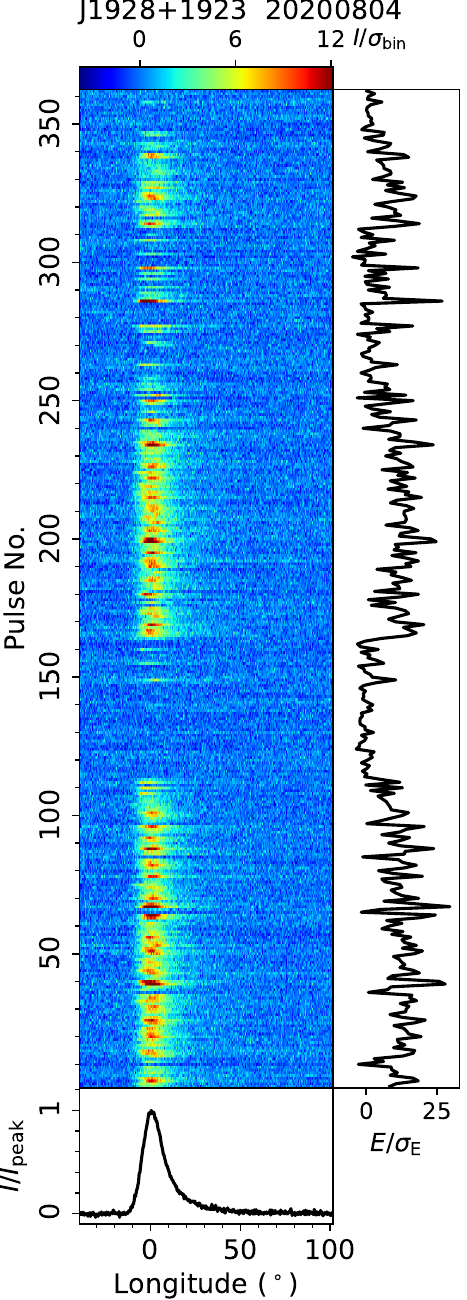}
\figcaption{Single pulse sequence of PSR J1928+1923 from the FAST observation on 20200804.
\label{subfig:TP:J1928+1923}}
\end{figure}

\begin{figure}[htpb]
\centering
\includegraphics[width=0.39\textwidth, angle=0]{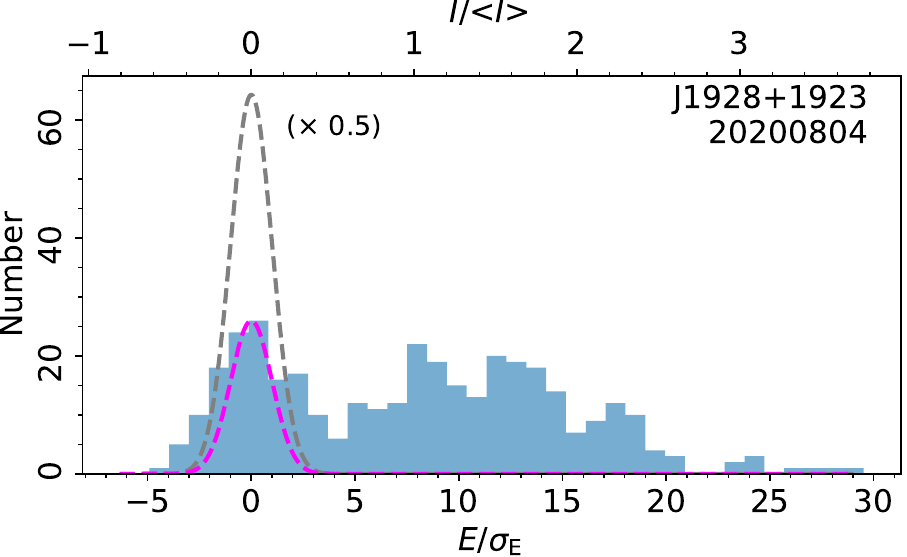}
\figcaption{On-pulse energy histogram of single pulses of PSR J1928+1923 from the FAST observation on 20200804.
\label{subfig:Hist:J1928+1923}}
\end{figure}

\subsection{J1927+2234}
\label{subsec:J1927+2234}

PSR J1927+2234 was found by \citet{Hulse1975} using the Arecibo telescope. \citep{Song2023} reported two and one drift features of the first and second components, respectively. 

This pulsar was observed by FAST on 20210608 for 5 minutes, deriving a rotation period $P=1.4309$~s and a dispersion measure $D\!M=185.7~{\rm cm^{-3}\,pc}$. 
Single pulse sequences are shown in Fig.~\ref{subfig:TP:J1927+2234}, illustrating the existence of nulling and subpulse drifting behaviors. From the on-pulse integral energy histogram (Fig.~\ref{subfig:Hist:J1927+2234}), the nulling fraction of this 5-minute observation is estimated to be 5$\pm$1\%. Fluctuation spectra of the leading and trailing parts in a mean pulse profile are displayed in Fig.~\ref{subfig:fluctu:J1927+2234}. 
2DFS of the leading profile part exhibits a positive drift feature, with the centroid frequencies estimated to be $1/P_3=0.166\pm0.001$ and $1/P_2=14\pm2$, that correspond to periodicities of $P_3=6.03\pm0.03$ periods and $P_2=26\pm4^\circ$. For 2DFS of the trailing part, the centroid of the drift feature is characterized by $1/P_3=0.182\pm0.001$ and $1/P_2=32\pm3$, yielding $P_3=5.51\pm0.02$ periods and $P_2=11\pm1^\circ$.

\subsection{J1928-0108}
\label{subsec:J1928-0108}

PSR J1928-0108 was discovered by \citet{Spiewak2020} using the Parkes Radio Telescope. 

The pulsar was observed by FAST on 20211216 for 5 minutes, deriving a rotation period $P=2.3660$~s and a dispersion measure $D\!M=124.1~{\rm cm^{-3}\,pc}$. 
From the single pulse sequence in Fig.~\ref{subfig:TP:J1928-0108} and the on-pulse integral energy histogram in Fig.~\ref{subfig:Hist:J1928-0108}, the pulsar has the nulling phenomenon, with a fraction of this observation estimated to be 24$\pm$4\%. Additionally, there are also bright subpulses sometimes appearing in the phase range of the central component.

\subsection{J1928+0840g}
\label{subsec:J1928+0840g}

PSR J1928+0840g was discovered in the FAST GPPS survey \citep{Han2021,han2025}.

This pulsar was observed by FAST on 20250810 for 5 minutes and 20251019 for 15 minutes. From the 15-minute data, a rotation period $P=1.7835$~s and a dispersion measure $D\!M=121.7~{\rm cm^{-3}\,pc}$ were determined. The single pulse sequence and a zoomed-in view of pulses No. 150-340 are shown in Figure~\ref{subfig:TP:J1928+0840g}. The distribution around zero energy in the on-pulse integral energy histogram (Fig.~\ref{subfig:Hist:J1928+0840g}) indicates the existence of nulls. The nulling fraction of this observation is estimated to be 16.8$\pm$1.1\%. 
More observations are necessary for further analysis of the nulling behavior.

\begin{figure}[htpb]
\centering
\includegraphics[width=0.44\textwidth, angle=0]{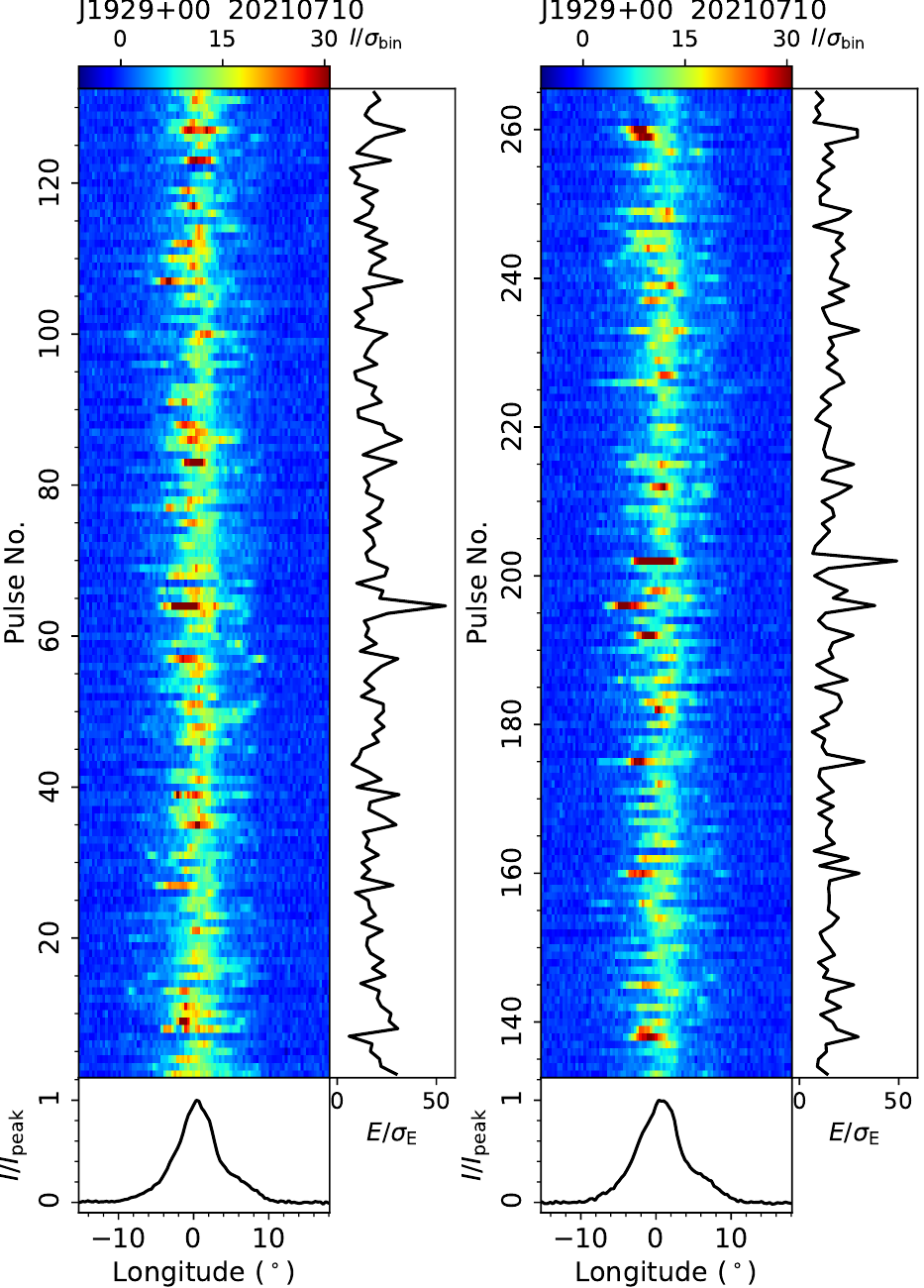}
\figcaption{Single pulse sequences of PSR J1929+00 from the FAST observation on 20210710.
\label{subfig:TP:J1929+00}}
\end{figure}

\begin{figure}[htpb]
\centering
\includegraphics[width=0.44\textwidth, angle=0]{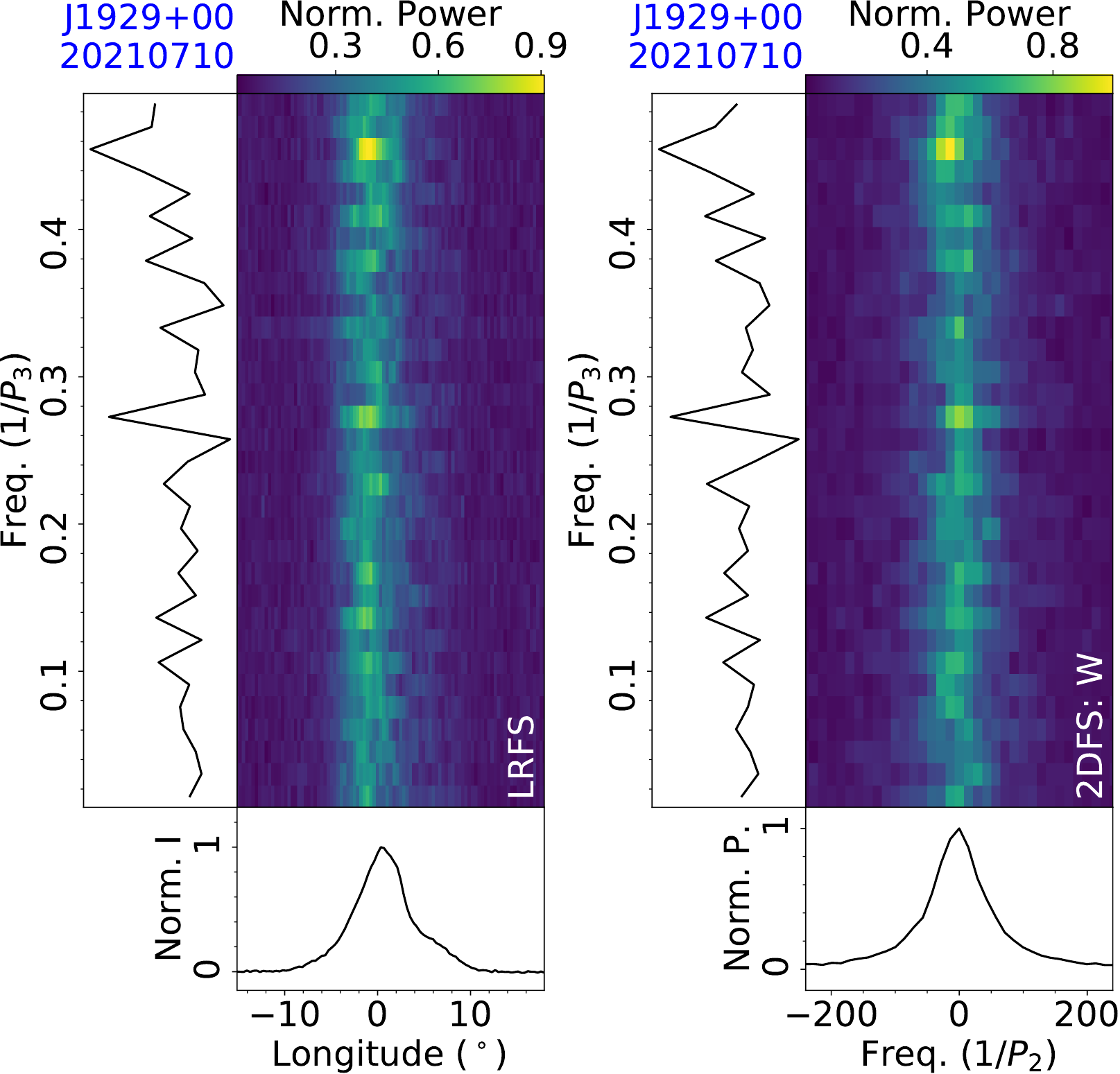}
\figcaption{Fluctuation analysis of PSR J1929+00 for the observation on 20210710, with LRFS and 2DFS for the on-pulse region of a mean pulse profile.
\label{subfig:fluctu:J1929+00}}
\end{figure}

\begin{figure}[htpb]
\centering
\includegraphics[width=0.22\textwidth, angle=0]{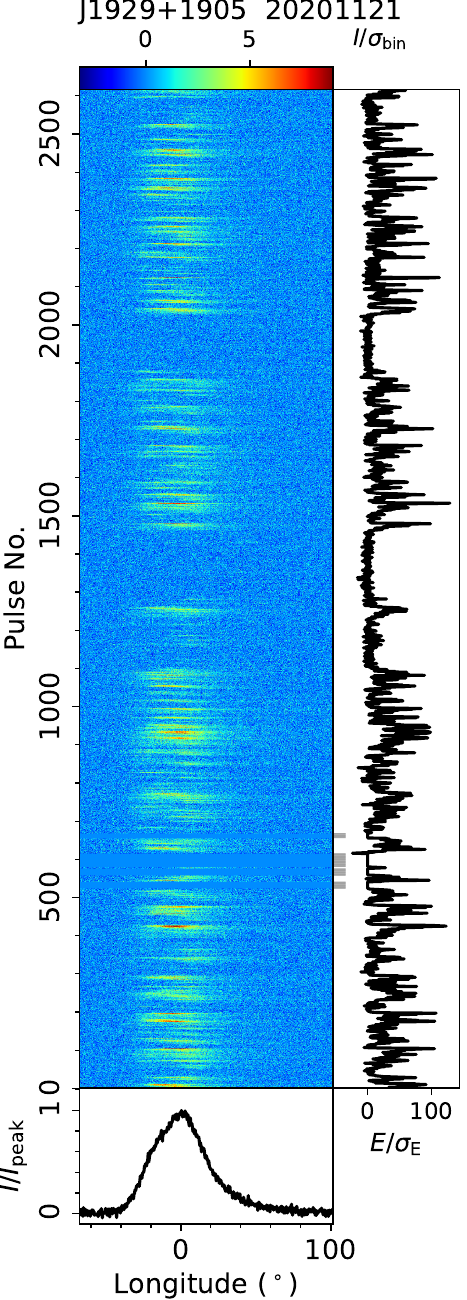}
\includegraphics[width=0.22\textwidth, angle=0]{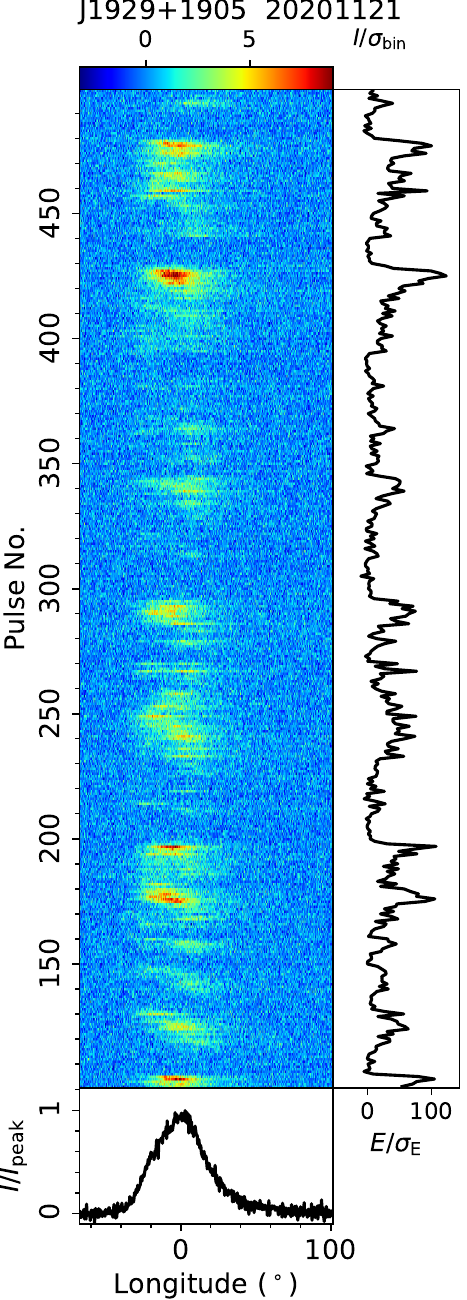}
\figcaption{Single pulse sequence of PSR J1929+1905 from the FAST observation on 20201121, and three zoomed-in views of pulses No. 100-500, 1200-1600 and 1800-2200. -- to be continued.
\label{subfig:TP:J1929+1905}}
\end{figure}

\begin{figure}[htpb]
\centering
\includegraphics[width=0.22\textwidth, angle=0]{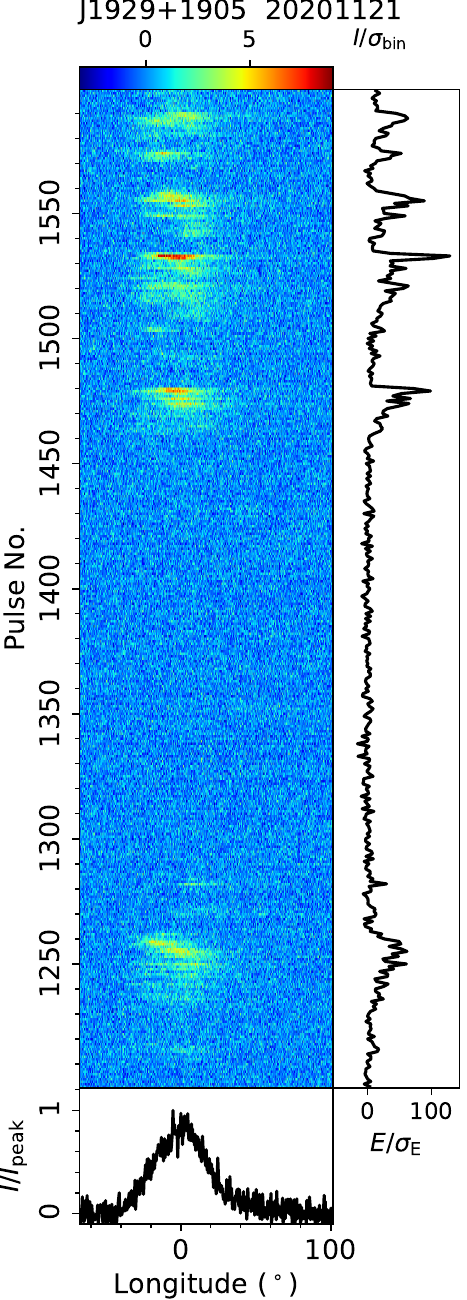}
\includegraphics[width=0.22\textwidth, angle=0]{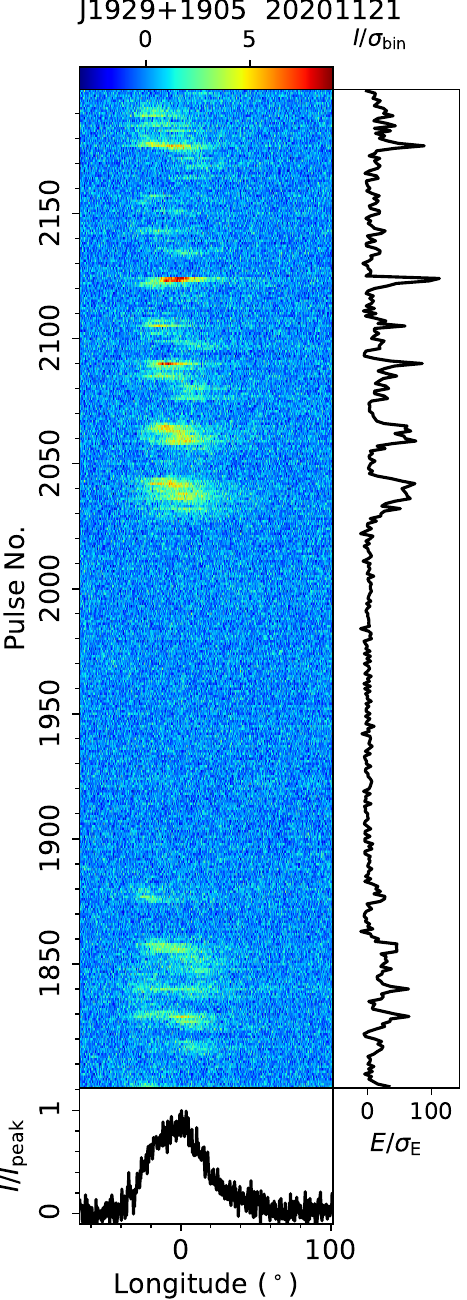}
\figcaption{Continued.}
\end{figure}


\begin{figure}[htpb]
\centering
\includegraphics[width=0.22\textwidth, angle=0]{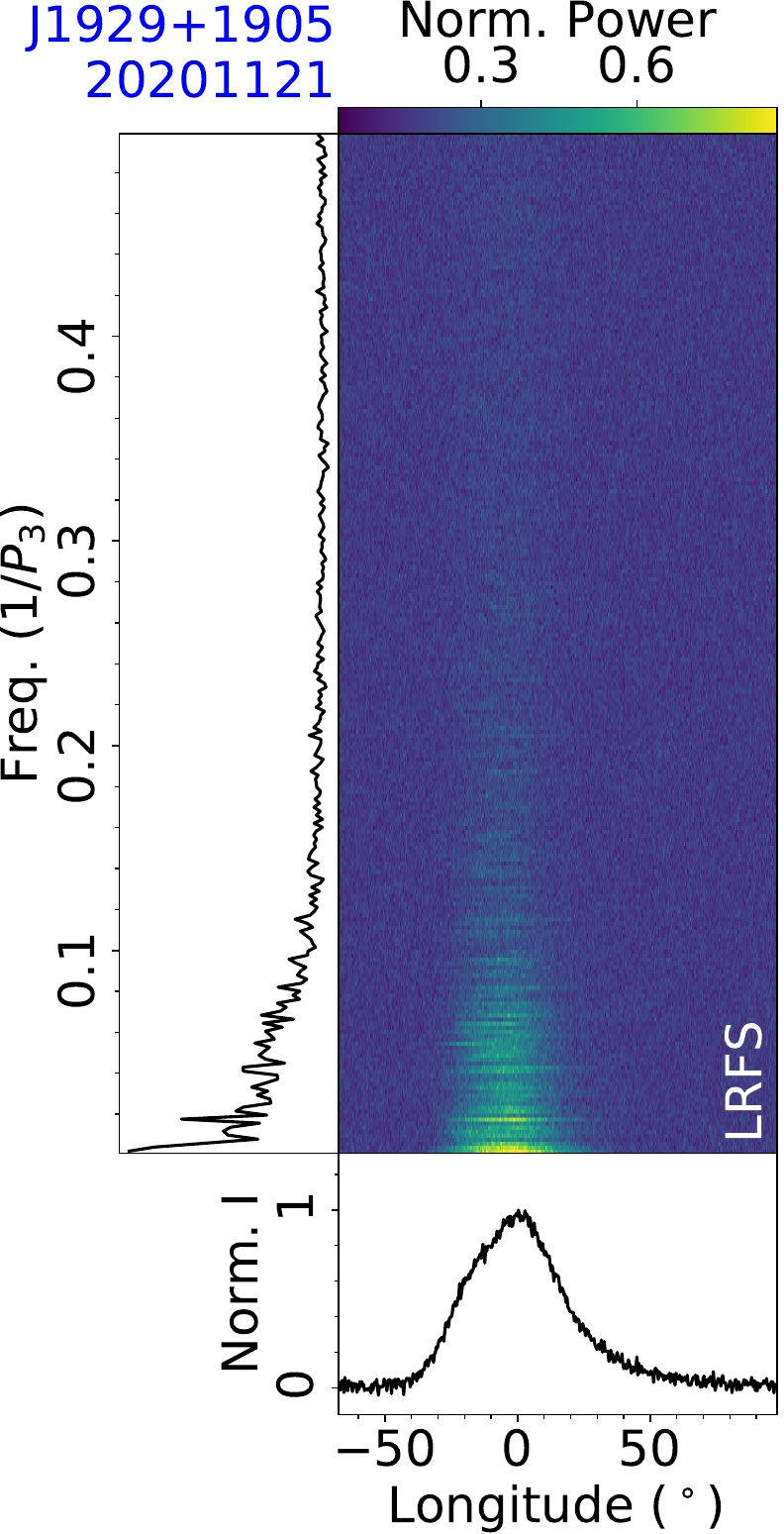}
\includegraphics[width=0.22\textwidth, angle=0]{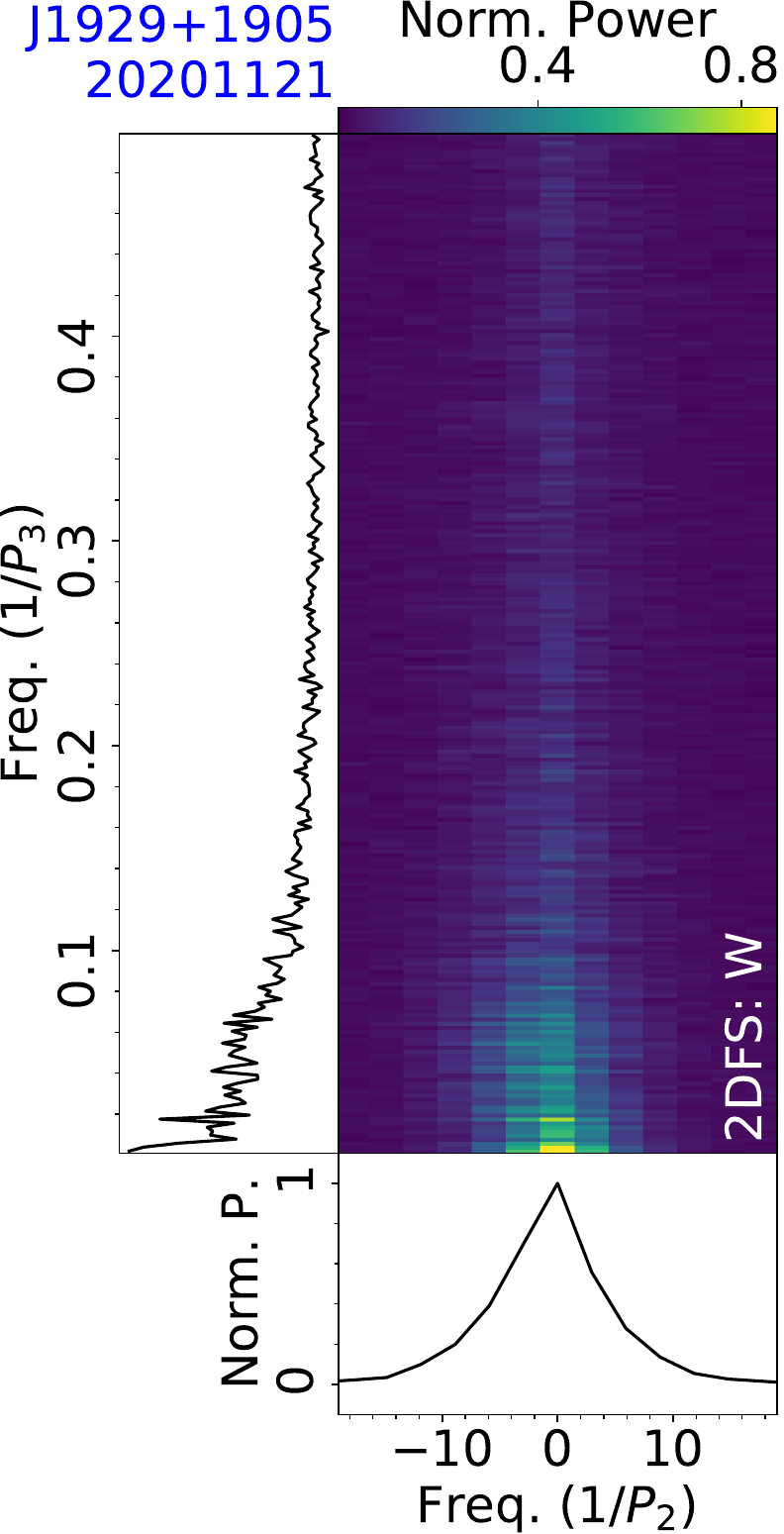}
\figcaption{Fluctuation analysis of PSR J1929+1905 for the observation on 20201121, with LRFS and 2DFS for the on-pulse region of a mean pulse profile.
\label{subfig:fluctu:J1929+1905}}
\end{figure}

\begin{figure}[htpb]
\centering
\includegraphics[width=0.225\textwidth, angle=0]{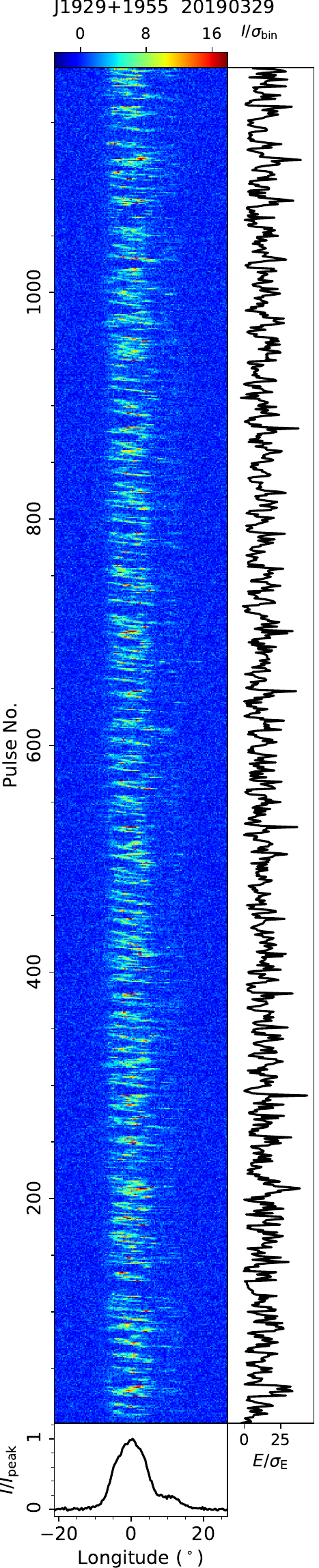}
\includegraphics[width=0.225\textwidth, angle=0]{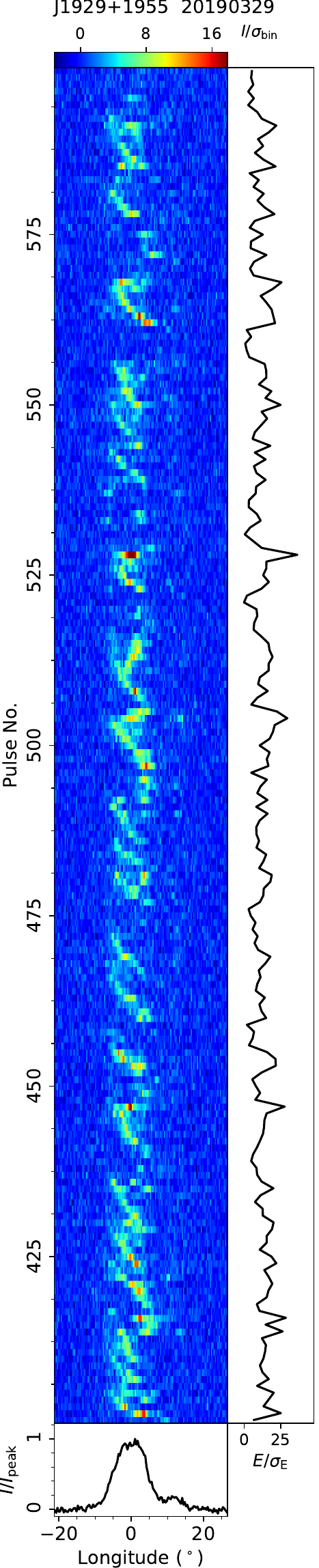}\\
\figcaption{Single pulse sequences of PSR J1929+1955 from the FAST observation on 20190329 -- to be continued.\\
\label{subfig:TP:J1929+1955}}
\addtocounter{figure}{-1}
\end{figure}

\begin{figure}[htpb]
\centering
\includegraphics[width=0.225\textwidth, angle=0]{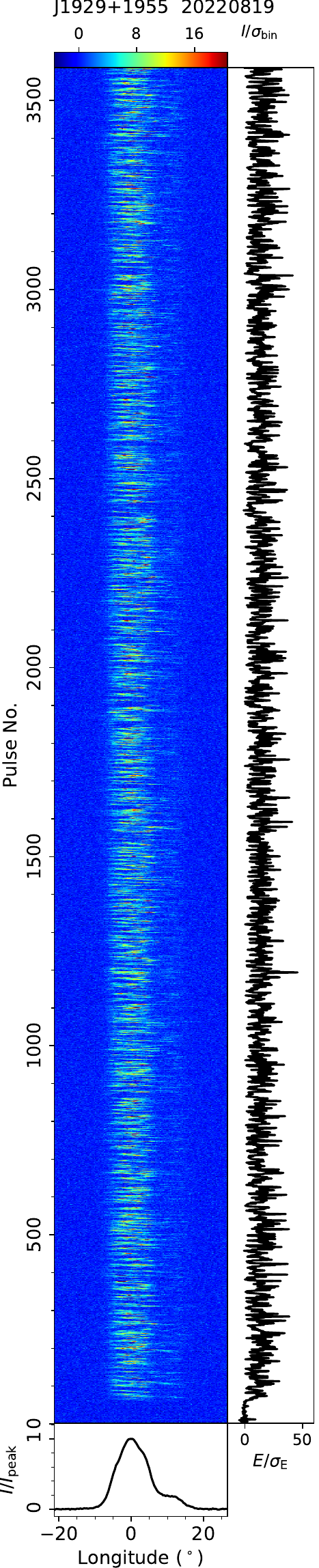}
\includegraphics[width=0.225\textwidth, angle=0]{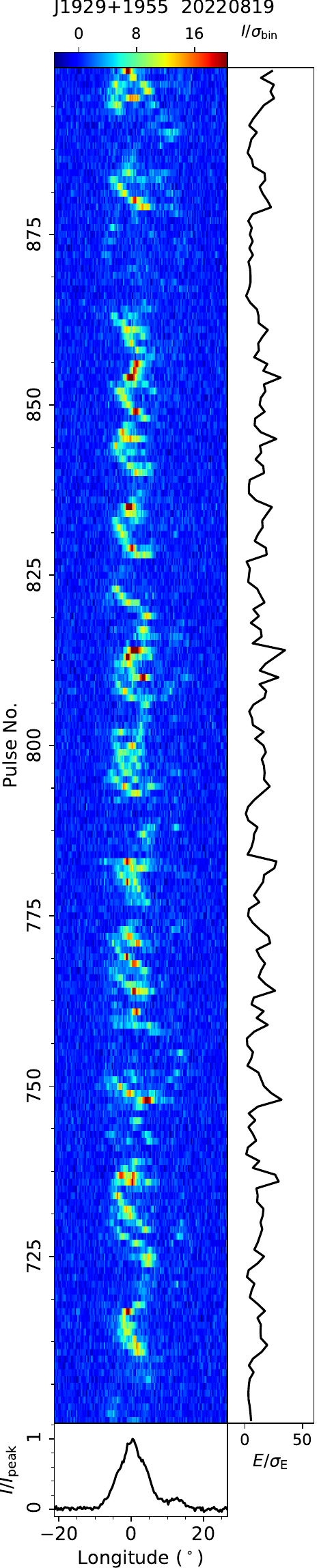}\\
\figcaption{Continued and ended.\\}
\end{figure}

\begin{figure}[htpb]
\centering
\includegraphics[width=0.22\textwidth, angle=0]{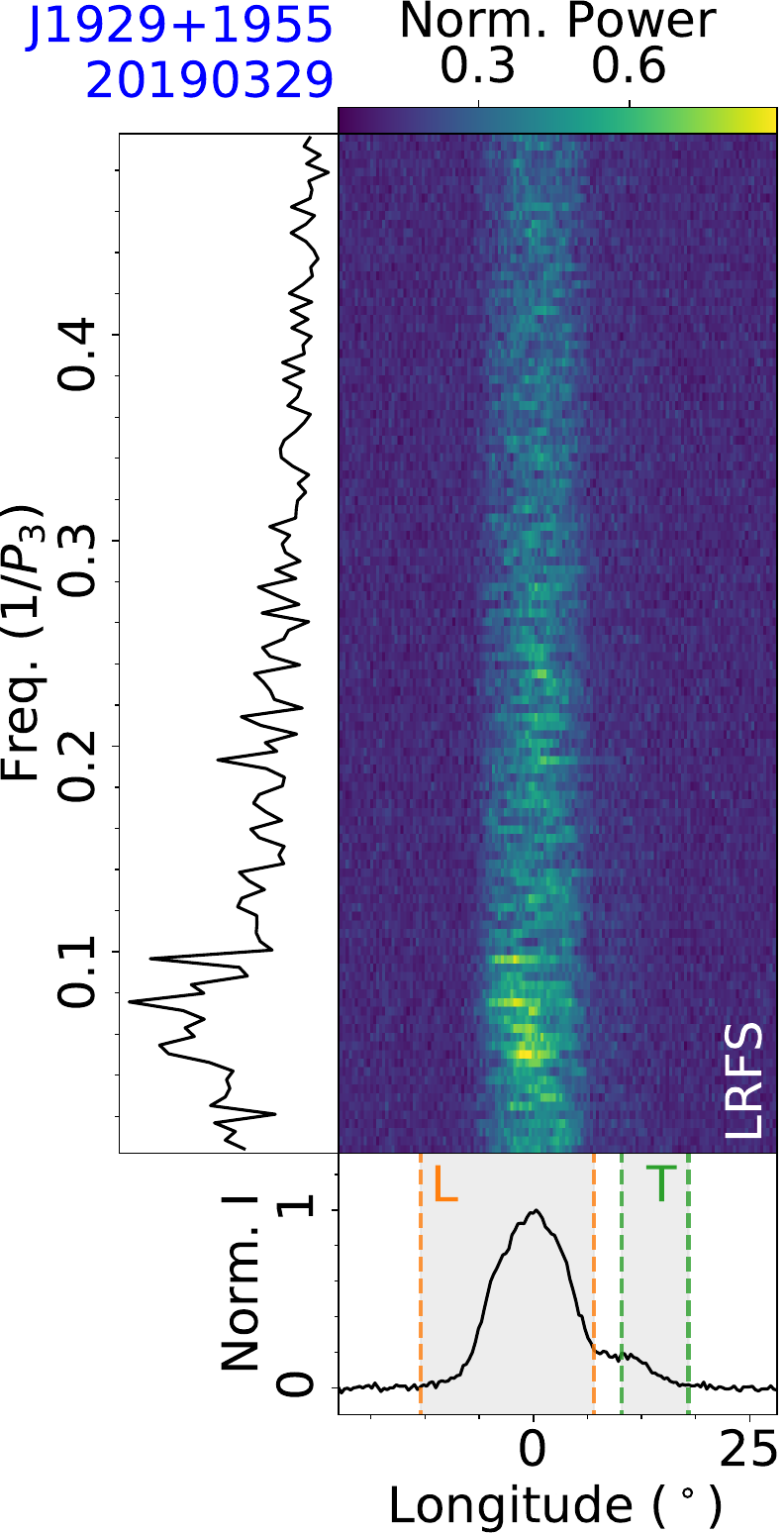}
\includegraphics[width=0.22\textwidth, angle=0]{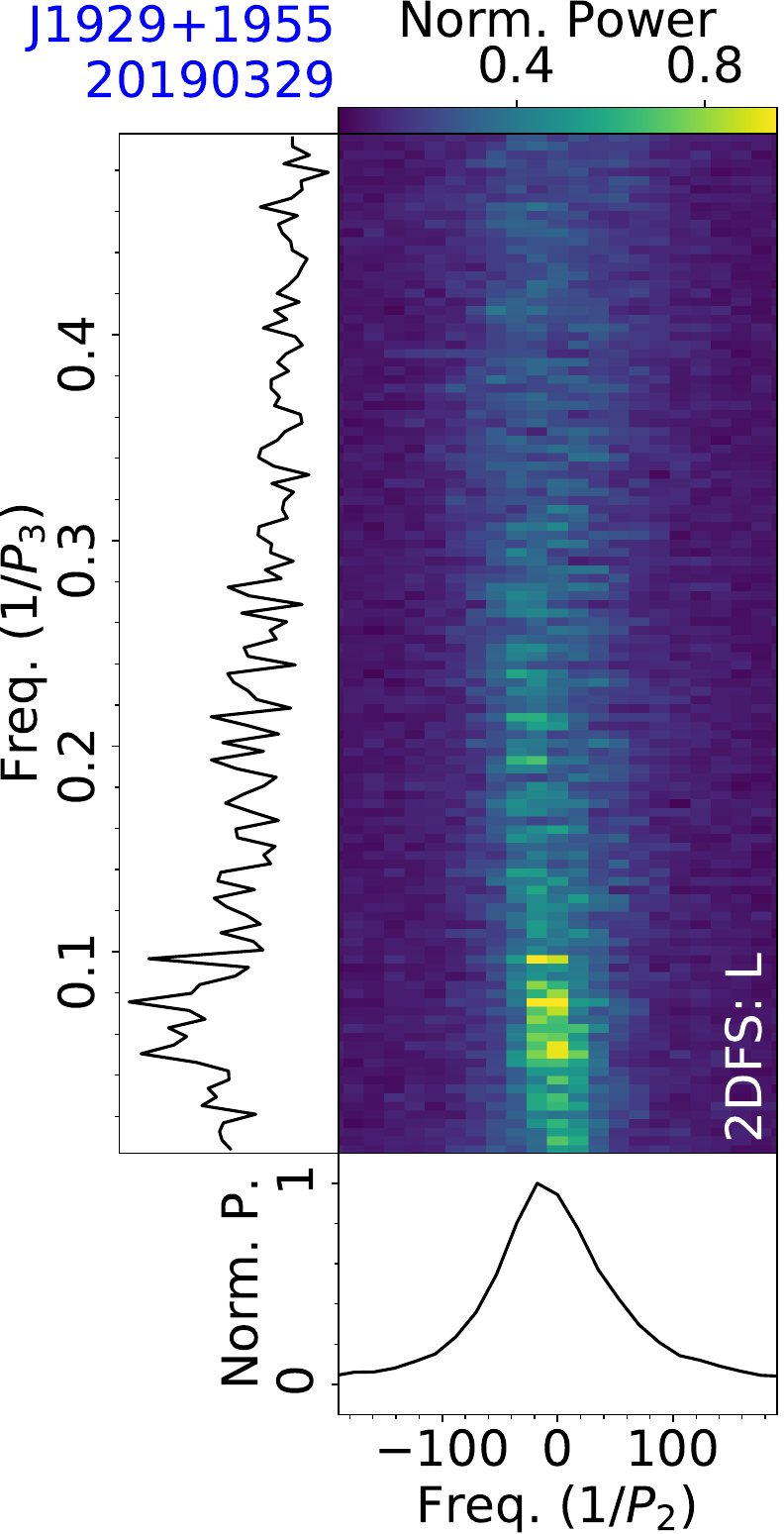}\\
\includegraphics[width=0.22\textwidth, angle=0]{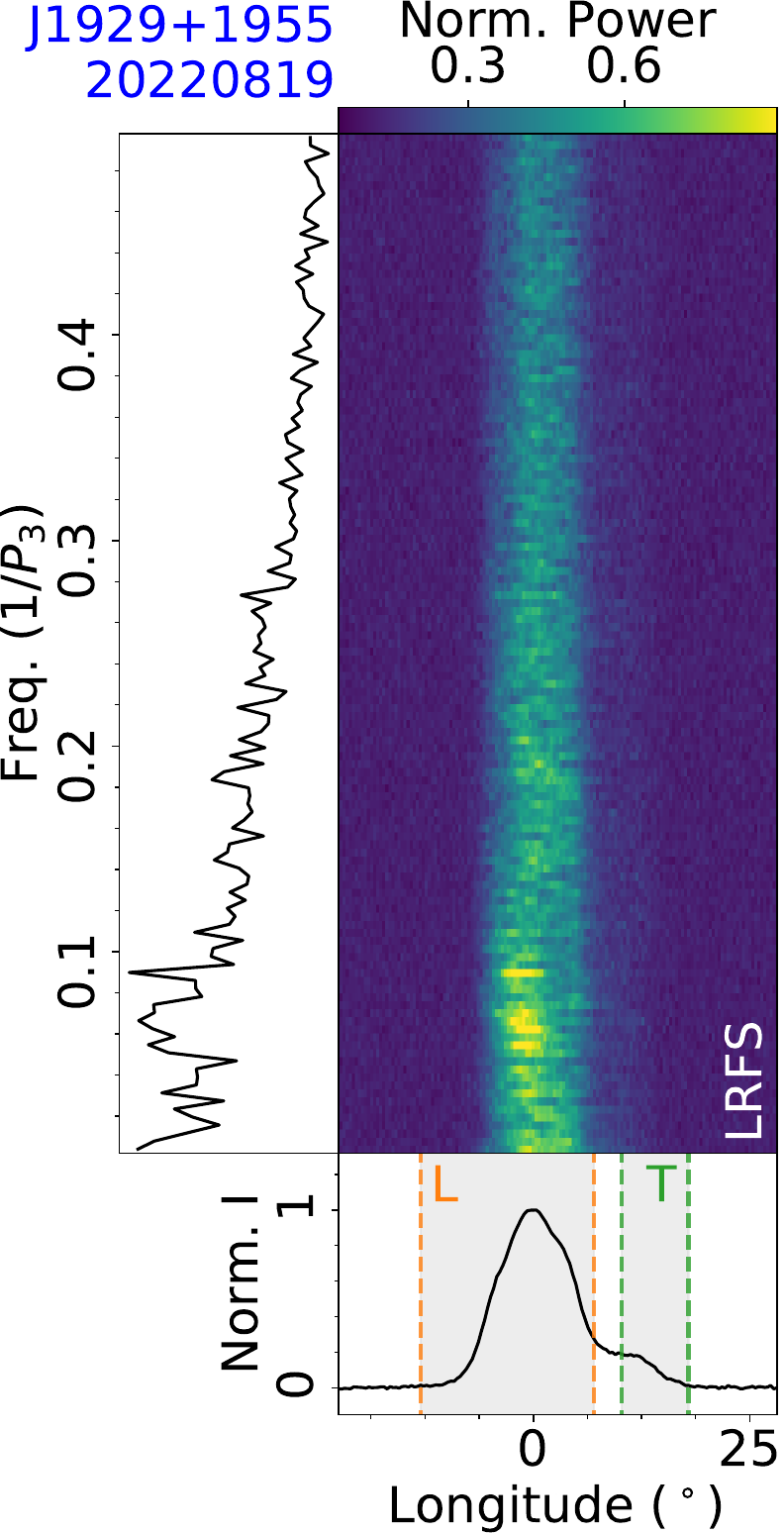}
\includegraphics[width=0.22\textwidth, angle=0]{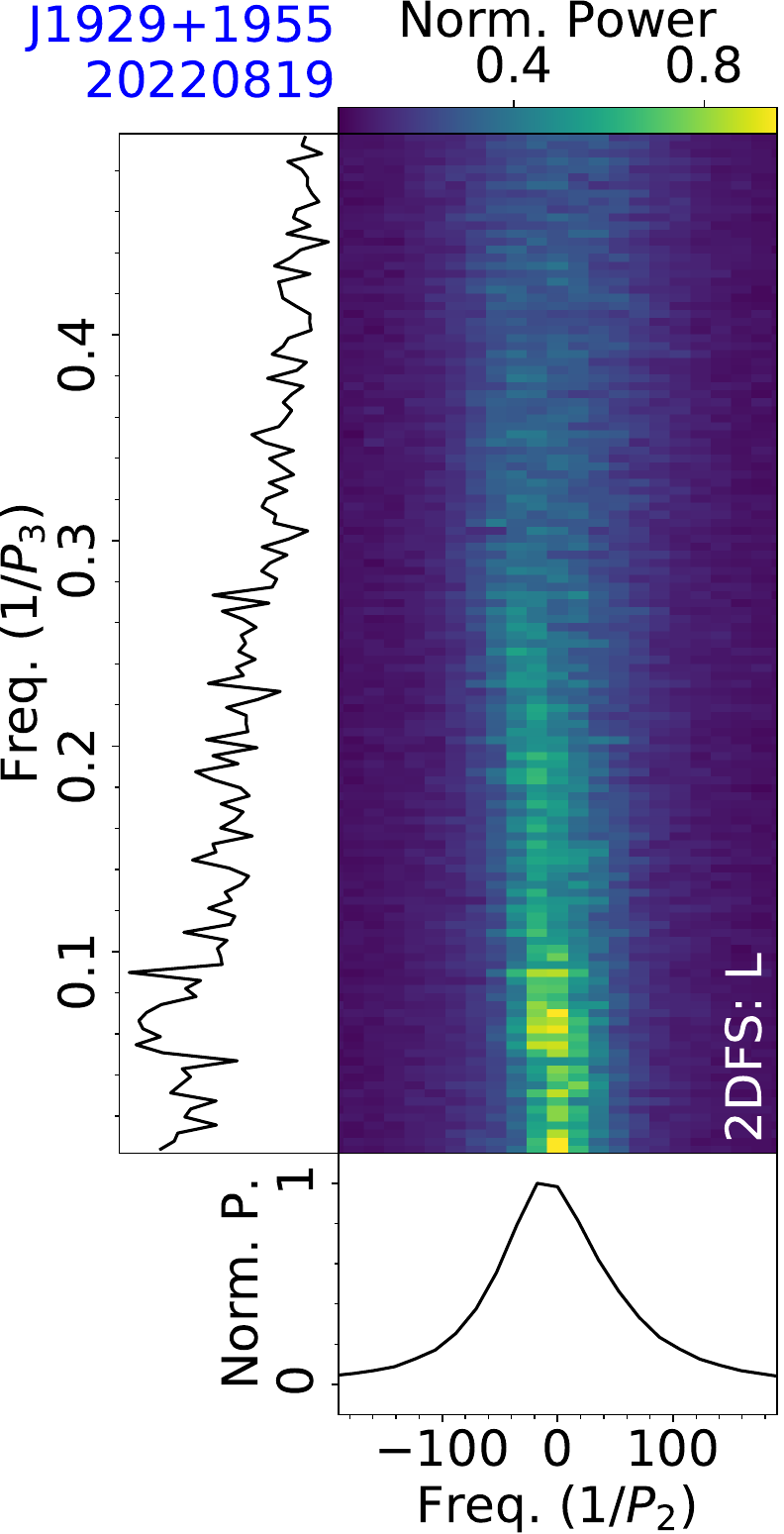}
\figcaption{Fluctuation analysis of PSR J1929+1955 from FAST observations on 20211123 and 20220819, with LRFS and 2DFS for the on-pulse region of mean pulse profiles.
\label{subfig:fluctu:J1929+1955}}
\end{figure}

\begin{figure}[htpb]
\centering
\includegraphics[width=0.22\textwidth, angle=0]{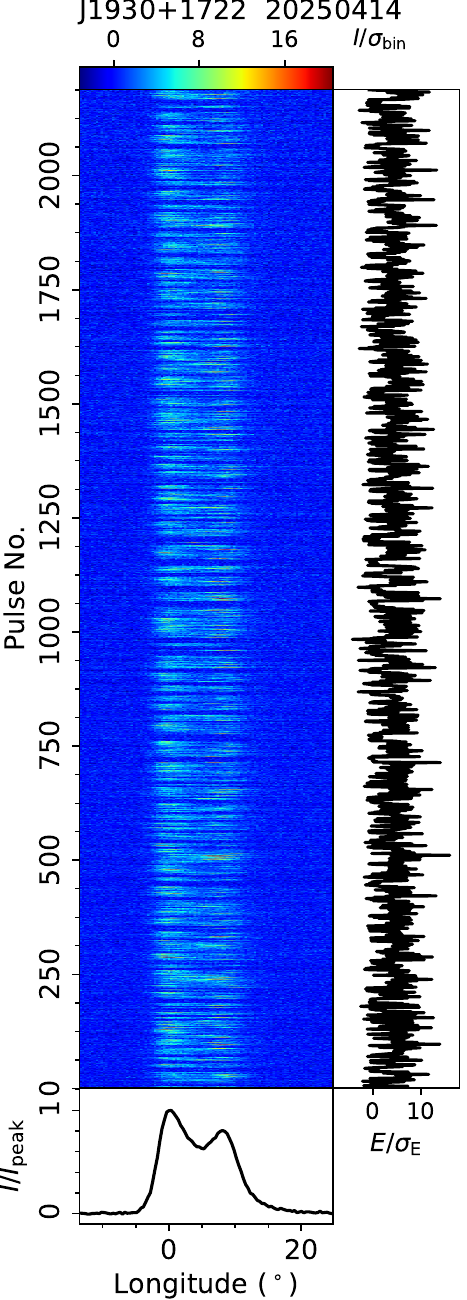}
\includegraphics[width=0.22\textwidth, angle=0]{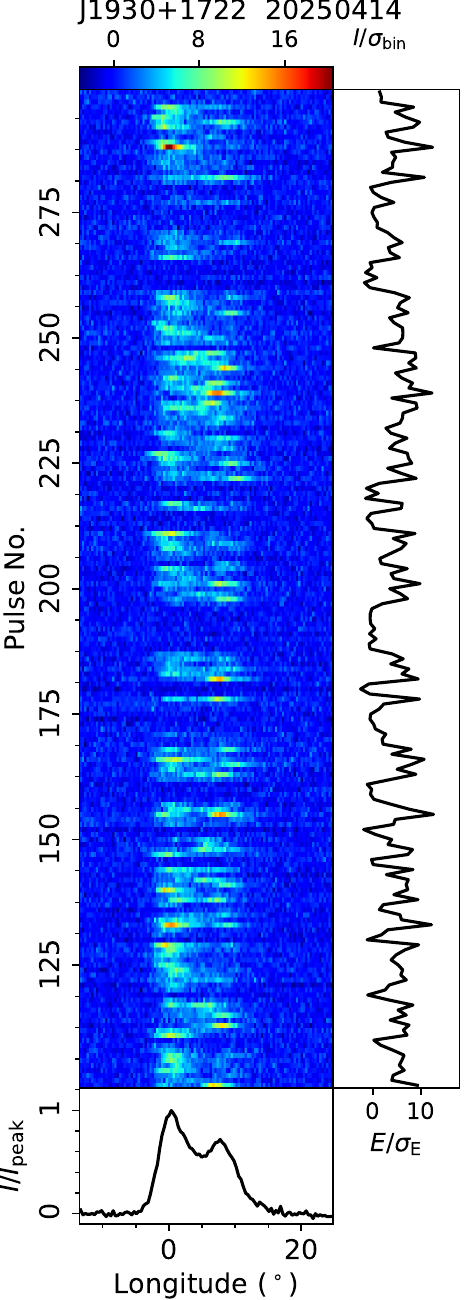}\\
\figcaption{Single pulse sequence of PSR J1930+1722 from the FAST observation on 20250414, and a zoomed-in view of pulses No. 100-300.
\label{subfig:TP:J1930+1722}}
\end{figure}

\begin{figure}[htpb]
\centering
\includegraphics[width=0.39\textwidth, angle=0]{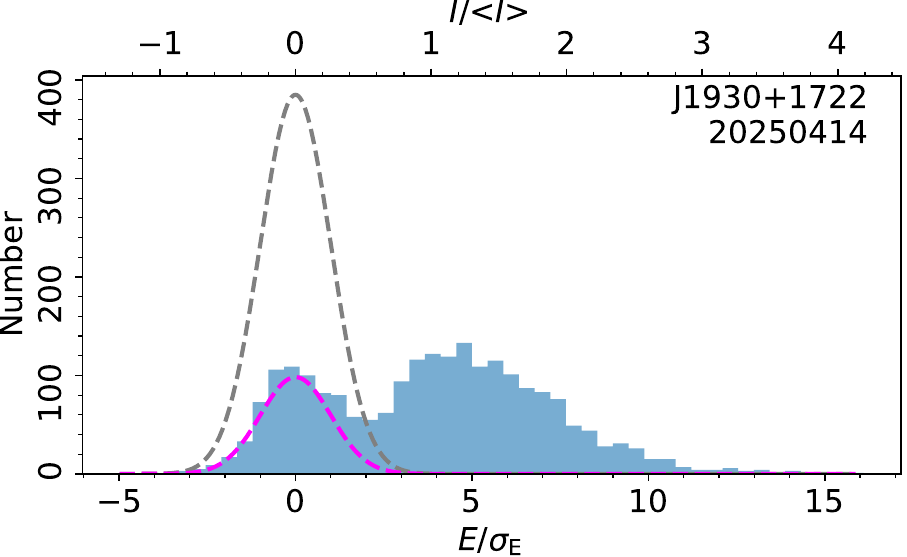}
\figcaption{On-pulse energy histogram of single pulses of PSR J1930+1722 from the FAST observation on 20250414.
\label{subfig:Hist:J1930+1722}}
\end{figure}

\begin{figure}[htpb]
\centering
\includegraphics[width=0.22\textwidth, angle=0]{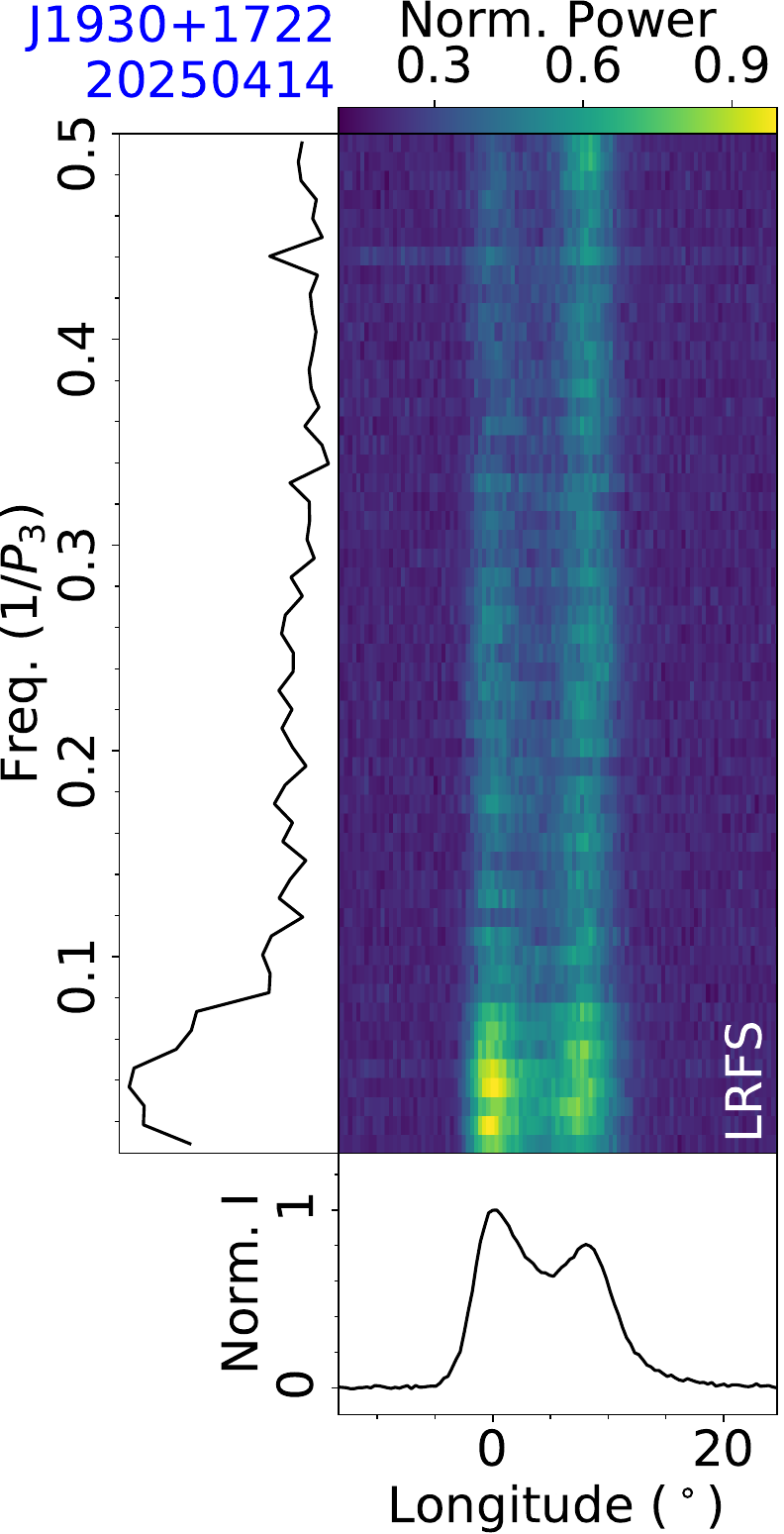}
\includegraphics[width=0.22\textwidth, angle=0]{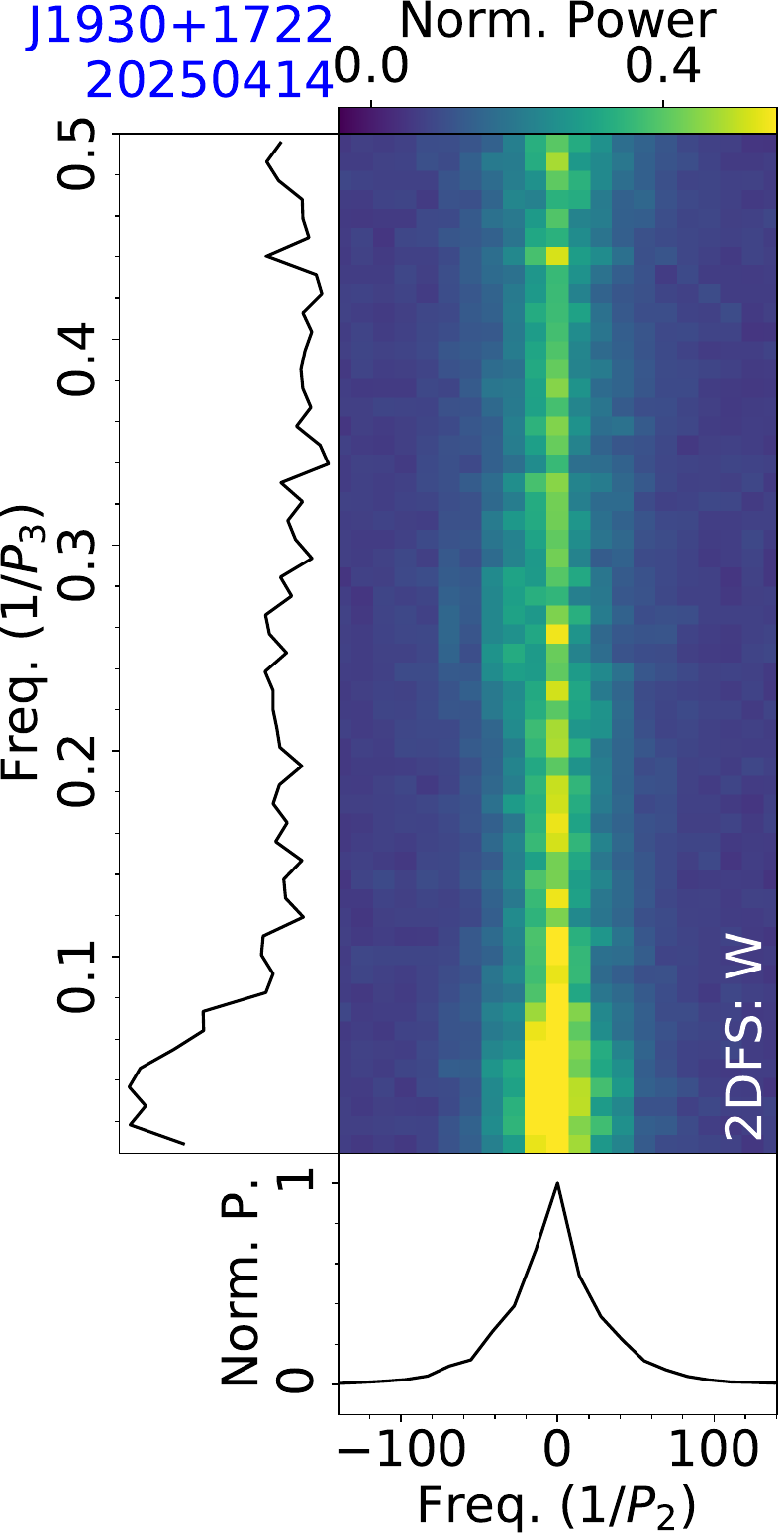}
\figcaption{Fluctuation analysis of PSR J1930+1722 for the observation on 20250414, with LRFS and 2DFS for the on-pulse region of a mean pulse profile.
\label{subfig:fluctu:J1930+1722}}
\end{figure}

\subsection{J1928+1923}
\label{subsec:J1928+1923}

PSR J1928+1923 was discovered in a deep Parkes multibeam survey \citep{Lorimer2013}. 

This pulsar was observed by FAST on 20200804 for 5 minutes, deriving a rotation period $P=0.8173$~s and a dispersion measure $D\!M=482.0~{\rm cm^{-3}\,pc}$. 
The single pulse sequence in Fig.~\ref{subfig:TP:J1928+1923} displays the nulling phenomenon. From the on-pulse integral energy histogram in Fig.~\ref{subfig:Hist:J1928+1923}, the nulling fraction of this observation is estimated to be 20.2$\pm$3\%.

\begin{figure}[htpb]
\centering
\includegraphics[width=0.22\textwidth, angle=0]{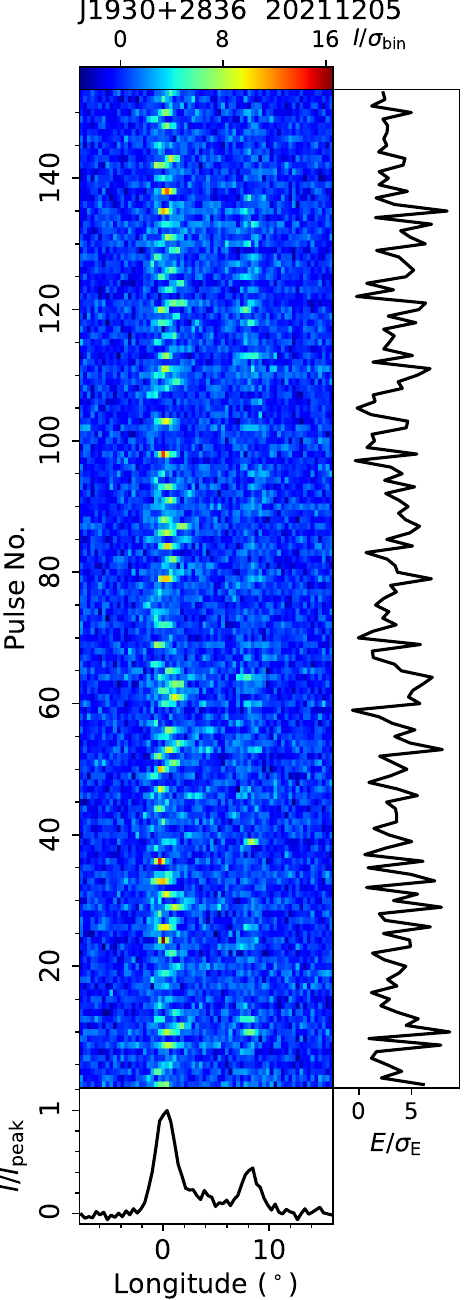}
\figcaption{Single pulse sequence of PSR J1930+2836 from the FAST observation on 20211205.
\label{subfig:TP:J1930+2836}}
\end{figure}

\begin{figure}[htpb]
\centering
\includegraphics[width=0.22\textwidth, angle=0]{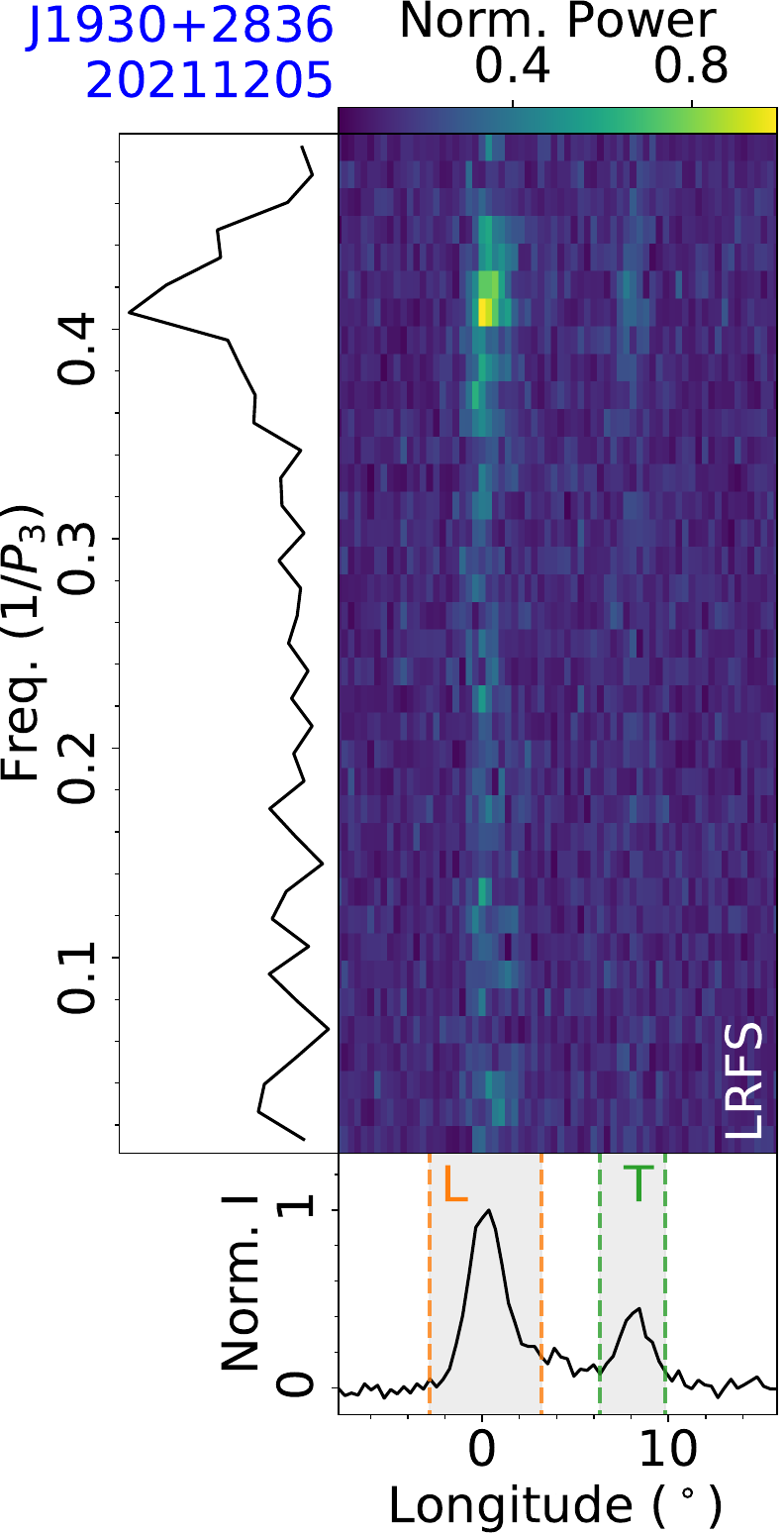}
\includegraphics[width=0.22\textwidth, angle=0]{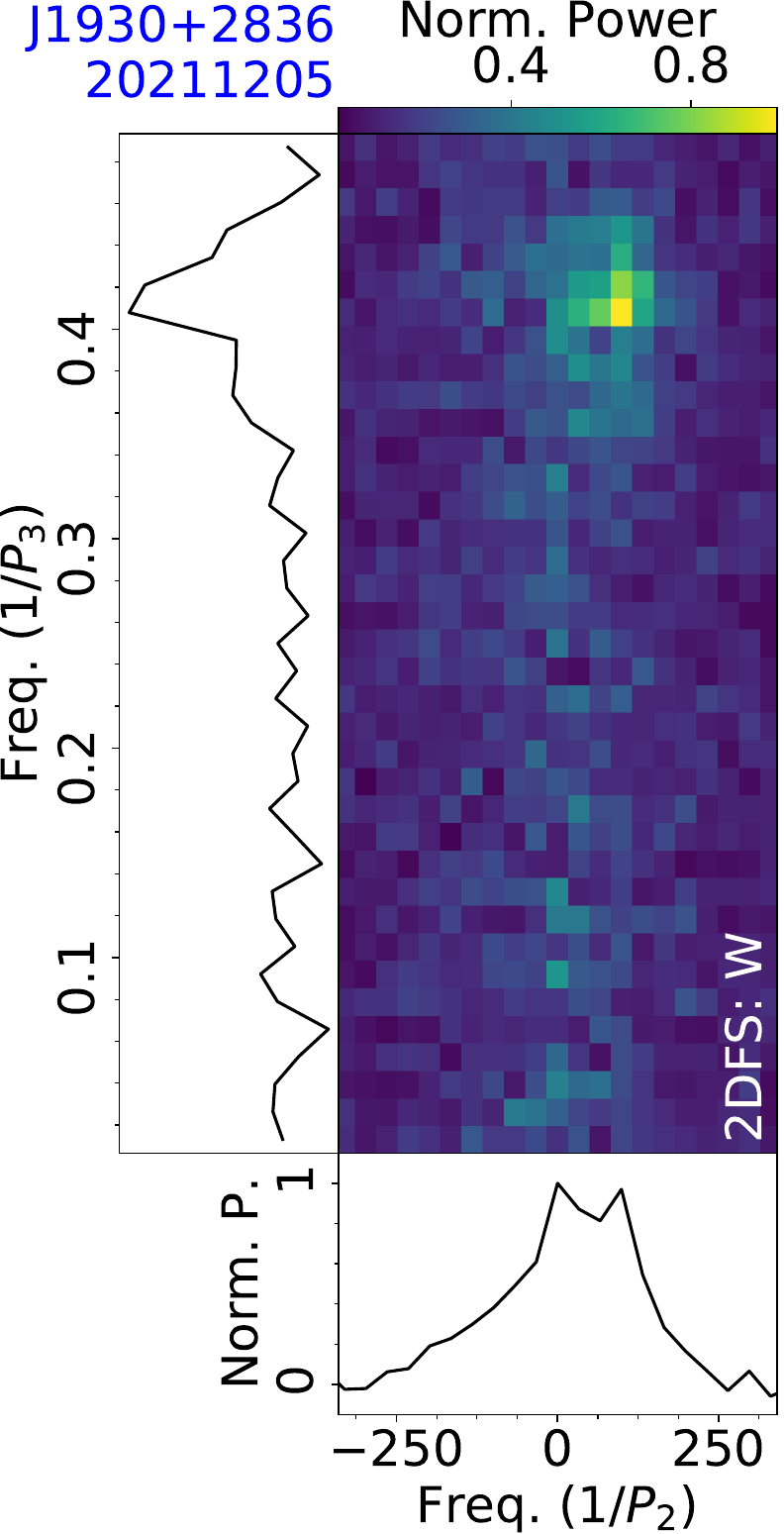}\\
\includegraphics[width=0.22\textwidth, angle=0]{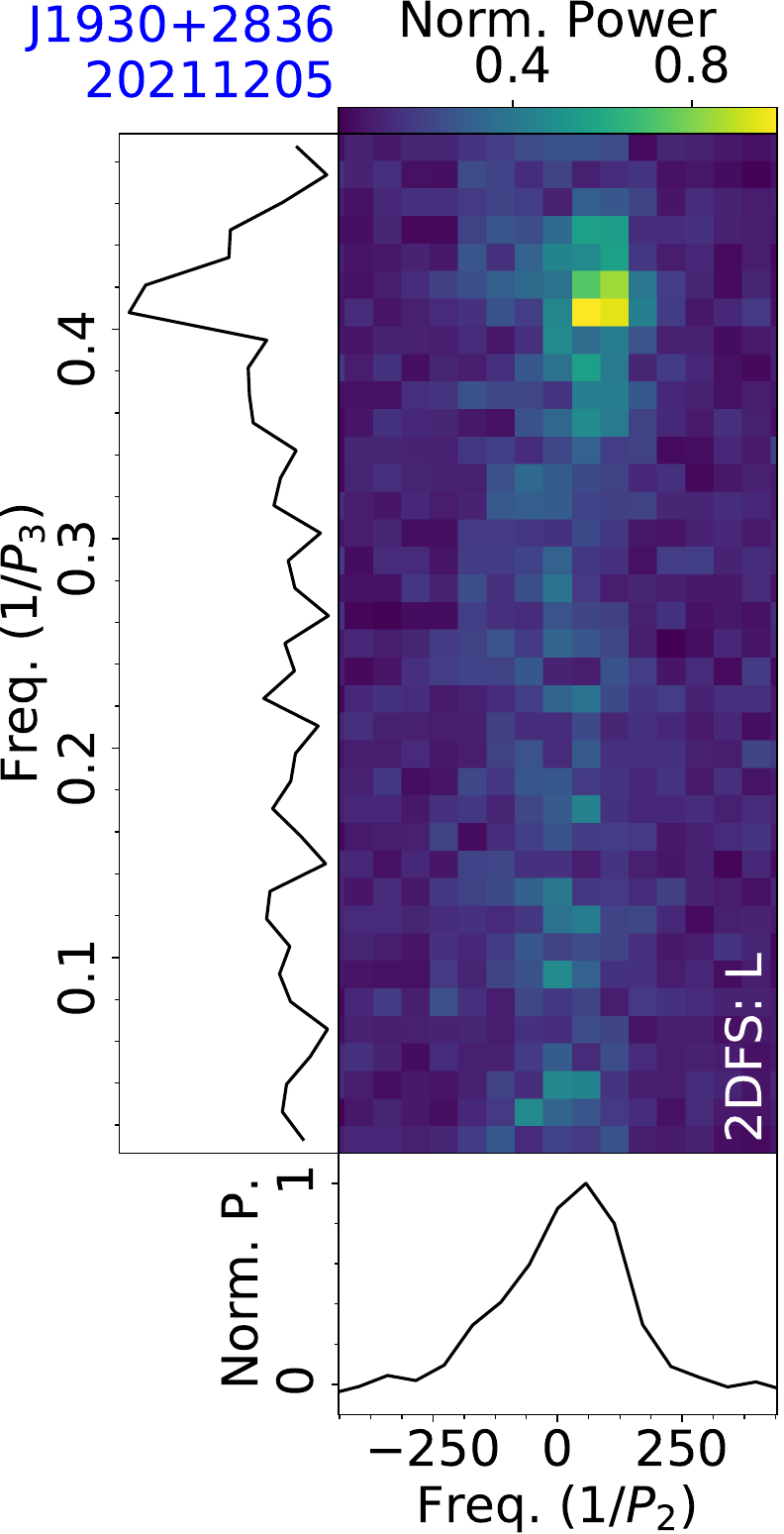}
\includegraphics[width=0.22\textwidth, angle=0]{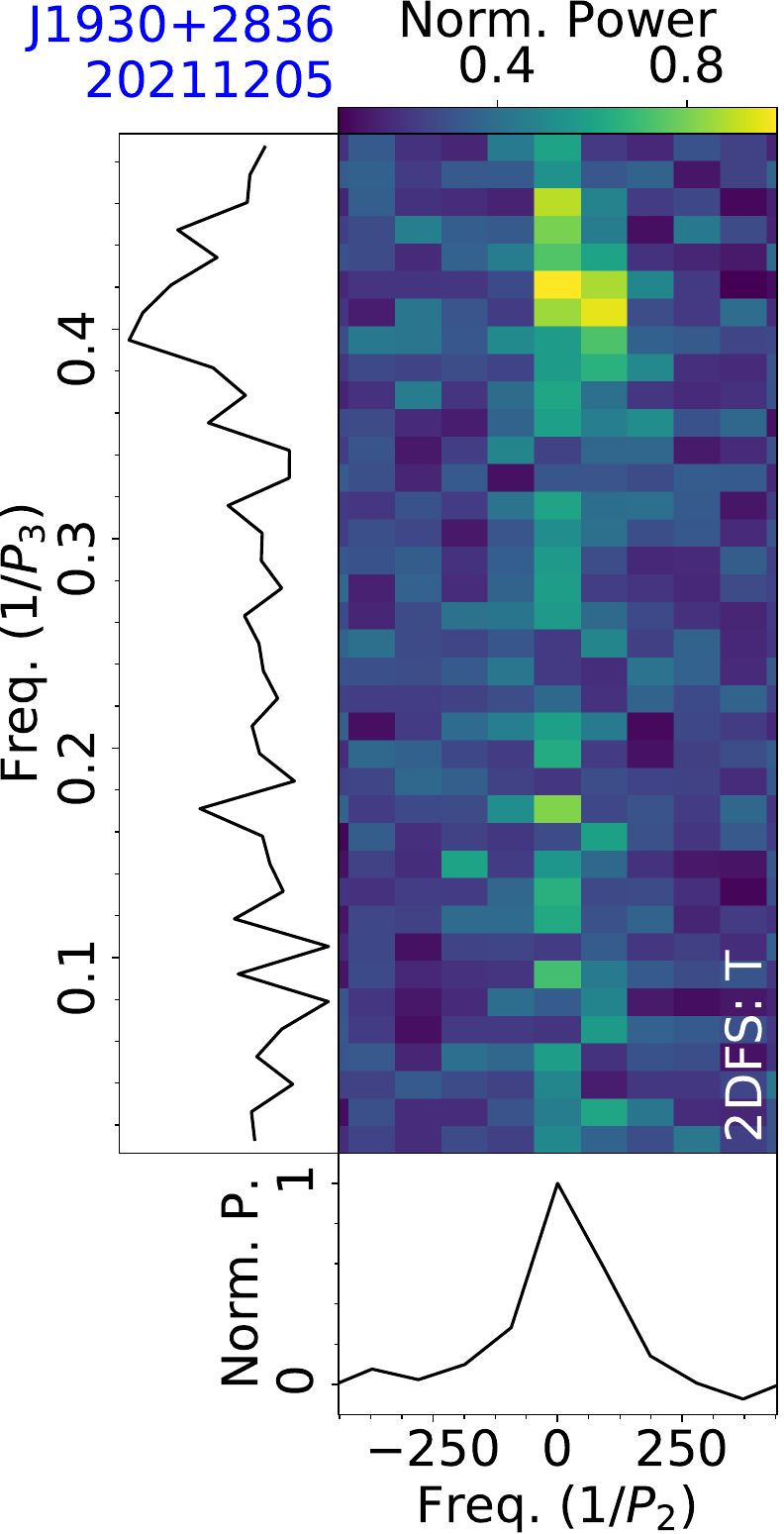}
\figcaption{Fluctuation analysis of PSR J1930+2836 for the observation on 20211205, with LRFS (top-left), and 2DFS for the on-pulse phase region (top-right), leading part (bottom-left) and trailing part (bottom-right) of a mean pulse profile.
\label{subfig:fluctu:J1930+2836}}
\end{figure}

\subsection{J1929+00}
\label{subsec:J1929+00}

PSR J1929+00 was discovered by \citet{Camilo1996} with the Arecibo telescope. 

This pulsar was observed by FAST on 20210710 for 5 minutes, deriving a rotation period $P=1.1669$~s and a dispersion measure $D\!M=43.5~{\rm cm^{-3}\,pc}$. Single pulse sequences in Fig.~\ref{subfig:TP:J1929+00} show a subpulse modulation phenomenon. The fluctuation spectra are shown in Fig.~\ref{subfig:fluctu:J1929+00}. A preferred negative drift feature is identified in the 2DFS with the centroid at $1/P_3=0.441\pm0.003$ and $1/P_2=-12\pm3$, corresponding to periodicities of $P_3=2.27\pm0.02$ periods and $P_2=-31\pm7$ degrees. 

\subsection{J1929+1905}
\label{subsec:J1929+1905}

PSR J1929+1905 was found in a deep Parkes multibeam survey \citep{Lorimer2013}. 

This pulsar was observed by FAST on 20201121 for 15 minutes, deriving a rotation period $P=0.3392$~s and a dispersion measure $D\!M=528.7~{\rm cm^{-3}\,pc}$. 
The single pulse sequence of this observation and three zoomed-in sequences are shown in Fig.~\ref{subfig:TP:J1929+1905}, which display drifting bands and intensity variations. 
Fluctuation spectra shown in Fig.~\ref{subfig:fluctu:J1929+1905} exhibit a negative drift feature, with the centroid frequencies of $1/P_3=0.0379\pm0.0002$ and $1/P_2=-1.2\pm0.1$, corresponding to periodicities of $P_3=26.4\pm0.1$ periods and $P_2=-311\pm20^\circ$.

\begin{figure}[htpb]
\centering
\includegraphics[width=0.22\textwidth, angle=0]{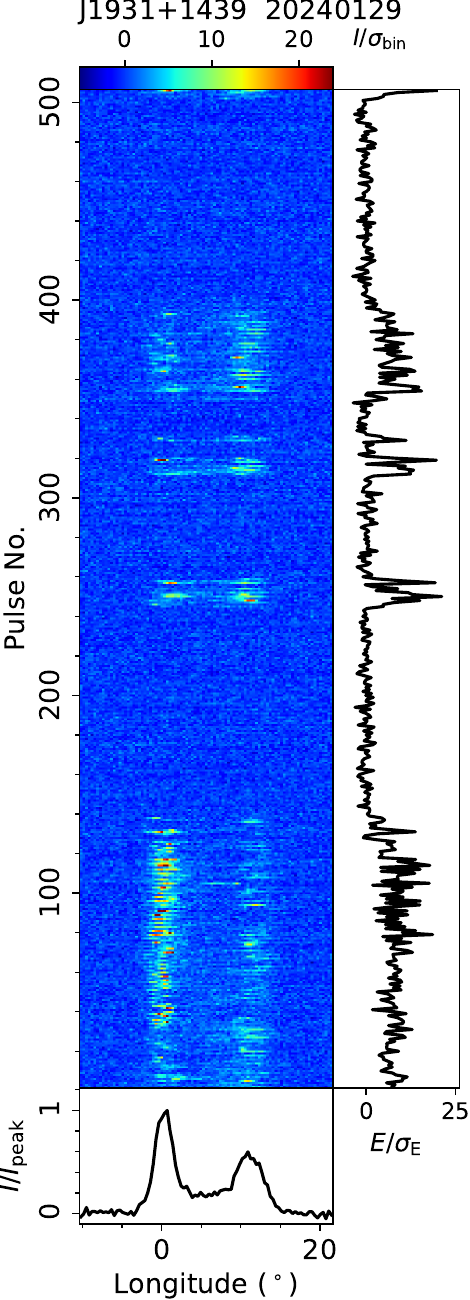}
\includegraphics[width=0.22\textwidth, angle=0]{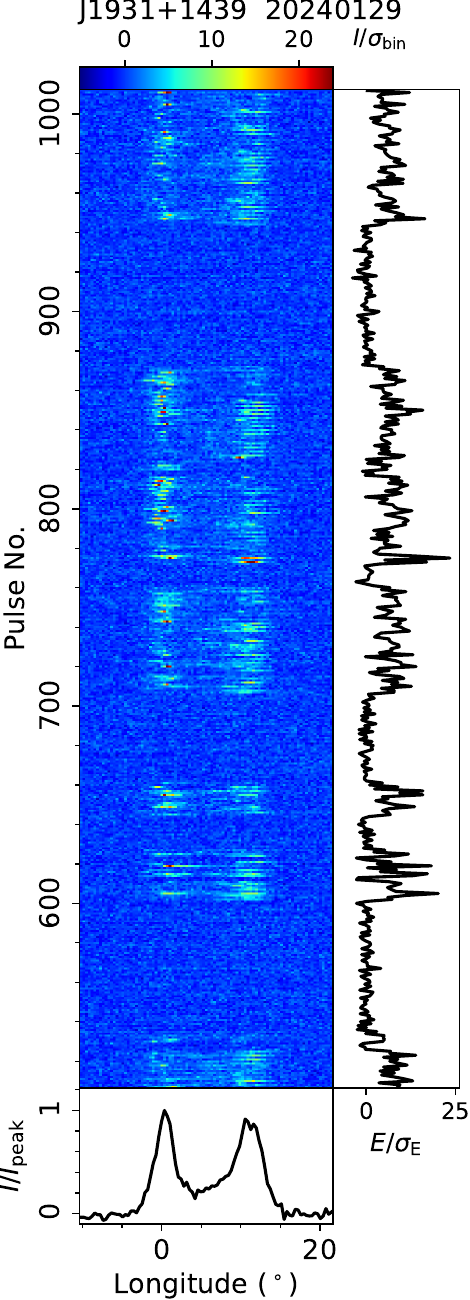}
\figcaption{Single pulse sequences of PSR J1931+1439 from the FAST observation on 20240129.
\label{subfig:TP:J1931+1439}}
\end{figure}

\begin{figure}[htpb]
\centering
\includegraphics[width=0.39\textwidth, angle=0]{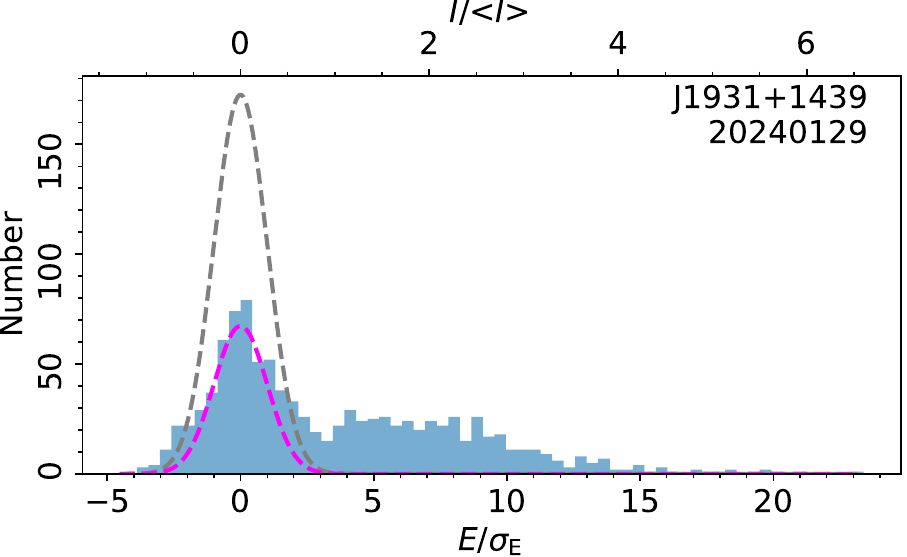}
\figcaption{On-pulse energy histogram of single pulses of PSR J1931+1439 from the FAST observation on 20240129.
\label{subfig:Hist:J1931+1439}}
\end{figure}

\begin{figure}[htpb]
\centering
\includegraphics[width=0.22\textwidth, angle=0]{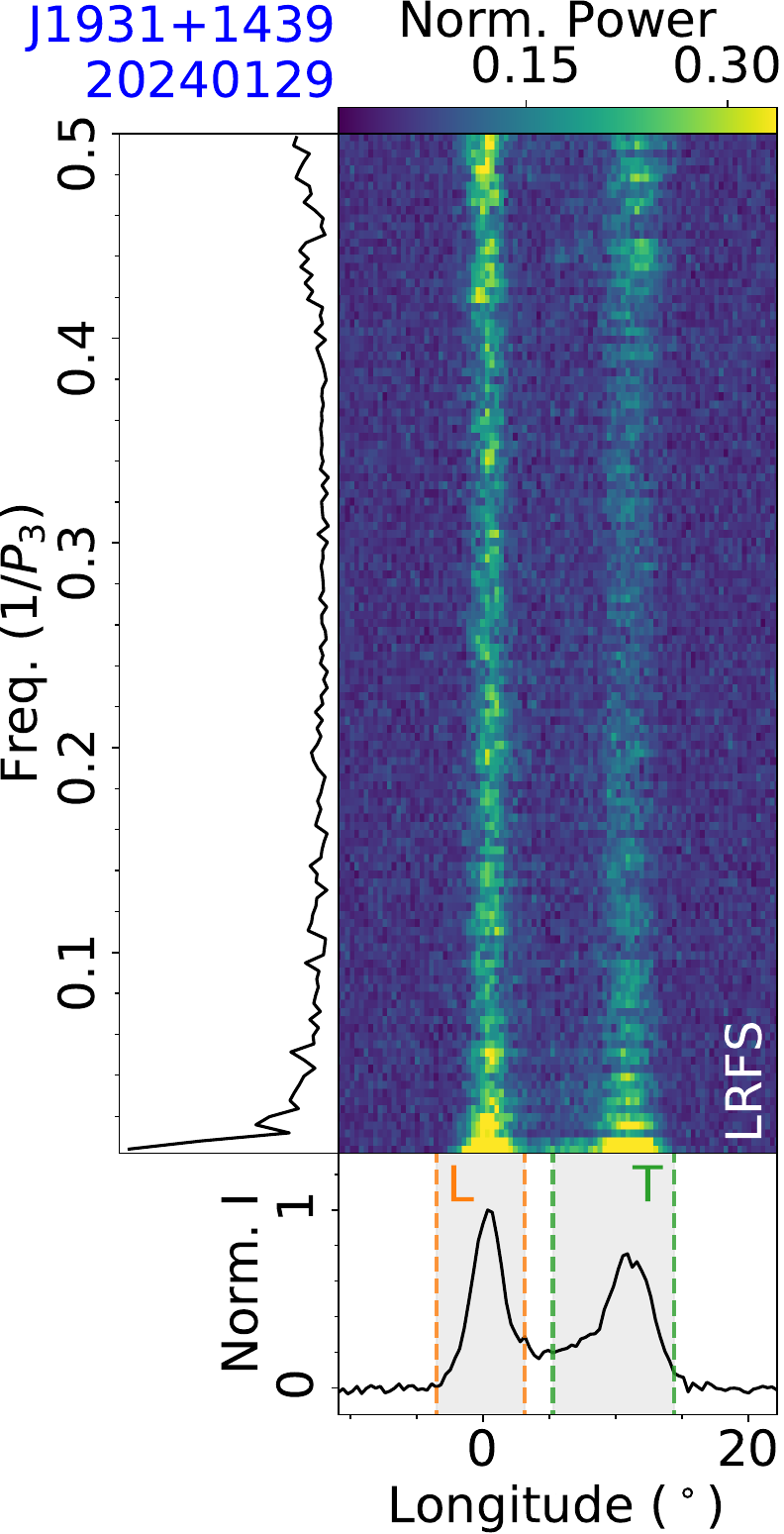}
\includegraphics[width=0.22\textwidth, angle=0]{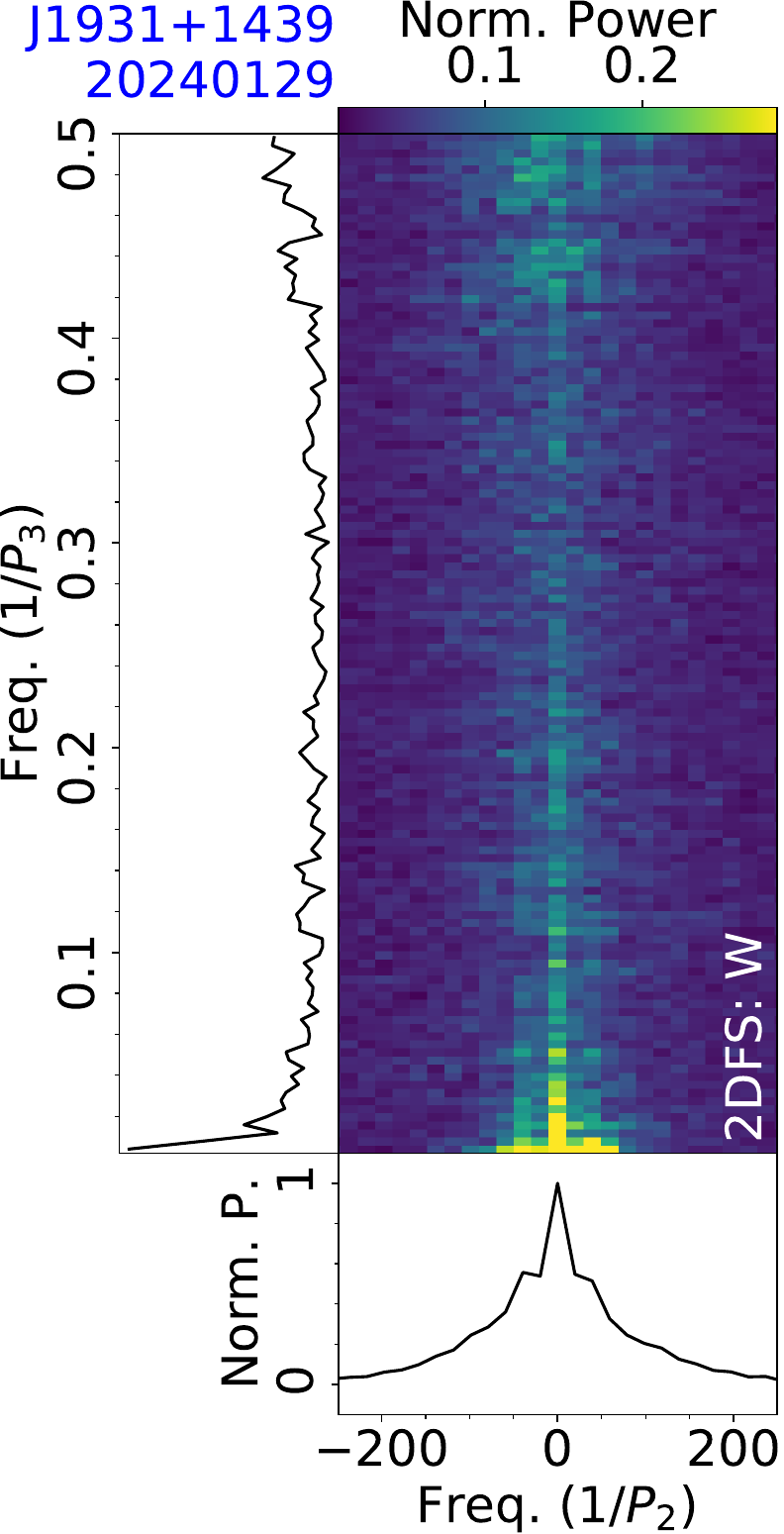}\\
\includegraphics[width=0.22\textwidth, angle=0]{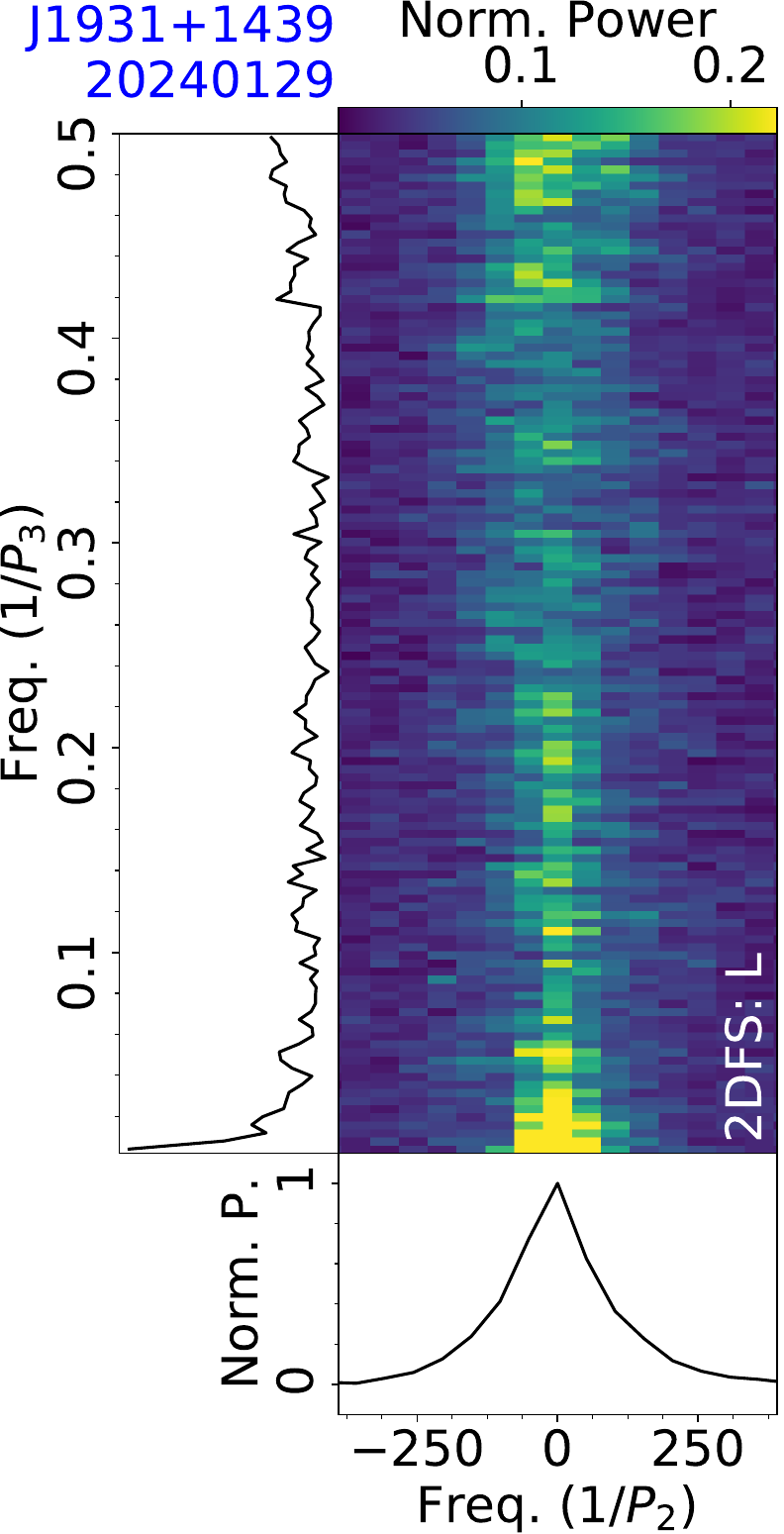}
\includegraphics[width=0.22\textwidth, angle=0]{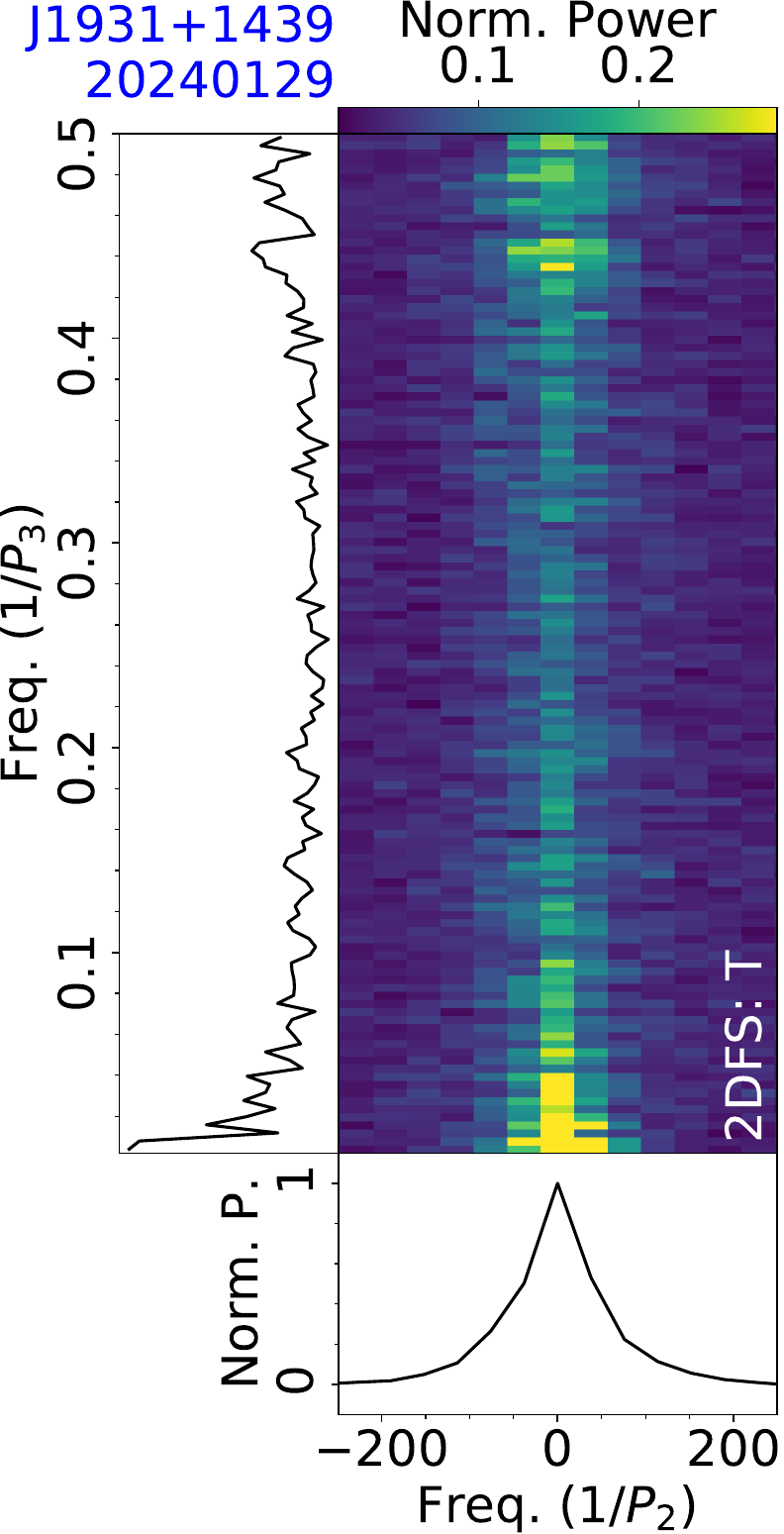}
\figcaption{Fluctuation analysis of PSR J1931+1439 for the observation on 20240129, with LRFS (top-left), and 2DFS for the on-pulse phase region (top-right), leading part (bottom-left) and trailing part (bottom-right) of a mean pulse profile.
\label{subfig:fluctu:J1931+1439}}
\end{figure}

\subsection{J1929+1955}
\label{subsec:J1929+1955}

PSR J1929+1955 was discovered by \citep{Lorimer2013} using the Parkes telescope. Negative drifting of $P_3=13\pm5$ periods and $P_2=-90^{+56}_{-122}$ degrees was reported by \citet{Song2023}. 

This pulsar was observed by FAST on 20190329 for 5 minutes and on 20220819 for 15 minutes. 
From the 15-minute data, a rotation period $P=0.2578$~s and a dispersion measure $D\!M=279.2~{\rm cm^{-3}\,pc}$ were derived. 
Single pulse sequences of these two observations in Fig.~\ref{subfig:TP:J1929+1955} illustrate that subpulses of this pulsar mainly negatively drift, sometimes changing drifting direction and displaying a "C"-shape pattern. Fluctuation spectra of these two observations are shown in Fig.~\ref{subfig:fluctu:J1929+1955}. 
%
The modulation frequency over time of the negatively drifting feature is widely distributed in 2DFS. 
For the main drift feature in 2DFS of the observation on 20190329, centroid frequencies are estimated to be $1/P_3=0.068\pm0.001$ and $1/P_2=-5.2\pm0.6$, yielding periodicities of $P_3=14.7\pm0.2$ periods and $P_2=-69\pm8^\circ$. 
For the observation on 20220819, 2DFS exhibits the main drift feature with the centroid of $1/P_3=0.0744\pm0.0004$ and $1/P_2=-5.8\pm0.6$, corresponding to $P_3=13.4\pm0.1$ periods and $P_2=-62\pm6^\circ$.

\begin{figure}[htpb]
\centering
\includegraphics[width=0.22\textwidth, angle=0]{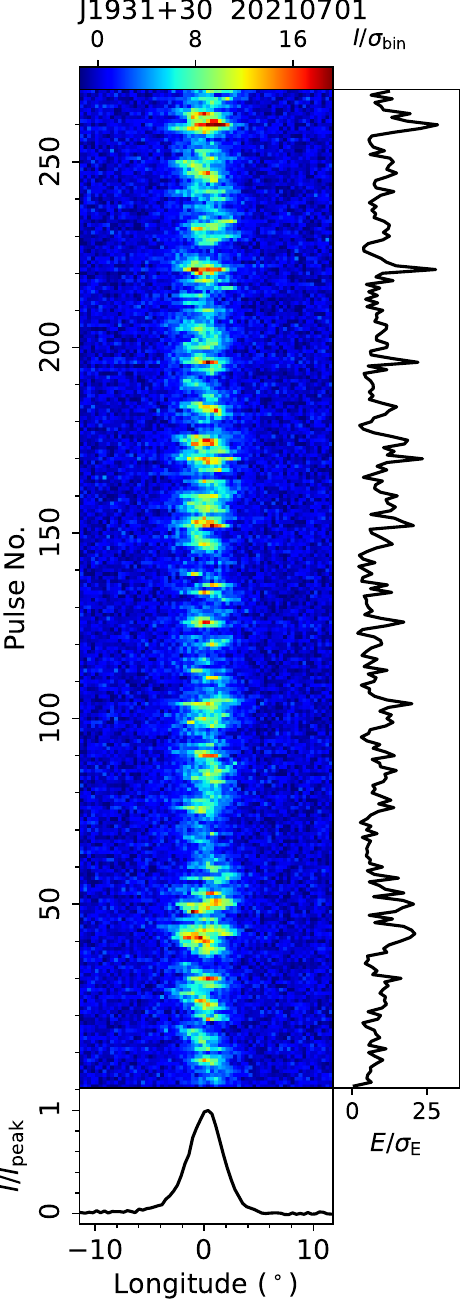}
\includegraphics[width=0.22\textwidth, angle=0]{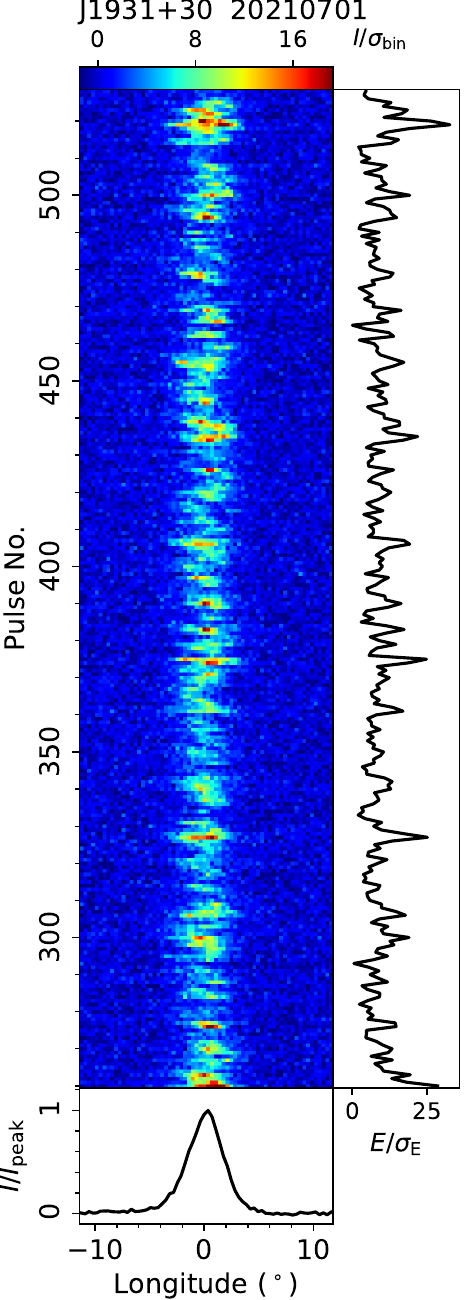}
\figcaption{Single pulse sequences of PSR J1931+30 from the FAST observation on 20210701.
\label{subfig:TP:J1931+30}}
\end{figure}

\begin{figure}[htpb]
\centering
\includegraphics[width=0.22\textwidth, angle=0]{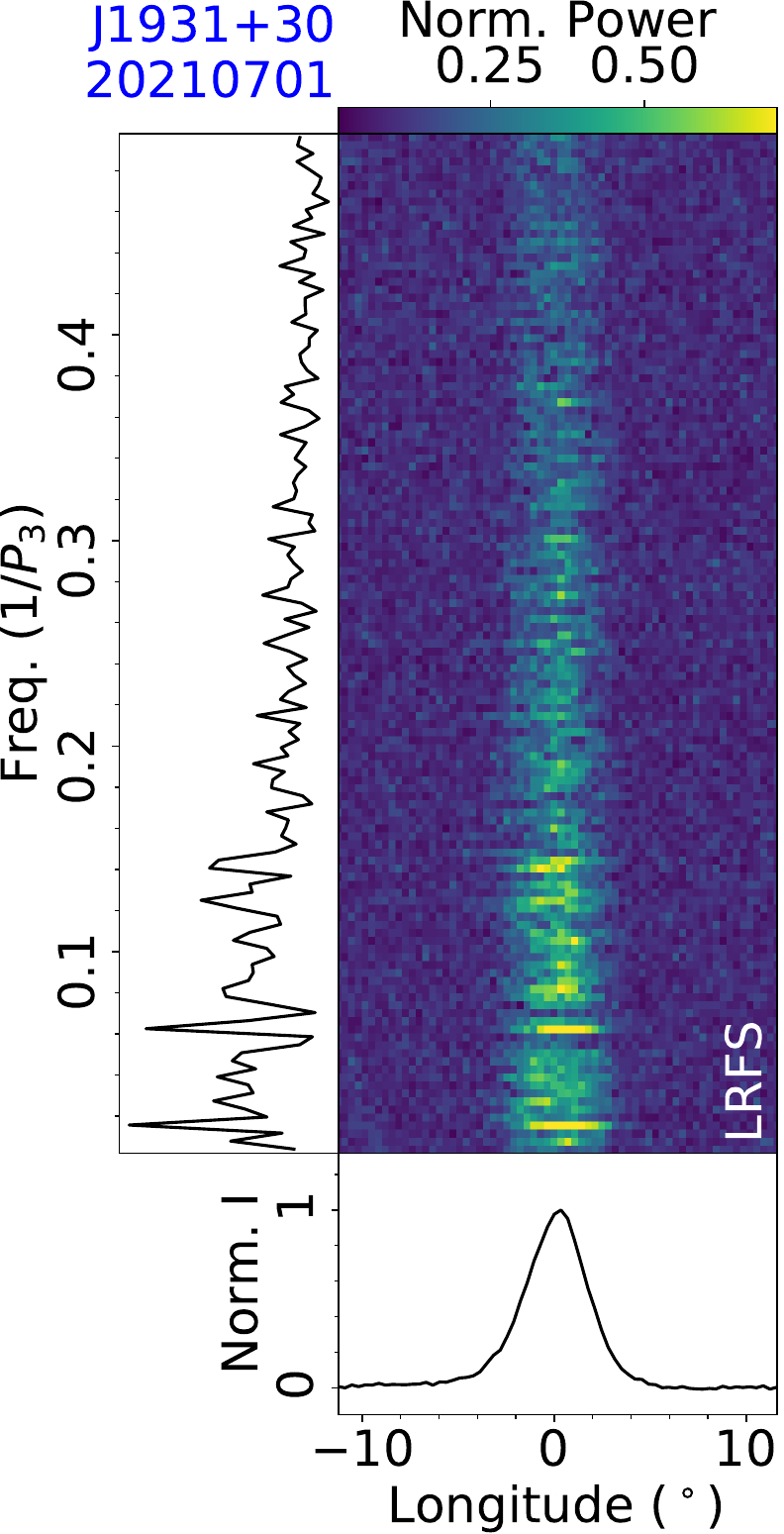}
\includegraphics[width=0.22\textwidth, angle=0]{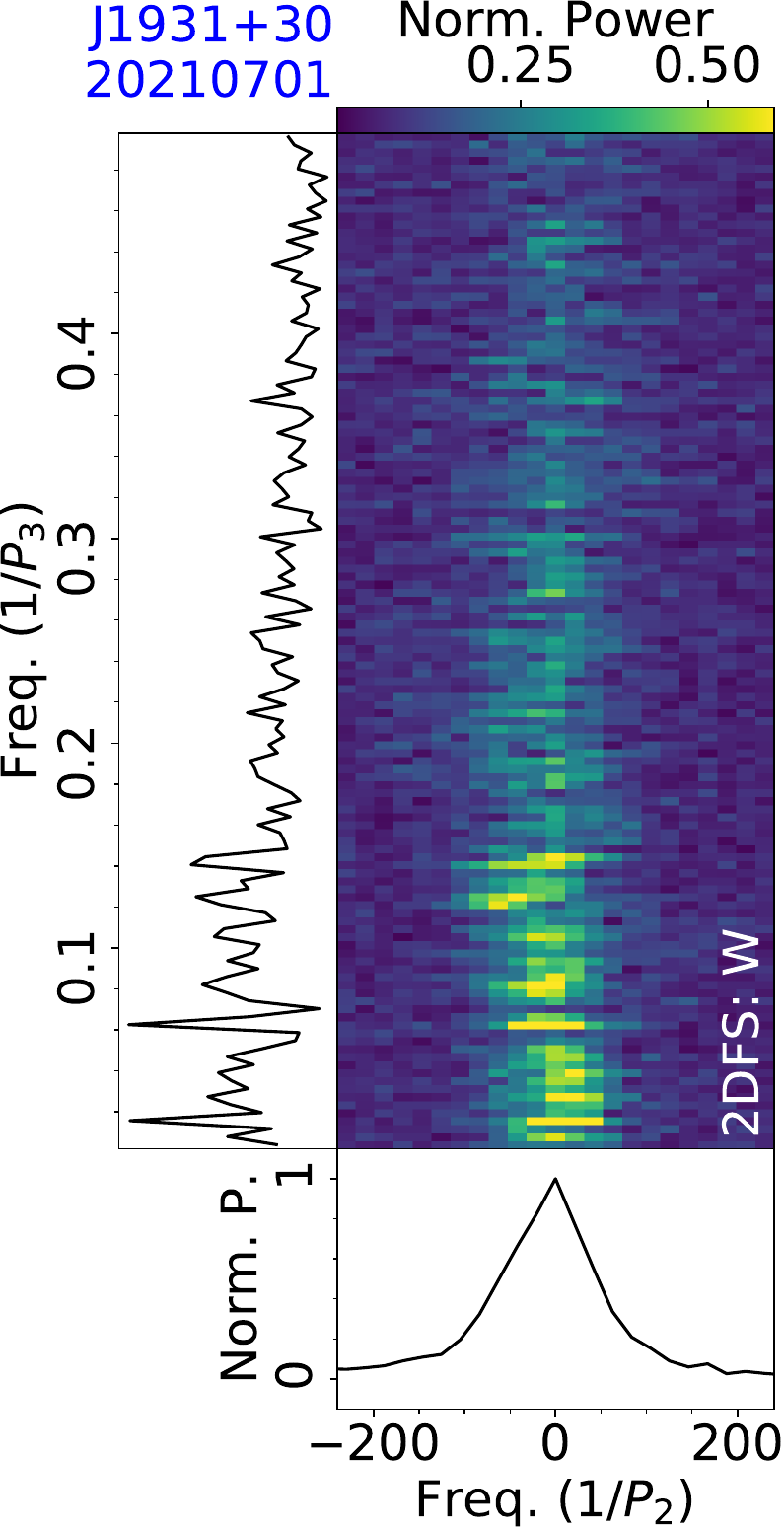}
\figcaption{Fluctuation analysis of PSR J1931+30 for the observation on 20210701, with LRFS and 2DFS for the on-pulse region of a mean pulse profile.
\label{subfig:fluctu:J1931+30}}
\end{figure}

\begin{figure}[htpb]
\centering
\includegraphics[width=0.22\textwidth, angle=0]{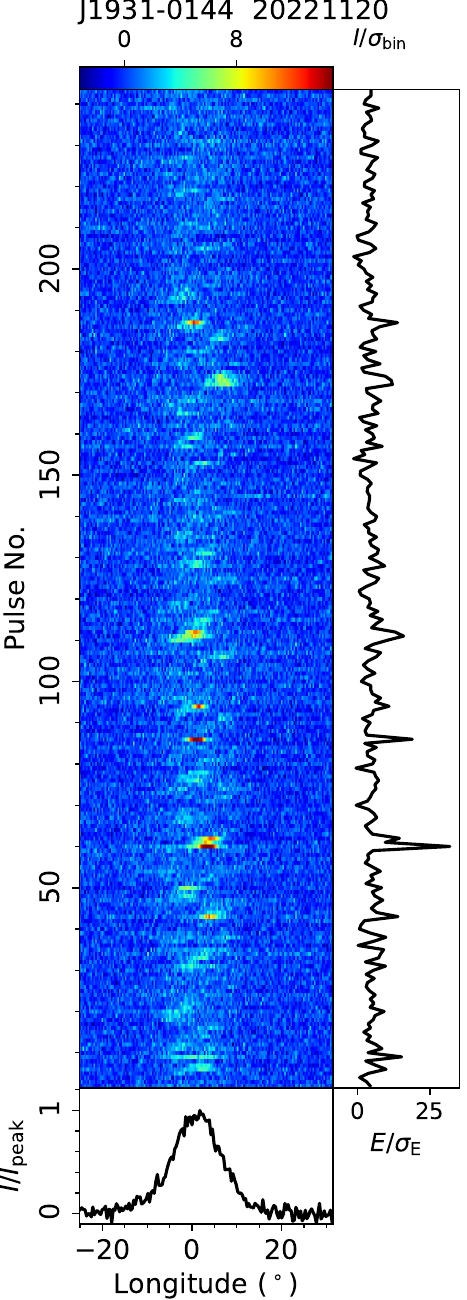}
\includegraphics[width=0.22\textwidth, angle=0]{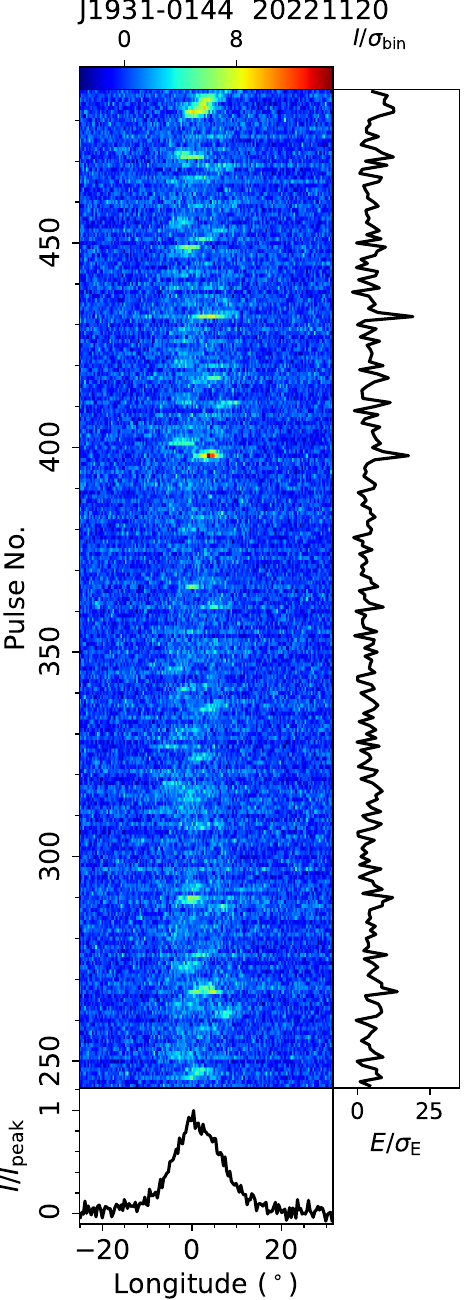}
\figcaption{Single pulse sequences of PSR J1931-0144 from the FAST observation on 20221120.
\label{subfig:TP:J1931-0144}}
\end{figure}

\begin{figure}[htpb]
\centering
\includegraphics[width=0.22\textwidth, angle=0]{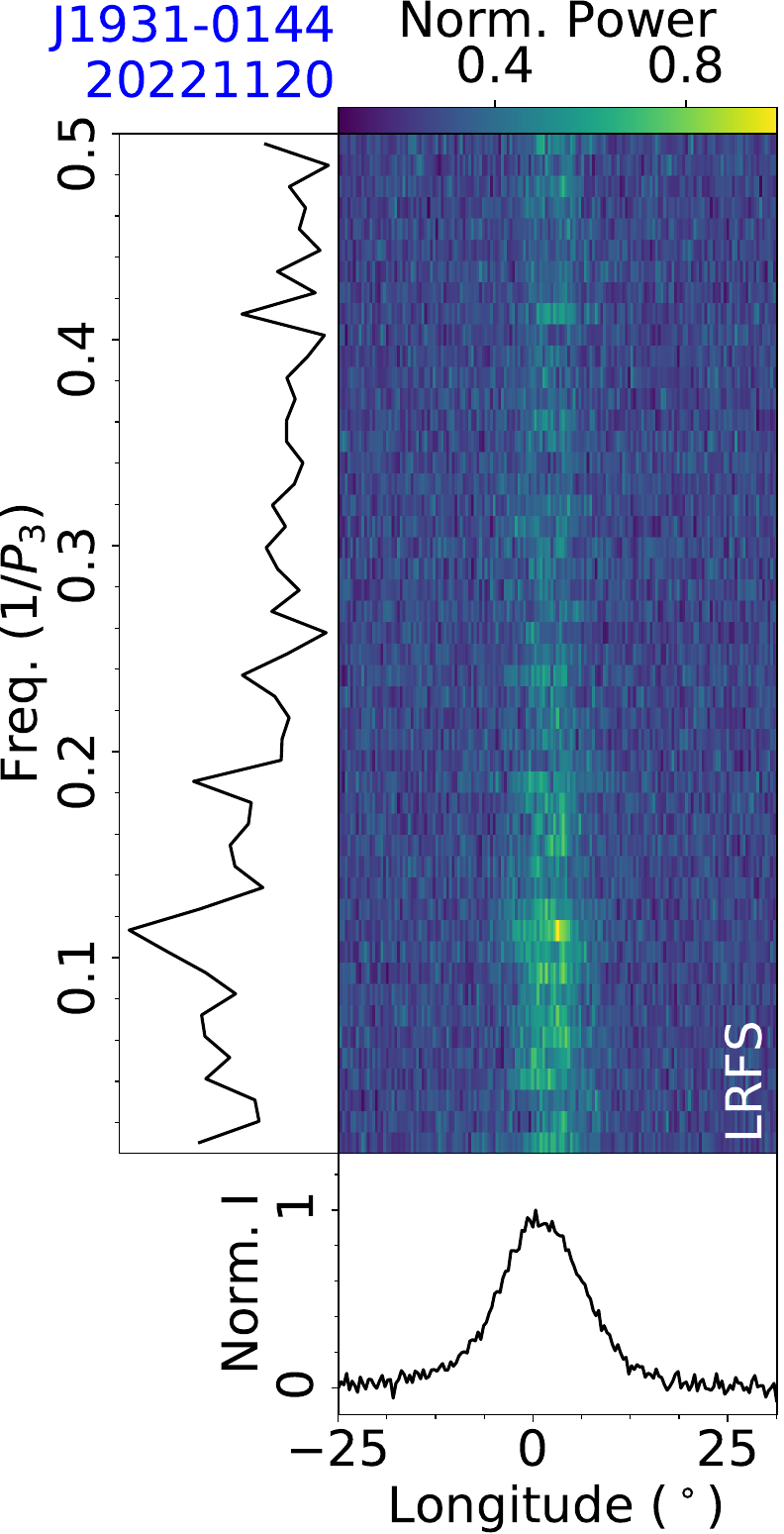}
\includegraphics[width=0.22\textwidth, angle=0]{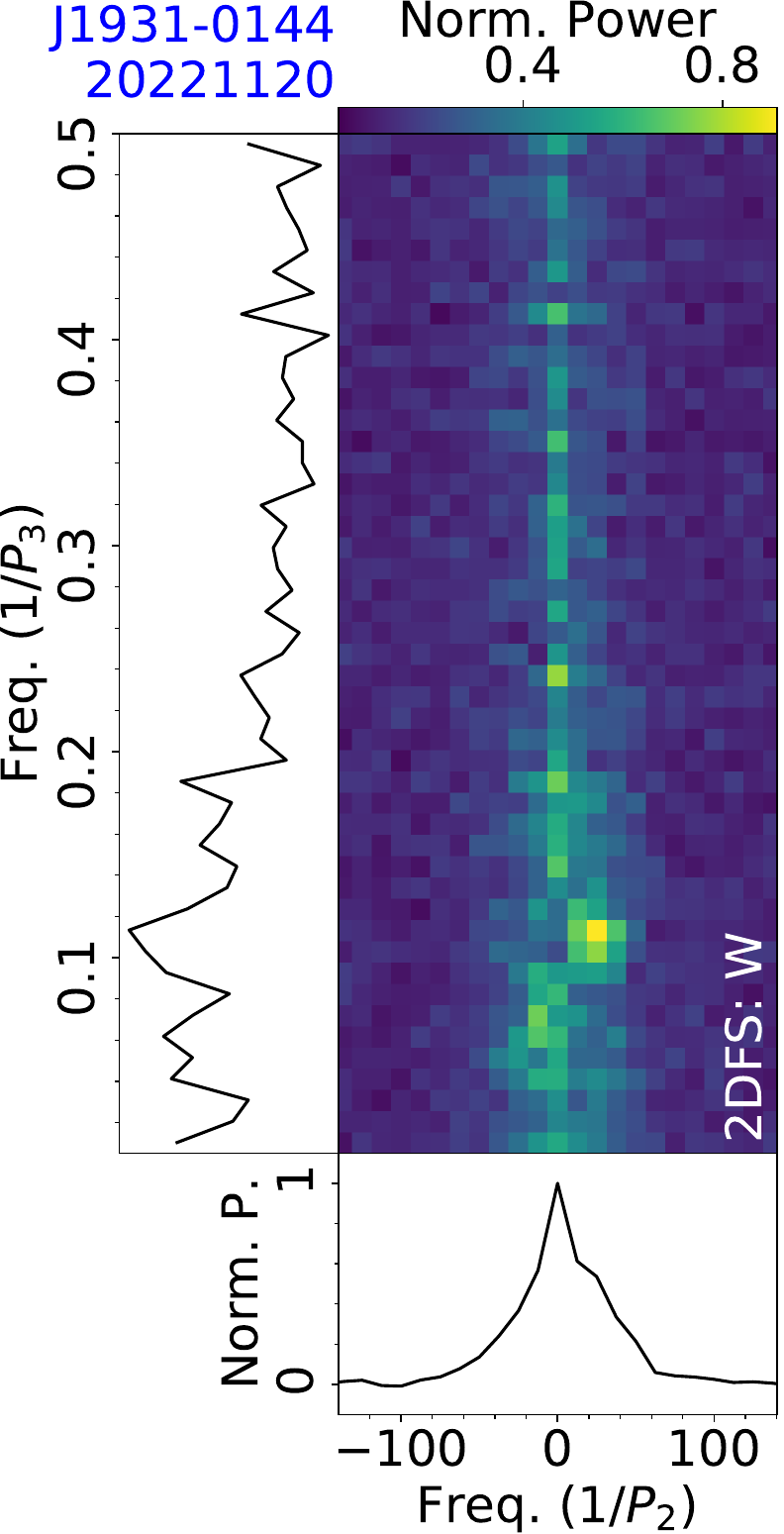}
\figcaption{Fluctuation analysis of PSR J1931-0144 for the observation on 20221120, with LRFS and 2DFS for the on-pulse region of a mean pulse profile.
\label{subfig:fluctu:J1931-0144}}
\end{figure}



\begin{figure}[htpb]
\centering
\includegraphics[width=0.22\textwidth, angle=0]{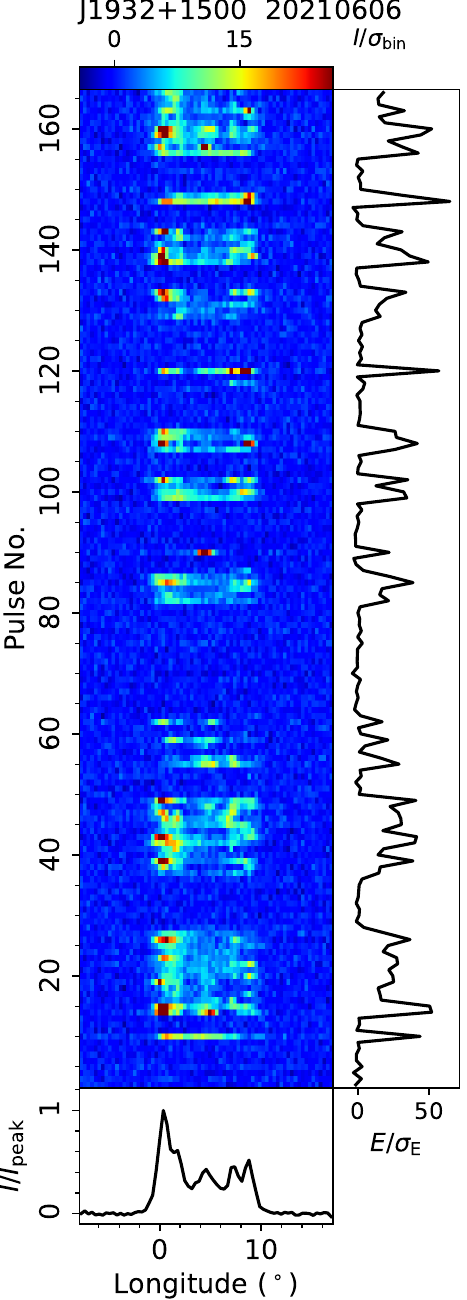}
\figcaption{Single pulse sequence of PSR J1932+1500 from the FAST observation on 20210606.
\label{subfig:TP:J1932+1500}}
\end{figure}

\begin{figure}[htpb]
\centering
\includegraphics[width=0.39\textwidth, angle=0]{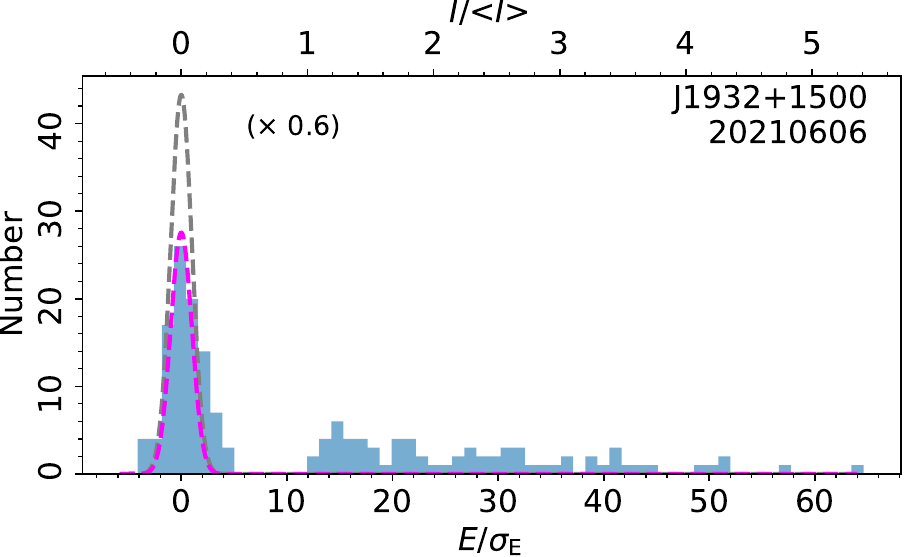}
\figcaption{On-pulse energy histogram of single pulses of PSR J1932+1500 from the FAST observation on 20210606.
\label{subfig:Hist:J1932+1500}}
\end{figure}

\subsection{J1930+1722}
\label{subsec:J1930+1722}

PSR J1930+1722 was discovered in a Parkes multibeam survey \citep{Lorimer2013}.

This pulsar was observed by FAST on 20250414 for 59 minutes, with a rotation period $P=1.6096$~s and a dispersion measure $D\!M=198.0~{\rm cm^{-3}\,pc}$ determined. The single pulse sequence and a zoomed-in view of pulses No. 100-300 are shown in Fig.~\ref{subfig:TP:J1930+1722}, illustrating the existence of nulling and subpulse drifting phenomena. The nulling fraction is estimated from the on-pulse integral energy histogram (Fig.~\ref{subfig:Hist:J1930+1722}) to be 26$\pm$2\%. From LRFS and 2DFS in Fig.~\ref{subfig:fluctu:J1930+1722}, the centroid of the negative drift feature is characterized by frequencies of $1/P_3=0.263\pm0.003$ and $1/P_2=-69\pm1$, corresponding to $P_3=3.81\pm0.04$ periods and $P_2=-5.2\pm0.1$ degrees. The low-frequency modulation feature in fluctuation spectra is from the nulling behavior, with a centroid periodicity of $25.8\pm0.5$ periods.

\begin{figure}[htpb]
\centering
\includegraphics[width=0.22\textwidth, angle=0]{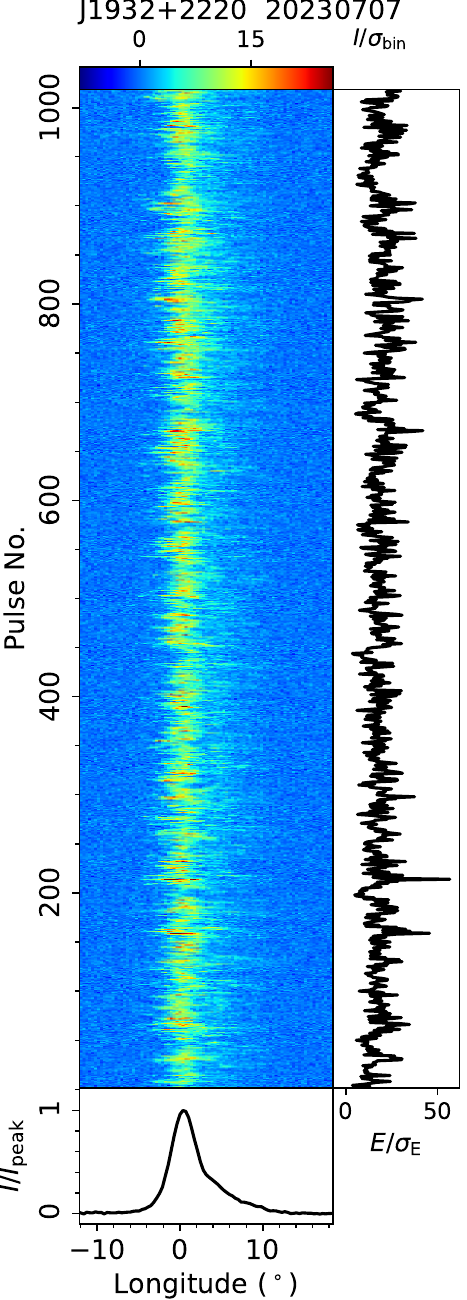}
\includegraphics[width=0.22\textwidth, angle=0]{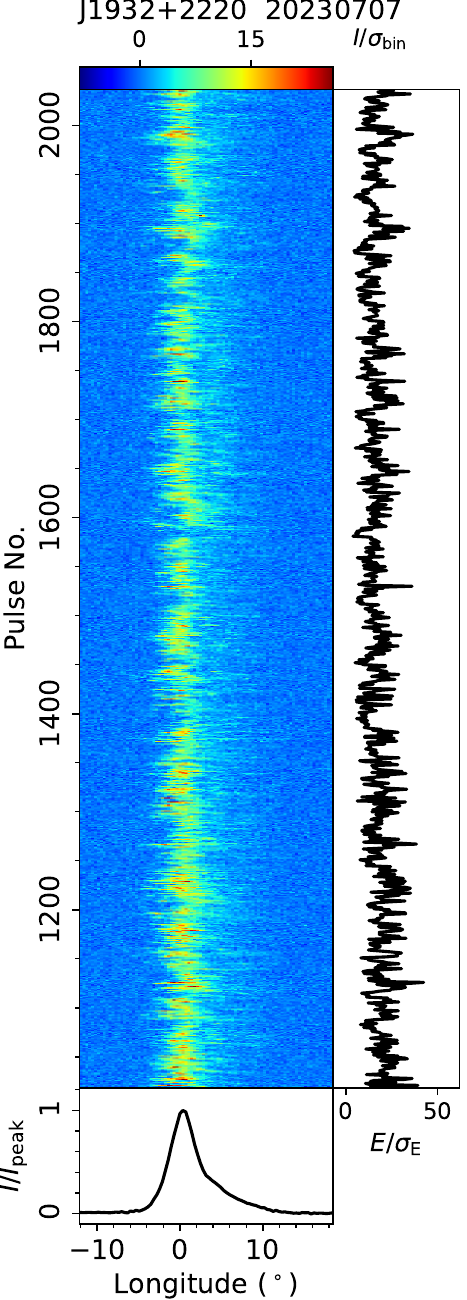}
\figcaption{Single pulse sequences of PSR J1932+2220 from the FAST observation on 20230707.
\label{subfig:TP:J1932+2220}}
\end{figure}

\begin{figure}[htpb]
\centering
\includegraphics[width=0.22\textwidth, angle=0]{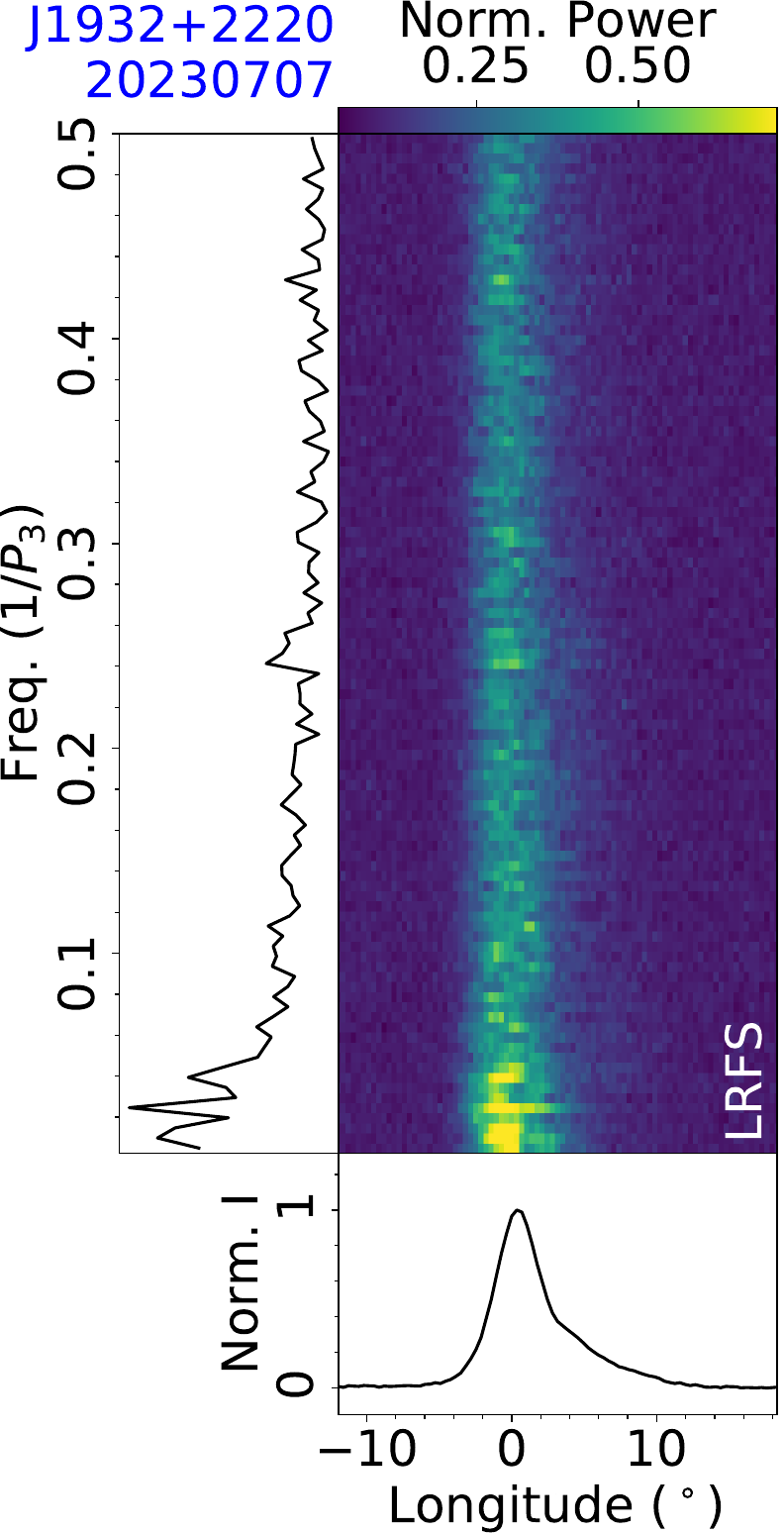}
\includegraphics[width=0.22\textwidth, angle=0]{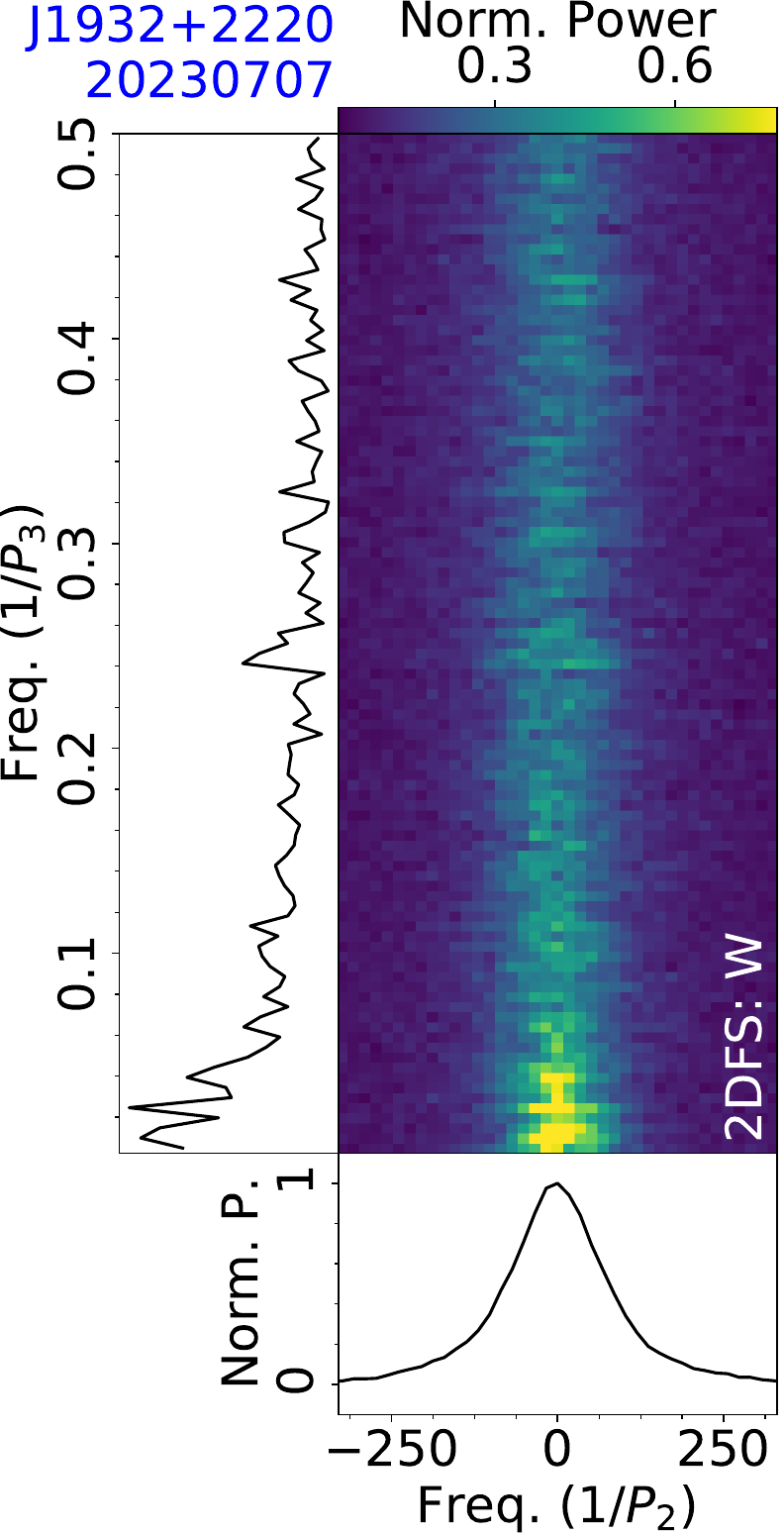}
\figcaption{Fluctuation analysis of PSR J1932+2220 for the observation on 20230707, with LRFS and 2DFS for the on-pulse region of the mean pulse profile.
\label{subfig:fluctu:J1932+2220}}
\end{figure}

\subsection{J1930+2836}
\label{subsec:J1930+2836}

PSR J1930+2836 was discovered by FAST in the Commensal Radio Astronomy FAST Survey (CRAFTS) (http://groups.bao.ac.cn/ism/CRAFTS/).

This pulsar was observed by FAST on 20211205 for 3 minutes, yielding a rotation period $P=1.1792$~s and a dispersion measure $D\!M=60.4~{\rm cm^{-3}\,pc}$. The single pulse sequence is displayed in Fig.~\ref{subfig:TP:J1930+2836}, illustrating the existence of subpulse drifting behavior. 
Drifting properties are derived from fluctuation spectra (Fig.~\ref{subfig:fluctu:J1930+2836}). The positive drift feature for the leading part in a mean pulse profile exhibits the centroid frequencies of $1/P_3=0.407\pm0.002$ and $1/P_2=88\pm5$, corresponding to $P_3=2.46\pm0.01$ periods and $P_2=4.1\pm0.2$ degrees. For the trailing profile part, the centroid of the drift feature is characterized by $1/P_3=0.416\pm0.005$ and $1/P_2=47\pm10$, yielding periodicities of $P_3=2.41\pm0.03$ periods and $P_2=8\pm2$ degrees.

\subsection{J1931+1439}
\label{subsec:J1931+1439}

PSR J1931+1439 was discovered in the Arecibo Pulsar-ALFA (PALFA) survey \citep{Lazarus2015}. 

This pulsar was observed by FAST on 20240129 for 30 minutes and 20240917 for 15 minutes. From the longer observation, a rotation period $P=1.7793$~s and a dispersion measure $D\!M=237.5~{\rm cm^{-3}\,pc}$ were derived. 
Here we present the analysis of the data on 20240129. Single pulse sequences are shown in Fig.~\ref{subfig:TP:J1931+1439}, displaying nulling as well as modulation behaviors. The nulling fraction is estimated from the on-pulse integral energy histogram (Fig.~\ref{subfig:Hist:J1931+1439}) to be 39$\pm$3\%. 
Fluctuation spectra are shown in Fig.~\ref{subfig:fluctu:J1931+1439}. 
For the leading part in the mean pulse profile, there is a preferred negative drift feature with centroid frequencies estimated to be $1/P_3=0.463\pm0.002$ and $1/P_2=-39\pm4$ in 2DFS, corresponding to periodicities of $P_3=2.16\pm0.01$ periods and $P_2=-9\pm1^\circ$. 
2DFS of the trailing profile part exhibits a modulation feature with the centroid of $1/P_3=0.462\pm0.001$, yielding $P_3=2.17\pm0.01$ periods.

\begin{figure}[htpb]
\centering
\includegraphics[width=0.22\textwidth, angle=0]{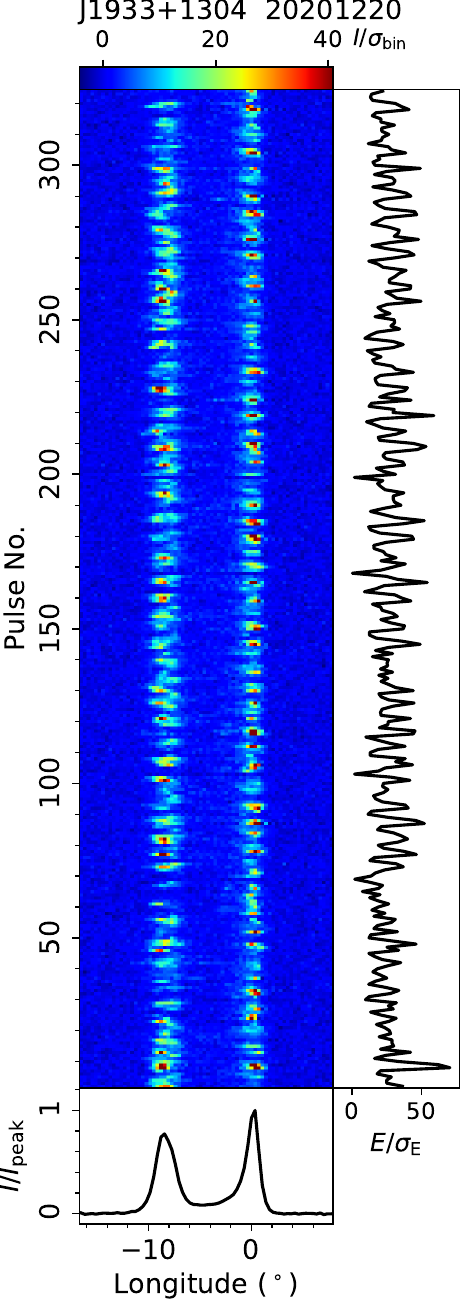}
\figcaption{Single pulse sequence of PSR J1933+1304 from the FAST observation on 20201220.
\label{subfig:TP:J1933+1304}}
\end{figure}

\begin{figure}[htpb]
\centering
\includegraphics[width=0.22\textwidth, angle=0]{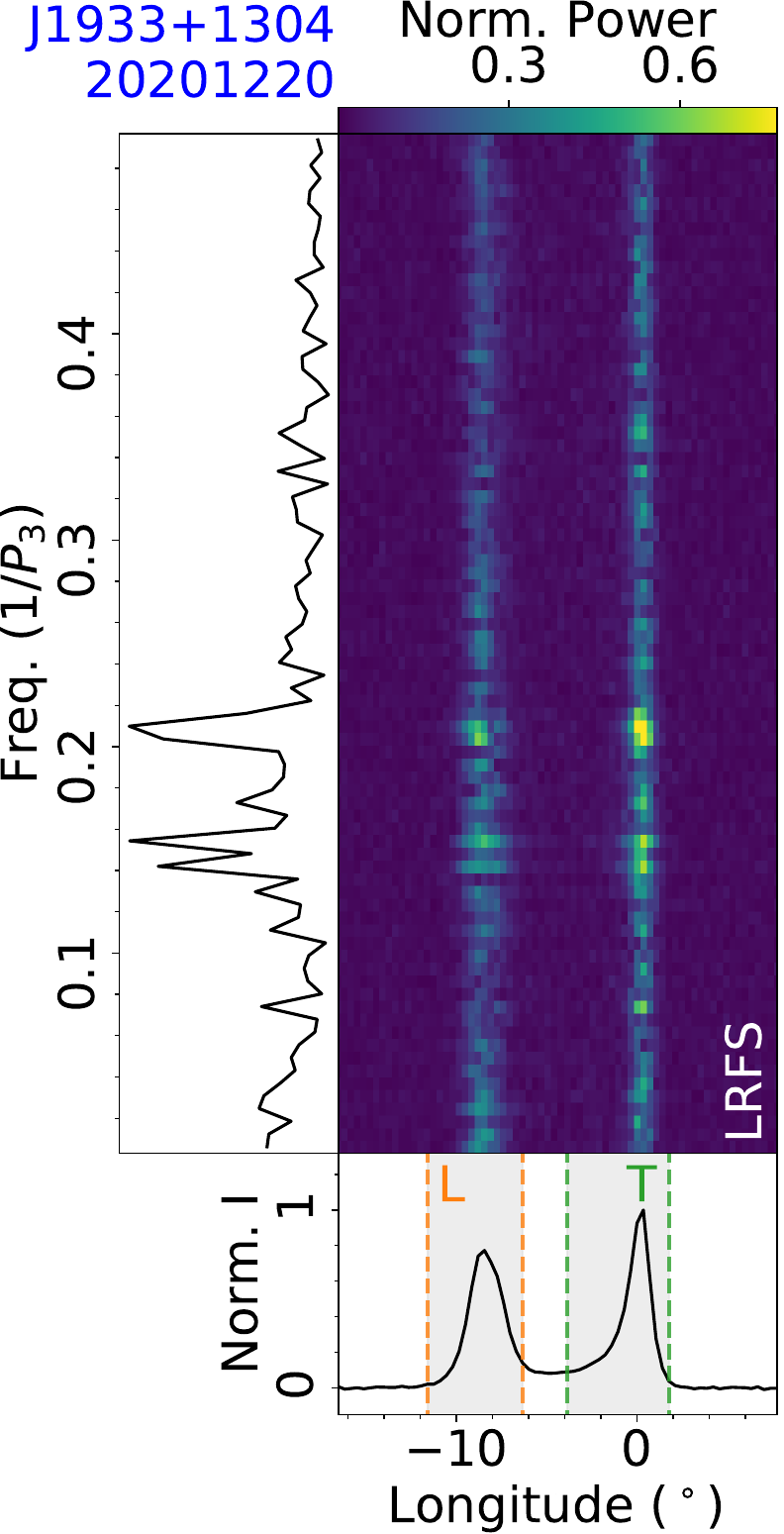}
\includegraphics[width=0.22\textwidth, angle=0]{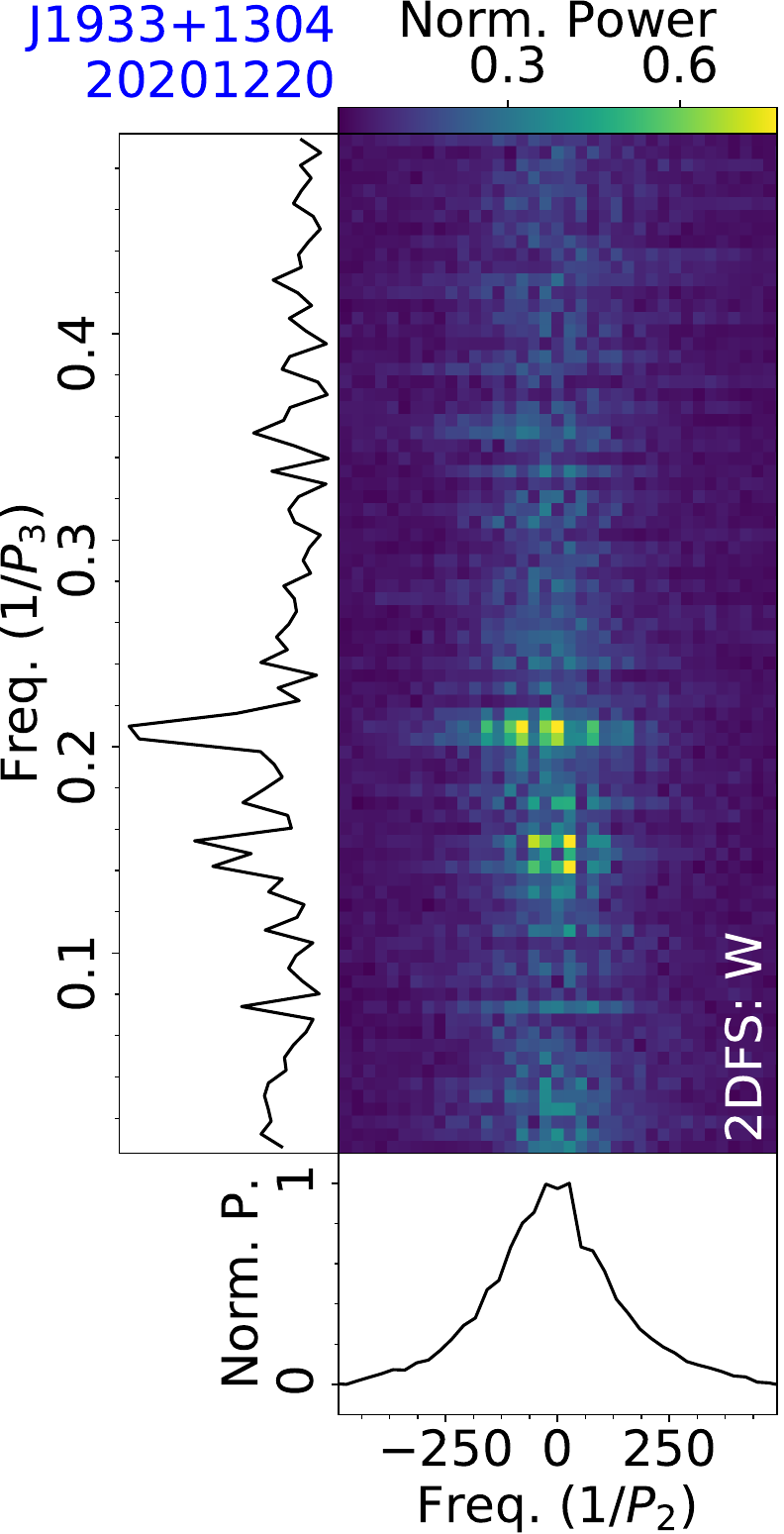}\\
\includegraphics[width=0.22\textwidth, angle=0]{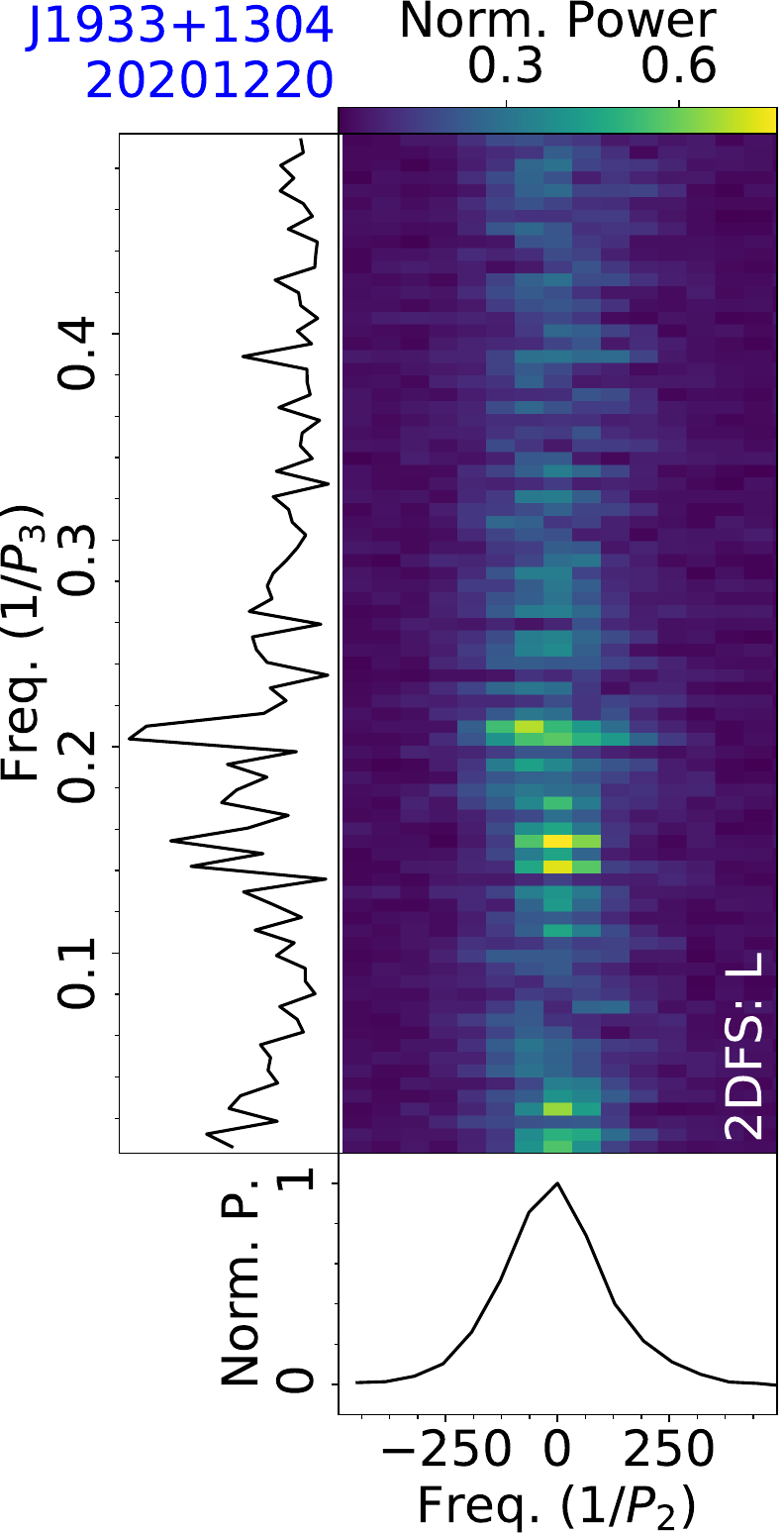}
\includegraphics[width=0.22\textwidth, angle=0]{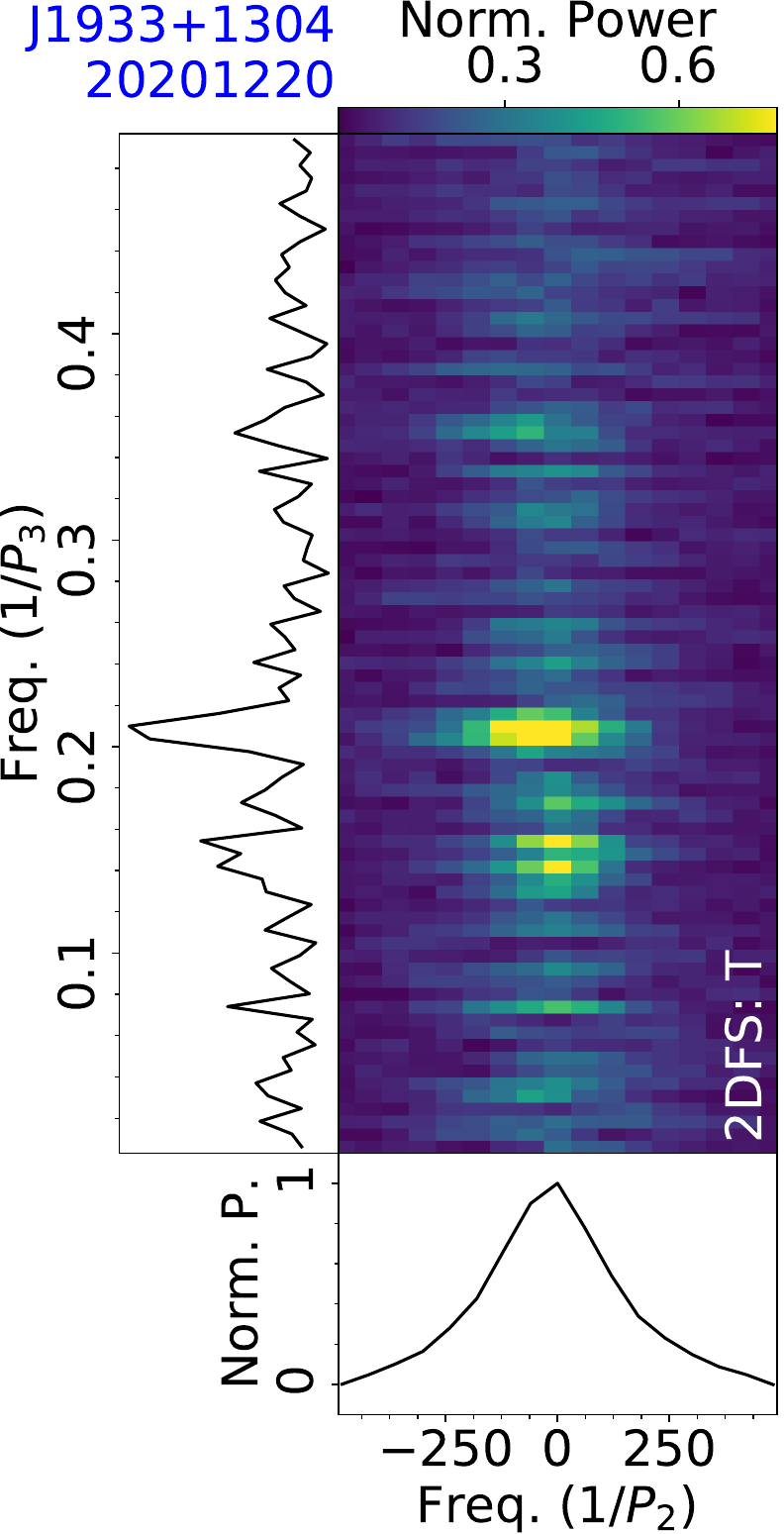}
\figcaption{Fluctuation analysis of PSR J1933+1304 from the FAST observation on 20201220, with LRFS (top-left), and 2DFS for the on-pulse phase region (top-right), leading part (bottom-left) and trailing part (bottom-right) of a mean pulse profile.
\label{subfig:fluctu:J1933+1304}}
\end{figure}

\subsection{J1931+30}
\label{subsec:J1931+30}

PSR J1931+30 was discovered by \citet{Camilo1996} using the Arecibo radio telescope at 430 MHz. 

This pulsar was observed by FAST on 20210701 for 5 minutes, deriving a rotation period $P=0.5821$~s and a dispersion measure $D\!M=53.3~{\rm cm^{-3}\,pc}$. 
Single pulse sequences are shown in Fig.~\ref{subfig:TP:J1931+30}, where the drift bands are not systematic. From LRFS and 2DFS in Fig.~\ref{subfig:fluctu:J1931+30}, the modulation frequency with time is wide, with the centroid of $1/P_3=0.077\pm0.001$ ($P_3=13.0\pm0.2$ periods) and $1/P_2=-14\pm1$ ($P_2=-26\pm2$ degrees).

\subsection{J1931-0144}
\label{subsec:J1931-0144}

PSR J1931-0144 was discovered by FAST \citep{Cameron2020,You2021}.

This pulsar was observed by FAST on 20221120 for 5 minutes, and a rotation period $P=0.5937$~s and a dispersion measure $D\!M=37.7~{\rm cm^{-3}\,pc}$ were derived. Single pulse sequences in Fig.~\ref{subfig:TP:J1931-0144} show the subpulse drifting phenomenon. From fluctuation spectra in Fig.~\ref{subfig:fluctu:J1931-0144}, centroid frequencies of the positive drift feature are $1/P_3=0.109\pm0.001$ and $1/P_2=24\pm1$, corresponding to periodicities of $P_3=9.2\pm0.1$ periods and $P_2=15\pm1$ degrees.



\subsection{J1932+1500}
\label{subsec:J1932+1500}

PSR J1932+1500 was discovered in the Arecibo L-band Feed Array pulsar survey \citet{Lyne2017}. 

This pulsar was observed by FAST on 20210606 for 5 minutes, deriving a rotation period $P=1.8641$~s and a dispersion measure $D\!M=91.3~{\rm cm^{-3}\,pc}$. 
The single pulse sequence shown in Fig.~\ref{subfig:TP:J1932+1500} displays the nulling phenomenon. The nulling fraction of this observation is estimated from the on-pulse integral energy histogram (Fig.~\ref{subfig:Hist:J1932+1500}) to be 38$\pm$3\%.

\subsection{J1932+2220}
\label{subsec:J1932+2220}

PSR J1932+2220 was discovered by \citet{Hulse1975} using the Arecibo telescope. \citet{Song2023} reported a $P_3$-only feature with $P_3=41\pm8$ periods. 

This pulsar was observed by FAST on 20230707 for 5 minutes, deriving a rotation period $P=0.1445$~s and a dispersion measure $D\!M=218.9~{\rm cm^{-3}\,pc}$. 
Single pulse sequences in Fig.~\ref{subfig:TP:J1932+2220} show a modulation signature. From the fluctuation in Fig.~\ref{subfig:fluctu:J1932+2220}, the low-frequency modulation feature has a centroid at $1/P_3=0.0261\pm0.0004$, corresponding to $P_3=38\pm1$ periods. This value is consistent with the $P_3$ reported by \citet{Song2023}.

\subsection{J1933+1304}
\label{subsec:J1933+1304}

PSR J1933+1304 was discovered by \citet{Hulse1975} using the Arecibo telescope.
From previous work, there are 4.92-period and 6.6-period modulations common to both components \citet{Rankin2023,Song2023}.

The observation of the FAST on 20201220 for 5 minutes, deriving a rotation period $P=0.9284$~s and a dispersion measure $D\!M=176.3~{\rm cm^{-3}\,pc}$. 
Single-pulse behavior of the FAST observation is consistent with previous studies. 
Single pulse sequences in Fig.~\ref{subfig:TP:J1933+1304} display the modulation behavior. The fluctuation spectra are shown in Fig.~\ref{subfig:fluctu:J1933+1304}, and there are two features in 2DFS for both the leading and trailing parts in the mean pulse profile. 
For 2DFS of the leading profile part, there is a negative subpulse drift feature with the centroid frequencies of $1/P_3=0.207\pm0.001$ ($P_3=4.83\pm0.02$ periods) and $1/P_2=-37\pm7$ ($P_2=-10\pm2^\circ$), as well as an amplitude variation feature without any significant phase modulation with the centroid frequency of $1/P_3=0.151\pm0.001$ ($P_3=6.61\pm0.03^\circ$). 
2DFS of the trailing part also exhibits a negative drift feature with the centroid of $1/P_3=0.2076\pm0.0004$ ($P_3=4.82\pm0.01$ periods) and $1/P_2=-43\pm5$ ($P_2=-8\pm1^\circ$), and a modulation feature with the centroid characterized by $1/P_3=0.145\pm0.001$ ($P_3=6.91\pm0.03$ periods). 
Additionally, there are 4 single pulses whose on-pulse integral energies are less than 3$\sigma_{\rm E}$, which seems to be related to short-duration nulling behavior.

\begin{figure}[htpb]
\centering
\includegraphics[width=0.22\textwidth, angle=0]{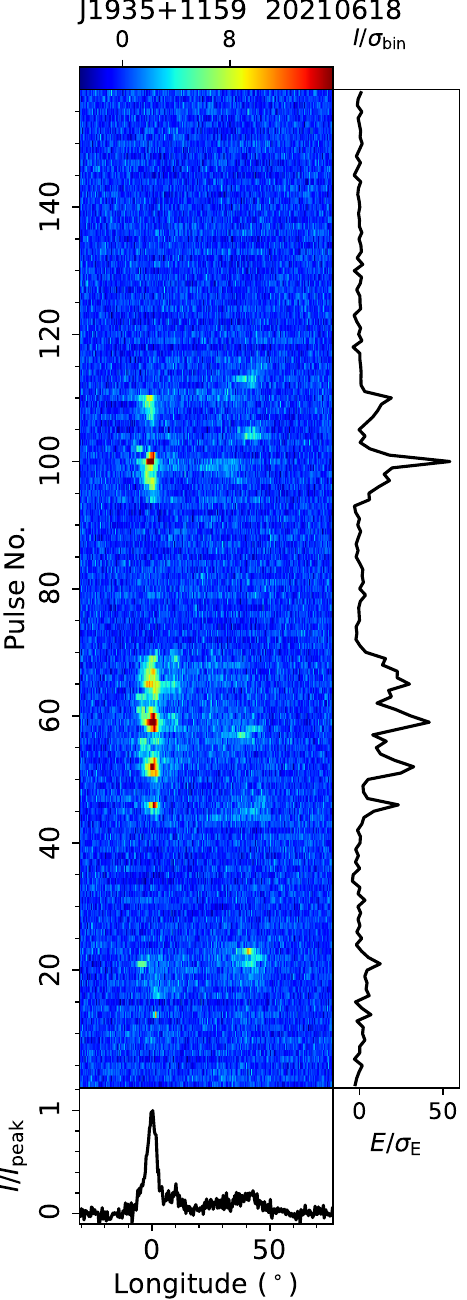}
\figcaption{Single pulse sequence of PSR J1935+1159 from the FAST observation on 20210618.
\label{subfig:TP:J1935+1159}}
\end{figure}

\begin{figure}[htpb]
\centering
\includegraphics[width=0.39\textwidth, angle=0]{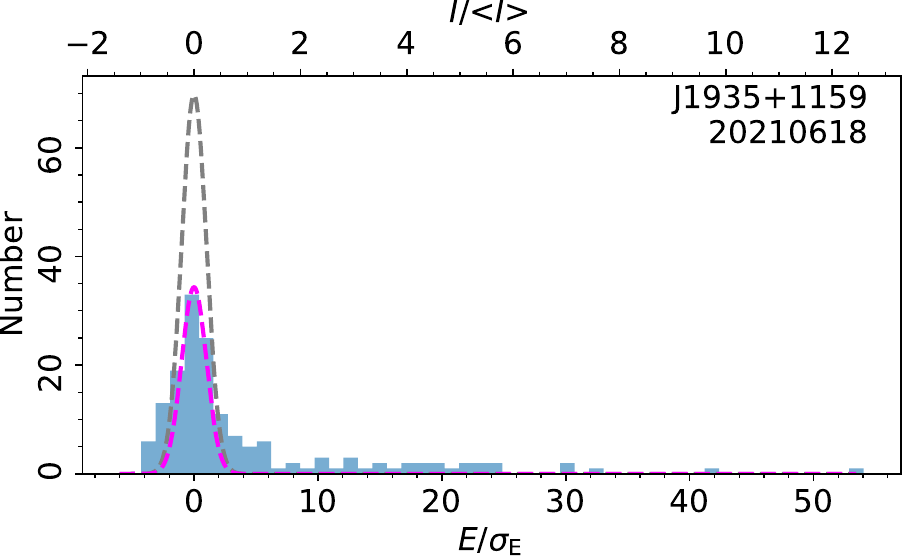}
\figcaption{Energy histogram of single pulses integrated in longitude range between -6.0$^\circ$ and 3.2$^\circ$ of PSR J1935+1159 from the FAST observation on 20210618.
\label{subfig:Hist:J1935+1159}}
\end{figure}

\begin{figure}[htpb]
\centering
\includegraphics[width=0.22\textwidth, angle=0]{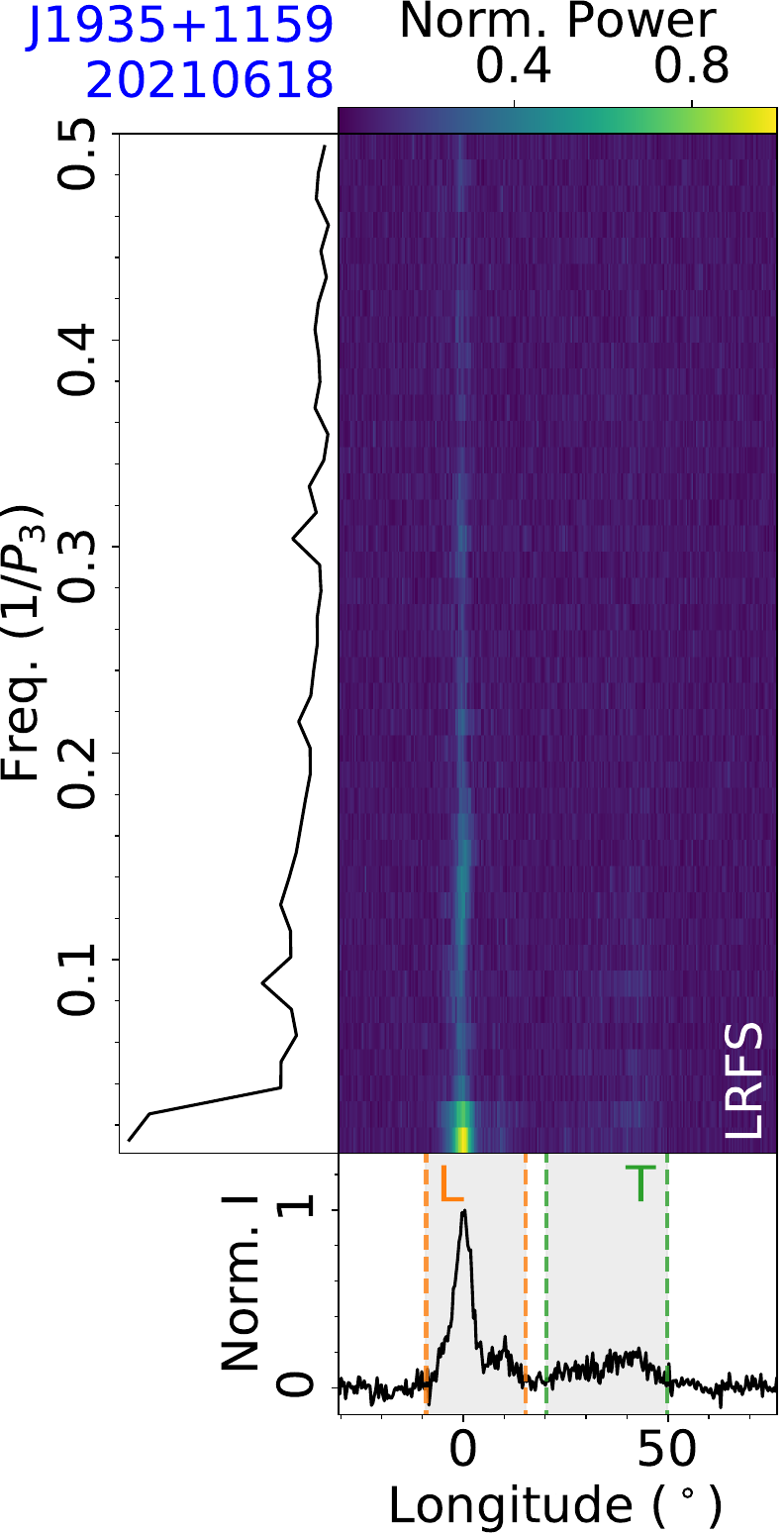}
\includegraphics[width=0.22\textwidth, angle=0]{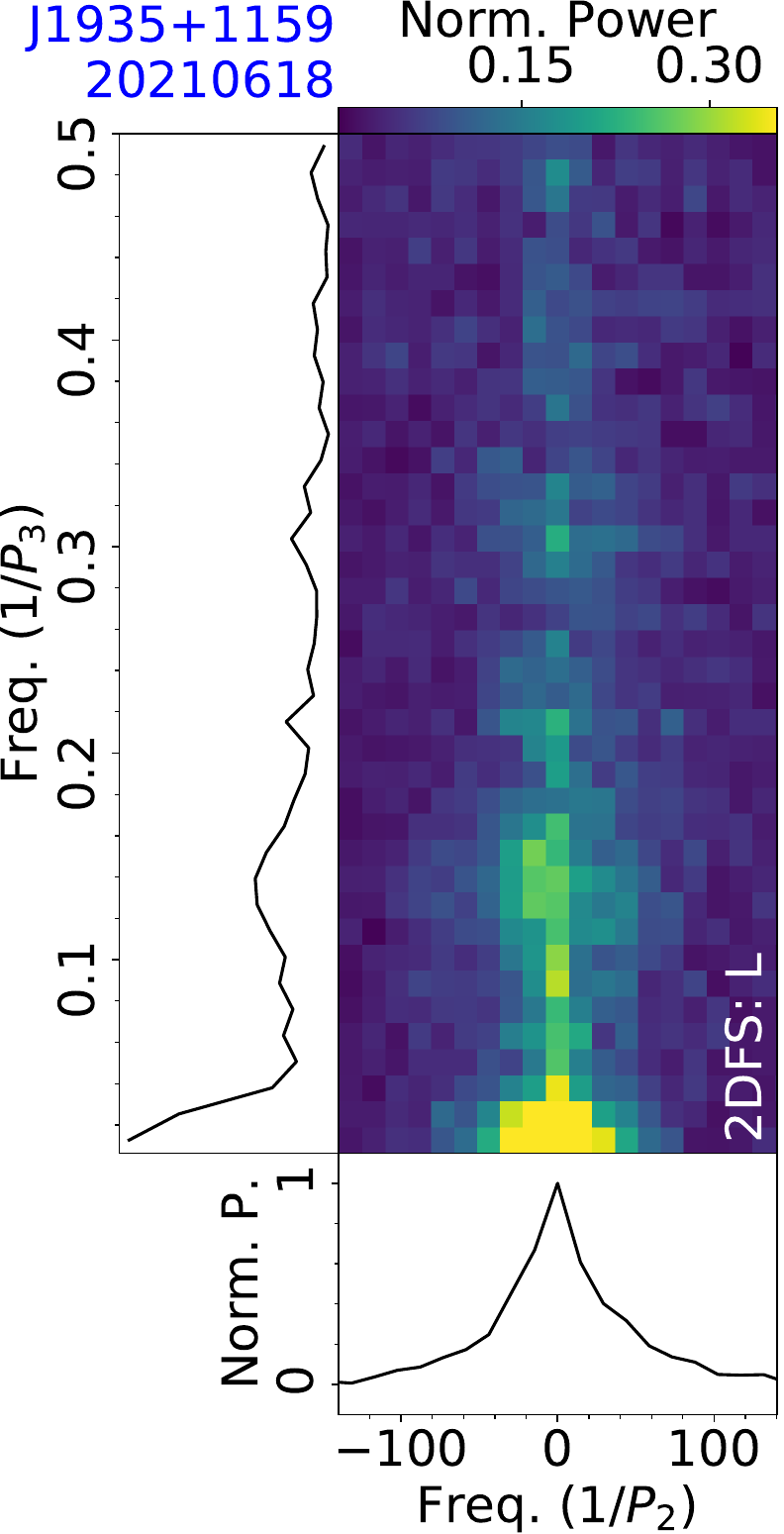}
\figcaption{Fluctuation analysis of PSR J1935+1159 for the observation on 20210618, with LRFS and 2DFS for the leading part of a mean pulse profile.
\label{subfig:fluctu:J1935+1159}}
\end{figure}

\begin{figure}[htpb]
\centering
\includegraphics[width=0.22\textwidth, angle=0]{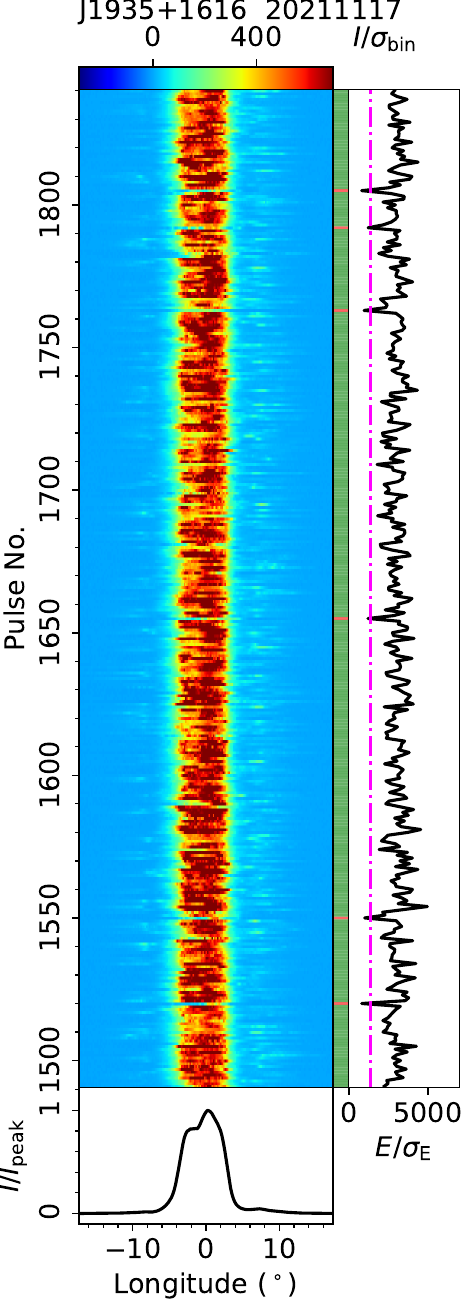}
\includegraphics[width=0.22\textwidth, angle=0]{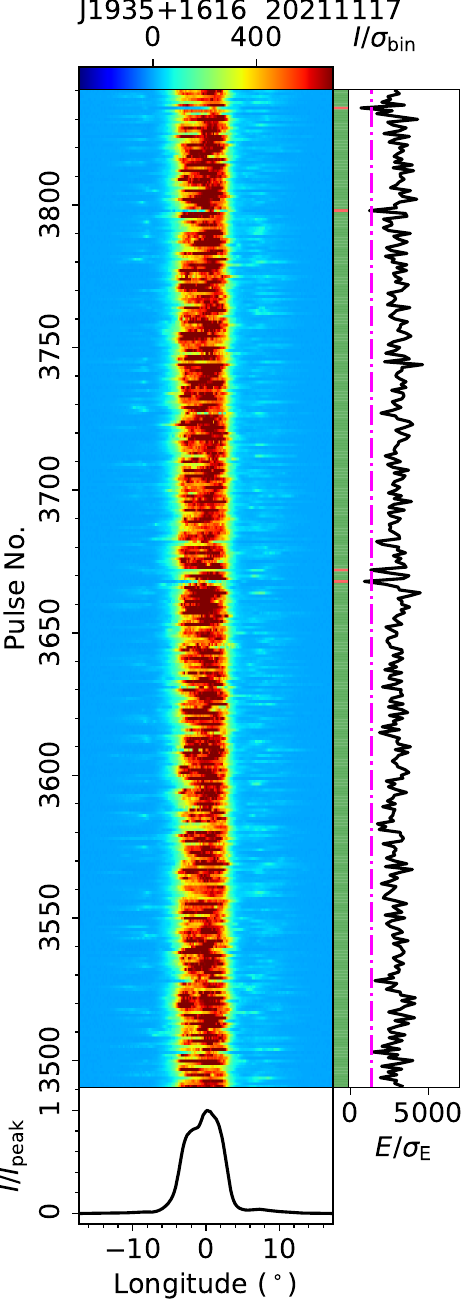}
\figcaption{Two single pulse segments of pulses No. 1491-1840 and 3491-3840 of PSR J1935+1616 from the FAST observation on 20211117. 
The green and red bars represent normal or weak emission modes. In the right subpanel, the on-pulse energy variation is plotted against period, with dashed lines for thresholds to distinguish two states.
\label{subfig:TP:J1935+1616}}
\end{figure}

\begin{figure}[htpb]
\centering
\includegraphics[width=0.39\textwidth, angle=0]{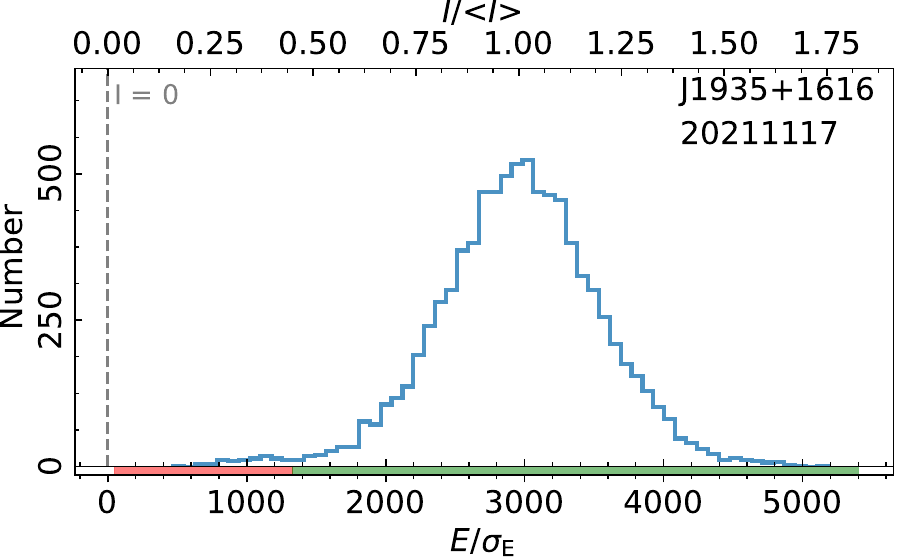}
\figcaption{Energy histogram of single pulses integrated over the central component of PSR J1935+1616, from the FAST observation on 20211117. The red and green bars indicate the weak and normal emission modes.
\label{subfig:Hist:J1935+1616}}
\end{figure}

\begin{figure}[htpb]
\centering
\includegraphics[width=0.42\textwidth, angle=0]{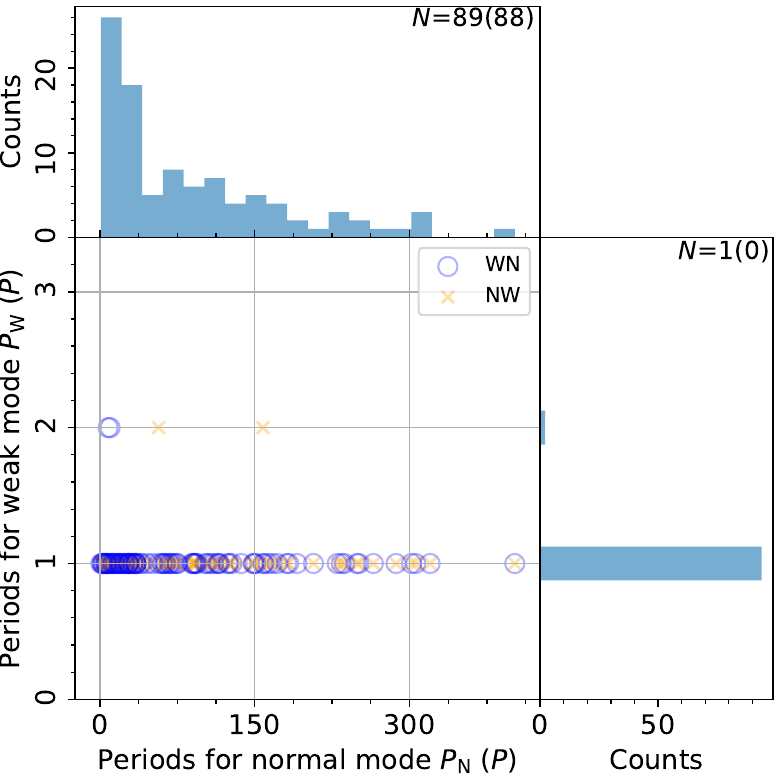}
\figcaption{Distribution of period numbers for continuous normal emission $P_{\rm N}$ against period numbers for adjacent weak emission $P_{\rm W}$ of PSR J1935+1616 observed by FAST on 20211117, as well as the duration histograms for the normal state and weak state shown in the top and right panels, respectively. 
\label{subfig:scaleHist:J1935+1616}}
\end{figure}

\begin{figure}[htpb]
\centering
\includegraphics[width=0.37\textwidth, angle=0]{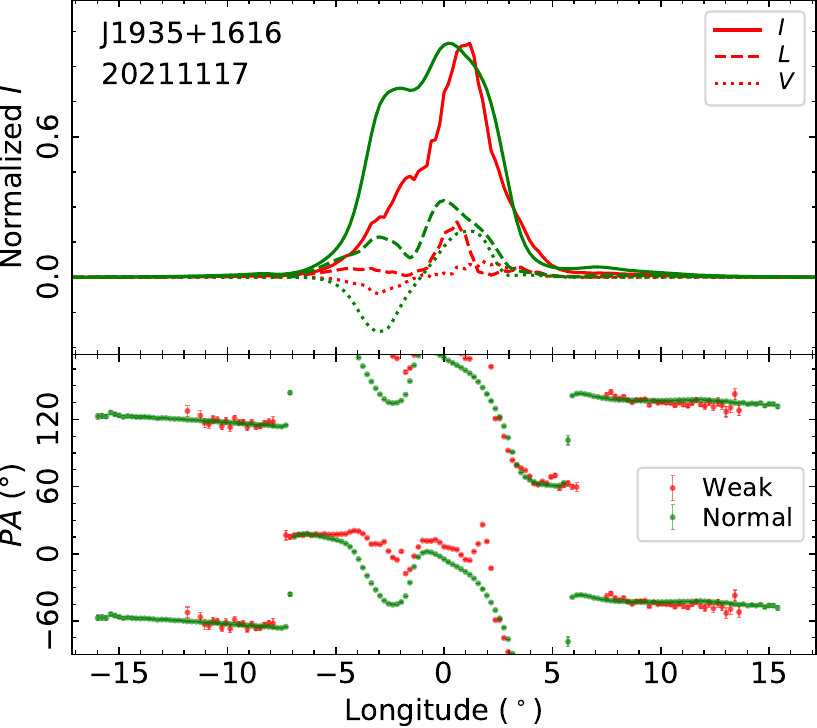}
\figcaption{Mean polarization profiles (the top panel) for the weak (red) and normal (green) emission modes of PSR J1935+1616 observed on 20211117, as well as the averaged PA curves (the bottom panel). Profiles in the top panel are normalized by their respective peaks. \label{subfig:PolModes:J1935+1616}}
\end{figure}

\begin{figure}[htpb]
\centering
\includegraphics[width=0.22\textwidth, angle=0]{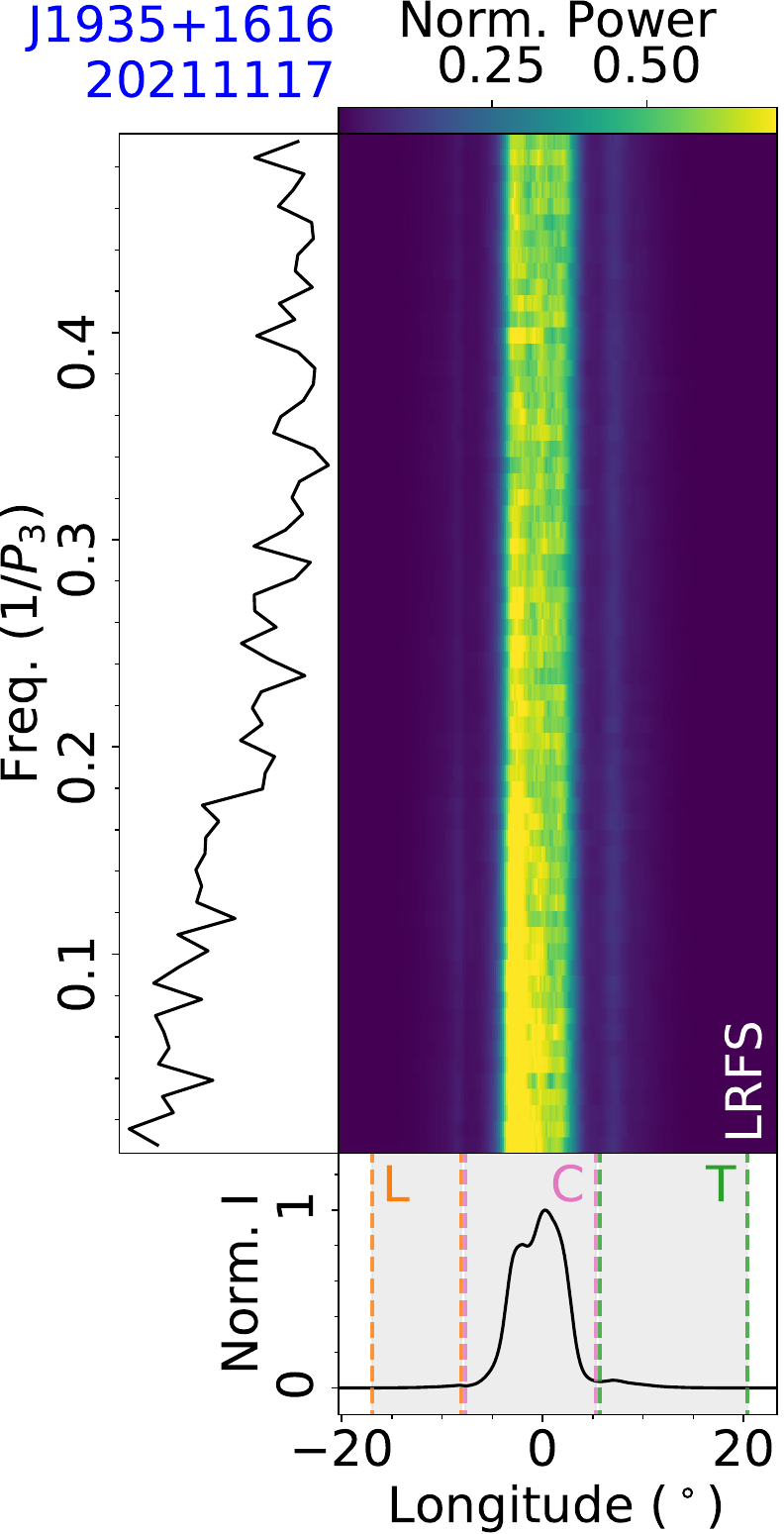}
\includegraphics[width=0.22\textwidth, angle=0]{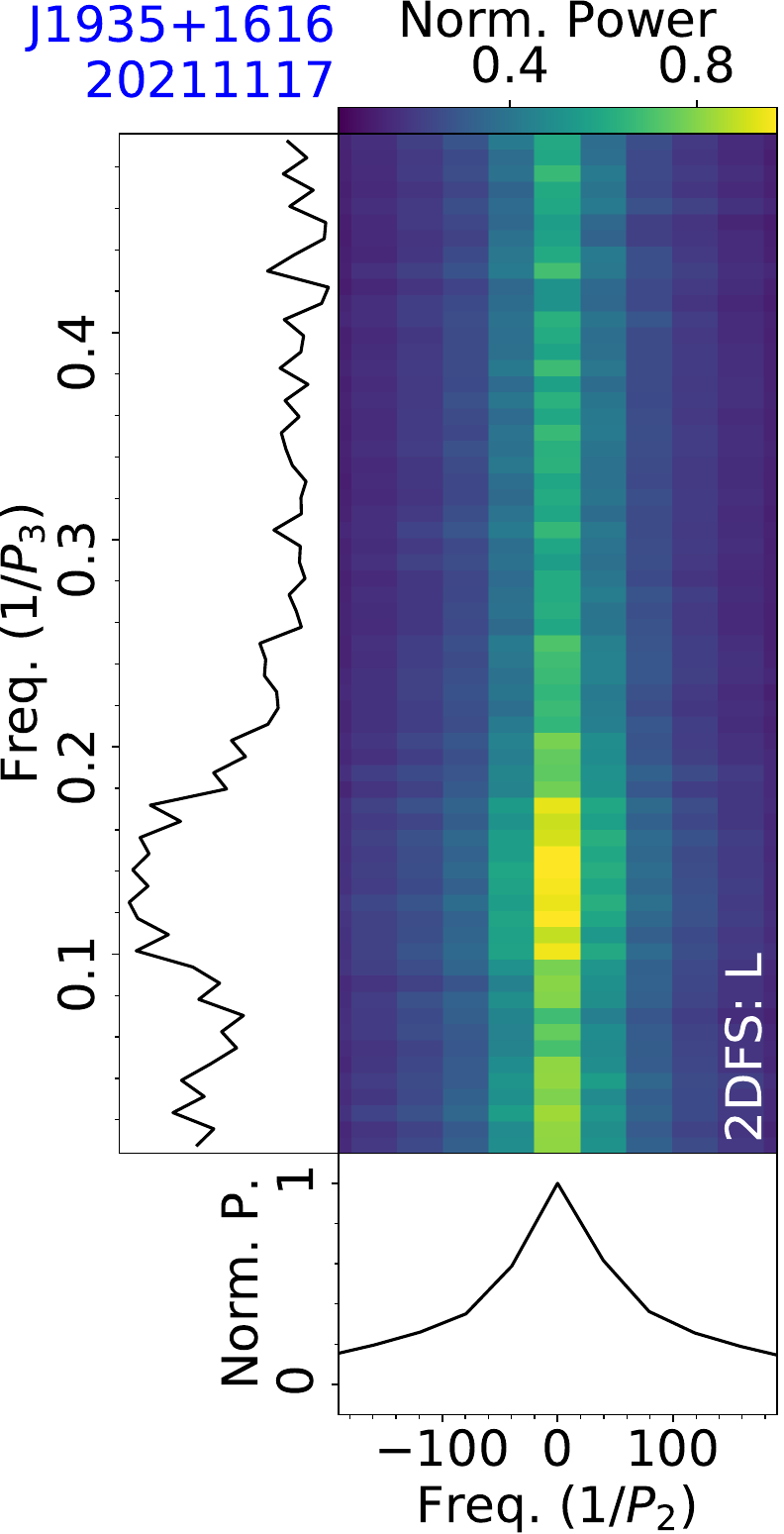}\\
\includegraphics[width=0.22\textwidth, angle=0]{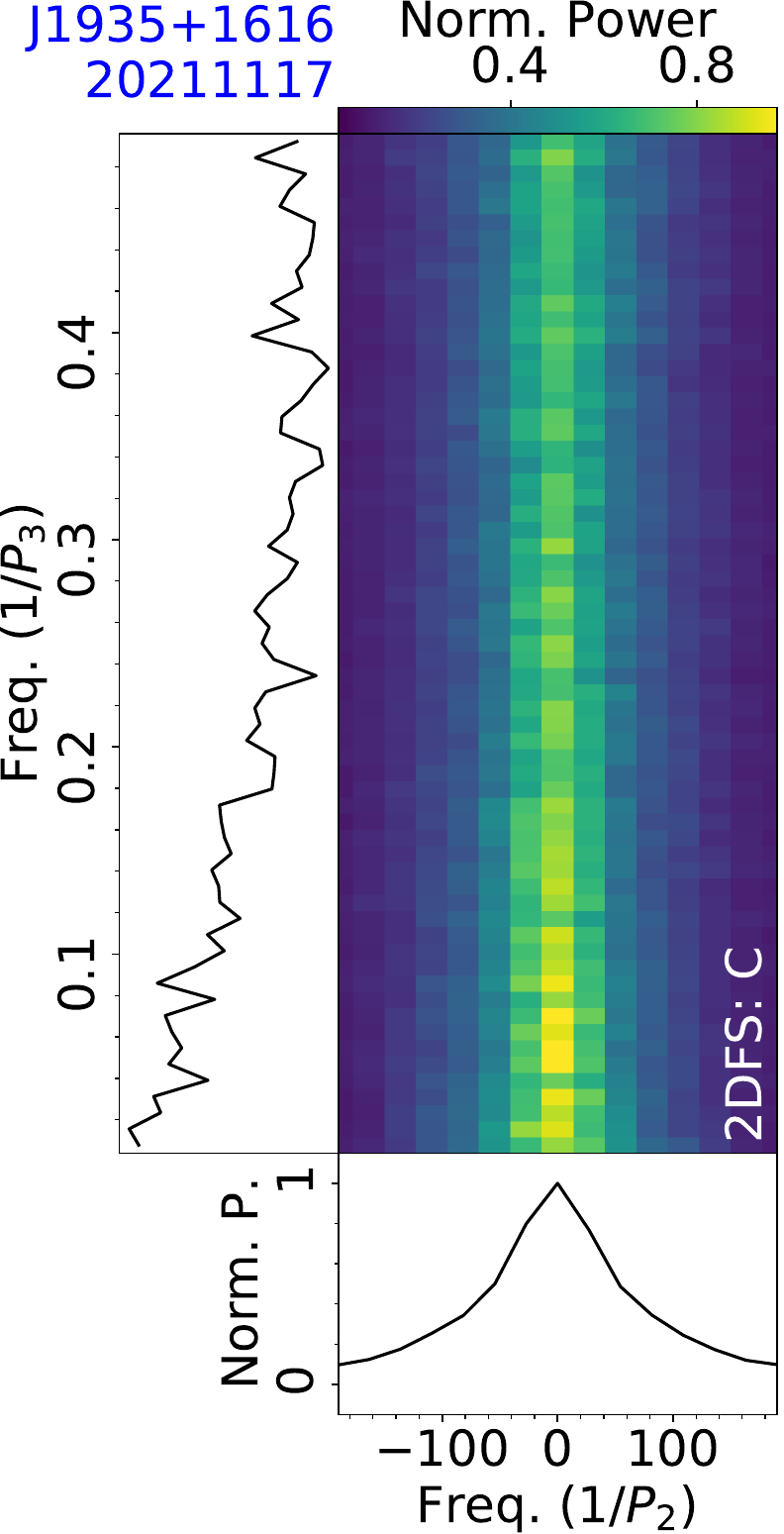}
\includegraphics[width=0.22\textwidth, angle=0]{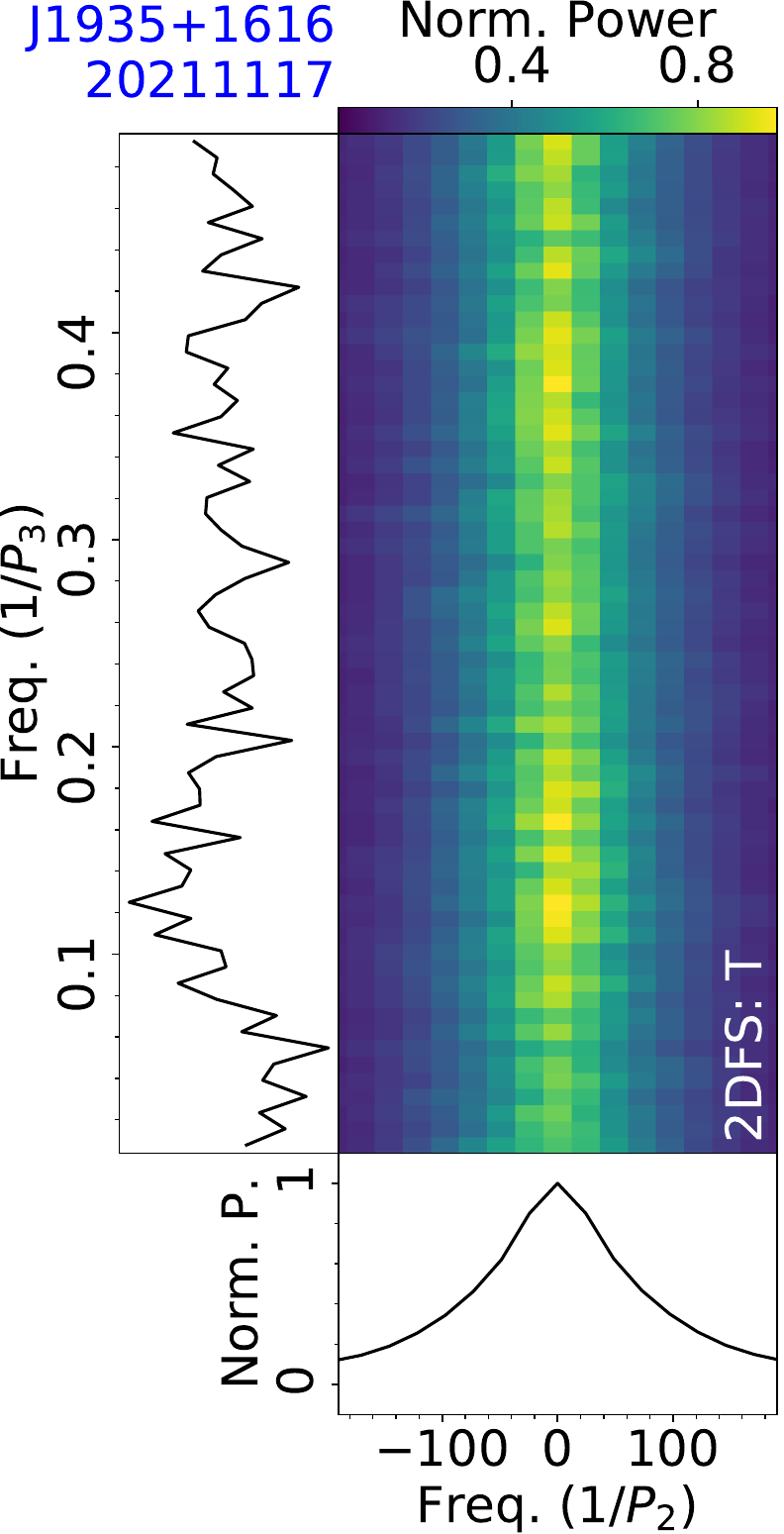}
\figcaption{Fluctuation analysis of PSR J1935+1616 for the observation on 20211117, with LRFS (top-left), and 2DFS for the leading part (top-right), central part (bottom-left) and trailing part (bottom-right) of a mean pulse profile.
\label{subfig:fluctu:J1935+1616}}
\end{figure}

\subsection{J1935+1159}
\label{subsec:J1935+1159}

PSR J1935+1159 was discovered in the Arecibo–Caltech drift-scan survey \citep{Chandler2003}. From previous studies, this pulsar has nulling and modulation phenomena. Long nulls were shown by \citet{Brinkman2018}. 
\citet{Song2023} reported drift feature of the leading component and $P_3$-only feature of the trailing component. 

This pulsar was observed by FAST on 20210618 for 5 minutes, deriving a rotation period $P=1.9395$~s and a dispersion measure $D\!M=189.6~{\rm cm^{-3}\,pc}$. 
Nulling behavior is apparent in the single pulse sequence (Fig.~\ref{subfig:TP:J1935+1159}), and the nulling fraction is estimated to be 38.2\% from the integral energy histogram of the longitude range between -6.0$^\circ$ and 3.2$^\circ$ (Fig.~\ref{subfig:Hist:J1935+1159}). In 2DFS of the leading part in the mean pulse profile (Fig.~\ref{subfig:fluctu:J1935+1159}), there is a negative drift feature with the centroid frequencies of $1/P_3=0.139\pm0.001$ and $1/P_2=-14\pm2$, corresponding to $P_3=7.2\pm0.1$ periods and $P_2=-25\pm3^\circ$.

\begin{figure}[htpb]
\centering
\includegraphics[width=0.22\textwidth, angle=0]{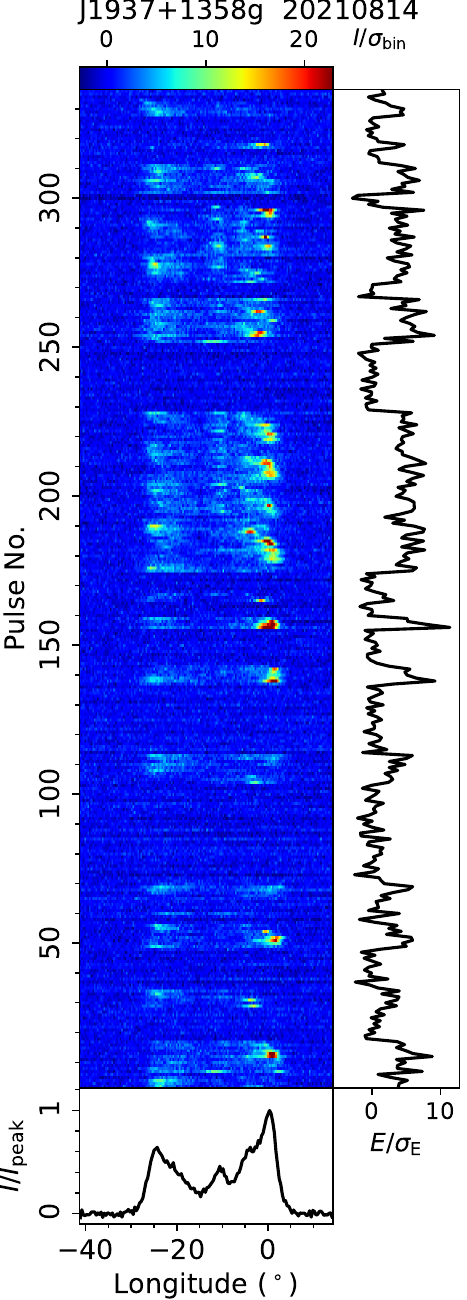}
\figcaption{Single pulse sequence of PSR J1937+1358g from the FAST observation on 20210814.
\label{subfig:TP:J1937+1358g}}
\end{figure}

\begin{figure}[htpb]
\centering
\includegraphics[width=0.39\textwidth, angle=0]{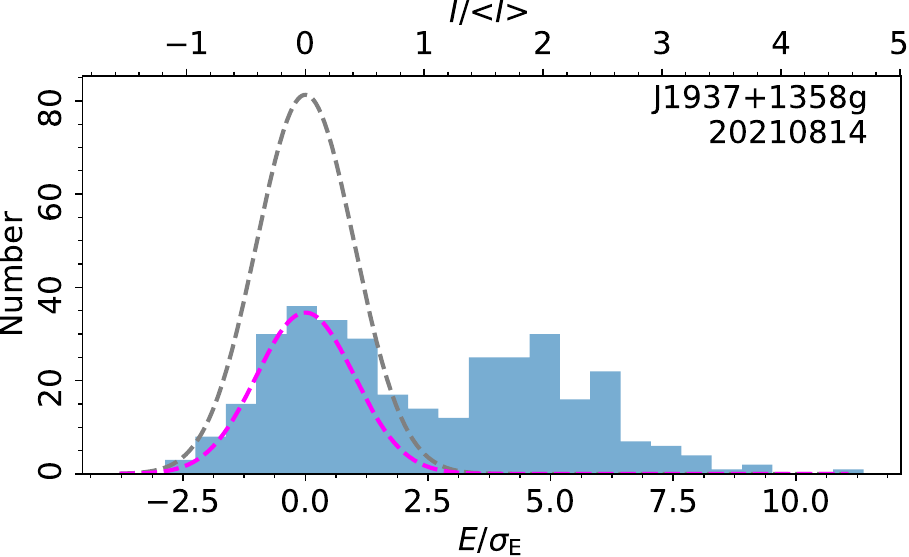}
\figcaption{On-pulse energy histogram of single pulses of PSR J1937+1358g from the FAST observation on 20210814.
\label{subfig:Hist:J1937+1358g}}
\end{figure}

\begin{figure}[htpb]
\centering
\includegraphics[width=0.22\textwidth, angle=0]{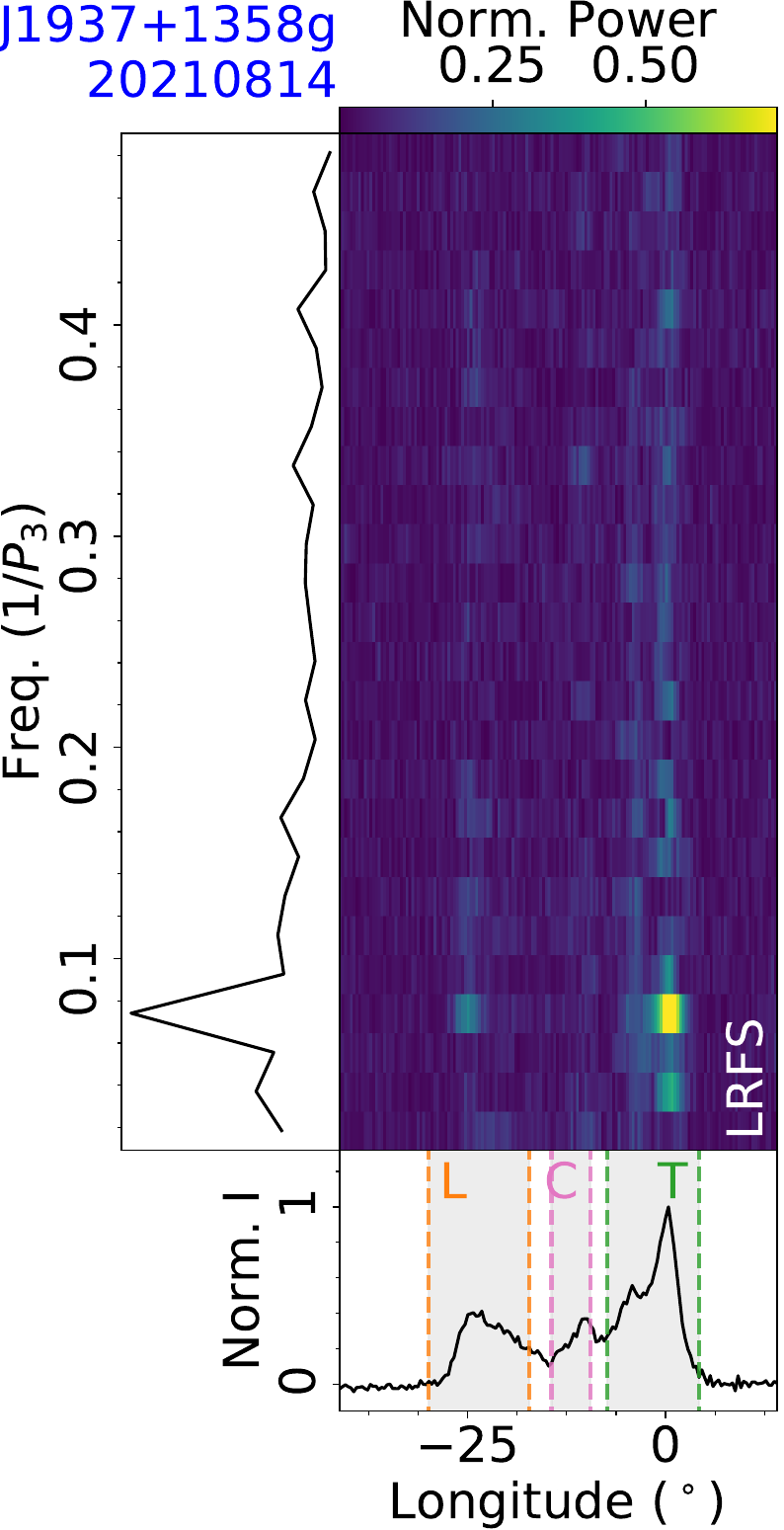}
\includegraphics[width=0.22\textwidth, angle=0]{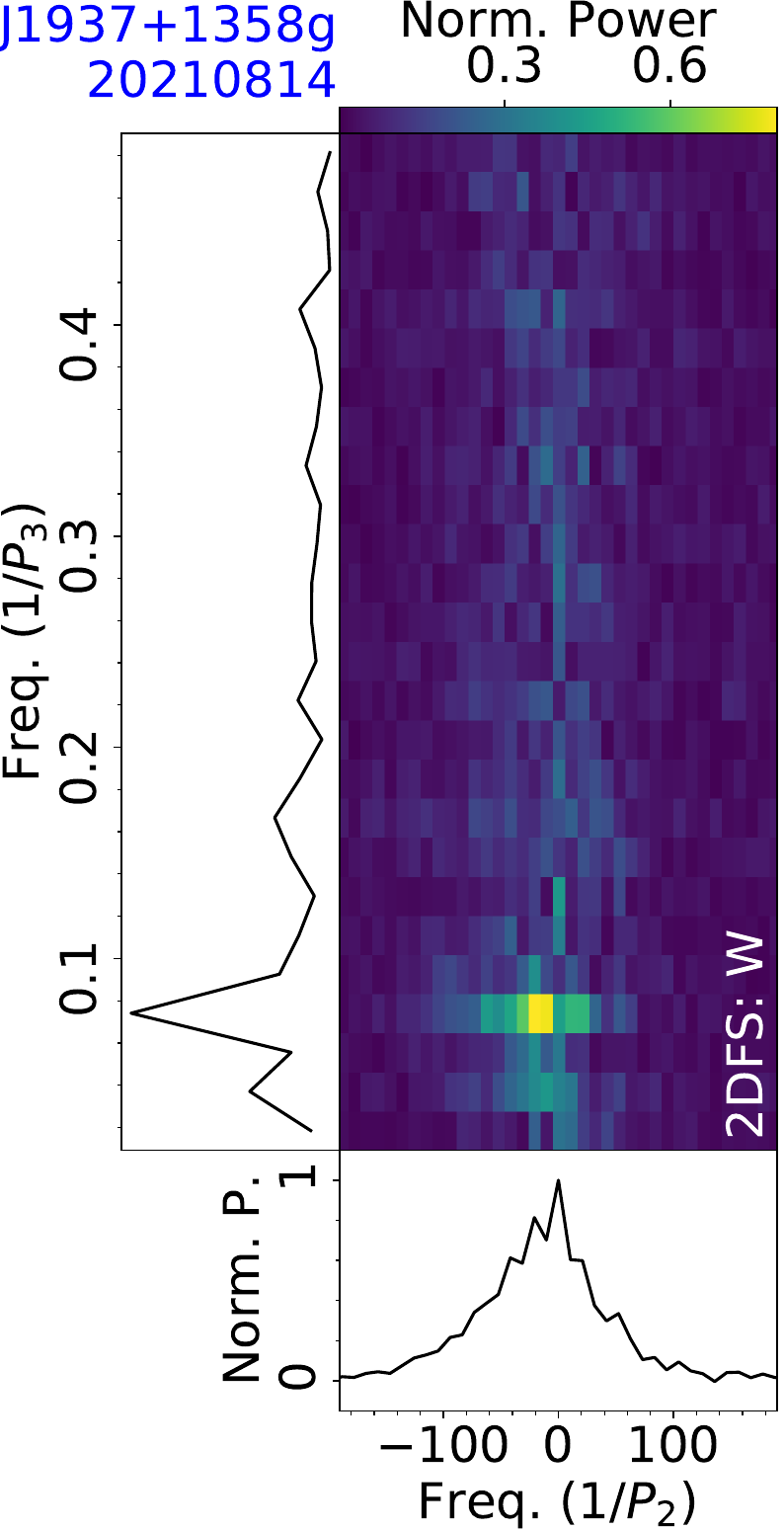}\\
\includegraphics[width=0.22\textwidth, angle=0]{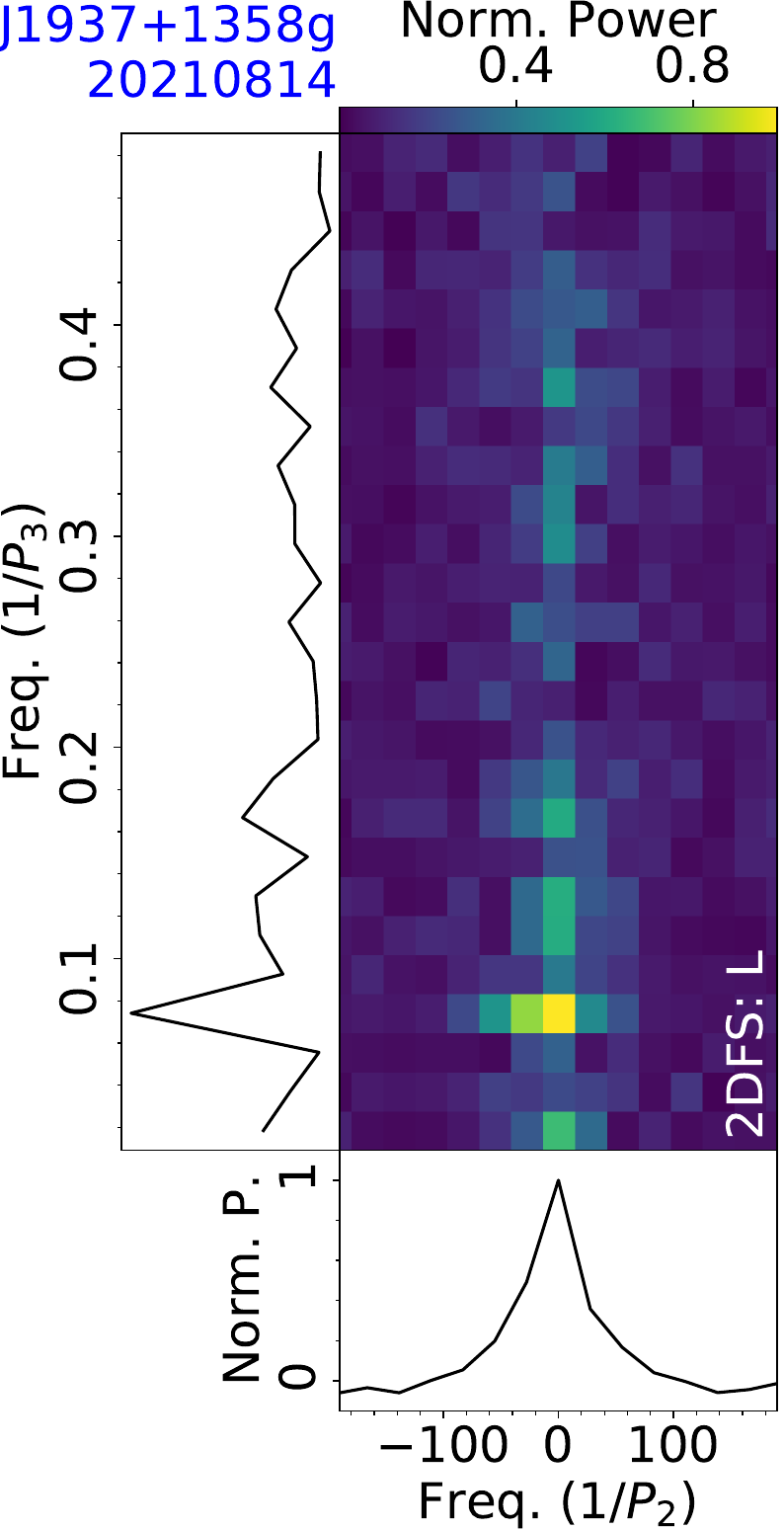}
\includegraphics[width=0.22\textwidth, angle=0]{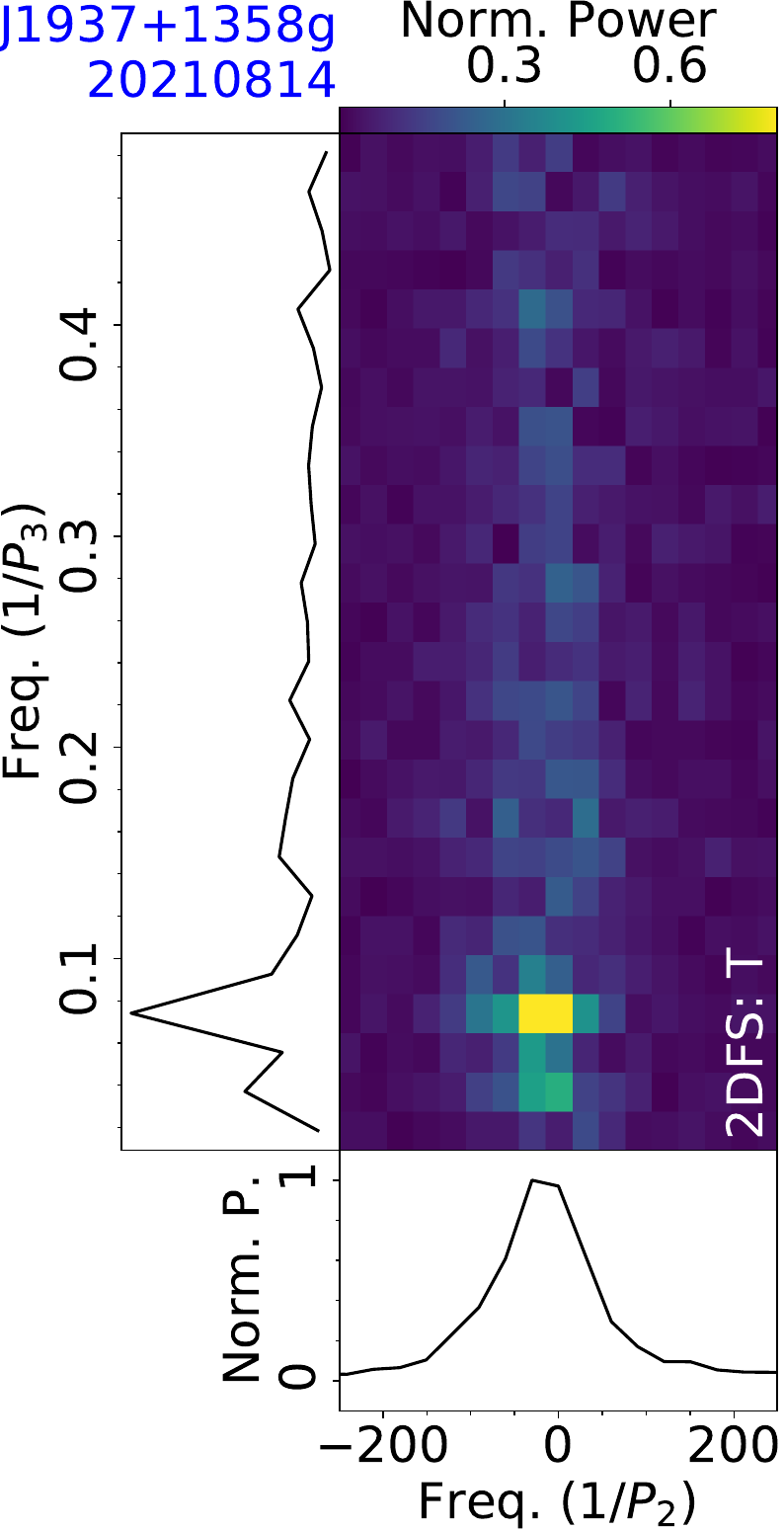}
\figcaption{Fluctuation analysis for pulses No.175-230 of PSR J1937+1358g from the FAST observation on 20210814, with LRFS (top-left), and 2DFS for the on-pulse region (top-right), leading part (bottom-left) and trailing part (bottom-right) of a mean pulse profile.
\label{subfig:fluctu:J1937+1358g}}
\end{figure}

\subsection{J1935+1616}
\label{subsec:J1935+1616}

PSR J1935+1616 was discovered by the Jodrell Bank Mk I telescope at 408 MHz \citep{Davies1970a}.  
The nulling fraction was estimated to be less than 0.25\% at 408 MHz by \citet{Ritchings1976} and less than 0.06\% at 645 MHz by \citet{Biggs1992}. From previous studies, this pulsar also has a modulation behavior. \citet{Backer1973} reported the pulsar with a broad short-period feature without a preferred drift direction at 430 MHz. \citet{Oster1977} showed drifting at 430 MHz, and preferred positive drifting in a broad feature near the $P_3$= 2 periods alias border of 2DFS was reported by \citet{Weltevrede2006,Weltevrede2007} from the 21-cm and 92-cm data. 

This pulsar has been observed by FAST on 20190919 for 53 minutes and 20211117 for 52 minutes.
From the data of 20190919, a rotation period $P=0.3588$~s and a dispersion measure $D\!M=158.5~{\rm cm^{-3}\,pc}$ were estimated. 
FAST data illustrate that this pulsar has mode changing and modulation behaviors, and single-pulse features between the two observations are similar. Here we give the analysis of the observation on 20211117.

\subsubsection{Mode changing}

Two single pulse segments of the observation on 20211117 are shown in Fig.~\ref{subfig:TP:J1935+1616}. From the integral energy histogram of the central component of the observation on 20211117 (Fig.~\ref{subfig:Hist:J1935+1616}), there is no distribution around zero corresponding to nulls. Instead, there is a distribution corresponding to the weak state, and the single pulses of the weak and normal modes are distinguished from the histogram. From the distribution of continuous period numbers for adjacent weak and normal modes (Fig.~\ref{subfig:scaleHist:J1935+1616}), the duration of the weak mode is very short of 1 or 2 rotation periods. While the duration of the normal mode is 89$\pm$88 periods. 

Averaged Polarization profiles as well as PA curves of the weak and normal emission modes are displayed in Fig.~\ref{subfig:PolModes:J1935+1616}. The profile peak of the weak emission mode is slightly lagged relative to the normal mode. For the weak mode, the part of the longitude between -5$^\circ$ and -1$^\circ$ in the profile is much weaker than the part of -1$^\circ$ to 5$^\circ$. PA of two wings is consistent between two emission states, while it is different in the longitude range between -5$^\circ$ and 2$^\circ$.

\subsubsection{Modulation}

Modulation features of the leading (-17.0$^\circ$,-8.1$^\circ$), central (-7.7$^\circ$,5.3$^\circ$) and trailing (5.7$^\circ$,20.3$^\circ$) parts in the profile are investigated respectively, and fluctuation spectra are displayed in Fig.~\ref{subfig:fluctu:J1935+1616}. 
For the leading part, there are two modulation features with centroid temporal modulation frequencies of $1/P_3=0.030\pm0.001$ ($P_3=34\pm1$ periods) and $1/P_3=0.135\pm0.001$ ($P_3=7.40\pm0.05$ periods). 
The temporal modulation frequency of the central part is low and wide, and the centroid modulation frequency of the feature is $1/P_3=0.094\pm0.002$, corresponding to $P_3=10.7\pm0.2$ periods. 
The trailing part also has a broad modulation frequency feature, and the frequency $1/P_3$ of the centroid is $0.268\pm0.003$ ($P_3=3.74\pm0.04$ periods).

\subsection{J1937+1358g}
\label{subsec:J1937+1358g}

PSR J1937+1358g was discovered by \citep{Han2021,han2025}. 

This pulsar was observed by FAST on 20210814 for 15 minutes. The single pulse sequence of this observation is shown in Fig.~\ref{subfig:TP:J1937+1358g}. The pulsar has both nulling and subpulse drifting behaviors. The nulling fraction of the observation is estimated from the on-pulse integral energy histogram (Fig.~\ref{subfig:Hist:J1937+1358g}) to be 43$\pm$0.02\%.
The LRFS and 2DFS for pulses 175-230 are presented in Fig.~\ref{subfig:fluctu:J1937+1358g} to do the fluctuation analysis and decrease the effect of nulls. 
2DFS of the leading part in the mean pulse profile has a negative drifting feature, with the centroid frequencies of $1/P_3=0.074\pm0.003$ and $1/P_2=-15\pm5$, corresponding to periodicities of $P_3=13.5\pm0.5$ periods and $P_2=-24\pm7^\circ$. For the trailing profile part, the centroid of the drift feature is characterized by $1/P_3=0.063\pm0.002$ and $1/P_2=-16\pm4$, yielding $P_3=16.0\pm0.6$ periods and $P_2=-22\pm5^\circ$.

\begin{figure}[htpb]
\centering
\includegraphics[width=0.22\textwidth, angle=0]{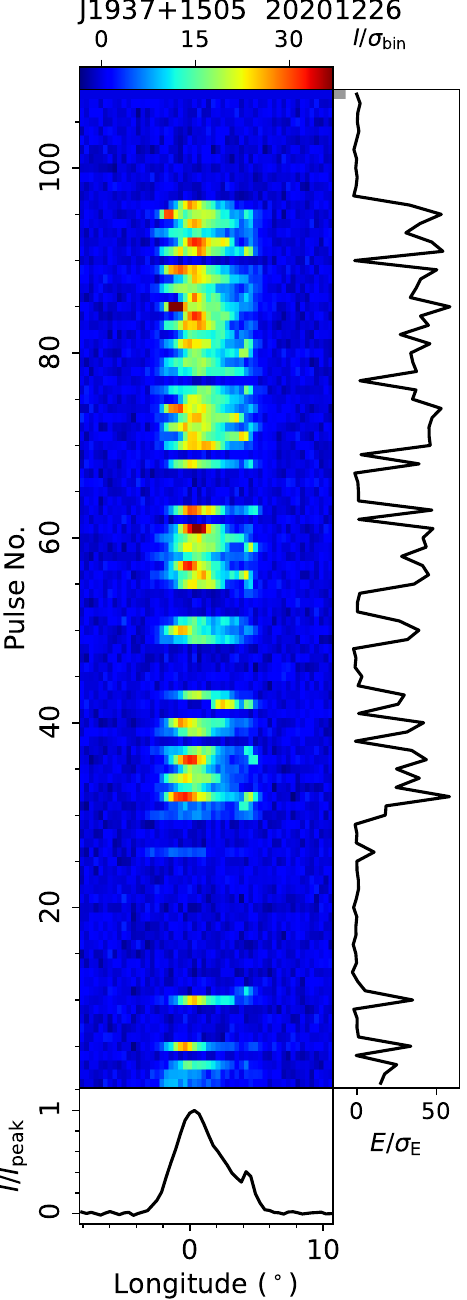}
\figcaption{Single pulse sequence of PSR J1937+1505 from the FAST observation on 20201226.
\label{subfig:TP:J1937+1505}}
\end{figure}

\begin{figure}[htpb]
\centering
\includegraphics[width=0.39\textwidth, angle=0]{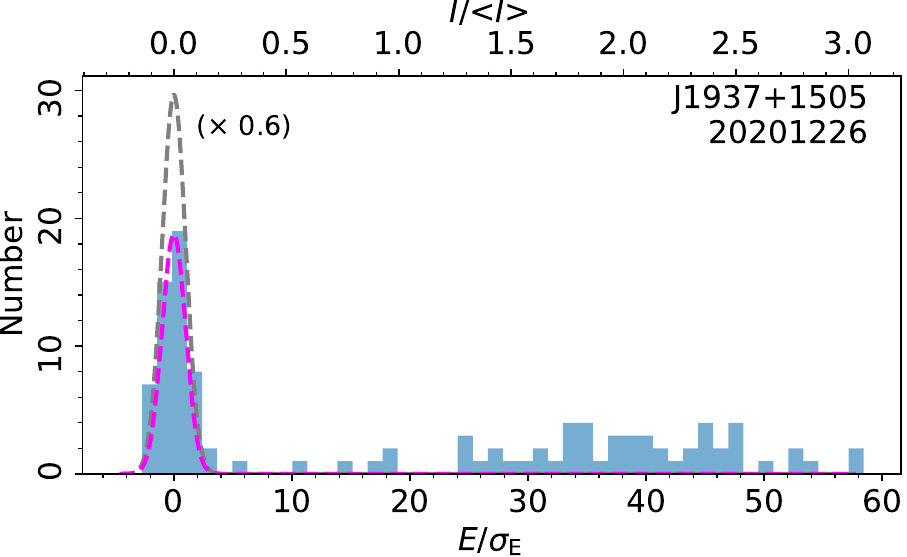}
\figcaption{On-pulse energy histogram of single pulses of PSR J1937+1505 for the observation on 20201226.
\label{subfig:Hist:J1937+1505}}
\end{figure}

\begin{figure}[htpb]
\centering
\includegraphics[width=0.22\textwidth, angle=0]{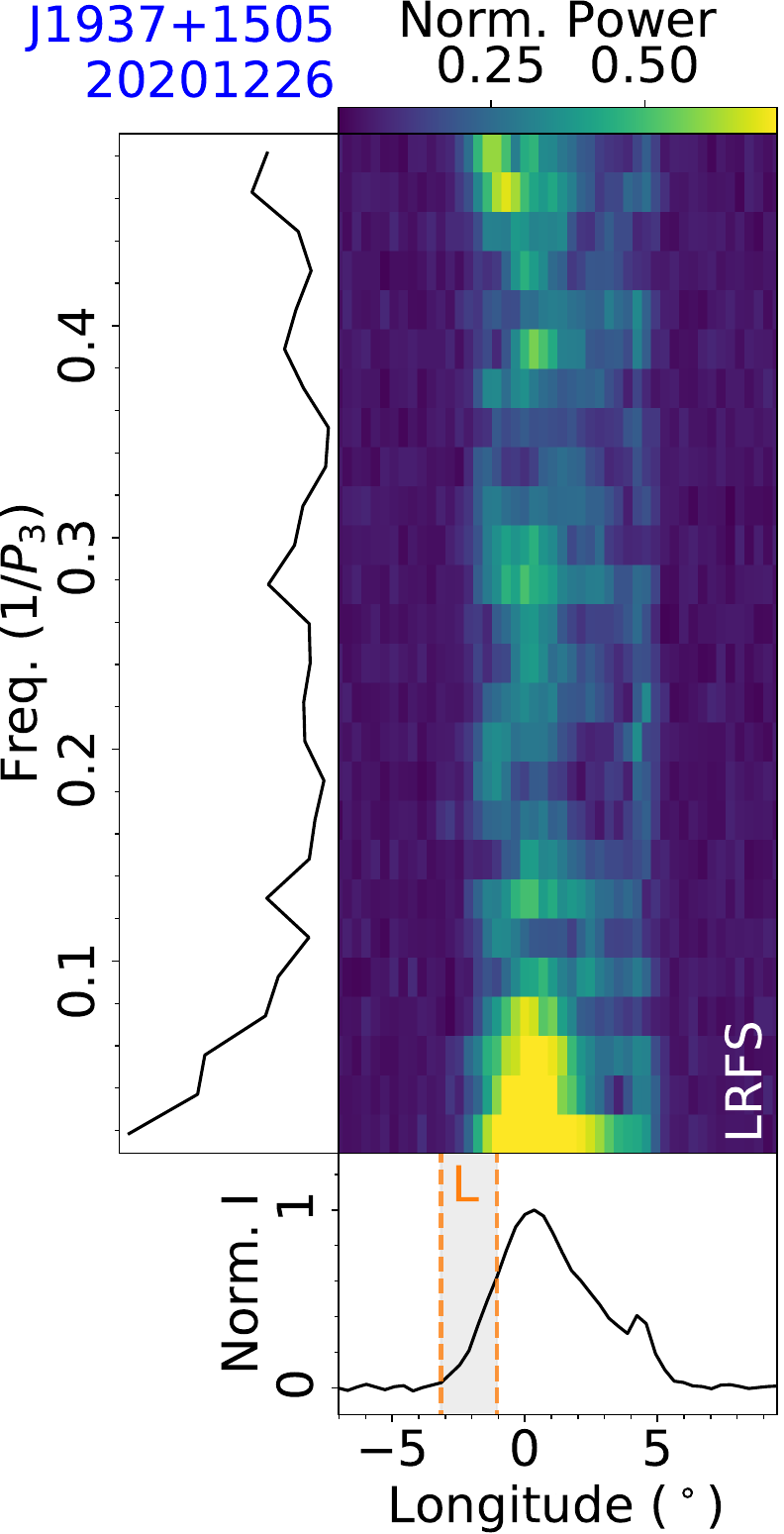}
\includegraphics[width=0.22\textwidth, angle=0]{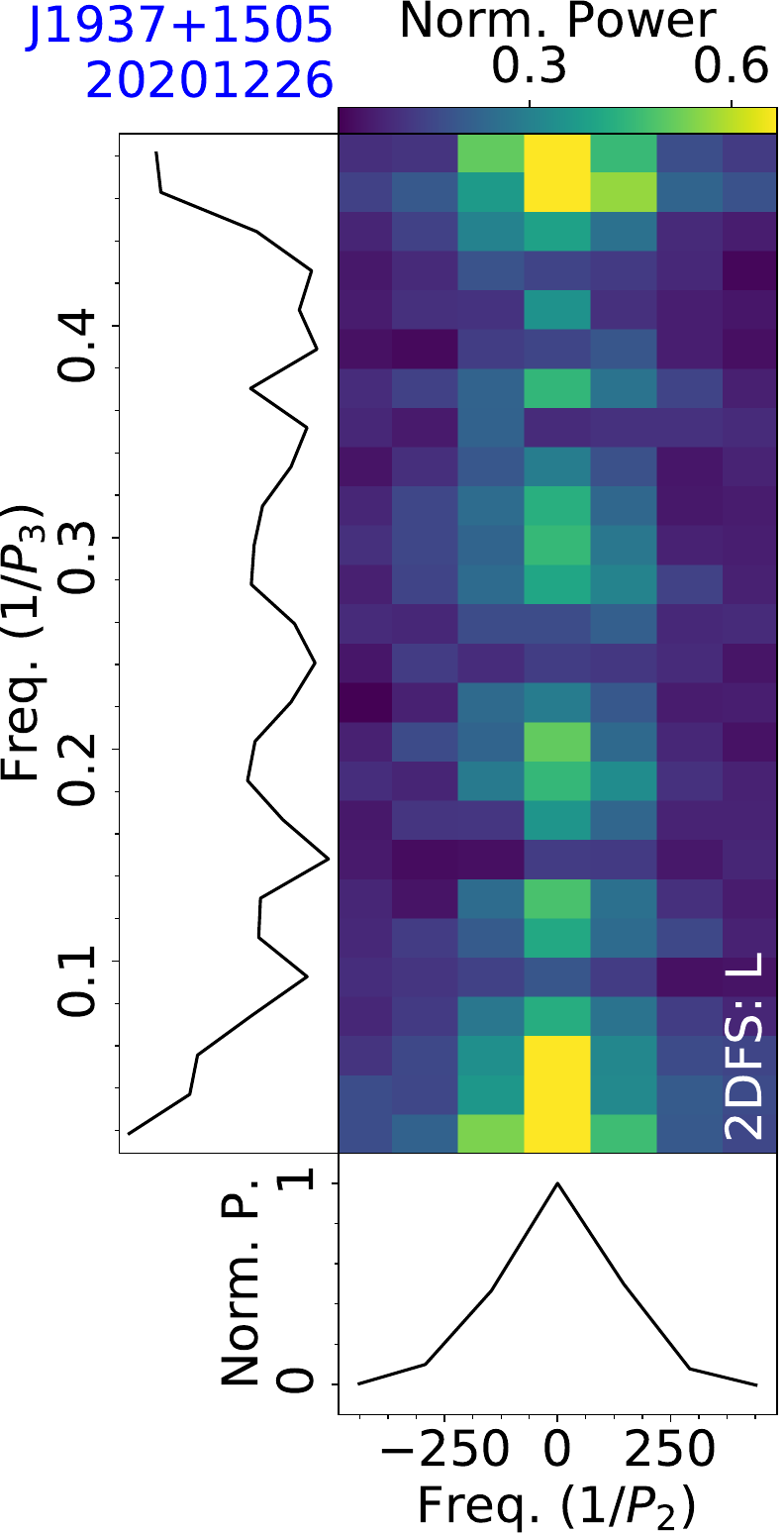}
\figcaption{Fluctuation analysis of PSR J1937+1505 from the FAST observation on 20201226, with LRFS and 2DFS for the leading part of a mean pulse profile.
\label{subfig:fluctu:J1937+1505}}
\end{figure}

\subsection{J1937+1505}
\label{subsec:J1937+1505}

PSR J1937+1505 was discovered by \citet{hfs+04} in the Parkes multibeam pulsar survey. 

This pulsar was observed by FAST on 20201226 for 5 minutes, deriving a rotation period $P=2.8730$~s and a dispersion measure $D\!M=228.1~{\rm cm^{-3}\,pc}$. 
The single pulse sequence is displayed in Fig.~\ref{subfig:TP:J1937+1505}, where the pulsar is newly found to have nulling and subpulse modulation behaviors. The nulling fraction is estimated to be 38$\pm$4\% from the on-pulse integral energy histogram. From the LRFS and 2DFS in Fig.~\ref{subfig:fluctu:J1937+1505}, there is a temporal modulation feature with the centroid frequency of $1/P_3=0.471\pm0.003$ for the leading part of the profile, which corresponds to $P_3=2.12\pm0.01$ periods. However, limited to the high-percent nulling and short pulse numbers, the accurate modulation parameters require further observation.

\begin{figure}[htpb]
\centering
\includegraphics[width=0.22\textwidth, angle=0]{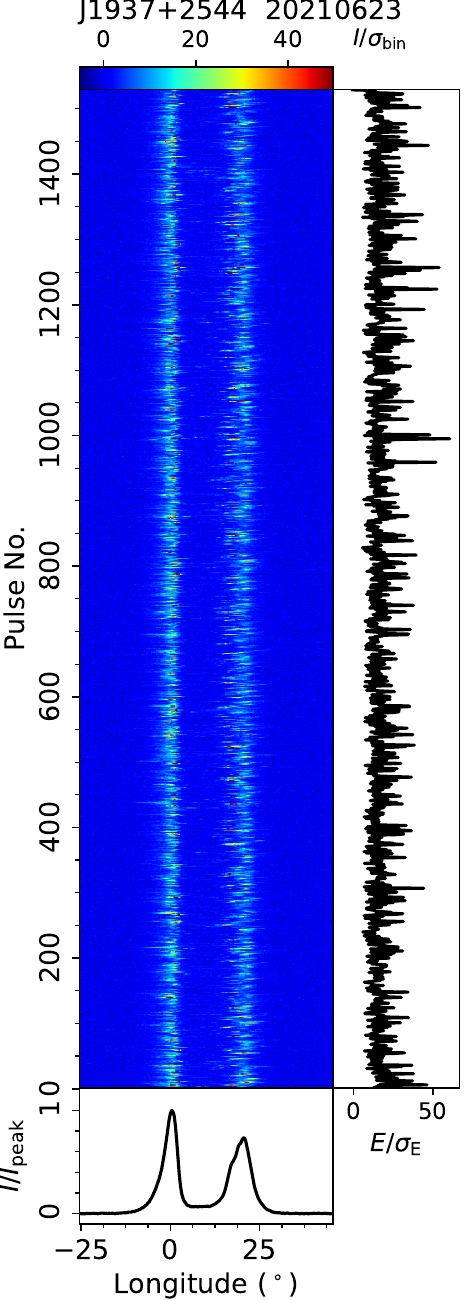}
\includegraphics[width=0.22\textwidth, angle=0]{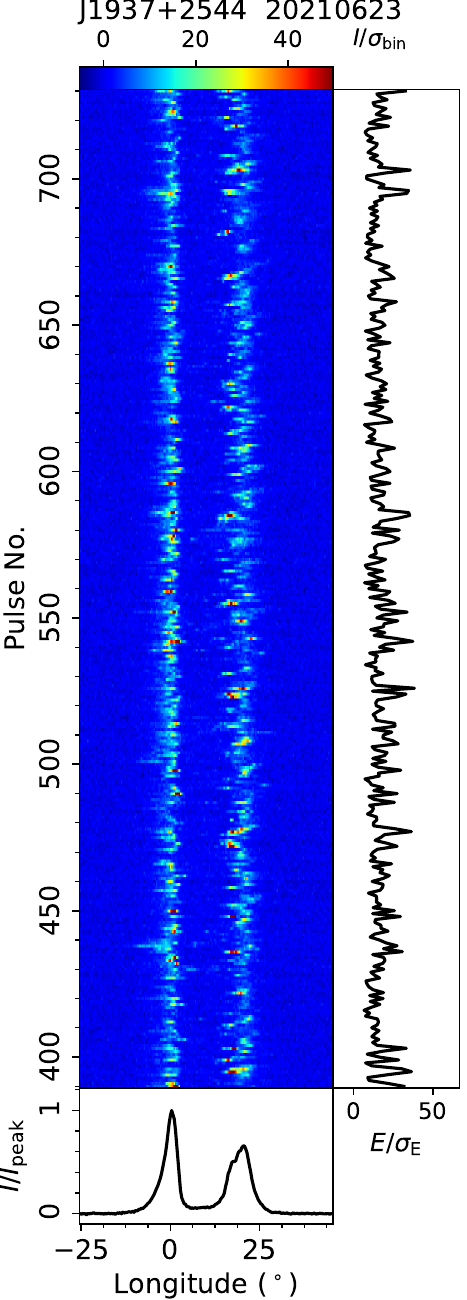}
\figcaption{Single pulse sequences of PSR J1937+2544 from the FAST observation on 20210623.
\label{subfig:TP:J1937+2544}}
\end{figure}

\begin{figure}[htpb]
\centering
\includegraphics[width=0.22\textwidth, angle=0]{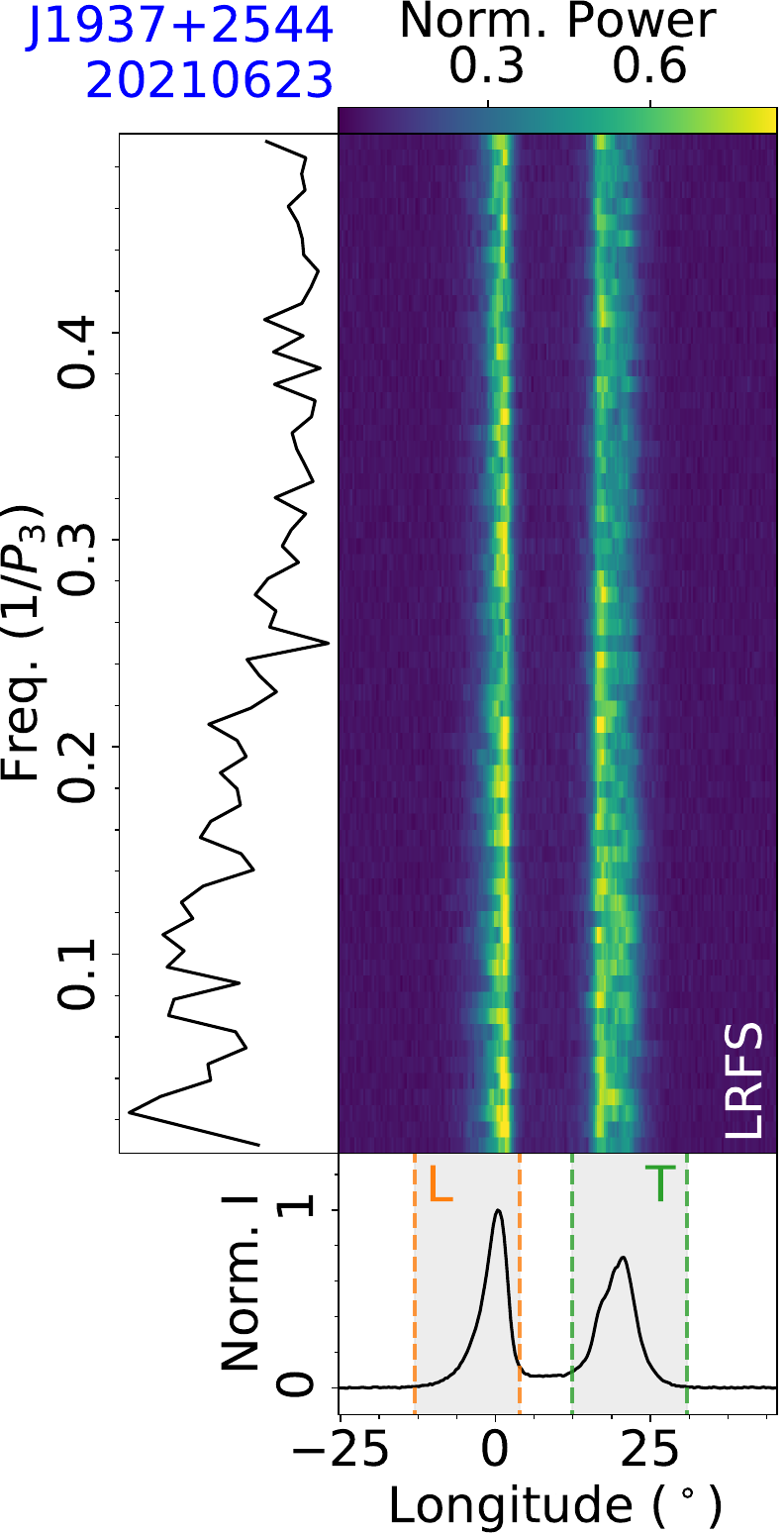}
\includegraphics[width=0.22\textwidth, angle=0]{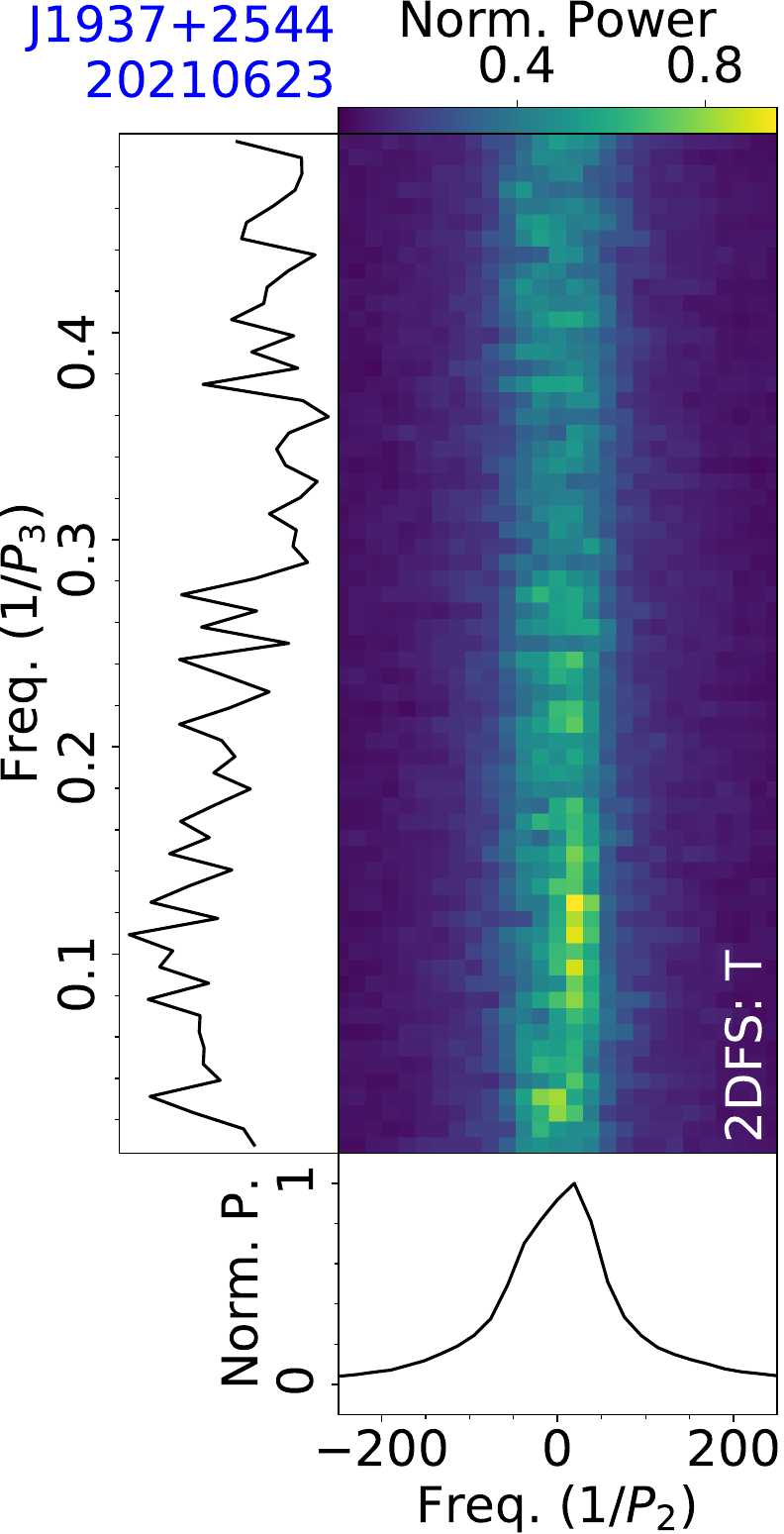}
\figcaption{Fluctuation analysis of PSR J1937+2544 for the observation on 20210623, with LRFS and 2DFS for the trailing part of a mean pulse profile.
\label{subfig:fluctu:J1937+2544}}
\end{figure}

\begin{figure}[htpb]
\centering
\includegraphics[width=0.22\textwidth, angle=0]{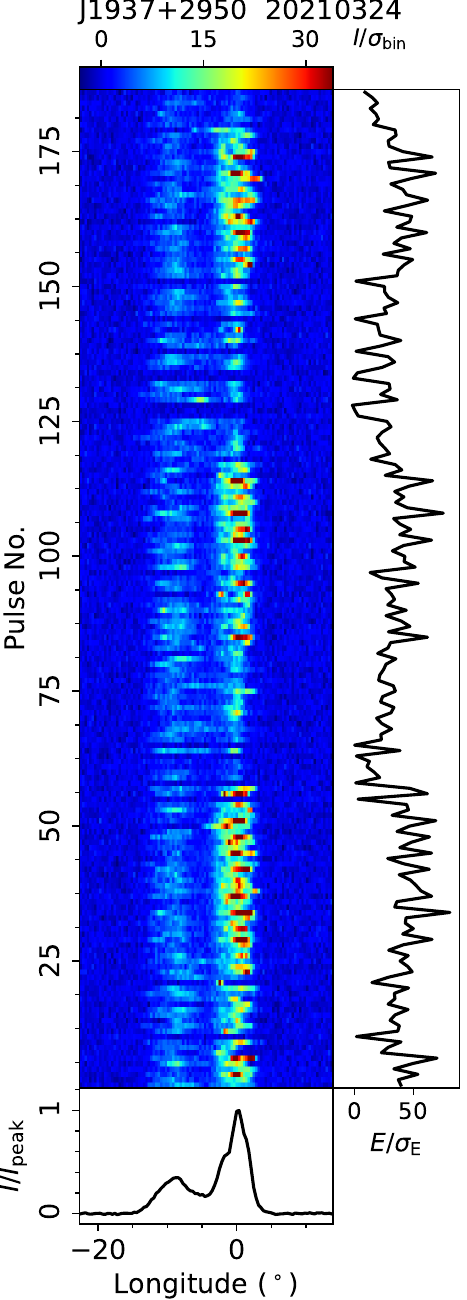}
\figcaption{Single pulse sequence of PSR J1937+2950 from the FAST observation on 20210324.
\label{subfig:TP:J1937+2950}}
\end{figure}

\begin{figure}[htpb]
\centering
\includegraphics[width=0.22\textwidth, angle=0]{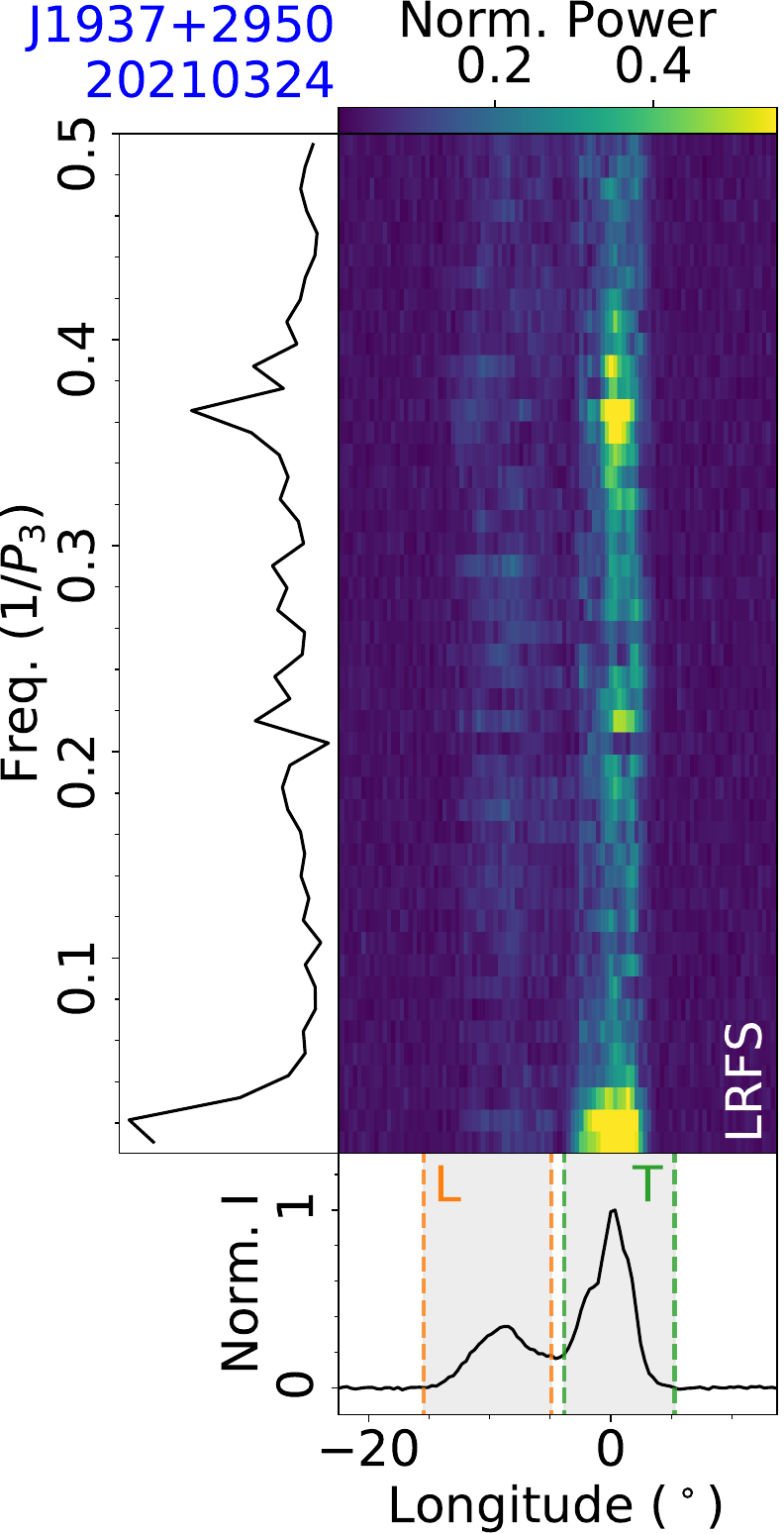}
\includegraphics[width=0.22\textwidth, angle=0]{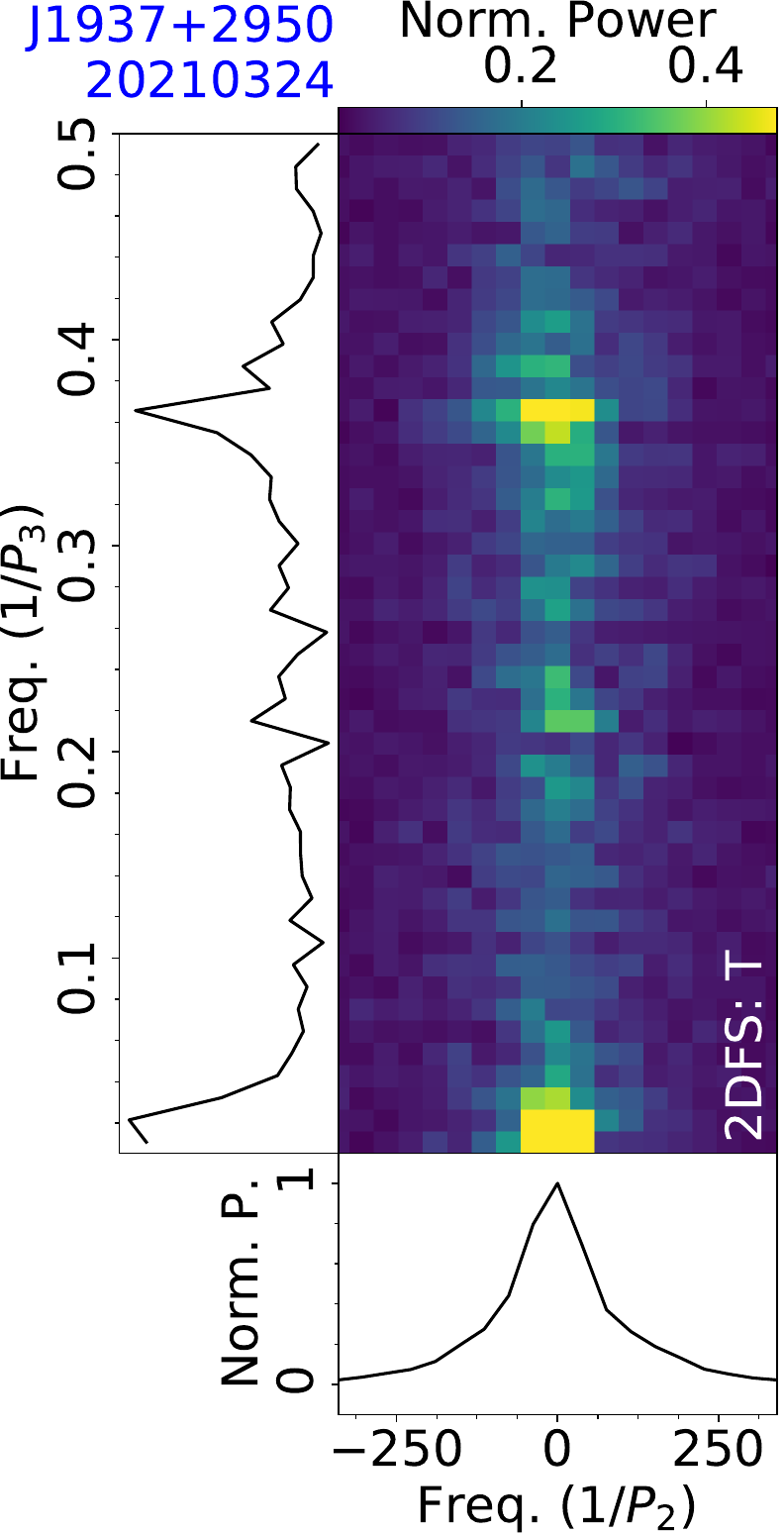}
\figcaption{Fluctuation analysis of PSR J1937+2950 for the observation on 20210324, with LRFS and 2DFS for the trailing part of a mean pulse profile.
\label{subfig:fluctu:J1937+2950}}
\end{figure}

\subsection{J1937+2544}
\label{subsec:J1937+2544}

PSR J1937+2544 was found by \citep{Dewey1985} using the 92 m telescope at Green Bank. 

The pulsar was observed by FAST on 20210623 for 5 minutes, deriving a rotation period $P=0.2010$~s and a dispersion measure $D\!M=53.2~{\rm cm^{-3}\,pc}$. 
The single pulse sequence of the observation and the zoomed view of pulses No. 390-730 are shown in Fig.~\ref{subfig:TP:J1937+2544}. 
In 2DFS of the trailing profile part (Fig.~\ref{subfig:fluctu:J1937+2544}), the preferred direction of the main drift feature is positive, with the centroid frequencies of $1/P_3=0.104\pm0.001$ and $1/P_2=21\pm1$, corresponding to $P_3=9.6\pm0.1$ periods and $P_2=18\pm1^\circ$.

\begin{figure}[htpb]
\centering
\includegraphics[width=0.22\textwidth, angle=0]{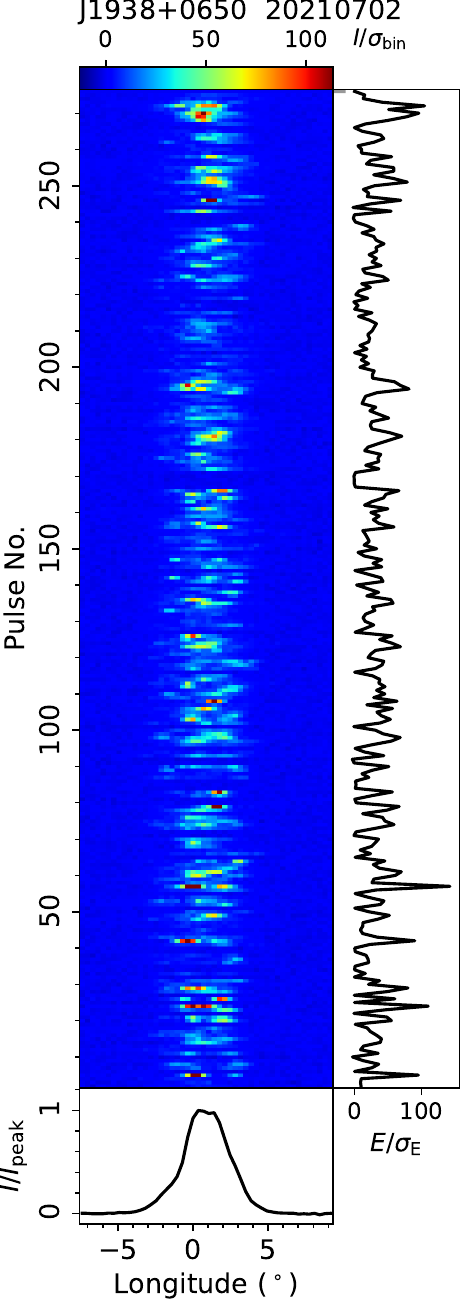}
\includegraphics[width=0.22\textwidth, angle=0]{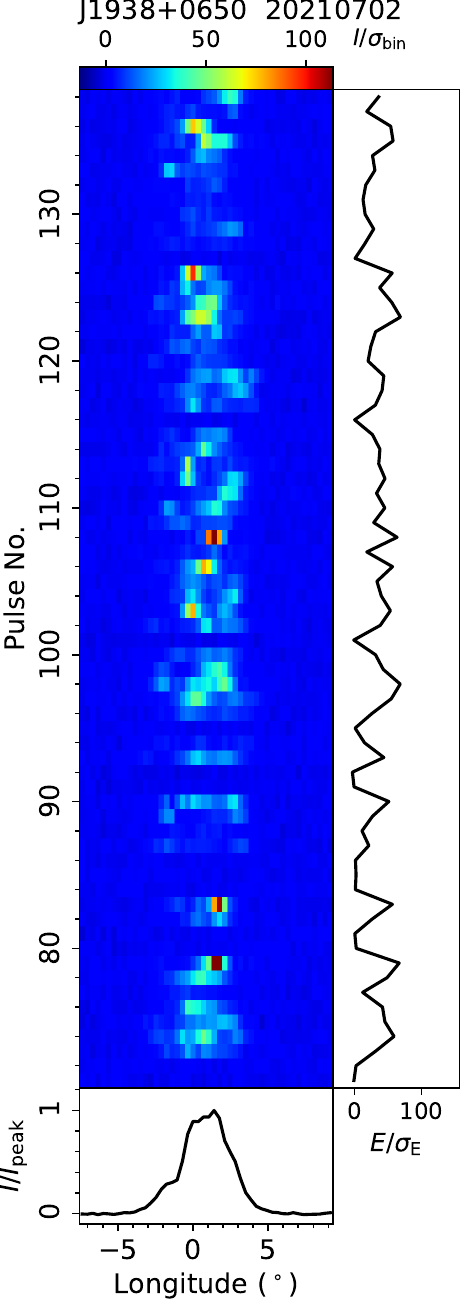}
\figcaption{Single pulse sequences of PSR J1938+0650 from the FAST observation on 20210702.
\label{subfig:TP:J1938+0650}}
\end{figure}

\begin{figure}[htpb]
\centering
\includegraphics[width=0.39\textwidth, angle=0]{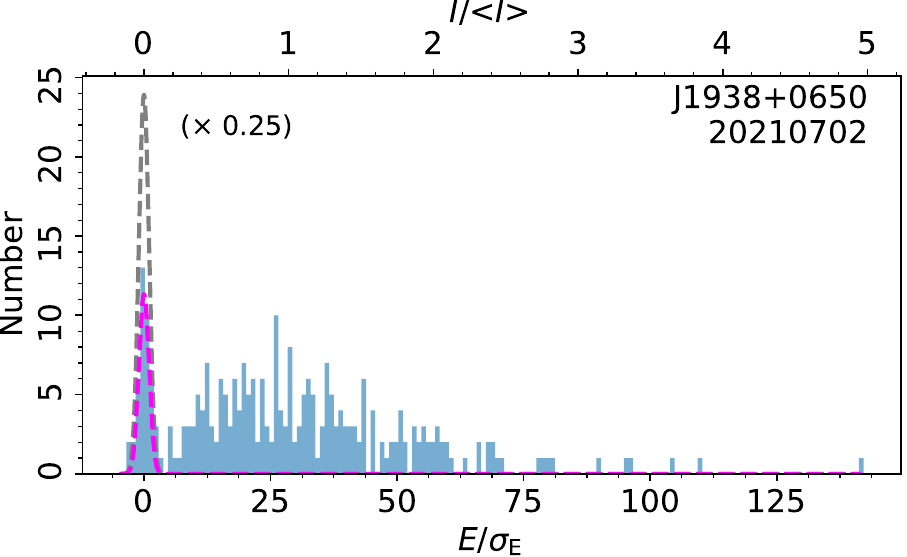}
\figcaption{On-pulse energy histogram of PSR J1938+0650 from the FAST observation on 20210702.
\label{subfig:Hist:J1938+0650}}
\end{figure}

\begin{figure}[htpb]
\centering
\includegraphics[width=0.39\textwidth, angle=0]{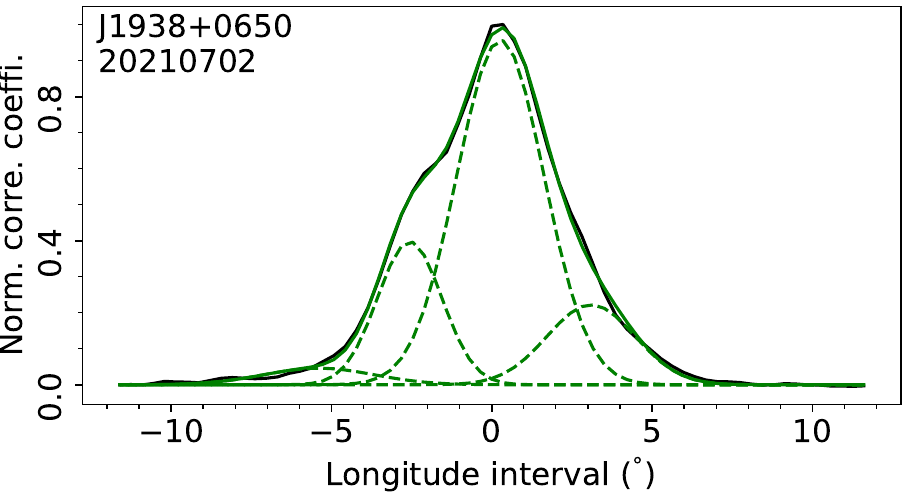}
\figcaption{Cross correlation of PSR J1938+0650 from the FAST observation on 20210702.
\label{subfig:Corre:J1938+0650}}
%
\centering
\includegraphics[width=0.22\textwidth, angle=0]{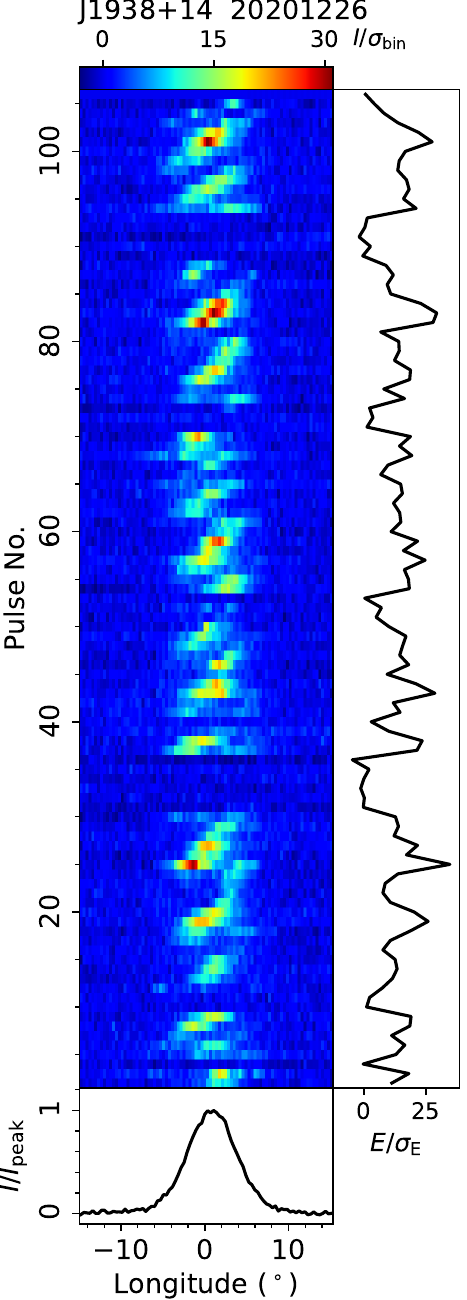}
\figcaption{Single pulse sequence of PSR J1938+14 from the FAST observation on 20201226.
\label{subfig:TP:J1938+14}}
\end{figure}

\begin{figure}[htpb]
\centering
\includegraphics[width=0.39\textwidth, angle=0]{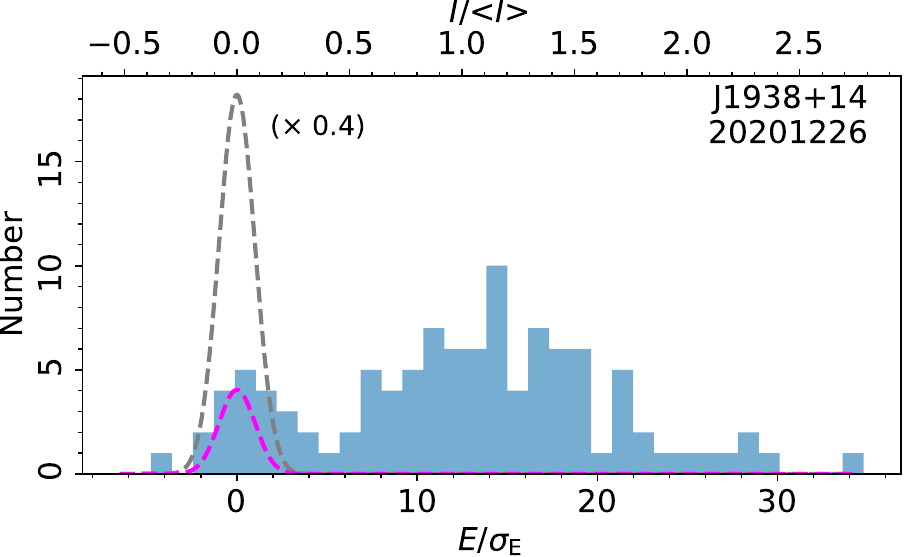}
\figcaption{On-pulse energy histogram of single pulses of PSR J1938+14 from the FAST observation on 20201226.
\label{subfig:Hist:J1938+14}}
\end{figure}

\begin{figure}[htpb]
\centering
\includegraphics[width=0.22\textwidth, angle=0]{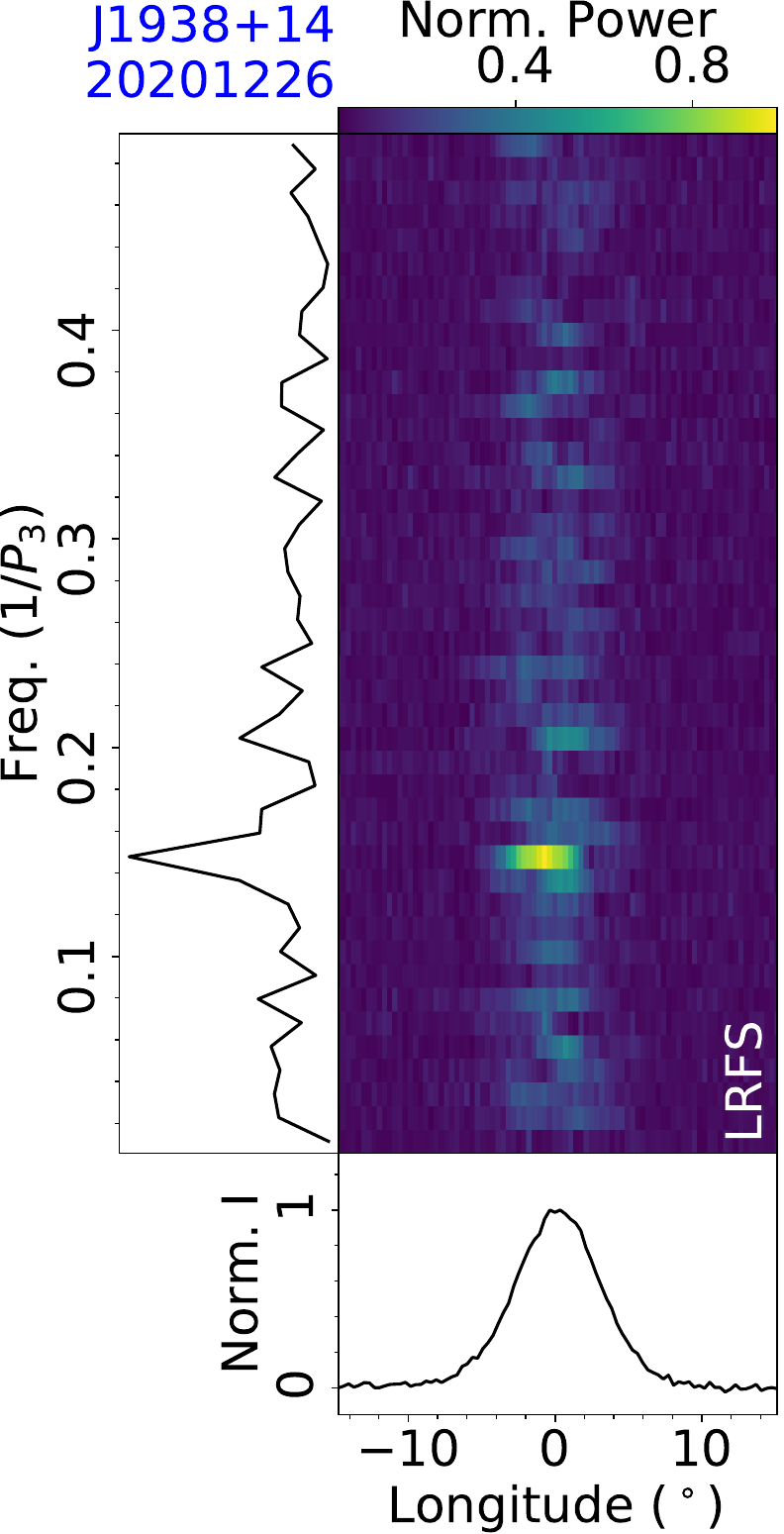}
\includegraphics[width=0.22\textwidth, angle=0]{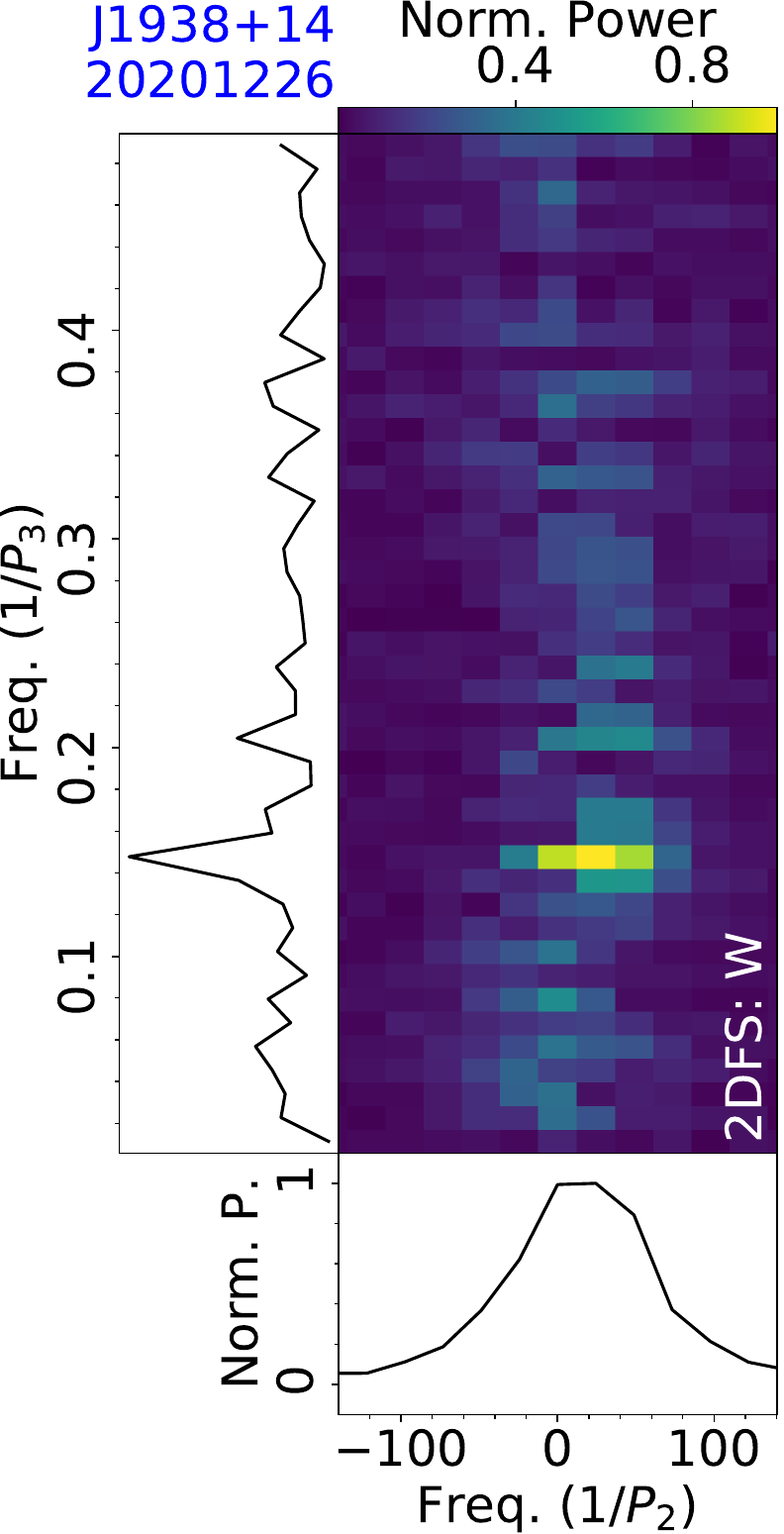}
\figcaption{Fluctuation analysis of PSR J1938+14 for the observation on 20201226, with LRFS and 2DFS for the on-pulse region of a mean pulse profile.
\label{subfig:fluctu:J1938+14}}
\end{figure}

\subsection{J1937+2950}
\label{subsec:J1937+2950}

PSR J1937+2950 was discovered by the Westerbork Synthesis Radio Telescope \citep{Janssen2009}. 

This pulsar was observed by FAST on 20210324 for 5 minutes, deriving a rotation period $P=1.6573$~s and a dispersion measure $D\!M=112.7~{\rm cm^{-3}\,pc}$. The single pulse sequence in Fig.~\ref{subfig:TP:J1937+2950} shows the subpulse drifting and low-frequency modulation for the trailing part in a mean pulse profile. LRFS and 2DFS of the trailing profile part are shown in Fig.~\ref{subfig:fluctu:J1937+2950}. There are two modulation features, with centroid frequencies of $1/P_3=0.357\pm0.001$ and $0.019\pm0.001$, corresponding to $P_3=2.80\pm0.01$ and $53\pm3$ periods. The 2.8-period modulation is nested within the low-frequency modulation. 
Longer observations are required for a detailed analysis of single pulse behaviors.

\begin{figure}[htpb]
\centering
\includegraphics[width=0.22\textwidth, angle=0]{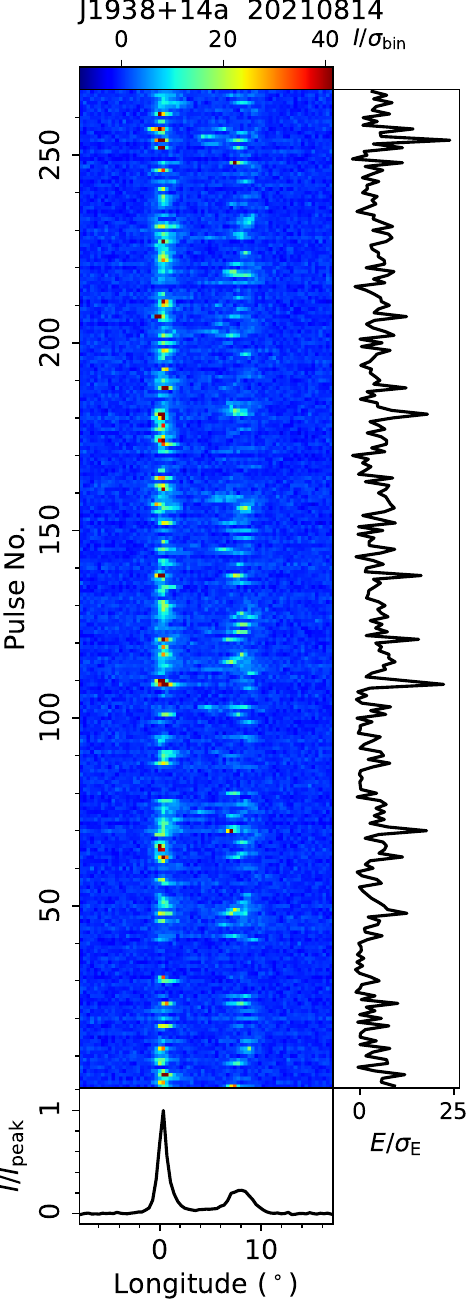}
\includegraphics[width=0.22\textwidth, angle=0]{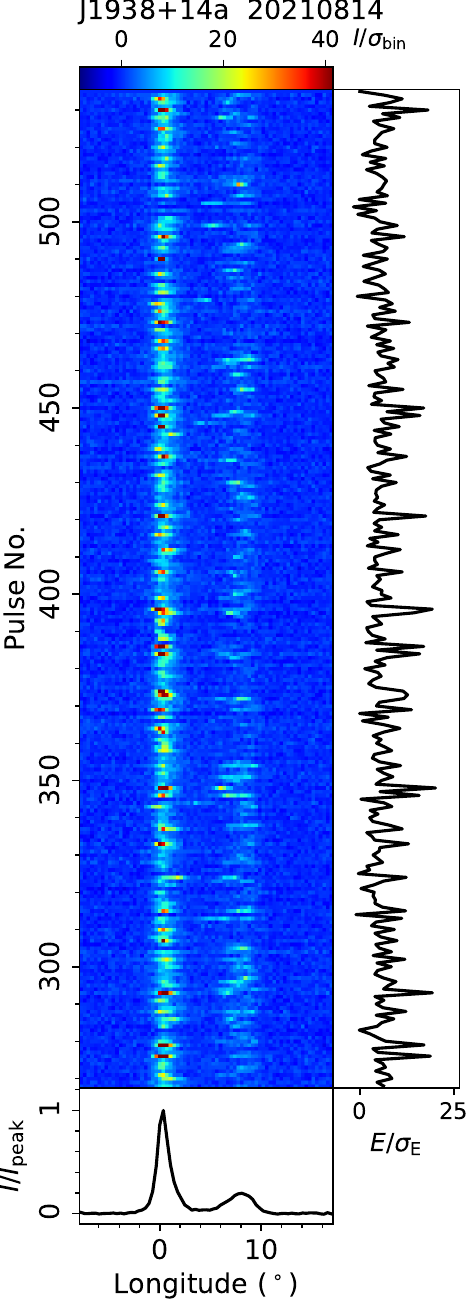}
\figcaption{Single pulse sequences of PSR J1938+14a from the FAST observation on 20210814.
\label{subfig:TP:J1938+14a}}
\end{figure}

\begin{figure}[htpb]
\centering
\includegraphics[width=0.39\textwidth, angle=0]{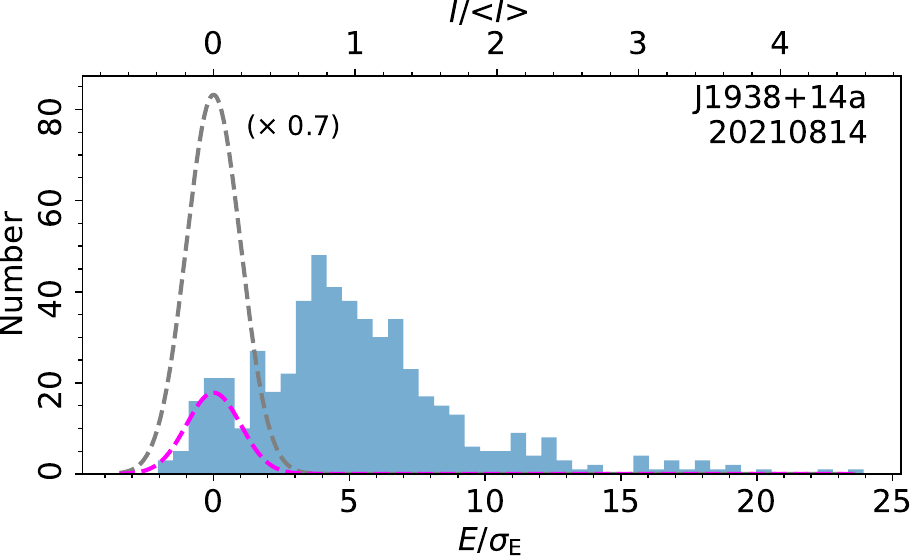}
\figcaption{Energy histogram of single pulses of PSR J1938+14a from the FAST observation on 20210814.
\label{subfig:Hist:J1938+14a}}
\end{figure}

\begin{figure}[htpb]
\centering
\includegraphics[width=0.22\textwidth, angle=0]{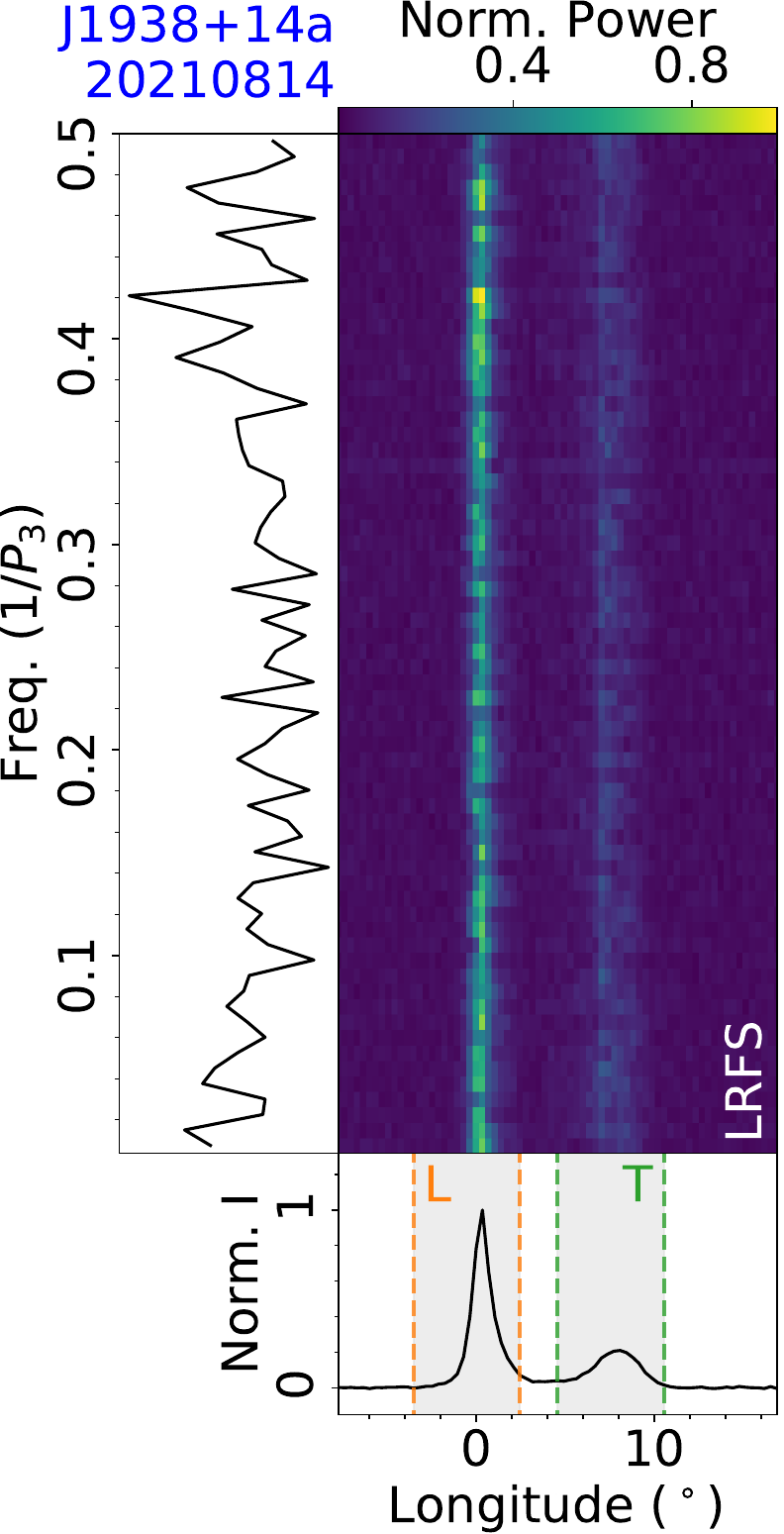}
\includegraphics[width=0.22\textwidth, angle=0]{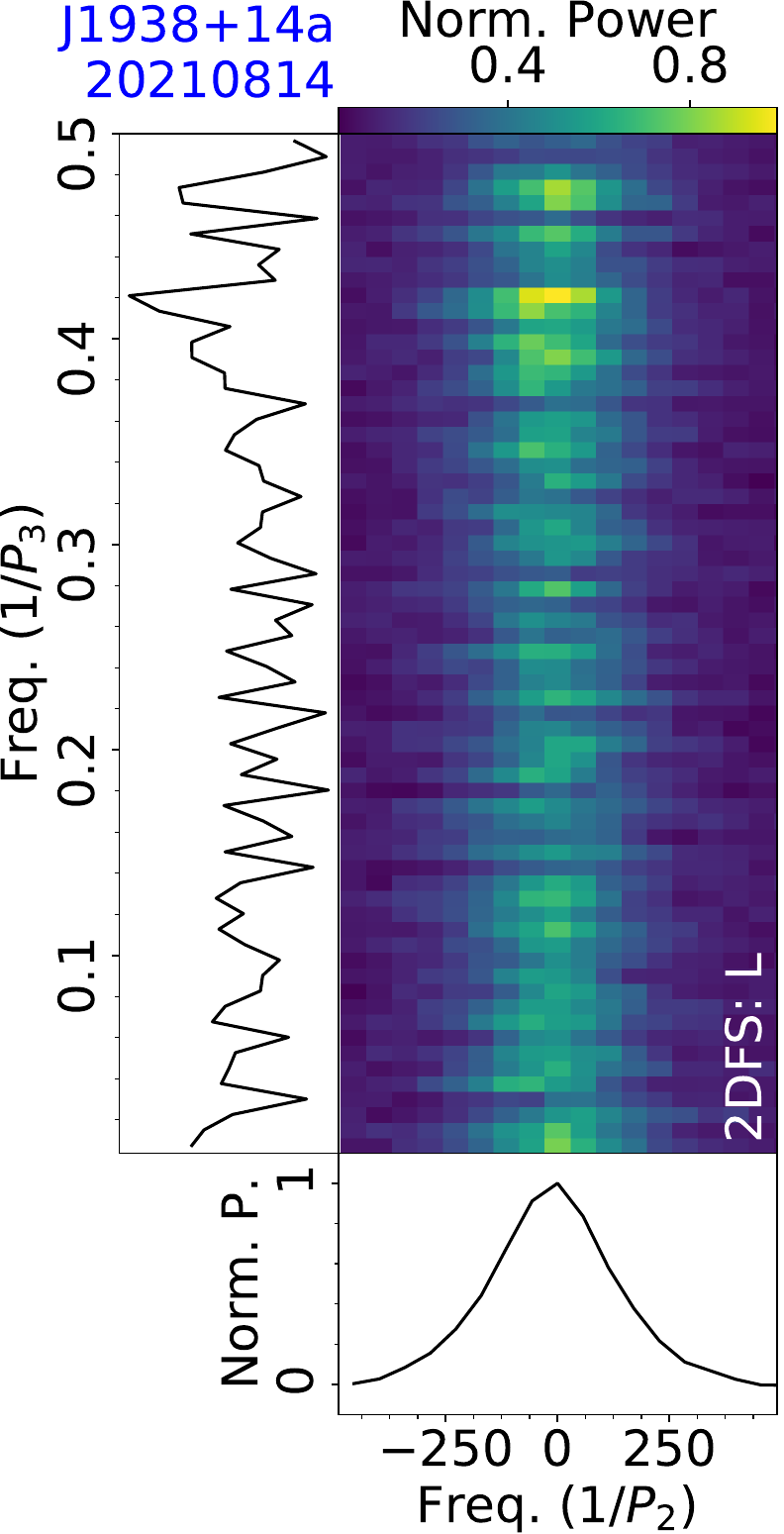}
\figcaption{Fluctuation analysis of PSR J1938+14a for the observation on 20210814, with LRFS and 2DFS for the leading part of a mean pulse profile.
\label{subfig:fluctu:J1938+14a}}
\end{figure}

\subsection{J1938+0650}
\label{subsec:J1938+0650}

PSR J1938+0650 was discovered by \citet{Foster1995} using the Arecibo radio telescope. 

This pulsar was observed by FAST on 20210702 for 5 minutes, deriving a rotation period $P=1.1215$~s and a dispersion measure $D\!M=74.9~{\rm cm^{-3}\,pc}$. 
Single pulse sequences are shown in Fig.~\ref{subfig:TP:J1938+0650}. The pulsar is found to have the nulling phenomenon, and the nulling fraction of this observation is estimated to be 12$\pm$1\% from the on-pulse integral energy histogram in Fig.~\ref{subfig:Hist:J1938+0650}. Single pulse sequences display some occasional drifting bands, such as pulse Nos. 102-115, although the drifting behavior is not systematic and always appears. Using the cross-correlation method (Fig.~\ref{subfig:Corre:J1938+0650}), the drifting parameters are estimated to be $D=0.27\pm0.18$ degrees per period and $P_2=2.83\pm0.04^\circ$ for these drifting bands.

\begin{figure}[htpb]
\centering
\includegraphics[width=0.22\textwidth, angle=0]{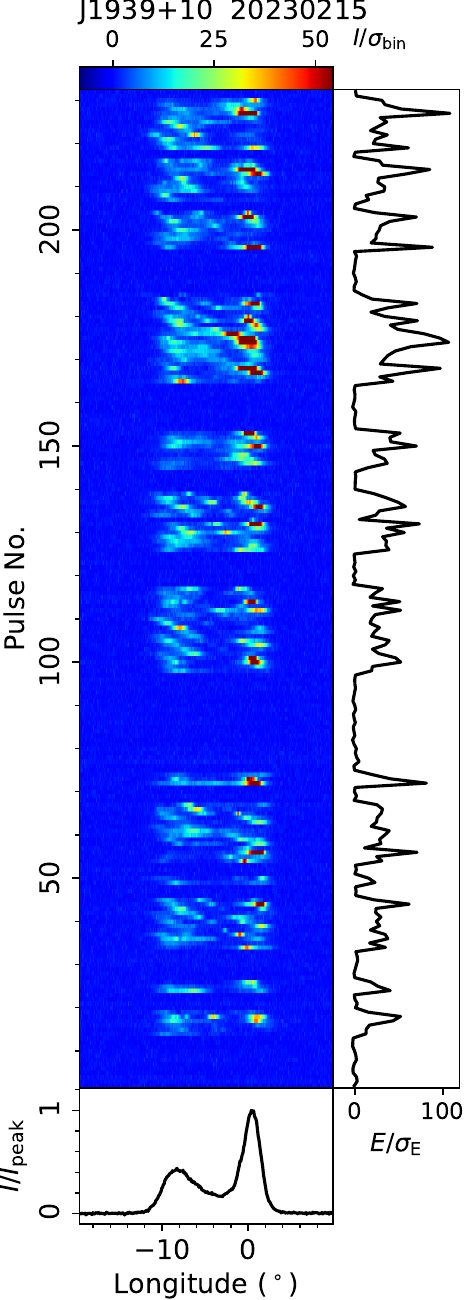}
\includegraphics[width=0.22\textwidth, angle=0]{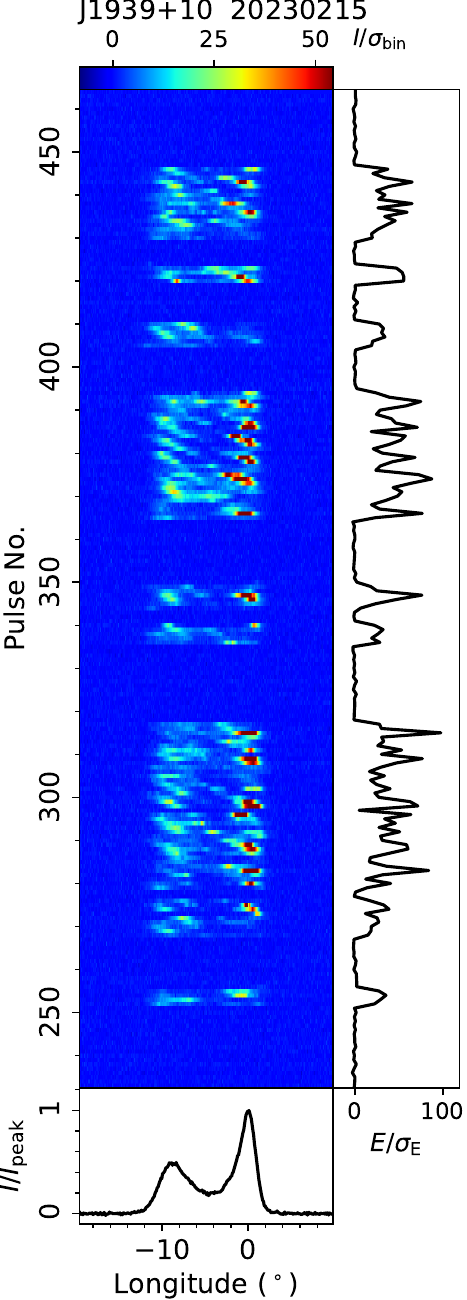}
\figcaption{Single pulse sequences of PSR J1939+10 from the FAST observation on 20230215.
\label{subfig:TP:J1939+10}}
\end{figure}

\begin{figure}[htpb]
\centering
\includegraphics[width=0.39\textwidth, angle=0]{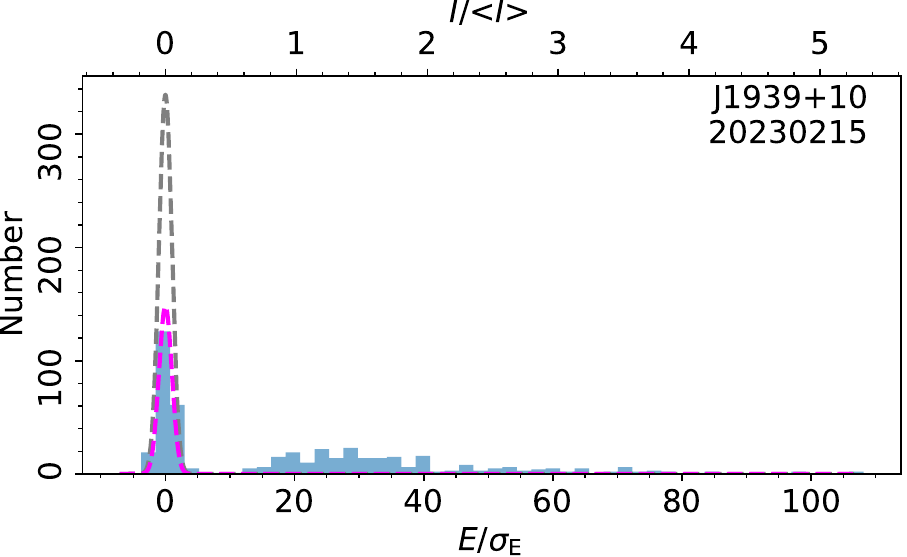}
\vspace{-0.3cm}
\figcaption{On-pulse energy histogram of single pulses of PSR J1939+10 from the FAST observation on 20230215.
\label{subfig:Hist:J1939+10}}
%
\centering
\includegraphics[width=0.44\textwidth, angle=0]{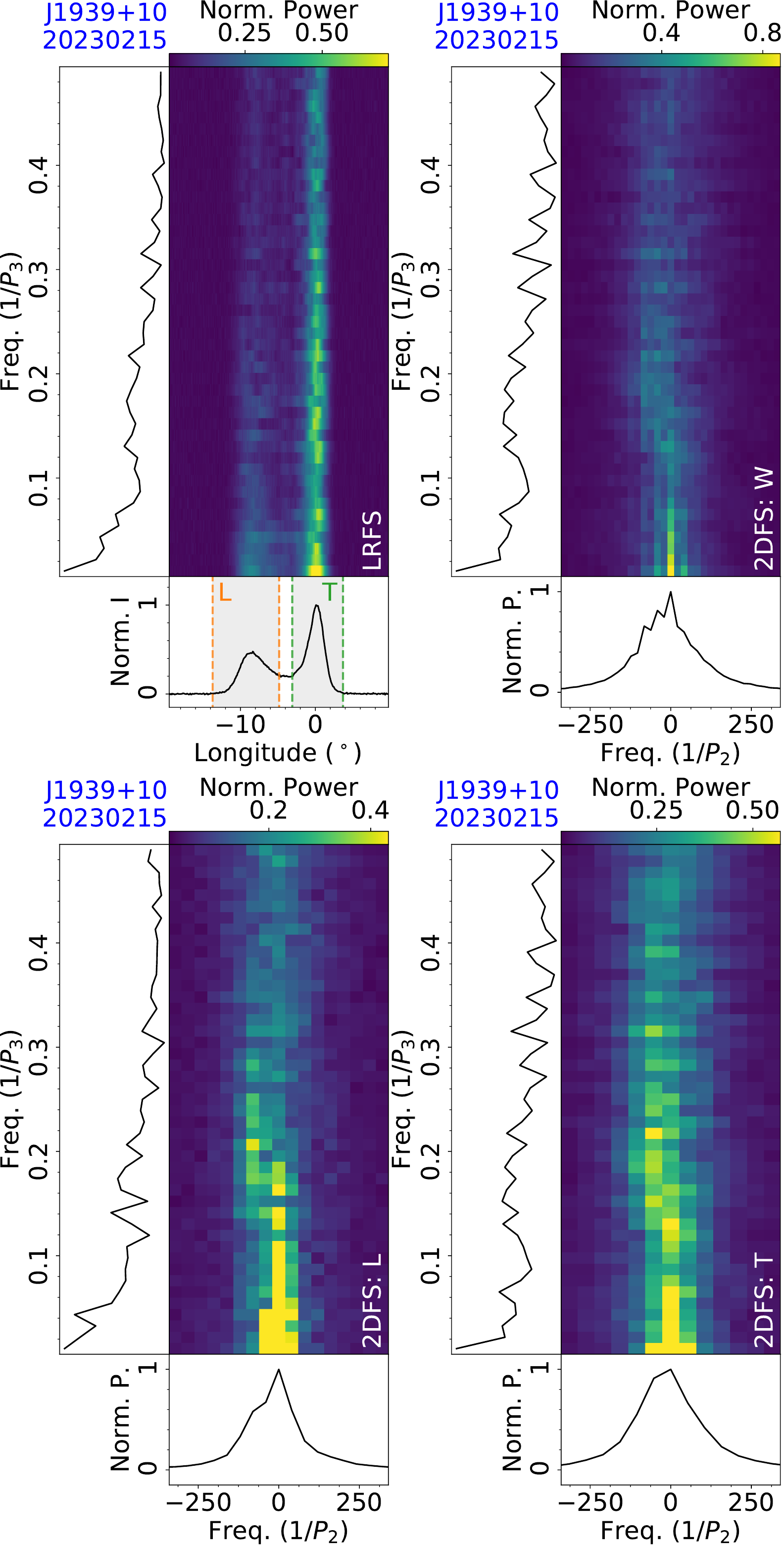}
\vspace{-0.3cm}
\figcaption{Fluctuation analysis of PSR J1939+10 for the observation on 20230215, with LRFS (top-left), and 2DFS for the on-pulse phase region (top-right), leading part (bottom-left) and trailing part (bottom-right) of a mean pulse profile.
\label{subfig:fluctu:J1939+10}}
\end{figure}

\begin{figure}[htpb]
\centering
\includegraphics[width=0.22\textwidth, angle=0]{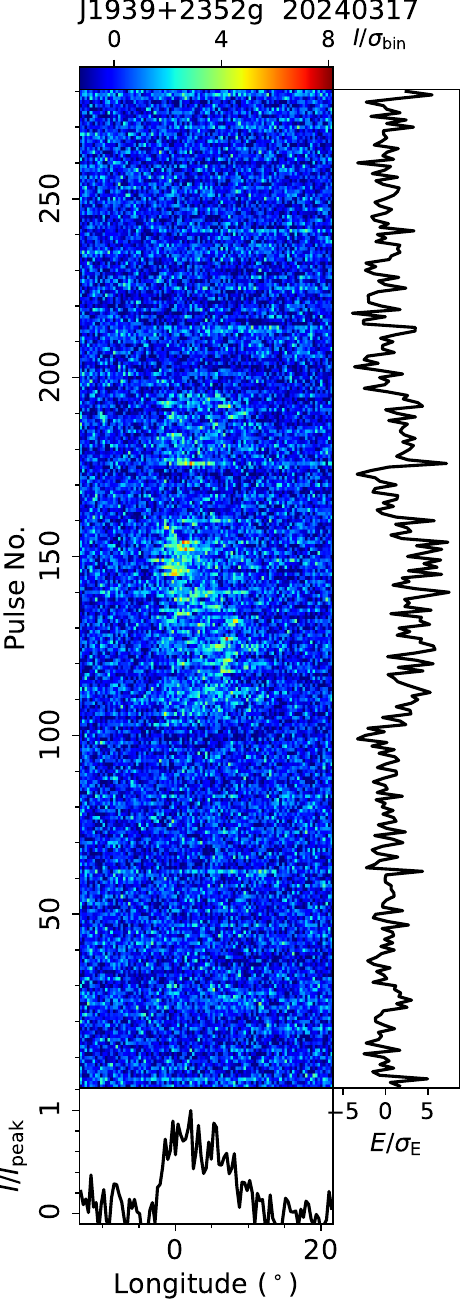}
\includegraphics[width=0.22\textwidth, angle=0]{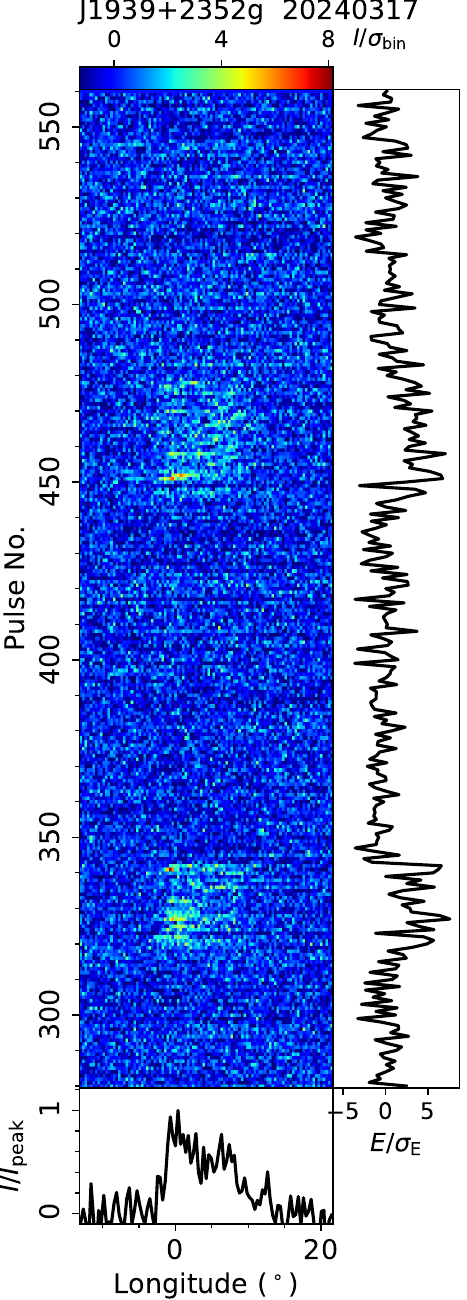}
\figcaption{Single pulse sequences of PSR J1939+2352g from the FAST observation on 20240317.
\label{subfig:TP:J1939+2352g}}
\end{figure}

\begin{figure}[htpb]
\centering
\includegraphics[width=0.39\textwidth, angle=0]{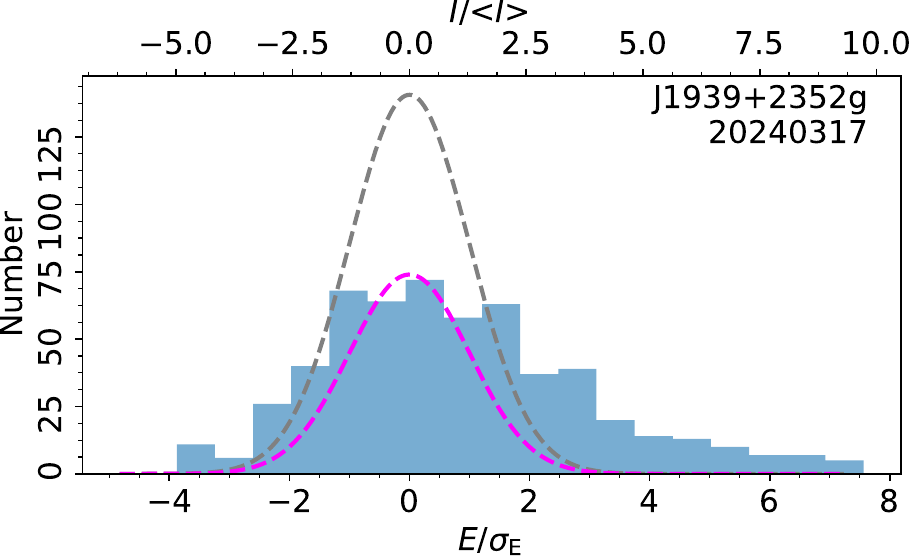}
\figcaption{On-pulse energy histogram of single pulses of PSR J1939+2352g from the FAST observation on 20240317. \label{subfig:Hist:J1939+2352g}}
\end{figure}

\begin{figure}[htpb]
\centering
\includegraphics[width=0.22\textwidth, angle=0]{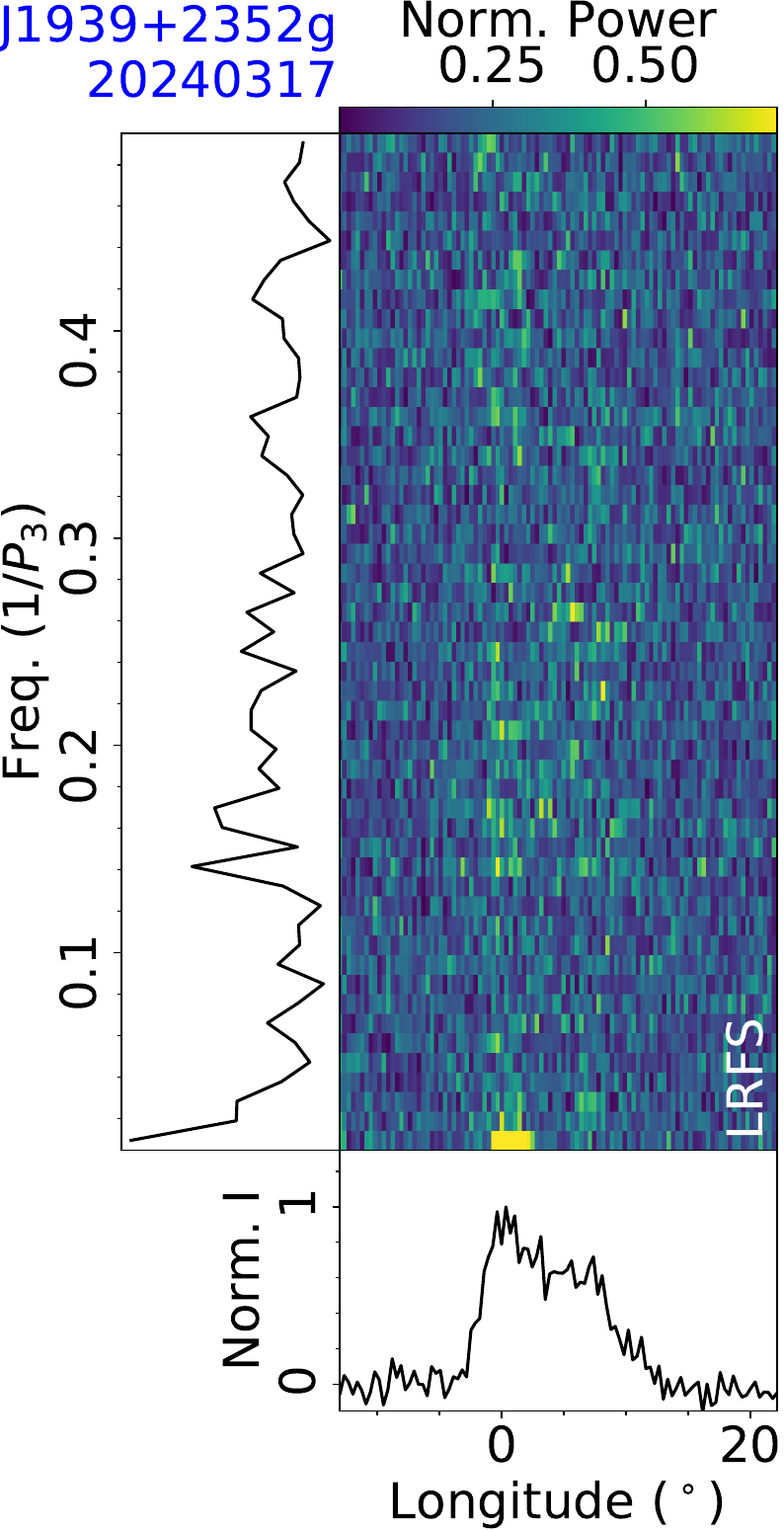}
\includegraphics[width=0.22\textwidth, angle=0]{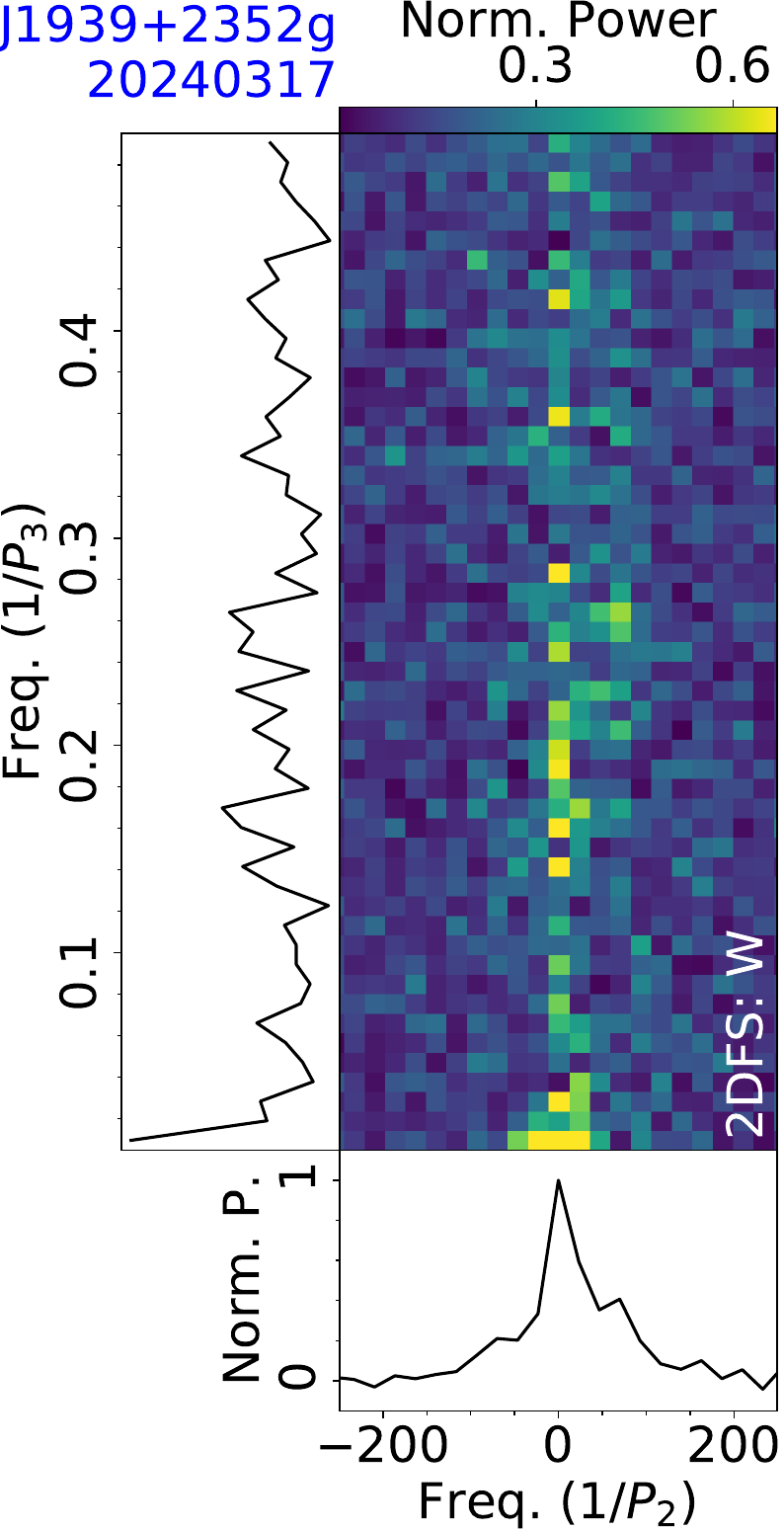}
\figcaption{Fluctuation analysis of PSR J1939+2352g for the observation on 20240317, with LRFS and 2DFS for the on-pulse region of a mean pulse profile.
\label{subfig:fluctu:J1939+2352g}}
\end{figure}

\subsection{J1938+14}
\label{subsec:J1938+14}

PSR J1938+14 was discovered in the Arecibo 327 MHz Drift Pulsar Survey via a single-pulse search \citep{Deneva2016}. 

J1938+14 was observed by FAST on 20201226 for 5 minutes, deriving a rotation period $P=2.9028$~s and a dispersion measure $D\!M=75.6~{\rm cm^{-3}\,pc}$. 
The single pulse sequence in Fig.~\ref{subfig:TP:J1938+14} indicates nulling and subpulse drifting of this pulsar. The nulling fraction is estimated to be 9$\pm$1\% from the on-pulse integral energy histogram in Fig.~\ref{subfig:Hist:J1938+14}. Centroid frequencies of the main feature in 2DFS (Fig.~\ref{subfig:fluctu:J1938+14}) are $1/P_3=0.148\pm0.002$ and $1/P_2=25\pm4$, which correspond to periodicities of $P_3=6.8\pm0.1$ periods and $P_2=15\pm3^\circ$.

\subsection{J1938+14a}
\label{subsec:J1938+14a}

PSR J1938+14a was discovered in the Green Bank North Celestial Cap (GBNCC) pulsar survey (http://astro.phys.wvu.edu/GBNCC/). 

The pulsar is observed by FAST on 20210814 for 15 minutes, deriving a rotation period $P=1.6615$~s and a dispersion measure $D\!M=114.4~{\rm cm^{-3}\,pc}$. 
Single pulse sequences in Fig.~\ref{subfig:TP:J1938+14a} display that this pulsar has the nulling phenomenon, and the leading component is modulated temporally. The nulling fraction of this observation is estimated to be 15$\pm$1\% from the on-pulse integral energy histogram (Fig.~\ref{subfig:Hist:J1938+14a}). 
For the leading part of the mean pulse profile, LRFS and 2DFS in Fig.~\ref{subfig:fluctu:J1938+14a} indicate a temporal modulation feature with the centroid of $1/P_3=0.424\pm0.003$, corresponding to the periodicity of $P_3=2.36\pm0.02$ periods.

\begin{figure}[htpb]
\centering
\includegraphics[width=0.22\textwidth, angle=0]{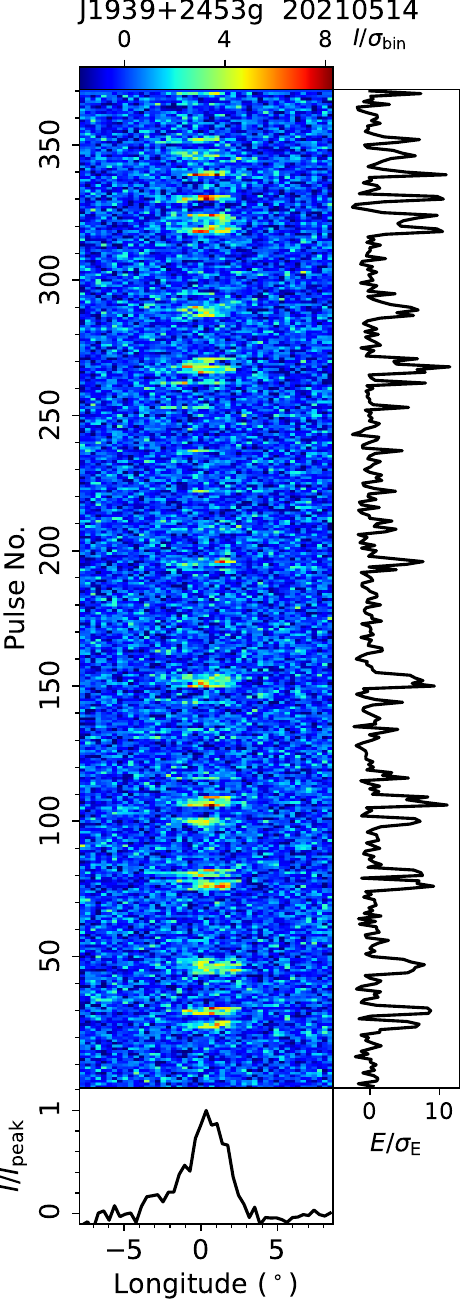}
\figcaption{Single pulse sequence of PSR J1939+2453g from the FAST observation on 20210514.
\label{subfig:TP:J1939+2453g}}
\end{figure}

\begin{figure}[htpb]
\centering
\includegraphics[width=0.39\textwidth, angle=0]{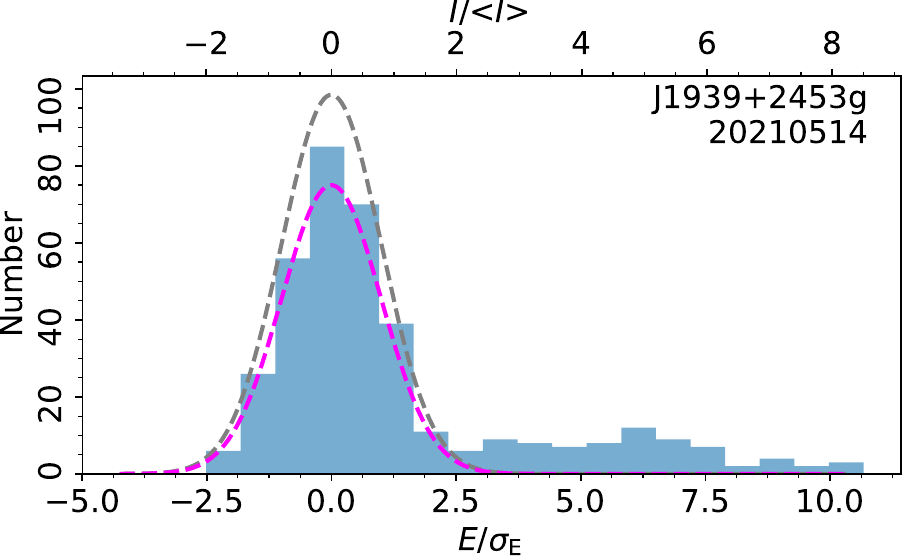}
\figcaption{On-pulse energy histogram of single pulses of PSR J1939+2453g from the FAST observation on 20210514.
\label{subfig:Hist:J1939+2453g}}
\end{figure}

\subsection{J1939+10}
\label{subsec:J1939+10}

PSR J1939+10 was discovered by the Arecibo telescope \citep{Camilo1996}, and reported to exhibit dwarf pulses during the nulling state from the sensitive FAST observations \citep{Yan2024}. 

The pulsar was observed by FAST on 20240317 for 18 minutes, deriving a rotation period $P=2.3087$~s and a dispersion measure $D\!M=73.6~{\rm cm^{-3}\,pc}$. 
Single pulse sequences are shown in Fig.~\ref{subfig:TP:J1939+10}, displaying nulling and subpulse drifting phenomena. From the on-pulse integral energy histogram (Fig.~\ref{subfig:Hist:J1939+10}), the nulling fraction of this observation is estimated to be 44$\pm$13\%. Fluctuation spectra of leading and trailing parts in the mean pulse profile are shown in Fig.~\ref{subfig:fluctu:J1939+10}, and negative drifting parameters are estimated from the centroid of the drift features in 2DFS. 
For 2DFS of the leading profile part, the centroid frequencies of the drift feature are $1/P_3=0.207\pm0.002$ and $1/P_2=-80\pm2$, corresponding to $P_3=4.84\pm0.05$ periods and $P_2=-4.5\pm0.1^\circ$. 
The drift feature in 2DFS of the trailing profile part exhibits the centroid of $1/P_3=0.208\pm0.003$ and $1/P_2=-48\pm3$, yielding $P_3=4.8\pm0.1$ periods and $P_2=-7.6\pm0.5^\circ$. 

\begin{figure}[htpb]
\centering
\includegraphics[width=0.22\textwidth, angle=0]{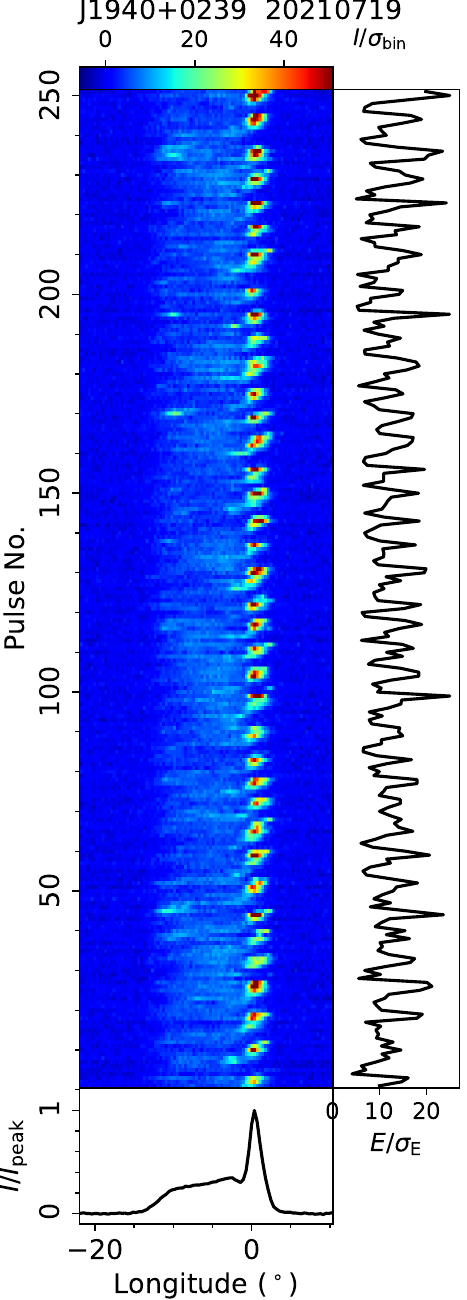}
\figcaption{Single pulse sequence of PSR J1940+0239 from the FAST observation on 20210719.
\label{subfig:TP:J1940+0239}}
\end{figure}

\begin{figure}[htpb]
\centering
\includegraphics[width=0.44\textwidth, angle=0]{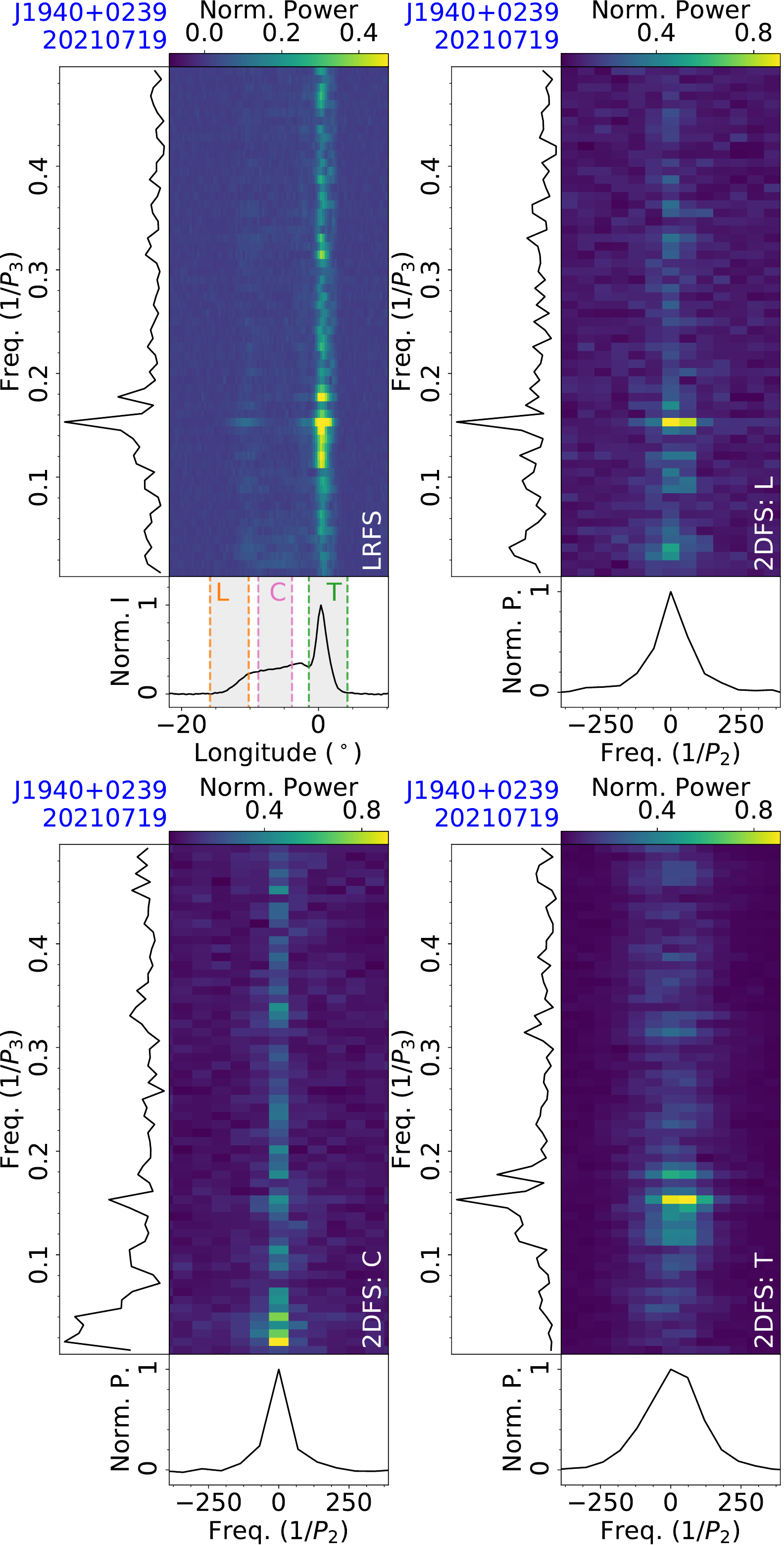}
\figcaption{Fluctuation analysis of PSR J1940+0239 for the observation on 20210719, with LRFS (top-left), and 2DFS for the leading part (top-right), central part (bottom-left) and trailing part (bottom-right) of a mean pulse profile.
\label{subfig:fluctu:J1940+0239}}
\end{figure}

\subsection{J1939+2352g}
\label{subsec:J1939+2352g}

PSR J1939+2352g was discovered in the FAST GPPS survey \citep{Han2021,han2025}.

This pulsar was observed by FAST on 20240317 for 20 minutes, deriving a rotation period $P=2.1451$~s and a dispersion measure $D\!M=413.5~{\rm cm^{-3}\,pc}$. 
The nulling fraction is estimated to be 52.6\% from the on-pulse integral energy histogram in Fig.~\ref{subfig:Hist:J1939+2352g}. Additionally, there is also drifting behavior from the single pulse (Fig.~\ref{subfig:TP:J1939+2352g}) and fluctuation spectra (Fig.~\ref{subfig:fluctu:J1939+2352g}). 
In 2DFS, there is a positive drift feature with the centroid of $1/P_3=0.245\pm0.004$ and $1/P_2=61\pm3$, corresponding to periodicities of $P_3=4.1\pm0.1$ periods and $P_2$=5.9$\pm0.3$ degrees.

\begin{figure}[htpb]
\centering
\includegraphics[width=0.22\textwidth, angle=0]{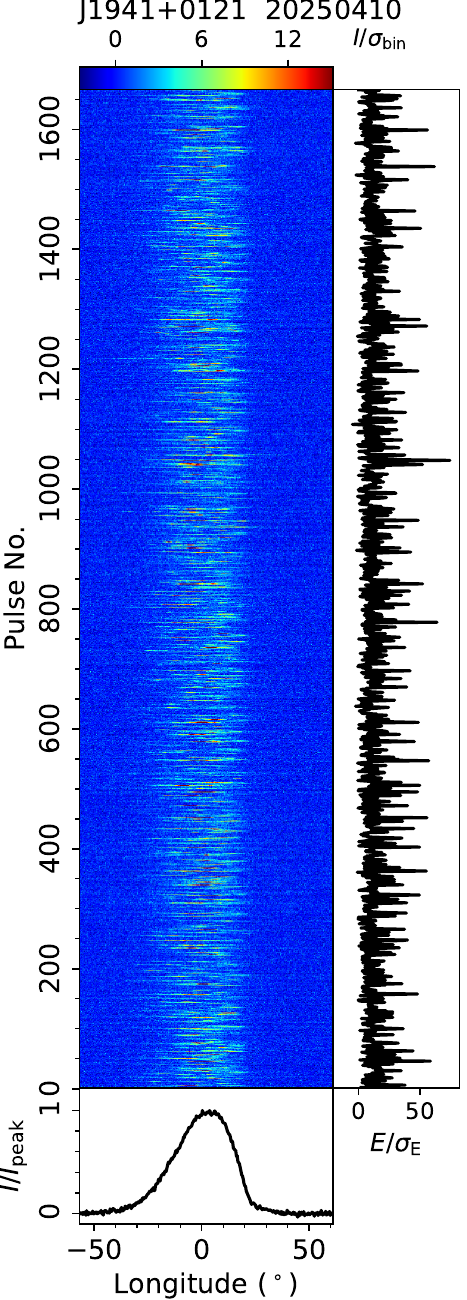}
\includegraphics[width=0.22\textwidth, angle=0]{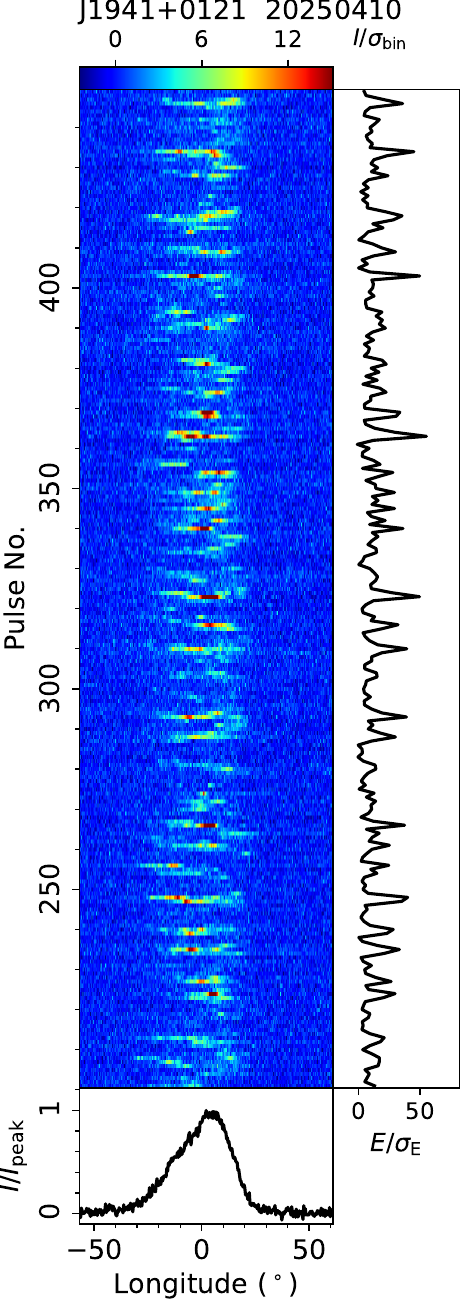}
\figcaption{Single pulse sequence of PSR J1941+0121 from the FAST observation on 20250410, and a zoomed-in view of pulses No. 200-450.
\label{subfig:TP:J1941+0121}}
\end{figure}

\begin{figure}[htpb]
\centering
\includegraphics[width=0.22\textwidth, angle=0]{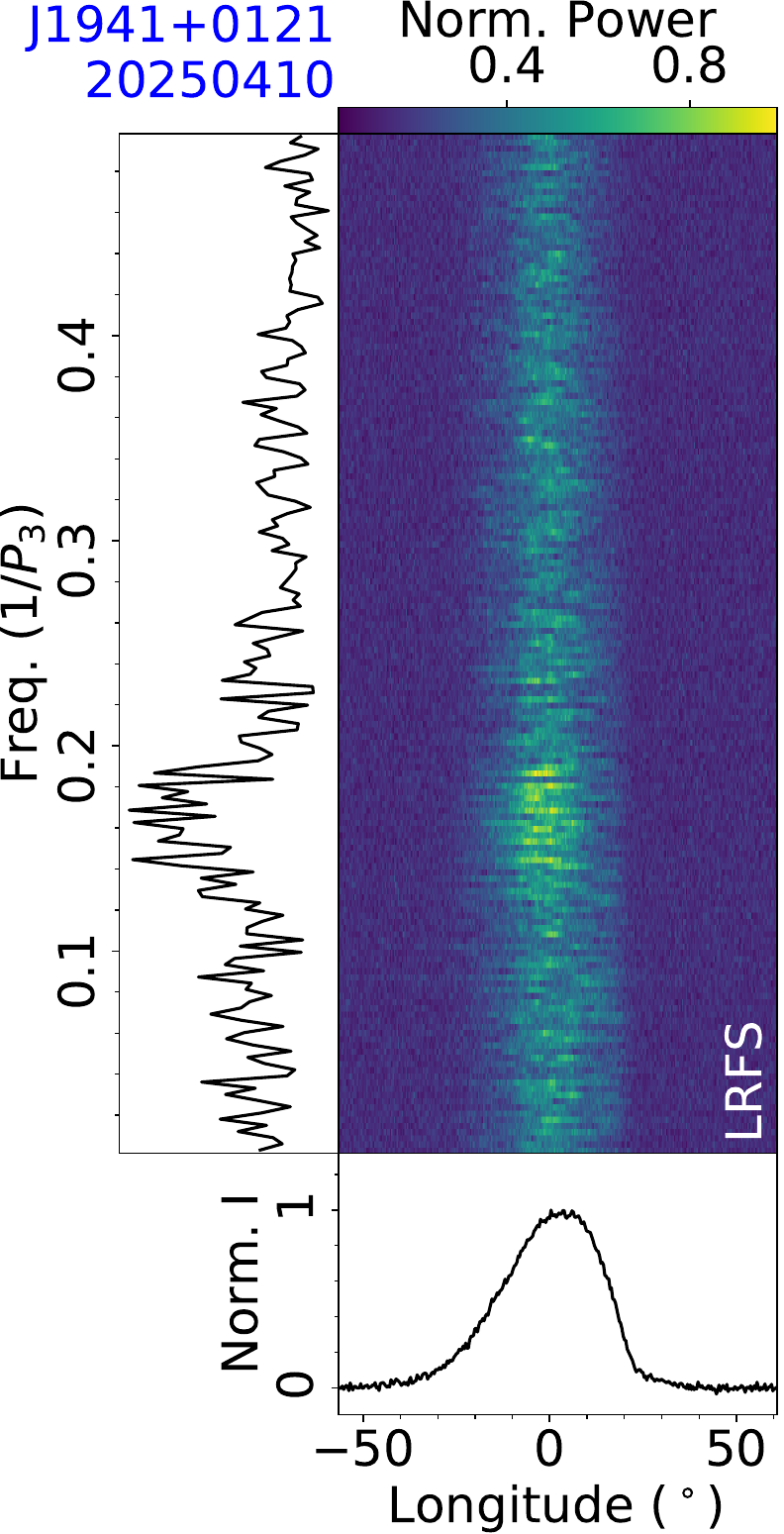}
\includegraphics[width=0.22\textwidth, angle=0]{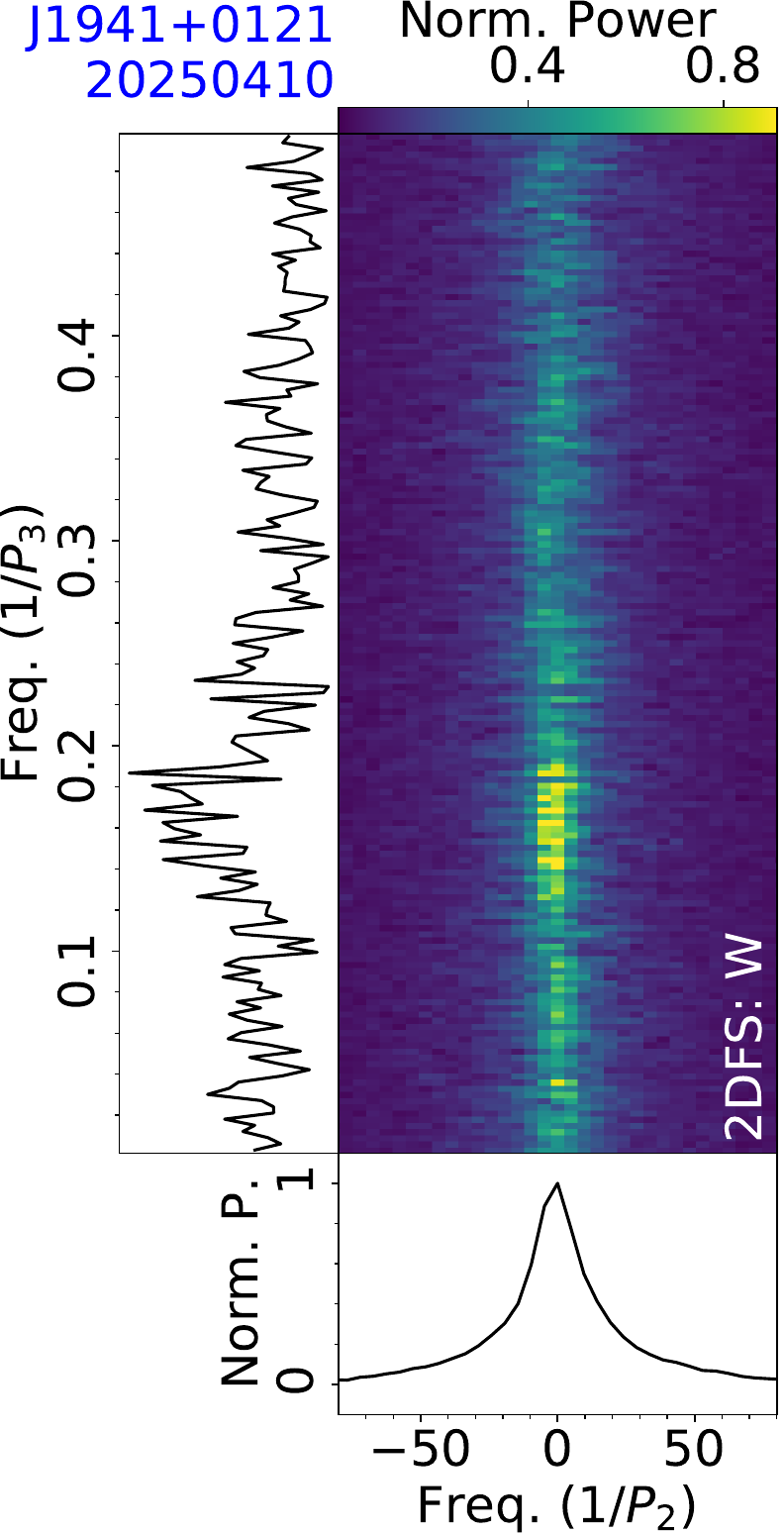}
\figcaption{Fluctuation analysis of PSR J1941+0121 for the observation on 20250410, with LRFS and 2DFS for the on-pulse region of a mean pulse profile.
\label{subfig:fluctu:J1941+0121}}
\end{figure}

\begin{figure}[htpb]
\centering
\includegraphics[width=0.22\textwidth, angle=0]{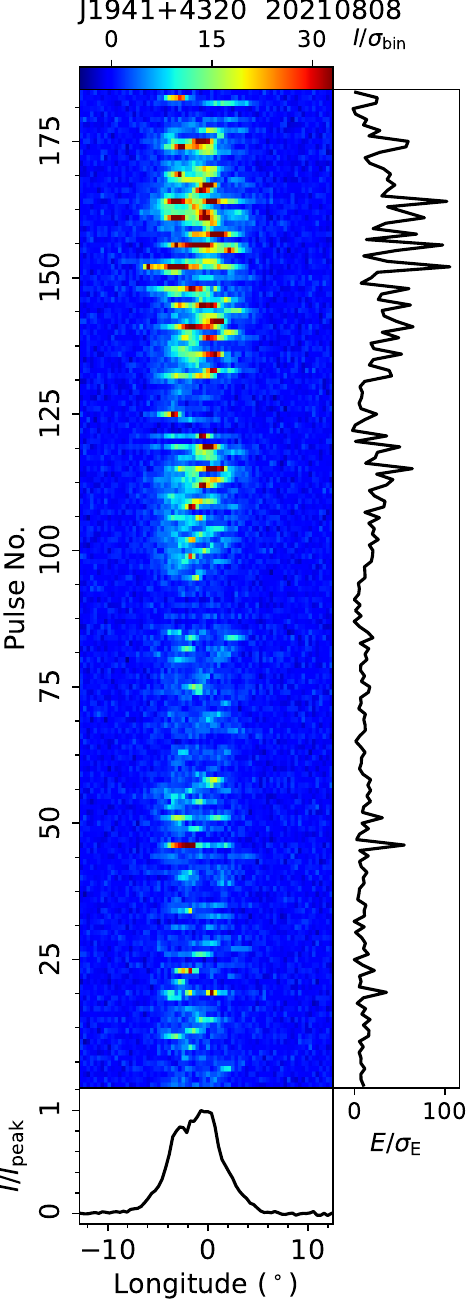}
\includegraphics[width=0.22\textwidth, angle=0]{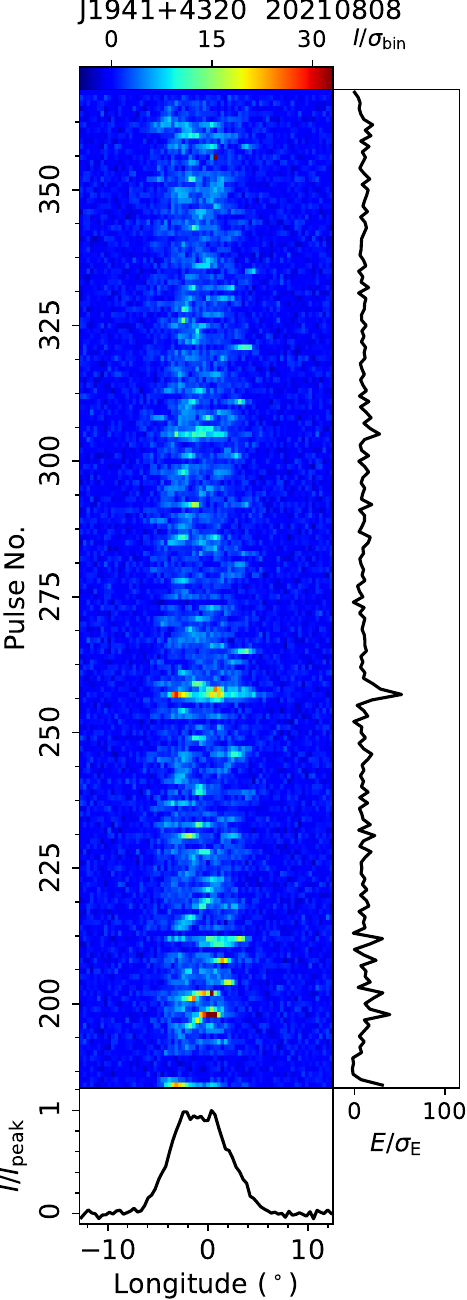}
\figcaption{Single pulse sequences of PSR J1941+4320 from the FAST observation on 20210808.
\label{subfig:TP:J1941+4320}}
\end{figure}

\begin{figure}[htpb]
\centering
\includegraphics[width=0.39\textwidth, angle=0]{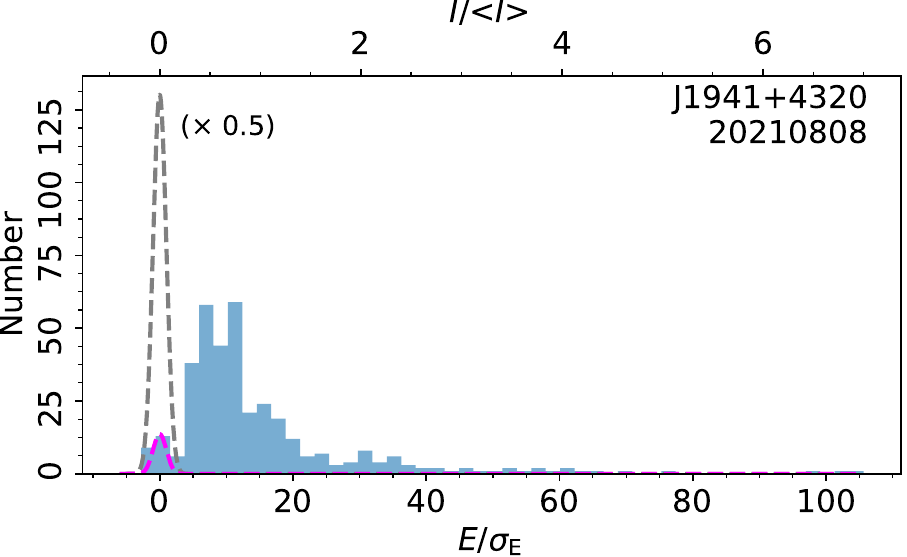}
\figcaption{On-pulse energy histogram of single pulses of PSR J1941+4320 from the FAST observation on 20210808.
\label{subfig:Hist:J1941+4320}}
\end{figure}

\begin{figure}[htpb]
\centering
\includegraphics[width=0.22\textwidth, angle=0]{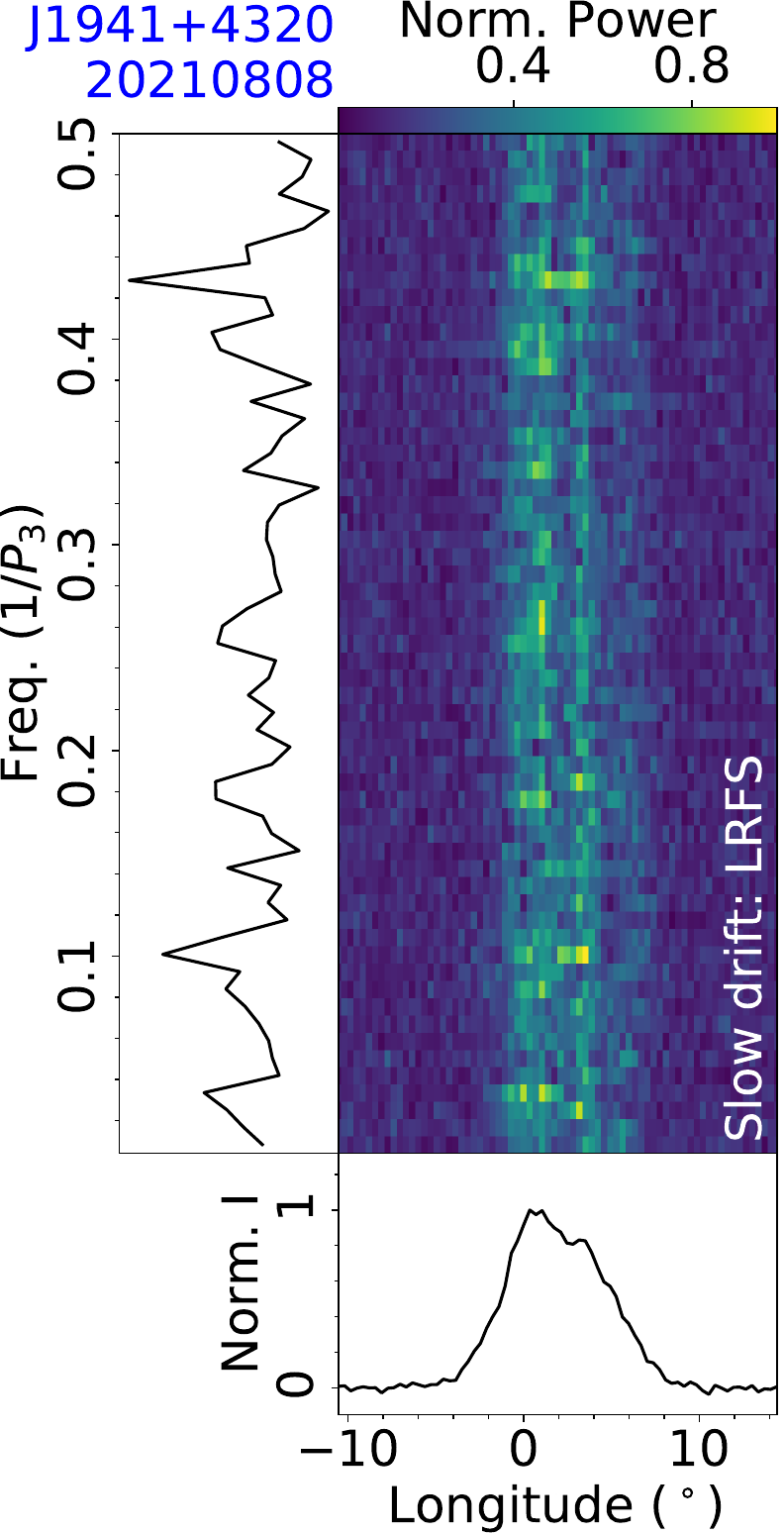}
\includegraphics[width=0.22\textwidth, angle=0]{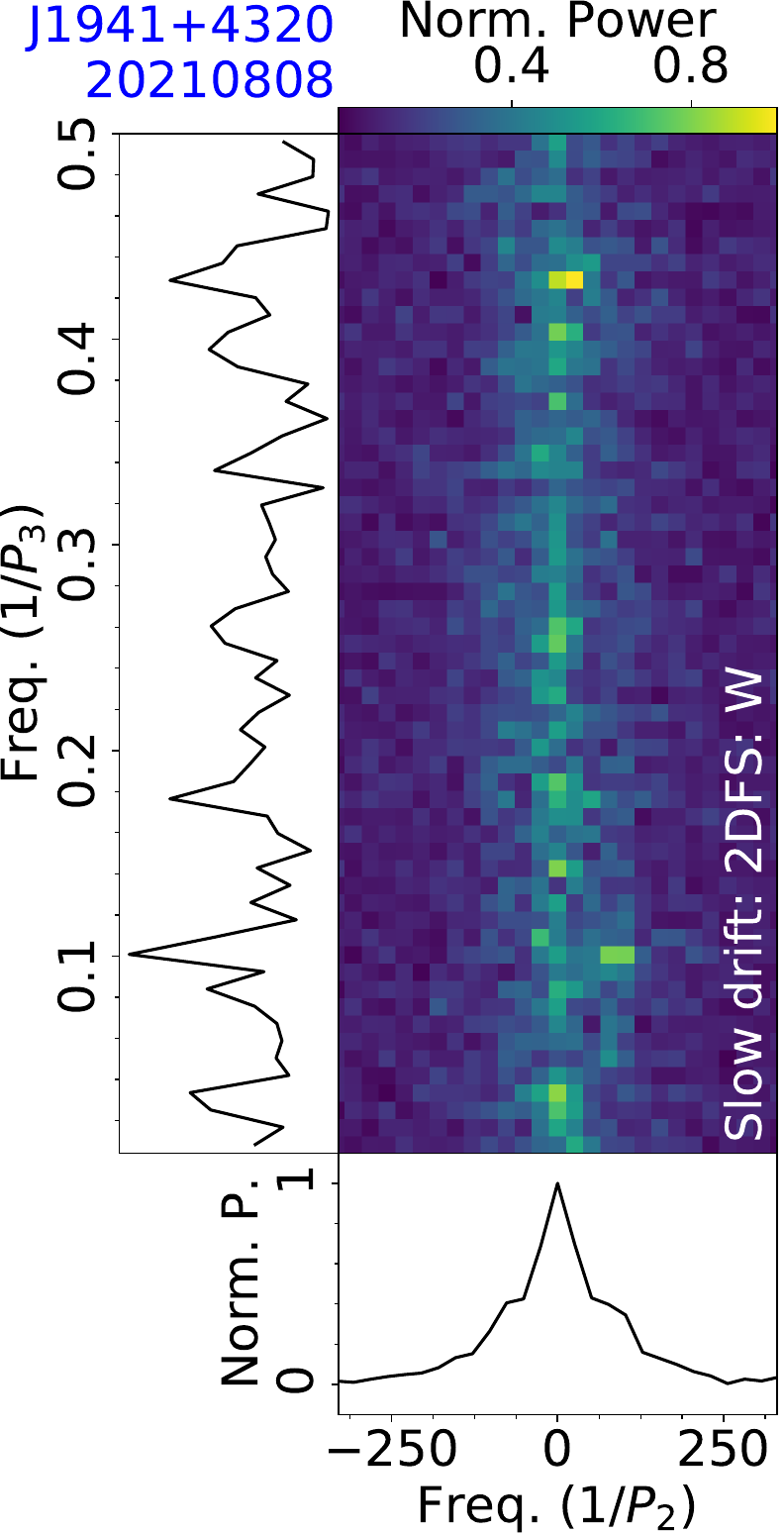}
\figcaption{Fluctuation analysis of the slow drifting mode (pulses No. 1-85 and 210-360) of PSR J1941+4320 from the FAST observation on 20210808, with LRFS and 2DFS for the on-pulse region of a mean pulse profile. \label{subfig:fluctuMode1:J1941+4320}}
\end{figure}

\begin{figure}[htpb]
\centering
\includegraphics[width=0.39\textwidth, angle=0]{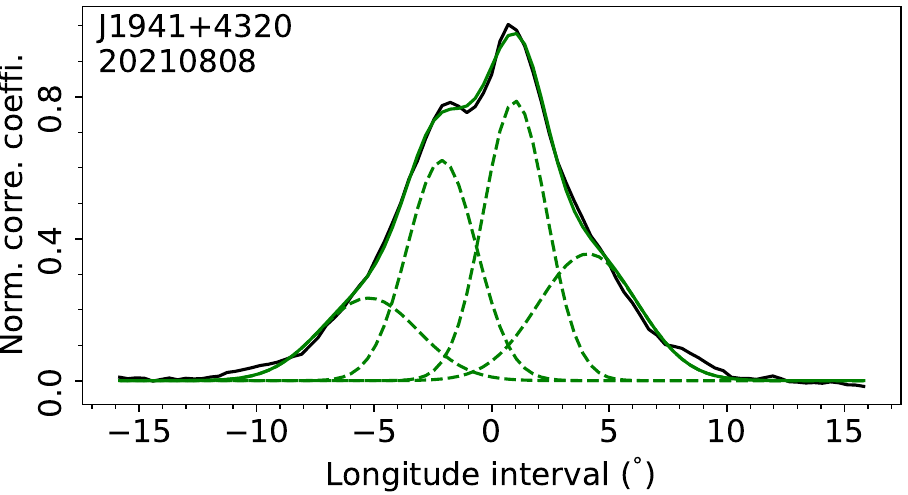}
\figcaption{Cross correlation for single pulses of the fast drifting mode (pulses No. 93-110) of the FAST observation on 20210808 of PSR J1941+4320.
\label{subfig:CorreMode2:J1941+4320}}
\end{figure}

\begin{figure}[htpb]
\centering
\includegraphics[width=0.22\textwidth, angle=0]{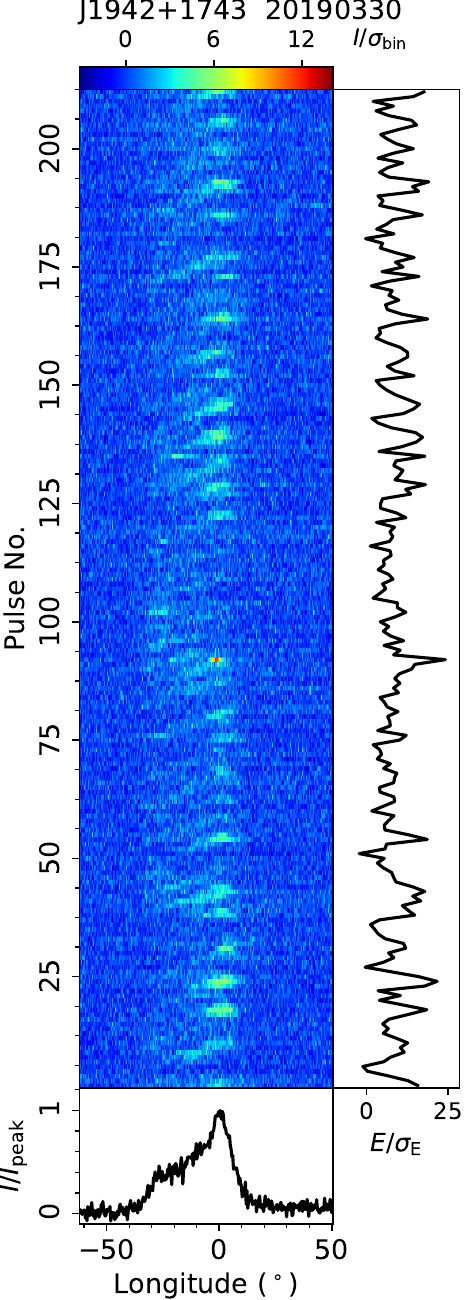}
\includegraphics[width=0.22\textwidth, angle=0]{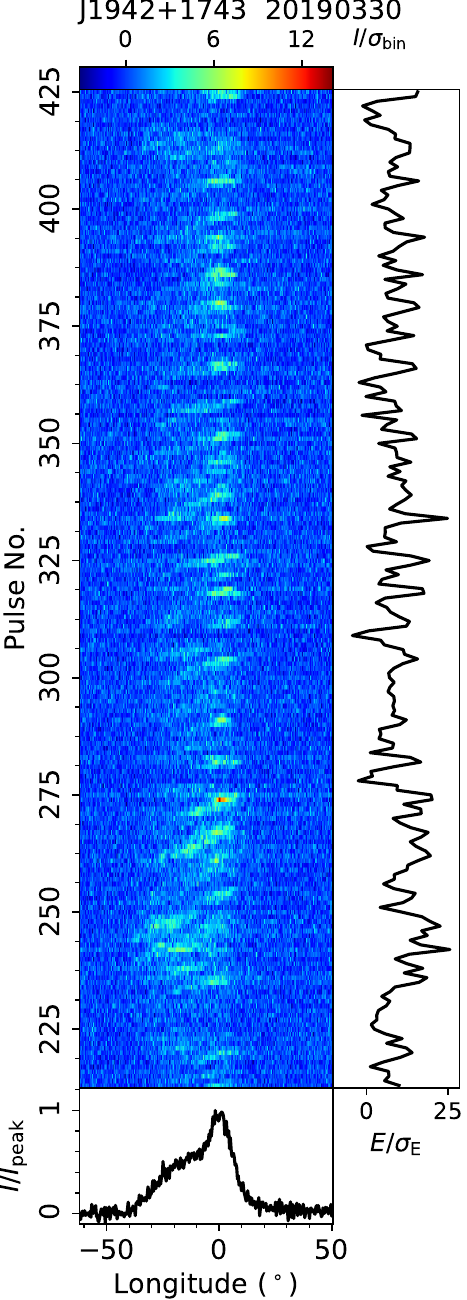}
\figcaption{Single pulse sequences of PSR J1942+1743 from the FAST observation on 20190330.
\label{subfig:TP:J1942+1743}}
\end{figure}

\begin{figure}[htpb]
\centering
\includegraphics[width=0.44\textwidth, angle=0]{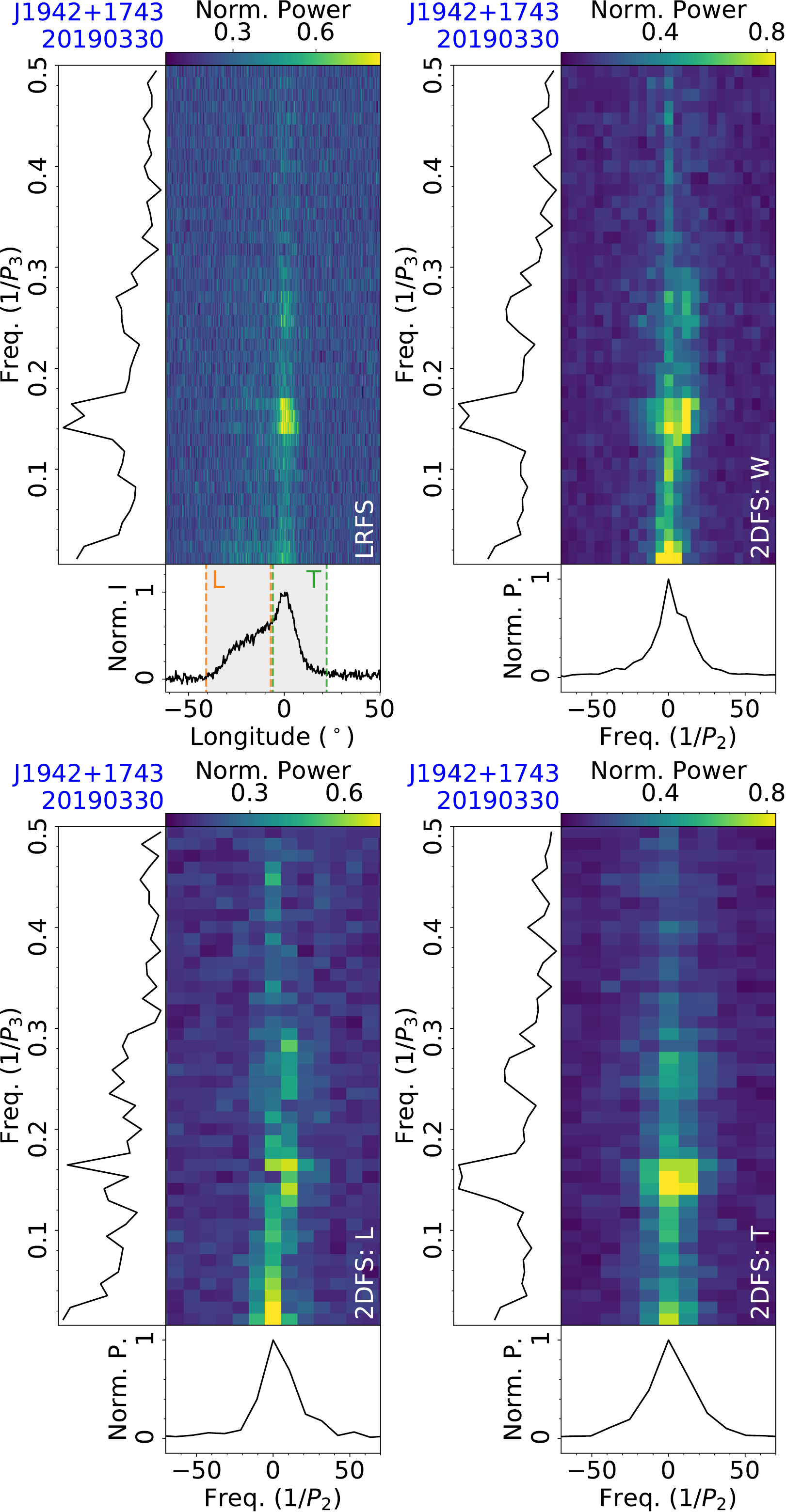}
\figcaption{Fluctuation analysis of PSR J1942+1743 for the observation on 20190330, with LRFS (top-left), and 2DFS for the on-pulse region (top-right), leading part (bottom-left) and trailing part (bottom-right) of a mean pulse profile.
\label{subfig:fluctu:J1942+1743}}
\end{figure}

\subsection{J1939+2453g}
\label{subsec:J1939+2453g}

PSR J1939+2453g was discovered in the FAST GPPS survey \citep{Han2021,han2025}. 

This pulsar was observed by FAST on 20210514 for 15 minutes, deriving a rotation period $P=2.4024$~s and a dispersion measure $D\!M=226.8~{\rm cm^{-3}\,pc}$. 
The single pulse sequence of this observation is displayed in Fig.~\ref{subfig:TP:J1939+2453g}, which shows changes between nulling and emission states. The nulling fraction of this observation is estimated to be 76$\pm$4\% from the on-pulse integral energy histogram in Fig.~\ref{subfig:Hist:J1939+2453g}.

\subsection{J1940+0239}
\label{subsec:J1940+0239}

PSR J1940+0239 was discovered in the High Time Resolution Universe survey by \citet{Morello2019}. 

The pulsar was observed by FAST on 20210719 for 5 minutes, deriving a rotation period $P=1.2322$~s and a dispersion measure $D\!M=86.1~{\rm cm^{-3}\,pc}$. 
We first report the subpulse behavior of the pulsar here. In single pulse sequences (Fig.~\ref{subfig:TP:J1940+0239}), drift bands of the trailing component are clear. LRFS and 2DFS of three longitude parts in the mean pulse profile are displayed in Fig.~\ref{subfig:fluctu:J1940+0239}. 
For the leading profile part, 2DFS exhibits a preferred positive drift feature with the centroid of $1/P_3=0.153\pm0.002$ and $1/P_2=28\pm12$, corresponding to $P_3=6.5\pm0.1$ periods and $P_2=13\pm6^\circ$. 
For the trailing profile part, there is also a positive drift feature in 2DFS, with the centroid characterized by $1/P_3=0.148\pm0.001$ and $1/P_2=33\pm4$, yielding $P_3=6.7\pm0.1$ periods and $P_2=11\pm1^\circ$. 
2DFS of the central phase part in the profile has a low-frequency modulation feature, with the centroid frequency of $1/P_3=0.032\pm0.002$, corresponding to $P_3=31\pm2$, that requires longer observations to confirm.

\begin{figure}[htpb]
\centering
\includegraphics[width=0.22\textwidth, angle=0]{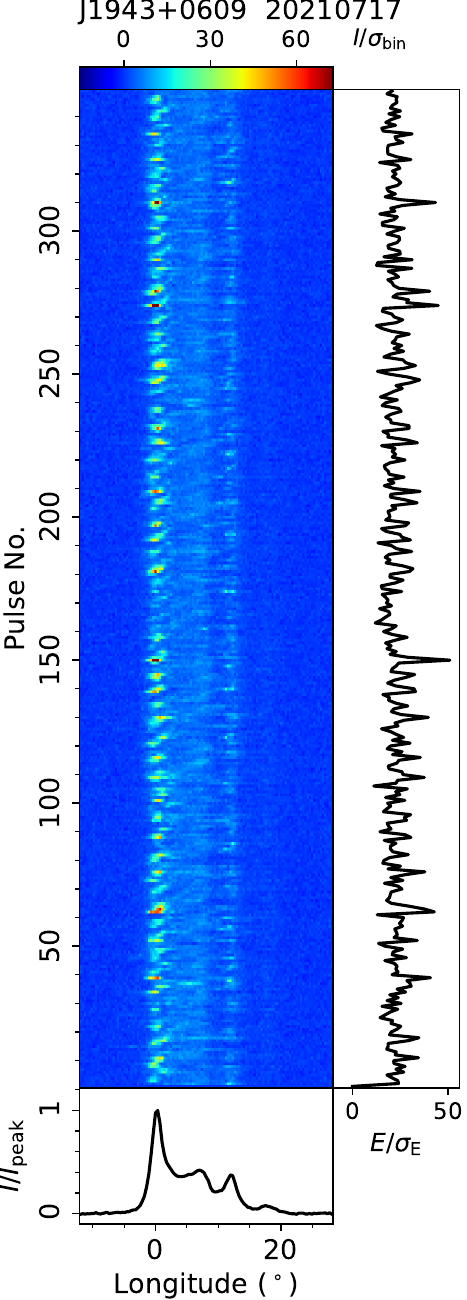}
\includegraphics[width=0.22\textwidth, angle=0]{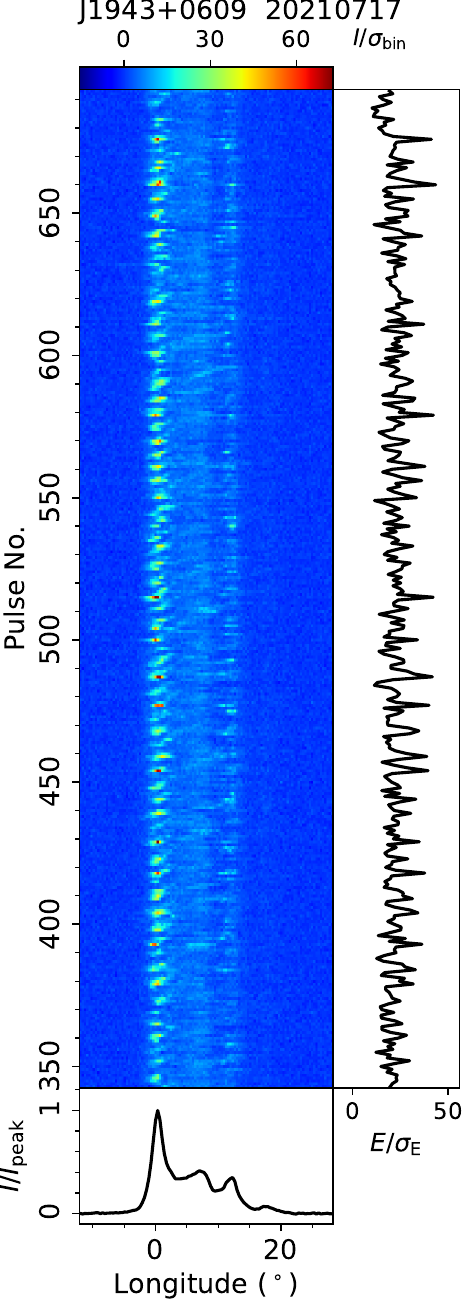}
\figcaption{Single pulse sequences of PSR J1943+0609 from the FAST observation on 20210717.
\label{subfig:TP:J1943+0609}}
\end{figure}

\begin{figure}[htpb]
\centering
\includegraphics[width=0.22\textwidth, angle=0]{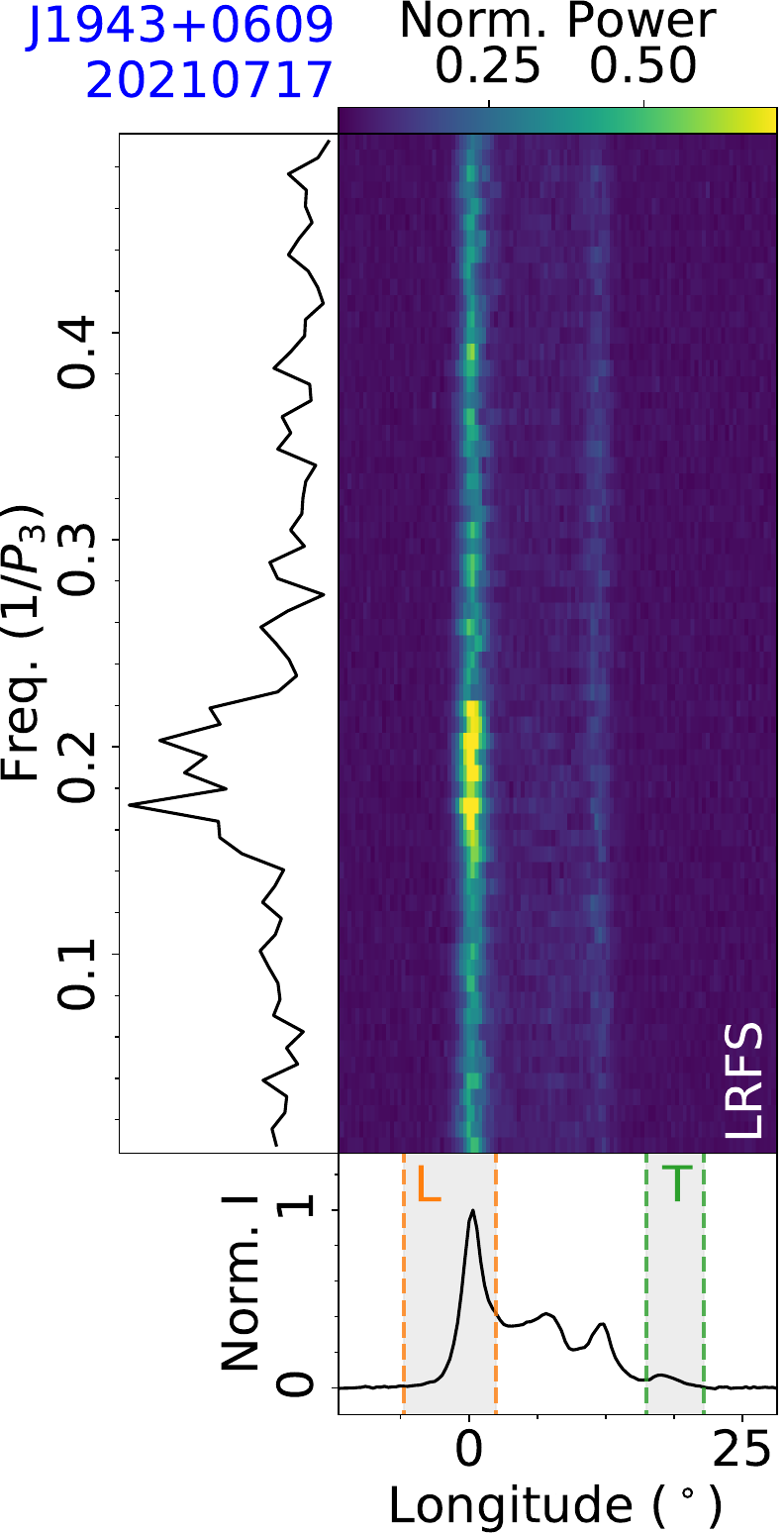}
\includegraphics[width=0.22\textwidth, angle=0]{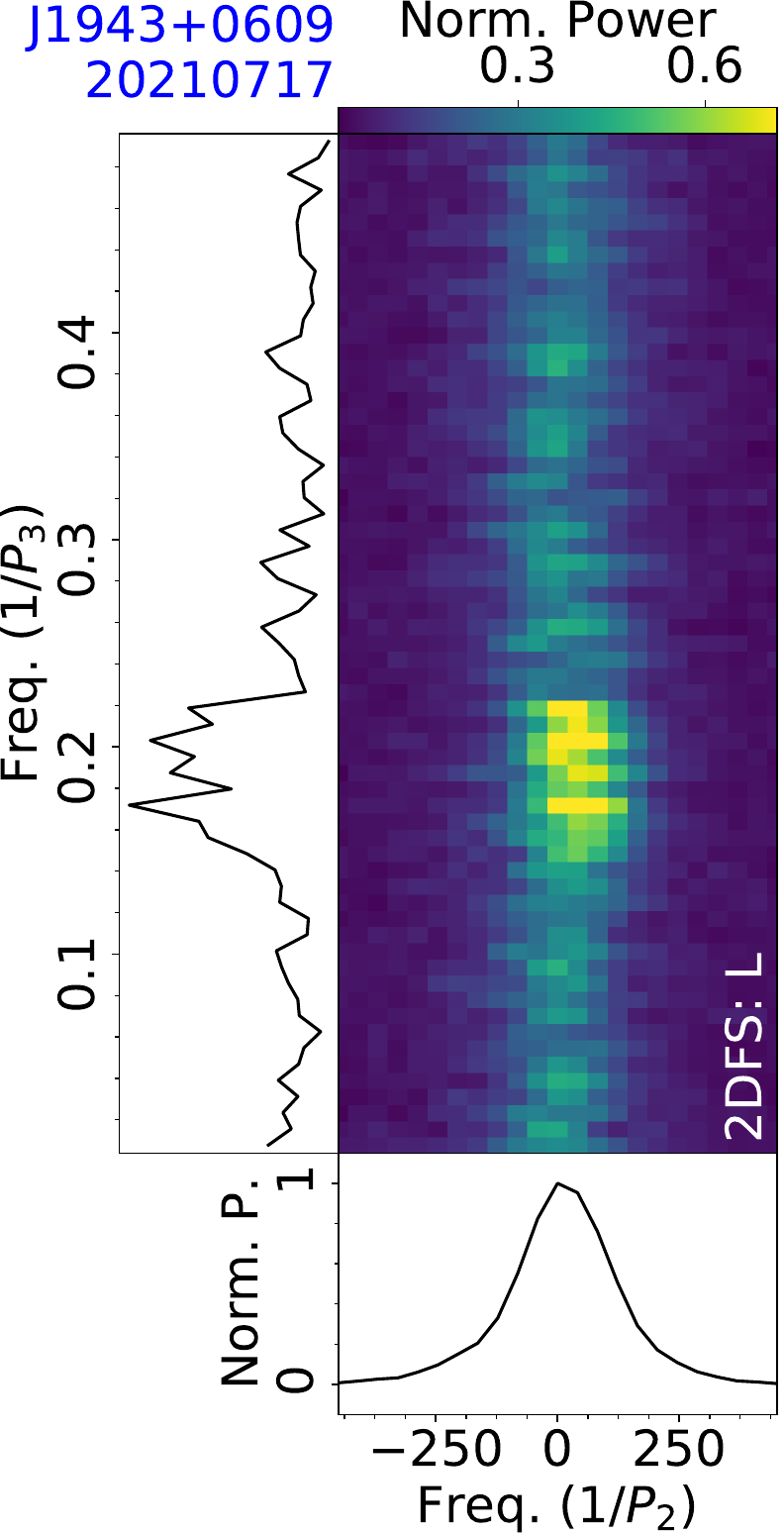}
\figcaption{Fluctuation analysis of PSR J1943+0609 for the observation on 20210717, with LRFS and 2DFS for the leading part of a mean pulse profile.
\label{subfig:fluctu:J1943+0609}}
\end{figure}

\begin{figure}[htpb]
\centering
\includegraphics[width=0.22\textwidth, angle=0]{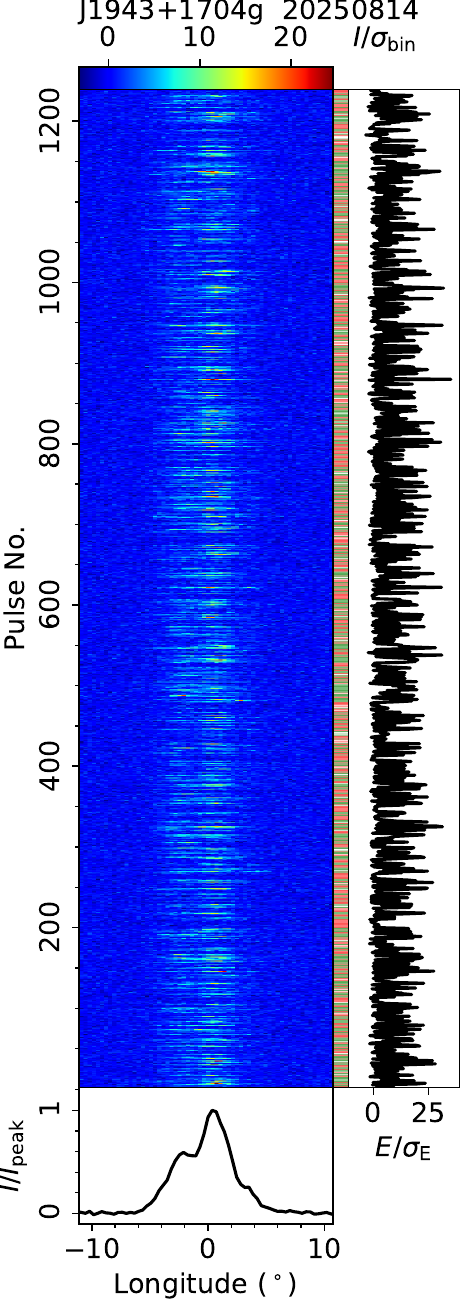}
\includegraphics[width=0.22\textwidth, angle=0]{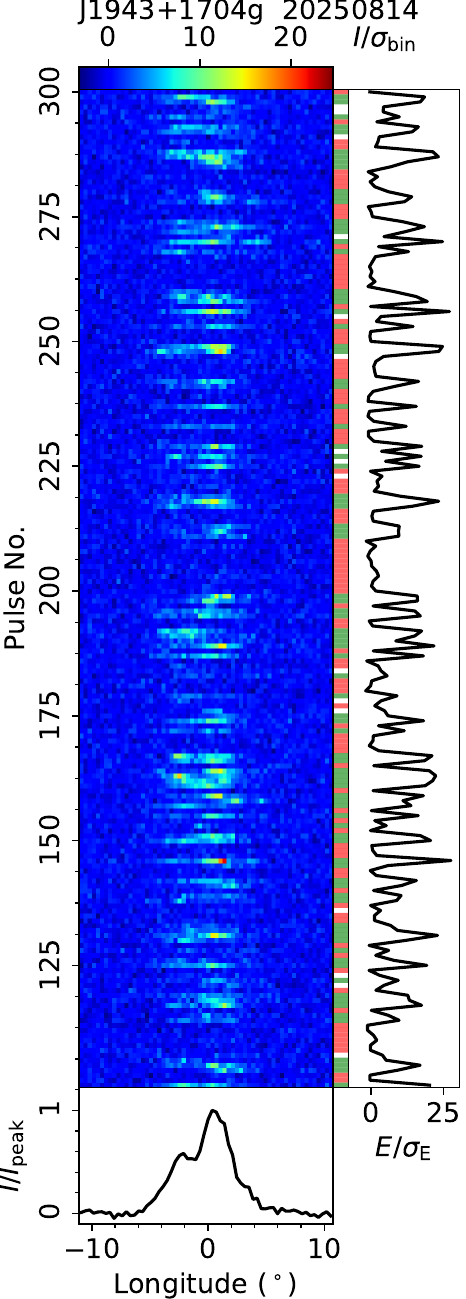}
\figcaption{Single pulse sequence of PSR J1943+1704g from the FAST observation on 20230214, and a zoomed-in view of pulses No. 100-300. 
The red and green bars represent weak or bright emission modes. In the right subpanel, the on-pulse energy variation is plotted against period.
\label{subfig:TP:J1943+1704g}}
\end{figure}

\begin{figure}[htpb]
\centering
\includegraphics[width=0.39\textwidth, angle=0]{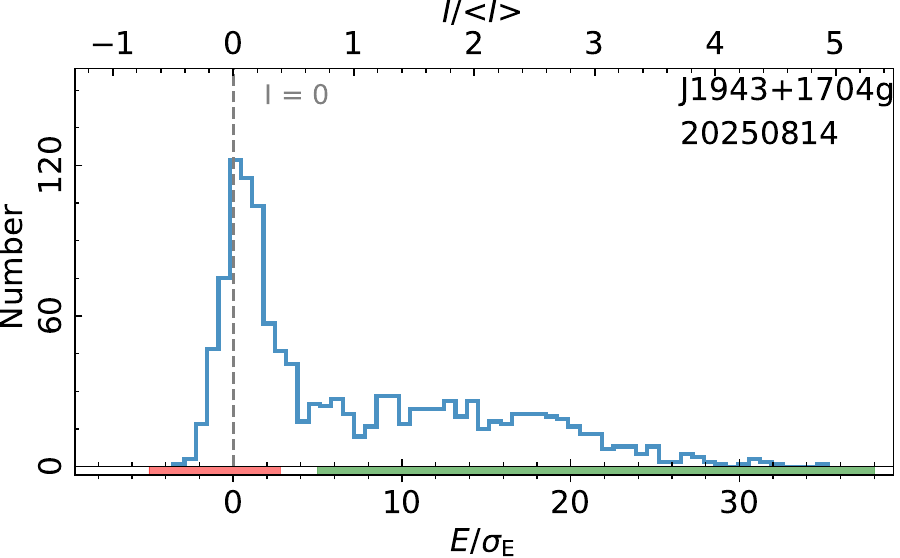}
\figcaption{On-pulse energy histogram of single pulses of PSR J1943+1704g from the FAST observation on 20230214. 
The red and green bars indicate the weak and bright emission modes.
\label{subfig:Hist:J1943+1704g}}
\end{figure}

\begin{figure}[htpb]
\centering
\includegraphics[width=0.39\textwidth, angle=0]{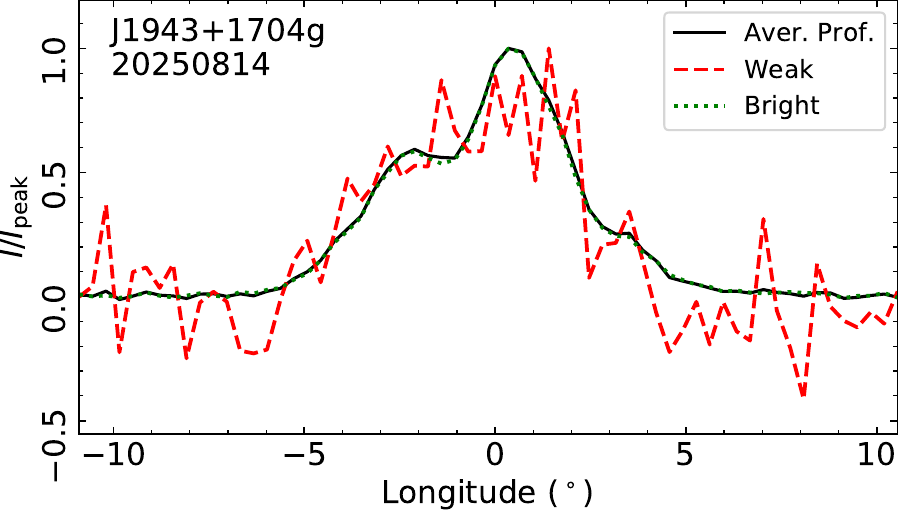}
\figcaption{Mean profiles for the weak (red dashed) and bright (green dotted) emission modes of PSR J1943+1704g observed on 20250814, which are normalized by their respective peaks of the mean pulse profile.
\label{subfig:ProfModes:J1943+1704g}}
\end{figure}

\subsection{J1941+0121}
\label{subsec:J1941+0121}

PSR J1941+0121 was discovered by \citet{Boyles2013} with the Green Bank Telescope. Drifting parameters of $P_3=6.2\pm0.5$ periods and $P_2=-87^{+21}_{-435}$ degrees were reported by \citet{Song2023}.

This pulsar was observed by FAST on 20250410 for 6 minutes, with a rotation period $P=0.2173$~s and a dispersion measure $D\!M=51.5~{\rm cm^{-3}\,pc}$ derived. The single pulse sequence and a zoomed-in view of pulses No. 200-450 in Fig.~\ref{subfig:TP:J1941+0121} show the subpulse drifting phenomenon. From LRFS and 2DFS in Fig.~\ref{subfig:fluctu:J1941+0121}, the centroid modulation frequencies of the negative drift feature are estimated to be $1/P_3=0.159\pm0.001$ and $1/P_2=-1.5\pm0.2$, corresponding to periodicities of $P_3=6.28\pm0.03$ periods and $P_2=-248\pm39$ degrees.

\subsection{J1941+4320}
\label{subsec:J1941+4320}

PSR J1941+4320 was discovered by \citet{Stovall2014} using the Green Bank Telescope.

The pulsar was observed by FAST on 20210808 for 5 minutes, deriving a rotation period $P=0.8409$~s and a dispersion measure $D\!M=79.2~{\rm cm^{-3}\,pc}$. 
Single pulse sequences of the observation are shown in Fig.~\ref{subfig:TP:J1941+4320}. 
J1941+4320 has nulling behavior, with the fraction of this observation estimated to be 5$\pm$1\% from the on-pulse integral energy histogram in Fig.~\ref{subfig:Hist:J1941+4320}. 
From single pulse sequences in Fig.~\ref{subfig:TP:J1941+4320}, the pulsar has mode changes. Single pulses between No. 130 and 175 have brighter emission, and there is no systematic modulation for this segment. There are systematic drift bands for single pulses of 1-85 and 210-360, and this drifting mode is defined as the Sd mode. Fluctuation spectra of this slow drifting mode, as shown in Fig.~\ref{subfig:fluctuMode1:J1941+4320}, illustrate that the centroid frequencies of the drift feature are $1/P_3=0.101\pm0.002$ ($P_3=9.9\pm0.2$ periods) and $P_2=90\pm5$ ($P_2=4.0\pm0.2^\circ$). 
$P_3=9.9\pm0.1$ periods and $P_2=4.0\pm0.2^\circ$. Additionally, there seems to be faster subpulse drifting in the single pulse segment between pulses Nos 93 and 110, which is defined to be Fd mode. From the cross-correlation method (Fig.~\ref{subfig:CorreMode2:J1941+4320}), the drifting parameters of the fast drifting mode are $D=1.0\pm0.4$ degrees per period and $P_2=3.11\pm0.08^\circ$.

\subsection{J1942+1743}
\label{subsec:J1942+1743}

PSR J1942+1743 was discovered by \citet{Hulse1975} using the Arecibo telescope.

This pulsar was observed by FAST on 20190330 for 5 minutes, and a rotation period $P=0.6962$~s and a dispersion measure $D\!M=186.6~{\rm cm^{-3}\,pc}$ were determined. 
Single pulse sequences in Fig.~\ref{subfig:TP:J1942+1743} display subpulse drifting behavior. 
Drifting properties of the leading and trailing parts in the mean pulse profile are estimated from the fluctuation spectra in Fig.~\ref{subfig:fluctu:J1942+1743}. The main drift feature for the leading profile part exhibits the centroid frequencies of $1/P_3=0.156\pm0.002$ and $1/P_2=11\pm1$, corresponding to $P_3=6.4\pm0.1$ periods and $P_2=34\pm5$ degrees. For the trailing profile part, the centroid of the main drift feature is characterized by frequencies of $1/P_3=0.153\pm0.001$ and $1/P_2=4\pm1$, yielding $P_3=6.6\pm0.1$ periods and $1/P_2=89\pm27$ degrees.

\begin{figure}[htpb]
\centering
\includegraphics[width=0.22\textwidth, angle=0]{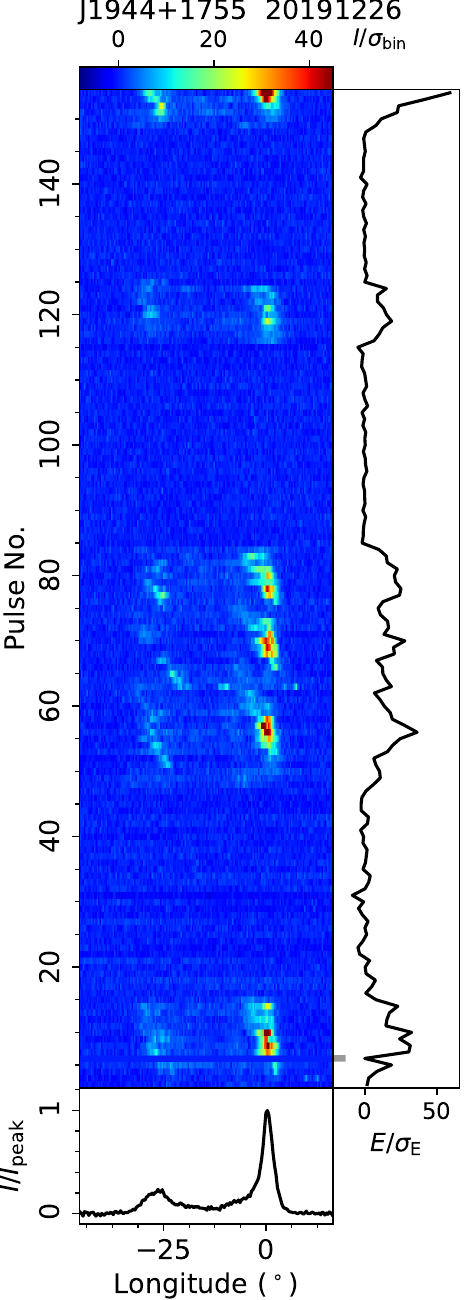}
\includegraphics[width=0.22\textwidth, angle=0]{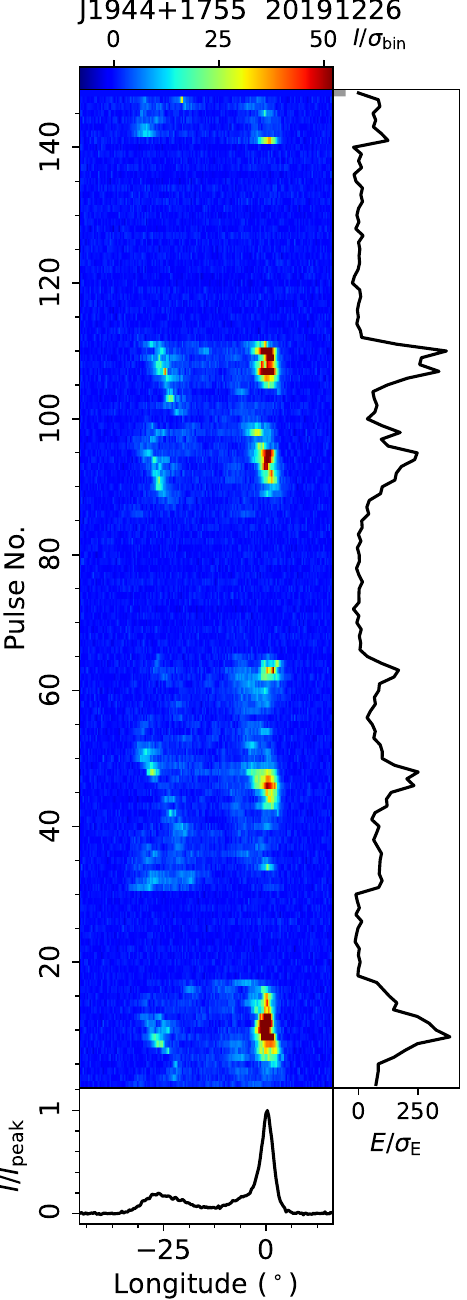}
\figcaption{Single pulse sequences of PSR J1944+1755 for two beams of P3M11 and P4M11 from the FAST observation on 20191226 . \label{subfig:TP:J1944+1755}}
\end{figure}

\begin{figure}[htpb]
\centering
\includegraphics[width=0.39\textwidth, angle=0]{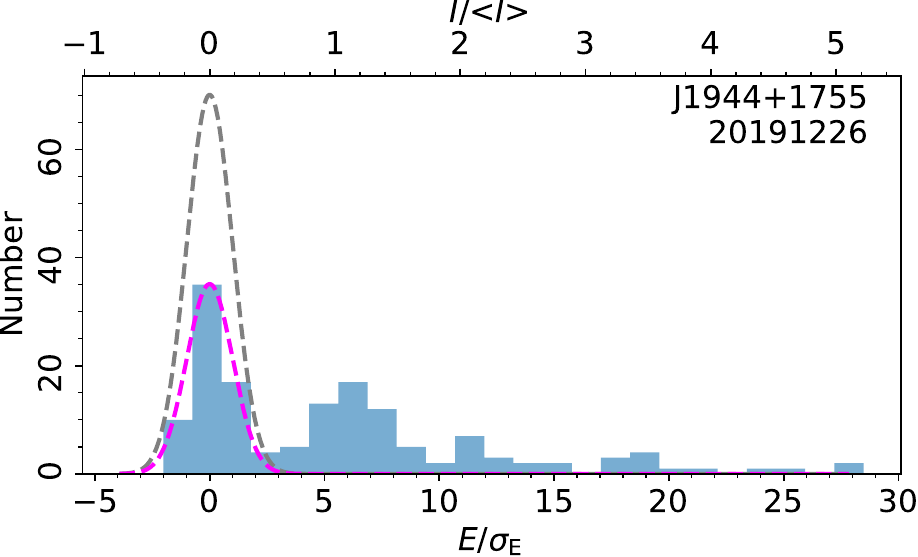}
\figcaption{On-pulse energy histogram of single pulses of PSR J1944+1755 from the FAST observation on 20191226.
\label{subfig:Hist:J1944+1755}}
\end{figure}

\begin{figure}[htpb]
\centering
\includegraphics[width=0.22\textwidth, angle=0]{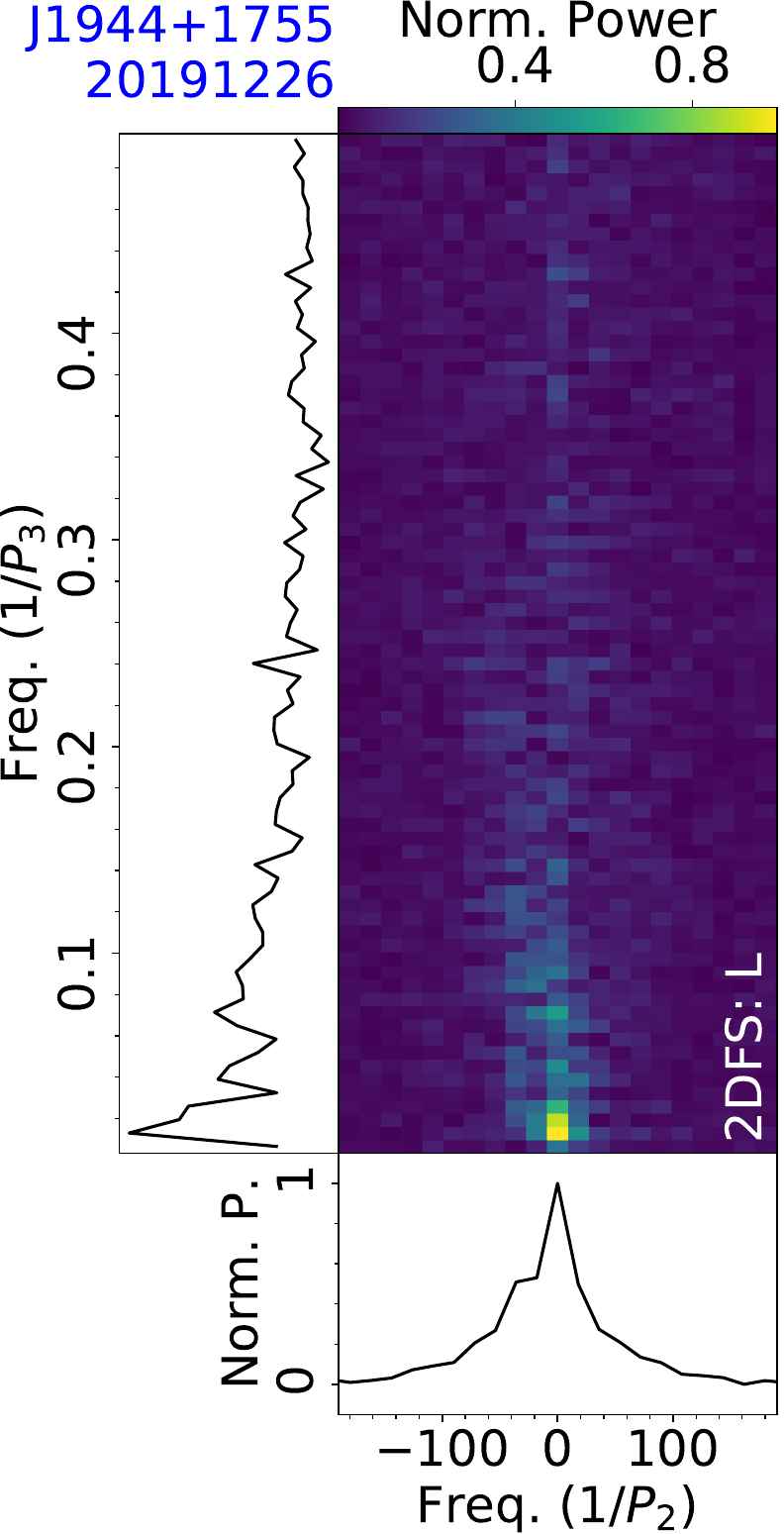}
\includegraphics[width=0.22\textwidth, angle=0]{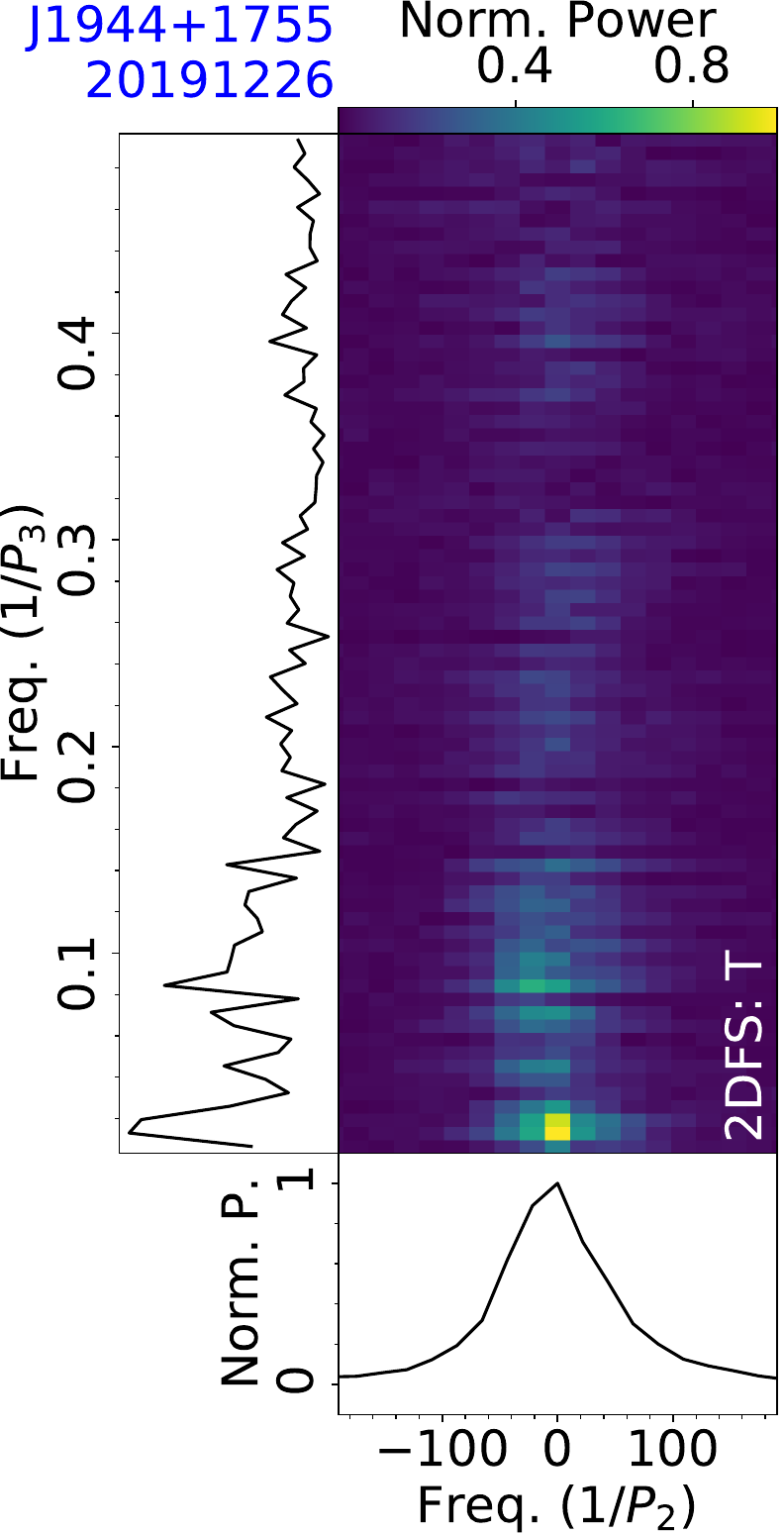}\\
\includegraphics[width=0.22\textwidth, angle=0]{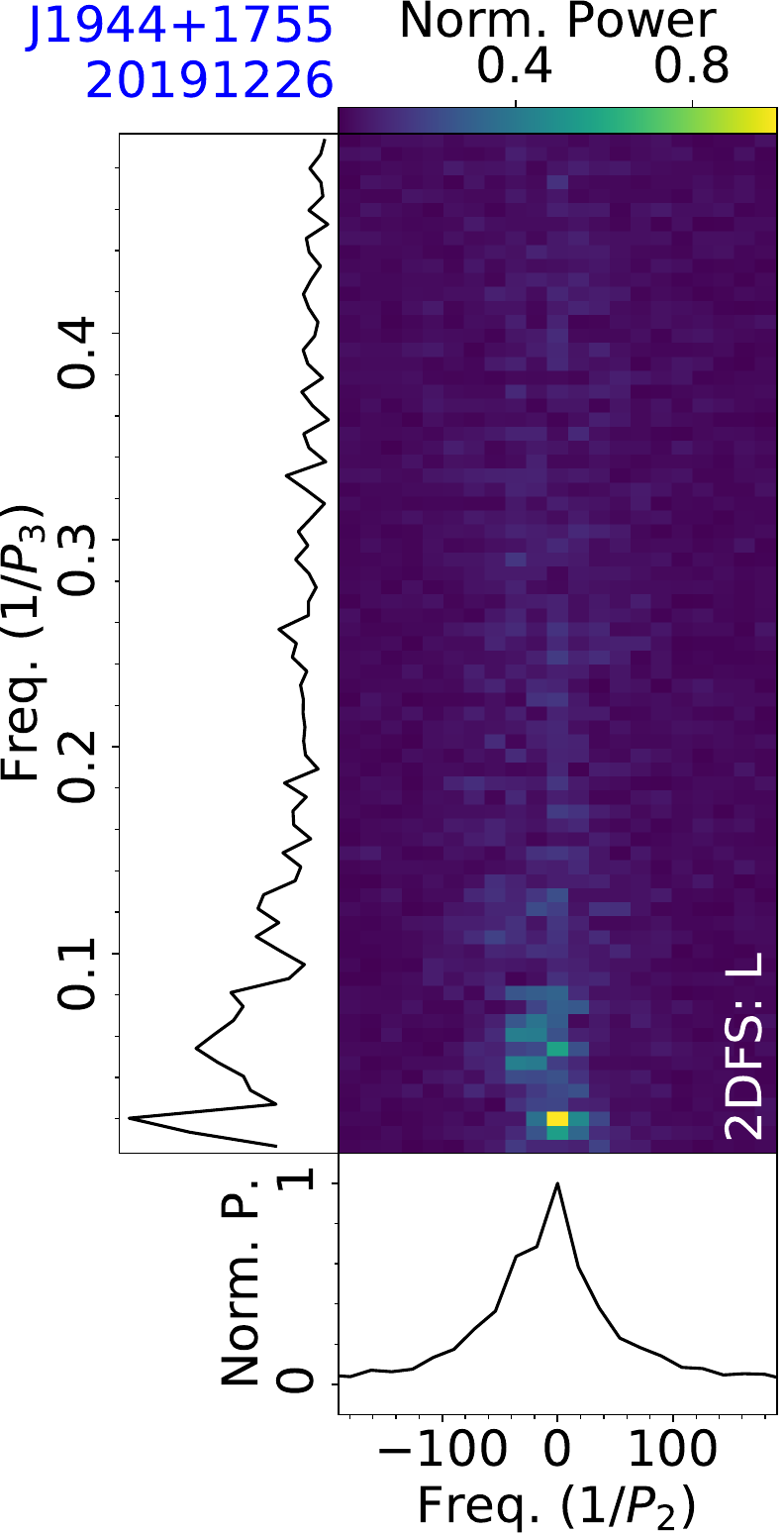}
\includegraphics[width=0.22\textwidth, angle=0]{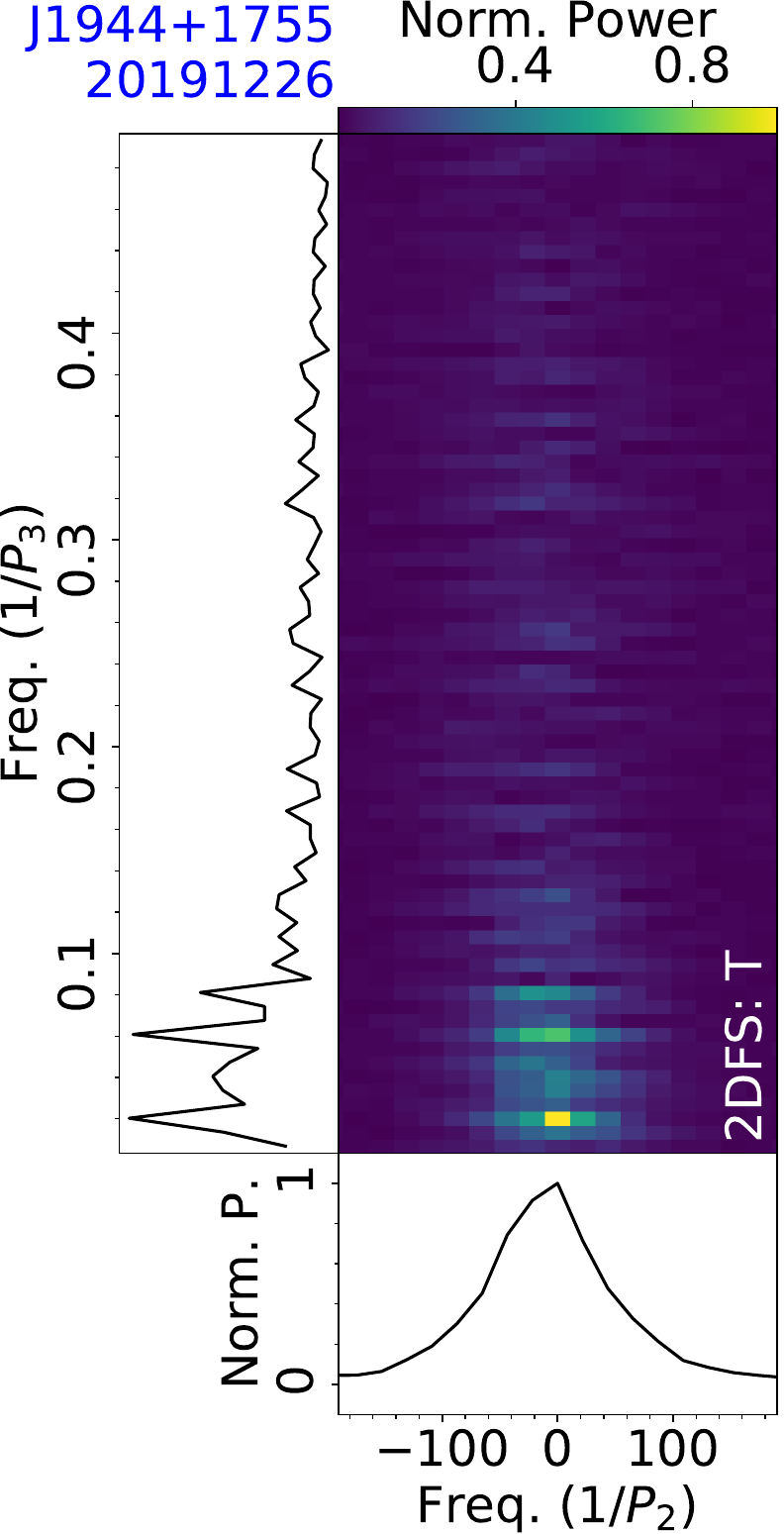}
\figcaption{Fluctuation analysis of PSR J1944+1755 from two beam data of the FAST observation on 20191226, with 2DFS for the leading and trailing parts of a mean pulse profile.
\label{subfig:fluctu:J1944+1755}}
\end{figure}

\subsection{J1943+0609}
\label{subsec:J1943+0609}

PSR J1943+0609 was discovered by \citet{Edwards2001} using the Parkes 64-m radio telescope at 1374 MHz. The drifting parameters have been reported by \citet{Song2023}. 

Here we show the result from the FAST observation conducted for 5 minutes on 20210717, deriving a rotation period $P=0.4462$~s and a dispersion measure $D\!M=70.5~{\rm cm^{-3}\,pc}$. 
Single pulse sequences are shown in Fig.~\ref{subfig:TP:J1943+0609}, where the leading profile part has systematic drifting behavior, while the modulation behavior of other phase parts is not obvious. 
LRFS and 2DFS of the leading part in the mean pulse profile in Fig.~\ref{subfig:fluctu:J1943+0609} exhibit a positive drift feature, with the centroid frequencies of $1/P_3=0.186\pm0.001$ and $1/P_2=48.7\pm2.3$, that correspond to periodicities of $P_3=5.37\pm0.02$ periods and $P_2=7.4\pm0.4^\circ$.

\begin{figure}[htpb]
\centering
\includegraphics[width=0.22\textwidth, angle=0]{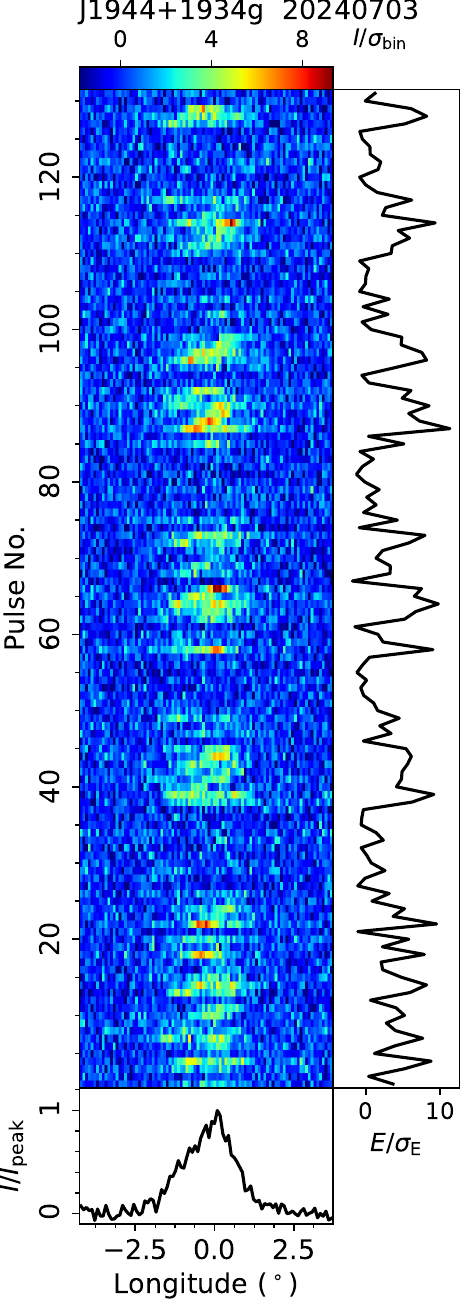}
\includegraphics[width=0.22\textwidth, angle=0]{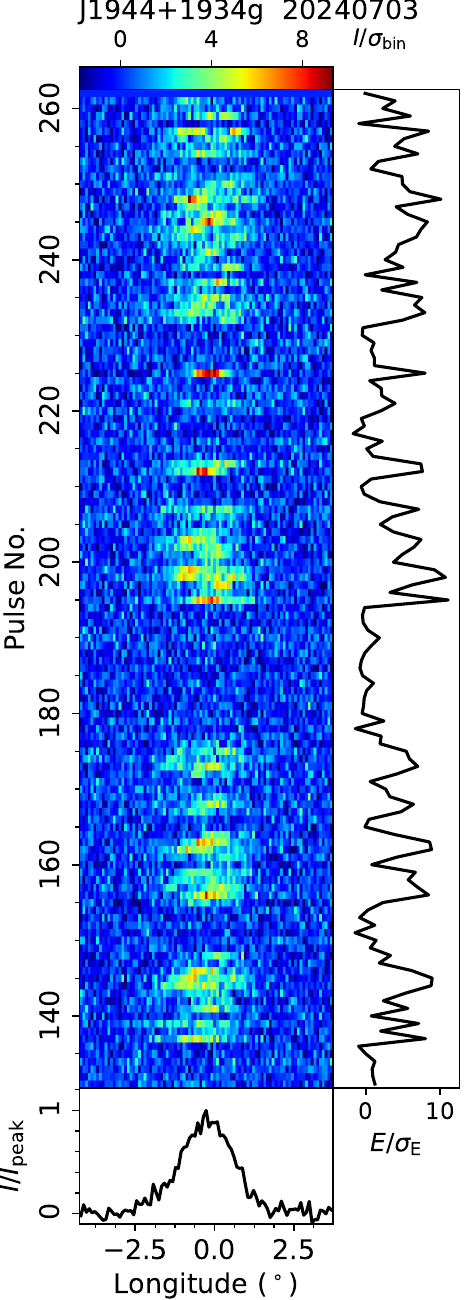}
\figcaption{Single pulse sequences of PSR J1944+1934g from the FAST observation on 20240703.
\label{subfig:TP:J1944+1934g}}
\end{figure}

\begin{figure}[htpb]
\centering
\includegraphics[width=0.39\textwidth, angle=0]{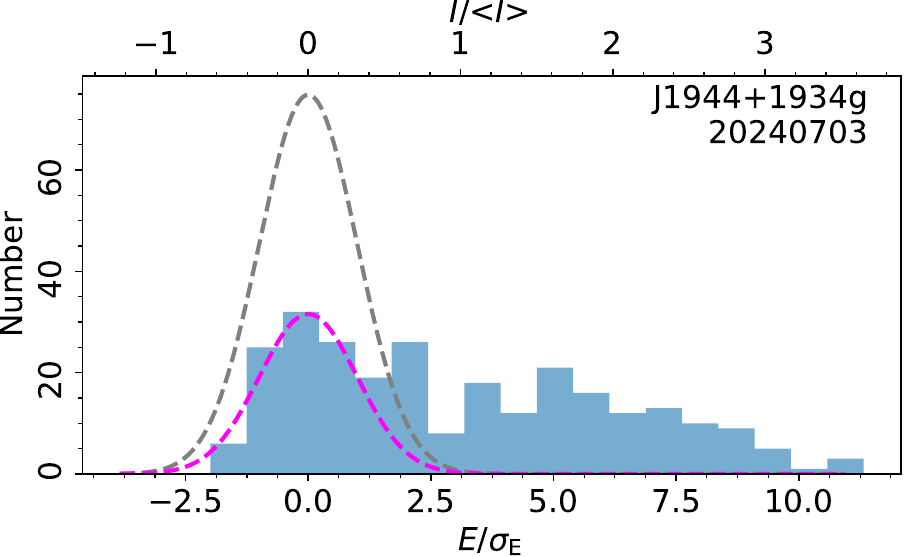}
\figcaption{On-pulse energy histogram of single pulses of PSR J1944+1934g from the FAST observation on 20240703.
\label{subfig:Hist:J1944+1934g}}
\end{figure}

\begin{figure}[htpb]
\centering
\includegraphics[width=0.22\textwidth, angle=0]{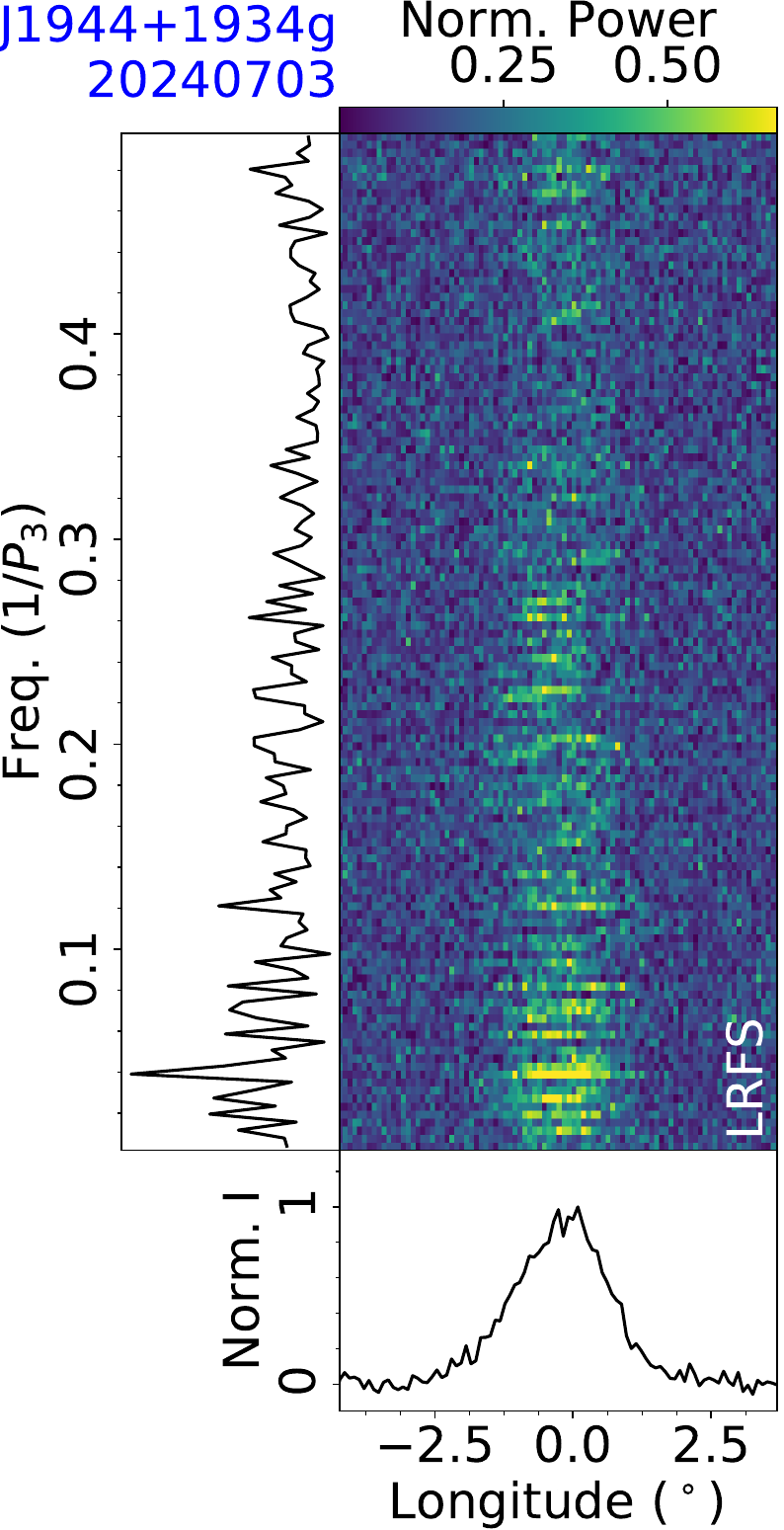}
\includegraphics[width=0.22\textwidth, angle=0]{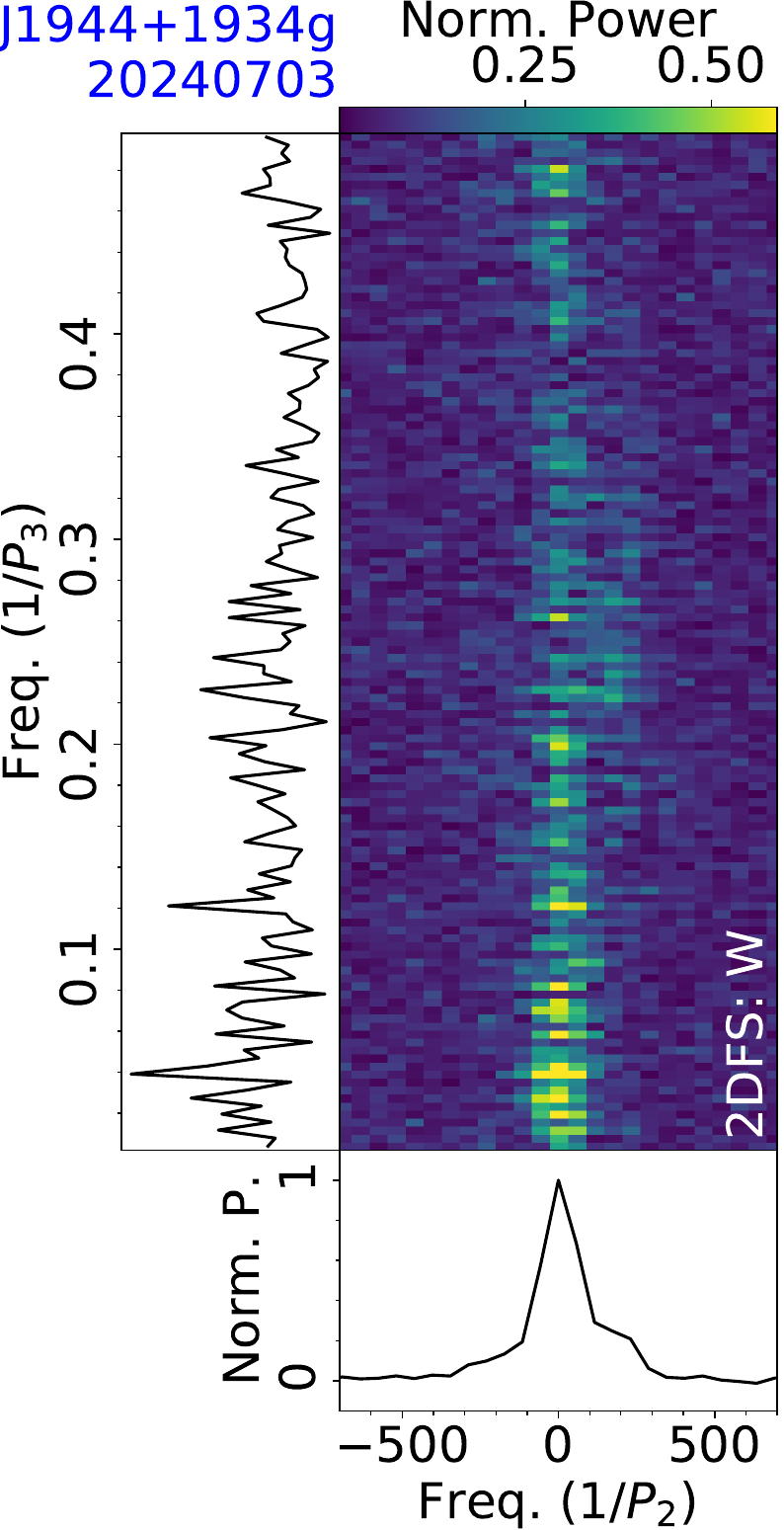}
\figcaption{Fluctuation analysis of PSR J1944+1934g for the observation on 20240703, with LRFS and 2DFS for the on-pulse region of a mean pulse profile.
\label{subfig:fluctu:J1944+1934g}}
\end{figure}

\subsection{J1943+1704g}
\label{subsec:J1943+1704g}

PSR J1943+1704g was discovered in the FAST GPPS survey \citep{Han2021,han2025}. 

This pulsar was observed by FAST on 20250814 for 15 minutes, with a rotation period $P=0.7251$~s and a dispersion measure $D\!M=157.1~{\rm cm^{-3}\,pc}$ determined.
The single pulse sequence and a zoomed-in view of pulses No. 100-300 are shown in Fig.~\ref{subfig:TP:J1943+1704g}, illustrating a wide variation in the integrated energy of single pulses. The on-pulse integral energy histogram of single pulses in Fig.~\ref{subfig:Hist:J1943+1704g} reveals that there is a distribution around a low energy value, corresponding to the weak emission mode. The weak and bright emission modes of single pulses are distinguished from this energy histogram, labeled in red and green, respectively. Fig.~\ref{subfig:ProfModes:J1943+1704g} shows the mean profiles of two modes.

\begin{figure}[htpb]
\centering
\includegraphics[width=0.22\textwidth, angle=0]{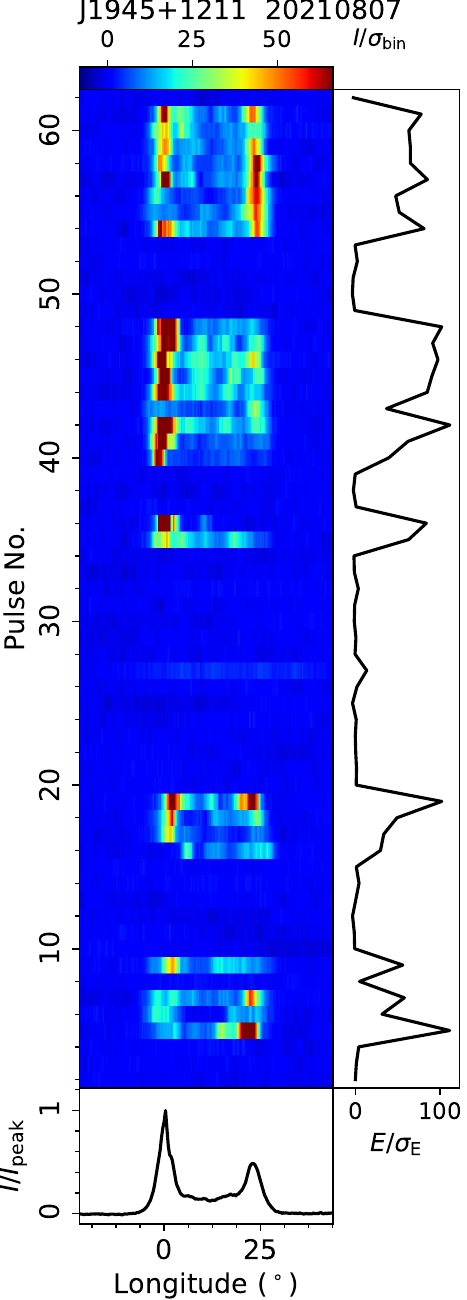}
\figcaption{Single pulse sequence of PSR J1945+1211 from the FAST observation on 20210807.
\label{subfig:TP:J1945+1211}}
\end{figure}

\begin{figure}[htpb]
\centering
\includegraphics[width=0.39\textwidth, angle=0]{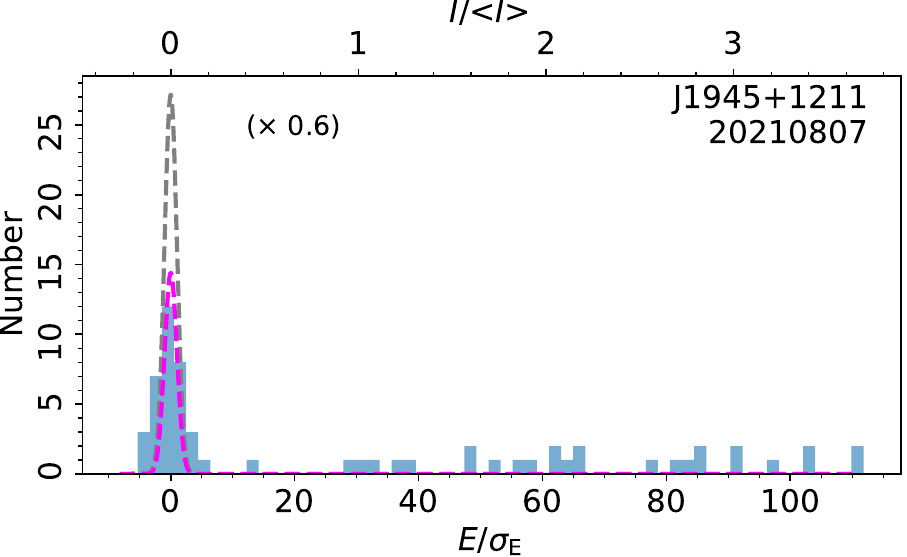}
\figcaption{On-pulse energy histogram of single pulses of PSR J1945+1211 from the FAST observation on 20210807.
\label{subfig:Hist:J1945+1211}}
\end{figure}

\begin{figure}[htpb]
\centering
\includegraphics[width=0.22\textwidth, angle=0]{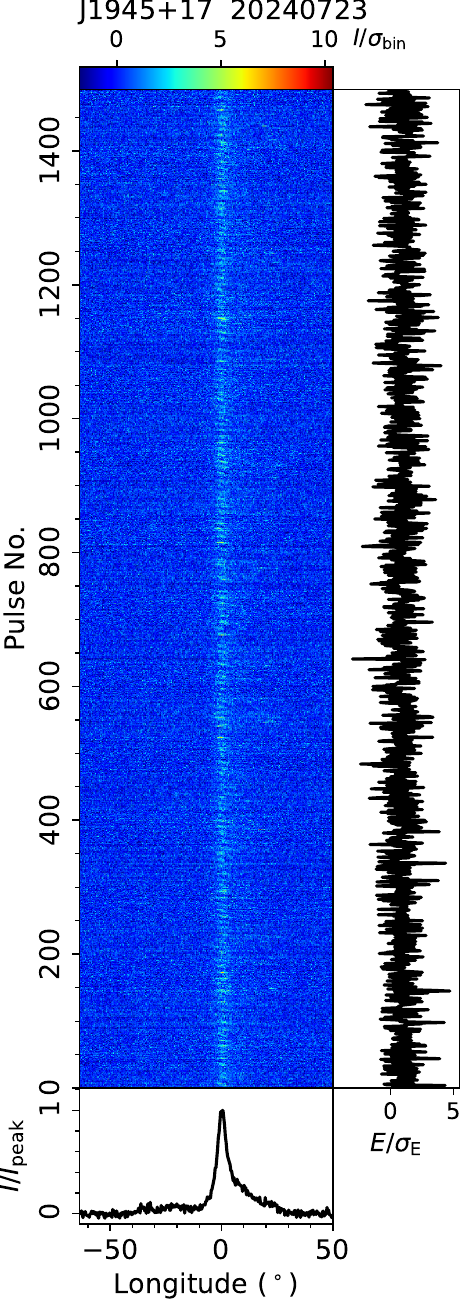}
\includegraphics[width=0.22\textwidth, angle=0]{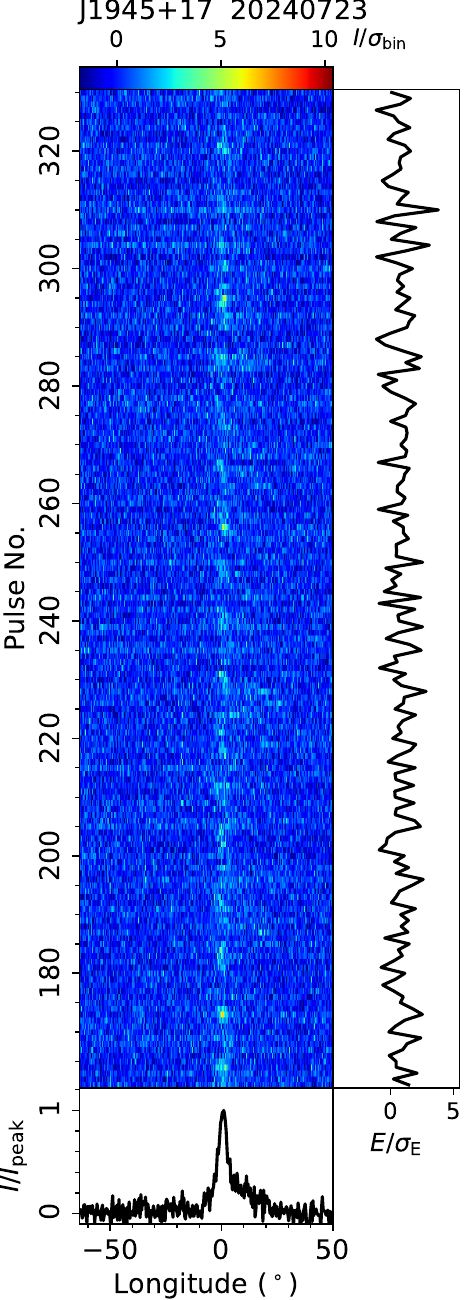}
\figcaption{Single pulse sequence of PSR J1945+17 from the FAST observation on 20240723, and a zoomed-in view of pulses No. 161-330.
\label{subfig:TP:J1945+17}}
\end{figure}

\begin{figure}[htpb]
\centering
\includegraphics[width=0.22\textwidth, angle=0]{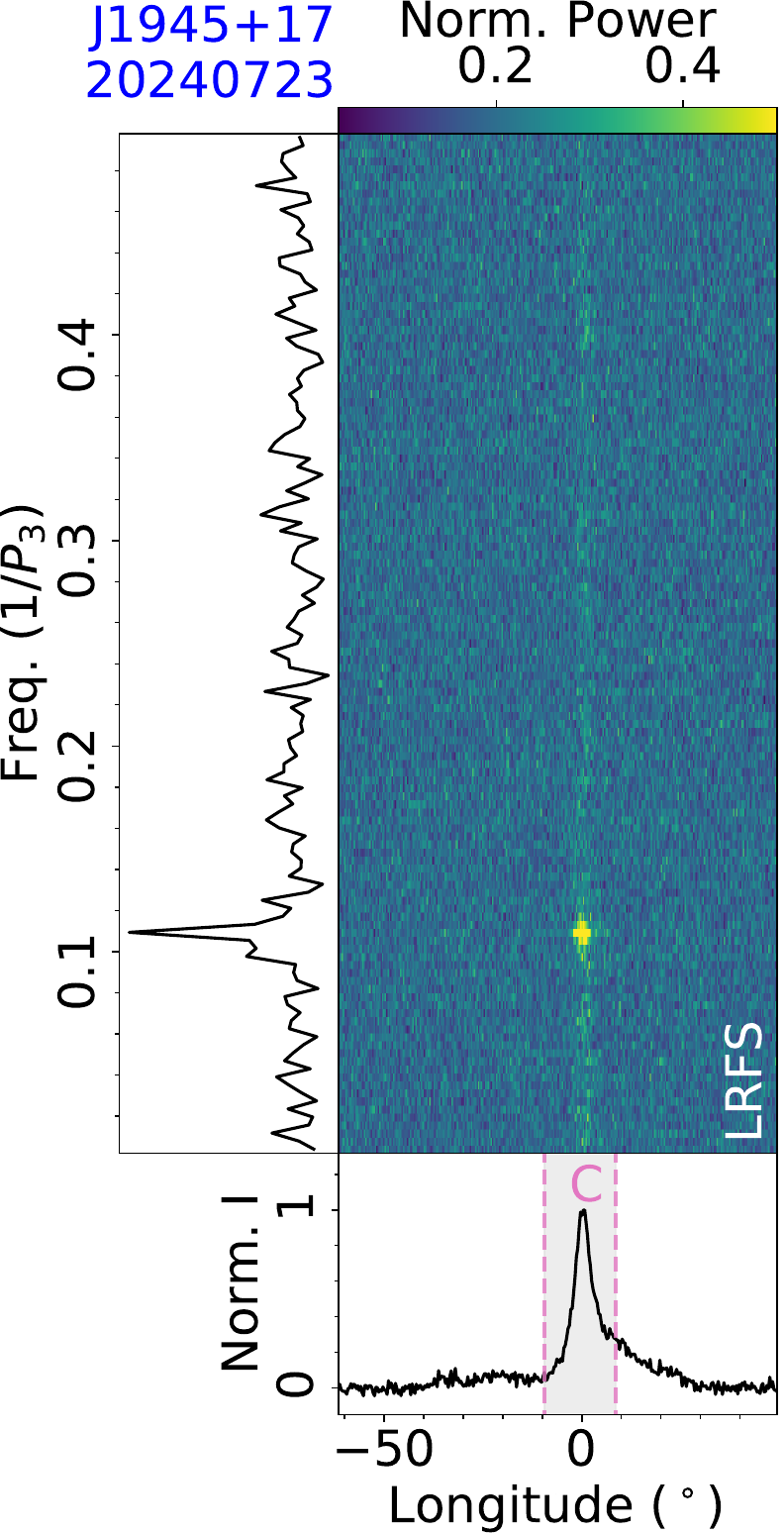}
\includegraphics[width=0.22\textwidth, angle=0]{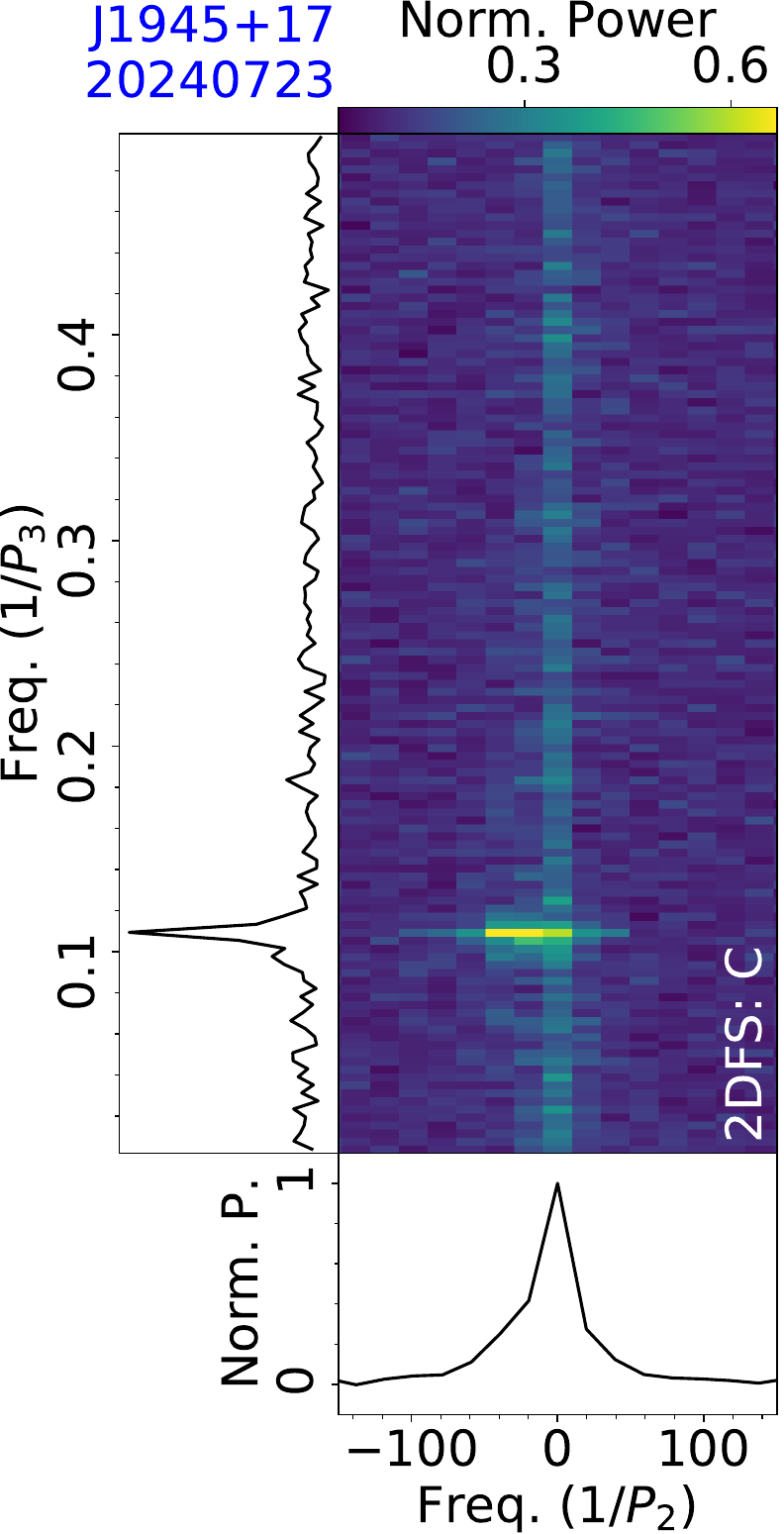}
\figcaption{Fluctuation analysis of PSR J1945+17 for the observation on 20240723, with LRFS and 2DFS for the central part of a mean pulse profile.
\label{subfig:fluctu:J1945+17}}
\end{figure}

\subsection{J1944+1755}
\label{subsec:J1944+1755}

PSR J1944+1755 was discovered by \citet{Hulse1975} using the Arecibo telescope. 
The pulsar has nulls with a fraction of more than 60\% at 430 MHz \citep{Lorimer2002}. \citet{Song2023} reported $P_3$=15$\pm$2 periods and $P_2$=-14$^{+1}_{-3}$ degrees, and $P_3$=15(1) periods and $P_2$=-29$^{+6}_{-5}$ degrees for two components. 

This pulsar was observed by FAST on 20191226 and 20240703, both for 5 minutes. From the observation on 20191226, a rotation period $P=1.9970$~s and a dispersion measure $D\!M=171.8~{\rm cm^{-3}\,pc}$ were derived. 
Single pulse sequences observed on 20191226 are shown in Fig.~\ref{subfig:TP:J1944+1755}, which display nulling and subpulse drifting behaviors. The nulling fractions are estimated from the on-pulse energy histogram (Fig.~\ref{subfig:Hist:J1944+1755}) to be 36$\pm$5\% and 50$\pm$8\% for beams P3M11 and P4M11 observed on 20191226, and 57$\pm$5\% for the observation on 20240703. 2DFS of the leading and trailing components for two data of 20191226 are shown in Fig.~\ref{subfig:fluctu:J1944+1755}, from which the drifting parameters are estimated. 
For the data of P3M11, centroid frequencies of the drift feature in 2DFS are $1/P_3=0.085\pm0.001$ ($P_3=11.8\pm0.1$ periods) and $1/P_2=-21\pm1$ ($P_2=-23\pm7^\circ$) for the leading component, and $1/P_3=0.089\pm0.001$ ($P_3=11.3\pm0.1$ periods) and $1/P_2=-21\pm1$ ($P_2=-17\pm1^\circ$) for the trailing component. 
While the centroid drifting parameters of the P4M11 data are relatively different. For two components, they are $1/P_3=0.061\pm0.001$ ($P_3=16.5\pm0.2$ periods) and $1/P_2=-17\pm1$ ($P_2=-21\pm2^\circ$), and $1/P_3=0.058\pm0.001$ ($P_3=17.4\pm0.2$ periods) and $1/P_2=-11\pm1$ ($P_2=-34\pm4^\circ$).
Differences in drifting between two data could also be perceived from single pulse sequences, and physical causes need more observations.

\begin{figure}[htpb]
\centering
\includegraphics[width=0.22\textwidth, angle=0]{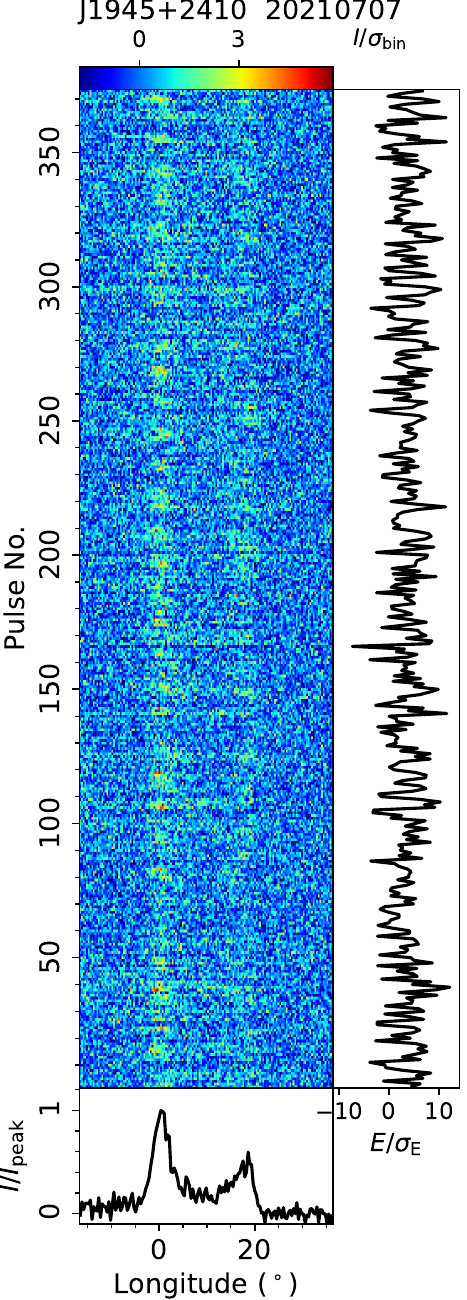}
\figcaption{Single pulse sequence of PSR J1945+2410 from the FAST observation on 20210707.
\label{subfig:TP:J1945+2410}}
\end{figure}

\begin{figure}[htpb]
\centering
\includegraphics[width=0.44\textwidth, angle=0]{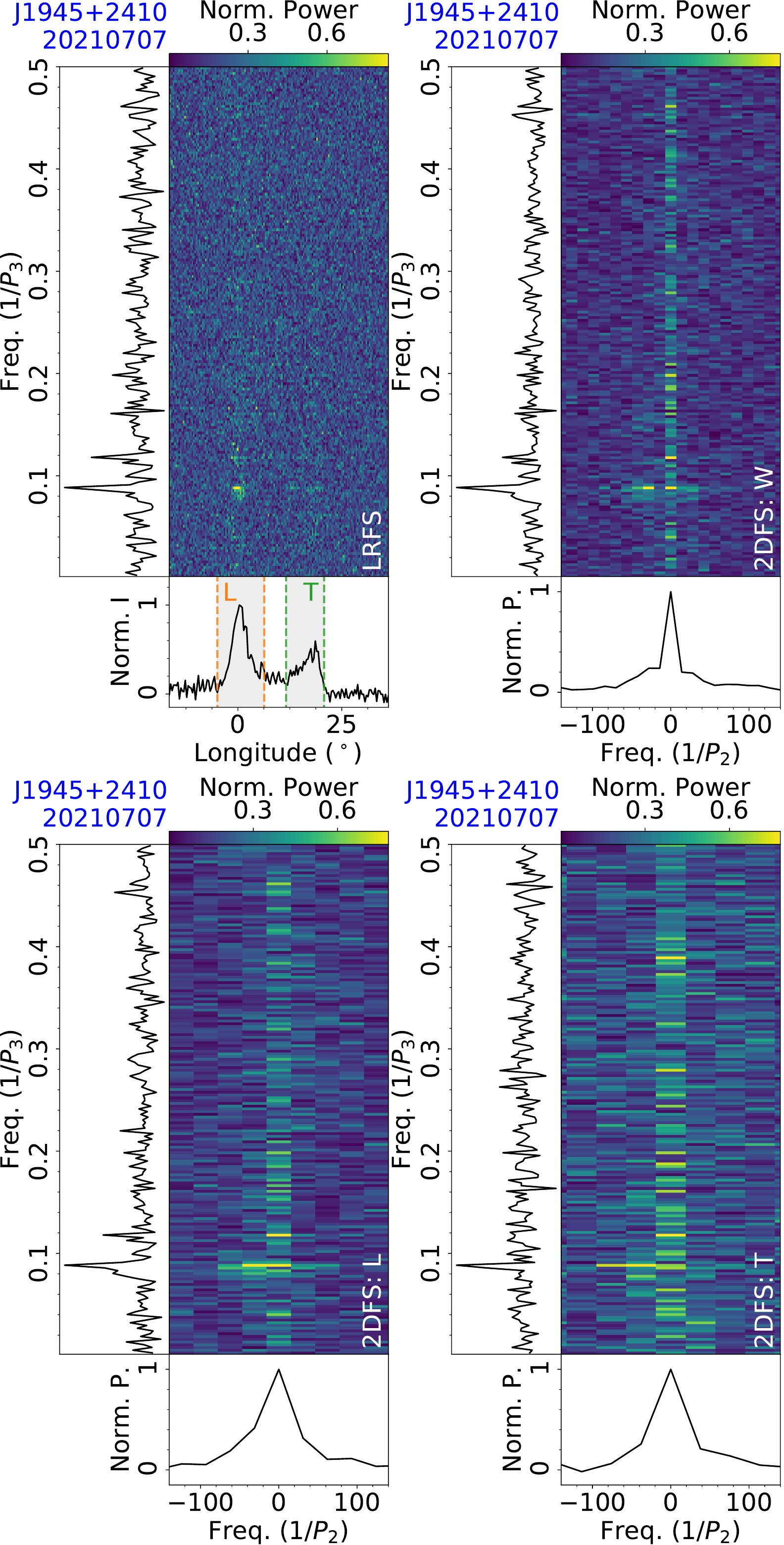}
\figcaption{Fluctuation analysis of PSR J1945+2410 for the observation on 20210707, with LRFS (top-left), and 2DFS for the on-pulse region (top-right), leading part (bottom-left) and trailing part (bottom-right) of a mean pulse profile.
\label{subfig:fluctu:J1945+2410}}
\end{figure}

\subsection{J1944+1934g}
\label{subsec:J1944+1934g}

PSR J1944+1934g was discovered in the FAST GPPS survey \citep{Han2021,han2025}. 

This pulsar was observed by FAST on 20240703 for 15 minutes, yielding a rotation period $P=3.4452$~s and a dispersion measure $D\!M=241.0~{\rm cm^{-3}\,pc}$. 
Single pulse sequences are displayed in Fig.~\ref{subfig:TP:J1944+1934g}, and the pulsar has nulling and subpulse drifting behaviors. From the on-pulse integral energy histogram in Fig.~\ref{subfig:Hist:J1944+1934g}, the nulling fraction of this observation is estimated to be 42$\pm$1\%. The drifting parameters are estimated from LRFS and 2DFS (Fig.~\ref{subfig:fluctu:J1944+1934g}). 
The centroid frequencies of the positive drift feature are $1/P_3=0.258\pm0.002$ and $1/P_2=187\pm4$, corresponding to periodicities of $P_3=3.88\pm0.03$ periods and $P_2=1.93\pm0.04^\circ$.

\begin{figure}[htpb]
\centering
\includegraphics[width=0.22\textwidth, angle=0]{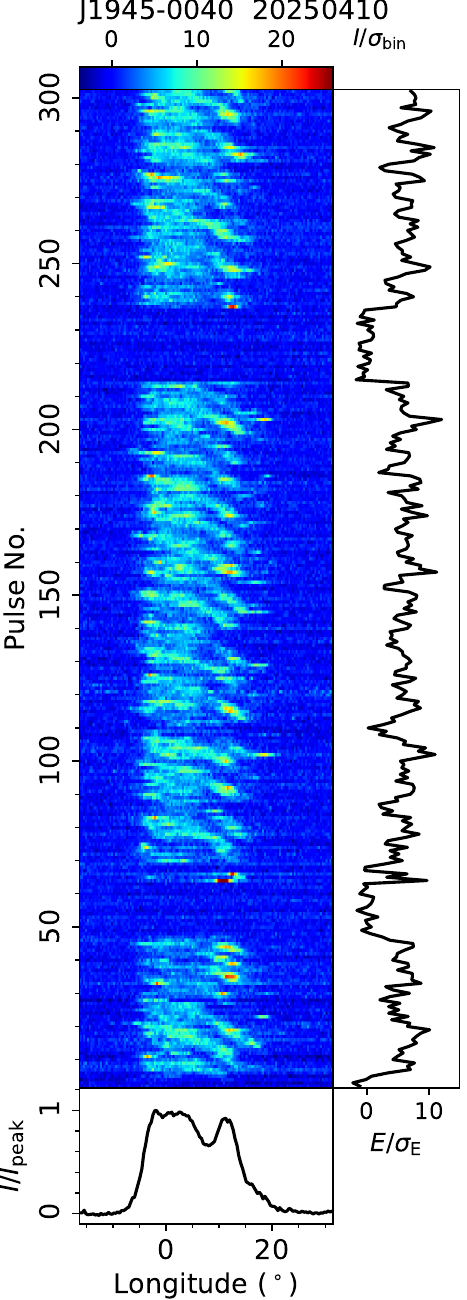}
\includegraphics[width=0.22\textwidth, angle=0]{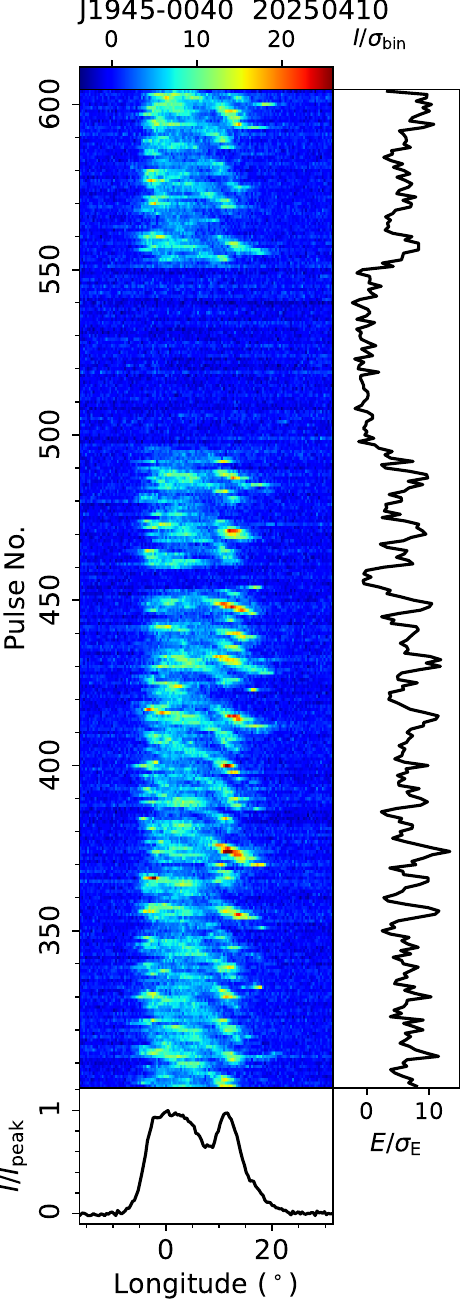}
\figcaption{Single pulse sequences of PSR J1945-0040 from the FAST observation on 20250410.
\label{subfig:TP:J1945-0040}}
\end{figure}

\begin{figure}[htpb]
\centering
\includegraphics[width=0.39\textwidth, angle=0]{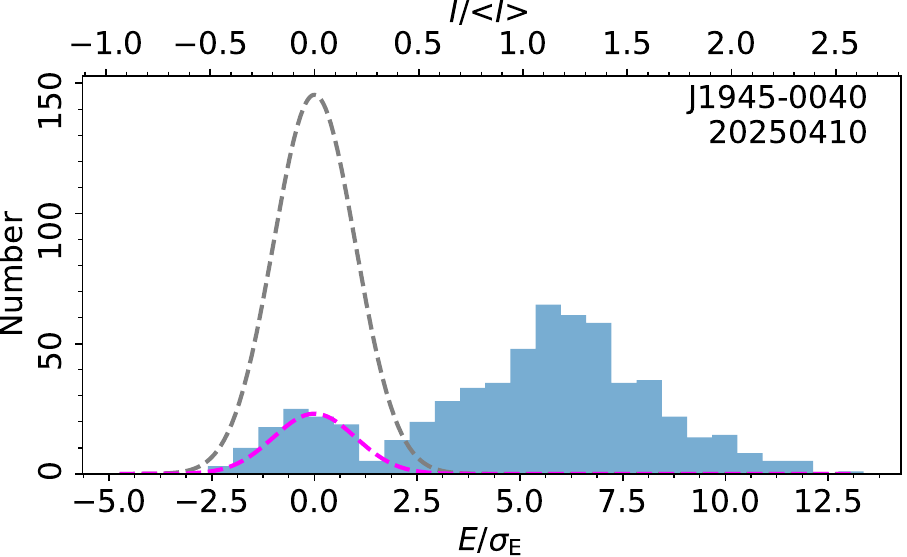}
\figcaption{On-pulse energy histogram of single pulses of PSR J1945-0040 from the FAST observation on 20250410.
\label{subfig:Hist:J1945-0040}}
\end{figure}

\begin{figure}[htpb]
\centering
\includegraphics[width=0.44\textwidth, angle=0]{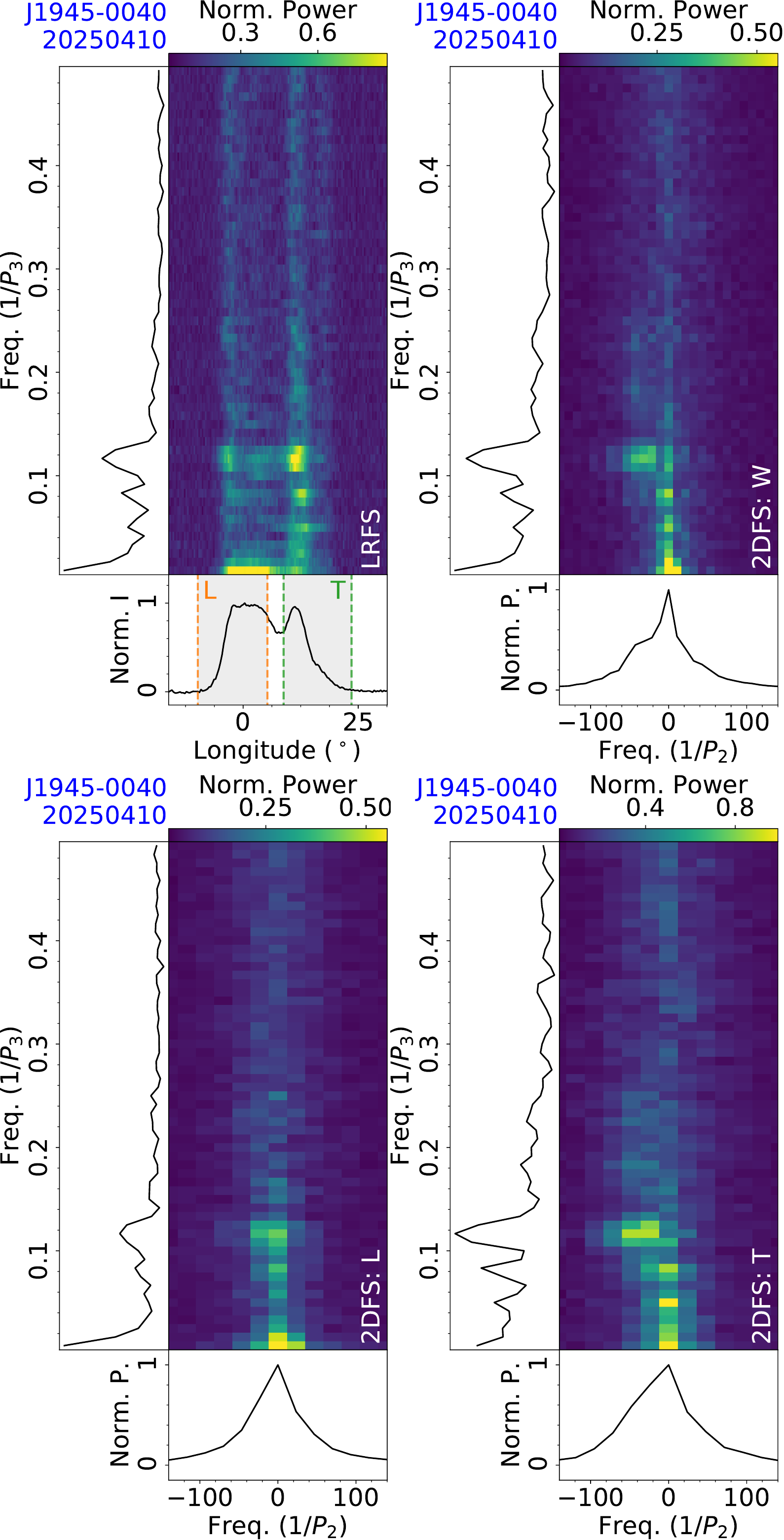}
\figcaption{Fluctuation analysis of PSR J1945-0040 for the observation on 20250410, with LRFS (top-left), and 2DFS for the on-pulse region (top-right), leading part (bottom-left) and trailing part (bottom-right) of a mean pulse profile.
\label{subfig:fluctu:J1945-0040}}
\end{figure}

\subsection{J1945+1211}
\label{subsec:J1945+1211}

PSR J1945+1211 was discovered by FAST \citep{Cameron2020}. 

The pulsar was also observed for 5 minutes on 20210807 by FAST, yielding a rotation period $P=4.7565$~s and a dispersion measure $D\!M=92.7~{\rm cm^{-3}\,pc}$. 
The single pulse sequence of this observation, shown in Fig.~\ref{subfig:TP:J1945+1211}, displays the nulling phenomenon. From the on-pulse energy histogram in Fig.~\ref{subfig:Hist:J1945+1211}, the nulling fraction is estimated to be 32$\pm$7\%.

\begin{figure}[htpb]
\centering
\includegraphics[width=0.22\textwidth, angle=0]{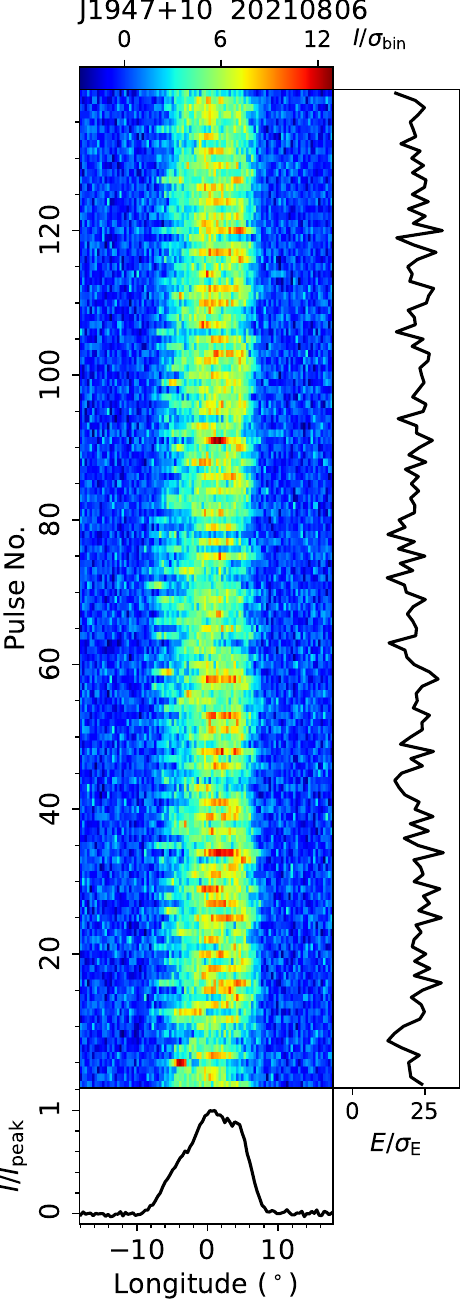}
\includegraphics[width=0.22\textwidth, angle=0]{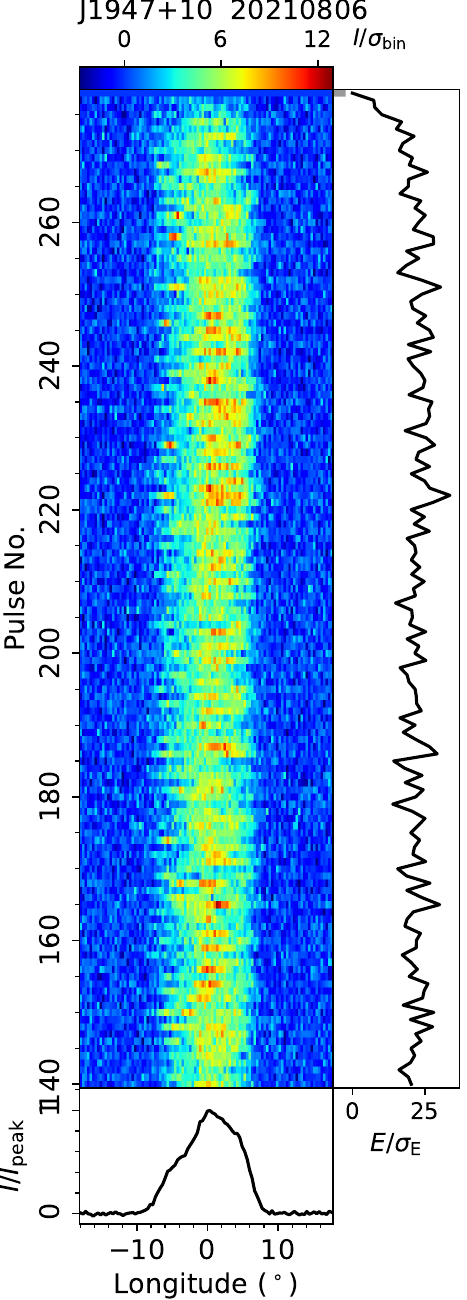}
\figcaption{Single pulse sequences of PSR J1947+10 from the FAST observation on 20210806.
\label{subfig:TP:J1947+10}}
\end{figure}

\begin{figure}[htpb]
\centering
\includegraphics[width=0.44\textwidth, angle=0]{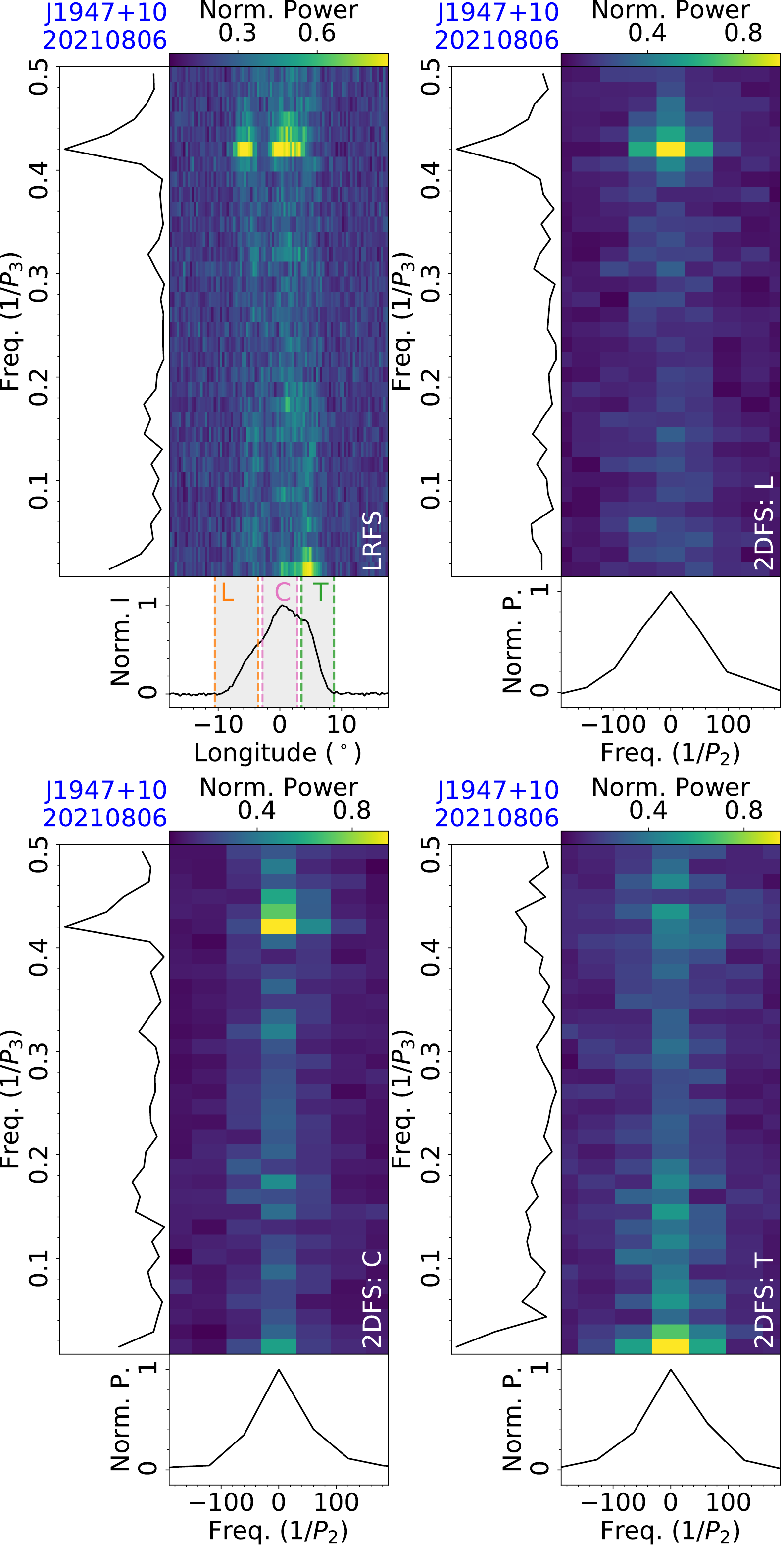}
\figcaption{Fluctuation analysis of PSR J1947+10 for the observation on 20210806, with LRFS (top-left), and 2DFS for the leading part (top-right), central part (bottom-left) and trailing part (bottom-right) of a mean pulse profile.
\label{subfig:fluctu:J1947+10}}
\end{figure}

\subsection{J1945+17}
\label{subsec:J1945+17}

PSR J1945+17 was discovered by \citet{Parent2022} in the Pulsar Arecibo L-band Feed Array (PALFA) survey.

This pulsar was observed by FAST on 20240723 for 15 minutes, deriving a rotation period $P=0.6041$~s and a dispersion measure $D\!M=173.0~{\rm cm^{-3}\,pc}$. 
The brightest component in the central part of the profile shows systematic drifting behavior. The fluctuation spectra are displayed in Fig.~\ref{subfig:fluctu:J1945+17}, where the centroid frequencies of the drift feature are estimated to be $1/P_3=0.1089\pm0.0004$ and $1/P_2=-22\pm2$, corresponding to drifting periodicities of $P_3=9.18\pm0.03$ periods and $P_2=-16\pm1^\circ$.

\begin{figure}[htpb]
\centering
\includegraphics[width=0.22\textwidth, angle=0]{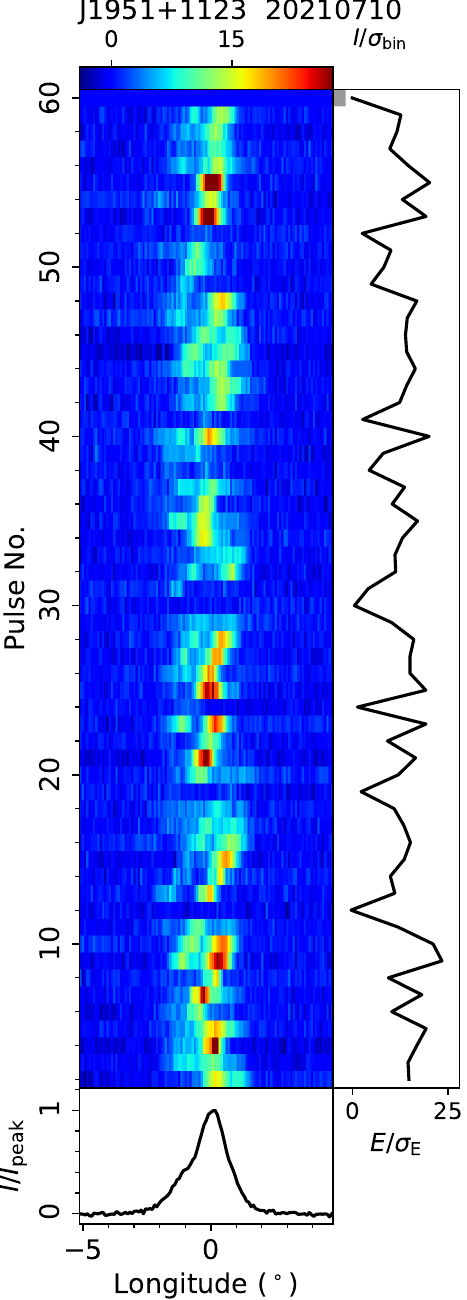}
\figcaption{Single pulse sequence of PSR J1951+1123 from the FAST observation on 20210710.
\label{subfig:TP:J1951+1123}}
\end{figure}

\begin{figure}[htpb]
\centering
\includegraphics[width=0.39\textwidth, angle=0]{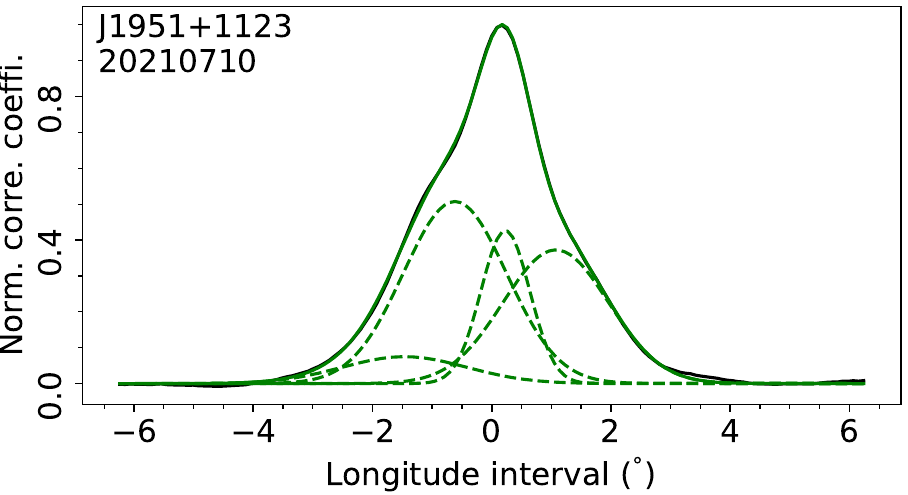}
\figcaption{Cross correlation of PSR J1951+1123 from the FAST observation on 20210710.
\label{subfig:Corre:J1951+1123}}
\end{figure}

\begin{figure}[htpb]
\centering
\includegraphics[width=0.22\textwidth, angle=0]{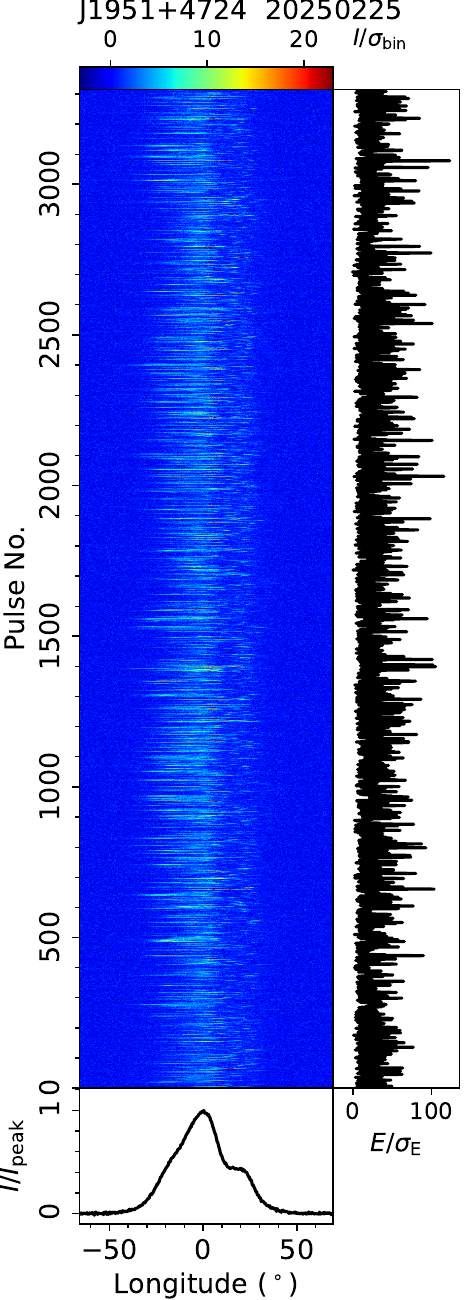}
\includegraphics[width=0.22\textwidth, angle=0]{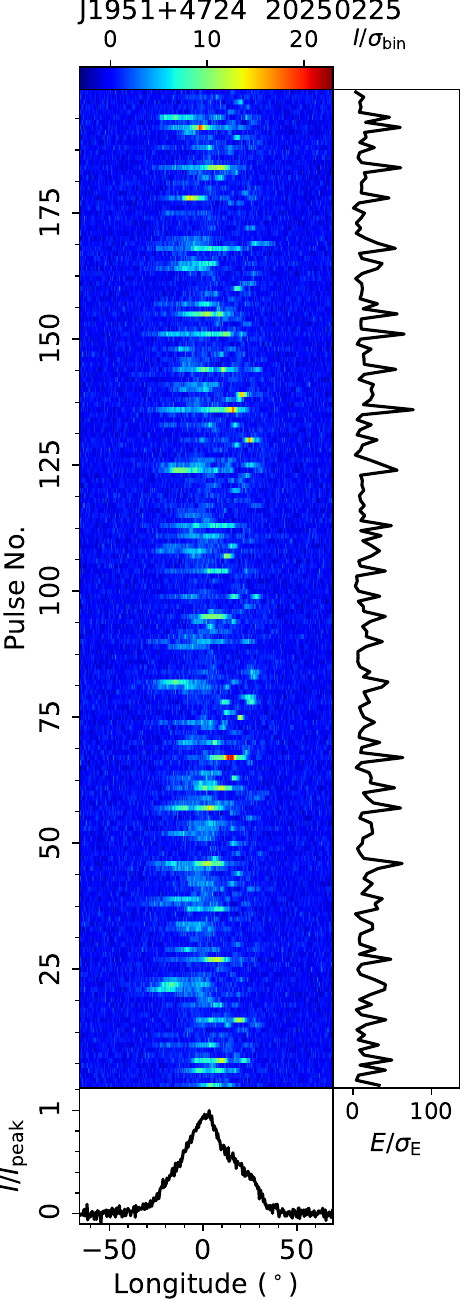}
\figcaption{Single pulse sequence of PSR J1951+4724 from the FAST observation on 20250225, and a zoomed-in view of pulses No. 1-200.
\label{subfig:TP:J1951+4724}}
\end{figure}

\begin{figure}[htpb]
\centering
\includegraphics[width=0.44\textwidth, angle=0]{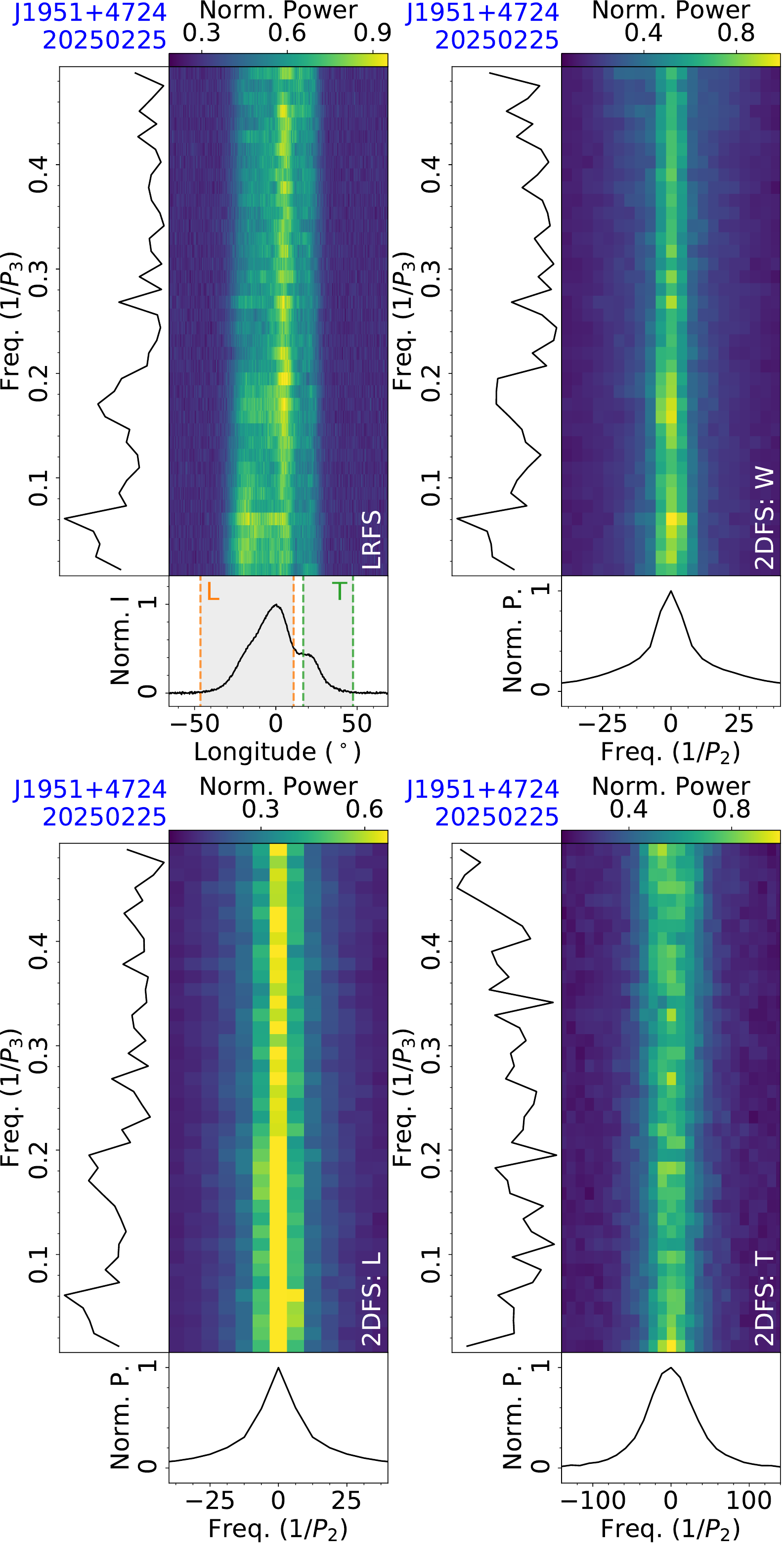}
\figcaption{Fluctuation analysis of PSR J1951+4724 for the observation on 20250225, with LRFS (top-left), and 2DFS for the on-pulse region (top-right), leading part (bottom-left) and trailing part (bottom-right) of the mean pulse profile.
\label{subfig:fluctu:J1951+4724}}
\end{figure}

\subsection{J1945+2410g}
\label{subsec:J1945+2410g}

PSR J1945+2410g was discovered in the FAST GPPS survey \citep{Han2021,han2025}. 

This pulsar was observed by FAST on 20210707 for 15 minutes, deriving a rotation period $P=2.3774$~s and a dispersion measure $D\!M=480.9~{\rm cm^{-3}\,pc}$. 
Single pulse sequences in Fig.~\ref{subfig:TP:J1945+2410} and fluctuation spectra in Fig.~\ref{subfig:fluctu:J1945+2410} illustrate that subpulses corresponding to leading and trailing components all have negative drifting with similar temporal modulation periodicities. 
In 2DFS of the leading component, the centroid frequencies of the negative drift feature are $1/P_3=0.085\pm0.001$ and $1/P_2=-22\pm3$, which correspond to $P_3=11.7\pm0.1$ periods and $P_2=-16\pm2^\circ$. 
2DFS of the trailing component exhibits a drift feature with the centroid characterized by $1/P_3=0.089\pm0.001$ and $1/P_2=-36\pm6$, yielding $P_3=11.3\pm0.1$ periods and $P_2=-10\pm2^\circ$.

\subsection{J1945-0040}
\label{subsec:J1945-0040}

PSR J1945-0040 was discovered in the second Molonglo pulsar survey \citep{Manchester1978}. This pulsar was reported to exhibit nulling behavior \citep{Weisberg1986,Redman2009}. The subpulse drifting parameters for two components are also shown by \citet{Song2023}, which are $P_3=8.5\pm0.4$ periods and $P_2=-34^{+6}_{-11}$ degrees, and $P_3=8.5\pm0.3$ periods and $P_2=-13^{+3}_{-2}$ degrees, respectively.

This pulsar was observed by FAST on 20250410 for 11 minutes, with a rotation period $P=1.0455$~s and a dispersion measure $D\!M=59.0~{\rm cm^{-3}\,pc}$ determined. Single pulse sequences in Fig.~\ref{subfig:TP:J1945-0040} display nulling and subpulse negatively drifting phenomena, and the trailing part in the mean pulse profile has a lower absolute drifting rate compared to the leading part. The nulling fraction of this observation is estimated from the on-pulse energy histogram in Fig.~\ref{subfig:Hist:J1945-0040}, which is 15.9$\pm$1.6\%. Fluctuation spectra are shown in Fig.~\ref{subfig:fluctu:J1945-0040}. For the leading profile part, the centroid frequencies of the negative drift feature are $1/P_3=0.114\pm0.001$ and $1/P_2=-10\pm2$, corresponding to periodicities of $P_3=8.8\pm0.1$ periods and $P_2=-34\pm8$ degrees. The drift feature in 2DFS of the trailing profile part is characterized by the centroid frequencies of $1/P_3=0.116\pm0.001$ and $1/P_2=-36\pm2$, yielding $P_3=8.6\pm0.1$ periods and $P_2=-10\pm1$ degrees.

\subsection{J1947+10}
\label{subsec:J1947+10}

PSR J1947+10 was discovered by the Arecibo telescope at 430 MHz \citep{Camilo1996}. 

This pulsar was observed by FAST on 20210806 for 5 minutes, deriving a rotation period $P=1.1109$~s and a dispersion measure $D\!M=128.7~{\rm cm^{-3}\,pc}$. 
Single pulse sequences of the observation are shown in Fig.~\ref{subfig:TP:J1947+10}. 
The LRFS and 2DFS for the leading, central, and trailing parts of the mean pulse profile are shown in Fig.~\ref{subfig:fluctu:J1947+10}, with three modulation features corresponding to these profile ranges.
2DFS of the leading profile part has a temporal modulation feature with the centroid of $1/P_3=0.427\pm0.002$, corresponding to $P_3=2.34\pm0.01$ periods. 
In 2DFS of the central profile part, the preferred positive drift feature has centroid frequencies of $1/P_3=0.431\pm0.002$ and $1/P_2=18\pm8$, yielding periodicities of $P_3=2.32\pm0.01$ periods and $P_2=20\pm10$ periods. 
There is a low-frequency modulation feature in 2DFS of the trailing profile part, while longer observations are required to confirm this feature.

\begin{figure}[htpb]
\centering
\includegraphics[width=0.44\textwidth, angle=0]{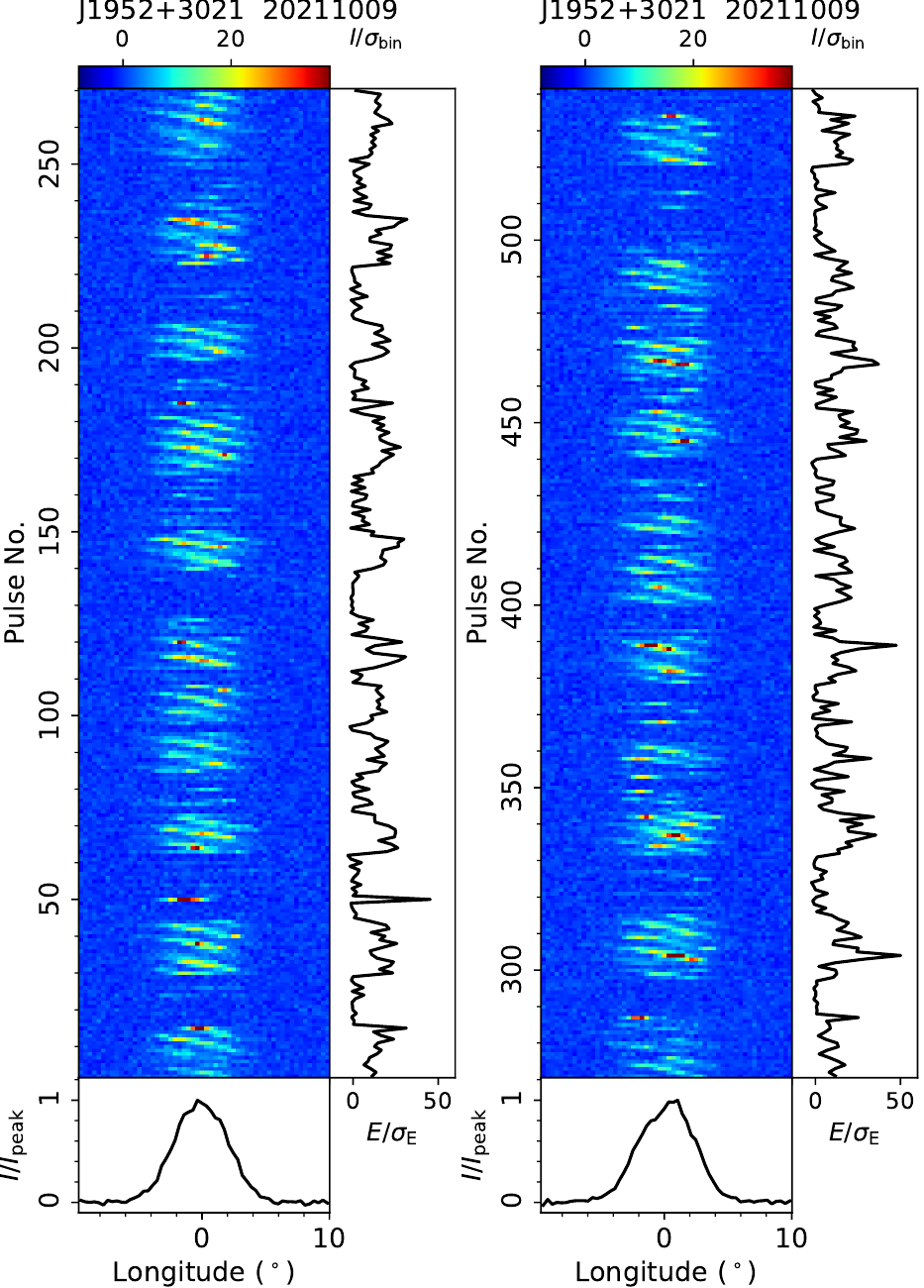}
\figcaption{Single pulse sequences of PSR J1952+3021 from the FAST observation on 20211009.
\label{subfig:TP:J1952+3021}}
\end{figure}

\begin{figure}[htpb]
\centering
\includegraphics[width=0.39\textwidth, angle=0]{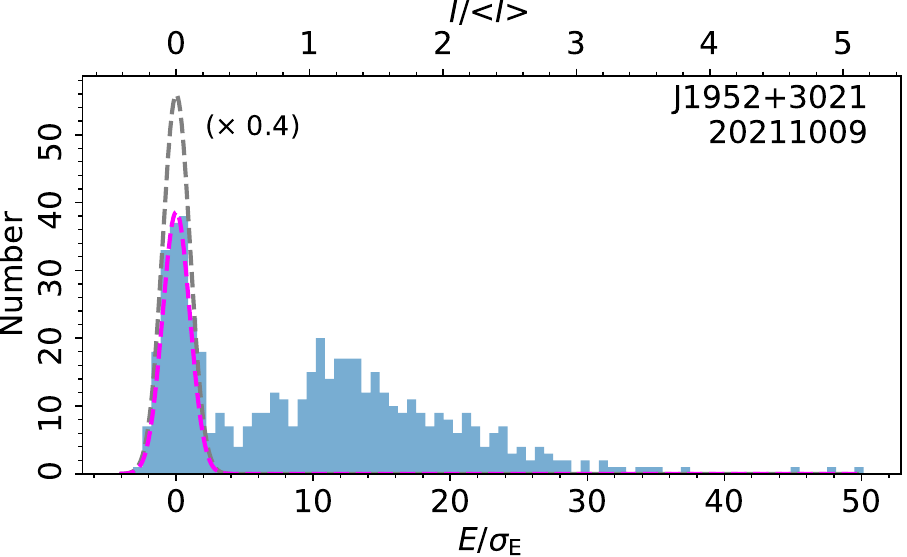}
\figcaption{On-pulse energy histogram of single pulses of PSR J1952+3021 from the FAST observation on 20211009.
\label{subfig:Hist:J1952+3021}}
\end{figure}

\begin{figure}[htpb]
\centering
\includegraphics[width=0.44\textwidth, angle=0]{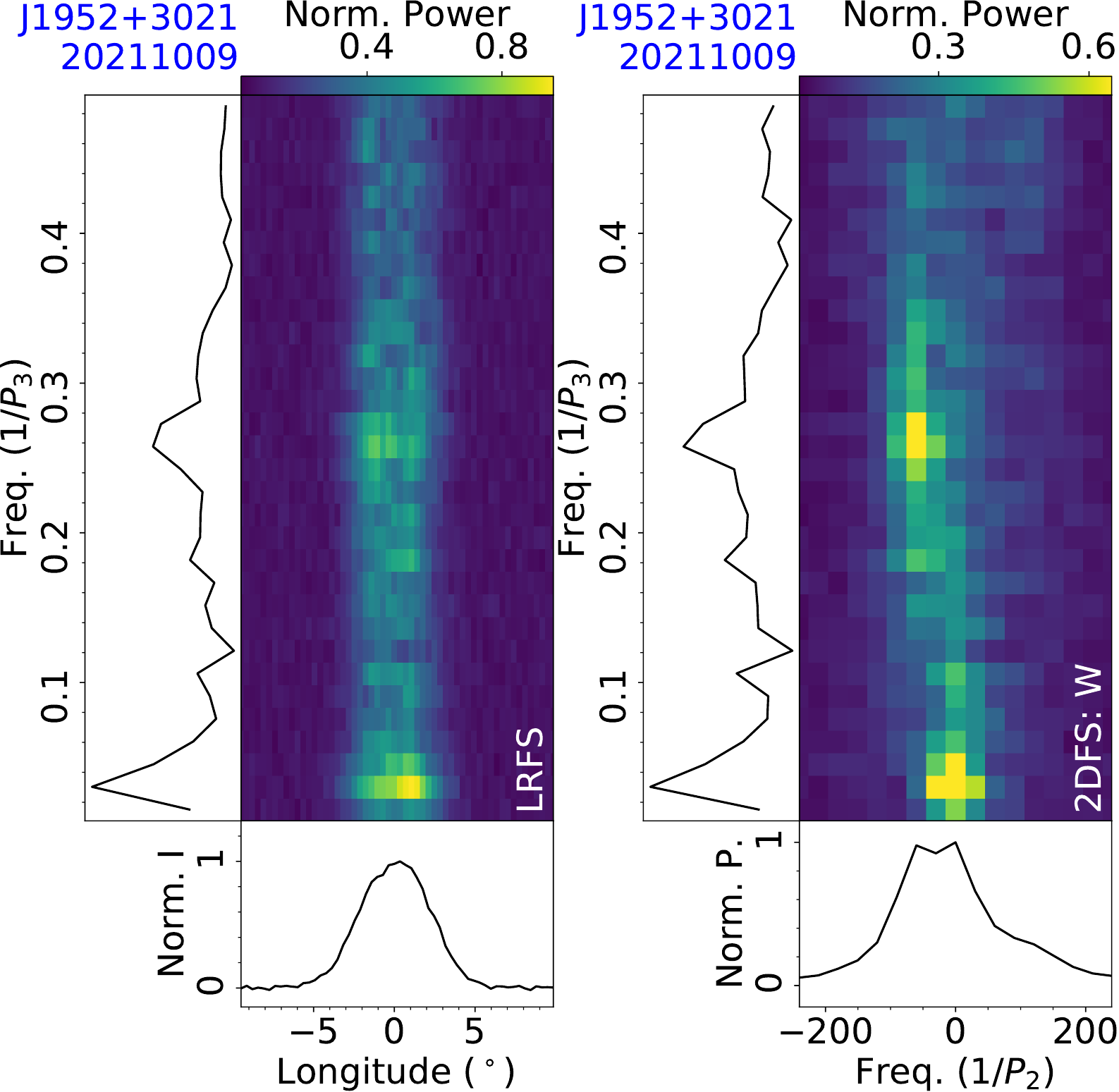}
\figcaption{Fluctuation analysis of PSR J1952+3021 for the observation on 20211009, with LRFS and 2DFS for the on-pulse region of a mean pulse profile.
\label{subfig:fluctu:J1952+3021}}
\end{figure}

\begin{figure}[htpb]
\centering
\includegraphics[width=0.22\textwidth, angle=0]{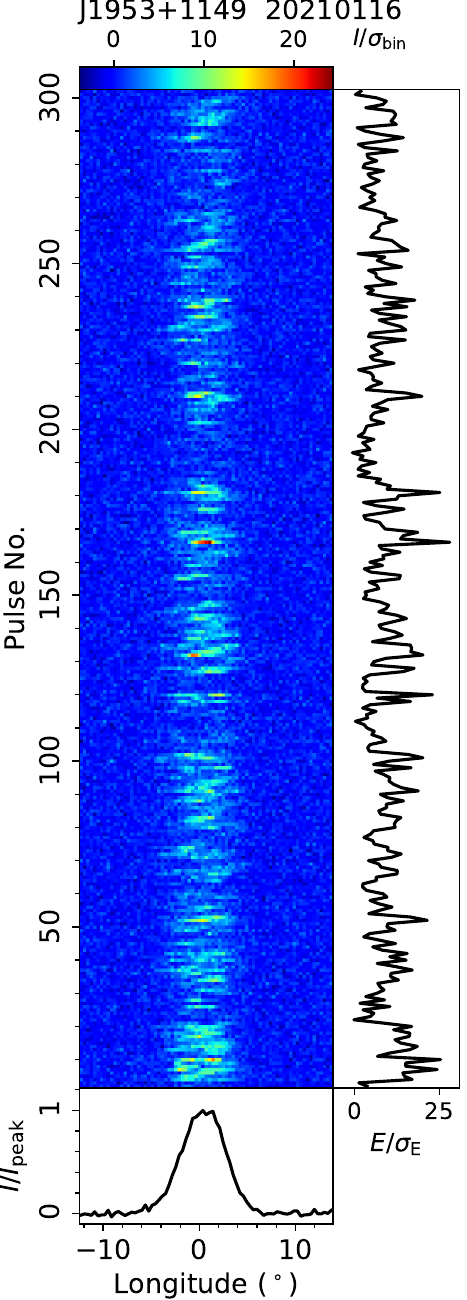}
\includegraphics[width=0.22\textwidth, angle=0]{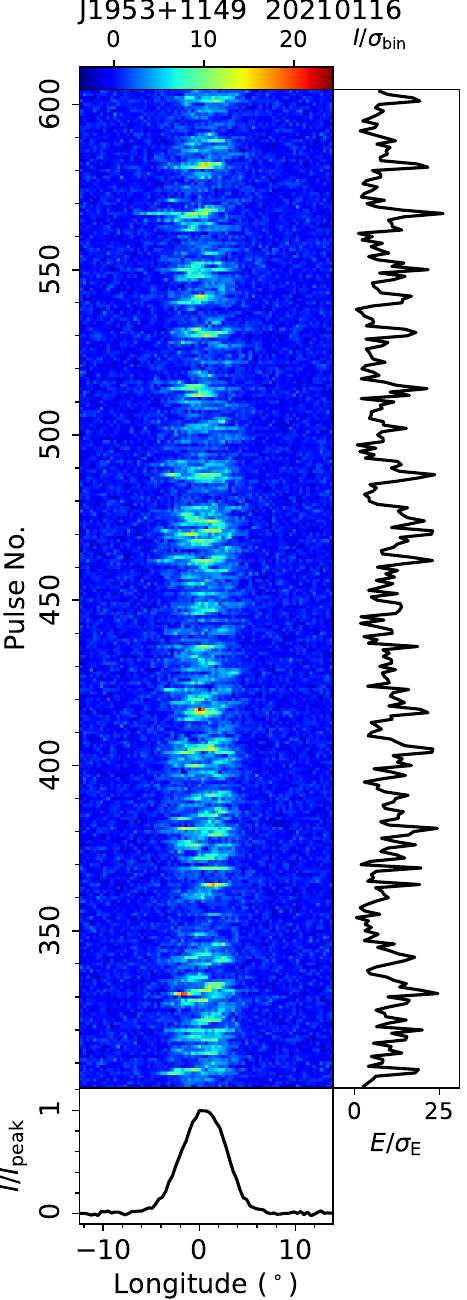}
\figcaption{Single pulse sequences of PSR J1953+1149 from the FAST observation on 20210116.
\label{subfig:TP:J1953+1149}}
\end{figure}

\begin{figure}[htpb]
\centering
\includegraphics[width=0.22\textwidth, angle=0]{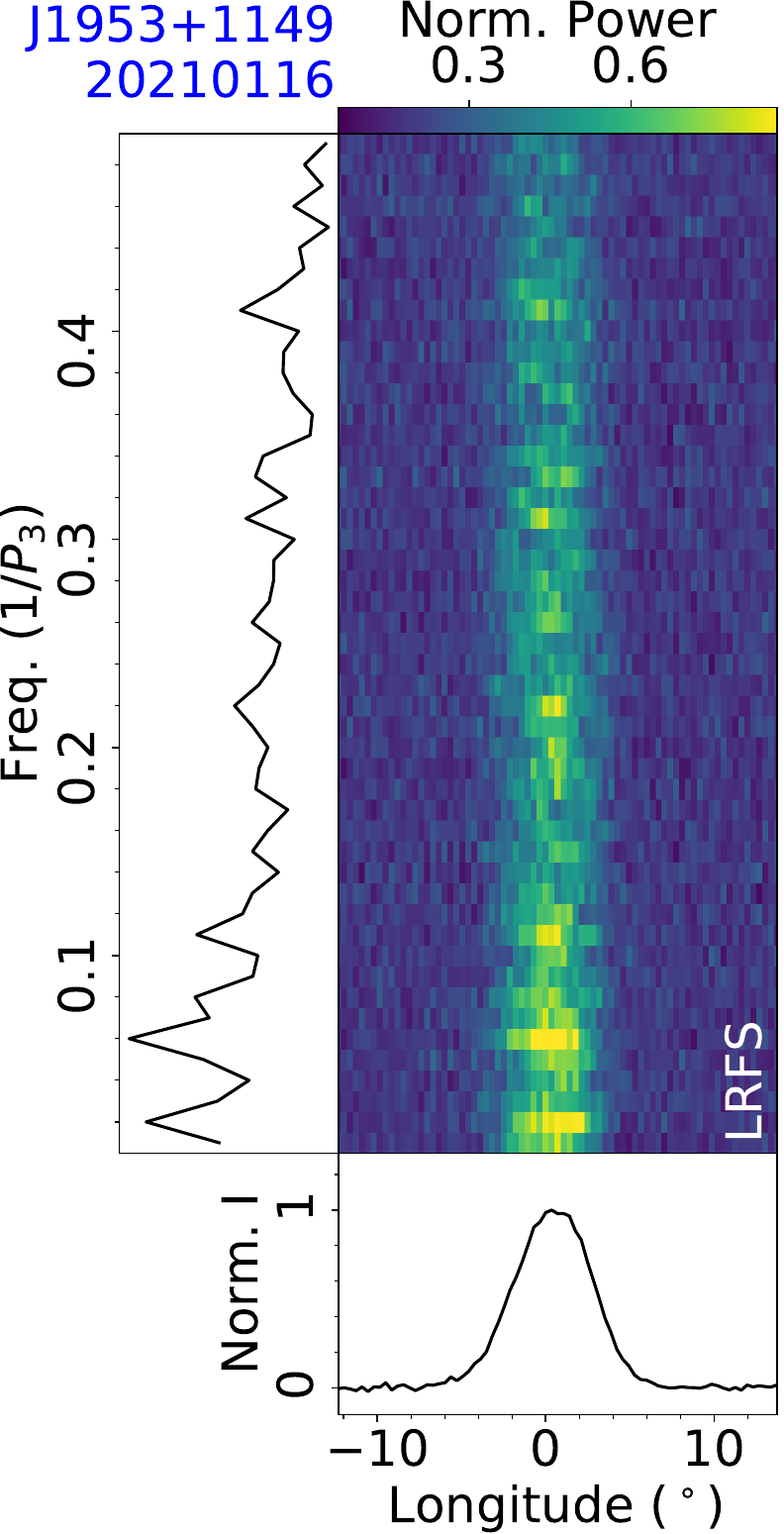}
\includegraphics[width=0.22\textwidth, angle=0]{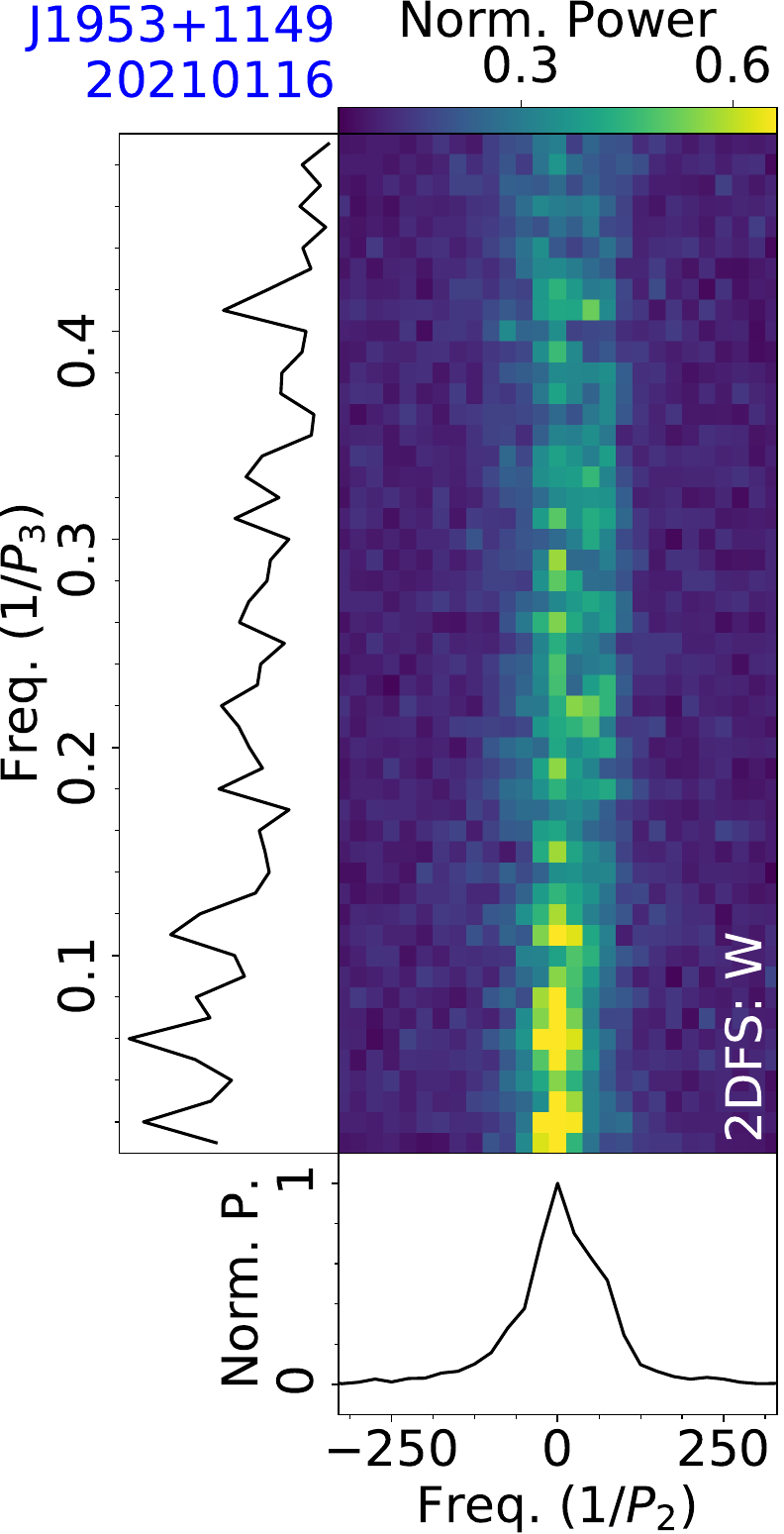}
\figcaption{Fluctuation analysis of PSR J1953+1149 for the observation on 20210116, with LRFS and 2DFS for the on-pulse region of a mean pulse profile.
\label{subfig:fluctu:J1953+1149}}
\end{figure}

\subsection{J1951+1123}
\label{subsec:J1951+1123}

PSR J1951+1123 was discovered using the 305 m radio telescope at Arecibo \citep{Nice1995}. Two drift features were reported by \citet{Song2023}, with temporal periodicities of 4.8$\pm$0.2 and 17$\pm$3 periods. 

The pulsar was observed by FAST on 20210710 for 5 minutes, deriving a rotation period $P=5.0936$~s and a dispersion measure $D\!M=27.3~{\rm cm^{-3}\,pc}$. 
The single pulse sequence of the observation in Fig.~\ref{subfig:TP:J1951+1123} displays that the drifting behavior is not systematic. From the cross-correlation method, the averaged drifting parameters are $D=0.2\pm0.1$ degrees per period and $P_2=0.85\pm0.03^\circ$.

\begin{figure}[htpb]
\centering
\includegraphics[width=0.22\textwidth, angle=0]{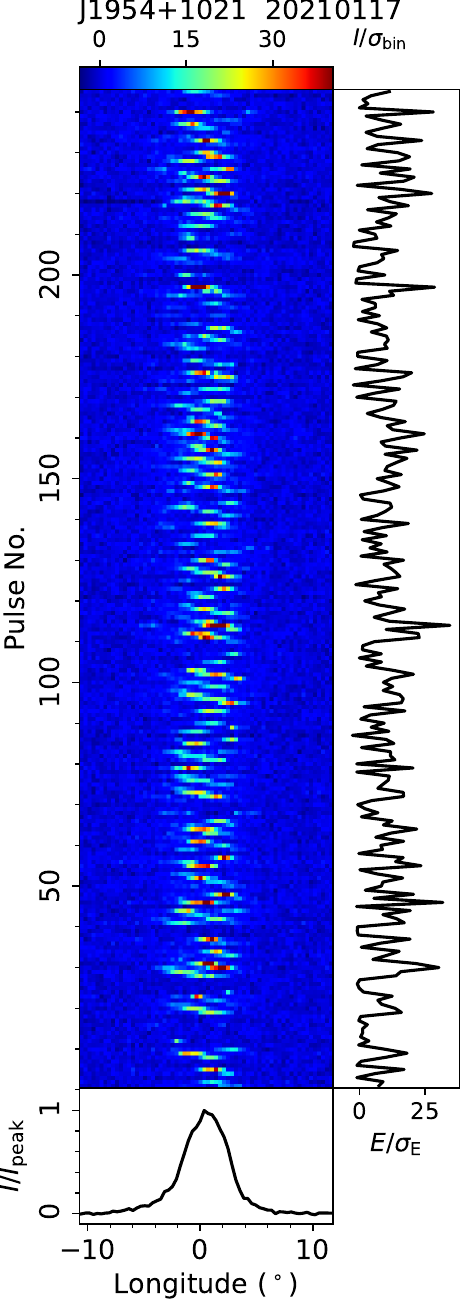}
\figcaption{Single pulse sequence of PSR J1954+1021 from the FAST observation on 20210117.
\label{subfig:TP:J1954+1021}}
\end{figure}

\begin{figure}[htpb]
\centering
\includegraphics[width=0.39\textwidth, angle=0]{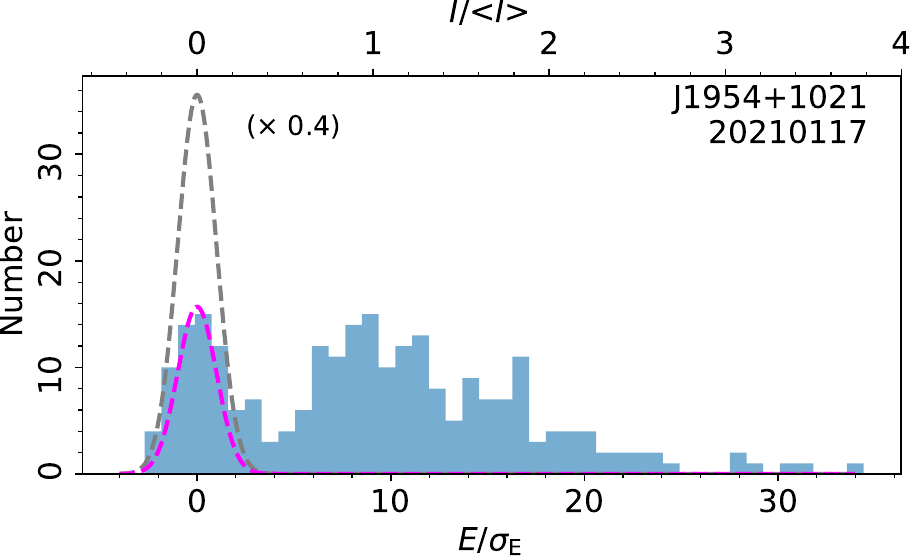}
\figcaption{On-pulse energy histogram of single pulses of PSR J1954+1021 from the FAST observation on 20210117.
\label{subfig:Hist:J1954+1021}}
\end{figure}

\begin{figure}[htpb]
\centering
\includegraphics[width=0.22\textwidth, angle=0]{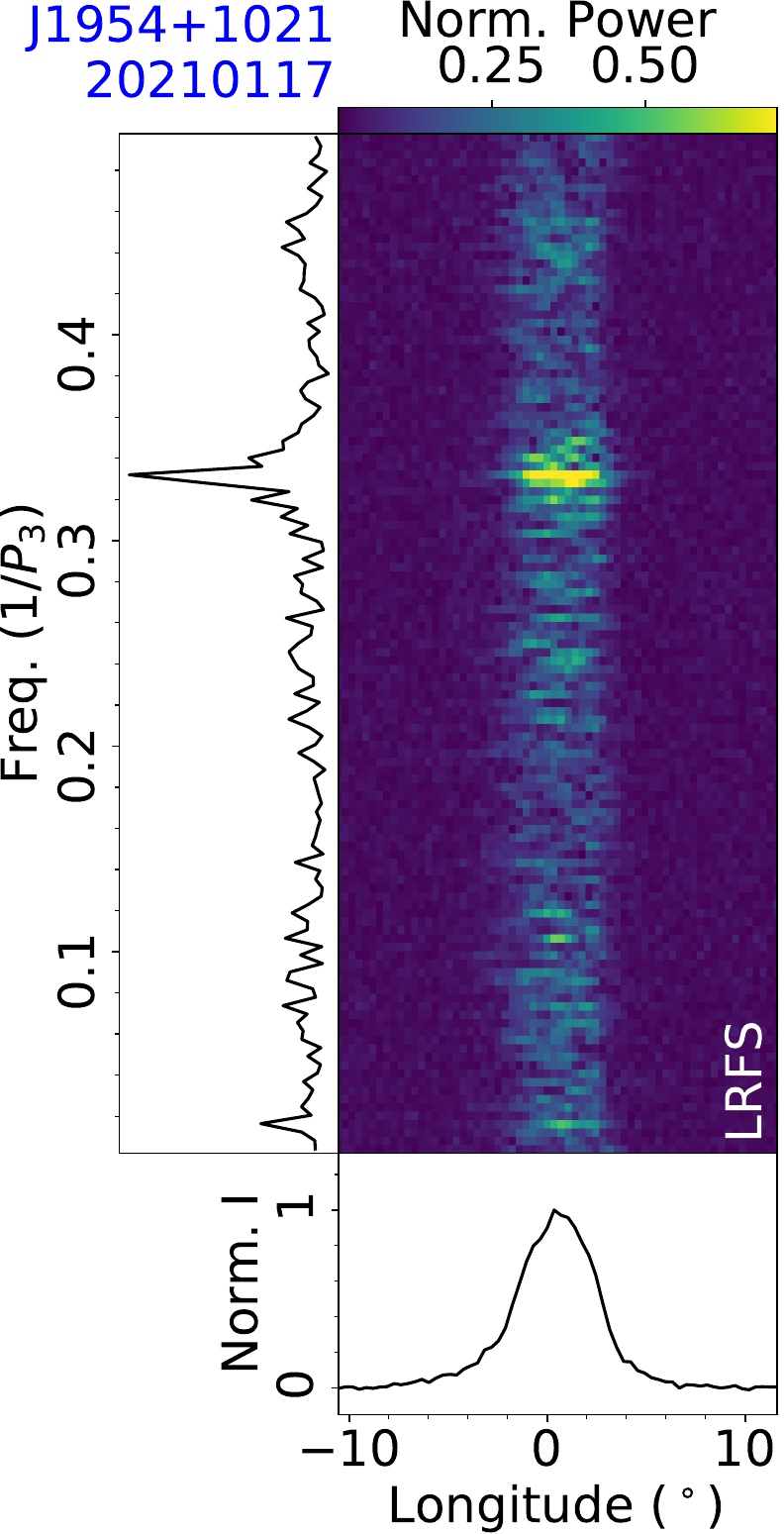}
\includegraphics[width=0.22\textwidth, angle=0]{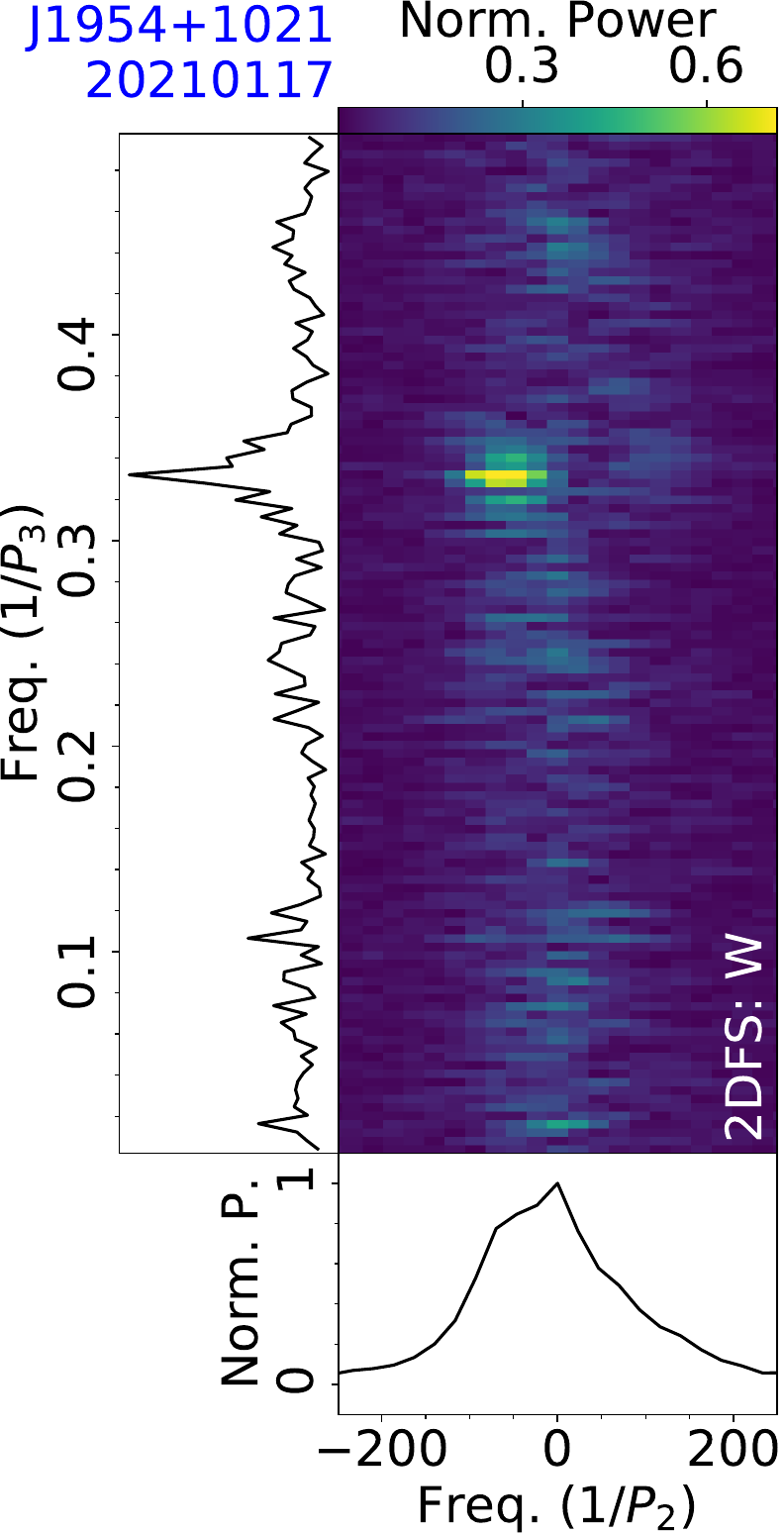}
\figcaption{Fluctuation analysis of PSR J1954+1021 for the observation on 20210117, with LRFS and 2DFS for the on-pulse region of a mean pulse profile.
\label{subfig:fluctu:J1954+1021}}
\end{figure}

\subsection{J1951+4724}
\label{subsec:J1951+4724}

PSR J1951+4724 was discovered by FAST in  the Commensal Radio Astronomy FAST Survey (CRAFTS) \citep{Cruces2021}.

This pulsar was observed by FAST on 20250225 for 10 minutes, deriving a rotation period $P=0.1819$~s and a dispersion measure $D\!M=107.2~{\rm cm^{-3}\,pc}$. 
The single pulse sequence and a zoomed-in view of pulses No. 1-200 in Fig.~\ref{subfig:TP:J1951+4724} indicate the existence of subpulse modulation behavior. Fluctuation spectra are shown in Fig.~\ref{subfig:fluctu:J1951+4724}. For the leading part of the mean pulse profile, two modulation features are present with centroid frequencies of $1/P_3=0.056\pm0.002$ and $0.168\pm0.002$, corresponding to $P_3=18\pm1$ and $5.9\pm0.1$ periods, respectively. For the trailing profile part, there is a preferred negative drift feature with a centroid at $1/P_3=0.433\pm0.005$ and $1/P_2=-3\pm1$, yielding $P_3=2.31\pm0.03$ periods and $P_2=-138\pm73$ degrees.

\begin{figure}[htpb]
\centering
\includegraphics[width=0.22\textwidth, angle=0]{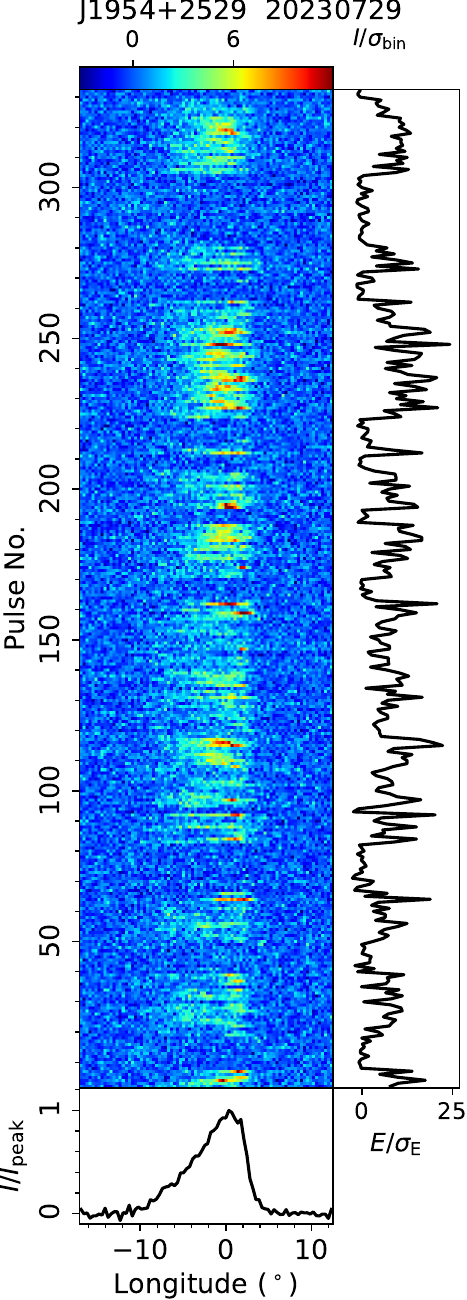}
\figcaption{Single pulse sequence of PSR J1954+2529 from the FAST observation on 20230729.
\label{subfig:TP:J1954+2529}}
\end{figure}

\begin{figure}[htpb]
\centering
\includegraphics[width=0.39\textwidth, angle=0]{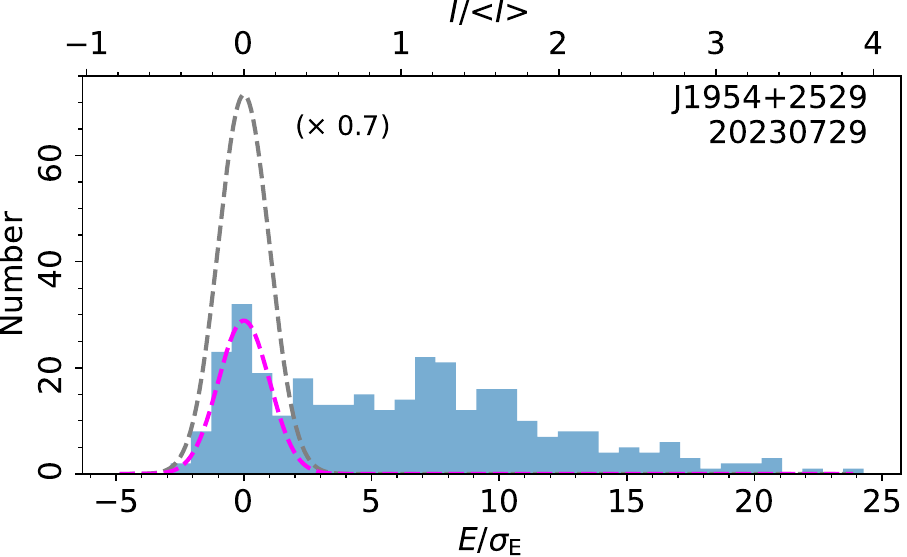}
\figcaption{On-pulse energy histogram of single pulses of PSR J1954+2529 from the FAST observation on 20230729.
\label{subfig:Hist:J1954+2529}}
\end{figure}

\subsection{J1952+3021}
\label{subsec:J1952+3021}

PSR J1952+3021 was discovered in the PALFA survey using the Arecibo telescope \citep{Patel2018}. Nulling behavior was displayed by \citep{Parent2022} at 1.4 GHz from the Arecibo observation. 

This pulsar was observed by FAST on 20211009 for 15 minutes, with a rotation period $P=1.6658$~s and a dispersion measure $D\!M=190.0~{\rm cm^{-3}\,pc}$ determined. 
Single pulse sequences in Fig.~\ref{subfig:TP:J1952+3021} show nulling and subpulse drifting phenomena. The nulling fraction of this observation is estimated to be 28$\pm$2\% from the single-pulse on-pulse energy histogram (Fig.~\ref{subfig:Hist:J1952+3021}). Fluctuation spectra are displayed in Fig.~\ref{subfig:fluctu:J1952+3021}, and the centroid of the negative drift feature is at $1/P_3=0.260\pm0.002$ and $1/P_2=-59\pm2$, corresponding to $P_3=3.84\pm0.03$ periods and $P_2=-6.1\pm0.2$ degrees. The low-frequency modulation periodicity of 31$\pm$2 periods in fluctuation spectra arises from nulls.

\begin{figure}[htpb]
\centering
\includegraphics[width=0.22\textwidth, angle=0]{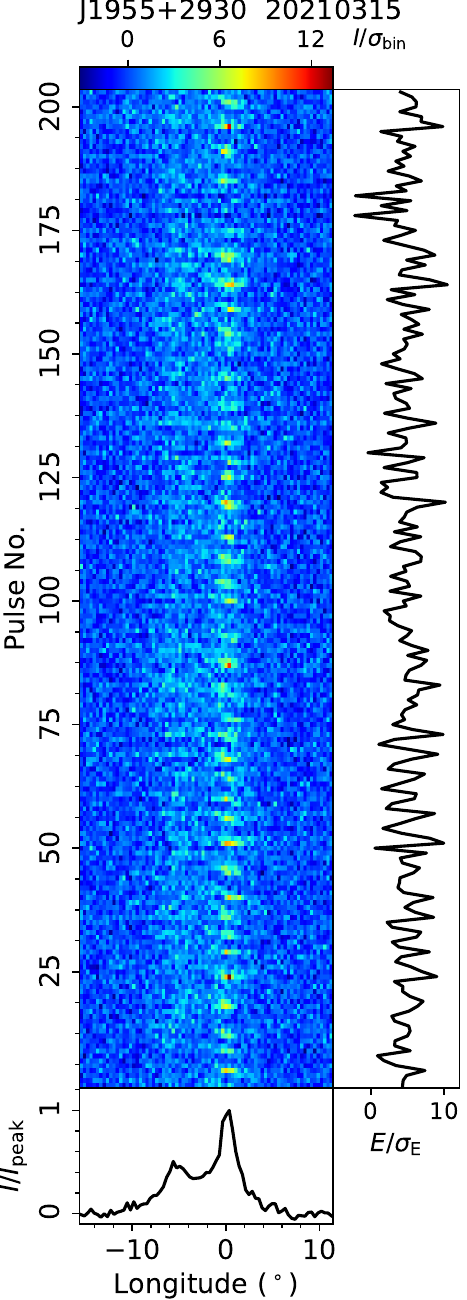}
\includegraphics[width=0.22\textwidth, angle=0]{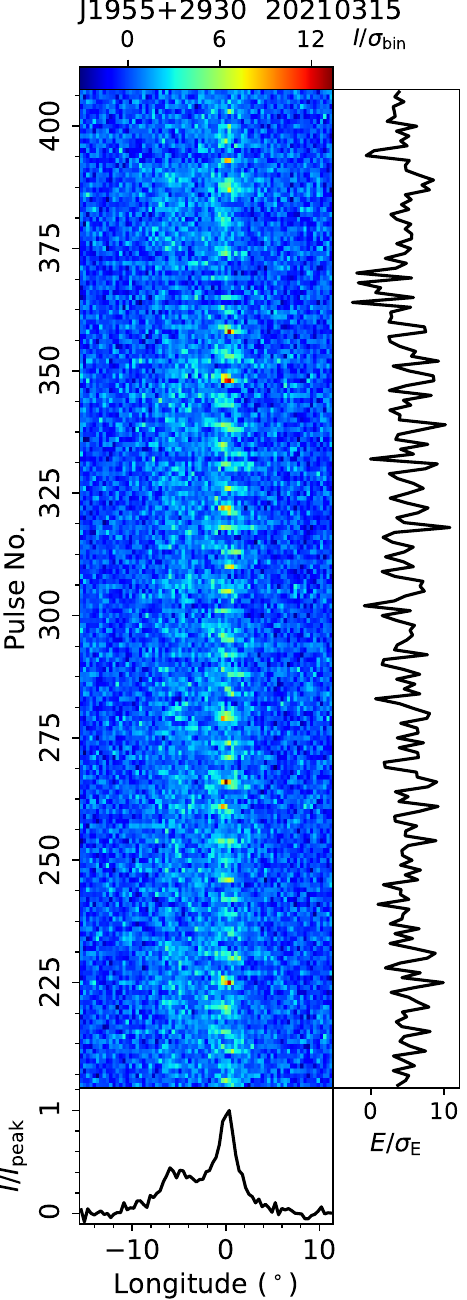}
\figcaption{Single-pulse sequences of PSR J1955+2930 from the FAST observation on 20210315.
\label{subfig:TP:J1955+2930}
}
\end{figure}

\begin{figure}[htpb]
\centering
\includegraphics[width=0.22\textwidth, angle=0]{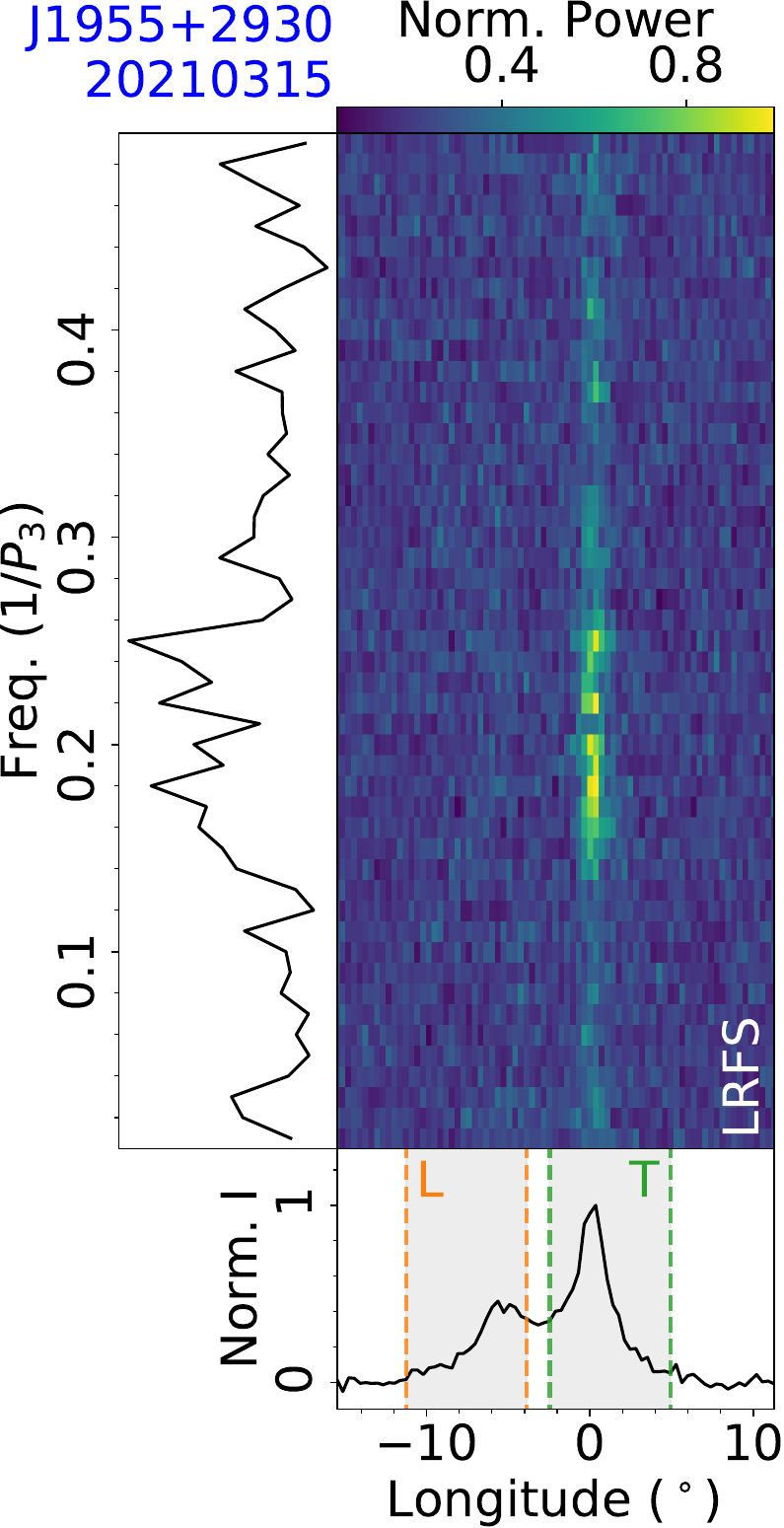}
\includegraphics[width=0.22\textwidth, angle=0]{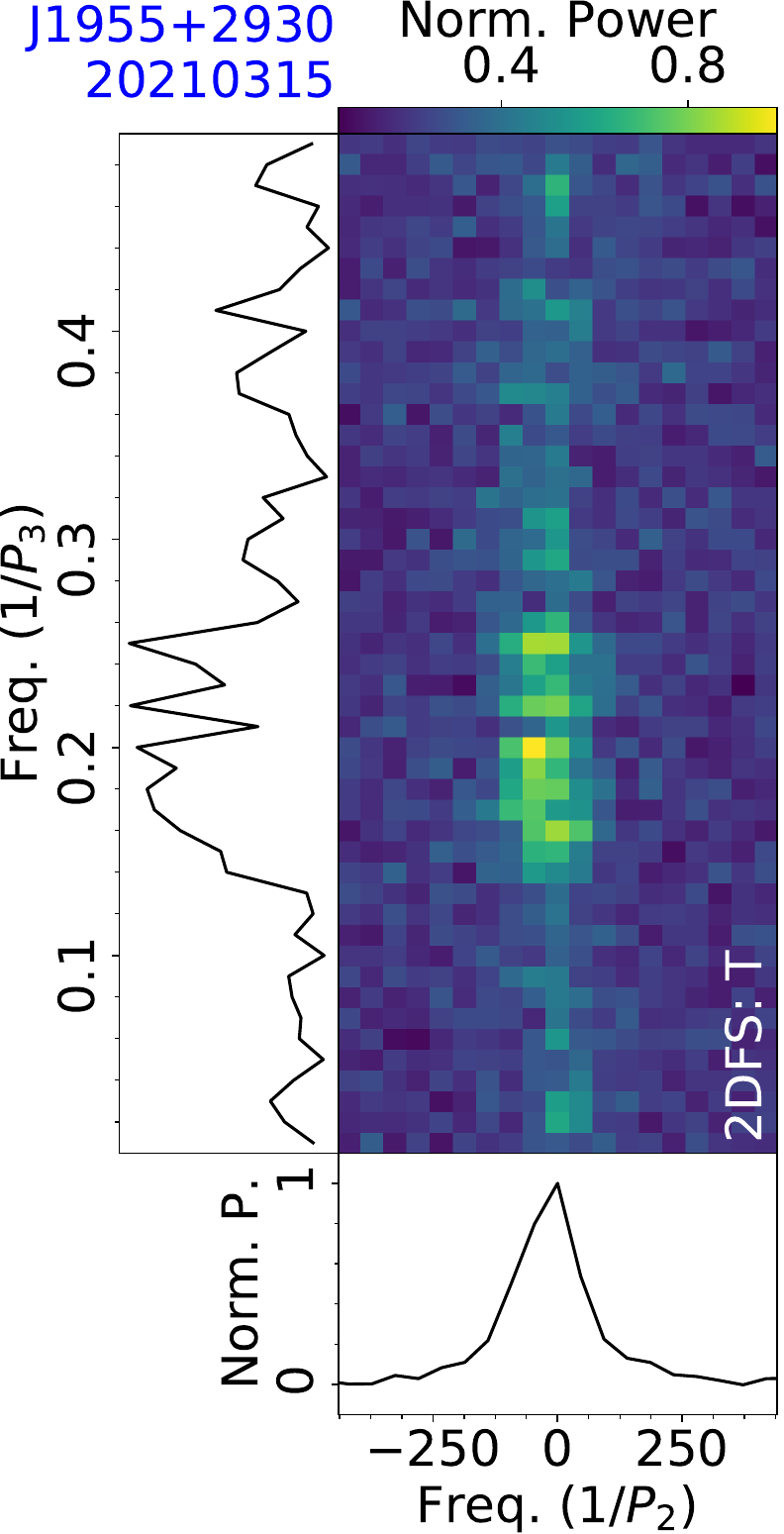}
\figcaption{Fluctuation analysis of PSR J1955+2930 for the observation on 20210315, with LRFS and 2DFS for the trailing part of a mean pulse profile.
\label{subfig:fluctu:J1955+2930}}
\end{figure}

\subsection{J1953+1149}
\label{subsec:J1953+1149}

PSR J1953+1149 was discovered by the 305 m radio telescope at Arecibo \citep{Nice1995}. Subpulse drifting of $P_3$=3.2$\pm$0.9 periods and $P_2$=31$^{+7}_{-24}$ degrees was reported by \citet{Song2023}.

The pulsar was also observed by FAST on 20210116 for 9 minutes, deriving a rotation period $P=0.8520$~s and a dispersion measure $D\!M=139.7~{\rm cm^{-3}\,pc}$. 
Single pulse sequences are shown in Fig.~\ref{subfig:TP:J1953+1149}, illustrating the unsystematic drifting behavior. The drift feature in 2DFS (Fig.~\ref{subfig:fluctu:J1953+1149}), whose centroid is $1/P_3=0.280\pm0.003$ ($P_3=3.57\pm0.04$ periods) and $1/P_2=49\pm1$ ($P_2=7.3\pm0.2^\circ$), is widely distributed across the temporal modulation frequency range. Additionally, a low-frequency modulation feature with the centroid of $1/P_3=0.059\pm0.002$ ($P_3=17.0\pm0.5$ periods) exists in 2DFS.

\begin{figure}[htpb]
\centering
\includegraphics[width=0.22\textwidth, angle=0]{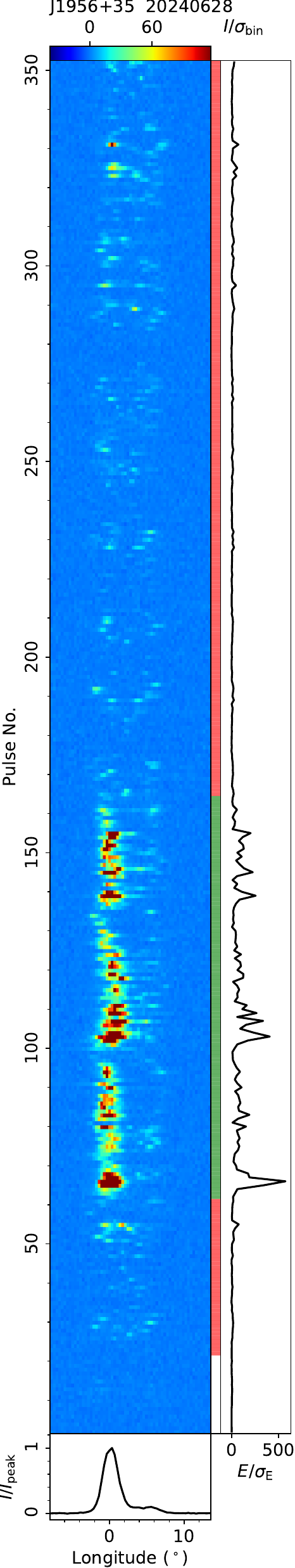}
\vspace{-0.2cm}
\figcaption{Single pulse sequence of PSR J1956+35 from the FAST observation on 20240628.
\label{subfig:TP:J1956+35}}
\end{figure}

\begin{figure}[htpb]
\centering
\includegraphics[width=0.37\textwidth, angle=0]{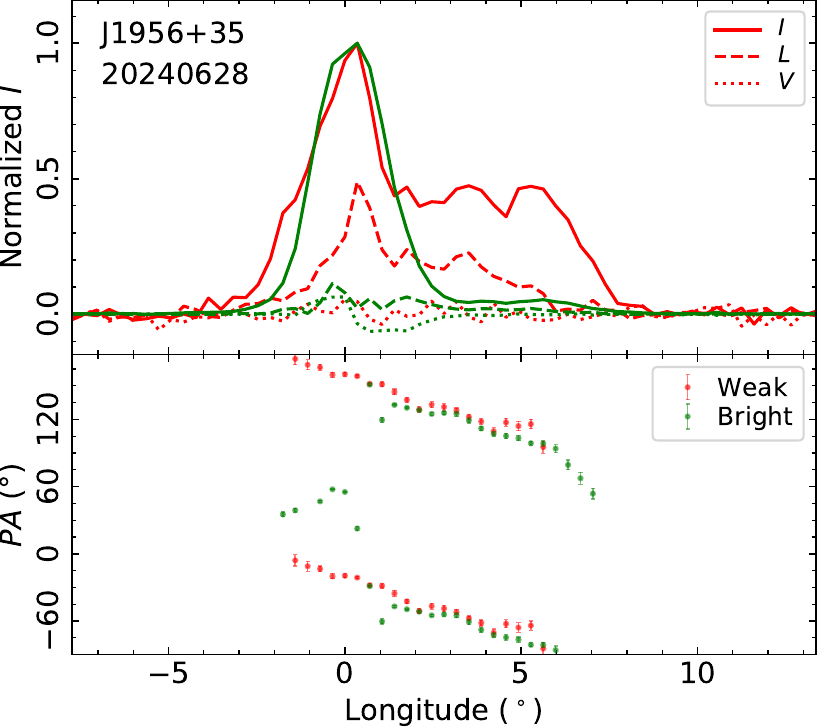}
\figcaption{Mean polarization profiles (the top panel) for weak and bright emission modes of PSR J1956+35 observed on 20240628, as well as the averaged PA curves (the bottom panel).
\label{subfig:PolModes:J1956+35}}
\end{figure}

\begin{figure}[htpb]
\centering
\includegraphics[width=0.44\textwidth, angle=0]{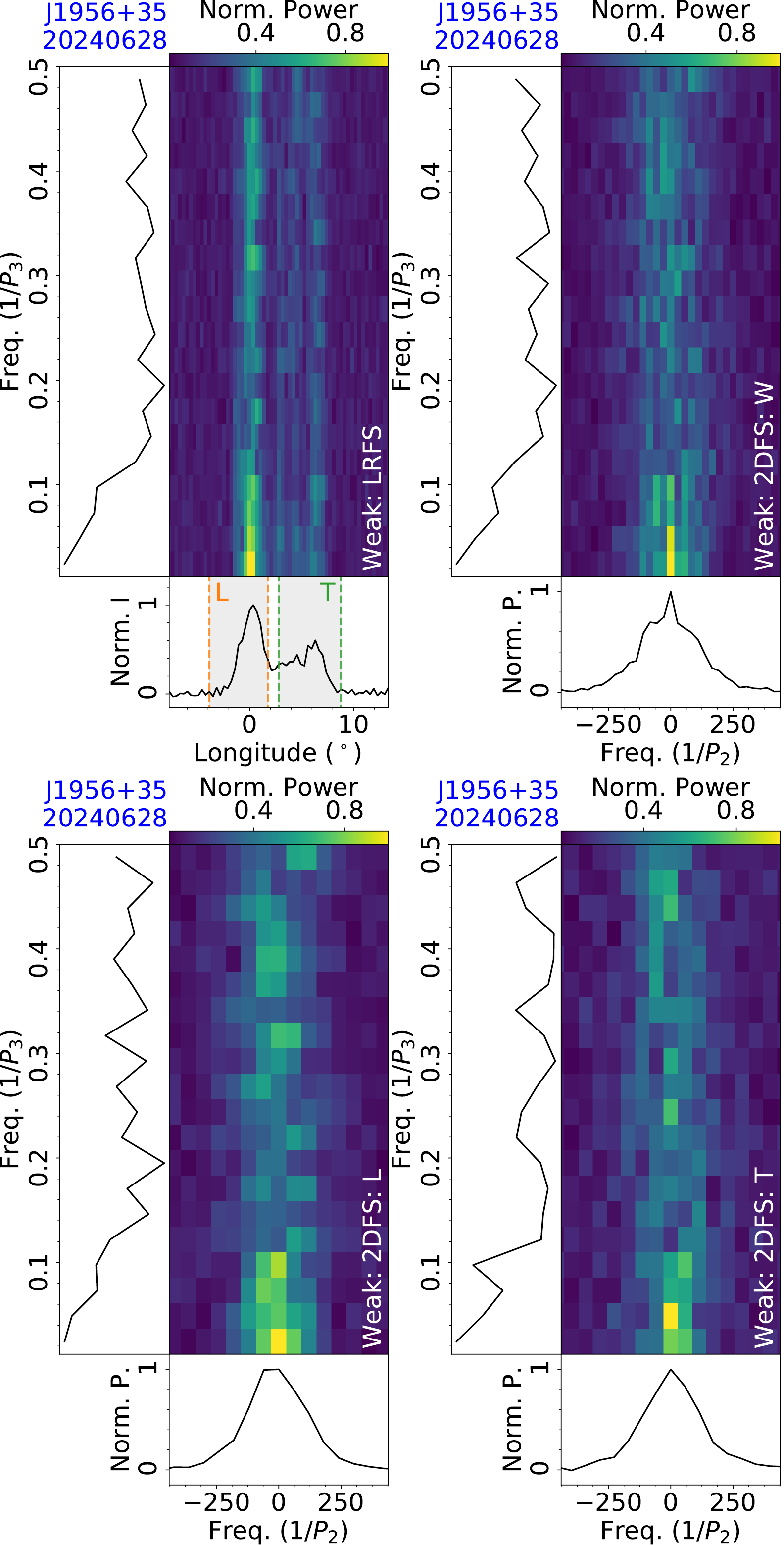}
\figcaption{Fluctuation analysis of PSR J1956+35 for the observation on 20240628, with LRFS (top-left), and 2DFS for the on-pulse region (top-right), leading part (bottom-left) and trailing part (bottom-right) of a mean pulse profile for the weak emission mode.
\label{subfig:fluctu:J1956+35}}
\end{figure}

\subsection{J1954+1021}
\label{subsec:J1954+1021}

PSR J1954+1021 was discovered in 350 MHz searches with the Green Bank Telescope \citet{Kawash2018}. 

This pulsar was observed by FAST on 20210117 for 9 minutes, yielding a rotation period $P=2.0996$~s and a dispersion measure $D\!M=81.6~{\rm cm^{-3}\,pc}$. 
The single pulse sequence in Fig.~\ref{subfig:TP:J1954+1021} displays nulling and subpulse drifting phenomena, which are reported for the first time. The nulling fraction is estimated to be 18$\pm$2\% from the on-pulse integral energy histogram in Fig.~\ref{subfig:Hist:J1954+1021}. 
Fluctuation spectra shown in Fig.~\ref{subfig:fluctu:J1954+1021} exhibit a negative drift feature, with the centroid frequencies estimated to be $1/P_3=0.3321\pm0.004$ and $1/P_2=-62\pm2$, corresponding to periodicities of $P_3=3.012\pm0.004$ periods and $P_2=-5.8\pm0.2^\circ$.

\begin{figure}[htpb]
\centering
\includegraphics[width=0.22\textwidth, angle=0]{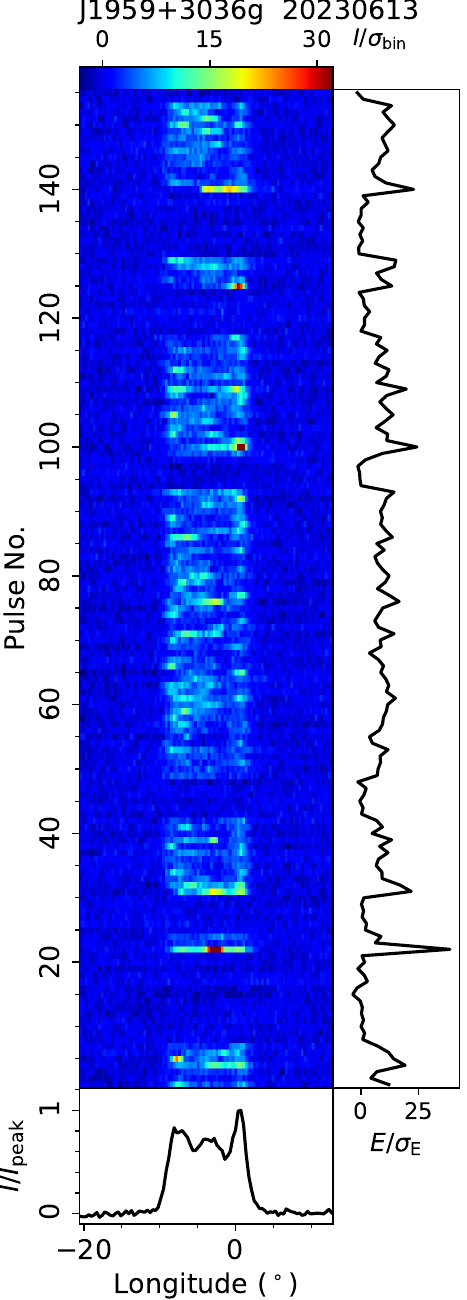}
\includegraphics[width=0.22\textwidth, angle=0]{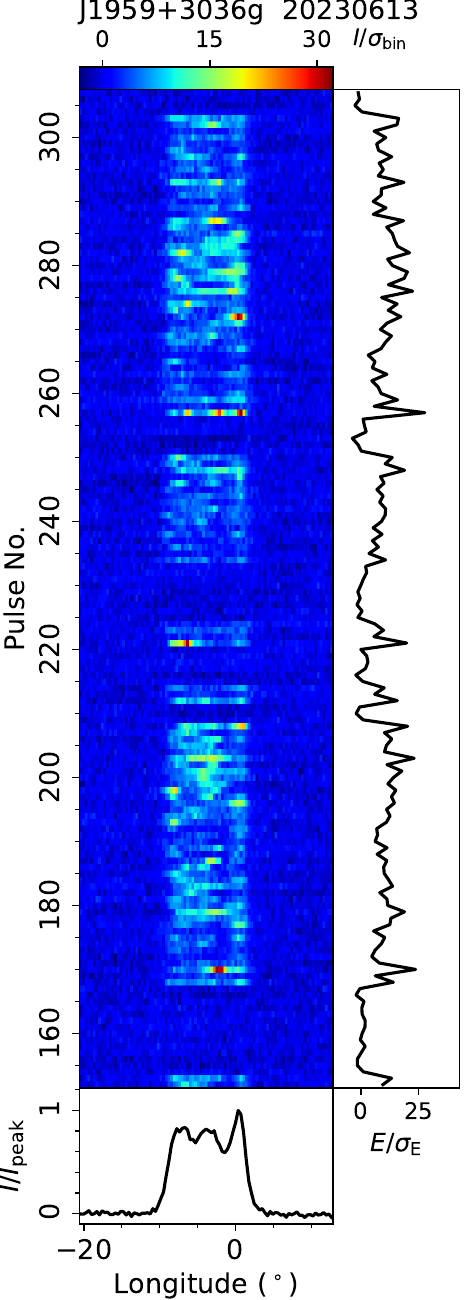}
\figcaption{Single-pulse sequences of PSR J1959+3036g from the FAST observation on 20230613.
\label{subfig:TP:J1959+3036g}
}
\end{figure}

\begin{figure}[htpb]
\centering
\includegraphics[width=0.39\textwidth, angle=0]{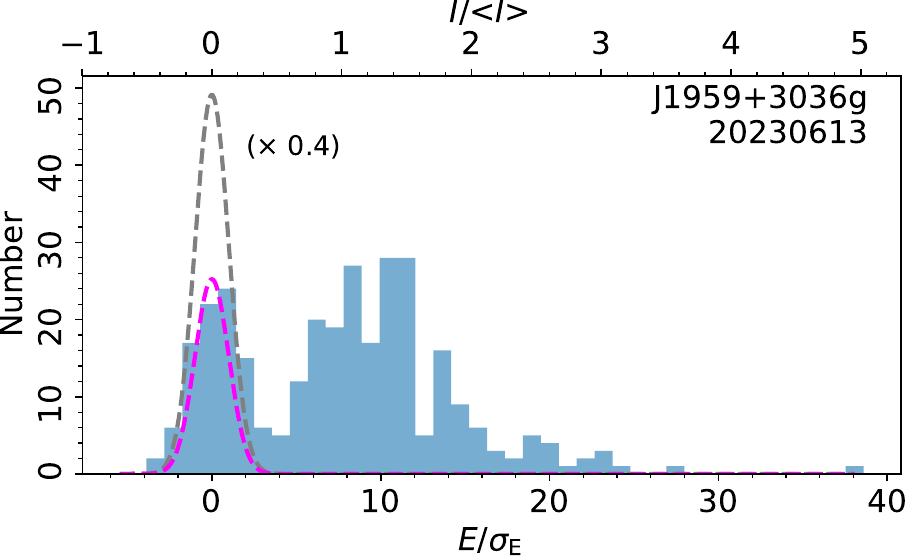}
\figcaption{On-pulse energy histogram of single pulses of PSR J1959+3036g from the FAST observation on 20230613.
\label{subfig:Hist:J1959+3036g}}
\end{figure}

\begin{figure}[htpb]
\centering
\includegraphics[width=0.22\textwidth, angle=0]{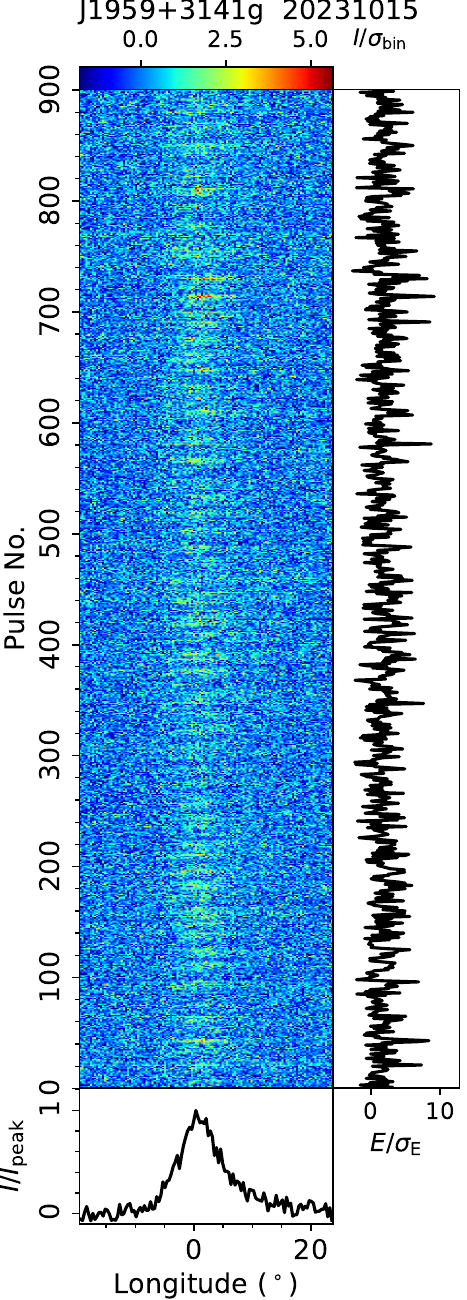}
\includegraphics[width=0.22\textwidth, angle=0]{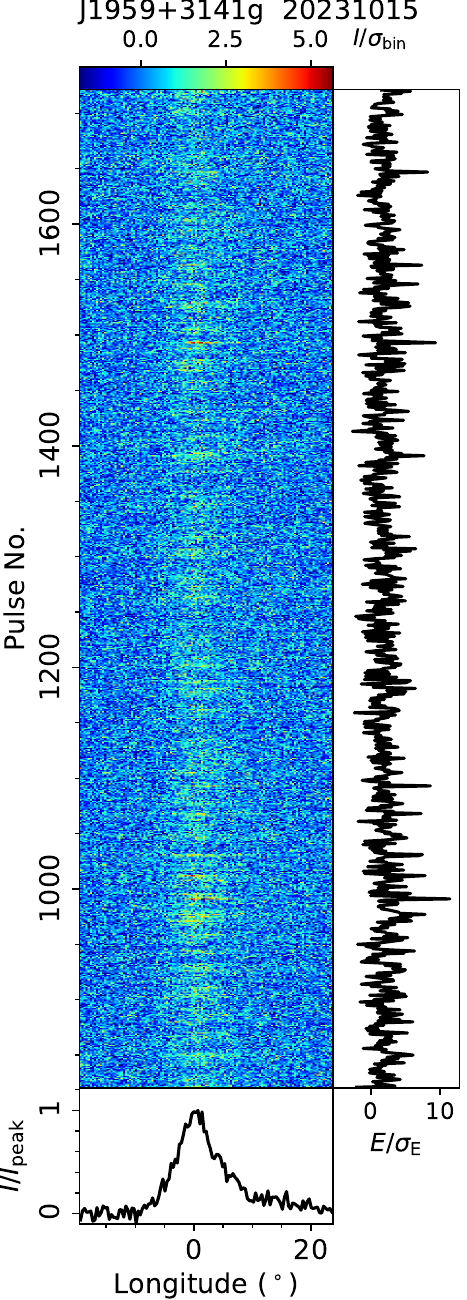}
\figcaption{Single pulse sequences of PSR J1959+3141g from the FAST observation on 20231015.
\label{subfig:TP:J1959+3141g}}
\end{figure}

\begin{figure}[htpb]
\centering
\includegraphics[width=0.22\textwidth, angle=0]{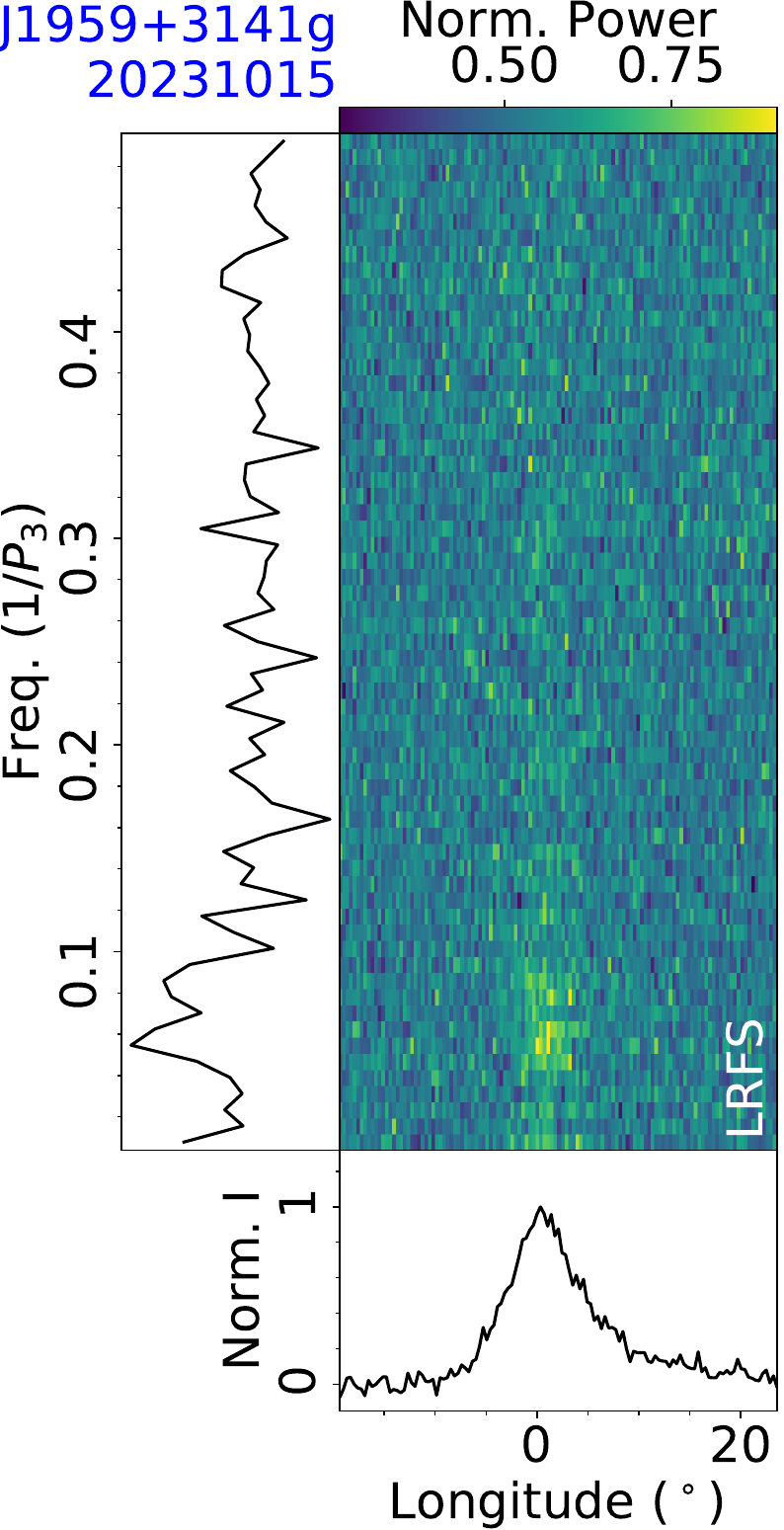}
\includegraphics[width=0.22\textwidth, angle=0]{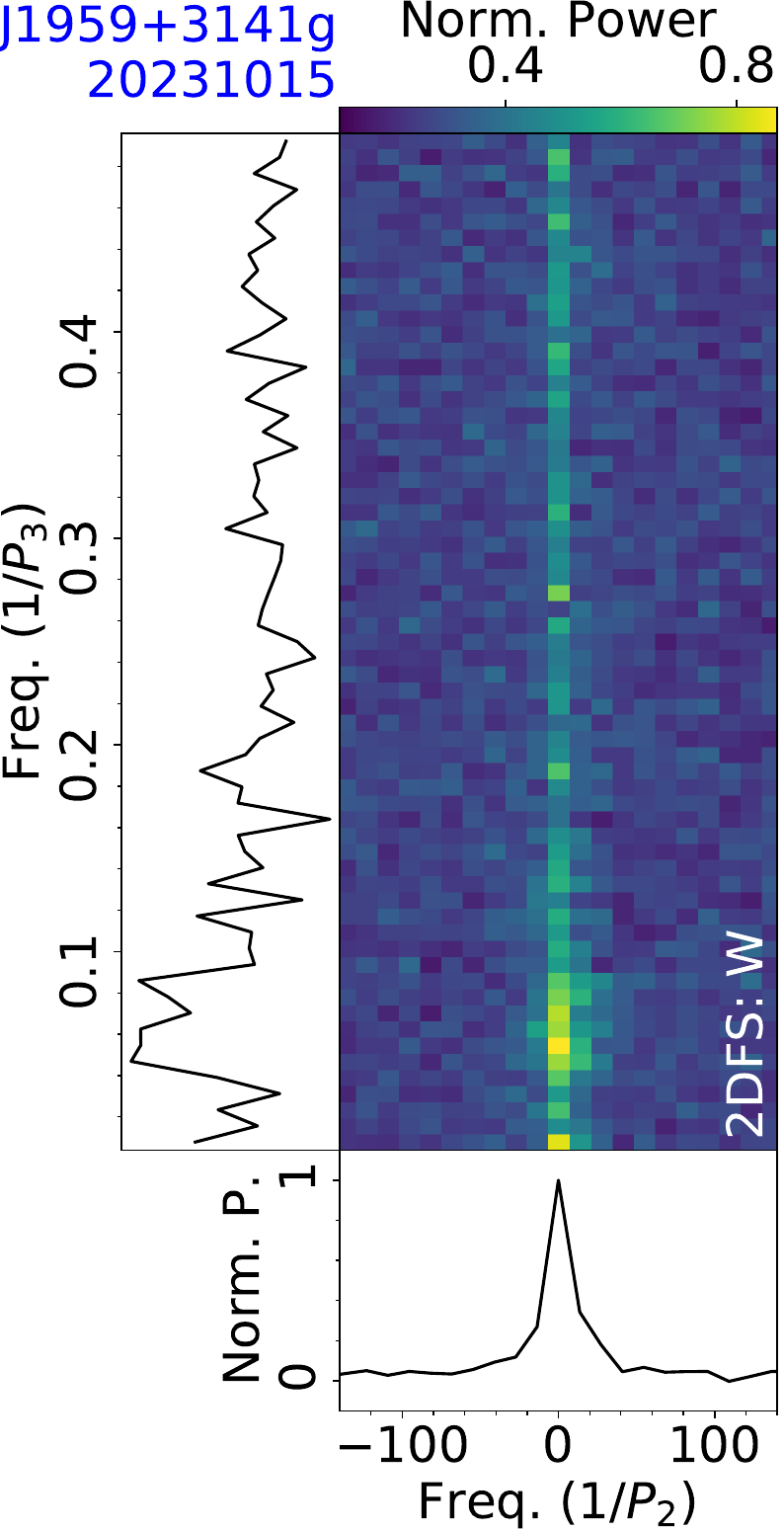}
\figcaption{Fluctuation analysis of PSR J1959+3141g for the observation on 20231015, with LRFS and 2DFS for the on-pulse region of a mean pulse profile.
\label{subfig:fluctu:J1959+3141g}}
\end{figure}

\subsection{J1954+2529}
\label{subsec:J1954+2529}

PSR J1954+2529 was discovered in the Pulsar Arecibo L-band Feed Array (PALFA) survey, which was reported to display nulling behavior \citep{Parent2022}.

This pulsar was observed by FAST on 20220829 and 20230729 for 59 and 5 minutes, respectively. From the longer observation, a rotation period $P=0.9313$~s and a dispersion measure $D\!M=182.7~{\rm cm^{-3}\,pc}$ were derived. The single pulse sequence of the observation on 20230729 in Fig.~\ref{subfig:TP:J1954+2529} shows the nulling phenomenon, and the nulling fraction is estimated from the on-pulse integral energy histogram (Fig.~\ref{subfig:Hist:J1954+2529}) to be 28.3$\pm$1.5\%.

\begin{figure}[htpb]
\gridline{
\includegraphics[width=0.22\textwidth, angle=0]{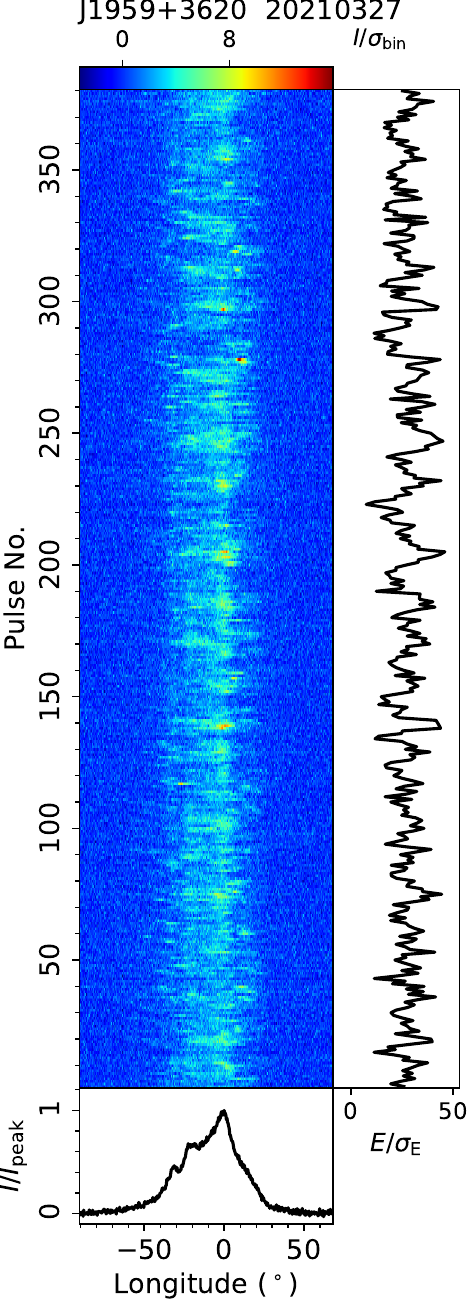}
\includegraphics[width=0.22\textwidth, angle=0]{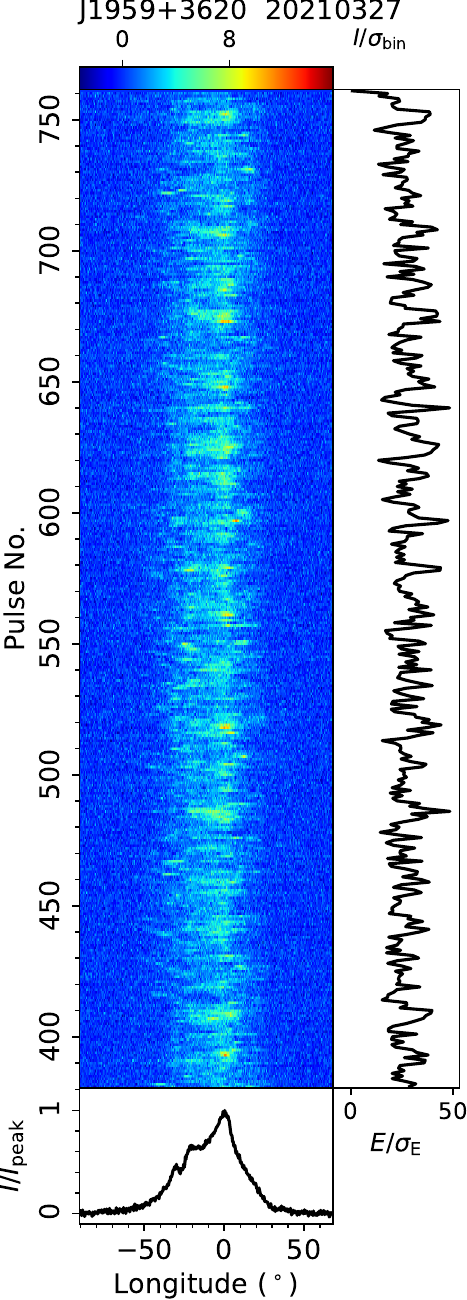}
}
\figcaption{Single pulse sequences of PSR J1959+3620 from the FAST observation on 20210327.
\label{subfig:TP:J1959+3620}}
\end{figure}

\begin{figure}[htpb]
\centering
\includegraphics[width=0.22\textwidth, angle=0]{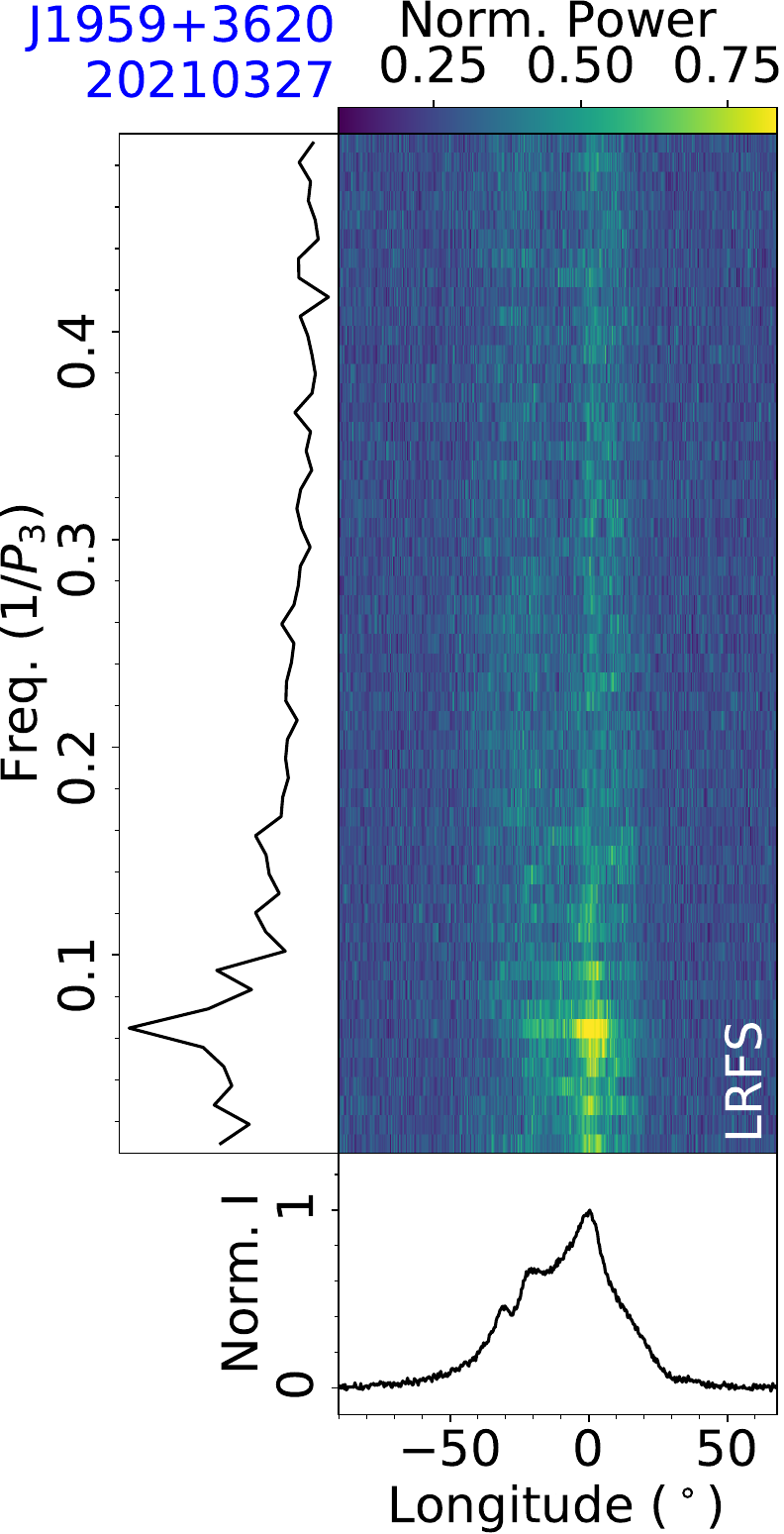}
\includegraphics[width=0.22\textwidth, angle=0]{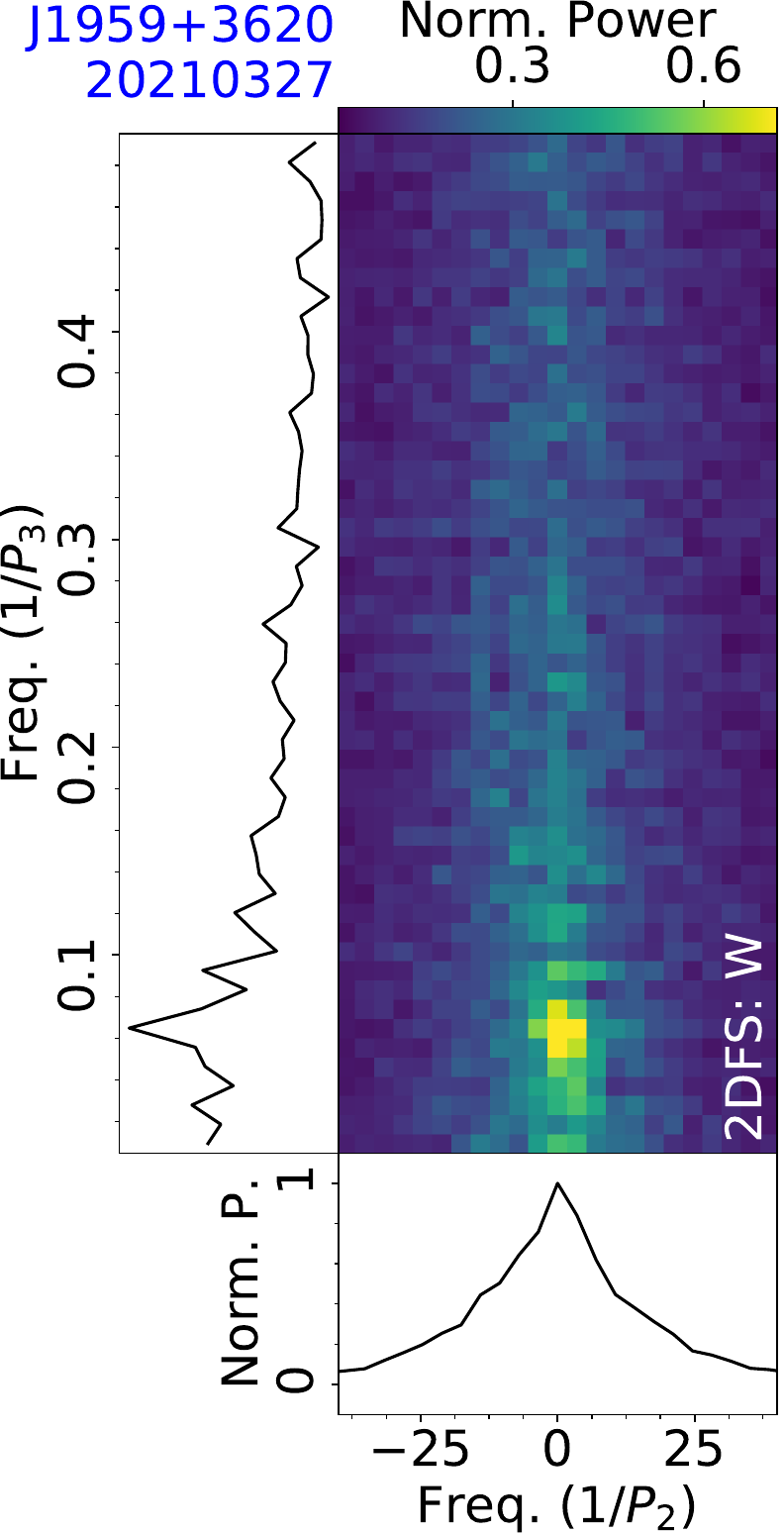}
\figcaption{Fluctuation analysis of PSR J1959+3620 for the observation on 20210327, with LRFS and 2DFS for the on-pulse region of a mean pulse profile.
\label{subfig:fluctu:J1959+3620}}
\end{figure}

\subsection{J1955+2930}
\label{subsec:J1955+2930}

PSR J1955+2930 was discovered by the Arecibo telescope \citep{Parent2022}. 

This pulsar was observed by FAST on 20210315 for 7 minutes, with a rotation period and a dispersion measure determined to be $P=1.0738$~s and a dispersion measure $D\!M=211.3~{\rm cm^{-3}\,pc}$. Single pulse sequences are shown in Fig.~\ref{subfig:TP:J1955+2930}, displaying the subpulse drifting phenomenon for the trailing part in a mean pulse profile. From the fluctuation spectra in Fig.~\ref{subfig:fluctu:J1955+2930}, centroid frequencies of the negative drift feature for the trailing profile part are estimated to be $1/P_3=0.200\pm0.002$ and $1/P_2=-25\pm3$, corresponding to periodicities of $P_3=5.00\pm0.05$ periods and $P_2=-15\pm2$ degrees.

\begin{figure}[htpb]
\gridline{
\includegraphics[width=0.22\textwidth, angle=0]{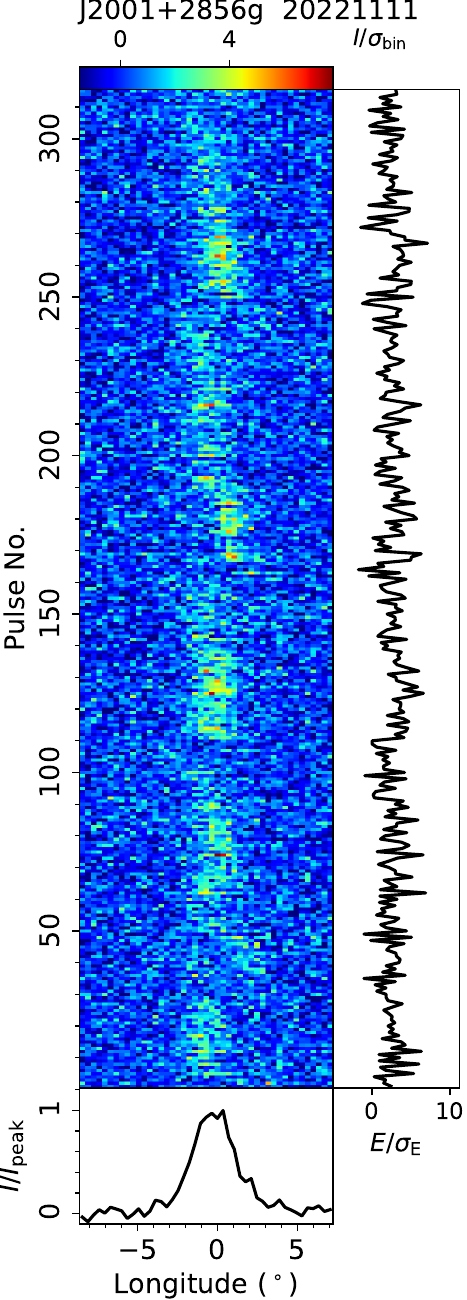}
\includegraphics[width=0.22\textwidth, angle=0]{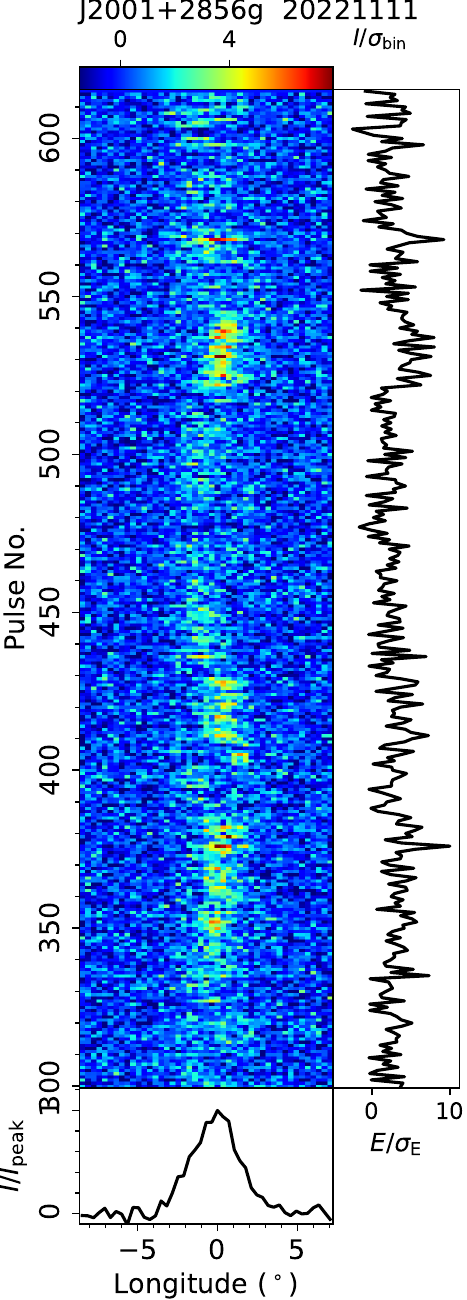}
}
\figcaption{Single pulse sequences of PSR J2001+2856g from the FAST observation on 20221111.
\label{subfig:TP:J2001+2856}}
\end{figure}

\begin{figure}[htpb]
\centering
\includegraphics[width=0.22\textwidth, angle=0]{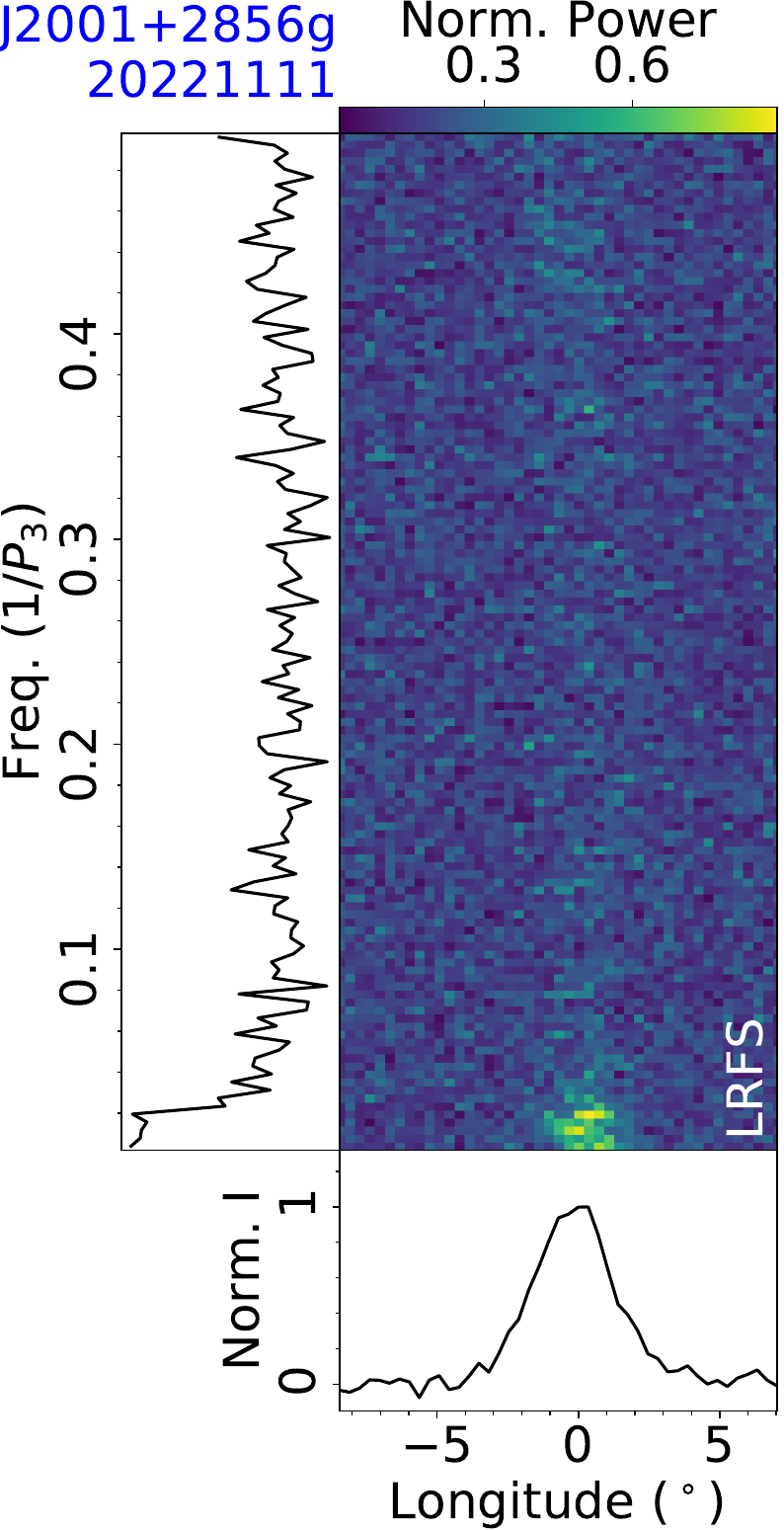}
\includegraphics[width=0.22\textwidth, angle=0]{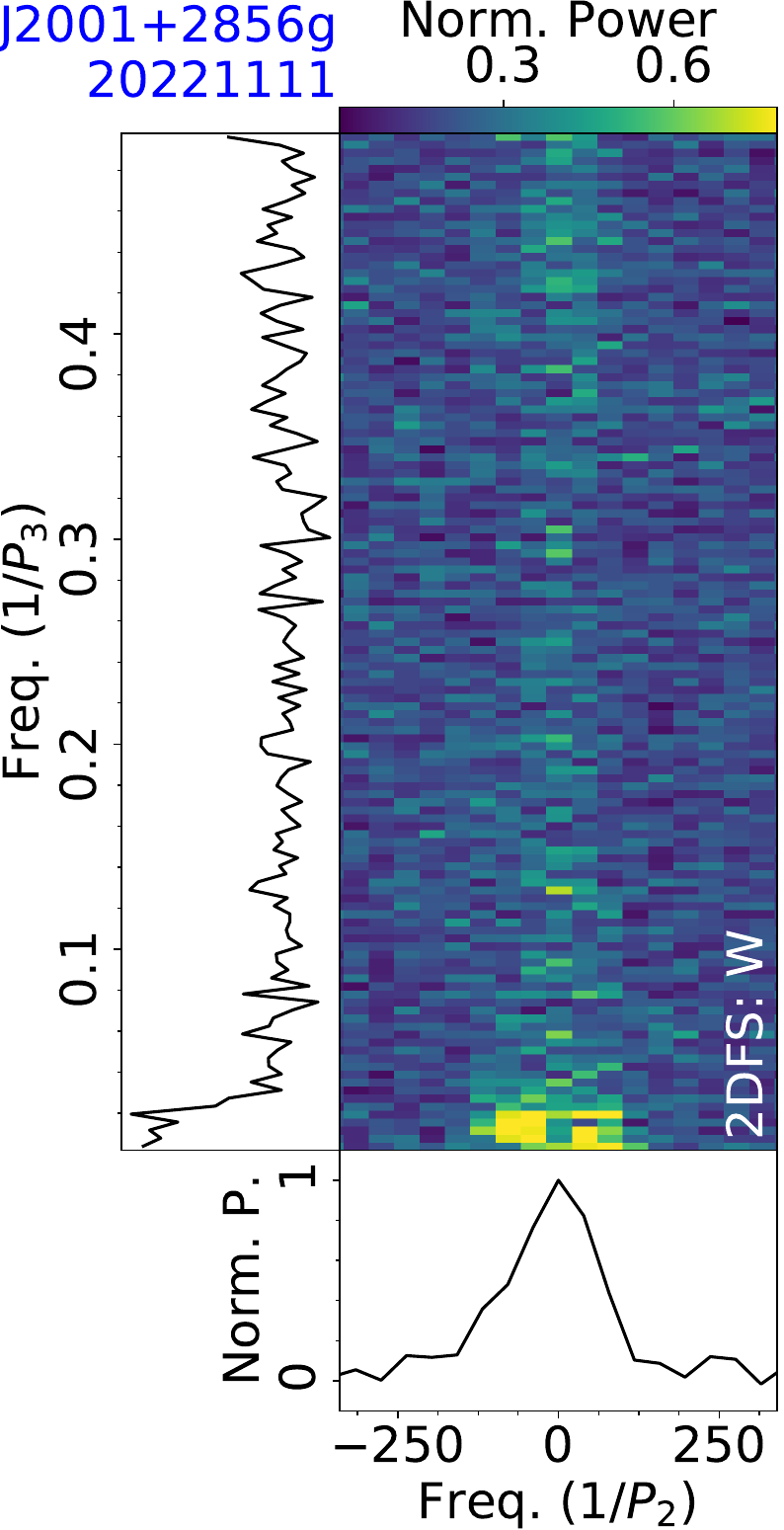}
\figcaption{Fluctuation analysis of PSR J2001+2856g for the observation on 20221111, with LRFS and 2DFS for the on-pulse region of a mean pulse profile.
\label{subfig:fluctu:J2001+2856}}
\end{figure}

\subsection{J1956+35}
\label{subsec:J1956+35}

PSR J1956+35 was discovered by CHIME (https://www.chime-frb.ca/galactic).

This pulsar was observed by FAST on 20240628 for 5 minutes, deriving a rotation period $P=0.8755$~s and a dispersion measure $D\!M=153.3~{\rm cm^{-3}\,pc}$. The single pulse sequence in Fig.~\ref{subfig:TP:J1956+35} displays mode changes between weak and bright emission modes, which are labeled by red and green colors, and there is also bi-drifting behavior for pulses No. 165-280 in the weak state. 
Mean polarization profiles and the average PA curves of two emission modes are shown in Fig.~\ref{subfig:PolModes:J1956+35}. The leading profile part is stronger than the trailing part for two modes, and the contrast between the two profile parts is much greater in the bright emission mode. The difference between the PA curves of the two emission modes over longitudes from $-1.5^\circ$ to 1$^\circ$ may be attributed to the orthogonal modes. For pulses No. 165-280 of the weak emission state, LRFS and 2DFS are shown in Fig.~\ref{subfig:fluctu:J1956+35}. 2DFS of the leading profile part exhibits a negative drift feature, with the centroid modulation frequencies estimated to be $1/P_3=0.055\pm0.004$ and $1/P_2=-26\pm10$, corresponding to periodicities of $P_3=18\pm1$ periods and $P_2=-14\pm5$ degrees. For the trailing profile part, the centroid of the positive drift feature is characterized by frequencies of $1/P_3=0.056\pm0.005$ and $1/P_2=27\pm7$, yielding $P_3=18\pm1$ periods and $P_2=13\pm4$ degrees.

Detailed single pulse properties require more observations.

\begin{figure}[htpb]
\centering
\includegraphics[width=0.22\textwidth, angle=0]{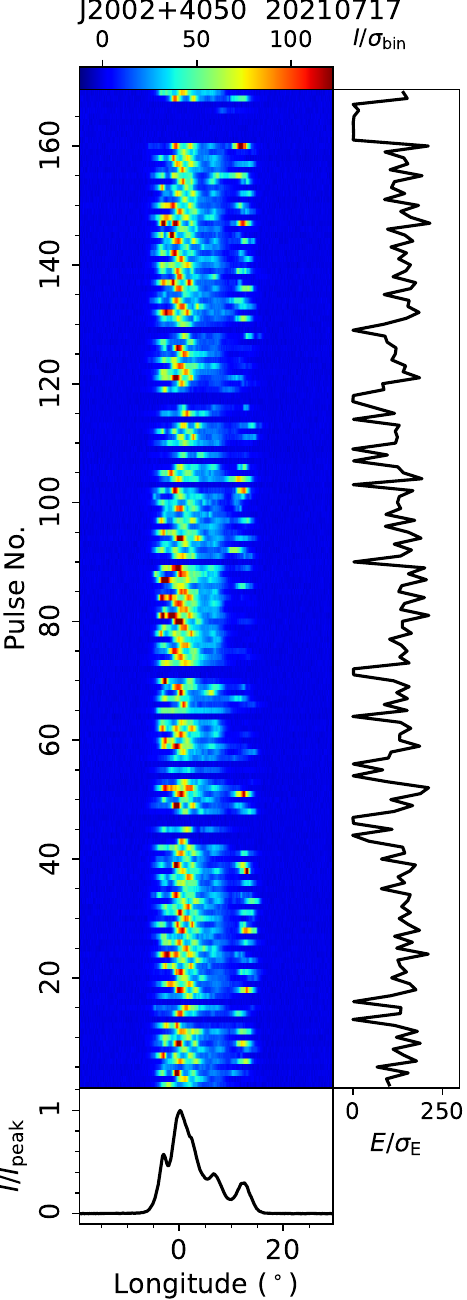}
\includegraphics[width=0.22\textwidth, angle=0]{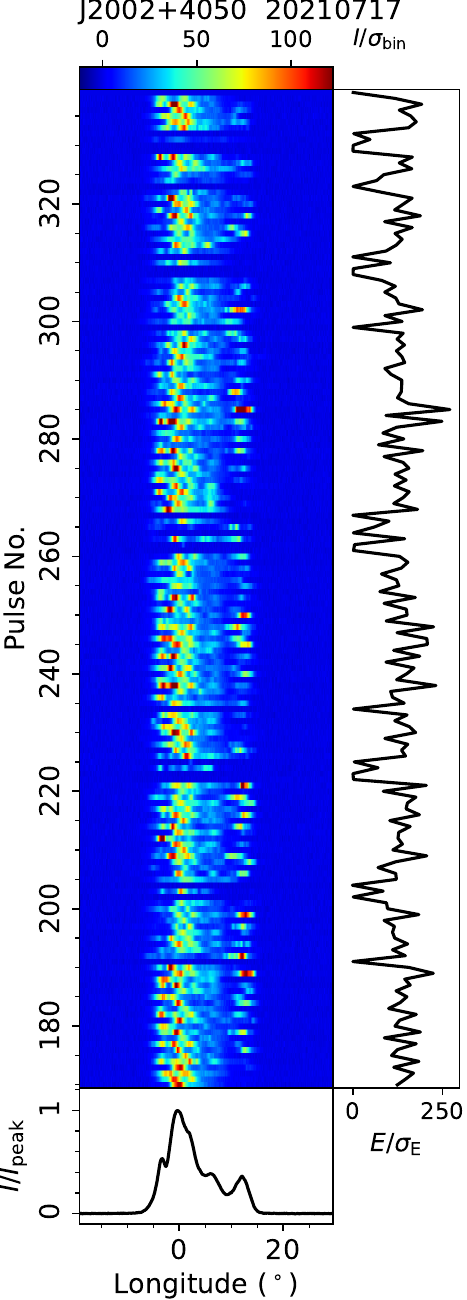}
\figcaption{Single pulse sequences of PSR J2002+4050 from the FAST observation on 20210717.
\label{subfig:TP:J2002+4050}}
\end{figure}

\begin{figure}[htpb]
\centering
\includegraphics[width=0.39\textwidth, angle=0]{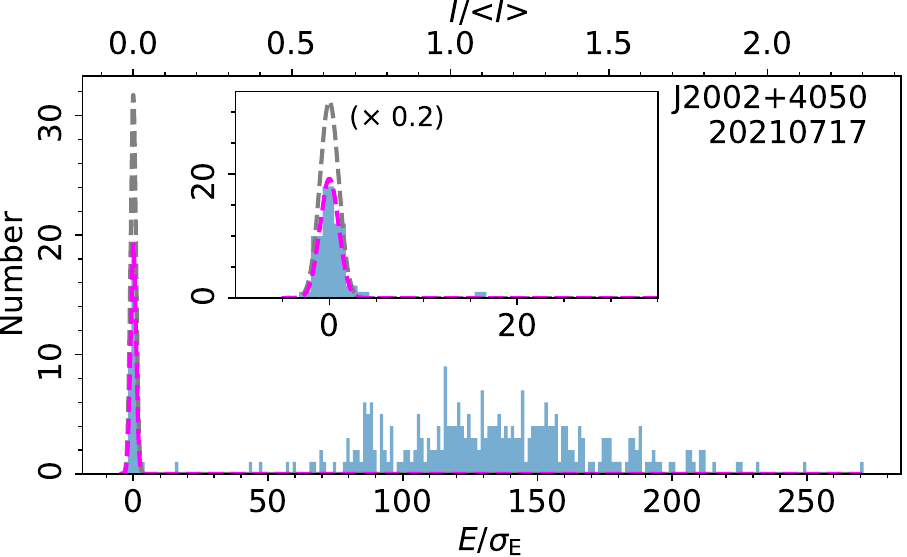}
\figcaption{On-pulse energy histogram of single pulses of PSR J2002+4050 from the FAST observation on 20210717. The inset provides a view of the x‑axis region from -10 to 35.
\label{subfig:Hist:J2002+4050}}
\end{figure}

\begin{figure}[htpb]
\centering
\includegraphics[width=0.44\textwidth, angle=0]{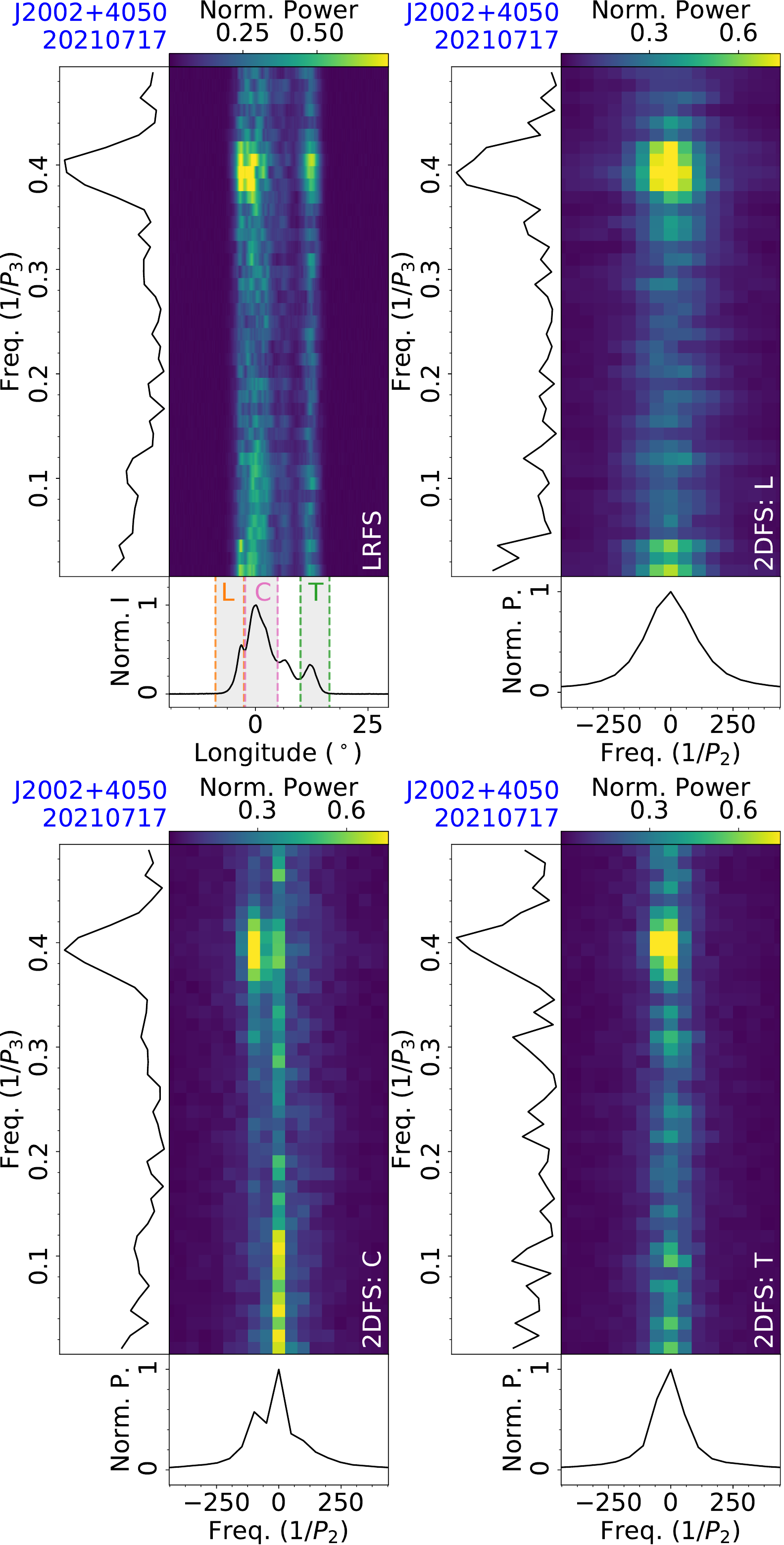}
\figcaption{Fluctuation analysis of PSR J2002+4050 from the FAST observation on 20210717, with LRFS (top-left), and 2DFS for the on-pulse region (top-right), leading part (bottom-left) and trailing part (bottom-right) of a mean pulse profile.
\label{subfig:fluctu:J2002+4050}}
\end{figure}

\subsection{J1959+3036g}
\label{subsec:J1959+3036g}

PSR J1959+3036g was discovered by \citet{Han2021} from the FAST GPPS survey. 

This pulsar was observed by FAST on 20230613 for 15 minutes, yielding a rotation period $P=2.8883$~s and a dispersion measure $D\!M=179.1~{\rm cm^{-3}\,pc}$. 
Single pulse sequences in Fig.~\ref{subfig:TP:J1959+3036g} show the nulling phenomenon. The on-pulse integral energy histogram is displayed in Fig.~\ref{subfig:Hist:J1959+3036g}, where the distribution around 0 energy confirms the existence of the nulling phenomenon. The nulling fraction of this observation is estimated to be 20.5$\pm$1.6\%.

\subsection{J1959+3141g}
\label{subsec:J1959+3141g}

PSR J1959+3141g was discovered in the FAST GPPS survey \citep{Han2021,han2025}.

This pulsar was observed by FAST on 20231015 for 15 minutes, yielding a rotation period $P=0.5146$~s and a dispersion measure $D\!M=342.8~{\rm cm^{-3}\,pc}$. 
Single pulse sequences are displayed in Fig.~\ref{subfig:TP:J1959+3141g}. From LRFS and 2DFS in Fig.~\ref{subfig:fluctu:J1959+3141g}, the pulsar tends to have a positive drifting behavior. The centroid frequencies of the drift feature in 2DFS are estimated to be $1/P_3=0.065\pm0.001$ and $1/P_2=6\pm1$, corresponding to $P_3=15.4\pm0.2$ periods and $P_2=58\pm11^\circ$.

\begin{figure}[htpb]
\centering
\includegraphics[width=0.22\textwidth, angle=0]{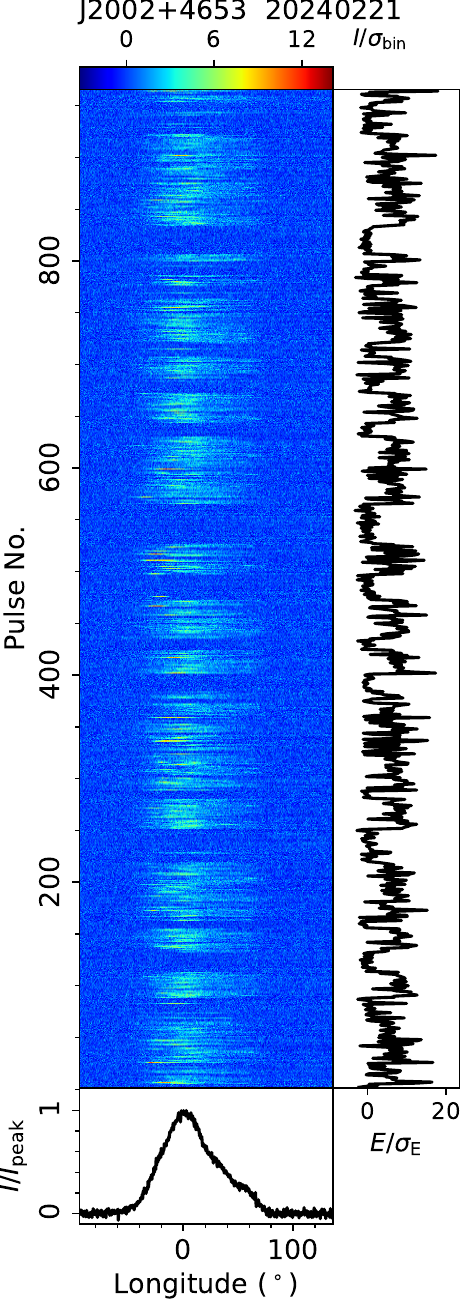}
\includegraphics[width=0.22\textwidth, angle=0]{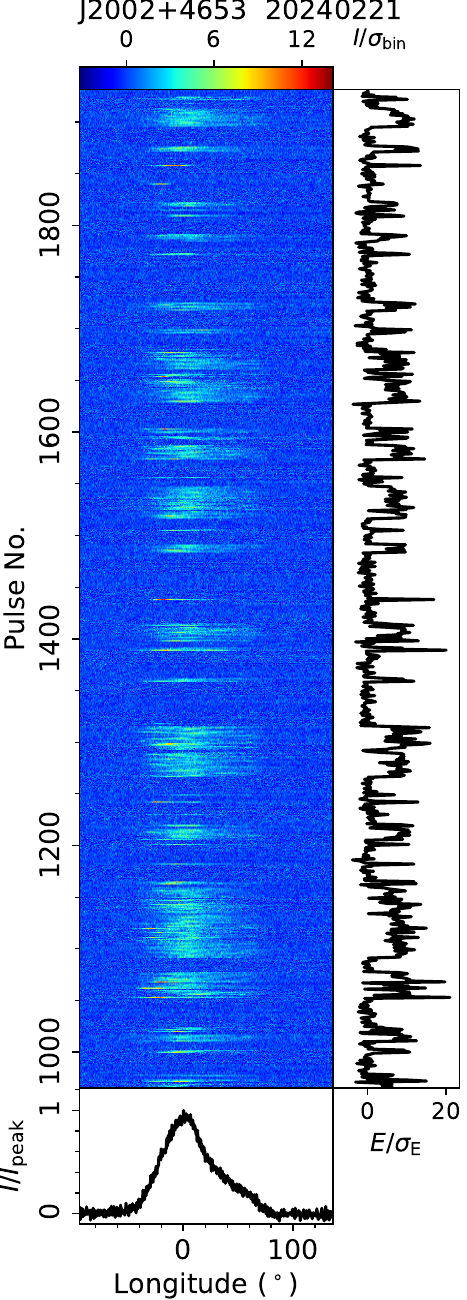}
\figcaption{Single pulse sequences of PSR J2002+4653 from the FAST observation on 20240221.
\label{subfig:TP:J2002+4653}}
\end{figure}

\begin{figure}[htpb]
\centering
\includegraphics[width=0.39\textwidth, angle=0]{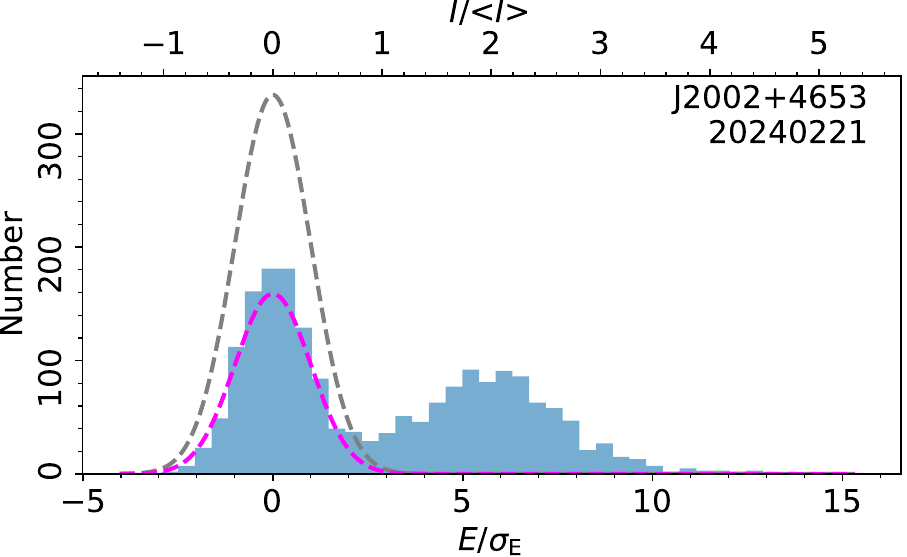}
\figcaption{On-pulse energy histogram of single pulses of PSR J2002+4653 from the FAST observation on 20240221.
\label{subfig:Hist:J2002+4653}}
\end{figure}

\subsection{J1959+3620}
\label{subsec:J1959+3620}

PSR J1959+3620 was discovered in the Northern High Time Resolution Universe survey with the 100-m Effelsberg radio telescope \citep{Barr2013}.

This pulsar was observed by FAST on 20210327 for 5 minutes, yielding a rotation period $P=0.4061$~s and a dispersion measure $D\!M=316.8~{\rm cm^{-3}\,pc}$. Single pulse sequences in Fig.~\ref{subfig:TP:J1959+3620} show the modulation phenomenon. For the on-pulse phase range, the centroid temporal modulation frequency in the fluctuation spectra (Fig.~\ref{subfig:fluctu:J1959+3620}) is estimated to be $1/P_3=0.054\pm0.001$, corresponding to the periodicity of $P_3=18.6\pm0.3$ periods.

\subsection{J2001+2856g}
\label{subsec:J2001+2856g}

PSR J2001+2856g was discovered in the FAST GPPS survey \citep{Han2021,han2025}.

This pulsar was observed by FAST on 20221004 for 5 minutes, on 20221111 for 15 minutes, and on 20250417 for 60 minutes. From the 15-minute data, a rotation period $P=1.4456$~s and a dispersion measure $D\!M=233.7~{\rm cm^{-3}\,pc}$ were determined. 
Single pulse sequences of the observation on 20221111, shown in Fig.~\ref{subfig:TP:J2001+2856}, display drifting bands of two drifting directions, and subpulses mainly drift negatively. In LRFS and 2DFS in Fig.~\ref{subfig:fluctu:J2001+2856}, the centroid frequencies of the negative drift feature are $1/P_3=0.017\pm0.001$ and $1/P_2=-59\pm3$, corresponding to periodicities of $P_3=58\pm2$ periods and $P_2=-6.1\pm0.3^\circ$. The centroid of the positive drift feature is characterized by $1/P_3=0.015\pm0.001$ and $1/P_2=55\pm4$, yielding $P_3=67\pm3$ periods and $P_2=6.5\pm0.5^\circ$.

\subsection{J2002+4050}
\label{subsec:J2002+4050}

PSR J2002+4050 was discovered by the 92-m telescope at Green Bank \citep{Stokes1985}. Subpulse drifting was reported by \citet{Weltevrede2006} at 21 cm, with $P_2=-7.3^{+0.8}_{-0.6}$ degrees for the leading component and $P_2=-17^{+3}_{-8}$ degrees for the rest of the pulse profile, and $P_3=2.5\pm0.1$ periods. \citet{Basu2020MNRAS} subsequently reported drift rates of $-3.5\pm0.4$ and $-5.1\pm0.8$ degrees per period for the leading conal component and the trailing component, respectively, with a $P_3$ of $\sim$2.5 periods at 1.6 GHz, and a nulling fraction of 11.5$\pm$0.4\%. Dwarf pulses were detected by \citet{Yan2024} with FAST. 

This pulsar was observed by FAST on 20210717 for 5 minutes, deriving a rotation period $P=0.9051$~s and a dispersion measure $D\!M=130.4~{\rm cm^{-3}\,pc}$. Single pulse sequences in Fig.~\ref{subfig:TP:J2002+4050} show nulling and subpulse modulation. 
The nulling fraction of this observation is estimated to be 12.1$\pm$0.9\% from the on-pulse energy histogram in Fig.~\ref{subfig:Hist:J2002+4050}. 
From the fluctuation spectra shown in Fig.~\ref{subfig:fluctu:J2002+4050}, the leading edge of the profile exhibits a subpulse modulation, while the central and trailing profile parts both have negative subpulse drifting. 
For the leading edge of the profile, the centroid frequency of the modulation feature in 2DFS is $1/P_3=0.395\pm0.001$, corresponding to $P_3=2.53\pm0.01$ periods. In the 2DFS of the central profile part, the negative drift feature has a centroid at $1/P_3=0.393\pm0.001$ and $1/P_2=-95\pm4$, yielding $P_3=2.55\pm0.01$ periods and $P_2=-3.8\pm0.2$ degrees. For the trailing profile part, the negative drift feature in the 2DFS exhibits a centroid at $1/P_3=0.397\pm0.002$ and $1/P_2=-26\pm5$, yielding $P_3=2.52\pm0.01$ periods and $P_2=-14\pm3$ degrees.

\begin{figure}[htpb]
\centering
\includegraphics[width=0.22\textwidth, angle=0]{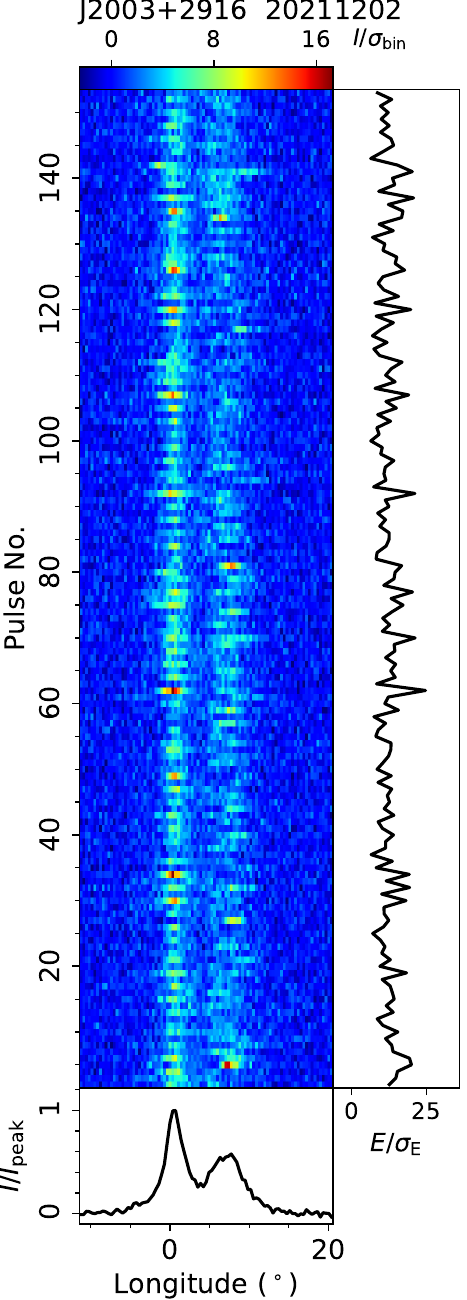}
\includegraphics[width=0.22\textwidth, angle=0]{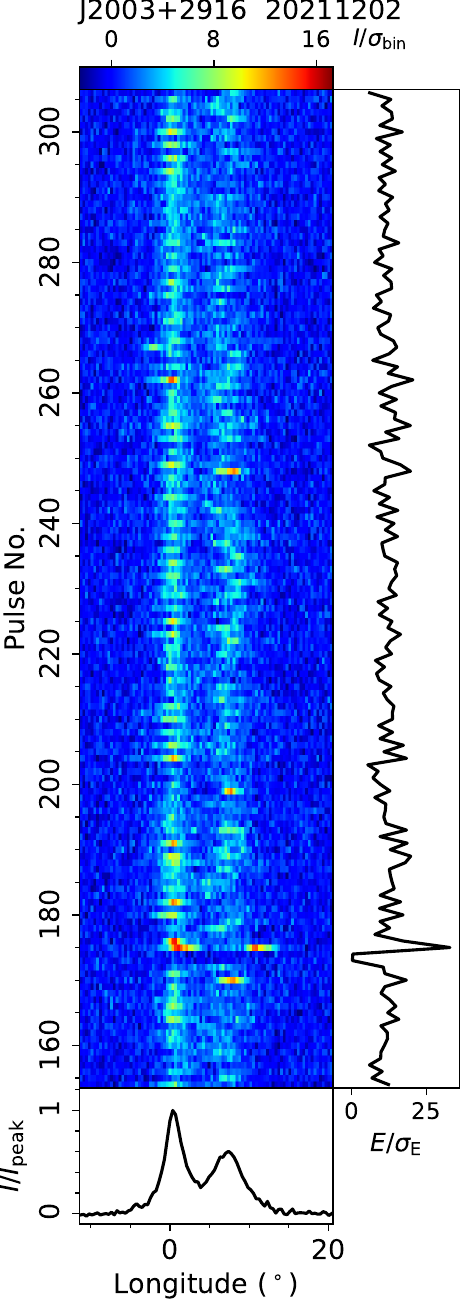}
\figcaption{Single pulse sequences of PSR J2003+2916 from the FAST observation on 20211202.
\label{subfig:TP:J2003+2916}}
\end{figure}

\begin{figure}[htpb]
\centering
\includegraphics[width=0.39\textwidth, angle=0]{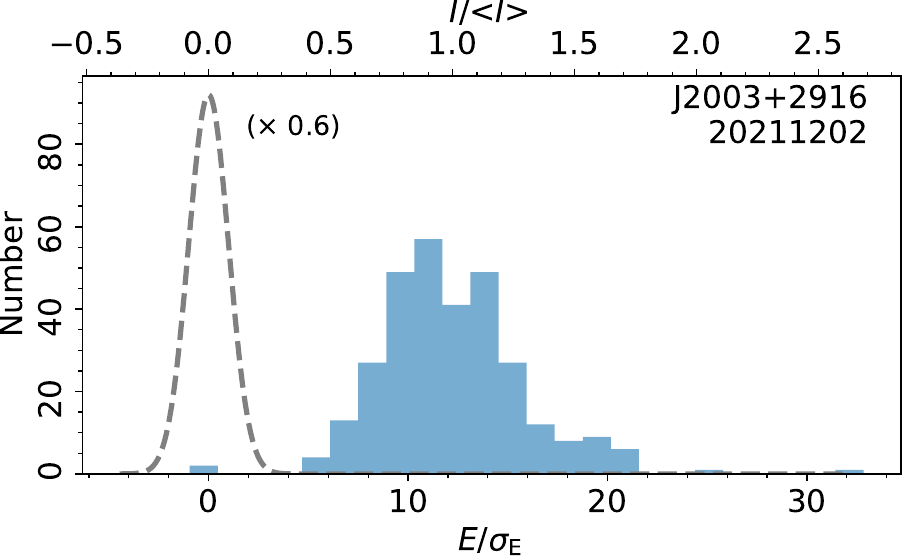}
\figcaption{On-pulse energy histogram of single pulses of PSR J2003+2916 from the FAST observation on 20211202. \label{subfig:Hist:J2003+2916}}
\end{figure}

\begin{figure}[htpb]
\centering
\includegraphics[width=0.44\textwidth, angle=0]{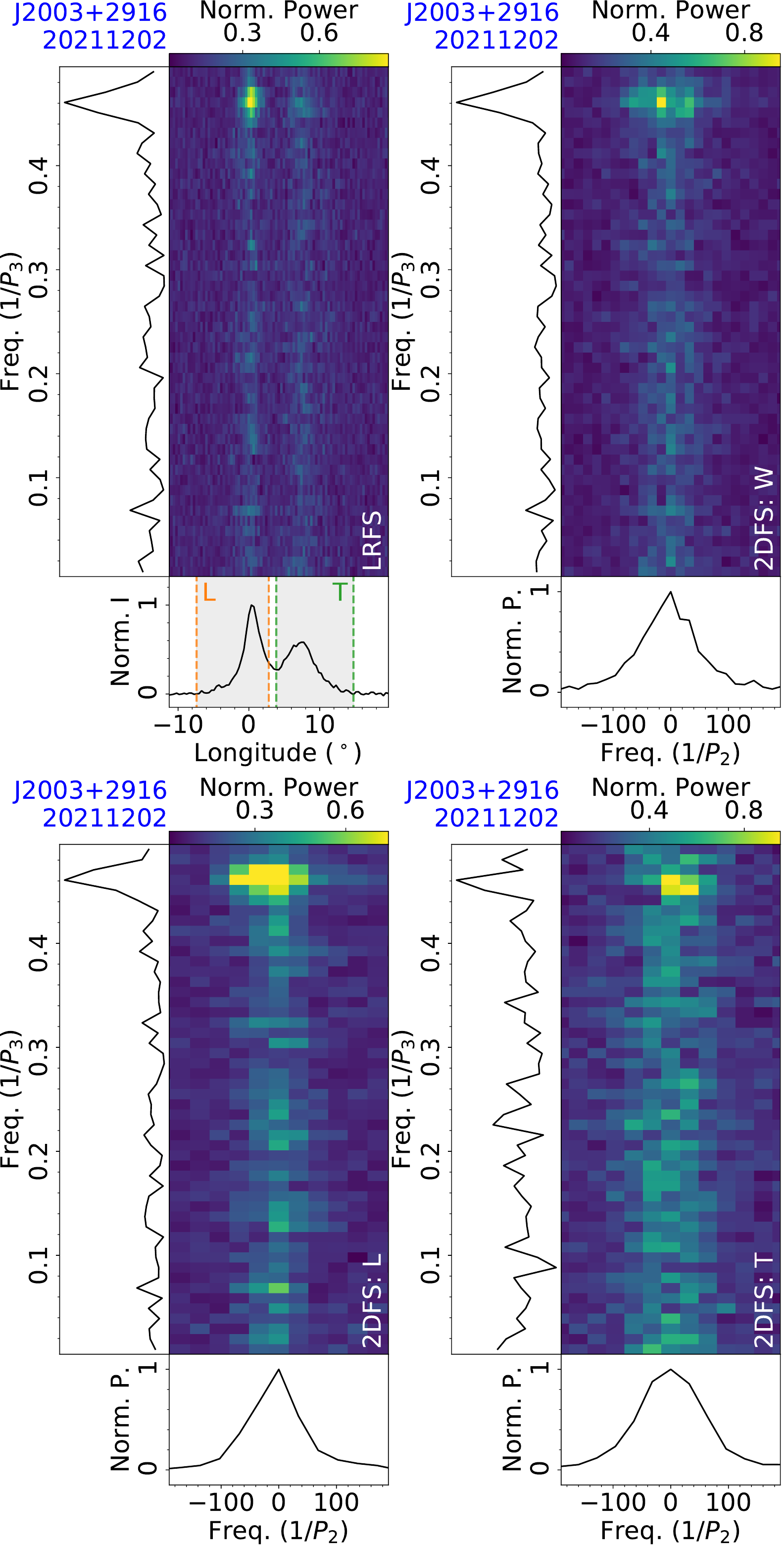}
\figcaption{Fluctuation analysis of PSR J2003+2916 from the FAST observation on 20211202, with LRFS (top-left), and 2DFS for the on-pulse region (top-right), leading part (bottom-left) and trailing part (bottom-right) of a mean pulse profile.
\label{subfig:fluctu:J2003+2916}}
\end{figure}

\begin{figure}[htpb]
\centering
\includegraphics[width=0.22\textwidth, angle=0]{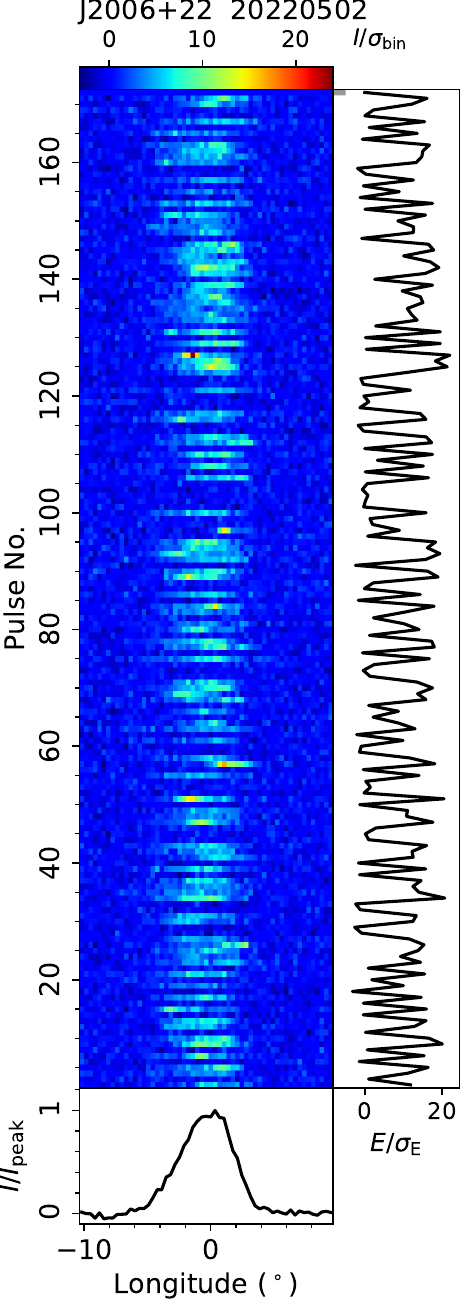}
\figcaption{Single pulse sequence of PSR J2006+22 from the FAST observation on 20220502.
\label{subfig:TP:J2006+22}}
\end{figure}

\begin{figure}[htpb]
\centering
\includegraphics[width=0.39\textwidth, angle=0]{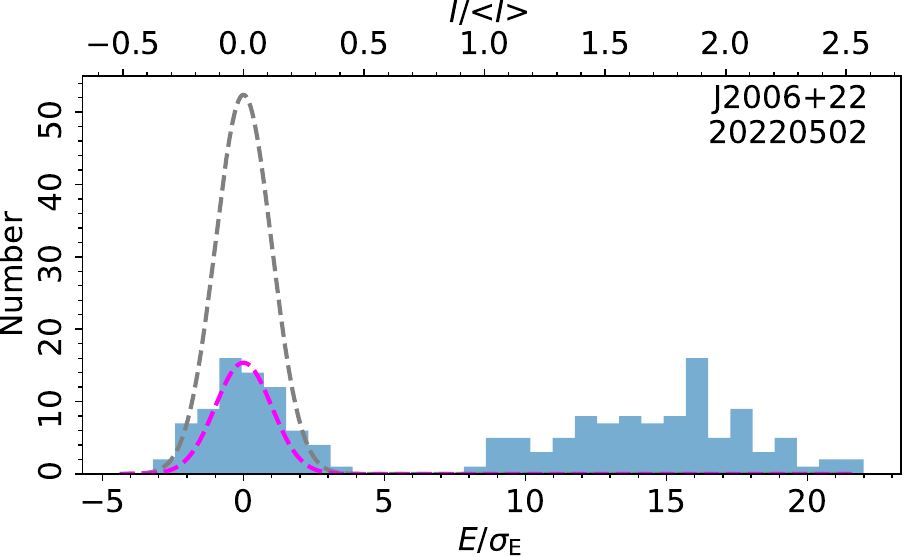}
\figcaption{On-pulse energy histogram of single pulses of PSR J2006+22 from the FAST observation on 20220502.
\label{subfig:Hist:J2006+22}}
\end{figure}

\subsection{J2002+4653}
\label{subsec:J2002+4653}

PSR J2002+4653 was discovered by FAST in the Commensal Radio Astronomy FAST Survey (CRAFTS) (http://groups.bao.ac.cn/ism/CRAFTS/).

This pulsar was observed by FAST on 20210707 for 4 minutes and 20240221 for 8 minutes. From the longer data, a rotation period $P=0.2483$~s and a dispersion measure $D\!M=144.8~{\rm cm^{-3}\,pc}$ were derived. Single pulse sequences of the observation on 20240221 in Fig.~\ref{subfig:TP:J2002+4653} show the nulling phenomenon. The nulling fraction of this observation is estimated to be 47.5$\pm$3.4\%.

\begin{figure}[htpb]
\centering
\includegraphics[width=0.22\textwidth, angle=0]{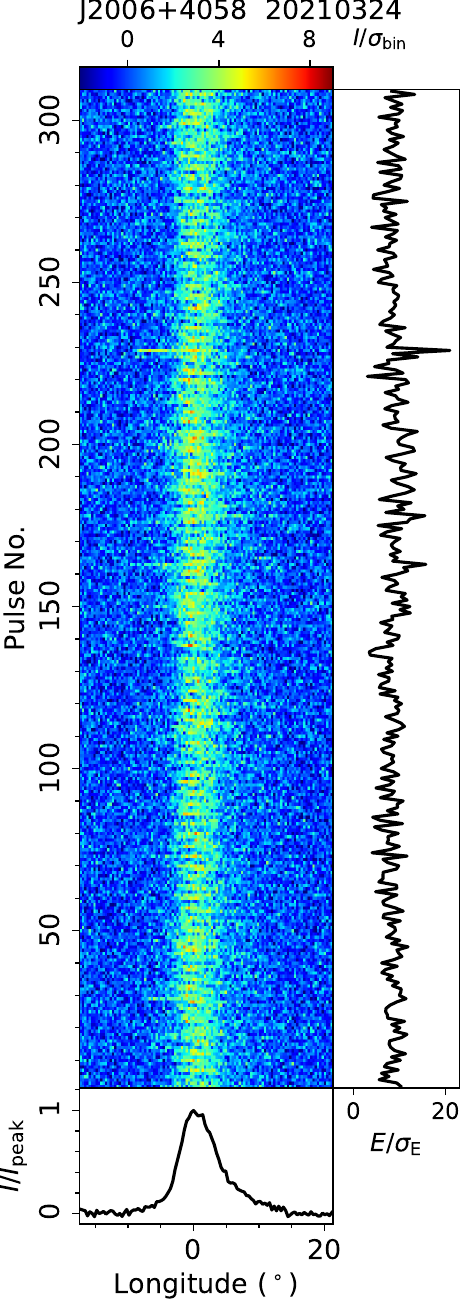}
\includegraphics[width=0.22\textwidth, angle=0]{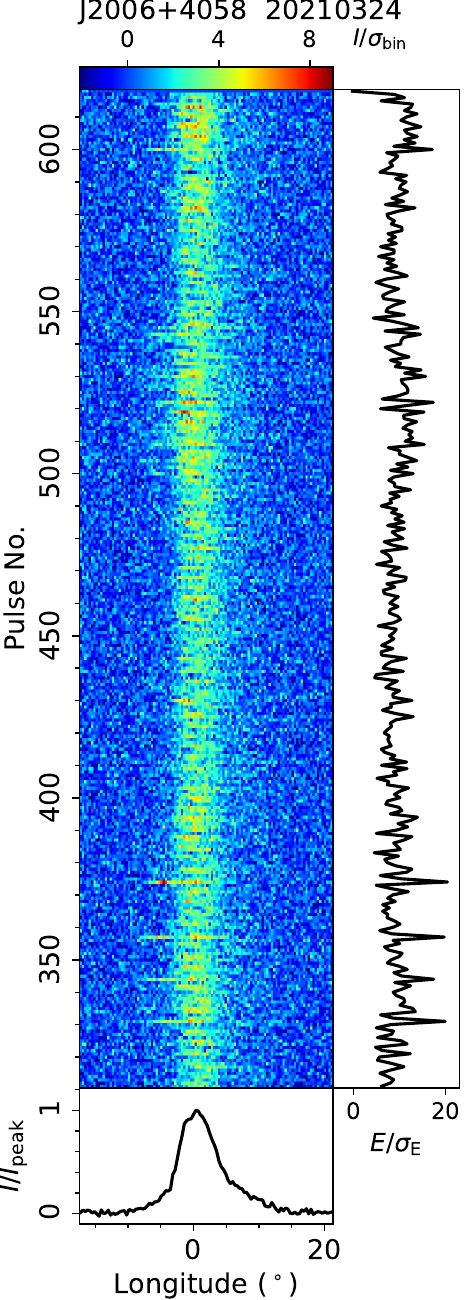}
\figcaption{Single pulse sequences of PSR J2006+4058 from the FAST observation on 20210324.
\label{subfig:TP:J2006+4058}}
\end{figure}

\begin{figure}[htpb]
\centering
\includegraphics[width=0.22\textwidth, angle=0]{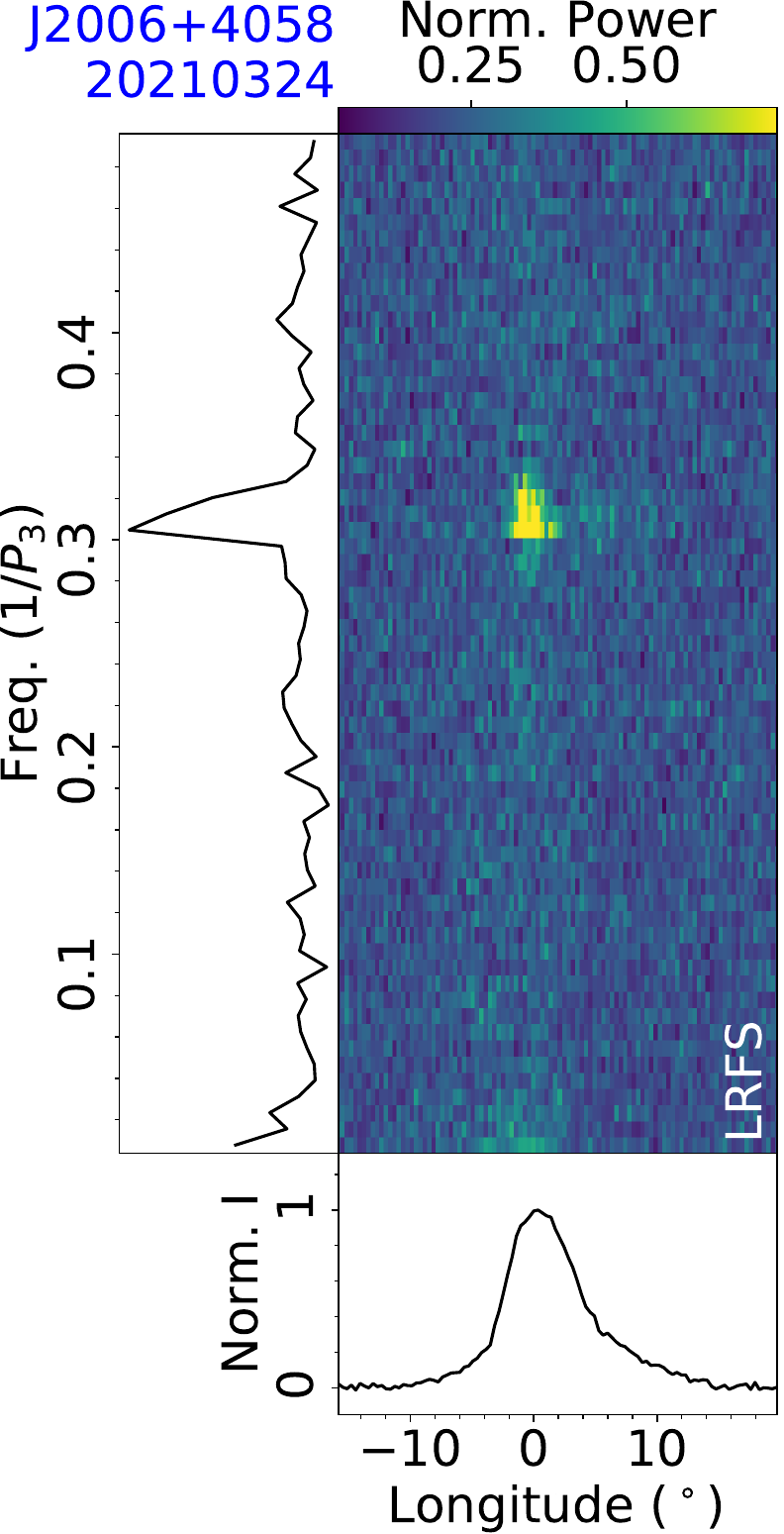}
\includegraphics[width=0.22\textwidth, angle=0]{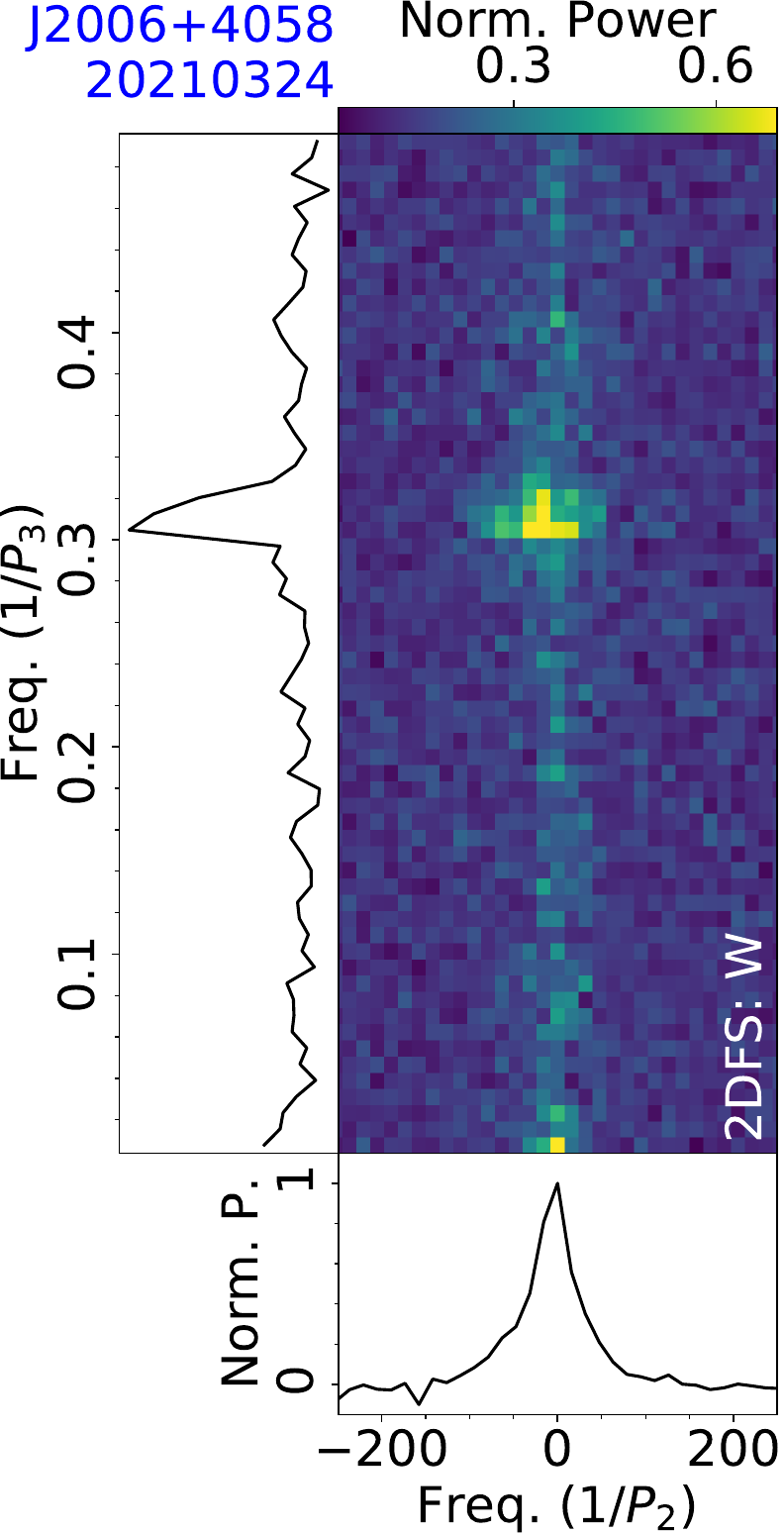}
\figcaption{Fluctuation analysis of PSR J2006+4058 for the observation on 20210324, with LRFS and 2DFS for the on-pulse region of a mean pulse profile.
\label{subfig:fluctu:J2006+4058}}
\end{figure}

\begin{figure}[htpb]
\centering
\includegraphics[width=0.22\textwidth, angle=0]{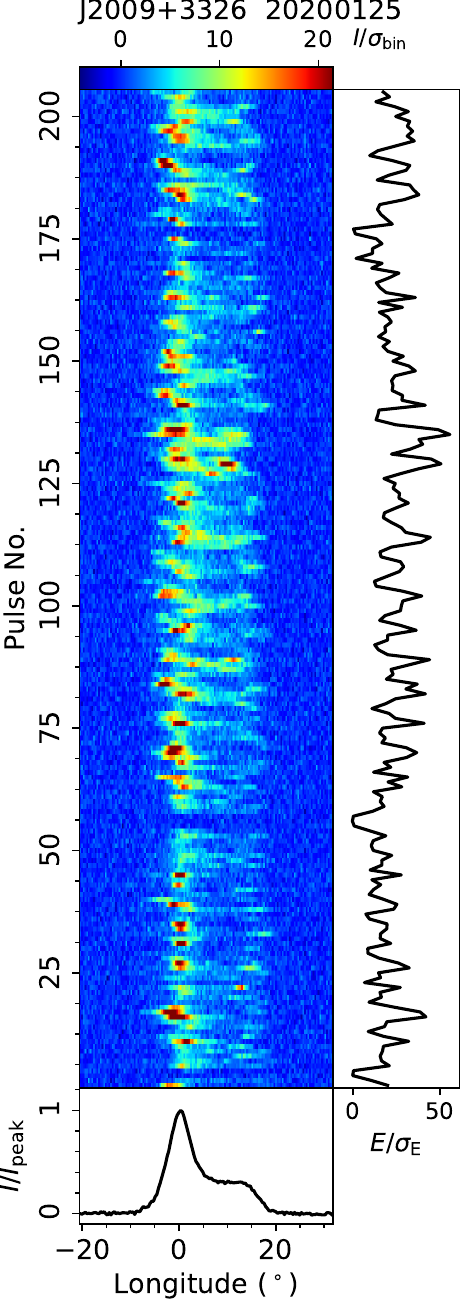}
\figcaption{Single pulse sequence of PSR J2009+3326 from the FAST observation on 20200125.
\label{subfig:TP:J2009+3326}}
\end{figure}

\begin{figure}[htpb]
\centering
\includegraphics[width=0.44\textwidth, angle=0]{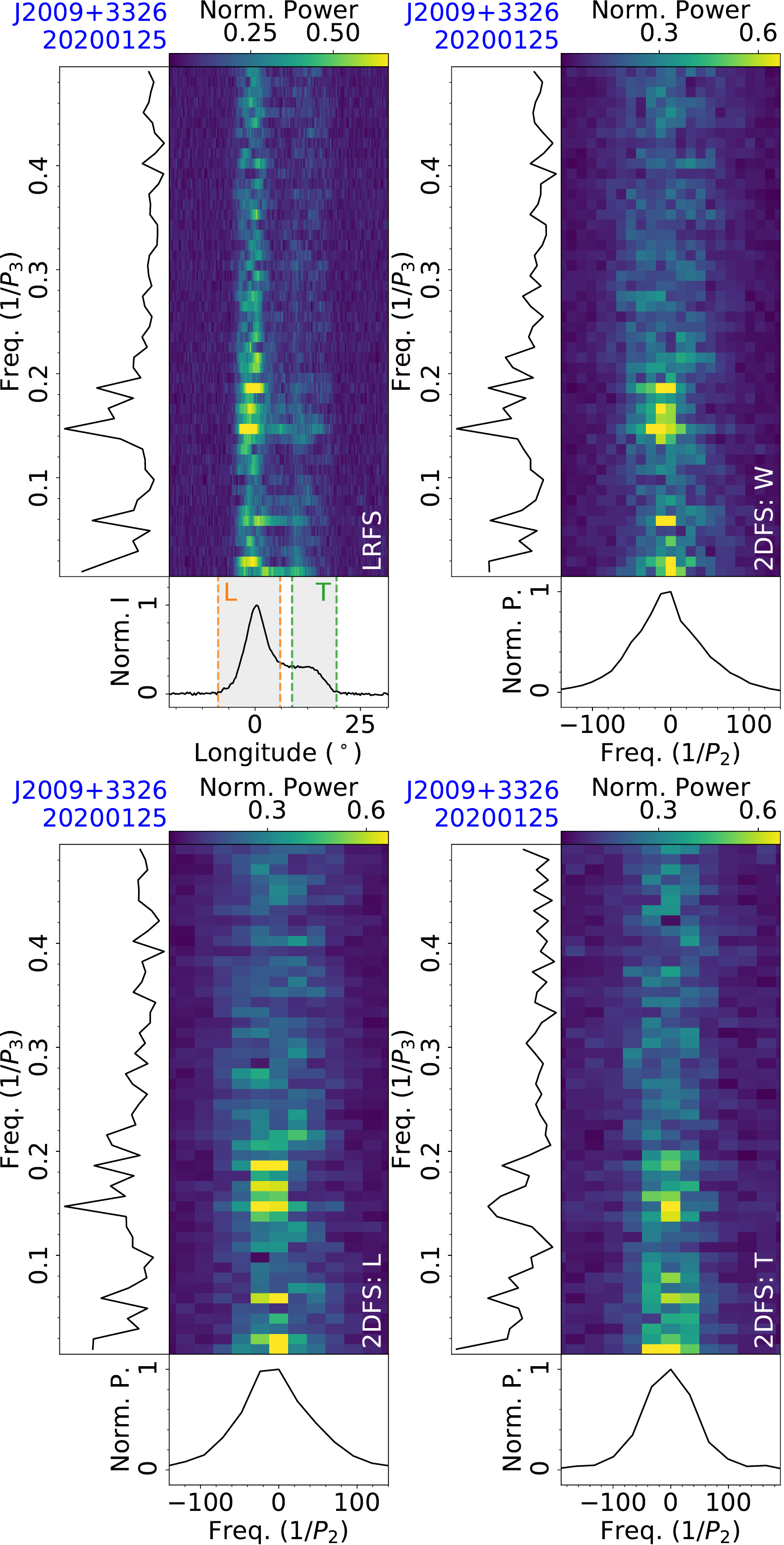}
\figcaption{Fluctuation analysis of PSR J2009+3326 from the FAST observation on 20200125, with LRFS (top-left), and 2DFS for the on-pulse region (top-right), leading part (bottom-left) and trailing part (bottom-right) of a mean pulse profile.
\label{subfig:fluctu:J2009+3326}}
\end{figure}

\subsection{J2003+2916}
\label{subsec:J2003+2916}

PSR J2003+2916 was discovered in the Pulsar Arecibo L-band Feed Array (PALFA) survey \citep{Parent2022}. 

This pulsar was observed by FAST on 20211202 for 5 minutes, deriving a rotation period $P=1.0100$~s and a dispersion measure $D\!M=208.0~{\rm cm^{-3}\,pc}$. 
Single pulse sequences in Fig.~\ref{subfig:TP:J2003+2916} display a subpulse modulation with a periodicity of about 2 periods, and two single pulses (Nos. 173 and 174) have energies less than 3 $\sigma_{\rm E}$. 
Fluctuation spectra are shown in Fig.~\ref{subfig:fluctu:J2003+2916}. 
In 2DFS of the leading profile part, the negative drift feature has a centroid at $1/P_3=0.462\pm0.001$ and $1/P_2=-19\pm3$, corresponding to $P_3=2.165\pm0.004$ periods and $P_2=-18\pm3^\circ$. The trailing profile part exhibits a positive drift feature in 2DFS, and the centroid is characterized by $1/P_3=0.455\pm0.002$ and $1/P_2=16\pm5$, yielding $P_3=2.20\pm0.01$ periods and $P_2=23\pm8^\circ$.

\begin{figure}[htpb]
\centering
\includegraphics[width=0.44\textwidth, angle=0]{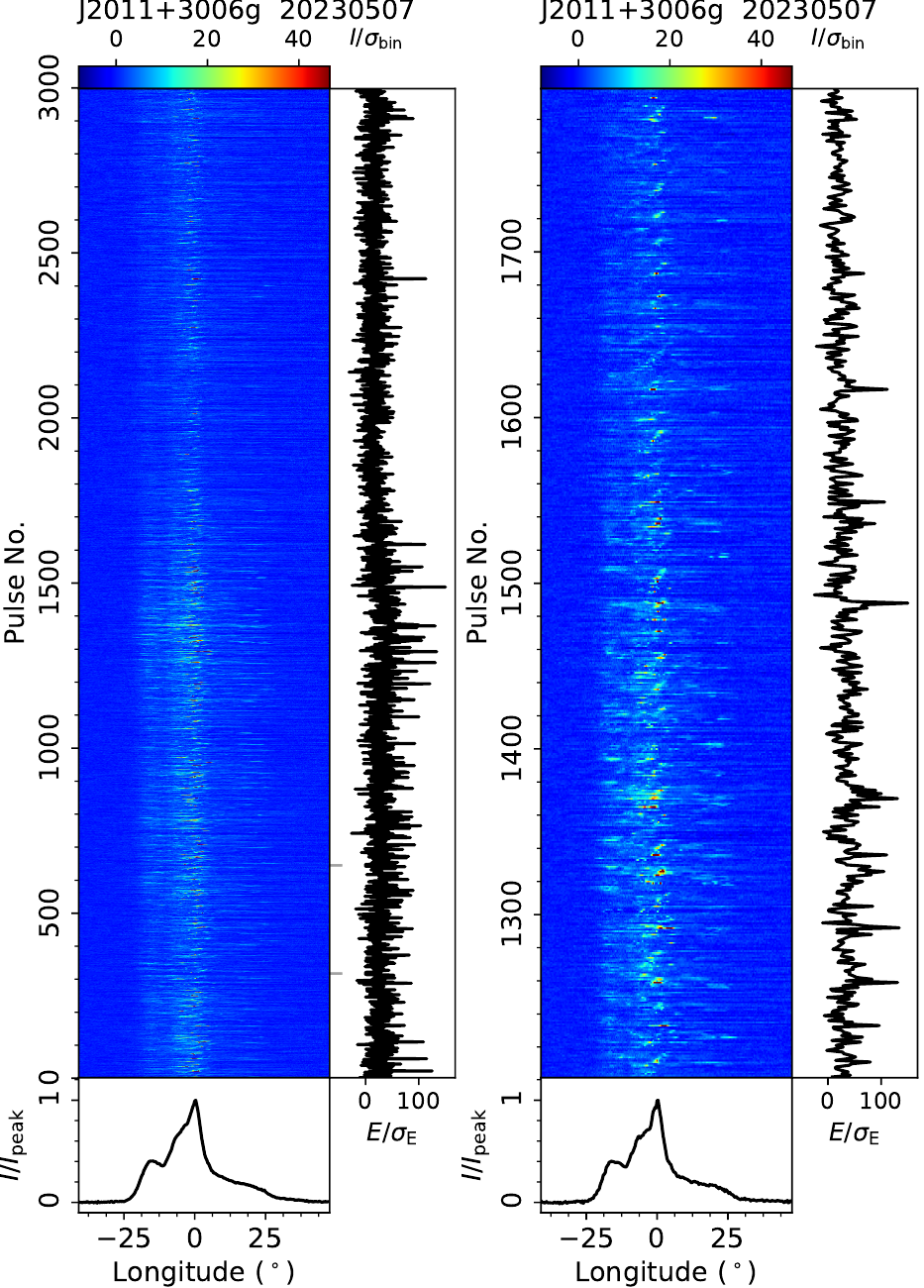}
\figcaption{Single pulse sequence of PSR J2011+3006g from the FAST observation on 20230507, and a zoomed-in view of pulses No. 1200-1800. 
\label{subfig:TP:J2011+3006g}}
\end{figure}

\begin{figure}[htpb]
\centering
\includegraphics[width=0.44\textwidth, angle=0]{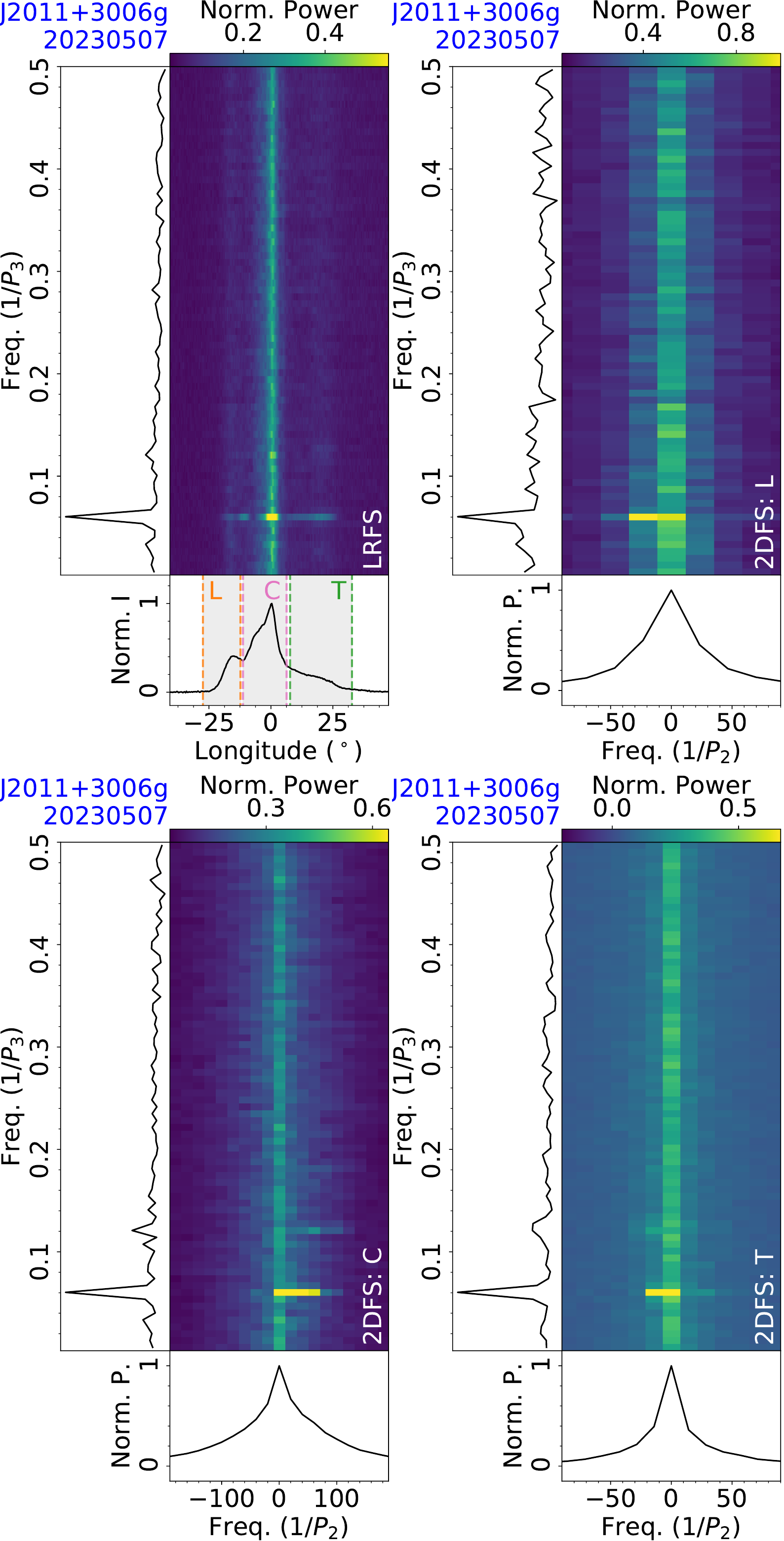}
\figcaption{Fluctuation analysis of PSR J2011+3006g from the FAST observation on 20230507, with LRFS (top-left), and 2DFS for the leading (top-right), central (bottom-left) and trailing parts (bottom-right) of the mean pulse profile.
\label{subfig:fluctu:J2011+3006g}}
\end{figure}

\begin{figure}[htpb]
\centering
\includegraphics[width=0.22\textwidth, angle=0]{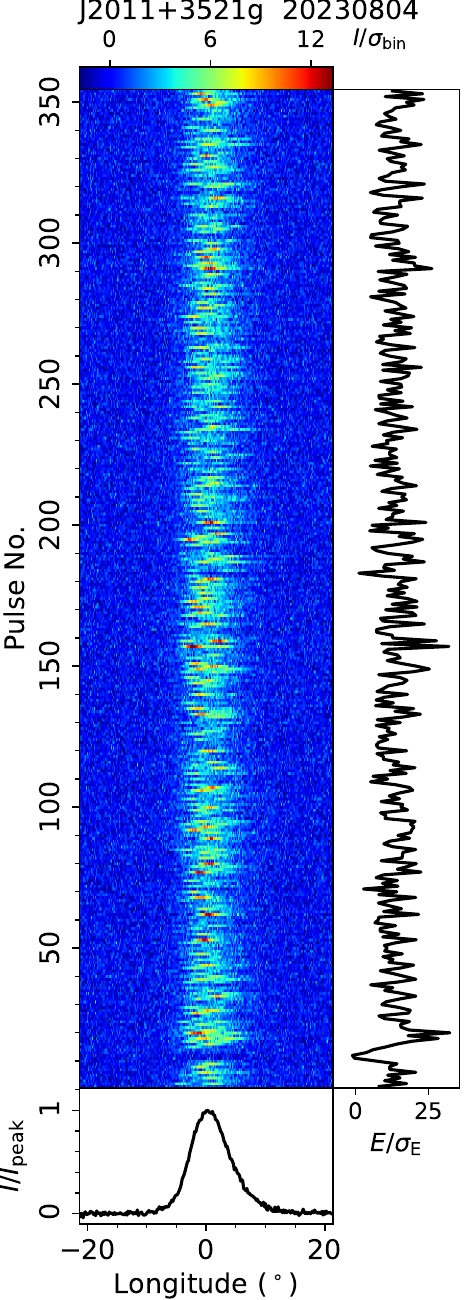}
\figcaption{Single pulse sequence of PSR J2011+3521g from the FAST observation on 20230804.
\label{subfig:TP:J2011+3521g}}
\end{figure}

\begin{figure}[htpb]
\centering
\includegraphics[width=0.39\textwidth, angle=0]{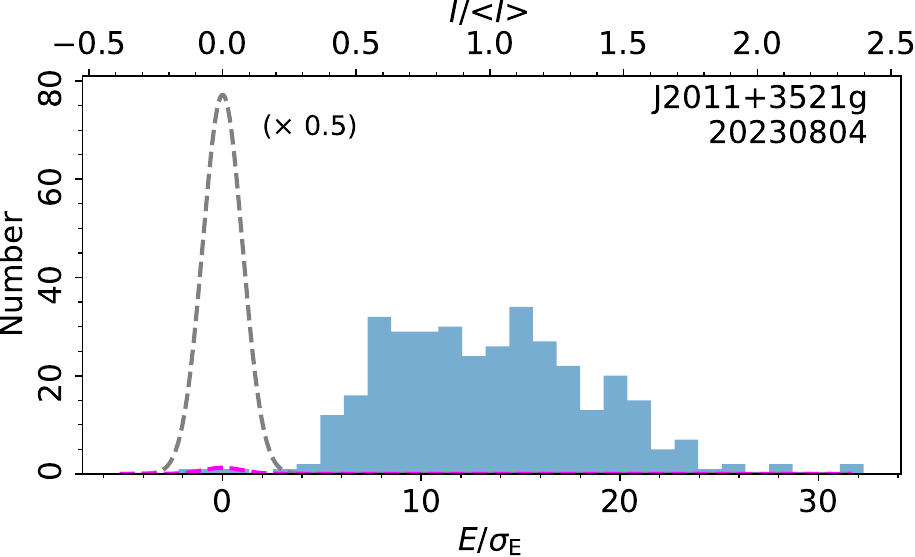}
\figcaption{On-pulse energy histogram of single pulses of PSR J2011+3521g from the FAST observation on 20230804.
\label{subfig:Hist:J2011+3521g}}
\end{figure}

\begin{figure}[htpb]
\centering
\includegraphics[width=0.22\textwidth, angle=0]{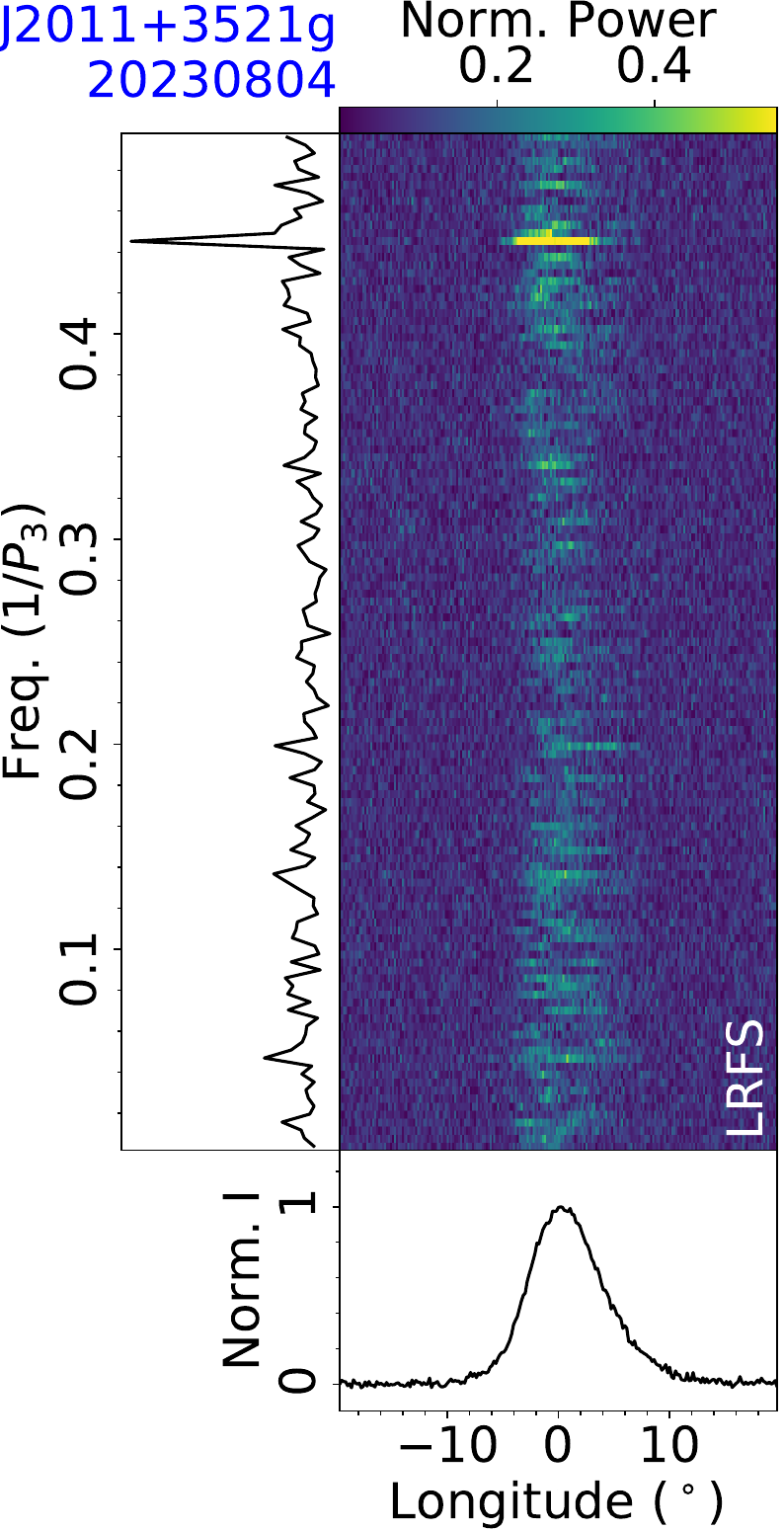}
\includegraphics[width=0.22\textwidth, angle=0]{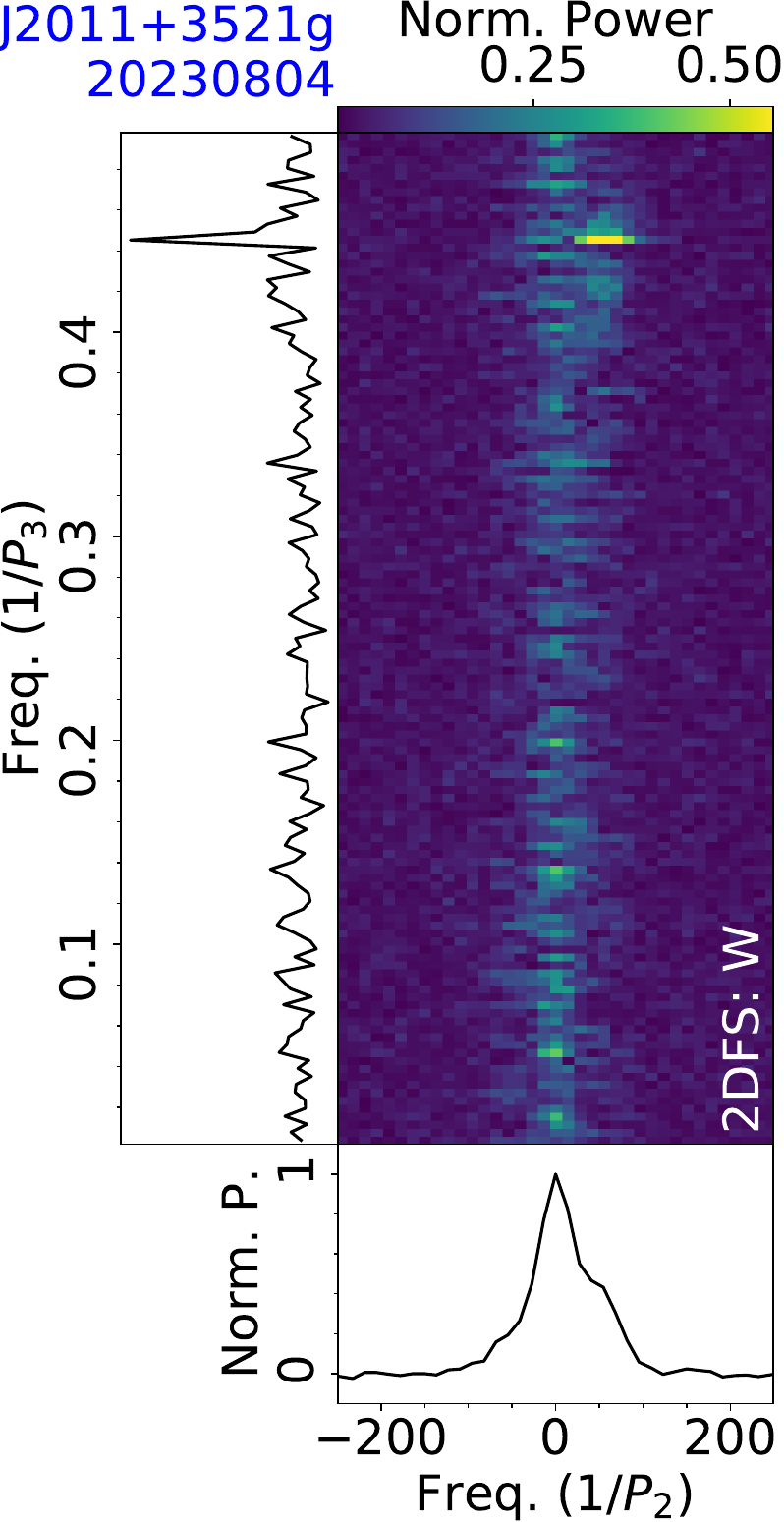}
\figcaption{Fluctuation analysis of PSR J2011+3521g for the observation on 20230804, with LRFS and 2DFS for the on-pulse region of a mean pulse profile.
\label{subfig:fluctu:J2011+3521g}}
\end{figure}

\subsection{J2006+22}
\label{subsec:J2006+22}

PSR J2006+22 was discovered in the LOFAR Tied-Array All-Sky Survey (LOTAAS) \citep{Sanidas2019}. 

This pulsar was observed by FAST on 20220502 for 5 minutes, yielding a rotation period $P=1.7417$~s and a dispersion measure $D\!M=130.3~{\rm cm^{-3}\,pc}$. The single pulse sequence (Fig.~\ref{subfig:TP:J2006+22}) and the on-pulse integral energy histogram (Fig.~\ref{subfig:Hist:J2006+22}) illustrate the nulling phenomenon. The nulling fraction of this observation is estimated to be 29.3$\pm$3.6\%.

\begin{figure}[htpb]
\centering
\includegraphics[width=0.22\textwidth, angle=0]{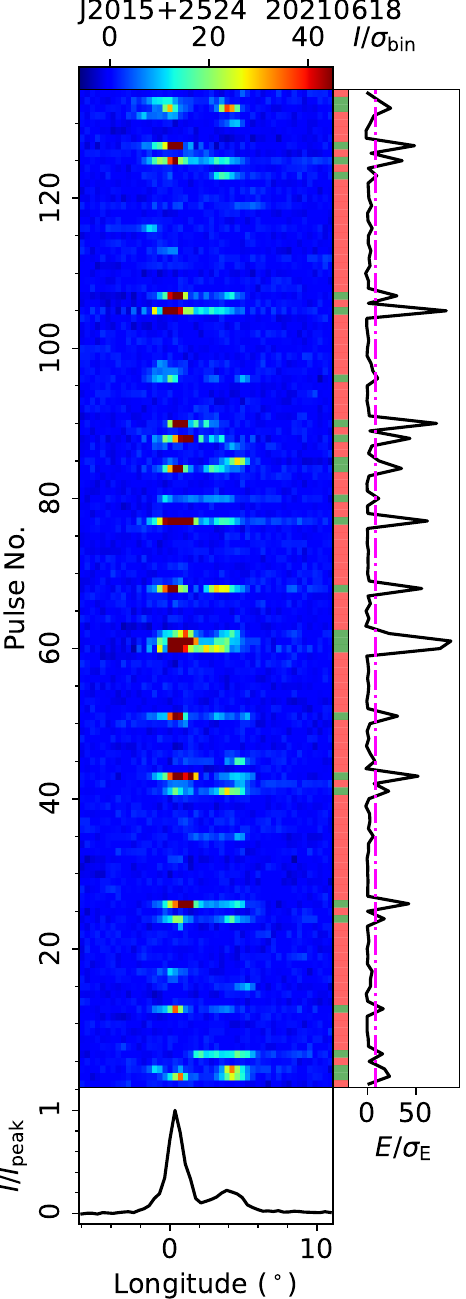}
\figcaption{Single pulse sequence of PSR J2015+2524 from the FAST observation on 20210618. 
The green and red bars represent bright or weak emission modes. In the right subpanel, the on-pulse energy variation is plotted against period, with dashed lines for thresholds to distinguish two states.
\label{subfig:TP:J2015+2524}}
\end{figure}

\begin{figure}[htpb]
\centering
\includegraphics[width=0.39\textwidth, angle=0]{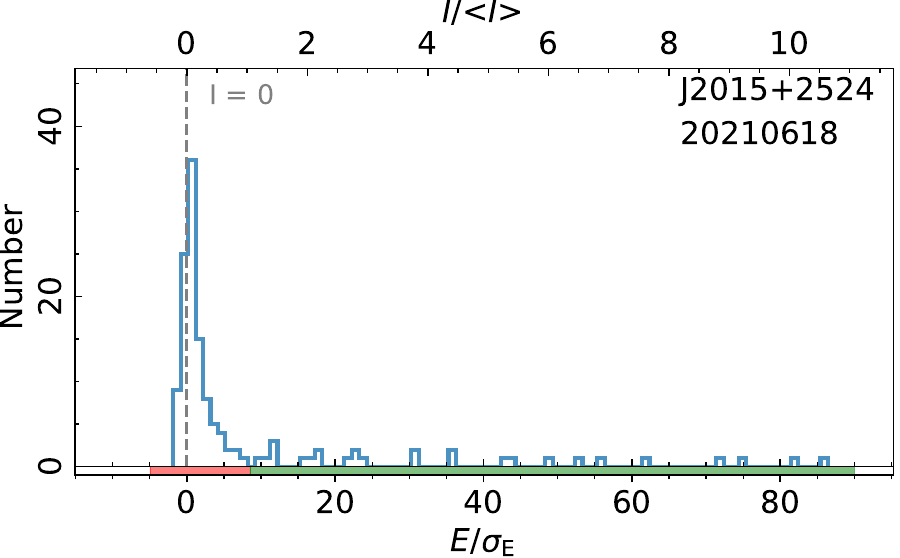}
\figcaption{On-pulse energy histogram of single pulses of PSR J2015+2524 from the FAST observation on 20210618. The red and green bars indicate the weak and bright emission modes.
\label{subfig:Hist:J2015+2524}}
\end{figure}

\subsection{J2006+4058}
\label{subsec:J2006+4058}

PSR J2006+4058 was discovered by the Five-hundred-metre Aperture Spherical radio-Telescope (FAST) during its commissioning \citep{Cruces2021}. 

The pulsar was also observed by FAST on 20210324 for 5 minutes, deriving a rotation period $P=0.4997$~s and a dispersion measure $D\!M=257.8~{\rm cm^{-3}\,pc}$. 
Single pulse sequences are shown in Fig.~\ref{subfig:TP:J2006+4058}. In the 2DFS (Fig.~\ref{subfig:fluctu:J2006+4058}), there is a negative drift feature, with the centroid frequencies estimated to be $1/P_3=0.311\pm0.001$ and $1/P_2=-21\pm2$, corresponding to $P_3=3.22\pm0.01$ periods and $P_2=-17\pm1^\circ$.



\subsection{J2009+3326}
\label{subsec:J2009+3326}

PSR J2009+3326 was discovered by \citep{Cordes2006} using the Arecibo telescope. 

This pulsar was observed by FAST on 20200125 for 5 minutes, deriving a rotation period $P=1.4384$~s and a dispersion measure $D\!M=264.8~{\rm cm^{-3}\,pc}$. The single pulse sequence in Fig.~\ref{subfig:TP:J2009+3326} displays negative subpulse drifting and temporal modulation for the leading and trailing profile parts, respectively. 
Fluctuation spectra are shown in Fig.~\ref{subfig:fluctu:J2009+3326}. For the leading profile part, the centroid of the negative drift feature in 2DFS is at $1/P_3=0.162\pm0.001$ and $1/P_2=-13\pm2$, corresponding to $P_3=6.16\pm0.04$ periods and $P_2=-28\pm5$ degrees. For the trailing profile part, the centroid frequency of the modulation feature in 2DFS is $1/P_3=0.162\pm0.002$, yielding $P_3=6.2\pm0.1$ periods.

\begin{figure}[htpb]
\centering
\includegraphics[width=0.22\textwidth, angle=0]{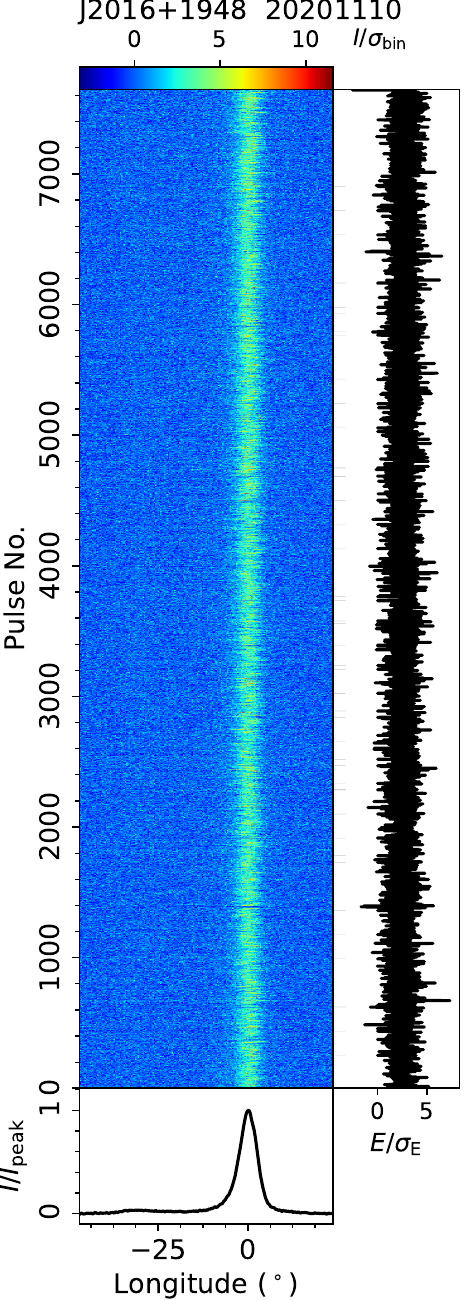}
\includegraphics[width=0.22\textwidth, angle=0]{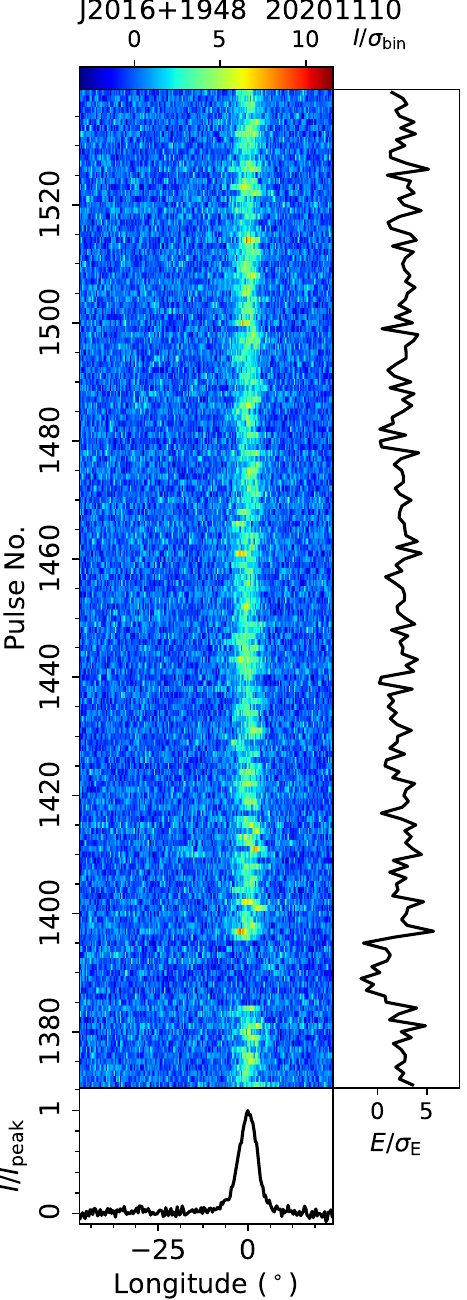}
\figcaption{Single pulse sequence of PSR J2016+1948 from the FAST observation on 20201110, and a zoomed-in view of pulses No. 1370-1539.
\label{subfig:TP:J2016+1948}}
\end{figure}

\begin{figure}[htpb]
\centering
\includegraphics[width=0.22\textwidth, angle=0]{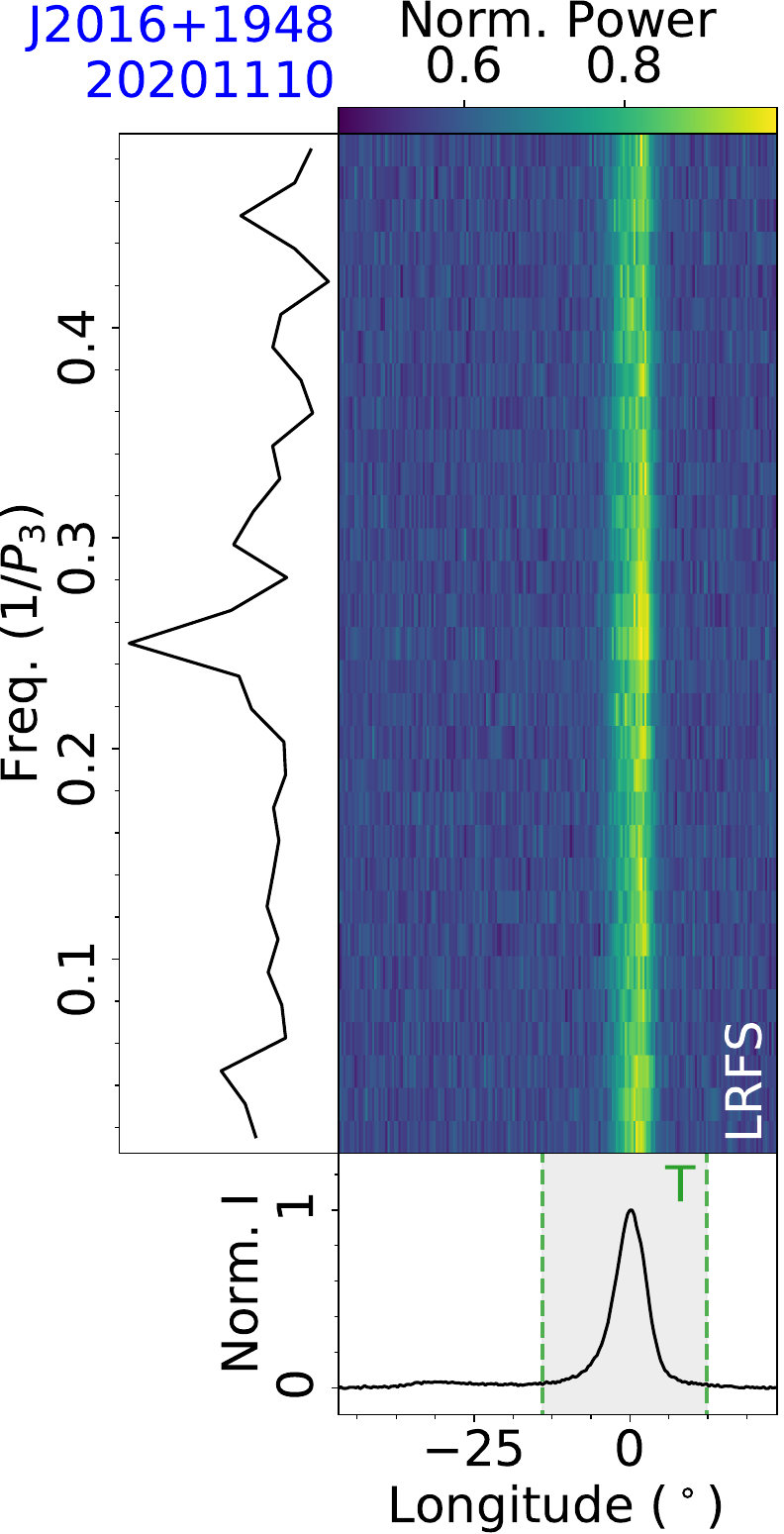}
\includegraphics[width=0.22\textwidth, angle=0]{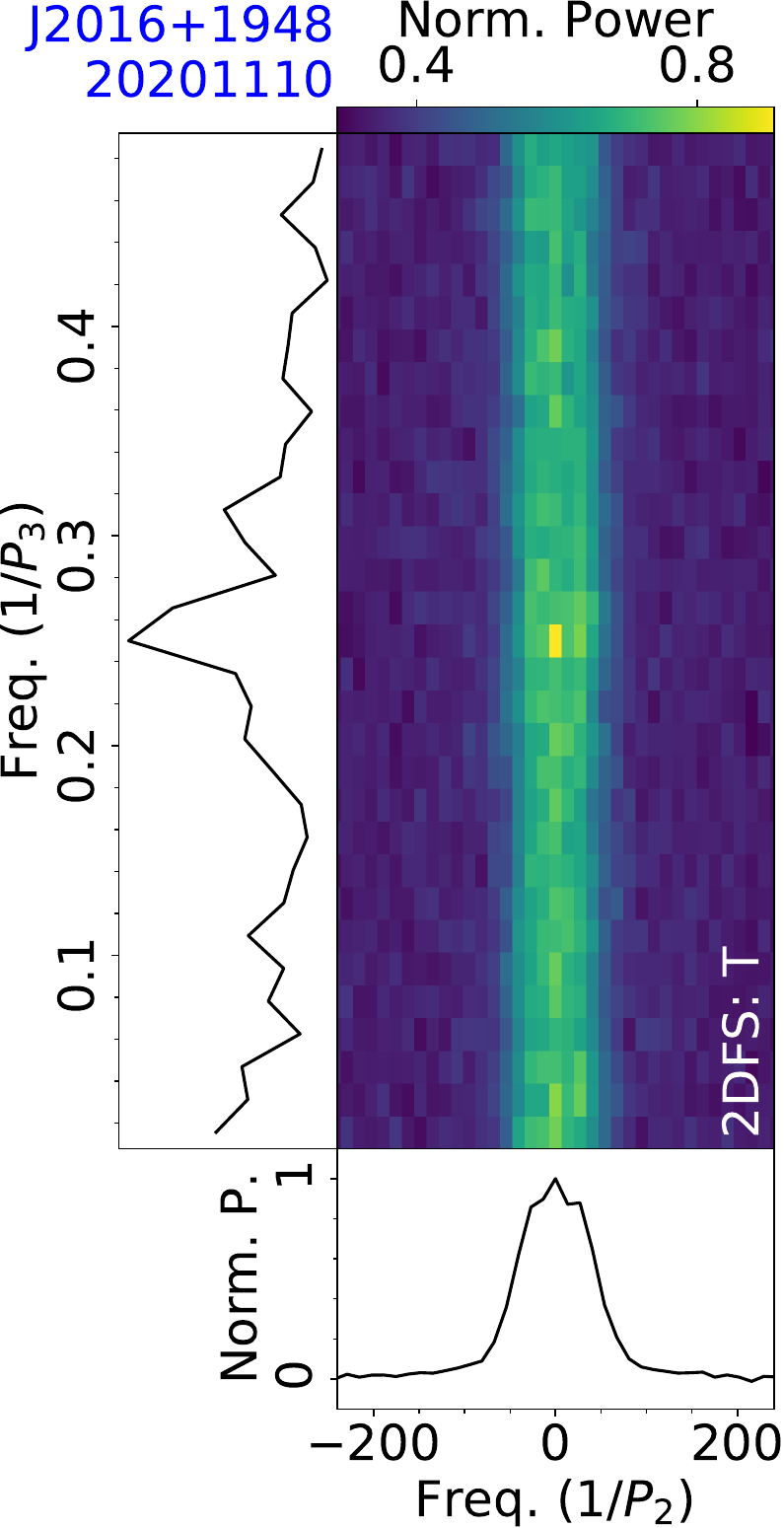}
\figcaption{Fluctuation analysis of PSR J2016+1948 for the FAST observation on 20201110, with LRFS and 2DFS for the trailing part of a mean pulse profile.
\label{subfig:fluctu:J2016+1948}}
\end{figure}

\begin{figure}[htpb]
\centering
\includegraphics[width=0.22\textwidth, angle=0]{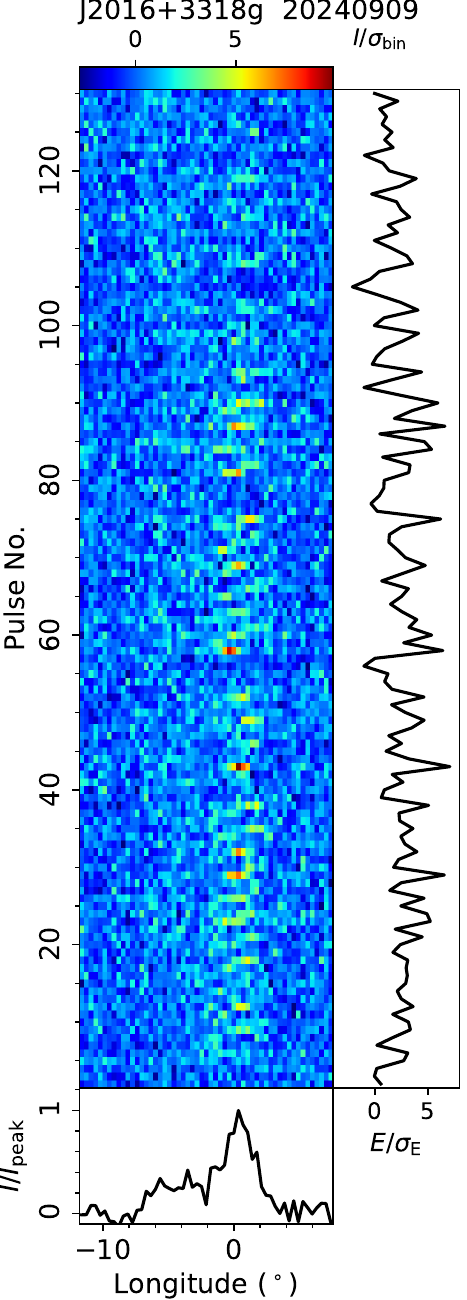}
\includegraphics[width=0.22\textwidth, angle=0]{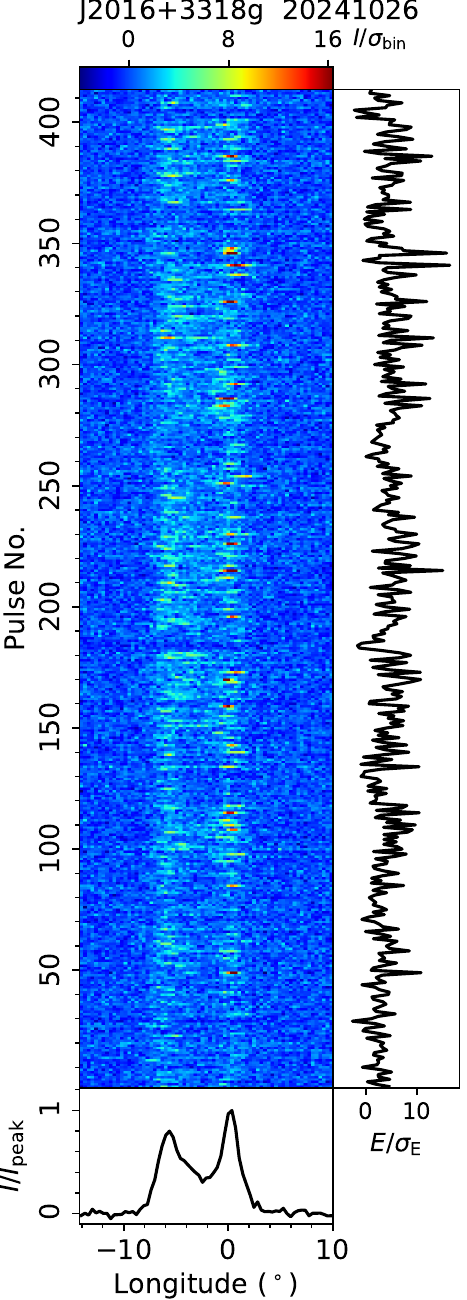}
\figcaption{Single pulse sequences of PSR J2016+3318g from FAST observations on 20240909 and 20241026.
\label{subfig:TP:J2016+3318g}}
\end{figure}

\begin{figure}[htpb]
\centering
\includegraphics[width=0.39\textwidth, angle=0]{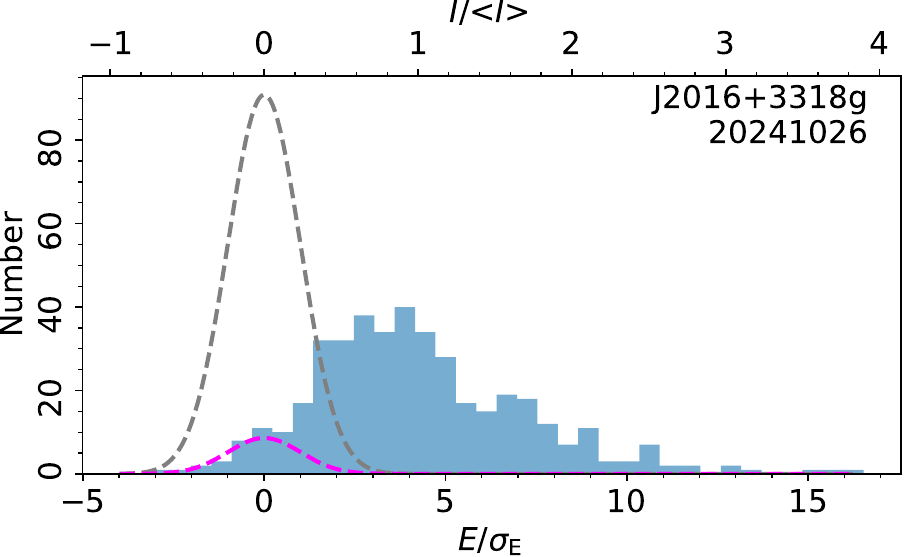}
\figcaption{On-pulse energy histogram of single pulses of PSR J2016+3318g from the FAST observation on 20241026.
\label{subfig:Hist:J2016+3318g}}
\end{figure}

\begin{figure}[htpb]
\centering
\includegraphics[width=0.22\textwidth, angle=0]{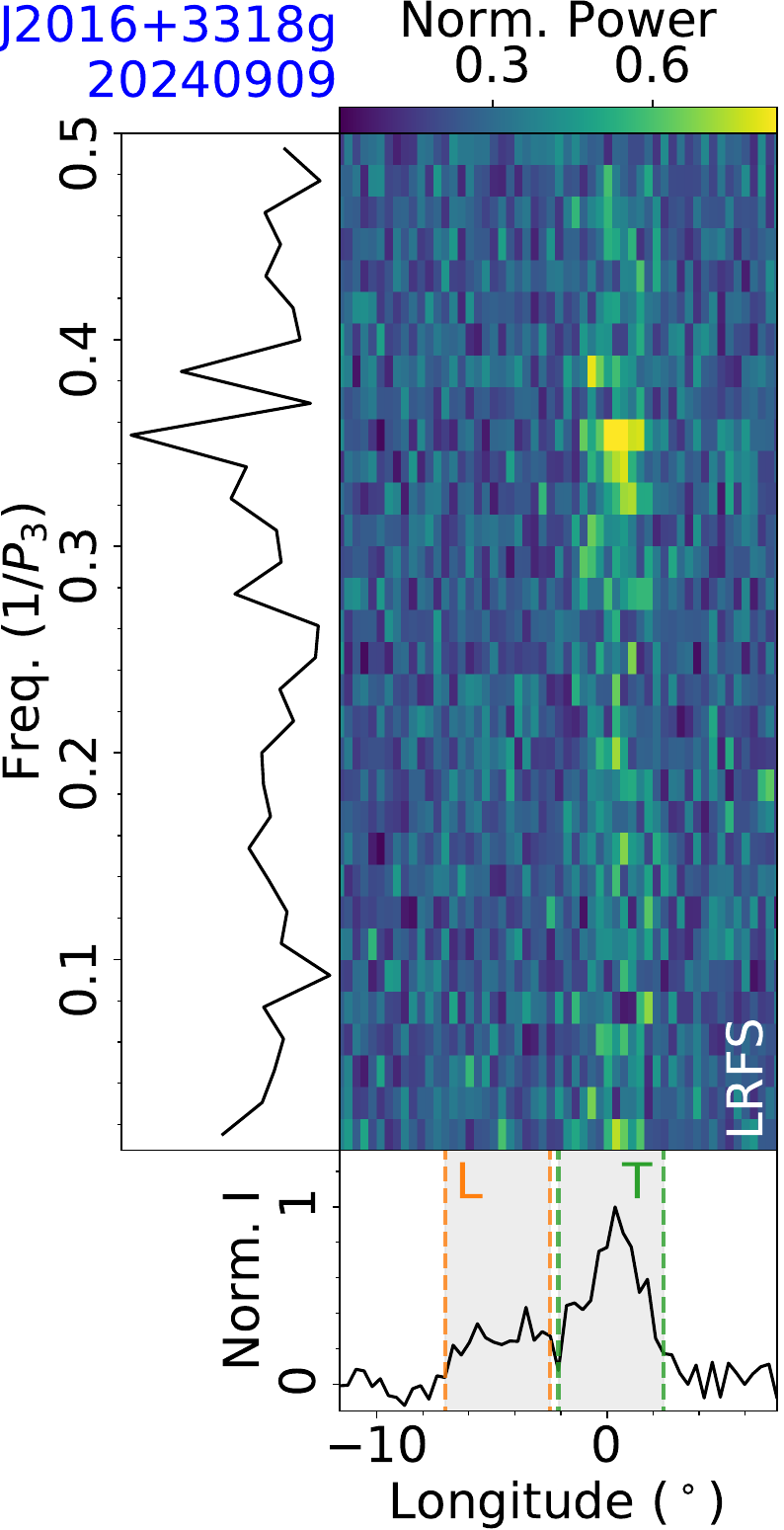}
\includegraphics[width=0.22\textwidth, angle=0]{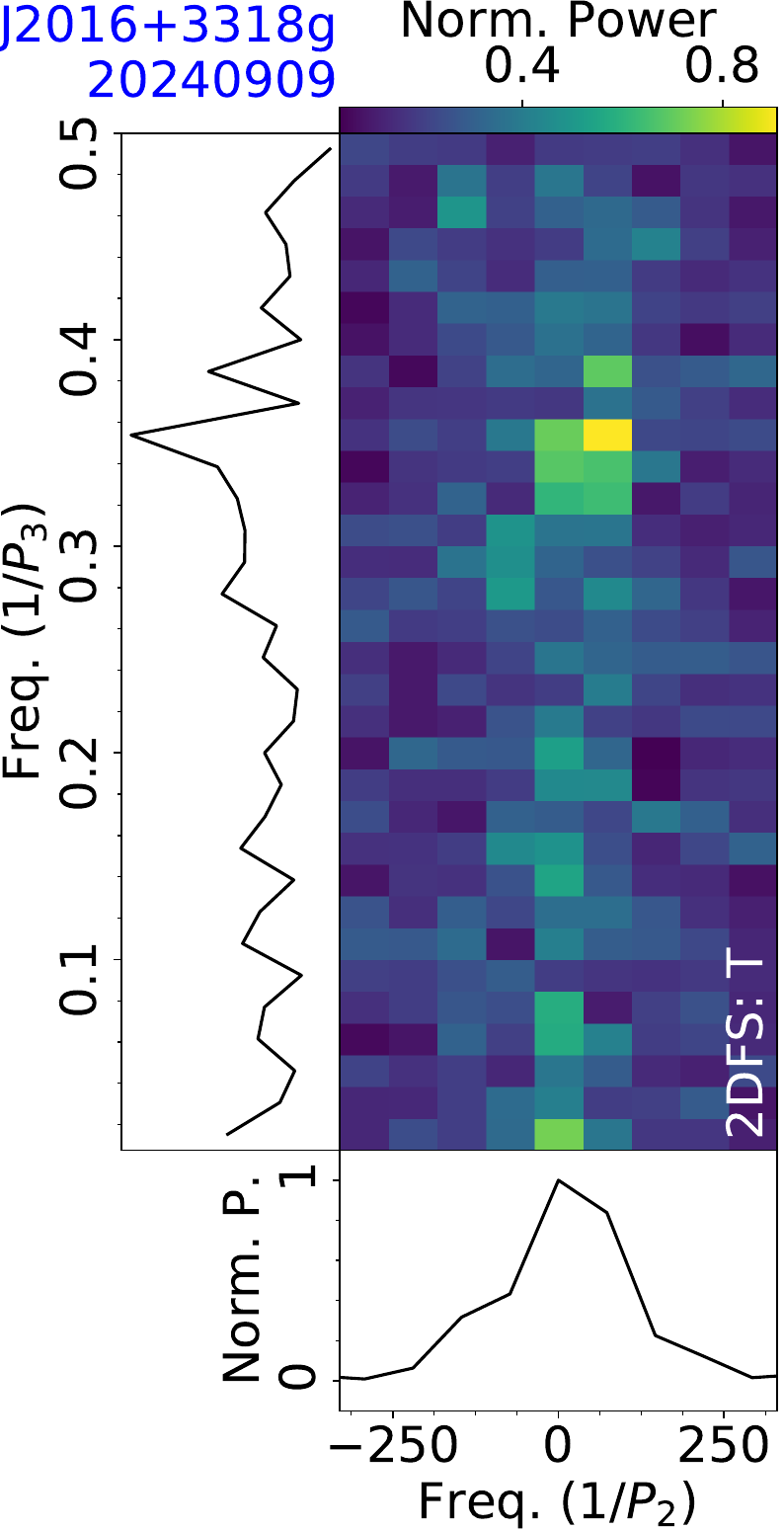}\\
\includegraphics[width=0.22\textwidth, angle=0]{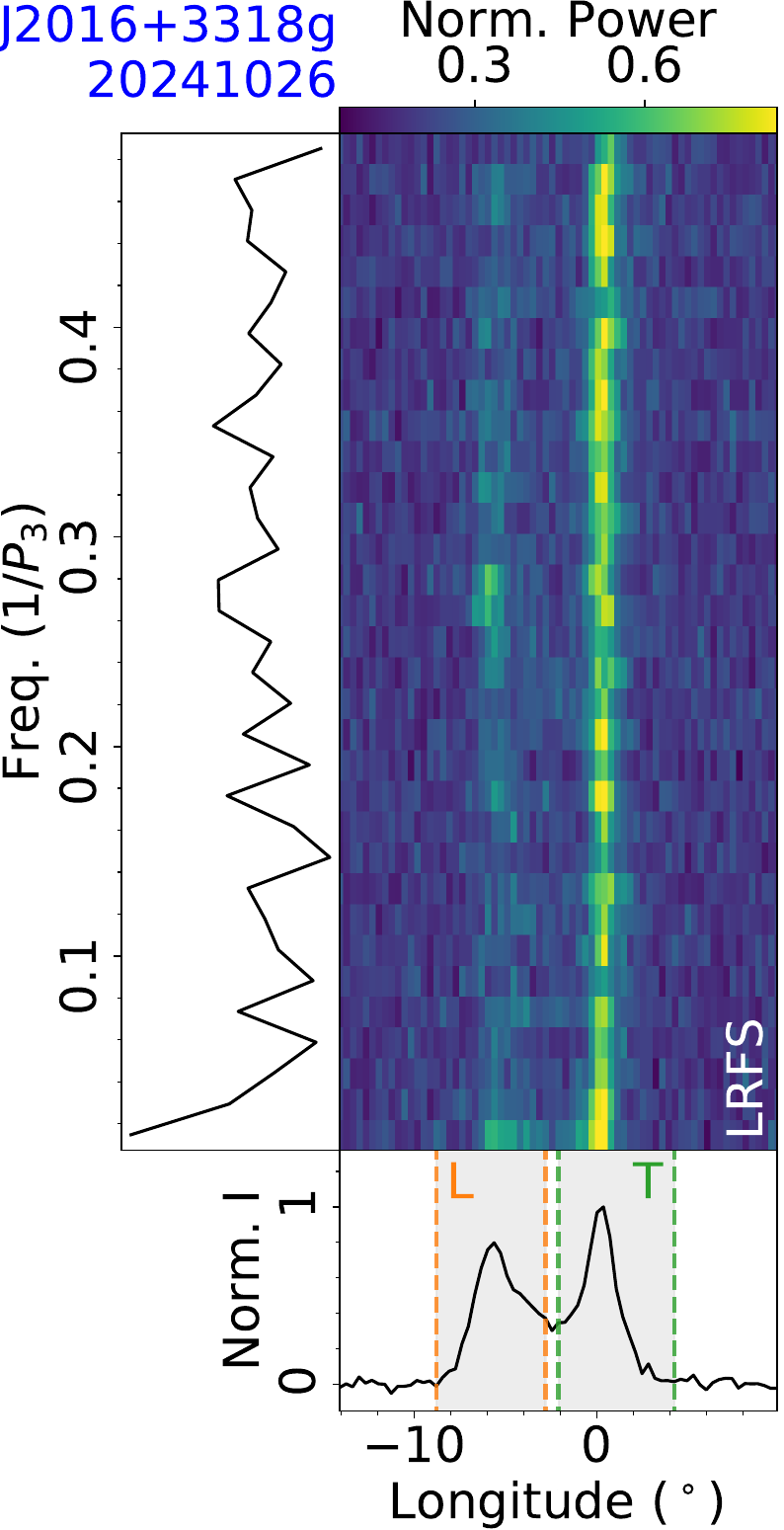}
\includegraphics[width=0.22\textwidth, angle=0]{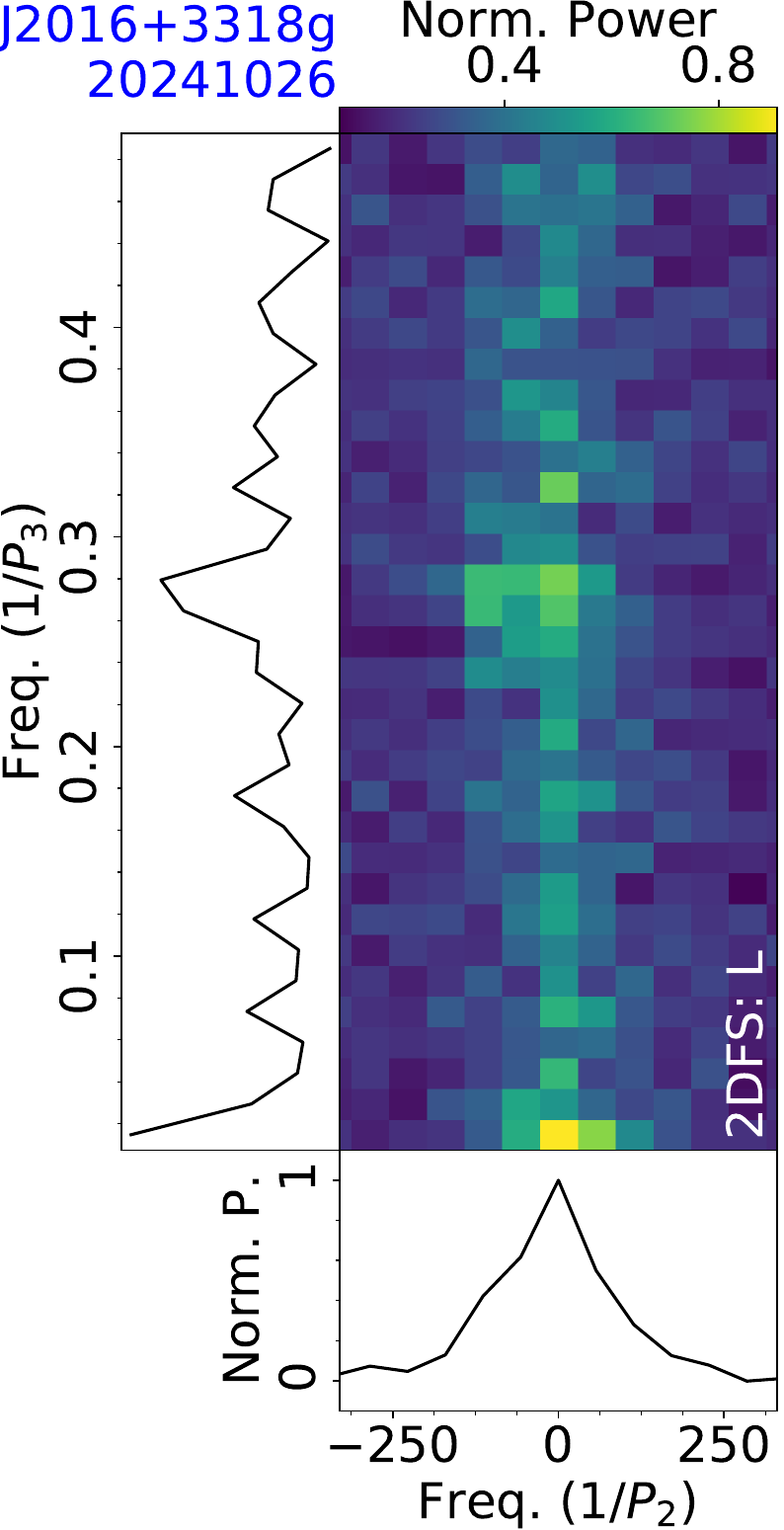}
\figcaption{Fluctuation analysis of PSR J2016+3318g for FAST observations on 20240909 and 20241026, with LRFS and 2DFS for the trailing and leading parts of mean pulse profiles.
\label{subfig:fluctu:J2016+3318g}}
\end{figure}

\subsection{J2011+3006g}
\label{subsec:J2011+3006g}

PSR J2011+3006g was discovered in the FAST GPPS survey \citep{Han2021,han2025}. 

This pulsar was observed by FAST on 20230507 for 125 minutes, deriving a rotation period $P=2.5054$~s and  a dispersion measure $D\!M=8.0~{\rm cm^{-3}\,pc}$. The single pulse sequence and a zoomed-in view of pulses No. 1200-1800 show that the leading and trailing profile parts have negative drifting, while the central part exhibits positive drifting. Fluctuation spectra are displayed in Fig.~\ref{subfig:fluctu:J2011+3006g}. In 2DFS of the leading profile part, the centroid of the negative drift feature is at $1/P_3=0.060\pm0.001$ ($P_3=16.6\pm0.4$ periods) and $1/P_2=-15\pm5$ ($P_2=-23\pm8$ degrees). The positive drift feature in 2DFS of the central profile part is at $1/P_3=0.060\pm0.001$ ($P_3=16.6\pm0.3$ periods) and $1/P_2=31\pm3$ ($P_2=12\pm1$ degrees). For the trailing part, the centroid of the negative drift feature is characterized by $1/P_3=0.060\pm0.001$ ($P_3=16.6\pm0.4$ periods) and $1/P_2=-7\pm3$ ($P_2=-55\pm25$ degrees).

\subsection{J2011+3521g}
\label{subsec:J2011+3521g}

PSR J2011+3521g was discovered in the FAST GPPS survey \citep{Han2021,han2025}. 

This pulsar was observed by FAST on 20230804 for 6 minutes, yielding a rotation period $P=0.9432$~s and a dispersion measure $D\!M=439.2~{\rm cm^{-3}\,pc}$. 
The single pulse sequence is shown in Fig.~\ref{subfig:TP:J2011+3521g}, that display short-duration nulling and a modulation of about 2 periods. There are 4 pulses whose on-pulse integral energies are less than 3 $\sigma_{\rm E}$ (Fig.~\ref{subfig:Hist:J2011+3521g}). LRFS and 2DFS are displayed in Fig.~\ref{subfig:fluctu:J2011+3521g}. 
The centroid of the mean drift feature in 2DFS is characterized by $1/P_3=0.4471\pm0.0004$ and $1/P_2=55\pm1$, corresponding to periodicities of $P_3=2.236\pm0.002$ periods and $P_2=6.6\pm0.2^\circ$.

\begin{figure}[htpb]
\centering
\includegraphics[width=0.22\textwidth, angle=0]{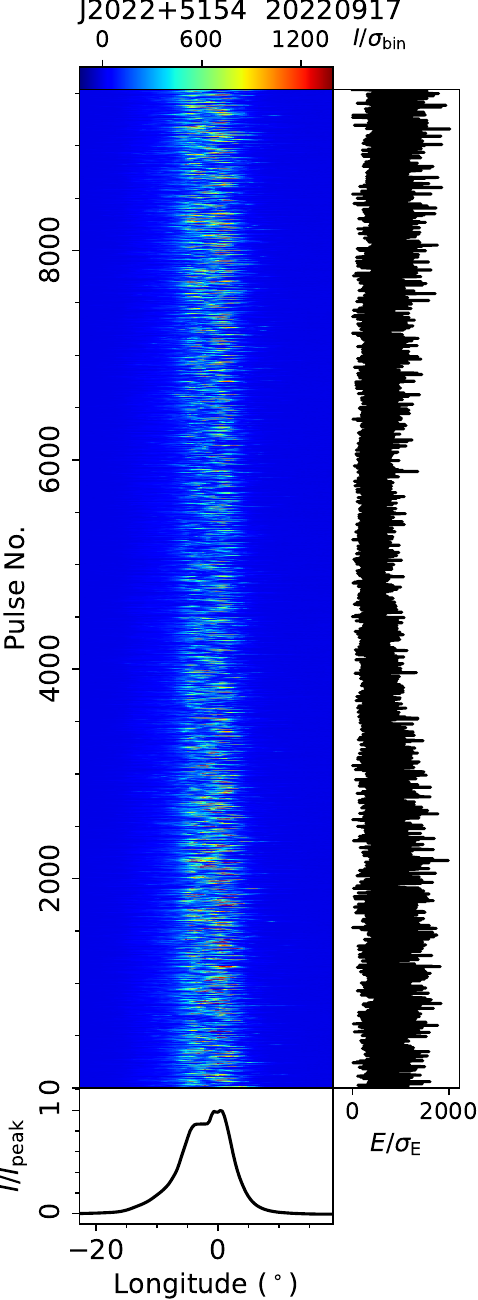}
\includegraphics[width=0.22\textwidth, angle=0]{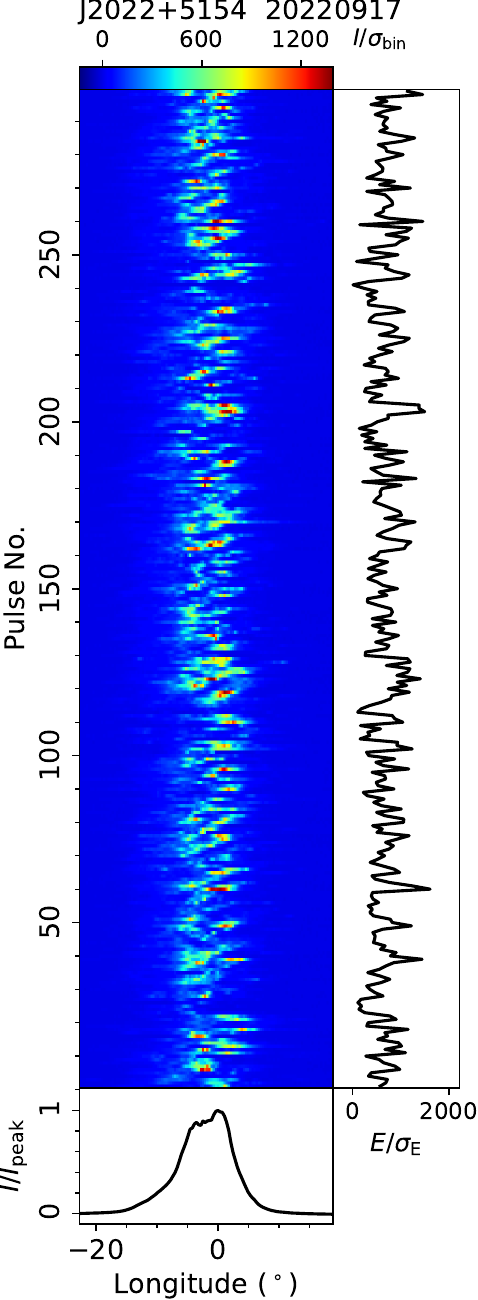}
\figcaption{Single pulse sequence of PSR J2022+5154 from FAST observations on 20220917, and a zoomed-in view of pulses No. 1-300. 
\label{subfig:TP:J2022+5154}}
\end{figure}

\begin{figure}[htpb]
\centering
\includegraphics[width=0.44\textwidth, angle=0]{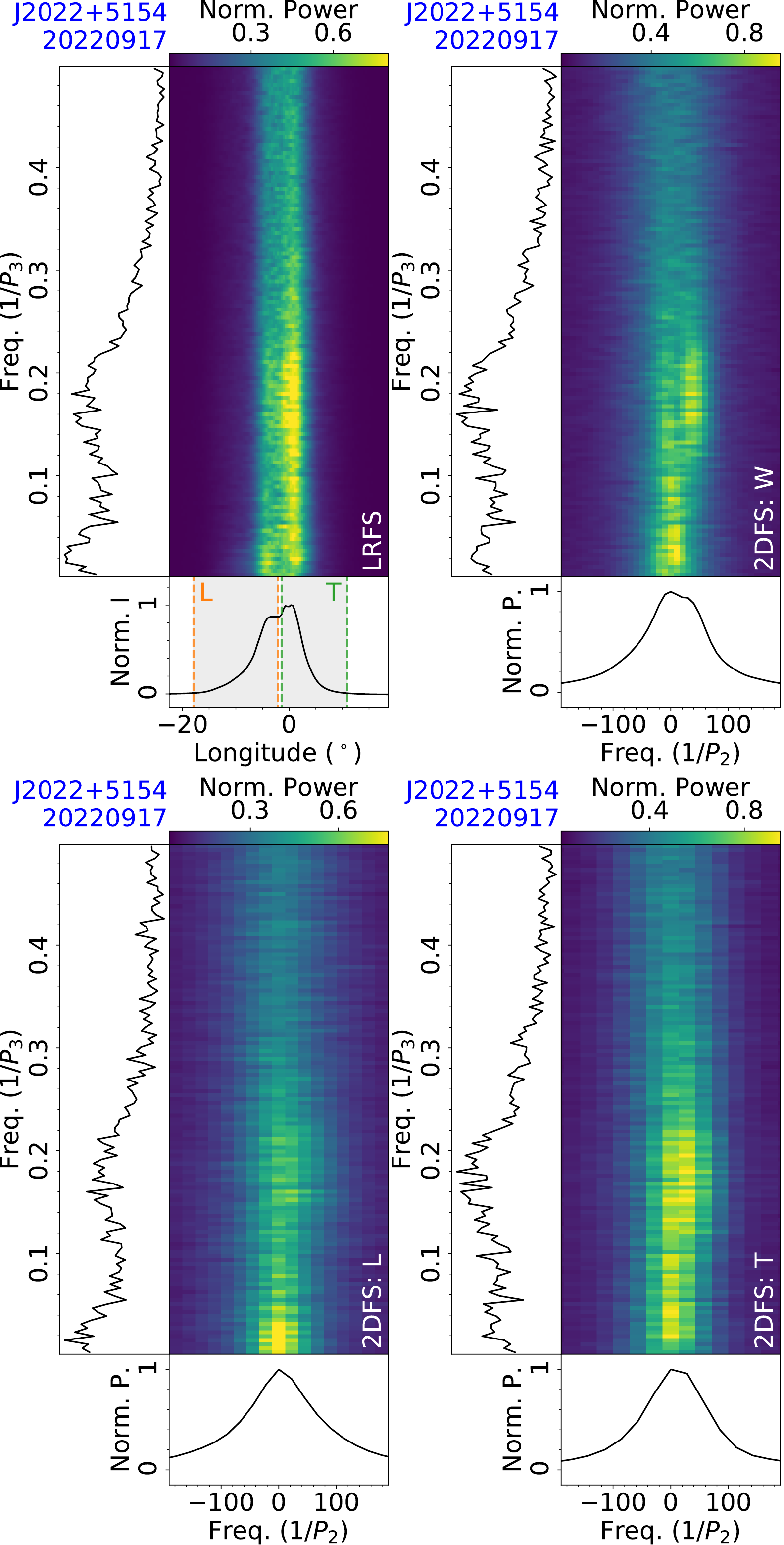}
\figcaption{Fluctuation analysis of PSR J2022+5154 from the FAST observation on 20220917, with LRFS (top-left), and 2DFS for the on-pulse region (top-right), leading part (bottom-left) and trailing part (bottom-right) of the mean pulse profile.
\label{subfig:fluctu:J2022+5154}}
\end{figure}

\subsection{J2015+2524}
\label{subsec:J2015+2524}

PSR J2015+2524 was found by the 305 m radio telescope at Arecibo \citep{Nice1995}. 

This pulsar was observed by FAST for 5 minutes on 20210618, yielding a rotation period $P=2.3033$~s and a dispersion measure $D\!M=11.0~{\rm cm^{-3}\,pc}$. 
The distribution around zero energy in the on-pulse integral energy histogram (Fig.~\ref{subfig:Hist:J2015+2524}) is not symmetric, which indicates the existence of the weak state. Based on the energy histogram, single pulses are distinguished into weak or bright emission modes, which correspond to two colors in the central bar of the single pulse sequence.




\begin{figure}[htpb]
\centering
\includegraphics[width=0.22\textwidth, angle=0]{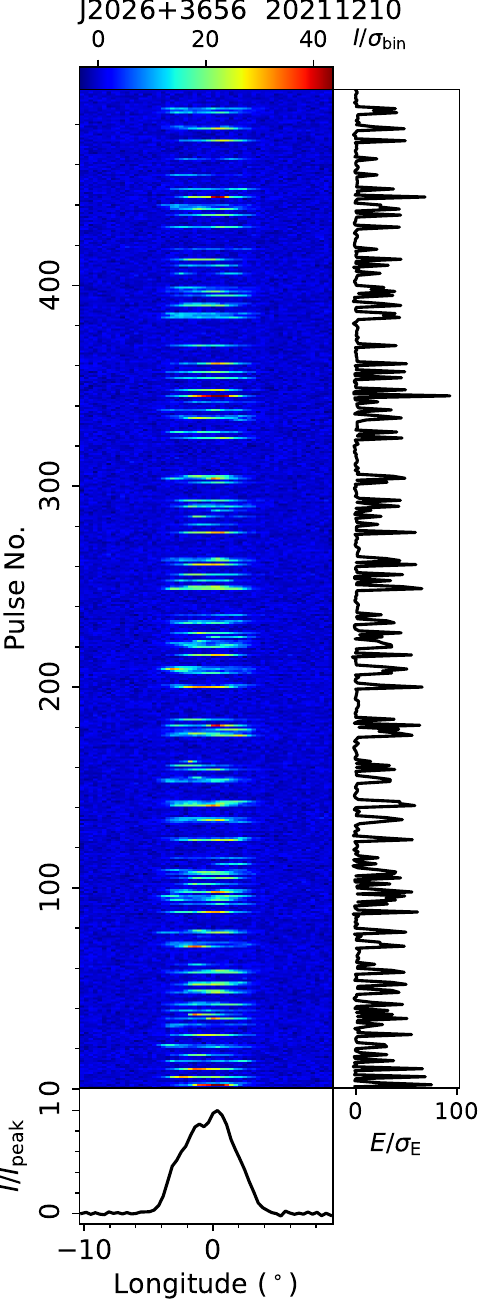}
\figcaption{Single pulse sequence of PSR J2026+3656 from the FAST observation on 20211210.
\label{subfig:TP:J2026+3656}}
\end{figure}

\begin{figure}[htpb]
\centering
\includegraphics[width=0.39\textwidth, angle=0]{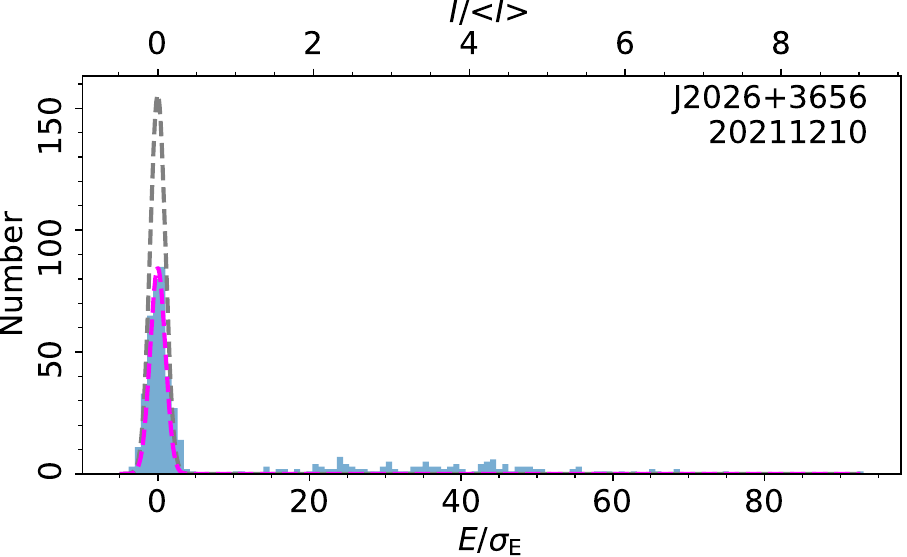}
\figcaption{On-pulse energy histogram of single pulses of PSR J2026+3656 from the FAST observation on 20211210.
\label{subfig:Hist:J2026+3656}}
\end{figure}

\begin{figure}[htpb]
\centering
\includegraphics[width=0.22\textwidth, angle=0]{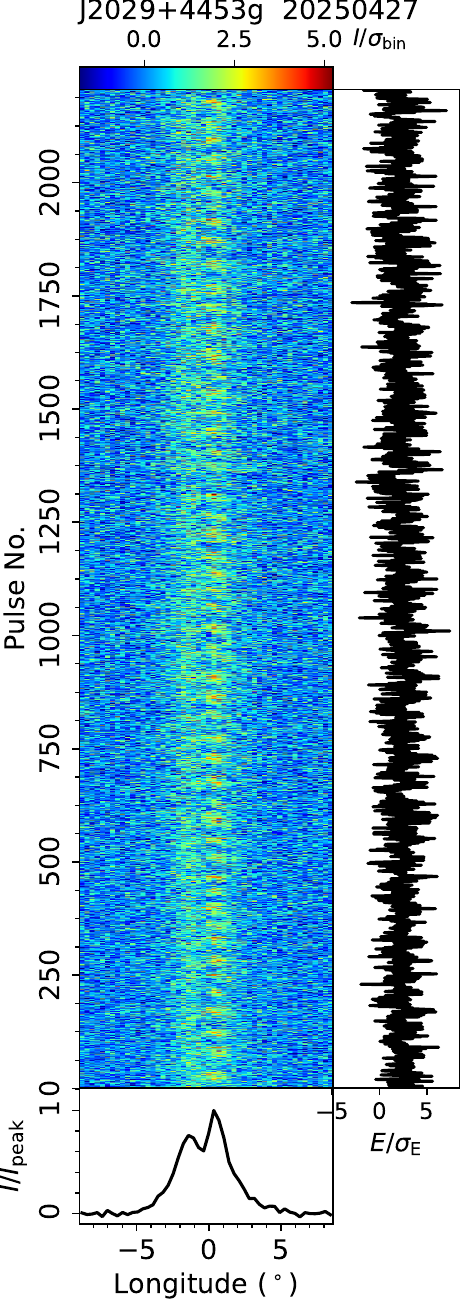}
\includegraphics[width=0.22\textwidth, angle=0]{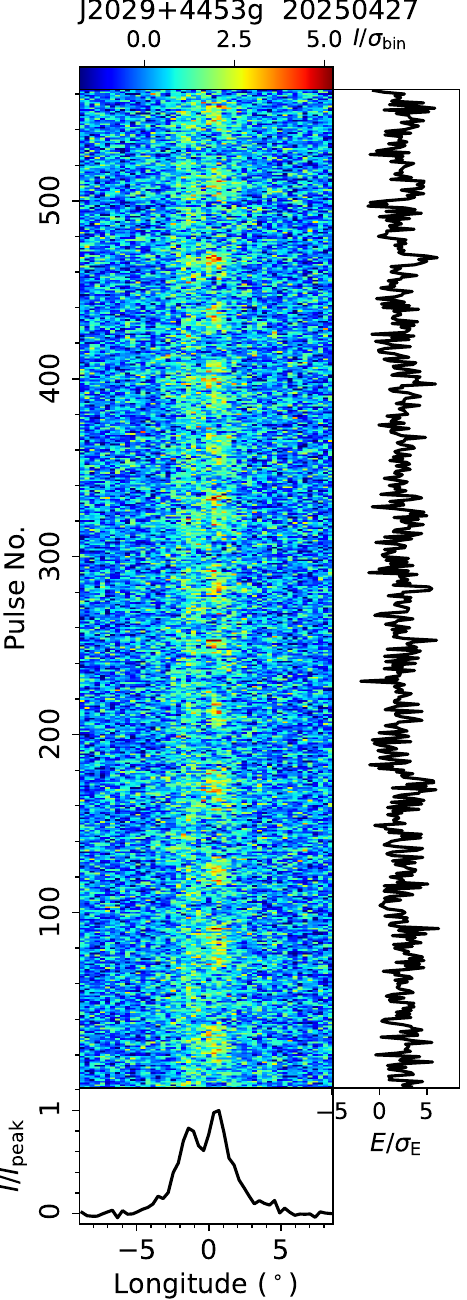}
\figcaption{Single pulse sequence of PSR J2029+4453g from the FAST observation on 20250427, and a zoomed-in view of pulses No. 1-560.
\label{subfig:TP:J2029+4453g}}
\end{figure}

\begin{figure}[htpb]
\centering
\includegraphics[width=0.22\textwidth, angle=0]{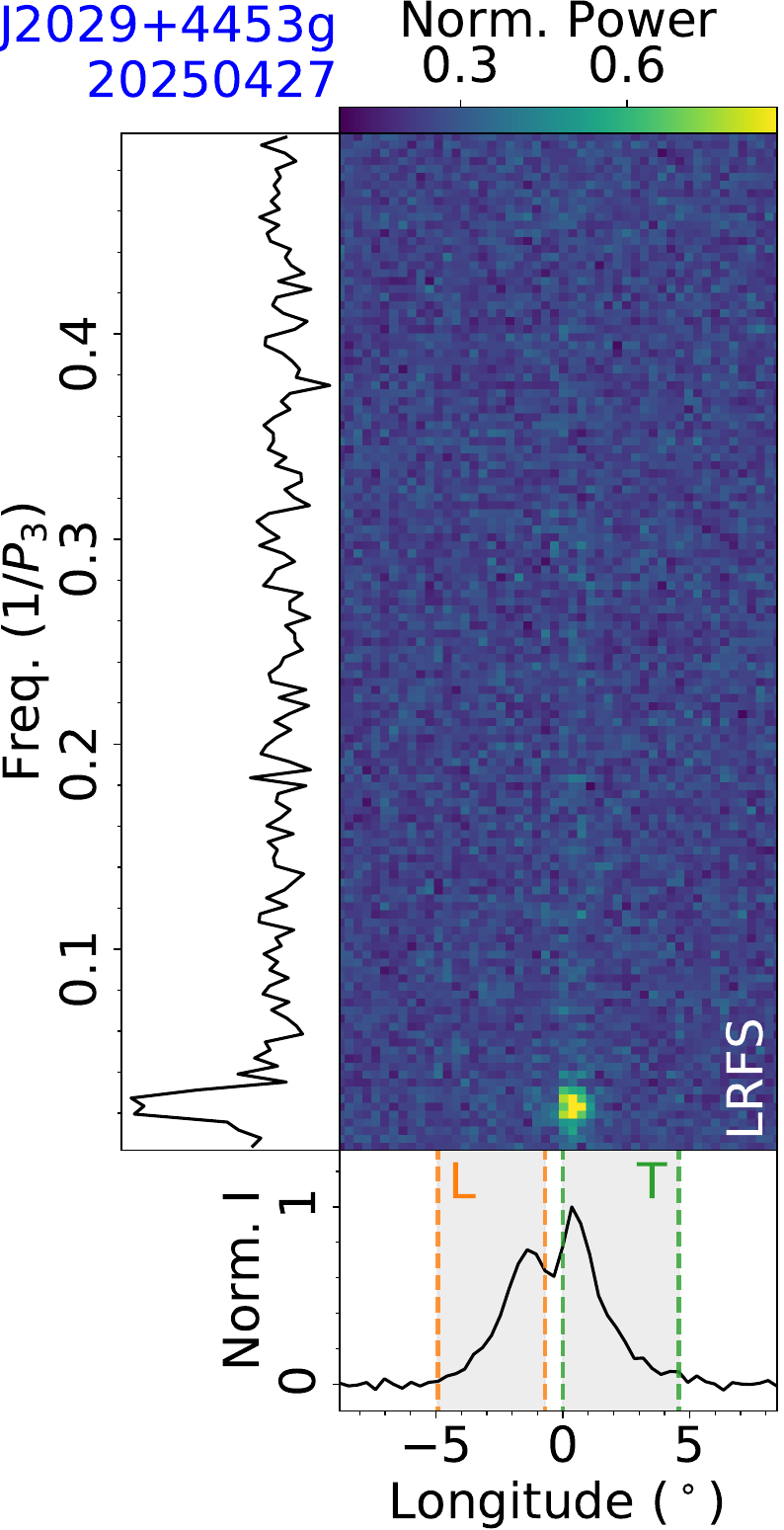}
\includegraphics[width=0.22\textwidth, angle=0]{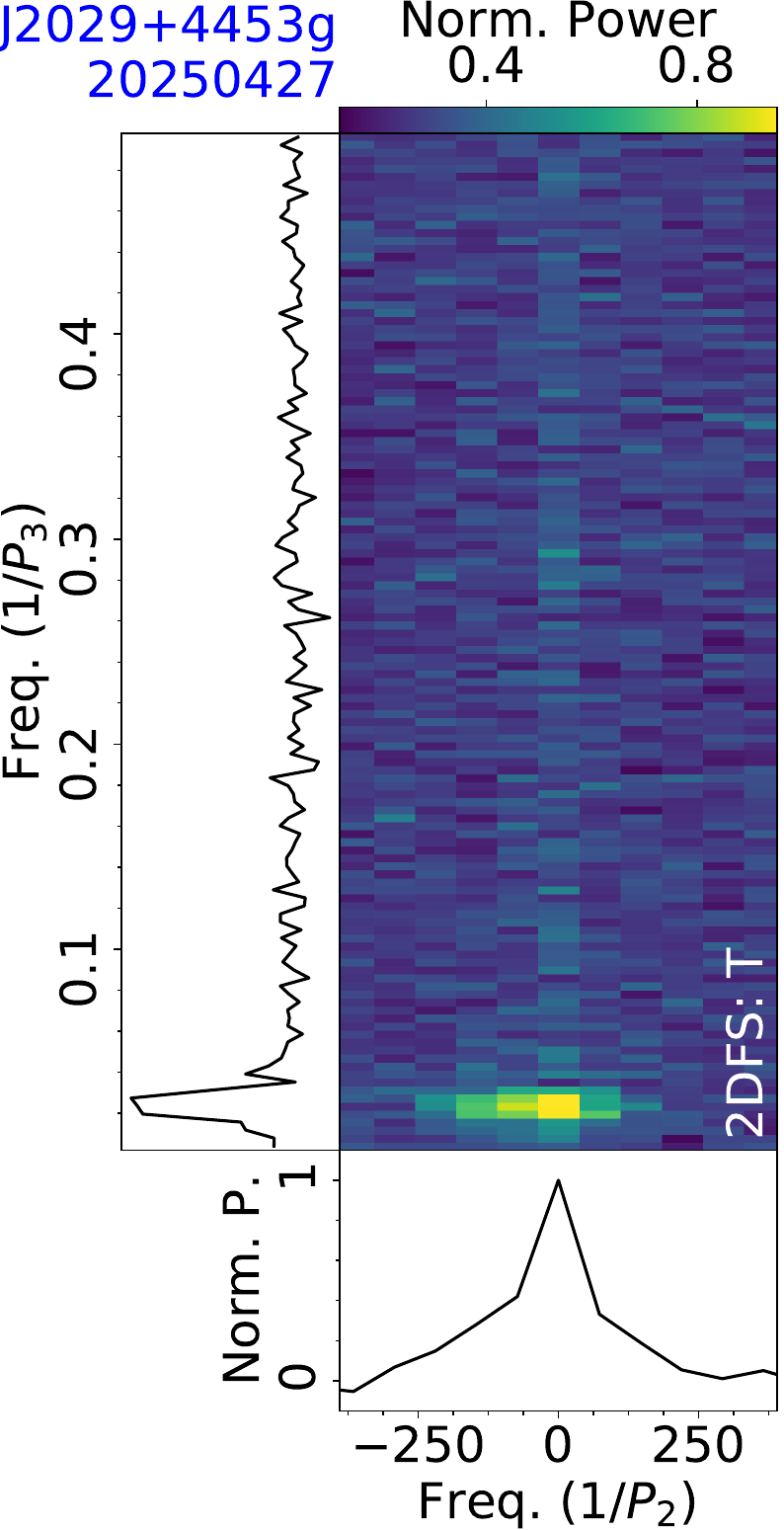}
\figcaption{Fluctuation analysis of PSR J2029+4453g for the observation on 20250427, with LRFS and 2DFS for the trailing part of a mean pulse profile.
\label{subfig:fluctu:J2029+4453g}}
\end{figure}

\subsection{J2016+1948}
\label{subsec:J2016+1948}

PSR J2016+1948 was discovered by the Arecibo Radio Telescope \citep{Navarro2003}. 

This pulsar was observed by FAST on 20201110 for 8 minutes, and a rotation period $P=0.06495$~s and a dispersion measure $D\!M=33.8~{\rm cm^{-3}\,pc}$ were determined. The single pulse sequence and a zoomed-in view are shown in Fig.~\ref{subfig:TP:J2016+1948}. A decrease in energy for pulses 1385 to 1395 is revealed. From fluctuation spectra in Fig.~\ref{subfig:fluctu:J2016+1948}, the 2DFS of the trailing profile part exhibits a feature with the centroid frequencies of $1/P_3=0.257\pm0.002$ and $1/P_2=10\pm2$, corresponding to periodicities of $P_3=3.89\pm0.03$ periods and $P_2=37\pm9$ degrees. 
More FAST observations are required for the confirmation of whether the energy decrease corresponds to the nulling state or the weak emission mode, as well as for a detailed analysis.

\begin{figure}[htpb]
\centering
\includegraphics[width=0.22\textwidth, angle=0]{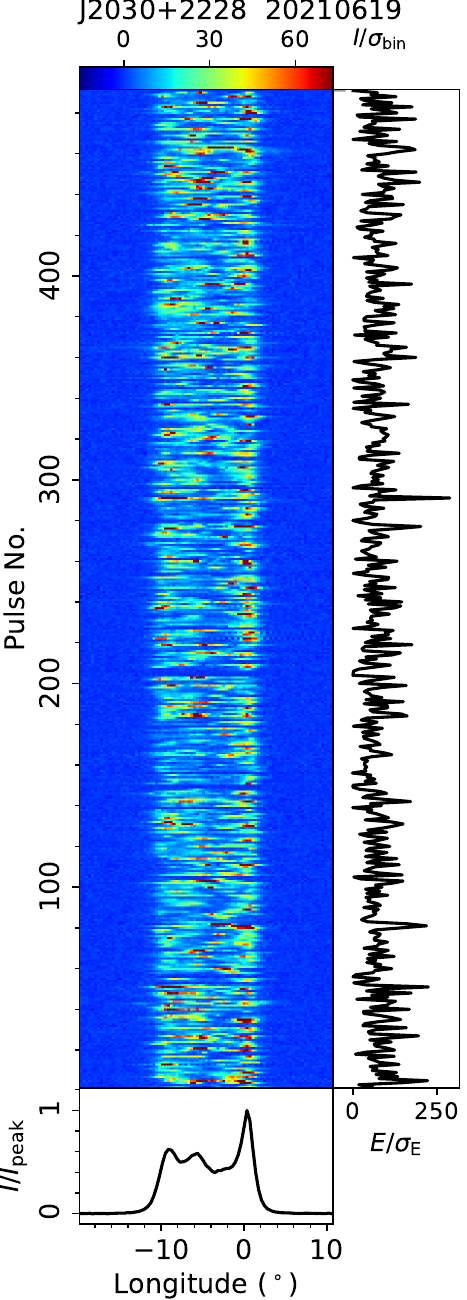}
\includegraphics[width=0.22\textwidth, angle=0]{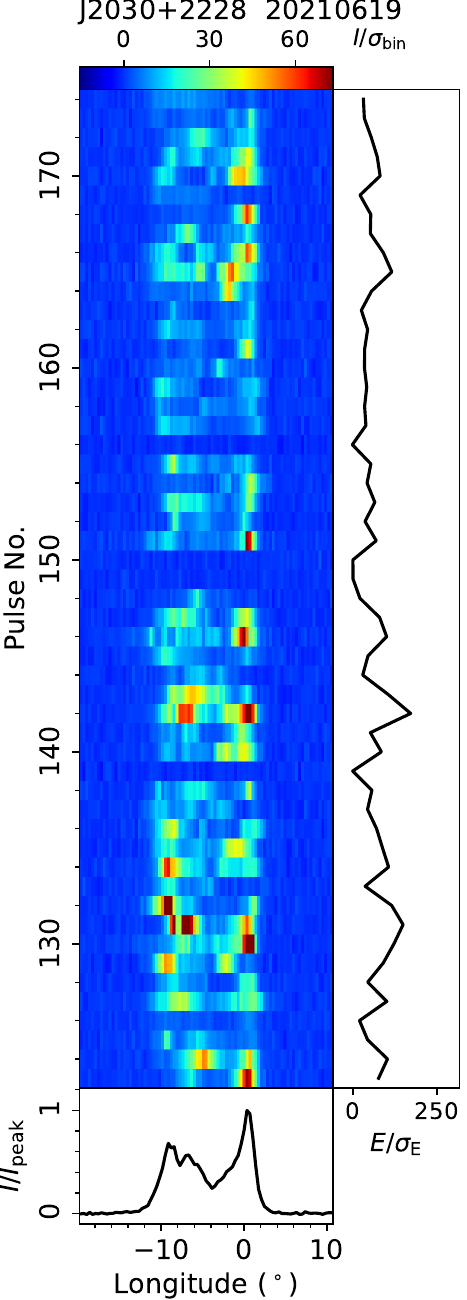}
\figcaption{Single pulse sequence of PSR J2030+2228 from the FAST observation on 20210619, with a zoomed-in view of pulses No. 123-174.
\label{subfig:TP:J2030+2228}}
\end{figure}

\begin{figure}[htpb]
\centering
\includegraphics[width=0.39\textwidth, angle=0]{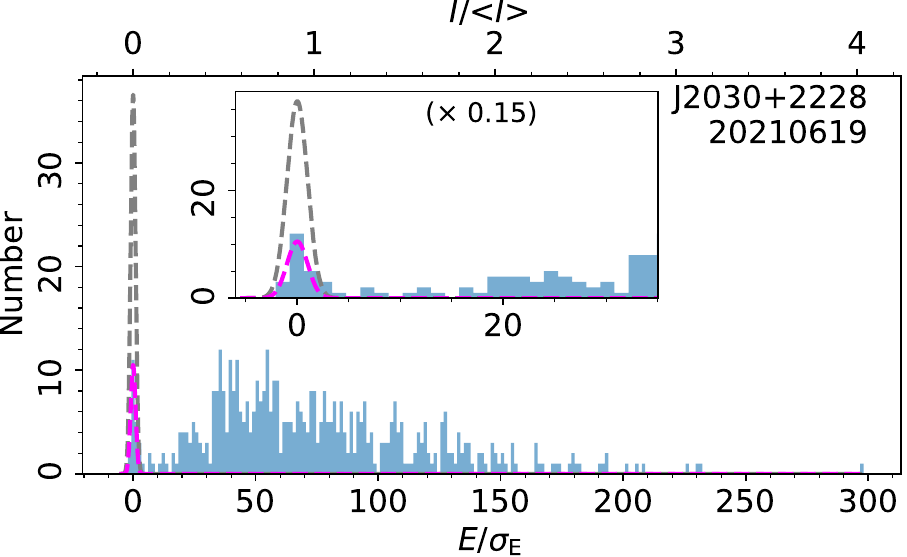}
\figcaption{On-pulse energy histogram of single pulses of PSR J2030+2228 from the FAST observation on 20210619.
\label{subfig:Hist:J2030+2228}}
\end{figure}

\subsection{J2016+3318g}
\label{subsec:J2016+3318g}

PSR J2016+3318g was discovered in the FAST GPPS survey \citep{Han2021,han2025}. 

This pulsar was observed by FAST on 20240909 for 5 minutes and 20241026 for 15 minutes. From the 15-minute observation, a rotation period and a dispersion measure are derived to be $P=2.1785$~s and a dispersion measure $D\!M=380.1~{\rm cm^{-3}\,pc}$. Single pulses of two observations in Figure~\ref{subfig:TP:J2016+3318g} reveal nulling and subpulse drifting behaviors. 
From the on-pulse integral energy distribution in Figure~\ref{subfig:Hist:J2016+3318g}, the nulling fraction for the data of 20241026 is estimated to be  9.5$\pm$0.8\%. 
LRFS and 2DFS of the trailing part in the mean pulse profile for the data on 20240909 and the leading profile part on 20241026 are shown in Figure~\ref{subfig:fluctu:J2016+3318g}, where the drift properties are different between the two observations. 
For the observation on 20240909, 2DFS of the trailing profile part has a positive drift feature, with the centroid frequencies of $1/P_3=0.342\pm0.002$ and $1/P_2=44\pm10$, corresponding to drifting periodicities of $P_3=2.93\pm0.02$ periods and $P_2=8\pm2^\circ$. However, for the data on 20241026, 2DFS of the leading profile part displays a negative drift feature. The centroid of this feature is characterized by $1/P_3=0.273\pm0.002$ and $1/P_2=-69\pm9$, yielding $P_3=3.67\pm0.03$ periods and $P_2=-5.2\pm0.7^\circ$. 
More FAST observations are required for the analysis of clear single pulse properties of J2016+3318g.

\subsection{J2022+5154}
\label{subsec:J2022+5154}

PSR J2022+5154 was discovered by the Jodrell Bank Mk I telescope \citep{Davies1970a}. Subpulse drifting with variable drift rate has been presented by \citet{Oster1977} at 1720 MHz and \citep{Weltevrede2006} at 21 cm. \citet{Chen2024} reported a 5-period subpulse drifting and a 40-period periodic amplitude modulation at 2250 MHz.

This pulsar was observed by FAST on 20220917 for 84 minutes, driving a rotation period $P=0.5292$~s and a dispersion measure $D\!M=22.9~{\rm cm^{-3}\,pc}$. The single pulse sequence and a zoomed-in view of pulses No. 1-300 in Fig.~\ref{subfig:TP:J2022+5154} reveal the unsystematic subpulse drifting behavior. Fluctuation spectra are shown in Fig.~\ref{subfig:fluctu:J2022+5154}, in which the leading and trailing parts of the mean pulse profile exhibit different modulation features. 
For the leading profile part, the low-frequency modulation feature is more prominent than the positive drift feature in 2DFS. The low-frequency feature has a centroid frequency of $0.0272\pm0.0003$ ($P_3=36.7\pm0.4$ periods), and the drift feature is centered at $1/P_3=0.172\pm0.001$ and $1/P_2=16\pm1$ ($P_3=5.80\pm0.02$ periods and $P_2=22\pm1$ degrees). 
For the trailing profile part, the centroid of the low-frequency modulation feature is at $1/P_3=0.0529\pm0.0004$ ($P_3=18.9\pm0.1$ periods), and that of the positive drift feature is at $1/P_3=0.1708\pm0.0004$ and $1/P_2=22\pm1$ ($P_3=5.85\pm0.01$ periods and $P_2=16\pm1$ degrees).

\begin{figure}[htpb]
\centering
\includegraphics[width=0.22\textwidth, angle=0]{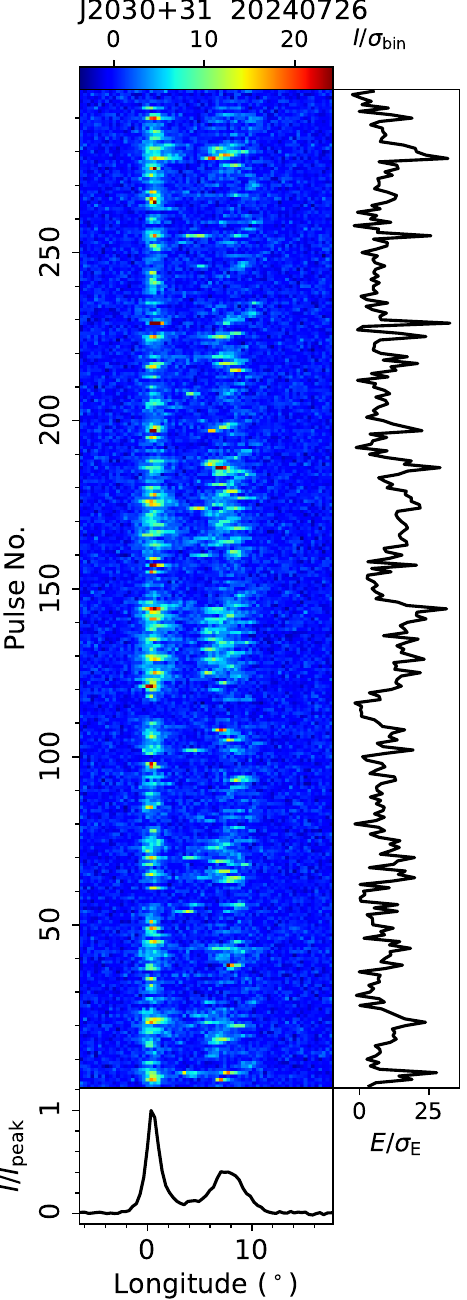}
\figcaption{Single pulse sequence of PSR J2030+31 from the FAST observation on 20240726.
\label{subfig:TP:J2030+31}}
\end{figure}

\begin{figure}[htpb]
\centering
\includegraphics[width=0.39\textwidth, angle=0]{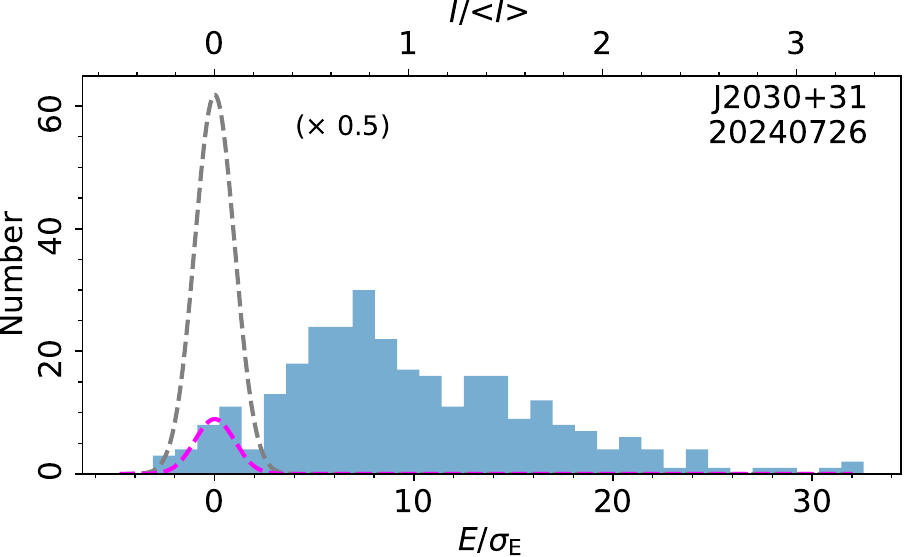}
\figcaption{On-pulse energy histogram of single pulses of PSR J2030+31 from the FAST observation on 20240726.
\label{subfig:Hist:J2030+31}}
\end{figure}

\begin{figure}[htpb]
\centering
\includegraphics[width=0.22\textwidth, angle=0]{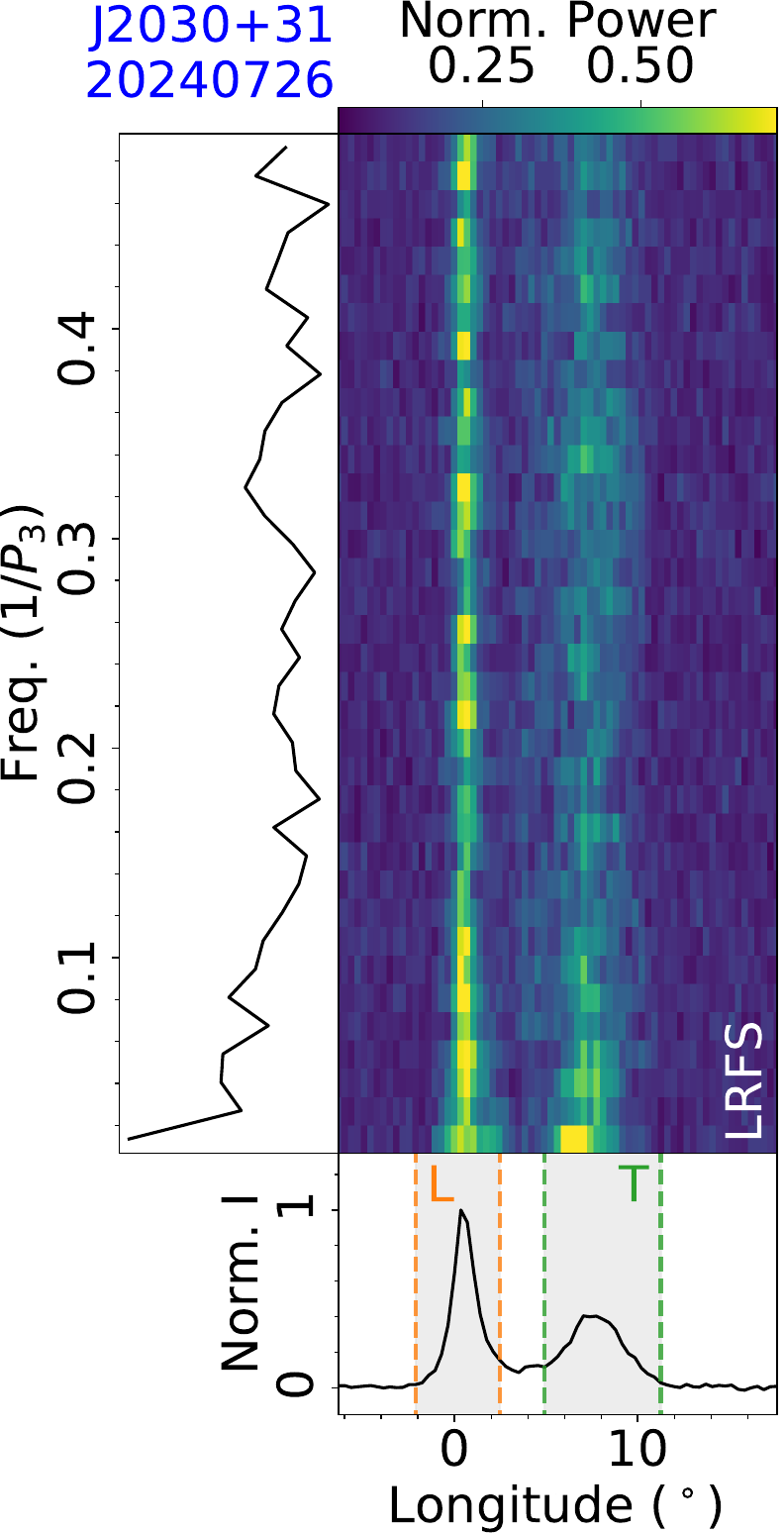}
\includegraphics[width=0.22\textwidth, angle=0]{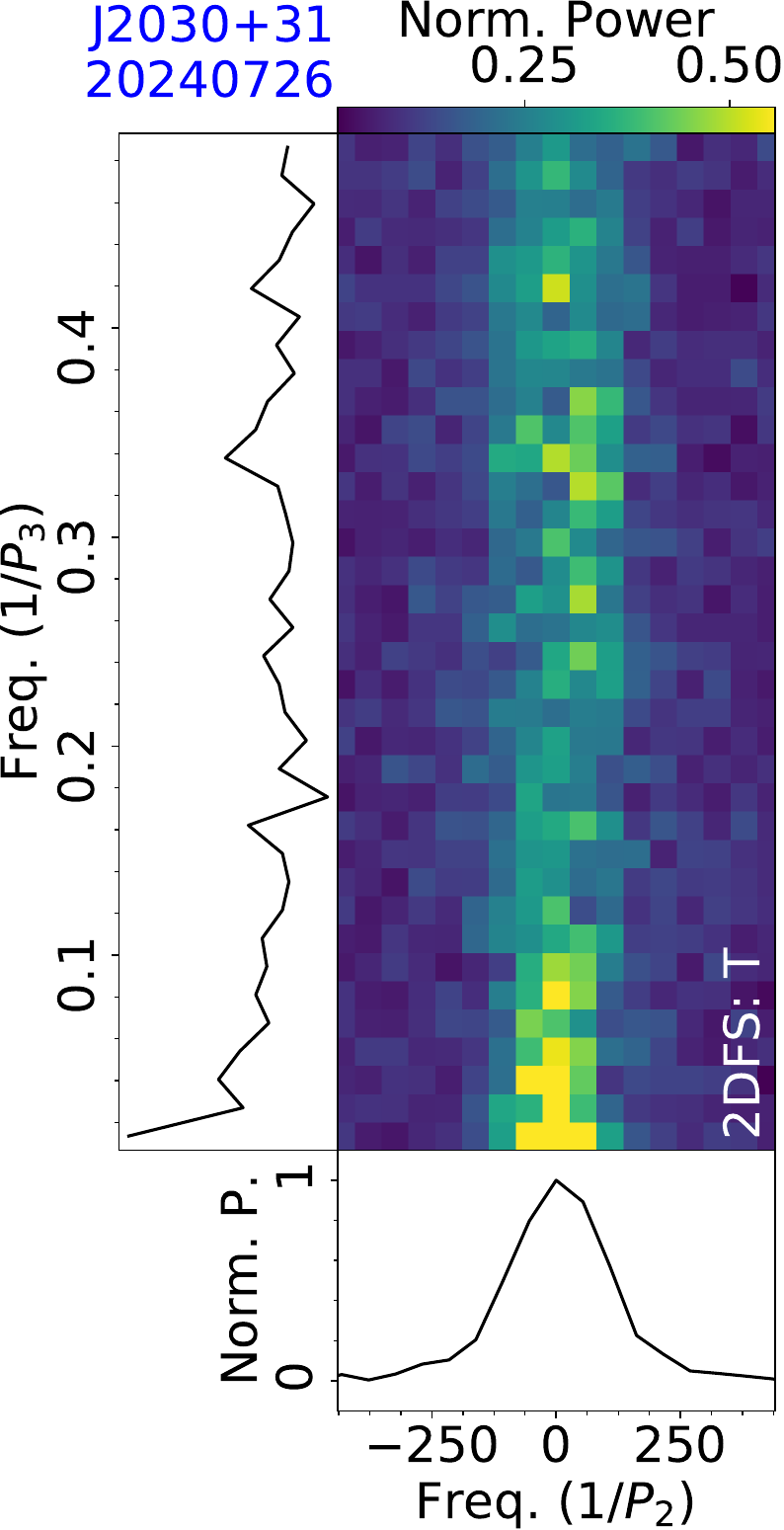}
\figcaption{Fluctuation analysis of PSR J2030+31 for the observation on 20240726, with LRFS and 2DFS for the trailing part of a mean pulse profile.
\label{subfig:fluctu:J2030+31}}
\end{figure}

\subsection{J2026+3656g}
\label{subsec:J2026+3656g}

The pulsar was discovered in the GPPS survey by FAST \citep{Han2021,han2025}. 

This pulsar was observed by FAST on 20211210 for 15 minutes, yielding a rotation period $P=1.7857$~s and a dispersion measure $D\!M=278.9~{\rm cm^{-3}\,pc}$. 
The single pulse sequence is displayed in Fig.~\ref{subfig:TP:J2026+3656}, indicating the nulling behavior of the pulsar. The nulling fraction is estimated to be 54$\pm$4\% from the on-pulse integral energy histogram in Fig.~\ref{subfig:Hist:J2026+3656}.

\subsection{J2029+4453g}
\label{subsec:J2029+4453g}

PSR J2029+4453g was discovered in the FAST GPPS survey \citep{Han2021,han2025}. 

This pulsar was observed by FAST on 20220606 for 15 minutes and 20250427 for 50 minutes. From the longer observation, a rotation period $P=1.3613$~s and a dispersion measure $D\!M=332.9~{\rm cm^{-3}\,pc}$ were determined. 
The single pulse sequence of the observation on 20250427 is shown in Fig.~\ref{subfig:TP:J2029+4453g}, where the trailing profile part displays good modulation behavior. From LRFS and 2DFS in Fig.~\ref{subfig:fluctu:J2029+4453g}, the trailing profile part has a negative drift feature, with centroid frequencies of the feature estimated to be $1/P_3=0.0246\pm0.0003$ and $1/P_2=-45\pm6$, corresponding to periodicities of $P_3=40.7\pm0.5$ periods and $P_2=-8\pm1^\circ$. 
The drifting parameters from the observation on 20220606 are consistent with 20250427, which are $P_3=48\pm2$ periods and $P_2=-10\pm2^\circ$.

\subsection{J2030+2228}
\label{subsec:J2030+2228}

PSR J2030+2228 was discovered by the Arecibo telescope \citep{Hulse1975}. 

The pulsar was observed by FAST on 20210619 for 5 minutes, deriving a rotation period $P=0.6305$~s and a dispersion measure $D\!M=71.9~{\rm cm^{-3}\,pc}$. 
The single pulse sequence and a zoomed-in view of pulses No. 123-174 in Fig.~\ref{subfig:TP:J2030+2228} show nulls with short durations. The nulling fraction of this observation is estimated to be 4.3$\pm$0.3\% from the on-pulse energy histogram in Fig.~\ref{subfig:Hist:J2030+2228}.




\begin{figure}[htpb]
\centering
\includegraphics[width=0.22\textwidth, angle=0]{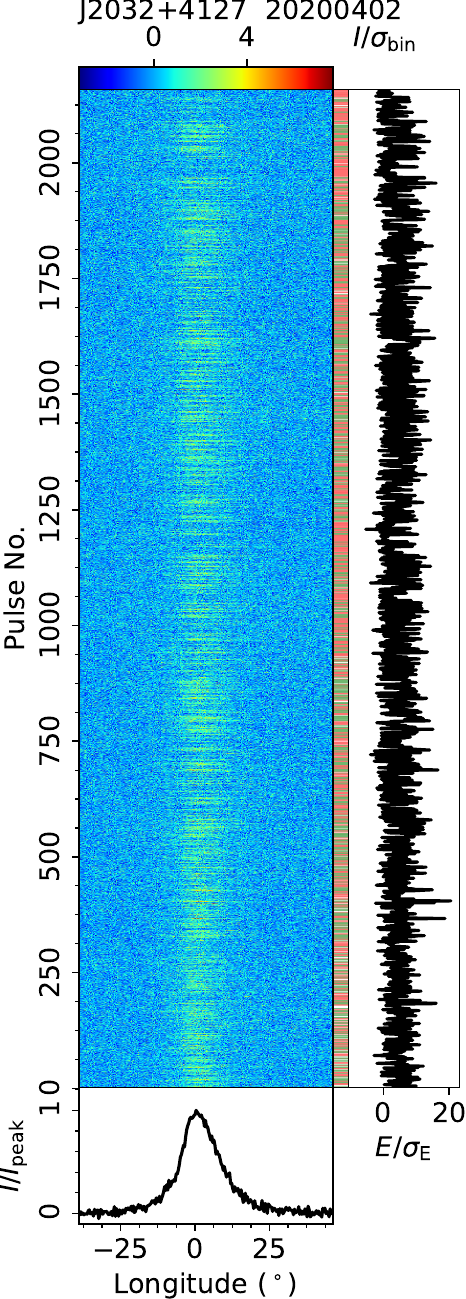}
\includegraphics[width=0.22\textwidth, angle=0]{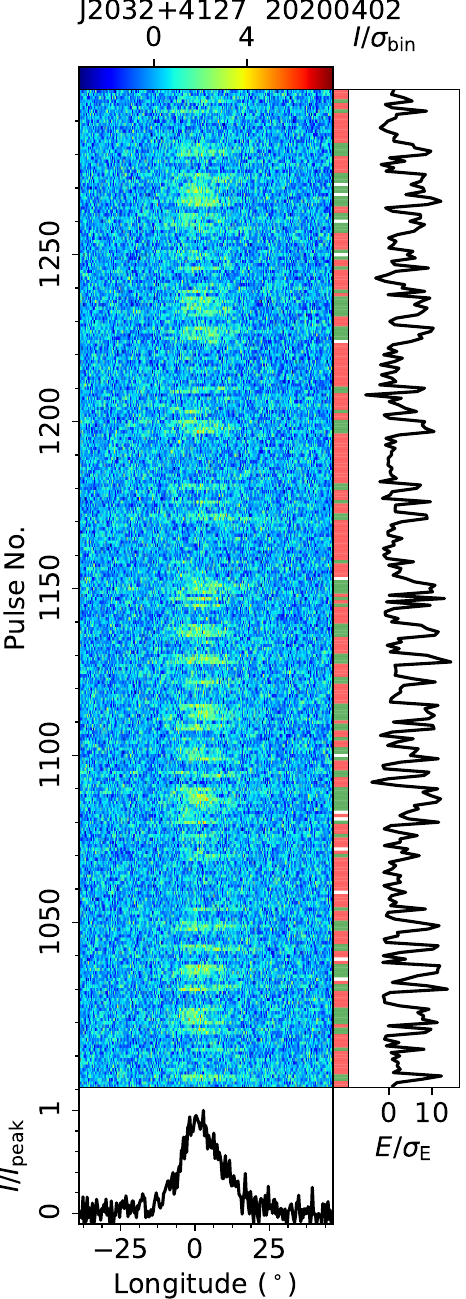}
\figcaption{Single pulse sequence of PSR J2032+4127 from the FAST observation on 20200402, with a zoomed-in view of pulses No. 1000-1300.
\label{subfig:TP:J2032+4127}}
\end{figure}

\begin{figure}[htpb]
\centering
\includegraphics[width=0.39\textwidth, angle=0]{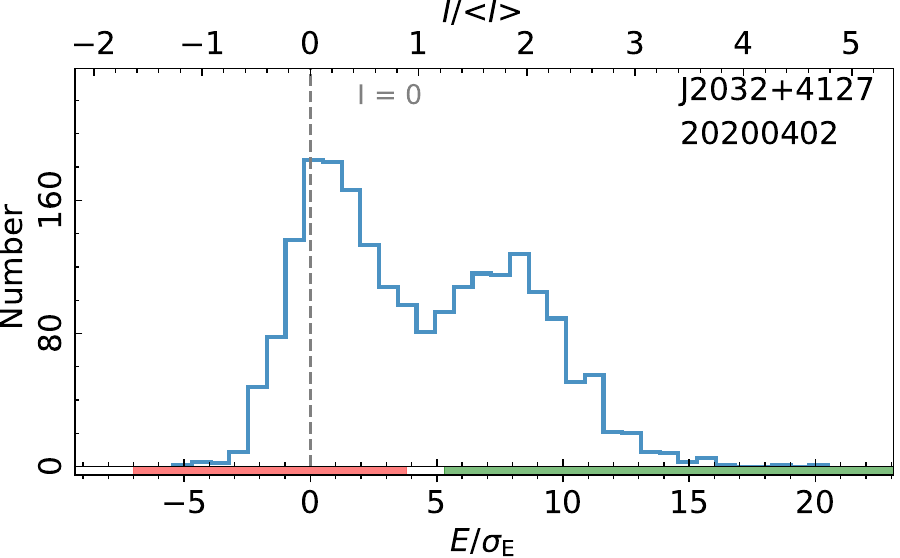}
\figcaption{On-pulse energy histogram of single pulses of PSR J2032+4127 from the FAST observation on 20200402. \label{subfig:Hist:J2032+4127}}
\end{figure}

\begin{figure}[htpb]
\centering
\includegraphics[width=0.39\textwidth, angle=0]{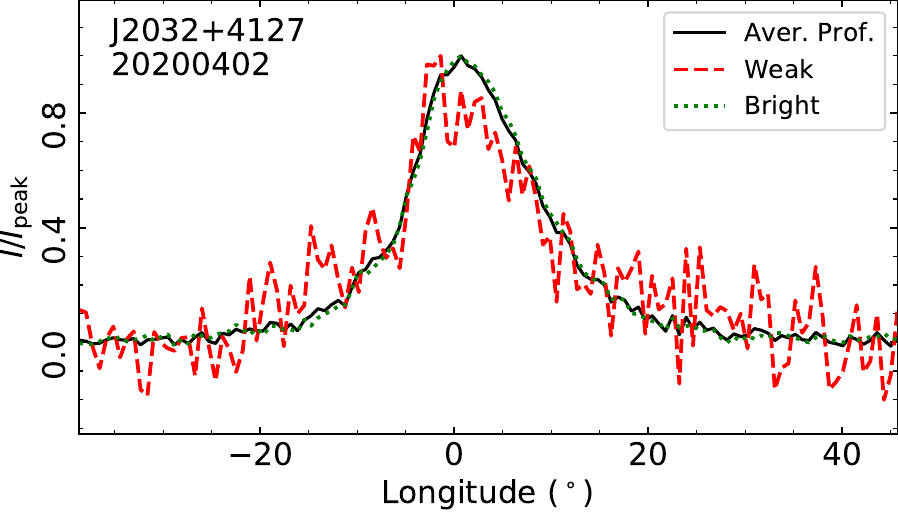}
\figcaption{Mean profiles of weak and bright emission modes of PSR J2032+4127 from the FAST observation on 20200402. \label{subfig:profModes:J2032+4127}}
\end{figure}

\begin{figure}[htpb]
\centering
\includegraphics[width=0.22\textwidth, angle=0]{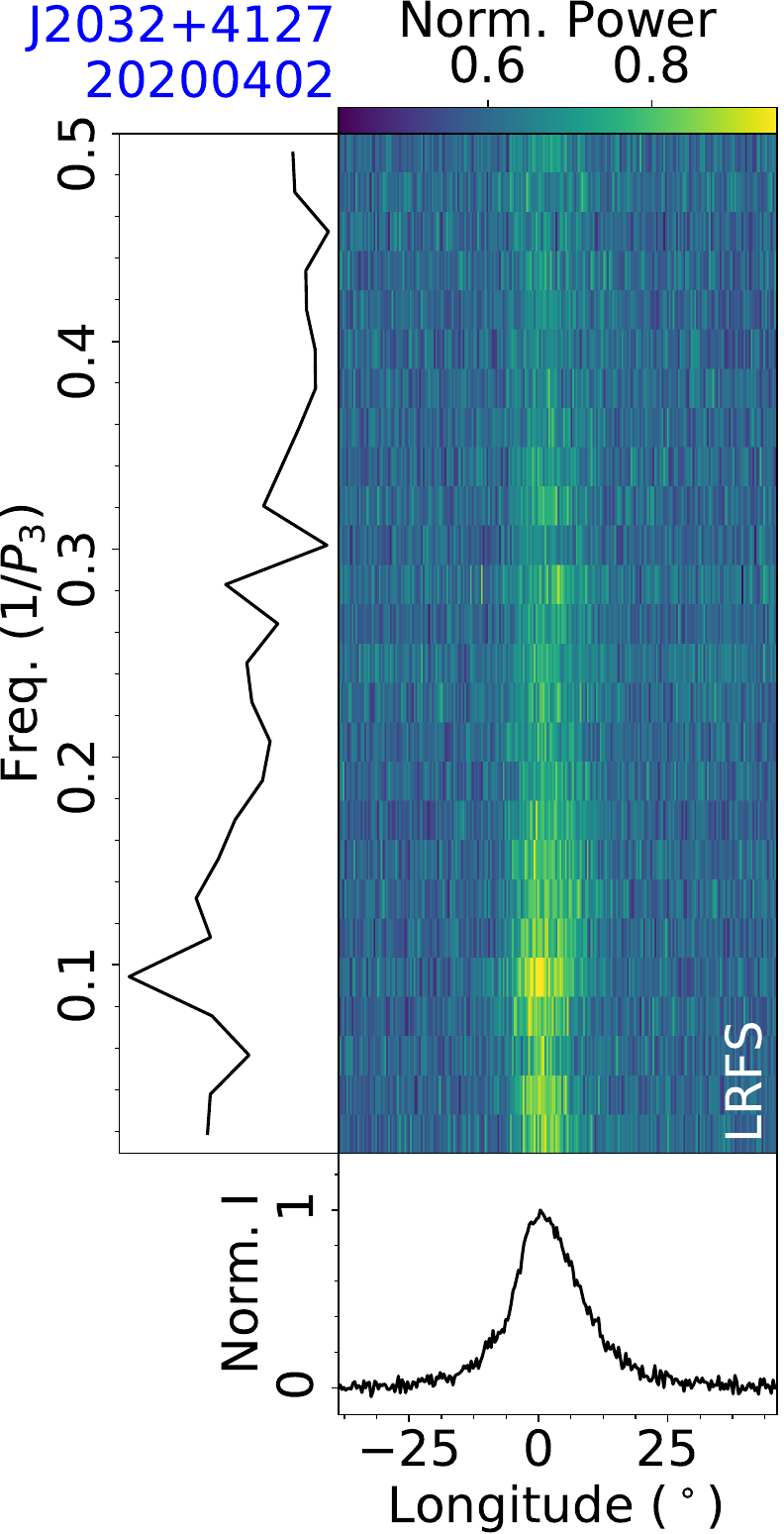}
\includegraphics[width=0.22\textwidth, angle=0]{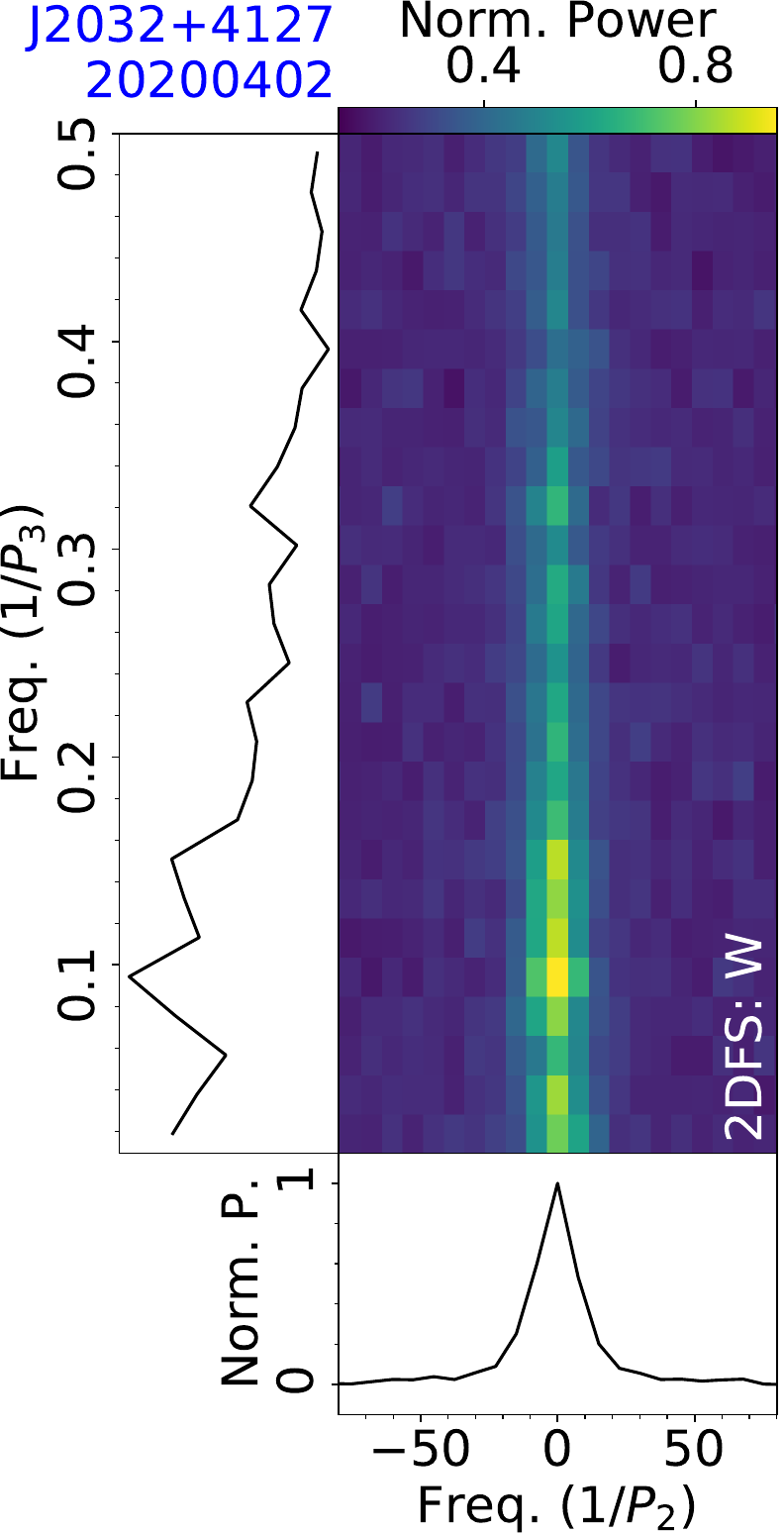}
\figcaption{Fluctuation analysis of PSR J2032+4127 for the FAST observation on 20200402, with LRFS and 2DFS for the on-pulse phase region of the mean pulse profile.
\label{subfig:fluctu:J2032+4127}}
\end{figure}

\begin{figure}[htpb]
\centering
\includegraphics[width=0.22\textwidth, angle=0]{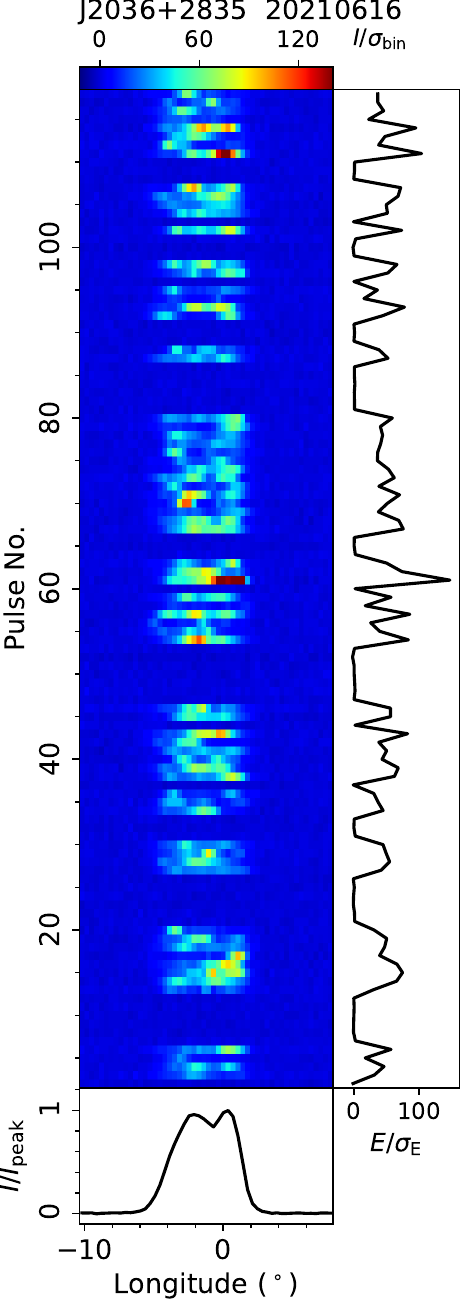}
\includegraphics[width=0.22\textwidth, angle=0]{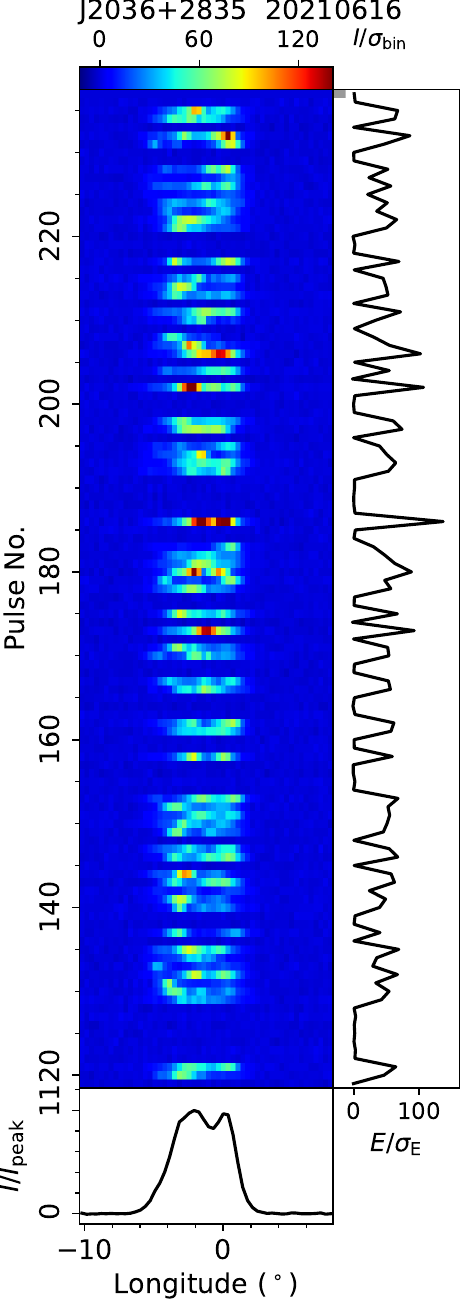}
\figcaption{Single pulse sequences of PSR J2036+2835 from the FAST observation on 20210616.
\label{subfig:TP:J2036+2835}}
\end{figure}

\begin{figure}[htpb]
\centering
\includegraphics[width=0.39\textwidth, angle=0]{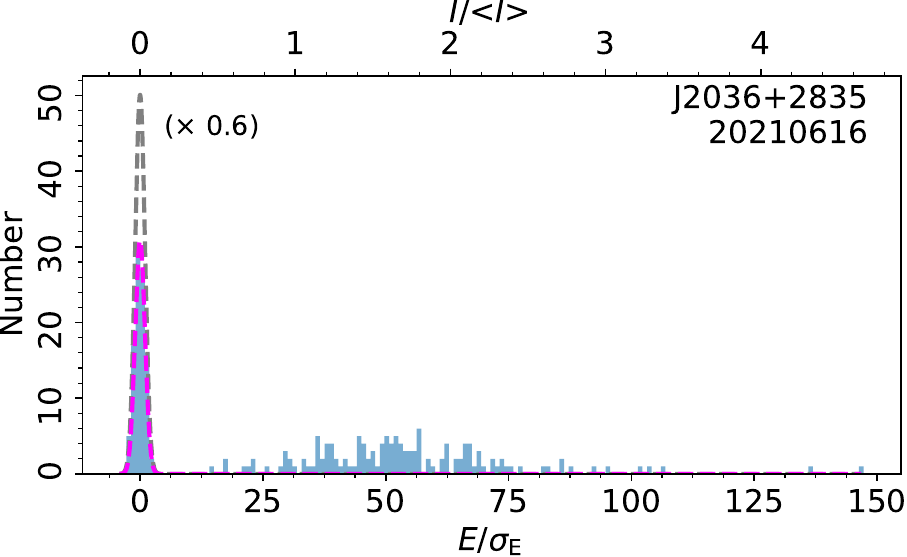}
\figcaption{On-pulse energy histogram of single pulses of PSR J2036+2835 from the FAST observation on 20210616.
\label{subfig:Hist:J2036+2835}}
\end{figure}

\subsection{J2030+31}
\label{subsec:J2030+31}

PSR J2030+31 was discovered by CHIME (https://www.chime-frb.ca/galactic).

This pulsar was observed by FAST on 20240726 for 5 minutes, and a rotation period $P=1.0147$~s and a dispersion measure $D\!M=132.2~{\rm cm^{-3}\,pc}$ were determined. The single pulse sequence in Fig.~\ref{subfig:TP:J2030+31} displays the existence of nulling and subpulse drifting phenomena. 
From the on-pulse integral energy histogram (Fig.~\ref{subfig:Hist:J2030+31}), the nulling fraction of this observation is estimated to be 7.2$\pm$0.7\%. 
From fluctuation spectra in Fig.~\ref{subfig:fluctu:J2030+31}, the centroid frequencies of the positive drift feature for the trailing profile part are $1/P_3=0.307\pm0.004$ and $1/P_2=65\pm5$, corresponding to periodicities of $P_3=3.25\pm0.05$ periods and $P_2=5.6\pm0.4$ degrees. 
Longer observations are required to characterize nulling and subpulse-drifting properties in more detail. 

\begin{figure}[htpb]
\centering
\includegraphics[width=0.22\textwidth, angle=0]{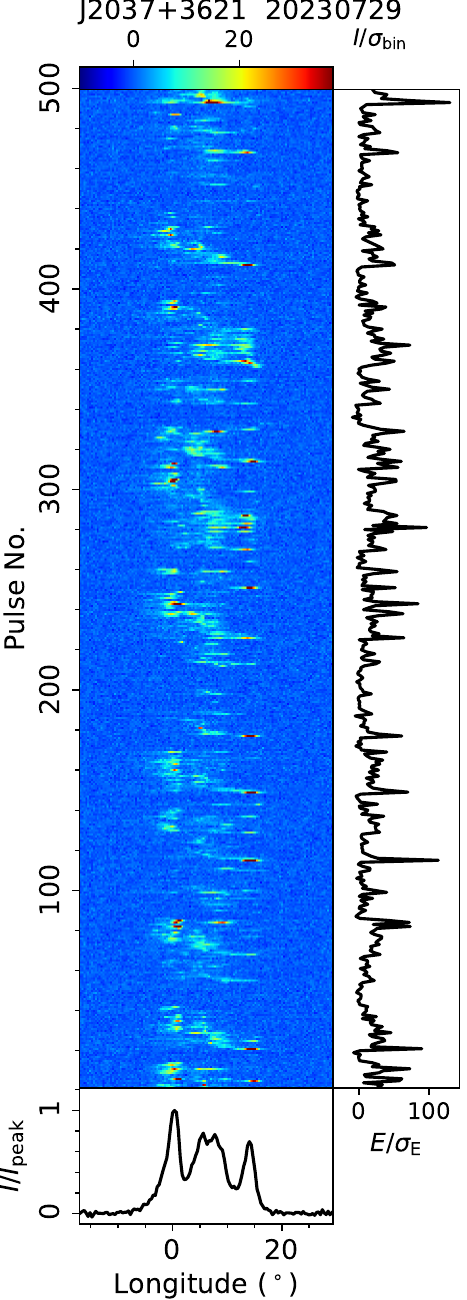}
\figcaption{Single pulse sequence of PSR J2037+3621 from the FAST observation on 20221122.
\label{subfig:TP:J2037+3621}}
\end{figure}

\begin{figure}[htpb]
\centering
\includegraphics[width=0.44\textwidth, angle=0]{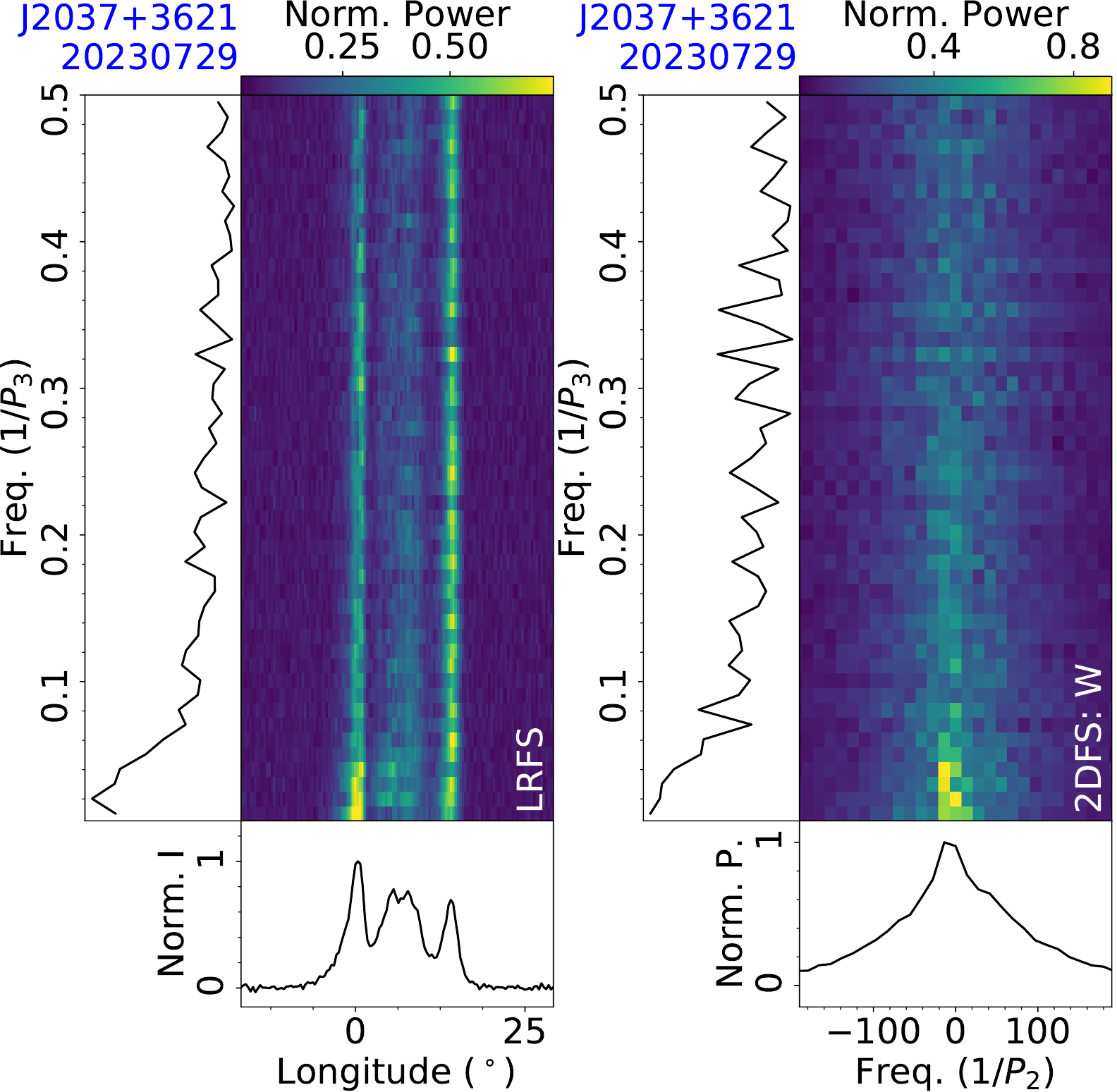}
\figcaption{Fluctuation analysis of PSR J2037+3621 for the observation on 20230729, with LRFS and 2DFS for the on-pulse phase region of a mean pulse profile.
\label{subfig:fluctu:J2037+3621}}
\end{figure}

\begin{figure}[htpb]
\centering
\includegraphics[width=0.22\textwidth, angle=0]{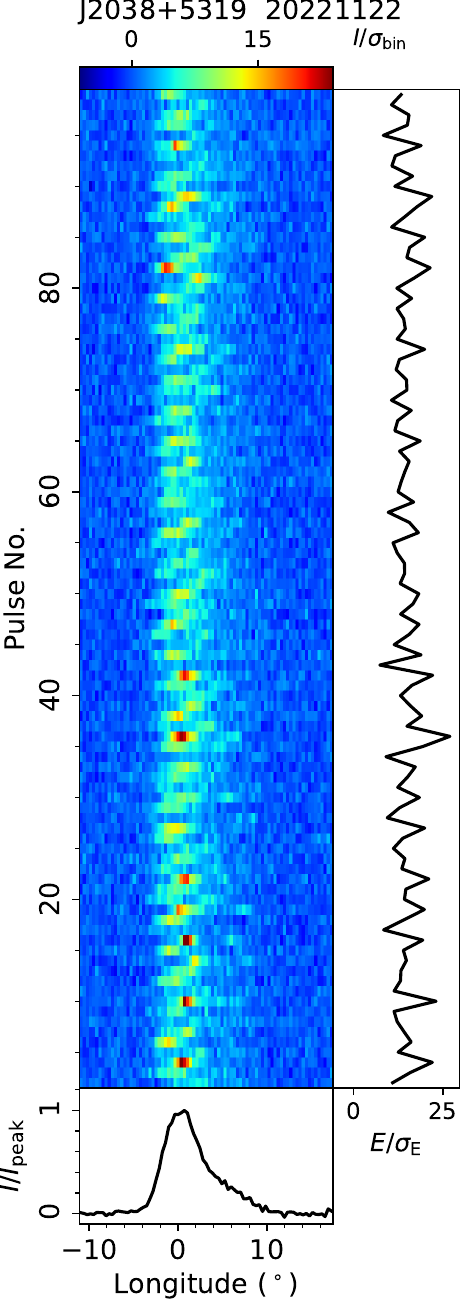}
\includegraphics[width=0.22\textwidth, angle=0]{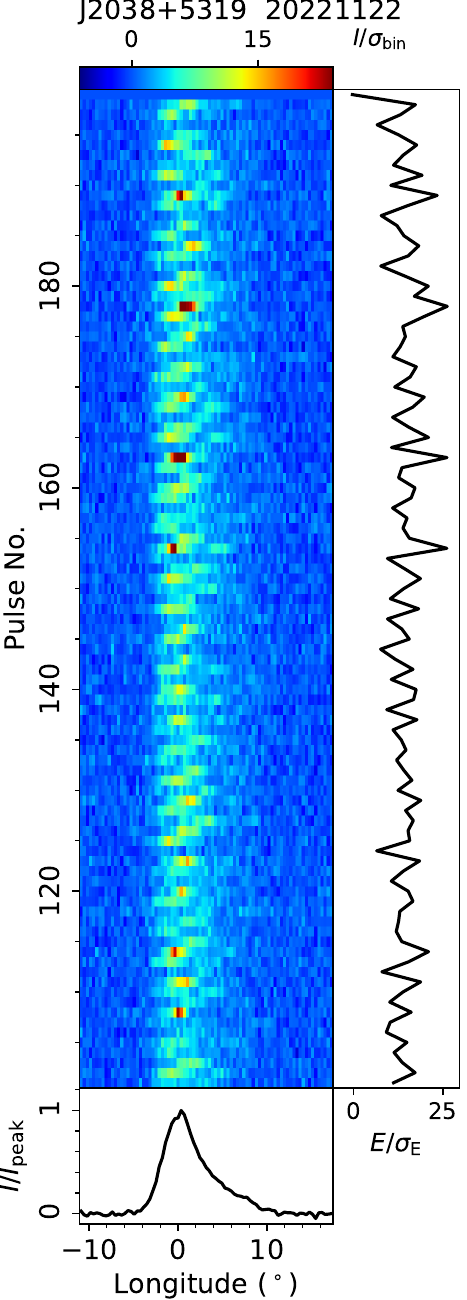}
\figcaption{Single pulse sequences of PSR J2038+5319 from the FAST observation on 20221122.
\label{subfig:TP:J2038+5319}}
\end{figure}

\begin{figure}[htpb]
\centering
\includegraphics[width=0.22\textwidth, angle=0]{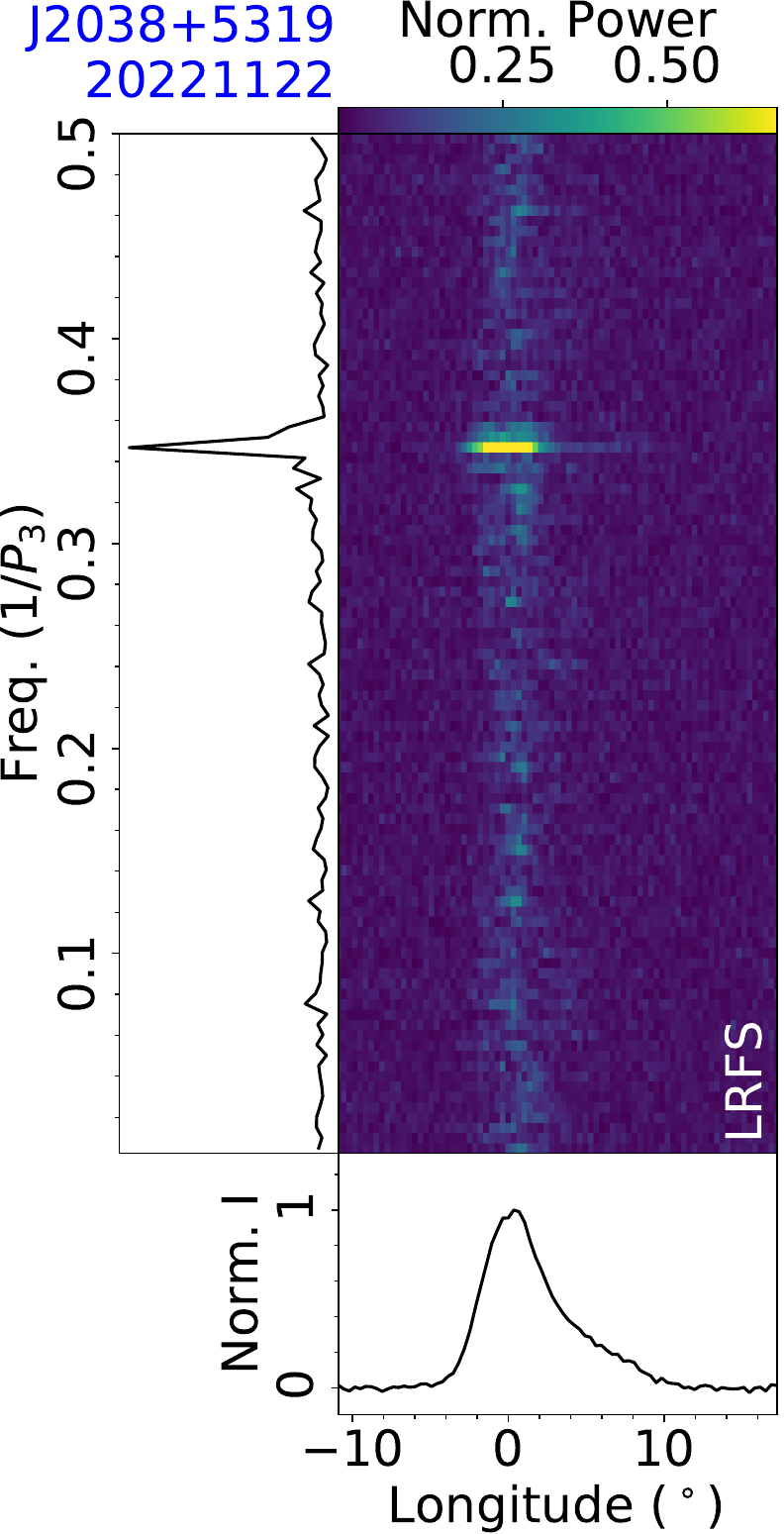}
\includegraphics[width=0.22\textwidth, angle=0]{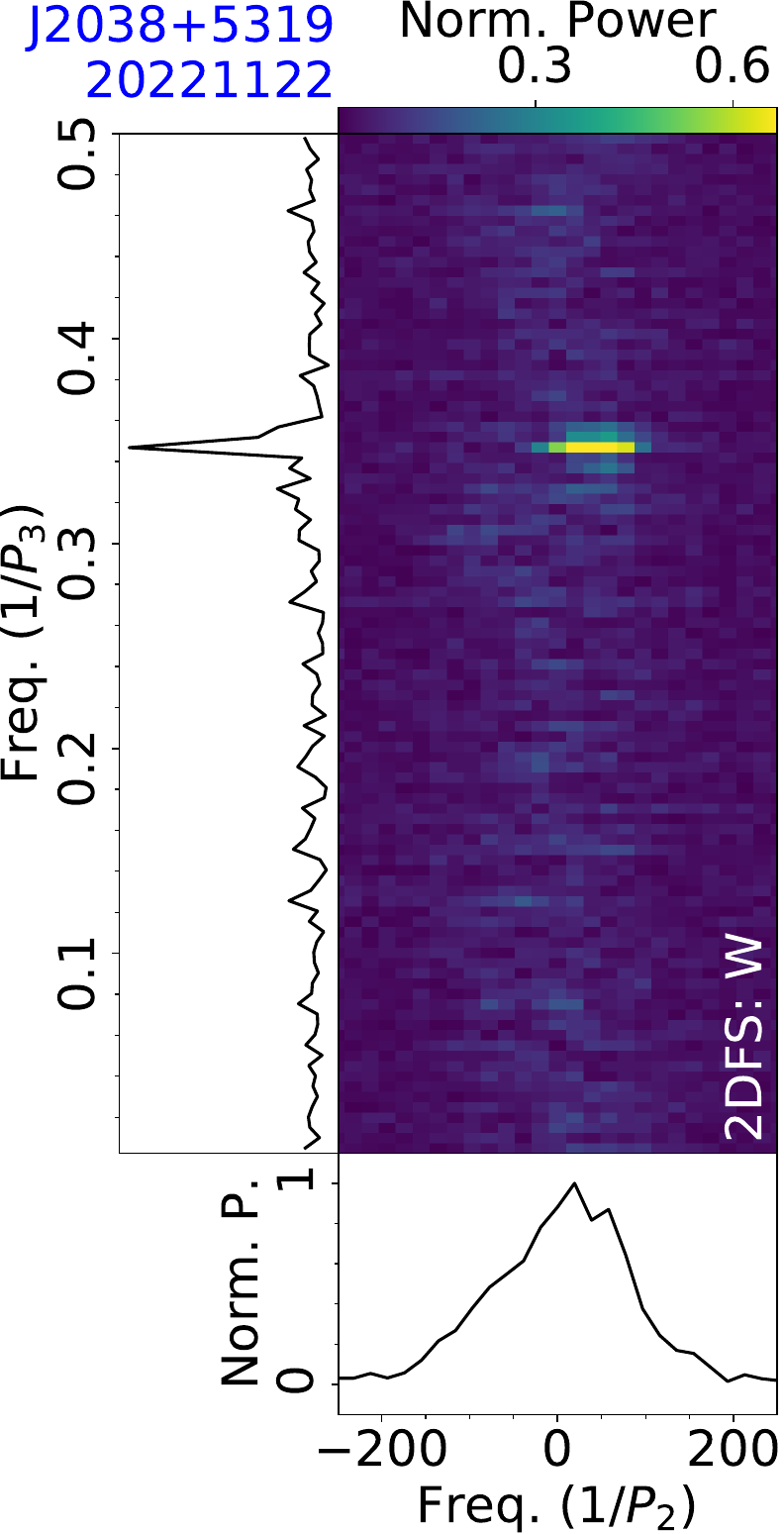}
\figcaption{Fluctuation analysis of PSR J2038+5319 for the observation on 20221122, with LRFS and 2DFS for the on-pulse phase region of a mean pulse profile.
\label{subfig:fluctu:J2038+5319}}
\end{figure}

\subsection{J2032+4127}
\label{subsec:J2032+4127}

PSR J2032+4127 was discovered by the Fermi LAT through blind frequency searches \citep{Abdo2009}.

This pulsar was observed by FAST on 20200402 for 5 minutes, deriving a rotation period $P=0.1432$~s and a dispersion measure $D\!M=114.6~{\rm cm^{-3}\,pc}$. The single pulse sequence and a zoomed-in view of pulses No. 1000-1300 in Fig.~\ref{subfig:TP:J2032+4127} reveal changes between weak and bright emission modes. These two modes are distinguished from the on-pulse energy histogram shown in Fig.~\ref{subfig:Hist:J2032+4127}, where they are labeled in red and green, respectively. The corresponding mean pulse profiles are displayed in Fig.~\ref{subfig:profModes:J2032+4127}. Fluctuation spectra in Fig.~\ref{subfig:fluctu:J2032+4127} show a quasi-periodic feature of mode changes, and the centroid of the modulation feature is at $1/P_3=0.103\pm0.003$, yielding $P_3=9.7\pm0.3$ periods.

\begin{figure}[htpb]
\centering
\includegraphics[width=0.22\textwidth, angle=0]{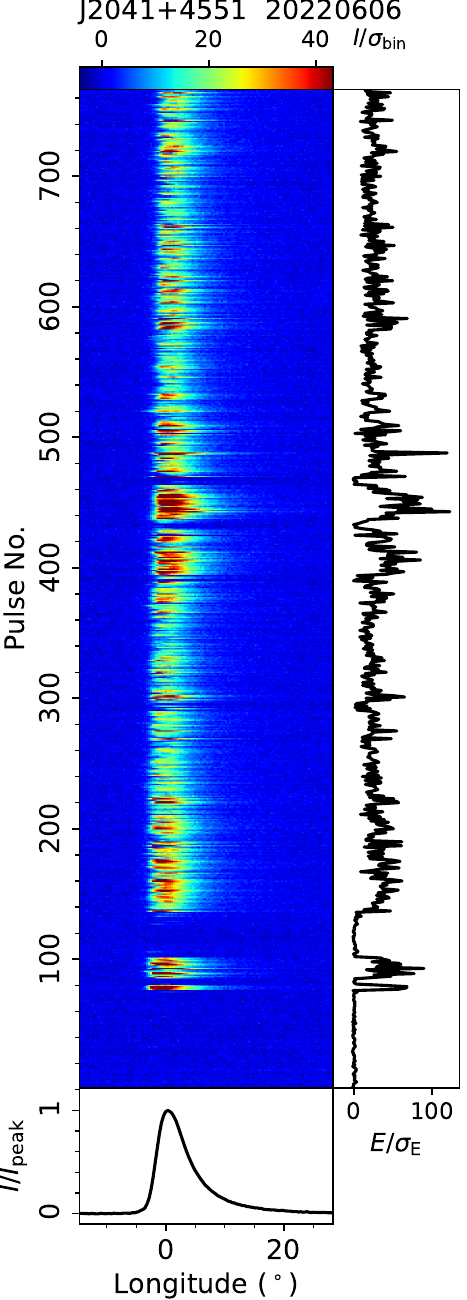}
\figcaption{Single pulse sequence of PSR J2041+4551 from the FAST observation on 20220606.
\label{subfig:TP:J2041+4551}}
\end{figure}

\begin{figure}[htpb]
\centering
\includegraphics[width=0.39\textwidth, angle=0]{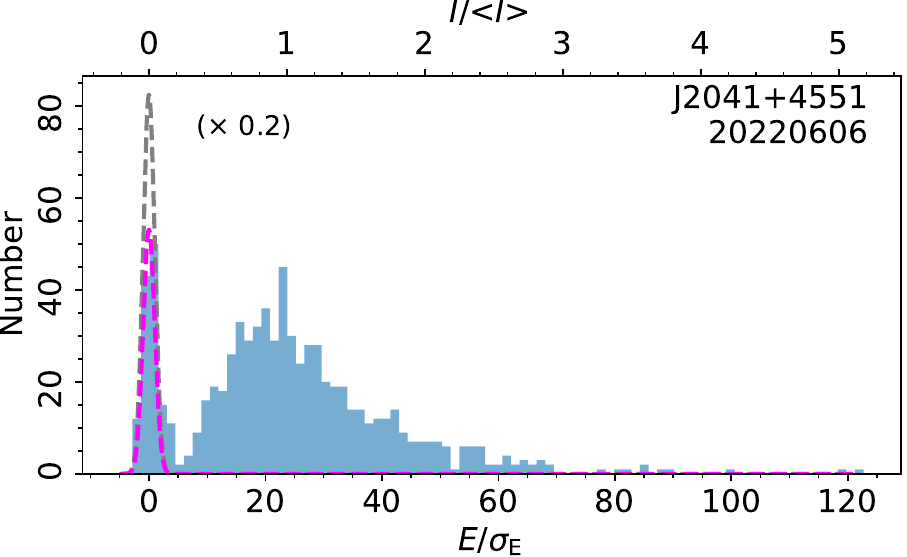}
\figcaption{On-pulse energy histogram of single pulses of PSR J2041+4551 from the FAST observation on 20220606.
\label{subfig:Hist:J2041+4551}}
\end{figure}

\begin{figure}[htpb]
\centering
\includegraphics[width=0.22\textwidth, angle=0]{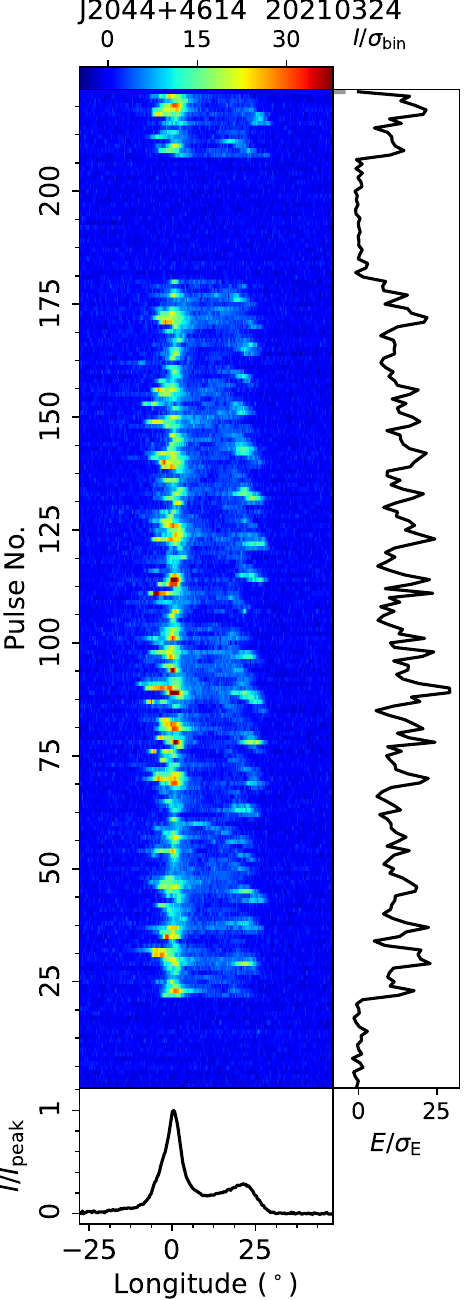}
\figcaption{Single pulse sequence of PSR J2044+4614 from the FAST observation on 20210324.
\label{subfig:TP:J2044+4614}}
\end{figure}

\begin{figure}[htpb]
\centering
\includegraphics[width=0.39\textwidth, angle=0]{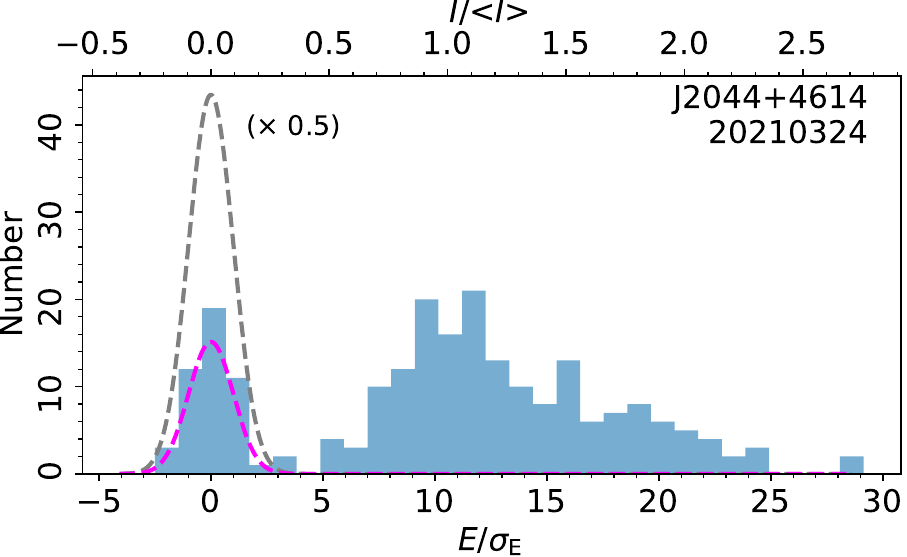}
\vspace{-0.2cm}
\figcaption{On-pulse energy histogram of single pulses of PSR J2044+4614 from the FAST observation on 20210324.
\label{subfig:fluctu:J2044+4614}}
\end{figure}

\begin{figure}[htpb]
\centering
\includegraphics[width=0.44\textwidth, angle=0]{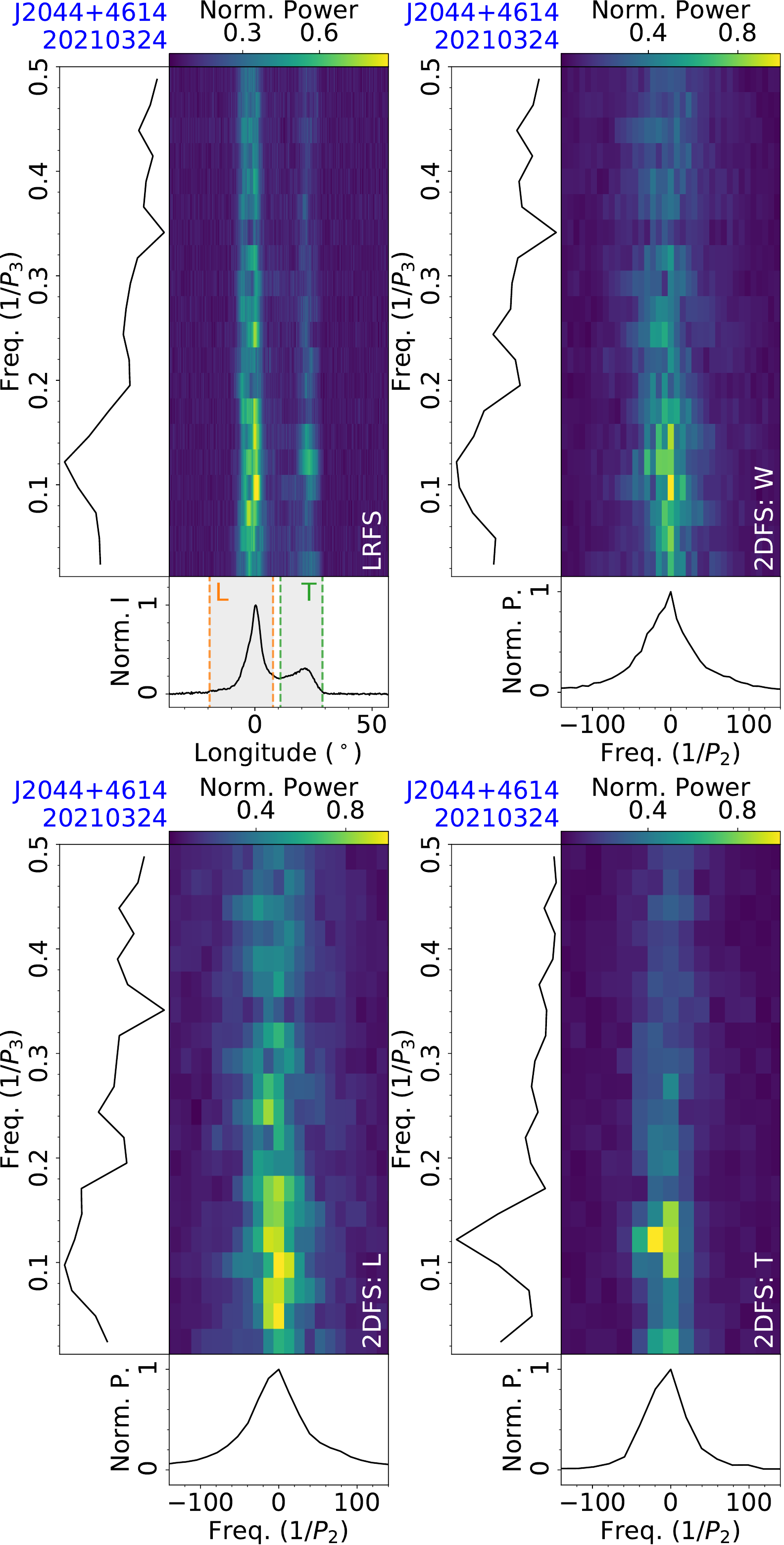}
\figcaption{Fluctuation analysis of PSR J2044+4614 from the FAST observation on 20210324, with LRFS (top-left), and 2DFS for the on-pulse region (top-right), leading part (bottom-left) and trailing part (bottom-right) of a mean pulse profile.
\label{subfig:Hist:J2044+4614}}
\end{figure}

\subsection{J2036+2835}
\label{subsec:J2036+2835}

PSR J2036+2835 was discovered in the Northern High Time Resolution Universe survey \citep{Barr2013}. 

This pulsar was observed by FAST on 20210616 for 5 minutes, deriving a rotation period $P=1.3586$~s and a dispersion measure $D\!M=83.5~{\rm cm^{-3}\,pc}$. 
Single pulse sequences are shown in Fig.~\ref{subfig:TP:J2036+2835}, which illustrates the existence of nulls. From the energy histogram in Fig.~\ref{subfig:Hist:J2036+2835}, the nulling fraction of this observation is estimated to be 37$\pm$3\%.

\begin{figure}[htpb]
\centering
\includegraphics[width=0.22\textwidth, angle=0]{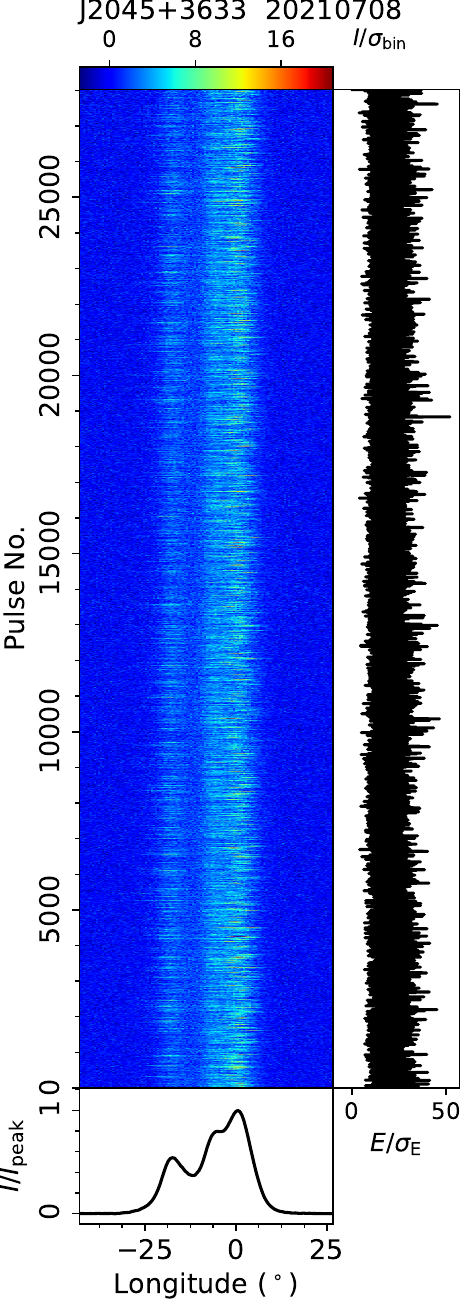}
\includegraphics[width=0.22\textwidth, angle=0]{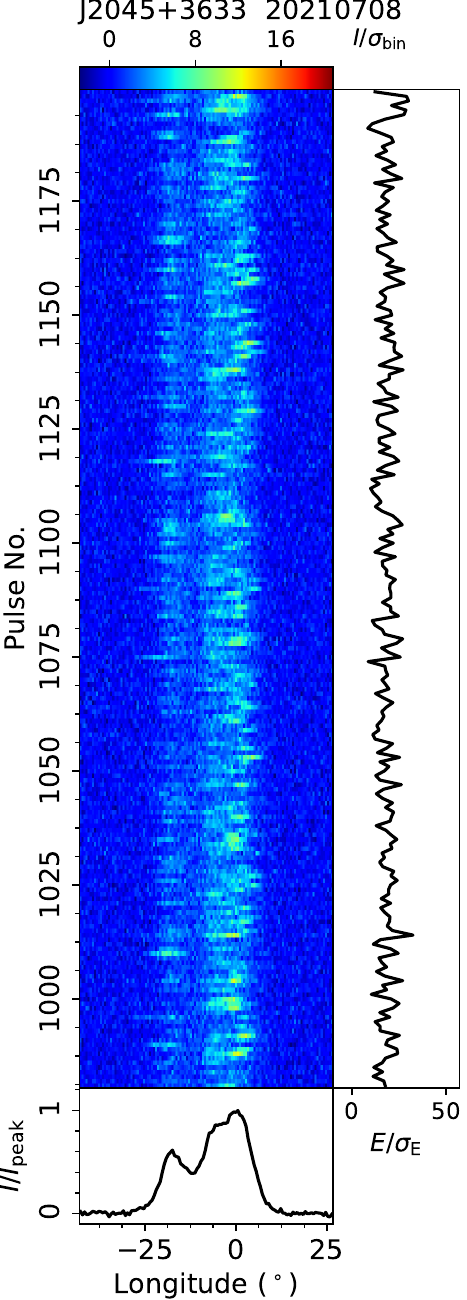}
\figcaption{Single pulse sequence of PSR J2045+3633 from the FAST observation on 20210708, and a zoomed-in view of pulses No. 980-1200.
\label{subfig:TP:J2045+3633}}
\end{figure}

\begin{figure}[htpb]
\centering
\includegraphics[width=0.44\textwidth, angle=0]{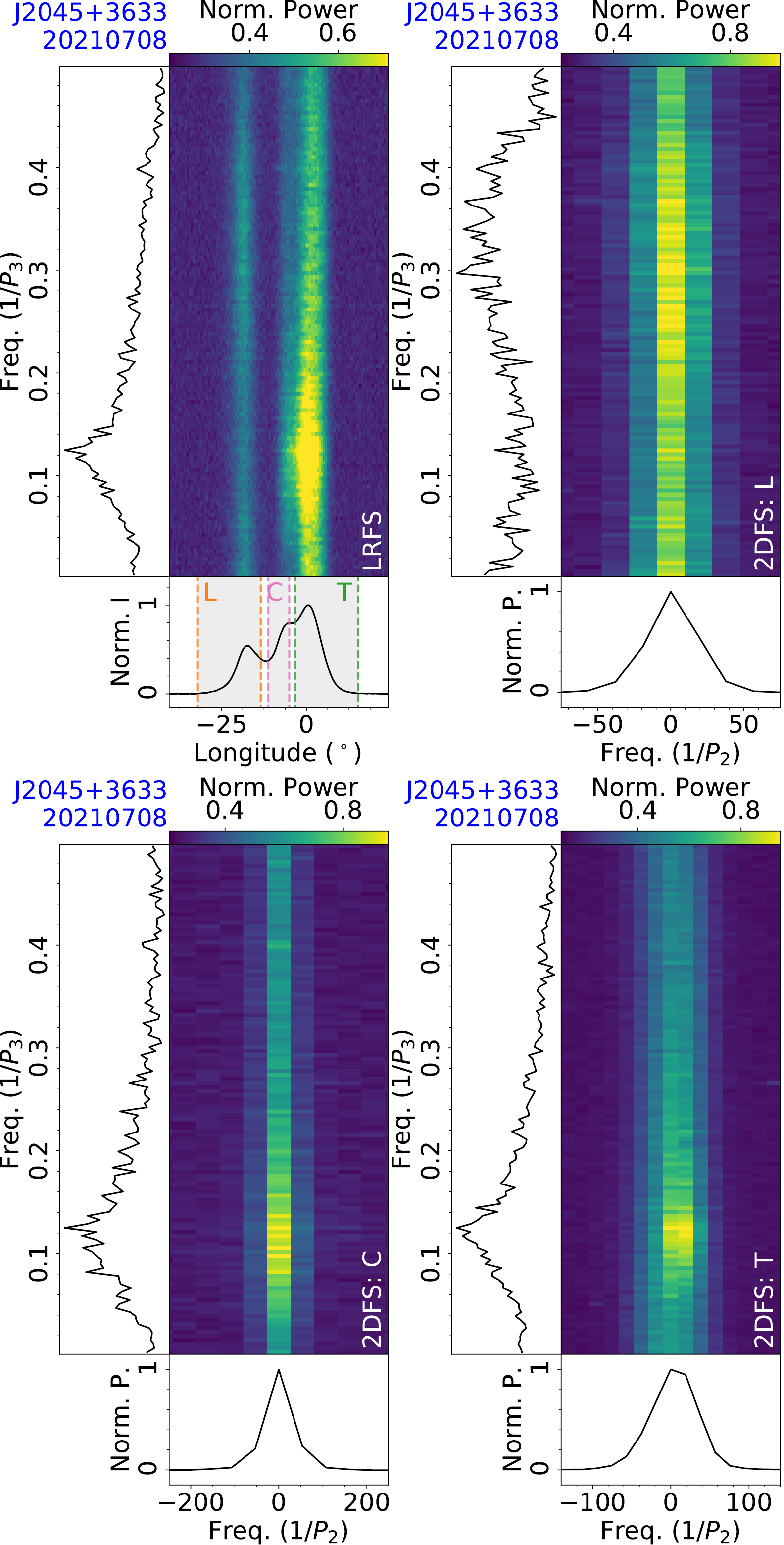}
\figcaption{Fluctuation analysis of PSR J2045+3633 from the FAST observation on 20210708, with LRFS (top-left), and 2DFS for the leading part (top-right), central part (bottom-left) and trailing part (bottom-right) of a mean pulse profile.
\label{subfig:fluctu:J2045+3633}}
\end{figure}

\begin{figure}[htpb]
\centering
\includegraphics[width=0.22\textwidth, angle=0]{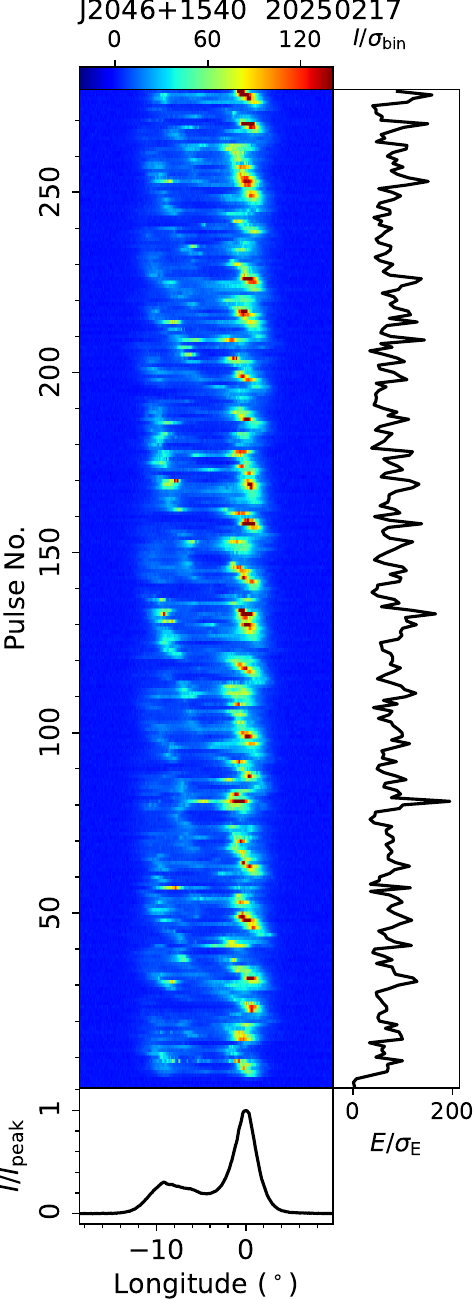}
\includegraphics[width=0.22\textwidth, angle=0]{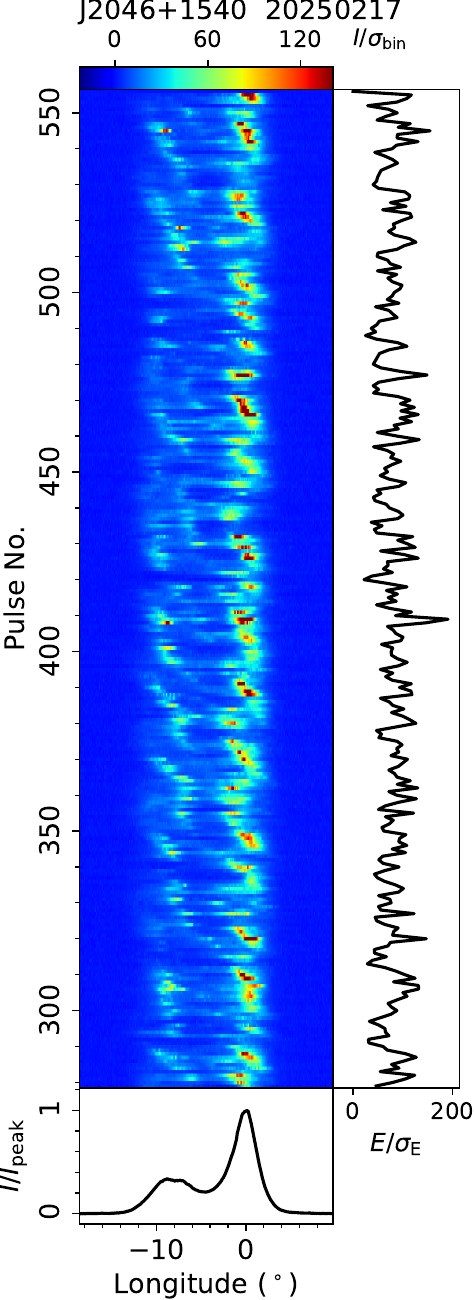}
\figcaption{Single pulse sequences of PSR J2046+1540 from FAST observations on 20250217.
\label{subfig:TP:J2046+1540}}
\end{figure}

\begin{figure}[htpb]
\centering
\includegraphics[width=0.44\textwidth, angle=0]{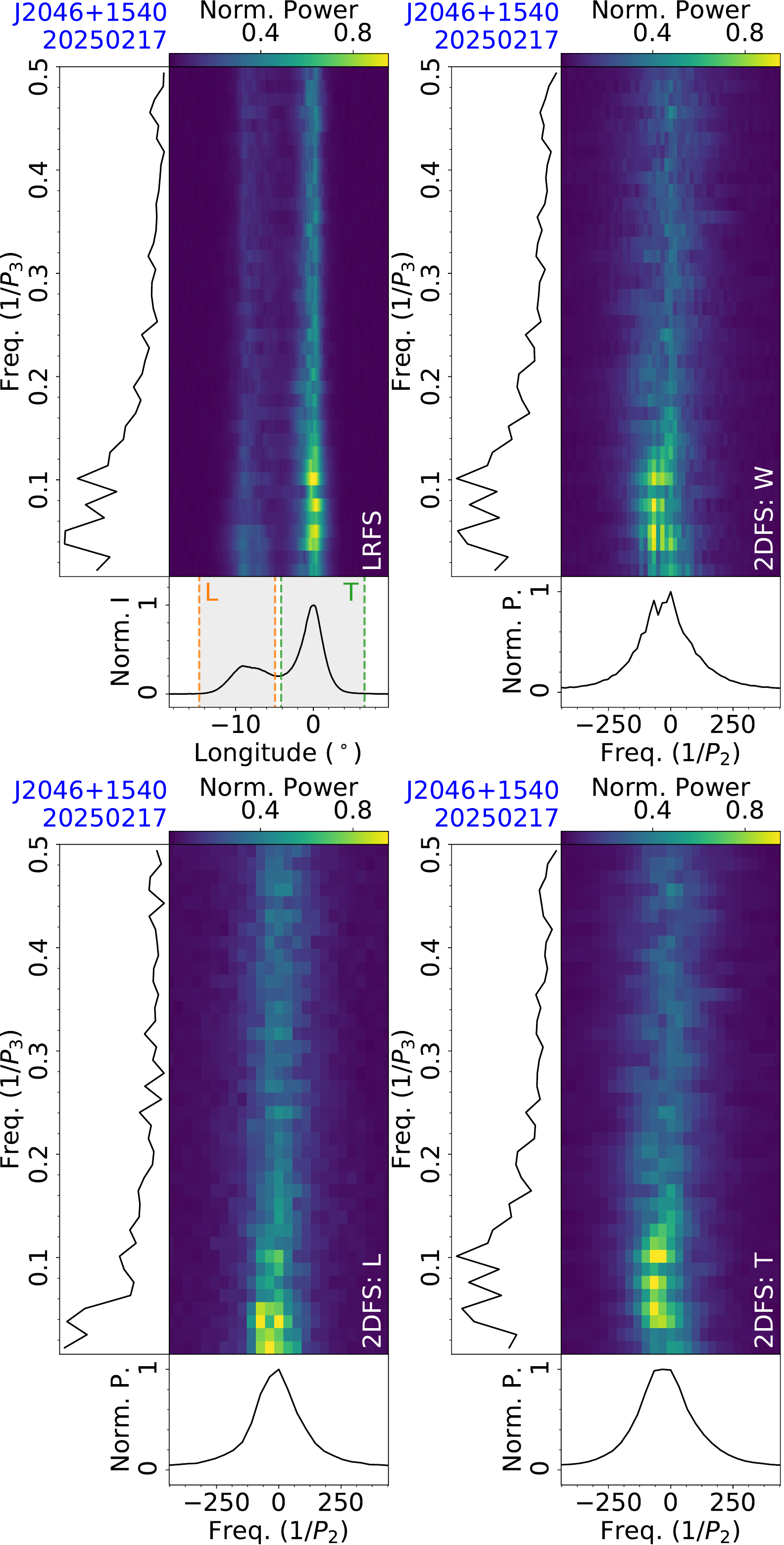}
\figcaption{Fluctuation analysis of PSR J2046+1540 from the FAST observation on 20250217, with LRFS (top-left), and 2DFS for the on-pulse region (top-right), leading part (bottom-left) and trailing part (bottom-right) of the mean pulse profile.
\label{subfig:fluctu:J2046+1540}}
\end{figure}

\begin{figure}[htpb]
\centering
\includegraphics[width=0.22\textwidth, angle=0]{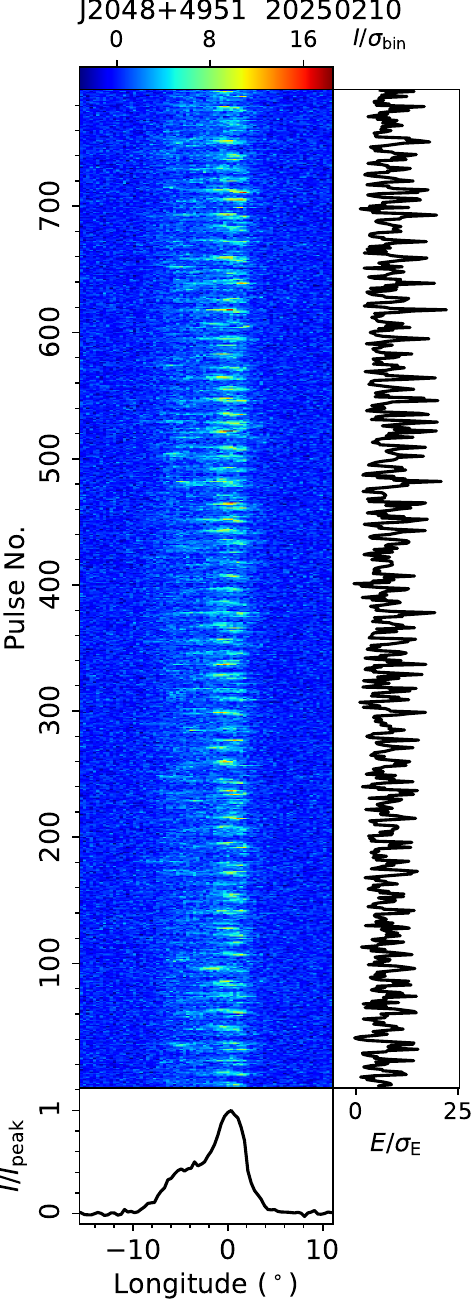}
\includegraphics[width=0.22\textwidth, angle=0]{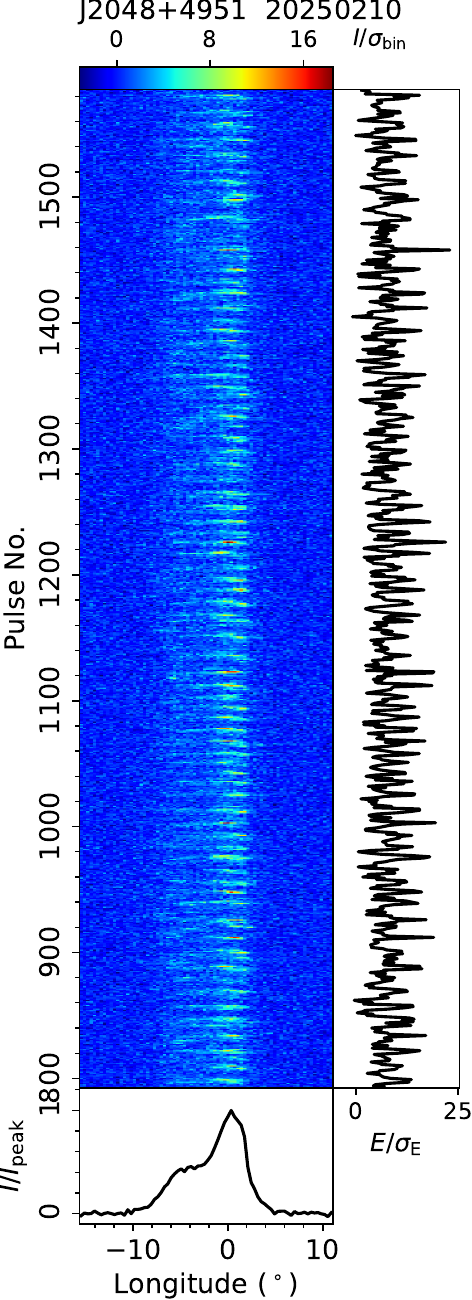}
\figcaption{Single pulse sequences of PSR J2048+4951 from the FAST observation on 20250210.
\label{subfig:TP:J2048+4951}}
\end{figure}

\begin{figure}[htpb]
\centering
\includegraphics[width=0.44\textwidth, angle=0]{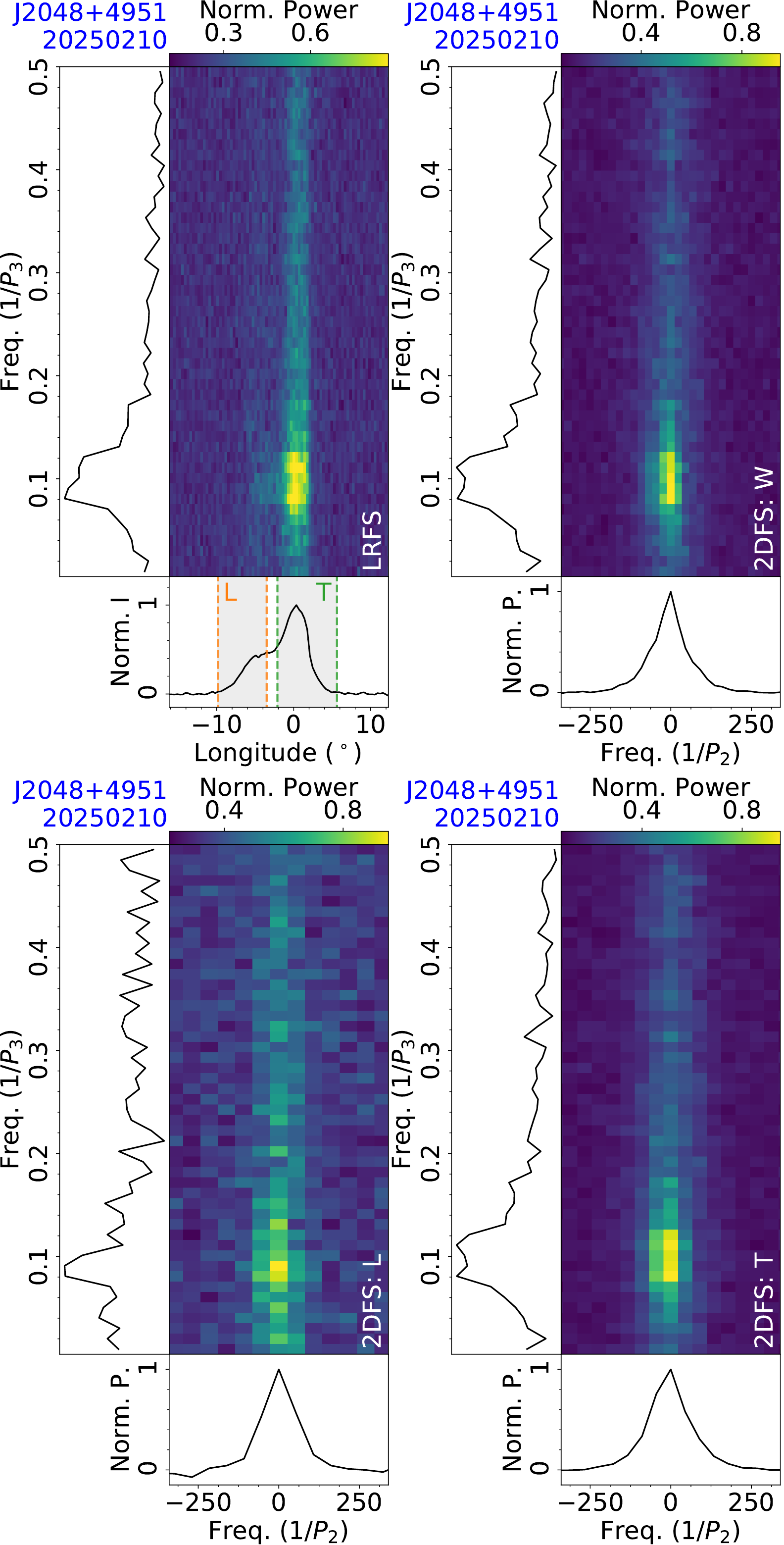}
\figcaption{Fluctuation analysis of PSR J2048+4951 for the observation on 20250210, with LRFS (top-left), and 2DFS for the on-pulse phase region (top-right), leading part (bottom-left) and trailing part (bottom-right) of a mean pulse profile.
\label{subfig:fluctu:J2048+4951}}
\end{figure}

\subsection{J2037+3621}
\label{subsec:J2037+3621}

PSR J2037+3621 was discovered in the Princeton-NRAO pulsar survey with the 92 m telescope at Green Bank \citep{Dewey1985}.

This pulsar was observed by FAST on 20230729 for 5 minutes, deriving a rotation period $P=0.6187$~s and a dispersion measure $D\!M=93.9~{\rm cm^{-3}\,pc}$. The single pulse sequence in Fig.~\ref{subfig:TP:J2037+3621} displays a negative drifting behavior. Fluctuation spectra are shown in Fig.~\ref{subfig:fluctu:J2037+3621}, and the centroid of the negative drift feature in 2DFS is at $1/P_3=0.031\pm0.001$ and $1/P_2=-3\pm1$, corresponding to periodicities of $P_3=32\pm1$ periods and $P_2=-109\pm40$ degrees.

\subsection{J2038+5319}
\label{subsec:J2038+5319}

PSR J2038+5319 was discovered in the Princeton-NRAO pulsar survey using the 92 m telescope at Green Bank \citep{Dewey1985}. 

This pulsar was observed by FAST on 20221122 for 5 minutes, yielding a rotation period $P=1.4246$~s and a dispersion measure $D\!M=160.0~{\rm cm^{-3}\,pc}$. Single pulse sequences in Fig.~\ref{subfig:TP:J2038+5319} display the subpulse drifting phenomenon in a positive direction. From LRFS and 2DFS in Fig.~\ref{subfig:fluctu:J2038+5319}, the centroid of the drift feature is characterized by $1/P_3=0.3476\pm0.0004$ and $1/P_2=44\pm2$, which correspond to $P_3=2.877\pm0.004$ periods and $P_2=8.3\pm0.3^\circ$.

\begin{figure}[htpb]
\centering
\includegraphics[width=0.22\textwidth, angle=0]{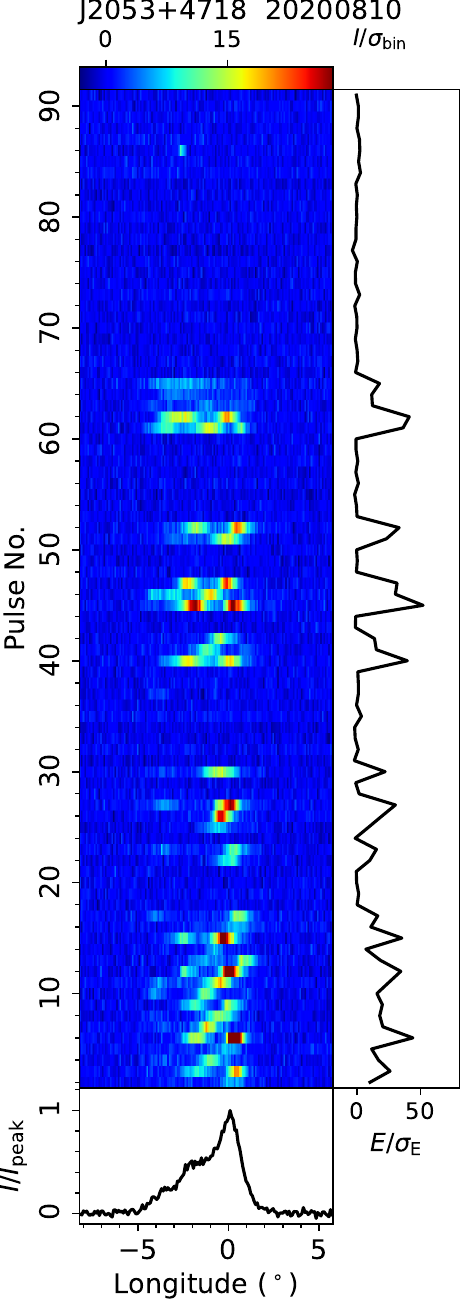}
\includegraphics[width=0.22\textwidth, angle=0]{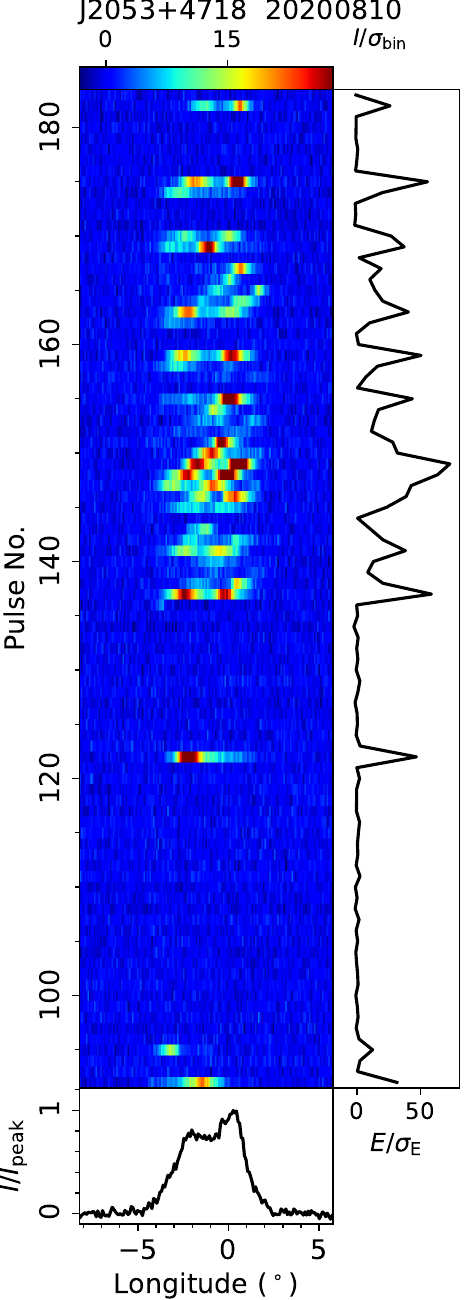}
\figcaption{Single pulse sequences of PSR J2053+4718 from the FAST observation on 20200810.
\label{subfig:TP:J2053+4718}}
\end{figure}

\begin{figure}[htpb]
\centering
\includegraphics[width=0.39\textwidth, angle=0]{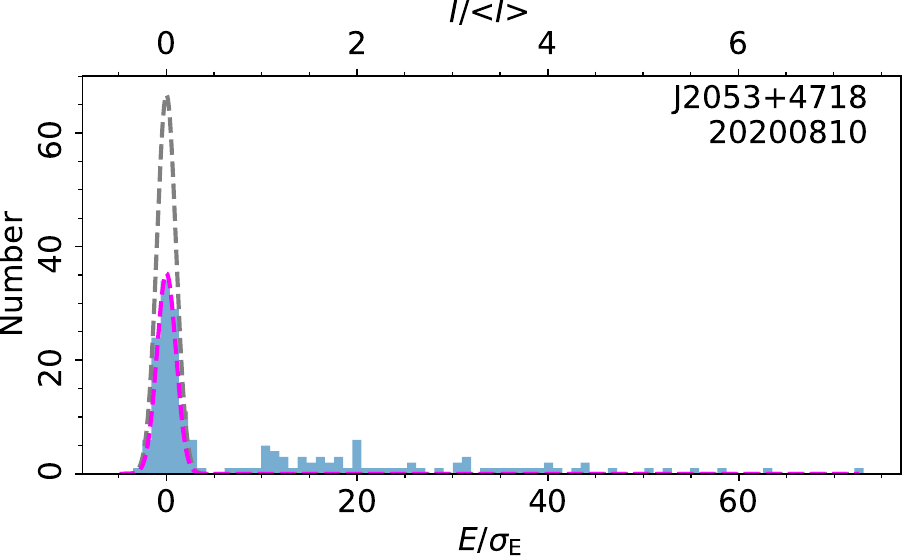}
\figcaption{On-pulse energy histogram of single pulses of PSR J2053+4718 from the FAST observation on 20210621.
\label{subfig:Hist:J2053+4718}}
%
\centering
\includegraphics[width=0.39\textwidth, angle=0]{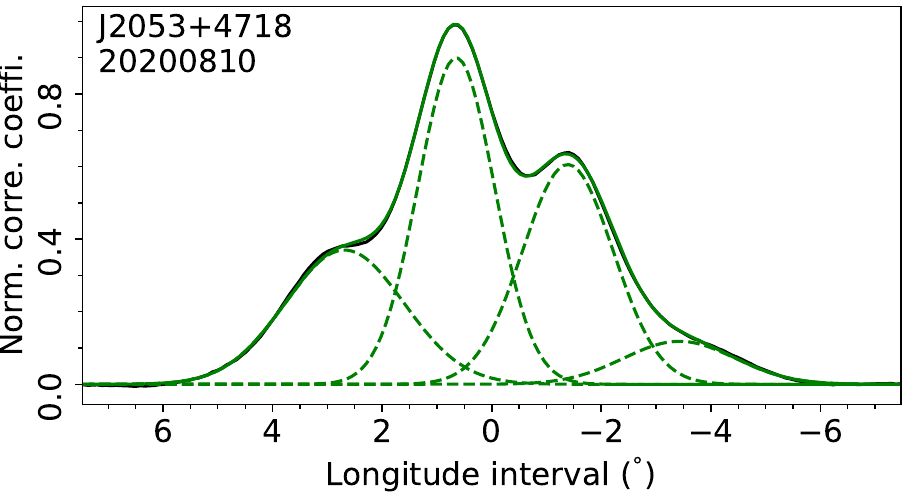}
\figcaption{Cross correlation of PSR J2053+4718 from the FAST observation on 20200810.
\label{subfig:Corre:J2053+4718}}
\end{figure}

\subsection{J2041+4551}
\label{subsec:J2041+4551}

PSR J2041+4551 was discovered in the Green Bank 820 MHz Pulsar Survey \citep{McEwen2024}. 

This pulsar was observed by FAST on 20220606 for 15 minutes, deriving a rotation period $P=1.1597$~s and a dispersion measure $D\!M=307.2~{\rm cm^{-3}\,pc}$. The single pulse sequence in Fig.~\ref{subfig:TP:J2041+4551} shows the nulling phenomenon, with the nulling fraction estimated to be 13$\pm$1\% from the on-pulse energy histogram (Fig.~\ref{subfig:Hist:J2041+4551}). 

The nulling phenomenon was also observed in FAST observations on 20240106 for 4 minutes and on 20240402 for 2 minutes. However, no nulling was observed in the data from 20220502 (5 minutes), 20240814 (2 minutes), and 20241026 (5 minutes).

\subsection{J2044+4614}
\label{subsec:J2044+4614}

PSR J2044+4614 was first reported by \citet{Sayer1996} from the observation of the 43 m telescope at Green Bank. \citet{Ng2020} observed the pulsar in the frequency range of 400-800 MHz and reported the nulling fraction of more than 9\%. 

This pulsar was observed by FAST on 20210324 for 5 minutes, deriving a rotation period $P=1.3927$~s and a dispersion measure $D\!M=316.0~{\rm cm^{-3}\,pc}$. 
The single pulse sequence in Fig.~\ref{subfig:TP:J2044+4614} shows the clear nulling phenomenon. There is a clear peak around zero in the on-pulse integral energy histogram (Fig.~\ref{subfig:Hist:J2044+4614}), indicating the existence of nulls with a fraction of 17$\pm$2\% for this observation. 
Fluctuation spectra are shown in Fig.~\ref{subfig:Hist:J2044+4614}. 
For the leading part of a mean pulse profile, the main drift feature is widely distributed in 2DFS, which exhibits a centroid of $1/P_3=0.110\pm0.003$ and $1/P_2=-3\pm1$, corresponding to periodicities of $P_3=9.1\pm0.2$ periods and $P_2=-103\pm32^\circ$. 
Subpulse drifting of the trailing profile part is more systematic than that of the leading part. 
2DFS of the trailing profile part has the centroid frequencies of $1/P_3=0.122\pm0.003$ and $1/P_2=-15\pm2$, yielding $P_3=8.2\pm0.2$ periods and $P_2=-24\pm3^\circ$.
In the single pulse sequence, there seems to be a quasi-periodic variation of the trailing longitude edge, which requires longer observation to confirm.

\begin{figure}[htpb]
\centering
\includegraphics[width=0.22\textwidth, angle=0]{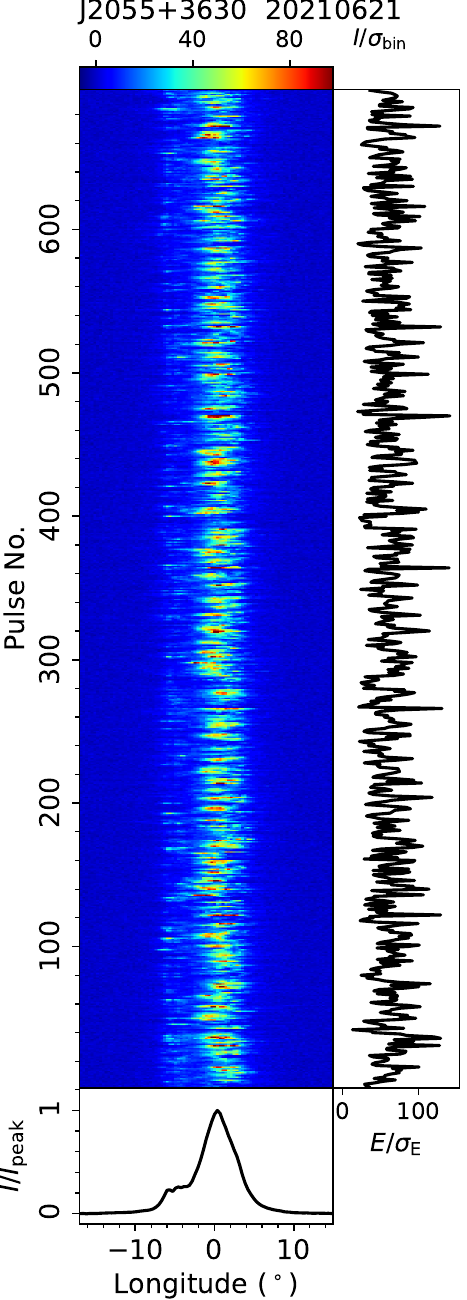}
\includegraphics[width=0.22\textwidth, angle=0]{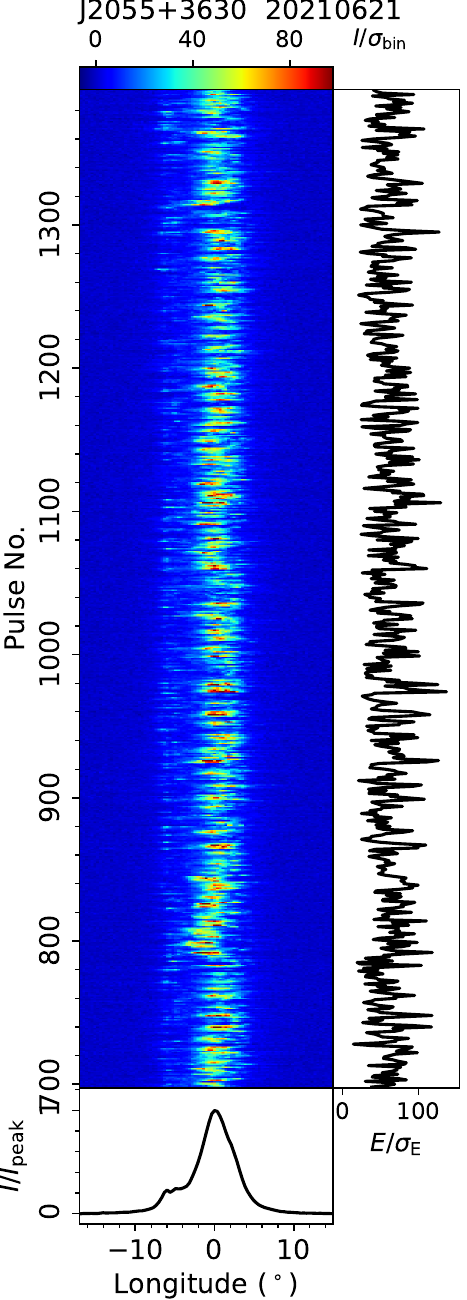}
\figcaption{Single pulse sequences of PSR J2055+3630 from the FAST observation on 20210621.
\label{subfig:TP:J2055+3630}}
\end{figure}

\begin{figure}[htpb]
\centering
\includegraphics[width=0.39\textwidth, angle=0]{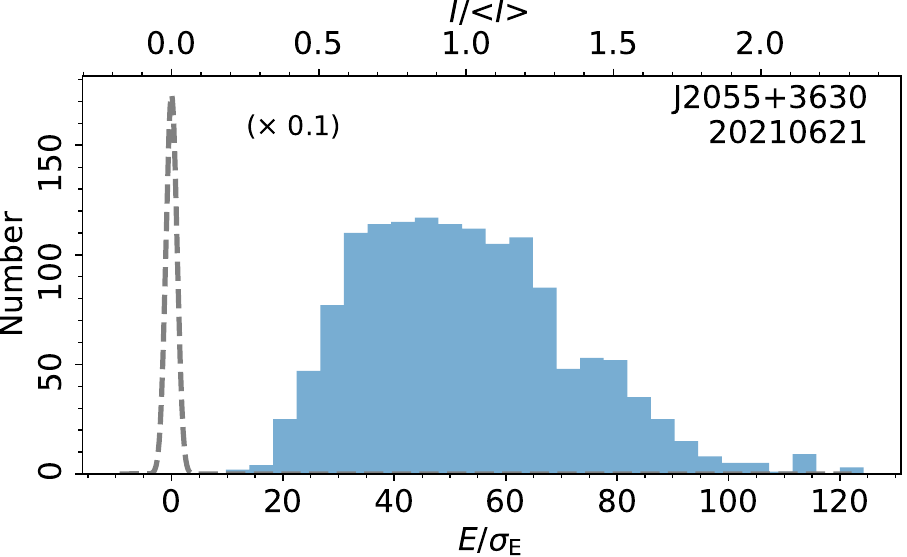}
\figcaption{On-pulse energy histogram of single pulses of PSR J2055+3630 from the FAST observation on 20210621.
\label{subfig:Hist:J2055+3630}}
\end{figure}

\begin{figure}[htpb]
\centering
\includegraphics[width=0.22\textwidth, angle=0]{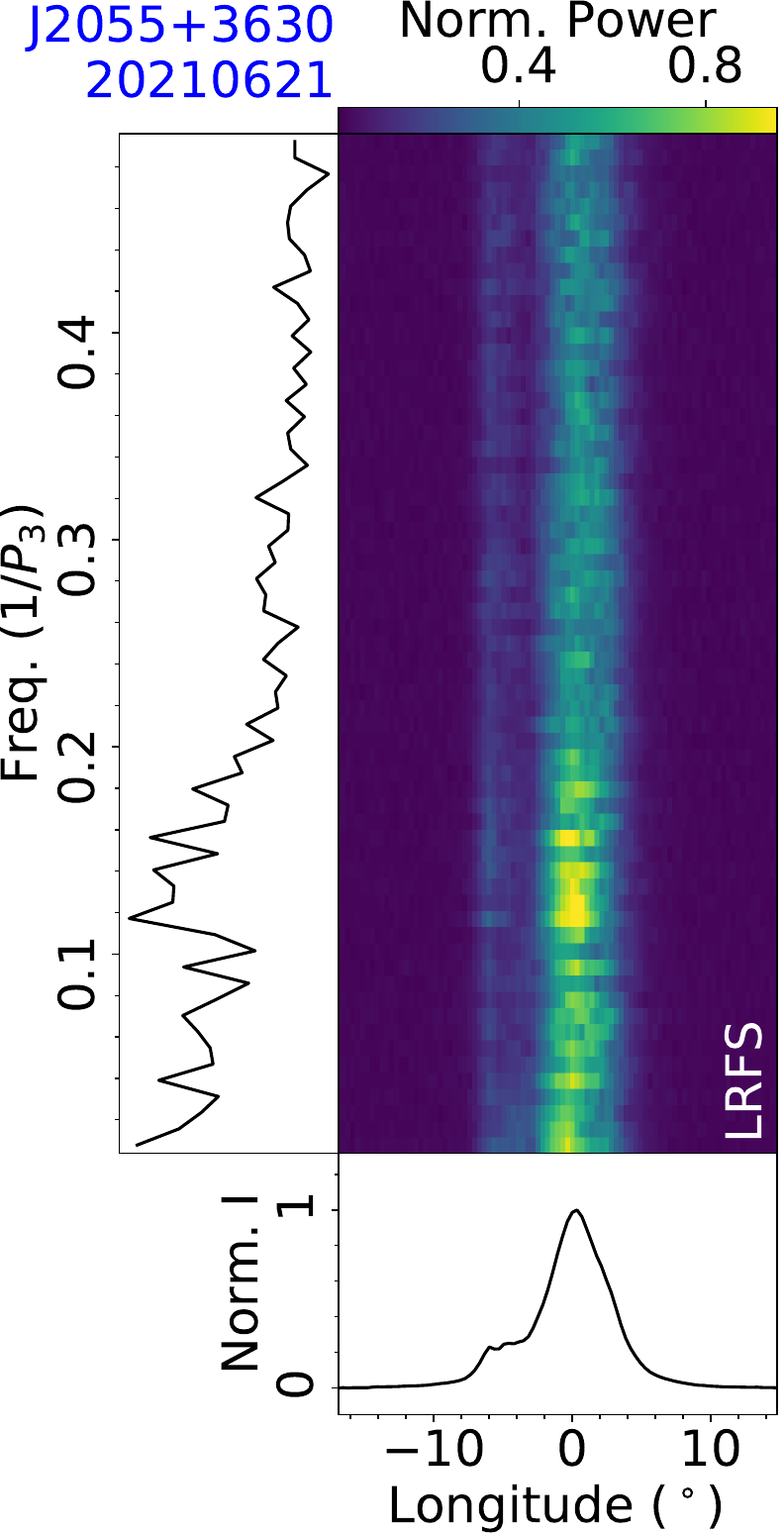}
\includegraphics[width=0.22\textwidth, angle=0]{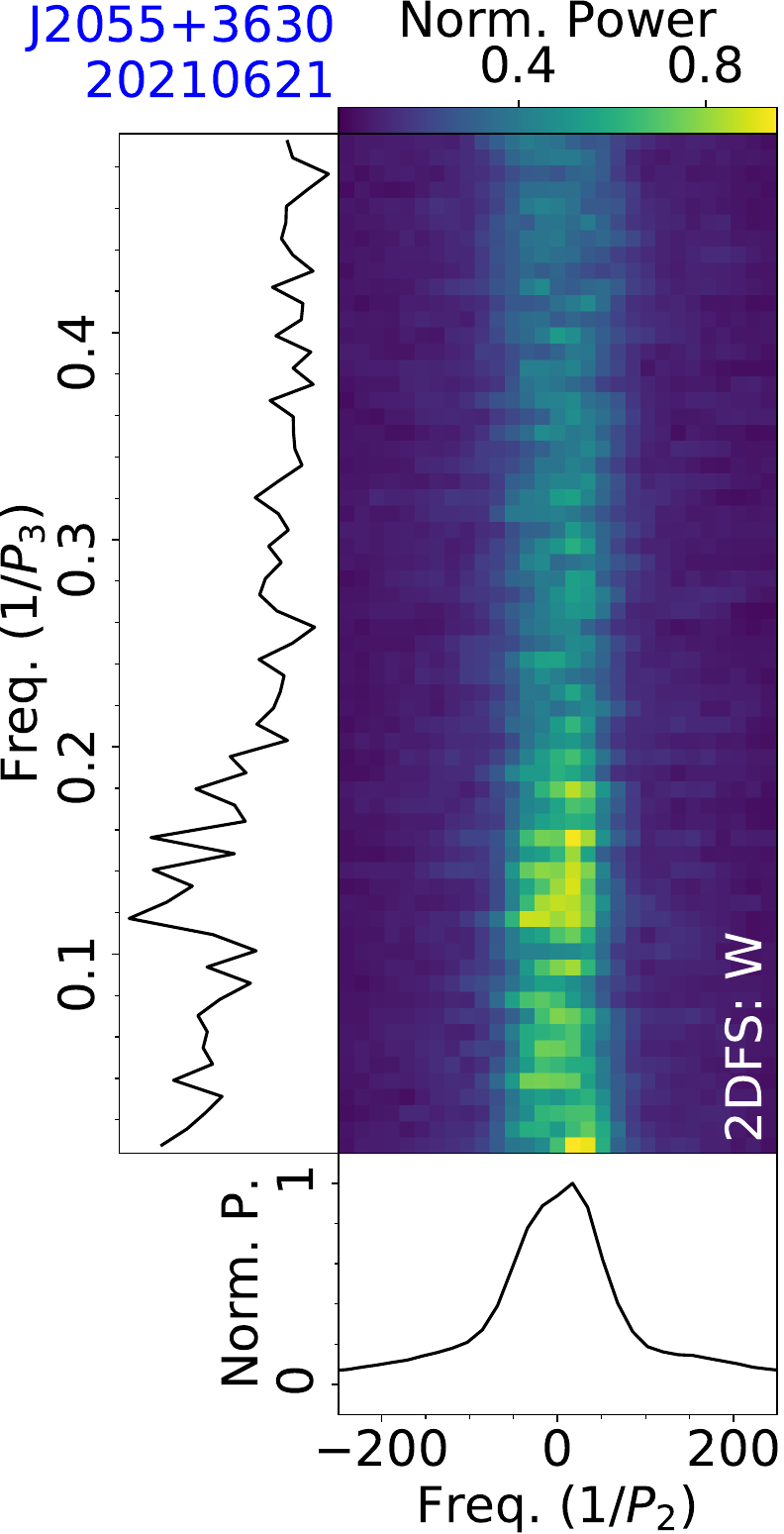}
\figcaption{Fluctuation analysis of PSR J2055+3630 for the observation on 20210621, with LRFS and 2DFS for the on-pulse region of a mean pulse profile.
\label{subfig:fluctu:J2055+3630}}
\end{figure}

\subsection{J2045+3633}
\label{subsec:J2045+3633}

PSR J2045+3633 was discovered by \citet{bcf+17} in the Northern High Time Resolution Universe pulsar survey using the Effelsberg radio telescope at 1.4 GHz. 

This pulsar was observed by FAST on 20200404 for 5 minutes and 20210708 for 15 minutes. From the 15-minute observation, a rotation period $P=0.0317$~s and a dispersion measure $D\!M=129.6~{\rm cm^{-3}\,pc}$ were determined. 
Single pulse sequences of the data on 20210708 are shown in Fig.~\ref{subfig:TP:J2045+3633}. This pulsar is found to have modulation behavior from the fluctuation spectra in Fig.~\ref{subfig:fluctu:J2045+3633}, and the temporal modulation periodicities are different for different longitude regions corresponding to different components. The modulation frequencies on time of three longitude parts in the profile are widely distributed in 2DFS. 
In 2DFS of the leading profile part, there seems a feature with the centroid modulation frequency of $1/P_3=0.307\pm0.001$, corresponding to $P_3=3.26\pm0.01$, as well as a low-frequency feature. However, the central and trailing longitude ranges have different properties with the leading range. For the central profile part, the centroid temporal modulation frequency is $1/P_3=0.120\pm0.001$, yielding $P_3=8.35\pm0.05$ periods. In 2DFS of the trailing profile part, there is a subpulse drift feature with the centroid frequencies of $1/P_3=0.1172\pm0.0003$ ($P_3=8.54\pm0.02$ periods) and $1/P_2=11\pm1$ ($P_2=33\pm2^\circ$).

\subsection{J2046+1540}
\label{subsec:J2046+1540}

PSR J2046+1540 was discovered in the Molonglo pulsar survey \citep{Manchester1978}. Negative drifting was reported in previous studies. At 21 cm, \citet{Weltevrede2006} measured the drifting of the trailing component with $P_3=18\pm6$ periods and $P_2=-7.1^{+0.5}_{-1.4}$ degrees. At 92 cm, \citet{Weltevrede2007} reported values of $P_3=18\pm4$ periods, $P_2=-25^{+10}_{-10}$ degrees and $P_3=13\pm1$ periods, $P_2=-16^{+3}_{-4}$ degrees for two components. \citet{Basu2016} presented negative drifting with $P_3=23.0\pm6.1$ periods at 618 MHz.  \citet{Song2023} recently reported that at 1280 MHz the leading component has $P_3=27\pm10$ periods and $P_2=-15^{+3}_{-8}$ degrees, while the trailing component shows two drift features: one with $P_3=23\pm4$ periods and $P_2=-6^{+1}_{-3}$ degrees, and another with $P_3=9.0\pm0.8$ periods and $P_2=-8^{+2}_{-2}$ degrees.

This pulsar was observed by FAST on 20250217 for 11 minutes, deriving a rotation period $P=1.1383$~s and a dispersion measure $D\!M=39.7~{\rm cm^{-3}\,pc}$. Single pulse sequences in Fig.~\ref{subfig:TP:J2046+1540} show that both the leading and trailing parts of the mean pulse profile exhibit negative drifting with variable drifting rates. Fluctuation spectra are shown in Fig.~\ref{subfig:fluctu:J2046+1540}, and drift features appear temporally broad, reflecting variable drifting behavior. In 2DFS of the leading profile part, the drift feature has a centroid at $1/P_3=0.049\pm0.001$ and $1/P_2=-38\pm3$, corresponding to $P_3=20.2\pm0.4$ periods and $P_2=-10\pm1$ degrees. The centroid of the drift feature is at $1/P_3=0.080\pm0.001$ and $1/P_2=-56\pm2$, yielding $P_3=12.5\pm0.1$ periods and $P_2=-6.5\pm0.2$ degrees.

\begin{figure}[htpb]
\centering
\includegraphics[width=0.22\textwidth, angle=0]{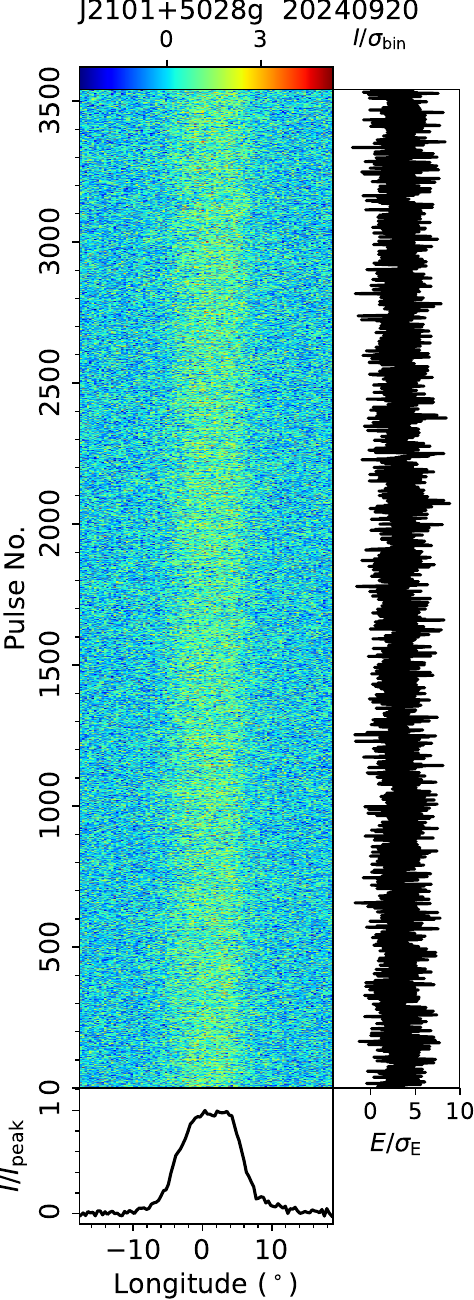}
\includegraphics[width=0.22\textwidth, angle=0]{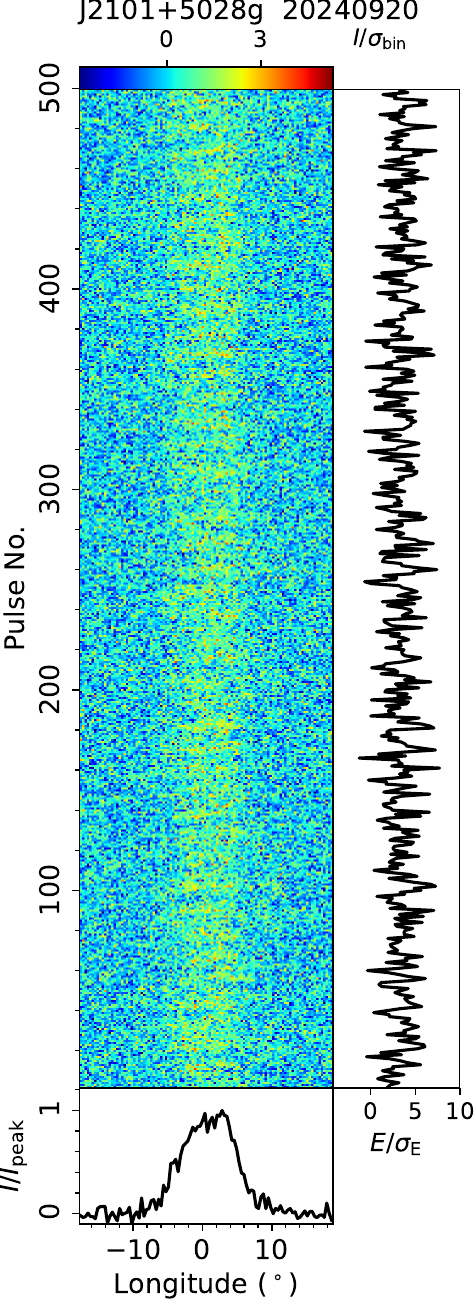}
\figcaption{Single pulse sequences of PSR J2101+5028g from the FAST observation on 20240920.
\label{subfig:TP:J2101+5028g}}
\end{figure}

\begin{figure}[htpb]
\centering
\includegraphics[width=0.22\textwidth, angle=0]{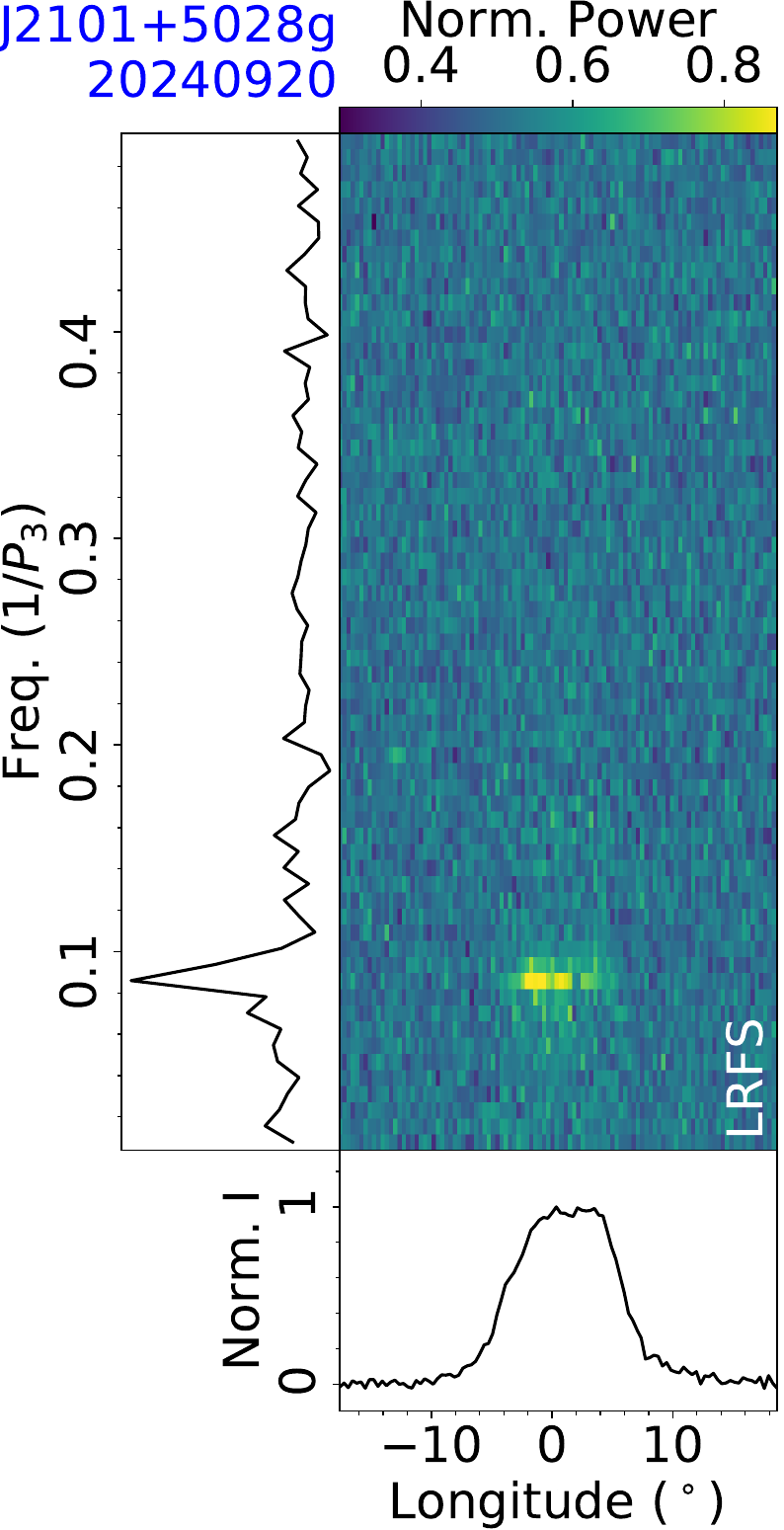}
\includegraphics[width=0.22\textwidth, angle=0]{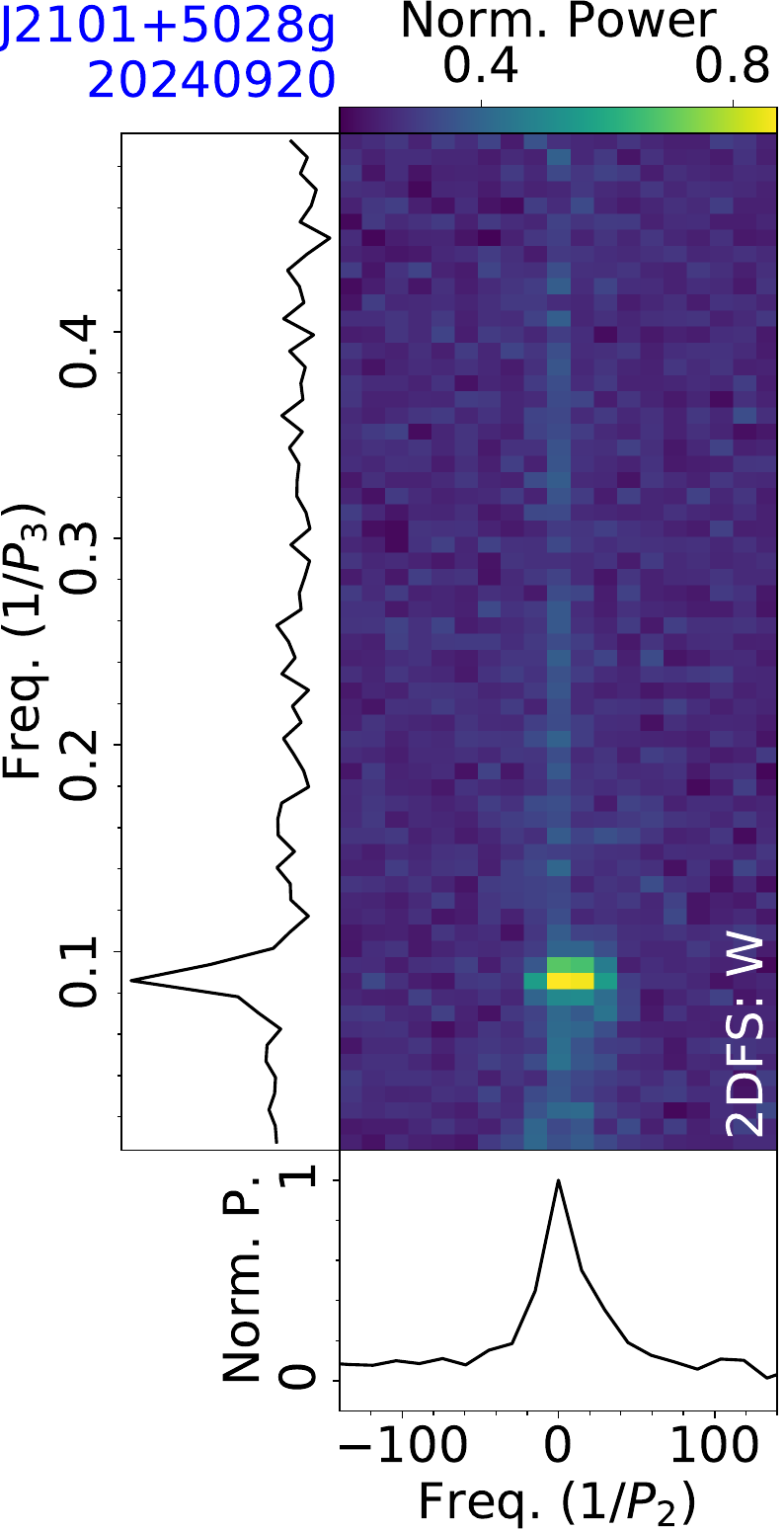}
\figcaption{Fluctuation analysis of PSR J2101+5028g for the observation on 20240920, with LRFS and 2DFS for the on-pulse region of a mean pulse profile.
\label{subfig:fluctu:J2101+5028g}}
\end{figure}

\subsection{J2048+4951}
\label{subsec:J2048+4951}

PSR J2048+4951 was discovered in the SPAN512 pulsar survey using the L-band receiver of the Nançay Radio Telescope \citep{Desvignes2022}. 

This pulsar was observed by FAST on 20240731 for 5 minutes and on 20250210 for 15 minutes. From the 15-minute data, a rotation period $P=0.5683$~s and a dispersion measure $D\!M=223.2~{\rm cm^{-3}\,pc}$ were derived. Single pulse sequences of the observation on 20250210 are shown in Fig.~\ref{subfig:TP:J2048+4951}, illustrating the existence of a temporal modulation phenomenon. LRFS and 2DFS are shown in Fig.~\ref{subfig:fluctu:J2048+4951}. 
2DFS of the leading part in the mean pulse profile has a temporal modulation feature with the centroid frequency of $1/P_3=0.085\pm0.002$, corresponding to $P_3=11.8\pm0.2$ periods. 
In the 2DFS of the trailing profile part, the negative drift feature exhibits the centroid of $1/P_3=0.097\pm0.001$ and $1/P_2=-20\pm4$, yielding $P_3=10.3\pm0.1$ periods and $P_2=-18\pm4^\circ$.
Single-pulse features of the observation on 20240731 are consistent with those on 20250210.

\begin{figure}[htpb]
\centering
\includegraphics[width=0.22\textwidth, angle=0]{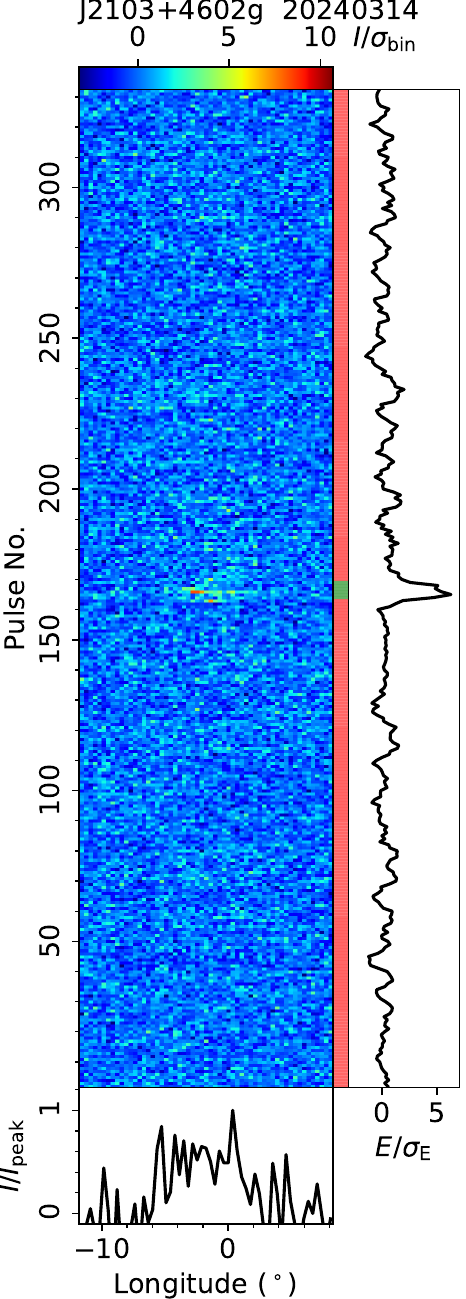}
\includegraphics[width=0.22\textwidth, angle=0]{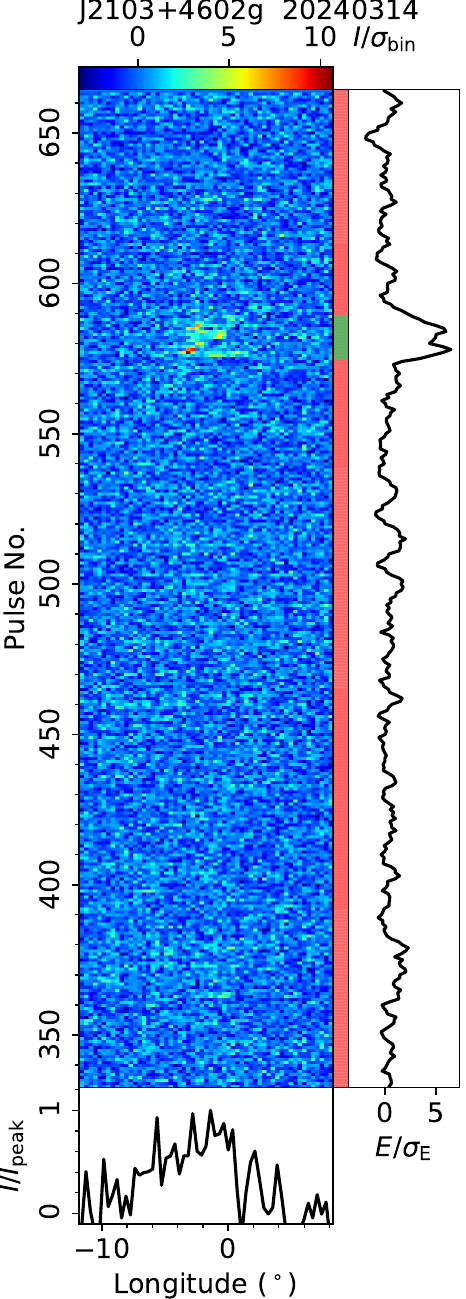}
\figcaption{Single pulse sequences of PSR J2103+4602g from the FAST observation on 20240314. The on-pulse energy variation versus period is smoothed over every 5 periods.
\label{subfig:TP:J2103+4602g}}
\end{figure}

\begin{figure}[htpb]
\centering
\includegraphics[width=0.39\textwidth, angle=0]{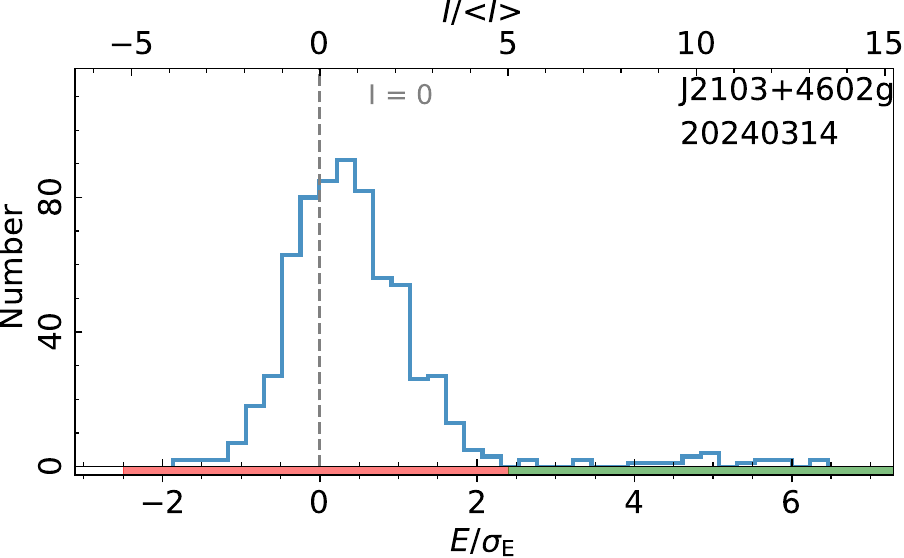}
\figcaption{Histogram of the smoothed on-pulse integral energy for single pulses of PSR J2103+4602g from the FAST observation on 20240314. The red and green bars indicate the weak and bright emission modes.
\label{subfig:Hist:J2103+4602g}}
\end{figure}

\begin{figure}[htpb]
\centering
\includegraphics[width=0.22\textwidth, angle=0]{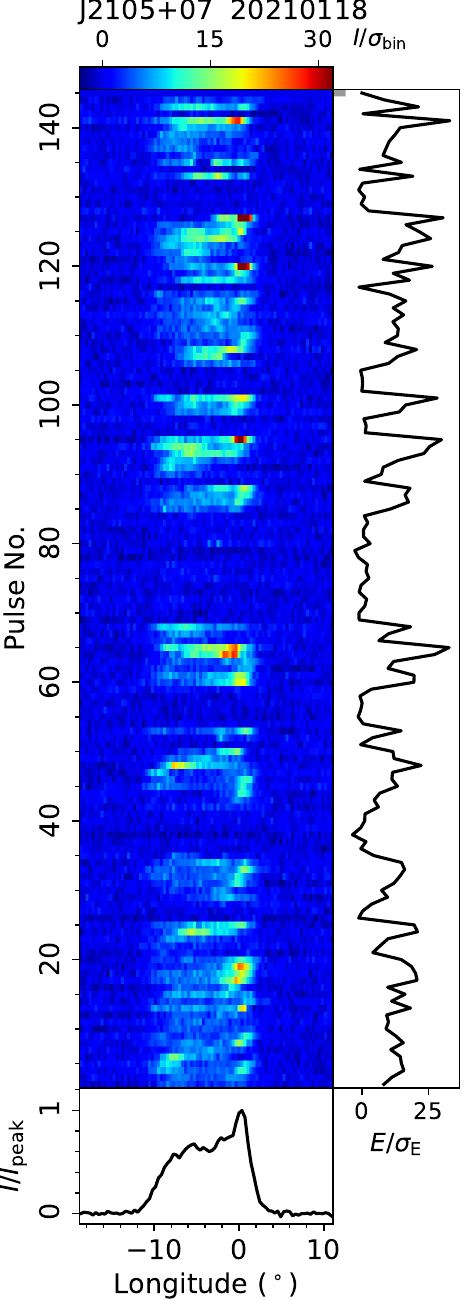}
\figcaption{Single pulse sequence of PSR J2105+07 from the FAST observation on 20210118.
\label{subfig:TP:J2105+07}}
\end{figure}

\begin{figure}[htpb]
\centering
\includegraphics[width=0.39\textwidth, angle=0]{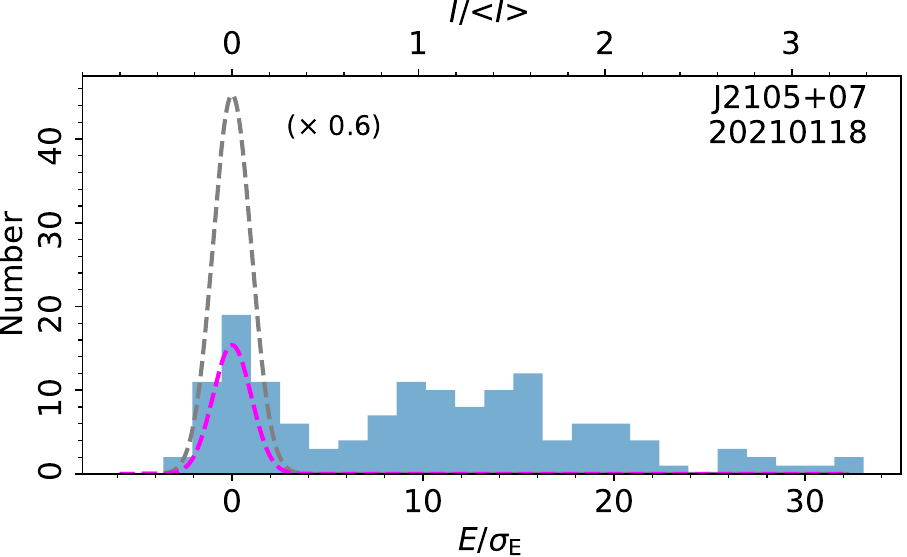}
\figcaption{On-pulse energy histogram of single pulses of PSR J2105+07 from the FAST observation on 20210118.
\label{subfig:Hist:J2105+07}}
\end{figure}

\begin{figure}[htpb]
\centering
\includegraphics[width=0.22\textwidth, angle=0]{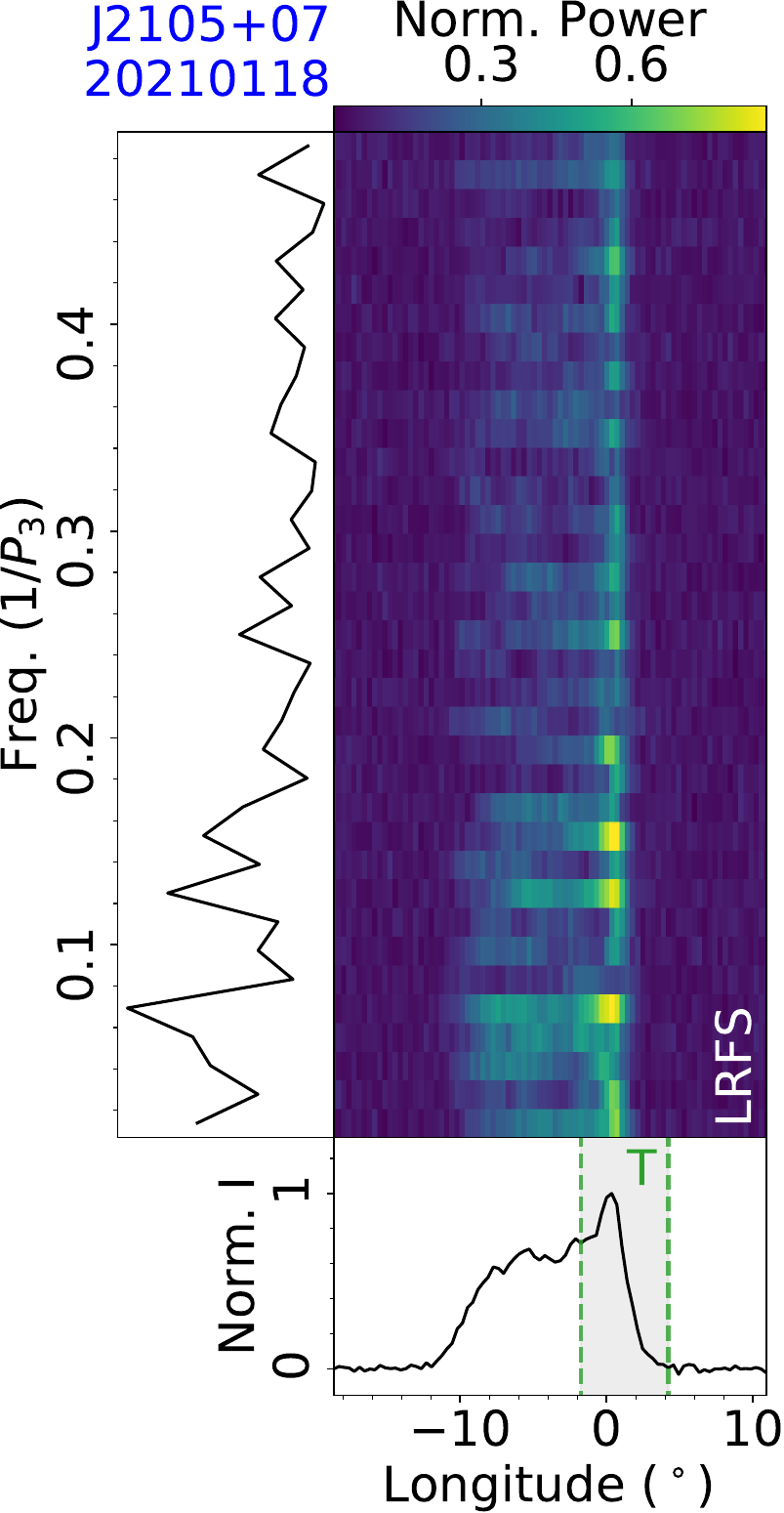}
\includegraphics[width=0.22\textwidth, angle=0]{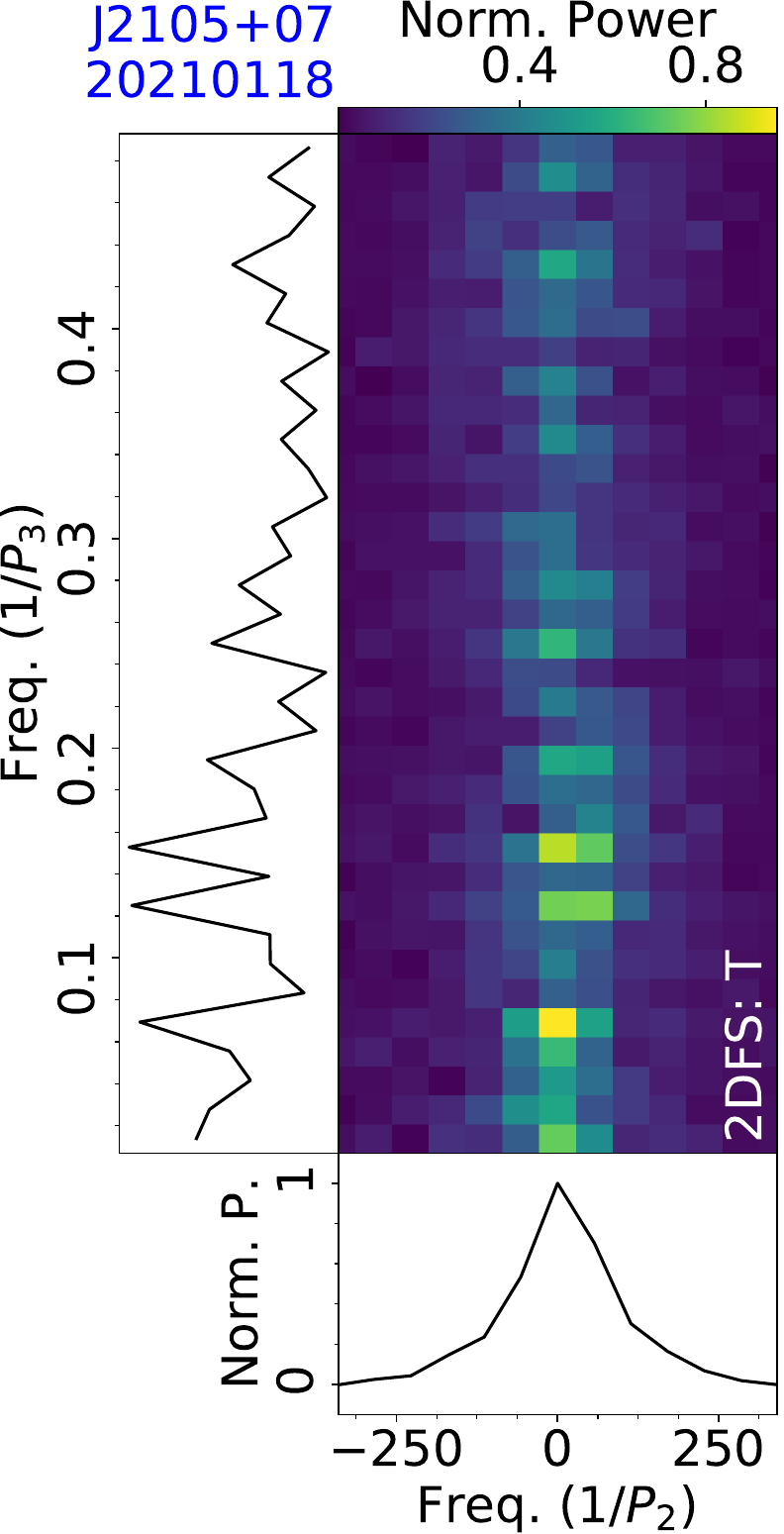}
\figcaption{Fluctuation analysis of PSR J2105+07 for the observation on 20210118, with LRFS and 2DFS for the leading part of a mean pulse profile.
\label{subfig:fluctu:J2105+07}}
\end{figure}

\subsection{J2053+4718}
\label{subsec:J2053+4718}

PSR J2053+4718 was discovered by FAST in the Commensal Radio Astronomy FAST Survey (CRAFTS) \citep{Cruces2021}.

The pulsar was observed by FAST on 20200810 for 15 minutes, deriving a rotation period $P=4.9102$~s and a dispersion measure $D\!M=322.3~{\rm cm^{-3}\,pc}$ were determined. 
Single pulse sequences in Fig.~\ref{subfig:TP:J2053+4718} display nulling and positive subpulse drifting behaviors. The nulling fraction of this observation is estimated to be 53$\pm$2\% from the on-pulse integral energy histogram (Fig.~\ref{subfig:Hist:J2053+4718}). Drifting parameters are obtained from the correlation method, which are $D=0.65\pm0.02$ degrees per period and $P_2=2.04\pm0.01^\circ$.

\begin{figure}[htbp]
\centering
\includegraphics[width=0.22\textwidth, angle=0]{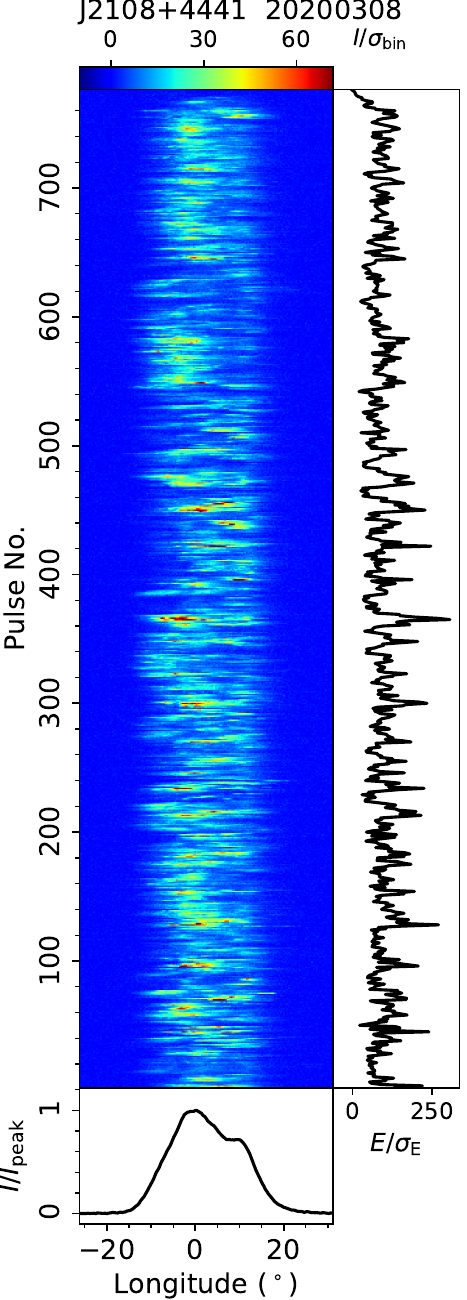}
\figcaption{Single pulse sequence of PSR J2108+4441 from the FAST observation on 20200308.
\label{subfig:TP:J2108+4441}}
\end{figure}

\begin{figure}[htbp]
\centering
\includegraphics[width=0.22\textwidth, angle=0]{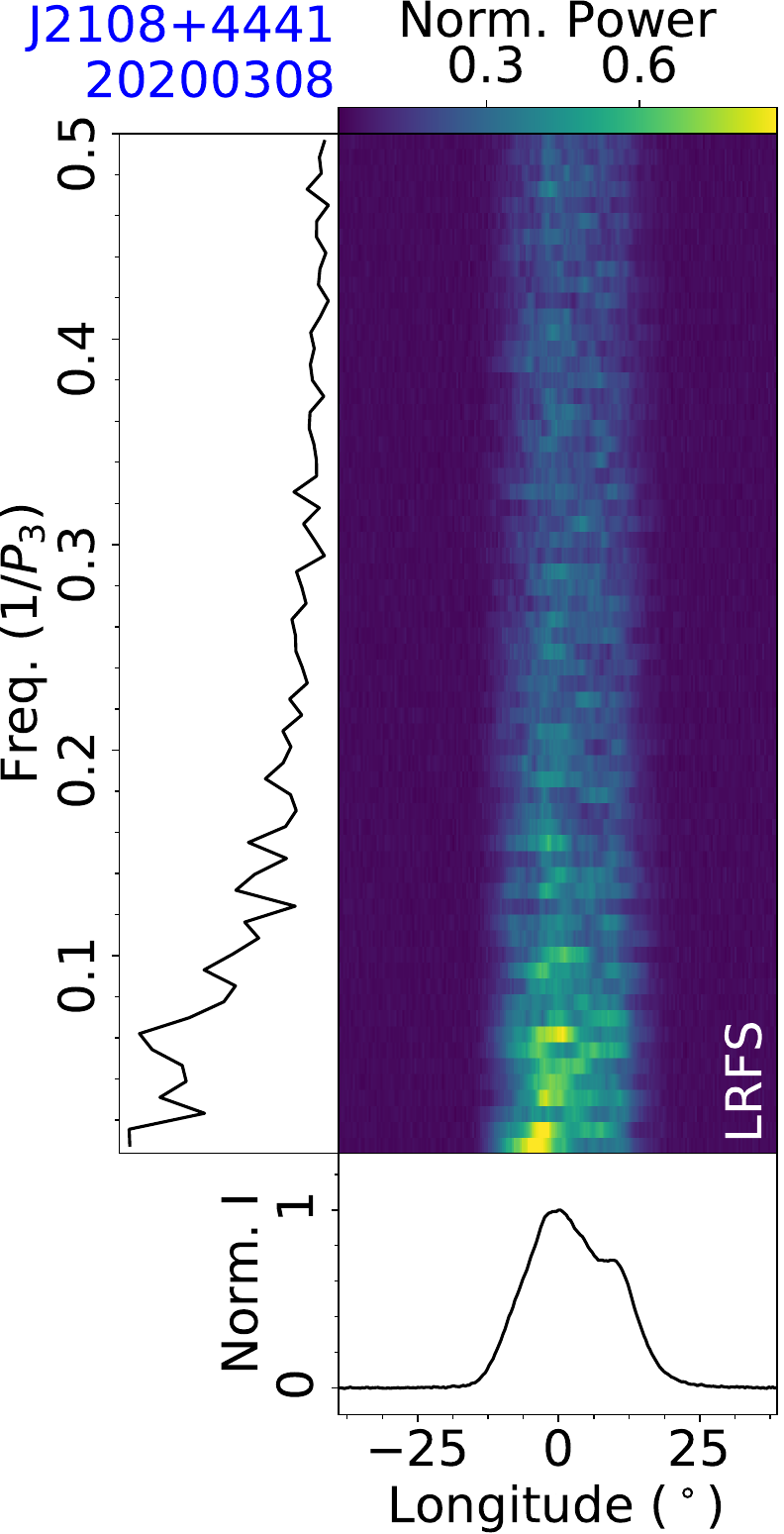}
\includegraphics[width=0.22\textwidth, angle=0]{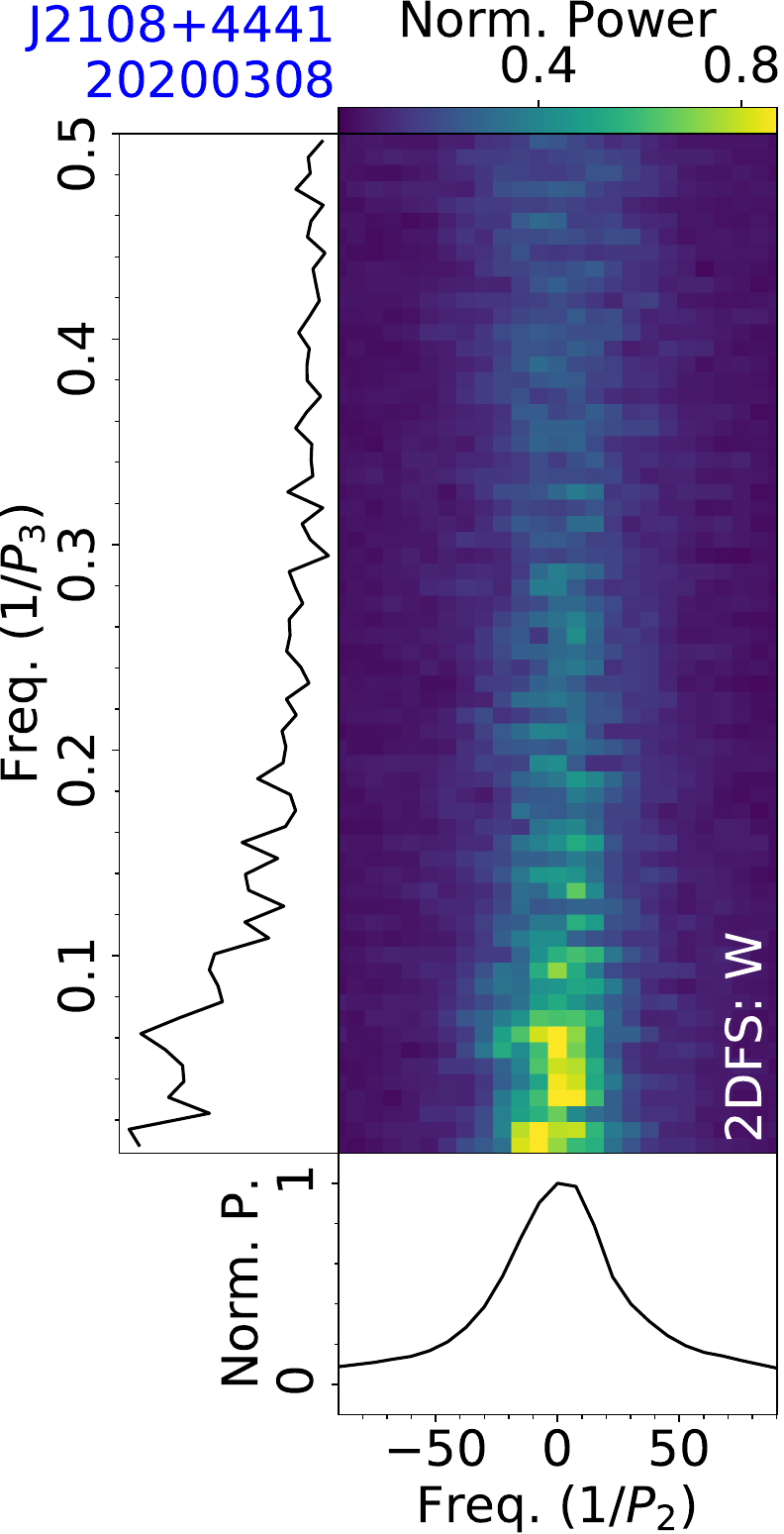}
\figcaption{Fluctuation analysis of PSR J2108+4441 for the observation on 20200308, with LRFS and 2DFS for the on-pulse region of a mean pulse profile.
\label{subfig:fluctu:J2108+4441}}
\end{figure}

\begin{figure}[htpb]
\centering
\includegraphics[width=0.22\textwidth, angle=0]{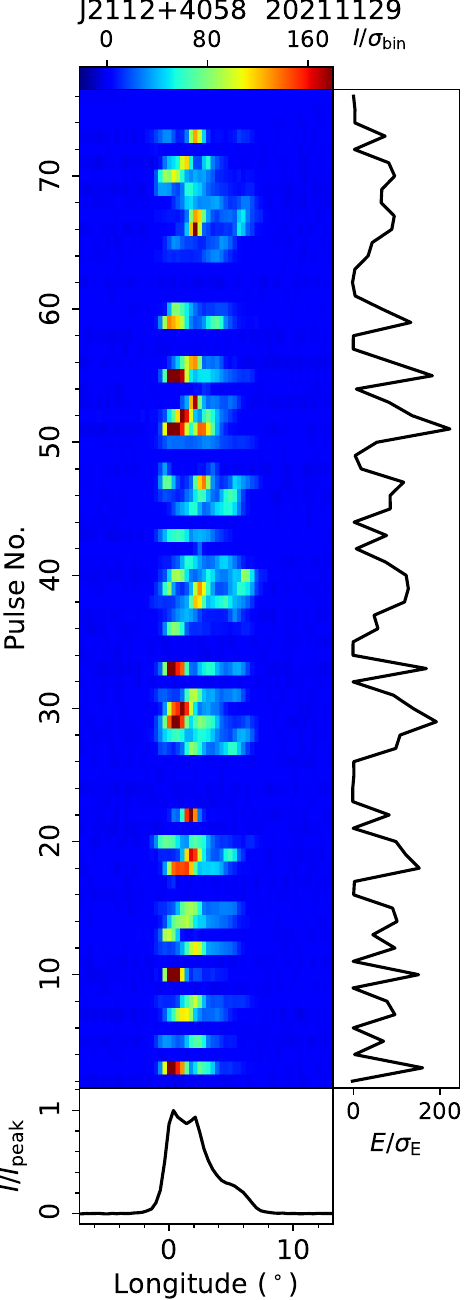}
\figcaption{Single pulse sequence of PSR J2112+4058 from the FAST observation on 20211129.
\label{subfig:TP:J2112+4058}}
\end{figure}

\begin{figure}[htpb]
\centering
\includegraphics[width=0.39\textwidth, angle=0]{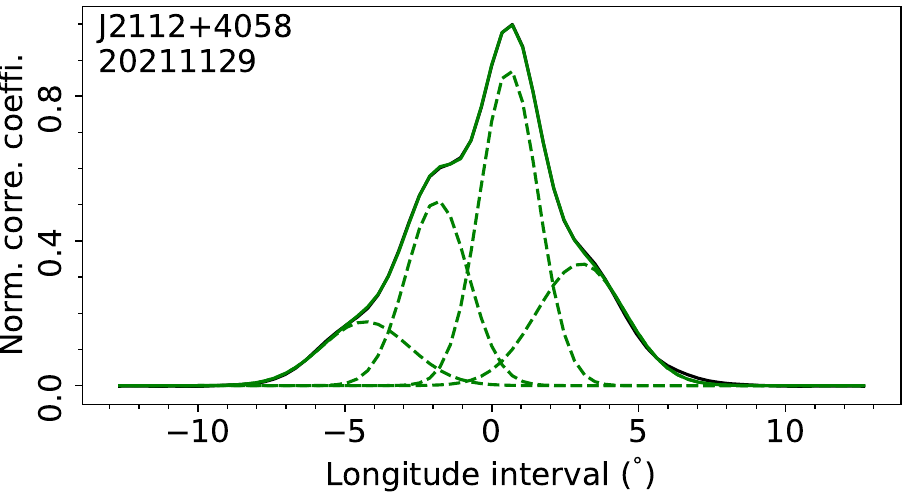}
\figcaption{Cross correlation of PSR J2112+4058 from the observation on 20211129.
\label{subfig:Corre:J2112+4058}}
\end{figure}

\begin{figure}[htpb]
\centering
\includegraphics[width=0.39\textwidth, angle=0]{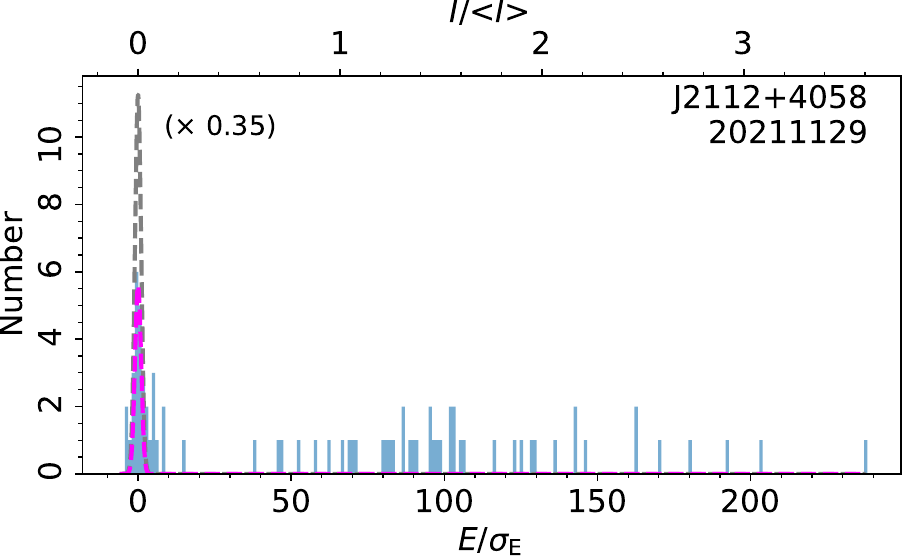}
\figcaption{On-pulse energy histogram of single pulses of PSR J2112+4058 from the FAST observation on 20211129.
\label{subfig:Hist:J2112+4058}}
\end{figure}

\subsection{J2055+3630}
\label{subsec:J2055+3630}

PSR J2055+3630 was discovered by the transit telescope of the National Radio Astronomy Observatory \citep{Damashek1982}. The fraction of null pulses was estimated to be less than 0.7\%, listed by \citet{Weisberg1986} at 430 MHz. 

This pulsar was observed by FAST on 20210621 for 5 minutes, yielding a rotation period $P=0.2215$~s and a dispersion measure $D\!M=97.4~{\rm cm^{-3}\,pc}$. 
From single pulse sequences and on-pulse integral energy histogram (Fig.~\ref{subfig:TP:J2055+3630} and Fig.~\ref{subfig:Hist:J2055+3630}), no nulling behavior exists in our observation. In 2DFS (Fig.~\ref{subfig:fluctu:J2055+3630}), there are broad drift features with a preferred positive drift sense, and the wide modulation frequency on phase is around zero, similar to the results of \citet{Weltevrede2006}. 
The centroid of the drift feature is characterized by frequencies of $1/P_3=0.097\pm0.001$ and $1/P_2=2.3\pm0.7$, corresponding to periodicities of $P_3=10.3\pm0.1$ periods and $P_2=158\pm47^\circ$. 
Additionally, subpulses occasionally drift to the earlier phase and then back to the normal longitude range, such as single pulses around No.800.

\begin{figure}[htpb]
\centering
\includegraphics[width=0.22\textwidth, angle=0]{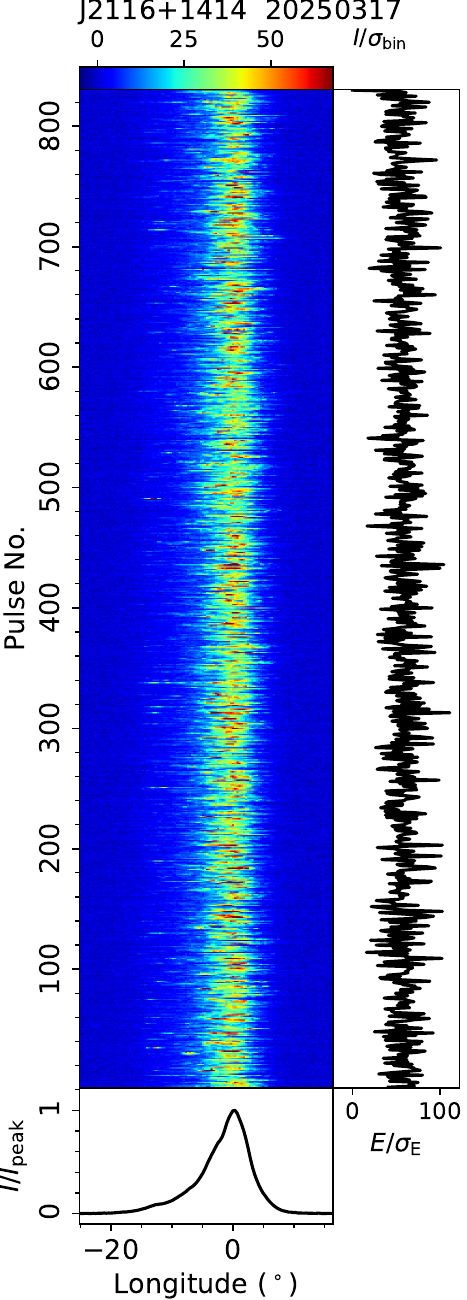}
\includegraphics[width=0.22\textwidth, angle=0]{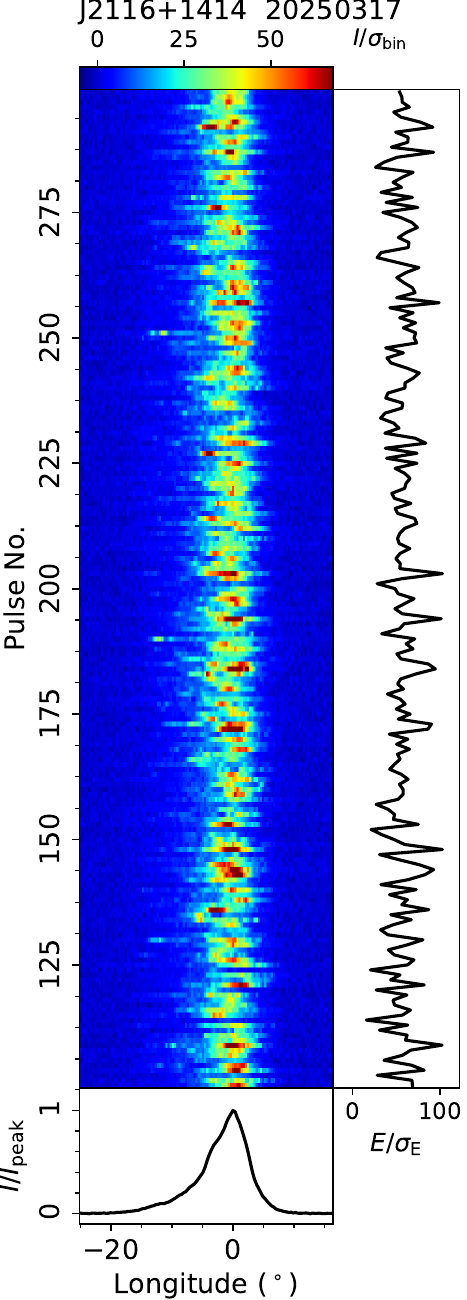}
\figcaption{Single pulse sequence of PSR J2116+1414 from the FAST observation on 20250317, as well as a zoomed-in view of pulses No. 100-300.
\label{subfig:TP:J2116+1414}}
\end{figure}

\begin{figure}[htpb]
\centering
\includegraphics[width=0.22\textwidth, angle=0]{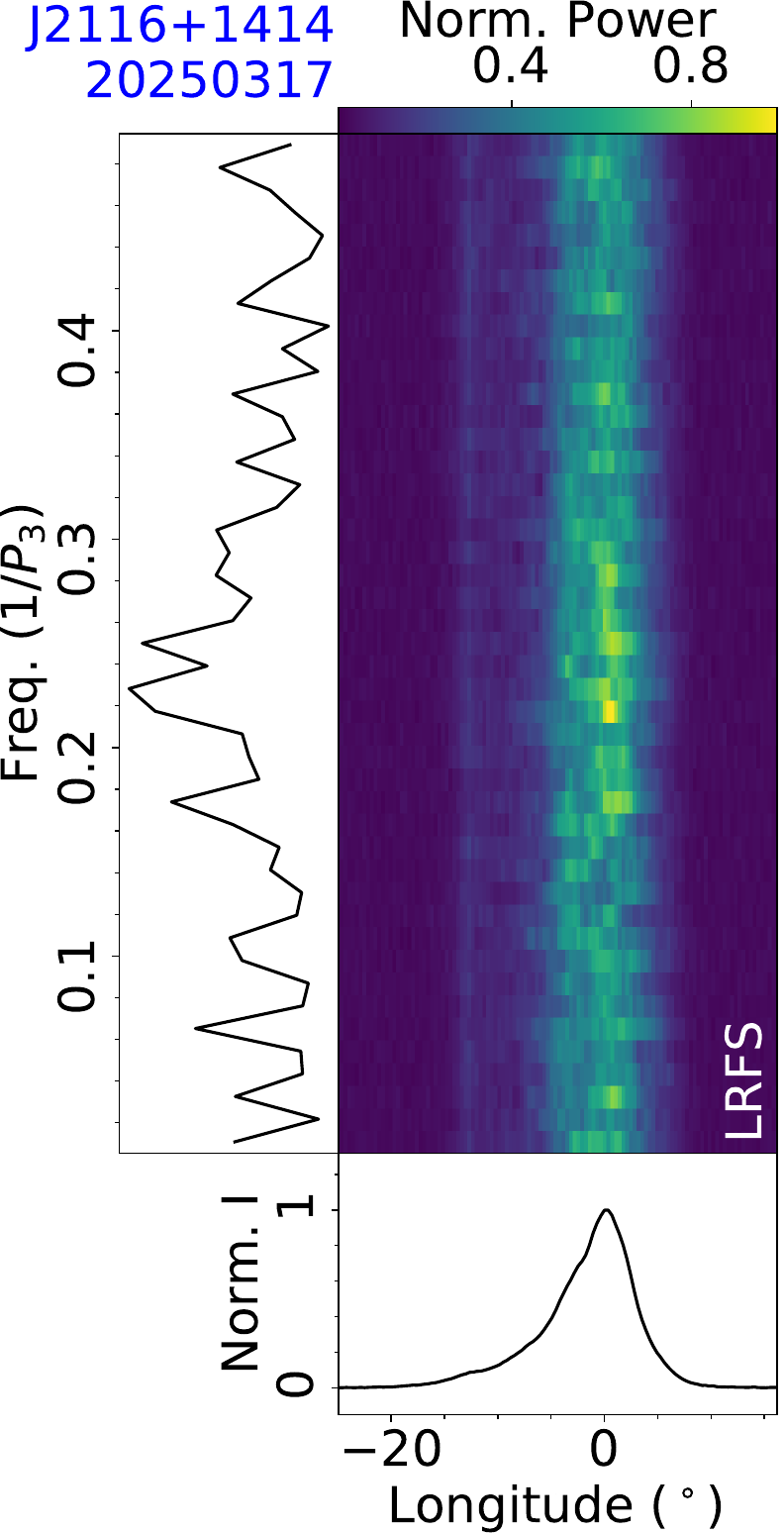}
\includegraphics[width=0.22\textwidth, angle=0]{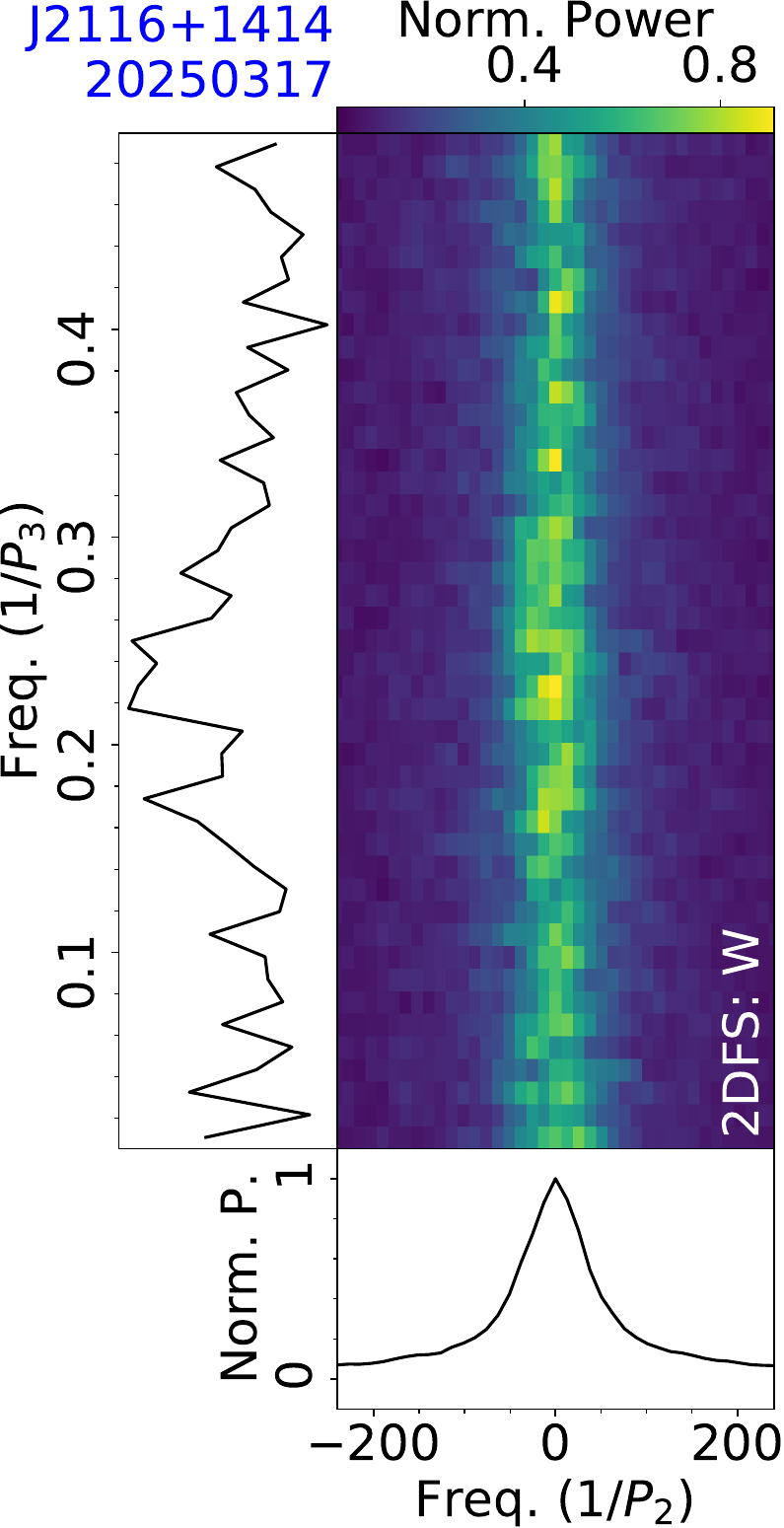}
\figcaption{Fluctuation analysis of PSR J2116+1414 for the observation on 20250317, with LRFS and 2DFS for the on-pulse region of a mean pulse profile.
\label{subfig:fluctu:J2116+1414}}
\end{figure}

\subsection{J2101+5028g}
\label{subsec:J2101+5028g}

PSR J2101+5028g was discovered in the FAST GPPS survey \citep{Han2021,han2025}. 

This pulsar was observed by FAST on 20240731 for 5 minutes and 20240920 for 15 minutes, with a rotation period of 0.2544~s and a dispersion measurement of 157.2 cm$^{-3}$pc from the observation on 20240920. 
Single pulse sequences and fluctuation spectra of the observation on 20240920 are shown in Fig.~\ref{subfig:TP:J2101+5028g} and \ref{subfig:fluctu:J2101+5028g}, respectively. There is a positive drift feature in 2DFS, with the centroid frequencies estimated to be $1/P_3=0.089\pm0.001$ and $1/P_2=8\pm2$, corresponding to periodicities of $P_3=11.3\pm0.1$ periods and $P_2=43\pm9^\circ$. The single pulse behavior of the observation on 20240731 is consistent with that on 20240920.

\begin{figure}[htpb]
\centering
\includegraphics[width=0.22\textwidth, angle=0]{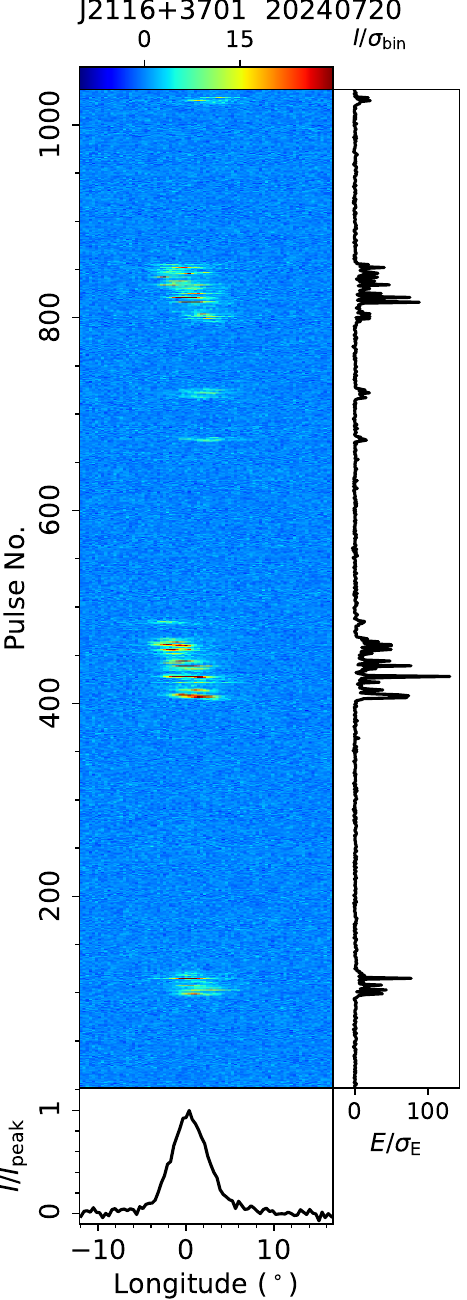}
\includegraphics[width=0.22\textwidth, angle=0]{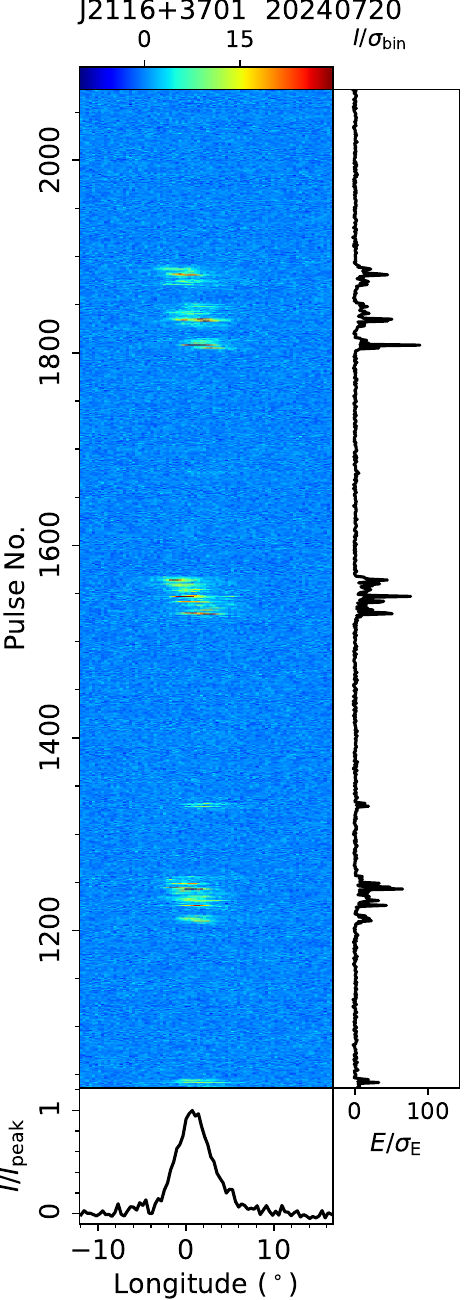}
\figcaption{Single pulse sequences of PSR J2116+3701 from the FAST observation on 20240720.
\label{subfig:TP:J2116+3701}}
\end{figure}

\begin{figure}[htpb]
\centering
\includegraphics[width=0.39\textwidth, angle=0]{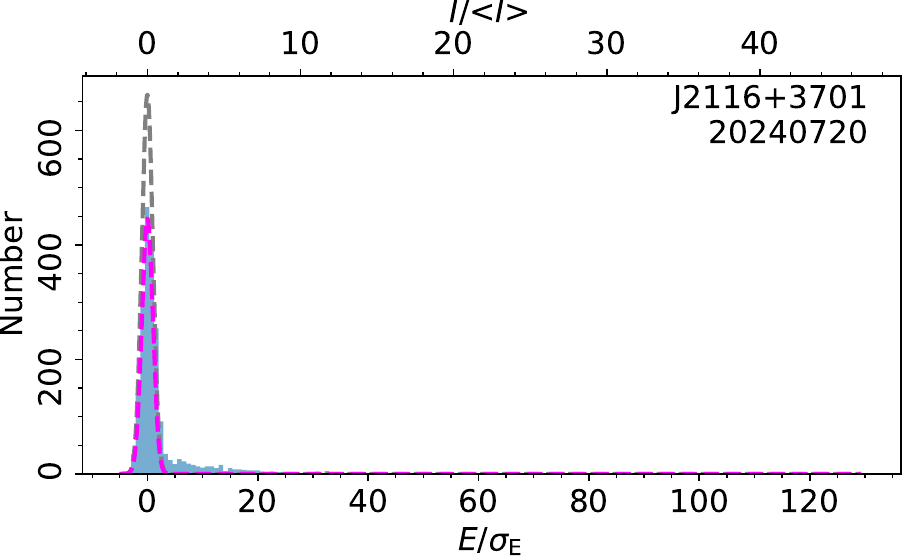}
\figcaption{On-pulse energy histogram of single pulses of PSR J2116+3701 from the FAST observation on 20240720.
\label{subfig:Hist:J2116+3701}}
\end{figure}

\begin{figure}[htpb]
\centering
\includegraphics[width=0.22\textwidth, angle=0]{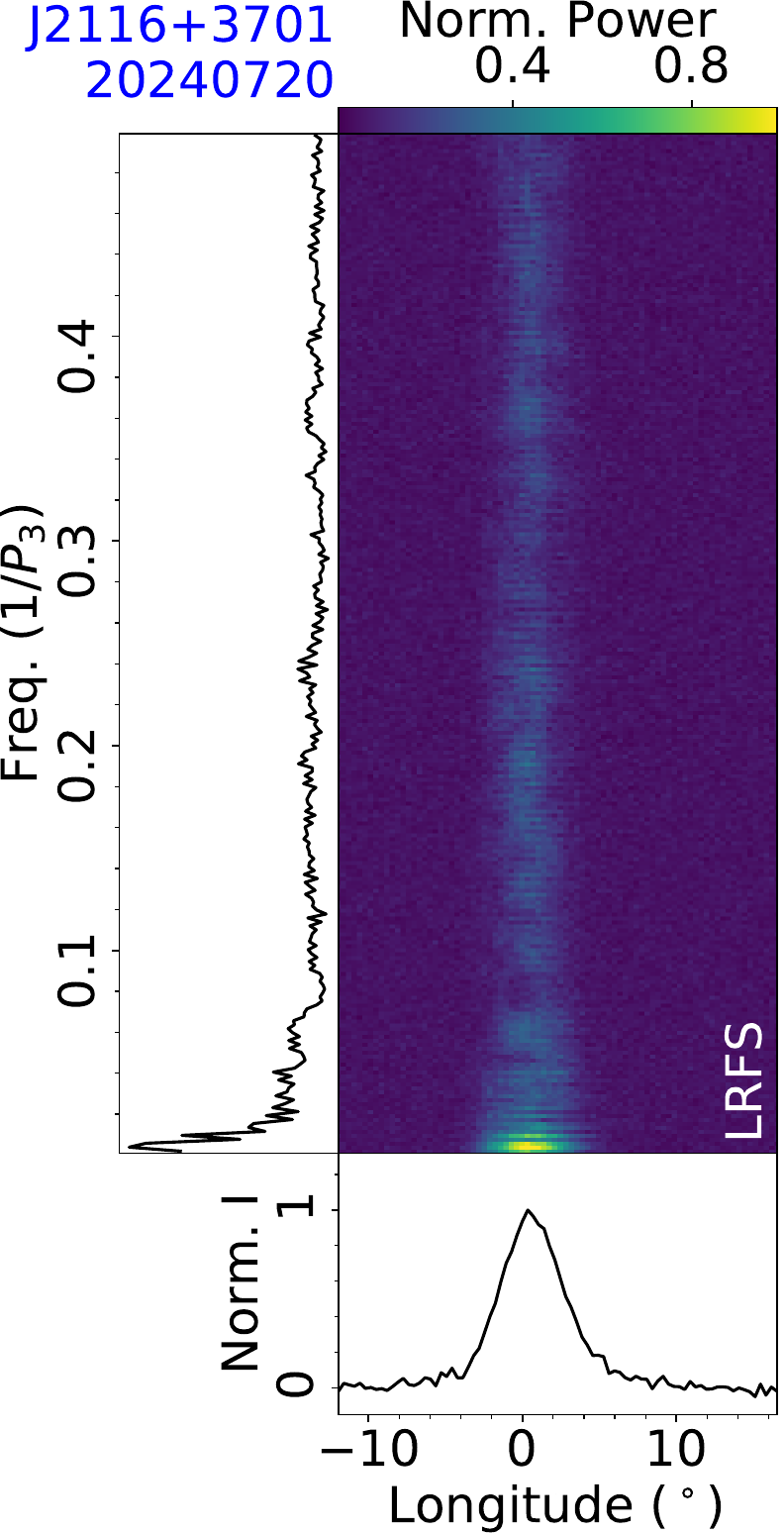}
\includegraphics[width=0.22\textwidth, angle=0]{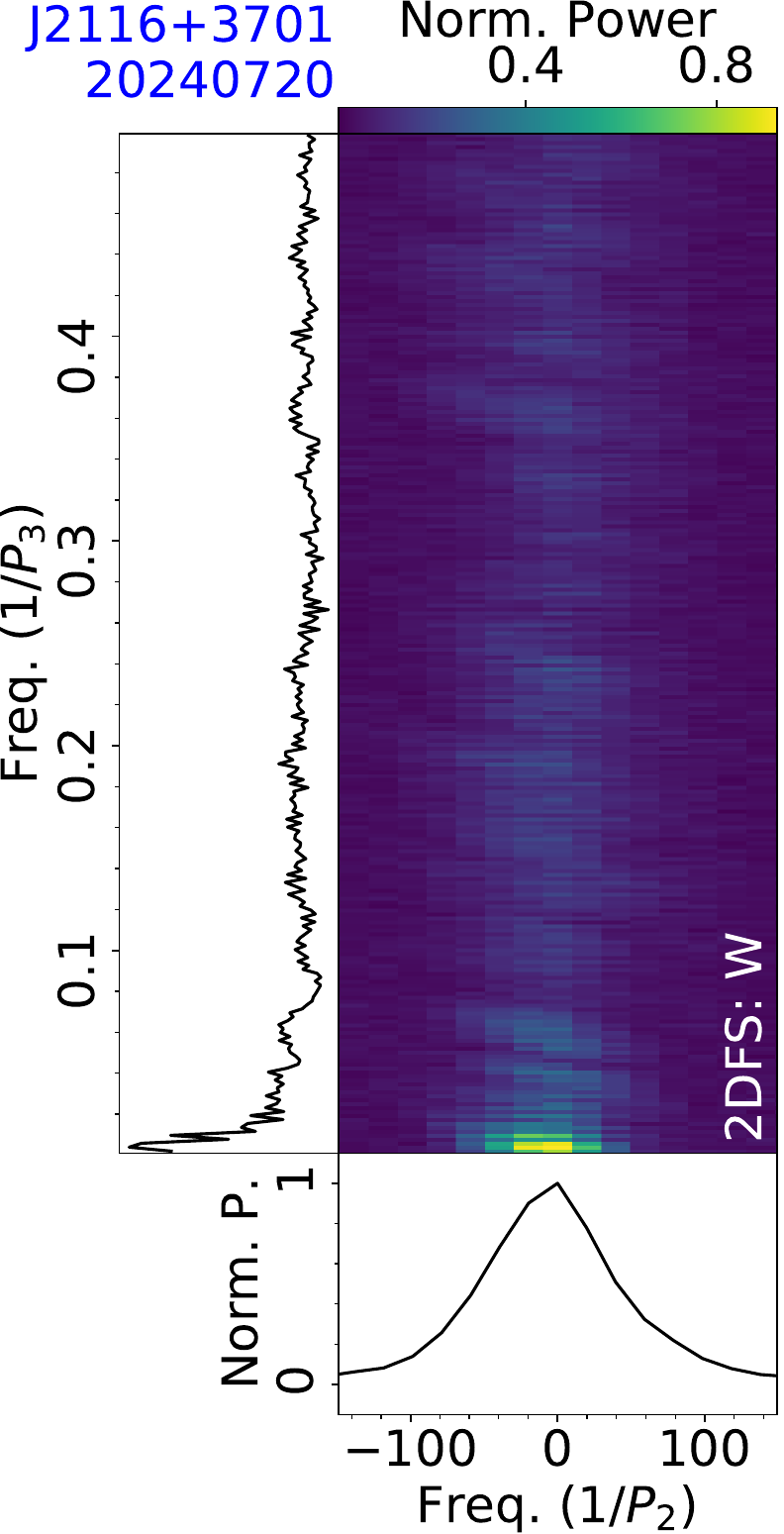}
\figcaption{Fluctuation analysis of PSR J2116+3701 for the observation on 20240720, with LRFS and 2DFS for the on-pulse region of a mean pulse profile.
\label{subfig:fluctu:J2116+3701}}
\end{figure}

\subsection{J2103+4602g}
\label{subsec:J2103+4602g}

PSR J2103+4602g was discovered in the FAST GPPS survey \citep{Han2021,han2025}.

The pulsar was observed by FAST on 20240314 for 15 minutes, and the rotation period and a dispersion measure are estimated to be $P=1.3599$~s and a dispersion measure $D\!M=207.2~{\rm cm^{-3}\,pc}$. Two segments of bright emission are exhibited in single pulse sequences shown in Fig.~\ref{subfig:TP:J2103+4602g}. The on-pulse integral energies versus period are smoothed over every 5 periods. From the smoothed energy histogram (Fig.~\ref{subfig:Hist:J2103+4602g}), the weak and bright emission modes of single pulses are distinguished, labeled using red and green colors.

\begin{figure}[htpb]
\centering
\includegraphics[width=0.22\textwidth, angle=0]{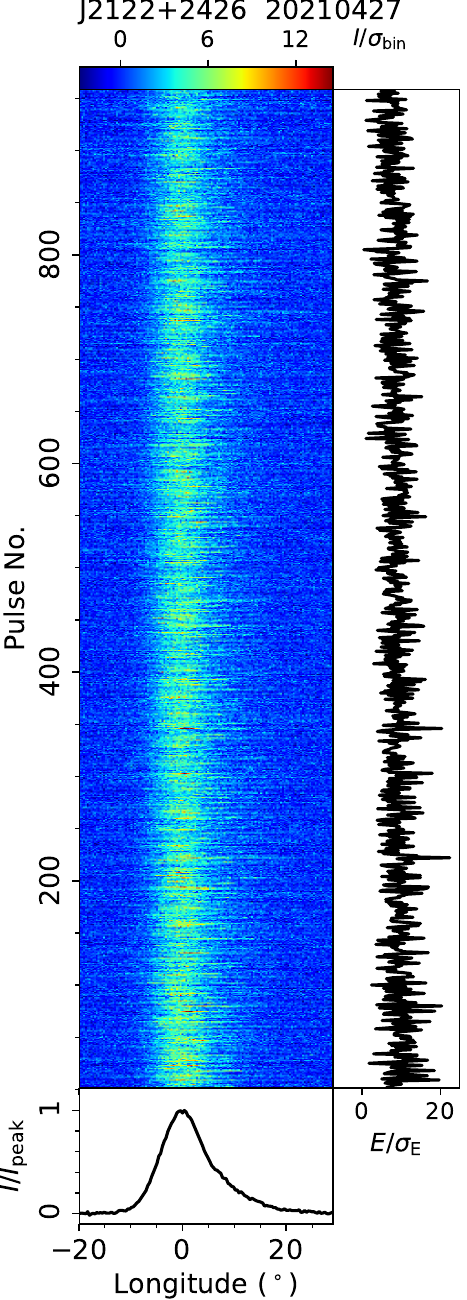}
\includegraphics[width=0.22\textwidth, angle=0]{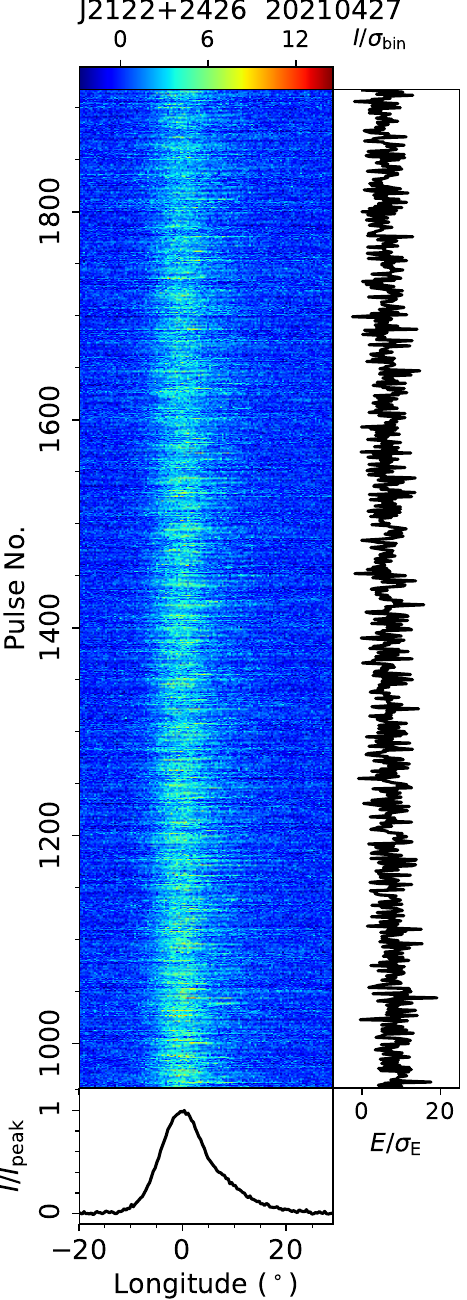}
\figcaption{Single pulse sequences of PSR J2122+2426 from the FAST observation on 20210427.
\label{subfig:TP:J2122+2426}}
\end{figure}

\begin{figure}[htpb]
\centering
\includegraphics[width=0.22\textwidth, angle=0]{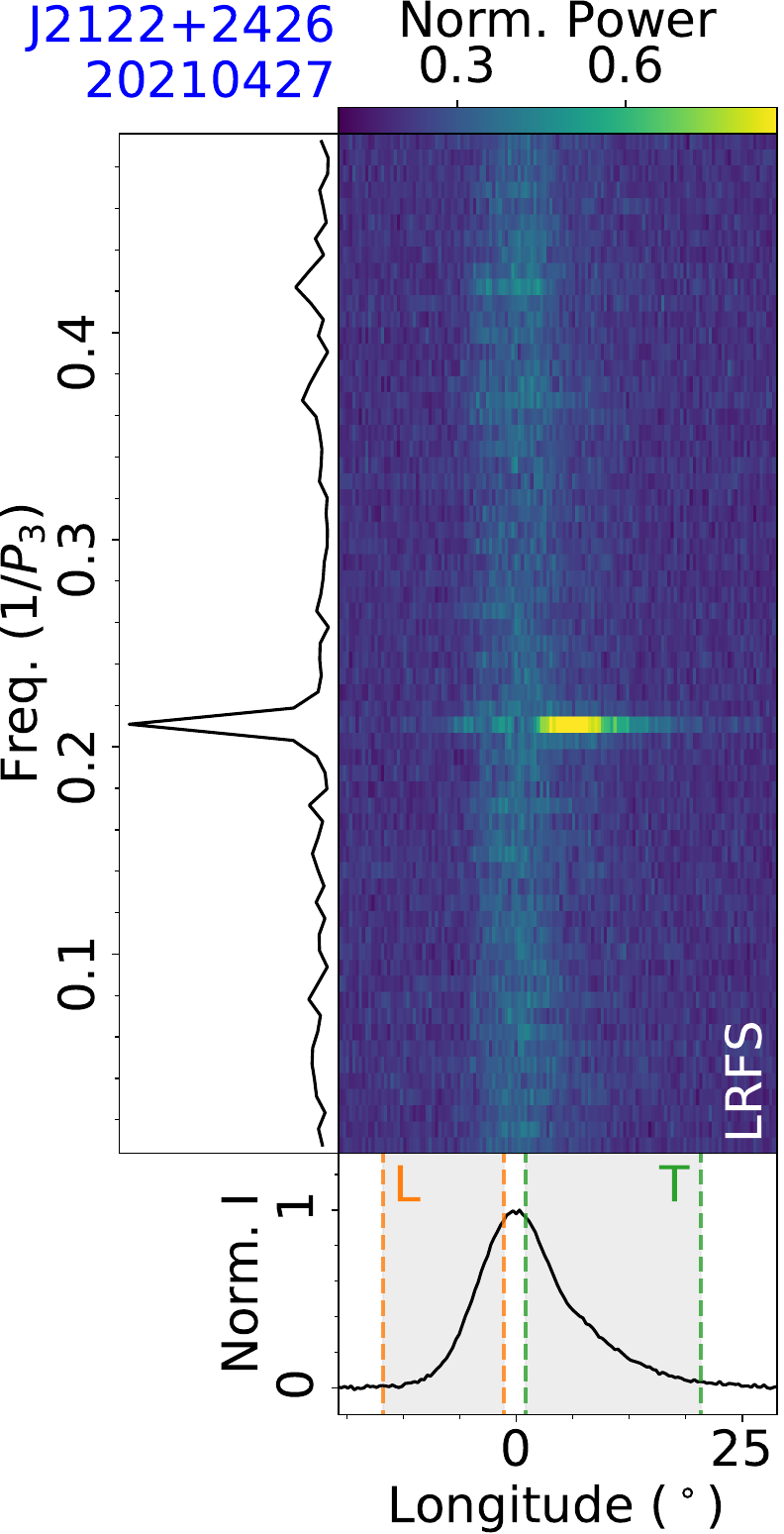}
\includegraphics[width=0.22\textwidth, angle=0]{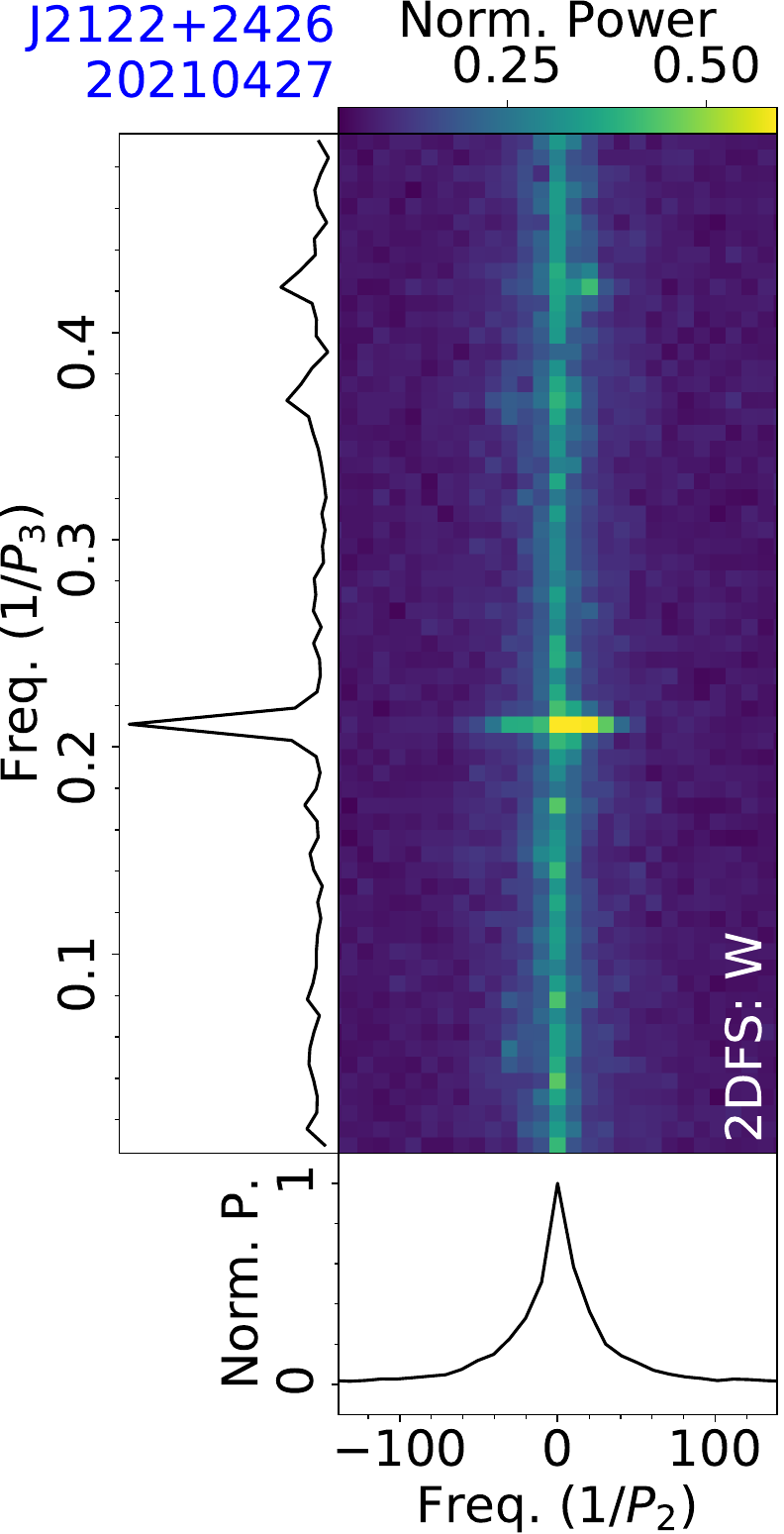}
\figcaption{Fluctuation analysis of PSR J2122+2426 for the observation on 20210427, with LRFS and 2DFS for the on-pulse region of a mean pulse profile.
\label{subfig:fluctu:J2122+2426}}
\end{figure}

\begin{figure}[htpb]
\centering
\includegraphics[width=0.22\textwidth, angle=0]{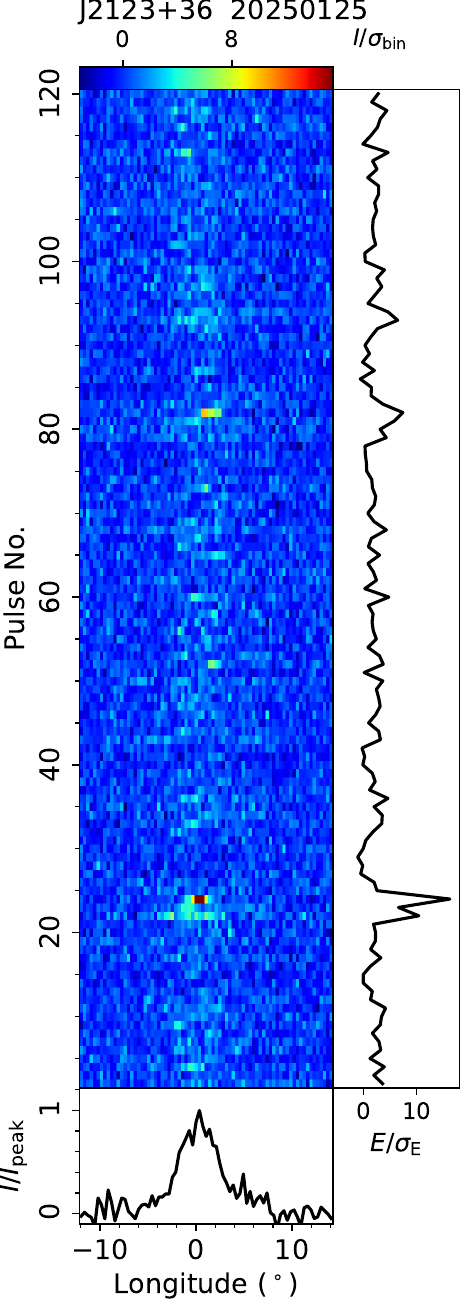}
\includegraphics[width=0.22\textwidth, angle=0]{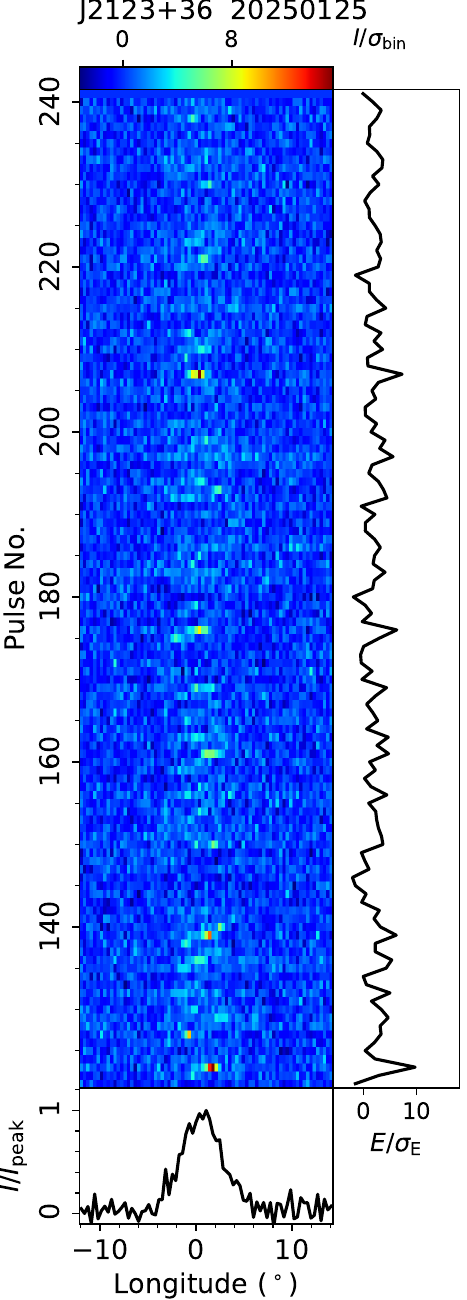}
\figcaption{Single pulse sequences of PSR J2123+36 from the FAST observation on 20250125.
\label{subfig:TP:J2123+36}}
\end{figure}

\begin{figure}[htpb]
\centering
\includegraphics[width=0.22\textwidth, angle=0]{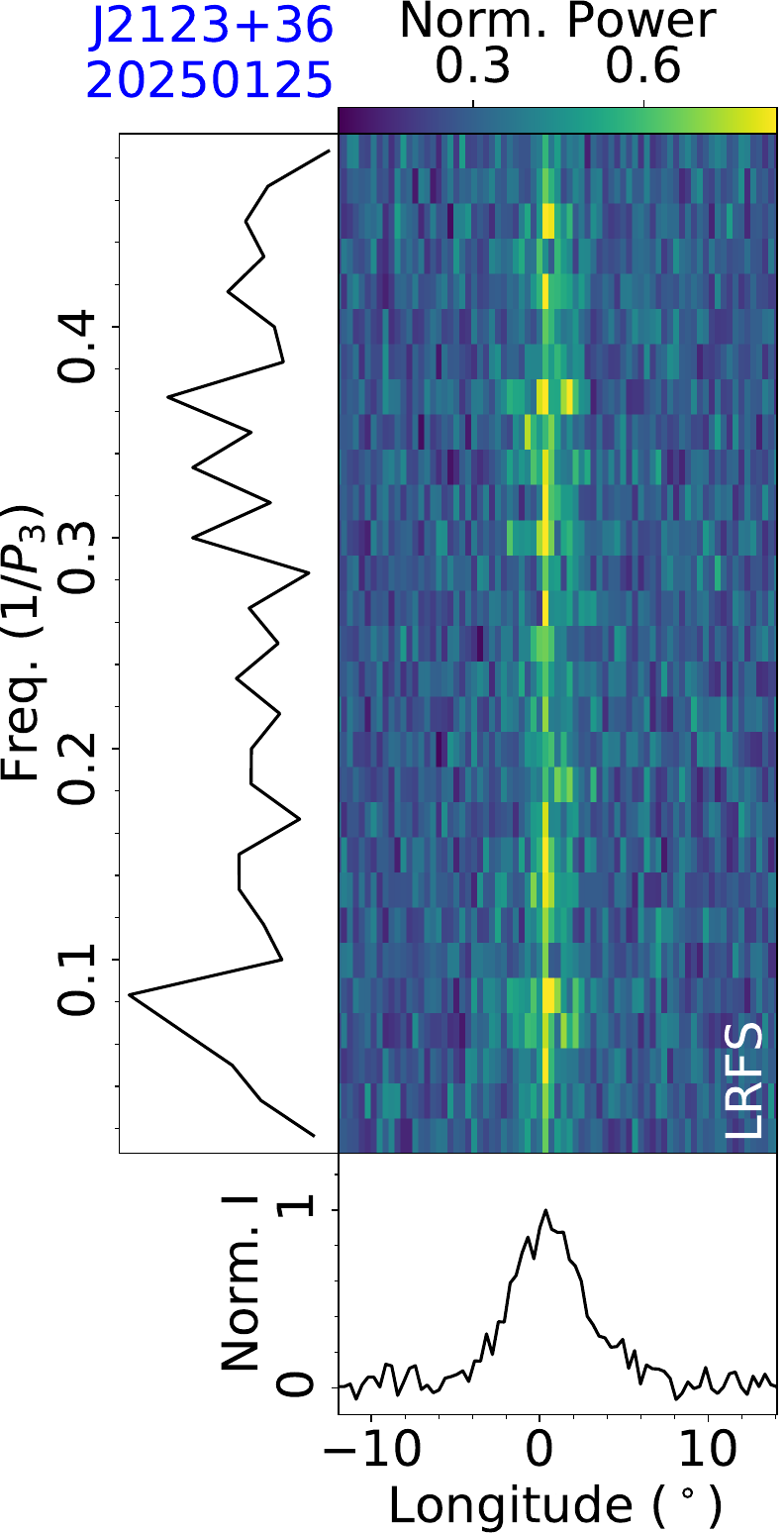}
\includegraphics[width=0.22\textwidth, angle=0]{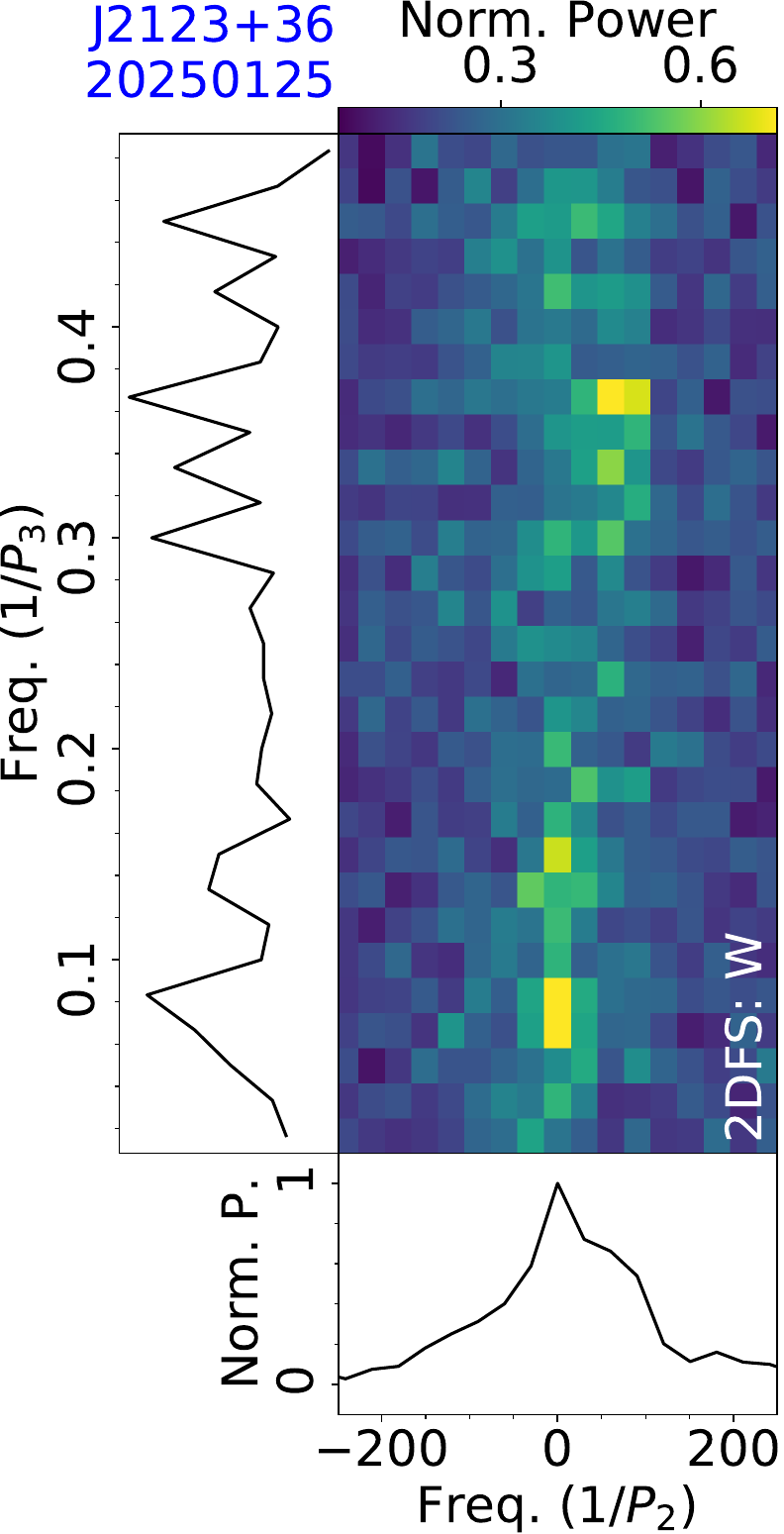}
\figcaption{Fluctuation analysis of PSR J2123+36 for the observation on 20250125, with LRFS and 2DFS for the on-pulse region of a mean pulse profile.
\label{subfig:fluctu:J2123+36}}
\end{figure}

\subsection{J2105+07}
\label{subsec:J2105+07}

PSR J2105+07 was discovered in the Arecibo 327 MHz Drift Pulsar Survey \citep{Deneva2016}. 

The FAST observed the pulsar on 20210118 for 9 minutes, yielding a rotation period $P=3.7468$~s and a dispersion measure $D\!M=50.0~{\rm cm^{-3}\,pc}$. 
The single pulse sequence is shown in Fig.~\ref{subfig:TP:J2105+07}. From the on-pulse integral energy histogram in Fig.~\ref{subfig:Hist:J2105+07}, the nulling fraction of this observation is fitted to be 20$\pm$4\%. 
LRFS and 2DFS of the trailing part in the mean pulse profile are shown in Fig.~\ref{subfig:fluctu:J2105+07}. 
2DFS of the trailing profile part exhibits a positive drift feature, with the centroid frequencies estimated to be $1/P_3=0.153\pm0.003$ and $1/P_2=34\pm7$, corresponding to periodicities of $P_3=6.5\pm0.1$ periods and $P_2=11\pm2^\circ$.

\begin{figure}[htpb]
\centering
\includegraphics[width=0.22\textwidth, angle=0]{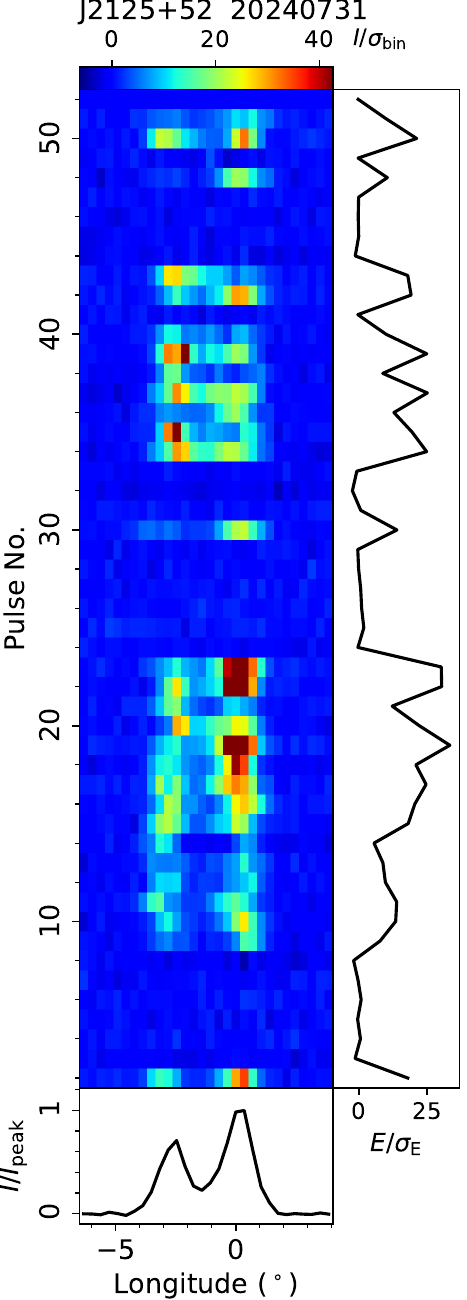}
\figcaption{Single pulse sequence of PSR J2125+52 from the FAST observation on 20240731.
\label{subfig:TP:J2125+52}}
\end{figure}

\begin{figure}[htpb]
\centering
\includegraphics[width=0.39\textwidth, angle=0]{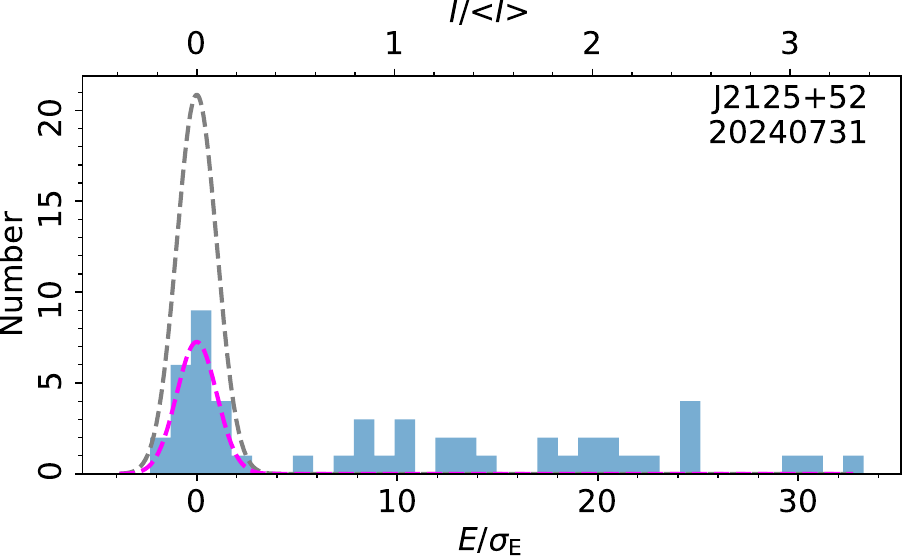}
\figcaption{On-pulse energy histogram of single pulses of PSR J2125+52 from the FAST observation on 20240731.
\label{subfig:Hist:J2125+52}}
\end{figure}

\subsection{J2108+4441}
\label{subsec:J2108+4441}

PSR J2108+4441 was discovered by \citet{Davies1973} using the Mark 1A radio telescope at Jodrell Bank. Low frequency modulation has been detected \citep{Weltevrede2006,Weltevrede2007}. 

The pulsar was observed by FAST on 20200308 for 5 minutes, deriving a rotation period $P=0.4149$~s and a dispersion measure $D\!M=140.1~{\rm cm^{-3}\,pc}$. 
The single pulse sequence in Fig.~\ref{subfig:TP:J2108+4441} illustrates that the drifting behavior is not systematic. From the 2DFS of the FAST observation in Fig.~\ref{subfig:fluctu:J2108+4441}, there are mainly two drift features. 
The positive drift feature exhibits the centroid frequencies of $1/P_3=0.063\pm0.001$ and $1/P_2=3.0\pm0.4$, corresponding to $P_3=15.9\pm0.1$ periods and $P_2=120\pm16^\circ$. 
There is also a negatively phase modulated feature in 2DFS, exhibiting in the leading profile part from LRFS, which may be caused by the leading phase variation of single pulses.

\subsection{J2112+4058}
\label{subsec:J2112+4058}

PSR J2112+4058 was discovered by FAST \citep{Cruces2021}. The pulsar was reported to have dwarf pulses by \citep{Yan2024}. 

This pulsar was observed by FAST on 20211129 for 5 minutes, deriving a rotation period $P=4.0612$~s and a dispersion measure $D\!M=125.7~{\rm cm^{-3}\,pc}$. 
The nulling fraction is estimated from the energy histogram in Fig.~\ref{subfig:Hist:J2112+4058} to be 0.17$\pm$0.02\%. 
From the single pulse sequence (Fig.~\ref{subfig:TP:J2112+4058}) of this observation, the pulsar also has subpulse drifting behavior. From the cross-correlation method (Fig.~\ref{subfig:Corre:J2112+4058}), we calculate the drifting parameters to be $D=0.59\pm0.07$ degrees per period and $P_2=2.44\pm0.02^\circ$.

\subsection{J2116+1414}
\label{subsec:J2116+1414}

PSR J2116+1414 was discovered by \citet{Manchester1978} in the second Molonglo pulsar survey. \citet{Song2023} reported the parameters of the negative drift feature: $P_3=3.7\pm0.4$ and $P_2=-57^{+35}_{-21}$.

This pulsar was observed by FAST on 20250317 for 6 minutes, deriving a rotation period $P=0.4401$~s and a dispersion measure $D\!M=56.0~{\rm cm^{-3}\,pc}$. 
The single pulse sequence and a zoomed-in view of pulses No. 100-300 are shown in Fig.~\ref{subfig:TP:J2116+1414}, exhibiting no systematic drifting behavior. From fluctuation spectra in Fig.~\ref{subfig:fluctu:J2116+1414}, there is a preferred negative drift feature, with the temporal fluctuation frequency widely distributed. The centroid frequencies of the drift feature are estimated to be $1/P_3=0.232\pm0.004$ and $1/P_2=-13\pm2$, corresponding to periodicities of $P_3=4.3\pm0.1$ periods and $P_2=-28\pm5$ degrees.

\subsection{J2116+3701}
\label{subsec:J2116+3701}

PSR J2116+3701 was discovered by the CHIME telescope \citep{Dong2023}. 

This pulsar was observed by FAST on 20240720 for 5 minutes, deriving a rotation period $P=0.1459$~s and a dispersion measure $D\!M=44.1~{\rm cm^{-3}\,pc}$. Single pulse sequences are shown in Fig.~\ref{subfig:TP:J2116+3701}, illustrating the existence of nulls. The nulling fraction is estimated from the on-pulse integral energy histogram in Fig.~\ref{subfig:Hist:J2116+3701}, which is 67$\pm$3\%. Interestingly, the pulse drifts from the later to the earlier phase during the emission state and displays the negative drifting phenomenon. Fluctuation spectra are displayed in Fig.~\ref{subfig:fluctu:J2116+3701}, where the centroid frequencies of the drift feature are $1/P_3=0.0055\pm0.0001$ and $1/P_2=-10\pm1$, corresponding to periodicities of $P_3=183\pm5$ periods and $P_2=-38\pm5^\circ$. 
$P_3=181\pm15$ periods and $P_2=-38\pm17^\circ$. The temporally modulated periodicity of 181 periods is from the occurrences of nulls.

\begin{figure}[htpb]
\centering
\includegraphics[width=0.22\textwidth, angle=0]{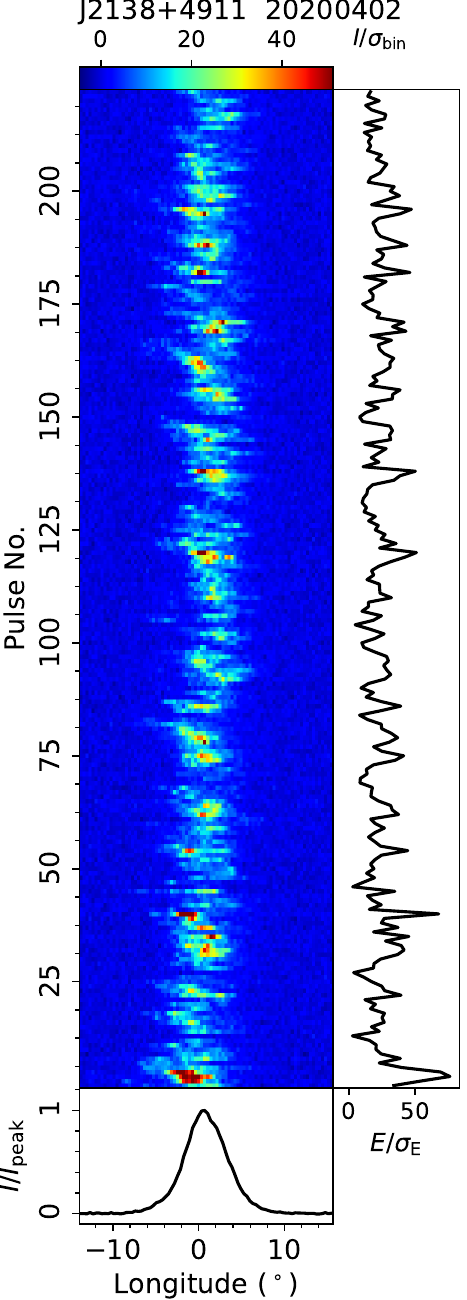}
\includegraphics[width=0.22\textwidth, angle=0]{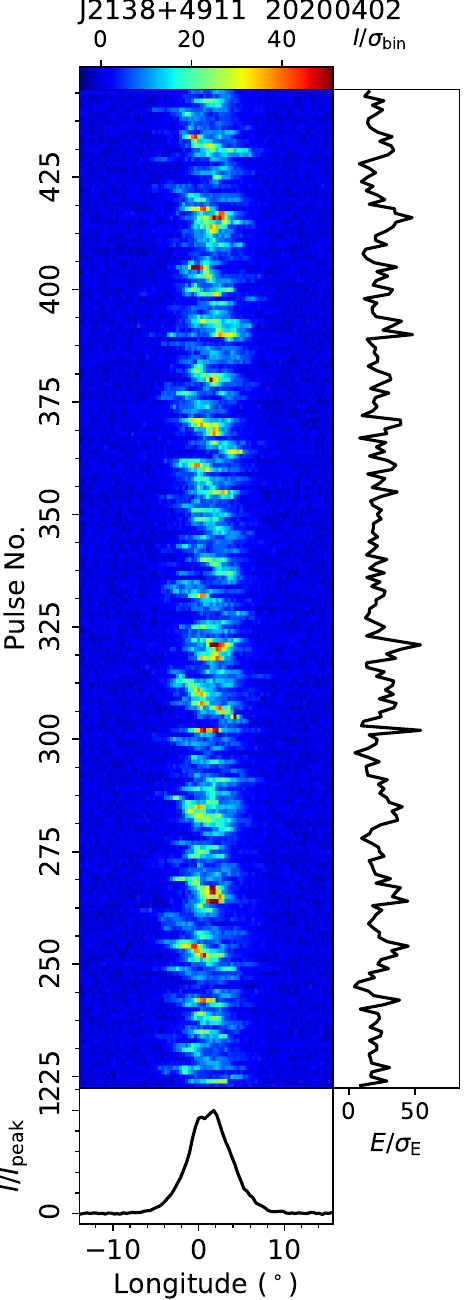}
\figcaption{Single pulse sequences of PSR J2138+4911 from the FAST observation on 20200402.
\label{subfig:TP:J2138+4911}}
\end{figure}

\begin{figure}[htpb]
\centering
\includegraphics[width=0.22\textwidth, angle=0]{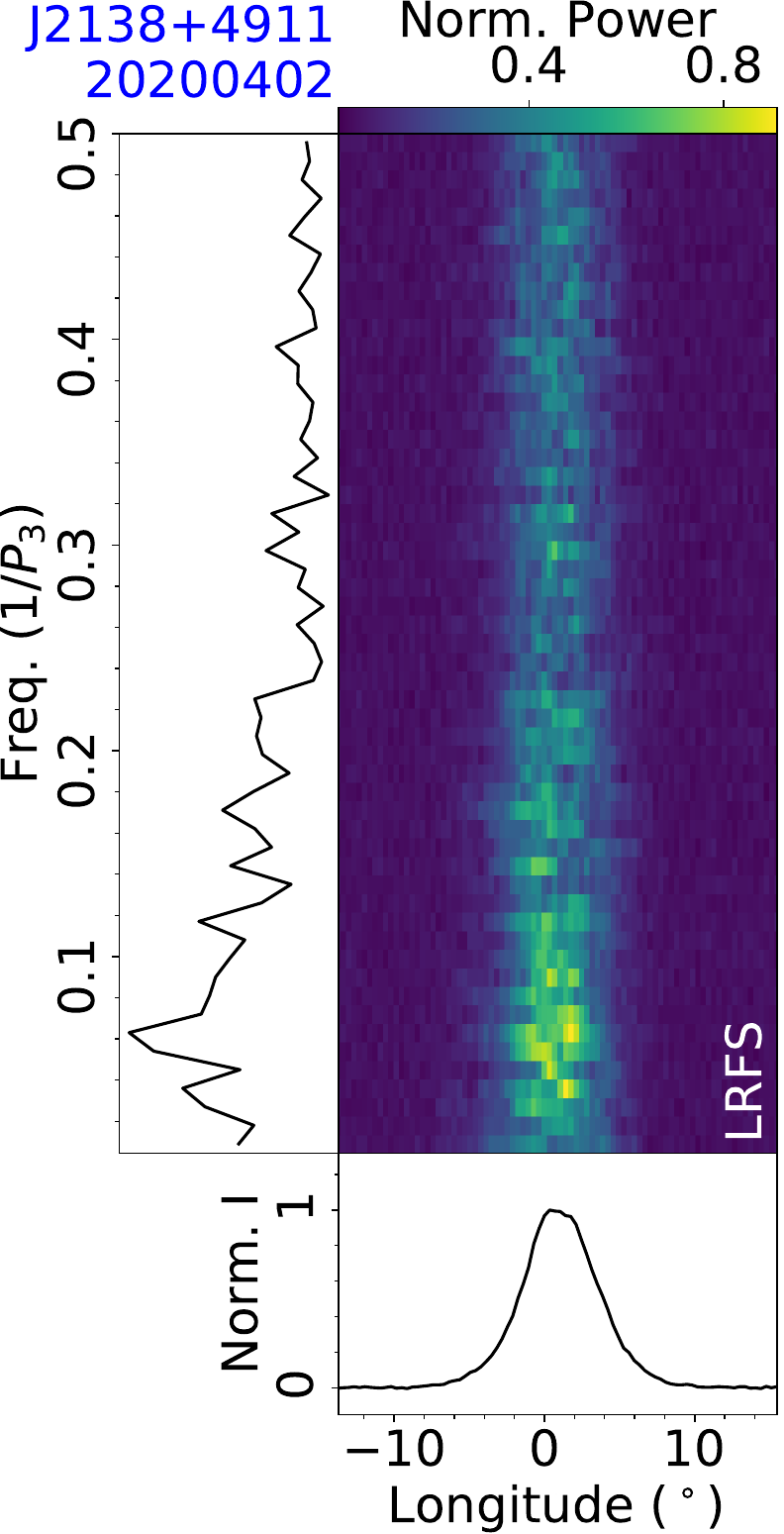}
\includegraphics[width=0.22\textwidth, angle=0]{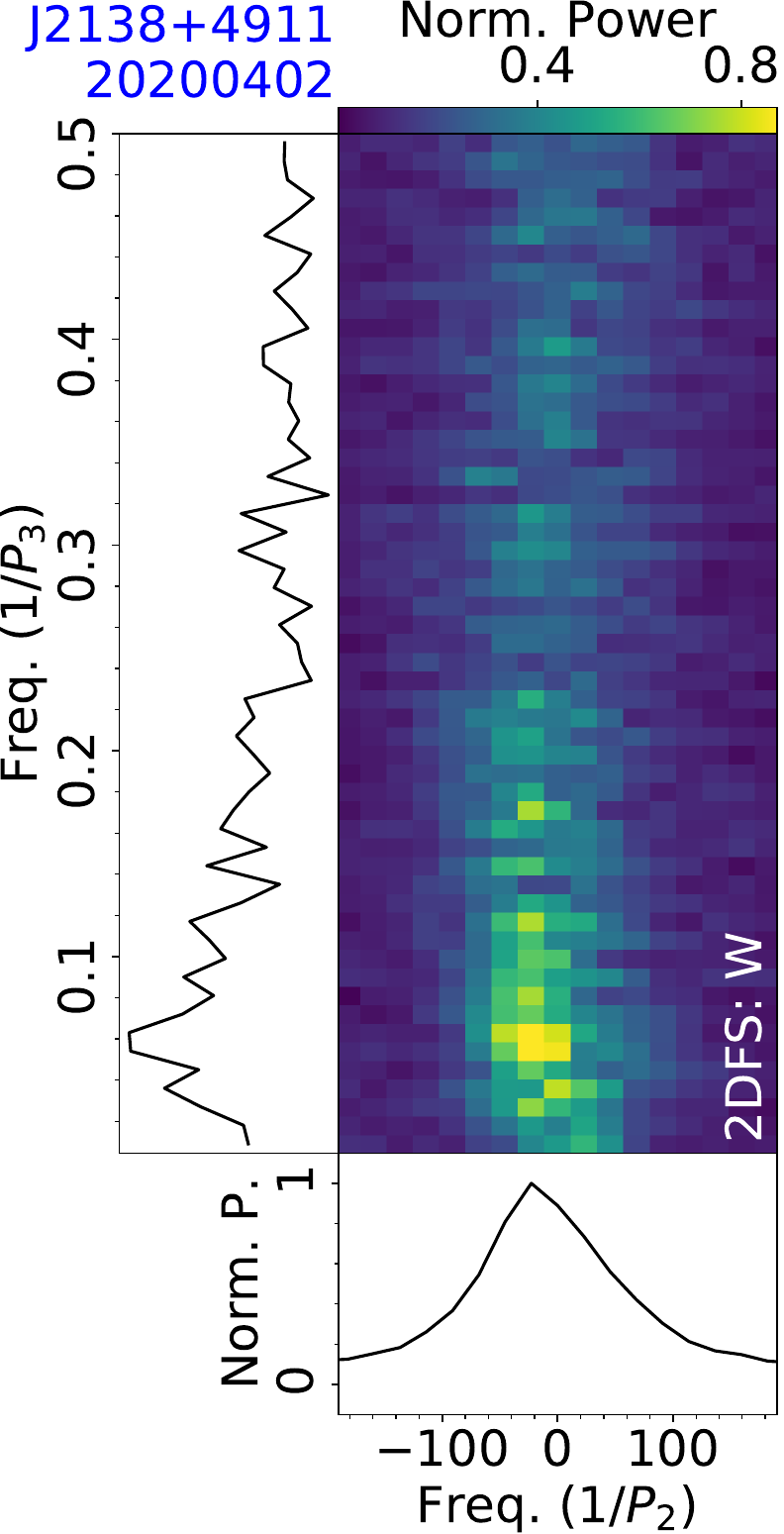}
\figcaption{Fluctuation analysis of PSR J2138+4911 for the observation on 20200402, with LRFS and 2DFS for the on-pulse region of a mean pulse profile.
\label{subfig:fluctu:J2138+4911}}
\end{figure}

\begin{figure}[htpb]
\centering
\includegraphics[width=0.22\textwidth, angle=0]{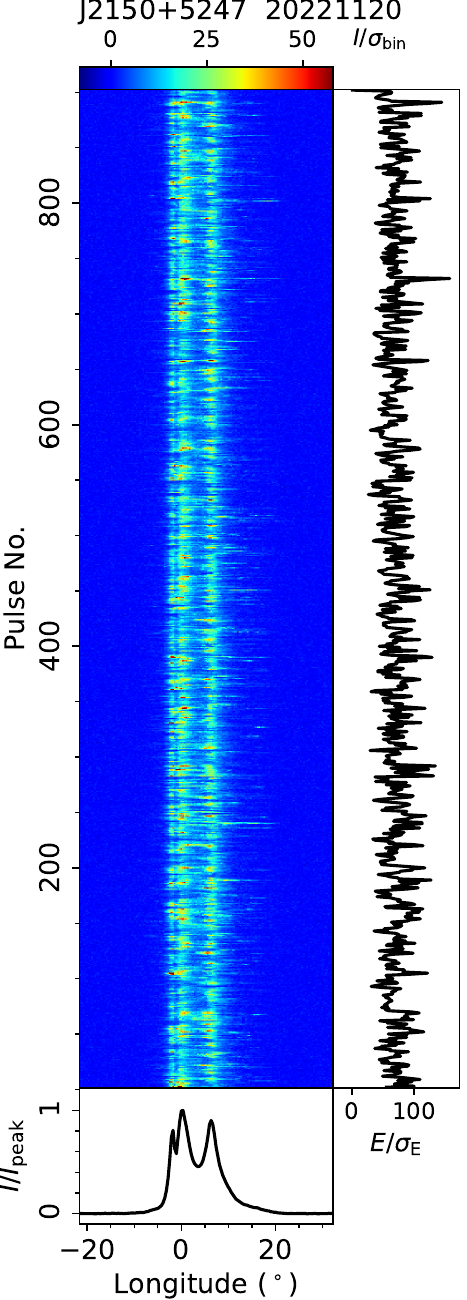}
\includegraphics[width=0.22\textwidth, angle=0]{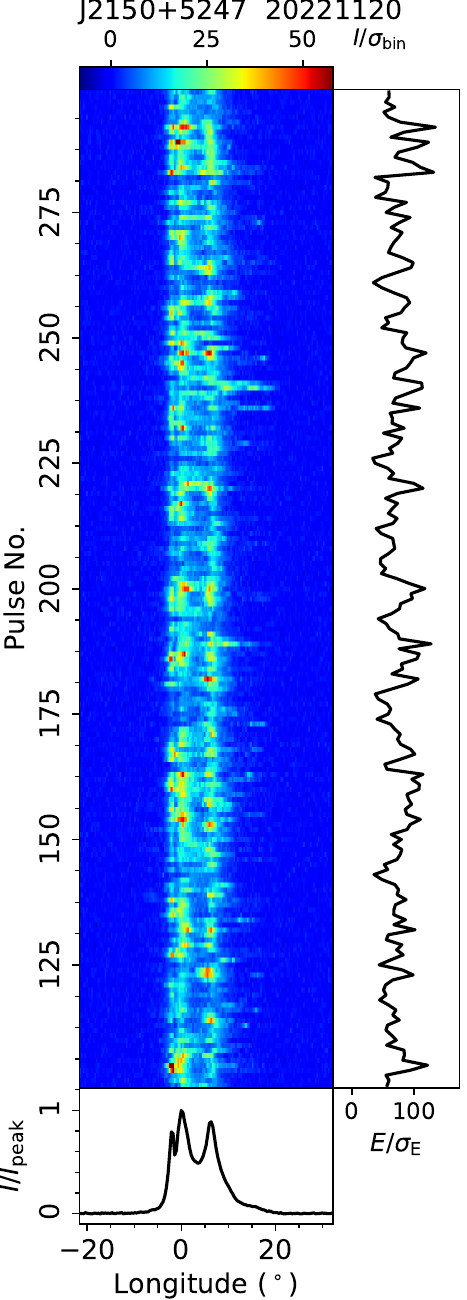}
\figcaption{The single pulse sequence of PSR J2150+5247 from the FAST observation on 20221120, and a zoomed-in view of pulses No.101-300.
\label{subfig:TP:J2150+5247}}
\end{figure}

\begin{figure}[htpb]
\centering
\includegraphics[width=0.44\textwidth, angle=0]{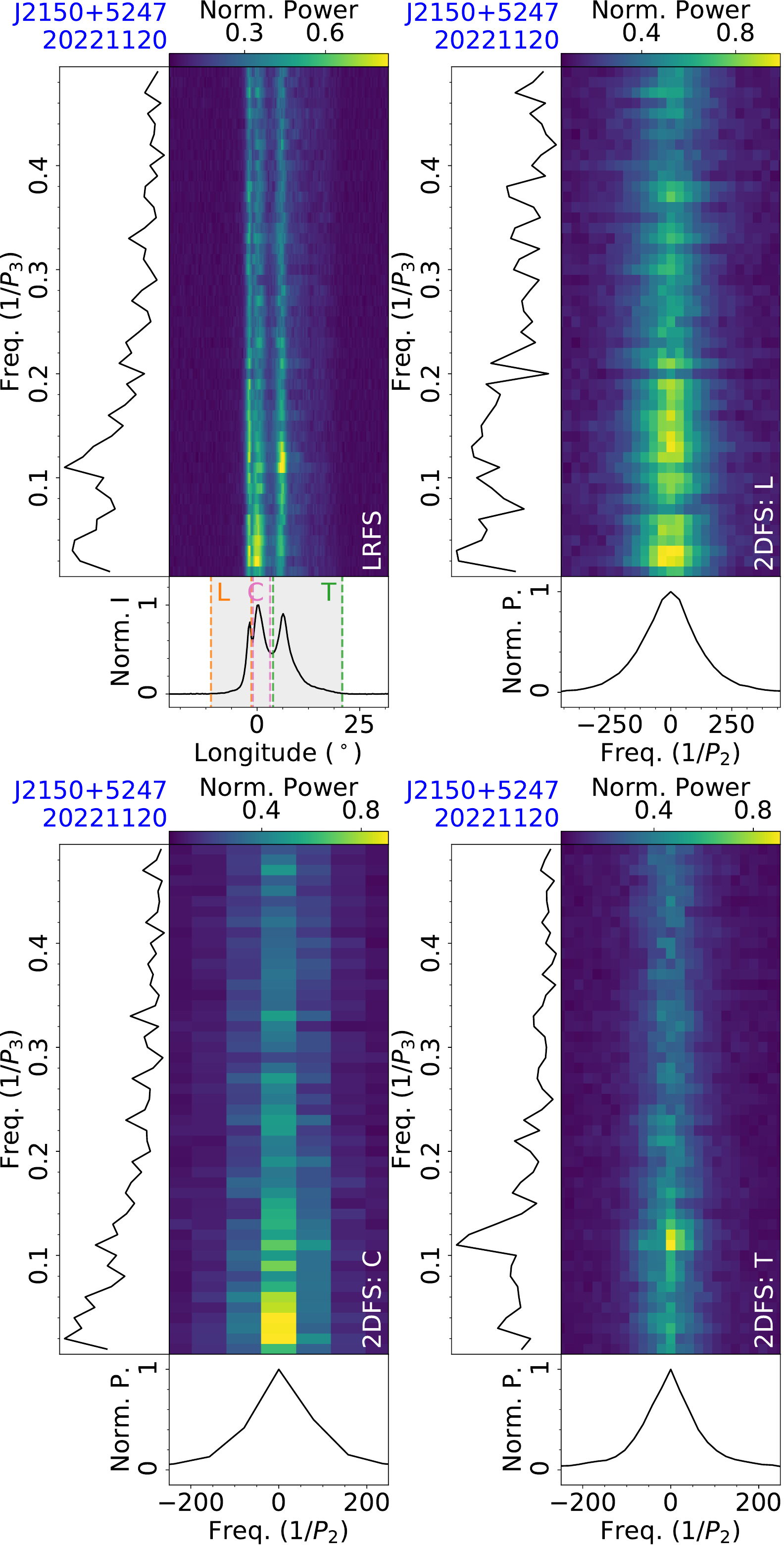}
\figcaption{Fluctuation analysis of PSR J2150+5247 for the observation on 20221120, with LRFS (top-left), and 2DFS for the leading part (top-right), central part (bottom-left) and trailing part (bottom-right) of a mean pulse profile.
\label{subfig:fluctu:J2150+5247}}
\end{figure}

\subsection{J2122+2426}
\label{subsec:J2122+2426}

PSR J2122+2426 was discovered in the LOFAR Tied-Array All-Sky Survey \citep{Sanidas2019}. 

The pulsar was observed by FAST on 20210427 for 17 minutes, deriving a rotation period $P=0.5414$~s and a dispersion measure $D\!M=8.5~{\rm cm^{-3}\,pc}$. 
Single pulse sequences are displayed in Fig.~\ref{subfig:TP:J2122+2426}. From the fluctuation spectra shown in Fig.~\ref{subfig:fluctu:J2122+2426}, the pulsar has a preferred positive drifting behavior. The drift feature in 2DFS exhibits centroid frequencies of $1/P_3=0.211\pm0.001$ and $1/P_2=5\pm1$, corresponding to periodicities of $P_3=4.74\pm0.02$ periods and $P_2=75\pm21^\circ$.

\begin{figure}[htpb]
\centering
\includegraphics[width=0.22\textwidth, angle=0]{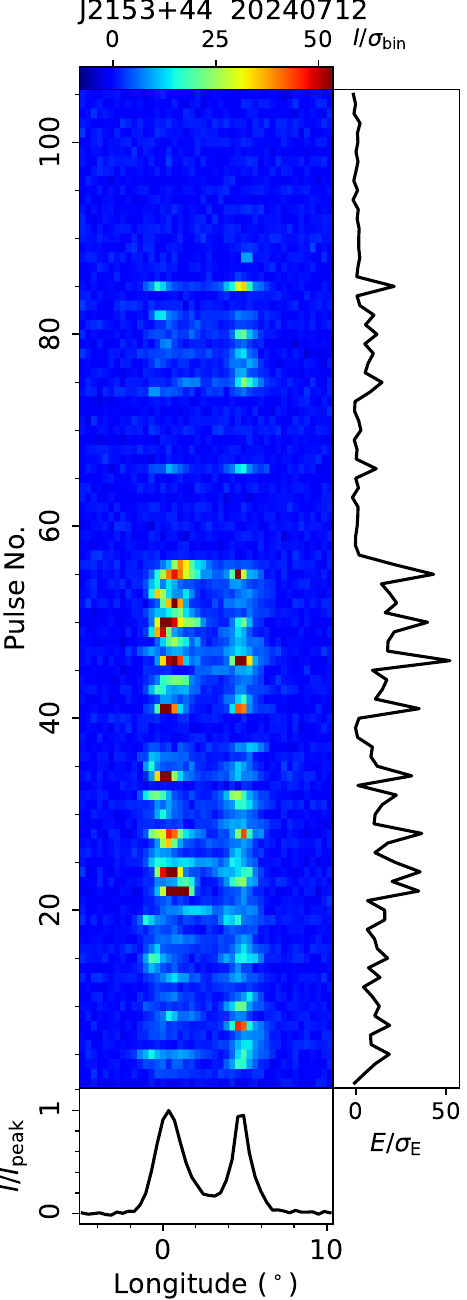}
\figcaption{Single pulse sequence of PSR J2153+44 from the FAST observation on 20240712.
\label{subfig:TP:J2153+44}}
\end{figure}

\begin{figure}[htpb]
\centering
\includegraphics[width=0.39\textwidth, angle=0]{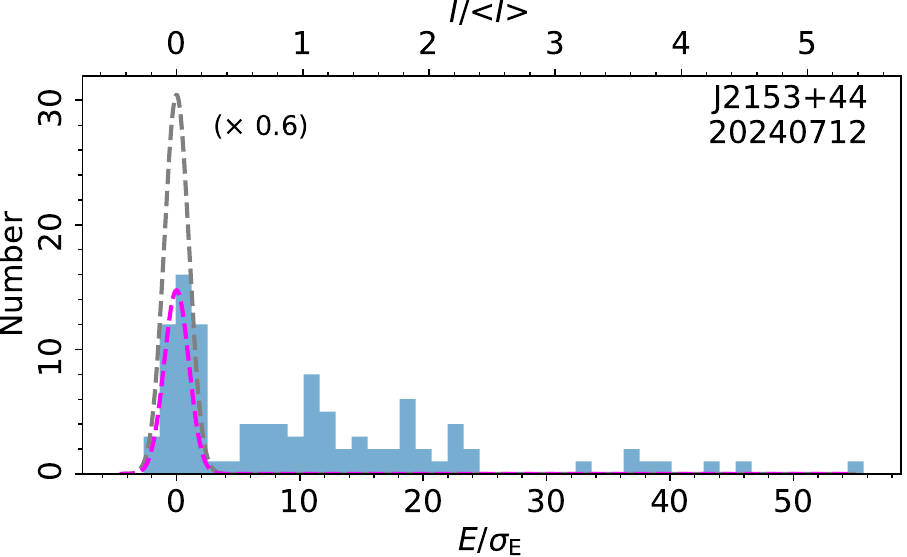}
\figcaption{On-pulse energy histogram of single pulses of PSR J2153+44 from the FAST observation on 20240712.
\label{subfig:Hist:J2153+44}}
\end{figure}

\begin{figure}[htpb]
\centering
\includegraphics[width=0.22\textwidth, angle=0]{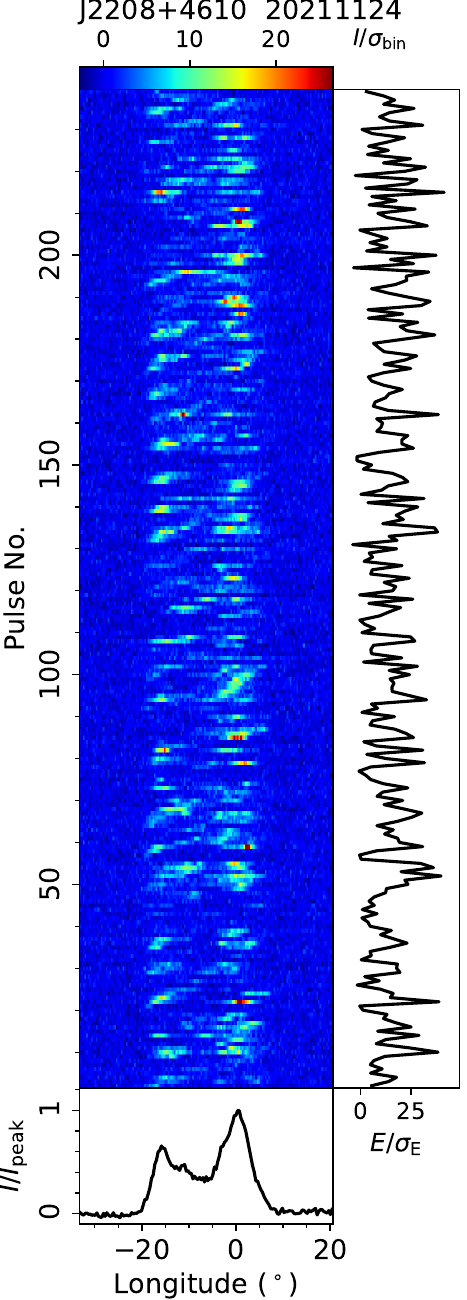}
\includegraphics[width=0.22\textwidth, angle=0]{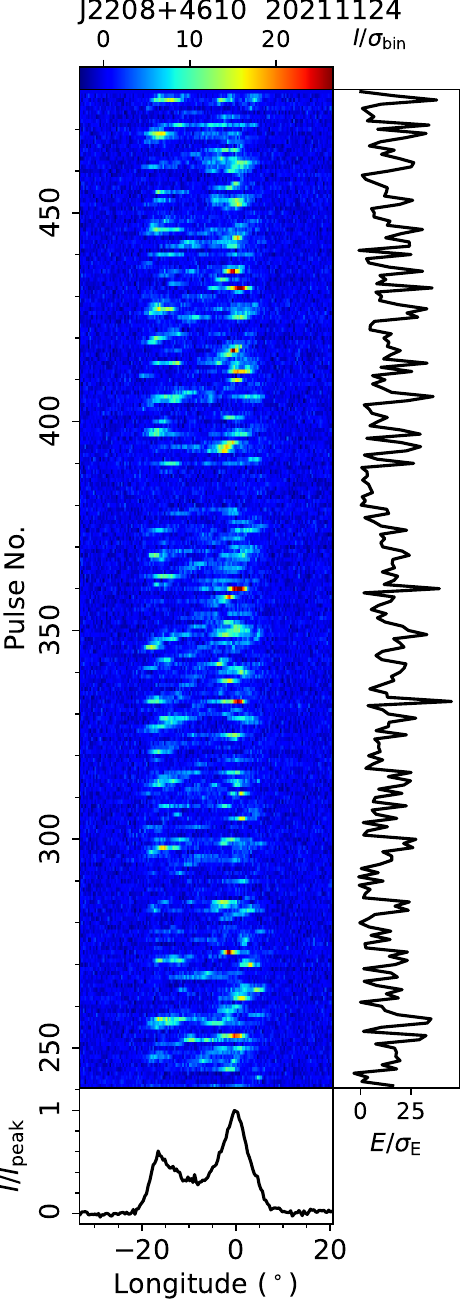}
\figcaption{Single pulse sequences of PSR J2208+4610 from the FAST observation on 20211124.
\label{subfig:TP:J2208+4610}}
\end{figure}

\begin{figure}[htpb]
\centering
\includegraphics[width=0.44\textwidth, angle=0]{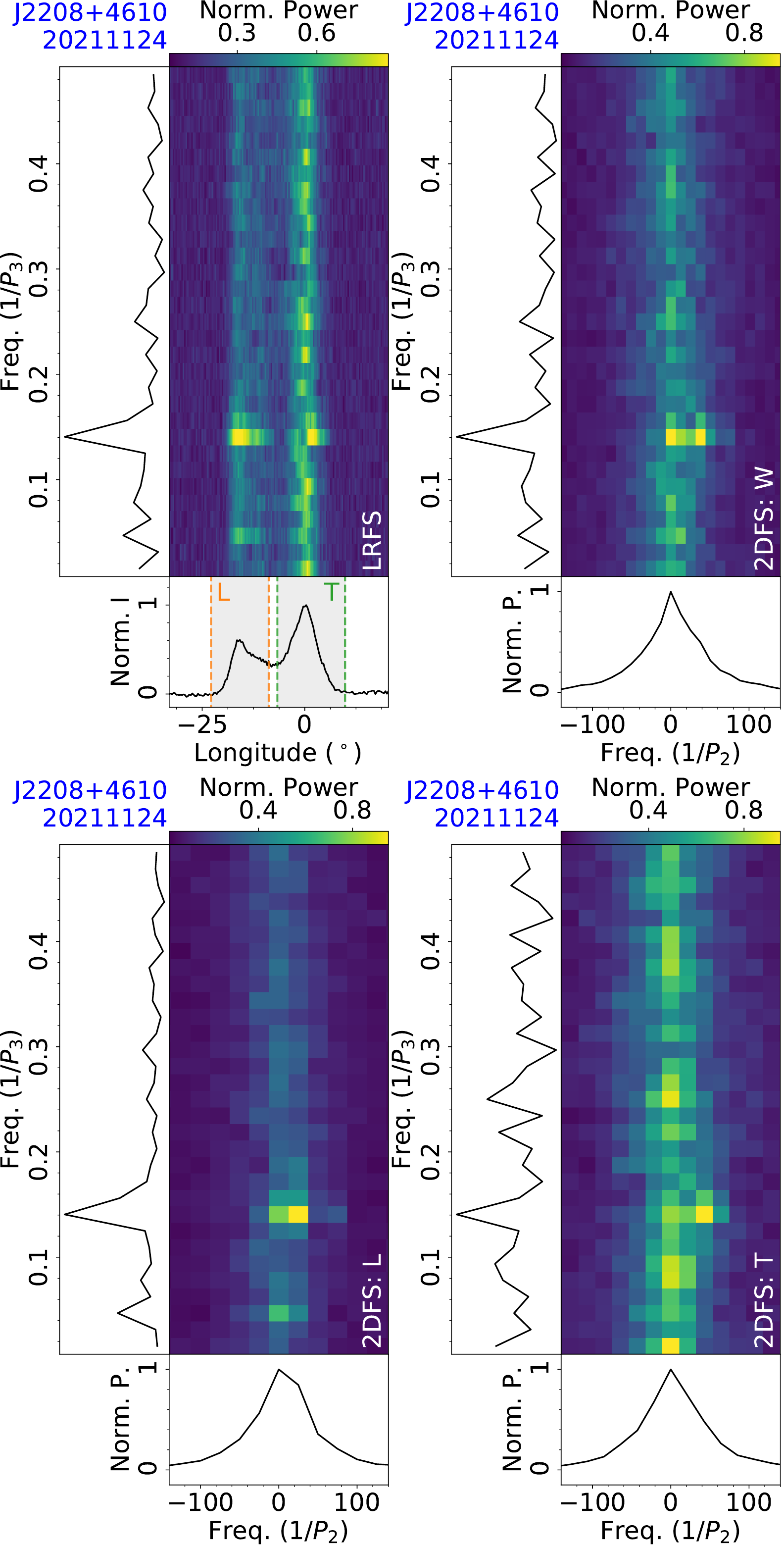}
\figcaption{Fluctuation analysis of PSR J2208+4610 for the observation on 20211124, with LRFS (top-left), and 2DFS for the on-pulse region (top-right), leading part (bottom-left) and trailing part (bottom-right) of a mean pulse profile.
\label{subfig:fluctu:J2208+4610}}
\end{figure}

\subsection{J2123+36}
\label{subsec:J2123+36}

PSR J2123+36 was discovered in the LOFAR Tied-Array All-Sky Survey (LOTAAS) \citep{Sanidas2019}. 

This pulsar was observed by FAST on 20250125 for 5 minutes, deriving a rotation period $P=1.2941$~s and a dispersion measure $D\!M=108.7~{\rm cm^{-3}\,pc}$.
Single pulse sequences in Fig.~\ref{subfig:TP:J2123+36} indicate the existence of the subpulse drifting phenomenon. Fluctuation spectra in Fig.~\ref{subfig:fluctu:J2123+36} show the weak positive drift feature, with the centroid frequencies of $1/P_3=0.355\pm0.003$ and $1/P_2=67\pm6$, corresponding to periodicities of $P_3=2.81\pm0.02$ periods and $P_2=5.4\pm0.5^\circ$.

\subsection{J2125+52}
\label{subsec:J2125+52}

PSR J2125+52 was discovered by CHIME (https://www.chime-frb.ca/galactic).

This pulsar was observed by FAST on 20240731 for 5 minutes, deriving a rotation period $P=5.8012$~s and a dispersion measure $D\!M=259.7~{\rm cm^{-3}\,pc}$. 
The single pulse sequence in Fig.~\ref{subfig:TP:J2125+52} displays the nulling phenomenon. The nulling fraction of this FAST observation is estimated from the on-pulse integral energy histogram to be 35$\pm$3\%.

\subsection{J2138+4911}
\label{subsec:J2138+4911}

PSR J2138+4911 was discovered by the Green Bank Telescope (GBT) at 350 MHz \citep{Hessels2008}. 

The FAST observed the pulsar on 20200402 for 5 minutes, deriving a rotation period $P=0.6962$~s and a dispersion measure $D\!M=167.8~{\rm cm^{-3}\,pc}$. 
Single pulse sequences are displayed in Fig.~\ref{subfig:TP:J2138+4911}. Fluctuation spectra in Fig.~\ref{subfig:fluctu:J2138+4911} illustrate that the pulsar has a negative drifting behavior, and the range of the temporally modulated frequency is wide. 
The centroid of the drift feature is characterized by $1/P_3=0.101\pm0.002$ and $1/P_2=-24\pm1$, corresponding to periodicities of $P_3=9.9\pm0.2$ periods and $P_2=-15\pm1^\circ$.

\begin{figure}[htpb]
\centering
\includegraphics[width=0.22\textwidth, angle=0]{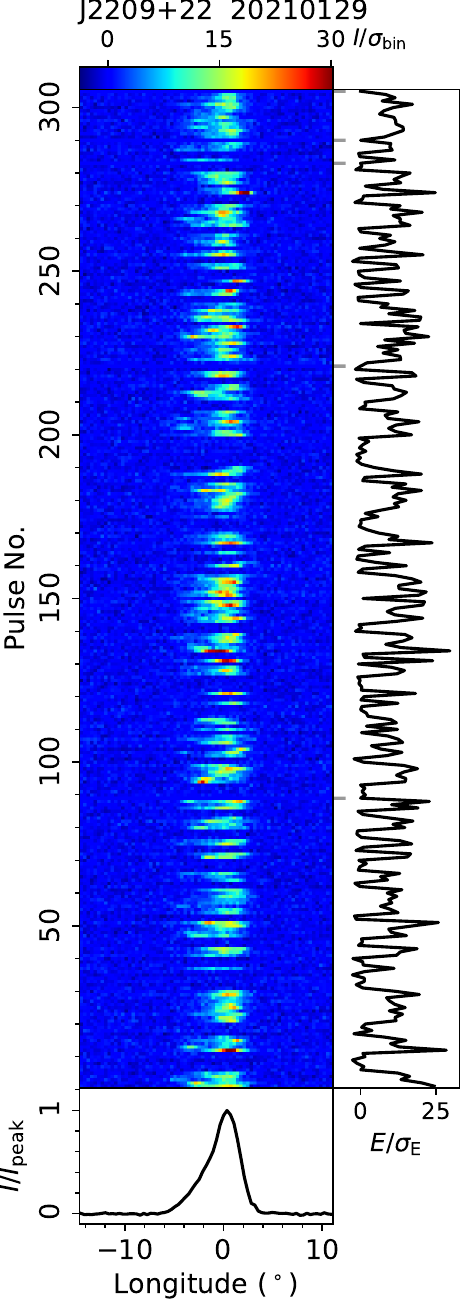}
\figcaption{Single pulse sequence of PSR J2209+22 from the FAST observation on 20210129.
\label{subfig:TP:J2209+22}}
\end{figure}

\begin{figure}[htpb]
\centering
\includegraphics[width=0.39\textwidth, angle=0]{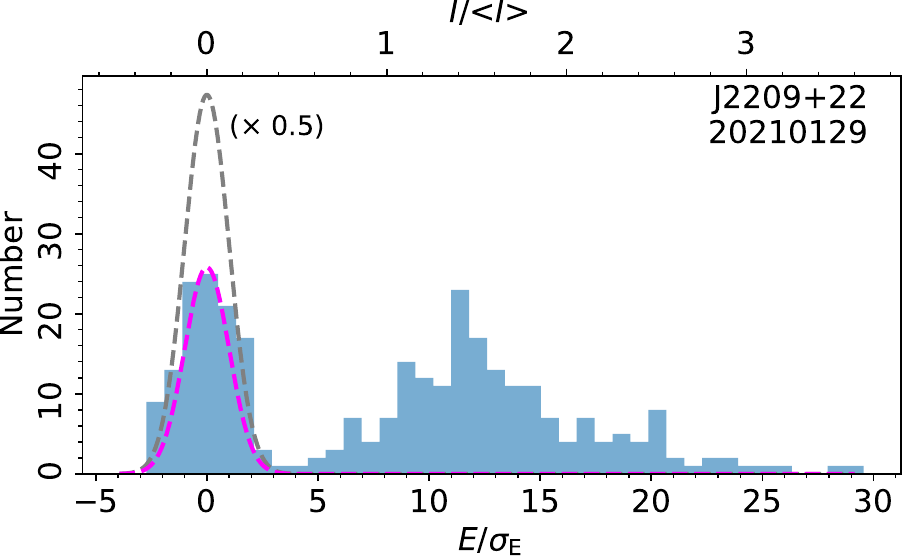}
\figcaption{On-pulse energy histogram of single pulses of PSR J2209+22 from the FAST observation on 20210129.
\label{subfig:Hist:J2209+22}}
\end{figure}

\subsection{J2150+5247}
\label{subsec:J2150+5247}

PSR J2150+5247 was discovered in the Princeton-NRAO pulsar survey, carried out at 390 MHz using the 92 m telescope at Green Bank \citep{Dewey1985}. \citet{Ng2020} reported the discovery of the mode-switching behavior with CHIME.

This pulsar was observed by FAST on 20221120 for 5 minutes, deriving a rotation period $P=0.3322$~s and a dispersion measure $D\!M=148.8~{\rm cm^{-3}\,pc}$. The single pulse sequence and a zoomed-in view are displayed in Fig.~\ref{subfig:TP:J2150+5247}. LRFS and 2DFS of the leading, central, and trailing parts in a mean pulse profile are shown in Fig.~\ref{subfig:fluctu:J2150+5247}. In 2DFS of the leading part in a mean pulse profile, there is a temporal modulation feature whose $f_3$ is widely distributed from 0 to 0.22, with the centroid of $1/P_3=0.104\pm0.002$ ($P_3=9.7\pm0.1$ periods). The centroid temporally modulated frequency related to the central profile part in a mean profile is $1/P_3=0.074\pm0.002$, corresponding to $P_3=13.6\pm0.3$ periods. For the trailing part in the profile, there is a narrower modulation feature with the centroid of $1/P_3=0.118\pm0.001$ ($P_3=8.5\pm0.1$ periods). The modulation properties vary across the profile.

\begin{figure}[htpb]
\centering
\includegraphics[width=0.22\textwidth, angle=0]{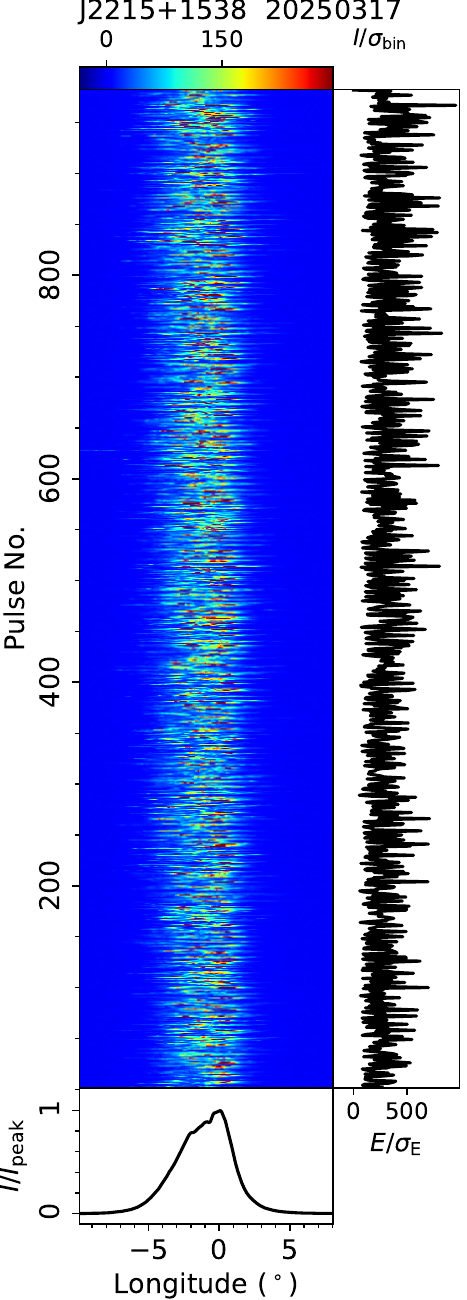}
\includegraphics[width=0.22\textwidth, angle=0]{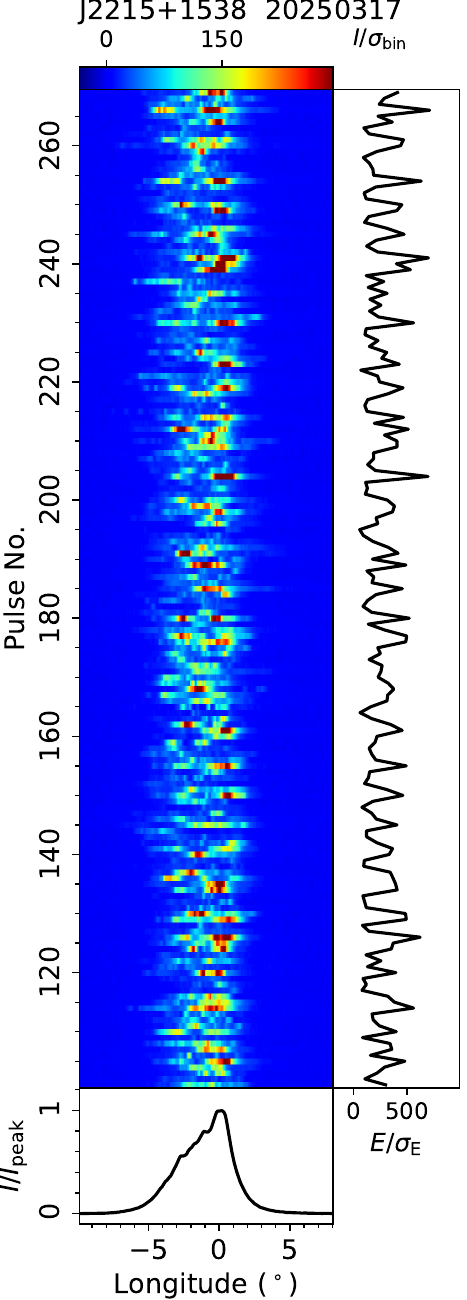}
\figcaption{Single pulse sequence of PSR J2215+1538 from the FAST observation on 20250317, and a zoomed-in view of pulses No. 100-270.
\label{subfig:TP:J2215+1538}}
\end{figure}

\begin{figure}[htpb]
\centering
\includegraphics[width=0.22\textwidth, angle=0]{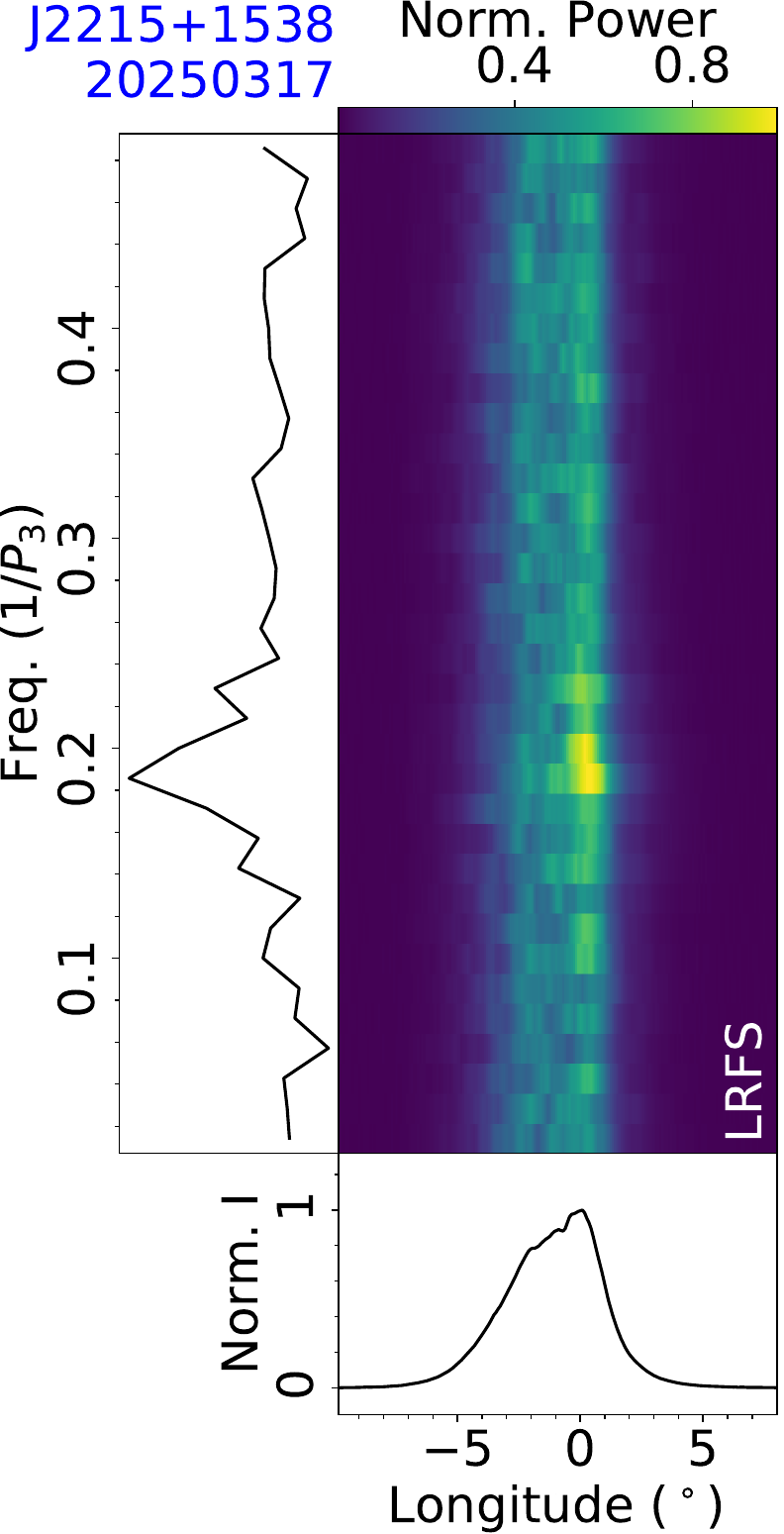}
\includegraphics[width=0.22\textwidth, angle=0]{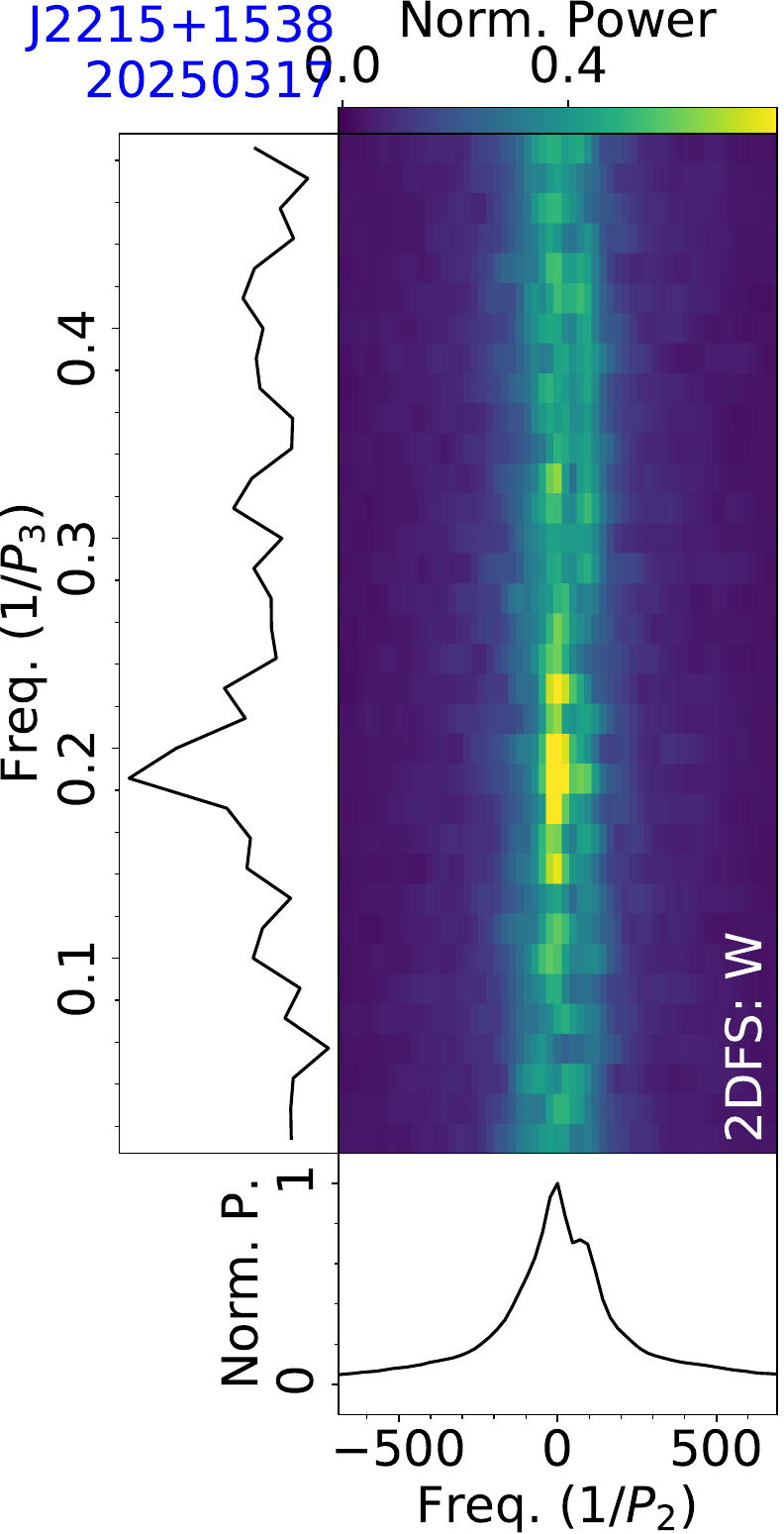}
\figcaption{Fluctuation analysis of PSR J2215+1538 for the observation on 20250317, with LRFS and 2DFS for the on-pulse region of a mean pulse profile.
\label{subfig:fluctu:J2215+1538}}
\end{figure}

\begin{figure}[htpb]
\centering
\includegraphics[width=0.44\textwidth, angle=0]{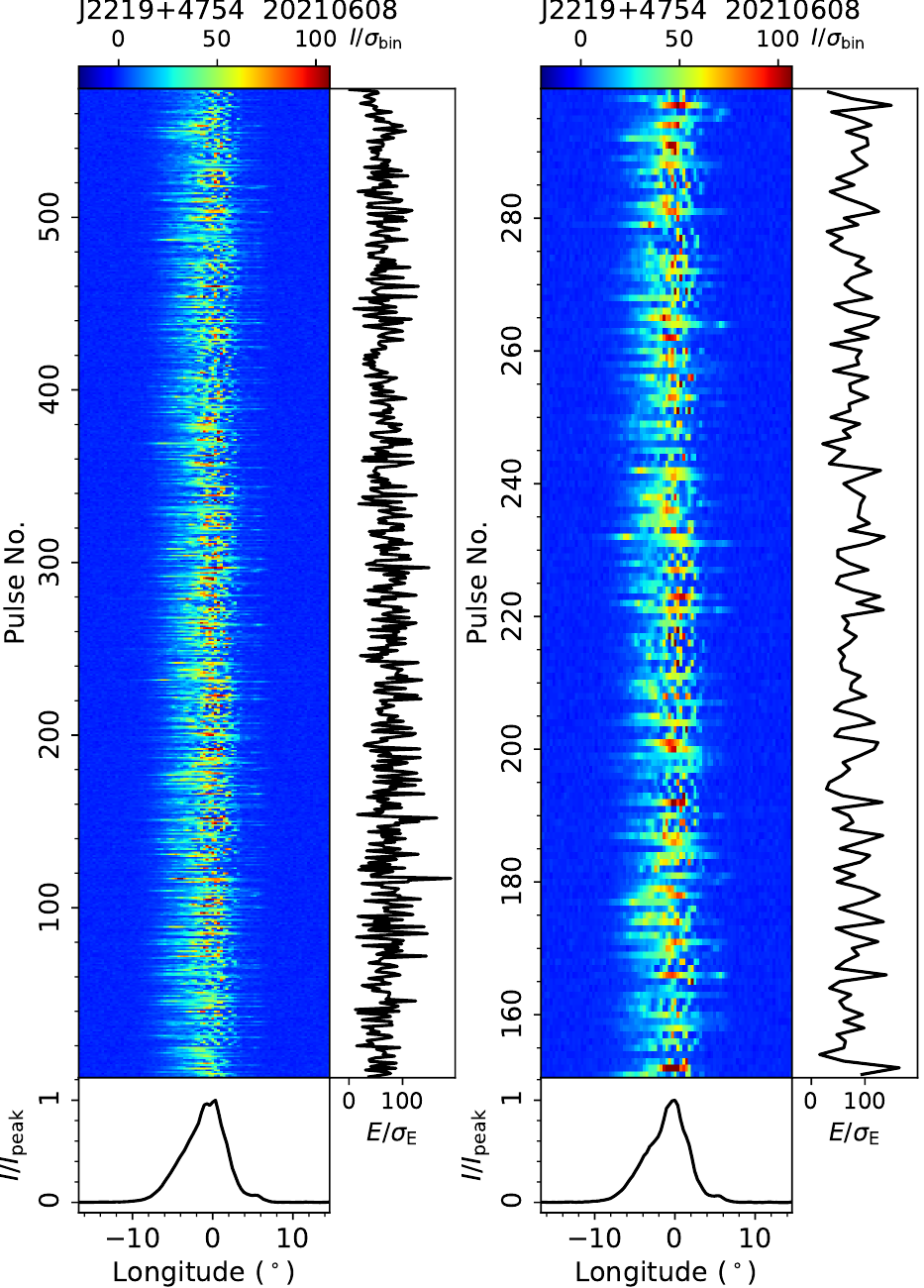}
\figcaption{Single pulse sequence of PSR J2219+4754 from the FAST observation on 20210608, and a zoomed-in view of pulses No.150-300. 
\label{subfig:TP:J2219+4754}}
\end{figure}

\begin{figure}[htpb]
\centering
\includegraphics[width=0.44\textwidth, angle=0]{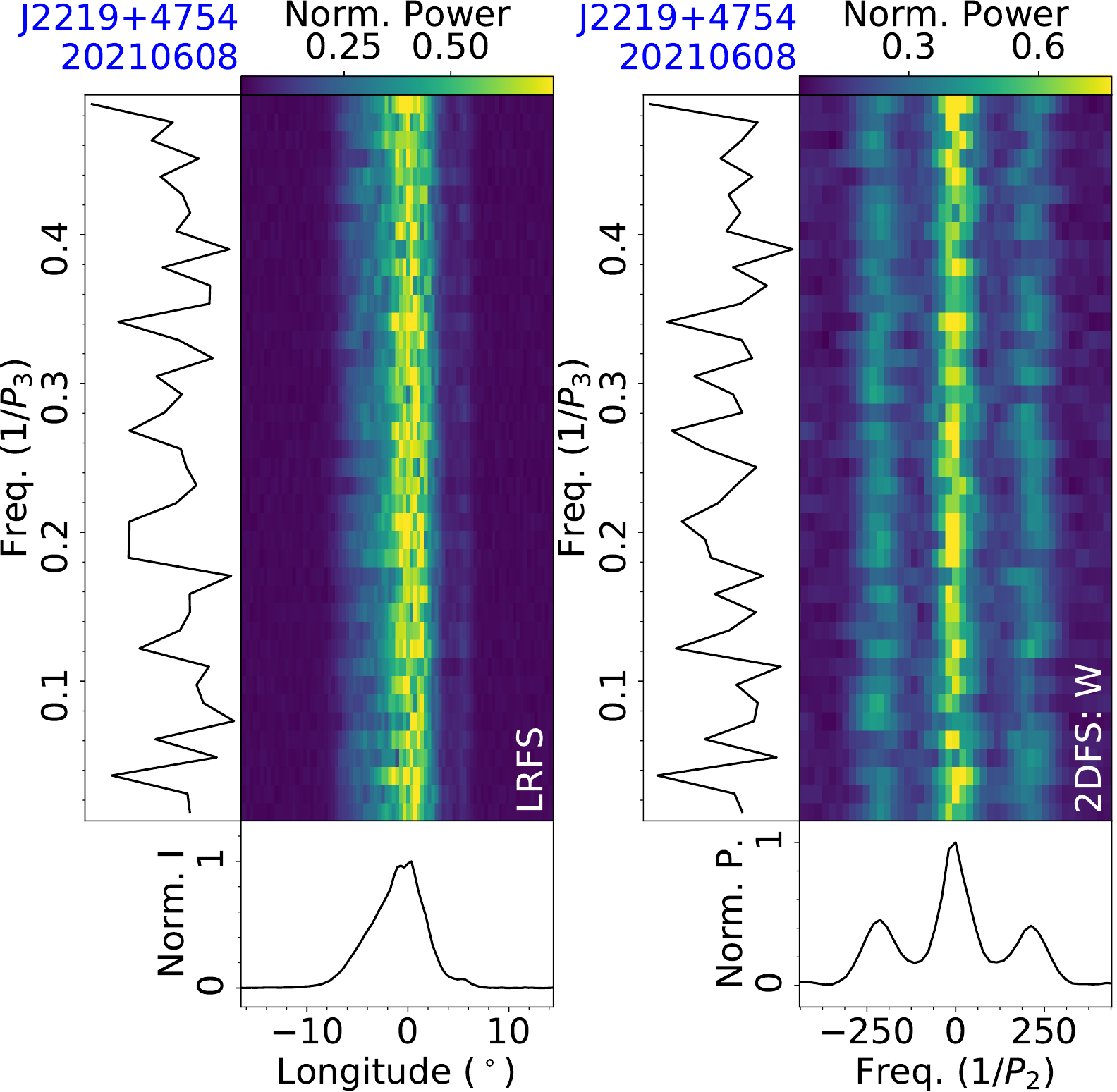}
\figcaption{Fluctuation analysis of PSR J2219+4754 for the observation on 20210608, with LRFS and 2DFS for the on-pulse region of a mean pulse profile.
\label{subfig:fluctu:J2219+4754}}
\end{figure}

\subsection{J2153+44}
\label{subsec:J2153+44}

PSR J2153+44 was discovered by CHIME (https://www.chime-frb.ca/galactic).

This pulsar was observed by FAST on 20240712 for 5 minutes, yielding a rotation period $P=2.8928$~s and  a dispersion measure $D\!M=141.7~{\rm cm^{-3}\,pc}$. 
The single pulse sequence in Fig.~\ref{subfig:TP:J2153+44} illustrates nulling behavior. From the on-pulse integral energy histogram in Fig.~\ref{subfig:Hist:J2153+44}, the nulling fraction is estimated to be 29$\pm$1\%.

\begin{figure}[htpb]
\centering
\includegraphics[width=0.22\textwidth, angle=0]{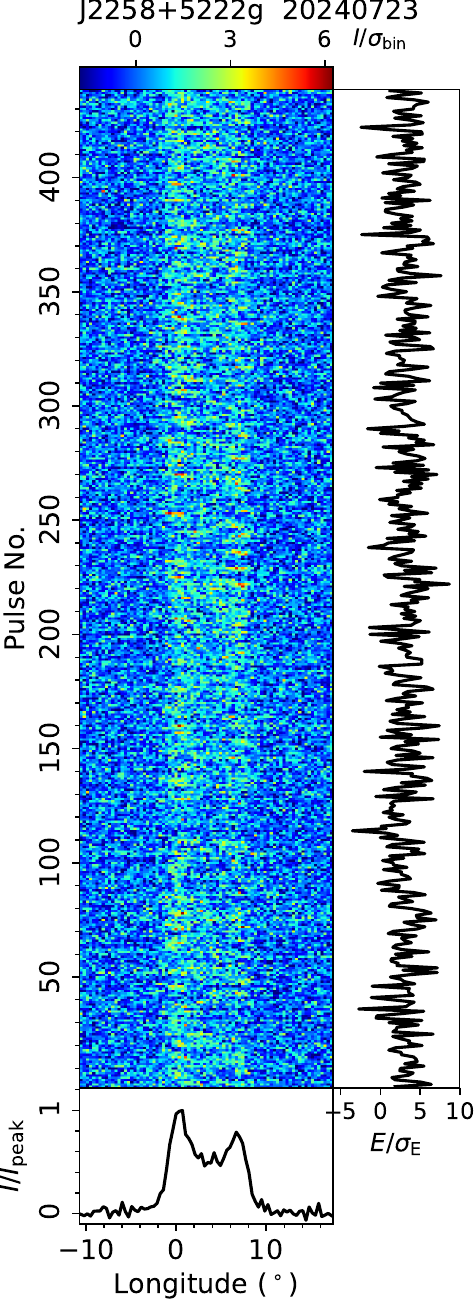}
\includegraphics[width=0.22\textwidth, angle=0]{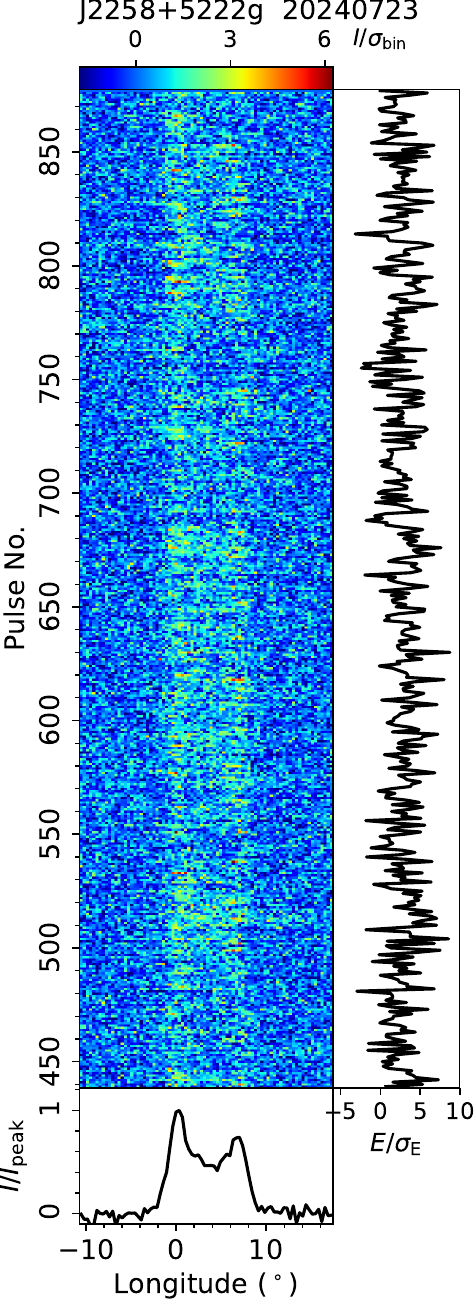}
\figcaption{Single pulse sequences of PSR J2258+5222g from the FAST observation on 20240723.
\label{subfig:TP:J2258+5222g}}
\end{figure}

\begin{figure}[htpb]
\centering
\includegraphics[width=0.44\textwidth, angle=0]{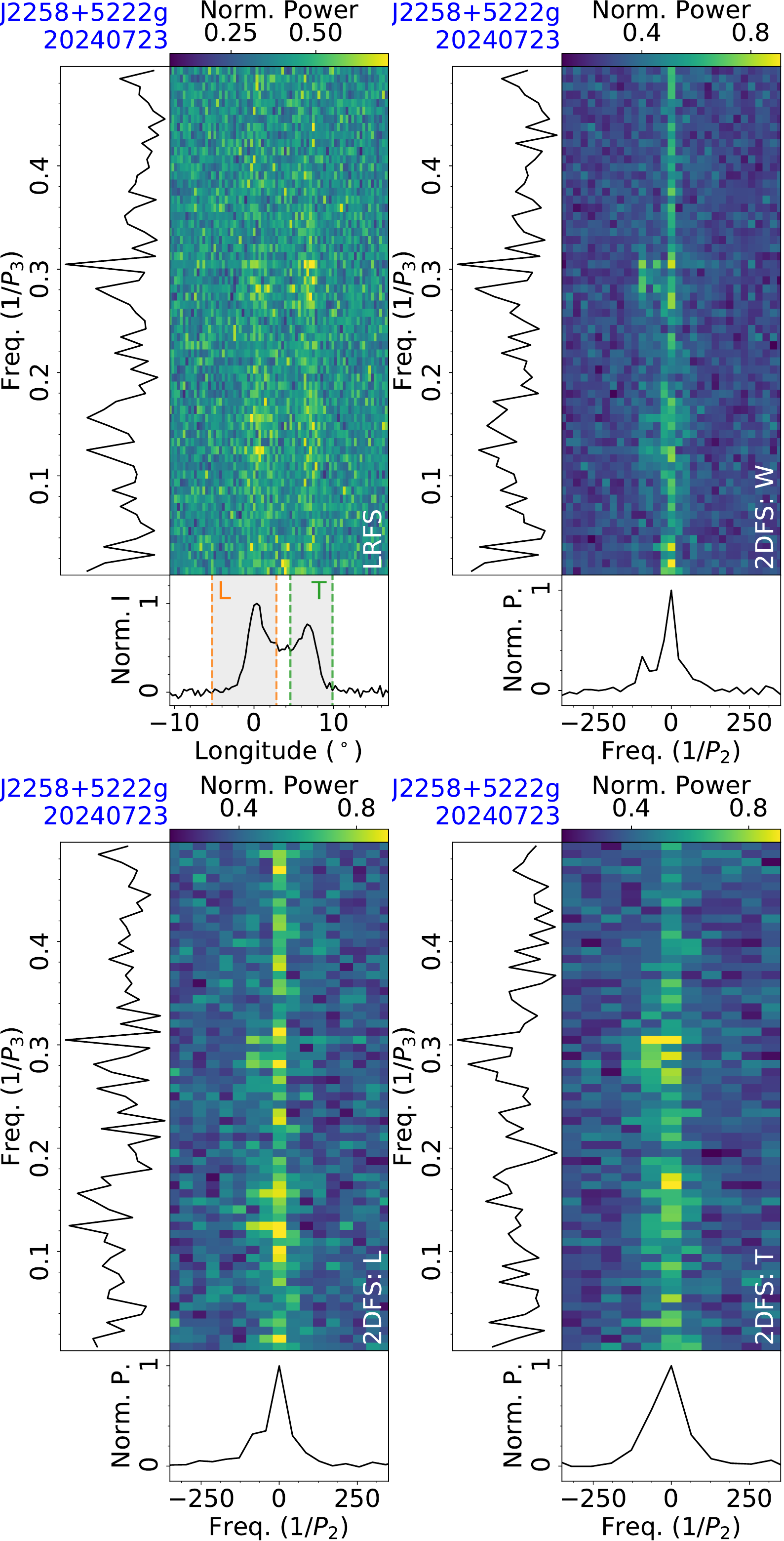}
\figcaption{Fluctuation analysis of PSR J2258+5222g for the observation on 20240723, with LRFS (top-left), and 2DFS for the on-pulse region (top-right), leading part (bottom-left) and trailing part (bottom-right) of a mean pulse profile.
\label{subfig:fluctu:J2258+5222g}}
\end{figure}

\subsection{J2208+4610}
\label{subsec:J2208+4610}

PSR J2208+4610 was discovered by \citet{Dong2023} using the CHIME telescope. 

The pulsar was observed by FAST on 20211124 for 5 minutes, deriving a rotation period $P=0.6426$~s and a dispersion measure $D\!M=62.1~{\rm cm^{-3}\,pc}$. 
The single pulse sequence in Fig.~\ref{subfig:TP:J2208+4610} and fluctuation spectra in Fig.~\ref{subfig:fluctu:J2208+4610} illustrate that the pulsar has subpulse drifting behavior, and the leading part in the mean pulse profile has a more systematic drifting than the trailing part. 
The centroid frequencies of the drift feature in 2DFS of the leading profile part are $1/P_3=0.145\pm0.003$ and $1/P_2=15\pm4$, corresponding to $P_3=6.9\pm0.1$ periods and $P_2=24\pm6^\circ$. 
For 2DFS of the trailing profile part, the drift feature exhibits centroid frequencies of $1/P_3=0.146\pm0.003$ and $1/P_2=43\pm5$, yielding $P_3=6.8\pm0.2$ periods and $P_2=8.4\pm0.9^\circ$.

\begin{figure}[htpb]
\centering
\includegraphics[width=0.22\textwidth, angle=0]{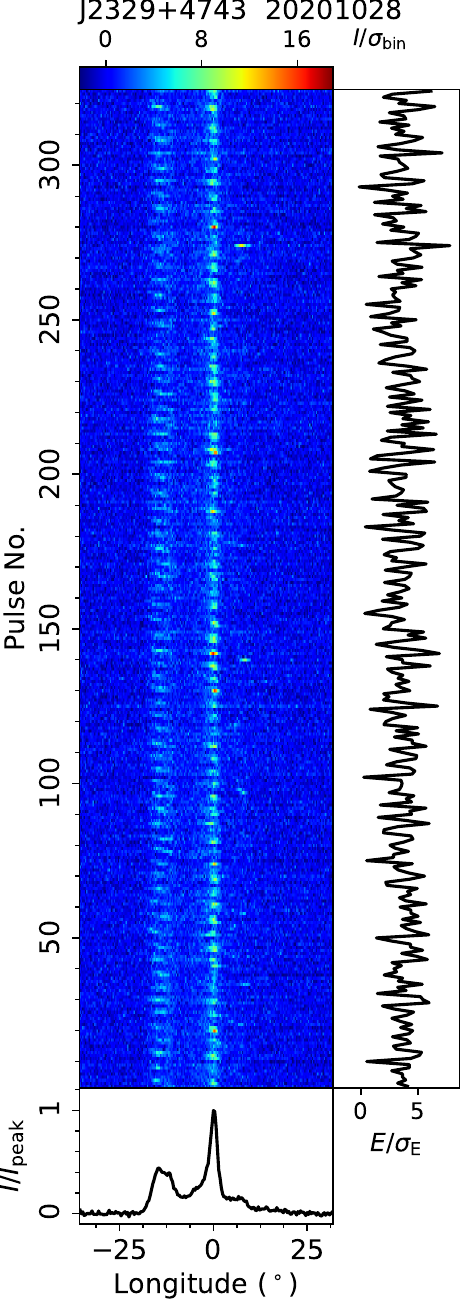}
\includegraphics[width=0.22\textwidth, angle=0]{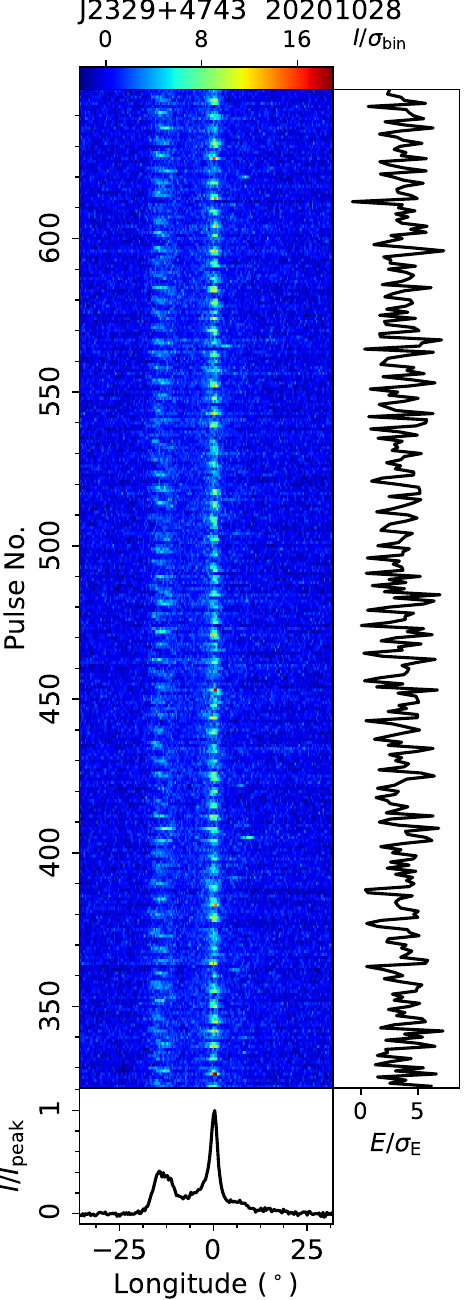}
\figcaption{Single pulse sequences of PSR J2329+4743 from the FAST observation on 20201028.
\label{subfig:TP:J2329+4743}}
\end{figure}

\begin{figure}[htpb]
\centering
\includegraphics[width=0.44\textwidth, angle=0]{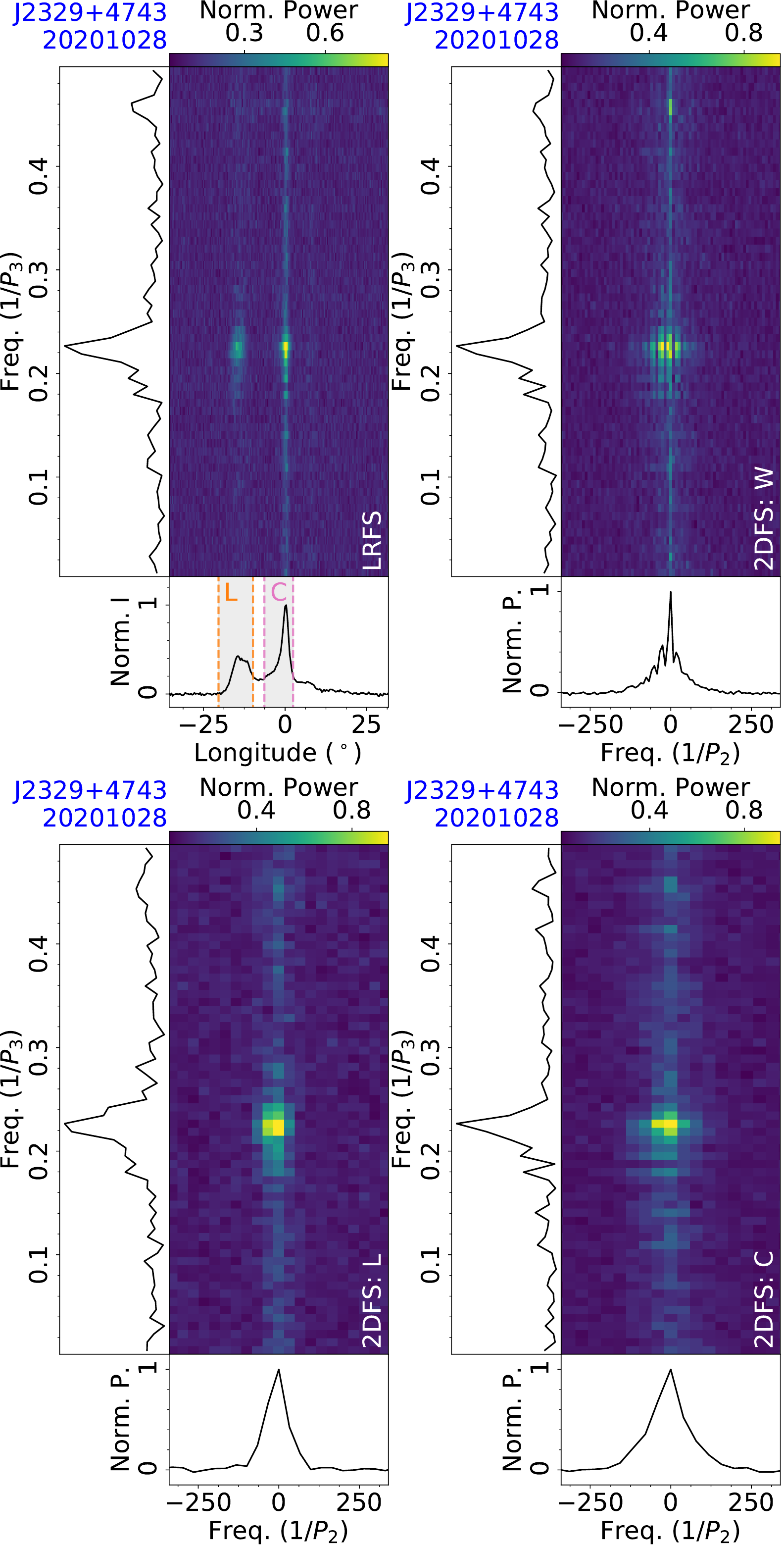}
\figcaption{Fluctuation analysis of PSR J2329+4743 for the observation on 20201028, with LRFS (top-left), and 2DFS for the on-pulse region (top-right), leading part (bottom-left) and central part (bottom-right) of a mean pulse profile.
\label{subfig:fluctu:J2329+4743}}
\end{figure}

\subsection{J2209+22}
\label{subsec:J2209+22}

PSR J2209+22 was reported as RRAT in the Pushchino survey \citep{Tyulbashev2018}. 
This pulsar was detected by FAST from the observation on 20210129 for 9 minutes, yielding a rotation period $P=1.7771$~s and a dispersion measure $D\!M=44.1~{\rm cm^{-3}\,pc}$. 
The single pulse sequence is displayed in Fig.~\ref{subfig:TP:J2209+22}. The distribution around zero energy in the on-pulse integral energy histogram of Fig.~\ref{subfig:Hist:J2209+22} indicates the existence of nulls. The nulling fraction of this observation is estimated to be 28$\pm$4\% from the histogram.

\begin{figure}[htpb]
\centering
\includegraphics[width=0.22\textwidth, angle=0]{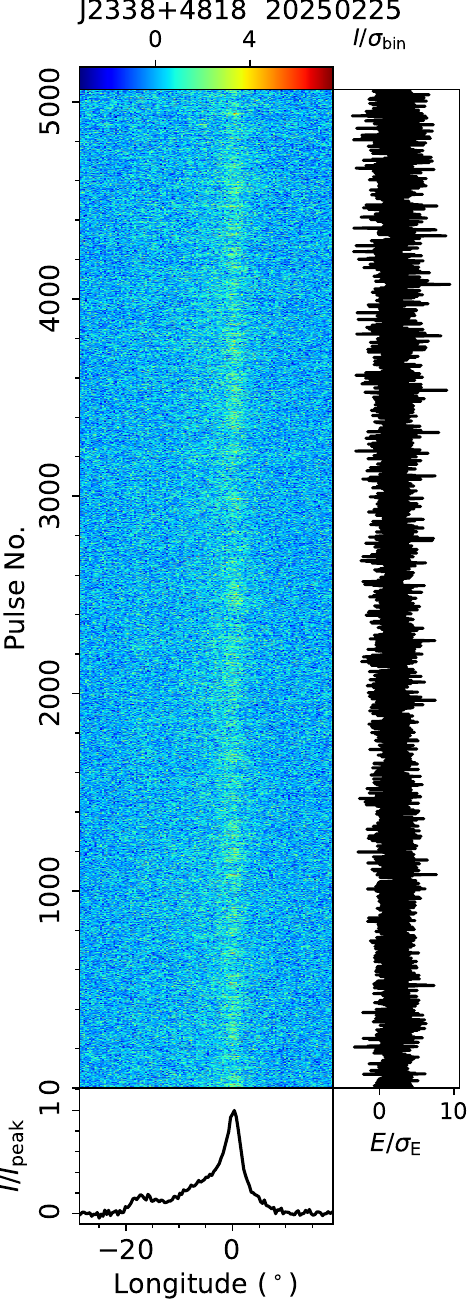}
\includegraphics[width=0.22\textwidth, angle=0]{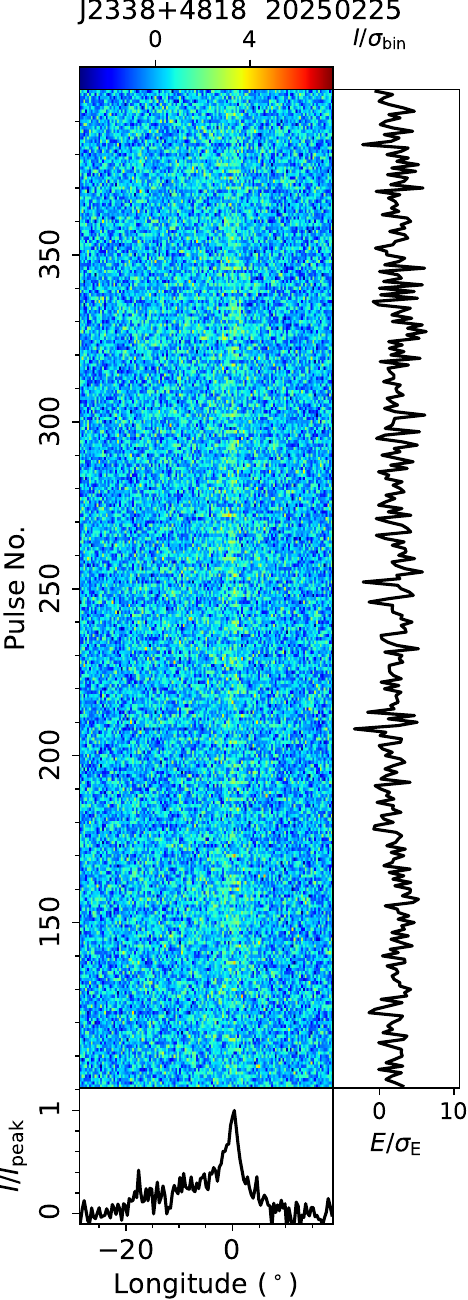}
\figcaption{The single pulse sequence of PSR J2338+4818 from the FAST observation on 20250225, and a zoomed-in view of pulses No. 101-400.
\label{subfig:TP:J2338+4818}}
\end{figure}

\begin{figure}[htpb]
\centering
\includegraphics[width=0.22\textwidth, angle=0]{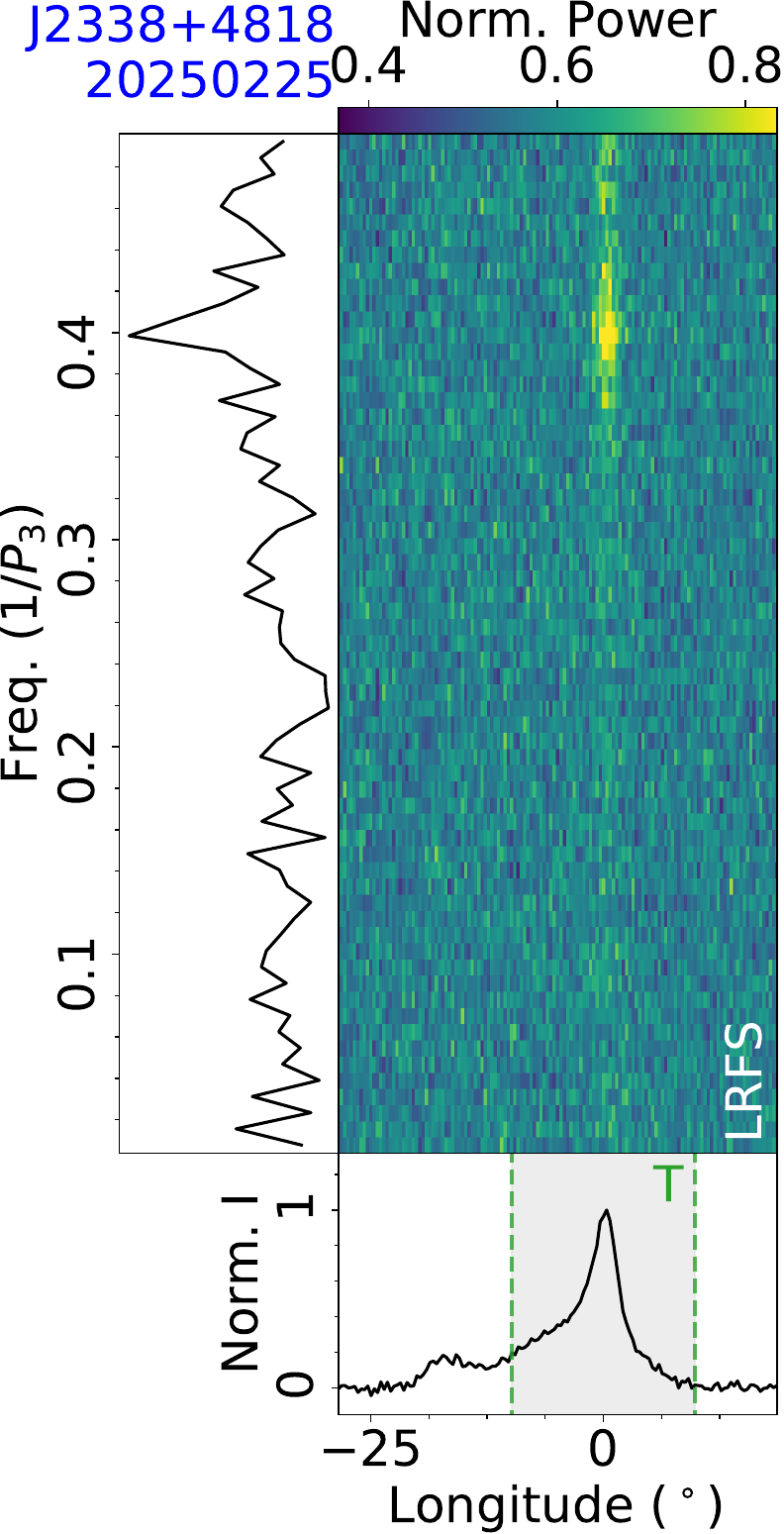}
\includegraphics[width=0.22\textwidth, angle=0]{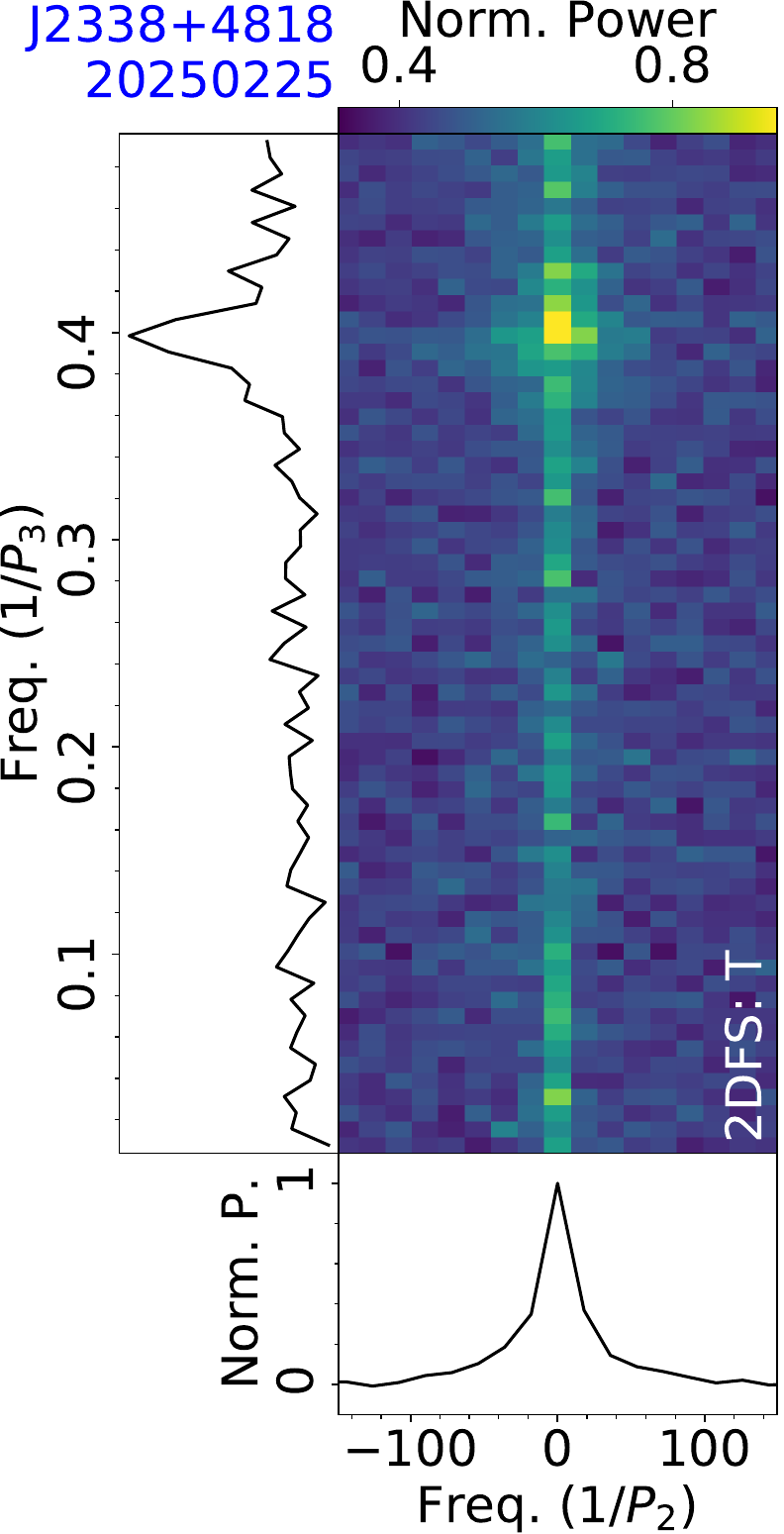}
\figcaption{Fluctuation analysis of PSR J2338+4818 for the observation on 20250225, with LRFS and 2DFS for the trailing part of a mean pulse profile.
\label{subfig:fluctu:J2338+4818}}
\end{figure}

\subsection{J2215+1538}
\label{subsec:J2215+1538}

PSR J2215+1538 was discovered by the Arecibo telescope \citep{CNT1996}. \citet{Song2023} reported the subpulse drifting parameters of $P_3=5.2\pm0.3$ periods and $P_2=38^{+14}_{-34}$ degrees. 

This pulsar was observed by FAST on 20250317 for 6 minutes, deriving a rotation period $P=0.3742$~s and a dispersion measure $D\!M=29.1~{\rm cm^{-3}\,pc}$. The single pulse sequence and a zoomed-in view of pulses No. 100-270 in Fig.~\ref{subfig:TP:J2215+1538} show the modulation phenomenon but no systematic drifting band. LRFS and 2DFS are displayed in Fig.~\ref{subfig:fluctu:J2215+1538}, and the main modulation feature exhibits a temporal modulation frequency of $1/P_3=0.189\pm0.002$, corresponding to $P_3=5.30\pm0.05$ periods. 


\begin{figure}[htpb]
\centering
\includegraphics[width=0.22\textwidth, angle=0]{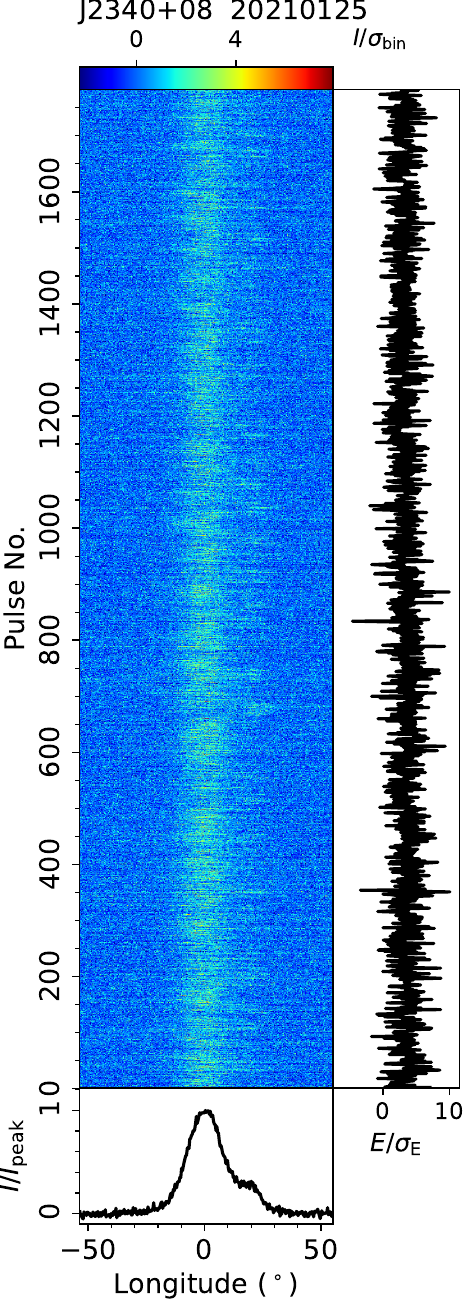}
\includegraphics[width=0.22\textwidth, angle=0]{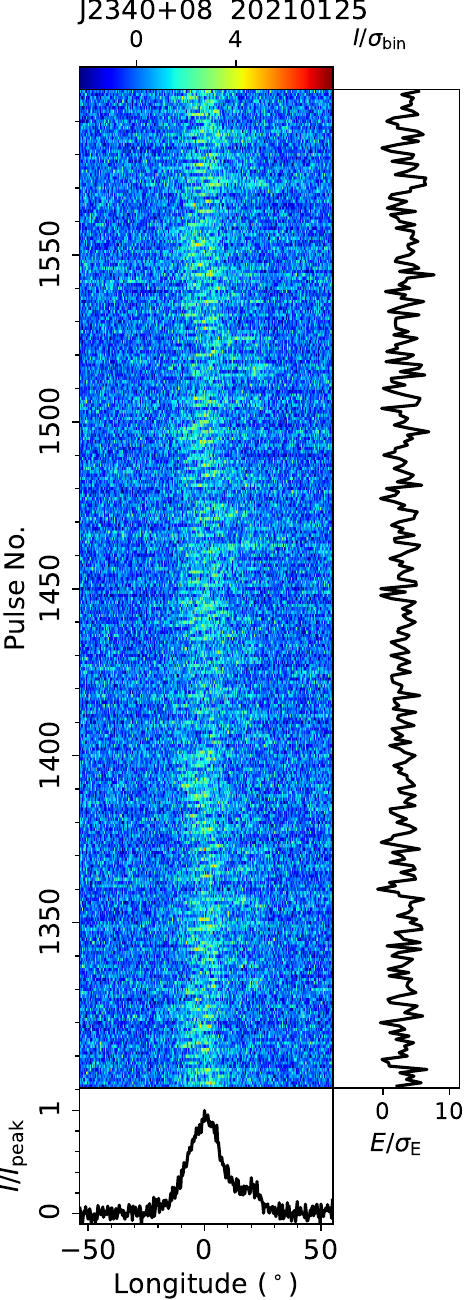}
\figcaption{Single pulse sequence of PSR J2340+08 from the FAST observation on 20210125, and a zoomed-in view of pulses No. 1301-1600.
\label{subfig:TP:J2340+08}}
\end{figure}

\begin{figure}[htpb]
\centering
\includegraphics[width=0.22\textwidth, angle=0]{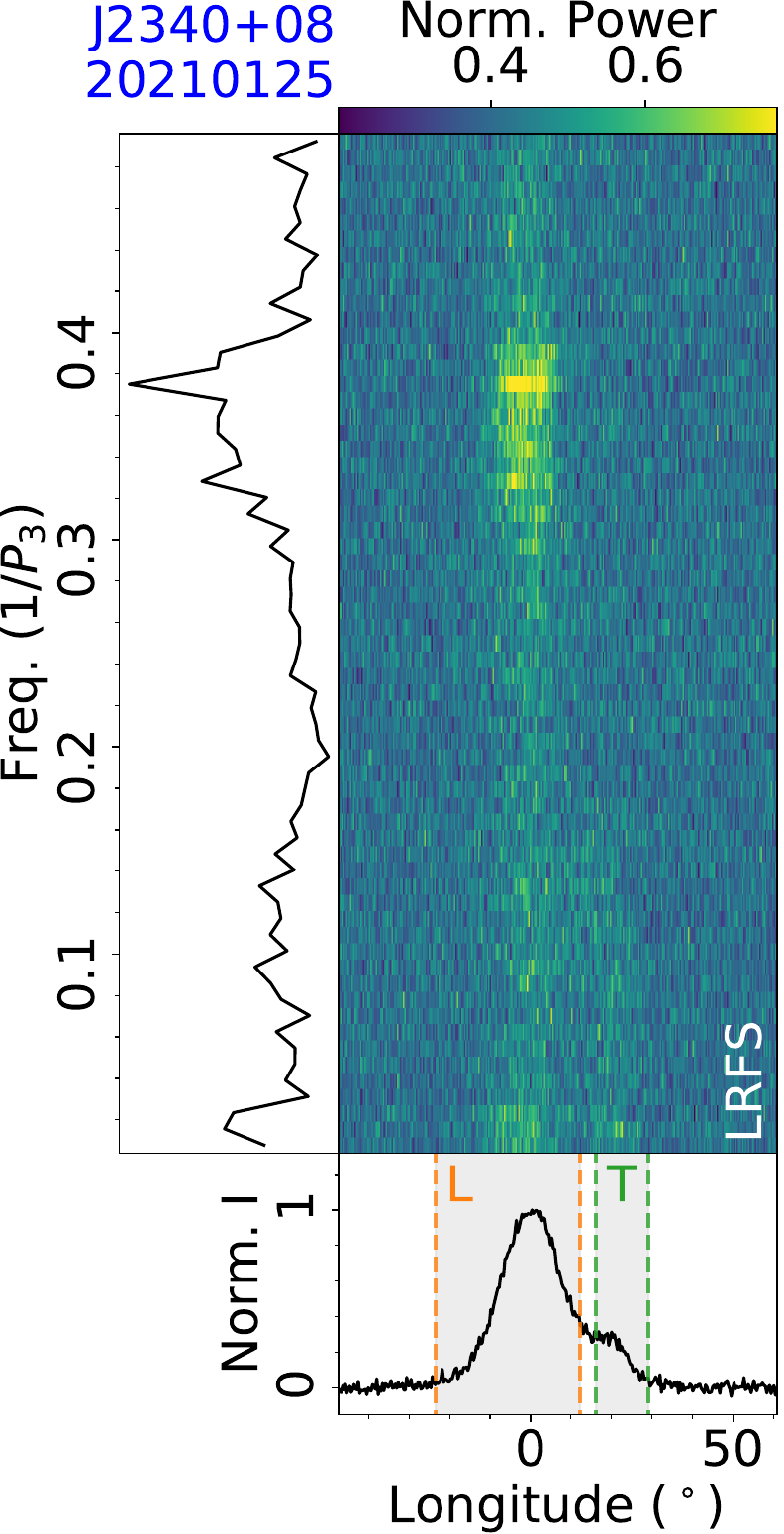}
\includegraphics[width=0.22\textwidth, angle=0]{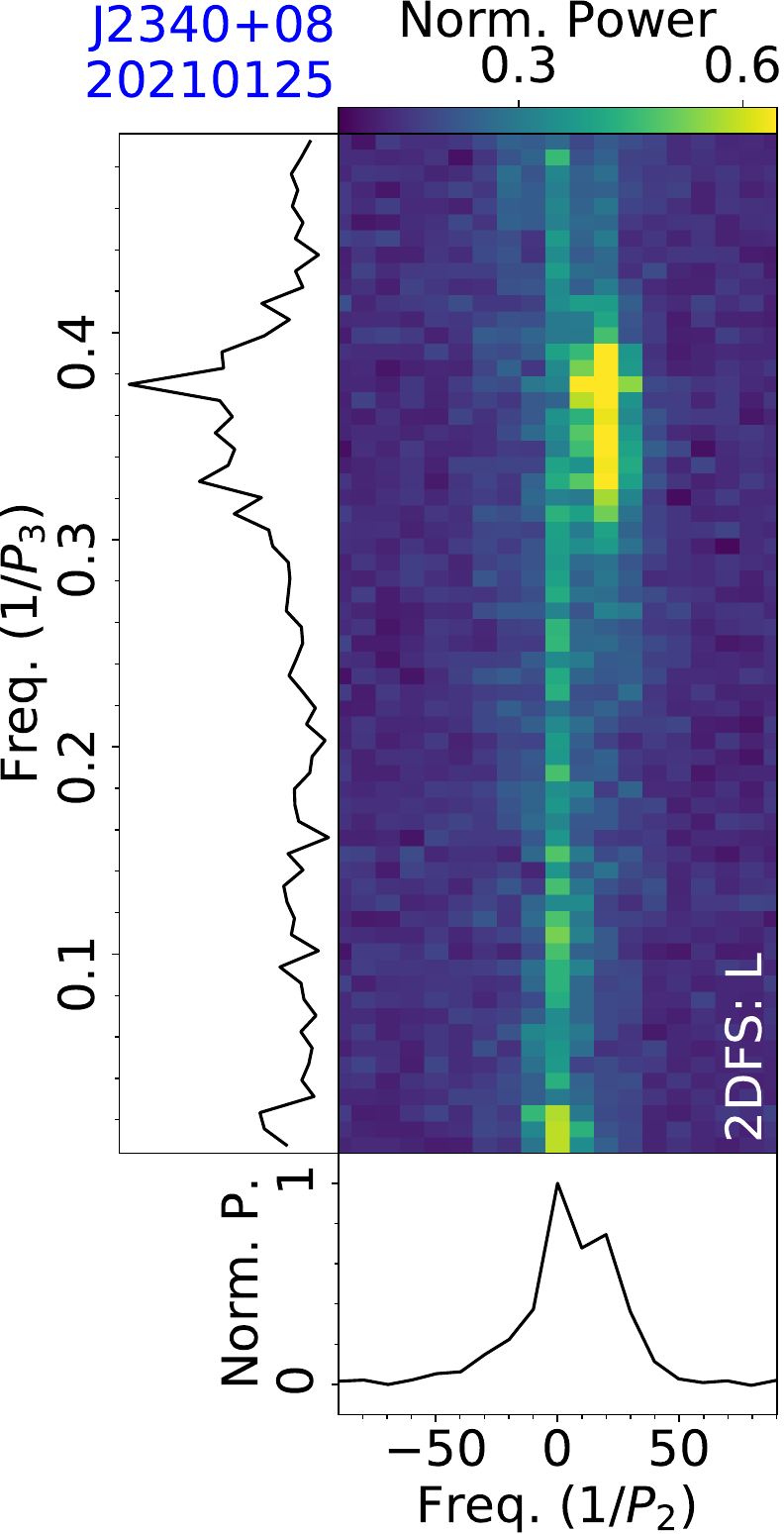}
\figcaption{Fluctuation analysis of PSR J2340+08 for the observation on 20210125, with LRFS and 2DFS for the leading part of a mean pulse profile.
\label{subfig:fluctu:J2340+08}}
\end{figure}

\subsection{J2219+4754}
\label{subsec:J2219+4754}

PSR J2219+4754 was discovered by \citet{Taylor1969} with the 300-foot transit telescope at the US National Radio Astronomy Observatory. \citet{Weltevrede2006} reported two vertical bands over the whole $P_3$ range in 2DFS observed at 21 cm. A preferred negative drift with $P_3=4.1\pm0.6$ periods and $P_2=-140^{+10}_{-100}$ degrees was detected by \citet{Weltevrede2007} at 92 cm, and the modulation of this pulsar was interpreted as a quasi-periodic intensity modulation in the pulses, but not much correlation between the subpulse positions from pulse to pulse. 

This pulsar was observed by FAST on 20210608 for 5 minutes, deriving a rotation period $P=0.5384$~s and a dispersion measure $D\!M=43.3~{\rm cm^{-3}\,pc}$. The single pulse sequence and a zoomed-in view of pulses No. 150-300 in Fig.~\ref{subfig:TP:J2219+4754} show that most single pulses consist of several subpulses. The 2DFS in Fig.~\ref{subfig:fluctu:J2219+4754} exhibits negative and positive drift features with constant $P_2$ that are widely distributed over $1/P_3$ from 0 to 0.5, indicating the stable phase interval between subpulses within a single pulse, but a variable drift rate over time. The centroid frequencies of two features are $1/P_2=-229\pm1$ and $231\pm2$, corresponding to $P_2=-1.57\pm0.01$ and $1.56\pm0.01$ degrees.

\subsection{J2258+5222g}
\label{subsec:J2258+5222g}

PSR J2258+5222g was discovered in the FAST GPPS survey \citep{Han2021,han2025}.

This pulsar was observed by FAST on 20240723 for 15 minutes and 20240908 for 14 minutes. From the data on 20240723, a rotation period $P=1.0305$~s and a dispersion measure $D\!M=125.3~{\rm cm^{-3}\,pc}$ were derived. Single pulse sequences of this observation are displayed in Fig.~\ref{subfig:TP:J2258+5222g}. 
From fluctuation spectra in Fig.~\ref{subfig:fluctu:J2258+5222g}, there are two drift features for both components, corresponding to two drifting modes. 
The modes with larger and smaller $P_3$ are referred to as drifting modes A and B, and their drifting parameters are derived from the centroid in 2DFS. For the drifting mode A, the drift feature of the leading profile part has centroid frequencies of $1/P_3=0.141\pm0.002$ ($P_3=7.1\pm0.1$ periods) and $1/P_2=-43\pm5$ ($P_2=-8\pm1^\circ$), and the trailing profile part has $1/P_3=0.141\pm0.002$ ($P_3=7.1\pm0.1$ periods) and $1/P_2=-32\pm5$ ($P_2=-11\pm2^\circ$). 
For the drifting mode B, the centroid frequencies of the feature in 2DFS are $1/P_3=0.292\pm0.002$ ($P_3=3.43\pm0.02$ periods) and $1/P_2=-66\pm6$ ($P_2=-5.5\pm0.5\circ$) for the leading profile part, and $1/P_3=0.288\pm0.001$ ($P_3=3.48\pm0.02$ periods) and $1/P_2=-41\pm6$ ($P_2=-9\pm1\circ$) for the trailing part.
Drifting parameters show large uncertainty and variations between two observations (See Table~\ref{Tab:moduPars-modeModu}), likely attributable to weak drift features in 2DFS. Accurate parameter determination requires longer observations.

\subsection{J2329+4743}
\label{subsec:J2329+4743}

PSR J2329+4743 was discovered in the LOFAR Tied-Array All-Sky Survey (LOTAAS) \citep{Sanidas2019}. 
The pulsar was observed by FAST on 20201028 for 8 minutes, yielding a rotation period $P=0.7285$~s and a dispersion measure $D\!M=43.4~{\rm cm^{-3}\,pc}$. 
There are systematic drift bands in single pulse sequences (Fig.~\ref{subfig:TP:J2329+4743}). LRFS and 2DFS in Fig.~\ref{subfig:fluctu:J2329+4743} show a clear drift feature, and the range of temporally modulated frequency is narrow, from which the leading and central parts in the mean pulse profile have similar temporal modulation periodicities.
In 2DFS of the leading profile part, the drift feature has the centroid of $1/P_3=0.218\pm0.001$ and $1/P_2=-14\pm2$, corresponding to periodicities of $P_3=4.56\pm0.02$ periods and $P_2=-26\pm4^\circ$. 2DFS of the central profile part exhibits the drift feature, with centroid frequencies of $1/P_3=0.223\pm0.001$ and $1/P_2=-16\pm3$, yielding $P_3=4.48\pm0.01$ periods and $P_2=-22\pm5^\circ$.

\begin{figure}[htpb]
\centering
\includegraphics[width=0.44\textwidth, angle=0]{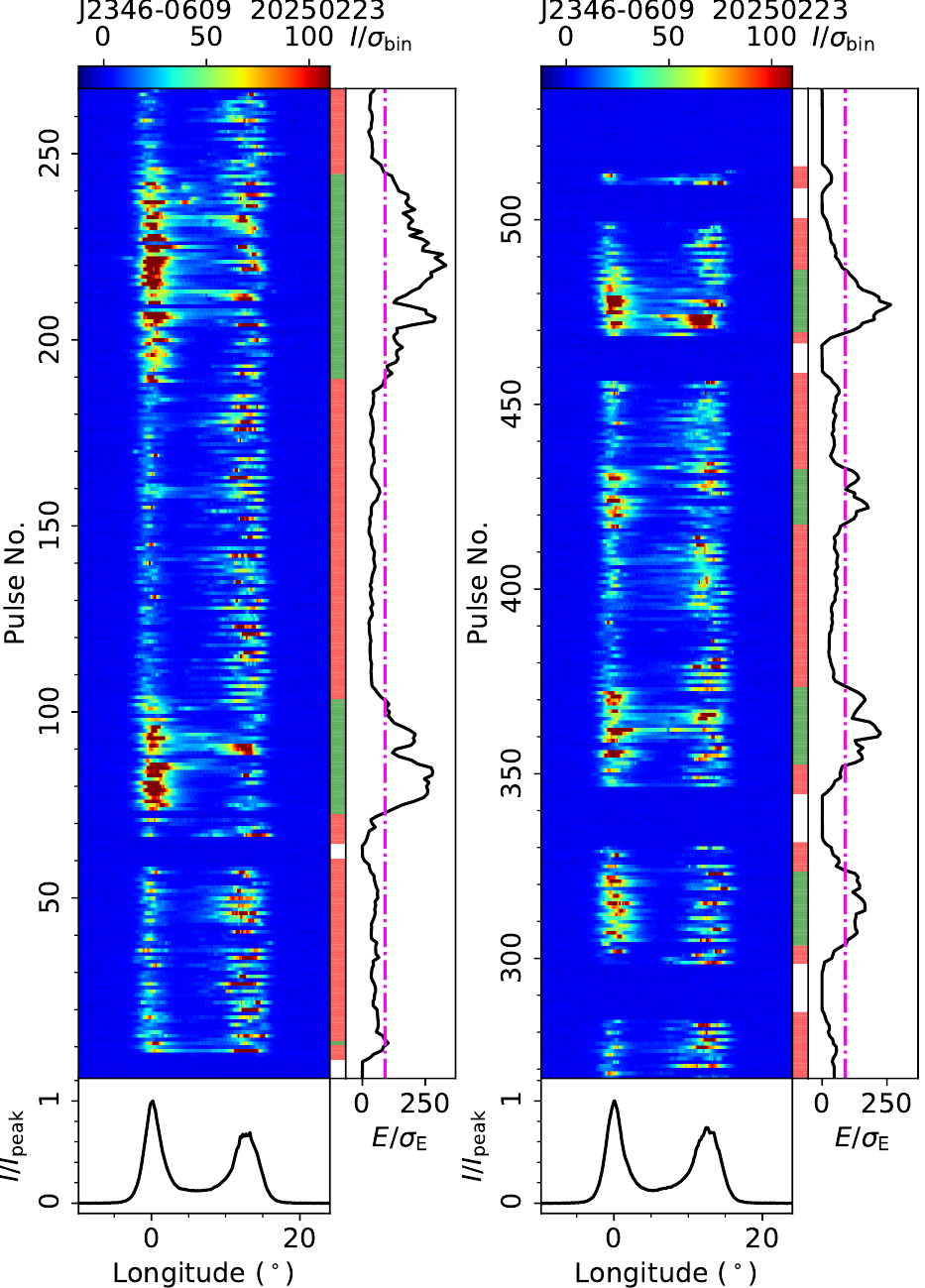}
\figcaption{Single pulse sequences of PSR J2346-0609 from the FAST observation on 20250223. The on-pulse energy variation versus period is smoothed over every 5 periods.
\label{subfig:TP:J2346-0609}}
\end{figure}

\begin{figure}[htpb]
\centering
\includegraphics[width=0.39\textwidth, angle=0]{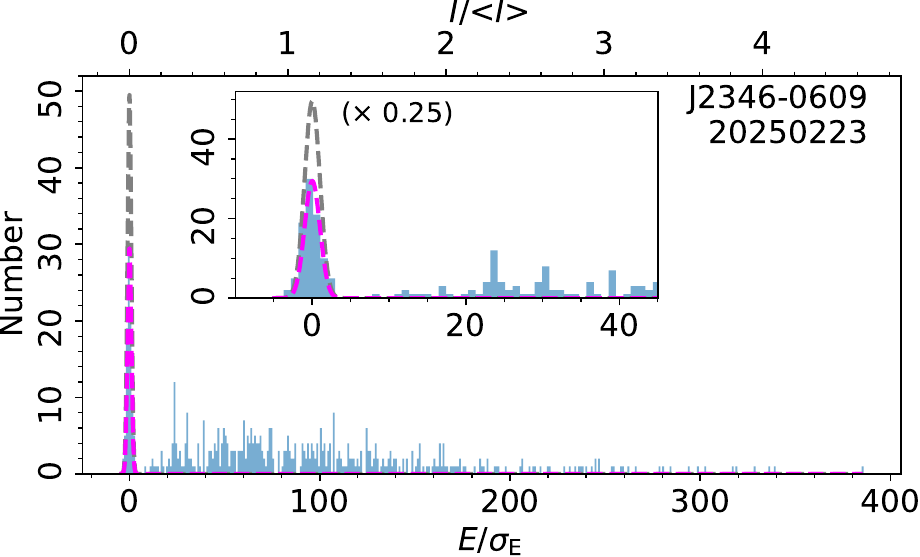}
\figcaption{On-pulse energy histogram of single pulses of PSR J2346-0609 from the FAST observation on 20250223. The inset provides a view of the x‑axis region from -10 to 45.
\label{subfig:Hist:J2346-0609}}
\end{figure}

\begin{figure}[htpb]
\centering
\includegraphics[width=0.39\textwidth, angle=0]{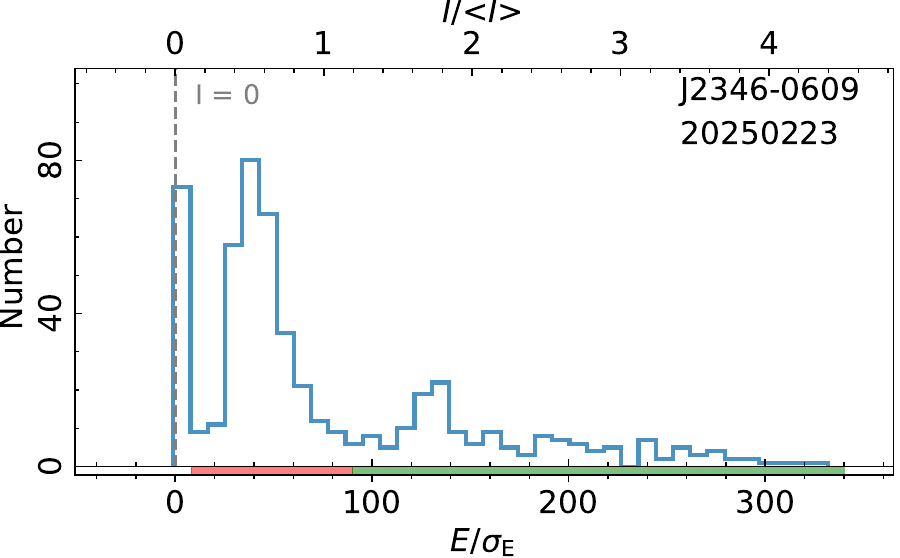}
\figcaption{Histogram of the smoothed integral energy for the leading profile part of single pulses of PSR J2346-0609. The red and green bars indicate the weak and bright emission modes. The data is from the FAST observation on 20250223.
\label{subfig:HistMode:J2346-0609}}
\end{figure}

\begin{figure}[htpb]
\centering
\includegraphics[width=0.37\textwidth, angle=0]{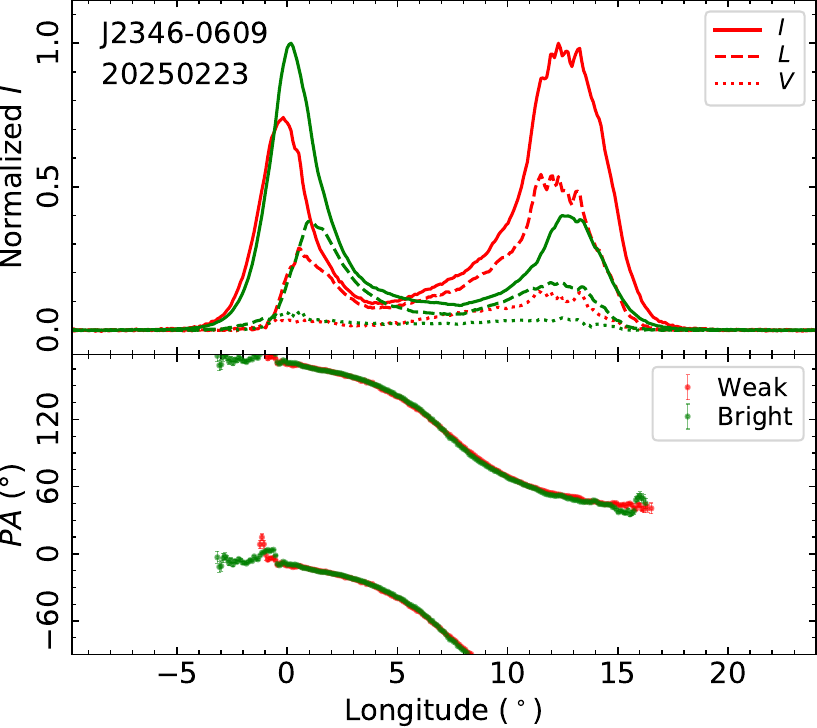}
\figcaption{Mean polarization profiles (the top panel) for weak and bright emission modes of PSR J2346-0609 observed on 20250223, as well as the averaged PA curves (the bottom panel).
\label{subfig:PolModes:J2346-0609}}
\end{figure}

\begin{figure}[htpb]
\centering
\includegraphics[width=0.44\textwidth, angle=0]{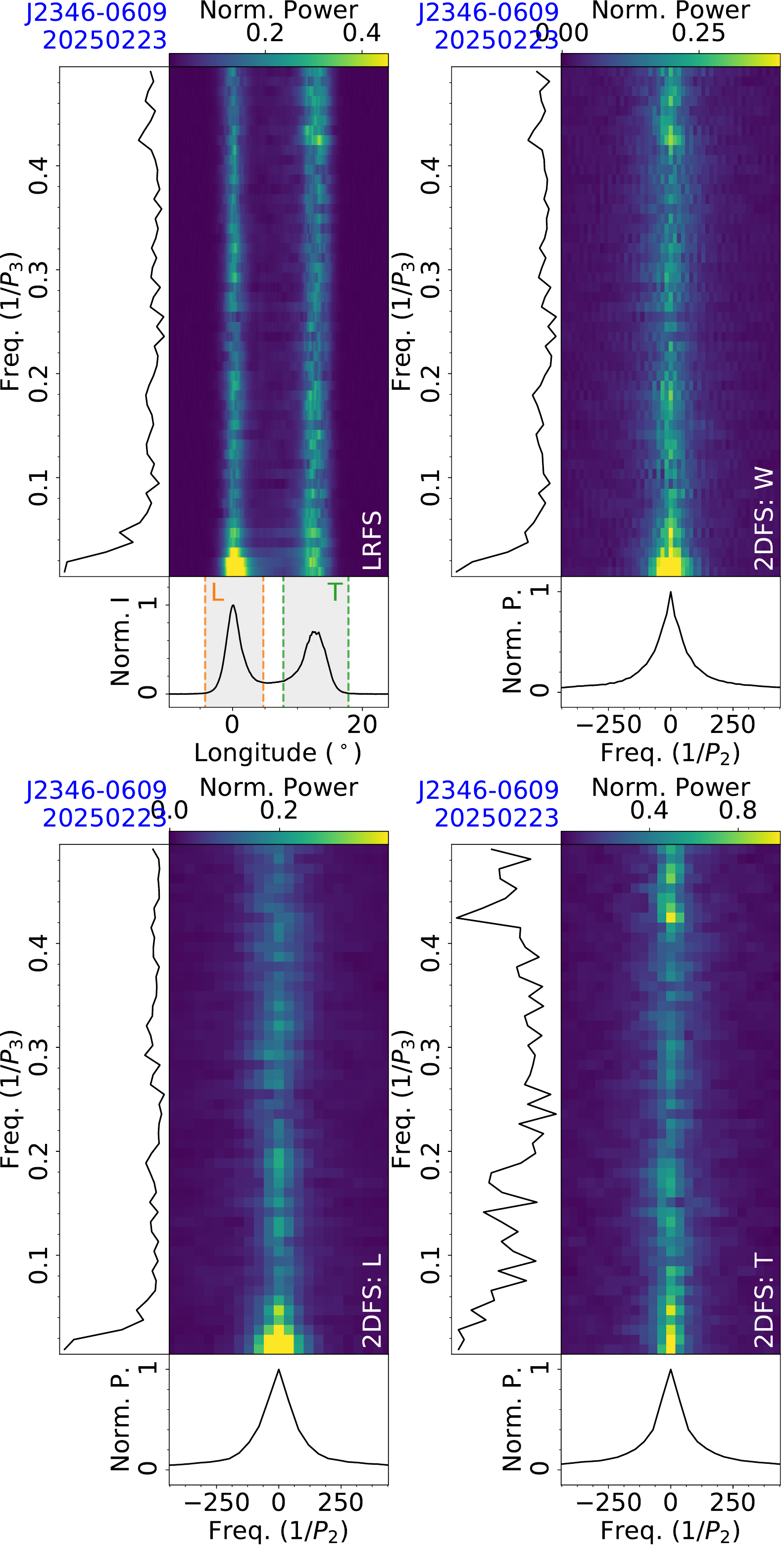}
\figcaption{Fluctuation analysis of PSR J2346-0609 for the observation on 20250223, with LRFS (top-left), and 2DFS for the on-pulse region (top-right), leading part (bottom-left) and trailing part (bottom-right) of a mean pulse profile.
\label{subfig:fluctu:J2346-0609}}
\end{figure}

\subsection{J2338+4818}
\label{subsec:J2338+4818}

PSR J2338+4818 was discovered by FAST \citep{Cruces2021}. 

This pulsar was also observed by FAST on 20250225 for 10 minutes, deriving a rotation period $P=0.1187$~s and a dispersion measure $D\!M=33.8~{\rm cm^{-3}\,pc}$. The single pulse sequence and a zoomed-in view are shown in Fig.~\ref{subfig:TP:J2338+4818}. From LRFS and 2DFS in Fig.~\ref{subfig:fluctu:J2338+4818}, the trailing part in a mean pulse profile has a temporal modulation, with the centroid frequency of $1/P_3=0.400\pm0.001$, corresponding to the periodicity of $P_3=2.50\pm0.01$ periods.

\subsection{J2340+08}
\label{subsec:J2340+08}

PSR J2340+08 was discovered by \citet{Deneva2013} using the Arecibo telescope at 327 MHz.

The pulsar was observed by FAST on 20210125 for 9 minutes, deriving a rotation period $P=0.3033$~s and a dispersion measure $D\!M=23.2~{\rm cm^{-3}\,pc}$. 
Single pulse sequences of the observation are shown in Fig.~\ref{subfig:TP:J2340+08}. 
There is a clear drift feature in 2DFS of the leading profile part with the centroid frequencies of $1/P_3=0.357\pm0.001$ and $1/P_2=20\pm1$, corresponding to $P_3=2.80\pm0.01$ periods and $P_2=18.2\pm0.5^\circ$. Additionally, there is also a low-frequency modulation in 2DFS with a centroid temporally modulated frequency of $1/P_3=0.016\pm0.001$, yielding the periodicity of $P_3=61\pm4$ periods. From the zoomed-in single pulse sequence in Fig.~\ref{subfig:TP:J2340+08}, the low-frequency modulation is superimposed on the subpulse drifting behavior.

\subsection{J2346-0609}
\label{subsec:J2346-0609}

PSR J2346-0609 was discovered by \citet{Manchester1996} in the Parkes Southern Pulsar Survey. Previous studies have reported the low-frequency modulation, especially in the leading component, as well as the subpulse drifting behavior of $P_3\sim2.3$ periods for the trailing component \citep{Weltevrede2006,Song2023}. In addition, the pulsar also has the nulling phenomenon, with a nulling fraction of 42.5$\pm$3.8\% at 333 MHz and 28.7$\pm$1.8\% at 618 MHz \citep{Basu2017}.

This pulsar was observed by FAST on 20250223 for 11 minutes, with a rotation period $P=1.1815$~s and a dispersion measure $D\!M=22.9~{\rm cm^{-3}\,pc}$ determined. Single pulse sequences in Fig.~\ref{subfig:TP:J2346-0609} illustrate that the pulsar has nulling, subpulse modulation, as well as mode-changing phenomena. The nulling fraction of this observation is estimated from the on-pulse integral energy histogram (Fig.~\ref{subfig:Hist:J2346-0609}), which is 14.9$\pm$1.3\%. From single pulse sequences, the emission is occasionally enhanced, especially for the leading part in the mean pulse profile. Weak and bright emission modes of single pulses are distinguished from the integral energy histogram of the leading profile part (Fig.\ref{subfig:HistMode:J2346-0609}). Mean polarization profiles and PA curves of two emission modes are shown in Fig.~\ref{subfig:PolModes:J2346-0609}. The weak emission mode exhibits a stronger trailing component relative to the leading component, while the bright emission mode has a stronger leading component. For the PA curves of the two modes, there is no obvious difference. 
From fluctuation spectra in Fig.~\ref{subfig:fluctu:J2346-0609}, the strong low-frequency modulation feature in the 2DFS of the leading profile part primarily results from mode changes. 2DFS of the trailing profile part exhibits a modulation feature with the centroid temporal modulation frequency of $1/P_3=0.448\pm0.002$, corresponding to the periodicity of $P_3=2.23\pm0.01$ periods.

\begin{figure*}[thbp]
\centering
\includegraphics[width=0.6\textwidth, angle=0]{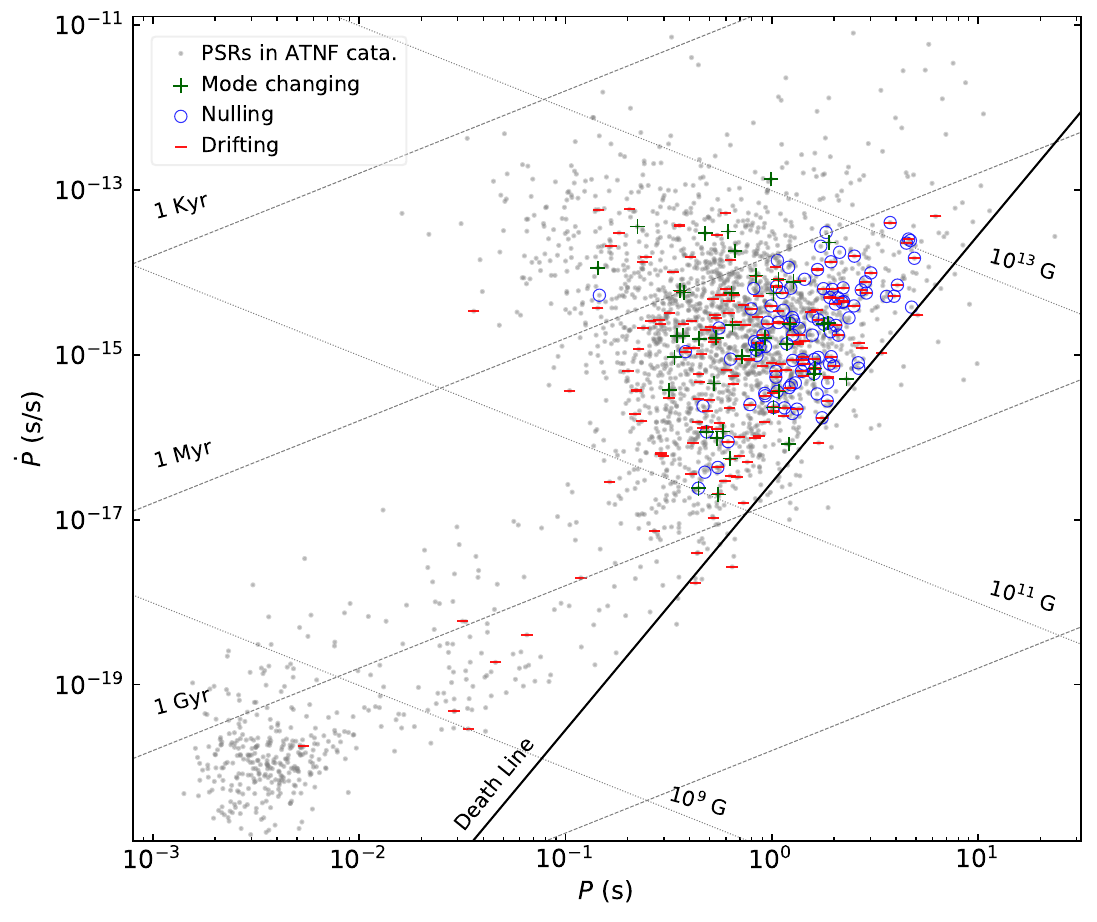}
\vspace{-3mm}
\caption{$P-\dot{P}$ diagram for known pulsars and pulsars in this paper. All known pulsars are the grey dots. Pulsars with nulling, mode changing, or subpulse modulation phenomena are represented by blue circles, green pluses, and red minus signs. 
The traditional death line is drawn for $BP^{-2}=0.17 \times 10^{12}{\rm G}{\rm s}^{-2}$ \citep{Ruderman1975}.}
\label{FigPPdot}
\end{figure*}

\begin{table*}[htbp]
\setlength\tabcolsep{13pt}
\flushleft
\caption{Statistics and p-values of the 1D Kolmogorov-Smirnov tests applied to fundamental pulsar parameters of pulsars with nulling, mode changing, or modulation behaviors, comparing with the distribution of known normal pulsars in the ATNF pulsar catalog \citep{Manchester+2005AJ....129.1993M}. For pulsars with subpulse-drifting, the Spearman correlation coefficients and p-values are obtained between the drifting parameters ($P_3$ and $P_2$ in degrees or milliseconds) and fundamental pulsar parameters. 
\label{Tab:KSValue}}
\footnotesize 
\begin{tabular}{lccccc}
\hline 
           & log($P$)          & log($\dot{P}$)  & log($\tau$)      & log($B$)  & log($\dot{E}$) \\
\hline 
\multicolumn{6}{c}{Statistics and p-values of K-S test.}\\
\hline
Nulling           & {\bf 0.48 / 1E-20} & 0.12 / 0.10        & {\bf 0.26 / 3E-06} & {\bf 0.18 / 2E-03} & {\bf 0.43 / 1E-16} \\
Mode-changing     & 0.17 / 0.19        & 0.12 / 0.59        & 0.07 / 0.98        & 0.08 / 0.96       & 0.10 / 0.83  \\
Subpulse Drifting & {\bf 0.13 / 5E-03} & {\bf 0.13 / 7E-03} & {\bf 0.19 / 1E-05} & 0.07 / 0.35       & {\bf 0.20 / 2E-06} \\
\hline
\multicolumn{6}{c}{Spearman correlation coefficients and p-values}\\
\hline
log($P_3$)           & {\bf -0.33 / 1E-09} & -0.10 / 0.14        & -0.02 / 0.80     & {\bf -0.22 / 6E-04} & {\bf 0.14 / 0.03}  \\
log($P_2$($^\circ$)) & {\bf -0.35 / 4E-10} & {\bf -0.22 / 2E-03} & 0.11 / 0.13      & {\bf -0.33 / 2E-06} & 0.09 / 0.21 \\
log($P_2$(ms))       & {\bf 0.32 / 3E-08}  & -0.01 / 0.94        & 0.10 / 0.16      & 0.07 / 0.32         & {\bf -0.27 / 1E-04} \\
\hline 
\end{tabular}
\end{table*}

\begin{figure*}[htpb]
\centering
\setlength\tabcolsep{1pt}
\renewcommand\arraystretch{0}
\begin{tabular}{ccccc}
\includegraphics[height=0.19\textwidth, angle=0]{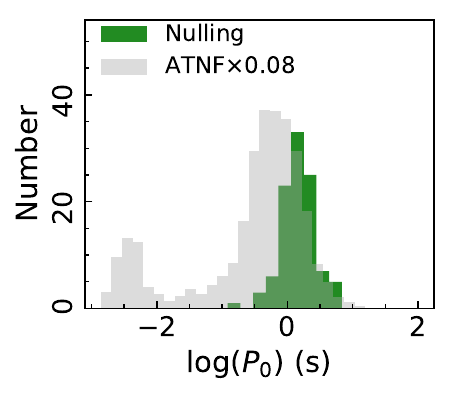}&
\includegraphics[height=0.19\textwidth, angle=0]{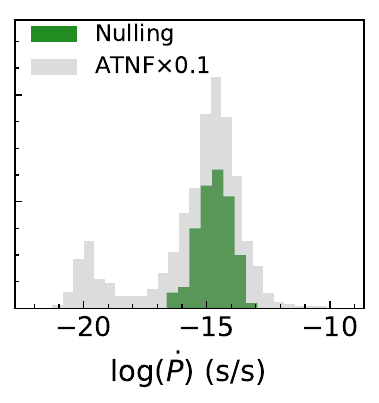}&
\includegraphics[height=0.19\textwidth, angle=0]{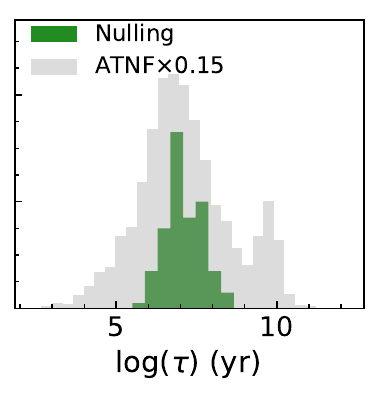}&
\includegraphics[height=0.19\textwidth, angle=0]{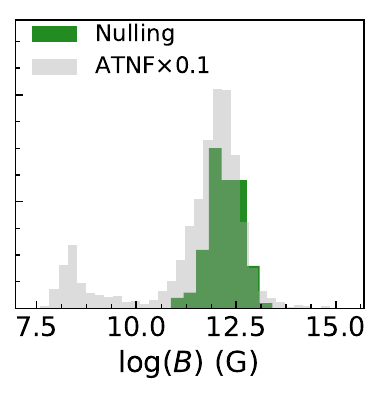}&
\includegraphics[height=0.19\textwidth, angle=0]{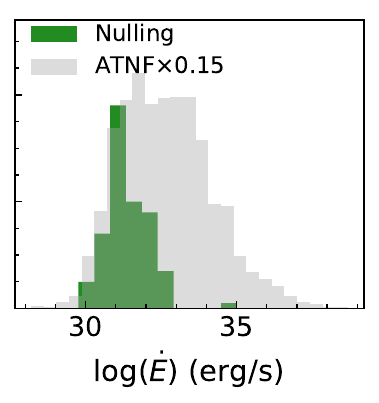}\\
\includegraphics[height=0.19\textwidth, angle=0]{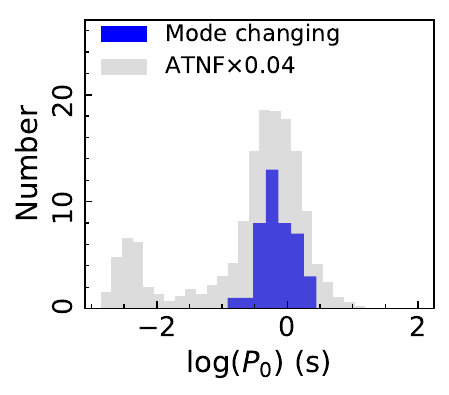}&
\includegraphics[height=0.19\textwidth, angle=0]{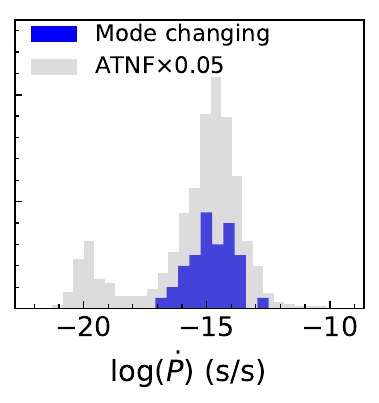}&
\includegraphics[height=0.19\textwidth, angle=0]{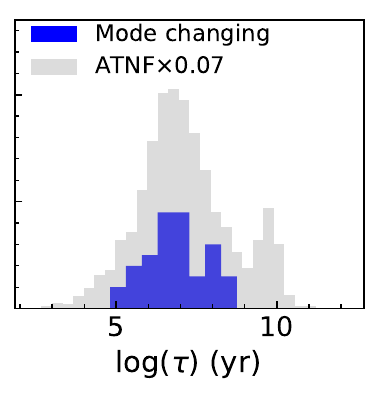}&
\includegraphics[height=0.19\textwidth, angle=0]{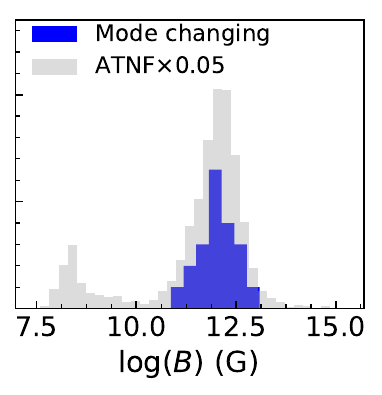}&
\includegraphics[height=0.19\textwidth, angle=0]{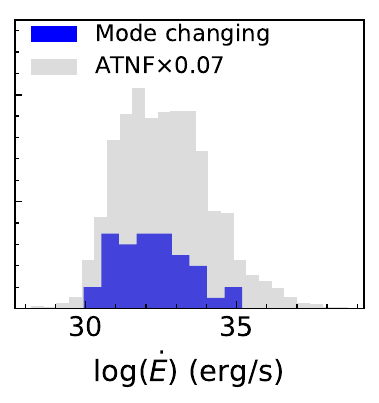}\\
\includegraphics[height=0.19\textwidth, angle=0]{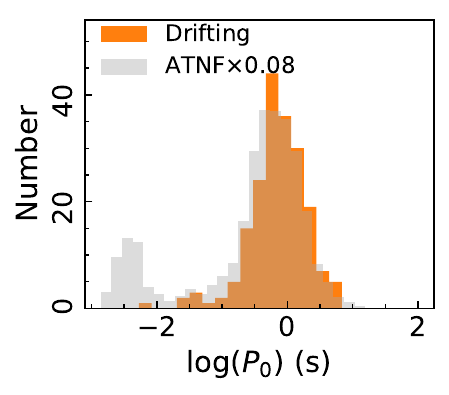}&
\includegraphics[height=0.19\textwidth, angle=0]{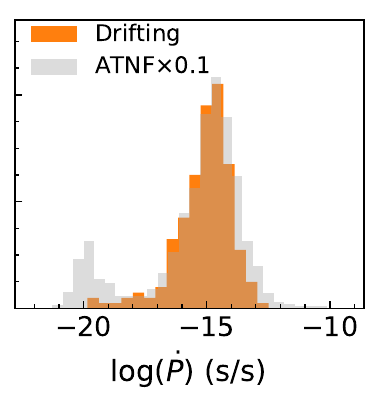}&
\includegraphics[height=0.19\textwidth, angle=0]{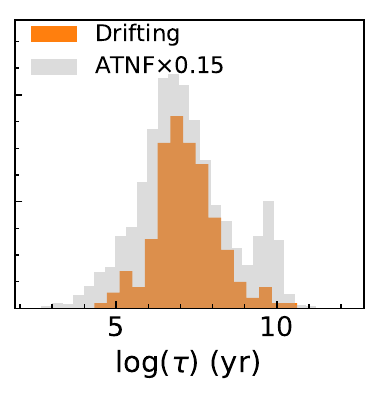}&
\includegraphics[height=0.19\textwidth, angle=0]{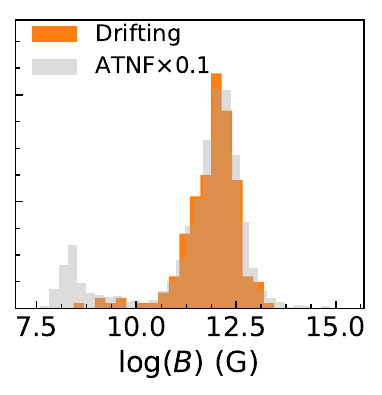}&
\includegraphics[height=0.19\textwidth, angle=0]{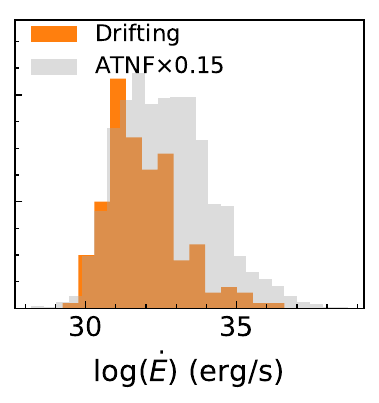}
\end{tabular}
\caption{The distribution of fundamental parameters (rotation period $P$, period derivation $\dot{P}$, characteristic age $\tau$, magnetic field strength $B$ and spin-down energy loss $\dot{E}$) for pulsars with single pulse behaviours (nulling, mode-changing or subpulse drifting), compared with the distribution of all known pulsars from the ATNF pulsar catalog \citep{Manchester+2005AJ....129.1993M}. 
\label{subfig:fluctu:FigDisNormalPSRPars}
}
%
\vspace{-0.3cm}
\centering
\setlength\tabcolsep{1pt}
\renewcommand\arraystretch{0}
\begin{tabular}{ccccc}
\includegraphics[height=0.19\textwidth, angle=0]{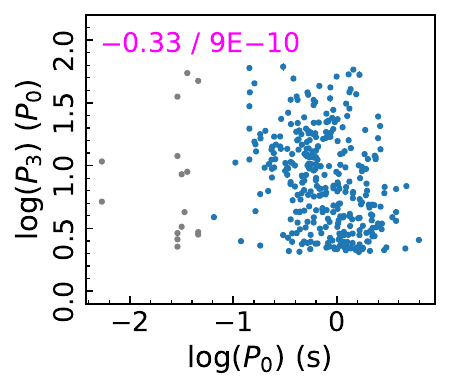}&
\includegraphics[height=0.19\textwidth, angle=0]{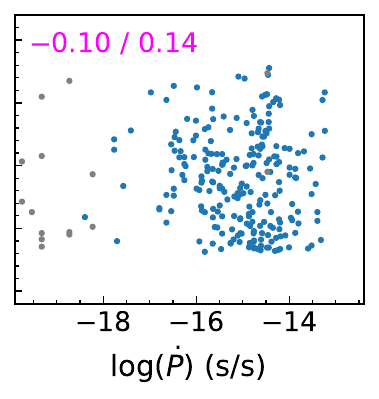}&
\includegraphics[height=0.19\textwidth, angle=0]{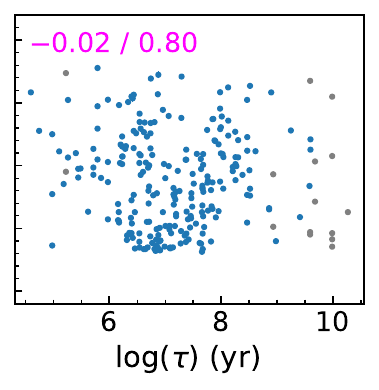}&
\includegraphics[height=0.19\textwidth, angle=0]{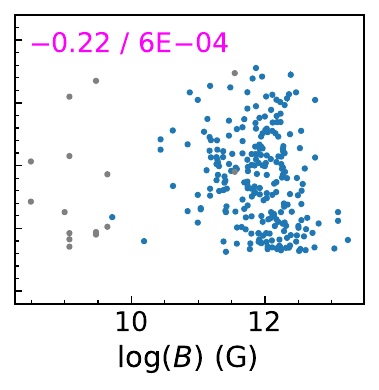}&
\includegraphics[height=0.19\textwidth, angle=0]{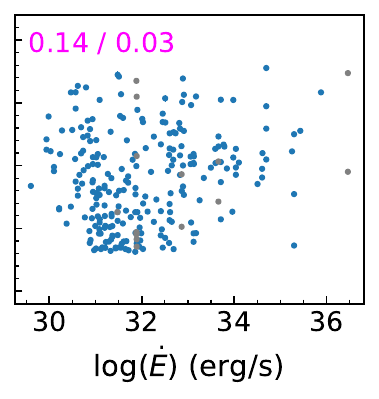}\\
\includegraphics[height=0.19\textwidth, angle=0]{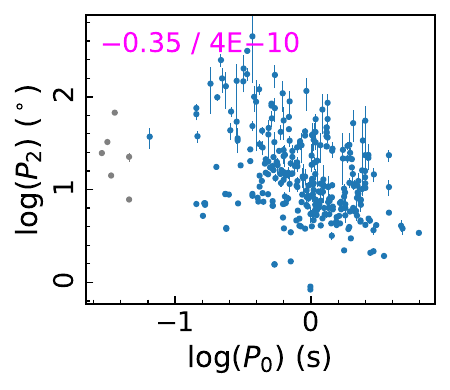}&
\includegraphics[height=0.19\textwidth, angle=0]{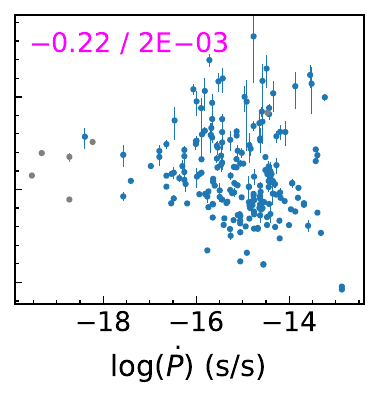}&
\includegraphics[height=0.19\textwidth, angle=0]{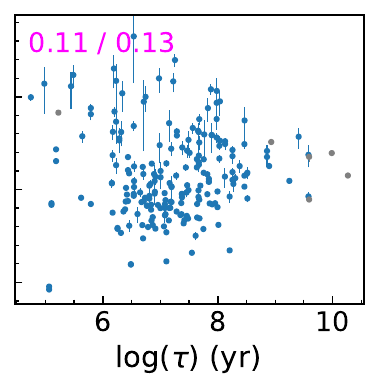}&
\includegraphics[height=0.19\textwidth, angle=0]{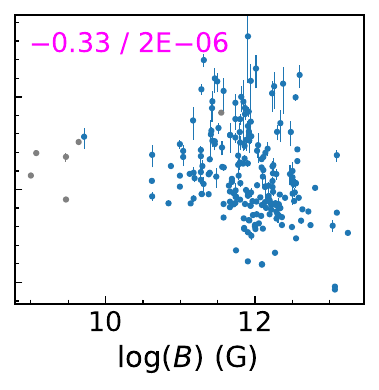}&
\includegraphics[height=0.19\textwidth, angle=0]{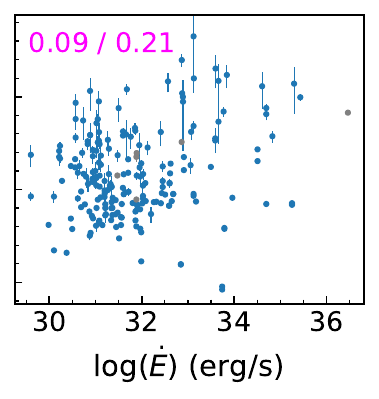}\\
\includegraphics[height=0.19\textwidth, angle=0]{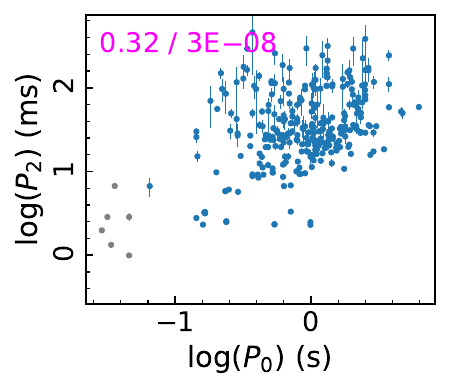}&
\includegraphics[height=0.19\textwidth, angle=0]{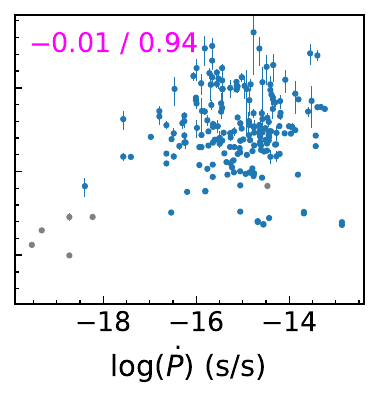}&
\includegraphics[height=0.19\textwidth, angle=0]{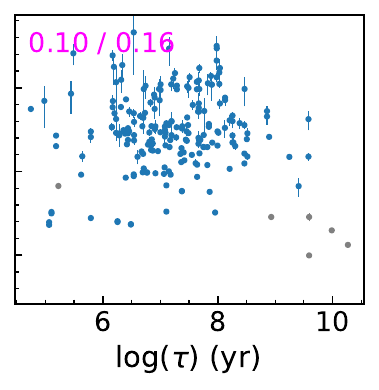}&
\includegraphics[height=0.19\textwidth, angle=0]{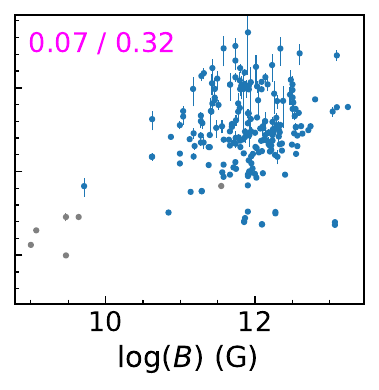}&
\includegraphics[height=0.19\textwidth, angle=0]{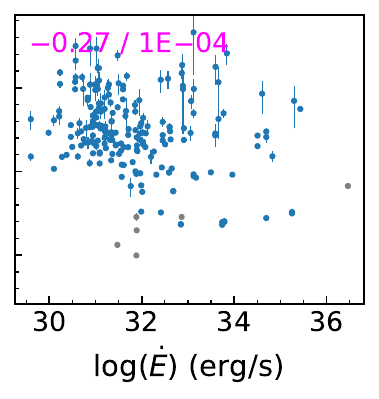}
\end{tabular}
\caption{Correlating subpulse drifting parameters ($P_3$, $P_2$ in degree or millisecond) with  fundamental pulsar parameters taken from the ATNF pulsar catalog \citep{Manchester+2005AJ....129.1993M}. The Spearman correlation coefficient and the p-value are labeled in each panel.}
\label{FigCorrP3Drift}
\end{figure*}



\section{Discussion and Conclusions}
\label{sect:Conc}

Using FAST single pulse time-series data obtained via the GPPS survey and other projects by the authors, we detect the nulling from 160 pulsars, mode changing from 52 pulsars, and subpulse modulation from 272 pulsars. Among them, these phenomena from 127, 51 and 180 pulsars are newly detected, respectively. This substantially expands the samples of pulsars with these single pulse behaviors. 

As we see above, some pulsars exhibit two or more of these features, for example, mode changes accompanied by distinct subpulse driftings, the nulling with mode changing. 
%
%
%
%
%
For some pulsars with a similar $P_3$, such as PSRs J1940+0239, J1954+2923 and J2051+4434g, the drifting bands may be discontinuous across different longitude ranges,  
%
which may be related to phase-locked drifting and phase steps for different components 
\citep{Bhattacharyya2009,Zhang2019}. 

Pulsars exhibiting single-pulse behaviors are investigated through two kinds of statistic analyses: the interplay with pulsar fundamental parameters -- rotation period ($P$), period derivation ($\dot{P}$), magnetic field strength ($B=3.2\times10^{19}{\rm G}\sqrt{P\dot{P}}$), characteristic age ($\tau=P/(2\dot{P})$) or spin-down energy loss ($\dot{E}\simeq3.95\times10^{31}{\rm erg/s}(\frac{\dot{P}}{10^{-15}})(\frac{P}{\rm s})^{-3}$) via histogram analyses, the correlations between these fundamental parameters and drifting subpulse properties, mainly the time interval $P_3$ and phase separation $P_2$ between drifting bands. One can see from the $P-\dot{P}$ diagram in Figure~\ref{FigPPdot} for known pulsars that pulsars exhibiting nulling are more concentrated to the death line \citep{Ruderman1975}.



Histograms of rotation period $P$, period derivation $\dot{P}$, magnetic field strength $B$, characteristic age $\tau$ or spin-down energy loss $\dot{E}$ for pulsars with nulling, mode-changing, and subpulse modulation are shown in Fig~\ref{subfig:fluctu:FigDisNormalPSRPars}. To compare whether the parameters of pulsars exhibiting these phenomena and those of all known normal pulsars (with $P>$50~ms) originate from the same distributions, Kolmogorov-Smirnov tests \citep{Hodges1958} are performed. Parameters of all known normal pulsars are taken from the ATNF pulsar catalog \citep{Manchester+2005AJ....129.1993M}, the most complete sample of known pulsars currently available. 
The null hypothesis states that both samples originate from the same distribution, and that the KS statistic is large and the p-value is less than 0.05 indicates that the distributions are statistically incompatible. From histograms in Figure~\ref{subfig:fluctu:FigDisNormalPSRPars} and statistic values and p-values of K-S tests are listed in Table~\ref{Tab:KSValue}.

For pulsars with nulling phenomenon, the p-values corresponding to $P$, $\tau$, $B$ and $\dot{E}$ are less than 0.05, and as a result, the null hypothesis is rejected. This means that these parameters of nulling pulsars with those of all known pulsars are from distinct distributions. Combined with analysis of distributions in Figure~\ref{subfig:fluctu:FigDisNormalPSRPars}, pulsars exhibiting nulling behavior are older and have longer periods, stronger magnetic field strengths, and lower spin-down energy loss, which agrees with previous studies \citep{Ritchings1976, Wang2007, Ng2020}.

Mode-changing pulsars were reported to have longer period \citep{Ng2020}. While p-values about all pulsar parameters corresponding to mode-changing behavior in Table~\ref{Tab:KSValue} are all more than 0.05. As a result, we cannot reject the null hypothesis that the parameters of pulsars with mode-changing phenomenon are from different distributions with that of all known pulsars.

Similarly, the quantities of drifting pulsars, including period, period derivation, characteristic age, and spin-down energy loss, are from different distributions with that of all known pulsars with low p-values. Therefore, in addition to older age as \citet{Weltevrede2006} reported, pulsars with subpulse drifting behavior also tend to have longer period, lower period deviation, and spin-down energy loss.

In conclusion, pulsars with nulling and subpulse modulation behaviors tend to have longer periods, older ages, and lower spin-down energy loss. Additionally, pulsars exhibiting subpulse modulation phenomena have lower period deviation, while those showing nulls tend to have stronger magnetic field strength.

To measure the correlation between drifting parameters $P_3$ and $P_2$ with the pulsar parameters for pulsars with the rotation period more than 50 ms, Spearman correlation coefficients and associated p-values \citep{Zwillinger2000,Kendall1979} are calculated and listed in Table~\ref{Tab:KSValue}, with the null hypothesis that two quantities are uncorrelated. Spearman correlation coefficients of -1 or +1 imply an exact negatively or positively monotonic relationship, and a p-value less than 0.05 indicates that two quantities are correlated. 


From Figure~\ref{FigCorrP3Drift} and Table~\ref{Tab:KSValue}, the correlation coefficients are relatively higher, and the p-values are far less than 0.05 for correlations of $P_3$ with $P$, $\dot{P}$, and $B$, which manifests that $P_3$ has higher negative correlations with rotation period, period derivation, and magnetic field strength. However, smaller correlation coefficients and higher p-values more than 0.1 of $P_3$ with pulsar period derivation $\dot{P}$ and characteristic age $\tau$ cannot indicate the correlations. $P_2$ in degrees is negatively correlated with rotation period $P$, period derivation $\dot{P}$ and magnetic field strength $B$. However, $P_2$ in millisecond units is positively correlated with rotation period $P$, and negatively correlated with spin-down energy loss $\dot{E}$. This is different from the correlations when $P_2$ is expressed in degrees.


\section*{Data availability}
FAST original observational data can be accessed through the FAST data center after a one-year priority protection for proposers. The data series for all individual pulse stacks of pulsars published in this 
paper are available at \url{https://psr-fast.zmtt.bao.ac.cn/} after publication of this paper.

\begin{acknowledgments}
We thank the referee for helpful comments.
This work made use of the data from FAST (Five-hundred-meter Aperture Spherical radio Telescope)(https://cstr.cn/31116.02.FAST).  FAST is a Chinese national mega-science facility, operated by the National Astronomical Observatories, Chinese Academy of Sciences. 
The authors were supported by the National Key R\&D Program of China No. 2025YFA161400, the National Natural Science Foundation of China (NSFC, grant numbers 12588202 and 12041303), and the Chinese Academy of Sciences via project JZHKYPT-2021-06, 
the Postdoctoral Fellowship Program of CPSF under Grant Number GZC20252098, 
and the National SKA Program of China grant 2020SKA0120100 and 2022SKA0120103.
\end{acknowledgments}

\bibliography{ref}{}
\bibliographystyle{aasjournal}

\end{document}